\documentclass[11pt,twoside=true,a4paper,titlepage,open=right,abstracton,numbers=noenddot]{scrartcl}

 \DeclareOldFontCommand{\rm}{\normalfont\rmfamily}{\mathrm}
 \DeclareOldFontCommand{\bf}{\normalfont\bfseries}{\mathbf}
 \DeclareOldFontCommand{\it}{\normalfont\itshape}{\mathit}

\usepackage[inner=3.0cm,outer=2.7cm,top=2cm,bottom=2cm,includeheadfoot]{geometry}

\usepackage[ngerman, english]{babel}

\usepackage{hyperref}

\usepackage[dvips]{graphicx}
\usepackage[percent]{overpic}
\usepackage{lscape}
\usepackage{amsmath}
\usepackage{wasysym}
\usepackage{bbm}
\usepackage{amssymb}
\usepackage{appendix}
\usepackage{rotating}
\usepackage[percent]{overpic}
\usepackage[usenames, dvipsnames]{color}
\usepackage{longtable}
\usepackage[verbose]{placeins}
\usepackage{pdfpages}
\usepackage{epsfig}
\usepackage{epstopdf}
\usepackage{tabularx}
\usepackage{upgreek}
\usepackage{bm}
\usepackage{sidecap}

\usepackage{amsthm}
\usepackage{algorithm}
\usepackage{algpseudocode}
\usepackage{extarrows}
\usepackage{slashed}
\usepackage{arydshln}
\usepackage{multirow}
\usepackage{multicol}

\newtheorem*{CompEx}{Problem}

\usepackage{tabularx}
\usepackage{chngcntr}
\usepackage{morefloats}
\usepackage{subcaption}

\usepackage{cite}
\usepackage{url}
\usepackage{tablefootnote}
\usepackage{float}
\usepackage{fancyhdr}
\usepackage{CJKutf8}

\newenvironment{rcases}
  {\left.\begin{aligned}}
  {\end{aligned}\right\rbrace}

\usepackage{fancyhdr}
\author{Y.Wunderlich}
\title{}
\date{08.01.2018}

\numberwithin{equation}{section}
\counterwithin{table}{section}
\counterwithin{figure}{section}

\begin{document}

\pagenumbering{roman}
\thispagestyle{plain}

\begin{titlepage}
\thispagestyle{empty}
\selectlanguage{ngerman}
\begin{center}
{
\rule{\textwidth}{1pt}\par
\vspace{0.5\baselineskip}
\huge\bf The complete experiment problem of pseudoscalar meson photoproduction in a truncated partial wave analysis\\
\rule{\textwidth}{1pt}\par

\vfill
{\Large
\vspace{1ex}
Dissertation\\
\vspace{1ex}
\vspace{1ex}
\vspace{1ex}
}
{\large
zur \\
\vspace{1ex}
\vspace{1ex}
\vspace{1ex}
\vspace{1ex}
Erlangung des Doktorgrades (Dr. rer. nat.) \\
\vspace{1ex}
\vspace{1ex}
\vspace{1ex}
\vspace{1ex}
der \\
\vspace{1ex}
\vspace{1ex}
Mathematisch-Naturwissenschaftlichen Fakult\"at\\
\vspace{1ex}
der\\
  \vspace{1ex}
Rheinischen Friedrich-Wilhelms-Universit\"at\\
 \vspace{1ex}
Bonn\\
\vfill vorgelegt von\\
 \vspace{1ex}
 \vspace{1ex}
 \vspace{1ex}
\begin{Large}Yannick Wunderlich\end{Large}\\
 \vspace{1ex}
aus\\
 \vspace{1ex}
Bonn-Duisdorf\\
 \vfill
Bonn 2018}}
\end{center}
\end{titlepage}

\thispagestyle{empty}

\selectlanguage{ngerman}
\cleardoublepage
\thispagestyle{empty} { \vspace*{0.025\textheight}
\parindent0pt
\centering Angefertigt mit Genehmigung der Mathematisch-Naturwissenschaftlichen Fakult\"at der Rheinischen Friedrich-Wilhelms-Universit\"at Bonn

\vspace*{0.1\textheight}

\vspace*{0.2\textheight}

\vspace*{0.4\textheight}

\begin{tabular}{ll}
1. Referent: &Prof.~Dr.~Reinhard~Beck\\[1ex]
2. Referent: &PD.~Dr.~Bernard~C.~Metsch\\[1.5ex]
Tag der Promotion: &04.12.2018\\[1.5ex]
Erscheinungsjahr: &2019\\[1.5ex]
\end{tabular}
}

\thispagestyle{empty}
\cleardoublepage
\thispagestyle{empty}

\selectlanguage{english}

\begin{abstract}
One of the remaining open challenges in Standard Model phenomenology is the formation of bound states of quarks in the strong coupling regime of Quantum Chromodynamics (QCD). This is true in particular for the excited states of baryons ($qqq$-states). There is still room for improvement of our understanding, which is seen once predictions from constituent quark models or lattice QCD are confronted with information on baryon resonances extracted from experiment.  \newline
From the experimental side, one studies baryon resonances by impinging strong or electromagnetic probes coming from manufacturable beams (pions, photons, electrons, $\ldots$) on target nucleons in order to produce excited states. Then, the decay products are studied in order to infer the resonances. Thus, generally many different reactions are important for baryon spectroscopy. \newline
The main currently accepted method to determine physical properties of resonances (masses, widths, quantum numbers, $\ldots$) from data are so-called energy-dependent (ED) fits. Here, reaction-theoretic models construct the amplitude as a function of energy, and the model-parameters are (loosely speaking) varied in a fit. Then, the resulting amplitude is analytically continued into the complex energy plane to search for the resonance-poles. In almost all ED approaches, many reactions are analyzed at once in so-called coupled-channel fits. \newline \newline
An alternative Ansatz consists of trying to extract maximal information on reaction-amp\-li\-tudes from the data, without introducing any kind of model-assumptions. For reactions involving particles with spin, e.g. $\pi N \rightarrow \pi N$, $\gamma N \rightarrow \pi N$, $e N \rightarrow e^{\prime} \pi N$, $\ldots$, generally $n$ complex spin amplitudes are necessary to model-independently constraint the full reaction $\mathcal{T}$-matrix. Furthermore, the {\it data} for such reactions consist of $n^{2}$ measurable so-called polarization observables (or -asymmetries), which in general have to be measured in order to obtain information on amplitude-interferences. Thus, the question for an optimization of the measurement effort arises and one searches for so-called {\it complete experiments}. Those are minimal subsets of all measurable polarization observables sufficient in order to maximally constrain the underlying amplitudes. The complete experiment problem is most commonly treated as a purely mathematical problem, i.e. for idealized data with infinite precision. \newline
This thesis treats the problem of complete experiments for the photoproduction of a single pseudoscalar meson $\varphi$, with a recoil-baryon $B$ in the final state: $\gamma N \rightarrow \varphi B$. In this case, $4$ complex amplitudes are accompanied by $16$ polarization observables. The observables are again divided into the unpolarized differential cross section $\sigma_{0}$, $3$ single-spin asymmetries and $12$ double-polarization observables which are again subdivided into the classes of beam-target ($\mathcal{BT}$), beam-recoil ($\mathcal{BR}$) and target-recoil ($\mathcal{TR}$) observales. \newline
In an earlier work, W.-T. Chiang and F. Tabakin deduced completeness-rules according to which {\it $8$ carefully selected observables} are sufficient in order to determine the $4$ amplitudes up to one overall phase. However, these rules are again only valid for idealized data. One point which is troubling about the complete experiments according to Chiang and Tabakin is that they enforce the measurement of the double-polarization observables with recoil polarization ($\mathcal{BR}$ and $\mathcal{TR}$), a task which is very hard to accomplish experimentally. \newline
The goal of this thesis was therefore to study the analogous problem, however not for extraction of the full spin amplitudes, but for the photoproduction partial waves ('multipoles') in a truncated partial wave analysis (TPWA) up to some maximal angular momentum cutoff $\ell_{\mathrm{max}}$. The extraction of partial waves in such analyses proceeds on each energy-bin individually, thus one refers to them as single-energy (SE) fits. The work has been triggered initially by a paper from the author V. F. Grushin, which investigates similar questions for quite low truncation orders ($\ell_{\mathrm{max}} = 1$). Here, a promising aspect of Grushin's work was that he has been able to infer (almost) unique multipole-solutions without using any double-polarization observables at all. \newline
The early concept for the thesis consisted of approaching the complete experiment problem for the TPWA from two sides. Those consist of the purely mathematical, or algebraic, side, which should then be complemented by numerical investigations on TPWAs. The present document collects the results of the project. \newline
A review and further development of an earlier work by A. S. Omelaenko, which discussed linear factor decompositions of the polynomial amplitude, partial wave zero's (akin to so-called {\it Barrelet zero's}) and discrete ambiguities in TPWAs, is shown. This approach allowed for an identification of candidates for complete sets in a TPWA, although it was not fully clear up to which $\ell_{\mathrm{max}}$ such candidates hold up. \newline
A welcome by-product of the formalization for the photoproduction TPWA performed in this work was the possibility of doing so-called {\it moment-analyses} on the angular distributions of the observables alone. A survey of such analyses for observables in $\pi^{0}$-photoproduction is presented. \newline
Lastly, the completeness-rules postulated in the algebraic part of the thesis had to be checked using model-independent numerical methods for the extraction of multipoles. Complete sets in TPWAs are thus studied numerically for synthetic idealized model-data, pseudo-data with errors of variable size and then, finally, also for real data. The influence of errors on the precision of extracted multipoles, as well as on the stability of the fits in general, is studied using the bootstrap.
\end{abstract}

\cleardoublepage
\selectlanguage{ngerman}

\begin{abstract}
Eine der noch verbleibenden Herausforderungen in der Ph\"anomenologie des Standardmodells ist das Verst\"andnis der Bildung von Bindungszust\"anden von Quarks, verursacht durch die Quantenchromodynamik (QCD) bei starker Kopplung. Dies trifft insbesondere zu auf die Anregungszust\"ande von Baryonen ($qqq$-Zust\"anden). Es wird klar, dass hier noch Bedarf an Verbesserungen besteht, sobald die Vorhersagen von Konstituenten-Quarkmodellen oder auch Gitter-QCD Rechnungen mit den bis heute aus Experimenten extrahierten Informationen \"uber Baryon-Resonanzen verglichen werden. \newline
Um Baryon-Resonanzen experimentell studieren zu k\"onnen, werden typischerweise stark oder elektromagnetisch wechselwirkende Sonden (Pionen, Photonen, Elektronen, $\ldots$) an Target-Nukleonen gestreut um Anregungszust\"ande zu erzeugen. Man analysiert dann die Zerfallsprodukte dieser Zust\"ande, um Informationen \"uber die auftretenden Resonanzen zu erhalten. Von daher sind im Allgemeinen mehrere verschiedene Reaktionen f\"ur die Baryonspektroskopie wichtig. \newline
Die heutzutage akzeptierte Methode zur Extraktion der physikalischen Eigenschaften von Resonanzen (Massen, Zerfallsbreiten, Quantenzahlen, $\ldots$) aus Streudaten sind die sogenannten Energie-abh\"angigen ({\it engl.: energy-dependent}, ED) Fits. In solchen Verfahren werden Reaktions-theoretische Modelle konstruiert um die Amplitude als Funktion der Energie zu erhalten. Die Modell-Parameter werden dann in numerischen Anpassungsverfahren variiert und bestimmt. Die resultierende Amplitude wird dann analytisch in die komplexe Energie-Ebene fortgesetzt um die Resonanz-Pole zu finden. In beinahe allen ED Ans\"atzen werden mehrere Reaktionen gleichzeitig analysiert, was auf sogenannte Fits f\"ur gekoppelte Kan\"ale ({\it engl.: coupled-channels analyses}) hinausl\"auft. \newline \newline
Ein alternativer Ansatz zu den oben genannten ED-Fits besteht aus dem Versuch, maximale Informationen \"uber die Reaktionsamplituden aus den Daten zu extrahieren, jedoch ohne dabei Modellannahmen zu machen. F\"ur Reaktionen an denen Teilchen mit nicht-verschwindendem Spin teilnehmen, also zum Beispiel $\pi N \rightarrow \pi N$, $\gamma N \rightarrow \pi N$, $e N \rightarrow e^{\prime} \pi N$, $\ldots$, sind im Allgemeinen $n$ komplexe sogenannte Spin-Amplituden notwendig, um die volle $\mathcal{T}$-Matrix der Reaktion modellunabh\"angig zu parametrisieren. Dar\"uber hinaus bestehen die Daten f\"ur solche Reaktionen typischerweise aus $n^{2}$ sogenannten Polarisationsobservablen (oder Polarisationsasymmetrien), welche im Allgemeinen auch gemessen werden m\"ussen um auf Beitr\"age von hinreichend vielen Interferenz-Termen sensitiv zu sein. Von daher entsteht schnell die Frage nach einer Optimierung des Messaufwandes und diese f\"uhrt zur Suche nach sogenannten {\it Vollst\"andigen Experimenten} ({\it engl.: complete experiments}). Dies sind minimale Teilmengen von Observablen, welche hinreichend sind, um maximale Zwangsbedingungen f\"ur die zu Grunde liegenden Spin-Amplituden zu liefern. Die Ermittlung von vollst\"andigen Experimenten wird h\"aufig als rein mathematisches Problem behandelt, also f\"ur idealisierte perfekte Daten mit unendlicher Pr\"azision. \newline
Diese Arbeit behandelt die vollst\"andigen Experimente f\"ur die Photoproduktion eines einzelnen pseudoskalaren Mesons $\varphi$, mit einem R\"ucksto\ss-Baryon $B$ im Endzustand: $\gamma N \rightarrow \varphi B$. In diesem Fall stehen $4$ komplexen Amplituden, $16$ Polarisationsobservablen gegen\"uber. Die Observablen sind unterteilt in den unpolarisierten differenziellen Wirkungsquerschnitt $\sigma_{0}$, $3$ Einfach-polarisations Asymmetrien und $12$ Doppelpolarisationsobservablen. Letztere sind wiederum unterteilt in Observablen vom Typ Strahl-Target ({\it engl.: beam-target} ($\mathcal{BT}$)), Strahl-R\"ucksto{\ss} ({\it engl.: beam-recoil} ($\mathcal{BR}$)) und Target-R\"ucksto{\ss} ({\it engl.: target-recoil} ($\mathcal{TR}$)). \newline
In einer Ver\"offentlichung von W.-T. Chiang und F. Tabakin wurden Vollst\"andigkeitsregeln aufgestellt, welche besagen, dass {\it $8$ sorgf\"altig ausgew\"ahlte Observablen} ausreichend sind, um die $4$ Amplituden bis auf eine globale Phase festzulegen. Jedoch sind diese Regeln wieder nur g\"ultig f\"ur idealisierte, unendlich pr\"azise Daten. Ein Aspekt der vollst\"andigen Experimente nach Chiang and Tabakin, welcher problematisch erscheint, besteht darin dass sie eine Messung der Doppelpolarisationsobservablen mit R\"ucksto{\ss}polarisation ($\mathcal{BR}$ und $\mathcal{TR}$) erzwingen, was sich experimentell als \"au{\ss}erst schwierig herausstellt. \newline
Das Ziel dieser Dissertation war daher, ein v\"ollig analoges Problem zu untersuchen, allerdings nicht f\"ur die Extraktion der vollen Spin-Amplituden, sondern der Partialwellen der Photoproduktion (sog. 'Multipolen') in einer trunkierten Partialwellenanalyse (TPWA), abgeschnitten bei der maximalen Drehimpuls-Quantenzahl $\ell_{\mathrm{max}}$. Die Bestimmung von Partialwellen in solchen Analysen geschieht in jedem Energie-Bin einzeln. Aus diesem Grunde werden sie oft auch als {\it single-energy} (SE) Fits bezeichnet. Ein weiterer Punkt, welcher das Interesse in das Thema dieser Arbeit geweckt hat, war eine Ver\"offentlichung von V. F. Grushin, in welcher der Autor sehr \"ahnliche Fragestellungen f\"ur niedrige Trunkierungs-Ordnungen ($\ell_{\mathrm{max}} = 1$) untersuchte. Dabei war es Grushin m\"oglich (beinahe) eindeutige Multipol-L\"osungen ausfindig zu machen, ohne \"uberhaupt Doppelpolarisationsobservablen zu untersuchen. \newline
In der fr\"uhzeitigen Konzeption dieser Arbeit wurde es als sinnvoll erachtet, das Problem der vollst\"andigen Experimente in der TPWA von zwei komplement\"aren Sichtweisen aus zu untersuchen. Die erste besteht hier aus einer rein mathematischen, oder algebraischen, Untersuchung. Parallel dazu sollten numerische Berechnungen angestellt werden, welche im Idealfall die mathematischen Erkenntnisse untermauern. Das vorliegende Dokument tr\"agt die Ergebnisse dieses Projektes zusammen. \newline
Zuerst wird die \"Uberpr\"ufing und Weiterf\"uhrung einer fr\"uheren Arbeit von A. S. Omelaenko gezeigt, welche die Linearfaktorzerlegung der polynomialen Amplitude mit den Partialwellen \"aquivalenten komplexen Nullstellen (\"ahnlich den sog. {\it Barrelet zero's}) und die daraus resultierenden diskreten Ambiguit\"aten in TPWAs behandelt. Dieser algebraische Zugang erlaubte es, erste Kandidaten f\"ur vollst\"andige Experimente aufzustellen. Jedoch war es alles andere als klar, bis zu welchen Ordnungen in $\ell_{\mathrm{max}}$ sich die Kandidaten-S\"atze in der Praxis tats\"achlich als vollst\"andig erweisen. \newline
Eine willkommene Folge der in dieser Arbeit betriebenen Formalisierung der Pho\-to\-pro\-duk\-tions-TPWA bestand aus der M\"oglichkeit, sog. {\it Moment-Analysen} auf die Winkelverteilungen der Observablen anzuwenden. Ein \"Uberblick solcher Untersuchungen f\"ur Observablen aus der $\pi^{0}$-Photoproduktion wird pr\"asentiert. \newline
Die durch die algebraischen Untersuchungen zu Tage getretenen Kandidaten f\"ur vollst\"andige Experimente werden schlussendlich mittels modellunabh\"angiger numerischer Methoden zur Extraktion von Multipolen getestet. Dies geschieht f\"ur synthetische Theorie-Daten aus ED Modell-L\"osungen, f\"ur Pseudodaten mit Fehlern von variabler Gr\"o{\ss}e und auch f\"ur echte Messdaten. Der Einfluss der experimentellen Fehler auf die Pr\"azision der extrahieren SE-Multipole, aber auch auf die Stabilit\"at der TPWA-Fits, wird mit Hilfe der Bootstrap-Methode studiert.
\end{abstract}

\cleardoublepage
\pagenumbering{arabic}
\selectlanguage{english}

\tableofcontents

\renewcommand{\pagestyle}{headings}

\clearpage

\section{Introduction} \label{chap:Intro}

\allowdisplaybreaks

\subsection{The Standard Model of elementary particle physics and strong interactions} \label{sec:StandardModel}

The two revolutions in physics at the beginning of the $20^{\mathrm{th}}$ century, namely the discovery of the theory of relativity and quantum mechanics, set into motion a development which lead, in the $1970$s, to the completion of the {\it Standard Model} of elementary particle physics (See \cite{Patrignani:2016xqp} for more detail. Textbook treatments is given, for instance, in \cite{DonoghueEtAlSM, AitchisonHey}.). This model represents, at the time of this writing, the experimentally well-tested and accepted fundamental description of nature. It is a relativistic gauge field theory based on the local gauge group
\begin{equation}
 G_{SM} = SU(3)_{c} \times SU(2)_{L} \times U(1)_{Y} \mathrm{.} \label{eq:StandardModelGaugeGroup}
\end{equation}
The particle content of the Standard Model is listed in Table \ref{tab:ParticleContentSM}. There are $12$ fundamental spin-$\frac{1}{2}$ fermions in the Standard Model, so-called {\it quarks} and {\it leptons}. They are further divided into three generations with two flavours of lepton and quark, each. These fermions represent, roughly speaking, the matter-part of the Standard Model. On the other hand, forces are mediated by spin-$1$ gauge bosons for the strong, weak and electromagnetic interactions. \newline
The $SU(3)_{c}$-factor in the direct product (\ref{eq:StandardModelGaugeGroup}) describes the strong interactions among quarks, the theory of { \it Quantum Chromodynamics} (QCD). Interactions in QCD are dictated by gauging a three-valued charge degree of freedom of the quarks called {\it color} (thus a subscript $c$). The strong interactions are most relevant in connection to the subject of this thesis and thus they are discussed in more detail below. \newline
Concerning the electroweak factor $SU(2)_{L} \times U(1)_{Y}$ in the gauge-group (\ref{eq:StandardModelGaugeGroup}), it should be mentioned that only left-chiral fermions transform as doublets under $SU(2)$ (thus the sub-script $L$). Therefore, the electroweak theory is chiral and totally violates parity. The sub-script $Y$ in $U(1)_{Y}$ refers to the weak hypercharge. \newline
Weak interactions are known to be short-ranged. Therefore, the gauge bosons have to be massive. In order to accomplish the introduction of gauge boson masses without spoiling the gauge-invariance of the original electroweak theory, the full electroweak symmetry of the Standard Model is spontaneously broken to the $U(1)$ of quantum electrodynamics (QED), symbolically $SU(2)_{L} \times U(1)_{Y} \rightarrow U(1)_{em}$, at an energy scale set by the vacuum expectation value (VEV) of the so-called Higgs field $v \sim 246 \hspace*{2pt} \mathrm{GeV}$. The Higgs-field is an $SU(2)_{L}$ doublet of complex scalar fields $\Phi = \left( \phi^{+}, \phi^{0} \right)^{T}$, which obtains a phenomenologically introduced potential that accommodates the symmetry-breaking. \newline
After electroweak symmetry-breaking, the weak gauge bosons $W^{\pm}$ and $Z^{0}$ acquire masses while the photon $\gamma$ remains massless. The Fermions also obtain mass-terms, via the breaking of Yukawa interaction-terms among the left-handed fermion doublets, the Higgs-doublet and the right-handed fermion-singlets. The form of the Yukawa-interactions is motivated by the full electroweak symmetry. This generation of mass, both for the gauge bosons as well as the fermions, is called the {\it Higgs mechanism}. As a consequence of the symmetry-breaking, a massive spin-$0$ boson, the {\it Higgs boson} $H^{0}$, remains in the spectrum. It is a triumph of experimental particle physics that a particle possessing all the properties of a Higgs-boson has been measured by the ATLAS- and CMS-collaborations at the Large Hadron Collider (LHC), during the time of preparation of this thesis \cite{ATLASHiggsPaper, CMSHiggsPaper}. \newline
The fourth known fundamental force of nature not mentioned up to now is gravity.

\begin{table}
 \begin{center}
  \boxed{\text{Standard Model (strong, weak and electromagnetic interactions)}} \\
 \end{center}
 \centering
 \begin{tabular}{ccccc}
  \hline
  \hline
  Generation & Fermions (Spin $1/2$) & $SU(3)_{c}$-rep. & $Q \left[ e \right]$ & Mass\tablefootnote{The determination/interpretation of quark-mass parameters is a non-trivial issue, since quarks are confined. For further discussion, see \cite{DonoghueEtAlSM} and reviews in \cite{Patrignani:2016xqp}.} $\left[ \mathrm{GeV}/c^{2} \right]$ \\
  \hline
  \multirow{2}{*}{$1^{\mathrm{st}}$}  & $u$ & $\bm{3}$ & $+2/3$ & $2.2 \times 10^{-3}$ \\
  & $d$ & $\bm{3}$ & $-1/3$ & $4.7 \times 10^{-3}$ \\
  \hdashline
  \multirow{2}{*}{$2^{\mathrm{nd}}$}  & $c$ & $\bm{3}$ & $+2/3$ & $1.28$ \\
   & $s$ & $\bm{3}$ & $-1/3$ & $0.096$ \\
  \hdashline
  \multirow{2}{*}{$3^{\mathrm{rd}}$}  & $t$ & $\bm{3}$ & $+2/3$ & $173.1 \pm 0.6$ \\
   & $b$ & $\bm{3}$ & $-1/3$ & $4.18$ \\
  \hline
  \multirow{2}{*}{$1^{\mathrm{st}}$}  & $\nu_{e}$ & $\bm{1}$ & $0$ & $< 0.2 \times 10^{-8}$ \\
   & $e^{-}$ & $\bm{1}$ & $-1$ & $5.10999 \times 10^{-4}$ \\
  \hdashline
  \multirow{2}{*}{$2^{\mathrm{nd}}$}  & $\nu_{\mu}$ & $\bm{1}$ & $0$ & $< 1.9 \times 10^{-4}$ \\
   & $\mu^{-}$ & $\bm{1}$ & $-1$ & $0.10566$ \\
  \hdashline
  \multirow{2}{*}{$3^{\mathrm{rd}}$}  & $\nu_{\tau}$ & $\bm{1}$ & $0$ & $< 0.0182$ \\
   & $\tau^{-}$ & $\bm{1}$ & $-1$ & $1.77686 \pm 0.00012$ \\
  \hline
  \hline
  \multicolumn{2}{c}{Gauge-Bosons (Spin $1$)} & $SU(3)_{c}$-rep. & $Q \left[ e \right]$ & Mass $\left[ \mathrm{GeV}/c^{2} \right]$ \\
  \hline
  \multicolumn{2}{c}{$\gamma$} & $\bm{1}$ & $0$ & $< 1 \times 10^{-27}$ \\
  \multicolumn{2}{c}{$W^{\pm}$} & $\bm{1}$ & $\pm 1$ & $80.385 \pm 0.015$ \\
  \multicolumn{2}{c}{$Z^{0}$} & $\bm{1}$ & $0$ & $91.1876 \pm 0.0021$ \\
  \multicolumn{2}{c}{$\mathrm{gluons}$ $\left( A_{\mu}^{a} \right)$} & $\bm{8}$ & $0$ & $0$ \\
  \hline
  \hline
  \multicolumn{2}{c}{Higgs (Spin $0$)} & $SU(3)_{c}$-rep. & $Q \left[ e \right]$ & Mass $\left[ \mathrm{GeV}/c^{2} \right]$ \\
  \hline
  \multicolumn{2}{c}{$H^{0}$} & $\bm{1}$ & $0$ & $125.09 \pm 0.24$ \\
  \hline
  \hline
 \end{tabular}
 \vspace*{5pt}
 \begin{center}
  \boxed{\text{General Relativity (gravity)}}
 \end{center}
  \begin{tabular}{ccc}
  \hline
  \hline
  Graviton (Spin $2$) & $Q \left[ e \right]$ & Mass $\left[ \mathrm{GeV}/c^{2} \right]$ \\
  \hline
  $h_{\mu \nu}$ & $0$ & $< 6 \times 10^{-41}$ \\
  \hline
  \hline
 \end{tabular}
\caption[The particle content of the Standard Model as well as Gravity.]{{\it Top:} The particle content for the Standard Model of elementary particle physics after electroweak symmetry breaking is given. For the three generations (or families) of quarks and leptons, the force-mediating gauge-bosons and for the only elementary scalar, the Higgs boson, the corresponding $SU(3)_{c}$-representation, electric charge and mass-estimate are quoted (with data taken from \cite{Patrignani:2016xqp}). Electric charges are given in units of the proton-charge $+e$. The quarks are in the fundamental, or triplet-, representation $\bm{3}$ of $SU(3)_{c}$ while gluons are in the adjoint, or octet-, representation $\bm{8}$. All remaining particles are color-singlets $\bm{1}$ and thus do not partake in strong interactions. In the Table, all further information about weak interactions and the corresponding quantum numbers has been suppressed, which would otherwise also require a distinction of chiralities. \newline
{\it Bottom:} In a small sub-Table, some information is given on the graviton. In treatments for quantum gravity such as \cite{DonoghueEtAlARTEFT}, the graviton is not charged electrically and furthermore, at least as a fundamental field quantum, massless. Still, the PDG \cite{Patrignani:2016xqp} gives a mass-bound. Further knowledge on the quantum numbers of the graviton would require a sensible scheme to unify gravity with the Standard Model. This author does not want to give the impression that he has any idea about such a unification.}
\label{tab:ParticleContentSM}
\end{table}
%


The currently accepted classical field theory for gravitation, the {\it General Theory of Relativity} (GR), was published by Einstein at the beginning of the previous century \cite{EinsteinART} (Reference \cite{WeinbergART} is a popular textbook.). According to this theory, any system which carries energy and momentum influences the geometry, in particular the curvature, of spacetime itself. The latter is encoded in the metric tensor $g_{\mu \nu} (x)$. Test-particles move then along geodesics in spacetime. In a simplified statement: "Matter tells spacetime how to curve and spacetime tells matter how to move." \cite{WheelerQuoteART}. It has long been known that in case one makes an Ansatz for the metric tensor as a flat Minkowski-background plus a small metric perturbation $h$, i.e. $g_{\mu \nu} (x) = \eta_{\mu \nu} + h_{\mu \nu} (x)$, then GR admits wave-solutions for the field $h_{\mu \nu}$ \cite{WeinbergART}. In another breakthrough of experimental physics which occurred during the writing of this thesis, the LIGO-collaboration measured for the first time gravitational waves stemming from a distant inward spiral- and merger-event of two black holes, a signal named {\it GW150914} \cite{LIGOGravWaves}. \newline
The quantization of gravity is one of the long-lasting challenges of theoretical physics. In case the metric perturbation $h_{\mu \nu}$ is treated as the relevant degree of freedom for the gravitational field, GR can be quantized consistently as an Effective Field Theory (EFT) \cite{DonoghueGREFT1994, DonoghueEtAlARTEFT}. The quantized perturbation is then called {\it graviton} and it is a spin-$2$ particle which mediates the gravitational interaction. The graviton is also included into Table \ref{tab:ParticleContentSM}. \newline
Quantum-GR as an Effective Field Theory is expected to break down no later than at the Planck scale \cite{DonoghueEtAlARTEFT}
\begin{equation}
 M_{\mathrm{P}} = \sqrt{ \frac{\hbar c}{ G_{N} } } = 1.22 \times 10^{19} \frac{\mathrm{GeV}}{c^{2}} \mathrm{,} \label{eq:PlanckScale}
\end{equation}
where $c$ is the velocity of light, $\hbar$ is Planck's constant and $G_{N}$ is Newton's constant. \newline
Furthermore, the quantization of gravity alone does not achieve a unification of GR with the Standard Model. A hypothetical suggestion of new ultraviolet physics for both the Standard Model and gravity which may accomplish the unification is string theory \cite{GSWStringTheory}. \newline
Now we turn again to QCD, in particular to its very important property of {\it asymptotic freedom} \cite{GrossWilczek1973, Politzer1973}. The following discussion can be found in many textbooks \cite{DonoghueEtAlSM, AitchisonHey} and, partly, in the PDG-review article \cite{Patrignani:2016xqp}. Suppose one has a Dirac-field $q_{i} (x)$ which describes a quark. Here, the index $i$ has been written explicitly and it specifies the color-charge state the quark is in. This index can take tree values $i = 1,2,3$ which are typically denoted as colors {\it red}, {\it green} and {\it blue}. The dynamics of QCD is now fixed by the requirement of having the theory invariant under {\it local} $SU(3)_{c}$-transformations, which act on the quark-fields as
\begin{equation}
q_{j} (x) \longrightarrow q_{j}^{\prime} (x) = \left[\bm{U} (x)\right]_{jk} q_{k} (x) = \exp\left[- i \alpha_{a} (x) \frac{\bm{\lambda}^{a}_{jk}}{2} \right] q_{k} (x) \mathrm{.} \label{eq:QuarkTrafoFundamentalRep}
\end{equation}
Here, the term 'local' means that the transformation-parameters $\alpha_{a}(x)$ depend on the space\-time-coordinate. The $\bm{\lambda}^{a}$ are called Gell-Mann matrices\footnote{Listings of the Gell-Mann matrices are given in many places, for instance \cite{DonoghueEtAlSM}.} and they form a representation of the generators of $SU(3)_{c}$. The quarks are said to transform in the fundamental $3$-dimensional, or triplet-, representation $\bm{3}$ of $SU(3)_{c}$. From here on, $SU(3)_{c}$-triplet indices are denoted $ j,k, \ldots $. The indices from the beginning of the alphabet $\left( a,b, \ldots \right)$ belong to another representation of $SU(3)_{c}$, the so-called adjoint or, in this case, octet-representation $\bm{8}$. Also, repeated indices are summed in equation (\ref{eq:QuarkTrafoFundamentalRep}) and in the following. \newline
In order to obtain a {\it lagrangian} which defines the theory and which is invariant under local $SU(3)_{c}$-transformations, the ordinary derivatives $\partial_{\mu} q_{i}$ of the quark-fields have to be mo\-di\-fied. Loosely speaking, one has to cancel the product-rule terms arising once the transformed field (\ref{eq:QuarkTrafoFundamentalRep}) is differentiated with respect to the space\-time-argument. Thus, a {\it gauge-covariant derivative} $\left(D_{\mu} \right)_{jk}$ is introduced such that a differentiated quark-field transforms covariantly, i.e. $\left(D_{\mu} q\right)_{i} \rightarrow \left(D_{\mu} q\right)^{\prime}_{i} = \left[ \bm{U} (x) \right]_{ij} \left( D_{\mu} q \right)_{j} $ \cite{DonoghueEtAlSM}. The gauge-covariant derivative for QCD is defined in terms of an ordinary derivative and a connection-term involving the gauge-, or {\it gluon}-, fields $A^{a}_{\mu} (x)$ as
\begin{equation}
 \left(D_{\mu} \right)_{jk} = \delta_{jk} \partial_{\mu} + i g_{3} A^{a}_{\mu} \frac{\bm{\lambda}^{a}_{jk}}{2} \mathrm{,} \label{eq:NonAbelianGaugeCovariantDerivative}
\end{equation}
with the $SU(3)_{c}$ gauge-coupling $g_{3}$. It can be shown that $D_{\mu}$ as given here is covariant if and only if the transformation-rule of the gluon-fields resembles that of an element of the adjoint representation $\bm{8}$ plus a correction-term depending in the derivative of the matrix $\bm{U} (x)$ \cite{AitchisonHey}. Thus, the fact that gluons transform in the $\bm{8}$ of $SU(3)_{c}$ is tied intimately to the demanded non-abelian gauge-invariance. \newline
We quote now the lagrangian of QCD already in a form which is suitable for perturbative quantization. It reads\footnote{Note that the $CP$-violating QCD $\theta$-term \cite{Patrignani:2016xqp} is neglected here.} \cite{DonoghueEtAlSM}
\begin{align}
 \mathcal{L}_{\mathrm{QCD}} &= \sum_{q} \bar{q}_{j} \left( i \slashed{D}_{jk} - m_{q} \delta_{jk} \right) q_{k} - \frac{1}{4} F^{a}_{\mu \nu} F^{a \hspace*{1pt} \mu \nu} - \frac{1}{2 \xi_{0}} \left( \partial_{\mu} A^{\mu}_{a} \right)^{2} \nonumber \\
 & \hspace*{11.5pt} + \partial_{\mu} \bar{c}_{a} \partial^{\mu} c_{a} + g_{3,0} f_{abe} A^{\mu}_{a} \left( \partial_{\mu} \bar{c}_{b} \right) c_{e} \mathrm{.} \label{eq:QCDLagrangian} 
\end{align}
The first, Dirac-bilinear-, term is here summed over all quark-flavours $q \in \left\{ u, d, c, s, b, t \right\}$, see Table \ref{tab:ParticleContentSM}. The Feynman-slashed covariant derivative is $\slashed{D}_{jk} = \gamma^{\mu} \left(D_{\mu} \right)_{jk}$, with Dirac-matrices $\gamma^{\mu}$. At second place there appears the pure gauge-kinetic (or Yang-Mills-) term defined in terms of the non-abelian gauge field-strength tensor $F^{a}_{\mu \nu}$, which in component-form is given by
\begin{equation}
 F^{a}_{\mu \nu} = \partial_{\mu} A^{a}_{\nu} - \partial_{\nu} A^{a}_{\mu} - g_{3,0} f^{abc} A^{b}_{\mu} A^{c}_{\nu} \mathrm{.} \label{eq:NonAbelianFieldStrength}
\end{equation}
The numbers $f^{abc}$ are called the {\it structure constants}\footnote{The structure constants define the Lie-algebra of $SU(3)_{c}$ via $\left[ \bm{\lambda}_{a}, \bm{\lambda}_{b} \right] = 2 i f_{abc} \bm{\lambda}_{c}$, with repeated indices summed. For an abelian gauge-group, these constants would all vanish.} of $SU(3)$. All the remaining terms in (\ref{eq:QCDLagrangian}) are present as a consequence of a Lorentz-invariant gauge-fixing procedure, which is itself necessary for quantization \cite{DonoghueEtAlSM}. The third term in the first line of (\ref{eq:QCDLagrangian}) is a so-called gauge-fixing term, while the entire second line belongs to Grassmann-valued scalar fields $\left\{ c_{a} (x), a=1,\ldots,8 \right\}$ called {\it ghosts}. The ghosts couple only to gluons and never appear in asymptotic states, but only within closed loops. Furthermore, all sub-scripts '$0$' in equation (\ref{eq:QCDLagrangian}) denote bare quantities. \newline
Once the Feynman-rules for perturbative QCD are derived \cite{Patrignani:2016xqp, DonoghueEtAlSM, BailinLove}, it is seen that the covariant derivative in the Dirac-bilinear terms in the lagrangian (\ref{eq:QCDLagrangian}) implies a quark-antiquark-gluon vertex which is proportional to $g_{3}$. Once the field-strength (\ref{eq:NonAbelianFieldStrength}) is inserted into the gauge-kinetic term of (\ref{eq:QCDLagrangian}), a $3$-gluon vertex (also proportional to $g_{3}$) and a $4$-gluon vertex (proportional to $g_{3}^{2}$) arise. The gluons carry color-charge degrees of freedom themselves and thus self-interact, as a consequence of the non-abelian gauge-structure of QCD. One of the many implications of this fact will be outlined in the following. \newline
Once perturbative calculations are preformed at an order where Feynman diagrams with a
\begin{center}
\begin{figure}[ht]
 \centering
\includegraphics[width=0.99\textwidth,trim=0 45 0 37.5,clip]{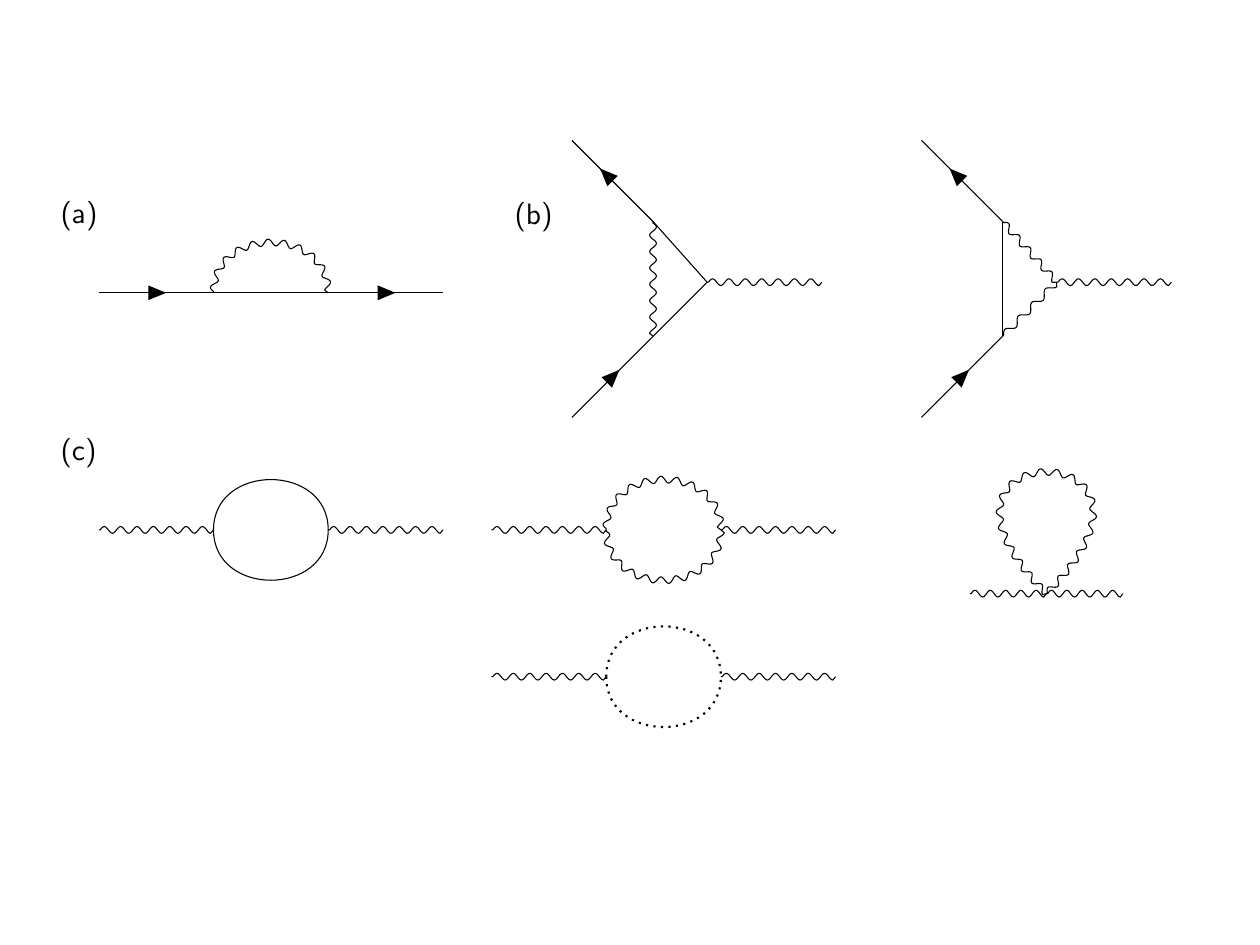}
 \caption[Diagrams contributing to the QCD $\beta$-function at one loop order.]{The Feynman diagrams contributing to QCD at $1$-loop order are shown sche\-ma\-ti\-cal\-ly (cf. \cite{DonoghueEtAlSM, BailinLove}). Solid lines are quarks, wavy lines represent gluons and dotted lines are ghosts. Indices (Lorentz-group, color) and $4$-momenta are not shown explicitly, as well as counterterm diagrams. \newline Terms are divided into: quark self-energy (a), corrections to the quark-antiquark-gluon vertex (b) and to the gluon-propagator (c). The second diagram of (b) and the second, third and fourth diagram of (c) involve gluon self-interactions as well as ghosts and thus would be absent in the abelian case (i.e. QED).}
\label{fig:OCDDiagramsOneLoop}
\end{figure}
\end{center}
non-vanishing number of closed {\it loops} are present, the procedure of {\it renormalization} \cite{PeskinSchroeder} becomes necessary. As an example, the topologies arising from quantum-corrections to QCD in the $1-$loop order are shown schematically in Figure \ref{fig:OCDDiagramsOneLoop}. \newline
Such loop amplitudes are momentum-space integrals which contain divergent parts. Thus one first has to regularize\footnote{I.e., the use of some prescription to split the loops additively into divergent and finite parts.} the loops and then absorb the divergent parts into so called {\it renormalization factors}. These factors relate the redefined, so-called {\it renormalized}, fields and physical constants to the bare quantities in the original lagrangian (\ref{eq:QCDLagrangian}) multiplicatively. Amplitudes are in the end expressed in terms of renormalized quantities and thus become finite. This whole process introduces a new energy scale, the so-called {\it renormalization scale} $\mu$, into the problem, which specifies the energy at which renormalized quantities are measured. \newline
Bare quantities on the other hand should not depend on the scale $\mu$ \cite{PeskinSchroeder, DonoghueEtAlSM}. This consistency condition leads to so-called {\it renormalization group equations}, which are dif\-fe\-rential equations for renormalized quantities as functions of $\mu$. \newline
The dependence of the coupling constant $g_{3}$ on the renormalization scale $\mu$ is described by the so-called {\it beta-function} of QCD. For $N_{c} = 3$ colors and $n_{f}$ active quark-flavours, this function reads in the $1$-loop approximation \cite{DonoghueEtAlSM, Patrignani:2016xqp}
\begin{equation}
 \beta_{QCD} = \mu \frac{\partial g_{3}}{\partial \mu} = - \left( 11 - \frac{2 n_{f}}{3} \right) \frac{g_{3}^{3}}{16 \pi^{2}} + \mathcal{O} \left( g_{3}^{5} \right) = - \beta_{0} \frac{g_{3}^{3}}{16 \pi^{2}} + \mathcal{O} \left( g_{3}^{5} \right) \mathrm{.} \label{eq:BetaQCD1LoopOrder}
\end{equation}
\newpage
\begin{figure}[ht]
\centering
 \includegraphics[width=0.69\textwidth]{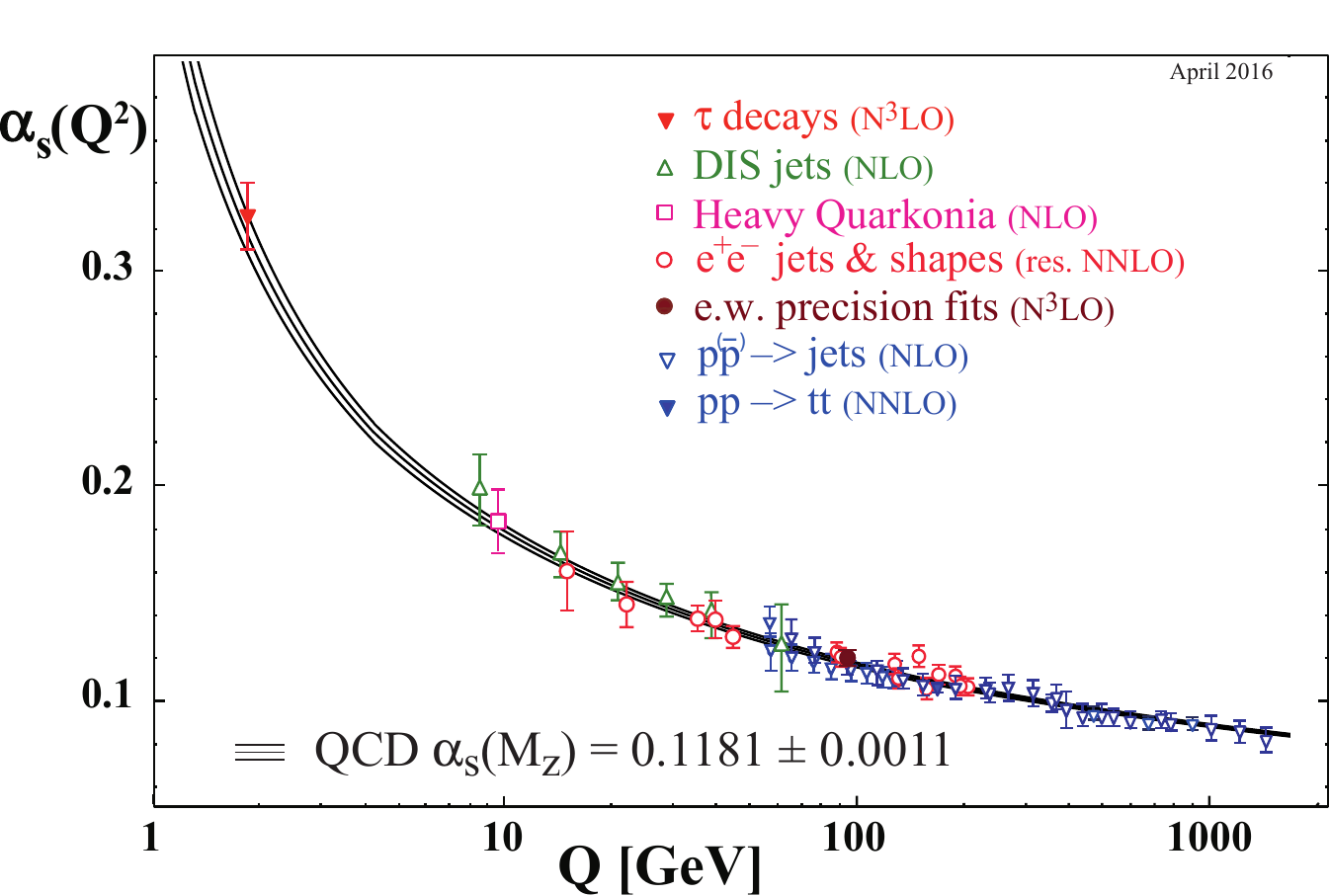}
 \caption[The running coupling constant of QCD as a function of $4$-momentum transfer.]{The strong coupling constant $\alpha_{s} (Q^{2})$ is shown as a function of energy \cite{Patrignani:2016xqp}. Experiments and systems used for the determination of $\alpha_{s}$ are indicated in the legend of the plot as well as by the datapoints.}
\label{fig:AlphaSRunning}
\end{figure}
It is seen that the leading coefficient $- \beta_{0}$ has a negative sign for $n_{f} < 17$. As far as we know, this is fulfilled by nature (see Table \ref{tab:ParticleContentSM}), a fact which leads to the so-called asymptotic freedom of QCD. The differential equation (\ref{eq:BetaQCD1LoopOrder}) can be integrated using elementary methods. Evaluating the result at $\mu = Q^{2}$, i.e. the $4$-momentum transfer of a particular reaction, gives a measure for the strength of the interaction at the energy of the respective process. Redefining $\alpha_{s}\left( Q^{2} \right) \equiv g_{3}^{2} \left( Q^{2} \right) / \left( 4 \pi \right)$, one obtains \cite{DonoghueEtAlSM}
\begin{equation}
 \alpha_{s}\left( Q^{2} \right) = \frac{4 \pi}{\left( 11 - \frac{2 n_{f}}{3} \right)} \frac{1}{\log \left( Q^{2} / \Lambda_{QCD}^{2} \right)} \mathrm{,} \label{eq:RunningCouplingToOneLoopOrder}
\end{equation}
Thus, for $Q^{2}\rightarrow \infty$, $\alpha_{s}$ vanishes and quarks are free. The constant $\Lambda_{QCD}$ is the scale at which $\alpha_{s}$ diverges. The QCD $\beta$-function can of course be evaluated at higher loop-orders. Up to now, precision calculations have been performed up to the $4$-loop level \cite{Patrignani:2016xqp}. \newline
 Defining the quantity $a_{s} := \alpha_{s}/(4 \pi)$, the authors of reference \cite{DonoghueEtAlSM} write the result as
\begin{equation}
 \mu^{2} \frac{\partial a_{s}}{\partial \mu^{2}} = - \beta_{0} a_{s}^{2} - \beta_{1} a_{s}^{3} - \beta_{2} a_{s}^{4} - \beta_{3} a_{s}^{5}  + \ldots \mathrm{,} \label{eq:BetaFunctionToFourLoops}
\end{equation}
also quoting approximate values for $\beta_{0}, \ldots, \beta_{3}$. With an expansion of the QCD beta-function up to $4$-loop order as a theoretical basis, the determination of $\alpha_{s}$ from experimental data is typically attempted at kinematics where the coupling is still so small that perturbative QCD can yield reasonable approximations. Collective attempts to do so have lead to a famous picture, shown in Figure \ref{fig:AlphaSRunning}, for the energy-dependence of the strong coupling. \newline

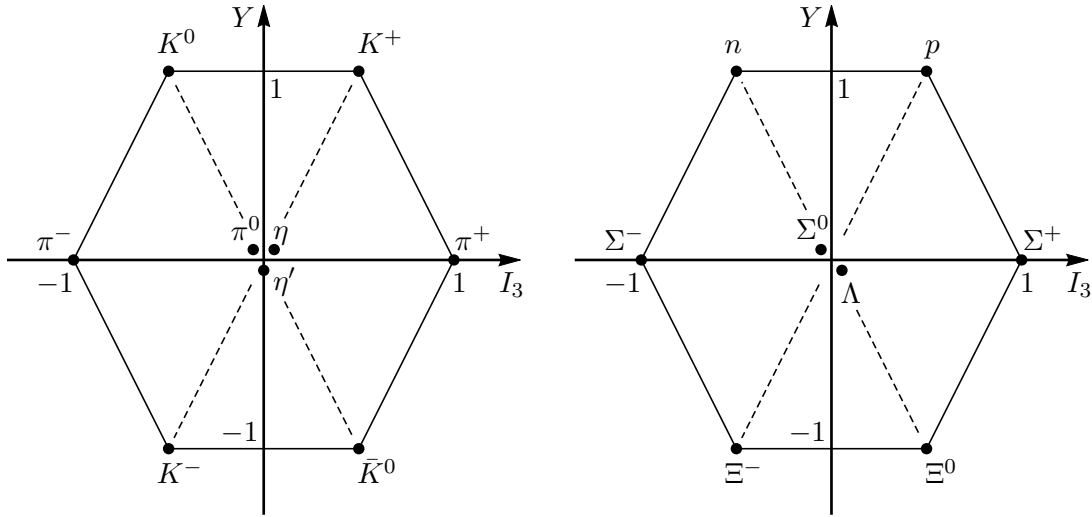
\begin{figure}[ht]
\centering
 \begin{overpic}[width=0.46\textwidth]%
      {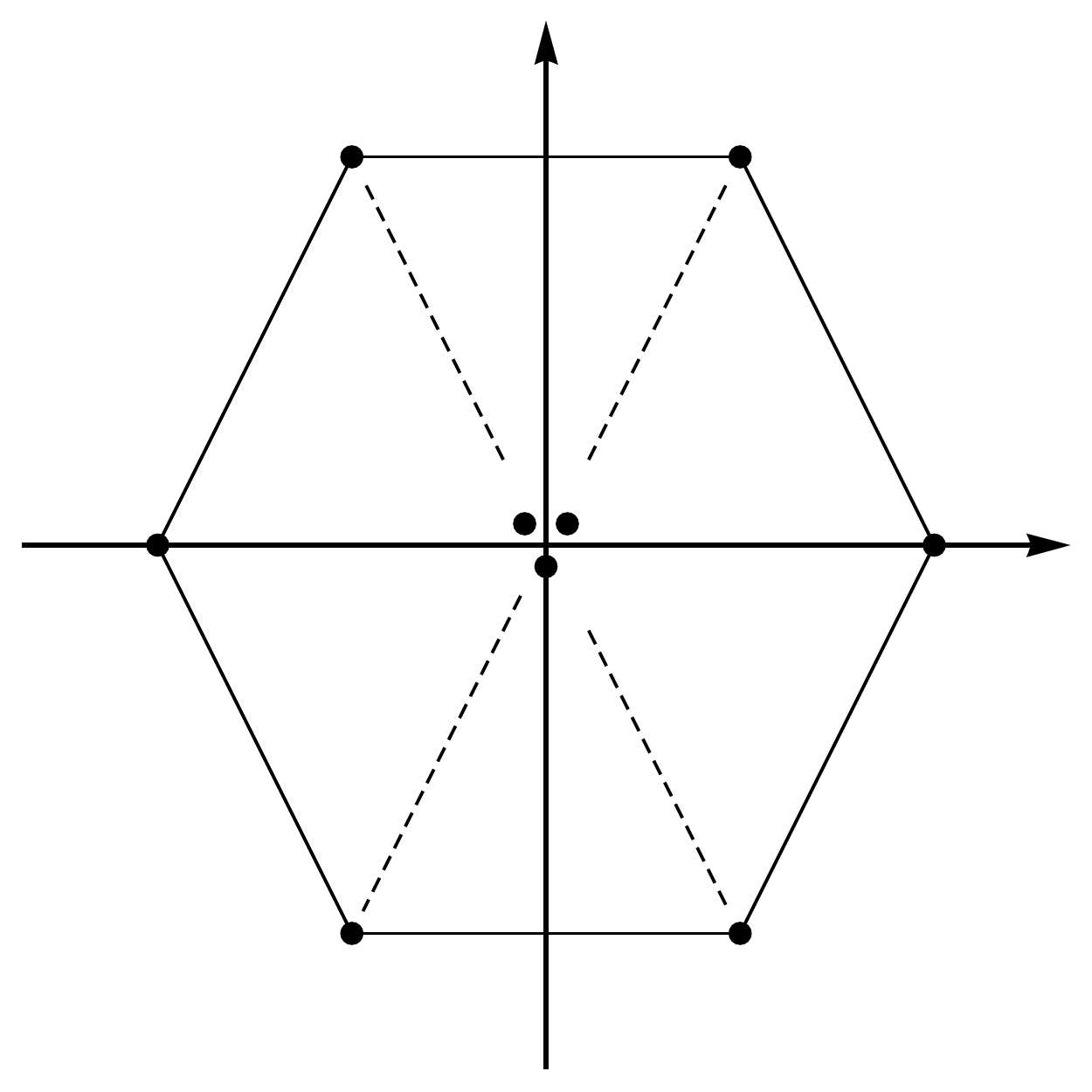}
  \put(94,44){$I_{3}$}
  \put(44,94){$Y$}
  \put(30,89){$K^{0}$}
  \put(67.5,89){$K^{+}$}
  \put(30,8){$K^{-}$}
  \put(67.5,8){$\bar{K}^{0}$}
  \put(85.25,44){$1$}
  \put(7.5,44){$-1$}
  \put(85.75,52){$\pi^{+}$}
  \put(7.5,52){$\pi^{-}$}
  \put(43.85,53.8){$\pi^{0}$}
  \put(51.75,54){$\eta$}
  \put(51.75,44.5){$\eta^{\prime}$}
  \put(51,80.5){$1$}
  \put(42,16){$-1$}
\end{overpic} \hspace*{5pt}
 \begin{overpic}[width=0.46\textwidth]%
      {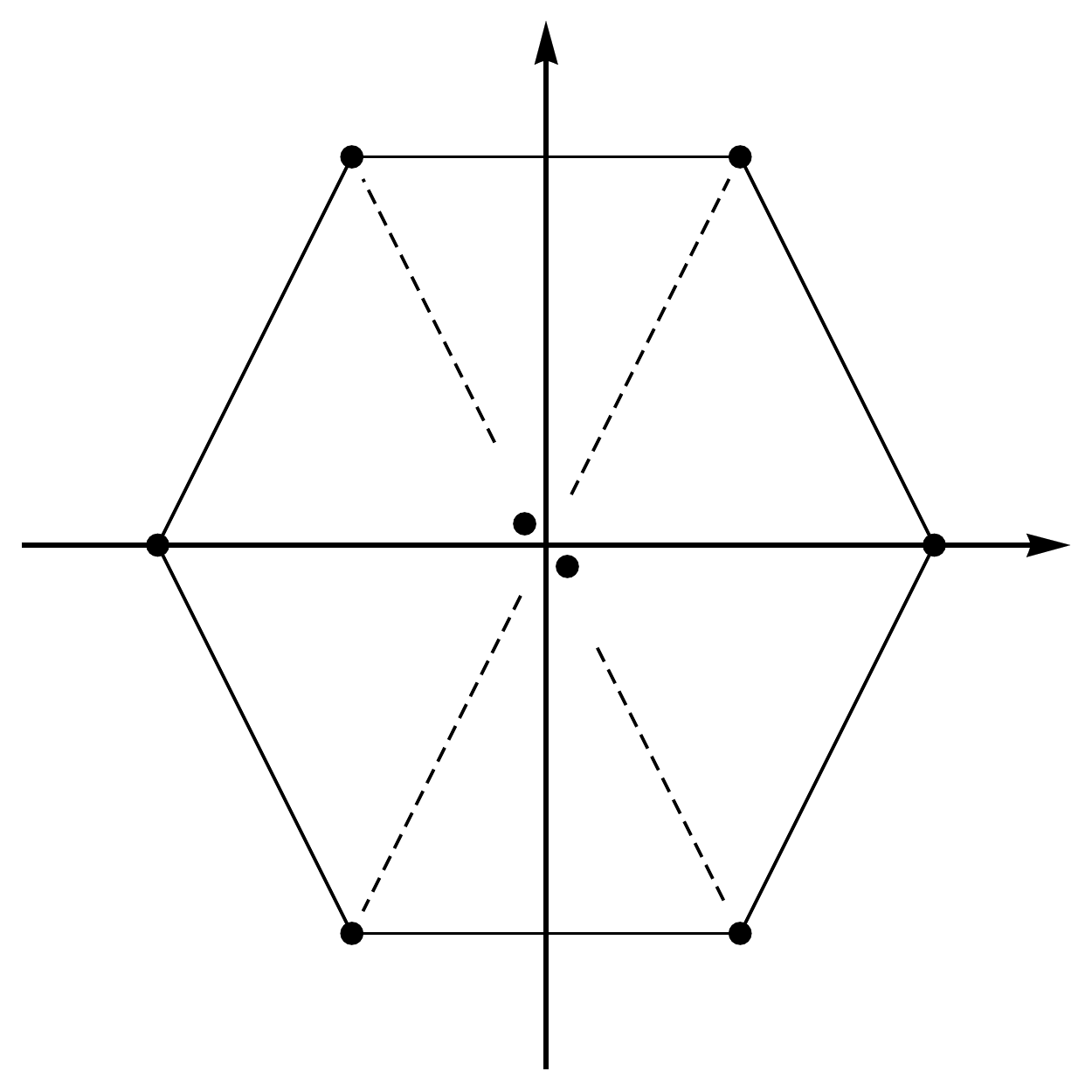}
  \put(94,44){$I_{3}$}
  \put(44,94){$Y$}
  \put(30,89){$n$}
  \put(67.5,89){$p$}
  \put(30,8){$\Xi^{-}$}
  \put(67.5,8){$\Xi^{0}$}
  \put(85.25,44){$1$}
  \put(7.5,44){$-1$}
  \put(85.75,52){$\Sigma^{+}$}
  \put(7.5,52){$\Sigma^{-}$}
  \put(43.35,54){$\Sigma^{0}$}
  \put(51.75,42){$\Lambda$}
  \put(51,80.5){$1$}
  \put(42,16){$-1$}
\end{overpic}
 \caption[Weight diagrams for the pseudoscalar meson nonet and baryon octet.]{Shown here are $Y$-$I_{3}$ diagrams (or weight diagrams) for the ground state pseudoscalar meson nonet (left) and baryon octet (right), with the strong hypercharge $Y$ and the third component of the strong isospin $I_{3}$. Quantum numbers are $J^{PC} = 0^{-+}$ for the mesons and $J^{P} = \frac{1}{2}^{+}$ for the baryons. \newline
 The nonet of vector mesons and the baryon decuplet are not shown here in favor of brevity (see reference \cite{Patrignani:2016xqp}). The pictures are facsimiles of plots which can be found in many places in the literature (see, for instance, \cite{HalzenMartin}).}
\label{fig:WeightDiagramsMesonsBaryons}
\end{figure}

An up to now mathematically unproven working hypothesis consists of the assumption that the strong coupling indeed continues to increase for lower and lower $Q^{2}$. This is then assumed to lead to the {\it confinement} of quarks. Thus, quarks only occur in bound states, the latter of which are called {\it hadrons} and by themselves interact strongly. However, hadrons do not carry any net color-charge. Quarks never appear as asymptotic states of any reaction, but instead they {\it hadronize}. \newline
Hadrons can be categorized in a group-theoretical construct known as the {\it quark model} \cite{GellMann1962QuarkModel, GellMann1964QuarkModel, Neeman1961QuarkModel, ZweigQuarkModel2}. Again, the PDG \cite{Patrignani:2016xqp} gives a good review of the subject. In the most basic form, the quark model classifies hadrons under so-called flavour-symmetries. However, it has to be noted that such flavour-symmetries are not exact, but are realized as broken symmetries in nature. This is contrary to the color-symmetry, which is exact in the QCD-lagrangian. The following facts are also discussed for instance in the book by Halzen and Martin \cite{HalzenMartin}. \newline
Considering only up, down and strange quarks $\left(u,d,s\right)^{T}$, an overarching flavour symmetry of $SU(3)_{\bm{f}}$ can be assumed. The simplest possible color-neutral bound states are either made up from a quark-antiquark pair $(q \bar{q})$, called {\it mesons}, or of three quarks $(qqq)$, in which case the states are called {\it baryons}. The full wave function is then defined on the direct product of the Hilbert-spaces belonging to the quarks and anti-quarks, respectively. 
Thus, one also has to form the direct product of the flavour-parts, with a quark $q$ transforming in the $\bm{3}$- and an anti-quark transforming in the $\bm{\bar{3}}$- representation of $SU(3)_{\bm{f}}$. The decomposition of such a direct product, for instance in case of mesons $(q \bar{q})$, into a direct sum of irreducible representations is conventionally written in highly condensed mathematical notation as  
\begin{equation}
 \bm{3} \otimes \bm{\bar{3}} = \bm{8} \oplus \bm{1} \mathrm{.} \label{eq:MesonIrrepDecomposition}
\end{equation}
Thus, the mesonic flavour-space decomposes into the direct sum of an octet $\bm{8}$ and a singlet $\bm{1}$, which do not transform into each other under a general $SU(3)_{\bm{f}}$-rotation. Irreducible representations are generally referred to as {\it multiplets}. \newline
For the flavour-part of a baryonic ($qqq$) wavefunction, the same arguments apply but now the direct product of three triplets $\bm{3}$ has to be taken. The resulting decomposition into irreducible representations is\footnote{The subscript denote symmetric ($S$), mixed-symmetric ($M$) and antisymmetric ($A$) states under interchange of any two quarks.}
\begin{equation}
 \bm{3} \otimes \bm{3} \otimes \bm{3} = \bm{10}_{S} \oplus \bm{8}_{M} \oplus \bm{8}_{M} \oplus \bm{1}_{A} \mathrm{,} \label{eq:BaryonIrrepDecomposition}
\end{equation}
i.e. one obtains a decuplet $\bm{10}$, two octets $\bm{8}$ and a singlet $\bm{1}$. Furthermore, in case one would apply the decompositions (\ref{eq:MesonIrrepDecomposition}) and (\ref{eq:BaryonIrrepDecomposition}) to color $SU(3)_{c}$, it is seen that both products admit a singlet and thus both are a sensible Ansatz for color-neutral hadrons. \newline
The octets for the ground state pseudoscalar mesons and spin-$\frac{1}{2}$ baryons are shown as so-called $Y$-$I_{3}$-diagrams, or {\it weight diagrams}, in Figure \ref{fig:WeightDiagramsMesonsBaryons}. Here, $I_{3}$ is the third component of the strong isospin and $Y = \mathcal{B} + \mathcal{S}$ is the so-called {\it hypercharge}\footnote{The given definition is only valid for $u$, $d$ and $s$ quarks, with the heavier flavors removed from the picture.}, with {\it baryon number} $\mathcal{B}$ and {\it strangeness} $\mathcal{S}$. One has $\mathcal{B} = 1$ for baryons and $\mathcal{B} = 0$ for mesons, while the total strangeness quantum-number receives a contribution of $\mathcal{S} = -1$ from every strange valence-quark present in the hadron. The charges of the hadrons are related to hypercharge and isospin $3$-component by the {\it Gell-Mann Nishijima relation} $Q = I_{3} + Y/2$ \cite{Patrignani:2016xqp}. The familiar nucleons $N = \left( p,n \right)^{T}$ are given as an $SU(2)$ sub-doublet in the baryon octet shown in Figure \ref{fig:WeightDiagramsMesonsBaryons}. \newline
It should be noted that there is no reason to forbid mixing of isoscalar states with the same $J^{PC}I$ quantum numbers. For ground state mesons, the $\eta$ and $\eta^{\prime}$ are mixtures of the isoscalar states from the octet and singlet and are thus both indicated in Figure \ref{fig:WeightDiagramsMesonsBaryons}. In the ground state baryon octet, the flavor singlet is forbidden via Fermi statistics \cite{Patrignani:2016xqp}. 
\clearpage
Hadrons are composite objects and therefore have an excitation spectrum. The spectroscopy of hadrons represents an important field and is relevant in oder to obtain a deeper understanding of the inner workings of QCD. Hadron spectroscopy, in particular baryon spectroscopy, is the field to which the subject of this thesis is most intimately related. Thus, a brief summary on the experimental and theoretical tools of baryon spectroscopy will be given in the next section. \newline
A further motivation for why the study of bound-state formation in QCD is interesting is given by the generation of hadronic masses, in particular now for the ground state hadrons. It can be shown that here the so-called {\it trace-anomaly} of QCD is relevant \cite{DonoghueEtAlSM}. The mass of a hadron is directly related to the zero momentum-transfer matrix element of the trace of the QCD energy-momentum tensor $\Theta^{\mu \nu}$. The fact that this trace does not vanish is an anomaly and therefore purely a quantum effect. For instance, for the nucleon the result for the relevant matrix-element is \cite{DonoghueEtAlSM}
\begin{align}
 m_{N} \hspace*{1pt} \bar{u} \left( \vec{p} \hspace*{1.5pt} \right) u \left( \vec{p} \hspace*{1.5pt} \right) &= \left< N \left( \vec{p} \hspace*{1.5pt} \right) \right| \Theta^{\mu}_{\hspace*{1.5pt} \mu} \left| N \left( \vec{p} \hspace*{1.5pt} \right) \right> \nonumber \\
 &= \left< N \left( \vec{p} \hspace*{1.5pt} \right) \right| \frac{\beta_{QCD}}{2 g_{3}} F^{a}_{\mu \nu} F^{a \hspace*{1pt} \mu \nu} + m_{u} \bar{u} u + m_{d} \bar{d} d + m_{s} \bar{s} s \left| N \left( \vec{p} \hspace*{1.5pt} \right) \right> \mathrm{.} \label{eq:NucleonTraceAnomaly}
\end{align}
Determinations of the so-called Pion Nucleon $\sigma$-term \cite{HoferichterEtAlPiNSigma} show that the contribution of the up- and down quarks to the quantity (\ref{eq:NucleonTraceAnomaly}) are small. Thus, the bulk comes from the strange quark and the gluon-term, where it is believed that the contribution of gluons dominates \cite{DonoghueEtAlSM}.
Therefore, although the mass of all elementary particles on the lagrangian-level is generated entirely by the Higgs-mechanism, the mass of the baryonic matter surrounding us, which is made up of protons and neutrons, is not. Instead, the bulk of the mass of the latter is generated by dynamical effects peculiar to QCD.

\clearpage

\subsection{Baryon spectroscopy} \label{sec:BaryonSpectroscopy}

Here we outline briefly the experimental and theoretical methods of hadron spectroscopy. Although in particular the mentioned experiments can be used for the spectroscopy of mesons as well, here everything will be presented with a slope to baryon spectroscopy. For the latter subject, we have to mention the very elaborate review-article by Klempt and Richard \cite{KlemptRichard}, which is fundamental for the ensuing discussion. \newline
Much in analogy with atomic emission spectroscopy (see for instance \cite{DemtroederIII}), the spectra of hadrons can be studied by first isolating the system under consideration and then injecting energy, or more precisely $4$-momentum, into it. This injection will cause the system to transition into some metastable state of higher energy, the state to be studied, which then decays via emission of particles that are characteristic for the physics in question. The detailed study of the decay-products can then yield information on the constituents of the system and their dynamics. \newline
For the case of atomic physics \cite{DemtroederIII}, the material under study is first fixed in either solid or gaseous form. The metastable intermediate states are just atoms with one or more electrons in a higher atomic shell. These then decay under emission of photons, which then yield an emission spectrum giving detailed information about the energy-level scheme of the respective element. The experimental methods of excitation adequate for characteristic energies of atomic physics ($E=\mathrm{eV},\ldots,\mathrm{keV}$)\; examples are here the use of flames, coupled plasmas, gas lamps, $\ldots$\; are still simple and small enough that they fit in a small to moderately sized laboratory. \newline
Here persists the difference to experiments in hadron spectroscopy. Hadronic physics takes place on length-scales of $1$ $\mathrm{fm}$ or less, which corresponds to characteristic energy-scales in the $\mathrm{MeV}$-range up to a few $\mathrm{GeV}$. Thus, one already has to conduct more sophisticated intermediate-energy experiments, typically scattering-experiments, using complicated detector-systems. \newline
From now on, we confine to fixed-target experiments. The target-particles are in most cases given by nucleons $N$. Then, either strong or electromagnetic probes are impinged on the target. The standard example for strong probes are pions $\pi$ (but also, for instance, kaons $K$), while pertinent electromagnetic probes are either photons $\gamma$ or electrons $e$. \newline
The initial state thus consists in each case of a two-particle system. However, since neither particle-type nor particle-number are conserved quantities for relativistic reactions\footnote{See any field theory textbook, for instance \cite{PeskinSchroeder}.}, it is possible to produce a multitude of different final states. This is of course only possible in case the respective final states are allowed kinematically, i.e. the total $4$-momentum is conserved, and certain important quantum numbers are unchanged as well\footnote{Kinematics and quantum numbers can serve as first guiding principles to judge the feasibility of a particular channel. However, specific channels may still be suppressed due to certain dynamical reasons, which can generally only emerge after a more detailed analysis.} (e.g. baryon number, total electric charge, strangeness, $\ldots$). Thus, different possible final states, or {\it channels}, arise which can be collectively denoted by the symbol $X$ for 'everything'. Thus, the conglomerate of allowed channels, for different probes, can be denoted collectively as
\begin{align}
 \pi N \longrightarrow X \mathrm{,} \label{eq:PionNToEverything} \\
 \gamma N \longrightarrow X \mathrm{,} \label{eq:PhotonNToEverything} \\
 e N \longrightarrow X \mathrm{,} \label{eq:ElectronNToEverything}
\end{align}
and a hierarchy of reaction arises which are ordered in energy. The excited states of hadrons to be studied then occur as metastable intermediate states, or {\it resonances}, in such (purely or partly) hadronic reactions. \newline
The most fundamental reactions for the study of baryon resonances are elastic pion-nucleon scattering and charge-exchange reactions $\pi N \longrightarrow \pi N$. Data generally consist of the ordinary differential cross section. However, since the spin degrees-of-freedom of the nucleons can be polarized, in a brief notation\footnote{Vectors denote polarizable spins.} $\pi \vec{N} \longrightarrow \pi \vec{N}$, one also has the possibility to measure so-called {\it polarization observables}. Important facilities taking polarized and unpolarized data for pion-nucleon reactions have been at CERN \cite{AlbrowEtAl1972}, Rutherford-lab \cite{BrownEtAl1978} and ITEP in Moskow \cite{AlekseevEtAl1989}, among others. More information on the data can be found in the SAID-database \cite{SAID} or the review \cite{KlemptRichard}. \newline
For higher energies, further inelastic channels become possible, for instance di-pion production $\pi N \longrightarrow \pi \pi N$. One should mention here the COMPASS-collaboration \cite{AbbonEtAl2007}, which studies highly inelastic pion-induced processes such as $\pi N \longrightarrow \pi \pi \pi N$ \cite{Mikhasenko:2017jtg}. The interest of these measurements lies however nowadays mainly within meson-spectroscopy. \newline
Over the most recent years, the photo-induced reactions (\ref{eq:PhotonNToEverything}) have contributed a lot of contraints on the baryon-spectrum \cite{KlemptRichard, CommonPaper}. The simplest possible reaction is the production of one pseudoscalar meson, the most pertinent version of which is photoproduction of pions $\vec{\gamma} \vec{N} \longrightarrow \pi \vec{N}$. Since the photon has spin-$1$, there are more polarization observables accessible in this case than for Pion-Nucleon scattering. Pseudoscalar meson photoproduction is at the center of attention in this work and therefore the details of this reactions are given in a lot more detail below (sections \ref{sec:Photoproduction} to \ref{sec:CompExpsTPWA}). Experimental facilities running dedicated programs on photoproduction are located at ELSA in Bonn \cite{Thiel:2012}, MAMI in Mainz \cite{Adlarson:2015} and JLab in Newport News \cite{Dugger:2013}. \newline

How exactly the properties of resonances can be extracted from data for hadronic reactions is a complicated matter by itself. Also, the subject of this thesis is a subtopic of this particular issue. Therefore, a survey of currently applied methods will be provided in section \ref{subsec:Energy-dependentModels} below. \newline
Having extracted the resonances from scattering data, one has to have some kind of prediction to compare to, which ideally should come from QCD itself. Since in the low-energy region QCD cannot be attacked using standard perturbative methods (cf. section \ref{sec:StandardModel}), alternative sophisticated theoretical approaches have to be used. Among them are {\it Constituent Quark Models}, {\it Lattice-QCD} and {\it Chiral Effective Field Theory}. These three main approaches shall be outlined briefly in the following. \newline

\textbf{Phenomenological Constituent Quark Models (CQM)} \newline
The quark models as initially proposed by Isgur and Karl \cite{IsgurKarl1, IsgurKarl2} model a hadron as a bound state of so-called {\it constituent quarks}. The latter are the current-quarks from the QCD-lagrangian (\ref{eq:QCDLagrangian}), but surrounded by a cloud of quarks and gluons. Sometimes the constituent quarks are also referred to as dressed quarks. \newline
Since QCD itself cannot be invoked for a description of the interactions among the constituent quarks from first principles, these interactions have to be modeled by some effective potentials. For the long-range part of the interactions, a linearly rising confinement-potential is assumed almost universally. Differences among models lie mostly within the phenomenological description of the moderate short-ranged interactions of QCD. \newpage

\begin{sidewaysfigure}[ht]
\centering
\vspace*{-10pt}
 \includegraphics[width=0.955\textwidth]{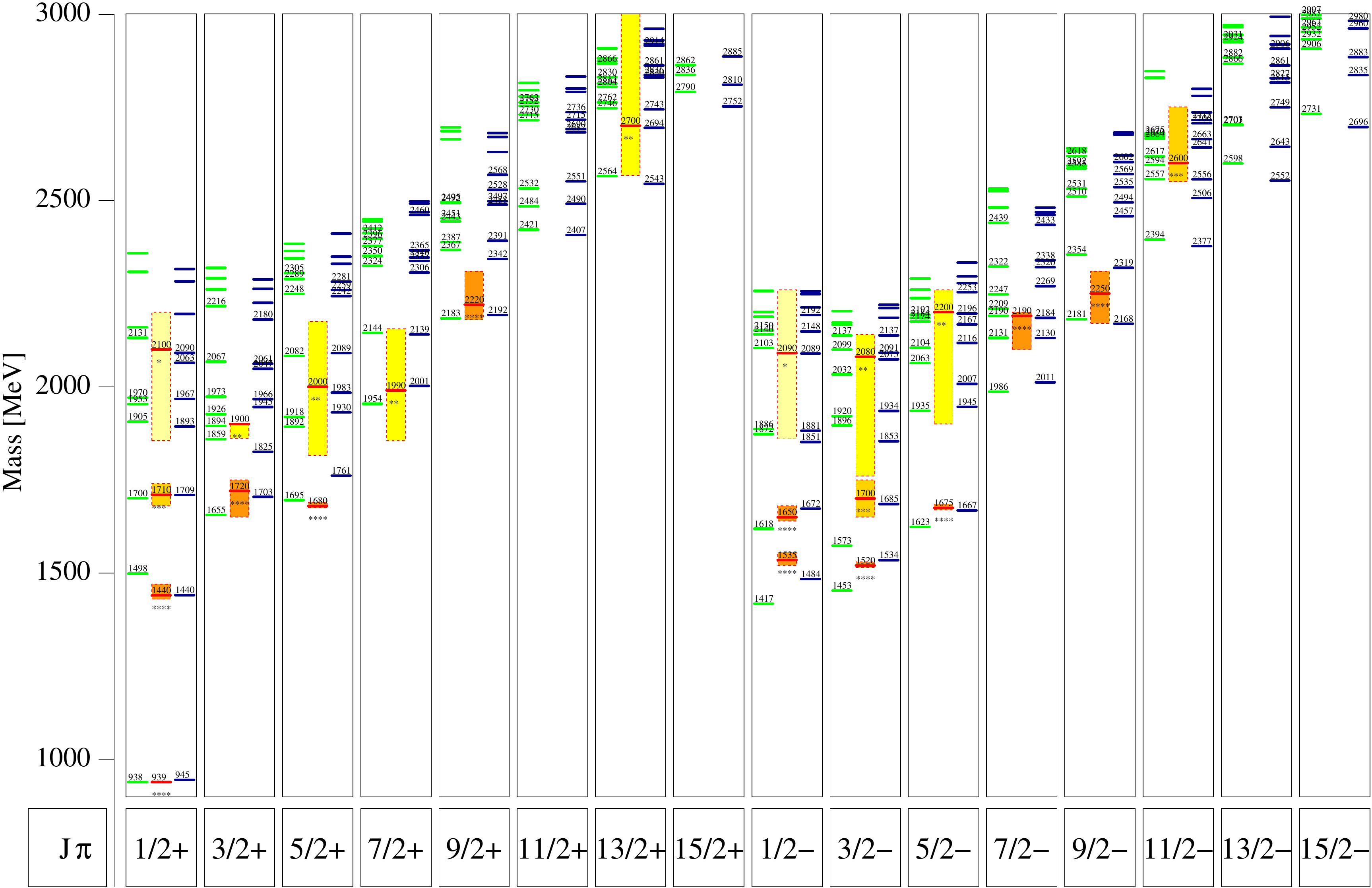} 
 \caption[Predictions for the nucleon excitation spectrum from the Bonn CQM.]{The spectra of nucleon-resonances (isospin $I = \frac{1}{2}$) calculated and published by Ronniger and Metsch \cite{Ronniger:2011td} are shown (Figure provided in high resolution with kind permission by B. Metsch.). For more explanations, see the main text.}
 \vspace*{-18pt}
\label{fig:NucleonSpectrum}
\end{sidewaysfigure}

\begin{sidewaysfigure}[ht]
\centering
\vspace*{-10pt}
 \includegraphics[width=0.955\textwidth]{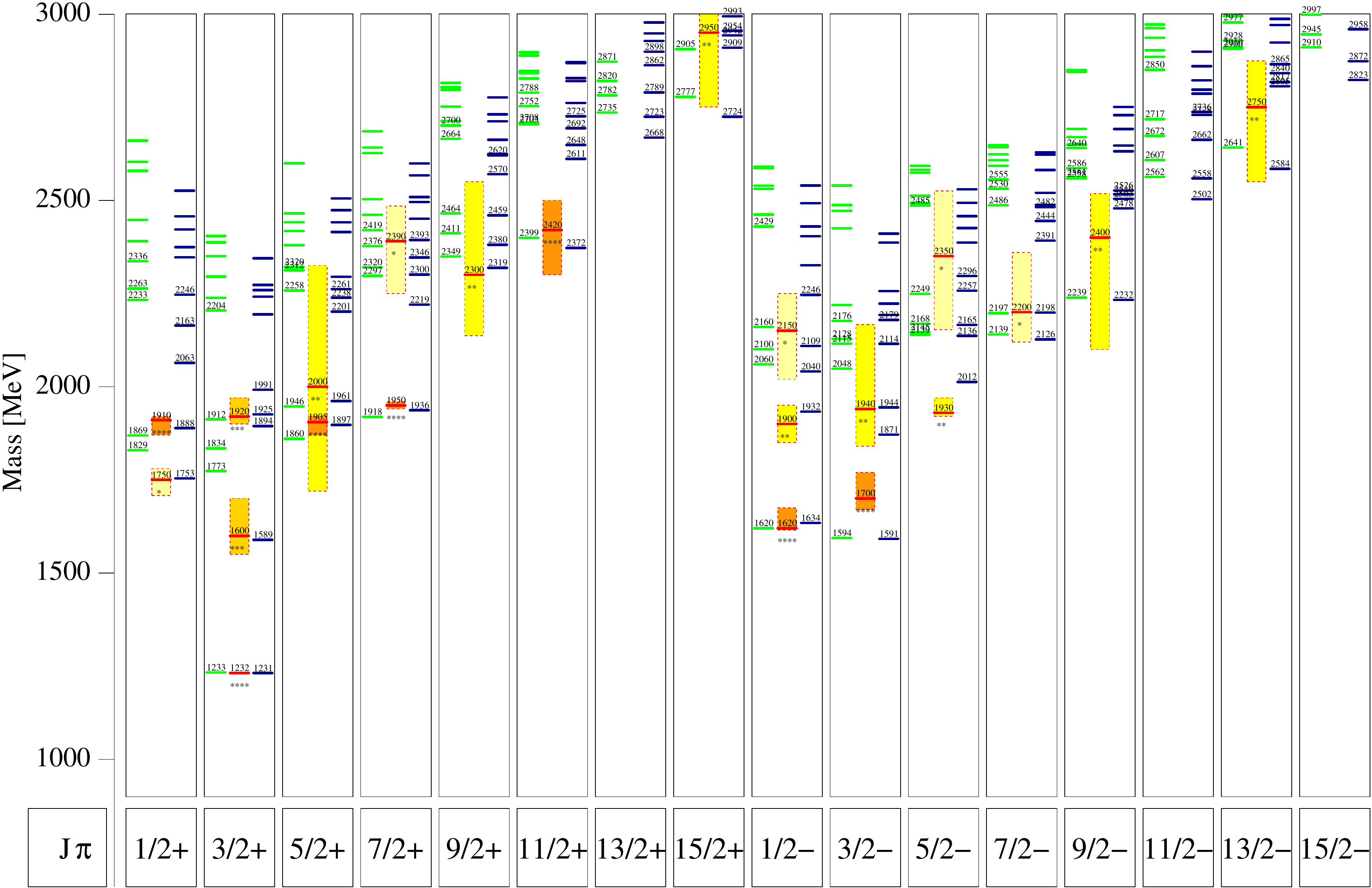}
 \caption[Predictions for the Delta excitation spectrum from the Bonn CQM.]{The spectra of Delta-resonances (isospin $I = \frac{3}{2}$) calculated and published by Ronniger and Metsch \cite{Ronniger:2011td} are shown (Figure provided in high resolution with kind permission by B. Metsch.). For more explanations, see the main text.}
 \vspace*{-18pt}
\label{fig:DeltaSpectrum}
\end{sidewaysfigure}

Once a form of interaction is fixed, some kind of either differential (i.e. Schr\"{o}dinger-type) or integral (Lippmann-Schwinger-/Bethe-Salpeter-type) equation has to be solved in order to yield the ground- and excited states of the whole spectrum as purely non-perturbative phenomena. At this point, considerable technical complications can arise and in most cases approximations have to be made. Also, solutions can be obtained practically only in a numerical way. \newline
As an illustration, we mention here the Bonn-model developed by L\"{o}ring, Kretzschmar, Metsch and Petry \cite{LoringEtAlCQM1} for the description of baryons. This model is formulated in a fully relativistically covariant formalism, using the Bethe-Salpeter equation for bound states of the three-quark system as a basis for all predictions. However, some simplifying approximations are then made \cite{LoringEtAlCQM1}. \newline
First results for the nucleon- and Delta-spectra were published in \cite{LoringEtAlCQM2}. Only up-, down- and strange quarks have been assumed as dynamical. In order to obtain these results, a three-body confinement-kernel has been fed into the Bethe-Salpeter equation with a local three-quark potential of the form $V_{\mathrm{conf.}}^{(3)} \left( \vec{x}_{1}, \vec{x}_{2}, \vec{x}_{3} \right) = 3 a \hat{\Gamma}_{o} + b \sum_{i<j} \left| \vec{x}_{i} - \vec{x}_{j} \right| \hat{\Gamma}_{s}$ \cite{LoringEtAlCQM2, Ronniger:2011td} in coordinate space. This potential is defined by two parameters, an offset $a$ and the slope $b$, the latter of which determines how strongly the potential rises linearly with the distance among the quarks. The operators $\hat{\Gamma}_{o}$ and $\hat{\Gamma}_{s}$ denote Dirac-structures for the offset- and slope part, which have been chosen phenomenologically based on the assumption that the confinement-forces are in a good approximation spin-independent. Two such choices have been employed in the original calculation \cite{LoringEtAlCQM2}, denoted as model $\mathcal{A}$ and $\mathcal{B}$, respectively. \newline
For the residual short-range QCD, a two-body potential was chosen based on 't Hooft's effective instanton-induced interaction \cite{tHooftInstantons1976}. This effective $2$-body interaction introduces three additional free parameters, two couplings and one effective range parameter which is introduced by the regularization of the potential. \newline
The original model \cite{LoringEtAlCQM2} has remarkably few free parameters. Aside from the five parameters introduced in the potentials, only the constituent quark masses $m_{n}$ and $m_{s}$ of the non-strange and strange quarks are input to the model. Fixing these parameters from the mass values and splittings of a few well-known resonances, L\"{o}ring {\it et al.} arrived at predictions for the whole remainder of the spectrum. \newline
The most recent results from the Bonn-model have been published by Ronniger and Metsch \cite{Ronniger:2011td}. In these calculations, the above mentioned model has been modified by an additional flavour-dependent interaction. This new interaction term has been parametrized purely phenomenologically, but it has been inspired by the known forms of the pseudoscalar and pseudovector coupling-terms for spin-$\frac{1}{2}$ fermions (quarks) to the flavour-nonet of pseudoscalar mesons (cf. Figure \ref{fig:WeightDiagramsMesonsBaryons}). The flavour-dependent potential introduces four additional free parameters, two couplings and two effective ranges, which would lead to a total of $11$ quantities to be determined. However, Ronniger and Metsch fixed the effective range of the 't Hooft force to the previous results \cite{LoringEtAlCQM2}, which makes the new calculation a $10$-parameter description of the baryon resonance spectra. \newline
In Figures \ref{fig:NucleonSpectrum} and \ref{fig:DeltaSpectrum}, results of model $\mathcal{A}$ from the original publication \cite{LoringEtAlCQM2} and the newer description involving flavour-dependent forces (model $\mathcal{C}$) of reference \cite{Ronniger:2011td} are compared to experimental values of baryon resonances from the $2010$ edition of the Review of particle physics \cite{Nakamura:2010zzi}. Nucleon resonances (isospin $I = \frac{1}{2}$) are shown in Figure \ref{fig:NucleonSpectrum}, while Figure \ref{fig:DeltaSpectrum} contains results on Delta-resonances (isospin $I = \frac{3}{2}$). Each column in the spectra belongs to specific spin-parity quantum numbers $J^{P}$.

\clearpage

\clearpage
The masses of the resonances are shown on the ordinate in units of MeV. Results from model $\mathcal{A}$ are plotted on the left side in each column, experimental results from the PDG \cite{Nakamura:2010zzi} are at the center and results from model $\mathcal{C}$ are on the right side of each column. Resonance masses are indicated by a line in each case. For the experimental values, shaded bars are drawn as well in order to represent the mass uncertainty. Furthermore, the PDG-rating \cite{Nakamura:2010zzi} is also indicated by stars. \newline
In the published spectra of Figures \ref{fig:NucleonSpectrum} and \ref{fig:DeltaSpectrum}, hints of the so-called {missing-resonance problem} can be seen. I.e., especially in the high mass region, many more states are generically predicted by quark models than have been measured until now\footnote{The experimental values in Figure \ref{fig:NucleonSpectrum} are not any more up-to-date. Still, the above mentioned problem remains essentially unresolved.}. This has been one of the main motivations to do photoproduction experiments after the shutdown of the Pion-beams. \newline
In addition, it should be mentioned that the additional assumption of a flavour-dependent interaction brought a significant improvement of the description of the excited negative-parity states in the Delta-spectrum (cf. Figure \ref{fig:DeltaSpectrum}) \cite{Ronniger:2011td}. \newline

\begin{figure}[ht]
\centering
\hspace*{20pt}
 \begin{overpic}[width=0.95\textwidth]%
      {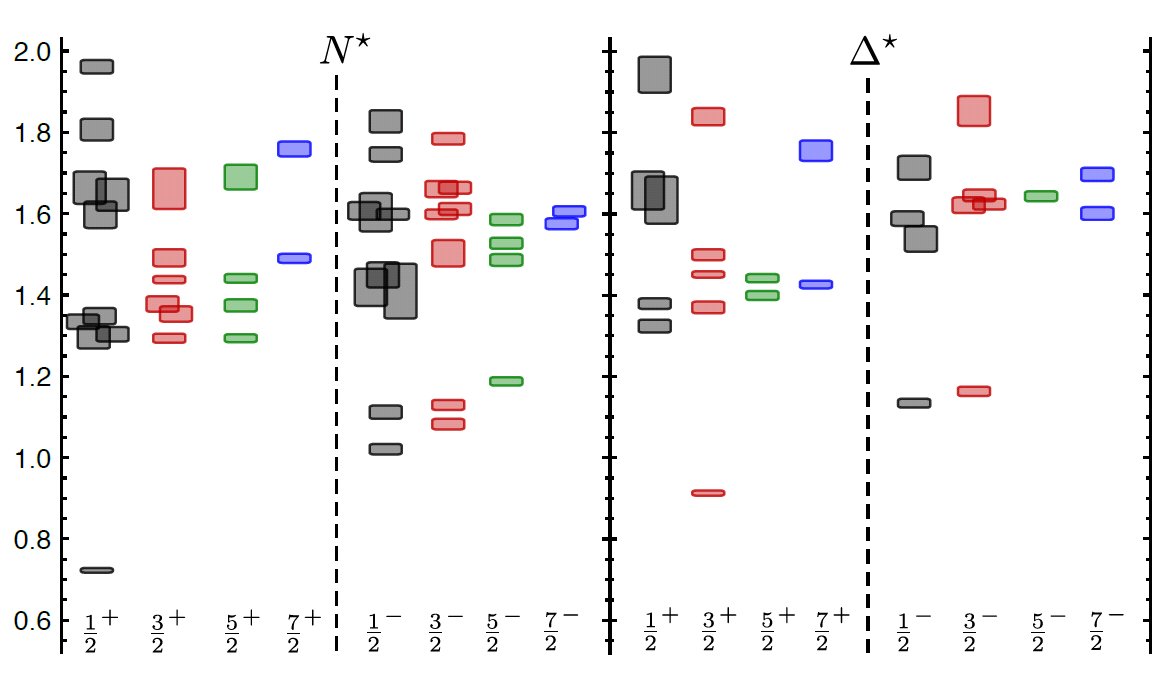}
  \put(51,-3){\begin{Large}$J^{P}$\end{Large}}
  \put(-4,28){\begin{Large}$\frac{m}{m_{\Omega}}$\end{Large}}
\end{overpic}
\vspace*{5pt}
 \caption[Predictions for the nucleon excitation spectrum from Lattice-QCD.]{The nucleon- ({\it Left}) and Delta-states ({\it Right}) from the Lattice-QCD calculation published by Edwards et al. \cite{EdwardsEtAl2011} are shown. Quantum numbers $J^{P}$ are given and masses are scaled in units of the mass of the $\Omega^{-}$-baryon calculated in the lattice-simulation (the PDG-value for this mass is $m_{\Omega} \simeq 1672$ $\mathrm{MeV}$ \cite{Patrignani:2016xqp}). The employed lattice corresponds here to an unphysical pion-mass of $m_{\pi}=396$ $\mathrm{MeV}$. The fact that energy-levels are only given as continuous bands of values is an artifact of the spin-identification procedure \cite{EdwardsEtAl2011}.}
\label{fig:LatticeNucleonSpectra}
\end{figure}

\textbf{Lattice QCD} \newline
This approach dates back to the original idea of lattice gauge-theory proposed first by Wilson \cite{Wilson1974}. A didactical account can be found in volume II of reference \cite{AitchisonHey}. In Lattice-QCD, the full theory of Quantum Chromodynamics is solved numerically on a discrete grid in euclidean spacetime\footnote{Euclidean spacetime is reached from Minkowski spacetime by the substitution $t \rightarrow \tau:= i t$, i.e. by going to imaginary time. In this way, the rapidly oscillating integrands of path-integrals become exponentially damped and the path-integrals themselves thus well-defined. After a calculation has been done, one can always analytically continue from euclidean back to Minkowski spacetime.}. Correlation-functions are set up for the physical problem under study and evaluated on the lattice using the Feynman path-integral. This represents a non-perturbative {\it ab initio} approach. \newline
Theoretically, in case arbitrarily large lattices were possible with arbitrarily fine lattice-spacing $a$, one would be able to solve QCD very close to the so-called {\it continuum-limit}. In practice lattices of, for instance, $16^{4}$ sites require already quite some calculational effort \cite{AitchisonHey}, such that in the most general cases supercomputers have to be employed. Thus, practical calculations are usually done away from the physical continuum-limit and subtle techniques have to be used, in the end, to extrapolate the results to the real world. \newline
Regarding the size of lattices, it should be said that they have to be large enough such that the objects under study, e.g. nucleons, which have themselves the spatial dimension $R$, fit well inside. Otherwise one would pick up so-called 'finite-size effects'. On the other hand, in case the lattice itself becomes too large, one would at some point get $R \simeq a$ and the discrete nature of the lattice would become apparent. The ideal situation would be, for an object of mass $m$ and a lattice of side-length $L$ with $N$ lattice-sites \cite{AitchisonHey}:
\begin{equation}
 a \ll R \sim \frac{1}{m} \ll L = N a \mathrm{.} \label{eq:IdealLatticeEstimate}
\end{equation}
This estimate also hints at the fact that finite non-zero lattice-spacings $a$ yield lower bounds on the masses of hadrons to be studied. A measure for this effect is often given by the pion-mass $m_{\pi}$. \newline
A first lattice-calculation of excited nucleon- and Delta-states has been published by Edwards and collaborators \cite{EdwardsEtAl2011} and the results are shown for an unphysical pion-mass of $m_{\pi}=396$ $\mathrm{MeV}$ in Figure \ref{fig:LatticeNucleonSpectra}. \clearpage
These results are not yet perfect. The pion-mass is still far from the physical value $m_{\pi} = 140$ $\mathrm{MeV}$ and also one should say that the resonances in this lattice-calculation are stable particles, i.e. not even the high-lying states can decay via the strong interactions. Furthermore, a comparison with analogous spectra for $m_{\pi}=524$ $\mathrm{MeV}$, which are shown in the same publication, shows still some modification of the respective energy-levels, depending on the pion-mass. \newline
But still, this result is the first of its kind and a comparison to the Bonn-model \cite{LoringEtAlCQM2}, Figures \ref{fig:NucleonSpectrum} and \ref{fig:DeltaSpectrum}, is meaningful. The resemblance between the lattice- and CQM-spectra is already quite striking. In particular for the low-lying states, for instance the whole negative-parity sector of the nucleon-spectra, patterns of levels look quite similar. However, it should be noted that the relative ordering of levels can be quite different between both methods of calculation. \newline

\textbf{Chiral effective field theory ($\chi$EFT)} \newline
This description of low-energy QCD is rooted in the Effective Field Theory (EFT)-techniques pioneered by Weinberg \cite{WeinbergEFT}. Here, only a brief outline can be given. More detailed accounts can be found in the literature \cite{DonoghueEtAlSM, Kubis2007}. Chiral EFT itself has been originated by Gasser and Leutwyler \cite{GasserLeutwyler1, GasserLeutwyler2}. \newline
The whole idea rests upon a particular symmetry-pattern in the QCD-lagrangian (\ref{eq:QCDLagrangian}). Suppose one has up-, down- and strange quarks. Left- and right-chiral projections of the quark fields are defined as $q_{L,R} := 1/2 \left( 1 \mp \gamma_{5} \right) q$, with the chirality-operator $\gamma_{5} = i \gamma^{0} \gamma^{1} \gamma^{2} \gamma^{3}$. The chiral projections are then organized into a flavor-vector of Dirac-fields $\Psi_{L,R} := \left( u_{L,R}, d_{L,R}, s_{L,R} \right)^{T}$. \newline
A chiral symmetry transformation now consists of an independent global $SU(3)$-rotation for the left- and right-handed components: $\Psi_{L,R} \longrightarrow \exp \left[ i \alpha^{a}_{L,R} \bm{\lambda}^{a} \right] \Psi_{L,R}$. It can be shown furthermore that the direct product of left- and right-chiral rotations is isomorphic to the product of so-called {\it vector- and axial}\footnote{A vector-transformation transforms both left- and right-handed components by the same phase. An axial-transformation transforms both components by the exact opposite phase.} $SU(3)$-transformations $SU(3)_{L} \times SU(3)_{R} \cong SU(3)_{V} \times SU(3)_{A}$. It is now crucial to observe that the kinetic energy terms of the quarks in the QCD-lagrangian, i.e.
\begin{equation}
 \bar{\Psi} \slashed{D} \Psi = \bar{\Psi}_{L} \slashed{D} \Psi_{L} + \bar{\Psi}_{R} \slashed{D} \Psi_{R} \mathrm{,} \label{eq:KineticEnergyTerms}
\end{equation}
are chirally invariant, while the mass-terms are not. This is the case, since a standard Dirac mass-term mixes the left- and right-handed field-components:
\begin{equation}
  \bar{\Psi} \bm{m} \Psi = \bar{\Psi}_{L} \bm{m} \Psi_{R} + \bar{\Psi}_{R} \bm{m} \Psi_{L} \mathrm{,} \label{eq:DiracMassTerm}
\end{equation}
with $\bm{m}$ the quark mass-matrix. \newline
Even in the case of vanishing quark-masses, the accepted picture is that chiral symmetry has to be broken spontaneously \cite{DonoghueEtAlSM, AitchisonHey}. This breaking is measured by the so-called quark-condensate $\left< 0 \right| \bar{q} q \left| 0 \right>$. In group-theoretical language, the whole group of chiral symmetry transformations is broken to the subgroup of pure vector rotations $SU(3)_{L} \times SU(3)_{R} \longrightarrow SU(3)_{V}$. Thus, the subgroup of axial $SU(3)$-rotations is not a symmetry any-more and there has to exist one massless scalar Goldstone-boson for each broken symmetry-generator. For the eight $SU(3)_{\bm{f}}$-generators, the eight ground-state pseudoscalar mesons (cf. Figure \ref{fig:WeightDiagramsMesonsBaryons}) neatly fit into the picture as Goldstone-bosons. However, chiral symmetry is broken explicitly in QCD as well, via the mass-term (\ref{eq:DiracMassTerm}), such that the pseudoscalar Goldstone bosons become massive \cite{DonoghueEtAlSM, Kubis2007}. \newline
Under guidance of the above mentioned considerations, so-called {\it chiral effective lagrangians} (or just {\it chiral lagrangians}) can be constructed. In case of pure Goldstone-boson dy\-na\-mics \cite{Kubis2007}, the fundamental field-variable used to construct such lagrangians is the unitary matrix $U(x) := \exp \left[ i \phi^{a} (x) \bm{\lambda}^{a} / F_{\pi} \right]$, with Goldstone-boson fields $\phi^{a}$, Gell-Mann matrices $\bm{\lambda}^{a}$ and Pion decay-constant $F_{\pi}$. Lagrangians constructed from this matrix are then built to be chirally invariant. Thus, the Golstone-bosons have replaced the quarks as dynamical degrees of freedom in the lagrangian, but the symmetries of QCD are kept intact. \newline
The chiral lagrangian contains then an infinite amount of appropriately contructed kinetic energy- and mass terms, which however can be ordered in the powers of momenta, or derivatives, and powers of the mass matrix which are present. Given this chiral ordering-scheme of terms, each operator is then accompanied by a so-called {\it Low-Energy Constant} (LEC), such that the number of LECs grows exponentially with the chiral term-order in the lagrangian. The LECs have to be determined either from a fit to data or from Lattice-QCD, for instance. Once this is done, one obtains a predictive and consistent theory of Goldstone-boson dynamics. \newline
Chiral EFT can be generalized to describe interactions among pions and nucleons \cite{BernardEtAl1995}, or even to processes like photoproduction \cite{BernardEtAl1991}. Furthermore, the general theory of nuclear forces can be rooted more deeply in the Standard Model using chiral EFT methods \cite{EpelbaumModernTheory}. Chiral effective field theory can even be applied to the calculation of individual isolated Baryon resonances, see for instance \cite{BrunsEtAl2010}. One can also introduce resonances as explicit higher-spin degrees of freedom, as is done e.g. for the $\Delta$-resonance in a recent PhD-thesis studying pion-induced reactions \cite{SiemensPhD}. \newline
Chiral dynamics have blossomed into a huge research-field, due to the fact that the quality of chiral approximations to hadronic amplitudes in the low-energy regime is generally excellent. However, the evaluation of a whole excitation spectrum, such as that of the nucleon, all at once is not practically feasible.


\subsubsection{Energy-dependent models for the extraction of baryon-resonances} \label{subsec:Energy-dependentModels}

In order to extract information on resonances from scattering-data, it is necessary to pa\-ra\-met\-rize the amplitude as a function of energy in a suitable way and then fit this parametrization to the data. The most important aspects of this method shall be outline in the following. Analyses of this kind fall under the umbrella term of {\it energy-dependent} (ED) fits. \newline
In the beginning of section \ref{sec:BaryonSpectroscopy}, the hierarchy of hadronic reactions suitable for the study of baryon resonances has been mentioned. Since a lot of strongly interacting particles are present in the particle-zoo \cite{Patrignani:2016xqp}, reactions are manifold and their kinematic thresholds are usually not very far separated in energy. Generically, a few thresholds exist over typical {\it resonance-regions} in energy. Moreover, the same resonance is usually allowed to couple to a multitude of channels. Therefore, in almost all cases a so-called {\it coupled-channels} approach is used in
ED fits. \newline
The principles and methods of analytic {\it S-Matrix theory}\footnote{S-Matrix theory has originally been proposed in the $50$s and $60$s as a candidate for a fundamental theory of the strong interactions. The advent of QCD has prevented that purpose, but the knowledge is still in frequent use, for instance in the energy-dependent reaction-analyses.} \cite{PolkinghorneEtAl} are often employed in such energy-dependent analyses. The S-matrix, or {\it scattering matrix} is an abstract operator in the Hilbert-space of asymptotic scattering-states, which defines the probability for transitioning from an initial state $\left| a \right>$ to a final state $\left| b \right>$ by its matrix-element\footnote{More precisely: the probability is given by the modulus-squared of the matrix element.} $\left< b \right| \hat{S} \left| a \right>$. This notation for state-vectors is highly condensed. We should mention that each vector carries discrete quantum numbers (spin, isospin, ...) and continuous quantum numbers, i.e. $4$-momenta, necessary to specify the state. Moreover, it should be mentioned explicitly that the S-Matrix can act in channel-space.\newline
Since the S-matrix contains interactions as well as processes with no interactions at all, it is customary to split-off the non-interactions additively, which can be done in terms of abstract operators as follows\footnote{Factors of '$2$' and '$i$' are conventional in this equation.} \cite{PeskinSchroeder,PolkinghorneEtAl,KlemptKMatrix}
\begin{equation}
 \hat{S} = \mathbbm{1} + 2 i \hat{T} \mathrm{.} \label{eq:SAndTMatrix}
\end{equation}
This introduces another operator, the so-called {\it T-matrix} (or {\it transition operator}) $\hat{T}$, which is the main object of interest. Matrix-elements of the $\hat{T}$-operator are then the amplitudes to be modeled. S-matrix theory sets the requirements \cite{PolkinghorneEtAl} of {\it Analyticity}, {\it Unitarity} and {\it Crossing}, about which we now provide some more detail. \newline

\begin{figure}[ht]
\centering
 \hspace*{10pt}  \includegraphics[width=0.65\textwidth]{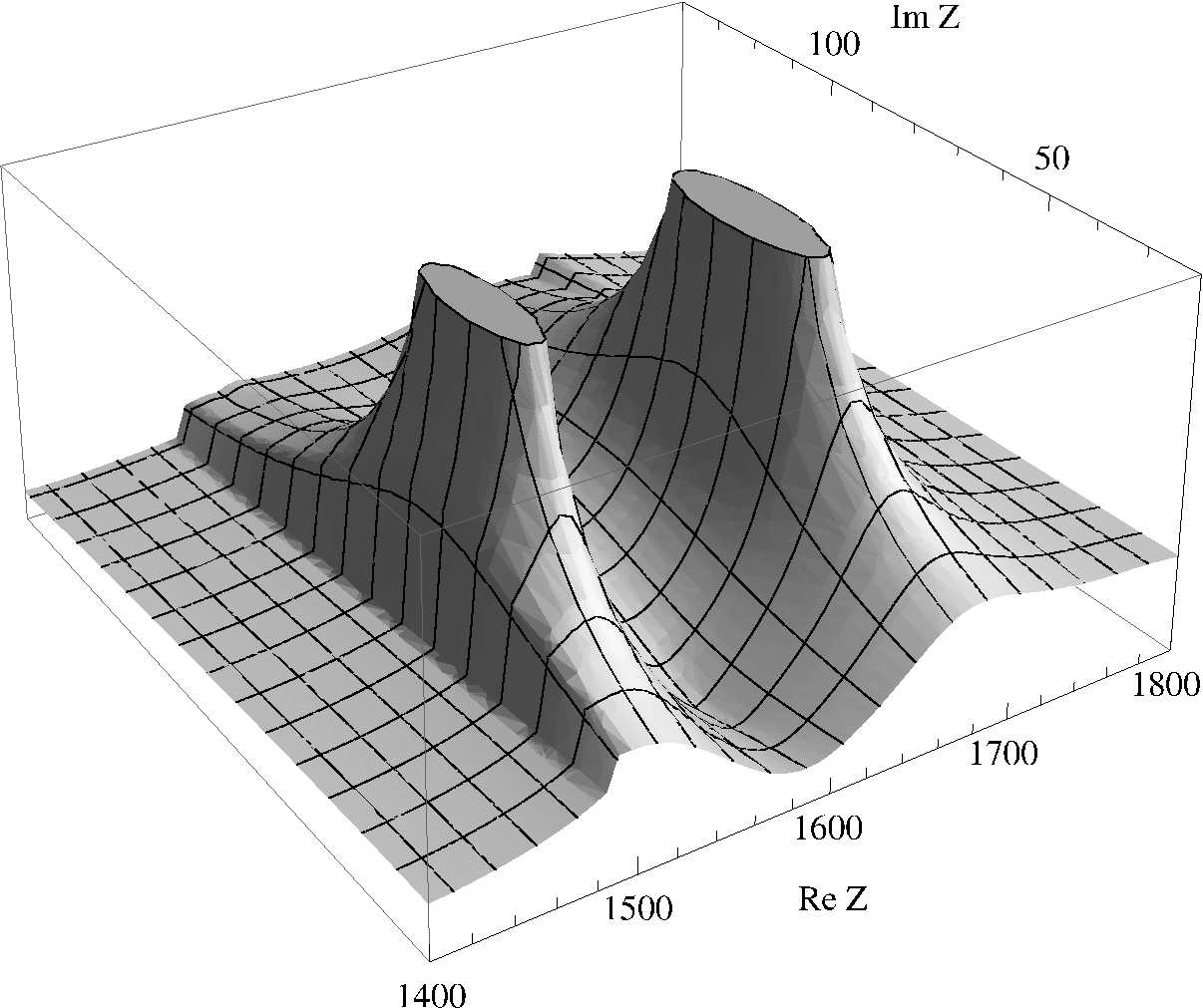}
\vspace*{10pt} 
 \caption[An illustration showing resonance-poles contributing to a particular partial wave from the J\"{u}lich-Bonn dynamical coupled-channels model.]{Shown here is the modulus $\left| T \right|$ of the partial wave $S_{11}$ (i.e. $\ell = 0$, $I = \frac{1}{2}$ and $J=\frac{1}{2}$) of the J\"{u}lich meson exchange model, as a function of complex energy $Z$ $\left[\mathrm{MeV}\right]$. The figure is taken over identically from reference \cite{MichaelAnalyticProps}. \newline
 The resonance-poles of the states $N^{\ast} (1535)$ and $N^{\ast} (1650)$ can be seen. Below the $N^{\ast} (1535)$-pole, the discontinuity belonging to the $\eta N$-cut is visible as a step. Behind the $N^{\ast} (1650)$, one can still see the $\rho N$-cut which has a complex branch-point.}
 \label{fig:DoeringEtAlPoles}
\end{figure}

\textbf{Analyticity}\newline
The amplitudes $\mathcal{T}_{ba} := \left< b \right| \hat{T} \left| a \right>$, describing the transition of an asymptotic 'in' state $\left| a \right>$ into an asymptotic 'out' state $\left| b \right>$, are functions of external Lorentz-invariants for the given process. For a $2\rightarrow 2$ reaction with all particles on the mass-shell, these would be just the Mandelstam-variables\footnote{For initial $4$-momenta $p_{1}$ and $p_{2}$, final $4$-momenta $p_{3}$ and $p_{4}$, one has: $s := \left( p_{1} + p_{2} \right)^{2}$, $t := \left( p_{1} - p_{3} \right)^{2}$, $u := \left( p_{1} - p_{4} \right)^{2}$. The variable $s$ is related to the total center-of-mass energy $W$ via: $W = \sqrt{s}$.} $s$ and $t$. \newline
The physical amplitudes, i.e. those defined for real physical values of the external invariants, have to be boundary values of multivariate {\it analytic functions} (tantamount to {\it complex dif\-fe\-ren\-ti\-ab\-le functions}) \cite{BehnkeSommer, CartanComplexAnalysis}, of the invariants. Again for the simplest example-case of a 2-body reaction, Mandelstam formulated this analyticity requirement as a double-dispersion relation \cite{Mandelstam1958DispRel}. In practice however, one typically encounters discussions of $2$-body channels with the variable $t$ held fixed. Then, the amplitude is an ordinary single-variable analytic function on the complex energy-plane. \newline
To be more precise, amplitudes are not required to be purely analytic functions, but rather {\it meromorphic} \cite{BehnkeSommer} one’s. This means that isolated singularities, in this case poles, are allowed to exist \cite{PolkinghorneEtAl}. Even extended singularities, i.e. branch-points and cuts, are not excluded. Moreover, physics dictates where and of which kind these singularities have to be. First-order poles are either stable bound states or resonances. It is now common ground that {\it only poles} are model-independent signals for resonances \cite{Patrignani:2016xqp}. Thus, pole-parameters (positions and residues) should be extracted once a model has been fitted. \newline
Branch points on the other hand correspond, roughly speaking, to thresholds. Thus, the analytic amplitudes are necessarily multivalued functions defined on multiple {\it Riemann sheets} \cite{BehnkeSommer}. Keeping the overview of the sheet-structure can become quite formidable for more complicated models (cf. comments made below on the J\"{u}lich-model). \newline
The property of analyticity is known to be linked to the fundamental principle of {\it microcausality}. This has been shown for specific examples \cite{GellMannEtAlAnalyticity}, but a general proof of the connection is lacking \cite{PolkinghorneEtAl}. \newline
As a simple example for an analytic amplitude, we quote here the relativistic {\it Breit-Wigner formula}, describing an isolated resonance in the $S$-wave\footnote{I.e. a partial wave for definite relative angular momentum $\ell = 0$. More on partial waves shall be elaborated below and in section \ref{sec:Photoproduction}.} of a $2$-body reaction of scalar particles with equal mass $m$ \cite{Aitchison2015}:
\begin{equation}
 \mathcal{T}_{0} (s) = \frac{g^{2}}{s_{\mathrm{R}} - s - i \rho (s) g^{2}} \mathrm{.} \label{eq:RelBreitWigner}
\end{equation}
Here, $g$ is the coupling of the resonance to the initial and final state, while $\rho(s)$ is the $2$-body phase-space factor, which for equal-mass particles reads $\rho(s) = \sqrt{1 - \frac{4 m^{2}}{s}}$ and $s_{\mathrm{R}}$ is the real CMS energy-squared of the resonance. In particular, this amplitude contains no further contributions from any non-resonant backgrounds. \newline
A lot of complications are absent from the simple Breit-Wigner formula (\ref{eq:RelBreitWigner}). Apart from the above-mentioned coupled-channels effects, in the individual channels one generally has to model the amplitude for a (possible) dense population of multiple resonances and also including the effects of non-resonant background-processes. These complications lead to the necessity for more intricate models such as those mentioned here. An example for the contribution of poles and cuts to a partial wave in a realistic model is shown in Figure \ref{fig:DoeringEtAlPoles}. \newline

\textbf{Unitarity} \newline
In the standard probabilistic interpretation of quantum mechanics, the conservation of probabilities is a meaningful and universally assumed constraint. Fixing a given initial state $\left| i \right>$, \clearpage
then the total probability for transitioning from this particular initial state to any allowed\footnote{I.e. allowed by kinematics and, if present, any further imposed internal symmetries.} final state {\it has to be} $1$. It can be shown \cite{PolkinghorneEtAl} that for the abstract S-Matrix operator $\hat{S}$, this requirement is equivalent to the unitarity of this operator: $\hat{S}^{\dagger} \hat{S} \equiv \mathbbm{1}$. \newline
It is furthermore interesting which kinds of constraints this simple unitarity equation implies for the transition operator $\hat{T}$. Inserting the general parametrization (\ref{eq:SAndTMatrix}) into the unitarity-relation for $\hat{S}$, one obtains the so-called {\it optical theorem} in operator-notation \cite{PolkinghorneEtAl, PeskinSchroeder}:
\begin{equation}
 \frac{1}{2i} \left( \hat{T} - \hat{T}^{\dagger} \right) = \hat{T}^{\dagger} \hat{T} \mathrm{.} \label{eq:UnitarityTMatrix} 
\end{equation}
%
Taking matrix-elements of this equation between states $\left| a \right>$ and $\left| b \right>$ yields the following result
\begin{equation}
 \frac{1}{2i} \left( \left< b \right| \hat{T} \left| a \right> - \left< b \right| \hat{T}^{\dagger} \left| a \right> \right) = \frac{1}{2i} \left( \left< b \right| \hat{T} \left| a \right> - \left< a \right| \hat{T} \left| b \right>^{\ast} \right) \equiv \left< b \right| \hat{T}^{\dagger} \hat{T} \left| a \right> \mathrm{,} \label{eq:UnitarityTMatrixStep2}
\end{equation}
for the process $a \rightarrow b$. In order to arrive at the desired unitarity-equation, a resolution of the identity\footnote{The sum over $n$ in this equation encompasses the sum over all kinematically allowed intermediate states, as well as all discrete and continuum quantum numbers. Thus, unitarity equations for full amplitudes such as (\ref{eq:UnitarityTMatrixStep3}) are generally integral-equations and thus a lot more complicated than the simple notation chosen here may suggest.} $\mathbbm{1} \equiv \sum_{n} \left| n \right> \left< n \right|$ has to be inserted into the operator-product on the right-hand-side. The sum is here restriced to states $\left| n \right>$ which are allowed by kinematics (as well as, if present, further internal symmetries) as intermediate states. This means that the further one ascends in the total CMS-energy, the more terms occur in this sum. \newline
The left-hand side of equation (\ref{eq:UnitarityTMatrixStep2}) would, for the simple case of a $2 \rightarrow 2$-reaction with scalar particles and time-reversal invariant interactions, reduce to the imaginary part $\mathrm{Im} \left[ \mathcal{T}_{ba}  \right]$. In more general cases, this is no longer true. However, what remains true is that this left-hand-side will in the end define a {\it discontinuity} of the amplitude across an appropriately defined point-region in the space of external invariants. Thus one can write symbolically
\begin{equation}
 \mathrm{Disc} \left[ \mathcal{T}_{ba} \right] = \sum_{n} \left< b \right| \hat{T}^{\dagger} \left| n \right> \left< n \right| \hat{T} \left| a \right> = \sum_{n} \mathcal{T}_{nb}^{\ast} \mathcal{T}_{na} \mathrm{.} \label{eq:UnitarityTMatrixStep3}
\end{equation}
It is seen that the branching behavior of the amplitude is essentially implied by unitarity-equations. In particular, if the energy is raised, additional terms are allowed on the right-hand-side of equation (\ref{eq:UnitarityTMatrixStep3}), leading to additional singularities. This is one particular example of physics implying singularities of the otherwise analytic amplitudes. \newline
Keeping unitarity-constraints intact in parametrizations of amplitudes is a very important, but not an easy task. For instance, while one single Breit-Wigner amplitude (\ref{eq:RelBreitWigner}) is unitary, the sum of two such amplitudes is not \cite{Aitchison2015}. Here, it can be of use to introduce the so-called {\it K-matrix} \cite{KlemptKMatrix}. \newline
The most important first step in the direction of the K-Matrix is observing that the operator-relation (\ref{eq:UnitarityTMatrix}), which is valid under unitarity, can be rewritten as
\begin{equation}
 \left( \hat{T}^{-1} + i \mathbbm{1} \right)^{\dagger} = \hat{T}^{-1} + i \mathbbm{1} \mathrm{.} \label{eq:OptTheoremRewritten}
\end{equation}
This expression motivates the definition of the K-matrix operator\footnote{More precisely, its inverse.}, which is \cite{KlemptKMatrix}
\begin{equation}
 \hat{K}^{-1} := \hat{T}^{-1} + i \mathbbm{1} \mathrm{,} \label{eq:KMatrixDef}
\end{equation}
and which may be inverted, for the T-Matrix, via \cite{KlemptKMatrix}:
\begin{equation}
 \hat{T} = \hat{K} \left( \mathbbm{1} - i \hat{K} \right)^{-1} \equiv \left( \mathbbm{1} - i \hat{K} \right)^{-1} \hat{K} \mathrm{.} \label{eq:TMatrixInTermsOfKMatrix}
\end{equation}
Furthermore, equation (\ref{eq:OptTheoremRewritten}) is equivalent to the statement that for a unitary T-Matrix, the K-operator is hermitean
\begin{equation}
 \hat{K}^{\dagger} = \hat{K} \mathrm{.} \label{eq:KMatrixHermitean}
\end{equation}
The argument can be reversed in the sense that it is possible to construct a reaction-theoretic model by first specifying the functional form of the K-Matrix in a suitable basis. Then, in case the postulated K-Matrix is hermitean, one automatically obtains unitary S- and T-Matrices. However, this tells nothing about how this specific K-Matrix should be parametrized and a lot of freedom still exists. \newline
In case the schematic operator-treatment given here is to be turned into a proper Lorentz-invariant amplitude-theory, more care has to be taken, in particular with appearing factors of phase-space functions. We make no attempt to do this here, but instead refer to \cite{KlemptKMatrix}. In this case, as always, the devil is in the detail. \newline
\clearpage

\textbf{Crossing} \newline
A property of scattering amplitudes deduced from the Feynman-rules in perturbation theory \cite{PeskinSchroeder} is that the {\it same} analytic funtions have to describe different physical processes, which are related to each other by so-called {\it crossing}-tranformations \cite{PolkinghorneEtAl}. Most generally, a crossing transformation moves one particle from the initial/final state to the final/initial state, conjugates it into its own antiparticle and replaces the particle's $4$-momentum $p$ by $-p$. In abbreviated notation \cite{PeskinSchroeder}, where '$S$' is just a matrix-element of the S-Matrix, symmetry under crossing of the particle $\phi(p)$ states that\footnote{In this brief discussion of crossing, $\phi(p)$ stands for any kind of particle.}:
\begin{equation}
 S \left( \phi(p) + \ldots \rightarrow \ldots \right) \equiv S \left( \ldots \rightarrow \ldots + \bar{\phi} (-p) \right) \mathrm{.} \label{eq:CrossingTrafoSMatrix}
\end{equation}
In most discussions \cite{PolkinghorneEtAl}, one encounters $2 \rightarrow 2$-reactions related by crossing. For instance, in case the direct production channel, or $s$-channel reads $\phi_{1} + \phi_{2} \longrightarrow \phi_{3} + \phi_{4}$, crossing-transformations lead to the so-called $t$-channel $\phi_{1} + \bar{\phi}_{3} \longrightarrow \bar{\phi}_{2} + \phi_{4}$ and $u$-channel $\phi_{1} + \bar{\phi}_{4} \longrightarrow \bar{\phi}_{2} + \phi_{3}$. As a concrete physical example, the processes $\pi^{+} p \longrightarrow \pi^{+} p$, $\pi^{+}  \pi^{-} \longrightarrow p \bar{p}$ and $\pi^{+} \bar{p} \longrightarrow \pi^{+} \bar{p}$ are crossing-related. It is however also perfectly feasible to relate $2$-body scattering processes to decays via crossing-transformations. \newline

It is clear that the models currently at the market fulfill the above-mentioned principles with a varying degree of rigor. Approximations have to be made and the exact fulfillment of analyticity, unitarity and crossing is an idealization which in most cases cannot be fully satisfied. The question is then which approximations can make the models tractable, while not distorting the physics in a too crude way. \newline
We outline four of the most prominent models in the following. We also refer to reference \cite{CommonPaper}, from which we will draw heavily and which contains a more complete and illustrative overview. Also, from now on we resort to the notations used by the respective authors in the original publications. \newline

\textbf{The Bonn-Gatchina model} \newline
The Bonn-Gatchina partial wave analysis represents a flexible analysis-tool for the simultaneous analysis of many different channels. At the time of this writing, the model is already roughly $15$ years in the making. The fundamentals of the approach have been published in many works over the years, see for instance \cite{AnisovichEtAlPWABook,Anisovich:2011fc}. The results can be viewed by anyone on the web-page \cite{BoGa}. In its most recent incarnation, this model uses a so-called '$N/D$-inspired approach', about which a good review can be found in \cite{CommonPaper}. We rely here mainly on this review. \newline
The model is a coupled-channel K-Matrix analysis with a multitude of channels. The transition-amplitude is written as\footnote{The Bonn-Gatchina group practically always uses the letter '$A$' for the T-Matrix.} $\hat{\bm{A}}(s)$ with matrix-elements $A_{ab}(s)$ for channels $a$ and $b$. The channel-indices run over $2$-body systems such as $\pi N$, $\eta N$, $K \Lambda$, $\pi \Delta$ and the electromagnetic channel $\gamma N$. However, also $3$-body states such as $\pi \pi N$ and $\pi \eta N$ are included into the analysis. \newline
The BnGa-group uses the so-called covariant tensor-formalism \cite{AnisovichEtAlPWABook} in order to define the partial-wave decompositions of their amplitudes. Generally, partial waves depend on the set of good quantum numbers made up of the total angular momentum $J$, parity $P$ and isospin $I$. However, the dependence on $J^{P} \hspace*{0.5pt} I$ is only implicit in the following formulas, such that the discussion becomes more digestible. In the chosen parametrization Ansatz, the transition-amplitude now reads in matrix-form \cite{CommonPaper}
\begin{equation}
 \hat{\bm{A}}(s) = \hat{\bm{K}} \left( \hat{\mathbbm{1}} - \hat{\bm{B}} \hat{\bm{K}} \right)^{-1} \mathrm{.} \label{eq:BnGaTransitionAmplitude}
\end{equation}
The diagonal matrix $\hat{\bm{B}}$ contains rescattering loop-diagrams. The imaginary part of each loop is generally given by the phase-space, i.e. $B_{j} = \mathrm{Re} \left[B_{j}\right] + i \rho_{j}$. Dropping the real parts would correspond to the usual 'K-Matrix approximation'. However, the BnGa-group has updated the model in such a way as to include more of such loops in the form of once subtracted dispersion-intergals, which can be inspected in reference \cite{CommonPaper}. \newline
At the heart of the analysis is a 'pole$+$background'-parametrization for the K-Matrix, i.e.
\begin{equation}
 K_{ab} (s) := \sum_{\alpha} \frac{g_{a}^{(\alpha)} g_{b}^{(\alpha)}}{M_{\alpha}^{2} - s} + f_{ab} (s) \mathrm{.} \label{eq:BonnGatchinaKMatrix}
\end{equation}
Here, $M_{\alpha}$ and $g_{a}^{(\alpha)}$ are the mass of the resonance '$\alpha$' and the coupling of the respective resonance to the channel '$a$'. The functions $f_{ab}(s)$ describe non-resonant transitions $a \rightarrow b$ and in most partial waves it is sufficient to assume them as constants. Partial waves where this is not the case are $S_{11}$ (i.e.: $\ell = 0$, $J=1/2$, $I=1/2$) and $S_{31}$ (i.e.: $\ell = 0$, $J=3/2$, $I=1/2$), where the Bonn-Gatchina group employs the functional form
\begin{equation}
 f_{ab} (s) = \frac{f^{(1)}_{ab} + f^{(2)}_{ab} \sqrt{s}}{s - s_{0}^{ab}} \mathrm{,} \label{eq:fabBackgroundSpecialCase}
\end{equation}
with $f_{ab}^{(i)}$ and $s_{0}^{ab}$ constants to be fitted to the data. \newline
Since the $\gamma N$-interaction has a relatively weakly coupled compared to the purely hadronic channels, it is sufficient to approximate the photoproduction-amplitudes in a so-called $P$-vector ('production-vector') approach. The photoproduction-amplitude for the process $a \rightarrow b$ is then \cite{CommonPaper}
\begin{equation}
 A_{a} = \hat{P}_{b} \left( \hat{\mathbbm{1}} - \hat{B} \hat{K} \right)^{-1}_{ba} \mathrm{.} \label{eq:AmplitudeInPVectorApproach}
\end{equation}
The parametrization of the $P$-vector looks quite similar to equation (\ref{eq:BonnGatchinaKMatrix}). It reads
\begin{equation} 
  P_{b} (s) := \sum_{\alpha} \frac{g_{\gamma N}^{(\alpha)} \hspace*{1pt} g_{b}^{(\alpha)}}{M_{\alpha}^{2} - s} + \tilde{f}_{\left( \gamma N \right)b} (s) \mathrm{.} \label{eq:BonnGatchinaPVector}
\end{equation}
Here, $g_{\gamma N}^{(\alpha)}$ are the photo-couplings of the resonances and $\tilde{f}_{\left( \gamma N \right)b} (s)$ are functions describing non-resonant transitions. In practice, the latter are again assumed to be constants. \newline
In case of decay-modes which are weakly coupled, like for instance the $3$-body final states, the BnGa-group employs a so-called $D$-vector approximation. For processes where both production and decay are weakly coupled, they use a $PD$-vector approach. More details on both approaches can be found in \cite{CommonPaper} and references therein. \newline
In order to describe the forward-peaks in the data present at high energies and which originate from $t$-channel exchanges, currently the BnGa-group uses so-called {\it reggeized} amplitudes. The invariant amplitude describing the $t$-channel exchange of particles that lie on a {\it Regge-trajectory} \cite{PolkinghorneEtAl} $\alpha (t)$ is then given as \cite{CommonPaper, AnisovichEtAlPWABook}
\begin{equation}
 A(s,t,u) = g(t) \hspace*{1pt} \frac{1 + \xi \exp \left[ - i \pi \alpha (t) \right]}{\sin \left[ \pi \alpha(t) \right]} \hspace*{1pt} \left( \frac{\nu}{\nu_{0}} \right)^{\alpha(t)} \mathrm{.} \label{eq:BnGaReggionAmplitude}
\end{equation}
In this expression, $\nu = (s - u)/2$ is the so-called crossing-variable, $\nu_{0}$ a normalization factor and $\xi$ the {\it signature} of the trajectory. The function $g(t) = c \exp (- b t)$ is both a vertex-function and a form-factor. In order to remove unwanted poles for negative $t$, additional factors of $\Gamma$-functions have to be introduced. For more details on the Regge-amplitudes, see \cite{AnisovichEtAlPWABook,CommonPaper}. \newline
The Bonn-Gatchina partial wave analysis code has a reputation for being well-optimized and thus for allowing fits of the above-described flexible reaction-theoretic model in short runtime. Therefore, a lot of different scenarios for resonances can be tested and compared. The analysis has, at the time of this writing, almost benchmark-status and a lot of the newly-included resonances in the baryon summary-Tables of the PDG \cite{Patrignani:2016xqp} stem from results worked out by the Bonn-Gatchina group. \newline

\textbf{The J\"{u}lich-Bonn model} \newline
The J\"{u}lich-Bonn model is a dynamical coupled-channels (DCC) approach, capable of performing a combined analysis of pion-induced
\cite{DeborahPion} and photon-induced \cite{DeborahPhotoProd, DeborahEtaPhotoProd} reactions. \newline
At the heart of this analysis lie recursive scattering-equations, in this case Lippmann-Schwinger equations. For the Pion-induced reactions, the respective equation for the T-Matrix reads in the partial-wave basis \cite{CommonPaper}:
\begin{equation}
 T_{\mu \nu} (q,p^{\prime};E) = V_{\mu \nu} (q,p^{\prime};E) + \sum_{\kappa} \int_{0}^{\infty} dp \hspace*{2pt} p^{2} \hspace*{2pt} V_{\mu \kappa} (q,p;W) G_{\kappa} (p;W) T_{\kappa \nu} (p,p^{\prime};W) \mathrm{.} \label{eq:JuelichModelPiNLippmannSchwinger}
\end{equation}
The greek indices $\mu$, $\nu$ and $\kappa$ denote here the initial, final and intermediate channels. They run over the explicit $2$-body channels $\pi N$, $\eta N$, $K \Sigma$ and $K \Lambda$ as well as the 'effective' $2$-body channels $\pi \Delta$, $\sigma N$ and $\rho N$. The latter are used to model the $3$-particle intermediate state $\pi \pi N$. Therefore, $2$-body unitarity is exact in the J\"{u}lich approach, while $3$-body unitarity is satisfied up to a good approximation \cite{DeborahPion, MichaelAnalyticProps}. \newline
The $G_{\kappa}$ denote $2$- and $3$-body propagators. Their form may be inspected in the publication \cite{MichaelAnalyticProps} and references therein. Kinematical variables in equation (\ref{eq:JuelichModelPiNLippmannSchwinger}) are the total center-of-mass energy $W$ as well as the moduli $q = \left| \vec{q} \right|$ and $p^{\prime} = \left| \vec{p}^{\prime} \right|$ of the outgoing and incoming $3$-momenta. The latter are allowed to violate the mass-shell condition \cite{CommonPaper}. \newline
Finally, the scattering-potential $V_{\mu \nu}$ in equation (\ref{eq:JuelichModelPiNLippmannSchwinger}) is constructed from effective lagrangians \cite{DeborahPion}. It includes (bare) $s$-channel poles as genuine resonances, as well as $t$- and $u$-channel exchanges of a relatively generous collection of light mesons and baryons. Explicit expressions for this potential can be found in \cite{DeborahPion, CommonPaper}. \newline
As special property of the Lippmann-Schwinger equation (\ref{eq:JuelichModelPiNLippmannSchwinger}) is that apart from the pre-included genuine resonances, it is capable to produce so-called {\it dynamically generated} resonances. The latter are poles in the T-Matrix that arise as a by-product of the solution of equation (\ref{eq:JuelichModelPiNLippmannSchwinger}) and have not been put in initially as genuine states. \newline
The J\"{u}lich-Bonn description of pseudoscalar meson photoproduction is built onto the had\-ro\-nic meson-baryon amplitudes in a phenomenoligical approach \cite{DeborahPhotoProd}. A photoproduction-kernel $V_{\mu \gamma}$ is defined, and the photoproduction amplitude consists of this kernel plus a Lippmann-Schwinger iterated term, which includes all rescattering processes generated by the meson-baryon model. The resulting equation for the photoproduction partial waves $M_{\mu \gamma}$, so-called {\it multipoles}, reads \cite{CommonPaper}
\begin{equation}
 M_{\mu \gamma} (q;W) = V_{\mu \gamma} (q;W) + \sum_{\kappa} \int_{0}^{\infty} dp \hspace*{2pt} p^{2} \hspace*{2pt} T_{\mu \kappa} (q,p;W) G_{\kappa} (p;W) V_{\kappa \gamma} (p;W) \mathrm{.} \label{eq:JuelichModelPhotoProdLippmannSchwinger}
\end{equation}
The index '$\gamma$' labels the state $\gamma N$, while the T-Matrix $T_{\mu \kappa}$ contains the full meson-baryon model obtained by solving equation (\ref{eq:JuelichModelPiNLippmannSchwinger}). \newline
Explicit expressions for the photoproduction-kernel can be found in \cite{DeborahPhotoProd, CommonPaper}. The kernel $V_{\mu \gamma}$ can again be decomposed into pole- and non-pole parts. It is important to mention that in this approach, the J\"{u}lich-Bonn group chose to parametrize the non-pole terms and the creation-vertex functions for the resonances as energy-dependent polynomials \cite{DeborahPhotoProd}. The hadronic annihilation-vertex functions have been taken over from the meson-baryon model. \newline
The J\"{u}lich-Bonn group is well-known for implementing and documenting unitarity- and, in particular, analyticity-constraints in a way that is as theoretically clean as possible. For instance, this group has published an account of the analytic continuation of the meson-baryon model over no less than $256$ Riemann-sheets\footnote{This number of sheets is true for the channel-space considered in reference \cite{MichaelAnalyticProps}. Since further channels have been included since then, one has an even more complicated sheet-structure.} (!) in painstaking detail \cite{MichaelAnalyticProps}. \newline

\textbf{The MAID-analysis} \newline
The \underline{M}ainz scattering-\underline{A}nalysis \underline{I}nteractive \underline{D}ial-in, or MAID, is a collection of ED analyses for Photo- and Electroproduction reactions, the results of which can be viewed at the web-page \cite{MAID}. We confine here to the so-called {\it 'unitary isobar-model'} \cite{MAID2007} applicable to the photoproduction reaction. Quite generally, MAID is not a fully coupled-channel model. The model defines the T-Matrix additively as a classic 'background$+$resonance'-Ansatz:
\begin{equation}
 t_{\gamma \pi} (W,Q^{2}) \equiv t_{\gamma \pi}^{B} \left(W,Q^{2}\right) + t_{\gamma \pi}^{R} \left(W,Q^{2}\right) \mathrm{.} \label{eq:MAIDTMatrixAnsatz}
\end{equation}
Here, $W$ is the total center-of-mass energy and $Q^{2}$ measures the virtuality of the photon. A specific partial wave is denoted by the following multi-index which collects good quantum numbers: $\alpha = \left\{ j,\ell,\ldots \right\}$. The background-part is defined in a unitarized K-Matrix inspired Ansatz via
\begin{equation}
 t_{\gamma \pi}^{B,\alpha} (W,Q^{2}) \equiv v_{\gamma \pi}^{B,\alpha} \left(W,Q^{2}\right) \left[ 1 + i t^{\alpha}_{\pi N} \left( W \right) \right]  \mathrm{,} \label{eq:MAIDPhenBackground}
\end{equation}
where the phenomenological potential $v_{\gamma \pi}^{B,\alpha}$ is defined by Born-Terms and the amplitude $t^{\alpha}_{\pi N}$ represents the on-shell part of the Pion-Nucleon rescattering. The Pion-Nucleon elastic amplitudes are parametrized as $t^{\alpha}_{\pi N} = \left[ \eta_{\alpha} \exp \left( 2 i \delta_{\alpha} \right) - 1 \right] /2i$, with phase-shifts and inelasticities imported from the GWU/SAID-analysis (see below). \newline
The resonant part of the T-Matrix Ansatz (\ref{eq:MAIDTMatrixAnsatz}) is modeled by a Breit-Wigner shape as follows
\begin{equation}
 t_{\gamma \pi}^{R} \left(W,Q^{2}\right) := \bar{\mathcal{A}}^{R}_{\alpha} \left(W,Q^{2}\right) \frac{f_{\gamma N} (W) \Gamma_{\mathrm{tot.}} (W) M_{\mathrm{R}} f_{\pi N} (W) }{M_{\mathrm{R}}^{2} - W^{2} - i M_{\mathrm{R}} \Gamma_{\mathrm{tot.}} (W)} \exp \left[ i \phi_{R} \left(W,Q^{2}\right) \right] \mathrm{.} \label{eq:MAIDResonanceAnsatz}
\end{equation}
Here, $M_{\mathrm{R}}$ is the resonance-mass, $\Gamma_{\mathrm{tot.}} (W)$ is the total width of the resonance, $f_{\gamma N}$ and $f_{\pi N}$ are Breit-Wigner factors relating total and partial decay-widths and finally, $\phi_{R}$ is a unitarization-phase. The electromagnetic coupling $\bar{\mathcal{A}}^{R}_{\alpha}$ for the resonance-excitation is for most resonances parametrized independently of $W$ as:
\begin{equation}
 \bar{\mathcal{A}}^{R}_{\alpha} \left(W,Q^{2}\right) \equiv \bar{\mathcal{A}}^{R}_{\alpha} \left(Q^{2}\right) = \mathcal{A}_{\alpha} (0) \left( 1 + a_{1} Q^{2} + a_{2} Q^{4} + \ldots \right) \exp \left[ - b_{1} Q^{2} \right] \mathrm{.} \label{eq:MAIDResonancecouplingQ2Dependence}
\end{equation}
The only exception is the $\Delta(1232) \frac{3}{2}^{+}$-resonance, where this quantity depends on the photon's $3$-momentum $k \left(W,Q^{2}\right)$. \newline
Recent activities of the MAID-group include analyses of $\eta$- and $\eta^{\prime}$-photoproduction \cite{TiatorEtAlEta} and the implementation of methods from Regge-phenomenology \cite{KashevarovEtAlRegge}. \newline

\textbf{The SAID-analysis} \newline
The \underline{S}cattering \underline{A}nalysis \underline{I}nteractive \underline{D}ial-in, or SAID, provided by the George Washington University (GWU) is historically the first ED analysis to be found on the web \cite{SAID}. \newline
In the most recently published fits, the SAID-group employs phenomenological K-Matrix approaches for the analysis of Pion-Nucleon elastic and Photoproduction reactions \cite{WorkmanEtAl2012ChewMPiN, WorkmanEtAl2012ChewMPhotoprod}. The Pion-Nucleon elastic T-Matrix is
\begin{equation}
 T_{\alpha \beta} = \left[ 1 - \bar{K} C \right]^{-1}_{\alpha \sigma} \bar{K}_{\sigma \beta} \mathrm{,} \label{eq:SAIDPiNElasticParametrization}
\end{equation}
where $C$ denotes a so-called Chew-Mandelstam function \cite{WorkmanEtAl2012ChewMPiN, WorkmanEtAl2012ChewMPhotoprod} and the greek indices run over the channels $\pi N$, $\pi \Delta$, $\rho N$ and $\eta N$, with the repeated index $\sigma$ summed. The photon-induced reactions are then built on top of this construction by defining
\begin{equation}
 T_{\alpha \gamma} = \left[ 1 - \bar{K} C \right]^{-1}_{\alpha \sigma} \bar{K}_{\sigma \gamma} \mathrm{,} \label{eq:SAIDPiNElasticParametrization}
\end{equation}
where the K-Matrix elements $\bar{K}_{\sigma \gamma}$ yield the couplings of the system $\gamma N$ to the purely hadronic channels. \newline
It is important to mention that the energy-dependence of the K-Matrix in the SAID-approach is defined purely in terms of phenomenological energy-polynomials. The only resonance which is put in 'by hand', as a genuine K-Matrix pole, is the $\Delta (1232) \frac{3}{2}^{+}$. All other T-Matrix poles are produced by the K-Matrix factor $\left[ 1 - \bar{K} C \right]^{-1}$. Therefore, the T-Matrices for the Pion-Nucleon elastic and the Photoproduction reactions can only have the same pole-structure in this approach. \newline

The list of models described here is by no means complete. Further ED fit approaches are for instance the DCC model by the ANL/Osaka group \cite{Kamano:2013iva}, works done at Kent State University by Manley and collaborators \cite{Shrestha:2012ep} or analyses performed by the Giessen-group \cite{Shklyar:2012zz, Shklyar:2012js}. Also, the classic Pion-Nucleon analyses by H\"{o}hler \cite{Hoehler84}, Cutkosky and collaborators \cite{CMB} should be mentioned. \newline
This concludes our compilation of the most important ED-models currently on the market. A complementary approach to such models consists of asking the question which maximal amount of information on the amplitudes of a particular process can be inferred from data alone, without any model-assumptions on the functional form of these amplitudes in terms of physical parameters. \newline
This Ansatz is less sophisticated than the methods outlined above, but still it has the advantage of, ideally, total model-independence. Also, it leads directly to the problem of so-called {\it complete experiments}. This thesis treats complete experiments for the special case of pseudoscalar meson photoproduction. Moreover, the better-known discussion for the full amplitudes is here adapted for an expansion into partial waves. Thus, we introduce the model-independent amplitude-formalism for this process in section \ref{sec:Photoproduction}.

\clearpage

\subsection{The reaction of pseudoscalar meson photoproduction} \label{sec:Photoproduction}

In the following we confine our attention to the production of a pseudoscalar meson $\varphi$ and recoil-baryon $B$ by impinging a real photon $\gamma$ on a target-nucleon $N$
\begin{equation}
 \gamma N \longrightarrow \varphi B \mathrm{.} \label{eq:PhotoprodReaction}
\end{equation}
The model-independent amplitude-formalism outlined in the following can account for the pertinent channel of pion photoproduction $\gamma N \rightarrow \pi N$. Furthermore, it may also be applied to the cases of production of eta (-prime) mesons $\gamma N \rightarrow \eta^{(\prime)} N$, or the strangeness-production channels with kaons and hyperons in the final state: $\gamma N \rightarrow K \Lambda$, $\gamma N \rightarrow K \Sigma$ (see Fig. \ref{fig:WeightDiagramsMesonsBaryons}). \newline
Before elaborating further on the amplitudes, a few more notes should be made on the reaction-kinematics. One introduces $4$-momentum vectors for the photon $p_{\gamma}$, the nucleon $P_{i}$, the produced meson $p_{\varphi}$ and the recoil-baryon $P_{f}$. These have to fulfill $4$-momentum conservation
\begin{equation}
 p_{\gamma} + P_{i} = p_{\varphi} + P_{f} \mathrm{.} \label{eq:4MomentumCons}
\end{equation}
It is customary to define the Lorentz-invariant {\it Mandelstam variables}
\begin{equation}
 s = \left( p_{\gamma} + P_{i} \right)^{2} \mathrm{,} \hspace*{5pt} t = \left( p_{\gamma} - p_{\varphi} \right)^{2} \mathrm{,} \hspace*{5pt} u = \left( p_{\gamma} - P_{f} \right)^{2} \mathrm{.} \label{eq:MandelstamVariables}
\end{equation}
Since all the initial and final particles are on their mass-shell, the Mandelstam variables are not independent, but fulfill the constraint $s+t+u = \sum_{j} m_{j}^{2} = m_{\varphi}^{2} + m_{N}^{2} + m_{B}^{2}$. Furthermore, it can be shown that for a $2 \rightarrow 2$-process of on-shell particles, only two kinematic variables are fully independent \cite{CGLN}. These can be chosen to be, for instance, the Mandelstam variables $s$ and $t$. \newline
For the most common choices of reference-frames, i.e. the laboratory (LAB-) frame and the center-of-mass (CMS-) frame, kinematics are illustrated in Figure \ref{fig:SandorfiCoordinates}. We refrain here from providing a more formal treatment of the reaction-kinematics in the Mandelstam-plane, which may be found for instance in reference \cite{BycklingK}. Instead, some results on energies and scattering-angles in the two systems and their relation to frame-independent quantities are listed (see reference \cite{Sandorfi:2010uv} and derivations in \cite{MyDiplomaThesis}):
\begin{itemize}
 \item[$\ast)$] Relations between the total center of mass (CMS) energy $W$ and the laboratory (LAB) energy $E_{\gamma}^{\mathrm{LAB}}$ of the photon:
 \begin{equation}
W = \sqrt{s} = \sqrt{m_{N} \left( m_{N} + 2 E_{\gamma}^{\mathrm{LAB}} \right)} \mathrm{,} \hspace*{5pt} E_{\gamma}^{\mathrm{LAB}} = \frac{s - m_{N}^{2}}{2 m_{N}} \mathrm{.} \label{eq:WconvertedToEGammaLAB}
\end{equation}
\item[$\ast)$] The magnitudes of the CMS 3-momenta for the photon $ k = | \vec{k} | = | \vec{p}_{\gamma}^{~\mathrm{CMS}} | $ and the produced meson $ q = \left| \vec{q} \right| = \left| \vec{p}_{\varphi}^{~\mathrm{CMS}} \right| $, written in terms of Lorentz invariants (furthermore: $\left| \vec{P}_{i}^{\mathrm{CMS}} \right| = | \vec{k} |$ and $\left| \vec{P}_{f}^{\mathrm{CMS}} \right| = | \vec{q} |$):
\begin{align}
 k &= E_{\gamma}^{\mathrm{CMS}} = \frac{s - m_{N}^{2}}{2 \sqrt{s}} \mathrm{,} \label{eq:CMS3MomentumPhoton} \\
 q &=  \frac{1}{2 \sqrt{s}} \sqrt{ \left\{ s - \left( m_{\varphi} + m_{B} \right)^{2} \right\} \left\{ s - \left( m_{\varphi} - m_{B} \right)^{2} \right\}} \mathrm{.} \label{eq:CMS3MomentumMeson}
\end{align}
\item[$\ast)$] The phase-space-factor becomes $\rho = q/k$ and is thus determined by Lorentz invariants.
\item[$\ast)$] The CMS-angle $\theta_{\varphi}^{\mathrm{CMS}}$ is fixed by invariants via the following expression
\begin{equation}
 \cos \theta_{\varphi}^{\mathrm{CMS}} = \frac{ t - m_{\varphi}^{2} + 2 k \sqrt{m_{\varphi}^{2} + q^{2}} }{ 2 q k } \mathrm{,} \label{eq:CosThetaCMSInvariants}
\end{equation}
while CMS- and LAB-angles can be related using the equation
\begin{equation}
\tan \left( \theta_{\varphi}^{\mathrm{LAB}} \right) = \frac{\sin \left( \theta_{\varphi}^{\mathrm{CMS}} \right)}{\gamma \left( \cos \left( \theta_{\varphi}^{\mathrm{CMS}} \right) + \frac{\beta}{\beta_{\varphi}^{\mathrm{CMS}}} \right)} \mathrm{.} \label{eq:thetaMLAB}
\end{equation}
Here, $\gamma = 1/\sqrt{1 - \beta^{2}}$, $\beta$ is the LAB-velocity of the center of mass and $\beta_{\varphi}^{\mathrm{CMS}}$ is the meson velocity in the CMS frame.
\end{itemize}
\begin{figure}[ht]
\centering
\begin{overpic}[width=1.00\textwidth]%
      {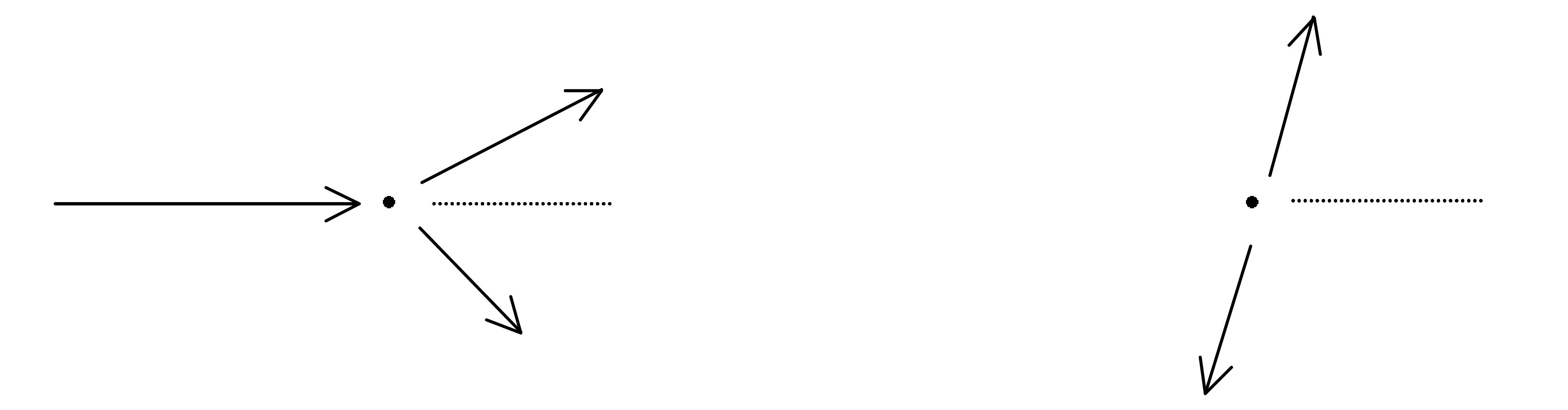}
 \put(0,15.0){\begin{tiny}$p_{\gamma}^{\mathrm{LAB}} = \left( E_{\gamma}^{\mathrm{LAB}}, \thinspace \vec{p}_{\gamma}^{~\mathrm{LAB}} \right)$\end{tiny}}
 \put(17.6,10.1){\begin{tiny}$ m_{T} = m_{N} $\end{tiny}}
 \put(11.5,7.0){\begin{tiny}$ P_{i}^{\mathrm{LAB}} = \left( m_{N}, \thinspace \vec{0} \right)  $\end{tiny}}
 \put(29,14.1){\begin{tiny}$ \theta_{\varphi}^{\mathrm{LAB}} $\end{tiny}}
 \put(33.5,7.0){\begin{tiny}$ P_{f}^{\mathrm{LAB}} = \left( E_{f}^{\mathrm{LAB}}, \thinspace \vec{p}_{f}^{~\mathrm{LAB}} \right) $\end{tiny}}
 \put(33.5,3.6){\begin{tiny}$ m_{R} = m_{B} $\end{tiny}}
 \put(39.25,21.0){\begin{tiny}$ m_{\varphi} $\end{tiny}}
 \put(17.0,22.0){\begin{tiny}$ p_{\varphi}^{\mathrm{LAB}} = \left( E_{\varphi}^{\mathrm{LAB}}, \thinspace \vec{p}_{\varphi}^{~\mathrm{LAB}} \right) $\end{tiny}}
 \put(82.0,14.6){\begin{tiny}$ \theta_{\varphi}^{\mathrm{CMS}} $\end{tiny}}
 \put(80.5,6.6){\begin{tiny}$ P_{f}^{\mathrm{CMS}} $\end{tiny}}
 \put(86.4,6.6){\begin{tiny}$ = \left( E_{f}^{\mathrm{CMS}}, \thinspace \vec{p}_{f}^{~\mathrm{CMS}} \right) $\end{tiny}}
 \put(86.4,2.6){\begin{tiny}$ = \left( E_{f}^{\mathrm{CMS}}, \thinspace - \vec{q} \right) $\end{tiny}}
 \put(60.0,20.1){\begin{tiny}$ p_{\varphi}^{\mathrm{CMS}} $\end{tiny}}
 \put(65.45,20.1){\begin{tiny}$ = \left( E_{\varphi}^{\mathrm{CMS}}, \thinspace \vec{p}_{\varphi}^{~\mathrm{CMS}} \right) $\end{tiny}}
 \put(65.45,16.1){\begin{tiny}$ = \left( E_{\varphi}^{\mathrm{CMS}}, \thinspace \vec{q} \right) $\end{tiny}}
 \put(77.5,0.0){\begin{tiny}$ m_{R} = m_{B} $\end{tiny}}
 \put(84.25,25.75){\begin{tiny}$ m_{\varphi} $\end{tiny}}
\end{overpic} \\
\caption[The LAB- and CMS-kinematics of photoproduction in one simple picture.]{The kinematics of pseudoscalar meson photoproduction in both the LAB- and CMS-frame. Individual particles are denoted by their masses. The picture, except for a relabelling, is taken over from ref. \cite{Sandorfi:2010uv}.}
\label{fig:SandorfiKinematics}
\end{figure}
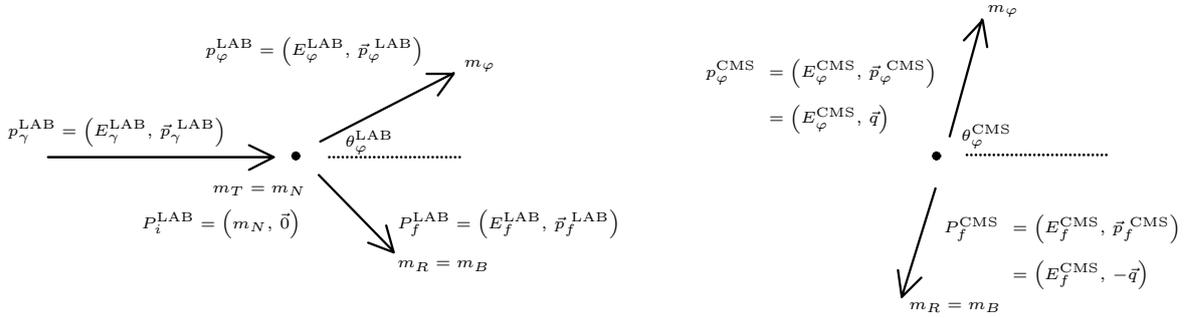

\subsubsection{Photoproduction amplitudes and multipoles} \label{subsec:PhotoproductionAmpl}

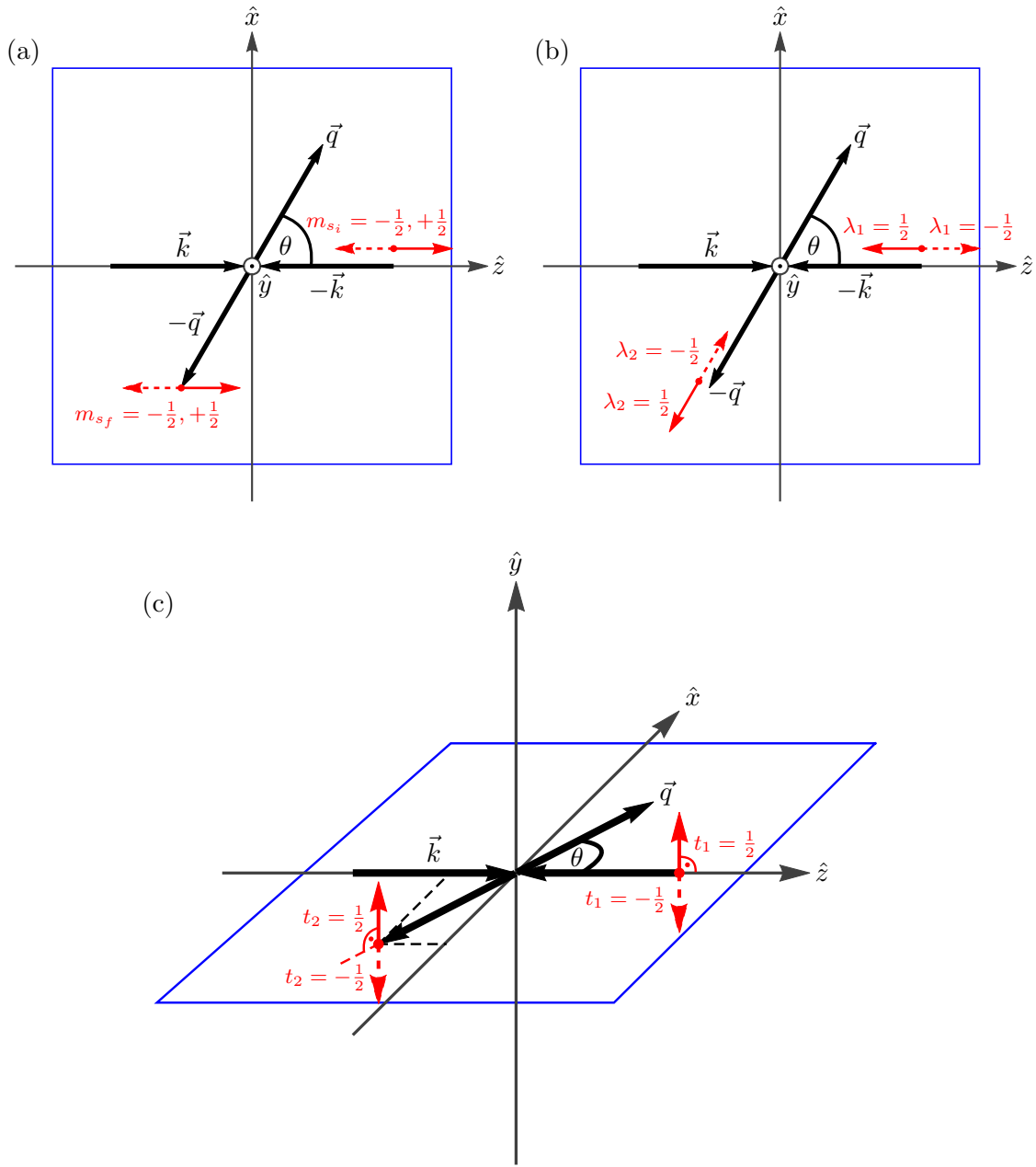
\begin{figure}[ht]
 \centering
 \begin{overpic}[width=0.46\textwidth]%
      {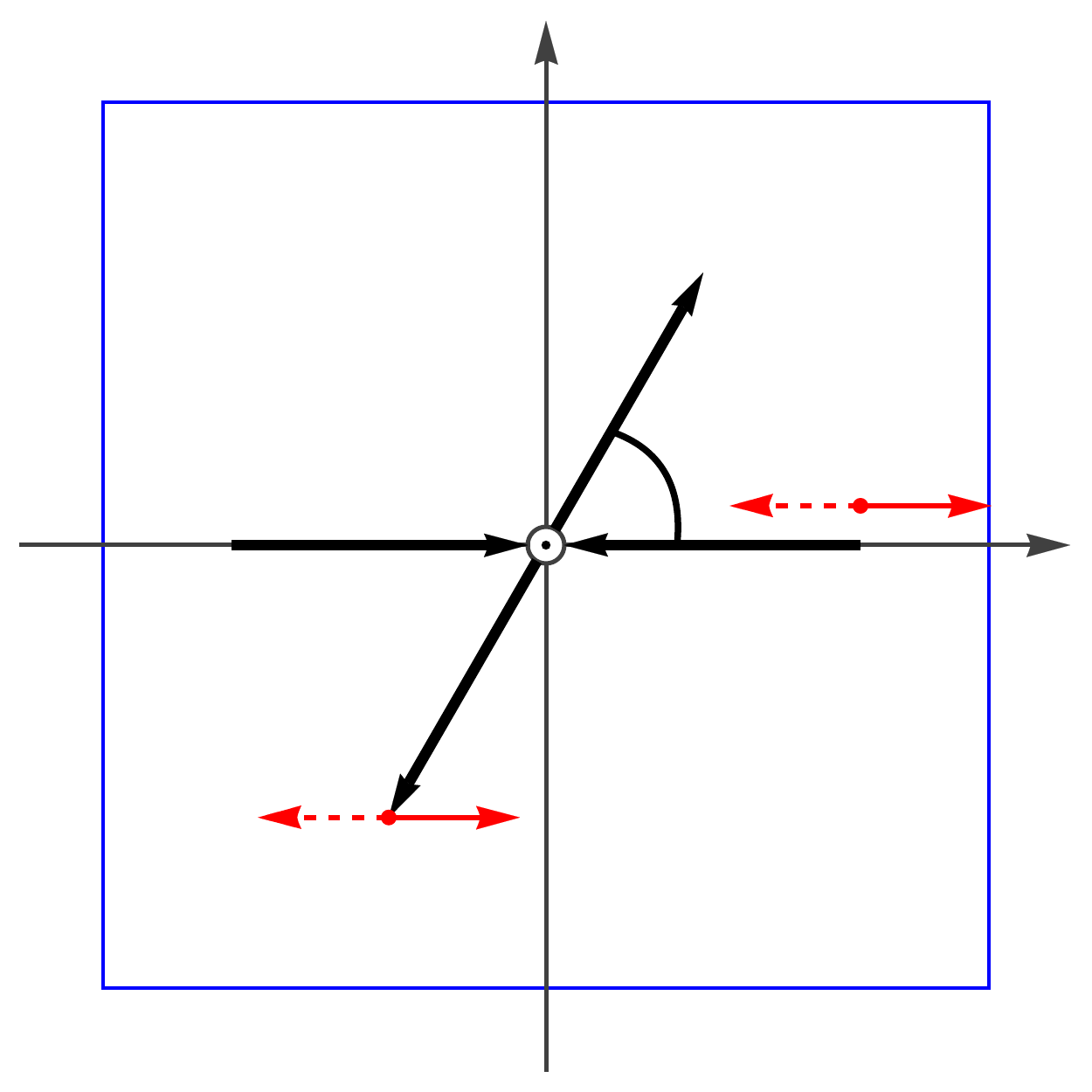}
      \put(0,92.5){(a)}
      \put(98.75,48.5){$\hat{z}$}
      \put(48.5,99.5){$\hat{x}$}
      \put(51.25,44.75){$\hat{y}$}
      \put(55.5,52.5){$\theta$}
      \put(61.5,43.5){$-\vec{k}$}
      \put(34.5,52){$\vec{k}$}
      \put(65,75.5){$\vec{q}$}
      \put(32.75,37){$-\vec{q}$}
      \put(61,58){\begin{footnotesize}\textcolor{red}{$m_{s_{i}} = - \frac{1}{2} , + \frac{1}{2}$}\end{footnotesize}}
      \put(14,18.5){\begin{footnotesize}\textcolor{red}{$m_{s_{f}} = - \frac{1}{2} , + \frac{1}{2}$}\end{footnotesize}}
\end{overpic} \hspace*{7.5pt}
 \begin{overpic}[width=0.46\textwidth]%
      {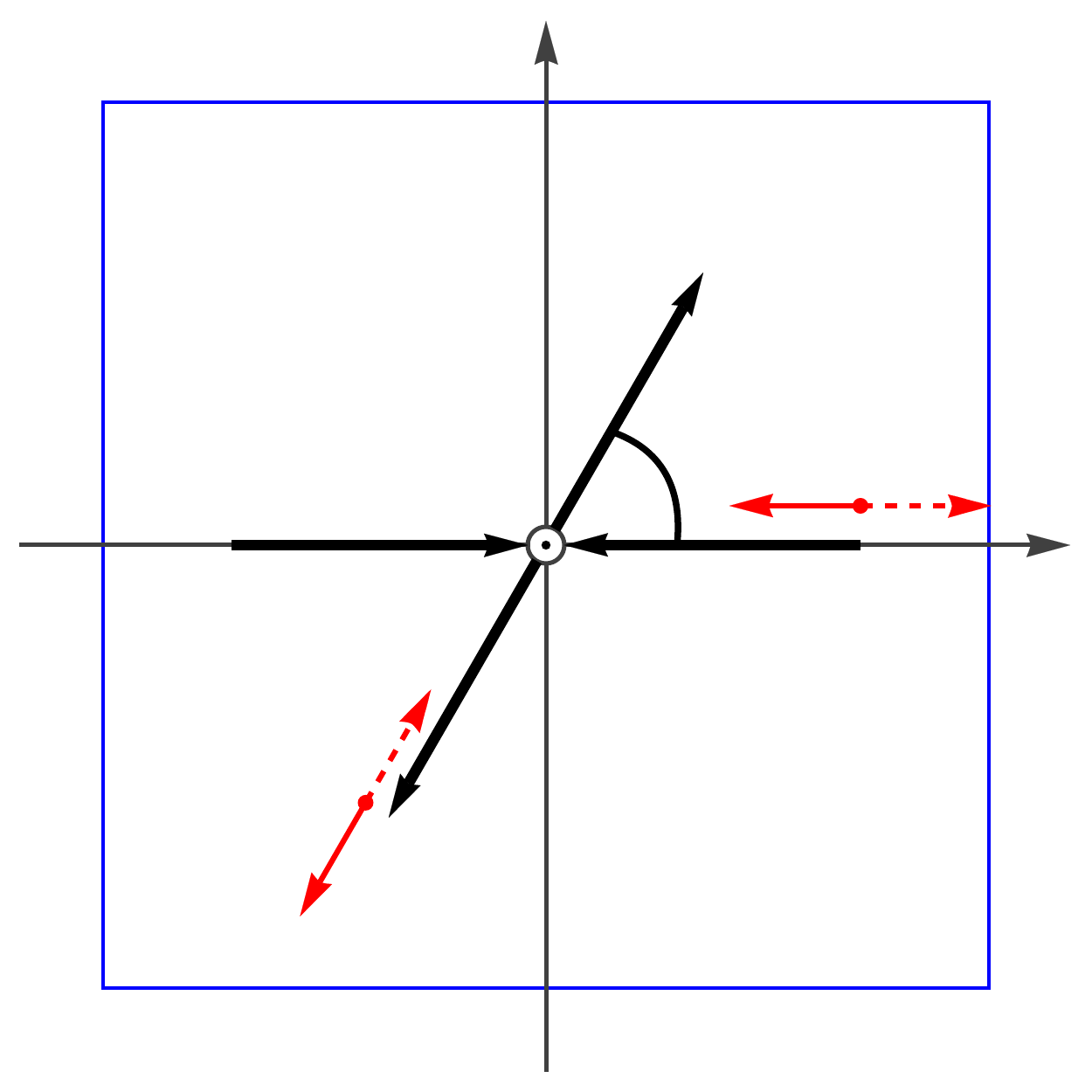}
      \put(0,92.5){(b)}
      \put(98.75,48.5){$\hat{z}$}
      \put(48.5,99.5){$\hat{x}$}
      \put(51.25,44.75){$\hat{y}$}
      \put(55.5,52.5){$\theta$}
      \put(61.5,43.5){$-\vec{k}$}
      \put(34.5,52){$\vec{k}$}
      \put(65,75.5){$\vec{q}$}
      \put(35.25,22.75){$-\vec{q}$}
      \put(80.25,57){\begin{footnotesize}\textcolor{red}{$\lambda_{1} = - \frac{1}{2}$}\end{footnotesize}}
      \put(62.75,57){\begin{footnotesize}\textcolor{red}{$\lambda_{1} = \frac{1}{2}$}\end{footnotesize}}
      \put(16.5,31.5){\begin{footnotesize}\textcolor{red}{$\lambda_{2} = - \frac{1}{2}$}\end{footnotesize}}
      \put(14,21){\begin{footnotesize}\textcolor{red}{$\lambda_{2} = \frac{1}{2}$}\end{footnotesize}}
\end{overpic} \\
\vspace*{10pt}
 \begin{overpic}[width=0.7\textwidth]%
      {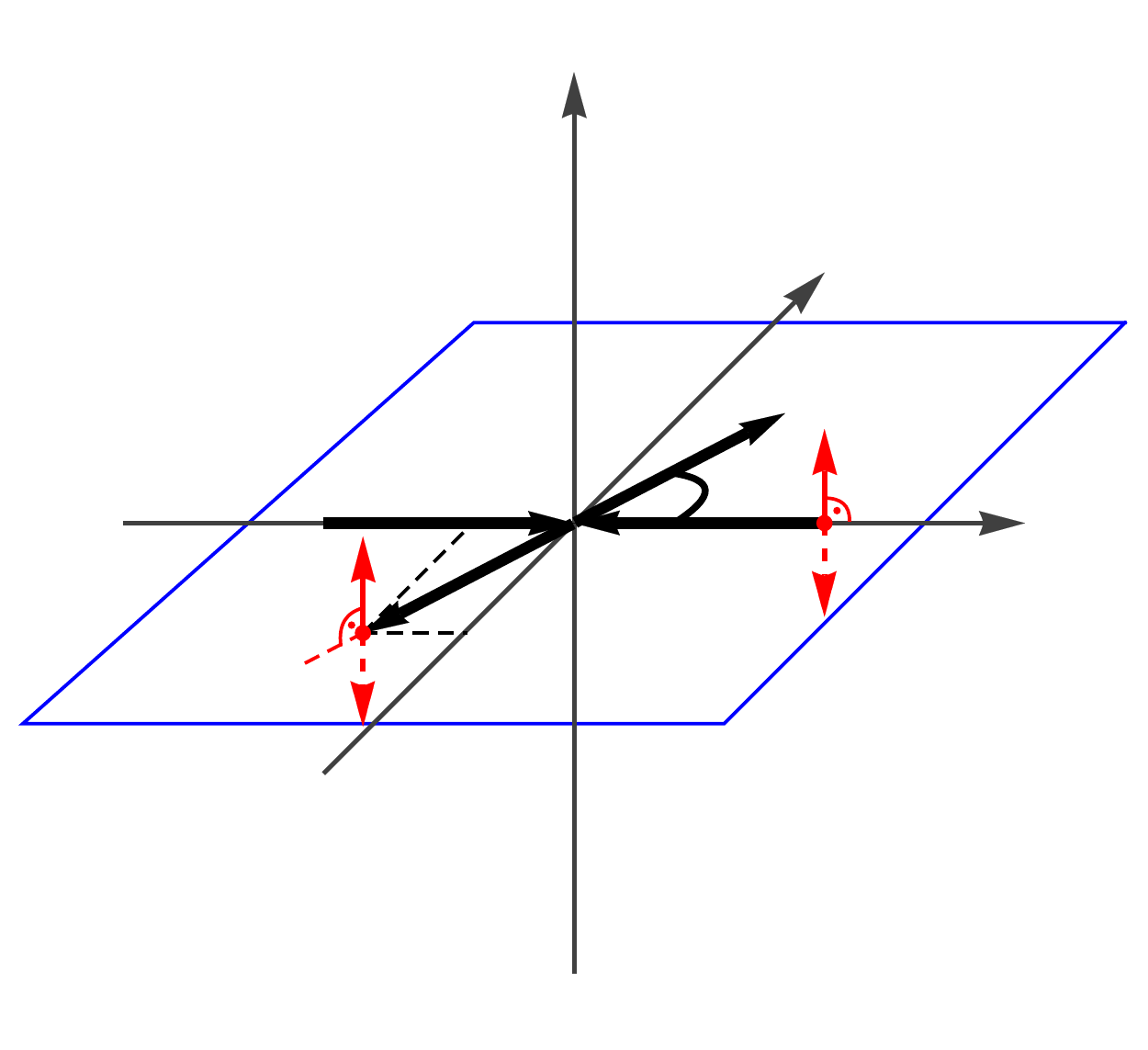}
      \put(0,81){(c)}
      \put(90,44.7){$\hat{z}$}
      \put(49,86.65){$\hat{y}$}
      \put(72.6,68.35){$\hat{x}$}
      \put(57.25,46.75){$\theta$}
      \put(38,47.35){$\vec{k}$}
      \put(69.2,55.65){$\vec{q}$}
      \put(73.75,49){\begin{footnotesize}\textcolor{red}{$t_{1} = \frac{1}{2}$}\end{footnotesize}}
      \put(59.25,41.5){\begin{footnotesize}\textcolor{red}{$t_{1} = - \frac{1}{2}$}\end{footnotesize}}
      \put(21.5,39){\begin{footnotesize}\textcolor{red}{$t_{2} = \frac{1}{2}$}\end{footnotesize}}
      \put(19.15,31){\begin{footnotesize}\textcolor{red}{$t_{2} = - \frac{1}{2}$}\end{footnotesize}}
\end{overpic}
\caption[Illustration of the axes of spin-quantization for the different bases of photoproduction spin-amplitudes.]{The sketches illustrate the different axes of spin-quantization corresponding to each system of $4$ complex spin-amplitudes introduced in this section. CMS-coordinates are used everywhere. The so-called {\it reaction plane}, which is spanned by the CMS $3$-momenta $\vec{k}$ and $\vec{q}$, is indicated as a blue quadrangle and always defined to coincide with the $\left< x - z \right>$-plane. \newline (a) Quantization using the eigenvalues of a conventional spin-$z$ eigenstate for the initial nucleon and final baryon. (b) Quantization via the helicity quantum numbers, i.e. the projection of the spin along the direction of motion. (c) The $y$-direction normal to the reaction plane is used for the definition of {\it transversity} quantum numbers (eq. (\ref{eq:TransversityRotationst12})).}
\label{fig:SPinAmplitudeIllustrations}
\end{figure}

In any order in perturbation theory, the complex transition-amplitude $\mathcal{T}_{fi}$ for photoproduction would be given by an invariant transition operator $\mathcal{T}$ in Dirac-spinor space, sandwiched between two spinors for the initial (target-) nucleon $N_{T}$ and final-state (recoil-) baryon $B_{R}$
\begin{equation}
\mathcal{T}_{fi} (s,t) = \bar{u}_{B_{R}} \left( P_{f},m_{s_{f}} \right) \hat{\mathcal{T}} u_{N_{T}} \left( P_{i},m_{s_{i}} \right) \mathrm{.} \label{eq:ScatteringMatrixElement}
\end{equation}
As shown by Chew, Goldberger, Low and Nambu (CGLN) \cite{CGLN}, the most general transition operator can be decomposed as a linear combination of four terms 
\begin{equation}
\hat{\mathcal{T}} = \sum \limits_{i = 1}^{4} A_{i} (s,t) \hat{M}_{i} \mathrm{,} \label{eq:MostGeneralT}
\end{equation}
where the complex functions $A_{i} (s,t)$ are called {\it invariant amplitudes} and are further specified by the process-dynamics, while the four spin-matrices $M_{i}$ need to be found using more general arguments. They have to be built from the $4$-momenta of the reaction, polarization $4$-vector $\epsilon$ of the photon, the Dirac-operator $\gamma^{\mu}$ and, since the produced meson is a pseudoscalar, the chirality operator $\gamma_{5}$. CGLN showed that $4$-momentum conservation combined with Lorentz- and gauge\footnote{Gauge-invariance can be checked by investigating whether the quantities (\ref{eq:M1Inv}) to (\ref{eq:M4Inv}) vanish once the replacement $\epsilon \rightarrow p_{\gamma}$ is applied. Then, the QED Ward-identity is satisfied.}-invariance suffice to determine the $\hat{M}_{i}$. A conventional choice is \cite{Hanstein}
\begin{align}
\hat{M}_{1} &= i \gamma_{5} \displaystyle{\not} \epsilon \thinspace \displaystyle{\not} p_{\gamma} \mathrm{,} \label{eq:M1Inv} \\
\hat{M}_{2} &= 2 i \gamma_{5} \left\{ \left( P \cdot p_{\gamma} \right) \left( p_{\varphi} \cdot \epsilon \right) - \left( P \cdot \epsilon \right) \left( p_{\gamma} \cdot p_{\varphi} \right) \right\} \mathrm{,} \label{eq:M2Inv} \\
\hat{M}_{3} &= i \gamma_{5} \left\{ \displaystyle{\not} \epsilon \left( p_{\gamma} \cdot p_{\varphi} \right) - \displaystyle{\not} p_{\gamma} \left( p_{\varphi} \cdot \epsilon \right) \right\} \mathrm{,} \label{eq:M3Inv} \\
\hat{M}_{4} &= 2 i \gamma_{5} \left\{ \displaystyle{\not} \epsilon \left( P \cdot p_{\gamma} \right) - \displaystyle{\not} p_{\gamma} \left( P \cdot \epsilon \right) - m_{N} \displaystyle{\not} \epsilon \thinspace \displaystyle{\not} p_{\gamma} \right\} \mathrm{,} \label{eq:M4Inv}
\end{align}
with slash-notation $\displaystyle{\not} u = u_{\mu} \gamma^{\mu}$ and $P:=P_{i} + P_{f}$. Now, upon adopting CMS-coordinates and explicit expressions of the Dirac-spinors $u_{N_{T}}$ and $u_{B_{R}}$, the most general tansition matrix element reduces further to the form \cite{CGLN,Sandorfi:2010uv}

\begin{align}
\bar{u}_{B_{R}} \left( P_{f}^{\mathrm{CMS}},m_{s_{f}} \right) \hat{\mathcal{T}} u_{N_{T}} \left( P_{i}^{\mathrm{CMS}},m_{s_{i}} \right) &= \frac{4 \pi W}{\sqrt{m_{N}m_{B}}} \chi_{m_{s_{f}}}^{\dagger} F_{\mathrm{CGLN}} \chi_{m_{s_{i}}} \nonumber \\
&= \frac{4 \pi W}{\sqrt{m_{N}m_{B}}} \big<m_{s_{f}}\big| F_{\mathrm{CGLN}} \big|m_{s_{i}}\big> \mathrm{,} \label{eq:DiractoCGLN}
\end{align}
with the $2 \times 2$-operator in Pauli-spinor space:
\begin{equation}
 F_{\mathrm{CGLN}} =  i \vec{\sigma} \cdot \hat{\epsilon} F_{1} + \vec{\sigma} \cdot \hat{q}
   \vec{\sigma} \cdot \left( \hat{k} \times \hat{\epsilon} \right)  F_{2}  + i \vec{\sigma} \cdot \hat{k}  \hat{q} \cdot \hat{\epsilon}  F_{3} + i \vec{\sigma} \cdot \hat{q}  \hat{q} \cdot \hat{\epsilon} F_{4} \mathrm{.} \label{eq:FCGLNOperator} 
\end{equation}
Here, normalized CMS $3$-momenta are written with a hat and the vector $\vec{\sigma} = \left[ \sigma_{x}, \sigma_{y}, \sigma_{z} \right]^{T}$ collects the usual Pauli-matrices. The four complex functions $F_{i}$ are typically written as depending on the total CMS-energy and scattering-angle, $F_{i} = F_{i} \left( W,\theta\right)$ and are termed, according to their inventors, as {\it CGLN-amplitudes} \cite{CGLN}. The relations among the invariant amplitudes $A_{i}$ and CGLN amplitudes $F_{i}$ are linear and invertible, but still quite complicated. Furthermore, they contain kinematical singularities \cite{Hanstein}.\newline
For the further discussion of the model-independent amplitude formalism, it will turn out fruitful to introduce new schemes of spin-quantization for the matrix-element in equation (\ref{eq:DiractoCGLN}), other than the eigenvalues $m_{s}$ of spin-$z$ eigenstates. Using suitable rotations to different axes of quantization, one can introduce alternatives to the CGLN amplitudes, called {\it helicity-} and {\it transversity-amplitudes}. The CMS frame of photoproduction and the relevant axes are shown in Figure \ref{fig:SPinAmplitudeIllustrations}. The change in axes affects here only the spins of the initial state nucleon and the final state baryon. \newline 
In the ensuing discussion, the general rotation-matrices\footnote{These are representations of a general rotation parametrized with Euler-angles: $D_{m_{1} m_{2}}^{\left(j\right)} (\alpha,\beta,\gamma) = \left< j,m_{1} \right| e^{-i \alpha  \hat{J}_{z}} e^{-i \beta \hat{J}_{y}} e^{i \gamma \hat{J}_{z}} \left| j,m_{2} \right>$ \cite{MartinSpearman}.} $D_{m_{1} m_{2}}^{\left(j\right)}(\alpha,\beta,\gamma)$ in a spin-$j$ representation will be needed. In order to rotate from a unit-vector pointing in the $+\hat{z}$ direction to an arbitrary unit-vector with polar coordinates $(\theta, \phi)$, the choice of Euler-angles $\left(\alpha,\beta,\gamma\right) = \left(\phi,\theta,-\phi\right)$ is conventional \cite{MartinSpearman}. Then, the rotation-matrices can be reduced, via 
\begin{equation}
D^{\left(j\right)}_{m_{1}, m_{2}} \left( \phi, \theta, - \phi \right) = e^{-i \left( m_{1} - m_{2} \right) \phi} d^{\left(j\right)}_{m_{1}, m_{2}} \left( \theta \right) \mathrm{,} \label{eq:DefGeneralSpinJRotationWignerDs}
\end{equation}
to the so-called {\it Wigner $d$-functions} $d_{m_{1} m_{2}}^{\left(j\right)}$. Further properties of the latter are listed by Martin and Spearman \cite{MartinSpearman}. \newline

\clearpage
For instance, the $4$ helicity amplitudes $H_{i}$ can be defined by a rotation from the $\hat{z}$ axis in the CMS frame to the $3$-momentum axes of the target nucleon and recoil baryon, respectively\footnote{Formally, the helicity $\lambda$ of a particle is the eigenvalue of the operator which measures the projection of the spin on the direction of motion: $\hat{\Lambda} := \left( \hat{\vec{S}} \cdot \hat{p} \right)$.}. Using helicity quantum numbers $\lambda_{1} = \pm 1/2$ for the nucleon and $\lambda_{2} = \pm 1/2$ for the baryon, the corresponding rotations in the spin-$\frac{1}{2}$ read (according to Fasano, Tabakin and Saghai \cite{FTS}):
\begin{equation}
\left| \lambda_{1} \right> = \sum \limits_{m_{s_{i}} = \pm 1/2} D^{\left( 1/2 \right) }_{m_{s_{i}}, \lambda_{1}} \left( 0, \pi, 0 \right) \left| m_{s_{i}} \right> \mathrm{,} \hspace*{5pt} \left| \lambda_{2} \right> = \sum \limits_{m_{s_{f}} = \pm 1/2} D^{\left( 1/2 \right) }_{m_{s_{f}}, \lambda_{2}} \left( 0, \pi + \theta, 0 \right) \left| m_{s_{f}} \right> \mathrm{.} \label{eq:HelicityRotationsLambda12}
\end{equation}
The angles $\pi$ and $\pi + \theta$ in the $d$-functions are correct since the CMS $3$-momenta for nucleon and baryon are $- \vec{k}$ and $-\vec{q}$. Using (\ref{eq:DefGeneralSpinJRotationWignerDs}) and standard expressions for the $d^{(1/2)}$ \cite{Sandorfi:2010uv, MartinSpearman}, it can be seen that both rotations become
\begin{align}
\left| \lambda_{1} \right> = \eta_{\lambda_{1}} \sum \limits_{m_{s_{i}} = \pm 1/2} d^{\left( 1/2 \right) }_{m_{s_{i}}, - \lambda_{1}} \left( 0 \right) \left| m_{s_{i}} \right> \mathrm{,} \hspace*{5pt} \left| \lambda_{2} \right> = \eta_{\lambda_{2}} \sum \limits_{m_{s_{f}} = \pm 1/2} d^{\left( 1/2 \right) }_{m_{s_{f}}, - \lambda_{2}} \left( \theta \right) \left| m_{s_{f}} \right> \mathrm{,} \label{eq:SimplifiedHelicityRotationsLambda12}
\end{align}
with $\eta_{\lambda} = \left( - 1 \right)^{1/2 + \lambda}$. Furthermore, knowing that the CGLN-operator $F_{\mathrm{CGLN}}$ (\ref{eq:FCGLNOperator}) can be written as a scalar-product of the photon-polarization and a hadronic transition-current $F_{\mathrm{CGLN}} = \vec{\epsilon} \cdot \hat{\vec{\bm{J}}}$, it is seen that this operator becomes a spherical vector operator for circular photon polarizations $\left(\hat{\epsilon}_{c}\right)_{\lambda_{\gamma} = \pm 1} = (\mp 1/\sqrt{2}) \left( \hat{e}_{x} \pm i \hat{e}_{y} \right)$:
\begin{equation}
\hat{\bm{J}}_{1, \lambda_{\gamma}} = \left( \hat{\epsilon}_{c} \right)_{\lambda_{\gamma}} \cdot \hat{\vec{\bm{J}}} \mathrm{.} \label{eq:DefJ_1Lambda}
\end{equation}
For this spherical vector, the relation between helicity- and spin-$\hat{z}$ matrix elements is
\begin{equation}
\left< \lambda_{2} \right| \hat{\bm{J}}_{1, \lambda_{\gamma}} \left| \lambda_{1} \right> = \eta_{\lambda_{1}} \eta_{\lambda_{2}} \sum \limits_{m_{s_{i}}, m_{s_{f}}} d^{\left( 1/2 \right) }_{m_{s_{f}}, - \lambda_{2}} \left( \theta \right) d^{\left( 1/2 \right) }_{m_{s_{i}}, - \lambda_{1}} \left( 0 \right) \big< m_{s_{f}} \big| \hat{\bm{J}}_{1, \lambda_{\gamma}} \big| m_{s_{i}} \big> \mathrm{.} \label{eq:MatrixElementJHelicityBasis}
\end{equation}
Thus, we introduce a set of $4$ non-redundant helicity amplitudes according to reference \cite{FTS}, using matrix elements of the spherical vector for\footnote{The $\lambda_{\gamma} = -1$ case is related to these amplitudes by parity and thus redundant.} $\lambda_{\gamma} = +1$:
\begin{align}
H_{1} \left( W, \theta \right) &\equiv  \left< \lambda_{2} = + 1/2 \right| \hat{\bm{J}}_{1, +1} \left| \lambda_{1} = - 1/2 \right> \mathrm{,} \label{eq:DefH1} \\
H_{2} \left( W, \theta \right) &\equiv  \left< \lambda_{2} = + 1/2 \right| \hat{\bm{J}}_{1, +1} \left| \lambda_{1} = + 1/2 \right> \mathrm{,} \label{eq:DefH2} \\
H_{3} \left( W, \theta \right) &\equiv  \left< \lambda_{2} = - 1/2 \right| \hat{\bm{J}}_{1, +1} \left| \lambda_{1} = - 1/2 \right> \mathrm{,} \label{eq:DefH3} \\
H_{4} \left( W, \theta \right) &\equiv  \left< \lambda_{2} = - 1/2 \right| \hat{\bm{J}}_{1, +1} \left| \lambda_{1} = + 1/2 \right> \mathrm{.} \label{eq:DefH4}
\end{align}
The linear transformation connecting CGLN- and helicity-amplitudes is invertible and $\theta$-dependent. For the sake of brevity, it is quoted in appendix \ref{subsec:CGLNMatrixReps} (equation (\ref{eq:HelCGLNTrafoAppendix})). \newline
Transversity amplitudes shall be introduced in Chiang/Tabakin-conventions \cite{ChTab}. According to this reference, for transversity quantum-numbers correspond to spin-projections along the normal of the so-called {\it reaction-plane}. This plane is spanned by $\vec{k}$ and $\vec{q}$ and is conventionally chosen to be the $\left< x-z \right>$-plane (see Figure \ref{fig:SPinAmplitudeIllustrations}). Therefore, one has to rotate to the unit-vector $\hat{e}_{y} \equiv \left(\theta = \pi/2, \phi = \pi/2\right)$ using the rotations
\begin{equation}
\left| t_{1} \right> = \sum \limits_{\lambda_{1} = \pm 1/2} D^{\left( 1/2 \right) }_{\lambda_{1}, t_{1}} \left( \frac{\pi}{2}, \frac{\pi}{2}, - \frac{\pi}{2} \right) \left| \lambda_{1} \right> \mathrm{,} \hspace*{5pt} \left| t_{2} \right> = \sum \limits_{\lambda_{2} = \pm 1/2} D^{\left( 1/2 \right) }_{\lambda_{2}, t_{2}} \left( \frac{\pi}{2}, \frac{\pi}{2}, - \frac{\pi}{2} \right) \left| \lambda_{2} \right> \mathrm{.} \label{eq:TransversityRotationst12}
\end{equation}
We now define a set of non-redundant transversity amplitudes but it should be mentioned that, in order to become consistent with the expressions of Chiang/Tabakin, we had to introduce additional phases of two of the amplitudes
\begin{align}
b_{1} \left( W,  \theta \right) &\equiv  \left< t_{2} = - 1/2 \right| \hat{\bm{J}}_{1, +1} \left| t_{1} = + 1/2 \right> \mathrm{,} \label{eq:Defb1} \\
b_{2} \left( W,  \theta \right) &\equiv  \left< t_{2} = + 1/2 \right| \hat{\bm{J}}_{1, +1} \left| t_{1} = - 1/2 \right> \mathrm{,} \label{eq:Defb2} \\
b_{3} \left( W,  \theta \right) &\equiv \left( + i \right) \left< t_{2} = - 1/2 \right| \hat{\bm{J}}_{1, +1} \left| t_{1} = - 1/2 \right> \mathrm{,} \label{eq:Defb3} \\
b_{4} \left( W,  \theta \right) &\equiv \left( - i \right) \left< t_{2} = + 1/2 \right| \hat{\bm{J}}_{1, +1} \left| t_{1} = + 1/2 \right> \mathrm{.} \label{eq:Defb4}
\end{align}
Then, the relation between helicity- and transversity amplitudes according to reference \cite{ChTab} becomes
\begin{align}
 b_{1} &= \frac{1}{2} \left[ H_{1} + H_{4} - i \left( H_{2} - H_{3} \right) \right] \mathrm{,} \label{eq:b1InTermsOfHelAmpls} \\
 b_{2} &= \frac{1}{2} \left[ H_{1} + H_{4} + i \left( H_{2} - H_{3} \right) \right] \mathrm{,} \label{eq:b2InTermsOfHelAmpls} \\
 b_{3} &= \frac{1}{2} \left[ H_{1} - H_{4} + i \left( H_{2} + H_{3} \right) \right] \mathrm{,} \label{eq:b3InTermsOfHelAmpls} \\
 b_{4} &= \frac{1}{2} \left[ H_{1} - H_{4} - i \left( H_{2} + H_{3} \right) \right] \mathrm{.} \label{eq:b4InTermsOfHelAmpls}
\end{align}
This linear relation is unitary, invertible and looks simple. Both the helicity- and transversity amplitudes are very important for the whole discussion in this thesis, so we quoted their definition here explicitly. Once the observables of photoproduction (to be introduced in the next section) are written using these amplitudes, they simplify significantly. Especially the transversity basis will become very important for the discussion of complete experiments later on. \newline

In case one wishes to write a partial wave expansion of the photoproduction amplitude, the expressions are most concisely written in the helicity-formalism. We show here expressions according to books by Martin and Spearman \cite{MartinSpearman}, as well as Bransden and Moorhouse \cite{BransdenMoorhouse}, but the idea dates back to the work by Jacob and Wick \cite{JacobWick}. \newline
According to these references, the general transition matrix elements for a $2$-body reaction $1+2\rightarrow3+4$ of particles with arbitrary spin are matrix elements of $\mathcal{T}$ in the Hilbert-space of $2$-particle helicity states and can be expanded into rotation-matrices. We fix the convention that the initial CMS-momentum $\vec{k}$ is pointing in the $\hat{z}$-direction, while the final state direction $\hat{q}$ is parametrized by polar angles $\left(\theta, \phi\right)$. The expansion then is \cite{MartinSpearman}
\begin{align}
 &\left< \theta \phi\mathrm{;} \mu_{1} \mu_{2} \mathrm{;} n_{\mathrm{out}} \right| \mathcal{T} \left| 0 0 \mathrm{;} \lambda_{1} \lambda_{2} \mathrm{;} n_{\mathrm{in}} \right> \nonumber \\
 & \hspace*{20pt} = \sum_{j} (2 j + 1) D_{\lambda \mu}^{\left(j\right) \ast} \left( \phi, \theta, -\phi \right) \left< \mu_{1} \mu_{2} \mathrm{;} n_{\mathrm{out}}  \right| \mathcal{T}^{j} \left| \lambda_{1} \lambda_{2} \mathrm{;} n_{\mathrm{out}}  \right> \mathrm{,} \label{eq:GeneralAngMomDecomposition}
\end{align}
where $\lambda_{1,2}$ and $\mu_{1,2}$ are initial and final helicities, $\lambda := \lambda_{1} - \lambda_{2}$ as well as $\mu := \mu_{1} - \mu_{2}$ and the quantum numbers $n_{\mathrm{in}/\mathrm{out}}$ can carry additional information to specify states (particle species, internal quantum numbers, $\ldots$). The partial wave amplitude $\left< \mu_{1} \mu_{2} \mathrm{;} n_{\mathrm{out}}  \right| \mathcal{T}^{j} \left| \lambda_{1} \lambda_{2} \mathrm{;} n_{\mathrm{out}}  \right>$ is a reduced matrix element which survives after the application of angular momentum conservation. \newline
Now, we specify the general series (\ref{eq:GeneralAngMomDecomposition}) to photoproduction. We fix the meson to be particle '$1$' in the final state, i.e. $\mu_{1} = 0$. Furthermore, we define
\begin{equation}
\mathcal{T}_{\mu_{2},\lambda_{1} \lambda_{2}} (s,t) := \left< \theta \phi\mathrm{;} \mu_{2} \mathrm{;} \varphi B \right| \mathcal{T} \left| 0 0 \mathrm{;} \lambda_{1} \lambda_{2} \mathrm{;} \gamma N \right> \hspace*{5pt} \mathrm{and} \hspace*{5pt} t_{\mu,\lambda}^{j} (s) := \left< \mu_{2} \mathrm{;} \varphi B  \right| \mathcal{T}^{j} \left| \lambda_{1} \lambda_{2} \mathrm{;} \gamma N \right> \mathrm{.} \label{eq:PhotoprodHelicityPWsDefinition} 
\end{equation}
Then, using equation (\ref{eq:DefGeneralSpinJRotationWignerDs}), the helicity partial wave series for photoproduction becomes
\begin{equation}
 \mathcal{T}_{\mu_{2},\lambda_{1} \lambda_{2}} (s,t) = e^{i\left(\lambda-\mu\right)\phi} \sum_{j=\mathrm{max}\left(\left|\lambda\right|,\left|\mu\right|\right)}^{\infty} \left(2 j + 1\right) t_{\mu,\lambda}^{j} (s) d_{\lambda \mu}^{\left(j\right)} \left( \theta \right)  \mathrm{.} \label{eq:HelicityPWExpPhotoproductionSpecified}
\end{equation}
For the conventional coordinate choice $\phi = 0$ (see Figure \ref{fig:SPinAmplitudeIllustrations}), the helicity transition amplitude on the left hand side coincides with the amplitudes $H_{i}$ defined above \cite{BransdenMoorhouse}. Furthermore, it is again customary to consider only photon-helicities $\lambda_{\gamma} = +1$, in order to arrive at parity-inequivalent amplitudes. Then, the relative helicities are
\begin{align}
 \lambda &= \lambda_{1} - \lambda_{2} = \left(\mathrm{photon} \hspace*{2.5pt} \mathrm{helicity}\right) - \left(\mathrm{initial}   \hspace*{2.5pt} \mathrm{nucleon} \hspace*{2.5pt} \mathrm{helicity}\right) \mathrm{,} \label{eq:InitialHelDifference} \\
 \mu &= - \mu_{2} = - \left(\mathrm{final} \hspace*{2.5pt} \mathrm{nucleon} \hspace*{2.5pt} \mathrm{helicity}\right) \mathrm{.} \label{eq:FinalHel}
\end{align}
They can take the values $\lambda \in \left\{\frac{1}{2},\frac{3}{2}\right\}$ and $\mu \in \left\{-\frac{1}{2},\frac{1}{2}\right\}$. One has $4$ complex partial wave amplitudes for every order in the $j$-expansion except for the lowest order, $j=1/2$, where only two are given. \newline
In the literature, different bases for the photoproduction partial waves are employed. We follow here the definitions of Bransden and Moorhouse \cite{BransdenMoorhouse}. One problem with the helicity partial waves $t_{\mu,\lambda}^{j}$ is that they are not parity-conserving. However, one can define parity-conserving helicity amplitudes $A_{\ell +}$, $B_{\ell+}$, $A_{(\ell+1) -}$ and $B_{(\ell+1)-}$. One should note that these partial waves carry the relative angular momentum quantum number $\ell$ in the final state as index and the $\pm$-signs distinguish the cases $j=\ell \pm 1/2$. \newline
However, the basis which is most commonly used is that of so-called {\it electric and magnetic multipoles} $E_{\ell \pm}$ and $M_{\ell \pm}$. Bransden and Moorhouse quote relations among the $A$- and $B$- amplitudes and the multipoles. From these relations, it is feasible to infer the following definitions of multipoles in terms of helicity partial waves (of course, again only valid within the conventions of reference \cite{BransdenMoorhouse}):
\begin{align}
 E_{\ell +} &:= \frac{1}{ \sqrt{2} (\ell + 1)} \left\{ \left( - t_{\frac{1}{2},\frac{1}{2}}^{j} + \sqrt{\frac{\ell}{\ell + 2}} \hspace*{2pt} t_{\frac{1}{2},\frac{3}{2}}^{j} \right) + \left( - t_{-\frac{1}{2},\frac{1}{2}}^{j} + \sqrt{\frac{\ell}{\ell + 2}} \hspace*{2pt} t_{-\frac{1}{2},\frac{3}{2}}^{j} \right) \right\} \mathrm{,} \label{eq:MultipoleDefsHelicityAmpls1} \\
 M_{\ell +} &:= \frac{(-1)}{ \sqrt{2} (\ell + 1)} \left\{ \left( t_{\frac{1}{2},\frac{1}{2}}^{j} + \sqrt{\frac{\ell+2}{\ell}} \hspace*{2pt} t_{\frac{1}{2},\frac{3}{2}}^{j} \right) + \left( t_{-\frac{1}{2},\frac{1}{2}}^{j} + \sqrt{\frac{\ell+2}{\ell}} \hspace*{2pt} t_{-\frac{1}{2},\frac{3}{2}}^{j} \right) \right\} \mathrm{,} \label{eq:MultipoleDefsHelicityAmpls2} \\
 E_{(\ell+1) -} &:= \frac{1}{ \sqrt{2} (\ell + 1)} \left\{ \left( - t_{\frac{1}{2},\frac{1}{2}}^{j} - \sqrt{\frac{\ell+2}{\ell}} \hspace*{2pt} t_{\frac{1}{2},\frac{3}{2}}^{j} \right) + \left( t_{-\frac{1}{2},\frac{1}{2}}^{j} + \sqrt{\frac{\ell+2}{\ell}} \hspace*{2pt} t_{-\frac{1}{2},\frac{3}{2}}^{j} \right) \right\} \mathrm{,} \label{eq:MultipoleDefsHelicityAmpls3} \\
 M_{(\ell+1) -} &:= \frac{1}{ \sqrt{2} (\ell + 1)} \left\{ \left( t_{\frac{1}{2},\frac{1}{2}}^{j} - \sqrt{\frac{\ell}{\ell+2}} \hspace*{2pt} t_{\frac{1}{2},\frac{3}{2}}^{j} \right) + \left( - t_{-\frac{1}{2},\frac{1}{2}}^{j} + \sqrt{\frac{\ell}{\ell+2}} \hspace*{2pt} t_{-\frac{1}{2},\frac{3}{2}}^{j} \right) \right\} \mathrm{.} \label{eq:MultipoleDefsHelicityAmpls4}
\end{align}
It should be mentioned that the definitions (\ref{eq:MultipoleDefsHelicityAmpls2}) and (\ref{eq:MultipoleDefsHelicityAmpls3}) are only valid provided $\ell \geq 1$. Thus, the two multipoles present for $j=1/2$ are $E_{0+}$ and $M_{1-}$. Then, one can attribute $4$ multipoles to each half-integer order $j \geq 3/2$. \newline
Finally, we quote without derivation the form of the photoproduction partial wave expansion which is commonly used in a lot of references (see, for instance \cite{Sandorfi:2010uv}) and which was first given by CGLN \cite{CGLN}. This is the multipole expansion for the $4$ CGLN-amplitudes, written here in terms of Legendre-polynomials and their derivatives:
\begin{align}
F_{1} \left( W, \theta \right) &= \sum \limits_{\ell = 0}^{\infty} \Big\{ \left[ \ell M_{\ell+} \left( W \right) + E_{\ell+} \left( W \right) \right] P_{\ell+1}^{'} \left( \cos \theta \right) \nonumber \\
 & \quad \quad \quad + \left[ \left( \ell+1 \right) M_{\ell-} \left( W \right) + E_{\ell-} \left( W \right) \right] P_{\ell-1}^{'} \left( \cos \theta \right) \Big\} \mathrm{,} \label{eq:MultExpF1} \\
F_{2} \left( W, \theta \right) &= \sum \limits_{\ell = 1}^{\infty} \left[ \left( \ell+1 \right) M_{\ell+} \left( W \right) + \ell M_{\ell-} \left( W \right) \right] P_{\ell}^{'} \left( \cos \theta \right) \mathrm{,} \label{eq:MultExpF2} \\
F_{3} \left( W, \theta \right) &= \sum \limits_{\ell = 1}^{\infty} \Big\{ \left[ E_{\ell+} \left( W \right) - M_{\ell+} \left( W \right) \right] P_{\ell+1}^{''} \left( \cos \theta \right) \nonumber \\
 & \quad \quad \quad + \left[ E_{\ell-} \left( W \right) + M_{\ell-} \left( W \right) \right] P_{\ell-1}^{''} \left( \cos \theta \right) \big\} \mathrm{,} \label{eq:MultExpF3} \\
F_{4} \left( W, \theta \right) &= \sum \limits_{\ell = 2}^{\infty} \left[ M_{\ell+} \left( W \right) - E_{\ell+} \left( W \right) - M_{\ell-} \left( W \right) - E_{\ell-} \left( W \right) \right] P_{\ell}^{''} \left( \cos \theta \right) \mathrm{.} \label{eq:MultExpF4}
\end{align}
This expansion is conventionally written as a series in $\ell$ instead of $j$. Furthermore, it can be inverted formally by a set of $4$ projection-integrals (see for instance Ball et al. \cite{JSBall})
\begin{align}
M_{\ell+} &= \frac{1}{2 \left( \ell+1 \right)} \int_{-1}^{1} dx \left[ F_{1} P_{\ell} \left( x \right) - F_{2} P_{\ell+1} \left( x \right) - F_{3} \frac{P_{\ell-1} \left( x \right) - P_{\ell+1} \left( x \right)}{ 2\ell + 1} \right] \mathrm{,} \label{eq:MlplusProjection} \\
E_{\ell+} &= \frac{1}{2 \left( \ell+1 \right)} \int_{-1}^{1} dx \bigg[ F_{1} P_{\ell} \left( x \right) - F_{2} P_{\ell+1} \left( x \right) + \ell F_{3} \frac{P_{\ell-1} \left( x \right) - P_{\ell+1} \left( x \right)}{ 2\ell + 1} \nonumber \\
 & \quad \quad \quad \quad \quad \quad \quad \quad + \left( \ell+1 \right) F_{4} \frac{P_{\ell} \left( x \right) - P_{\ell+2} \left( x \right)}{ 2\ell + 3} \bigg] \mathrm{,} \label{eq:ElplusProjection} \\
M_{\ell-} &= \frac{1}{2 \ell} \int_{-1}^{1} dx \left[ - F_{1} P_{\ell} \left( x \right) + F_{2} P_{\ell-1} \left( x \right) + F_{3} \frac{P_{\ell-1} \left( x \right) - P_{\ell+1} \left( x \right)}{ 2\ell + 1} \right] \mathrm{,} \label{eq:MlminusProjection} \\
E_{\ell-} &= \frac{1}{2 \ell} \int_{-1}^{1} dx \bigg[ F_{1} P_{\ell} \left( x \right) - F_{2} P_{\ell-1} \left( x \right) - \left( \ell+1 \right) F_{3} \frac{P_{\ell-1} \left( x \right) - P_{\ell+1} \left( x \right)}{ 2\ell + 1} \nonumber \\
 & \quad \quad \quad \quad \quad \enspace - \ell F_{4} \frac{P_{\ell-2} \left( x \right) - P_{\ell} \left( x \right)}{ 2\ell - 1} \bigg] \mathrm{.} \label{eq:ElminusProjection}
\end{align}
This form of the partial wave expansion will be employed in the remainder of this thesis. One of its advantages consists of the fact that the expansion (\ref{eq:MultExpF1}) to (\ref{eq:MultExpF4}) has been found to be consistent over the whole investigated literature, while other forms of the partial wave expansion may very well change depending on the reference. \newline
We close this section by providing some information of the physical meaning of the multipoles. Brief treatments of the following facts can be found in the paper by Drechsel and Tiator \cite{DrechselTiator} and the thesis by Leukel \cite{LeukelPhD}. An extensive formal treatment of photoproduction state-vectors with a certain multipolarity is given in the PhD-thesis by Alharbi \cite{AlharbiPhD}. \newline \newline
In the initial $\gamma N$-state, photon and nucleon have a relative orbital angular momentum, quantized by total angular momentum quantum number $g$. Then, a conventional coupling-scheme to write initial state partial wave basis vectors is to first couple $g$ and the photon spin $S_{\gamma} = 1$ to a total angular momentum quantum number $L_{\gamma}$ \cite{AlharbiPhD}. \newpage
\begin{table}[h]
\centering
\begin{tabular}{|c|c|c|c|c|c||c|c|c|c|c|c|}
\hline
\multicolumn{2}{|c|}{$ \gamma N $-system} & \multicolumn{4}{c||}{$ \varphi B $-system} & \multicolumn{2}{c}{$ \gamma N $-system} & \multicolumn{4}{|c|}{$ \varphi B $-system} \\
\hline
$ L_{\gamma} $ & $ \mathcal{M} L $ & $ J $ & $ \ell $ & $ \mathcal{M}_{\ell \pm} $ & $ P $ & $ L_{\gamma} $ & $ \mathcal{M} L $ & $ J $ & $ \ell $ & $ \mathcal{M}_{\ell \pm} $ & $ P $ \\
\hline
\hline
 $ 1 $ & $ E1 $ & $ 1/2 $ & $ 0 $ & $ E_{0+} $ & $ - $ &  $ 2 $ & $ E2 $ & $ 3/2 $ & $ 1 $ & $ E_{1+} $ & $ + $ \\
\cline{4 - 6} \cline{10 - 12}
  &  &  &  $ \displaystyle{\not} 1 $ &  &  &  &  &  &  $ \displaystyle{\not} 2 $ &  &  \\
\cline{3 - 6} \cline{9 - 12}
  &  & $ 3/2 $ &  $ \displaystyle{\not} 1 $ &  &  &  &  & $ 5/2 $ &  $ \displaystyle{\not} 2 $ &  &  \\
\cline{4 - 6} \cline{10 - 12}
  &  &  & $ 2 $ & $ E_{2-} $ & $ - $ &  &  &  & $ 3 $ & $ E_{3-} $ & $ + $ \\
\cline{2 - 6} \cline{8 - 12}
  & $ M1 $ & $ 1/2 $ & $ \displaystyle{\not} 0 $ &  &  &  & $ M2 $ & $ 3/2 $ & $ \displaystyle{\not} 1 $ &  &  \\
\cline{4 - 6} \cline{10 - 12}
  &  &  &  $ 1 $ & $ M_{1-} $ & $ + $ &  &  &  & $ 2 $ & $ M_{2-} $ & $ - $ \\
\cline{3 - 6} \cline{9 - 12}
  &  & $ 3/2 $ &  $ 1 $ & $ M_{1+} $ & $ + $ &  &  & $ 5/2 $ & $ 2 $ & $ M_{2+} $ & $ - $ \\
\cline{4 - 6} \cline{10 - 12}
  &  &  & $ \displaystyle{\not} 2 $ &  &  &  &  &  & $ \displaystyle{\not} 3 $ &  &  \\
\hline
\end{tabular}
\caption[Correspondence between photoproduction multipoles and quantum numbers.]{The final state multipoles $ \mathcal{M}_{\ell \pm} $ corresponding to electromagnetic dipole ($ L_{\gamma} = 1 $) and quadrupole ($ L_{\gamma} = 2 $) excitations, which can be inferred from the selection rules mentioned in the main text, are tabulated here. Parity conservation forbids some values for the quantum number $\ell$ of the final angular momentum's magnitude (those are the slashed ones). This table is taken over, up to slight modifications, from reference \cite{LeukelPhD}.}
\label{tab:MultipolesMesonToPhotonCorrespondence}
\end{table}
In case photon and nucleon were in a relative $S$-wave $g=0$, one would have $L_{\gamma} = 1$. Otherwise, for $g \geq 1$ only the possibilities
\begin{equation}
 L_{\gamma} = g+1,g,g-1 \mathrm{,} \label{eq:TotalPhotonSpinCouplingsNonSWave}
\end{equation}
remain. In order to obtain a total angular momentum for the initial state, one then couples $L_{\gamma}$ and $S_{N}=\frac{1}{2}$ to $j = \left|L_{\gamma} \pm \frac{1}{2}\right|$. Once the coupling is performed, Alharbi \cite{AlharbiPhD} again defines new linear combinations of the partial wave basis states which, according to his statements, help to ensure gauge invariance. These new partial wave states turn out to correspond to electric and magnetic photons. The parity eigenvalue for these initial state partial wave states then becomes $\bm{P}_{i} = \eta_{\gamma} \eta_{N} (-1)^{n+L_{\gamma}+1}$, where $n=1$ corresponds to excitations by magnetic photons ($ML$) and $n=2$ to electric photons ($EL$). The quantities $\eta_{\gamma}$ and $\eta_{N}$ are intrinsic parities. \newline
In the final state, the coupling of angular momenta is simpler and one obtains $j= \left|\ell \pm \frac{1}{2}\right|$ from the relative angular momentum $\ell$ between meson and baryon and the baryon spin $S_{B} = \frac{1}{2}$. The parity eigenvalue of the final partial wave state is $\bm{P}_{f} = \eta_{\varphi} \eta_{B} (-1)^{\ell}$. Setting the known intrinsic parities $\eta_{\gamma} = \eta_{\varphi} = -1$ and $\eta_{N} = \eta_{B} = +1$ (for an octet-baryon), as well as demanding conservation of $\bm{P}$ over the whole process, one arrives at well known selection rules for electric and magnetic photons \cite{DrechselTiator, LeukelPhD}
\begin{align}
EL \thinspace &\mathrm{:} \enspace \left( -1 \right)^{L} = \bm{P} = \left( -1 \right)^{\ell+1} \thinspace \Rightarrow \thinspace \left| L - \ell \right| = 1 \mathrm{,} \label{eq:ParityElectricPhoton} \\
ML \thinspace &\mathrm{:} \enspace \left( -1 \right)^{L+1} = \bm{P} = \left( -1 \right)^{\ell+1} \thinspace \Rightarrow \thinspace L = \ell  \mathrm{.} \label{eq:ParityMagneticPhoton}
\end{align}
These rules, combined with the conservation of $j$, allow to assign the final state multipole-amplitudes $E_{\ell \pm}$ and $M_{\ell \pm}$ to electromagnetic excitations $\mathcal{M} L$, as well as to definite conserved quantum numbers $J^{P}$. Therefore, resonances with certain spin-parity quantum numbers can only contribute to maximally two different multipoles. Results are shown in Table \ref{tab:MultipolesMesonToPhotonCorrespondence} for the lowest angular momenta. \newline
This concludes our introduction of photoproduction amplitudes. It is now of interest how they are linked to measurable quantities.

\subsubsection{Polarization observables} \label{subsec:PhotoproductionObs}

\begin{figure}[ht]
\centering
\vspace*{-7.5pt}
\includegraphics[width=0.75\textwidth]{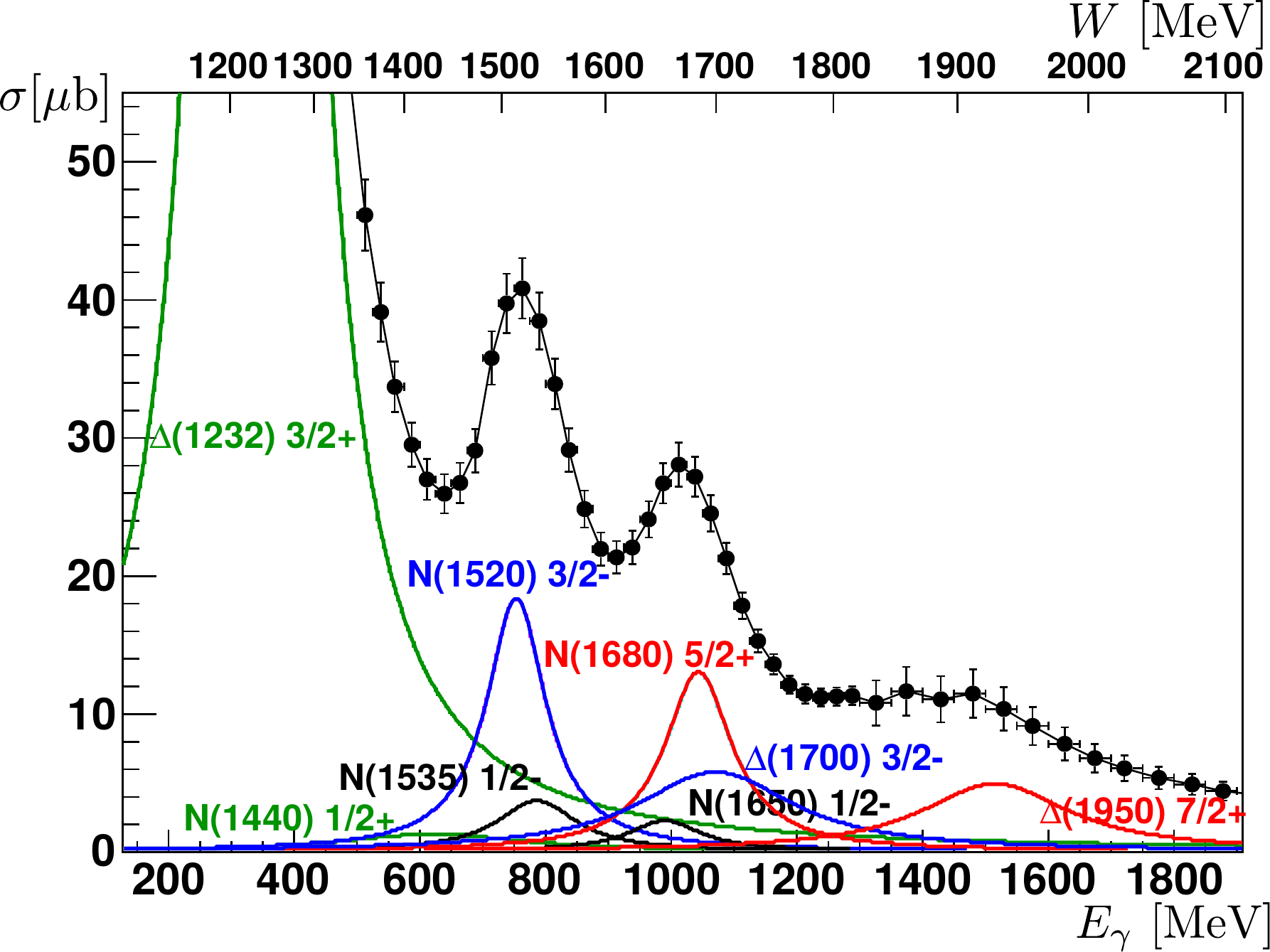} \\
\vspace*{-5pt}
\caption[The total cross section for $\gamma p \rightarrow \pi^{0} p$ and schematic Breit-Wigner resonances.]{The total cross section for $\gamma p \rightarrow \pi^{0} p$ is shown and compared to peaks which correspond to squared Breit-Wigner functions for well-established nucleon resonances \cite{Patrignani:2016xqp}. Bumps in the measured cross section receive here contributions from multiple overlapping resonances. $S$- and $P$-wave resonances ($\ell = 0,1$) are drawn in green, $D$-waves ($\ell = 2$) in blue, $F$-waves ($\ell = 3$) are red and $G$-waves ($\ell = 4$) black. The figure is taken from references \cite{Thiel:2016,LFitPaper}.}
\label{fig:WQResonances2015}
\end{figure}

The most basic expression for the photoproduction cross section, with spins specified in the CMS-frame, is given by the product of the phase space factor $\rho = q/k$ and the square of the matrix element (\ref{eq:DiractoCGLN}) \cite{Sandorfi:2010uv}
\begin{align}
\left( \frac{d \sigma}{d \Omega} \right) \left( \hat{\epsilon}, m_{s_{i}}, m_{s_{f}} \right) &= \frac{q}{k} \left| \big<m_{s_{f}}\big| F_{\mathrm{CGLN}} \big|m_{s_{i}}\big> \right|^{2} \mathrm{.} \label{eq:DCSfromCGLNSpinz}
\end{align}
However, in case an experiment is performed which features neither a manipulation of the spins of the initial state particles, nor a measurement of the final state spins, one can only measure the unpolarized differential cross section $\sigma_{0}$. This quantity follows from (\ref{eq:DCSfromCGLNSpinz}) by averaging over the initial state polarizations and summing over those in the final state:
\begin{equation}
\sigma_{0} (W,\theta) \equiv \left( \frac{d \sigma}{d \Omega} \right)_{0} (W,\theta) = \frac{1}{2} \sum \limits_{N\mathrm{-spins}} \thinspace \frac{1}{2} \sum \limits_{B\mathrm{-spins}} \thinspace \sum \limits_{\gamma \mathrm{-spins}} \left( \frac{d \sigma}{d \Omega} \right) \left( \hat{\epsilon}, m_{s_{i}}, m_{s_{f}} \right) \mathrm{.} \label{eq:UnpolarizedCSDefinition}
\end{equation}
In case one integrates this observable over the scattering angle, the total cross section $\bar{\sigma} := \int d\Omega \sigma_{0}$ is even easier to obtain from an experiment. It can be shown that, when written in terms of electric and magnetic multipoles, the total cross section $\bar{\sigma}$ becomes strictly a sum over squared moduli of multipoles (or partial waves) \cite{AlharbiPhD, LeukelPhD}:
\begin{align}
 \bar{\sigma} (W) &= 2 \pi \frac{q}{k} \sum_{\ell = 0}^{\infty} \Big\{ (\ell + 1)^{2} (\ell + 2) \left| E_{\ell+} \right|^{2} + (\ell - 1) \ell^{2} \left| E_{\ell -} \right|^{2}    \nonumber \\
 & \hspace*{62.5pt} + \ell (\ell + 1)^{2} \left| M_{\ell+} \right|^{2} + \ell^{2} (\ell + 1) \left| M_{\ell-} \right|^{2} \Big\} \mathrm{.} \label{eq:TCSInTermsOfMultsLZeroExpansionIntroduction}
\end{align}
In case the extraction of multipoles (and later: resonance parameters) from data for the total cross section $\bar{\sigma}$ would be the goal, a problem would emerge which is illustrated in Figure \ref{fig:WQResonances2015}. There, the measured cross section for the process $\gamma p \rightarrow \pi^{0} p$ is compared to squared Breit-Wigner functions for the well-established nucleon resonances in this energy-region. Nucleon resonances are all relatively short lived. Thus, they have a large width which, combined with the fact that the low-energy region is relatively densely populated with resonances, leads to the phenomenon that states overlap strongly. For a fixed energy $W$, it is therefore possible that many different resonances contribute to the total cross section (\ref{eq:TCSInTermsOfMultsLZeroExpansionIntroduction}) as a sum of squares and thus also to the bumps visible in the data in Figure \ref{fig:WQResonances2015}. Then, it is intuitively clear that further disentanglement of the resonance-contributions from $\bar{\sigma}$ alone is just impossible. \newline
In case the unpolarized differential cross section (\ref{eq:UnpolarizedCSDefinition}) were measured, the information from the angular distribution would provide access to some interference-terms among multipoles. Interference-terms will turn out to be just what is needed in order to achieve a unique partial wave decomposition, but it will seen later (section \ref{sec:CompExpsTPWA}) that the unpolarized cross section alone cannot provide enough such terms. \newline
From another point of view, in case one would wish to extract not the multipoles, but the full production amplitudes ($F_{i}$, $H_{i}$ or $b_{i}$) from the data, it again becomes clear quickly that the unpolarized cross section cannot suffice. As seen in section \ref{subsec:PhotoproductionAmpl}, up to an arbitrary choice of basis, $4$ complex amplitudes would have to be extracted.

\begin{figure}[ht]
\centering
\vspace*{-12.5pt}
\begin{overpic}[width=0.8\textwidth]%
      {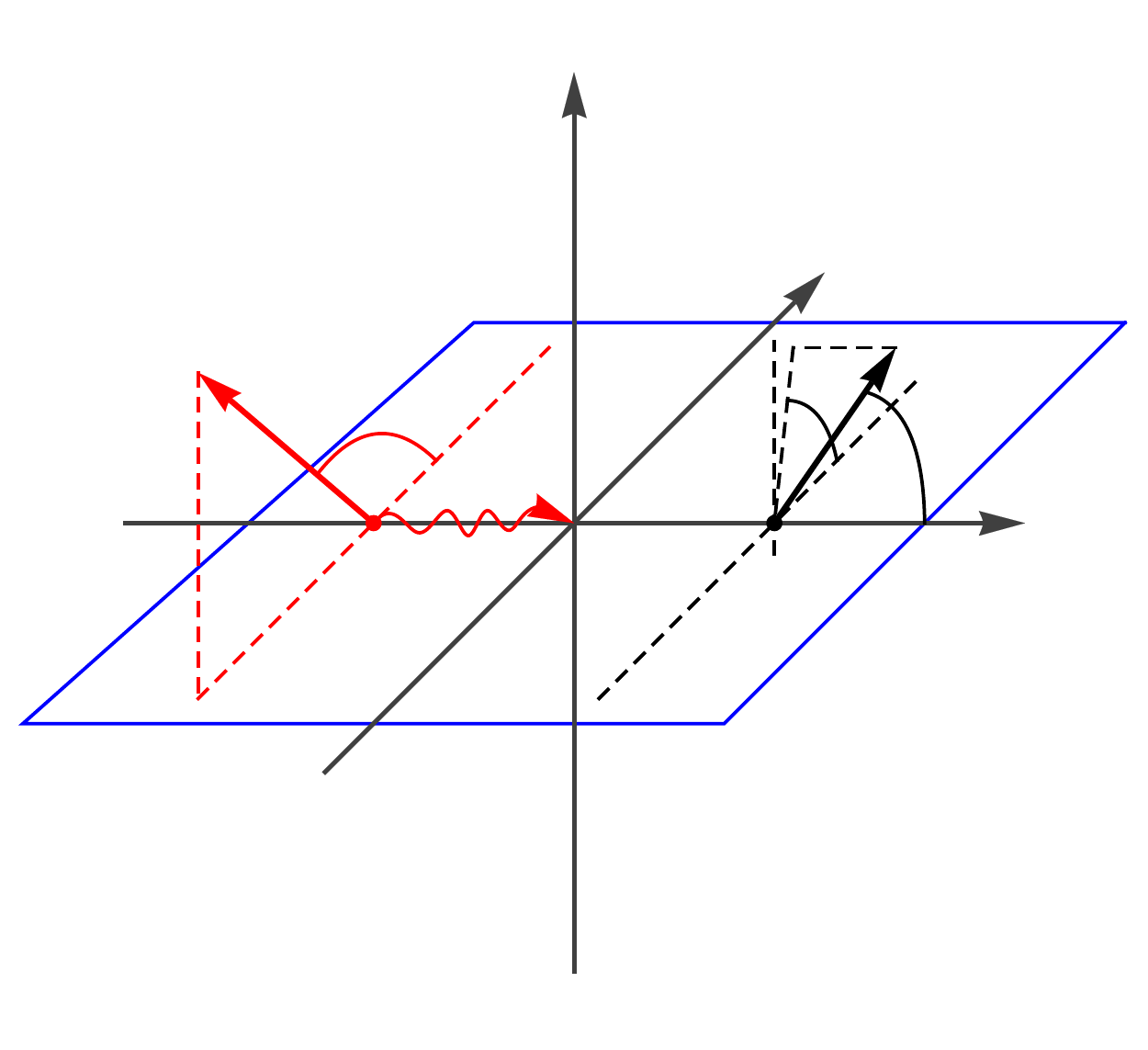}
      \put(0,84){$\boxed{\mathrm{initial}}$}
      \put(90,44.7){$\hat{z}$}
      \put(49,86.65){$\hat{y}$}
      \put(72.6,68.35){$\hat{x}$}
      \put(30,40.5){\begin{small}\textcolor{red}{$\gamma \left( \vec{k} \right)$}\end{small}}
      \put(31.5,49.75){\textcolor{red}{$\phi_{\gamma}$}}
      \put(14,60.5){\textcolor{red}{$\vec{\epsilon}_{L}$}}
      \put(62.45,39){\begin{small}$N \left( - \vec{k} \right)$\end{small}}
      \put(76,48.85){$\theta_{p}$}
      \put(68.6,53.05){$\phi_{p}$}
      \put(78.35,59.75){$\vec{P}^{T}$}
\end{overpic} \\ 
\vspace*{-5pt} \hspace*{2pt}
\begin{overpic}[width=0.8\textwidth]%
      {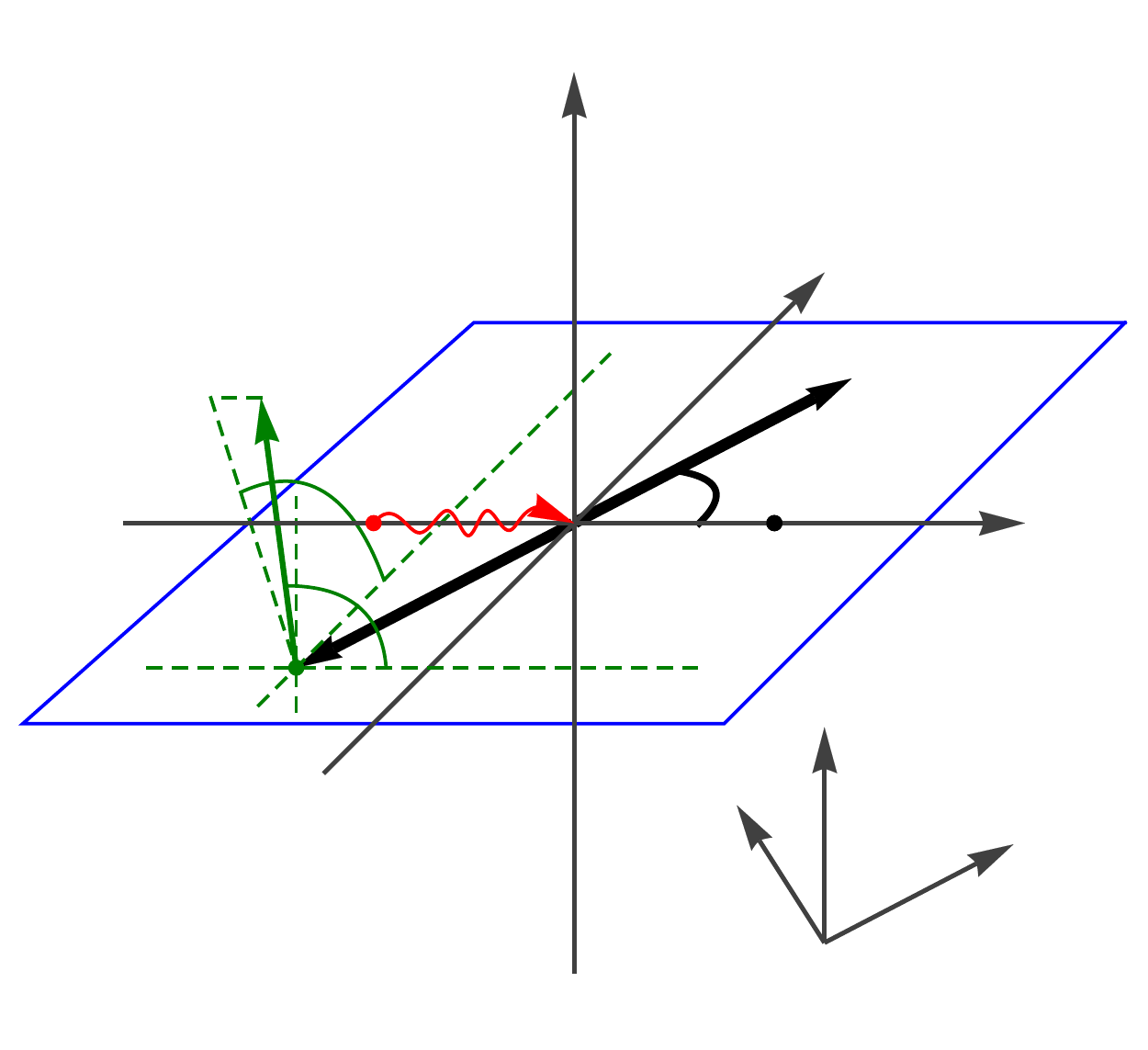}
      \put(0,84){$\boxed{\mathrm{final}}$}
      \put(90,44.7){$\hat{z}$}
      \put(49,86.65){$\hat{y}$}
      \put(72.6,68.35){$\hat{x}$}
      \put(58.15,46.75){$\theta$}
      \put(30,49.25){\begin{small}\textcolor{red}{$\gamma \left( \vec{k} \right)$}\end{small}}
      \put(62.45,41){\begin{small}$N \left( - \vec{k} \right)$\end{small}}
      \put(74.95,58.5){\begin{small}$\varphi \left( \vec{q} \hspace*{1pt} \right)$\end{small}}
      \put(12.95,29.95){\begin{small}$B \left( - \vec{q} \hspace*{1pt} \right)$\end{small}}
      \put(20.95,58){\textcolor{ForestGreen}{$\vec{P}^{R}$}}
      \put(26.35,37.3){\textcolor{ForestGreen}{$\theta_{p'}$}}
      \put(27.35,42.5){\textcolor{ForestGreen}{$\phi_{p'}$}}
      \put(89,17.4){$\hat{z}^{\prime} \upuparrows \vec{q}$}
      \put(70.85,29.45){$\hat{y}^{\prime} \equiv \hat{y}$}
      \put(61.7,21.35){$\hat{x}^{\prime}$}
\end{overpic}
\vspace*{-15pt}
\caption[Vectors and angles used to specify polarizations for pseudoscalar meson photoproduction in the CMS frame.]{The figure depicts vectors and angles used to specify polarizations for pseudoscalar meson photoproduction in the CMS frame. For a description of the angles $ \phi_{\gamma} $, $ \theta_{p} $, $ \phi_{p} $, $ \theta_{p'} $ and $ \phi_{p'} $ see the main text. Moduli of vectors are strongly exaggerated. The figures are very similar to those given by Sandorfi et al. \cite{Sandorfi:2010uv}.}
\label{fig:SandorfiCoordinates}
\end{figure}
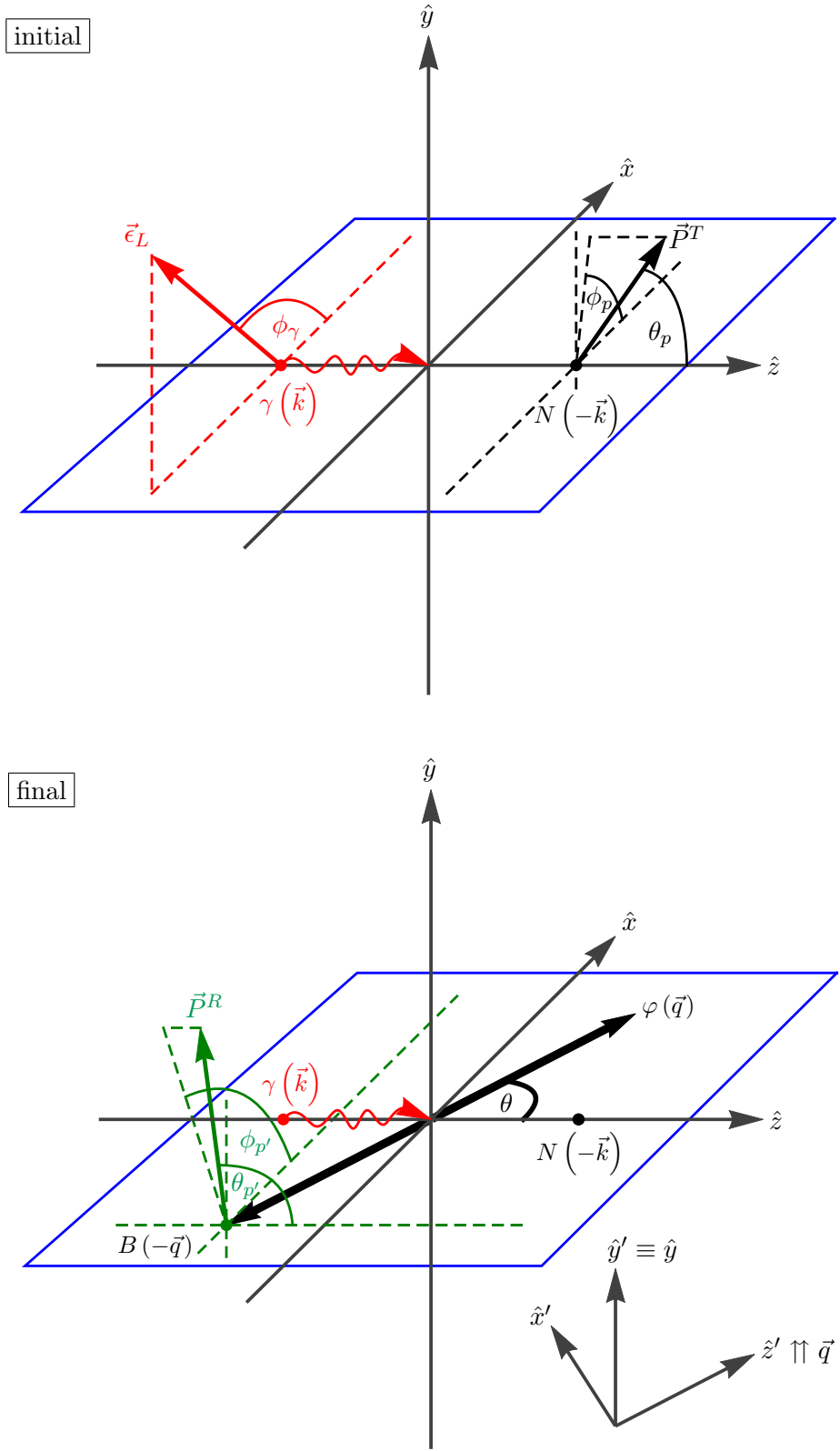

These correspond to $8$ real numbers or, upon fixing an arbitrary overall phase (more on this issue will follow later), to $7$ real quantities. These real degrees of freedom would have to be determined for every energy and angle. The unpolarized cross section $\sigma_{0}$ can only provide one real number and therefore is again insufficient. \newline
These considerations serve as enough motivation to define and measure so-called {\it polarization observables} in photoproduction. Definitions are simplified by writing the CGLN matrix-element for arbitrary directional spin-states for the target-nucleon $\left| \hat{P}^{T} \right>$ and recoil-baryon $\left| \hat{P}^{R} \right>$, respectively\footnote{A spin-state $\left| \hat{s} \right>$ having spin-projection $+1/2$ in the direction of an arbitrary unit-vector $\hat{s} = (\theta, \phi)$, i.e. $\left( \hat{\vec{S}} \cdot \hat{s} \right) \left| \hat{s} \right> = + \frac{1}{2} \left| \hat{s} \right>$, follows from an ordinary spin-$z$ state $\left| m \right>$ by use of rotation-matrices \cite{Sandorfi:2010uv}, similarly to section \ref{subsec:PhotoproductionAmpl}: $\left| \hat{s} \right> = \sum_{m} D^{(1/2)}_{m, + 1/2} \left( \phi, \theta, - \phi \right) \left| m \right> $.} \cite{Sandorfi:2010uv,MyDiplomaThesis}:
\begin{equation}
\left< \hat{P}^{R} \right| F_{\mathrm{CGLN}} \left( \hat{\epsilon} \right) \left| \hat{P}^{T} \right> \mathrm{.} \label{eq:CGLNMatrixElementSpinArbitrary}
\end{equation}
Figure \ref{fig:SandorfiCoordinates} provides a consistent polar-angle parametrization of the different polarization vectors within the CMS-frame $\left( \hat{x}, \hat{y}, \hat{z} \right)$. The $\hat{z}$-direction is chosen to coincide with the initial photon momentum $\vec{k}$. The Figures show the case of linear photon-polarization, with a polarization-vector $\vec{\epsilon}_{L}$ specified by an angle $\phi_{\gamma}$ rotating from the $\hat{x}$- to the $\hat{y}$-direction. The polarization-vectors $\vec{P}^{T}$ and $\vec{P}^{R}$ of the target-nucleon and recoil-baryon are shown as well, together with their respective polar- and azimuthal angles. \newline
The importance of the 'primed' coordinate system $\left( \hat{x}^{\prime}, \hat{y}^{\prime}, \hat{z}^{\prime} \right)$, visible explicitly in the lower half of Figure \ref{fig:SandorfiCoordinates}, should be stressed here. It arises from the system $\left( \hat{x}, \hat{y}, \hat{z} \right)$ by a rotation around the $\hat{y}$-axis by the scattering-angle $\theta$ (i.e., $\hat{e}_{z} \upuparrows \vec{k}$ and $\hat{e}_{z^{\prime}} \upuparrows \vec{q}$).
\newpage
\clearpage
This primed coordinate system is important for the names and definitions of polarization observables with recoil-polarization (cf. Table \ref{tab:DefObservables} below). \newline 
Using the appropriate polarization unit-vectors, the differential cross section for arbitrary beam-, target-, and recoil polarizations $(B,T,R)$ becomes
\begin{equation}
\left( \frac{d \sigma}{d \Omega} \right)^{\left( B, T, R \right)} \left( \hat{\epsilon}, \hat{P}^{T}, \hat{P}^{R} \right) = \frac{q}{k} \left| \left< \hat{P}^{R} \right| F_{\mathrm{CGLN}} \left( \hat{\epsilon} \right) \left| \hat{P}^{T} \right> \right|^{2} \mathrm{.} \label{eq:DCSDefinitionArbitrarySpinProjection}
\end{equation}
A polarization observable\footnote{We will follow here a notation according to Chiang and Tabakin \cite{ChTab}.} $\Omega$ is then defined as an asymmetry among differential cross sections for different polarization states $\left( B_{1}, T_{1}, R_{1} \right)$ and $\left( B_{2}, T_{2}, R_{2} \right)$, normalized to the unpolarized cross section \cite{Sandorfi:2010uv,MyDiplomaThesis}:
\begin{equation}
\Omega = \frac{\beta \left[ \left( \frac{d \sigma}{d \Omega} \right)^{\left( B_{1}, T_{1}, R_{1} \right)} - \left( \frac{d \sigma}{d \Omega} \right)^{\left( B_{2}, T_{2}, R_{2} \right)} \right]}{\left( \frac{d \sigma}{d \Omega} \right)_{0}} \mathrm{.} \label{eq:ObservableDefinitionGeneric}
\end{equation}
Irrespective of how both polarization configurations are chosen, the unpolarized cross section is always given by the sum
\begin{equation}
\enspace \left( \frac{d \sigma}{d \Omega} \right)_{0} = \beta \left[ \left( \frac{d \sigma}{d \Omega} \right)^{\left( B_{1}, T_{1}, R_{1} \right)} + \left( \frac{d \sigma}{d \Omega} \right)^{\left( B_{2}, T_{2}, R_{2} \right)} \right] \mathrm{.} \label{eq:UnpolarizedCrossSectionDenominator}
\end{equation}
The non-normalized asymmetry is often called a {\it profile function} \cite{ChTab,MyDiplomaThesis} and distinguished by a check-mark on the $\Omega$:
\begin{equation}
\check{\Omega} = \beta \left[ \left( \frac{d \sigma}{d \Omega} \right)^{\left( B_{1}, T_{1}, R_{1} \right)} - \left( \frac{d \sigma}{d \Omega} \right)^{\left( B_{2}, T_{2}, R_{2} \right)} \right] \mathrm{.} \label{eq:ProfileFunctionDefinition}
\end{equation}
This quantity of course carries the physical dimension of a cross section, while the observable (\ref{eq:ObservableDefinitionGeneric}) is dimension-less. The parameter $\beta$ is a consistency-factor introduced by Sandorfi et al. \cite{Sandorfi:2010uv}, which takes the value $\beta = 1/2$ for observables with only beam- and target polarization and $\beta = 1$ for quantities with recoil polarization. \newline
As it turns out, for the photoproduction of a single pseudoscalar meson, a total of $16$ non-redundant polarization asymmetries can be defined \cite{Barker75,ChTab}. They are conventionally given as the unpolarized cross section $\sigma_{0}$, three single-spin observables $\Sigma$, $T$ and $P$ as well as three more classes of observables with double-polarization of type beam-target ($\mathcal{BT}$), beam-recoil ($\mathcal{BR}$) and target-recoil ($\mathcal{TR}$). \newline
These classes of double-polarization observables contain $4$ quantities each. Furthermore, triple-polarization measurements may be defined, but they turn out to be related to the single- and double-polarization observables and are thus redundant. Sandorfi and collaborators show this fact in a rigorous derivation within the spin-density matrix formalism \cite{Sandorfi:2010uv}. \newline
The $16$ observables of pseudoscalar meson photoproduction are summarized in Table \ref{tab:DefObservables} and \ref{tab:ChTabHelTrObs}. The definitions of the corresponding asymmetries (\ref{eq:ObservableDefinitionGeneric}) and (\ref{eq:ProfileFunctionDefinition}) are listed in Table \ref{tab:DefObservables} and should be consistent with the CMS-coordinates shown in Figure \ref{fig:SandorfiCoordinates}. Table \ref{tab:ChTabHelTrObs} on the other hand collects expressions for the $16$ observables in terms of helicity- and transversity amplitudes $H_{i}$ and $b_{i}$. \newline
\begin{table}[ht]
\centering
\caption[Definitions of polarization observables in the CMS-frame.]{Overview of the definitions for the $16$ observables, taken from \cite{Sandorfi:2010uv}. "unp" indicates the need to average over initial, or to sum over final polarization-states (cf. \cite{Sandorfi:2010uv, MyDiplomaThesis}). For the definition of the appearing angles, cf. Figure \ref{fig:SandorfiCoordinates}. \newline A definite value for $\lambda_{\gamma}$ refers here to a state of circular photon polarization, where one has $\left( \hat{\epsilon}_{c} \right)_{\lambda_{\gamma} = \pm 1} = \mp \frac{1}{\sqrt{2}} \left( \hat{e}_{x} \pm i \hat{e}_{y} \right)$ \cite{Sandorfi:2010uv}. The variable $\phi_{\gamma}$ describes states of linear photon polarization, with $\hat{\epsilon}_{L} = \left( \cos \left(\phi_{\gamma}\right), \sin \left(\phi_{\gamma}\right), 0 \right)$.}
\label{tab:DefObservables}
\begin{tabular}{cccccccc}
  &  &  &  &  &  &  &  \\
\hline
\hline
\hline
  &  & \multicolumn{2}{c}{Beam} & \multicolumn{2}{c}{Target} & \multicolumn{2}{c}{Recoil} \\
 Observable & $ \left( \sigma_{1} - \sigma_{2} \right) $ & $ \lambda_{\gamma} $ & $ \phi_{\gamma} $ & $ \theta_{p} $ & $ \phi_{p} $ & $ \theta_{p'} $ & $ \phi_{p'} $ \\
\hline
\hline
\vspace*{-12pt} \\
 $ \sigma_{0} \left(\theta\right) $ &  & unp & unp & unp & unp & unp & unp \\
\hline
\vspace*{-10pt} \\
 $ 2 \check{\Sigma} $ & $ \sigma_{1} = \sigma \left( \perp, 0, 0 \right) $ & - & $ \pi/2 $ & unp & unp & unp & unp \\
  & $ \sigma_{2} = \sigma \left( \parallel, 0, 0 \right) $ & - & $ 0 $ & unp & unp & unp & unp \\
\hline
\vspace*{-10pt} \\
 $ 2 \check{T} $ & $ \sigma_{1} = \sigma \left( 0, +y, 0 \right) $ & unp & unp & $ \pi/2 $ & $ \pi/2 $ & unp & unp \\
  & $ \sigma_{2} = \sigma \left( 0, -y, 0 \right) $ & unp & unp & $ \pi/2 $ & $ 3 \pi/2 $ & unp & unp \\
\hline
\vspace*{-10pt} \\
 $ \check{P} $ & $ \sigma_{1} = \sigma \left( 0, 0, +y' \right) $ & unp & unp & unp & unp & $ \pi/2 $ & $ \pi/2 $ \\
  & $ \sigma_{2} = \sigma \left( 0, 0, -y' \right) $ & unp & unp & unp & unp & $ \pi/2 $ & $ 3 \pi/2 $ \\
\hline
\hline
\vspace*{-10pt} \\
 $ 2 \check{E} $ & $ \sigma_{1} = \sigma \left( +1, +z, 0 \right) $ & $ +1 $ & - & $ 0 $ & $ 0 $ & unp & unp \\
  & $ \sigma_{2} = \sigma \left( +1, -z, 0 \right) $ & $ +1 $ & - & $ \pi $ & $ 0 $ & unp & unp \\
\hline
\vspace*{-10pt} \\
 $ 2 \check{G} $ & $ \sigma_{1} = \sigma \left( + \pi/4, +z, 0 \right) $ & - & $ \pi/4 $ & $ 0 $ & $ 0 $ & unp & unp \\
  & $ \sigma_{2} = \sigma \left( + \pi/4, -z, 0 \right) $ & - & $ \pi/4 $ & $ \pi $ & $ 0 $ & unp & unp \\
\hline
\vspace*{-10pt} \\
 $ 2 \check{H} $ & $ \sigma_{1} = \sigma \left( + \pi/4, +x, 0 \right) $ & - & $ \pi/4 $ & $ \pi/2 $ & $ 0 $ & unp & unp \\
  & $ \sigma_{2} = \sigma \left( + \pi/4, -x, 0 \right) $ & - & $ \pi/4 $ & $ \pi/2 $ & $ \pi $ & unp & unp \\
\hline
\vspace*{-10pt} \\
 $ 2 \check{F} $ & $ \sigma_{1} = \sigma \left( +1, +x, 0 \right) $ & $ +1 $ & - & $ \pi/2 $ & $ 0 $ & unp & unp \\
  & $ \sigma_{2} = \sigma \left( +1, -x, 0 \right) $ & $ +1 $ & - & $ \pi/2 $ & $ \pi $ & unp & unp \\
\hline
\hline
\vspace*{-10pt} \\
 $ \check{C}_{x'} $ & $ \sigma_{1} = \sigma \left( +1, 0, +x' \right) $ & $ +1 $ & - & unp & unp & $ \pi/2 + \theta $ & $ 0 $ \\
  & $ \sigma_{2} = \sigma \left( +1, 0, -x' \right) $ & $ +1 $ & - & unp & unp & $ 3 \pi/2 + \theta $ & $ 0 $ \\
\hline
\vspace*{-10pt} \\
 $ \check{C}_{z'} $ & $ \sigma_{1} = \sigma \left( + 1, 0, +z' \right) $ & $ +1 $ & - & unp & unp & $ \theta $ & $ 0 $ \\
  & $ \sigma_{2} = \sigma \left( + 1, 0, -z' \right) $ & $ +1 $ & - & unp & unp & $ \pi + \theta $ & $ 0 $ \\
\hline
 \vspace*{-10pt} \\
 $ \check{O}_{x'} $ & $ \sigma_{1} = \sigma \left( + \pi/4, 0, +x' \right) $ & - & $ \pi/4 $ & unp & unp & $ \pi/2 + \theta $ & $ 0 $ \\
  & $ \sigma_{2} = \sigma \left( + \pi/4, 0, -x' \right) $ & - & $ \pi/4 $ & unp & unp & $ 3 \pi/2 + \theta $ & $ 0 $ \\
\hline
\vspace*{-10pt} \\
 $ \check{O}_{z'} $ & $ \sigma_{1} = \sigma \left( +\pi/4, 0, +z' \right) $ & - & $ \pi/4 $ & unp & unp & $ \theta $ & $ 0 $ \\
  & $ \sigma_{2} = \sigma \left( +\pi/4, 0, -z' \right) $ & - & $ \pi/4 $ & unp & unp & $ \pi + \theta $ & $ 0 $ \\
  \hline
  \hline
\vspace*{-10pt} \\
 $ \check{L}_{x'} $ & $ \sigma_{1} = \sigma \left( 0, +z, +x' \right) $ & unp & unp & $ 0 $ & $ 0 $ & $ \pi/2 + \theta $ & $ 0 $ \\
  & $ \sigma_{2} = \sigma \left( 0, +z, -x' \right) $ & unp & unp & $ 0 $ & $ 0 $ & $ 3\pi/2 + \theta $ & $ 0 $ \\
\hline
\vspace*{-10pt} \\
 $ \check{L}_{z'} $ & $ \sigma_{1} = \sigma \left( 0, +z, +z' \right) $ & unp & unp & $ 0 $ & $ 0 $ & $ \theta $ & $ 0 $ \\
  & $ \sigma_{2} = \sigma \left( 0, +z, -z' \right) $ & unp & unp & $ 0 $ & $ 0 $ & $ \pi + \theta $ & $ 0 $ \\
\hline
\vspace*{-10pt} \\
 $ \check{T}_{x'} $ & $ \sigma_{1} = \sigma \left( 0, +x, +x' \right) $ & unp & unp & $ \pi/2 $ & $ 0 $ & $ \pi/2 + \theta $ & $ 0 $ \\
  & $ \sigma_{2} = \sigma \left( 0, +x, -x' \right) $ & unp & unp & $ \pi/2 $ & $ 0 $ & $ 3 \pi/2 + \theta $ & $ 0 $ \\
\hline
\vspace*{-10pt} \\
 $ \check{T}_{z'} $ & $ \sigma_{1} = \sigma \left( 0, +x, +z' \right) $ & unp & unp & $ \pi/2 $ & $ 0 $ & $ \theta $ & $ 0 $ \\
  & $ \sigma_{2} = \sigma \left( 0, +x, -z' \right) $ & unp & unp & $ \pi/2 $ & $ 0 $ & $ \pi + \theta $ & $ 0 $ \\
\hline
\hline
\hline
\end{tabular}
\end{table}
\clearpage
One motivation for the detailed treatment of helicity- and transversity amplitudes in section \ref{subsec:PhotoproductionAmpl} is the fact that they have the special property of making the expressions for the polarization observables, shown in Table \ref{tab:ChTabHelTrObs}, look simple. Furthermore, the expressions seem to exhibit a certain regularity. Generally, it is seen that observables are bilinear forms\footnote{I.e., linear combinations of {\it bilinear products} $H_{i}^{\ast} H_{j}$ or $b_{i}^{\ast} b_{j}$.} defined by hermitean\footnote{Any matrix $\bm{\hat{M}}$ with complex entries is called {\it hermitean} if and only if: $\bm{\hat{M}}^{\dagger} = \left(\bm{\hat{M}}^{\ast} \right)^{T} = \left(\bm{\hat{M}}^{T} \right)^{\ast} \equiv \bm{\hat{M}}$ \cite{FalkoLorenz2}.} matrices. \newline
This fact is formalized by writing\footnote{The abbreviations $\check{\Omega}^{\alpha} = \frac{1}{2} \left< H \right| \Gamma^{\alpha} \left| H \right>$ and $\check{\Omega}^{\alpha} = \frac{1}{2} \left< b \right| \tilde{\Gamma}^{\alpha} \left| b \right>$ are used as well, cf. Table \ref{tab:ChTabHelTrObs}.} \cite{ChTab}
\begin{equation}
 \check{\Omega}^{\alpha} (W,\theta) = \frac{1}{2} \sum_{i,j=1}^{4} H_{i}^{\ast} (W,\theta) \Gamma^{\alpha}_{ij} H_{j} (W,\theta) \mathrm{,} \hspace*{5pt} \check{\Omega}^{\alpha} (W,\theta) = \frac{1}{2} \sum_{i,j=1}^{4} b_{i}^{\ast} (W,\theta) \tilde{\Gamma}^{\alpha}_{ij} b_{j} (W,\theta) \mathrm{,} \label{eq:ObservablesBilHelTrProductForm}
\end{equation}
where the index $\alpha$ counts observables: $\alpha = 1,\ldots,16$. Furthermore, the regularity of the expressions comes from the defining algebras of hermitean and unitary Dirac $\Gamma$-matrices, or $\tilde{\Gamma}$-matrices, respectively (cf. reference \cite{ChTab}). The algebra defined by such matrices will turn out to be of fundamental importance to this work and therefore expressions for the matrices, as well as their most important properties, are given in appendix \ref{subsec:HelTrGammaReps}. \newline
In case the observables are written in the CGLN-amplitudes, the resulting expressions are still bilinear but are defined by more complicated, $\theta$-dependent matrices $\check{\Omega}^{\alpha} = \frac{1}{2} \left< F \right| \hat{A}^{\alpha} (\theta) \left| F \right>$. The matrices can be found in appendix \ref{subsec:CGLNMatrixReps}, while fully written-out expressions can be seen in references \cite{Sandorfi:2010uv, FTS}. \newline
For more general $2\rightarrow2$ reactions involving particles with spin, at least for cases with an even number of $n$ amplitudes, it turns out that observables are written in the transversity basis by a generalization of the $\Gamma$-algebra to $n \times n$-matrices \cite{ChTab}. Examples would be the electroproduction of pseudoscalar mesons $e N \rightarrow e^{\prime} \varphi N$ ($6$ amplitudes) \cite{LotharEtAlElectro}, the photoproduction of two pseudoscalar mesons, for instance pions $\gamma N \rightarrow \pi \pi N$ \cite{RobertsOed} or even the simpler reaction of pion-nucleon scattering $\pi N \rightarrow \pi N$ \cite{Hoehler84,DeanLee}. Therefore, one may expect that some of the results obtained in this work for pseudoscalar meson photoproduction may hold in more general circumstances. \newline
The $16$ observables listed in Table \ref{tab:ChTabHelTrObs} turn out to be not independent, but instead quadratic equations among the quantities $\check{\Omega}^{\alpha}$ exist \cite{Sandorfi:2010uv}. The quadratic constraints come from so-called {\it Fierz-identities} for the $\Gamma$-matrices \cite{ChTab}. These identities are explained in more detail in appendix \ref{sec:ObservableAlgebra}, while a consistent set of $37$ quadratic constraints are listed, for quick reference, in appendix \ref{subsec:FierzIdentities}. \newline
Two facts are striking about the observable algebra: first of all, the $7$ real numbers defining a set of $4$ phase-constrained full spin-amplitudes are now accompanied by $16$ measurable quantities. Therefore, a measurement of all $16$ observables is likely to result in an over-constraint on the amplitudes. Secondly, the observables themselves have been proven to be mathematically dependent on each other in non-trivial ways. Both observations motivate the search of so-called {\it complete experiments} \cite{Barker75,ChTab}. We describe this problem for full spin-amplitudes in the next section and generalize it to the extraction of multipoles in section \ref{sec:CompExpsTPWA}.

\begin{sidewaystable}[h]
\centering
\begin{tabular}{cccccc}
\hline
\hline \\
Observable & Helicity-representation & Transversity-representation & BHP-form & BTP-form & Type \\
\hline \\
$ \check{\Omega}^{1} = \sigma_{0} \left(\theta\right) $ & $ \frac{1}{2} \left( \left| H_{1} \right|^{2} + \left| H_{2} \right|^{2} + \left| H_{3} \right|^{2} + \left| H_{4} \right|^{2} \right) $ & $ \frac{1}{2} \left( \left| b_{1} \right|^{2} + \left| b_{2} \right|^{2} + \left| b_{3} \right|^{2} + \left| b_{4} \right|^{2} \right) $ & $ \frac{1}{2} \left< H \right| \Gamma^{1} \left| H \right> $ & $ \frac{1}{2} \left< b \right| \tilde{\Gamma}^{1} \left| b \right> $ & \\
$ \check{\Omega}^{4} = - \check{\Sigma} $ & $ \mathrm{Re} \left[ H_{1} H_{4}^{\ast} - H_{2} H_{3}^{\ast} \right] $ & $ \frac{1}{2} \left( \left| b_{1} \right|^{2} + \left| b_{2} \right|^{2} - \left| b_{3} \right|^{2} - \left| b_{4} \right|^{2} \right) $ & $ \frac{1}{2} \left< H \right| \Gamma^{4} \left| H \right> $ & $ \frac{1}{2} \left< b \right| \tilde{\Gamma}^{4} \left| b \right> $ & $\mathcal{S}$ \\
$ \check{\Omega}^{10} = - \check{T} $ & $ \mathrm{Im} \left[ H_{1} H_{2}^{\ast} + H_{3} H_{4}^{\ast} \right] $ & $ \frac{1}{2} \left( - \left| b_{1} \right|^{2} + \left| b_{2} \right|^{2} + \left| b_{3} \right|^{2} - \left| b_{4} \right|^{2} \right) $ & $ \frac{1}{2} \left< H \right| \Gamma^{10} \left| H \right> $ & $ \frac{1}{2} \left< b \right| \tilde{\Gamma}^{10} \left| b \right> $ & \\
$ \check{\Omega}^{12} = \check{P} $ & $ \mathrm{Im} \left[ - H_{1} H_{3}^{\ast} - H_{2} H_{4}^{\ast} \right] $ & $ \frac{1}{2} \left( - \left| b_{1} \right|^{2} + \left| b_{2} \right|^{2} - \left| b_{3} \right|^{2} + \left| b_{4} \right|^{2} \right) $ & $ \frac{1}{2} \left< H \right| \Gamma^{12} \left| H \right> $ & $ \frac{1}{2} \left< b \right| \tilde{\Gamma}^{12} \left| b \right> $ & \\
 & & & & \\
$ \check{\Omega}^{3} = \check{G} $ & $ \mathrm{Im} \left[ H_{1} H_{4}^{\ast} - H_{3} H_{2}^{\ast} \right] $ & $ \mathrm{Im} \left[ - b_{1} b_{3}^{\ast} - b_{2} b_{4}^{\ast} \right] $ & $ \frac{1}{2} \left< H \right| \Gamma^{3} \left| H \right> $ & $ \frac{1}{2} \left< b \right| \tilde{\Gamma}^{3} \left| b \right> $ & \\
$ \check{\Omega}^{5} = \check{H} $ & $ \mathrm{Im} \left[ - H_{2} H_{4}^{\ast} + H_{1} H_{3}^{\ast} \right] $ & $ \mathrm{Re} \left[ b_{1} b_{3}^{\ast} - b_{2} b_{4}^{\ast} \right] $ & $ \frac{1}{2} \left< H \right| \Gamma^{5} \left| H \right> $ & $ \frac{1}{2} \left< b \right| \tilde{\Gamma}^{5} \left| b \right> $ & $\mathcal{BT}$ \\
$ \check{\Omega}^{9} = - \check{E} $ & $ \frac{1}{2} \left( \left| H_{1} \right|^{2} - \left| H_{2} \right|^{2} + \left| H_{3} \right|^{2} - \left| H_{4} \right|^{2} \right) $ & $ \mathrm{Re} \left[ b_{1} b_{3}^{\ast} + b_{2} b_{4}^{\ast} \right] $ & $ \frac{1}{2} \left< H \right| \Gamma^{9} \left| H \right> $ & $ \frac{1}{2} \left< b \right| \tilde{\Gamma}^{9} \left| b \right> $ & \\
$ \check{\Omega}^{11} = \check{F} $ & $ \mathrm{Re} \left[ - H_{2} H_{1}^{\ast} - H_{4} H_{3}^{\ast} \right] $ & $ \mathrm{Im} \left[ b_{1} b_{3}^{\ast} - b_{2} b_{4}^{\ast} \right] $ & $ \frac{1}{2} \left< H \right| \Gamma^{11} \left| H \right> $ & $ \frac{1}{2} \left< b \right| \tilde{\Gamma}^{11} \left| b \right> $ & \\
 & & & & \\
$ \check{\Omega}^{14} = \check{O}_{x'} $ & $ \mathrm{Im} \left[ - H_{2} H_{1}^{\ast} + H_{4} H_{3}^{\ast} \right] $ & $ \mathrm{Re} \left[ - b_{1} b_{4}^{\ast} + b_{2} b_{3}^{\ast} \right] $ & $ \frac{1}{2} \left< H \right| \Gamma^{14} \left| H \right> $ & $ \frac{1}{2} \left< b \right| \tilde{\Gamma}^{14} \left| b \right> $ & \\
$ \check{\Omega}^{7} = - \check{O}_{z'} $ & $ \mathrm{Im} \left[ H_{1} H_{4}^{\ast} - H_{2} H_{3}^{\ast} \right] $ & $ \mathrm{Im} \left[ - b_{1} b_{4}^{\ast} - b_{2} b_{3}^{\ast} \right] $ & $ \frac{1}{2} \left< H \right| \Gamma^{7} \left| H \right> $ & $ \frac{1}{2} \left< b \right| \tilde{\Gamma}^{7} \left| b \right> $ & $\mathcal{BR}$ \\
$ \check{\Omega}^{16} = - \check{C}_{x'} $ & $ \mathrm{Re} \left[ H_{2} H_{4}^{\ast} + H_{1} H_{3}^{\ast} \right] $ & $ \mathrm{Im} \left[ b_{1} b_{4}^{\ast} - b_{2} b_{3}^{\ast} \right] $ & $ \frac{1}{2} \left< H \right| \Gamma^{16} \left| H \right> $ & $ \frac{1}{2} \left< b \right| \tilde{\Gamma}^{16} \left| b \right> $ & \\
$ \check{\Omega}^{2} = - \check{C}_{z'} $ & $ \frac{1}{2} \left( \left| H_{1} \right|^{2} + \left| H_{2} \right|^{2} - \left| H_{3} \right|^{2} - \left| H_{4} \right|^{2} \right) $ & $ \mathrm{Re} \left[ b_{1} b_{4}^{\ast} + b_{2} b_{3}^{\ast} \right] $ & $ \frac{1}{2} \left< H \right| \Gamma^{2} \left| H \right> $ & $ \frac{1}{2} \left< b \right| \tilde{\Gamma}^{2} \left| b \right> $ & \\
 & & & & \\
$ \check{\Omega}^{6} = - \check{T}_{x'} $ & $ \mathrm{Re} \left[ - H_{1} H_{4}^{\ast} - H_{2} H_{3}^{\ast} \right] $ & $ \mathrm{Re} \left[ - b_{1} b_{2}^{\ast} + b_{3} b_{4}^{\ast} \right] $ & $ \frac{1}{2} \left< H \right| \Gamma^{6} \left| H \right> $ & $ \frac{1}{2} \left< b \right| \tilde{\Gamma}^{6} \left| b \right> $ & \\
$ \check{\Omega}^{13} = - \check{T}_{z'} $ & $ \mathrm{Re} \left[ - H_{1} H_{2}^{\ast} + H_{4} H_{3}^{\ast} \right] $ & $ \mathrm{Im} \left[ b_{1} b_{2}^{\ast} - b_{3} b_{4}^{\ast} \right] $ & $ \frac{1}{2} \left< H \right| \Gamma^{13} \left| H \right> $ & $ \frac{1}{2} \left< b \right| \tilde{\Gamma}^{13} \left| b \right> $ & $\mathcal{TR}$ \\
$ \check{\Omega}^{8} = \check{L}_{x'} $ & $ \mathrm{Re} \left[ H_{2} H_{4}^{\ast} - H_{1} H_{3}^{\ast} \right] $ & $ \mathrm{Im} \left[ - b_{1} b_{2}^{\ast} - b_{3} b_{4}^{\ast} \right] $ & $ \frac{1}{2} \left< H \right| \Gamma^{8} \left| H \right> $ & $ \frac{1}{2} \left< b \right| \tilde{\Gamma}^{8} \left| b \right> $ & \\
$ \check{\Omega}^{15} = \check{L}_{z'} $ & $ \frac{1}{2} \left( - \left| H_{1} \right|^{2} + \left| H_{2} \right|^{2} + \left| H_{3} \right|^{2} - \left| H_{4} \right|^{2} \right) $ & $ \mathrm{Re} \left[ - b_{1} b_{2}^{\ast} - b_{3} b_{4}^{\ast} \right] $ & $ \frac{1}{2} \left< H \right| \Gamma^{15} \left| H \right> $ & $ \frac{1}{2} \left< b \right| \tilde{\Gamma}^{15} \left| b \right> $ & \\
 & & & & \\
\hline
\hline
\end{tabular}
\caption[Polarization observables in terms of helicity- and transversity amplitudes.]{ This Table collects the definitions of the $16$ polarization observables in terms of the helicity- ($H_{i}$) as well as transversity amplitudes ($b_{i}$) defined in section \ref{subsec:PhotoproductionAmpl} (see appendix \ref{sec:ObservableAlgebra} for matrix representations of the $\Gamma$- and $\tilde{\Gamma}$-matrices). The non-normalized asymmetries (or: profile functions) (\ref{eq:ProfileFunctionDefinition}) are shown here and therefore a factor of $\rho = q/k$ should appear in front of each quantity which is here, however, suppressed. The Table is taken over (up to sign-changes) from reference \cite{ChTab}.}
\label{tab:ChTabHelTrObs}
\end{sidewaystable}

\clearpage

\subsection{Complete experiments: full production amplitudes} \label{sec:CompExpsFullAmp}

\begin{figure}[ht]
\begin{overpic}[width=0.499\textwidth,trim=0 125 0 0,clip]%
      {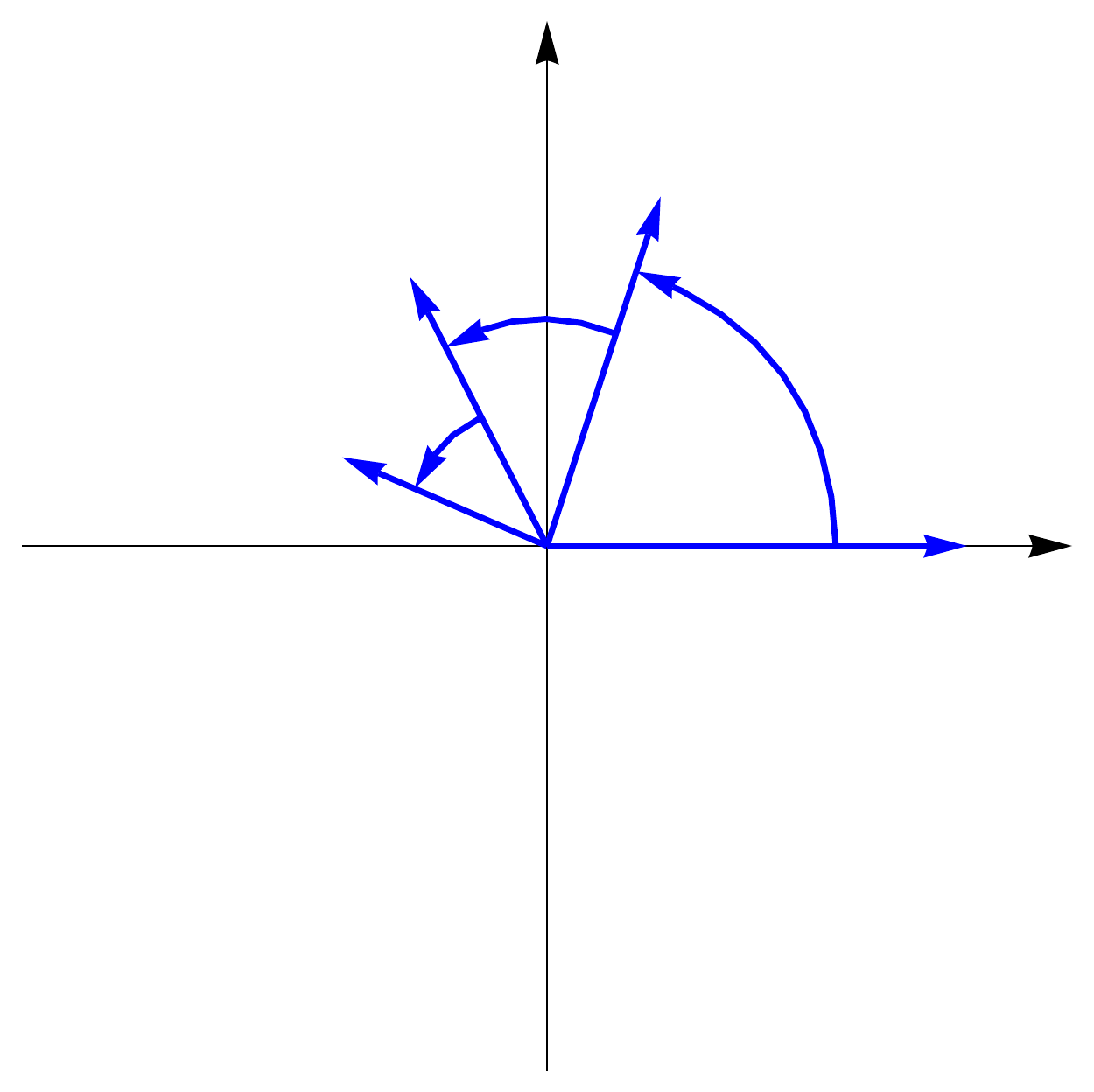}
  \put(92.6,8.5){Re}
  \put(52.65,60.25){Im}
  \put(84.1,18.0){\textcolor{blue}{$ \tilde{b_{1}} $}}
  \put(59.6,49.0){\textcolor{blue}{$ \tilde{b_{2}} $}}
  \put(32.1,43.0){\textcolor{blue}{$ \tilde{b_{3}} $}}
  \put(25.0,23.5){\textcolor{blue}{$ \tilde{b_{4}} $}}
  \put(73.1,33.5){\textcolor{blue}{$\phi^{b}_{21}$}}
  \put(44.6,40.0){\textcolor{blue}{$\phi^{b}_{32}$}}
  \put(32.0,27.0){\textcolor{blue}{$\phi^{b}_{43}$}}
\end{overpic}
\begin{overpic}[width=0.499\textwidth,trim=0 125 0 0,clip]%
      {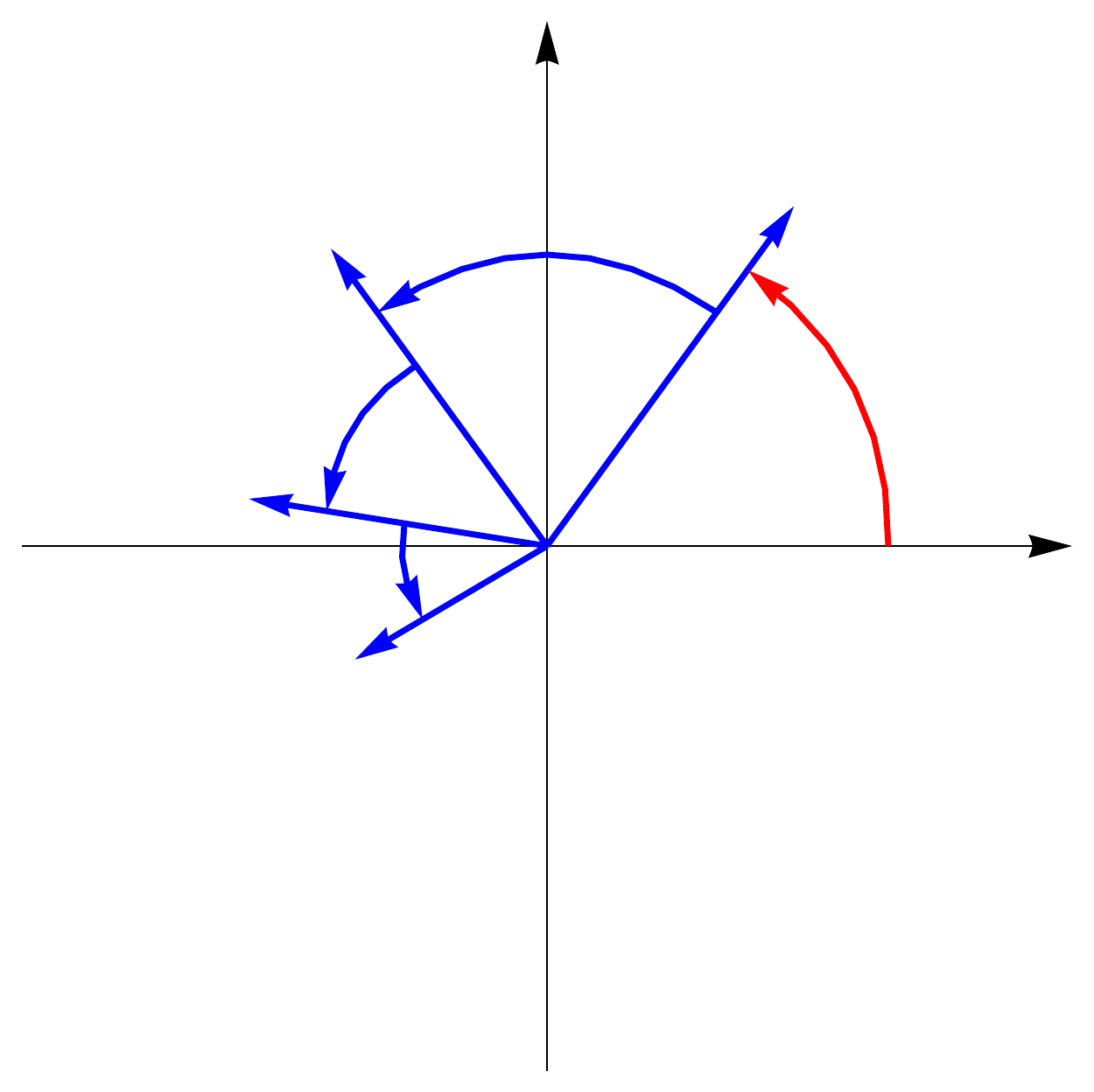}
  \put(92.6,8.5){Re}
  \put(52.65,60.25){Im}
  \put(80.6,28.5){\textcolor{red}{$ \phi^{b} \left( W, \theta \right) $}}
  \put(72.6,49.0){\textcolor{red}{$ b_{1} $}}
  \put(25.1,45.0){\textcolor{red}{$ b_{2} $}}
  \put(17.0,19.5){\textcolor{red}{$ b_{3} $}}
  \put(26.5,2.5){\textcolor{red}{$ b_{4} $}}
  \put(44.6,45.5){\textcolor{blue}{$\phi^{b}_{21}$}}
  \put(23.5,28.5){\textcolor{blue}{$\phi^{b}_{32}$}}
  \put(28.0,10.5){\textcolor{blue}{$\phi^{b}_{43}$}}
\end{overpic}
\caption[Transversity amplitudes $b_{i}$ in the complex plane, with and without phase-constraint.]{The figures illustrate the arrangement of the full transversity amplitudes in the complex plane. This schematic is taken from the thesis \cite{MyDiplomaThesis}, where it was drawn for CGLN-amplitudes. Left: Phase-constrained amplitudes $ \tilde{b}_{i} $, defined by the property that $\tilde{b}_{1}$ has to be real and positive, are plotted. This set of amplitudes can be reconstructed from a complete set of measurements. Three relative phase angles are indicated. Right: The assumed \textit{true} solution for the actual transversity amplitudes $ b_{i} $ is shown, which is obtained from the $ \tilde{b}_{i} $ via a rotation by a connecting overall phase $ \phi^{b} \left(W, \theta \right) $. All quantities that are drawn in red, i.e. $ \phi^{b} \left(W, \theta \right) $ as well as the $ b_{i} $, cannot be calculated directly from a complete experiment.}
\label{fig:CompareCompExAmplitudesToActualAmplitudes}
\end{figure}
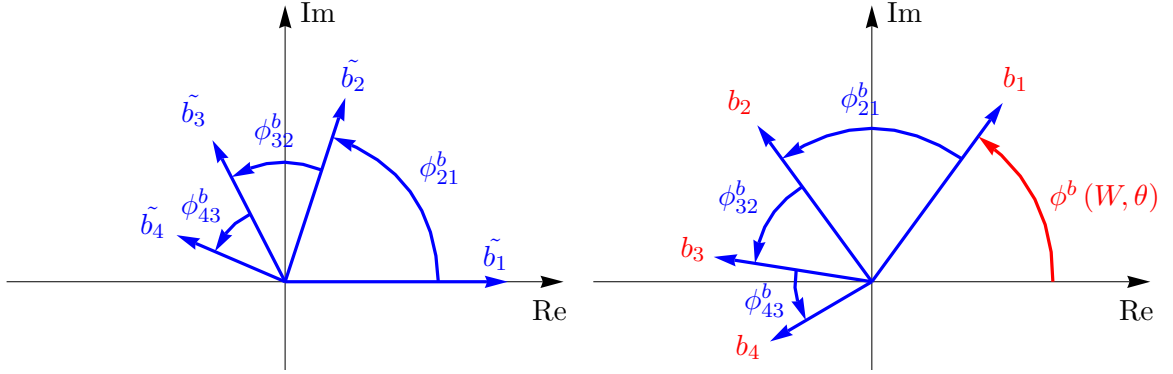

Now, we consider the problem of how many and which of the $16$ polarization observables $\check{\Omega}^{\alpha}$ need to be measured in order to uniquely extract the 
full spin-amplitudes of photoproduction, i.e. for instance the helicity amplitudes $\left\{ H_{i} \right\}$ or transversity amplitudes $\left\{ b_{i} \right\}$. 
This issue is called \textit{complete experiment problem} and received quite some attention in the literature \cite{Barker75, KeatonWorkman1, ChTab, Ireland:2010bi, Vrancx:2013pza, Nys:2015kqa, Nys:2016uel, WorkmanEtAl2017}. Another definition 
of a complete experiment is that of a \textit{minimal set of measurements capable of predicting all other possible experiments} \cite{Tiator:2011tu}. It can be seen that complete experiments defined in the latter terms are equivalent to those defined via amplitude extraction. \newline
It should be mentioned that here we describe this problem only in its academic, or mathematically precise, version. This means we assume ideal data without any
uncertainties. Then, the extraction of amplitudes is a mathematically well-defined problem. \newline
In a very recent publication \cite{WorkmanEtAl2017}, the extraction of full spin amplitudes out of polarization asymmetries has been investigated
again and was there termed a \textit{Complete Experiment Analysis (CEA)}, thus we adopt this term here. Furthermore, for simplicity, from now on we consider the problem 
for transversity amplitudes. \newline
It should be stated that the term \textit{uniquely} for the extraction of amplitudes above means, more precisely, unique up to a phase. The observables are invariant
under a very general rotation of all transversity amplitudes by an energy- and angle-dependent phase\footnote{This transformation is actually, in the literature on partial
wave analysis, called a \textit{continuum ambiguity} \cite{BowcockBurkhardt} (more precisely, in this case, a \textit{1-fold} continuum ambiguity) and it plays an important
role in the topic of ambiguities in the extraction of partial waves from data (see sections \ref{sec:CompExpsTPWA} and \ref{sec:OmelaenkoApproachIntro}).}
\begin{equation}
 b_{i} (W,\theta) \longrightarrow e^{i \Phi (W,\theta)} b_{i} (W,\theta) \mathrm{,} \label{eq:BiTrAmplitudesPhaseTrafo}
\end{equation}
because of their bilinear structure (see eq. (\ref{eq:ObservablesBilHelTrProductForm}) and Table \ref{tab:ChTabHelTrObs}). Due to this invariance, the overall phase
in return cannot be obtained from the data \cite{MyDiplomaThesis}. \newline
The complete experiment problem can be reformulated more precisely as a question for which minimum subsets of all $16$ observables are sufficient in order to extract 
the moduli $\left| b_{i} \right|$ and relative phases $\phi^{b}_{kj}$ (i.e., a phase-angle between two transversity amplitudes $b_{k}$ and $b_{j}$) uniquely. Figure 
\ref{fig:CompareCompExAmplitudesToActualAmplitudes} gives an illustration. \newline
Chiang and Tabakin published a solution to this problem in 1997 \cite{ChTab}. They find that generally \textit{8 carefully selected observables} can constitute a complete experiment.
We outline their findings here in a bit more detail for two reasons:
\begin{itemize}
 \item[(i)] The result has lead directly to the inception of this thesis, where the same question will be asked for the extraction of partial waves (see section \ref{sec:CompExpsTPWA}). 
 Therefore, the result by Chiang and Tabakin will serve as an important reference point to what is found later. 
 \item[(ii)] It is important to outline the logic behind the approach of Chiang and Tabakin, since we will follow a very similar one for the extraction of partial waves. 
\end{itemize}
The first step Chiang and Tabakin undertook consisted of a systematic study of discrete linear and antilinear ambiguities of the observables. This was motivated by an earlier study
from roughly the same time \cite{KeatonWorkman1}, which constructed special cases of such kinds of ambiguities and showed that certain complete sets of 9 observables claimed earlier by
Barker et al. \cite{Barker75} were in fact incomplete. However, first the concepts of the above mentioned ambiguities should be defined more precisely. \newline

\clearpage

Linear and antilinear ambiguities are transformations of the four amplitudes of the following form:
\begin{equation}
 b_{i} \rightarrow b_{i}^{L} = \sum_{j=1}^{4} L_{ij} b_{j} \mathrm{,} \hspace*{10pt} b_{i} \rightarrow b_{i}^{A} = \sum_{j=1}^{4} A_{ij} b_{j}^{\ast} \mathrm{.} \label{eq:LanAntiLinAmbChTabChapter2}
\end{equation}
Here, the generally complex matrices $L$ and $A$ are unitary and generate the linear- and antilinear ambiguity, respectively. Unitarity of the matrices is required because the cross section $\sigma_{0} = 1/2 \frac{q}{k} \sum_{i} \left| b_{i} \right|^{2}$ has to be left invariant. More generally, a subset of observables $\check{\Omega}^{\alpha_{n}}$ represented by Gamma-matrices $\tilde{\Gamma}^{\alpha_{n}}$ is invariant under a linear ambiguity in case the condition \cite{ChTab}
\begin{equation}
 L^{\dagger} \tilde{\Gamma}^{\alpha_{n}} L = \tilde{\Gamma}^{\alpha_{n}} \mathrm{,} \label{eq:LinearAmbTest1}
\end{equation}
is fulfilled for all $\alpha_{n}$. Because of the assumed unitarity of $L$, this condition is just equivalent to
the vanishing of the commutator
\begin{equation}
 \left[ L, \tilde{\Gamma}^{\alpha_{n}} \right] = 0 \mathrm{.} \label{eq:LinearAmbTest2}
\end{equation}
For the antilinear ambiguity, the invariance condition reads \cite{ChTab}
\begin{equation}
 \left( A^{\dagger} \tilde{\Gamma}^{\alpha_{n}} A \right)^{T} = \tilde{\Gamma}^{\alpha_{n}} \mathrm{.} \label{eq:AntiLinearAmbTest}
\end{equation}
Chiang and Tabakin now made an argument that, in order to test for (anti-) linear ambiguities generated by the most general matrices $L$ and $A$, the conditions (\ref{eq:LinearAmbTest1}) and (\ref{eq:AntiLinearAmbTest}) have to be checked only for the $\tilde{\Gamma}$-matrices. This is justified by the fact that every complex $4 \times 4$ matrix can be expanded into $\tilde{\Gamma}$ matrices (cf. appendix \ref{subsec:HelTrGammaReps}). The group $\mathcal{S}$ observables are always assumed to be measured, since they fix the moduli of the $b_{i}$ (see the discussion further below). Thus, only those $\tilde{\Gamma}$-matrices have to be tested which are either linear or antilinear ambiguities of the four observables $\left\{\sigma_{0}, \Sigma, T, P\right\}$. Chiang and Tabakin performed this task and came up with interesting ambiguity-patterns for the remaining $12$ observables. In particular, subsets of $8$ or even more observables (sometimes up to $10$) were identified which all share a particular ambiguity. Examples of subsets fulfilling the completeness-rules proposed earlier by Barker et al. \cite{Barker75} have been disproved in this way. \newline
All subsets of observables that do not share any discrete ambiguity come up automatically as candidates for complete experiments. However, the studies of ambiguities could only serve as a first hint at complete sets. Chiang and Tabakin assumed that ambiguity-transformations with a more general mathematical structure than that of the discrete ambiguities discussed above may exist, i.e. some non-linear transformations
\begin{equation}
 b_{i} \longrightarrow b_{i}^{\prime} \left( b_{1}, b_{2}, b_{3}, b_{4} \right) \mathrm{.} \label{eq:TransversityAmplitudesGeneralTransformation}
\end{equation}
Therefore, a sufficient condition for each complete set consists only of a solution of the inverse problem posed by the extraction of moduli and relative-phases of the transversity-amplitudes. This explicit solution represents the second important step in the argument. \newline
An elegant way towards this goal consists of employing the completeness relation of the $\tilde{\Gamma}$-matrices (see appendix \ref{subsec:HelTrGammaReps})
\begin{equation}
\frac{1}{4} \sum_{\alpha = 1}^{16} \tilde{\Gamma}^{\alpha}_{ba} \tilde{\Gamma}^{\alpha}_{st} = \delta_{as} \delta_{bt} \mathrm{,} \label{eq:CompletenessRelGammaTildeMatrices}
\end{equation}
in order to formally invert the bilinear forms (\ref{eq:ObservablesBilHelTrProductForm}) posed by the observables themselves for the so-called bilinear products of the transversity amplitudes:
\begin{equation}
 b_{i}^{\ast} b_{j} = \frac{1}{2} \sum_{\alpha=1}^{16} \left( \tilde{\Gamma}^{\alpha}_{ij} \right)^{\ast} \check{\Omega}^{\alpha} \mathrm{.} \label{eq:BilProductTransversityExplicitExpression}
\end{equation}
It is seen quickly that this expression yields the moduli $\left| b_{i} \right| = \sqrt{b_{i}^{\ast} b_{i}}$ and relative phases $e^{i \phi^{b}_{kj}} = b_{j}^{\ast} b_{k} / \left( \left| b_{j} \right| \left| b_{k} \right| \right)$. In particular, it is found that the four group $\mathcal{S}$ observables completely fix the moduli $\left| b_{i} \right|$ (cf. Table \ref{tab:ChTabHelTrObs}). Reference \cite{MyDiplomaThesis} contains a listing of the resulting formulas. \newline
The extraction of analytic expressions for the bilinear products represents the easy part of the CEA. The much more difficult next step is the reduction of the right-hand-side of equation (\ref{eq:BilProductTransversityExplicitExpression}) to a minimum number of $8$ observables. In order to do this, one has to employ the so-called \textit{Fierz-identities} for the $\tilde{\Gamma}$ matrices, which hold as a result of their fundamental algebraic properties (see appendix \ref{subsec:FierzIdentities})
\begin{equation}
 \tilde{\Gamma}^{\alpha}_{ij} \tilde{\Gamma}^{\beta}_{st} = \sum \limits_{\delta, \eta}^{} C_{\delta \eta}^{\alpha \beta} \tilde{\Gamma}^{\delta}_{it} \tilde{\Gamma}^{\eta}_{sj} \quad \mathrm{where} \quad C_{\delta \eta}^{\alpha \beta} = \frac{1}{16} \mathrm{Tr} \left[ \tilde{\Gamma}^{\delta} \tilde{\Gamma}^{\alpha} \tilde{\Gamma}^{\eta} \tilde{\Gamma}^{\beta} \right] \mathrm{.} \label{eq:FierzrelationsIntro}
\end{equation}
These relations imply constraints among products of the observables $\check{\Omega}^{\alpha}$, defined by the Fierz-traces $C_{\delta \eta}^{\alpha \beta}$:
\begin{equation}
\check{\Omega}^{\alpha} \check{\Omega}^{\beta} = \sum \limits_{\delta, \eta}^{} C^{\alpha \beta}_{\delta \eta} \check{\Omega}^{\delta} \check{\Omega}^{\eta} \mathrm{.} \label{eq:FierzRelObservablesIntro}
\end{equation}
In total, $37$ such identities have been given by Chiang and Tabakin \cite{ChTab}. Appendix \ref{subsec:FierzIdentities} provides a listing, which is consistent with the sign-conventions used in this work.

\begin{sidewaystable}[h]
\caption[Example for one of the Tables by Chiang and Tabakin, listing complete sets for the CEA in photoproduction.]{A list of some complete experiments according to Chiang and Tabakin (ref. \cite{ChTab}) is given. The nomenclature is as follows: The unpolarized cross section as well as Group S observables $\left\{\sigma_{0}, \Sigma, T, P\right\}$ are already assumed to be measured in each case. The $ X $-symbols mark three measurements that are additionally selected from the double-polarization observables, while the $ O $'s indicate the choices for a fourth observable that yields a complete set (i.e., resolves all ambiguities). This Table is taken over identically from reference \cite{ChTab}.}
\label{tab:CompExExampleList}
\begin{tabular}{cccccccccccccccccccccccccccccc}
  &  &  &  &  &  &  &  &  &  &  &  &  &  &  &  &  &  &  &  &  &  &  &  &  &  &  &  &  &  \\
\hline
\hline
  &  &  &  &  &  &  &  &  &  &  &  &  &  &  &  &  &  &  &  &  &  &  &  &  &  &  &  &  &  \\
$ G $ &  &  $ X $ & $ X $ & $ X $ & $ X $ & $ X $ & $ X $ & $ X $ & $ X $ &  & $ X $ & $ X $ & $ X $ & $ X $ & $ X $ & $ X $ & $ X $ & $ X $ &  & $ X $ & $ X $ & $ X $ & $ X $ & $ X $ & $ X $ & $ X $ & $ X $ &  & \\
$ H $ &  &  $ X $ & $ X $ & $ X $ & $ X $ & $ X $ & $ X $ & $ X $ & $ X $ &  &  &  &  &  &  &  &  &  &  &  &  &  &  &  &  &  &  &  & $\mathcal{BT}$ \\
$ E $ &  &  &  &  &  &  &  &  &  &  & $ X $ & $ X $ & $ X $ & $ X $ & $ X $ & $ X $ & $ X $ & $ X $ &  &  &  &  &  &  &  &  &  &  &  \\
$ F $ &  &  &  &  &  &  &  &  &  &  &  &  &  &  &  &  &  &  &  & $ X $ & $ X $ & $ X $ & $ X $ & $ X $ & $ X $ & $ X $ & $ X $ &  & \\
  &  &  &  &  &  &  &  &  &  &  &  &  &  &  &  &  &  &  &  &  &  &  &  &  &  &  &  &  &  \\
  &  &  &  &  &  &  &  &  &  &  &  &  &  &  &  &  &  &  &  &  &  &  &  &  &  &  &  &  &  \\
$ O_{x'} $ &  & $ X $ &  &  $ O $ &  &  $ O $ & $ O $ & $ O $ & $ O $ &  & $ X $ & $ O $ & $ O $ & $ O $ & $ O $ & $ O $ & $ O $ & $ O $ &  & $ X $ &  & $ O $ &  & $ O $ &  &  & $ O $ &  &  \\
$ O_{z'} $ &  &  &  $ X $ &  &  $ O $ & $ O $ & $ O $ & $ O $ & $ O $ &  & $ O $ & $ X $ & $ O $ & $ O $ & $ O $ & $ O $ & $ O $ & $ O $ &  &  & $ X $ &  & $ O $ &  & $ O $ & $ O $ &  &  & $\mathcal{BR}$ \\
$ C_{x'} $ &  & $ O $ &  &  $ X $ &  &  $ O $ & $ O $ & $ O $ & $ O $ &  & $ O $ & $ O $ & $ X $ & $ O $ & $ O $ & $ O $ & $ O $ & $ O $ &  & $ O $ &  &  $ X $ &  &  & $ O $ & $ O $ &  &  &  \\
$ C_{z'} $ &  &  &  $ O $ &  &  $ X $ & $ O $ & $ O $ & $ O $ & $ O $ &  & $ O $ & $ O $ & $ O $ & $ X $ & $ O $ & $ O $ & $ O $ & $ O $ &  &  & $ O $ &  &  $ X $ & $ O $ &  &  & $ O $ &  & \\
  &  &  &  &  &  &  &  &  &  &  &  &  &  &  &  &  &  &  &  &  &  &  &  &  &  &  &  &  &  \\
  &  &  &  &  &  &  &  &  &  &  &  &  &  &  &  &  &  &  &  &  &  &  &  &  &  &  &  &  &  \\
$ T_{x'} $ &  & $ O $ & $ O $ & $ O $ & $ O $ & $ X $ & $ O $ & $ O $ & $ O $ &  & $ O $ & $ O $ & $ O $ & $ O $ & $ X $ &  &  $ O $ &  &  & $ O $ &  &  & $ O $ & $ X $ &  & $ O $ &  &  &  \\
$ T_{z'} $ &  & $ O $ & $ O $ & $ O $ & $ O $ & $ O $ & $ X $ & $ O $ & $ O $ &  & $ O $ & $ O $ & $ O $ & $ O $ &  &  $ X $ &  &  $ O $ &  &  & $ O $ & $ O $ &  &  & $ X $ &  &  $ O $ &  & $\mathcal{TR}$ \\
$ L_{x'} $ &  & $ O $ & $ O $ & $ O $ & $ O $ & $ O $ & $ O $ & $ X $ & $ O $ &  & $ O $ & $ O $ & $ O $ & $ O $ & $ O $ &  & $ X $ &  &  &  & $ O $ & $ O $ &  & $ O $ &  & $ X $ &  &  &  \\
$ L_{z'} $ &  & $ O $ & $ O $ & $ O $ & $ O $ & $ O $ & $ O $ & $ O $ & $ X $ &  & $ O $ & $ O $ & $ O $ & $ O $ &  &  $ O $ &  & $ X $ &  & $ O $ &  &  & $ O $ &  & $ O $ &  &  $ X $ &  & \\
  &  &  &  &  &  &  &  &  &  &  &  &  &  &  &  &  &  &  &  &  &  &  &  &  &  &  &  &  &  \\
\hline
\hline
\end{tabular}
\end{sidewaystable}

\clearpage

It is now a central claim of reference \cite{ChTab}, that the Fierz-relations are enough to show that the complete sets of $8$ observables are in fact complete. This means, it should always be possible to use the Fierz-identities in order to express the asymmetries which are not contained in the complete set by those who are. In the end, there turn out to exist quite a lot of complete sets. They have all been tabulated by Chiang/Tabakin and we provide here only one of their tables as an example (Table \ref{tab:CompExExampleList}). \newline
A few general features of the complete experiments for the extraction of the full spin amplitudes are \cite{ChTab}:
\begin{itemize}
 \item[1.)] The group $\mathcal{S}$ observables $\left\{\sigma_{0}, \Sigma, T, P\right\}$ have to be measured, since they fix the moduli $\left|b_{i}\right|$ of the four transversity amplitudes. This is also convenient, since the single-spin obervables are most easily measured. However, the single-spin observables are not the only set of quantities which can be expressed purely by moduli in a certain spin-amplitude basis. Other sets can be brought to such a form as well (cf. appendix \ref{subsec:AmbiguitiesSimDiagObservablesI}), such that starting with the group $\mathcal{S}$ is in some sense a convention and not mandatory.
 \item[2.)] From the remaining $4$ double polarization measurements, no more than two observables are allowed to be picked from the same class (i.e. $\mathcal{BT}$, $\mathcal{BR}$ and $\mathcal{TR}$). This fact directly implies that the double polarization observables with recoil polarization ($\mathcal{BR}$ and $\mathcal{TR}$) have to be measured for a unique extraction of the full spin amplitudes. This is troubling since these kinds of observables are extremely difficult to obtain in an experiment.
\end{itemize}
It is actually not difficult to see that one of the implications of point 2.), namely that single-spin and $\mathcal{BT}$-observables alone cannot yield all moduli and relative phases of the $b_{i}$, is true. The group $\mathcal{S}$ fixes the moduli $\left| b_{i} \right|$, while an explicit evaluation of equation (\ref{eq:BilProductTransversityExplicitExpression}) (or alternatively, inspection of Table \ref{tab:ChTabHelTrObs}) shows that the $\mathcal{BT}$-observables yield only two relative phases, $\phi^{b}_{13}$ and $\phi^{b}_{24}$. One needed relative phase (for instance $\phi^{b}_{23}$) is undetermined, which leads to a continuum of ambiguous solutions. \newline
This concludes the discussion of the CEA according to Chiang and Tabakin \cite{ChTab}. We commence with establishing the central issue of this thesis, which consists of the analogous problem, but for the extraction of multipoles.

\subsection{Complete experiments in a truncated partial wave analysis (TPWA)} \label{sec:CompExpsTPWA}

While an extraction of the full spin-amplitudes from real data up to an energy- and angle dependent phase would by itself already represent an impressive achievement, it would unfortunately be insufficient to uniquely constrain the multipoles. This is true due to the following argument pointed out by Sarantsev \cite{AndreyPrivComm} during the writing of the thesis \cite{MyDiplomaThesis}. \newline
For simplicity, we assume that CGLN amplitudes $F_{i}$ have been extracted uniquely from a complete experiment up to an overall phase. This assumption is justified, since the amplitude extraction formulated for transversity amplitudes in section \ref{sec:CompExpsFullAmp} is in fact possible for any other amplitude basis (see reference \cite{MyDiplomaThesis} for more details). This means, we again assume that the {\it true} solution $F_{i}$ is connected to phase-constrained amplitudes $\tilde{F}_{i}$ by an energy- and angle-dependent phase: $F_{i} (W, \theta) = e^{i \phi^{F} (W, \theta)} \tilde{F}_{i} (W,\theta)$. Furthermore, let the $\tilde{F}_{i}$ be constrained in such a way that $\tilde{F}_{1}$ is real and positive. Then, the multipole-projection formula for e.g. $M_{\ell +}$ becomes:
\begin{align}
M_{\ell+} &= \frac{1}{2 \left( \ell+1 \right)} \int_{-1}^{1} dx \left[ F_{1} P_{\ell} \left( x \right) - F_{2} P_{\ell+1} \left( x \right) - F_{3} \frac{P_{\ell-1} \left( x \right) - P_{\ell+1} \left( x \right)}{ 2\ell + 1} \right] \nonumber \\
 &= \frac{1}{2 \left( \ell+1 \right)} \int_{-1}^{1} d\left(\cos \theta\right) e^{i \phi^{F}\left( W, \theta \right)} \bigg[ \left| F_{1} \right| P_{\ell} \left( \cos \theta \right) - \left| F_{2} \right| e^{i \phi^{F}_{21}} P_{\ell+1} \left( \cos \theta \right) \nonumber \\
 & \quad \quad \quad \quad \quad \quad  \quad  \quad \quad \quad  \quad \quad - \left| F_{3} \right| e^{i \phi^{F}_{31}} \frac{P_{\ell-1} \left( \cos \theta \right) - P_{\ell+1} \left( \cos \theta \right)}{ 2\ell + 1} \bigg] \mathrm{.} \label{eq:OverallPhaseKillerArgument}
\end{align}
Here, the moduli and relative phases (or generally the term in square brackets in the expression above) are fixed by the complete experiment\footnote{A subtle point: The CEA operates on a discrete grid in $(W,\theta)$, defined by wherever the idealized data are given. However, the multipole-projection needs the moduli and relative phases at least as continuous functions in the angular variable $\cos \theta$. Thus, some kind of interpolation of the CEA-results would be needed. How the latter is performed (splines, fits of high-order polynomials, ...), is in principle a source of systematic uncertainty. Also, the TPWA discussed in this section is in principle a method of angular interpolation. However, here the cutoff of the polynomials $\ell_{\mathrm{max}}$ has a physical significance.}. However, the angular-dependence of $\phi^{F}\left( W, \theta \right)$ is {\it unknown} and cannot be inferred from the CEA. Therefore, the projection to the {\it true} multipole solution cannot be performed without knowledge of the phase\footnote{Generally, partial waves defining the decompositions of amplitudes that are related by an angle-dependent phase (or continuum ambiguity transformation, cf. section \ref{sec:CompExpsFullAmp}) are admixtures of one another. The mixing-relations are linear but infinite series. For the $2 \rightarrow 2$-scattering of scalar particles, the mixing-formula is simple and will be elaborated a bit more in section \ref{sec:OmelaenkoApproachIntro}. For photoproduction, mixing-relations become a lot more involved and generally things become more subtle, since one has $4$ amplitudes which can be rotated generally by different phases ($4$-fold continuum ambiguity). This issue is not touched further here. We just mention that mixing-phenomena have been mentioned in earlier works \cite{DeanLee, Omelaenko} and received some more recent attention \cite{AlfredPhasePaper, MyPhasePaper}.}. For the remaining multipoles, the situation is similar. \newline

The problem of the unknown overall phase therefore blocks the way from the CEA to multipoles. A possible way out of this issue is to not choose the way of the full amplitude extraction, but instead to try and {\it 'solve'} for the multipoles directly. In this Ansatz, one truncates the infinite multipole-expansions of the full (e.g. CGLN) amplitudes, equations (\ref{eq:MultExpF1}) to (\ref{eq:MultExpF4}), at some maximal angular momentum quantum number $\ell_{\mathrm{max}}$, inserts this finite expansion into the observables $\check{\Omega}^{\alpha}$ and then extracts the multipoles directly from the data. This procedure defines the so-called {\it truncated partial wave analysis (TPWA)} for photoproduction and it is the central object of interest in this thesis. The associated numerical fit-procedure is generally referred to as a {\it single-energy (SE)} fit (or energy-independent fit), as opposed to the energy-dependent (ED) methods summarized in section \ref{subsec:Energy-dependentModels}. \newline
One can then ask the question for an algebraic uniqueness-problem for the multipole-solutions (a detailed definition is given below). The problem arises essentially as the synthesis of the algebra of the photoproduction-observables described in section \ref{subsec:PhotoproductionObs} (specifically, see Table \ref{tab:ChTabHelTrObs}), as well as appendix \ref{sec:ObservableAlgebra}, and a truncated form of the partial wave expansion (i.e. the assumption of a polynomial amplitude). \newline
However, in order to really treat the uniqueness-, or completeness-, problem mathematically exact, one needs to assume the existence of exact solutions for truncated partial wave analyses. I.e. one assumes that the polynomials arising from the truncation are an exact representation of the $F_{i}$ with only a finite number of multipoles. This assumption is idealized and never realized in nature (this is discussed further below). Thus, while one can study the TPWA as a purely algebraic uniqueness problem, one needs to bear in mind that the underlying assumptions are always academic. Before commencing with a more detailed statement of the complete experiment problem for the TPWA, it is useful to systematize the expansions a bit more. Here, we follow the scheme and notation introduced by Tiator in his NSTAR-proceeding \cite{Tiator:2011tu}. \newline
First of all, it is useful to determined the maximal orders of $\cos \theta$ appearing in the polynomial-amplitudes $F_{i}$ for any specific truncation order $\ell_{\mathrm{max}}$. These highest orders can, in principle, be simply read off from the multipole expansions (\ref{eq:MultExpF1}) to (\ref{eq:MultExpF4}). They are:
\begin{equation}
F_{1} \sim (\cos \theta)^{\ell_{\mathrm{max}}}\mathrm{,} \hspace*{2pt} F_{2} \sim (\cos \theta)^{\ell_{\mathrm{max}} - 1}\mathrm{,} \hspace*{2pt} F_{3} \sim (\cos \theta)^{\ell_{\mathrm{max}} - 1}\mathrm{,} \hspace*{2pt} F_{4} \sim (\cos \theta)^{\ell_{\mathrm{max}} - 2} \mathrm{.} \label{eq:CGLNLeadingPowers}
\end{equation}
Since observables are just bilinear forms of amplitudes, it is clear that they turn out to be polynomials as well. Using the leading powers (\ref{eq:CGLNLeadingPowers}), the polynomial-orders of the $\check{\Omega}^{\alpha}$ can be inferred from their definitions (see Tables \ref{tab:A1toA8} and \ref{tab:A9toA16} in appendix \ref{subsec:CGLNMatrixReps}). As an example, the beam asymmetry $\check{\Sigma}$ shall be treated here. For real $\cos \theta$, one can just count the maximal power yielded by every bilinear product in the definition:
\begin{align} 
\check{\Sigma} &= - \frac{q}{k} \sin^{2} \theta \hspace*{1pt} \mathrm{Re} \Big[ \left| F_{3} \right|^{2} + \left| F_{4} \right|^{2} + 2 \big\{ F_{1}^{\ast} F_{4} +  F_{2}^{\ast} F_{3} + \cos \theta  F_{3}^{\ast} F_{4} \big\} \Big] \nonumber \\ &= - \frac{q}{k} \sin^{2} \theta \hspace*{1pt} \mathrm{Re} \Big[ \underbrace{\left| F_{3} \right|^{2}}_{\textcolor{black}{\sim \cos \theta ^{2 \ell_{\mathrm{max}} - 2}}} + \underbrace{\left| F_{4} \right|^{2}}_{\textcolor{black}{\sim \cos \theta ^{2 \ell_{\mathrm{max}}-4}}} + 2 \big\{ \underbrace{F_{1}^{\ast} F_{4}}_{\textcolor{black}{\sim \cos \theta ^{2 \ell_{\mathrm{max}}-2}}} \nonumber \\ & \hspace*{77pt}  + 2 \underbrace{F_{2}^{\ast} F_{3}}_{\textcolor{black}{\sim \cos \theta ^{2 \ell_{\mathrm{max}}-2}}} + 2 \underbrace{\cos (\theta) F_{3}^{\ast} F_{4}}_{\textcolor{black}{\sim \cos \theta ^{2 \ell_{\mathrm{max}}-2}}} \big\} \Big] \mathrm{.} \label{eq:SigmaCountingTPWAStandardForm}
\end{align} 
Thus, it is seen that the profile function $\check{\Sigma}$ divided by $\sin^{2} \theta$ has highest power
\begin{equation}
 \frac{\check{\Sigma}}{\sin^{2} \theta} \sim \cos \theta^{2 \ell_{\mathrm{max}} - 2} \mathrm{,} \label{eq:SigmaPolynomialOrder}
\end{equation}
and that therefore the angular distribution of this asymmetry can be written, for any order $\ell_{\mathrm{max}}$, as a finite polynomial:
\begin{equation}
 \check{\Sigma}\left( W, \theta \right) = \frac{q}{k} \hspace*{3pt} \sin^{2} \theta \sum \limits_{n = 0}^{2 \ell_{\mathrm{max}} - 2} a_{n}^{\check{\Sigma}} \left( W \right) \cos^{n} \theta \mathrm{.} \label{eq:SigmaCosPolynomialForm}
\end{equation}
It is clear that the coefficients $a_{n}^{\check{\Sigma}} \left( W \right)$ depend on the multipoles present in the truncation. Before elaborating more on this dependence, we generalize the expression (\ref{eq:SigmaCosPolynomialForm}) to all observables. The counting illustrated for the beam asymmetry in equation (\ref{eq:SigmaCountingTPWAStandardForm}) may be performed for any of the $16$ observables, considering their definitions in terms of CGLN-amplitudes (cf. appendix \ref{subsec:CGLNMatrixReps}). Choosing to expand the observables into $\cos \theta$-monomials (just as in equation (\ref{eq:SigmaCosPolynomialForm})), all observables can be cast in a kind of standard for in the TPWA, as noted in the proceeding \cite{Tiator:2011tu}. \newline
We state here the first version of this standard form, already noting that the energy-dependent expansion coefficients analogous to $a_{n}^{\check{\Sigma}} \left( W \right)$ have to be bilinear in the multipoles. This has to be expected, since the observables themselves are bilinear. The TPWA expressed in powers of $\cos \theta$ reads
\begin{align}
\check{\Omega}^{\alpha} \left( W, \theta \right) &= \frac{q}{k} \hspace*{3pt} \sin^{\beta_{\alpha}} \theta \sum \limits_{n = 0}^{2 \ell_{\mathrm{max}} + \gamma_{\alpha}} a_{n}^{\check{\Omega}^{\alpha}} \left( W \right) \cos^{n} \theta \mathrm{,}  \label{eq:LowECosStandardParametrization1} \\
a_{n}^{\check{\Omega}^{\alpha}} \left( W \right) &= \left< \mathcal{M}_{\ell_{\mathrm{max}}} \left( W \right) \right| \mathcal{C}_{n}^{\check{\Omega}^{\alpha}} \left| \mathcal{M}_{\ell_{\mathrm{max}}} \left( W \right) \right> \mathrm{,} \label{eq:LowECosStandardParametrization2}
\end{align}
\begin{table}[h]
\centering
\begin{tabular}{ccc|cccccc||ccc|ccccccc}
\hline
\hline
  &  &  &  &  &  &  &  &  &  &  &  &  &  &  &  &  &  \\
Type & $\alpha$ & $ \check{\Omega}^{\alpha} $ &  & $ \beta_{\alpha} $ &  & $ \gamma_{\alpha} $ &  & $ \delta_{\alpha} $ & Type & $\alpha$ & $ \check{\Omega}^{\alpha} $ &  & $ \beta_{\alpha} $ &  & $ \gamma_{\alpha} $ &  & $ \delta_{\alpha} $  \\
\hline
  &  &  &  &  &  &  &  &  &  &  &  &  &  &  &  &  &  \\
  & $1$ & $ I \left( \theta \right) $ &  & $ 0 $ &  & $ 0 $ &  & $ -2 $ & & $14$ & $ \check{O}_{x'} $ &  & $ 1 $ &  & $ 0 $ &  & $ -1 $   \\
 $\mathcal{S}$ & $4$ & $ \check{\Sigma} $ &  & $ 2 $ &  & $ -2 $ &  & $ -2 $ & $\mathcal{BR}$ & $7$ & $ \check{O}_{z'} $ &  & $ 2 $ &  & $ -1 $ &  & $ -1 $   \\
  & $10$ & $ \check{T} $ &  & $ 1 $ &  & $ -1 $ &  & $ -1 $ &  & $16$ & $ \check{C}_{x'} $ &  & $ 1 $ &  & $ 0 $ &  & $ -1 $   \\
  & $12$ & $ \check{P} $ &  & $ 1 $ &  & $ -1 $ &  & $ -1 $ &  & $2$ & $ \check{C}_{z'} $ &  & $ 0 $ &  & $ +1 $ &  & $ -1 $   \\
\hline
  &  &  &  &  &  &  &  &  &  &  &  &  &  &  &  &  &  \\
  & $9$ & $ \check{E} $ &  & $ 0 $ &  & $ 0 $ &  & $ -1 $ &  & $6$ & $ \check{T}_{x'} $ &  & $ 2 $ &  & $ -1 $ &  & $ -2 $   \\
 $\mathcal{BT}$ & $3$ & $ \check{G} $ &  & $ 2 $ &  & $ -2 $ &  & $ -1 $ &  $\mathcal{TR}$ & $13$ & $ \check{T}_{z'} $ &  & $ 1 $ &  & $ 0 $ &  & $ -2 $   \\
  & $5$ & $ \check{H} $ &  & $ 1 $ &  & $ -1 $ &  & $ -1 $ &  & $8$ & $ \check{L}_{x'} $ &  & $ 1 $ &  & $ 0 $ &  & $ -2 $   \\
  & $11$ & $ \check{F} $ &  & $ 1 $ &  & $ -1 $ &  & $ -1 $ &  & $15$ & $ \check{L}_{z'} $ &  & $ 0 $ &  & $ +1 $ &  & $ -2 $   \\
\hline
\hline
\end{tabular}
\caption[Definition of the angular parametrization in a TPWA, for all 16 observables.]{The parameters listed here describe the angular parametrizations for the $16$ polarization observables given in equations (\ref{eq:LowECosStandardParametrization1}), (\ref{eq:LowEAssocLegStandardParametrization1}) and (\ref{eq:ProfileFunctionPoleExpansion}). The numbers are taken over (up to slight modifications) from Tiator \cite{Tiator:2011tu}. \newline For ease of reference, the index $\alpha$, labelling all observables according to the defining $4 \times 4$ $\Gamma$-matrices, is also given, cf. Table \ref{tab:ChTabHelTrObs}.}
\label{tab:AngularDistributionsParameters}
\end{table}
with one index labeling all the observables: $\alpha = 1,\ldots,16$. The constants $\beta_{\alpha}$ and $\gamma_{\alpha}$, which specify the expansion of the angular distribution for any observable, are listed in Table \ref{tab:AngularDistributionsParameters}. We sort the multipoles into a vector\footnote{In later chapters, the same vector is mostly denoted as $\left| \mathcal{M}_{\ell} \right>$ instead of $\left| \mathcal{M}_{\ell_{\mathrm{max}}} \right>$.}
\begin{equation}
 \left| \mathcal{M}_{\ell_{\mathrm{max}}} \right> = \left[ E_{0+}, E_{1+}, M_{1+}, M_{1-}, E_{2+}, E_{2-}, M_{2+}, M_{2-}, \ldots , M_{\ell_{\mathrm{max}}-} \right]^{T} \mathrm{,} \label{eq:MultipoleVectorIntro}
\end{equation}
with adjont:
\begin{equation}
 \left< \mathcal{M}_{\ell_{\mathrm{max}}} \right| = \left[ E^{\ast}_{0+}, E^{\ast}_{1+}, M^{\ast}_{1+}, M^{\ast}_{1-}, E^{\ast}_{2+}, E^{\ast}_{2-}, M^{\ast}_{2+}, M^{\ast}_{2-}, \ldots , M^{\ast}_{\ell_{\mathrm{max}}-} \right] \mathrm{.} \label{eq:MultipoleVectorAdjointIntro}
\end{equation}
Thus, the expansion coefficients $a_{n}^{\check{\Omega}^{\alpha}} \left( W \right)$ are defined in terms of multipoles by certain $(4 \ell_{\mathrm{max}})\times(4 \ell_{\mathrm{max}})$-matrices $\mathcal{C}_{n}^{\check{\Omega}^{\alpha}}$. Since the coefficients have to be real, these matrices are either symmetric or hermitean\footnote{Actually, observables which can, in the CGLN-basis, be written as a real part have real symmetric matrices, while those written as an imaginary part have complex hermitean matrices.}. \newline
Expanding the observables into powers of $\cos \theta$ (equation (\ref{eq:LowECosStandardParametrization1})) is just a convention and not mandatory. One can choose any polynomial-basis. In this work, we found it convenient to use associated Legendre polynomials $P_{\ell}^{m} (\cos \theta)$ \cite{Abramowitz}. The corresponding form of the TPWA, which is fully equivalent to equations (\ref{eq:LowECosStandardParametrization1}) and (\ref{eq:LowECosStandardParametrization2}), reads (for the constants $\beta_{\alpha}$ and $\gamma_{\alpha}$, see again Table \ref{tab:AngularDistributionsParameters})
\begin{align}
\check{\Omega}^{\alpha} \left( W, \theta \right) &= \frac{q}{k} \hspace*{3pt} \sum \limits_{n = \beta_{\alpha}}^{2 \ell_{\mathrm{max}} + \beta_{\alpha} + \gamma_{\alpha}} \left(a_{L}\right)_{n}^{\check{\Omega}^{\alpha}} \left( W \right) P^{\beta_{\alpha}}_{n} \left( \cos \theta \right) \mathrm{,}  \label{eq:LowEAssocLegStandardParametrization1} \\
\left(a_{L}\right)_{n}^{\check{\Omega}^{\alpha}} \left( W \right) &= \left< \mathcal{M}_{\ell_{\mathrm{max}}} \left( W \right) \right| \left( \mathcal{C}_{L}\right)_{n}^{\check{\Omega}^{\alpha}} \left| \mathcal{M}_{\ell_{\mathrm{max}}} \left( W \right) \right> \mathrm{.} \label{eq:LowEAssocLegStandardParametrization2}
\end{align}
This associated Legendre-form of the TPWA has been found to be beneficial to work with. The $\sin \theta$-factors in front of the observables are absorbed into the angular fitting functions and furthermore, the $P_{\ell}^{m}$ are orthogonal, a fact which attenuates the size of correlations between the extracted Legendre coefficients. Another point in favor of the Legendre form is that the multipole-bilinear forms (\ref{eq:LowEAssocLegStandardParametrization2}) turn out to be somewhat simplified compared to those present in the $\cos \theta$-form (\ref{eq:LowECosStandardParametrization2}). \newline 
As an example, we now consider the lowest Legendre coefficient of the beam asymmetry $\left( a_{L}  \right)_{2}^{\check{\Sigma}}$, evaluated in the order $\ell_{\mathrm{max}} = 2$. Fully written out in terms of multipoles, it reads:
\begin{align}
 \textcolor{black}{\left( a_{L}  \right)_{2}^{\check{\Sigma}}} &= \frac{1}{14} \Big[E_{2-}^{\ast} \Big(-7 E_{2-}+7 E_{0+}-2 E_{2+}+7 M_{2-}-7 M_{2+}\Big)+7 E_{0+}^{\ast} \Big(E_{2-}+E_{2+}+M_{2-} \nonumber \\ & \hspace*{14.5pt} -M_{2+}\Big)  + E_{2+}^{\ast} \Big(-2 E_{2-}+7 E_{0+}-18 (4 E_{2+}+M_{2-}-M_{2+})\Big)+M_{2-}^{\ast} \Big(7 E_{2-}+7 E_{0+} \nonumber \\ & \hspace*{14.5pt} -18 E_{2+}
  +21 M_{2-}+9 M_{2+}\Big) + M_{2+}^{\ast} \Big(-7 E_{2-}-7 E_{0+}+9 (2 E_{2+}+M_{2-}+4 M_{2+})\Big) \nonumber \\ 
 & \hspace*{14.5pt} +7 \Big(E_{1+}^{\ast} \Big(-3 E_{1+}-M_{1-}+M_{1+}\Big) + M_{1-}^{\ast} \Big(M_{1+}-E_{1+}\Big) \nonumber \\ & \hspace*{14.5pt} +M_{1+}^{\ast} \Big(E_{1+}+M_{1-}+M_{1+}\Big)\Big)\Big] \mathrm{.} \label{eq:a2SigmaCoeffWrittenOut} 
\end{align}
This formula looks somewhat confusing. It turns out to be more instructive to look at the matrix $\left(\mathcal{C}_{L}\right)_{2}^{\check{\Sigma}}$ which defines it:
\begin{equation}
 \left(\mathcal{C}_{L}\right)_{2}^{\check{\Sigma}} = \left[
\begin{array}{c|ccc|cccc|c}
 0 & 0 & 0 & 0 & \frac{1}{2} & \frac{1}{2} & -\frac{1}{2} & \frac{1}{2} & \ldots \\
\hline
 0 & -\frac{3}{2} & \frac{1}{2} & -\frac{1}{2} & 0 & 0 & 0 & 0 & \\
 0 & \frac{1}{2} & \frac{1}{2} & \frac{1}{2} & 0 & 0 & 0 & 0 & \ldots \\
 0 & -\frac{1}{2} & \frac{1}{2} & 0 & 0 & 0 & 0 & 0 & \\
\hline
 \frac{1}{2} & 0 & 0 & 0 & -\frac{36}{7} & -\frac{1}{7} & \frac{9}{7} & -\frac{9}{7} & \\
 \frac{1}{2} & 0 & 0 & 0 & -\frac{1}{7} & -\frac{1}{2} & -\frac{1}{2} & \frac{1}{2} & \ldots \\
 -\frac{1}{2} & 0 & 0 & 0 & \frac{9}{7} & -\frac{1}{2} & \frac{18}{7} & \frac{9}{14} & \\
 \frac{1}{2} & 0 & 0 & 0 & -\frac{9}{7} & \frac{1}{2} & \frac{9}{14} & \frac{3}{2} & \ldots \\
\hline
 \vdots &  & \vdots &  & \vdots &  & \vdots &  &  \ddots
\end{array}
\right] \mathrm{.} \label{eq:C2SigmaCoeffMatrixIntro}
\end{equation}

Here, the dots indicate contributions from partial wave interferences in higher truncation-orders $\ell_{\mathrm{max}} > 2$. For $\ell_{\mathrm{max}} \rightarrow \infty$, the dimension of the matrices defining the multipole-bilinear forms (\ref{eq:LowEAssocLegStandardParametrization2}) would of course also become infinite. \newline Furthermore, from equation (\ref{eq:C2SigmaCoeffMatrixIntro}), we see in comparison to the multipole-vectors (\ref{eq:MultipoleVectorIntro}) and (\ref{eq:MultipoleVectorAdjointIntro}) that up to $\ell_{\mathrm{max}} = 2$, the coefficient $\left( a_{L}  \right)_{2}^{\check{\Sigma}}$ is made up of interference contributions between $S$- and $D$-waves (upper right and lower left box in (\ref{eq:C2SigmaCoeffMatrixIntro})), of $P$-waves with themselves (central diagonal block in (\ref{eq:C2SigmaCoeffMatrixIntro})) and of $D$-waves with themselves (lower right diagonal block in (\ref{eq:C2SigmaCoeffMatrixIntro})). Such block-structures have been found for all Legendre-coefficients and all observables during the investigations of this thesis. Since the matrices $\left( \mathcal{C}_{L}\right)_{n}^{\check{\Omega}^{\alpha}}$ define the algebraic problem at the center of this work, a listing of such matrices is provided for the group $\mathcal{S}$- and $\mathcal{BT}$-observables in appendix \ref{sec:TPWAFormulae}, for the truncation order $\ell_{\mathrm{max}} = 5$. There, we also outline how to extract them using computer algebra. \newline
Another important consequence of the bilinear structure of the equations (\ref{eq:LowEAssocLegStandardParametrization2}) present in a TPWA is that, at least in case no model-assumptions are made, multipoles can only be extracted up to an energy-dependent overall phase. This is just rooted in the fact that, once all multipoles are rotated by the {\it same} phase $\phi_{\mathcal{M}} (W)$
\begin{equation}
 E_{\ell \pm} (W) \longrightarrow e^{i \phi_{\mathcal{M}} (W)} E_{\ell \pm} (W) \mathrm{,} \hspace*{5pt} M_{\ell \pm} (W) \longrightarrow e^{i \phi_{\mathcal{M}} (W)} M_{\ell \pm} (W) \mathrm{,} \label{eq:EnergyDependentPhaseTrafoMultipoles}
\end{equation}
and then inserted into the bilinear forms (\ref{eq:LowEAssocLegStandardParametrization2}), the latter remain invariant. Thus, at least one phase-convention has to be fixed in a fully model-independent TPWA. One can then either try to extract real- and imaginary parts of phase-constrained multipoles, or one could directly attempt to determine their moduli and relative-phases. This point is again picked up once actual TPWA-fits are discussed in this work (see chapter \ref{chap:TPWA}). \newline
Before stating the complete experiment problem formally, another point should be mentioned regarding the truncation-error present in the TPWA. In consistent relativistic (field-theoretical) descriptions of reactions such as photoproduction (see the models described in section \ref{subsec:Energy-dependentModels}, as well as the references cited there), there are not only resonant contributions in the direct production-channel ($s$-channel) present, which would only contribute to isolated multipoles. 

In addition, one also has non-resonant contributions, for instance from particle exchanges in the so-called {\it crossed channels} \cite{PolkinghorneEtAl,PeskinSchroeder}. For example, so-called $t$-channel diagrams bring (roughly speaking) the Mandelstam-variable $t$, and therefore also $\cos \theta$, within energy-denominators. Thus, such diagrams always project on all partial waves. \newline
The TPWA is therefore in all cases only an approximation of the {\it true} physics and it depends, in principle, on the considered channel and energy-region how quickly it converges towards the true amplitude. In case of pion photoproduction, the neutral production channel $\gamma p \rightarrow \pi^{0} p$ is known to be described quite well by few multipoles within the low-energy region, while for the charged channel $\gamma p \rightarrow \pi^{+} n$ this is not the case \cite{Grushin}. In the latter case, a $\pi^{+}$-meson can be exchanged in the $t$-channel. This $t$-channel pole is then, even for low energies, close to the physical region and makes the TPWA converge slowly. \newline
For the sake of completeness, we quote here an alternative Ansatz for the presence of such a strong pole-contribution \cite{Tiator:2011tu, Grushin, BaldinThresholdPP}. It is then possible to cast the TPWA as an expansion around the pole:
\begin{equation}
 \check{\Omega}^{\alpha} \left( W, \theta \right) = \frac{q}{k} \sin^{\beta_{\alpha}} \theta \sum_{n=\delta_{\alpha}}^{2 \ell_{\mathrm{max}} + \gamma_{\alpha}} a^{\check{\Omega}^{\alpha}}_{n} (W) \kappa(\theta)^{n} \mathrm{.} \label{eq:ProfileFunctionPoleExpansion}
\end{equation}
The additional set of parameters $\delta_{\alpha}$ describing this parametrization for all observables are given in Table \ref{tab:AngularDistributionsParameters} as well. The expansion-parameter is here $\kappa\left(\theta\right) = \frac{m_{\pi^{+}}^{2} - t}{2 k E^{\mathrm{CMS}}_{\pi^{+}}} = \left\{ 1 - \beta \cos \left( \theta \right) \right\} $ and $\beta = q_{\pi^{+}}/E^{\mathrm{CMS}}_{\pi^{+}}$ is the CMS-velocity of the $\pi^{+}$. Considering the numbers in Table \ref{tab:AngularDistributionsParameters}, it is seen that some observables contain, in addition to the single pole term $ \sim 1/\kappa(\theta) $, a contribution of the double pole term $ \sim 1/\kappa^{2} (\theta) $. These observables are the unpolarized cross section $ \sigma_{0} $, the beam asymmetry $ \check{\Sigma} $ and all $\mathcal{TR}$ observables. The double pole term has the special attribute of being calculable exactly (cf. \cite{Grushin}). Thus, only the coefficients higher than $ a_{-1}^{\check{\Omega}^{\alpha}} $ have to be extracted from angular distributions of the observables \cite{Tiator:2011tu}. Another convenient property of the $t$-channel pole contribution, which should be mentioned, is the fact that the coefficients of the single pole-terms, $ a_{-1}^{\check{\Omega}^{\alpha}} $, are linear in the multipoles and therefore can resolve the problem of the unknown overall phase (equation (\ref{eq:EnergyDependentPhaseTrafoMultipoles})).\newline
We note that Grushin and collaborators \cite{Grushin} have been able to perform an analysis of the $\pi^{+}$-production channel with minimal model-dependence using this Ansatz. Furthermore, we assume that parametrizations such as (\ref{eq:ProfileFunctionPoleExpansion}) can be generalized to other channels when a similar situation is present, even in case the exchanged particle and the meson in the final state are not the same. \newline
Having introduced a systematic way to write the TPWA for all $16$ photoproduction observables, we now state the complete experiment problem in a more formal way (cf. \cite{MyDiplomaThesis}):
\newpage
\begin{CompEx}[Complete Experiment: TPWA] \textcolor{white}{Hallo Welt ;-)} \\
In order to construct a complete experiment, the problem is to find and utilize a (possibly minimum) set of expansion coefficients $ a_{k}^{\check{\Omega}^{\alpha}} \left( W \right) $ (or $ \left(a_{L}\right)_{k}^{\check{\Omega}^{\alpha}} \left( W \right) $), determined from the angular distributions of a corresponding (also possibly minimum) set of observables via equations (\ref{eq:LowECosStandardParametrization1}), (\ref{eq:LowEAssocLegStandardParametrization1}) or (\ref{eq:ProfileFunctionPoleExpansion}) (depending on the kind of channel considered), that permits an unambiguous determination of the multipoles $ \left\{ E_{\ell \pm}, M_{\ell \pm} \right\} $ contributing in the considered energy area bound by a value $ W_{\mathrm{max}} $ (i.e. up to some $ \ell = \ell_{\mathrm{max}} $). \newline
In case the parametrization (\ref{eq:LowEAssocLegStandardParametrization1}) (no t-channel pole expansion) is used, unambiguous means only unambiguous up to an overall phase, whereas under usage of (\ref{eq:ProfileFunctionPoleExpansion}) (t-channel pole present) it means free of all ambiguities including the continuous phase ambiguity.
\end{CompEx}
Though one can ask the question for complete experiments in the context of slightly model-dependent parametrizations such as (\ref{eq:ProfileFunctionPoleExpansion}), we will set the focus to the fully model-independent form (equations (\ref{eq:LowECosStandardParametrization1}) to  (\ref{eq:LowEAssocLegStandardParametrization2})) in the remainder of this thesis. This is just a matter of philosophy, since complete experiment should ideally be valid without any model-dependence. Of course, the fact cannot be disregarded that in practical applications, some model-dependence may have to enter the multipole-extraction in the end.
\vfill
\begin{table}[h]
\centering
\begin{tabular}{cc|ccc||cc|ccc}
\hline
\hline
  &  &  &  &  &  &  &  &  &  \\
Type & $ \check{\Omega}^{\alpha} $ &  & $ N_{a^{\alpha}_{k}} $ &  & Type & $ \check{\Omega}^{\alpha} $ &  & $ N_{a^{\alpha}_{k}} $ &  \\
\hline
\hline
  &  &  &  &  &  &  &  &  &  \\
  & $ \sigma_{0} $ &  & $ (2 \ell_{\mathrm{max}} + 1) $ &  &  & $ \check{O}_{x'} $ &  & $ (2 \ell_{\mathrm{max}} + 1) $ &   \\
 $\mathcal{S}$ & $ \check{\Sigma} $ &  & $ (2 \ell_{\mathrm{max}} - 1) $ &  & $\mathcal{BR}$ & $ \check{O}_{z'} $ &  & $ 2 \ell_{\mathrm{max}} $ &   \\
  & $ \check{T} $ &  & $ 2 \ell_{\mathrm{max}} $ &  &  & $ \check{C}_{x'} $ &  & $ (2 \ell_{\mathrm{max}} + 1) $ &   \\
  & $ \check{P} $ &  & $ 2 \ell_{\mathrm{max}} $ &  &  & $ \check{C}_{z'} $ &  & $ (2 \ell_{\mathrm{max}} + 2) $ &   \\
\hline
 Sum of $N_{a^{\alpha}_{k}}$ &  &  & $ 8 \ell_{\mathrm{max}} $ &  & Sum of $N_{a^{\alpha}_{k}}$ &  &  & $ (8 \ell_{\mathrm{max}} + 4) $ & \\
\hline
\hline
  &  &  &  &  &  &  &  &  &  \\
  & $ \check{E} $ &  & $ (2 \ell_{\mathrm{max}} + 1) $ &  &  & $ \check{T}_{x'} $ &  & $ 2 \ell_{\mathrm{max}} $ &   \\
 $\mathcal{BT}$ & $ \check{G} $ &  & $ (2 \ell_{\mathrm{max}} - 1) $ &  & $\mathcal{TR}$ & $ \check{T}_{z'} $ &  & $ (2 \ell_{\mathrm{max}} + 1) $ &   \\
  & $ \check{H} $ &  & $ 2 \ell_{\mathrm{max}} $ &  &  & $ \check{L}_{x'} $ &  & $ (2 \ell_{\mathrm{max}} + 1) $ &  \\
  & $ \check{F} $ &  & $ 2 \ell_{\mathrm{max}} $ &  &  & $ \check{L}_{z'} $ &  & $ (2 \ell_{\mathrm{max}} + 2) $ &  \\
\hline
 Sum of $N_{a^{\alpha}_{k}}$ &  &  & $ 8 \ell_{\mathrm{max}} $ &  & Sum of $N_{a^{\alpha}_{k}}$ &  &  & $ (8 \ell_{\mathrm{max}} + 4) $ & \\
\hline
\hline
\end{tabular}
\caption[Counting the number of angular fit-parameters in a TPWA.]{For the angular parametrizations without $t$-channel pole, i.e. equations (\ref{eq:LowECosStandardParametrization1}) and (\ref{eq:LowEAssocLegStandardParametrization1}), the numbers of angular fit parameters yielded by any observables are listed here for an arbitrary truncation order $\ell_{\mathrm{max}}$. Also, numbers of parameters are summed up for every group of observables (i.e. type $\mathcal{S}$, $\mathcal{BT}$, $\mathcal{BR}$ and $\mathcal{TR}$). \newline It can be seen, that the single-spin and type $\mathcal{BT}$-observables are equivalent in regard to their capability to yield angular fit coefficients. The remaining double-polarization observables of type $\mathcal{BR}$ and $\mathcal{TR}$ yield more coefficients, thus they may be better capable of constraining multipole-solutions. \newline However, the group $\mathcal{S}$ observables alone already yield more fit coefficients than the free parameters present in the TPWA ($8 \ell_{\mathrm{max}}$ compared to $(8 \ell_{\mathrm{max}} - 1)$). The fit coefficients of type $\mathcal{S}$ and $\mathcal{BT}$-observables combined exceed free parameters by more than a factor of two ($16 \ell_{\mathrm{max}}$ vs. $(8 \ell_{\mathrm{max}} - 1)$).}
\label{tab:NumberOfLegendreCoefficients}
\end{table}
\clearpage
As a next step, it is interesting to compare the number of constraining quantities, i.e. the Legendre coefficients, to the number of free multipole parameters present in a TPWA. This will be done directly for arbitrary $\ell_{\mathrm{max}}$. In a fully model-independent TPWA, there are $4 \ell_{\mathrm{max}}$ complex multipoles that have to be extracted from the data, which correspond to $8 \ell_{\mathrm{max}}$ real variables. However, one overall phase remains undetermined (due to the invariance under the rotation (\ref{eq:EnergyDependentPhaseTrafoMultipoles})) and therefore, once a phase-constraint is fixed, there remain
\begin{equation}
 8 \ell_{\mathrm{max}} - 1 \mathrm{,} \label{eq:FreeParametersModelIndepTPWA}
\end{equation}
real numbers\footnote{Speaking of real numbers in this context is of course a mathematical idealization which, in a practical numerical situation, cannot be fulfilled.} that have to be determined. For comparison, the number of Legendre-coefficients provided by the angular parametrizations of every observable are listed in Table \ref{tab:NumberOfLegendreCoefficients} (These numbers may just be read off from equation (\ref{eq:LowECosStandardParametrization1}) or (\ref{eq:LowEAssocLegStandardParametrization1}).). \newline
A first direct naive comparison of the numbers seems very promising. The group $\mathcal{S}$ observables alone for instance yield $8 \ell_{\mathrm{max}}$ Legendre coefficients, which already exceeds the number of free parameters (\ref{eq:FreeParametersModelIndepTPWA}) by one.  Complementing the group $\mathcal{S}$ by only one $\mathcal{BT}$-observable, for instance $\check{F}$, would already yield $10 \ell_{\mathrm{max}}$ coefficients. Picking all group $\mathcal{S}$- and $\mathcal{BT}$-observables would still avoid the experimentally almost inaccessible recoil double-polarization observables, while introducing a total of $16 \ell_{\mathrm{max}}$ Legendre coefficients, thus exceeding the number of free parameters (\ref{eq:FreeParametersModelIndepTPWA}) by a factor of two, a fact that remains true for any truncation order $\ell_{\mathrm{max}}$. \newline
However, a word of warning should be expressed in regard of the bilinear nature of the equation-systems constructed from the forms (\ref{eq:LowEAssocLegStandardParametrization2}) (or (\ref{eq:LowECosStandardParametrization2})). Such equation systems are, if viewed in terms of the real and imaginary parts of (phase-constrained) multipoles, just multivariate polynomials of order $2$ with generally complex coefficients. Then, we are looking for the simultaneous roots of such polynomials. A less rigorous use of language would be to say that one is just looking at {\it 'quadratic equations'} in multiple variables. However, as is well-known, the most basic quadratic equation in one variable, i.e.  $x^{2} + a x + b = 0$, already generally possesses two solutions. The bilinear equations considered in this work are, in some sense, just a generalization of this simple equation. Therefore, based purely on their mathematical nature, one may anticipate multiple solutions, or so-called {\it ambiguities}, to exist. Then, in case one considers a set of observables which still has ambiguities, a complete experiment would be an enlarged set containing additional observables capable of resolving these multiple solutions. \newline
Furthermore, since the complete experiment problem for the TPWA is again a uniqueness problem for the solution of bilinear forms, one cannot help but to acknowledge the resemblance it has to the CEA for the full spin amplitudes discussed in section \ref{sec:CompExpsFullAmp}. For instance, for a truncation at $\ell_{\mathrm{max}}=1$, the group $\mathcal{S}$- and $\mathcal{BT}$-observables would already yield $16$ Legendre coefficients (the counterparts of the observables $\check{\Omega}^{\alpha}$ in the CEA) for the extraction of $4$ multipoles (analogues of the CEA's full spin-amplitudes $b_{i}$) and both problems would look very similar, with the dimensions being the same. Then, one could hope to be successful with an {\it effective linearization} strategy, which in case of the CEA allowed for the expression of the bilinear products $b_{i}^{\ast} b_{j}$ as a linear combination of observables (see equation (\ref{eq:BilProductTransversityExplicitExpression})). Thus, one could {\it hope} to be able to derive equations similar to, for instance
\begin{equation}
 E_{0+}^{\ast} M_{1-} = \sum_{\alpha,n} \bm{k}^{\alpha}_{n} \left(a_{L}\right)_{n}^{\check{\Omega}^{\alpha}} \left( W \right) \mathrm{,} \label{eq:HypotheticalBilProductInverted}
\end{equation}
where the product $E_{0+}^{\ast} M_{1-}$ may be defined in terms of some complex coefficients $\bm{k}^{\alpha}_{n}$. Once all bilinear products are known, moduli and relative phases would follow readily. \newline
However, it has to be reported that all attempts to obtain an algebraic inversion with expressions similar to equation (\ref{eq:HypotheticalBilProductInverted}) did not succeed in the course of this work. The reason is that, from all possible combinations of matrices $ \left( \mathcal{C}_{L}\right)_{n}^{\check{\Omega}^{\alpha}}$ present in some low truncation order, with the case $\ell_{\mathrm{max}} = 1$ investigated explicitly, none were found which were linearly independent. However, the linear independence of the $\tilde{\Gamma}$-matrices has been the most important fact leading to the inversion (\ref{eq:BilProductTransversityExplicitExpression}) in section \ref{sec:CompExpsFullAmp}. \newline
Furthermore, a simple estimate can show that the Ansatz of effective linearization, even in case it would work for low truncations, becomes impossible for higher truncation orders. For this calculation to work, one would need a full basis of the vectorspace of hermitean $(4 \ell_{\mathrm{max}})\times(4 \ell_{\mathrm{max}})$-matrices. The real dimension of the latter is $(4 \ell_{\mathrm{max}})^{2}$. However, all $16$ observables can yield only $(32 \ell_{\mathrm{max}} + 8)$ Legendre coefficients in any truncation order $\ell_{\mathrm{max}}$ (see Table \ref{tab:NumberOfLegendreCoefficients}). Since we need (at least) one Legendre coefficient for every basis-vector, the inequality $(4 \ell_{\mathrm{max}})^{2} < (32 \ell_{\mathrm{max}} + 8)$ has to be fulfilled. This is only possible for $\ell_{\mathrm{max}} \leq 2$. \newline

\begin{figure}[ht]
\centering
\includegraphics[width=0.8\textwidth,trim=0 0 0 0,clip]{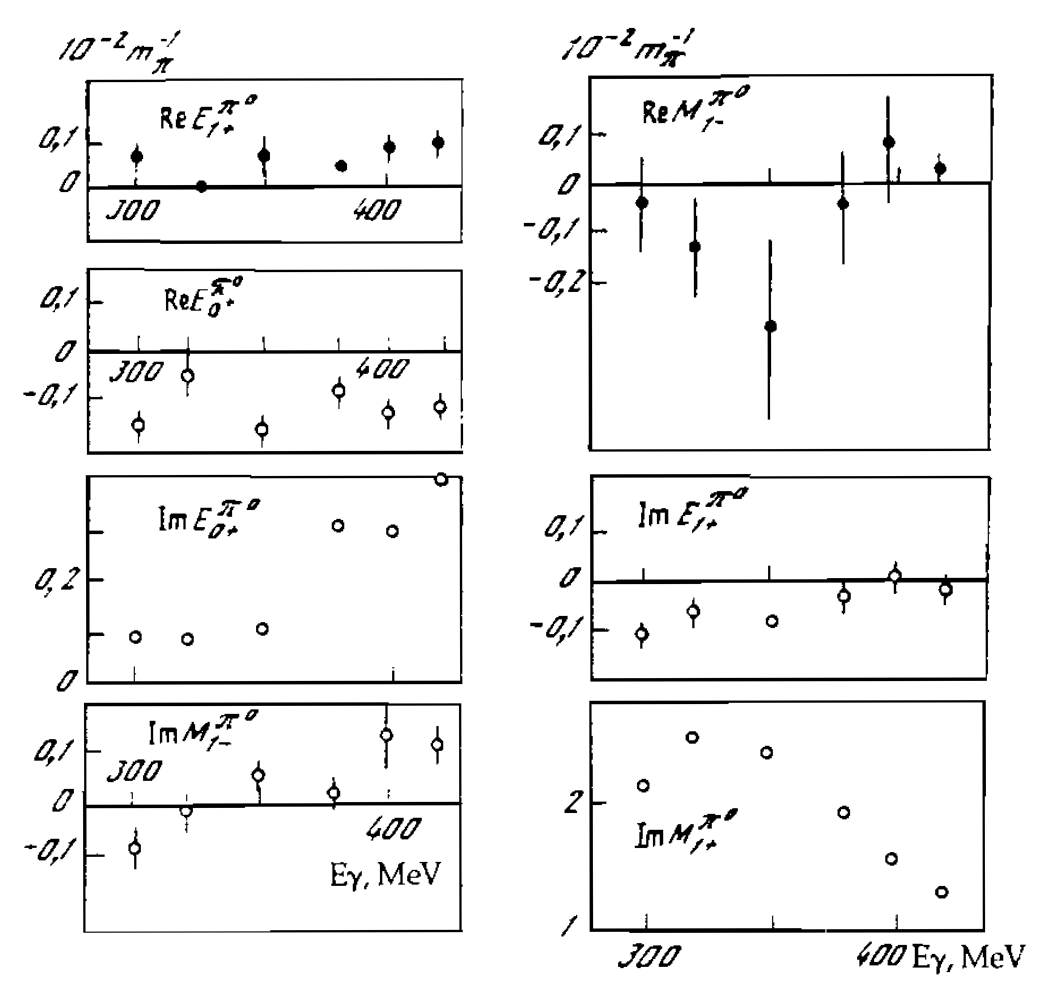}
\caption[Results of Grushin's fit to single-spin observables for $\gamma p \rightarrow \pi^{0} p$ in the $\Delta$-region.]{Shown are the results of Grushin's fit \cite{Grushin} of a TPWA truncated at $\ell_{\mathrm{max}} = 1$, performed on $6$ energies within the low-energy region of the reaction $\gamma p \rightarrow \pi^{0} p$. Data for the group $\mathcal{S}$ observables $\left\{\sigma_{0}, \check{\Sigma}, \check{T}, \check{P} \right\}$ \cite{PionPPDataCompilation, OldPionDataAleksandrov, Belyaev:1983} have been fitted. Real and imaginary parts of all multipoles for $\ell = 0$ and $1$ are shown in units of $10^{-2} m_{\pi}$, except for the real part of $M_{1+}$, which has been fixed to the result of a dispersive analysis \cite{SchwelaWeizel} in order to remove the overall phase ambiguity. The original figure has been taken over from reference \cite{Grushin}.}
\label{fig:GrushinGroupSFitResults}
\end{figure}

We conclude this introductory chapter by highlighting some results from the literature on multipole analyses which have, for one reason or another, been important for the inception of this thesis. The first and most important one is the work by Grushin \cite{Grushin}, which has already been mentioned before. Here, the author analyzed the channel $\gamma p \rightarrow \pi^{0} p$ on an energy-grid of $6$ points in the low-energy region. Data for the group $\mathcal{S}$ observables $\left\{\sigma_{0}, \check{\Sigma}, \check{T}, \check{P} \right\}$ \cite{PionPPDataCompilation, OldPionDataAleksandrov, Belyaev:1983} were analyzed which are, at least from today's perspective, quite out of date. Grushin fitted a TPWA truncated at $\ell_{\mathrm{max}} = 1$ and was, although only $4$ observables were analyzed, able to extract a unique best estimate for the $S$- and $P$-wave multipoles. The results are shown in Figure \ref{fig:GrushinGroupSFitResults}. \newline
It is worthwhile to quote more details on the ambiguities the author encountered and how he resolved them. First, Grushin performed a numerical search of multipole solutions without imposing any phase constraint. For this, all $S$- and $P$-waves were varied freely and initial conditions for the minimizations were drawn around the results of a dispersive analysis by Schwela and Weizel \cite{SchwelaWeizel}. The real and imaginary parts of the $4$ non phase-constrained multipoles were organized into a parameter vector $\left\{y_{n}, \hspace*{1.5pt} n=1,\ldots,8 \right\}$ and initial conditions generated according to the prescription \cite{Grushin}
\begin{equation}
y_{n,0} = y_{n}^{\mathrm{th.}} (3 \gamma - 1)\mathrm{,} \hspace*{5pt} n=1,\ldots,8 \mathrm{,} \label{eq:GrushinsRandomSampling}
\end{equation}
where $y_{n}^{\mathrm{th.}}$ are the dispersive results and $\gamma$ is a random variable drawn from the interval $\left(0,1\right)$. In this way, Grushin generated between $100$ and $150$ initial parameter configurations in a mildly model-dependent way and then performed a fit for each one of them. He then encountered a degeneracy in the solution which was a superposition of two effects \cite{Grushin}. \newline
The first one was a discrete (binary) ambiguity \cite{Omelaenko}, which has been known to exist at the time, for the $4$ group $\mathcal{S}$ observables. The second one was the continuous overall phase ambiguity\footnote{Both notions, i.e. discrete and continuous ambiguities, are clarified further in the beginning of chapter \ref{chap:Omelaenko}.}. In order to resolve the latter, Grushin introduced two methods, each one making further physical assumptions:
\begin{itemize}
 \item[a.)] Setting $\mathrm{Im} \left[ M_{1-} \right] = 0$, i.e. introducing an assumption that a small, non-resonant amplitude is real.
 \item[b.)] Fixing $\mathrm{Re} \left[ M_{1+} \right]$ to the result of dispersive calculations.
\end{itemize}

\clearpage

Both methods above have been able to resolve the continuous phase ambiguity. Furthermore, demanding agreement between the results of both methods a.) and b.), Grushin was able to remove the remaining discrete ambiguity. The unique result is then shown, for method b.), in Figure \ref{fig:GrushinGroupSFitResults}. \newline
We again would like to stress the importance of the Grushin-fits for the motivation of this thesis. They showed that within a TPWA, it can be possible to arrive at a unique amplitude extraction introducing minimal model assumptions, even in cases where the completeness rules for the extraction of the full spin-amplitudes (see section \ref{sec:CompExpsFullAmp}) are violated. The Grushin-fits have been recently revisited and confirmed by Workman \cite{WorkmanGrushinFits}. However, one has to keep in mind that while the result is interesting, it was obtained in a quite low truncation order. Still, the data base of $\pi^{0}$-photoproduction has been improved a lot in the recent years, a fact which spawns a lot of hope for unique solutions. \newline
\begin{figure}[ht]
\hspace*{-14pt}
\includegraphics[width=0.56\textwidth,trim=0 0 0 0,clip]{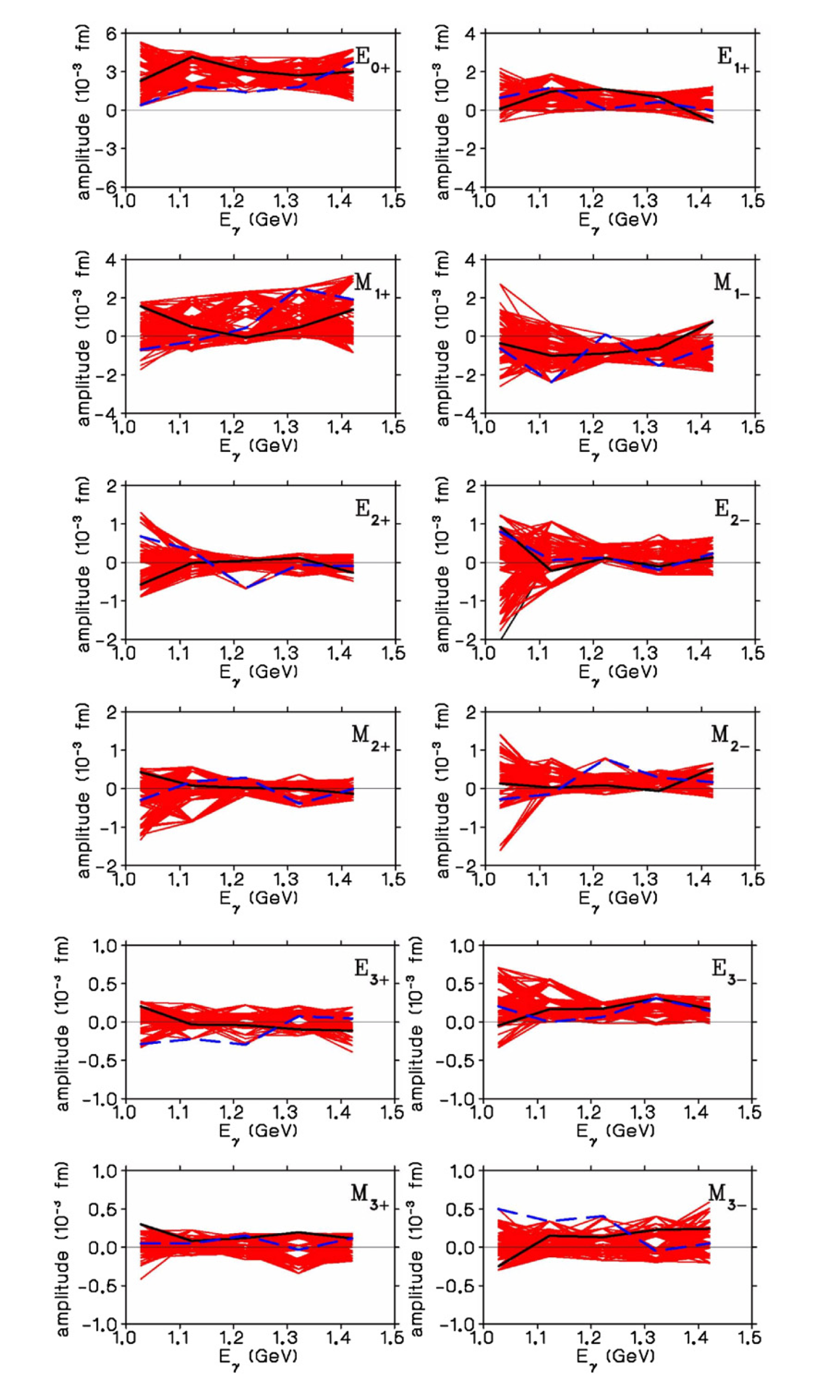} \hspace*{-25pt}
\includegraphics[width=0.555\textwidth,trim=0 0 0 0,clip]{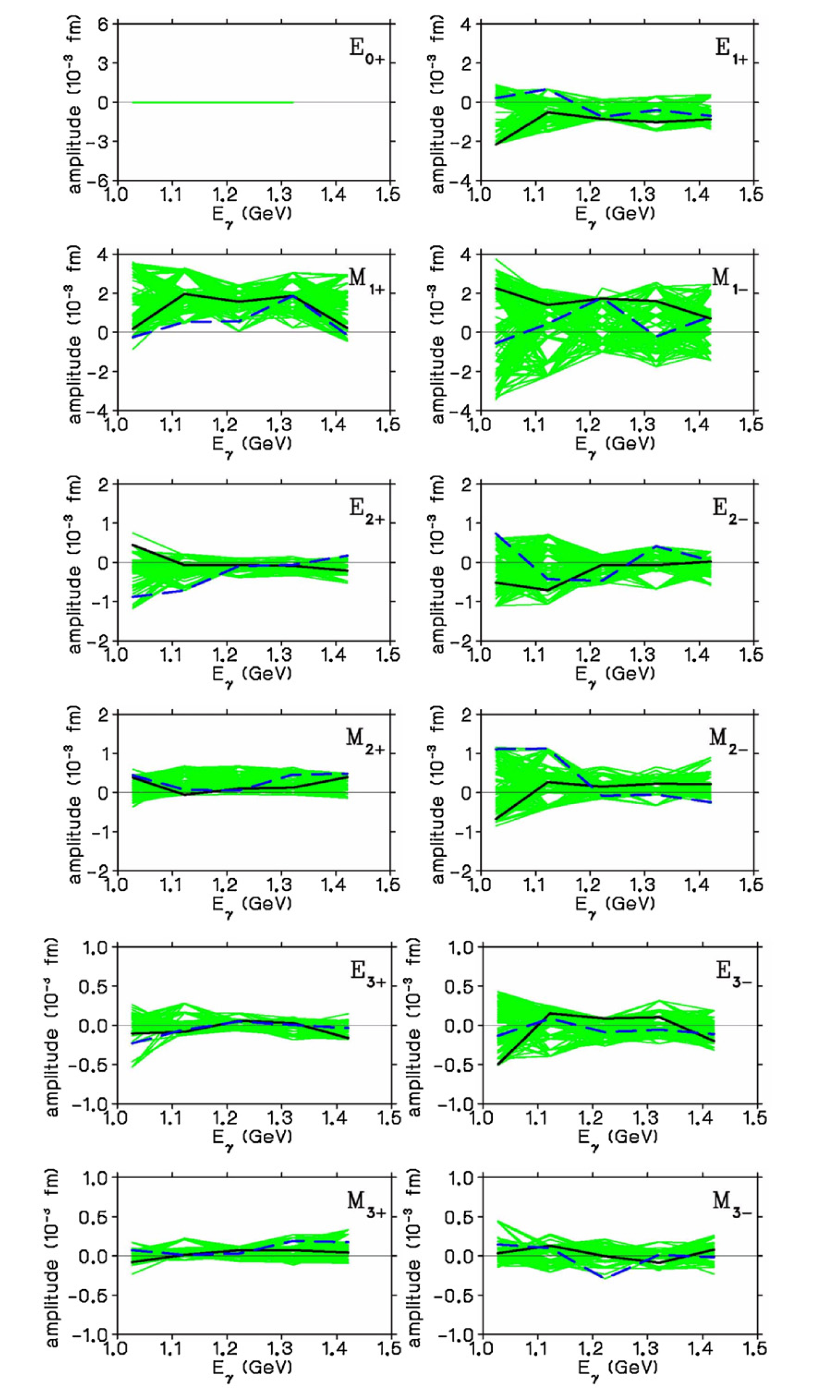}
\caption[Results of fits by Sandorfi {\it et al.} to a complete set of observables for $\gamma p \rightarrow K \Lambda$.]{The results for the real- (red lines, left column) and imaginary parts (green lines, right column) of the electric and magnetic multipoles $\left\{ E_{\ell \pm}, M_{\ell \pm} \right\}$ for $\ell = 0$ to $3$ are shown for a fit of the photoproduction channel $\gamma p \rightarrow K^{+} \Lambda$. The fit is described in the paper of Sandorfi, Hoblit, Kamano and Lee \cite{Sandorfi:2010uv}, from which these figures are taken over identically. \newline The authors analyzed a relatively large dataset, comprised of the observables $\sigma_{0}$, $\Sigma$, $T$, $P$, $C_{x^{\prime}}$, $C_{z^{\prime}}$, $O_{x^{\prime}}$ and $O_{z^{\prime}}$. All multipoles from $\ell = 0$ to $3$ have been varied freely, while higher partial waves from $\ell = 4$ to $8$ were fixed to phenomenological Born-amplitudes. Initial parameter-configurations for the fits were determined from a Monte Carlo sampling of the amplitude space, and the $S$-wave multipole $E_{0+}$ has been fixed to be real in all fits. \newline The bands show all solutions from the global minimum in $\chi^{2}/\mathrm{point}$ (black solid line), up to the solution located within an interval of $0.2$ above this global minimum (blue dashed line).}
\label{fig:SandorfiEtAlFitResults}
\end{figure}

\clearpage

Another interesting multipole analysis with a higher truncation order was published by Sandorfi, Hoblit, Kamano and Lee \cite{Sandorfi:2010uv}. This work presents a fairly model-independent analysis of the channel $\gamma p \rightarrow K^{+} \Lambda$. Some results are shown in Figure \ref{fig:SandorfiEtAlFitResults}. The authors analyzed data for the complete\footnote{Complete in the sense of Chiang and Tabakin \cite{ChTab}.} set of observables $\sigma_{0}$, $\Sigma$, $T$, $P$, $C_{x^{\prime}}$, $C_{z^{\prime}}$, $O_{x^{\prime}}$ and $O_{z^{\prime}}$, varying all $S$-, $P$-, $D$- and $F$-wave multipoles freely. One restriction was made for the $S$-wave, which has been fixed to be real and positive. Furthermore, all multipoles from $\ell=4$ to $8$ were fixed to phenomenological Born-amplitudes, which the authors also show in some detail. \newline
The fitting-procedure for the multipoles from $\ell=0$ to $3$ in this analysis is less model-dependent than in Grushin's case. The authors employed large random samples of parameter configurations in amplitude space (i.e. of real and imaginary parts of all multipoles from $\ell = 0$ to $3$), up to $10^{7}$ for each energy. Minimizations of $\chi^{2}$ were carried out, but only in case the $\chi^{2}$ of a random sample was within a range of $10^{4}$ times the current best value.
Furthermore, it should be mentioned that an attempt has been made to correct systematic errors in the data in this fit, by using the Fierz-identities (see section \ref{sec:CompExpsFullAmp} and appendix \ref{subsec:FierzIdentities}). The fulfillment of the latter served as a criterion to deduce correcting overall scale factors for the different datasets. In this way, Sandorfi {\it et al.} \cite{Sandorfi:2010uv} arrived at bands of multipole-solutions which can be seen in Figure \ref{fig:SandorfiEtAlFitResults}. All solutions in the bands describe the (re-scaled) data equally well, since they are local minima in $\chi^{2}/\mathrm{points}$ within a region of $0.2$ above the global minimum. The spread in these bands of solutions is quite large, in particular for the $S$- and $P$-wave multipoles. Thus, from the fits by Sandorfi {\it et al.} it was clear in the beginning of this work that single-energy fits can become quite unstable. \newline
However, it was not clear whether this fact is just rooted in the larger truncation order, when compared to the Grushin-fits, or also in the statistical precision of the data. Sandorfi {\it et al.} actually followed the latter idea and performed studies of mock-data generated from Bonn-Gatchina multipoles \cite{BoGa} with variable errors assigned. They showed that in case of $5 \%$ statistical precision and furthermore for a fit of {\it all} $16$ observables, the stability of the multipole fit would increase significantly \cite{Sandorfi:2010uv}. \newline

Lastly, it should be mentioned that during the writing of this thesis, another paper has been published \cite{CommonPaper} which investigates similar questions for the uniqueness of multipole solutions, but in the context of the energy-dependent PWAs outlined in section \ref{subsec:Energy-dependentModels}. There, the impact of new data, measured mostly for $\pi^{0}$-photoproduction, on the solutions of the analyses SAID, BnGa and J\"uBo is studied. The new data for $\pi^{0}$-production contain the observables $\Sigma$ \cite{Hornidge:2013,Dugger:2013}, $P$, $T$ and $H$ \cite{Hartmann:2015}, $G$ \cite{Thiel:2012, Thiel:2016} and $E$ \cite{Gottschall:2014, Gottschall:2015}. For the reaction $\gamma p \rightarrow \pi^{+} n$, new data were included for the beam asymmetry $\Sigma$ \cite{Dugger:2013}. \newline
Solutions for the three energy-dependent PWAs mentioned above were compared before and after the inclusion of the new polarization data. It was then seen that, for most multipoles, the PWA-model solutions {\it converge} towards each other, as a result of the new data. This result is very important philosophically even for this thesis, since it implies that an underlying {\it true} multipole-solution provided by nature seems to exist. The different ED PWAs tend towards this solution, even though they are constructed in, partly, very different ways (see section \ref{subsec:Energy-dependentModels}). \newline
The authors of reference \cite{CommonPaper} even quantified this convergence-effect, by defining a variance between two multipole-solutions from two different PWA-models $1$ and $2$. This quantity is defined by the following sum over the $16$ multipoles of the process $\gamma p \rightarrow \pi^{0} p$ from $\ell=0$
\begin{figure}[ht]
\centering
 \begin{minipage}[t]{0.496\textwidth}
        \includegraphics[width=\textwidth]{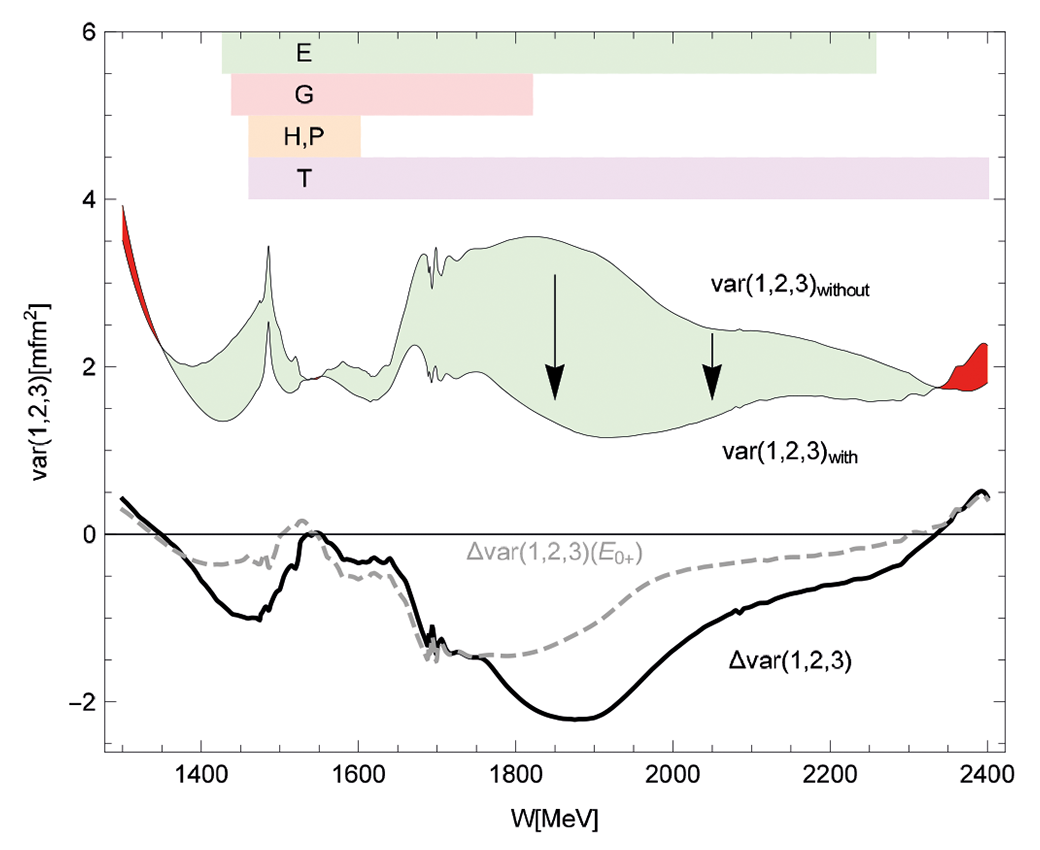}
        \vspace*{-20pt}
        \caption[Comparisons of variances between energy-dependent PWA models, before and after the inclusion of new polarization data.]{{\it Right:} The variance (\ref{eq:MultipoleVarianceMichael}) evaluated between two PWAs each and summed over all $\gamma p \rightarrow \pi^{0} p$-multipoles up to $\ell = 4$, is shown. $(a)$: before inclusion of the new data; $(b)$: after inclusion of the new data; $(c)$: difference between $(a)$ and $(b)$. {\it Left:} The variance between all three PWAs SAID, BnGa and J\"uBo is shown. The improvement of the variance due to the $S$-wave $E_{0+}$ alone is shown as a dashed curve. \newline The pictures are taken over identically from reference \cite{CommonPaper}.}
        \label{fig:CommonPaperShot1}
    \end{minipage}
 \begin{minipage}[t]{0.496\textwidth}
 \vspace*{-6.25cm}
        \includegraphics[width=\textwidth]{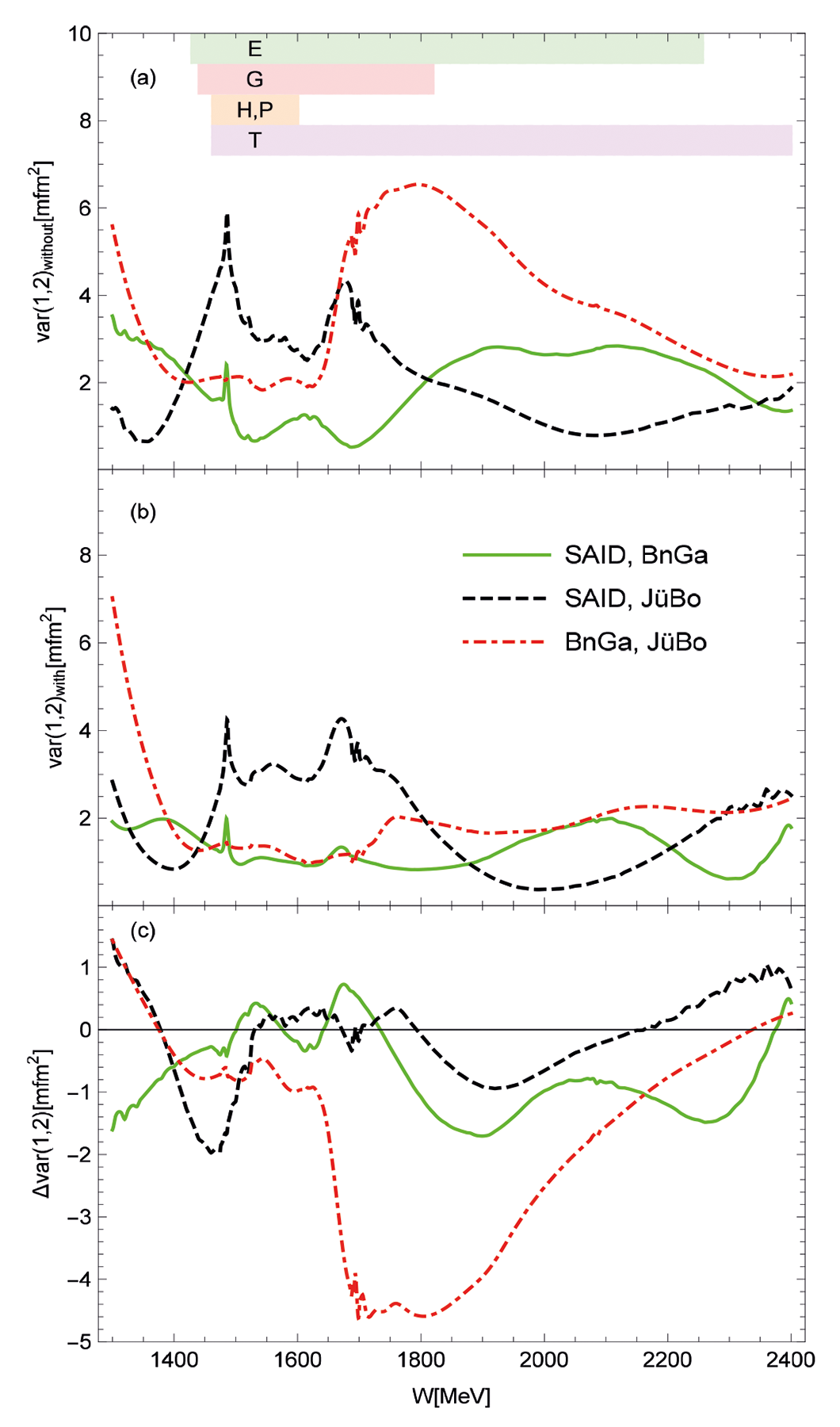}
        \label{fig:CommonPaperShot2}
    \end{minipage}
\end{figure}
 (i.e. $\mathcal{M} (1) = E^{\pi^{0}}_{0+}$) to $\ell = 4$ (i.e. $\mathcal{M} (16) = M^{\pi^{0}}_{4-}$): 
\begin{equation}
 \mathrm{var}(1,2) = \frac{1}{2} \sum_{i=1}^{16} \left[ \mathcal{M}_{1} (i) - \mathcal{M}_{2} (i) \right]  \left[ \mathcal{M}^{\ast}_{1} (i) - \mathcal{M}^{\ast}_{2} (i) \right] \mathrm{.} \label{eq:MultipoleVarianceMichael}
\end{equation}
In Figure \ref{fig:CommonPaperShot1}, such variances are plotted as pair-wise comparisons of SAID vs. BnGa, SAID vs. J\"uBo and BnGa  vs. J\"uBo. A variance between all three PWAs is shown as well. Furthermore, the kinematic ranges of the new polarization data are indicated as colored bars in the plots. Indeed, variances are reduced significantly after inclusion of the new polarization data into the fits. This effect is seen first and foremost in the energy regions where new data are given. \newline

This concludes the introduction and motivation of the complete experiment problem in a TPWA. The following thesis is organized according to the original idea to approach this problem, namely to investigate both from the mathematical (algebraic) side and from the numerical side. \newline
Mathematical considerations deal mainly with the ambiguities present in this problem, as well as with the capability of observables to resolve them. They are presented in chapter \ref{chap:Omelaenko}. \newline

One of the benefits of the formalization for the TPWA presented in this section is that the angular parametrizations (\ref{eq:LowECosStandardParametrization1}) and (\ref{eq:LowEAssocLegStandardParametrization1}) permit one to perform a very simple numerical procedure known as {\it moment analysis} (cf. \cite{MikhasenkoProceeding2014,LFitPaper}). This kind of analysis is still quite useful to throw a first glimpse into a (newly measured) dataset. The results of moment analyses are presented in chapter \ref{chap:LFits}. \newline
Finally, the main numerical part of this work is shown in chapter \ref{chap:TPWA} and it consists of practical truncated partial wave analyses. A model-independent scheme for the extraction of multipoles is explained and then applied to (ideal) model-data, pseudo-data with variable precision and then finally also to real data. Approaches for treating both statistical and systematic errors in the fits are discussed as well. \newline

The thesis is written in a partly cumulative form. Results which have been published in the course of this work are thus presented in the form of the original papers in their respective (sub-) chapters. The publications are:
\begin{itemize}
 \item Y.~Wunderlich, R.~Beck and L.~Tiator,
  ``The complete-experiment problem of photoproduction of pseudoscalar mesons in a truncated partial wave analysis,''
  Phys.\ Rev.\ C {\bf 89}, no. 5, 055203 (2014).\newline - This paper forms the most important part of chapter \ref{chap:Omelaenko} and is presented in section \hspace*{3.85pt} \ref{sec:WBTpaper}.
 \item Y.~Wunderlich, F.~Afzal, A.~Thiel and R.~Beck,
  ``Determining the dominant partial wave contributions from angular distributions of single- and double-polarization observables in pseudoscalar meson photoproduction,''
  Eur.\ Phys.\ J.\ A {\bf 53}, no. 5, 86 (2017). \newline - \underline{Remark:} For copyright reasons, an earlier version of this paper, which has been \\ \hspace*{3.9pt} generated toward the end of the refereeing process, is included as the main part of \\ \hspace*{3.9pt} chapter \ref{chap:LFits}, in section \ref{sec:LFitsPaper}. The published version can be obtained from the journal \\ \hspace*{3.9pt} \textit{Eur.\ Phys.\ J.\ A} and it has the DOI: 10.1140/epja/i2017-12255-0.
\end{itemize}
Each paper is thus presented as its own chapter, together with an introduction and a brief summary. \newline
Material which has been developed in the course this thesis, but has not been published until now, is contained in chapters \ref{chap:CompleteSetsOf4}, \ref{chap:TPWA} and in the appendices, with all of them written in conventional form. \newline

The mathematical considerations described in chapter \ref{chap:Omelaenko} have lead to another publication, namely
\begin{itemize}
 \item Y.~Wunderlich, A.~{\v{S}}varc, R. L.~Workman, L.~Tiator and R.~Beck,
  ``Toward an understanding of discrete ambiguities in truncated partial wave analyses,''
  Phys.\ Rev.\ C {\bf 96}, no. 6, 065202 (2017).
\end{itemize}
However, since the latter paper does not directly address the process of photoproduction, which is at the center of attention in this thesis, we chose not to re-print the full publication here. Some of its content is cited at the end of the introduction to chapter \ref{chap:Omelaenko}, in section \ref{sec:OmelaenkoApproachIntro}.

\clearpage



\clearpage
\thispagestyle{empty}
\textcolor{white}{Hallo Welt :-)}
\clearpage

\section{Mathematical approach to the TPWA problem} \label{chap:Omelaenko}

\subsection{Introduction} \label{sec:OmelaenkoApproachIntro}

\begin{figure}[h]
 \centering
 \begin{overpic}[width=0.325\textwidth]{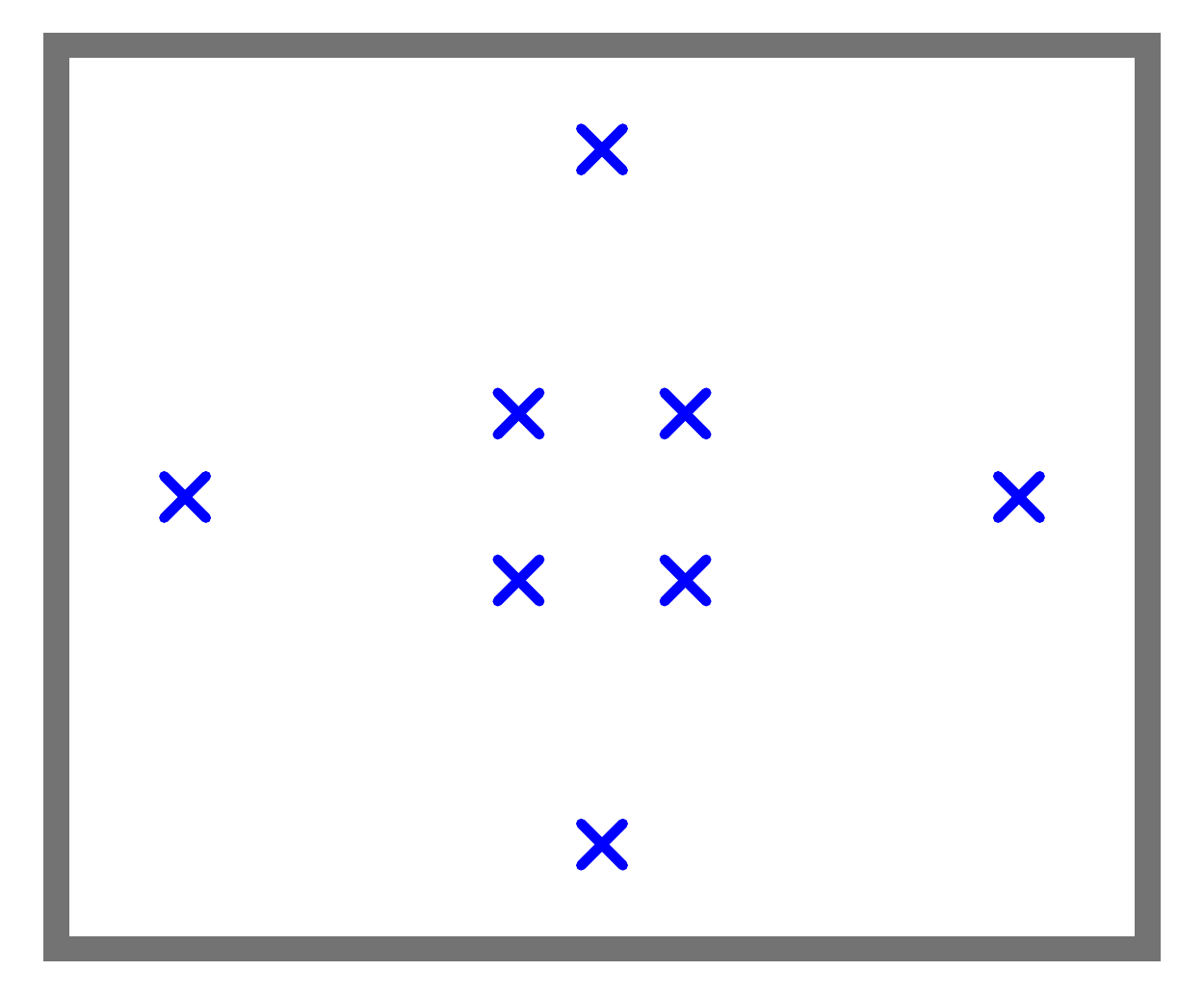}
 \put(8,66){(i)}
 \end{overpic}
 \hspace*{-10pt}
 \begin{overpic}[width=0.325\textwidth]{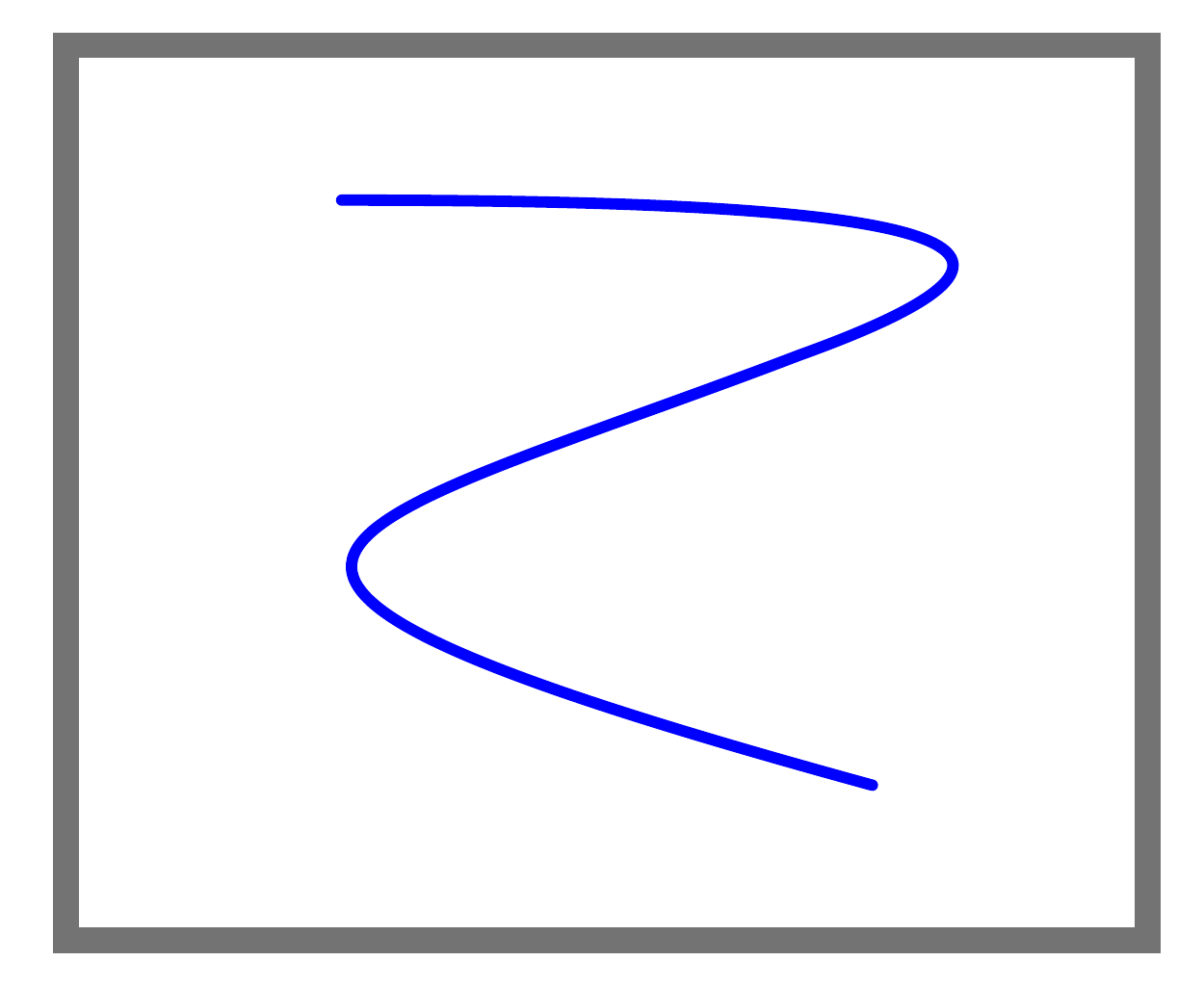}
 \put(8,66){(ii)}
 \end{overpic}
 \hspace*{-10pt}
 \begin{overpic}[width=0.325\textwidth]{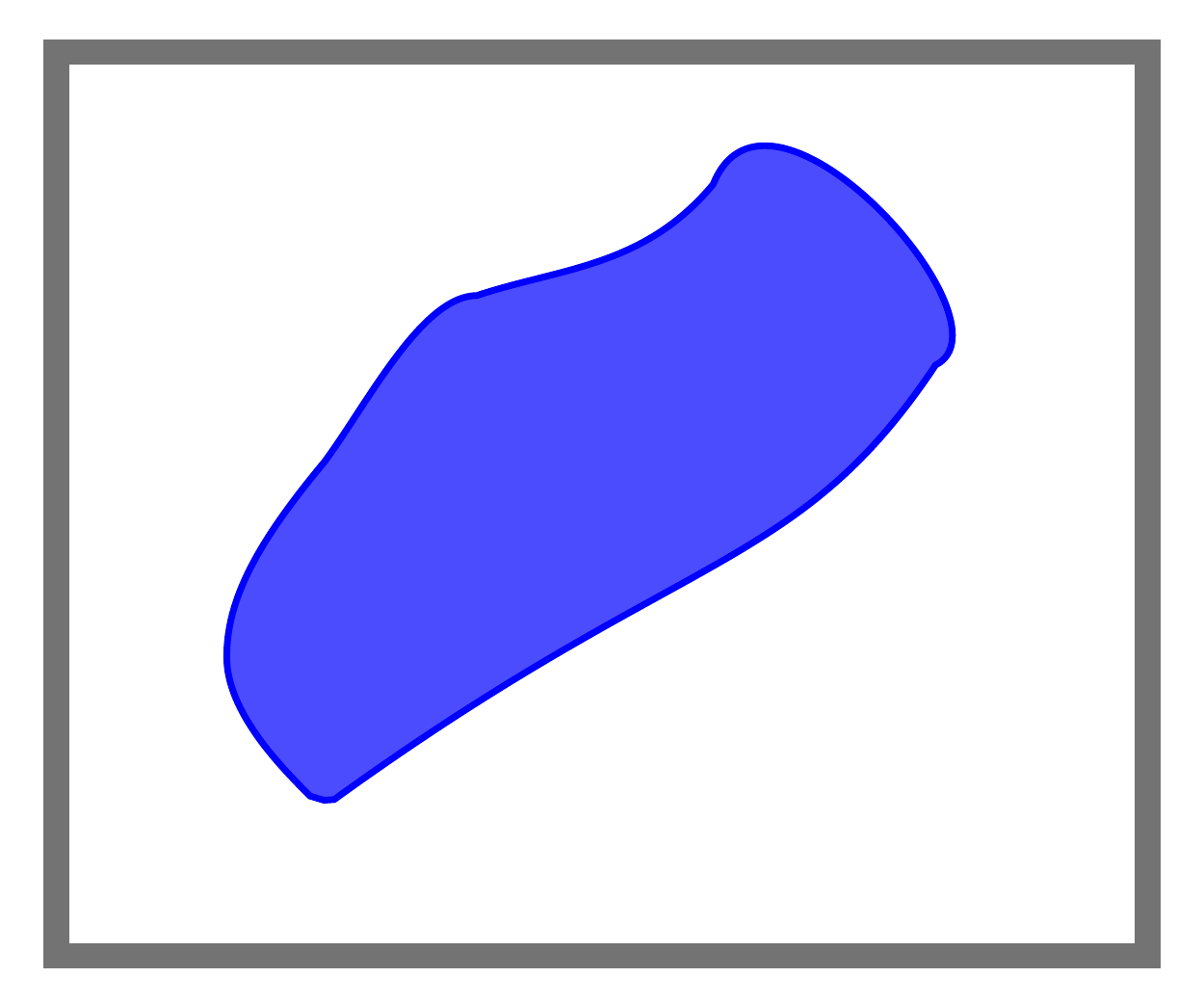}
 \put(8,66){(iii)}
 \end{overpic}
\caption[Sketches illustrating the concepts of discrete and continuum ambiguities.]{The concepts of discrete and continuum ambiguities are illustrated in three sketches. The grey boxes are $2$-dimensional representations of the full $(8 \ell_{\mathrm{max}} - 1)$-dimensional parameter-space of the multipoles in a truncated PWA-model, or the infinite-dimensional parameter space in case of an infinite model, respectively. \newline Discrete ambiguities are shown in (i), where the markers correspond to single points in amplitude-space that yield the same values for an incomplete set of observables. The pictures (ii) and (iii) illustrate lower-dimensional as well as $(8 \ell_{\mathrm{max}} - 1)$-dimensional regions in parameter space (shown as a line and a shaded area respectively), that represent point-sets on which a continuum-ambiguity \cite{BowcockBurkhardt} exists. This means that there exist continuous mappings connecting each point in the respective set and for each value of the mapping, the observables do not change. For a TPWA, the latter kind of ambiguities can in principle be ruled out by the theorem of local invertibility, as described in the main text. Furthermore, the case of mixed discrete and continuum ambiguities, leading to disconnected "islands of ambiguity" (cf. Bowcock and Burkhardt \cite{BowcockBurkhardt}), is not shown in the sketches.}
\label{fig:AmbiguityTypeExplanationPlots}
\end{figure}

In this chapter, we wish to study complete experiments in a TPWA mathematically. This means that we mostly consider the academic case of data with infinite precision,
which in a certain truncation order $L = \ell_{\mathrm{max}}$ admits mathematically exact solutions for the multipoles. \newline
Logically, the aim is to follow the structure of the discussion of complete experiments for the full spin amplitudes performed by Chiang and Tabakin \cite{ChTab} as closely as possible. Therefore, it is useful to briefly recapitulate their discussion which is also depicted in section \ref{sec:CompExpsTPWA}. It is divided into the steps
\begin{itemize}
 \item[1.)] \underline{Treatment of ambiguities:} \newline First of all, Chiang and Tabakin investigated the discrete ambiguities of po\-la\-ri\-za\-tion observables in a detailed way. \newline For this purpose, the transversity-representation of the observables can be chosen: $\check{\Omega}^{\alpha} = \frac{1}{2} \sum_{i,j} b_{i}^{\ast} \tilde{\Gamma}^{\alpha}_{ij} b_{j}$ (see section \ref{subsec:PhotoproductionObs} and appendix \ref{sec:ObservableAlgebra}). The approach consists of investigating the linear ($L$) and antilinear ($A$) ambiguities generated by the transformations
\begin{equation}
 b_{i} \rightarrow b_{i}^{L, \alpha} = \sum_{j=1}^{4} \tilde{\Gamma}^{\alpha}_{ij} b_{j} \mathrm{,} \hspace*{10pt} b_{i} \rightarrow b_{i}^{A, \alpha} = \sum_{j=1}^{4} \tilde{\Gamma}^{\alpha}_{ij} b_{j}^{\ast} \mathrm{,} \hspace*{10pt} \alpha = 1,\ldots,16 \mathrm{,} \label{eq:LanAntiLinAmbChTabChapter2}
\end{equation}
and determining the subset of these transformations that leaves the four group $\mathcal{S}$ observables $\left\{\sigma_{0},\Sigma,T,P\right\}$ invariant (the latter being sums of moduli-squared of the $b_{i}$). Then, the remaining $12$ observables are tested regarding whether or not they are invariant under the found discrete ambiguities of the group $\mathcal{S}$. The subsets of observables which are found to be non-invariant under all of the candidate ambiguities, then give a first hint at the complete experiments. Generally, the ambiguity-study provides a relatively quick and elegant way to identify candidates for complete sets of observables, without having to fully solve the inverse problem for the amplitudes.
 \item[2.)] \underline{Explicit (algebraic) solution of the inverse problem:} \newline In order to investigate the correctness of the results found in step 1.), the bilinear forms defining possible complete sets of observables have to be solved explicitly in order to obtain the amplitudes up to an overall phase. Utilizing special properties of the Dirac $\tilde{\Gamma}$-matrices (appendix \ref{subsec:HelTrGammaReps}), Chiang and Tabakin \cite{ChTab} found the following algebraic expression yielding the bilinear products of the $b_{i}$
\begin{equation}
 b_{i}^{\ast} b_{j} = \frac{1}{2} \sum_{\alpha} \left( \tilde{\Gamma}^{\alpha}_{ij} \right)^{\ast} \check{\Omega}^{\alpha} \mathrm{.} \label{eq:BiStarBjInverted}
\end{equation}
Using the Fierz-identities (see \cite{ChTab} and appendix \ref{subsec:FierzIdentities}) for the observables, it is possible to verify the complete experiments postulated in step 1.) by explicitly showing that the right hand side of (\ref{eq:BiStarBjInverted}) is calculable in terms of $8$ well-selected observables.
\end{itemize}
This chapter will be an attempt at the TPWA-analogue of point 1.), namely the study of ambiguities. Regarding the procedure corresponding to point 2.) in a TPWA, it has to be said that it was not possible (as stated in section \ref{sec:CompExpsTPWA}) to derive algebraic equations that facilitate the explicit inversion of the central bilinear equation systems in a TPWA, namely equation (\ref{eq:LowEAssocLegStandardParametrization2})
\begin{equation}
\left(a_{L}\right)_{k}^{\check{\Omega}^{\alpha}} = \left< \mathcal{M}_{\ell} \right| \left( \mathcal{C}_{L}\right)_{k}^{\check{\Omega}^{\alpha}} \left| \mathcal{M}_{\ell} \right> \mathrm{.} \label{eq:BilinearEqSystemIntroduction}
\end{equation}
Therefore, the analogue to the inversion (\ref{eq:BiStarBjInverted}), which in case of the complete experiments for full amplitudes (section \ref{sec:CompExpsFullAmp}) was still possible analytically, will in the TPWA-case consist of the numerical minimization of "$\chi^{2}$-like" discrepancy functions for exactly solvable data, where the word 'exact' is meant up to a small numerical error
\begin{equation}
\hspace*{-5pt}\Phi \left( \left\{ \mathcal{M}_{\ell} \right\} \right) := \sum_{\alpha, c_{k_{\alpha}}} \left[ \check{\Omega}^{\alpha}_{\mathrm{Data}} \left(c_{k_{\alpha}}\right) - \check{\Omega}^{\alpha}_{\mathrm{Fit}} \left(c_{k_{\alpha}}, \left\{ \mathcal{M}_{\ell} \right\}\right) \right]^{2} \mathrm{.} \label{eq:DefDiscrFunctIntroduction}
\end{equation}
Here, the summation runs over the indices of all the tested observables, as well as over discrete points $c_{k_{\alpha}} \equiv \cos \left( \theta_{k^{\alpha}} \right)$ where the solvable ideal data exist. \newline
Logically, the approach in this thesis will be chosen very much in accord with Chiang and Tabakin \cite{ChTab}. This means we will first of all try to learn as much as possible about the nature of the ambiguities occurring in the TPWA. This will lead to the postulating of (at least mathematically) complete sets of observables. Then, the completeness of the respective sets will have to be verified by numerical solutions, obtained by minimizing functions such as (\ref{eq:DefDiscrFunctIntroduction}). We refer to chapter \ref{chap:TPWA} for details on the results of the second step. \newline
In order to find all the relevant ambiguities, we first have to fully specify the Ansatz for the TPWA that is assumed to be solved. The equation system $\left(a_{L}\right)_{k}^{\check{\Omega}^{\alpha}} = \left< \mathcal{M}_{\ell} \right| \left( \mathcal{C}_{L}\right)_{k}^{\check{\Omega}^{\alpha}} \left| \mathcal{M}_{\ell} \right>$, which defines the procedure, shall be solved for a finite truncation at $L = \ell_{\mathrm{max}}$. Furthermore, we assume a convention that fixes one energy-dependent overall-phase $\phi_{\mathcal{M}} (W)$ for all multipoles. We outline now how this Ansatz leads to the fact that all ambiguities occurring in the TPWA are at most {\it discrete}, i.e. occurring at discrete points in amplitude space and that no so-called "continuum ambiguities" exist any more. Figure \ref{fig:AmbiguityTypeExplanationPlots} further illustrates the different kinds of ambiguities. \newline
According to Bowcock and Burkhardt \cite{BowcockBurkhardt}, all continuum ambiguities originate from an energy- and angle-dependent phase rotation acting on the full amplitudes, which as mentioned in section \ref{sec:CompExpsFullAmp} leaves all observables invariant
\begin{equation}
 F_{j} (W,\theta) \rightarrow e^{i \phi (W, \theta)} F_{j} (W,\theta) \mathrm{,} \hspace*{5pt} j = 1,\ldots,4 \mathrm{.} \label{eq:ContAmbIntroduction}
\end{equation}
Ambiguities coming from both the energy- and angle-dependence of the phase function $\phi (W, \theta)$ are effectively restricted in the TPWA problem defined above. The reasons are:
\begin{itemize}
 \item[(i)] Assuming that the phase-function $\phi$ in equation (\ref{eq:ContAmbIntroduction}) is only energy-dependent, it is seen that multiplying this phase-rotation into the full CGLN-amplitudes is, in case a truncation is assumed, fully equivalent to a multiplication of all multipoles by the same energy-dependent phase (cf. equations (\ref{eq:MultExpF1}) to (\ref{eq:MultExpF4})). However, this freedom was explicitly fixed in the TPWA by assuming some convention for the overall phase. Therefore, this kind of ambiguity cannot occur any more in the fit.
 \item[(ii)] The $\theta$-dependence of the continuum ambiguity transformation (\ref{eq:ContAmbIntroduction}) is a lot more subtle. It can be shown \cite{DeanLee,AlfredPhasePaper,MyPhasePaper}, that a rotation of reaction amplitudes by an angle-dependent phase leads to the mixing of multipoles (or, more generally, partial waves). This is demonstrated in the context of simple scalar scattering amplitudes in an appendix at the end of this section, but holds also for more complicated reactions like photoproduction. The full functional freedom to choose $\phi (W, \theta)$ as an arbitrary, continuous, $\theta$-dependent function is troublesome here, since the number of possibilities to construct such functions is infinite (the problem is infinite-dimensional). \newline
 However, it can be seen that if one starts with a PWA-model truncated at some $L$, rotates it and then demands the rotated model to have the same truncation order $L$, the full functional freedom in the choice of $\phi (W, \theta)$ breaks down \cite{MyPhasePaper}. What remains are a finite, countable set of rotations $\varphi_{p} (W, \theta)$, each of them generating a specific discrete ambiguity. Thus, the severeness of the full angle-dependent continuum ambiguity gets alleviated by restricting the fit to models truncated at a specific order $L$. All statements made here have as of now been best understood in the context of toy models for scalar $2 \rightarrow 2$ scattering. For this case, explicit derivations can be found in the appendix at the end of this section, but they are not essential for the main investigations in this chapter. \newline
 Further investigations \cite{AlfredPhasePaper,MyPhasePaper} have shown that similar facts are highly likely to hold for photoproduction as well. 
\end{itemize}
Another strong motivation to expect the {\it discrete} nature of the ambiguities in the TPWA-problem as defined above comes from the usage of the theorem of local invertibility, cf. for example the book by Forster \cite{ForsterII}.

\vfill

\clearpage
This theorem states that for a differentiable map among spaces with the same real dimension, for example the $(8L-1)$-dimensional parameter space of real- and imaginary-parts of the phase-constrained multipoles and $(8L-1)$ well-selected Legendre-coefficients $\left(a_{L}\right)_{k}^{\check{\Omega}^{\alpha}}$, there exist local regions around points in either space where the respective map is uniquely invertible, if and only if at these points the so-called Jacobian ($\equiv$ determinant of the Jacobian matrix) does not vanish. \newline
In the course of this work, we performed tests using this theorem for some low truncation angular momenta $L$. For this purpose, $(8L-1)$ Legendre coefficients were selected from the group $\mathcal{S}$ observables $\left\{\sigma_{0},\Sigma,T,P\right\}$, denoted now as $a_{i}$ (with the two indices $(\alpha,k)$ merged into the "multi-index" $i$). Then, Jacobians were evaluated by their definition
\begin{equation}
 \bm{\mathcal{J}} := \mathrm{det} \left( \frac{\partial a_{i}}{\partial y_{j}} \right) \mathrm{,} \label{eq:JacobianDefIntroduction}
\end{equation}
where the real- and imaginary parts of the phase constrained multipoles are now collected into the $(8L-1)$-dimensional real parameter vector $(y_{1},\ldots,y_{8L-1})$. We have to report here, without showing the results explicitly, that the Jacobians were clearly generally non-vanishing. Rather, they turned out to be quite complicated algebraic functions of the parameters $y_{j}$. Therefore, by means of the above mentioned theorem, at every point in amplitude space where $\bm{\mathcal{J}}$ does not vanish, there exists a local neighbourhood where the TPWA is uniquely invertible. Continuum ambiguities are then excluded in this local region. \newline
Having motivated the absence of continuum ambiguities in the TPWA as defined above, we now turn to the Ansatz that facilitates the investigation of the discrete ambiguities. \newline
The procedure was first elaborated by Gersten \cite{Gersten}. To illustrate the most important points, we consider the simple example case of $2 \rightarrow 2$ scattering of spinless particles, described by an amplitude $A (s,t) \equiv A (W,\theta)$. This amplitude can be conventionally expanded into partial waves according to
\begin{equation}
 A \left( W, \theta \right) = \sum_{\ell = 0}^{\infty} (2 \ell + 1) A_{\ell} (W) P_{\ell} (\cos \theta) \mathrm{.} \label{eq:StandardSpinlessPWExp}
\end{equation}
If this expansion is now truncated at some finite $L = \ell_{\mathrm{max}}$, then the differential cross section which is (ignoring pre-factors) defined as $\sigma_{0} \left( W, \theta \right) = \left| A \left( W, \theta \right) \right|^{2}$, can be written as a Legendre-expansion
\begin{equation}
 \sigma_{0} \left( W, \theta \right) = \sum_{n = 0}^{2L} (2 n + 1) \bm{c}_{n} (W) P_{n} (\cos \theta) \mathrm{,} \label{eq:StandardSpinlessCrossSectionExp}
\end{equation}
where the energy-dependent real expansion coefficients $\bm{c}_{n}$ are bilinear hermitean forms in the partial waves $A_{\ell}$ (cf. Bowcock and Burkhardt \cite{BowcockBurkhardt})
\allowdisplaybreaks
\begin{align}
 \bm{c}_{n} (W) &= \sum_{\ell, \ell^{\prime}} \left<  \ell, 0 ; \ell^{\prime}, 0 | n, 0 \right>^{2} \frac{(2 \ell + 1) (2 \ell^{\prime} + 1)}{(2 n + 1)} A_{\ell} (W) A_{\ell^{\prime}}^{\ast} (W) \nonumber \\ 
 &\equiv \sum_{\ell, \ell^{\prime}} \bar{\mathcal{C}}^{n}_{\ell, \ell^{\prime}} A_{\ell} (W) A_{\ell^{\prime}}^{\ast} (W) \mathrm{.} \label{eq:SpinlessExpCoeffForm}
\end{align}
In this expansion, one finds the usual Clebsch-Gordan coefficients $\left<  \ell, m ; \ell^{\prime}, m^{\prime} | n, M \right>$ (cf. \cite{ThompsonAngMom}). The question is now how many and which discrete ambiguities exist in the procedure of solving the $\bm{c}_{n}$ for the partial waves $A_{\ell}$ (up to an overall phase). Gersten \cite{Gersten} noticed that a truncated version of the expansion (\ref{eq:StandardSpinlessPWExp}), being just a polynomial of order $L$ in $\cos \theta$, can by means of the fundamental theorem of algebra be decomposed into linear factors as follows

\begin{equation}
 A \left( W, \theta \right) = \lambda \prod_{i = 1}^{L} \left( \cos \theta - \alpha_{i} (W) \right) \mathrm{,} \label{eq:LinFactDecompIntroduction}
\end{equation}
provided that the highest partial wave $A_{L}$ has been absorbed into the overall complex normalization-factor $\lambda$. The scattering-amplitude is now mainly given in terms of $L$ complex roots $\left\{\alpha_{i}\right\}$. When written in terms of the decomposition (\ref{eq:LinFactDecompIntroduction}), the differential cross section takes the form
\begin{equation}
 \sigma_{0} \left( W, \theta \right) = \left| \lambda \right|^{2} \prod_{i=1}^{L} \left( \cos \theta - \alpha_{i} \right) \left( \cos \theta - \alpha^{\ast}_{i} \right) \mathrm{,} \label{eq:DCSLinFactDecomp}
\end{equation}
where $\cos \theta$ was assumed to be real. It is now quickly seen that the complex conjugation of either all the roots $\left\{\alpha_{i}\right\}$
\begin{equation}
 \alpha_{i} \rightarrow \alpha_{i}^{\ast} \mathrm{,} \hspace*{5pt} \forall i = 1,\ldots,L \mathrm{,} \label{eq:ConjugateAllRootsIntroduction}
\end{equation}
or only subsets of them, leaves the differential cross section (\ref{eq:DCSLinFactDecomp}) unchanged. The conjugation of different combinations of roots generates precisely the discrete partial-wave-ambiguities that were searched for. A simple inductive counting argument shows that there can be at most $\mathcal{N} = 2^{L}$ of them for each truncation order $L$ (a proof of this statement is given at the end of appendix \ref{subsec:AccidentalAmbProofsI}). \newline
The exponentially rising number of ambiguities seems troubling and the issue arises about whether or not one can demand additional constraints that can reduce the number of ambiguities, without introducing model-dependent amplitudes. In the energy region where the spinless scattering $a + b \rightarrow a + b$ is purely elastic, a very powerful constraint is provided by the unitarity of the $S$-matrix, corresponding to the fundamental principle of conservation of probabilities (cf. section \ref{subsec:Energy-dependentModels}). We repeat here an argument by Berends and Ruijsenaars \cite{BerendsEtAl} in the version given by Bowcock and Burkhardt \cite{BowcockBurkhardt}. Although the latter reference calls the argument non-conclusive, it is still interesting to see how it hints at the fact that unitarity can reduce the number of ambiguities. \newline
As stated already in more detail within section \ref{subsec:Energy-dependentModels}, the unitarity relation $\hat{S}^{\dagger} \hat{S} = \mathbbm{1}$ for the $S$-matrix $\hat{S} \equiv \mathbbm{1} + 2 i \hat{T}$ is well-known to lead to unitarity equations such as (\ref{eq:UnitarityTMatrixStep2}) and (\ref{eq:UnitarityTMatrixStep3}) for the matrix-elements of the transition-operator $\hat{T}$. Defining matrix elements of $\hat{T}$ between the initial state $\left| i \right>$ and final state $\left| f \right>$ and factoring out a $4$-momentum conserving delta-function defines the spinless scattering amplitude called '$A$': $\left< f \right| \hat{T} \left| i \right> \equiv \mathcal{T}_{fi} =: (2 \pi)^{4} \delta^{4} \left( \tilde{P}_{f} - P_{i} \right) A_{fi}$. Then, for time-reversal invariant interactions, the unitarity-equation (\ref{eq:UnitarityTMatrixStep3}) reads for purely elastic scattering \cite{BowcockBurkhardt}:
\begin{equation}
 \mathrm{Im} \left[ A_{fi} \right] = \int d \Phi_{\bm{n}} A_{f\bm{n}}^{\ast} A_{\bm{n}i} \mathrm{.} \label{eq:UnitarityFullAmplitude}
\end{equation}
The integral on the right-hand-side reaches over the full invariant phase space of the only possible $2$-particle intermediate state $\left| \bm{n} \right>$\footnote{For the general case of a total initial $4$-momentum $P_{i}$ leading to an intermediate state $\left| \bm{n} \right>$ containing $N_{\bm{n}}$ particles, the phase-space differential is defined as \cite{BowcockBurkhardt, PeskinSchroeder}: \newline $d \Phi_{\bm{n}} := \prod_{j = 1}^{N_{\bm{n}}} \frac{d^{3} p_{j}}{(2 \pi)^{3} 2 E_{j}} (2 \pi)^{4} \delta^{(4)} \left( P_{i} - \sum_{j=1}^{N_{\bm{n}}} p_{j} \right)$, with $E_{j}$ the energy of the respective intermediate particle. For the special case of the $2$-body system $\bm{n}$ present in equation (\ref{eq:UnitarityFullAmplitude}), this becomes: $d \Phi_{\bm{n}} \equiv \frac{d^{3} p_{1}}{(2 \pi)^{3} 2 E_{1}} \frac{d^{3} p_{2}}{(2 \pi)^{3} 2 E_{2}} (2 \pi)^{4} \delta^{(4)} \left( P_{i} - p_{1} - p_{2} \right)$.}. If now the partial wave expansion (\ref{eq:StandardSpinlessPWExp}) is applied to the full amplitudes appearing on the left- and right-hand-side of equation (\ref{eq:UnitarityFullAmplitude}) and furthermore, a lot of care is taken with the intermediate phase-space-integral, it is seen that in the region of elastic scattering, unitarity takes the following particularly simple form for the partial waves \cite{BowcockBurkhardt}
\begin{equation}
 \mathrm{Im} \left[ A_{\ell} \right] = \left| A_{\ell} \right|^{2} \mathrm{,} \hspace*{5pt} \ell = 0, \ldots , L \mathrm{.} \label{eq:SpinlessPWUnitarityOneEquation}
\end{equation}
This is equivalent to the parametrization of the partial waves in terms of phase-shifts
\begin{equation}
 A_{\ell} = \frac{1}{2 i q} \left( e^{2 i \delta_{\ell}} - 1 \right) = \frac{1}{q} e^{i \delta_{\ell}} \sin \delta_{\ell} \mathrm{,} \label{eq:PWElasticUnitaryParametrization}
\end{equation}
where $q$ is the modulus of the CMS $3$-momentum in the initial state. Since from equation (\ref{eq:SpinlessExpCoeffForm}) one always has $\bm{c}_{2L} = \bar{\mathcal{C}}^{2L}_{L, L} \left| A_{L} \right|^{2}$, it is seen that the data and unitarity directly fix $\sin^{2} \delta_{L}$ \cite{BowcockBurkhardt}. One may thus just choose $\sin \delta_{L} > 0$. \newline \newline
It is interesting that the powerful constraints imposed by the elastic unitarity equations (\ref{eq:SpinlessPWUnitarityOneEquation}), which in their general solution (\ref{eq:PWElasticUnitaryParametrization}) are capable of reducing the number of free parameters for each partial wave from two (for instance real- and imaginary part) to one phase-shift $\delta_{\ell}$, still do not resolve all ambiguities, even for truncated analyses. Crichton \cite{Crichton} first found an example for two ambiguous solutions, in a TPWA truncated at $L = 2$ which imposed elastic unitarity. The fact that for $L=2$, only a two-fold discrete ambiguity can exist under elastic unitarity has been established \cite{Martin1969} and later generalized to some higher truncation orders \cite{AtkinsonEtAl1973a, AtkinsonEtAl1974, CornilleDrouffe1974}. It is believed, but (to our knowledge) not proven, that at most a two-fold Crichton-type ambiguity exists in the elastic energy-region for any truncation order or even infinite partial wave expansions \cite{BowcockBurkhardt}. Having mentioned the fact that even the powerful constraint of elastic unitarity allows for some residual discrete ambiguities, we proceed by outlining the argument by  Berends and Ruijsenaars \cite{BerendsEtAl,BowcockBurkhardt}, illustrating the capability of unitarity-arguments to remove discrete ambiguities. \newline
Unitarity shall be imposed for all partial waves except the highest one. It is assumed in the beginning that the $\mathcal{N} = 2^{L}$ possible ambiguous solutions contain $N_{\mathrm{unit.}} < \mathcal{N}$ solutions that obey unitarity. Then, the partial wave amplitudes corresponding to these solutions have to satisfy the unitarity equations
\begin{equation}
 \mathrm{Im} \left[ A_{\ell}^{k} \right] = \left| A_{\ell}^{k} \right|^{2} \mathrm{,} \hspace*{5pt} k = 1, \ldots, N_{\mathrm{unit.}} \mathrm{;} \hspace*{5pt} \ell = 0, \ldots , L - 1 \mathrm{.} \label{eq:SpinlessPWUnitarity}
\end{equation}
With the phase $\delta_{L}$ fixed, the amplitudes are still determined by $2L+1$ quantities. These are $\delta_{L}$, the real parts of the zeros $\left\{\alpha_{i}\right\}$ and the moduli of their imaginary parts. The set of $N_{\mathrm{unit.}}$ unitarity equations (\ref{eq:SpinlessPWUnitarity}) on the other hand represents in total $N_{\mathrm{unit.}} L$ constraints. In case one assumes the constraints to be independent, then in order for a solution to exist at all, there are not allowed to be more constraining relations than free variables in the problem. This directly leads to \cite{BowcockBurkhardt}
\begin{equation}
 N_{\mathrm{unit.}} L \leq 2 L + 1 \mathrm{,} \label{eq:UnitaritySolutionEstimate}
\end{equation}
which is equivalent to $N_{\mathrm{unit.}} \leq 2$. This estimate motivates that for spinless elastic scattering, unitarity provides a powerful constraint that is very likely to reduce the number of ambiguities. This constraint is of course lost as soon as one crosses the first inelastic threshold. The occurrence of the full set of $\mathcal{N} = 2^{L}$ discrete ambiguities does not depend on the energy region under consideration, since no assumption about unitarity was made in its derivation. It is valid for partial waves generally given in terms of two real variables. The latter could be either the real- and imaginary part of each wave, or a phase-shift $\delta_{\ell}$ and a general elasticity $\eta_{\ell} \leq 1$. \newline \newline
In the case of pseudoscalar meson photoproduction, the problem is that the reaction itself is already a production and not an elastic scattering, i.e. there cannot be a fully elastic energy region. However, for the special case of photoproduction of pions
\begin{equation}
 \gamma N \rightarrow \pi N \mathrm{,} \label{eq:PionPhProdIntroduction}
\end{equation}
unitarity in the guise of the Watson-theorem \cite{KMWatson} can have a similar effect of removing discrete ambiguities \cite{Grushin,WorkmanGrushinFits}. This is of course only true as long as one does not exceed the threshold for the $2\pi$ production process $\gamma N \rightarrow \pi \pi N $. For general CMS-energies, one has to live with the photoproduction-analogue of the $2^{L}$ discrete Gersten-ambiguities derived above, which will amount to an even larger number of discrete partial-wave ambiguities. \newline
One final remark has to be made about the discrete ambiguities derived above. Gersten \cite{Gersten} states, without proof, that the $2^{L}$ possibilities arising from the complex conjugation of different combinations of roots already give {\it all} discrete ambiguities that can possibly be there. This means that the $2^{L}$ ambiguities already fully exhaust the maximal possible set of simultaneous solutions of the bilinear equation system composed out of the equations (\ref{eq:SpinlessExpCoeffForm}). No more solutions can exist, which may be built by manipulating the roots in a more complicated way than just complex conjugation. This seems like a very strong statement and in the course of this work, no proof of it was found. However, since every reference dealing with discrete partial-wave ambiguities (cf. \cite{BowcockBurkhardt}, \cite{BerendsEtAl} among others) indeed states that all non-trivial discrete ambiguities can be reached via the Gersten-procedure (provided the energy-dependent overall phase $\phi(W)$ is fixed), we just assume this statement to be true in the ensuing discussion. This question has also been re-investigated in a recent paper \cite{MyPhasePaper}. \newline \newline
In section \ref{sec:WBTpaper}, a publication \cite{MyCompExTPWAPaper} written in the course of this thesis is contained which deals with the application of Gersten's method to photoproduction. This approach was first published by Omelaenko \cite{Omelaenko}. However, in section \ref{sec:WBTpaper}, first conclusions regarding completeness-rules for polarization observables in a TPWA are drawn from this previously published work. \newline
Appendix \ref{sec:AdditionsChapterII} contains supplementary as well as extending material that is not contained in the original publication, due to the fact that either it would have made the paper too extensive, or just because the respective results were not known and worked out at the time section \ref{sec:WBTpaper} was published. Appendix \ref{sec:AccidentalAmbProofs} in particular further explores the nature and relevance of different types of discrete ambiguities in a TPWA. \newline \newpage

\textit{Appendix: The angular dependence of the phase-rotation function $\phi (W, \theta)$ causes the mixing of partial waves. Furthermore, discrete ambiguities correspond to a unique and countable set of angle-dependent phase rotations $\varphi_{p} (W, \theta)$, which leave the truncation order $L$ of an original truncated model unchanged:} \newline

We point out the fact that an angular-dependent phase-function in the con\-ti\-nu\-um-ambiguity transformation can lead to the mixing of partial waves. All (!) of the formal developments shown in this section-appendix can be found in another publication \cite{MyPhasePaper} written in the course of this work, which however contains more details. \newline We assume infinite partial wave series (\ref{eq:StandardSpinlessPWExp}) for both the original amplitude $A (W,\theta)$ as well as the transformed one $\tilde{A} (W,\theta) := e^{i \phi(W,\theta)} A (W,\theta)$. This means that, in reverse, the partial waves of both amplitudes can be extracted by means of standard projection integrals
\allowdisplaybreaks
\begin{align}
 A_{\ell} (W) &= \frac{1}{2} \int_{-1}^{1} d (\cos \theta) A (W,\theta) P_{\ell} (\cos \theta) \mathrm{,} \label{eq:OriginalAmplProjection} \\
 \tilde{A}_{\ell} (W) &= \frac{1}{2} \int_{-1}^{1} d (\cos \theta) \tilde{A} (W,\theta) P_{\ell} (\cos \theta) \mathrm{.} \label{eq:TransformedAmplProjection}
\end{align}
Furthermore, the analytic angular-dependence of the phase function $\phi (W,\theta)$ that was assumed above, makes it possible to write an infinite Legendre-series for the phase-\textit{rotation} function
\begin{equation}
 e^{i \phi (W, \theta)} = \sum_{\ell^{\prime} = 0}^{\infty} L_{\ell^{\prime}} (W) P_{\ell^{\prime}} (\cos \theta) \mathrm{,} \label{eq:PhaseRotTrafoLegendreSeries}
\end{equation}
which is fully specified by the complex energy-dependent expansion coefficients $L_{\ell^{\prime}}$. Utilizing this expansion, it is seen that the projection of the transformed partial wave (\ref{eq:TransformedAmplProjection}) becomes
\allowdisplaybreaks
\begin{align}
 \tilde{A}_{\ell} (W) &= \frac{1}{2} \int_{-1}^{1} d (\cos \theta) \tilde{A} (W,\theta) P_{\ell} (\cos \theta) \nonumber \\
   &= \frac{1}{2} \int_{-1}^{1} d (\cos \theta) e^{i \phi (W, \theta)} A (W,\theta) P_{\ell} (\cos \theta) \nonumber \\
   &= \frac{1}{2} \int_{-1}^{1} d (\cos \theta) \sum_{\ell^{\prime} = 0}^{\infty} L_{\ell^{\prime}} (W) P_{\ell^{\prime}} (\cos \theta) A (W,\theta) P_{\ell} (\cos \theta) \nonumber \\
  &= \sum_{\ell^{\prime} = 0}^{\infty} L_{\ell^{\prime}} (W) \hspace*{2pt} \frac{1}{2} \int_{-1}^{1} d (\cos \theta) A (W,\theta) P_{\ell^{\prime}} (\cos \theta) P_{\ell} (\cos \theta) \mathrm{.} \label{eq:MixingDerivation1}
\end{align}
The product of two Legendre-polynomials under the integral can be dealt with by applying a known formula \cite{Adams, ThompsonAngMom} for the expansion of this product into, again, Legendre polynomials
\begin{align}
 P_{k} (x) P_{\ell} (x) &= \sum_{m = \left| k - \ell \right|}^{ k + \ell } \left( \begin{array}{ccc} k & l & m \\ 0 & 0 & 0 \end{array} \right)^{2} (2m+1) P_{m} (x) \nonumber \\
 &= \sum_{m = \left| k - \ell \right|}^{ k + \ell } \left< k,0 ; \ell,0 | m , 0 \right>^{2} P_{m} (x) \mathrm{.} \label{eq:AdamsFormula}
\end{align}
This formula can be written by either using squared $3j$-symbols or squared Clebsch-Gordan coefficients. Using the knowledge of this expansion, the expression (\ref{eq:MixingDerivation1}) becomes
\allowdisplaybreaks
\begin{align}
 \tilde{A}_{\ell} (W) &= \sum_{\ell^{\prime} = 0}^{\infty} L_{\ell^{\prime}} (W) \hspace*{2pt} \frac{1}{2} \int_{-1}^{1} d (\cos \theta) A (W,\theta) \sum_{m = \left| \ell^{\prime} - \ell \right|}^{ \ell^{\prime} + \ell } \left< \ell^{\prime},0 ; \ell,0 | m , 0 \right>^{2} P_{m} (\cos \theta) \nonumber \\
 &= \sum_{\ell^{\prime} = 0}^{\infty} L_{\ell^{\prime}} (W) \hspace*{2pt} \sum_{m = \left| \ell^{\prime} - \ell \right|}^{ \ell^{\prime} + \ell } \left< \ell^{\prime},0 ; \ell,0 | m , 0 \right>^{2}  \hspace*{2pt}  \frac{1}{2} \int_{-1}^{1} d (\cos \theta) A (W,\theta) P_{m} (\cos \theta) \nonumber \\
 &= \sum_{\ell^{\prime} = 0}^{\infty} L_{\ell^{\prime}} (W) \hspace*{2pt} \sum_{m = \left| \ell^{\prime} - \ell \right|}^{ \ell^{\prime} + \ell } \left< \ell^{\prime},0 ; \ell,0 | m , 0 \right>^{2} A_{m} (W) \mathrm{,} \label{eq:MixingDerivation2Final}
\end{align}
where in the last step the projection formula (\ref{eq:OriginalAmplProjection}) was used. It is seen that the partial waves of the transformed amplitude are generally a complicated admixture of an infinite tower of partial waves from the original amplitude, which illustrates just how severe the ambiguity really is. \newline
Now, we establish the connection of the above mentioned general continuum ambiguities and the discrete ambiguities of Gersten \cite{Gersten}. Recalling the linear factor decomposition (\ref{eq:LinFactDecompIntroduction}) from the main text, i.e.
\begin{equation} 
A (W, \theta)  = \sum_{\ell = 0}^{L} (2 \ell + 1) A_{\ell} (W) P_{\ell} (\cos \theta) \equiv \lambda \prod_{i = 1}^{L} \left( \cos \theta - \alpha_{i} \right) \mathrm{,} \label{eq:GerstenDecomposedScalarAmplitude} 
\end{equation}
the discrete ambiguity transformations which leave the differential cross section $\sigma_{0}$ invariant shall now be established more formally. As mentioned in the introduction, those transformations consist of the complex conjugations of either all or subsets of the roots $\left\{ \alpha_{i} \right\}$. We use here the formal scheme of Gersten \cite{Gersten}, who counted all ambiguities by labelling a set of $2^{L}$ maps $\bm{\uppi}_{\hspace*{0.035cm}p}$, for $p = 0,\ldots,(2^{L} - 1)$, defined by:
\begin{equation}\bm{\uppi}_{\hspace*{0.035cm}p} \left(\alpha_{i}\right) := \begin{cases}
                    \alpha_{i} &\mathrm{,} \hspace*{3pt} \mu_{i} \left(p\right) = 0 \\
                    \alpha_{i}^{\ast} &\mathrm{,} \hspace*{3pt} \mu_{i} \left(p\right) = 1 
                   \end{cases}\mathrm{,} \hspace*{2.5pt} \mathrm{using} \hspace*{2.5pt} p = \sum_{i = 1}^{L} \mu_{i} \left( p \right) 2^{(i-1)}\mathrm{.} \label{eq:GerstenMapsScalarExample}
\end{equation}
This kind of binary counting of ambiguities is generalized to photoproduction in section \ref{subsec:TheoryDataFitsLmax1} and appendix \ref{subsec:AccidentalAmbProofsI}. Provided an energy-dependent phase-convention, conventionally for instance $A_{0} = \mathrm{Re} \left[ A_{0} \right] \geq 0$, has been established, the ambiguities (\ref{eq:GerstenMapsScalarExample}) exhaust all possibilities for discrete symmetries in the TPWA problem truncated at $L$. As stated in the main text, this fact is not proven in the original publication by Gersten \cite{Gersten}. Further below, another method will be outlined which does not depend explicitly on the Gersten-roots $\left\{ \alpha_{i} \right\}$ and still leads to the conclusion that this uniqueness of the Gersten-ambiguities is true. \newline
However, first of all the formal definition of the ambiguity maps (\ref{eq:GerstenMapsScalarExample}) allows one to introduce transformed, truncated amplitudes
\begin{equation} 
A^{(p)} (W, \theta) = \lambda \prod_{i = 1}^{L} \left( \cos \theta - \bm{\uppi}_{\hspace*{0.035cm}p} \left[\alpha_{i}\right] \right) \equiv \sum_{\ell = 0}^{L} (2 \ell + 1) A^{(p)}_{\ell} (W) P_{\ell} (\cos \theta) \mathrm{,} \label{eq:GerstenTransformedScalarAmplitude} 
\end{equation} 
for $p = 0,\ldots,(2^{L} - 1)$ which, by way of their definition, yield the exact same cross section as $A (W, \theta)$. They are seen to have the same truncation order $L$ as the original amplitude and can be decomposed into partial waves $A^{(p)}_{\ell} (W)$.
Using the formal expressions (\ref{eq:GerstenDecomposedScalarAmplitude}) and (\ref{eq:GerstenTransformedScalarAmplitude}), a countable set of phase-functions $\varphi_{p} (W, \theta)$, capable of rotating the original amplitude $A$ to the discrete ambiguity $A^{(p)}$, can be defined easily by just dividing both amplitudes 
\begin{equation}
e^{i \varphi_{p} (W, \theta)} = \frac{A^{(p)} (W, \theta)}{A(W,\theta)} = \frac{\left( \cos \theta - \bm{\uppi}_{\hspace*{0.035cm}p} \left[\alpha_{1}\right] \right) \ldots \left( \cos \theta - \bm{\uppi}_{\hspace*{0.035cm}p} \left[\alpha_{L}\right] \right)}{\left( \cos \theta - \alpha_{1} \right) \ldots \left( \cos \theta - \alpha_{L} \right)} \mathrm{.} \label{eq:GerstenFormalismPhaseRotation} 
\end{equation}
Therefore, the rotations $e^{i \varphi_{p} (W, \theta)}$, belonging to $\varphi_{p} (W, \theta)$ for $p = 0,\ldots,(2^{L} - 1)$, can be quickly extracted from the Gersten-formalism. Multiplying the right hand side of equation (\ref{eq:GerstenFormalismPhaseRotation}) by its own complex conjugate, it can be seen (remembering the definition (\ref{eq:GerstenMapsScalarExample})) that it has modulus $1$ for all real $\theta$. Furthermore, for any non-trivial ambiguity $\bm{\uppi}_{\hspace*{0.035cm}p}$, i.e. any apart from the identity $\bm{\uppi}_{\hspace*{0.035cm}0}$, the rotation (\ref{eq:GerstenFormalismPhaseRotation}) is angle-dependent for real values of $\theta$. Therefore, when written as a Legendre-series as in equation (\ref{eq:PhaseRotTrafoLegendreSeries}), the non-trivial ambiguities (\ref{eq:GerstenFormalismPhaseRotation}) will be infinite series in $\cos \theta$. The partial waves $A^{(p)}_{\ell} (W)$ of the rotated amplitude will, by means of the mixing formula (\ref{eq:MixingDerivation2Final}), be complicated admixtures of partial waves from the original amplitude. However, since from the definition (\ref{eq:GerstenFormalismPhaseRotation}) it can be seen that the rotated amplitude is manifestly truncated as well, the admixture caused by the infinite number of non-vanishing Legendre-coefficients $L_{\ell^{\prime}} (W)$ has to be such that all partial waves $A^{(p)}_{\ell} (W)$ above the cutoff $L$ have to vanish exactly. This can only be accomplished by an exact cancellation-effect in all higher partial waves above $L$. \newline
Having obtained the expression (\ref{eq:GerstenFormalismPhaseRotation}) for the phase-rotations generating the discrete Gersten-ambiguities, a further interesting result found in the course of this work is the fact that they are the {\it only} angle-dependent phases that can rotate the original truncated amplitude $A(W, \theta)$ again into a truncated one $A^{(p)} (W, \theta)$. This fact is closely connected to the statement by Gersten mentioned at the end of this introduction, namely that the maps $\bm{\uppi}_{\hspace*{0.035cm}p}$ already fully exhaust all discrete ambiguities that can exist in a TPWA. In order to establish this unique feature of the phases $\varphi_{p} (W, \theta)$ as correct, it would be desirable to have an alternative method which does not make explicit use of the Gersten-formalism. The remainder of this appendix shall be used to describe a scheme developed for this purpose in the course of this work (see reference \cite{MyPhasePaper} for more details).\newline
One starts out with a truncated amplitude $A(W,x) = \sum_{\ell \leq L} (2 \ell + 1) A_{\ell} (W) P_{\ell} (x)$, where from now on the notation $x = \cos \theta$ is used. Then, a scheme is constructed to systematically search for all possible phase-rotation functions $F(W,x) := e^{i \varphi (W,x)}$ leading to a rotated model $\tilde{A} (W,x)$, such that the following two conditions are satisfied
\begin{itemize}
 \item[(i)] The rotated amplitude $\tilde{A}$, coming out of an amplitude $A$ truncated at $L$, has to be truncated as well, i.e.
 \begin{equation}
 \tilde{A}_{L + k} (W) = 0 \mathrm{,} \hspace*{5pt} \forall k = 1,\ldots,\infty \mathrm{.} \label{eq:RotAmplTruncatedAsWellRequireThis}
 \end{equation}
 \item[(ii)] The complex solution-function $F (W, x)$ has to have modulus 1 at each angle $x$
 \begin{equation}
 \left| F (W, x) \right|^{2} = 1 \mathrm{,} \hspace*{5pt} \forall x \in \left[ -1, 1 \right] \mathrm{.} \label{eq:PhaseModulus1}
 \end{equation}
\end{itemize}
The proposed problem is a functional-analysis (or ``functional'' for short) problem, since one tries to scan a full vectorspace, consisting out of an infinte numer of possible functions $\varphi (W,x)$, for solutions of the problem. The goal is to obtain a complex function $F (W, x)$ of modulus $1$ everywhere, which is a solution of the infinite set of functional conditions
\allowdisplaybreaks
\begin{equation}
 \tilde{A}_{L + k} (W) = \frac{1}{2} \int_{-1}^{+1} d x  F (W, x)  A (W, x) P_{L + k} (x) \equiv 0 \mathrm{,} \hspace*{5pt} \forall \hspace*{2pt} k = 1, \ldots, \infty \mathrm{.} \label{eq:FunctProblem}
\end{equation}
Since in all practical examples solution functions were only possible to be found numerically, it is clear that in all realistic cases, the infinite tower of conditions (\ref{eq:FunctProblem}) has to be implemented for a finite set of indices
\begin{equation}
 k = 1, \ldots, K_{\mathrm{cut}} \mathrm{,} \label{eq:NIndexRestriction}
\end{equation}
with a cutoff $K_{\mathrm{cut}}$ chosen as large as possible. Solutions to the conditions (i) and (ii) above can now be found numerically, by means of the minimization of a suitably chosen functional. A simple definition \cite{MyPhasePaper}, which implements both constraints, can be written symbolically as (with the energy-dependences suppressed)
\begin{align}
 &\bm{W} \left[ F(x) \right] := \sum_{x} \left(  \mathrm{Re}\left[F(x)\right]^{2}  + \mathrm{Im}\left[F(x)\right]^{2} - 1 \right)^{2}  + \mathrm{Im} \left[  \frac{1}{2} \int_{-1}^{+1} dx \hspace*{1pt} F(x) A(x) \right]^{2}  \nonumber \\
  & \hspace*{1pt} + \sum_{k \geq 1} \Bigg\{ \mathrm{Re} \left[ \frac{1}{2} \int_{-1}^{+1} dx \hspace*{1pt} F(x) A(x) P_{L + k} (x) \right]^{2} + \mathrm{Im} \left[ \frac{1}{2} \int_{-1}^{+1} dx \hspace*{1pt} F(x) A(x) P_{L + k} (x) \right]^{2}  \Bigg\} \mathrm{.} \label{eq:FunctProblemMinimizationFunctionalDef}
\end{align}
The first term in this expression guarantees the unimodularity of $F(x)$. It is here written as a sum $\sum_{x}$, since in all practical examples it has been implemented as a sum over a discrete, finely sown grid of equidistant points $\left\{x_{n}\right\} \in \left[ -1,1 \right]$. The convention $A_{0} = \mathrm{Re} \left[ A_{0} \right]$ for the solely energy-dependent overall phase is incorporated in the definition (\ref{eq:FunctProblemMinimizationFunctionalDef}) by the second term in the first line. This means that here, a sign-ambiguity $A_{\ell} \rightarrow - A_{\ell}$ is still possible. Finally, the condition (\ref{eq:FunctProblem}) for all partial waves above $L$ to vanish is implemented by the infinite sum in the second line of equation (\ref{eq:FunctProblemMinimizationFunctionalDef}). It is clear that, for practical calculations, this sum over $k$ has to be truncated at a suitable $K_{\mathrm{cut}}$. \newline 
In order to accomplish the numerical search for functions minimizing the functional $\bm{W} \left[ F(x) \right]$, a general way of parametrizing the sought-after functions has to be chosen (cf. \cite{MyPhasePaper}). A Legendre-series can be used just as in expression (\ref{eq:PhaseRotTrafoLegendreSeries}) above, which in practical cases however can only be a finite sum
\begin{equation}
 F(W,x) = \sum_{\ell^{\prime} = 0}^{\mathcal{L}_{\mathrm{cut}}} L_{\ell^{\prime}} (W) P_{\ell^{\prime}} (x) \mathrm{.} \label{eq:PhaseRotTrafoLegendreSeriesTruncated}
\end{equation}
Another possibility is to parametrize $F(W,x)$ by its values on a finite grid of points $\left\{x_{n}\right\} \in \left[ -1,1 \right]$. Both methods, i.e. Legendre-expansions and function-discretization, have yielded consistent results in practical examples \cite{MyPhasePaper}. \newline
Finally, we have to state here the important fact that for all examples considered in this work, which were all given by Toy-models, the numerical minimization of the functional (\ref{eq:FunctProblemMinimizationFunctionalDef}) has, for large ensembles of randomly chosen initial functions, yielded only the discrete Gersten-ambiguites (\ref{eq:GerstenFormalismPhaseRotation}) as solutions. This means, from all investigations done up to this point, one can proclaim the behavior \cite{MyPhasePaper}
\begin{align}
 \bm{W} \left[ F (W,\theta) \right] &\longrightarrow \mathrm{min.} \equiv 0 \mathrm{,} \label{eq:MinimFunctionalDef} \\
 \mathrm{for} \hspace*{2.5pt}  F (W,\theta) &\longrightarrow F_{p} (W,\theta) = e^{i \varphi_{p} (W, \theta)} \mathrm{,} \hspace*{2.5pt} p = 0,\ldots,(2^{L} - 1) \mathrm{.} \label{eq:FunctionLimit}
\end{align}

\subsection{The complete-experiment problem of photoproduction of pseudoscalar mesons in a truncated partial wave analysis} \label{sec:WBTpaper}

\includepdf[pages={1}, frame=false, noautoscale=true, scale=0.86, pagecommand={}]{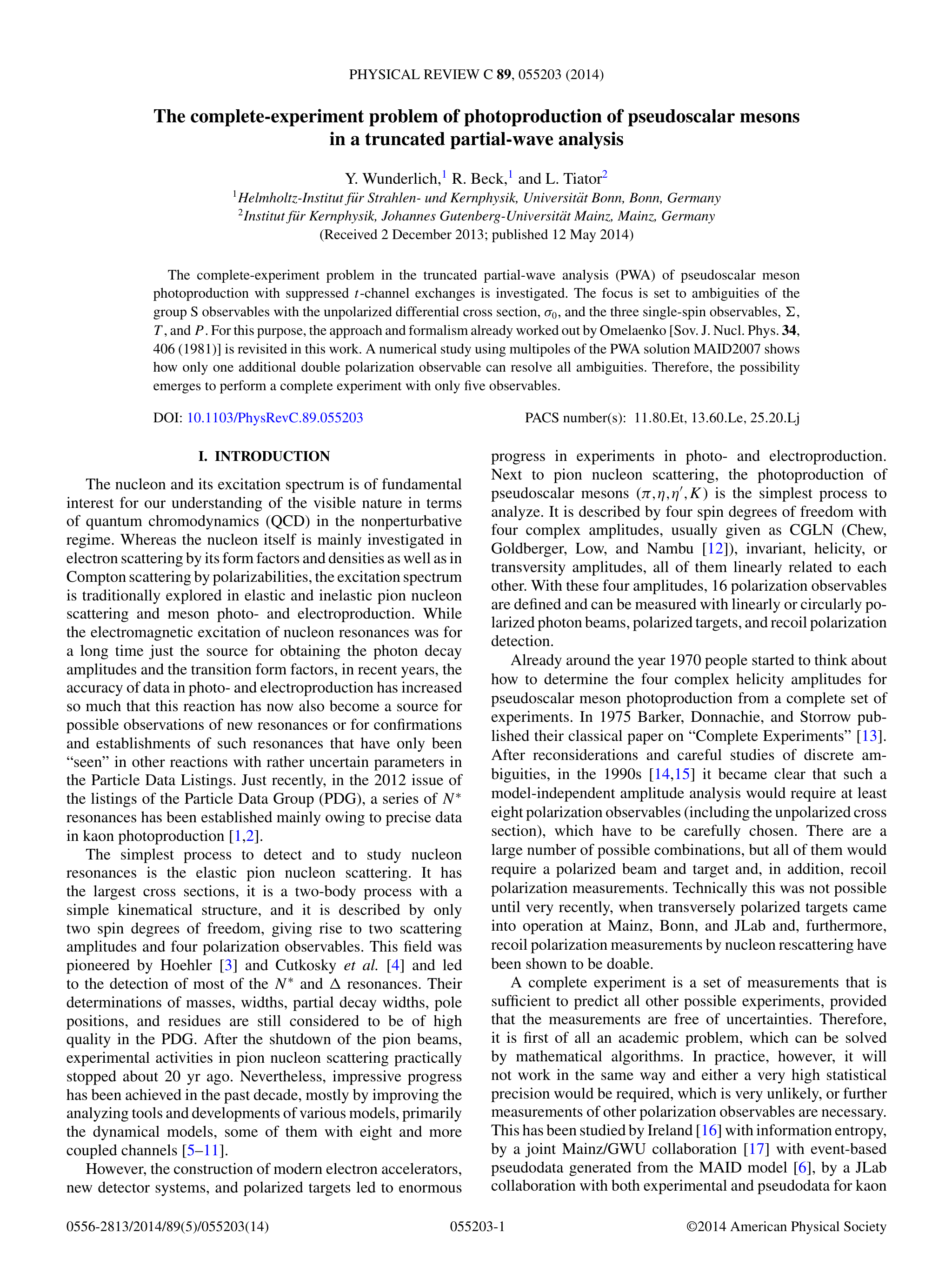}

\newpage

\includepdf[pages={2}, frame=false, noautoscale=true, scale=0.86, pagecommand={}]{Paper_PhysRevC89_055203}

\newpage

\includepdf[pages={3}, frame=false, noautoscale=true, scale=0.86, pagecommand={}]{Paper_PhysRevC89_055203}

\newpage

\includepdf[pages={4}, frame=false, noautoscale=true, scale=0.86, pagecommand={}]{Paper_PhysRevC89_055203}

\newpage

\includepdf[pages={5}, frame=false, noautoscale=true, scale=0.86, pagecommand={}]{Paper_PhysRevC89_055203}

\newpage

\includepdf[pages={6}, frame=false, noautoscale=true, scale=0.86, pagecommand={}]{Paper_PhysRevC89_055203}

\newpage

\includepdf[pages={7}, frame=false, noautoscale=true, scale=0.86, pagecommand={}]{Paper_PhysRevC89_055203}

\newpage

\includepdf[pages={8}, frame=false, noautoscale=true, scale=0.86, pagecommand={}]{Paper_PhysRevC89_055203}

\newpage

\includepdf[pages={9}, frame=false, noautoscale=true, scale=0.86, pagecommand={}]{Paper_PhysRevC89_055203}

\newpage

\includepdf[pages={10}, frame=false, noautoscale=true, scale=0.86, pagecommand={}]{Paper_PhysRevC89_055203}

\newpage

\includepdf[pages={11}, frame=false, noautoscale=true, scale=0.86, pagecommand={}]{Paper_PhysRevC89_055203}

\newpage

\includepdf[pages={12}, frame=false, noautoscale=true, scale=0.86, pagecommand={}]{Paper_PhysRevC89_055203}

\newpage

\includepdf[pages={13}, frame=false, noautoscale=true, scale=0.86, pagecommand={}]{Paper_PhysRevC89_055203}

\newpage

\includepdf[pages={14}, frame=false, noautoscale=true, scale=0.86, pagecommand={}]{Paper_PhysRevC89_055203}

\clearpage

\subsection{Summary of chapter \ref{chap:Omelaenko}} \label{sec:ChapIISummary}

The preceding section contains a paper \cite{MyCompExTPWAPaper} published in the course of this work, on the application of Gersten's method \cite{Gersten} to the problem of discrete ambiguities in the case of a TPWA in photoproduction. The results have been worked out previously by Omelaenko \cite{Omelaenko}, in a paper which is however not widely distributed and also not easy to read. Thus, gaps had to be filled in by hand in order to re-derive some of Omelaenko's results, which has however been invaluable for the numerical TPWA-fits, which then had to follow (see chapter \ref{chap:TPWA}). Furthermore, the paper shown in section \ref{sec:WBTpaper} contains applications of Omelaenko's formalism to partial waves from recent PWA-models \cite{MAID,SAID,BoGa}, which were new, as well as an interpretation of the results oriented towards the goal of a complete experiment. \newline
Surprisingly, the simple algebraic Ansatz employed in order to study discrete ambiguities has lead to the postulation of {\it complete sets of $5$ observables} in a TPWA. This has been possible at least under the highly idealized circumstances that the TPWA possesses an exact solution. Thus, the minimum number of observables required by Chiang and Tabakin \cite{ChTab} for the extraction of the full spin amplitudes, e.g. helicity amplitudes $H_{i}$ or transversity amplitudes $b_{i}$ (cf. section \ref{sec:CompExpsFullAmp}), has been undercut. Furthermore, the complete sets of $5$ quantities contain cases that do not require any double-polarization observables with recoil polarization, namely
\begin{equation}
 \left\{ \sigma_{0}, \Sigma, T, P, F \right\} \hspace*{2.5pt} \mathrm{and} \hspace*{2.5pt} \left\{ \sigma_{0}, \Sigma, T, P, G \right\} \mathrm{.} \label{eq:CompleteSetsOf5ExamplesSummary}
\end{equation}
However, the work presented in section \ref{sec:WBTpaper} does not contain all results of the more mathematically oriented investigations done in the course of this thesis. Elaborations on further ideas can be found in appendix \ref{sec:AdditionsChapterII}. \newline
Among these extended treatments are further mathematical details on and classifications of the different kinds of ambiguities, i.e. the {\it double ambiguity} vs. {\it accidental ambiguities}, and under which circumstances the latter can become dangerous (see appendices \ref{sec:DoubleAmbiguityTrafoActingOnBi} and \ref{sec:AccidentalAmbProofs}). Some rough estimates on the abundance of such accidental ambiguities are made (cf. appendices \ref{subsec:AccidentalAmbProofsIII} and \ref{sec:ProjectionIntegrals}). Furthermore, the interesting possibility is investigated of making algebraic investigations analogous to Omelaenko \cite{Omelaenko}, without necessarily starting from the group $\mathcal{S}$ observables, but any other possible set of simultaneously diagonalizable observables (see appendix \ref{sec:AmbiguitiesSimDiagObservables}). \newline
None of the above-mentioned additional investigations are essential for an understanding of the basic message of this thesis. Some of the material will be referenced in the analyses of theory-data in section \ref{sec:TheoryDataFits}, where the necessary results will also be recited in the main text. \newline

Before commencing with the results of the numerical part of this thesis, i.e. the analyses with Legendre moments contained in chapter \ref{chap:LFits} and the actual multipole-fits (i.e. TPWAs) discussed in detail in chapter \ref{chap:TPWA}, we comment on an interesting result obtained by L. Tiator towards the end of the processing time of this thesis \cite{LotharPrivateComm2016,WorkmanEtAl2017}.

\clearpage

\section{Comment on a recent result: complete sets of 4 observables} \label{chap:CompleteSetsOf4}

This work followed until now the basic logic of Omelaenko's procedure \cite{Omelaenko} (cf. chapter \ref{chap:Omelaenko} and Appendix \ref{sec:AdditionsChapterII}). This means that first of all, the discrete ambiguities of the four group $\mathcal{S}$ observables $\left\{\sigma_{0}, \check{\Sigma}, \check{T}, \check{P}\right\}$ have been worked out, using a suitbale set of root-variables $\left\{\alpha_{k}, \beta_{k^{\prime}}\right\}$. Then, the remaining observables were investigated in order to find quantities capable of resolving all exact discrete ambiguities, which has lead to candidates for mathematically complete sets composed of $5$ observables. \newline
The procedure has been elaborated in detail during the main part of chapter \ref{chap:Omelaenko} and has been the direction followed in the whole time of this thesis project. As a result, the algebraic investigations, as well as all numerical tests of them (cf. parts of chapter \ref{chap:TPWA}, especially \ref{sec:TheoryDataFits}), have been fully consistent. This means, at least in the idealized case of fits to truncated model-data, the mathematical completeness of Omelaenko's sets of $5$ has been fully confirmed. \newline
However, during the writing of this document, L. Tiator \cite{LotharPrivateComm2016} has found that mathematically complete sets composed of just $4$ observables are possible as well. The discovery was made in the context of numerical simulations in MATHEMATICA. \newline
This poses an exciting result, since it has not been known until now and was not to be anticipated when following the Omelaenko-paradigm outlined above. Some mentioning of these new complete sets of $4$ has already been published in a recent work on complete experiments in photoproduction \cite{WorkmanEtAl2017}. Though Tiator found $196$ distinct complete sets of $4$ using his own simulations, some examples have been confirmed using the numerical codes written for this work. In particular, those codes use the same method as the fits to truncated MAID model-data outlined in section \ref{sec:TheoryDataFits}. All complete sets of $4$ composed of just group $\mathcal{S}$ and $\mathcal{BT}$ observables, which amount to $6$ sets and are listed exclusively in Table \ref{tab:LotharsGroupSAndBTCompleteSetsOf4Listing} below, were confirmed to be mathematically complete using the codes written for this thesis. The remaining $190$ sets found by Tiator have not been checked. \newline
The problem with the newly discovered complete sets of $4$ is that, while they have been found and tested for their completeness using numerical simulations, they are not yet well-understood algebraically, on the contrary to Omelaenko's complete sets of $5$. Some pre\-limi\-nary attempts have been made to solve this problem, using the known parametrization of observables in terms of Omelaenko-roots $\left\{\alpha_{k}, \beta_{k^{\prime}}\right\}$, however they have not lead to a complete understanding. Thus, further investigations have not been included in this thesis.
\vfill
\begin{longtable}{r|cccc}
\caption[Subset of all $196$ complete sets of $4$ listed in Table \ref{tab:Lothars196CompleteSetsOf4Listing}, containing all possibilities to form such sets out of just the group $\mathcal{S}$ and $\mathcal{BT}$ observables.]{Here, a subset of all $196$ complete sets of $4$ listed in Table \ref{tab:Lothars196CompleteSetsOf4Listing} is given, which contains all possibilities to form such sets out of just the group $\mathcal{S}$ and $\mathcal{BT}$ observables. All of these sets have been verified to be mathematically complete in the course of this work, within the context of analyses of model-data. For more details, see the main text.}
\label{tab:LotharsGroupSAndBTCompleteSetsOf4Listing}
\endfirsthead
\endhead
Set-Nr. & \multicolumn{4}{c}{Observables} \\
\hline                                    
 1 & $\sigma_{0}$ & $\check{\Sigma}$ & $\check{P}$ & $\check{F}$   \\
 2 & $\sigma_{0}$ & $\check{\Sigma}$ & $\check{F}$ & $\check{H}$  \\
 3 & $\sigma_{0}$ & $\check{T}$ & $\check{P}$ & $\check{F}$  \\
 4 & $\sigma_{0}$ & $\check{T}$ & $\check{P}$ & $\check{G}$  \\ 
 5 & $\sigma_{0}$ & $\check{T}$ & $\check{F}$ & $\check{H}$ \\
 6 & $\sigma_{0}$ & $\check{T}$ & $\check{G}$ & $\check{H}$ \\
\end{longtable}

\clearpage

It is not even clear whether the parametrization in terms of roots $\left\{\alpha_{k}, \beta_{k^{\prime}}\right\}$ is the correct one to discuss the complete sets of $4$ algebraically. In case it is used, several interesting new ways to form discrete ambiguities can be found, apart from conjugations of roots. This is because in all cases, not all observables in the respective sets are 'diagonal' in the same sense as the group $\mathcal{S}$ observables, i.e. they cannot all be written as a linear combination of modulus-squared spin amplitudes (e.g., in the case of transversity amplitudes, $\left| b_{i} \right|^{2}$). However, it may very well be that another, as of yet unknown, parametrization exists, which is better suited to shed light on the problem. \newline
With kind permission by L. Tiator \cite{LotharPrivateComm2016}, we list all the $196$ complete sets of $4$ found for photoproduction in Table \ref{tab:Lothars196CompleteSetsOf4Listing} below.

\clearpage

\begin{longtable}{r|cccc||r|cccc}
\caption[The $196$ distinct possibilities to form mathematically complete sets out of $4$ observables.]{The $196$ distinct possibilities to form mathematically complete sets out of $4$ observables are listed here. They have been found by L. Tiator. Some of them have been verified to be complete in the course of this work. For more details, see the main text. The Table is continued on the following pages. \\}
\label{tab:Lothars196CompleteSetsOf4Listing}
\endfirsthead
\endhead
Set-Nr. & \multicolumn{4}{c}{Observables} & Set-Nr. & \multicolumn{4}{c}{Observables} \\
\hline                                    
 1 & $\sigma_{0}$ & $\check{\Sigma}$ & $\check{T}$ & $\check{C}_{x^{\prime}}$ &                   41 & $\sigma_{0}$ & $\check{T}$ & $\check{F}$ & $\check{C}_{z^{\prime}}$  \\ 
 2 & $\sigma_{0}$ & $\check{\Sigma}$ & $\check{T}$ & $\check{C}_{z^{\prime}}$  &                  42 & $\sigma_{0}$ & $\check{T}$ & $\check{F}$ & $\check{L}_{x^{\prime}}$  \\ 
 3 & $\sigma_{0}$ & $\check{\Sigma}$ & $\check{T}$ & $\check{L}_{x^{\prime}}$  &                  43 & $\sigma_{0}$ & $\check{T}$ & $\check{F}$ & $\check{L}_{z^{\prime}}$  \\ 
 4 & $\sigma_{0}$ & $\check{\Sigma}$ & $\check{T}$ & $\check{L}_{z^{\prime}}$  &                  44 & $\sigma_{0}$ & $\check{T}$ & $\check{G}$ & $\check{H}$ \\               
 5 & $\sigma_{0}$ & $\check{\Sigma}$ & $\check{P}$ & $\check{F}$  &                               45 & $\sigma_{0}$ & $\check{T}$ & $\check{G}$ & $\check{T}_{x^{\prime}}$ \\
 6 & $\sigma_{0}$ & $\check{\Sigma}$ & $\check{P}$ & $\check{O}_{x^{\prime}}$  &                  46 & $\sigma_{0}$ & $\check{T}$ & $\check{G}$ & $\check{T}_{z^{\prime}}$ \\
 7 & $\sigma_{0}$ & $\check{\Sigma}$ & $\check{P}$ & $\check{O}_{z^{\prime}}$  &                  47 & $\sigma_{0}$ & $\check{T}$ & $\check{H}$ & $\check{C}_{x^{\prime}}$ \\
 8 & $\sigma_{0}$ & $\check{\Sigma}$ & $\check{P}$ & $\check{T}_{x^{\prime}}$  &                  48 & $\sigma_{0}$ & $\check{T}$ & $\check{H}$ & $\check{C}_{z^{\prime}}$ \\
 9 & $\sigma_{0}$ & $\check{\Sigma}$ & $\check{P}$ & $\check{T}_{z^{\prime}}$  &                  49 & $\sigma_{0}$ & $\check{T}$ & $\check{H}$ & $\check{O}_{x^{\prime}}$ \\
 10 & $\sigma_{0}$ & $\check{\Sigma}$ & $\check{F}$ & $\check{H}$  &                              50 & $\sigma_{0}$ & $\check{T}$ & $\check{H}$ & $\check{O}_{z^{\prime}}$ \\
 11 & $\sigma_{0}$ & $\check{\Sigma}$ & $\check{F}$ & $\check{O}_{x^{\prime}}$  &                 51 & $\sigma_{0}$ & $\check{T}$ & $\check{H}$ & $\check{T}_{x^{\prime}}$ \\
 12 & $\sigma_{0}$ & $\check{\Sigma}$ & $\check{F}$ & $\check{O}_{z^{\prime}}$  &                 52 & $\sigma_{0}$ & $\check{T}$ & $\check{H}$ & $\check{T}_{z^{\prime}}$ \\
 13 & $\sigma_{0}$ & $\check{\Sigma}$ & $\check{H}$ & $\check{O}_{x^{\prime}}$  &                 53 & $\sigma_{0}$ & $\check{T}$ & $\check{H}$ & $\check{L}_{x^{\prime}}$ \\
 14 & $\sigma_{0}$ & $\check{\Sigma}$ & $\check{H}$ & $\check{O}_{z^{\prime}}$  &                 54 & $\sigma_{0}$ & $\check{T}$ & $\check{H}$ & $\check{L}_{z^{\prime}}$ \\
 15 & $\sigma_{0}$ & $\check{\Sigma}$ & $\check{H}$ & $\check{T}_{x^{\prime}}$  &                 55 & $\sigma_{0}$ & $\check{T}$ & $\check{C}_{x^{\prime}}$ & $\check{T}_{x^{\prime}}$ \\
 16 & $\sigma_{0}$ & $\check{\Sigma}$ & $\check{H}$ & $\check{T}_{z^{\prime}}$  &                 56 & $\sigma_{0}$ & $\check{T}$ & $\check{C}_{x^{\prime}}$ & $\check{T}_{z^{\prime}}$ \\
 17 & $\sigma_{0}$ & $\check{\Sigma}$ & $\check{C}_{x^{\prime}}$ & $\check{O}_{x^{\prime}}$ &     57 & $\sigma_{0}$ & $\check{T}$ & $\check{C}_{z^{\prime}}$ & $\check{T}_{x^{\prime}}$ \\
 18 & $\sigma_{0}$ & $\check{\Sigma}$ & $\check{C}_{x^{\prime}}$ & $\check{O}_{z^{\prime}}$ &     58 & $\sigma_{0}$ & $\check{T}$ & $\check{C}_{z^{\prime}}$ & $\check{T}_{z^{\prime}}$ \\
 19 & $\sigma_{0}$ & $\check{\Sigma}$ & $\check{C}_{z^{\prime}}$ & $\check{O}_{x^{\prime}}$ &     59 & $\sigma_{0}$ & $\check{T}$ & $\check{O}_{x^{\prime}}$ & $\check{T}_{x^{\prime}}$ \\
 20 & $\sigma_{0}$ & $\check{\Sigma}$ & $\check{C}_{z^{\prime}}$ & $\check{O}_{z^{\prime}}$ &     60 & $\sigma_{0}$ & $\check{T}$ & $\check{O}_{x^{\prime}}$ & $\check{T}_{z^{\prime}}$ \\
 21 & $\sigma_{0}$ & $\check{\Sigma}$ & $\check{O}_{x^{\prime}}$ & $\check{T}_{x^{\prime}}$ &     61 & $\sigma_{0}$ & $\check{T}$ & $\check{O}_{z^{\prime}}$ & $\check{T}_{x^{\prime}}$ \\
 22 & $\sigma_{0}$ & $\check{\Sigma}$ & $\check{O}_{x^{\prime}}$ & $\check{T}_{z^{\prime}}$ &     62 & $\sigma_{0}$ & $\check{T}$ & $\check{O}_{z^{\prime}}$ & $\check{T}_{z^{\prime}}$ \\
 23 & $\sigma_{0}$ & $\check{\Sigma}$ & $\check{O}_{x^{\prime}}$ & $\check{L}_{x^{\prime}}$ &     63 & $\sigma_{0}$ & $\check{T}$ & $\check{T}_{x^{\prime}}$ & $\check{L}_{x^{\prime}}$ \\
 24 & $\sigma_{0}$ & $\check{\Sigma}$ & $\check{O}_{x^{\prime}}$ & $\check{L}_{z^{\prime}}$ &     64 & $\sigma_{0}$ & $\check{T}$ & $\check{T}_{x^{\prime}}$ & $\check{L}_{z^{\prime}}$ \\
 25 & $\sigma_{0}$ & $\check{\Sigma}$ & $\check{O}_{z^{\prime}}$ & $\check{T}_{x^{\prime}}$ &     65 & $\sigma_{0}$ & $\check{T}$ & $\check{T}_{z^{\prime}}$ & $\check{L}_{x^{\prime}}$ \\
 26 & $\sigma_{0}$ & $\check{\Sigma}$ & $\check{O}_{z^{\prime}}$ & $\check{T}_{z^{\prime}}$ &     66 & $\sigma_{0}$ & $\check{T}$ & $\check{T}_{z^{\prime}}$ & $\check{L}_{z^{\prime}}$ \\
 27 & $\sigma_{0}$ & $\check{\Sigma}$ & $\check{O}_{z^{\prime}}$ & $\check{L}_{x^{\prime}}$ &     67 & $\sigma_{0}$ & $\check{P}$ & $\check{E}$ & $\check{O}_{x^{\prime}}$ \\
 28 & $\sigma_{0}$ & $\check{\Sigma}$ & $\check{O}_{z^{\prime}}$ & $\check{L}_{z^{\prime}}$ &     68 & $\sigma_{0}$ & $\check{P}$ & $\check{E}$ & $\check{O}_{z^{\prime}}$ \\
 29 & $\sigma_{0}$ & $\check{T}$ & $\check{P}$ & $\check{F}$  &                                   69 & $\sigma_{0}$ & $\check{P}$ & $\check{E}$ & $\check{T}_{x^{\prime}}$ \\
 30 & $\sigma_{0}$ & $\check{T}$ & $\check{P}$ & $\check{G}$  &                                   70 & $\sigma_{0}$ & $\check{P}$ & $\check{E}$ & $\check{T}_{z^{\prime}}$ \\
 31 & $\sigma_{0}$ & $\check{T}$ & $\check{P}$ & $\check{C}_{x^{\prime}}$  &                      71 & $\sigma_{0}$ & $\check{P}$ & $\check{F}$ & $\check{C}_{x^{\prime}}$ \\
 32 & $\sigma_{0}$ & $\check{T}$ & $\check{P}$ & $\check{C}_{z^{\prime}}$  &                      72 & $\sigma_{0}$ & $\check{P}$ & $\check{F}$ & $\check{C}_{z^{\prime}}$ \\
 33 & $\sigma_{0}$ & $\check{T}$ & $\check{P}$ & $\check{O}_{x^{\prime}}$  &                      73 & $\sigma_{0}$ & $\check{P}$ & $\check{F}$ & $\check{O}_{x^{\prime}}$ \\
 34 & $\sigma_{0}$ & $\check{T}$ & $\check{P}$ & $\check{O}_{z^{\prime}}$  &                      74 & $\sigma_{0}$ & $\check{P}$ & $\check{F}$ & $\check{O}_{z^{\prime}}$ \\
 35 & $\sigma_{0}$ & $\check{T}$ & $\check{P}$ & $\check{T}_{x^{\prime}}$  &                      75 & $\sigma_{0}$ & $\check{P}$ & $\check{F}$ & $\check{T}_{x^{\prime}}$ \\
 36 & $\sigma_{0}$ & $\check{T}$ & $\check{P}$ & $\check{T}_{z^{\prime}}$  &                      76 & $\sigma_{0}$ & $\check{P}$ & $\check{F}$ & $\check{T}_{z^{\prime}}$ \\
 37 & $\sigma_{0}$ & $\check{T}$ & $\check{P}$ & $\check{L}_{x^{\prime}}$  &                      77 & $\sigma_{0}$ & $\check{P}$ & $\check{F}$ & $\check{L}_{x^{\prime}}$ \\
 38 & $\sigma_{0}$ & $\check{T}$ & $\check{P}$ & $\check{L}_{z^{\prime}}$  &                      78 & $\sigma_{0}$ & $\check{P}$ & $\check{F}$ & $\check{L}_{z^{\prime}}$ \\
 39 & $\sigma_{0}$ & $\check{T}$ & $\check{F}$ & $\check{H}$  &                                   79 & $\sigma_{0}$ & $\check{P}$ & $\check{G}$ & $\check{O}_{x^{\prime}}$ \\
 40 & $\sigma_{0}$ & $\check{T}$ & $\check{F}$ & $\check{C}_{x^{\prime}}$  &                      80 & $\sigma_{0}$ & $\check{P}$ & $\check{G}$ & $\check{O}_{z^{\prime}}$ \\
 \end{longtable}
 
\begin{longtable}{r|cccc||r|cccc}
\label{tab:Lothars196CompleteSetsOf4Listing2}
 Set-Nr. & \multicolumn{4}{c}{Observables} & Set-Nr. & \multicolumn{4}{c}{Observables} \\
 \endfirsthead
\hline  
 81 & $\sigma_{0}$ & $\check{P}$ & $\check{G}$ & $\check{T}_{x^{\prime}}$ &                  125 & $\sigma_{0}$ & $\check{F}$ & $\check{H}$ & $\check{L}_{x^{\prime}}$  \\             
 82 & $\sigma_{0}$ & $\check{P}$ & $\check{G}$ & $\check{T}_{z^{\prime}}$ &                  126 & $\sigma_{0}$ & $\check{F}$ & $\check{H}$ & $\check{L}_{z^{\prime}}$  \\              
 83 & $\sigma_{0}$ & $\check{P}$ & $\check{H}$ & $\check{O}_{x^{\prime}}$ &                  127 & $\sigma_{0}$ & $\check{F}$ & $\check{C}_{x^{\prime}}$ & $\check{O}_{x^{\prime}}$  \\ 
 84 & $\sigma_{0}$ & $\check{P}$ & $\check{H}$ & $\check{T}_{z^{\prime}}$ &                  128 & $\sigma_{0}$ & $\check{F}$ & $\check{C}_{x^{\prime}}$ & $\check{O}_{z^{\prime}}$  \\
 85 & $\sigma_{0}$ & $\check{P}$ & $\check{C}_{x^{\prime}}$ & $\check{O}_{x^{\prime}}$ &     129 & $\sigma_{0}$ & $\check{F}$ & $\check{C}_{z^{\prime}}$ & $\check{O}_{x^{\prime}}$  \\ 
 86 & $\sigma_{0}$ & $\check{P}$ & $\check{C}_{x^{\prime}}$ & $\check{O}_{z^{\prime}}$ &      130 & $\sigma_{0}$ & $\check{F}$ & $\check{C}_{z^{\prime}}$ & $\check{O}_{z^{\prime}}$  \\ 
 87 & $\sigma_{0}$ & $\check{P}$ & $\check{C}_{x^{\prime}}$ & $\check{T}_{x^{\prime}}$ &      131 & $\sigma_{0}$ & $\check{F}$ & $\check{O}_{x^{\prime}}$ & $\check{T}_{x^{\prime}}$  \\ 
 88 & $\sigma_{0}$ & $\check{P}$ & $\check{C}_{x^{\prime}}$ & $\check{T}_{z^{\prime}}$ &      132 & $\sigma_{0}$ & $\check{F}$ & $\check{O}_{x^{\prime}}$ & $\check{T}_{z^{\prime}}$  \\ 
 89 & $\sigma_{0}$ & $\check{P}$ & $\check{C}_{z^{\prime}}$ & $\check{O}_{x^{\prime}}$  &     133 & $\sigma_{0}$ & $\check{F}$ & $\check{O}_{x^{\prime}}$ & $\check{L}_{x^{\prime}}$  \\ 
 90 & $\sigma_{0}$ & $\check{P}$ & $\check{C}_{z^{\prime}}$ & $\check{O}_{z^{\prime}}$  &     134 & $\sigma_{0}$ & $\check{F}$ & $\check{O}_{x^{\prime}}$ & $\check{L}_{z^{\prime}}$  \\                                                                                  
 91 & $\sigma_{0}$ & $\check{P}$ & $\check{C}_{z^{\prime}}$ & $\check{T}_{x^{\prime}}$  &     135 & $\sigma_{0}$ & $\check{F}$ & $\check{O}_{z^{\prime}}$ & $\check{T}_{x^{\prime}}$  \\ 
 92 & $\sigma_{0}$ & $\check{P}$ & $\check{C}_{z^{\prime}}$ & $\check{T}_{z^{\prime}}$  &   136 & $\sigma_{0}$ & $\check{F}$ & $\check{O}_{z^{\prime}}$ & $\check{T}_{z^{\prime}}$  \\ 
 93 & $\sigma_{0}$ & $\check{P}$ & $\check{O}_{x^{\prime}}$ & $\check{O}_{z^{\prime}}$  &   137 & $\sigma_{0}$ & $\check{F}$ & $\check{O}_{z^{\prime}}$ & $\check{L}_{x^{\prime}}$  \\
 94 & $\sigma_{0}$ & $\check{P}$ & $\check{O}_{x^{\prime}}$ & $\check{T}_{x^{\prime}}$  &   138 & $\sigma_{0}$ & $\check{F}$ & $\check{O}_{z^{\prime}}$ & $\check{L}_{z^{\prime}}$ \\
 95 & $\sigma_{0}$ & $\check{P}$ & $\check{O}_{x^{\prime}}$ & $\check{T}_{z^{\prime}}$  &   139 & $\sigma_{0}$ & $\check{G}$ & $\check{H}$ & $\check{O}_{x^{\prime}}$ \\
 96 & $\sigma_{0}$ & $\check{P}$ & $\check{O}_{x^{\prime}}$ & $\check{L}_{x^{\prime}}$  &   140 & $\sigma_{0}$ & $\check{G}$ & $\check{H}$ & $\check{O}_{z^{\prime}}$ \\
 97 & $\sigma_{0}$ & $\check{P}$ & $\check{O}_{x^{\prime}}$ & $\check{L}_{z^{\prime}}$  &   141 & $\sigma_{0}$ & $\check{G}$ & $\check{H}$ & $\check{T}_{x^{\prime}}$ \\
 98 & $\sigma_{0}$ & $\check{P}$ & $\check{O}_{z^{\prime}}$ & $\check{T}_{x^{\prime}}$  &   142 & $\sigma_{0}$ & $\check{G}$ & $\check{H}$ & $\check{T}_{z^{\prime}}$ \\
 99 & $\sigma_{0}$ & $\check{P}$ & $\check{O}_{z^{\prime}}$ & $\check{T}_{z^{\prime}}$  &   143 & $\sigma_{0}$ & $\check{G}$ & $\check{C}_{x^{\prime}}$ & $\check{T}_{x^{\prime}}$ \\
 100 & $\sigma_{0}$ & $\check{P}$ & $\check{O}_{z^{\prime}}$ & $\check{L}_{x^{\prime}}$  &  144 & $\sigma_{0}$ & $\check{G}$ & $\check{C}_{x^{\prime}}$ & $\check{T}_{z^{\prime}}$ \\
 101 & $\sigma_{0}$ & $\check{P}$ & $\check{O}_{z^{\prime}}$ & $\check{L}_{z^{\prime}}$  &  145 & $\sigma_{0}$ & $\check{G}$ & $\check{C}_{z^{\prime}}$ & $\check{T}_{x^{\prime}}$ \\
 102 & $\sigma_{0}$ & $\check{P}$ & $\check{T}_{x^{\prime}}$ & $\check{T}_{z^{\prime}}$  &  146 & $\sigma_{0}$ & $\check{G}$ & $\check{C}_{z^{\prime}}$ & $\check{T}_{z^{\prime}}$ \\
 103 & $\sigma_{0}$ & $\check{P}$ & $\check{T}_{x^{\prime}}$ & $\check{L}_{x^{\prime}}$  &  147 & $\sigma_{0}$ & $\check{G}$ & $\check{O}_{x^{\prime}}$ & $\check{T}_{x^{\prime}}$ \\
 104 & $\sigma_{0}$ & $\check{P}$ & $\check{T}_{x^{\prime}}$ & $\check{L}_{z^{\prime}}$  &  148 & $\sigma_{0}$ & $\check{G}$ & $\check{O}_{x^{\prime}}$ & $\check{T}_{z^{\prime}}$ \\
 105 & $\sigma_{0}$ & $\check{P}$ & $\check{T}_{z^{\prime}}$ & $\check{L}_{x^{\prime}}$  &  149 & $\sigma_{0}$ & $\check{G}$ & $\check{O}_{z^{\prime}}$ & $\check{T}_{x^{\prime}}$ \\
 106 & $\sigma_{0}$ & $\check{P}$ & $\check{T}_{z^{\prime}}$ & $\check{L}_{z^{\prime}}$  &  150 & $\sigma_{0}$ & $\check{G}$ & $\check{O}_{z^{\prime}}$ & $\check{T}_{z^{\prime}}$ \\
 107 & $\sigma_{0}$ & $\check{E}$ & $\check{H}$ & $\check{O}_{x^{\prime}}$  &               151 & $\sigma_{0}$ & $\check{G}$ & $\check{T}_{x^{\prime}}$ & $\check{L}_{x^{\prime}}$ \\
 108 & $\sigma_{0}$ & $\check{E}$ & $\check{H}$ & $\check{O}_{z^{\prime}}$  &               152 & $\sigma_{0}$ & $\check{G}$ & $\check{T}_{x^{\prime}}$ & $\check{L}_{z^{\prime}}$ \\
 109 & $\sigma_{0}$ & $\check{E}$ & $\check{H}$ & $\check{T}_{x^{\prime}}$  &               153 & $\sigma_{0}$ & $\check{G}$ & $\check{T}_{z^{\prime}}$ & $\check{L}_{x^{\prime}}$ \\
 110 & $\sigma_{0}$ & $\check{E}$ & $\check{H}$ & $\check{T}_{z^{\prime}}$  &               154 & $\sigma_{0}$ & $\check{G}$ & $\check{T}_{z^{\prime}}$ & $\check{L}_{z^{\prime}}$ \\
 111 & $\sigma_{0}$ & $\check{E}$ & $\check{O}_{x^{\prime}}$ & $\check{T}_{x^{\prime}}$  &  155 & $\sigma_{0}$ & $\check{H}$ & $\check{C}_{x^{\prime}}$ & $\check{O}_{x^{\prime}}$ \\
 112 & $\sigma_{0}$ & $\check{E}$ & $\check{O}_{x^{\prime}}$ & $\check{T}_{z^{\prime}}$  &  156 & $\sigma_{0}$ & $\check{H}$ & $\check{C}_{x^{\prime}}$ & $\check{O}_{z^{\prime}}$ \\
 113 & $\sigma_{0}$ & $\check{E}$ & $\check{O}_{z^{\prime}}$ & $\check{T}_{x^{\prime}}$  &  157 & $\sigma_{0}$ & $\check{H}$ & $\check{C}_{x^{\prime}}$ & $\check{T}_{x^{\prime}}$ \\
 114 & $\sigma_{0}$ & $\check{E}$ & $\check{O}_{z^{\prime}}$ & $\check{T}_{z^{\prime}}$  &  158 & $\sigma_{0}$ & $\check{H}$ & $\check{C}_{x^{\prime}}$ & $\check{T}_{z^{\prime}}$ \\
 115 & $\sigma_{0}$ & $\check{F}$ & $\check{G}$ & $\check{C}_{x^{\prime}}$  &               159 & $\sigma_{0}$ & $\check{H}$ & $\check{C}_{z^{\prime}}$ & $\check{O}_{x^{\prime}}$ \\
 116 & $\sigma_{0}$ & $\check{F}$ & $\check{G}$ & $\check{C}_{z^{\prime}}$  &               160 & $\sigma_{0}$ & $\check{H}$ & $\check{C}_{z^{\prime}}$ & $\check{O}_{z^{\prime}}$ \\
 117 & $\sigma_{0}$ & $\check{F}$ & $\check{G}$ & $\check{L}_{x^{\prime}}$  &               161 & $\sigma_{0}$ & $\check{H}$ & $\check{C}_{z^{\prime}}$ & $\check{T}_{x^{\prime}}$ \\
 118 & $\sigma_{0}$ & $\check{F}$ & $\check{G}$ & $\check{L}_{z^{\prime}}$  &               162 & $\sigma_{0}$ & $\check{H}$ & $\check{C}_{z^{\prime}}$ & $\check{T}_{z^{\prime}}$ \\
 119 & $\sigma_{0}$ & $\check{F}$ & $\check{H}$ & $\check{C}_{x^{\prime}}$  &               163 & $\sigma_{0}$ & $\check{H}$ & $\check{O}_{x^{\prime}}$ & $\check{O}_{z^{\prime}}$ \\
 120 & $\sigma_{0}$ & $\check{F}$ & $\check{H}$ & $\check{C}_{z^{\prime}}$  &               164 & $\sigma_{0}$ & $\check{H}$ & $\check{O}_{x^{\prime}}$ & $\check{T}_{x^{\prime}}$ \\
 121 & $\sigma_{0}$ & $\check{F}$ & $\check{H}$ & $\check{O}_{x^{\prime}}$  &               165 & $\sigma_{0}$ & $\check{H}$ & $\check{O}_{x^{\prime}}$ & $\check{T}_{z^{\prime}}$ \\
 122 & $\sigma_{0}$ & $\check{F}$ & $\check{H}$ & $\check{O}_{z^{\prime}}$  &               166 & $\sigma_{0}$ & $\check{H}$ & $\check{O}_{x^{\prime}}$ & $\check{L}_{x^{\prime}}$ \\
 123 & $\sigma_{0}$ & $\check{F}$ & $\check{H}$ & $\check{T}_{x^{\prime}}$  &               167 & $\sigma_{0}$ & $\check{H}$ & $\check{O}_{x^{\prime}}$ & $\check{L}_{z^{\prime}}$ \\
 124 & $\sigma_{0}$ & $\check{F}$ & $\check{H}$ & $\check{T}_{z^{\prime}}$  &               168 & $\sigma_{0}$ & $\check{H}$ & $\check{O}_{z^{\prime}}$ & $\check{T}_{x^{\prime}}$ \\
\pagebreak Set-Nr. & \multicolumn{4}{c}{Observables} & Set-Nr. & \multicolumn{4}{c}{Observables} \\
\hline   
169 & $\sigma_{0}$ & $\check{H}$ & $\check{O}_{z^{\prime}}$ & $\check{T}_{z^{\prime}}$ &                 183 & $\sigma_{0}$ & $\check{C}_{z^{\prime}}$ & $\check{O}_{z^{\prime}}$ & $\check{T}_{x^{\prime}}$  \\ 
170 & $\sigma_{0}$ & $\check{H}$ & $\check{O}_{z^{\prime}}$ & $\check{L}_{x^{\prime}}$ &                 184 & $\sigma_{0}$ & $\check{C}_{z^{\prime}}$ & $\check{O}_{z^{\prime}}$ & $\check{T}_{z^{\prime}}$  \\ 
171 & $\sigma_{0}$ & $\check{H}$ & $\check{O}_{z^{\prime}}$ & $\check{L}_{z^{\prime}}$ &                 185 & $\sigma_{0}$ & $\check{O}_{x^{\prime}}$ & $\check{O}_{z^{\prime}}$ & $\check{T}_{x^{\prime}}$  \\ 
 172 & $\sigma_{0}$ & $\check{H}$ & $\check{T}_{x^{\prime}}$ & $\check{T}_{z^{\prime}}$ &                186 & $\sigma_{0}$ & $\check{O}_{x^{\prime}}$ & $\check{O}_{z^{\prime}}$ & $\check{T}_{z^{\prime}}$  \\ 
 173 & $\sigma_{0}$ & $\check{H}$ & $\check{T}_{x^{\prime}}$ & $\check{L}_{x^{\prime}}$ &                187 & $\sigma_{0}$ & $\check{O}_{x^{\prime}}$ & $\check{T}_{x^{\prime}}$ & $\check{T}_{z^{\prime}}$  \\ 
 174 & $\sigma_{0}$ & $\check{H}$ & $\check{T}_{x^{\prime}}$ & $\check{L}_{z^{\prime}}$ &                188 & $\sigma_{0}$ & $\check{O}_{x^{\prime}}$ & $\check{T}_{x^{\prime}}$ & $\check{L}_{x^{\prime}}$  \\ 
 175 & $\sigma_{0}$ & $\check{H}$ & $\check{T}_{z^{\prime}}$ & $\check{L}_{x^{\prime}}$ &                189 & $\sigma_{0}$ & $\check{O}_{x^{\prime}}$ & $\check{T}_{x^{\prime}}$ & $\check{L}_{z^{\prime}}$  \\ 
 176 & $\sigma_{0}$ & $\check{H}$ & $\check{T}_{z^{\prime}}$ & $\check{L}_{z^{\prime}}$ &                 190 & $\sigma_{0}$ & $\check{O}_{x^{\prime}}$ & $\check{T}_{z^{\prime}}$ & $\check{L}_{x^{\prime}}$ \\
 177 & $\sigma_{0}$ & $\check{C}_{x^{\prime}}$ & $\check{O}_{x^{\prime}}$ & $\check{T}_{x^{\prime}}$ &    191 & $\sigma_{0}$ & $\check{O}_{x^{\prime}}$ & $\check{T}_{z^{\prime}}$ & $\check{L}_{z^{\prime}}$ \\
 178 & $\sigma_{0}$ & $\check{C}_{x^{\prime}}$ & $\check{O}_{x^{\prime}}$ & $\check{T}_{z^{\prime}}$ &    192 & $\sigma_{0}$ & $\check{O}_{z^{\prime}}$ & $\check{T}_{x^{\prime}}$ & $\check{T}_{z^{\prime}}$ \\
 179 & $\sigma_{0}$ & $\check{C}_{x^{\prime}}$ & $\check{O}_{z^{\prime}}$ & $\check{T}_{x^{\prime}}$ &    193 & $\sigma_{0}$ & $\check{O}_{z^{\prime}}$ & $\check{T}_{x^{\prime}}$ & $\check{L}_{x^{\prime}}$ \\
 180 & $\sigma_{0}$ & $\check{C}_{x^{\prime}}$ & $\check{O}_{z^{\prime}}$ & $\check{T}_{z^{\prime}}$ &    194 & $\sigma_{0}$ & $\check{O}_{z^{\prime}}$ & $\check{T}_{x^{\prime}}$ & $\check{L}_{z^{\prime}}$ \\
 181 & $\sigma_{0}$ & $\check{C}_{z^{\prime}}$ & $\check{O}_{x^{\prime}}$ & $\check{T}_{x^{\prime}}$ &    195 & $\sigma_{0}$ & $\check{O}_{z^{\prime}}$ & $\check{T}_{z^{\prime}}$ & $\check{L}_{x^{\prime}}$ \\
 182 & $\sigma_{0}$ & $\check{C}_{z^{\prime}}$ & $\check{O}_{x^{\prime}}$ & $\check{T}_{z^{\prime}}$ &    196 & $\sigma_{0}$ & $\check{O}_{z^{\prime}}$ & $\check{T}_{z^{\prime}}$ & $\check{L}_{z^{\prime}}$ \\
\end{longtable}

\clearpage
\thispagestyle{empty}
\textcolor{white}{Hallo Welt :-)}
\clearpage

\section{Determining the partial wave content of polarization observables} \label{chap:LFits}

\subsection{Introduction} \label{sec:LFitsIntro}

In case a set of polarization observables is newly measured and one wishes to gain first information and an intuition about the dominant contributing partial waves, it is good to have a method which achieves just that, while avoiding the complications and calculational efforts of a full TPWA or even the fit of an energy-dependent model. \newline
Such a method can be deduced by using just the first ingredient of the parametrization of observables in a TPWA detailed in section \ref{sec:CompExpsTPWA}, i.e. the angular parametrization in terms of e.g. associated Legendre polynomials $P_{\ell}^{m} (\cos \theta)$ \cite{Abramowitz}. The formula to be used for the truncation angular momentum $\ell_{\mathrm{max}}$ reads (cf. equation (\ref{eq:LowEAssocLegStandardParametrization1}))
\begin{equation}
\check{\Omega}^{\alpha} \left( W, \theta \right) = \rho \hspace*{3pt} \sum \limits_{k = \beta_{\alpha}}^{2 \ell_{\mathrm{max}} + \beta_{\alpha} + \gamma_{\alpha}} \left(a_{L}\right)_{k}^{\check{\Omega}^{\alpha}} \left( W \right) P^{\beta_{\alpha}}_{k} \left( \cos \theta \right) \mathrm{,}  \label{eq:LowEAssocLegStandardParametrization1Chapter3}
\end{equation}
for an arbitrary profile function $\check{\Omega}^{\alpha}$. In case of for example the beam asymmetry $\Sigma$, the general parametrization (\ref{eq:LowEAssocLegStandardParametrization1Chapter3}) reads
\begin{equation}
\check{\Sigma} \left( W, \theta \right) = \rho \hspace*{3pt} \sum \limits_{k = 2}^{2 \ell_{\mathrm{max}}} \left(a_{L}\right)_{k}^{\check{\Sigma}} \left( W \right) P^{2}_{k} \left( \cos \theta \right) \mathrm{,}  \label{eq:LowEAssocLegBeamAsymmetry1Chapter3}
\end{equation}
which for the particularly simple example of an $S$- and $P$-wave truncation ($\ell_{\mathrm{max}} = 1$) reduces to
\begin{equation}
\check{\Sigma} \left( W, \theta \right) = \rho \hspace*{3pt} \left(a_{L}\right)_{2}^{\check{\Sigma}} \left( W \right) P^{2}_{2} \left( \cos \theta \right) \mathrm{.}  \label{eq:LowEAssocLegBeamAsymmetry1SPWaves}
\end{equation}
By fitting parametrizations such as these to angular distributions of data, it is possible to infer the content of the dominant contributing partial waves by looking at the $\chi^{2}/\mathrm{ndf}$ of fits with different truncation orders $\ell_{\mathrm{max}}$. The dominant multipoles can be inferred reliably, while certain partial-wave interferences can remain hidden (cf. section \ref{sec:LFitsPaper}). The procedure is in other contexts sometimes referred to as \textit{moment analysis} \cite{MikhasenkoProceeding2014}. \newline
Although the method is simple, it can also be very useful. It can be performed without a lot of additional advice and is less calculationally "costly" and complicated than a full TPWA or even fitting an energy-dependent model. Furthermore, it can serve as a first preparatory guideline for these more complicated procedures. \newline
Details on the analysis scheme as well as a survey of its application to recent polarization measurements are given in the next section, where a paper \cite{LFitPaper} on the subject written in the course of this thesis is shown. \newline
\underline{Remark:} For copyright reasons, an earlier version of reference \cite{LFitPaper}, which has been generated toward the end of the refereeing process, is included in section \ref{sec:LFitsPaper}. The version published by \textit{Eur.\ Phys.\ J.\ A} has the DOI: 10.1140/epja/i2017-12255-0.

\vspace*{40pt}

\subsection{Determining the dominant partial wave contributions from angular distributions of single- and double-polarization observables in pseudoscalar meson photoproduction} \label{sec:LFitsPaper}

\includepdf[pages={1}, frame=false, noautoscale=true, scale=0.86, pagecommand={}]{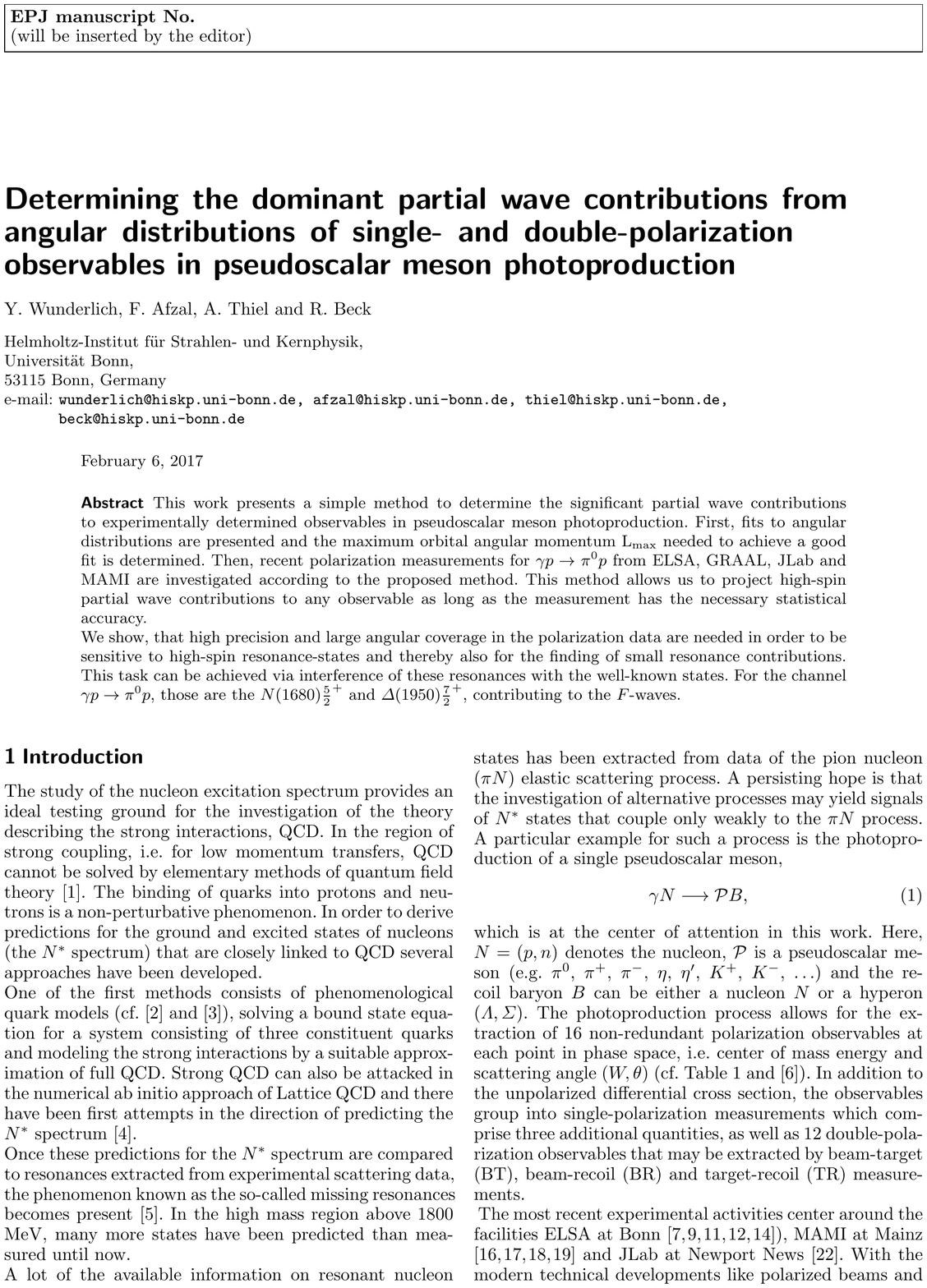}

\newpage

\includepdf[pages={2}, frame=false, noautoscale=true, scale=0.86, pagecommand={}]{LFit_paper_long_rev4-1}

\newpage

\includepdf[pages={3}, frame=false, noautoscale=true, scale=0.86, pagecommand={}]{LFit_paper_long_rev4-1}

\newpage

\includepdf[pages={4}, frame=false, noautoscale=true, scale=0.86, pagecommand={}]{LFit_paper_long_rev4-1}

\newpage

\includepdf[pages={5}, frame=false, noautoscale=true, scale=0.86, pagecommand={}]{LFit_paper_long_rev4-1}

\newpage

\includepdf[pages={6}, frame=false, noautoscale=true, scale=0.86, pagecommand={}]{LFit_paper_long_rev4-1}

\newpage

\includepdf[pages={7}, frame=false, noautoscale=true, scale=0.86, pagecommand={}]{LFit_paper_long_rev4-1}

\newpage

\includepdf[pages={8}, frame=false, noautoscale=true, scale=0.86, pagecommand={}]{LFit_paper_long_rev4-1}

\newpage

\includepdf[pages={9}, frame=false, noautoscale=true, scale=0.86, pagecommand={}]{LFit_paper_long_rev4-1}

\newpage

\includepdf[pages={10}, frame=false, noautoscale=true, scale=0.86, pagecommand={}]{LFit_paper_long_rev4-1}

\newpage

\includepdf[pages={11}, frame=false, noautoscale=true, scale=0.86, pagecommand={}]{LFit_paper_long_rev4-1}

\newpage

\includepdf[pages={12}, frame=false, noautoscale=true, scale=0.86, pagecommand={}]{LFit_paper_long_rev4-1}

\newpage

\includepdf[pages={13}, frame=false, noautoscale=true, scale=0.86, pagecommand={}]{LFit_paper_long_rev4-1}

\newpage

\includepdf[pages={14}, frame=false, noautoscale=true, scale=0.86, pagecommand={}]{LFit_paper_long_rev4-1}

\newpage

\includepdf[pages={15}, frame=false, noautoscale=true, scale=0.86, pagecommand={}]{LFit_paper_long_rev4-1}

\newpage

\includepdf[pages={16}, frame=false, noautoscale=true, scale=0.86, pagecommand={}]{LFit_paper_long_rev4-1}

\newpage

\includepdf[pages={17}, frame=false, noautoscale=true, scale=0.86, pagecommand={}]{LFit_paper_long_rev4-1}

\newpage

\includepdf[pages={18}, frame=false, noautoscale=true, scale=0.86, pagecommand={}]{LFit_paper_long_rev4-1}

\newpage

\includepdf[pages={19}, frame=false, noautoscale=true, scale=0.86, pagecommand={}]{LFit_paper_long_rev4-1}

\newpage

\includepdf[pages={20}, frame=false, noautoscale=true, scale=0.86, pagecommand={}]{LFit_paper_long_rev4-1}

\newpage

\includepdf[pages={21}, frame=false, noautoscale=true, scale=0.86, pagecommand={}]{LFit_paper_long_rev4-1}

\newpage

\includepdf[pages={22}, frame=false, noautoscale=true, scale=0.86, pagecommand={}]{LFit_paper_long_rev4-1}

\newpage

\includepdf[pages={23}, frame=false, noautoscale=true, scale=0.86, pagecommand={}]{LFit_paper_long_rev4-1}

\newpage

\includepdf[pages={24}, frame=false, noautoscale=true, scale=0.86, pagecommand={}]{LFit_paper_long_rev4-1}

\newpage

\includepdf[pages={25}, frame=false, noautoscale=true, scale=0.86, pagecommand={}]{LFit_paper_long_rev4-1}

\newpage

\includepdf[pages={26}, frame=false, noautoscale=true, scale=0.86, pagecommand={}]{LFit_paper_long_rev4-1}

\newpage

\includepdf[pages={27}, frame=false, noautoscale=true, scale=0.86, pagecommand={}]{LFit_paper_long_rev4-1}

\newpage

\includepdf[pages={28}, frame=false, noautoscale=true, scale=0.86, pagecommand={}]{LFit_paper_long_rev4-1}

\newpage

\includepdf[pages={29}, frame=false, noautoscale=true, scale=0.86, pagecommand={}]{LFit_paper_long_rev4-1}

\newpage

\includepdf[pages={30}, frame=false, noautoscale=true, scale=0.86, pagecommand={}]{LFit_paper_long_rev4-1}

\newpage

\includepdf[pages={31}, frame=false, noautoscale=true, scale=0.86, pagecommand={}]{LFit_paper_long_rev4-1}

\newpage

\includepdf[pages={32}, frame=false, noautoscale=true, scale=0.86, pagecommand={}]{LFit_paper_long_rev4-1}

\newpage

\includepdf[pages={33}, frame=false, noautoscale=true, scale=0.86, pagecommand={}]{LFit_paper_long_rev4-1}

\newpage

\includepdf[pages={34}, frame=false, noautoscale=true, scale=0.86, pagecommand={}]{LFit_paper_long_rev4-1}

\newpage

\includepdf[pages={35}, frame=false, noautoscale=true, scale=0.86, pagecommand={}]{LFit_paper_long_rev4-1}

\newpage

\includepdf[pages={36}, frame=false, noautoscale=true, scale=0.86, pagecommand={}]{LFit_paper_long_rev4-1}

\newpage

\includepdf[pages={37}, frame=false, noautoscale=true, scale=0.86, pagecommand={}]{LFit_paper_long_rev4-1}

\newpage

\includepdf[pages={38}, frame=false, noautoscale=true, scale=0.86, pagecommand={}]{LFit_paper_long_rev4-1}

\newpage

\includepdf[pages={39}, frame=false, noautoscale=true, scale=0.86, pagecommand={}]{LFit_paper_long_rev4-1}

\newpage

\includepdf[pages={40}, frame=false, noautoscale=true, scale=0.86, pagecommand={}]{LFit_paper_long_rev4-1}

\newpage

\includepdf[pages={41}, frame=false, noautoscale=true, scale=0.86, pagecommand={}]{LFit_paper_long_rev4-1}

\newpage

\includepdf[pages={42}, frame=false, noautoscale=true, scale=0.86, pagecommand={}]{LFit_paper_long_rev4-1}

\newpage

\includepdf[pages={43}, frame=false, noautoscale=true, scale=0.86, pagecommand={}]{LFit_paper_long_rev4-1}

\newpage

\includepdf[pages={44}, frame=false, noautoscale=true, scale=0.86, pagecommand={}]{LFit_paper_long_rev4-1}

\newpage

\includepdf[pages={45}, frame=false, noautoscale=true, scale=0.86, pagecommand={}]{LFit_paper_long_rev4-1}

\newpage

\includepdf[pages={46}, frame=false, noautoscale=true, scale=0.86, pagecommand={}]{LFit_paper_long_rev4-1}

\newpage

\includepdf[pages={47}, frame=false, noautoscale=true, scale=0.86, pagecommand={}]{LFit_paper_long_rev4-1}

\newpage

\includepdf[pages={48}, frame=false, noautoscale=true, scale=0.86, pagecommand={}]{LFit_paper_long_rev4-1}

\newpage

\includepdf[pages={49}, frame=false, noautoscale=true, scale=0.86, pagecommand={}]{LFit_paper_long_rev4-1}

\newpage

\includepdf[pages={50}, frame=false, noautoscale=true, scale=0.86, pagecommand={}]{LFit_paper_long_rev4-1}

\newpage

\includepdf[pages={51}, frame=false, noautoscale=true, scale=0.86, pagecommand={}]{LFit_paper_long_rev4-1}

\newpage

\includepdf[pages={52}, frame=false, noautoscale=true, scale=0.86, pagecommand={}]{LFit_paper_long_rev4-1}

\newpage

\includepdf[pages={53}, frame=false, noautoscale=true, scale=0.86, pagecommand={}]{LFit_paper_long_rev4-1}

\newpage

\includepdf[pages={54}, frame=false, noautoscale=true, scale=0.86, pagecommand={}]{LFit_paper_long_rev4-1}

\newpage

\includepdf[pages={55}, frame=false, noautoscale=true, scale=0.86, pagecommand={}]{LFit_paper_long_rev4-1}

\newpage

\includepdf[pages={56}, frame=false, noautoscale=true, scale=0.86, pagecommand={}]{LFit_paper_long_rev4-1}

\newpage

\includepdf[pages={57}, frame=false, noautoscale=true, scale=0.86, pagecommand={}]{LFit_paper_long_rev4-1}

\newpage

\includepdf[pages={58}, frame=false, noautoscale=true, scale=0.86, pagecommand={}]{LFit_paper_long_rev4-1}

\newpage

\includepdf[pages={59}, frame=false, noautoscale=true, scale=0.86, pagecommand={}]{LFit_paper_long_rev4-1}

\newpage

\includepdf[pages={60}, frame=false, noautoscale=true, scale=0.86, pagecommand={}]{LFit_paper_long_rev4-1}

\newpage

\includepdf[pages={61}, frame=false, noautoscale=true, scale=0.86, pagecommand={}]{LFit_paper_long_rev4-1}

\newpage

\includepdf[pages={62}, frame=false, noautoscale=true, scale=0.86, pagecommand={}]{LFit_paper_long_rev4-1}

\newpage

\includepdf[pages={63}, frame=false, noautoscale=true, scale=0.86, pagecommand={}]{LFit_paper_long_rev4-1}

\newpage

\includepdf[pages={64}, frame=false, noautoscale=true, scale=0.86, pagecommand={}]{LFit_paper_long_rev4-1}

\newpage

\includepdf[pages={65}, frame=false, noautoscale=true, scale=0.86, pagecommand={}]{LFit_paper_long_rev4-1}

\newpage

\includepdf[pages={66}, frame=false, noautoscale=true, scale=0.86, pagecommand={}]{LFit_paper_long_rev4-1}

%
%
%
%
%
%
%


%
%
%

\subsection{Summary of chapter \ref{chap:LFits}} \label{sec:ChapIVSummary}

In the preceding section, a publication \cite{LFitPaper} has been presented on the subject of Legendre moment analyses in pseudoscalar meson photoproduction. The required formalism was presented in some detail and then applied to a relatively large collection of single- and double-polarization data for the reaction $\gamma p \longrightarrow \pi^{0} p$. \newline
The performance of a moment analysis represents a useful preparatory step towards the extraction of multipoles in a full TPWA. The latter will be the subject of chapter \ref{chap:TPWA}. \newline
Within the investigations shown in section \ref{sec:LFitsPaper}, the possibility has shown up to fit the set of seven observables $\left\{ \sigma_{0}, \Sigma, T, P, E, G, H  \right\}$ \cite{Adlarson:2015,GRAAL,Hartmann:2014,Hartmann:2015,Gottschall:2014,Thiel:2012,Thiel:2016} in the second resonance region, employing a TPWA with a truncation at $\ell_{\mathrm{max}} = 2$ or $3$. This dataset will be analyzed further within section \ref{subsec:2ndResRegionDataFits} of chapter \ref{chap:TPWA}. But first, the necessary formal and numerical machinery of the TPWA has to be introduced. \newline
Lastly, the paper in section \ref{sec:LFitsPaper} introduced a novel way to illustrate the partial-wave in\-ter\-fe\-ren\-ces present in the Legendre coefficients as colored plots. The matrices shown in appendix \ref{sec:TPWAFormulae}, which have been worked out as a part of this thesis, have been the basis of these graphical representations.

\clearpage

\section{Numerical truncated partial wave analyses} \label{chap:TPWA}

This chapter treats the methods and shows the results for full numerical multipole-fits done in the course of this work. This comprises the ``fits'' to exactly solvable model data, as well as real fits of actually measured data. \newline
The chapter therefore stands in symbiosis to chapter \ref{chap:Omelaenko} which treated the mathematical ambiguities of the TPWA. Furthermore, results will be shown that were announced in the latter chapter. Also, in quite a lot of places, we will refer to the results and intuitions obtained in chapter \ref{chap:Omelaenko}. As mentioned in sections \ref{sec:CompExpsTPWA} and \ref{sec:OmelaenkoApproachIntro}, an analytic solution of the equation systems appearing in a TPWA is generally not possible. Therefore, the numerical solution of particular model-TPWA's provides the only way to check the proposed complete experiments of chapter \ref{chap:Omelaenko}, the latter coming just out of the study of ambiguities. \newline
Furthermore, this chapter can be seen as a logical continuation of chapter \ref{chap:LFits}, which showed results obtained by fitting the associated Legendre expansion to angular distributions of observables.

\subsection{Outline of numerical TPWA fits} \label{sec:TPWAFitsIntro}

We use here the standard form of the TPWA, using an associated Legendre expansion for the angular distributions of observables (See equations (\ref{eq:LowEAssocLegStandardParametrization1}) to (\ref{eq:LowEAssocLegStandardParametrization2}) and Table \ref{tab:AngularDistributionsParameters} of section \ref{sec:CompExpsTPWA}), which is quoted here for convenience
\allowdisplaybreaks
\begin{align}
\check{\Omega}^{\alpha} \left( W, \theta \right) &= \rho \hspace*{3pt} \sum \limits_{k = \beta_{\alpha}}^{2 \ell_{\mathrm{max}} + \beta_{\alpha} + \gamma_{\alpha}} \left(a_{L}\right)_{k}^{\check{\Omega}^{\alpha}} \left( W \right) P^{\beta_{\alpha}}_{k} \left( \cos \theta \right) \mathrm{,}  \label{eq:LowEAssocLegStandardParametrization1TPWAChap} \\
\left(a_{L}\right)_{k}^{\check{\Omega}^{\alpha}} \left( W \right) &= \left< \mathcal{M}_{\ell_{\mathrm{max}}} \left( W \right) \right| \left( \mathcal{C}_{L}\right)_{k}^{\check{\Omega}^{\alpha}} \left| \mathcal{M}_{\ell_{\mathrm{max}}} \left( W \right) \right> \mathrm{.} \label{eq:LowEAssocLegStandardParametrization2TPWAChap}
\end{align}
Suppose now that some set of observables has been measured. These would correspond to some subset of indicex $\alpha$ in equation (\ref{eq:LowEAssocLegStandardParametrization1TPWAChap}). The set $\left\{ \check{\Omega}^{1},  \check{\Omega}^{4},  \check{\Omega}^{10},  \check{\Omega}^{12},  \check{\Omega}^{11} \right\}$ would for example correspond to the observables $\left\{ \sigma_{0}, \check{\Sigma}, \check{T}, \check{P}, \check{F} \right\}$. \newline
A TPWA shall be fitted to these observables corresponding to some fixed truncation order $L = \ell_{\mathrm{max}}$. Multipoles are the goal of this fit, at each energy of the dataset individually. Adjacent energy bins do not know about each other in this kind of analysis. Therefore, the kind of fit described in the following is also called a ``single energy''- (SE-) fit. \newline  One possibility to accomplish this task, which has been pursued in this work, consists of a strategy akin to the work Grushin \cite{Grushin}, which consists of first extracting the Legendre coefficients from the angular distributions of profile functions (cf. chapter \ref{chap:LFits}) and then obtaining multipoles in a second step by fitting them to these coefficients. This can be written in brief form as
\begin{equation}
 \check{\Omega}^{\alpha} \overset{\mathrm{I}}{\longrightarrow} \left( a_{L} \right)^{\alpha}_{k} \overset{\mathrm{II}}{\longrightarrow} \mathcal{M}_{\ell} = \left\{ E_{\ell \pm}, M_{\ell \pm} \right\} \mathrm{.} \label{eq:FitWayOverLegCoeffsIntermediateStep}
\end{equation}
In regard to the numerical estimation of multipoles, one can choose from several different methods \cite{BlobelLohrmann}. We resort here to the method of least squares \cite{BlobelLohrmann} for both steps I and II. The expressions employed in each step are now elaborated in more detail. \newline \clearpage

\underline{Step I:} \newline

In case exact theoretical model data are solved, we minimize the simple sum of squares
\begin{equation}
 \Phi_{a}^{\alpha} \left( \left\{ a_{L} \right\} \right) = \sum_{c_{k_{\alpha}}} \left[ \check{\Omega}^{\alpha}_{\mathrm{Data}} (c_{k_{\alpha}}) - \check{\Omega}^{\alpha}_{\mathrm{Fit}} \left( c_{k_{\alpha}}, \left\{\left(a_{L}\right)^{\alpha}_{k}\right\} \right) \right]^{2} \mathrm{,} \label{eq:LegCoeffFitChi2ModelFit}
\end{equation}
where $c_{k_{\alpha}} := \cos \left( \theta_{k_{\alpha}} \right)$ denotes the (possibly varying) angular grid of each observable. The fit is done for each observable ($\equiv \alpha$-value) individually. In case real data are fitted, with each datapoint $\check{\Omega}^{\alpha}_{\mathrm{Data}} (c_{k_{\alpha}})$ endowed with a statistical error $\Delta \check{\Omega}^{\alpha}_{\mathrm{Data}} (c_{k_{\alpha}})$ assumed to be a normal standard error, the error-weighted version of (\ref{eq:LegCoeffFitChi2ModelFit}) has to be minimized, which is equal to
\begin{equation}
 \left(\chi^{2}_{a}\right)^{\alpha} \left( \left\{ a_{L} \right\} \right) = \sum_{c_{k_{\alpha}}} \left[ \frac{ \check{\Omega}^{\alpha}_{\mathrm{Data}} (c_{k_{\alpha}}) - \check{\Omega}^{\alpha}_{\mathrm{Fit}} \left( c_{k_{\alpha}}, \left\{\left(a_{L}\right)^{\alpha}_{k}\right\} \right)}{\Delta \check{\Omega}^{\alpha}_{\mathrm{Data}} (c_{k_{\alpha}})} \right]^{2} \mathrm{.} \label{eq:LegCoeffFitChi2DataFit}
\end{equation}
In both cases, the value of the model function $\check{\Omega}^{\alpha}_{\mathrm{Fit}} \left( c_{k_{\alpha}}, \left\{\left(a_{L}\right)^{\alpha}_{k}\right\} \right)$ is given by evaluating the angular parametrization (\ref{eq:LowEAssocLegStandardParametrization1TPWAChap}) at $c_{k_{\alpha}}$. \newline
A few comments are in order regarding the form (\ref{eq:LegCoeffFitChi2DataFit}) used in the fits to data. In this approach here, the profile functions $\check{\Omega}^{\alpha} = \Omega^{\alpha} \ast \sigma_{0}$ are fitted. Therefore, in case one is provided with a dataset for the dimensionless asymmetry $\Omega^{\alpha}$, the process of calculating the profile function prior to the fit may induce correlations among the unpolarized differential cross section and the profile functions of all other observables. These correlations are not taken into account in equation (\ref{eq:LegCoeffFitChi2DataFit}). The reason for this is that in all datasets investigated for this thesis, the errors for the differential cross sections are practically negligible compared to the errors of the asymmetries. The errors of the profile functions were calculated using standard Gaussian error propagation in all cases. \newline
This first step of the analysis is easily implemented in standard fitting-routines as for example given in ROOT. In our case, we employ the routine \textit{NonLinearModelFit} in MATHEMATICA \cite{Mathematica8,Mathematica11,MathematicaLanguage,MathematicaBonnLicense}. The respective fit-routine used then yields in the end the numbers $\left(a_{L}^{\mathrm{Fit}}\right)^{\alpha}_{k}$ for the Legendre coefficients. In case of the data-fit (\ref{eq:LegCoeffFitChi2DataFit}), standard errors as well as covariance- and correlation-matrices of the Legendre coefficients are also interesting, which can be extracted from \textit{NonLinearModelFit} as well. Alternative fit-routines normally also provide this service. \newline

\underline{Step II:} \newline

This is the actual multipole-fit and the more difficult part of the analysis. In the case of the exactly solvable data, the parameters $\left(a_{L}^{\mathrm{Fit}}\right)^{\alpha}_{k}$ coming from each individual profile function are generally non-correlated to a good approximation. Furthermore, the elements of the covariance matrix $\mathrm{\textbf{C}}$ of the $\left(a_{L}^{\mathrm{Fit}}\right)^{\alpha}_{k}$ are too small to be numerically tractable. Therefore, we drop any covariance terms and minimize the least squares discrepancy function
\begin{equation}
\Phi_{\mathcal{M}} \left( \left\{ \mathcal{M}_{\ell} \right\} \right) = \sum_{\alpha, k} \left[ \left(a_{L}^{\mathrm{Fit}}\right)^{\alpha}_{k} -  \left< \mathcal{M}_{\ell} \right| \left( \mathcal{C}_{L}\right)_{k}^{\check{\Omega}^{\alpha}} \left| \mathcal{M}_{\ell} \right> \right]^{2} \mathrm{,} \label{eq:PhiTheoryDataFitStep2}
\end{equation}
as a function of the real- and imaginary parts of the multipoles. In case of a data-fit however, it is generally not a good approximation any more to disregard the correlations among the resulting parameters $\left(a_{L}^{\mathrm{Fit}}\right)^{\alpha}_{k}$ from step I. They are clearly non-vanishing and especially for higher truncations $L$ may significantly influence the result of the fit. Instead, one should fit a correlated $\chi^{2}$ \cite{Grushin,BlobelLohrmann}, which by collecting the indices $\alpha$ and $k$ into multi-indices $i = \left( \alpha, k \right)$ can be written as
\begin{equation}
\chi^{2}_{\mathcal{M}} \left( \left\{ \mathcal{M}_{\ell} \right\} \right) = \sum_{i,j} \Big[ \left(a_{L}^{\mathrm{Fit}}\right)_{i} - \left< \mathcal{M}_{\ell} \right| \left(\mathcal{C}_{L}\right)_{i} \left| \mathcal{M}_{\ell} \right> \Big] \mathrm{\textbf{C}}^{-1}_{ij} \Big[ \left(a_{L}^{\mathrm{Fit}}\right)_{j} - \left< \mathcal{M}_{\ell} \right| \left(\mathcal{C}_{L}\right)_{j} \left| \mathcal{M}_{\ell} \right> \Big] \mathrm{.} \label{eq:CorrelatedChisquare}
\end{equation}
This function is specified by the fitted Legendre coefficients $\left(a_{L}^{\mathrm{Fit}}\right)_{i}$ as well as their covariance matrix $\mathrm{\textbf{C}}_{ij}$, which is also a result of step I. The function $\chi^{2}_{\mathcal{M}} = \chi^{2}_{\mathcal{M}} \left(  \left\{ \mathcal{M}_{\ell} \right\} \right)$ is again minimized as a function of the real- and imaginary parts of the multipoles. \newline
The latter enter the functions (\ref{eq:PhiTheoryDataFitStep2}) and (\ref{eq:CorrelatedChisquare}) via the multipole vector
\begin{equation}
 \left| \mathcal{M}_{\ell} \right> = \left[ E_{0+}, E_{1+}, M_{1+}, M_{1-}, E_{2+}, E_{2-}, M_{2+}, M_{2-}, \ldots , M_{L-}  \right]^{T} \mathrm{.} \label{eq:MultipoleVectorChap4}
\end{equation}
and its hermitean adjoint $\left< \mathcal{M}_{\ell} \right|$. It is furthermore worth mentioning that the correlated $\chi^{2}$ (\ref{eq:CorrelatedChisquare}) can be reduced to a function that has a non-correlated diagonal shape similar to (\ref{eq:LegCoeffFitChi2DataFit}) by means of an orthogonal transformation. This is standard knowledge \cite{BlobelLohrmann}, but still the important points are elaborated in appendix \ref{sec:CorrelatedChi2Reduction}. \newline
Since $\mathrm{\textbf{C}}$ is real and symmetric, so is its inverse. Therefore, the latter can be diagonalized by an orthogonal matrix $\mathrm{O}$ according to $\mathrm{O}^{T} \left( \mathrm{\textbf{C}}^{-1} \right) \mathrm{O} =: \bm{D}$, such that $\bm{D}$ is a diagonal matrix. Then, the function (\ref{eq:CorrelatedChisquare}) is equivalent to
\allowdisplaybreaks
\begin{equation}
 \chi^{2}_{\mathcal{M}} \left(  \left\{ \mathcal{M}_{\ell} \right\} \right) = \sum_{i} \bm{c}_{i} \left[ \left(a_{L}^{\mathrm{Fit}}\right)_{i}^{R} - \left< \mathcal{M}_{\ell} \right| \left(\mathcal{C}_{L}\right)_{i}^{R} \left| \mathcal{M}_{\ell} \right>  \right]^{2} \mathrm{,} \label{eq:ReductionToDiagonalChisquareChap4}
\end{equation}
where $\bm{c}_{i}$ are the eigenvalues of $\mathrm{\textbf{C}}^{-1}$ and the rotated Legendre coefficients and TPWA fit-matrices have been defined according to
\begin{equation}
 \left(a_{L}^{\mathrm{Fit}}\right)_{i}^{R} := \sum_{j} \mathrm{O}^{T}_{ij} \left(a_{L}^{\mathrm{Fit}}\right)_{j} \mathrm{,} \hspace*{5pt} \left(\mathcal{C}_{L}\right)_{i}^{R} := \sum_{j} \mathrm{O}^{T}_{ij} \left(\mathcal{C}_{L}\right)_{j} \mathrm{.} \label{eq:DefRotatedFitParametersAndMatricesChap4}
\end{equation}
In this way, the correlated $\chi^{2}$ can also be implemented in algorithms that can only handle functions that are strictly sums of squares. The Levenberg-Marquardt algorithm employed in the routine \textit{FindMinimum} of MATHEMATICA \cite{Mathematica8,Mathematica11,MathematicaLanguage,MathematicaBonnLicense}, which was mainly employed in this work, is an example for such a case. \newline
Furthermore, the diagonalization that lead from the expression (\ref{eq:CorrelatedChisquare}) to (\ref{eq:ReductionToDiagonalChisquareChap4}) is a proof of the fact that the function $\chi^{2}_{\mathcal{M}}$ defined in (\ref{eq:CorrelatedChisquare}) follows a chisquare-distribution \cite{BlobelLohrmann}. \newline

In contrast to Grushin's \cite{Grushin} method of breaking the TPWA into steps $\mathrm{I}$ and $\mathrm{II}$ as described above, we also implemented the possibility of fitting the multipoles directly to the data as an alternative. This means we insert the Legendre coefficients in terms of multipoles (\ref{eq:LowEAssocLegStandardParametrization2TPWAChap}) directly into the non-correlated chisquare, i.e. a sum over individual terms such as given in equation (\ref{eq:LegCoeffFitChi2DataFit}), upon which the following quantity is minimized
\begin{align}
 \chi^{2}_{\mathrm{data}} \left(  \left\{ \mathcal{M}_{\ell} \right\} \right) &= \sum_{\alpha, c_{k_{\alpha}}} \left[ \frac{ \check{\Omega}^{\alpha}_{\mathrm{Data}} (c_{k_{\alpha}}) - \check{\Omega}^{\alpha}_{\mathrm{Fit}} \left( c_{k_{\alpha}}, \left\{ \mathcal{M}_{\ell} \right\} \right) }{\Delta \check{\Omega}^{\alpha}_{\mathrm{Data}} (c_{k_{\alpha}})} \right]^{2} \mathrm{,} \label{eq:ChiSquareDirectFit} \\
 \mathrm{with} \hspace*{5pt} \check{\Omega}^{\alpha}_{\mathrm{Fit}} \left( c_{k_{\alpha}}, \left\{ \mathcal{M}_{\ell} \right\} \right) &:= \frac{q}{k} \sum_{n = \beta_{\alpha}}^{2 L + \beta_{\alpha} + \gamma_{\alpha}} \left< \mathcal{M}_{\ell} \right| \left( \mathcal{C}_{L}\right)_{n}^{\check{\Omega}^{\alpha}}  \left| \mathcal{M}_{\ell} \right> \hspace*{2pt} P^{\beta_{\alpha}}_{n} \left(  c_{k_{\alpha}} \right) \mathrm{.} \label{eq:FitFunctionDirectFit}
\end{align}
At first glance, both ways of fitting look fully equivalent and we have to state that in all cases where both were tried, they yielded consistent results. Still, we regard the direct fit to the data as a {\it safer} option in case datasets are analyzed which are dominated by systematic errors. In section \ref{subsec:DeltaRegionDataFits}, one such case will be shown. \newline

The parameters to vary in fit step II of Grushin's method (\ref{eq:FitWayOverLegCoeffsIntermediateStep}), as well as in the direct fit to the data ((\ref{eq:ChiSquareDirectFit}) $\&$ (\ref{eq:FitFunctionDirectFit})), are in any case the real- and imaginary parts of the multipoles. A word of caution is in order in regard of how to vary them. \newline
As is well known and described in sections \ref{sec:CompExpsTPWA} and \ref{sec:OmelaenkoApproachIntro}, a fully model-independent TPWA truncated at some finite $L$ is only capable of determining the multipoles up to an unknown energy-dependent overall phase. This originates from the fact that multipoles enter the functions equations (\ref{eq:PhiTheoryDataFitStep2}), (\ref{eq:CorrelatedChisquare}) and (\ref{eq:ReductionToDiagonalChisquareChap4}) through bilinear forms. The latter are completely blind to a rotation of all multipoles by the same overall phase. \newline
One way out of this issue consists of fixing some convention for the overall phase, effectively removing one real degree of freedom from the problem but also removing the above mentioned energy-dependent rest of the continuum-ambiguity. In this work, whenever multipoles are extracted from the maximally model-independent TPWA formulated above, we extracted ``phase-constrained'' multipoles $\mathcal{M}_{\ell}^{C}$, defined by setting the phase of $E_{0+}$ to zero
\begin{equation}
 \mathrm{Re} \left[ E_{0+}^{C} \right] > 0 \hspace*{3.75pt} \& \hspace*{3.75pt} \mathrm{Im} \left[ E_{0+}^{C} \right] = 0 \mathrm{.} \label{eq:PhaseConstraintMultipoles}
\end{equation}
There exists of course an infinity of different possibilities to fix a convention for the phase. Another possibility would be to regard the phase of a particular multipole to be well known (e.g. the phase of $M_{1+}$ in the $\Delta$-region for $\pi^{0}$ photoproduction) and fix it to the already pre-determined value. \newline
In our case, the phase-constrained multipoles enter the bilinears via their multipole-vector
\begin{equation}
 \left| \mathcal{M}_{\ell}^{C} \right> = \left( \mathrm{Re} \left[ E_{0+}^{C} \right], \mathrm{Re} \left[ E_{1+}^{C} \right] + i \hspace*{1.5pt} \mathrm{Im} \left[ E_{1+}^{C} \right], \mathrm{Re} \left[ M_{1+}^{C} \right] + i \hspace*{1.5pt} \mathrm{Im} \left[ M_{1+}^{C} \right], \ldots  \right)^{T} \mathrm{.} \label{eq:MultipoleVectorConstrainedChap4}
\end{equation}
The real- and imaginary parts of the $\mathcal{M}_{\ell}^{C}$ are then varied and the problem of the remaining ambiguity is effectively removed. \newline
Another possibility would be to re-phrase the minimization problem in terms of moduli and relative-phases of the multipoles
\begin{equation}
 \left( \left| E_{0+} \right|, \ldots , \left| M_{L-} \right|\right), \hspace*{2pt} \mathrm{and} \hspace*{2pt} \left( \phi_{E_{1+}} - \phi_{E_{0+}},  \phi_{M_{1+}} - \phi_{E_{1+}}, \ldots, \phi_{M_{L-}} - \phi_{M_{L+}} \right) \mathrm{.} \label{eq:ModRelPhasesListing}
\end{equation}
These parameters effectively contain all the information on multipoles up to an overall phase and can then, upon re-formulation, be extracted in a fit. Though this second method is certainly used in a lot of places, in this work we only fit real- and imaginary parts. \newline
Concerning the overall phase, it should be mentioned that it is possible to fit it out of the data in a TPWA, at the price of introducing some additional theoretical assumptions or model-dependencies into the fit. A good example is the analysis of Grushin \cite{Grushin}, who introduced the Fermi-Watson theorem \cite{KMWatson} into a simultaneous fit of $(\pi^{+} n)$- and $(\pi^{0} p)$-photoproduction in the $\Delta$-region. It was then possible to perform an analysis in which the overall phase of the multipoles in the $(\pi^{+} n)$-channel could be fixed by introducting the exactly calculable pion-pole contribution explicitly, which determines the initially unknown phase in the fit via interference-effects. Then, the Fermi-Watson theorem is capable to carry this knowledge of the phase in the $(\pi^{+} n)$-channel over to the $(\pi^{0} p)$-channel, also determining the overall phase there. \newline
The above mentioned phenomenon, that knowledge of some, possibly higher partial waves can determine the overall phase in a fit deserves some more elaboration. For a truncation at $L = \ell_{\mathrm{max}}$, one can decompose a generic fit-matrix entering equation (\ref{eq:PhiTheoryDataFitStep2}), (\ref{eq:CorrelatedChisquare}) or (\ref{eq:ReductionToDiagonalChisquareChap4}) into blocks as follows
\begin{equation}
 \left( \mathcal{\mathcal{C}}_{L} \right)_{i} =  \left[ \begin{array}{ccc|ccc}  &  &  &  &  &  \\  & \left( \bar{\mathcal{C}}_{\ell \leq L} \right)_{i} &  &  & \left( \tilde{\mathcal{C}}_{\ell} \right)_{i}  &  \\  &  &  &  &  &  \\ \hline  &  &  &  &  &  \\ & \left[\left( \tilde{\mathcal{C}}_{\ell} \right)_{i}\right]^{\dagger} &  &  & \left( \hat{\mathcal{C}}_{\ell > L} \right)_{i}  &  \\  &  &  &  &  &  \end{array} \right] \mathrm{.} \label{eq:FitMatrixDecomposition}
\end{equation}
Here, the matrices $\left( \bar{\mathcal{C}}_{\ell \leq L} \right)$ and $\left( \hat{\mathcal{C}}_{\ell > L} \right)$ only mix lower and higher multipoles among themselves, respectively. The off-diagonal matrix $\left( \tilde{\mathcal{C}}_{\ell} \right)$ however generates interference-terms among higher and lower multipoles. \newline
Now, suppose that all multipoles with $\ell \leq L$ are running freely in a fit, without introducing a phase-constraint and furthermore a good model or estimate for the higher partial waves is known. In case the higher multipoles are held fixed at the particular model, then the off-diagonal matrices generate contributions to $\chi^{2}$ that are only of linear and not bilinear nature and through which the overall phase can be determined (sub-scripts on sub-matrices are dropped here):
\begin{small}
\allowdisplaybreaks
\begin{align}
\left( a_{L} \right)_{i} &= \left[ \begin{array}{c|c}  \mathcal{M}^{\ast}_{\ell \leq L}  & \mathcal{M}^{\ast}_{\ell > L} \end{array} \right]  \left[ \begin{array}{ccc|ccc}  &  &  &  &  &  \\  & \left( \bar{\mathcal{C}} \right)_{i} &  &  & \left( \tilde{\mathcal{C}} \right)_{i}  &  \\  &  &  &  &  &  \\ \hline  &  &  &  &  &  \\ & \left[\left( \tilde{\mathcal{C}} \right)_{i}\right]^{\dagger} &  &  & \left( \hat{\mathcal{C}} \right)_{i}  &  \\  &  &  &  &  &  \end{array} \right] \left[ \begin{array}{c} \\ \mathcal{M}_{\ell \leq L} \\  \\ \hline  \\  \mathcal{M}_{\ell > L} \\ \textcolor{white}{\mathrm{Hi}} \end{array} \right] \nonumber \\
 &= \sum_{m, n} \left( \mathcal{M}_{\ell \leq} \right)^{\ast}_{m}  \left[ \left( \bar{\mathcal{C}} \right)_{i} \right]_{m,n}  \left( \mathcal{M}_{\ell \leq} \right)_{n} + \sum_{p, q} \left( \mathcal{M}_{\ell \leq} \right)^{\ast}_{p}  \left[ \left( \tilde{\mathcal{C}} \right)_{i} \right]_{p,q}  \left( \mathcal{M}_{\ell >} \right)_{q} \nonumber \\
 & \quad + \sum_{r, l} \left( \mathcal{M}_{\ell >} \right)^{\ast}_{r}  \left[ \left( \tilde{\mathcal{C}} \right)_{i} \right]^{\dagger}_{r,l}  \left( \mathcal{M}_{\ell \leq} \right)_{l} + \sum_{b, c} \left( \mathcal{M}_{\ell >} \right)^{\ast}_{b}  \left[ \left( \hat{\mathcal{C}} \right)_{i} \right]_{b,c}  \left( \mathcal{M}_{\ell >} \right)_{c}  \nonumber \\
 &\simeq \sum_{m, n} \left( \mathcal{M}_{\ell \leq} \right)^{\ast}_{m}  \left[ \left( \bar{\mathcal{C}} \right)_{i} \right]_{m,n}  \left( \mathcal{M}_{\ell \leq} \right)_{n} + \underbrace{2 \hspace*{1.5pt} \mathrm{Re} \left[ \sum_{p, q} \left( \mathcal{M}_{\ell \leq} \right)^{\ast}_{p}  \left[ \left( \tilde{\mathcal{C}} \right)_{i} \right]_{p,q}  \left( \mathcal{M}_{\ell >} \right)_{q} \right]}_{\displaystyle \mathrm{Interference} \hspace*{2pt} \mathrm{term:} \hspace*{2pt} \mathrm{can} \hspace*{2pt} \mathrm{fix} \hspace*{2pt} \mathrm{the} \hspace*{2pt} \mathrm{phase}} \mathrm{.} \label{eq:PhaseFixingInterferenceTermCalculation}
\end{align}
\end{small}
\noindent
This method works only under the conditions that first of all, the known and fixed multipoles are relatively large, and second that the data are precise enough to resolve the interference. Thereby they can then also determine the overall phase \cite{Grushin, WorkmanGrushinFits}. \newline
For the sake of completeness, we mention here also the possibility of introducing model-dependence via penalty terms, binding the single-energy fit to a particular energy-dependent model, to the chisquare (\ref{eq:CorrelatedChisquare}) \cite{SchumannEtAl}. This is also elaborated more at the end of appendix \ref{subsec:AccidentalAmbProofsIII}. Although this method is certainly also a possibility to introduce some information on the phase, we feel that it introduces quite a strong model-dependence. Since the fits in this work are preformed more in the spirit of extracting amplitudes from complete experiments as model-independently as possible, we do not pursue it further. \newline

We close this section with a discussion of some of the possible benefits of the two-step fit-method (\ref{eq:FitWayOverLegCoeffsIntermediateStep}) as opposed to the direct fit ((\ref{eq:ChiSquareDirectFit}) $\&$ (\ref{eq:FitFunctionDirectFit})), which were however both employed in this work. \newline
Step I, the extraction of Legendre coefficients from data of profile functions, is still quite simple and can be implemented easily in any standard fit-routine. Furthermore, since it consists just of linear polynomial-fits, it is generally not plagued with ambiguities. The Legendre coefficients $\left(a_{L}^{\mathrm{Fit}}\right)_{i}$ can therefore be considered as an equivalent representation of the data in an irreducible form. \newline
Chapter \ref{chap:LFits} has also exemplified the usefulness of such Legendre-fits to obtain a first guess of the order $L$ of multipoles relevant in a full TPWA-fit. This first guess can however still be too low due to interference effects caused by off-diagonal blocks similar to those indicated in equation (\ref{eq:FitMatrixDecomposition}) and (\ref{eq:PhaseFixingInterferenceTermCalculation}). More details can be found in chapter \ref{chap:LFits}. \newline
Fit step II represents the much more difficult part of the analysis. All the possible difficulties due to discrete ambiguities of the bilinear equation systems
\begin{equation}
 \left(a_{L}^{\mathrm{Fit}}\right)_{i} = \left< \mathcal{M}_{\ell}^{C} \right| \left( \mathcal{C}_{L} \right)_{i} \left| \mathcal{M}_{\ell}^{C} \right> \mathrm{,} \label{eq:BilEqSystemChap4ConstrMultipoles}
\end{equation}
which were discussed at length in chaper \ref{chap:Omelaenko}, enter in this second stage of the analysis. The implementation of the minimized chisquare-functions is also generally not so easily done. The respective functions have to be programmed by hand, then passed to a minimization routine. \newline
In case Legendre-coefficients are extracted first in the additional step, one obtains the opportunity to study ambiguities of certain bilinear forms more 'selectively' in case one so desires. This possibility does not exist in the direct fit to the data. However, the two-step method does not bring an advantage in computational speed, which seems at first counterintuitive since the minimized function in the direct fit ((\ref{eq:ChiSquareDirectFit}) $\&$ (\ref{eq:FitFunctionDirectFit})) is a lot bigger than the correlated chisquare (\ref{eq:CorrelatedChisquare}) of fit step $\mathrm{II}$. \newline

In regard to the results from chapter \ref{chap:Omelaenko}, one can anticipate the chisquare functions in fit step II to have, especially for higher orders in $L$, a large number of local minima. This may even be the case when a well-separated global minimum exists. The question arises how to search for all the relevant minima in a model-independent way. This will be the subject of the next section. \newline
In a fit to data, one can expect the multipole-parameters to be extractable only with a finite statistical accuracy. Furthermore, as discussed in appendix \ref{subsec:AccidentalAmbProofsIII}, data with finite precision are generally expected to cause more problems with ambiguities. A computationally intensive method to both obtain a robust error estimate for the multipoles as well as to check for ambiguities will be the subject of section \ref{sec:BootstrappingIntroduction}. \newline
The largest part of the remainder of this chapter has been written under the paradigm of Grushin's method of two-step fitting (\ref{eq:FitWayOverLegCoeffsIntermediateStep}). The direct fit to data is invoked explicitly only in section \ref{subsec:DeltaRegionDataFits}. But again, we have to stress that in all cases where both methods have been compared, they did yield equivalent results.

\clearpage

\subsection{Optimization method by Monte Carlo sampling of multipole spaces} \label{sec:MonteCarloSampling}

We consider now the general problem of the minimization of the functions $\Phi (\mathcal{M}_{\ell}^{C})$ and $\chi^{2} (\mathcal{M}_{\ell})$ defined in fit step II of the previous section\footnote{Or alternatively, the minimization of $\chi^{2}_{\mathrm{data}} (\mathcal{M}_{\ell})$ for a direct fit to the data, see equations (\ref{eq:ChiSquareDirectFit}) and (\ref{eq:FitFunctionDirectFit}) of section \ref{sec:TPWAFitsIntro}.}, while the real- and imaginary parts of the phase constrained multipoles $(\mathcal{M}_{\ell}^{C}) = (\left\{E_{\ell \pm}, M_{\ell \pm}\right\})$ are varied. \newline
We use the MATHEMATICA-routine \textit{FindMinimum} \cite{Mathematica8,Mathematica11,MathematicaLanguage,MathematicaBonnLicense}. This method, like a large class of alternative minimization routines, requires as input a vector that specifies the covariantes as well as their start-parameters, which is in this case
\begin{equation}
 \left\{ \left( \mathrm{Re} E_{0+}^{C}, \left[ \mathrm{Re} E_{0+}^{C} \right]^{0} \right), \left( \mathrm{Re} E_{1+}^{C}, \left[ \mathrm{Re} E_{1+}^{C} \right]^{0} \right), \ldots \left( \mathrm{Im} M_{L-}^{C}, \left[ \mathrm{Im} M_{L-}^{C} \right]^{0} \right) \right\} \mathrm{.} \label{eq:FitInputVector}
\end{equation}
A subscript ``0'' indicates a start-parameter here. The issue is now how to choose these numerical start-configurations. \newline
Many groups, like MAID \cite{SchumannEtAl} or SAID \cite{WorkmanGrushinFits}, publish fits that use multipoles from an energy-dependent (ED) model fit as start-parameters. This procedure is acceptable if one wants to test the ED model with the single-energy fit, thereby ``refining'' the multipoles obtained in the ED model. There could be structures in the data of a single channel that an ED model, especially for a coupled-channel fit, may not reproduce accurately or be insensitive to. \newline

In this thesis however, the goal is to extract multipoles from data in a way that is maximally model-independent. Then, introducing a dependence on a particular model from the beginning is no meaningful practice. \newline
Instead, we adopt here the idea used in the work of Sandorfi, Hoblit, Kamano and Lee \cite{Sandorfi:2010uv}. There, the parameter space of the multipoles is sampled by random numbers, followed by minimizations of $\chi^{2}$ using each of the sampled configurations as start-parameters. In this way, a \textit{pool} of start-parameters is generated, leading then to a \textit{pool of solutions}. These solutions then ideally contain the global minimum as well as all relevant local minima. \newline
The only open question that remains is how to pre-constrain the relevant part of the multipole parameter-space prior to sampling. Clearly, the interval $\left[ - \infty, \infty \right]$ for each real- and imaginary part is non-acceptable. \newline
In this work, a geometrical method utilizing the total cross section is employed, which will be outlined in the following. Originally, it was believed that this was precisely the method used by Sandorfi \textit{et al.}, but a private inquiry \cite{SandorfiPrivateComm} clarified that this is not the case. In the literature that was investigated for this work, the following method did not appear. This should however not be viewed as an exclusion of the fact that some reference may already exist that uses the method. \newline
As quoted in section \ref{subsec:PhotoproductionObs}, the total cross section is written in terms of multipoles as
\begin{align}
 \bar{\sigma} (W) &:= \int d \Omega \hspace*{1.75pt} \sigma_{0} \left( W, \theta \right) = 2 \pi \int_{-1}^{+1} d \cos \theta \hspace*{1.75pt} \sigma_{0} \left( W, \cos \theta \right) \nonumber \\
  &= 2 \pi \frac{q}{k} \sum_{\ell = 0}^{\infty} \Big\{ (\ell + 1)^{2} (\ell + 2) \left| E_{\ell+} \right|^{2} + (\ell - 1) \ell^{2} \left| E_{\ell -} \right|^{2}    \nonumber \\
 & \hspace*{62.5pt} + \ell (\ell + 1)^{2} \left| M_{\ell+} \right|^{2} + \ell^{2} (\ell + 1) \left| M_{\ell-} \right|^{2} \Big\} \mathrm{.} \label{eq:TCSInTermsOfMultsLZeroExpansion}
\end{align}
In a TPWA, one uses the version of this expression truncated at some suitable $L$ and assumes that it is a good approximation for the full infinite sum. This approximation can always be assumed to be good, provided one truncates at an $L$ that is motivated by fits to polarization observables. This is true since $\bar{\sigma}$ does not contain any interference terms. \newline
The fact that the total cross section is strictly a sum of moduli squared of multipoles can be exploited in two ways. To exemplify the method, we write (\ref{eq:TCSInTermsOfMultsLZeroExpansion}) for a truncation at $L = 1$ and furthermore directly in terms real- and imaginary parts of phase-constrained multipoles
\allowdisplaybreaks
\begin{align}
 \bar{\sigma} (W) &\approx 4 \pi \frac{q}{k} \Big( \mathrm{Re}\left[E^{C}_{0+}\right]^{2} + 6 \hspace*{1.5pt} \mathrm{Re}\left[E^{C}_{1+}\right]^{2} + 6 \hspace*{1.5pt} \mathrm{Im}\left[E^{C}_{1+}\right]^{2} + 2 \hspace*{1.5pt} \mathrm{Re}\left[M^{C}_{1+}\right]^{2} \nonumber \\
 & \hspace*{38pt} + 2 \hspace*{1.5pt} \mathrm{Im}\left[M^{C}_{1+}\right]^{2} + \mathrm{Re}\left[M^{C}_{1-}\right]^{2} + \mathrm{Im}\left[M^{C}_{1-}\right]^{2}  \Big) \mathrm{.} \label{eq:TCSTruncatedAtL1PhaseConstrainedMults}
\end{align}
The mathematical form of this expression can now be exploited in two ways:
\begin{itemize}
 \item[(i)] The quantity $\bar{\sigma} (W)$ constrains parameter-intervals for the multipoles. To see this, one can first consider the hypothetic case that all parameters of the phase-constrained multipoles vanish except for $\mathrm{Re}\left[E^{C}_{0+}\right]$. Then equation (\ref{eq:TCSTruncatedAtL1PhaseConstrainedMults}) becomes (now written as a strict equality, disregarding the small approximation-error)
 \begin{equation}
  \bar{\sigma} (W) = 4 \pi \frac{q}{k} \mathrm{Re}\left[E^{C}_{0+}\right]^{2} \mathrm{,} \label{eq:MultIntDerivationPart1}
 \end{equation}
 which, remembering that our phase-convention (\ref{eq:PhaseConstraintMultipoles}) forces the $S$-wave to have a positive real part, is equal to
 \begin{equation}
  \mathrm{Re}\left[E^{C}_{0+}\right] = \sqrt{\frac{k}{q} \frac{\bar{\sigma} (W)}{4 \pi}} \mathrm{.} \label{eq:MultIntDerivationPart2}
 \end{equation}
 Now, in case the contribution from the remaining multipoles on the right hand side of (\ref{eq:TCSTruncatedAtL1PhaseConstrainedMults}) is non-vanishing but equal to a finite remainder, here denoted as $\bar{\sigma}_{r}$, then it can be seen that the real part of the phase-constrained $S$-wave becomes
 \begin{equation}
  \mathrm{Re}\left[E^{C}_{0+}\right] = \sqrt{\frac{k}{q} \frac{\left[\bar{\sigma} (W) - \bar{\sigma}_{r} (W)\right]}{4 \pi}} < \sqrt{\frac{k}{q} \frac{\bar{\sigma} (W)}{4 \pi}} \mathrm{.} \label{eq:MultIntDerivationPart3}
 \end{equation}
 Equivalently, the parameter $\mathrm{Re}\left[E^{C}_{0+}\right]$ is seen to be confined to an interval:
 \begin{equation}
 \mathrm{Re}\left[E^{C}_{0+}\right] \in \left[ 0, \sqrt{\frac{k}{q} \frac{\bar{\sigma} (W)}{4 \pi}} \right] \mathrm{.} \label{eq:SWaveParInt}
 \end{equation}
 In an analogous way, parameter intervals can be derived for the remaining multipoles, where it has to be remembered that since the phase-convention (\ref{eq:PhaseConstraintMultipoles}) does not require them to be positive, the possibility of a negative sign for the square-root can not be disregarded. For the truncation at $L=1$, the resulting parameter-intervals for the fit parameters are
 \allowdisplaybreaks
 \begin{align}
  \mathrm{Re}\left[E^{C}_{1+}\right] \in \left[ - \sqrt{\frac{k}{q} \frac{\bar{\sigma} }{24 \pi}} , \sqrt{\frac{k}{q} \frac{\bar{\sigma} }{24 \pi}} \right] &\mathrm{,} \hspace*{3pt} \mathrm{Im}\left[E^{C}_{1+}\right] \in \left[ - \sqrt{\frac{k}{q} \frac{\bar{\sigma} }{24 \pi}} , \sqrt{\frac{k}{q} \frac{\bar{\sigma} }{24 \pi}} \right] \mathrm{,} \label{eq:MultParIntsLmax1E1Plus} \\
  \mathrm{Re}\left[M^{C}_{1+}\right] \in \left[ - \sqrt{\frac{k}{q} \frac{\bar{\sigma} }{8 \pi}} , \sqrt{\frac{k}{q} \frac{\bar{\sigma} }{8 \pi}} \right] &\mathrm{,} \hspace*{3pt} \mathrm{Im}\left[M^{C}_{1+}\right] \in \left[ - \sqrt{\frac{k}{q} \frac{\bar{\sigma} }{8 \pi}} , \sqrt{\frac{k}{q} \frac{\bar{\sigma} }{8 \pi}} \right] \mathrm{,} \label{eq:MultParIntsLmax1M1Plus} \\
  \mathrm{Re}\left[M^{C}_{1-}\right] \in \left[ - \sqrt{\frac{k}{q} \frac{\bar{\sigma} }{4 \pi}} , \sqrt{\frac{k}{q} \frac{\bar{\sigma} }{4 \pi}} \right] &\mathrm{,} \hspace*{3pt} \mathrm{Im}\left[M^{C}_{1-}\right] \in \left[ - \sqrt{\frac{k}{q} \frac{\bar{\sigma} }{4 \pi}} , \sqrt{\frac{k}{q} \frac{\bar{\sigma} }{4 \pi}} \right] \mathrm{.} \label{eq:MultParIntsLmax1M1Minus}
 \end{align}
 It is therefore seen that the total cross section constrains the multipole solution to lie in a $7$-dimensional hypercube for $L = 1$, which is at each energy specified by the single number $\bar{\sigma}$. The procedure described here is seen to be directly generalizable to any finite truncation order $L$, where the respective intervals for each multipole can be read of from the general expansion (\ref{eq:TCSInTermsOfMultsLZeroExpansion}). \newline
 Therefore, the total cross section $\bar{\sigma}$ generally confines the parameters of the model-independent TPWA to an $(8L - 1)$-dimensional hypercube.
 \item[(ii)] The total cross section (\ref{eq:TCSInTermsOfMultsLZeroExpansion}) truncated at some $L$ furthermore has the mathematical form of a constraint that defines an $(8L - 2)$-dimensional ellipsoid in the $(8L - 1)$-dimensional hypercube of the multipole-parameters. To see this, it is useful to recall the basic mathematical definition of an $(n-1)$-dimensional sphere in an $n$-dimensional real embedding space \cite{ForsterIII}. The basic example is the $1$-dimensional circle, or $1$-sphere $S_{1}$. In the $2$-dimensional plane spanned by the vectors $(x,y)^{T}$, this point set is defined by all the values of $x$ and $y$ that fulfill the constraint
 \begin{equation}
  x^{2} + y^{2} = 1 \mathrm{.} \label{eq:CircleEquation}
 \end{equation}
 The $(n-1)$-sphere $S_{(n-1)}$, defined as a sub-manifold embedded in the $n$-dimensional real parameter space $\mathbbm{R}^{n}$, is defined analogously by the equation
 \begin{equation}
  x_{1}^{2} + x_{2}^{2} + \ldots + x_{n}^{2} = 1 \mathrm{,} \label{eq:Nminus1SphereEquation}
 \end{equation}
 which has to hold for every point lying on the sub-manifold. It can now be seen that equation (\ref{eq:TCSTruncatedAtL1PhaseConstrainedMults}) defines a $6$-sphere in the space of suitably re-scaled phase-constrained multipoles. Upon rescaling the $\mathcal{M}_{\ell}$ according to
 \allowdisplaybreaks
 \begin{align}
  E^{C}_{\ell+} \rightarrow \sqrt{\frac{k}{q} \frac{\bar{\sigma}}{2 \pi (\ell + 1)^{2} (\ell + 2)}} E^{C}_{\ell+} &\mathrm{,} \hspace*{5pt} E^{C}_{\ell-} \rightarrow \sqrt{\frac{k}{q} \frac{\bar{\sigma}}{2 \pi (\ell - 1) \ell^{2}}} E^{C}_{\ell-} \mathrm{,} \label{eq:E^{C}M^{C}ultsRescaling} \\
  M^{C}_{\ell+} \rightarrow \sqrt{\frac{k}{q} \frac{\bar{\sigma}}{2 \pi \ell (\ell + 1)^{2}}} M^{C}_{\ell+} &\mathrm{,} \hspace*{5pt} M^{C}_{\ell-} \rightarrow \sqrt{\frac{k}{q} \frac{\bar{\sigma}}{2 \pi \ell^{2} (\ell + 1)}} M^{C}_{\ell-} \mathrm{,} \label{eq:MultsRescaling} 
 \end{align}
 it can be verified that the defining equation for $\bar{\sigma}$ (\ref{eq:TCSTruncatedAtL1PhaseConstrainedMults}) reproduces equation (\ref{eq:Nminus1SphereEquation}) for $n = 7$, written now in terms of the re-scaled multipoles. Since an $(n-1)$-sphere and an $(n-1)$-ellipsoid are generally equivalent up to an independent re-scaling of the coordinate axes \cite{ForsterIII}, we see by an inversion of (\ref{eq:MultsRescaling}) that actually the total cross section $\bar{\sigma}$ defines a $6$-ellipsoid in the parameter space of the phase-constrained multipoles. \newline
 The generalization to arbitrary $L$ is straightforward, illustrating the fact that indeed the total cross section defines an $(8 L - 2)$-ellipsoid.
\end{itemize}
We now want to exploit the two points elaborated above for the sampling procedure. Since our scheme then needs the number $\bar{\sigma} (W)$ as input at every energy where it is applied, it is not a prior sampling that does not need any information from the data, but instead develops a \textit{half-posterior} character. The advantage is that the total cross section is of course easily extracted from the data. In case a good distinct measurement for $\bar{\sigma}$ already has been done, one can just use this number. If not, the total cross section is most easily extracted from a truncated Legendre-fit to the angular distribution of the unpolarized differential cross section $\sigma_{0}$. Up to a pre-factor, $\bar{\sigma}$ is equal to the zeroth coefficient $\left(a_{L}\right)_{0}^{\sigma_{0}}$. To see this, one can just compare some finite truncations of equation (\ref{eq:TCSInTermsOfMultsLZeroExpansion}) with the matrix defining $\left(a_{L}\right)_{0}^{\sigma_{0}}$ as a hermitean form. The latter is printed for $L=5$ in appendix \ref{sec:TPWAFormulae}. \newline
Instead, one can just use the fact that $P_{0} (x) = 1$ and therefore do the well-known proof
\begin{equation}
 \int_{-1}^{1} d x \hspace*{2pt} P_{\ell} (x) = \int_{-1}^{1} d x \hspace*{2pt} P_{\ell} (x) \hspace*{1.75pt} P_{0} (x) = 2 \delta_{\ell 0} \mathrm{.} \label{eq:LegPolyIntKroneckerDelta}
\end{equation}
With this, the above mentioned connection is quickly established
\begin{align}
 \bar{\sigma} &= 2 \pi \int_{-1}^{1} d \cos \theta \hspace*{2pt} \sigma_{0} = 2 \pi \frac{q}{k} \int_{-1}^{1} d \cos \theta \hspace*{2pt} \sum_{n = 0}^{2L} \left(a_{L}\right)_{n}^{\sigma_{0}} P_{n} (\cos \theta) \nonumber \\
 &= 2 \pi \frac{q}{k} \sum_{n = 0}^{2L} \left(a_{L}\right)_{n}^{\sigma_{0}} \int_{-1}^{1} d \cos \theta \hspace*{2pt} P_{n} (\cos \theta) = 4 \pi \frac{q}{k} \left(a_{L}\right)_{0}^{\sigma_{0}} \mathrm{.} \label{eq:TCSCoeffA0Derivation}
\end{align}
Having extracted $\bar{\sigma}$, there is already considerable knowledge about the relevant part of the multipole parameter space available at each energy. We now describe the sampling method used in this work. It is illustrated in Figure \ref{fig:MultSpaceSamplingMethod}. \newline

The following search method is employed at a given energy.

\begin{itemize}
 \item[1.)] Extract $\bar{\sigma} (W)$ from the data.
 \item[2.)] Use Monte Carlo methods to generate a pool consisting of $N_{MC}$ start-con\-fi\-gu\-ra\-tions lying on the $(8L-2)$-ellipsoid defined by $\bar{\sigma}$. The result of this step is a set of vectors (cf. equation (\ref{eq:FitInputVector}))
 \begin{equation}
  \left( \left[ \mathrm{Re} E_{0+}^{C} \right]^{0}_{i}, \left[ \mathrm{Re} E_{1+}^{C} \right]^{0}_{i}, \ldots , \left[ \mathrm{Im} M_{L-}^{C} \right]^{0}_{i} \right) \mathrm{,} \hspace*{2pt} \mathrm{for} \hspace*{2pt} i = 1,\ldots,N_{MC} \mathrm{.} \label{eq:MCPointsVector}
 \end{equation}
 A suitable choice of the number $N_{MC}$ is expected to depend heavily on the truncation order $L$. A good choice consists in part of intuition about ambiguities and in part of guesswork. More details on this problem are given below. \newline
 Furthermore, there exists an infinity of possible algorithms to generate a set of random numbers that fulfill the $\bar{\sigma}$-constraint. A particularly simple and efficient one, which was chosen for the results shown in this thesis, is described in appendix \ref{sec:MCSamplingAlgorithms}. We however note that the algorithm chosen here does not generate a distribution of points that is flat, or uniform, on the $\bar{\sigma}$-ellipsoid.
 \item[3.)] Run a \textit{FindMinimum}-minimization of $\Phi_{\mathcal{M}}$ or $\chi^{2}_{\mathcal{M}}$ (or alternatively $\chi^{2}_{\mathrm{data}}$, cf. section \ref{sec:TPWAFitsIntro}) for each element in the pool of start-con\-fi\-gu\-ra\-tions. \newline
 This leads, in the ideal case that each minimization converges without issues, to a \textit{solution-pool} consisting of values for parameters in the minimum as well as the minimum of the function itself, i.e.
 \begin{equation}
  \left( \mathcal{M}_{\ell}^{C} \right)_{i} \hspace*{2pt} \mathrm{and} \hspace*{2pt} \left(\Phi_{\mathcal{M}}\right)_{i} \hspace*{2pt} \mathrm{/} \hspace*{2pt} \left(\chi^{2}_{\mathcal{M}}\right)_{i} \hspace*{2pt} \mathrm{for} \hspace*{2pt} i = 1,\ldots,N_{MC} \mathrm{.} \label{eq:SolutionPool}
 \end{equation}
\begin{figure}[ht]
 \centering
 \vspace*{-7.5pt}
 \hspace*{15pt} \begin{overpic}[width=0.725\textwidth]{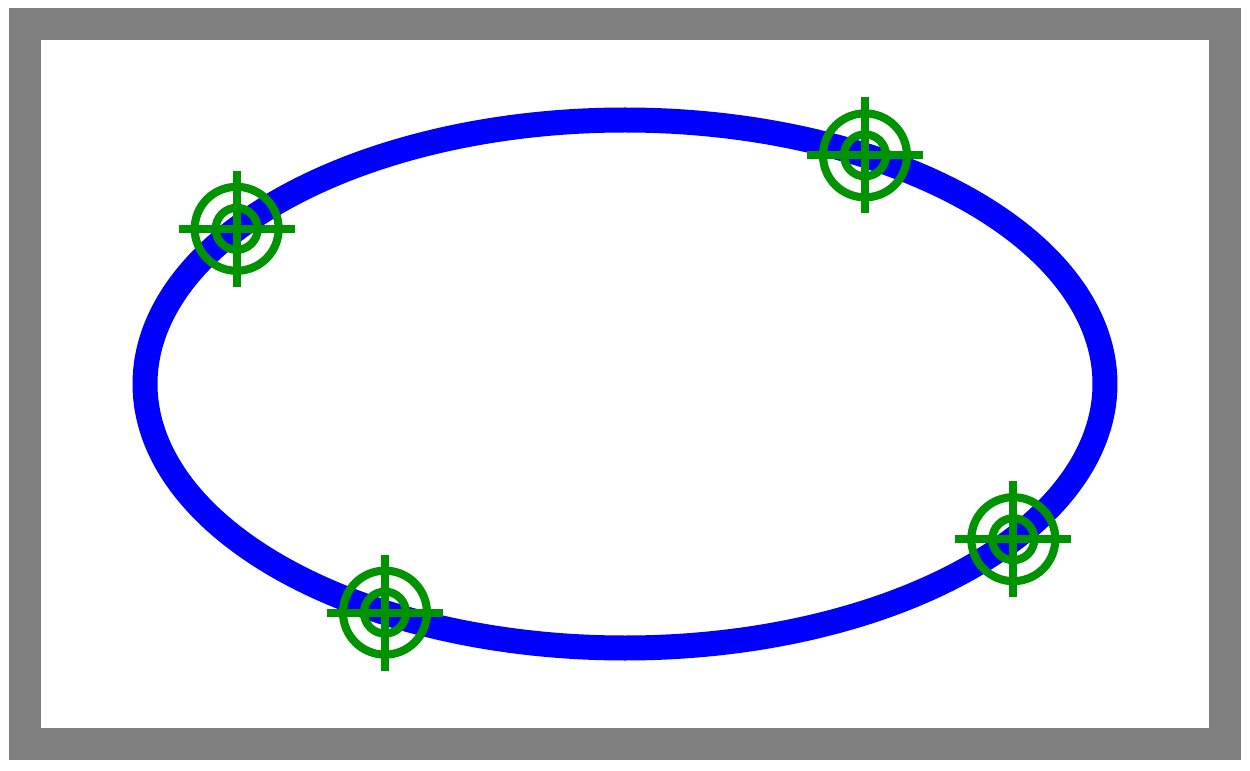}
\put(-10,30.3){\begin{Large}1.)\end{Large}}
 \end{overpic} \\
 \hspace*{15pt} \begin{overpic}[width=0.725\textwidth]{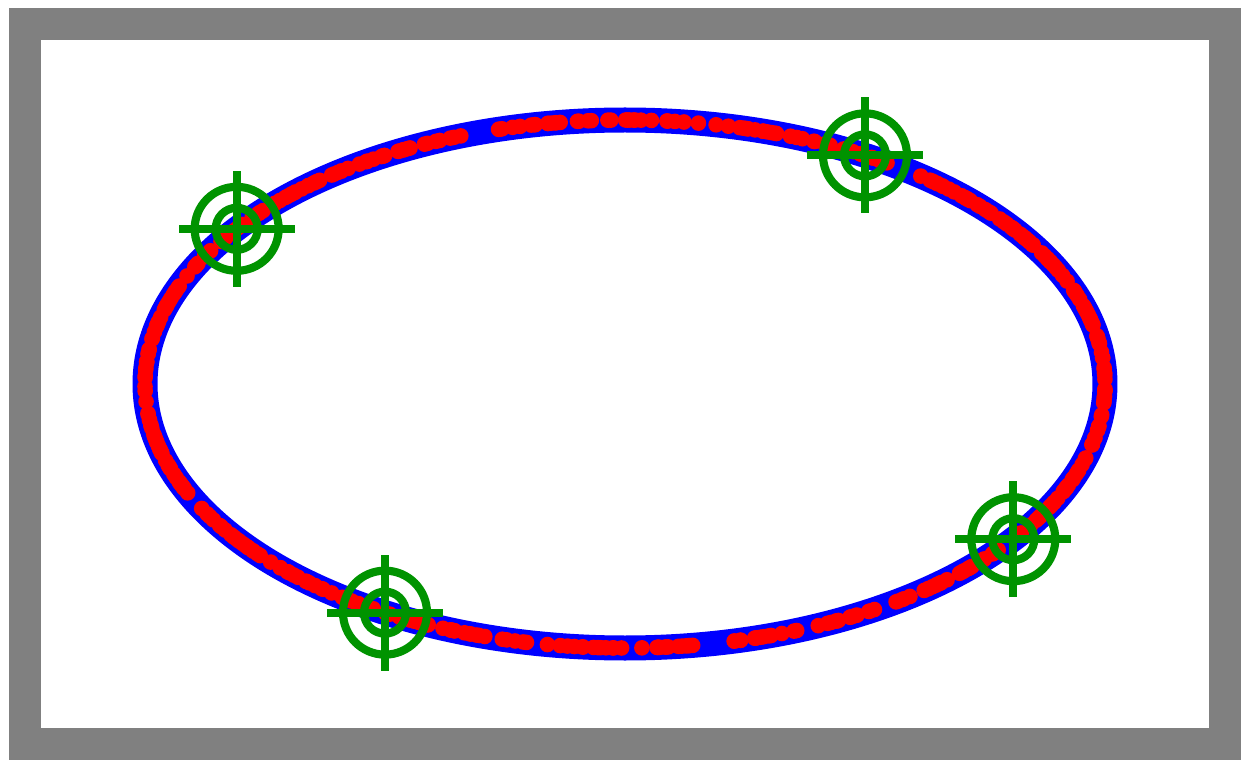}
\put(-10,30.3){\begin{Large}2.)\end{Large}}
 \end{overpic} \\
 \hspace*{15pt} \begin{overpic}[width=0.725\textwidth]{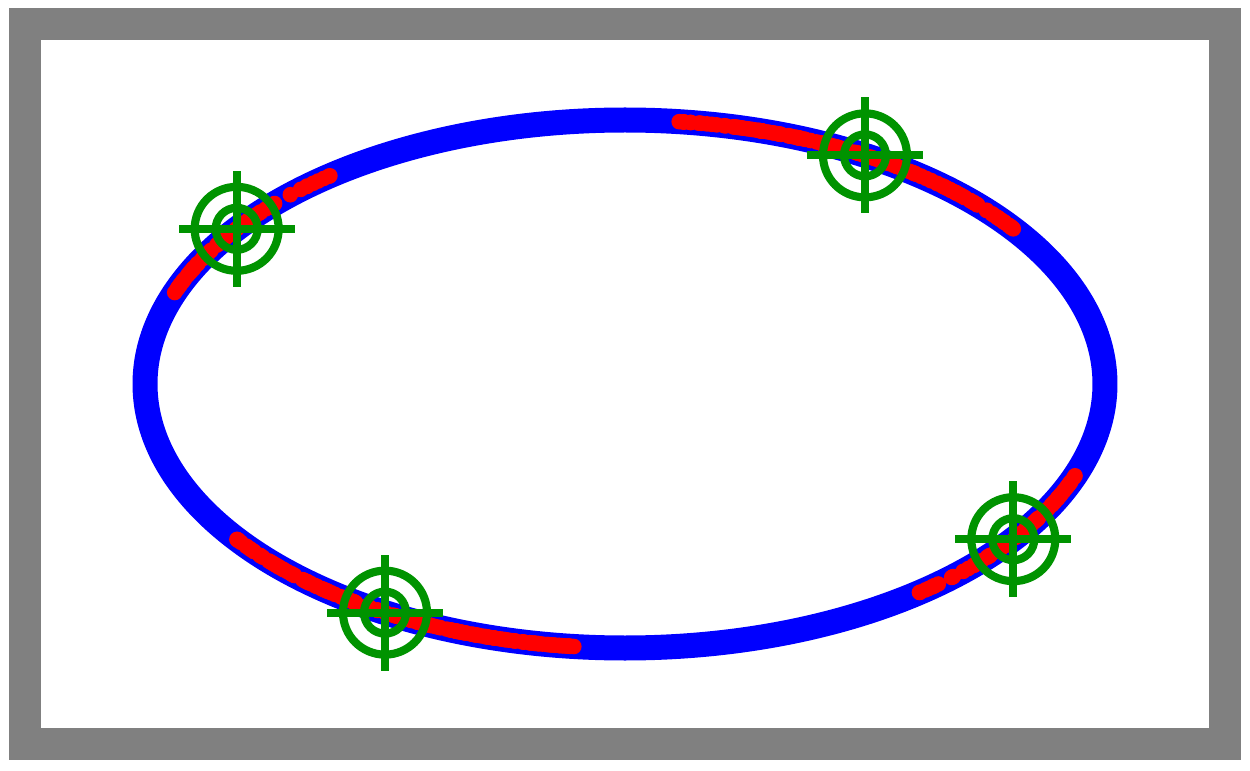}
\put(-10,30.3){\begin{Large}3.)\end{Large}}
 \end{overpic} \\
 \caption[Schematics illustrating the Monte Carlo search method for a fully model-independent TPWA.]{Three schematics are shown to illustrate the Monte Carlo search method outlined in the main text. The box indicates the $(8 L - 1)$-hypercube defined by the total cross section $\bar{\sigma}$, while the corresponding ellipsoid is drawn inside the box. \newline Upon extraction of the total cross section from the data, step 1.), it is only known that the solutions to the TPWA problem (crosshairs) have to lie somewhere on the ellipsoid. Then, in step 2.), the relevant part of the parameter-space is sampled with randomly chosen start-con\-fi\-gu\-ra\-tions. During the minimizations performed in step 3.), each sampled set of start-parameters should ideally converge towards one of the solutions.}
 \label{fig:MultSpaceSamplingMethod}
\end{figure}

\clearpage

 \item[4.)] Store and compare the solutions in the pool with respect to resulting $\Phi_{\mathcal{M}}$ or $\chi^{2}_{\mathcal{M}}$. The smallest value of the minimized function in the pool, denoted here as
 \begin{equation}
 \left(\Phi_{\mathcal{M}}\right)_{j}^{\mathrm{best}} \hspace*{2pt} \mathrm{/} \hspace*{2pt} \left(\chi^{2}_{\mathcal{M}}\right)_{j}^{\mathrm{best}} \mathrm{,} \label{eq:ChiSquareBest}
 \end{equation}
 then defines the global minimum. This will then inevitably yield the global minimum for the given $N_{MC}$-pool, which need of course not be the true global minimum. \newline
 Also, mathematical ambiguities, i.e. configurations of different resulting fit-parameters $\left( \mathcal{M}_{\ell}^{C} \right)_{i}$ that nonetheless yield the exact same value for the minimum up to a small numerical error, may exist. This case is to be expected for example, when the observables $\left\{\sigma_{0}, \check{\Sigma}, \check{T}, \check{P}\right\}$ are fitted (cf. chapter \ref{chap:Omelaenko}). Nonetheless, the appearance of mathematical ambiguities should always be checked by hand. \newline
 Another problem consists of the fact that a numerical minimization, even if two start-con\-fi\-gu\-ra\-tions run into a minimum that is one and the same mathematically, will only yield the same numbers for the fit-parameters up to the chosen numerical precision. In other words, solutions will have a tiny scatter around con\-fi\-gu\-ra\-tions that are mathematically indifferent. In case one wishes to purge the solution-pool of these redundant solutions, a suitable clustering-algorithm should be applied. Such algorithms are already pre-implemented in MATHEMATICA in the method \textit{FindClusters} \cite{MathematicaLanguage}. However, we also quote a particularly simple version of an algorithm that tries to sort out the non-redundant solutions, using the idea of a pre-adjusted \textit{$\epsilon$-ball} around any mathematical solution, in appendix \ref{sec:MCSamplingAlgorithms}. Both versions have been found to work in the course of this thesis. However, they should be trusted only in situations where the structure of values in the solution-pool is reasonably well-behaved. If in doubt, we omit the usage of a clustering-algorithm and keep the full solution-pool, since there is always the danger of throwing away non-redundant solutions. \newline
 Finally, all the non-redundant solutions found in the pool can be compared and selected with respect to the global pool-minimum. This can be done via well-known rules of thumb such as to keep all solutions in the range $\chi^{2}_{\mathrm{best}} / \mathrm{ndf} + 1$. Also, one can consider the actual probability-theoretical $\chi^{2}$-distributions \cite{BlobelLohrmann} for the given degrees of freedom and make decisions based on the grounds of it. For the fit step II as defined in section \ref{sec:TPWAFitsIntro}, the number of degrees of freedom is estimated as
 \begin{equation}
 \mathrm{ndf} = N_{a^{\alpha}_{k}} - (8 L - 1) \mathrm{,} \label{eq:NDOFFitStep2}
 \end{equation}
 with $N_{a^{\alpha}_{k}}$ being the number of Legendre coefficients provided by the particular set of fitted observables, in the employed truncation order $L$. In case the data are fitted directly, without division into steps $\mathrm{I}$ and $\mathrm{II}$, we estimate $\mathrm{ndf}$ as
 \begin{equation}
 \mathrm{ndf} = N_{\mathrm{data}} - (8 L - 1) \mathrm{.} \label{eq:NDOFDirectFit}
 \end{equation}
 Here, $N_{\mathrm{data}}$ is just the total number of datapoints from all observables included into the fit.
\end{itemize}
We now come to a central issue in regard to the Monte Carlo search-method proposed here, which is the question about how one can be sure to first of all have the correct global minimum as the global pool-minimum and furthermore, if one has really mapped out all the relevant local minima of the $\chi^{2}$-function. The intuitive answer would be that one can be sure enough, provided $N_{MC}$ has been chosen \textit{large enough}. But how large is this exactly? \newline
A good intuition for the appropriate number $N_{MC}$ can be obtained by a consideration of the upper bound for possible accidental ambiguities in the TPWA problem, derived in chapter \ref{chap:Omelaenko} and with more details in appendix \ref{subsec:AccidentalAmbProofsI}. Here, we base our suggestions for the number of sampling points on the number $N_{\mathrm{AC}}^{\mathrm{total}}$ of maximally possible distinct multipole solutions originating from accidental symmetries, which is quoted in equation (\ref{eq:AccAmbCountingTotalN}) of appendix \ref{subsec:AccidentalAmbProofsI} as $N_{\mathrm{AC}}^{\mathrm{total}}=2^{4 L} - 2$. Our estimates are summarized in Table \ref{tab:AccAmbPossibilityNumberMotivatesNMonteCarlo}. \newline
Provided are suggested numbers $N_{MC}$ for the lowest truncation orders $L$ ranging from $1$ to $6$. Also, the dimensions of the corresponding parameter-spaces $\mathcal{M}_{\ell}^{C}$ as well as the values for $N_{\mathrm{AC}}^{\mathrm{total}}$ are listed. The estimates for $N_{MC}$ were based in particular on the considerations of appendix \ref{subsec:AccidentalAmbProofsIII}, where it was found that for $L = 1$, only a low fraction of $2$ to $5$ $\%$ of all accidental ambiguities was expected to be dangerous. Provided that this behaviour carries over to the higher $L$, a hypothesis for which no proof was found here, the numbers for $N_{MC}$ were chosen such that, on average, every local minimum corresponding to an accidental symmetry is found at least $50$ to $100$ times. \newline
These are just suggestions for $N_{MC}$ and the numbers in Table \ref{tab:AccAmbPossibilityNumberMotivatesNMonteCarlo} were already chosen quite generously. However, in practical analyses, sampling pools of comparable sizes have generally yielded good results. What is particularly startling is that since the ambiguities rise exponentially in number, so should the number of sampling points. Therefore, for higher $L$ one can in our view not avoid sampling millions of points. This makes the suggested search method very costly for fitting higher truncation orders. \newline

Another good question is how one can be sure that the solution-pool resulting from a particular $N_{MC}$ really contains the correct global minimum as well as all relevant local minima. The only good answer we can give here is again to apply brute force, e.g. by re-fitting with $2 \times N_{MC}$ sampling points and comparing the solutions of both pools. In case no new distinct solutions are found, this is of course not a mathematical proof of their non-existence. Still, it would be a very strong signal that all solutions indeed have been mapped out. \newline
Lastly, we would like to mention the fact that the Ansatz for a model independent search-method discussed in this section is not only applicable in a TPWA, but for the numerical solution of the traditional complete experiment \cite{ChTab} as well, this time solving for the full spin amplitudes (cf. section \ref{sec:CompExpsFullAmp}). 
\vfill
\begin{table}[h]
\centering
\begin{tabular}{c|r|r|r}
\hline
\hline
 $L$ & $\mathrm{dim}\left( \mathcal{M}_{\ell}^{C} \right) = (8L - 1)$ &$N_{\mathrm{AC}}^{\mathrm{total}}$ & Suggested $N_{MC}$ \\
\hline
 $1$ & $7$ & $14$ & $100$ - $1\hspace*{1pt}000$ \\
 $2$ & $15$ & $254$ & $1\hspace*{1pt}000$ - $10\hspace*{1pt}000$ \\
 $3$ & $23$ & $4\hspace*{1pt}094$ & $10\hspace*{1pt}000$ - $50\hspace*{1pt}000$ \\
 $4$ & $31$ & $65\hspace*{1pt}534$ & $50\hspace*{1pt}000$ - $100\hspace*{1pt}000$ \\
 $5$ & $39$ & $1\hspace*{1pt}048\hspace*{1pt}574$ & $1\hspace*{1pt}000\hspace*{1pt}000$ - $5\hspace*{1pt}000\hspace*{1pt}000$ \\
 $6$ & $47$ & $16\hspace*{1pt}777\hspace*{1pt}214$ & $20\hspace*{1pt}000\hspace*{1pt}000$ - $50\hspace*{1pt}000\hspace*{1pt}000$ \\
\hline
\hline
\end{tabular}
\caption[Suggested ranges for the start-configurations $N_{MC}$ in the fully model-independent TPWA for the lowest orders $L$.]{This Table shows suggested ranges for the start-configurations $N_{MC}$ in the fully model-independent TPWA for the lowest orders $L$. The estimated values are motivated by the power-law in equation (\ref{eq:AccAmbCountingTotalN}) of appendix \ref{subsec:AccidentalAmbProofsI}, the results of which are also listed for comparison. The dimension of the parameter space of phase-constrained multipoles is displayed as well for every order.}
\label{tab:AccAmbPossibilityNumberMotivatesNMonteCarlo}
\end{table}
\clearpage

If for example the goal is the extraction of helicity amplitudes $\left\{ H_{i} \right\}$ up to an energy- and angle-dependent overall phase $\phi^{H} (W,\theta)$, then the quantity that constrains the amplitude-space in this case is the differential cross section measured at the appropriate energy and angle (see Table \ref{tab:ChTabHelTrObs}, section \ref{subsec:PhotoproductionObs})
\begin{equation}
 \sigma_{0} \left( W, \theta \right) =  \frac{1}{2} \left( \left| H_{1} \right|^{2} + \left| H_{2} \right|^{2} + \left| H_{3} \right|^{2} + \left| H_{4} \right|^{2} \right) \mathrm{.} \label{eq:DCSExprForMotivationOfCompExFitChap4}
\end{equation}
This observable is again strictly a sum of squares and can be utilized to generate Monte Carlo start-con\-fi\-gu\-ra\-tions for the reduced helicity amplitudes $\left\{ \tilde{H}_{i} \right\}$, where $\tilde{H}_{1}$ is required to be purely real (as well as $\tilde{H}_{1} \equiv \mathrm{Re}\left[\tilde{H}_{1}\right] > 0$). Then, fits can be performed utilizing the remaining $7$ carefully chosen observables required by Chiang and Tabakin \cite{ChTab} and the helicity amplitudes recovered up to a phase. \newline
We have tested this procedure also in the course of this thesis for observables obtained from randomly generated helicity amplitudes, and the search method also works in this case. \newline

\clearpage

\subsection{Fitting theoretical model data} \label{sec:TheoryDataFits}

The purpose of this section is to illustrate the numerical search method for minima of the $\Phi$- or $\chi^{2}$- function in a TPWA which was presented in section \ref{sec:MonteCarloSampling}. Furthermore, connections will be drawn between the numerical investigations of this present chapter and the more abstract investigations on discrete ambiguities from chapter \ref{chap:Omelaenko}, which have been put in a more formal and precise shape in appendix \ref{subsec:AccidentalAmbProofsIII}. \newline
Both of these goals can be accomplished by the numerical solution of \textit{theory-data} stemming directly from a model. In this case, synthetic data generated from MAID 2007 \cite{MAID2007,MAID} were provided specifically for this work by the MAID-group \cite{LotharPrivateComm}. Such theoretical data are exact up to a significant number of digits reflecting the numerical precision of the model calculation, in this case around $10^{-10}$. Therefore it is clear that such data are a lot more precise than any dataset that may be extracted from the real world. Studies of theory-data serve first of all a purely academic purpose. \newline
Some of the provided MAID-datasets are even more artificial in the sense that they are \textit{truncated data}. This means that the data are not stemming from the full model, but were generated using MAID multipoles up to a certain order. Truncated theory-data for the orders $\ell_{\mathrm{max}} = 1,2,3$ and $4$ were provided \cite{LotharPrivateComm} and investigated in this work. These data are what, in the context of numerical studies, can be considered closest to the academic case of exactly solvable data in a TPWA, since they do not contain interferences from higher partial waves. They serve as an ideal testing ground for the proposed Monte Carlo search-method as well as the ambiguity statements from chapter \ref{chap:Omelaenko} and appendix \ref{sec:AccidentalAmbProofs}, since the latter were derived also in the context of \textit{exact} truncations at $L = \ell_{\mathrm{max}}$. \newline
Moreover, theory-data are of course also given as extracted from the full MAID model. They formally contain contributions from all higher partial waves for $L \rightarrow \infty$ ($t$-channel diagrams, $\ldots$). Investigations of these data and comparisons to the above mentioned truncated data are well suited to investigate a point of critique described in more detail in appendix \ref{subsec:AccidentalAmbProofsIII}, namely that the truncation itself inherently has an approximation error. \newline
In the following, we will cover the results for the truncated theory-data at $\ell_{\mathrm{max}} = 1$ in quite some detail. This is the simplest possible case and the degree of complexity is still manageable. For truncated theory-data at higher orders, only the most important results are described. Results for the full model theory-data are then again studied in more detail and the (de-) stabilization of TPWA-fits is discussed.

\subsubsection{MAID2007 theory-data up to $\ell_{\mathrm{max}} = 1$} \label{subsec:TheoryDataFitsLmax1}

The theory-data considered in this section have been generated using just the $S$- and $P$- wave multipoles from the MAID2007 analysis for the channel $\gamma p \rightarrow \pi^{0} p$ \cite{MAID2007,MAID}, i.e. explicitly the complex partial waves
\begin{equation}
 \left\{  E_{0+}^{p \pi^{0}}, E_{1+}^{p \pi^{0}}, M_{1+}^{p \pi^{0}}, M_{1-}^{p \pi^{0}} \right\} \mathrm{.} \label{eq:SPWavesMAIDPi0}
\end{equation}
No higher partial waves contribute to these data. Therefore, one can anticipate a mathematically exact solution of the TPWA problem to exist. Such solutions are of course exact only within a finite, but very good, numerical precision. The task is now to fit the multipoles (\ref{eq:SPWavesMAIDPi0}) out of the theoretical data, imposing an overall phase constraint (see equation (\ref{eq:PhaseConstraintMultipoles})) on the respective fit parameters.

\newpage

\begin{figure}[ht]
 \centering
\begin{overpic}[width=0.485\textwidth]{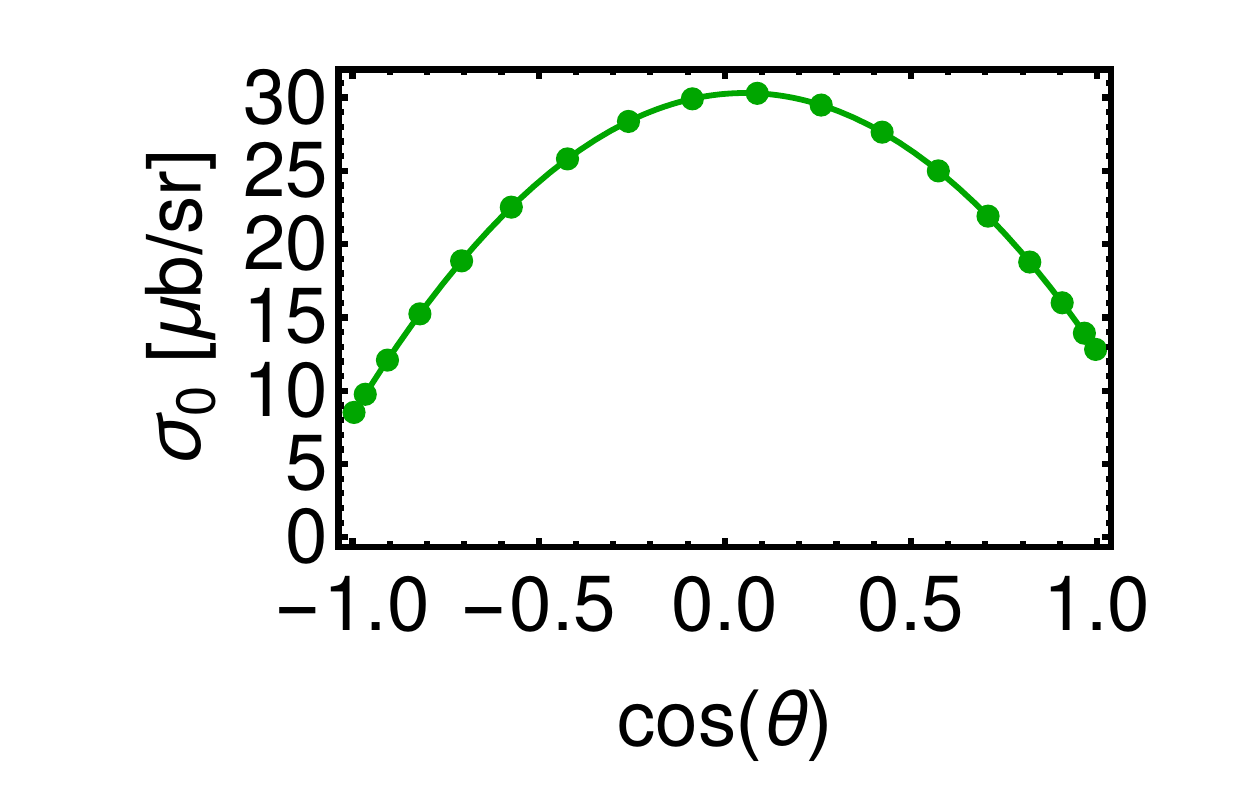}
 \put(85,65){\begin{Large}$E_{\gamma} = 330 \hspace*{2pt} \mathrm{MeV}$\end{Large}}
 \end{overpic}
\begin{overpic}[width=0.485\textwidth]{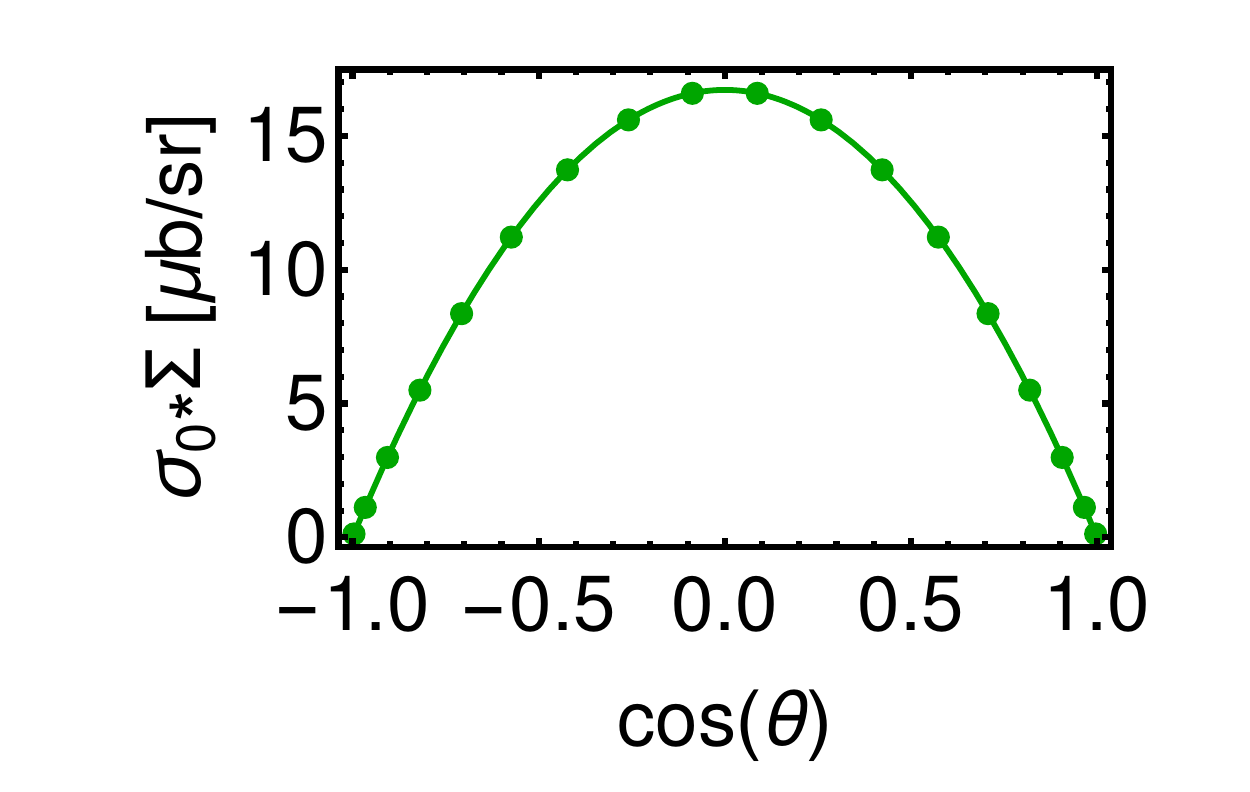}
 \end{overpic} \\
\begin{overpic}[width=0.485\textwidth]{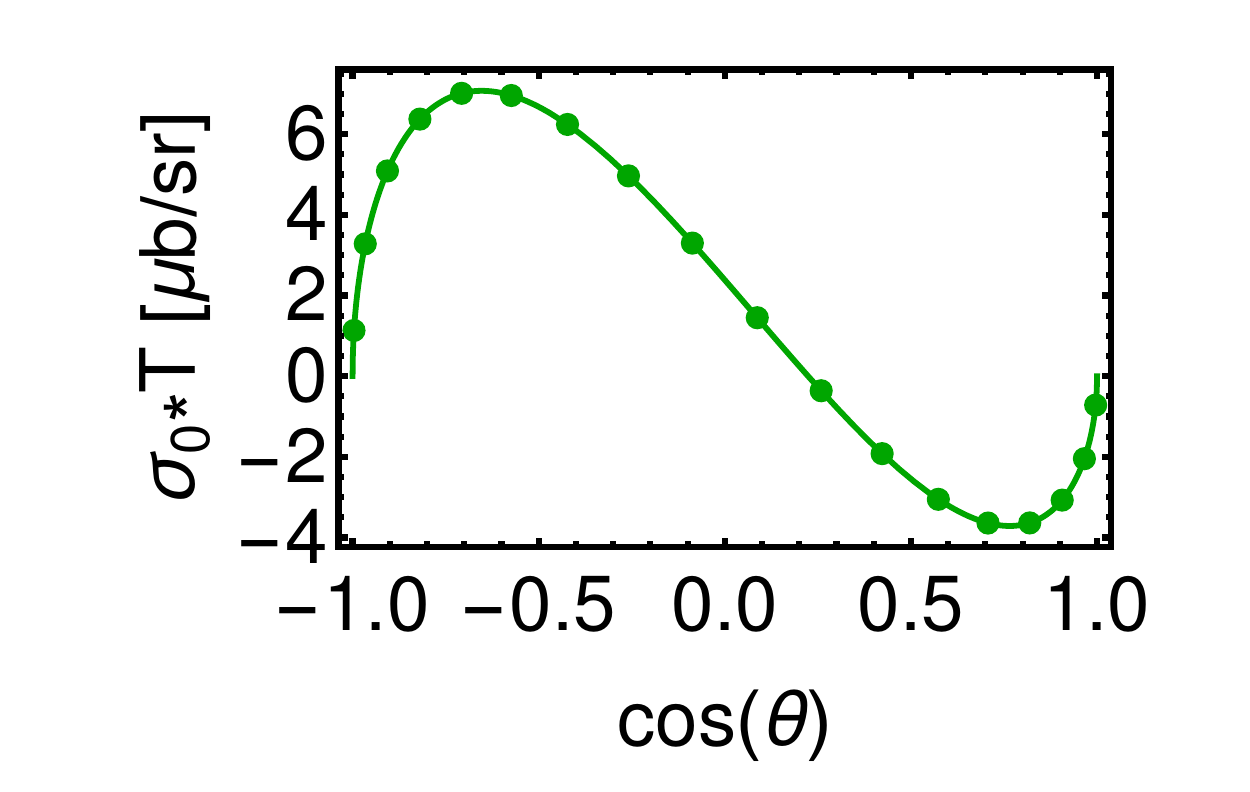}
 \end{overpic}
\begin{overpic}[width=0.485\textwidth]{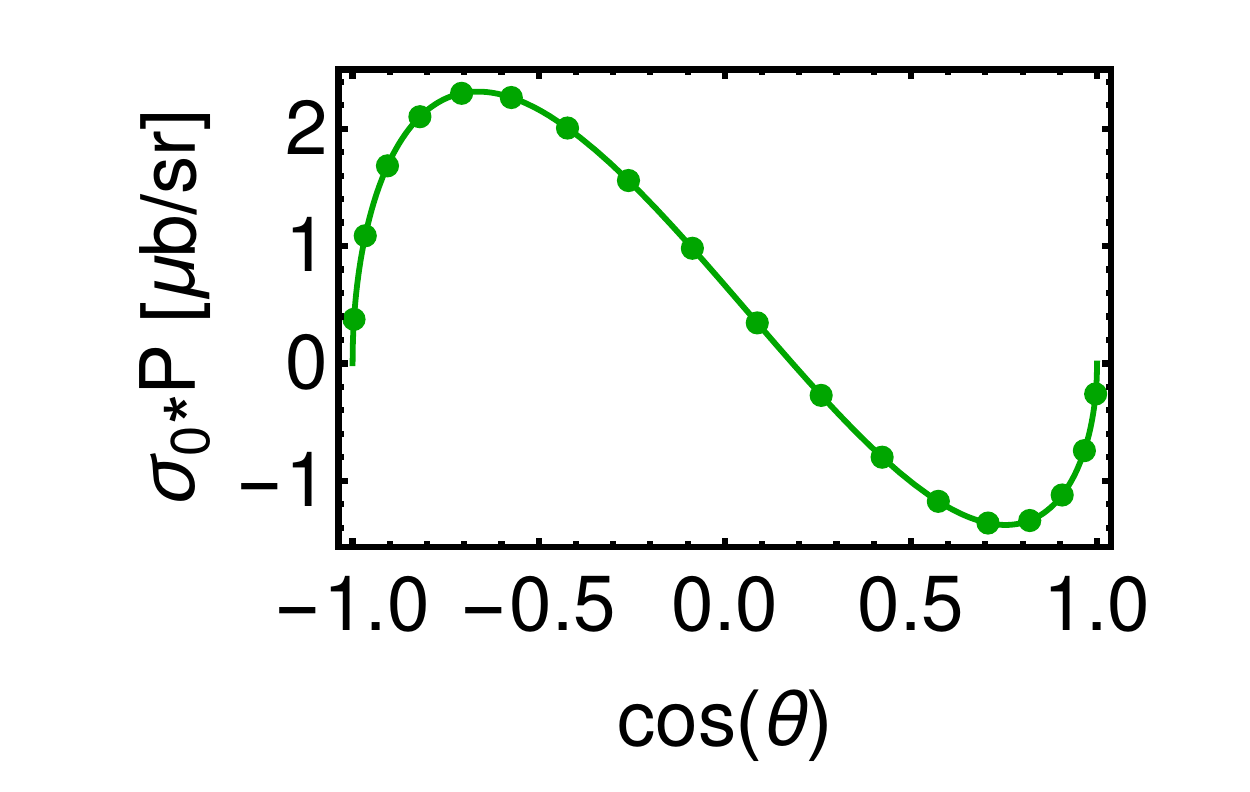}
 \end{overpic} \\
\begin{overpic}[width=0.485\textwidth]{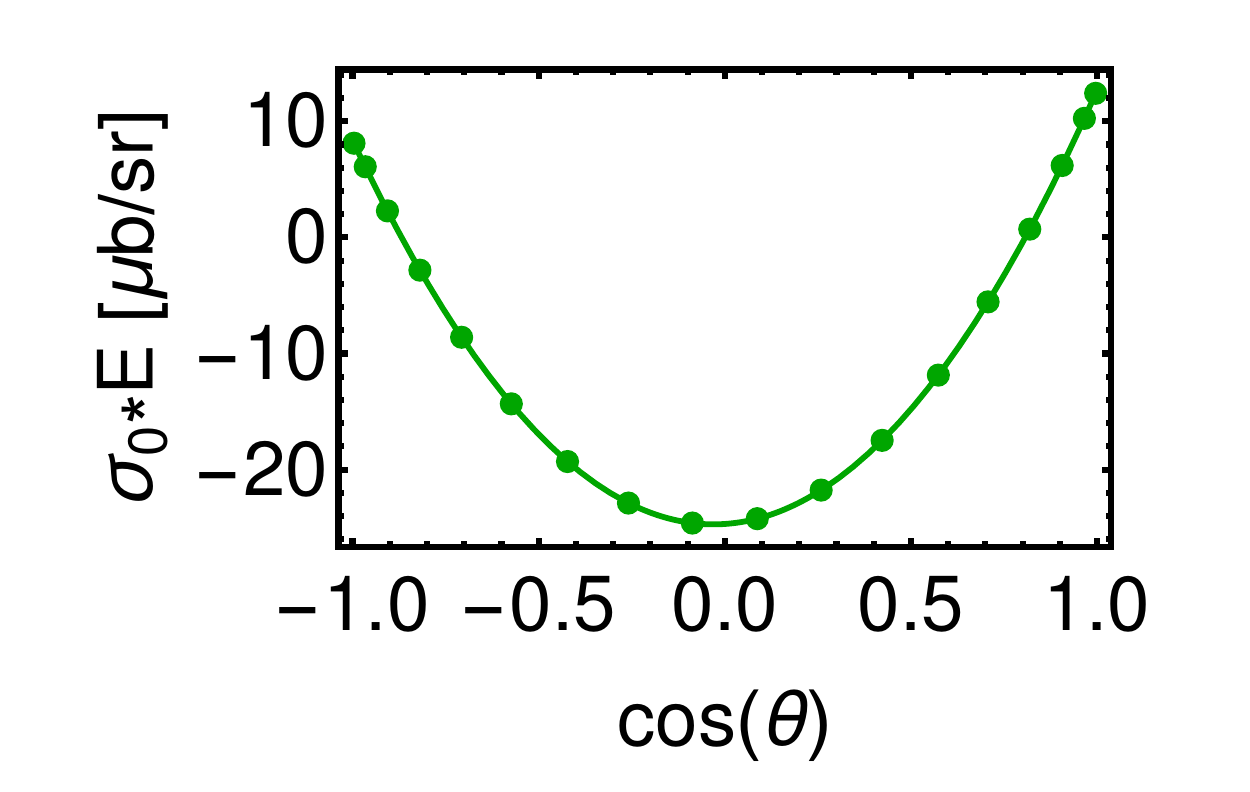}
 \end{overpic}
\begin{overpic}[width=0.485\textwidth]{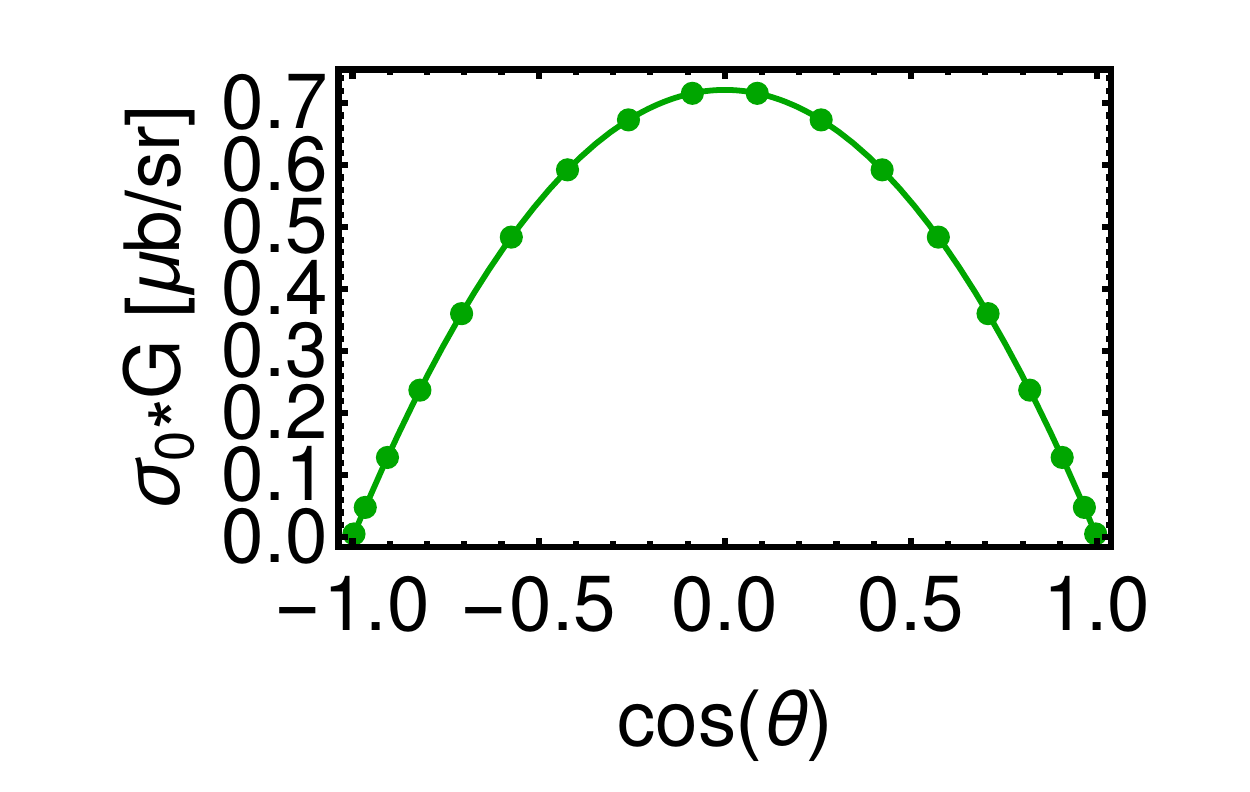}
 \end{overpic} \\
\begin{overpic}[width=0.485\textwidth]{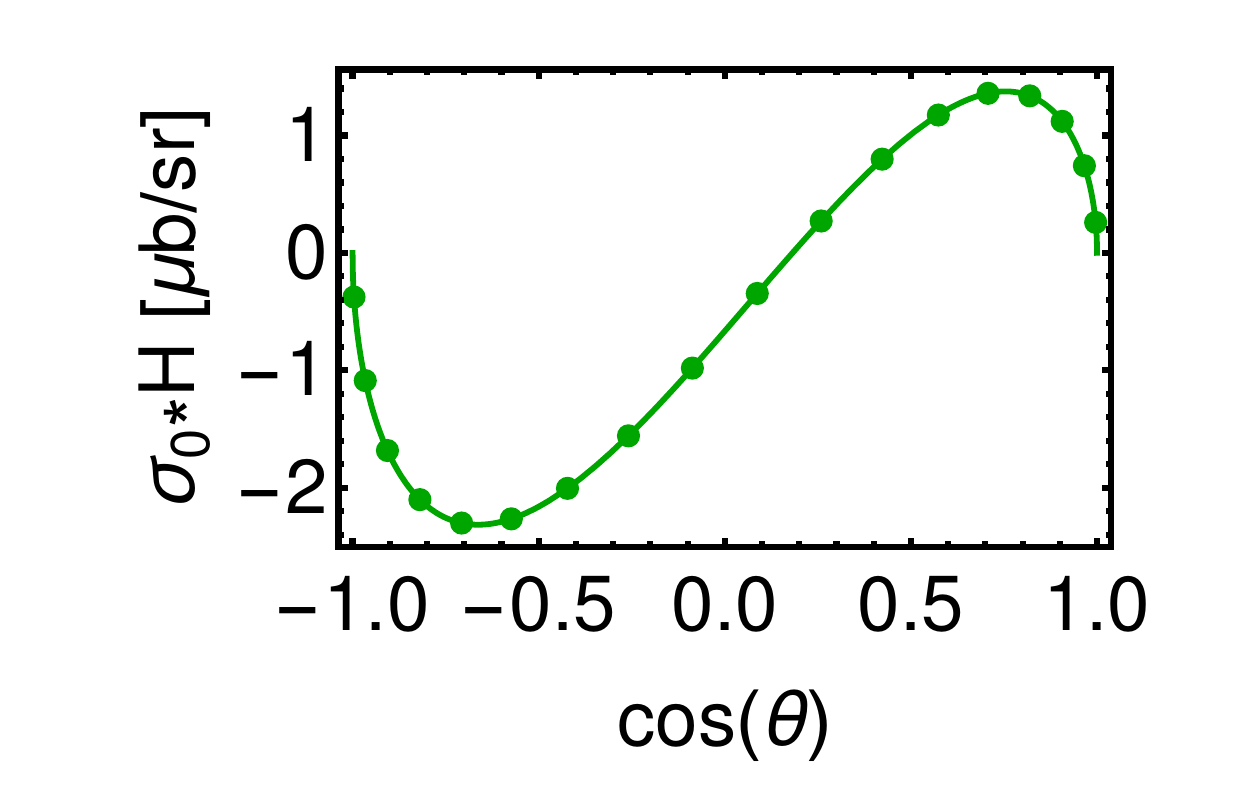}
 \end{overpic}
\begin{overpic}[width=0.485\textwidth]{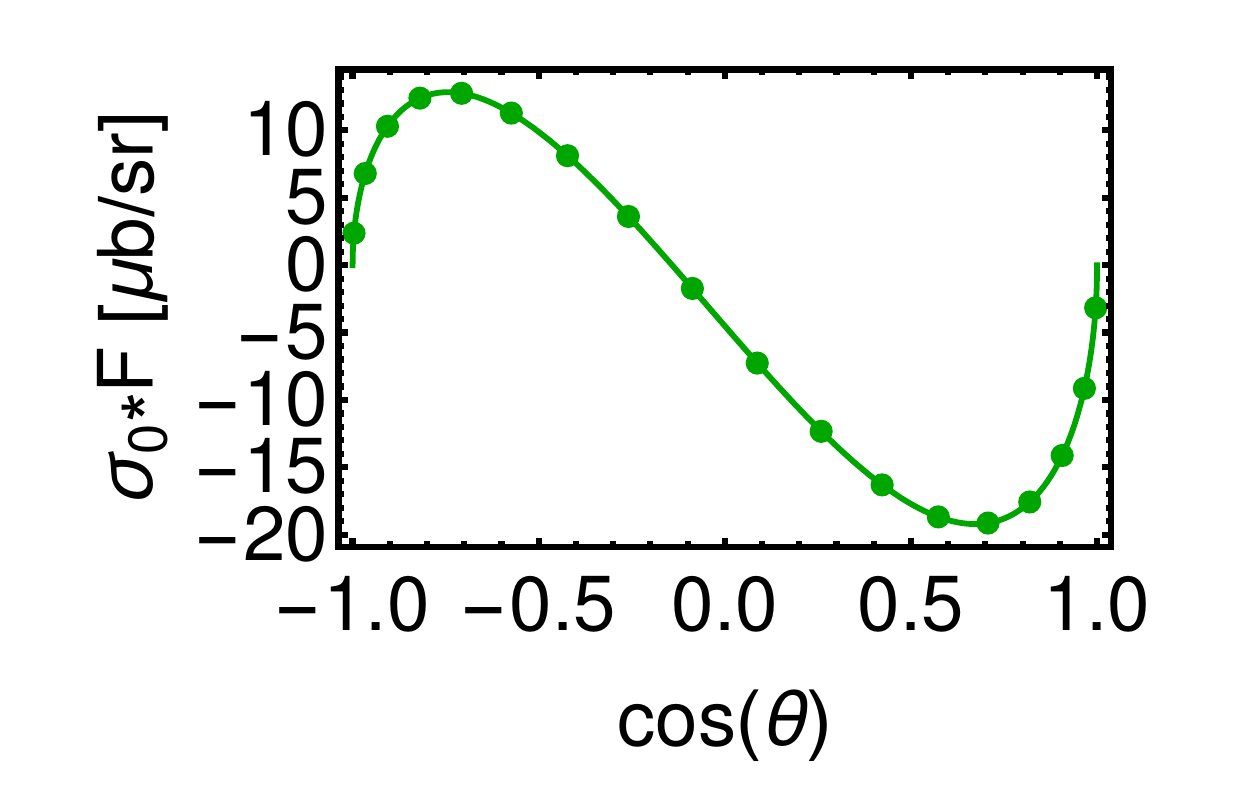}
 \end{overpic}
\caption[Angular distributions of MAID2007 theory-data truncated at $L=1$, for $E_{\gamma} = 330 \hspace*{2pt} \mathrm{MeV}$.]{MAID2007 theory-data \cite{LotharPrivateComm,MAID2007} truncated at $\ell_{\mathrm{max}} = 1$ are shown here. Depicted are the angular distributions of the profile functions belonging to the group $\mathcal{S}$ and $\mathcal{BT}$ observables at an example photon energy of $E_{\gamma} = 330 \hspace*{2pt} \mathrm{MeV}$. The TPWA Legendre fit for $\ell_{\mathrm{max}} = 1$ is also shown, which naturally describes the data perfectly.}
\label{fig:Lmax1ThDataFitGroupSBTObservablesExampleEnergy}
\end{figure}

\clearpage

\begin{figure}[ht]
 \centering
\begin{overpic}[width=0.32\textwidth]{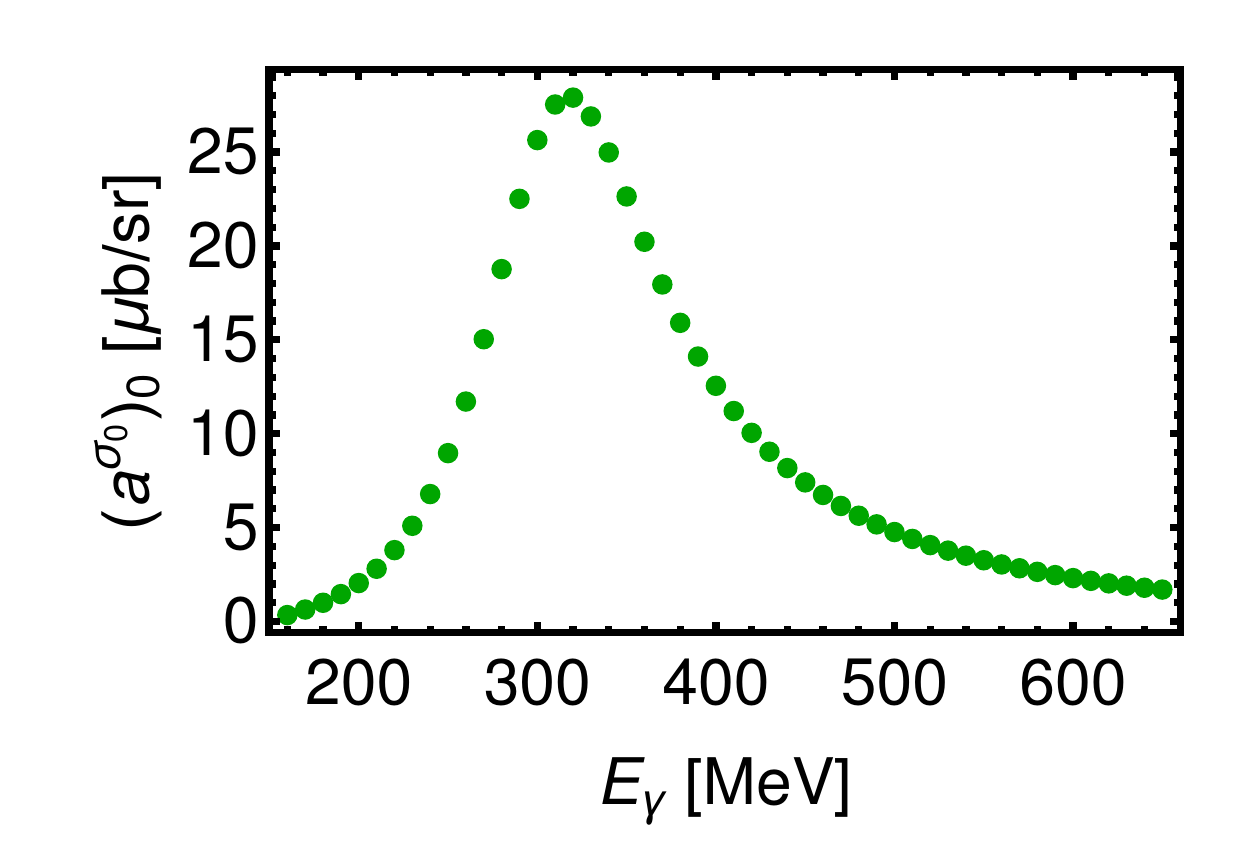}
 \end{overpic}
\begin{overpic}[width=0.32\textwidth]{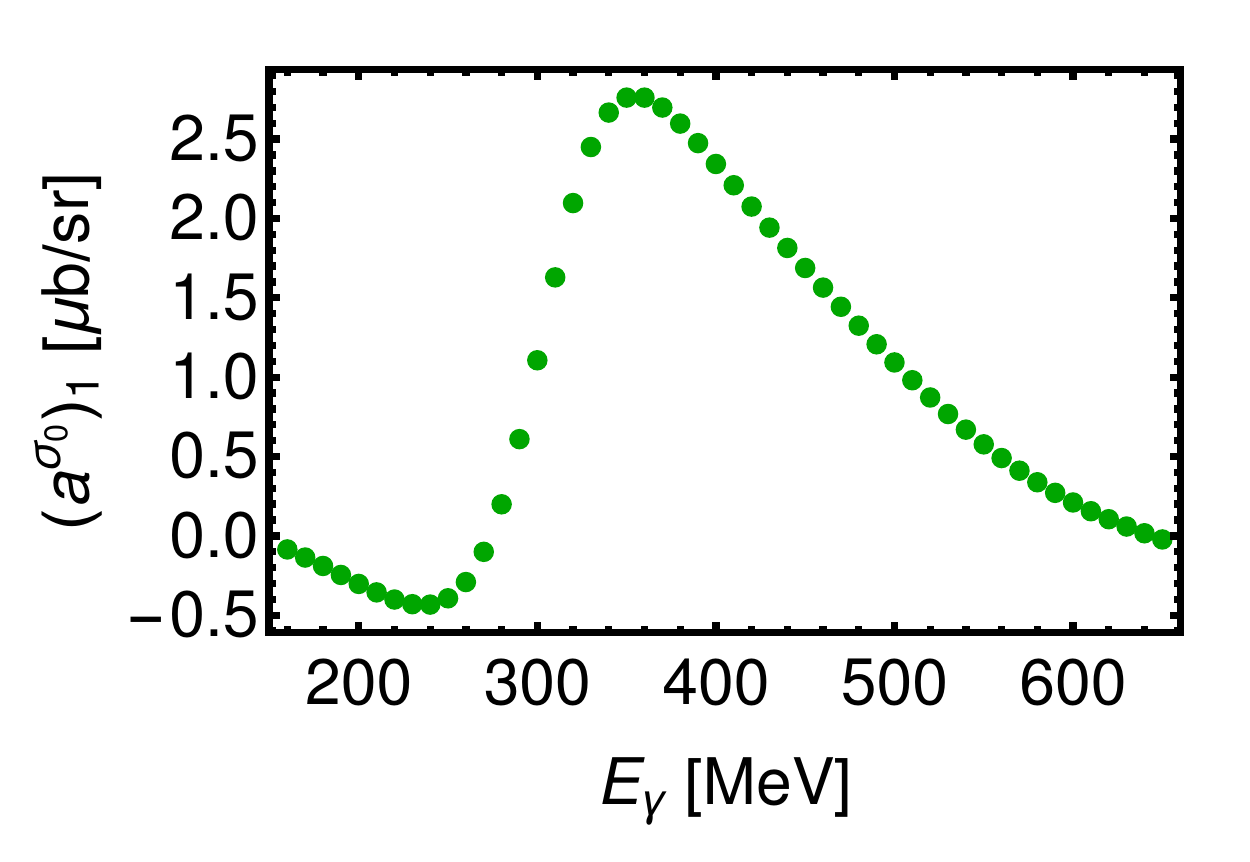}
 \end{overpic}
\begin{overpic}[width=0.32\textwidth]{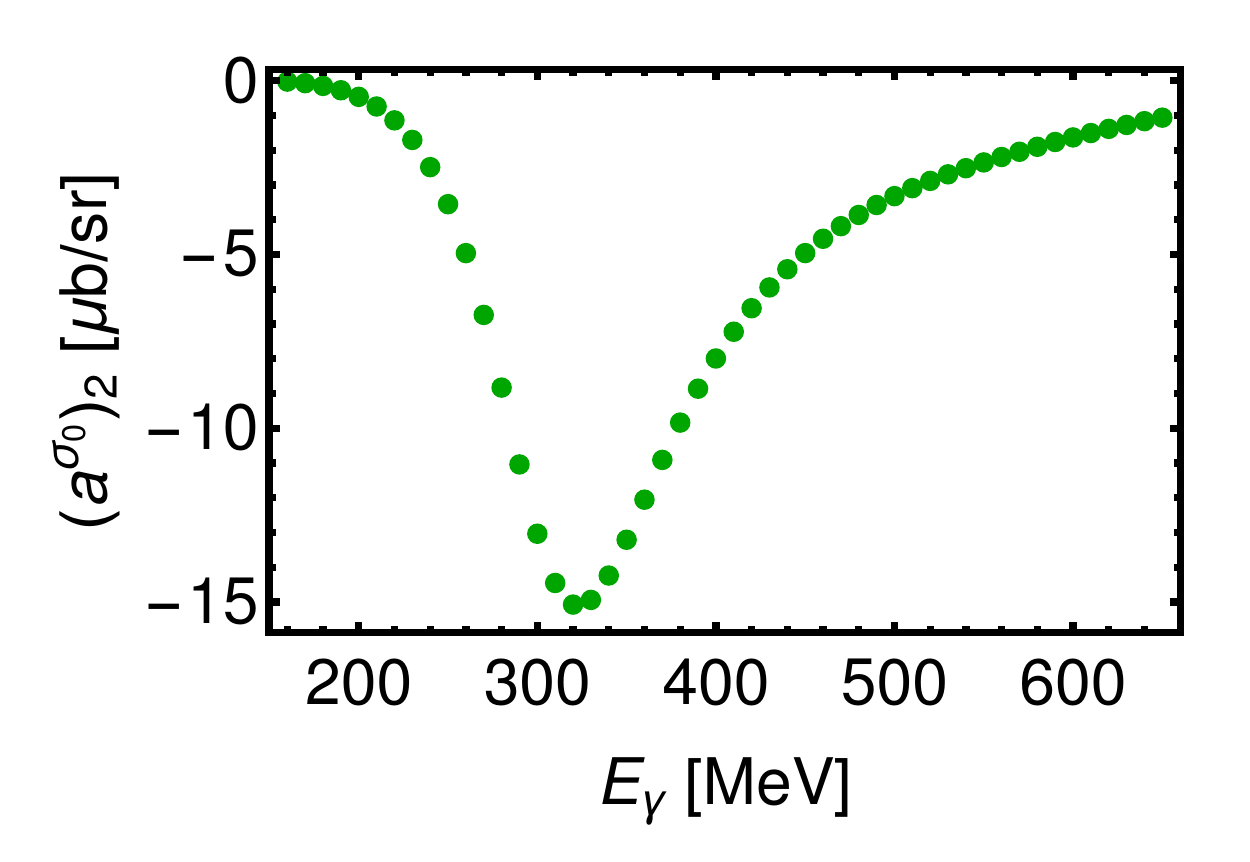}
 \end{overpic} \\
\begin{overpic}[width=0.32\textwidth]{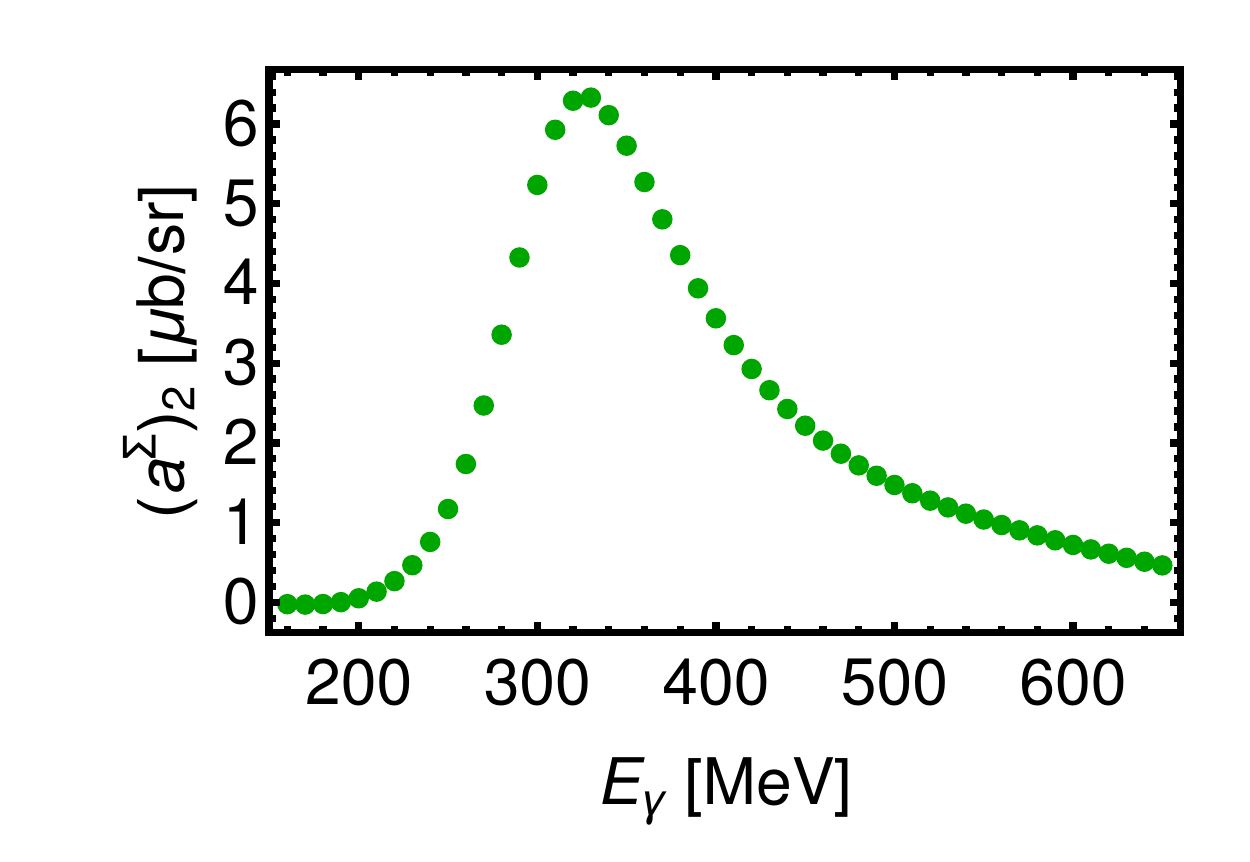}
 \end{overpic}
\begin{overpic}[width=0.32\textwidth]{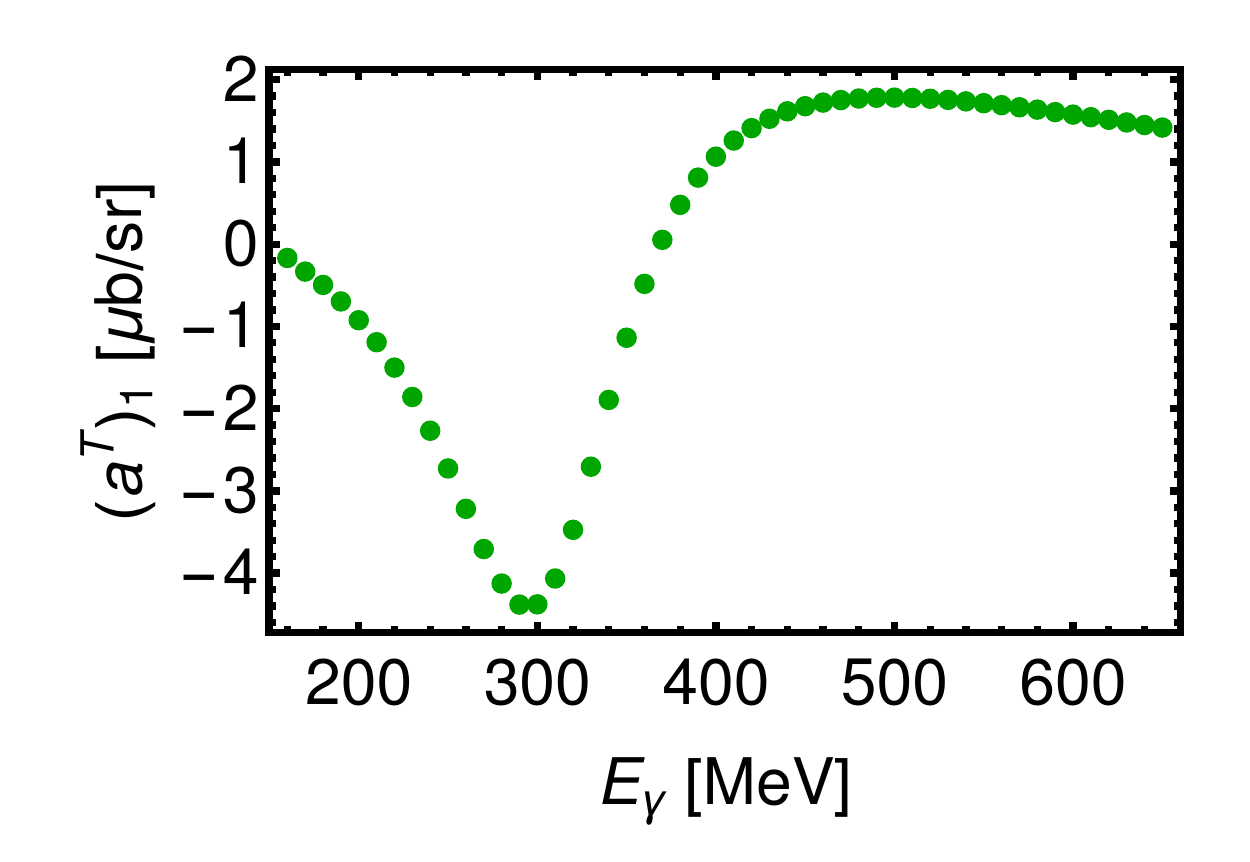}
 \end{overpic}
\begin{overpic}[width=0.32\textwidth]{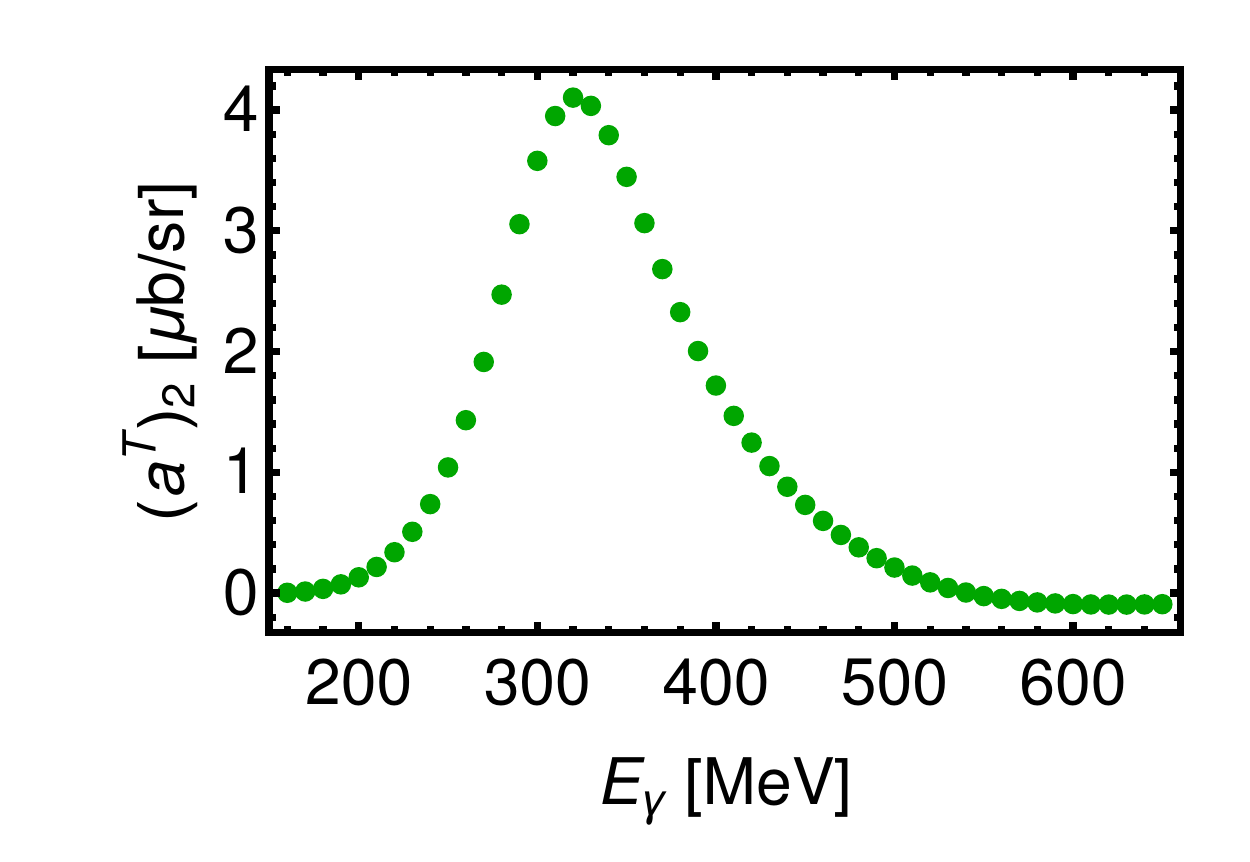}
 \end{overpic}  \\
\begin{overpic}[width=0.32\textwidth]{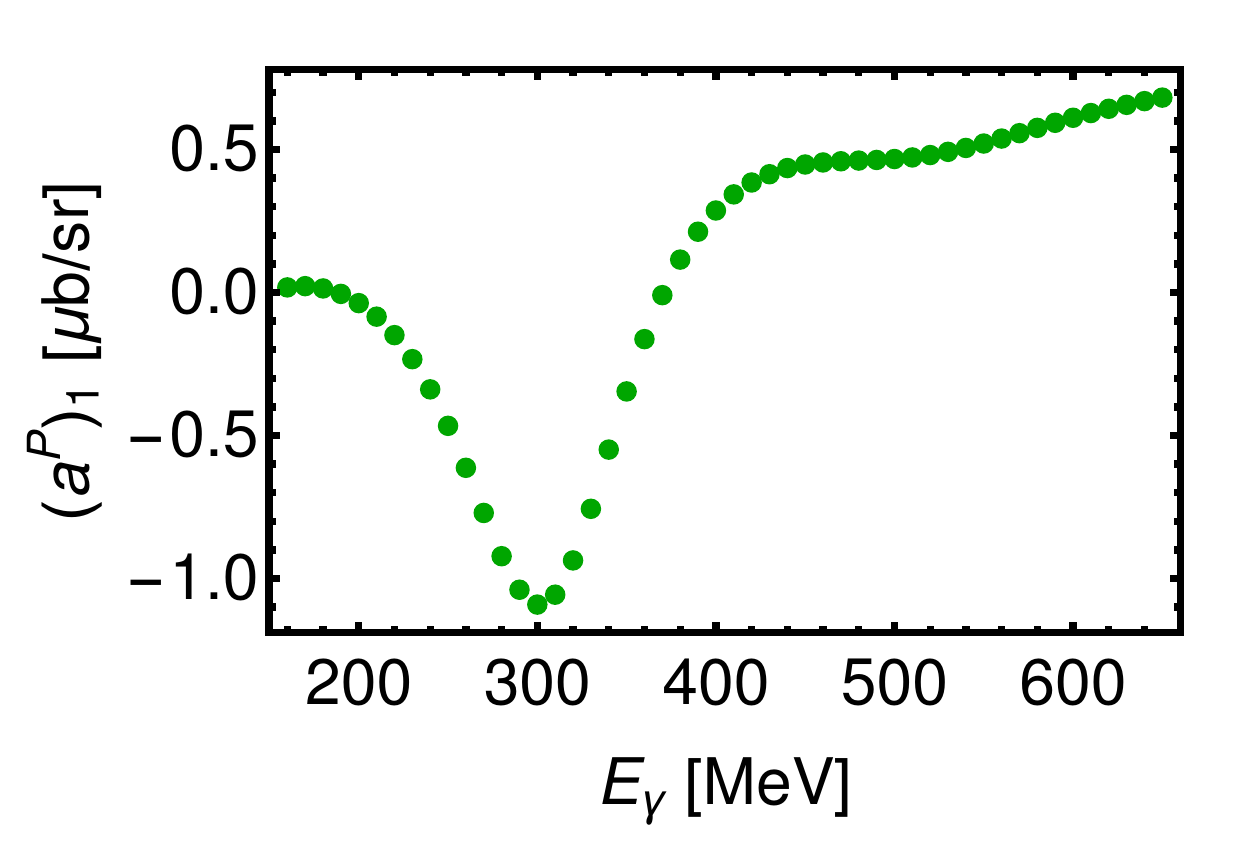}
 \end{overpic}
\begin{overpic}[width=0.32\textwidth]{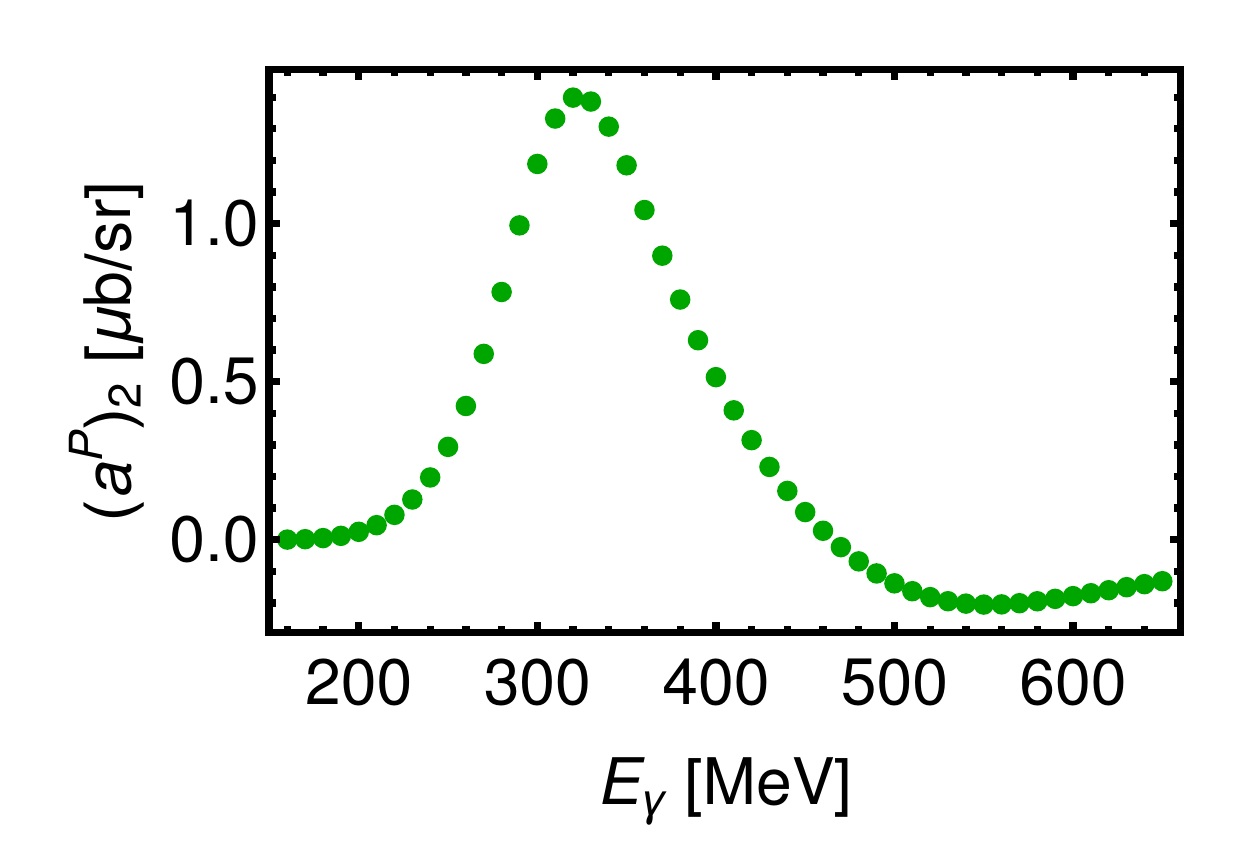}
 \end{overpic}
\caption[Legendre coefficients of the group $\mathcal{S}$ observables extracted from the truncated theory-data at $\ell_{\mathrm{max}} = 1$.]{Shown here are the Legendre coefficients of the group $\mathcal{S}$ observables extracted from the truncated theory-data at $\ell_{\mathrm{max}} = 1$.}
\label{fig:Lmax1ThDataFitLegCoeffsGroupS}
\end{figure}

\begin{figure}[ht]
 \centering
\begin{overpic}[width=0.32\textwidth]{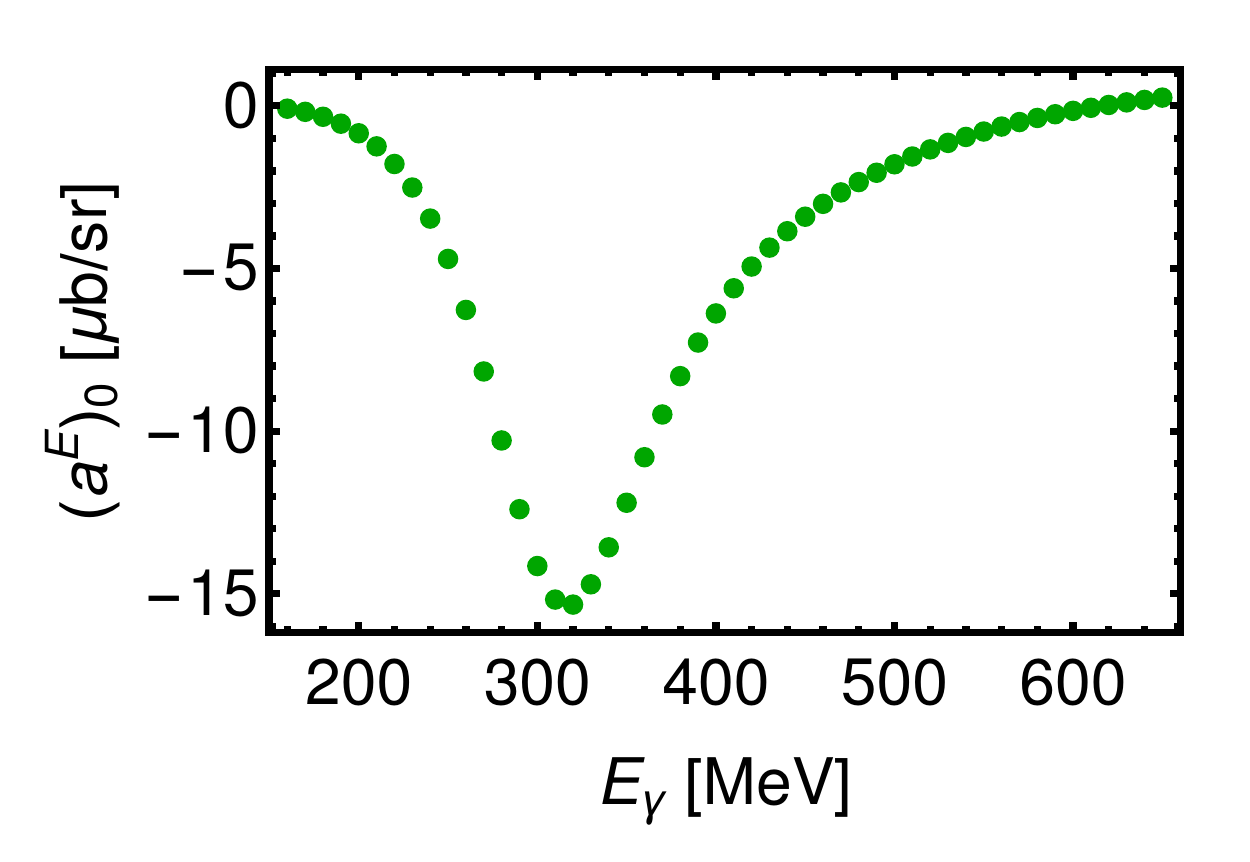}
 \end{overpic}
\begin{overpic}[width=0.32\textwidth]{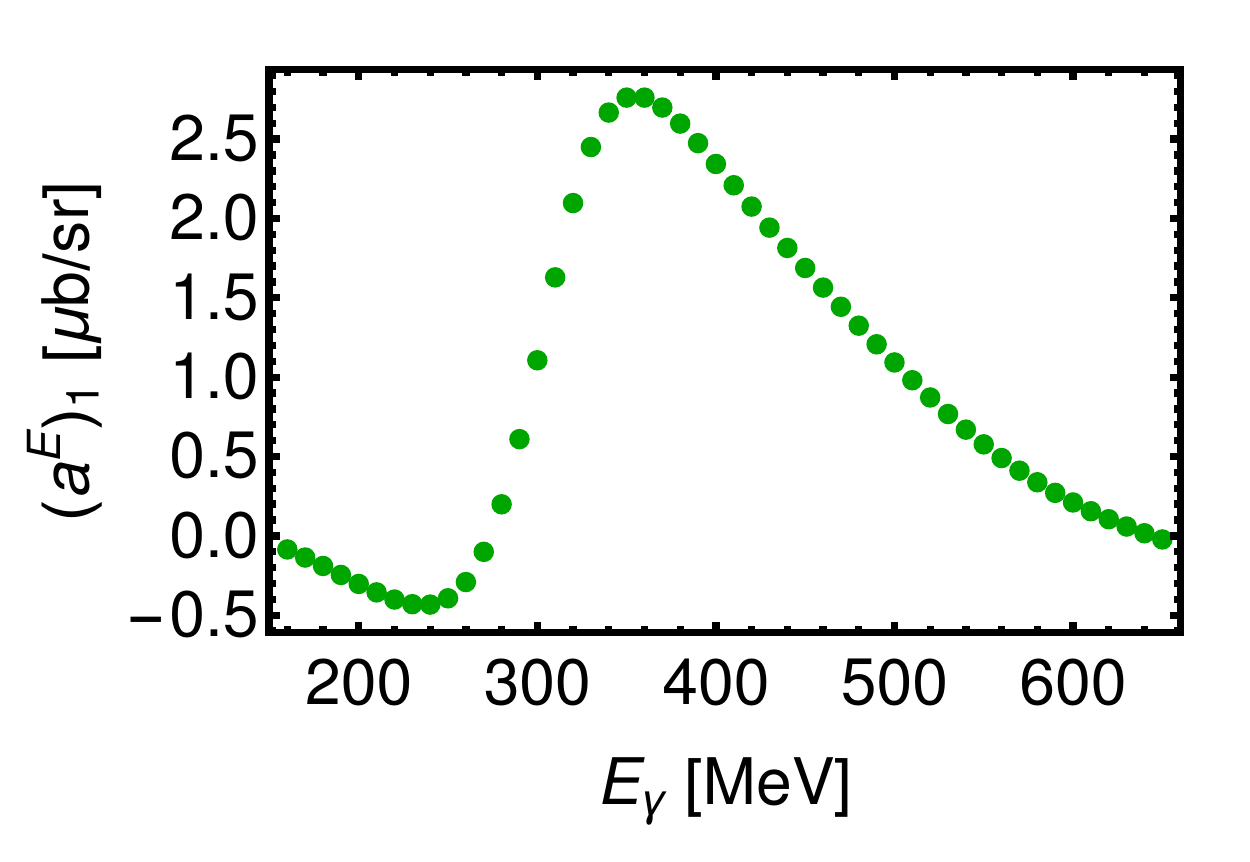}
 \end{overpic}
\begin{overpic}[width=0.32\textwidth]{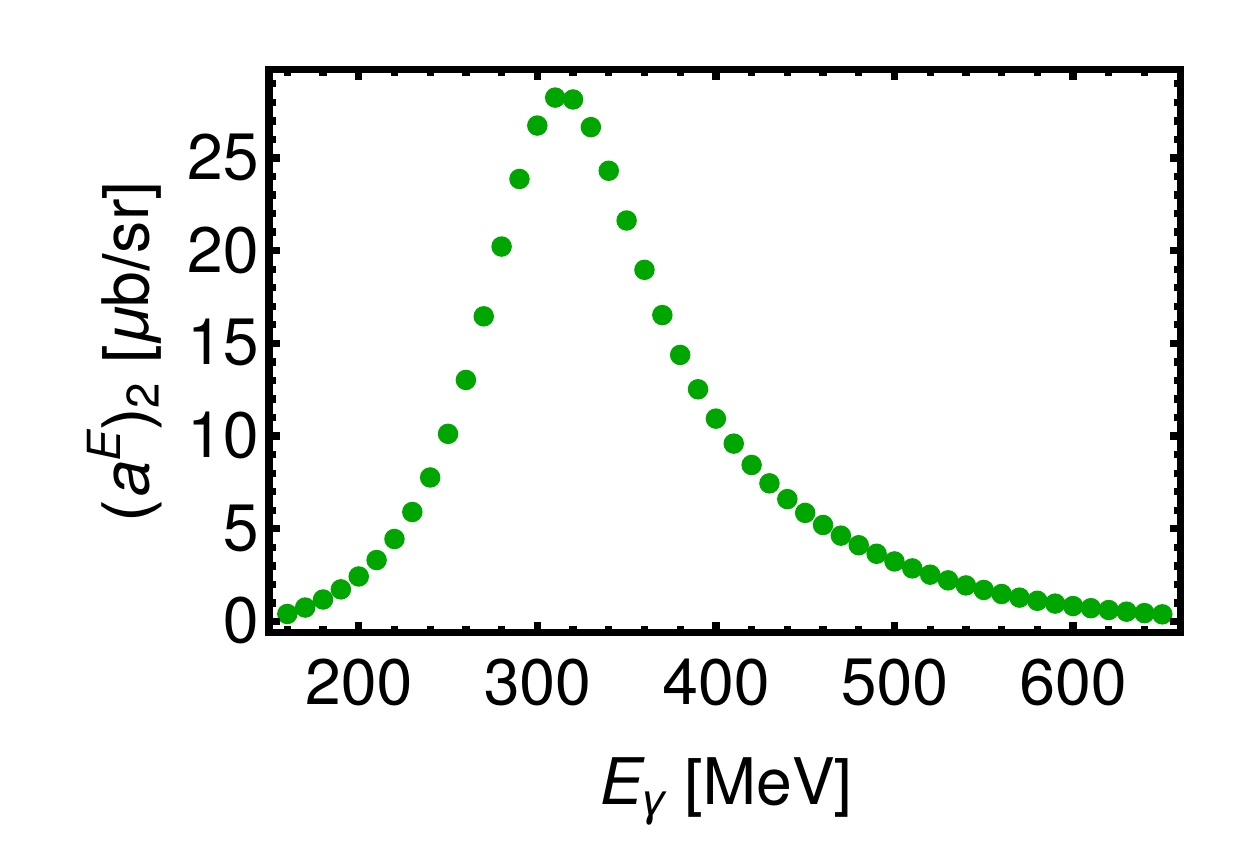}
 \end{overpic} \\
\begin{overpic}[width=0.32\textwidth]{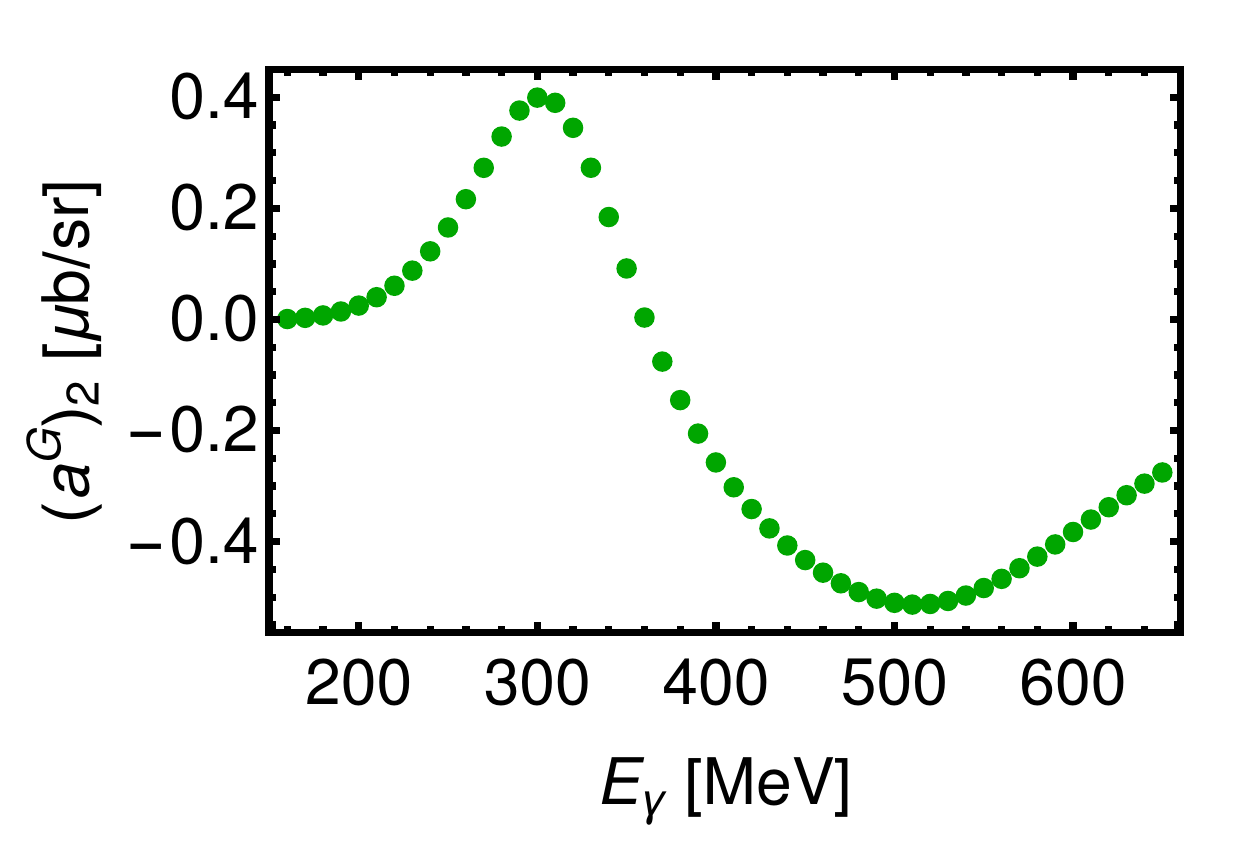}
 \end{overpic}
\begin{overpic}[width=0.32\textwidth]{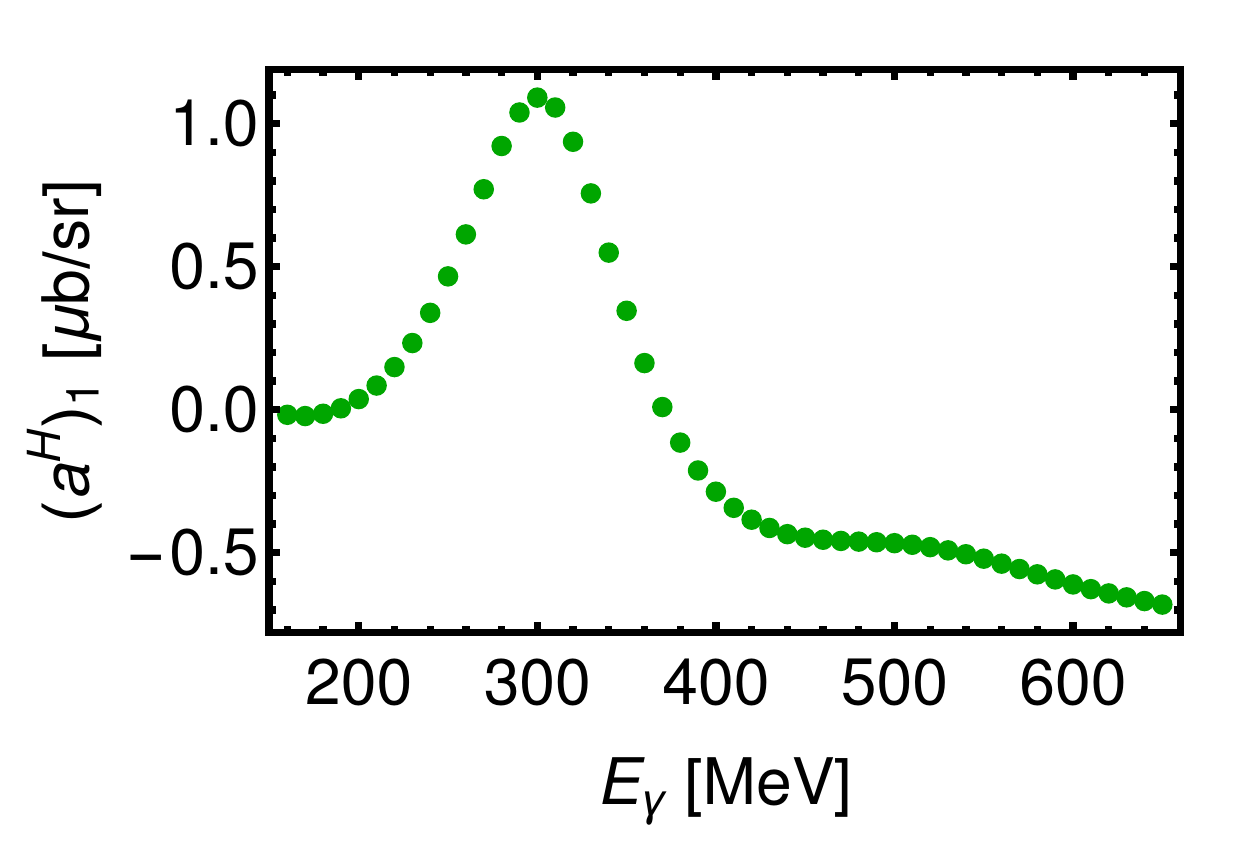}
 \end{overpic}
\begin{overpic}[width=0.32\textwidth]{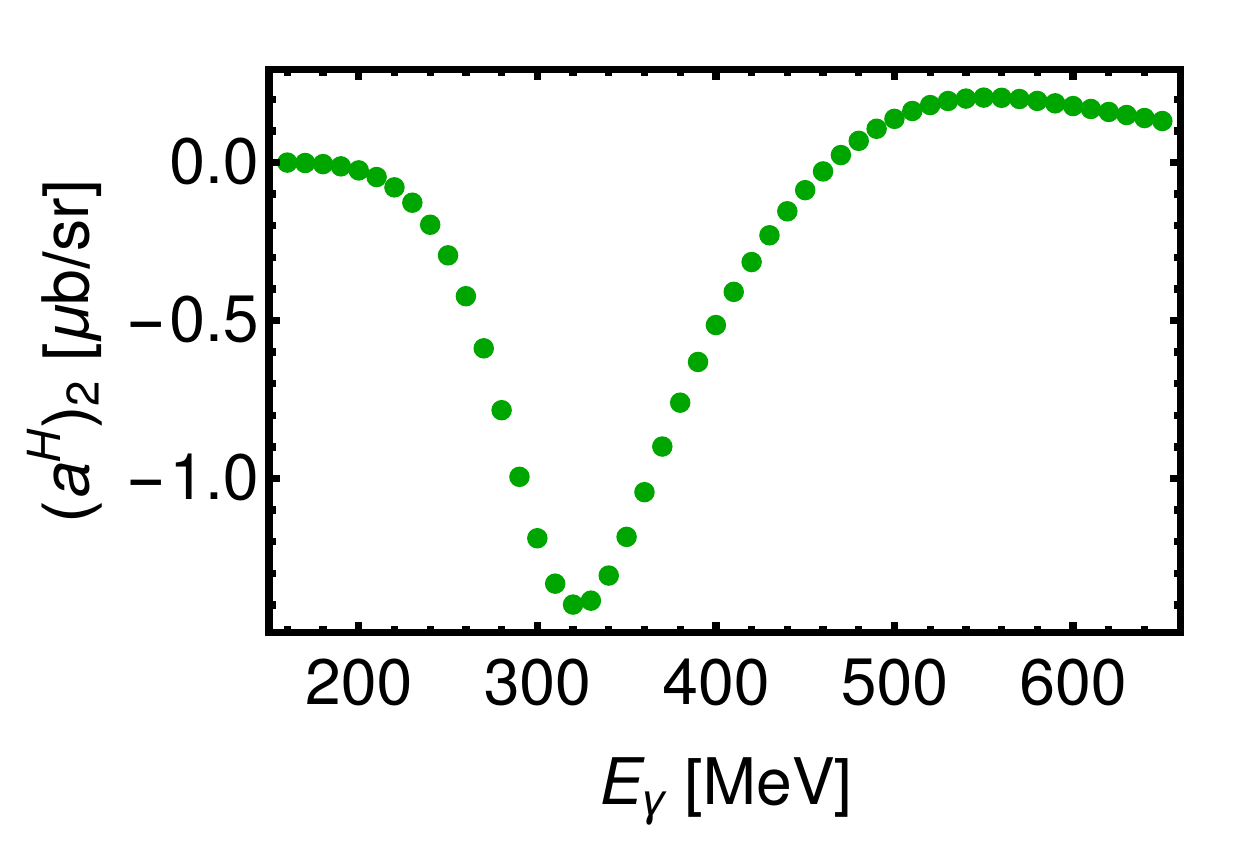}
 \end{overpic}  \\
\begin{overpic}[width=0.32\textwidth]{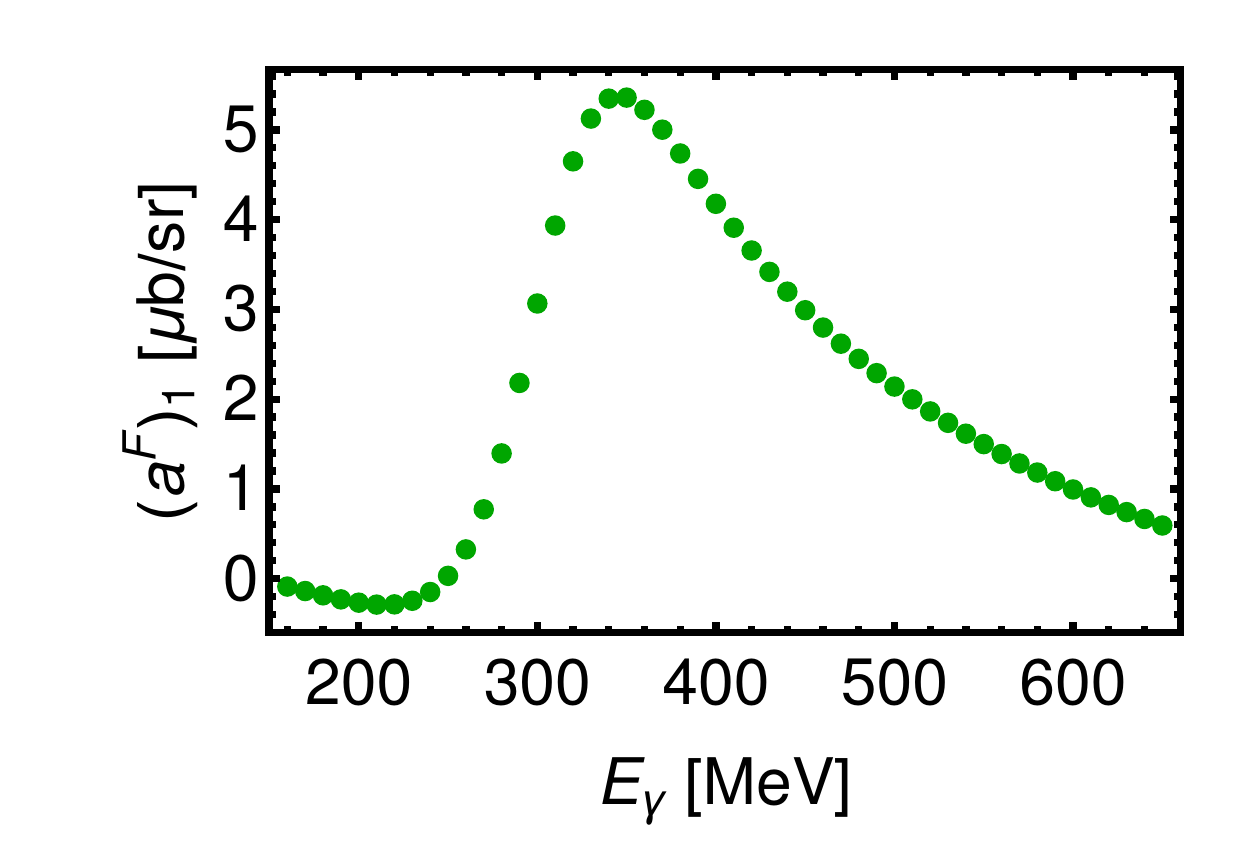}
 \end{overpic}
\begin{overpic}[width=0.32\textwidth]{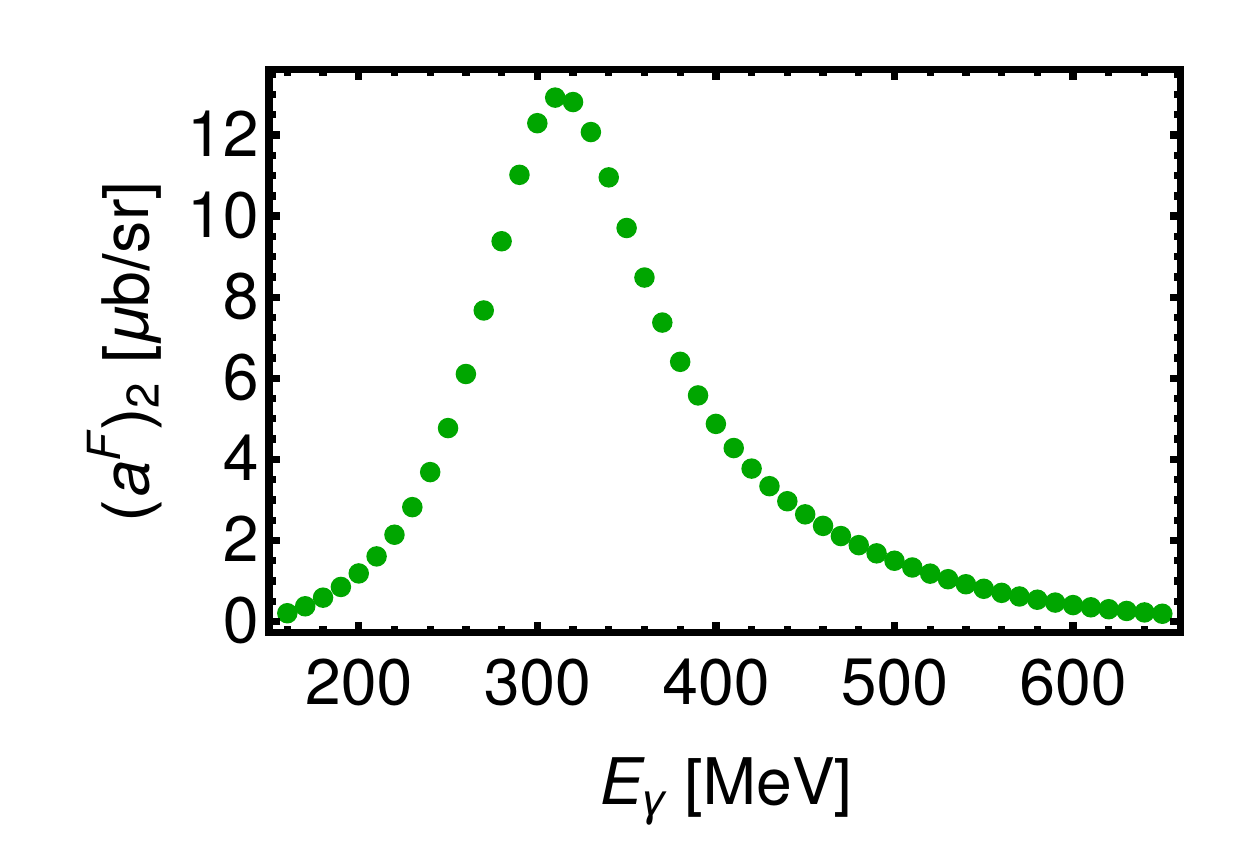}
 \end{overpic}
\caption[Legendre coefficients of the $\mathcal{BT}$ observables extracted from the truncated theory-data at $\ell_{\mathrm{max}} = 1$.]{Here, the Legendre coefficients of the $\mathcal{BT}$ observables fitted out of the truncated theory-data at $\ell_{\mathrm{max}} = 1$ are shown.}
\label{fig:Lmax1ThDataFitLegCoeffsBT}
\end{figure}

The MAID data are provided in the photon energy interval $E_{\gamma} \in \left[ 160, 1500 \right] \hspace*{2pt} \mathrm{MeV}$, with $10$ $\mathrm{MeV}$-steps. In every energy bin, the angular distributions are given in terms of the variable $\theta$, ranging from $5^{\circ}$ to $175^{\circ}$ in $5^{\circ}$-steps. These specifications are valid for the truncated MAID theory-data considered in this section, as well as for the remaining datasets used in sections \ref{subsec:TheoryDataFitsLmax2} and \ref{subsec:TheoryDataFitsLmaxInfinite}. \newline
We choose here to consider the theory-data in the energy range $E_{\gamma} \in \left[ 160, 650 \right] \hspace*{2pt} \mathrm{MeV}$, i.e. testing just the $\Delta$-region\footnote{I.e., the region where the resonance $\Delta(1232) \frac{3}{2}^{+}$ \cite{Patrignani:2016xqp} dominates the multipole $M_{1+}$ in $\gamma p \longrightarrow \pi^{0} p$.}. In this way, numerical efforts are still not too extreme and one has a good comparison to the ambiguity diagrams in Figure II of section \ref{sec:WBTpaper} and Figure \ref{fig:ExampleEpsilonPiParametersInAmbiguityDiagram} of appendix \ref{subsec:AccidentalAmbProofsI}. \newline
Furthermore, we transform the angular variable from $\theta$ to $\cos \theta$, since the TPWA problem is most conveniently phased in terms of the latter. This has the consequence that angular distributions of observables exist now on $18$ non-equidistant points. \newline
First, Legendre coefficients are fitted out of the $\ell_{\mathrm{max}} = 1$ theory-data according to fit step I outlined in section \ref{sec:TPWAFitsIntro}, i.e. by minimizing the functions $\Phi_{a}^{\alpha}$. Figure \ref{fig:Lmax1ThDataFitGroupSBTObservablesExampleEnergy} shows the angular distributions of the theory-data for the group $\mathcal{S}$ and $\mathcal{BT}$ observables. The energy $E_{\gamma} = 330 \hspace*{2pt} \mathrm{MeV}$ is chosen as an example. Also, the Legendre fits for $\ell_{\mathrm{max}} = 1$ are shown in the Figure. The extracted Legendre coefficients for all the considered energy bins of the theory-data are plotted in Figures \ref{fig:Lmax1ThDataFitLegCoeffsGroupS} and \ref{fig:Lmax1ThDataFitLegCoeffsBT}. \newline
The Legendre coefficients are then used as input to minimizations of functions of the form $\Phi_{\mathcal{M}}$ (\ref{eq:PhiTheoryDataFitStep2}) according to fit step II described in section \ref{sec:TPWAFitsIntro}. In all cases considered here, minimizations were performed using the Monte Carlo method described in 

\newpage

section \ref{sec:MonteCarloSampling}, employing a pool of $N_{MC} = 1500$ start configurations.
The minimizations then lead to a full pool of $1500$ solutions clustered around the minima, out of which the non-redundant one's are sorted using the methods described in section \ref{sec:MonteCarloSampling} and appendix \ref{sec:MCSamplingAlgorithms}. The number of non redundant solutions $N_{\mathrm{nonred}}$ will of course change from energy bin to energy bin and reflect the full solution structure of the problem. One can now test different combinations of observables regarding their capability to uniquely solve the TPWA problem for the theory-data. Different choices of observables correspond to different index-sets for $\alpha$ on the right hand side of equation (\ref{eq:PhiTheoryDataFitStep2}). \newline

As a first example, we consider a fit of just the group $\mathcal{S}$ observables, or better said the profile functions of them, composed out of $\left\{ \sigma_{0}, \check{\Sigma}, \check{T}, \check{P} \right\}$. In this case, the index $\alpha$ in the discrepancy function has to run over the values $\left(1,4,10,12\right)$. Fully written out in terms of observable-names, it reads
\allowdisplaybreaks
\hspace*{-5pt}
\begin{align}
 \Phi_{\mathcal{M}} \left( \left\{ \mathcal{M}_{\ell}^{C} \right\} \right) &= \sum_{k = 0}^{2} \left[ \left(a_{L}\right)^{\sigma_{0}}_{k} -  \left< \mathcal{M}_{\ell}^{C} \right| \left( \mathcal{C}_{L}\right)_{k}^{\sigma_{0}} \left| \mathcal{M}_{\ell}^{C} \right> \right]^{2}  \nonumber \\
 & + \left[ \left(a_{L}\right)^{\check{\Sigma}}_{2} -  \left< \mathcal{M}_{\ell}^{C} \right| \left( \mathcal{C}_{L}\right)_{2}^{\check{\Sigma}} \left| \mathcal{M}_{\ell}^{C} \right> \right]^{2} + \sum_{m = 1}^{2} \left[ \left(a_{L}\right)^{\check{T}}_{m} -  \left< \mathcal{M}_{\ell}^{C} \right| \left( \mathcal{C}_{L}\right)_{m}^{\check{T}} \left| \mathcal{M}_{\ell}^{C} \right> \right]^{2} \nonumber \\
  & + \sum_{n = 1}^{2} \left[ \left(a_{L}\right)^{\check{P}}_{n} -  \left< \mathcal{M}_{\ell}^{C} \right| \left( \mathcal{C}_{L}\right)_{n}^{\check{P}} \left| \mathcal{M}_{\ell}^{C} \right> \right]^{2} \mathrm{.} \label{eq:ExamplePhiFitFunction}
\end{align}
The phase-constrained $S$- and $P$-wave multipoles (cf. equation (\ref{eq:MultipoleVectorConstrainedChap4})) are varied as free
\newpage
\begin{figure}[ht]
 \centering
 \vspace*{-10pt}
\begin{overpic}[width=0.475\textwidth]{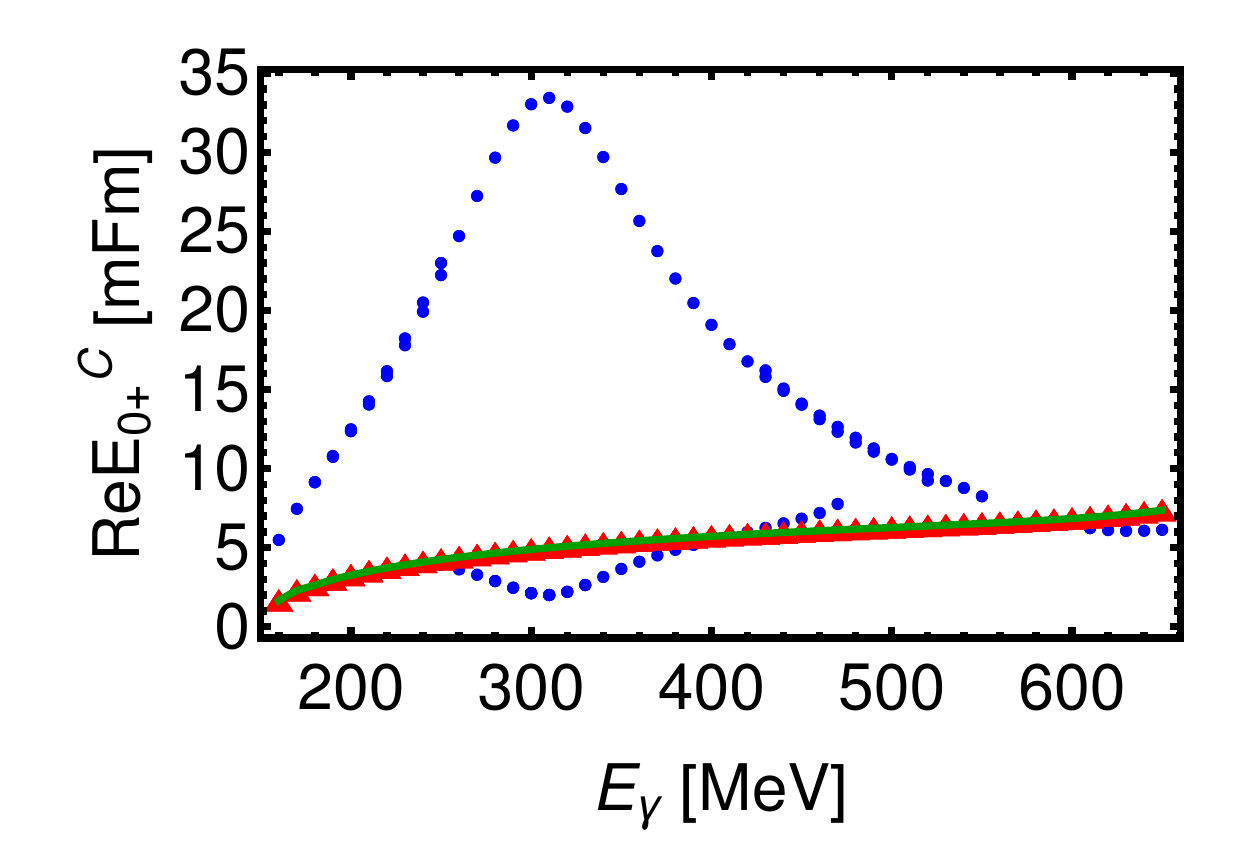}
 \end{overpic} \\ 
 \vspace*{-5pt}
\begin{overpic}[width=0.475\textwidth]{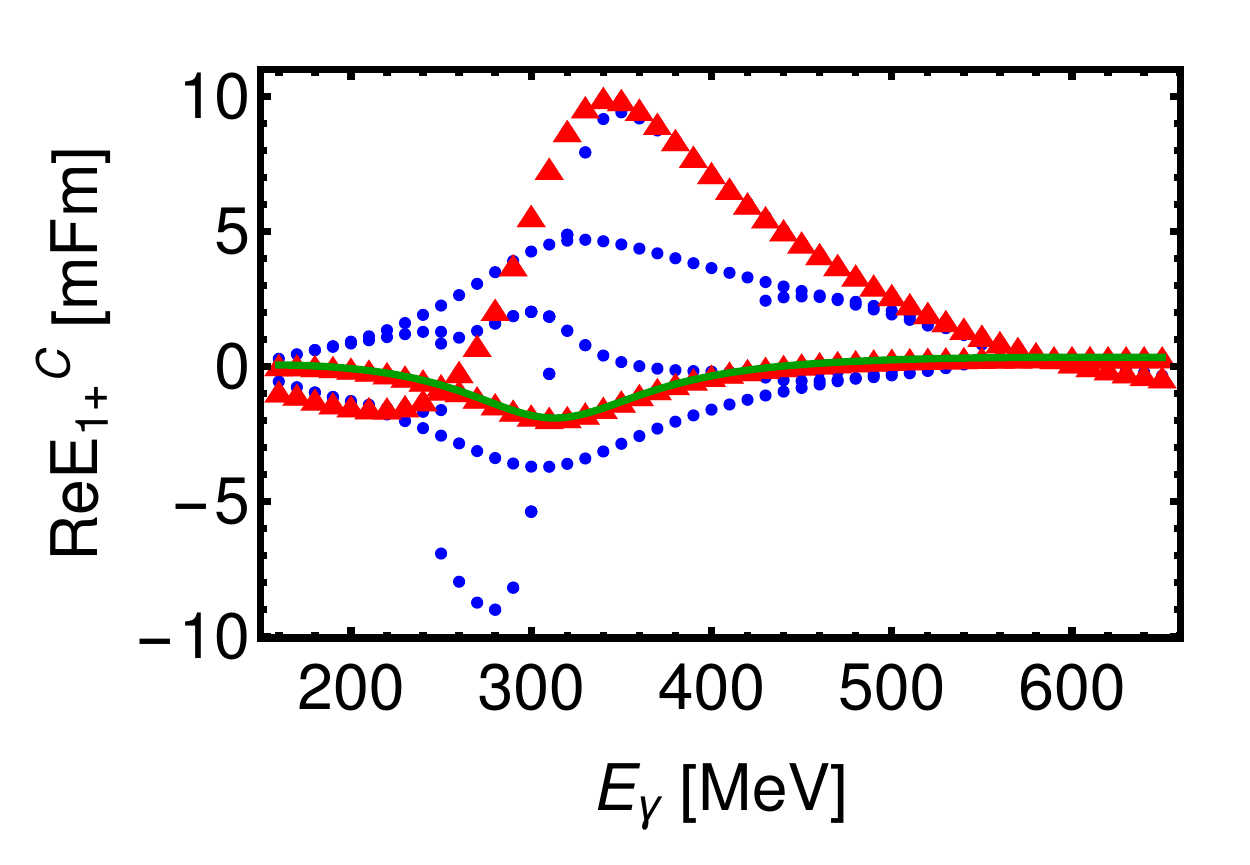}
 \end{overpic} \hspace*{5pt}
\begin{overpic}[width=0.475\textwidth]{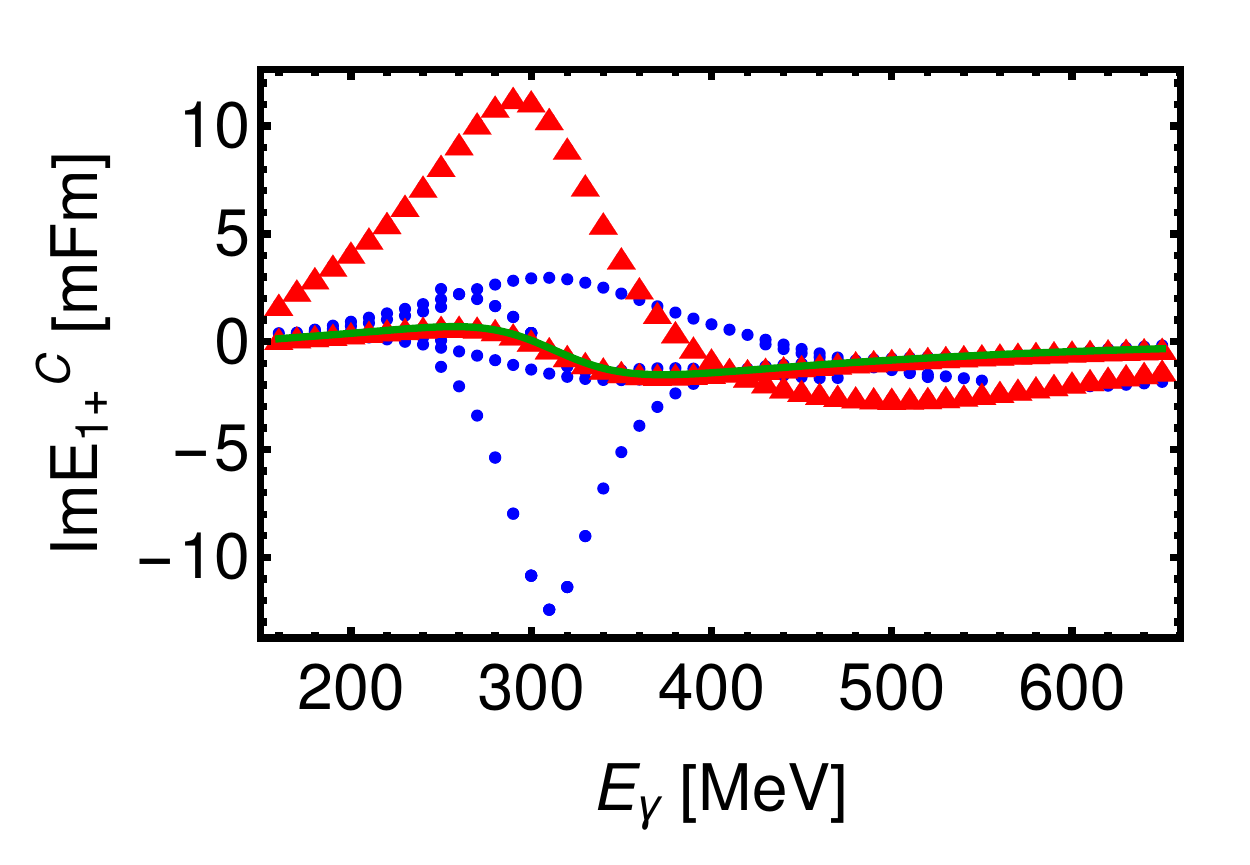}
 \end{overpic} \\
\begin{overpic}[width=0.475\textwidth]{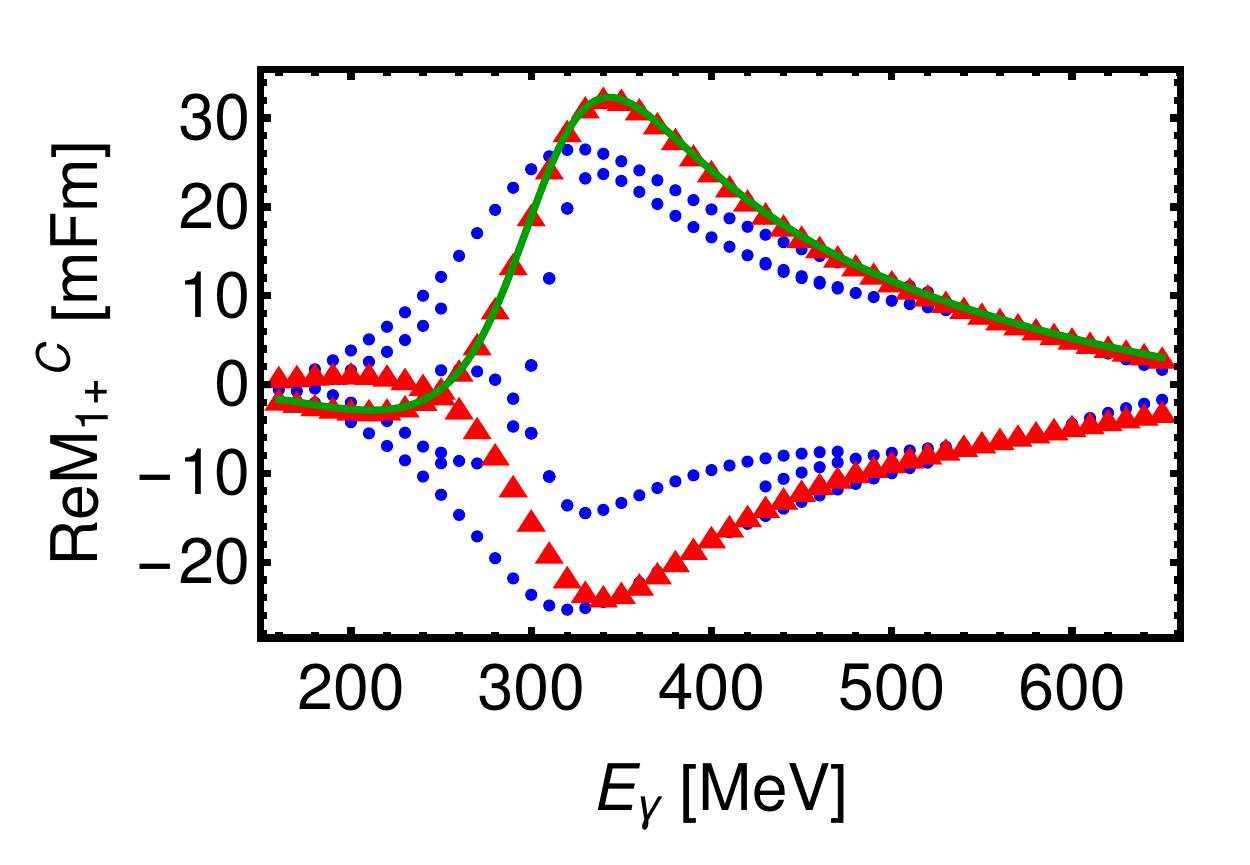}
 \end{overpic} \hspace*{5pt}
\begin{overpic}[width=0.475\textwidth]{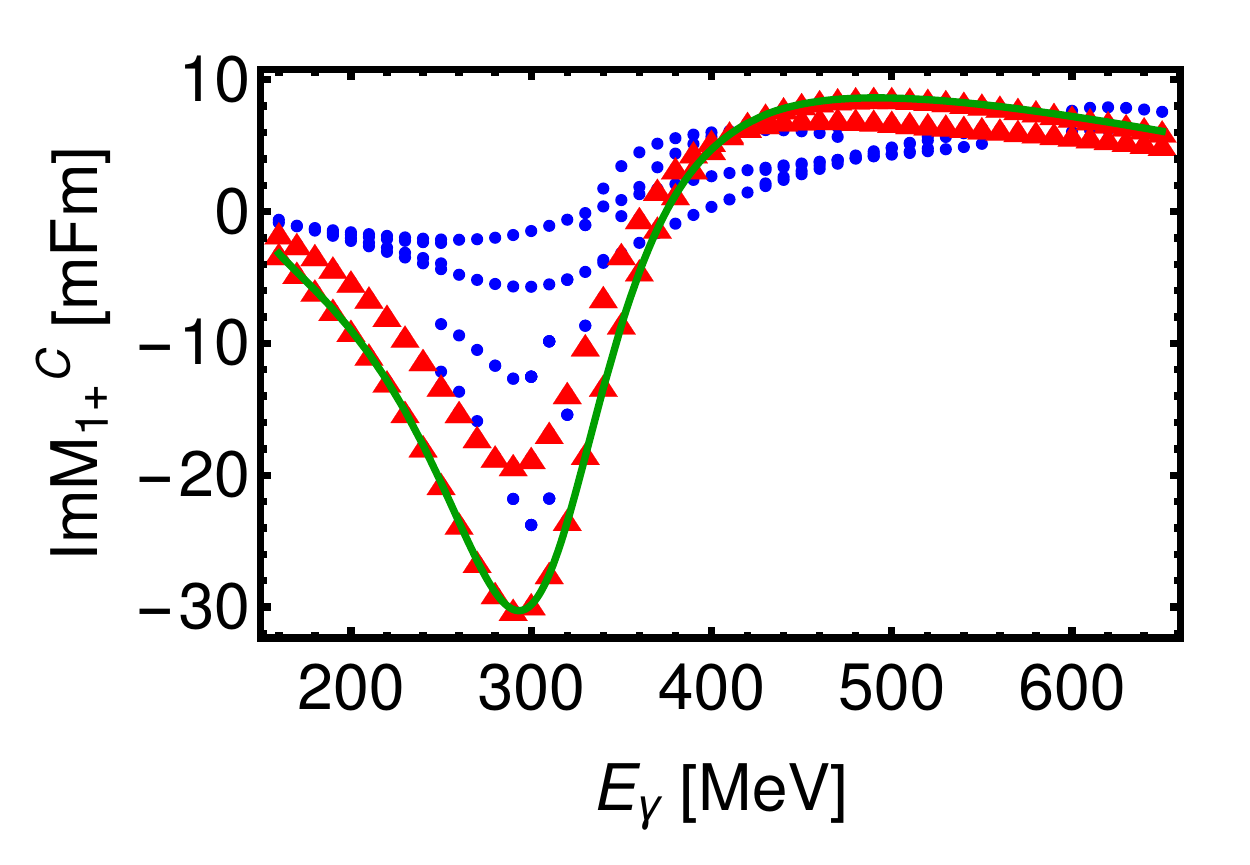}
 \end{overpic} \\
\begin{overpic}[width=0.475\textwidth]{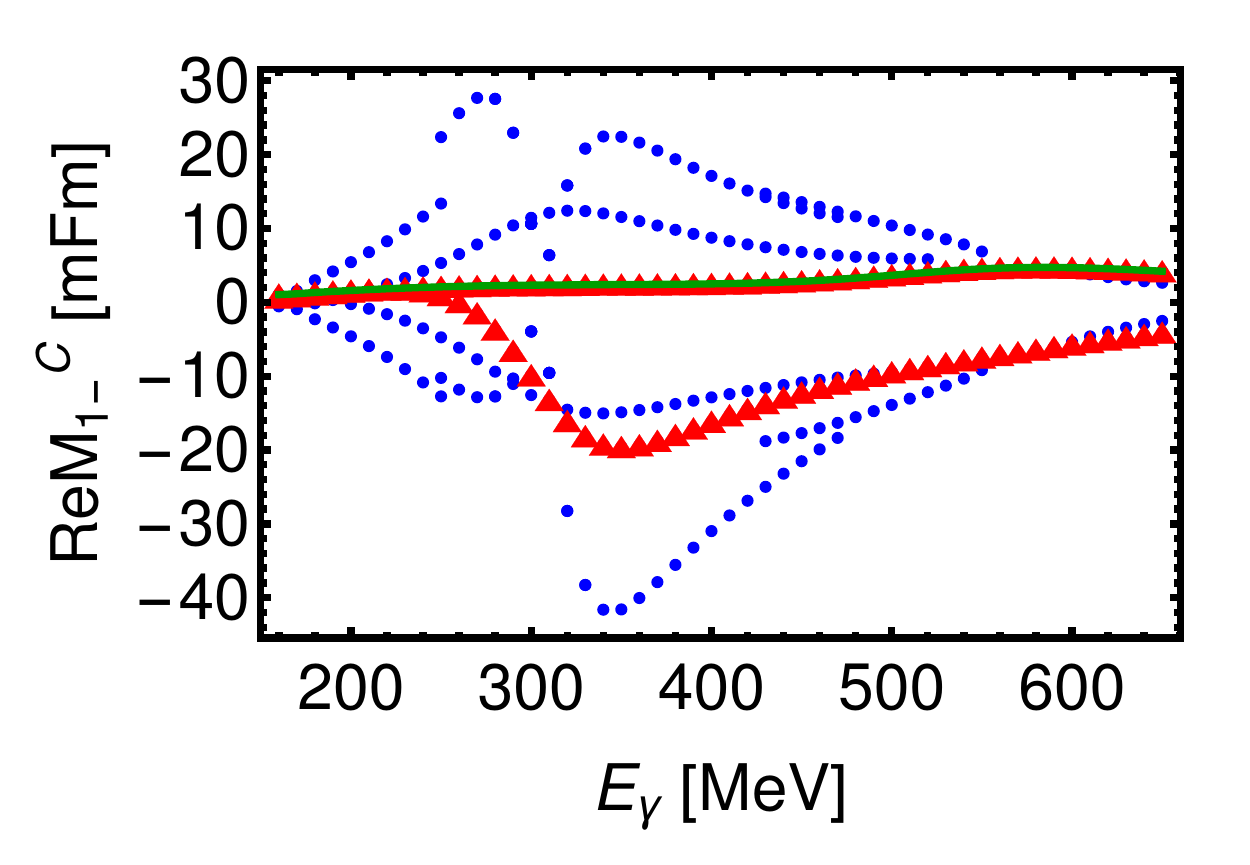}
 \end{overpic} \hspace*{5pt}
\begin{overpic}[width=0.475\textwidth]{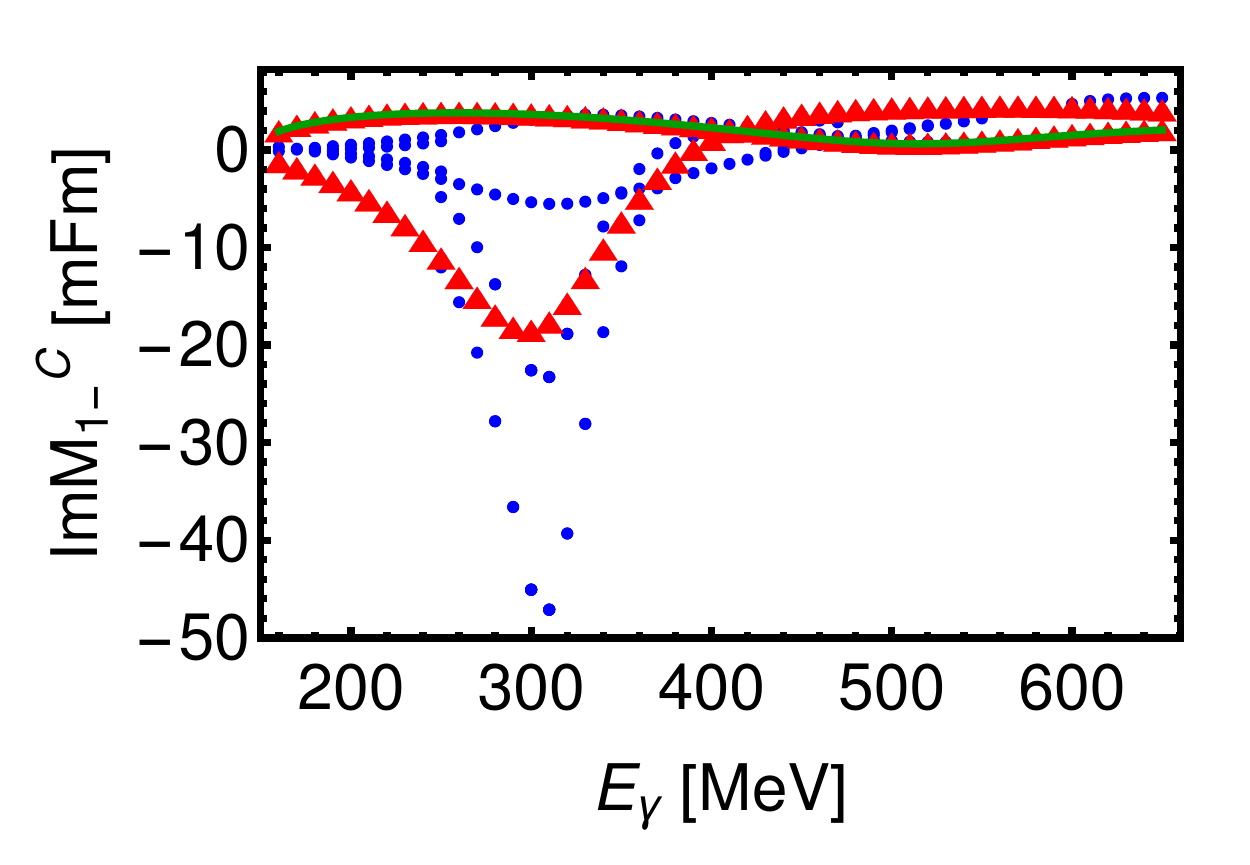}
 \end{overpic}
\vspace*{-5pt}
\caption[Solutions found in a TPWA fit to truncated MAID theory-data of the group $\mathcal{S}$ observables $\left\{ \sigma_{0}, \check{\Sigma}, \check{T}, \check{P} \right\}$, for $\ell_{\mathrm{max}}=1$.]{These figures depict the solutions found in the TPWA fit of the group $\mathcal{S}$ observables $\left\{ \sigma_{0}, \check{\Sigma}, \check{T}, \check{P} \right\}$. In each energy bin, two solutions are found that have an equally good value of $\Phi_{\mathcal{M}}$ (red triangles), typically in the parameter region of $10^{-15}$ to $10^{-16}$. Therefore, these solutions may be considered exact within a good numerical precision. One of them is equal to the original MAID2007 solution \cite{MAID2007,MAID} (green continuous line). The other one can be shown to equal the double ambiguity of the MAID multipoles. \newline
All remaining non redundant solutions are indicated by blue dots. Their value of $\Phi_{\mathcal{M}}$ is typically around $10$ orders of magnitude larger than the best solution.}
\label{fig:Lmax1ThDataFitBestSols}
\end{figure}
\clearpage

parameters in the minimization of this function. The matrices $\left( \mathcal{C}_{L}\right)_{k}^{\Omega^{\alpha}}$ can be read off from the expressions given in appendix \ref{sec:TPWAFormulae}.

The resulting solutions are shown in Figure \ref{fig:Lmax1ThDataFitBestSols}. Two equivalent best solutions, with values of the minimization function $\Phi_{\mathcal{M}}^{\mathrm{Best}}$ around $10^{-15}$ to $10^{-16}$, are found in each energy bin. For two example bins, numerical values of the non-redundant solutions are included in Table \ref{tab:Lmax1TheoryDataExampleBinsResults} in the appendix at the end of this section\footnote{In order to help the quickness of reference, important numerical results of the analyses discussed here have been collected in an appendix at the end of {\it this} section, in Tables \ref{tab:Lmax1TheoryDataExampleBinsResults} to \ref{tab:Lmax1TheoryDataAccAmbiguitiesNumbers2}.}. There, the two best solutions can also be seen. \newline

\begin{figure}[ht]
 \centering
 \vspace*{-10pt}
\begin{overpic}[width=0.475\textwidth]{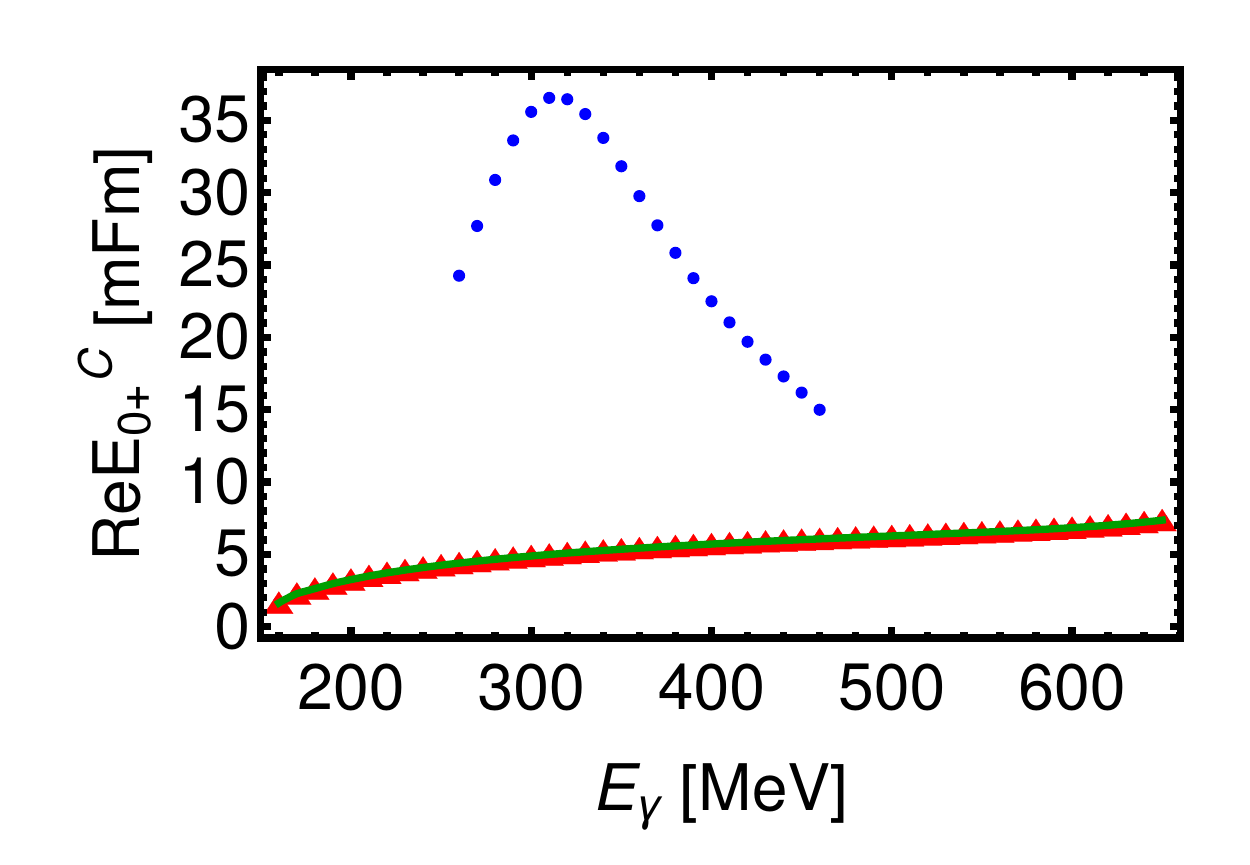}
 \end{overpic} \\
 \vspace*{7.5pt}
\begin{overpic}[width=0.475\textwidth]{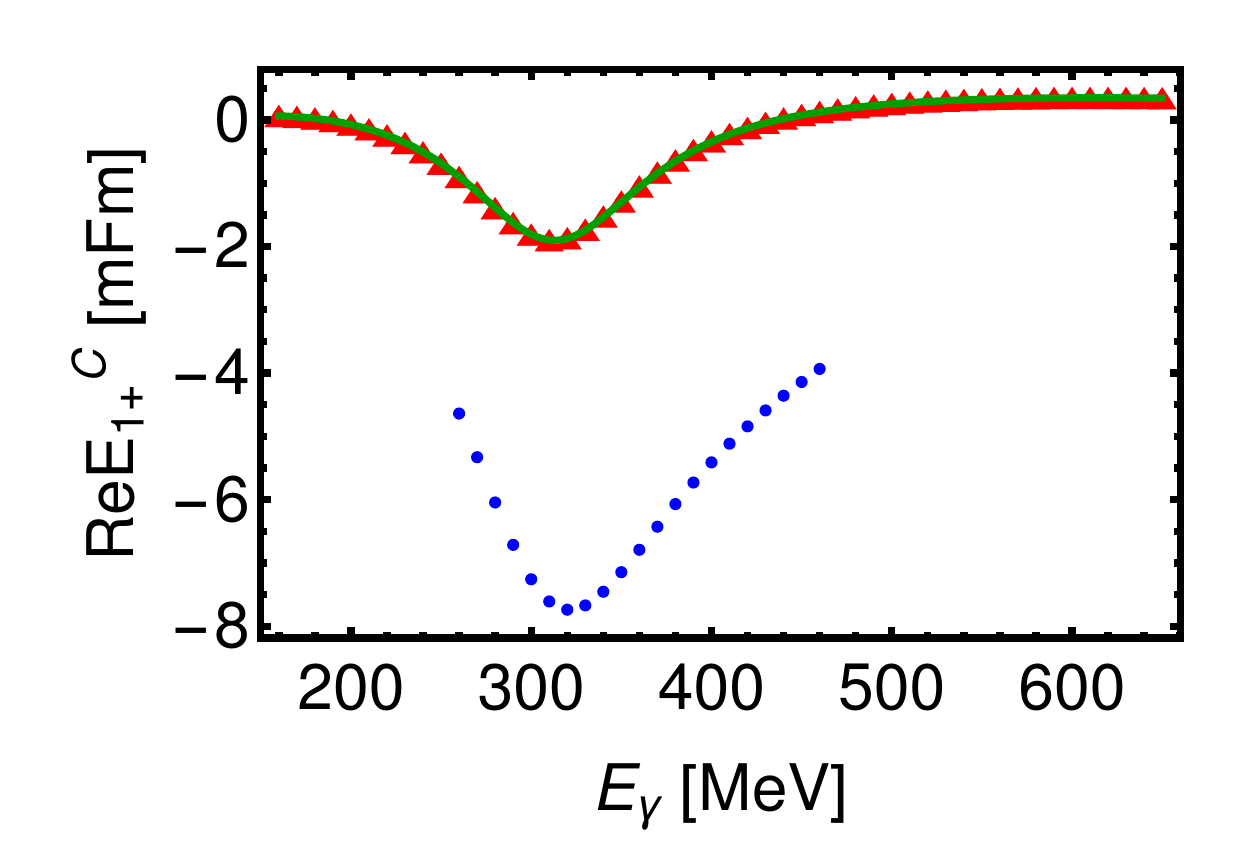}
 \end{overpic} \hspace*{5pt}
\begin{overpic}[width=0.475\textwidth]{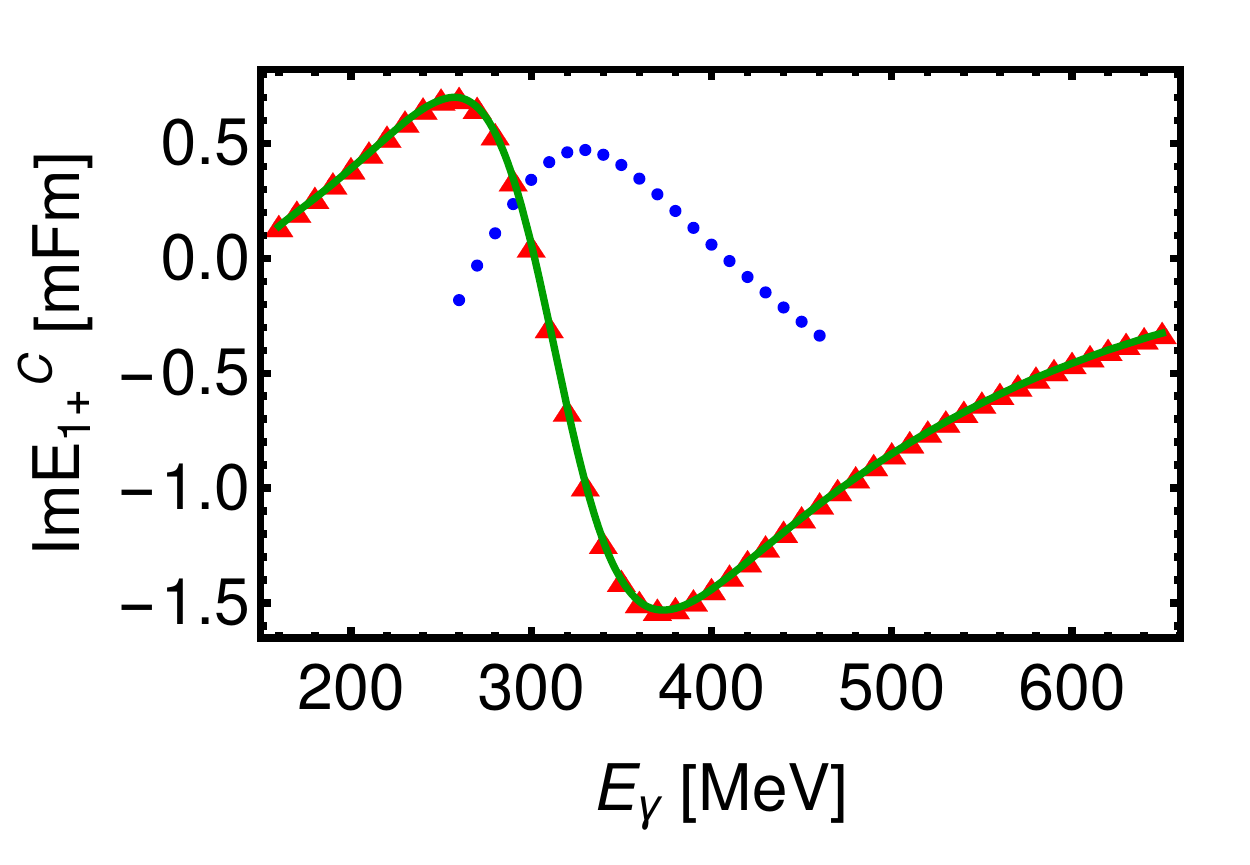}
 \end{overpic} \\
\begin{overpic}[width=0.475\textwidth]{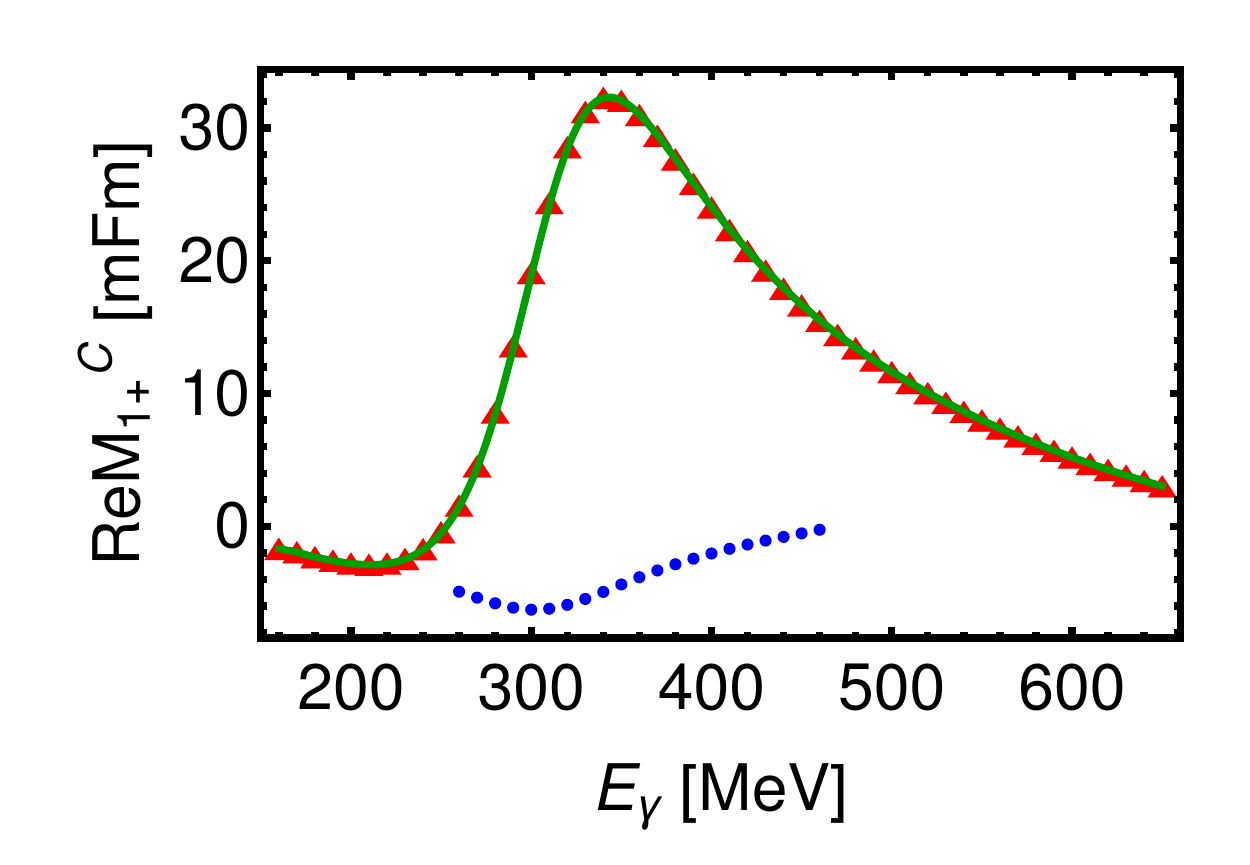}
 \end{overpic} \hspace*{5pt}
\begin{overpic}[width=0.475\textwidth]{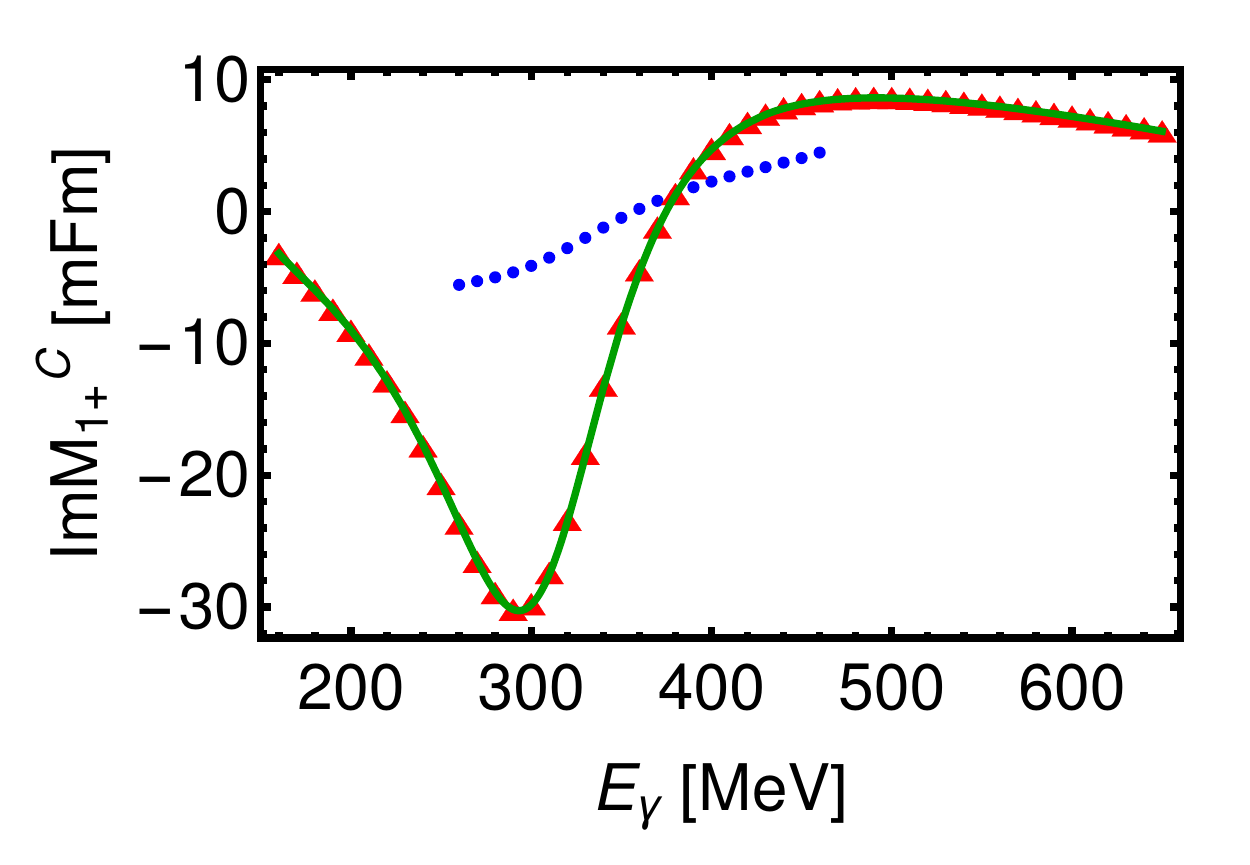}
 \end{overpic} \\
\begin{overpic}[width=0.475\textwidth]{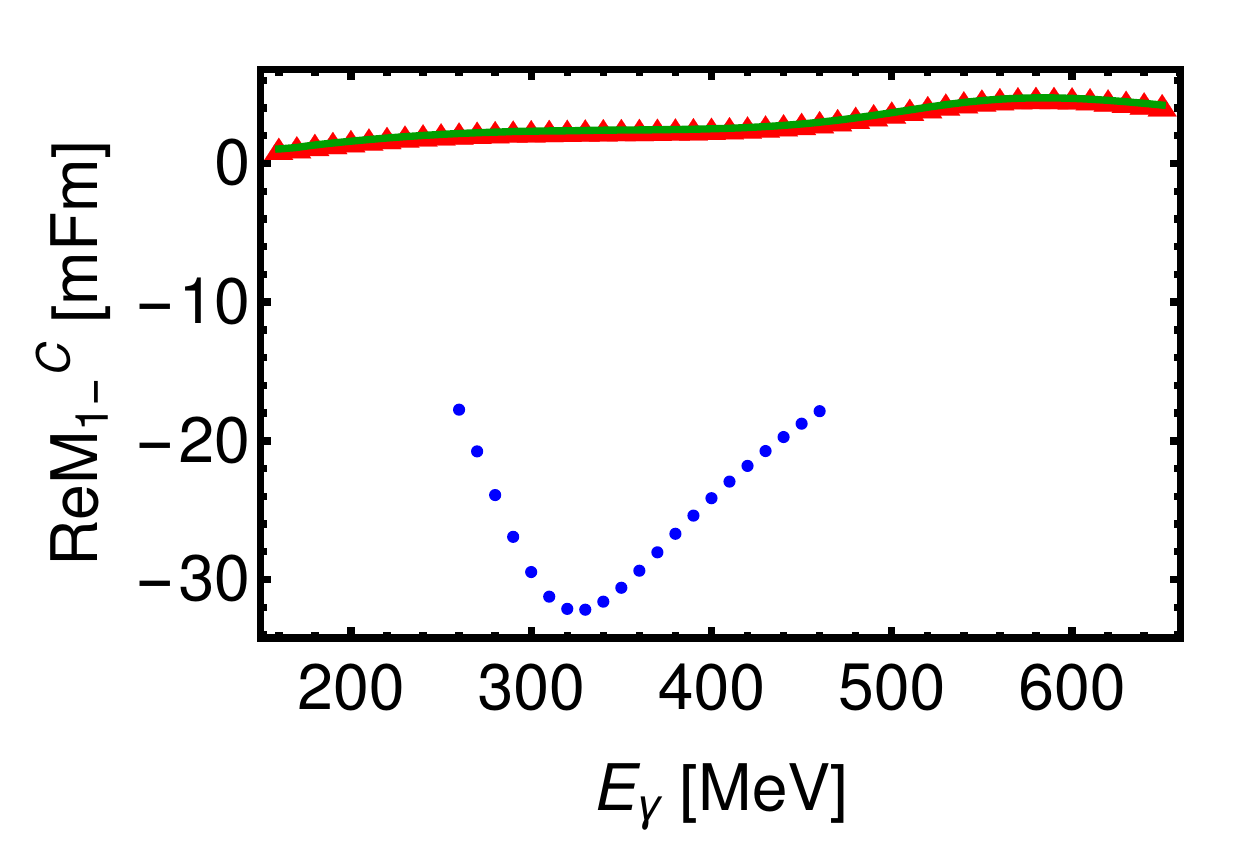}
 \end{overpic} \hspace*{5pt}
\begin{overpic}[width=0.475\textwidth]{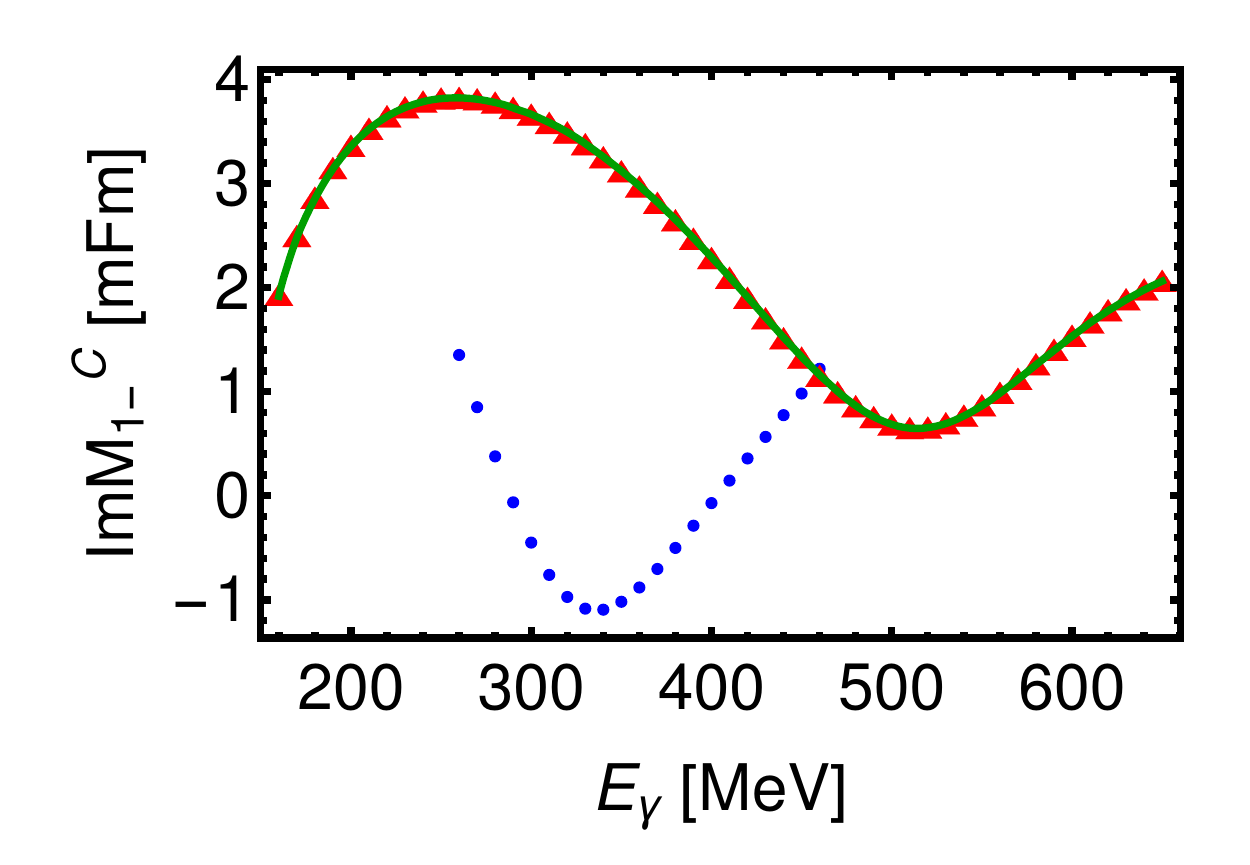}
 \end{overpic}
\vspace*{7.5pt}
\caption[Solutions found in a TPWA fit to truncated MAID theory-data of the observables $\left\{ \sigma_{0}, \check{\Sigma}, \check{T}, \check{P}, \check{F} \right\}$, for $\ell_{\mathrm{max}}=1$.]{Here, results of TPWA fits to truncated MAID theory-data \cite{LotharPrivateComm,MAID2007} are shown which include the observable $\check{F}$ in addition to the single spin observables. The latter have lead to the results shown in Figure \ref{fig:Lmax1ThDataFitBestSols}. A unique solution, corresponding precisely to the MAID multipoles \cite{MAID2007,MAID}, is recovered. Moreover, many local minima have been purged by inclusion of the $F$-observable.}
\label{fig:Lmax1ThDataFitBestSolsGroupSAndF}
\end{figure}

\begin{figure}[ht]
 \centering
 \vspace*{-10pt}
\begin{overpic}[width=0.475\textwidth]{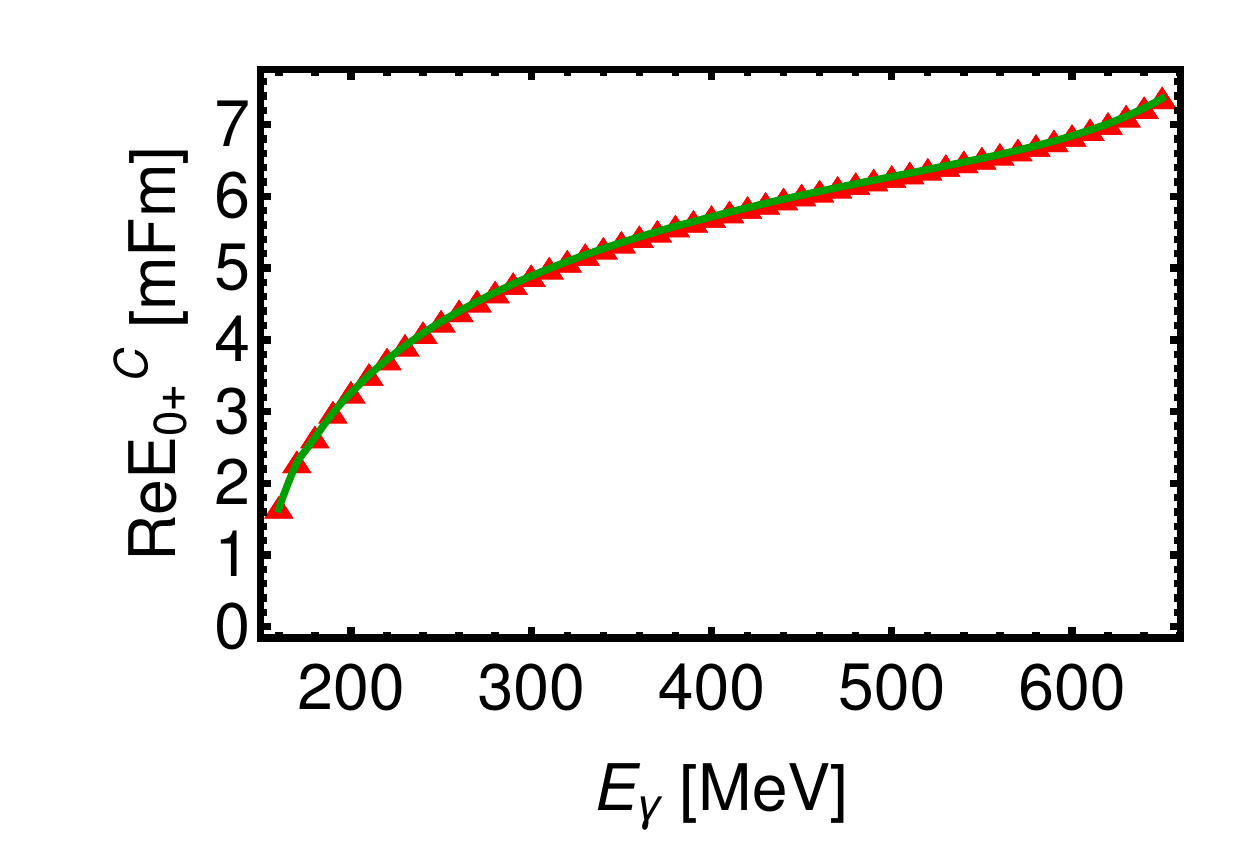}
 \end{overpic} \\
 \vspace*{7.5pt}
\begin{overpic}[width=0.475\textwidth]{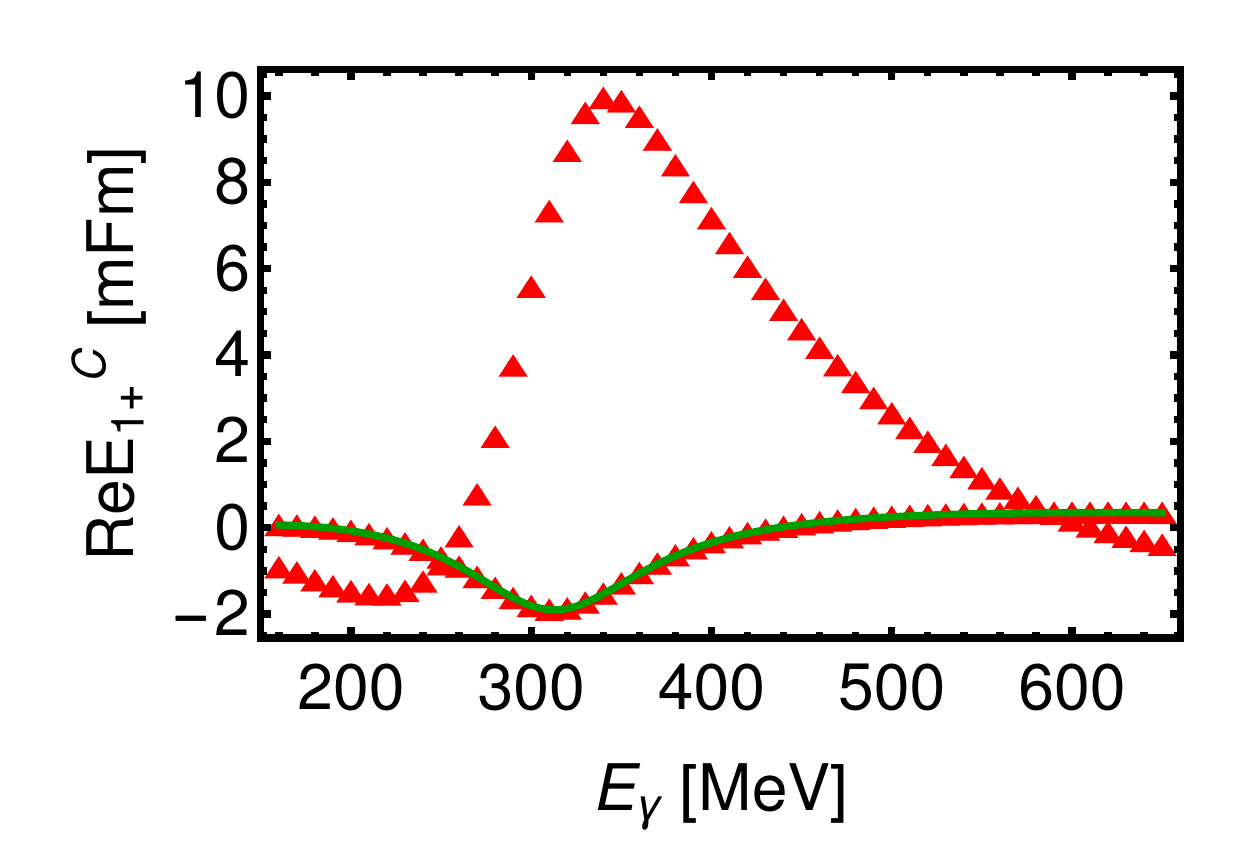}
 \end{overpic} \hspace*{5pt}
\begin{overpic}[width=0.475\textwidth]{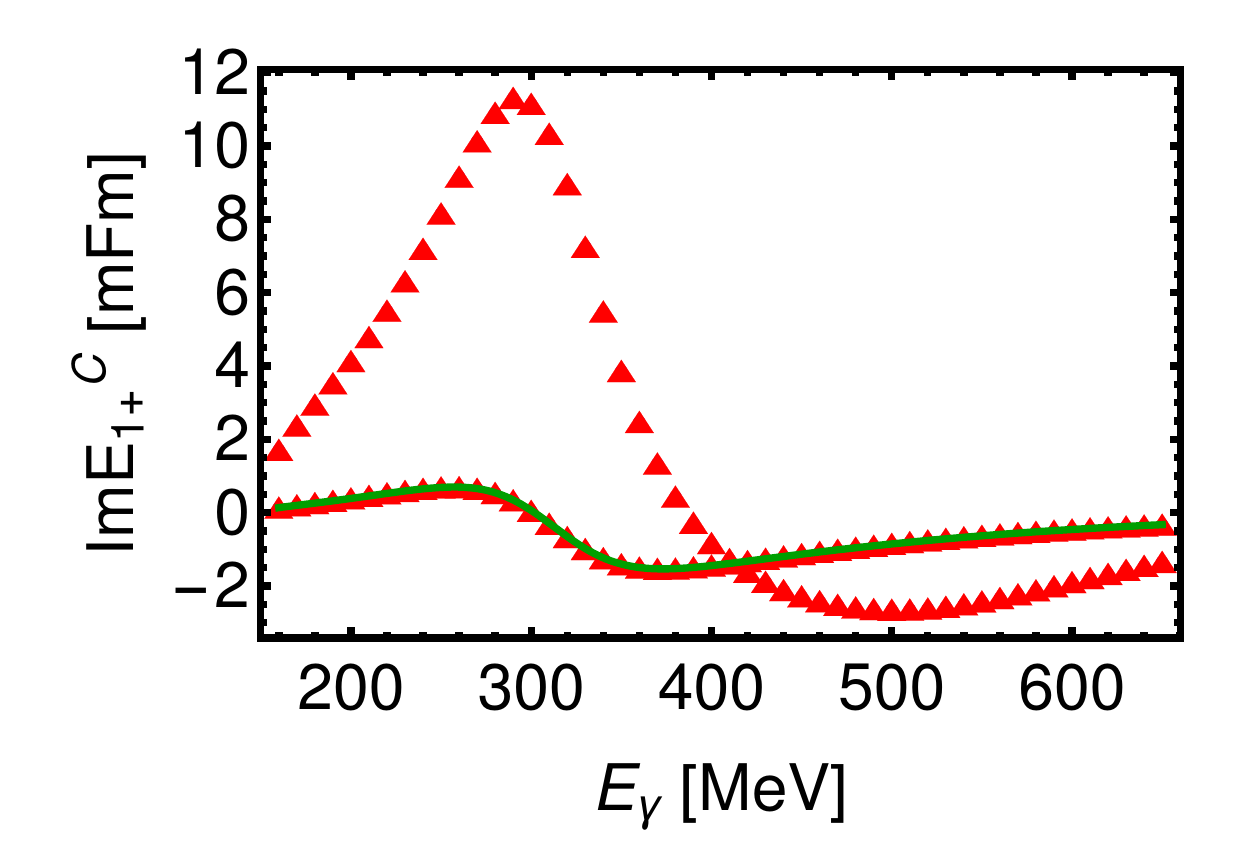}
 \end{overpic} \\
\begin{overpic}[width=0.475\textwidth]{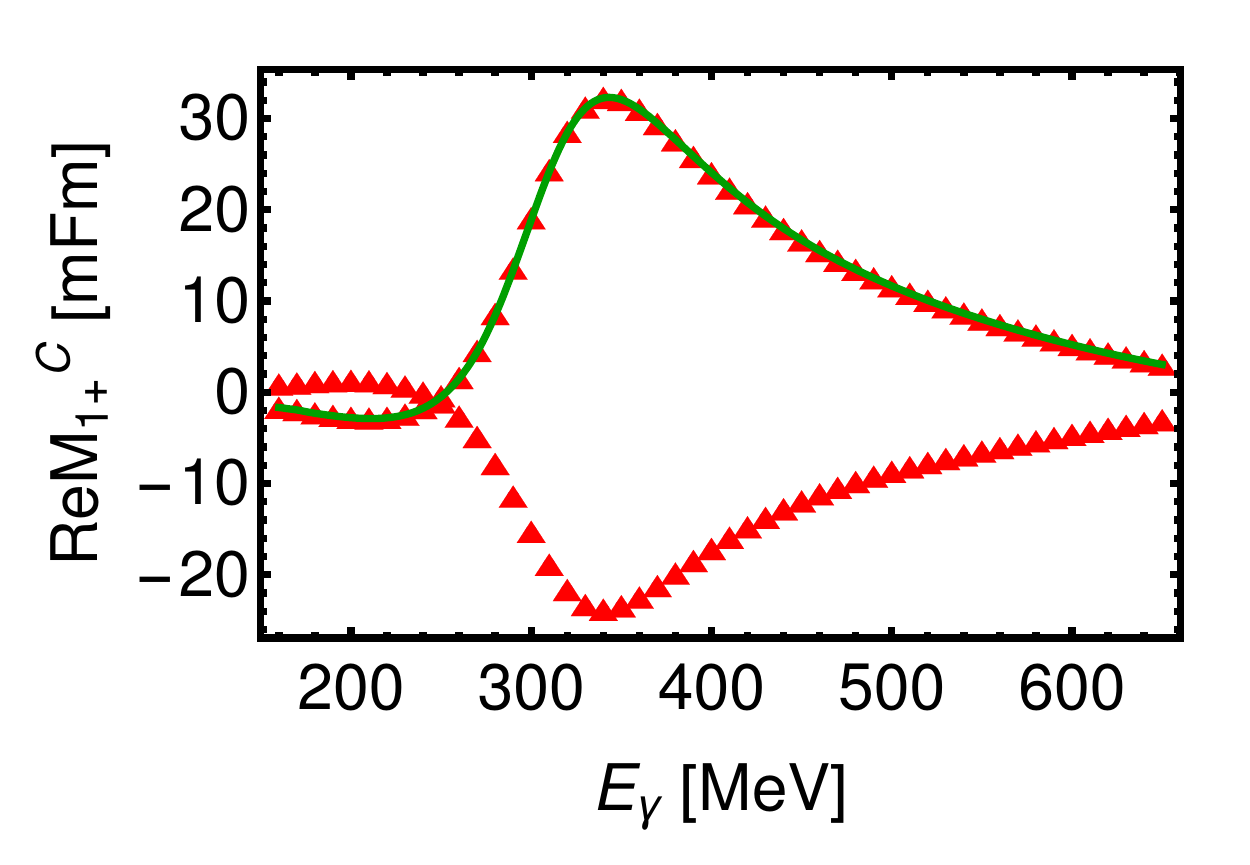}
 \end{overpic} \hspace*{5pt}
\begin{overpic}[width=0.475\textwidth]{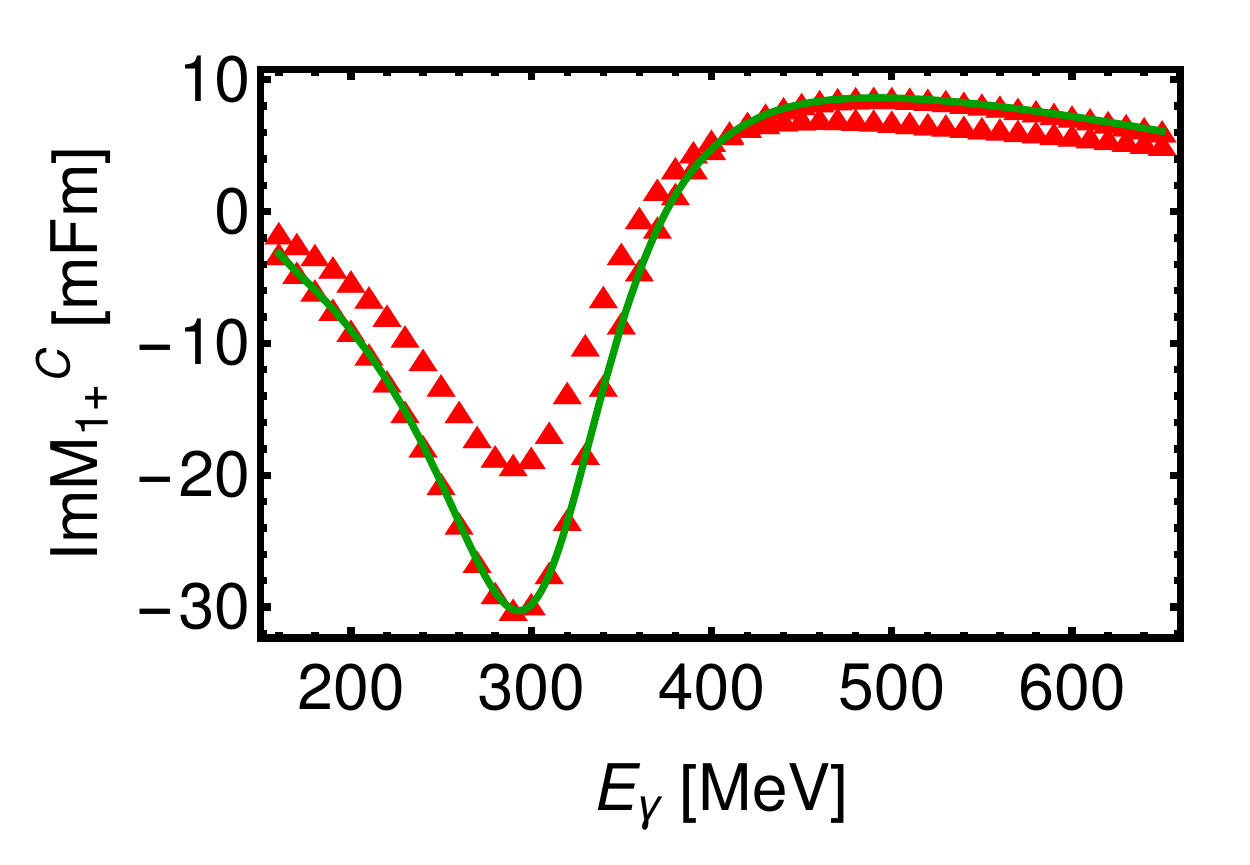}
 \end{overpic} \\
\begin{overpic}[width=0.475\textwidth]{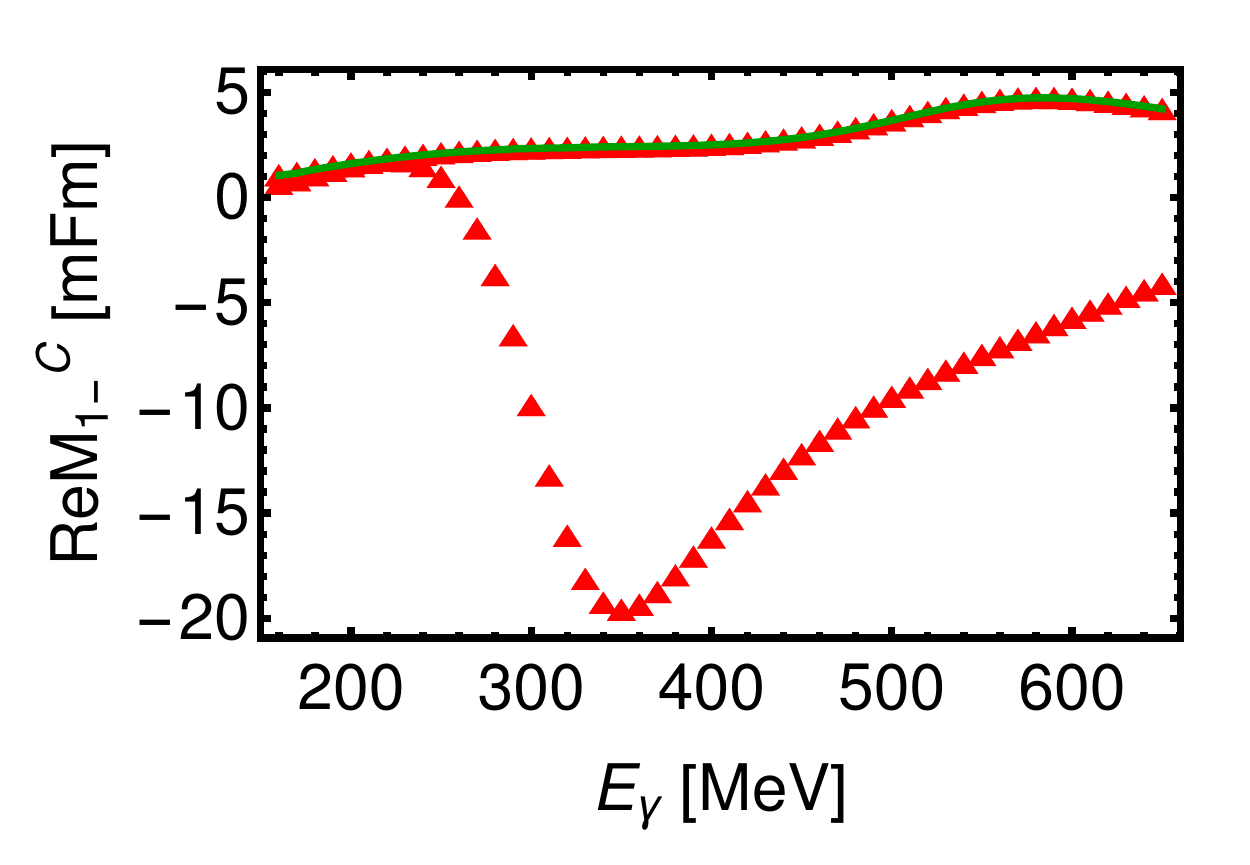}
 \end{overpic} \hspace*{5pt}
\begin{overpic}[width=0.475\textwidth]{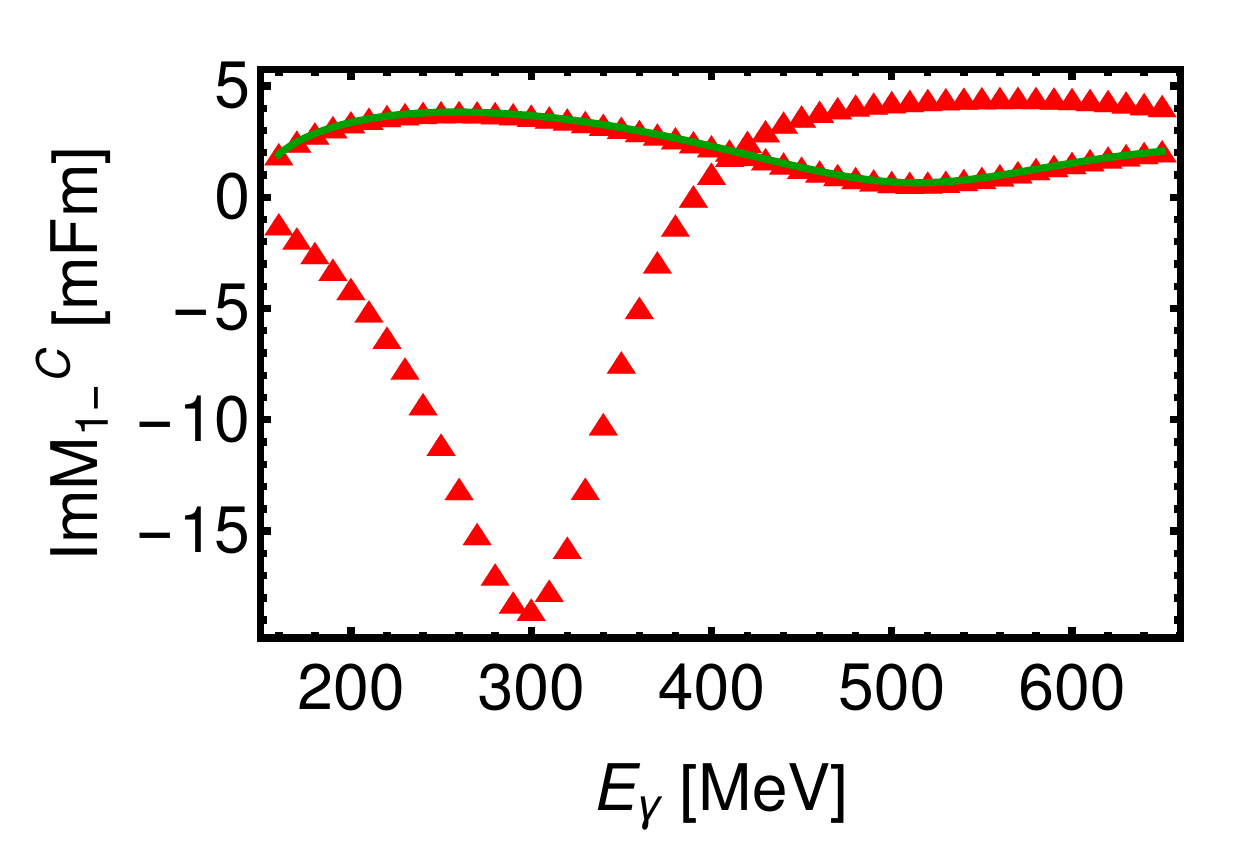}
 \end{overpic}
\vspace*{7.5pt}
\caption[Solutions found in a TPWA fit to truncated MAID theory-data of the observables $\left\{ \sigma_{0}, \check{\Sigma}, \check{T}, \check{P}, \check{E}, \check{H} \right\}$, for $\ell_{\mathrm{max}}=1$.]{These pictures show results of TPWA-fits to truncated MAID theory-data \cite{LotharPrivateComm,MAID2007}, for the case in which the $4$ single spin observables have been complemented by the $\mathcal{BT}$-observables $\check{E}$ and $\check{H}$. All local minima present as blue dots in Figure \ref{fig:Lmax1ThDataFitBestSols} are removed. Yet, the double ambiguity belonging to the true MAID solution remains, in accordance to the discussion in chapter \ref{chap:Omelaenko} and appendix \ref{sec:AdditionsChapterII}.}
\label{fig:Lmax1ThDataFitBestSolsGroupSAndEH}
\end{figure}

These parameter configurations are 'exact' solutions of the problem, but only within a finite numerical precision. One of those two solutions can be shown to coincide with the multipoles of the MAID model \cite{MAID2007,MAID} used to generate the theory-data. The other one is equal to the double ambiguity of the MAID solution. We will point out in more detail below how this can be proven. One can just pass from the resulting multipoles to Omelaenko-roots, conjugate all roots and then transform back to multipoles. \newline
Furthermore, several local minima are found in the group $\mathcal{S}$ fit, typically corresponding to a $\Phi_{\mathcal{M}}$ value at least $10$ orders of magnitude larger than $\Phi_{\mathcal{M}}^{\mathrm{Best}}$. Some of these solutions are already quite good. However, in case we require an exact solution to result in a minimization function of the order of $10^{-15}$, they are clearly not exact. Several of these local minima will be linked to accidental symmetries in a more detailed discussion of the group $\mathcal{S}$ fit later in this section. In addition, it has to be said that all minima, also the local ones, occur in pairs of equal $\Phi_{\mathcal{M}}$, due to the fact that the single spin observables are invariant under the double ambiguity transformation (discussed in chapter \ref{chap:Omelaenko} and appendix \ref{sec:AdditionsChapterII}). This latter fact leads to a doubling of all minima, not only the global minima. \newline \newline
As a next task, it is interesting to investigate whether or not one of the complete sets composed out of $5$ observables, suggested in section \ref{sec:WBTpaper}, can make this model-TPWA unique. We choose here to include the profile function $\check{F}$ in addition to the single spin observables and run a fit in complete analogy to the previous case. Here, also $N_{MC} = 1500$ was chosen. The results are shown in Figure \ref{fig:Lmax1ThDataFitBestSolsGroupSAndF}. \newline
It is seen that the inclusion of $\check{F}$ completely removes the double ambiguity. Furthermore, this $\mathcal{BT}$ observable has the capability to almost completely purge the resulting non redundant solutions of the local minima visible in the group $\mathcal{S}$ fit, cf. Figure \ref{fig:Lmax1ThDataFitBestSols}. Very few local minima remain and their values of $\Phi_{\mathcal{M}}$ are quite large, rendering them completely negligible. Therefore, we conclude that this (admittedly simple) model-TPWA is solved uniquely and the set $\left\{ \sigma_{0}, \check{\Sigma}, \check{T}, \check{P}, \check{F} \right\}$ is complete in this case. \newline \newline
As a final example, the $\mathcal{BT}$ observables $\check{E}$ and $\check{H}$ have been added to the single spin asymmetries. The results of the TPWA fit are shown in Figure \ref{fig:Lmax1ThDataFitBestSolsGroupSAndEH}. \newline
In this case, the true MAID solution and the double ambiguity remain as equally good solutions, almost indistinguishable in $\Phi_{\mathcal{M}}$. This could be expected, since $\check{E}$ and $\check{H}$ are invariant under the double ambiguity transformation as well. However, it can also be seen that all the local minima present in the group $\mathcal{S}$ fit have been removed by inclusion of the two $\mathcal{BT}$ asymmetries. This fact can tell something about the ability the observables $\check{H}$ and $\check{E}$ have in stabilizing TPWA fits, although both cannot resolve the double ambiguity.

\newpage


\clearpage

The remaining part of this section will be used to further explore the results of the fit to the group $\mathcal{S}$ observables $\left\{ \sigma_{0}, \check{\Sigma}, \check{T}, \check{P} \right\}$. Here, one has the chance to draw connections to the mathematical discussion on ambiguities contained in chapter \ref{chap:Omelaenko} and appendix \ref{sec:AdditionsChapterII}, in particular the statements about accidental ambiguities in appendix \ref{sec:AccidentalAmbProofs}. Since the model-TPWA here is still quite simple, the degree of complexity in the discussion remains manageable. \newline
As a further motivation, we again refer to the example solutions shown in Table \ref{tab:Lmax1TheoryDataExampleBinsResults} towards the end of this section. \newline
Here, the energy bins $E_{\gamma} = 510 \hspace*{2pt} \mathrm{MeV}$ and $E_{\gamma} = 610 \hspace*{2pt} \mathrm{MeV}$ are shown. All solutions occur in pairs, as mentioned above. In both examples, two best solutions are found, having values of $\Phi_{\mathcal{M}}$ compatible with zero (i.e. around $10^{-15}$ - $10^{-16}$). Furthermore, in each case a pair of solutions occurs that has a non-vanishing, but still quite small value for the discrepancy function $\Phi_{\mathcal{M}}$. To be more precise, the value is $\Phi_{\mathcal{M}} = 9.7 \times 10^{-5} \left( \mu b /\mathrm{sr} \right)^{2}$ for the $510 \hspace*{2pt} \mathrm{MeV}$ bin and $\Phi_{\mathcal{M}} = 1.0 \times 10^{-6} \left( \mu b /\mathrm{sr} \right)^{2}$ in the $610 \hspace*{2pt} \mathrm{MeV}$ bin. Two more solutions were even found at $E_{\gamma} = 510 \hspace*{2pt} \mathrm{MeV}$, but due to their relatively large value of $\Phi_{\mathcal{M}}$ they will be considered no longer. \newline
Both example energy bins of Table \ref{tab:Lmax1TheoryDataExampleBinsResults} were not chosen purely randomly. Rather, they have been picked in view to the discussion of accidental ambiguities in chapter \ref{chap:Omelaenko} and specifically in appendix \ref{subsec:AccidentalAmbProofsI}. \newline

For a further understanding of this section, some definitions are needed from a more precise formalization of the main results found in chapter \ref{chap:Omelaenko}. However, this formalization has been included in appendix \ref{sec:AccidentalAmbProofs}. In order to keep the discussion in this section self-contained, we cite the most important mathematical objects and statements made in appendices \ref{subsec:AccidentalAmbProofsI} and \ref{subsec:AccidentalAmbProofsII}, but for the special case $L \equiv \ell_{\mathrm{max}} = 1$. In the respective appendices, generalizations to arbitrary finite $L$ can be found. \newline
We assume a set of Omelaenko-roots $\left( \alpha_{k}, \beta_{k^{\prime}} \right)$ to correspond to one existing exact solution of the TPWA problem. For $L=1$, the indices $k$ and $k^{\prime}$ run from $1$ to $2$. In section \ref{sec:WBTpaper}, possible ambiguity transformations are described as 'complex conjugation of either all roots, or any subset of them'. This description can be brought into more precise terms using the numbering-scheme due to Gersten \cite{Gersten}, which in case of $L=1$ would introduce $16$ 'maps' $\bm{\uppi}_{\hspace*{0.035cm}n}$ for the respective ambiguity transformations, which act according to (see equations (\ref{eq:BinaryRepresentationOfN}) and (\ref{eq:PiNMapsDefinition}) of appendix \ref{subsec:AccidentalAmbProofsI})
\begin{equation}
 \bm{\uppi}_{\hspace*{0.035cm}n} \left(\alpha_{k}\right) := \begin{cases}
                    \alpha_{k} &\mathrm{,} \hspace*{3pt} \mu_{k} \left(n\right) = 0 \\
                    \alpha_{k}^{\ast} &\mathrm{,} \hspace*{3pt} \mu_{k} \left(n\right) = 1 
                   \end{cases}  \hspace*{5pt} \mathrm{and} \hspace*{5pt}
 \bm{\uppi}_{\hspace*{0.035cm}n} \left(\beta_{k^{\prime}}\right) := \begin{cases}
                    \beta_{k^{\prime}} &\mathrm{,} \hspace*{3pt} \nu_{k^{\prime}} \left(n\right) = 0 \\
                    \beta_{k^{\prime}}^{\ast} &\mathrm{,} \hspace*{3pt} \nu_{k^{\prime}} \left(n\right) = 1 
                   \end{cases} \mathrm{.} \label{eq:PiNMapsDefinitionMainText}
\end{equation}
All maps $\bm{\uppi}_{\hspace*{0.035cm}n}$ get assigned an index $n \in \left\{ 0,1,\ldots,15 \right\}$ unambiguously, which has the binary representation
\begin{equation}
 n = \sum_{k = 1}^{2} \mu_{k} \left( n \right) 2^{(k-1)} + \sum_{k^{\prime} = 1}^{2} \nu_{k^{\prime}} \left( n \right) 2^{(k^{\prime} + 1)} \mathrm{.} \label{eq:BinaryRepresentationOfNMainText}
\end{equation}
As an illustration, the assignments of indices $n$ to combinatorial possibilities of root-con\-ju\-ga\-tions can be seen in Table \ref{tab:AllCCPossibilitiesLEquals1CaseMainText}.\newline
It has been useful to introduce notations for two important different subsets of all ambiguity transformations. We have chosen the symbol $\mathcal{P}$ for all possible conjugations (\ref{eq:PiNMapsDefinitionMainText}) and the symbol $\hat{\mathcal{P}}$ for all transformations $\mathcal{P}$ except for the identity $\bm{\uppi}_{\hspace*{0.035cm}0}$ and the double ambiguity (i.e. conjugation of all roots) $\bm{\uppi}_{\hspace*{0.035cm}15}$, i.e. $\hat{\mathcal{P}} := \mathcal{P} \setminus \left\{ \bm{\uppi}_{\hspace*{0.035cm} 0 }, \bm{\uppi}_{\hspace*{0.035cm}15} \right\}$ (see appendix \ref{subsec:AccidentalAmbProofsI}).
\clearpage
\begin{table}[ht]
\centering
\begin{tabular}{c|cccc||c|cccc}
\hline
\hline
$n$ & $\bm{\uppi}_{\hspace*{0.035cm}n}\left(\alpha_{1}\right)$ & $\bm{\uppi}_{\hspace*{0.035cm}n}\left(\alpha_{2}\right)$ & $\bm{\uppi}_{\hspace*{0.035cm}n}\left(\beta_{1}\right)$ & $\bm{\uppi}_{\hspace*{0.035cm}n}\left(\beta_{2}\right)$ & $n$ & $\bm{\uppi}_{\hspace*{0.035cm}n}\left(\alpha_{1}\right)$ & $\bm{\uppi}_{\hspace*{0.035cm}n}\left(\alpha_{2}\right)$ & $\bm{\uppi}_{\hspace*{0.035cm}n}\left(\beta_{1}\right)$ & $\bm{\uppi}_{\hspace*{0.035cm}n}\left(\beta_{2}\right)$  \\
\hline
 $0$ & $\alpha_{1}$ & $\alpha_{2}$ & $\beta_{1}$ & $\beta_{2}$ & $8$ & $\alpha_{1}$ & $\alpha_{2}$ & $\beta_{1}$ & $\beta_{2}^{\ast}$ \\
 $1$ & $\alpha_{1}^{\ast}$ & $\alpha_{2}$ & $\beta_{1}$ & $\beta_{2}$ & $9$ & $\alpha_{1}^{\ast}$ & $\alpha_{2}$ & $\beta_{1}$ & $\beta_{2}^{\ast}$ \\
 $2$ & $\alpha_{1}$ & $\alpha_{2}^{\ast}$ & $\beta_{1}$ & $\beta_{2}$ & $10$ & $\alpha_{1}$ & $\alpha_{2}^{\ast}$ & $\beta_{1}$ & $\beta_{2}^{\ast}$ \\
 $3$ & $\alpha_{1}^{\ast}$ & $\alpha_{2}^{\ast}$ & $\beta_{1}$ & $\beta_{2}$ & $11$ & $\alpha_{1}^{\ast}$ & $\alpha_{2}^{\ast}$ & $\beta_{1}$ & $\beta_{2}^{\ast}$ \\
\hline
 $4$ & $\alpha_{1}$ & $\alpha_{2}$ & $\beta_{1}^{\ast}$ & $\beta_{2}$ & $12$ & $\alpha_{1}$ & $\alpha_{2}$ & $\beta_{1}^{\ast}$ & $\beta_{2}^{\ast}$ \\
 $5$ & $\alpha_{1}^{\ast}$ & $\alpha_{2}$ & $\beta_{1}^{\ast}$ & $\beta_{2}$ & $13$ & $\alpha_{1}^{\ast}$ & $\alpha_{2}$ & $\beta_{1}^{\ast}$ & $\beta_{2}^{\ast}$ \\
 $6$ & $\alpha_{1}$ & $\alpha_{2}^{\ast}$ & $\beta_{1}^{\ast}$ & $\beta_{2}$ & $14$ & $\alpha_{1}$ & $\alpha_{2}^{\ast}$ & $\beta_{1}^{\ast}$ & $\beta_{2}^{\ast}$ \\
 $7$ & $\alpha_{1}^{\ast}$ & $\alpha_{2}^{\ast}$ & $\beta_{1}^{\ast}$ & $\beta_{2}$ & $15$ & $\alpha_{1}^{\ast}$ & $\alpha_{2}^{\ast}$ & $\beta_{1}^{\ast}$ & $\beta_{2}^{\ast}$ \\
\hline
\hline
\end{tabular}
\caption[All possibilities for complex conjugation of Omelaenko-roots, for $L=1$, shown in the main text.]{All possibilities for complex conjugations of roots, numbered according to the binary representation (\ref{eq:BinaryRepresentationOfNMainText}) and the rule (\ref{eq:PiNMapsDefinitionMainText}), for a truncation at $L = 1$. This Table can be found as well in appendix \ref{subsec:AccidentalAmbProofsI}, where transformations are shown as acting of the phases of the roots and not the roots themselves.}
\label{tab:AllCCPossibilitiesLEquals1CaseMainText}
\end{table}
The set $\hat{\mathcal{P}}$ contains the candidates for what have been called {\it accidental ambiguities} in section \ref{sec:WBTpaper}. \newline
Omelaenko's multiplicative consistency-constraint among the roots (see reference \cite{Omelaenko} and equation (40) in section \ref{sec:WBTpaper}) reads in this case here, i.e. for $L=1$:
\begin{equation}
 \alpha_{1} \alpha_{2} = \beta_{1} \beta_{2} \mathrm{.} \label{eq:OmelaenkoConstConstraintMainTextChap5}
\end{equation}
As argued in more detail in appendix \ref{subsec:AccidentalAmbProofsII}, a numerical TPWA-fit which is performed using multipoles, and {\it not} the roots $\left( \alpha_{k}, \beta_{k^{\prime}} \right)$, has this constraint already implemented implicitly and is {\it not} able to violate it. \newline
However, as argued with some rough probabilistic arguments in appendix \ref{subsec:AccidentalAmbProofsIII}, it is very likely that every $\bm{\uppi} \in \hat{\mathcal{P}}$, when applied to the roots $\left( \alpha_{k}, \beta_{k^{\prime}} \right)$, violates Omelaenko's constraint (\ref{eq:OmelaenkoConstConstraintMainTextChap5}). In case this violation is 'small', we choose to parametrize it by a {\it violation parameter} $\epsilon_{\bm{\uppi}}$, with $\epsilon_{\bm{\uppi}} > 0$, $\epsilon_{\bm{\uppi}} \ll 1$, and $\bm{\uppi} \left( \alpha_{1} \right) \hspace*{1pt} \bm{\uppi} \left( \alpha_{2} \right) = \bm{\uppi} \left( \beta_{1} \right) \hspace*{1pt} \bm{\uppi} \left( \beta_{2} \right) e^{i \epsilon_{\bm{\uppi}}}$ (cf. appendices \ref{subsec:AccidentalAmbProofsI} and \ref{subsec:AccidentalAmbProofsII}, in particular equation (\ref{eq:GeneralizedViolatedCRWithNIndex})). \newline
Now, in case a $\bm{\uppi} \in \hat{\mathcal{P}}$ with $\epsilon_{\bm{\uppi}} \ll 1$ exists, the following behavior of the TPWA-fit has been postulated heuristically in appendix \ref{subsec:AccidentalAmbProofsII}. Since the fit itself cannot violate Omelaenko's constraint (\ref{eq:OmelaenkoConstConstraintMainTextChap5}), it instead adopts a multipole-solution which corresponds to a set of roots $\bm{\tilde{\uppi}} \left( \alpha_{k} \right)$ and $\bm{\tilde{\uppi}} \left( \beta_{k^{\prime}} \right)$, such that the $\bm{\tilde{\uppi}} \left( \alpha_{k}, \beta_{k^{\prime}} \right)$ are {\it close} to the $\bm{\uppi} \left( \alpha_{k}, \beta_{k^{\prime}} \right)$ in root-space and the constraint (\ref{eq:OmelaenkoConstConstraintMainTextChap5}) is {\it restored}, i.e.:
\begin{equation}
 \bm{\tilde{\uppi}} \left( \alpha_{1} \right) \bm{\tilde{\uppi}} \left( \alpha_{2} \right) = \bm{\tilde{\uppi}} \left( \beta_{1} \right) \bm{\tilde{\uppi}} \left( \beta_{2} \right) \mathrm{.} \label{eq:OmelCOnstrRestoredMainText}
\end{equation}
As mentioned above, this behavior had to be postulated and has up to now not been substantiated by a proof, but it reflects accurately what is found in generic TPWAs, at least for theory-data. \newline
In order to quantify the notion of 'closeness' between the roots $\bm{\tilde{\uppi}} \left( \alpha_{k}, \beta_{k^{\prime}} \right)$ and $\bm{\uppi} \left( \alpha_{k}, \beta_{k^{\prime}} \right)$, we chose in appendix \ref{subsec:AccidentalAmbProofsII} to introduce an exponential parametrization by writing $\bm{\tilde{\uppi}} \left( \alpha_{k} \right) \equiv e^{\xi_{k}} \bm{\uppi} \left( \alpha_{k} \right)$ and $\bm{\tilde{\uppi}} \left( \beta_{k^{\prime}} \right) \equiv e^{\zeta_{k^{\prime}}} \bm{\uppi} \left( \beta_{k^{\prime}} \right)$, using infinitesimal complex parameters $\xi_{k}$ and $\zeta_{k^{\prime}}$ (cf. equation (\ref{eq:ClosenessOfRootsExpressed3}) in appendix \ref{subsec:AccidentalAmbProofsII}). Since (\ref{eq:OmelCOnstrRestoredMainText}) has to hold, the $\left( \xi_{k}, \zeta_{k^{\prime}} \right)$ are {\it not} independent, but satisfy the constraint (cf. appendix \ref{subsec:AccidentalAmbProofsII}, especially equation (\ref{eq:RestoringParametersConstraint}))
\begin{equation}
- \xi_{1} - \xi_{2} + \zeta_{1} + \zeta_{2} = i \epsilon_{\bm{\uppi}} \mathrm{.}
\end{equation}
As argued in a lot more detail in appendix \ref{subsec:AccidentalAmbProofsII}, the roots $\bm{\tilde{\uppi}} \left( \alpha_{k}, \beta_{k^{\prime}} \right)$ adopted by the TPWA-fit are not an exact symmetry of the group $\mathcal{S}$ observables any more! An equivalent statement of this fact is to say that, assuming that the roots $\left( \alpha_{k}, \beta_{k^{\prime}} \right)$, which represent the assumed 'exact' TPWA-solution, lead to a minimum in the discrepancy function $\Phi_{\mathcal{M}}$ of $\Phi_{\mathcal{M}}^{\mathrm{Best}} \simeq 10^{-16} \equiv 0$ within a fit to the group $\mathcal{S}$ observables, this minimum receives a positive correction in case the solution corresponding to the roots $\bm{\tilde{\uppi}} \left( \alpha_{k}, \beta_{k^{\prime}} \right)$ is found: $\Phi_{\mathcal{M}}^{\bm{\tilde{\uppi}}} = \Phi_{\mathcal{M}}^{\mathrm{Best}} + \delta \Phi_{\mathcal{M}}^{\bm{\tilde{\uppi}}}$ (see equation (\ref{eq:DeltaDiscrFunct}), Table \ref{tab:ThreeCases} and Figure \ref{fig:ChiSquareValleyCartoon} in appendix \ref{subsec:AccidentalAmbProofsII}). \newline
An approximation-formula for the correction $\delta \Phi_{\mathcal{M}}^{\bm{\tilde{\uppi}}}$, or just $\delta \Phi$ for short, has been derived in appendix \ref{subsec:AccidentalAmbProofsII}, which is linear in the moduli $\left| \xi_{k} \right|$ and $\left| \zeta_{k^{\prime}} \right|$. This approximation will be denoted as $\delta \Phi^{\mathrm{calc.}}_{\mathrm{lin.}}$ in the following. \newline 
This concludes the introduction of the required additional formalities. We continue our discussion of the fit to the MAID theory-data \cite{LotharPrivateComm,MAID2007} for the group $\mathcal{S}$ observables, in particular for the energies of interest, $E_{\gamma} = 510 \hspace*{2pt} \mathrm{MeV}$ and $E_{\gamma} = 610 \hspace*{2pt} \mathrm{MeV}$, mentioned above. \newline

Examples for accidental symmetries at the above mentioned two energies can be inferred from Figure 2 in section \ref{sec:WBTpaper} and they are further clarified in Figure \ref{fig:ExampleEpsilonPiParametersInAmbiguityDiagram} of appendix \ref{subsec:AccidentalAmbProofsI}, where the corresponding small violation parameters $\epsilon_{\bm{\uppi}}$ are illustrated as well. Specifically, one encounters here the ambiguities $\bm{\uppi}_{10} = (+,-,+,-)$ $(510 \hspace*{2pt} \mathrm{MeV})$ and $\bm{\uppi}_{9} = (-,+,+,-)$ $(610 \hspace*{2pt} \mathrm{MeV})$, defined by their action on the phases of the Omelaenko-roots (cf. Table \ref{tab:AllCCPossibilitiesLEquals1CaseMainText} above and Table \ref{tab:AllCCPossibilitiesLEquals1Case} of appendix \ref{subsec:AccidentalAmbProofsI}). These two ambiguities are very likely related to the second best minima listed in Table \ref{tab:Lmax1TheoryDataExampleBinsResults}. But how can the relations between local minima and accidental ambiguities be sorted out {\it numerically} for all relevant cases, at least in the context of an exactly solvable model-TPWA? \newline
As a first step towards an answer, it is useful to recapitulate the conditions under which the connection between multipole parameters and Omelaenko-roots (as well as the normalization coefficient $\left| a_{2L} \right|$) is really bijective. The first set of parameters are what we fit here, while the second set is better suited for a discussion of the ambiguity problem. The transition from multipoles to Omelaenko-parameters is always possible, i.e. one can always go in the direction
\begin{equation}
 \left( \mathcal{M}_{\ell}^{C} \right) \rightarrow  \left( \left| a_{2L} \right| ; \alpha_{k}, \beta_{k} \right) \mathrm{.} \label{eq:PossibleToPassFromMultsToRoots}
\end{equation}
The equations to use for this step are given in the appendix of the paper \cite{MyCompExTPWAPaper} shown in section \ref{sec:WBTpaper}. We quote them here again for convenience. The normalized coefficients $(\hat{a}_{0}, \hat{a}_{1})$ and $(\hat{b}_{0}, \hat{b}_{1})$, defining the normalized polynomials $A_{2L} (t)$ and $B_{2L} (t)$ in Omelaenko's reformulation of the TPWA problem \cite{Omelaenko} for $L = 1$, as well as the overall normalization coefficient $a_{2}$, can be evaluated in terms of phase constrained multipoles according to the definitions (see also \cite{Omelaenko})
\allowdisplaybreaks
\begin{align}
 a_{2} = E_{0+}^{C} - 3 E_{1+}^{C} - M_{1+}^{C} + M_{1-}^{C} &\mathrm{,} \hspace*{5pt} \hat{a}_{1} = 2 i \frac{2 M_{1+}^{C} + M_{1-}^{C}}{ E_{0+}^{C} - 3 E_{1+}^{C} - M_{1+}^{C} + M_{1-}^{C} }\mathrm{,} \label{eq:CoeffsDefs1} \\
 \hat{a}_{0} = \hat{b}_{0} = \frac{E_{0+}^{C} + 3 E_{1+}^{C} + M_{1+}^{C} - M_{1-}^{C}}{ E_{0+}^{C} - 3 E_{1+}^{C} - M_{1+}^{C} + M_{1-}^{C} } &\mathrm{,} \hspace*{5pt} \hat{b}_{1} = 2 i \frac{ 3 E_{1+}^{C} - M_{1+}^{C} + M_{1-}^{C}}{ E_{0+}^{C} - 3 E_{1+}^{C} - M_{1+}^{C} + M_{1-}^{C} } \mathrm{.} \label{eq:CoeffsDefs2} 
\end{align}
From these definitions, it is immediately apparent that $a_{2}$ carries the full information on the energy dependent overall phase. A rotation by this phase would just drop out of the remaining definitions of the normalized polynomial coefficients. Therefore, $a_{2}$ may be chosen purely real. Only its modulus is important, which can be extracted from the definition above. The Omelaenko-roots $\left(\alpha_{k}, \beta_{k}\right)$ can now be calculated as roots of the polynomials $A_{2L} (t)$ and $B_{2L} (t)$:
\allowdisplaybreaks
\begin{align}
 \left\{ \alpha_{1}, \alpha_{2} \right\} &\equiv \mathrm{Roots} \left( t^{2} + \hat{a}_{1} t + \hat{a}_{0} = 0 \right) \mathrm{,} \label{eq:AlphaRootDefChapter4} \\
 \left\{ \beta_{1}, \beta_{2} \right\} &\equiv \mathrm{Roots} \left( t^{2} + \hat{b}_{1} t + \hat{b}_{0} = 0 \right) \mathrm{,} \label{eq:BetaRootDefChapter4}
\end{align}
which is a task that is quickly performed here by MATHEMATICA \cite{Mathematica8,Mathematica11,MathematicaLanguage,MathematicaBonnLicense}. \newline
The inverse direction, i.e. going from Omelaenko parameters to multipoles, can in the most general case not be performed, i.e.
\begin{equation}
 \left( \left| a_{2L} \right| ; \alpha_{k}, \beta_{k} \right) \not\rightarrow \left( \mathcal{M}_{\ell}^{C} \right) \mathrm{.} \label{eq:NotPossibleToPassFromRootsToMults}
\end{equation}
A necessary and sufficient condition for this to be possible is given by the fulfillment of the multiplicative constraint for the roots (cf. equation (40) in section \ref{sec:WBTpaper})
\begin{equation}
 \left( \left| a_{2L} \right| ; \alpha_{k}, \beta_{k} \right) \rightarrow \left( \mathcal{M}_{\ell}^{C} \right) \mathrm{,} \hspace*{2pt} \mathrm{if} \hspace*{2pt} \mathrm{and} \hspace*{2pt} \mathrm{only} \hspace*{2pt} \mathrm{if} \hspace*{2pt} \prod_{i} \alpha_{i} = \prod_{j} \beta_{j} \mathrm{.} \label{eq:SpecialPossibilityToPassFromRootsToMults}
\end{equation}
Since the direction (\ref{eq:PossibleToPassFromMultsToRoots}) can always be followed and the TPWA fit procedure for the group $\mathcal{S}$ fit has produced a pool of non redundant solutions for the phase-constrained multipoles, it is always possible to evaluate the parameters
\begin{equation}
 \left( \left| a_{2}^{(i)} \right| ; \alpha_{1}^{(i)}, \alpha_{2}^{(i)}, \beta_{1}^{(i)}, \beta_{2}^{(i)}  \right) \mathrm{,} \hspace*{2pt} i = 1,\ldots,N_{\mathrm{nonred}} \mathrm{,} \label{eq:OmelParametersSolutionPool}
\end{equation}
from all solutions in the pool. Once this is done for all solutions, it is possible to uniquely identify the roots corresponding to the best solution (i.e. with smallest $\Phi_{\mathcal{M}}$, which is known), i.e.
\begin{equation}
 \left( \left| a_{2}^{\mathrm{Best}} \right| ; \alpha_{1}^{\mathrm{Best}}, \alpha_{2}^{\mathrm{Best}}, \beta_{1}^{\mathrm{Best}}, \beta_{2}^{\mathrm{Best}}  \right) \mathrm{.} \label{eq:BestSolRoots}
\end{equation}
Both steps are performed using the solution pool obtained in the group $\mathcal{S}$ fit. Then, it is possible to obtain all {\it exact} ambiguities of the single spin observables by acting on the roots of the best solution (\ref{eq:BestSolRoots}) with all relevant ambiguity transformations listed in Table \ref{tab:AllCCPossibilitiesLEquals1CaseMainText}. In the numbering scheme introduced above (equations (\ref{eq:PiNMapsDefinitionMainText}) and (\ref{eq:BinaryRepresentationOfNMainText})) as well as in appendix \ref{subsec:AccidentalAmbProofsI}, this procedure results in ambiguity-roots
\begin{equation}
 \bm{\uppi}_{\hspace*{0.035cm}n} \left( \alpha_{1}^{\mathrm{Best}}, \alpha_{2}^{\mathrm{Best}}, \beta_{1}^{\mathrm{Best}}, \beta_{2}^{\mathrm{Best}} \right) \mathrm{,} \hspace*{2pt} n = 1,\ldots,14 \mathrm{,} \label{eq:AccAmbTrafosBestSolution}
\end{equation}
in each energy bin. The modulus $\left| a_{2}^{\mathrm{Best}} \right|$ is unaffected by these ambiguity transformations. \newline
Having obtained all the candidates for exact accidental symmetries it is, as stated above, very likely that they violate the multiplicative constraint in equation (\ref{eq:SpecialPossibilityToPassFromRootsToMults}) by a small violation parameter $\epsilon_{\bm{\uppi}}$ (see the discussion above, as well as appendices \ref{subsec:AccidentalAmbProofsI} and \ref{subsec:AccidentalAmbProofsII}). \newline
Since the ambiguity transformations of the best solution are known (\ref{eq:AccAmbTrafosBestSolution}), violation parameters can be extracted via the argument function
\begin{equation}
 \epsilon_{\bm{\uppi}} \equiv \mathrm{Arg} \left[ \bm{\uppi} \left( \alpha_{1} \right) \hspace*{1pt} \bm{\uppi} \left( \alpha_{2} \right) \hspace*{1pt} \bm{\uppi} \left( \beta_{1} \right)^{-1} \hspace*{1pt} \bm{\uppi} \left( \beta_{2} \right)^{-1} \right] \mathrm{.} \label{eq:EpsPiDef}
\end{equation}
Here, we utilize the convenient fact that MATHEMATICA extracts the arguments of exponentials as values on the interval $\left[ - \pi, \pi \right]$ \cite{MathematicaLanguage}, which is helpful since parameters $\epsilon_{\bm{\uppi}}$ with modulus close to zero will now be searched. \newline

In appendix \ref{subsec:AccidentalAmbProofsIII}, a good upper bound for small values of the violation parameter was guessed to be given by $5^{\circ}$. Also, in the same appendix section the probability for accidental symmetries with such a small $\epsilon_{\bm{\uppi}}$ to occur was estimated to be quite small, around a few percent for $\ell_{\mathrm{max}} = 1$. \newline
Therefore, we loop here through all possible ambiguity roots (\ref{eq:AccAmbTrafosBestSolution}) and systematically store, for all energy bins, all cases where $\epsilon_{\bm{\uppi}}$ is smaller or equal to $5^{\circ}$ or, formulated in radian
\begin{equation}
 \epsilon_{\bm{\uppi}} \leq \left( 5^{\circ} \frac{2 \pi}{360^{\circ}} \right) \mathrm{rad} \simeq 0.0873 \hspace*{2pt} \mathrm{rad} \mathrm{.} \label{eq:EpsPiAmbBoundary}
\end{equation}
The question remains how to draw connections between the solutions in the pool (\ref{eq:OmelParametersSolutionPool}) other than the best solution and those ambiguities $\bm{\uppi} (\alpha_{k}, \beta_{k})$ for which $\epsilon_{\bm{\uppi}}$ fulfills (\ref{eq:EpsPiAmbBoundary}). The answer is closely related to the discussions in appendix \ref{subsec:AccidentalAmbProofsII}, which have been cited above. The fit-routine cannot violate Omelaenko's constraint, having it built in implicitly. As postulated in this thesis, the solutions found should be close to the exact ambiguities in root space, if and only if $\epsilon_{\bm{\uppi}} \ll 1$. \newline The distance in root-space between the exact ambiguity and the TPWA-solution in the pool has to be minimized. Therefore, for every possible ambiguity $\bm{\uppi} \in \hat{\mathcal{P}}$, we search for the TPWA-solution contained in (\ref{eq:OmelParametersSolutionPool}) that minimizes the quantity
\begin{equation}
 d_{i}^{\bm{\uppi}} := \sqrt{\left| \bm{\uppi} \left(\alpha_{1}^{\mathrm{Best}}\right) - \alpha_{1}^{(i)} \right|^{2} + \left| \bm{\uppi}\left( \alpha_{2}^{\mathrm{Best}} \right) - \alpha_{2}^{(i)} \right|^{2} + \left| \bm{\uppi} \left( \beta_{1}^{\mathrm{Best}} \right) - \beta_{1}^{(i)} \right|^{2} + \left| \bm{\uppi} \left( \beta_{2}^{\mathrm{Best}} \right) - \beta_{2}^{(i)} \right|^{2}} \mathrm{.} \label{eq:RootSolDistanceDefinition}
\end{equation}
The result is, for each $\bm{\uppi} \in \hat{\mathcal{P}}$, the number
\begin{equation}
 d_{\mathrm{min.}}^{\bm{\uppi}} := \mathrm{min} \left\{ d_{i}^{\bm{\uppi}} \Big| i = 1,\ldots,N_{\mathrm{nonred}} \right\} \mathrm{,} \label{eq:RootSolMinDistanceDefinition}
\end{equation}
as well as the corresponding solution index $j \in \left\{ 1,\ldots,N_{\mathrm{nonred}} \right\}$.
Naturally, for every candidate ambiguity that satisfies (\ref{eq:EpsPiAmbBoundary}), we expect small values of (\ref{eq:RootSolMinDistanceDefinition}). However, it is very important that, in this procedure, $d_{\mathrm{min.}}^{\bm{\uppi}}$ is not minimized by the true solution or the double ambiguity, but really by a solution corresponding to a $\bm{\uppi} \in \hat{\mathcal{P}}$, i.e. an accidental symmetry. This has in all cases been checked by hand. The true solution and the double ambiguity can then be dealt with separately. \newline
It has to be stated that in view of the parameter $\left| a_{2} \right|$, which specifies multipole solutions as well, the possibility that $\left| a_{2}^{\mathrm{Best}} \right| \not= \left| a_{2}^{(i)} \right|$ is disregarded here. This inequality is found to be in the sub-sub-percent range for all cases with $\epsilon_{\bm{\uppi}} \ll 1$ and may therefore be safely neglected. \newline
We report that $12$ candidates for ambiguities satisfying (\ref{eq:EpsPiAmbBoundary}) have been found in the case at hand. For two of those cases, the closest possible solution has been found to be the MAID solution, i.e. the fitter has not adopted a minimum corresponding to an accidental symmetry. For the remaining $10$ cases however, minima corresponding to ambiguities have been found as anticipated. They are listed, in conjunction with the corresponding values for $ \epsilon_{\bm{\uppi}}$, in Table \ref{tab:Lmax1TheoryDataAccAmbiguitiesNumbers}. Numbers for the found cases are provided in the appendix at the end of this section. Table \ref{tab:Lmax1TheoryDataAccBestRoots} lists the roots from the best solution found in all relevant cases. The corresponding ambiguity-transformed roots, as well as the roots of the obtained closest solutions from the pool, are given in the Tables \ref{tab:Lmax1TheoryDataAccAmbiguitiesRoots} and \ref{tab:Lmax1TheoryDataAccAmbiguitiesRoots2}. \newline

\begin{figure}[hb]
 \centering
\begin{overpic}[width=0.49\textwidth]{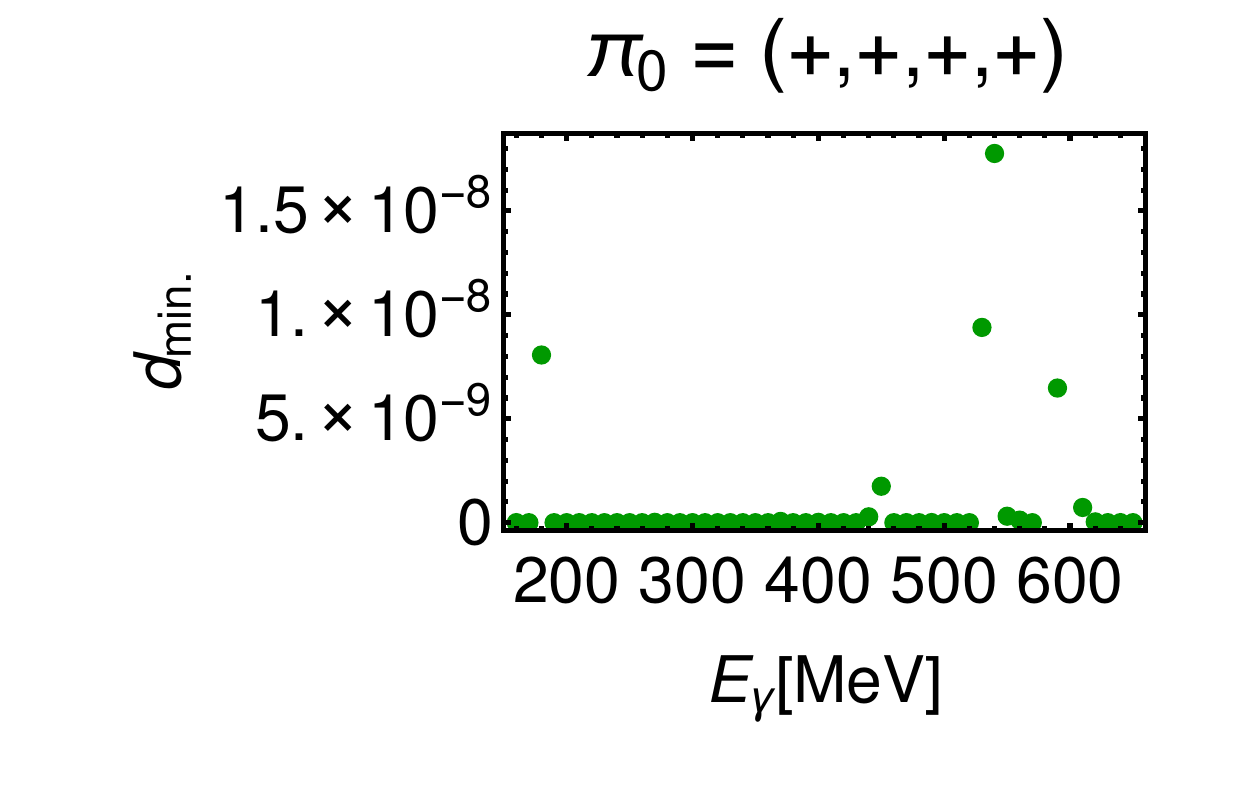}
 \end{overpic}
\begin{overpic}[width=0.49\textwidth]{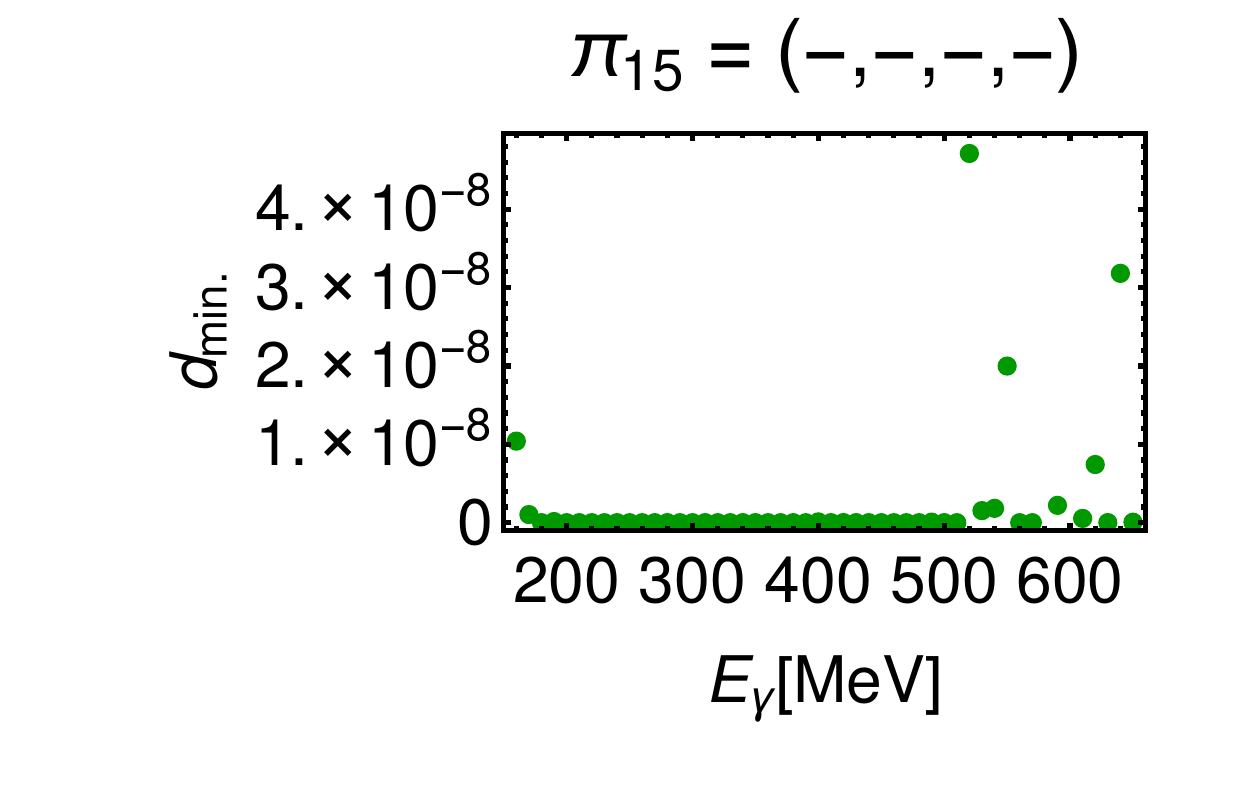}
 \end{overpic}
\caption[Plots of the dimensionless distance parameter to the closest fit solution $d_{\mathrm{min.}}^{\bm{\uppi}}$, for both the identity and the double ambiguity.]{The pictures show plots of the dimensionless distance parameter to the closest fit solution $d_{\mathrm{min.}}^{\bm{\uppi}}$, for both the identity $\bm{\uppi}_{0}$ and the double ambiguity $\bm{\uppi}_{15}$. The notation for the ambiguities is explained in the main text.}
\label{fig:Lmax1ThDataFitClosestSolutionPlot1}
\end{figure}

\begin{figure}[ht]
 \centering
\begin{overpic}[width=0.325\textwidth]{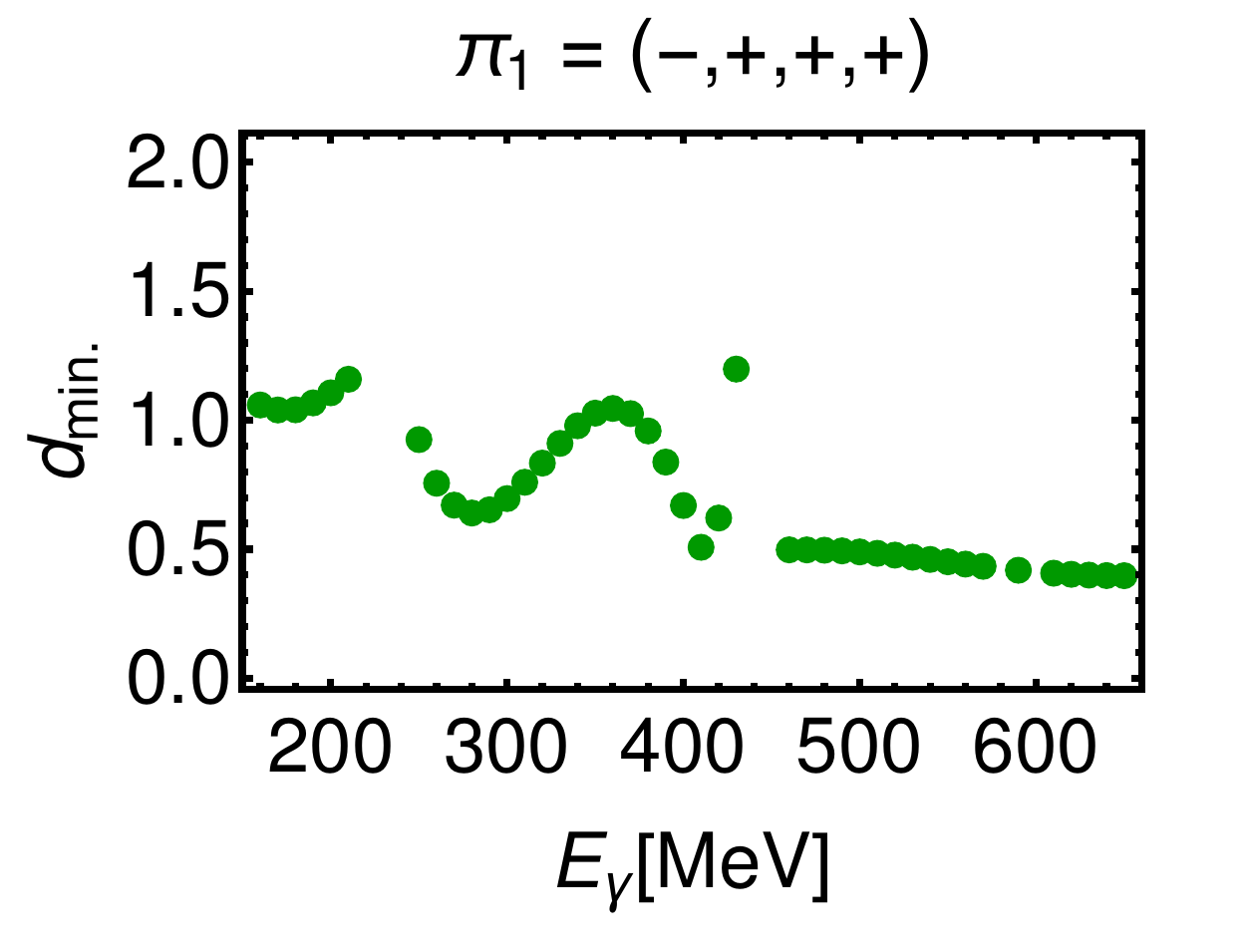}
 \end{overpic}
\begin{overpic}[width=0.325\textwidth]{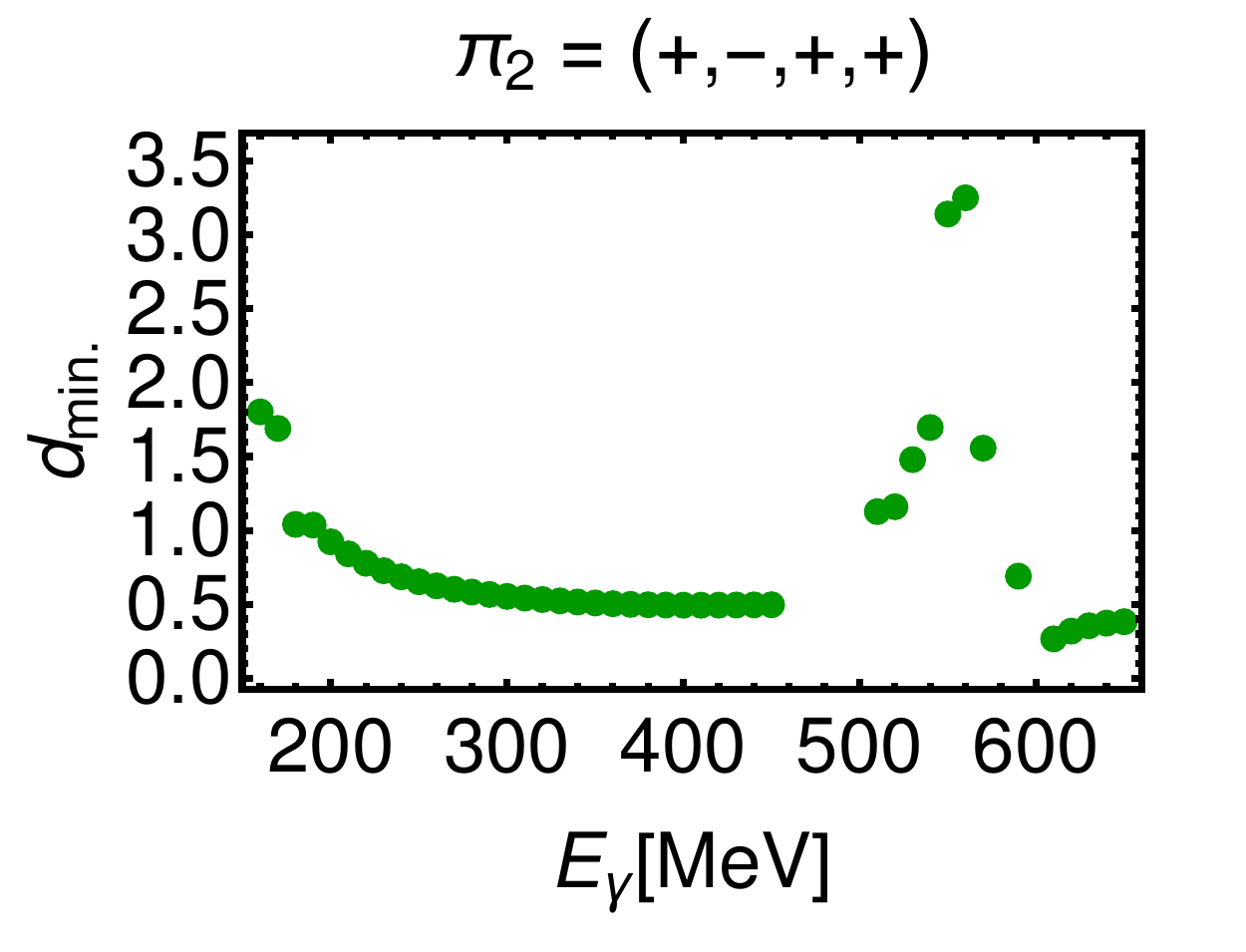}
 \end{overpic}
\begin{overpic}[width=0.325\textwidth]{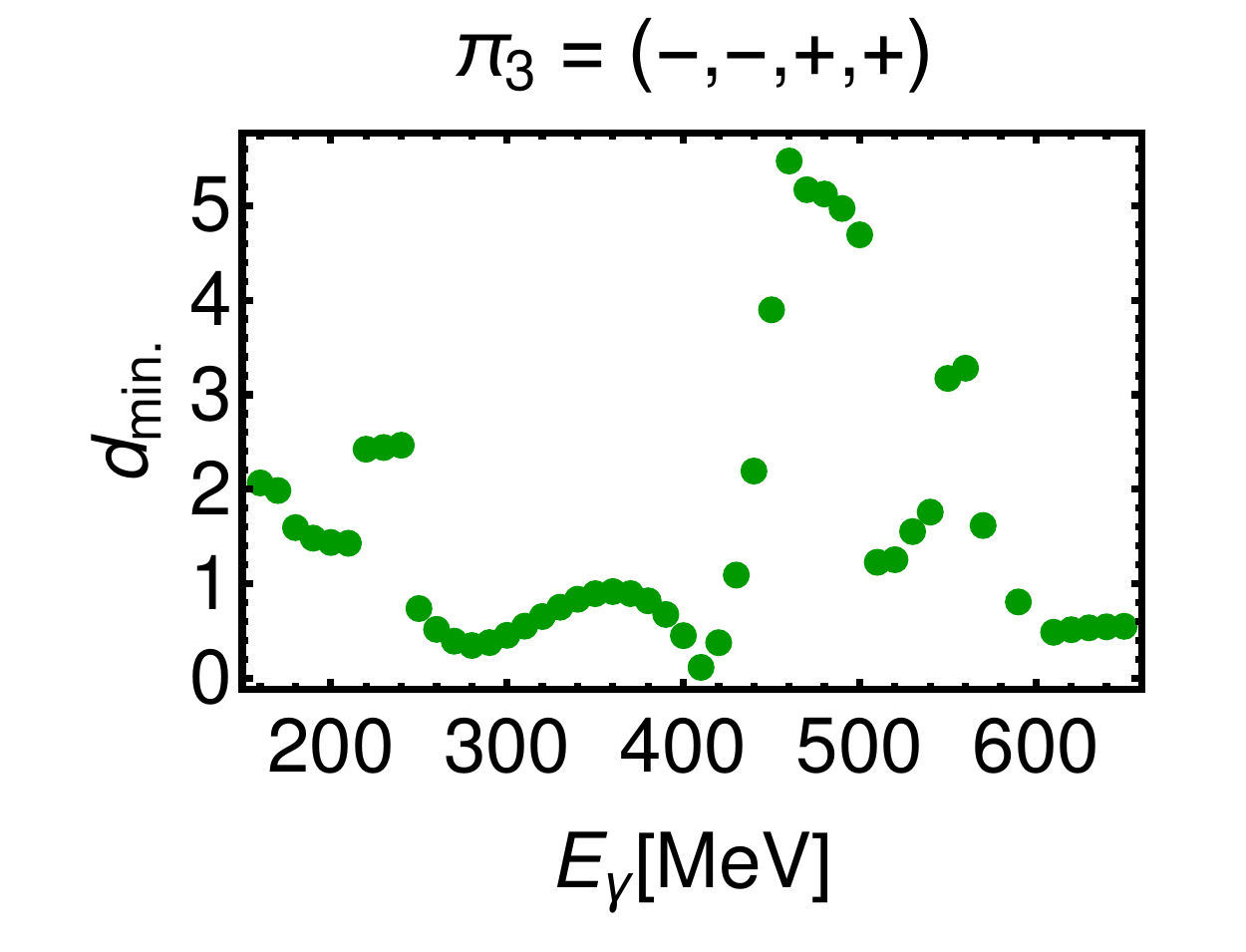}
 \end{overpic} \\
\begin{overpic}[width=0.325\textwidth]{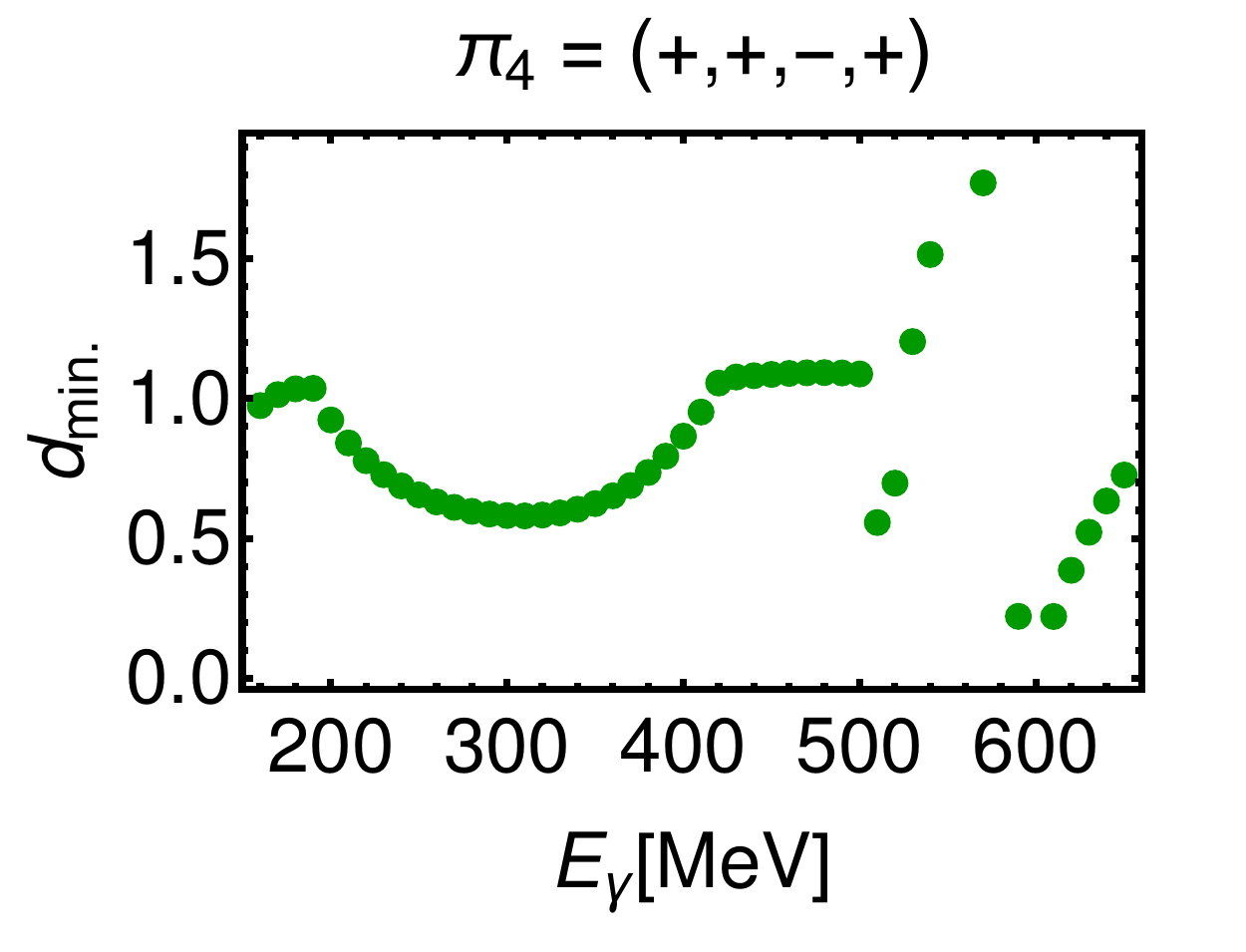}
 \end{overpic}
\begin{overpic}[width=0.325\textwidth]{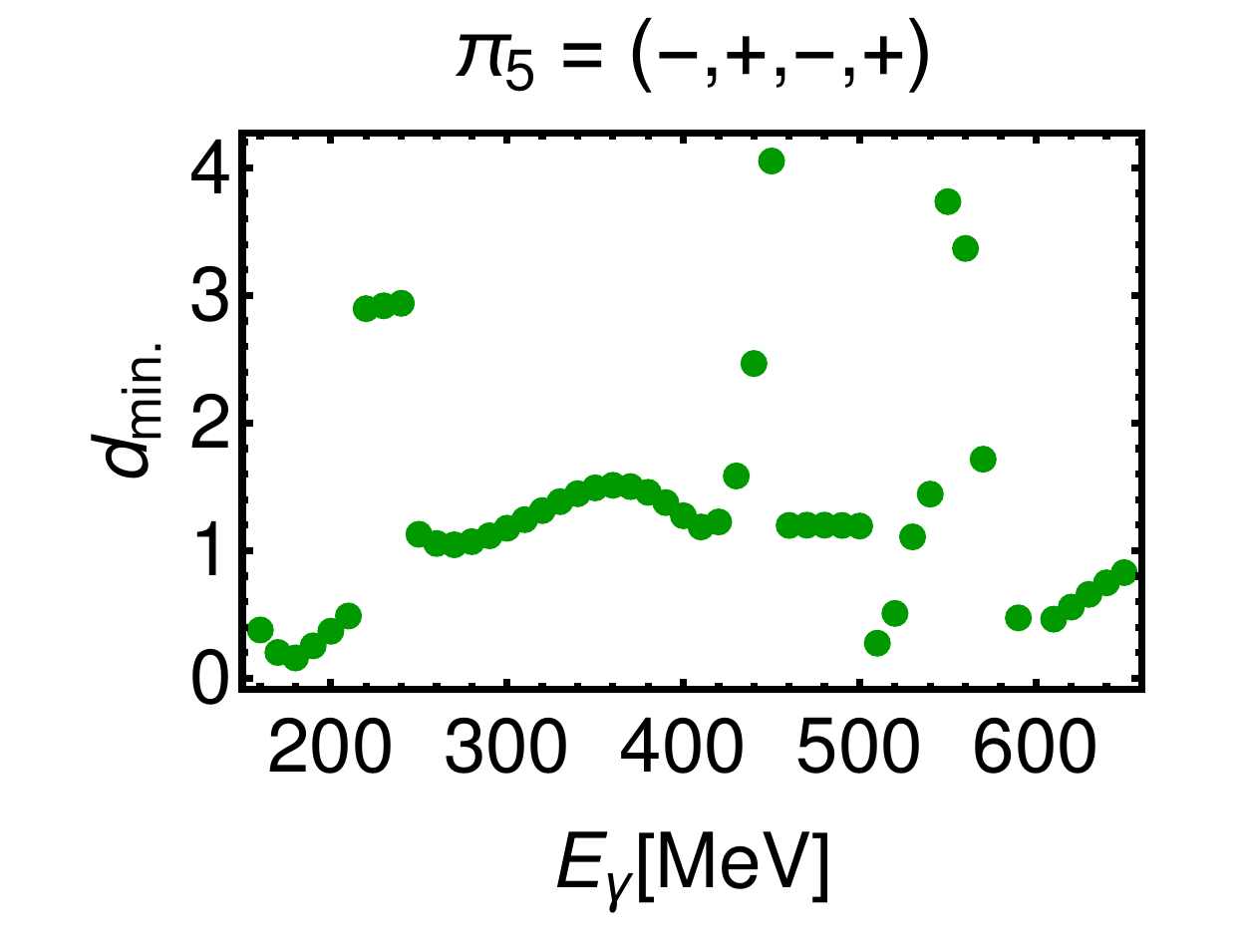}
 \end{overpic}
\begin{overpic}[width=0.325\textwidth]{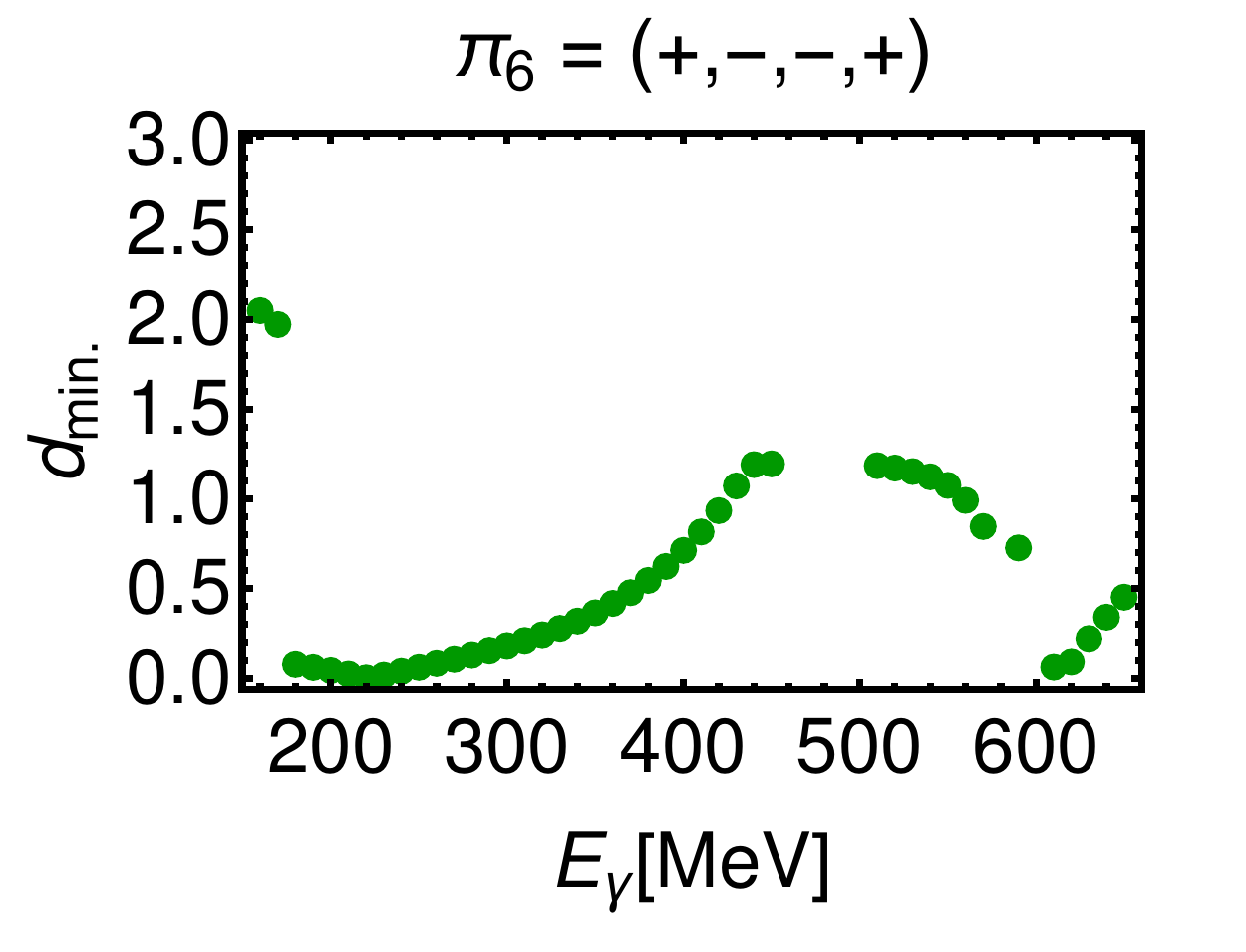}
 \end{overpic}  \\
\begin{overpic}[width=0.325\textwidth]{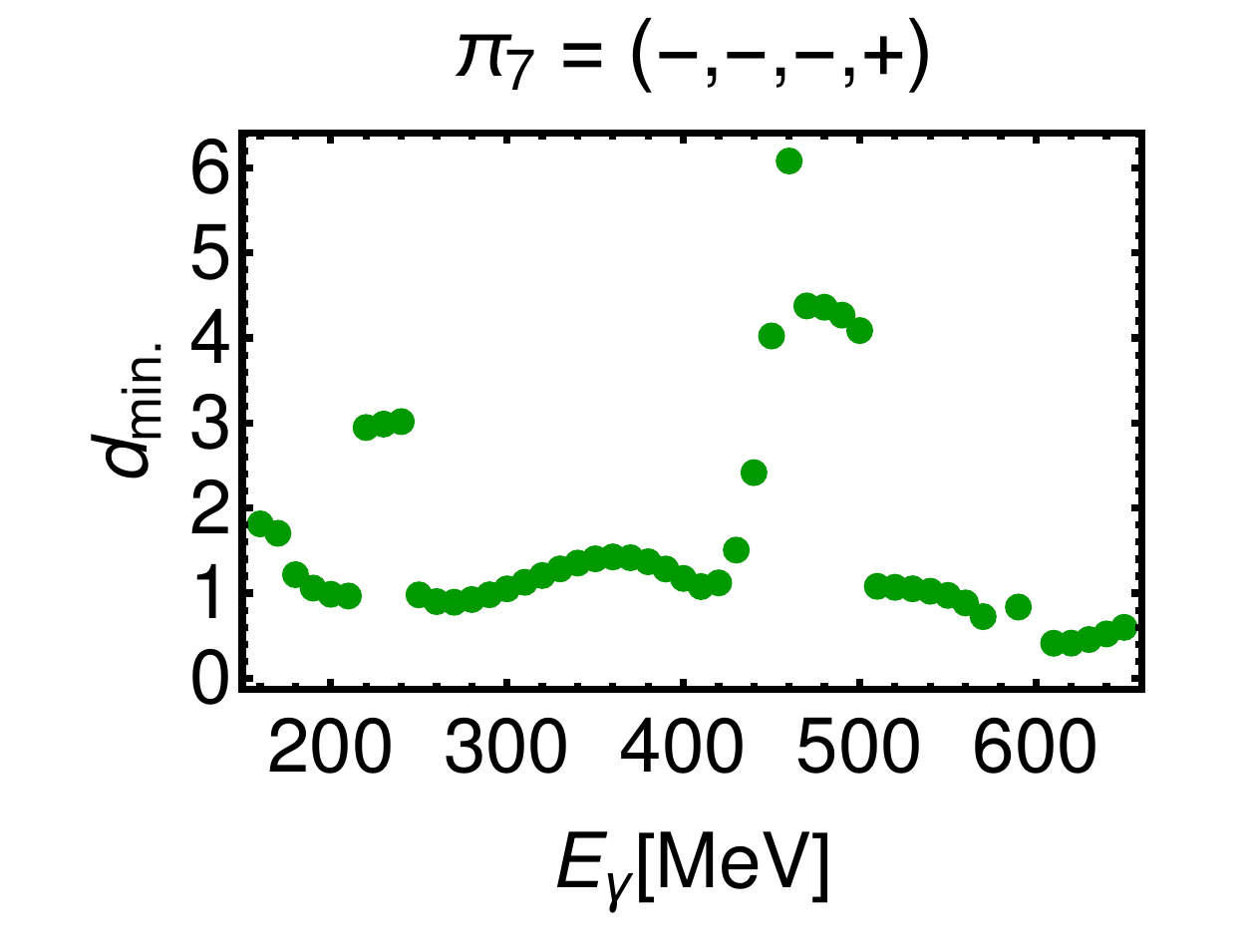}
 \end{overpic}
\begin{overpic}[width=0.325\textwidth]{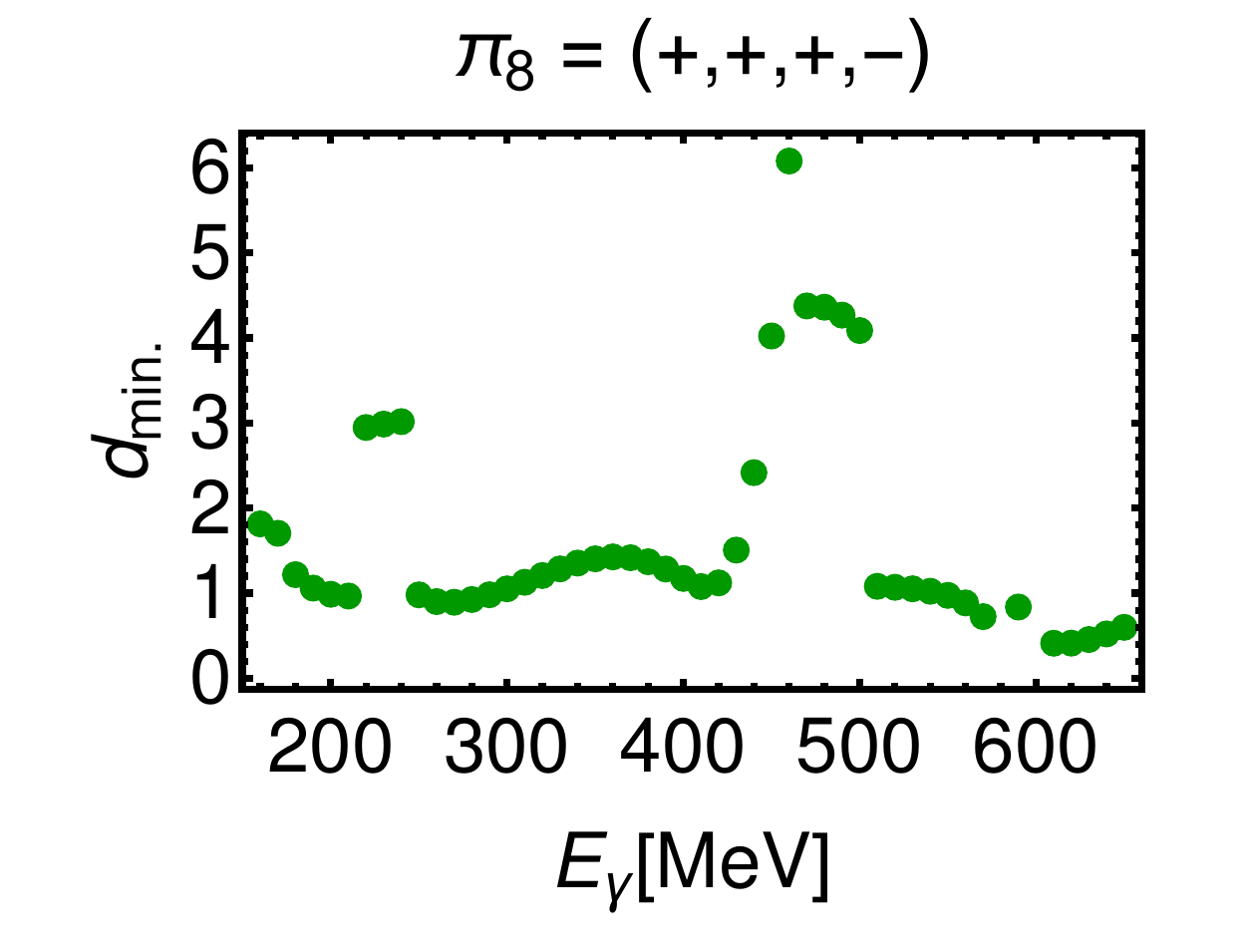}
 \end{overpic}
 \begin{overpic}[width=0.325\textwidth]{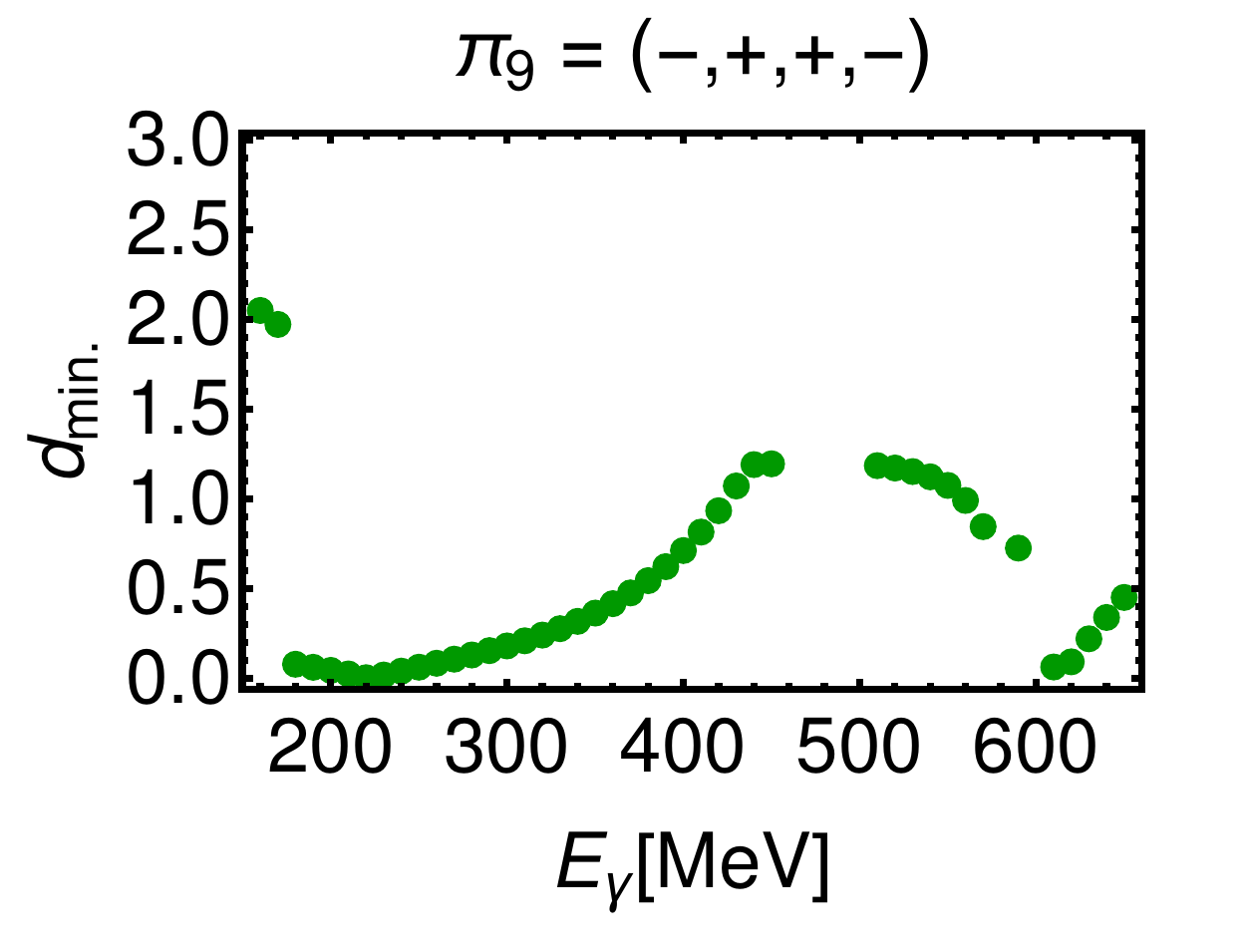}
 \end{overpic}  \\
 \begin{overpic}[width=0.325\textwidth]{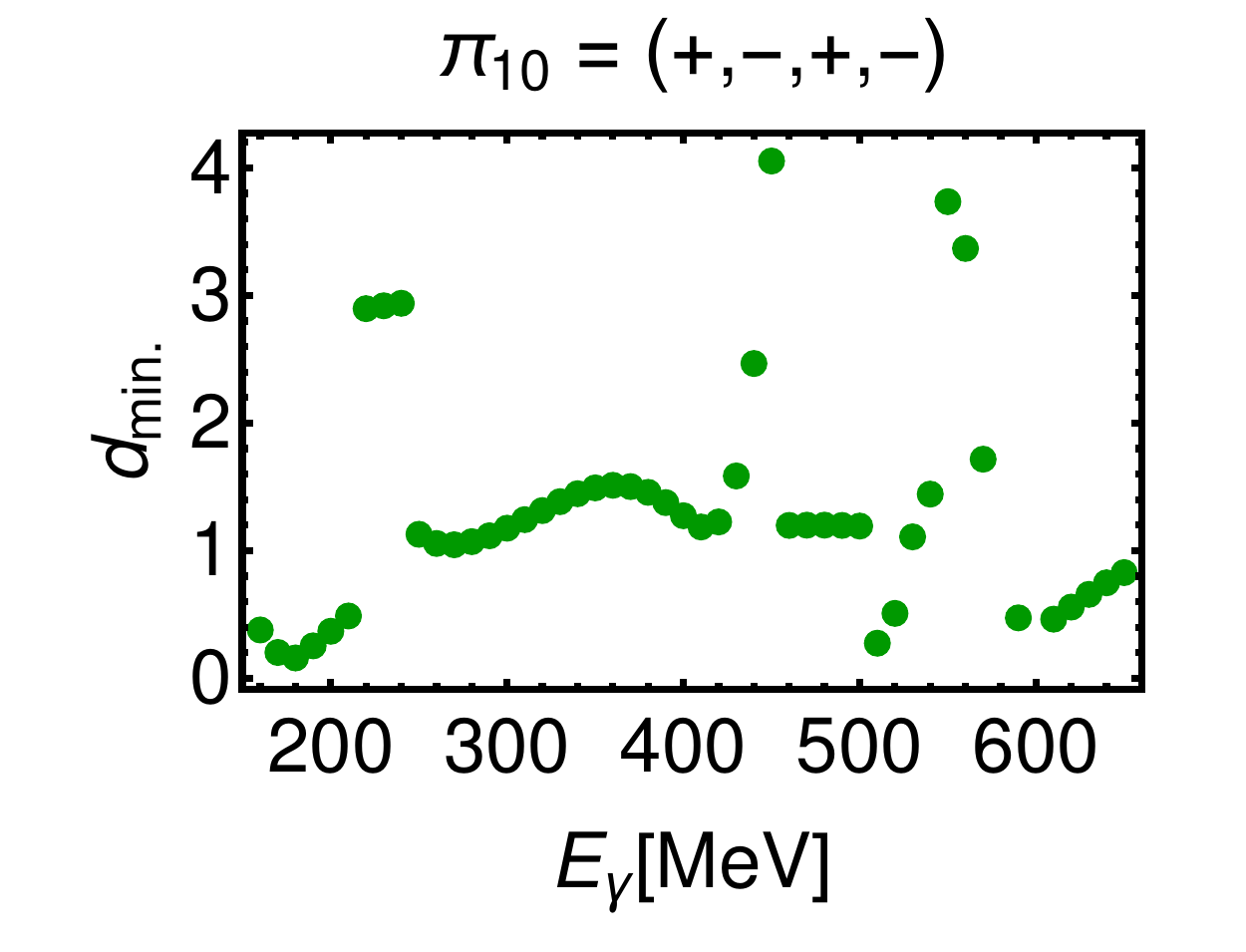}
 \end{overpic}
\begin{overpic}[width=0.325\textwidth]{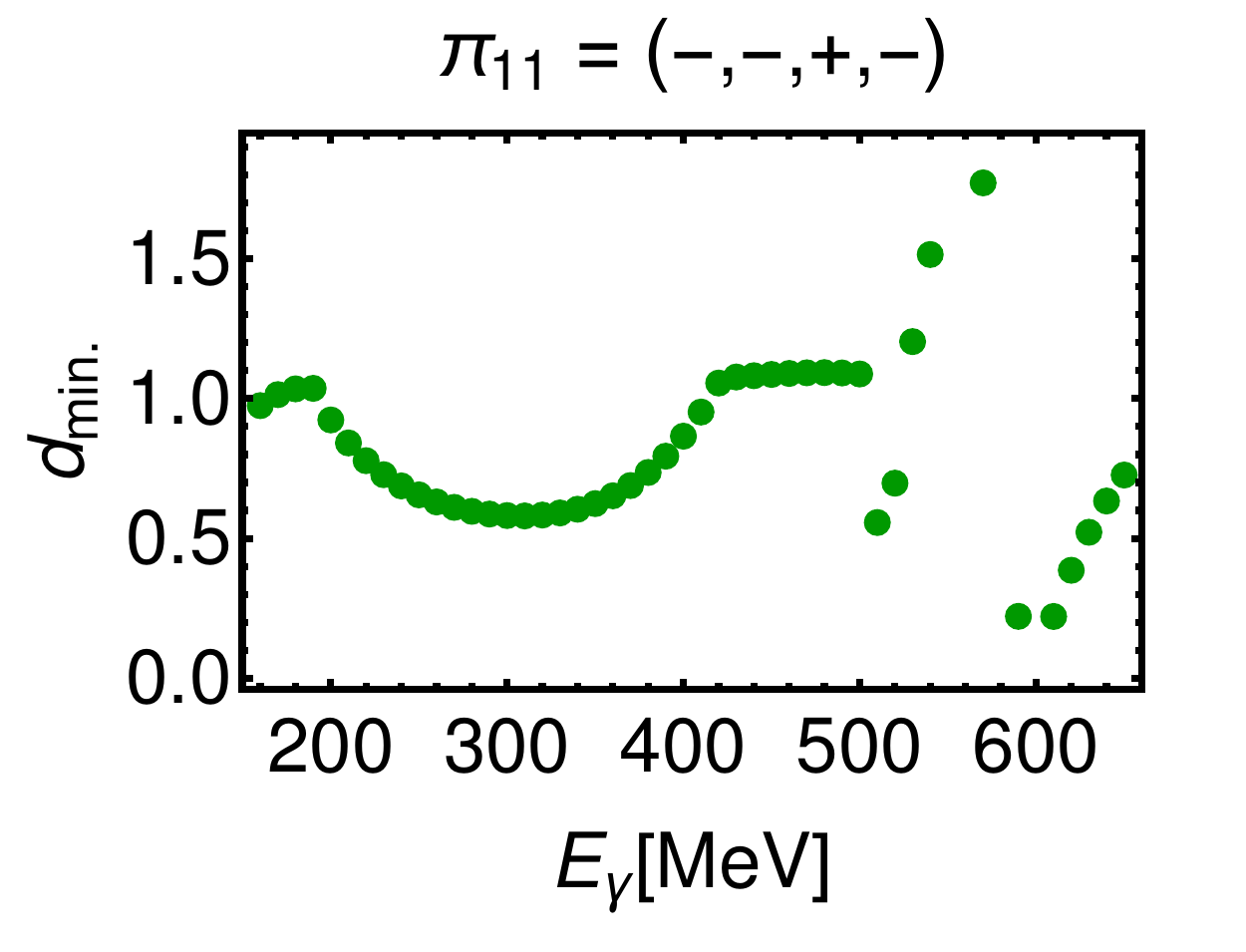}
 \end{overpic}  
 \begin{overpic}[width=0.325\textwidth]{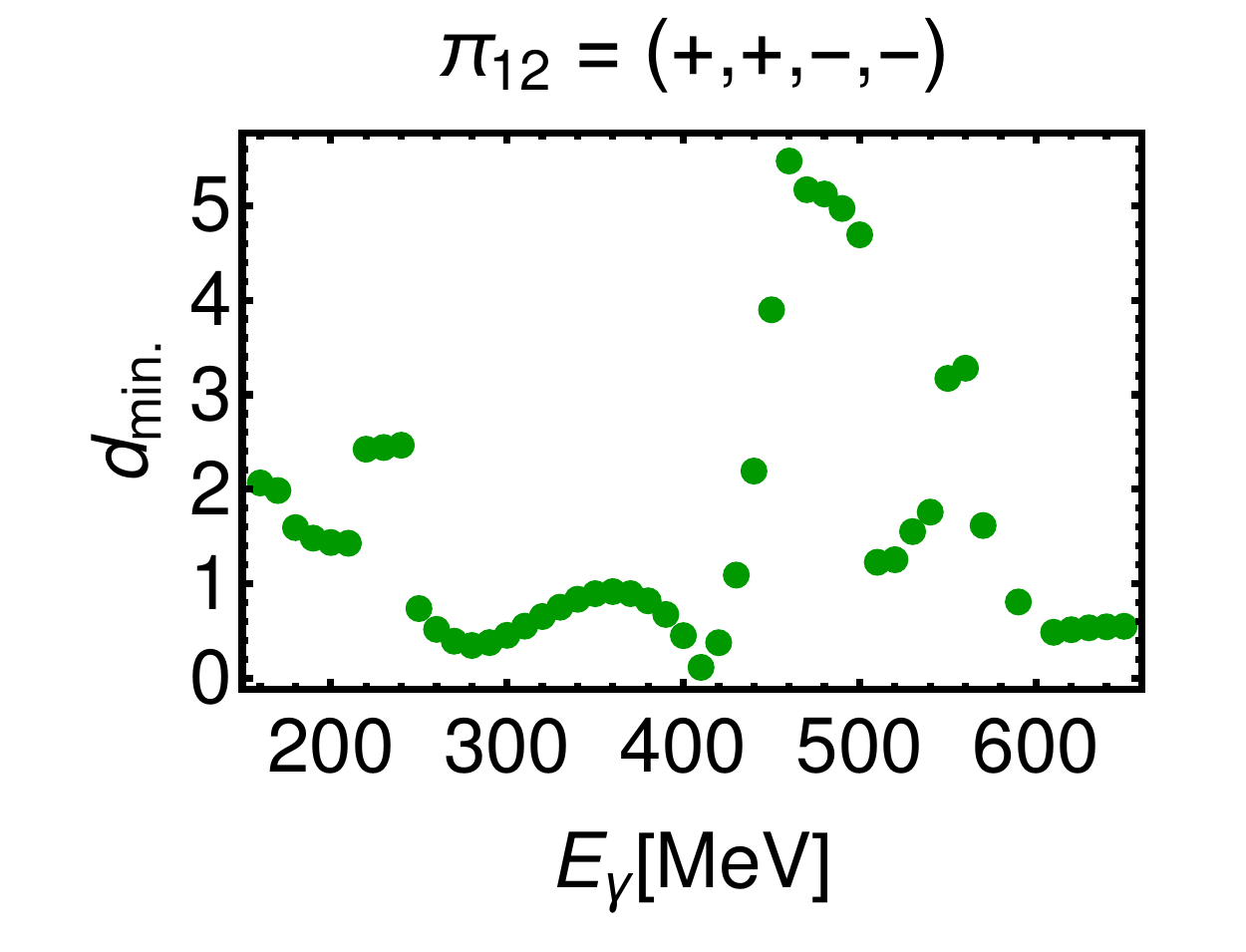}
 \end{overpic}  \\
  \begin{overpic}[width=0.325\textwidth]{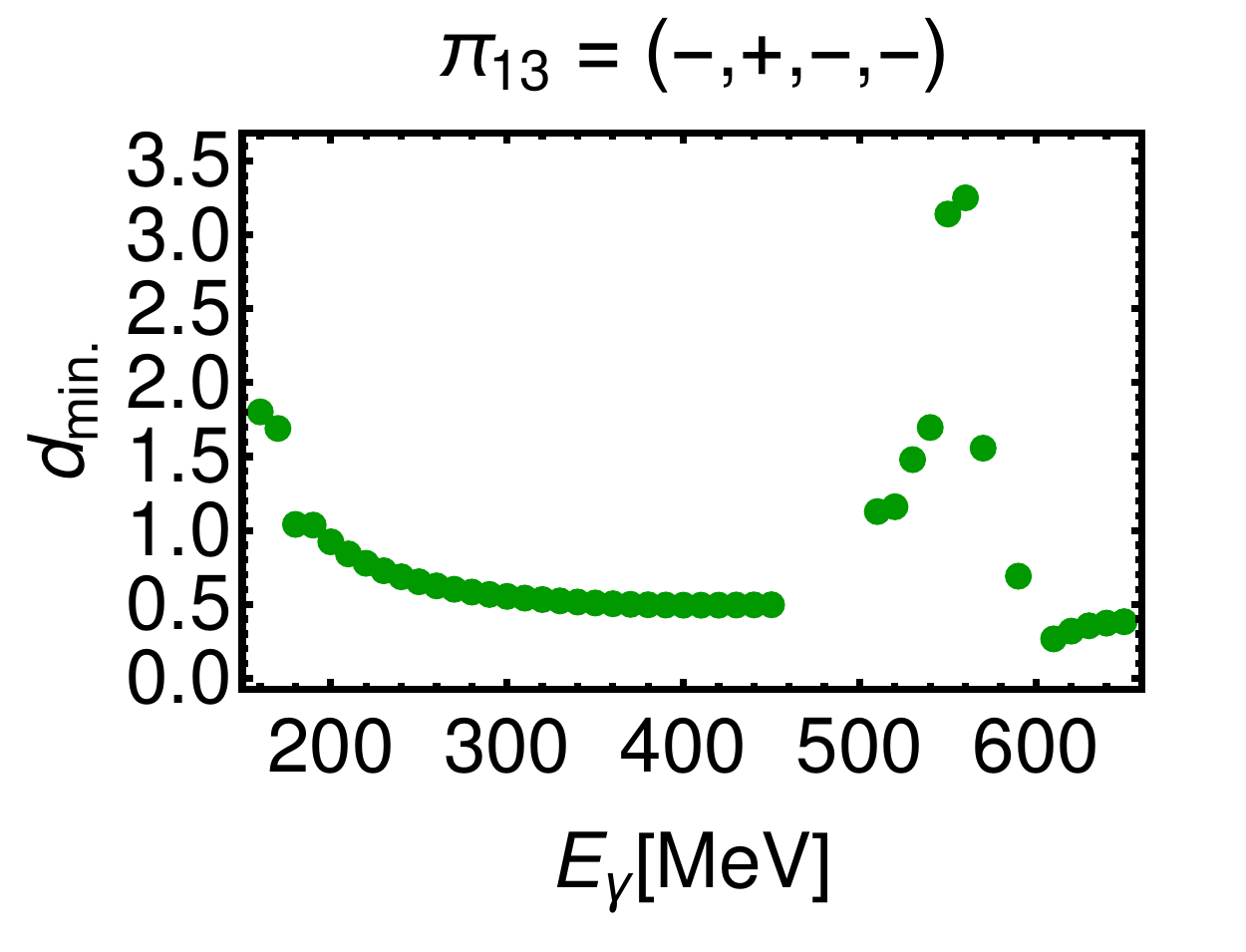}
 \end{overpic}
 \begin{overpic}[width=0.325\textwidth]{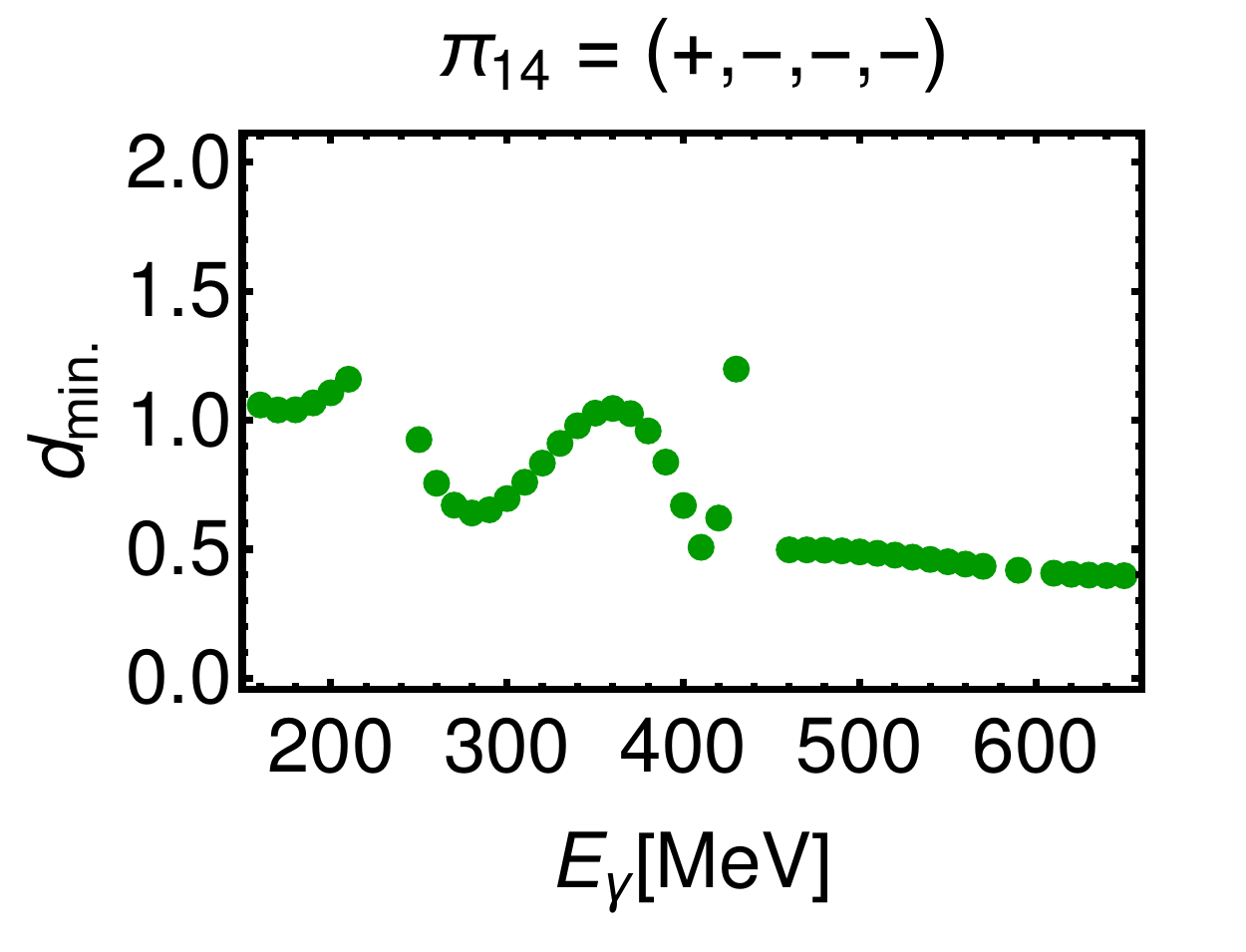}
 \end{overpic}
\caption[Plots of the dimensionless distance parameter to the closest fit solution $d_{\mathrm{min.}}^{\bm{\uppi}}$, for all possible accidental ambiguities.]{Shown here are plots of the dimensionless parameter $d_{\mathrm{min.}}^{\bm{\uppi}}$, measuring the distance to the closest fit solution for all possible accidental ambiguities and for all energies. The notation for the ambiguities is explained in the main text. \newline
For all accidental symmetries with a violation parameter $\epsilon_{\bm{\uppi}}$ smaller or equal to $5^{\circ}$, listed in Table \ref{tab:Lmax1TheoryDataAccAmbiguitiesNumbers}, a fit solution is found that is quite close in root space.}
\label{fig:Lmax1ThDataFitClosestSolutionPlot2}
\end{figure}

There is a good way to represent the minimal distances obtained graphically and link them to the accidental ambiguities. For this purpose, in Figures \ref{fig:Lmax1ThDataFitClosestSolutionPlot1} and \ref{fig:Lmax1ThDataFitClosestSolutionPlot2} the quantity $d_{\mathrm{min.}}^{\bm{\uppi}}$ has been plotted against energy for all ambiguities, also including the best solution corresponding to $\bm{\uppi}_{0} = (+,+,+,+)$ as well as its double ambiguity $\bm{\uppi}_{15} = (-,-,-,-)$. The notation chosen here denotes each ambiguity by its action on the phases $\left( \varphi_{1}, \varphi_{2}, \psi_{1}, \psi_{2} \right)$ of the Omelaenko roots $\alpha_{k} = \left| \alpha_{k} \right| e^{i \varphi_{k}}$ and $\beta_{k} = \left| \beta_{k} \right| e^{i \psi_{k}}$ corresponding to the best solution. Sign combinations such as $(+,-,+,-)$ then give a quick reference to which roots have been conjugated, and which not. \newline
For $\bm{\uppi}_{0}$ and $\bm{\uppi}_{15}$, it is natural that the distance $d_{\mathrm{min.}}^{\bm{\uppi}}$ to the closest solution found in the fit to the theory-data is vanishing up to a high numerical precision, as shown in Figure \ref{fig:Lmax1ThDataFitClosestSolutionPlot1}. The dimensionless distance parameter is here in the order of $10^{-8}$ to $10^{-9}$. This has to be expected since whenever a solution fulfills Omelaenko's constraint exactly, then so does the associated double ambiguity (see appendix \ref{sec:AccidentalAmbProofs}). \newline
For the remaining ambiguities however, all of them being accidental symmetries, the corresponding distances to the closest fit solution are generally non-vanishing, as can be seen in Figure \ref{fig:Lmax1ThDataFitClosestSolutionPlot2}. However, whenever an exact accidental symmetry with a violation parameter $\epsilon_{\bm{\uppi}}$ satisfying the bound (\ref{eq:EpsPiAmbBoundary}) exists, it is observed that the distance to the closest fit solution becomes small. Therefore, one can directly compare Table \ref{tab:Lmax1TheoryDataAccAmbiguitiesNumbers} to Figure \ref{fig:Lmax1ThDataFitClosestSolutionPlot2}. \newline
Particularly good examples are given by the first five ambiguities listed in Table \ref{tab:Lmax1TheoryDataAccAmbiguitiesNumbers}, corresponding to the ambiguity transformations $\bm{\uppi}_{6}$ and $\bm{\uppi}_{9}$. For both those ambiguities, a valley of small values is observed within the correct energy region in Figure \ref{fig:Lmax1ThDataFitClosestSolutionPlot2}. \newline
Furthermore, one has to mention the fact that all ambiguities listed in Table \ref{tab:Lmax1TheoryDataAccAmbiguitiesNumbers}, having corresponding fit solutions with small distance as seen in Figure \ref{fig:Lmax1ThDataFitClosestSolutionPlot2}, can be linked to ambiguities that can be read off from the ambiguity diagrams shown in Figure II of section \ref{sec:WBTpaper} and Figure \ref{fig:ExampleEpsilonPiParametersInAmbiguityDiagram}, appendix \ref{subsec:AccidentalAmbProofsI}. \newline \newline 

\vfill

\clearpage

\clearpage

The results obtained until now make it feasible to test some of the findings obtained in appendix \ref{subsec:AccidentalAmbProofsII}, in particular the linear approximations of the correction $\delta \Phi$ of the overall discrepancy function. This approximation is given in detail by equations (\ref{eq:DeltaDiscrFunct}), (\ref{eq:GroupSObsCorrComplete}), (\ref{eq:B1SquaredCorrFunct}), (\ref{eq:B2SquaredCorrFunct}), (\ref{eq:B3SquaredCorrFunct1}) and (\ref{eq:B4SquaredCorrFunct}). \newline
Furthermore, we evaluate the correction $\delta \Phi^{\mathrm{Fit}}$ by re-evaluating and summing up the functions $\Phi_{a}^{\alpha}$ (equation (\ref{eq:LegCoeffFitChi2ModelFit}), section \ref{sec:TPWAFitsIntro}), using Legendre-coefficients calculated from the mul\-ti\-pole-fit results that correspond to the accidental symmetries. \newline
In order to evaluate the linear approximation however, the parameters $\xi_{k}$ and $\zeta_{k}$ introduced in appendix \ref{subsec:AccidentalAmbProofsII} and cited in the discussion above, have to be extracted. Since we now know both the exact accidental symmetries $\bm{\uppi} (\alpha_{k}, \beta_{k})$, as well as the corresponding fit solutions $\bm{\tilde{\uppi}} (\alpha_{k}, \beta_{k})$, these parameters can be calculated via the definitions
\begin{equation}
 \xi_{k} = \log \left[ \frac{\bm{\tilde{\uppi}} \left(\alpha_{k}\right)}{\bm{\uppi} \left(\alpha_{k}\right)} \right] \mathrm{,} \hspace*{5pt} \zeta_{k^{\prime}} = \log \left[ \frac{\bm{\tilde{\uppi}} \left(\beta_{k^{\prime}}\right)}{\bm{\uppi} \left(\beta_{k^{\prime}}\right)} \right] \mathrm{.} \label{eq:XikZetakExtraction}
\end{equation}
We are content with the principal branch of the logarithm in each case, as returned by MATHEMATICA \cite{Mathematica8,Mathematica11,MathematicaLanguage,MathematicaBonnLicense}. Table \ref{tab:Lmax1TheoryDataAccAmbiguitiesNumbers2} stores the corresponding numbers for the relevant cases. The parameters (\ref{eq:XikZetakExtraction}) can now be used to evaluate the linear approximation $\delta \Phi^{\mathrm{calc.}}_{\mathrm{lin.}}$ as well as the full correction $\delta \Phi^{\mathrm{calc.}}$. The second task is achieved by inserting full exponentials $e^{\xi_{k}}$ and $e^{\zeta_{k}}$ in expressions such as equation (\ref{eq:B4ModSquareModifiedAmbTrafoI}) in appendix \ref{subsec:AccidentalAmbProofsII}. \newline
Numbers resulting from all calculations are given in Table \ref{tab:Lmax1TheoryDataAccAmbiguitiesNumbers}. It is seen that the linear calculation $\delta \Phi^{\mathrm{calc.}}_{\mathrm{lin.}}$ in some cases is an excellent approximation to the full $\delta \Phi^{\mathrm{calc.}}$. In other cases it fails rather badly. However, the re-calculated fit-offset $\delta \Phi^{\mathrm{Fit}}$ in all cases does not exactly coincide with $\delta \Phi^{\mathrm{calc.}}$. It is in most cases in the right order of magnitude. One can only guess that this discrepancy is rooted in numerical errors of the fitting or other processes such as the evaluation of roots. In case numbers carrying such small errors are combined in a complicated way to evaluate a precision quantity such as $\delta \Phi$, discrepancies may occur.
\begin{table}[hb]
\centering
 \begin{tabular}{c|cc|c|ccc}
  $E_{\gamma} \hspace*{2pt} \left[ \mathrm{MeV} \right]$ & $\bm{\uppi}_{n}$ & $\bm{\uppi}_{n} \left( \varphi_{1}, \varphi_{2}, \psi_{1}, \psi_{2} \right)$ & $\epsilon_{\bm{\uppi}}$ $\left[\mathrm{rad}\right]$ & $\delta \Phi^{\mathrm{calc.}}_{\mathrm{lin.}}$ & $\delta \Phi^{\mathrm{calc.}}$ & $\delta \Phi^{\mathrm{Fit}}$ \\ \hline \hline
  $200$ & $\bm{\uppi}_{9}$ & $\left( - , + , + , - \right)$ & $0.0721$ & $0.00305$ & $0.00308$ & $0.00138$ \\
  $210$ & $\bm{\uppi}_{9}$ & $\left( - , + , + , - \right)$ & $0.0371$ & $0.00169$ & $0.00171$ & $0.00085$ \\
  $220$ & $\bm{\uppi}_{6}$ & $\left( + , - , - , + \right)$ & $0.0047$ & $0.00005$ & $0.00005$ & $0.00003$ \\
  $230$ & $\bm{\uppi}_{9}$ & $\left( - , + , + , - \right)$ & $0.0247$ & $0.00276$ & $0.00277$ & $0.00159$ \\  
  $240$ & $\bm{\uppi}_{9}$ & $\left( - , + , + , - \right)$ & $0.0518$ & $0.02185$ & $0.02181$ & $0.01318$ \\
  $250$ & $\bm{\uppi}_{6}$ & $\left( + , - , - , + \right)$ & $0.0768$ & $0.08365$ & $0.08332$ & $0.05268$ \\  
  $410$ & $\bm{\uppi}_{3}$ & $\left( - , - , + , + \right)$ & $0.0248$ & $0.69395$ & $0.01832$ & $0.01547$ \\  
  $420$ & $\bm{\uppi}_{12}$ & $\left( + , + , - , - \right)$ & $0.0725$ & $6.06532$ & $0.10409$ & $0.08858$ \\
  $510$ & $\bm{\uppi}_{10}$ & $\left( + , - , + , - \right)$ & $0.0784$ & $0.30070$ & $0.00190$ & $0.00170$ \\
  $610$ & $\bm{\uppi}_{9}$ & $\left( - , + , + , - \right)$ & $0.0517$ & $0.00010$ & $0.00006$ & $0.00006$ \\  
 \end{tabular}
 \caption[Summary of candidates for accidental ambiguities in the fit of MAID2007 theory-data for $L=1$.]{This Table summarizes all the candidate ambiguities $\bm{\uppi}$ with $\epsilon_{\bm{\uppi}} \leq 5^{\circ}$ (\ref{eq:EpsPiAmbBoundary}) at their respective energies. Each ambiguity is represented by its action on the phases $\left( \varphi_{1}, \varphi_{2}, \psi_{1}, \psi_{2} \right)$ of the Omelaenko-roots, i.e. if it conjugates the roots $\alpha_{k} = \left| \alpha_{k} \right| e^{i \varphi_{k}}$ and $\beta_{k} = \left| \beta_{k} \right| e^{i \psi_{k}}$, or not. Furthermore, the respective violation parameters $\epsilon_{\bm{\uppi}}$ are given. Corrections $\delta \Phi$ to the minimum of the discrepancy function, which occur due to the finite violation parameter $\epsilon_{\bm{\uppi}} > 0$, are given as well. Methods used to estimate these corrections are described in the main text.}
 \label{tab:Lmax1TheoryDataAccAmbiguitiesNumbers}
\end{table}

\clearpage

We come to the conclusion on the fits to MAID theory-data truncated at $\ell_{\mathrm{max}} = 1$. The most important facts are that indeed this model-TPWA is complete with $5$ observables, e.g. the fitted example set $\left\{ \sigma_{0}, \check{\Sigma}, \check{T}, \check{P}, \check{F} \right\}$. Furthermore, local minima were found mainly in the fit to the group $\mathcal{S}$ observables. In case these local minima are good, they can in a lot of cases be attributed to accidental ambiguities with a small violation parameter $\epsilon_{\bm{\uppi}}$. \newline

For MAID theory-data truncated at a higher $L \equiv \ell_{\mathrm{max}}$, one can expect the same qualitative behaviour. It is however very likely that the number of ambiguities increases, resulting in a larger numerical effort to fit out the true MAID solution. The theory-data for truncations at $\ell_{\mathrm{max}} = 2,3,4$ will be the subject of section \ref{subsec:TheoryDataFitsLmax2}. \newline \newline \newline

\textit{Appendix: Additional Tables for the discussion of the group $\mathcal{S}$ theory-data fit} \newline

\begin{sidewaystable}[ht]
\centering
 \begin{tabular}{c|c|ccccccc}
  $E_{\gamma}$ & $\Phi_{\mathcal{M}} \left[ \left( \frac{\mu b}{\mathrm{sr}} \right)^{2} \right]$ & $\mathrm{Re}  E_{0+}^{C} $ & $\mathrm{Re}  E_{1+}^{C} $ & $\mathrm{Im}  E_{1+}^{C} $ & $\mathrm{Re}  E_{0+}^{C} $ & $\mathrm{Re}  E_{0+}^{C} $  &  $\mathrm{Re} E_{0+}^{C} $  & $\mathrm{Re}  E_{0+}^{C} $ \\ \hline \hline
  $510 \hspace*{1.5pt} \mathrm{MeV} $ & $1.9754\times10^{-15} \equiv 0$ & 0.632646 & 0.027314 & -0.080021 & 1.07991 & 0.856518 & 0.386767 & 0.065036  \\
   & $1.9754\times10^{-15} \equiv 0$ &  0.632646 & 0.231048 & -0.263827 & -0.821549 & 0.672712 & -0.903491 & 0.432648 \\
   & $0.000097$ & 1.00996 & -0.024845 & -0.142103 & -0.742484 & 0.444835 & -1.30413 & 0.224477  \\
   & $0.000097$ & 1.00996 & 0.187216 & -0.073452 & 0.904855 & 0.513485 & 0.97939 & 0.087176  \\
   & $0.014365$ &  0.99445 & -0.005875 & -0.079594 & 1.10743 & 0.521855 & 0.587794 & 0.085002 \\
   & $0.014365$ &  0.99445 & 0.173211 & -0.145618 & -0.940092 & 0.455831 & -0.922467 & 0.21705 \\
\hline \hline
  $610 \hspace*{1.5pt} \mathrm{MeV}$ & $3.4172\times10^{-16} \equiv 0$ & 0.69284 & 0.002489 & -0.177848 & -0.433243 & 0.564934 & -0.53778 & 0.437157 \\ 
    & $3.4172\times10^{-16} \equiv 0$ &   0.69284 & 0.034846 & -0.042592 & 0.470577 & 0.700189 & 0.463111 & 0.166645 \\
    & $0.000001$ & 0.623078 & 0.027933 & -0.038591 & 0.419156 & 0.787214 & 0.378222 & 0.168184 \\
    & $0.000001$ & 0.623078 & 0.013645 & -0.206344 & -0.377578 & 0.619462 & -0.461379 & 0.503688 \\
\hline \hline
 \end{tabular}
 \caption[The non-redundant from a fit of MAID2007 theory-data for $L=1$. Two energy-bins shown.]{This Table shows, as an example, the non-redundant solutions occurring as a result of the TPWA fit to the group $\mathcal{S}$ theory-data, for two energy bins. The parameters of the phase constrained multipoles are here given in the unit of $\left[10 \hspace*{1pt} \mathrm{mFm} \right]$. This is the unit in which they result out of fits to observables given in $\left[ \mu b / \mathrm{sr} \right]$. \newline All parameters shown are rounded to $6$ decimals. The resulting value of $\Phi_{\mathcal{M}}$ for the best solutions is well compatible with zero in this rounding approximation, but still the original resulting value is also given.}
 \label{tab:Lmax1TheoryDataExampleBinsResults}
\end{sidewaystable}
\begin{sidewaystable}[ht]
\centering
 {\renewcommand{\arraystretch}{1.25} \begin{tabular}{c|cccc}
  $E_{\gamma} \hspace*{2pt} \left[ \mathrm{MeV} \right]$ & $\alpha_{1}^{\mathrm{Best}}$ & $\alpha_{2}^{\mathrm{Best}}$ & $\beta_{1}^{\mathrm{Best}}$ & $\beta_{2}^{\mathrm{Best}}$ \\ \hline \hline
  $200$ &  $ -0.7255-1.6255 \hspace*{1pt} i $ & $ -0.0335-0.4616 \hspace*{1pt} i $ & $ 0.0574 +0.5265 \hspace*{1pt} i $ & $ 0.5823 +1.4422 \hspace*{1pt} i $  \\[1pt]
  $210$ &  $ -0.8393-1.8116 \hspace*{1pt} i $ & $ -0.0132-0.4206 \hspace*{1pt} i $ & $ 0.0264 +0.5272 \hspace*{1pt} i $ & $ 0.6422 +1.4564 \hspace*{1pt} i $ \\[1pt]
  $220$ &  $ -0.9512+1.993 \hspace*{1pt} i $ & $ 0.0029 +0.389 \hspace*{1pt} i $ & $ -0.0027-0.5265 \hspace*{1pt} i $ & $ 0.6993 -1.474 \hspace*{1pt} i $  \\[1pt]
  $230$ &  $ -1.0611+2.1722 \hspace*{1pt} i $ & $ 0.0157 +0.3638 \hspace*{1pt} i $ & $ -0.0292-0.5253 \hspace*{1pt} i $ & $ 0.7529 -1.4941 \hspace*{1pt} i $  \\[1pt]
  $240$ &  $ -1.1691-2.3508 \hspace*{1pt} i $ & $ 0.0261 -0.3433 \hspace*{1pt} i $ & $ -0.0535+0.5241 \hspace*{1pt} i $ & $ 0.8036 +1.516 \hspace*{1pt} i $  \\[1pt]
  $250$ &  $ -1.2741-2.5297 \hspace*{1pt} i $ & $ 0.0345 -0.3264 \hspace*{1pt} i $ & $ -0.0757+0.5231 \hspace*{1pt} i $ & $ 0.8511 +1.5397 \hspace*{1pt} i $  \\[1pt]
  $410$ &  $ -2.3515-6.7753 \hspace*{1pt} i $ & $ 0.0823 -0.2471 \hspace*{1pt} i $ & $ -0.3326+0.5332 \hspace*{1pt} i $ & $ 1.6044 +2.5025 \hspace*{1pt} i $  \\[1pt]
  $420$ &  $ -2.1076-7.2993 \hspace*{1pt} i $ & $ 0.0812 -0.2473 \hspace*{1pt} i $ & $ -0.3459+0.5359 \hspace*{1pt} i $ & $ 1.586 +2.6641 \hspace*{1pt} i $  \\[1pt]
  $510$ &  $ 0.0555 -0.2421 \hspace*{1pt} i $ & $ 7.0415 -8.0403 \hspace*{1pt} i $ & $ -0.66+3.5037 \hspace*{1pt} i $ & $ -0.5121+0.5405 \hspace*{1pt} i $  \\[1pt]
  $610$ &  $ 0.0714 +0.2036 \hspace*{1pt} i $ & $ 5.8247 -0.9199 \hspace*{1pt} i $ & $ -0.8396+0.1105 \hspace*{1pt} i $ & $ -0.5337-1.4041 \hspace*{1pt} i $  \\[1pt] \hline \hline
 \end{tabular}}
 \caption[For MAID2007 theory-data with $L=1$, the Omelaenko roots of the best solutions are shown for all energy bins where an accidental ambiguity can occur.]{Here, the Omelaenko roots of the best solutions (i.e. those with lowest value of $\Phi_{\mathcal{M}}$) are shown for all energy bins where an accidental ambiguity $\bm{\uppi}$ with $\epsilon_{\bm{\uppi}} \leq 5^{\circ}$ was found (cf. the discussion the main text). These roots serve as input for the calculation of the Omelaenko-roots belonging to the exact accidental symmetries $\bm{\uppi}_{\hspace*{0.035cm}n}$, see equation (\ref{eq:AccAmbTrafosBestSolution}). In the Tables \ref{tab:Lmax1TheoryDataAccAmbiguitiesRoots} and \ref{tab:Lmax1TheoryDataAccAmbiguitiesRoots2}, such ambiguity-transformed roots are printed in comparison to the closest solutions found by the fit-routine.}
 \label{tab:Lmax1TheoryDataAccBestRoots}
\end{sidewaystable}
\begin{sidewaystable}[ht]
\centering
 {\renewcommand{\arraystretch}{1.25} \begin{tabular}{c|cc|cccc|c}
  $E_{\gamma} \hspace*{2pt} \left[ \mathrm{MeV} \right]$ & $\bm{\uppi}_{n}$ & $\bm{\uppi}_{n} \left(\alpha_{1},\alpha_{2},\beta_{1},\beta_{2}\right)$ & $1.)$ $\alpha_{1}^{\mathrm{Amb.}}$ / $2.)$ $\alpha_{1}^{\mathrm{Fit}}$ & $1.)$ $\alpha_{2}^{\mathrm{Amb.}}$ / $2.)$ $\alpha_{2}^{\mathrm{Fit}}$ & $1.)$ $\beta_{1}^{\mathrm{Amb.}}$ / $2.)$ $\beta_{1}^{\mathrm{Fit}}$ & $1.)$ $\beta_{2}^{\mathrm{Amb.}}$ / $2.)$ $\beta_{2}^{\mathrm{Fit}}$ & $\epsilon_{\bm{\uppi}}$ \\ \hline \hline
  $200$ & $\bm{\uppi}_{9}$ & $\left(\alpha_{1}^{\ast},\alpha_{2},\beta_{1},\beta_{2}^{\ast}\right)$ & $ -0.7255+1.6255 \hspace*{1pt} i $ & $ -0.0335-0.4616 \hspace*{1pt} i $ & $ 0.0574 +0.5265 \hspace*{1pt} i $ & $ 0.5823 -1.4422 \hspace*{1pt} i $  &  $0.0721$ \\[1pt]
   &  &  &  $ -0.7031+1.636 \hspace*{1pt} i $ & $-0.0401-0.4609 \hspace*{1pt} i $ & $0.0456 +0.5275 \hspace*{1pt} i $ & $0.6137 -1.4299 \hspace*{1pt} i $   &   \\
   &  &   &   &   &   &   &   \\
  $210$ & $\bm{\uppi}_{9}$ & $\left(\alpha_{1}^{\ast},\alpha_{2},\beta_{1},\beta_{2}^{\ast}\right)$ & $ -0.8393+1.8116 \hspace*{1pt} i $ & $-0.0132-0.4206 \hspace*{1pt} i $ & $0.0264 +0.5272 \hspace*{1pt} i $ & $0.6422 -1.4564 \hspace*{1pt} i $  &  $0.0371$ \\[1pt]
   &   &  &  $ -0.8283+1.8175 \hspace*{1pt} i $ & $-0.0158-0.4204 \hspace*{1pt} i $ & $0.0198 +0.5273 \hspace*{1pt} i $ & $0.6604 -1.4492 \hspace*{1pt} i $   &   \\
   &   &  &   &   &   &   &   \\
  $220$ & $\bm{\uppi}_{6}$ & $\left(\alpha_{1},\alpha_{2}^{\ast},\beta_{1}^{\ast},\beta_{2}\right)$ & $ -0.9512+1.993 \hspace*{1pt} i $ & $0.0029 -0.389 \hspace*{1pt} i $ & $-0.0027+0.5265 \hspace*{1pt} i $ & $0.6993 -1.474 \hspace*{1pt} i $  &  $0.0047$\\[1pt]
   &   &  &  $ -0.9498+1.9937 \hspace*{1pt} i $ & $0.0026 -0.389 \hspace*{1pt} i $ & $-0.0035+0.5265 \hspace*{1pt} i $ & $0.7018 -1.473 \hspace*{1pt} i $   &   \\
   &   &  &   &   &   &   &   \\
  $230$ & $\bm{\uppi}_{9}$ & $\left(\alpha_{1}^{\ast},\alpha_{2},\beta_{1},\beta_{2}^{\ast}\right)$ &  $ -1.0611-2.1722 \hspace*{1pt} i $ & $0.0157 +0.3638 \hspace*{1pt} i $ & $-0.0292-0.5253 \hspace*{1pt} i $ & $0.7529 +1.4941 \hspace*{1pt} i $  &  $0.0247$ \\[1pt]
   &   &  &  $ -1.0684-2.1679 \hspace*{1pt} i $ & $0.0169 +0.3638 \hspace*{1pt} i $ & $-0.0246-0.5258 \hspace*{1pt} i $ & $0.7388 +1.4999 \hspace*{1pt} i $   &   \\
   &   &  &   &   &   &   &   \\
  $240$ & $\bm{\uppi}_{9}$ & $\left(\alpha_{1}^{\ast},\alpha_{2},\beta_{1},\beta_{2}^{\ast}\right)$  & $ -1.1691-2.3508 \hspace*{1pt} i $ & $0.0261 +0.3433 \hspace*{1pt} i $ & $-0.0535-0.5241 \hspace*{1pt} i $ & $0.8036 +1.516 \hspace*{1pt} i $  &  $0.0518$ \\[1pt]
   &   &  &  $ -1.1845-2.3414 \hspace*{1pt} i $ & $0.0282 +0.3433 \hspace*{1pt} i $ & $-0.0436-0.5257 \hspace*{1pt} i $ & $0.7723 +1.5291 \hspace*{1pt} i $   &   \\ \hline \hline
 \end{tabular}}
 \caption[For all the energies where exact accidental symmetries $\bm{\uppi}$ satisfying $\epsilon_{\bm{\uppi}} \leq 5^{\circ}$ were found, this Table lists corresponding sets of Omelaenko-roots.]{For all the energies where exact accidental symmetries $\bm{\uppi}$ satisfying $\epsilon_{\bm{\uppi}} \leq 5^{\circ}$ were found, this Table lists the roots corresponding to the exact symmetry called $\left( \alpha_{k}^{\mathrm{Amb.}}, \beta_{k}^{\mathrm{Amb.}} \right)$ (printed as upper parameter configuration), as well as the closest solution found by the fit-routine $\left( \alpha_{k}^{\mathrm{Fit.}}, \beta_{k}^{\mathrm{Fit.}} \right)$ (lower parameters). The fitted parameters $\left( \alpha_{k}^{\mathrm{Fit.}}, \beta_{k}^{\mathrm{Fit.}} \right)$ represent the ambiguities $\bm{\tilde{\uppi}}$ introduced in the main text of this section and in appendix \ref{subsec:AccidentalAmbProofsII}. The violation parameters $\epsilon_{\bm{\uppi}}$ of the respective symmetries are shown as well. (Continued in Table \ref{tab:Lmax1TheoryDataAccAmbiguitiesRoots2})}
 \label{tab:Lmax1TheoryDataAccAmbiguitiesRoots}
\end{sidewaystable}
\begin{sidewaystable}[ht]
\centering
 {\renewcommand{\arraystretch}{1.25} \begin{tabular}{c|cc|cccc|c}
  $E_{\gamma} \hspace*{2pt} \left[ \mathrm{MeV} \right]$ & $\bm{\uppi}_{n}$ &  $\bm{\uppi}_{n} \left(\alpha_{1},\alpha_{2},\beta_{1},\beta_{2}\right)$ & $1.)$ $\alpha_{1}^{\mathrm{Amb.}}$ / $2.)$ $\alpha_{1}^{\mathrm{Fit}}$ & $1.)$ $\alpha_{2}^{\mathrm{Amb.}}$ / $2.)$ $\alpha_{2}^{\mathrm{Fit}}$ & $1.)$ $\beta_{1}^{\mathrm{Amb.}}$ / $2.)$ $\beta_{1}^{\mathrm{Fit}}$ & $1.)$ $\beta_{2}^{\mathrm{Amb.}}$ / $2.)$ $\beta_{2}^{\mathrm{Fit}}$ & $\epsilon_{\bm{\uppi}}$ \\ \hline \hline
  $250$ & $\bm{\uppi}_{6}$ &   $\left(\alpha_{1},\alpha_{2}^{\ast},\beta_{1}^{\ast},\beta_{2}\right)$ &  $ -1.2741-2.5297 \hspace*{1pt} i $ & $0.0345 +0.3264 \hspace*{1pt} i $ & $-0.0757-0.5231 \hspace*{1pt} i $ & $ 0.8511 +1.5397 \hspace*{1pt} i $ &  $0.0768$ \\[1pt]
   &   &  &  $ -1.2975-2.5153 \hspace*{1pt} i $ & $ 0.0373 +0.3263 \hspace*{1pt} i $ & $ -0.0609-0.5263 \hspace*{1pt} i $ & $ 0.8023 +1.5605 \hspace*{1pt} i $   &   \\
   &  &   &   &   &   &   &   \\
  $410$ & $\bm{\uppi}_{3}$ &  $\left(\alpha_{1}^{\ast},\alpha_{2}^{\ast},\beta_{1},\beta_{2}\right)$ &  $ -2.3515+6.7753 \hspace*{1pt} i $ & $ 0.0823 +0.2471 \hspace*{1pt} i $ & $ -0.3326+0.5332 \hspace*{1pt} i $ & $ 1.6044 +2.5025 \hspace*{1pt} i $  &  $0.0248$ \\[1pt]
   &   &  &  $ -2.2864+6.7341 \hspace*{1pt} i $ & $ 0.0821 +0.2473 \hspace*{1pt} i $ & $ -0.3315+0.5385 \hspace*{1pt} i $ & $ 1.5189 +2.5057 \hspace*{1pt} i $   &   \\
   &  &   &   &   &   &   &   \\
  $420$ & $\bm{\uppi}_{12}$ &  $\left(\alpha_{1},\alpha_{2},\beta_{1}^{\ast},\beta_{2}^{\ast}\right)$ &  $ -2.1076-7.2993 \hspace*{1pt} i $ & $ 0.0812 -0.2473 \hspace*{1pt} i $ & $ -0.3459-0.5359 \hspace*{1pt} i $ & $ 1.586 -2.6641 \hspace*{1pt} i $  &  $0.0725$ \\[1pt]
   &   &  &  $ -2.3036-7.4637 \hspace*{1pt} i $ & $ 0.0819 -0.2467 \hspace*{1pt} i $ & $ -0.3494-0.5206 \hspace*{1pt} i $ & $ 1.861 -2.6506 \hspace*{1pt} i $   &   \\
   &  &   &   &   &   &   &   \\
  $510$ & $\bm{\uppi}_{10}$ &  $\left(\alpha_{1},\alpha_{2}^{\ast},\beta_{1},\beta_{2}^{\ast}\right)$ &  $ 0.0555 -0.2421 \hspace*{1pt} i $ & $ 7.0415 +8.0403 \hspace*{1pt} i $ & $ -0.66+3.5037 \hspace*{1pt} i $ & $ -0.5121-0.5405 \hspace*{1pt} i $  &  $0.0784$ \\[1pt]
   &   &  &  $ 0.0552 -0.2421 \hspace*{1pt} i $ & $ 7.2722 +8.0247 \hspace*{1pt} i $ & $ -0.808+3.5032 \hspace*{1pt} i $ & $ -0.5036-0.5531 \hspace*{1pt} i $   &   \\
   &  &   &   &   &   &   &   \\
  $610$ & $\bm{\uppi}_{9}$ &  $\left(\alpha_{1}^{\ast},\alpha_{2},\beta_{1},\beta_{2}^{\ast}\right)$ &  $ 0.0714 +0.2036 \hspace*{1pt} i $ & $ 5.8247 +0.9199 \hspace*{1pt} i $ & $ -0.8396-0.1105 \hspace*{1pt} i $ & $ -0.5337-1.4041 \hspace*{1pt} i $  &  $0.0517$ \\[1pt]
   &   &  &  $ 0.0714 +0.2036 \hspace*{1pt} i $ & $ 5.8321 +0.8758 \hspace*{1pt} i $ & $ -0.8293-0.1518 \hspace*{1pt} i $ & $ -0.5447-1.4078 \hspace*{1pt} i $   &   \\ \hline \hline
 \end{tabular}}
 \caption{This Table is the continuation of Table \ref{tab:Lmax1TheoryDataAccAmbiguitiesRoots} for higher energies.}
 \label{tab:Lmax1TheoryDataAccAmbiguitiesRoots2}
\end{sidewaystable}
\begin{sidewaystable}[ht]
\centering
 \begin{tabular}{c|ccccc|c}
  $E_{\gamma} \hspace*{2pt} \left[ \mathrm{MeV} \right]$ & $\bm{\uppi}_{n}$ & $\xi_{1}$ & $ \xi_{2} $ & $ \zeta_{1}$ & $\zeta_{2}$ & $\epsilon_{\bm{\uppi}}$ \\ \hline \hline
  $200$ & $\bm{\uppi}_{9}$ & $0.0004\, -0.0139 \hspace*{1pt} i$ & $-0.0002 - 0.0141 \hspace*{1pt} i$ & $-0.0004 + 0.0224 \hspace*{1pt} i$ & $0.0005\, +0.0217 \hspace*{1pt} i$ & $0.0721$ \\
  $210$ & $\bm{\uppi}_{9}$ & $0.0003\, -0.0062 \hspace*{1pt} i$ & $-0.0003 - 0.0062 \hspace*{1pt} i$ & $- 0.0004 + 0.0124 \hspace*{1pt} i$ & $0.0005\, + 0.0122 \hspace*{1pt} i$ & $0.0371$ \\
  $220$ & $\bm{\uppi}_{6}$ & $0.0001\, -0.0007 \hspace*{1pt} i$ & $- 0.0007 \hspace*{1pt} i$ & $- 0.0001 + 0.0017 \hspace*{1pt} i$ & $0.0001\, + 0.0017 \hspace*{1pt} i$ & $0.0047$ \\
  $230$ & $\bm{\uppi}_{9}$ & $-0.0003 - 0.0035 \hspace*{1pt} i$ & $0.0002\, -0.0033 \hspace*{1pt} i$ & $0.0006\, +0.0089 \hspace*{1pt} i$ & $-0.0006+0.0091 \hspace*{1pt} i$ & $0.0247$ \\  
  $240$ & $\bm{\uppi}_{9}$ & $-0.0006 - 0.0069 \hspace*{1pt} i$ & $0.0004\, -0.0063 \hspace*{1pt} i$ & $0.0014\, +0.0190 \hspace*{1pt} i$ & $-0.0016+0.0197 \hspace*{1pt} i$ & $0.0518$ \\
  $250$ & $\bm{\uppi}_{6}$ & $-0.0008-0.0097 \hspace*{1pt} i$ & $0.0006\, -0.0086 \hspace*{1pt} i$ & $0.0024\, + 0.0285 \hspace*{1pt} i$ & $-0.0026+0.0301 i$ & $0.0768$ \\  
  $410$ & $\bm{\uppi}_{3}$ & $-0.0084-0.0067 \hspace*{1pt} i$ & $0.0003\, +0.0011 \hspace*{1pt} i$ & $0.0063\, -0.0059 \hspace*{1pt} i$ & $-0.0144+0.0251 i$ & $0.0248$ \\  
  $420$ & $\bm{\uppi}_{12}$ & $0.0277\, -0.0183 \hspace*{1pt} i$ & $-0.0013 + 0.0031 \hspace*{1pt} i$ & $- 0.0172 - 0.0178 \hspace*{1pt} i$ & $0.0436\, + 0.0752 i$ & $0.0725$ \\
  $510$ & $\bm{\uppi}_{10}$ & $-0.0003-0.0011 \hspace*{1pt} i$ & $0.0132\, -0.0170 \hspace*{1pt} i$ & $0.0083\, + 0.0405 \hspace*{1pt} i$ & $0.0046\, +0.0199 i$ & $0.0784$ \\
  $610$ & $\bm{\uppi}_{9}$ & $0.0003\, +0.0002 \hspace*{1pt} i$ & $0.0001\, -0.0076 \hspace*{1pt} i$ & $-0.0044 + 0.0502 \hspace*{1pt} i$ & $0.0049\, - 0.0059 i$ & $0.0517$ \\  
 \end{tabular}
 \caption[This Table shows the quantities $\xi_{k}$ and $\zeta_{k^{\prime}}$, extracted via equation (\ref{eq:XikZetakExtraction}) for all relevant accidental ambiguities in the group $\mathcal{S}$ fit, for MAID theory-data truncated at $L=1$.]{This Table shows the quantities $\xi_{k}$ and $\zeta_{k^{\prime}}$, extracted via equation (\ref{eq:XikZetakExtraction}) for all relevant accidental ambiguities in the group $\mathcal{S}$ fit. Numbers have been rounded to $4$ decimals. The violation parameters $\epsilon_{\bm{\uppi}}$ are printed as well. It can be seen that the obtained numbers indeed have small modulus, as claimed in appendix \ref{subsec:AccidentalAmbProofsII}, and furthermore that they satisfy the constraint $i \epsilon_{\bm{\uppi}} = - \xi_{1} - \xi_{2} + \zeta_{1} + \zeta_{2}$ (at least within rounding-errors).}
 \label{tab:Lmax1TheoryDataAccAmbiguitiesNumbers2}
\end{sidewaystable}

\clearpage

\subsubsection{MAID2007 theory-data up to $\ell_{\mathrm{max}} = 2, 3, 4$} \label{subsec:TheoryDataFitsLmax2}

Here, we consider fits of MAID2007 theory-data \cite{LotharPrivateComm,MAID2007} for higher truncation orders than those investigated in the previous section, where theoretical data containing contributions from just the $S$- and $P$-waves were at the center of attention. Specifically, the cases $\ell_{\mathrm{max}} = 2, 3, 4$ are studied. For this purpose, the MAID-group \cite{LotharPrivateComm} has provided theory-data for the channel $\gamma p \longrightarrow \pi^{0} p$ containing contributions from all multipoles up to the respective truncation order. For instance, the truncated data for the case $\ell_{\mathrm{max}} = 2$ were generated using the MAID-multipoles
\begin{equation}
 \left\{  E_{0+}^{p \pi^{0}}, E_{1+}^{p \pi^{0}}, M_{1+}^{p \pi^{0}}, M_{1-}^{p \pi^{0}},  E_{2+}^{p \pi^{0}}, E_{2-}^{p \pi^{0}}, M_{2+}^{p \pi^{0}}, M_{2-}^{p \pi^{0}} \right\} \mathrm{.} \label{eq:SPDWavesMAIDPi0}
\end{equation}
Theory-data with higher truncation orders were generated in an exactly analogous way. Furthermore, irrespective of the particular dataset under consideration, we again restrict the energy range of the theory-data to the interval $E_{\gamma} \in \left[ 160, 650 \right] \hspace*{1pt} \mathrm{MeV}$. \newline
The goal of these investigations is to study the anticipated higher degree of complexity in the solution behavior of the theory-data, which is expected to occur once $\ell_{\mathrm{max}}$ is raised (cf. chapter \ref{chap:Omelaenko}, specifically section \ref{sec:WBTpaper}, as well as appendix \ref{sec:AccidentalAmbProofs}). We choose here the truncated theory-data, since it is in this case known that an 'exact' solution of the TPWA-problem exists, which was also a central assumption for the ambiguity theorems in chapter \ref{chap:Omelaenko}. \newline
Correspondingly, the study is performed by always fitting the \textit{correct} truncation order corresponding to the $\ell_{\mathrm{max}}$ used to generate the data. Again, the focus is set on potentially complete sets of observables composed of group $\mathcal{S}$ and $\mathcal{BT}$ observables only, since the complete sets proposed for a TPWA generally have the attractive feature of avoiding the double polarization observables with recoil polarization \cite{MyCompExTPWAPaper,Omelaenko}. \newline

\begin{figure}[ht]
 \centering
\vspace*{10pt}
\begin{overpic}[width=0.485\textwidth]{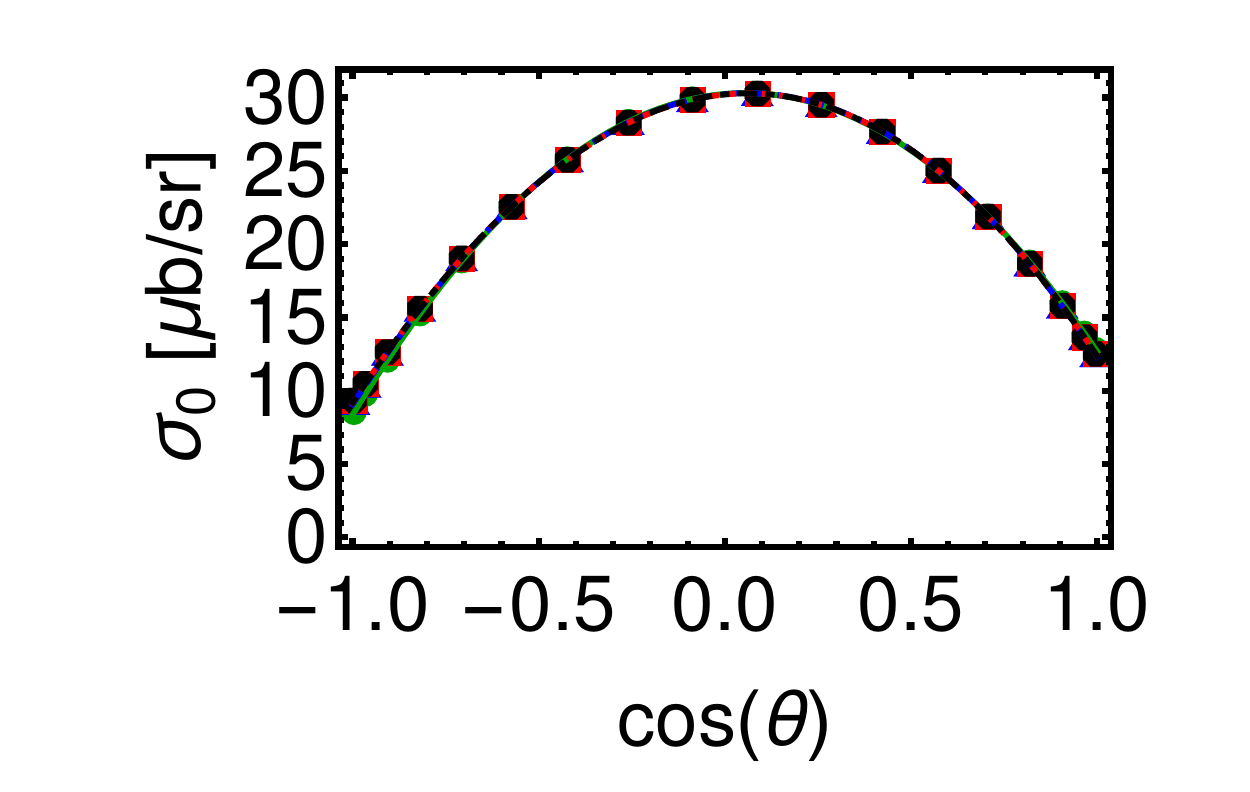}
 \put(85,65){\begin{Large}$E_{\gamma} = 330 \hspace*{2pt} \mathrm{MeV}$\end{Large}}
 \end{overpic}
\begin{overpic}[width=0.485\textwidth]{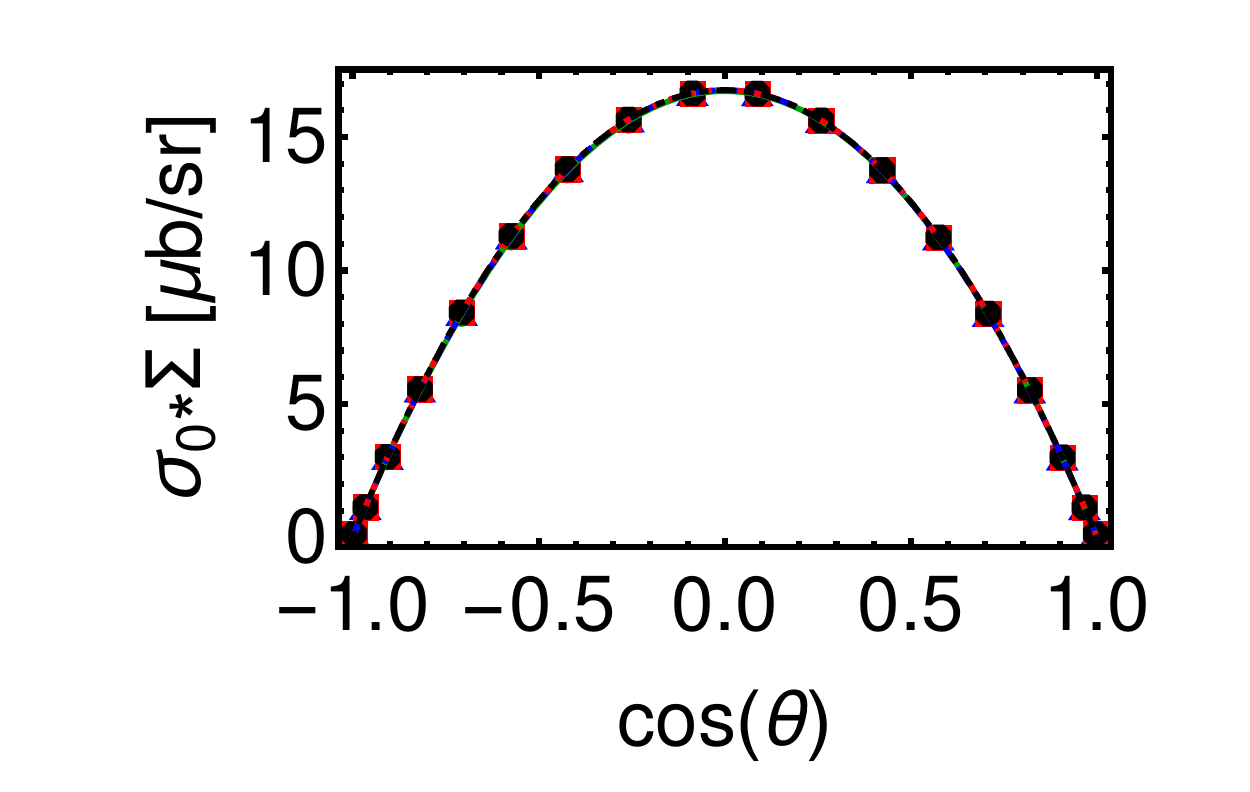}
 \end{overpic} \\
\begin{overpic}[width=0.485\textwidth]{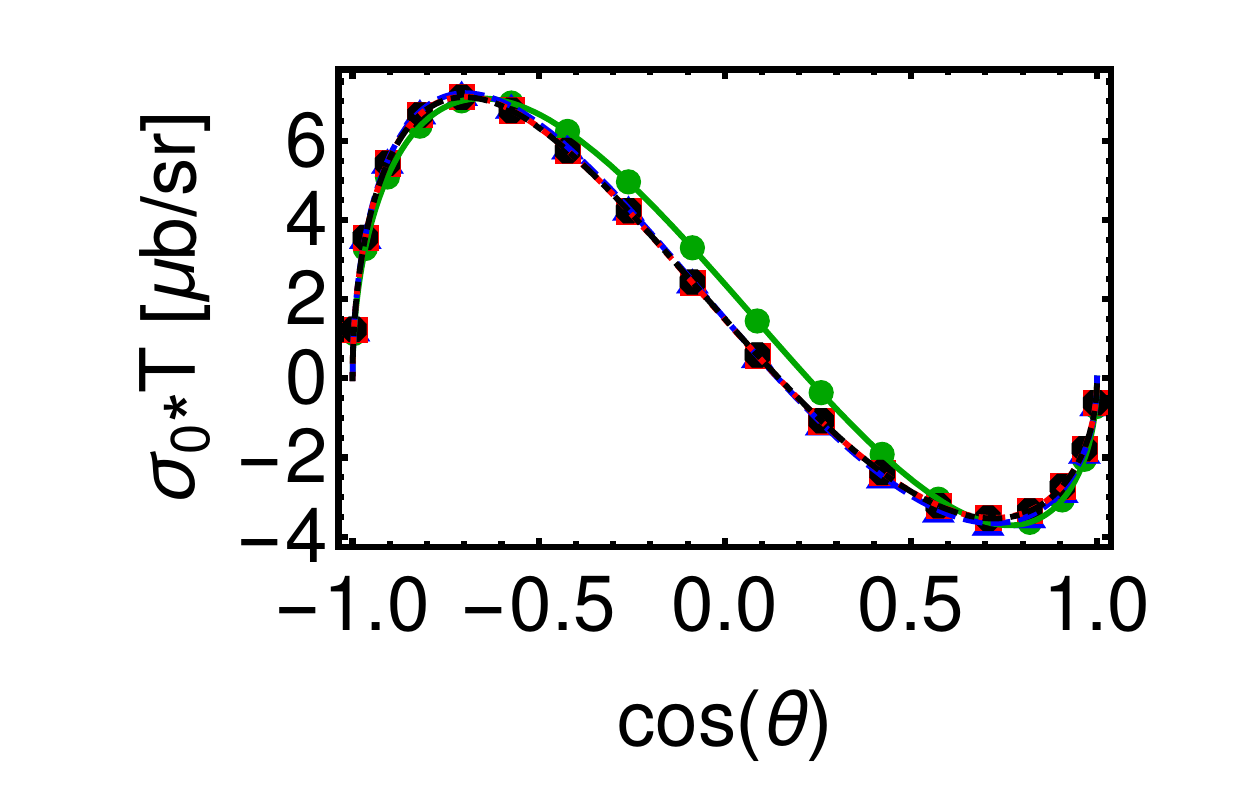}
 \end{overpic}
\begin{overpic}[width=0.485\textwidth]{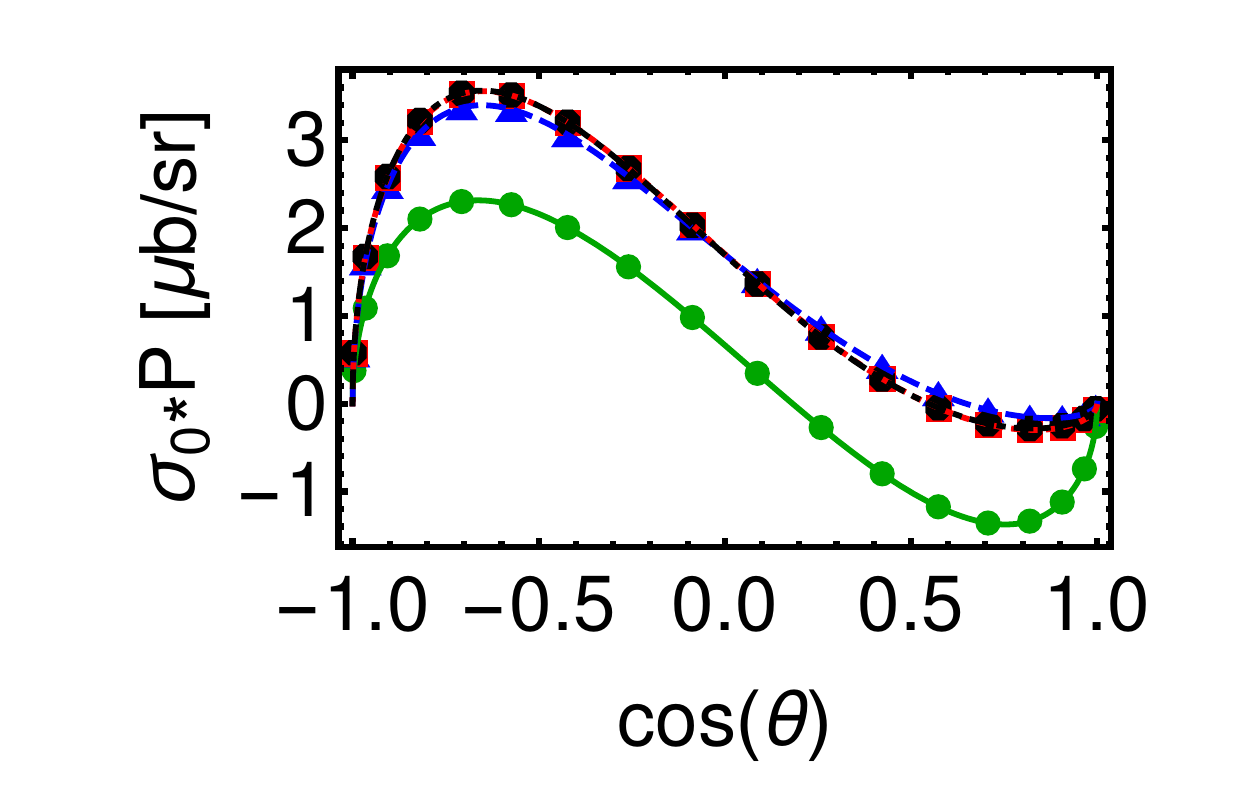}
 \end{overpic} \\
\begin{overpic}[width=0.485\textwidth]{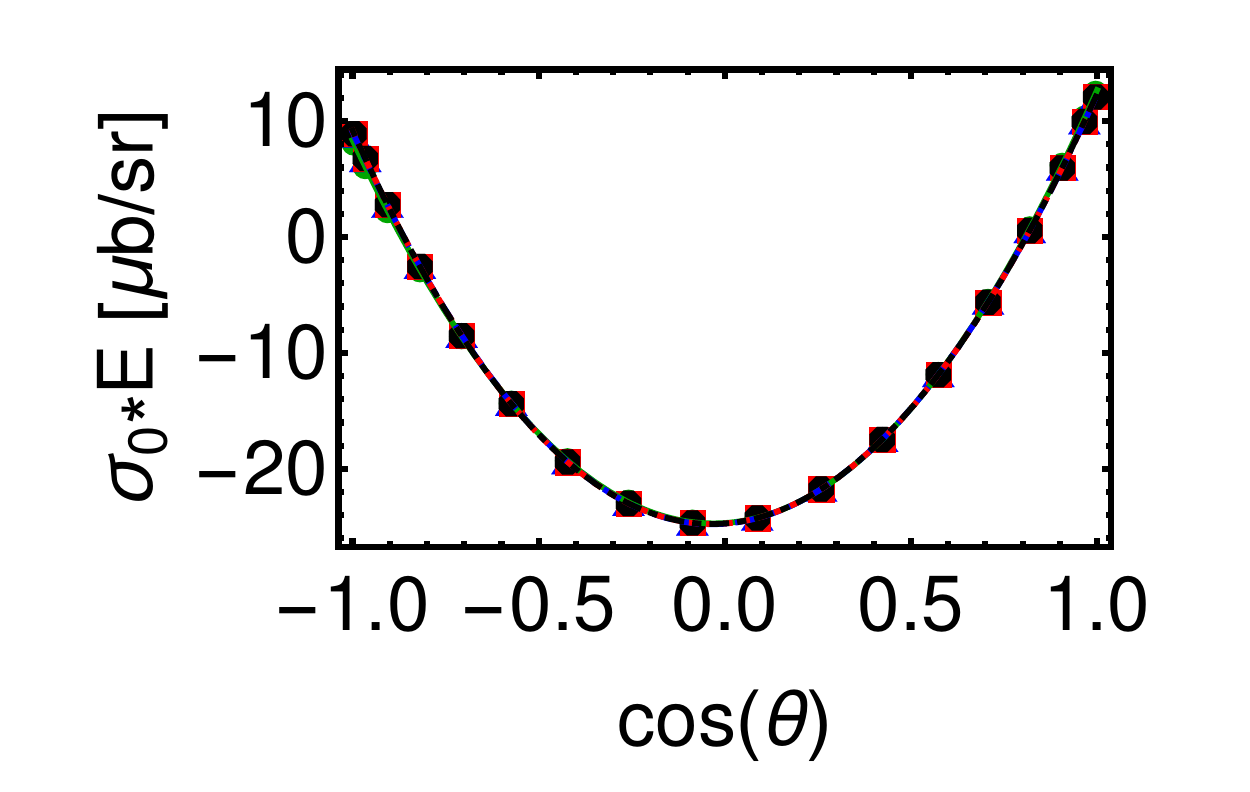}
 \end{overpic}
\begin{overpic}[width=0.485\textwidth]{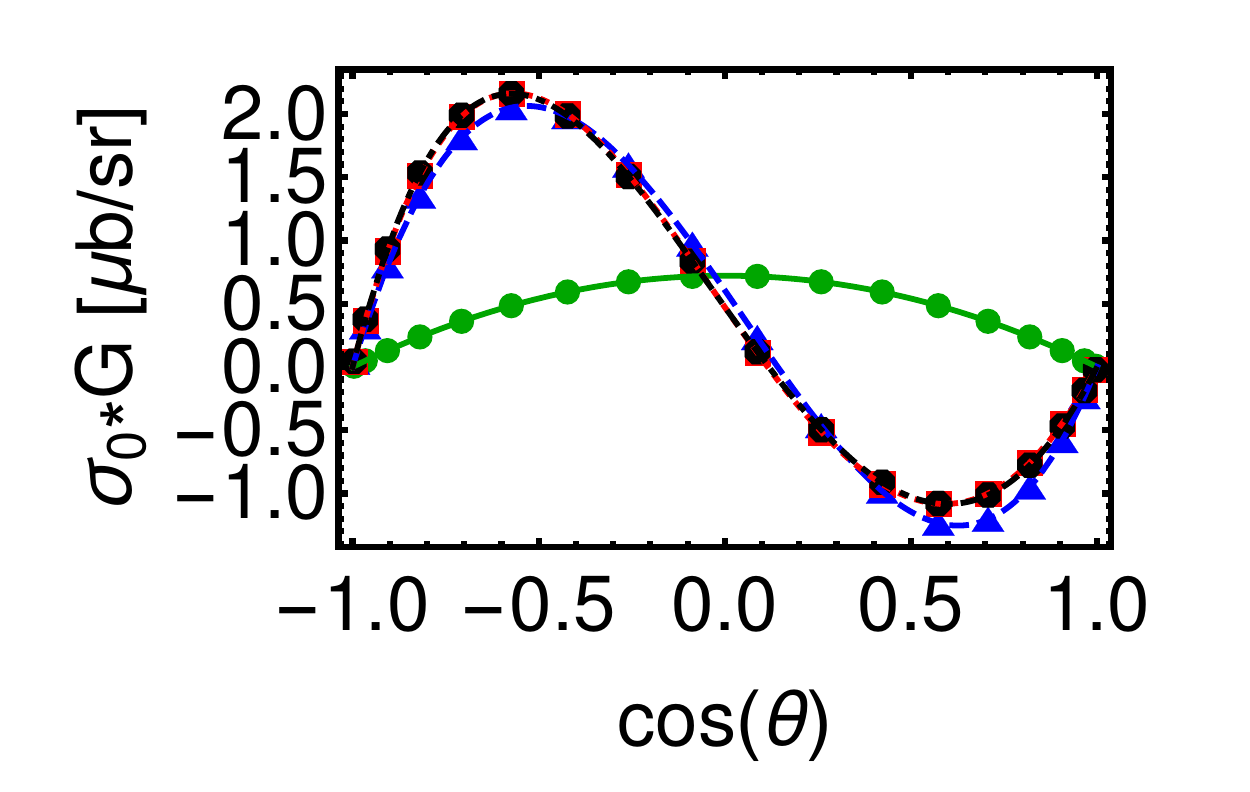}
 \end{overpic} \\
\begin{overpic}[width=0.485\textwidth]{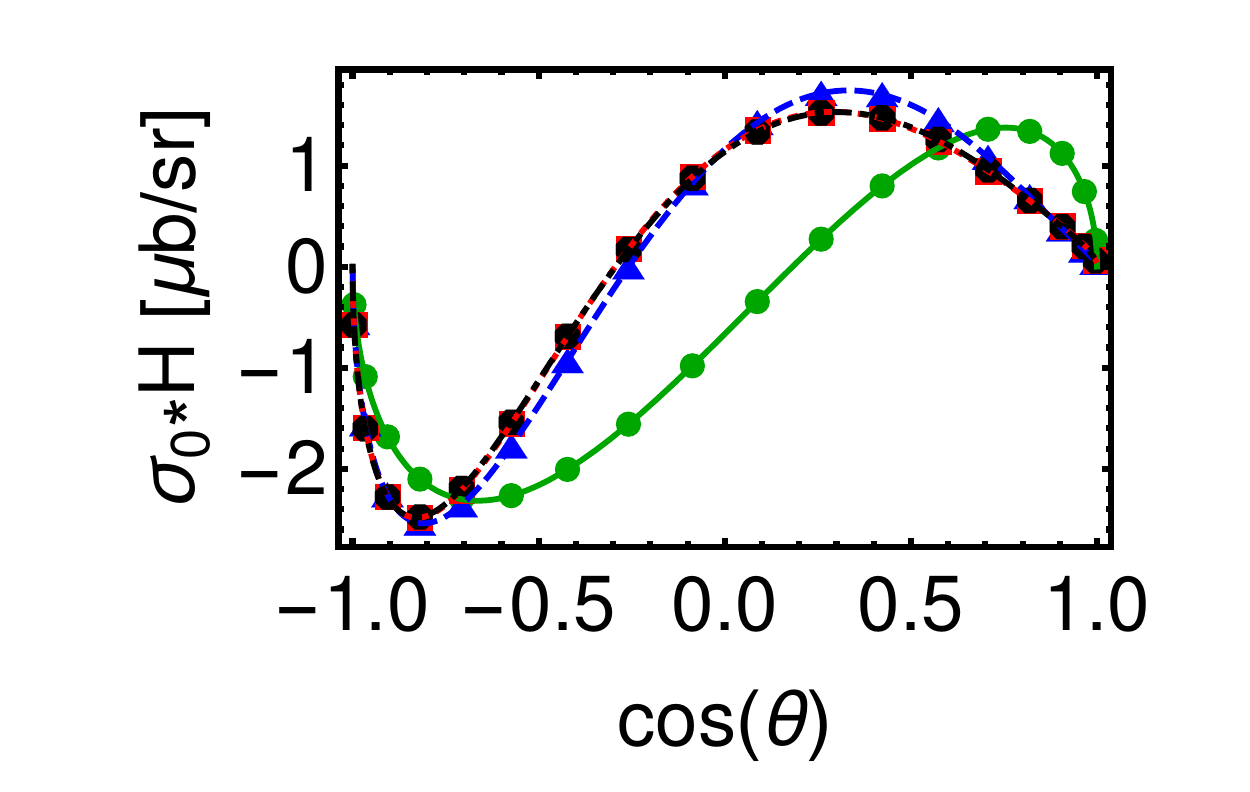}
 \end{overpic}
\begin{overpic}[width=0.485\textwidth]{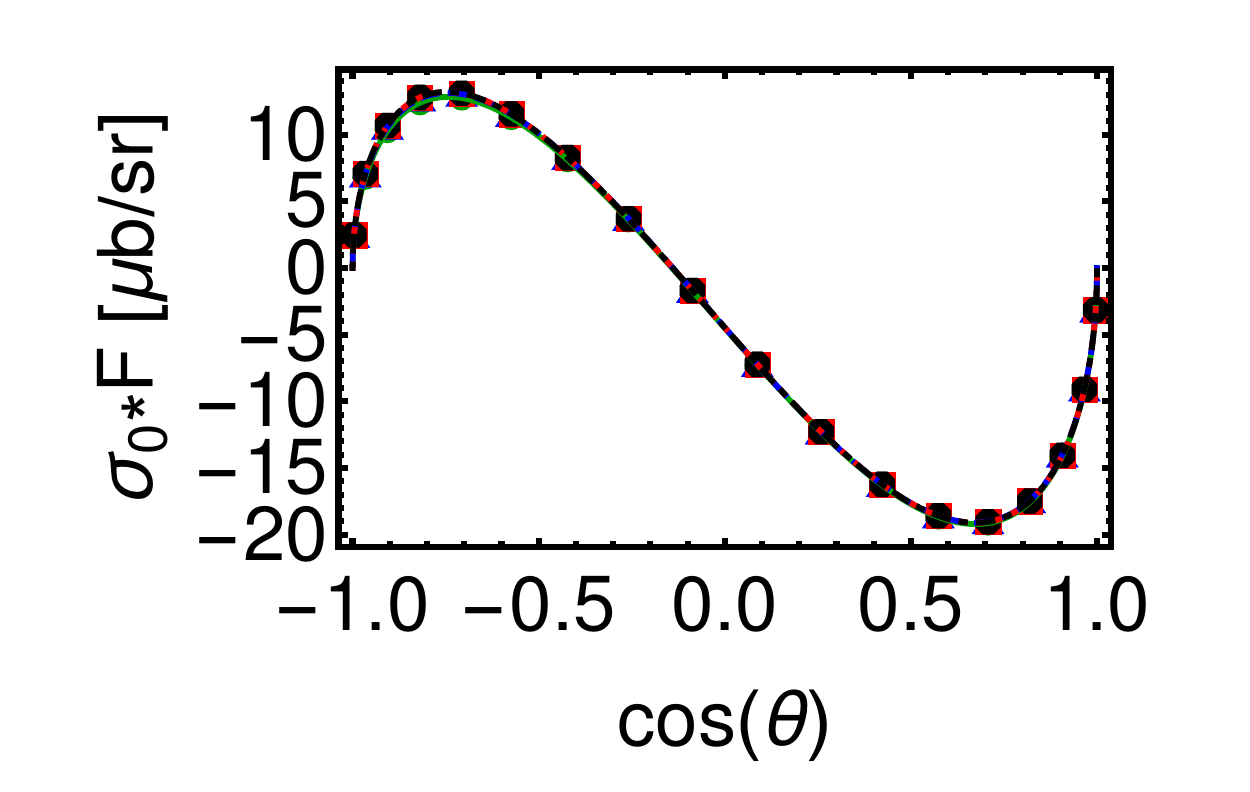}
 \end{overpic}
\caption[Angular distributions of MAID2007 theory-data truncated at $L=1,2,3,4$. Photon energy: $E_{\gamma} = 330 \hspace*{2pt} \mathrm{MeV}$.]{Truncated MAID2007 theory-data \cite{LotharPrivateComm,MAID2007} for different orders $\ell_{\mathrm{max}}$ are shown here. Depicted are the angular distributions of the profile functions belonging to the group $\mathcal{S}$ and $\mathcal{BT}$ observables at an example photon energy of $E_{\gamma} = 330 \hspace*{2pt} \mathrm{MeV}$. Explicitly shown are theory-data truncated at $\ell_{\mathrm{max}} = 1$ (green dots), $\ell_{\mathrm{max}} = 2$ (blue triangles), $\ell_{\mathrm{max}}= 3$ (red squares) and $\ell_{\mathrm{max}} = 4$ (black octagons). \newline The TPWA Legendre fits are also shown, where every theory-data set was fitted in the corresponding truncation order. Fit curves are plotted for $\ell_{\mathrm{max}} = 1$ (green solid line), $\ell_{\mathrm{max}} = 2$ (blue dashed line), $\ell_{\mathrm{max}} = 3$ (red dashed line) and $\ell_{\mathrm{max}} = 4$ (black dashed line).}
\label{fig:Lmax1234ThDataFitGroupSBTObservablesExampleEnergy}
\end{figure}

At first, it is advisable to compare the angular distributions of observables for different theory-data sets and study the effects of the introduction of higher partial waves. The corresponding plots can be seen in Figure \ref{fig:Lmax1234ThDataFitGroupSBTObservablesExampleEnergy}. Shown there are angular distributions of all fitted observables at the example energy $E_{\gamma} = 330 \hspace*{2pt} \mathrm{MeV}$, for all theory-data truncation orders from $\ell_{\mathrm{max}}= 1$ up to $4$. Also, a corresponding fit curve is shown for every observable, where the fit was done in the specific truncation order which matches that of the theory-data. \newline
Several interesting features can be observed. First, the difference of data and fit functions between the fit orders $\ell_{\mathrm{max}}= 3$ and $4$ are practically invisible. This fact just reflects the small size of the $G$-waves in the $\Delta$-energy region, which leads to generally tiny modifications once one truncates beyond the $F$-waves. However, in regard of the differences between high and low truncation orders, it is seen that the observables drastically differ from each other concerning the modifications they receive from the introduction of higher partial waves. The cross section $\sigma_{0}$, beam asymmetry $\Sigma$ and double polarization observable $E$ for instance show absolutely no perceivable difference between the theory-data truncated at $\ell_{\mathrm{max}}= 1$ or $\ell_{\mathrm{max}}= 4$. Quantities which are highly sensitive to such corrections are the asymmetries $P$, $G$ and $H$. These differences in the behavior of observables are true over the whole energy region of the theory-data, though minor variations can occur. \newline

\begin{figure}[ht]
 \centering
\begin{overpic}[width=0.48\textwidth]{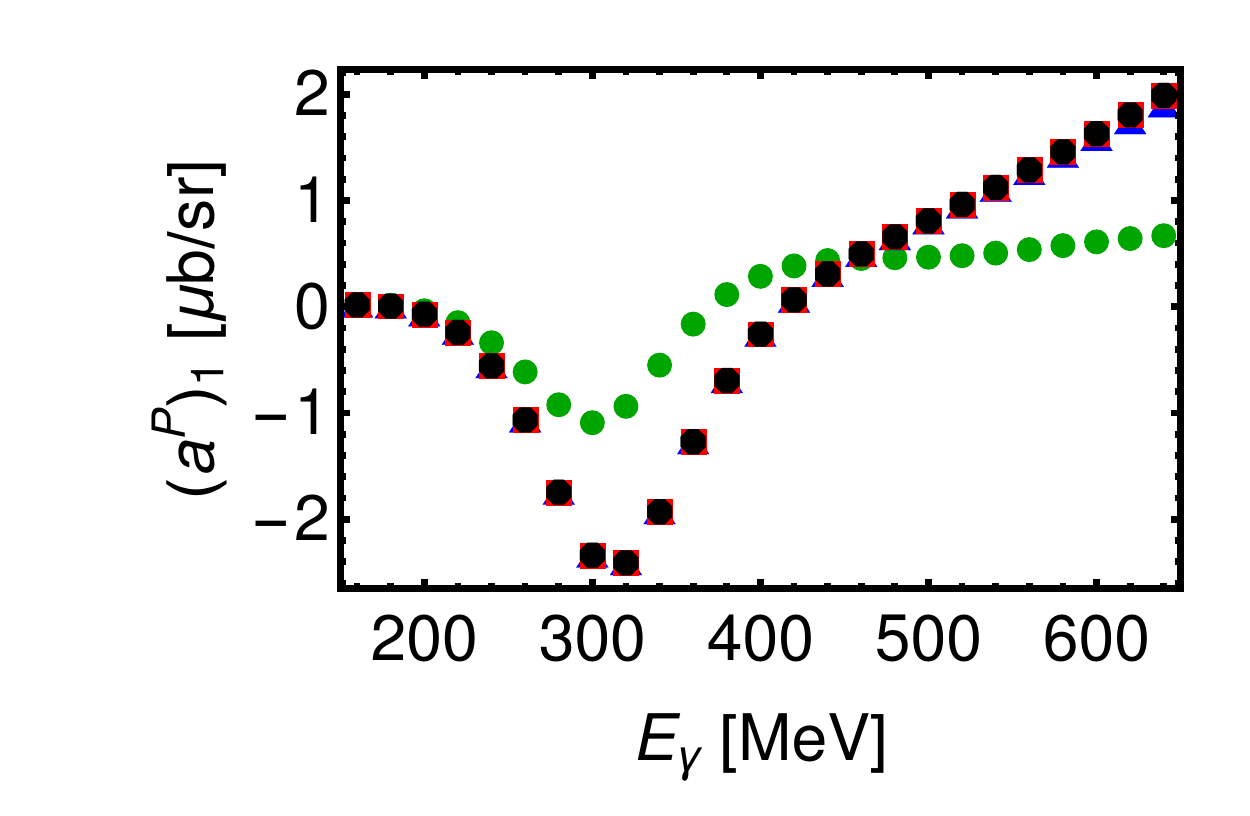}
 \end{overpic}
\begin{overpic}[width=0.48\textwidth]{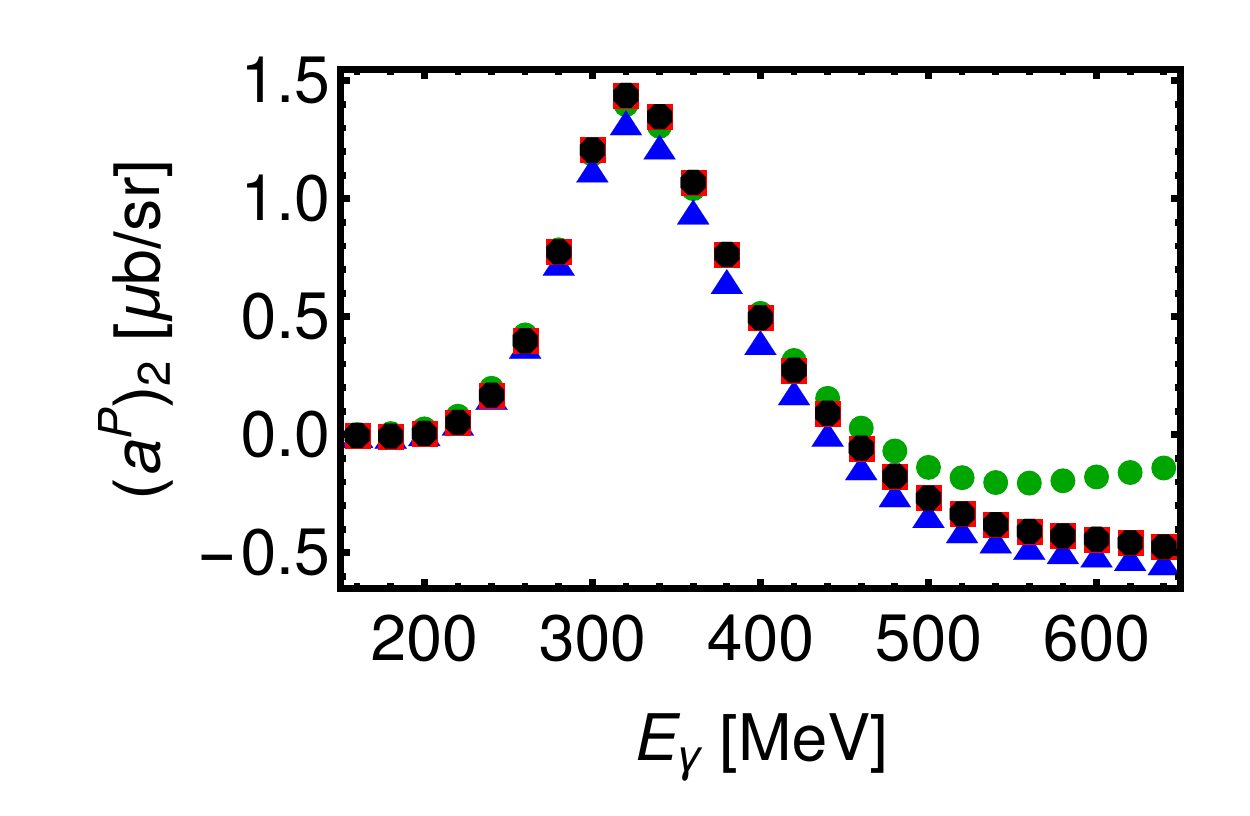}
 \end{overpic} \\
\begin{overpic}[width=0.48\textwidth]{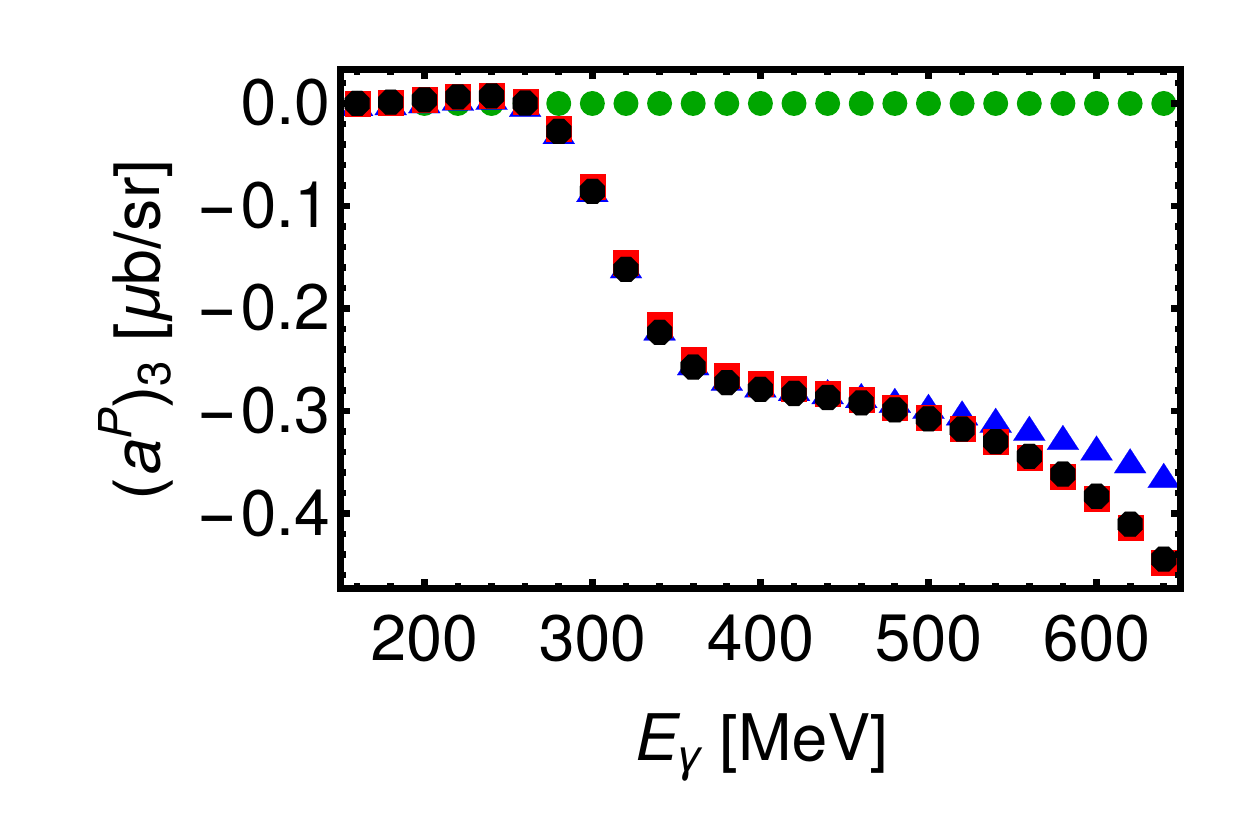}
 \end{overpic}
\begin{overpic}[width=0.48\textwidth]{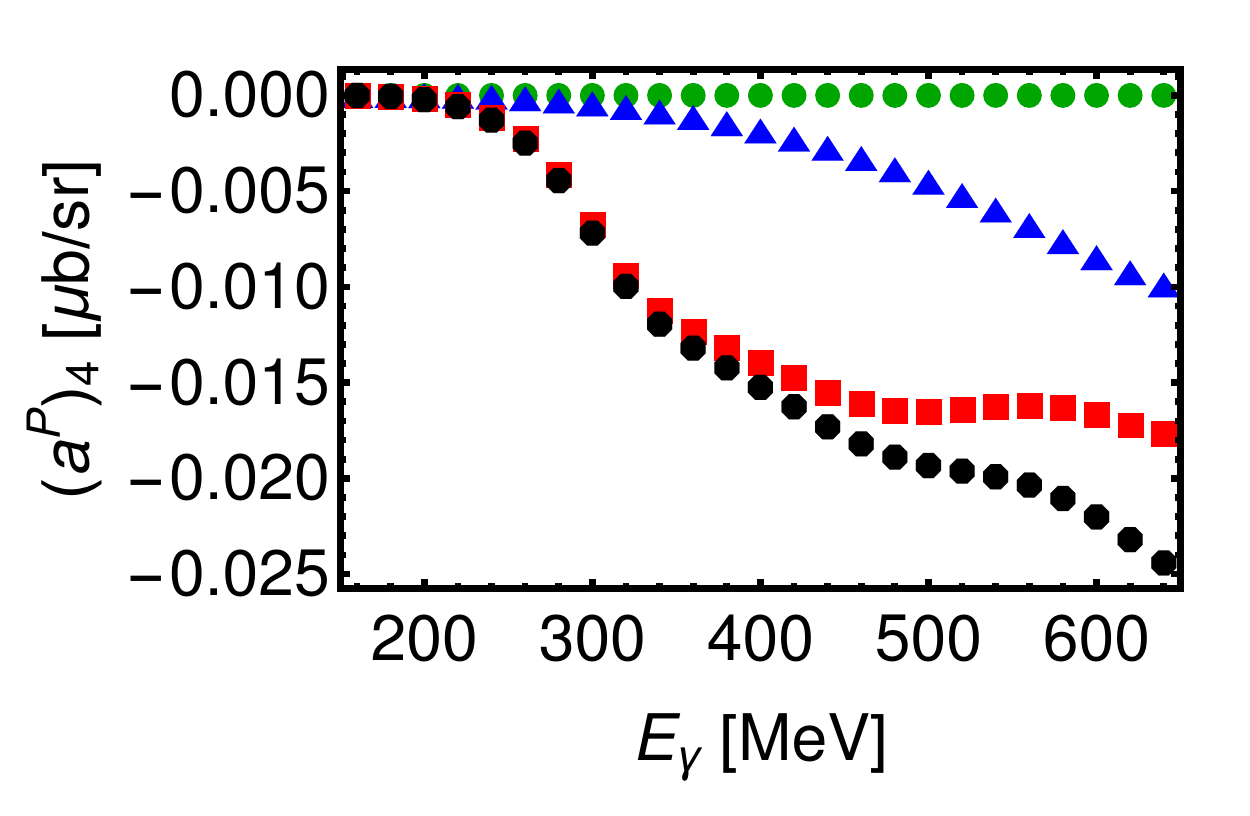}
 \end{overpic}
\caption[Legendre coefficients $\left(a^{P}\right)_{1,\ldots,4}$ of the observable $P$ extracted from the truncated theory-data at $\ell_{\mathrm{max}} = 1,2,3,4$.]{Shown here are the Legendre coefficients $\left(a^{P}\right)_{1,\ldots,4}$ of the observable $P$ extracted from the truncated theory-data \cite{LotharPrivateComm,MAID2007} with $\ell_{\mathrm{max}} = 1$ (green dots), $\ell_{\mathrm{max}} = 2$ (blue triangles), $\ell_{\mathrm{max}} = 3$ (red squares) and $\ell_{\mathrm{max}} = 4$ (black octagons).}
\label{fig:Lmax1234ThDataFitLegCoeffsP1}
\end{figure}

\begin{figure}[ht]
 \centering
\begin{overpic}[width=0.48\textwidth]{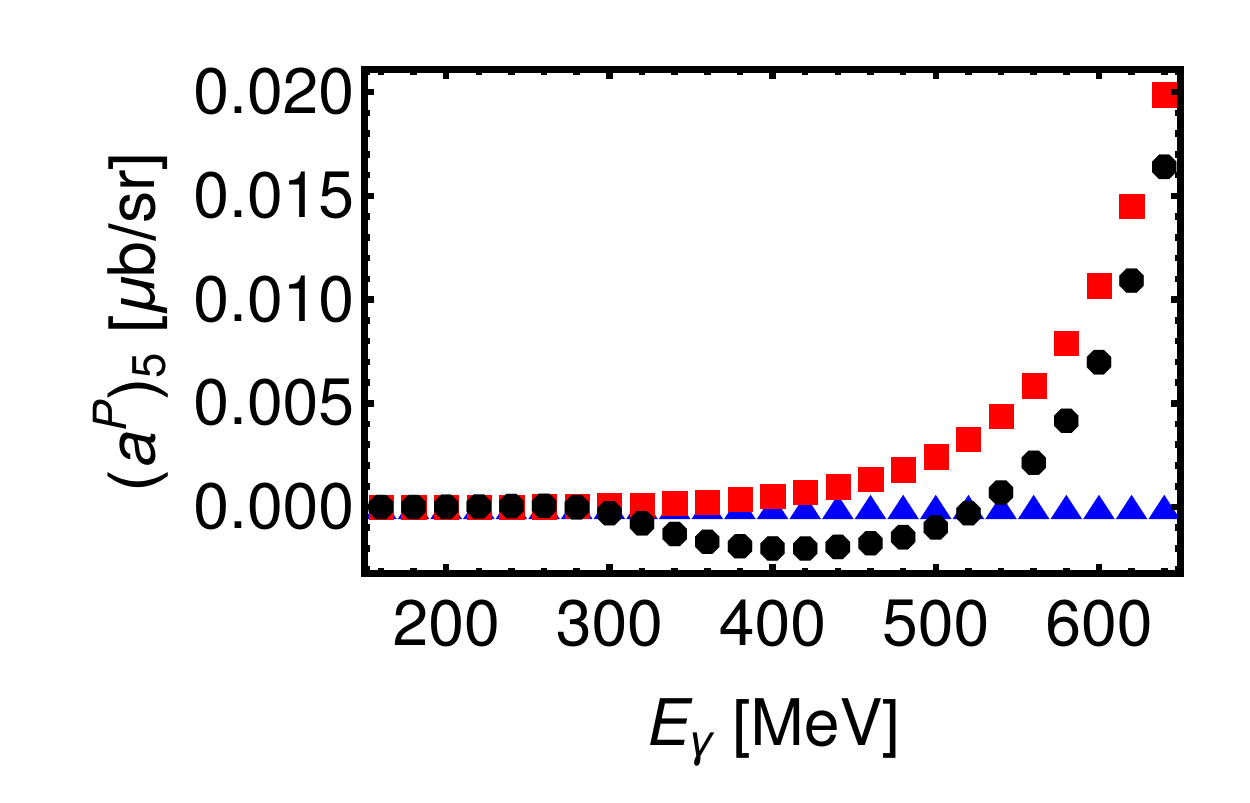}
 \end{overpic}
\begin{overpic}[width=0.48\textwidth]{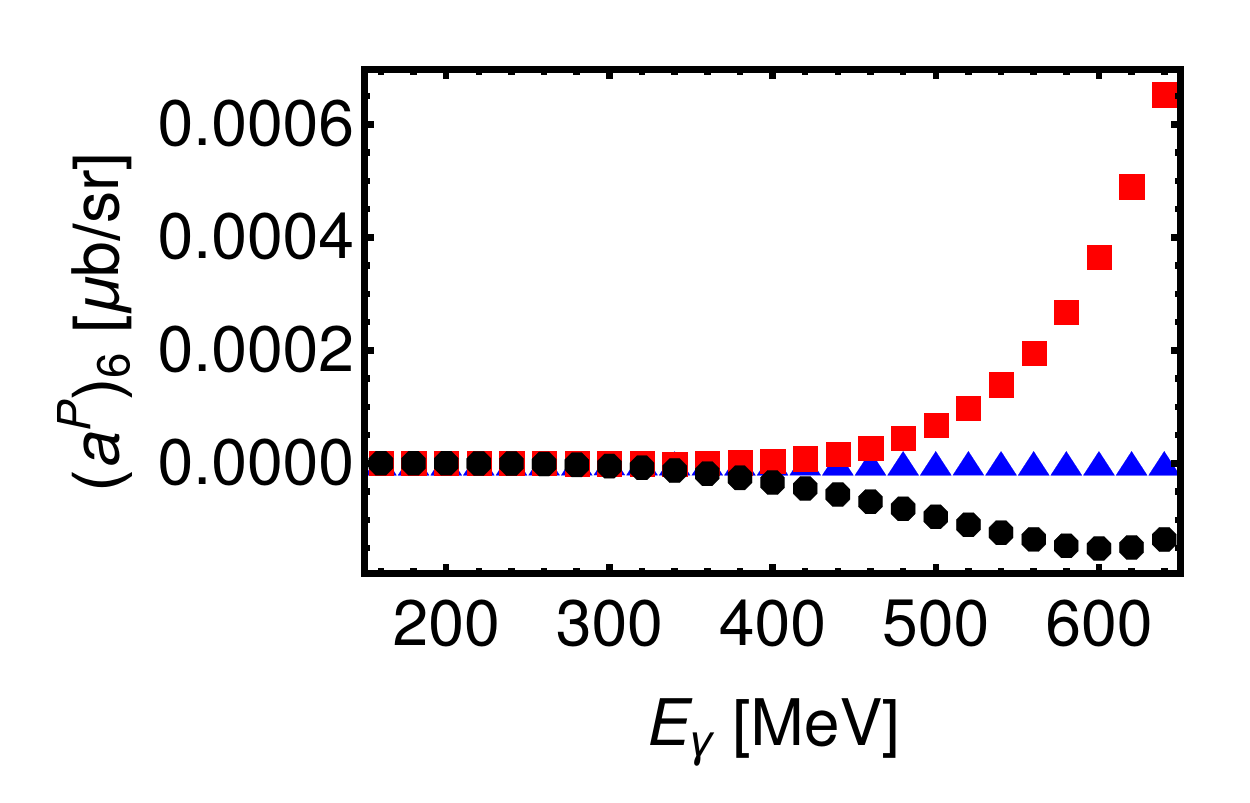}
 \end{overpic}
\caption[Legendre coefficients $\left(a^{P}\right)_{5,6}$ of the observable $P$ extracted from the truncated theory-data at $\ell_{\mathrm{max}} = 1,2,3,4$.]{This figure depicts the Legendre coefficients $\left(a^{P}\right)_{5,6}$ of the observable $P$ extracted from the truncated theory-data \cite{LotharPrivateComm,MAID2007} with $\ell_{\mathrm{max}} = 2$ (blue triangles), $\ell_{\mathrm{max}} = 3$ (red squares) and $\ell_{\mathrm{max}} = 4$ (black octagons).}
\label{fig:Lmax1234ThDataFitLegCoeffsP2}
\end{figure}

\begin{figure}[hb]
 \centering
 \begin{overpic}[width=0.48\textwidth]{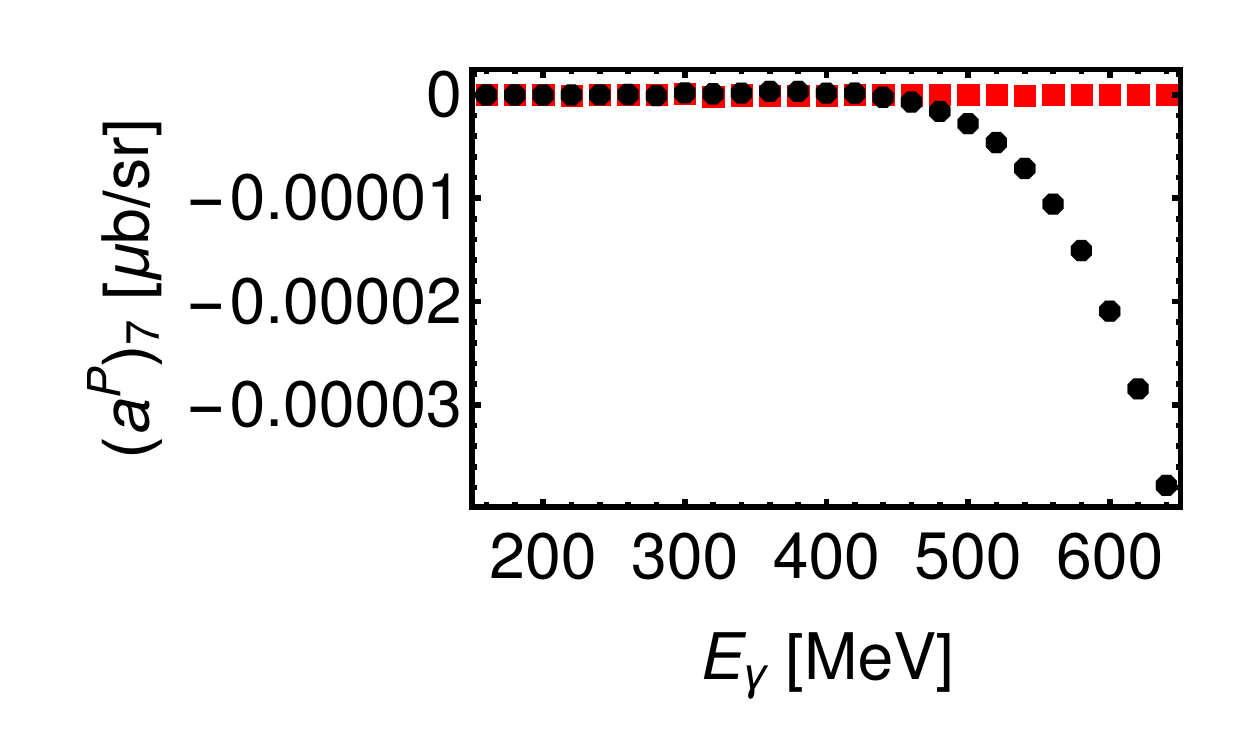}
 \end{overpic}
\begin{overpic}[width=0.48\textwidth]{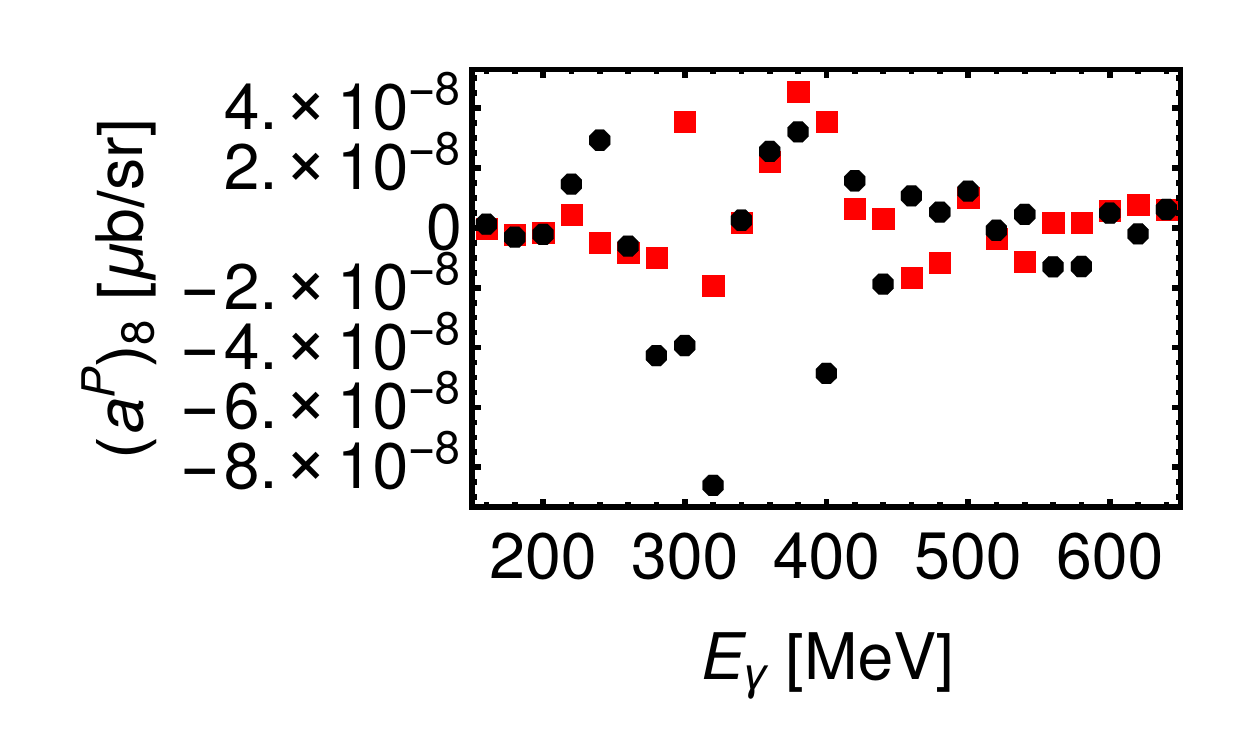}
 \end{overpic}
\caption[Legendre coefficients $\left(a^{P}\right)_{7,8}$ of the observable $P$ extracted from the truncated theory-data at $\ell_{\mathrm{max}} = 1,2,3,4$.]{Shown here are the Legendre coefficients $\left(a^{P}\right)_{7,8}$ of the observable $P$ extracted from the truncated theory-data \cite{LotharPrivateComm,MAID2007} with $\ell_{\mathrm{max}} = 3$ (red squares) and $\ell_{\mathrm{max}} = 4$ (black octagons).}
\label{fig:Lmax1234ThDataFitLegCoeffsP3}
\end{figure}

The observations made above are reflected in the behavior of the extracted Legendre coefficients. The observable $P$ is chosen as an example for a quantity with significant dependence on higher partial waves. Figures \ref{fig:Lmax1234ThDataFitLegCoeffsP1}, \ref{fig:Lmax1234ThDataFitLegCoeffsP2} and \ref{fig:Lmax1234ThDataFitLegCoeffsP3} show the coefficients fitted out of theory-data having different truncation orders, over the whole considered energy $E_{\gamma}$-region. \newpage

\clearpage

The coefficient $a^{P}_{1}$ has the largest absolute values and shows no visible modifications for truncations with $\ell_{\mathrm{max}} \geq 2$. For $a^{P}_{2}$, very small modifications due to $F$-waves are visible almost over the whole energy region. In the coefficient $a^{P}_{3}$, $F$-wave modifications are again small but become better visible for the higher energies. The quantities which show the influence of interference-terms between higher and lower partial waves a lot better are $a^{P}_{(4,5,6)}$. Moreover, such modifications can be seen over the full considered energy region. However, one has to observe that the absolute values of these three coefficients become smaller. They range from $10^{-2}$ to around $10^{-3}$. This can be another reason for the fact that modifications due to small partial wave interferences become better visible. \newline
The last two Legendre coefficients present in a truncation at the $G$-waves are $a^{P}_{7}$ and $a^{P}_{8}$. They have tiny absolute values, ranging from around $10^{-5}$ for $a^{P}_{7}$ to $10^{-8}$ in case of $a^{P}_{8}$. The quantity $a^{P}_{8}$ only receives, in a truncation at $\ell_{\mathrm{max}} = 4$, contributions from interferences of $G$-wave multipoles among themselves (i.e., it is a pure $\left<G,G\right>$-term, cf. chapter \ref{chap:LFits}). Thus, its scattering around zero and absence of any structure shows that the $G$-waves are practically vanishing for the MAID-model in this energy region (however, MAID $G$-waves were used explicitly to generate the theory-data). The coefficient $a^{P}_{7}$ however, which is just an $\left<F,G\right>$-term, shows modifications due to interferences especially for the higher energies. Thus, the tiny $G$-waves can still influence the second highest Legendre coefficient via interferences with the, relatively larger, $F$-waves. Although such modifications are invisible if one observes the angular distributions of the observables by eye, they can still influence the outcome of a multipole-fit. \newline

The statements made about the observable $P$ carry over, in principle, to all remaining observables. Even in the low energy region where, in this model, the resonance $\Delta (1232) \frac{3}{2}^{+}$ \cite{Patrignani:2016xqp} dominates, small higher partial waves can modify the Legendre coefficients (and thereby also the angular distributions of observables) via interferences with the larger low partial waves. On the one hand this is a blessing. Polarization observables are defined and measured in order to be able to detect such interferences at all. \newline 
However, for a multipole-fit in a TPWA, this blessing can turn into problems. Even for the academic case of data that are themselves truncated, interference terms are needed in order to solve the inverse problem for a sensible multipole solution. However, the compatibility of the bilinear equation systems to solve becomes, for the higher truncation orders $\ell_{\mathrm{max}}$, more fragile. The estimate for the number of possible ambiguities $N = 4^{2 \ell_{\mathrm{max}}}$ (cf. section \ref{sec:WBTpaper}) illustrates this fact. With these rising complications due to ambiguities, the stability of the fits and the prospect to solve for the correct MAID solution turns out to also depend crucially on even very small interference contributions, such as those in the observable $P$ above. We use the remainder of this section to illustrate the rise in complexity with $\ell_{\mathrm{max}}$ using the, in principle 'exactly' solvable, fits to theory-data as examples. Analyses of theory-data generated without a cutoff in $\ell_{\mathrm{max}}$ will be the subject of section \ref{subsec:TheoryDataFitsLmaxInfinite}. \newline

\clearpage

\begin{figure}[ht]
 \centering
\begin{overpic}[width=0.49\textwidth]{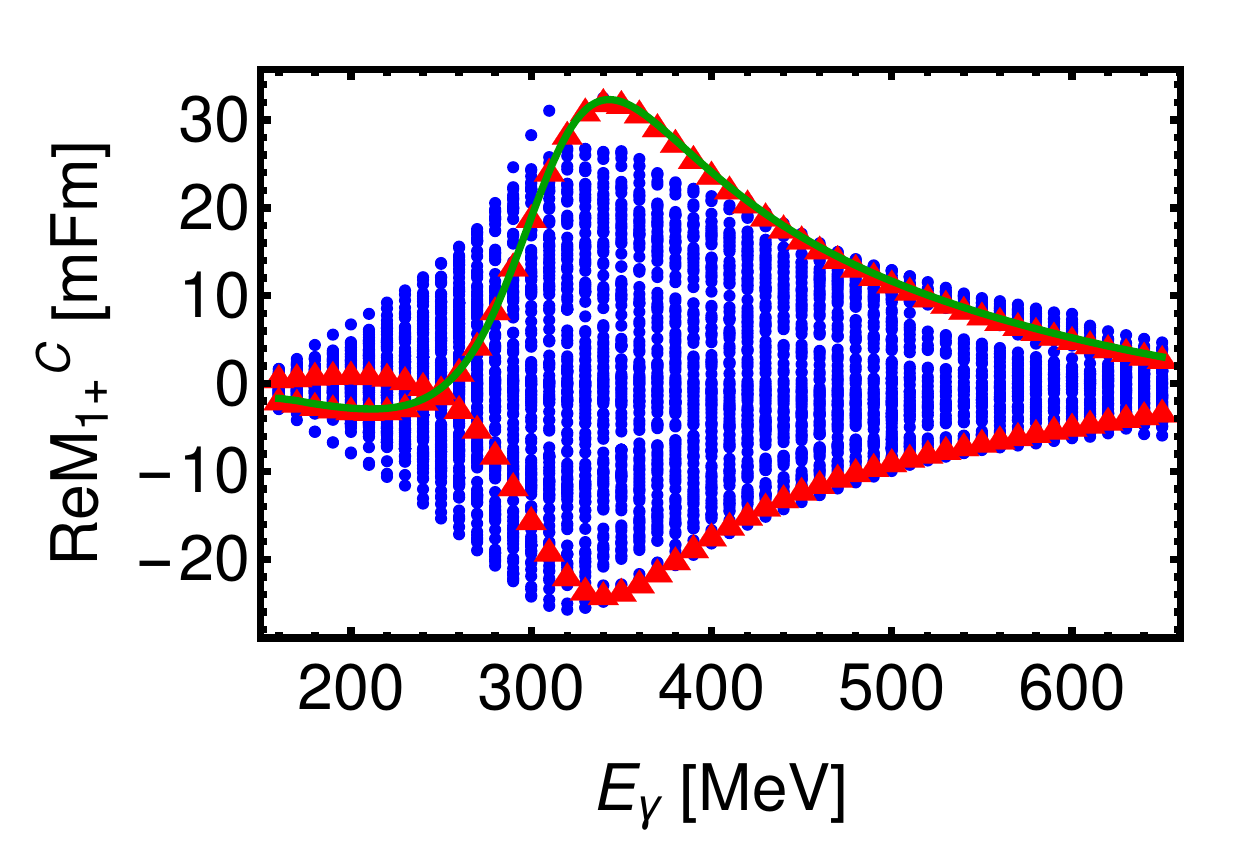}
 \end{overpic} \hspace*{-15pt}
\begin{overpic}[width=0.49\textwidth]{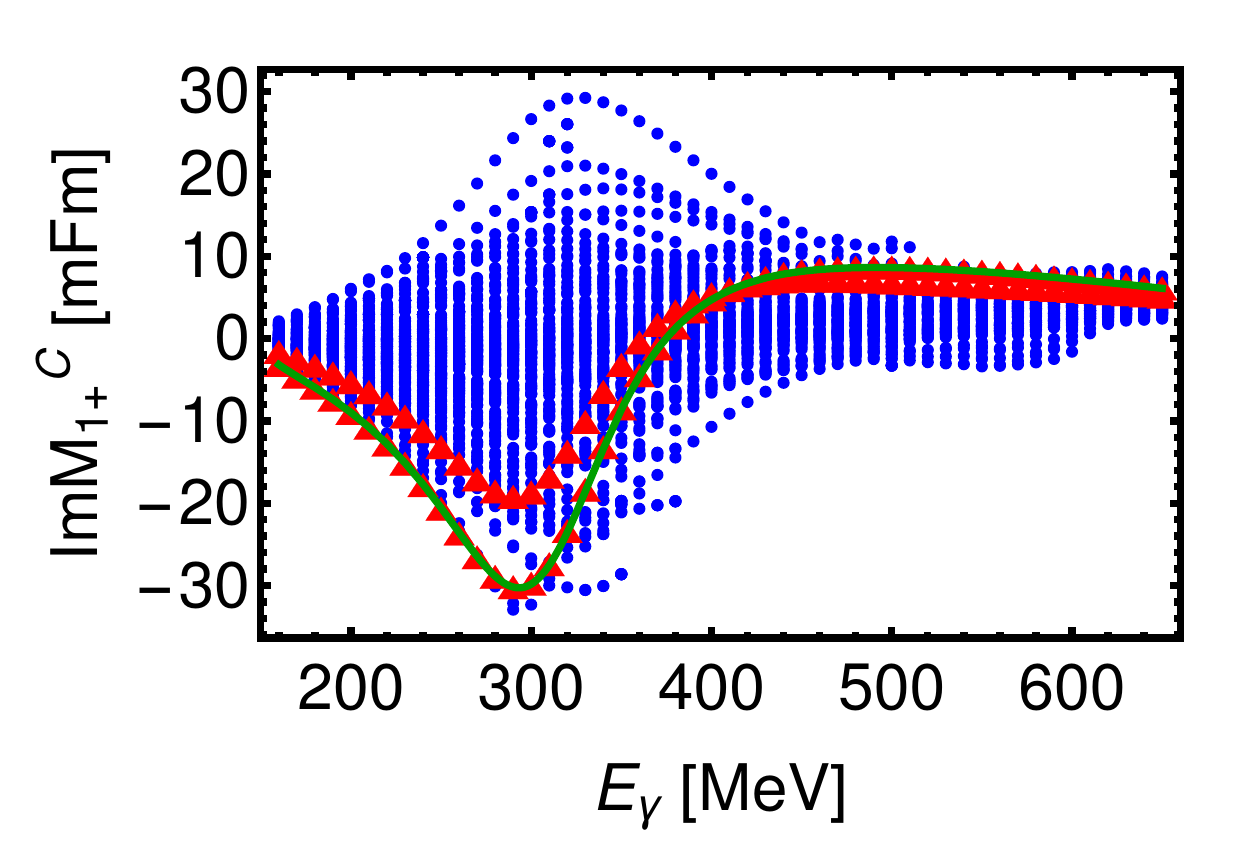}
 \end{overpic} \hspace*{-15pt}
\vspace*{0pt}
\caption[Solutions found in a TPWA fit to truncated MAID theory-data of the observables $\left\{ \sigma_{0}, \check{\Sigma}, \check{T}, \check{P} \right\}$, for $\ell_{\mathrm{max}}=2$. Only the $M_{1+}$-multipole is shown.]{This figure depicts the results of a TPWA-fit to MAID-theory-data \cite{LotharPrivateComm,MAID2007} for $\left\{ \sigma_{0}, \check{\Sigma}, \check{T}, \check{P} \right\}$ truncated at $\ell_{\mathrm{max}} = 2$, using a truncation at the $D$-waves in the fit. The attention is restricted to the $M_{1+}$-multipole. Blue dots show all solutions arising from a pool of $N_{MC} = 3000$ initial parameter configurations. Two solutions are found with best and nearly equal $\Phi_{\mathcal{M}}$, which are shown as red triangles. The MAID-solution \cite{MAID2007,MAID} used to generate the theory-data is plotted as a green solid line.}
\label{fig:Lmax2ThDataFitBestSolsGroupSExample}
\end{figure}

Similarly to the fits of MAID theory-data truncated at $\ell_{\mathrm{max}} = 1$ in section \ref{subsec:TheoryDataFitsLmax1}, phase-constrained multipoles have been extracted in the analyses here, employing the restriction $E_{0+} = \mathrm{Re} \left[ E_{0+} \right] \geq 0$. Furthermore, all fits were performed as outlined in section \ref{sec:TPWAFitsIntro}, using minimization functions $\Phi_{\mathcal{M}}$ as defined there and employing the Monte Carlo sampling technique described in section \ref{sec:MonteCarloSampling}. \newline
From the discussion in the latter section, especially from Table \ref{tab:AccAmbPossibilityNumberMotivatesNMonteCarlo}, it should be clear that the size $N_{MC}$ of the pool of Monte Carlo start configurations has to be increased if higher orders in $\ell_{\mathrm{max}}$ are fitted. For the analyses of theory-data truncated at $\ell_{\mathrm{max}} = 1$, $N_{MC} = 1500$ has been more than enough (cf. section \ref{subsec:TheoryDataFitsLmax1}). For all ensuing analyses, larger numbers were employed, which in order to keep the fits numerically tractable in an acceptable time have been chosen still smaller than the quite extreme estimates provided in Table \ref{tab:AccAmbPossibilityNumberMotivatesNMonteCarlo}. \newline

In case of MAID theory-data truncated at $\ell_{\mathrm{max}} = 2$ for instance, we have used a pool of $N_{MC} = 3000$ in every fit. Before presenting the fit of a complete experiment, some examples shall be treated which did not allow a unique solution for the MAID multipoles used to generate the theory-data. Still, the following examples allow for interesting comparisons to the results in section \ref{subsec:TheoryDataFitsLmax1}. \newline
The first example is a fit to the group $\mathcal{S}$ observables $\left\{ \sigma_{0}, \check{\Sigma}, \check{T}, \check{P} \right\}$. The minimization function $\Phi_{\mathcal{M}}$ is in this case just a straightforward generalization of the expression (\ref{eq:ExamplePhiFitFunction}) (section \ref{subsec:TheoryDataFitsLmax1}), with all summations running up to $2 \ell_{\mathrm{max}} = 4$ in this particular case. The results for the multipole $M_{1+}$ can be seen in Figure \ref{fig:Lmax2ThDataFitBestSolsGroupSExample}. It can be observed directly that the number of different solutions found in the fit is increased drastically compared to the analogous fit for $\ell_{\mathrm{max}} = 1$ (cf. Figure \ref{fig:Lmax1ThDataFitBestSols}, section \ref{subsec:TheoryDataFitsLmax1}). The number of non-redundant solutions obtained from the employed pool amounts to more than $100$ in the lower half of the considered energy region, while they become around $50$ - $70$ in the upper half. This increase means roughly a factor of $10$ compared to the $2$ - $6$ non-redundant final parameter configurations obtained in the simplest example $\ell_{\mathrm{max}} = 1$, section \ref{subsec:TheoryDataFitsLmax1}. \newpage

\begin{figure}[ht]
 \centering
\begin{overpic}[width=0.49\textwidth]{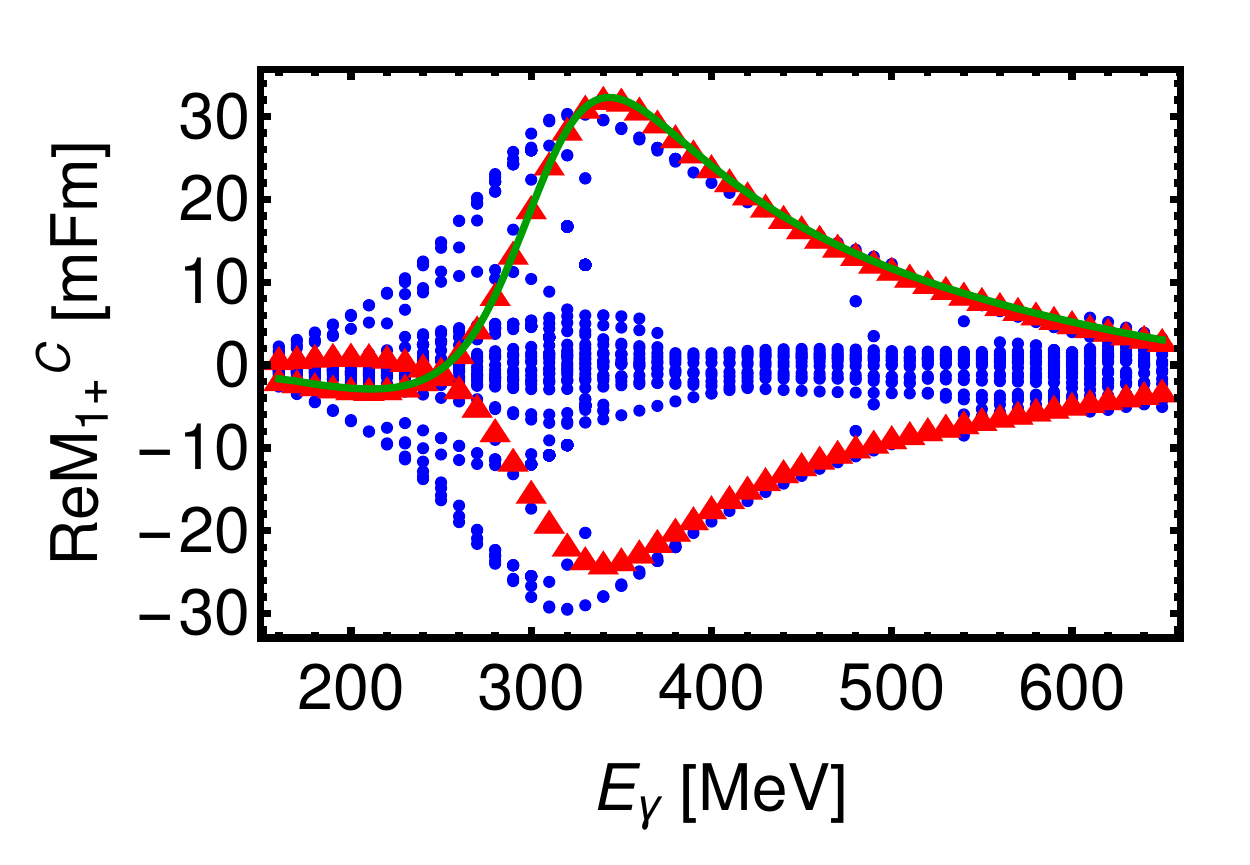}
 \end{overpic} \hspace*{-15pt}
\begin{overpic}[width=0.49\textwidth]{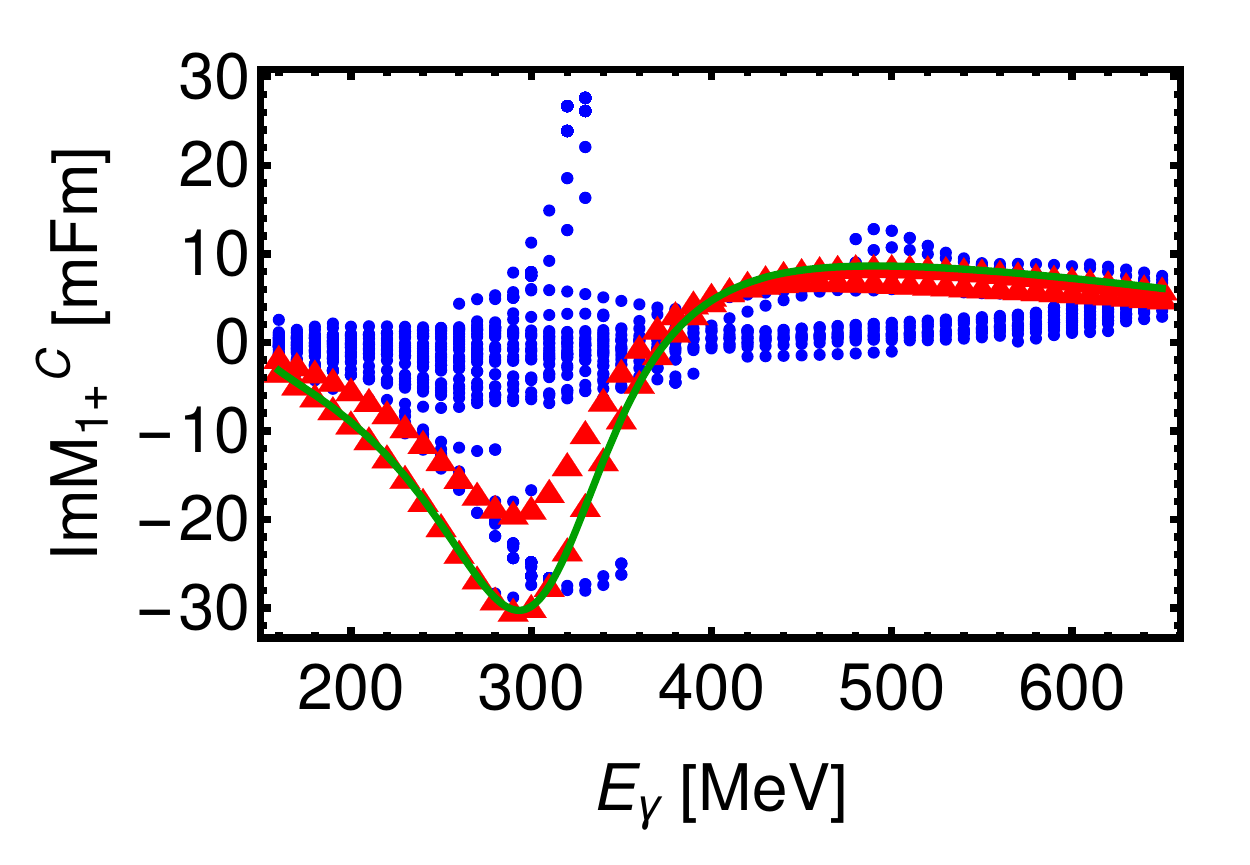}
 \end{overpic} \hspace*{-15pt}
\vspace*{0pt}
\caption[Solutions found in a TPWA fit to truncated MAID theory-data of the observables $\left\{ \sigma_{0}, \check{\Sigma}, \check{T}, \check{P}, \check{E}, \check{H} \right\}$, for $\ell_{\mathrm{max}}=2$. Only the $M_{1+}$-multipole is shown.]{The plots shown here are exactly analogous to those in Figure \ref{fig:Lmax2ThDataFitBestSolsGroupSExample}, except for the fact that here a larger set of observables was fitted, namely $\left\{ \sigma_{0}, \check{\Sigma}, \check{T}, \check{P}, \check{E}, \check{H} \right\}$. Again, a pool of $N_{MC} = 3000$ randomly generated Monte Carlo start configurations was used. The final configurations obtained from all minimizations are plotted as blue dots, while the best solutions are represented by red triangles. The MAID-solution \cite{MAID2007,MAID} used to generate the theory-data is plotted as a green solid line.}
\label{fig:Lmax2ThDataFitBestSolsGroupSAndEHExample}
\end{figure}

Furthermore, it is not possible to fit a global minimum out of the group $\mathcal{S}$ alone. Rather, two solutions exist with roughly equal value of $\Phi_{\mathcal{M}}$, which are well separated from the remaining local minima. One of them is found to be the true MAID solution, while the other can be identified with the double ambiguity. Therefore, the outcome of this fit is in accord with the treatment of ambiguities of chapter \ref{chap:Omelaenko} and appendix \ref{sec:AdditionsChapterII}. For this particular fit, a detailed assignment of multipole solutions with a particularly small $\Phi_{\mathcal{M}}$ to Omelaenko-ambiguities (i.e. those stemming from complex conjugation of the roots $\left\{ \alpha_{k}, \beta_{k} \right\}$) was performed in the same way as for $\ell_{\mathrm{max}} = 1$ in the preceding section. These detailed comparisons are not included here in favor of brevity. However, it should be stated that results were fully consistent with the expectations from a detailed treatment of discrete ambiguities. In particular, it should be stressed that all local minima in the present fit which happen to have a small value for $\Phi_{\mathcal{M}}$ can be put in a one-to-one correspondence to accidental ambiguities. Solutions related to such ambiguities are certainly there and can therefore be expected to exist for all higher truncation orders in $\ell_{\mathrm{max}}$. This fact really marks the present fit for $\ell_{\mathrm{max}} = 2$ as just a more complicated but mathematically fully analogous version of the simpler problem of $\ell_{\mathrm{max}} = 1$ treated at the end of section \ref{subsec:TheoryDataFitsLmax1}. \newline
Next, the effect of including the observables $E$ and $H$, each of them \textit{not} being able to resolve the double ambiguity (cf. section \ref{sec:WBTpaper}), is studied. Here, the minimization function $\Phi_{\mathcal{M}}$ receives seven additional constraining terms. Those are however defined by bilinear coefficient matrices $\left(\mathcal{C}_{L}\right)_{k}^{\check{E}}$ and $\left(\mathcal{C}_{L}\right)_{k}^{\check{H}}$ that are precisely invariant under the double ambiguity transformation. \newline Explicitly (and as an example), the minimization function reads in this case
\allowdisplaybreaks
\begin{align}
 \Phi_{\mathcal{M}} \left( \left\{ \mathcal{M}_{\ell}^{C} \right\} \right) &= \sum_{k = 0}^{4} \left[ \left(a_{L}\right)^{\sigma_{0}}_{k} -  \left< \mathcal{M}_{\ell}^{C} \right| \left( \mathcal{C}_{L}\right)_{k}^{\sigma_{0}} \left| \mathcal{M}_{\ell}^{C} \right> \right]^{2}  + \sum_{m = 2}^{4} \left[ \left(a_{L}\right)^{\check{\Sigma}}_{m} -  \left< \mathcal{M}_{\ell}^{C} \right| \left( \mathcal{C}_{L}\right)_{m}^{\check{\Sigma}} \left| \mathcal{M}_{\ell}^{C} \right> \right]^{2} \nonumber \\
 & + \sum_{n = 1}^{4} \left[ \left(a_{L}\right)^{\check{T}}_{n} -  \left< \mathcal{M}_{\ell}^{C} \right| \left( \mathcal{C}_{L}\right)_{n}^{\check{T}} \left| \mathcal{M}_{\ell}^{C} \right> \right]^{2} + \ldots \nonumber \\
 \ldots & + \sum_{q = 1}^{4} \left[ \left(a_{L}\right)^{\check{P}}_{q} -  \left< \mathcal{M}_{\ell}^{C} \right| \left( \mathcal{C}_{L}\right)_{q}^{\check{P}} \left| \mathcal{M}_{\ell}^{C} \right> \right]^{2}  + \sum_{r = 0}^{4} \left[ \left(a_{L}\right)^{\check{E}}_{r} -  \left< \mathcal{M}_{\ell}^{C} \right| \left( \mathcal{C}_{L}\right)_{r}^{\check{E}} \left| \mathcal{M}_{\ell}^{C} \right> \right]^{2} \nonumber \\
  & + \sum_{s = 1}^{4} \left[ \left(a_{L}\right)^{\check{H}}_{s} -  \left< \mathcal{M}_{\ell}^{C} \right| \left( \mathcal{C}_{L}\right)_{s}^{\check{H}} \left| \mathcal{M}_{\ell}^{C} \right> \right]^{2} \mathrm{.} \label{eq:ExamplePhiFitFunctionLmax2FitGroupSAndEH}
\end{align}
Results for the fits are shown in Figure \ref{fig:Lmax2ThDataFitBestSolsGroupSAndEHExample}. The true MAID solution as well as the double ambiguity are still found as two well-separated ``global'' minima, as one could expect. However, the number of distinct local minima has been reduced drastically and visibly, as compared to Figure \ref{fig:Lmax2ThDataFitBestSolsGroupSExample}. This again illustrates, similarly to the same example for the simpler case $\ell_{\mathrm{max}} = 1$ shown in section \ref{subsec:TheoryDataFitsLmax1} (cf. Figure \ref{fig:Lmax1ThDataFitBestSolsGroupSAndEH}), the capability of the observables $E$ and $H$ to resolve accidental ambiguities and stabilize a TPWA-fit, while not being able to resolve the mathematical double ambiguity. \newline

\begin{figure}[ht]
 \centering

\begin{overpic}[width=0.346\textwidth]{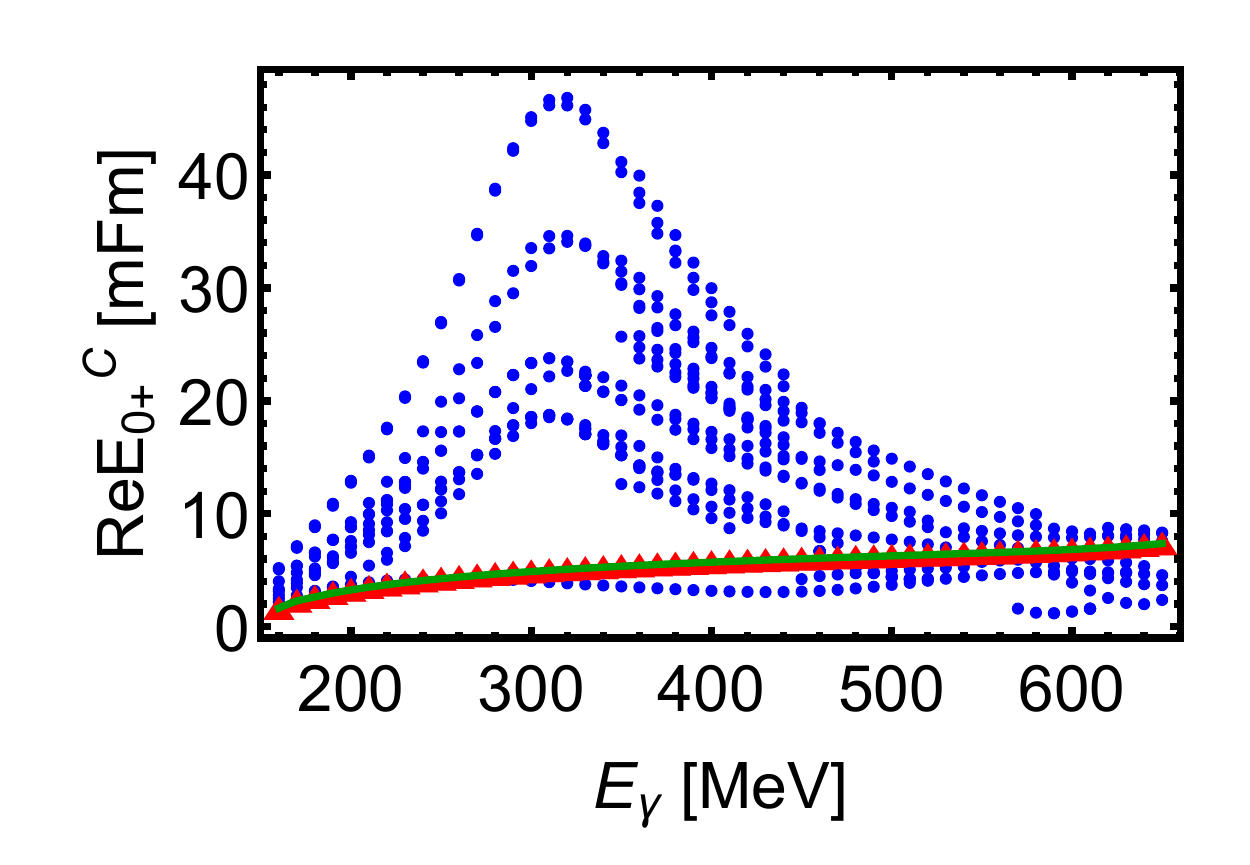}
 \end{overpic} \hspace*{-15pt}
\begin{overpic}[width=0.346\textwidth]{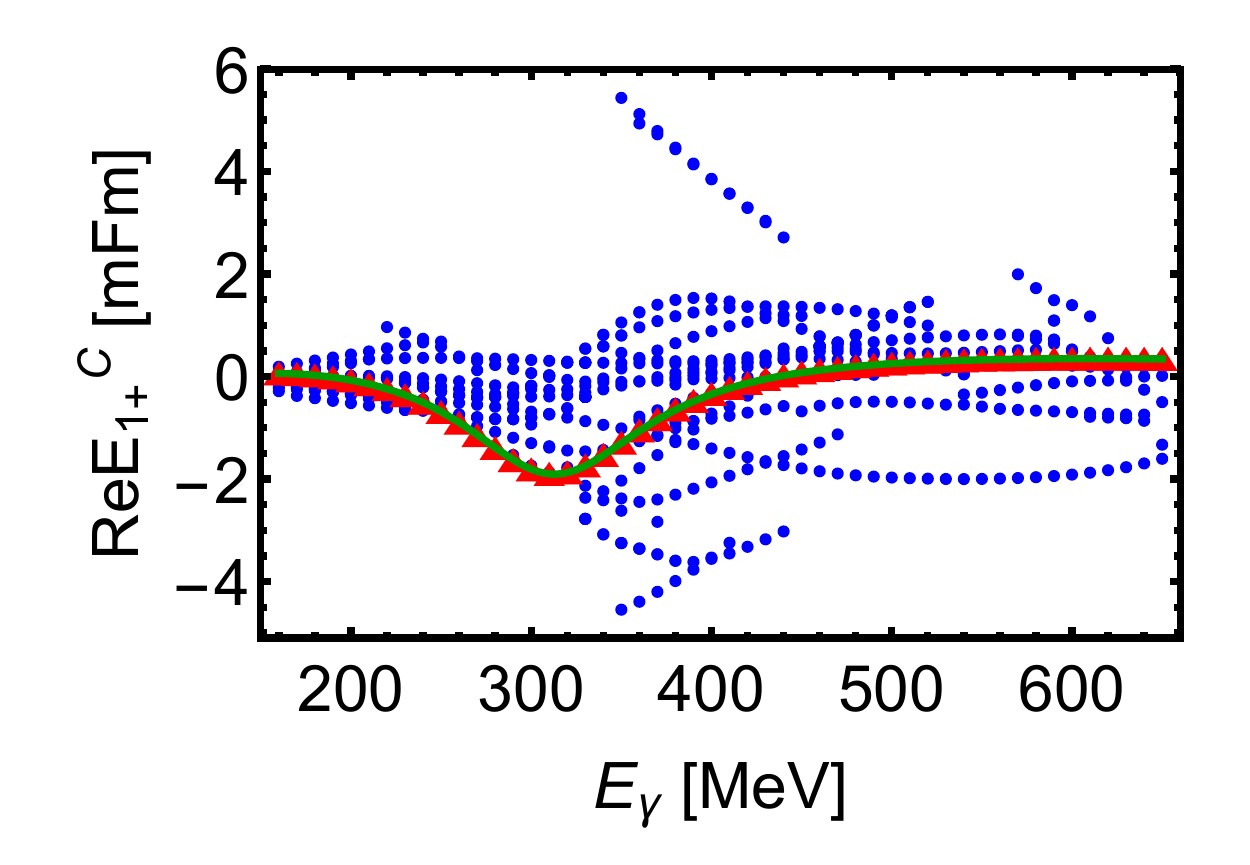}
 \end{overpic} \hspace*{-15pt}
\begin{overpic}[width=0.346\textwidth]{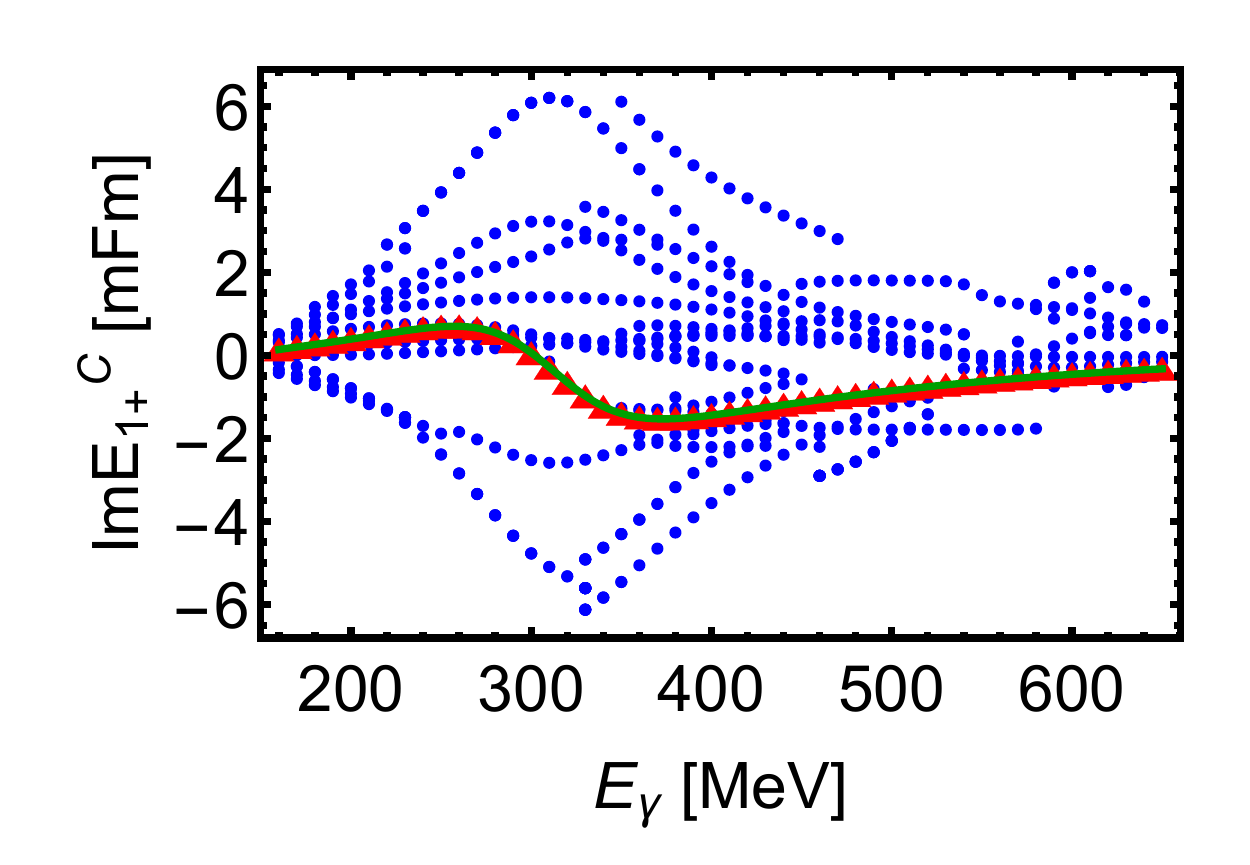}
 \end{overpic}  \\
\begin{overpic}[width=0.346\textwidth]{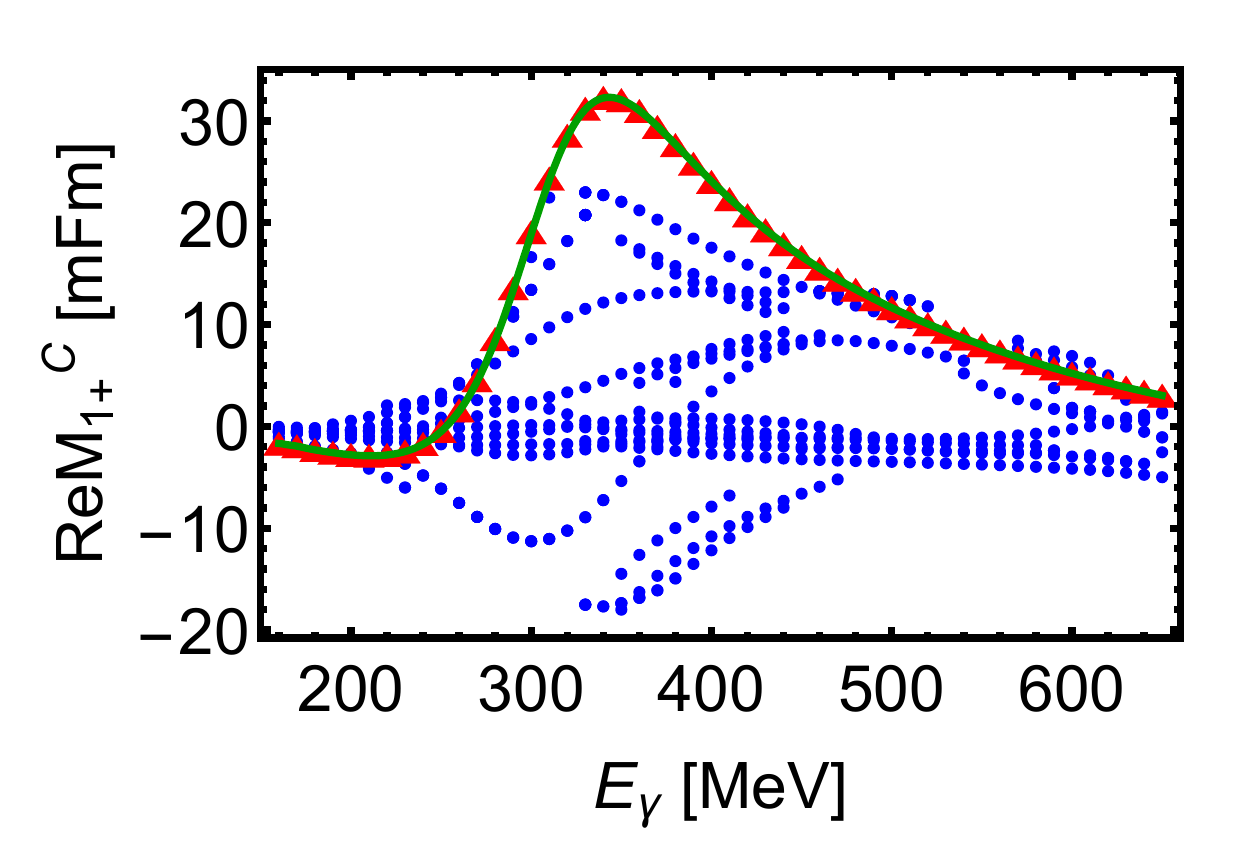}
 \end{overpic} \hspace*{-15pt}
\begin{overpic}[width=0.346\textwidth]{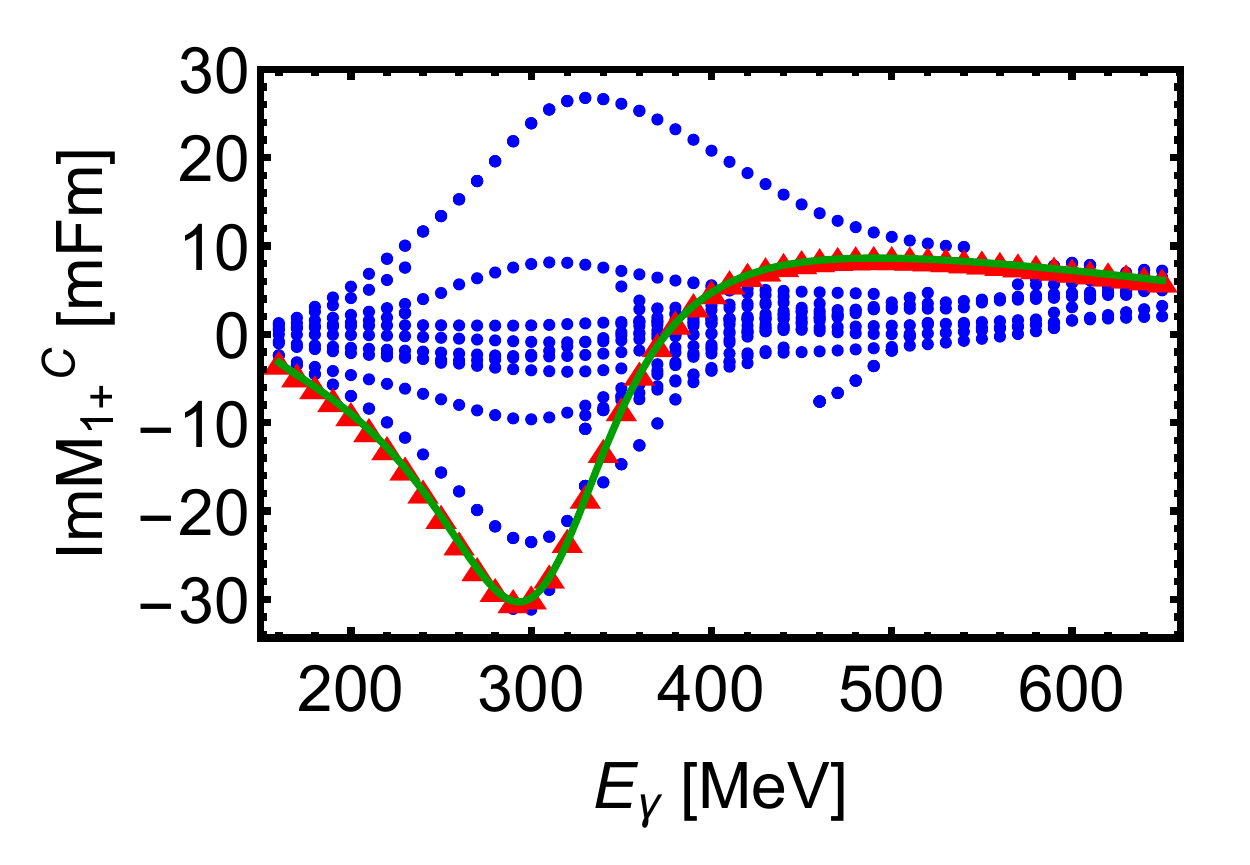}
 \end{overpic} \hspace*{-15pt}
\begin{overpic}[width=0.346\textwidth]{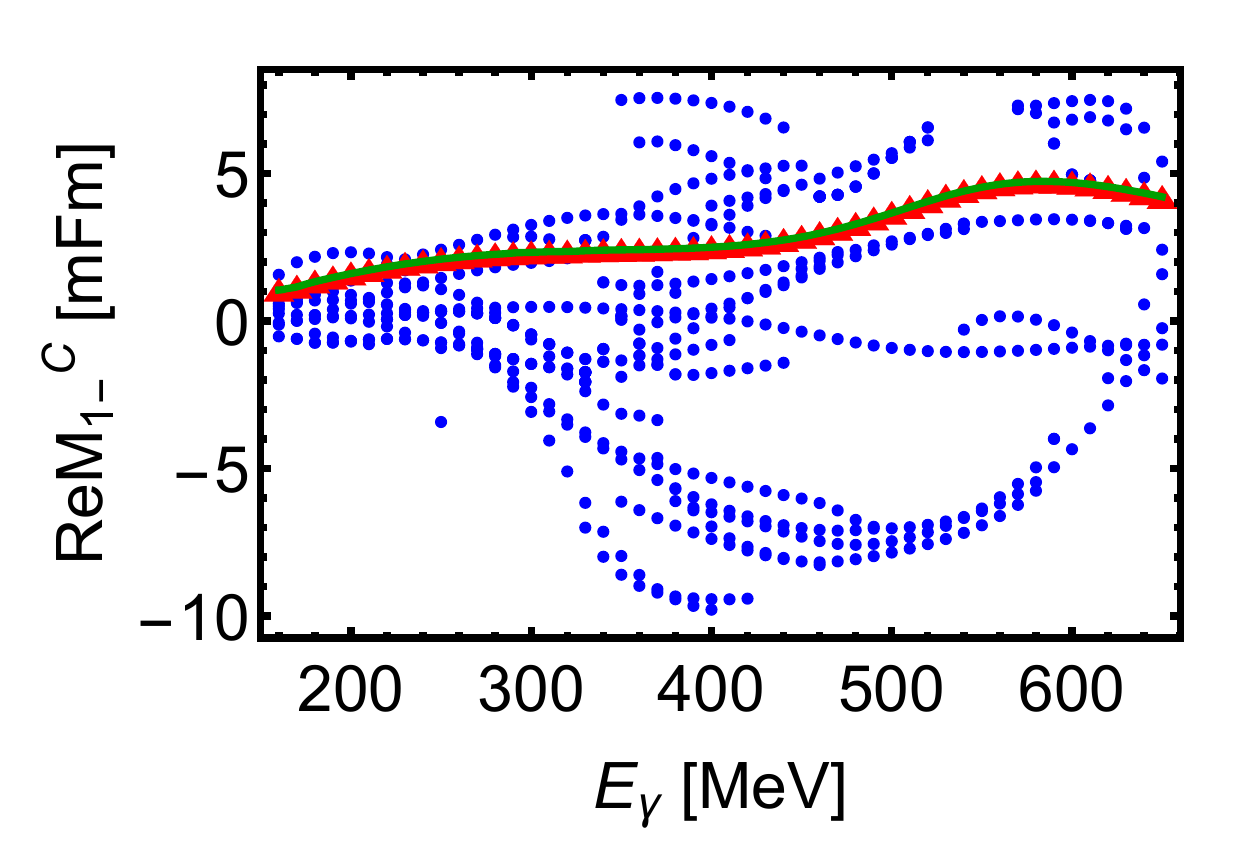}
 \end{overpic} \\
\begin{overpic}[width=0.346\textwidth]{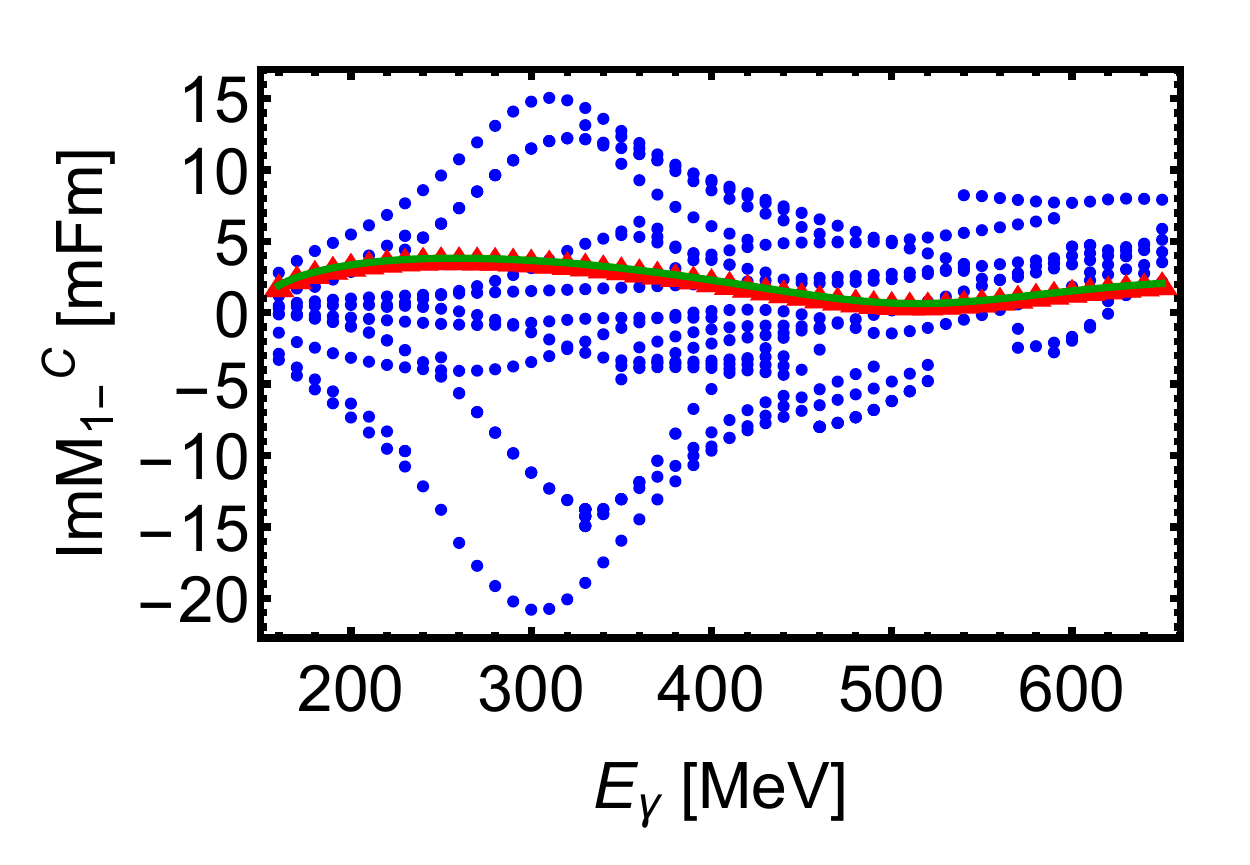}
 \end{overpic} \hspace*{-15pt}
\begin{overpic}[width=0.346\textwidth]{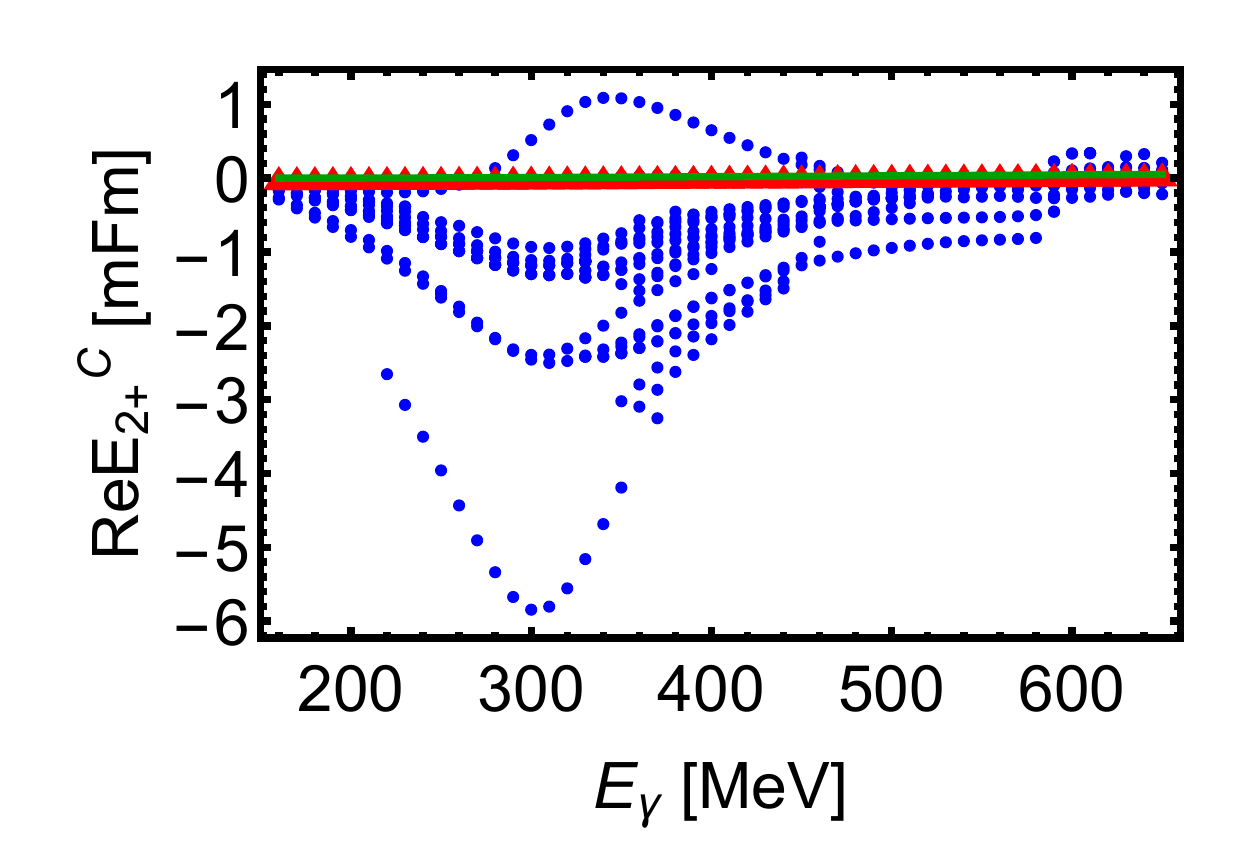}
 \end{overpic} \hspace*{-15pt}
\begin{overpic}[width=0.346\textwidth]{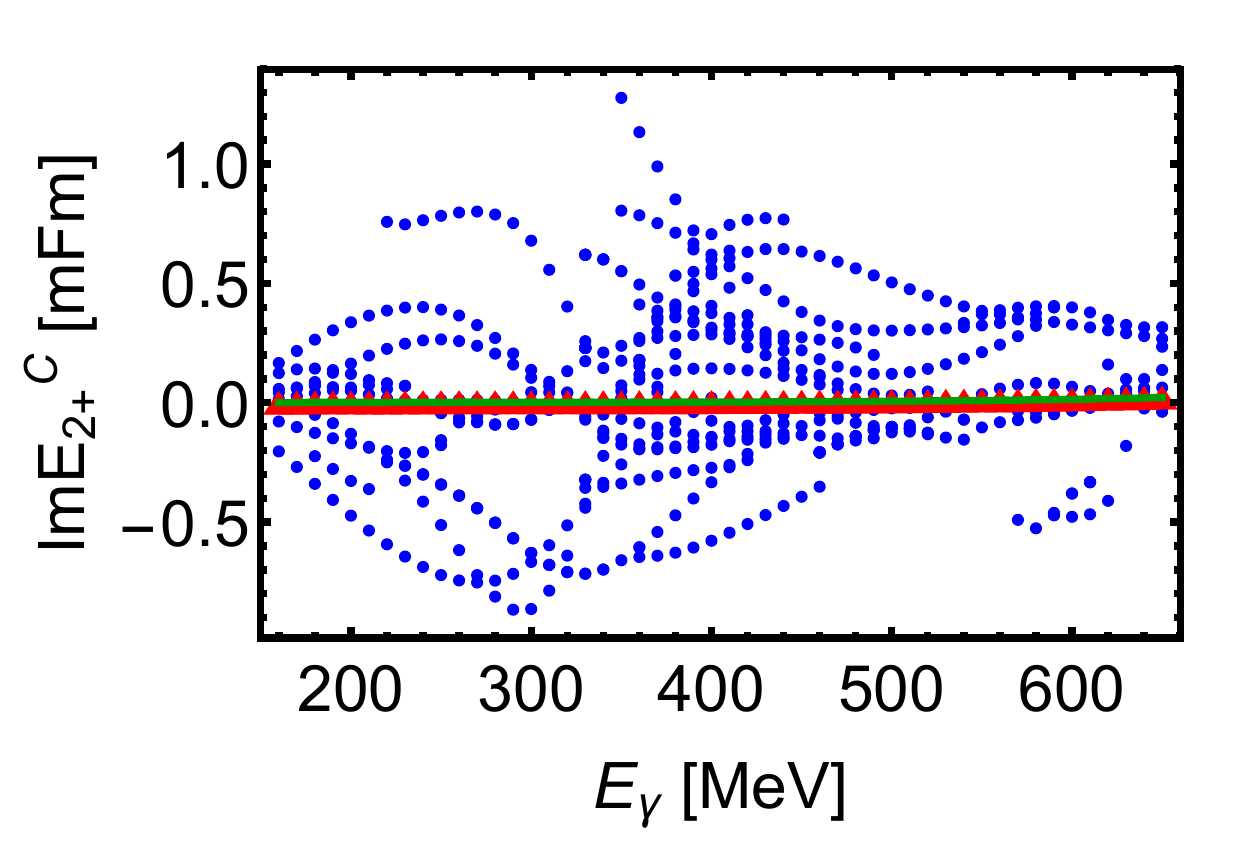}
 \end{overpic} \\
\begin{overpic}[width=0.346\textwidth]{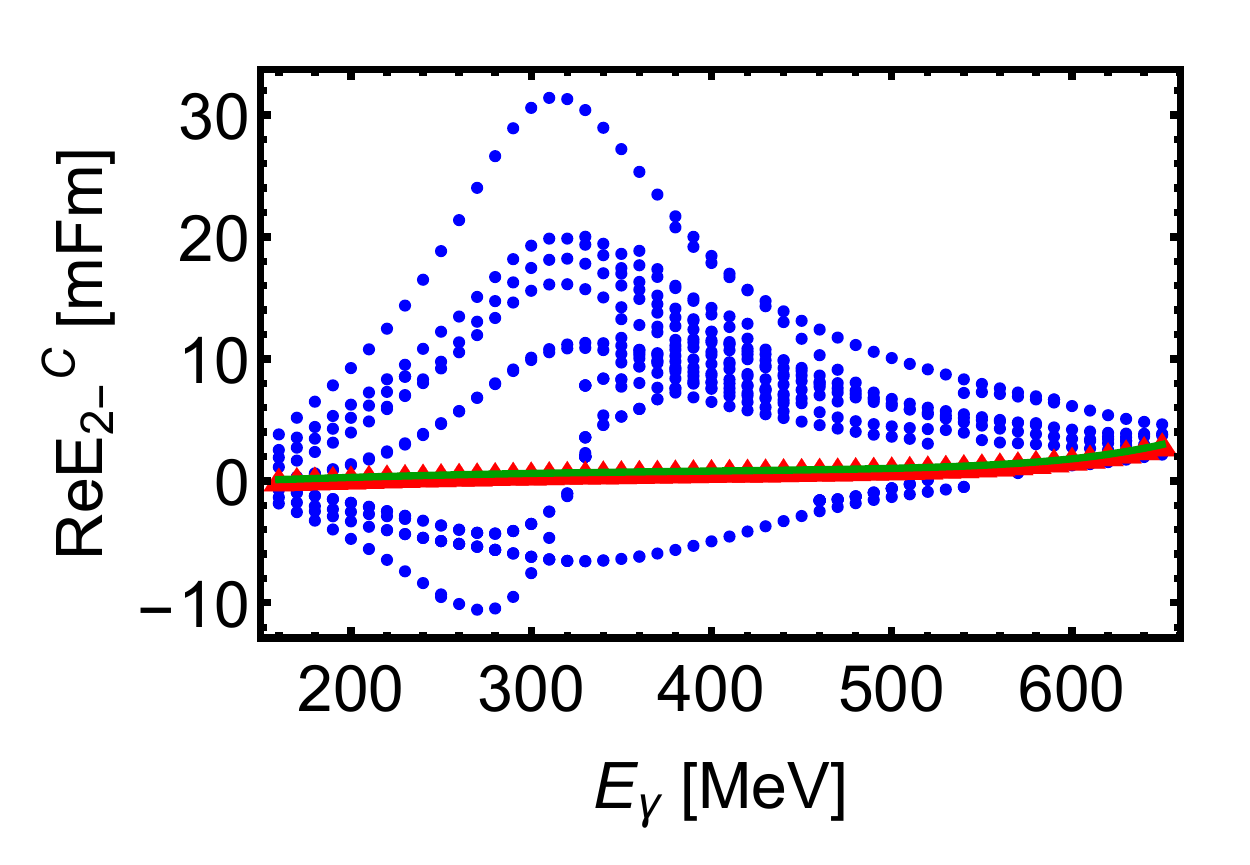}
 \end{overpic} \hspace*{-15pt}
\begin{overpic}[width=0.346\textwidth]{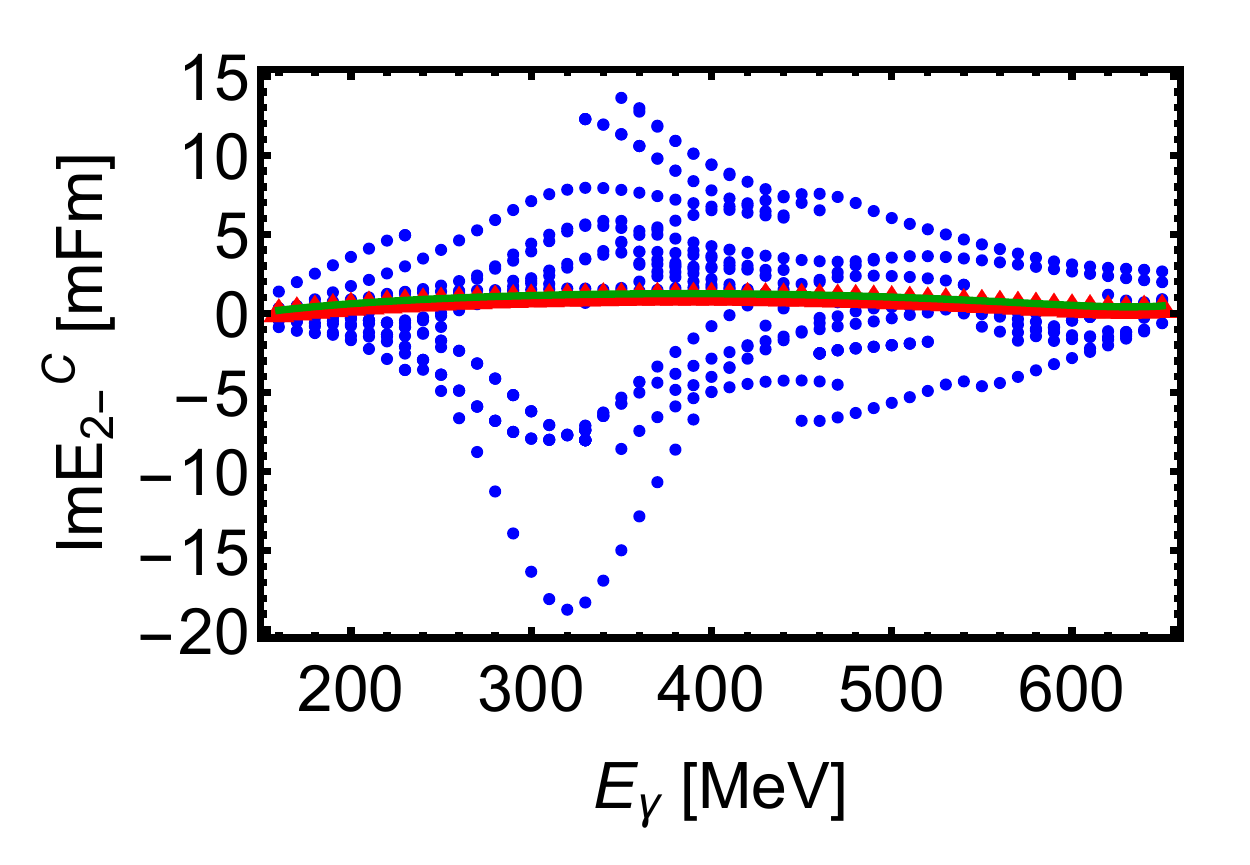}
 \end{overpic} \hspace*{-15pt}
\begin{overpic}[width=0.346\textwidth]{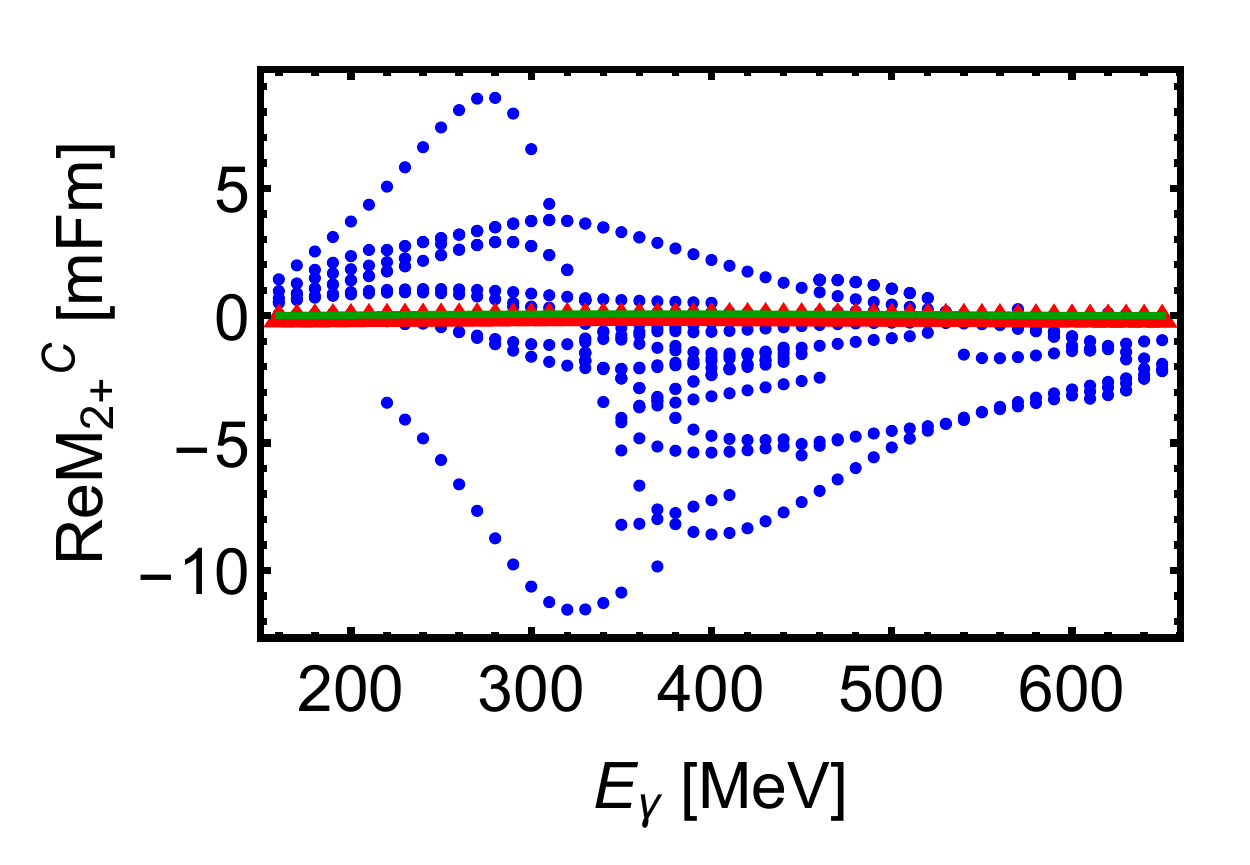}
 \end{overpic} \\
\begin{overpic}[width=0.3457\textwidth]{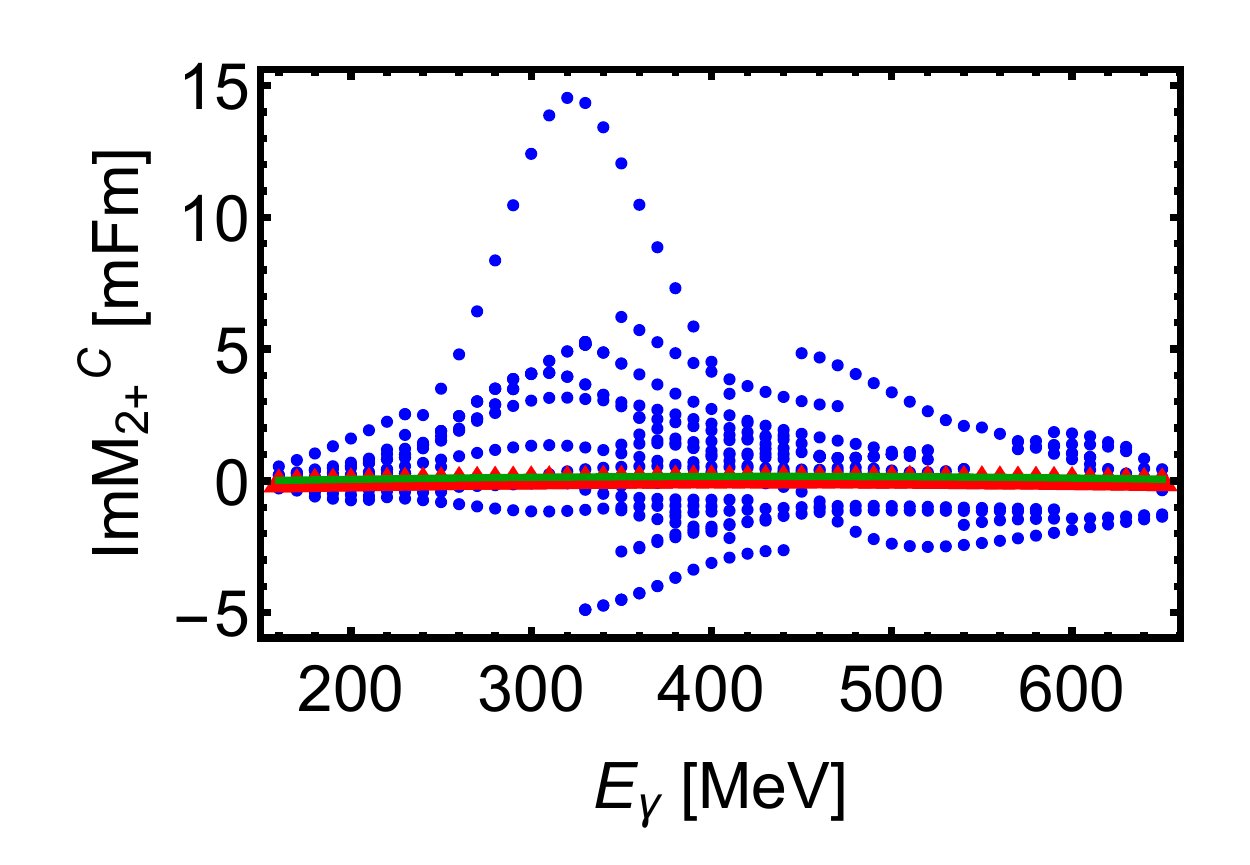}
 \end{overpic} \hspace*{-15pt}
\begin{overpic}[width=0.3457\textwidth]{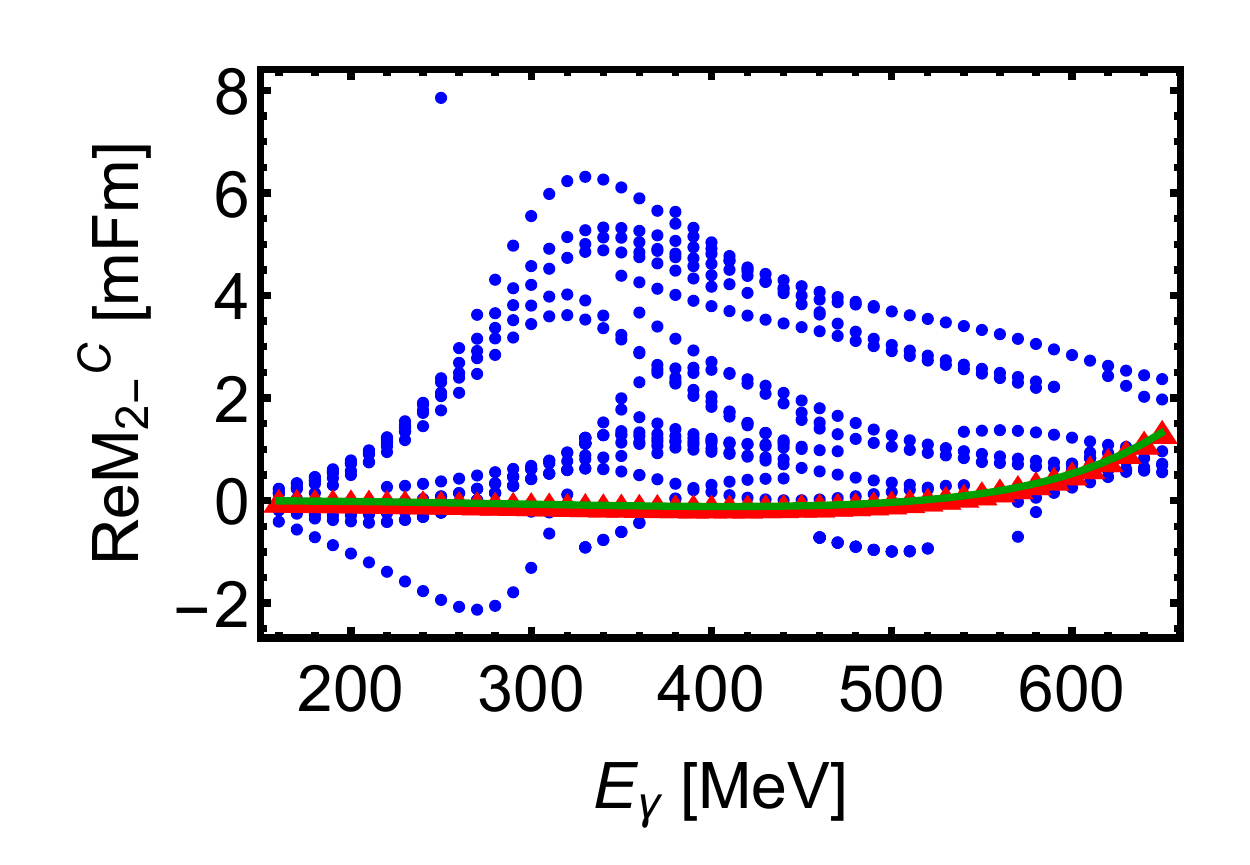}
 \end{overpic} \hspace*{-15pt}
\begin{overpic}[width=0.3457\textwidth]{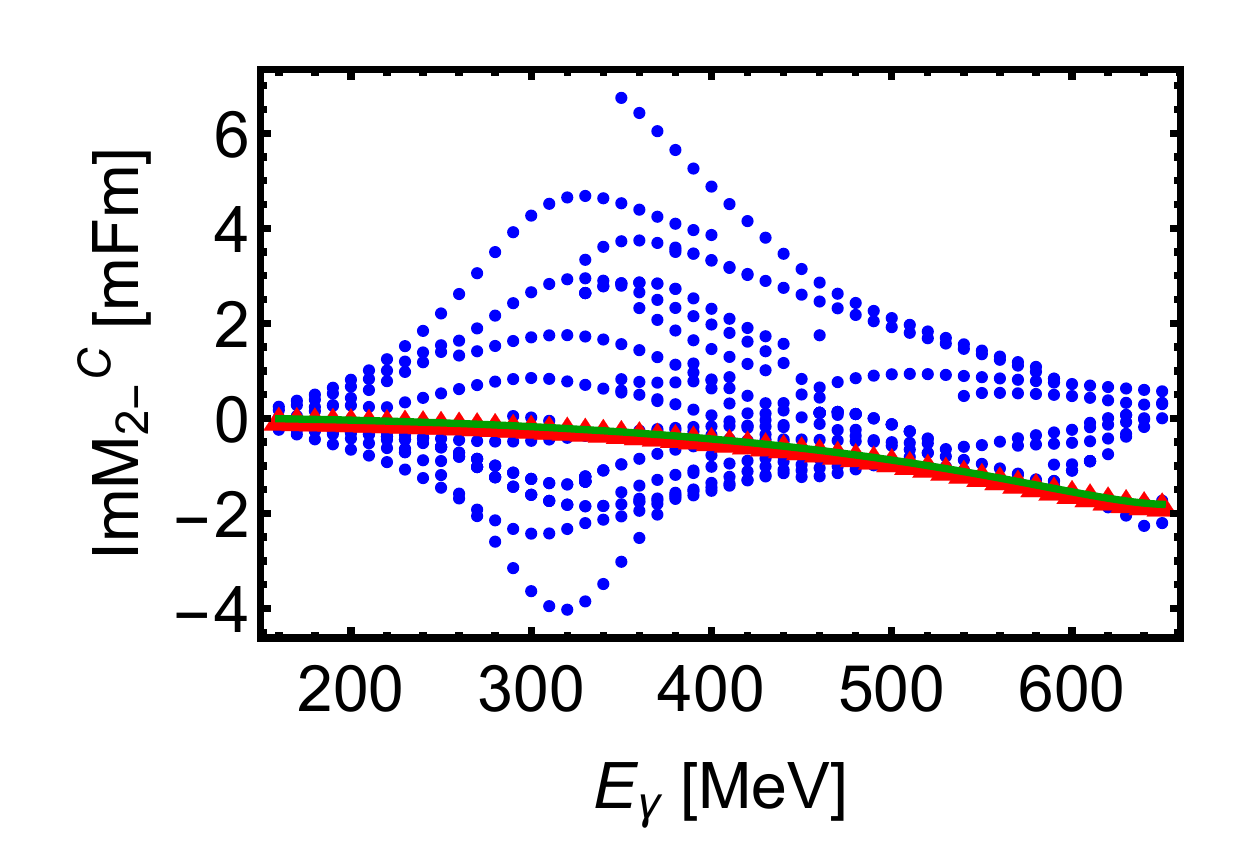}
 \end{overpic}
\vspace*{0pt}
\caption[Solutions found in a TPWA fit to truncated MAID theory-data of the observables $\left\{ \sigma_{0}, \check{\Sigma}, \check{T}, \check{P}, \check{F} \right\}$, for $\ell_{\mathrm{max}}=2$.]{Shown here are the results of a fit of an $S$-, $P$- and $D$-waves truncation to MAID theory-data \cite{LotharPrivateComm,MAID2007} truncated at $\ell_{\mathrm{max}} = 2$. Thus, the real and imaginary parts of all phase-constrained multipoles ($E_{0+} \equiv \mathrm{Re} \left[ E_{0+} \right] \geq 0$) for all partial waves from $E_{0+}$ up to $M_{2-}$ are plotted. \newline The observables $\left\{ \sigma_{0}, \check{\Sigma}, \check{T}, \check{P}, \check{F} \right\}$ were used to obtain the multipole solutions and a pool of $N_{MC} = 3000$ randomly chosen initial conditions yielded the fitted values. The whole solution pool is denoted by blue dots. A unique and well separated global minimum exists, which is represented by red triangles.}
\label{fig:Lmax2ThDataFitBestSols1}
\end{figure}

Following the study of some examples for non-complete sets of observables and illustrating the analogies and differences between truncations at $\ell_{\mathrm{max}} = 1$ and $\ell_{\mathrm{max}} = 2$, the remainder of this section shall be devoted solely to a particular candidate set for a complete experiment. The same set is chosen as in section \ref{subsec:TheoryDataFitsLmax1}, namely
\begin{equation}
 \left\{ \sigma_{0}, \check{\Sigma}, \check{T}, \check{P}, \check{F} \right\} \mathrm{.} \label{eq:ExampleCompleteSet}
\end{equation}
The unique solvability of this particular set, as well as the expected rise in the number of ambiguities, shall now be studied. In order to do this, MAID theory-data \cite{LotharPrivateComm,MAID2007} for the truncations $\ell_{\mathrm{max}} = 2,3,4$ are fitted. In case of $\ell_{\mathrm{max}} = 2$, we again employ a start configuration pool of $N_{MC} = 3000$. The results of the minimizations of $\Phi_{\mathcal{M}}$ are presented in Figure \ref{fig:Lmax2ThDataFitBestSols1} for all the fitted $S$-, $P$- and $D$-wave multipoles. \newline
In this particular case, a unique and well-separated global minimum exists, which exactly corresponds to the correct MAID solution. Therefore, the observable $\check{F}$ is capable of resolving the double ambiguity, which has been present in the preceding examples. For the unique MAID solution, one typically has values of $\Phi_{\mathcal{M}} \simeq (10^{-15} - 10^{-16}) \hspace*{2pt} \left( \mu b / sr \right)^{2}$ in the minimum. For the second best solutions, i.e. the local minima closest to the global one, the discrepancy function $\Phi_{\mathcal{M}}$ typically attains values around $10^{-4}$ to $10^{-7}$ $\left( \mu b / sr \right)^{2}$. The global minimum is thus separated by ten orders of magnitude. \newline
Apart from the ability to make this model-TPWA unique, the observable $\check{F}$ is also seen capable to reduce the total number of local minima in the fit, corresponding (in most cases) to accidental ambiguities. This fact is directly visible by comparing the results of this fit (Figure \ref{fig:Lmax2ThDataFitBestSols1}) to those including just the group $\mathcal{S}$ observables, in Figure \ref{fig:Lmax2ThDataFitBestSolsGroupSExample}. When fitting the complete experiment (\ref{eq:ExampleCompleteSet}), the total number of non-redundant solutions originating from the initial pool of $N_{MC} = 3000$ amounts, in most energy bins, to values between $10$ and $20$. For the higher energies, some $E_{\gamma}$-bins with a total number of solutions below $10$, around $5$ to $7$, exist. Comparing to the fit of the complete set (\ref{eq:ExampleCompleteSet}) for the simplest case $\ell_{\mathrm{max}} = 1$, where the TPWA had mostly only one unique minimum and for some energies a maximum of $2$ different solutions (cf. Figure \ref{fig:Lmax1ThDataFitBestSolsGroupSAndF}, section \ref{subsec:TheoryDataFitsLmax1}), it is seen that the number of distinct minima has grown (very) roughly by a factor of $10$. \newline

\begin{figure}[ht]
 \centering
\begin{overpic}[width=0.49\textwidth]{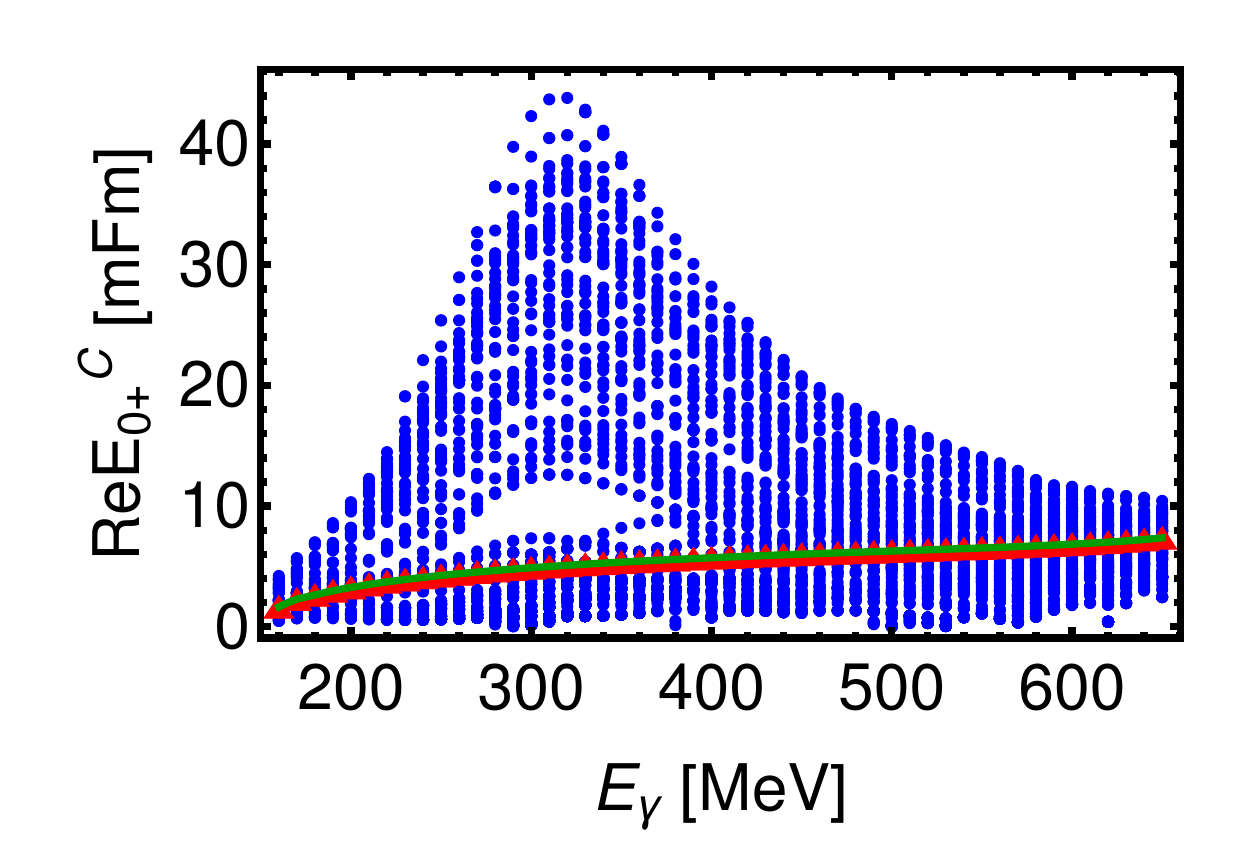}
 \end{overpic} \\
\begin{overpic}[width=0.49\textwidth]{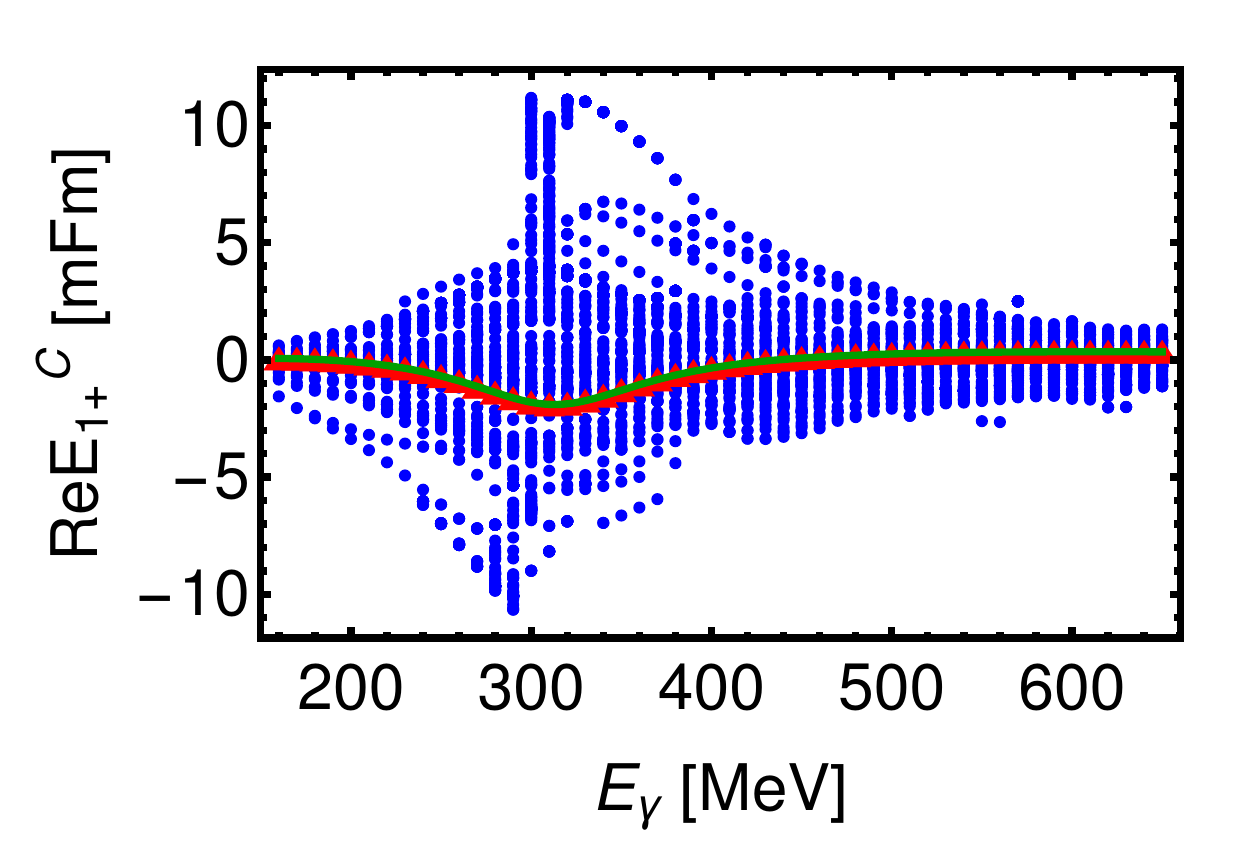}
 \end{overpic} \hspace*{-15pt}
\begin{overpic}[width=0.49\textwidth]{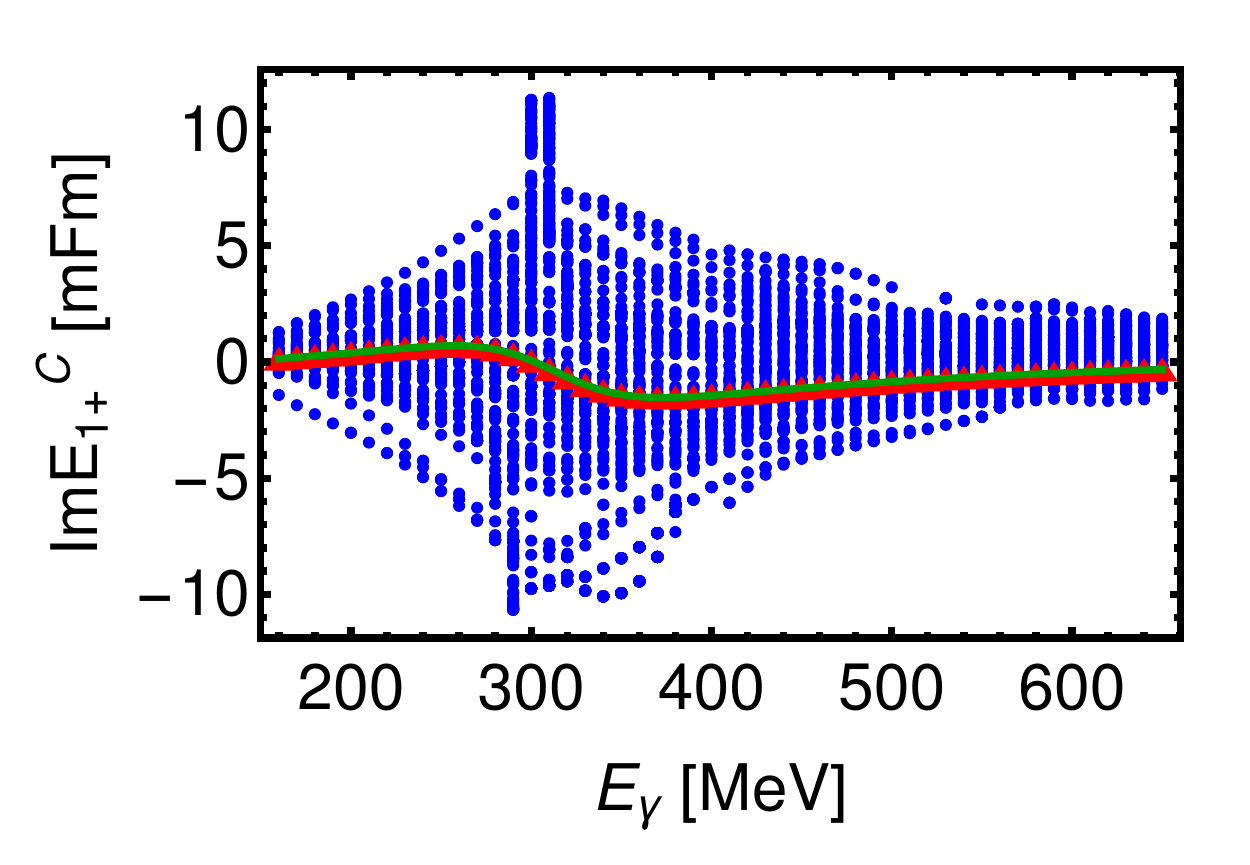}
 \end{overpic} \\
\begin{overpic}[width=0.49\textwidth]{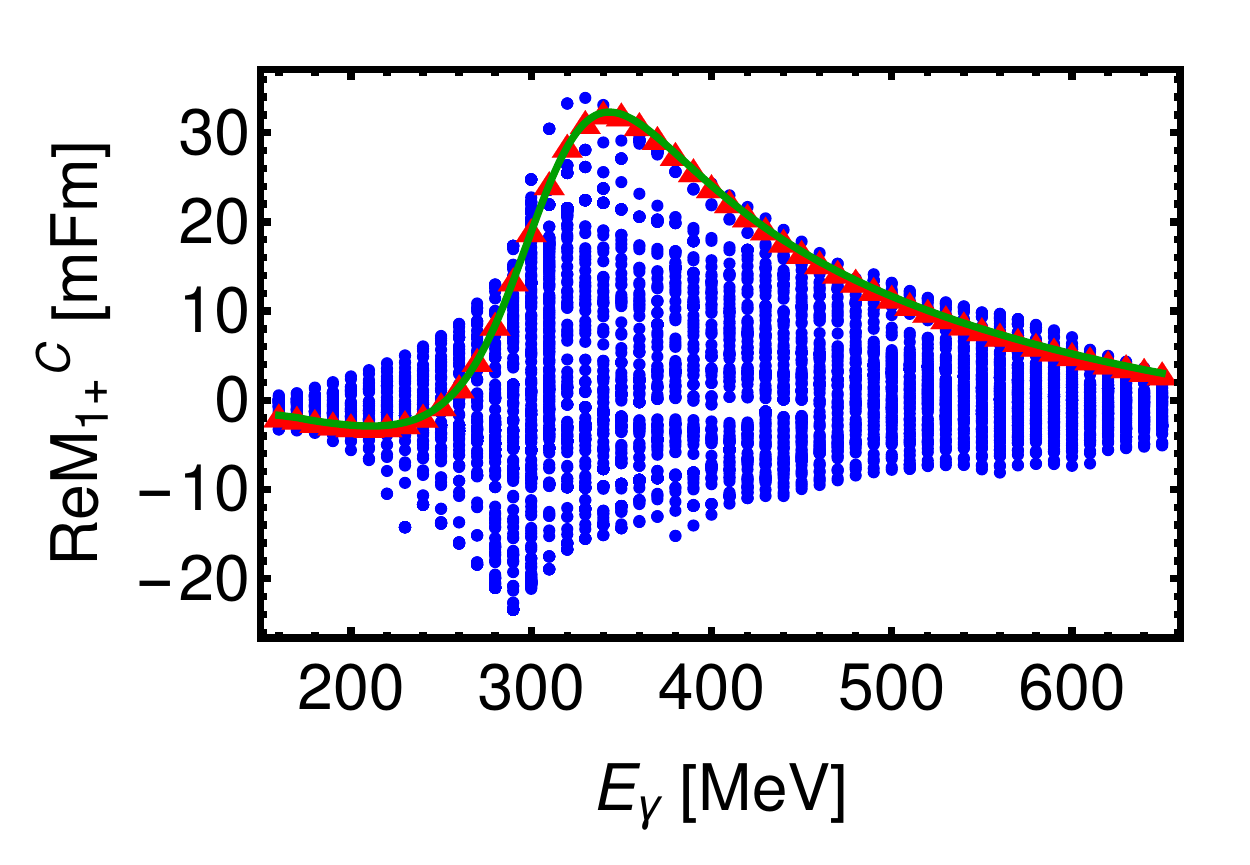}
 \end{overpic} \hspace*{-15pt}
\begin{overpic}[width=0.49\textwidth]{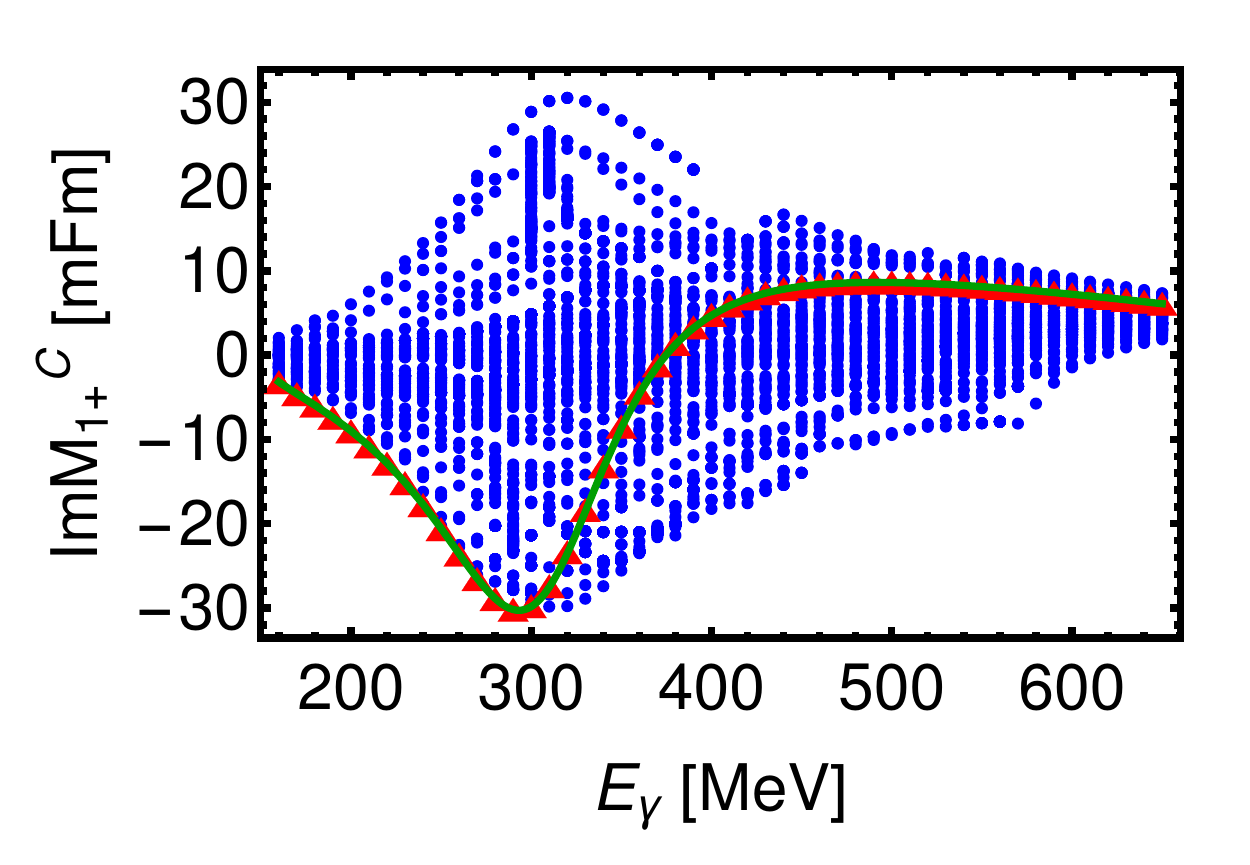}
 \end{overpic} \\
\begin{overpic}[width=0.49\textwidth]{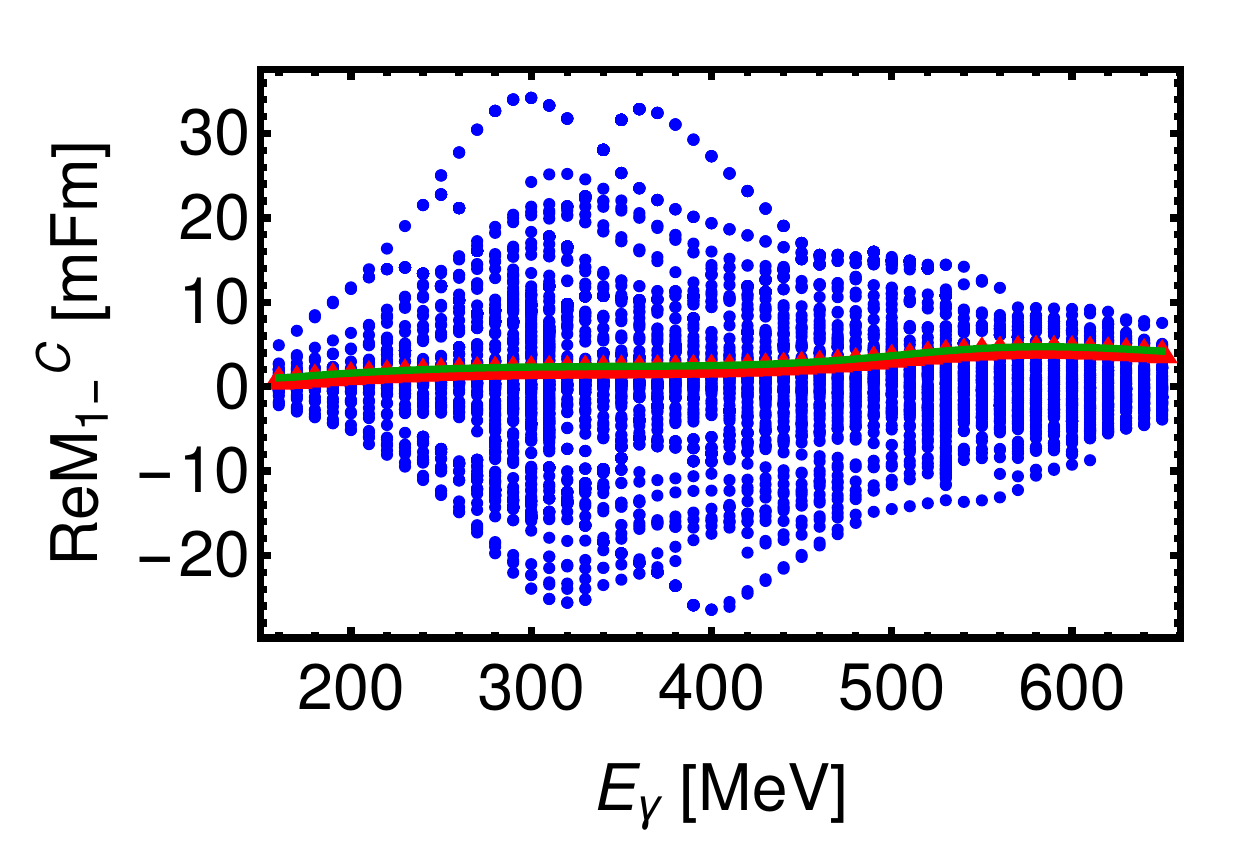}
 \end{overpic} \hspace*{-15pt}
\begin{overpic}[width=0.49\textwidth]{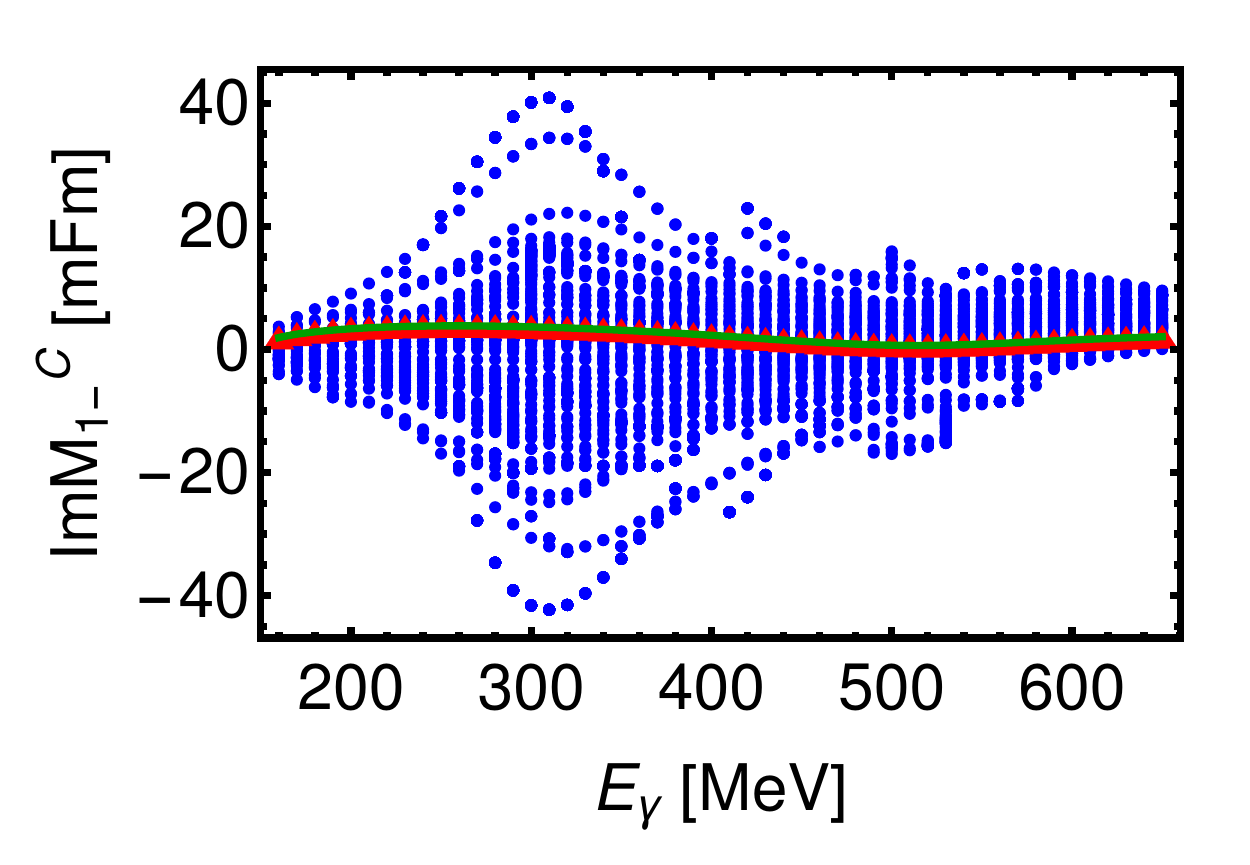}
 \end{overpic}
\vspace*{0pt}
\caption[Solutions found in a TPWA fit to truncated MAID theory-data of the observables $\left\{ \sigma_{0}, \check{\Sigma}, \check{T}, \check{P}, \check{F} \right\}$, for $\ell_{\mathrm{max}}=3$. The $S$- and $P$- wave multipoles are shown.]{Here, the results of an analysis of MAID theory-data \cite{LotharPrivateComm,MAID2007} truncated at $\ell_{\mathrm{max}} = 3$ are shown, using an $F$-wave truncation in the fit. The real and imaginary parts of phase-constrained multipoles for $E_{0+}$ up to $M_{1-}$ are plotted. \newline The observables $\left\{ \sigma_{0}, \check{\Sigma}, \check{T}, \check{P}, \check{F} \right\}$ were used to obtain the multipole solutions from a pool of $N_{MC} = 10000$ randomly chosen initial conditions. All solutions thus obtained are shown as blue dots. A global minimum exists, represented by the red triangles.}
\label{fig:Lmax3ThDataFitBestSols1}
\end{figure}

\begin{figure}[ht]
 \centering
\begin{overpic}[width=0.49\textwidth]{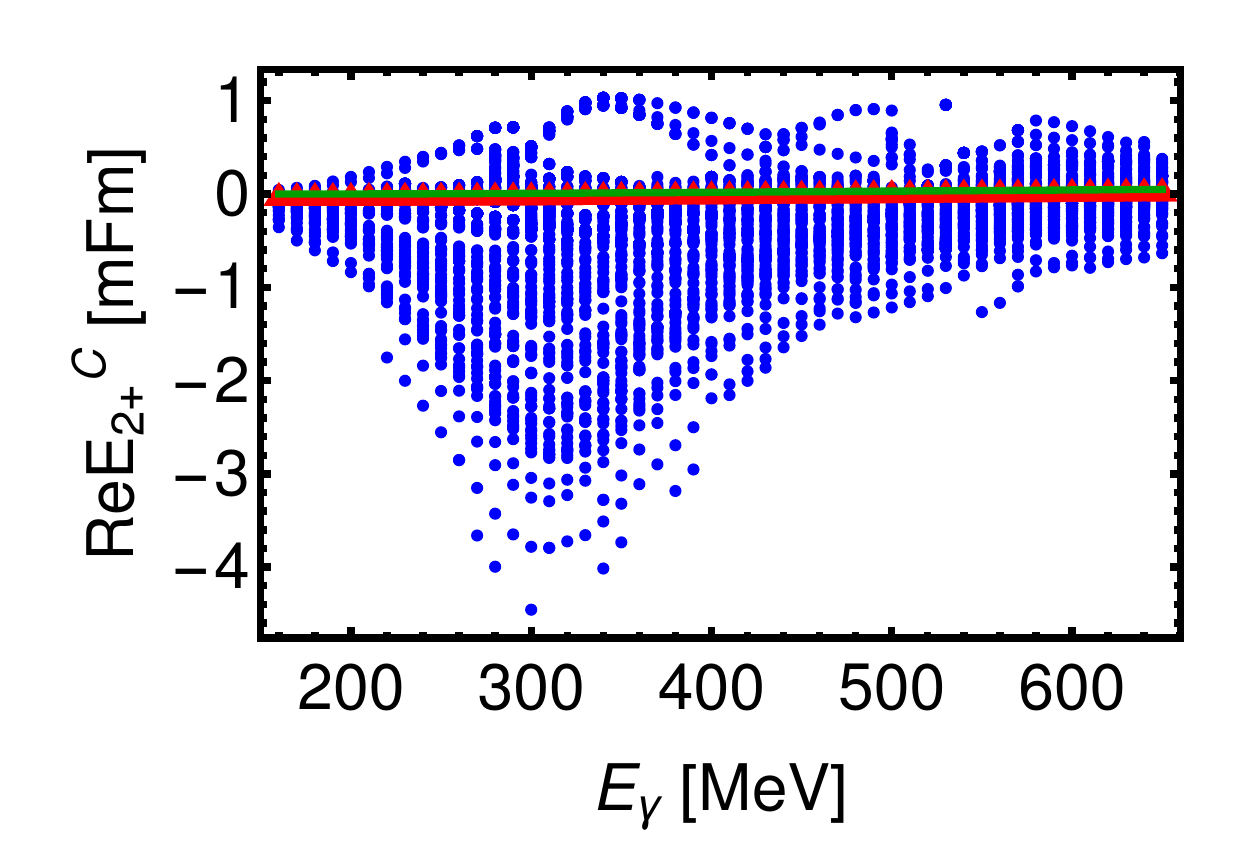}
 \end{overpic} \hspace*{-15pt}
\begin{overpic}[width=0.49\textwidth]{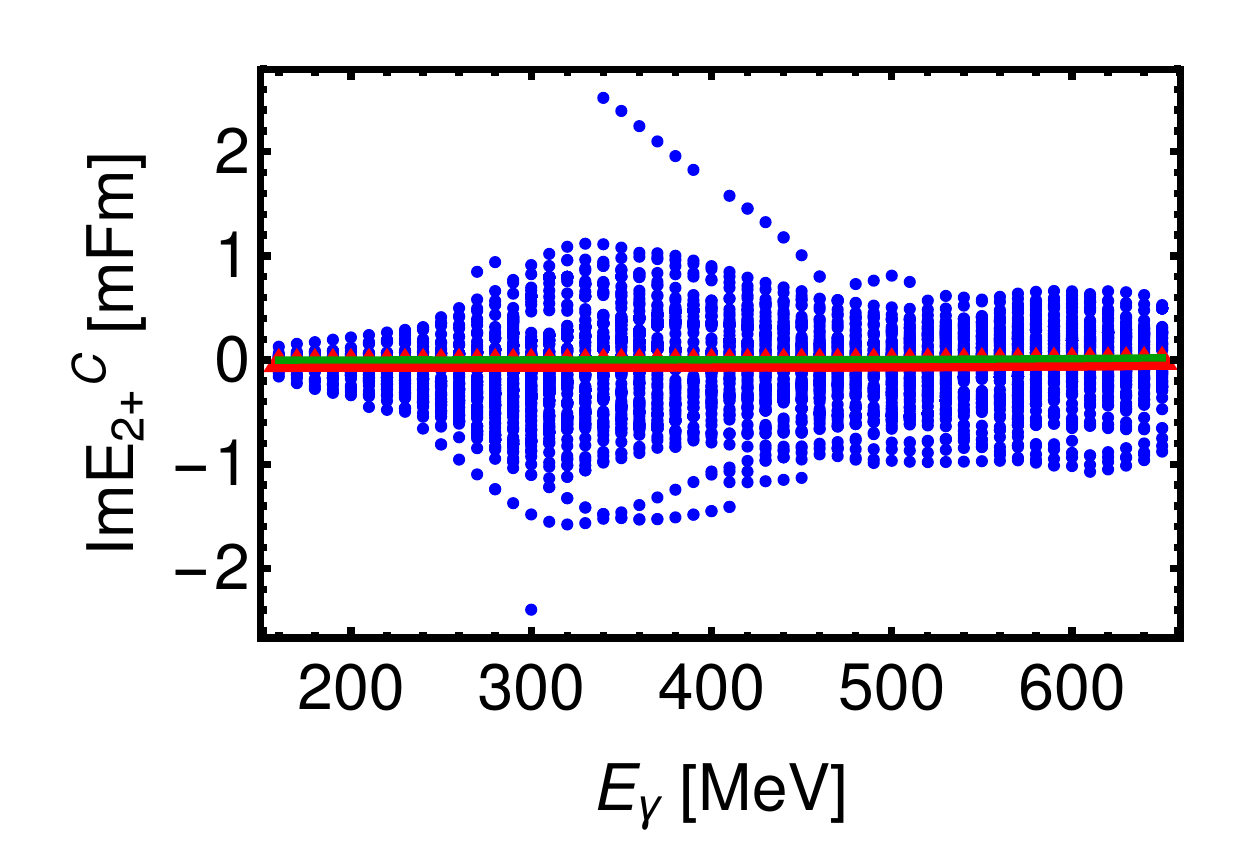}
 \end{overpic}  \\
\begin{overpic}[width=0.49\textwidth]{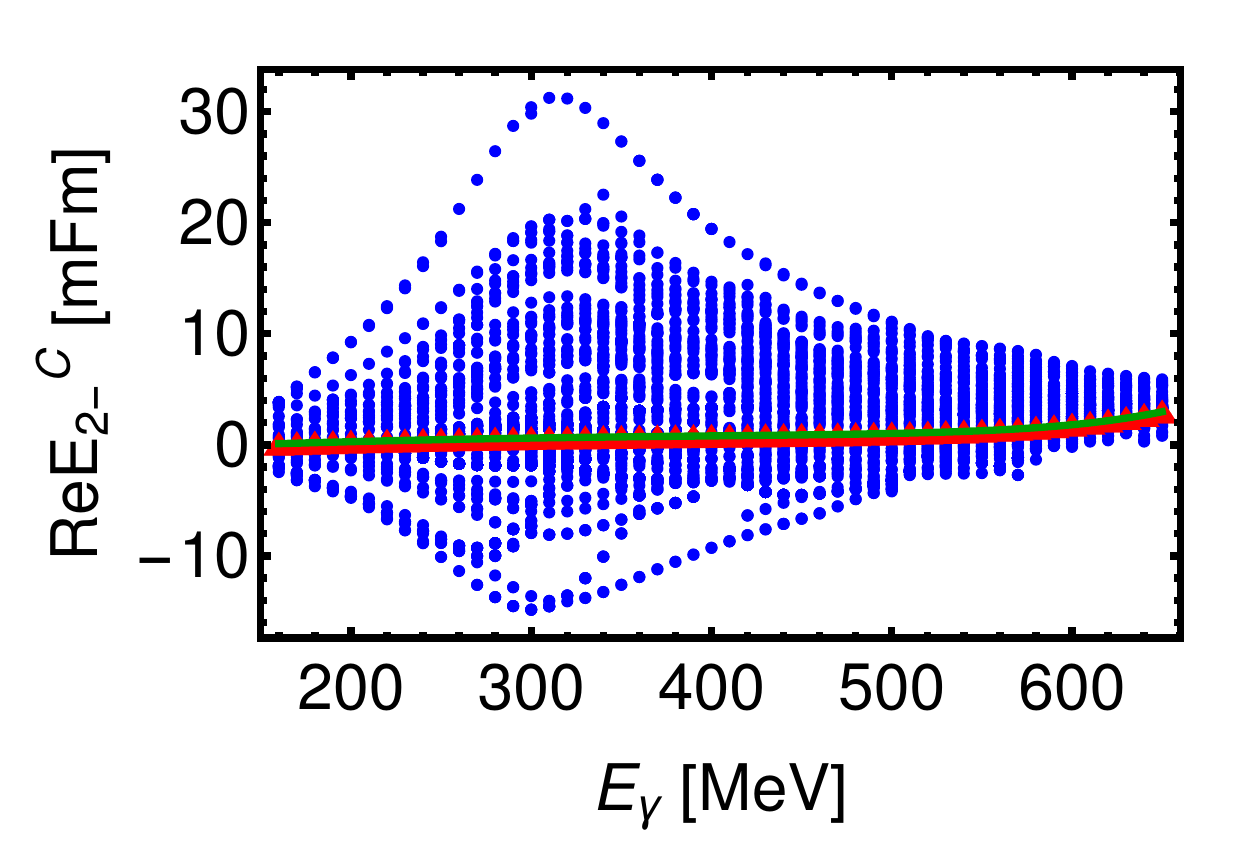}
 \end{overpic}  \hspace*{-15pt}
\begin{overpic}[width=0.49\textwidth]{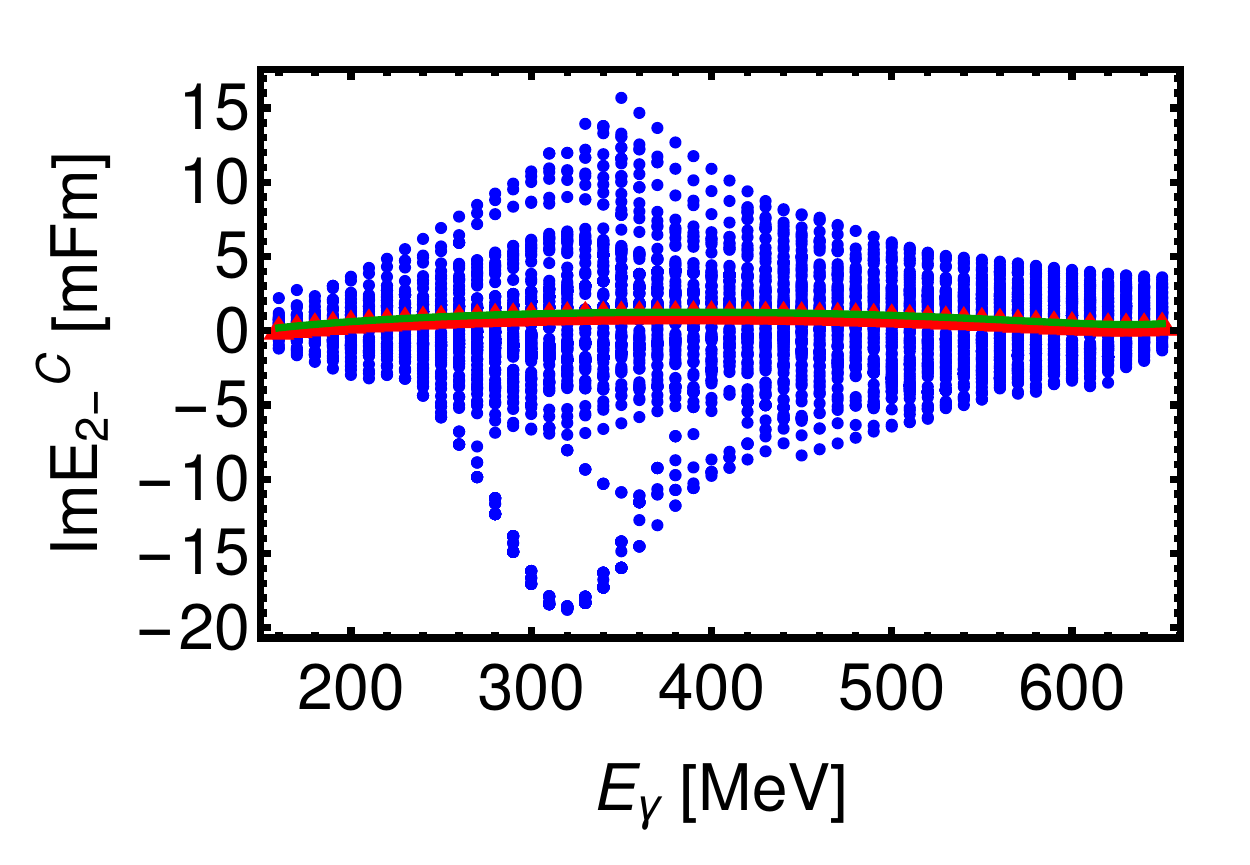}
 \end{overpic} \\
\begin{overpic}[width=0.49\textwidth]{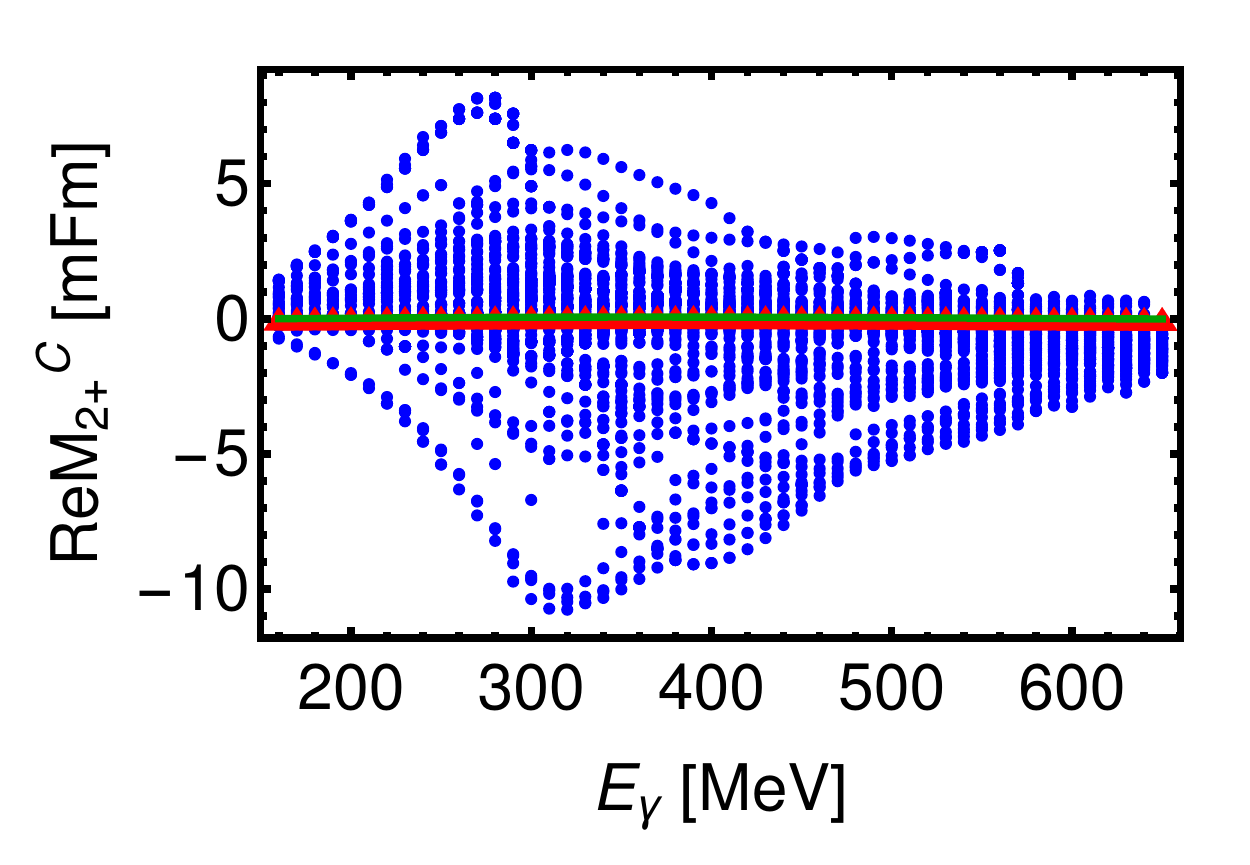}
 \end{overpic} \hspace*{-15pt}
\begin{overpic}[width=0.49\textwidth]{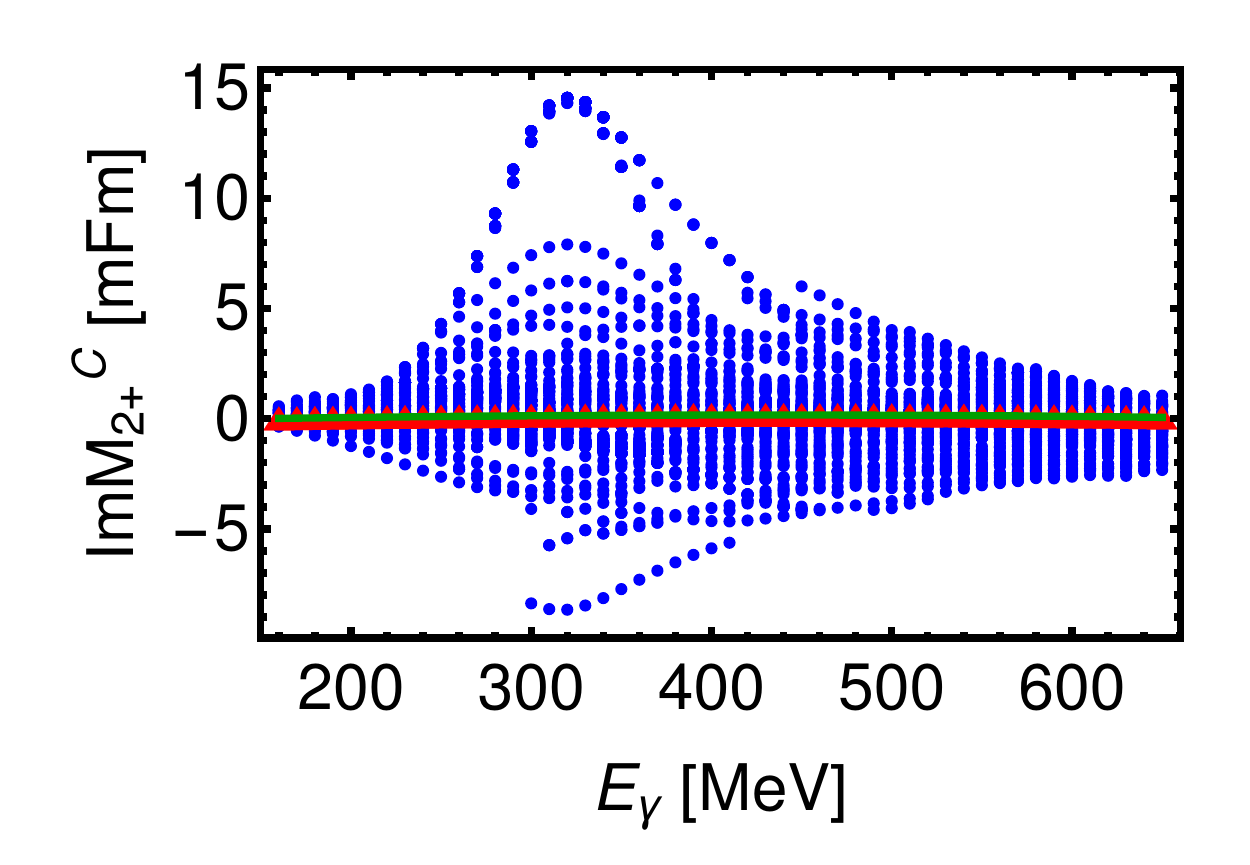}
 \end{overpic} \\
\begin{overpic}[width=0.49\textwidth]{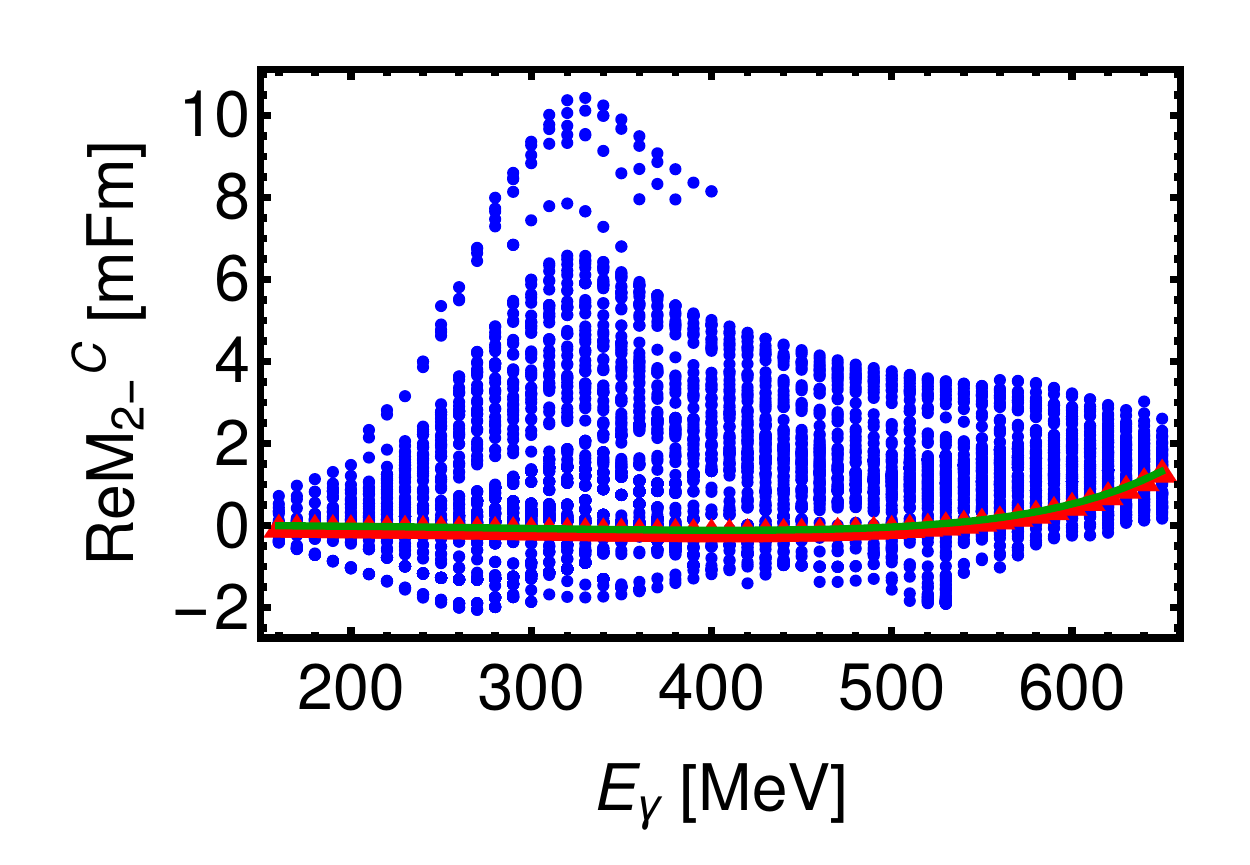}
 \end{overpic} \hspace*{-15pt}
\begin{overpic}[width=0.49\textwidth]{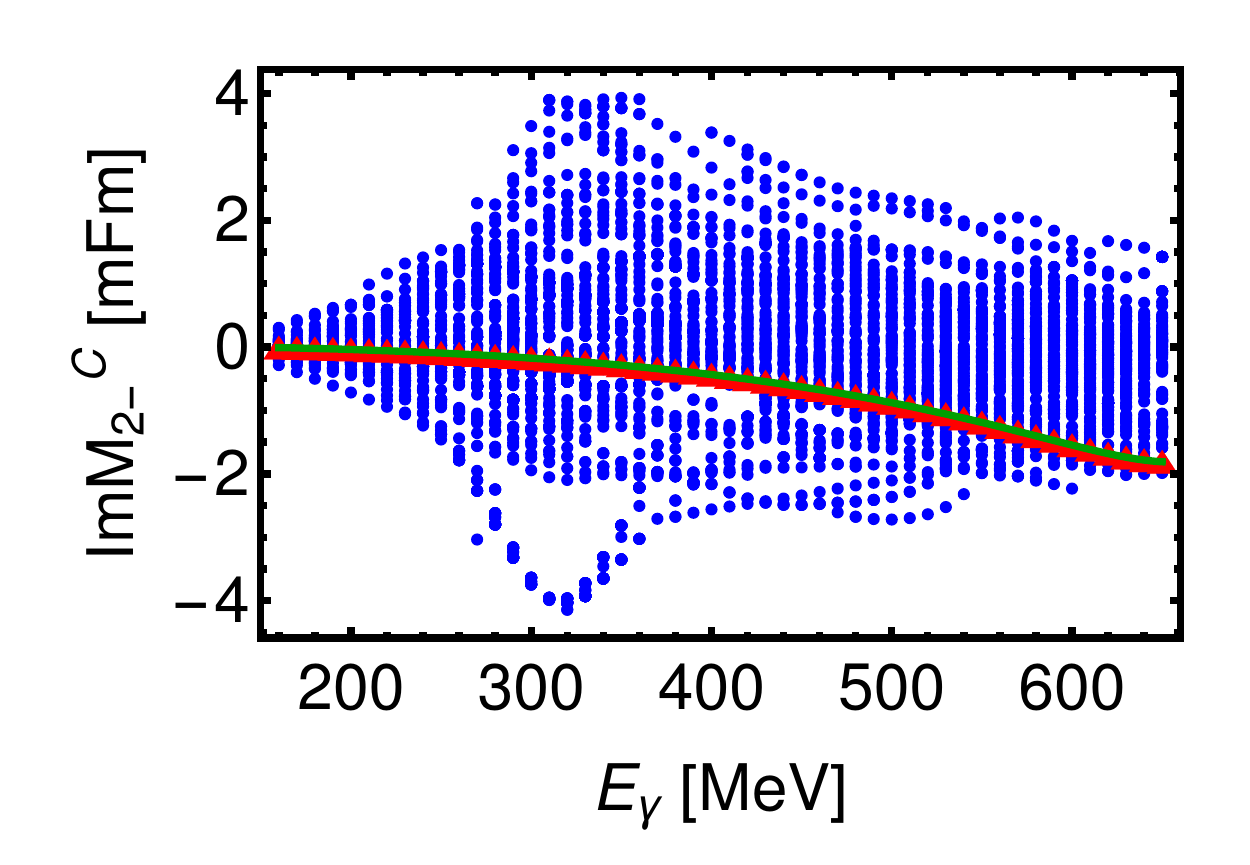}
 \end{overpic}
\vspace*{0pt}
\caption[Solutions found in a TPWA fit to truncated MAID theory-data of the observables $\left\{ \sigma_{0}, \check{\Sigma}, \check{T}, \check{P}, \check{F} \right\}$, for $\ell_{\mathrm{max}}=3$. The $D$- wave multipoles are shown.]{These plots are a continuation of Figure \ref{fig:Lmax3ThDataFitBestSols1}. Shown are the $D$-waves, i.e. all results for the multipoles $E_{2+}$ up to $M_{2-}$.}
\label{fig:Lmax3ThDataFitBestSols2}
\end{figure}

\begin{figure}[ht]
 \centering
\begin{overpic}[width=0.49\textwidth]{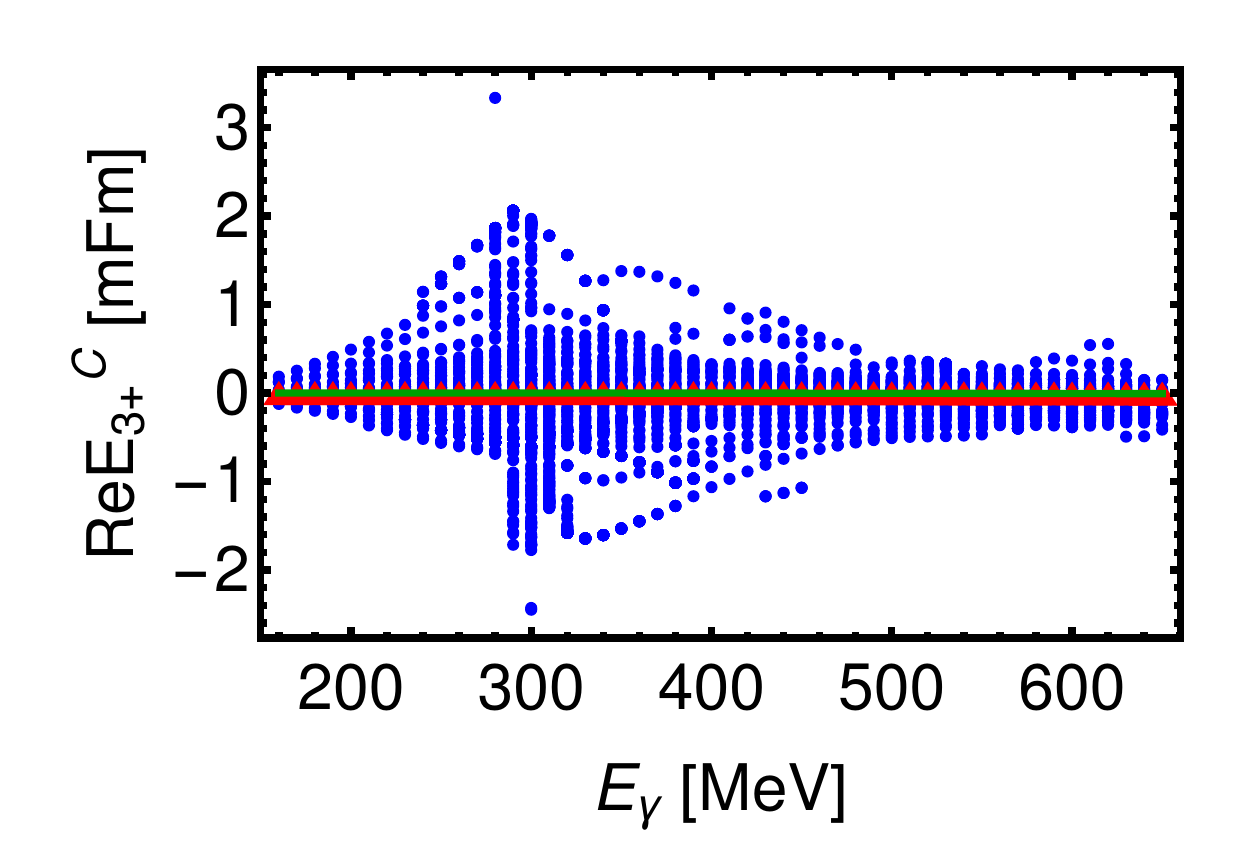}
 \end{overpic} \hspace*{-15pt}
\begin{overpic}[width=0.49\textwidth]{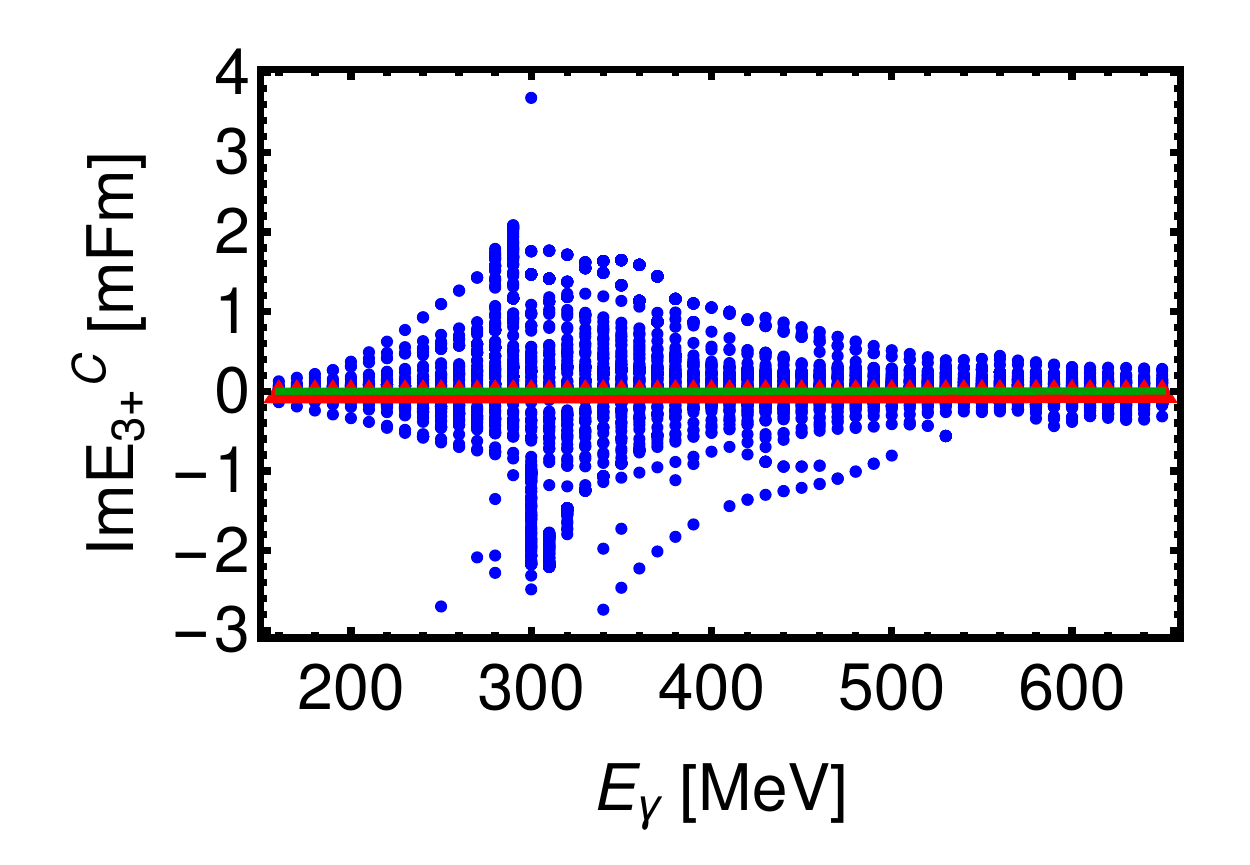}
 \end{overpic}  \\
\begin{overpic}[width=0.49\textwidth]{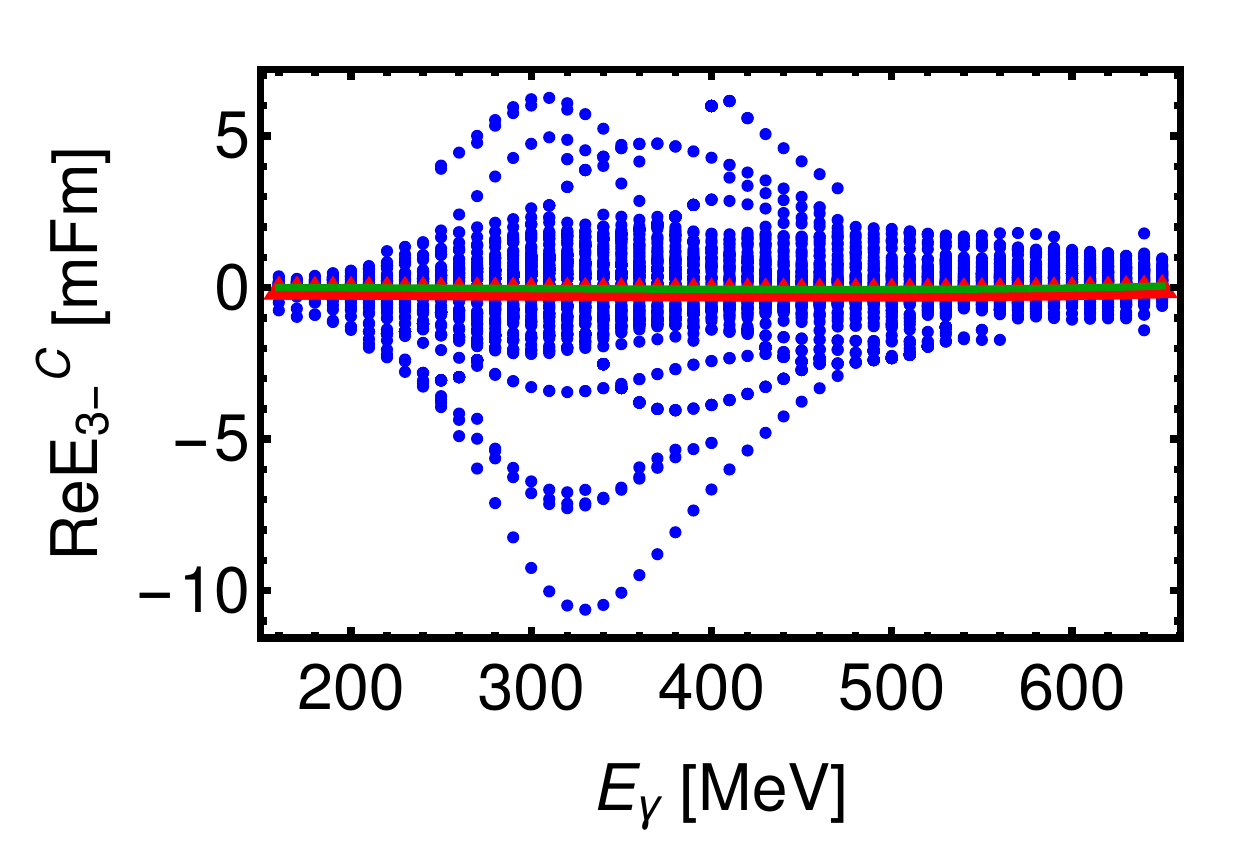}
 \end{overpic}  \hspace*{-15pt}
\begin{overpic}[width=0.49\textwidth]{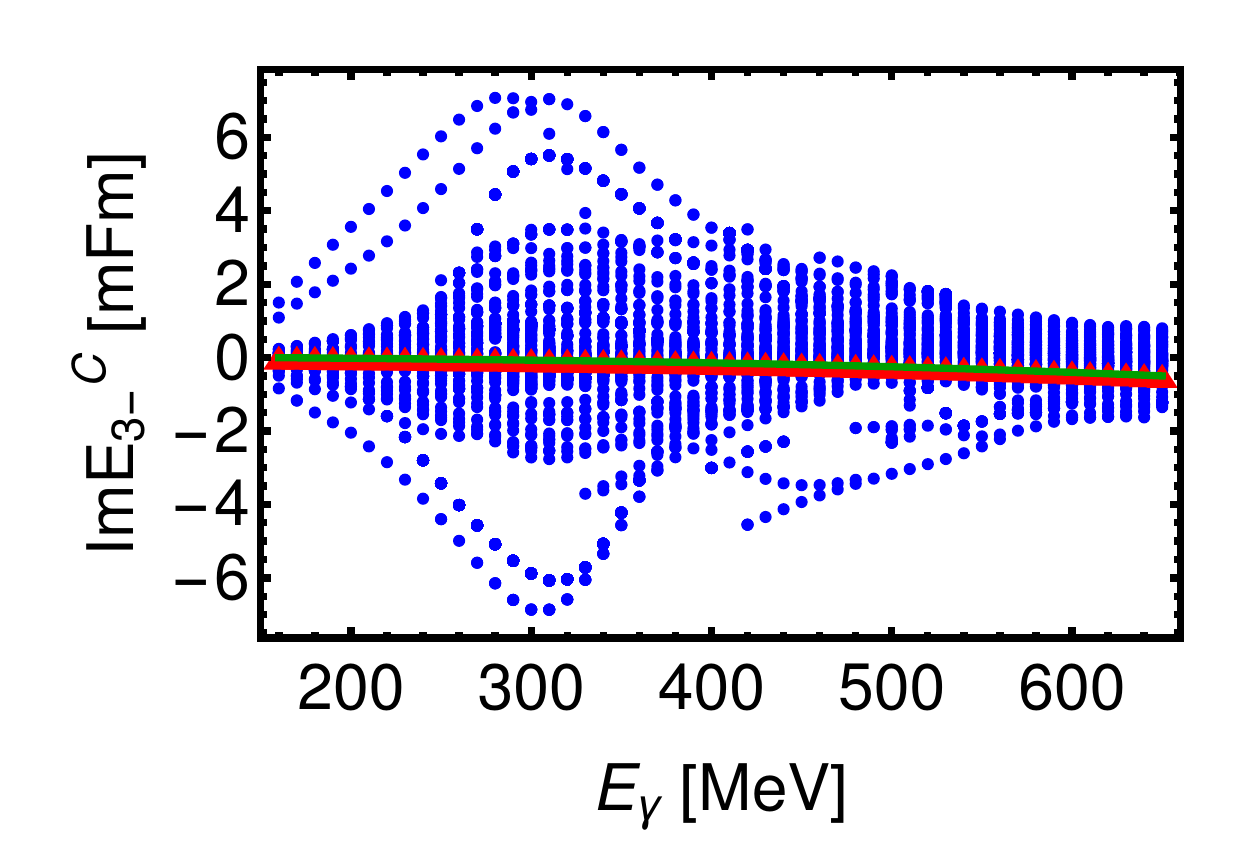}
 \end{overpic} \\
\begin{overpic}[width=0.49\textwidth]{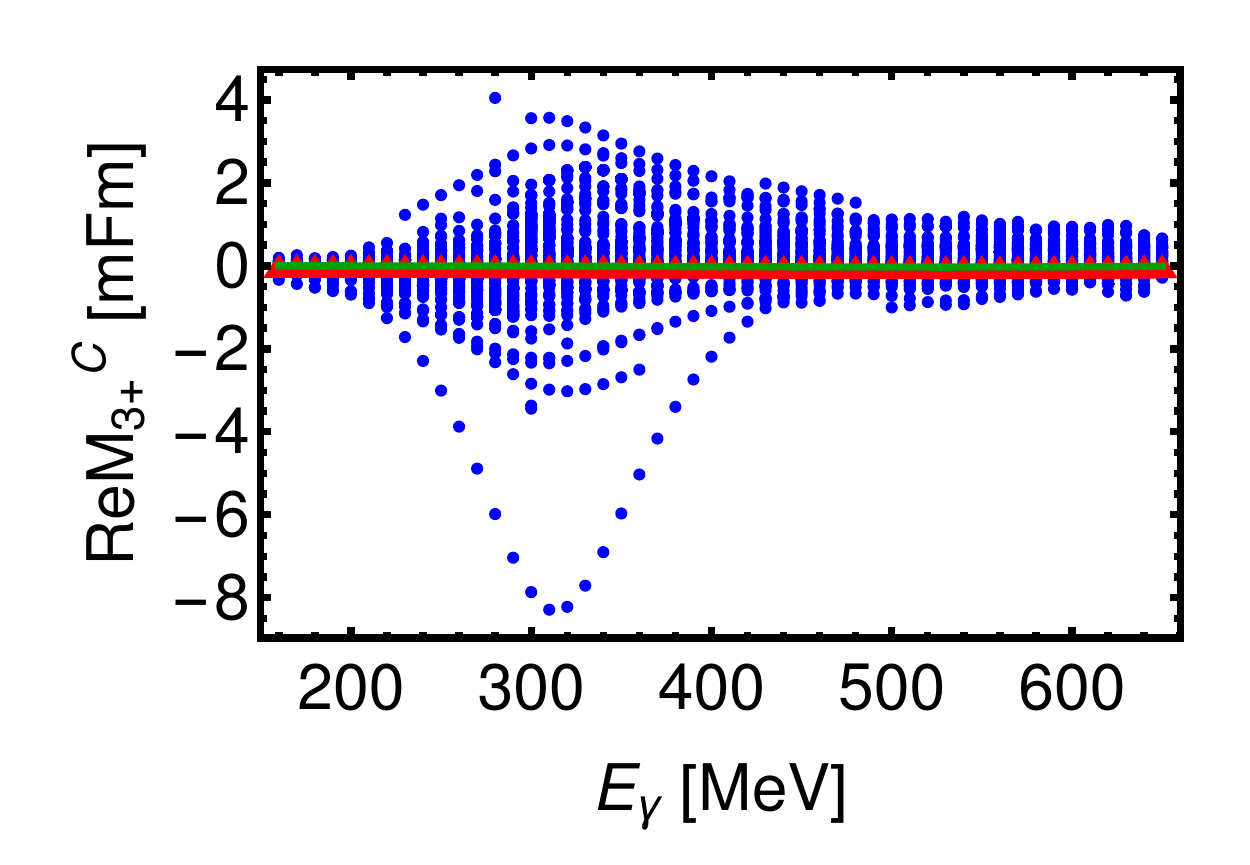}
 \end{overpic} \hspace*{-15pt}
\begin{overpic}[width=0.49\textwidth]{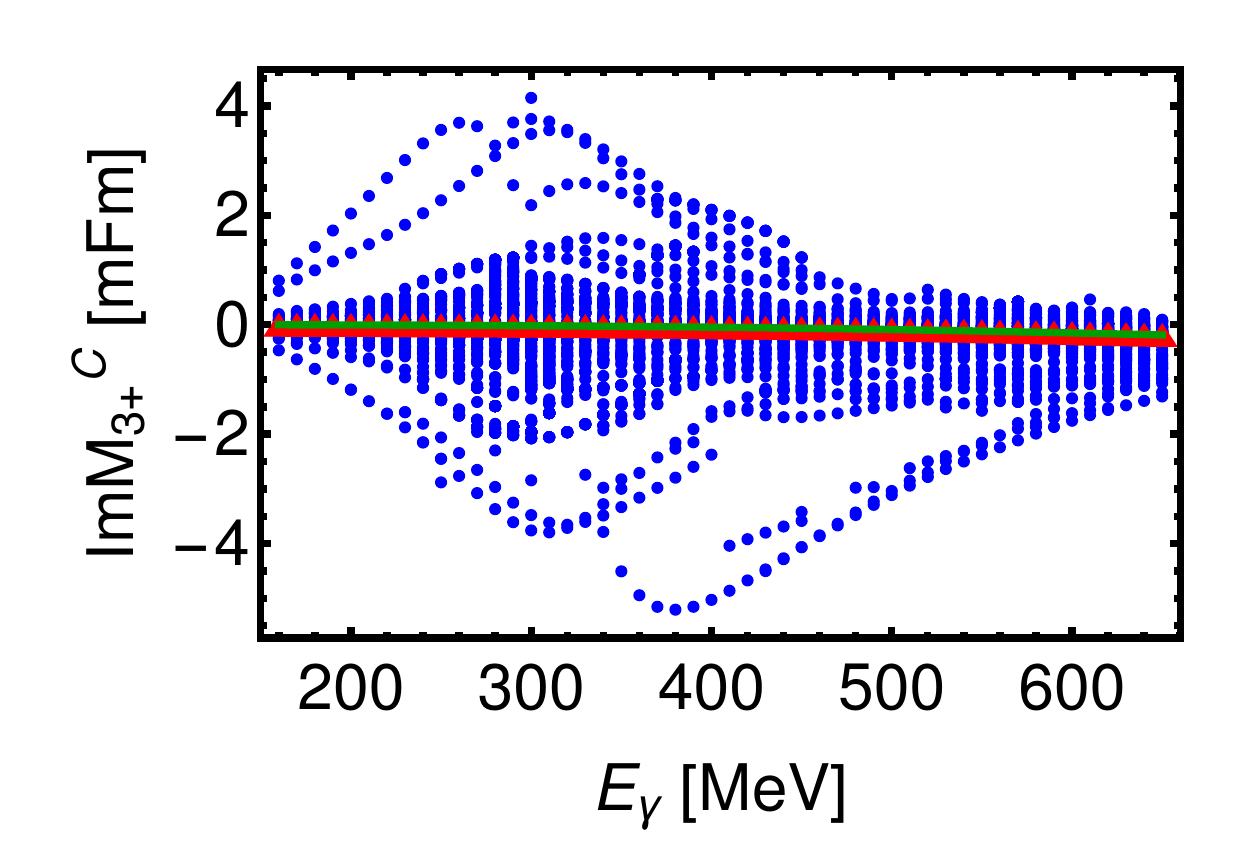}
 \end{overpic} \\
\begin{overpic}[width=0.49\textwidth]{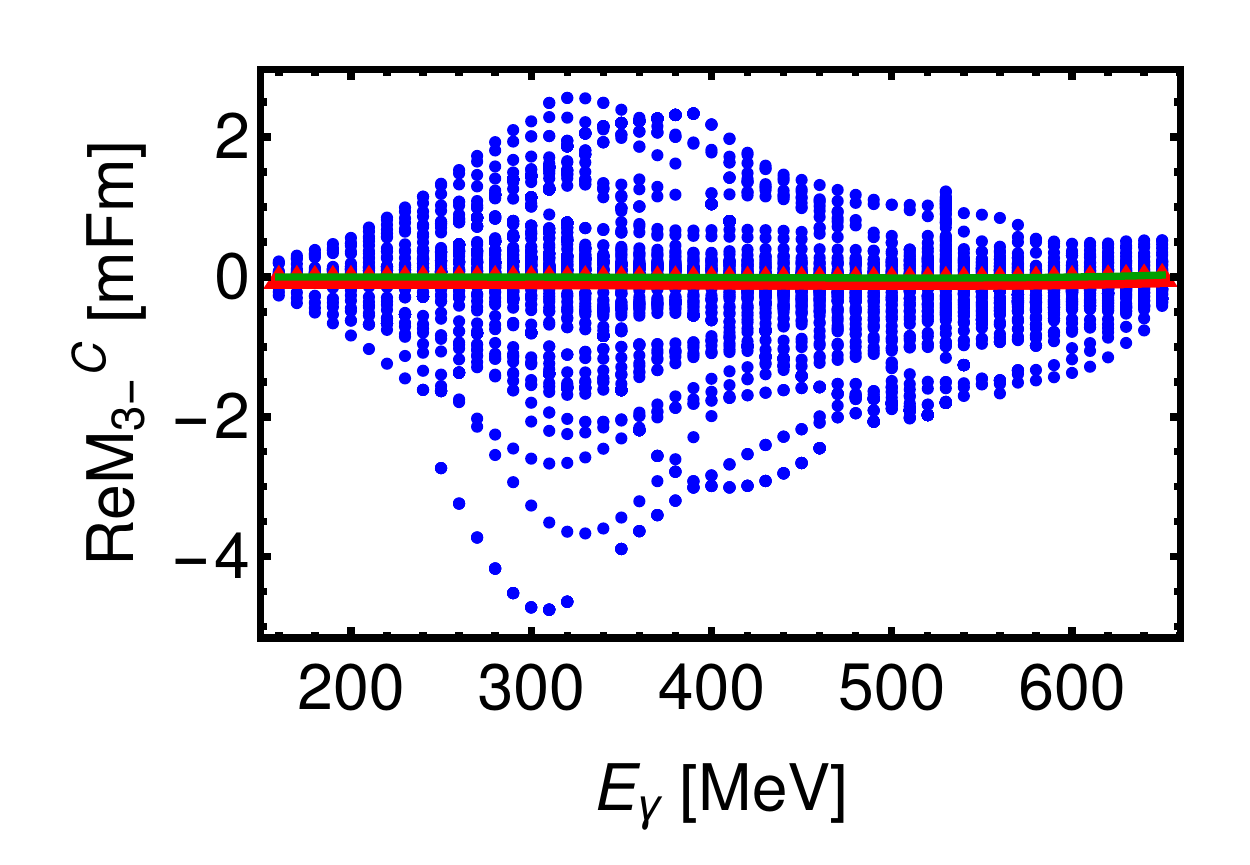}
 \end{overpic} \hspace*{-15pt}
\begin{overpic}[width=0.49\textwidth]{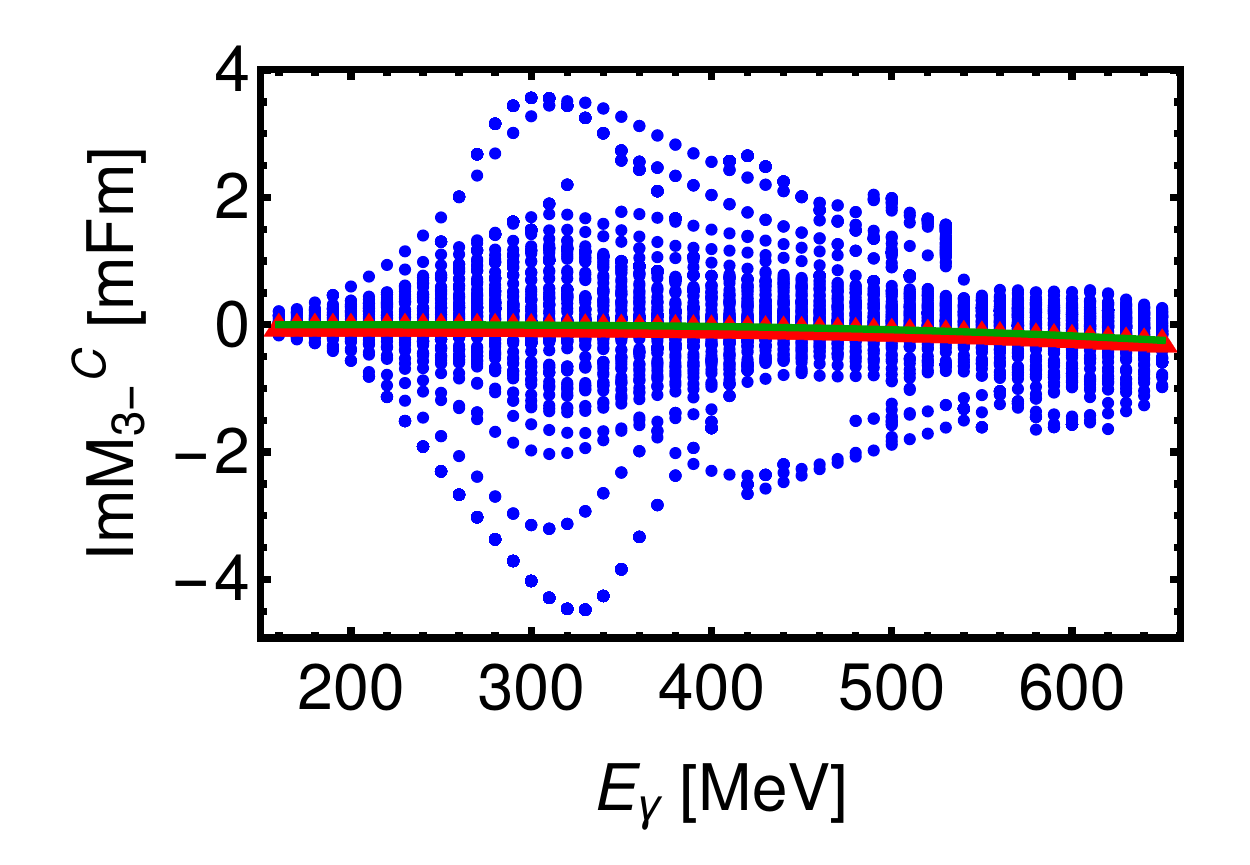}
 \end{overpic}
\vspace*{0pt}
\caption[Solutions found in a TPWA fit to truncated MAID theory-data of the observables $\left\{ \sigma_{0}, \check{\Sigma}, \check{T}, \check{P}, \check{F} \right\}$, for $\ell_{\mathrm{max}}=3$. The $F$- wave multipoles are shown.]{The diagrams shown here belong to the results in Figures \ref{fig:Lmax3ThDataFitBestSols1} and \ref{fig:Lmax3ThDataFitBestSols2}. Here, the $F$-waves, i.e. all multipoles from $E_{3+}$ up to $M_{3-}$, are plotted.}
\label{fig:Lmax3ThDataFitBestSols3}
\end{figure}

As a next step, the complete example set (\ref{eq:ExampleCompleteSet}) has been fitted for $\ell_{\mathrm{max}} = 3$, i.e. MAID theory-data truncated at exactly this order have been solved in a TPWA varying all multipoles up to the $F$-waves. For this purpose, the pool of Monte Carlo initial conditions has been raised to $N_{MC} = 10000$. Results for this fit are shown in Figures \ref{fig:Lmax3ThDataFitBestSols1}, \ref{fig:Lmax3ThDataFitBestSols2} and \ref{fig:Lmax3ThDataFitBestSols3}. \newpage

\clearpage

Again, a unique global minimum was adopted by the fit. This unique solution corresponds to the MAID multipoles used to generate the theory-data. Therefore, the example set (\ref{eq:ExampleCompleteSet}) is also complete for a fit of MAID theory-data truncated at $\ell_{\mathrm{max}} = 3$. However, as can be observed clearly, the total number of local minima has increased vastly. \newline
The global minimum has again a numerical value of roughly $\Phi_{\mathcal{M}} \simeq 10^{-16} \left( \mu b / sr \right)^{2}$ over the whole considered energy range. The local minima tend to get closer to the unique solution compared to the cases $\ell_{\mathrm{max}} = 2$ and $1$, especially for the lower energy bins. Values for the second best minima range around $\Phi_{\mathcal{M}} \simeq 10^{-11} \left( \mu b / sr \right)^{2}$ at the lower energies and $\Phi_{\mathcal{M}} \simeq 10^{-5} \left( \mu b / sr \right)^{2}$ at the highest energies. \newline
The number of non-redundant solutions of the problem also has increased compared to the previous example, as one could expect from the exponential law $N = 4^{2 \ell_{\mathrm{max}}}$ for the upper bound of discrete ambiguities. The number of all attained, global and local, minima takes values between $45$ and $457$ for all energy bins. However, numbers as large as the latter value are only found in a few singular energy bins. Mostly, one has between $100$ and $250$ minima. Again, compared to the previous order $\ell_{\mathrm{max}} = 2$, an increase by roughly a factor of $10$ can be observed. \newline

\begin{figure}[ht]
 \centering
\begin{overpic}[width=0.49\textwidth]{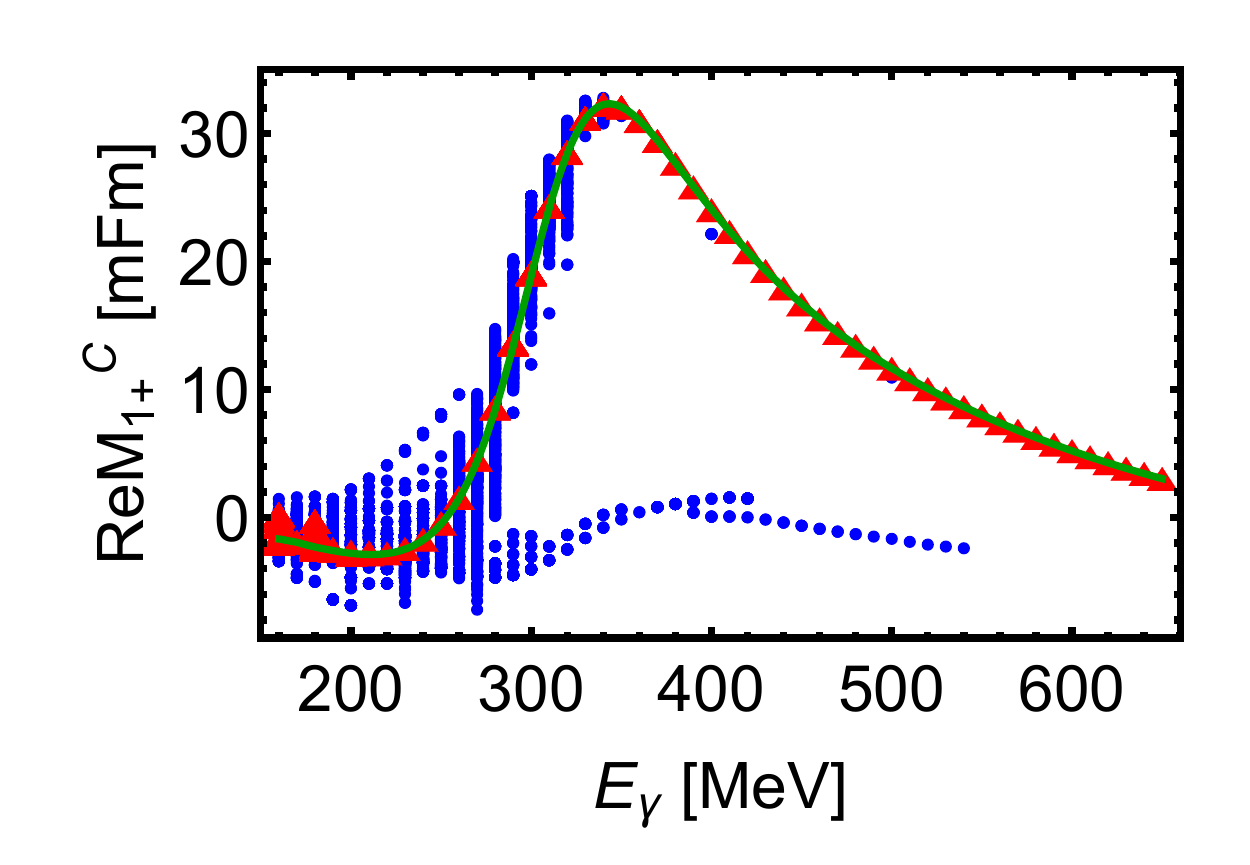}
 \end{overpic} \hspace*{-15pt}
\begin{overpic}[width=0.49\textwidth]{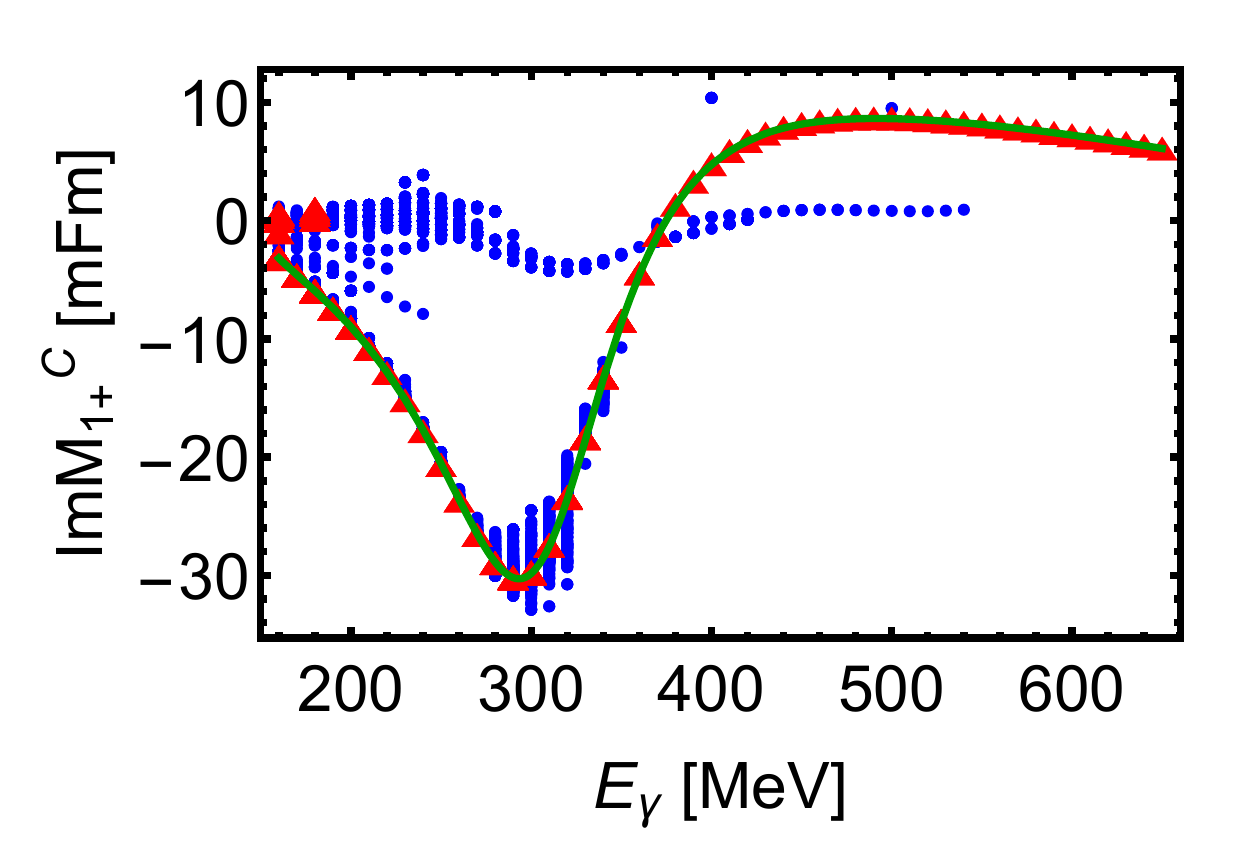}
 \end{overpic} \hspace*{-15pt}
\vspace*{0pt}
\caption[Solutions found in a TPWA fit to truncated MAID theory-data of the observables $\left\{ \sigma_{0}, \check{\Sigma}, \check{T}, \check{P}, \check{F} \right\}$, for $\ell_{\mathrm{max}}=4$. Only the $M_{1+}$-multipole is shown.]{This figure depicts the results of a TPWA-fit to MAID-theory-data \cite{LotharPrivateComm,MAID2007} for $\left\{ \sigma_{0}, \check{\Sigma}, \check{T}, \check{P}, \check{F} \right\}$ truncated at $\ell_{\mathrm{max}} = 4$, using a truncation at the $G$-waves in the fit. The attention is restricted to the $M_{1+}$-multipole. Blue dots show all solutions with $\Phi_{\mathcal{M}} < 10^{-9} \left( \mu b / sr \right)^{2}$ that arise from a pool of $N_{MC} = 10000$ initial parameter configurations. The solutions found with smallest values for $\Phi_{\mathcal{M}}$ are shown as red triangles. For the lowest energies, degenerate global minima arise. The MAID-solution \cite{MAID2007,MAID} used to generate the theory-data is plotted as a green solid line.}
\label{fig:Lmax4ThDataFitBestSolsGroupSAndFExample}
\end{figure}

The last explicit example considered in this section is the case of truncated MAID theory-data for $\ell_{\mathrm{max}} = 4$. It is the most numerically demanding fit and in some ways also the least well-behaved. Again, the postulated complete set (\ref{eq:ExampleCompleteSet}) has been fitted in a fully model independent TPWA truncated at the $G$-waves. In order to obtain solutions in an acceptable time, the size of the start configuration pool was not increased any further as compared to the previous example, i.e. still $N_{MC} = 10000$ point configurations have been used. Results are shown for the $M_{1+}$ multipole in Figure \ref{fig:Lmax4ThDataFitBestSolsGroupSAndFExample}. Due to the large number of local minima (see discussion below), not all solutions have been plotted there. Rather, we restricted this Figure to all solutions obeying the bound $\Phi_{\mathcal{M}} < 10^{-9} \left( \mu b / sr \right)^{2}$. \newline
In the lowest energy bins, the fit does not yield a well-separated global minimum. Several degenerate minima exist having discrepancy function values around $\Phi_{\mathcal{M}} \simeq 10^{-15}  \left( \mu b / sr \right)^{2}$. This situation starts to improve for energies larger than $300 \hspace*{2pt} \mathrm{MeV}$. From there on, a global minimum is found which corresponds to the correct MAID solution and is separated from the second best local minimum by about $4$ orders of magnitude in $\Phi_{\mathcal{M}}$. This separation becomes larger when going to the highest energies, where it is given mostly by $10$ orders of magnitude. Still, the behavior of the fits is quite singular in the lower energies and an explanation has to be sought. \newline
One can get further towards an understanding by considering the Legendre coefficients entering the TPWA fit, specifically the last two corresponding to orders $7$ and $8$. Those can be seen for the observable $P$, for instance, in Figure \ref{fig:Lmax1234ThDataFitLegCoeffsP3}. It is seen that the second to last Legendre coefficient, which is given in all cases as an $\left< F, G \right>$-term, shows values different from zero only in the second half of the considered energy region. This is fully consistent with the $G$-waves contained in the MAID-model. They all are strictly zero for the lowest energies and show small but rising moduli when ascending in energy. However, for the highest energies the moduli of MAID $G$-waves are still in regions around $10^{-6} \hspace*{2pt} \mathrm{mFm}$. Thus, in the low energy region the fit for $\ell_{\mathrm{max}} = 4$ \textit{overfits}, i.e.  it tries to fit out partial waves which are actually exactly zero in the theory-data. Once the $G$-waves develop some (small) strength, it is again feasible to extract them.


\clearpage

The higher degree of complication of this fit is also illustrated by the increase in the overall number of non-redundant solutions. For the most degenerate (low) energies, those can very well amount to over a thousand solutions, in the most extreme case counting $1933$ of them. When ascending in energy, these numbers get reduced somewhat and one typically gets a few hundred solutions, somewhere between $500$ and $1000$. One obtains again roughly a factor of $10$ for the increase in overall solutions, compared to the example for $\ell_{\mathrm{max}} = 3$. The solutions are in fact so many that in Figure \ref{fig:Lmax4ThDataFitBestSolsGroupSAndFExample}, although a lot of local minima with $\Phi_{\mathcal{M}} \geq 10^{-9} \left( \mu b / sr \right)^{2}$ have already been cut away in the plot, it is hard to tell whether the ambiguities that occur are still discrete or belong to a continuum of solutions. For the lowest energies, the latter scenario may very well be the case and it can originate from overfitting. However, above $300 \hspace*{2pt} \mathrm{MeV}$ we are, due to more detailed investigations of the resulting values of $\Phi_{\mathcal{M}}$, quite sure that ambiguities are in fact discrete. Still, there are many of them and they tend again to be less separated, in $\Phi_{\mathcal{M}}$, from the global minimum as compared to the $\ell_{\mathrm{max}} = 3$ case. \newline
In order to help the summary of all the fits performed in this section, some interesting metadata in the results have been collected in Table \ref{tab:Lmax1234TheoryDataFitsDataTable}. Shown are, for three exemplary energies and all fitted truncation orders, the total number of non-redundant solutions as well as the $\Phi_{\mathcal{M}}$-values for the global minimum (if attained) and the second best local one. \newline
As mentioned many times before, the number of local minima, or potential ambiguities, of the fit rises with higher $\ell_{\mathrm{max}}$. It can be seen that one gains (very) roughly a factor of $10$ in the number of solutions. Therefore, one observes an exponential growth behavior. This can however actually be expected from the formal treatment of the discrete partial wave ambiguities in chapter \ref{chap:Omelaenko} and appendix \ref{sec:AdditionsChapterII}. For a higher truncation order, there are more partial waves and therefore also more Omelaenko roots $\left\{ \alpha_{k}, \beta_{k} \right\}$. With this rise in the number of variables in the problem, the possibilities to form ambiguities grow exponentially, having an upper bound of $N = 4^{2 \ell_{\mathrm{max}}}$ (see section \ref{sec:WBTpaper}). \newline

\begin{table}[ht]
 \centering
 \begin{tabular}{cr|rcc|rcc}
  &   & \multicolumn{3}{c|}{ $E_{\gamma} = 190 \hspace*{2pt} \mathrm{MeV}$} & \multicolumn{3}{c}{  $E_{\gamma} = 400 \hspace*{2pt} \mathrm{MeV}$ } \\
\hline
 $\ell_{\mathrm{max}}$  & $N_{MC}$ & $N_{\mathrm{n.r.}}$ & $\Phi_{\mathcal{M}}^{\mathrm{Best}}$ & $\Phi_{\mathcal{M}}^{2^\mathrm{nd}}$ & $N_{\mathrm{n.r.}}$ & $\Phi_{\mathcal{M}}^{\mathrm{Best}}$ & $\Phi_{\mathcal{M}}^{2^\mathrm{nd}}$ \\
\hline
 $1$ & $1500$  & $1$  & $1.3 \times 10^{-15}$ & - & $2$ & $2.2 \times 10^{-15}$ & $50.29$ \\
 $2$ & $3000$  & $12$ & $1.6 \times 10^{-15}$ & $4.9 \times 10^{-7}$ & $20$ & $1.8 \times 10^{-14}$ & $1.4 \times 10^{-4}$ \\
 $3$ & $10000$ & $47$ & $3.6 \times 10^{-16}$ & $2.5 \times 10^{-11}$ & $124$ & $1.5 \times 10^{-14}$ & $4.4 \times 10^{-7}$ \\
 $4$ & $10000$ & $1321$ &  \multicolumn{2}{c|}{many loc. min. w. $\Phi \simeq 10^{-15}$}  & $493$ & $4.7 \times 10^{-15}$ & $7.2 \times 10^{-11}$ \\
\hline
\hline
  &   & \multicolumn{3}{c|}{ $E_{\gamma} = 610 \hspace*{2pt} \mathrm{MeV}$} &  &  &  \\
\hline
 $\ell_{\mathrm{max}}$  & $N_{MC}$ & $N_{\mathrm{n.r.}}$ & $\Phi_{\mathcal{M}}^{\mathrm{Best}}$ & $\Phi_{\mathcal{M}}^{2^\mathrm{nd}}$ &  &  &  \\
\hline
 $1$ & $1500$  & $1$ & $5.3 \times 10^{-16}$ & - &   &  &  \\
 $2$ & $3000$  & $12$ & $4.2 \times 10^{-16}$ & $6.5 \times 10^{-5}$ &  &  &  \\
 $3$ & $10000$ & $104$ & $1.1 \times 10^{-16}$ & $3.0 \times 10^{-5}$ &  &  &  \\
 $4$ & $10000$ & $759$ & $1.4 \times 10^{-15}$ & $5.9 \times 10^{-9}$ &  &  &  \\
 \end{tabular}
 \caption[Properties on the solutions obtained in the analyses of MAID theory-data. Three exemplary energies are shown for illustration.]{This Table shows some properties on the solutions obtained in the analyses of MAID theory-data, for the purpose of illustration. Three exemplary energies were selected from the whole considered energy region. All results stem from fits of the complete set $\left\{ \sigma_{0}, \check{\Sigma}, \check{T}, \check{P}, \check{F} \right\}$. Fits have always been performed in the same truncation order that has been employed in the generation of the theory-data (see the main text). \newline For every TPWA order $\ell_{\mathrm{max}}$, the following information on the obtained solutions is shown: the size of the employed pool of initial configurations $N_{MC}$, number of non-redundant solutions $N_{\mathrm{n.r.}}$ sorted out of the full pool of $N_{MC}$ final configurations, value $\Phi_{\mathcal{M}}^{\mathrm{Best}}$ of the minimization function in the global minimum, value $\Phi_{\mathcal{M}}^{2^\mathrm{nd}}$ of the same quantity in the second best minimum (i.e. the best minimum which is only local). Values for obtained minima are given in units of $\Big[ \left( \mu b / sr \right)^{2} \Big]$.}
 \label{tab:Lmax1234TheoryDataFitsDataTable}
\end{table}

Moreover, it can be observed in Table \ref{tab:Lmax1234TheoryDataFitsDataTable} that, while the global minimum is attained at discrepancy function values of about $\Phi_{\mathcal{M}} \simeq 10^{-16} \left( \mu b / sr \right)^{2}$, local minima tend to have $\Phi_{\mathcal{M}}$-values that come closer to, or are less well separated from, the global optimum whenever $\ell_{\mathrm{max}}$ is increased. Possible explanations for this behavior are manifold. However, it could very well be that, with the above mentioned rise in possibilities, one also obtains a larger probability to generate discrete ambiguities $\bm{\uppi} \in \hat{\mathcal{P}}$ (symbols are explained in section \ref{subsec:TheoryDataFitsLmax1} and appendix \ref{subsec:AccidentalAmbProofsI}) that can have a smaller violation parameter $\epsilon_{\bm{\uppi}}$ in Omelaenko's constraint equation (cf. expression (\ref{eq:GeneralizedViolatedCR}) in appendix \ref{subsec:AccidentalAmbProofsII})
\begin{equation}
\bm{\uppi} \left(\alpha_{1}\right) \ast \ldots \ast \bm{\uppi} \left(\alpha_{2L}\right) = \bm{\uppi} \left(\beta_{1}\right) \ast \ldots \ast \bm{\uppi} \left(\beta_{2L}\right) \ast e^{i \epsilon_{\bm{\uppi}}} \mathrm{.} \label{eq:GeneralizedViolatedCRChapter432}
\end{equation}
It has been explored in section \ref{subsec:TheoryDataFitsLmax1} and appendix \ref{subsec:AccidentalAmbProofsII} that ambiguities with smaller $\epsilon_{\bm{\uppi}}$ tend to be related to fit solutions with a better (i.e. smaller) value for $\Phi_{\mathcal{M}}$. \newline
The connection between local fit minima of good quality and discrete partial wave ambiguities derived from the Omelaenko formalism has been numerically rigorously established for $\ell_{\mathrm{max}} = 1$ and $2$ in the course of this work. There it has been confirmed to be correct, with results for $\ell_{\mathrm{max}} = 1$ shown in detail at the end of section \ref{subsec:TheoryDataFitsLmax1}. For the higher truncation orders, the connection has not been checked explicitly. However, these cases for higher $\ell_{\mathrm{max}}$ are very likely just mathematically analogous, but more complicated, versions of the $\ell_{\mathrm{max}} = 1,2$. This can be expected at least in the idealized case of theory-data analyses. \newline
Therefore, a strong case is made for the existence of ambiguous solutions in fully model-independent TPWA's which are related to Omelaenko's ambiguities. The well-known re\-fe\-rence on multipole analyses by Grushin \cite{Grushin} contains, at one place, a statement contrary to our result. The author of this work arrives at the conclusion to be able to generally completely disregard accidental ambiguities, in the context of a multipole analysis for pion data in the truncation order $\ell_{\mathrm{max}} = 1$. Explicitly, Grushin writes, on page 72 of \cite{Grushin}, lines 21 - 30:
\begin{quotation}
 \textit{We note in this context that a pressing need was noted in${^\mathrm{14}}$} (i.e., Omelaenko, ref. \cite{Omelaenko}) \textit{for the conduct of double-polarization experiments to achieve unambiguous determination of amplitudes in the channel $\gamma p \longrightarrow \pi^{0} p$ even at lower energies. ($\ldots$) However, in a real energy-independent analysis with an energy-interval of $\sim (20 - 30) \hspace*{1pt} \mathrm{MeV}$, these ``point-sources'' of ambiguity cannot appear as continuous branches of solutions. In other words, the concerns of the author in${^\mathrm{14}}$ are in practice unfounded, which was demonstrated by the multipole analysis here presented.}
\end{quotation}
However, the results of this section illustrated that Grushin probably drew these conclusions since his analysis was done in the simplest possible case, i.e. a truncation at the $P$-waves. In this order, fits are still quite well-behaved, as shown in section \ref{subsec:TheoryDataFitsLmax1}. The result for the higher truncation orders presented here enforce the idea that accidental ambiguities according to Omelaenko become important once $\ell_{\mathrm{max}}$ is increased and that they may very well form 'continuous branches'. \newline
An interesting question about theory-data analyses is whether or not complete TPWA experiments with $5$ observables can still yield the correct, unique multipole solution once the fitted MAID data are not truncated themselves, i.e. contain contributions from all partial waves up to infinity. The investigation of such data will be discussed in the next section.

\subsubsection{MAID2007 theory-data from the full model ($\ell_{\mathrm{max}} \rightarrow \infty$)} \label{subsec:TheoryDataFitsLmaxInfinite}

The investigations of the idealized situations provided by analyses of perfect data, i.e. data without errors, reach their conclusion in this section. Such idealized data were, for the investigations performed in this work, provided by the model MAID2007 \cite{LotharPrivateComm,MAID2007} for the reaction $\gamma p \longrightarrow \pi^{0} p$. In this section, perfect data are considered that stem from the full MAID model. MAID2007 does contain various kinds of $t$-channel exchanges \cite{MAID2007} and may, therefore, be regarded formally as a model including contributions from all partial waves up to infinity. \newline
It is interesting to investigate whether or not the higher partial waves can endanger the solvability of the proposed complete sets of $5$ observables found from the study of discrete ambiguities in chapter \ref{chap:Omelaenko}. These complete sets have passed the test in fits to truncated data for various cases of truncation and thus complication, in both section \ref{subsec:TheoryDataFitsLmax1} and \ref{subsec:TheoryDataFitsLmax2}. \newline

\begin{figure}[ht]
 \centering
\vspace*{10pt}
\begin{overpic}[width=0.485\textwidth]{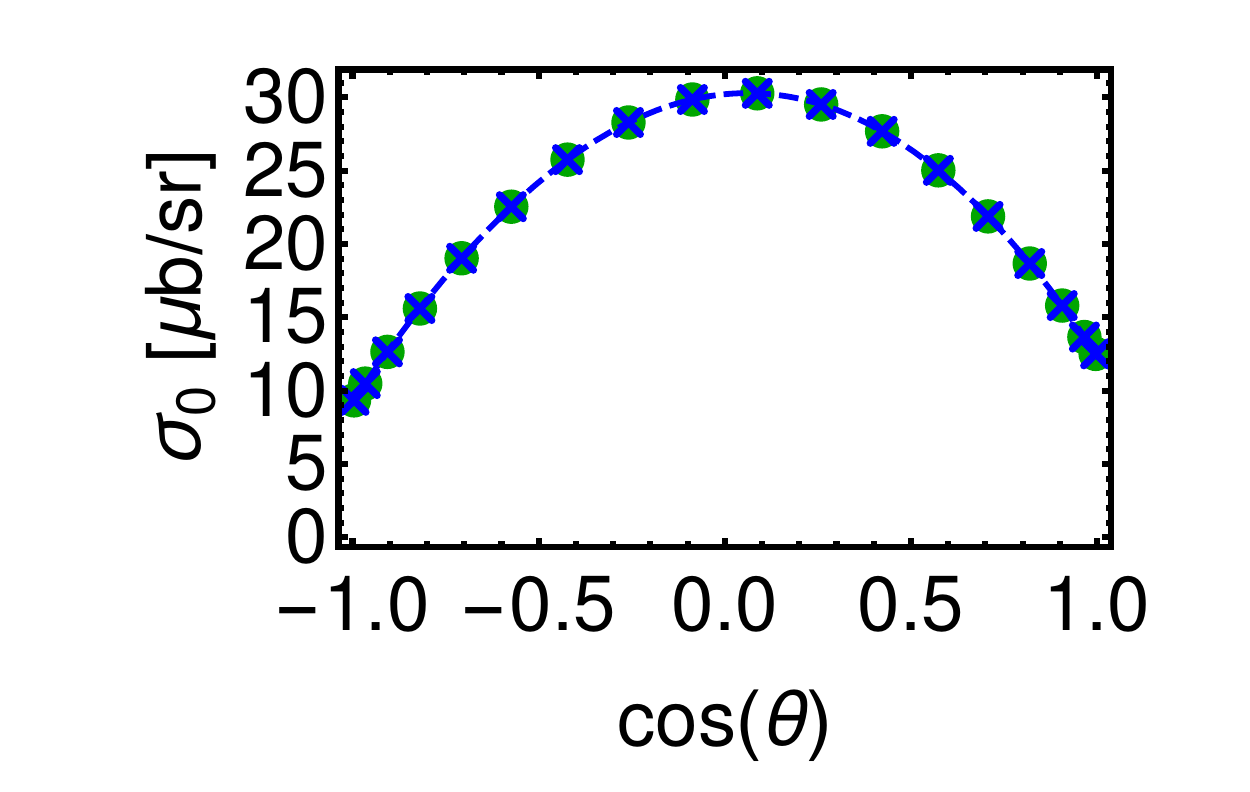}
 \put(85,65){\begin{Large}$E_{\gamma} = 330 \hspace*{2pt} \mathrm{MeV}$\end{Large}}
 \end{overpic}
\begin{overpic}[width=0.485\textwidth]{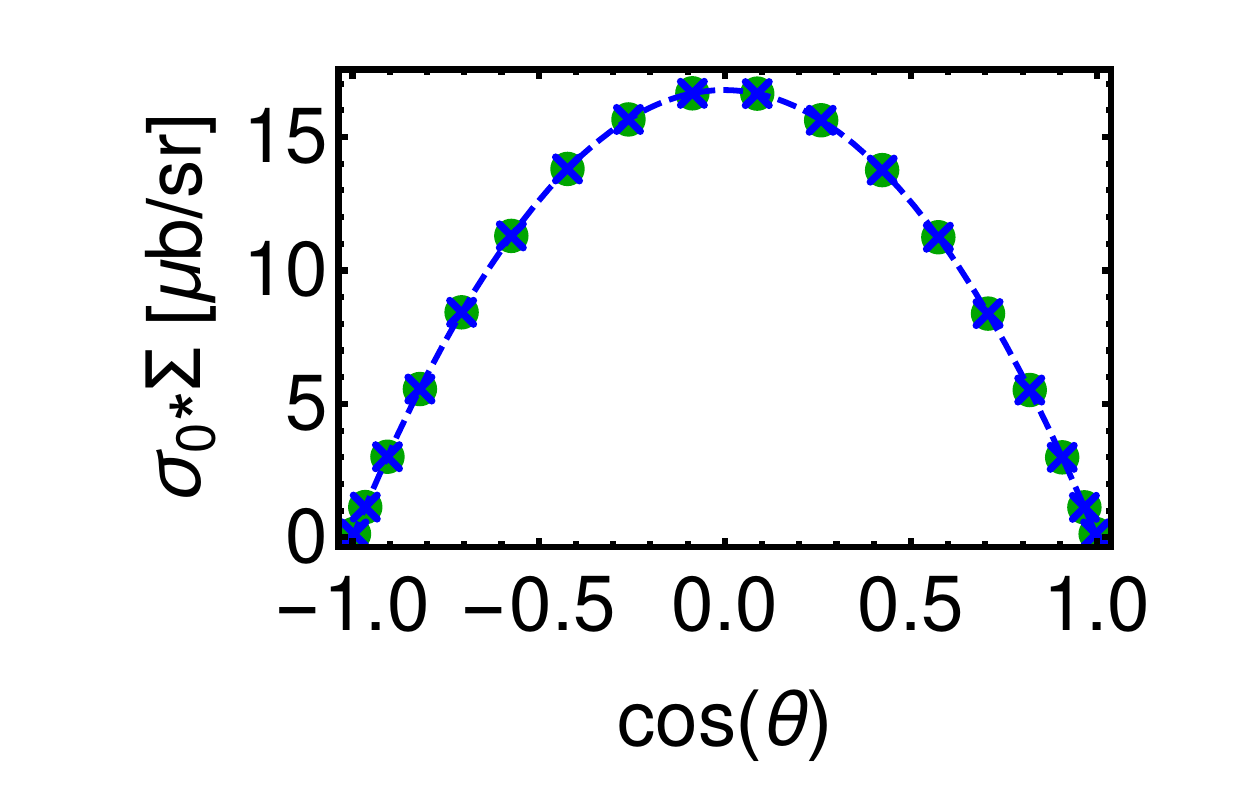}
 \end{overpic} \\
\begin{overpic}[width=0.485\textwidth]{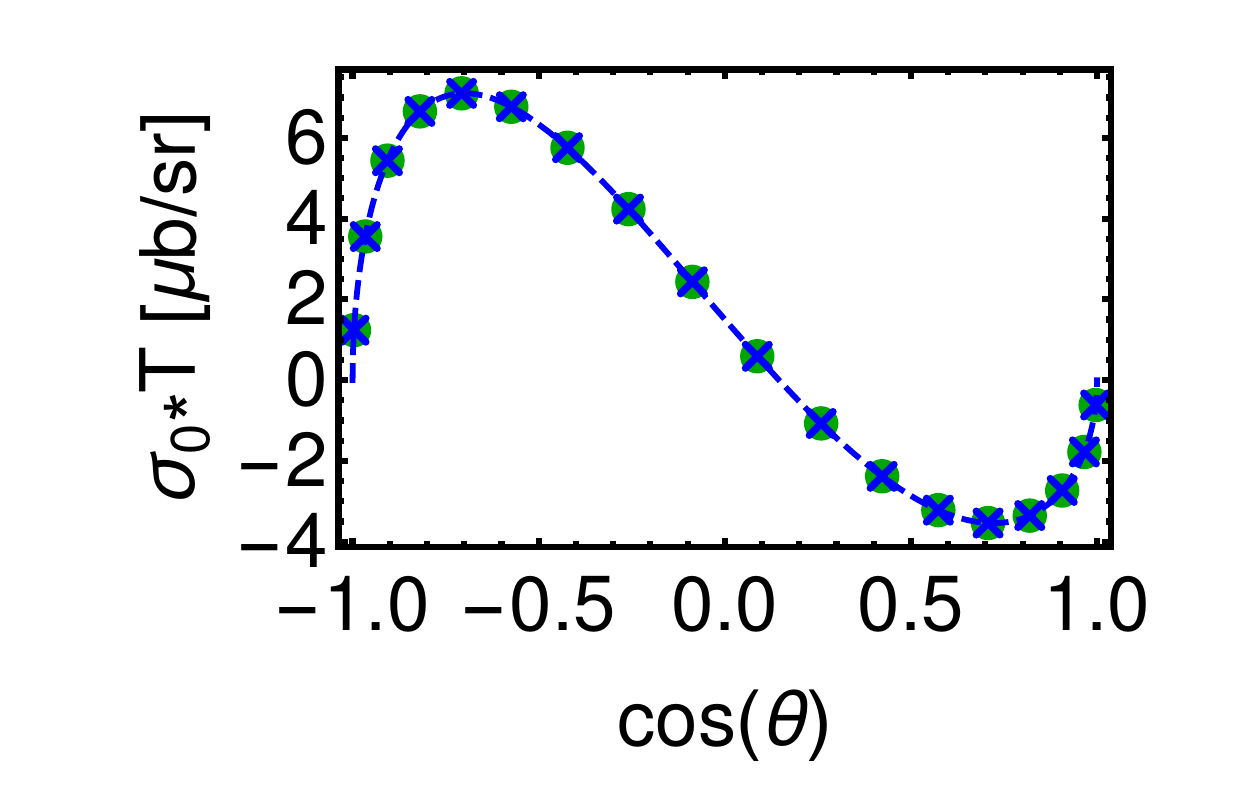}
 \end{overpic}
\begin{overpic}[width=0.485\textwidth]{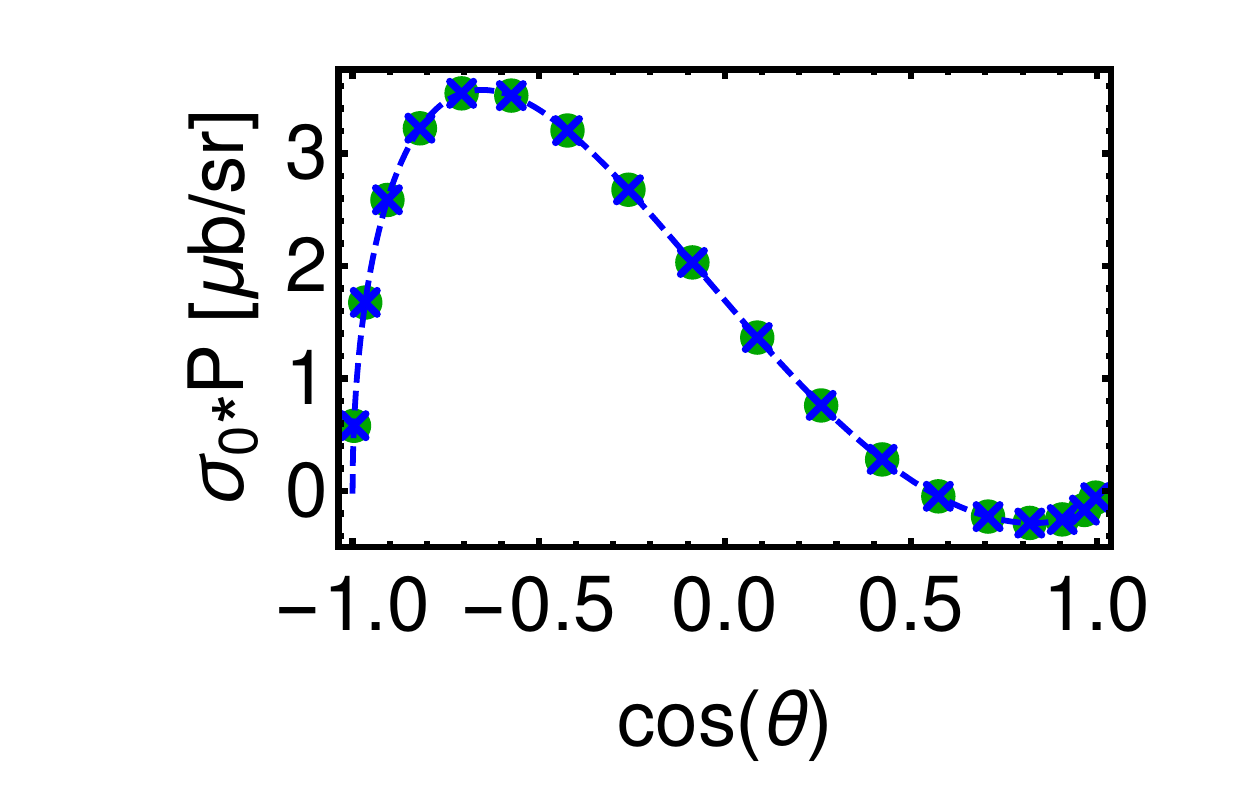}
 \end{overpic} \\
\begin{overpic}[width=0.485\textwidth]{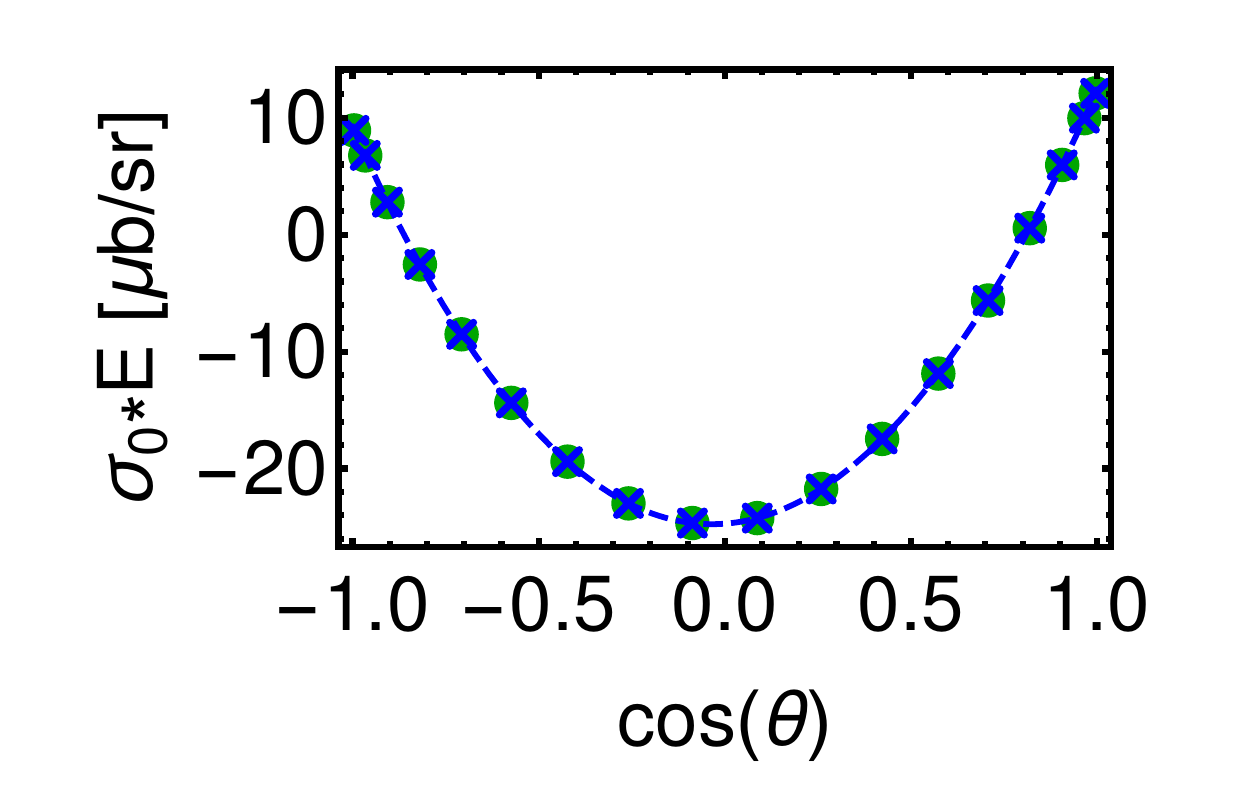}
 \end{overpic}
\begin{overpic}[width=0.485\textwidth]{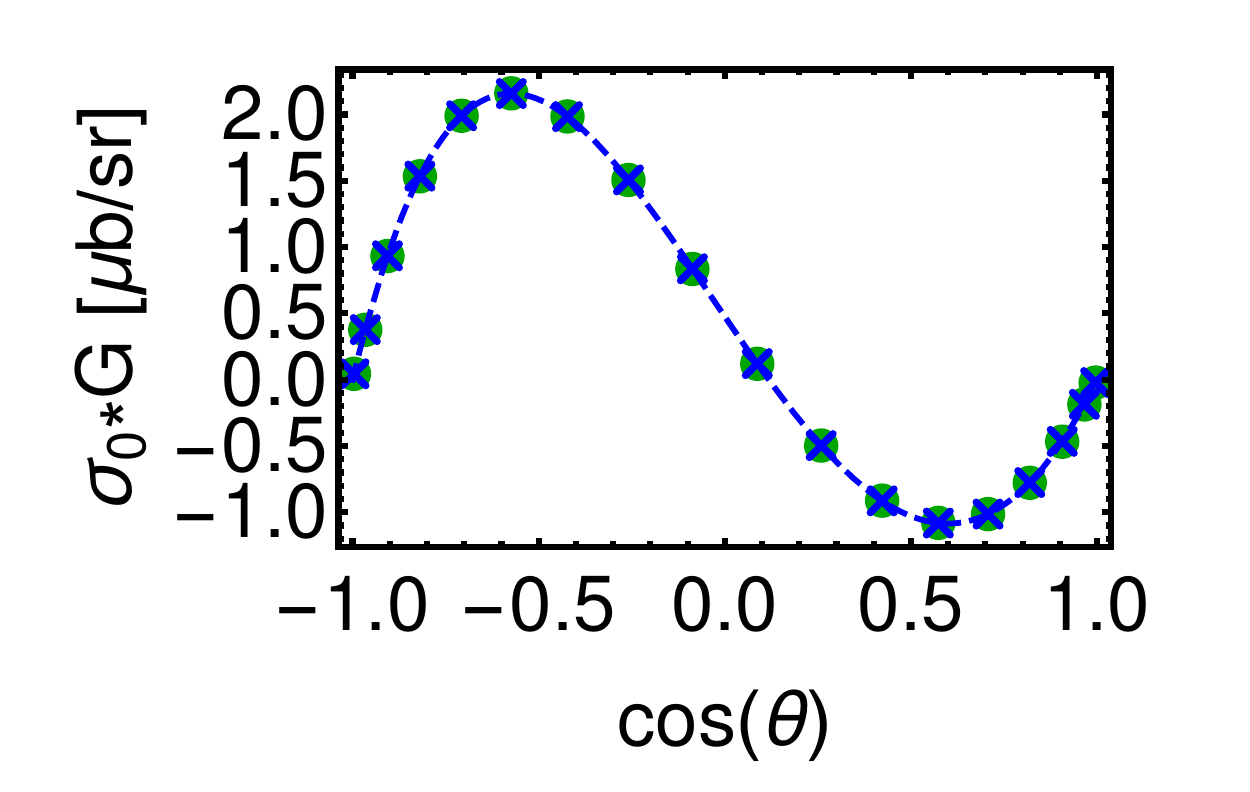}
 \end{overpic} \\
\begin{overpic}[width=0.485\textwidth]{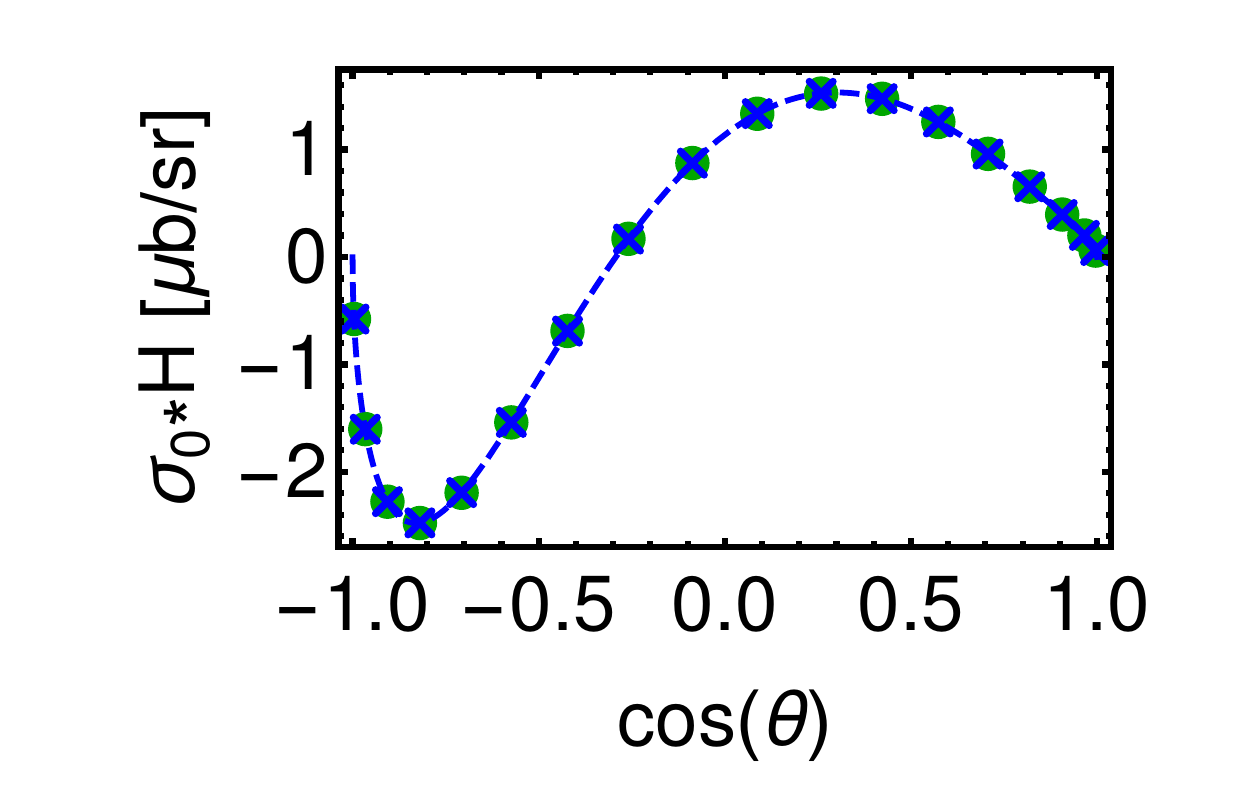}
 \end{overpic}
\begin{overpic}[width=0.485\textwidth]{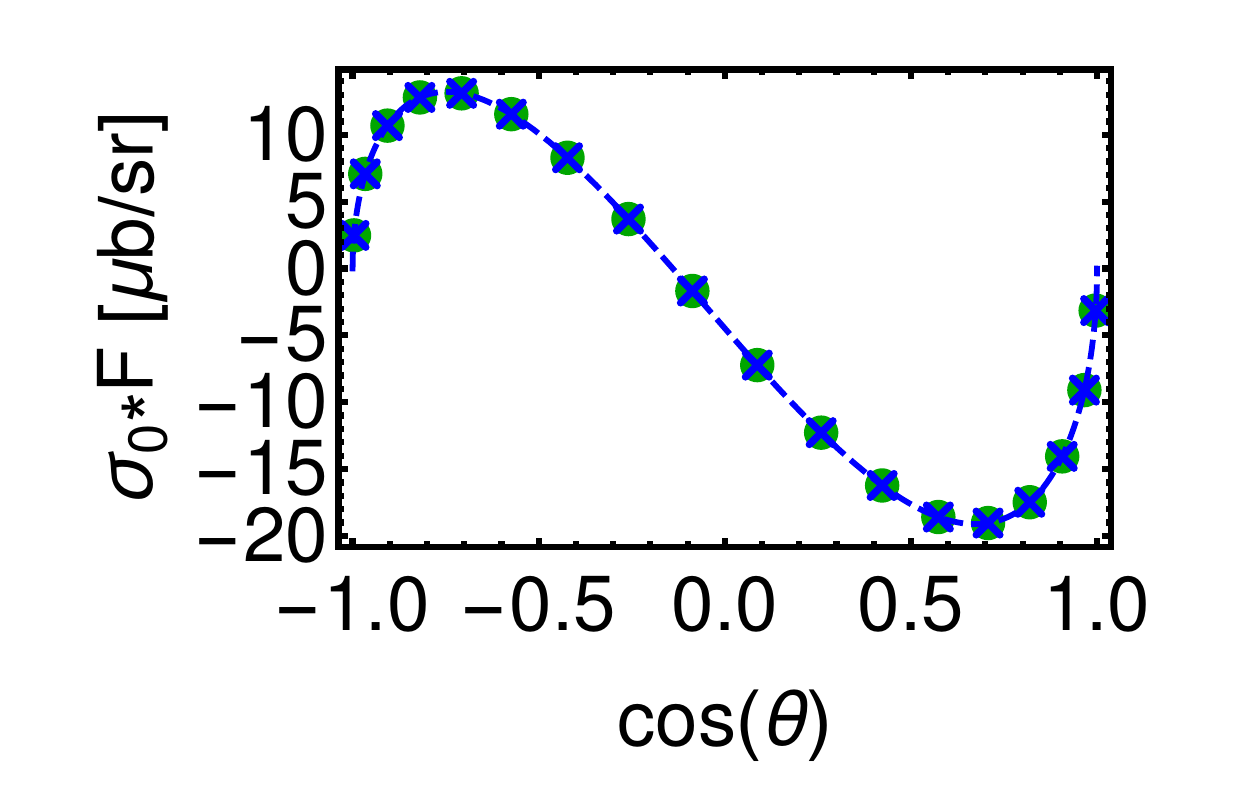}
 \end{overpic}
\caption[Angular distributions for MAID theory-data truncated at $\ell_{\mathrm{max}} = 4$ vs. the full model, for the energy $E_{\gamma} = 330 \hspace*{2pt} \mathrm{MeV}$.]{Truncated MAID2007 theory-data \cite{LotharPrivateComm,MAID2007} are shown here from the truncated dataset for $\ell_{\mathrm{max}} = 4$ (green dots), as well as from the full model (blue crosses). Depicted are the angular distributions of the profile functions belonging to the group $\mathcal{S}$ and $\mathcal{BT}$ observables at an example photon energy of $E_{\gamma} = 330 \hspace*{2pt} \mathrm{MeV}$. \newline A fit of an $\ell_{\mathrm{max}} = 4$ truncation to the \textit{full} model data is drawn as a blue dashed line, in order to illustrate that both sets of theory-data cannot be distinguished by eye, at least not in the $\Delta$-region. Modifications due to higher partial waves in the model become visible once Legendre coefficients are considered (see Figures \ref{fig:LmaxInfinityThDataFitLegCoeffsDCS1}, \ref{fig:LmaxInfinityThDataFitLegCoeffsDCS2} and \ref{fig:LmaxInfinityThDataFitLegCoeffsDCS3}).}
\label{fig:Lmax4VsInfinityThDataFitGroupSObservablesExampleEnergy}
\end{figure}

Before continuing with a description of the TPWA fits, the MAID data shall be examined in a bit more detail. It is interesting that when comparing the angular distributions of polarization observables from the MAID theory-data truncated at $\ell_{\mathrm{max}} = 4$ and the full model, no modifications due to higher partial waves are visible by eye. As an illustration, both solutions are plotted in Figure \ref{fig:Lmax4VsInfinityThDataFitGroupSObservablesExampleEnergy}. This is true especially in the $\Delta$-region considered here, since for such low energies the higher partial waves receive contributions mainly from Born terms which are, in fact, small (cf. results by MAID \cite{MAID2007,MAID}, SAID \cite{SAID} or Bonn-Gatchina \cite{BoGa}).

\clearpage

\begin{figure}[ht]
 \centering
\begin{overpic}[width=0.345\textwidth]{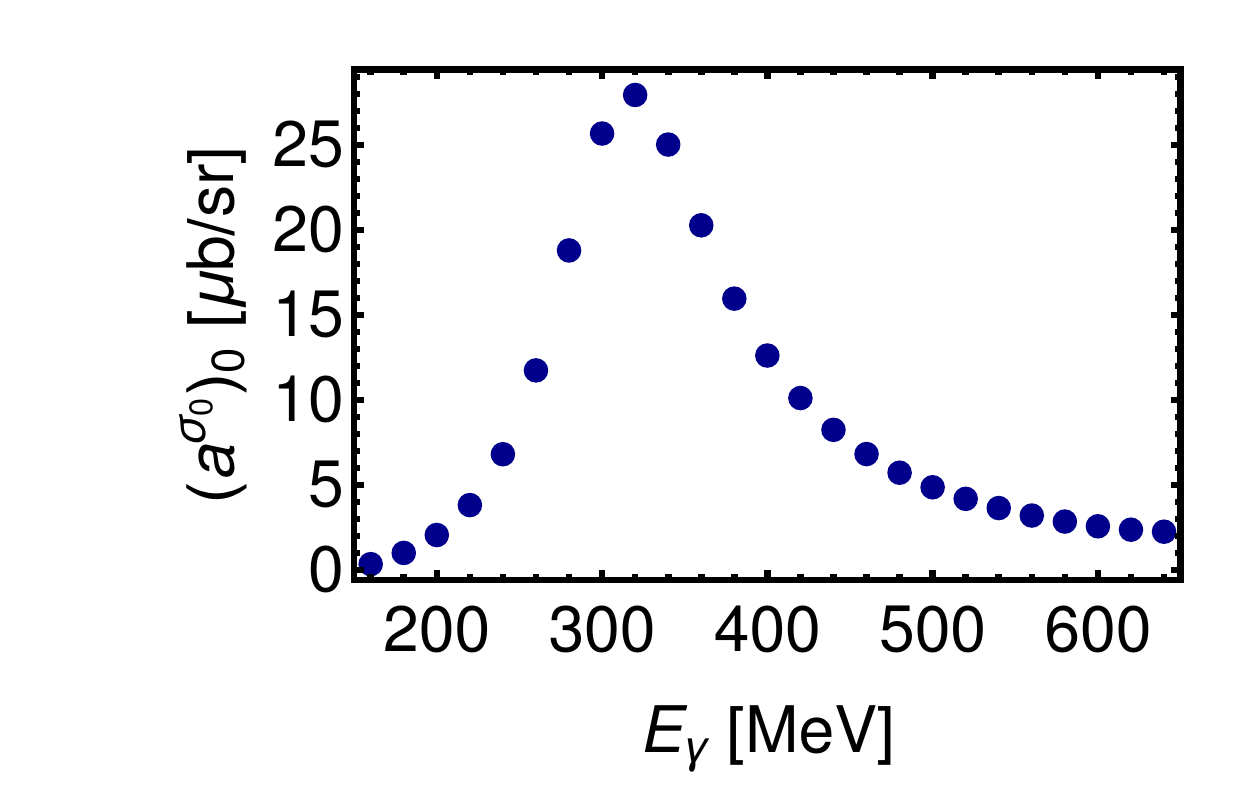}
 \end{overpic} \hspace*{-15pt}
\begin{overpic}[width=0.345\textwidth]{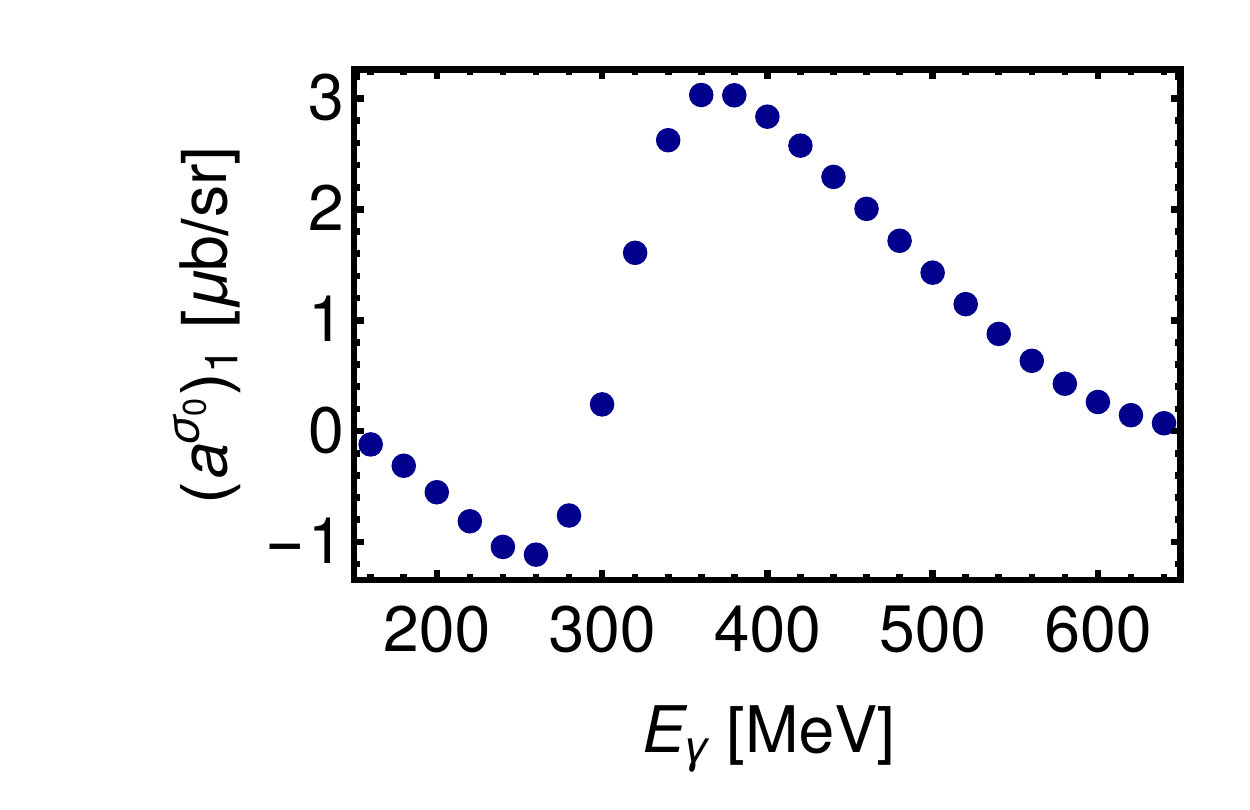}
 \end{overpic} \hspace*{-15pt}
\begin{overpic}[width=0.345\textwidth]{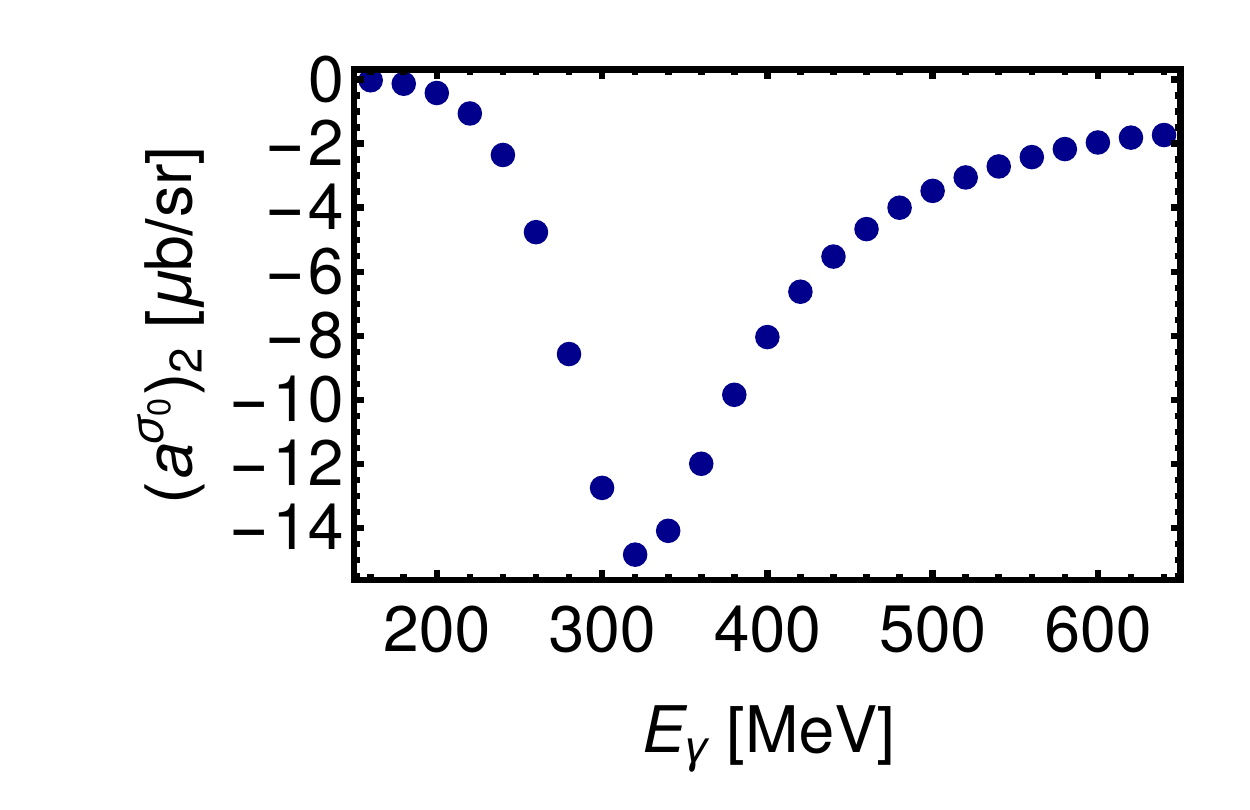}
 \end{overpic} \\
\begin{overpic}[width=0.345\textwidth]{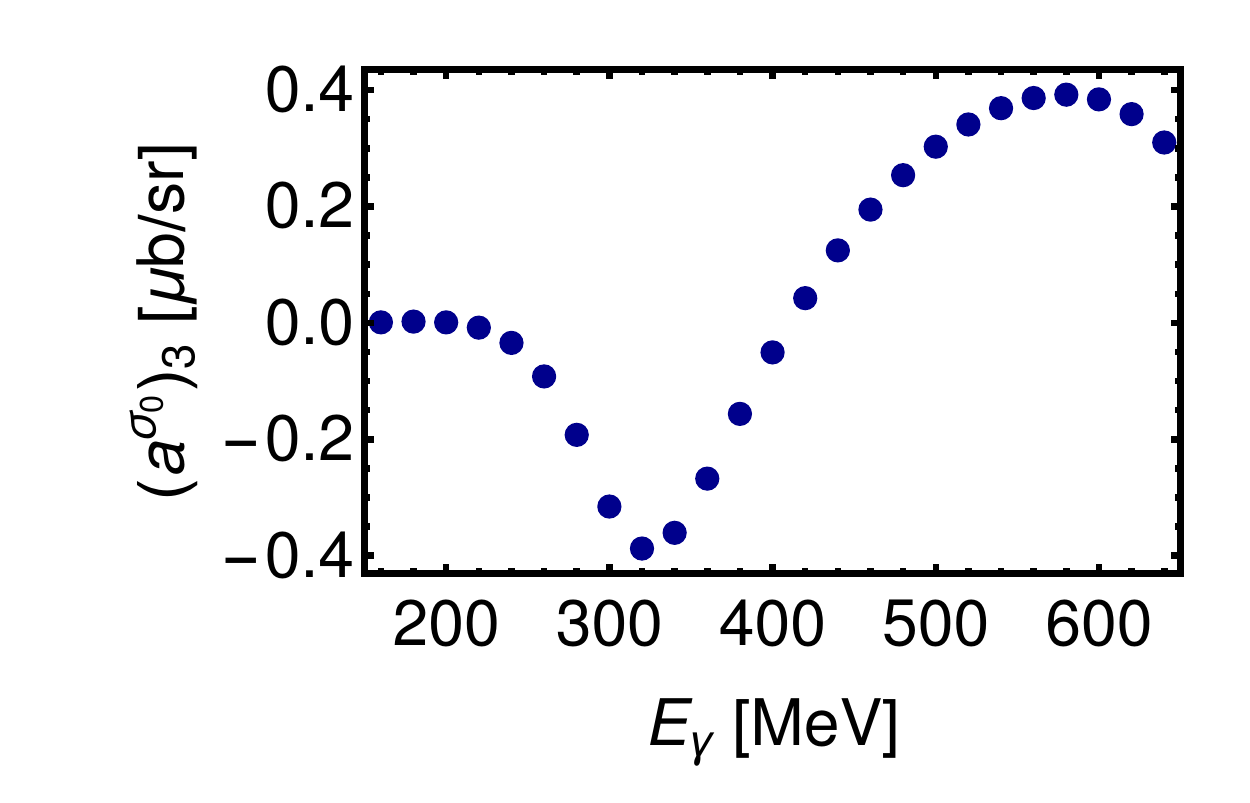}
 \end{overpic} \hspace*{-15pt}
 \begin{overpic}[width=0.345\textwidth]{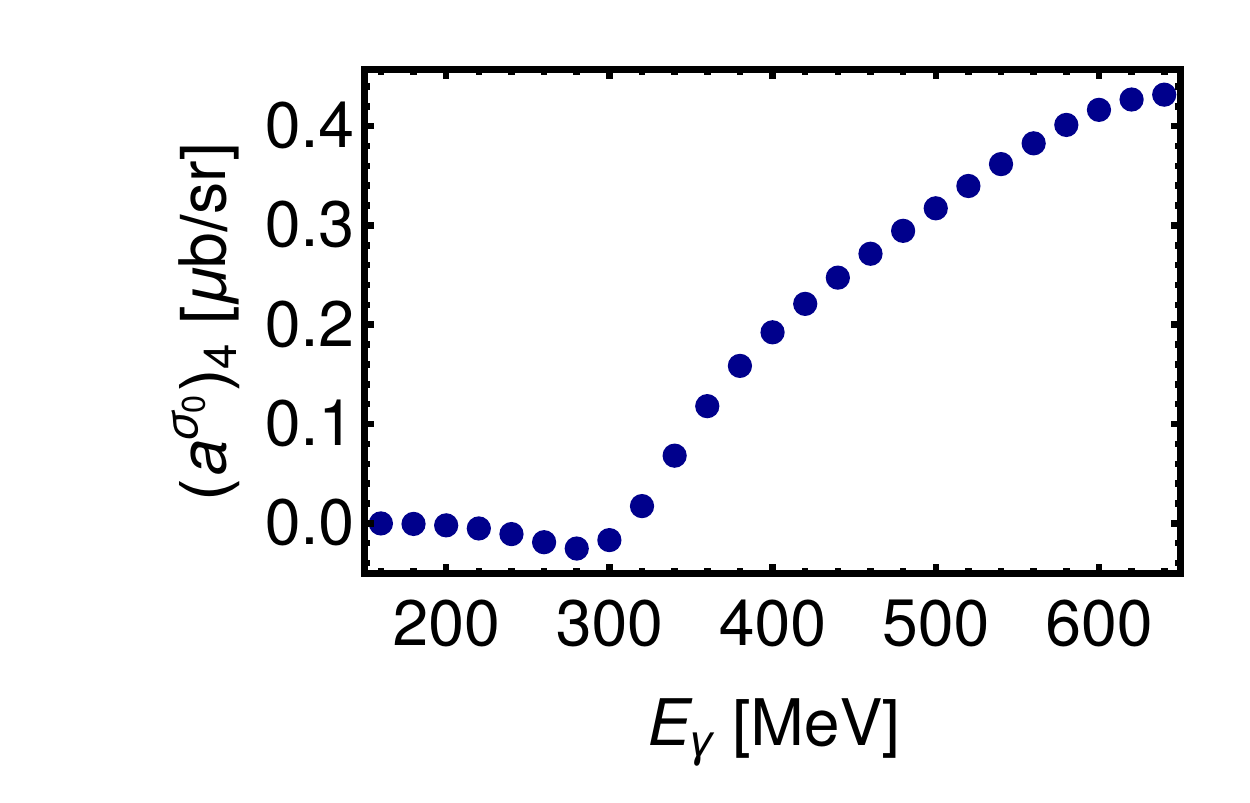}
 \end{overpic}
\caption[Legendre coefficients $\left(a^{\sigma_{0}}\right)_{0,\ldots,4}$ of the unpolarized cross secion $\sigma_{0}$ extracted from an $\ell_{\mathrm{max}} = 7$ TPWA to the full (non-truncated) MAID theory-data.]{Shown here are the Legendre coefficients $\left(a^{\sigma_{0}}\right)_{0,\ldots,4}$ of the unpolarized cross secion $\sigma_{0}$ extracted from an $\ell_{\mathrm{max}} = 7$ TPWA to the full MAID theory-data (blue dots).}
\label{fig:LmaxInfinityThDataFitLegCoeffsDCS1}
\end{figure}

\begin{figure}[ht]
 \centering
 \begin{overpic}[width=0.42\textwidth]{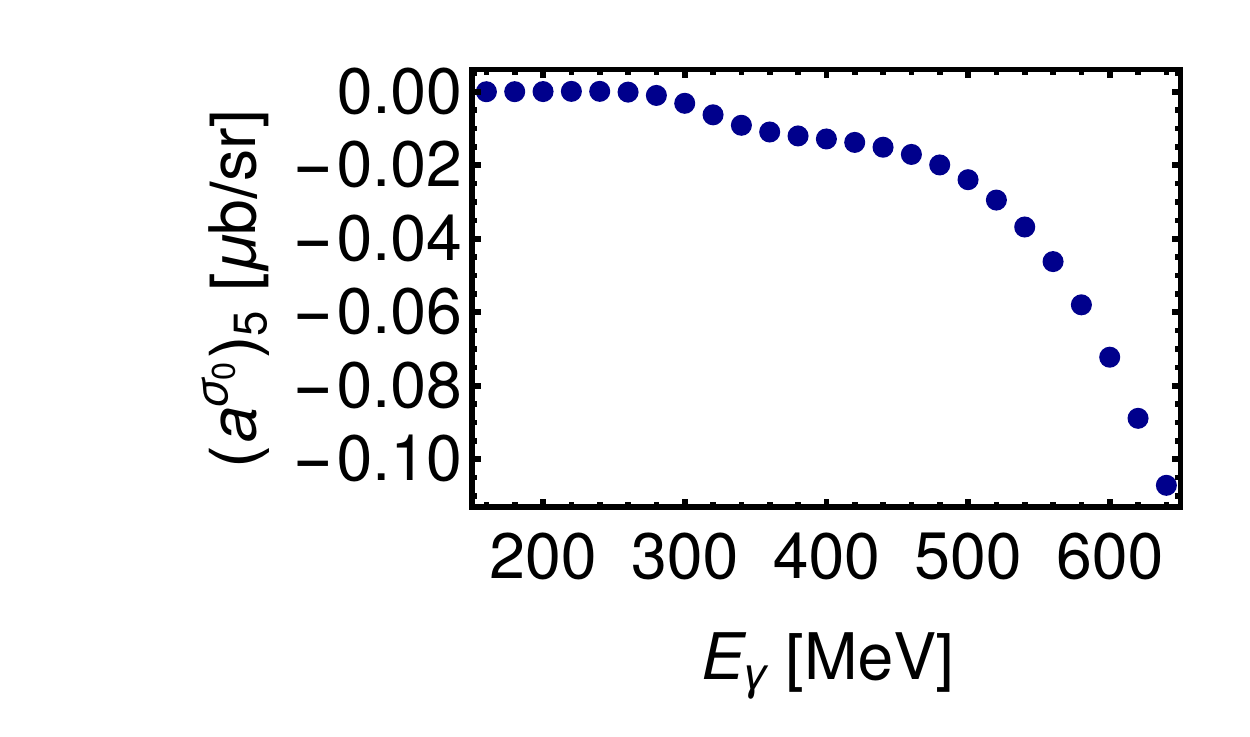}
 \end{overpic}
  \begin{overpic}[width=0.42\textwidth]{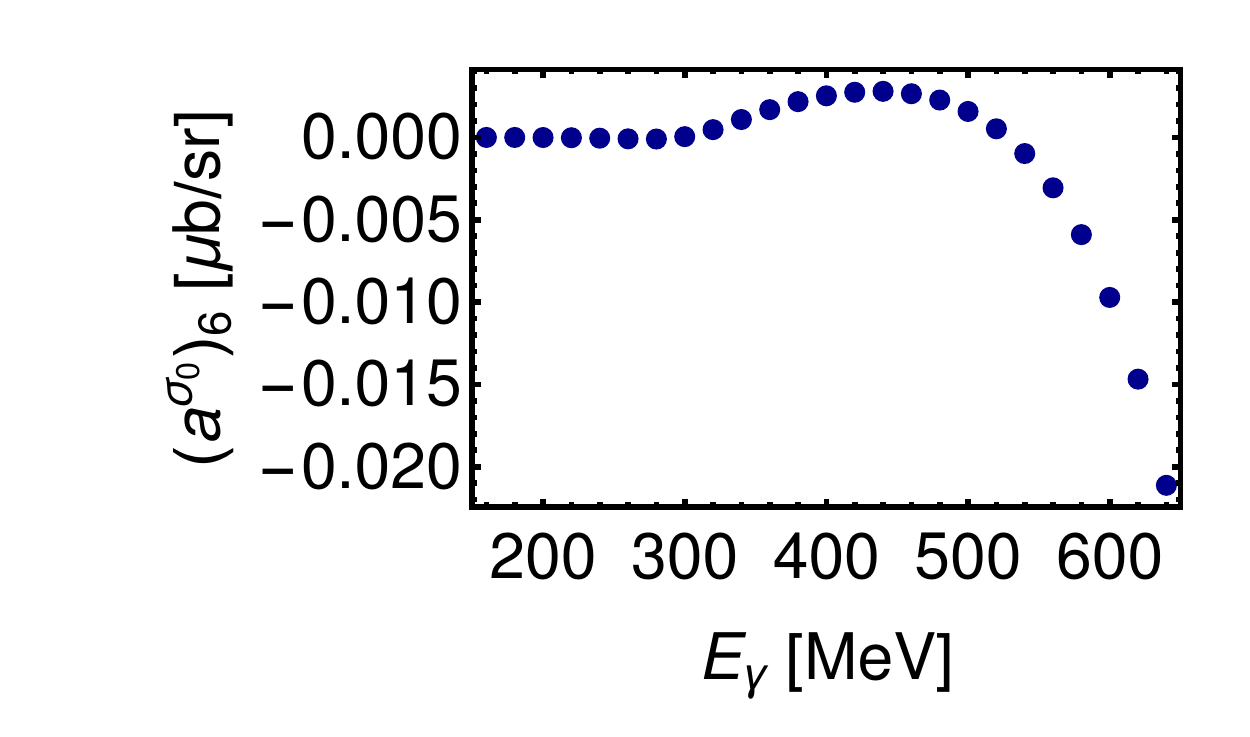}
 \end{overpic} \\
 \begin{overpic}[width=0.42\textwidth]{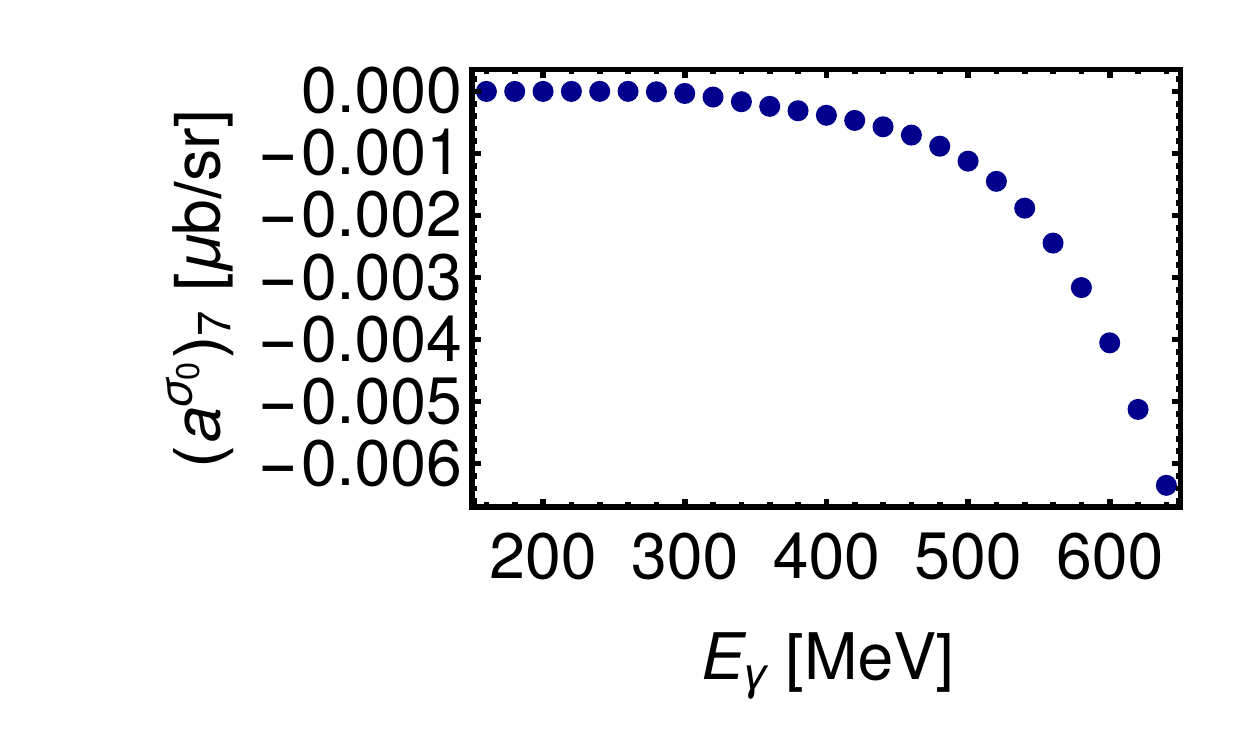}
 \end{overpic}
  \begin{overpic}[width=0.42\textwidth]{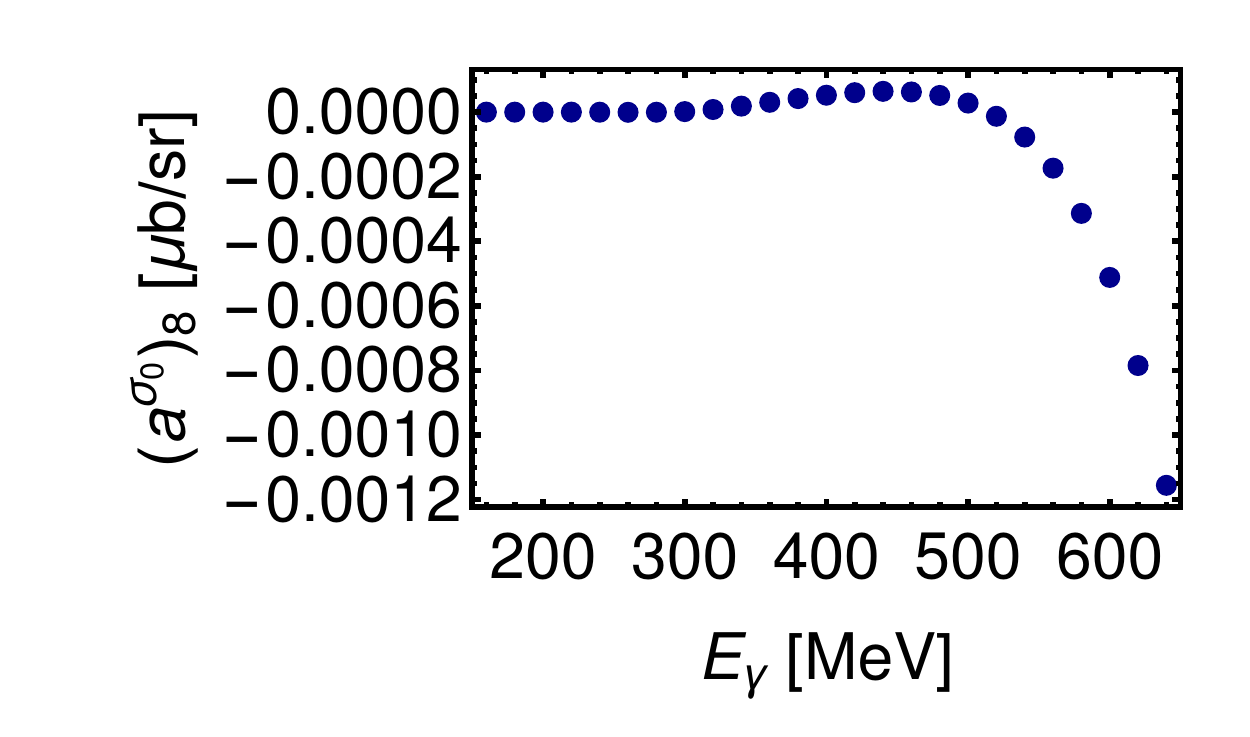}
 \end{overpic} \vspace*{-7pt}
\caption[Legendre coefficients $\left(a^{\sigma_{0}}\right)_{5,\ldots,8}$ of the unpolarized cross secion $\sigma_{0}$ extracted from an $\ell_{\mathrm{max}} = 7$ TPWA to the full (non-truncated) MAID theory-data.]{Shown here are the Legendre coefficients $\left(a^{\sigma_{0}}\right)_{5,\ldots,8}$ of the unpolarized cross secion $\sigma_{0}$ extracted from an $\ell_{\mathrm{max}} = 7$ TPWA to the full MAID theory-data (blue dots).}
\label{fig:LmaxInfinityThDataFitLegCoeffsDCS2}
\end{figure}

\begin{figure}[h]
 \centering
 \begin{overpic}[width=0.345\textwidth]{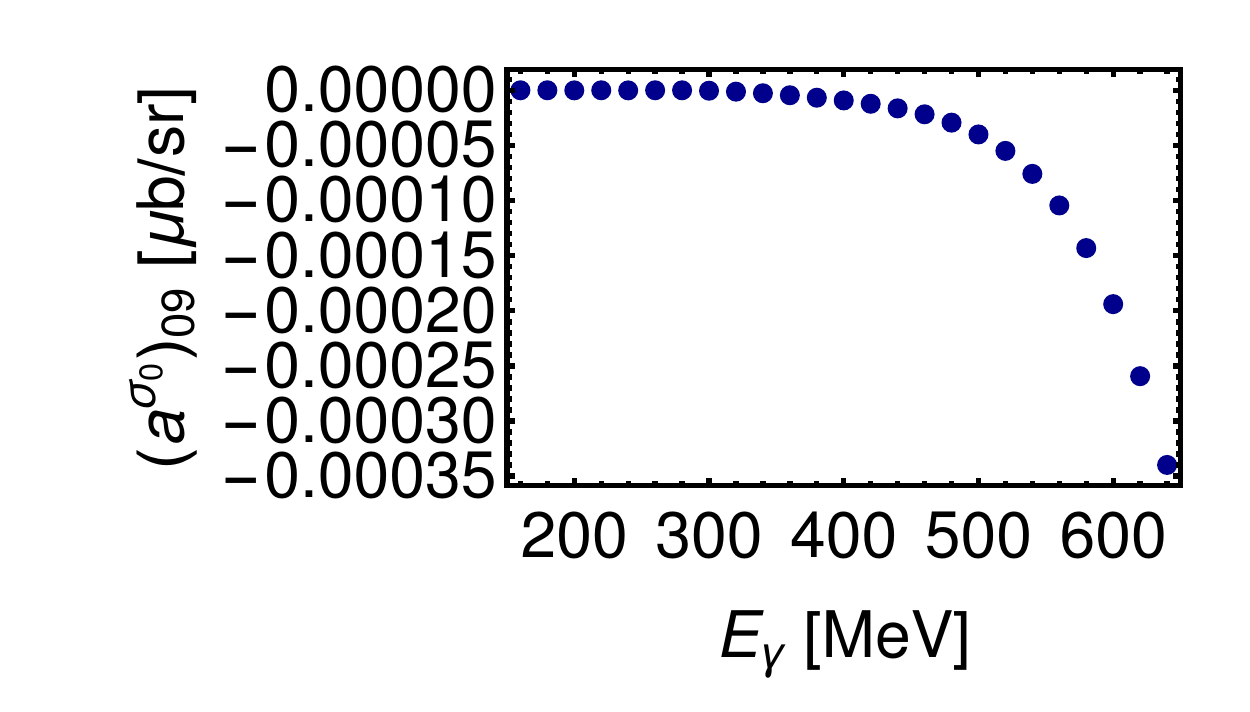}
 \end{overpic} \hspace*{-20pt}
  \begin{overpic}[width=0.345\textwidth]{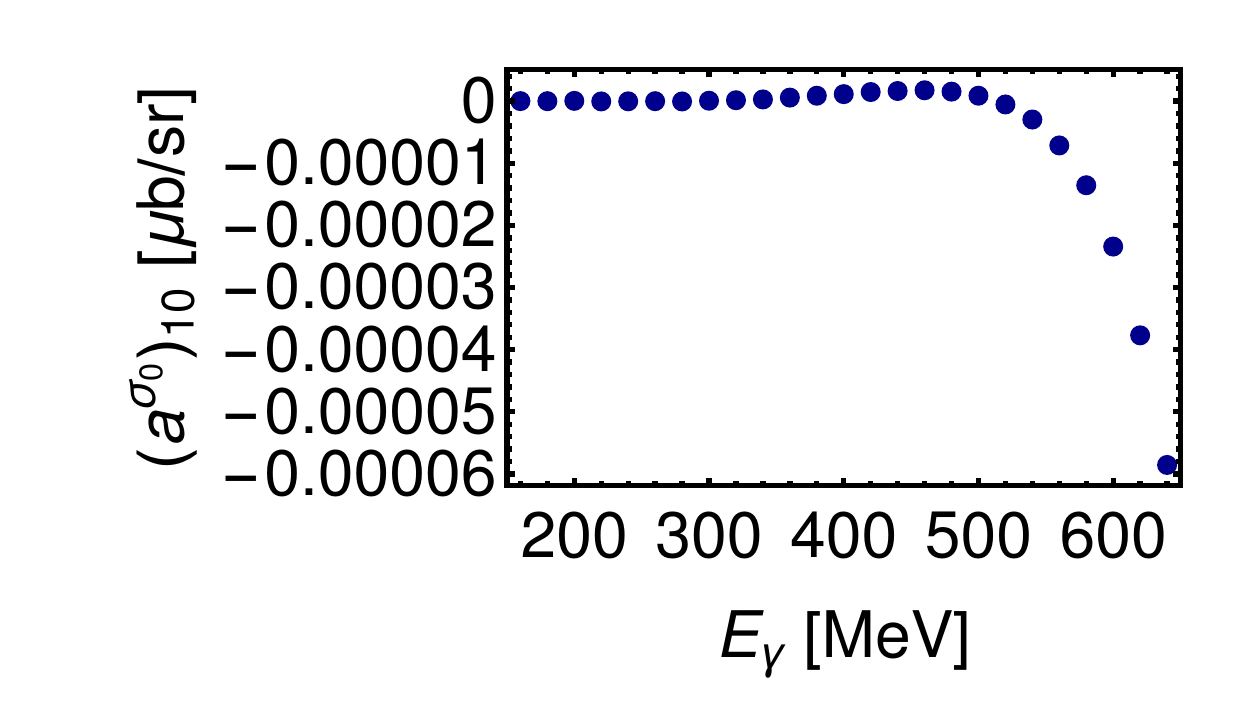}
 \end{overpic} \hspace*{-20pt}
  \begin{overpic}[width=0.345\textwidth]{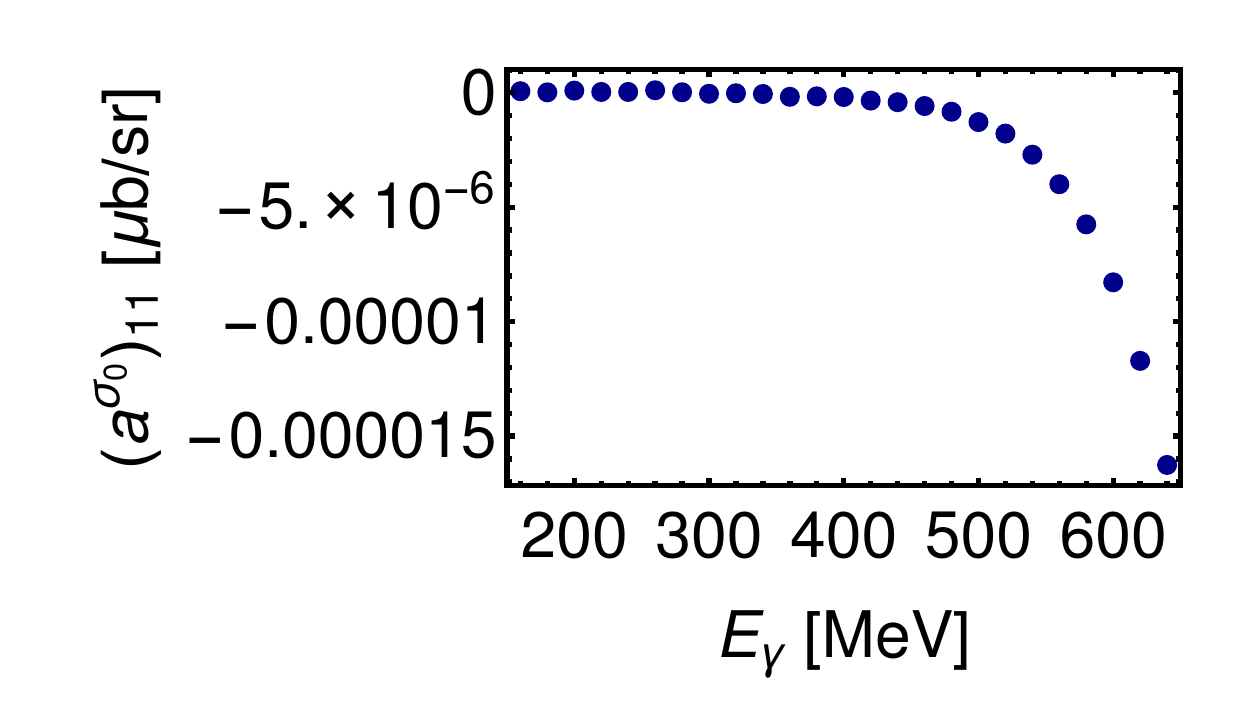}
 \end{overpic} \\
  \begin{overpic}[width=0.345\textwidth]{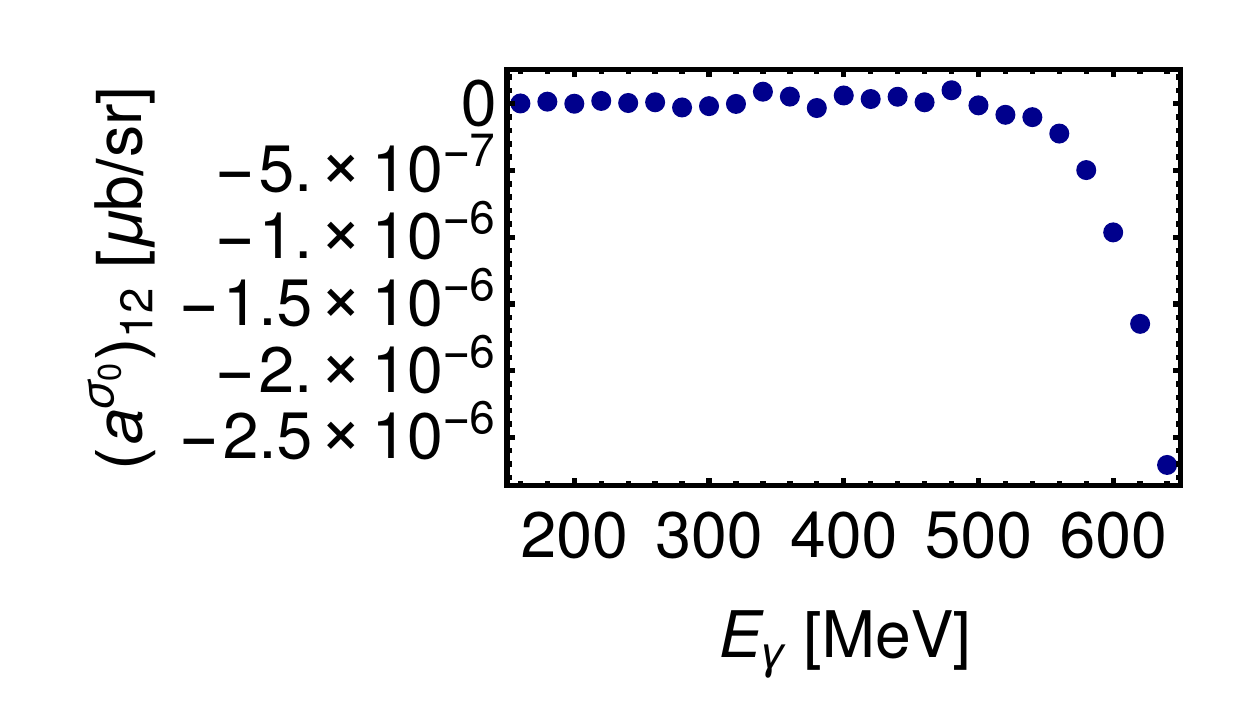}
 \end{overpic} \hspace*{-20pt} 
 \begin{overpic}[width=0.345\textwidth]{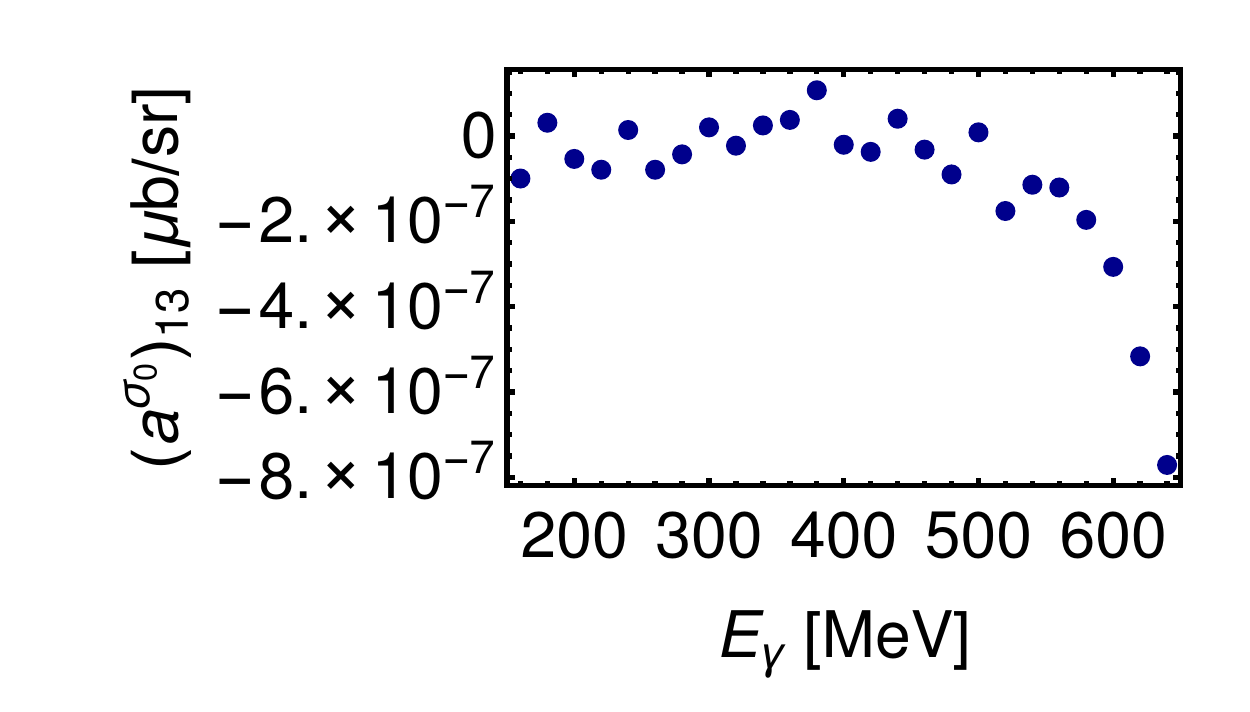}
 \end{overpic} \hspace*{-20pt}
  \begin{overpic}[width=0.345\textwidth]{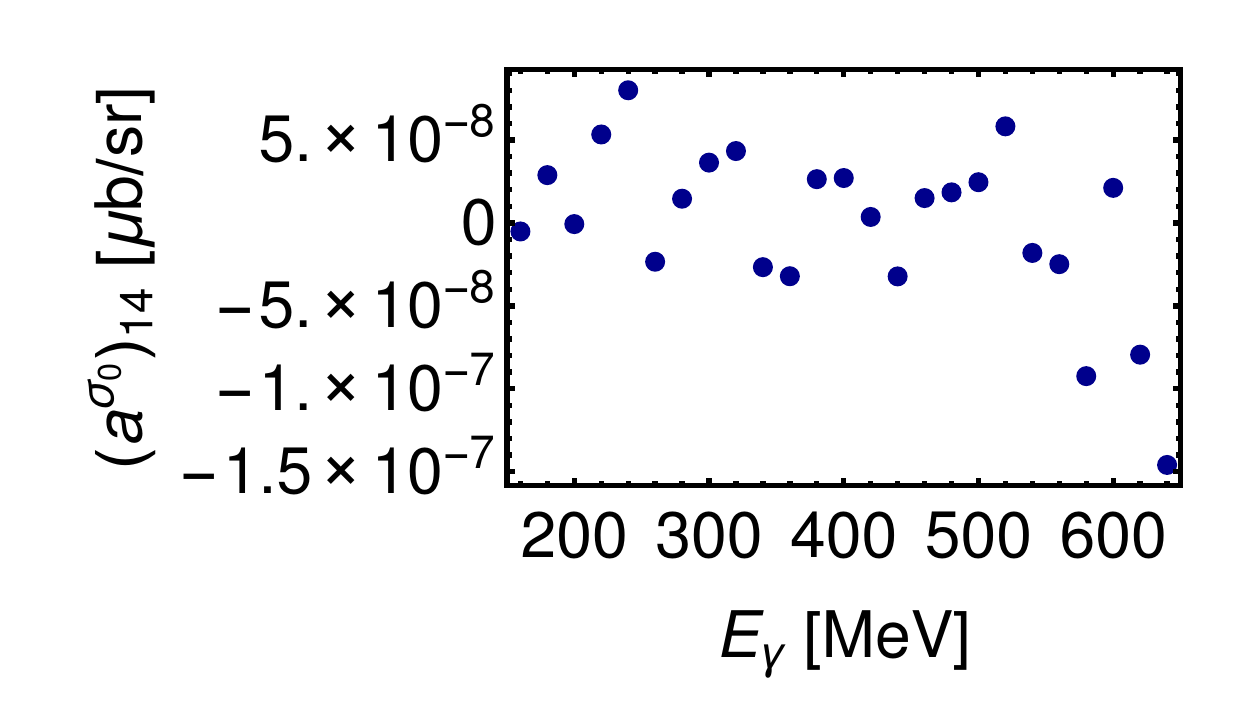}
 \end{overpic} \vspace*{-7pt}
\caption[Legendre coefficients $\left(a^{\sigma_{0}}\right)_{9,\ldots,14}$ of the unpolarized cross secion $\sigma_{0}$ extracted from an $\ell_{\mathrm{max}} = 7$ TPWA to the full (non-truncated) MAID theory-data.]{Shown here are the Legendre coefficients $\left(a^{\sigma_{0}}\right)_{9,\ldots,14}$ of the unpolarized cross secion $\sigma_{0}$ extracted from an $\ell_{\mathrm{max}} = 7$ TPWA to the full MAID theory-data (blue dots). For MAID theory-data truncated at $\ell_{\mathrm{max}} = 4$, these coefficients would be consistent with zero }
\label{fig:LmaxInfinityThDataFitLegCoeffsDCS3}
\end{figure}

The situation changes once Legendre coefficients are investigated. Here, modifications due to higher partial waves become visible even in the low energy region. Coefficients for the unpolarized cross section $\sigma_{0}$ are shown in Figures \ref{fig:LmaxInfinityThDataFitLegCoeffsDCS1}, \ref{fig:LmaxInfinityThDataFitLegCoeffsDCS2} and \ref{fig:LmaxInfinityThDataFitLegCoeffsDCS3}. This observable was chosen, since it has the highest non-vanishing Legendre coefficients within the $\Delta$-region, from all quantities belonging to group $\mathcal{S}$ and $\mathcal{BT}$. \newline
Here, the order $\ell_{\mathrm{max}} = 7$ has been fitted to obtain the results shown in the figures and it is seen that the highest coefficient $\left(a^{\sigma_{0}}\right)_{14}$ does not show a definite trend in energy and is consistent with zero (Figure \ref{fig:LmaxInfinityThDataFitLegCoeffsDCS3}). The next lower coefficient $\left(a^{\sigma_{0}}\right)_{13}$ already shows some tiny structure and strength for the highest energies. When descending through all Legendre coefficients from there, they all are seen to have a modulus that is non-zero and rising towards the higher energies. In case the cross section from the MAID data truncated at $\ell_{\mathrm{max}} = 4$ would have been fitted, all coefficients starting from $\left(a^{\sigma_{0}}\right)_{9}$ would have vanished. This illustrated the influence of the higher partial waves. Also, tiny modifications of the lower Legendre coefficients compared to the $\ell_{\mathrm{max}} = 4$ theory-data may not be excluded. \newline

It is very important to note that even tiny modifications due to higher partial waves such as those seen above can endanger the solvability of a TPWA even for perfect theory-data. It can be anticipated directly that, once a TPWA is fitted to data that contain all partial waves up to infinity, the bilinear equation systems solved in the model analysis, i.e.
\begin{equation}
\left(a_{L}\right)_{k}^{\check{\Omega}^{\alpha}} = \left< \mathcal{M}_{\ell} \right| \left( \mathcal{C}_{L}\right)_{k}^{\check{\Omega}^{\alpha}} \left| \mathcal{M}_{\ell} \right> \mathrm{,} \label{eq:BilinearEqSystemSec4Dot43}
\end{equation}
are generally not compatible any more (see reference \cite{Grushin} and appendix \ref{subsec:AccidentalAmbProofsIII}). This means one cannot expect the existence of an exact solution any more which, in a numerical context, would manifest itself as a minimum in $\Phi_{\mathcal{M}}$ at values around $10^{-16} (\mu b / sr)^{2}$, just as it did for the cases studies in the previous two sections. Then, one can anticipate that for cases where the analysis of the truncated theory-data was already susceptible to ambiguities, such as in those cases discussed in section \ref{subsec:TheoryDataFitsLmax2}, the readily abundant ambiguities could destroy the possibility to uniquely solve for the correct MAID multipoles. \newpage

\begin{figure}[ht]
 \centering
\begin{overpic}[width=0.475\textwidth]{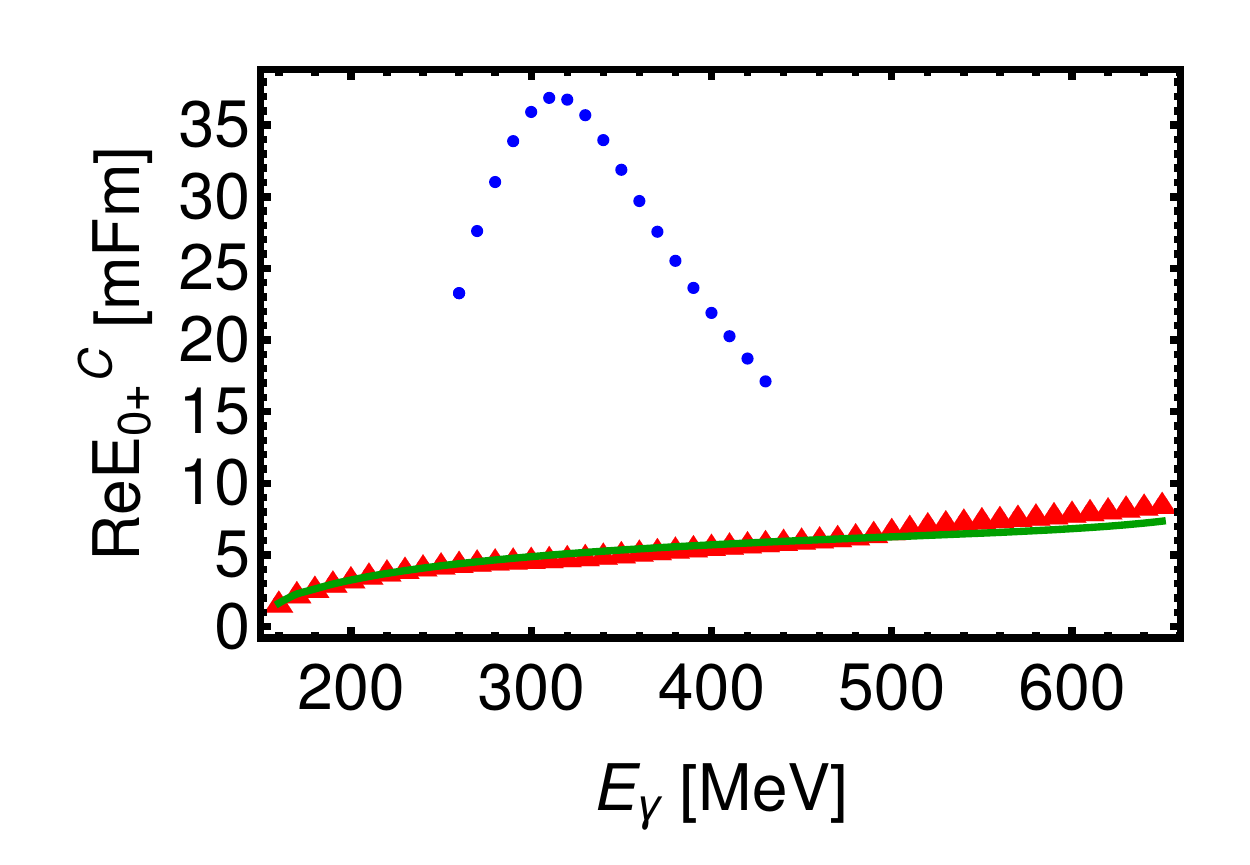}
 \end{overpic} \\
 \vspace*{-5pt}
\begin{overpic}[width=0.475\textwidth]{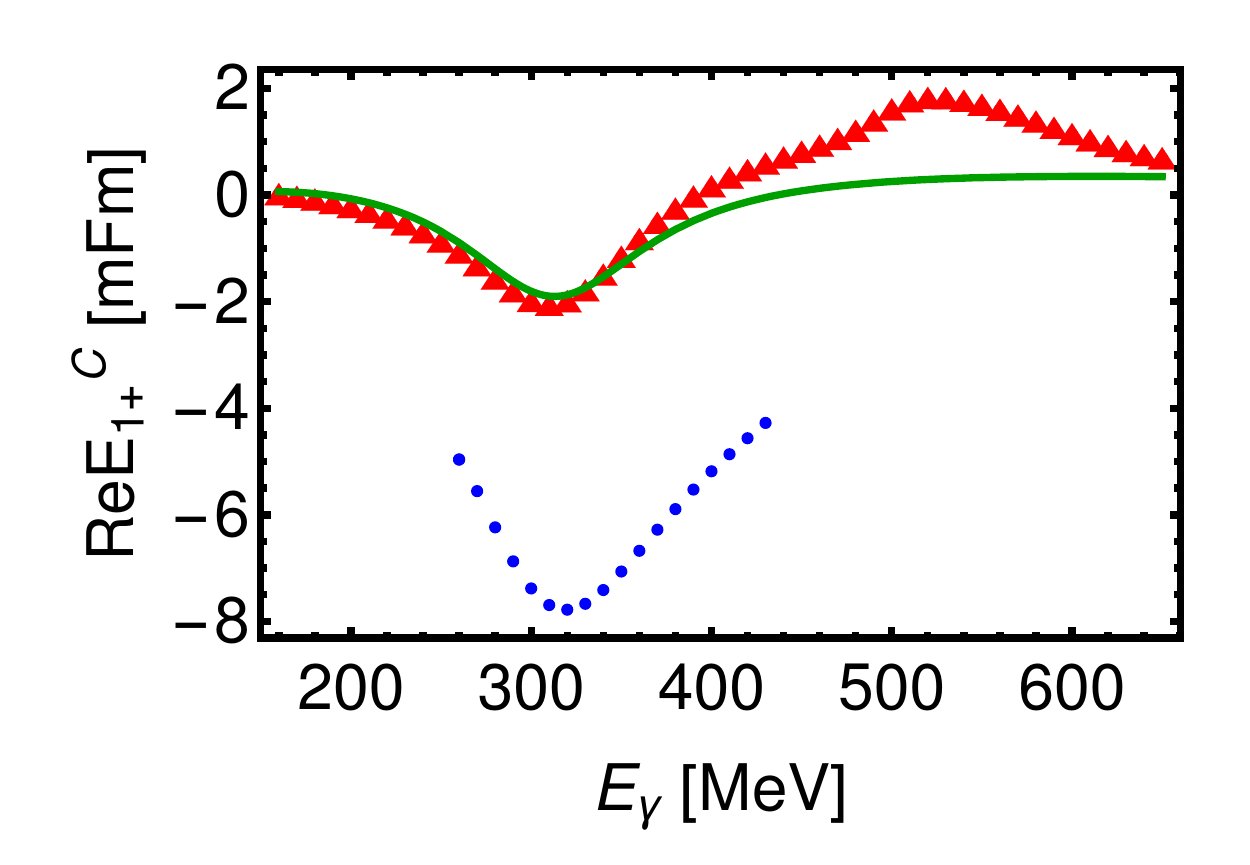}
 \end{overpic} \hspace*{5pt}
\begin{overpic}[width=0.475\textwidth]{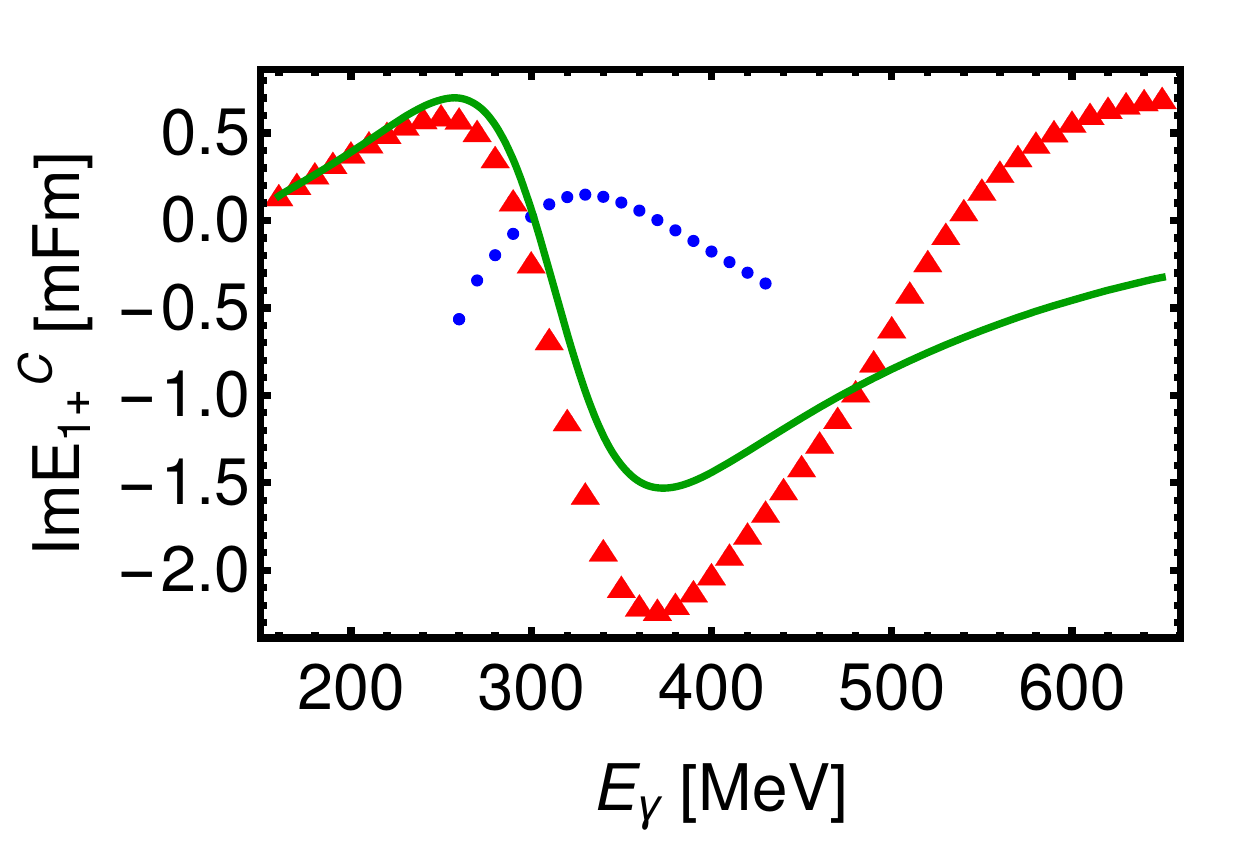}
 \end{overpic} \\
\begin{overpic}[width=0.475\textwidth]{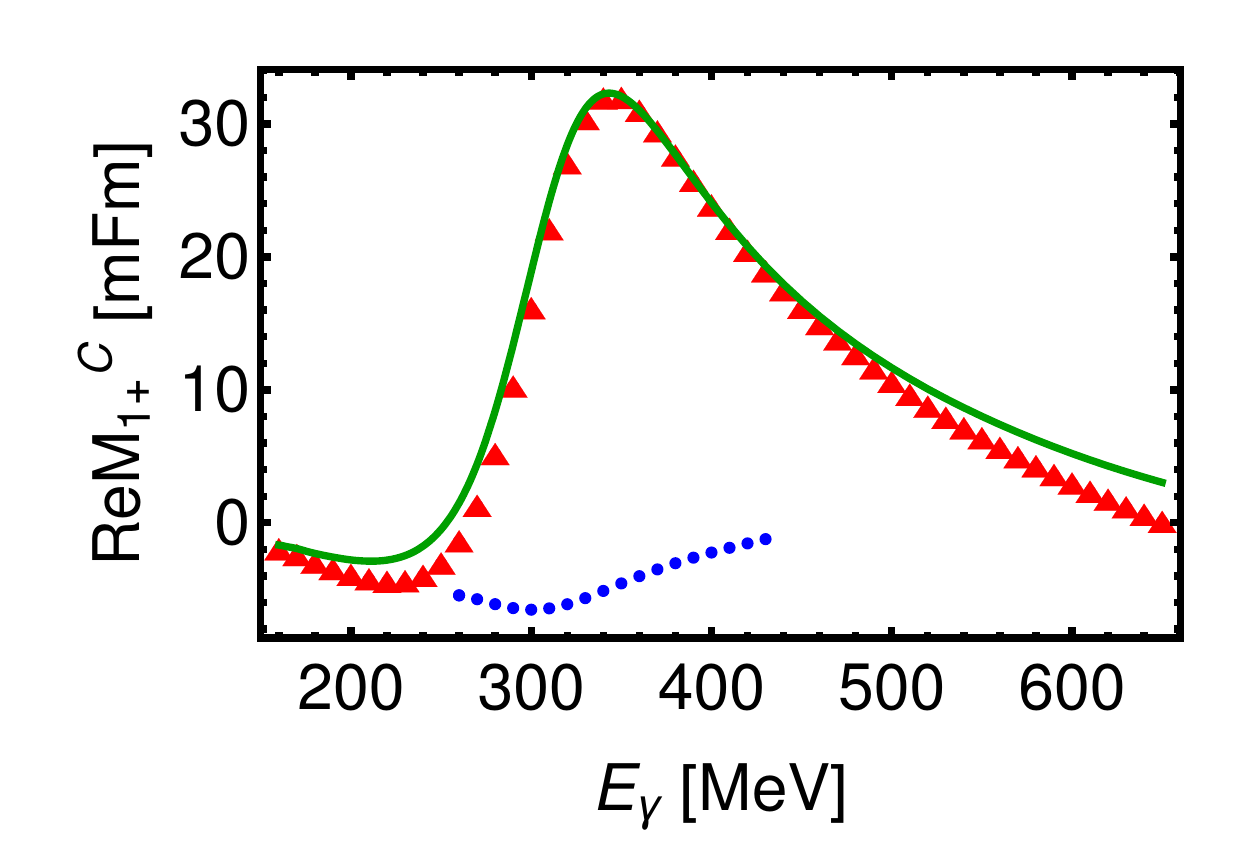}
 \end{overpic} \hspace*{5pt}
\begin{overpic}[width=0.475\textwidth]{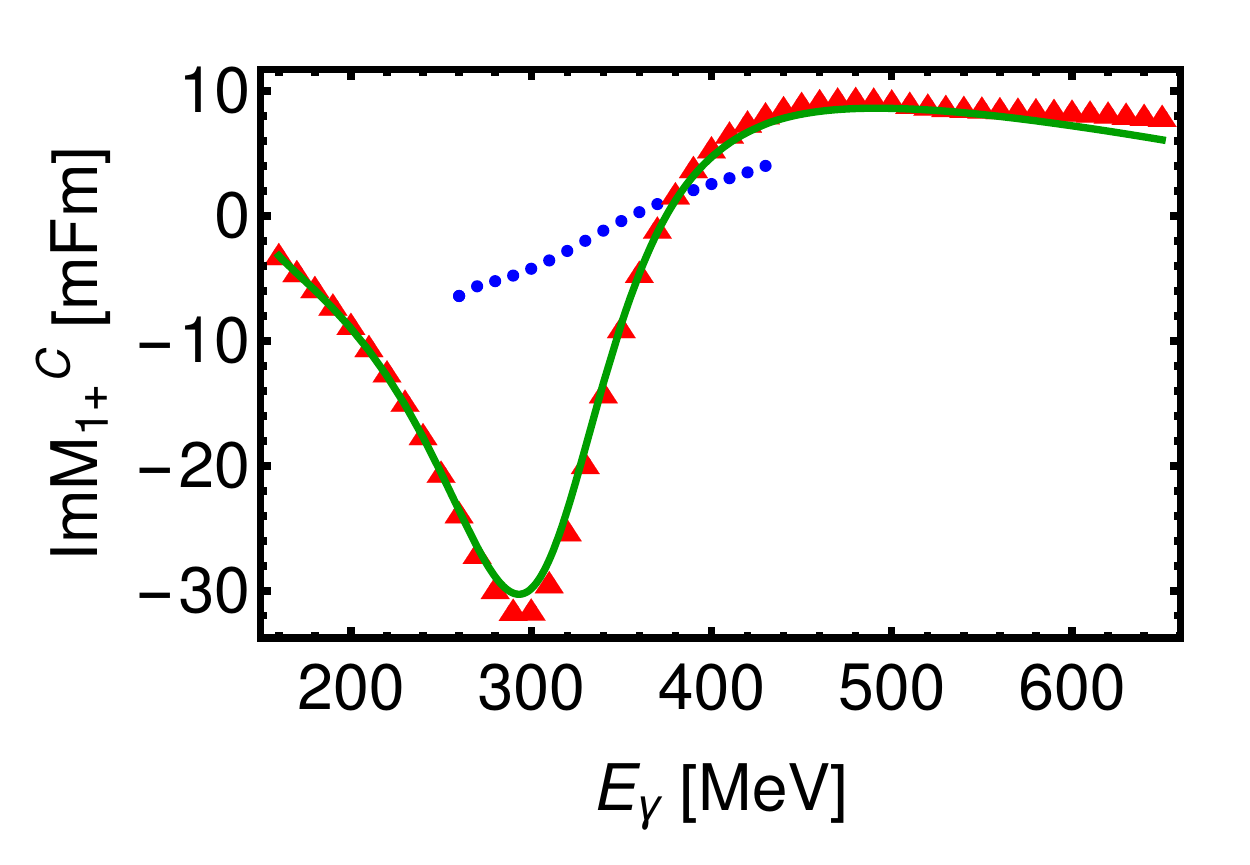}
 \end{overpic} \\
\begin{overpic}[width=0.475\textwidth]{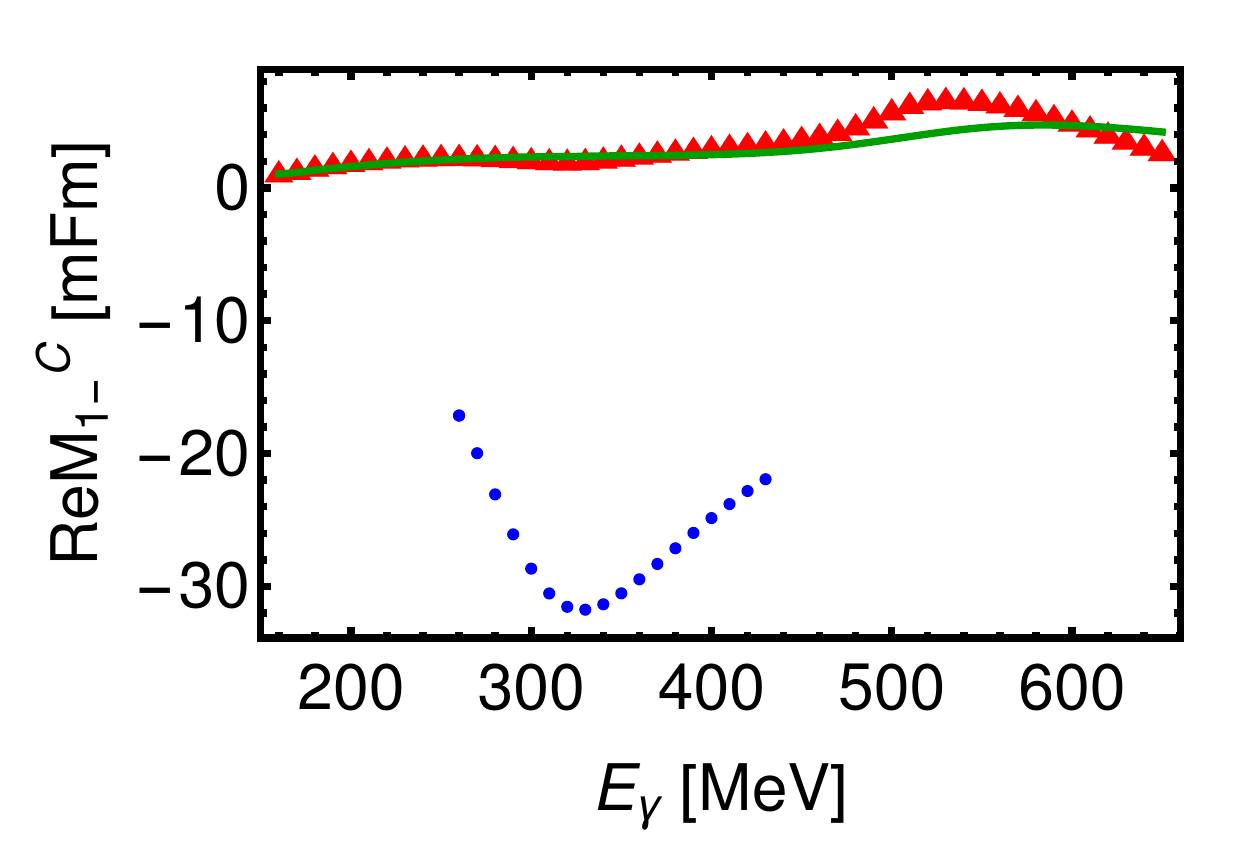}
 \end{overpic} \hspace*{5pt}
\begin{overpic}[width=0.475\textwidth]{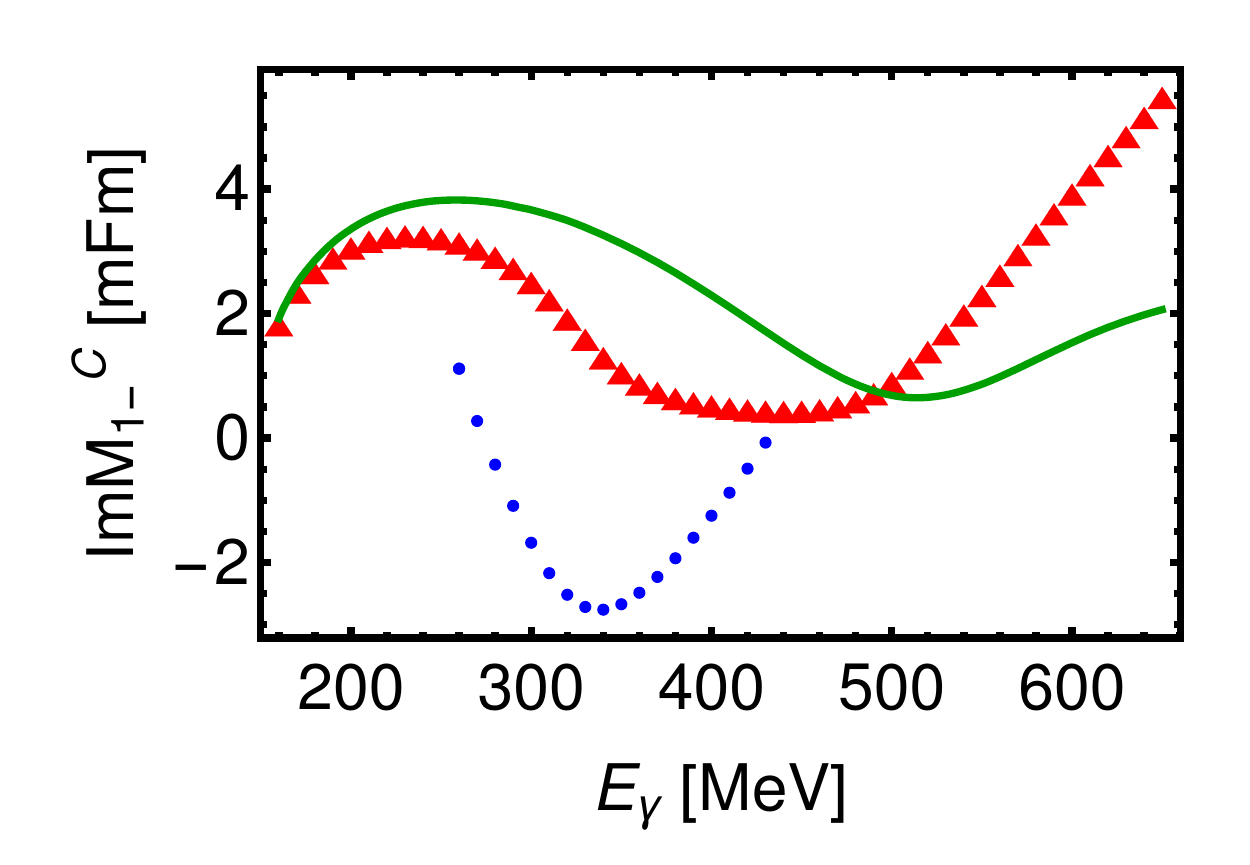}
 \end{overpic}
\vspace*{-5pt}
\caption[Multipole-results for a TPWA-fit of an $S$- and $P$-wave truncation to MAID theory-data from the full (non-truncated) model. The set of observables $\left\{ \sigma_{0}, \check{\Sigma}, \check{T}, \check{P}, \check{F} \right\}$ was analyzed.]{The figures depict results of a TPWA-fit of an $S$- and $P$-wave truncation to MAID theory-data \cite{LotharPrivateComm,MAID2007} from the full, non-truncated, model. Real and imaginary parts of all phase-constrained multipoles ($E_{0+} \equiv \mathrm{Re} \left[ E_{0+} \right] \geq 0$) for all partial waves from $E_{0+}$ up to $M_{1-}$ are shown. \newline The set of observables $\left\{ \sigma_{0}, \check{\Sigma}, \check{T}, \check{P}, \check{F} \right\}$ was used to obtain the multipole solutions from a pool of $N_{MC} = 1500$ randomly chosen initial configurations. The whole solution pool is denoted by blue dots. A unique and well separated global minimum exists, shown by the red triangles. The correct MAID solution \cite{MAID2007,MAID} is drawn as a green solid line.}
\label{fig:LmaxInfinityThDataFitBestSolsGroupSAndF}
\end{figure}

The following example cases will illustrate when and how this comes about. \newline
Just as in the previous studies of model data, we follow the model-independent TPWA scheme outlined in section \ref{sec:TPWAFitsIntro}, using pools of initial conditions of size $N_{MC}$ which are generated according to the Monte Carlo methods discussed in section \ref{sec:MonteCarloSampling}. This means in particular that in step $\mathrm{I}$ of the fit, Legendre coefficients are extracted in the \textit{same} order as in the subsequent multipole-fit, i.e. TPWA fit step $\mathrm{II}$. For the results shown in the following, orders of Legendre coefficients and multipoles always exactly matched. One could also think about the strategy to fit Legendre coefficients in quite high orders, followed by minimizations of $\Phi_{\mathcal{M}}$ involving only the lower multipoles. \newline
The hope in the latter case could be that the higher Legendre coefficients influence the lower ones in a manner that is consistent with the modifications due to small high partial waves, which would then have a correctional effect on the resulting multipoles. However, in practice such fits did generally not yield satisfactory results. \newline
In the first example, theory-data from the full MAID model are fitted for the set
\begin{equation}
 \left\{ \sigma_{0}, \check{\Sigma}, \check{T}, \check{P}, \check{F} \right\} \mathrm{,} \label{eq:CompleExampleSetGroupSAndFChapter4}
\end{equation}
which has been postulated as a complete experiment (section \ref{sec:WBTpaper}). We use here a truncation at $\ell_{\mathrm{max}} = 1$ and a pool of $N_{MC} = 1500$ start configurations. Results are shown in Figure \ref{fig:LmaxInfinityThDataFitBestSolsGroupSAndF}. \newline
For this example fit, a global minimum is found, which is still quite well separated from the few remaining local minima. Thus, in terms of stability this fit is still quite well behaved. However, once the global minimum is compared to the true MAID mutlipoles, discrepancies become apparent. Those are generated solely from the higher partial wave contributions present in the theory-data. \newline
 Quantities that turn out to be quite susceptible to this effect are the imaginary parts of $E_{1+}$ and $M_{-1}$, as well as the real part of $E_{1+}$, especially for the higher energies. The remaining parameters actually show a quite good agreement between the unique solution and the MAID model. Especially the dominating wave $M_{1+}$ comes out already quite well.
\clearpage

\begin{figure}[ht]
 \centering
\begin{overpic}[width=0.475\textwidth]{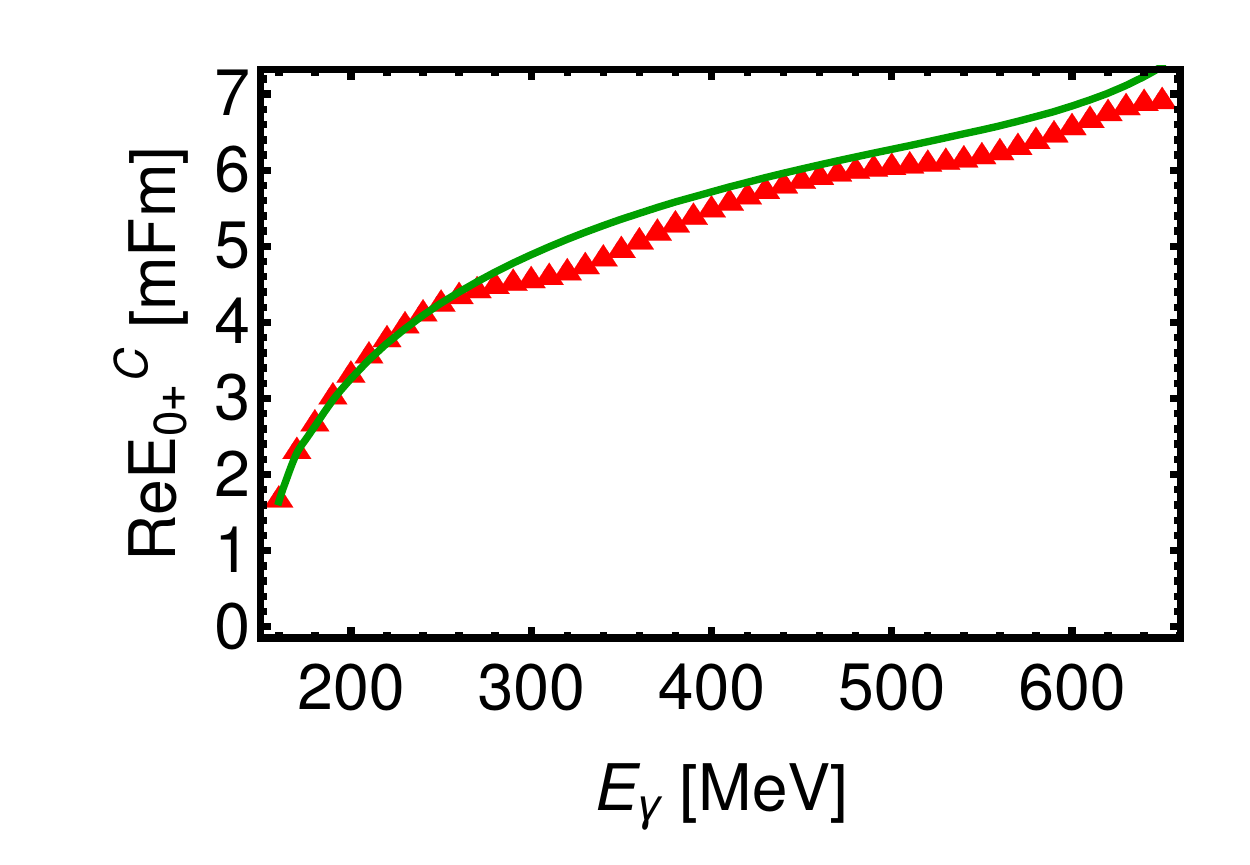}
 \end{overpic} \\
\vspace*{-5pt}
\begin{overpic}[width=0.475\textwidth]{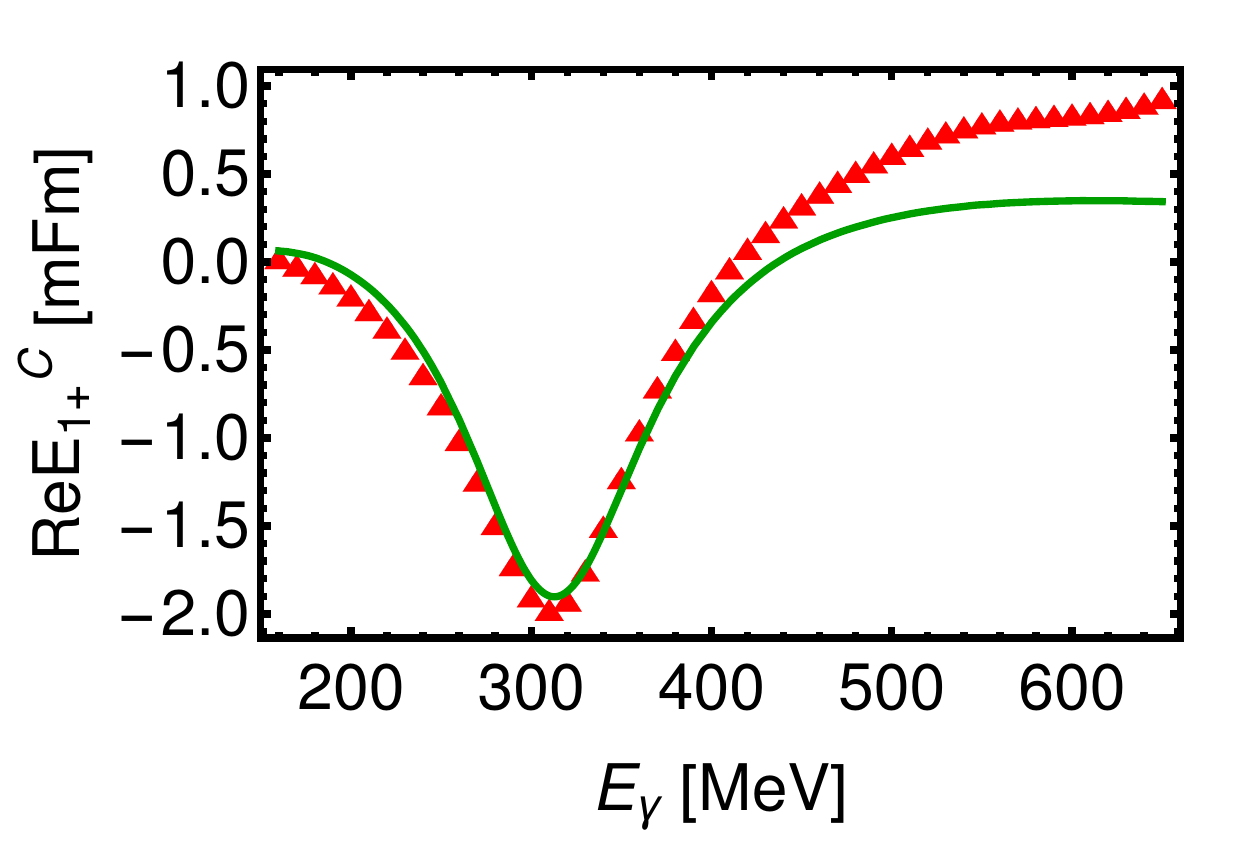}
 \end{overpic} \hspace*{5pt}
\begin{overpic}[width=0.475\textwidth]{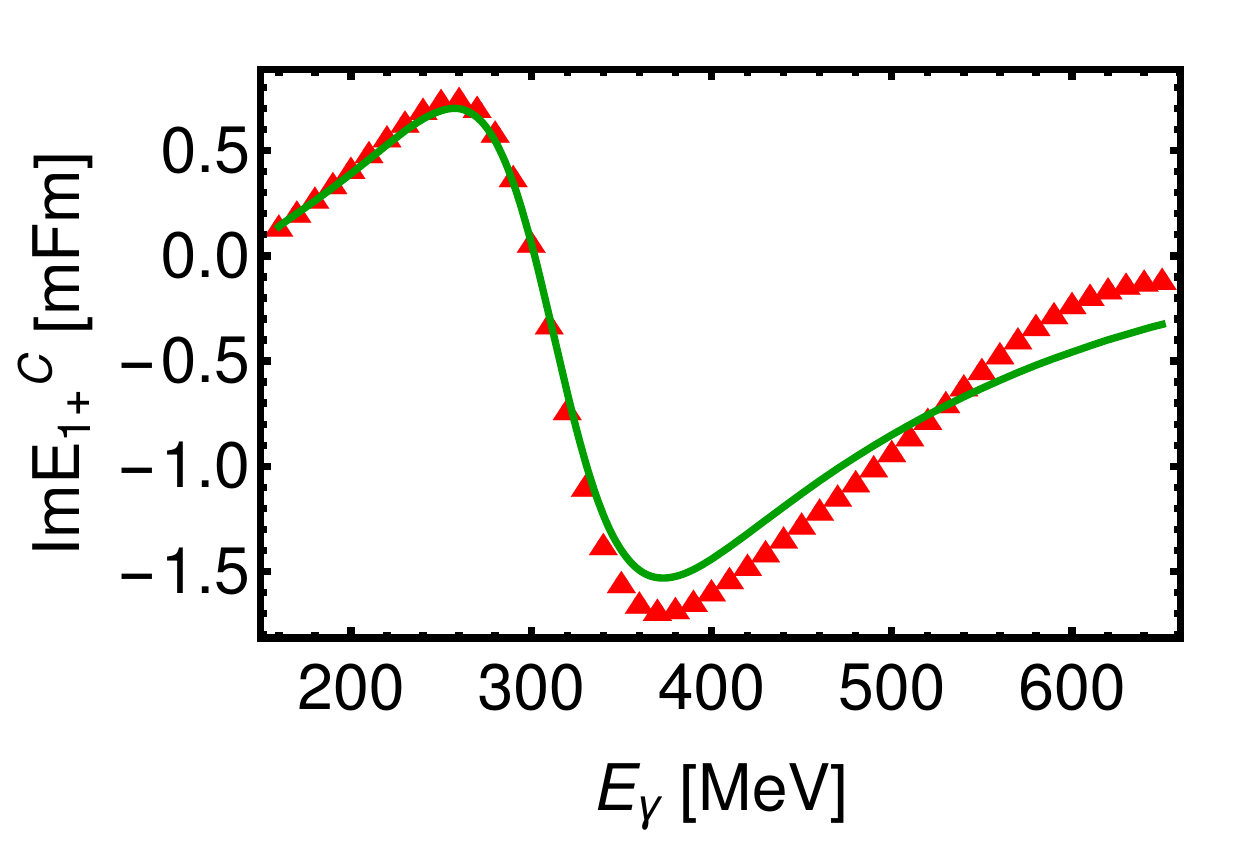}
 \end{overpic} \\
\begin{overpic}[width=0.475\textwidth]{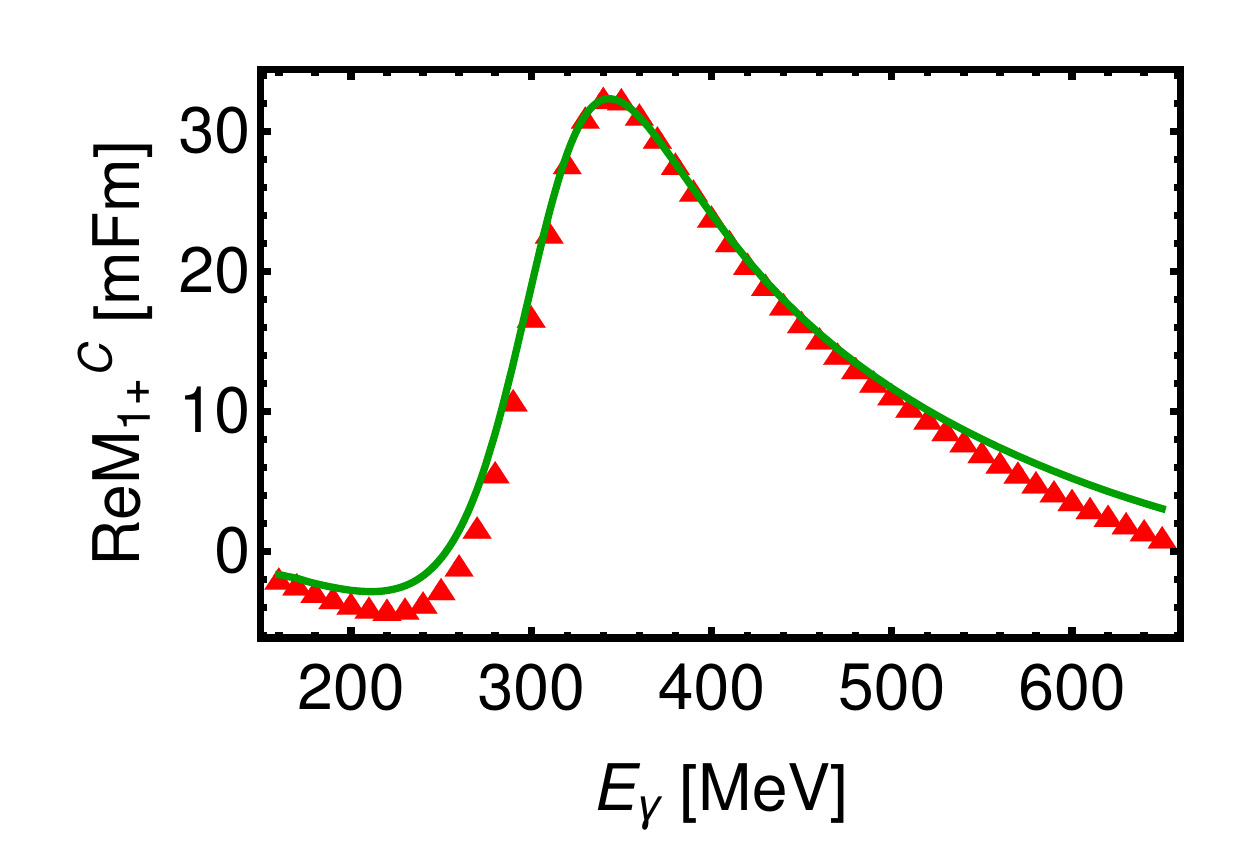}
 \end{overpic} \hspace*{5pt}
\begin{overpic}[width=0.475\textwidth]{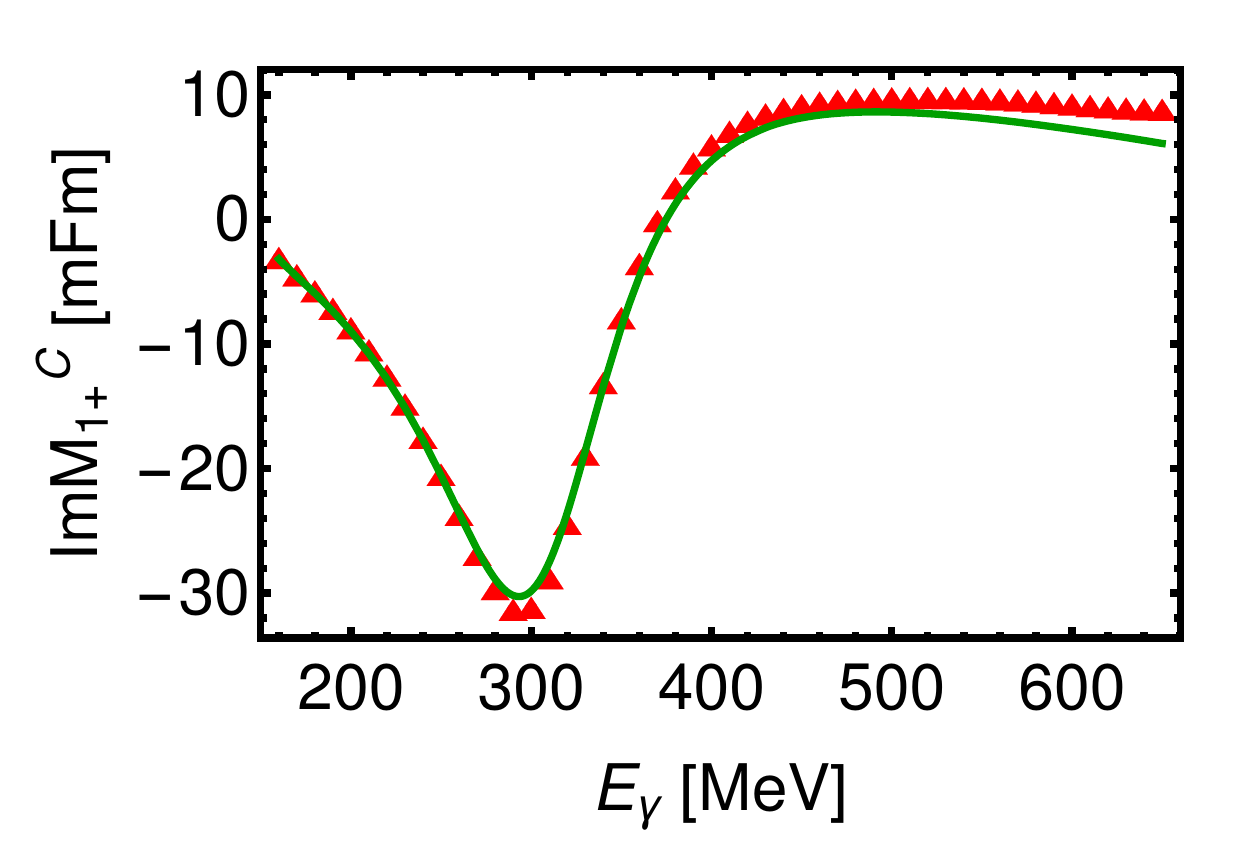}
 \end{overpic} \\
\begin{overpic}[width=0.475\textwidth]{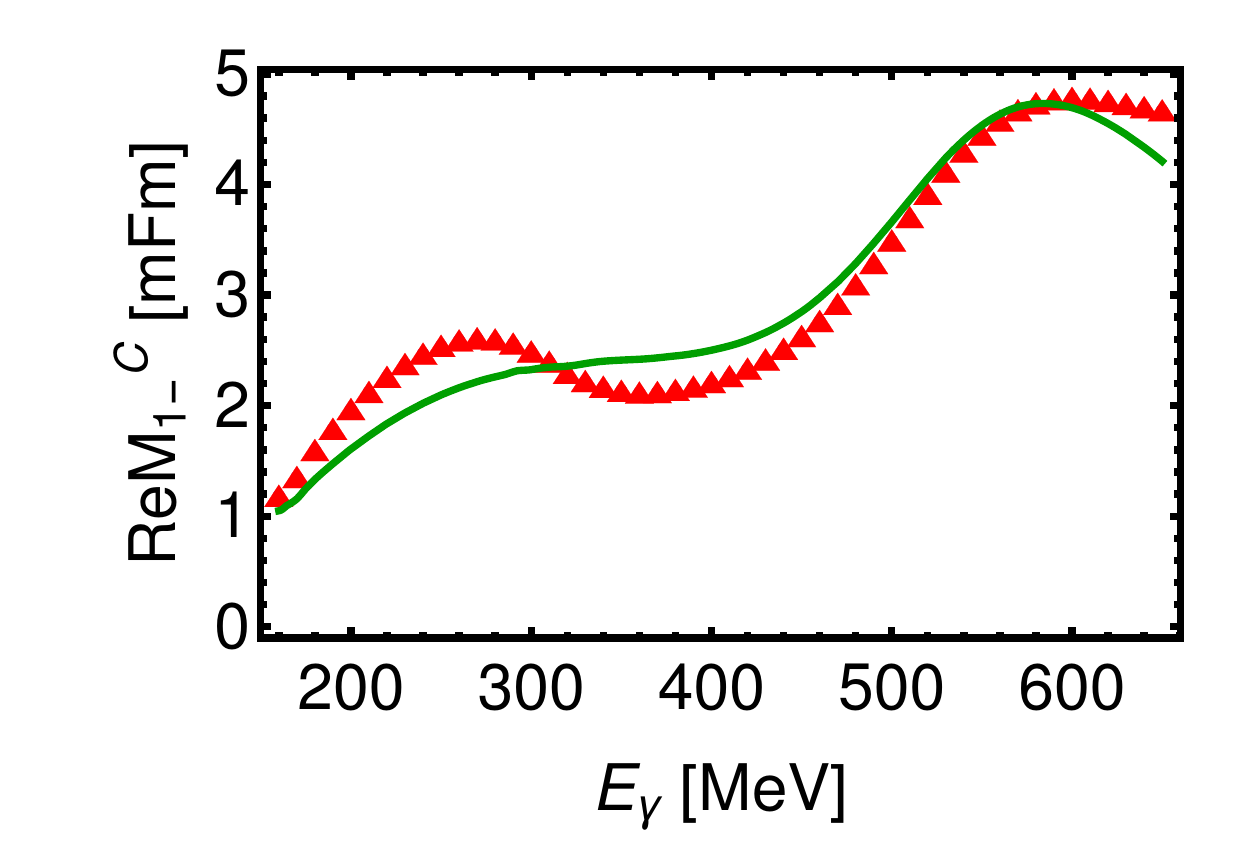}
 \end{overpic} \hspace*{5pt}
\begin{overpic}[width=0.475\textwidth]{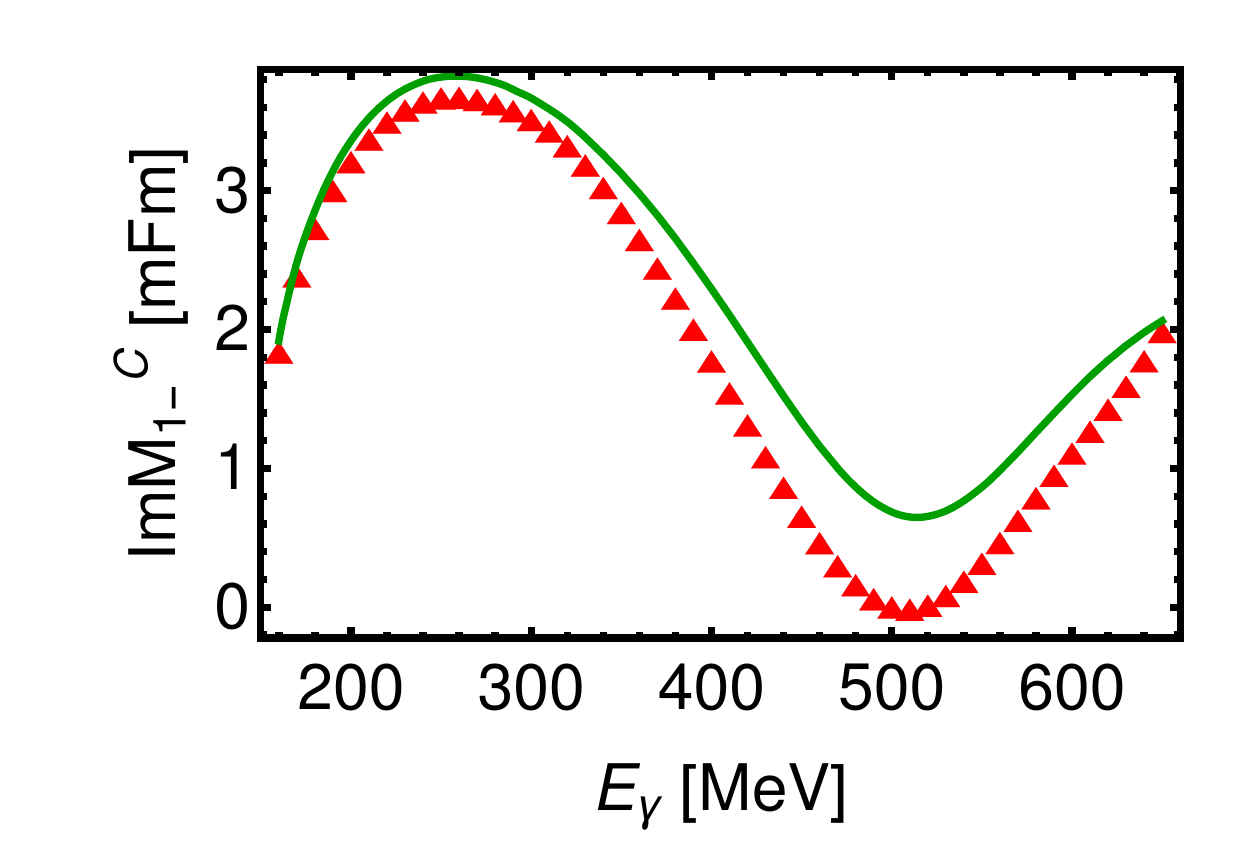}
 \end{overpic}
\vspace*{-5pt}
\caption[Multipole-results for a TPWA-fit of an $S$- and $P$-wave truncation to MAID theory-data from the full (non-truncated) model. All group $\mathcal{S}$ and $\mathcal{BT}$ were analyzed.]{Here, results are shown of a TPWA to the full MAID theory-data \cite{LotharPrivateComm,MAID2007}, fitted using $\ell_{\mathrm{max}} = 1$. The fit is fully analogous to the analysis that lead to what is seen in Figure \ref{fig:LmaxInfinityThDataFitBestSolsGroupSAndF}, except for the fact that here all group $\mathcal{S}$ and $\mathcal{BT}$ observables, i.e. $\left\{ \sigma_{0}, \check{\Sigma}, \check{T}, \check{P}, \check{E}, \check{G}, \check{H}, \check{F} \right\}$, were used to constrain the minimizations.}
\label{fig:LmaxInfinityThDataFitBestSolsGroupSAndBT}
\end{figure}

Before going to a higher order in $\ell_{\mathrm{max}}$, a valid means to drive the fit closer to the true solution would be to include more observables. Here, we have chosen to enlarge to the maximal set that does however still avoid recoil polarization observables, i.e. the set composed of the full classes group $\mathcal{S}$ and $\mathcal{BT}$
\begin{equation}
 \left\{ \sigma_{0}, \check{\Sigma}, \check{T}, \check{P}, \check{E}, \check{G}, \check{F}, \check{H} \right\} \mathrm{.} \label{eq:CompleExampleSetGroupSAndFChapter4_2}
\end{equation}
The fit is performed in otherwise exactly the same way as before. Results can be seen in Figure \ref{fig:LmaxInfinityThDataFitBestSolsGroupSAndBT}. The constrainng effect of the additional data can be seen in two ways. First, the fit is even more stable than before, since no local minima have been found and a unique global minimum exists. Furthermore, discrepancies to the true MAID multipoles have been reduced considerably compared to the previous fit, seen in Figure \ref{fig:LmaxInfinityThDataFitBestSolsGroupSAndF}. This is true especially for the multipoles $E_{1+}$ and $M_{1-}$. Therefore, it is possible again to get quite far while completely avoiding the double polarization observables with recoil polarization. The remaining discrepancies remain present especially for the higher energies. This is only logical, since there the higher partial waves become larger and their contributrions to the theory-data more significant, as can be also seen in Figures \ref{fig:LmaxInfinityThDataFitLegCoeffsDCS2} and \ref{fig:LmaxInfinityThDataFitLegCoeffsDCS3}. \newline
However, the TPWA truncated at $\ell_{\mathrm{max}} = 1$, even if it were constrained more by enlarging the set (\ref{eq:CompleExampleSetGroupSAndFChapter4_2}) to include all $16$ observables, could never converge exactly towards the MAID model. This is because all the partial wave interferences present in the full MAID model data, specifically those arising from the far off-diagonal entries of the hermitean matrices that define the Legendre coefficients (see section \ref{sec:TPWAFitsIntro} and appendix \ref{sec:TPWAFormulae}) for higher orders in $\ell_{\mathrm{max}}$, are not parametrized explicitly in the TPWA model used to fit the data, which is just truncated at the $S$- and $P$-waves. The TPWA, as a model, does not \textit{know} these terms which are however essential for a correct description of the data. The fit then compensates for this fact by converging to a minimum that is slightly off the correct MAID solution. \newline
In case one desires to get the fit results closer to the MAID multipoles, the only way out consists of raising $\ell_{\mathrm{max}}$ in the TPWA. This however introduces more multipoles, more Omelaenko-roots and thus also more possibilities to form ambiguities, as has also been illustrated by the results in section \ref{subsec:TheoryDataFitsLmax2}.

\clearpage

\clearpage

\begin{figure}[ht]
 \centering
\begin{overpic}[width=0.346\textwidth]{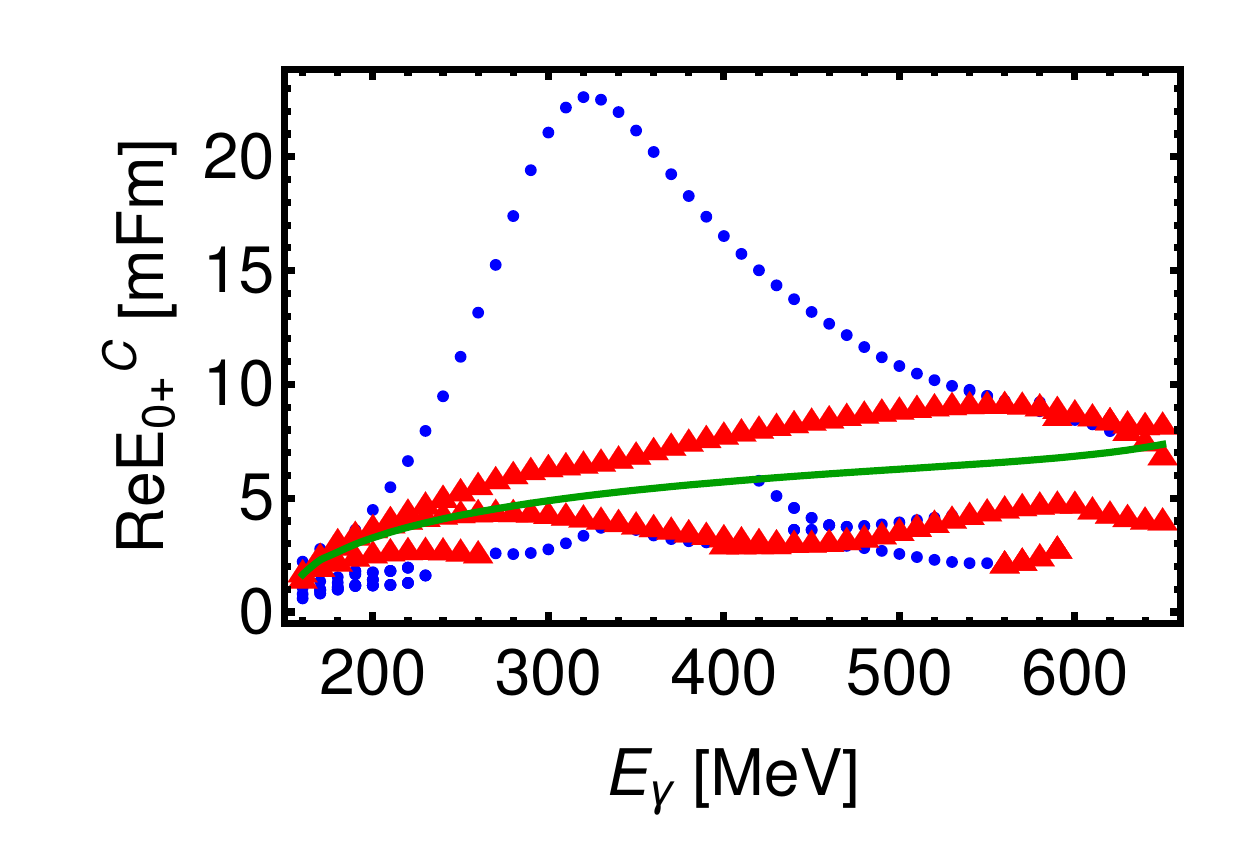}
 \end{overpic} \hspace*{-15pt}
\begin{overpic}[width=0.346\textwidth]{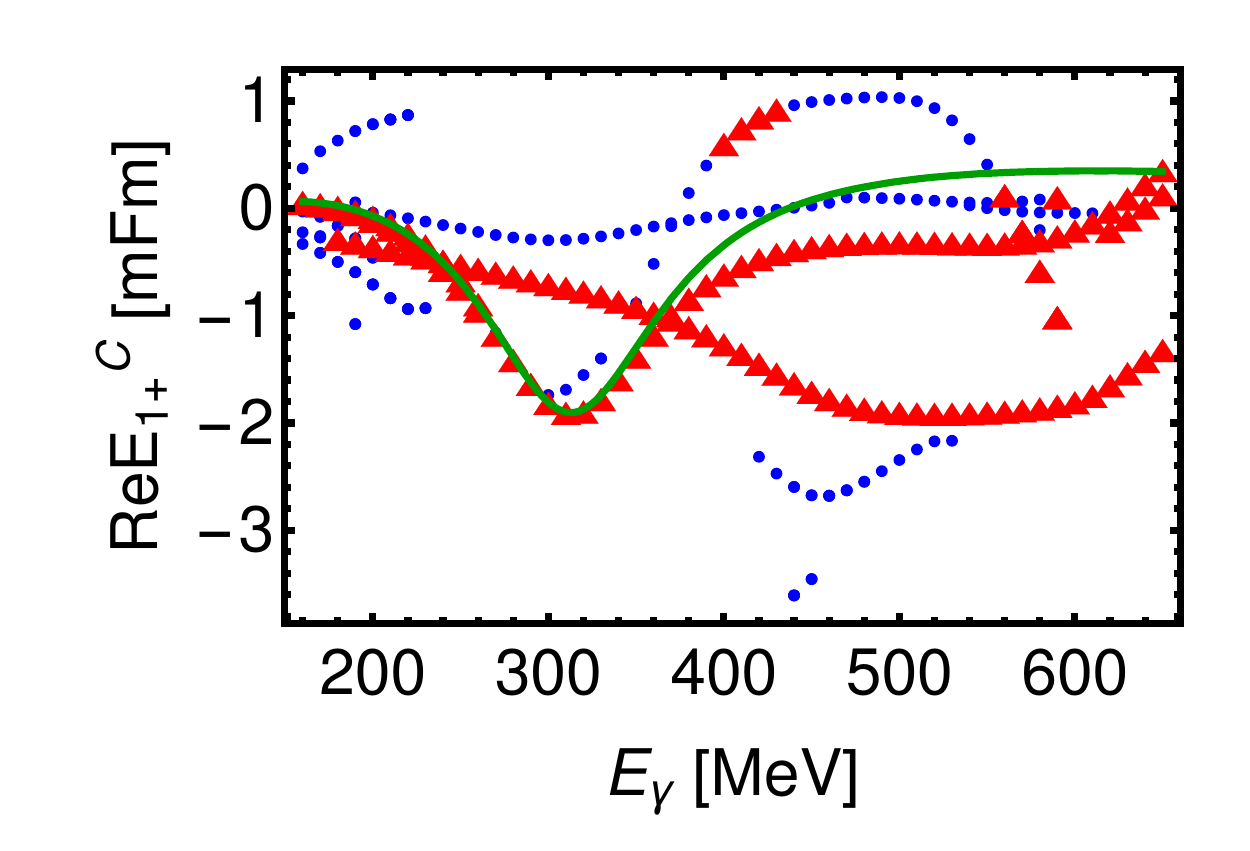}
 \end{overpic} \hspace*{-15pt}
\begin{overpic}[width=0.346\textwidth]{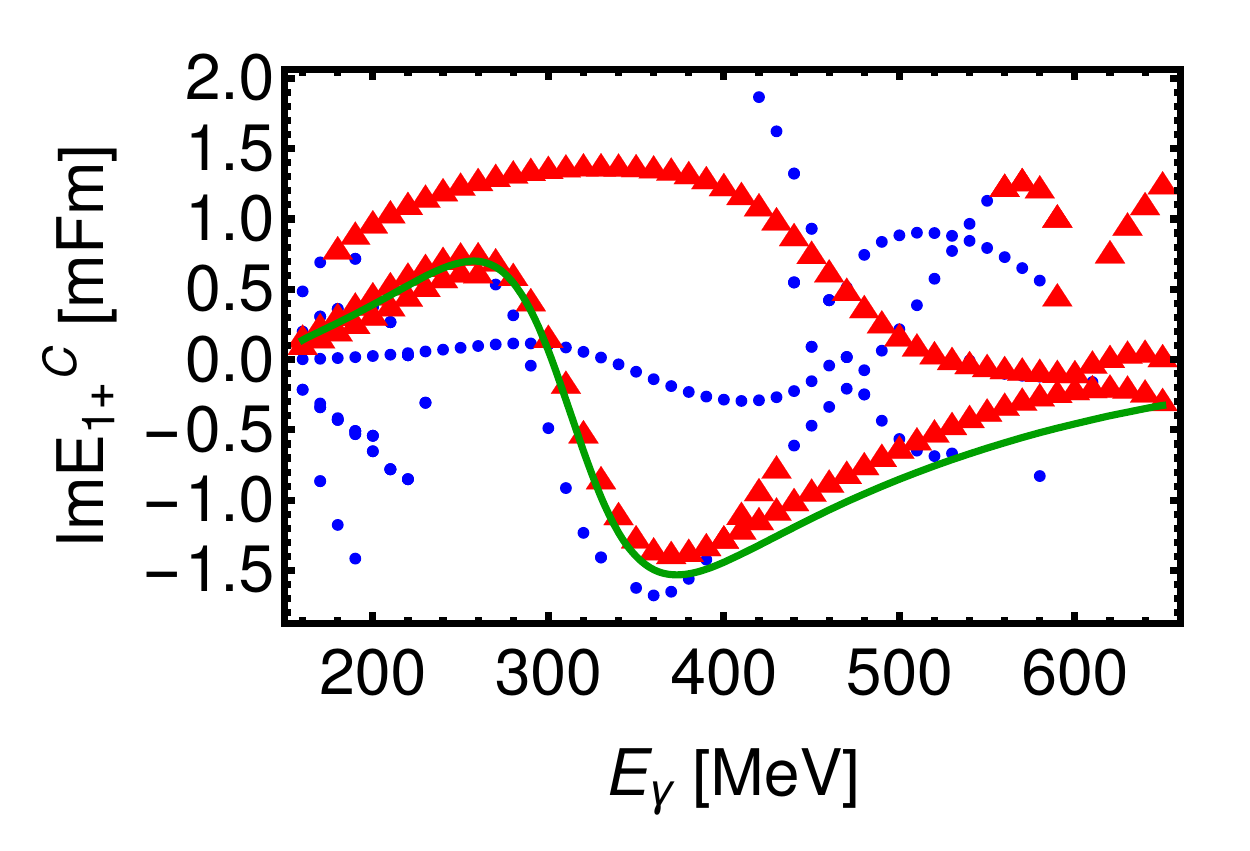}
 \end{overpic}  \\
\begin{overpic}[width=0.346\textwidth]{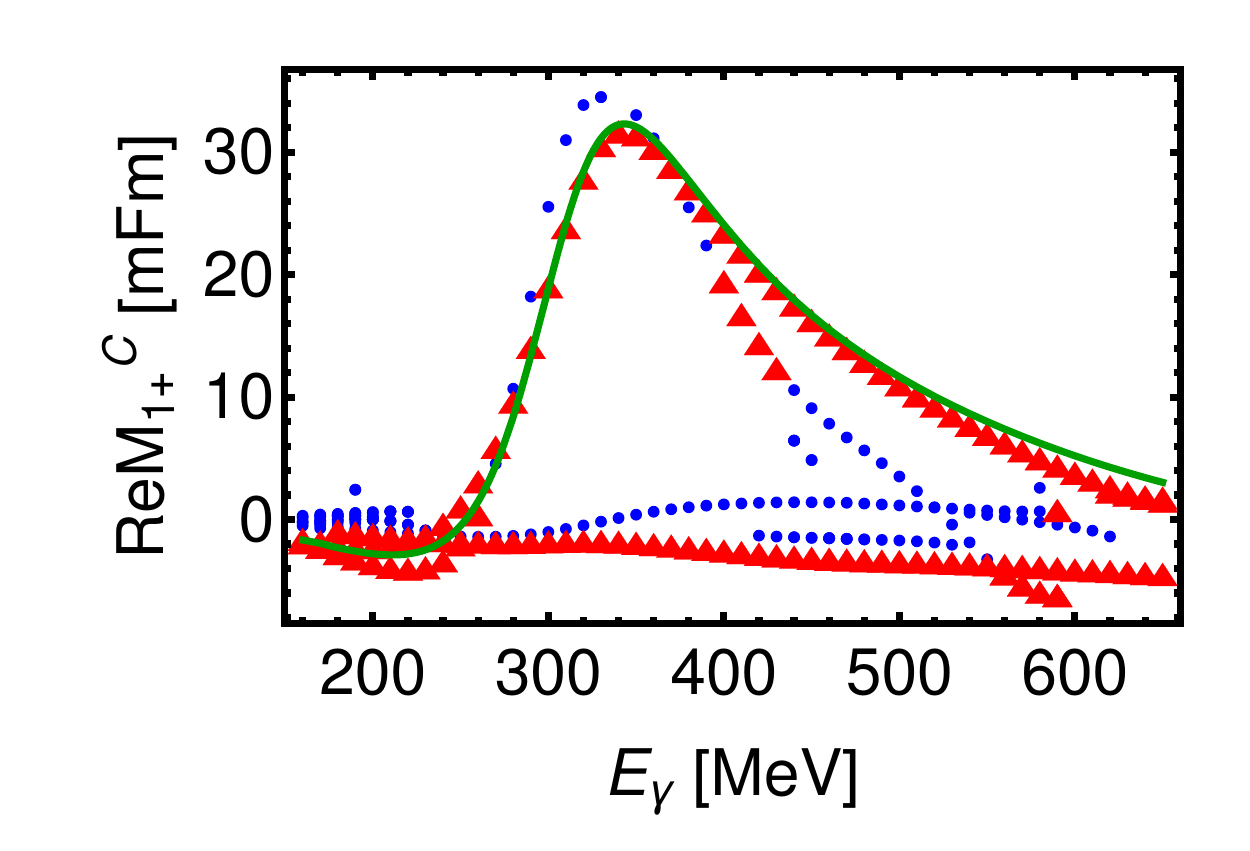}
 \end{overpic} \hspace*{-15pt}
\begin{overpic}[width=0.346\textwidth]{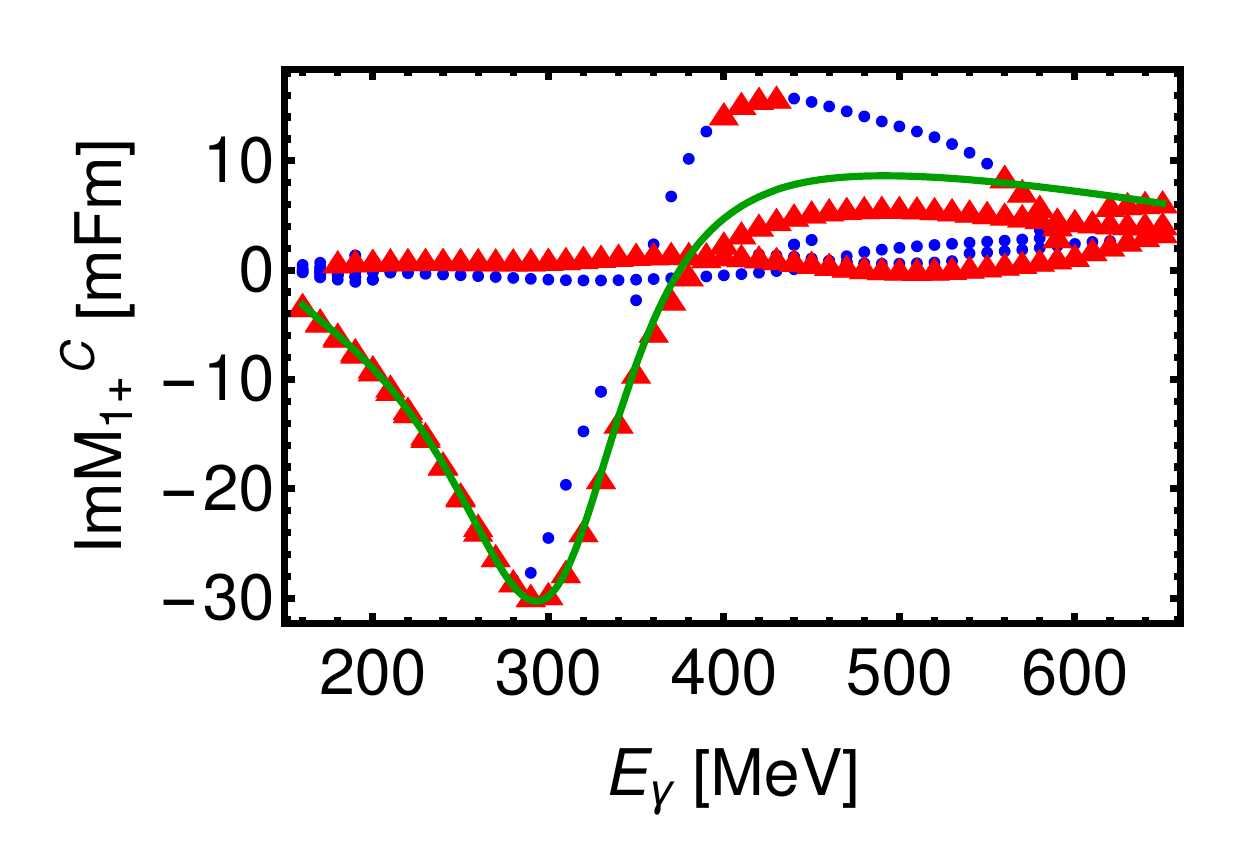}
 \end{overpic} \hspace*{-15pt}
\begin{overpic}[width=0.346\textwidth]{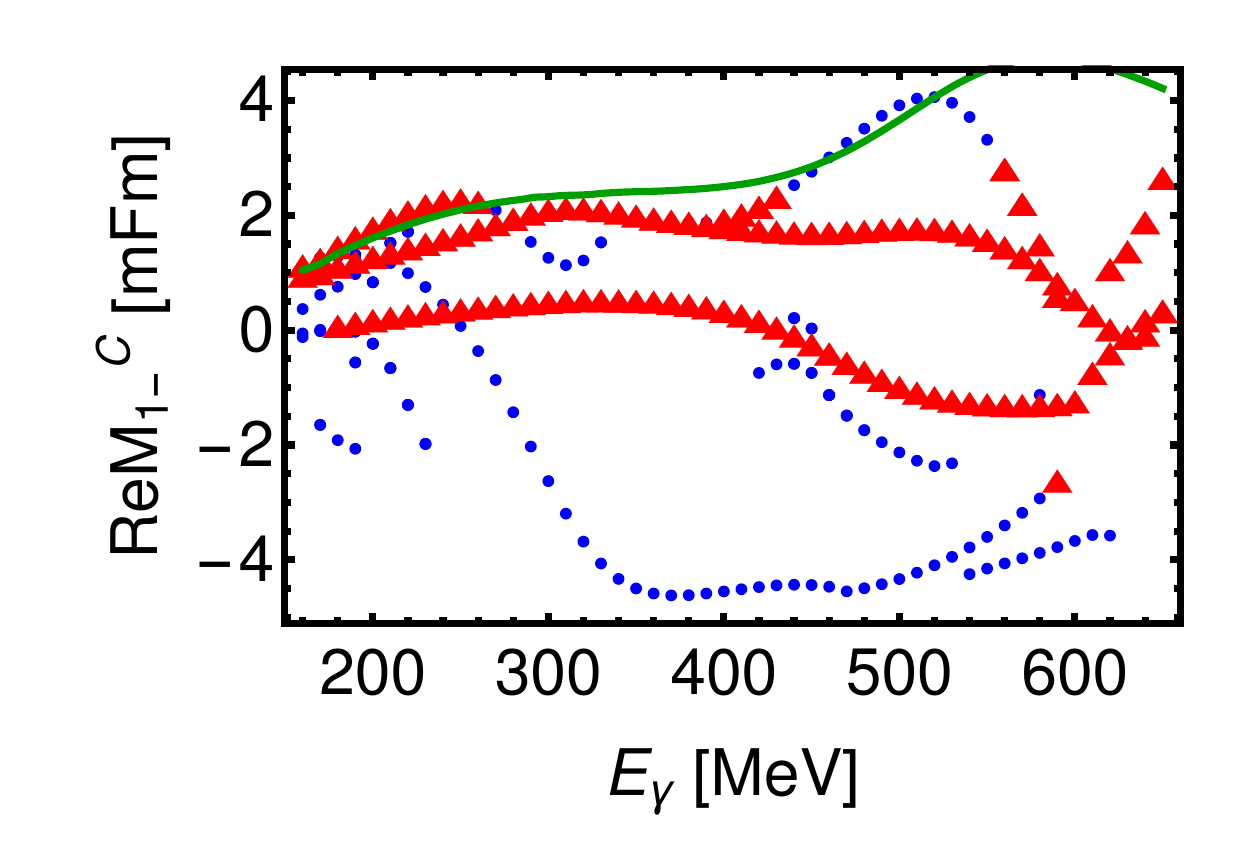}
 \end{overpic} \\
\begin{overpic}[width=0.346\textwidth]{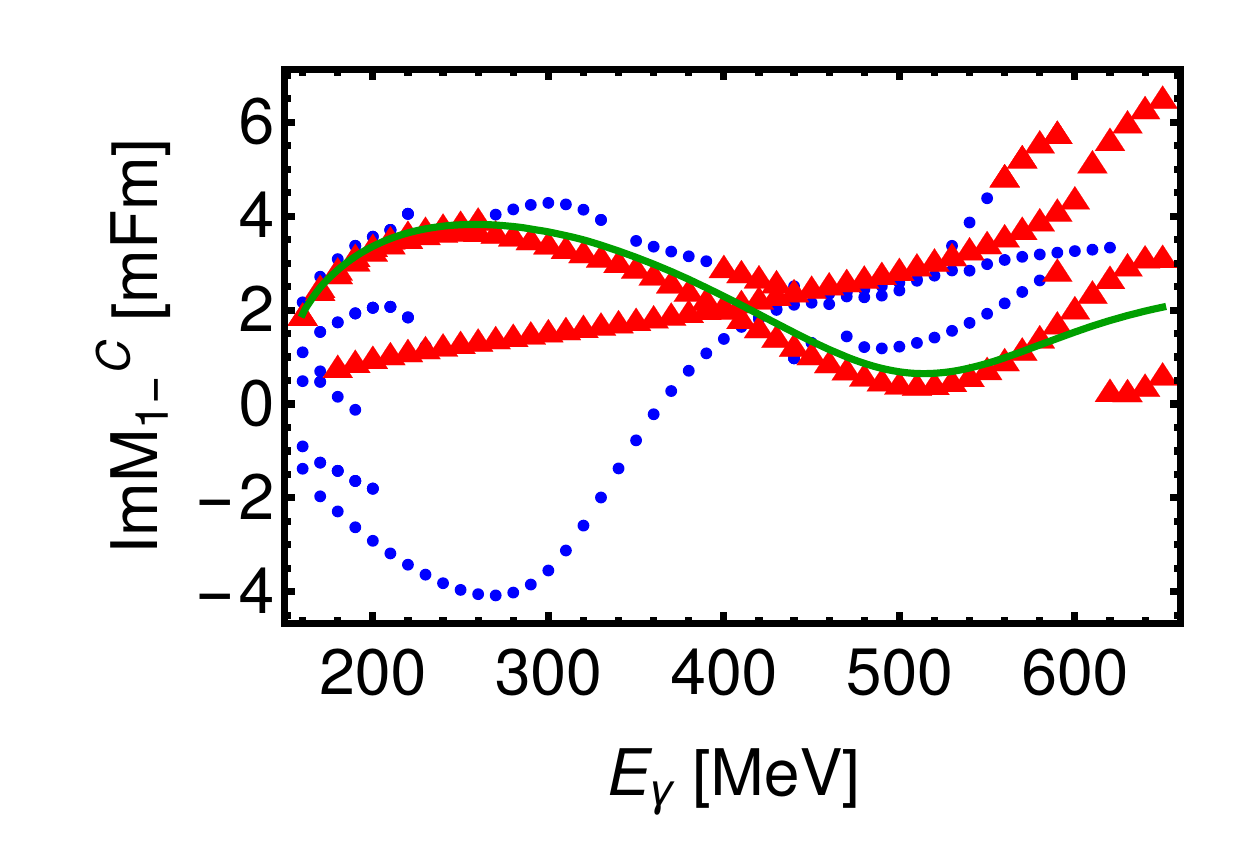}
 \end{overpic} \hspace*{-15pt}
\begin{overpic}[width=0.346\textwidth]{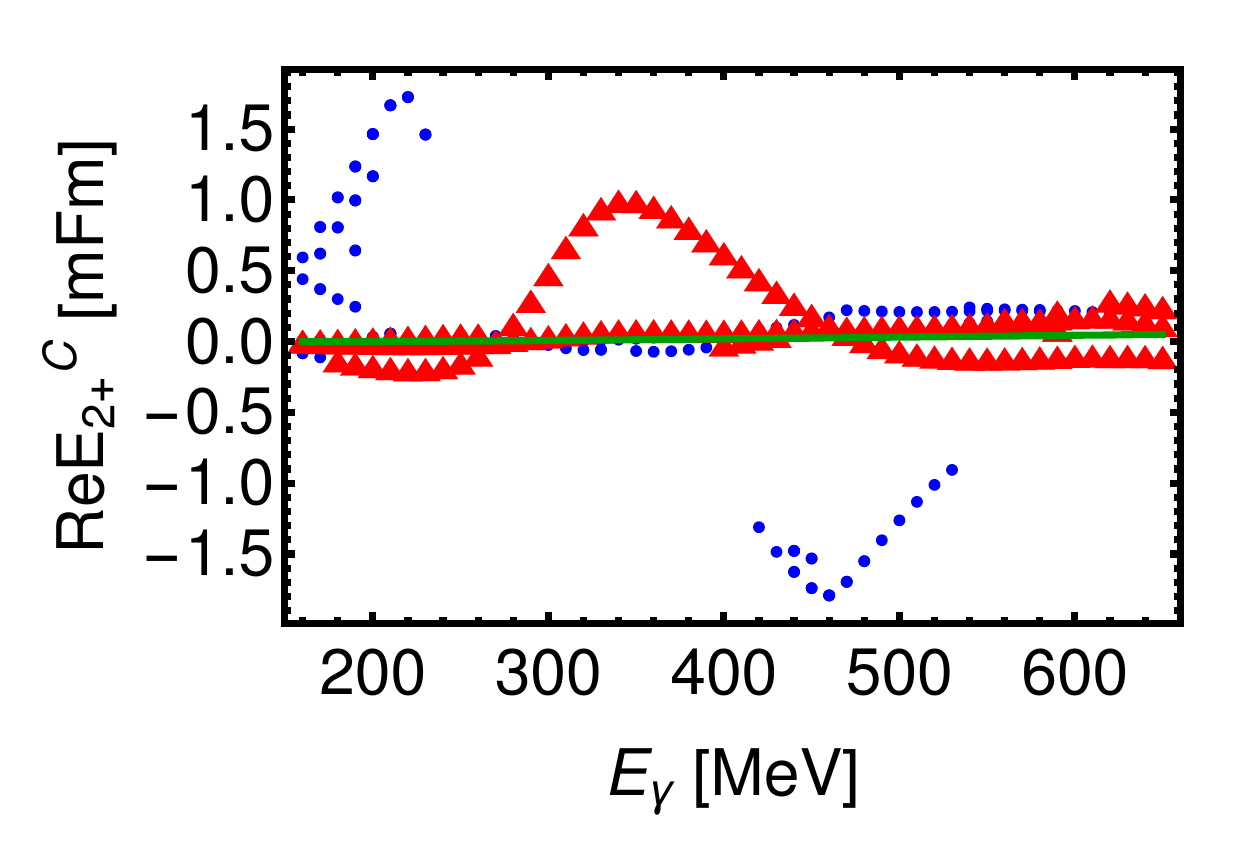}
 \end{overpic} \hspace*{-15pt}
\begin{overpic}[width=0.346\textwidth]{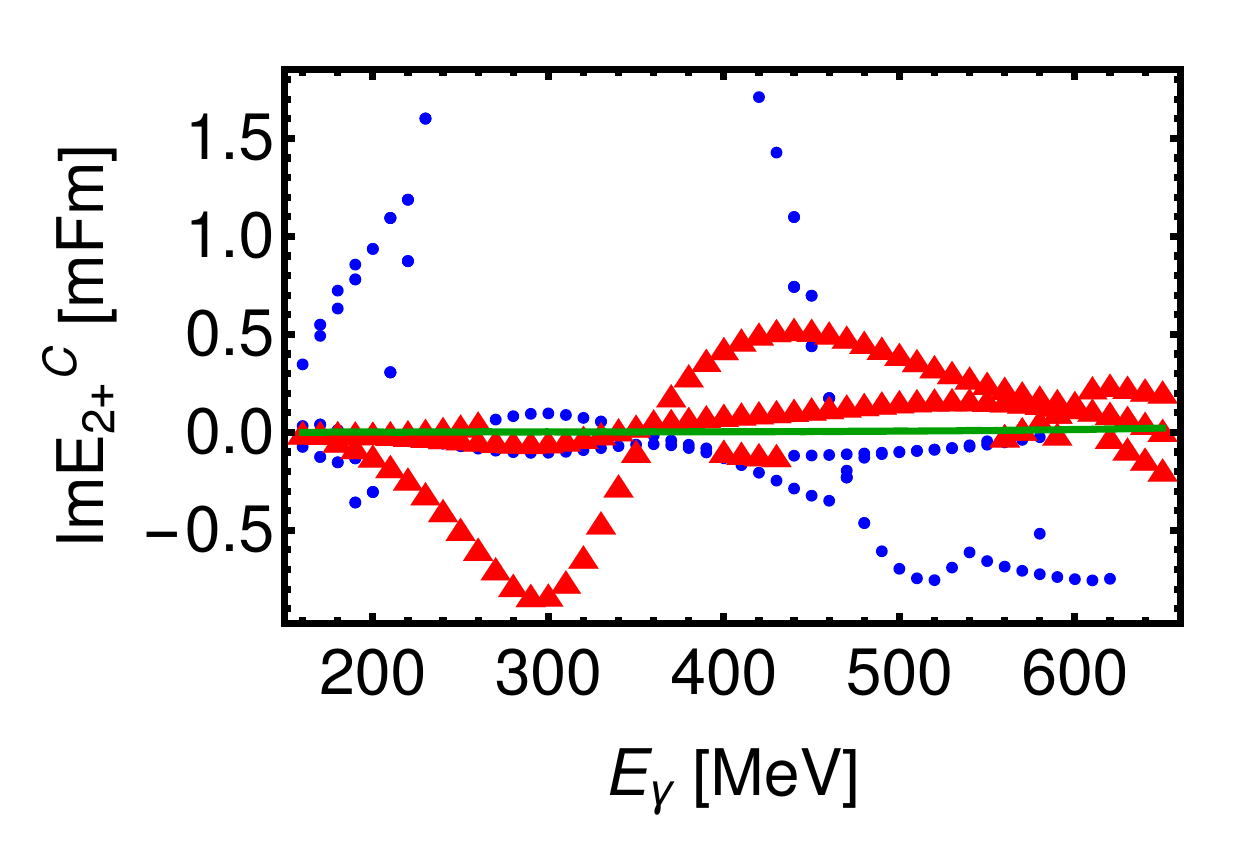}
 \end{overpic} \\
\begin{overpic}[width=0.346\textwidth]{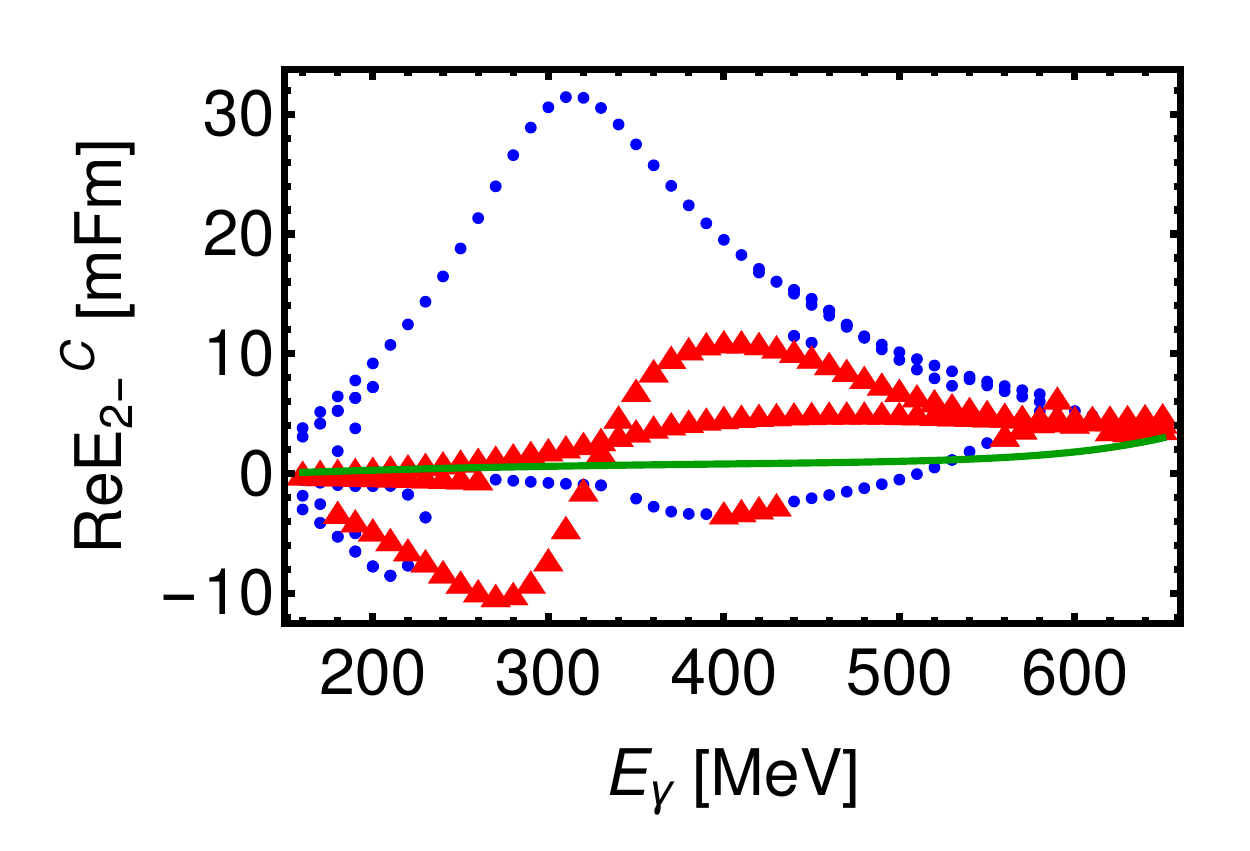}
 \end{overpic} \hspace*{-15pt}
\begin{overpic}[width=0.346\textwidth]{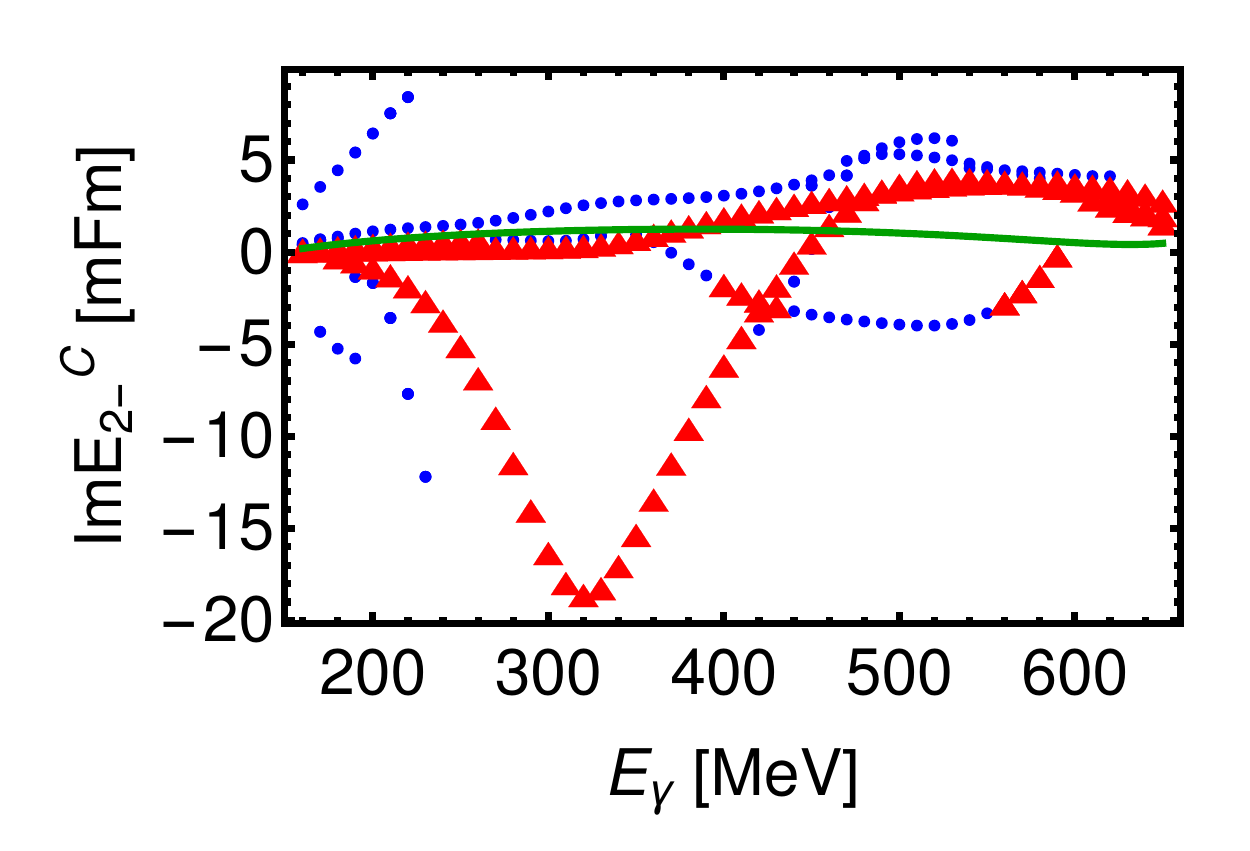}
 \end{overpic} \hspace*{-15pt}
\begin{overpic}[width=0.346\textwidth]{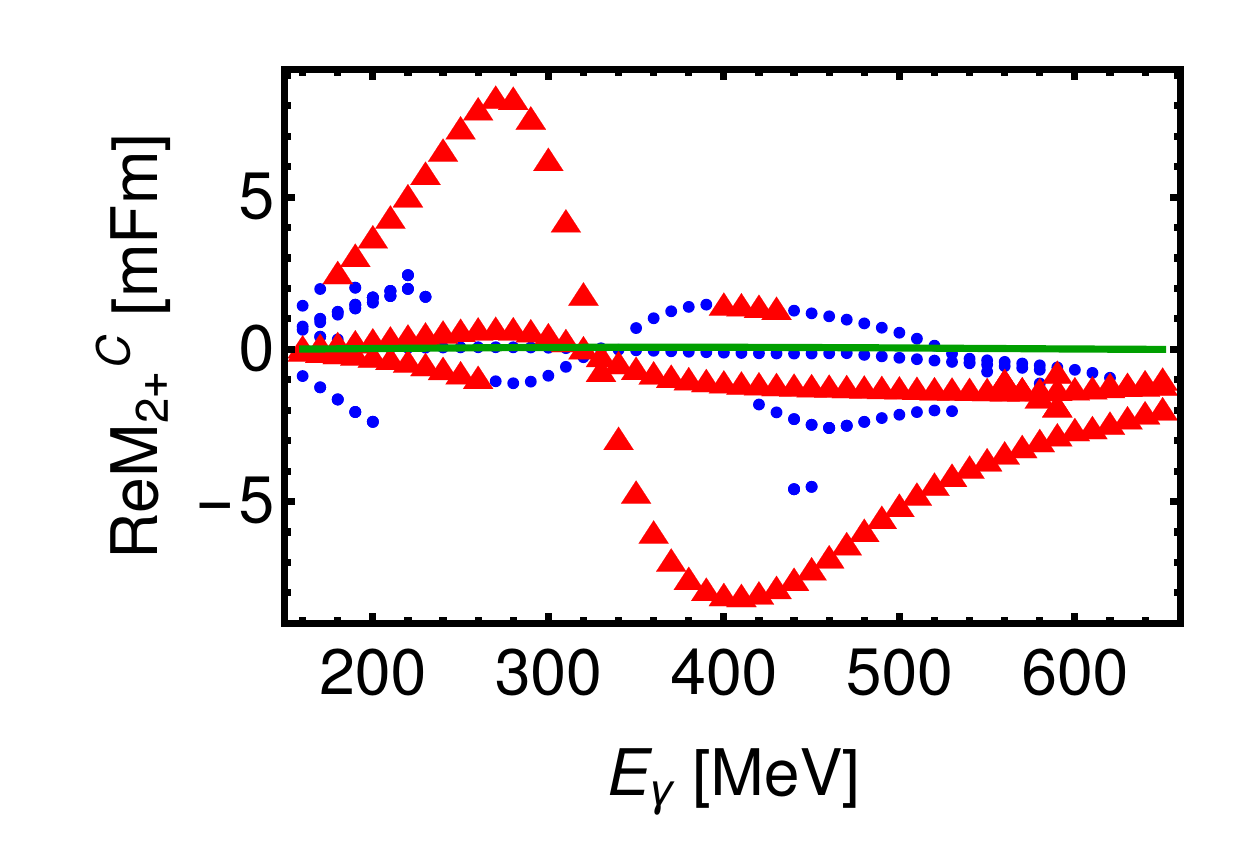}
 \end{overpic} \\
\begin{overpic}[width=0.3457\textwidth]{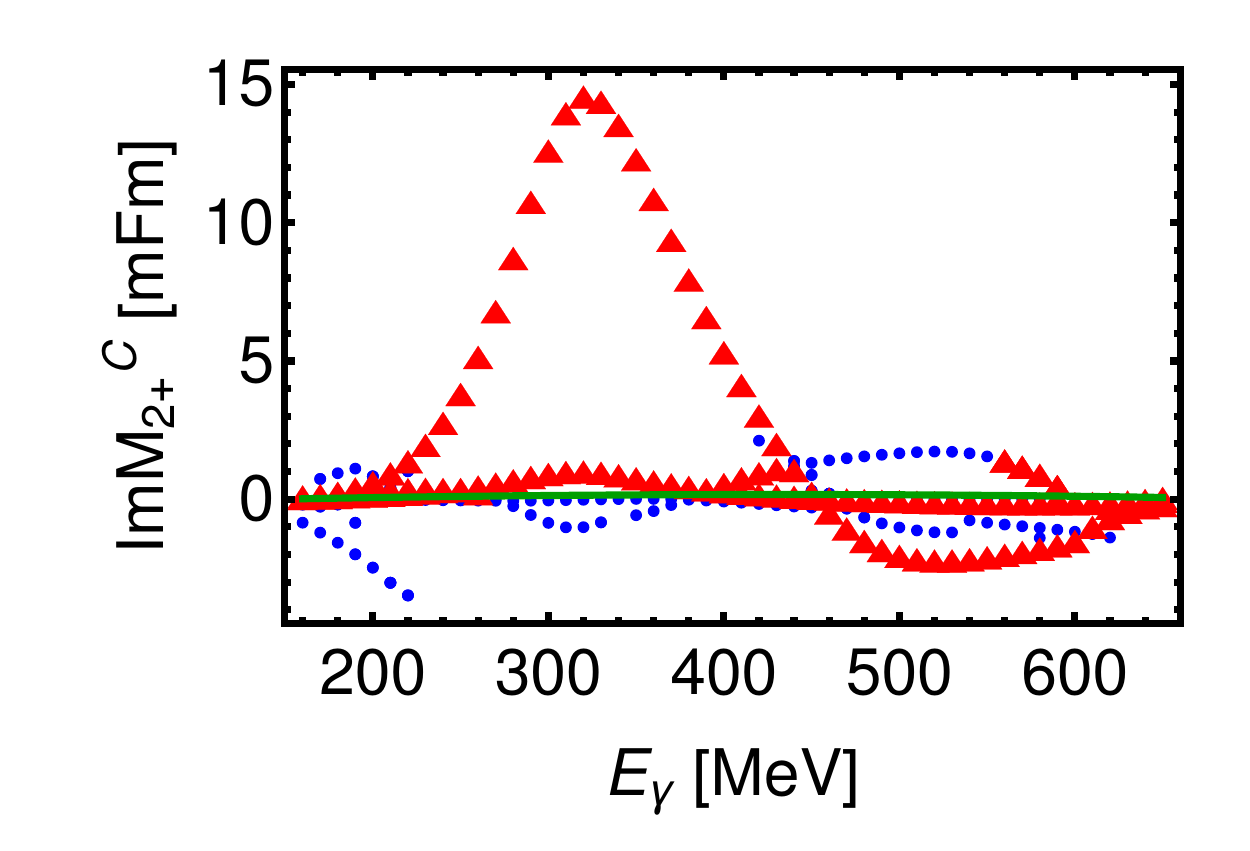}
 \end{overpic} \hspace*{-15pt}
\begin{overpic}[width=0.3457\textwidth]{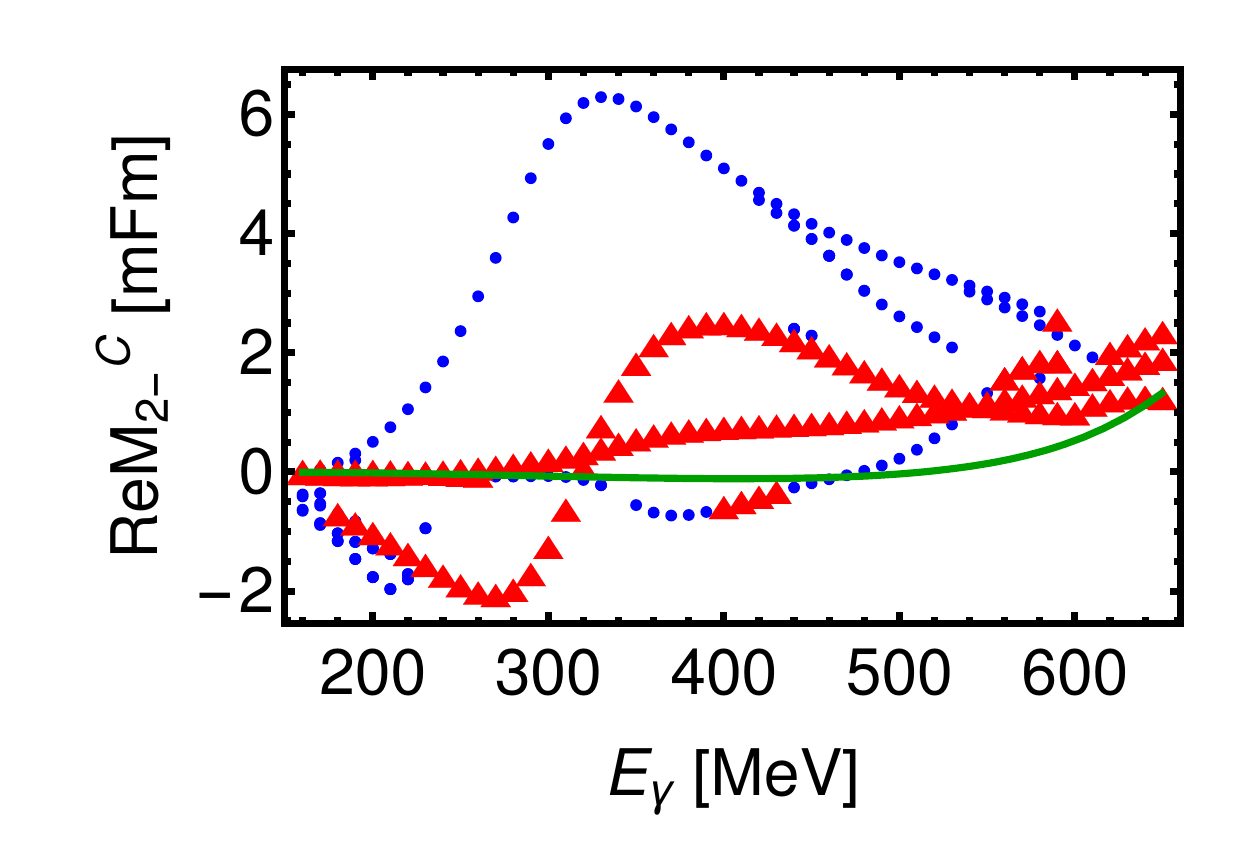}
 \end{overpic} \hspace*{-15pt}
\begin{overpic}[width=0.3457\textwidth]{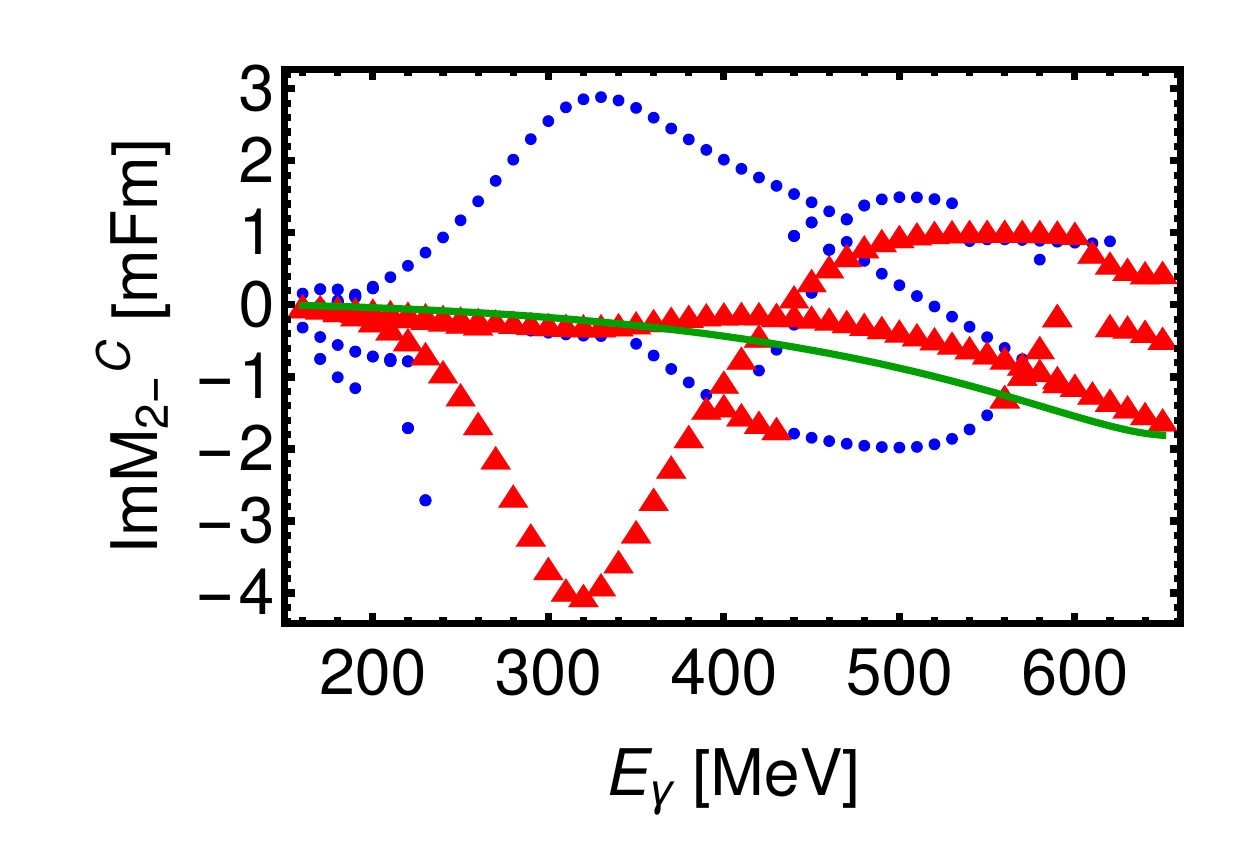}
 \end{overpic}
\vspace*{0pt}
\caption[Multipole-results for a TPWA-fit of an $S$-, $P$- and $D$-wave truncation to MAID theory-data from the full (non-truncated) model. All group $\mathcal{S}$ and $\mathcal{BT}$ were analyzed.]{Shown here are the results of a TPWA truncated at $\ell_{\mathrm{max}} = 2$ to MAID theory-data \cite{LotharPrivateComm,MAID2007} from the full model. Results for real and imaginary parts of all phase-constrained multipoles for all partial waves from $E_{0+}$ up to $M_{2-}$ are plotted. \newline The observables $\left\{ \sigma_{0}, \check{\Sigma}, \check{T}, \check{P}, \check{E}, \check{G}, \check{H}, \check{F} \right\}$ were used to constrain the multipole solutions and a pool of $N_{MC} = 3000$ randomly chosen initial conditions yielded the fitted values. A global minimum does exist, but it is not well-separated to other local minima. Rather, multiple final parameter configurations exist having the same value for $\Phi_{\mathcal{M}}$ within one order of magnitude. These solutions are plotted as red triangles. The remaining minima are shown as blue dots. The MAID solution \cite{MAID2007,MAID} is drawn as a green solid line.}
\label{fig:LmaxInfinityThDataLmax2FitBestSolsGroupSAndBT}
\end{figure}

The next example is a TPWA fit to the full model using $\ell_{\mathrm{max}} = 2$. In view of the above mentioned issues, the mathematically overcomplete set (\ref{eq:CompleExampleSetGroupSAndFChapter4_2}), i.e. all group $\mathcal{S}$ and $\mathcal{BT}$ observables, has been fitted. The parameter space of the multipoles has been scanned using a pool of $N_{MC} = 3000$ Monte Carlo start configurations. The results are shown in Figure \ref{fig:LmaxInfinityThDataLmax2FitBestSolsGroupSAndBT}. \newline

In this case, only one truncation order higher than in the previous examples, the stability of the fit is already lost. A global minimum exists and the number of obtained non-redundant solutions is not large. There are mostly up to $4$ in every energy bin. However, in addition to the global minimum, additional solutions are found which are located far away in amplitude space, but are closely clustered in the value of the discrepancy function $\Phi_{\mathcal{M}}$. While the global minimum takes values around $\Phi_{\mathcal{M}} \simeq 10^{-8} (\mu b / sr)$, the equivalent local minima are not far away in $\Phi_{\mathcal{M}}$ and at most one order of magnitude larger. However, among the multitude of solutions, there still exist some which are located quite closely to the MAID solution. \newline
The basic issue now becomes fully apparent. In order to have a chance to fit the true MAID multipoles out of the full theory-data, the truncation order of the TPWA has to be raised in order to properly take into account all relevant partial wave interference terms. This however makes the fit more unstable and susceptible to ambiguities. In order to make an estimate for the appropriate order in $\ell_{\mathrm{max}}$ at which the correct MAID solution \textit{might} re-emerge as the global minimum, the best is to look at the Legendre coefficients (Figures \ref{fig:LmaxInfinityThDataFitLegCoeffsDCS1} to \ref{fig:LmaxInfinityThDataFitLegCoeffsDCS3}). There, it was seen that the highest coefficient for $\ell_{\mathrm{max}} = 7$ was consistent with zero. However, trying a TPWA at such a high order would involve a quite extreme amount of Monte Carlo sampling due to the anticipated exponential rise in the number of ambiguities. Probably, hundreds of thousands up to millions of sampling points would have to be used in every energy bin in order to map out all solutions, a feat which seems numerically impractical and was therefore not pursued further. \newline

In summary, TPWA fits to the full MAID model showed instabilities for the higher truncation orders due to the appearance of ambiguities. The basic reason is the incompatibility of the TPWA equation systems (\ref{eq:BilinearEqSystemSec4Dot43}) due to all higher partial waves contributing in the theory-data. The effect is illustrated by the sketches in Figure \ref{fig:CompletenessChi2Cases} of appendix \ref{subsec:AccidentalAmbProofsIII}. \newline
Another basic feature shown by these analyses of perfect data is that it seems particularly difficult to fit small higher partial wave contributions out of data in a fully model independent way. This issue will emerge later in the analyses of real data. One possible way out of this issue would be to fix the higher waves to model values. Although this has been helpful in stabilizing theory-data fits such as the ones shown here, we will discuss results of this approach later, when analyses of real world data will be shown.

\clearpage

\subsection{Fitting heteroscedastic data: bootstrap methods} \label{sec:BootstrappingIntroduction}

Data for polarization observables taken in the real world can be expected to carry at least a statistical uncertainty. Systematic errors are not considered in this section. They are only expected to make the following matters worse. \newline
The statistical errors provided with a readily analyzed dataset of a polarization asymmetry typically vary from point to point. The technical term for this phenomenon is {\it heteroscedasticity} (cf. reference \cite{DavisonHinkley}). \newline
This means that, even if one assumes an exact solution of the TPWA-problem to exist in a world without statistical uncertainties, for actual data the equation systems
\begin{equation}
\left(a_{L}\right)_{k}^{\check{\Omega}^{\alpha}} = \left< \mathcal{M}_{\ell} \right| \left( \mathcal{C}_{L}\right)_{k}^{\check{\Omega}^{\alpha}} \left| \mathcal{M}_{\ell} \right> \mathrm{,} \label{eq:BilinearEqSystemSec4Dot4}
\end{equation}
used in the TPWA-step $\mathrm{II}$ of section \ref{sec:TPWAFitsIntro}, are not mathematically exactly solvable any more. In the words of Grushin \cite{Grushin}, they are {\it incompatible}. Put in another way, one does not get solutions in a TPWA-fit to real world data, but only best parameter estimates from e.g. statistical $\chi^{2}$-fits such as those outlined on section \ref{sec:TPWAFitsIntro}. \newline
Of course, as seen in section \ref{subsec:TheoryDataFitsLmaxInfinite}, the fact that the partial wave series is infinite can endanger the compatibility of the systems (\ref{eq:BilinearEqSystemSec4Dot4}) and the stability of the fits just as well. This point is however not the main concern of this section. \newline
The problem can now be formulated like this: Is there a numerical method capable of detecting whether or not ambiguities in a TPWA-fit exist and furthermore of mapping out where the ambiguities are situated? \newline
A second problem closely connected to the first one is given by the fact that the TPWA-fit, as a consequence of the bilinear forms $\left< \mathcal{M}_{\ell} \right| \left( \mathcal{C}_{L}\right) \left| \mathcal{M}_{\ell} \right>$ appearing in fit step $\mathrm{II}$, is a non-linear problem. Therefore, the standard error estimates returned by generic fit routines, which can be very good for fits of linear models, are generally not expected to be reliable any more in case of the TPWA. The second problem can therefore be phrased as: Is there a way to obtain a robust estimate for the statistical uncertainty of the multipole-parameters in case a TPWA is uniquely solvable? \newline

As an approach to both problems, we propose in this work the use of the so-called {\it bootstrap}. This is a statistical analysis method falling in the class of resampling techniques. It was first introduced by Efron in a seminal paper \cite{EfronOriginal}. There exists also an introductory textbook, co-written by the same author \cite{EfronTibshiraniBook}. Much of the following discussion as well as the applied notations are based on the latter reference. A rather technical and more elaborate introduction can be found in the book by Davison and Hinkley \cite{DavisonHinkley}. \newline

The method shall be illustrated briefly on the simplest example of a one-sample problem. Here, the data-structure is given by a tupel of $n$ measured values of one particular type of data, which can be organized in the vector
\begin{equation}
 \bm{x} = \left( x_{1}, x_{2}, \ldots , x_{n} \right) \mathrm{.} \label{eq:BooststrapExampleOneSample}
\end{equation}
The fundamental assumption is here, as always in statistics, that the measurement of the data is equivalent to drawing them from an in principle unknown underlying probability distribution $F$, written as
\begin{equation}
 F \overset{\mathrm{i.i.d.}}{\xrightarrow{\hspace*{0.75cm}}} \bm{x} = \left( x_{1}, x_{2}, \ldots , x_{n} \right) \mathrm{.} \label{eq:BooststrapExampleOneSampleDrawnFromF}
\end{equation}
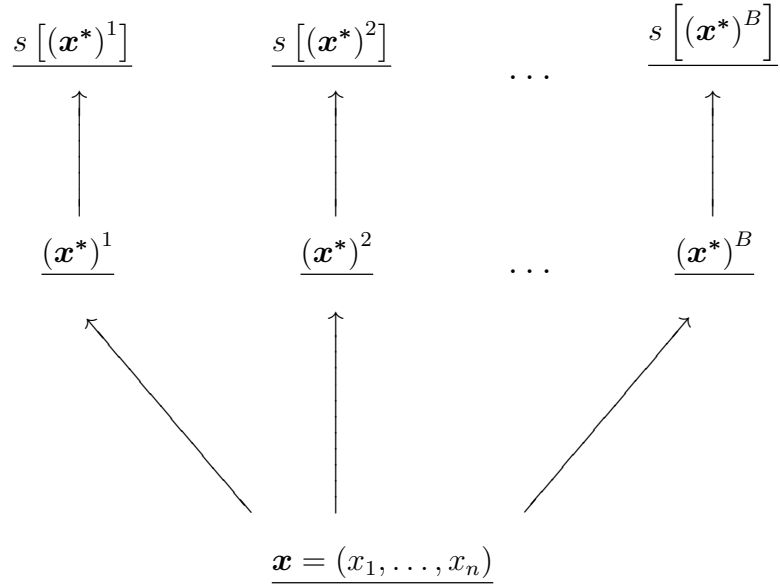
\begin{figure}[ht]
 \centering
\begin{overpic}[width=0.99\textwidth, trim = 0 175 0 0, clip]{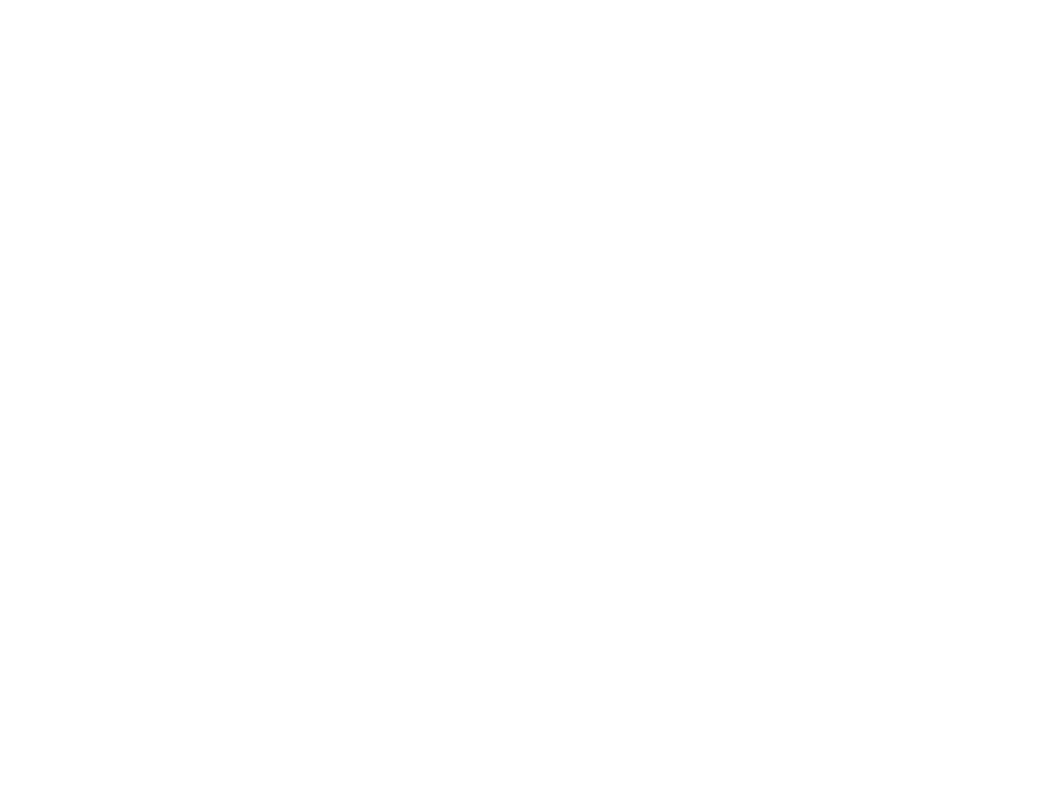}
\put(39,2.0){\begin{large}\underline{$\bm{x} = \left( x_{1},\ldots,x_{n} \right)$}\end{large}}
\put(22,7.25){\rotatebox{130}{$\xrightarrow{\hspace*{3.2cm}}$}}
\put(43.5,6.75){\rotatebox{90}{$\xrightarrow{\hspace*{2.5cm}}$}}
\put(60,6.5){\rotatebox{50}{$\xrightarrow{\hspace*{3.2cm}}$}}
\put(19.0,29){\begin{large}\underline{$\left( \bm{x^{\ast}} \right)^{1}$}\end{large}}
\put(41.5,29){\begin{large}\underline{$\left( \bm{x^{\ast}} \right)^{2}$}\end{large}}
\put(59.5,28){\begin{Large}$\ldots$\end{Large}}
\put(59.5,45){\begin{Large}$\ldots$\end{Large}}
\put(74,29){\begin{large}\underline{$\left( \bm{x^{\ast}} \right)^{B}$}\end{large}}
\put(21.25,32.75){\rotatebox{90}{$\xrightarrow{\hspace*{1.5cm}}$}}
\put(43.5,32.75){\rotatebox{90}{$\xrightarrow{\hspace*{1.5cm}}$}}
\put(76.25,32.75){\rotatebox{90}{$\xrightarrow{\hspace*{1.5cm}}$}}
\put(16.5,47.5){\begin{large}\underline{$s \left[\left( \bm{x^{\ast}} \right)^{1} \right]$}\end{large}}
\put(39,47.5){\begin{large}\underline{$s \left[\left( \bm{x^{\ast}} \right)^{2} \right]$}\end{large}}
\put(71.7,48.4){\begin{large}\underline{$s \left[\left( \bm{x^{\ast}} \right)^{B} \right]$}\end{large}}
 \end{overpic}
\caption[Basic schematic on the methodology of the bootstrap.]{The schematic illustrates the methodology of the bootstrap (Similar figures can be found in reference \cite{EfronTibshiraniBook}.). From the original data $\bm{x}$, an ensemble of $B$ bootstrap replications $\bm{x^ {\ast}}$ is generated by methods described in the main text. On each of the replicates, the statistic $s \left( \bm{x^ {\ast}} \right)$ can be evaluated just as on the original data. This generates an ensemble of bootstrap replicates of the statistic which can be processed for further analysis.}
\label{fig:BootstrapDrawingPic1}
\end{figure}

Here, the abbreviation {\it i.i.d.} stands for {\it independent and identically distributed sample drawn from $F$} \cite{EfronTibshiraniBook}. Whenever an arrow points from a probability distribution to some dataset in the following discussion, this is what is meant, without printing the abbreviation {\it i.i.d.} each time. \newline
Now, the bootstrap procedure for the estimation of the standard error of a statistical function, or just {\it statistic} $s (\bm{x})$ evaluated on the data, shall be described (cf. \cite{EfronTibshiraniBook}). A simple special case is given in case the statistic is the sample mean
\begin{equation}
 s \left( \bm{x} \right) = \bar{x} = \frac{1}{n} \sum_{i=1}^{n} x_{i} \mathrm{,} \label{eq:MeanOfXDefinition}
\end{equation}
but the procedure is more general. Of course, if (\ref{eq:MeanOfXDefinition}) is the statistic of interest, the correct formula for the standard error is well known \cite{EfronTibshiraniBook,DavisonHinkley}
\begin{equation}
 \widehat{\mathrm{se}} \left( \bar{x} \right) = \sigma_{\hat{F}} / \sqrt{n} = \sqrt{ \frac{1}{n^{2}} \sum \limits_{i=1}^{n} \left( x_{i} - \bar{x} \right)^{2} } \mathrm{.} \label{eq:StDevOfMeanOfXDefinition}
\end{equation}
The ``hat''-notation above indicates that $\widehat{\mathrm{se}} \left( \bar{x} \right)$ is a so called {\it plug-in estimator} \cite{EfronTibshiraniBook}, obtained by plugging the empirical distribution function $\hat{F}$ (obtained in the one-sample problem by applying the probabiltiy $1/n$ to each datapoint) into the definition of $\sigma_{F}$ as defined via the true underlying PDF\footnote{The abbreviation 'PDF' means 'probability distribution function'.} $F$. 
\begin{figure}[ht]
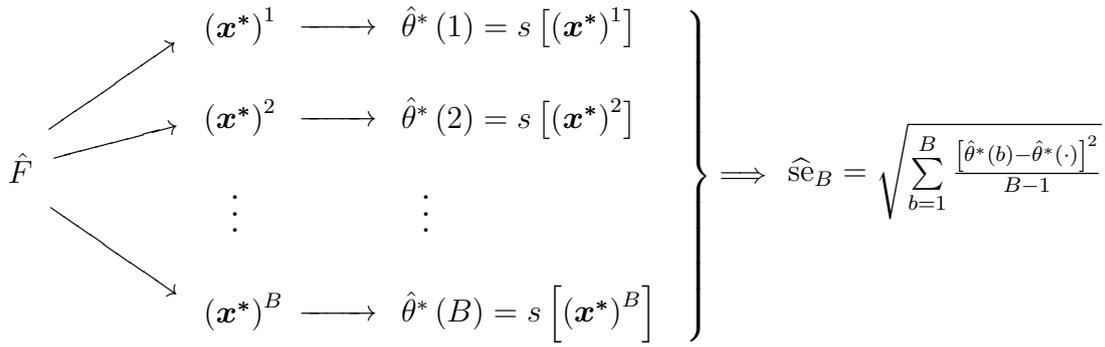

 \centering
\begin{overpic}[width=0.99\textwidth, trim = 0 310 0 0, clip]{blank_page.jpg}
\put(2,15.5){\begin{large}$\hat{F}$\end{large}}
\put(5.5,17){\rotatebox{15}{$\xrightarrow{\hspace*{1.5cm}}$}}
\put(4.75,19.5){\rotatebox{35}{$\xrightarrow{\hspace*{1.85cm}}$}}
\put(5.15,13){\rotatebox{325}{$\xrightarrow{\hspace*{1.85cm}}$}}
\put(19,28.28){\begin{large}$\left( \bm{x^{\ast}} \right)^{1}$\end{large}}
\put(19,20.28){\begin{large}$\left( \bm{x^{\ast}} \right)^{2}$\end{large}}
\put(19,3.28){\begin{large}$\left( \bm{x^{\ast}} \right)^{B}$\end{large}}
\put(21.55,14.75){\rotatebox{270}{\begin{Large}$\ldots$\end{Large}}}
\put(27.25,28.5){\rotatebox{0}{$\xrightarrow{\hspace*{0.8cm}}$}}
\put(27.25,20.5){\rotatebox{0}{$\xrightarrow{\hspace*{0.8cm}}$}}
\put(27.25,3.5){\rotatebox{0}{$\xrightarrow{\hspace*{0.8cm}}$}}
\put(36.15,28.35){\begin{large}$\hat{\theta}^{\ast} \left(1\right) = s \left[\left( \bm{x^{\ast}} \right)^{1}\right]$\end{large}}
\put(36.15,20.35){\begin{large}$\hat{\theta}^{\ast} \left(2\right) = s \left[\left( \bm{x^{\ast}} \right)^{2}\right]$\end{large}}
\put(36.15,3.35){\begin{large}$\hat{\theta}^{\ast} \left(B\right) = s \left[\left( \bm{x^{\ast}} \right)^{B}\right]$\end{large}}
\put(38.15,14.75){\rotatebox{270}{\begin{Large}$\ldots$\end{Large}}}
\put(60.15,15.5){$\begin{rcases}
   \\ \\ \\ \\ \\ \\ \\ \\
\end{rcases}\Longrightarrow $}
\put(69.75,15.65){\begin{large}$\widehat{\mathrm{se}}_{B} = \sqrt{ \sum \limits_{b=1}^{B} \frac{\left[ \hat{\theta}^{\ast} \left(b\right) - \hat{\theta}^{\ast} \left(\cdot\right) \right]^{2}}{B-1} }$\end{large}}
 \end{overpic}
\caption[Diagram illustrating the bootstrap method to estimate the standard error of a random sample, according to Efron and Tibshirani.]{This diagram (taken from \cite{EfronTibshiraniBook}) illustrates the bootstrap method to estimate the standard error of a random sample $\bm{x}$ drawn from an unknown probability distribution $F$. The emprical distribution function $\hat{F}$, obtained by applying the probability $1/n$ to each datapoint $x_{i}$, is used to generate an ensemble of $B$ booststrap replications $\bm{x^{\ast}}$ of the original data. \newline On each of the bootstrap replicates, the statistical function $s (\bm{x})$ is evaluated. Then, the standard error $\mathrm{se}_{B}$ is estimated by the standard deviation of the $B$ replications, with $\hat{\theta}^{\ast} \left(\cdot\right) = \sum_{b=1}^{B} \hat{\theta}^{\ast} \left(b\right)/B$.}
\label{fig:BootstrapDrawingPic2}
\end{figure}
The bootstrap can be used to get approximations for plug-in estimators such as (\ref{eq:StDevOfMeanOfXDefinition}), which in this example-case need not be done since the correct formula is already known. This will be elaborated further below. \newline
The basic method of the bootstrap is illustrated in Figure \ref{fig:BootstrapDrawingPic1}. Starting from the measured data $\bm{x}$, an ensemble of $B$ bootstrap replications $\left( \bm{x^{\ast}} \right)^{b}$ of the same size is generated, labeled by an index $b \in \left\{ 1, \ldots, B \right\}$. The statistic $s \left( \bm{x} \right)$ (in this case the mean (\ref{eq:MeanOfXDefinition})) can now be evaluated on each replicate in the same way as on the original data. \newline
Of course, the precise method of such a resampling of data should be elaborated. In case of the one-sample problem, the empirical PDF $\hat{F}$ can be implemented in a random number generator in order to yield the bootstrap replications. This means, one draws the bootstrap data $\bm{x^ {\ast}}$ numerically, according to
\begin{equation}
 \hat{F} \xrightarrow{\hspace*{0.75cm}} \bm{x^{\ast}} = \left( x^{\ast}_{1}, x^{\ast}_{2}, \ldots , x^{\ast}_{n} \right) \mathrm{.} \label{eq:BooststrapExampleOneSampleDrawnFromFHat}
\end{equation}
This is done $B$ times in order to yield the ensemble of bootstrap datasets. It is important that the bootstrap is defined by drawing with replacement. Therefore, it is possible that the same data-value appears multiple times in the bootstrap dataset as a result of (\ref{eq:BooststrapExampleOneSampleDrawnFromFHat}). \newline
The evaluation of the statistic $s (\bm{x})$ on each of the bootstrap datasets yields a distribution of values comprised by the $B$ bootstrap estimators $\hat{\theta}^{\ast} \left(b\right) = s \left[\left( \bm{x^{\ast}} \right)^{b}\right]$, $b = 1,\ldots,B$. The value of the plug-in estimator $\widehat{\mathrm{se}} \left( \bar{x} \right)$ can now be approximated via the evaluation of the standard deviation of the $B$ bootstrap replications \cite{EfronTibshiraniBook}
\begin{equation}
 \widehat{\mathrm{se}}_{B} = \sqrt{ \sum \limits_{b=1}^{B} \frac{\left[ \hat{\theta}^{\ast} \left(b\right) - \hat{\theta}^{\ast} \left(\cdot\right) \right]^{2}}{B-1} } \mathrm{,} \label{eq:StDevOfMeanOfBootstrapDistribution}
\end{equation}
with $\hat{\theta}^{\ast} \left(\cdot\right) = \sum_{b=1}^{B} \hat{\theta}^{\ast} \left(b\right)/B$. The procedure is illustrated in Figure \ref{fig:BootstrapDrawingPic2}. \newpage
\begin{figure}[ht]
 \centering
 \vspace*{-15pt}
\begin{overpic}[width=0.476\textwidth, trim = 0 30 0 0, clip]{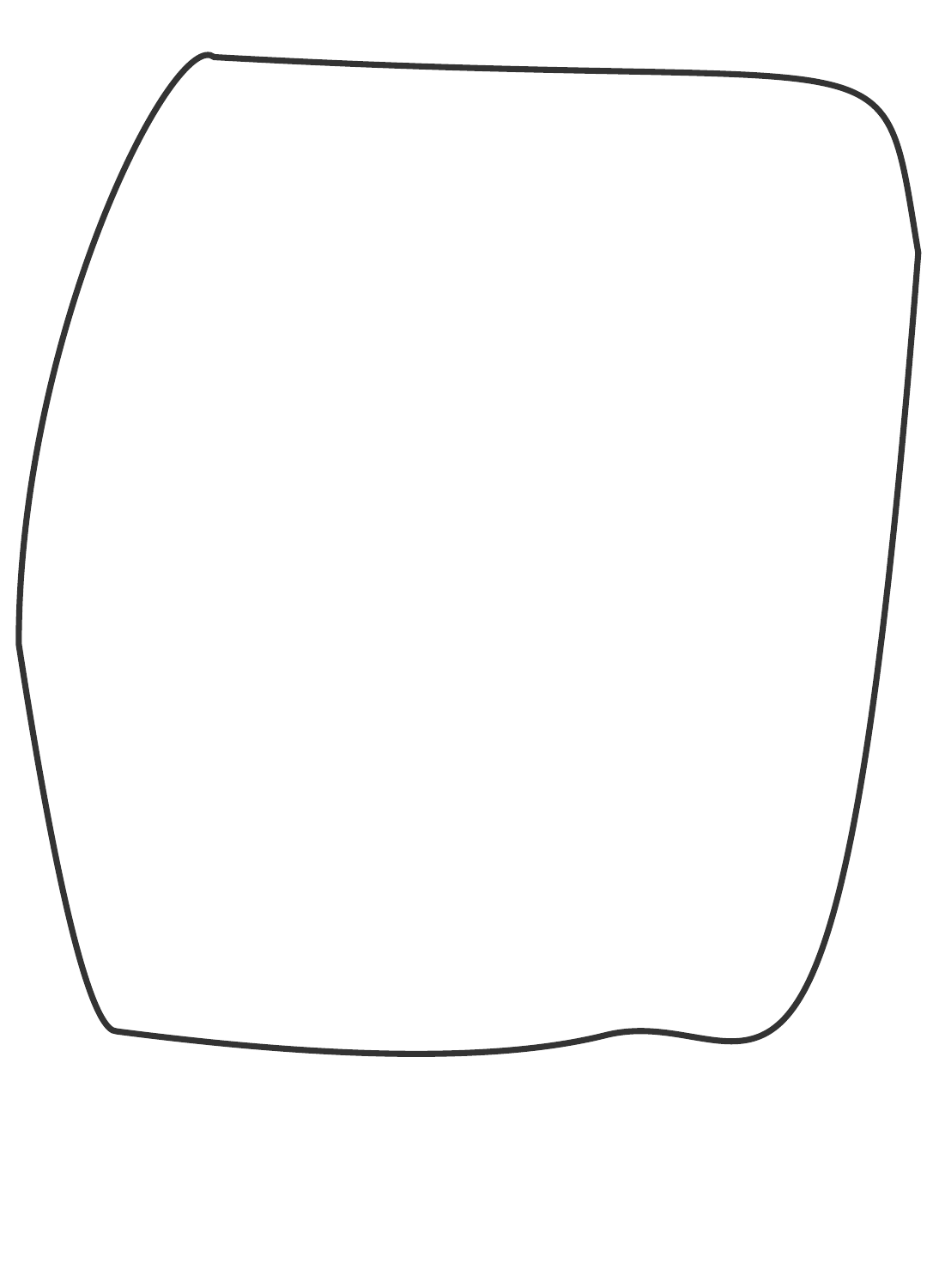}
\put(25,81.885){\begin{large}REAL WORLD\end{large}}
\put(14,51.65){$\mathrm{P}$}
\put(19.75,52.0){$\xrightarrow{\hspace*{1cm}}$}
\put(37,52.0){$\bm{x} = \left( x_{1}, \ldots, x_{n} \right)$}
\put(71.15,52.0){$\xLongrightarrow{\hspace*{1.55cm}}$}
\put(37.5,49){\rotatebox{270}{$\xrightarrow{\hspace*{0.8cm}}$}}
\put(33.35,30.9){$\hat{\theta} = s \left( \bm{x} \right)$}
\put(25,22.5){Statistic of interest}
\put(40,65.5){Observed Data}
\put(11.5,71.0){Unknown}
\put(9.5,65.5){Probability}
\put(11.5,60.0){Model}
 \end{overpic} \hspace*{0pt}
\begin{overpic}[width=0.510\textwidth, trim = 0 30 0 0, clip]{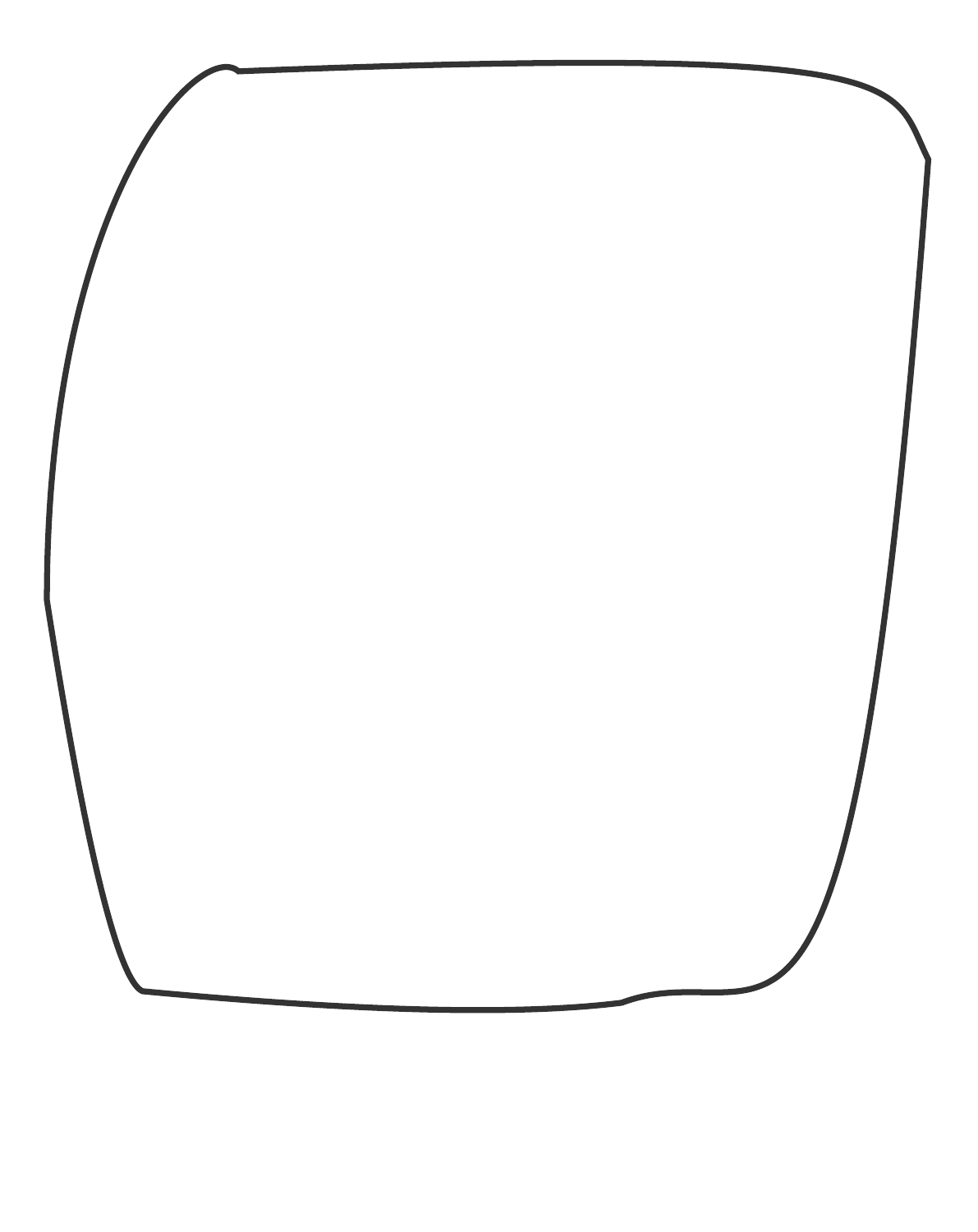}
\put(21.5,83.3){\begin{large}BOOTSTRAP WORLD\end{large}}
\put(15,52.5){$\hat{\mathrm{P}}$}
\put(21,53.2){$\xrightarrow{\hspace*{1.1cm}}$}
\put(40,53.25){$\bm{x^{\ast}} = \left( x^{\ast}_{1}, \ldots, x^{\ast}_{n} \right)$}
\put(40.5,51){\rotatebox{270}{$\xrightarrow{\hspace*{0.8cm}}$}}
\put(34.15,31.9){$\hat{\theta}^{\ast} = s \left( \bm{x^{\ast}} \right)$}
\put(23.5,22.5){Bootstrap Replication}
\put(40,67.2){Bootstrap Sample}
\put(12.5,73.0){Estimated}
\put(10.5,67.5){Probability}
\put(12.5,62.0){Model}
 \end{overpic} \\ \vspace*{-0.65cm}
\caption[Schematic picture illustrating the general philosophy of the bootstrap. The figure is a facsimile of a picture in the book by Efron and Tibshirani.]{This diagram, being a replica of Figure 8.3 in reference \cite{EfronTibshiraniBook}, illustrates the philosophy and wide range of applicability of the bootstrap. \newline A general probability model P can generate real world data $\bm{x}$ of general data-structure. In case the probability mechanism P can be estimated based solely from the observed data, in the crucial step ``$\Rightarrow$'' leading to $\hat{\mathrm{P}}$, bootstrap datasets can be generated and a resampling analysis performed. The arrow pointing upside down, i.e. the evaluation of the statistic, proceeds on the bootstrap data in exactly the same way as it does on the real data.} 
\label{fig:BootstrapPhilosophyPics}
\end{figure}
As the size of the bootstrap ensemble $B$ is increased, the estimate (\ref{eq:StDevOfMeanOfBootstrapDistribution}) will approach the result of the known formula (\ref{eq:StDevOfMeanOfXDefinition}). This is an important point: in the limit $B \rightarrow \infty$, the bootstrap can approximate plug-in estimators arbitrarily well, provided that all the underlying assumptions are correct. In other words, plug-in estimates are ideal bootstrap estimates. However, resampling with very large $B$ can not make the data itself better, by lowering for example the exact plug-in standard error $\widehat{\mathrm{se}} \left( \bar{x} \right)$. This is only logical. A higher resampling effort can make bootstrap estimates based on the observed data better. Since the bootstrap itself is a procedure based on the observed data, it cannot make the data better. \newline
The virtue of the bootstrap is that, in case of one-sample problems, it can be applied to the estimation of standard errors even in cases where no exact formula such as (\ref{eq:StDevOfMeanOfXDefinition}) is known \cite{EfronOriginal,EfronTibshiraniBook}. Furthermore, it can also be used in case the statistic $s \left( \bm{x} \right)$ is not evaluated by an analytic formula, but may be a result of complicated numerical calculations. \newline

The bootstrap itself is applicable to very general situations, ranging beyond the one-sample problems discussed until now. Shown here is a facsimile of a picture by Efron and Tibshirani \cite{EfronTibshiraniBook} in Figure \ref{fig:BootstrapPhilosophyPics}, which is well-suited to make this fact clear. \newline
One can think of data $\bm{x}$ given in very general data-structures, which could be a list of $n$ $m$-tupels, e.g. in the simplest case a list of $n$ pairs $\left( x_{i}, y_{i} \right)$, or even more complicated structures. These data are then thought of as being the result of sampling once from an underlying general probability-theoretic model P. In most cases discussed in books \cite{EfronTibshiraniBook}, the model P may be a collection of multiple PDFs of possibly different kind, giving rise to some subset of the general data-structure via random sampling. \newline
The crucial step in the bootstrap analysis scheme is now the inference of the empirical probability model $\hat{\mathrm{P}}$, solely on the basis of the observed data. This is illustrated as
\begin{equation}
 \bm{x} \Longrightarrow \hat{\mathrm{P}} \mathrm{,} \label{eq:DataYieldProbModel}
\end{equation}
in Figure \ref{fig:BootstrapPhilosophyPics}. For this step, no general procedure can be found. However, on the basis of the observed data, it is in most cases quickly performed by imposing quite natural assumptions, like picking the empirical PDF $\hat{F}$ in case of the one-sample problem. \newline
However, one has to mention here that in the step (\ref{eq:DataYieldProbModel}), there lies some source of non-uniqueness of the bootstrap. This means that, especially in the case of the application to fits (see below), there is not one unique way to bootstrap from observed data. However, once a sensible Ansatz for solving the step (\ref{eq:DataYieldProbModel}) is found, the remaining steps on how to do a bootstrap analysis are standardized (cf. Figures \ref{fig:BootstrapDrawingPic1}, \ref{fig:BootstrapDrawingPic2} and \ref{fig:BootstrapPhilosophyPics}). \newline
This illustrates the main strength of the bootstrap, which is its wide range of applicability, independent of any specific probability-theoretical assumptions. However, the procedure also has its drawbacks. \newline
The most obvious one is the increase of the computational cost to do an analysis. The time needed to do the analysis of the original data needs to be multiplied by $B$ (modulo possible gains by parallelization-techniques) for the bootstrap-analysis. Furthermore, there are many situations where a bootstrap analysis can go wrong \cite{EfronTibshiraniBook}, for instance when auto-correlations are present in the data, but underestimated or not taken into account properly in the bootstrap analysis. \newline
For the sake of fairness, it should be stated here that if auto-correlations in the data are known and quantified, there exist approaches for working such correlations directly into the probability model $\hat{\mathrm{P}}$ used for the bootstrap \cite{EfronTibshiraniBook, DavisonHinkley}. \newline

As mentioned above, the bootstrap can be applied even in cases where no theoretical formula for the standard error is known and the statistic of interest $s (\bm{x})$ is obtained from the data by means of complicated numerical calculations. An important example of such cases is given by fits of non-linear models to data, of which the TPWA-fit discussed here is again a special case. \newline

Two main approaches for the application of the bootstrap to fits are the so-called {\it residual-resampling} and {\it case-resampling} techniques \cite{EfronTibshiraniBook, DavisonHinkley}. \newline
When a model function $f \left( x ; \beta \right)$, defined by some parameter-vector $\beta$, is fitted to a distribution of data $\left\{ \left( x_{i}, y_{i} \right) \Big| i = 1,\ldots,n \right\}$ (disregarding errors and weighted fits for the moment), the so-called raw residual-estimates
\begin{equation}
 \hat{\epsilon}_{i} = y_{i} - f \left( x_{i} ; \hat{\beta} \right) \mathrm{,} \label{}
\end{equation}
can be evaluated. Here, $\hat{\beta}$ denotes the estimates for the fit-parameters returned by the fit-routine. Residual-resampling would now proceed by applying the probability $1/n$ to every residual and drawing with replacement $n$ residuals $\left\{ \hat{\epsilon}^{\ast}_{1}, \ldots, \hat{\epsilon}^{\ast}_{n} \right\}$ from the residual-vector. Then, the bootstrap-data would be generated by adding each drawn residual $\hat{\epsilon}^{\ast}_{i}$ (or a pro\-per\-ly re-normalized version thereof) to the fit-function $f (x_{i}; \hat{\beta})$. Re-fits are then performed and bootstrap distributions for e.g. the fit-parameters extracted. \newline
The case-resampling proceeds by drawing directly from the data. The probability $1/n$ is given to each datapoint and then bootstrap datasets are generated by drawing with replacement from the original data, just as in the one-sample example discussed above. \newline

The issue with the application of both the above mentioned methods to the TPWA-problem lies in the heteroscedasticity. The variances of the polarization data are certainly not constant over the angular distributions. It is generally difficult, if not impossible, to describe the variation of their variances by some estimated function. Furthermore, weighted fits are performed to the data, just as defined in equations (\ref{eq:LegCoeffFitChi2DataFit}) and (\ref{eq:ChiSquareDirectFit}) of section \ref{sec:TPWAFitsIntro}. This makes the application of both above mentioned textbook-methods troublesome. \newline

\begin{figure}[ht]
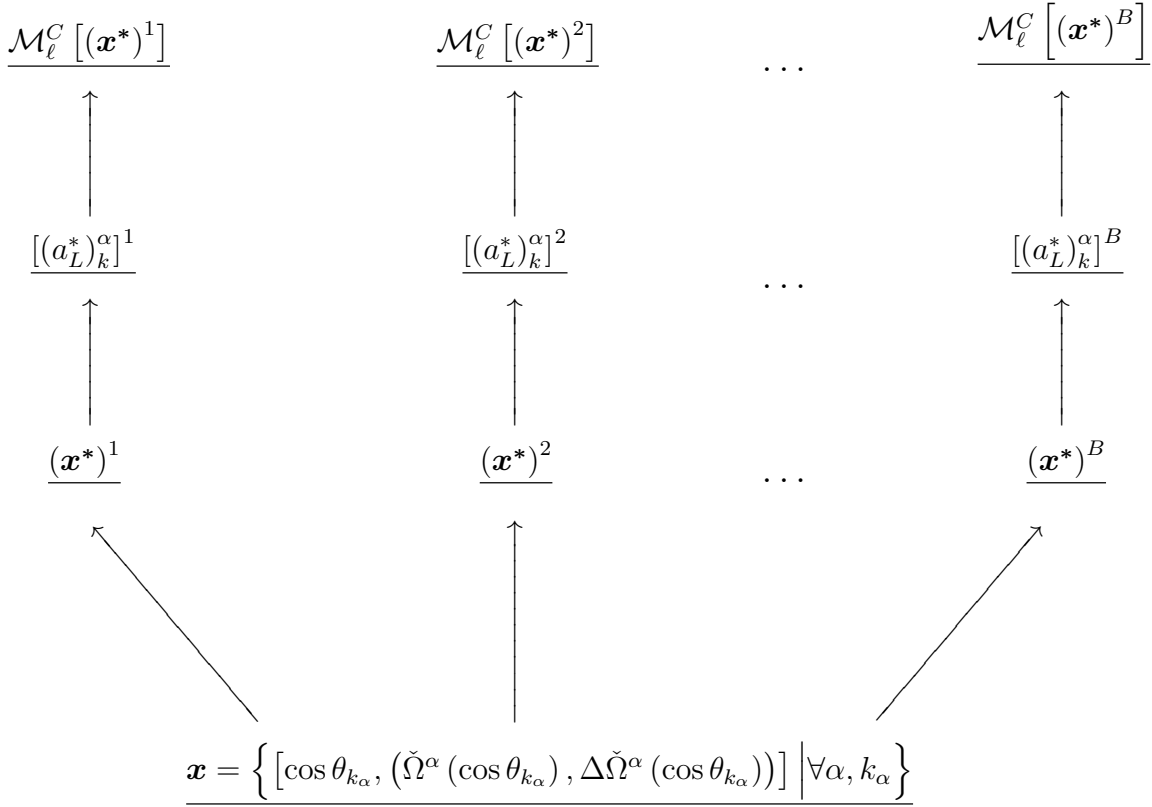

 \centering
\begin{overpic}[width=0.99\textwidth, trim = 0 20 0 0, clip]{blank_page.jpg}
\put(16,4.0){\begin{large}\underline{$\bm{x} = \left\{ \left[ \cos \theta_{k_{\alpha}}, \left( \check{\Omega}^{\alpha} \left( \cos \theta_{k_{\alpha}} \right), \Delta \check{\Omega}^{\alpha} \left( \cos \theta_{k_{\alpha}} \right) \right) \right] \Big| \forall \alpha , k_{\alpha} \right\}$}\end{large}}
\put(7,9.25){\rotatebox{130}{$\xrightarrow{\hspace*{3.2cm}}$}}
\put(43.5,8.75){\rotatebox{90}{$\xrightarrow{\hspace*{2.5cm}}$}}
\put(75,8.5){\rotatebox{50}{$\xrightarrow{\hspace*{3.2cm}}$}}
\put(4.0,31){\begin{large}\underline{$\left( \bm{x^{\ast}} \right)^{1}$}\end{large}}
\put(41.5,31){\begin{large}\underline{$\left( \bm{x^{\ast}} \right)^{2}$}\end{large}}
\put(66,30){\begin{Large}$\ldots$\end{Large}}
\put(66,47){\begin{Large}$\ldots$\end{Large}}
\put(66,66){\begin{Large}$\ldots$\end{Large}}
\put(89,31){\begin{large}\underline{$\left( \bm{x^{\ast}} \right)^{B}$}\end{large}}
\put(6.6,34.75){\rotatebox{90}{$\xrightarrow{\hspace*{1.5cm}}$}}
\put(43.5,34.75){\rotatebox{90}{$\xrightarrow{\hspace*{1.5cm}}$}}
\put(91,34.75){\rotatebox{90}{$\xrightarrow{\hspace*{1.5cm}}$}}
\put(2.5,49.5){\begin{large}\underline{$ \left[ \left( a_{L}^{\ast} \right)^{\alpha}_{k} \right]^{\hspace*{0.5pt}1}$}\end{large}}
\put(40,49.5){\begin{large}\underline{$ \left[ \left( a_{L}^{\ast} \right)^{\alpha}_{k} \right]^{\hspace*{0.5pt}2}$}\end{large}}
\put(87.7,49.5){\begin{large}\underline{$ \left[ \left( a_{L}^{\ast} \right)^{\alpha}_{k} \right]^{\hspace*{0.5pt}B}$}\end{large}}
\put(6.6,53){\rotatebox{90}{$\xrightarrow{\hspace*{1.5cm}}$}}
\put(43.5,53){\rotatebox{90}{$\xrightarrow{\hspace*{1.5cm}}$}}
\put(91,53){\rotatebox{90}{$\xrightarrow{\hspace*{1.5cm}}$}}
\put(0.5,67.75){\begin{large}\underline{$\mathcal{M}_{\ell}^{C} \left[ \left( \bm{x^{\ast}} \right)^{1} \right]$}\end{large}}
\put(37.75,67.75){\begin{large}\underline{$\mathcal{M}_{\ell}^{C} \left[ \left( \bm{x^{\ast}} \right)^{2} \right]$}\end{large}}
\put(84.75,68.75){\begin{large}\underline{$\mathcal{M}_{\ell}^{C} \left[ \left( \bm{x^{\ast}} \right)^{B} \right]$}\end{large}}
 \end{overpic}
\caption[The bootstrap-methodology, applied to the TPWA-problem.]{This diagram is the generalization of Figure \ref{fig:BootstrapDrawingPic1} to the case of a TPWA. The data-structure met in the TPWA is shown at the bottom. From these data, an ensemble of $B$ bootstrap replications is generated as described in the main text. For each bootstrap dataset, the TPWA step $\mathrm{I}$ leads to $B$ bootstrap replications of the Legendre-coefficients $\left( a_{L}^{\ast} \right)^{\alpha}_{k}$ (and covariance-matrices, $\ldots$) needed as input for fit step $\mathrm{II}$. Then, the performance of TPWA fit step $\mathrm{II}$ leads to $B$ replications of the statistics of interest, in this case the components of the multipole parameter-vector resulting from the fit.}
\label{fig:BootstrapDrawingPic3}
\end{figure}

\begin{figure}[ht]
 \centering
 \begin{minipage}{0.499\textwidth}
\begin{overpic}[width=1.0\textwidth]{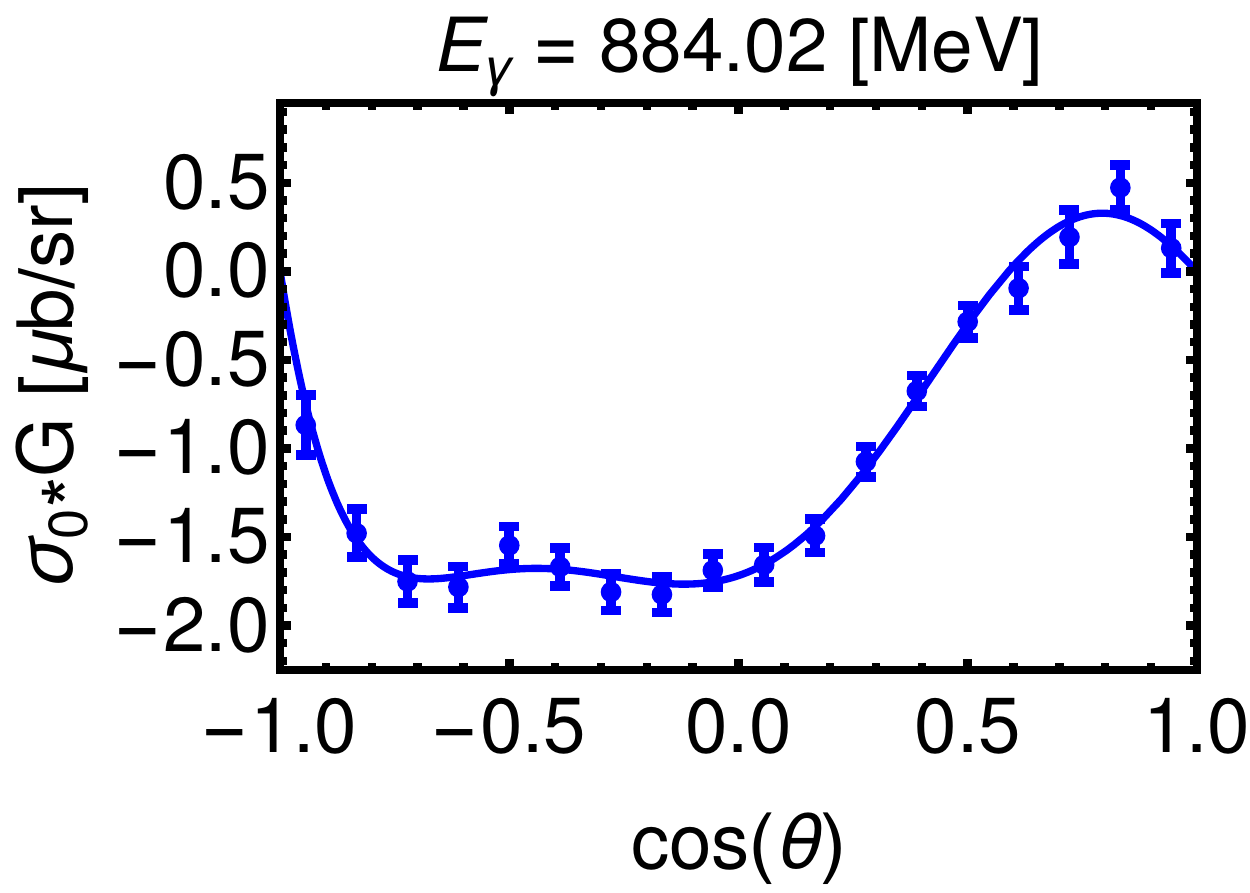}
 \end{overpic} \hspace*{20pt}
 \end{minipage}%
 \begin{minipage}{0.499\textwidth} \vspace*{-20pt}
 \hspace*{35pt}\begin{overpic}[width=0.800\textwidth]{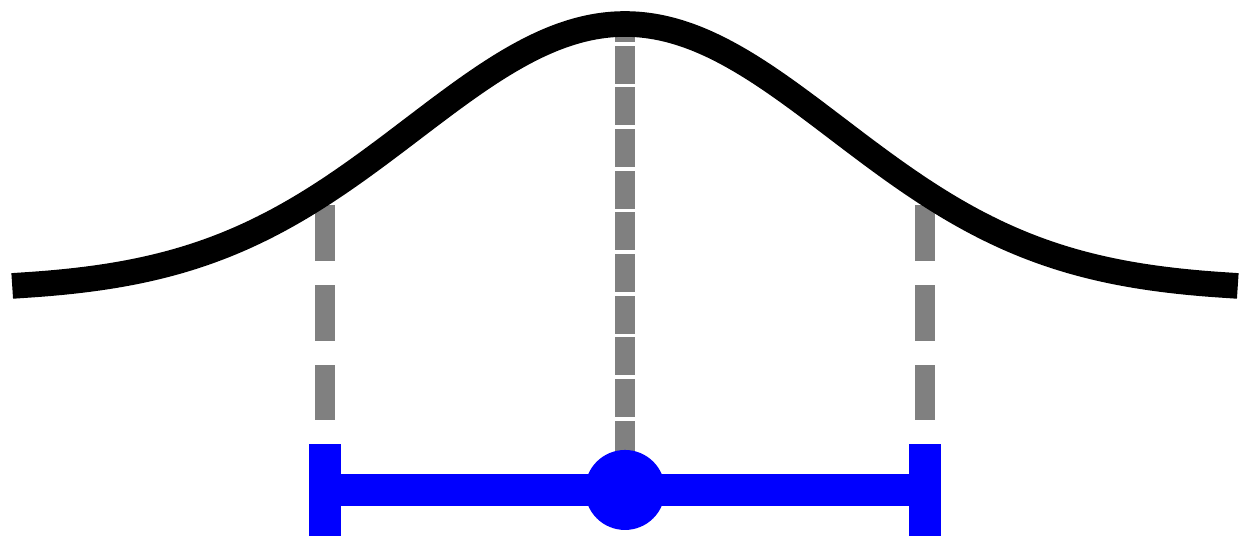}
 \end{overpic} \vspace*{10pt}
 \end{minipage} \\
 \begin{minipage}{0.499\textwidth}
\begin{overpic}[width=1.0\textwidth]{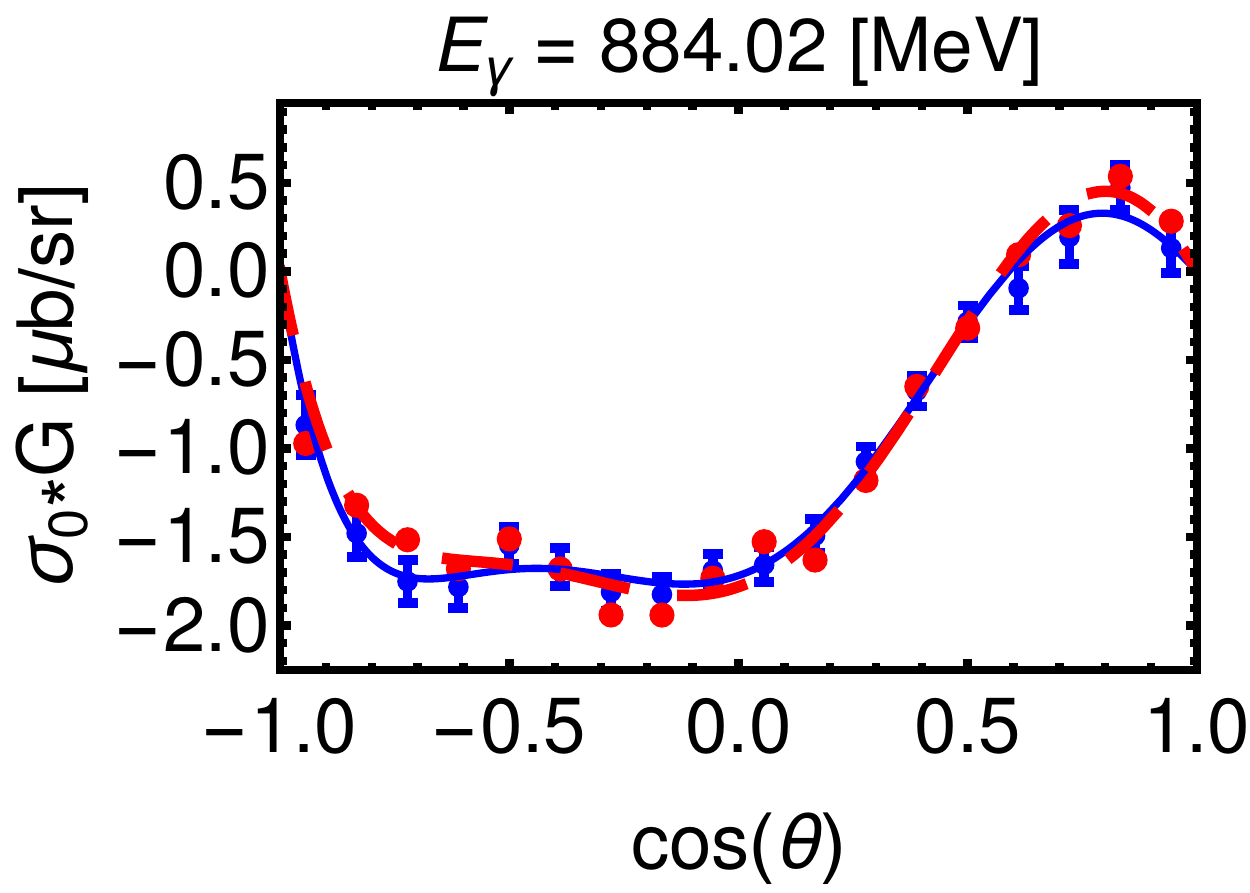}
 \end{overpic}
 \end{minipage}%
 \begin{minipage}{0.499\textwidth}
\begin{overpic}[width=1.0\textwidth]{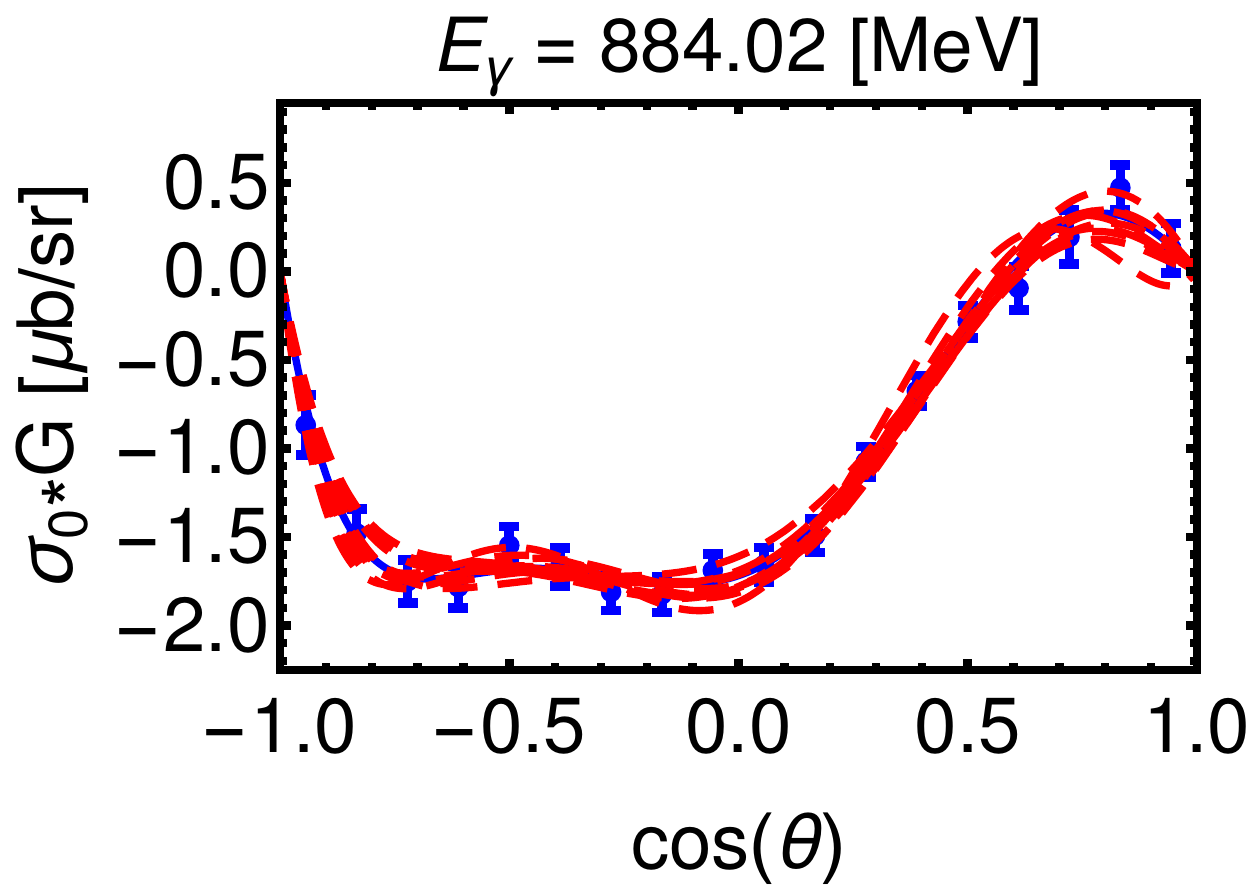}
 \end{overpic}
  \end{minipage}
\caption[The parametric bootstrap-Ansatz, illustrated on a recent measurement of the double-polarization observable $G$.]{The parametric bootstrap outlined in the main text is illustrated here. Shown are fits to a recent measurement of the polarization observable $G$ by the CBELSA/TAPS collaboration \cite{Thiel:2012,Thiel:2016}. Values for the profile function $\check{G} = \sigma_{0} \hspace*{1.5pt} G$ have been evaluated using a very recent cross section measurement from MAMI \cite{Adlarson:2015}. \newline The plot in the upper left shows the fit to the original data. A Gaussian distribution function is defined at each data-point, as illustrated in the upper right. The plot in the lower left shows the original data, plus the first bootstrap replicate (errors for the latter are not shown), as well as the corresponding first bootstrap fit (red dashed line). A similar plot, showing fits to the first $10$ bootstrap replicates (no bootstrap data shown this time), can be seen in the lower right.} 
\label{fig:BootstrapIllustrationUsingGObservable}
\end{figure}

Instead, we resort here to a version of the bootstrap employed often by physicists, but for which no reference seems to exist in the statistics literature \cite{UrbachPrivComm}. \newline
We carry the bootstrap methodology over to the TPWA-problem in a {\it parametric bootstrap} approach \cite{UrbachScript}. The application of similar, but in detail not exactly the same, methods to the TPWA has been discussed by Sandorfi in a talk at a recent $\mathrm{ECT}^{\ast}$-conference \cite{SandorfiTrentoTalk}. \newline
In order to employ the parametric bootstrap, it is useful to recapitulate the precise data-structure given in the TPWA. At each fixed energy, we have a data-vector of the following form
\begin{equation}
 \bm{x} = \left\{ \left[ \cos \theta_{k_{\alpha}}, \left( \check{\Omega}^{\alpha} \left( \cos \theta_{k_{\alpha}} \right), \Delta \check{\Omega}^{\alpha} \left( \cos \theta_{k_{\alpha}} \right) \right) \right] \Big| \forall \alpha , k_{\alpha} \right\} \mathrm{.} \label{eq:TPWADataStructure}
\end{equation}
Here, the index $\alpha$ runs over all observables included into the fit and different indices $k_{\alpha}$ for the angular points accommodate the fact that the statistics of the angular distributions can be quite different from measurement to measurement. \newline
Note that in the vector (\ref{eq:TPWADataStructure}), the observables have already been written in the form of profile functions. The error $\Delta \check{\Omega}^{\alpha}$ denotes the purely statistical error. In case the data are provided as dimensioned profile functions $\check{\Omega}^{\alpha}$, one can just pick the given error for the latter. If one is provided with data for a dimensionless asymmetry $\Omega^{\alpha}$, a suitable (in this case: gaussian, cf. comments in appendix \ref{sec:BootstrapAnsatzComments}) error propagation has to be performed in order to arrive at the error in (\ref{eq:TPWADataStructure}). \newline
To perform a bootstrap analysis to the TPWA, it is clear that again an ensemble of bootstrap replicates has to be generated from the original data and the TPWA fit steps $\mathrm{I}$ and $\mathrm{II}$ have to be applied to each replicate individually. Figure \ref{fig:BootstrapDrawingPic3} illustrates this for the TPWA data-structure. \newline
Typically, a dataset is provided with one number for the statistical error for each datapoint. This is the case for all datasets considered in this thesis. The standard-assumption one can make, in case no different information is given in conjunction with the dataset, is that of a standard normal distribution for each datapoint, with standard deviation given by the statistical error.
Therefore in this work, the Ansatz for the solution of (\ref{eq:DataYieldProbModel}), i.e. the inference of a suitable probability model $\hat{\mathrm{P}}$ from the observed data, is to parametrize a PDF $\hat{F}$ at each datapoint in the following way
\begin{equation}
 \hat{F} := \mathcal{N} \left[\check{\Omega}^{\alpha} \left( \cos \theta_{k_{\alpha}} \right) , \Delta \check{\Omega}^{\alpha} \left( \cos \theta_{k_{\alpha}} \right) \right] \mathrm{,} \label{eq:TPWAEmpDistrFunctDefinition}
\end{equation}
where, in order to keep the discussion self contained, the well-known definition of a normalized gaussian distribution is
\begin{equation}
 \mathcal{N} \left( \mu, \sigma \right) = \frac{1}{\sqrt{2 \pi} \sigma} \exp \left[ - \frac{1}{2} \frac{\left( x - \mu \right)^{2}}{\sigma^{2}} \right] \mathrm{.} \label{eq:NormalDistDef}
\end{equation}
To be specific, the bootstrap datasets generated from this distribution have the form
\begin{equation}
 \bm{x^{\ast}} = \left\{ \left[ \cos \theta_{k_{\alpha}}, \left( \check{\Omega}^{\alpha}_{\ast} \left( \cos \theta_{k_{\alpha}} \right), \Delta \check{\Omega}^{\alpha}_{\ast} \left( \cos \theta_{k_{\alpha}} \right) \right) \right] \Big| \forall \alpha , k_{\alpha} \right\} \mathrm{.} \label{eq:TPWABootstrapDataset}
\end{equation}
Each bootstrap datapoint is drawn from the normal distribution
\begin{equation}
 \mathcal{N} \left[ \check{\Omega}^{\alpha} \left( \cos \theta_{k_{\alpha}} \right) , \Delta \check{\Omega}^{\alpha} \left( \cos \theta_{k_{\alpha}} \right) \right] \longrightarrow \check{\Omega}^{\alpha}_{\ast} \left( \cos \theta_{k_{\alpha}} \right) \mathrm{,} \label{eq:BootstrapDataPointDef}
\end{equation}
while the standard error $\Delta \check{\Omega}^{\alpha}$ of the original datapoint is just re-applied to the bootstrap datapoint
\begin{equation}
 \Delta \check{\Omega}^{\alpha}_{\ast} \left( \cos \theta_{k_{\alpha}} \right) \equiv \Delta \check{\Omega}^{\alpha} \left( \cos \theta_{k_{\alpha}} \right) \mathrm{.} \label{eq:BootstrapDataPointErrorDef}
\end{equation}
In this way, an ensemble of $B$ bootstrap datasets is obtained, i.e. $\left( \bm{x^{\ast}} \right)^{b}$ for $b \in \left\{ 1, \ldots, B \right\}$. The TPWA is performed for each one of them (see Figure \ref{fig:BootstrapIllustrationUsingGObservable} for an illustration of the procedure on a particular dataset)
\begin{equation}
 \left( \bm{x^{\ast}} \right)^{b} \overset{\mathrm{I}}{\longrightarrow} \left[ \left( a_{L}^{\ast} \right)^{\alpha}_{k} \right]^{\hspace*{0.5pt}b} \overset{\mathrm{II}}{\longrightarrow} \left[ \left( \mathcal{M}_{\ell}^{C} \right)^{\ast} \right]^{b} \equiv \mathcal{M}_{\ell}^{C} \left[ \left( \bm{x^{\ast}} \right)^{b} \right] \mathrm{.} \label{eq:BootstrapFitStepOneAndTwo}
\end{equation}
\newpage

All this leads to parameter-distributions for the statistics of interest, which are in this case given by the multipole-fit-parameters. The bootstrap distributions are built from the B numbers $\hat{\theta}^{\ast}_{i} \left( b \right) = \left[\left(\mathcal{M}_{\ell}^{C} \right)^{\ast} \right]^{b}_{i}$, where each index $i$ labels one component of the multipole parameter-vector. For a standard TPWA-fit with phase-constrained multipoles, it therefore takes the values $i = 1,\ldots,(8 L - 1)$ (cf. sections \ref{sec:CompExpsTPWA} and \ref{sec:TPWAFitsIntro}). \newline
It is now advisable to investigate and further process the multipole parameter-distributions by use of a suitable histogram-class. MATHEMATICA \cite{Mathematica8,Mathematica11,MathematicaLanguage,MathematicaBonnLicense} provides a good collection of such classes. ROOT would be another applicable software framework. Further comments on the validity of this particular Ansatz for the bootstrap probability model are given in appendix \ref{sec:BootstrapAnsatzComments}. 
\clearpage

\begin{figure}[hb]
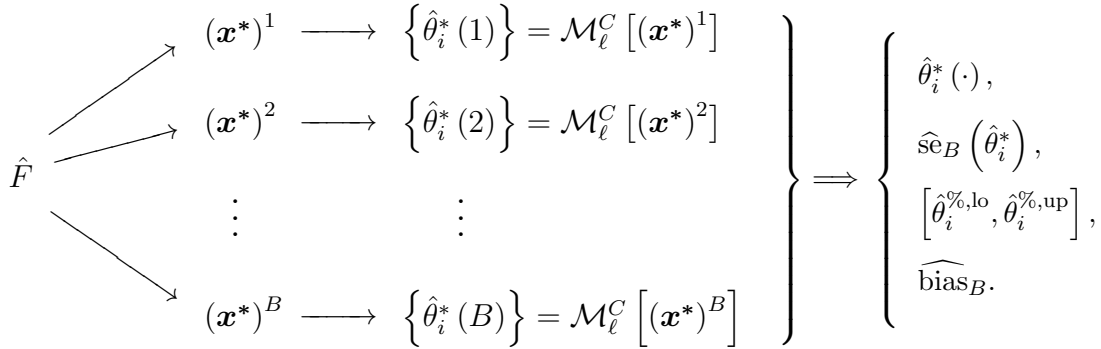

 \centering
\begin{overpic}[width=0.99\textwidth, trim = 0 290 0 0, clip]{blank_page.jpg}
\put(3,15.5){\begin{large}$\hat{F}$\end{large}}
\put(6.5,17){\rotatebox{15}{$\xrightarrow{\hspace*{1.5cm}}$}}
\put(5.75,19.5){\rotatebox{35}{$\xrightarrow{\hspace*{1.85cm}}$}}
\put(6.15,13){\rotatebox{325}{$\xrightarrow{\hspace*{1.85cm}}$}}
\put(20,28.18){\begin{large}$\left( \bm{x^{\ast}} \right)^{1}$\end{large}}
\put(20,20.35){\begin{large}$\left( \bm{x^{\ast}} \right)^{2}$\end{large}}
\put(20,3.35){\begin{large}$\left( \bm{x^{\ast}} \right)^{B}$\end{large}}
\put(22.55,14.75){\rotatebox{270}{\begin{Large}$\ldots$\end{Large}}}
\put(28.25,28.5){\rotatebox{0}{$\xrightarrow{\hspace*{0.8cm}}$}}
\put(28.25,20.5){\rotatebox{0}{$\xrightarrow{\hspace*{0.8cm}}$}}
\put(28.25,3.5){\rotatebox{0}{$\xrightarrow{\hspace*{0.8cm}}$}}
\put(37.15,28.35){\begin{large}$\left\{\hat{\theta}^{\ast}_{i} \left(1\right)\right\} = \mathcal{M}_{\ell}^{C} \left[ \left( \bm{x^{\ast}} \right)^{1} \right]$\end{large}}
\put(37.15,20.35){\begin{large}$\left\{\hat{\theta}^{\ast}_{i} \left(2\right)\right\} = \mathcal{M}_{\ell}^{C} \left[ \left( \bm{x^{\ast}} \right)^{2} \right]$\end{large}}
\put(37.15,3.35){\begin{large}$\left\{\hat{\theta}^{\ast}_{i} \left(B\right)\right\} = \mathcal{M}_{\ell}^{C} \left[ \left( \bm{x^{\ast}} \right)^{B} \right]$\end{large}}
\put(42.15,14.75){\rotatebox{270}{\begin{Large}$\ldots$\end{Large}}}
\put(69.15,15.5){$\begin{rcases}
   \\ \\ \\ \\ \\ \\ \\ \\
\end{rcases}\Longrightarrow $}
\put(78.15,15.5){$\begin{cases}
   \\ \\ \\ \\ \\ \\ \\
\end{cases} $}
\put(82,24.5){$\hat{\theta}^{\ast}_{i} \left( \cdot \right) \mathrm{,}$}
\put(82,18.5){$\widehat{\mathrm{se}}_{B} \left( \hat{\theta}^{\ast}_{i} \right) \mathrm{,}$}
\put(82,12.5){$\left[  \hat{\theta}_{i}^{\%, \mathrm{lo}}, \hat{\theta}_{i}^{\%, \mathrm{up}} \right] \mathrm{,}$}
\put(82,6.5){$\widehat{\mathrm{bias}}_{B} \mathrm{.}$}
 \end{overpic}
\caption[Diagram illustrating the general TPWA bootstrap-procedure.]{This diagram describes the TPWA bootstrap-procedure. The parametrization of the empirical PDF $\hat{F} = \mathcal{N} \left( \check{\Omega}^{\alpha}, \Delta \check{\Omega}^{\alpha} \right)$ is applied to each datapoint and used to generate bootstrap datasets. The TPWA is then applied to each one of them. The resulting distributions for the statistics of interest, in this case the multipole-parameters, are used to extract statistical quantities. The latter are listed on the right of the picture. They represent the final result of the bootstrap.}
\label{fig:BootstrapDrawingPic4}
\end{figure}

The bootstrap applied to the TPWA-problem is illustrated in Figure \ref{fig:BootstrapDrawingPic4}. This picture also denotes some quantities of interest derivable from the resulting parameter distributions. Those are:
\begin{itemize}
 \item[(i)] The mean $\hat{\theta}^{\ast}_{i} \left( \cdot \right)$ of each parameter distribution.
 \item[(ii)] The standard error $\widehat{\mathrm{se}}_{B} \left( \hat{\theta}^{\ast}_{i} \right)$, derived from each parameter distribution. This quantity can be used as a first measure for the statistical uncertainty of each multipole-parameter.
 \item[(iii)] Uncertainties may also be quantified by defining confidence intervals from the bootstrap distributions of the fit parameters. These are defined from the percentiles of the corresponding bootstrap histograms \cite{EfronTibshiraniBook,DavisonHinkley}. The so called {\it approximate $(1 - 2 \alpha)$ percentile interval} is defined in the following way \cite{EfronTibshiraniBook}
 \begin{equation}
  \left[  \hat{\theta}_{i}^{\%, \mathrm{lo}}, \hat{\theta}_{i}^{\%, \mathrm{up}} \right] = \left[  \left(\hat{\theta}^{\ast}_{i}\right)^{(\alpha)}_{B}, \left(\hat{\theta}^{\ast}_{i}\right)^{(1 - \alpha)}_{B} \right] \mathrm{.} \label{eq:PercentileConfidenceIntervalDef}
 \end{equation}
 Here, the quantity $\left(\hat{\theta}^{\ast}_{i}\right)^{(\alpha)}_{B}$ is the 100 $\cdot$ $\alpha$th percentile of the distribution provided by the $B$ numbers $\hat{\theta}^{\ast}_{i} \left( b \right)$, i.e. the $B$ $\cdot$ $\alpha$th value in an ordered list of the $B$ replications of $\hat{\theta}^{\ast}_{i}$ \cite{EfronTibshiraniBook}. In analogy, $\left(\hat{\theta}^{\ast}_{i}\right)^{(1 - \alpha)}_{B}$ denotes the 100 $\cdot$ $(1 - \alpha)$th percentile. In MATHEMATICA \cite{MathematicaLanguage}, good measures for these numbers can be directly extracted as the quantiles\footnote{For the formal definition of a quantile, see section \ref{sec:PseudoDataWithErrorsFitted}.} of normalized {\it HistogramDistributions}, provided the bootstrap distribution is stored in the latter.
 \item[(iv)] The mean value extracted in (i) also allows for the extraction of a bootstrap estimate of the so-called {\it bias} \cite{EfronTibshiraniBook}. The bias of $\hat{\theta} = s \left( \bm{x} \right)$ as an estimator for the quantity $\theta$ is defined as the difference between the expectation value of $\hat{\theta}$, evaluated using the true underlying probability distribution $F$ and the value of $\theta$ as a parameter of $F$
 \begin{equation}
  \mathrm{bias}_{F} := \mathrm{E}_{F} \left[ s \left( \bm{x} \right) \right] - s \left( \bm{x} \right) \mathrm{.} \label{eq:DefBias}
 \end{equation}
 The bootstrap estimate of this quantity is, in its elementary form, given by \cite{EfronTibshiraniBook}
 \begin{equation}
  \widehat{\mathrm{bias}}_{B} = \hat{\theta}^{\ast} (\cdot) - \hat{\theta} \mathrm{,} \label{eq:DefBiasBootstrap}
 \end{equation}
 and it approximates the ideal bootstrap ($\equiv$ plug-in) estimate $\mathrm{bias}_{\hat{F}}$. For an application of the formula (\ref{eq:DefBiasBootstrap}) to a TPWA, the mean $\hat{\theta}^{\ast} (\cdot)$ corresponds to the mean $\hat{\theta}^{\ast}_{i} \left( \cdot \right)$ for each multipole parameter-distribution as defined above. The plug-in estimate $\hat{\theta}$ corresponds here to the multipole-parameters $\hat{\theta}_{i} =\left(\mathcal{M}_{\ell}^{C} \right)_{i}$ resulting from a fit to the {\it original} data. Therefore, in order to estimate $\widehat{\mathrm{bias}}_{B}$ for a multipole parameter, one has to subtract the parameter value of the original fit from the mean of the corresponding bootstrap distribution. \newline
 The bias also provides a cross check of the TPWA bootstrap analysis. In order for the general scheme here to be valid, i.e. extracting parameters from a fit to the original data and then endowing these results with a bootstrap-error, the bias should be negligible. As a rule-of-thumb, Efron and Tibshirani \cite{EfronTibshiraniBook} state that a bias of less than $.25$ standard errors can be safely ignored
 \begin{equation}
 \frac{\left|\widehat{\mathrm{bias}}_{B} \left[ \hat{\theta}_{i} \right] \right|}{\widehat{\mathrm{se}}_{B} \left[ \hat{\theta}_{i} \right]} < 0.25 \mathrm{.} \label{eq:BootstrapRuleOfThumb}
 \end{equation}
 We will use this rule to cross check our bootstrap analyses. However, in case the bootstrap yields good results in the guise of normal shaped parameter histograms, then the rule of thumb (\ref{eq:BootstrapRuleOfThumb}) will be generally fulfilled. Whenever (\ref{eq:BootstrapRuleOfThumb}) is violated, the investigation of the corresponding histograms can generally reveal problems with the analysis, in this case mostly given by ambiguities (see the discussion further below).
\end{itemize}
In case the histograms resulting from the TPWA-bootstrap show well formed normal distributions, it is safe to conclude that the analysis worked without any problems. Then, we give the following numbers as results of the fit. \newline
For the parameters themselves, we take the results from the fit to the original data
\begin{equation}
 \left\{\hat{\theta}_{i}\right\} = \mathcal{M}_{\ell}^{C} \mathrm{.} \label{eq:BootstrapFitParameterResult}
\end{equation}
Errors are derived from the confidence intervals, where for a standard error we pick the approximate $.68$ percentile interval
 \begin{equation}
  \left[  \hat{\theta}_{i}^{\%, \mathrm{lo}}, \hat{\theta}_{i}^{\%, \mathrm{up}} \right] = \left[  \left(\hat{\theta}^{\ast}_{i}\right)^{.16}_{B}, \left(\hat{\theta}^{\ast}_{i}\right)^{.84}_{B} \right] \mathrm{.} \label{eq:PercentileConfidenceIntervalDefForStandardError}
 \end{equation}
\vfill
\clearpage
Upper and lower errors are then given as the distances of the fit result (\ref{eq:BootstrapFitParameterResult}) to the boun\-da\-ries of this interval. Furthermore, for comparison, we also quote the standard deviations of the bootstrap distributions of the multipole parameters. \newline
It is clear how the procedure described until now can help with the second problem mentioned in the beginning of this section, i.e. with the obtainment of good error estimates for the single energy partial waves. The question remains about what it can do about the first problem, the ambiguities. \newline
In the course of this work, no literature was found dealing specifically with bootstrap-analyses using models afflicted with ambiguities, such as the TPWA can very well be. However we state here as a sensible speculation, without mathematical proof, that while the bootstrap cannot help with the resolving of ambiguities, it can make them visible. Therefore, the proposal is
\begin{itemize}
 \item[$\ast)$] \underline{Claim:} A parametric bootstrap is applied to the TPWA-problem in such way that all possible ambiguities are discrete. Then, the ambiguous solutions allowed by the data show up as multiple normal-shaped peaks, either connected or disjoint, in the bootstrap-histograms composed of the values $\hat{\theta}^{\ast}_{i} \left( b \right)$. \newline 
 In this way, the bootstrap can directly map out all the ambiguities that exist with the approximately same probability based on the original data.
\end{itemize}
To be more precise, a bootstrap can of course only show all the ambiguities if a full minimum search as described in sections \ref{sec:TPWAFitsIntro} and \ref{sec:MonteCarloSampling}, using a pool of $N_{MC}$ start-configurations, is done for each bootstrap-dataset. \newline
For practical fits, especially using higher truncation orders, this mentioned procedure can become too expensive. Therefore, we present also an alternative, slightly reduced, procedure. The types of bootstrap-analyses performed in this work are:
\begin{itemize}
 \item[1.)] \underline{Full bootstrap-TPWA:} This procedure consists of first generating, in addition to the original data, an ensemble of $B$ bootstrap datasets
 \begin{equation}
  \left( \bm{x^{\ast}} \right)^{b} \mathrm{,} \hspace*{5pt} b = 1,\ldots,B \mathrm{,} \label{eq:BootstrapDatasets}
 \end{equation}
 and then performing a full minimum-search, using a pool of $N_{MC}$ start-pa\-ra\-me\-ter sets, for each dataset. This amounts to a total of $(B+1) N_{MC}$ numerical minimizations. \newline
 The details of this procedure lead to an important technical point, which has been disregarded until now. The bootstrap-TPWA as written above accounts for the case where a well-separated global minimum, having a chisquare $\left(\chi^{2}_{\mathcal{M}}\right)^{\mathrm{best}}$ out of the whole employed pool with size $N_{MC}$, is found for each search performed with every bootstrap dataset. However, as already seen in the studies of theory-data in section \ref{sec:TheoryDataFits}, it is quite possible that local minima, having chisquare very close to $\left(\chi^{2}_{\mathcal{M}}\right)^{\mathrm{best}}$, may exist. \newline
 Therefore, for the construction of bootstrap parameter distributions, it can become necessary to include all the non-redundant solutions within an interval
 \begin{equation}
  \left[ \left(\chi^{2}_{\mathcal{M}}\right)^{\mathrm{best}}_{b} , \hspace*{2pt} \left(\chi^{2}_{\mathcal{M}}\right)^{\mathrm{best}}_{b} + \Delta \chi^{2} \right] \mathrm{,} \label{eq:ChisquareIntervalBootstrap}
 \end{equation}
 coming from the full minimum search for every bootstrap dataset. For the choice of the cut-range $\Delta \chi^{2}$, which itself embodies the used notion of \textit{closeness}, one has to find some sensible criterion for each considered case. Theoretical $\chi^{2}$-distributions and their quantiles can help with this task. More details on such choices will be elaborated later, when actual datasets are fitted. \newline
 If ambiguities are found and made visible by the procedure considered here, the extraction of one well-defined error for each statistic of interest if of course impossible. If the normal-shaped peaks in the histograms are disjoint, one can do the error analysis for each one individually. \newline
 It should be mentioned here that even if the cut for local minima, equation (\ref{eq:ChisquareIntervalBootstrap}), is not built into the bootstrap-TPWA and instead the global minimum is picked from each bootstrap fit, ambiguities can still occur. \newline
 As a general result of the full bootstrap-TPWA, all non-redundant solutions coming from the $(B+1)$ minimum-searches, which survive the cut (\ref{eq:ChisquareIntervalBootstrap}), are compared and further processed in the bootstrap-histograms.
 \item[2.)] \underline{Reduced bootstrap-TPWA:} Here, one full minimum search is performed for the original data. In case a well-separated global minimum is obtained from this procedure, i.e. out of the pool of $N_{MC}$ Monte Carlo start configurations, then a new full search need not be done for each of the bootstrap datasets (\ref{eq:BootstrapDatasets}). Instead, one single fit is run for each dataset $\left( \bm{x^{\ast}} \right)^{b}$, each time starting at the global minimum obtained in the full fit to the original data.
\end{itemize}
Clearly, only the full bootstrap-TPWA is really capable of reliably mapping out the ambiguity structure allowed by the data. However, even in case the number $B$ of bootstrap replications is chosen relatively low (more information in how to choose it will follow), then the $(B+1) N_{MC}$ minimizations can make this method very costly. This is true since, for the higher truncation orders, the size $N_{MC}$ of the necessary pool grows exponentially (see section \ref{sec:MonteCarloSampling}). \newline
In case a good global minimum is found in the fit to the original data and only an error estimate for the multipoles is the goal, one can just do the reduced version of the bootstrap-TPWA. This way, of course a lot of calculation-time is saved and relatively large values of $B$ can be chosen. As can be seen later in particular examples (section \ref{subsec:2ndResRegionDataFits}), ambiguities can also occur in the reduced method. \newline \newline
This leads to the still open question on how to choose the number of bootstrap replicates $B$ in an analysis. Again, we resort to the original literature, in particular the book by Efron and Tibshirani \cite{EfronTibshiraniBook}. \newline
In order to get the ideal bootstrap estimate, corresponding to the plug-in estimate $\mathrm{se}_{\hat{F}} \left( \hat{\theta} \right)$ for the standard error of any statistic $\hat{\theta} = s \left( \bm{x} \right)$, one needs to take the limit $B \rightarrow \infty$. For practical analyses, this is clearly impossible. \newline
In order to obtain estimates for a good choice of $B$, we follow in some detail the discussion by Efron and Tibshirani. The argument is made strictly speaking only for one-sample problems and not for more complicated fits. However, since the same authors also employ similar numbers for $B$ in case fits are discussed \cite{EfronTibshiraniBook}, we take the thus obtained rules-of-thumb as a guideline for the TPWA-fits as well. \newline
The bootstrap estimate $\widehat{\mathrm{se}}_{B}$ for the standard error of a statistic $s (\bm{x})$ has a variance. A good approximation for this variance can be given by \cite{EfronTibshiraniBook}
\begin{equation}
 \mathrm{var} \left( \widehat{\mathrm{se}}_{B} \right) \simeq \frac{c_{1}}{n^{2}} + \frac{c_{2}}{n B} \mathrm{,} \label{eq:VarStErrorEfronTibshirani}
\end{equation}
where the constants $c_{1}$ and $c_{2}$ depend on the underlying true distribution function $F$. The number $n$ denotes here the number of datapoints in $\bm{x}$ drawn from $F$ via the measurement, while $B$ is the number of bootstrap replications generated from the dataset $\bm{x}$. It is important that neither $c_{1}$, nor $c_{2}$, depends on $n$ or $B$. \newline
Both terms on the right hand side of (\ref{eq:VarStErrorEfronTibshirani}) occur due to two in principle distinct sources of the variance. The first term exists solely due to the \textit{sampling variability}, i.e. the variance generated by the fact that only a finite number of datapoints $n$ can be drawn from the unknown true distribution $F$. The second term in equation (\ref{eq:VarStErrorEfronTibshirani}) represents the \textit{bootstrap resampling variability} and occurs from the above mentioned restriction of having only a finite number $B$ of bootstrap replicates available in a practical analysis. In case $n$ is held fixed, the second term $c_{2}/nB$ vanishes in the limit $B \rightarrow \infty$, as it should. \newline
In order to infer the necessary number of bootstrap datasets $B$, it is not advisable to use the variance (\ref{eq:VarStErrorEfronTibshirani}), but instead to define and investigate the \textit{coefficient of variation} of $\widehat{\mathrm{se}}_{B}$ \cite{EfronTibshiraniBook}
\begin{equation}
  \mathrm{cv}\left( \widehat{\mathrm{se}}_{B} \right) := \frac{\mathrm{var}\left(\widehat{\mathrm{se}}_{B}\right)^{1/2}}{\mathrm{E}\left[\widehat{\mathrm{se}}_{B}\right]} \mathrm{.} \label{eq:CoeffVariationEfronTibshirani}
\end{equation}
Here, both the variance and the expectation value exist due to the variability in the data $\bm{x}$. Next, we have to define the $k$-th moment of the bootstrap distribution of $\hat{\theta}^{\ast} = s \left(\bm{x^{\ast}}\right)$
\begin{equation}
  \hat{m}_{k} := \mathrm{E}_{B} \left[ \left( \hat{\theta}^{\ast} - \hat{\mu} \right)^{k} \right] \mathrm{,} \hspace*{5pt} \hat{\mu} = \mathrm{E}_{B} \left[ \hat{\theta}^{\ast} \right] \mathrm{,} \label{eq:MomentDef} 
\end{equation}
as well as the so-called \textit{kurtosis} \cite{EfronTibshiraniBook,DavisonHinkley} of the same distribution
\begin{equation}
 \hat{\Delta} := \frac{\hat{m}_{4}}{\left(\hat{m}_{2}\right)^{2}} - 3 \mathrm{.}\label{eq:KurtosisDef}
\end{equation}
Then, it is possible to derive a kind of master-formula connecting the coefficients of variation for the bootstrap estimate $\widehat{\mathrm{se}}_{B}$, as well as the full plug-in estimate $\widehat{\mathrm{se}}_{\infty}$, according to \cite{EfronTibshiraniBook}
\begin{equation}
 \mathrm{cv}\left( \widehat{\mathrm{se}}_{B} \right) = \left\{ \mathrm{cv}\left( \widehat{\mathrm{se}}_{\infty} \right)^{2} + \frac{\mathrm{E}(\hat{\Delta}) + 2}{4 B} \right\}^{\frac{1}{2}} \mathrm{.} \label{eq:EfronTibshiraniCoeffVarEstimate}
\end{equation}
Here, the expectation of the kurtosis $\mathrm{E}(\hat{\Delta})$ occurs again due to sampling variability. It measures how long-tailed the distribution of $\hat{\theta}^{\ast} = s \left( \bm{x^{\ast}} \right)$ is. \newline
Efron and Tibshirani \cite{EfronTibshiraniBook} state that for practical applications, $\mathrm{E}(\hat{\Delta})$ is rarely larger than $10$. Therefore, equation (\ref{eq:EfronTibshiraniCoeffVarEstimate}) helps to infer that in order to obtain a good bootstrap estimate for the standard error of a statistic, values of $B = 200$ or more are practically safe. \newline
However, the same authors argue that in order to get robust estimates for detailed local properties of bootstrap distributions, such as the percentiles used above to construct the confidence-intervals, one should choose larger values for $B$, at least within the range $B = 500,\ldots,1000$. \newline
We therefore resort to utilize a few hundred bootstrap datasets, typically around $B = 300$, for the more expensive case of the full bootstrap-TPWA. In case detailed properties of bootstrap-distributions shall be determined by the less expensive reduced method, we typically use numbers around $B = 1000,\ldots,2000$. It should also be noted that in order to avoid inconsistencies due to a possible periodicity of the employed random number generator, the number $B$ should not be chosen too large. \newline \newline

\subsection{Analyses of pseudodata} \label{sec:PseudoDataWithErrorsFitted}

As a first demonstration of the fit methods for heteroscedastic data described in the previous section, we present in the following analyses of synthetic pseudodata. \newline
In order to study the effects of a finite statistical precision, pseudodata can be a useful tool. One usually starts with a set of precisely solvable model-data, like the MAID07 theory-data used in section \ref{sec:TheoryDataFits}, for instance. The model-data should be such that their solution-behavior is well-known in case of vanishing errors. \newline
One would then proceed by applying a statistical model-error to each datapoint of the chosen theory-data set. The size and angular behavior of this model-error should be generated in such a way that it resembles a realistic situation as good as possible. A simple possible Ansatz on how to do this will be given below. \newline
Then, a set of pseudodata can be generated, having a purely statistical error of variable size. These pseudodata can then be fitted again using the Monte Carlo-sampling methods outline in section \ref{sec:MonteCarloSampling} as well as, for instance, resampling-techniques like the bootstrap described in section \ref{sec:BootstrappingIntroduction}. \newline
Since the size of the errors can be freely adjusted prior to fitting for such a pseudodataset, the door is opened to the study of many different scenarios. It is then easy to generate pseudodata for already measured observables, where the current statistical precision is known and then study the effect of improving the statistical precision.
Or, it would also be possible to study the impact of a newly measured observable, implementing a statistical error that is realistic for a first measurement. \newline

Clearly, one can study infinitely many scenarios using pseudodata. Since here we wish to shift the focus to analyses of real data soon, we choose not to go into detail about every pseudodata-study done in the course of this work. Instead, we focus on the demonstration of one fact: the decrease of the statistical precision can generally cause ambiguities to appear. This result can also be anticipated from the discussion in appendix \ref{subsec:AccidentalAmbProofsIII}, as well as remarks made by Grushin \cite{Grushin}. Of course, the advantage of the pseudodata-study now lies in the ability to systematically increase the errors of certain observables and scan for ambiguous solutions. Therefore, one has a methodical approach to investigate the effect a good statistical precision, or lack thereof, has on the solubility of the TPWA. \newline
To start, we choose a set of theory-data which is known to be well-solvable, i.e. for which the precise theory-data fits are still well-behaved. In view to the results shown in section \ref{sec:TheoryDataFits}, the MAID07-data \cite{LotharPrivateComm,MAID2007} truncated at $\ell_{\mathrm{max}} = 1$ ($S$- and $P$-waves) (cf. section \ref{subsec:TheoryDataFitsLmax1}) are picked as the simplest candidate. In order to investigate the increase in $\ell_{\mathrm{max}}$, we choose also to include the theory-data truncated at $\ell_{\mathrm{max}} = 2$ (see section \ref{subsec:TheoryDataFitsLmax2}) into the study and compare results. \newline

Firstly, one needs a method to apply at least somewhat realistic statistical model-errors to the pseudodata. Assuming Poisson-statistics, the standard-error of the count rate $N$ is assumed to be the square-root of $N$ in a kinematic bin, i.e. $\Delta N = \sqrt{N}$ \cite{ThomsonTalk}. Since furthermore the unpolarized differential cross section is itself proportional to the rate, $\sigma_{0} \propto N$, we assume the following for the error of the cross sections
\begin{equation}
 \Delta \sigma_{0} = c \sqrt{\sigma_{0}} \mathrm{,} \hspace*{5pt} \left[ c \right] = \sqrt{\frac{\mu b}{sr}} \mathrm{.} \label{eq:DCSModelDataError}
\end{equation}
The quantity $c$ is just a constant with which to tune the statistical error of $\sigma_{0}$ prior to fitting. It has the same value in each kinematic bin, i.e. for each energy $W$ and angle $\theta$. \newline
For a dimension-less asymmetry $\Omega^{\alpha}$ on the other hand, the scaling (\ref{eq:DCSModelDataError}) is not applicable since this quantity itself is a ratio of cross sections, i.e. $\Omega^{\alpha} = \left(\sigma^{(1)} - \sigma^{(2)}\right)/\sigma_{0}$ ($\sigma^{(1)}$ and $\sigma^{(2)}$ are polarized cross sections, cf. section \ref{subsec:PhotoproductionObs}). Instead, from this definition one would expect roughly\footnote{An analysis with gaussian error-propagation \cite{DemtroederI} can show under which circumstances the assumed approximation can become bad. Writing $\Omega^{\alpha} = \frac{\sigma^{(1)} - \sigma^{(2)}}{\sigma_{0}} \equiv \frac{\sigma^{(1)} - \sigma^{(2)}}{\sigma^{(1)} + \sigma^{(2)}}$ and propagating both the errors of $\sigma^{(1)}$ and $\sigma^{(2)}$, i.e. $\Delta \Omega^{\alpha} = \sqrt{\left( \frac{\partial \Omega^{\alpha}}{\partial \sigma^{(1)}} \Delta \sigma^{(1)} \right)^{2} + \left( \frac{\partial \Omega^{\alpha}}{\partial \sigma^{(2)}} \Delta \sigma^{(2)} \right)^{2} }$, the derivatives $\frac{\partial \Omega^{\alpha}}{\partial \sigma^{(1)}}$ and $\frac{\partial \Omega^{\alpha}}{\partial \sigma^{(2)}}$ become relevant. One obtains $\frac{\partial \Omega^{\alpha}}{\partial \sigma^{(1,2)}} = \frac{1}{\sigma^{(1)} + \sigma^{(2)}} \left[ \left( \pm 1 \right) - \frac{\sigma^{(1)} - \sigma^{(2)}}{\sigma^{(1)} + \sigma^{(2)}} \right]$. Since the derivatives are squared in the propagation-formula, the applied approximation becomes good in case one has $ \left[ \left( \pm 1 \right) - \frac{\sigma^{(1)} - \sigma^{(2)}}{\sigma^{(1)} + \sigma^{(2)}} \right]^{2} \simeq 1$, or equivalently $\left| \frac{\sigma^{(1)} - \sigma^{(2)}}{\sigma^{(1)} + \sigma^{(2)}} \right| \equiv \left| \Omega^{\alpha} \right| \ll 1$. Thus, whenever the dimensionless asymmetry approaches $\pm 1$, the approximation becomes bad. Upon some simple tests, we expect a relative approximation-error for $\Delta \Omega^{\alpha}$ of roughly $10\%$ for $\left|\Omega^{\alpha} \right| \leq 0.5$ and of roughly $(30-35)\%$ for $\left|\Omega^{\alpha} \right| \longrightarrow 1.0$. A further assumption necessary in order to arrive at $\Delta \Omega^{\alpha} \propto \sqrt{N}^{-1}$ is: $\sqrt{\left( \sigma^{(1)} \right)^{2} + \left( \sigma^{(2)} \right)^{2}} \propto \sqrt{N}$. \newline Thus, we acknowledge that under certain conditions, the approximation can become quite rough. Still, since here we are mainly interested in the behavior of the TPWA under some 'smearing' of the model-data, we are content with it. However, one can definitely invest more time and effort in order to model more realistic statistical errors for pseudodata.} a behavior of $\Delta \Omega^{\alpha} \propto \sqrt{N}^{-1}$. Therefore, we choose the following Ansatz for the statistical error of the pseudodata for $\Omega^{\alpha}$
\begin{equation}
 \Delta \Omega^{\alpha} (W,\theta) := \frac{1}{\sqrt{\epsilon^{\alpha}(W) \times \sigma_{0} (W,\theta)}} \mathrm{,} \hspace*{5pt} \left[ \epsilon^{\alpha}(W) \right] = \frac{sr}{\mu b} \mathrm{.} \label{eq:AsymmetryModelDataError}
\end{equation}
Here, $\epsilon^{\alpha}$ is again a scaling factor which determines the precision of the pseudodata. We assume it here as constant over each angular distribution, but it can generally not be assumed to be constant for all energies, and therefore carries a dependency on $W$. In practical studies of pseudodata, we fix the scaling-factor $\epsilon^{\alpha}$ as follows. For every polarization observable, labeled by the index $\alpha$, we define a factor $N_{\%}^{\alpha} \in \left[ 0, 1 \right]$ which specifies a certain \textit{percentage-error} for this quantity. We declare the convention that $N_{\%}^{\alpha}$ defines the error as a fraction of the maximal modulus of the asymmetry in each angular distribution, situated at $\theta_{\mathrm{max}}$
\begin{equation}
 \Delta \Omega^{\alpha}_{\mathrm{max}} (W,\theta_{\mathrm{max}}) = N_{\%}^{\alpha} \hspace*{1.5pt} \left| \Omega^{\alpha}_{\mathrm{max}} \left(W,\theta_{\mathrm{max}}\right) \right| \mathrm{,} \label{eq:MaxErrorConvention}
\end{equation}
With the error $\Delta \Omega^{\alpha}_{\mathrm{max}} (W,\theta_{\mathrm{max}})$ fixed by convention, the scaling factor $\epsilon^{\alpha}$ can be evaluated quickly
\begin{equation}
 \epsilon^{\alpha} (W) = \frac{1}{\sigma_{0}(W,\theta_{\mathrm{max}}) \times \left( N_{\%}^{\alpha} \right)^{2} \times \left[ \Omega^{\alpha}_{\mathrm{max}} \left(W,\theta_{\mathrm{max}}\right) \right]^{2}} \mathrm{.} \label{eq:EvaluationOfEpsilonAlpha}
\end{equation}
Thus, all the remaining errors in the angular distribution are given by the scaling behavior (\ref{eq:AsymmetryModelDataError}), once the factor $\epsilon^{\alpha}$ is determined via the convention (\ref{eq:MaxErrorConvention}). \newline
One has to mention that the above outline method for modeling statistical errors for pseudodata is only a first approximation. In order to arrive at more realistic pseudodata, one could think about the implementation of an energy-dependence of the errors which, in realistic experiments, enters due to the photon-flux. Moreover, it is also possible to design the pseudodata with regard to the detector, for instance by implementing known acceptance-gaps into the modeling of the errors. However, the construction of such realistic pseudodata is not the aim of this section. \newline 
Once the errors (\ref{eq:DCSModelDataError}) and (\ref{eq:AsymmetryModelDataError}) are fixed, pseudodata are generated by first evaluating $\check{\Omega}^{\alpha} = \sigma_{0} \Omega^{\alpha}$ and $\Delta \check{\Omega}^{\alpha} = \sigma_{0} \Delta \Omega^{\alpha}$ (The latter relation only holds for small $\Delta \sigma_{0}$, which will always be assumed in the following, cf. appendix \ref{sec:BootstrapAnsatzComments}.). Then, the pseudodata points are generated by drawing once from a normal-distribution centered at the model-value of the respective data point, i.e.
\begin{equation}
 \mathcal{N} \left( \check{\Omega}^{\alpha}_{\mathrm{model}}, \Delta \check{\Omega}^{\alpha} \right) \longrightarrow \check{\Omega}^{\alpha}_{\mathrm{p.d.}} \mathrm{.} \label{eq:PseudoDataGenerationDraw}
\end{equation}
Here, the subscripts \textit{model} and \textit{p.d.}, i.e. pseudodata, were added for clarification. Finally, the error $\Delta \check{\Omega}^{\alpha}$ is appended to each point $\check{\Omega}^{\alpha}_{\mathrm{p.d.}}$ and the pseudodata are ready for fitting. \newline 

\begin{figure}[ht]
 \centering
\vspace*{10pt}
\begin{overpic}[width=0.485\textwidth]{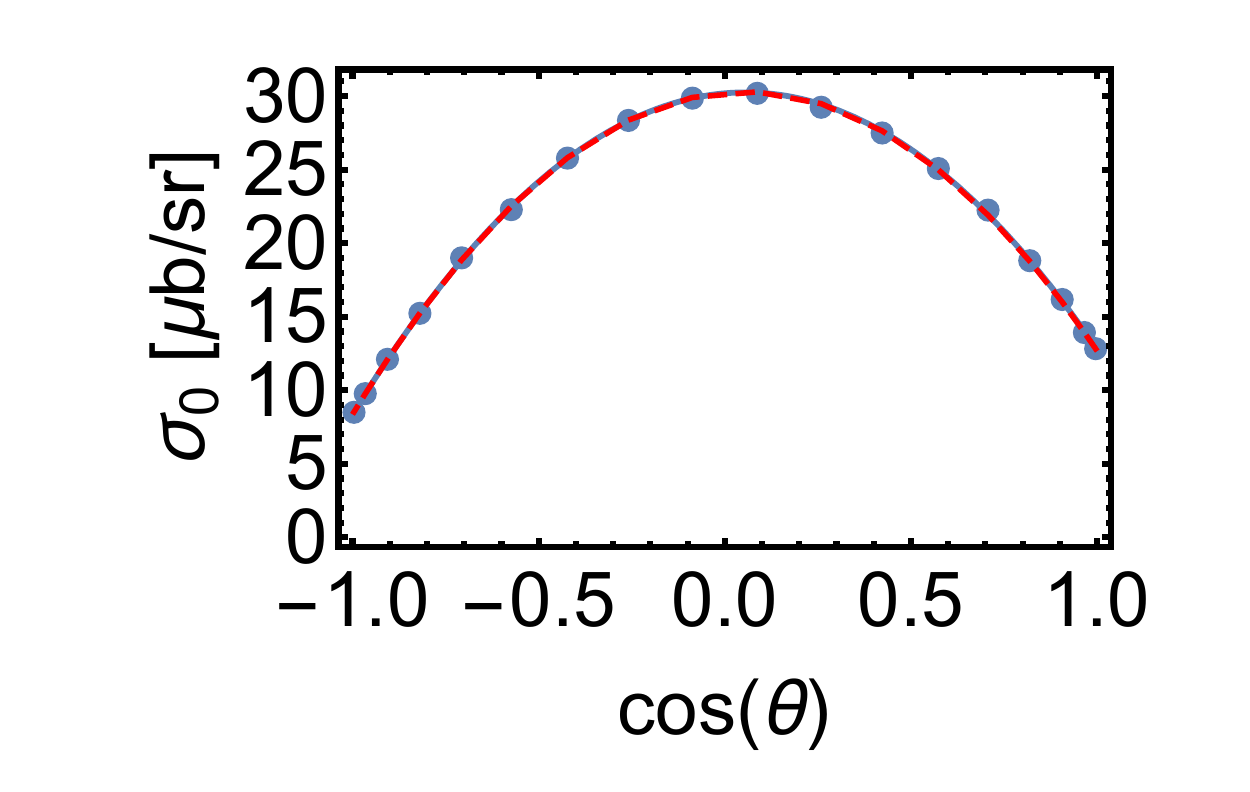}
 \put(85,65){\begin{Large}$E_{\gamma} = 330 \hspace*{2pt} \mathrm{MeV}$\end{Large}}
 \end{overpic}
\begin{overpic}[width=0.485\textwidth]{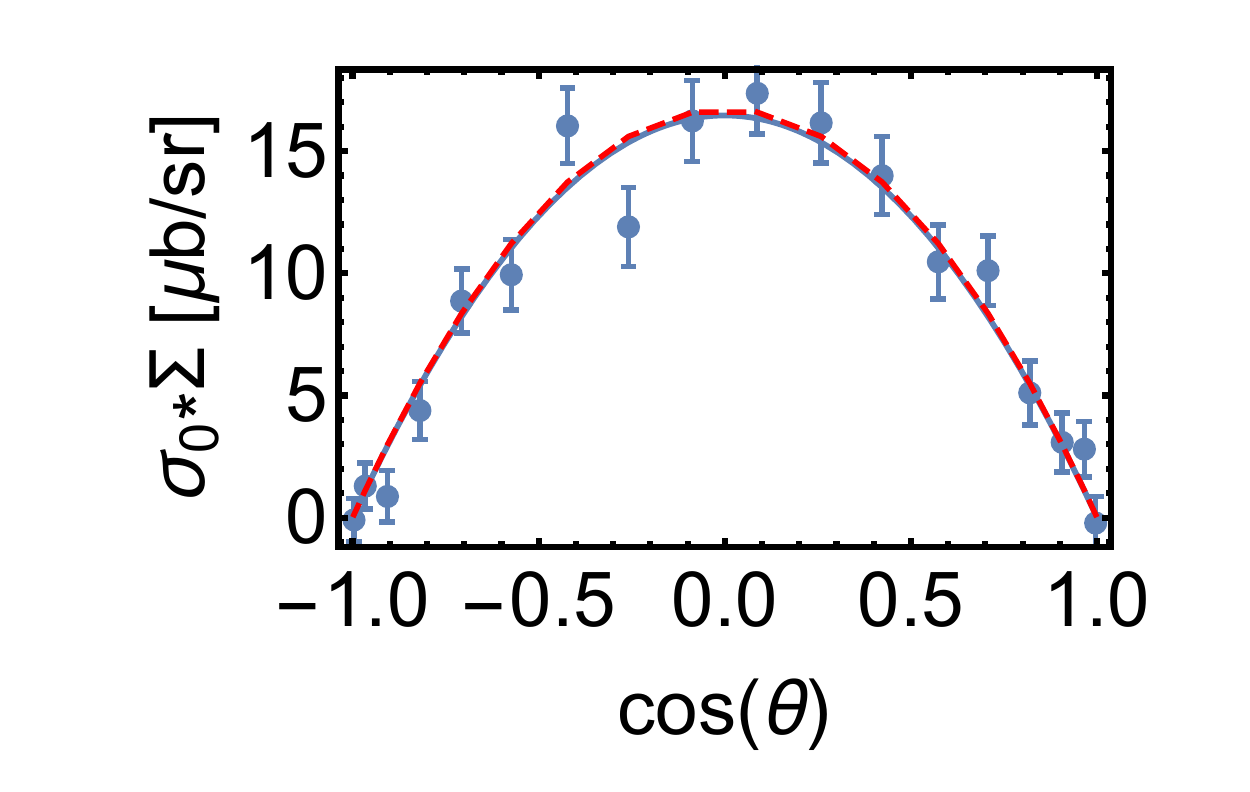}
 \end{overpic} \\
\begin{overpic}[width=0.485\textwidth]{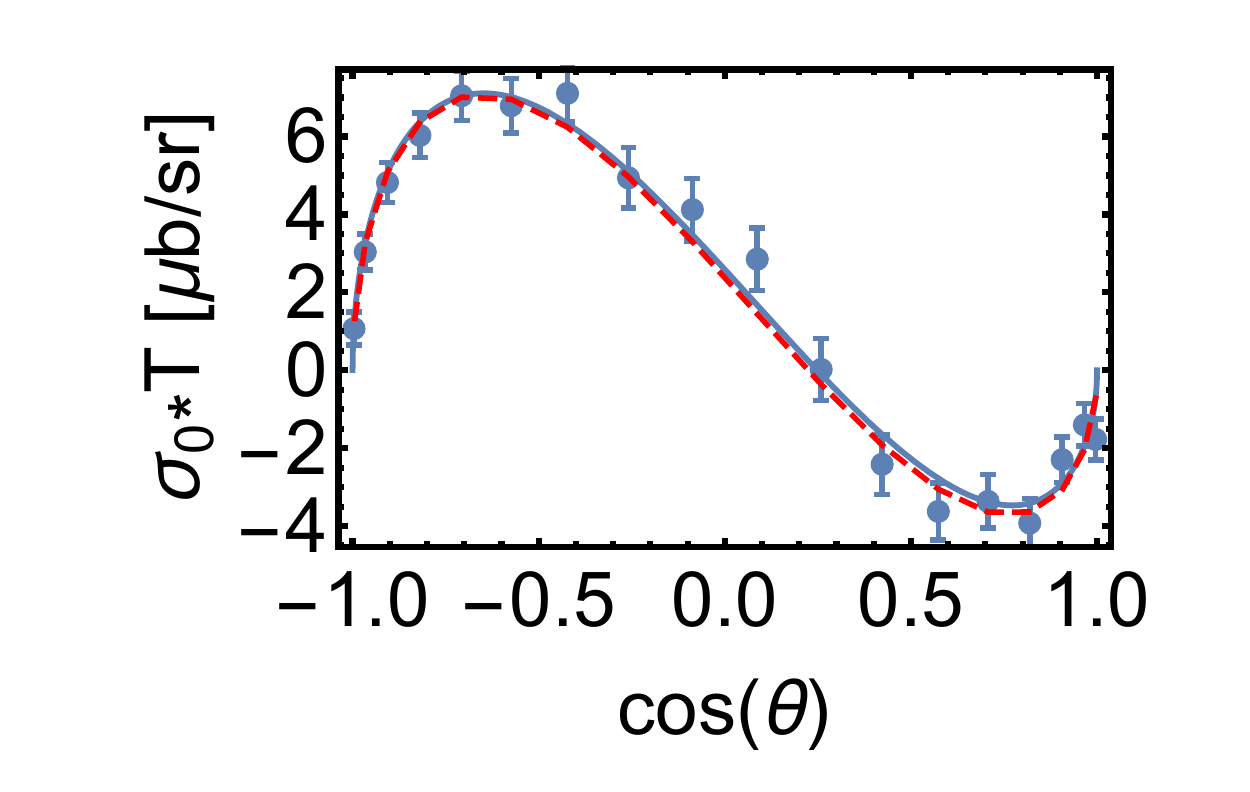}
 \end{overpic}
\begin{overpic}[width=0.485\textwidth]{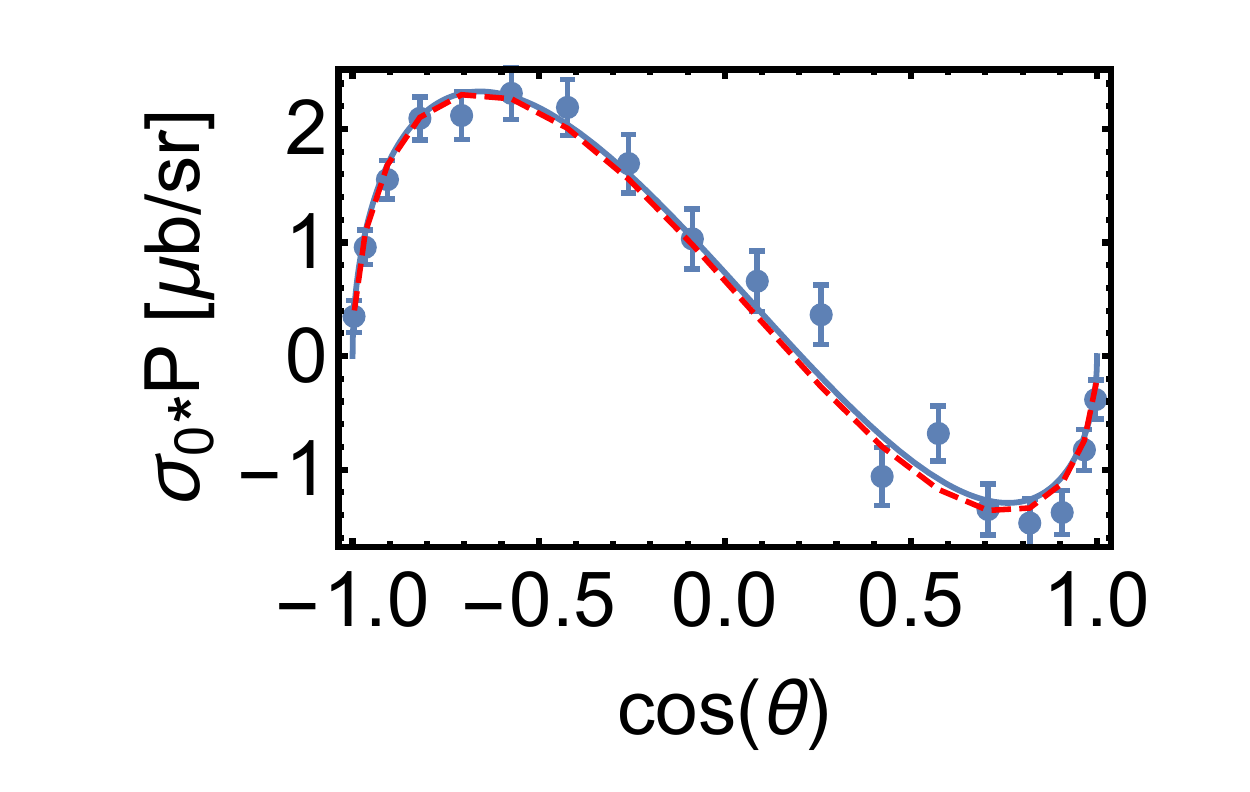}
 \end{overpic} \\
\begin{overpic}[width=0.485\textwidth]{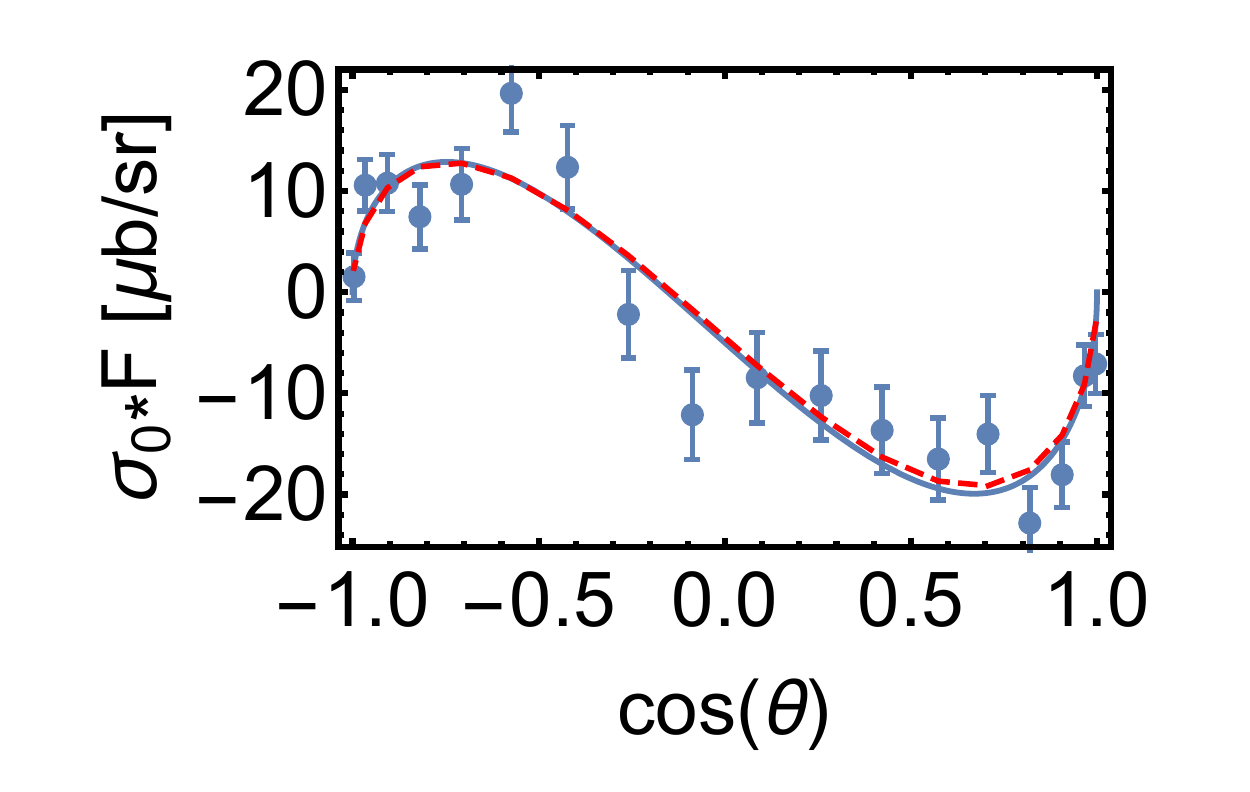}
 \end{overpic}
\caption[Angular distributions of the profile functions for pseudodata generated from the MAID2007-model.]{Shown here are angular distributions of the cross section $\sigma_{0}$ as well as the profile-functions $\check{\Omega}^{\alpha} = \sigma_{0} \Omega^{\alpha}$ belonging to $\Sigma$, $T$, $P$ and $F$. Blue points represent the pseudodata $\check{\Omega}^{\alpha}_{\mathrm{p.d.}}$ with errors $\Delta \check{\Omega}^{\alpha} = \sigma_{0} \Delta \Omega^{\alpha}$. Pseudodata points have been drawn away from the MAID model-data (cf. equation (\ref{eq:PseudoDataGenerationDraw})), which are shown as a red dashed line. The error-scenario $(iii)$ from Table \ref{tab:Lmax1PercentageErrorScenarios} has been utilized for the generation of the pseudodata. \newline
The fit to the pseudodata $\check{\Omega}^{\alpha}_{\mathrm{p.d.}}$ is plotted as a blue solid line. As can be observed, it may not exactly reproduce the MAID-model, as the randomization can introduce a small bias.}
\label{fig:Lmax1PseudoDataFitGroupSFObservablesExampleDeltaEnergy}
\end{figure}

\begin{figure}[hb]
 \centering
\begin{overpic}[width=0.775\textwidth]{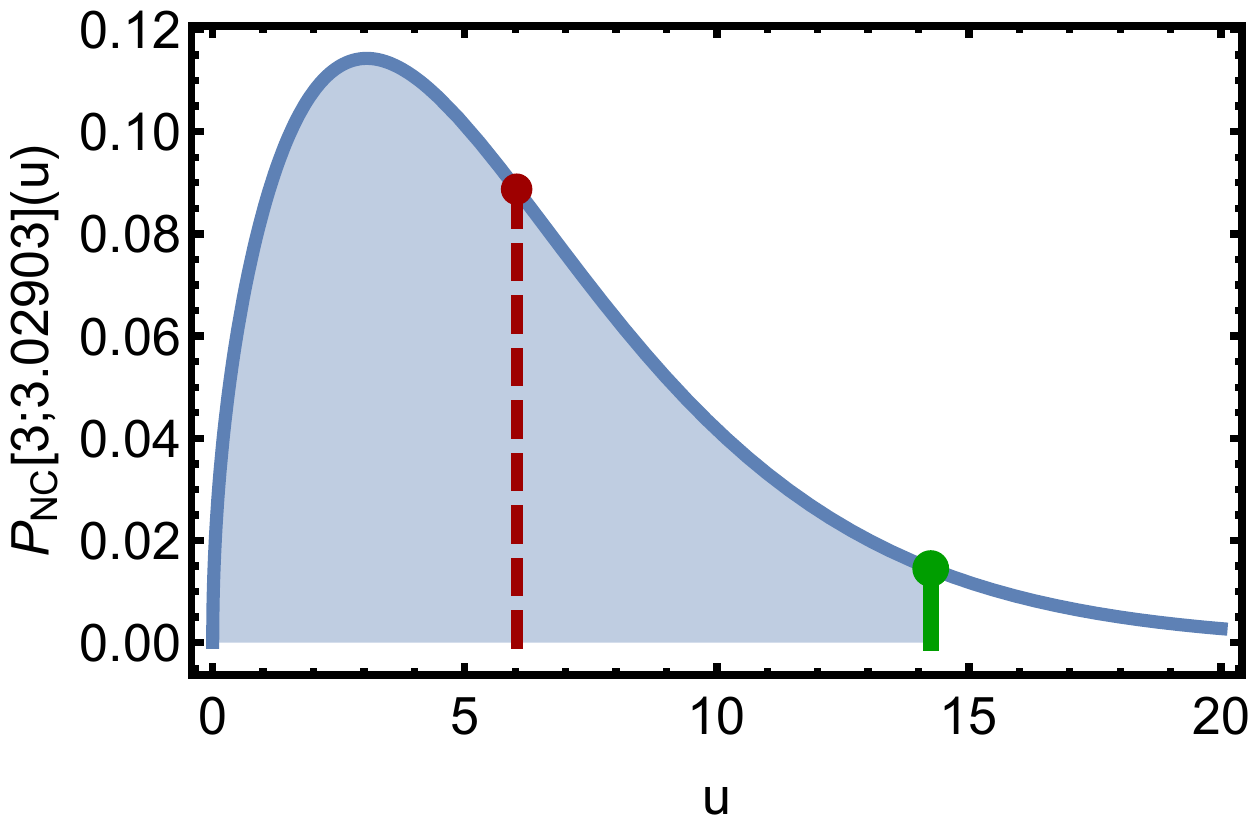}
 \end{overpic}
\caption[Example for a non-central chisquare-distribution.]{This plot shows an example for a non-central chisquare distribution $P_{NC} [r; \lambda] (u)$ (equation (\ref{eq:NonCentralChisquareDist})), the variable $u$ taking values of non-normalized $\chi^{2}$, with $r = \mathrm{ndf} = 3$ and decentralization-parameter $\lambda = \chi^{2}_{\mathrm{orig.}} = 3.02903$. It belongs to a particular pseudodata fit among those discussed in the main text. The mean of this distribution is indicated by the red dashed line, the $0.95$-quantile $u^{\left(P_{NC}\right)}_{0.95}$ by the green solid line. Thus, the shaded area under the curve shows the probability-fraction defining the quantile.}
\label{fig:ExampleNonCentralChiSquareDistribution}
\end{figure}

Some examples for analyses of pseudodata shall be considered in the following. We begin with the MAID-theory-data \cite{LotharPrivateComm,MAID2007} truncated at $\ell_{\mathrm{max}} = 1$, which are known to be exactly solvable (section \ref{subsec:TheoryDataFitsLmax1}). We investigate the mathematically complete set of observables
\begin{equation}
 \left\{ \sigma_{0}, \Sigma, T, P, F \right\} \mathrm{,} \label{eq:MathCompleteSetSec4Dot5}
\end{equation}
which are endowed with a statistical error and then used as the basis for pseudodata according to the description above. Furthermore, we want to focus here on an energy-bin close to the delta-resonance, $E_{\gamma} = 330 \hspace*{1pt} \mathrm{MeV}$. \newline
In order to study the precision of the multipole-fits to the pseudodata, as well as checking the stability of the latter, three scenarios for the errors of the fitted observables (\ref{eq:MathCompleteSetSec4Dot5}) have been studied. These are listed under $(i)$, $(ii)$ and $(iii)$ in Table \ref{tab:Lmax1PercentageErrorScenarios}. For the error of the cross section (\ref{eq:DCSModelDataError}), the assumption of a very precise dataset has been made in each case, using a $2\%$-error (i.e. $c = 0.02 \sqrt{\mu b / sr}$). This assumption is not unrealistically far away from modern measurements of this quantity \cite{Adlarson:2015,Hornidge:2013}. Furthermore, it allows for the study of the influence of the precision in the polarization-data alone. \newline
For the percentage-errors of the polarization observables, we always made the assumption that single-spin observables can be measured with twice the precision of the double-spin observables. Values range from the highly idealized case of $1\%$-errors for the group $\mathcal{S}$ observables and a $2\%$-error for $F$ (case $(i)$), up to the more realistic configuration of $10\%$- and $20\%$-errors. Figure \ref{fig:Lmax1PseudoDataFitGroupSFObservablesExampleDeltaEnergy} depicts as an example the angular distributions of the profile functions belonging to the pseudodata $\check{\Omega}^{\alpha}_{\mathrm{p.d.}}$ versus the curve of the original MAID theory-data $\check{\Omega}^{\alpha}_{\mathrm{model}}$, for a case generated using the error-scenario $(iii)$.
\vfill
\begin{table}[h]
 \centering
\begin{tabular}{r|c|cccc}
 Scenario-no. & $P_{c}$ & $P_{\%}^{\Sigma}$ & $P_{\%}^{T}$ & $P_{\%}^{P}$ &  $P_{\%}^{F}$  \\
\hline
$(i)$ & $2$ & $1$ & $1$ & $1$ & $2$ \\
$(ii)$ & $2$ & $5$ & $5$ & $5$ & $10$ \\
$(iii)$ & $2$ & $10$ & $10$ & $10$ & $20$ \\
$(iii)^{\prime}$ & $2$ & $30$ & $30$ & $95$ & $95$
\end{tabular}
\caption[Error-scenarios for analyses of pseudodata truncated at $L=1$.]{Shown here are the three scenarios $(i)$, $(ii)$ and $(iii)$ for the percentages $P_{c}$ and $P_{\%}^{\alpha}$ employed for the generation of errors in the pseudodata fits of MAID model data truncated at $\ell_{\mathrm{max}} = 1$. A fourth scenario $(iii)^{\prime}$ has been invoked in order to test at which point the $\ell_{\mathrm{max}}=1$-fit becomes unstable. \newline The factors $c$ and $N_{\%}^{\alpha}$ used in equations (\ref{eq:DCSModelDataError}) and (\ref{eq:MaxErrorConvention}) are obtained via $c = P_{c}/100$ and $N_{\%}^{\alpha}=P_{\%}^{\alpha}/100$.} 
\label{tab:Lmax1PercentageErrorScenarios}
\end{table}
\newpage
Once the pseudodata are prepared, we proceed by doing a full bootstrap-TPWA on them (cf. section \ref{sec:BootstrappingIntroduction}). This means we draw $B = 500$ bootstrap-replicates starting from the pseudodata $\check{\Omega}^{\alpha}_{\mathrm{p.d.}}$ (not from the MAID-model!) and then do a model-independent TPWA (cf. sections \ref{sec:TPWAFitsIntro} and \ref{sec:MonteCarloSampling}), using a pool of $N_{MC} = 250$ randomly chosen initial parameter configurations, on each of the bootstrap-replicates. Of course, the \textit{original data} $\check{\Omega}^{\alpha}_{\mathrm{p.d.}}$ are fitted using a model-independent TPWA as well. The truncation order in the TPWA matches that of the theory-data, i.e. the $S$- and $P$-waves are fitted. Furthermore, in TPWA fit step $\mathrm{I}$ (see section \ref{sec:TPWAFitsIntro}), Legendre-coefficients are extracted in error-weighted fits (equation (\ref{eq:LegCoeffFitChi2DataFit})), while for the extraction of the multipoles themselves (TPWA fit step $\mathrm{II}$), a correlated chisquare-function $\chi^{2}_{\mathcal{M}}$ (equation (\ref{eq:CorrelatedChisquare})) is minimized.  \newline
Once for each of the $(B+1)$ cases, the pool of $N_{MC}$ initial configurations has resulted in a TPWA solution-pool, the non-redundant solutions are sorted out of the pool according to the methods described in section \ref{sec:MonteCarloSampling} and appendix \ref{sec:MCSamplingAlgorithms}. Inevitably, a global minimum is found in each case. However, in addition a probabilistic criterion has to be used in order to decide which of the additional local minima to keep in the bootstrap-analysis, based on their value for $\chi^{2}$ in the minimum. \newline
Here we utilize the fact, which is based on experience with the bootstrap-analyses gained in the course of this work, that the bootstrap-values of the correlated chisquare (\ref{eq:CorrelatedChisquare}) follow very well a so-called non-central chisquare distribution. The latter is defined by a normalized probability distribution function $P_{NC} [r; \lambda] (u)$, which can be expressed as\footnote{Here, $\Gamma (z)$ is the {\it Gamma-function}: $\Gamma (z) := \int_{0}^{\infty} t^{z - 1} e^{-t} dt$ for $\mathrm{Re}[z] > 0$ \cite{Abramowitz}.} \cite{StuartOrdNonCentralChisquare,Abramowitz} 
\begin{equation}
 P_{NC} [r; \lambda] (u) = \frac{ e^{- (u + \lambda)/2} \hspace*{1pt} u^{r/2 - 1} }{2^{r/2}} \sum_{k = 0}^{\infty} \frac{ (\lambda u)^{k}}{ 2^{2 k} \hspace*{1pt} k! \hspace*{1pt} \Gamma \left( k + \frac{1}{2} r \right) } \mathrm{,} \hspace*{5pt} \int_{-\infty}^{\infty} du P_{NC} [r; \lambda] (u) = 1 \mathrm{.} \label{eq:NonCentralChisquareDist}
\end{equation}
The parameter $r$ counts the number of degrees of freedom, the other variable $\lambda$ is referred to as the {\it decentralization-parameter} and the variable $u$ takes here values of non-normalized chisquare. For $\lambda \rightarrow 0$, the distribution (\ref{eq:NonCentralChisquareDist}) turns into an ordinary chisquare distribution. \newline
Appendix \ref{sec:CorrelatedChi2Reduction} contains a proof of the fact that the correlated chisquare (\ref{eq:CorrelatedChisquare}) is always equivalent to a function which is explicitly chisquare-distributed. However, due to the fact that in the bootstrap-scheme here the replicate-data are always drawn from the original datapoints, a decentralization is introduced. The function (\ref{eq:NonCentralChisquareDist}) was found to generally reproduce the bootstrap-distribution of $\chi^{2}_{\mathcal{M}}$ very well in case the parameter $r$ is adjusted to the number of degrees of freedom in the original fit, which we estimate as (cf. section \ref{sec:MonteCarloSampling})
 \begin{equation}
 \mathrm{ndf} = N_{a^{\alpha}_{k}} - (8 \ell_{\mathrm{max}} - 1) = 10 - 7 = 3 \mathrm{.} \label{eq:NDOFFitStep2CitationChapter4Dot5}
 \end{equation}
Furthermore, the decentralization-parameter $\lambda$ has to be adjusted to the non-normalized chisquare of the fit to the original data (here, the pseudodata). Figure \ref{fig:ExampleNonCentralChiSquareDistribution} shows a plot of an example for a non-central chisquare distribution. \newline
For the bootstrap-distributions of our analysis, we fix the probabilistic convention to keep, for each bootstrap-replica $b \in \left\{ 1,\ldots,B \right\}$, the global minimum as well as all non-redundant solutions below the so-called $0.95$-quantile of the non-central chisquare distribution \newline $P_{NC} [\mathrm{ndf}; \chi^{2}_{orig.}] (u)$ from the fit to the original data. \newline
For any smooth probability distribution function $p(x)$, one can define the so-called cumulative probability distribution \cite{EfronTibshiraniBook}
\begin{equation}
 G_{p} (y) = \int_{-\infty}^{y} dx \hspace*{1pt} p(x) \mathrm{.} \label{eq:CumulativeProbDistributionGeneral} 
\end{equation}
\clearpage
\clearpage

\begin{figure}[ht]
 \centering
 \vspace*{-15pt}
\begin{overpic}[width=0.485\textwidth]{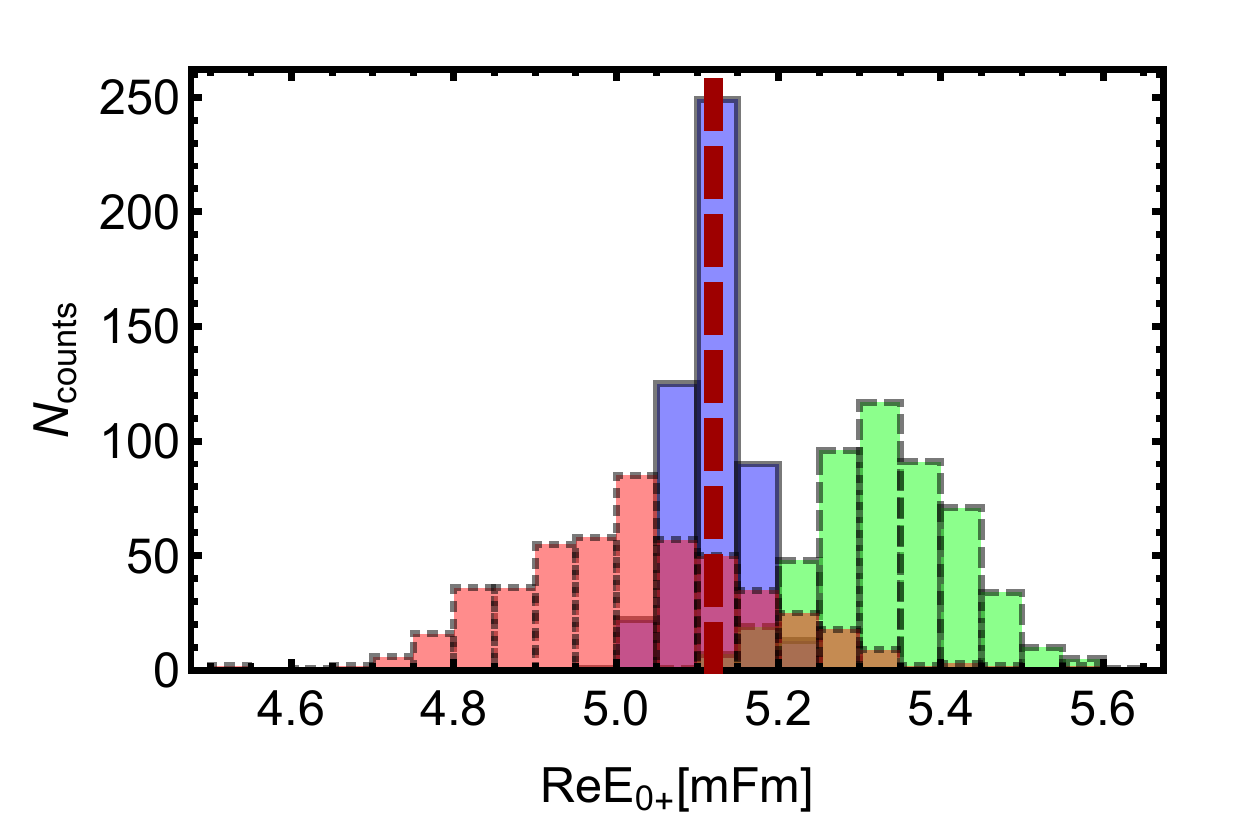}
 \end{overpic} \\
\begin{overpic}[width=0.485\textwidth]{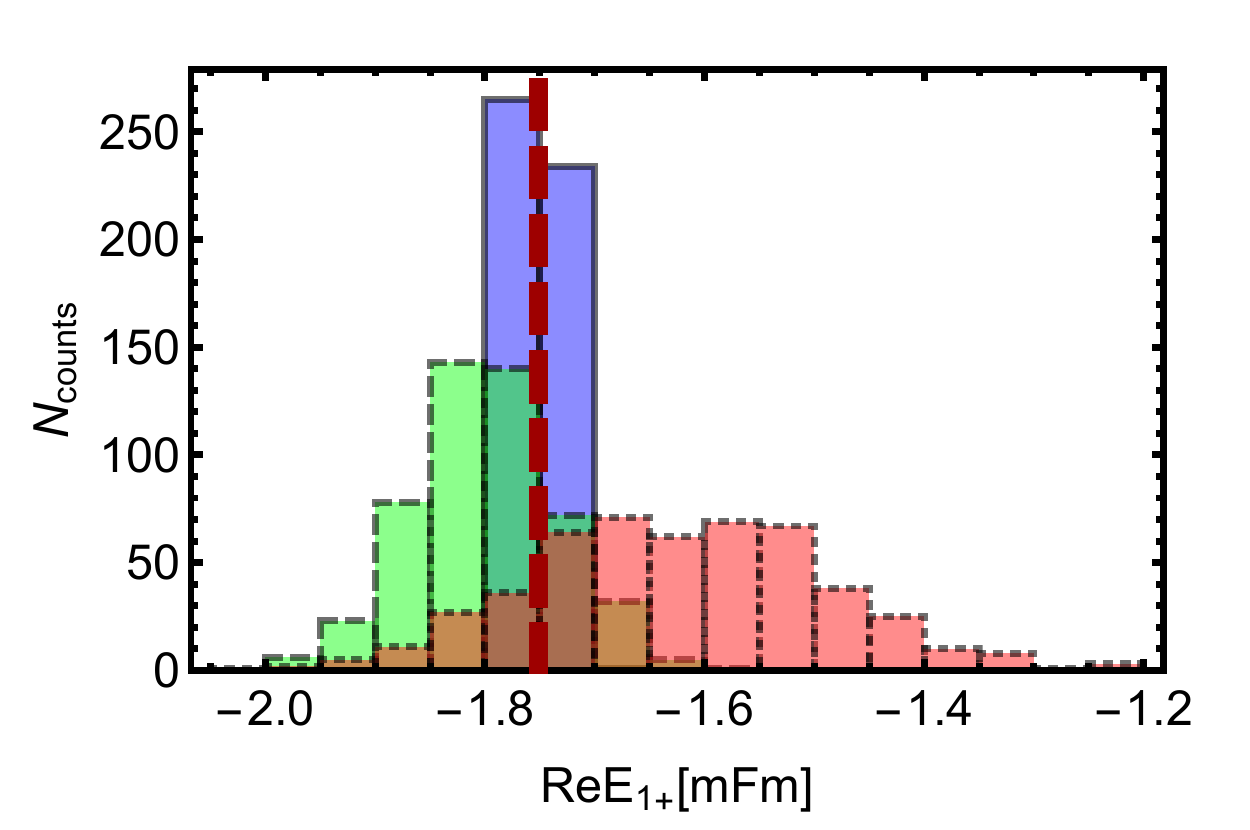}
 \end{overpic}
\begin{overpic}[width=0.485\textwidth]{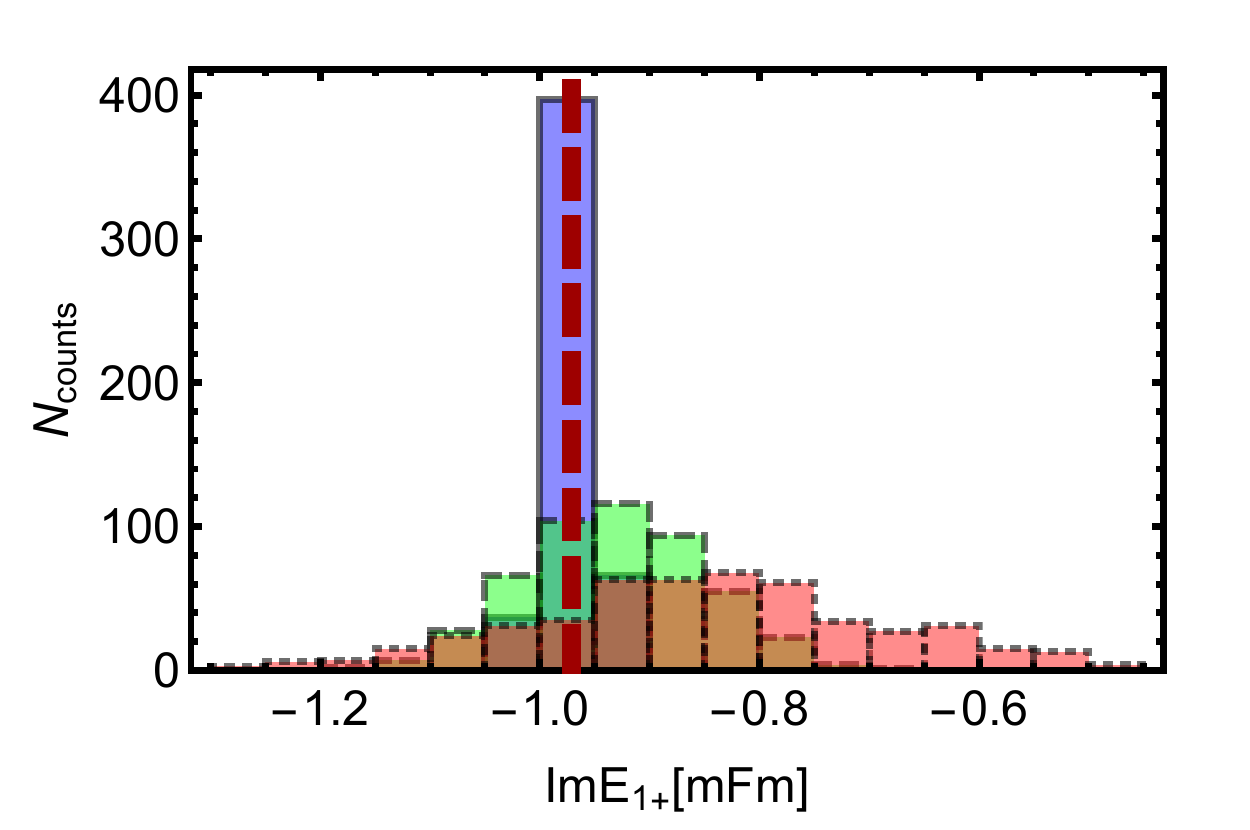}
 \end{overpic} \\
\begin{overpic}[width=0.485\textwidth]{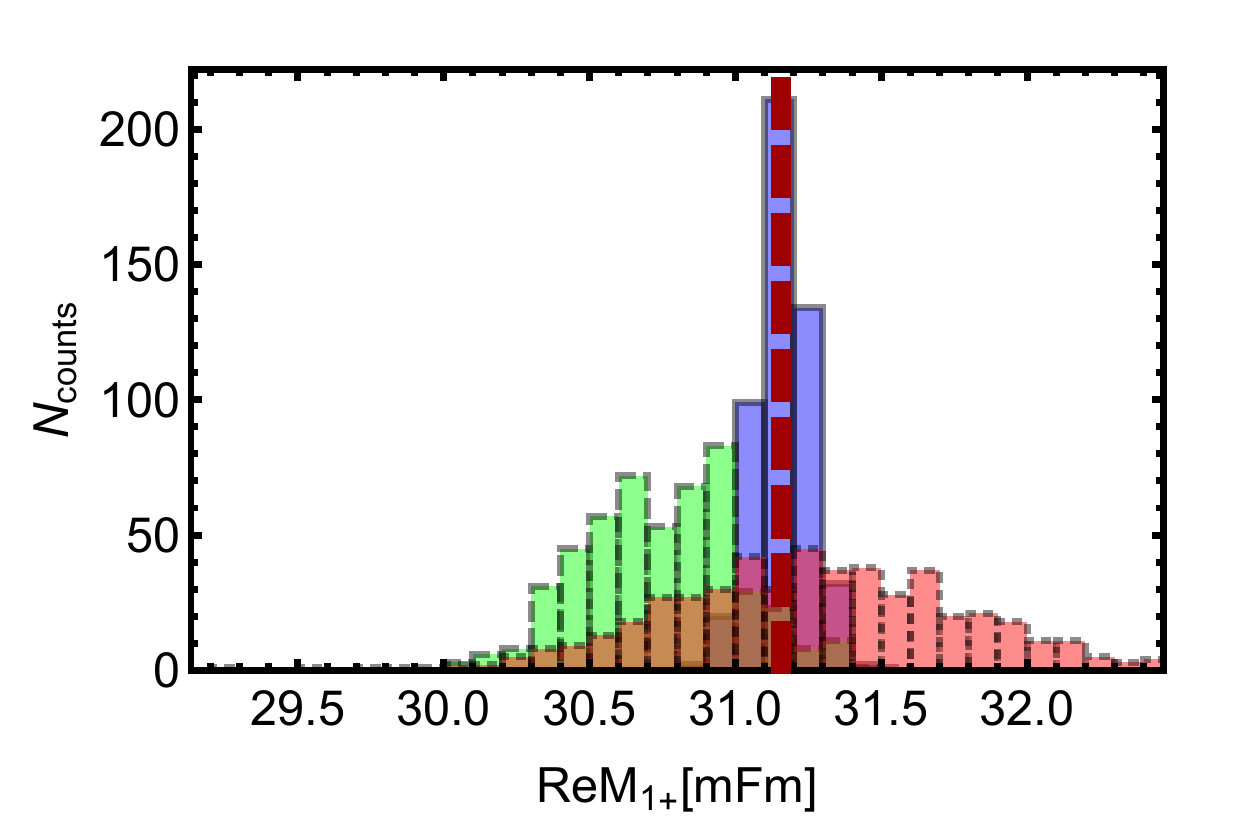}
 \end{overpic} 
 \begin{overpic}[width=0.485\textwidth]{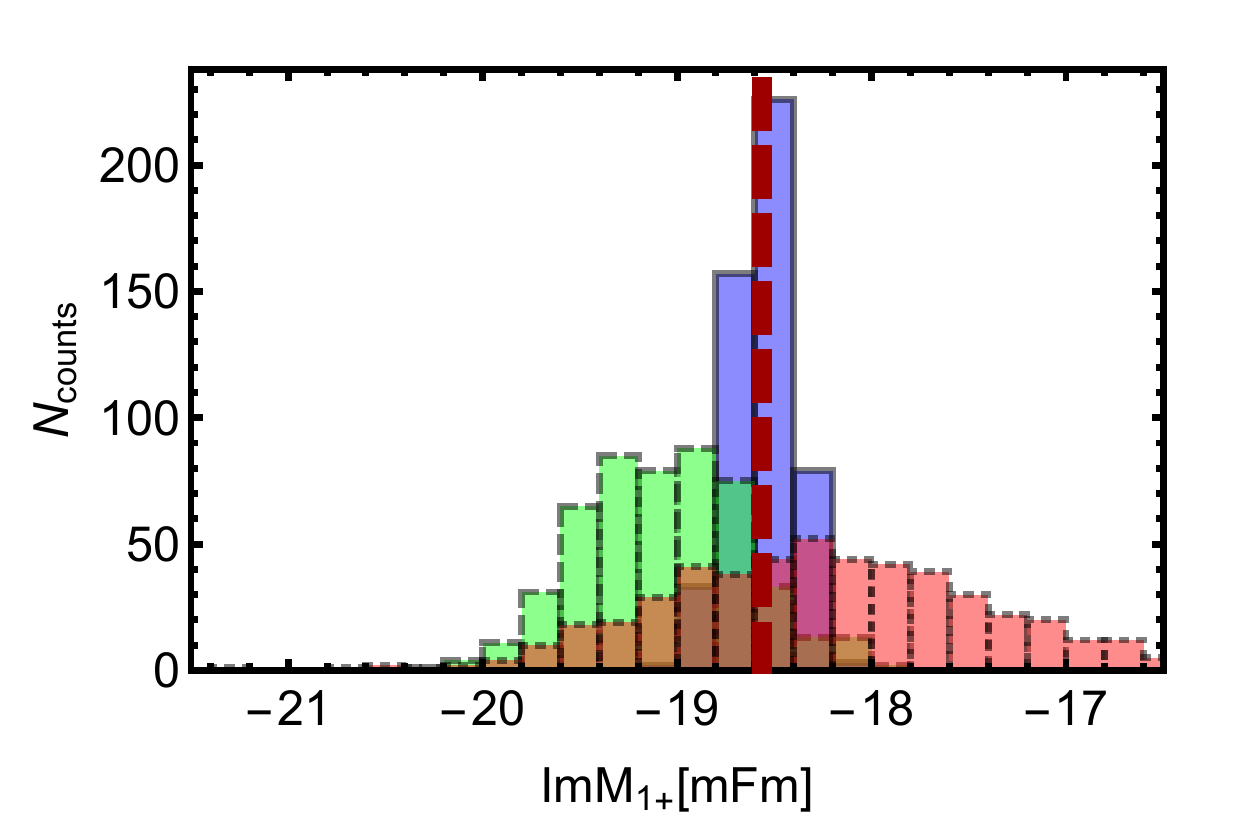}
 \end{overpic}\\
\begin{overpic}[width=0.485\textwidth]{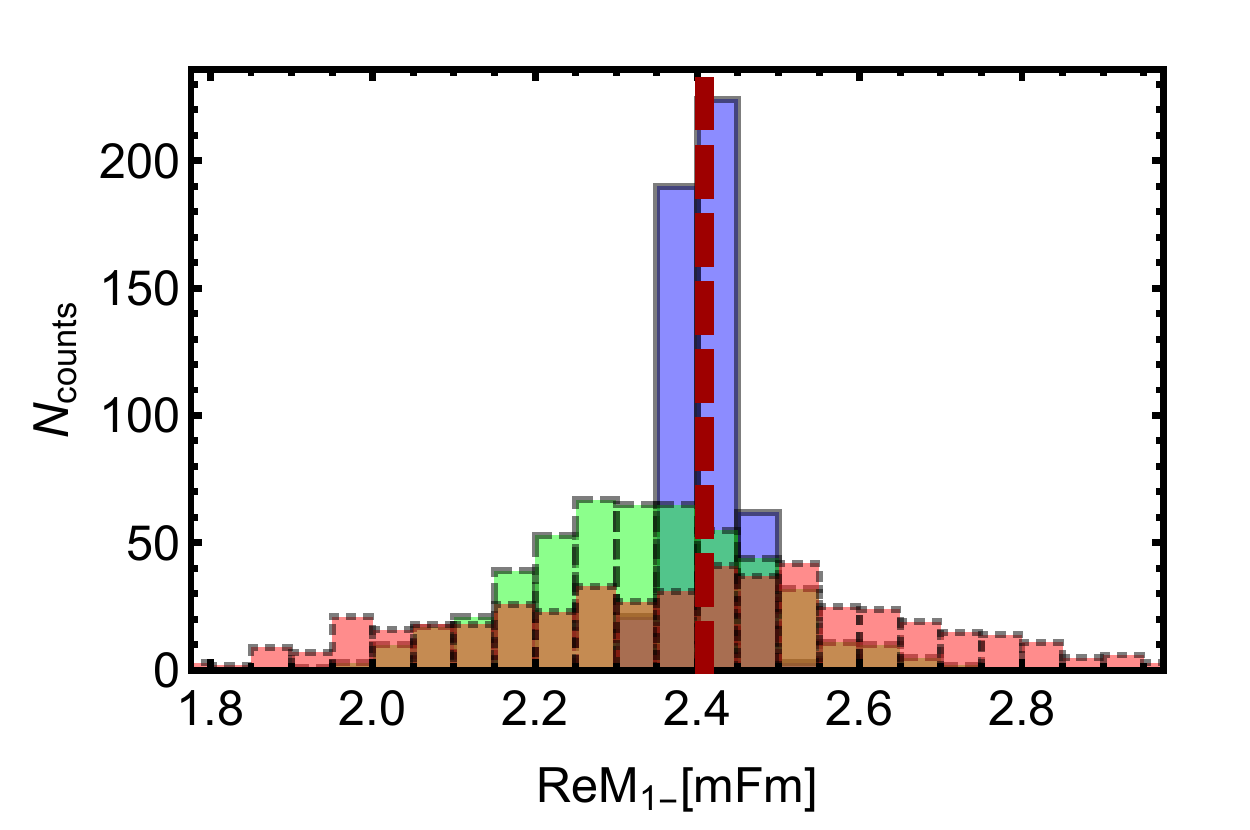}
 \end{overpic} 
\begin{overpic}[width=0.485\textwidth]{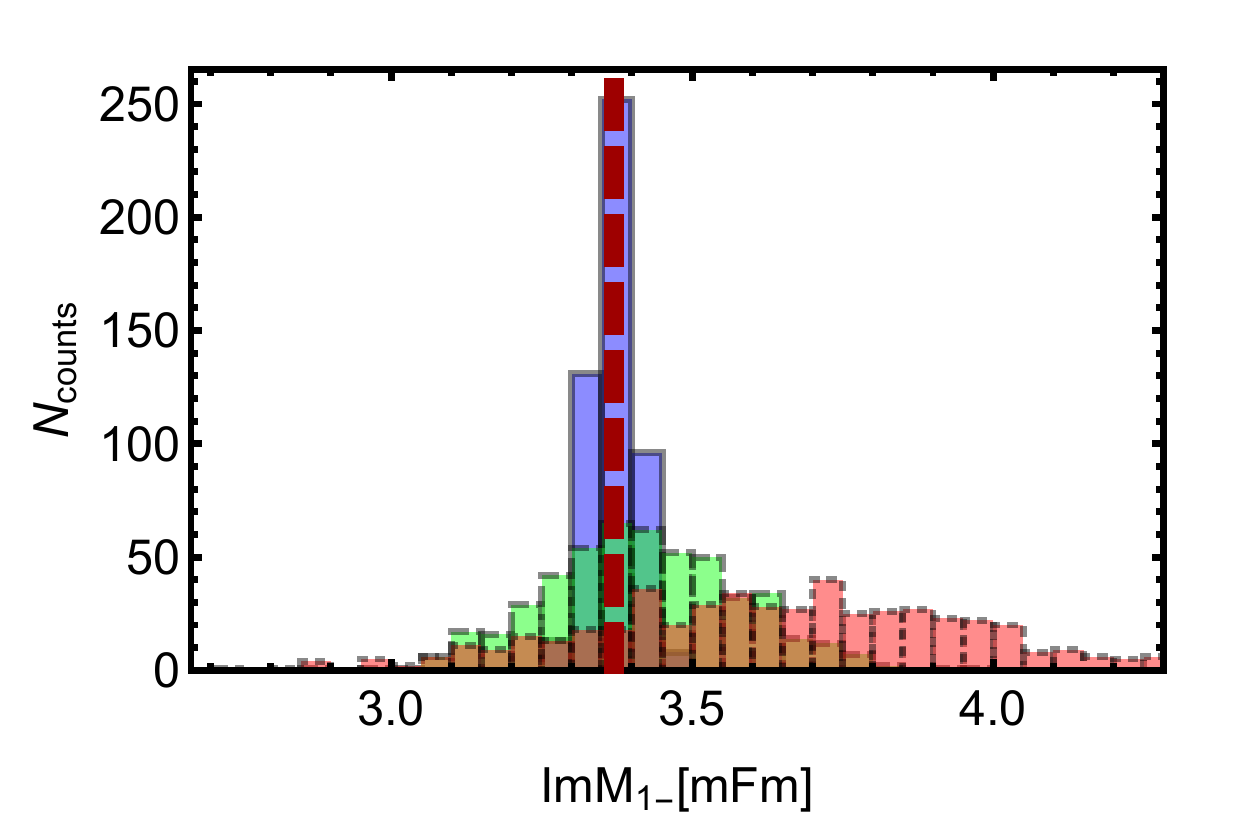}
 \end{overpic}
\caption[Bootstrap-histograms for $S$- and $P$-wave multipoles, resulting from analyses of pseudodata generated from the MAID2007 model. Results are shown for three different error-scenarios.]{The pictures show histograms of the bootstrap-distributions for the real- and imaginary parts of phase-constrained multipoles, resulting from the analyses of pseudodata generated from the MAID theory-data truncated at $\ell_{\mathrm{max}} = 1$. All solutions below the quantile $u^{\left(P_{NC}\right)}_{0.95}$ are included. Results for the error-scenarios $(i)$ (blue bars with solid boundary), $(ii)$ (green bars with dashed boundary) and $(iii)$ (red bars with dotted boundary) are plotted (cf. Table \ref{tab:Lmax1PercentageErrorScenarios}). \newline
A thick red dashed vertical line indicates the result of the fit to the original data (i.e. pseudodata) for error-scenario $(i)$. Those are included for orientation since they are close to the original MAID-model multipoles. Note that the corresponding original fit-results for scenarios $(ii)$ and $(iii)$ have {\it not} been plotted.}
\label{fig:Lmax1PseudoDataFitGroupSFObservablesMultHistograms}
\end{figure}

A quantile of the distribution $p(x)$, belonging to some probability fraction $q \in [0,1]$, can now be defined formally as the inverse of the cumulative distribution \cite{EfronTibshiraniBook}
\begin{equation}
 x^{(p)}_{q} = G_{p}^{-1} (q) \mathrm{.} \label{eq:QuantileGeneralDef}
\end{equation}
In our very broad probabilistic criterion, we therefore use the following quantile of the non-central chisquare distribution
\begin{equation}
 u^{\left(P_{NC}\right)}_{0.95} \mathrm{.} \label{eq:95PercentQuantileNCChisquareDef}
\end{equation}
Figure \ref{fig:ExampleNonCentralChiSquareDistribution} contains an illustration of this quantile. Using the global minimum as well as all solutions below the $0.95$-quantile in the cut, the parameters of the surviving solutions can be plotted into histograms which then show the resulting bootstrap-distributions. \newline
For the error-scenarios $(i)$, $(ii)$ and $(iii)$ and the MAID-theory-data truncated at $\ell_{\mathrm{max}} = 1$, the resulting histograms are shown in Figure \ref{fig:Lmax1PseudoDataFitGroupSFObservablesMultHistograms}. All parameter-distributions are \textit{unimodal} up to a very good approximation, i.e. they show only one peak. \newline
The observed peaks resemble the shape of (sometimes slightly asymmetric) gaussians. Furthermore, the bootstrap-distributions are broadened by an increase of the statistical error, which is fully expected. However, the relative increase of the width is different for different multipoles, i.e. different partial waves have a varying sensitivity to the decrease of the precision in the data. The $S$-wave multipole $E_{0+}$ for instance seems relatively stable, while the $P$-wave $M_{1-}$ is quite sensitive.
\vfill

\clearpage

\clearpage

\begin{figure}[ht]
 \centering
 \vspace*{-15pt}
\begin{overpic}[width=0.485\textwidth]{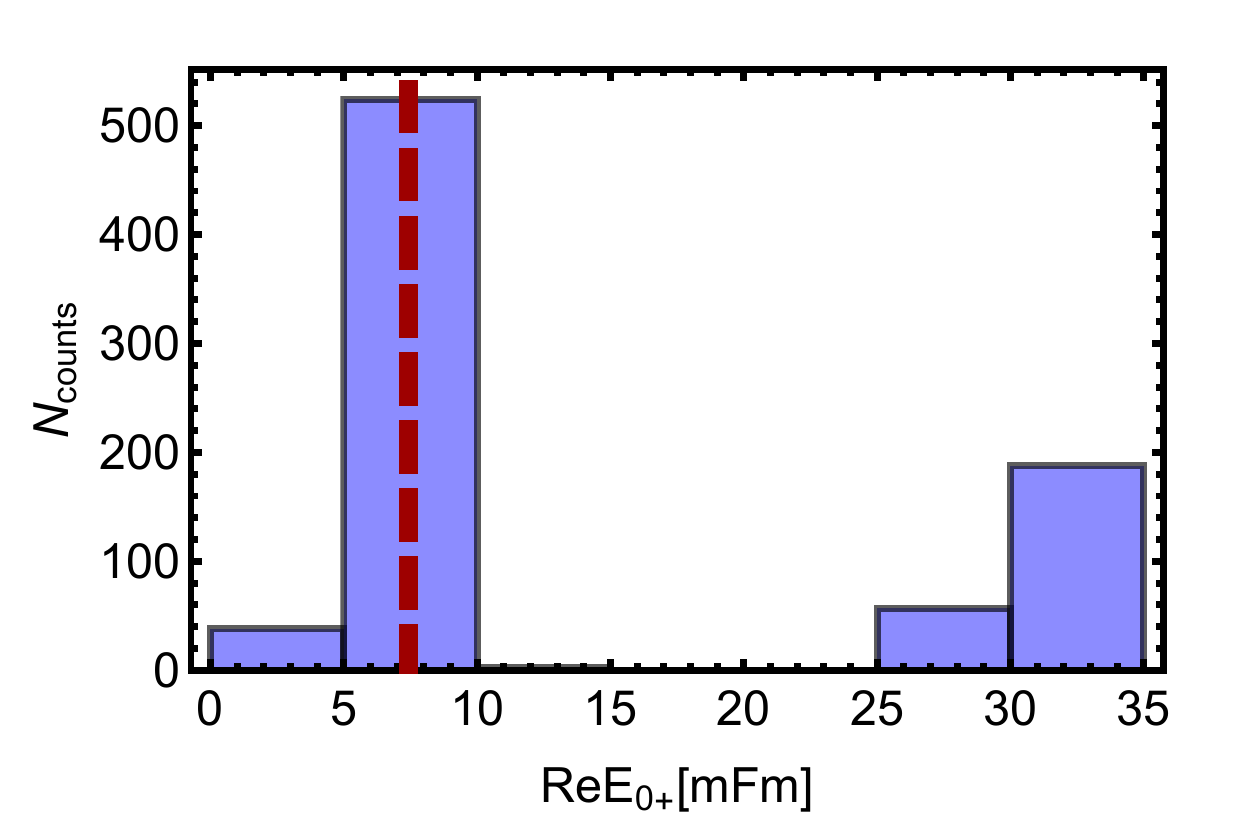}
 \end{overpic} \\
\begin{overpic}[width=0.485\textwidth]{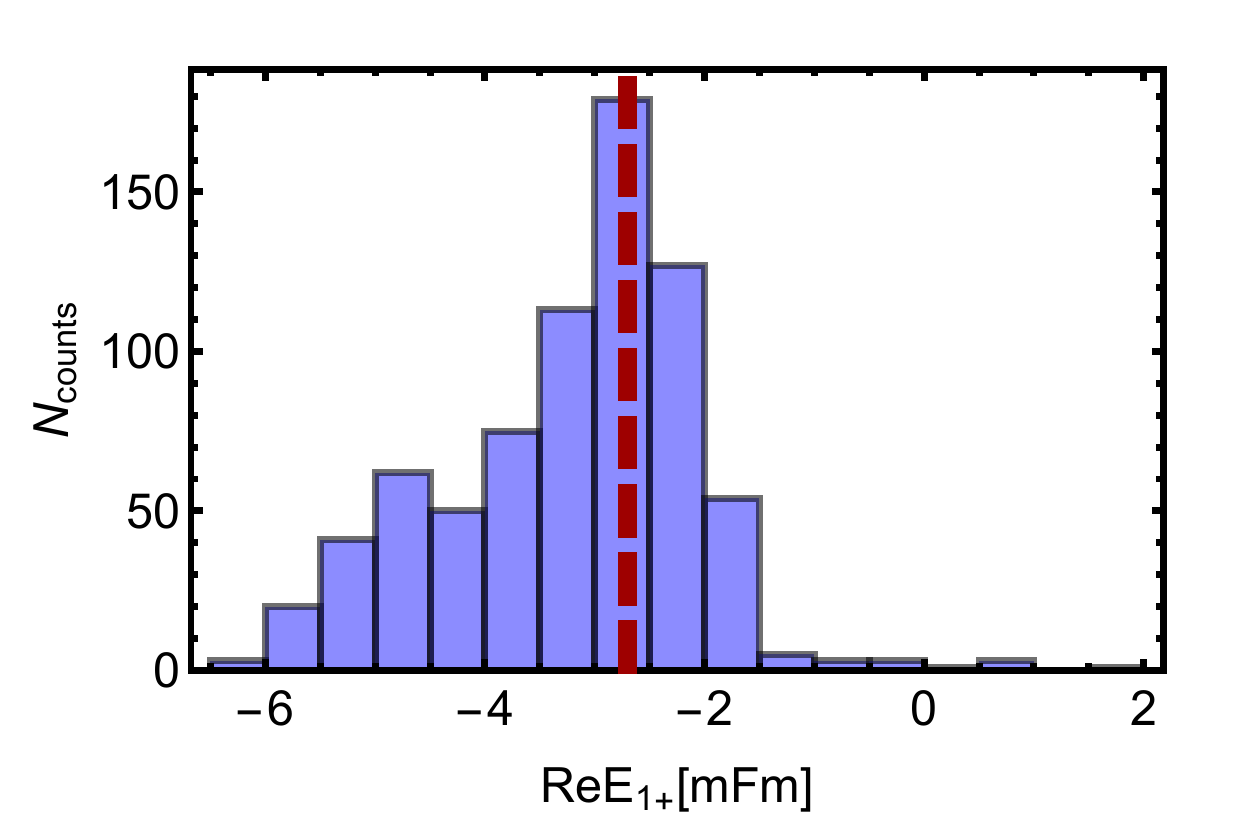}
 \end{overpic} 
\begin{overpic}[width=0.485\textwidth]{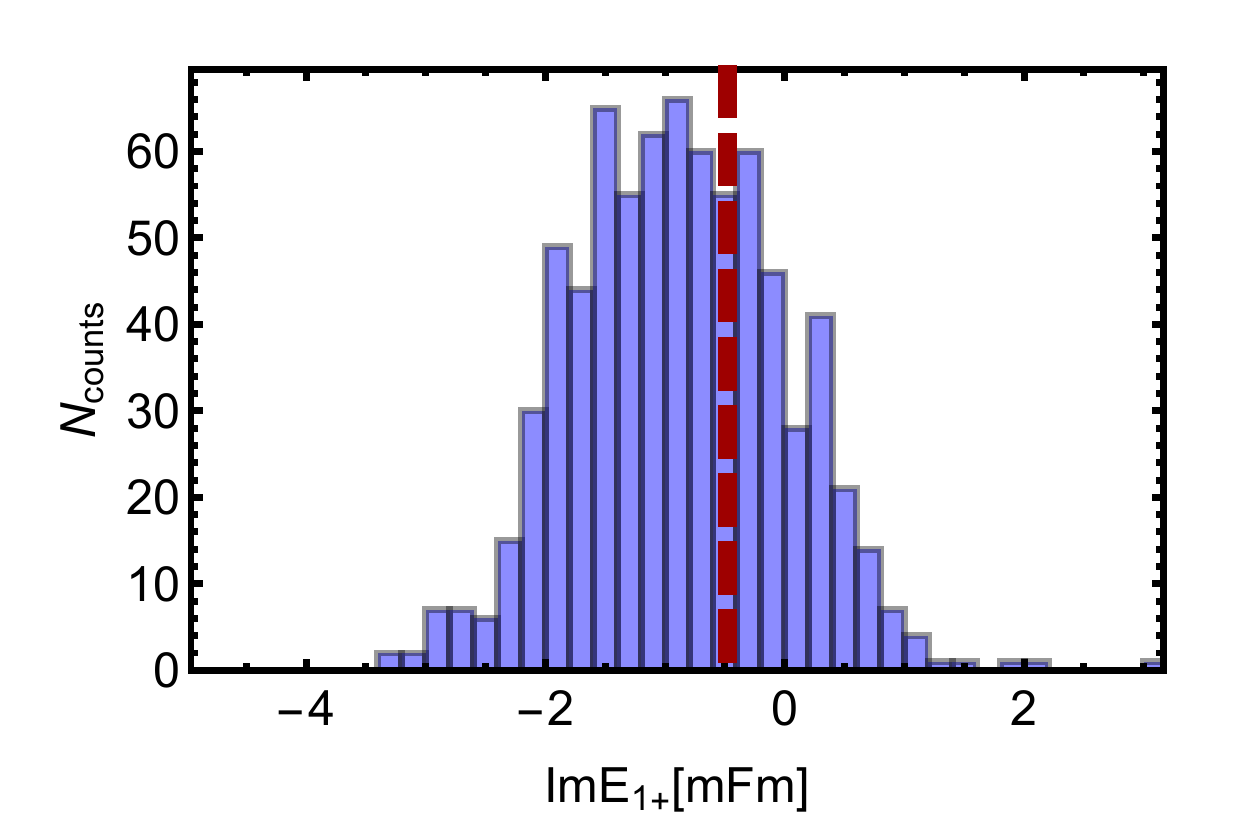}
 \end{overpic} \\
\begin{overpic}[width=0.485\textwidth]{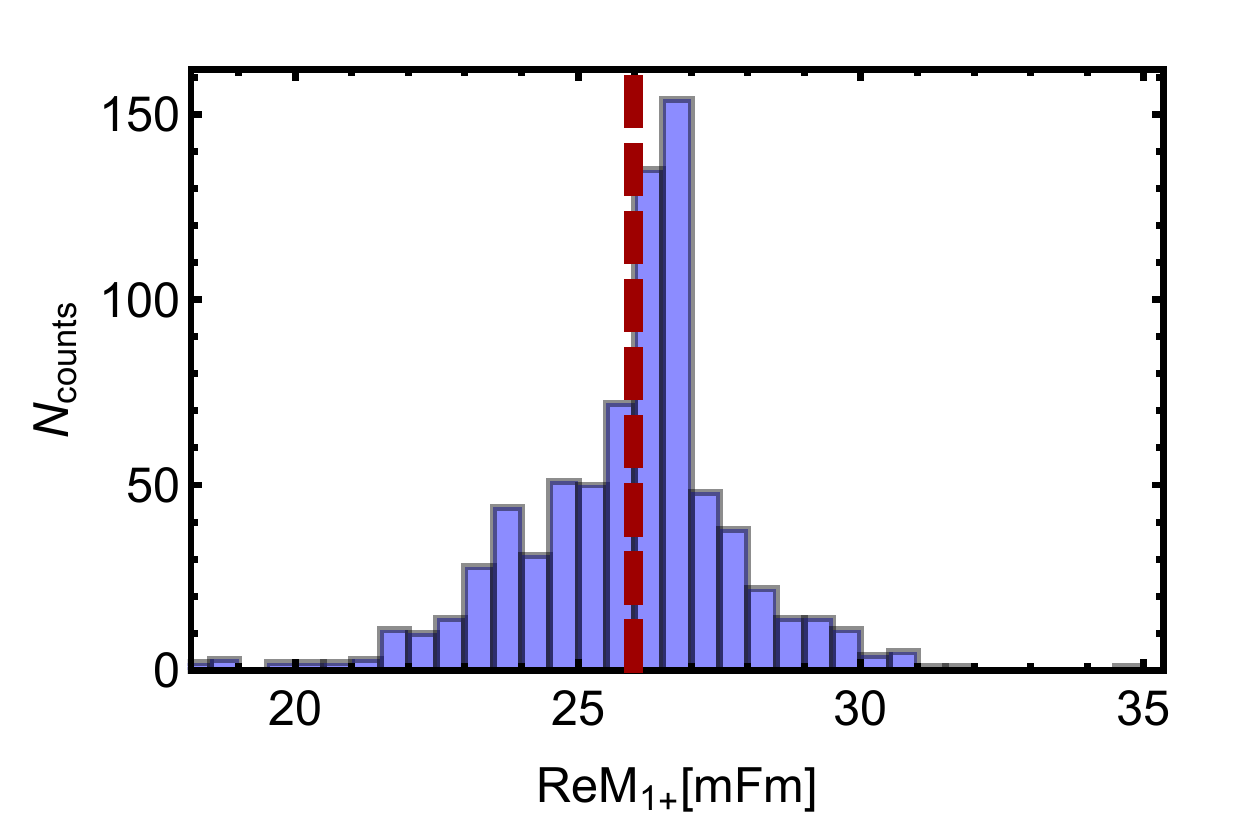}
 \end{overpic} 
 \begin{overpic}[width=0.485\textwidth]{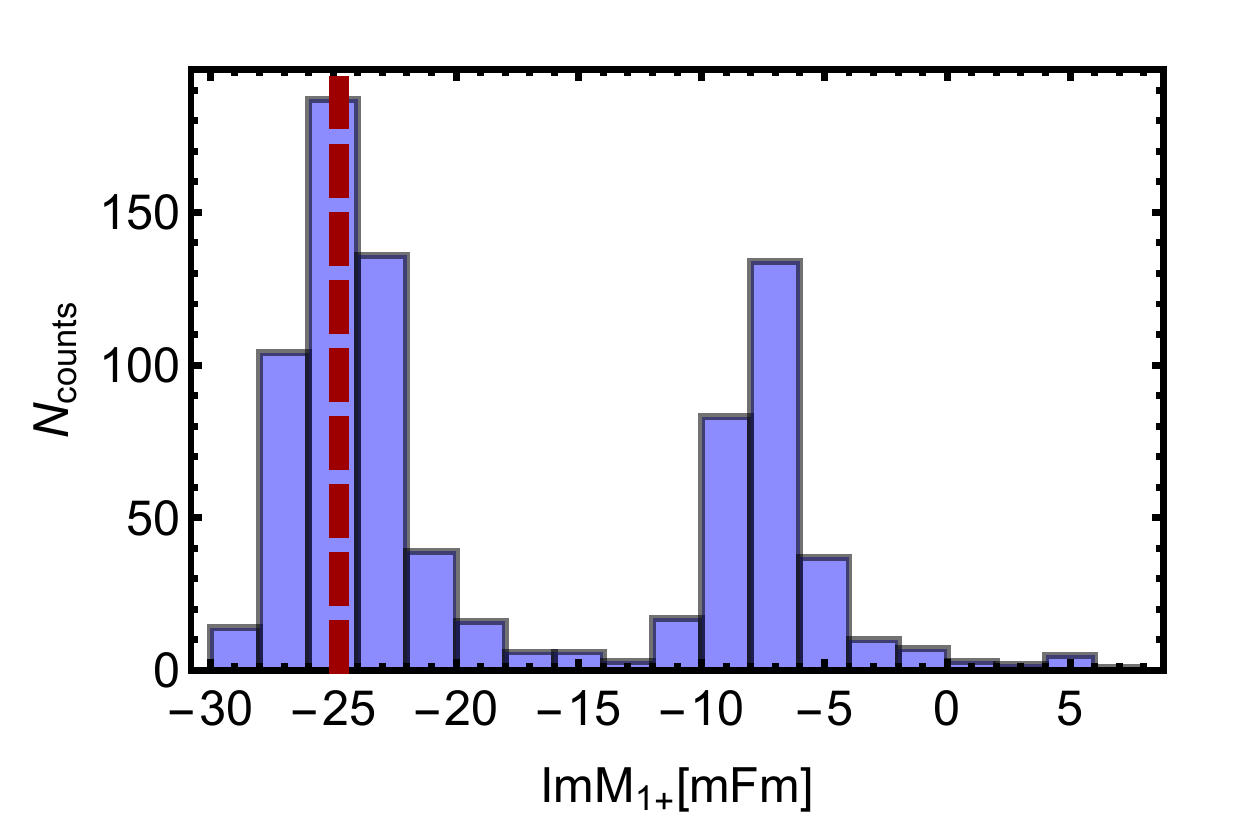}
 \end{overpic} \\
\begin{overpic}[width=0.485\textwidth]{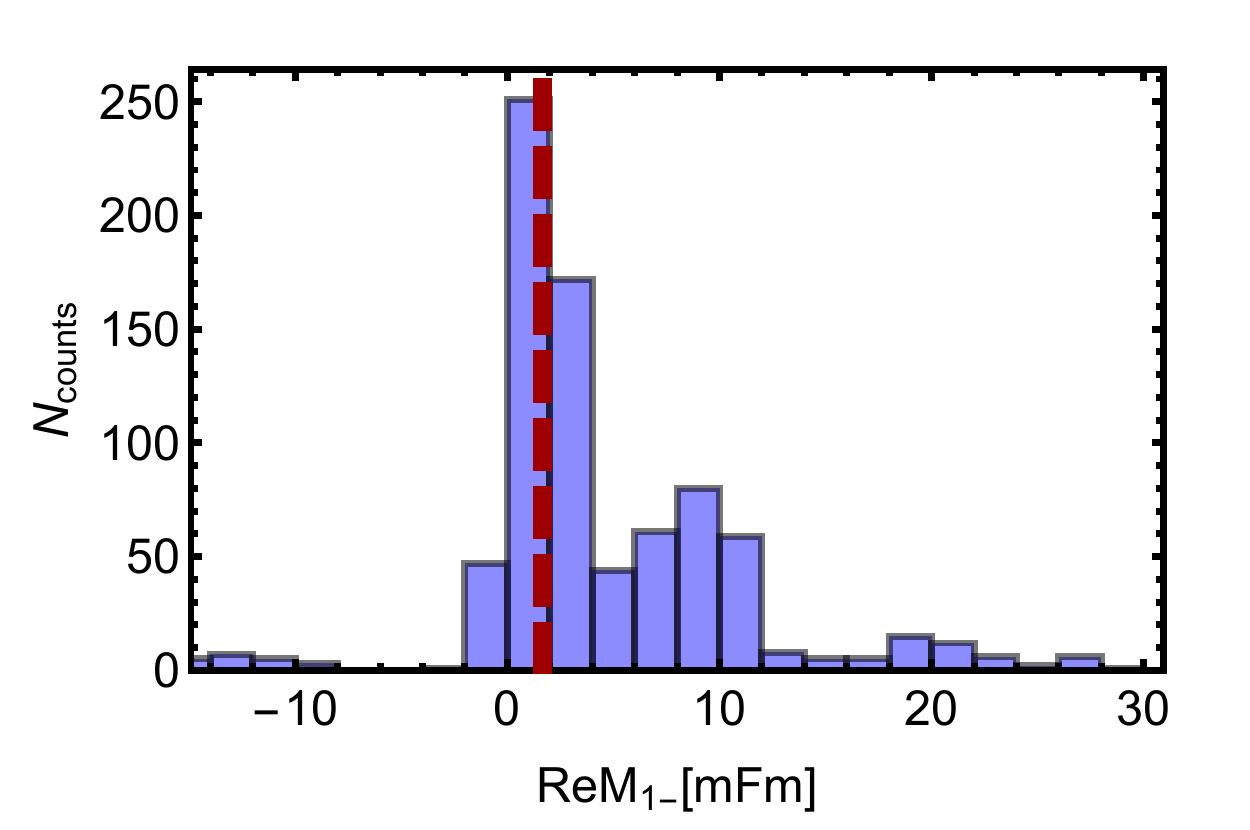}
 \end{overpic} 
\begin{overpic}[width=0.485\textwidth]{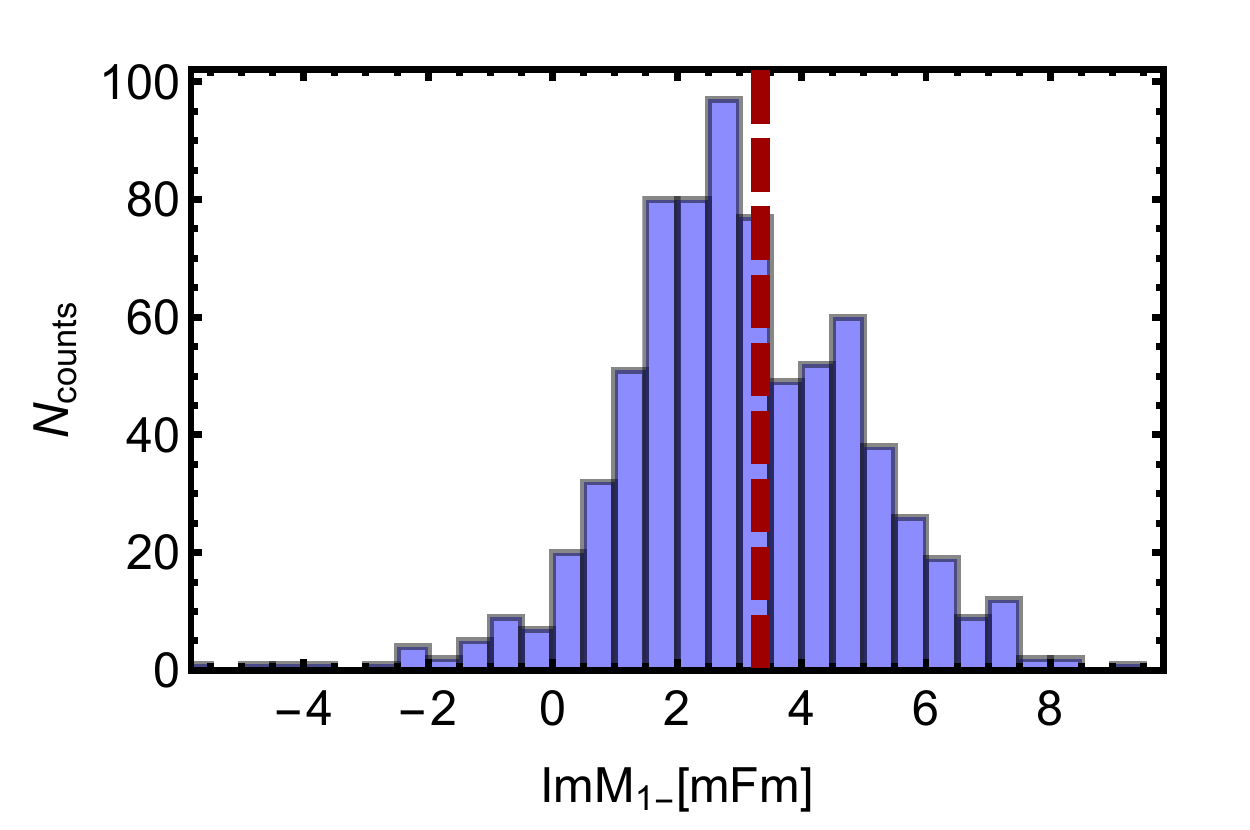}
 \end{overpic}
\caption[Bootstrap-histograms for $S$- and $P$-wave multipoles, resulting from analyses of pseudodata generated from the MAID2007 model. Results are shown for a fourth scenario with extremely large errors.]{The pictures show histograms of the bootstrap-distributions for the real- and imaginary parts of phase-constrained multipoles, resulting from the analyses of the MAID theory-data truncated at $\ell_{\mathrm{max}} = 1$. All solutions below the quantile $u^{\left(P_{NC}\right)}_{0.95}$ are included. Results for the error-scenario $(iii)^{\prime}$ (Table \ref{tab:Lmax1PercentageErrorScenarios}) are plotted.  \newline
A thick red dashed vertical line indicates the result of the fit to the original data (i.e. pseudodata) for scenario $(iii)^{\prime}$. This fit shows quite some bias compared to the fit of the original data for scenario $(i)$, which is shown in Figure \ref{fig:Lmax1PseudoDataFitGroupSFObservablesMultHistograms}, as may be expected.}
\label{fig:Lmax1PseudoDataFitGroupSFObservablesMultHistogramsCrazyErrors}
\end{figure}

\clearpage
Nonetheless, for the quite moderate error-scenarios considered up to now, the bootstrap distributions render the fits to pseudodata generated from the MAID-model for $\ell_{\mathrm{max}} = 1$ stable, since no ambiguities have shown up so far. \newline
In order to investigate the range of the stability for these fits, a quite extreme error-scenario listed as $(iii)^{\prime}$ in Table \ref{tab:Lmax1PercentageErrorScenarios} has been tested. Here, both the observables $P$ and $F$ have had a $95\%$-error applied to them, i.e. they barely add any more information to the fit. A particular real dataset to be discussed later (cf. section \ref{subsec:DeltaRegionDataFits}) will contain data for $P$ with a comparable lack in precision. The full bootstrap-TPWA (section \ref{sec:BootstrappingIntroduction}) has been applied with the same parameters $B = 500$ and $N_{MC} = 250$, as before.\newline
Bootstrap distributions, containing all non-redundant solutions below the quantile $u^{\left(P_{NC}\right)}_{0.95}$, are histogrammed in Figure \ref{fig:Lmax1PseudoDataFitGroupSFObservablesMultHistogramsCrazyErrors}. For most of the multipole-parameters, multimodal distributions can be observed. Exceptions are the imaginary part of $E_{1+}$ and the real part of $M_{1+}$. Fully disconnected peaks can be seen in the $S$-wave $\mathrm{Re}\left[ E_{0+} \right]$, almost disconnected one's for the parameter $\mathrm{Im}\left[ M_{1+} \right]$. For the parameter $\mathrm{Re}\left[ E_{1+} \right]$, a multimodality is suggested but not fully resolved. In summary, it is safe to conclude that for the extreme error-scenario $(iii)^{\prime}$, the stability of the TPWA-fit to the pseudodata is broken. The bootstrap-procedure has made the ambiguities visible in the parameter distributions. \newline

\begin{figure}[ht]
 \centering
\begin{overpic}[width=0.325\textwidth]{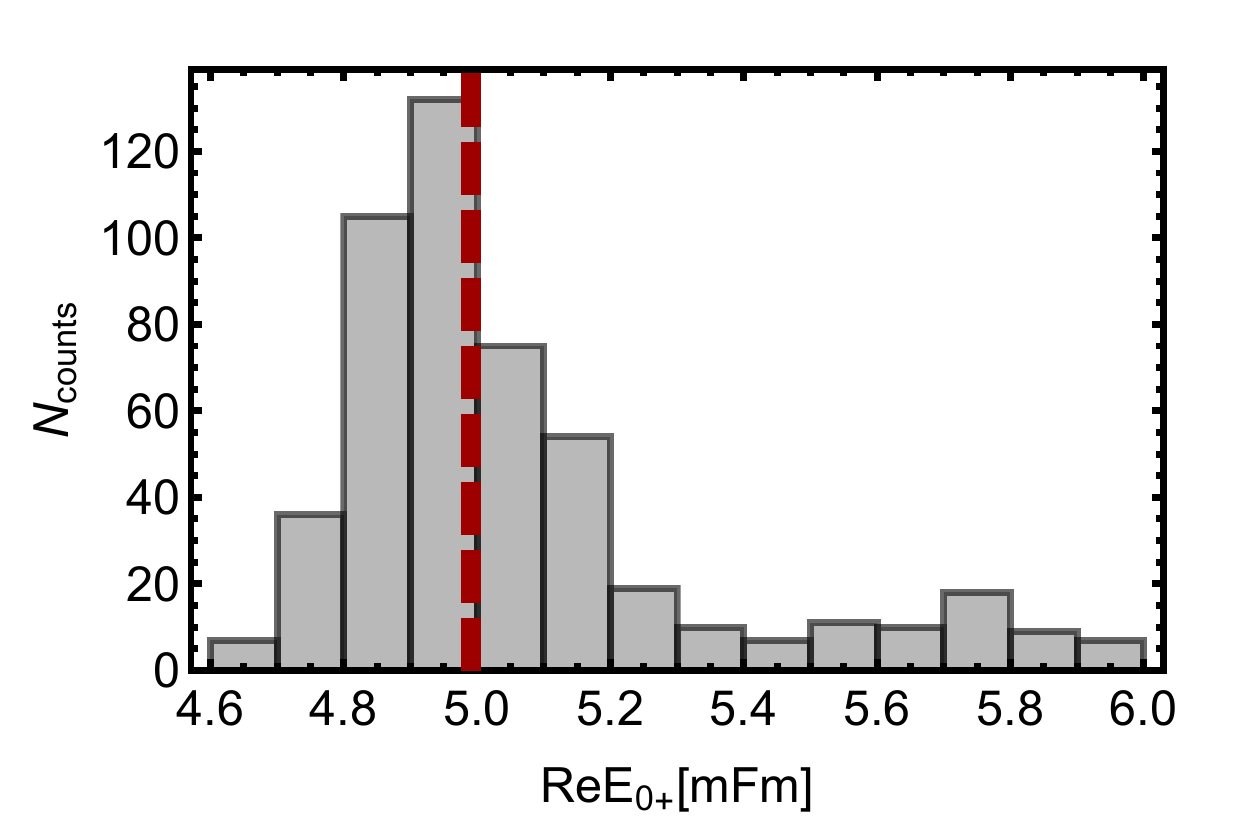}
 \end{overpic}
\begin{overpic}[width=0.325\textwidth]{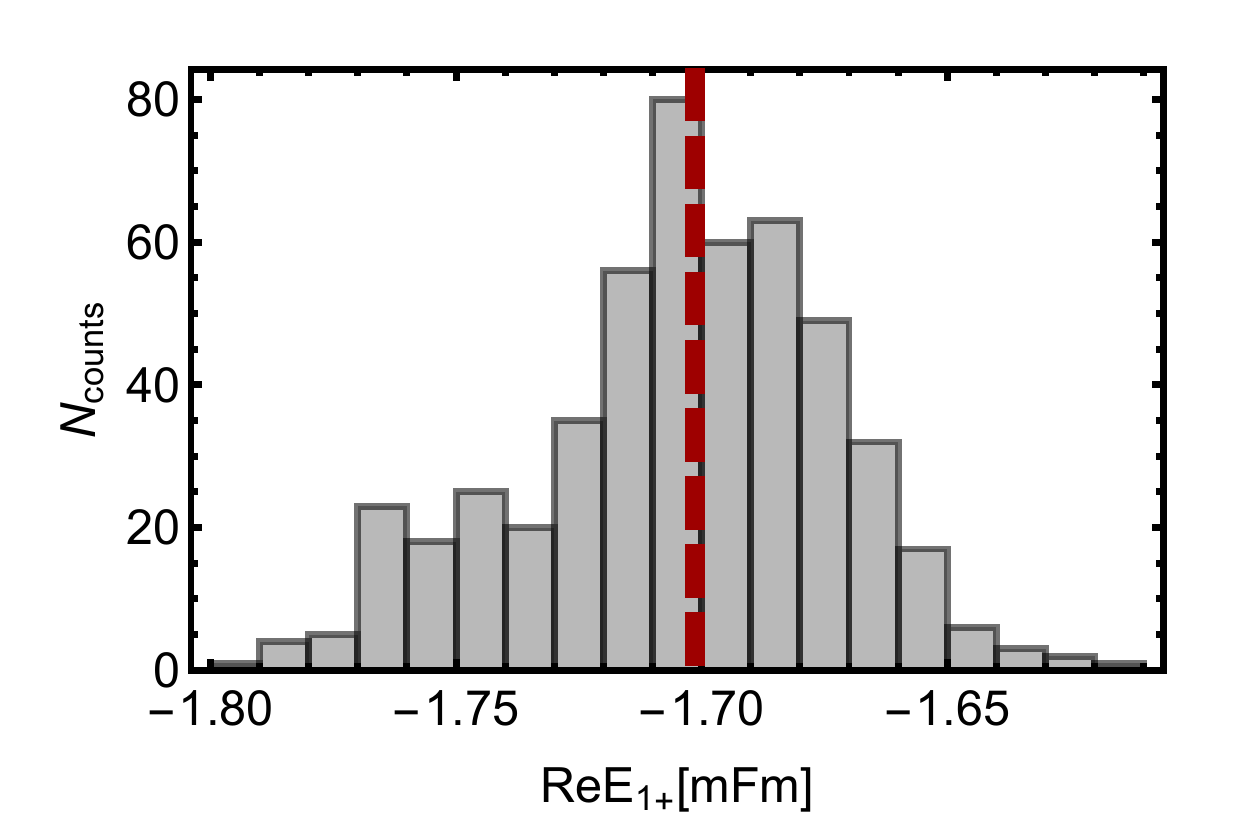}
 \end{overpic}
\begin{overpic}[width=0.325\textwidth]{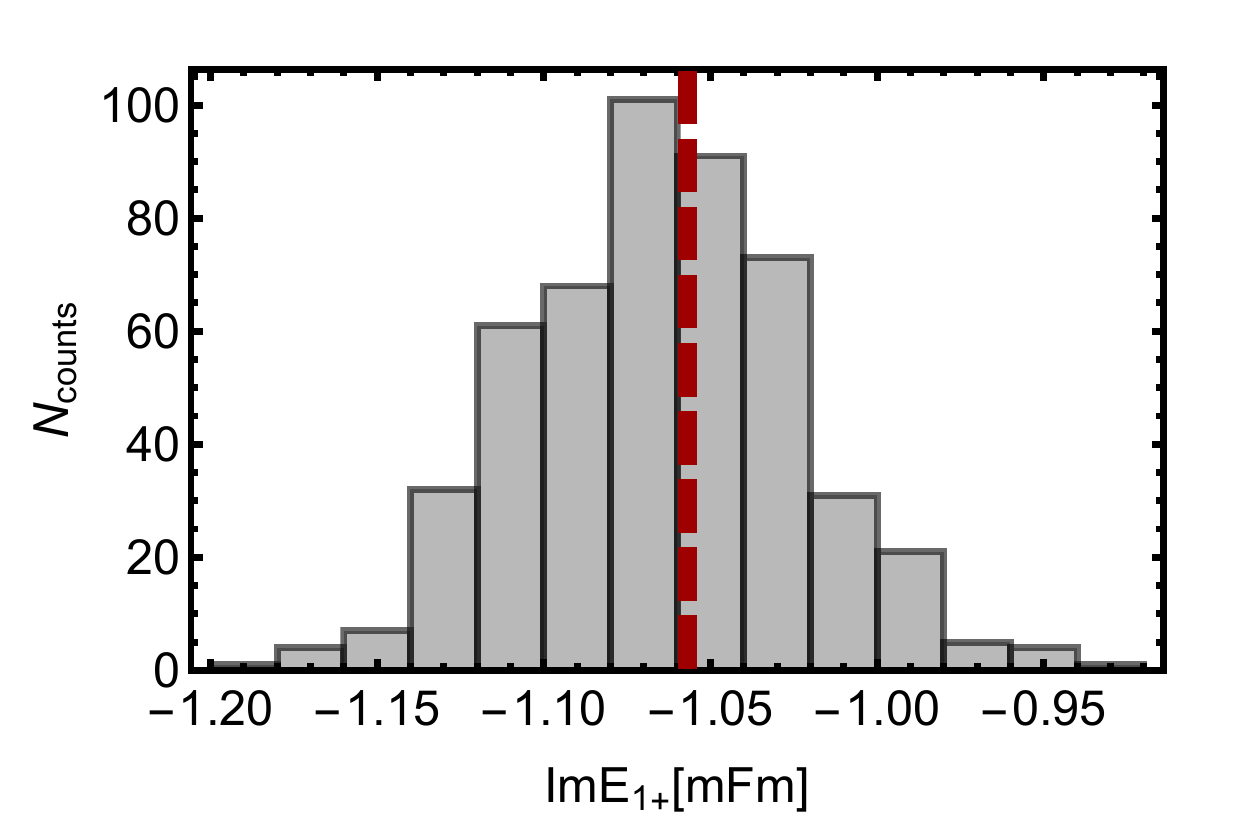}
 \end{overpic} \\
\begin{overpic}[width=0.325\textwidth]{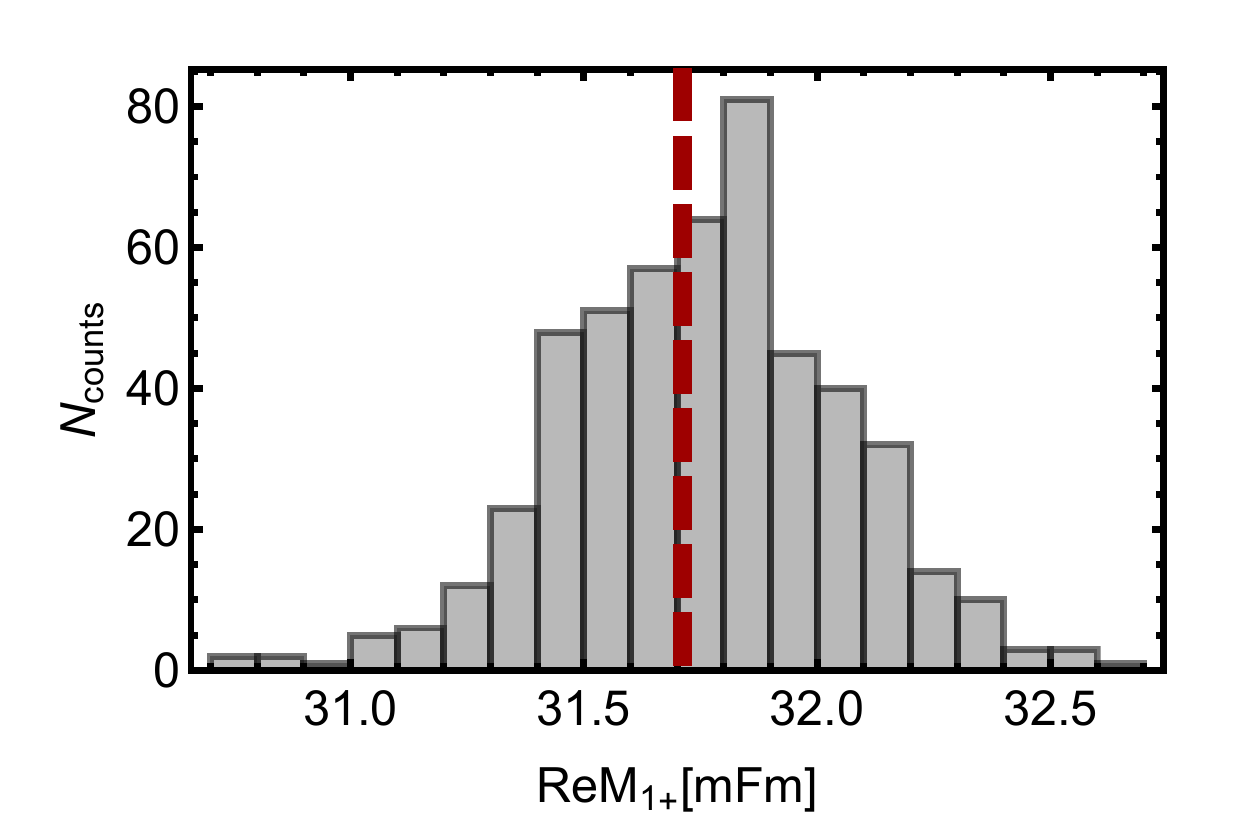}
 \end{overpic}
 \begin{overpic}[width=0.325\textwidth]{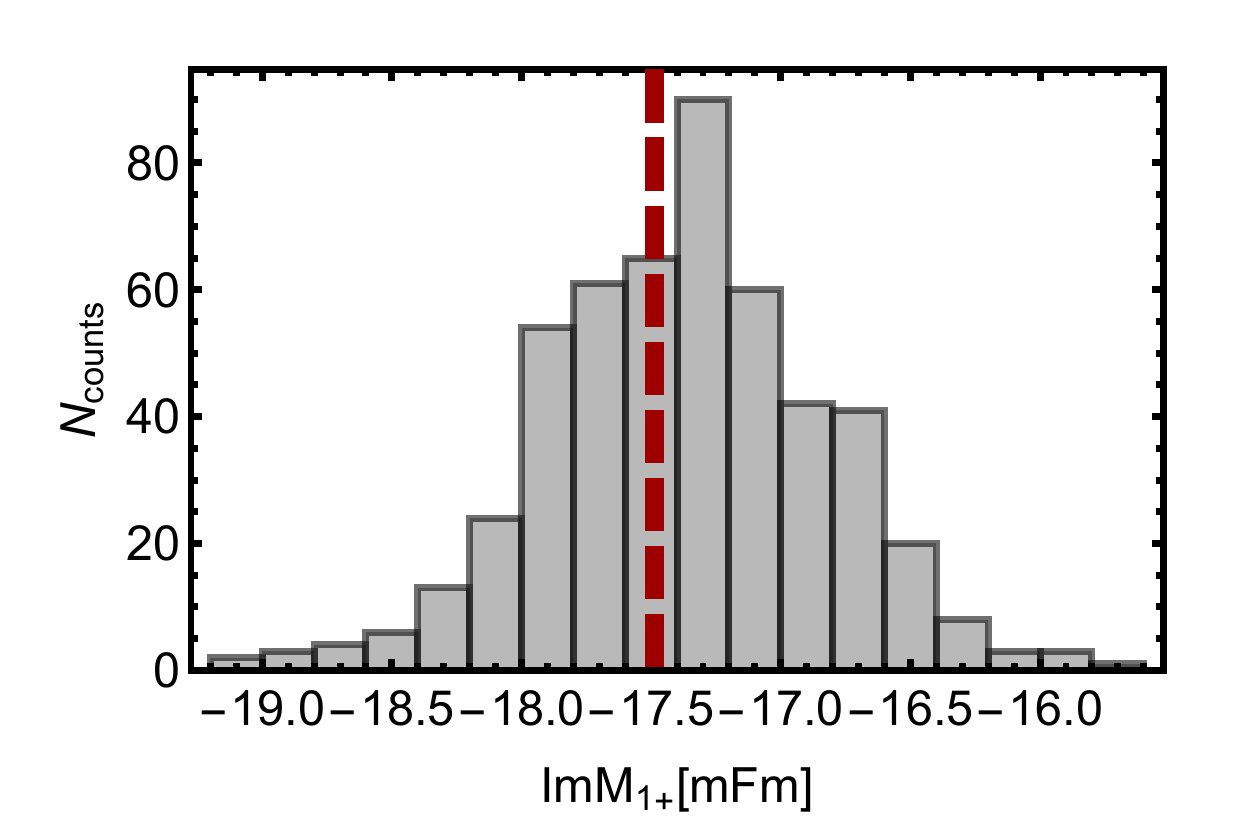}
 \end{overpic}
\begin{overpic}[width=0.325\textwidth]{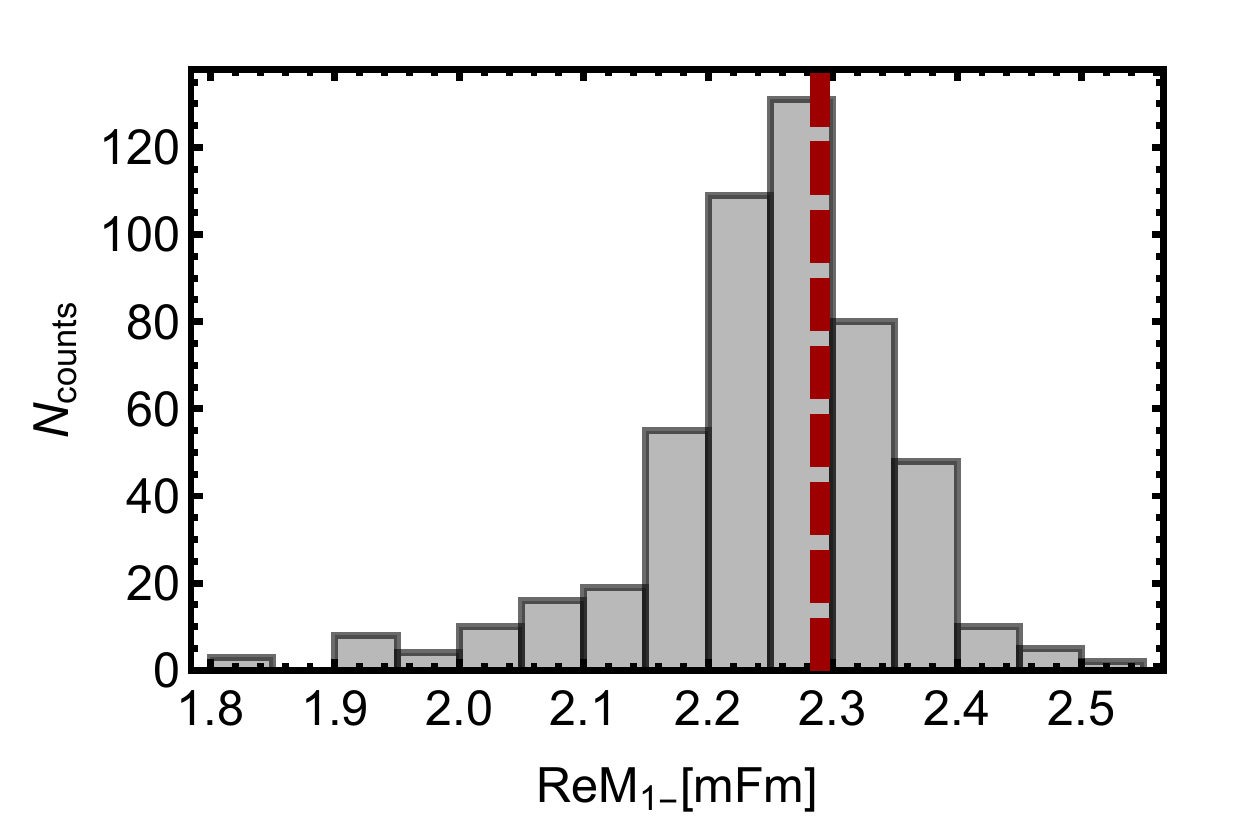}
 \end{overpic} \\
 \begin{overpic}[width=0.325\textwidth]{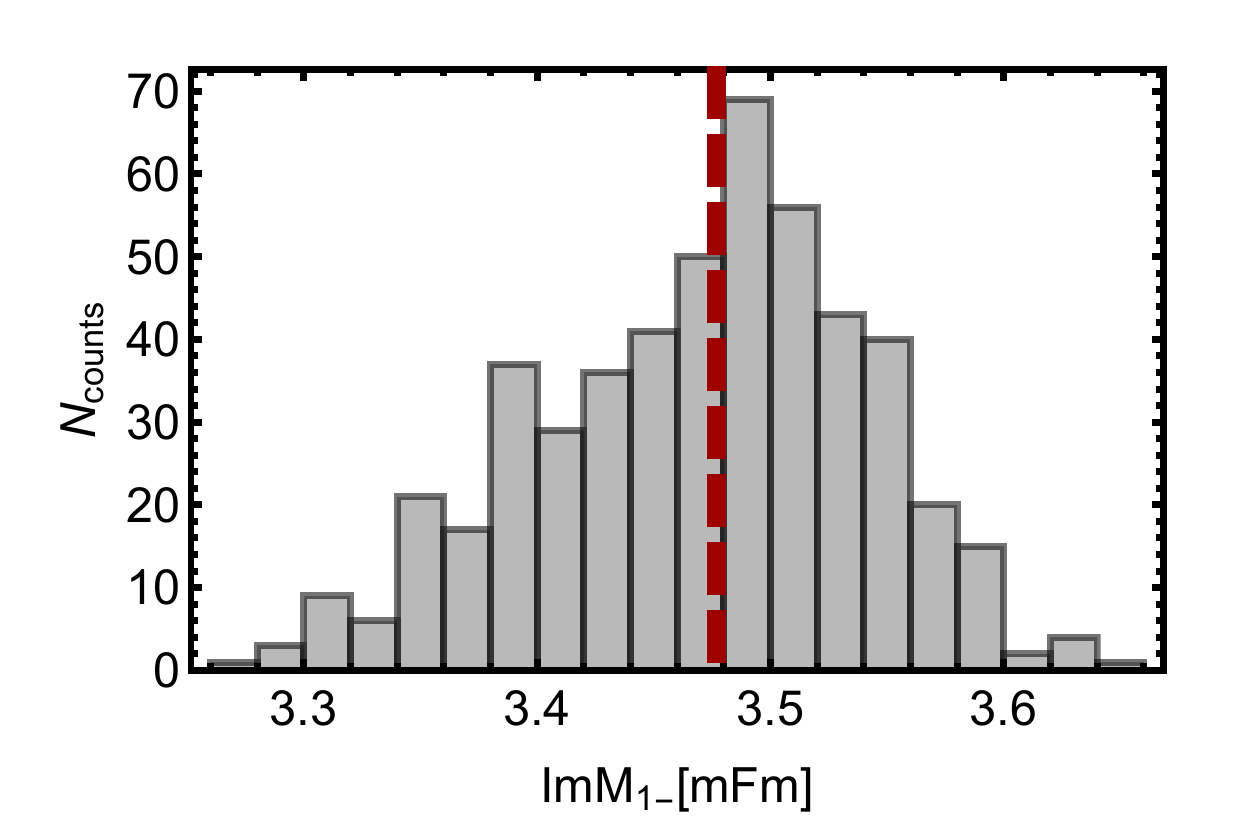}
 \end{overpic}
 \begin{overpic}[width=0.325\textwidth]{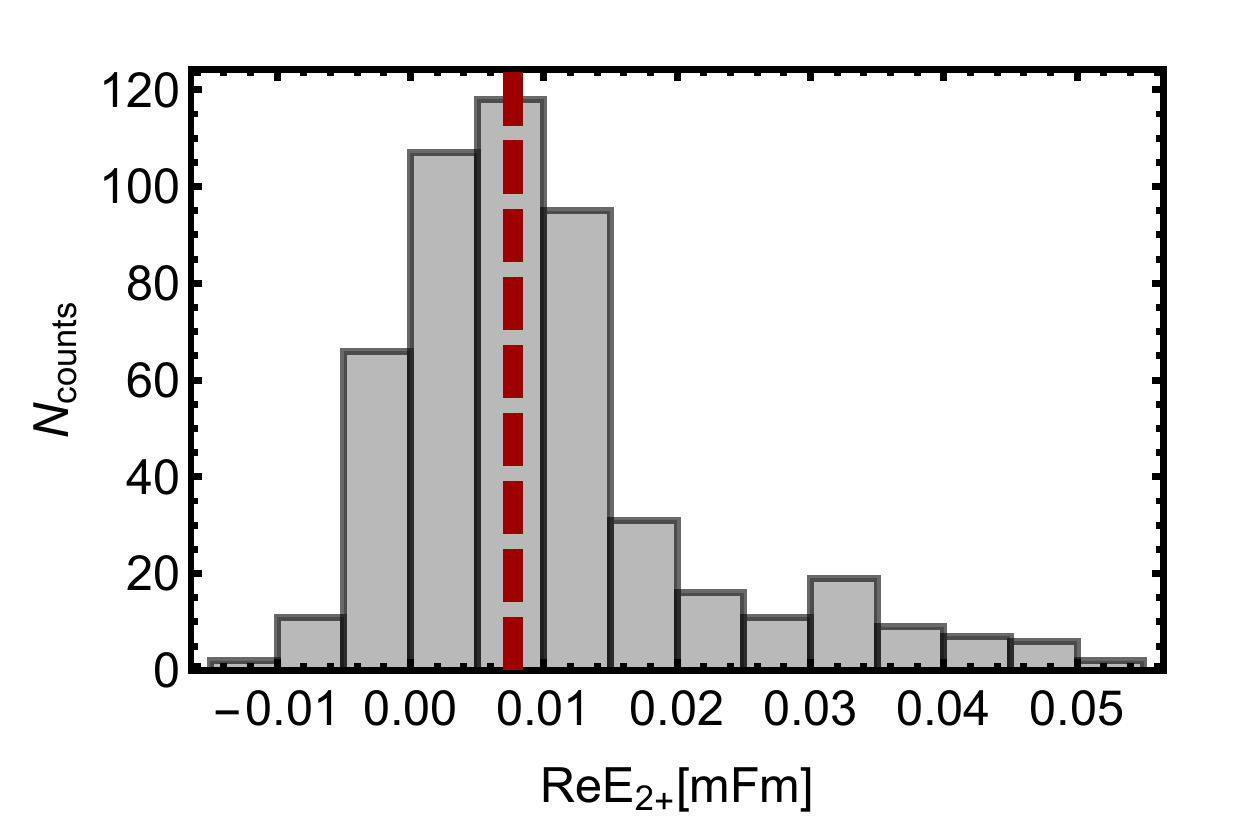}
 \end{overpic}
\begin{overpic}[width=0.325\textwidth]{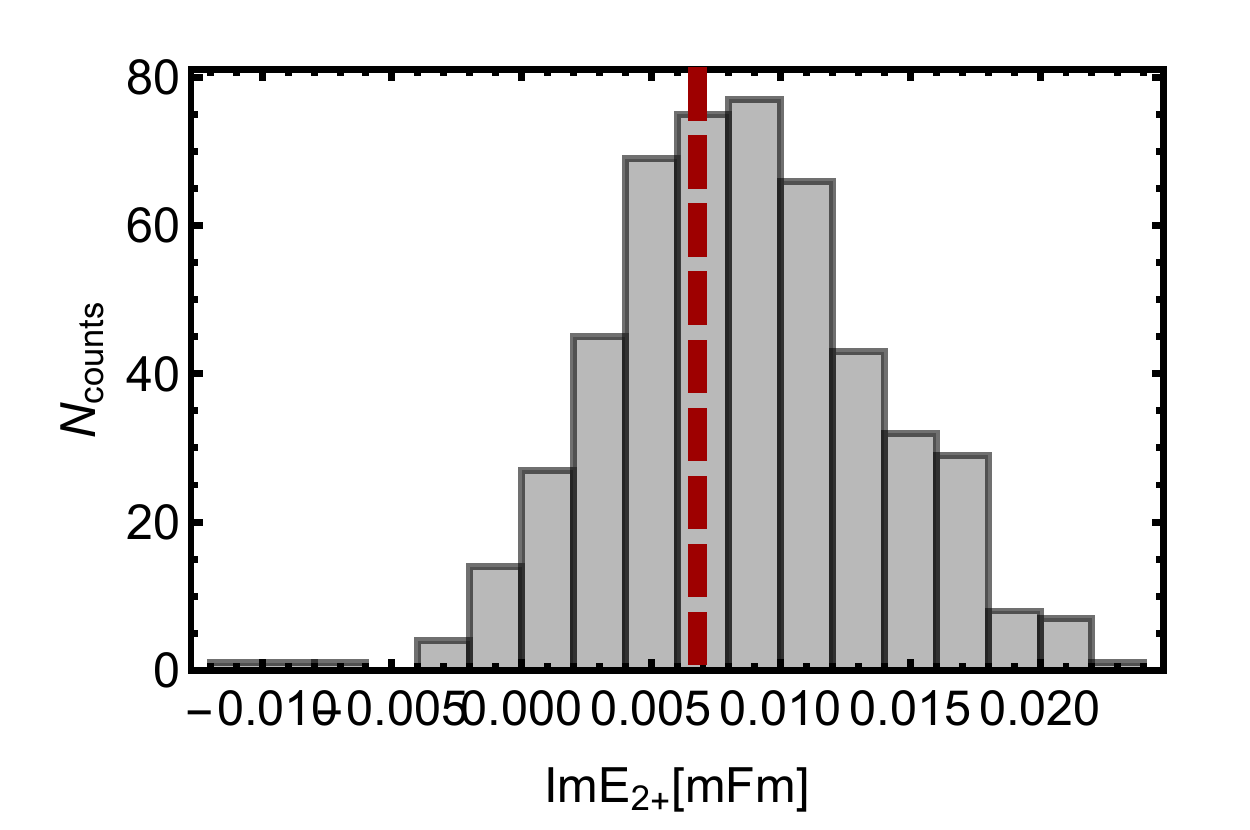}
 \end{overpic} \\
 \begin{overpic}[width=0.325\textwidth]{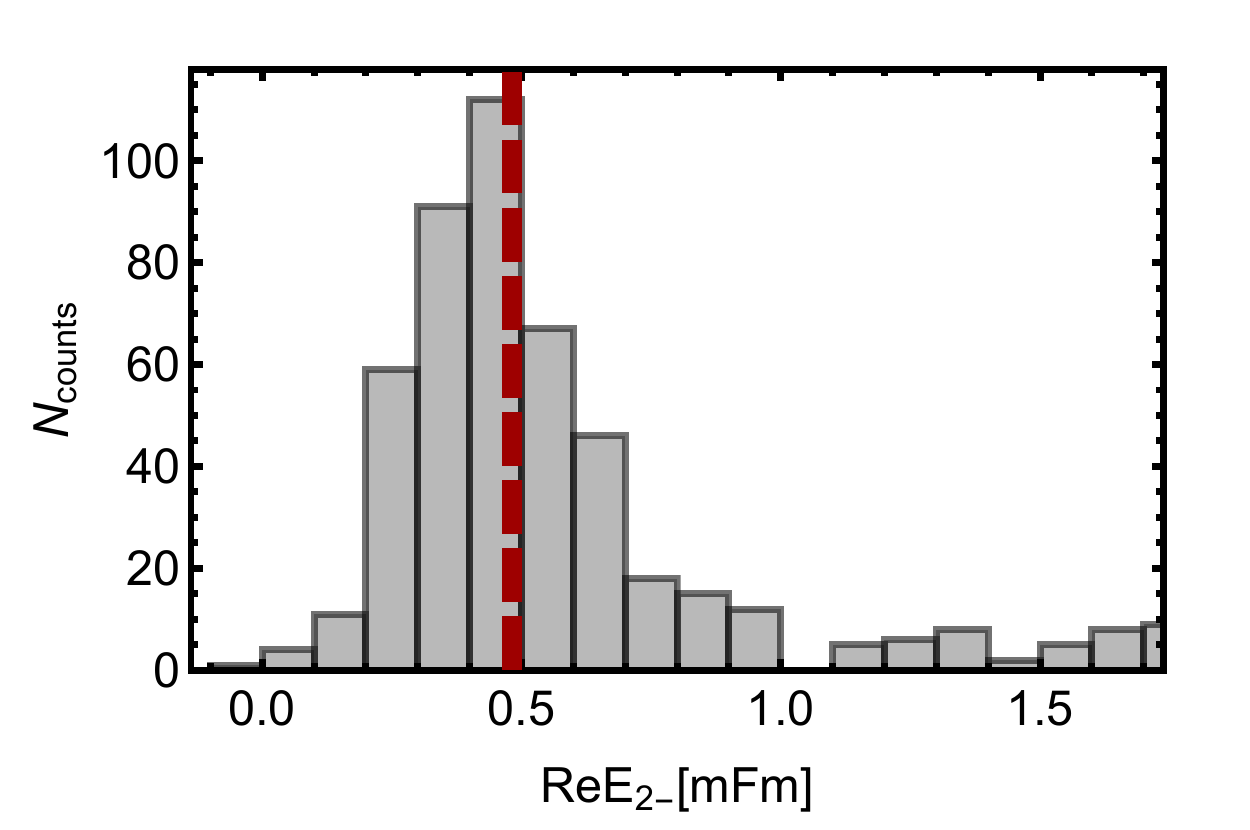}
 \end{overpic}
 \begin{overpic}[width=0.325\textwidth]{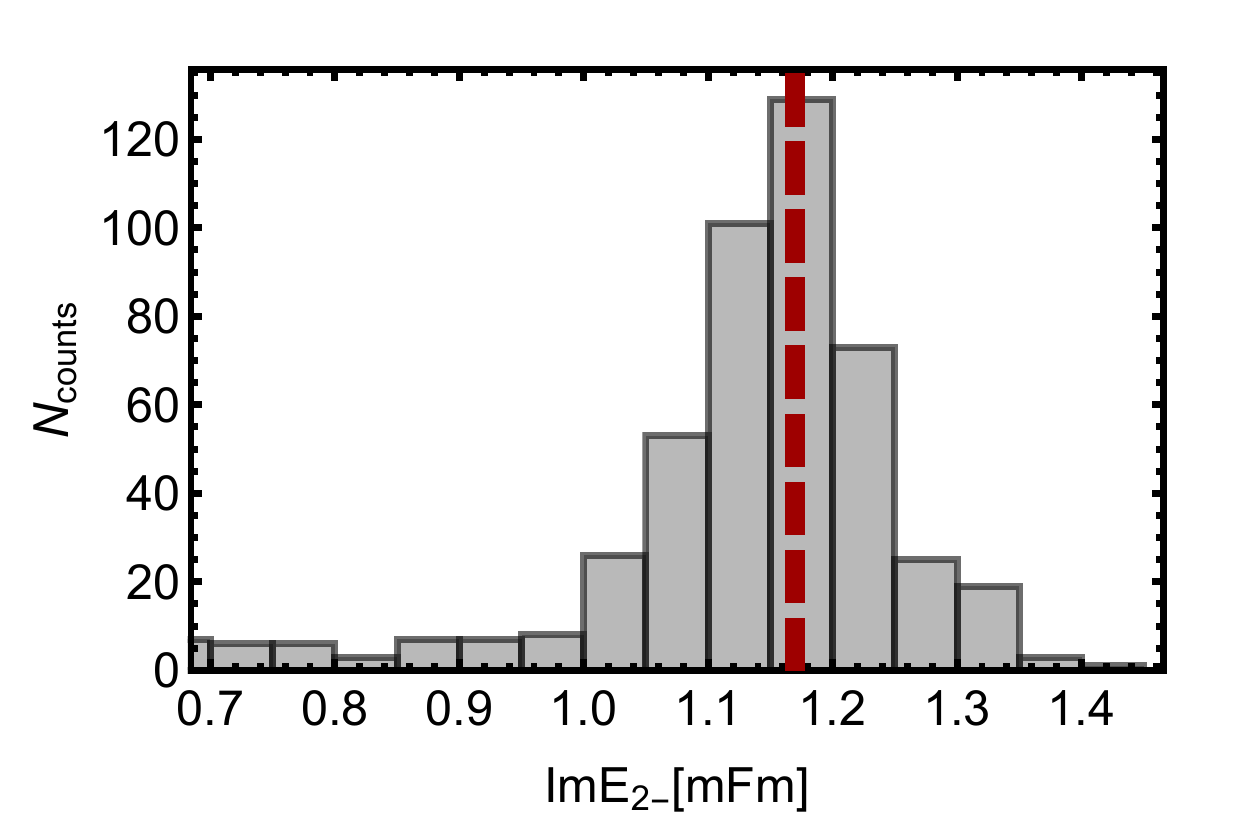}
 \end{overpic}
\begin{overpic}[width=0.325\textwidth]{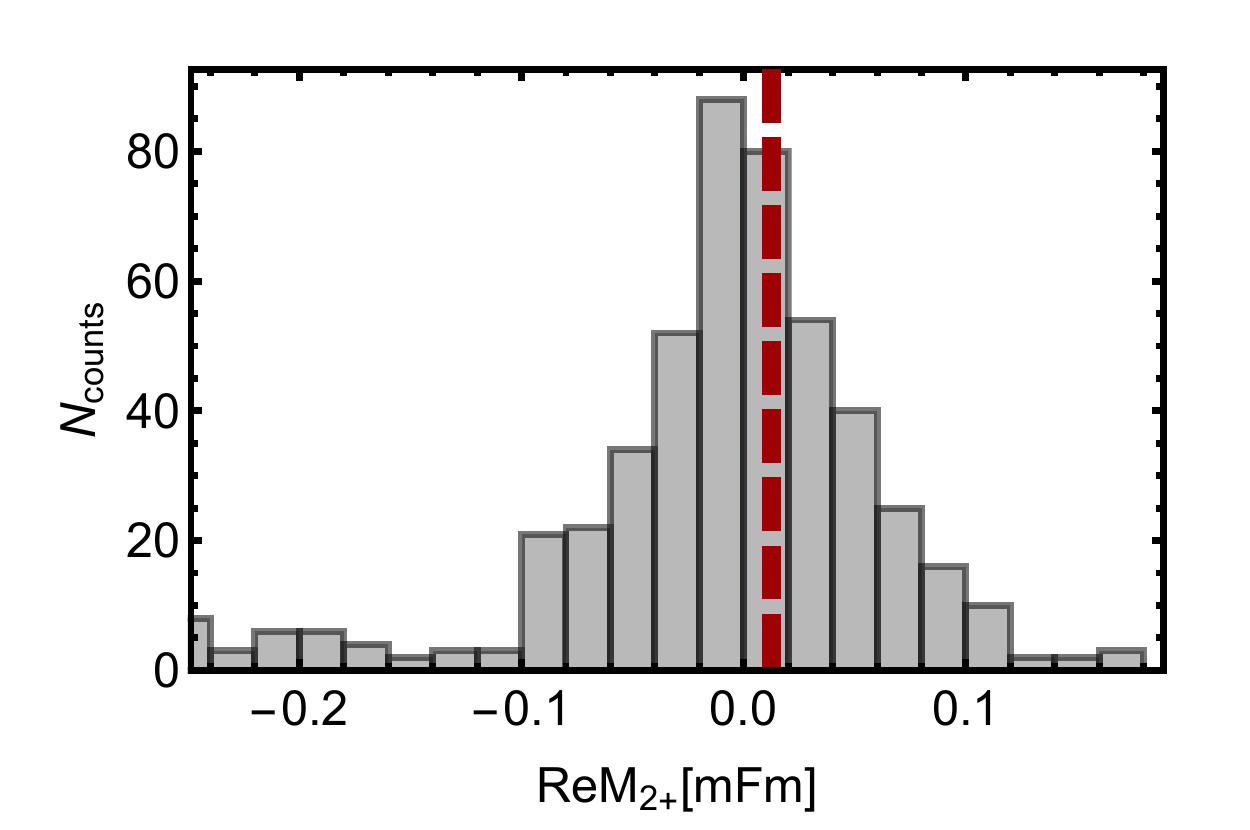}
 \end{overpic} \\
 \begin{overpic}[width=0.325\textwidth]{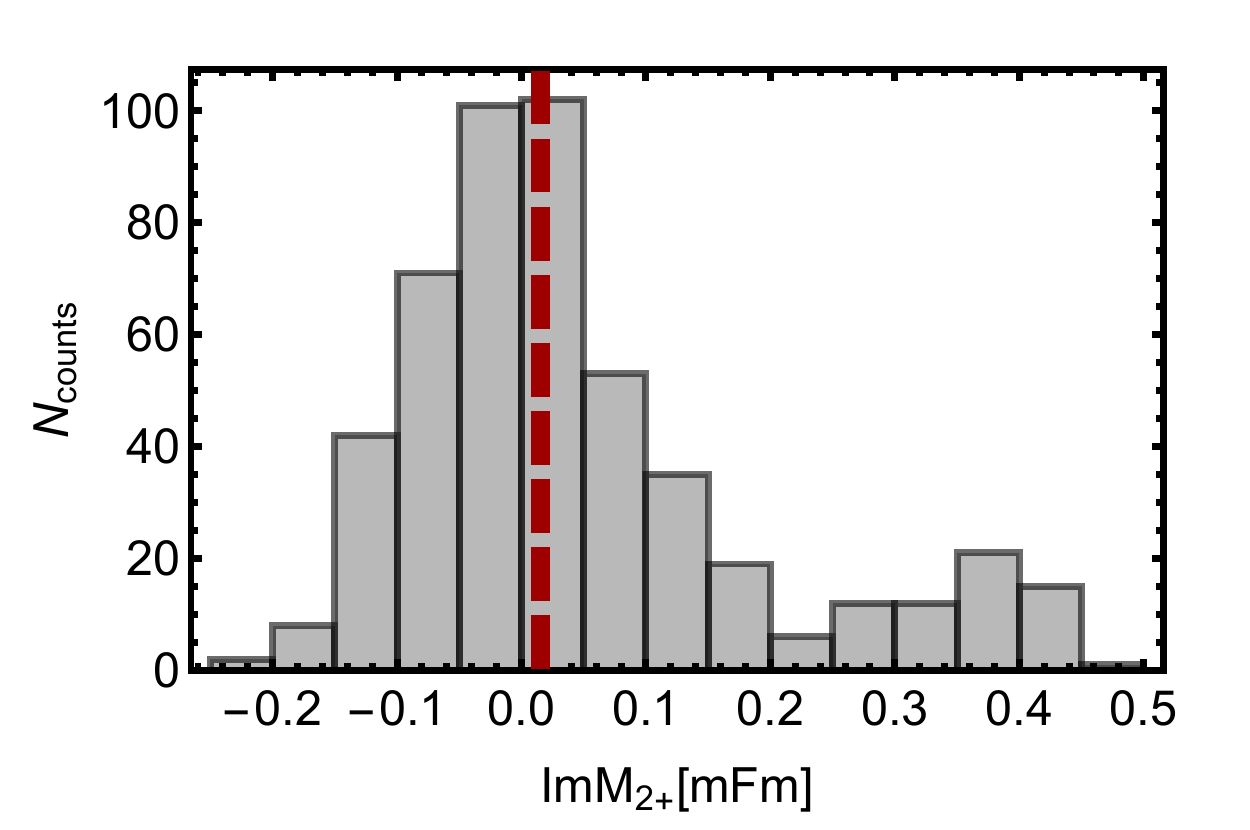}
 \end{overpic}
 \begin{overpic}[width=0.325\textwidth]{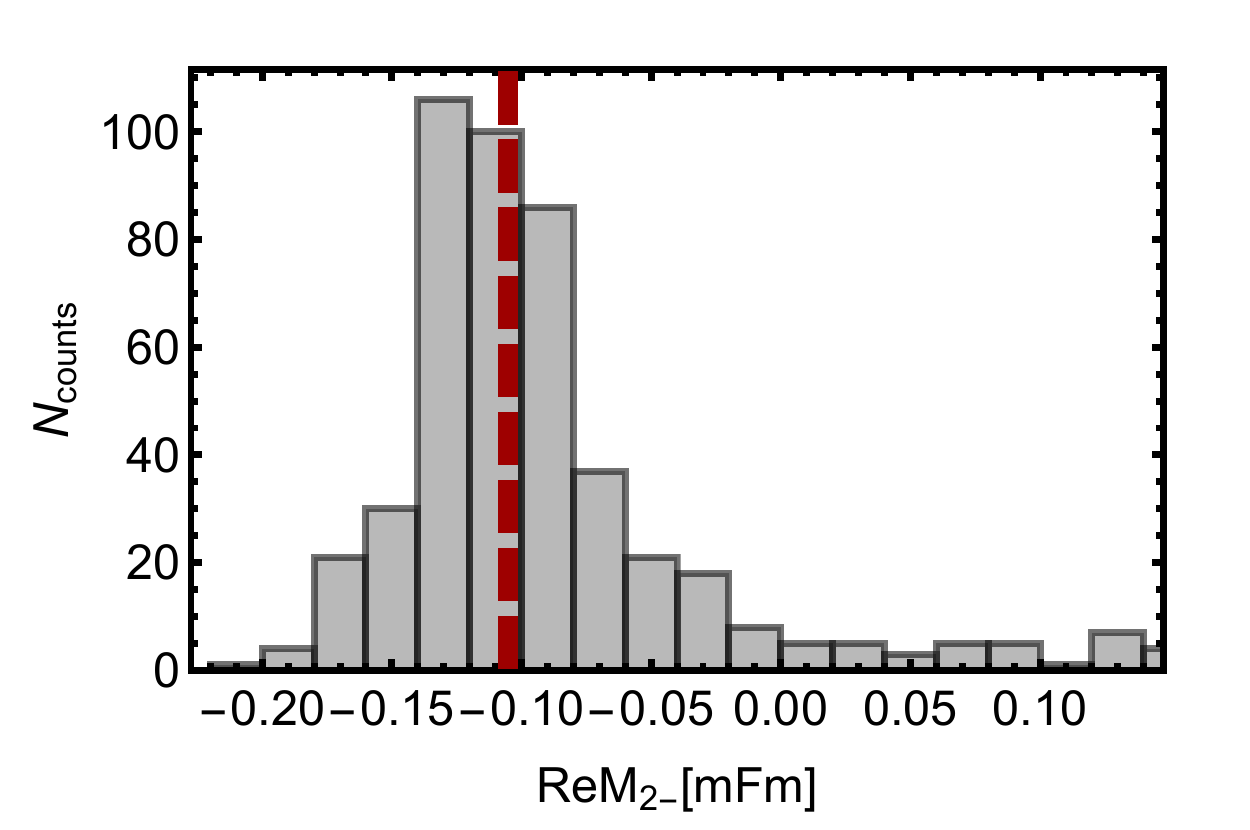}
 \end{overpic}
\begin{overpic}[width=0.325\textwidth]{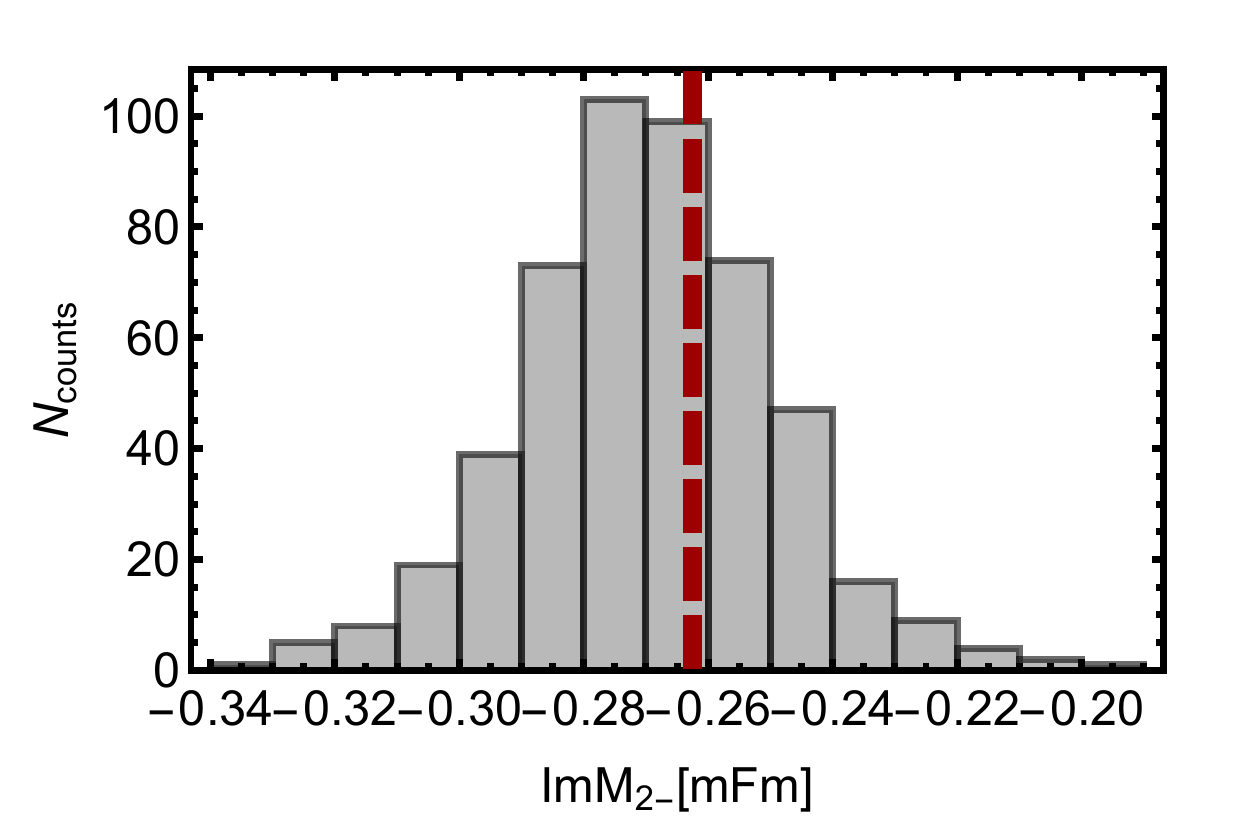}
 \end{overpic}
\caption[Bootstrap-histograms containing the global minima found in each fit of the $B=500$ bootstrap-replications of the MAID pseudodata truncated at $\ell_{\mathrm{max}} = 2$. Results are shown for the first error-scenario.]{The histograms shown here contain (only!) the global minima found in each fit of the $B=500$ bootstrap-replications of the MAID pseudodata truncated at $\ell_{\mathrm{max}} = 2$. Global minima were found by fitting with a pool of $N_{MC} = 400$ initial conditions for each replicate. For the results shown here, the error-scenario $(i)$ from Table \ref{tab:Lmax2PercentageErrorScenarios} has been analyzed. \newline
The global minimum of the fit to the original MAID pseudodata is indicated by a red dashed vertical line. Since the errors in scenario $(i)$ are very small, the original fit-result still gives quite a good indication of where the original MAID-multipoles are situated.}
\label{fig:Lmax2PseudoDataFitGroupSFObservablesMultHistograms1}
\end{figure}

\begin{figure}[ht]
 \centering
\begin{overpic}[width=0.325\textwidth]{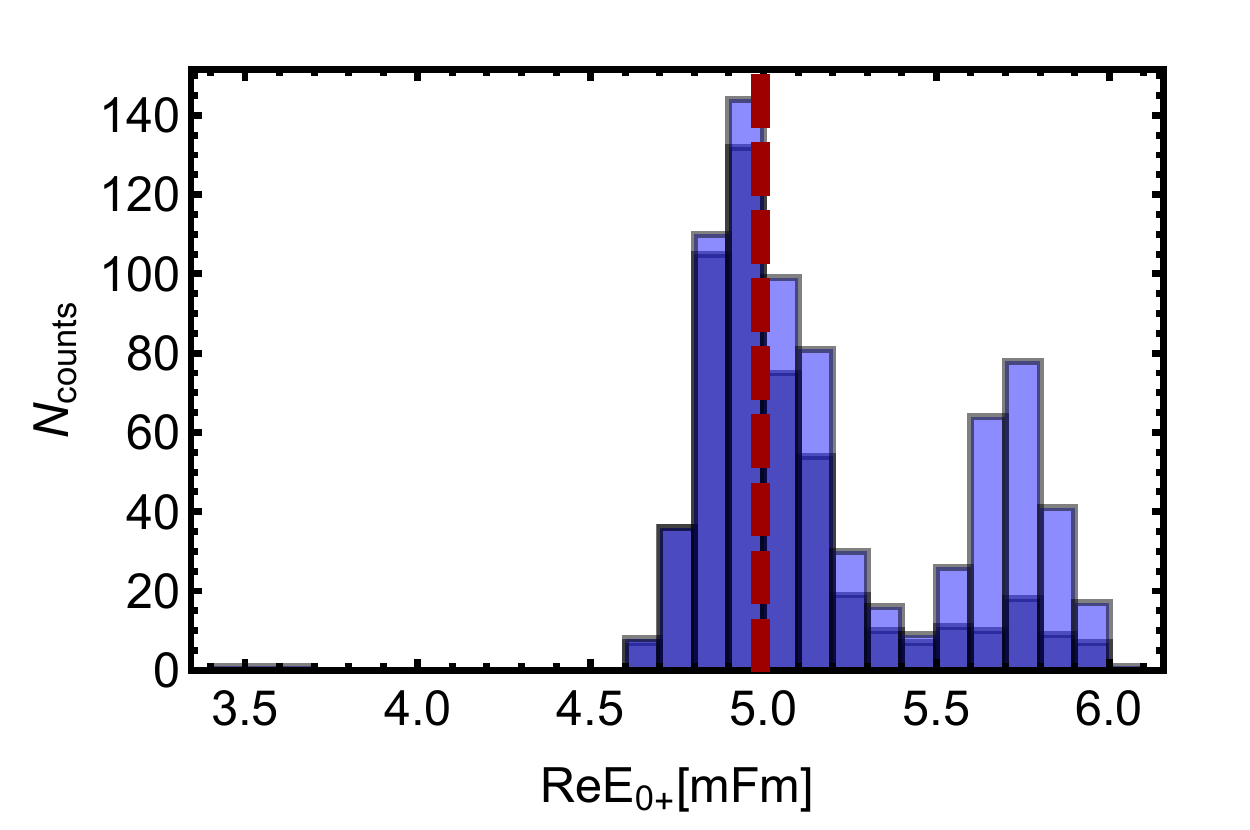}
 \end{overpic}
\begin{overpic}[width=0.325\textwidth]{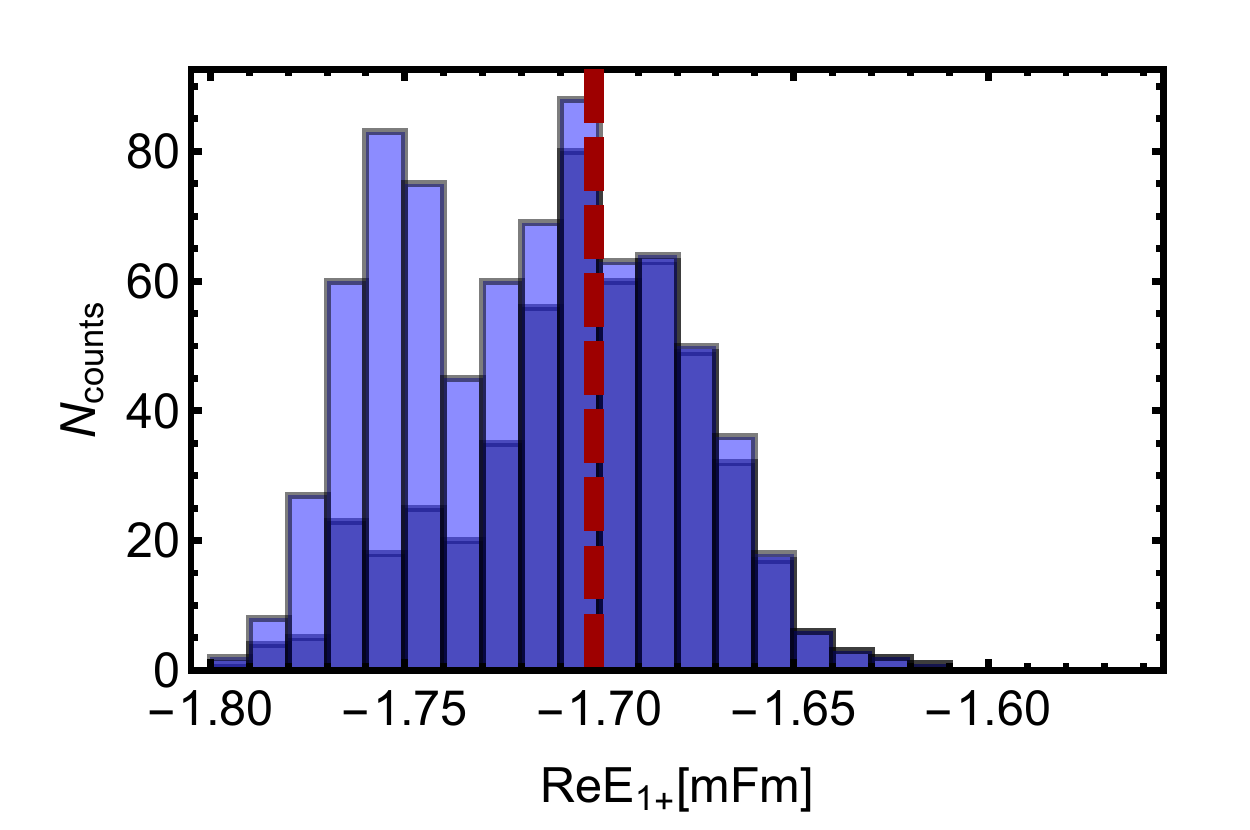}
 \end{overpic}
\begin{overpic}[width=0.325\textwidth]{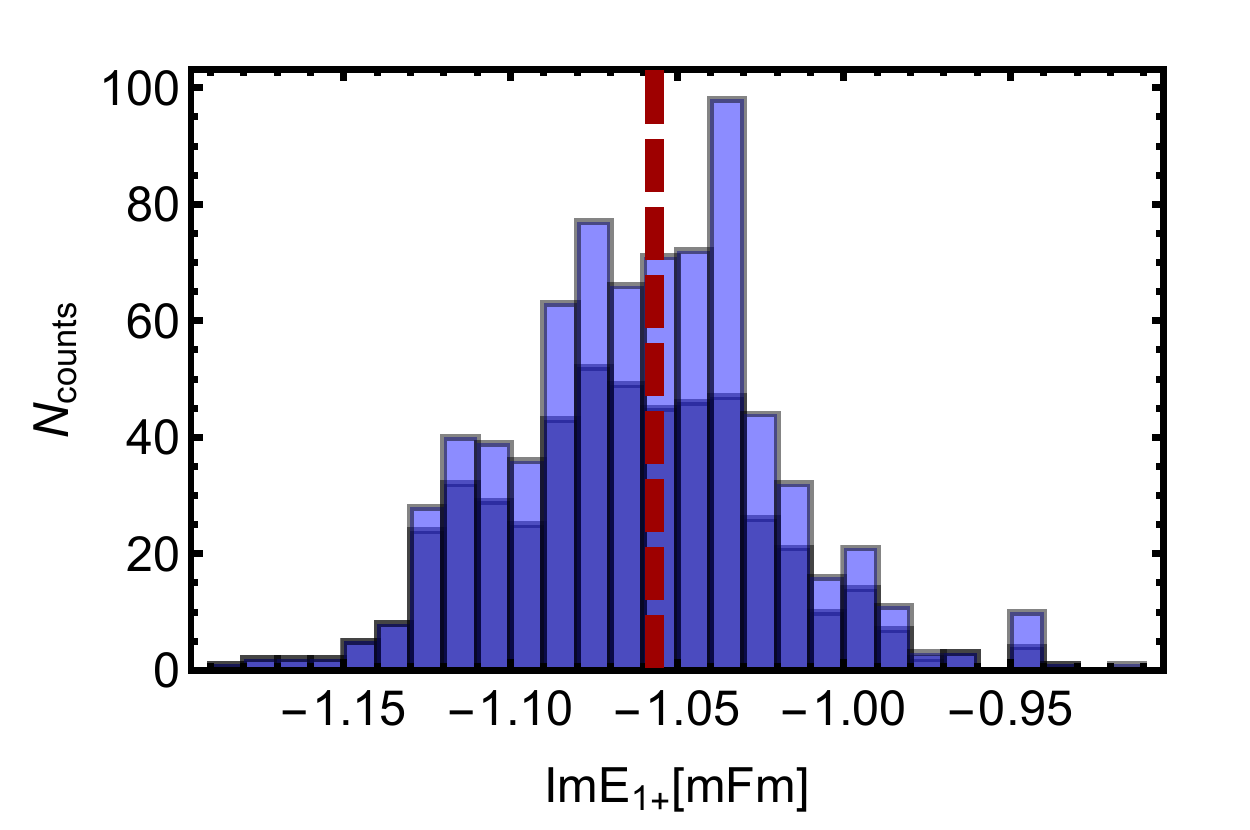}
 \end{overpic} \\
\begin{overpic}[width=0.325\textwidth]{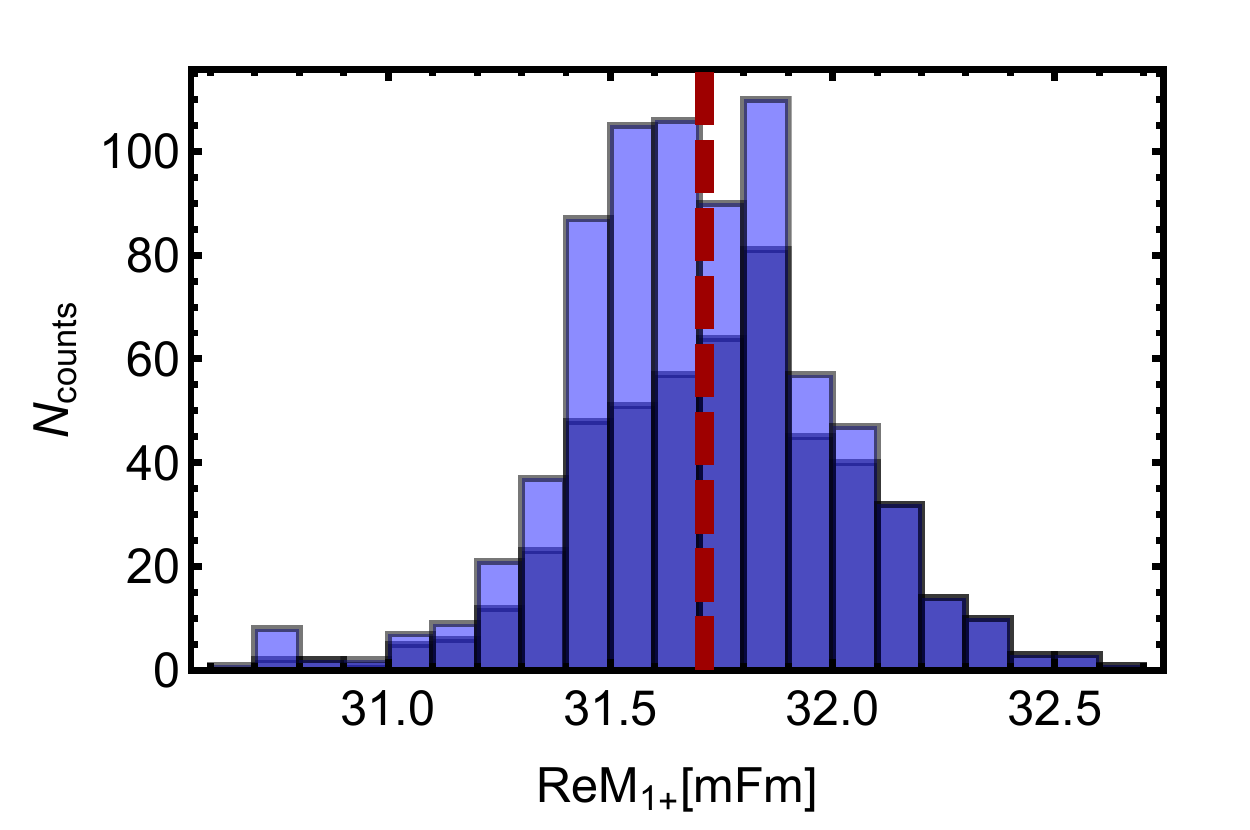}
 \end{overpic}
 \begin{overpic}[width=0.325\textwidth]{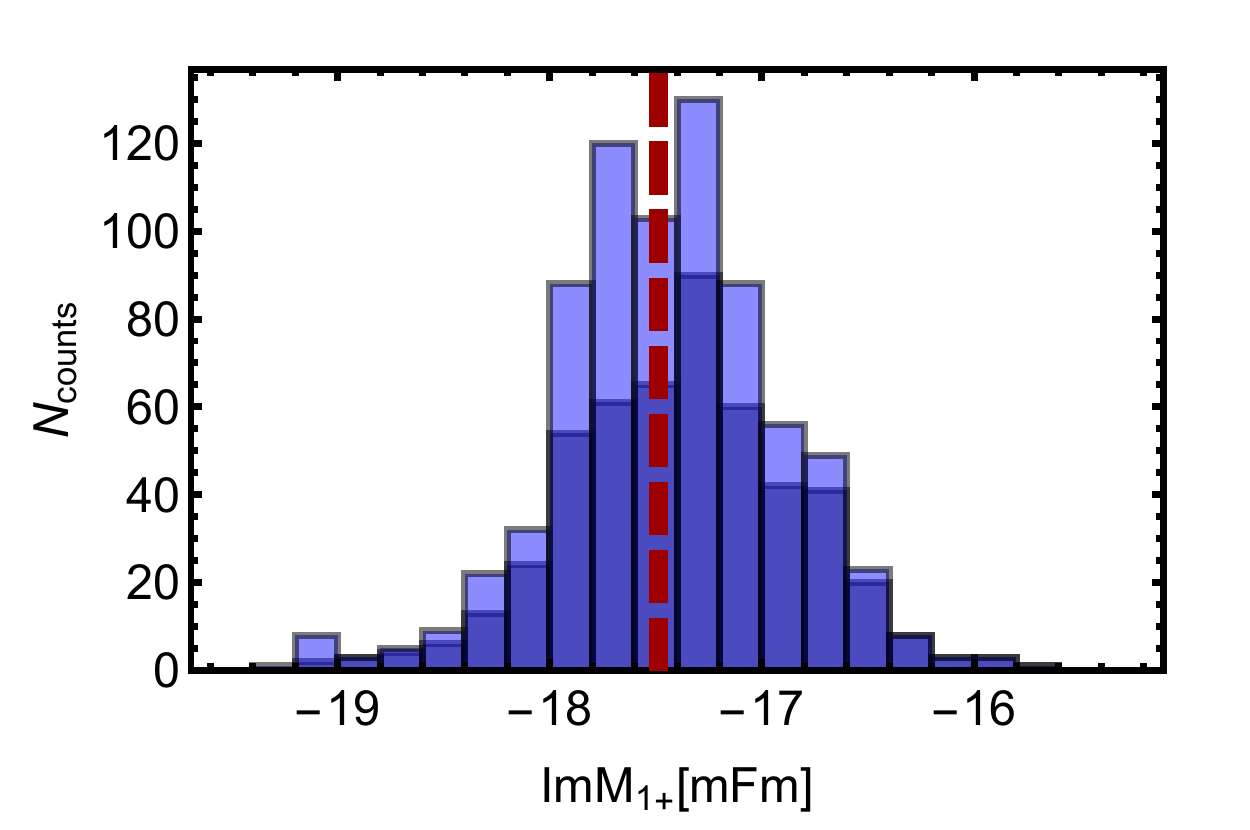}
 \end{overpic}
\begin{overpic}[width=0.325\textwidth]{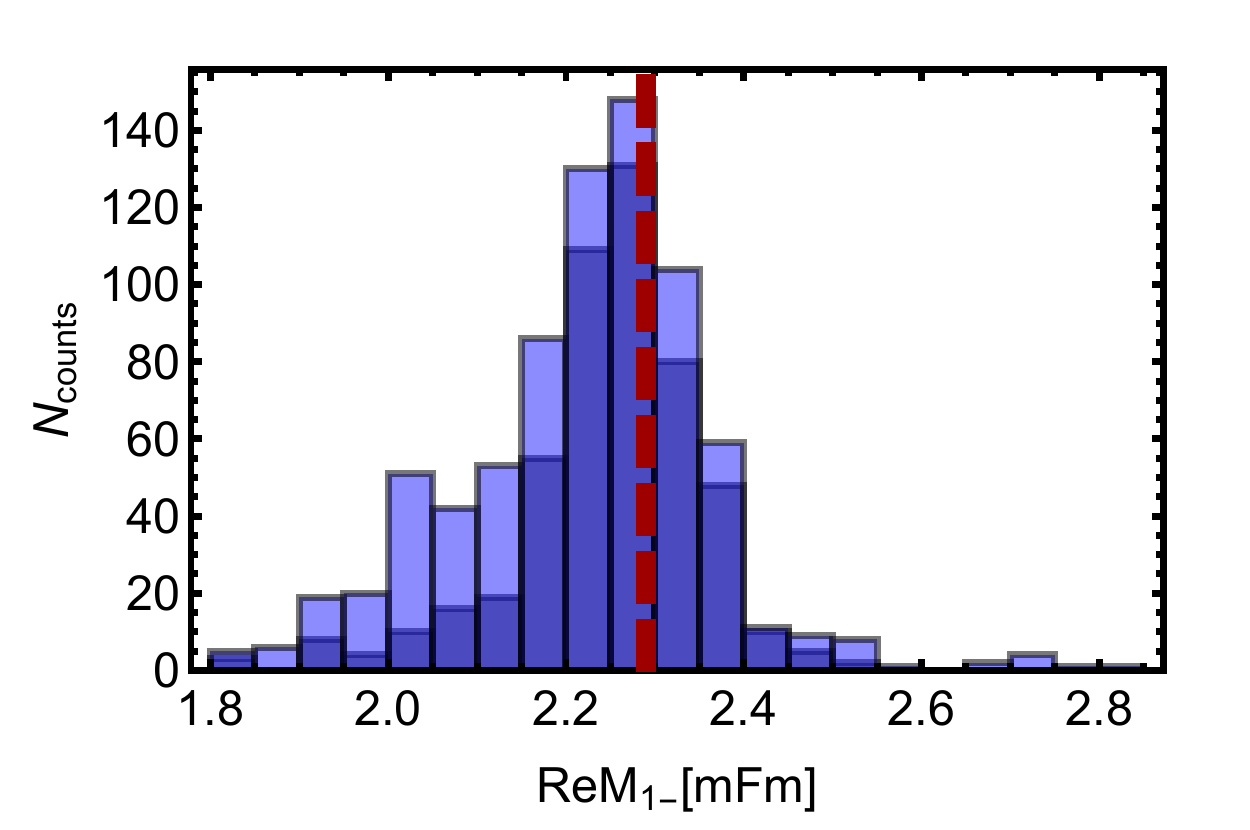}
 \end{overpic} \\
 \begin{overpic}[width=0.325\textwidth]{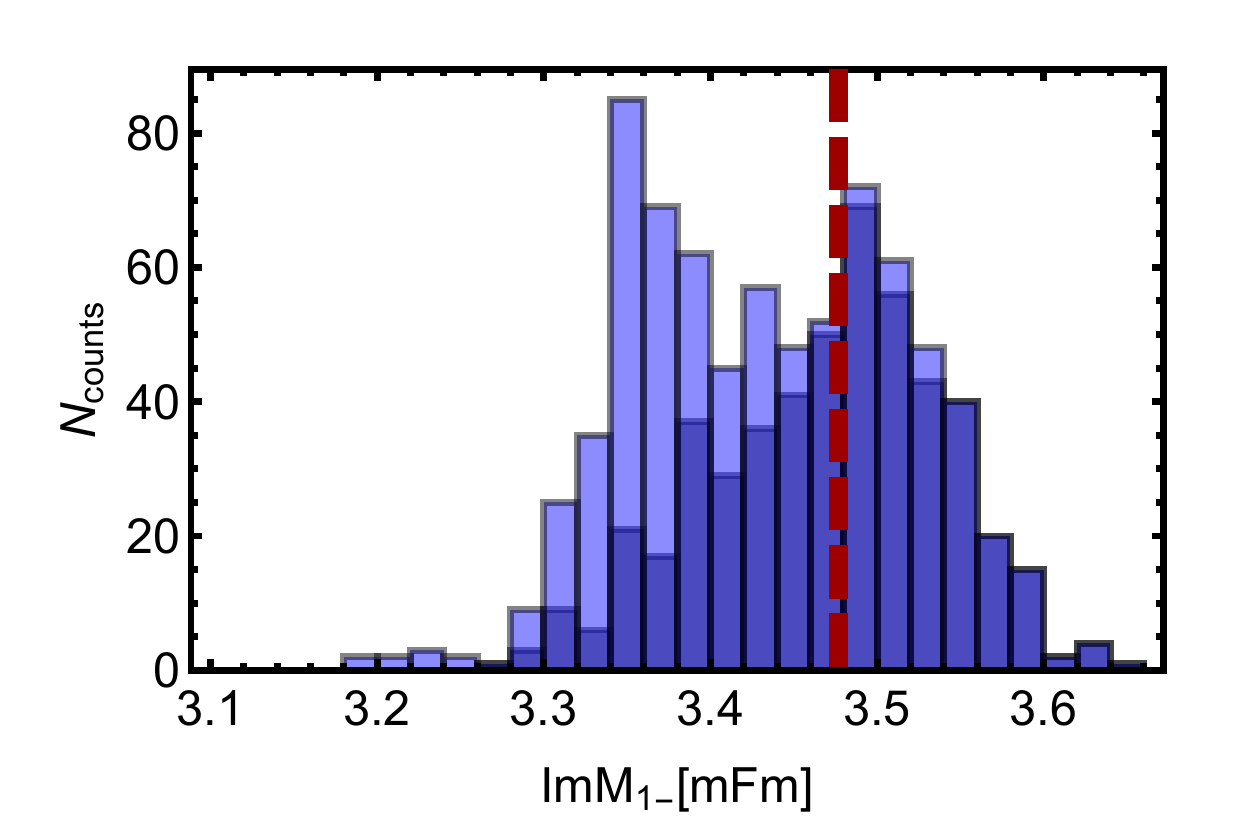}
 \end{overpic}
 \begin{overpic}[width=0.325\textwidth]{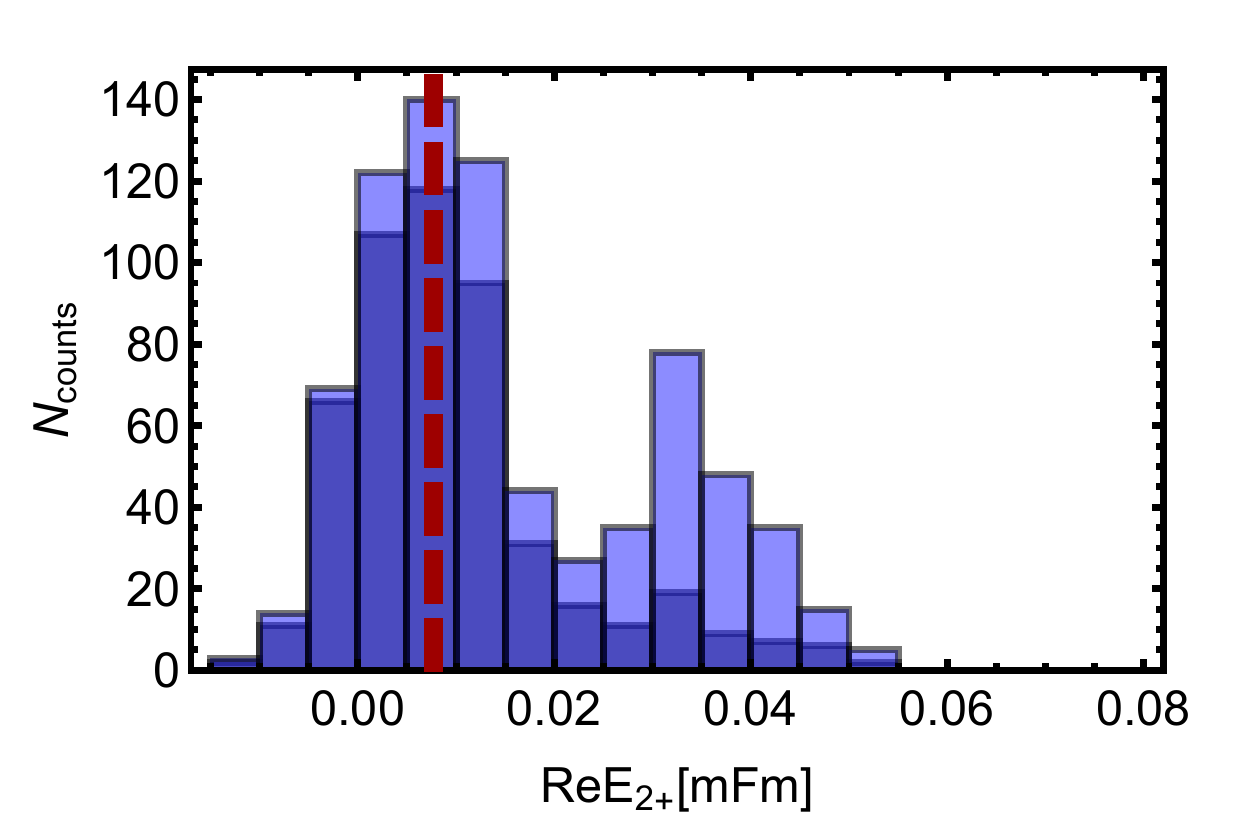}
 \end{overpic}
\begin{overpic}[width=0.325\textwidth]{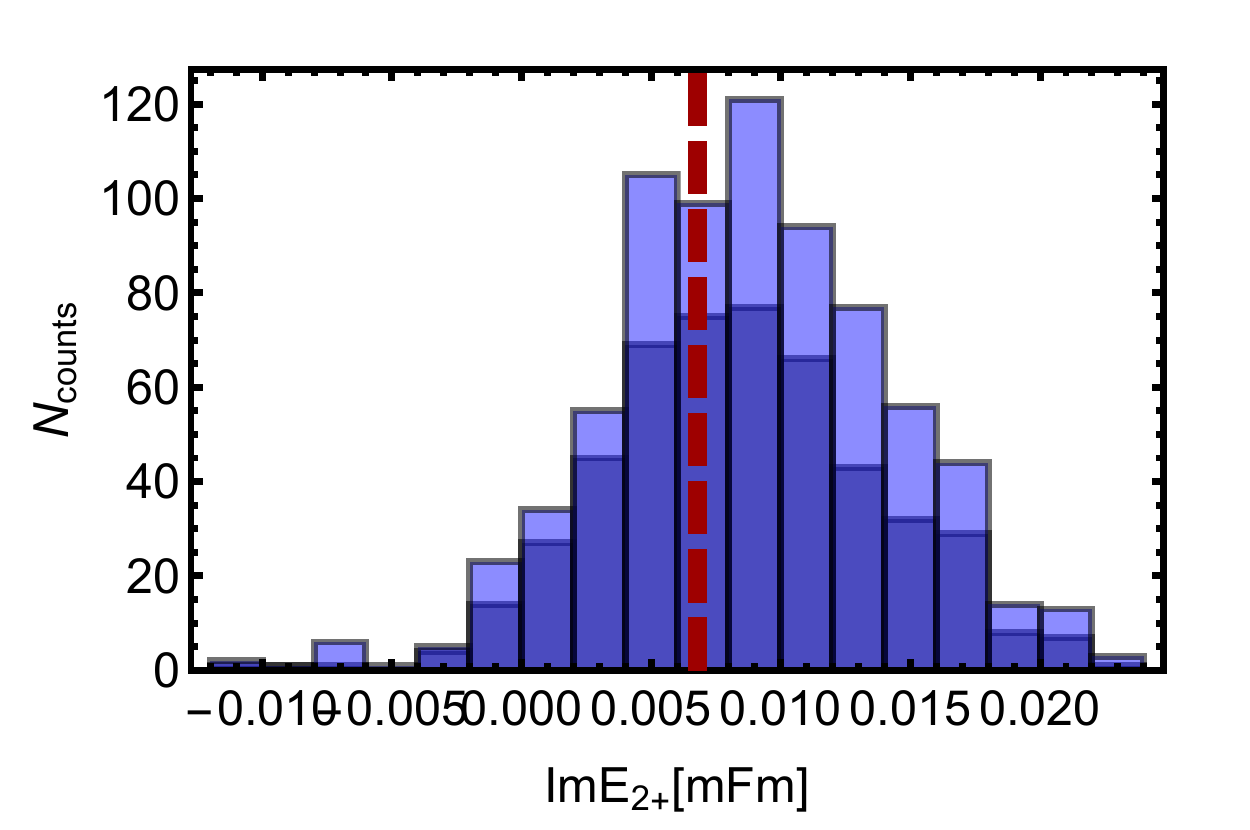}
 \end{overpic} \\
 \begin{overpic}[width=0.325\textwidth]{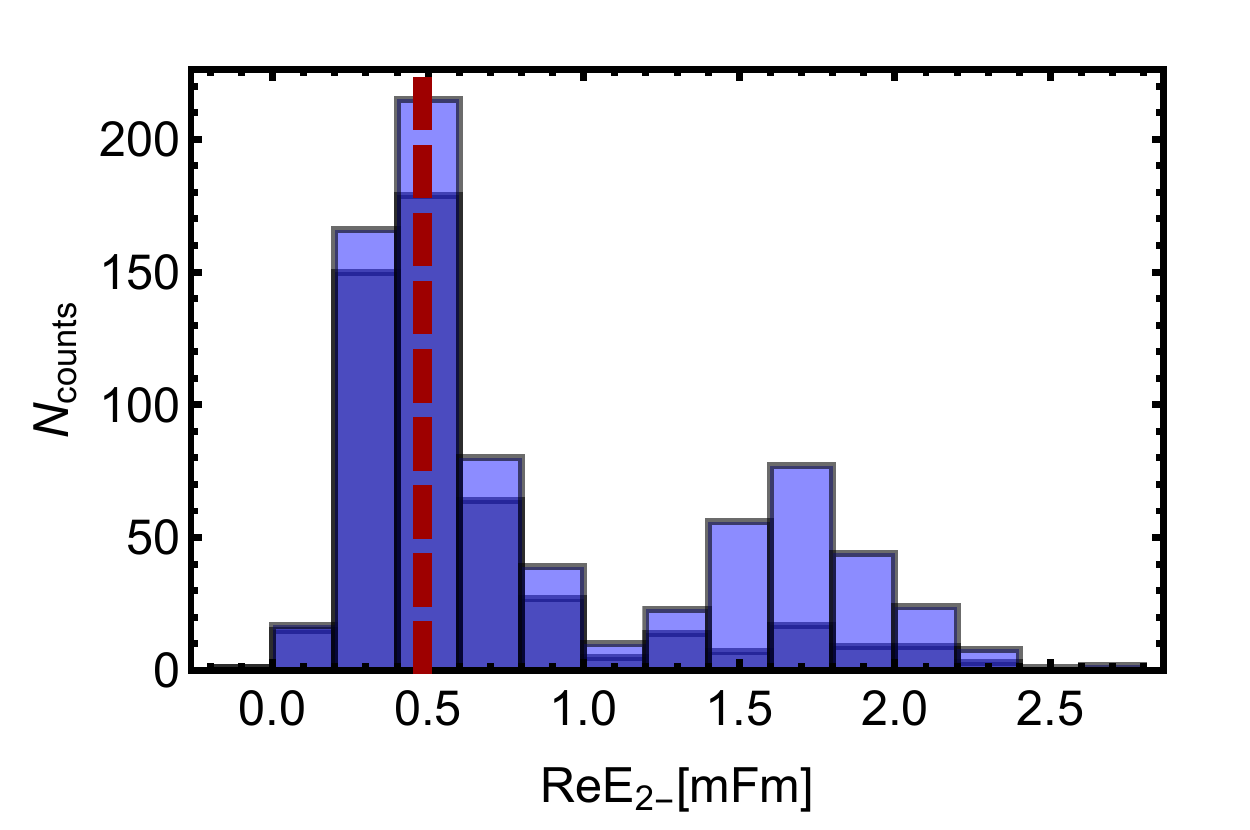}
 \end{overpic}
 \begin{overpic}[width=0.325\textwidth]{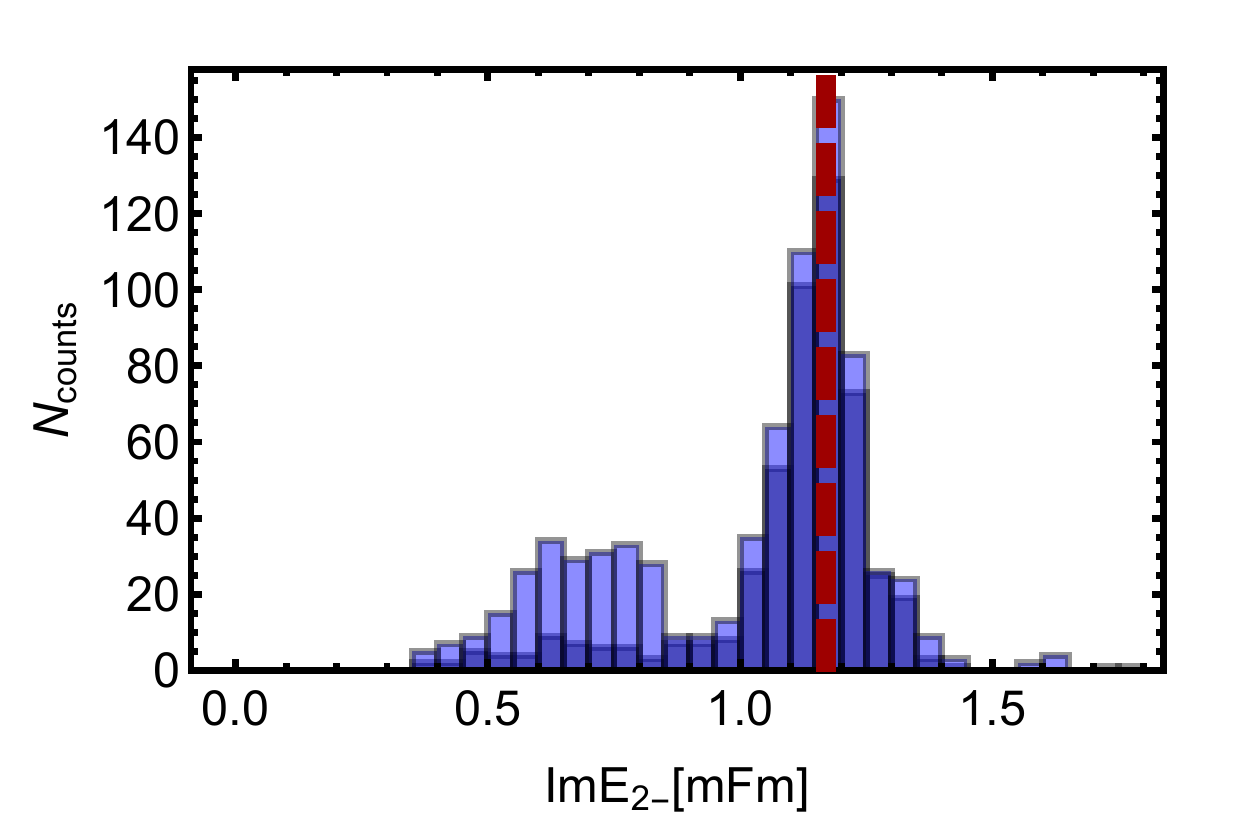}
 \end{overpic}
\begin{overpic}[width=0.325\textwidth]{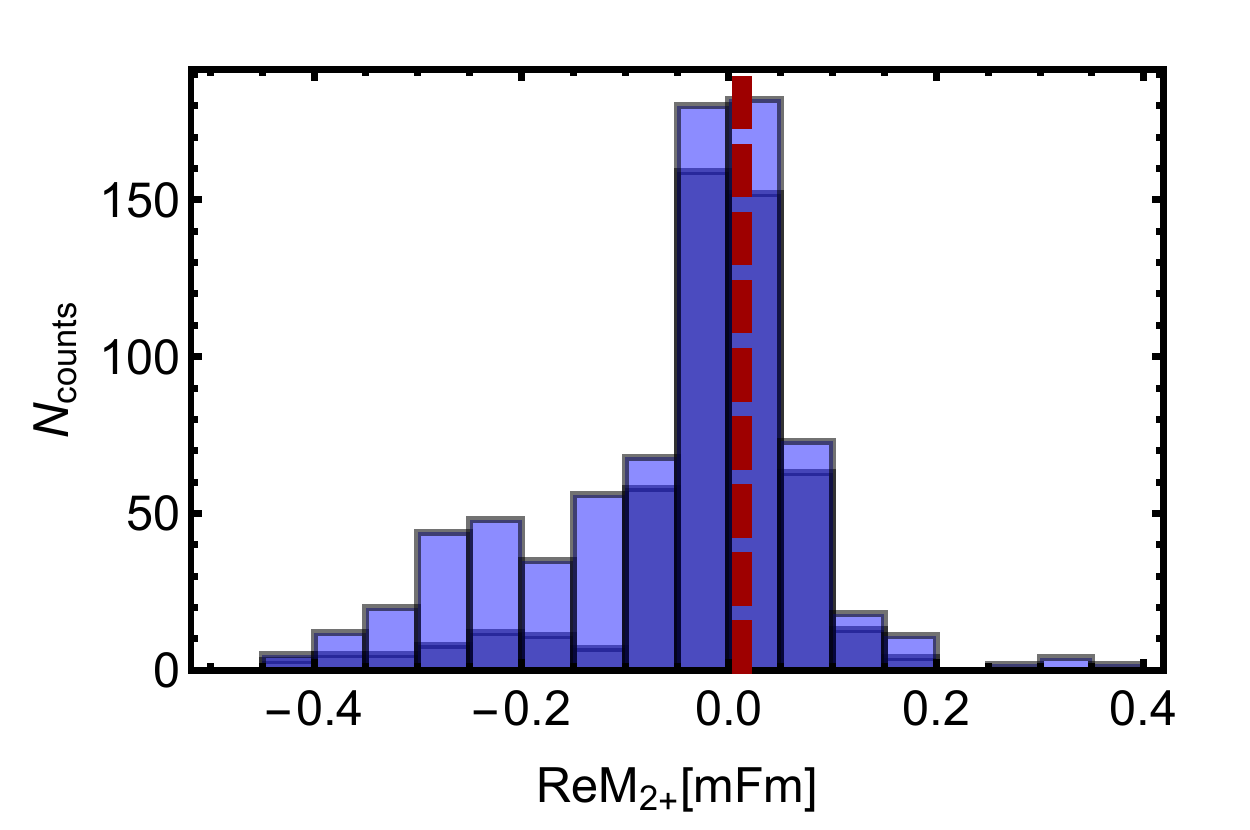}
 \end{overpic} \\
 \begin{overpic}[width=0.325\textwidth]{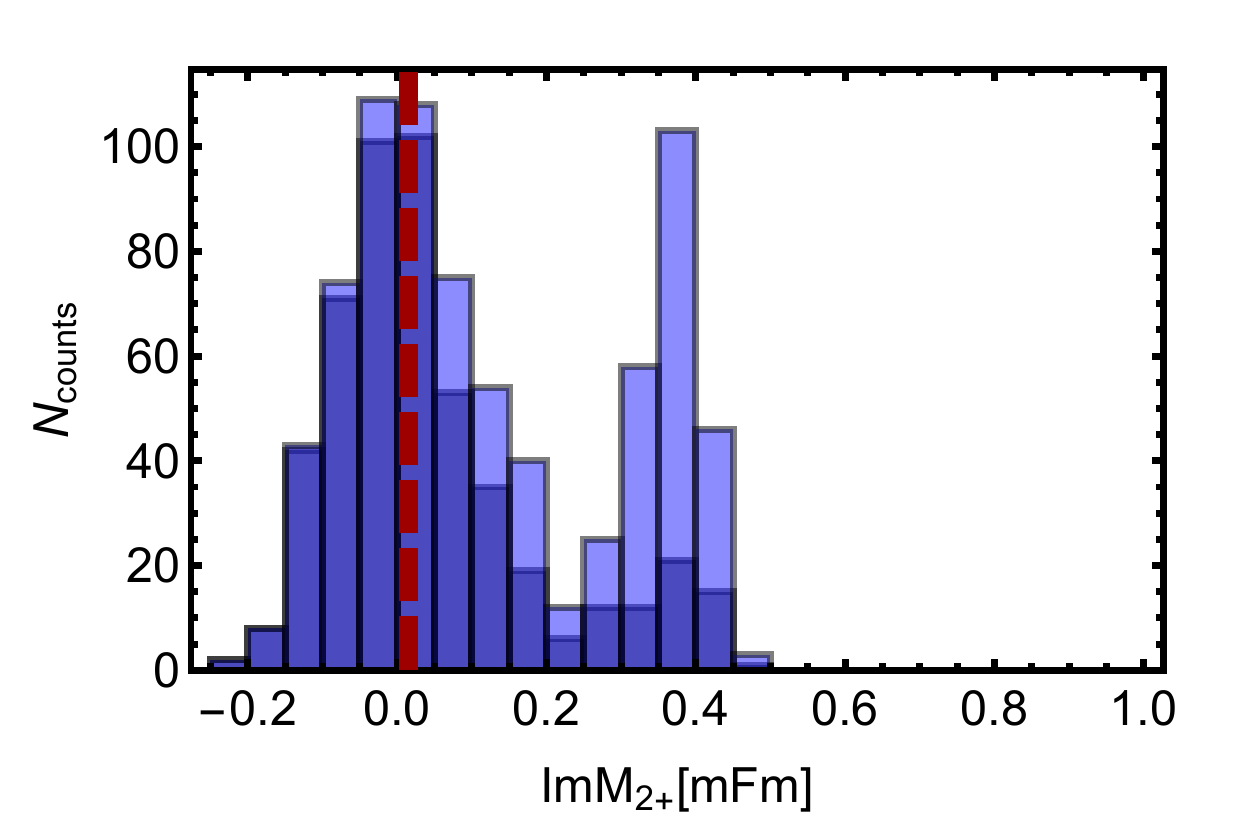}
 \end{overpic}
 \begin{overpic}[width=0.325\textwidth]{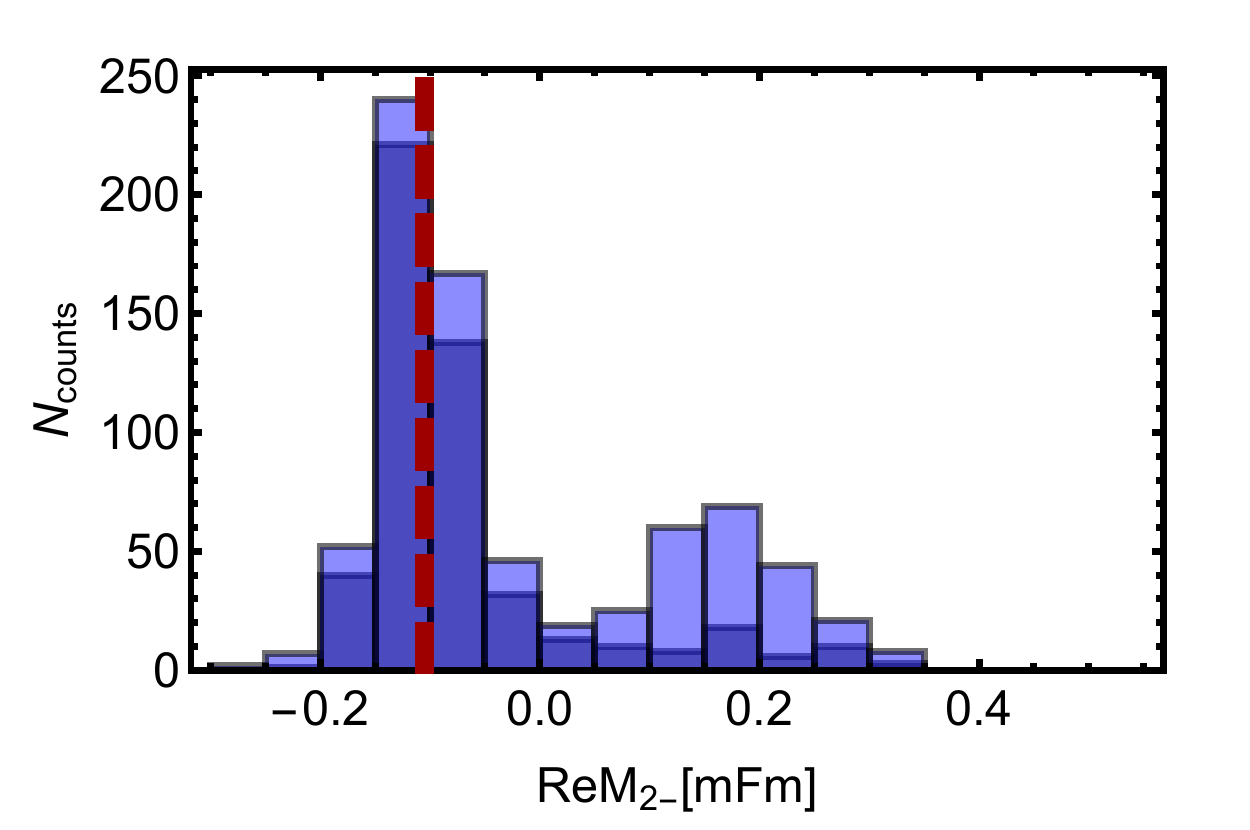}
 \end{overpic}
\begin{overpic}[width=0.325\textwidth]{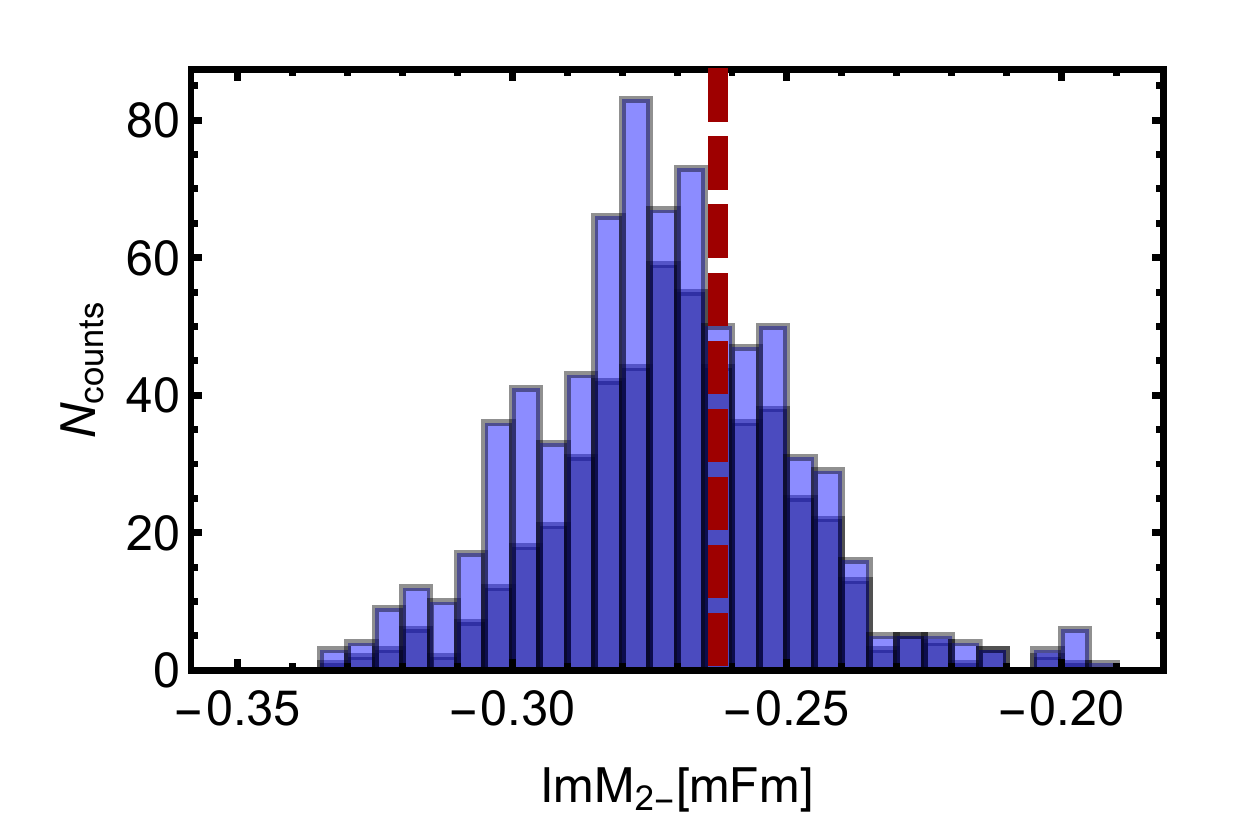}
 \end{overpic}
\caption[Bootstrap-histograms containing the global minima, as well as all non-redundant solutions having chisquare below the $0.95$-quantile of the corresponding non-central $\chi^{2}$-distribution, found in each fit of the $B=500$ bootstrap-replications of the MAID pseudodata truncated at $\ell_{\mathrm{max}} = 2$. Results are shown for the first error-scenario.]{The plots shown here are a continuation of Figure \ref{fig:Lmax2PseudoDataFitGroupSFObservablesMultHistograms1}. Again, bootstrap-distributions coming from an analysis of error-scenario $(i)$ are shown (cf. Table \ref{tab:Lmax2PercentageErrorScenarios}). The dark-blue bars show again only the global minima found for each one of $B=500$ bootstrap-replicates. The light-blue colored bars indicate the global minima, as well as all non-redundant solutions having chisquare below the $0.95$-quantile $u^{\left(P_{NC}\right)}_{0.95}$. \newline
The global minimum found in the fit of the original data is represented by red dashed vertical lines.}
\label{fig:Lmax2PseudoDataFitGroupSFObservablesMultHistograms2}
\end{figure}

\begin{figure}[ht]
 \centering
\begin{overpic}[width=0.325\textwidth]{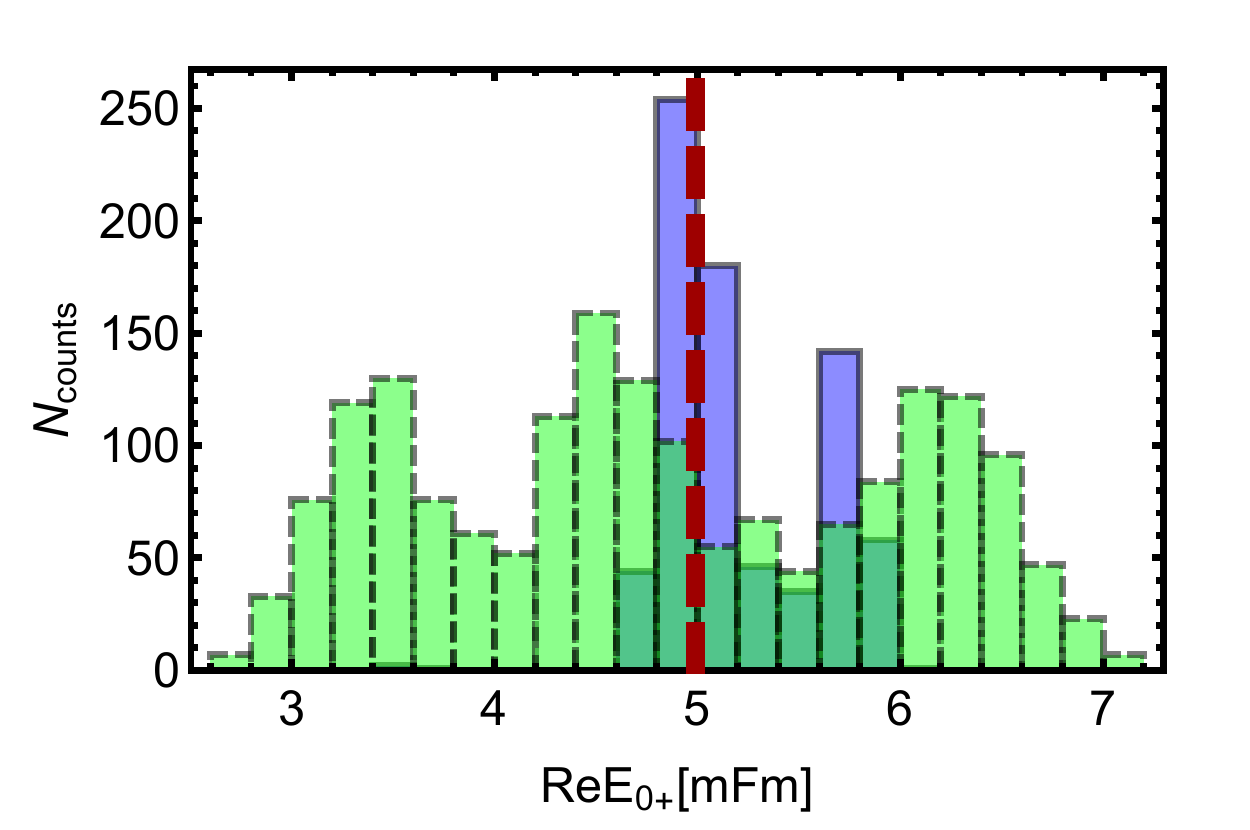}
 \end{overpic}
\begin{overpic}[width=0.325\textwidth]{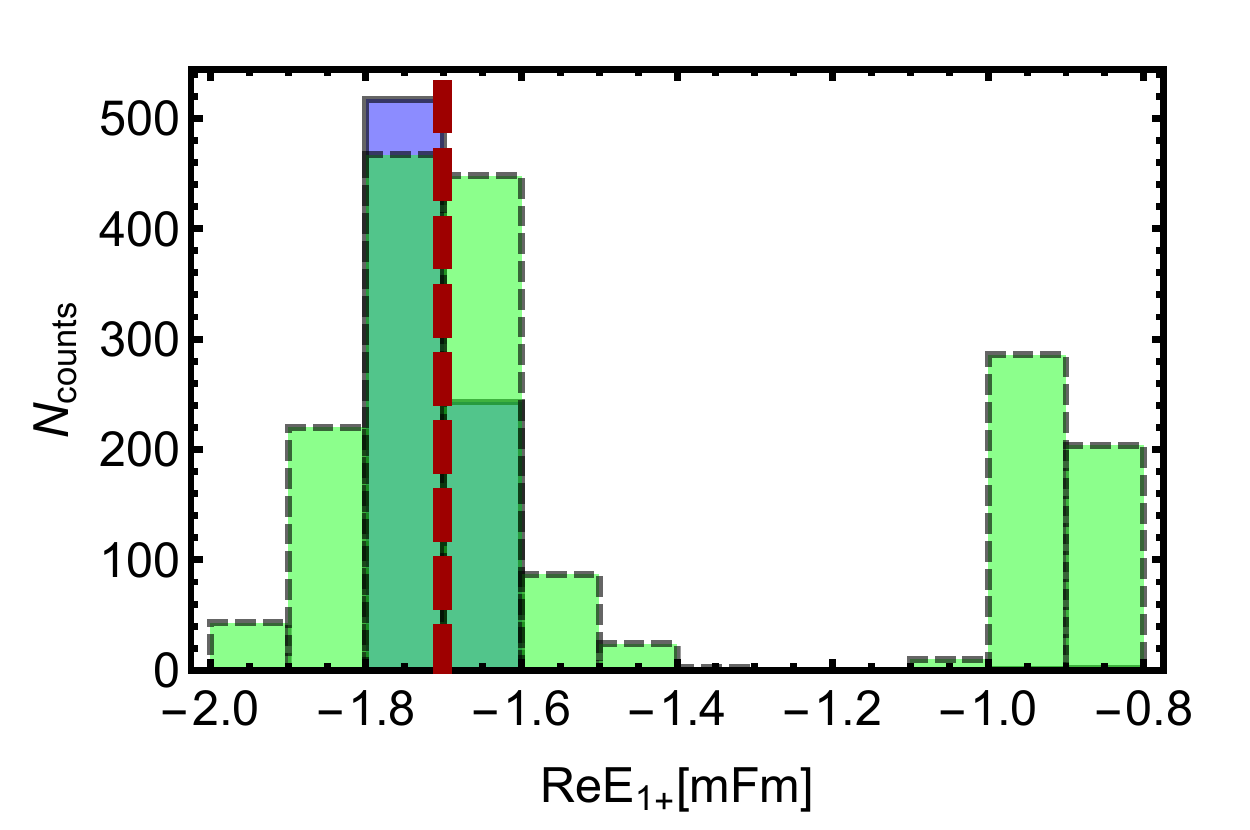}
 \end{overpic}
\begin{overpic}[width=0.325\textwidth]{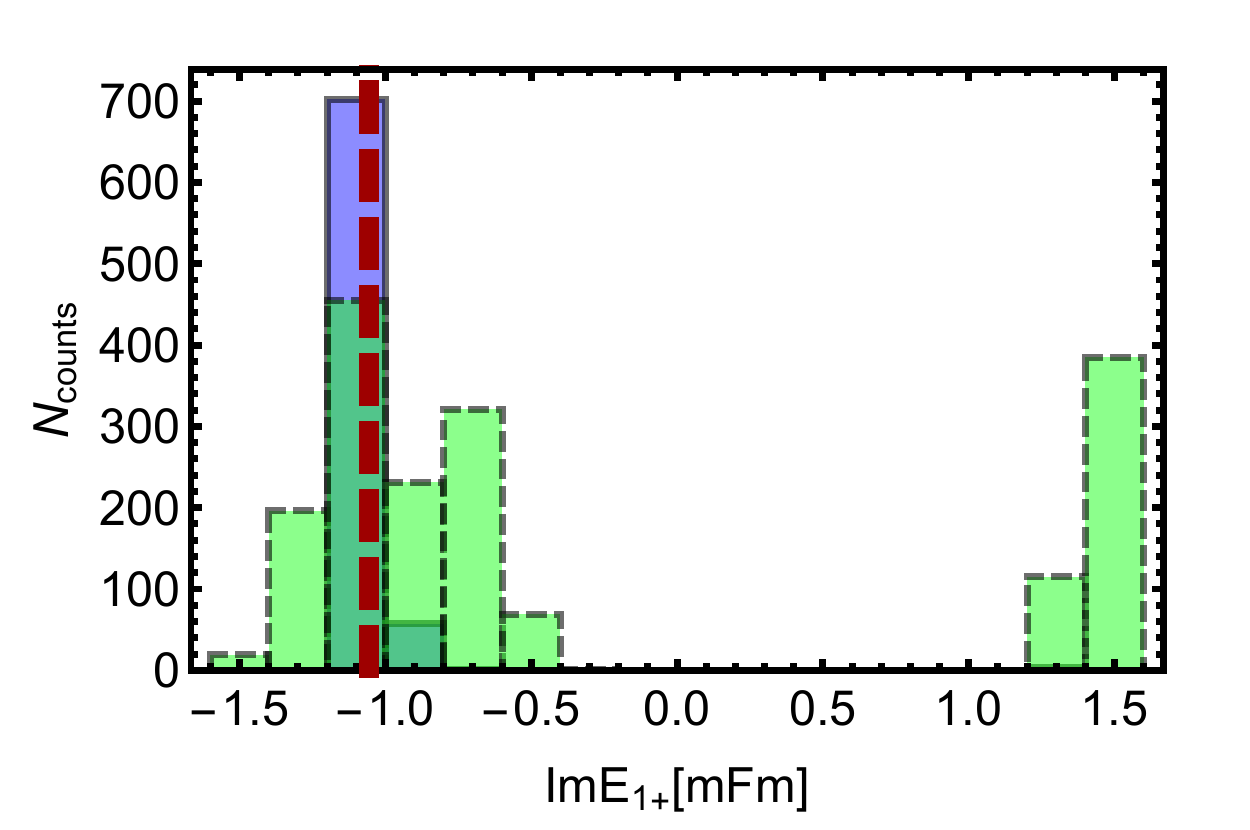}
 \end{overpic} \\
\begin{overpic}[width=0.325\textwidth]{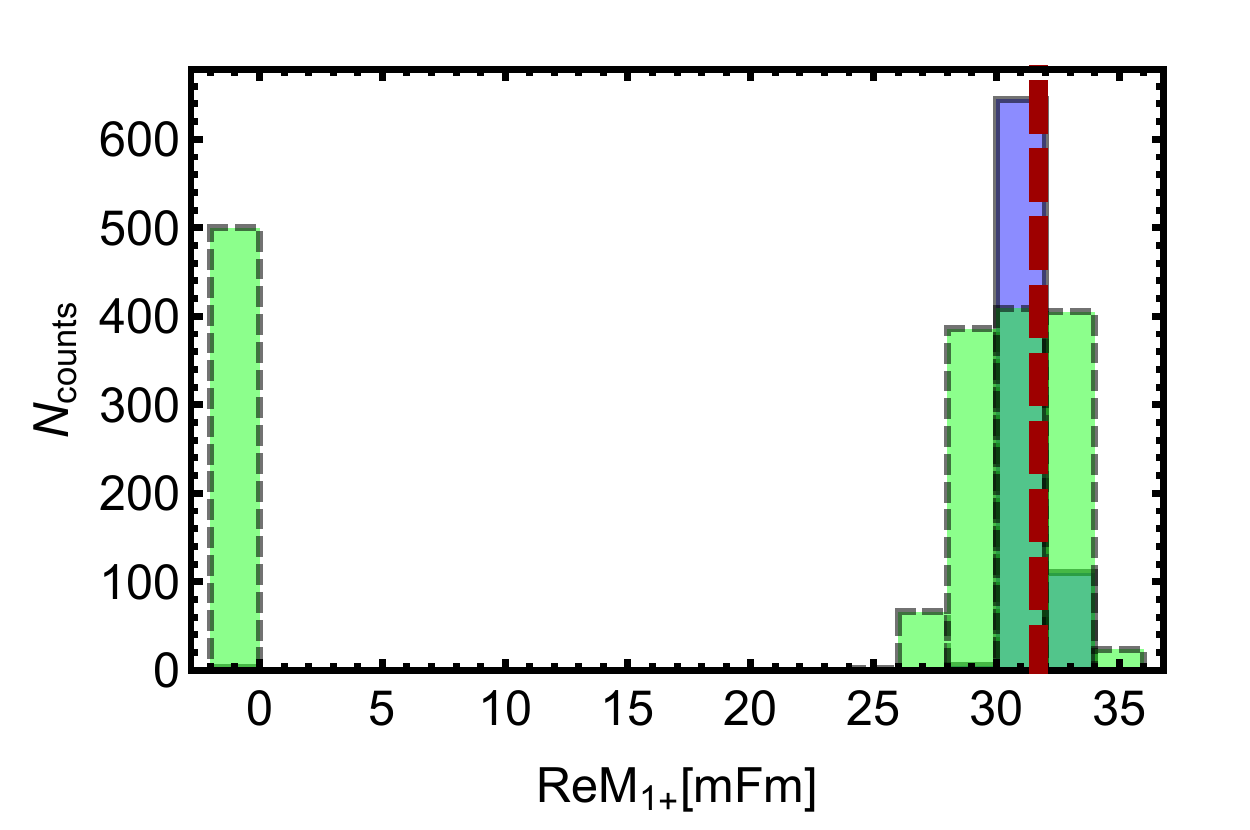}
 \end{overpic}
 \begin{overpic}[width=0.325\textwidth]{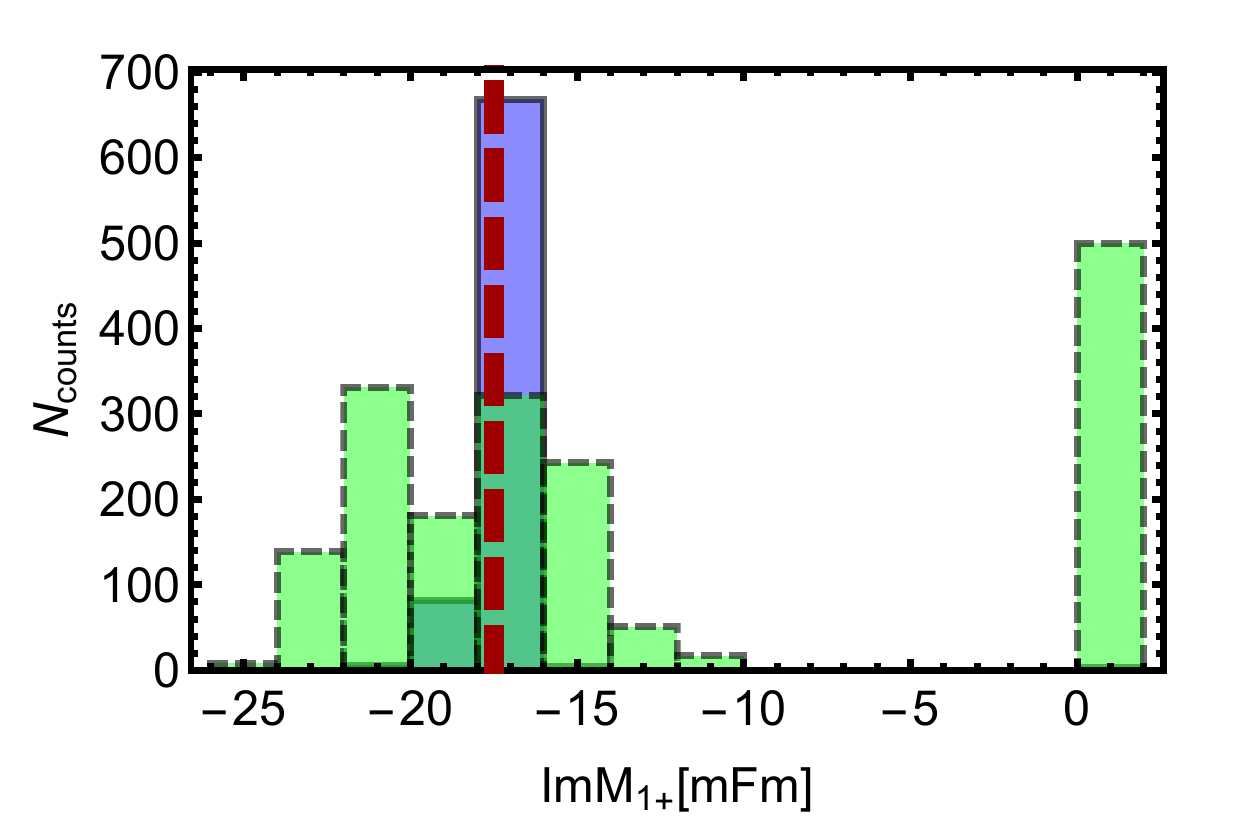}
 \end{overpic}
\begin{overpic}[width=0.325\textwidth]{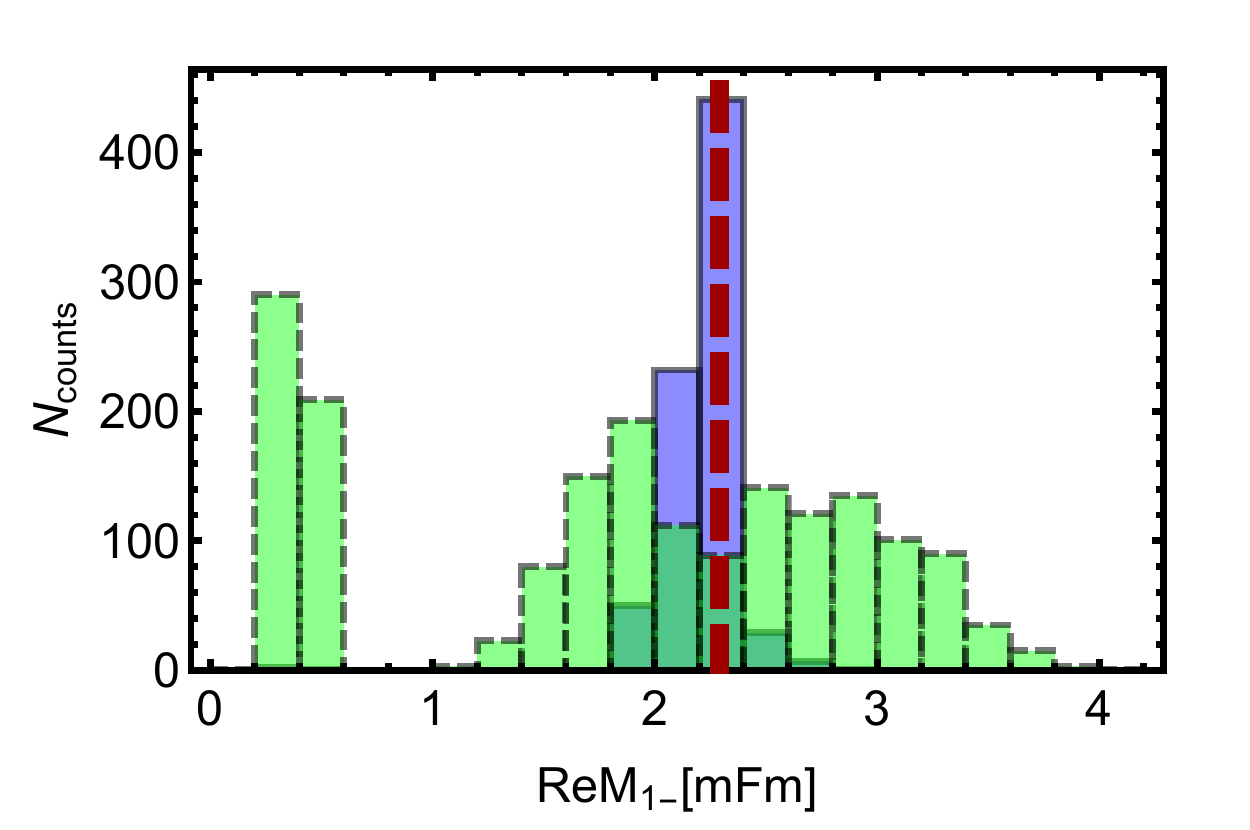}
 \end{overpic} \\
 \begin{overpic}[width=0.325\textwidth]{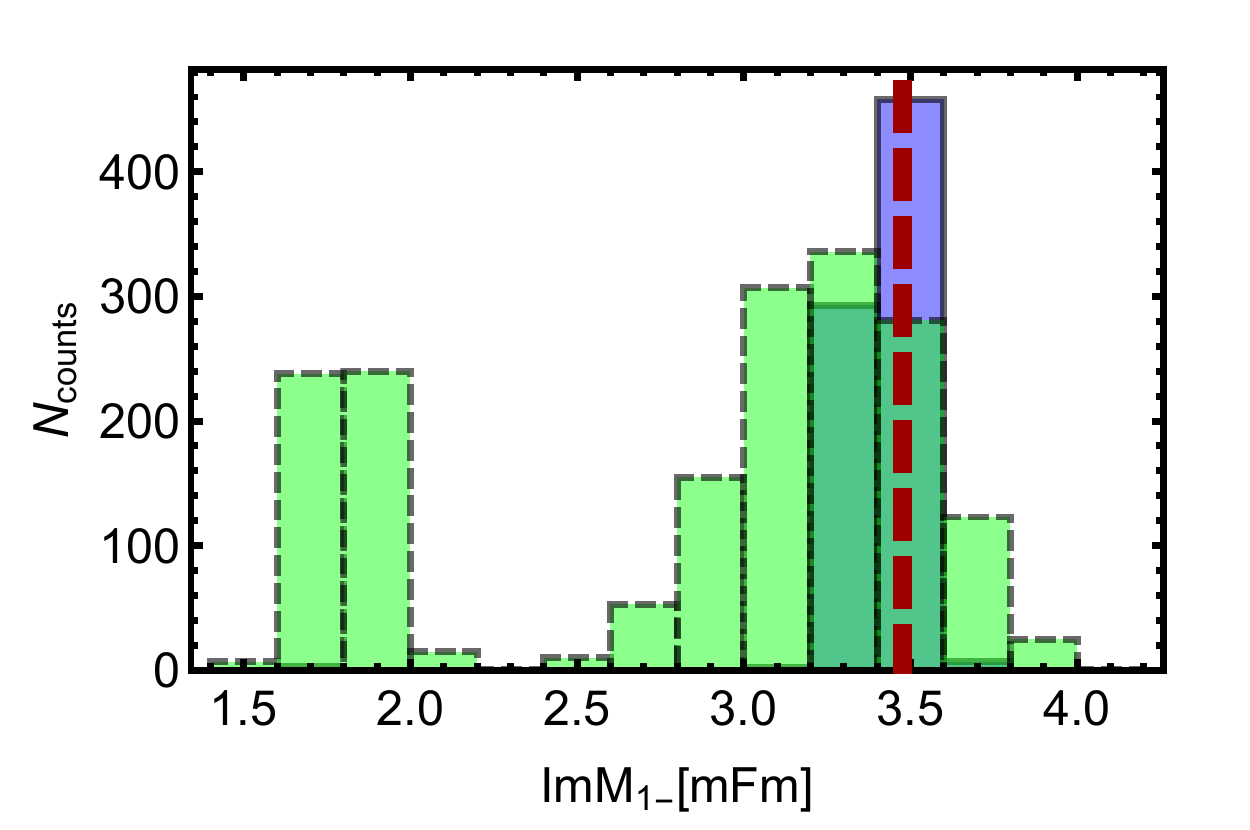}
 \end{overpic}
 \begin{overpic}[width=0.325\textwidth]{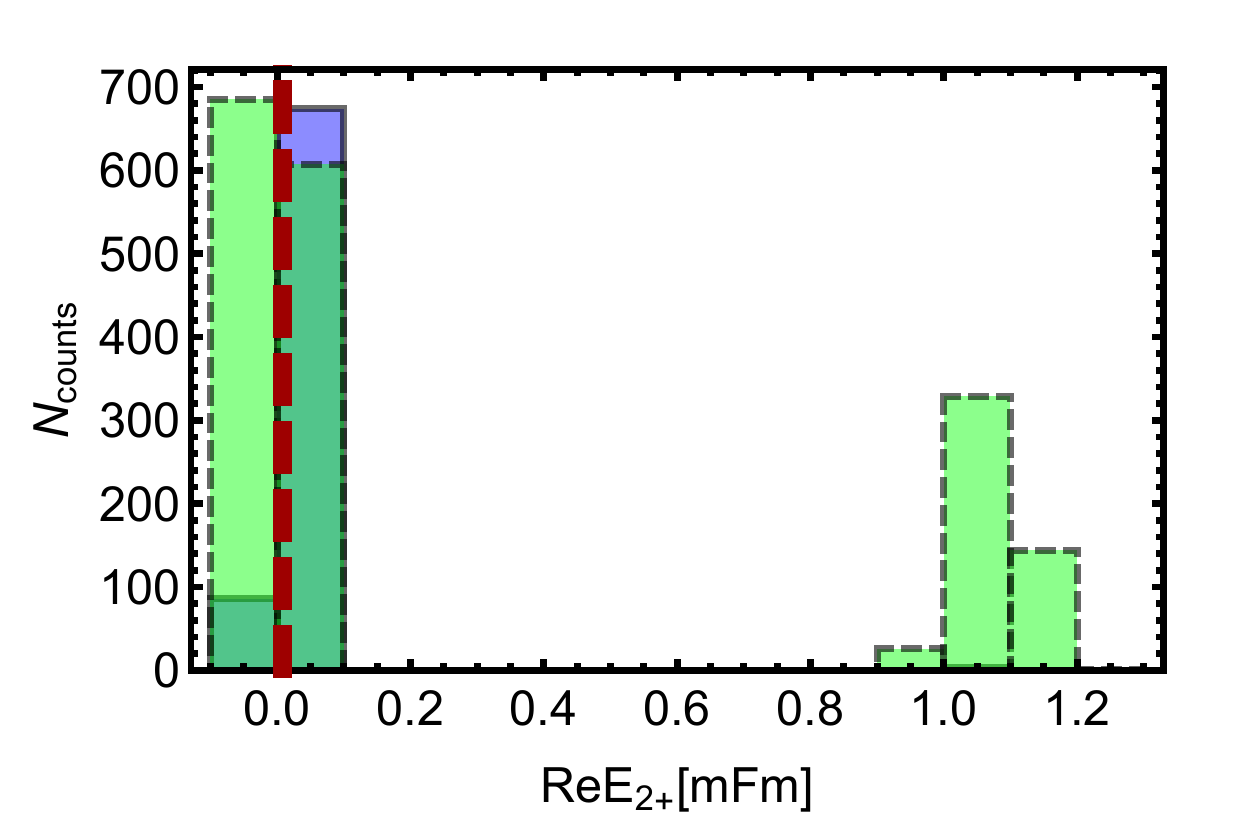}
 \end{overpic}
\begin{overpic}[width=0.325\textwidth]{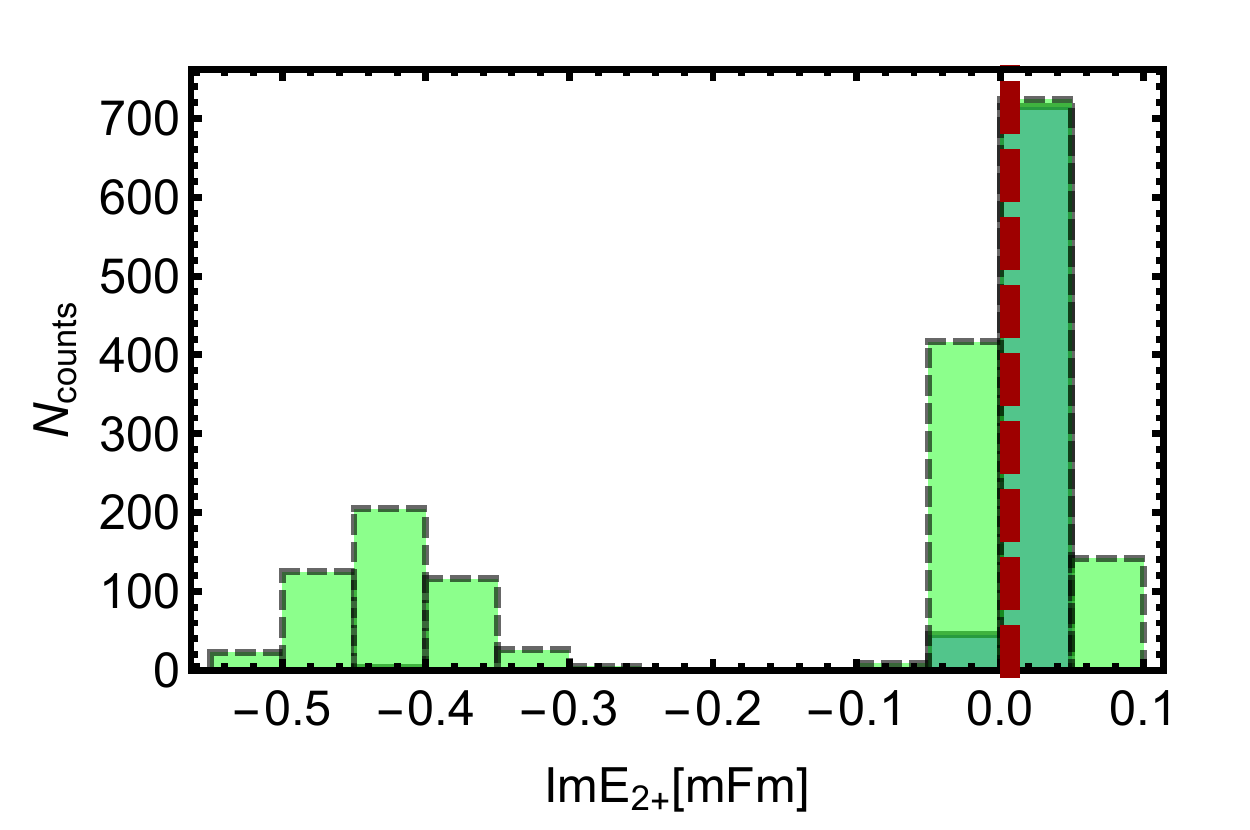}
 \end{overpic} \\
 \begin{overpic}[width=0.325\textwidth]{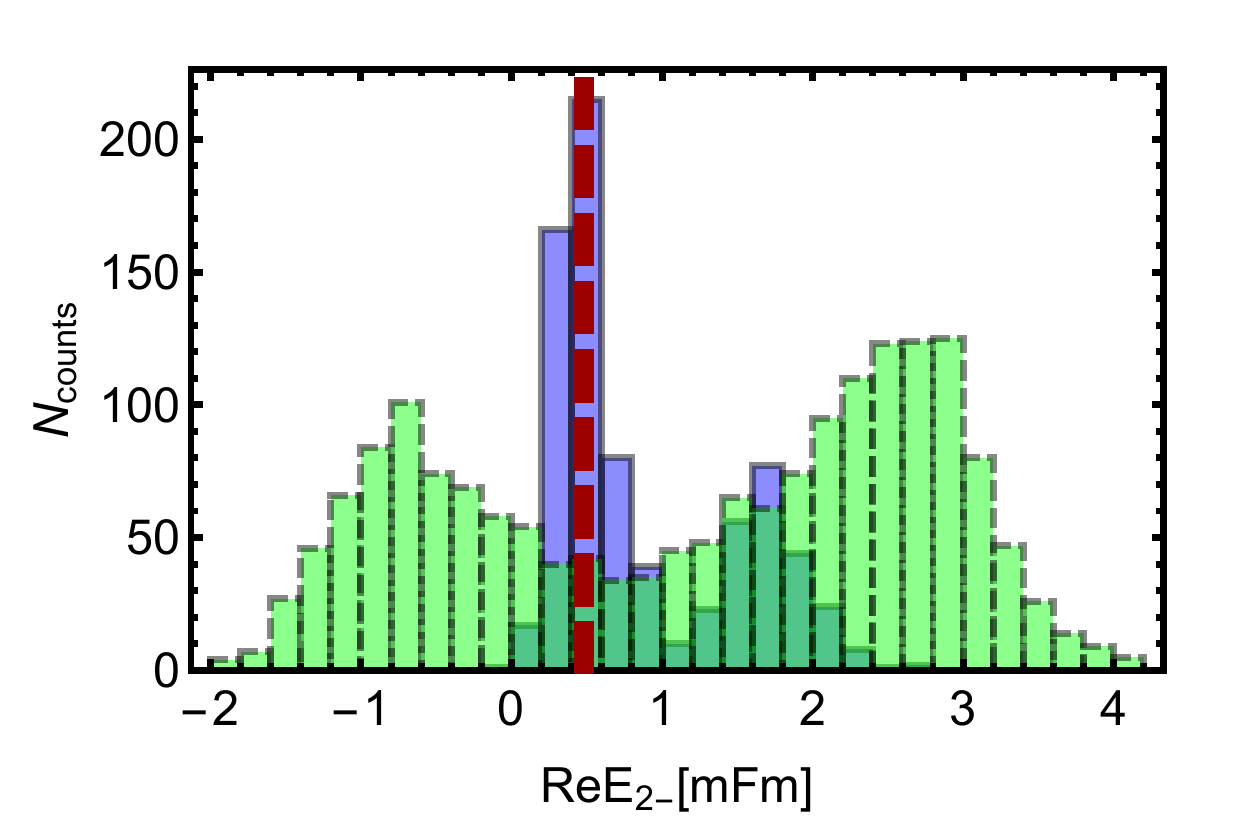}
 \end{overpic}
 \begin{overpic}[width=0.325\textwidth]{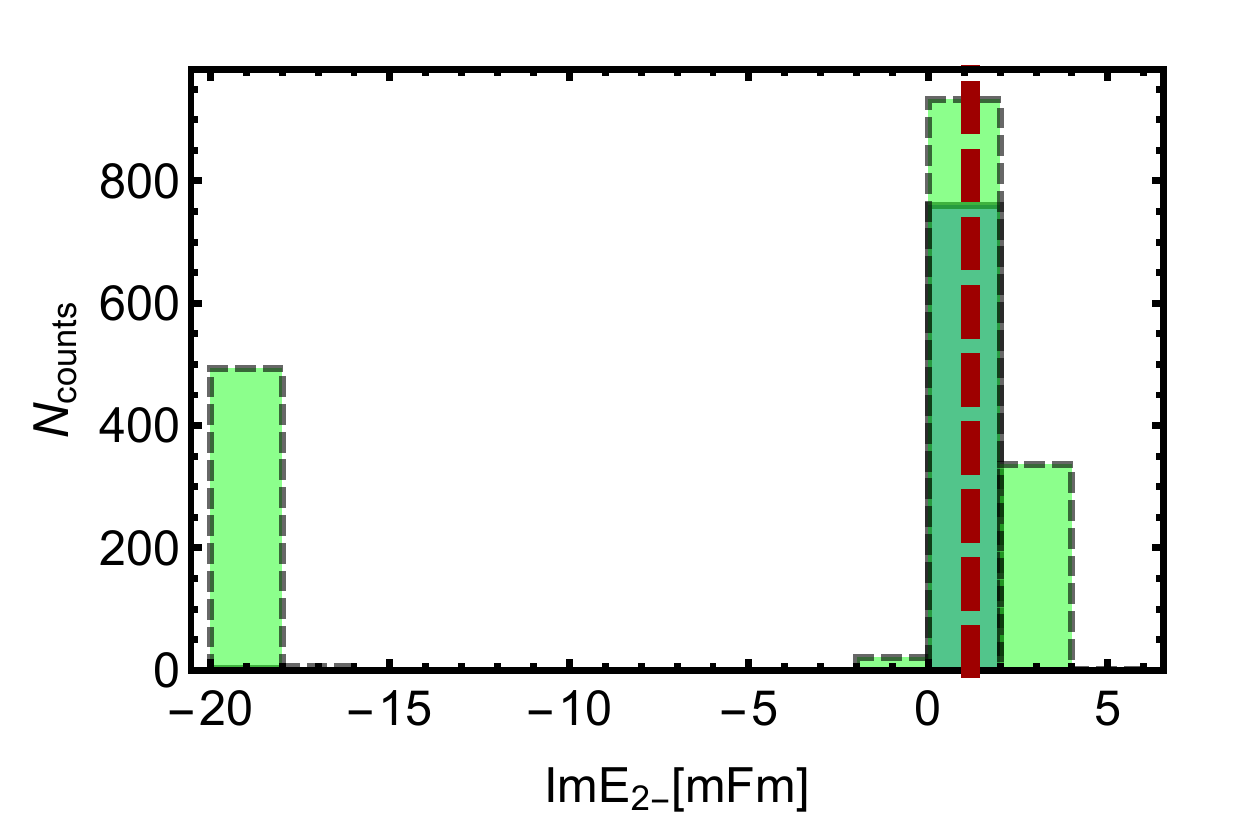}
 \end{overpic}
\begin{overpic}[width=0.325\textwidth]{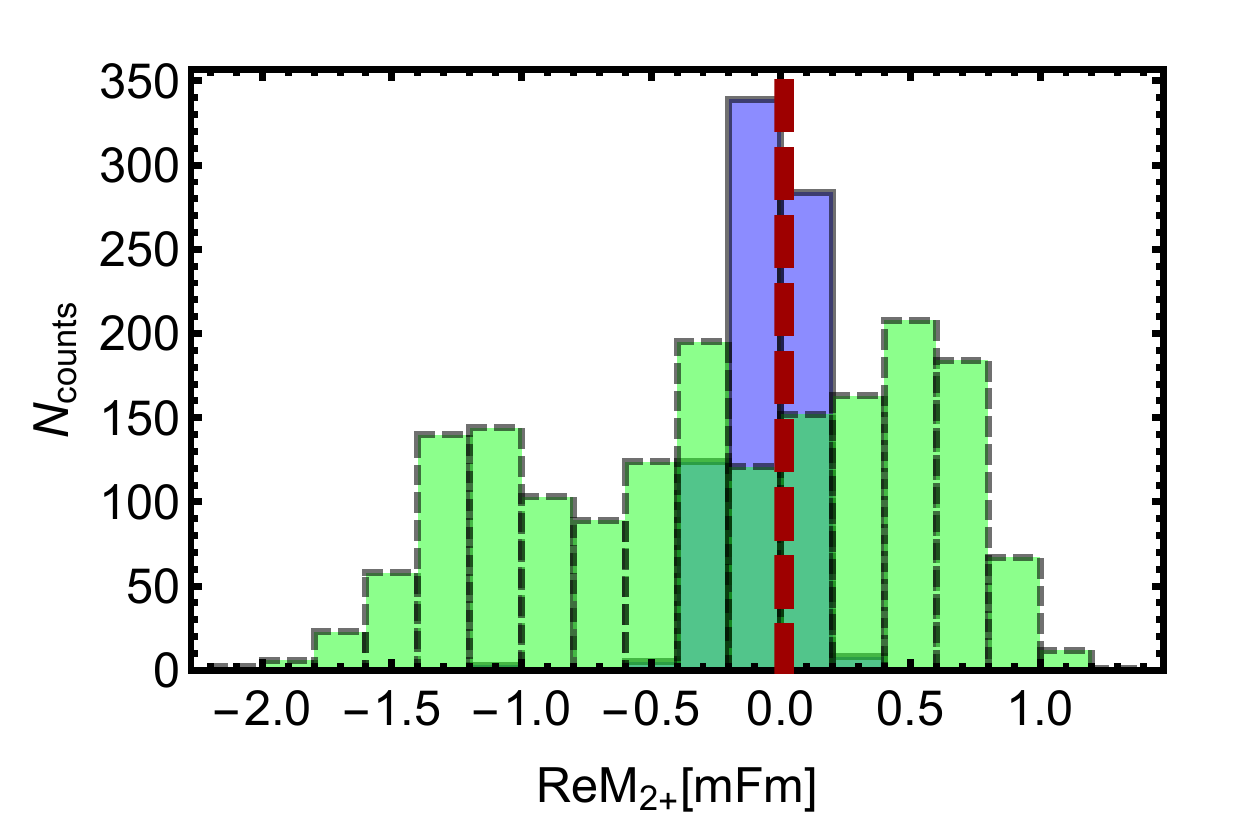}
 \end{overpic} \\
 \begin{overpic}[width=0.325\textwidth]{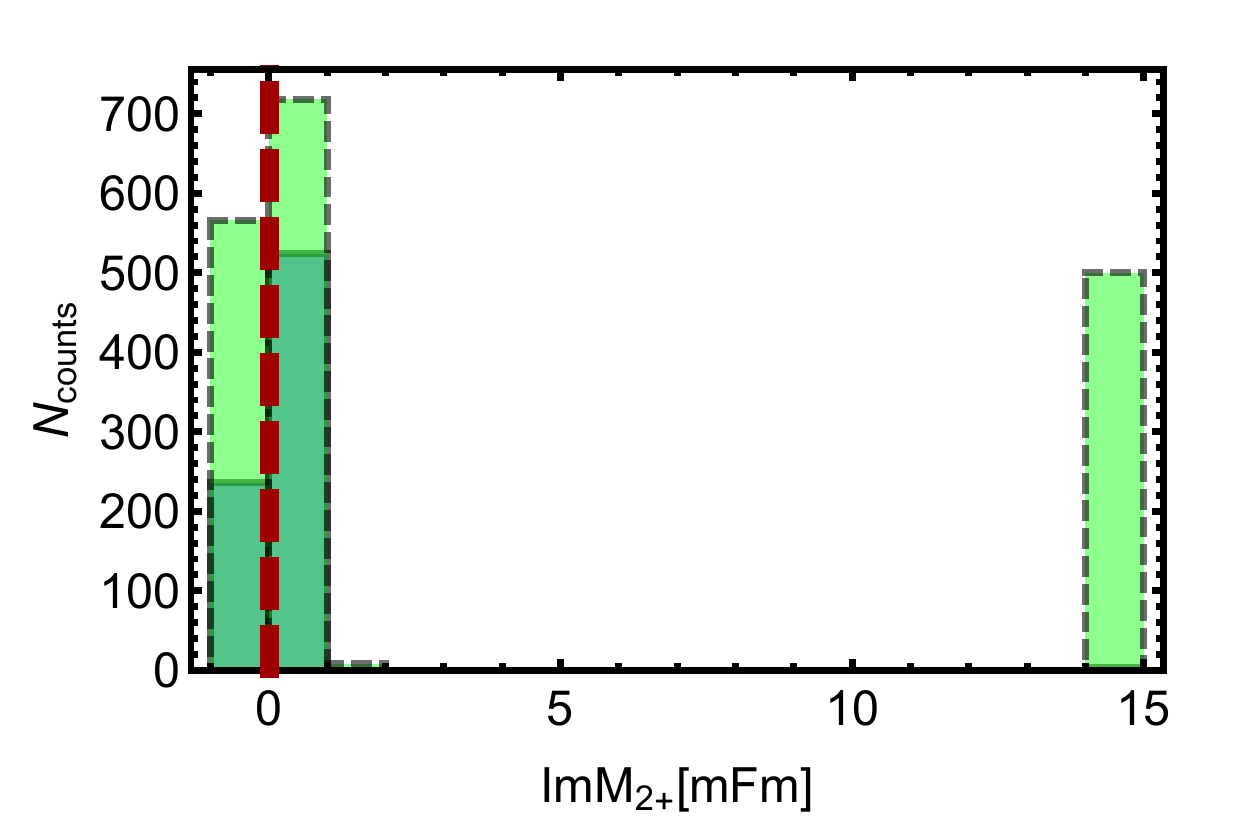}
 \end{overpic}
 \begin{overpic}[width=0.325\textwidth]{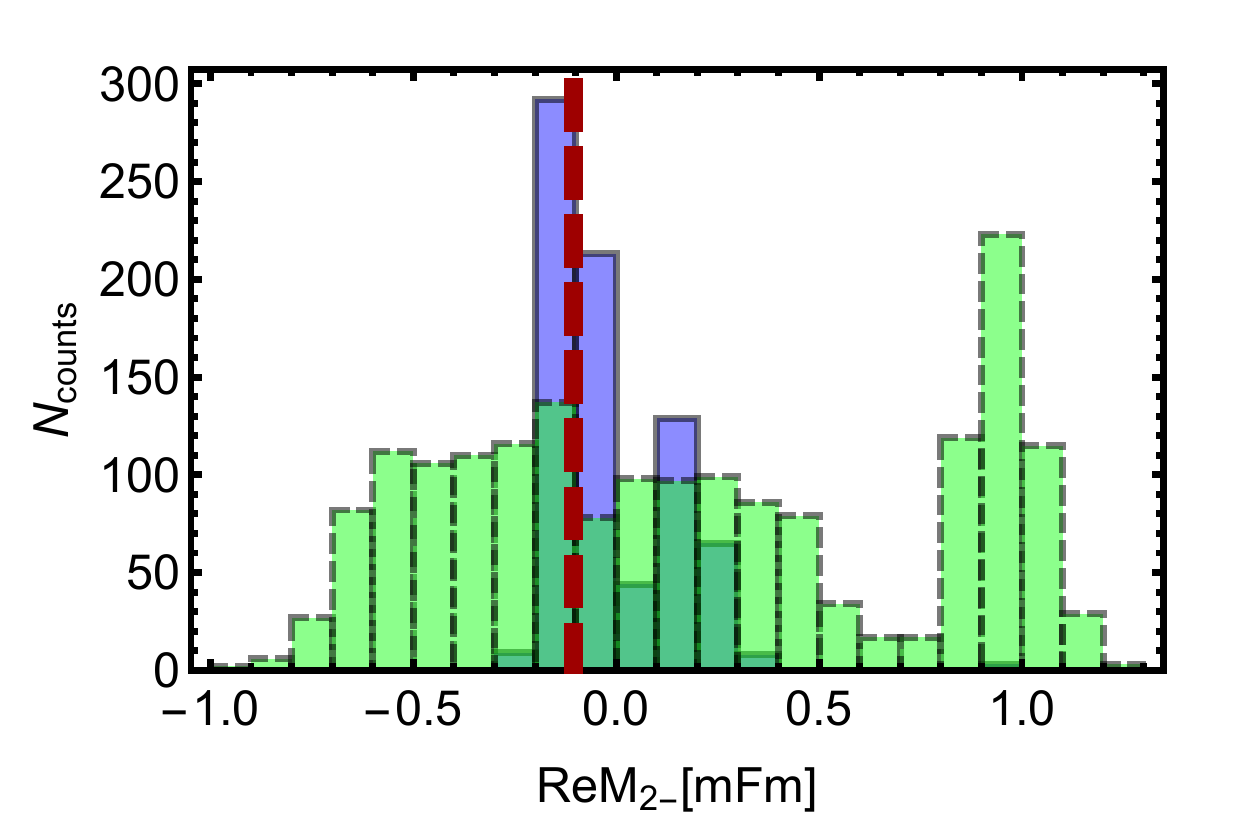}
 \end{overpic}
\begin{overpic}[width=0.325\textwidth]{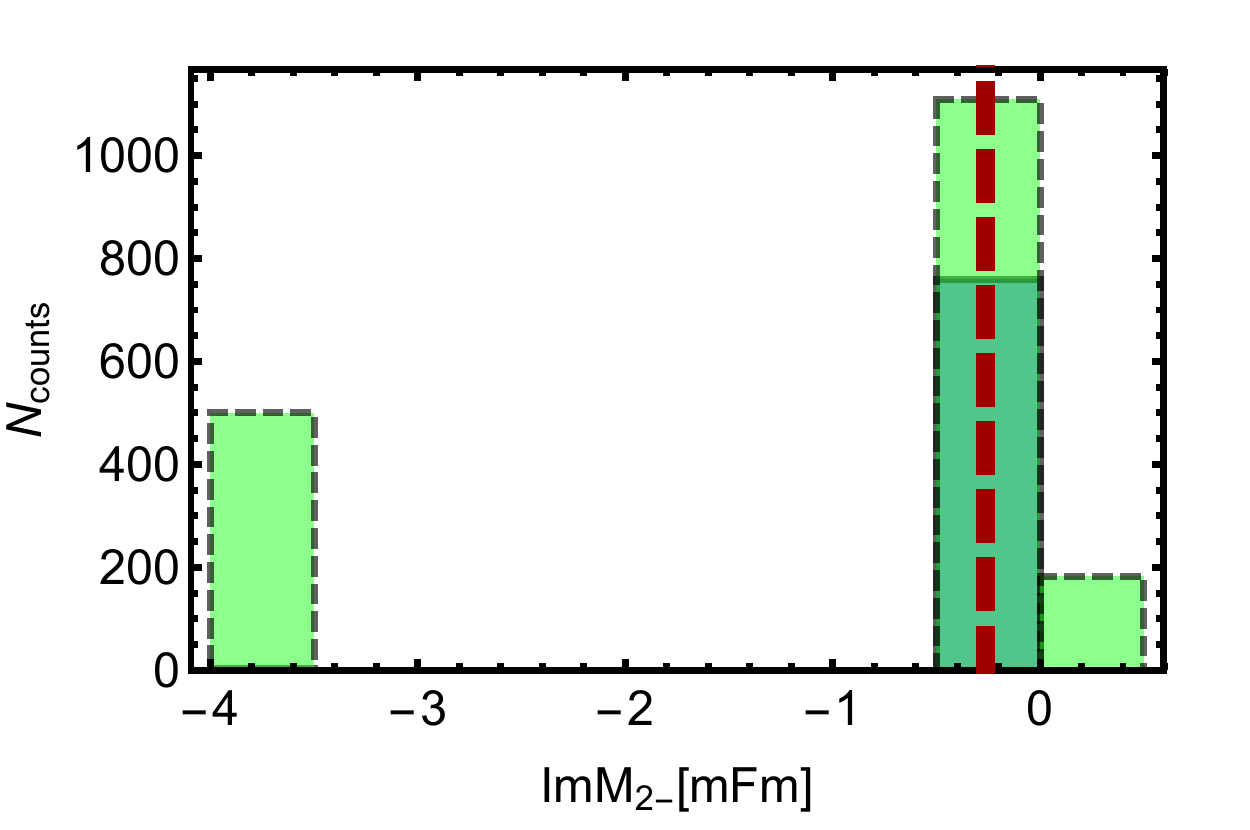}
 \end{overpic}
\caption[Bootstrap-histograms containing the global minima, as well as all non-redundant solutions having chisquare below the $0.95$-quantile of the corresponding non-central $\chi^{2}$-distribution, found in each fit of the $B=500$ bootstrap-replications of the MAID pseudodata truncated at $\ell_{\mathrm{max}} = 2$. Results are shown for the first and second error-scenario.]{The histograms show bootstrap-distributions coming from analyses of the error-scenarios $(i)$ (blue bars with solid boundary) and $(ii)$ (green bars with dashed boundary) (cf. Table \ref{tab:Lmax2PercentageErrorScenarios}) (MAID pseudodata truncated at $\ell_{\mathrm{max}} = 2$). All non-redundant solutions below the respective $0.95$-quantiles $u^{\left(P_{NC}\right)}_{0.95}$ have been included in the histograms. \newline
The global minimum of the fit to the original data of error-scenario $(i)$ is indicated by the red dashed vertical lines.}
\label{fig:Lmax2PseudoDataFitGroupSFObservablesMultHistograms3}
\end{figure}

\begin{figure}[ht]
 \centering
\begin{overpic}[width=0.325\textwidth]{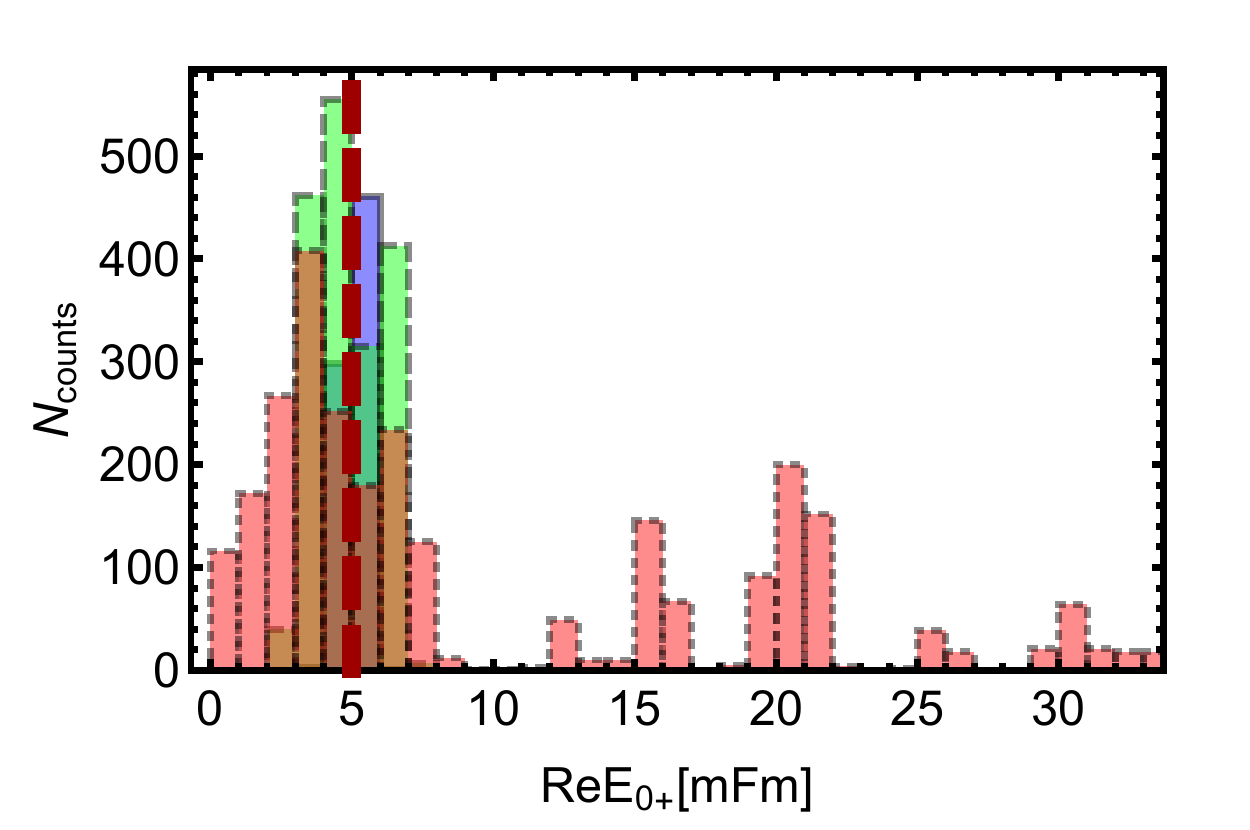}
 \end{overpic}
\begin{overpic}[width=0.325\textwidth]{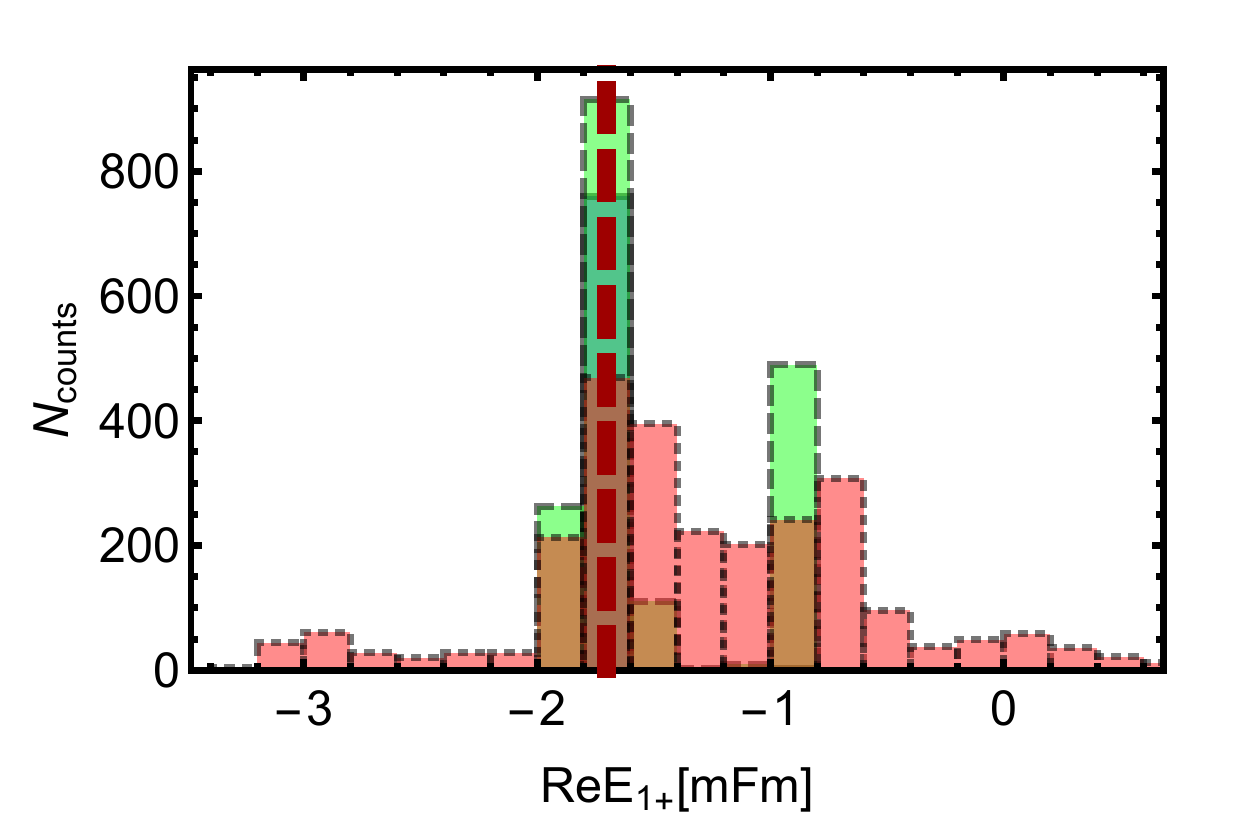}
 \end{overpic}
\begin{overpic}[width=0.325\textwidth]{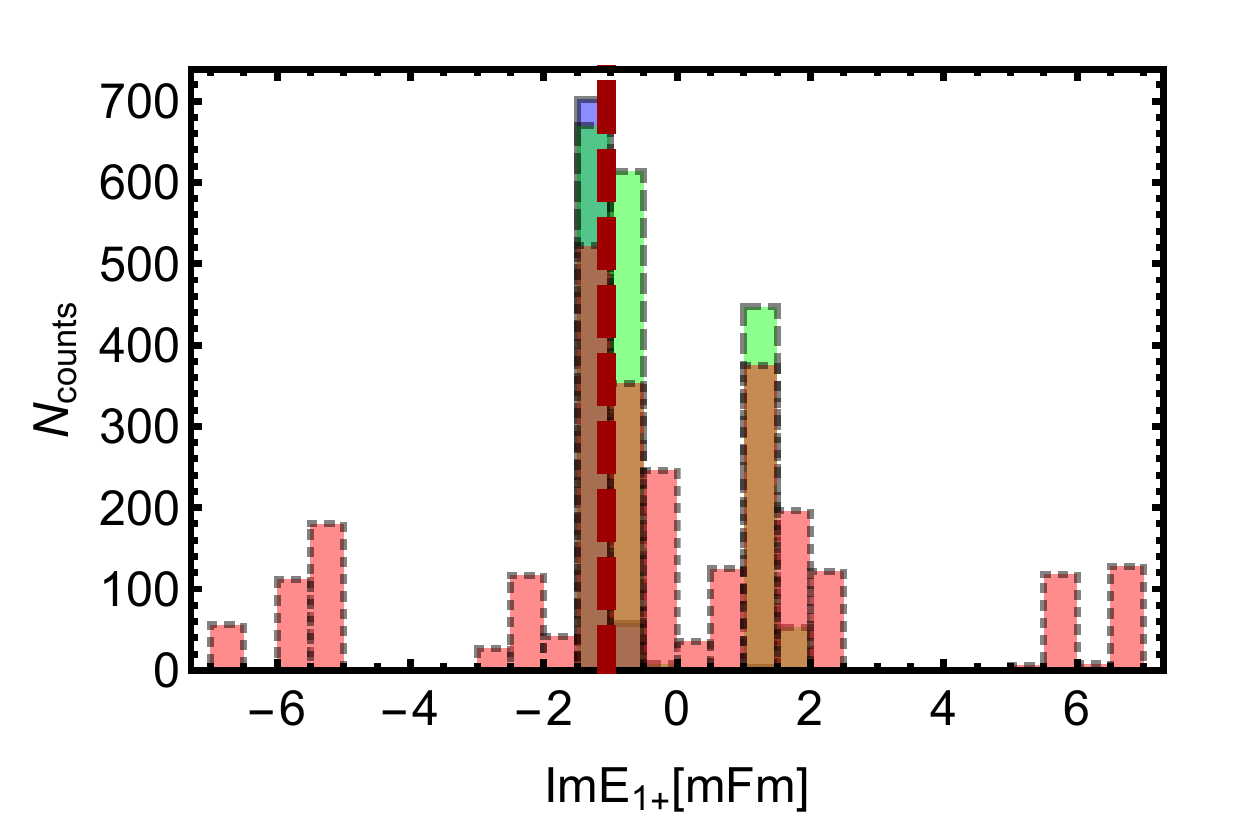}
 \end{overpic} \\
\begin{overpic}[width=0.325\textwidth]{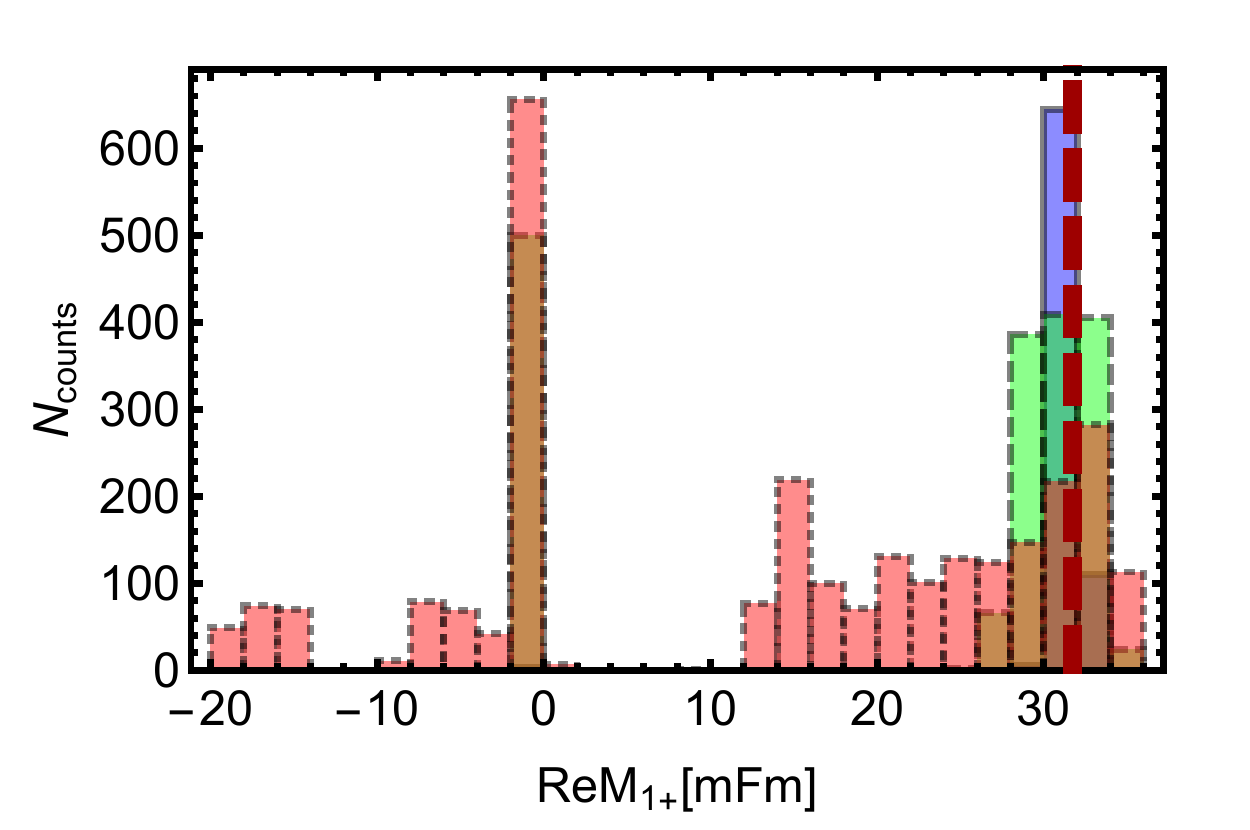}
 \end{overpic}
 \begin{overpic}[width=0.325\textwidth]{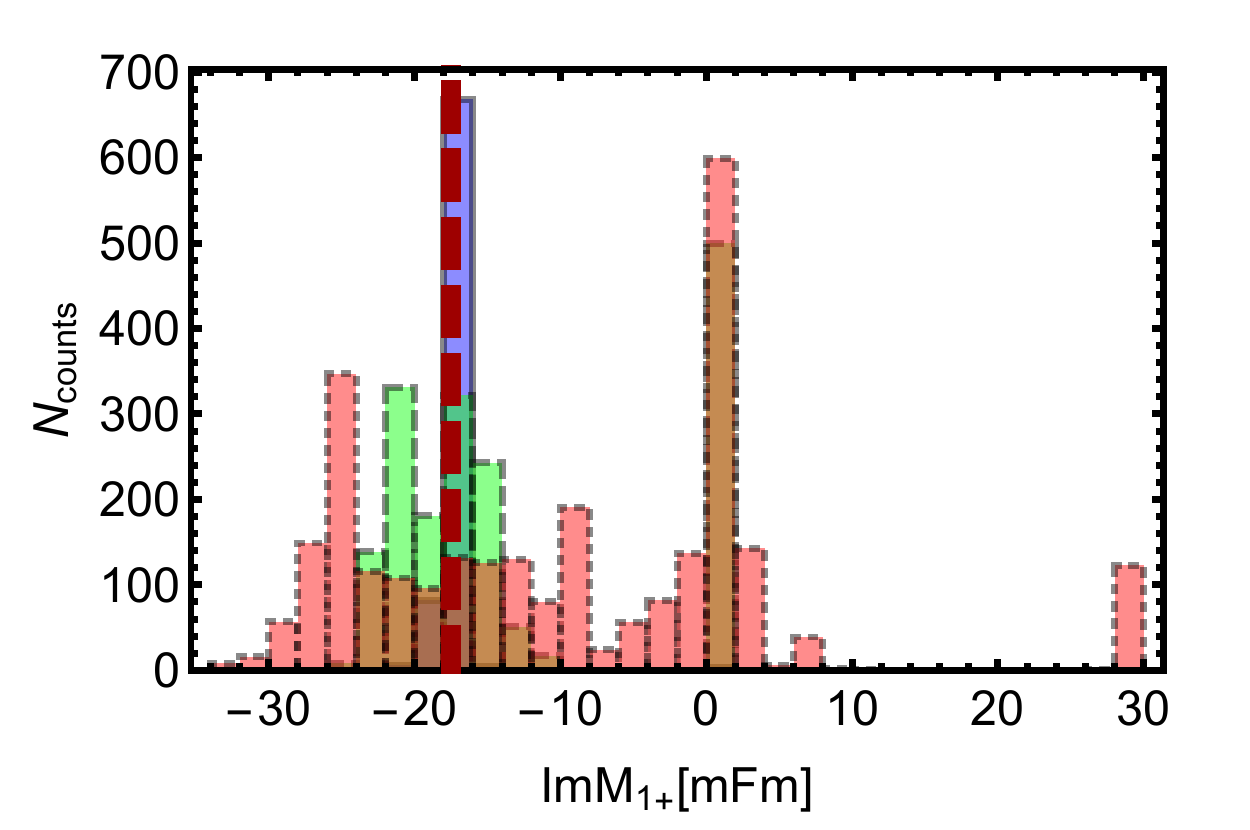}
 \end{overpic}
\begin{overpic}[width=0.325\textwidth]{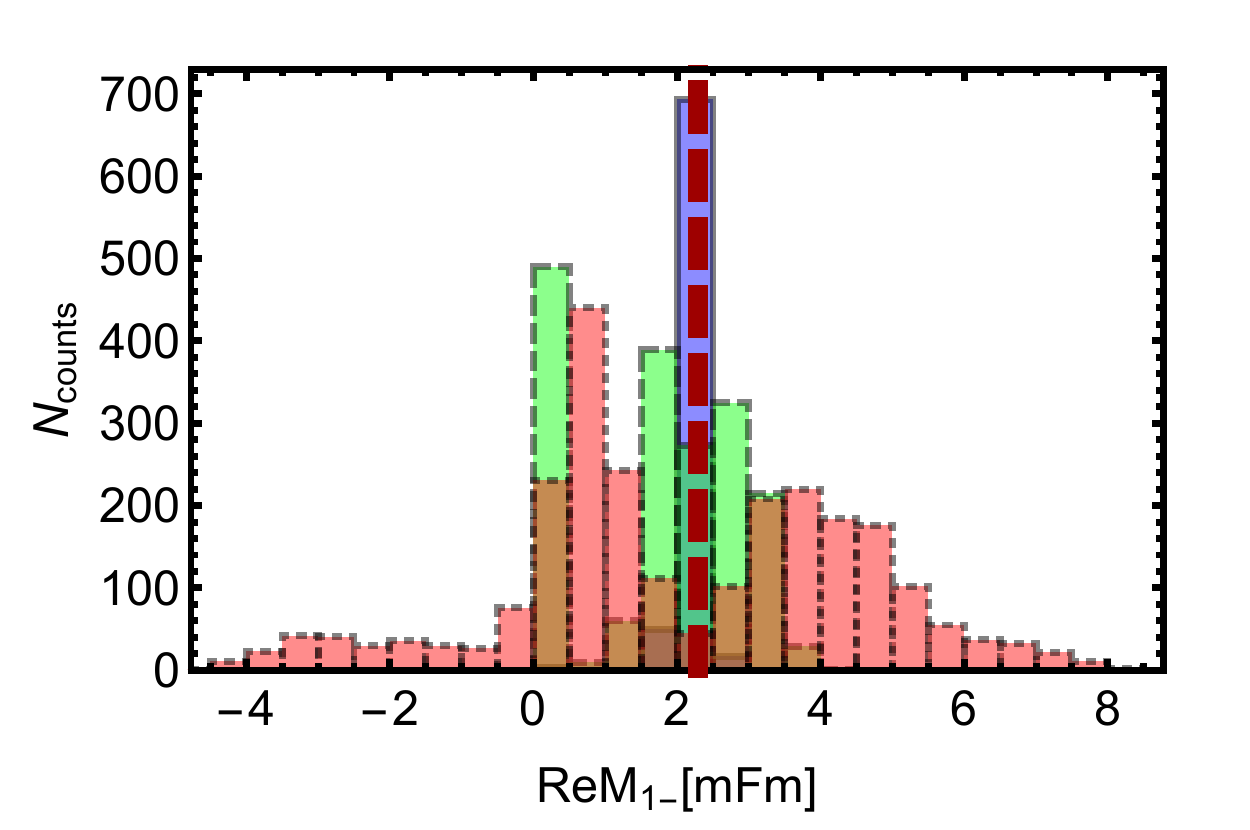}
 \end{overpic} \\
 \begin{overpic}[width=0.325\textwidth]{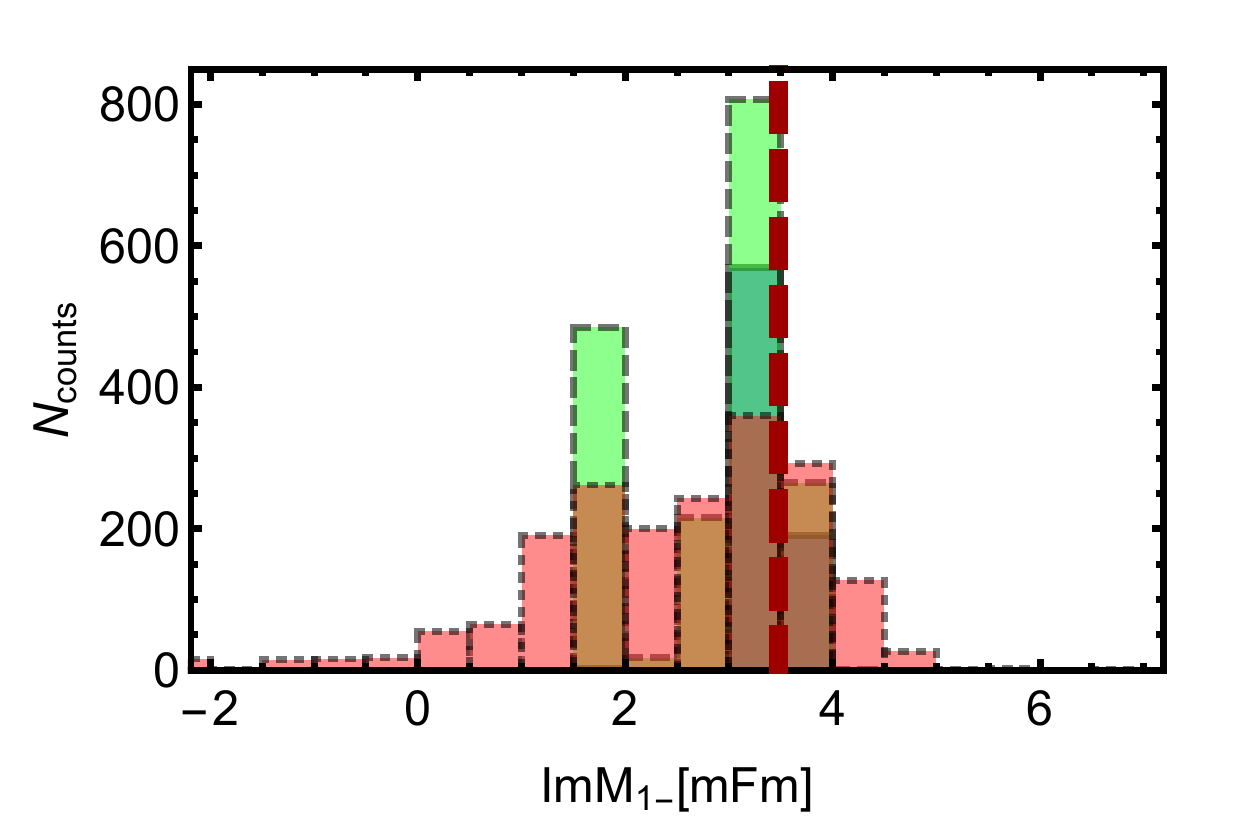}
 \end{overpic}
 \begin{overpic}[width=0.325\textwidth]{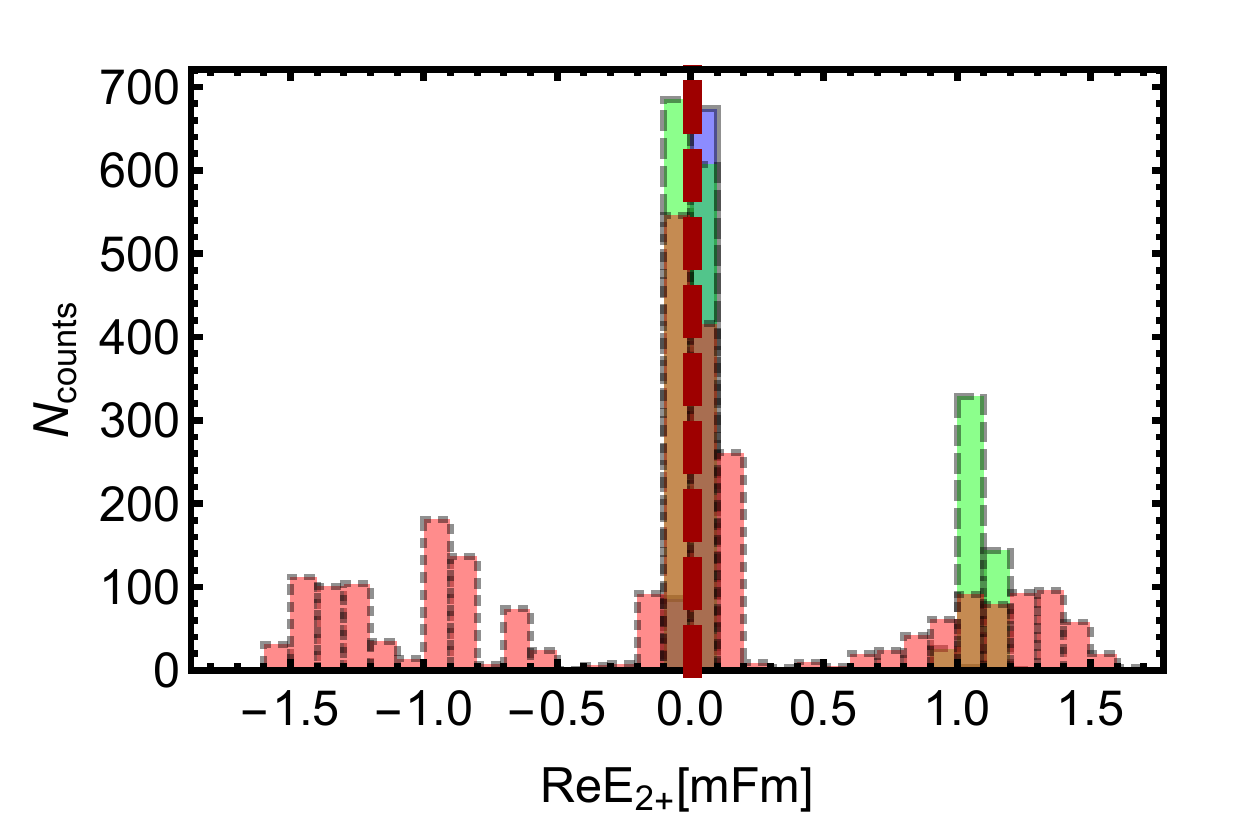}
 \end{overpic}
\begin{overpic}[width=0.325\textwidth]{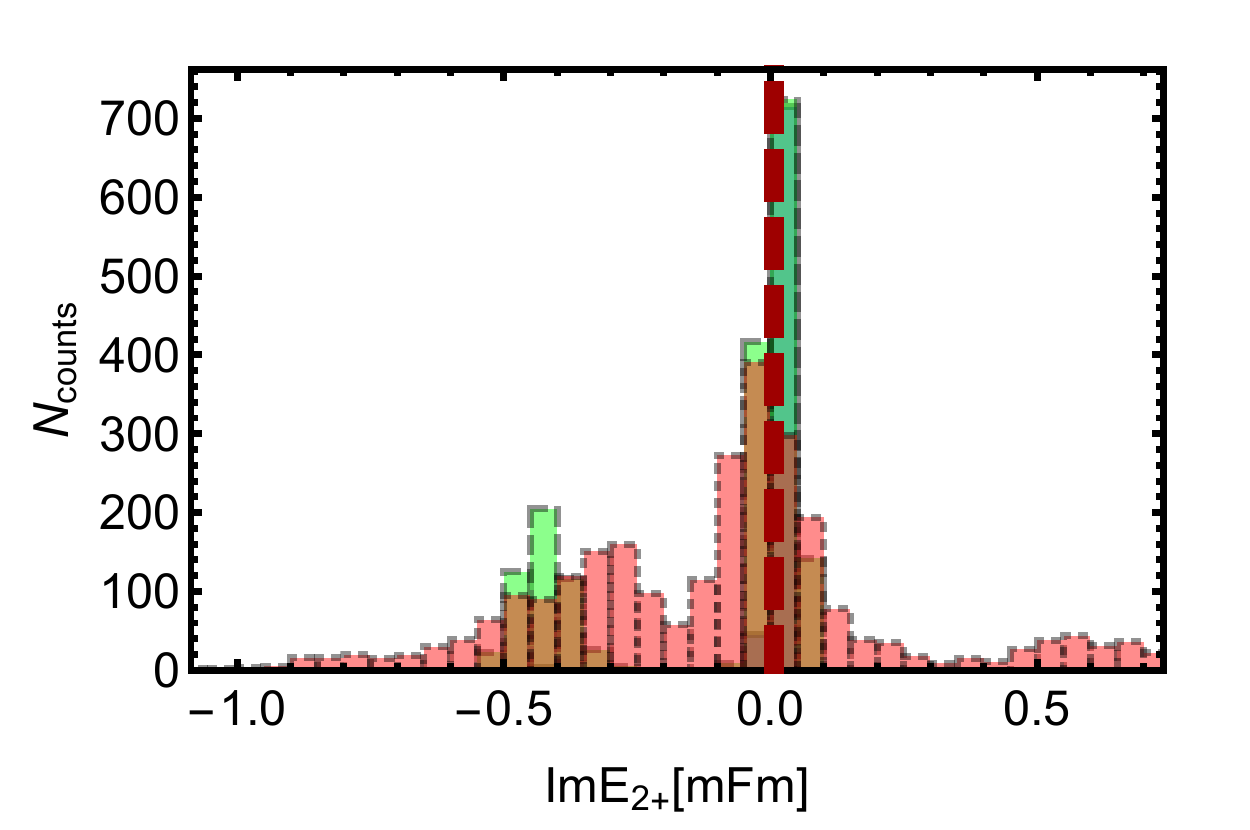}
 \end{overpic} \\
 \begin{overpic}[width=0.325\textwidth]{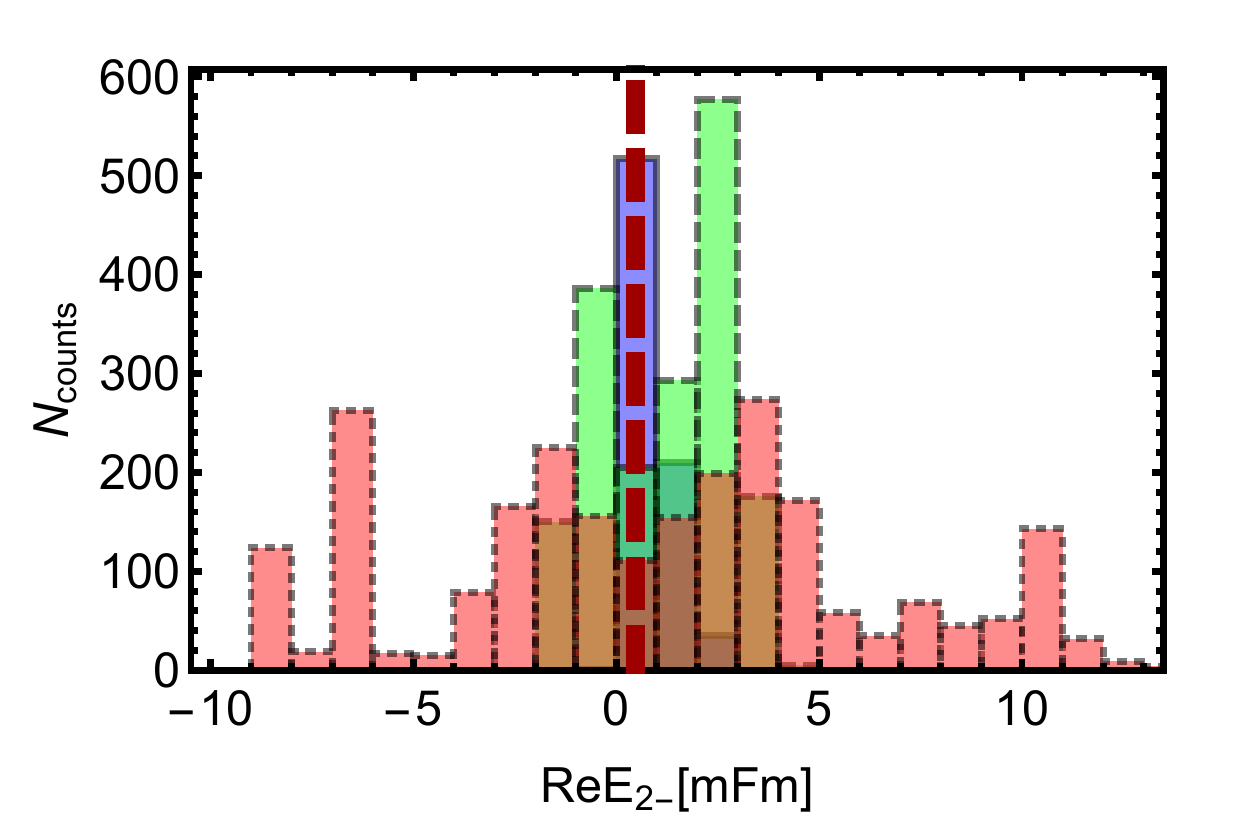}
 \end{overpic}
 \begin{overpic}[width=0.325\textwidth]{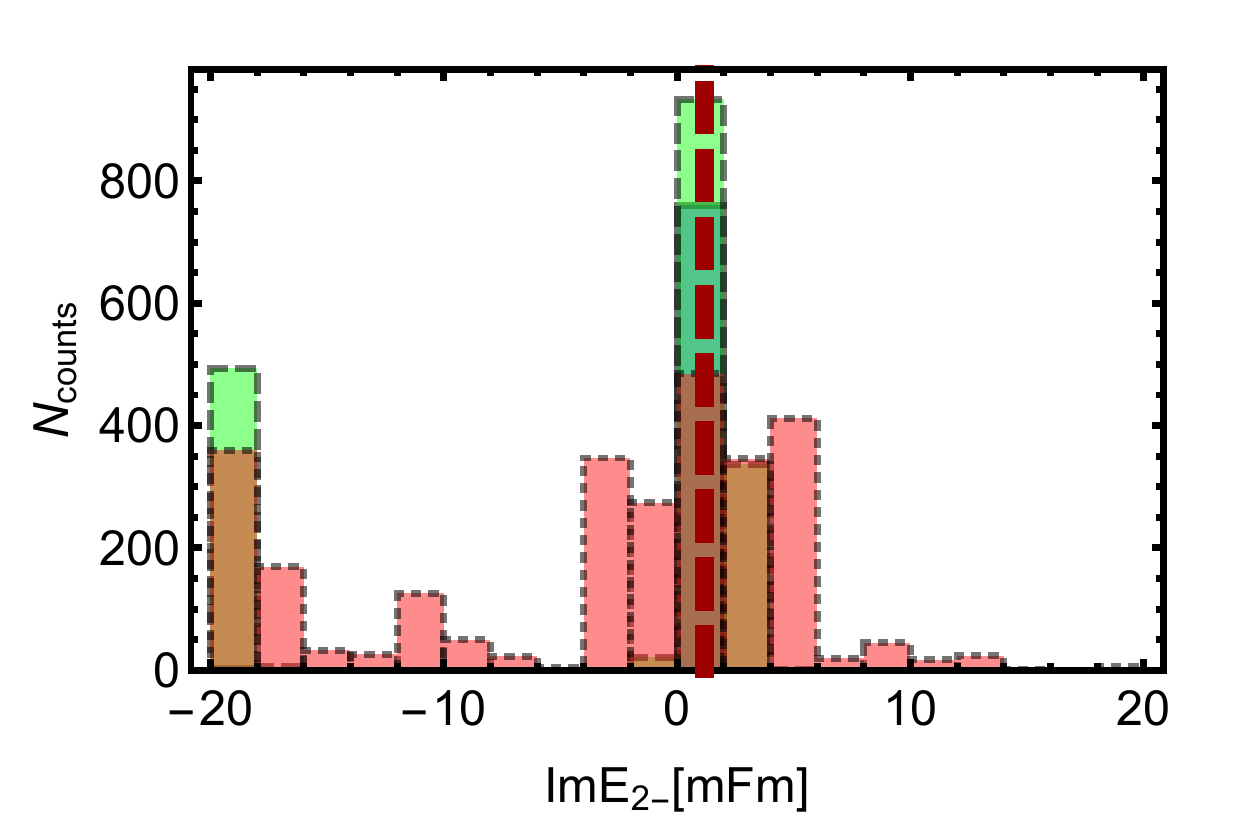}
 \end{overpic}
\begin{overpic}[width=0.325\textwidth]{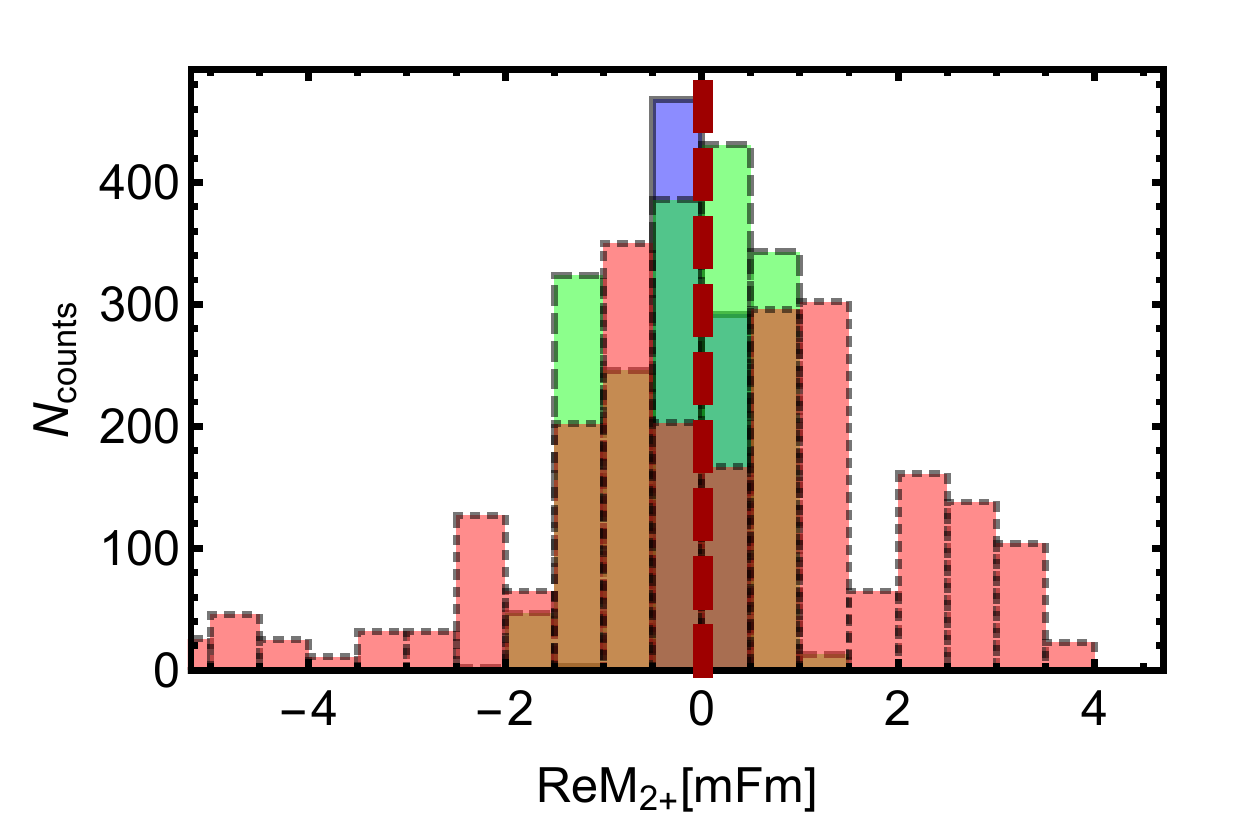}
 \end{overpic} \\
 \begin{overpic}[width=0.325\textwidth]{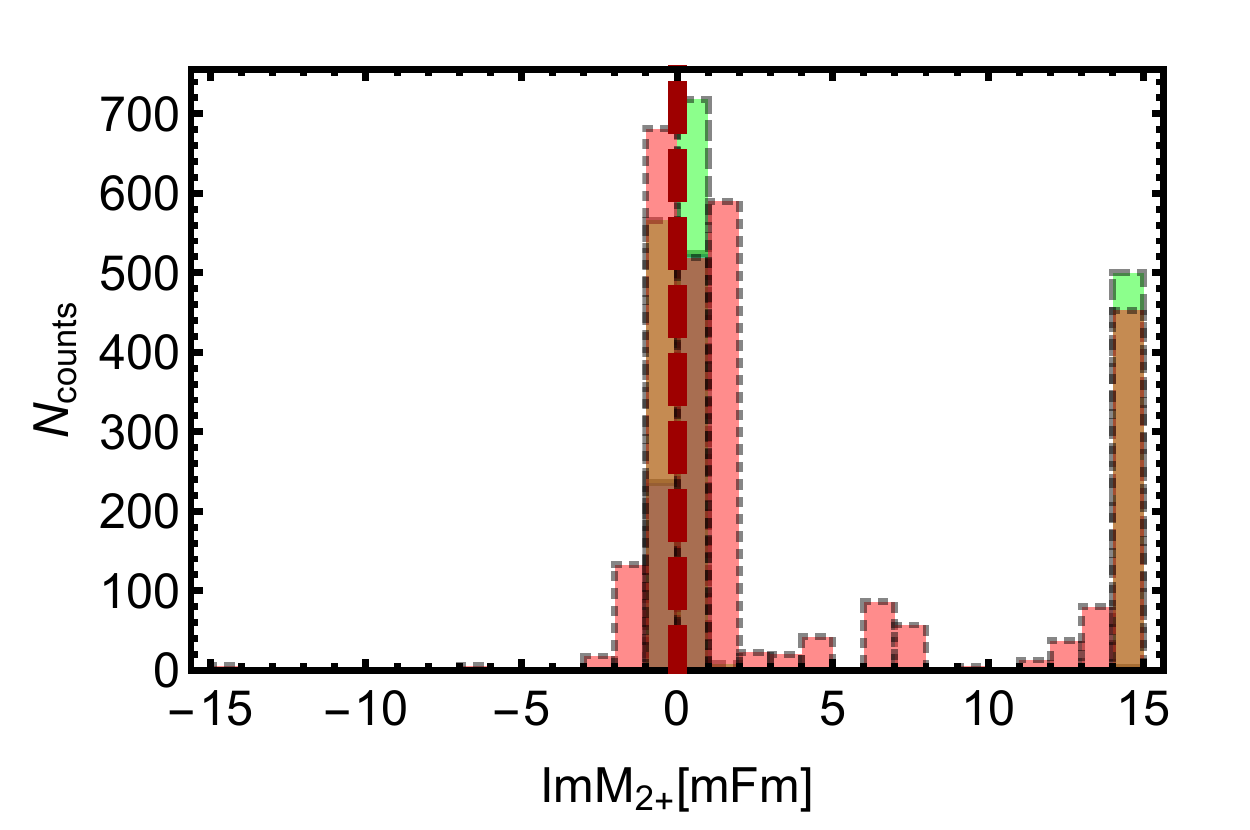}
 \end{overpic}
 \begin{overpic}[width=0.325\textwidth]{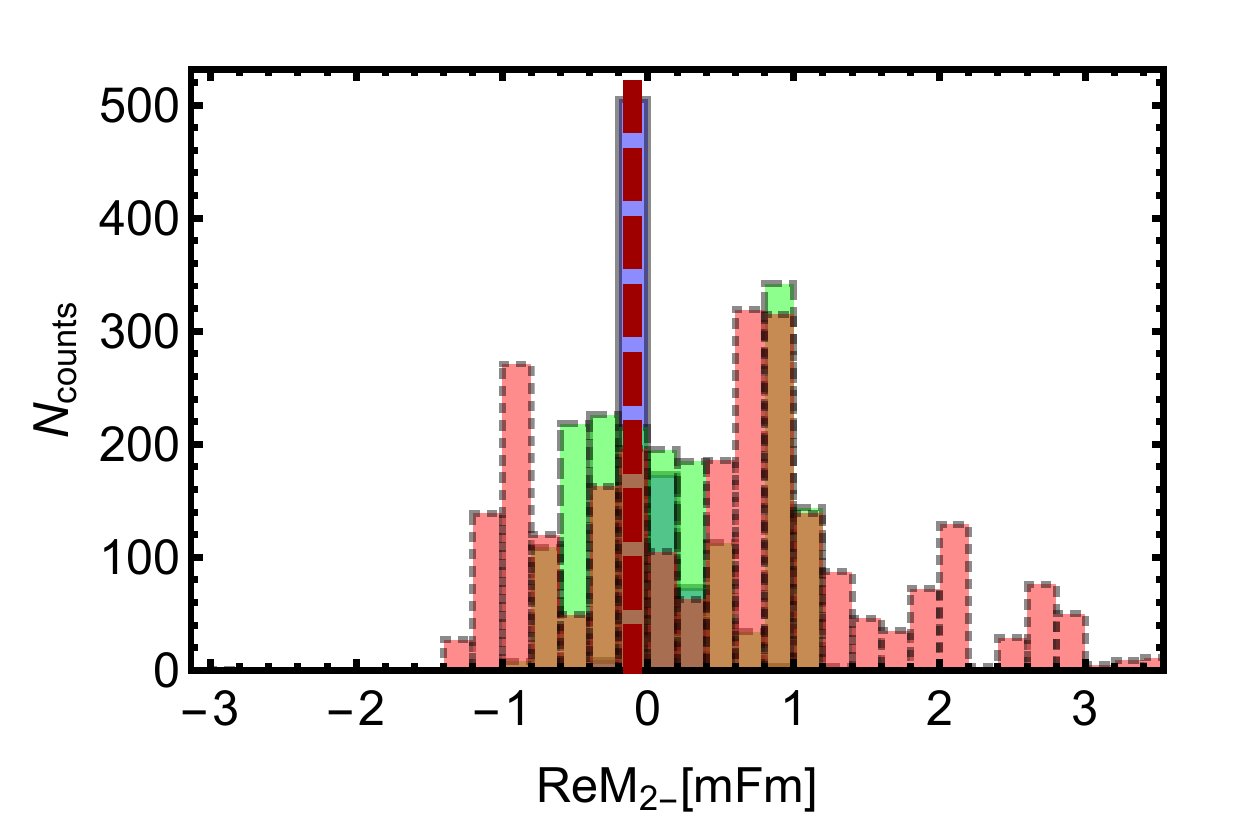}
 \end{overpic}
\begin{overpic}[width=0.325\textwidth]{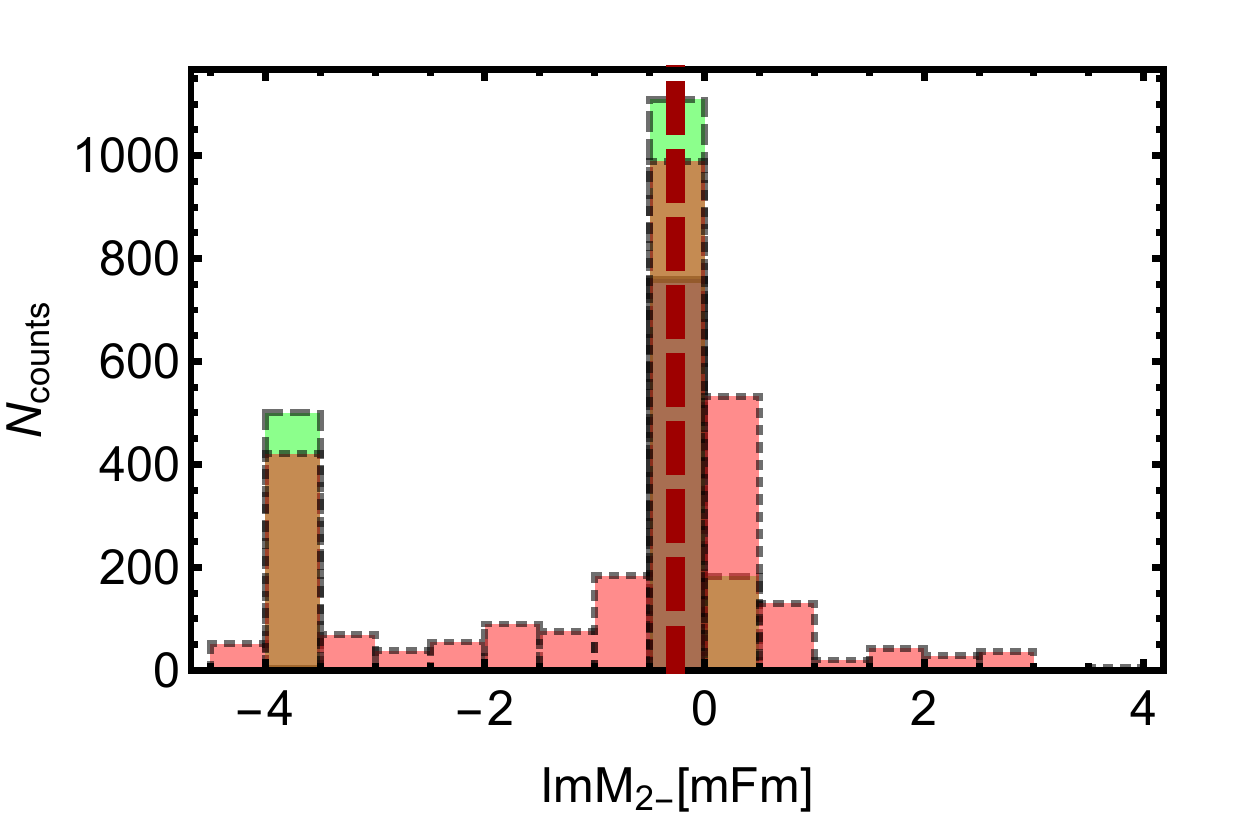}
 \end{overpic}
\caption[Bootstrap-histograms containing the global minima, as well as all non-redundant solutions having chisquare below the $0.95$-quantile of the corresponding non-central $\chi^{2}$-distribution, found in each fit of the $B=500$ bootstrap-replications of the MAID pseudodata truncated at $\ell_{\mathrm{max}} = 2$. Results are shown for the first, second and third error-scenario.]{The histograms show bootstrap-distributions coming from analyses of the error-scenarios $(i)$ (blue bars with solid boundary), $(ii)$ (green bars with dashed boundary) and $(iii)$ (red bars with dotted boundary) (cf. Table \ref{tab:Lmax2PercentageErrorScenarios}) (MAID pseudodata truncated at $\ell_{\mathrm{max}} = 2$). All non-redundant solutions below the respective $0.95$-quantiles $u^{\left(P_{NC}\right)}_{0.95}$ have been included in the histograms. \newline
The global minimum of the fit to the original data of error-scenario $(i)$ is indicated by the red dashed vertical lines.}
\label{fig:Lmax2PseudoDataFitGroupSFObservablesMultHistograms4}
\end{figure}

As a second example, which turns out a lot less stable when fitting pseudodata of comparable errors, we now consider MAID theory-data \cite{LotharPrivateComm} truncated at $\ell_{\mathrm{max}} = 2$ instead of $1$. These model data are also known to allow for 'exact' solutions (cf. section \ref{subsec:TheoryDataFitsLmax2}). Again, we focus on the Delta-energy $E_{\gamma} = 330 \hspace*{1pt} \mathrm{MeV}$ and consider the mathematically complete set of $5$ observables given by (\ref{eq:MathCompleteSetSec4Dot5}). \newline
Again, three scenarios for the errors are analyzed and compared, which are listed in Table \ref{tab:Lmax2PercentageErrorScenarios}. In view of the possible instability of the considered example, the first scenario $(i)$ has been chosen with extremely small errors, in the hope of obtaining at least one somewhat well-behaved fit. Scenarios $(ii)$ and $(iii)$ here correspond precisely to the errors in $(i)$ and $(ii)$ from the previous example (Table \ref{tab:Lmax1PercentageErrorScenarios}). In this way, fits for $\ell_{\mathrm{max}} = 1$ and $2$ can be compared. \newline
We performed a full bootstrap-TPWA on the pseudodata for all three error-scenarios, truncating the analysis at $\ell_{\mathrm{max}} = 2$ and utilizing $B=500$ bootstrap-replicates, as well as a pool of $N_{MC} = 400$ initial conditions for each replicate. \newline
For the generation of bootstrap-distributions for the fit-parameters, we again cut on the $0.95$-quantile $u^{\left(P_{NC}\right)}_{0.95}$ of the non-central chisquare distribution $P_{NC}\left[\mathrm{ndf}; \chi^{2}_{\mathrm{orig.}}\right] (u)$ coming from the fit to the original data. \newline
\vfill
\begin{table}[hb]
 \centering
\begin{tabular}{r|c|cccc}
 Scenario-no. & $P_{c}$ & $P_{\%}^{\Sigma}$ & $P_{\%}^{T}$ & $P_{\%}^{P}$ &  $P_{\%}^{F}$  \\
\hline
$(i)$ & $2$ & $0.1$ & $0.1$ & $0.1$ & $0.2$ \\
$(ii)$ & $2$ & $1$ & $1$ & $1$ & $2$ \\
$(iii)$ & $2$ & $5$ & $5$ & $5$ & $10$
\end{tabular}
\caption[Error-scenarios for analyses of pseudodata truncated at $L=2$.]{Shown here are three scenarios, for the percentages $P_{c}$ and $P_{\%}^{\alpha}$ employed for the generation of errors, in the pseudodata fits of MAID model data truncated at $\ell_{\mathrm{max}} = 2$. The factors $c$ and $N_{\%}^{\alpha}$ used in equations (\ref{eq:DCSModelDataError}) and (\ref{eq:MaxErrorConvention}) are obtained via $c = P_{c}/100$ and $N_{\%}^{\alpha}=P_{\%}^{\alpha}/100$.} 
\label{tab:Lmax2PercentageErrorScenarios}
\end{table}
\clearpage
Multimodal parameter distributions are generally seen to appear. Here, we choose to expose the instabilities of the fit in a sequence of plots of bootstrap-histograms. The first one is given in Figure \ref{fig:Lmax2PseudoDataFitGroupSFObservablesMultHistograms1}, where for the scenario $(i)$ of incredibly precise pseudodata, only the global minima found for each bootstrap-replicate are included. Here, the cut on the $0.95$-quantile has not yet been applied. The distributions of almost all multipole-parameters are unimodal and gaussian up to a (sometimes very good) approximation. Some parameters have distributions with quite elongated tails though, for instance $\mathrm{Re}\left[E_{0+}\right]$ or $\mathrm{Re}\left[E_{2+}\right]$. \newline
In case one still only considers error-scenario $(i)$, but now includes all non-redondant solutions below the $0.95$-quantile into the histograms, distributions are found which can be seen in Figure \ref{fig:Lmax2PseudoDataFitGroupSFObservablesMultHistograms2}. Multi-peak structures start to appear in the distributions for many of the parameters. In particular, the tails of some asymmetric distributions visible in Figure \ref{fig:Lmax2PseudoDataFitGroupSFObservablesMultHistograms1} have turned into peaks, once all solutions surviving the probabilistic cut are included. Furthermore, all peaks in the observed distributions are connected. \newline
A comparison of the bootstrap distributions for error-scenarios $(i)$ and $(ii)$ is plotted for all multipole-parameters in Figure \ref{fig:Lmax2PseudoDataFitGroupSFObservablesMultHistograms3}. Again, all solutions below the $0.95$-quantile are included in the histograms. Here, structures of multiple peaks can be observed, with individual peaks broadened for scenario $(ii)$. Also, for a lot of parameters peaks are observed which are fully disconnected in parameter-space, but no more than two such peaks are seen for each parameter individually. Some distributions, for instance those of $\mathrm{Re}\left[E_{0+}\right]$ or $\mathrm{Re}\left[E_{2-}\right]$, become diluted for error-scenario $(ii)$. They show still quite smooth distributions, with merged multi-modalities. A vast decrease of the bootstrap-TPWA's stability is observed when comparing both error-scenarios. However, one should note that for the errors present in scenario $(ii)$, the fit of the pseudodata for $\ell_{\mathrm{max}} = 1$ was still very stable (cf. Figure \ref{fig:Lmax1PseudoDataFitGroupSFObservablesMultHistograms}). All this illustrates, for the considered energy-bin of course, the possibility of a rapid increase of problems with ambiguities, once the truncation order of a TPWA is raised by one. \newline
Finally, a comparison of the bootstrap-distributions for all three error-scenarios is plotted in the histograms of Figure \ref{fig:Lmax2PseudoDataFitGroupSFObservablesMultHistograms4}. For error-scenario $(iii)$, the stability of the bootstrap-TPWA is now fully broken. Distributions are now diluted even more. For some parameters, multiple fully disconnected regions are given where solutions exist. Also, the shape of individual peaks can in many cases not be identified with a gaussian any more. \newline
The comparison between fits to pseudodata truncated at $\ell_{\mathrm{max}} = 1$ and $\ell_{\mathrm{max}} = 2$ has shown a feature which, empirically, has shown up as more general in the course of this work: once higher partial waves which are small, such as the $D$-waves in this example here at the $\Delta$-energy, are fitted out of the data and treated as completely free parameters, ambiguity-problems tend to appear. Thus, for higher energies where the $D$-waves become stronger, pseudodata-fits such as those performed here may turn out to be more stable. A way out of this issue consists of fixing the small higher partial waves to values from a theoretical model. In this way, the full model-independence of the TPWA is lost. Still, knowledge of small higher partial waves can be essential for obtaining better values for the fitted lower waves. Examples of such mildly model-dependent TPWAs will be given in section \ref{sec:RealWorldDataFits}. \newline
It is clear that an unlimited amount of further studies could be done on pseudodata. However, since in the course of this work, results were also obtained in analyses of real data, we continue in section \ref{sec:RealWorldDataFits} with a discussion of the latter. The machinery of bootstrap-TPWAs as presented here will be applied there as well.

\clearpage


\clearpage

\subsection{Results of fits to real photoproduction data} \label{sec:RealWorldDataFits}

In order to illustrate the techniques introduced for model-independent TPWA-fits using Monte Carlo-sampling of parameter spaces (sections \ref{sec:TPWAFitsIntro} and \ref{sec:MonteCarloSampling}), as well as the application of the bootstrap to the problem (section \ref{sec:BootstrappingIntroduction}), we consider here analyses of data for the reaction
\begin{equation}
 \gamma p \longrightarrow \pi^{0} p \mathrm{.} \label{eq:Pi0ProductionReactionSec4Dot6}
\end{equation}
Within the $\Delta$-region\footnote{The energy region where the resonance $\Delta(1232)\frac{3}{2}^{+}$ \cite{Patrignani:2016xqp} dominates the $M_{1+}$-multipole in the channel (\ref{eq:Pi0ProductionReactionSec4Dot6}) \cite{Grushin}. This is sometimes also referred to as the 'first resonance region' (section \ref{sec:LFitsPaper}).}, i.e. for energies from threshold up to approximately $E_{\gamma} = 500 \hspace*{1pt} \mathrm{MeV}$ (cf. section \ref{sec:LFitsPaper}), a set of combined measurement will be considered which forms a mathematically complete set, at least according to the rules of Omelaenko outlined in chapter \ref{chap:Omelaenko}. The observables are
\begin{equation}
 \left\{ \sigma_{0}, \Sigma, T, P, F \right\} \mathrm{.} \label{eq:DeltaRegionFitObservables}
\end{equation}
The majority of these data have been taken at the MAMI-facility \cite{Hornidge:2013,Leukel:2001,Schumann:2015}, with only the $P$-observable stemming from the Kharkov-data published by Belyaev et al. \cite{Belyaev:1983}. The data in the $\Delta$-region allow for first simple tests of the TPWA-machinery. \newline
Then, again for the reaction of $\pi^{0}$-photoproduction (\ref{eq:Pi0ProductionReactionSec4Dot6}), we will consider a set of combined measurements which have a kinematic overlap in the so-called second resonance region, i.e. from $E_{\gamma} = 500 \hspace*{1pt} \mathrm{MeV}$ up to roughly $E_{\gamma} = 900 \hspace*{1pt} \mathrm{MeV}$ (see section \ref{sec:LFitsPaper}). Here, we have a set of seven observables which is, according to Omelaenko, already mathematically over-complete:
\begin{equation}
 \left\{ \sigma_{0}, \Sigma, T, P, E, G, H \right\} \mathrm{.} \label{eq:2ndResRegionFitObservables}
\end{equation}
In this case, the experiment providing the majority of observables, namely the quantities $T$, $P$, $E$, $G$ and $H$, is the CBELSA/TAPS-experiment \cite{Hartmann:2014, Hartmann:2015, Gottschall:2014, Gottschall:2015, Thiel:2012, Thiel:2016}. The set (\ref{eq:2ndResRegionFitObservables}) is supplemented by data for the beam-asymmetry $\Sigma$ from the GRAAL-facility \cite{GRAAL}, as well as a very recent new measurement for the differential cross section $\sigma_{0}$ performed at MAMI \cite{Adlarson:2015}. The fits in the second resonance region can be expected to be be more numerically demanding and difficult to handle, since the truncation order $\ell_{\mathrm{max}}$ required will be larger than in the $\Delta$-region.

\subsubsection{$\gamma p \rightarrow \pi^{0} p$ in the $\Delta$-region} \label{subsec:DeltaRegionDataFits}

\textbf{Description of the datasets} \newline

The foundation of the analysis is provided by the measurement of the differential cross section $\sigma_{0}$ published by Hornidge et al. \cite{Hornidge:2013}. Data are given over the energy range from $E_{\gamma} = 146.95 \hspace*{1pt} \mathrm{MeV}$ up to $420.27 \hspace*{1pt} \mathrm{MeV}$ in $114$ almost equidistant energy-bins. Every energy-bin contains an angular distribution of $20$ datapoints with equal spacing in $\cos \theta$, covering the full angular interval. The statistical precision of this dataset is excellent, with errors totaling to only a fraction of few percent of the respective cross section. \newline
For the beam-asymmetry $\Sigma$, we pick the well-known results by Leukel et al. \cite{Leukel:2001, LeukelPhD}. Here, an energy-binning of constant separation $\Delta E_{\gamma} = 10 \hspace*{1pt} \mathrm{MeV}$ is provided, covering a region from $E_{\gamma} = 240 \hspace*{1pt} \mathrm{MeV}$ to $440 \hspace*{1pt} \mathrm{MeV}$. Angular distributions cover the whole range in $17$ bins each with equal spacing in $\theta$. \newline
\begin{figure}[h]
 \centering
 \begin{overpic}[width=0.475\textwidth]{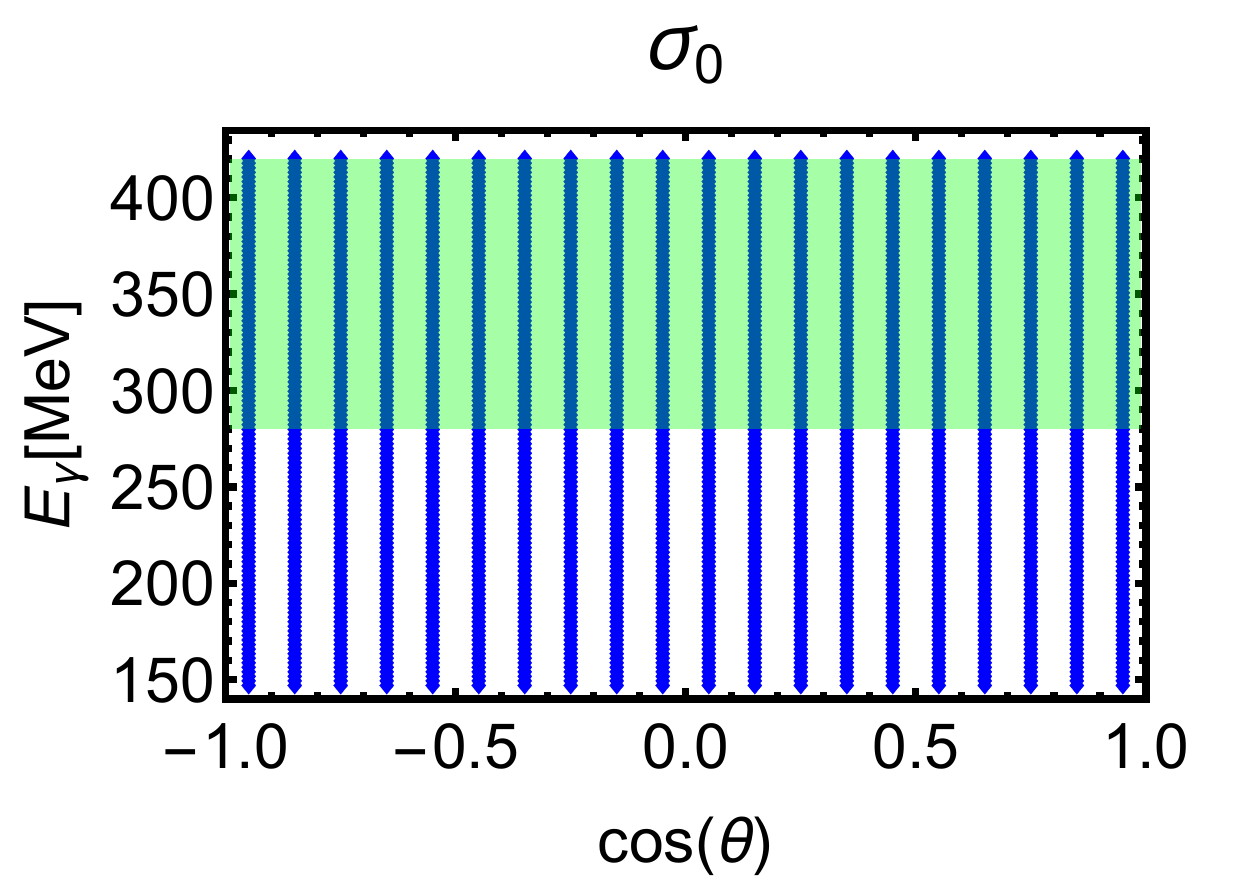}
 \end{overpic}
  \begin{overpic}[width=0.475\textwidth]{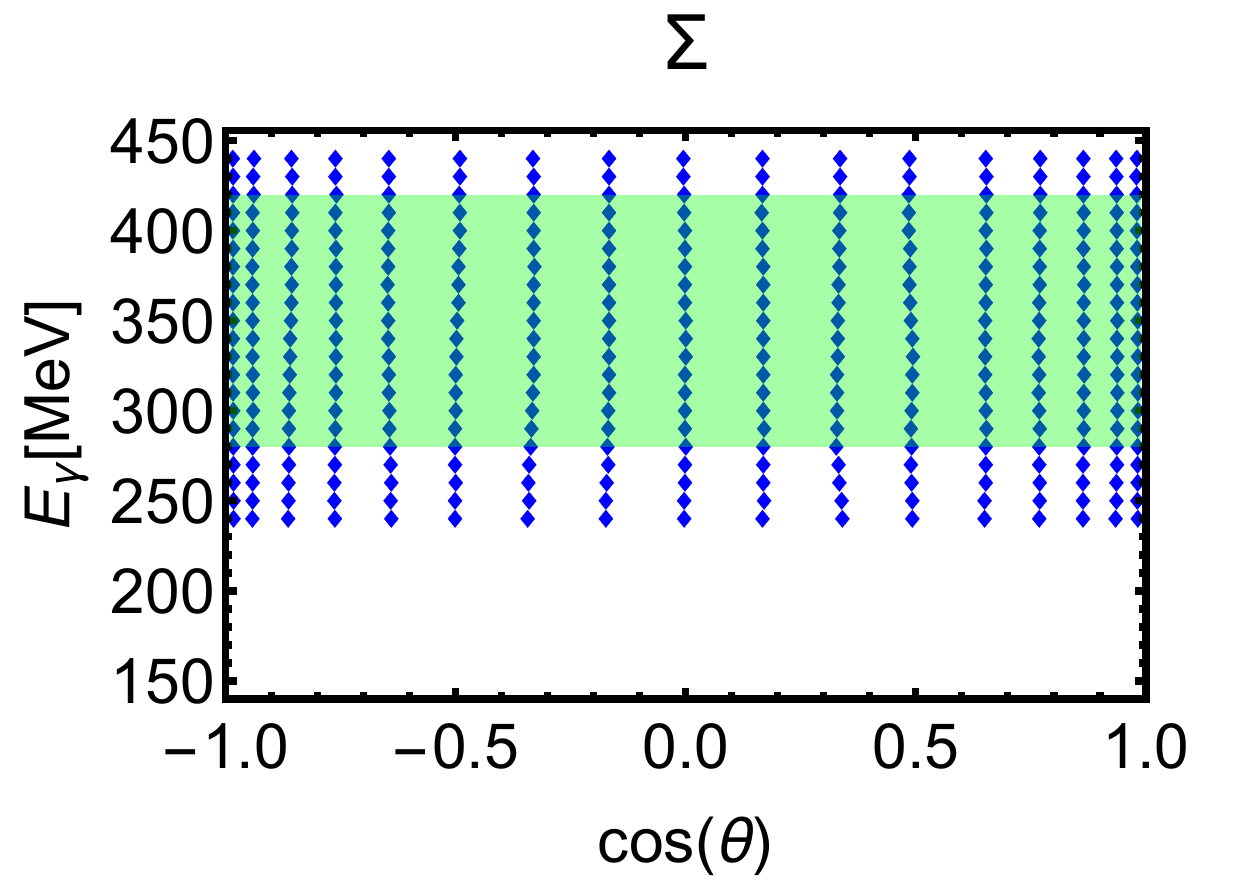}
 \end{overpic} \\
 \begin{overpic}[width=0.475\textwidth]{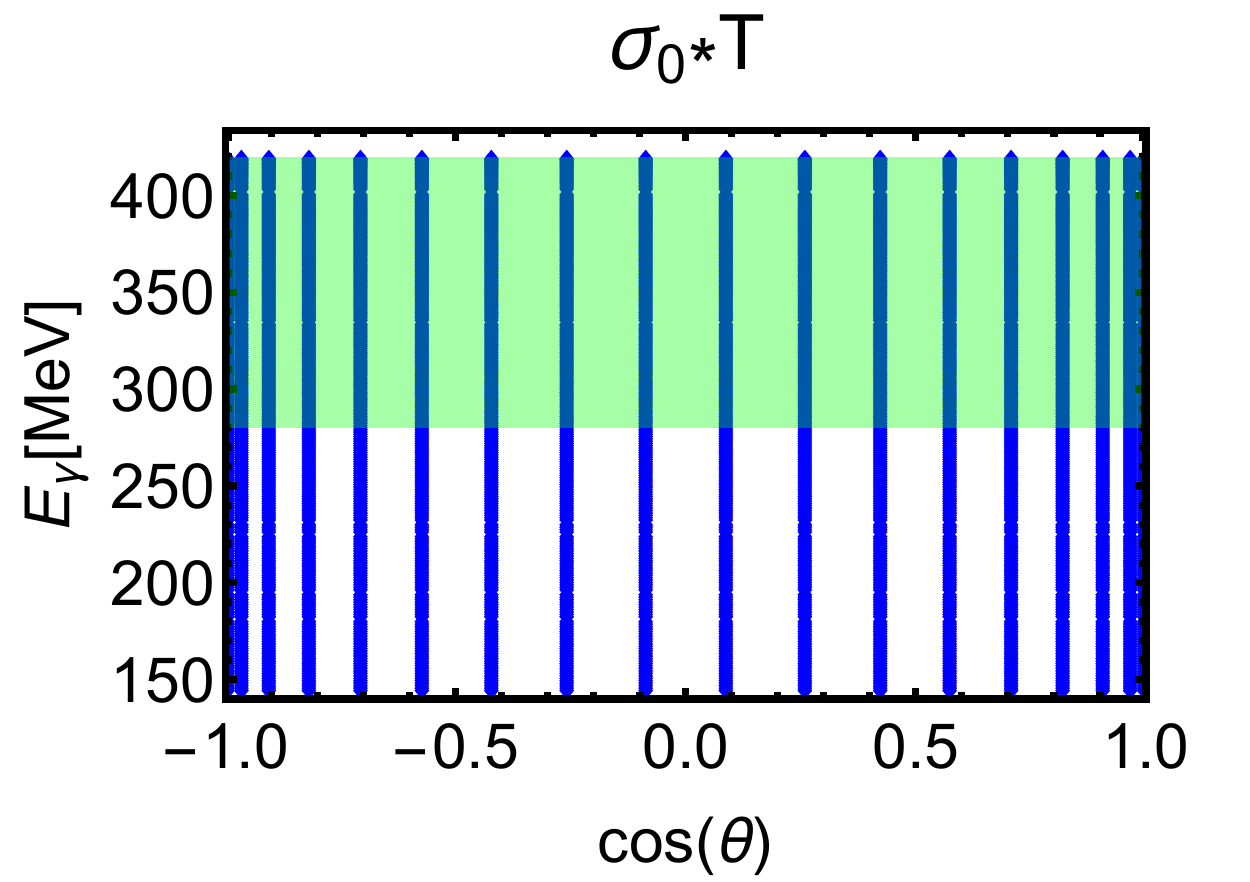}
 \end{overpic}
  \begin{overpic}[width=0.475\textwidth]{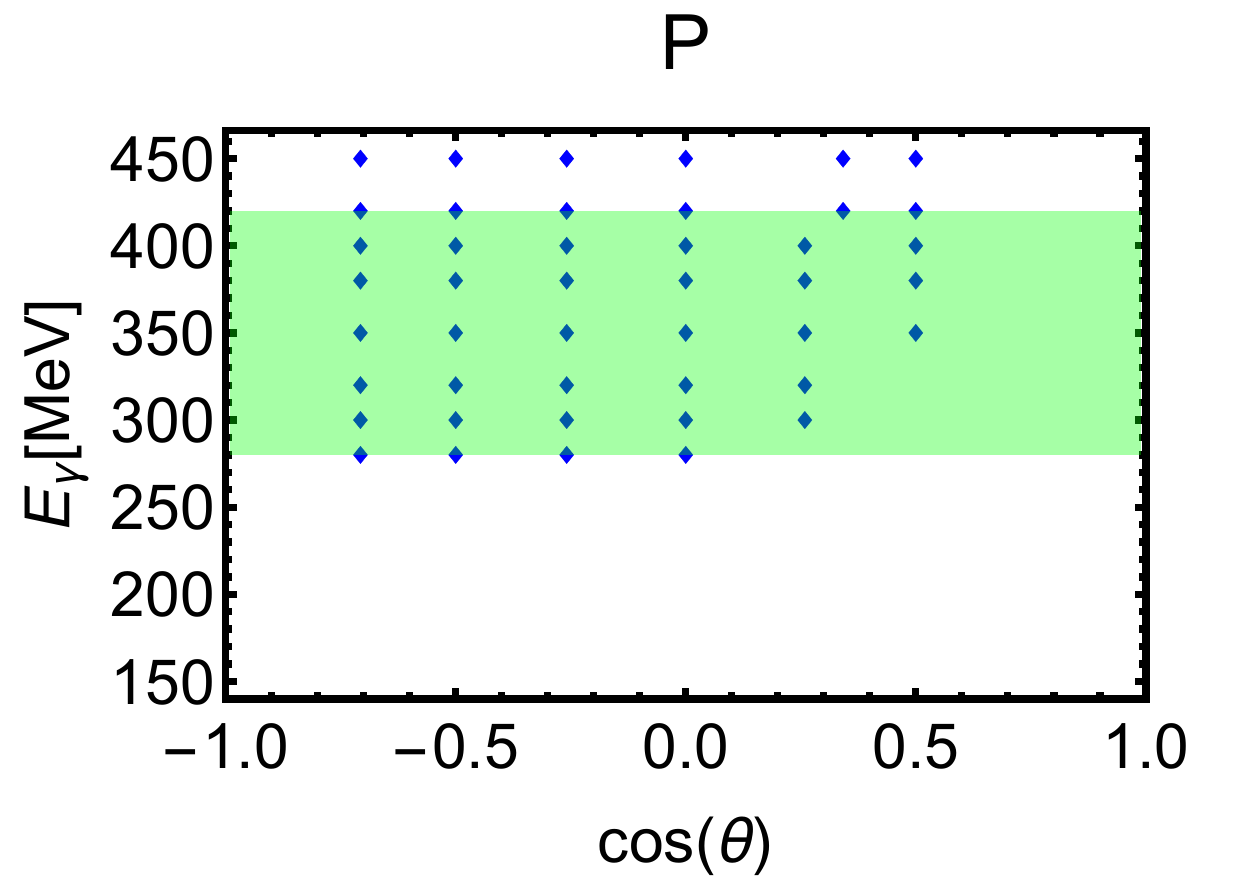}
 \end{overpic} \\
 \begin{overpic}[width=0.475\textwidth]{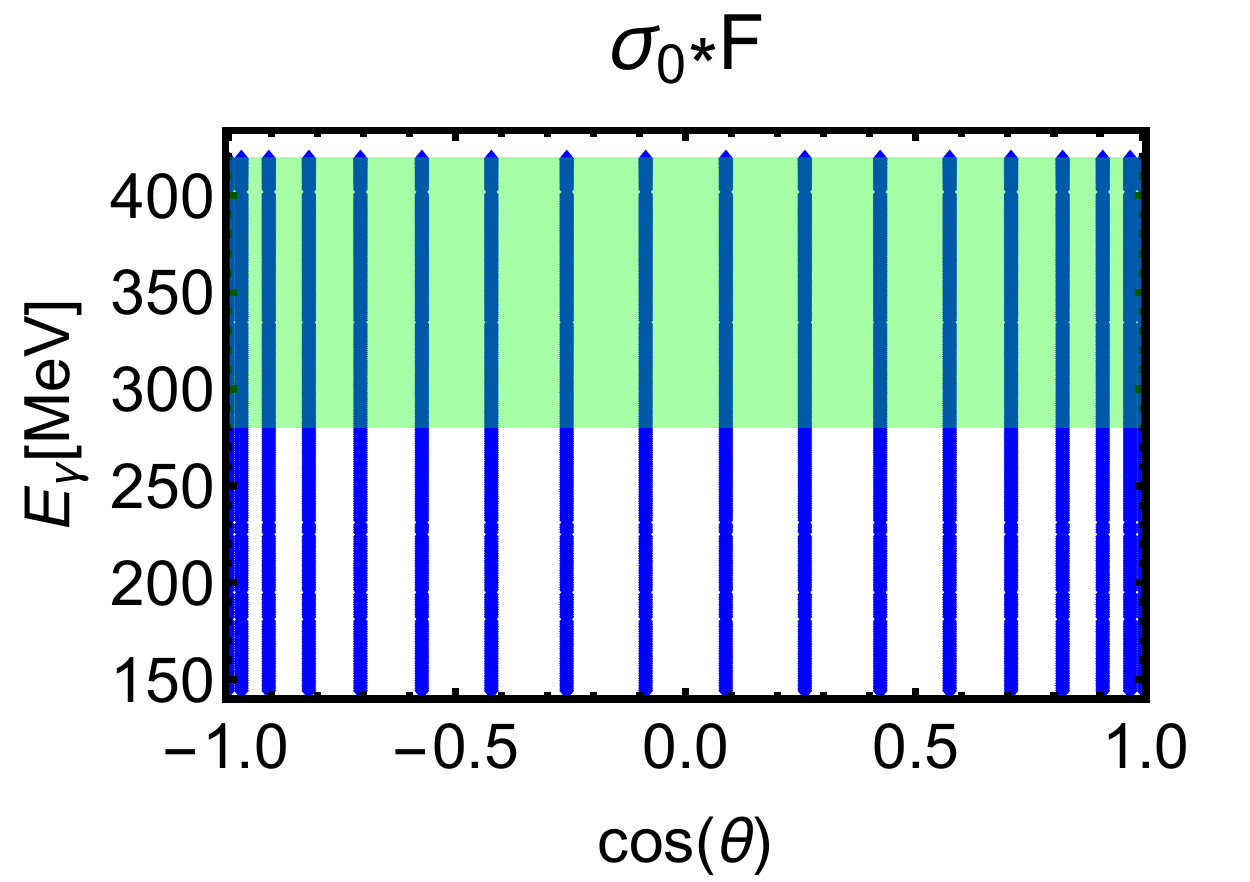}
 \end{overpic}
  \caption[The regions in phase space $(E_{\gamma}, \cos (\theta))$ covered by the five polarization-datasets analyzed in the $\Delta$-resonance region.]{The plots illustrate the regions in phase space $(E_{\gamma}, \cos (\theta))$ covered by the datasets for the cross section $\sigma_{0}$ \cite{Hornidge:2013}, the dimensionless asymmetries $\Sigma$ and $P$ \cite{Leukel:2001,Belyaev:1983} and the dimensioned profile functions $\sigma_{0} T = \check{T}$ and $\sigma_{0} F = \check{F}$ \cite{Schumann:2015}. Blue markers give the location of individual datapoints. The green-shaded region illustrates the energy-range where all observables overlap, i.e. the area on which the TPWA can be performed (cf. the main text).}
 \label{fig:DeltaRegionKinematicPlots}
\end{figure}

\clearpage
For the target-asymmetry $T$ and the double-polarization observable $F$, a very recently published dataset by Schumann, Otte et al. \cite{Schumann:2015} is utilized. Here, data for the dimensioned asymmetries $\check{T} = \sigma_{0} T$ and $\check{F} = \sigma_{0} F$ are provided, with the exact same kinematic binning. \newline
To be more precise, reference \cite{Schumann:2015} is a publication of only the $\check{T}$-data, but the results for $\check{F}$ used here come from the same cluster of beamtimes. They have been provided with permission by the CrystalBall-collaboration at MAMI. \newline
The data for $\check{T}$ and $\check{F}$ cover an energy-region from $E_{\gamma} = 144.29 \hspace*{1pt} \mathrm{MeV}$ to $419 \hspace*{1pt} \mathrm{MeV}$, with energies spaced only by $1$ to $2 \hspace*{1pt} \mathrm{MeV}$ in a very fine granularity. Angular distributions are fully covered with $18$ bins of equal spacing in $\theta$. The precision of these very recent CB-data is also acceptable. \newline
The dataset for the single-polarization observable $P$ by Belyaev et al. \cite{Belyaev:1983} represents, from a data-perspective, the limiting factor of the analysis in the $\Delta$-region described here. A total of $8$ energy-bins are given for $P$, over a range from $E_{\gamma} = 280 \hspace*{1pt} \mathrm{MeV}$ to $450 \hspace*{1pt} \mathrm{MeV}$. Spacings of points vary from $ \Delta E_{\gamma} = 20 \hspace*{1pt} \mathrm{MeV}$ to $30 \hspace*{1pt} \mathrm{MeV}$. Angular distributions contain between $4$ and $6$ non-equidistant points, with the forward-scattering region non-covered for all energies. Most importantly, the precision of the Belyaev-data is lacking, with statistical errors ranging between $40$ and $100 \hspace*{1pt} \%$ of the asymmetry itself. Therefore, of all the considered datasets, the Belyaev-data are those with lowest statistics as well as the smallest phase-space coverage. \newline
It is not clear a priori, whether or not the $P$-data can add any useful information to the fit. However, we have to stress that for a mathematically complete set of $5$ observables according to Omelaenko \cite{Omelaenko}, the observable $P$ is essential (see chapter \ref{chap:Omelaenko}). Furthermore, none of the complete sets of $4$ found by Tiator \cite{LotharPrivateComm2016} (listed in chapter \ref{chap:CompleteSetsOf4}) contains the combination $\left\{ \sigma_{0}, \Sigma, T, F \right\}$ alone. This behavior has been confirmed in simple theory-data fits performed in the course of this work, where a double ambiguity has been seen for the set $\left\{ \sigma_{0}, \Sigma, T, F \right\}$. Therefore, data for $P$ have to be included. \newline
The kinematic properties of the investigated datasets are summarized in the plots shown in Figure \ref{fig:DeltaRegionKinematicPlots}. Table \ref{tab:DeltaRegionSandorfiDataTable} contains condensed information on the fitted datasets. In summary, the provided datasets pose a total of $11 \hspace*{1pt} 681$ points of experimental information. \newline
\begin{table}[h]
 \centering
\begin{tabular}{rccccr}
\hline
\hline
 Group & Experiment & Observables & $E_{\gamma}$-range [MeV] & $\Delta E_{\gamma}$ [MeV] & av. stat. errors \\
\hline
  &  &  &  &  &  \\
 $1$ & MAMI/CB & $\sigma_{0}$ & $(146.95 - 420.27)$ & $\sim 2.5$ & $\sim (1 - 5) \%$ \\
  &  &  &  &  &  \\
 $2$ & MAMI/CB & $\left\{\check{T}, \check{F}\right\}$ & $(144.29 - 419.)$ & $\sim (1. - 2.)$ & $\sim (5 - 25) \%$ \\
  &  &  &  &  &  \\
 $3$ & MAMI/TAPS & $\Sigma$ & $(240. - 440.)$ & $10.$ & $\sim (5 - 15) \%$ \\
  &  &  &  &  &  \\
 $4$ & Kharkov & $P$ & $(280. - 450.)$ & $\sim (10. - 30.)$ & $\sim (40 - 100) \%$ \\
  &  &  &  &  &  \\
\hline
\hline
\end{tabular}
 \caption[Some specifics of the polarization datasets fitted in the $\Delta$-region.]{Some specifics of the polarization datasets fitted in the $\Delta$-region \cite{Hornidge:2013, Leukel:2001, Schumann:2015, Belyaev:1983} are collected. Datasets are divided into groups according to their kinematic binning. Energy-ranges as well as some information on the energy-binning $\Delta E_{\gamma}$ are given. Approximate quantities are marked with a $\sim$. Ranges for binnings as well as averages are provided, whenever appropriate. \newline
 Also, some rough estimate of the precision of the respective datasets is given, by providing an estimate for the average size of the statistical errors, relative to the data, in percent. For similar Tables (but different data), see reference \cite{Sandorfi:2010uv}.}
 \label{tab:DeltaRegionSandorfiDataTable}
\end{table}
%


\textbf{Kinematic re-binning of data and evaluation of the profile-functions} \newline

At this point, a problem is encountered which has not been present in the analyses of model-data presented in sections \ref{sec:TheoryDataFits} and \ref{sec:PseudoDataWithErrorsFitted}. The theory-data coming from the MAID-model have all been given with the exact same kinematic binning for all observables. However, as can be seen in Figure \ref{fig:DeltaRegionKinematicPlots}, this situation is generally not given for realistic datasets. \newline
Since the TPWA operates on adjacent energy-bins individually and furthermore incorporates information from all observables at each energy, it is clear that bins have to be brought to match in the energy-region covered by all observables. The angular distributions on the other hand are fitted in the TPWA, which makes it unnecessary to bring them to the same binning. \newline
As a general principle, the dataset with the weakest statistics has to dictate the overall energy-binning in the TPWA \cite{Sandorfi:2010uv}. For the datasets fitted in the $\Delta$-region, one would have to anhere to the $P$-dataset by Belyaev et al. \cite{Belyaev:1983} (see Figure \ref{fig:DeltaRegionKinematicPlots}). Thus, the analysis will be performed, in the end, on $7$ non-equidistant energy-bins in the interval from $E_{\gamma} = 280 \hspace*{1pt} \mathrm{MeV}$ to $420 \hspace*{1pt} \mathrm{MeV}$. \newline
For the method of re-binning we choose here the simplest possibility of taking the data as they are, at the closest points in energy. For every energy $E_{\gamma}^{P}$ dictated by $P$, we therefore loop through the full dataset of the particular observable (other than $P$) and then shift the whole angular distributions at the point in energy closest to $E_{\gamma}^{P}$, to the respective $P$-energy $E_{\gamma}^{P}$. \newline
Another feature of the TPWA as practitioned in this work consists of the fact that it is performed on dimensioned profile functions $\check{\Omega}^{\alpha}$, not dimensionless observables $\Omega^{\alpha}$. For the quantities $\check{T}$ and $\check{F}$ in the $\Delta$-region, which are already provided as profile functions, this poses no additional problem. For the remaining quantities $\Sigma$ and $P$, at some point the asymmetry has to be multiplied by the differential cross section in order to evaluate a value $\check{\Omega}^{\alpha} = \sigma_{0} \Omega^{\alpha}$ for the profile function (as well as the error $\Delta \check{\Omega}^{\alpha} = \sigma_{0} \Delta \Omega^{\alpha}$, see appendix \ref{sec:BootstrapAnsatzComments}). We again adhere to the simple method of kinematically closest points, searching the closest $\sigma_{0}$-value for every asymmetry-datapoint and then forming the profile function. This method again poses no additional inconsistency, since the differential cross section is in any case the observable with best statistics and thus the most datapoints in each angular distribution (cf. Figure \ref{fig:DeltaRegionKinematicPlots}). Again, no 'new' datapoints are generated this way. \newline
Plots of the resulting datapoints for the profile functions can be considered for one example-energy in Figure \ref{fig:DeltaRegionObsFitAngDistPlots} below. After the kinematic re-binning and the evaluation of the profile-functions, there remain $549$ datapoints as input for the multipole analysis. \newline
The method of kinematically closest points chosen here is the simplest possible one, but it should not disguise the fact that generally the preparation of data for the TPWA is already a complicated problem. The solution to this problem does not have a lot to do with he TPWA itself and we have found the simplest possible one here already sufficient to yield satisfactory analyses. \newline
Nonetheless, one may of course choose more sophisticated methods for the preparation of the data. We mention here only two possibilities: \newpage

\begin{itemize}
 \item[(i)] \underline{Model-dependent shifting:} \newline \newline
 We suppose here that the datapoint of a particular observable $\Omega_{\mathrm{exp.}}$ is to be shifted from its original energy $E_{\gamma}$ to a closeby energy $E_{\gamma}^{\prime}$ dictated by the overall binning for the TPWA. Provided that the distance between energies is small, one can drop all but the linear terms in the Taylor-series (see \cite{LefterisTalk})
 \begin{equation}
  \Omega_{\mathrm{exp.}} \left(E_{\gamma}^{\prime}, \cos (\theta)\right) = \Omega_{\mathrm{exp.}} \left(E_{\gamma}, \cos (\theta)\right) + \frac{\partial \Omega_{\mathrm{model}} \left(E_{\gamma}, \cos (\theta)\right)}{\partial E_{\gamma}} \left( E_{\gamma}^{\prime} - E_{\gamma}  \right) + \ldots \mathrm{,} \label{eq:ModelDependentShiftTaylorExpansion}
 \end{equation}
 assuming $\cos \theta$ to take some constant value. The estimate for the first derivative of $\Omega$ with respect to energy can of course not be extracted reliably from the data. However, as indicated in equation (\ref{eq:ModelDependentShiftTaylorExpansion}), it can be taken from a suitable model. Thus, the proposed method can be capable of correcting small systematic errors introduced by the simple method of kinematically closest points, but does this at the price of making the analysis model-dependent from the start. Result obtained from such model-dependent shifts have, for instance, been presented at a recent conference \cite{LefterisTalk}. 
 \item[(ii)] \underline{Two-dimensional interpolation:} \newline \newline
 It is also possible to use multi-variate interpolation techniques (for instance splines) and apply them to the $2$-dimensional problem posed by a particular dataset in $(E_{\gamma},\theta)$-space. One may try to interpolate the data $\Omega^{\alpha}$, as well as the values for the error-interval $\Omega^{\alpha} + \Delta \Omega^{\alpha}$ and $\Omega^{\alpha} - \Delta \Omega^{\alpha}$. The resulting two-dimensional function can then be sampled freely for the preparation of the TPWA, in whatever energy- and angle-binning one desires. \newline
 Such procedures can result in more smoothed-out results for the multipoles, but they require a lot of sophisticated implementation before the actual analysis has even started. Furthermore, one has to remember that the final result is still a fit to the original data, thus requiring a consistent definition for a test of the goodness of fit, such as $\chi^{2}$. The re-binning of data using spline-interpolations is employed, for instance, by members of the Mainz/Tuzla PWA-collaboration \cite{MainzTuzlaCollaboration}.
\end{itemize}

\textbf{Determination of an $\ell_{\mathrm{max}}$-estimate from angular distributions of the profile functions} \newline

Before continuing with the extraction of multipoles, a useful preparatory step consists of estimating a suitable truncation order $\ell_{\mathrm{max}}$ just from the angular distributions of the profile functions. This method has already been outlined and applied in detail in chapter \ref{chap:LFits}. Here, we only briefly summarize the results for the data in the $\Delta$-region. The suitable parametrizations in terms of associated Legendre polynomials $P_{\ell}^{m}$, to be fitted to the angular distributions, have been elaborated in both the introduction \ref{sec:CompExpsTPWA} and in section \ref{sec:LFitsPaper}. Here, we quote the necessary expressions for quick reference

\begin{align}
 \sigma_{0} (W,\theta) &= \frac{q}{k} \sum_{k=0}^{2 \ell_{\mathrm{max}}} \left( a_{L} \right)_{k}^{\sigma_{0}} P_{k} (\cos \theta) \mathrm{,} \label{eq:DCSAngDistChapter4} \\
 \check{\Sigma} (W,\theta) &= \frac{q}{k} \sum_{k=2}^{2 \ell_{\mathrm{max}}} \left( a_{L} \right)_{k}^{\check{\Sigma}} P^{2}_{k} (\cos \theta) \mathrm{,} \label{eq:SigmaAngDistChapter4}
\end{align}
\begin{figure}[h]
 \centering
 \begin{overpic}[width=0.475\textwidth]{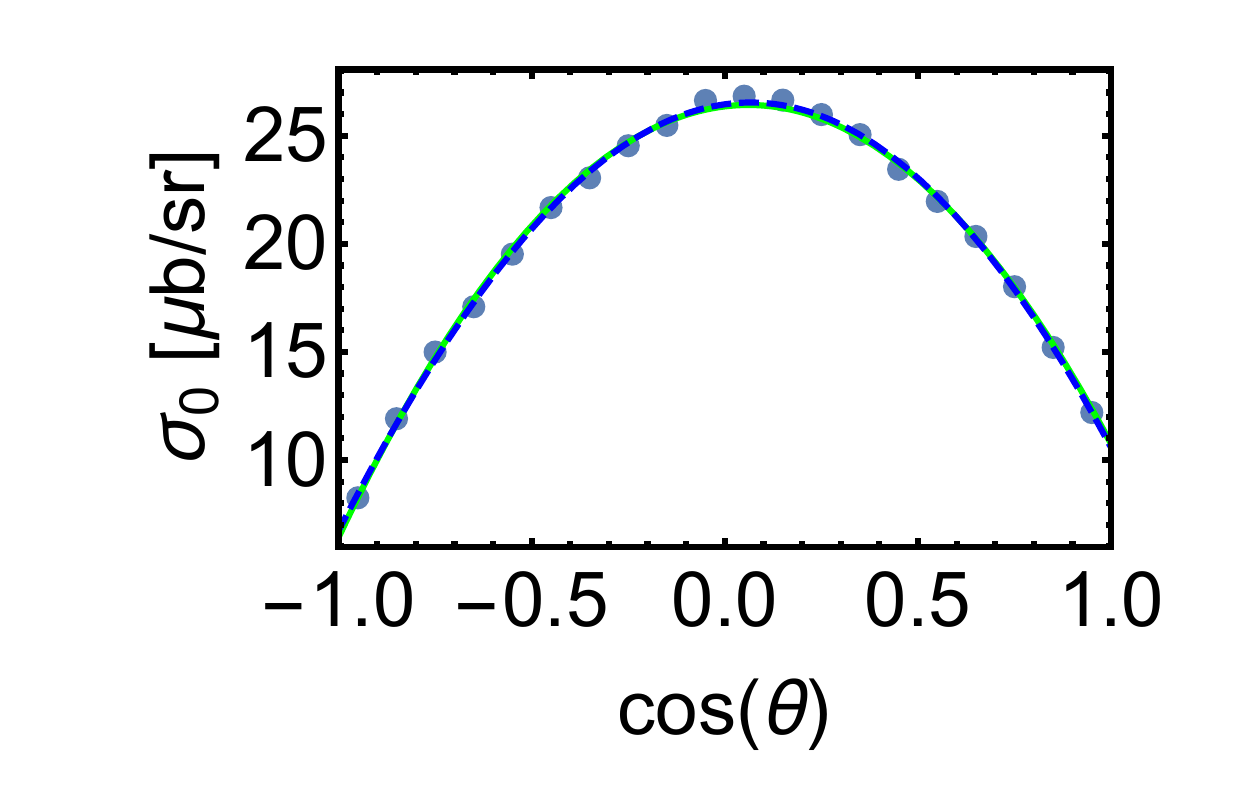}
 \put(81.5,68){\begin{Large}$E_{\gamma} = 350. \hspace*{2pt} \mathrm{MeV}$\end{Large}}
 \end{overpic}
 \begin{overpic}[width=0.475\textwidth]{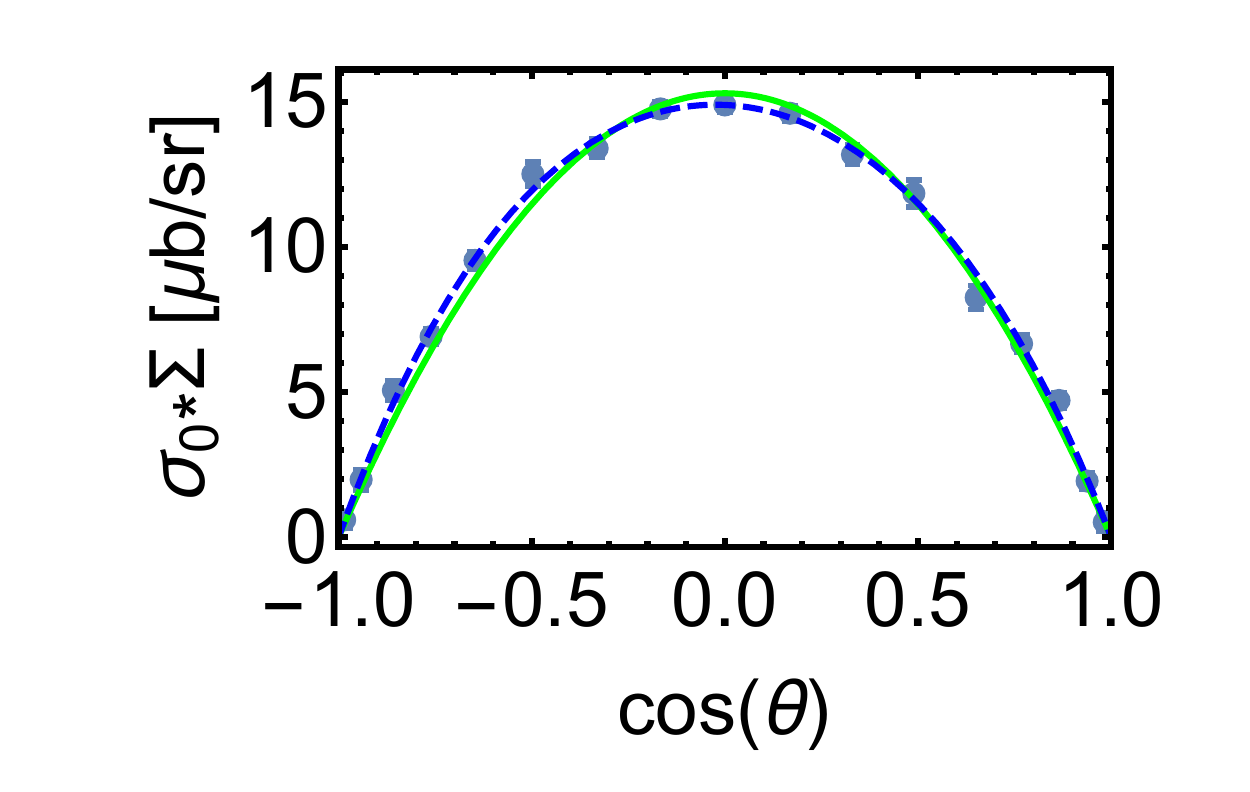}
 \end{overpic} \\
 \begin{overpic}[width=0.475\textwidth]{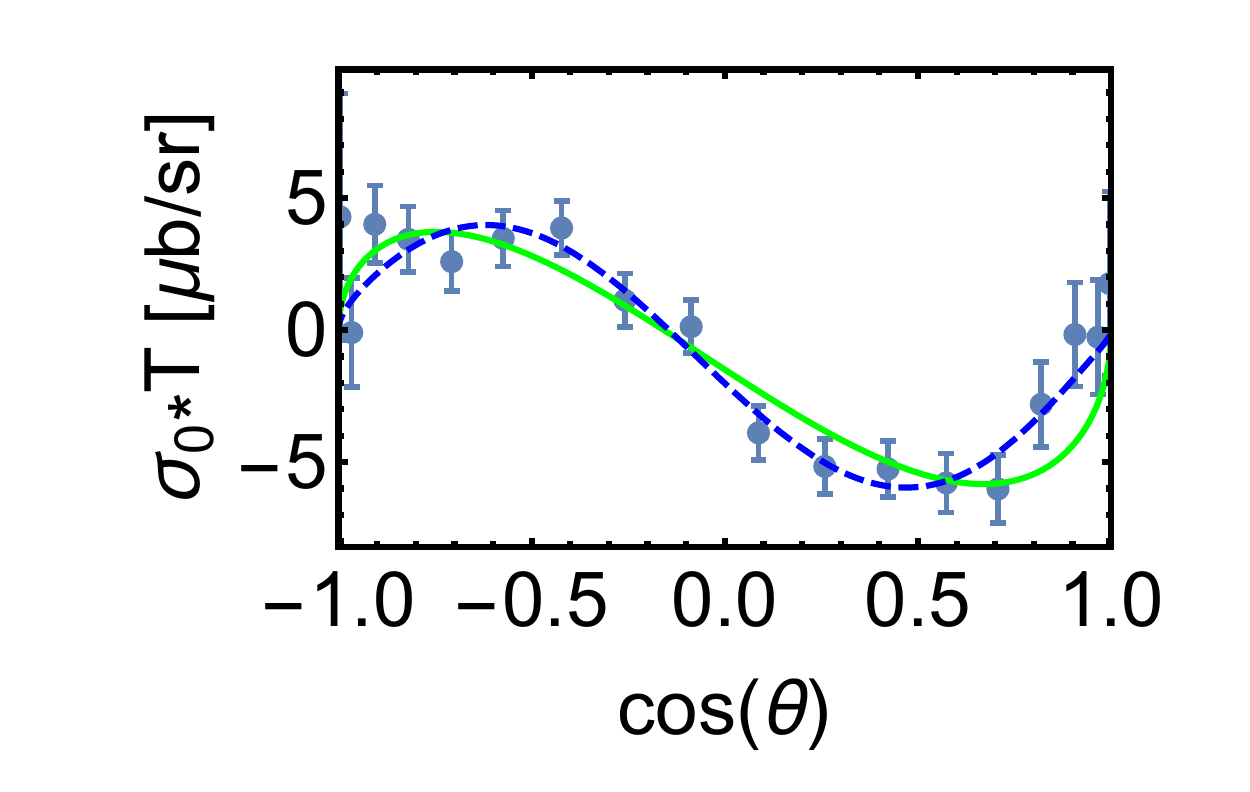}
 \end{overpic}
 \begin{overpic}[width=0.475\textwidth]{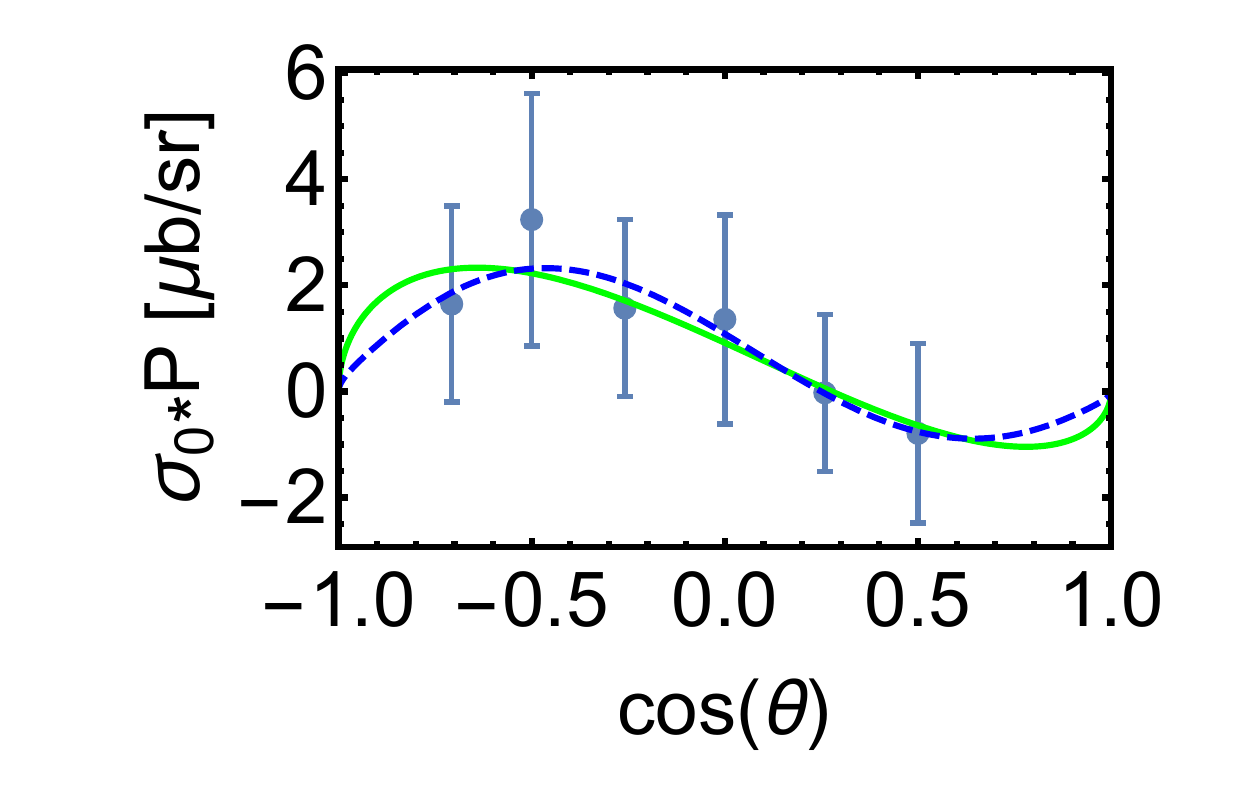}
 \end{overpic} \\
 \begin{overpic}[width=0.475\textwidth]{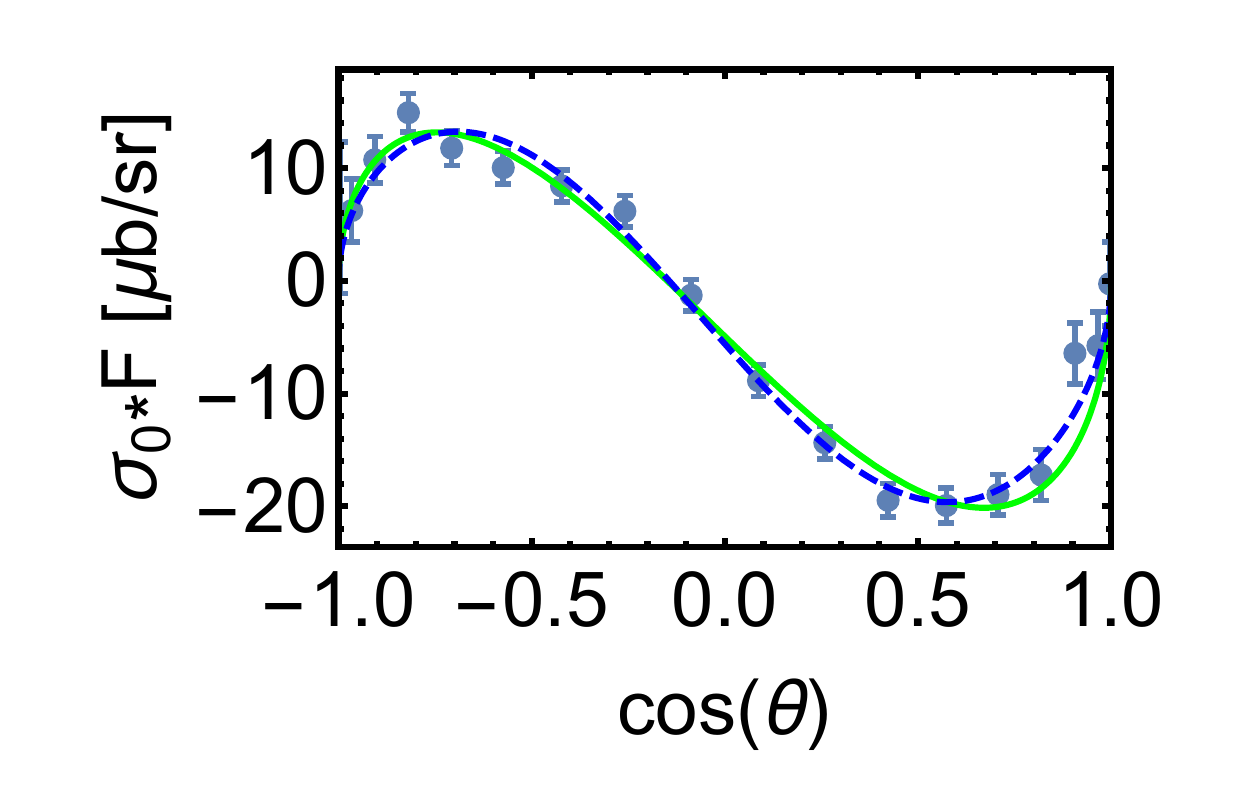}
 \end{overpic}
  \caption[Angular distributions of profile functions for the five polarization-datasets analyzed in the $\Delta$-resonance region. Data are plotted at a particular energy, $E_{\gamma} = 350. \hspace*{1pt} \mathrm{MeV}$.]{Shown here are angular distributions for the differential cross section $\sigma_{0}$ and the profile functions $\check{\Sigma} = \sigma_{0} \Sigma$, $\check{T} = \sigma_{0} T$, $\check{P} = \sigma_{0} P$ and $\check{F} = \sigma_{0} F$ of all polarization observables fitted in the $\Delta$-region. Data are shown at a particular energy, $E_{\gamma} = 350 \hspace*{1pt} \mathrm{MeV}$. Data coincide at this energy as a result of the kinematic re-binning elaborated in the main text. \newline
  Fits to the angular distributions can be seen as well, employing the truncation angular momenta $\ell_{\mathrm{max}} = 1$ (green solid line) and $\ell_{\mathrm{max}} = 2$ (blue dashed line). The fit-parametrizations follow the modulations quoted in equatons (\ref{eq:DCSAngDistChapter4}) to (\ref{eq:FAngDistChapter4}).}
 \label{fig:DeltaRegionObsFitAngDistPlots}
\end{figure}

\clearpage

\begin{figure}[h]
 \centering
 \begin{overpic}[width=0.475\textwidth]{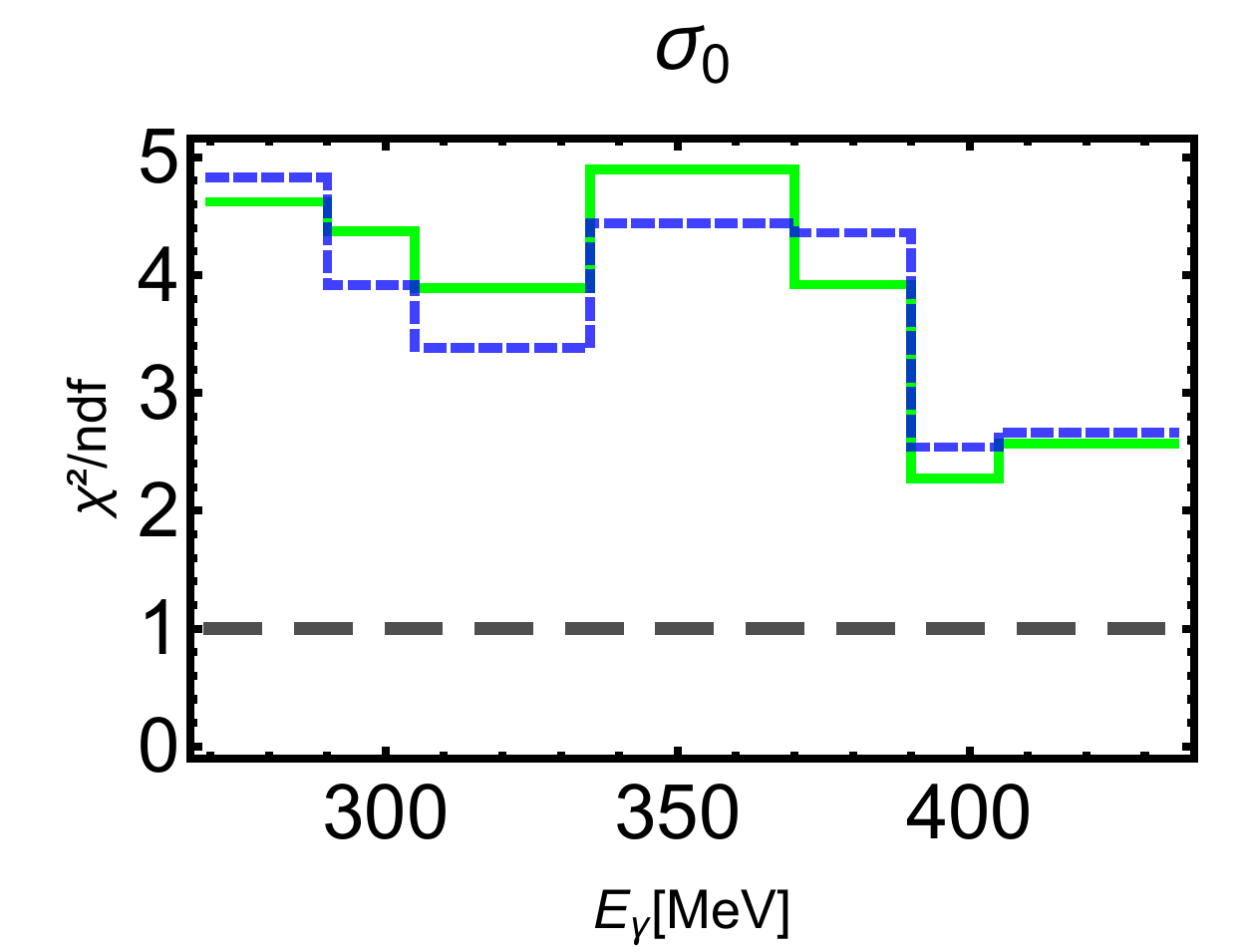}
 \end{overpic}
 \begin{overpic}[width=0.475\textwidth]{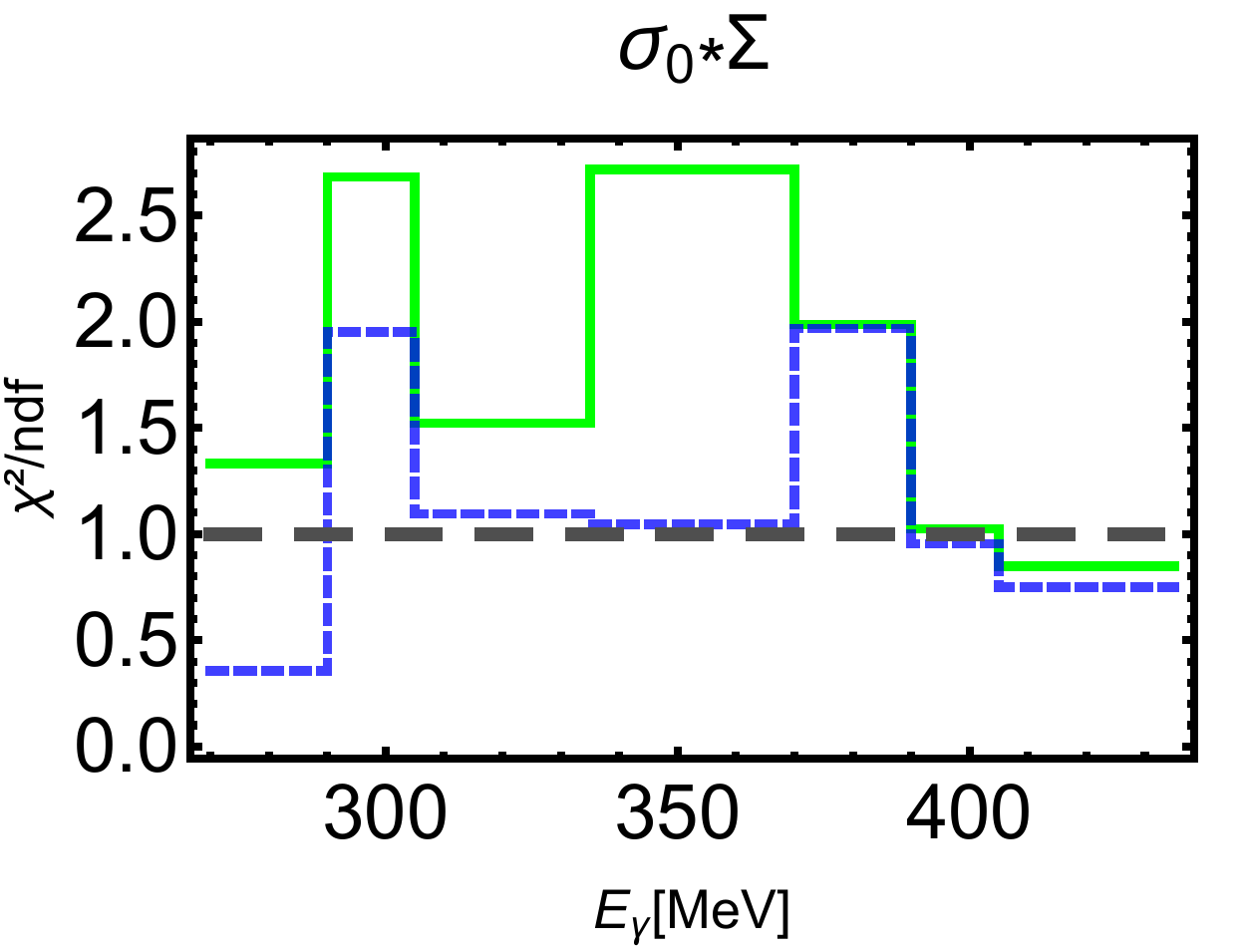}
 \end{overpic} \\
 \begin{overpic}[width=0.475\textwidth]{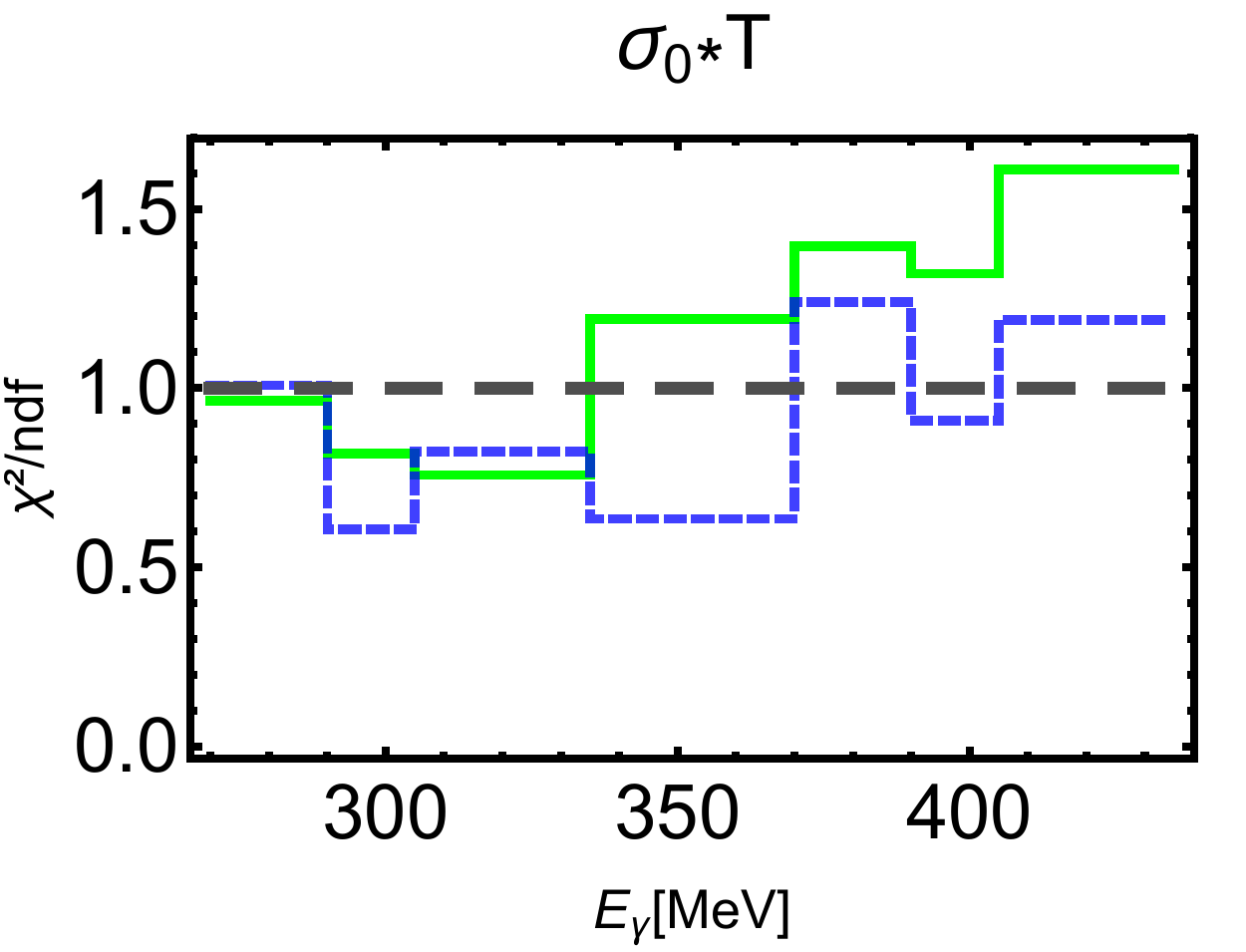}
 \end{overpic}
 \begin{overpic}[width=0.475\textwidth]{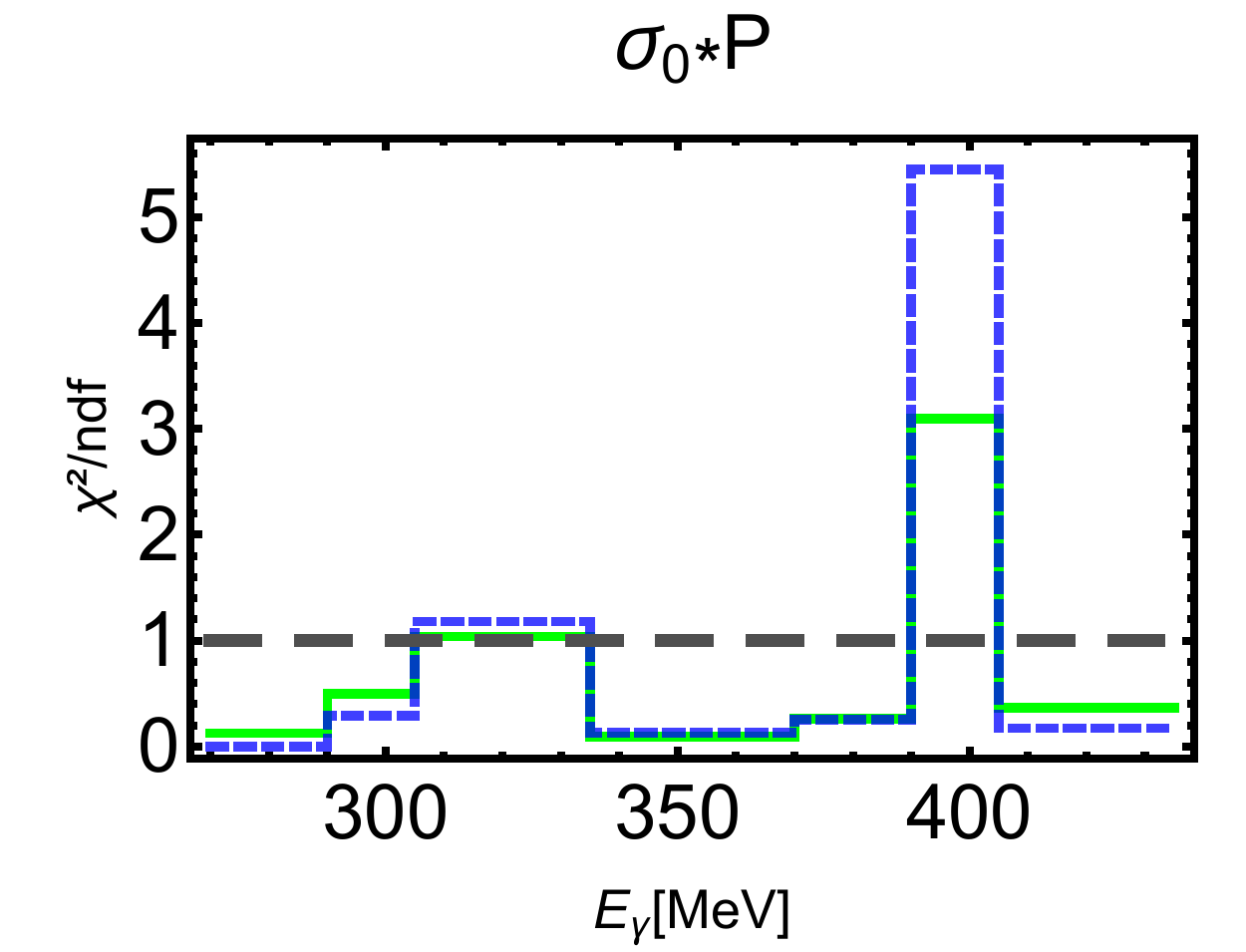}
 \end{overpic} \\
 \begin{overpic}[width=0.475\textwidth]{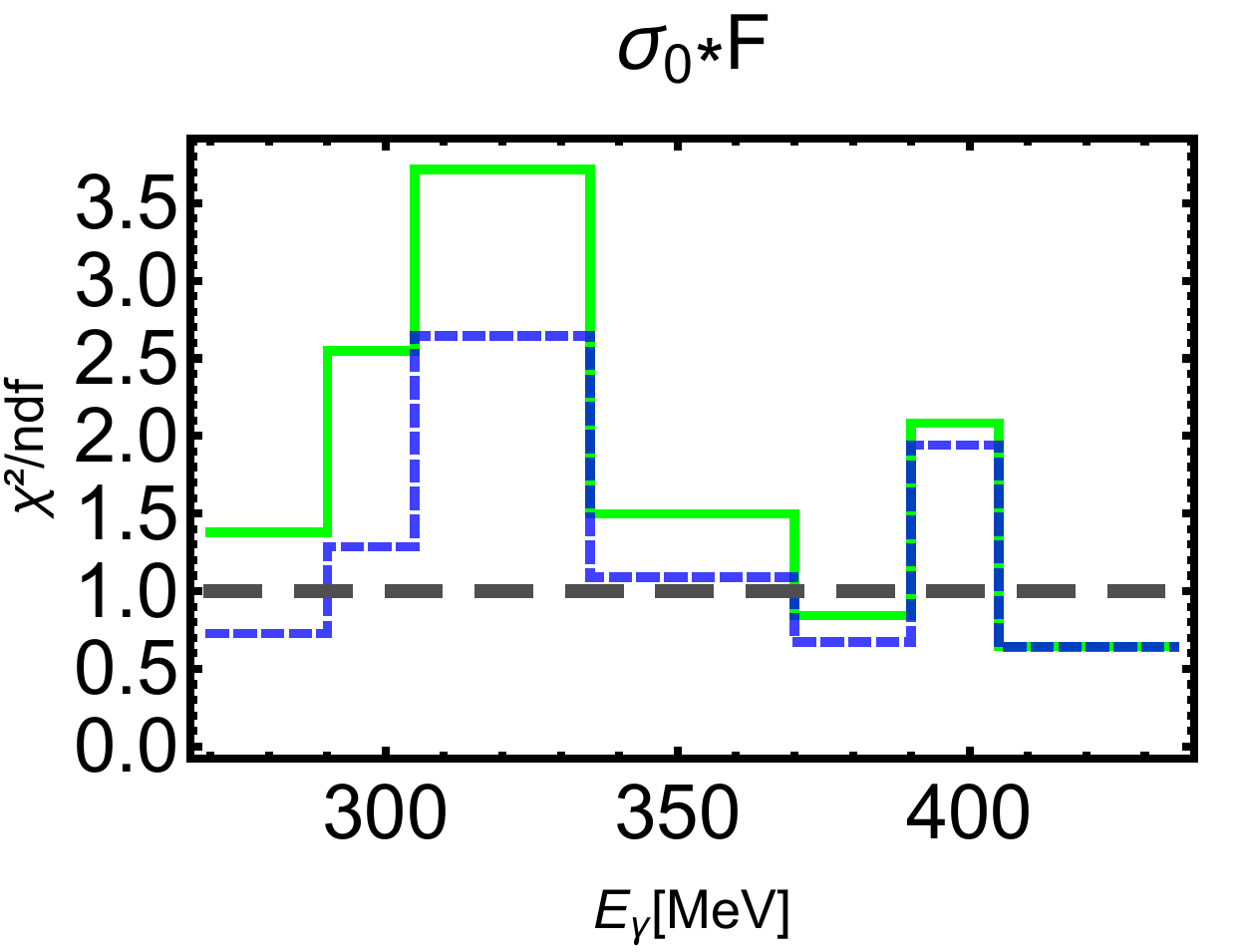}
 \end{overpic}
  \caption[The $\chi^{2}/\mathrm{ndf}$ resulting from fits to the angular distributions of the differential cross section and the four profile functions in the $\Delta$-resonance region, plotted vs. energy.]{The plots show the $\chi^{2}/\mathrm{ndf}$ resulting from fits of the parametrizations (\ref{eq:DCSAngDistChapter4}) to (\ref{eq:FAngDistChapter4}) to the angular distributions of the differential cross section and the four profile functions. Values are shown for all $7$ energy bins dictated by the $P$-dataset, on which the TPWA will be performed. Results are shown for a truncation at $\ell_{\mathrm{max}} = 1$ (green solid line) and $\ell_{\mathrm{max}} = 2$ (blue dashed line). The optimal value $\chi^{2}/\mathrm{ndf} = 1$ is indicated by a grey dashed horizontal line.\newline
  Note that in the first energy bin of the $P$-dataset, the number of degrees of freedom in a truncation at $\ell_{\mathrm{max}} = 2$ equals zero. Therefore, the value of $\chi^{2}/\mathrm{ndf}$ has been put to zero in the plot. However, it is clear that this first energy-bin cannot support a multipole-analysis for all truncation orders larger than $\ell_{\mathrm{max}} = 1$.}
 \label{fig:DeltaRegionChisquareLmaxPlots}
\end{figure}

\clearpage

\begin{align}
 \check{T} (W,\theta) &= \frac{q}{k} \sum_{k=1}^{2 \ell_{\mathrm{max}}} \left( a_{L} \right)_{k}^{\check{T}} P^{1}_{k} (\cos \theta) \mathrm{,} \label{eq:TAngDistChapter4} \\
 \check{P} (W,\theta) &= \frac{q}{k} \sum_{k=1}^{2 \ell_{\mathrm{max}}} \left( a_{L} \right)_{k}^{\check{P}} P^{1}_{k} (\cos \theta) \mathrm{,} \label{eq:PAngDistChapter4} \\
 \check{F} (W,\theta) &= \frac{q}{k} \sum_{k=1}^{2 \ell_{\mathrm{max}}} \left( a_{L} \right)_{k}^{\check{F}} P^{1}_{k} (\cos \theta) \mathrm{.} \label{eq:FAngDistChapter4}
\end{align}

\begin{figure}[h]
 \centering
 \begin{overpic}[width=0.325\textwidth]{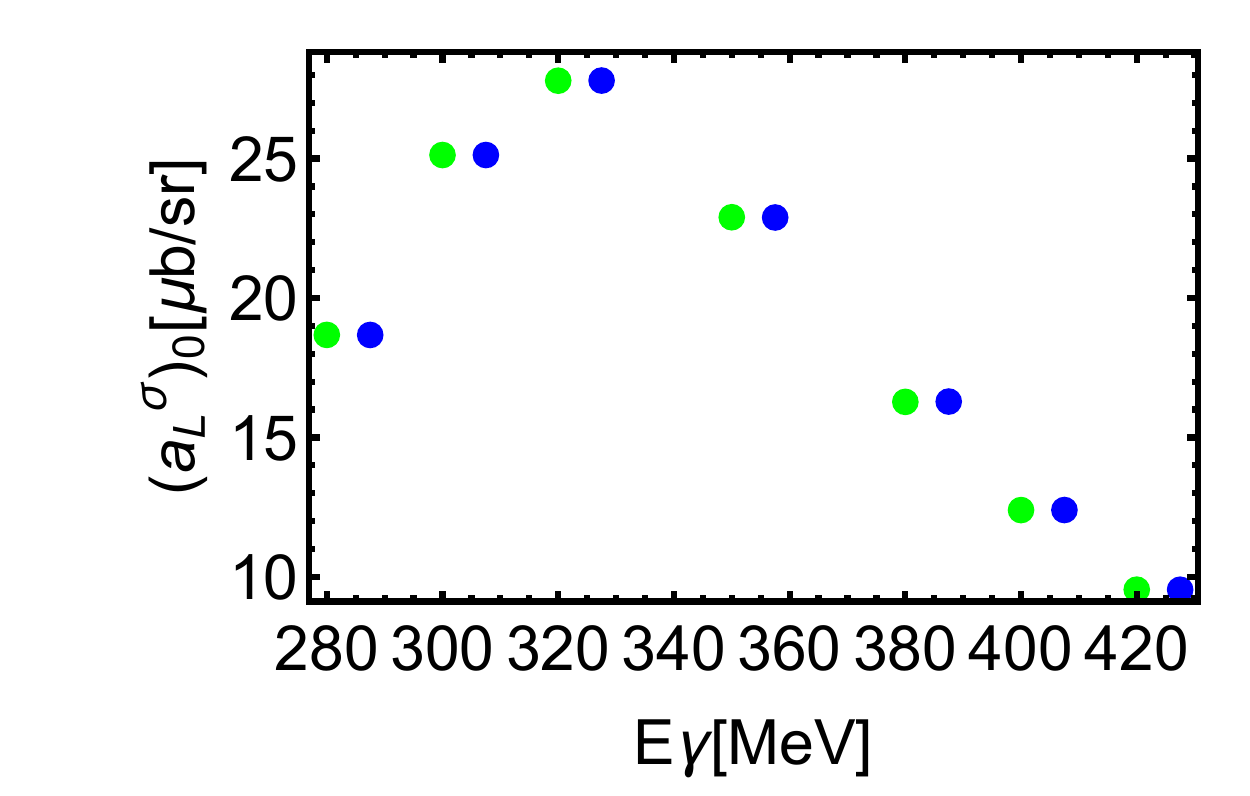}
 \end{overpic}
 \begin{overpic}[width=0.325\textwidth]{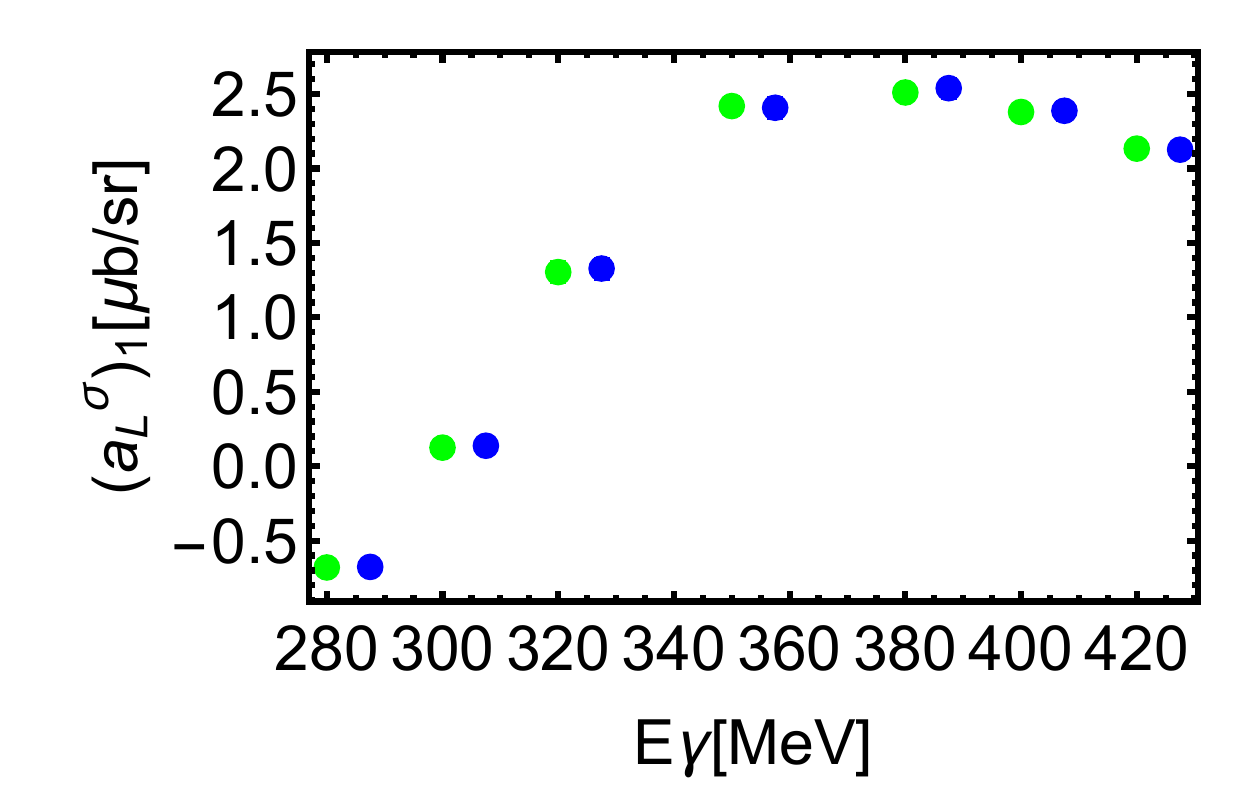}
 \end{overpic}
 \begin{overpic}[width=0.325\textwidth]{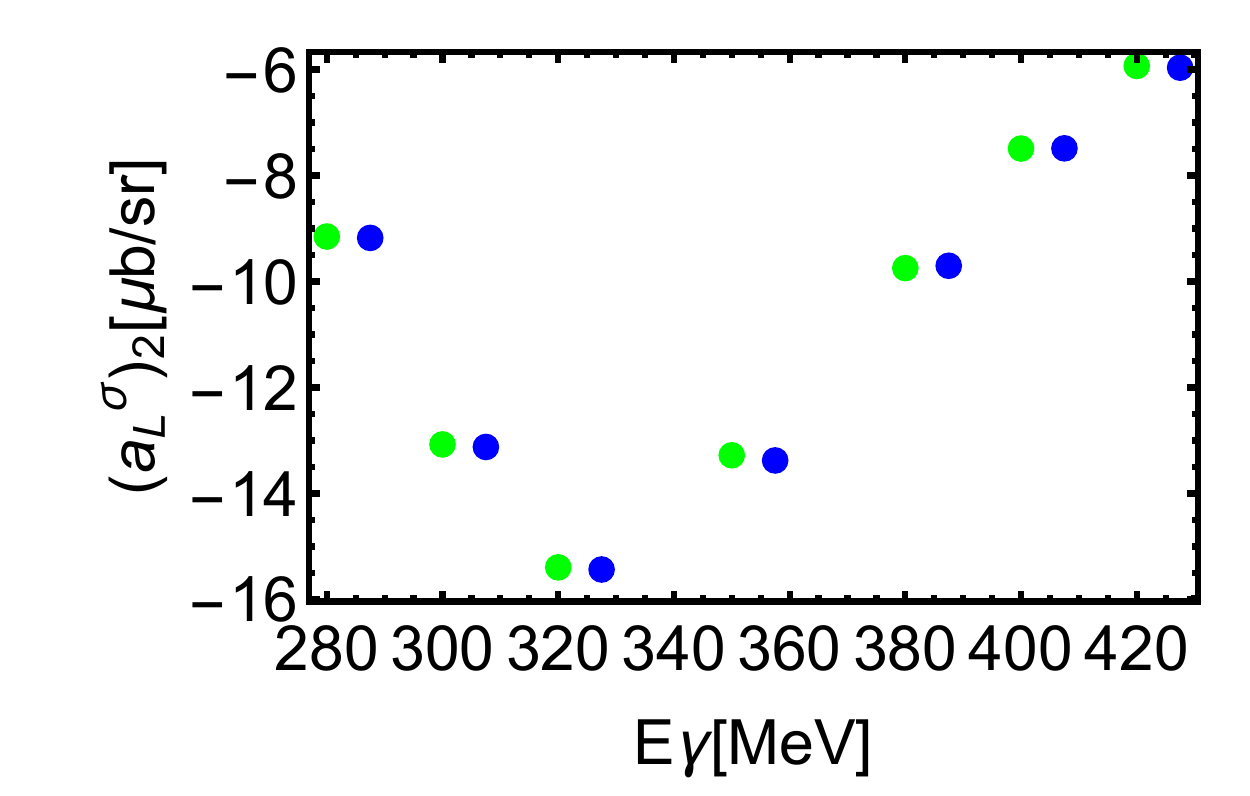}
 \end{overpic} \\
 \begin{overpic}[width=0.325\textwidth]{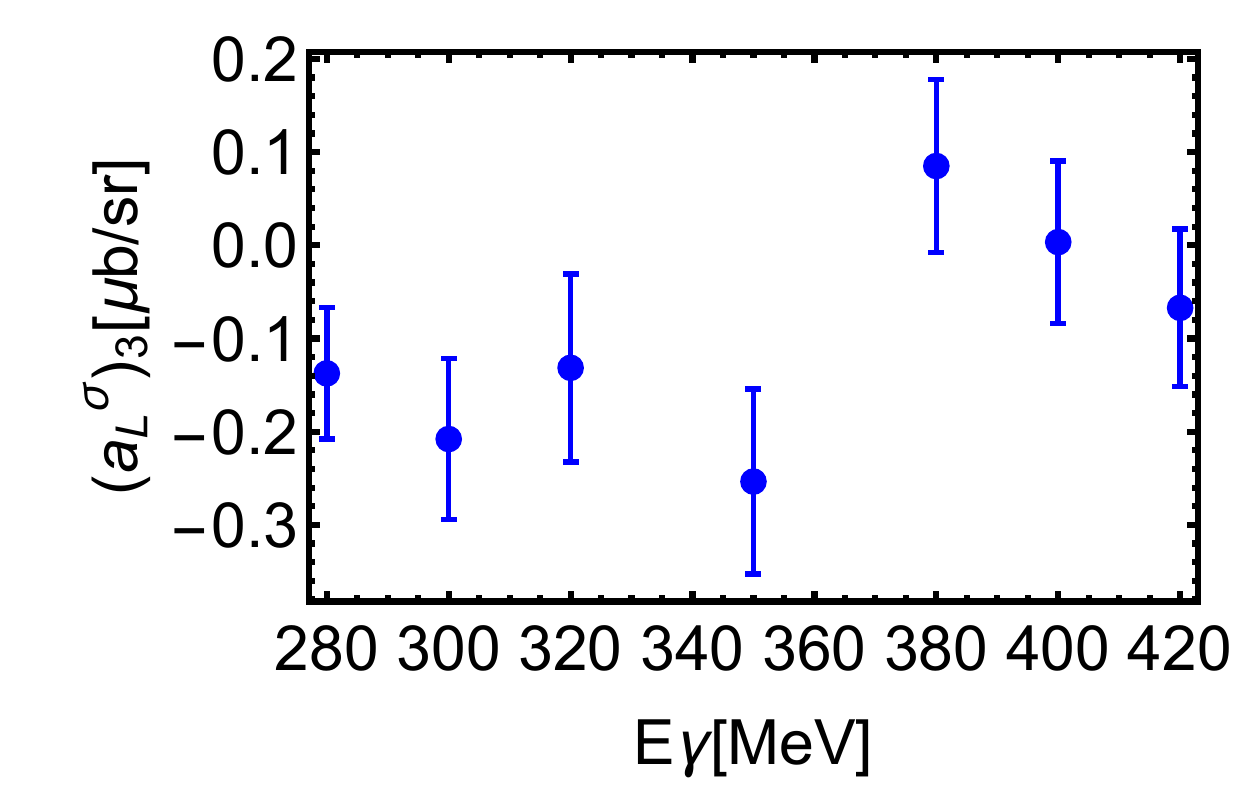}
 \end{overpic}
 \begin{overpic}[width=0.325\textwidth]{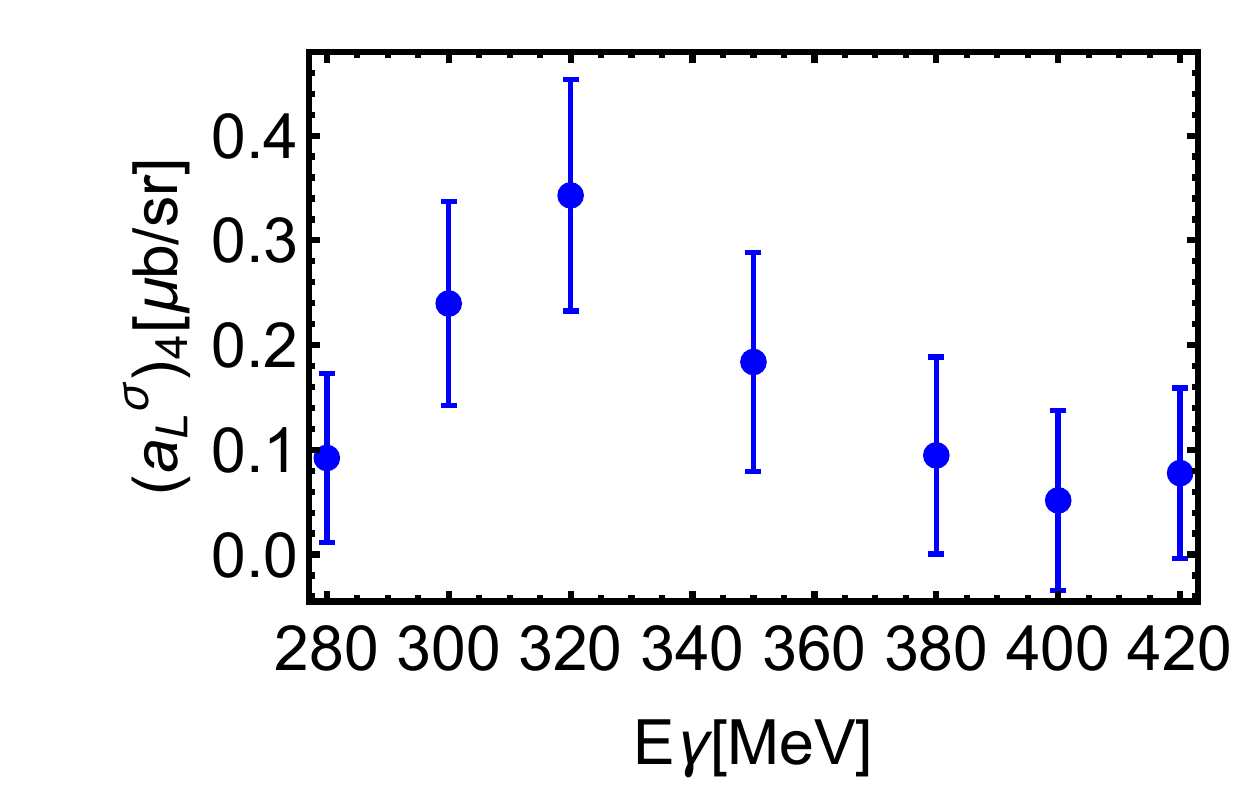}
 \end{overpic} \\
 \begin{overpic}[width=0.325\textwidth]{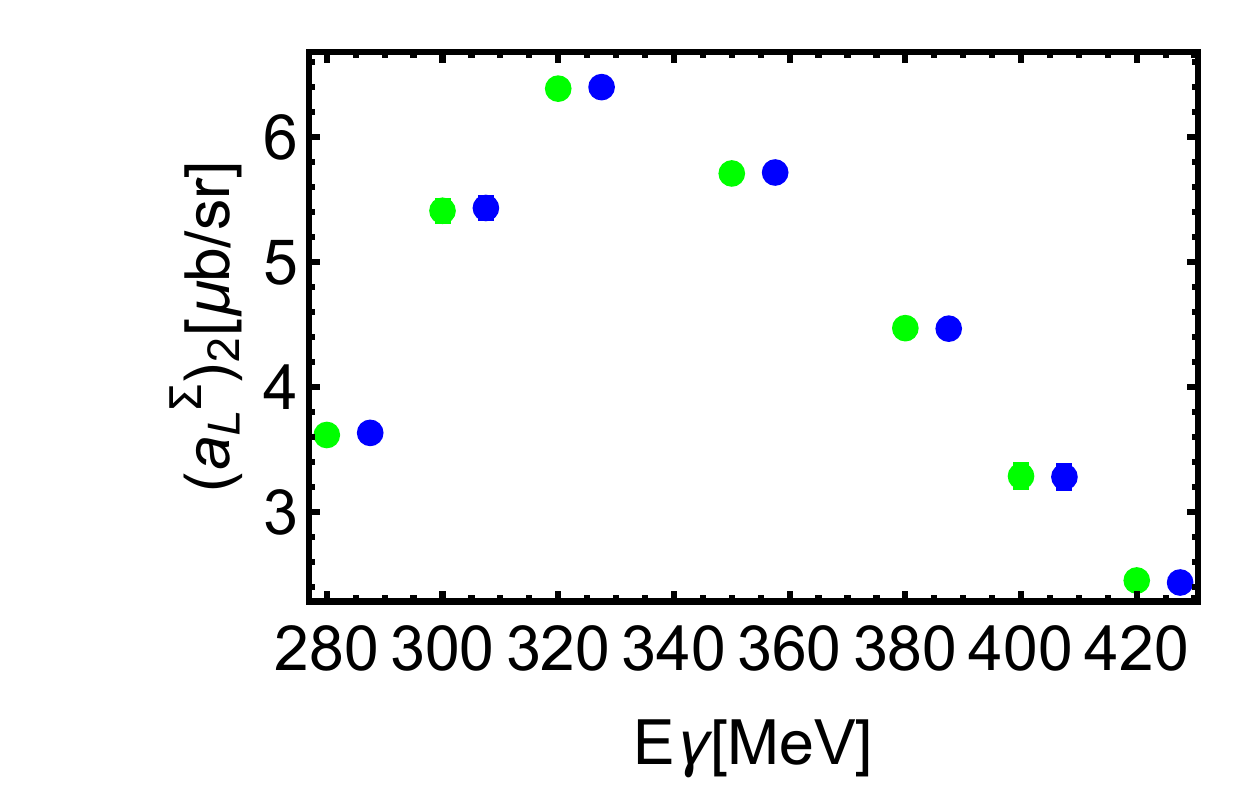}
 \end{overpic}
 \begin{overpic}[width=0.325\textwidth]{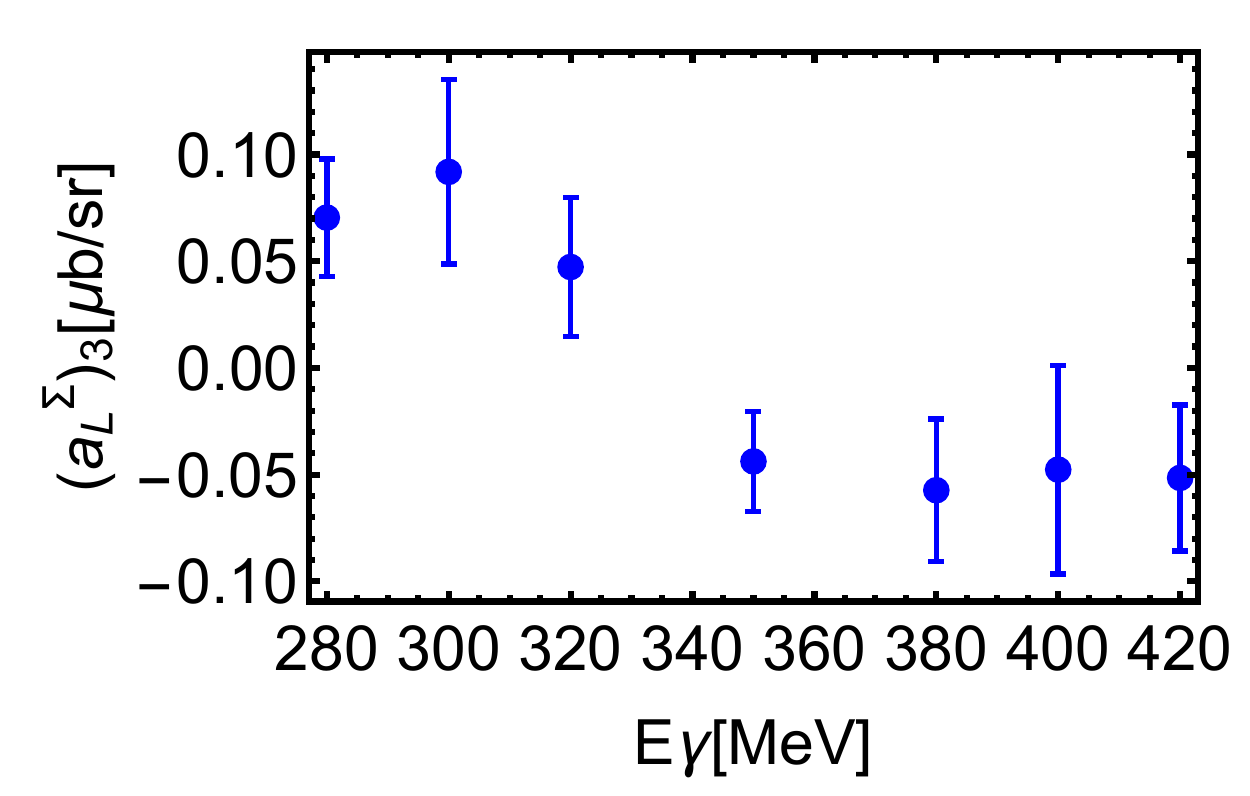}
 \end{overpic}
  \begin{overpic}[width=0.325\textwidth]{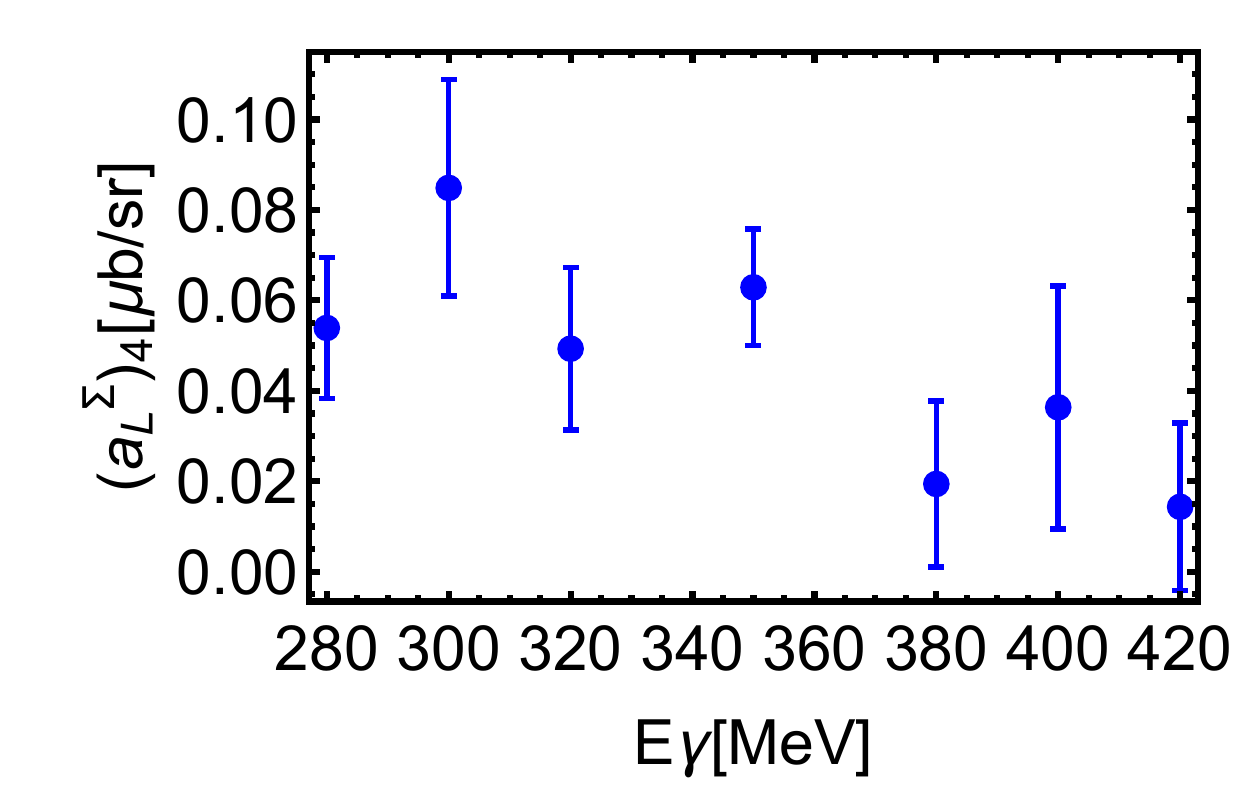}
 \end{overpic} \\
 \begin{overpic}[width=0.325\textwidth]{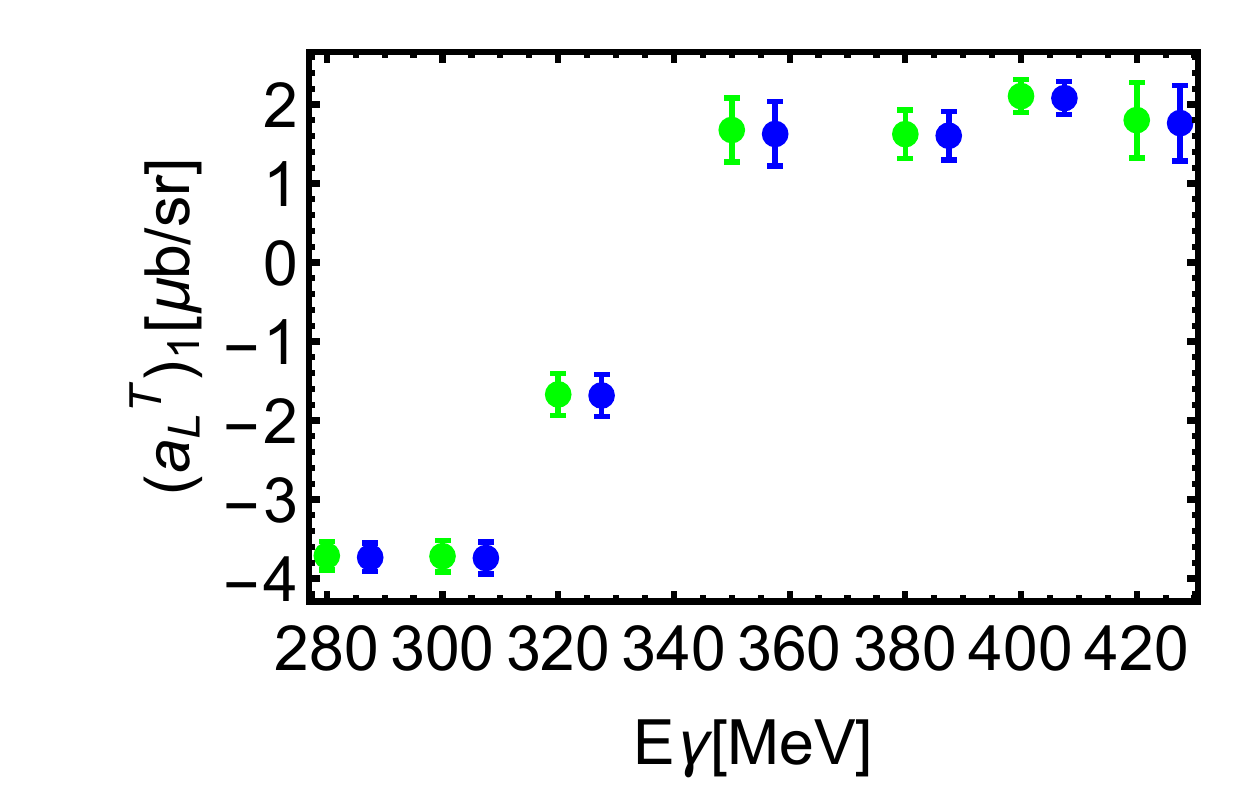}
 \end{overpic}
  \begin{overpic}[width=0.325\textwidth]{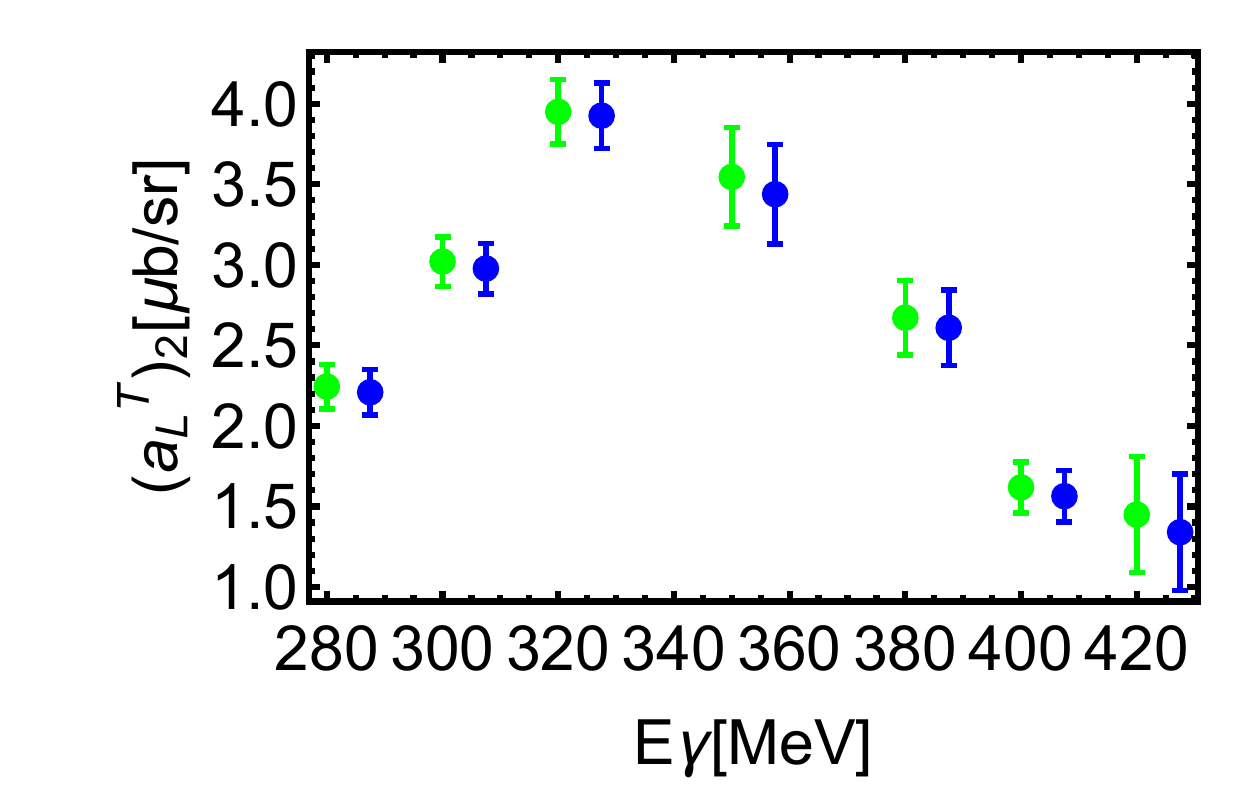}
 \end{overpic}
  \begin{overpic}[width=0.325\textwidth]{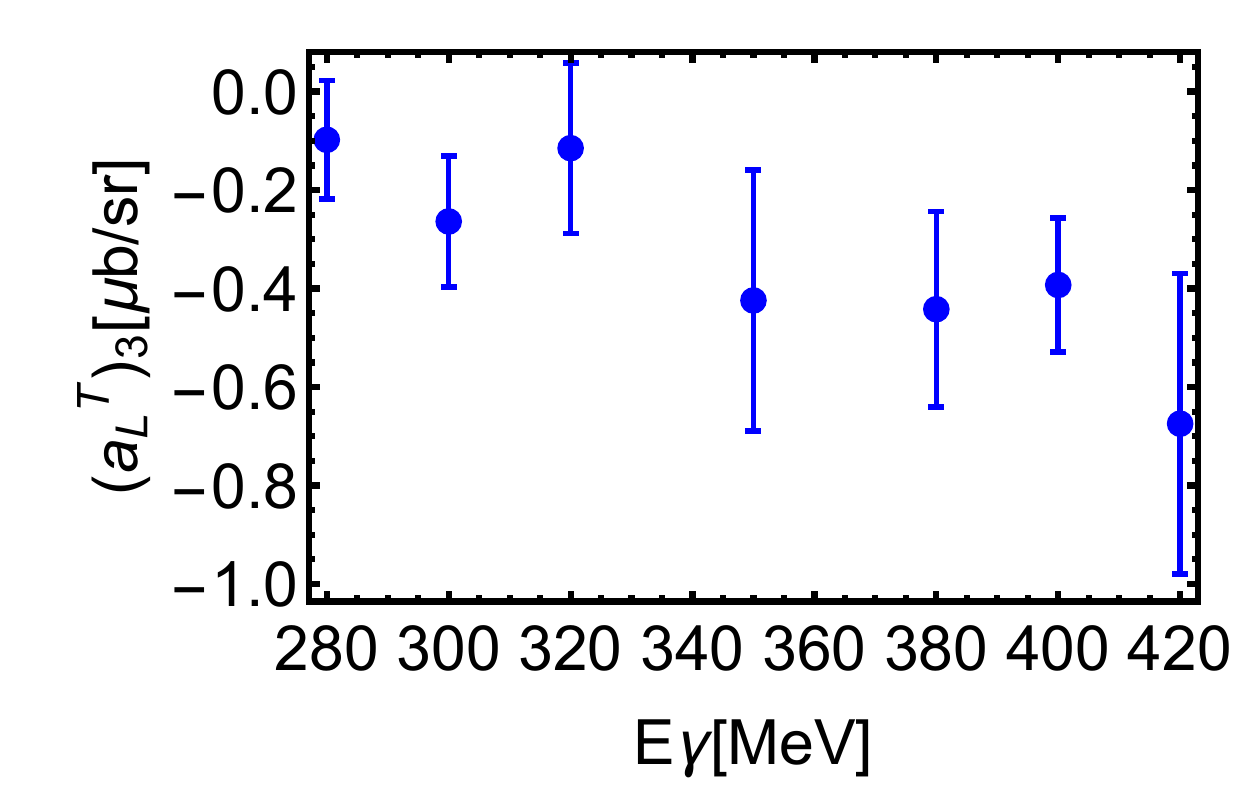}
 \end{overpic} \\
 \begin{overpic}[width=0.325\textwidth]{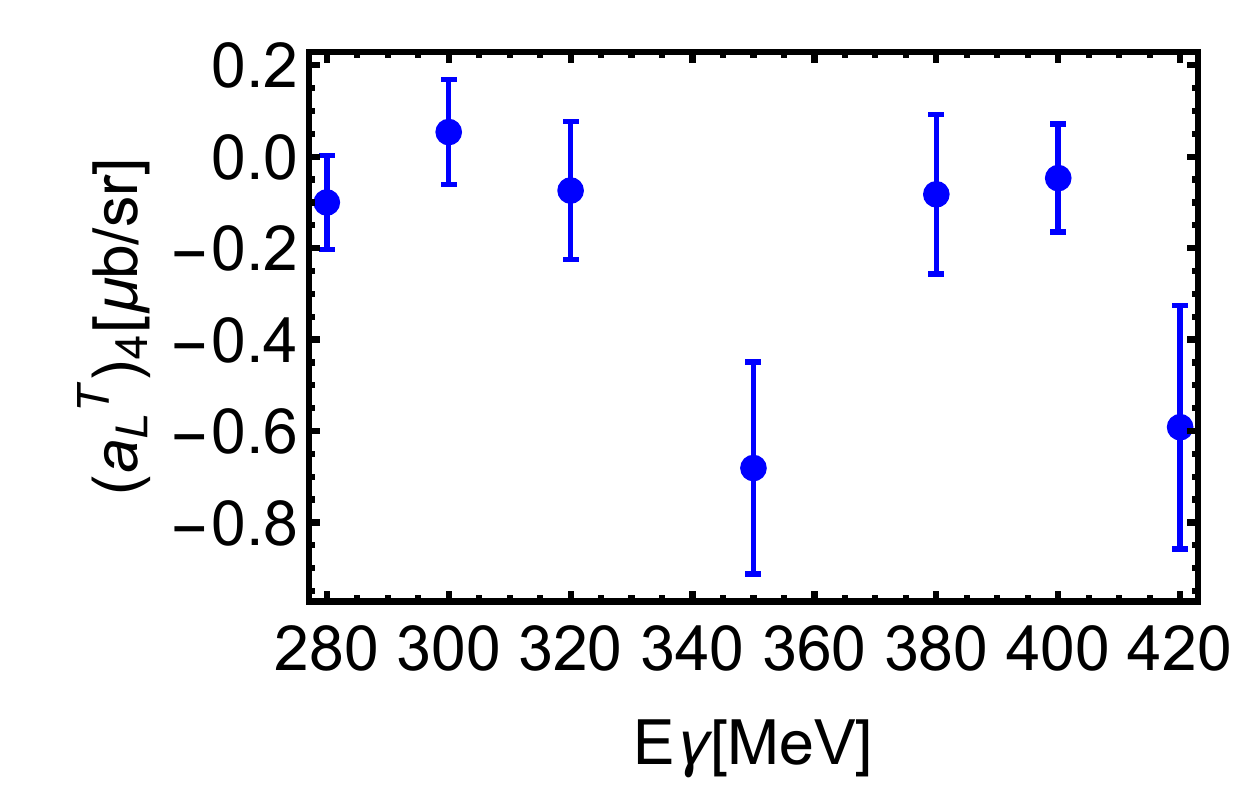}
 \end{overpic}
 \caption[Legendre-coefficients for the differential cross section $\sigma_{0}$ and the profile functions $\check{\Sigma}$ and $\check{T}$, extracted by fitting truncations at $\ell_{\mathrm{max}} = 1$ and $\ell_{\mathrm{max}} = 2$, in the $\Delta$-resonance region.]{The pictures show all Legendre-coefficients for the differential cross section $\sigma_{0}$ and the profile functions $\check{\Sigma}$ and $\check{T}$, extracted by fitting truncations at $\ell_{\mathrm{max}} = 1$ (green points) and $\ell_{\mathrm{max}} = 2$ (blue points). In comparison-plots, the results of the $D$-wave truncation have been shifted slightly to higher energies, in order to increase visibility. Errors have been extracted from the fits themselves and are, in this case, not bootstrapped.}
 \label{fig:DeltaRegionFittedLegCoeffsPlotsI}
\end{figure}

\begin{figure}[h]
 \centering
 \begin{overpic}[width=0.325\textwidth]{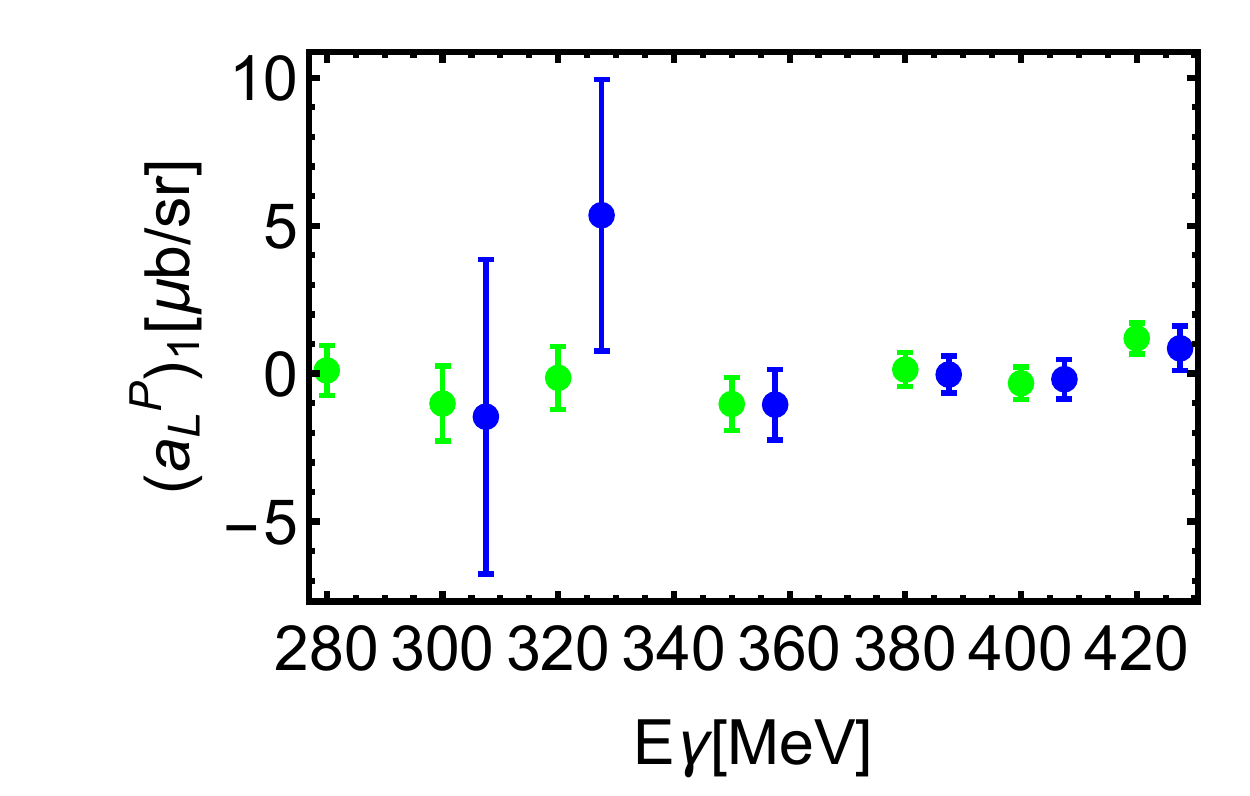}
 \end{overpic}
 \begin{overpic}[width=0.325\textwidth]{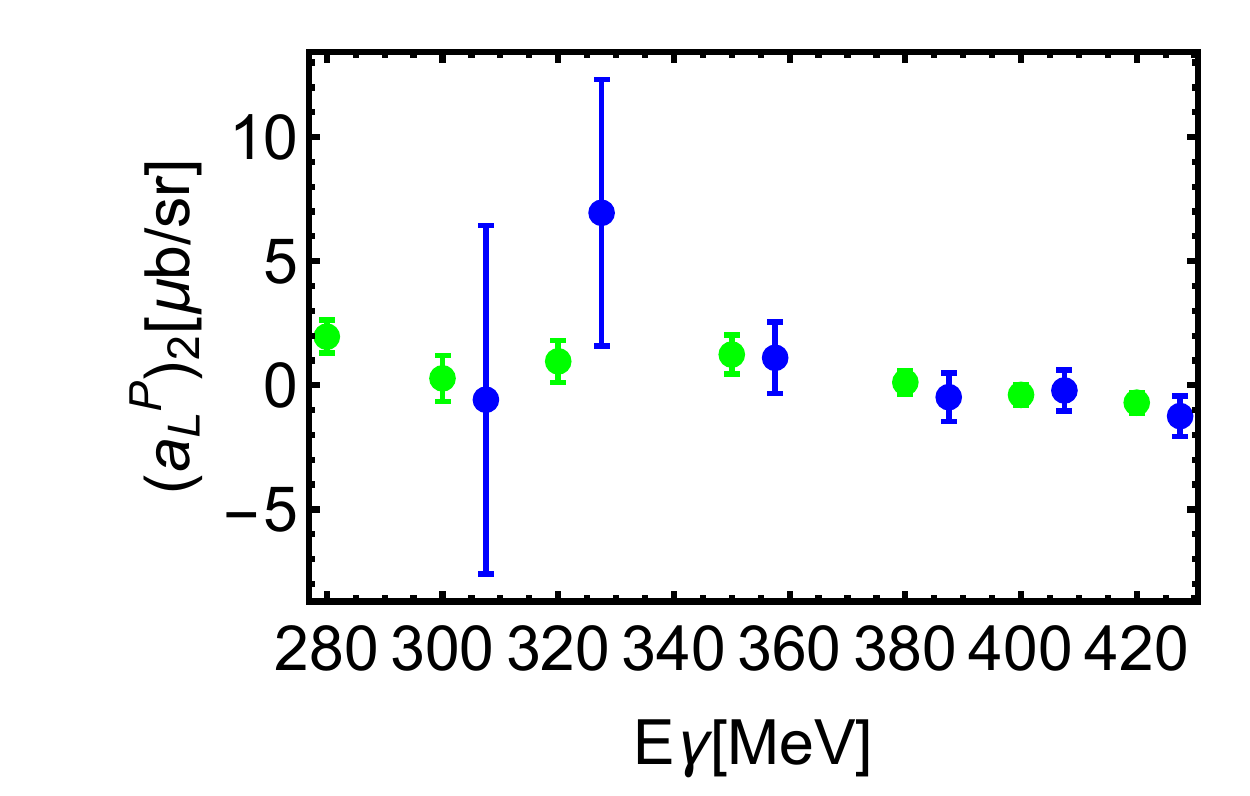}
 \end{overpic}
  \begin{overpic}[width=0.325\textwidth]{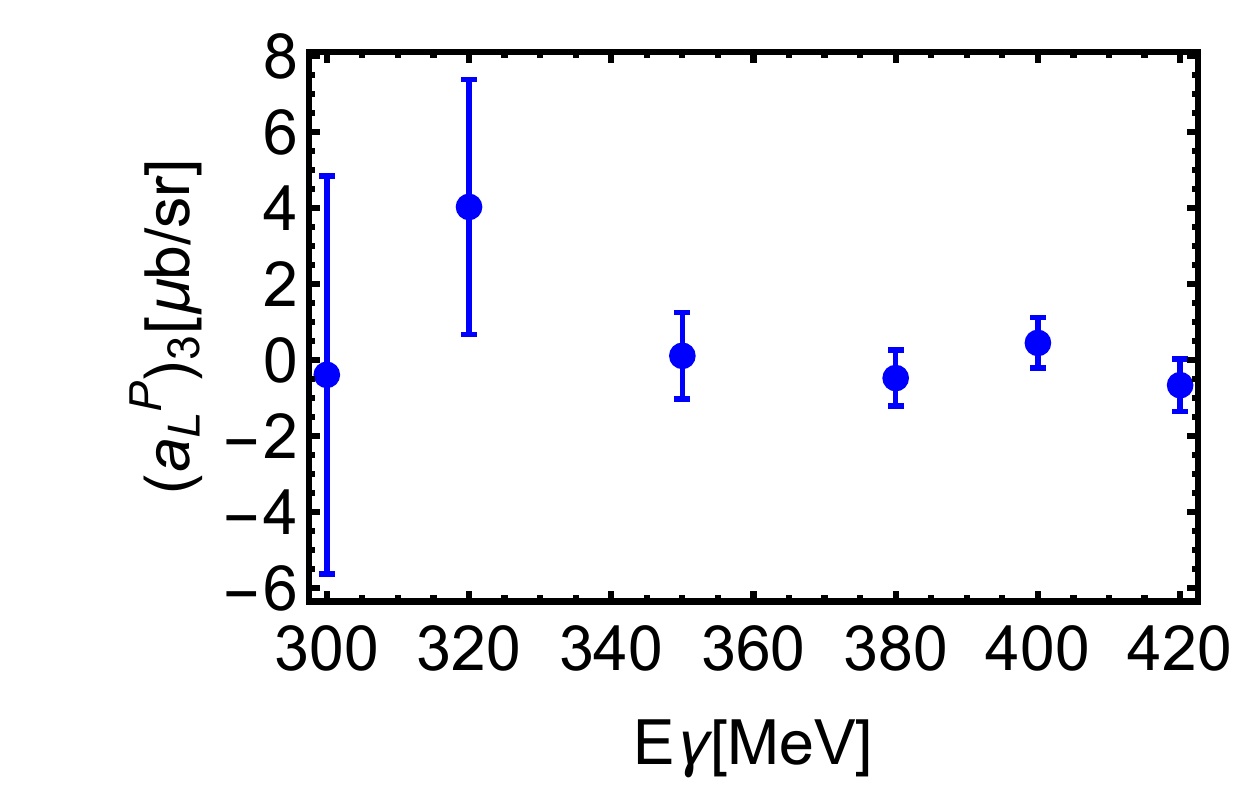}
 \end{overpic} \\
  \begin{overpic}[width=0.325\textwidth]{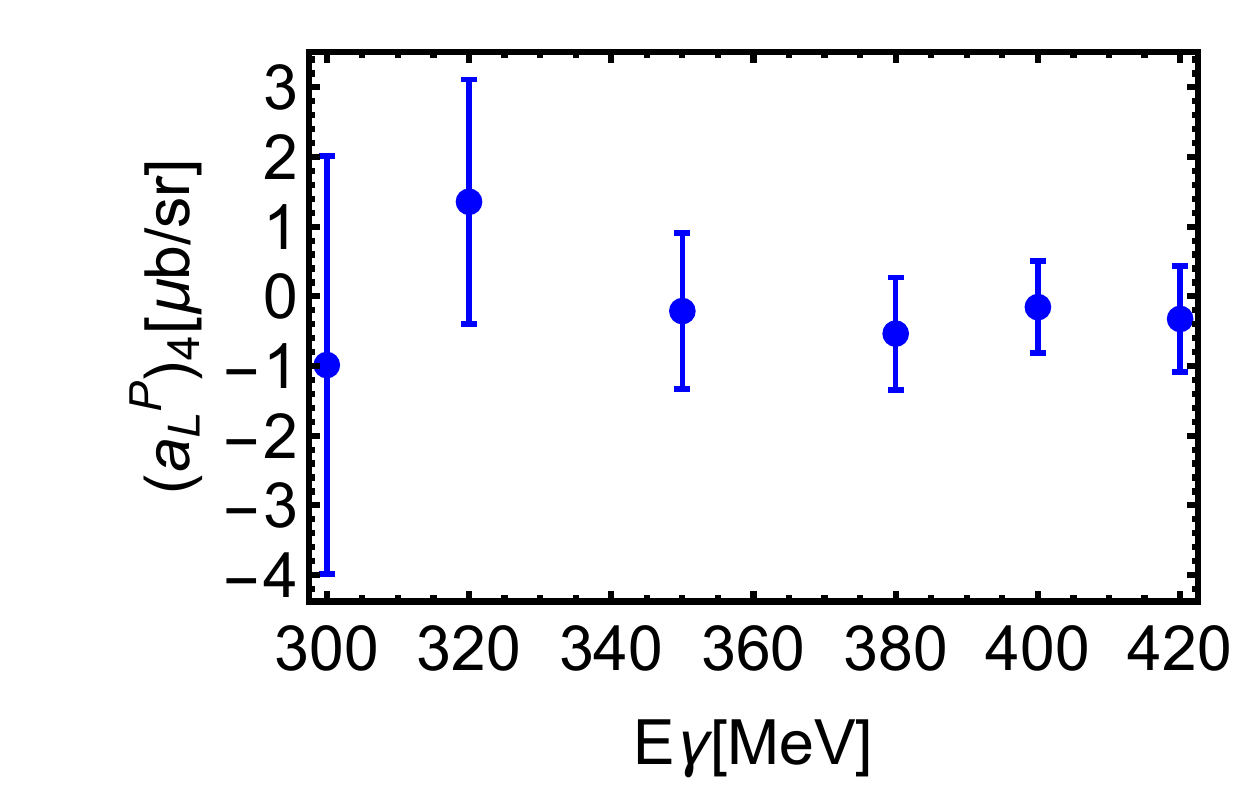}
 \end{overpic} \\
  \begin{overpic}[width=0.325\textwidth]{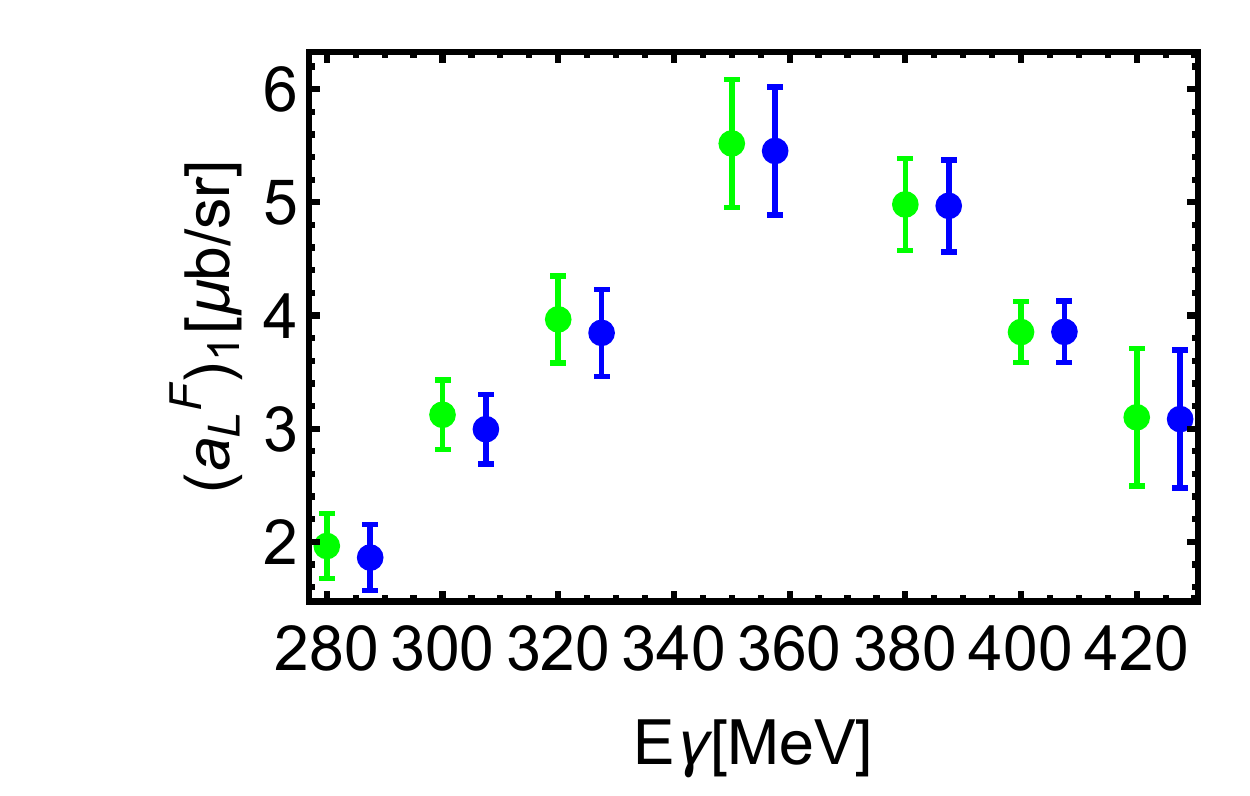}
 \end{overpic}
 \begin{overpic}[width=0.325\textwidth]{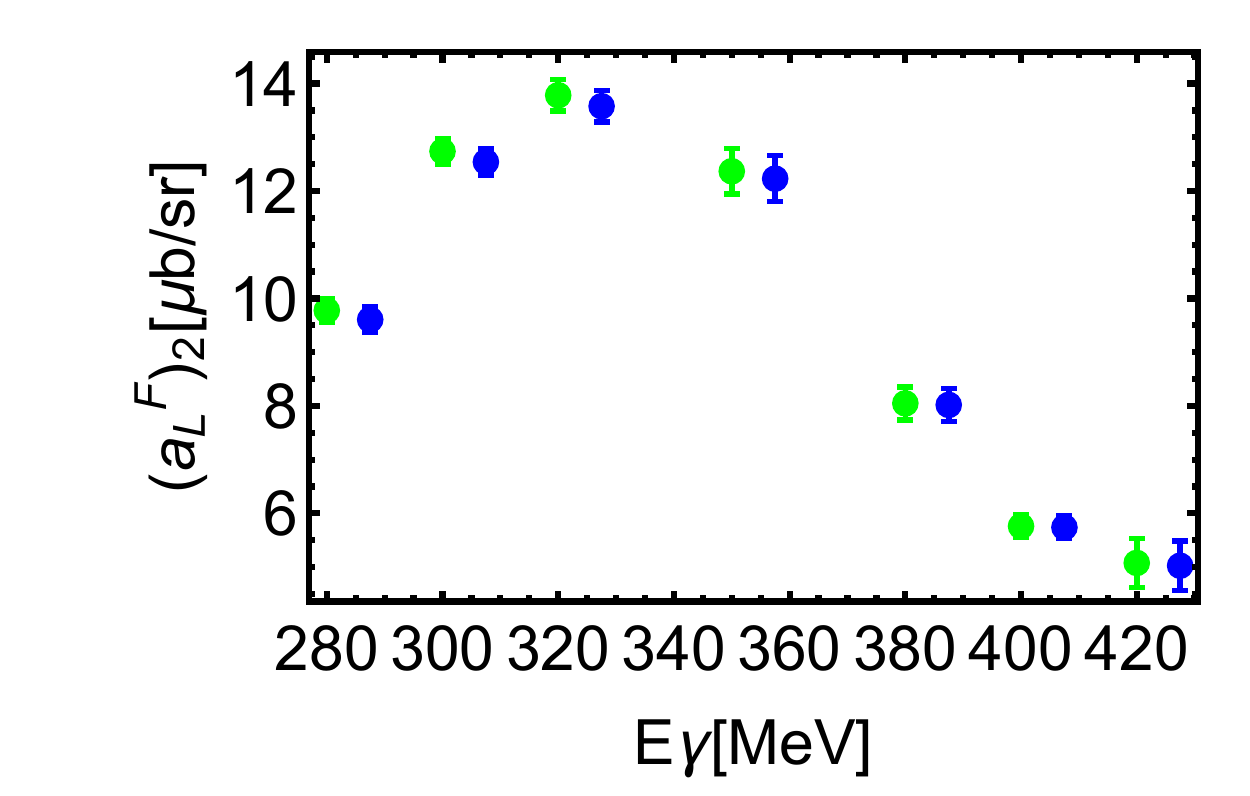}
 \end{overpic}
  \begin{overpic}[width=0.325\textwidth]{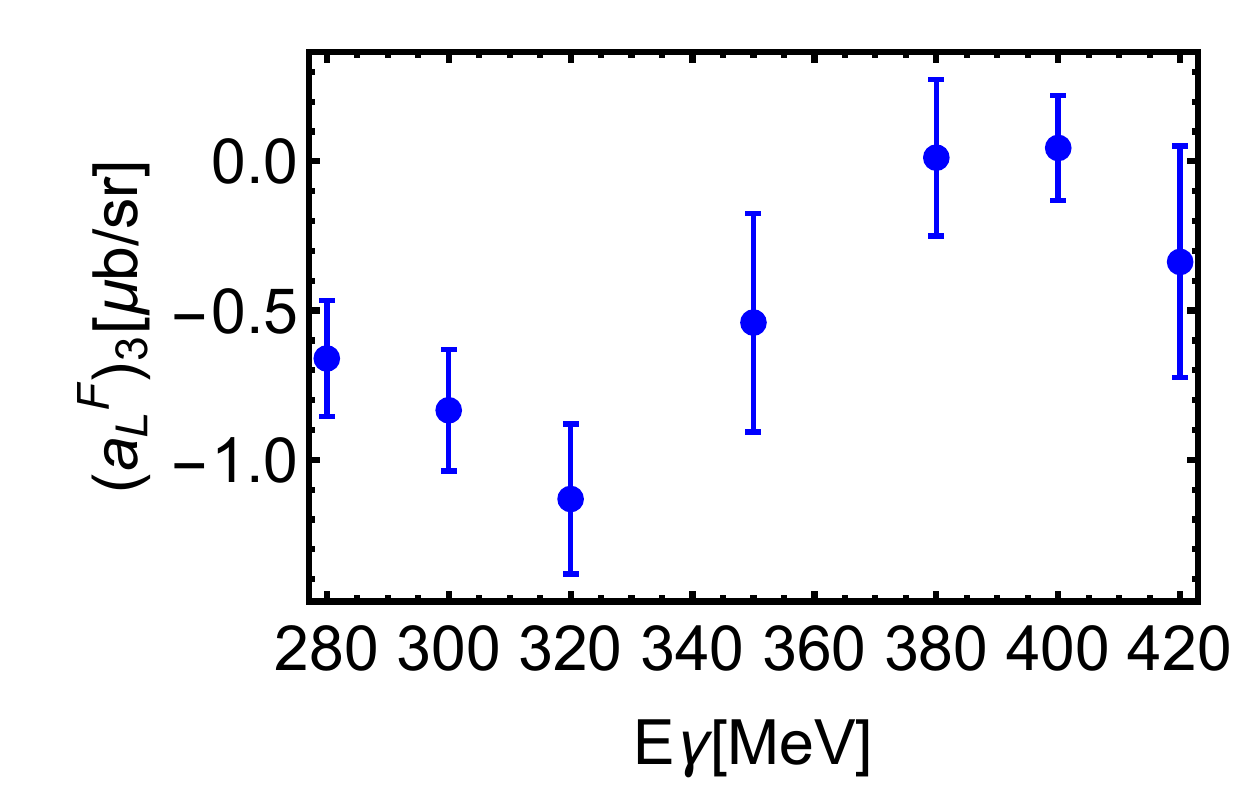}
 \end{overpic} \\
  \begin{overpic}[width=0.325\textwidth]{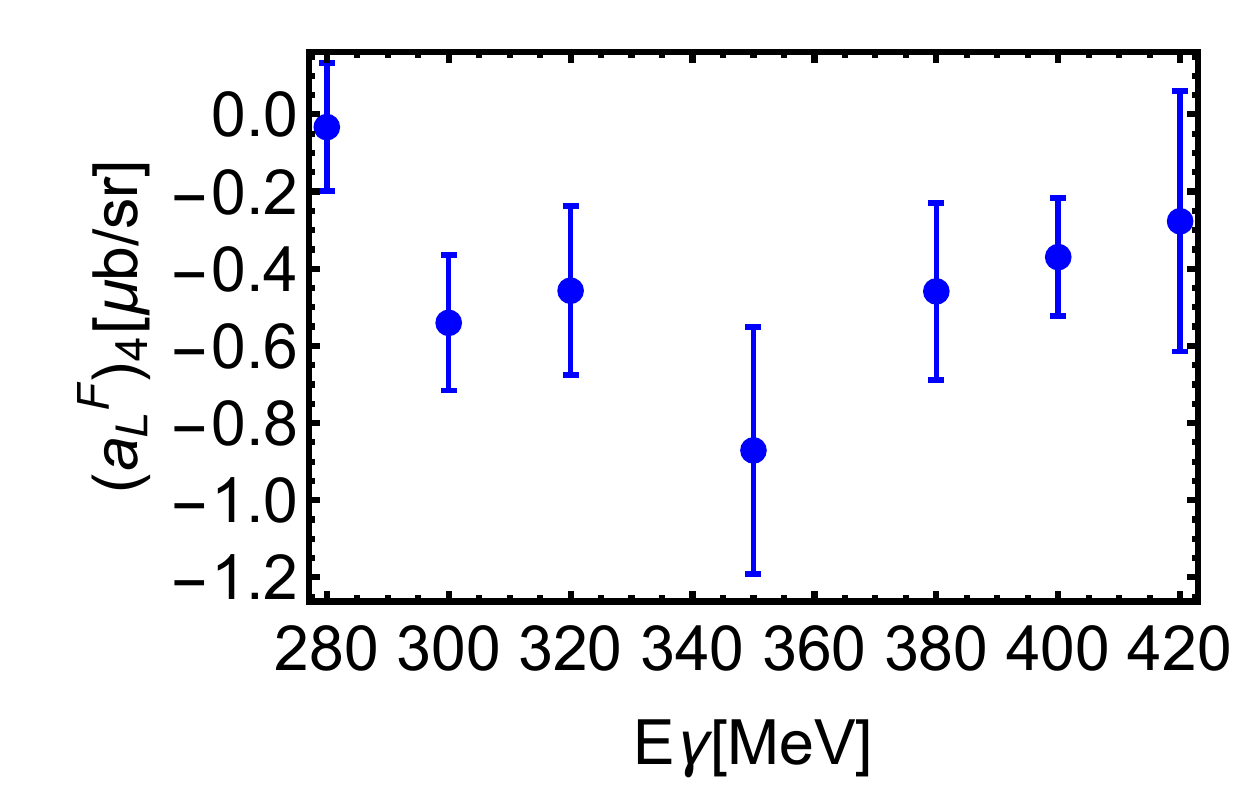}
 \end{overpic}
 \caption[The plots represent a continuation of Figure \ref{fig:DeltaRegionFittedLegCoeffsPlotsI} and the show extracted Legendre coefficients for the profile functions $\check{P}$ and $\check{F}$.]{These plots represent a continuation of Figure \ref{fig:DeltaRegionFittedLegCoeffsPlotsI} and they show extracted Legendre coefficients for the profile functions $\check{P}$ and $\check{F}$. \newline
 Note that for the $\check{P}$-observable, Legendre coefficients cannot be extracted in the first energy-bin (cf. the main text). }
 \label{fig:DeltaRegionFittedLegCoeffsPlotsII}
\end{figure}

\begin{figure}[h]
 \centering
 \begin{overpic}[width=0.325\textwidth]{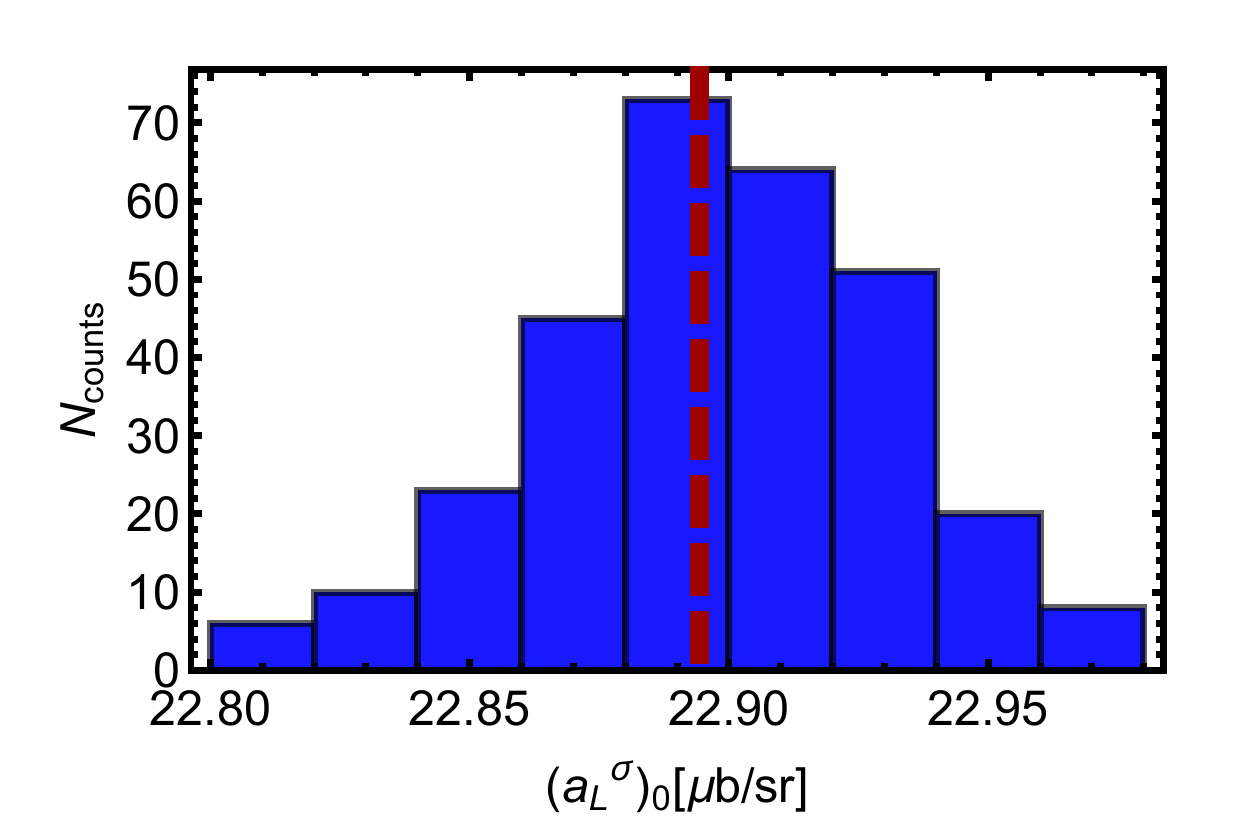}
  \put(123.75,75){\begin{Large}$E_{\gamma} = 350. \hspace*{2pt} \mathrm{MeV}$\end{Large}}
 \end{overpic}
 \begin{overpic}[width=0.325\textwidth]{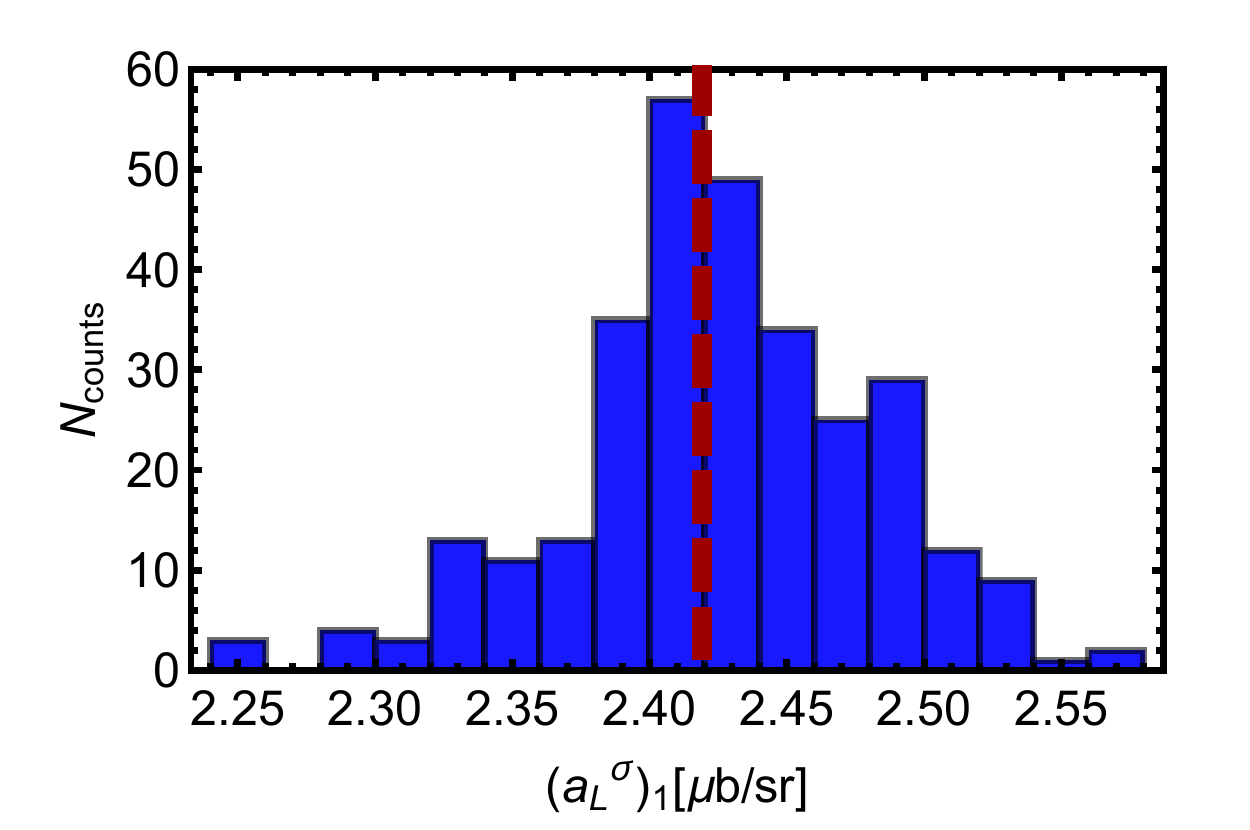}
 \end{overpic}
 \begin{overpic}[width=0.325\textwidth]{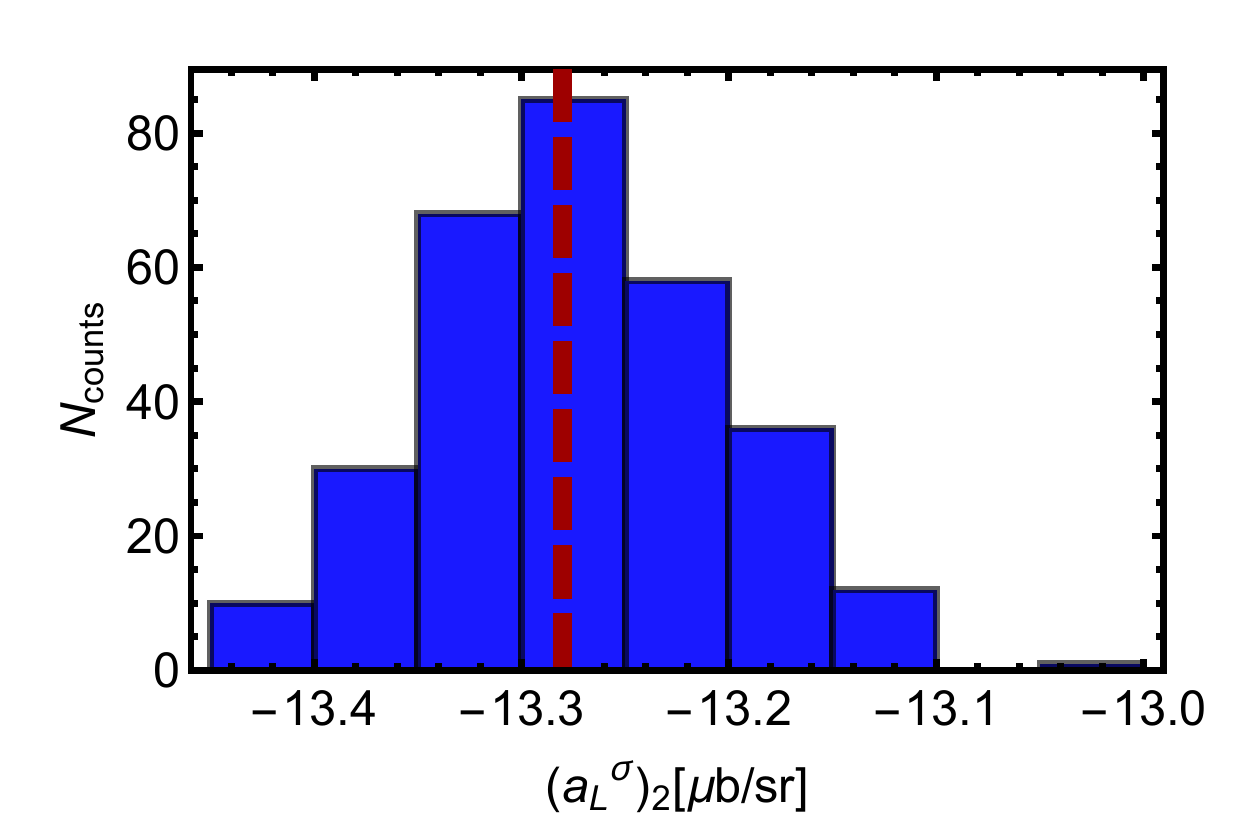}
 \end{overpic} \\
 \begin{overpic}[width=0.325\textwidth]{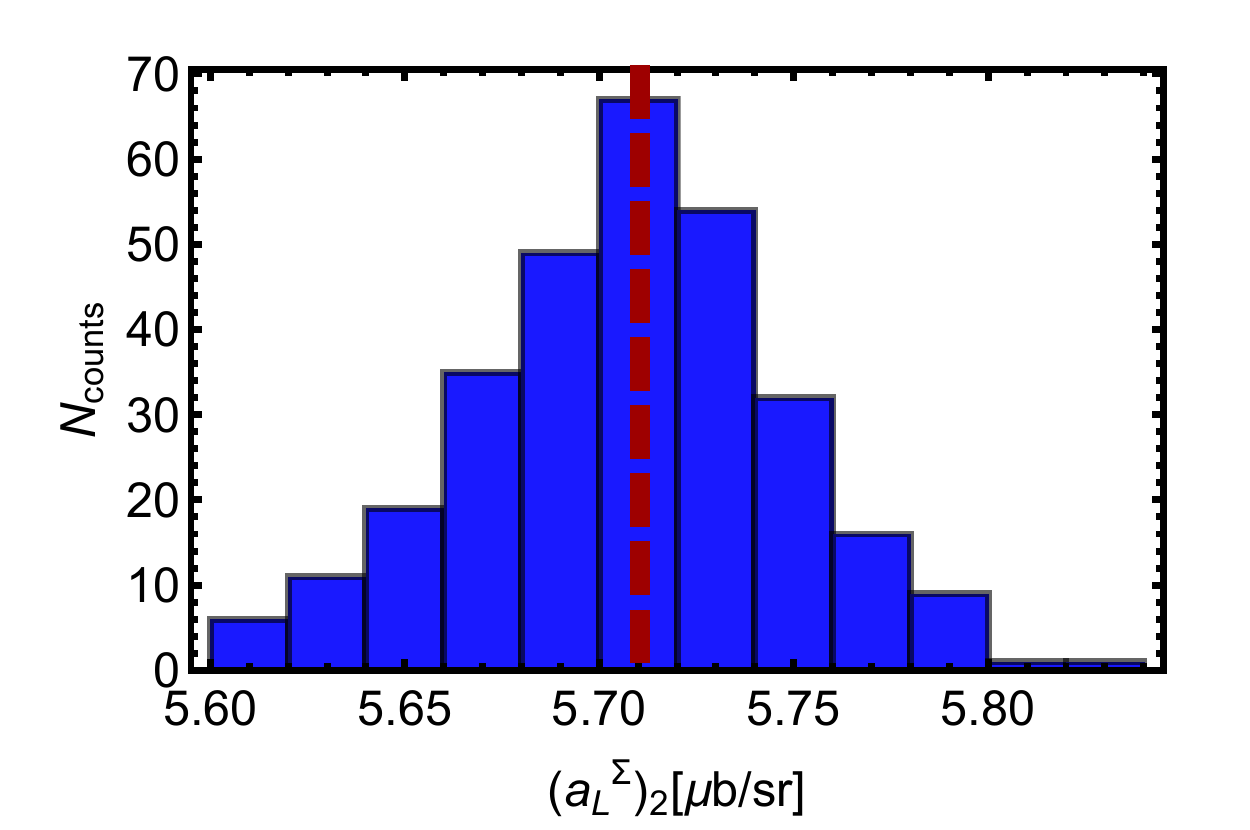}
 \end{overpic} \\
 \begin{overpic}[width=0.325\textwidth]{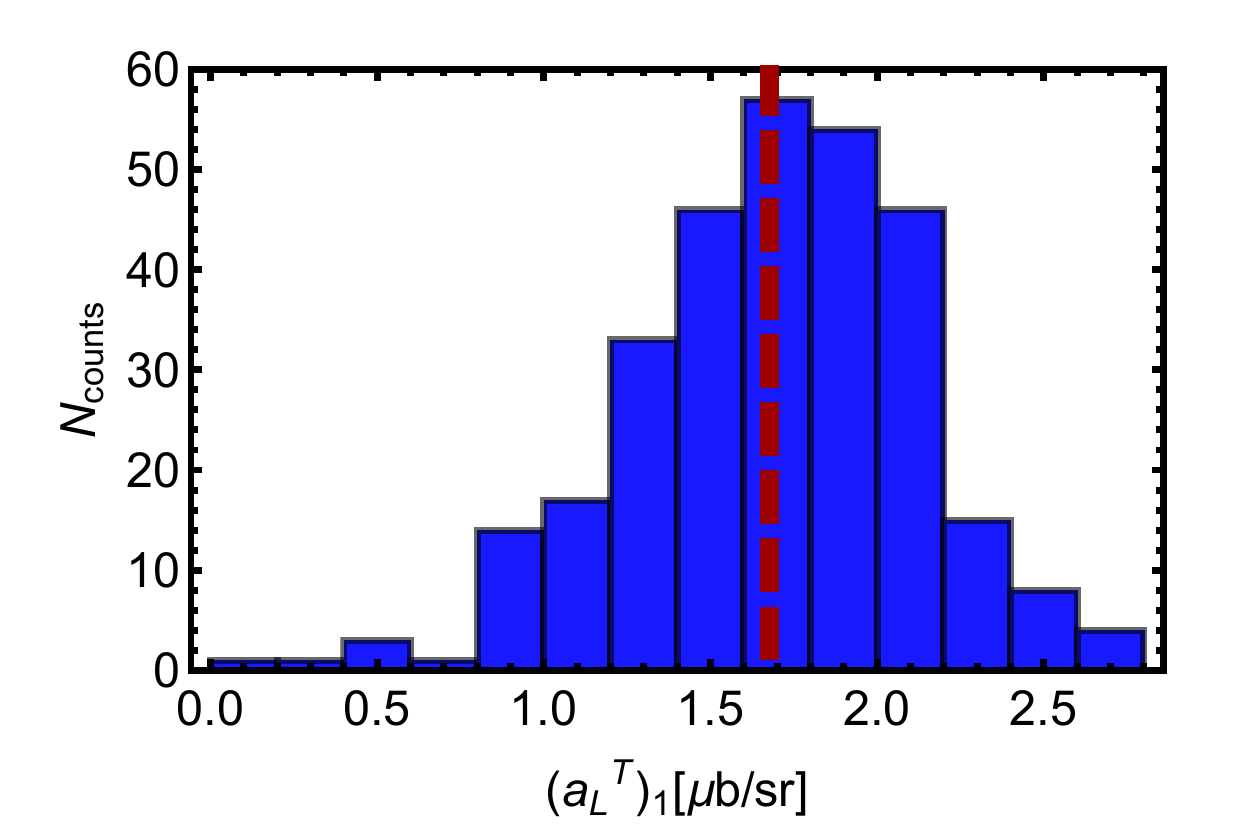}
 \end{overpic}
 \begin{overpic}[width=0.325\textwidth]{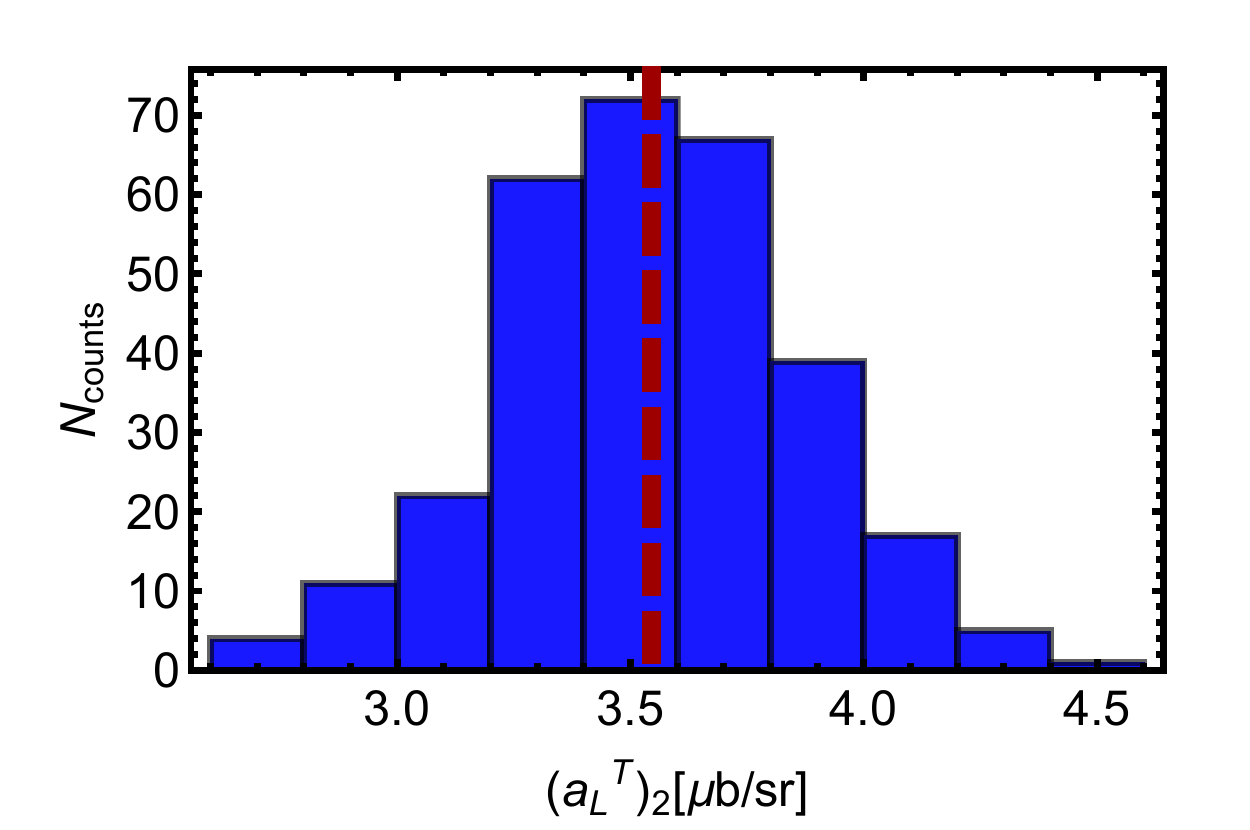}
 \end{overpic} \\
 \begin{overpic}[width=0.325\textwidth]{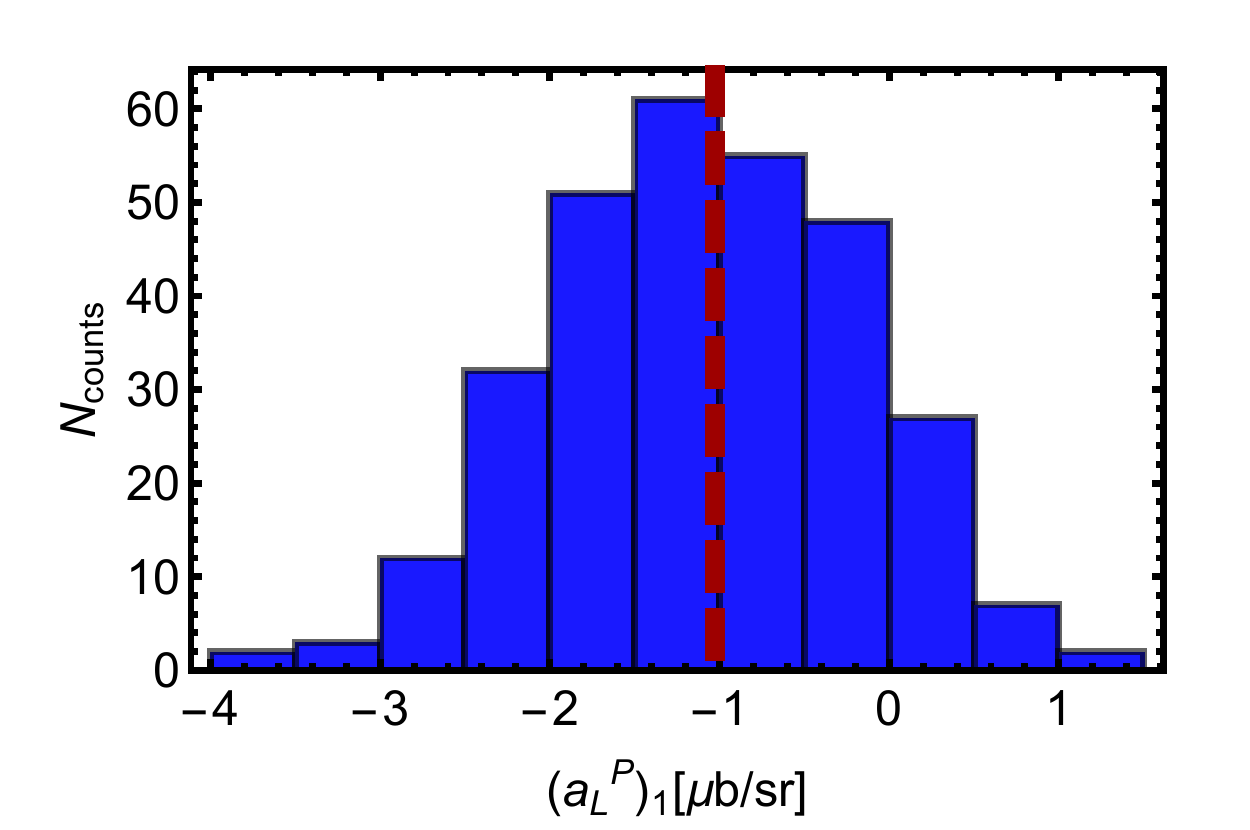}
 \end{overpic}
  \begin{overpic}[width=0.325\textwidth]{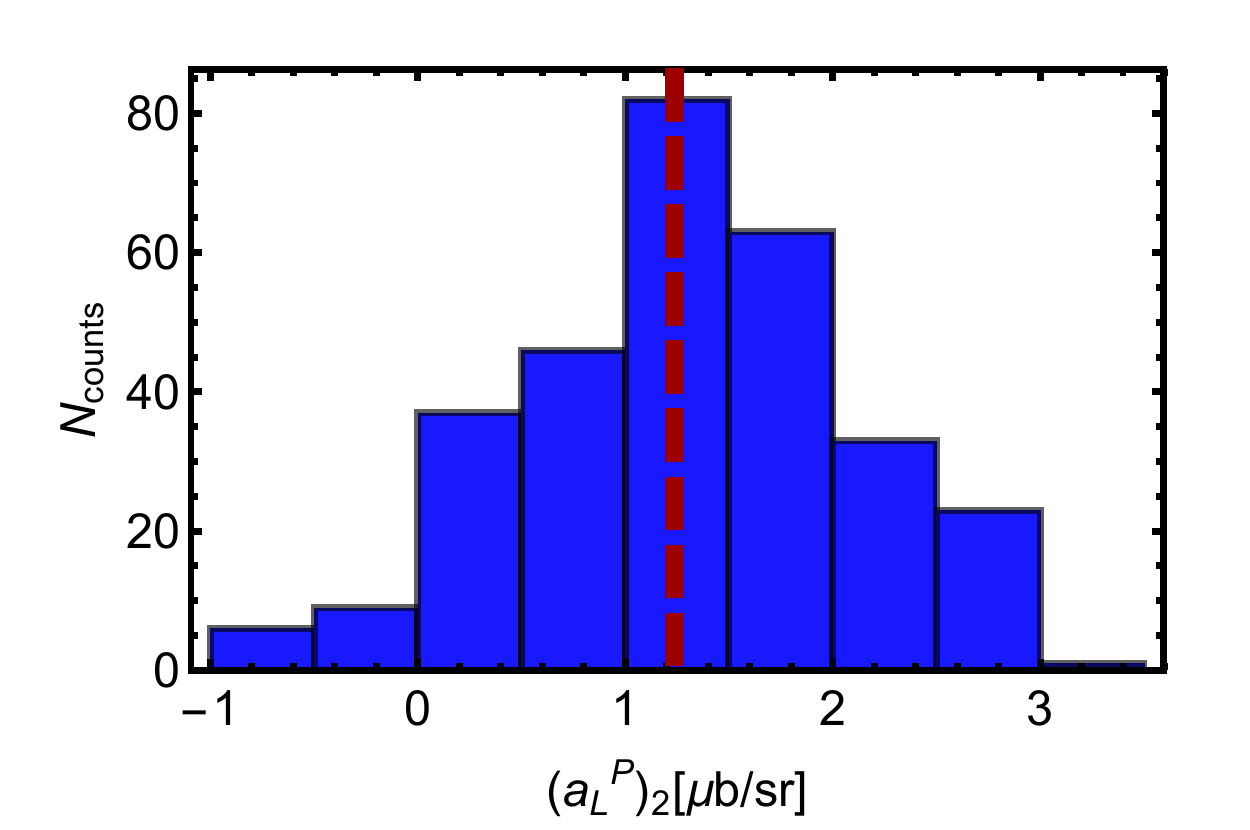}
 \end{overpic} \\
 \begin{overpic}[width=0.325\textwidth]{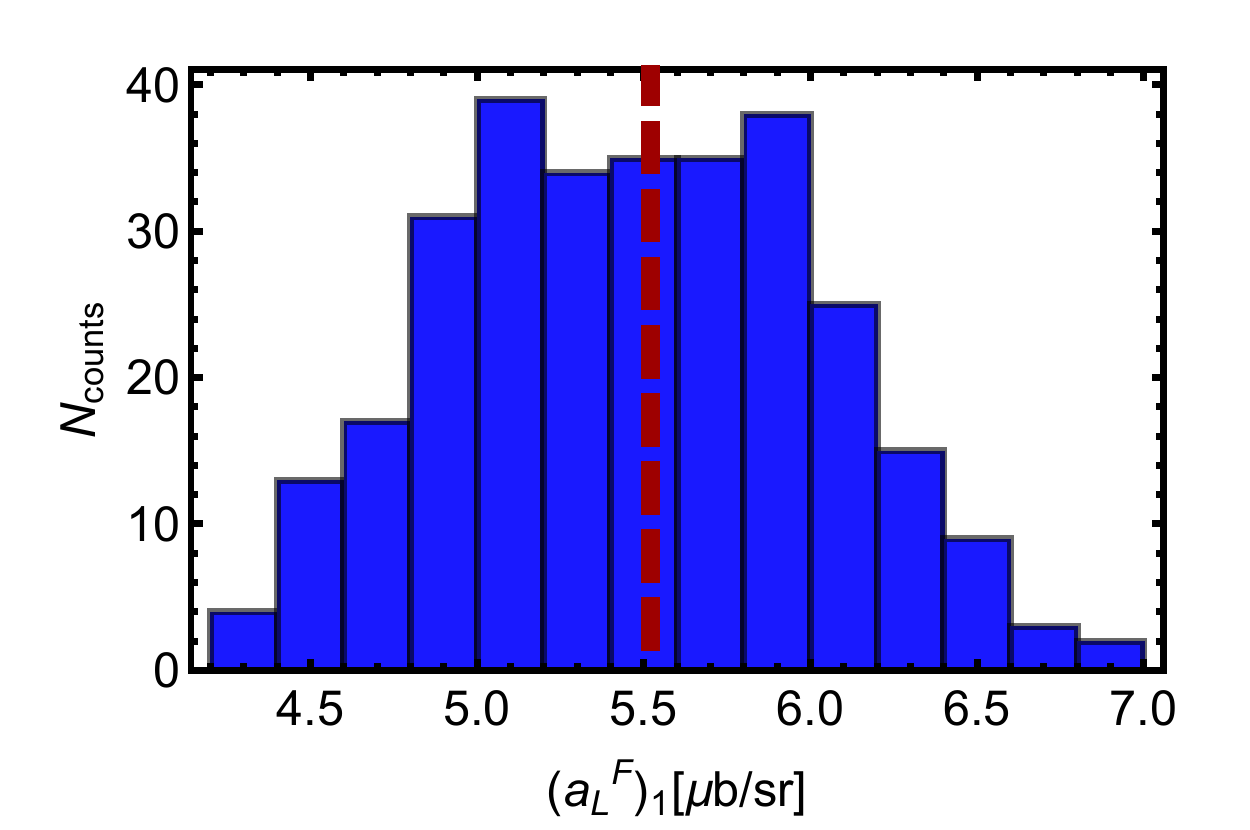}
 \end{overpic}
  \begin{overpic}[width=0.325\textwidth]{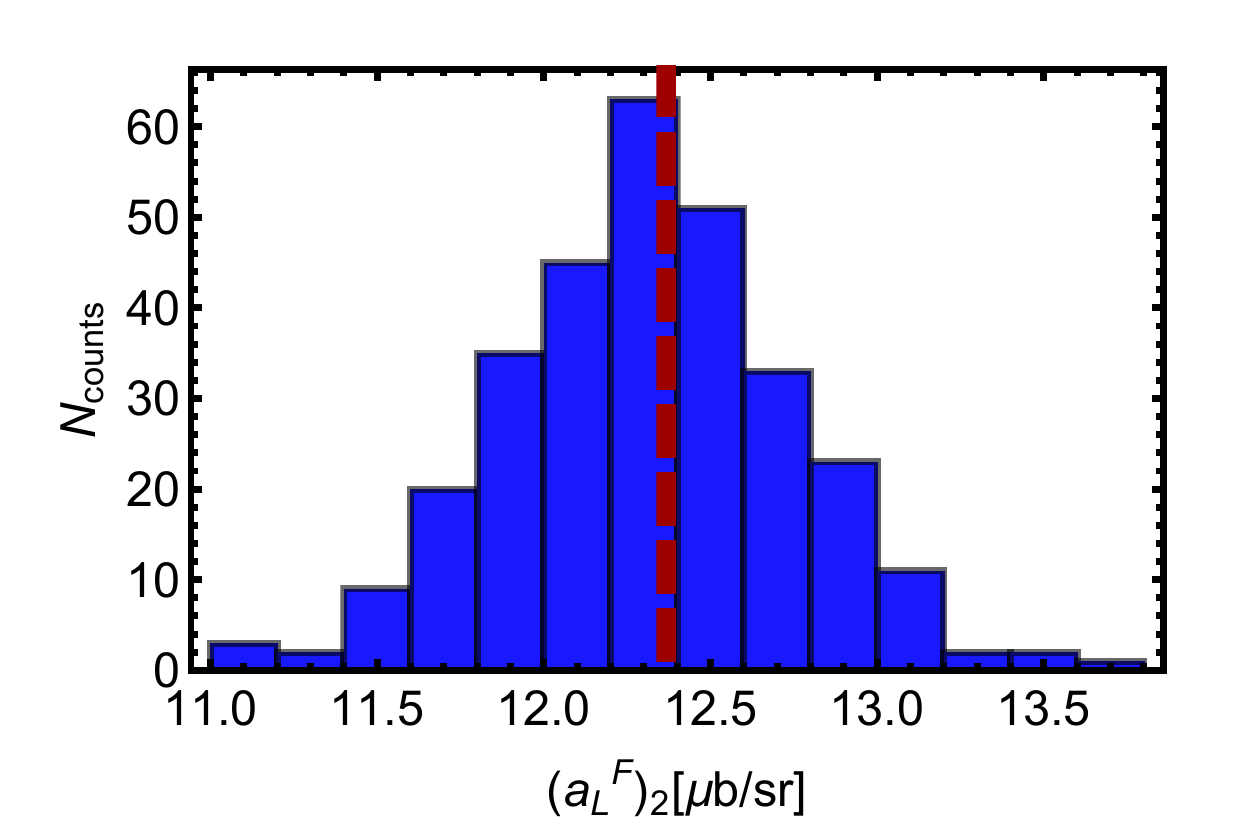}
 \end{overpic}
 \caption[Histograms showing bootstrap-distributions for the Legendre-coefficients of all observables fitted in the $\Delta$-region, extracted in a truncation at $\ell_{\mathrm{max}} = 1$, shown for the example-energy $E_{\gamma} = 350 \hspace*{1pt} \mathrm{MeV}$]{The histograms show bootstrap-distributions for the Legendre-coefficients of all observables fitted in the $\Delta$-region, extracted in a truncation at $\ell_{\mathrm{max}} = 1$. A relatively moderate number of $B = 300$ bootstrap-replicates has been generated from the original data and the particular example-energy $E_{\gamma} = 350 \hspace*{1pt} \mathrm{MeV}$ has been chosen. The result of the fit to the original data is indicated by a red dashed vertical line in each histogram.}
 \label{fig:DeltaRegionFittedLegCoeffsBootstrapHistos}
\end{figure}

Using such parametrizations for different $\ell_{\mathrm{max}}$, Legendre-coefficients can be extracted from the data. Therefore, this preparatory investigation of the angular distributions coincides with what has been called 'TPWA fit-step $\mathrm{I}$' in section \ref{sec:TPWAFitsIntro}. \newline
Figure \ref{fig:DeltaRegionObsFitAngDistPlots} contains the angular distributions of all observables considered in the $\Delta$-region, as well as fits of the parametrizations (\ref{eq:DCSAngDistChapter4}) to (\ref{eq:FAngDistChapter4}) for $\ell_{\mathrm{max}} = 1$ and $\ell_{\mathrm{max}} = 2$, all shown for the example of the fourth energy-bin $E_{\gamma} = 350 \hspace*{1pt} \mathrm{MeV}$. Only the statistical errors are plotted. The discrepancies in statistics among the $5$ datasets are seen immediately. The cross section $\sigma_{0}$ and beam asymmetry $\Sigma$ have been measured so precisely that error-bars can barely be seen in the plots. For the $P$-data especially, errors are so large that clearly these data do not permit the extraction of any meaningful Legendre coefficients for any higher $\ell_{\mathrm{max}}$. Furthermore, another problem posed by the statistics is the small number of datapoints in every angular distribution of $P$. In Figure \ref{fig:DeltaRegionObsFitAngDistPlots}, $P$ has $6$ points and thus, by inspection of the form of the expansion (\ref{eq:PAngDistChapter4}), would permit a fit with a positive number of degrees of freedom $\mathrm{ndf}$ for $\ell_{\mathrm{max}} = 1$ and $\ell_{\mathrm{max}} = 2$. In case of $\ell_{\mathrm{max}} = 3$, equation (\ref{eq:PAngDistChapter4}) would yield $6$ Legendre-coefficients, thus resulting in $\mathrm{ndf} = 0$. For the lowest energy, $P$-data are only given at four points in angle (see Figure \ref{fig:DeltaRegionKinematicPlots}) and therefore do not enable any meaningful fits above $\ell_{\mathrm{max}} = 1$. \newline
The same discrepancy in the statistical precision of the data can be observed in the distributions of the $\chi^{2}/\mathrm{ndf}$ for fits of different $\ell_{\mathrm{max}}$, plotted against energy. Such $\chi^{2}$-plots can be seen in Figure \ref{fig:DeltaRegionChisquareLmaxPlots}, where results of fits using $\ell_{\mathrm{max}} = 1$ and $2$ are included. As mentioned before, the $P$-data do not support an $\ell_{\mathrm{max}} = 2$-fit at the lowest energy, such that this value has been put to a default value zero in the plots. Furthermore, the optimal value of $\chi^{2}/\mathrm{ndf} = 1$ is indicated in the figures. \newline
All observables show improvements upon introducing the $D$-wave truncation $\ell_{\mathrm{max}} = 2$, compared to the $P$-wave truncation $\ell_{\mathrm{max}} = 1$. For the differential cross section $\sigma_{0}$ however, even the fit of a $D$-wave truncation does not yield a satisfactory $\chi^{2}$. For all the remaining observables and for most energies, it does. The measurements for the beam asymmetry $\Sigma$ and beam-target observable $F$ also indicate a slight need for $D$-waves for some energies, having a not quite satisfactory description in a $P$-wave truncation. Data for $T$ and especially $P$ are already described well in a $P$-wave approximation, with no real need to introduce the $D$-waves. \newline
This apparent discrepancy in precision makes it necessary to carefully weigh the order $\ell_{\mathrm{max}}$ to choose for the extraction of the multipoles, i.e. TPWA fit-step $\mathrm{II}$ (cf. section \ref{sec:TPWAFitsIntro}), to be performed in the ensuing discussion. Data for the differential cross section would, due to the large statistical precision in the measurement, even facilitate the extraction of meaningful information on Legendre-coefficients for $\ell_{\mathrm{max}}$-orders even larger than $2$. However, for these higher orders, it remains questionable if the remaining measurements yield any useful information. First and foremost, the $P$-data do not support any meaningful fits at all for all orders $\ell_{\mathrm{max}} \geq 3$. For these reasons,  we resort to TPWA-approaches with a maximum truncation of $\ell_{\mathrm{max}} = 2$ in the next paragraph. \newline
Apart from the $\ell_{\mathrm{max}}$-analysis, Legendre-coefficients are the most important result of the TPWA step $\mathrm{I}$ described in this paragraph. In Figures \ref{fig:DeltaRegionFittedLegCoeffsPlotsI} and \ref{fig:DeltaRegionFittedLegCoeffsPlotsII}, the coefficients resulting from fits of the parametrizations (\ref{eq:DCSAngDistChapter4}) to (\ref{eq:FAngDistChapter4}) to the data are shown for the orders $\ell_{\mathrm{max}} = 1$ and $\ell_{\mathrm{max}} = 2$. \newline
The coefficients which enter newly in the order $\ell_{\mathrm{max}} = 2$, i.e. those $\left(a_{L}\right)^{\check{\Omega}^{\alpha}}_{k}$ with lower index $k=3$ and $k=4$, take non-zero but small values for almost all observables except $P$. This can be seen as a small influence of $D$-waves already in the low-energy $\Delta$-region. The lower Legendre-coefficients, i.e. those already present in a $P$-wave truncation, are seen to be quite stable against increase of $\ell_{\mathrm{max}}$. The only exception is also here provided by the lower coefficients of the $P$-observable, which tend to be more unstable for the lower energies due to overfitting of the angular distributions (cf. Figure \ref{fig:DeltaRegionChisquareLmaxPlots}). The instability of Legendre-coefficients in similar situations will also be encountered later in the analysis of data within the second resonance region (cf. section \ref{subsec:2ndResRegionDataFits}). \newline
The $\ell_{\mathrm{max}}$-estimates and results for Legendre-coefficients (as well as their errors, covariance matrices, ...) are important results of TPWA fit step $\mathrm{I}$ and will serve as input in the actual multipole-analysis described below.  Furthermore, the data for the profile functions such as those shown in Figure \ref{fig:DeltaRegionObsFitAngDistPlots} are starting-points for bootstrap-analyses according to the description in section \ref{sec:BootstrappingIntroduction}, i.e. fits of bootstrap-replicates of the original data, generated by drawing from a normal distribution centered at the original datapoint: 
\begin{equation}
\mathcal{N} \left( \check{\Omega}^{\alpha}, \Delta \check{\Omega}^{\alpha} \right) \longrightarrow \check{\Omega}_{\ast}^{\alpha} \mathrm{.} \label{eq:BootstrapDefinitionDeltaRegionFit}
\end{equation}
It is clear that bootstrap-datasets $\left\{ \check{\Omega}_{\ast}^{\alpha} \right\}$ for the profile functions $\check{\Omega}^{\alpha}$ also generate bootstrap-ensembles for the Legendre coefficients (see Figures \ref{fig:BootstrapDrawingPic3} and \ref{fig:BootstrapIllustrationUsingGObservable} in section \ref{sec:BootstrappingIntroduction}).
An example for bootstrap-distributions of the Legendre coefficients is shown in Figure \ref{fig:DeltaRegionFittedLegCoeffsBootstrapHistos}. Such distributions exist as a part of the analysis, but will not be shown explicitly any more in the following. \newline

\clearpage


\clearpage

\textbf{Results of TPWA-fits performed to the selected data} \newline

For the extraction of multipoles, we utilize both fitting methods outlined in section \ref{sec:TPWAFitsIntro}. The first of both consists of fits according to Grushin \cite{Grushin}, i.e. extracting parameter estimates for the multipoles out of the Legendre coefficients determined previously. \newline
Upon labeling the analyzed observables $\left\{ \sigma_{0}, \check{\Sigma}, \check{T}, \check{P}, \check{F} \right\}$ by their index $\alpha$ and furthermore wrapping indices of observables and Legendre-coefficients into 'multi-indices', i.e. $i \equiv \left( \alpha, k \right)$, $\ldots$, the expression for the correlated chisquare is (see eq. (\ref{eq:CorrelatedChisquare})) 
\begin{equation}
\chi^{2}_{\mathcal{M}} \left(  \left\{ \mathcal{M}_{\ell} \right\} \right) = \sum_{i,j} \Big[ \left(a_{L}^{\mathrm{Fit}}\right)_{i} - \left< \mathcal{M}_{\ell} \right| \left(\mathcal{C}_{L}\right)_{i} \left| \mathcal{M}_{\ell} \right> \Big] \mathrm{\textbf{C}}^{-1}_{ij} \Big[ \left(a_{L}^{\mathrm{Fit}}\right)_{j} - \left< \mathcal{M}_{\ell} \right| \left(\mathcal{C}_{L}\right)_{j} \left| \mathcal{M}_{\ell} \right> \Big] \mathrm{.} \label{eq:CorrelatedChisquareQuotedForAnalysis}
\end{equation}
Furthermore, we also utilize the option to fit directly to the data. In the course of the discussion, it will become clear why this method is maybe even a bit better suited for this particular dataset. For quick reference, we quote the corresponding form of the chisquare (see equation (\ref{eq:ChiSquareDirectFit}))
\begin{equation}
 \chi^{2}_{\mathrm{data}} \left(  \left\{ \mathcal{M}_{\ell} \right\} \right) = \sum_{\check{\Omega}^{\alpha}, c_{k_{\alpha}}} \left[ \frac{ \check{\Omega}^{\alpha}_{\mathrm{Data}} (c_{k_{\alpha}}) - \check{\Omega}^{\alpha}_{\mathrm{Fit}} \left( c_{k_{\alpha}}, \left\{ \mathcal{M}_{\ell} \right\} \right) }{\Delta \check{\Omega}^{\alpha}_{\mathrm{Data}} (c_{k_{\alpha}})} \right]^{2} \mathrm{,} \label{eq:ChiSquareDirectFitRealDataFitSection}
\end{equation}
where $\check{\Omega}^{\alpha}_{\mathrm{Fit}} \left( \cos \theta, \left\{ \mathcal{M}_{\ell} \right\} \right) := \frac{q}{k} \sum_{n = \beta_{\alpha}}^{2 L + \beta_{\alpha} + \gamma_{\alpha}} \left< \mathcal{M}_{\ell} \right| \left( \mathcal{C}_{L}\right)_{n}^{\check{\Omega}^{\alpha}}  \left| \mathcal{M}_{\ell} \right> \hspace*{2pt} P^{\beta_{\alpha}}_{n} \left(  \cos \theta \right)$ as in equation (\ref{eq:FitFunctionDirectFit}) and the variables $c_{k_{\alpha}} \equiv \cos \left( \theta_{k_{\alpha}} \right)$ define the angular grid for the quantity $\check{\Omega}^{\alpha}$. The sum over $\alpha$ in equation (\ref{eq:ChiSquareDirectFitRealDataFitSection}) only includes the five observables analyzed in this section. The minima of either (\ref{eq:CorrelatedChisquareQuotedForAnalysis}) or (\ref{eq:ChiSquareDirectFitRealDataFitSection}) are searched according to the Monte Carlo strategies outlined in section \ref{sec:MonteCarloSampling}. \newline
In the following, the quality of fits is judged according to probability-theoretical chisquare-distributions. The {\it chisquare-distribution} for $r$ degrees of freedom reads \cite{BlobelLohrmann,Abramowitz}
\begin{equation}
 P \left[ r \right] (u) = \frac{ u^{\frac{r}{2} - 1} \exp \left( - \frac{u}{2} \right) }{\Gamma \left( \frac{r}{2} \right) 2^{r/2}} \mathrm{.} \label{eq:ChisquareDistDefinition}
\end{equation}
It can be obtained from the non-central chisquare-distribution, mentioned in section \ref{sec:PseudoDataWithErrorsFitted} (equation (\ref{eq:NonCentralChisquareDist})), by setting the decentralization parameter $\lambda$ to zero. However, the variable $u$ in function (\ref{eq:ChisquareDistDefinition}) corresponds to a non-normalized total test-statistic $\chi^{2}$. In case one wishes to write this probability-distribution for the normalized quantity $v := u/r \equiv \chi^{2}/\mathrm{ndf}$, the correct distribution-function becomes
\begin{equation}
 P \left[ r \right] (v) = r \frac{ (rv)^{\frac{r}{2} - 1} \exp \left( - \frac{rv}{2} \right) }{\Gamma \left( \frac{r}{2} \right) 2^{r/2}} \mathrm{.} \label{eq:ChisquareDistDefinitionInTermsOfNormalizedChi2}
\end{equation}
Now, to be more precise, we utilize the quantiles\footnote{For quick reference, we cite the definition here: given a smooth, normalized probability distribution function $p(x)$, the corresponding so-called cumulative distribution function is defined as $G_{p} (y) = \int_{-\infty}^{y} dx \hspace*{0.5pt} p(x)$. Then, the {\it quantile} $x^{(p)}_{q}$ of the distribution $p(x)$ belonging to the probability-fraction $q \in \left[ 0,1 \right]$ is formally the inverse: $x^{(p)}_{q} := G_{p}^{-1} (q)$ (see \cite{EfronTibshiraniBook}).} of the distribution (\ref{eq:ChisquareDistDefinitionInTermsOfNormalizedChi2}) for the normalized chisquare.
\begin{figure}[ht]
 \centering
 \vspace*{10pt}
\begin{overpic}[width=0.495\textwidth]{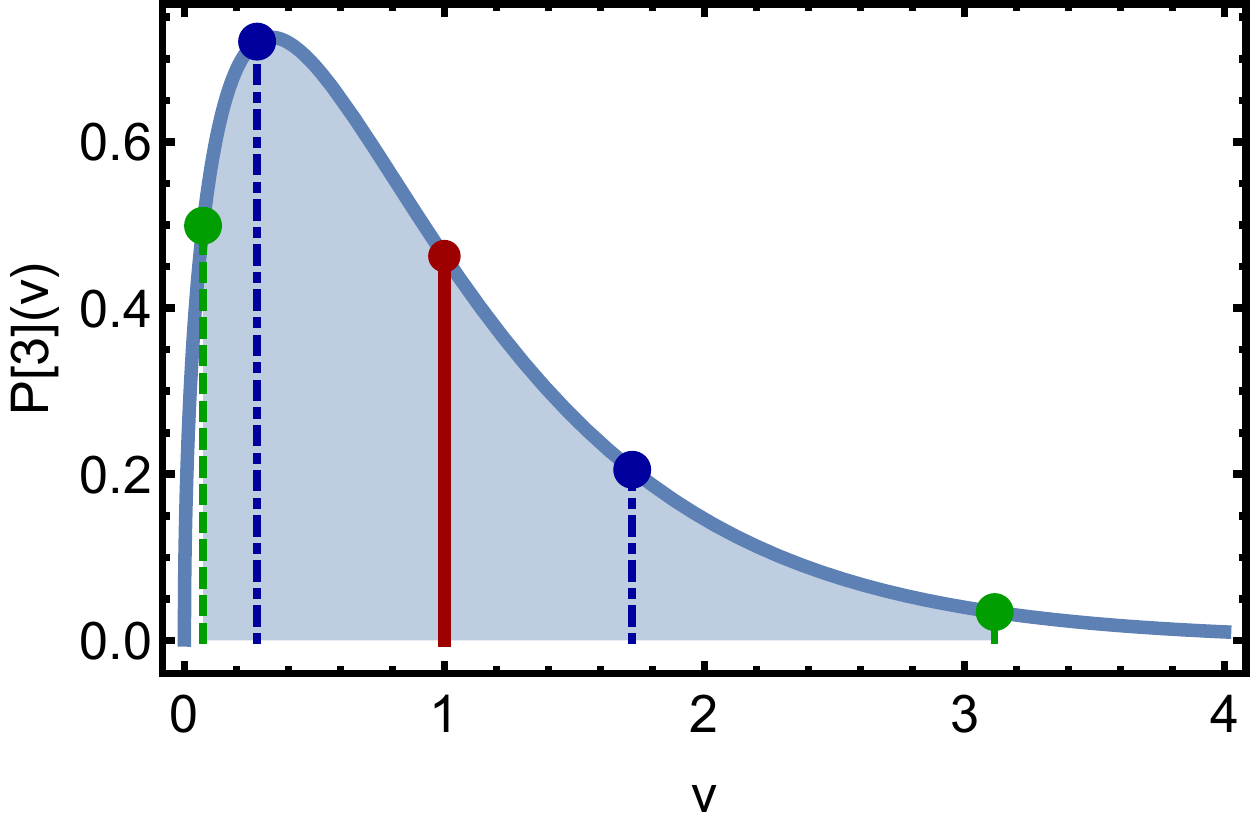}
 \put(0.5,70){\underline{Method:} fit Legendre coefficients, eq. (\ref{eq:CorrelatedChisquareQuotedForAnalysis}).}
 \end{overpic}
\begin{overpic}[width=0.495\textwidth]{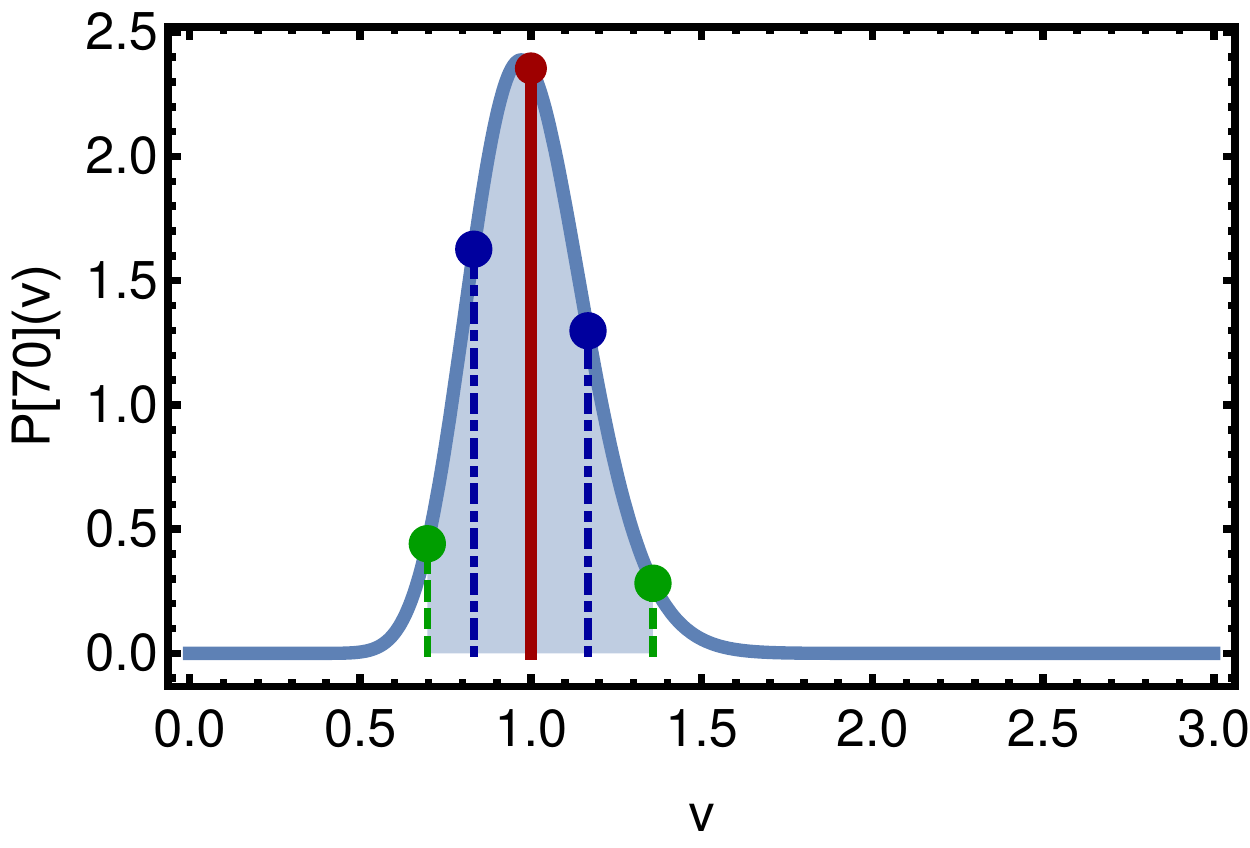}
 \put(14,70){\underline{Method:} direct fit to data, eq. (\ref{eq:ChiSquareDirectFitRealDataFitSection}).}
 \end{overpic}
\caption[Two relevant examples for $\chi^{2}$-distributions in the context of fits within the $\Delta$-resonance region.]{Two relevant examples for theoretical $\chi^{2}$-distributions $P[r](v)$ (\ref{eq:ChisquareDistDefinitionInTermsOfNormalizedChi2}), plotted against the nor\-ma\-lized chisquare $v = u/r \equiv \chi^{2}/\mathrm{ndf}$, are shown. {\it Left:} The distribution for the estimate of the number of degrees of freedom for an unconstrained fit at $\ell_{\mathrm{max}}=1$ is shown, using the method of fitting to Legendre-coefficients, or Grushin's \cite{Grushin} method, equation (\ref{eq:CorrelatedChisquareQuotedForAnalysis}). In this case, one typically has $\mathrm{ndf} \equiv r= N_{a^{\alpha}_{k}} - \left( 8 \ell_{\mathrm{max}} - 1 \right) = 10 - 7 =3$. {\it Right:} Here, a typical distribution for an unconstrained fit at $\ell_{\mathrm{max}}=1$, but for the method of directly fitting the data (\ref{eq:ChiSquareDirectFitRealDataFitSection}), is shown. An estimate of $\mathrm{ndf} \equiv r = N_{\mathrm{data}} - \left( 8 \ell_{\mathrm{max}} - 1 \right) = 77 - 7 =70$ was made (see discussion in the main text). Both distributions of course correspond to the special case that the given datasets for $\left\{ \sigma_{0}, \Sigma, T, P, F \right\}$ are fitted. \newline
The red solid line indicates the mean of the distributions, which is always just $1$. Some quantiles are shown as well. The green dashed lines mark the $0.025$- and $0.975$-quantiles. The shaded area is the fraction of probability encompassed by these two outer quantiles. In case of the gaussian distribution, this area would correspond to a $2\sigma$ confidence-interval. For comparison, the $0.16$- and $0.84$-quantiles are indicated as blue dash-dotted lines. They define the analogue of a $1\sigma$ confidence interval.}
\label{fig:ExampleChiSquareDistributionDeltaRegion}
\end{figure}
The pair of $0.16$- and $0.84$-quantiles of (\ref{eq:ChisquareDistDefinitionInTermsOfNormalizedChi2}) defines a $68\%$ confidence-interval, in analogy with the '$1 \sigma$' confidence interval of a gaussian probability distribution function. The pair of $0.025$- and $0.975$-quantiles then defines the analogue of a '$2 \sigma$' confidence-interval, i.e. a $95\%$ confidence-interval. We choose this latter interval as a region of $\chi^{2}/\mathrm{ndf}$ in which fits should be accepted. Whenever the result of a minimization falls outside of this region, it {\it should} be rejected. \newline
Depending on $r$, the shape of the chisquare distribution (\ref{eq:ChisquareDistDefinitionInTermsOfNormalizedChi2}) can change drastically, which also affects the positions of the quantiles. This fact is illustrated in Figure \ref{fig:ExampleChiSquareDistributionDeltaRegion}. \newline
The number of degrees of freedom 'ndf' is estimated differently for both fit methods, see sections \ref{sec:TPWAFitsIntro} and \ref{sec:MonteCarloSampling}. For a truncation at some $\ell_{\mathrm{max}}$, the estimate for the fit to Legendre-coefficients (\ref{eq:CorrelatedChisquareQuotedForAnalysis}) becomes $\mathrm{ndf} \equiv r= N_{a^{\alpha}_{k}} - \left( 8 \ell_{\mathrm{max}} - 1 \right)$, with $N_{a^{\alpha}_{k}}$ the number of coefficients coming from all five observables. In case of the direct fit to data (\ref{eq:ChiSquareDirectFitRealDataFitSection}), the estimate is $\mathrm{ndf} \equiv r= N_{\mathrm{data}} - \left( 8 \ell_{\mathrm{max}} - 1 \right)$, where $N_{\mathrm{data}}$ denotes the number of datapoints from all observables in a particular energy bin. Since $N_{\mathrm{data}}$ is typically a lot larger than $N_{a^{\alpha}_{k}}$, at least for low truncation orders $\ell_{\mathrm{max}}$, the chisquare distributions of direct fits (\ref{eq:ChiSquareDirectFitRealDataFitSection}) are typically a lot more slender that for Grushin's method (\ref{eq:CorrelatedChisquareQuotedForAnalysis}). Consequently, the quantiles also move closer together. This fact is illustrated by the comparison in Figure \ref{fig:ExampleChiSquareDistributionDeltaRegion}. These facts should be kept in mind when comparing results from both methods. \newline

We continue with the results of a TPWA in the lowest sensible truncation order suggested by the angular distributions of the observables, see Figure \ref{fig:DeltaRegionChisquareLmaxPlots}, i.e. $\ell_{\mathrm{max}} = 1$. \newline

\begin{figure}[ht]
 \centering
 \vspace*{-7.5pt}
\begin{overpic}[width=0.495\textwidth]{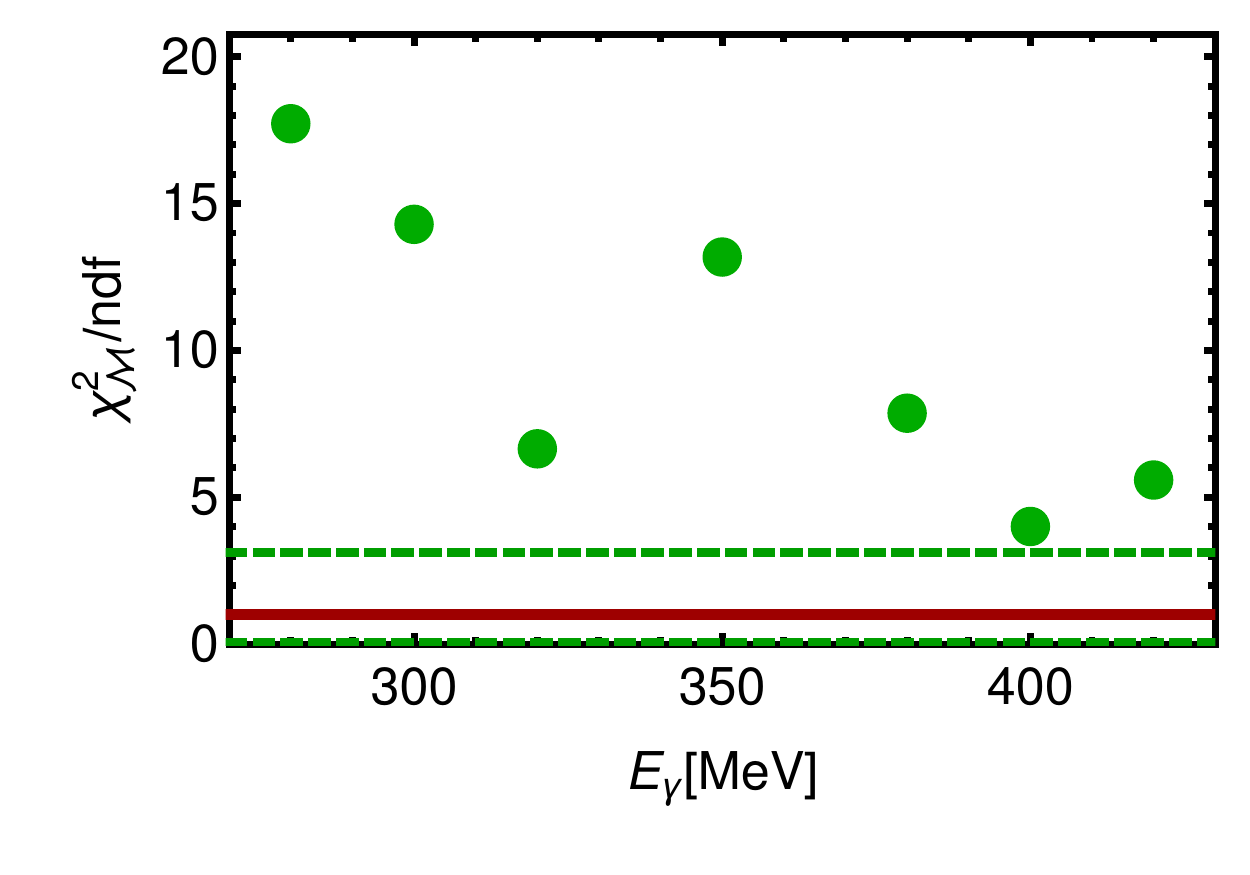}
 \put(0.5,66){a.)}
 \end{overpic}
\begin{overpic}[width=0.495\textwidth]{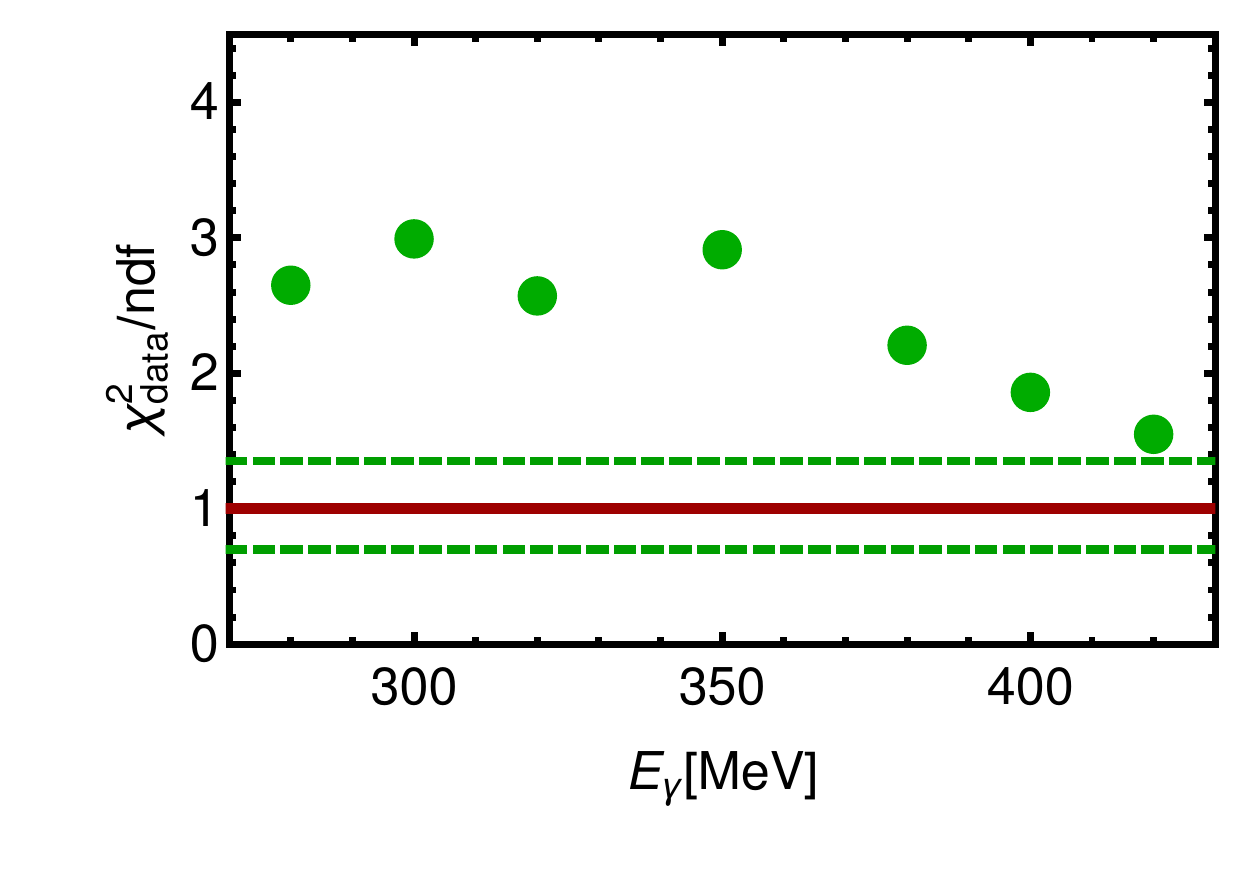}
 \end{overpic} \\
 \vspace*{2pt}
\begin{overpic}[width=0.325\textwidth]{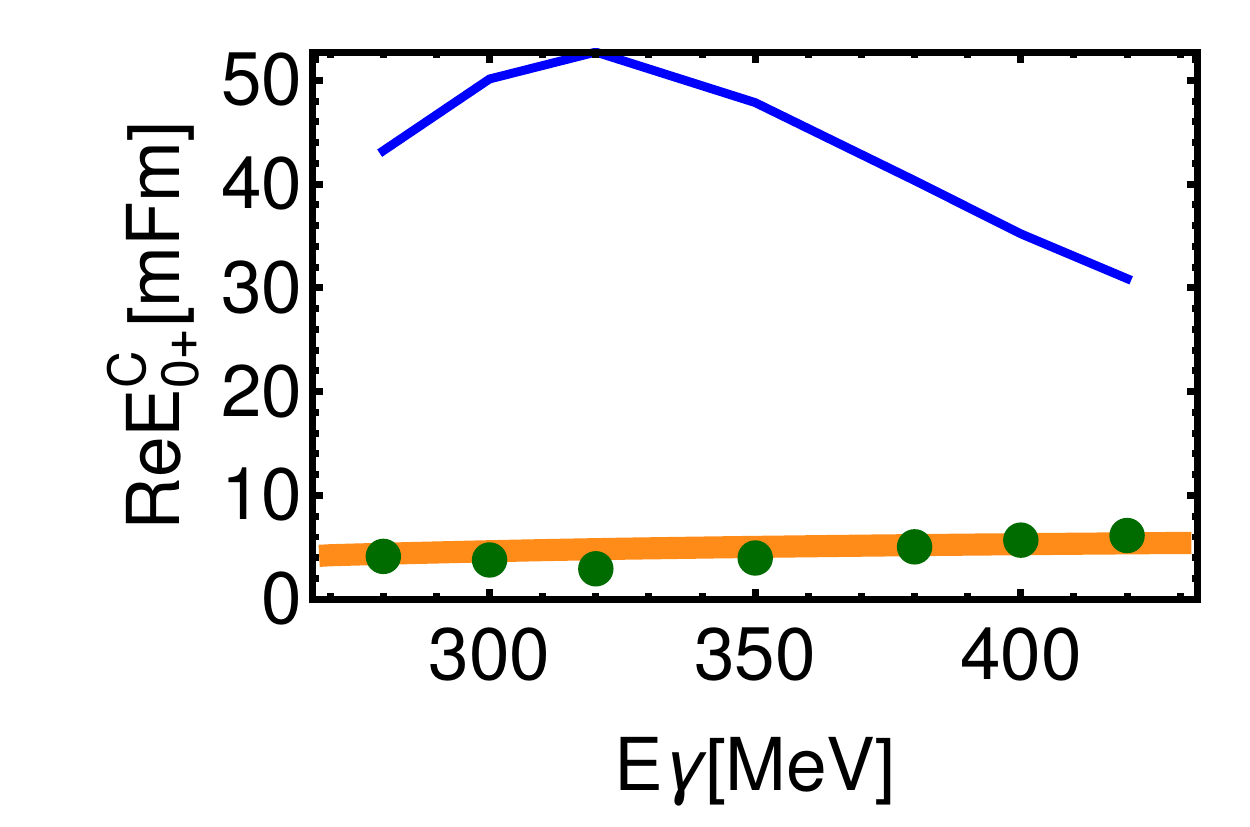}
 \put(0.5,65){b.)}
 \end{overpic}
\begin{overpic}[width=0.325\textwidth]{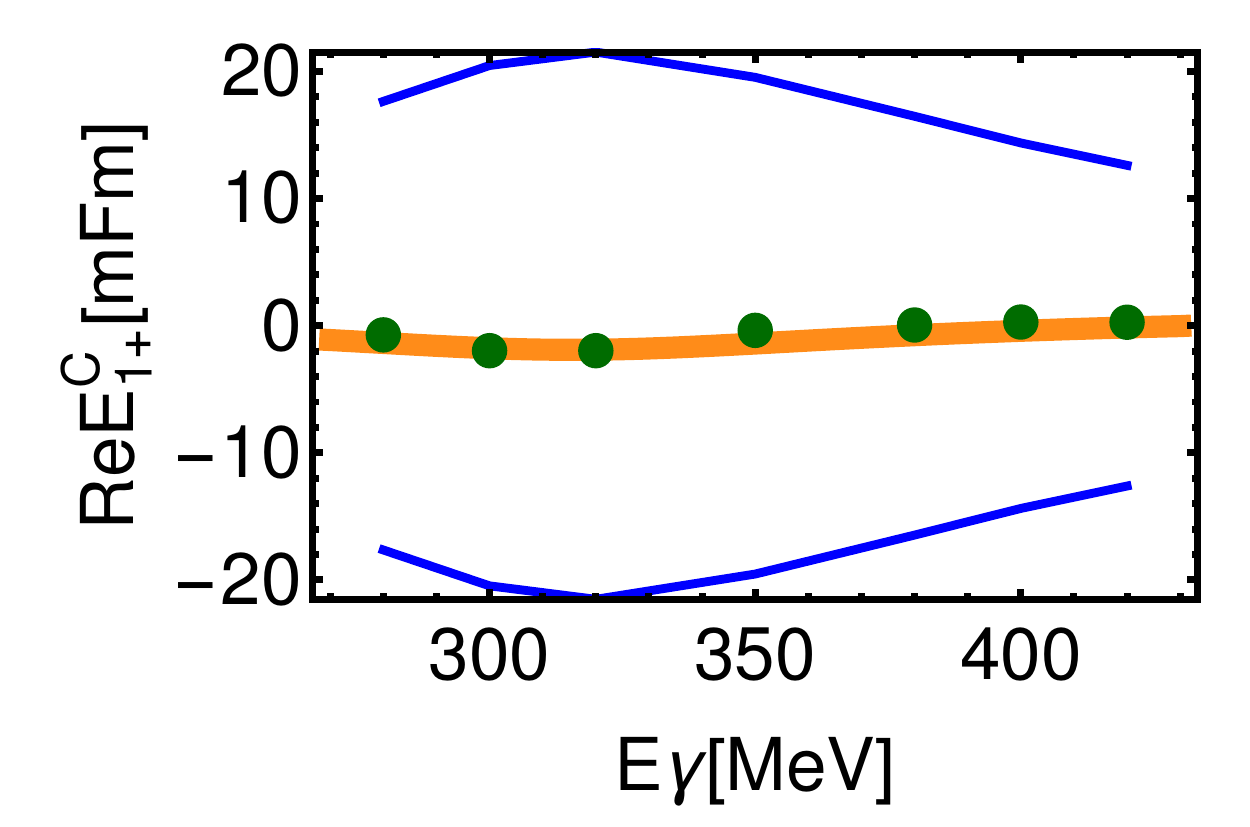}
 \end{overpic}
\begin{overpic}[width=0.325\textwidth]{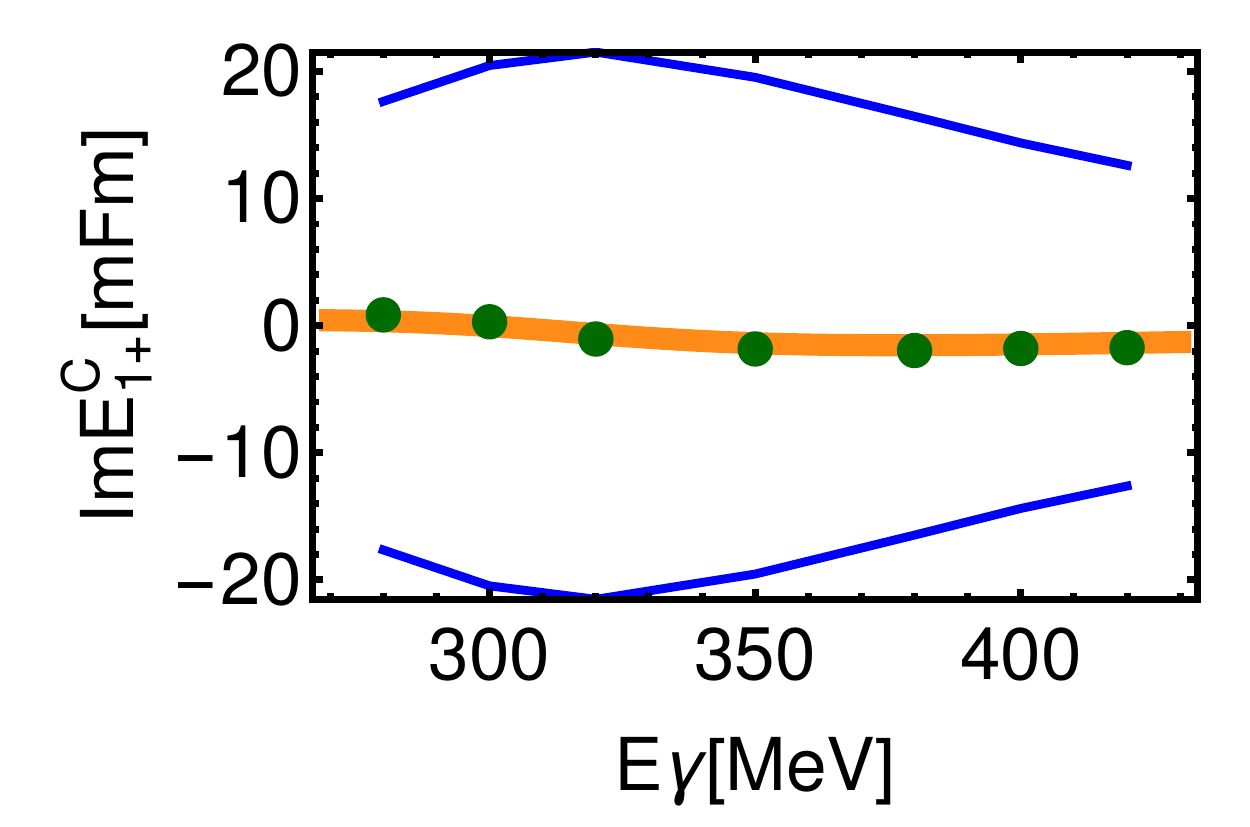}
 \end{overpic} \\
\begin{overpic}[width=0.325\textwidth]{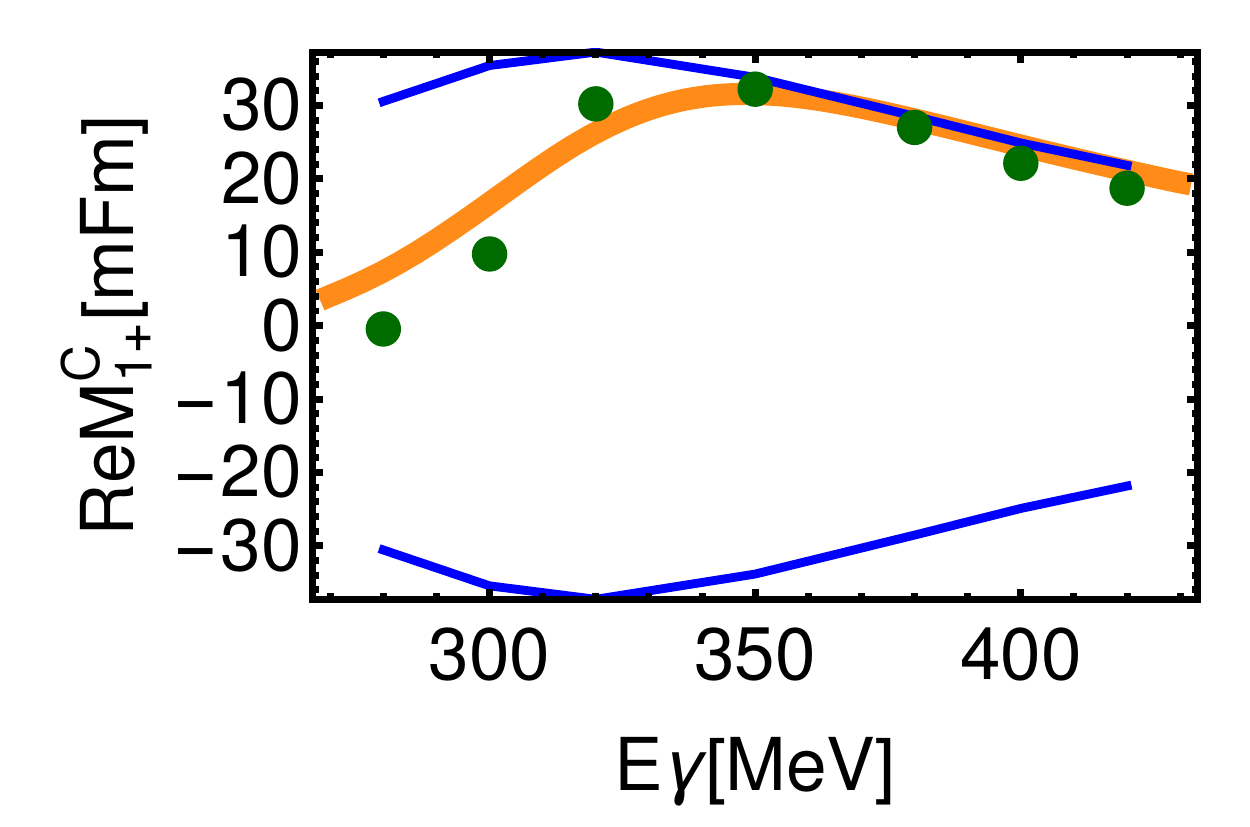}
 \end{overpic}
\begin{overpic}[width=0.325\textwidth]{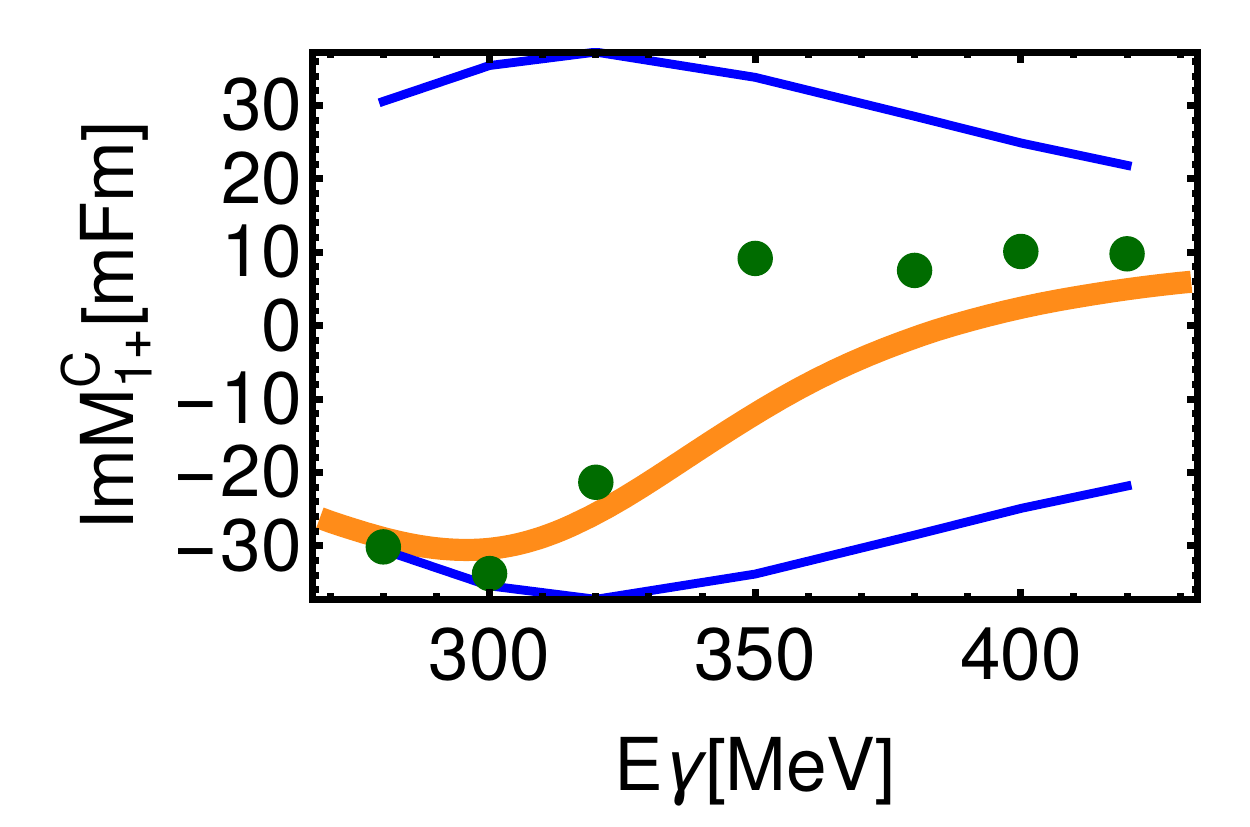}
 \end{overpic}
\begin{overpic}[width=0.325\textwidth]{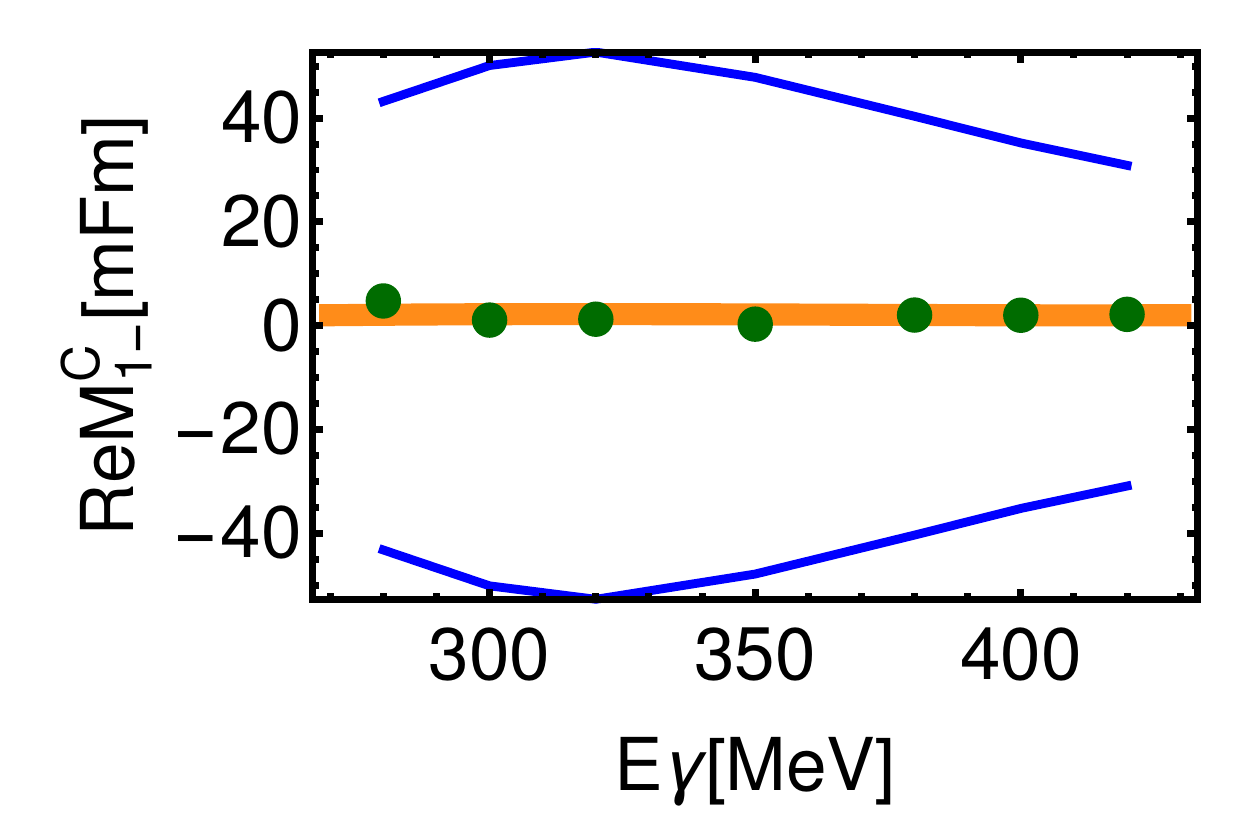}
 \end{overpic} \\
\begin{overpic}[width=0.325\textwidth]{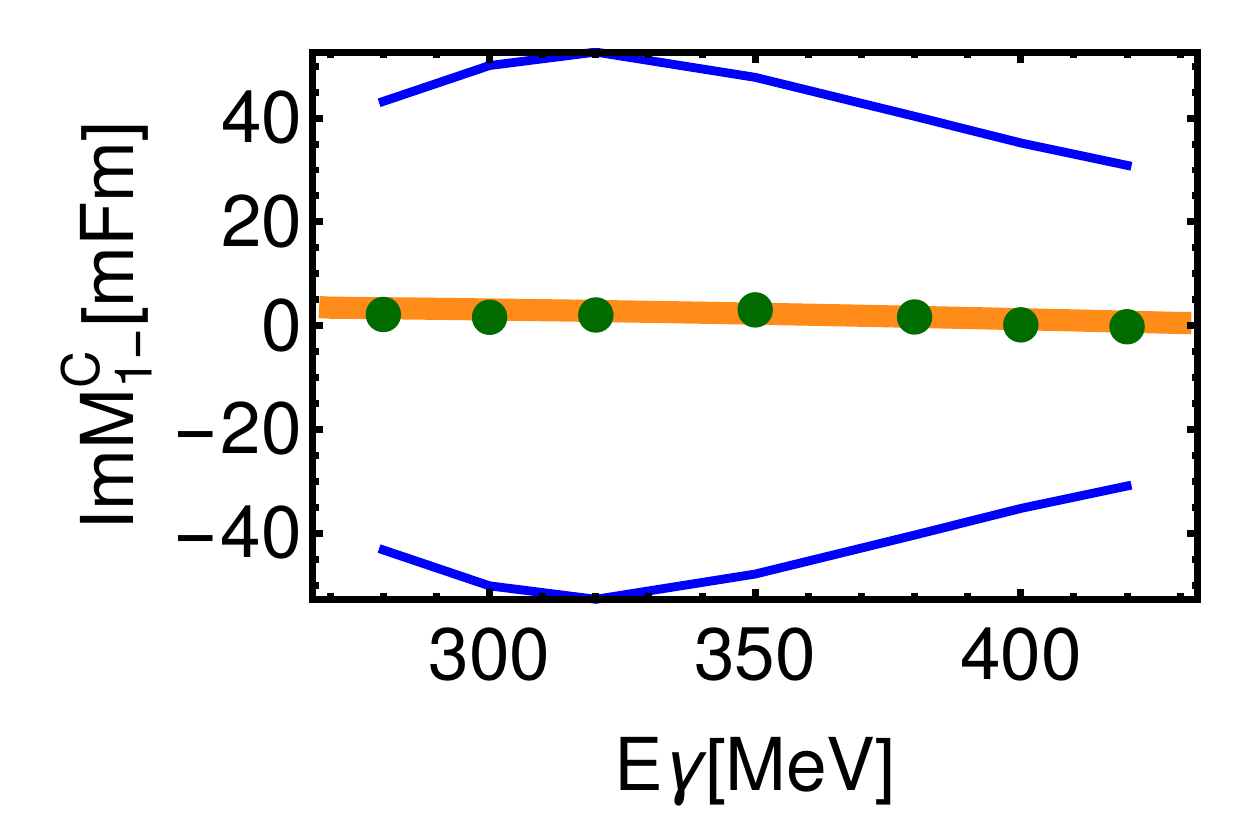}
 \end{overpic}
\caption[Results for the $7$ fit-parameters comprised of the real- and imaginary parts of phase-constrained $S$- and $P$-wave multipoles, as well as values for $\chi^{2} / \mathrm{ndf}$, for a model-independent TPWA with $\ell_{\mathrm{max}} = 1$ within the $\Delta$-resonance region.]{The given plots summarize the results of the fully unconstrained TPWA-fit with $\ell_{\mathrm{max}} = 1$ to the selected data in the $\Delta$-region. The fits employed a pool of $N_{MC} = 1000$ initial parameter configurations (cf. section \ref{sec:MonteCarloSampling}). \newline a.) {\it Left:} The obtained values for $\chi^{2} / \mathrm{ndf}$ are plotted vs. energy for the fit method of Grushin (\ref{eq:CorrelatedChisquareQuotedForAnalysis}); {\it Right:} The same plot for a direct fit to the data (\ref{eq:ChiSquareDirectFitRealDataFitSection}). In both cases, one global minimum is found (green dots). From the corresponding theoretical chisquare distributions, the mean (red line) as well as the pair of $0.025$- and $0.975$-quantiles (green dashed lines) are shown. For the estimates of $\mathrm{ndf}$ in both cases, as well as the meaning of the quantiles, see the main text (especially Figure \ref{fig:ExampleChiSquareDistributionDeltaRegion}). \newline b.) Results are shown for the fit-parameters, i.e. the real- and imaginary parts of the phase-constrained $S$- and $P$-wave multipoles. The global minimum is indicated by green dots and compared to the energy-dependent model-solution SAID CM12 (orange colored curves) \cite{WorkmanEtAl2012ChewMPhotoprod, SAID} (The SAID-solution has been rotated to the same phase-constraint!). For each fit-parameter, the maximal range set by the total cross section (cf. discussion in section \ref{sec:MonteCarloSampling}), as well as its energy-variation, is indicated by the blue solid lines. Results are shown for the fit to Legendre-coefficients (\ref{eq:CorrelatedChisquareQuotedForAnalysis}), but for a direct fit to the data (\ref{eq:ChiSquareDirectFitRealDataFitSection}), the plots look exactly the same.
}
\label{fig:Lmax1UnconstrainedFitResultsDeltaRegion}
\end{figure}

In the first attempt to analyze the data, we varied the $S$- and $P$-wave multipoles freely in the fit while setting all higher multipoles to zero. Furthermore, the constraint $\mathrm{Im} \left[E_{0+}\right] = 0$ $\&$ $\mathrm{Re}\left[E_{0+}\right] > 0$ has been employed in order to fix the overall phase\footnote{Thus the superscript '$C$' is written on all multipoles in all the following plots (meaning 'constrained').}, just as suggested in section \ref{sec:TPWAFitsIntro}. The fully model-independent Monte Carlo fit-method outlined in section \ref{sec:MonteCarloSampling} has been applied, utilizing a pool of $N_{MC} = 1000$ randomly chosen initial parameter configurations. Furthermore, we performed the analysis using both fit-approaches specified above, i.e. Grushin's method of fitting the Legendre-coefficients (equation (\ref{eq:CorrelatedChisquareQuotedForAnalysis})) as well as a direct fit to the data (equation (\ref{eq:ChiSquareDirectFitRealDataFitSection})). The results are summarized in Figure \ref{fig:Lmax1UnconstrainedFitResultsDeltaRegion}, where plots for $\chi^{2}_{\mathcal{M}}/\mathrm{ndf}$, $\chi^{2}_{\mathrm{data}}/\mathrm{ndf}$ and the multipoles coming out of Grushin's method (\ref{eq:CorrelatedChisquareQuotedForAnalysis}) are shown. It should be said that we estimated $\mathrm{ndf} = 10 - (8 - 1) = 7$ for Grushin's method and $\mathrm{ndf} = 79 - (8 - 1) = 72$ for the direct fit\footnote{Not every energy-bin has 79 data-points here. Rather, one has $77$ to $79$ points, depending on the energy. This small variation has been neglected in {\it all} $\chi^{2}$-quantiles evaluated for {\it all} figures shown in this section.}. \newline
An encouraging aspect about the results is given by the fact that this fit has only one global minimum and not any local minima anywhere close in $\chi^{2}$. This is true for both the fit to Legendre coefficients and the direct fit to data. Furthermore, once this global minimum is compared to the SAID-solution CM12\footnote{This particular model has been chosen since, according to a recommendation by Tiator \cite{LotharPrivateComm}, it is particularly trustworthy in the $\Delta$-region.} \cite{WorkmanEtAl2012ChewMPhotoprod, SAID}, it is seen that there is reasonable agreement in most multipoles. A noteable exception is here given by $\mathrm{Im} \left[M_{1+} \right]$. The set $\left\{ \sigma_{0}, \Sigma, T, P, F \right\}$ is seen to be complete, at least for this order in $\ell_{\mathrm{max}}$. \newline
Furthermore, we have to state that the resulting multipole-parameters for both fit-approaches (i.e. (\ref{eq:CorrelatedChisquareQuotedForAnalysis}) and (\ref{eq:ChiSquareDirectFitRealDataFitSection})) are the same up to the first $5$ non-vanishing digits. Thus, the plots for the direct fit to the data look exactly the same as the results shown for Grushin's method in Figure \ref{fig:Lmax1UnconstrainedFitResultsDeltaRegion}. \newline
Unfortunately, the encouragement is dampened by the fact that for all energies, as well as for both employed fit-approaches, the resulting solution is nowhere near acceptable when compared to theoretical chisquare-distributions. Every minimum is outside of the $95\%$-confidence interval set by the pair of $0.025$- and $0.975$-quantiles of the corresponding distributions (see Figure \ref{fig:Lmax1UnconstrainedFitResultsDeltaRegion}). Thus, while highly successful in regard of the unique solvability, this fit would have to be rejected completely. \newline
When searching for a reason why the fit-quality is so bad, two possible main reasons come to mind. The first would be the fact that the chosen model does not fully capture all the physics hidden in the data. In case of a fully model-independent TPWA, this could mean for instance that the truncation order has to be raised. Another reason could be that the data are in some sense {\it defective}, or at least not well suited for the applied analysis-scheme. \newline
When considering the $\chi^{2}$-plots for Legendre-fits shown in Figure \ref{fig:DeltaRegionChisquareLmaxPlots} above, it is clear that the differential cross section $\sigma_{0}$ measured by Hornidge et al. \cite{Hornidge:2013} demands much higher orders than $\ell_{\mathrm{max}} = 1$ or even $2$ to be fitted correctly. This is consistent with the average size of the statistical errors, which are extremely tiny, mostly within the range of $1\%$ of the magnitude of the cross section (Table \ref{tab:DeltaRegionSandorfiDataTable}). \newline
Regarding the sytematic errors of this dataset, reference \cite{Hornidge:2013} cites an upper limit of $4\%$ for the relative size of such uncertainties.

\clearpage

\clearpage

 Thus, here one has a dataset where {\it the systematic error dominates the statistical uncertainties}. This is a by-product of the immense statistics that can be gathered with today's experimental setups. For our analysis, the fact is now questioned whether the higher partial waves demanded by the Legendre-fits shown in Figure \ref{fig:DeltaRegionChisquareLmaxPlots} are actually physical, or just an artifact of large systematic uncertainties, which are not taken into account correctly. The latter scenario seems more probable, since according to older single-energy analyses \cite{Grushin} or modern energy-dependent models \cite{SAID,MAID,BoGa}, all multipoles with $\ell_{\mathrm{max}} \geq 2$ are known to be small, at least for the $\pi^{0}$-production channel and in the lower energy region. \newline
 Thus, the suspicion arises that the systematic errors, in particular for the differential cross section, have an important role to play. Therefore, in order to be safe we now continue solely with the Ansatz of fitting directly to data (equation (\ref{eq:ChiSquareDirectFitRealDataFitSection})). Grushin's method of fitting to the Legendre-coefficients has yielded fully equivalent results in all cases up to this point, still we abandon it here based on the assumption that the method, and especially the resulting values of $\chi^{2}_{\mathcal{M}}$, are more susceptible to systematic errors (cf. section \ref{sec:TPWAFitsIntro}). \newline
 
 For the moment, we assume the first reason for bad fit-quality given above to be true, i.e. that our model in the previous fit-attempt was not able to extract all the physics hidden in the data. Thus, we raise the truncation order to $\ell_{\mathrm{max}} = 2$ and make a fully model-independent fit, varying all $S$-, $P$- and $D$-waves under the same phase-constraint as used above. The Monte Carlo sampling is performed using $N_{MC} = 8000$ and the direct fit, equation (\ref{eq:ChiSquareDirectFitRealDataFitSection}), is used. We incorporate only the statistical errors of the data, as in the previous fit. Results can be seen in Figure \ref{fig:Lmax2UnconstrainedFitResultsDeltaRegion}. \newline
 Again, a global minimum is obtained, but now many local minima are found as well, many of them having a comparable quality in $\chi^{2}$. In all energy-bins, we count at least $10$ non-redundant minima in total, sometimes even more, up to $20$. The value of $\chi^{2}/\mathrm{ndf}$ has improved for the global minimum, but is for most bins not yet in perfect accordance with the expectations from theoretical chisquare-distributions. Only for the highest energy does the best solution fall into the $95\%$ confidence-interval. \newline
 A shocking observation is made once the obtained solutions are compared to the SAID-model (Figure \ref{fig:Lmax2UnconstrainedFitResultsDeltaRegion}): The global minimum shows large deviations to the model. It has a relatively small $S$-wave $E_{0+}$, which is fine, but the resonant multipole $M_{1+}$ comes out too small for most energies. In return, most of the $D$-waves are much too large! \newline
 In an attempt to find out whether the fit makes any sense at all, we looped through all non-redundant solutions in order to find the one which is most reasonably in accord with the SAID-solution. In order to do this, the solution was searched which describes the $E_{0+}$- and the resonant $M_{1+}$-multipole the best, i.e. for which the distance-quantity
 \begin{equation}
  \left| E_{0+}^{\mathrm{SAID}} - E_{0+}^{\mathrm{sol.}}  \right|^{2} + \left| M_{1+}^{\mathrm{SAID}} - M_{1+}^{\mathrm{sol.}} \right|^{2} \mathrm{,} \label{eq:}
 \end{equation}
becomes minimal. Indeed, such a solution exists and it turns out not only to match the course of the above-used multipoles from SAID pretty well, but also almost all the remaining partial waves. The particular solution is shown in Figure \ref{fig:Lmax2UnconstrainedFitResultsDeltaRegion}, where it is also seen to have a quality in $\chi^{2} / \mathrm{ndf}$ which is not a lot worse than that of the global minimum. We have chosen not to plot the plethora of multipole-solutions in Figure \ref{fig:Lmax2UnconstrainedFitResultsDeltaRegion}, for reasons of clarity. Instead, they can be viewed in Figure \ref{fig:Lmax2UnconstrainedFitMultipoleResultsDeltaRegionPlethora} of Appendix \ref{sec:NumericalTPWAFitResults}.\newline
The global minimum is, from a physical standpoint, unacceptable, even though it may be the best solution in $\chi^{2}$. We have encountered a phenomenon about multipole-fits also mentioned by Tiator in his proceeding \cite{Tiator:2011tu}:

\begin{figure}[ht]
 \centering
 \vspace*{-17.5pt}
\begin{overpic}[width=0.465\textwidth]{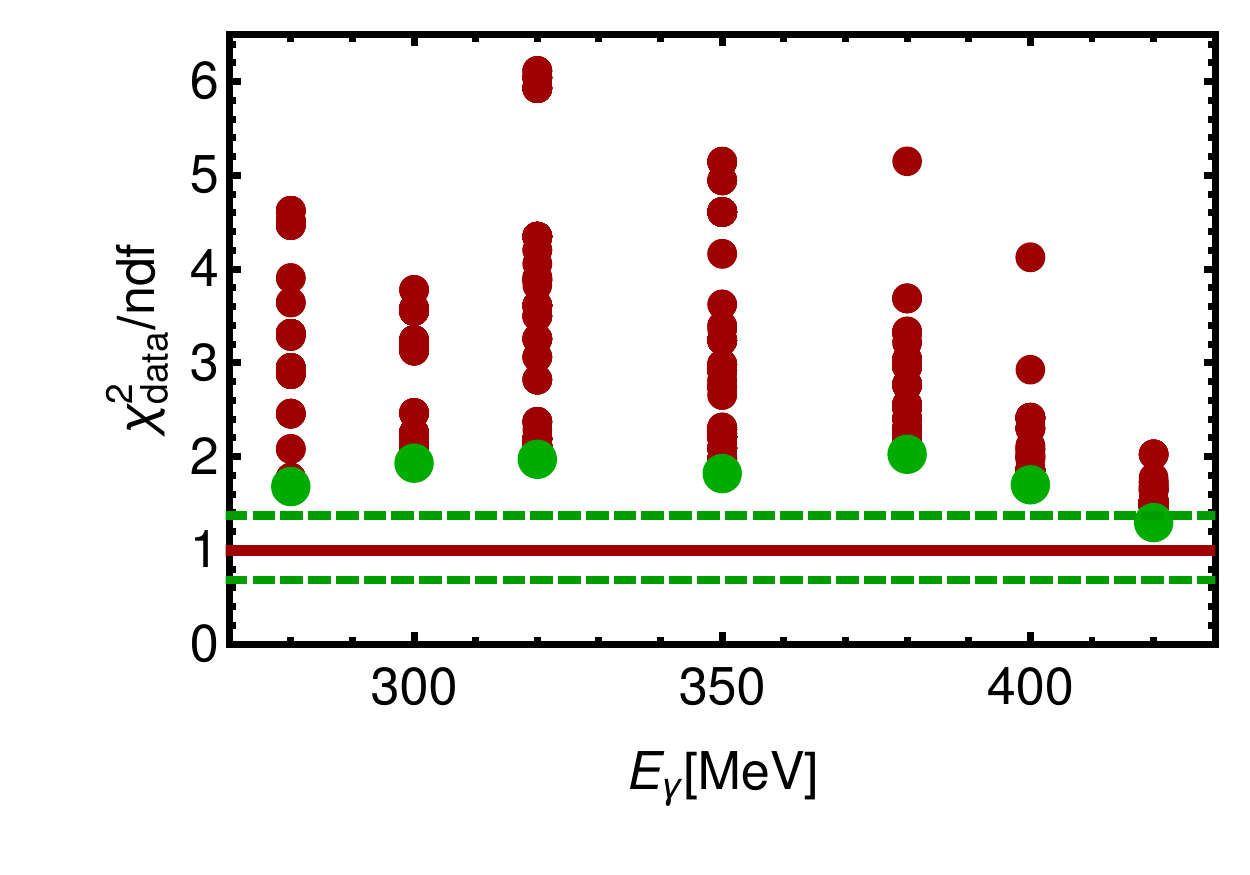}
 \put(-5,64){a.)}
 \end{overpic}
\begin{overpic}[width=0.465\textwidth]{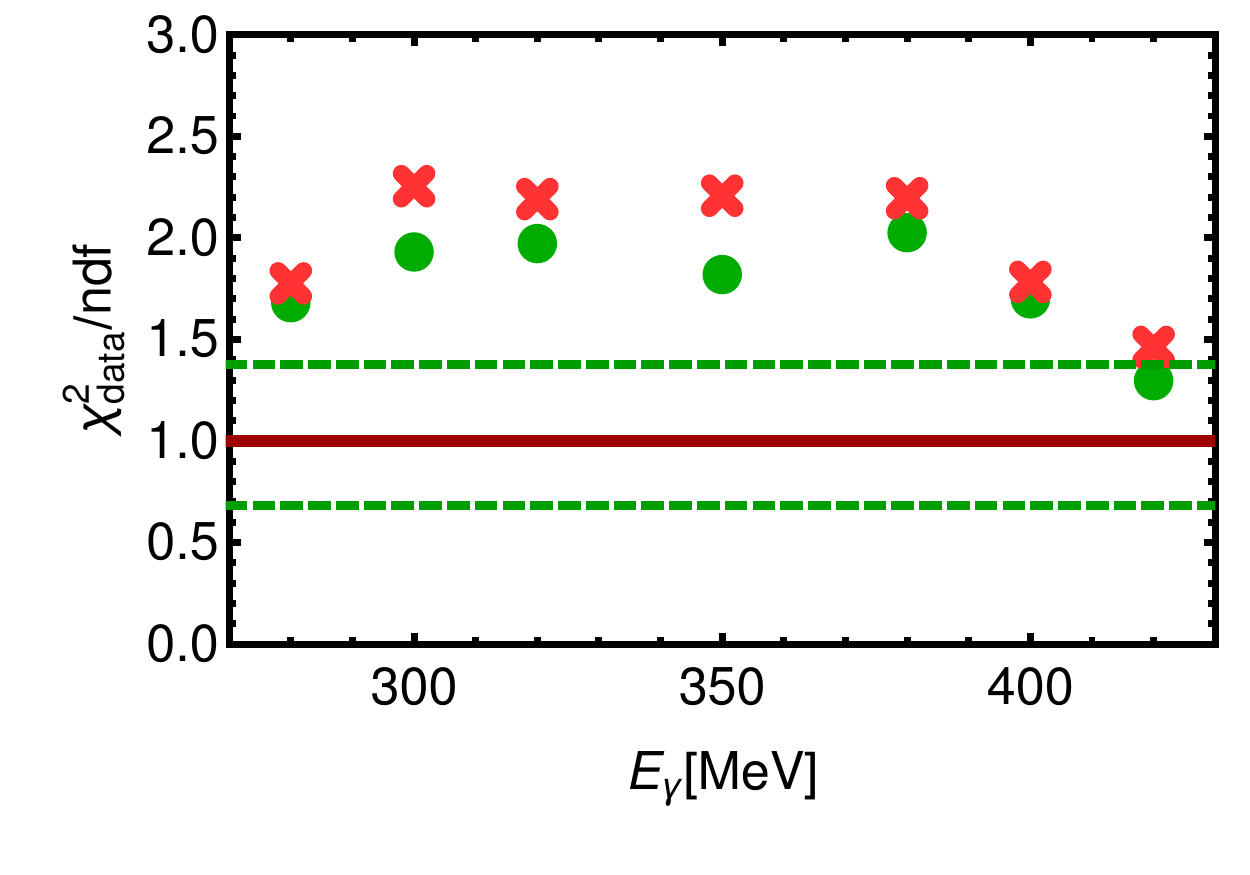}
 \end{overpic} \\
 \vspace*{-10pt}
\begin{overpic}[width=0.325\textwidth]{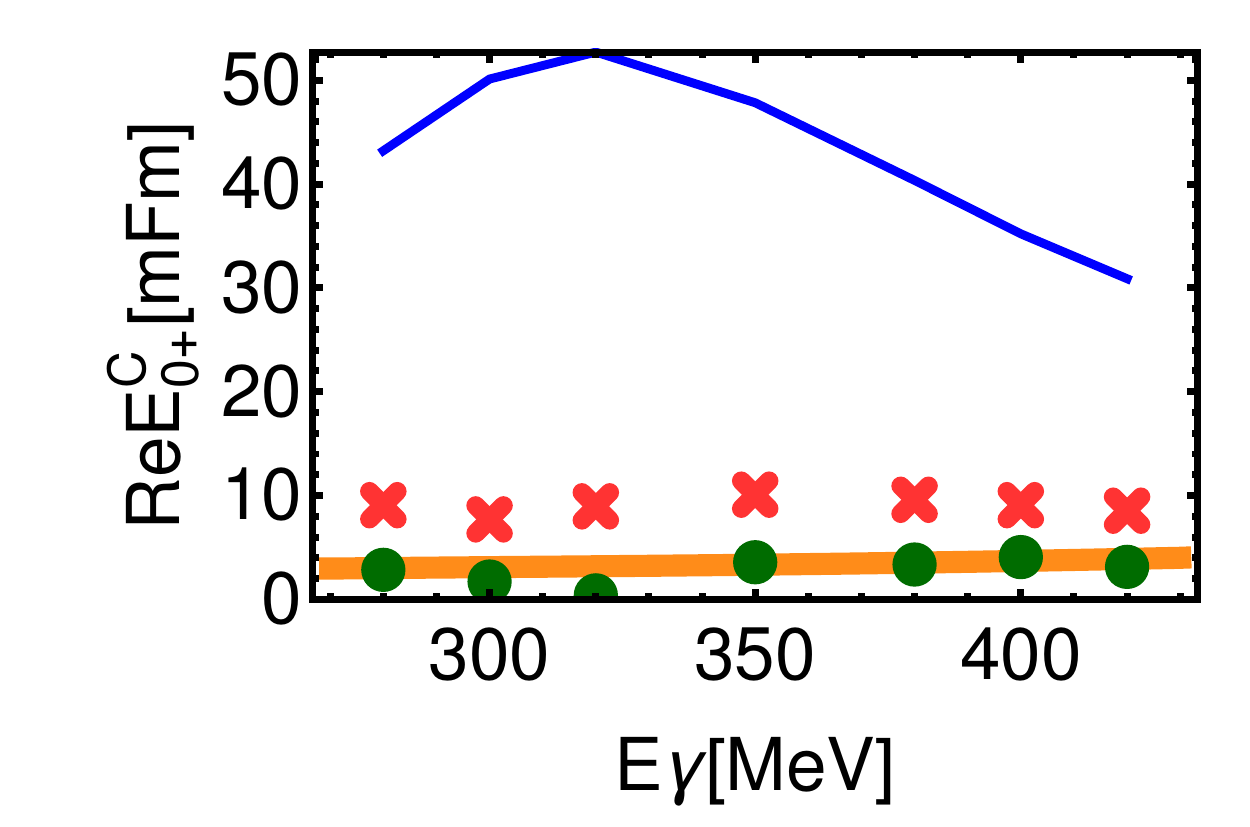}
 \put(0.5,60){b.)}
 \end{overpic}
\begin{overpic}[width=0.325\textwidth]{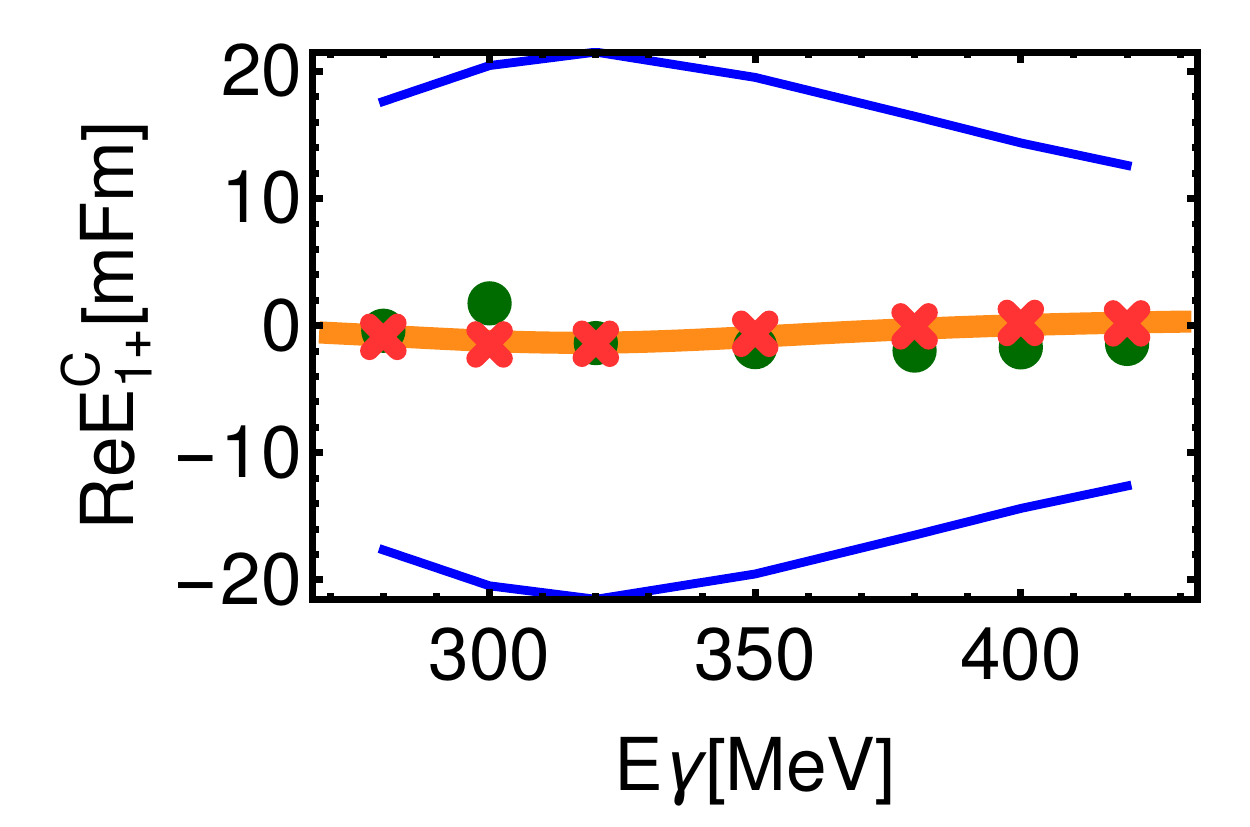}
 \end{overpic}
\begin{overpic}[width=0.325\textwidth]{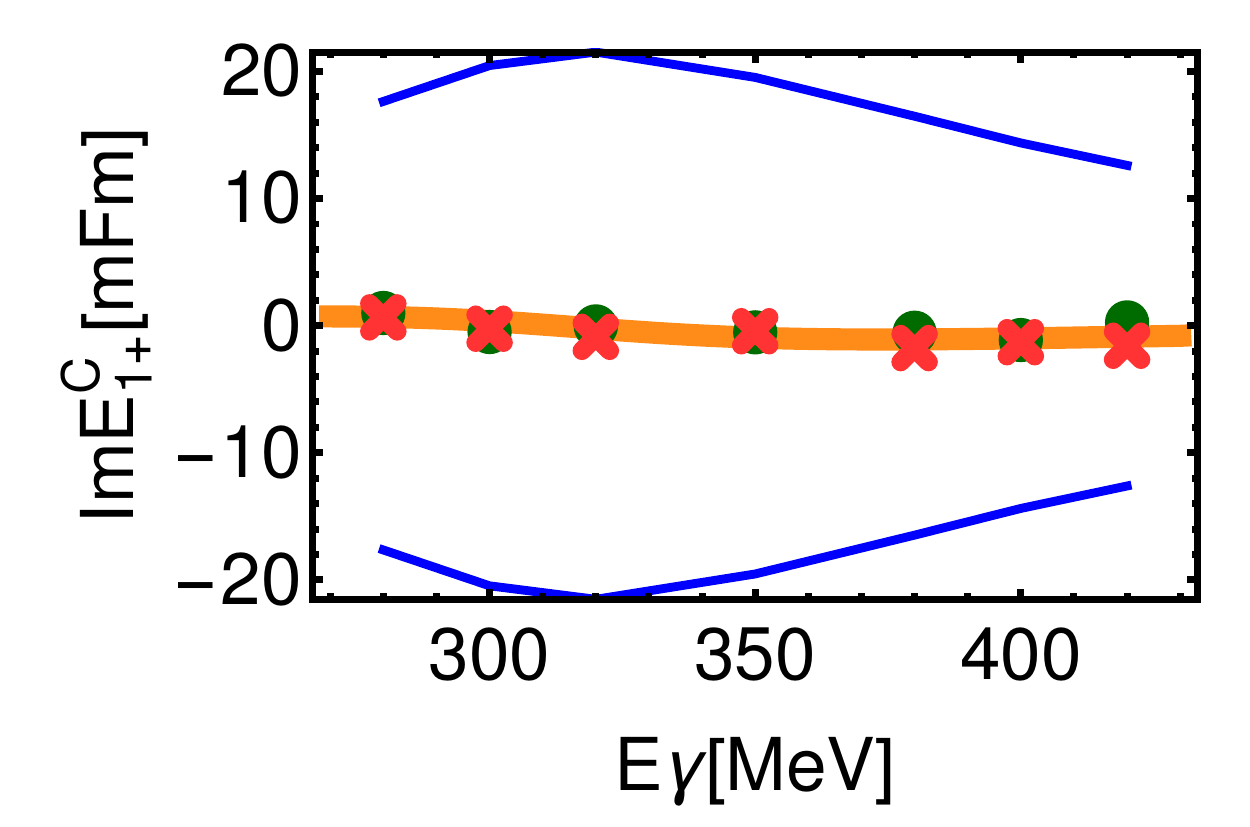}
 \end{overpic} \\
\begin{overpic}[width=0.325\textwidth]{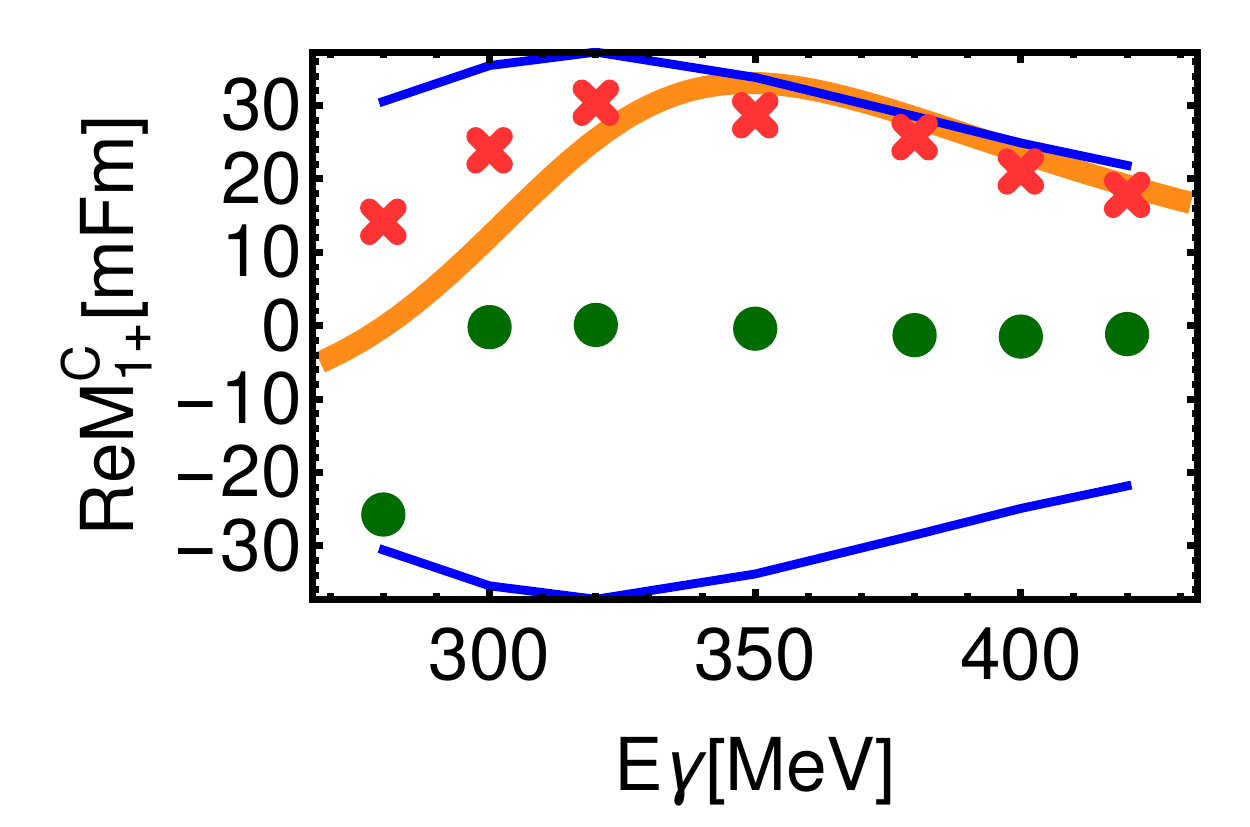}
 \end{overpic}
\begin{overpic}[width=0.325\textwidth]{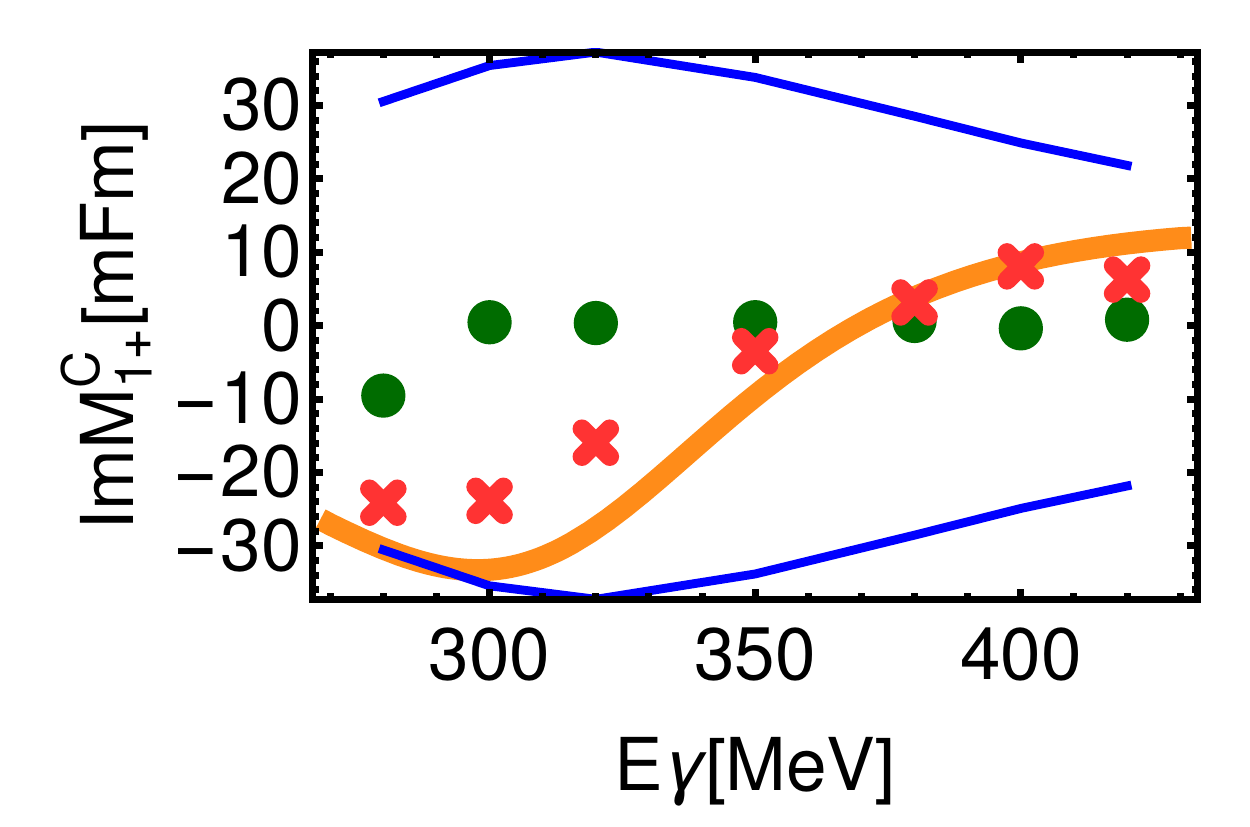}
 \end{overpic}
\begin{overpic}[width=0.325\textwidth]{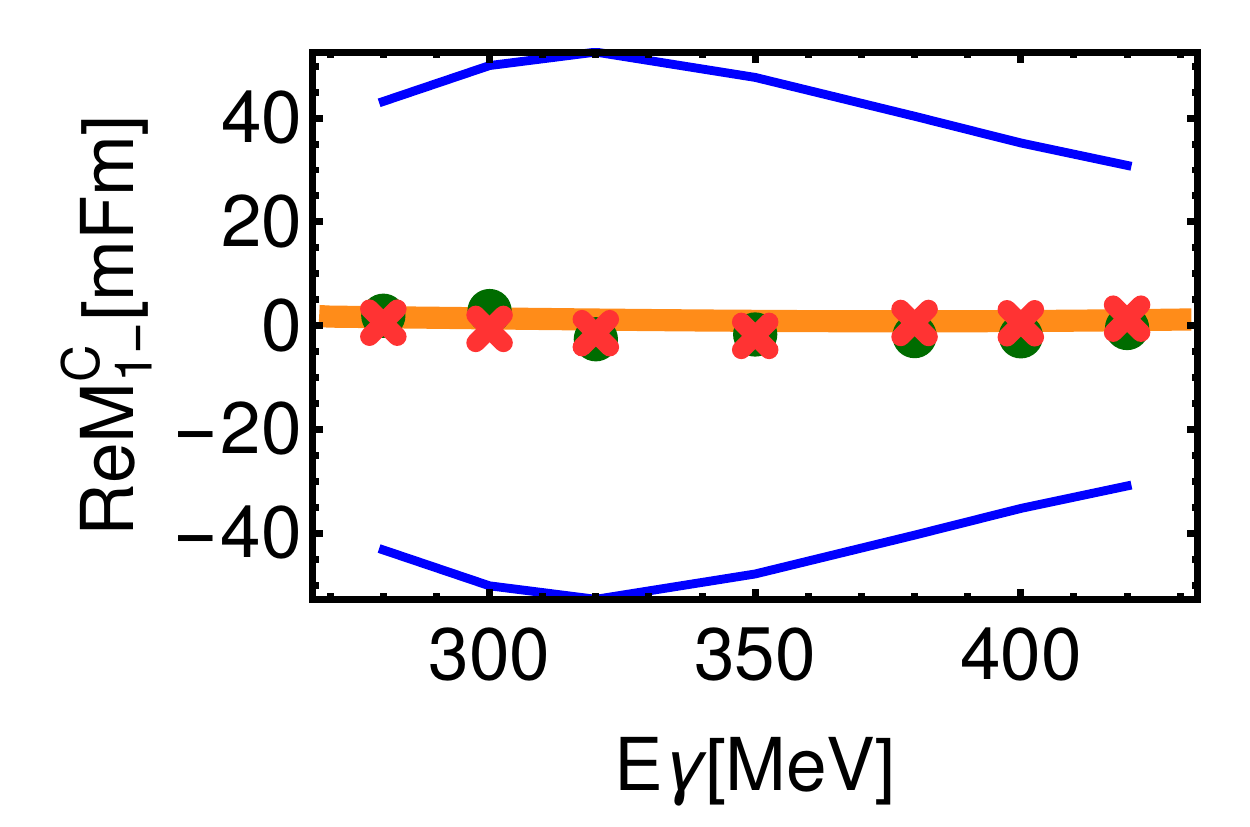}
 \end{overpic} \\
\begin{overpic}[width=0.325\textwidth]{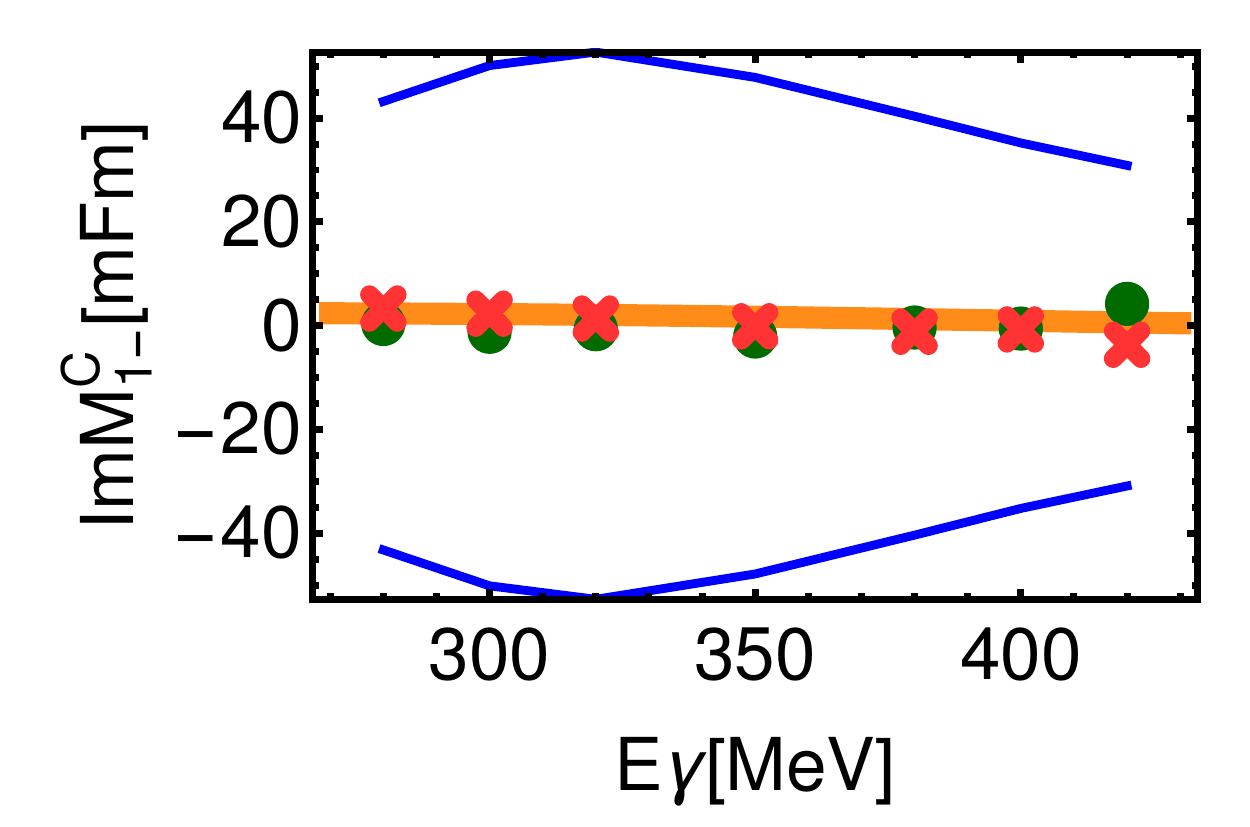}
 \end{overpic}
\begin{overpic}[width=0.325\textwidth]{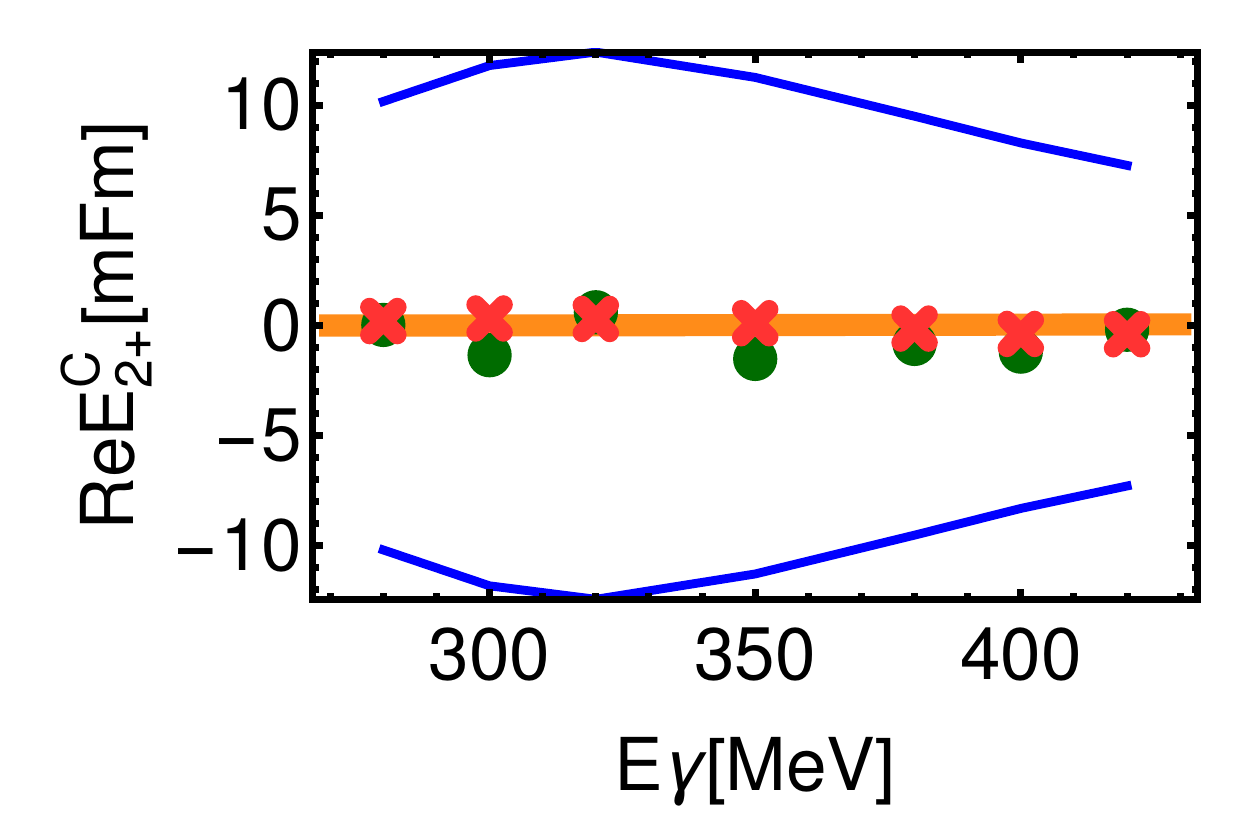}
 \end{overpic}
\begin{overpic}[width=0.325\textwidth]{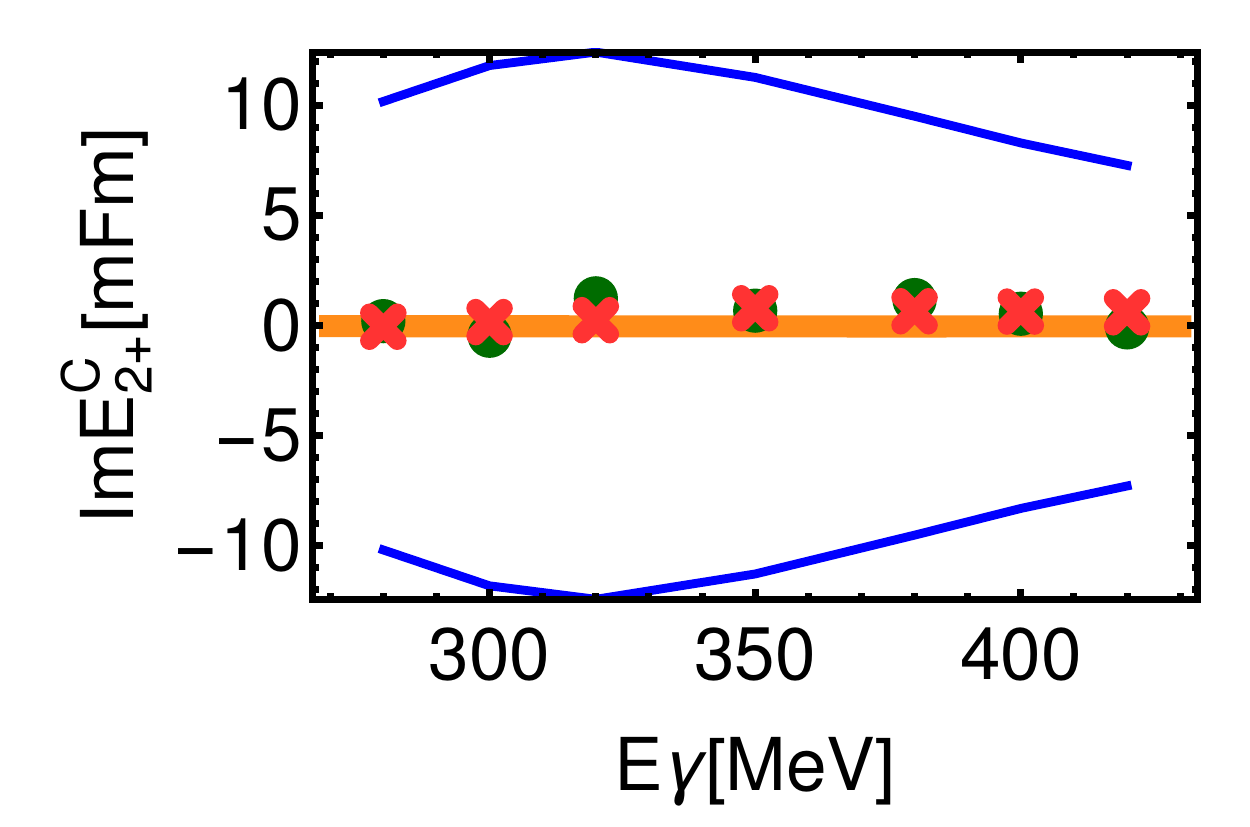}
 \end{overpic} \\
\begin{overpic}[width=0.325\textwidth]{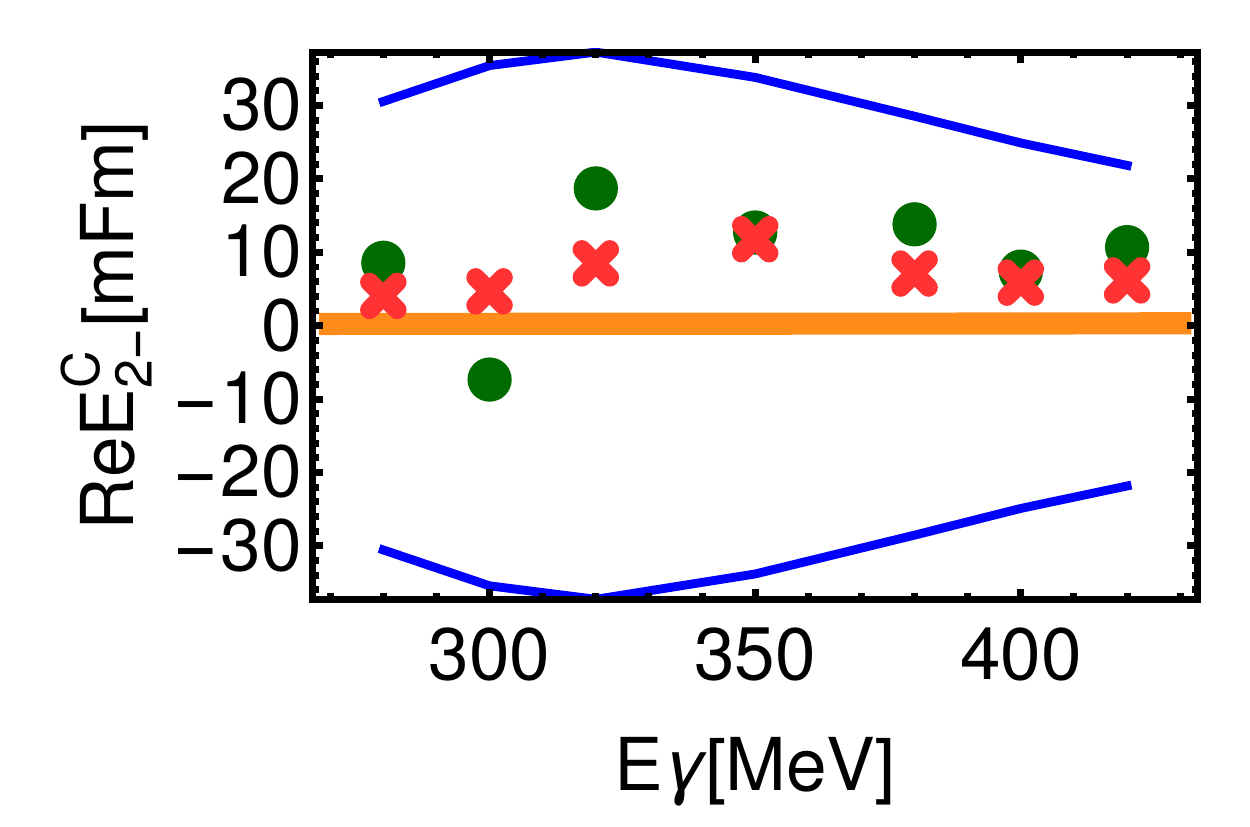}
 \end{overpic}
\begin{overpic}[width=0.325\textwidth]{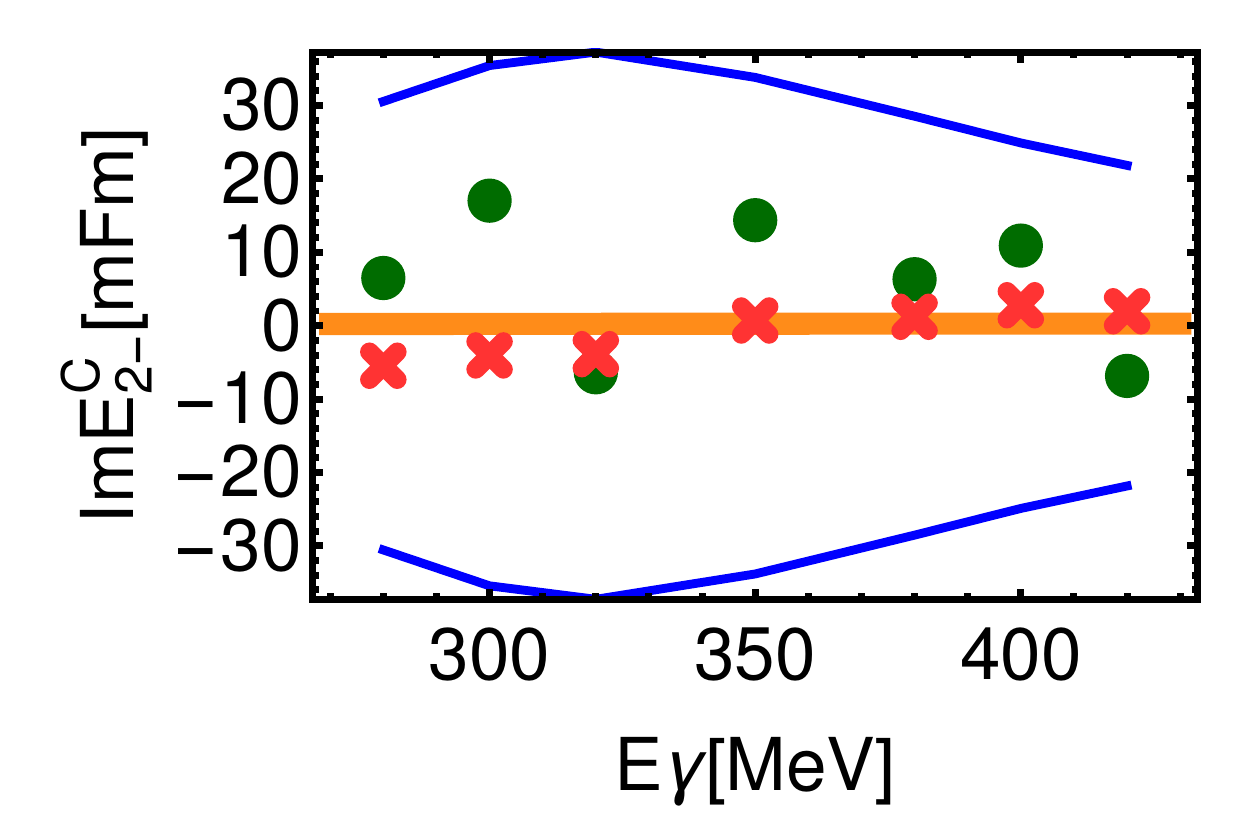}
 \end{overpic}
\begin{overpic}[width=0.325\textwidth]{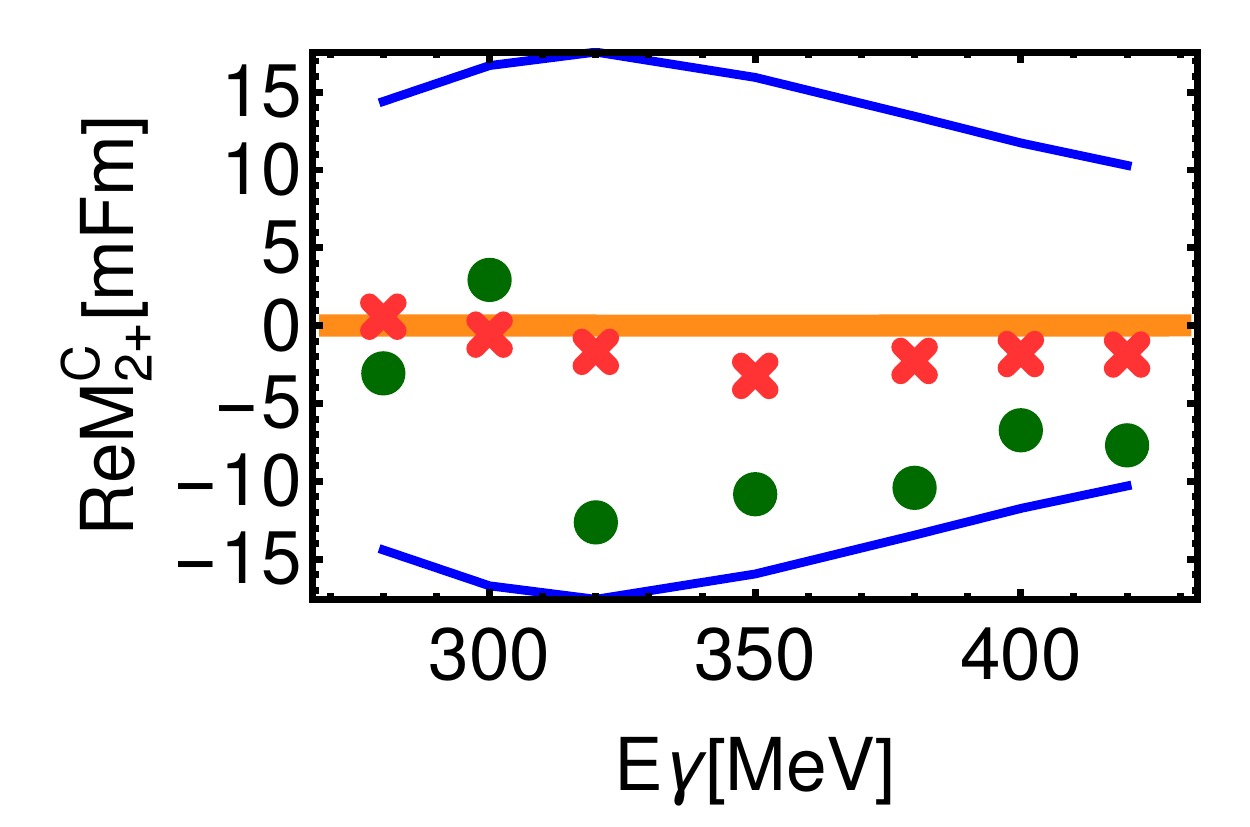}
 \end{overpic} \\
\begin{overpic}[width=0.325\textwidth]{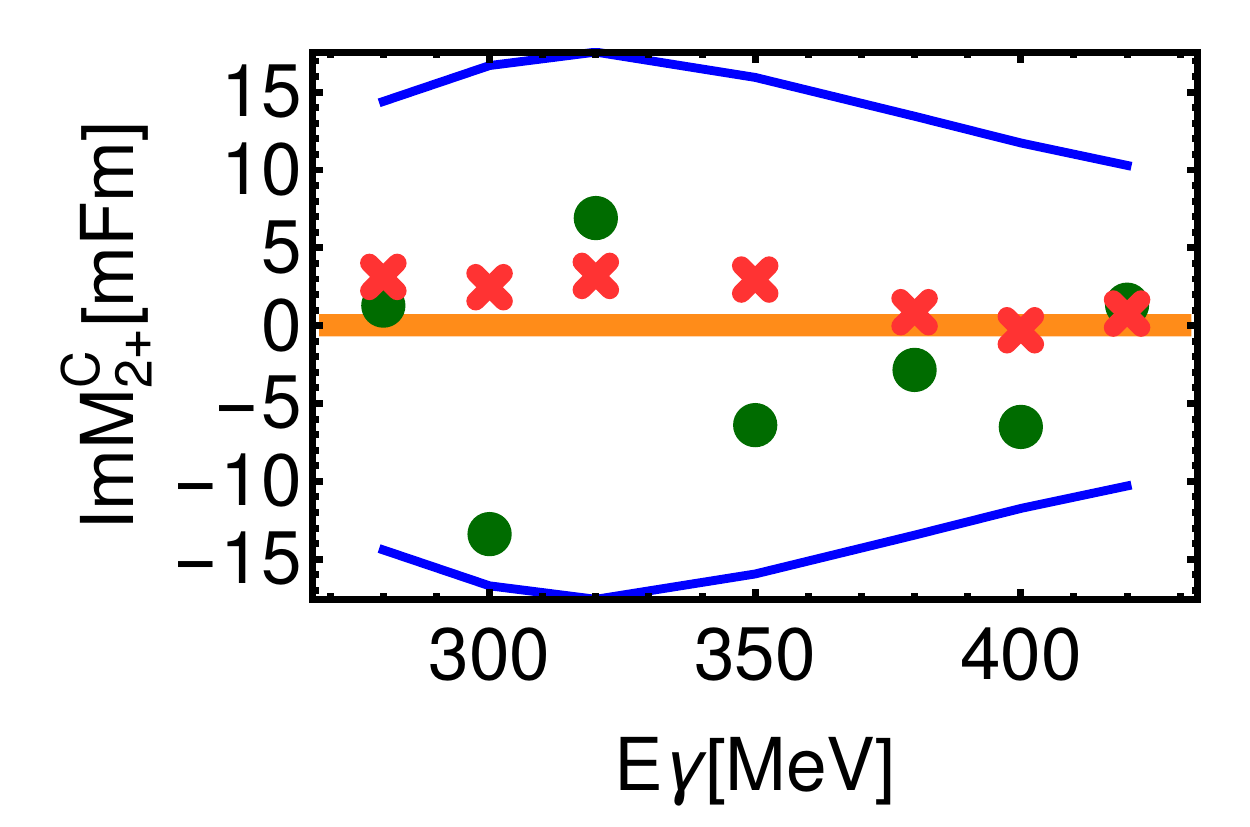}
 \end{overpic}
\begin{overpic}[width=0.325\textwidth]{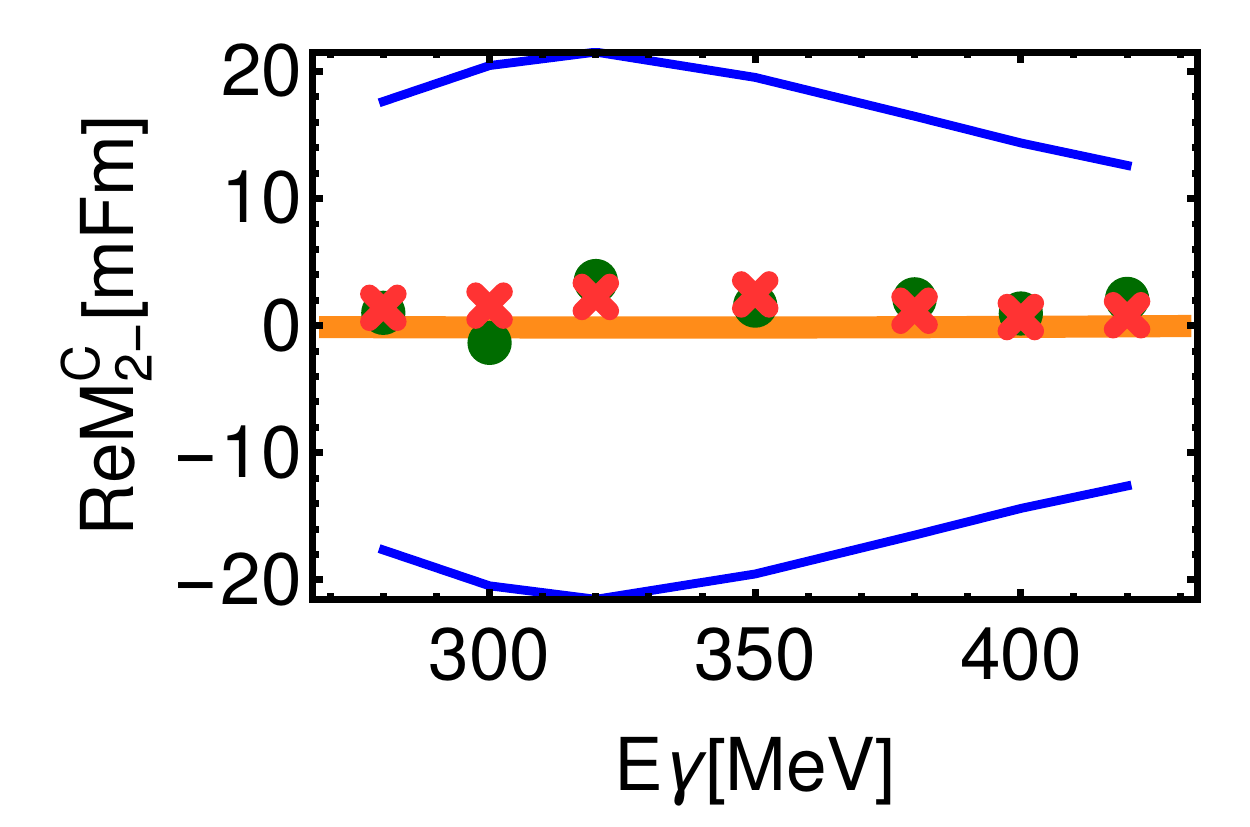}
 \end{overpic}
\begin{overpic}[width=0.325\textwidth]{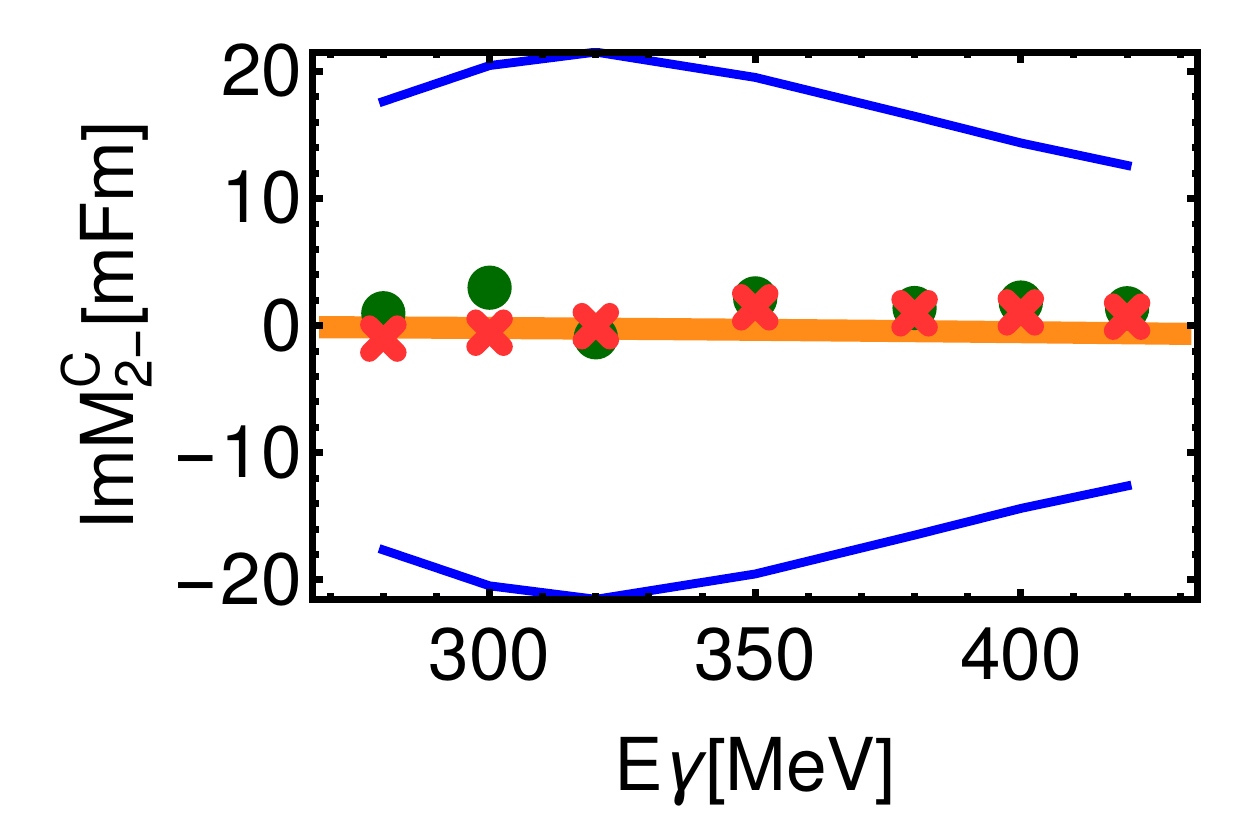}
 \end{overpic} \vspace*{-2pt}
\caption[Results for the $\chi^{2} / \mathrm{ndf}$ plotted vs. energy, as well as multipoles, are shown for the unconstrained TPWA with $\ell_{\mathrm{max}} = 2$ to a set of $5$ observables in the $\Delta$-region]{Results for an unconstrained TPWA with: $\ell_{\mathrm{max}} = 2$, $N_{MC} = 8000$. \newline a.) {\it Left:} The $\chi^{2} / \mathrm{ndf}$ is plotted vs. energy for the global minimum (large green dots), as well as all remaining non-redundant solutions obtained from the pool (smaller red dots). {\it Right:} The same plot, but only containing the global minimum (large green dots) and a particular solution closest to the SAID CM12 $E_{0+}$- and $M_{1+}$-multipoles \cite{SAID} (red crosses). \newline b.) Multipole-solutions are shown. The global minimum (green dots) and the local minimum closest to SAID CM12 (red crosses) are compared to the SAID CM12-solution \cite{SAID} (solid orange line).} 
\label{fig:Lmax2UnconstrainedFitResultsDeltaRegion}
\end{figure}

\clearpage

\begin{figure}[h]
 \centering
 \vspace*{0pt}
 \begin{overpic}[width=0.475\textwidth]{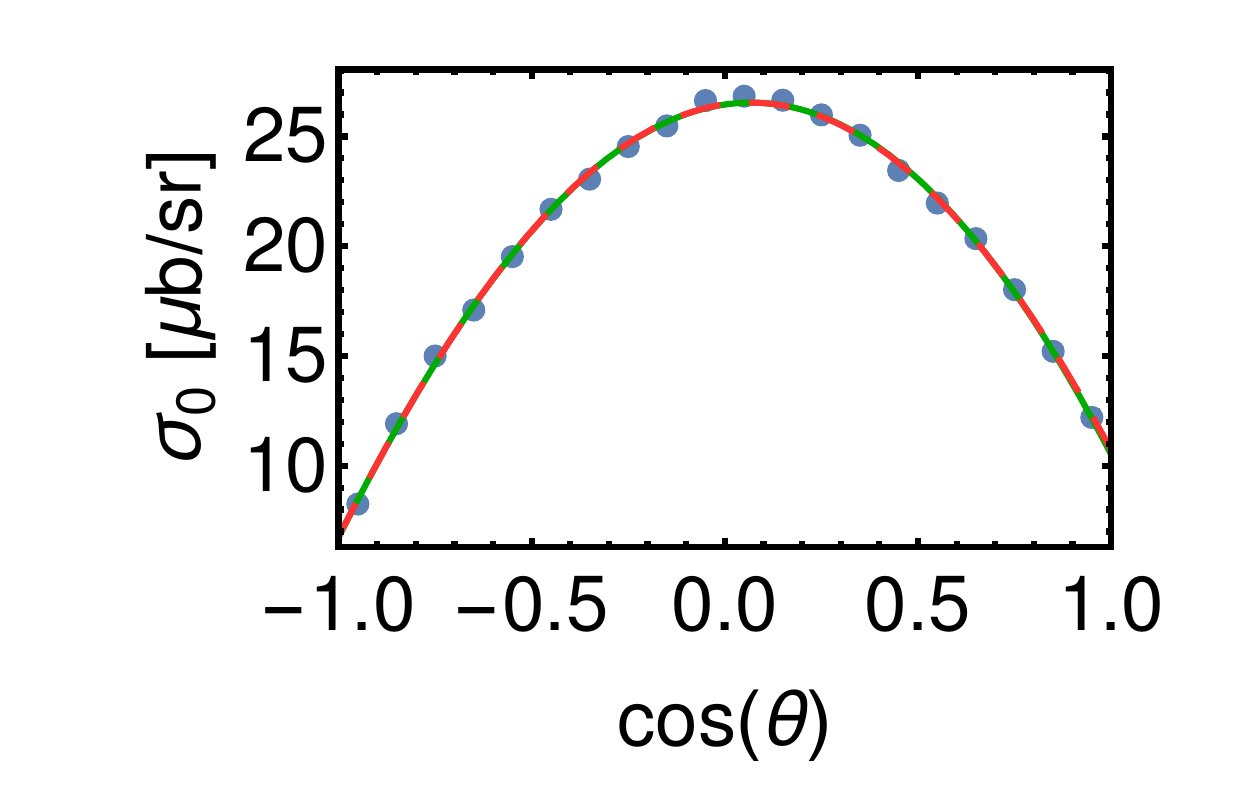}
 \put(81.5,62.5){\begin{Large}$E_{\gamma} = 350. \hspace*{2pt} \mathrm{MeV}$\end{Large}}
 \end{overpic}
 \begin{overpic}[width=0.475\textwidth]{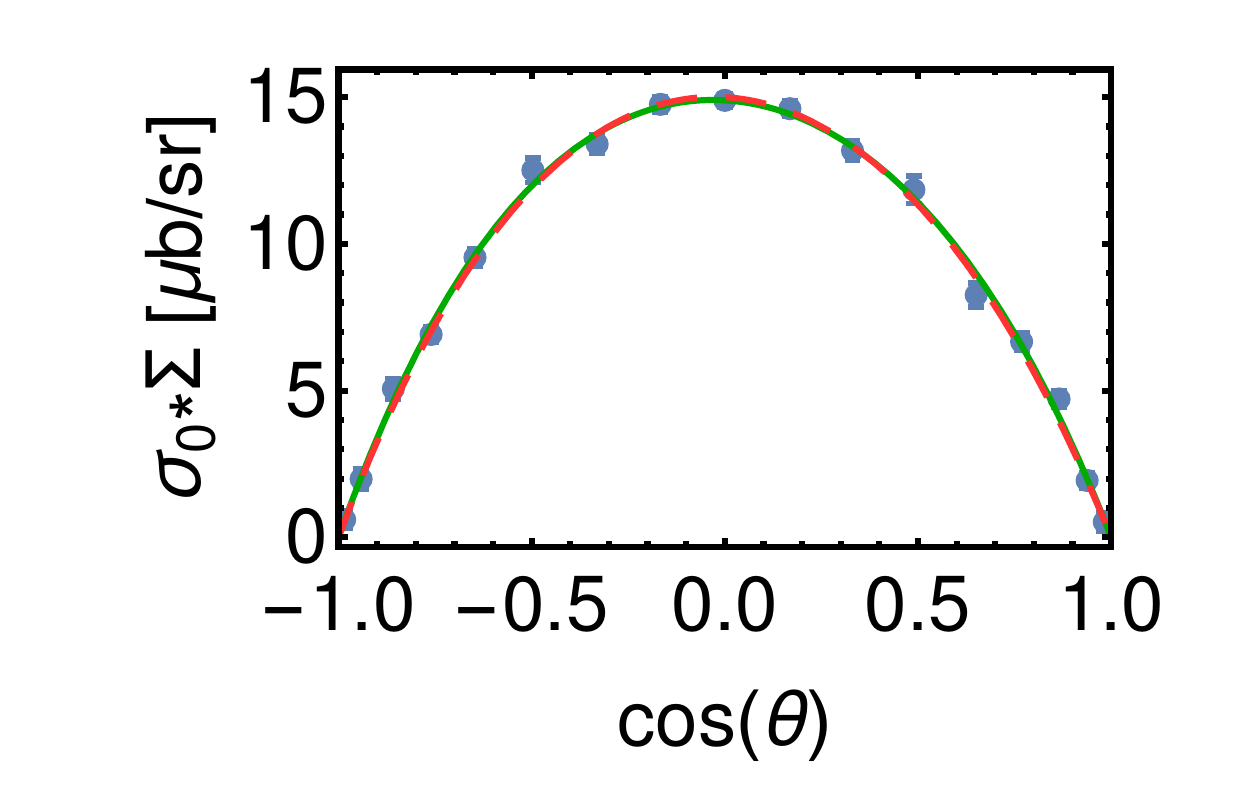}
 \end{overpic} \\
 \begin{overpic}[width=0.475\textwidth]{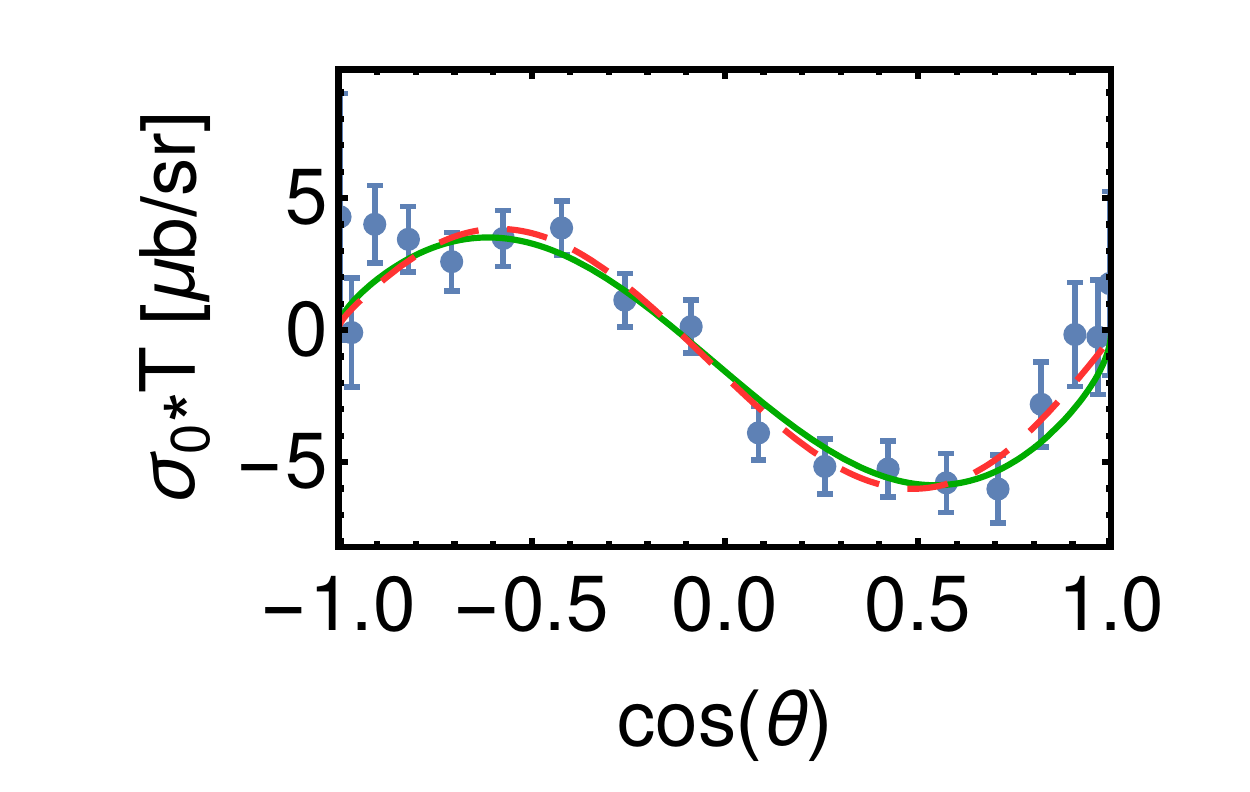}
 \end{overpic}
 \begin{overpic}[width=0.475\textwidth]{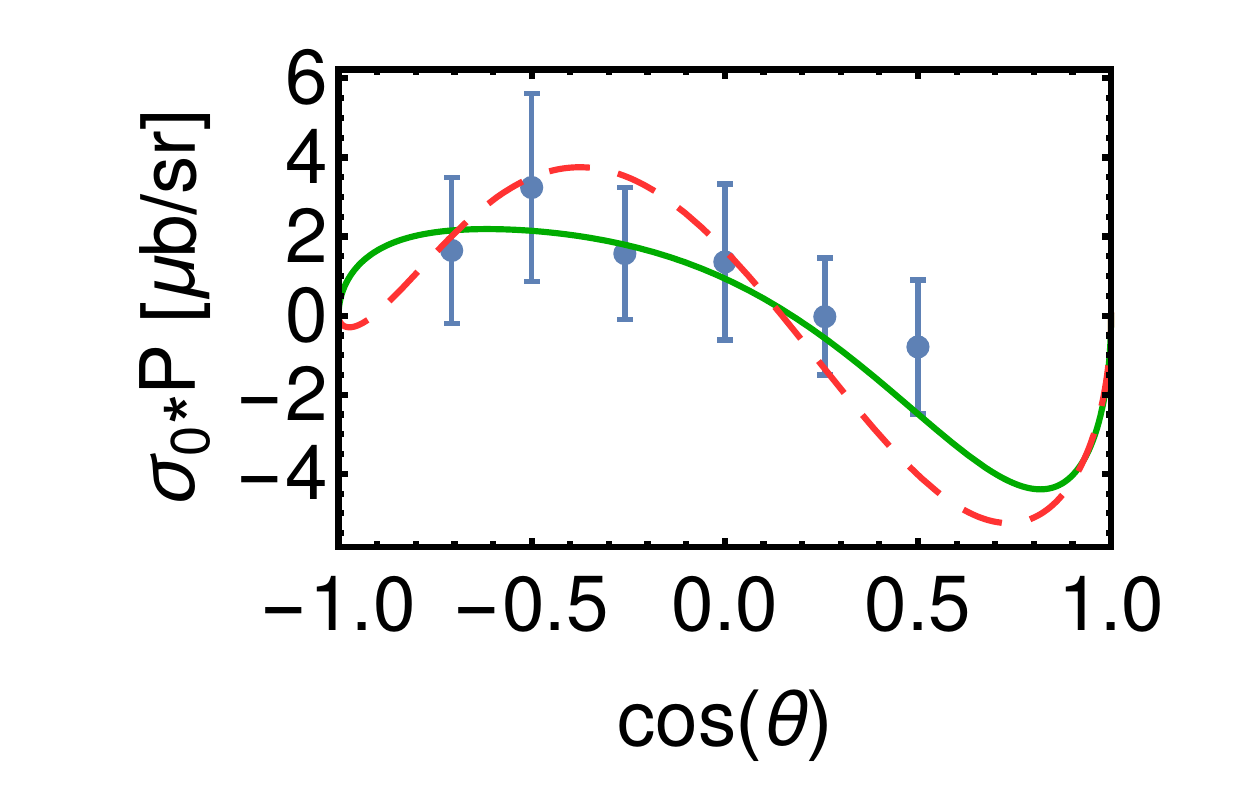}
 \end{overpic} \\
 \begin{overpic}[width=0.475\textwidth]{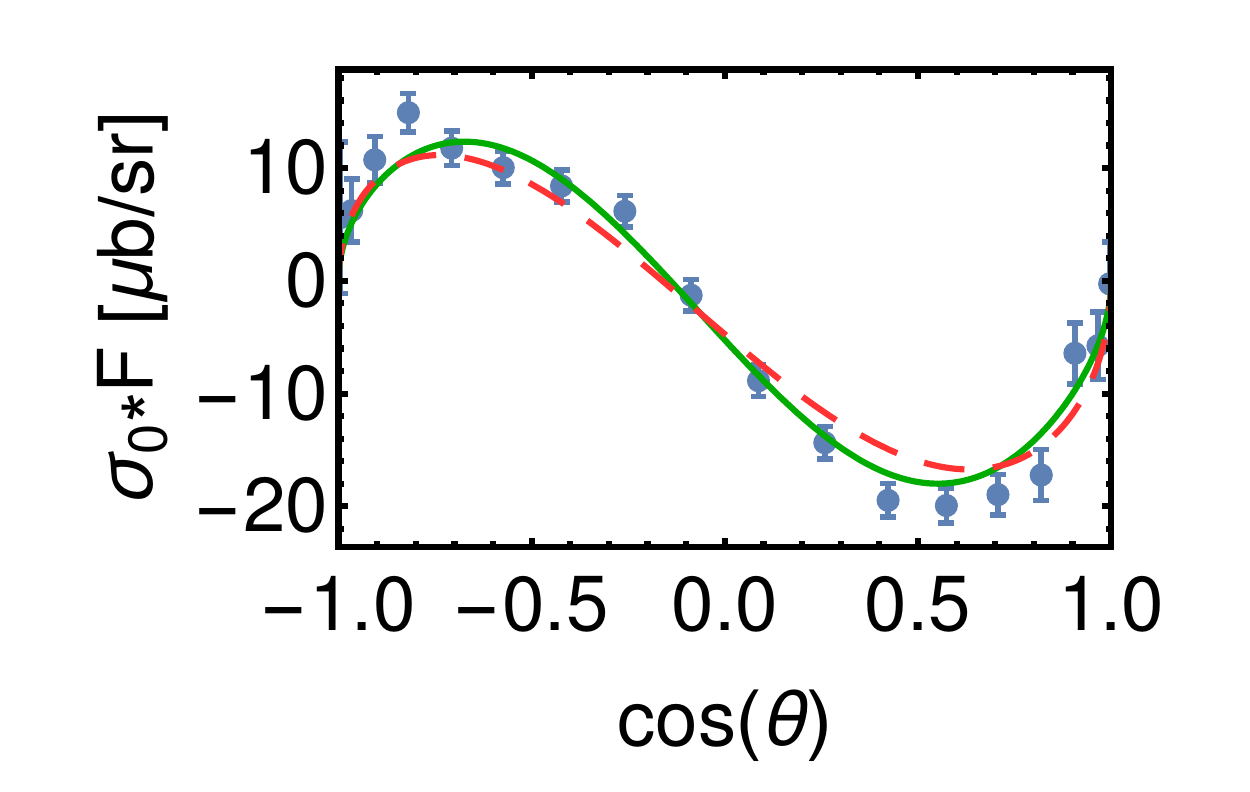}
 \end{overpic}
  \caption[Angular distributions of profile functions for the five polarization-datasets analyzed in the $\Delta$-resonance region. Data are plotted at a particular energy, $E_{\gamma} = 350. \hspace*{1pt} \mathrm{MeV}$. The global minimum for an unconstrained TPWA at $\ell_{\mathrm{max}} = 2$ is shown, as well as one ambiguity.]{Angular distributions for the profile functions of all $5$ polarization observables fitted in the $\Delta$-region are shown. The particular energy $E_{\gamma} = 350 \hspace*{1pt} \mathrm{MeV}$ has been chosen. \newline In addition, solutions from the unconstrained TPWA-fit with $\ell_{\mathrm{max}} = 2$ are plotted, namely the global minimum (green solid line) and the local minimum closest to the SAID CM12 $E_{0+}$- and $M_{1+}$-multipoles \cite{SAID} (red dashed line). Compare to Figure \ref{fig:Lmax2UnconstrainedFitResultsDeltaRegion} for a meaning of the solutions.
  }
 \label{fig:FourthEnergyBinGlobMinAndAmbiguity}
\end{figure}

\begin{quotation}
 \textit{`` For any finite precision of data, the solution of the problem becomes more involved and instead of exact solutions
of the sets of quadratic equations one has to search for a minimum similar to a chisquare
fit. Then the global minimum, which may already be difficult to find does not necessarily
give the correct solution. Therefore also local minima have to be considered and further
techniques are needed to arrive at the correct solution.''}
\end{quotation}
One can get one step closer to seeing what happened by considering both above-mentioned solutions plotted against the data itself, as is done in Figure \ref{fig:FourthEnergyBinGlobMinAndAmbiguity}\footnote{Care should be taken in comparing the results of Figure \ref{fig:FourthEnergyBinGlobMinAndAmbiguity} to those shown in Figure \ref{fig:DeltaRegionObsFitAngDistPlots}. The latter come from a Legendre-fit to each observable {\it individually}, while the former stem from the direct multipole-fit to {\it all} observables at the same time.}. \newline
For both observables $\sigma_{0}$ and $\check{\Sigma}$, the global minimum and the physically reasonable solution cannot be distinguished. Some discrepancies are seen in both $\check{T}$ and $\check{F}$, with only the $\check{F}$-data seeming capable to distinguish both solutions. Both solutions differ the most in the $\check{P}$-observable, which is however lacking the statistics to really help the decision. \newline
We are left with a situation which promises to get only worse once we raise the order in $\ell_{\mathrm{max}}$: the number of available discrete ambiguities will rise exponentially (cf. chapter \ref{chap:Omelaenko} and appendix \ref{sec:AdditionsChapterII}), but no observable except from maybe the cross section $\sigma_{0}$ has the statistical precision to really make a distinction. The cross section on the other hand is completely dominated by systematic effects. In our view, great danger lies here in the possible appearance of so-called 'outlier-points', i.e. data-points which may have a tiny statistical error while at the same time suffering a large systematic deviance. Such points give very large contributions to the final $\chi^{2}/\mathrm{ndf}$, worsening the apparent final quality of the fit. Furthermore, during the fit, such points have a huge weight in the minimized chisquare-function (\ref{eq:ChiSquareDirectFitRealDataFitSection}) and thus mainly determine the destination of the fit. \newline
In a scenario where $\ell_{\mathrm{max}}$ would be raised even more, the TPWA-fit would try to find a particular ambiguity which best describes even the smallest 'kink' in the angular distribution of the cross section, which then would probably be yielded as the final global minimum, since none of the remaining observables have even remotely the precision needed to resolve the ambiguity. In this way, one can obtain {\it spurious waves}. Thus, for instance, the 'unreasonable' global minimum seen in Figure \ref{fig:Lmax2UnconstrainedFitResultsDeltaRegion} {\it may} already be an artifact of the dominant systematic error of the cross section. \newline

In the end we are left with two problems: the appearance of unreasonable ambiguous minima and the unsatisfactory fit-quality for the global minimum. In an attempt to resolve the first issue, we choose to introduce a model-dependence into the procedure by fixing the $D$-wave multipoles to the SAID-solution CM12 \cite{WorkmanEtAl2012ChewMPhotoprod}. \newline
Thus, the truncation order remains $\ell_{\mathrm{max}} = 2$, but only the $S$- and $P$-waves are varied in a direct fit to the data (eq. (\ref{eq:ChiSquareDirectFitRealDataFitSection})). Still, we perform a full Monte Carlo sampling (cf. section \ref{sec:MonteCarloSampling}) for the varied multipoles. In case the $D$-waves are fixed to (for instance) SAID, one should correct the total cross section for the sampling-procedure according to
\begin{equation}
 \bar{\sigma}_{\bm{c}} := \bar{\sigma} - \Big\{ 36 \left| E^{\mathrm{SAID}}_{2+} \right|^{2} + 4 \left| E^{\mathrm{SAID}}_{2 -} \right|^{2} + 18 \left| M^{\mathrm{SAID}}_{2+} \right|^{2} + 12 \left| M^{\mathrm{SAID}}_{2-} \right|^{2} \Big\} \mathrm{.} \label{eq:TCSCorrectedDeltaRegionFit}
\end{equation}
Apart from that, the search-procedure does not change. However, it has to be mentioned that the SAID $D$-waves are inserted under the {\it same} phase-constraint as the remaining multipoles in the fit, i.e. they have been rotated with the inverse $E_{0+}^{\mathrm{SAID}}$-phase prior to fitting. \newpage

\begin{figure}[ht]
 \centering
 \vspace*{-2.5pt}
\begin{overpic}[width=0.495\textwidth]{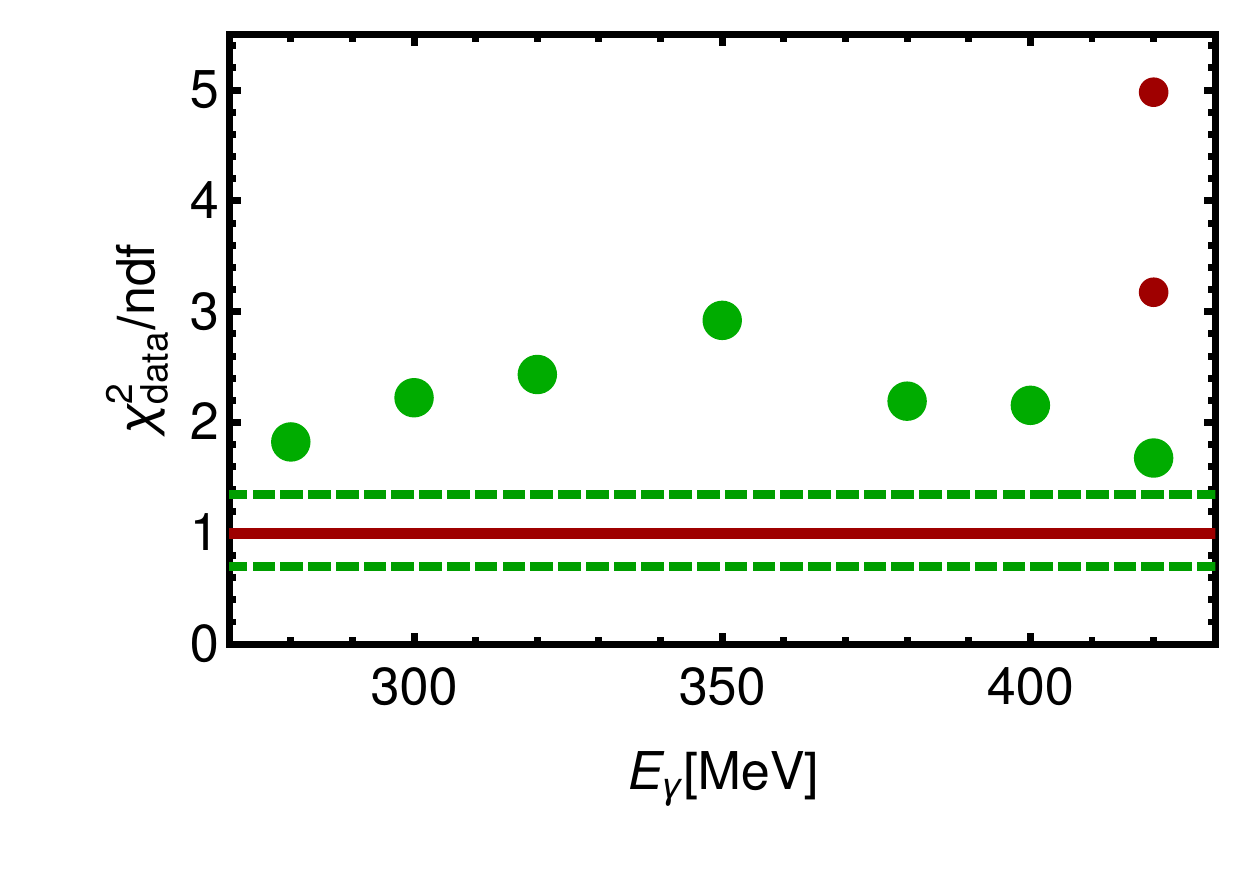}
 \put(0.5,66){a.)}
 \end{overpic} \\
 \vspace*{-1pt}
\begin{overpic}[width=0.325\textwidth]{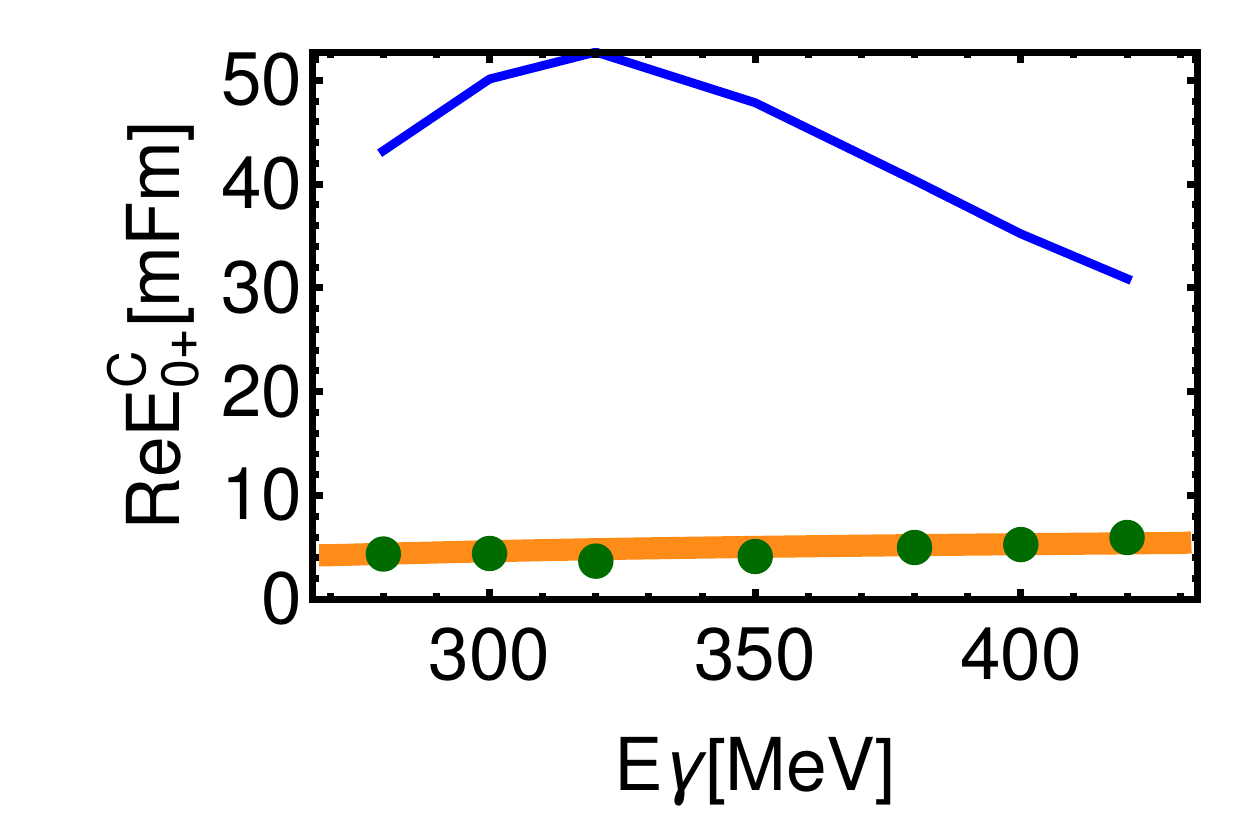}
 \put(0.5,65){b.)}
 \end{overpic}
\begin{overpic}[width=0.325\textwidth]{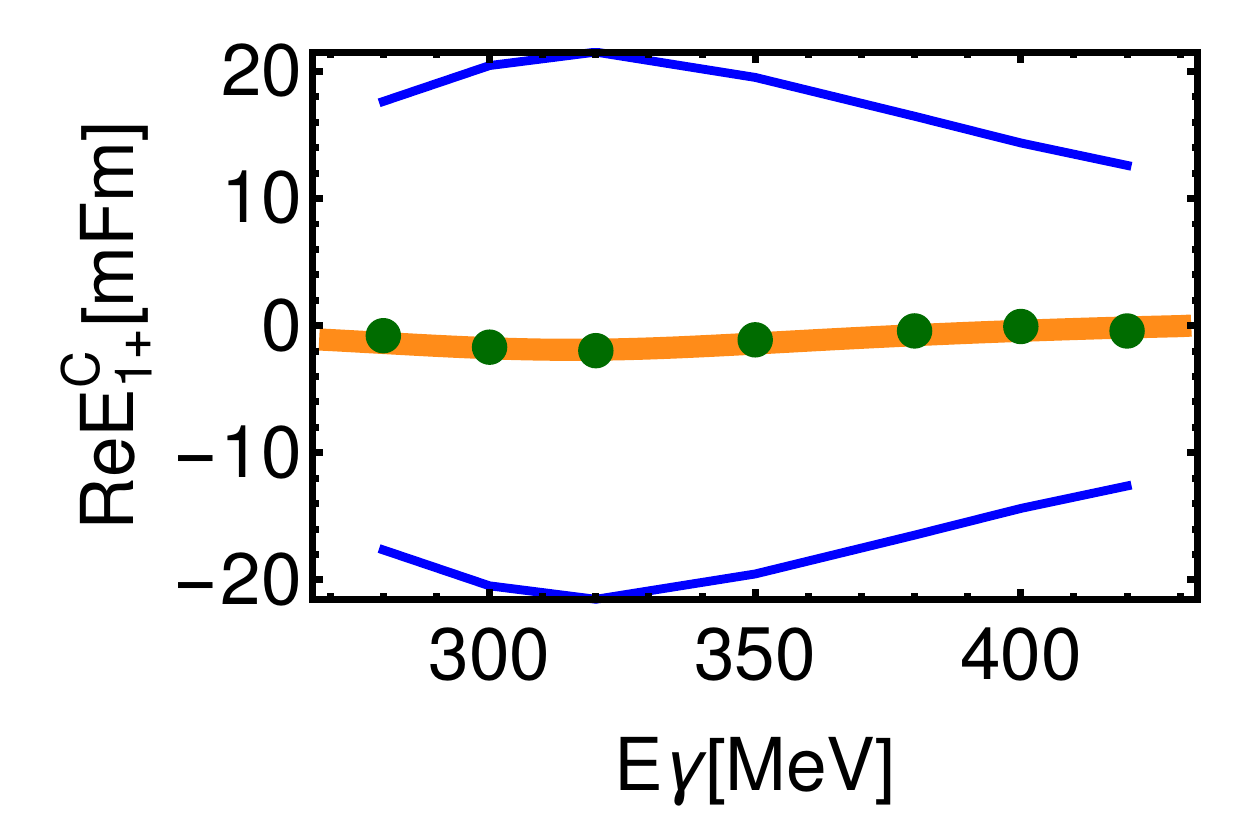}
 \end{overpic}
\begin{overpic}[width=0.325\textwidth]{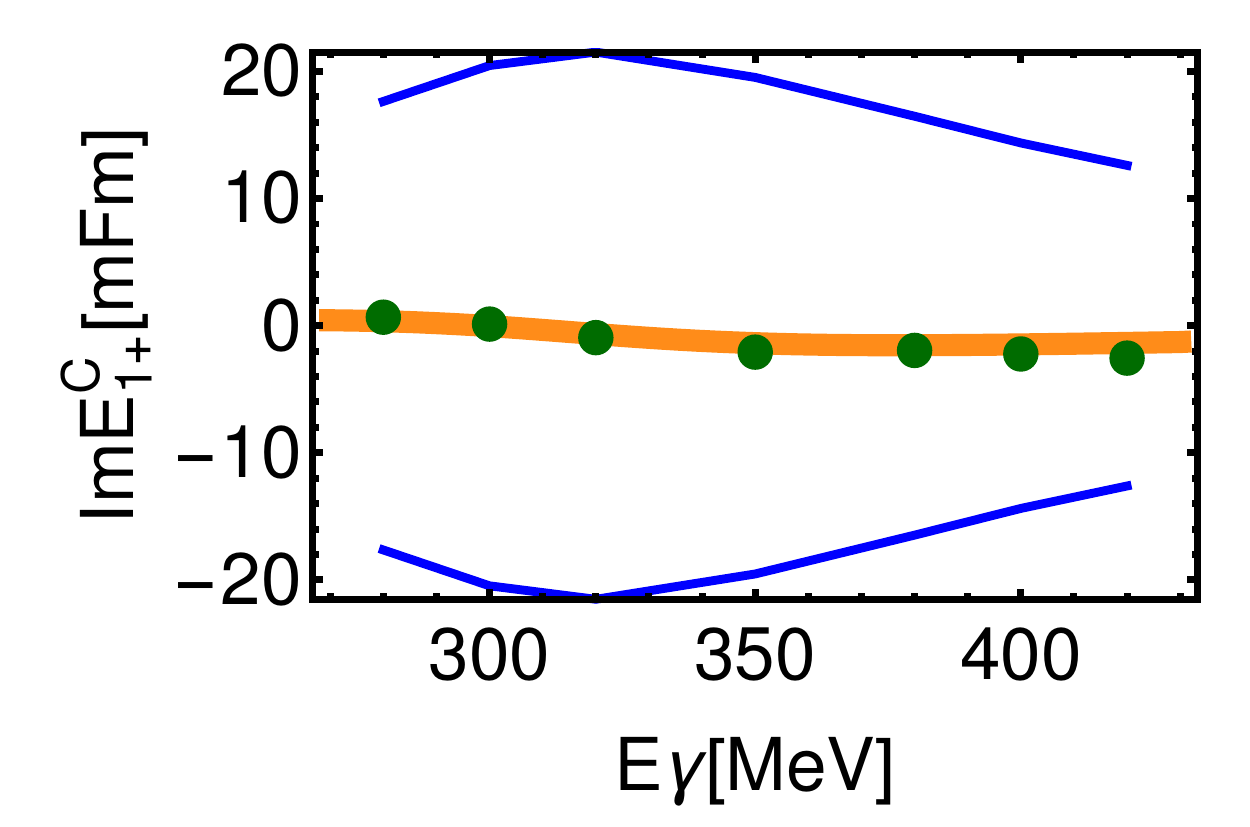}
 \end{overpic} \\
\begin{overpic}[width=0.325\textwidth]{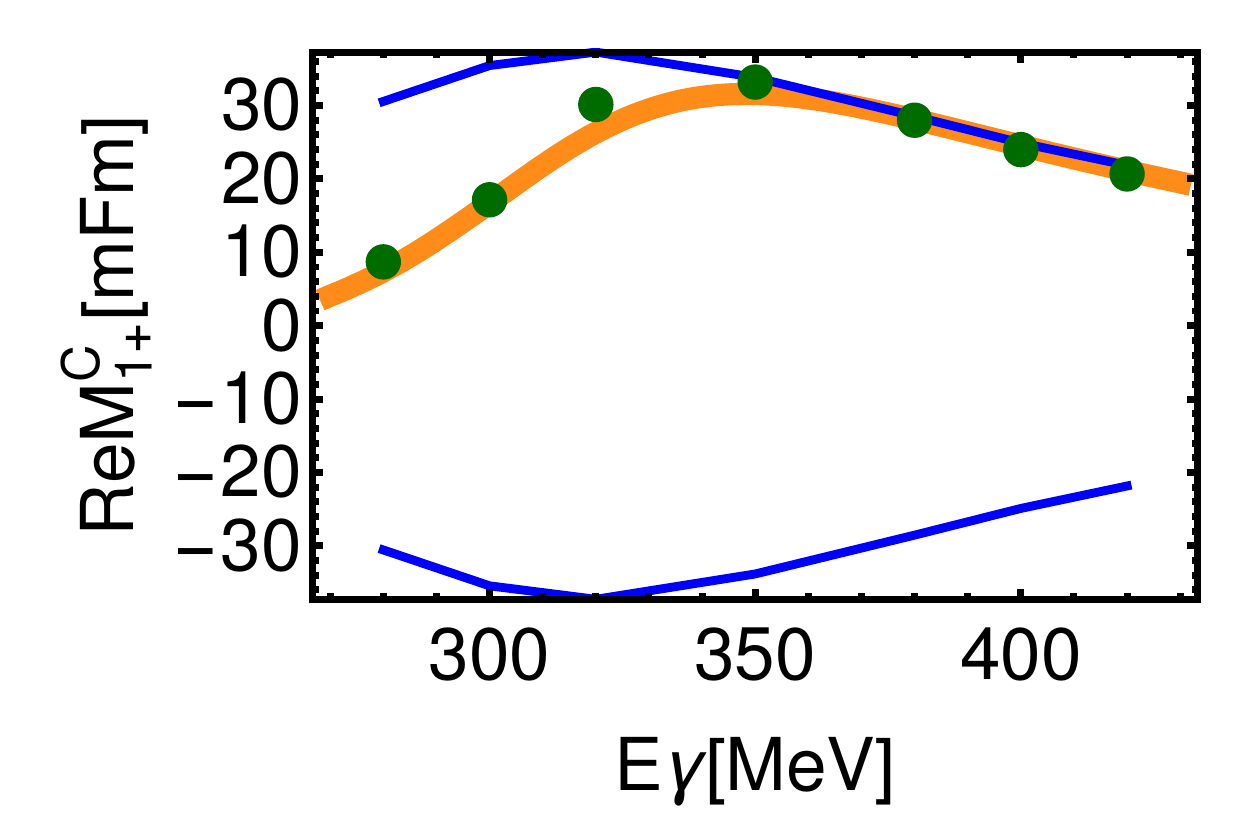}
 \end{overpic}
\begin{overpic}[width=0.325\textwidth]{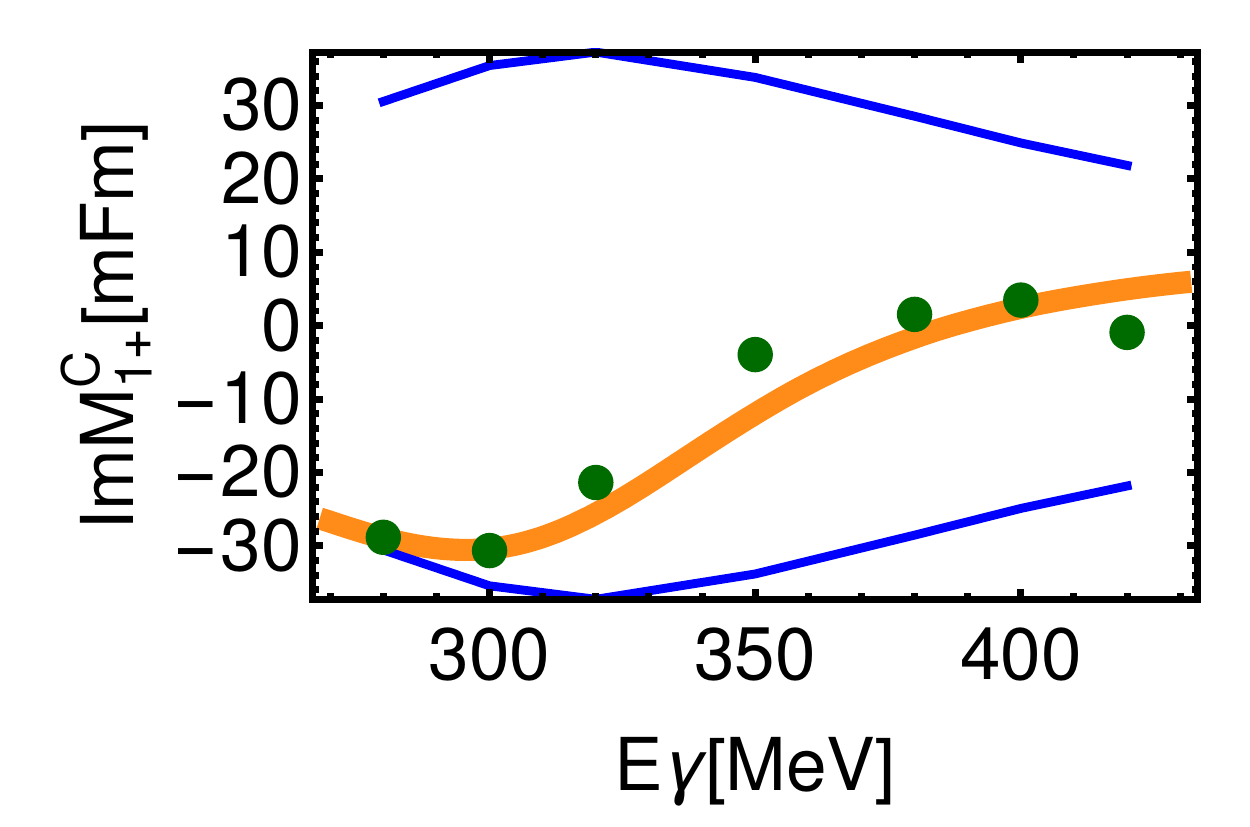}
 \end{overpic}
\begin{overpic}[width=0.325\textwidth]{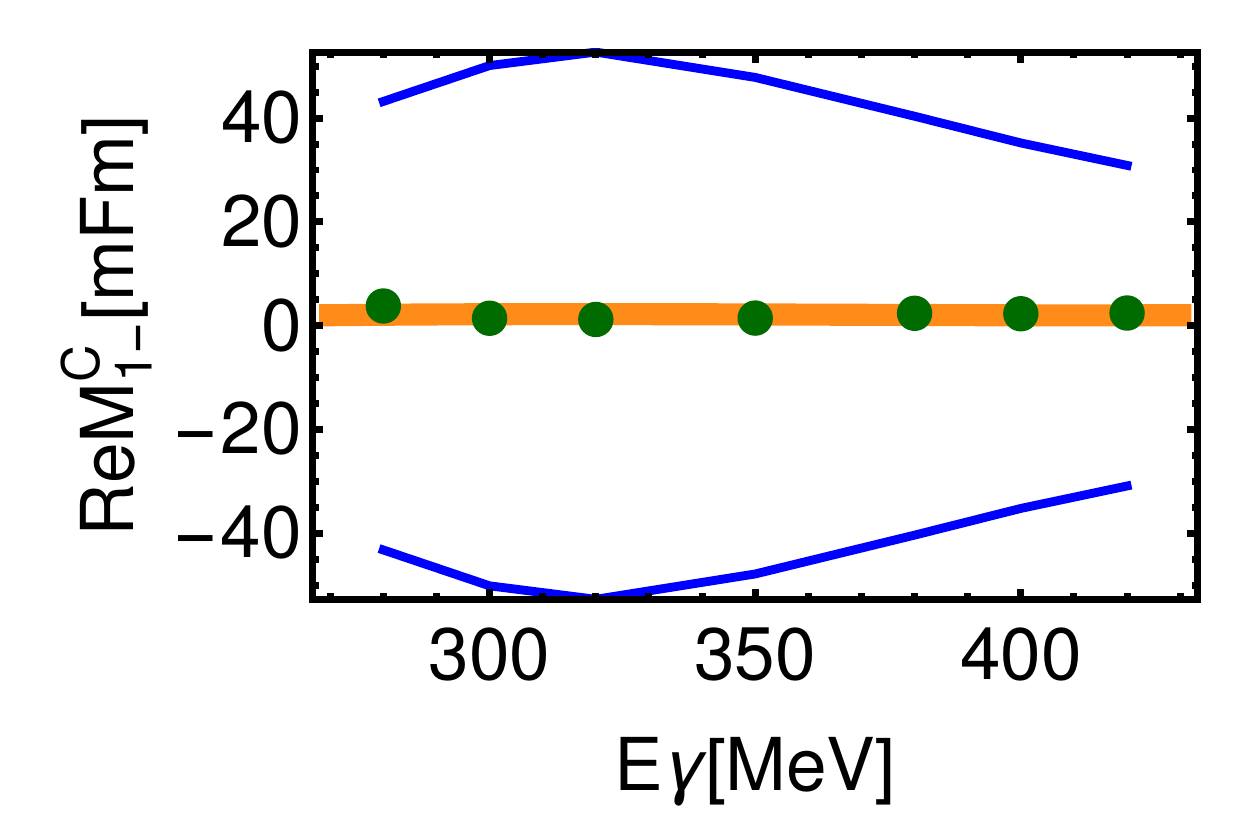}
 \end{overpic} \\
\begin{overpic}[width=0.325\textwidth]{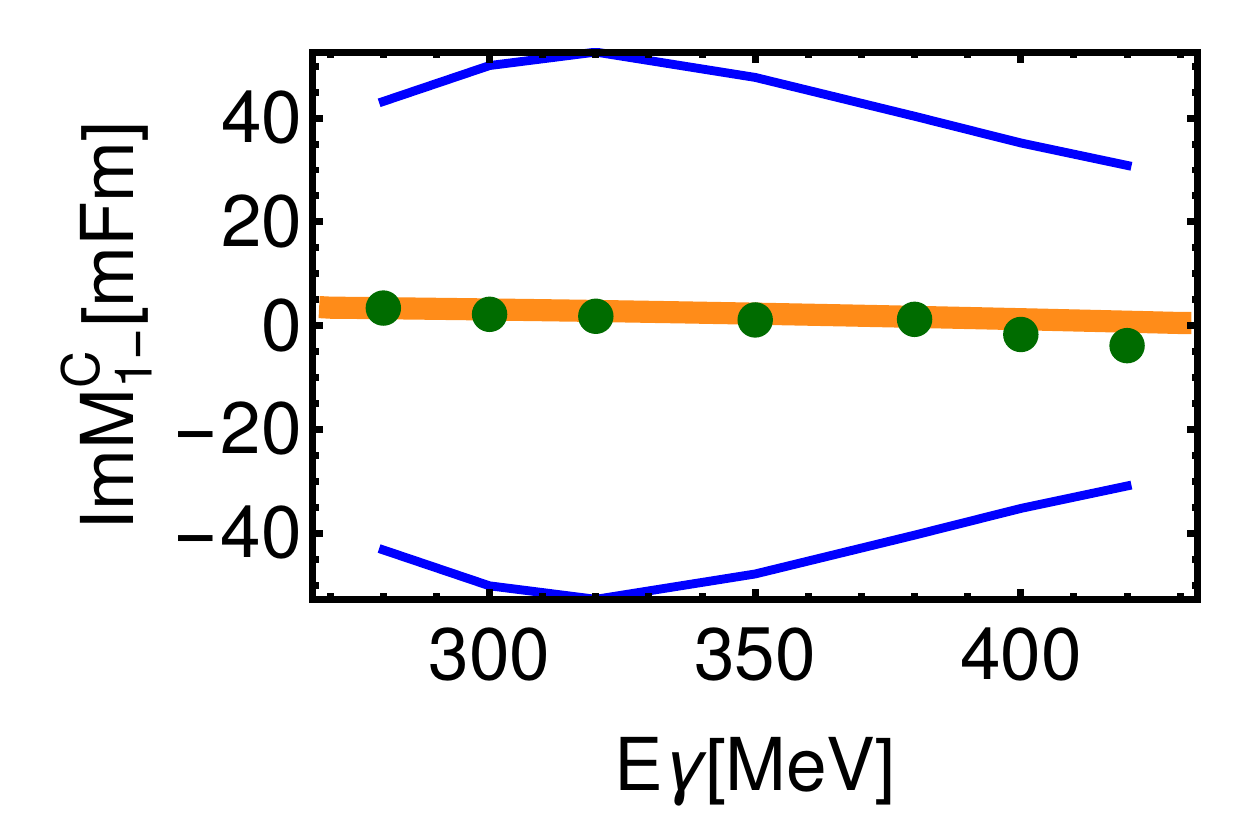}
 \end{overpic}
\caption[Results for the $7$ fit-parameters comprised of the real- and imaginary parts of phase-constrained $S$- and $P$-wave multipoles, as well as values for $\chi^{2} / \mathrm{ndf}$, for a TPWA using $\ell_{\mathrm{max}} = 2$ within the $\Delta$-resonance region. $D$-waves were fixed to the SAID-solution CM12.]{Results are summarized for a TPWA-fit with $\ell_{\mathrm{max}} = 2$ to the selected data in the $\Delta$-region. The fits employed a pool of $N_{MC} = 1000$ initial parameter configurations (cf. section \ref{sec:MonteCarloSampling}) and the $D$-wave multipoles have been fixed to the SAID-solution CM12 \cite{WorkmanEtAl2012ChewMPhotoprod}. \newline a.) The obtained values for $\chi^{2} / \mathrm{ndf}$ are plotted against energy for a direct fit to the data (\ref{eq:ChiSquareDirectFitRealDataFitSection}). One global minimum is obtained (green dots), which is fairly well separated against any local minima (red dots). From the corresponding theoretical chisquare distributions, the mean (red line) as well as the pair of $0.025$- and $0.975$-quantiles (green dashed lines) are shown. The number of degrees of freedom has been estimated to be $\mathrm{ndf} = 79-7 = 72$. \newline b.) The global minimum (green dots) is shown for the $7$ fit-parameters, i.e. the real- and imaginary parts of the phase-constrained $S$- and $P$-wave multipoles. The energy-dependent SAID-solution CM12 \cite{WorkmanEtAl2012ChewMPhotoprod, SAID} is shown as a solid orange colored curve. For each fit-parameter, the maximal range set by the total cross section (cf. discussion in section \ref{sec:MonteCarloSampling}), as well as its energy-variation, is indicated by the blue solid lines. Note that here, this maximal parameter range has been evaluated using the corrected cross section (\ref{eq:TCSCorrectedDeltaRegionFit}), since the fit is model-dependent.
}
\label{fig:Lmax2DWavesSAIDFitResultsDeltaRegionPurelyStat}
\end{figure}

The results of this procedure, obtained from a pool of $N_{MC} = 1000$ initial parameter configurations, can be inspected in Figure \ref{fig:Lmax2DWavesSAIDFitResultsDeltaRegionPurelyStat}. \newline
Exactly as in the case of the fit using $\ell_{\mathrm{max}} = 1$ (Figure \ref{fig:Lmax1UnconstrainedFitResultsDeltaRegion}), a nicely separated global minimum is found. There do not appear any dangerous local minima at all. However, the fit-quality of the best solution is still not in accord with chisquare-theory. Still, noticeable improvement of the $\chi^{2}/\mathrm{ndf}$-values has been obtained when compared to the results in Figure \ref{fig:Lmax1UnconstrainedFitResultsDeltaRegion}. \newline
The multipole-parameters show again a reasonable agreement with SAID CM12. Some (small) improvement has been achieved compared to the TPWA seen in Figure \ref{fig:Lmax1UnconstrainedFitResultsDeltaRegion}. 

\clearpage

\clearpage

Thus, one spots a dilemma: $D$-waves are necessary in order to correct the $S$- and $P$-waves via $\left< S, D \right>$- and $\left< P, D \right>$-interference terms. However, taking the $D$-waves into account fully model-independently has lead to considerable issues with ambiguities. \newline
Right now, the problem of a non-satisfactory fit-quality remains. As assumed above, this may be caused by the fact that in the data for the cross section $\sigma_{0}$ \cite{Hornidge:2013}, the systematic uncertainties dominate. \newline
Investigating further on this claim, we performed the following steps: inspired by the upper bound of a relative systematic error of $4\%$ given by Hornidge et al. \cite{Hornidge:2013}, we applied to each data-point for the cross section an uncertainty given by
\begin{equation}
 \Delta \sigma_{0, \mathrm{sys.}} (W,\theta) := 0.04 \times \sigma_{0} (W,\theta) \mathrm{.} \label{eq:SystematicErrorDCSHornidgeDataset}
\end{equation}
Then, in a first rough approximation, the statistical and systematic errors of the differential cross section were added in quadrature
\begin{equation}
 \Delta \sigma_{0} := \sqrt{ \left( \Delta \sigma_{0, \mathrm{stat.}} \right)^{2} + \left( \Delta \sigma_{0, \mathrm{sys.}} \right)^{2}} \mathrm{.} \label{eq:AddQuadratureDCSErrorsHordnidge}
\end{equation}
It has to be mentioned that this step may very well not be fully correct. First of all, the systematic error given by reference \cite{Hornidge:2013} may not parametrize a Gaussian probability distribution function at all, in case of which (\ref{eq:AddQuadratureDCSErrorsHordnidge}) would be false. However, the paper \cite{Hornidge:2013} itself contains no further statements in this direction. Even in case the systematic errors were Gaussian distributed, equation (\ref{eq:AddQuadratureDCSErrorsHordnidge}) would still be incorrect in case of correlated systematic errors among neighboring kinematic bins. \newline
Thus, the assumption (\ref{eq:AddQuadratureDCSErrorsHordnidge}) may be seen as a rough estimate in order to investigate the effect of the systematics on the fit-quality. A more sophisticated treatment of systematics will be given further below. \newline

\begin{figure}[ht]
 \centering
 \vspace*{-2.5pt}
\begin{overpic}[width=0.495\textwidth]{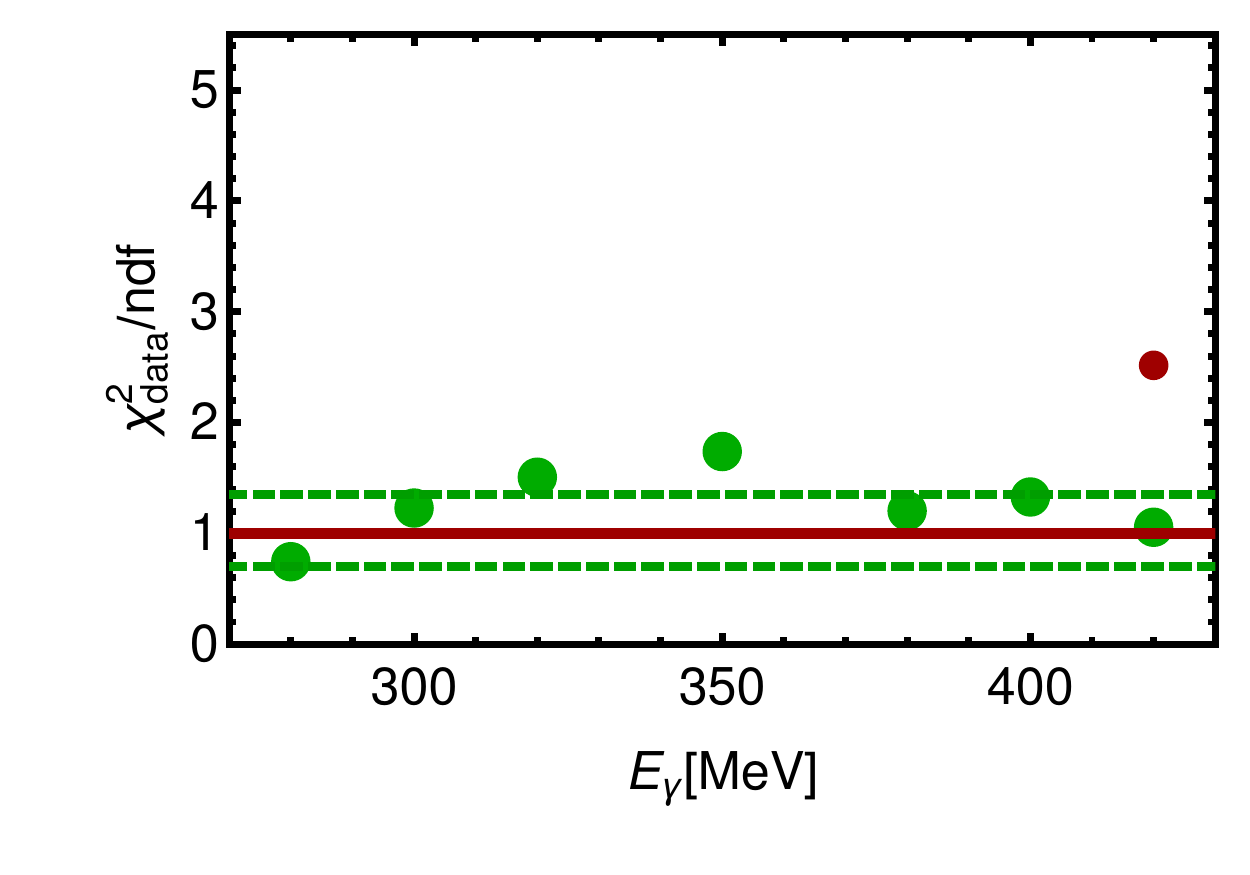}
 \put(0.5,66){a.)}
 \end{overpic} \\
 \vspace*{-1pt}
\begin{overpic}[width=0.325\textwidth]{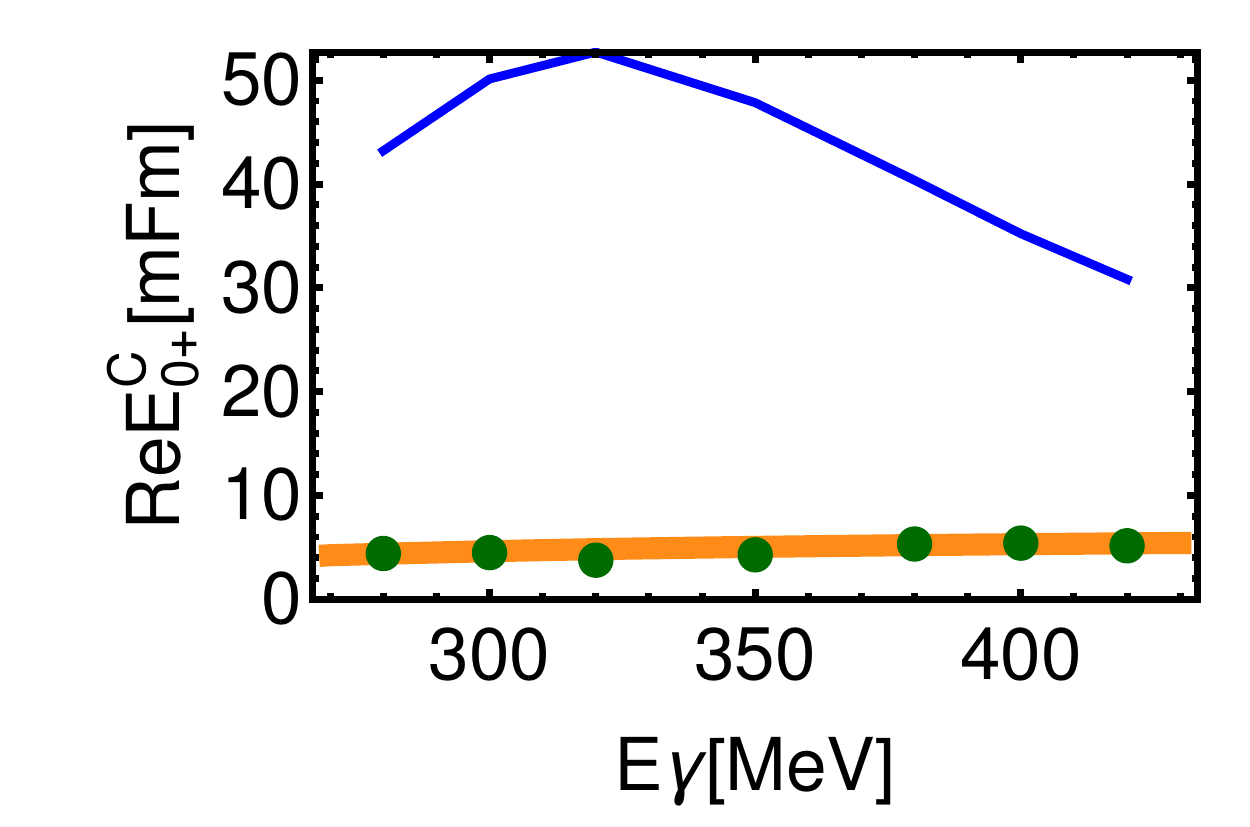}
 \put(0.5,65){b.)}
 \end{overpic}
\begin{overpic}[width=0.325\textwidth]{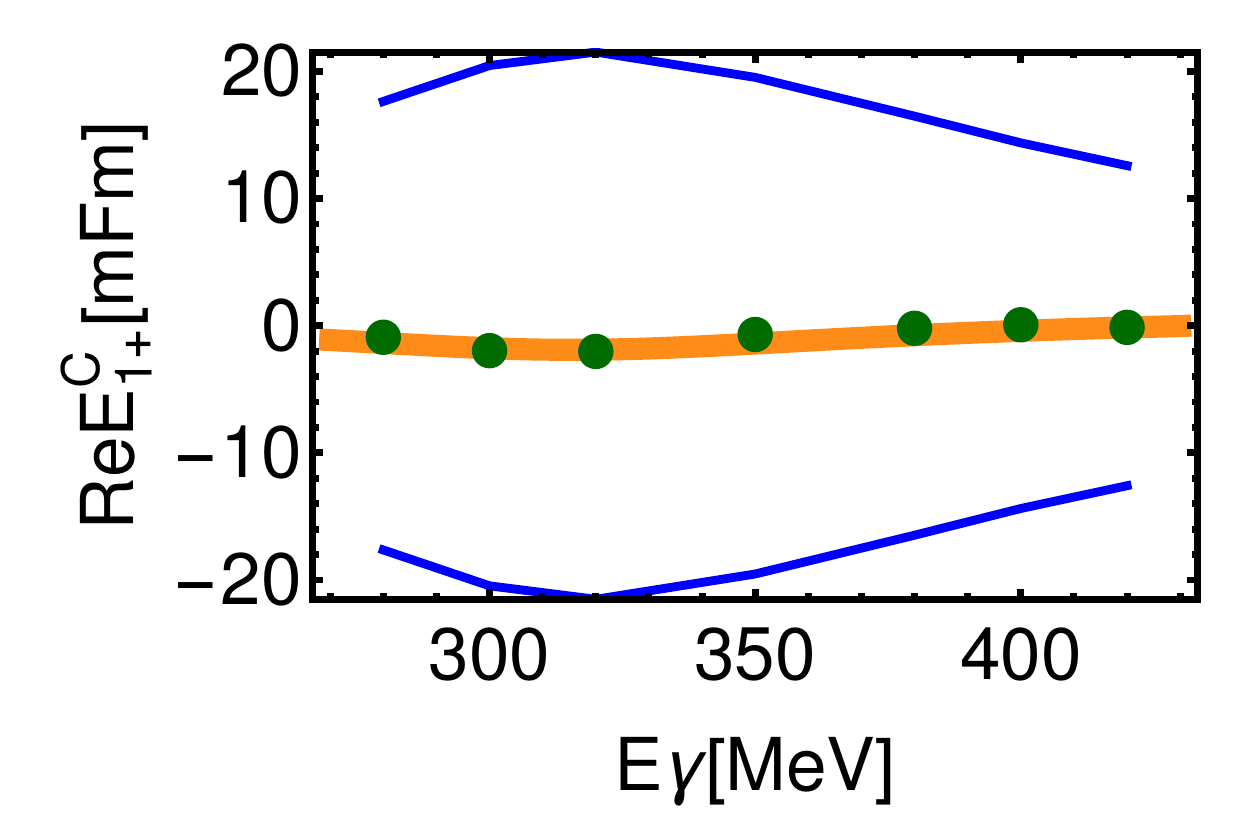}
 \end{overpic}
\begin{overpic}[width=0.325\textwidth]{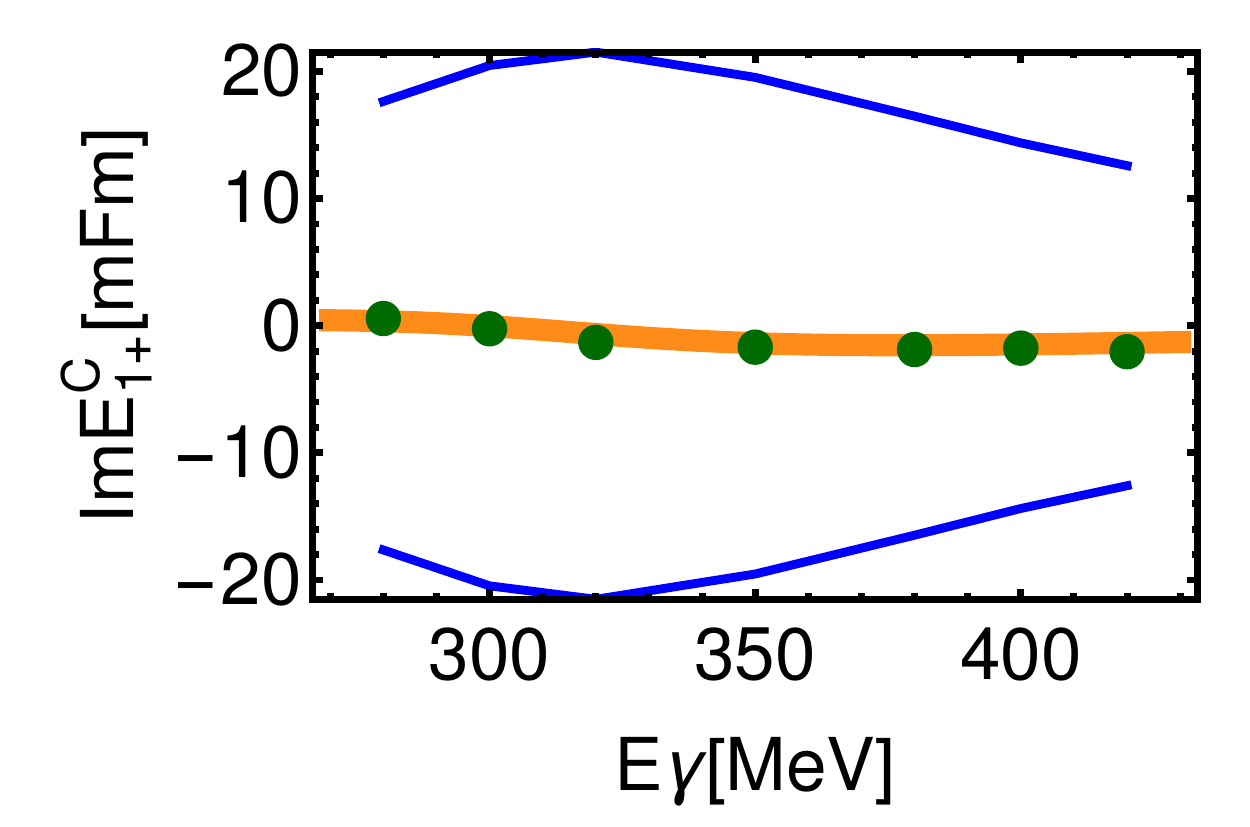}
 \end{overpic} \\
\begin{overpic}[width=0.325\textwidth]{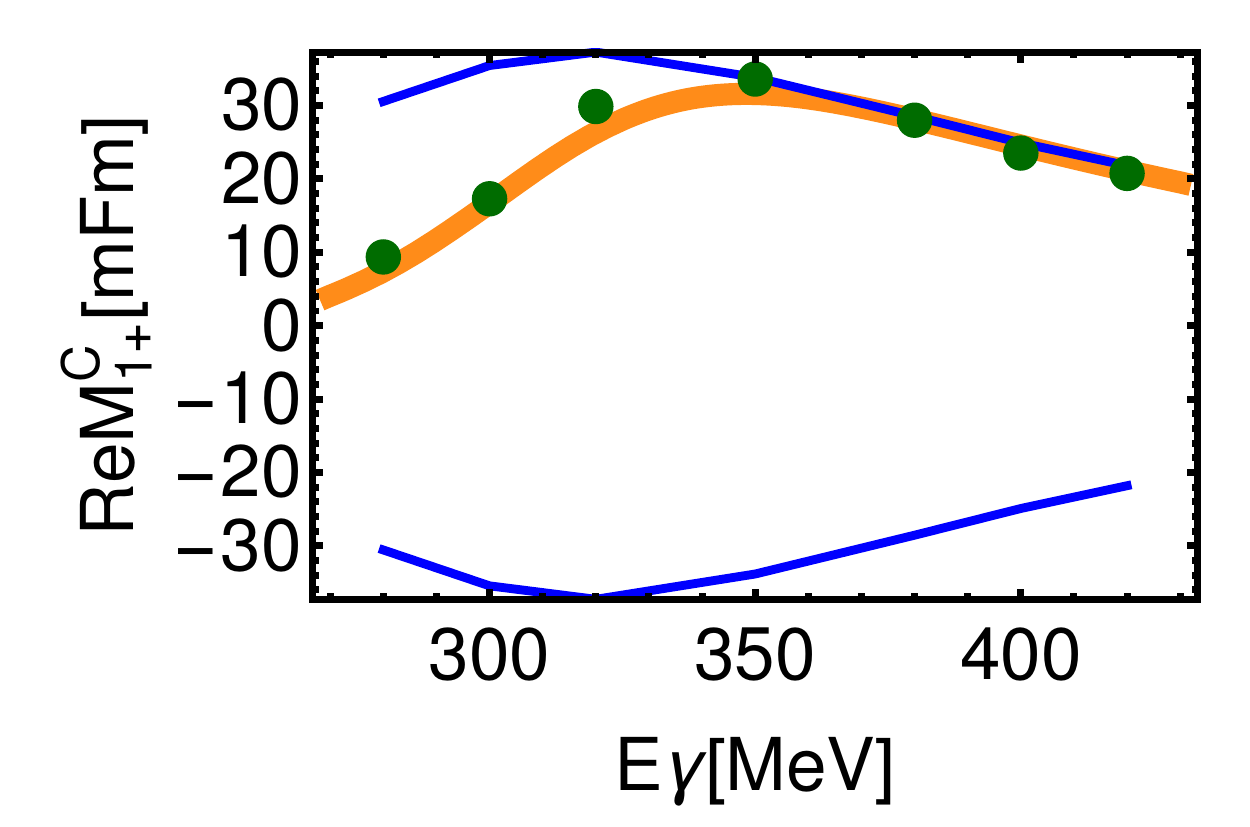}
 \end{overpic}
\begin{overpic}[width=0.325\textwidth]{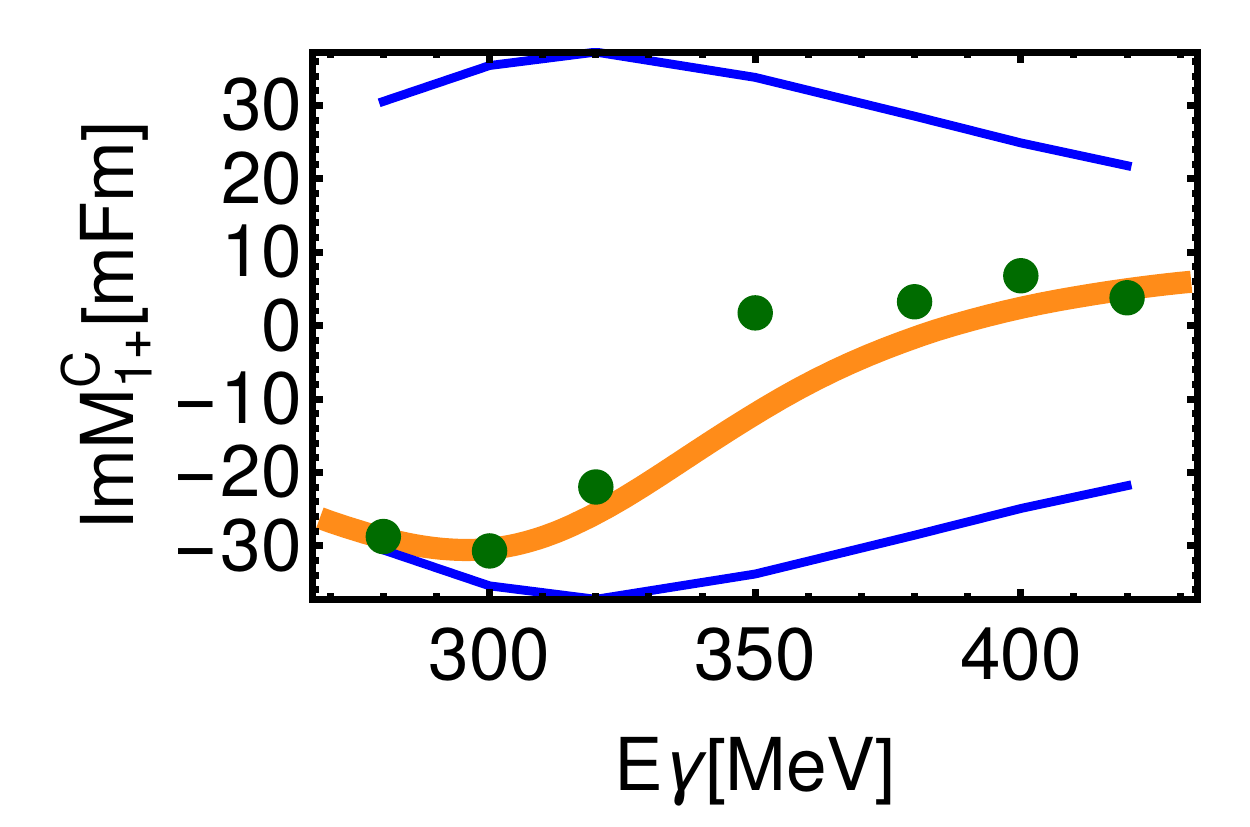}
 \end{overpic}
\begin{overpic}[width=0.325\textwidth]{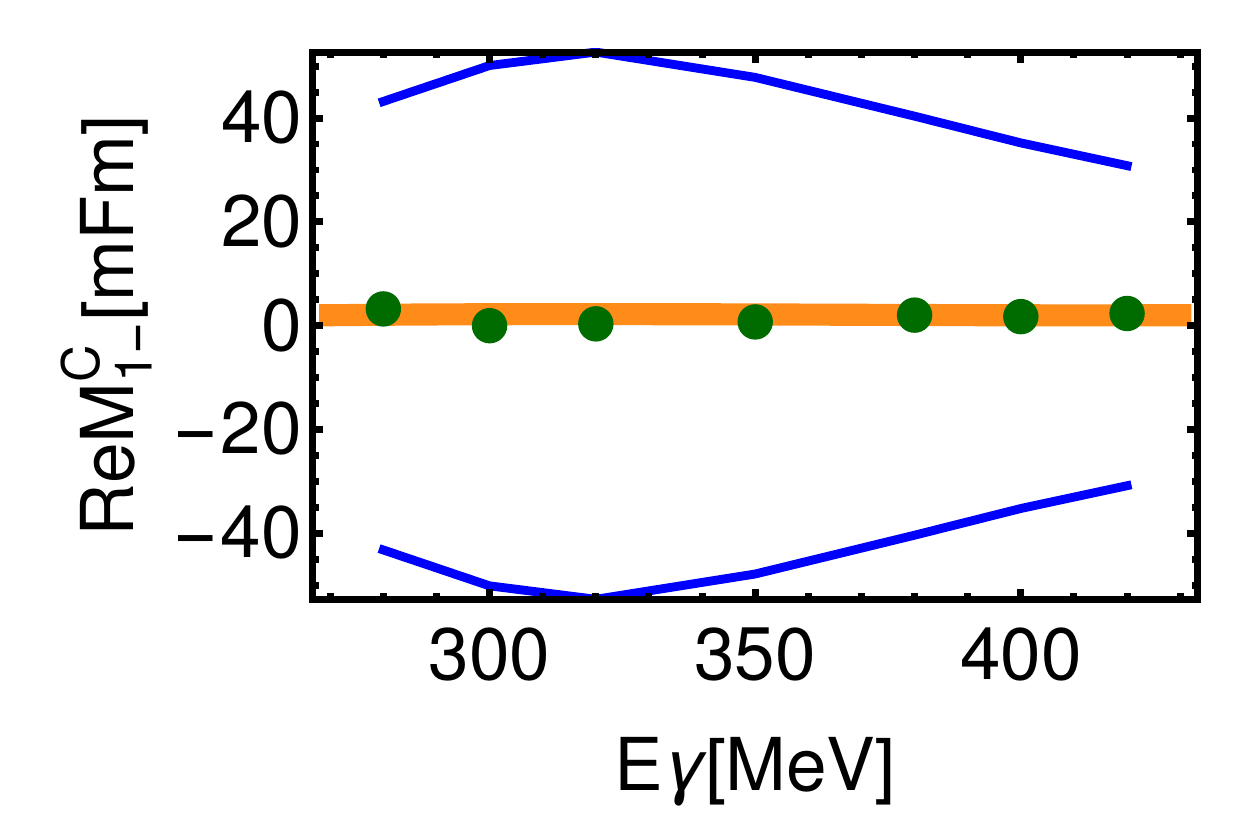}
 \end{overpic} \\
\begin{overpic}[width=0.325\textwidth]{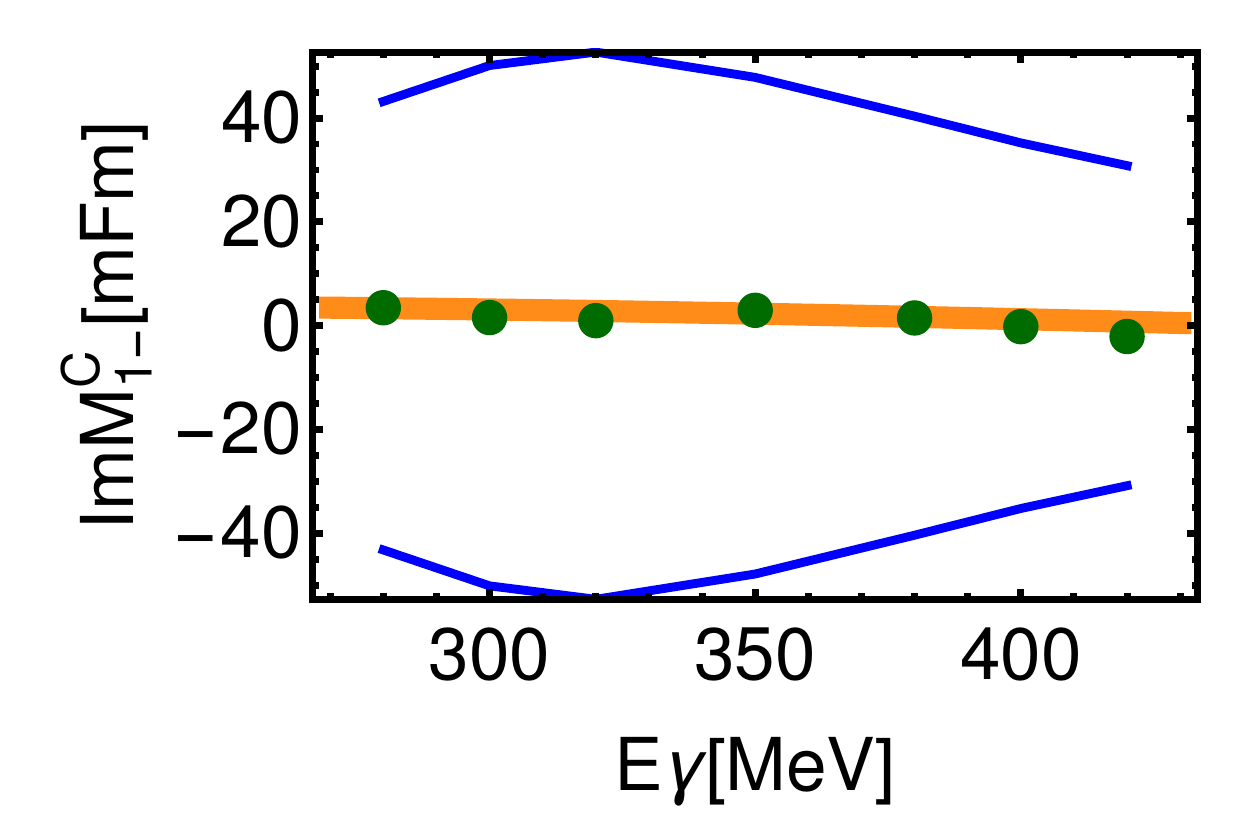}
 \end{overpic}
\caption[Results for the $7$ fit-parameters comprised of the real- and imaginary parts of phase-constrained $S$- and $P$-wave multipoles, as well as values for $\chi^{2} / \mathrm{ndf}$, for a TPWA using $\ell_{\mathrm{max}} = 2$ within the $\Delta$-resonance region. $D$-waves were fixed to the SAID-solution CM12 and statistical as well as systematic errors have been added in quadrature for the differential cross section $\sigma_{0}$.]{Results are summarized for a TPWA-fit with $\ell_{\mathrm{max}} = 2$ to the selected data in the $\Delta$-region. The fits employed a pool of $N_{MC} = 1000$ initial parameter configurations (cf. section \ref{sec:MonteCarloSampling}); $D$-wave multipoles have been set to the SAID-solution CM12 \cite{WorkmanEtAl2012ChewMPhotoprod}. Furthermore, a $4\%$ systematic error has been added in quadrature to the statistical error for each datapoint of the differential cross section $\sigma_{0}$ (see equation (\ref{eq:AddQuadratureDCSErrorsHordnidge})). \newline a.) The obtained values for $\chi^{2} / \mathrm{ndf}$ are plotted against energy for a direct fit to the data (\ref{eq:ChiSquareDirectFitRealDataFitSection}). The global minimum is shown (green dots), as well as other local minima (red dots). From the corresponding chisquare distribution, the mean (red line) as well as the pair of $0.025$- and $0.975$-quantiles (green dashed lines) are shown. The number of degrees of freedom has been estimated to be $\mathrm{ndf} = 79-7 = 72$. \newline b.) The global minimum (green dots) is shown for all fit-parameters (real- and imaginary parts of the phase-constrained $S$- and $P$-wave multipoles). The SAID-solution CM12 \cite{WorkmanEtAl2012ChewMPhotoprod, SAID} is shown as a solid orange colored curve. For each fit-parameter, the maximal range set by the total cross section (cf. discussion in section \ref{sec:MonteCarloSampling}), as well as its energy-variation, is indicated by the blue solid lines.
}
\label{fig:Lmax2DWavesSAIDFitResultsDeltaRegionAddQuadrature}
\end{figure}

The exact same fit as shown before in Figure \ref{fig:Lmax2DWavesSAIDFitResultsDeltaRegionPurelyStat} has been repeated, but employing the transformations (\ref{eq:SystematicErrorDCSHornidgeDataset}) and (\ref{eq:AddQuadratureDCSErrorsHordnidge}) to the data of the unpolarized cross section. The results are shown in Figure \ref{fig:Lmax2DWavesSAIDFitResultsDeltaRegionAddQuadrature}. \newline
The overall solution-structure has not changed compared to the case before, i.e we find a well-separated global minimum and only one harmless local minimum in the highest energy-bin. However, the resulting values of $\chi^{2} / \mathrm{ndf}$ are now of interest and it turns out that now, pleasantly, the addition in quadrature (equation (\ref{eq:AddQuadratureDCSErrorsHordnidge})) resulted in a global minimum which is well within the $95\%$ confidence-interval suggested by the theoretical chisquare-distribution, at least for $5$ of the $7$ energies! \newline
The obtained values for the multipole-fit-parameters show again a good agreement with SAID. When compared to the results of the previous fit (Figure \ref{fig:Lmax2DWavesSAIDFitResultsDeltaRegionPurelyStat}), the multipoles have varied somewhat, but mostly in the sub-percent to few percent range. The only quantity where changes are more noticeable in the plots is $\mathrm{Im} \left[ M_{1+} \right]$. \newline

This result solidifies the suspicion that the data for the unpolarized cross section have been responsible for the bad fit-qualities encountered in the first three fits. However, the estimate (\ref{eq:AddQuadratureDCSErrorsHordnidge}) used to obtain this result may be seen as somewhat crude. Thus, we present in the following an alternative method to take into account the systematic errors which has also been employed by researchers of The Cyprus Institute, C. N. Papanicolas and L. Markou \cite{LefterisPrivateComm}. This is the method of using {\it nuisance parameters} and it will be employed in an upcoming publication from the aforementioned researchers \cite{LefterisPaper}. 

\clearpage

The only alternative publication detailing the method of nuisance parameters, which we found in the course of this work, is a quite explicit talk by V. Blobel \cite{BlobelTalk}. \newline

The basic idea is the following: suppose a combination of measurements is to be fitted, where each publication belonging to a certain dataset quotes a relative, or 'few percent'-, estimate for the systematic error. In our case, for the dataset $\left\{ \check{\Omega}^{\alpha} \right\} = \left\{ \sigma_{0}, \Sigma, T, P, F \right\}$, this is actually the case for {\it every} observable. Apart from the $4\%$-estimate given by Hornidge and collaborators \cite{Hornidge:2013} in case of the cross section $\sigma_{0}$, Leukel gives an estimate of $4\%$ for $\Sigma$ \cite{LeukelPhD}. The most recent MAMI-data published by Schumann et al. \cite{Schumann:2015} estimate the relative size of the systematic error for $\check{T}$ to be maximally $8\%$. Since the $\check{F}$-data come from the same beamtime and analysis, we use the same estimate also in case of this double-polarization observable. Finally, for the $\check{P}$-data by Belyaev et al. \cite{Belyaev:1983}, a range from $2\%$ to $10\%$ is given for the systematic error, depending on energy and angle. Since in the provided data-{\it file} for $P$, an estimate of generally $5\%$ is given, we apply this latter number. \newline
Now, interpreting the percent-estimates given in the papers generally as normalization-, or scaling-, uncertainties, the approach is to introduce one dimensionless {\it nuisance parameter}\footnote{Or proportionality-factor, thus called '$\bm{p}$'.} $\bm{p}_{\alpha}$ for each measurement $\check{\Omega}^{\alpha}$ \cite{LefterisPrivateComm, LefterisPaper, BlobelTalk}. One possible approach, according to Blobel \cite{BlobelTalk}, is to let this nuisance parameter act as a modification of the fitting-functions and to define an 'effective chisquare-function', which is a modification of equation (\ref{eq:ChiSquareDirectFitRealDataFitSection}), as follows (cf. similar definitions in \cite{BlobelTalk})
\begin{align}
 \Phi_{\mathrm{data}} \left(  \left\{ \mathcal{M}_{\ell}  \right\}; \left\{ \bm{p}_{\alpha} \right\} \right) &:= \sum_{\check{\Omega}^{\alpha}, c_{k_{\alpha}}} \left( \frac{ \check{\Omega}^{\alpha}_{\mathrm{Data}} (c_{k_{\alpha}}) - \bm{p}_{\alpha} \hspace*{1.5pt} \check{\Omega}^{\alpha}_{\mathrm{Fit}} \left( c_{k_{\alpha}}, \left\{ \mathcal{M}_{\ell} \right\} \right) }{ \Delta \check{\Omega}^{\alpha}_{\mathrm{Data}} (c_{k_{\alpha}})} \right)^{2} \nonumber \\
 & \hspace*{12.5pt} + \sum_{\alpha} \left( \frac{\bm{p}_{\alpha} - 1}{\sigma_{\bm{p},\alpha}} \right)^{2} \mathrm{.} \label{eq:SystScalingFitFunction}
\end{align}
Here, the fit-functions $\check{\Omega}^{\alpha}_{\mathrm{Fit}} \left( \cos \theta, \left\{ \mathcal{M}_{\ell} \right\} \right)$ have been given above, directly below equation (\ref{eq:ChiSquareDirectFitRealDataFitSection}). The nuisance parameters $\left\{ \bm{p}_{\alpha} \right\}$ are varied as parameters in the fit, together with the real- and imaginary parts of the multipoles $\left\{ \mathcal{M}_{\ell}  \right\}$. However, the $\left\{ \bm{p}_{\alpha} \right\}$ are not varied completely freely, but instead are bound to the case of 'no re-scaling', i.e. $\bm{p}_{\alpha} = 1$, by penalty-terms. The normalization uncertainties $\sigma_{\bm{p},\alpha}$ are defined prior to fitting and are adjusted to the information contained in the publications of the data \cite{Hornidge:2013, LeukelPhD, Schumann:2015, Belyaev:1983}. \newline
Thus, the idea is, in a way, to 'fit-out the systematics'. In finding solutions for the $\left\{ \bm{p}_{\alpha} \right\}$, the fit determines in which direction the assumed scaling-errors act. Moreover, once a good minimum is found and the penalty-terms, as given in the minimum, are subtracted again from (\ref{eq:SystScalingFitFunction}), one obtains the equivalent to $\chi^{2}_{\mathrm{data}}$ from the fit without nuisance parameters. Then, one can get an idea about whether or not the introduction of the nuisance parameters has improved the quality of the description of the data. \newline
A further strength of the proposed method is that systematic errors are treated as {\it different} scalings for different datasets. Generally, in case one combines data coming from different experiments, one should take into account the fact that systematic effects manifest themselves differently for each individual measurement. The proposed method can accomplish this task. Lastly, we mention another method to take into account normalization errors, which has been proposed by Sandorfi et al. \cite{Sandorfi:2010uv}. Here, the authors used the Fierz-identities (cf. appendix \ref{subsec:FierzIdentities}) for a very similar purpose. \newpage

\begin{figure}[ht]
 \centering
 \vspace*{-5pt}
\begin{overpic}[width=0.495\textwidth]{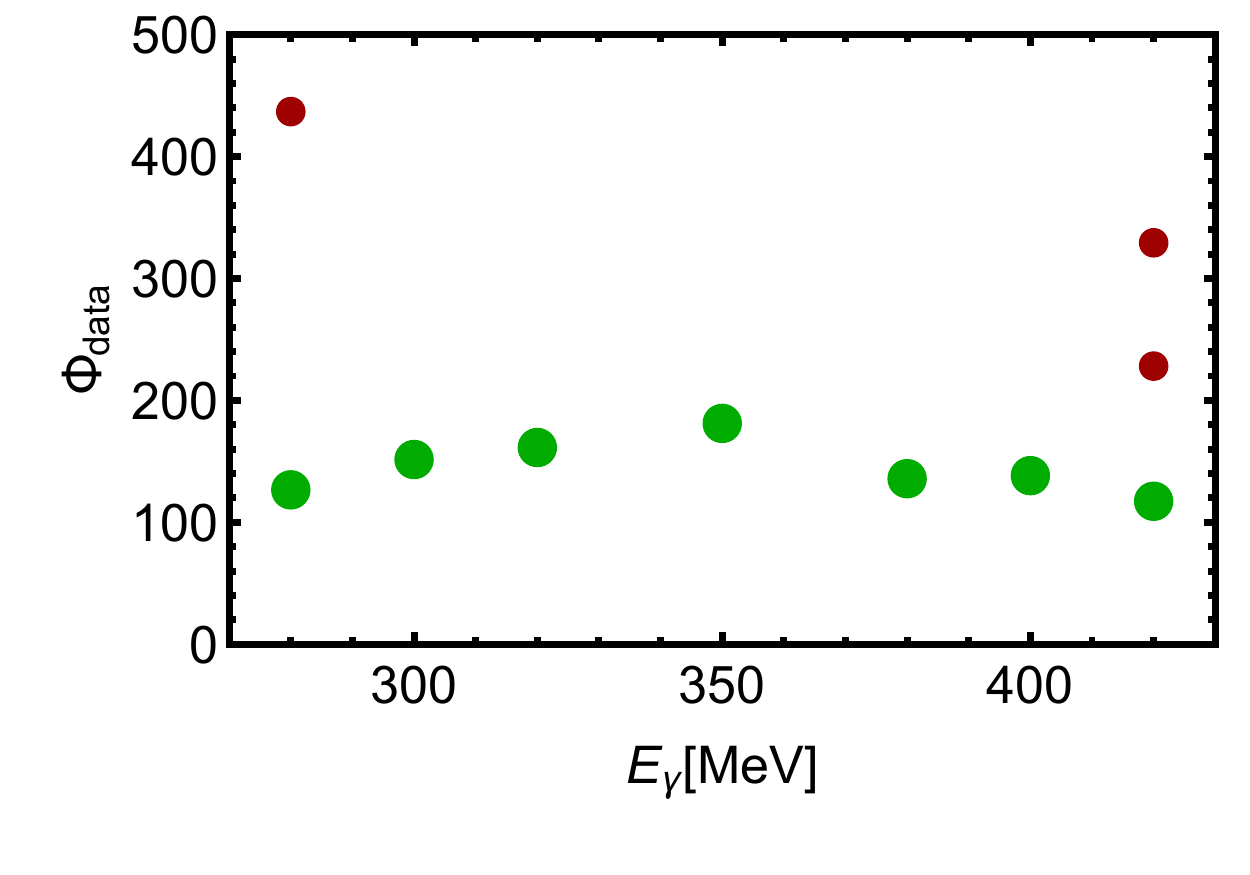}
 \put(0.5,65){a.)}
 \end{overpic}
\begin{overpic}[width=0.495\textwidth]{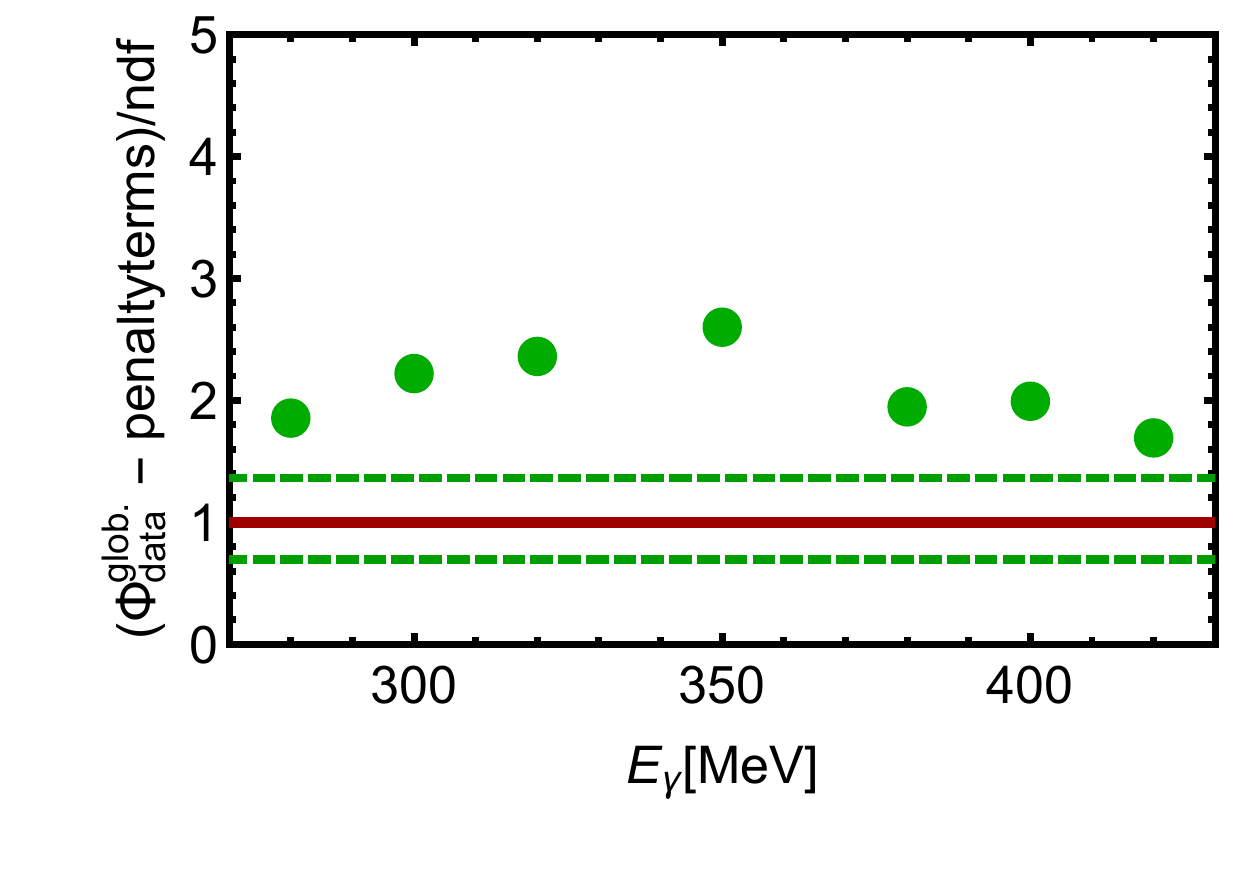}
 \end{overpic} \\
 \vspace*{-10pt}
\begin{overpic}[width=0.325\textwidth]{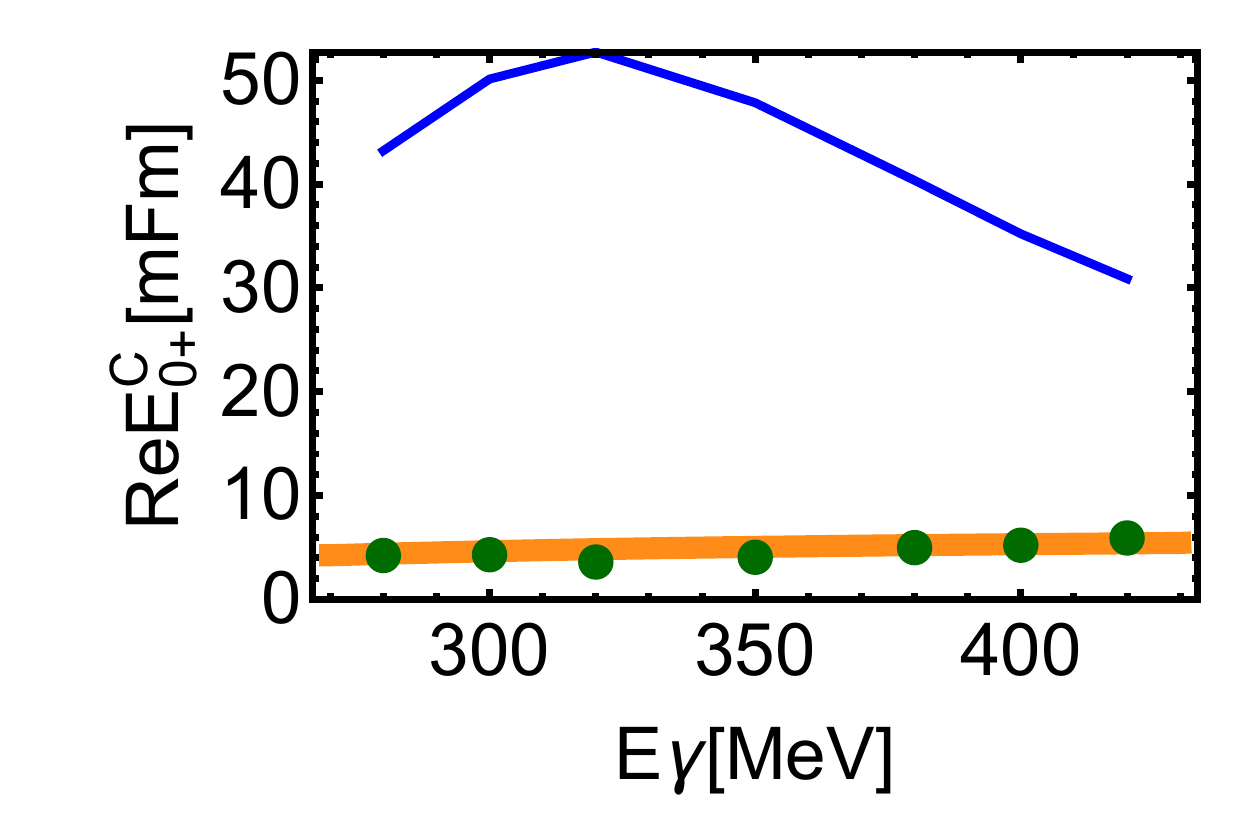}
 \put(0.5,64){b.)}
 \end{overpic}
\begin{overpic}[width=0.325\textwidth]{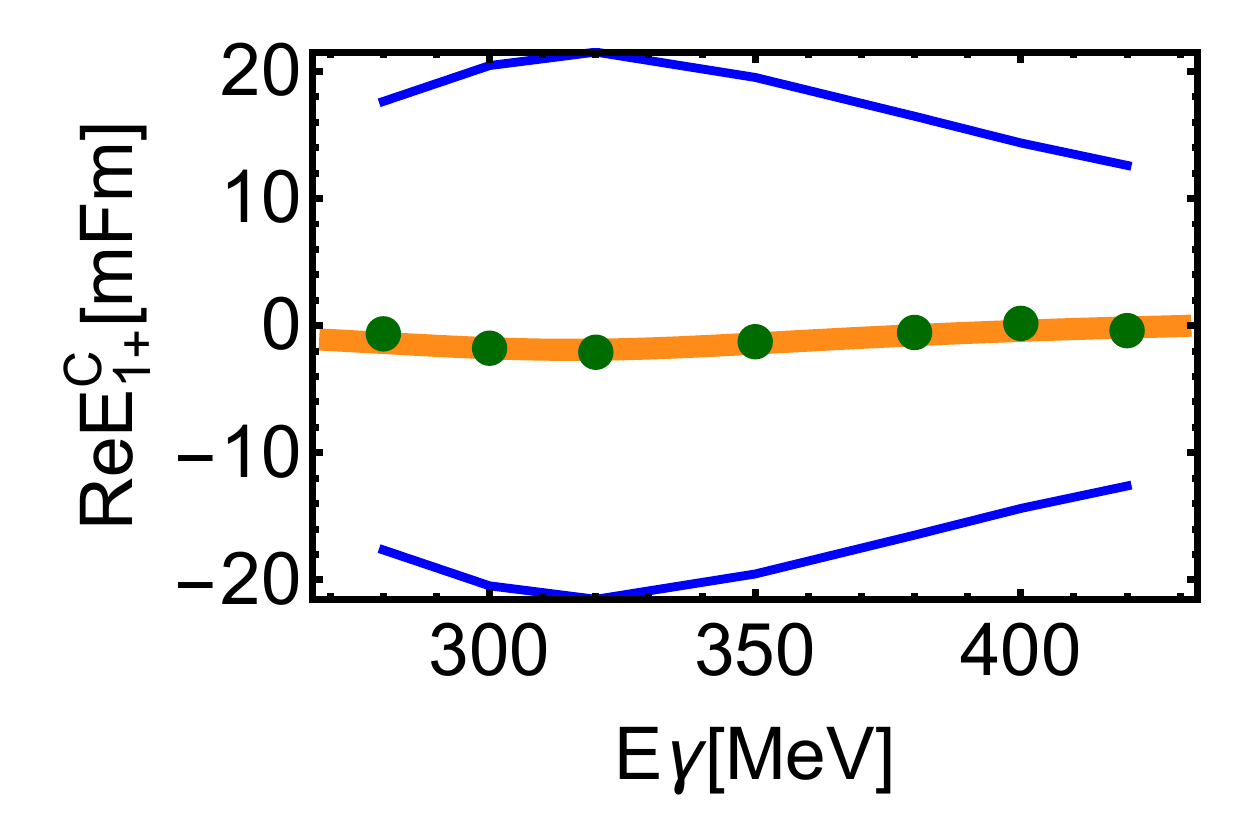}
 \end{overpic}
\begin{overpic}[width=0.325\textwidth]{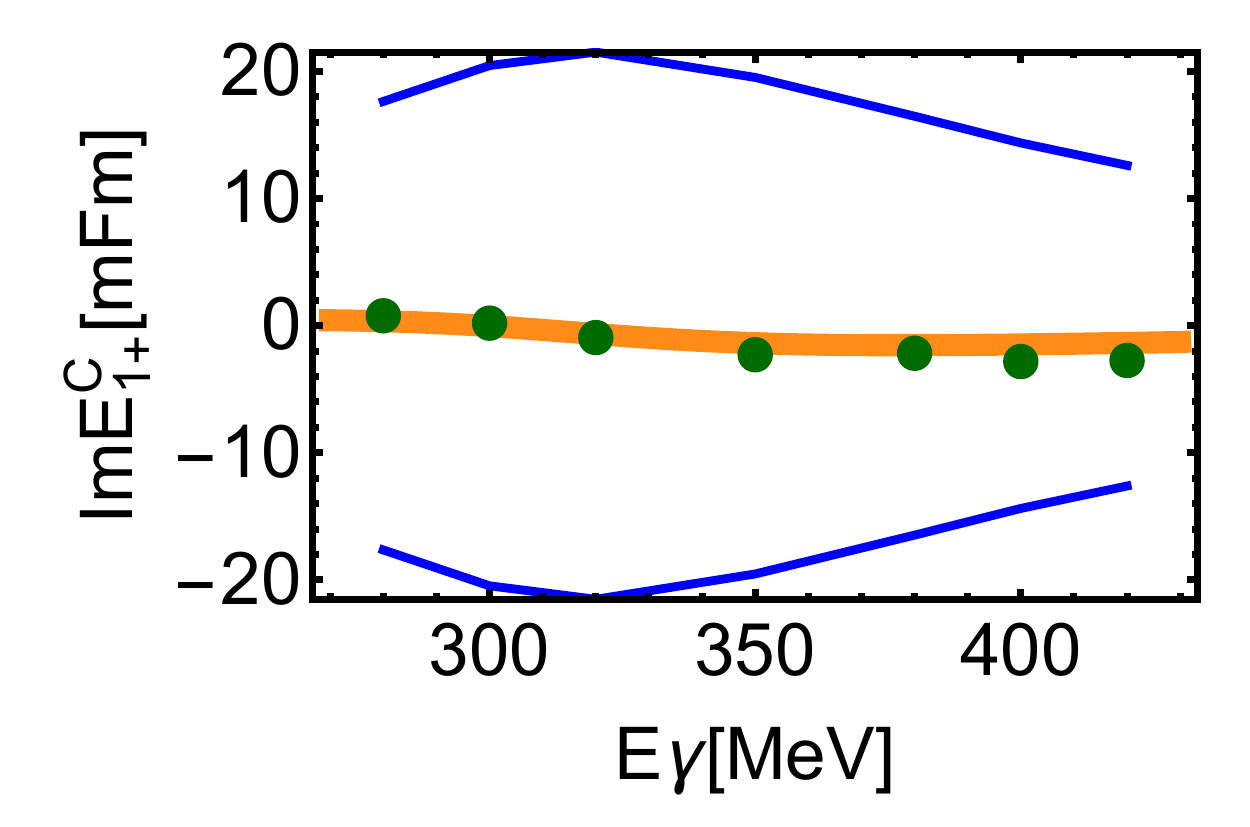}
 \end{overpic} \\
\begin{overpic}[width=0.325\textwidth]{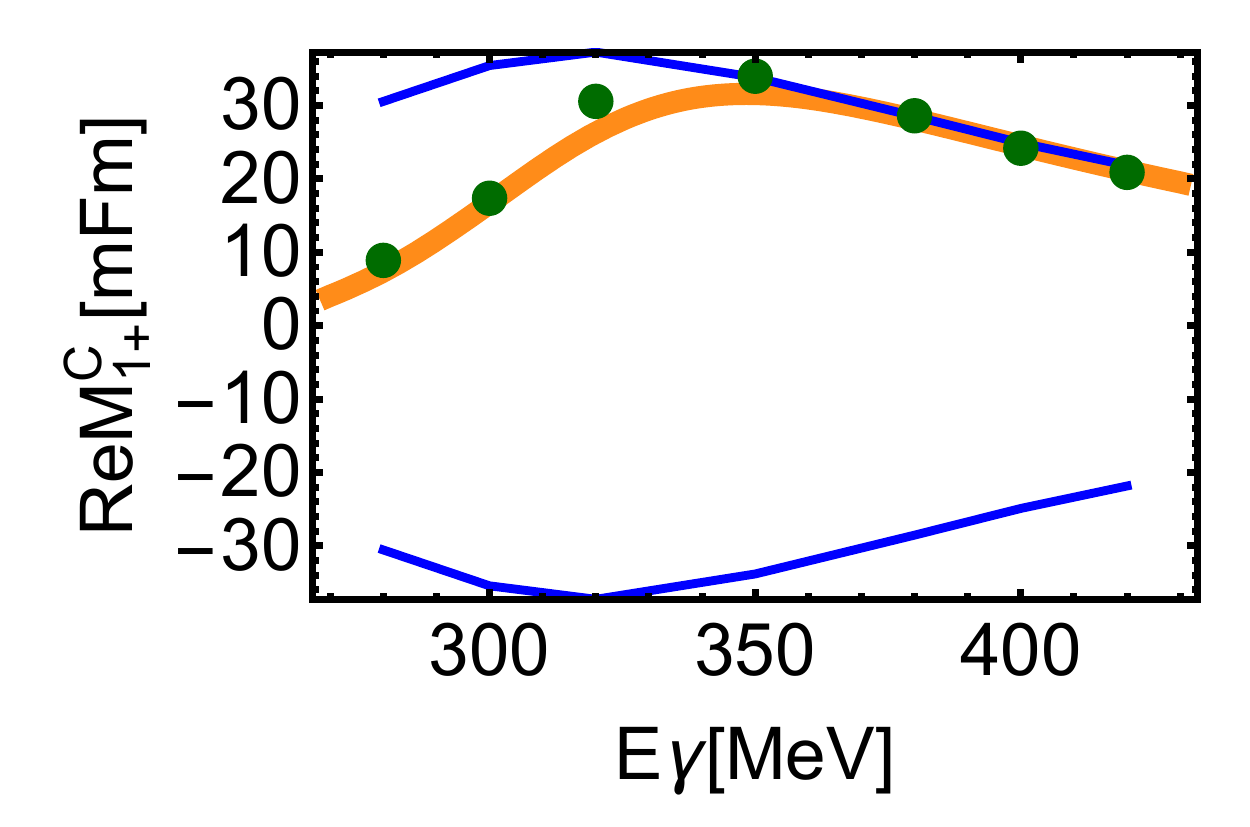}
 \end{overpic}
\begin{overpic}[width=0.325\textwidth]{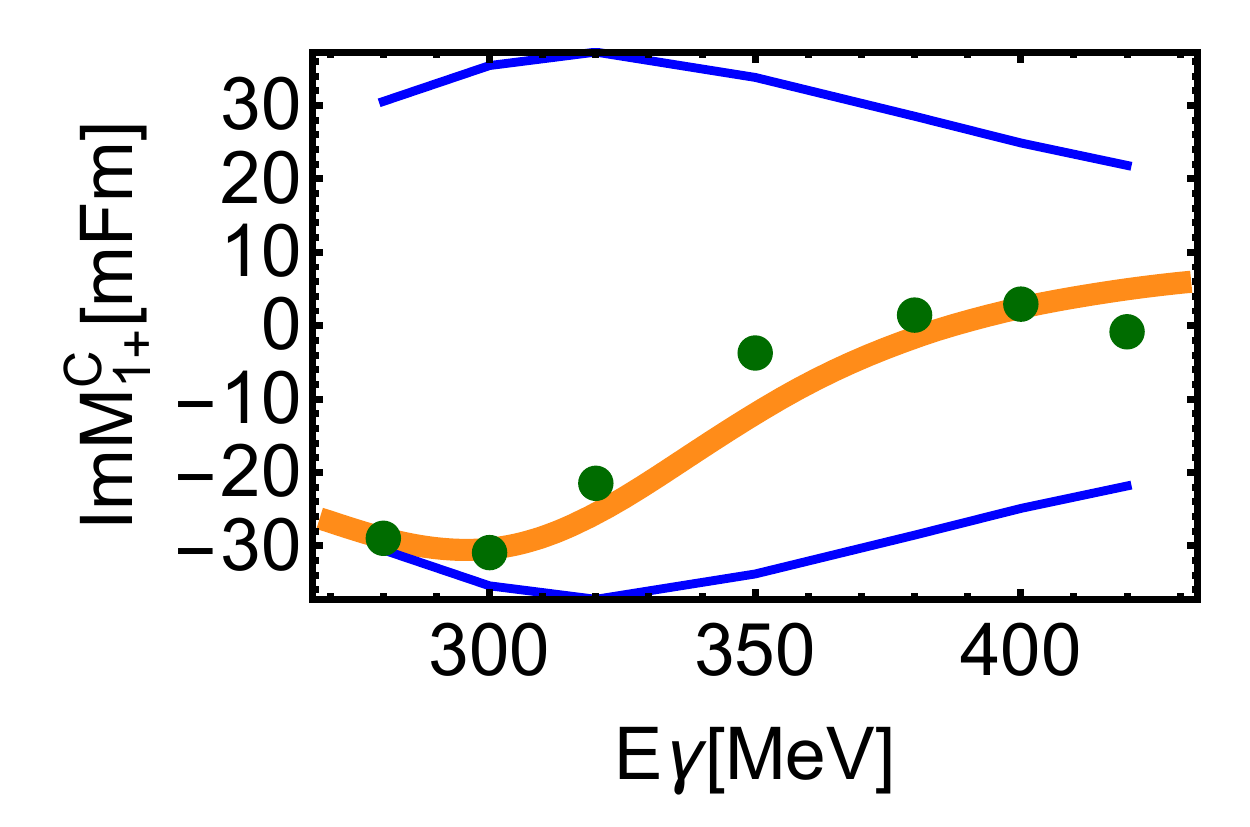}
 \end{overpic}
\begin{overpic}[width=0.325\textwidth]{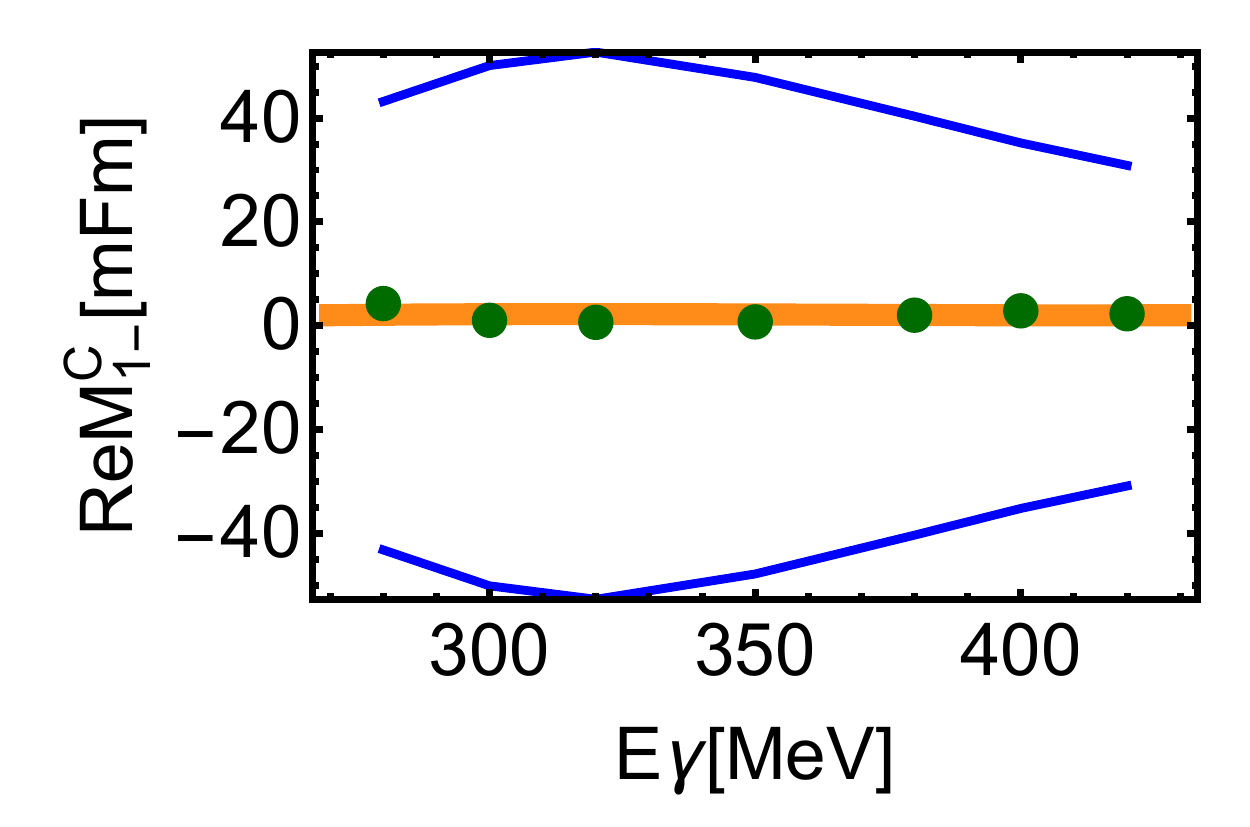}
 \end{overpic} \\
\begin{overpic}[width=0.325\textwidth]{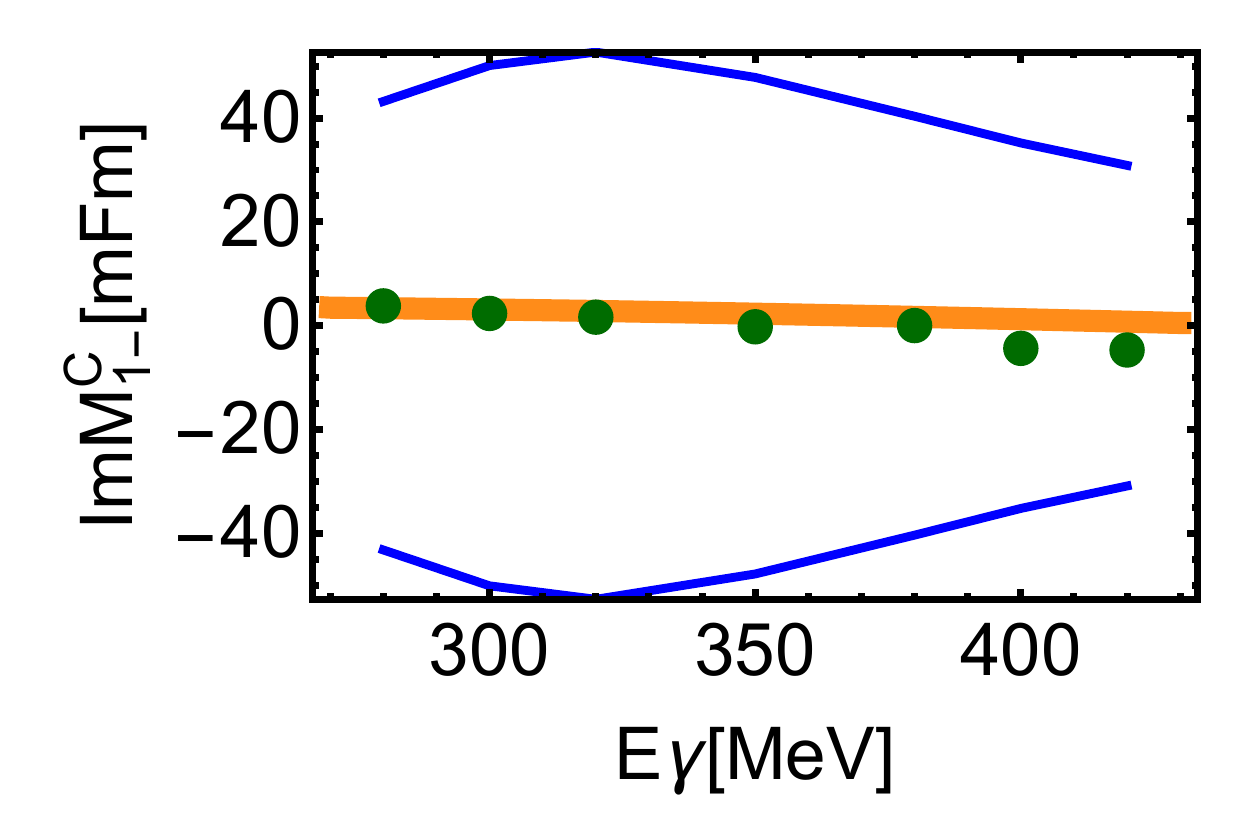}
 \end{overpic}
\caption[Results for the $7$ fit-parameters comprised of the real- and imaginary parts of phase-constrained $S$- and $P$-wave multipoles, as well as values for $\chi^{2} / \mathrm{ndf}$, for a TPWA employing $\ell_{\mathrm{max}} = 1$ within the $\Delta$-resonance region. The fits use scaling nuisance parameters for all $5$ observables.]{The given plots summarize the results of a TPWA-fit with $\ell_{\mathrm{max}} = 2$ to the selected data in the $\Delta$-region. $D$-wave multipoles have been fixed to the SAID-solution CM12 \cite{WorkmanEtAl2012ChewMPhotoprod}. The fits employed a pool of $N_{MC} = 1000$ initial parameter configurations (cf. section \ref{sec:MonteCarloSampling}). Furthermore, {\it nuisance parameters} have been employed as described in the main text. In particular, the 'effective chisquare-function' (\ref{eq:SystScalingFitFunction}) was minimized. \newline a.) {\it Left:} The obtained values for $\Phi_{\mathrm{data}}$ are plotted vs. energy for the fit method with nuisance parameters (\ref{eq:SystScalingFitFunction}). The global minimum (green dots) and local minima (red dots) are plotted. {\it Right:} The penalty-terms have been subtracted from the function $\Phi_{\mathrm{data}}$ in the global minimum (cf. the main text) and the resulting quantity was normalized to $\mathrm{ndf}$. This has been done only for the global minimum $\Phi_{\mathrm{data}}^{\mathrm{glob.}}$ (green dots). From the corresponding theoretical chisquare distributions, the mean (red line) as well as the pair of $0.025$- and $0.975$-quantiles (green dashed lines) are shown. For the estimate of the number of degrees of freedom, we used $\mathrm{ndf} = 79 - (7+4) = 68$. \newline b.) Results are shown for the real- and imaginary parts of the phase-constrained $S$- and $P$-wave multipoles. The global minimum (green dots) is compared to the model-solution SAID CM12 (orange colored curve) \cite{WorkmanEtAl2012ChewMPhotoprod, SAID}. For each fit-parameter, the maximal range set by the total cross section (cf. equation (\ref{eq:TCSCorrectedDeltaRegionFit})), as well as its energy-variation, is indicated by the blue solid lines.
}
\label{fig:Lmax2DWaveSAIDFitResultsDeltaRegionNuisanceParameters}
\end{figure}

Regarding the minimization of the effective chisquare (\ref{eq:SystScalingFitFunction}), we again do a Monte Carlo sampling of the initial conditions for the multipole-parameters $\left\{ \mathcal{M}_{\ell}  \right\}$. The $S$- and $P$-wave multipoles are varied, while the $D$-waves are again fixed to SAID CM12 \cite{WorkmanEtAl2012ChewMPhotoprod}. For each Monte Carlo fit, the nuisance parameters start at the value $\bm{p}_{\alpha} = 1$. The parameters $\sigma_{\bm{p},\alpha}$ are held fixed at the relative uncertainties quoted in the references \cite{Hornidge:2013, LeukelPhD, Schumann:2015, Belyaev:1983}, i.e. $\sigma_{\bm{p},\sigma_{0}} = 0.04$, $\sigma_{\bm{p},\check{\Sigma}} = 0.04$, $\sigma_{\bm{p},\check{P}} = 0.05$, and $\sigma_{\bm{p},\check{T}} = \sigma_{\bm{p},\check{F}} = 0.08$. Furthermore, we set $\bm{p}_{\check{T}} \equiv \bm{p}_{\check{F}}$, since both datasets stem from the same beam time. \newline
We employed the Monte Carlo fit method (section \ref{sec:MonteCarloSampling}) using a pool of $N_{MC} = 1000$ initial parameter configurations. The results are shown in Figure \ref{fig:Lmax2DWaveSAIDFitResultsDeltaRegionNuisanceParameters}. \newline
Again, a well-separated global minimum is found. The nuisance parameters attained in this global minimum are listed in Table \ref{tab:NuisanceParameterTable}. Nice features of the analysis are that first of all, the nuisance parameters come out mostly within the bounds demanded by the given systematic uncertainties $\sigma_{\bm{p},\alpha}$ and secondly that the same tendencies are found for the parameters irrespective of the energy. The fit tends to scale the fitting-functions of $\check{\Sigma}$, $\check{T}$ and $\check{F}$ up, while the functions for $\sigma_{0}$ and $\check{P}$ are scaled down. The fact that these general tendencies do not depend on the energy may be seen as a signal that something works out correctly. \newline
Subtracting the penalty-terms from the effective chisquare (\ref{eq:SystScalingFitFunction}), for parameters found in the global minimum and normalizing the resulting quantity to $\mathrm{ndf}$, one obtains a plot which is marked by '$\left(\Phi_{\mathrm{data}}^{\mathrm{glob.}} - \mathrm{penalty} \hspace*{1pt} \mathrm{terms}\right) / \mathrm{ndf}$' in Figure \ref{fig:Lmax2DWaveSAIDFitResultsDeltaRegionNuisanceParameters}. The resulting curve indeed shows small improvements compared to the $\chi^{2}_{\mathrm{data}}/\mathrm{ndf}$ seen in the equivalent fit without nuisance parameters, Figure \ref{fig:Lmax2DWavesSAIDFitResultsDeltaRegionPurelyStat}. However, the resulting numbers are still not in accordance with what is expected from theoretical chisquare-distributions. \newline
The multipole-fit-parameters again show good agreement with SAID CM12 \cite{WorkmanEtAl2012ChewMPhotoprod} in the global minimum. The change in parameter when compared to the previous two fits is again not drastic. It is mostly within the sub-percent to few-percent range (cf. Figure \ref{fig:Lmax2DWavesSAIDFitResultsDeltaRegionPurelyStat} or \ref{fig:Lmax2DWavesSAIDFitResultsDeltaRegionAddQuadrature}). An interesting observation is that the nuisance-corrected fit shown here has the capability to push the values for fit-parameters past the mathematical boundary set by total cross section (cf. section \ref{sec:MonteCarloSampling}), as can be seen for the parameter $\mathrm{Re} \left[ M_{1+} \right]$ at the fourth energy-bin (see Figure \ref{fig:Lmax2DWaveSAIDFitResultsDeltaRegionNuisanceParameters}). \newline

\vfill
\begin{table}[hb]
 \centering
 \begin{tabular}{l|ccccccc}
  $E_{\gamma}$ $\mathrm{[MeV]}$ & $280.$ & $300.$ & $320.$ & $350.$ & $380.$ & $400.$ & $420.$ \\
  \hline
  $\bm{p}_{\sigma_{0}}$ & $0.98$ & $0.98$ & $0.98$ & $0.95$ & $0.96$ & $0.96$ & $0.97$ \\
  $\bm{p}_{\check{\Sigma}}$ & $1.02$ & $1.01$ & $1.01$ & $1.01$ & $1.01$ & $1.01$ & $1.01$ \\
  $\bm{p}_{\check{P}}$ & $0.99$ & $0.99$ & $0.99$ & $1.00$ & $1.00$ & $0.99$ & $1.00$ \\
  $\bm{p}_{\check{T}} \equiv \bm{p}_{\check{F}}$ & $1.03$ & $1.03$ & $1.05$ & $1.14$ & $1.12$ & $1.11$ & $1.10$ \\
 \end{tabular}
 \caption[Results are shown for scaling nuisance parameters for a multipole-analysis in the $\Delta$-region.]{Nuisance parameters $\bm{p}_{\alpha}$ are shown resulting from the minimization of the effective chisquare-function (\ref{eq:SystScalingFitFunction}). Normalization uncertainties $\sigma_{\bm{p},\alpha}$ have been chosen in the minimizations as described in the main text. \newline
 The given numbers have been found in a well-separated global minimum. This minimum is the result of a search using the Monte Carlo methods described in section \ref{sec:MonteCarloSampling} for the $S$- and $P$-wave multipoles, while letting the nuisance parameters start from $\bm{p}_{\alpha} = 1$ for each fit. $D$-wave multipoles have been fixed to SAID CM12 \cite{WorkmanEtAl2012ChewMPhotoprod} in the analysis.}
\label{tab:NuisanceParameterTable}
\end{table}

\clearpage

The still unsatisfactory fit-quality for the fit using scaling-nuisance parameters may be traced back to the fact that, while the introduction of proportionality-factors to model systematic errors may already be a good start, the real potential damage caused to TPWA-fits comes from the fact that the systematic deviances are generally {\it angle-dependent}! In particular, the appearance of 'outlier-points' changes from angle to angle. Constant nuisance parameters such as those used in equation (\ref{eq:SystScalingFitFunction}) clearly cannot account for this effect. \newline
One of the most self-evident modifications that one could try, would be to turn the constant nuisance parameters into functions:
\begin{equation}
 \bm{p}_{\alpha} \longrightarrow \bm{p}_{\alpha} \left( \cos \theta \right) \mathrm{.} \label{eq:NuisanceParameterBecomesAFunction}
\end{equation}
Sadly, we cannot support this Ansatz by a citation at this moment. However, it seems a very nearby idea to try this. Thus, we suspect that it is not a new idea in any way. \newline
One, of course, would have to figure out a good suitable parametrization of the {\it nuisance-functions} (\ref{eq:NuisanceParameterBecomesAFunction}). These parametrizations should ideally not introduce too many additional parameters. Truncated Legendre-series or just polynomials in $\cos \theta$ might be a good idea. In any case, the quantities defining the parametrizations of the functions (\ref{eq:NuisanceParameterBecomesAFunction}) form then the new, enlarged set of nuisance parameters to be determined from the fit. It is also an interesting question which initial conditions should be chosen for these additional nuisance parameters, or how their penalty-terms should be defined. \newline
We have to state that we did not undertake additional attempts using functions such as (\ref{eq:NuisanceParameterBecomesAFunction}) in the course of this work. Therefore, at this point the method is just a speculation. \newline

As a last point regarding the method of nuisance parameters, we mention the fact that the parameters do not need to act on the fitting-functions as in done in equation (\ref{eq:SystScalingFitFunction}). Instead, one can also try to change the {\it data}. This is mentioned by Blobel \cite{BlobelTalk} and it is also in this form that Papanicolas and Markou employ the method \cite{LefterisPrivateComm,LefterisPaper}. \newline
In the latter case, the authors propose to apply a dimensionless scaling-factor $\bm{p}_{\alpha}$ and an offset $\bm{c}_{\alpha}$ with the dimension of a differential cross section to each measurement $\check{\Omega}^{\alpha}$, i.e. to replace \cite{LefterisPaper}
\begin{equation}
 \check{\Omega}^{\alpha}_{\mathrm{Data}} \longrightarrow \bm{c}_{\alpha} + \bm{p}_{\alpha} \hspace*{1.5pt}  \check{\Omega}^{\alpha}_{\mathrm{Data}} \mathrm{.} \label{eq:NuisanceManipulationOfData}
\end{equation}
Then, the effective chisquare-function becomes, including the required penalty-terms
\begin{align}
 \Phi_{\mathrm{data}} \left(  \left\{ \mathcal{M}_{\ell}  \right\}; \left\{ \bm{c}_{\alpha} \right\} , \left\{ \bm{p}_{\alpha} \right\} \right) &:= \sum_{\check{\Omega}^{\alpha}, c_{k_{\alpha}}} \left( \frac{ \left[ \bm{c}_{\alpha} + \bm{p}_{\alpha} \hspace*{1.5pt}  \check{\Omega}^{\alpha}_{\mathrm{Data}} (c_{k_{\alpha}}) \right] - \check{\Omega}^{\alpha}_{\mathrm{Fit}} \left( c_{k_{\alpha}}, \left\{ \mathcal{M}_{\ell} \right\} \right) }{ \bm{p}_{\alpha} \hspace*{1.5pt} \Delta \check{\Omega}^{\alpha}_{\mathrm{Data}} (c_{k_{\alpha}})} \right)^{2} \nonumber \\
 & \hspace*{12.5pt} + \sum_{\alpha} \left[ \left( \frac{\bm{c}_{\alpha}}{\sigma_{\bm{c},\alpha}} \right)^{2} + \left( \frac{\bm{p}_{\alpha} - 1}{\sigma_{\bm{p},\alpha}} \right)^{2} \right] \mathrm{.} \label{eq:SystScalingFitFunction2}
\end{align}
A comment is in order: As remarked by Blobel \cite{BlobelTalk}, whenever the data are rescaled in a nuisance-corrected fit, one has to apply the same transformation to the statistical errors for the datapoints, as done above. Otherwise one introduces an additional normalization bias! \newline

Concerning the normalization-uncertainties $\sigma_{\bm{p},\alpha}$, their values are fixed just as described above. The uncertainties $\sigma_{\bm{c},\alpha}$ for the offset-parameters have to be determined by some alternative consistent scheme. In this thesis, some attempts were done using the Ansatz (\ref{eq:SystScalingFitFunction2}) and in this case, we multiplied the dimensionless number $\sigma_{\bm{p},\alpha}$ to the maximum of the respective observable in a given angular distribution, in order to determine a reasonable estimate for $\sigma_{\bm{c},\alpha}$, i.e. $\sigma_{\bm{c},\alpha} = \sigma_{\bm{p},\alpha} \hspace*{1pt} \left( \check{\Omega}^{\alpha}_{\mathrm{Data}} \right)_{\mathrm{max}}$. \newline
Then, some attempts were done by minimizing the quantity (\ref{eq:SystScalingFitFunction2}), fitting the offsets $\left\{ \bm{c}_{\alpha} \right\}$ together with multipoles and scaling-factors. As initial conditions, we employed $\bm{c}_{\alpha} = 0$ in every fit. \newline
The results of attempts to change the data are not included here explicitly. It should however be stated that they were consistent to what has been found before, using the Ansatz of scaling the fit-functions (\ref{eq:SystScalingFitFunction}). This means, whenever the previous method (\ref{eq:SystScalingFitFunction}) tended to scale the fit-functions up, the present method (\ref{eq:SystScalingFitFunction2}) decreased the size of the data and vice versa. Moreover, it has not been possible to obtain a significant improvement of the fit-qualities using the Ansatz of changing the data. In order to improve on this, one could again try to let the angle-independent nuisance-parameters become functions of $\cos \theta$, but this is again speculation and goes beyond the investigations performed in the course of this work. \newline

For the final error-analysis, it is suitable to use the bootstrap (see section \ref{sec:BootstrappingIntroduction}). However, the question remains which of the above-described fits to select for such a resampling analysis. We choose here to show the results for the third fit in more detail, i.e. a direct fit of a truncation at $\ell_{\mathrm{max}} = 2$ with $D$-waves fixed to SAID CM12 \cite{WorkmanEtAl2012ChewMPhotoprod,SAID}, all while only using the statistical errors as input into the analysis and not treating systematic errors in any specific way. The global minimum found in this fit was shown in Figure \ref{fig:Lmax2DWavesSAIDFitResultsDeltaRegionPurelyStat}. \newline
It should be stated in the beginning, that two important global minima shown here, namely those resulting from the first fit shown in Figure \ref{fig:Lmax1UnconstrainedFitResultsDeltaRegion} and the third fit summarized in Figure \ref{fig:Lmax2DWavesSAIDFitResultsDeltaRegionPurelyStat}, have been checked to be stable even in the context of a {\it full} bootstrap-TPWA (cf. section \ref{sec:BootstrappingIntroduction}). This means that for each bootstrap-replicate of the data, where $B=400$ such replicates were employed, the complete Monte Carlo fit-procedure (section \ref{sec:MonteCarloSampling}) was done, using $N_{MC} = 500$ randomly generated initial conditions. Then, once the global minima of each fit to every replicate were plotted in histograms, the resulting distributions showed only one (sometimes asymmetric) gaussian peak. This means, no alternative solutiones 'turned up' in the bootstrap-distributions as a result of the statistical resampling. However, looking at how well the global minima in Figures \ref{fig:Lmax1UnconstrainedFitResultsDeltaRegion} and \ref{fig:Lmax2DWavesSAIDFitResultsDeltaRegionPurelyStat} are separated, this result does not surprise. In section \ref{subsec:2ndResRegionDataFits}, a fit will be shown where ambiguities actually did occur in bootstrap-distributions. \newline

\begin{figure}[ht]
\centering
\vspace*{-10pt}
\begin{overpic}[width=0.325\textwidth]{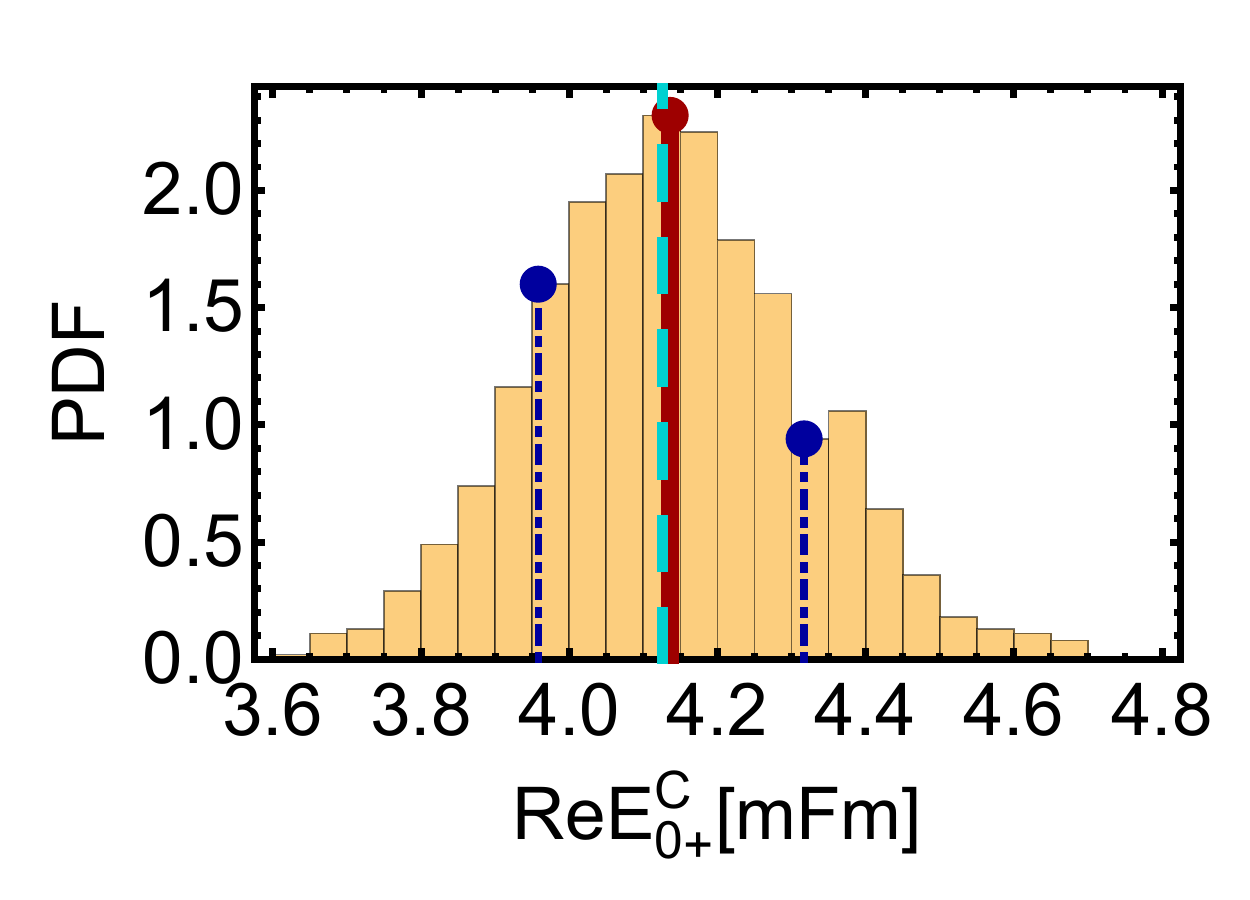}
 \end{overpic}
\begin{overpic}[width=0.325\textwidth]{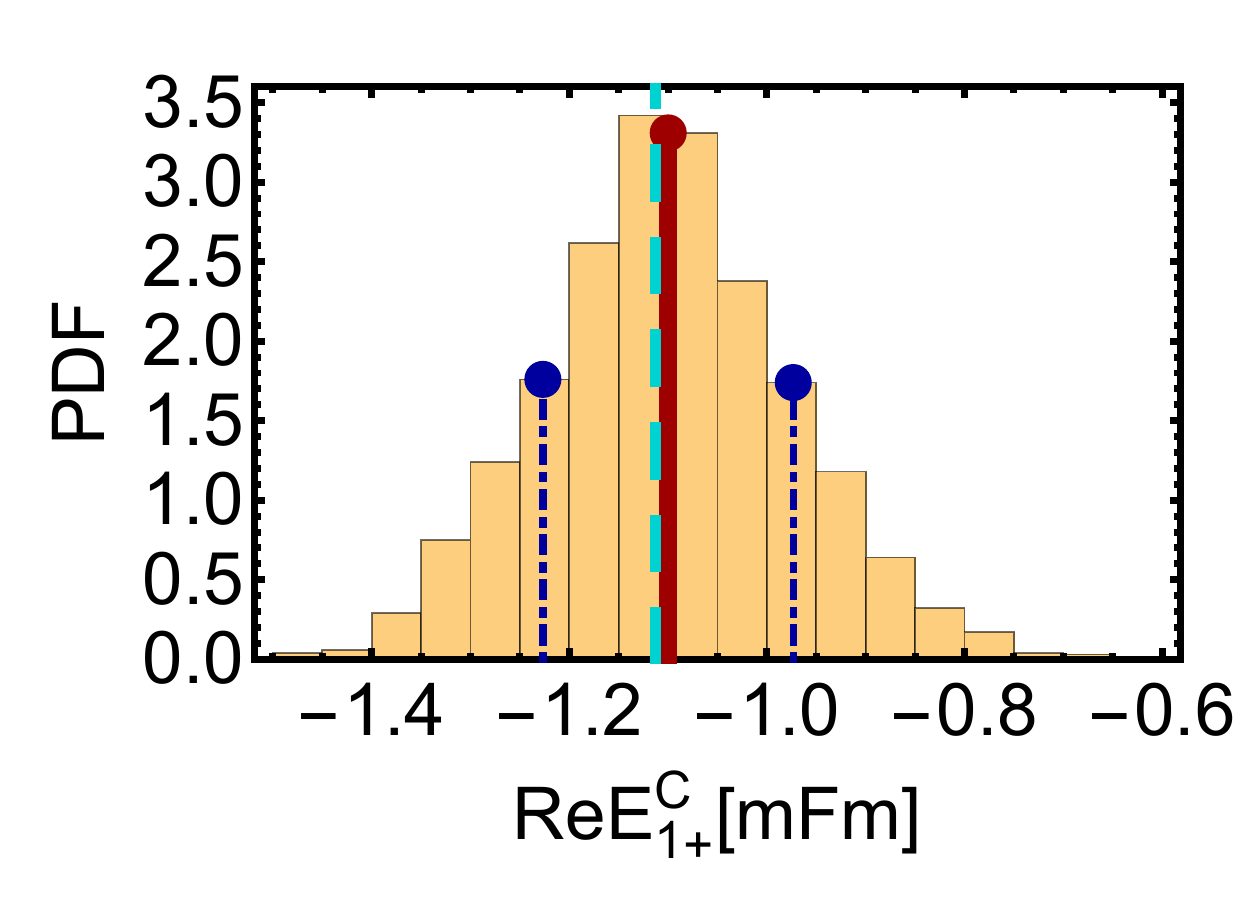}
 \end{overpic}
\begin{overpic}[width=0.325\textwidth]{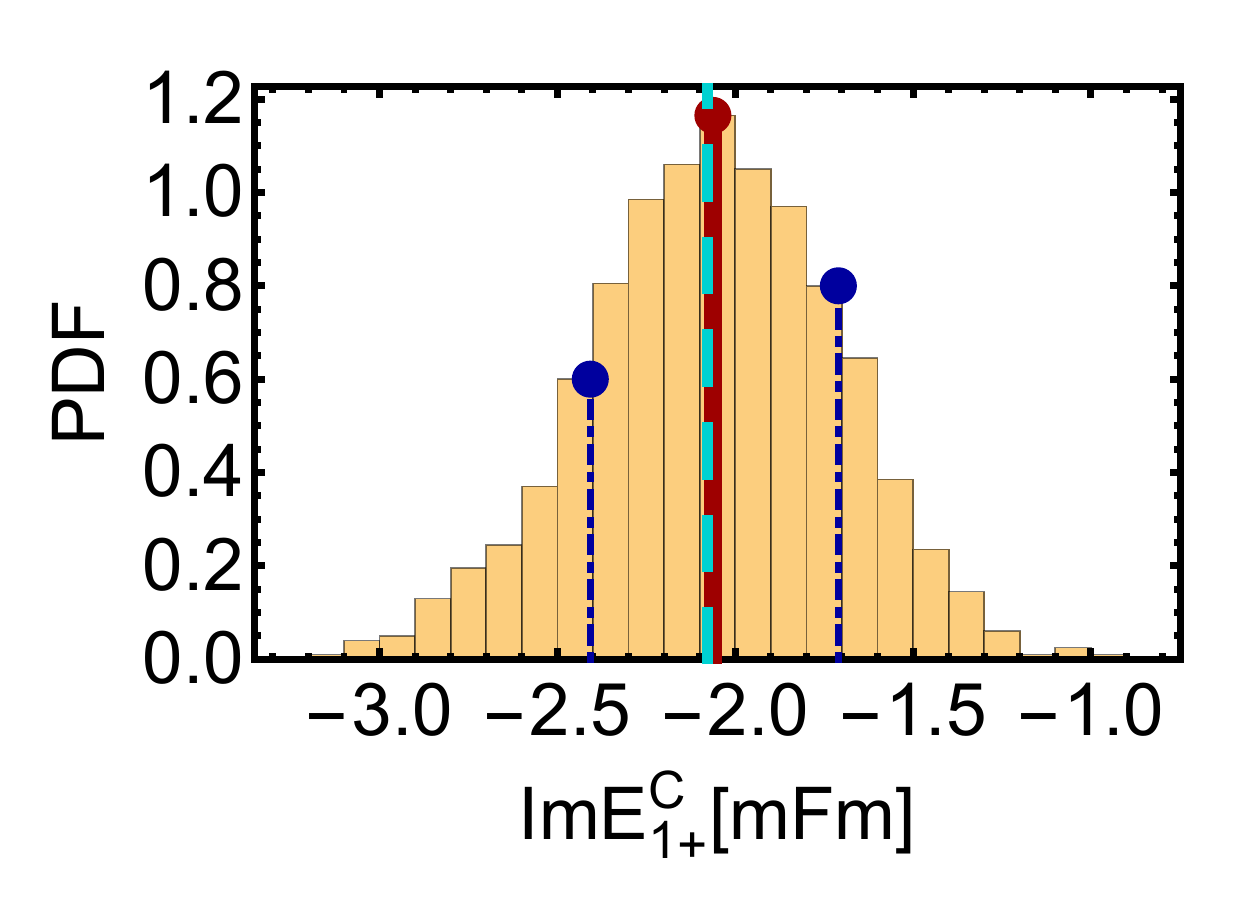}
 \end{overpic} \\ \vspace*{-8pt}
\begin{overpic}[width=0.325\textwidth]{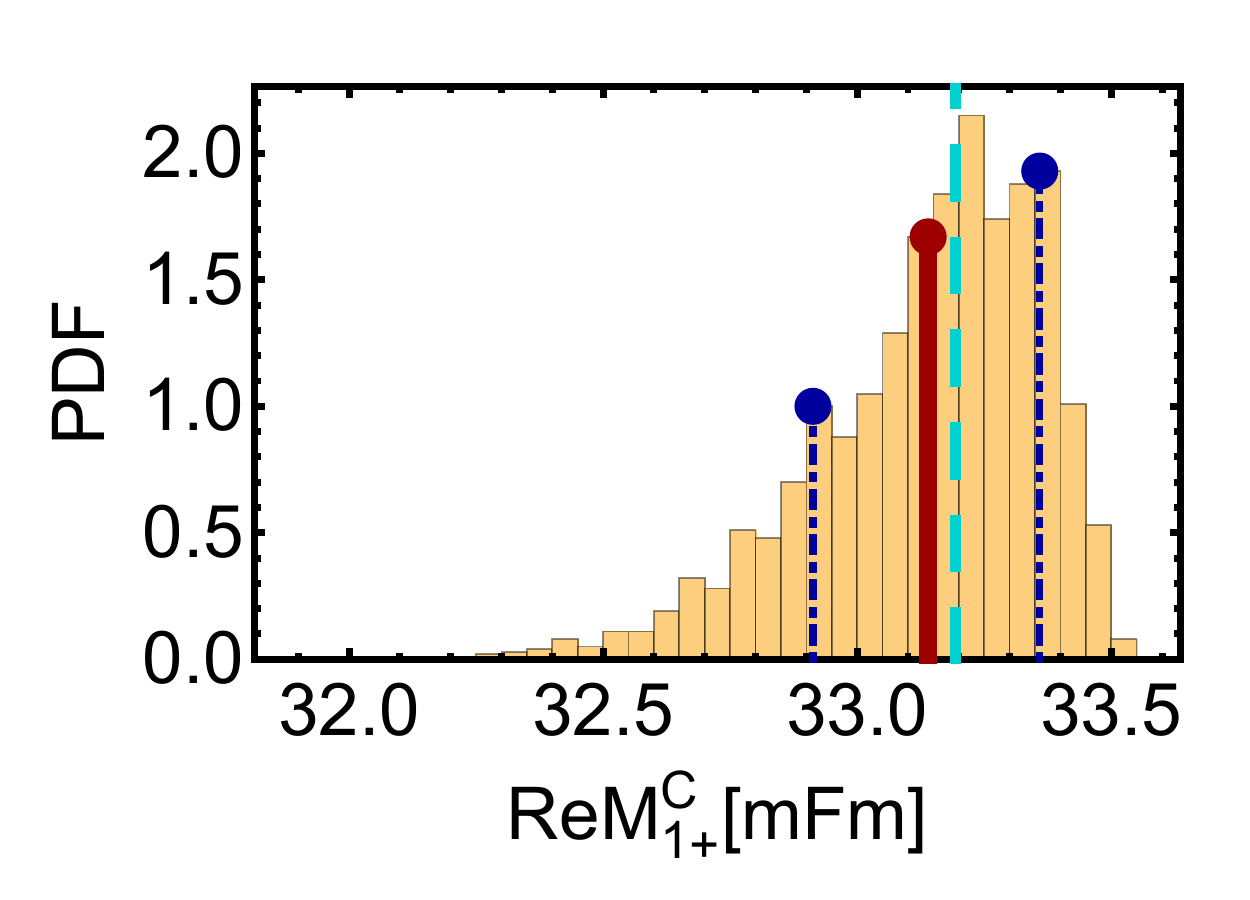}
 \end{overpic}
\begin{overpic}[width=0.325\textwidth]{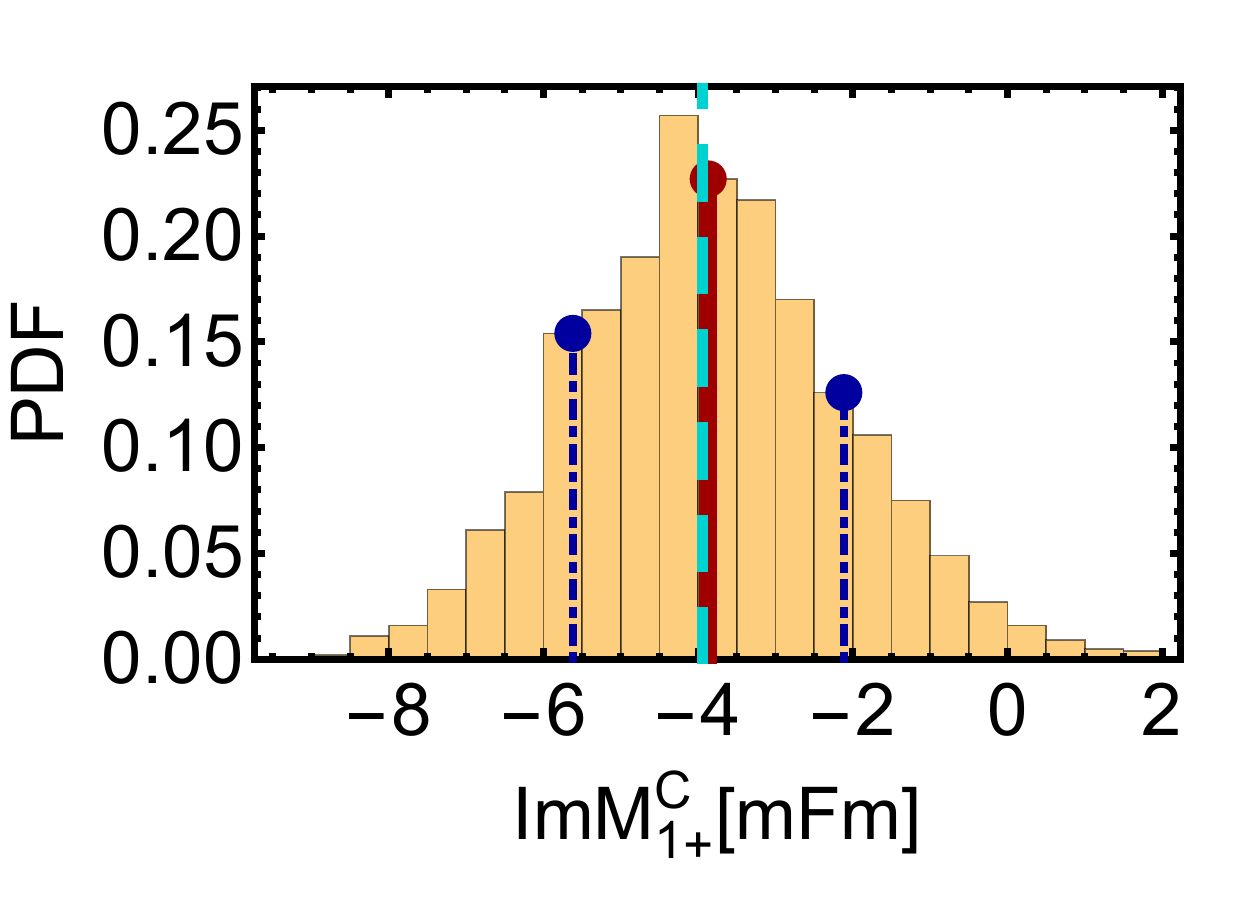}
 \end{overpic}
\begin{overpic}[width=0.325\textwidth]{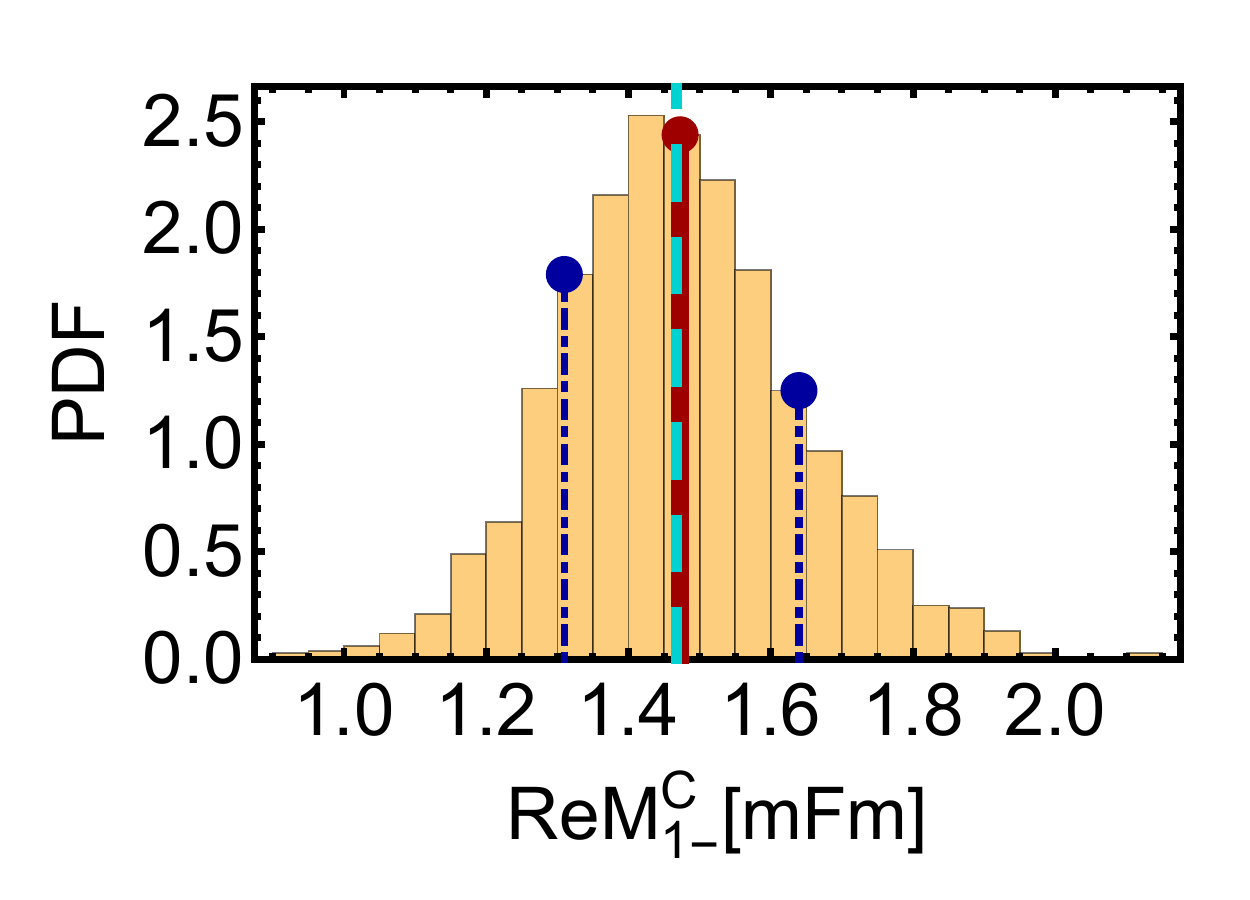}
 \end{overpic} \\ \vspace*{-8pt}
\begin{overpic}[width=0.325\textwidth]{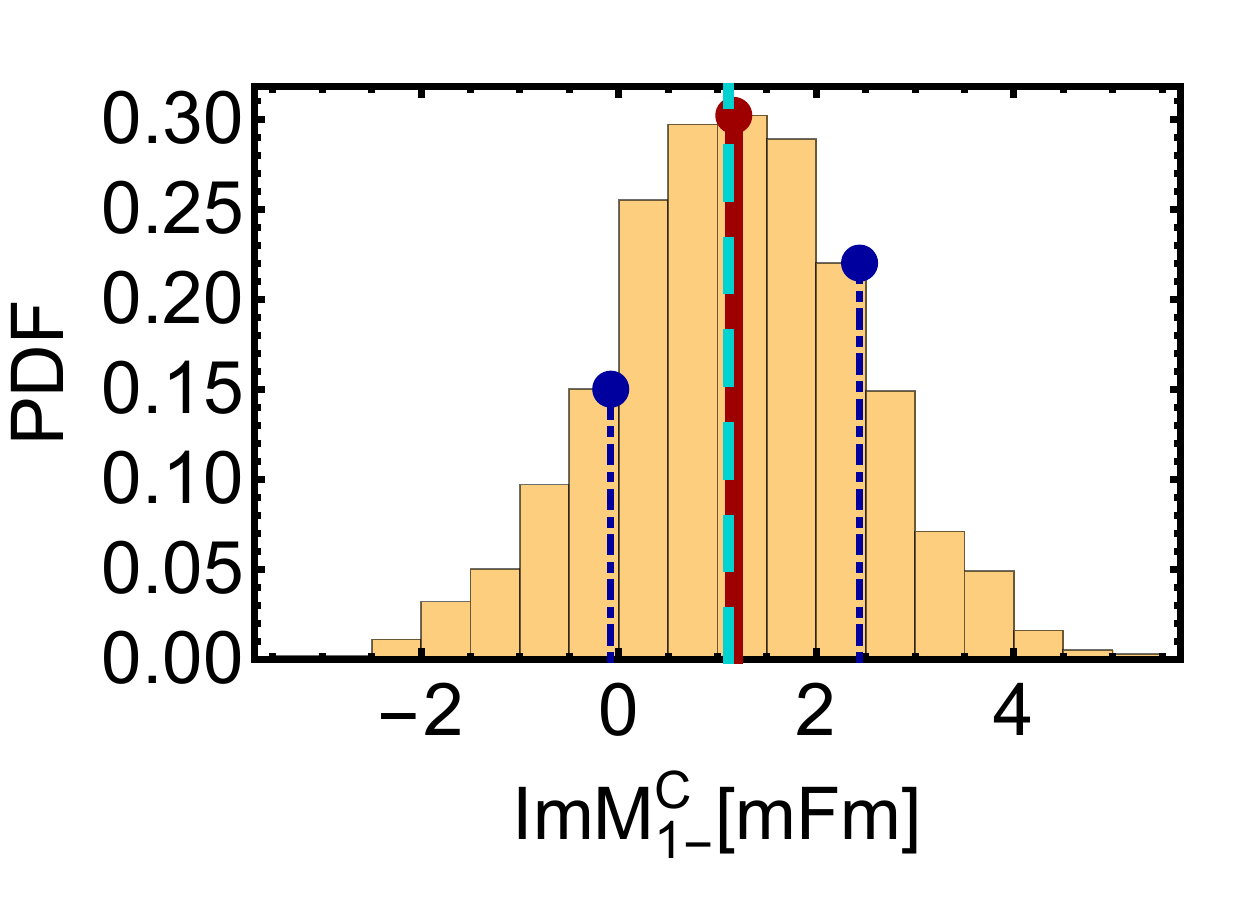}
 \end{overpic}
\caption[Bootstrap-distributions for multipole-fit-parameters in an analysis of photoproduction data on the $\Delta$-resonance region. The fourth energy-bin, \newline $E_{\gamma }\text{ = 350.0 MeV}$, is shown.]{The histograms show bootstrap-distributions for the real- and imaginary parts of phase-constrained $S$- and $P$-wave multipoles, for a TPWA bootstrap-analysis of photoproduction data in the $\Delta$-resonance region. The fourth energy-bin, $E_{\gamma }\text{ = 350.0 MeV}$, is shown. An ensemble of $B=2000$ bootstrap-replicates has been the basis of these results, using solely the statistical errors of the original datasets. The $D$-waves were held fixed to the SAID-solution CM12 \cite{WorkmanEtAl2012ChewMPhotoprod,SAID} during the entire analysis. The fit to the original data is shown in Figure \ref{fig:Lmax2DWavesSAIDFitResultsDeltaRegionPurelyStat}. Statistical quantities derived from these bootstrap-distributions are contained in Table \ref{tab:BootstrapFitResultsDeltaRegionFourthEnergyMainText}. \newline
The distributions have been normalized to $1$ via use of the object \textit{HistogramDistribution} in MATHEMATICA \cite{MathematicaLanguage,MathematicaBonnLicense}. Thus, $y$-axes are labeled as \textit{PDF}. The mean of each distribution is shown as a red solid line, while the $0.16$- and $0.84$-quantiles are indicated by blue dash-dotted lines. The global minimum of the fit to the original data is plotted as a cyan-colored dashed horizontal line.}
\label{fig:BootstrapHistosDeltaRegionEnergy4MainText}
\end{figure}

We commence with more details of a {\it reduced} bootstrap-TPWA (section \ref{sec:BootstrappingIntroduction}) for the third fit. In order to do this, bootstrap-replicates $\left\{\check{\Omega}_{\ast}^{\alpha}\right\}$ for the fitted profile functions were drawn out of normal distributions as $\mathcal{N} \left( \check{\Omega}^{\alpha}, \Delta \check{\Omega}^{\alpha} \right) \longrightarrow \check{\Omega}_{\ast}^{\alpha}$ (see equation (\ref{eq:BootstrapDefinitionDeltaRegionFit}) and section \ref{sec:BootstrappingIntroduction}). An ensemble of $B=2000$ bootstrap-datasets has been generated in this way. For each of these replicate-datasets, a full Monte Carlo minimum-search is omitted and instead a single re-fit is done, minimizing the chisquare-function (\ref{eq:ChiSquareDirectFitRealDataFitSection}), starting at the global minimum shown in Figure \ref{fig:Lmax2DWavesSAIDFitResultsDeltaRegionPurelyStat}. The $S$- and $P$-wave multipoles are varied, with $D$-waves fixed to SAID CM12 \cite{WorkmanEtAl2012ChewMPhotoprod,SAID}. The whole analysis operates under the overall phase-constraint $\mathrm{Im} \left[ E_{0+} \right] = 0$ $\&$ $\mathrm{Re} \left[ E_{0+} \right] > 0$. \newline
The resulting bootstrap-distributions for the multipole-fit-parameters can then be stored and used for further analysis (section \ref{sec:BootstrappingIntroduction}). As an example, histograms with bootstrap-distributions for the fourth energy-bin $E_{\gamma} = 350 \hspace*{1.5pt} \mathrm{MeV}$ can be seen in Figure \ref{fig:BootstrapHistosDeltaRegionEnergy4MainText}. Distributions for all the remaining energies are contained in appendix \ref{sec:NumericalTPWAFitResults} (see Figures \ref{fig:BootstrapHistosDeltaRegionEnergy1} to \ref{fig:BootstrapHistosDeltaRegionEnergies6and7}). \clearpage

 The bootstrap-distributions for the fit-parameters $\hat{\theta}_{i} = \left( \mathcal{M}_{\ell}^{C} \right)_{i}$ then yield statistical quantities of interest. For instance, the pair of $0.16$- and $0.84$-quantiles can be extracted, which define a $68\%$ confidence-interval for the individual parameter. The upper and lower bootstrap-errors $\Delta_{\pm}$ are then defined as the modulus of the distance of the fit-result for the global minimum (Figure \ref{fig:Lmax2DWavesSAIDFitResultsDeltaRegionPurelyStat}) to the lower quantile, which yields $\Delta_{-}$ and to the upper quantile, which defines $\Delta_{+}$. The main result of the fit is then given as the global minimum found in the fit to the original data, together with the asymmetric errors inferred from fits to the bootstrap-replicates (section \ref{sec:BootstrappingIntroduction}). \newline
 However, further quantities of interest are of course quickly extracted as well. We determine the mean $\hat{\theta}_{i}^{\ast} (\cdot)$, standard error $\widehat{\mathrm{se}}_{B} \left( \hat{\theta}^{\ast}_{i} \right)$ and bias-estimate $\widehat{\mathrm{bias}}_{B}$ from the bootstrap-distributions (cf. section \ref{sec:BootstrappingIntroduction}). \newpage
 \begin{table}[h]
\centering
\begin{tabular}{c|c|c|c|c|c}
\multicolumn{2}{l|}{$ E_{\gamma }\text{ = 350.0 MeV} $} & \multicolumn{2}{c|}{ $ \text{ndf = 72} $} & \multicolumn{2}{c}{ $ \chi^{2}_{\mathrm{data}}\text{/ndf = 2.92216} $ } \\
\hline
\hline
$ \hat{\theta}_{i} = \left( \mathcal{M}_{\ell}^{C} \right)_{i} \text{[mFm]} $ & $ \left(\hat{\theta}_{i}^{\mathrm{Best}}\right)_{- \Delta_{-}}^{+ \Delta_{+}} $ & $ \hat{\theta}_{i}^{\ast} (\cdot) $ & $ \widehat{\mathrm{se}}_{B} \left( \hat{\theta}^{\ast}_{i} \right) $ & $ \widehat{\mathrm{bias}}_{B} $ & $ \delta_{\mathrm{bias}} $\\
\hline
 $ \mathrm{Re} \left[ E_{0+}^{C} \right] $ & $ 4.12632_{-0.16819}^{+0.1907} $ & $ 4.13615 $ & $ 0.18187 $ & $ 0.00983 $ & $ 0.05407 $ \\
 $ \mathrm{Re} \left[ E_{1+}^{C} \right] $ & $ -1.11254_{-0.11417}^{+0.13926} $ & $ -1.09985 $ & $ 0.12748 $ & $ 0.01269 $ & $ 0.09951 $ \\
 $ \mathrm{Im} \left[ E_{1+}^{C} \right] $ & $ -2.07794_{-0.32956}^{+0.36794} $ & $ -2.0628 $ & $ 0.35378 $ & $ 0.01514 $ & $ 0.0428 $ \\
 $ \mathrm{Re} \left[ M_{1+}^{C} \right] $ & $ 33.193_{-0.28003}^{+0.1663} $ & $ 33.1397 $ & $ 0.22645 $ & $ -0.05335 $ & $ 0.2356 $ \\
 $ \mathrm{Im} \left[ M_{1+}^{C} \right] $ & $ -3.93992_{-1.68021}^{+1.82484} $ & $ -3.8695 $ & $ 1.74887 $ & $ 0.07042 $ & $ 0.04026 $ \\
 $ \mathrm{Re} \left[ M_{1-}^{C} \right] $ & $ 1.4676_{-0.15782}^{+0.172} $ & $ 1.4725 $ & $ 0.17162 $ & $ 0.0049 $ & $ 0.02856 $ \\
 $ \mathrm{Im} \left[ M_{1-}^{C} \right] $ & $ 1.10874_{-1.18874}^{+1.3299} $ & $ 1.1655 $ & $ 1.29632 $ & $ 0.05676 $ & $ 0.04379 $
\end{tabular}
\caption[Numerical results of a bootstrap-analysis are collected for a TPWA-fit of photoproduction data within the $\Delta$-resonance region. The $D$-waves were fixed to SAID CM12 \cite{WorkmanEtAl2012ChewMPhotoprod}. Shown are results for the fourth energy-bin, \newline $E_{\gamma }\text{ = 350.0 MeV}$.]{Numerical results of a bootstrap-analysis (cf. section \ref{sec:BootstrappingIntroduction}) are collected for a TPWA-fit of photoproduction data within the $\Delta$-resonance region, with $S$- and $P$-wave multipoles varied in the fit, while the $D$-waves were fixed to SAID CM12 \cite{WorkmanEtAl2012ChewMPhotoprod}. An ensemble of $B = 2000$ bootstrap-replicates has been applied. The global minimum found in the fit of the original data, which is itself shown in Figure \ref{fig:Lmax2DWavesSAIDFitResultsDeltaRegionPurelyStat}, has been used as initial parameter configuration in each of the bootstrap-fits. Shown are results for the fourth energy-bin, $E_{\gamma }\text{ = 350.0 MeV}$. Here, a global minimum has been found with $\chi^{2}_{\mathrm{data}}\text{/ndf = 2.92216}$. \newline
From the bootstrap-distributions of the fit-parameters, we extract quantiles which then define a confidence-interval for the individual parameter, composed of upper and lower bootstrap-errors $\Delta_{\pm}$. The global minimum is quoted in conjunction with these asymmetric errors (for more details, see the main text). Furthermore, the mean $\hat{\theta}_{i}^{\ast} (\cdot)$, standard error $\widehat{\mathrm{se}}_{B} \left( \hat{\theta}^{\ast}_{i} \right)$ and bias-estimate $\widehat{\mathrm{bias}}_{B}$ are extracted from the bootstrap-distributions. Lastly, we define and extract a bias test-parameter defined as $\delta_{\mathrm{bias}} := \left| \widehat{\mathrm{bias}}_{B} \right|/\widehat{\mathrm{se}}_{B}$ (equation (\ref{eq:BiasTestParameterDefinition})). \newline
All numbers are given in milli-Fermi, except for $\delta_{\mathrm{bias}}$ which does not carry dimension.}
\label{tab:BootstrapFitResultsDeltaRegionFourthEnergyMainText}
\end{table}
 The $\widehat{\mathrm{bias}}_{B}$-parameter is particularly useful to check the internal consistency of the analysis (cf. section \ref{sec:BootstrappingIntroduction} and reference \cite{EfronTibshiraniBook}). According to Efron and Tibshirani \cite{EfronTibshiraniBook}, biases of less than $25\%$ of the standard error $\widehat{\mathrm{se}}_{B} \left( \hat{\theta}^{\ast}_{i} \right)$ can be safely ignored, while for values substantially above this boundary, the analysis should be carefully reconsidered. Thus, we define a bias test-parameter
\begin{equation}
 \delta_{\mathrm{bias}} := \left| \widehat{\mathrm{bias}}_{B} \right|/\widehat{\mathrm{se}}_{B} \mathrm{,} \label{eq:BiasTestParameterDefinition}
\end{equation}
which is extracted for each bootstrap-distribution and given in the result-tables for quick reference. The numerical results stemming from the bootstrap-distribution for the fourth energy-bin $E_{\gamma} = 350 \hspace*{1.5pt} \mathrm{MeV}$ are shown in Table \ref{tab:BootstrapFitResultsDeltaRegionFourthEnergyMainText}. Derived statistical quantities for all the energies can be found in appendix \ref{sec:NumericalTPWAFitResults} (see Tables \ref{tab:DeltaRegionResultsFirstEnergy} to \ref{tab:DeltaRegionResultsFifthSixthSeventhEnergy}). \newline
It has to be stated that for almost all energies and fit-parameters, the quantity $\delta_{\mathrm{bias}}$ is a lot smaller than $0.25$. An exception to this rule is made by the parameter $\mathrm{Re}\left[ M_{1+}^{C} \right]$ and it turns out to be noticeably biased in the fourth, fifth, sixth and seventh energy-bin. This can be seen even in the example-Table \ref{tab:BootstrapFitResultsDeltaRegionFourthEnergyMainText}. A glance into the corresponding bootstrap-histograms (cf. Figure \ref{fig:BootstrapHistosDeltaRegionEnergy4MainText}) reveals that the corresponding distributions have a strongly asymmetric gaussian shape. This is probably caused by strong correlations between the respective parameter and one or multiple other fit-parameters. \newline
In case a noticeable bias is introduced by very asymmetric parameter distributions as in this case, we just ignore it. It could however also be that in cases of $\delta_{\mathrm{bias}} > 0.25$, the surpassing of the $25\%$ boundary is indicative of some deeper pathologies of the fit, most noticeably ambiguities (see section \ref{subsec:2ndResRegionDataFits}). Thus, the bias-check is in general a good method in order determine which bootstrap-distributions have to be investigated in more detail. \newline

\begin{figure}[h]
\centering
\begin{overpic}[width=0.325\textwidth]{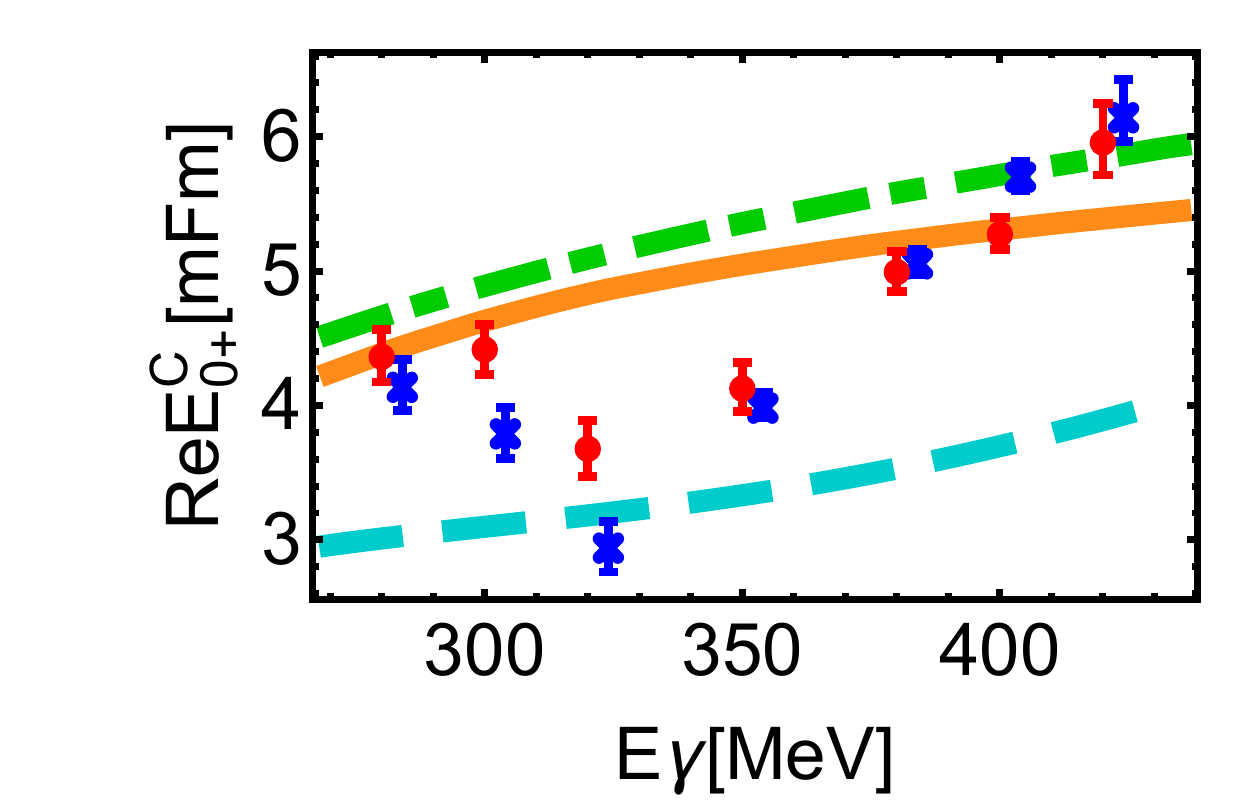}
 \end{overpic}
\begin{overpic}[width=0.325\textwidth]{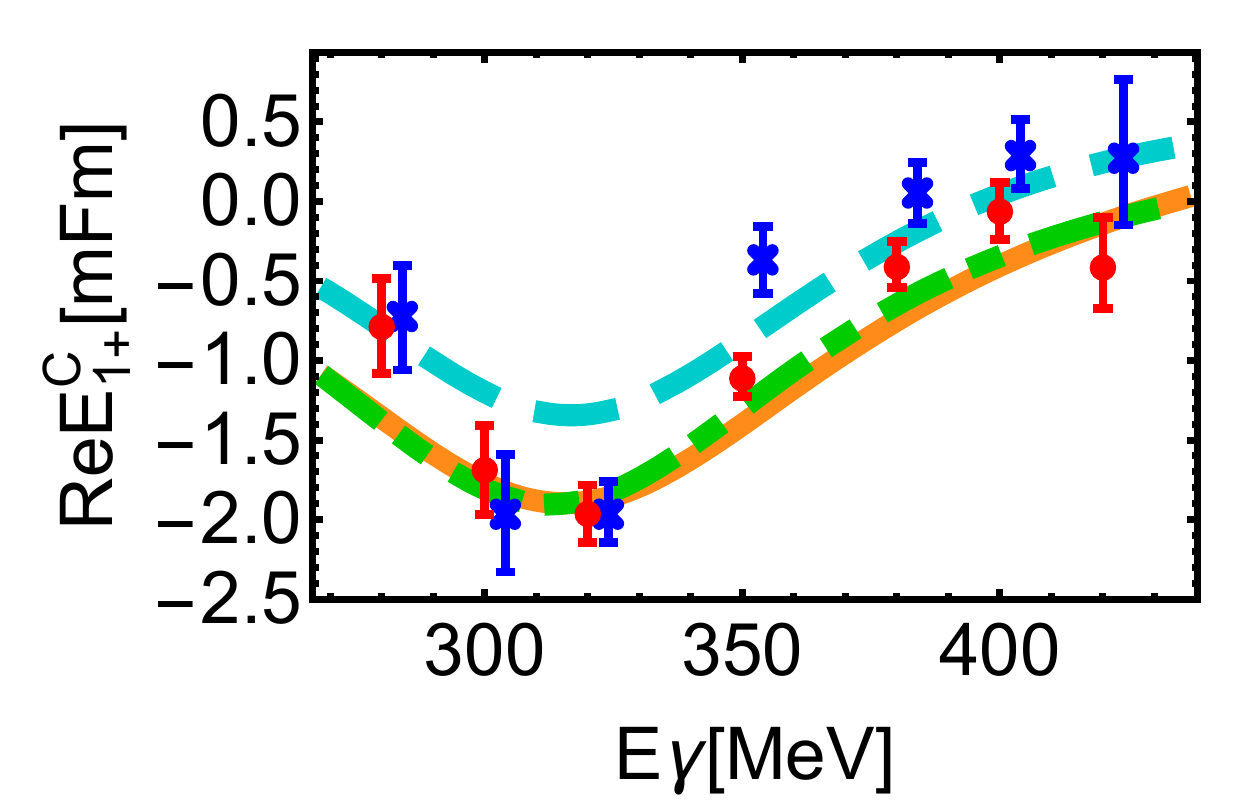}
 \end{overpic}
\begin{overpic}[width=0.325\textwidth]{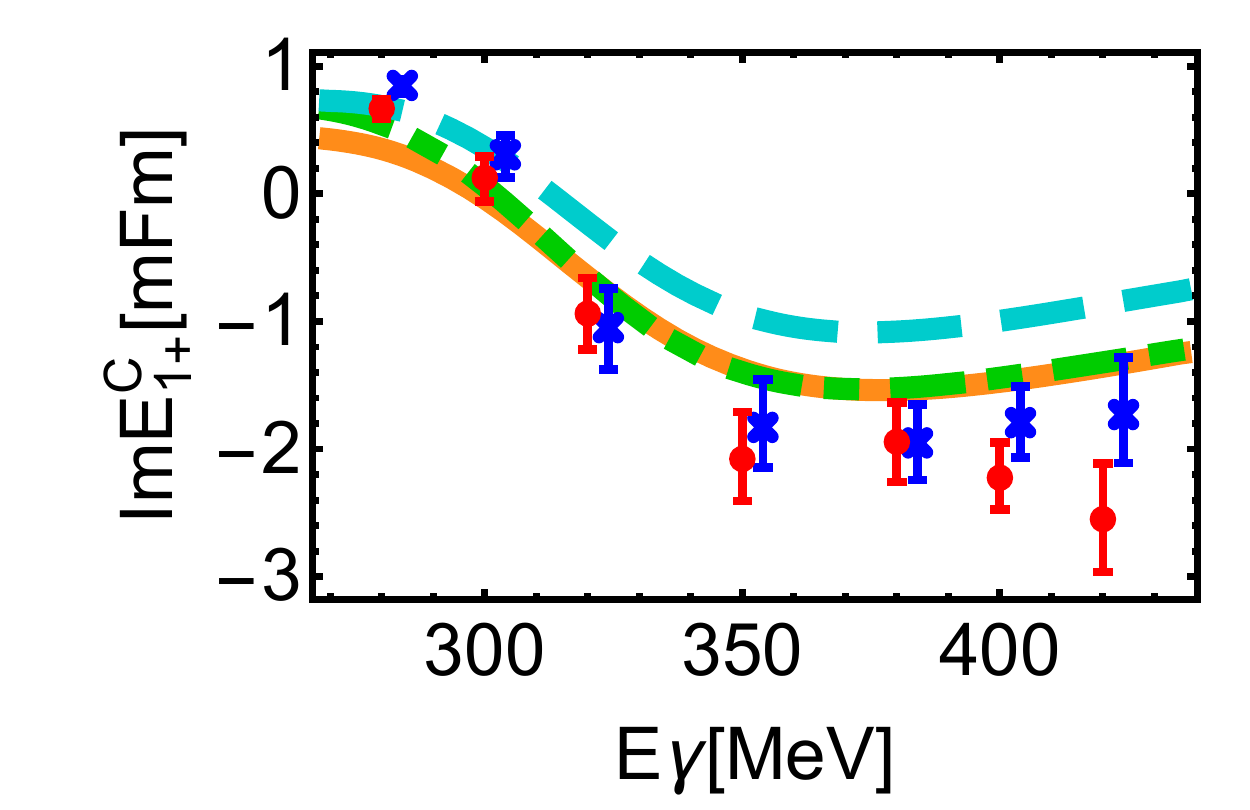}
 \end{overpic} \\
\begin{overpic}[width=0.325\textwidth]{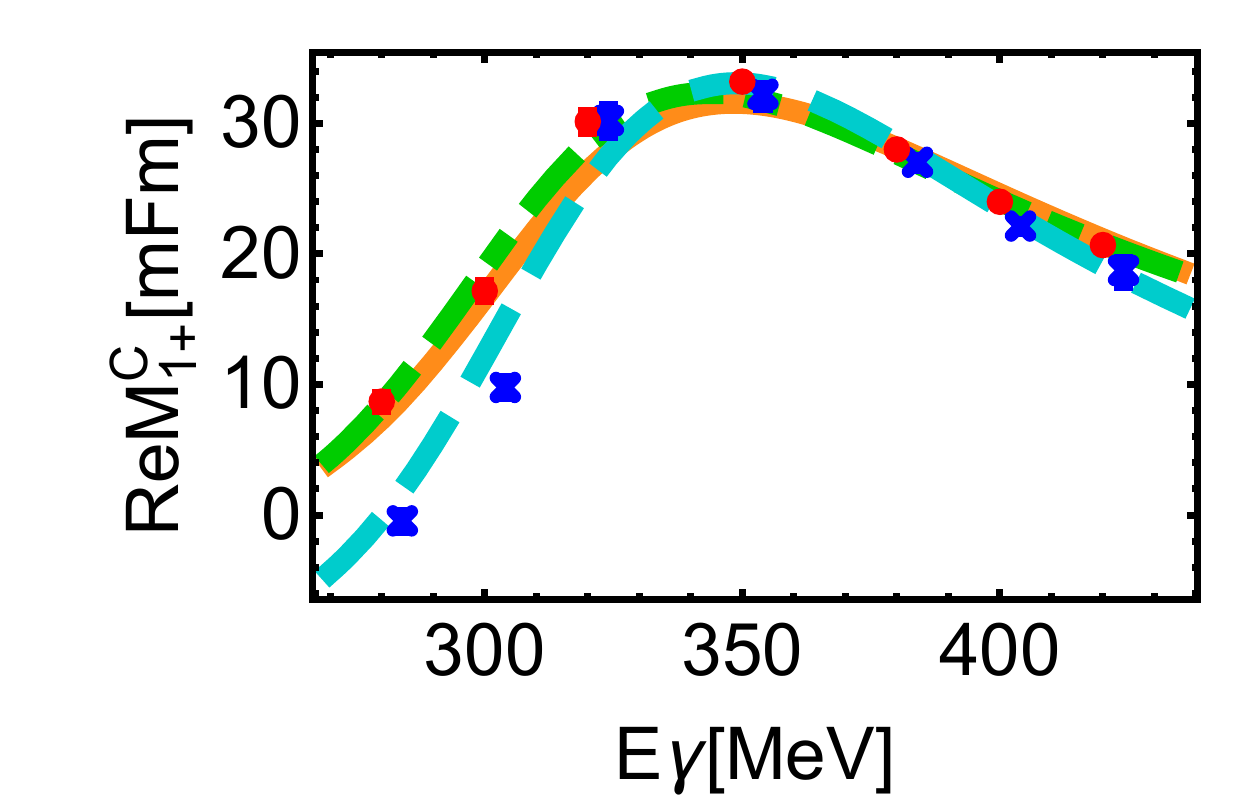}
 \end{overpic}
\begin{overpic}[width=0.325\textwidth]{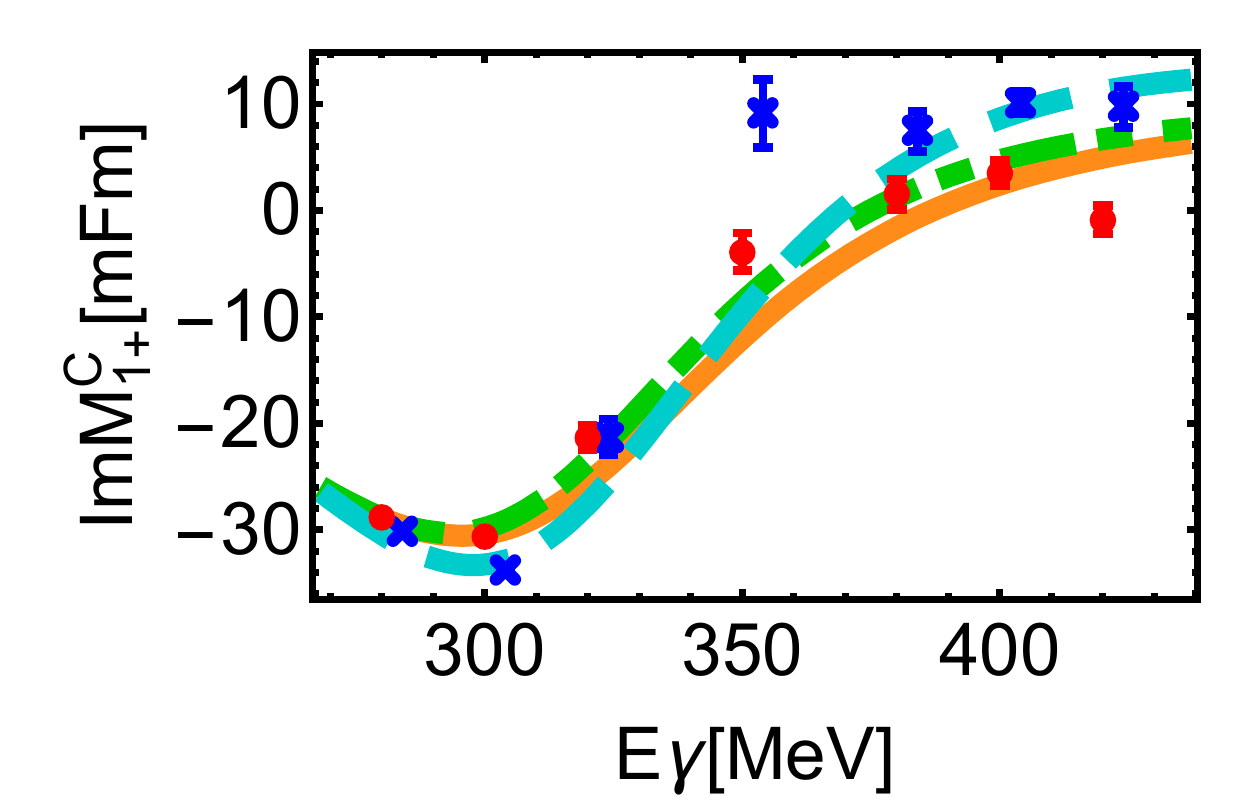}
 \end{overpic}
\begin{overpic}[width=0.325\textwidth]{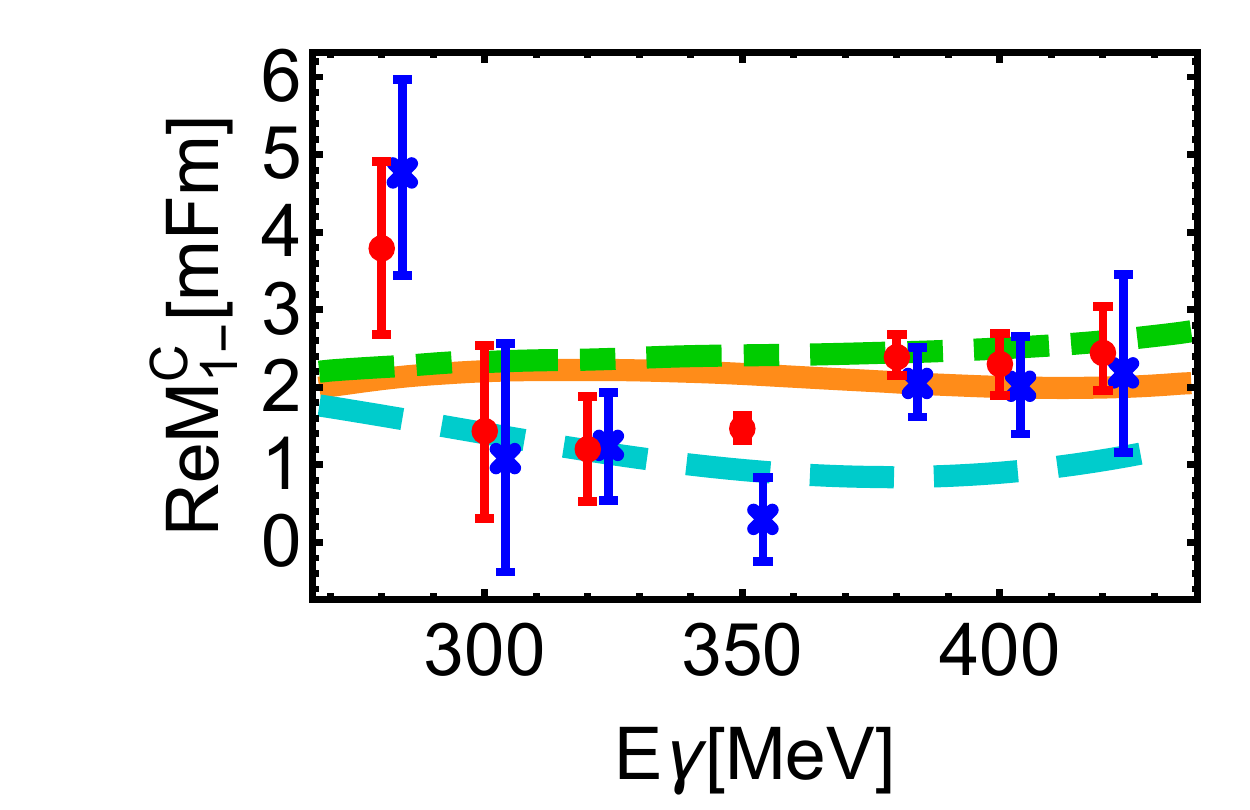}
 \end{overpic} \\
\begin{overpic}[width=0.325\textwidth]{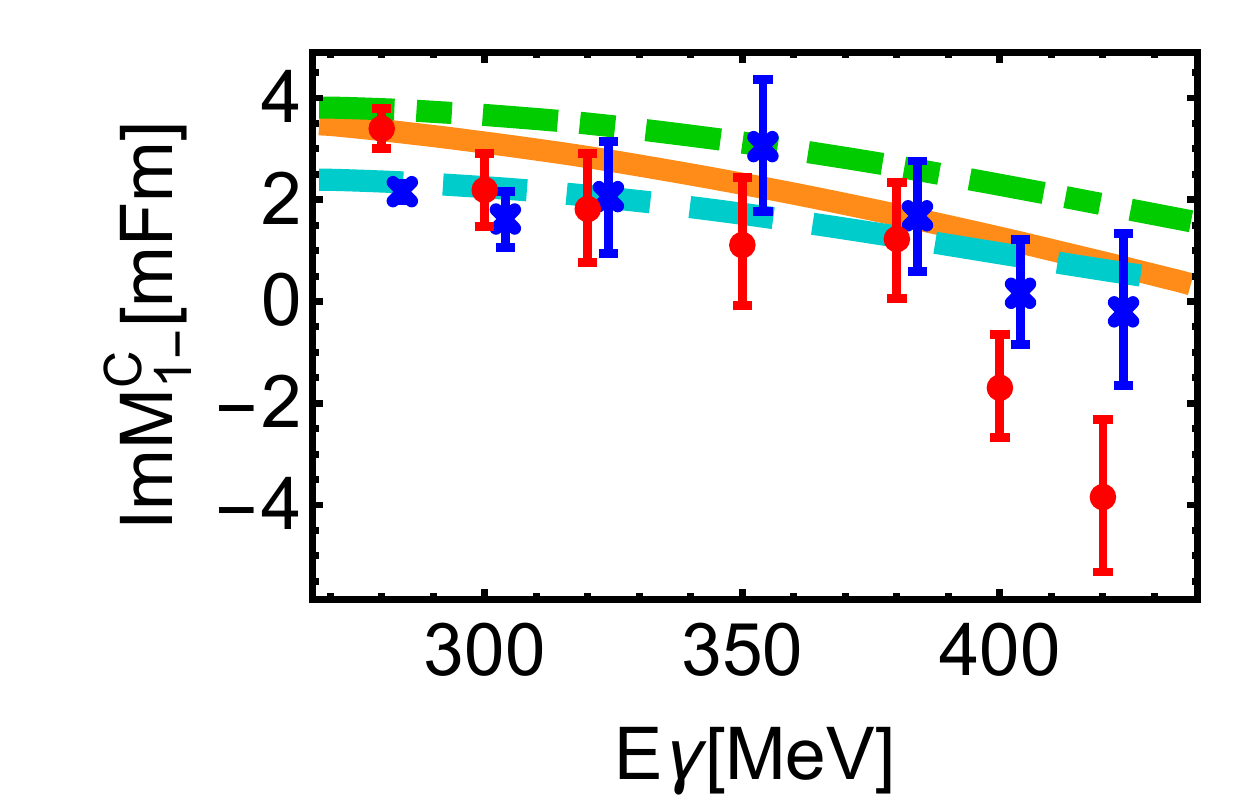}
 \end{overpic}
\caption[A comparison of the results of the bootstrap-analysis performed in the $\Delta$-resonance region, with and without inclusion $\&$ fixing of the $D$-waves, to energy-dependent PWA-models.]{The plots show a comparison of the results of bootstrap-analyses to energy-dependent PWA-models. The figures contain results of the fit with $\ell_{\mathrm{max}} = 2$ and $D$-waves fixed to SAID CM12 \cite{SAID} (red dots), as well as for the fit strictly truncated at $\ell_{\mathrm{max}} = 1$ (blue crosses, slightly shifted to the right). For the former case, the fit to the original data can be inspected in Figure \ref{fig:Lmax2DWavesSAIDFitResultsDeltaRegionPurelyStat}, while in the latter case the original fit was shown in Figure \ref{fig:Lmax1UnconstrainedFitResultsDeltaRegion}. Error-bars indicate statistical uncertainties determined from the bootstrapping-procedure (cf. the main text). \newline The results are compared to the PWA-solutions SAID CM12 (orange solid line) \cite{SAID}, BnGa 2014\_02 (cyan dashed line) \cite{BoGa} and MAID2007 (green dash-dotted line) \cite{MAID}.
}
\label{fig:BootstrapDeltaRegionResultsComparedToPWAI}
\end{figure}

Following the above-given detailed treatment of one particular bootstrap-analysis, we now continue with a comparison of general fit-results for the single-energy multipoles to energy-dependent PWA-models. A first interesting example is given by Figure \ref{fig:BootstrapDeltaRegionResultsComparedToPWAI}, where the above-described bootstrap-TPWA for the third fit (Figure \ref{fig:Lmax2DWavesSAIDFitResultsDeltaRegionPurelyStat}) is compared to MAID2007 \cite{MAID}, SAID CM12 \cite{SAID} and Bonn-Gatchina 2014\_02 \cite{BoGa}. Furthermore, results of a second bootstrap-analysis are shown in the Figure, namely for the first fit (Fig. \ref{fig:Lmax1UnconstrainedFitResultsDeltaRegion}), which did not include any $D$-waves at all. In this way, possible effects of $\left< S,D \right>$- and $\left<P,D\right>$-interferences can be seen.
\vfill

\clearpage
First, one should say that the agreement between the single-energy multipoles and the energy-dependent models is at least fair in the tendency of the energy-variation of the phase-constrained multipoles. However, energy-bins exist where the single-energy results do not agree with any model within the errors determined from the bootstrap. Examples for such cases are the parameter $\mathrm{Re}\left[ M_{1-}^{C} \right]$ in the first energy-bin or $\mathrm{Im}\left[ M_{1+}^{C} \right]$ in the fourth energy-bin. Interestingly however, not even the energy-dependent models agree among each other once the overall phase has been fixed to the same constraint. Thus, in most cases at least one model-curve passes through the bootstrap error-bar of at least one of the two fits. \newline
The sizes of the bootstrap-errors for both fits are in agreement. In almost all energy-bins, the large multipole $M_{1+}$, which contains contributions from the $\Delta (1232)$-resonance, is the most well-determined. The relative size of errors is smallest for this multipole. The multipole $M_{1-}$ on the other hand suffers the most variation due to statistical uncertainties for all energies. \newline
Comparing the two single-energy fits with and without SAID $D$-waves, it is seen that the biggest effect due to interferences is seen in the multipoles $E_{0+}$ and $M_{1+}$. The $S$-wave $E_{0+}$ is shifted a little bit away from the Bonn-Gatchina curve towards MAID and SAID in the lower half of the considered energy-range. However, a kind of 'dip'-structure is seen in $E_{0+}$ for both single-energy fits. In the first two energy-bins, it is seen in the quantity $\mathrm{Re}\left[M_{1+}^{C}\right]$ that the unconstrained fit without $D$-waves sits quite close to Bonn-Gatchina, while upon the inclusion of SAID $D$-waves is is shifted towards the MAID- and SAID-curves. \newline

\begin{figure}[h]
\centering
\begin{overpic}[width=0.325\textwidth]{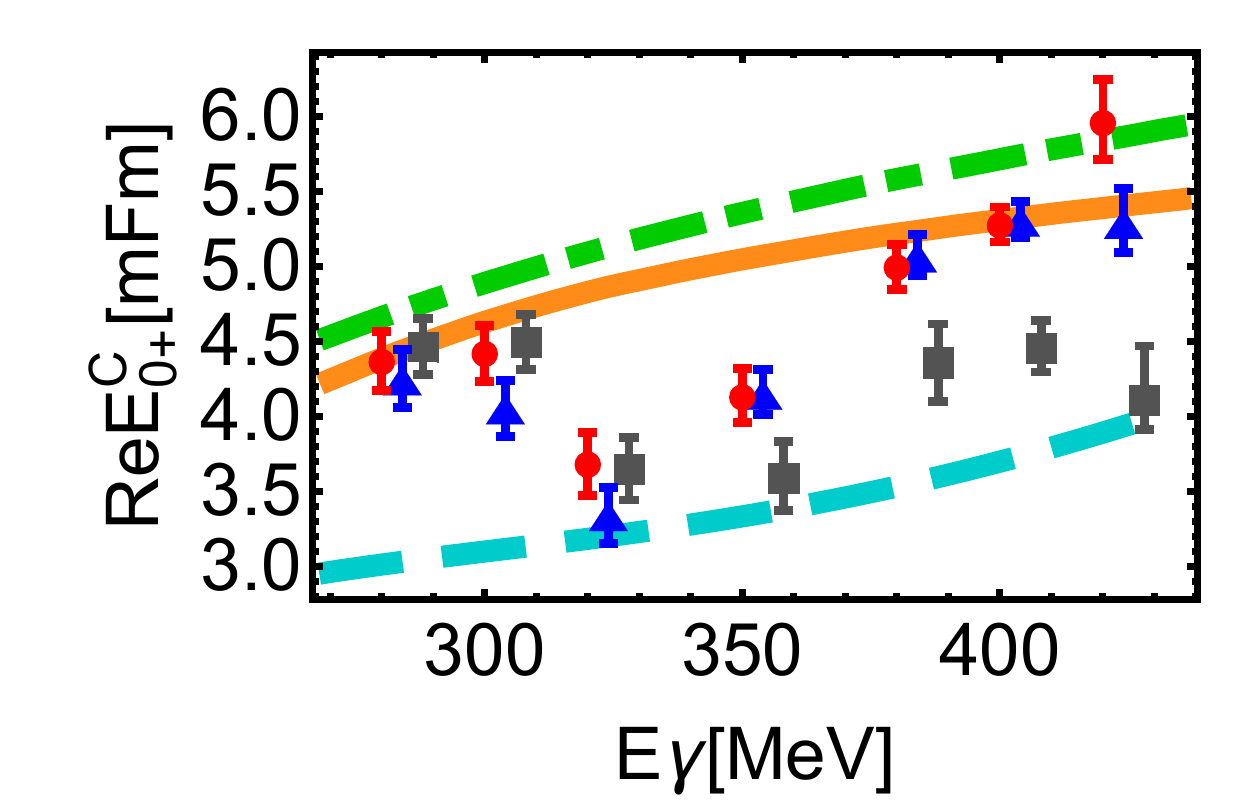}
 \end{overpic}
\begin{overpic}[width=0.325\textwidth]{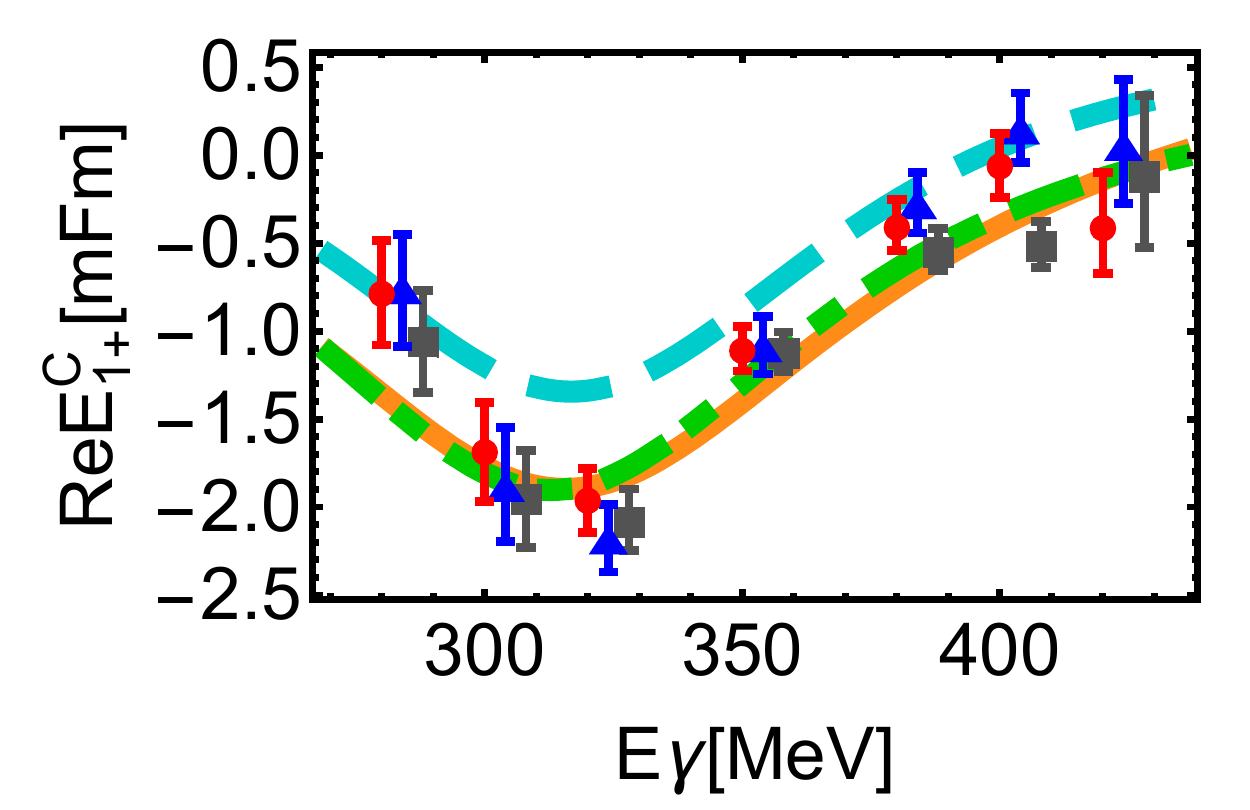}
 \end{overpic}
\begin{overpic}[width=0.325\textwidth]{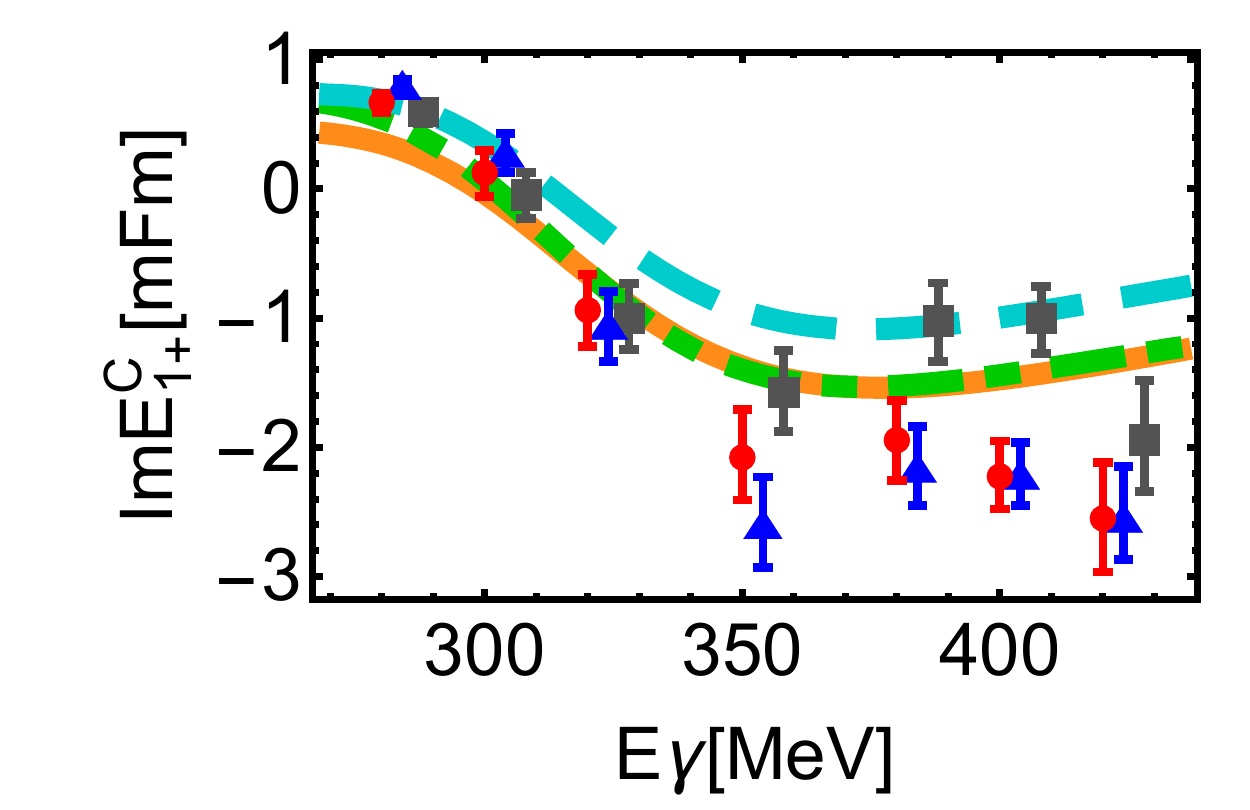}
 \end{overpic} \\
\begin{overpic}[width=0.325\textwidth]{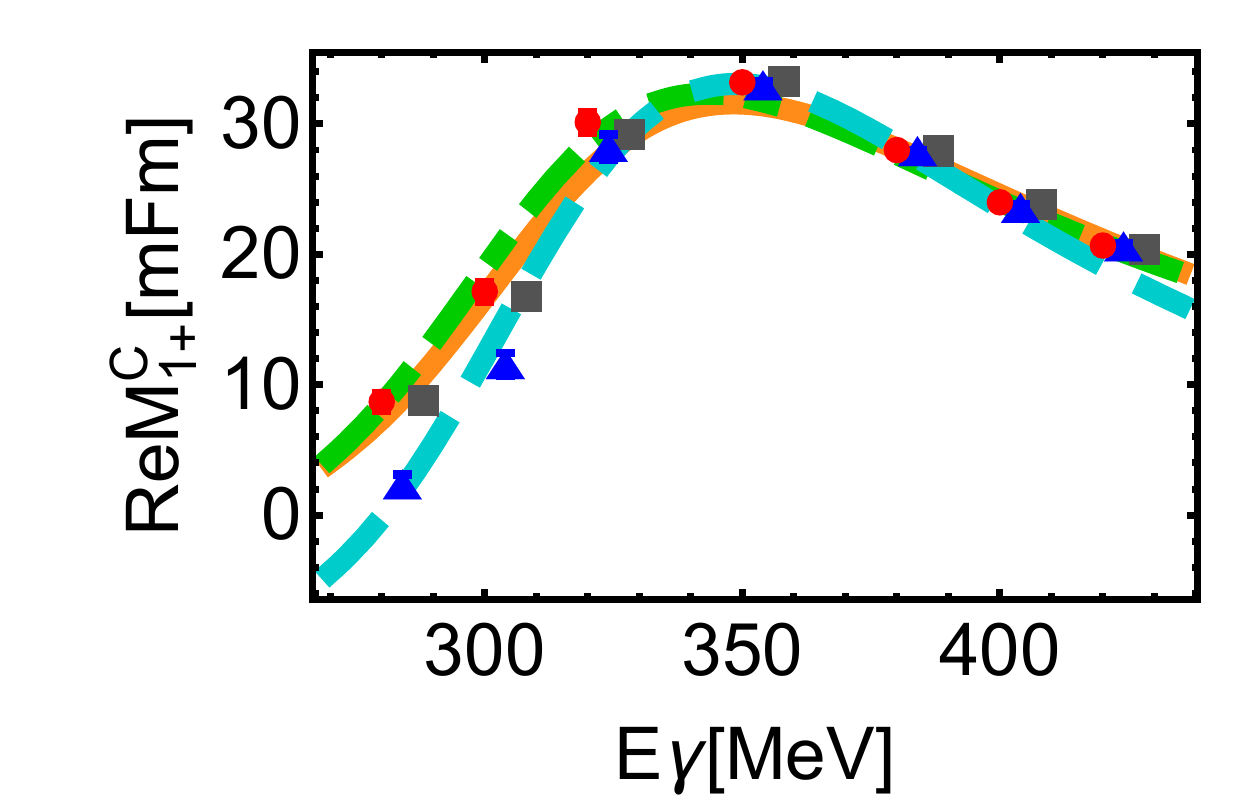}
 \end{overpic}
\begin{overpic}[width=0.325\textwidth]{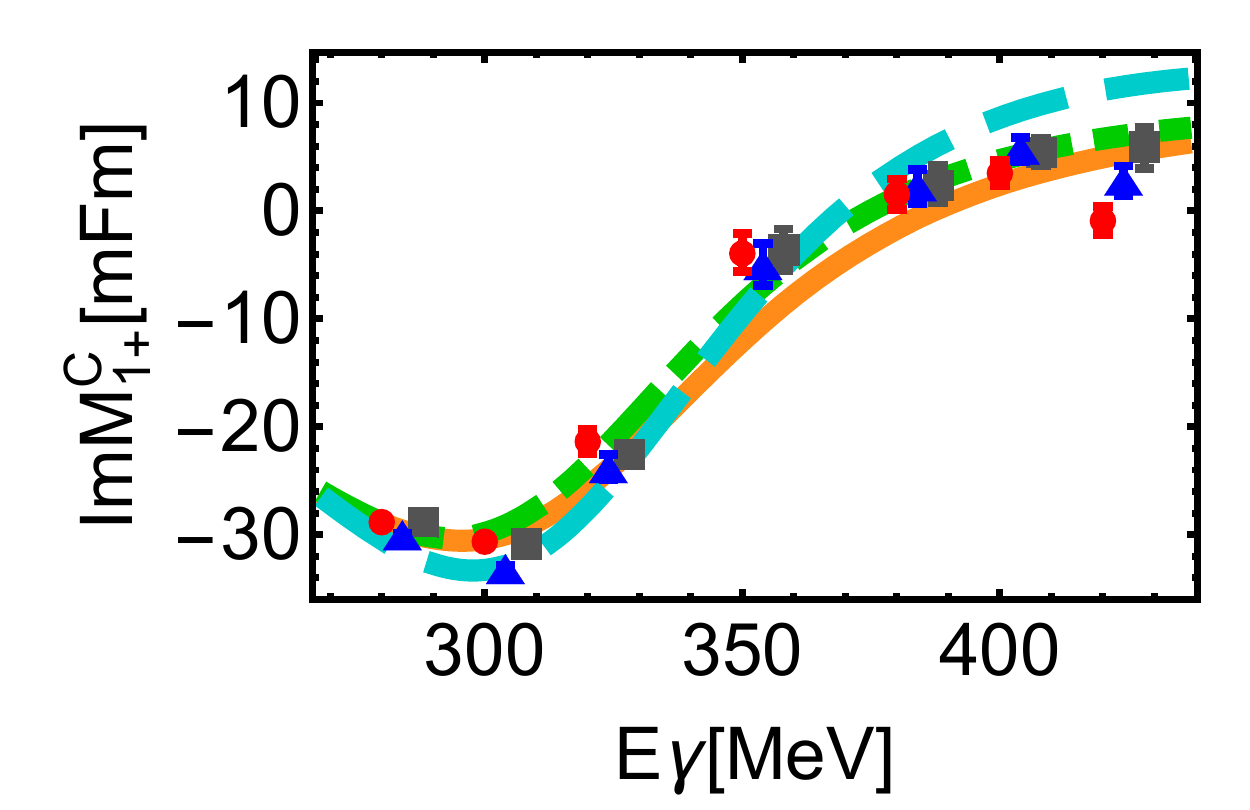}
 \end{overpic}
\begin{overpic}[width=0.325\textwidth]{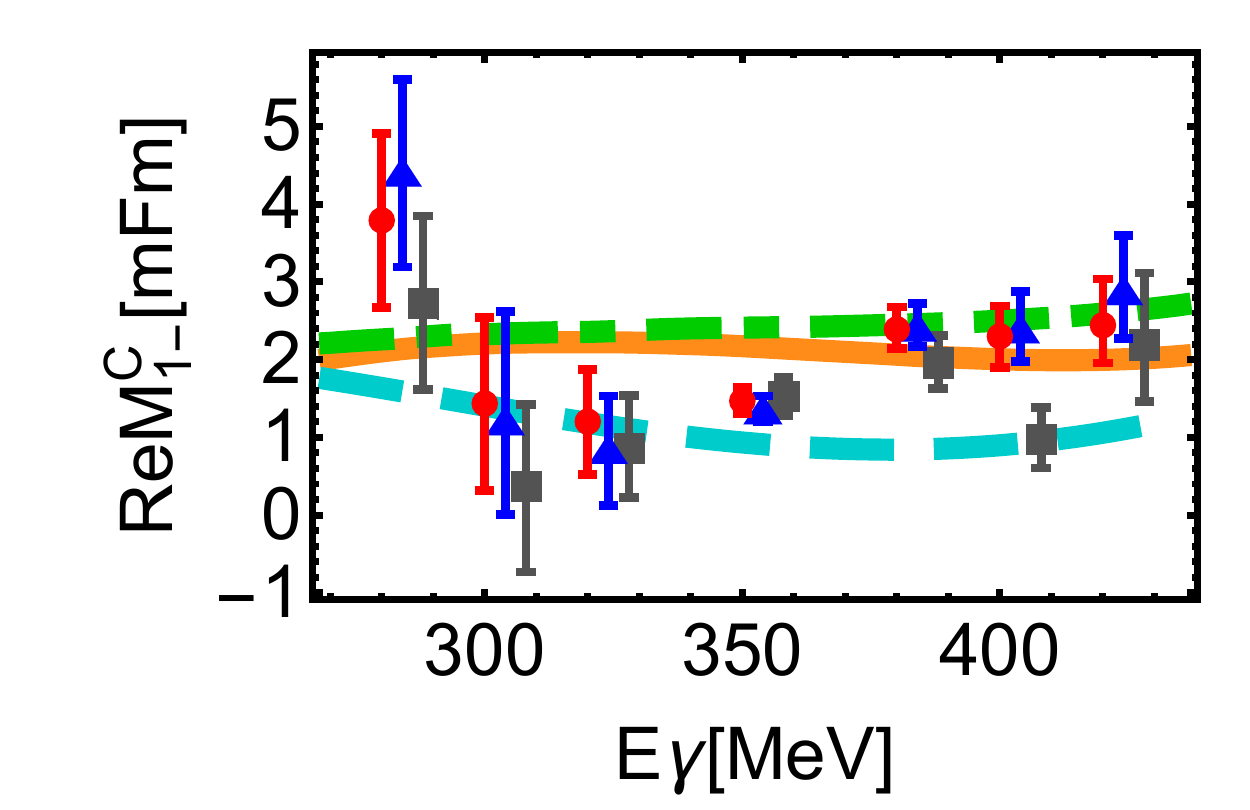}
 \end{overpic} \\
\begin{overpic}[width=0.325\textwidth]{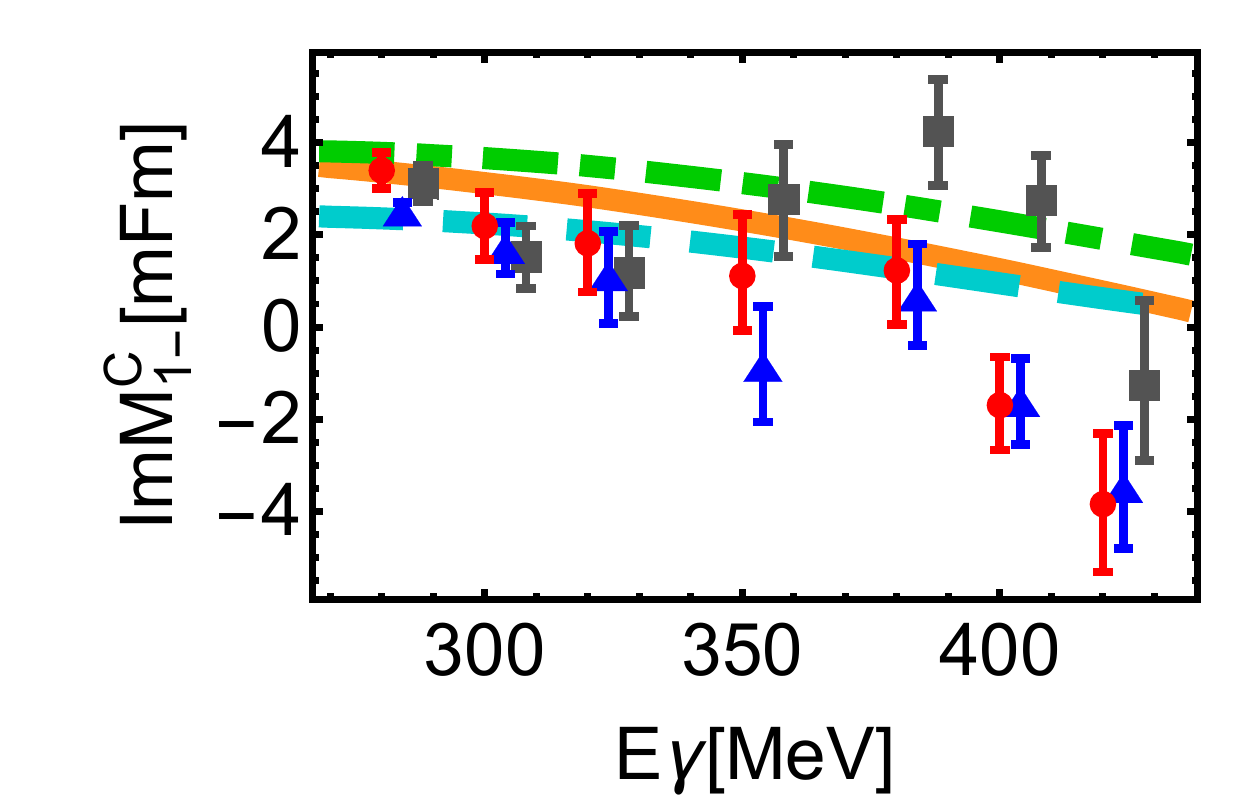}
 \end{overpic}
\caption[A comparison of the results of the bootstrap-analysis performed in the $\Delta$-resonance region, with inclusion $\&$ fixing of the $D$-waves to different PWA-models.]{The figures represent a comparison of the results of bootstrap-analyses to energy-dependent PWA-models. Shown are three fits using $\ell_{\mathrm{max}} = 2$, with $D$-waves fixed to three different energy-dependent PWA-models. The three cases comprise $D$-waves fixed to SAID CM12 \cite{WorkmanEtAl2012ChewMPhotoprod,SAID} (red dots), to Bonn-Gatchina 2014\_02 \cite{BoGa} (blue triangles, slightly shifted to the right) and to MAID2007 \cite{MAID2007,MAID} (grey boxes, slightly shifted to the right). Error-bars indicate statistical uncertainties determined from the bootstrapping-procedure (cf. the main text). \newline The results are compared to the PWA-solutions SAID CM12 (orange solid line) \cite{SAID}, BnGa 2014\_02 (cyan dashed line) \cite{BoGa} and MAID2007 (green dash-dotted line) \cite{MAID}.}
\label{fig:BootstrapDeltaRegionResultsComparedToPWAII}
\end{figure}

This last point could of course have come to pass simply due to the fact that the single-energy fit with $D$-waves fixed to SAID contains an intrinsic bias towards this particular model. In order to explore, or estimate, the size of this systematic 'model-uncertainty' of the single-energy fits, we repeated the whole analysis including a bootstrap-TPWA while keeping the $D$-wave multipoles fixed to the other two energy-dependent models, MAID \cite{MAID} and Bonn-Gatchina \cite{BoGa}. A comparative plot of the results is shown in Figure \ref{fig:BootstrapDeltaRegionResultsComparedToPWAII}. \newline
The systematic effects due to $D$-waves from different models are notable in, either real- or imaginary parts, of all multipoles. The shift mentioned above, which occurs for the parameter $\mathrm{Re}\left[M_{1+}^{C}\right]$ at low energies, really turns out to be a model-artifact. As seen in Figure \ref{fig:BootstrapDeltaRegionResultsComparedToPWAII}, the fit result indeed tends to match the PWA to which the $D$-waves have been fixed. Furthermore, the most dramatic deviances are observed for the fit with $D$-waves fixed to MAID2007, as opposed to the other two PWA-models. Drastic effects are seen in the higher energies for the parameters $\mathrm{Re}\left[E_{0+}^{C}\right]$, $\mathrm{Im}\left[E_{1+}^{C}\right]$ and $\mathrm{Im}\left[M_{1-}^{C}\right]$. \newline
The MATHEMATICA-codes \cite{Mathematica8,Mathematica11,MathematicaBonnLicense} which lead to these results have been double-checked, but it is not probable that some huge mistake has been done in the programming, since literally the same codes were used for fitting, just the inserted $D$-waves have been adapted. \newline
Apart from these most prominent visible deviances, systematic effects are not really dramatic when compared to the size of the bootstrapped error-bars. Actually, the results of fits with $D$-waves fixed to SAID CM12 and Bonn-Gatchine 2014\_02 agree quite well for most multipoles.

\clearpage

%

\textbf{Results of TPWA-fits performed to real data for $\left\{ \sigma_{0}, \Sigma, T, F \right\}$, supplemented by SAID-pseudodata for the recoil polarization asymmetry $P$} \newline

\begin{figure}[h]
 \centering
 \begin{overpic}[width=0.475\textwidth]{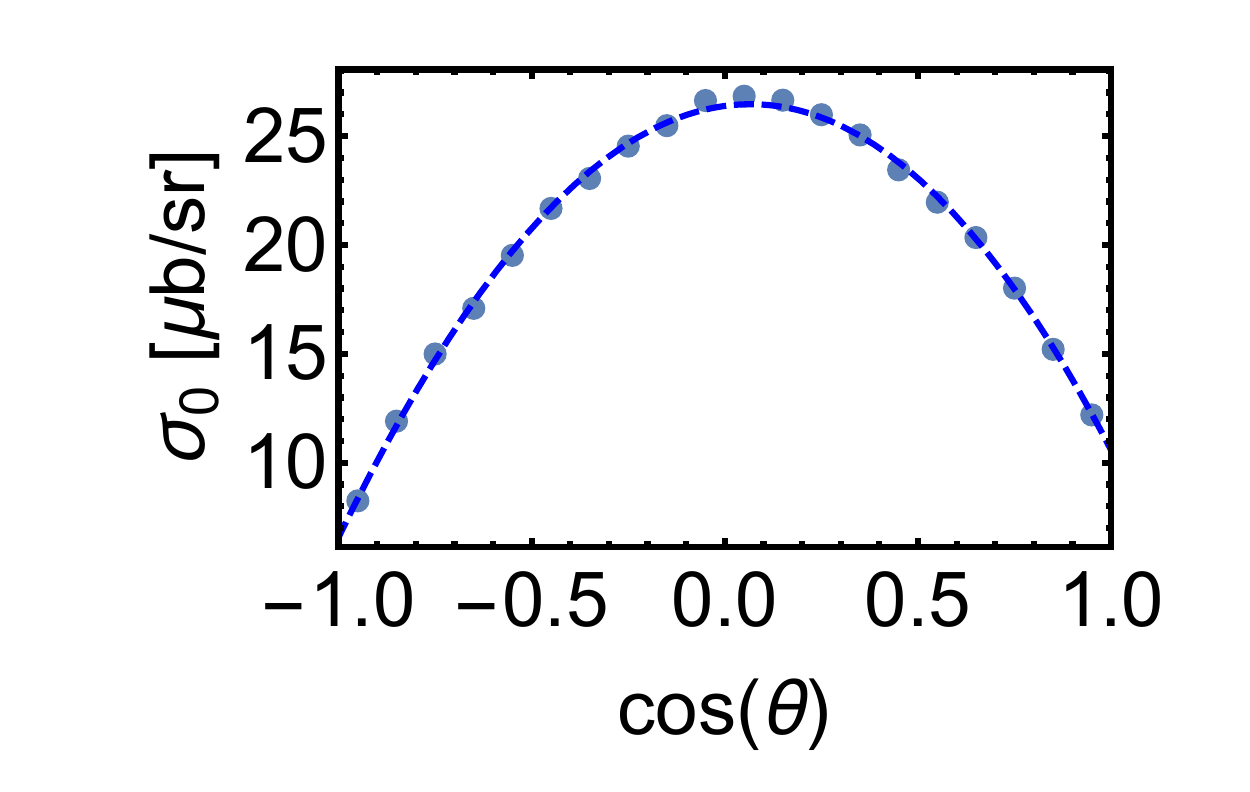}
 \put(81.5,68){\begin{Large}$E_{\gamma} = 350. \hspace*{2pt} \mathrm{MeV}$\end{Large}}
 \end{overpic}
 \begin{overpic}[width=0.475\textwidth]{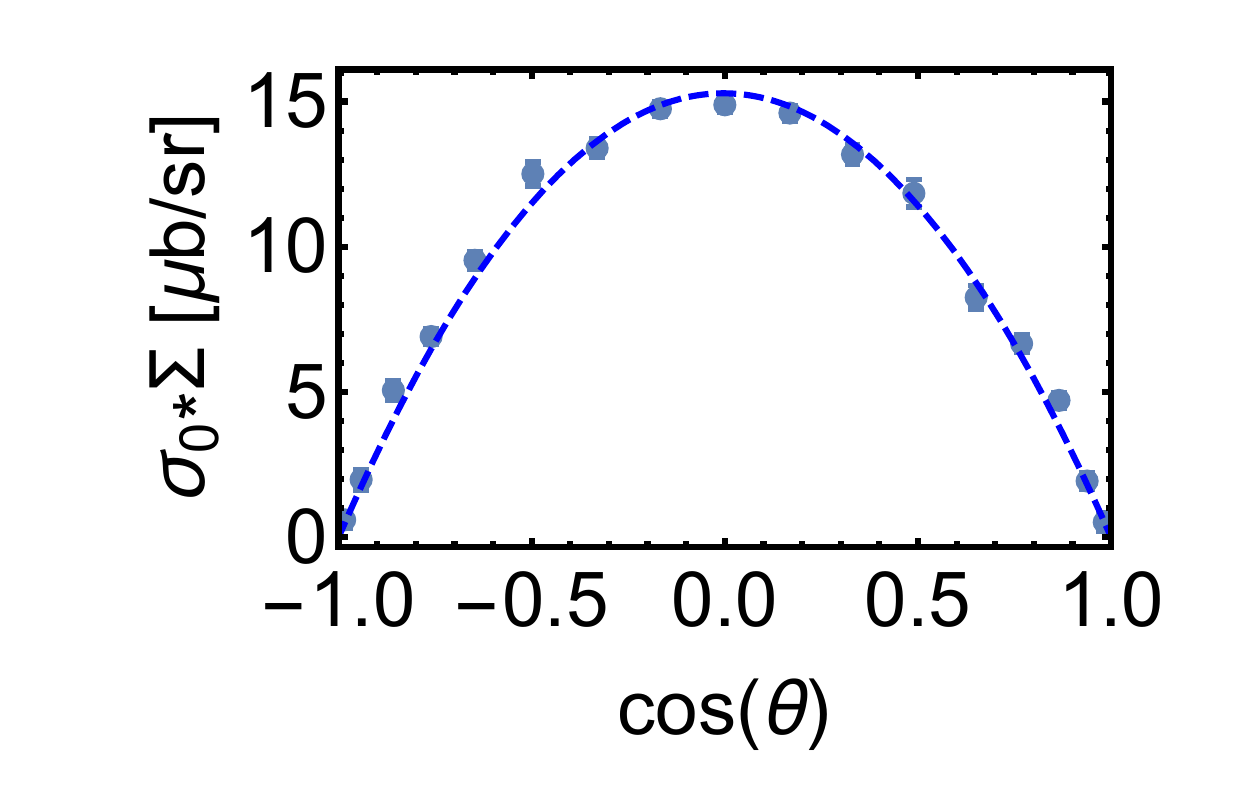}
 \end{overpic} \\
 \begin{overpic}[width=0.475\textwidth]{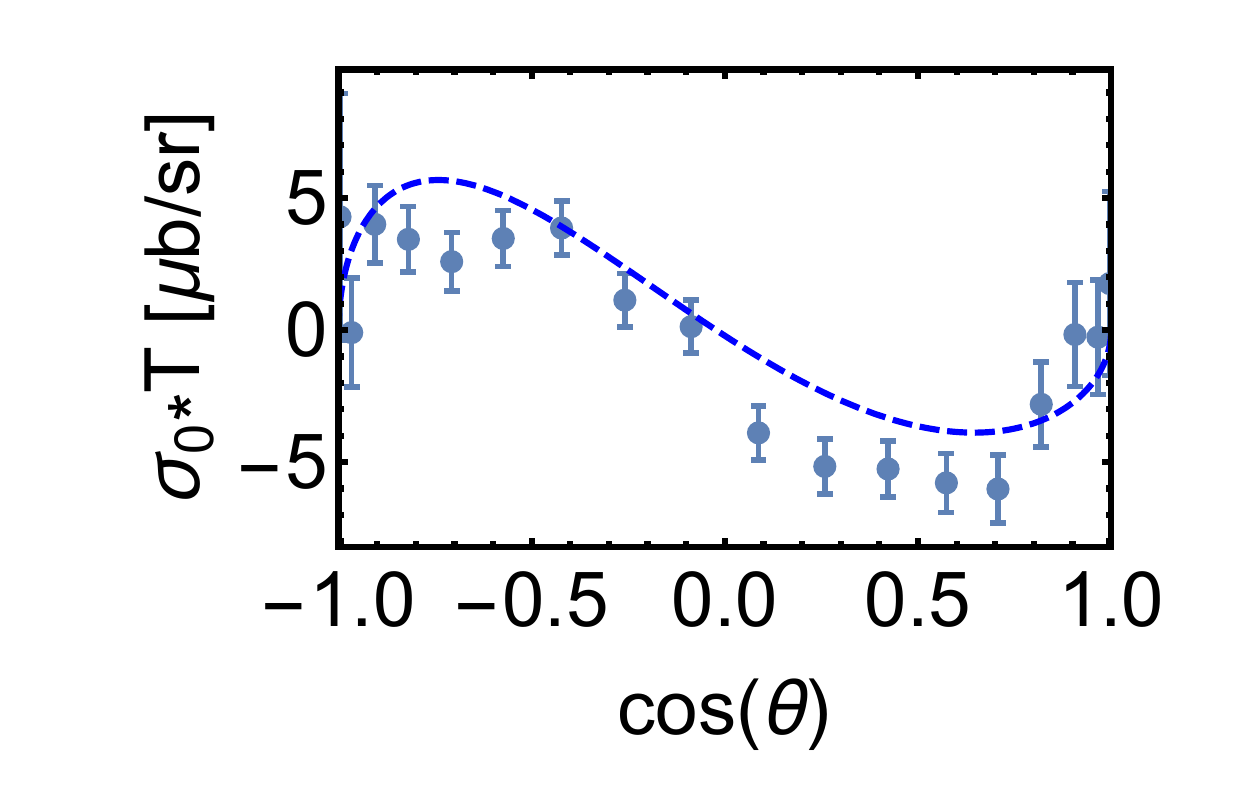}
 \end{overpic}
 \begin{overpic}[width=0.475\textwidth]{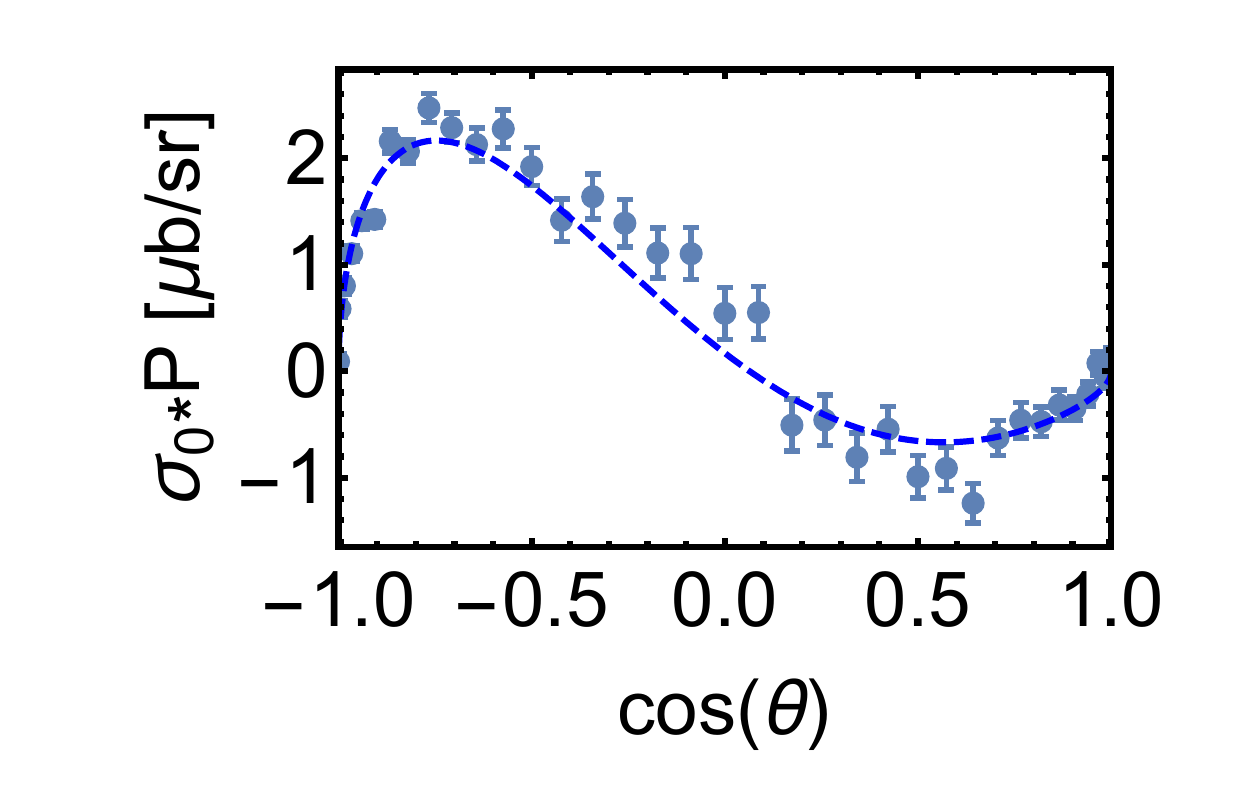}
 \end{overpic} \\
 \begin{overpic}[width=0.475\textwidth]{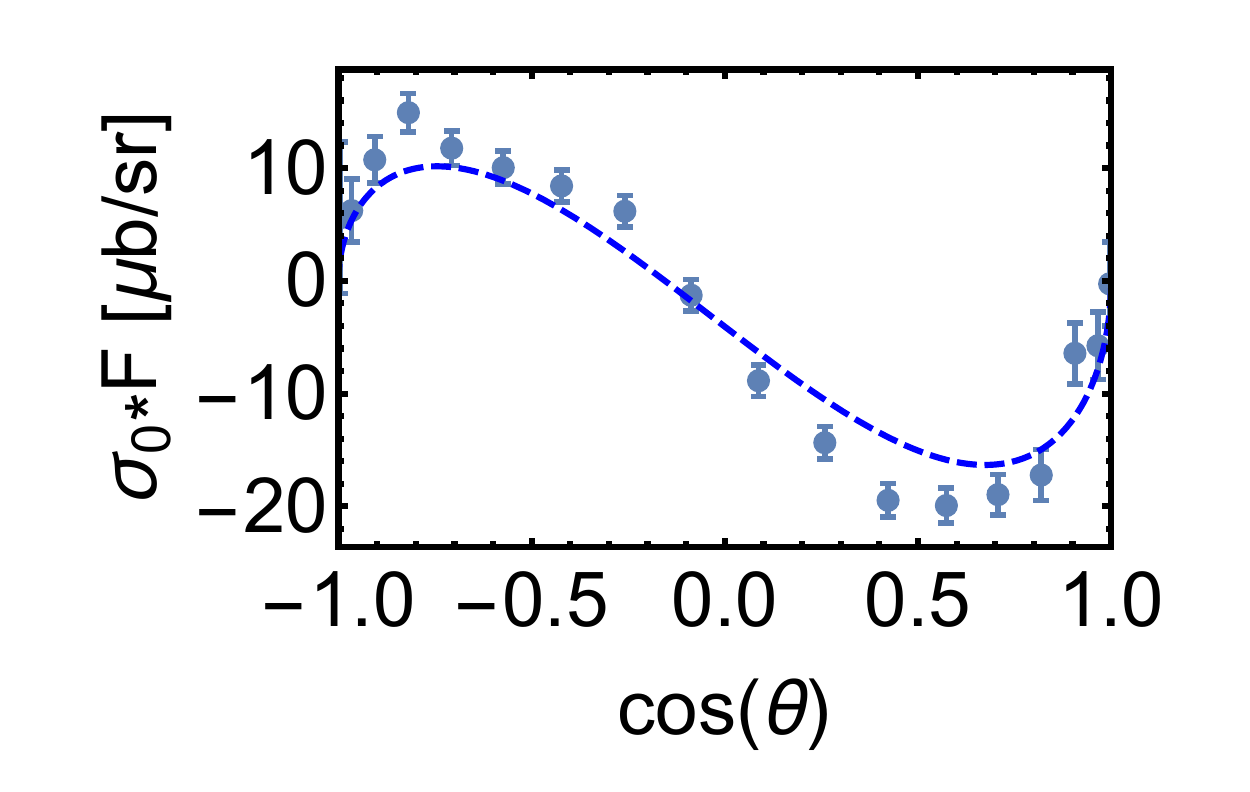}
 \end{overpic}
  \caption[Angular distributions of profile functions for the four measured polarization-datasets $\left\{\sigma_{0}, \Sigma, T, F\right\}$ analyzed in the $\Delta$-resonance region, complemented by SAID-pseudodata for the observable $P$ with $5\%$-errors. Data are plotted at a particular energy, $E_{\gamma} = 350. \hspace*{1pt} \mathrm{MeV}$.]{Angular distributions for the differential cross section $\sigma_{0}$ and the profile functions $\check{\Sigma} = \sigma_{0} \Sigma$, $\check{T} = \sigma_{0} T$ and $\check{F} = \sigma_{0} F$ stemming from actual measurements are shown. These real data were complemented with pseudodata for the $P$-observable created from the model SAID CM12 \cite{RonPrivComm,WorkmanEtAl2012ChewMPhotoprod}. The pseudodata were prepared with a $5\%$-error according to the method outlined in section \ref{sec:PseudoDataWithErrorsFitted}. Data are shown at a particular energy, $E_{\gamma} = 350 \hspace*{1pt} \mathrm{MeV}$. Data coincide at this energy as a result of the kinematic re-binning elaborated in the main text. \newline
  The result of a direct multipole-fit to the angular distributions can be seen as well, employing the truncation angular momentum $\ell_{\mathrm{max}} = 2$ and fixing the $D$-waves to SAID CM12 (blue dashed line).}
 \label{fig:DeltaRegionObsFitAngDistsPSAIDObservable}
\end{figure}

One obvious fact which imparied the precision and quality of the fit-results shown up to now consists of the comparatively weak statistics of the Kharkov-data for $P$ \cite{Belyaev:1983}. This becomes apparent by considering the plots of kinematical coverages of the data shown in Figure \ref{fig:DeltaRegionKinematicPlots} and the information contained in Table \ref{tab:DeltaRegionSandorfiDataTable}. \newline
In order to play through a hypothetical scenario in which high-statistics data for $P$ were available, we decided to combine the measured data for $\left\{\sigma_{0}, \Sigma, T, F\right\}$ \cite{Hornidge:2013, LeukelPhD, Leukel:2001, Schumann:2015} with pseudodata stemming from an energy-dependent model. R. Workman from the SAID-group has been friendly enough to provide model-data for this thesis-project \cite{RonPrivComm}, which have been extracted in a fine kinematical coverage from the SAID-solution CM12 \cite{WorkmanEtAl2012ChewMPhotoprod,SAID}. To be more precise, in these SAID-data for $P$ the energy-region from $E_{\gamma} = 145 \hspace*{1.5pt} \mathrm{MeV}$ to $421 \hspace*{1.5pt} \mathrm{MeV}$ is covered by $138$ equidistant points, with a spacing of $2 \hspace*{1.5pt} \mathrm{MeV}$. For each energy-point, the full angular region is covered by $37$ equidistant points in $\theta$, from $0 \hspace*{1.5pt}^{\circ}$ to $180 \hspace*{1.5pt}^{\circ}$. \newline
In order to create somewhat realistic pseudodata, we employed the method described in section \ref{sec:PseudoDataWithErrorsFitted}, which generates errors that scale with the inverse square-root of the unpolarized differential cross section, $1/\sqrt{\sigma_{0}}$. Within the conventions valid in section \ref{sec:PseudoDataWithErrorsFitted}, the pseudodata were endowed with a statistical precision corresponding to a $5\%$-error. These pseudodata were combined with real data for the remaining observables and then fitted. Needless to say, this makes the resulting fit even more model-biased towards SAID than the model-dependent fits shown up to this point (cf. Figures \ref{fig:Lmax2DWavesSAIDFitResultsDeltaRegionPurelyStat}, \ref{fig:Lmax2DWavesSAIDFitResultsDeltaRegionAddQuadrature} and \ref{fig:Lmax2DWaveSAIDFitResultsDeltaRegionNuisanceParameters}). However, this is not a big disadvantage since here we mainly wish to exemplify the influence of more precise $P$-data on the size of the statistical errors of the fit-parameters. An exemplary angular distribution of the $P$ pseudodata can be inspected in Figure \ref{fig:DeltaRegionObsFitAngDistsPSAIDObservable}, where the result of a fit is shown as well. \newline

\begin{figure}[ht]
 \centering
 \vspace*{-5pt}
\begin{overpic}[width=0.495\textwidth]{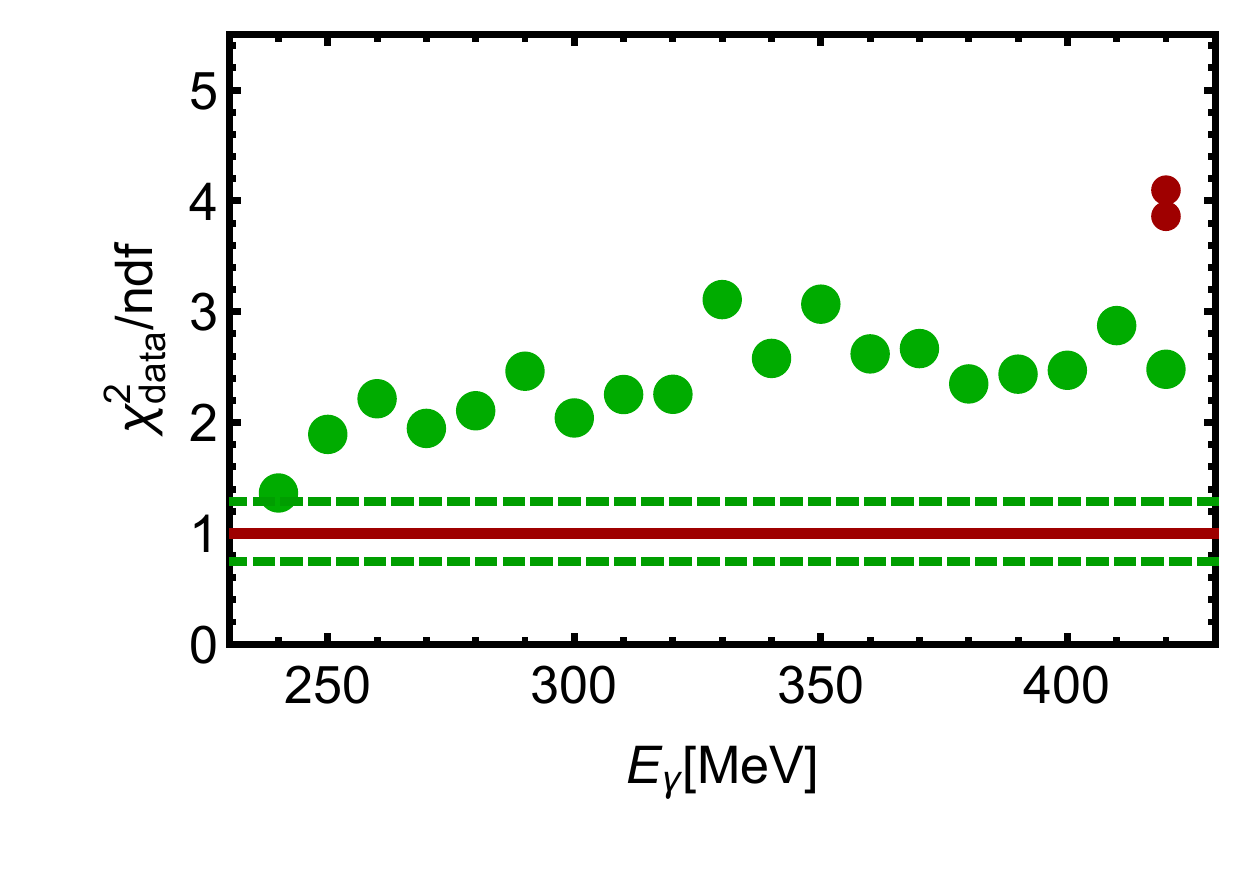}
 \put(0.5,66){a.)}
 \end{overpic} \\
 \vspace*{-5pt}
\begin{overpic}[width=0.325\textwidth]{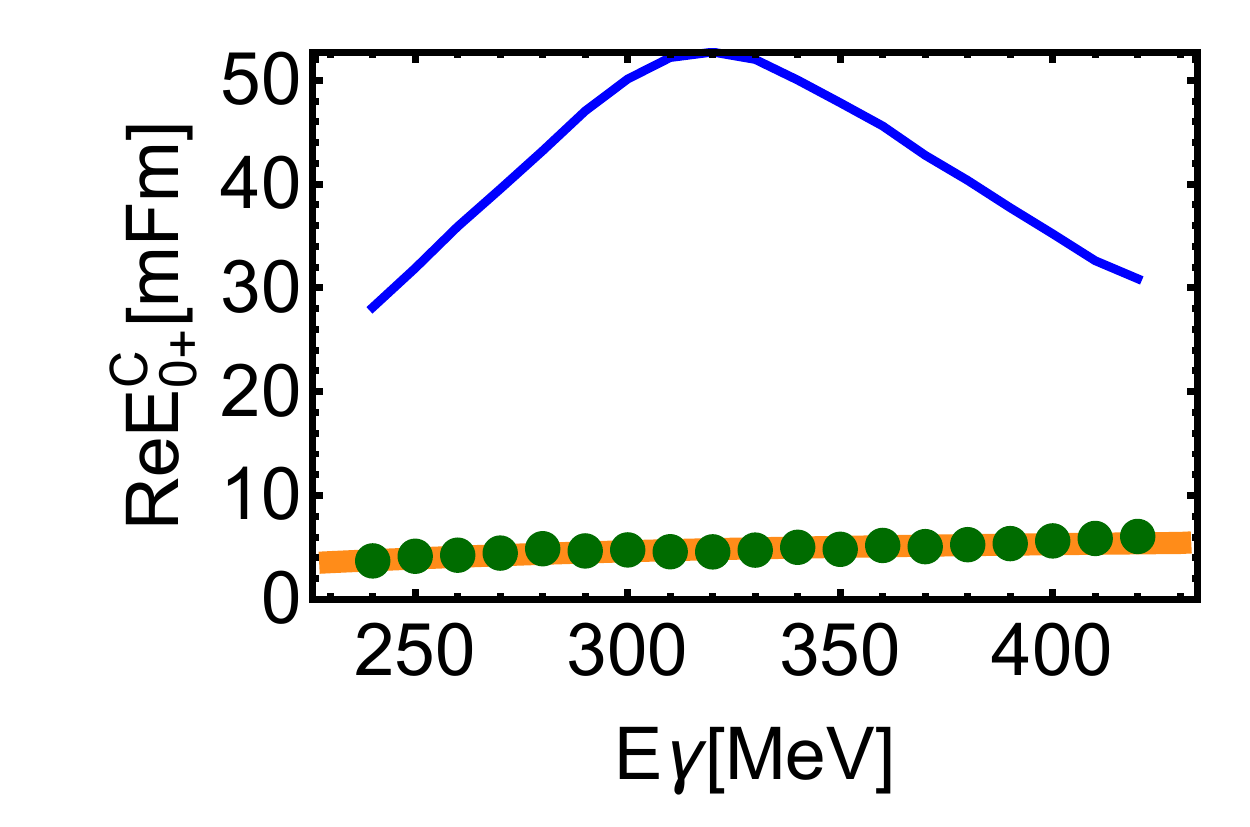}
 \put(0.5,65){b.)}
 \end{overpic}
\begin{overpic}[width=0.325\textwidth]{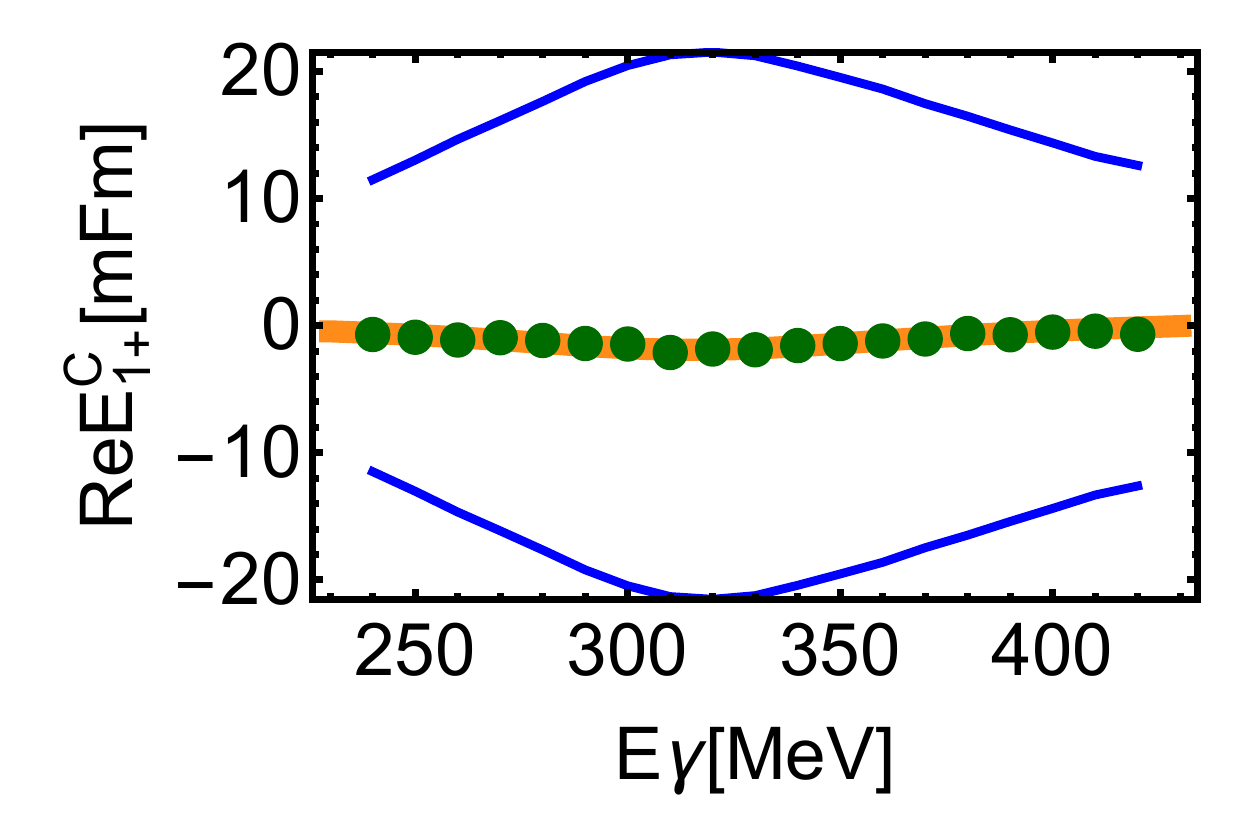}
 \end{overpic}
\begin{overpic}[width=0.325\textwidth]{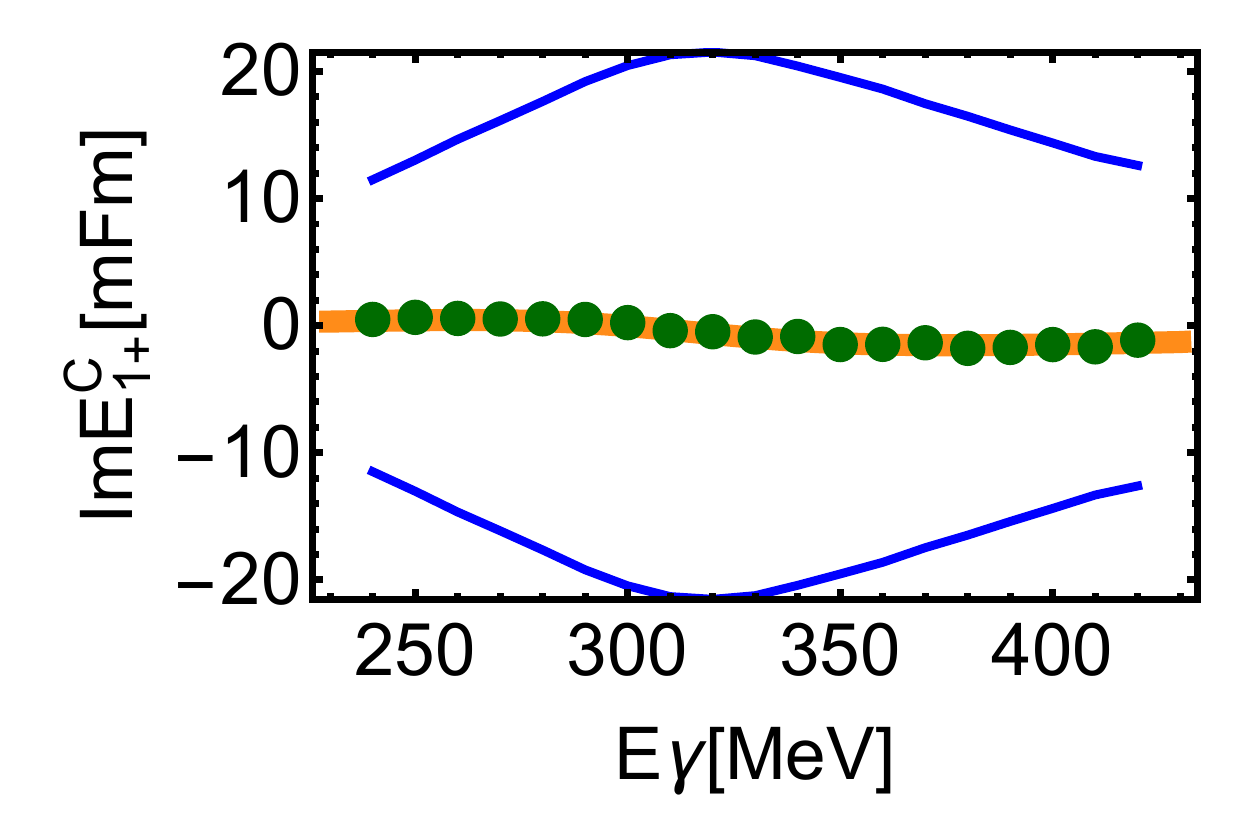}
 \end{overpic} \\
\begin{overpic}[width=0.325\textwidth]{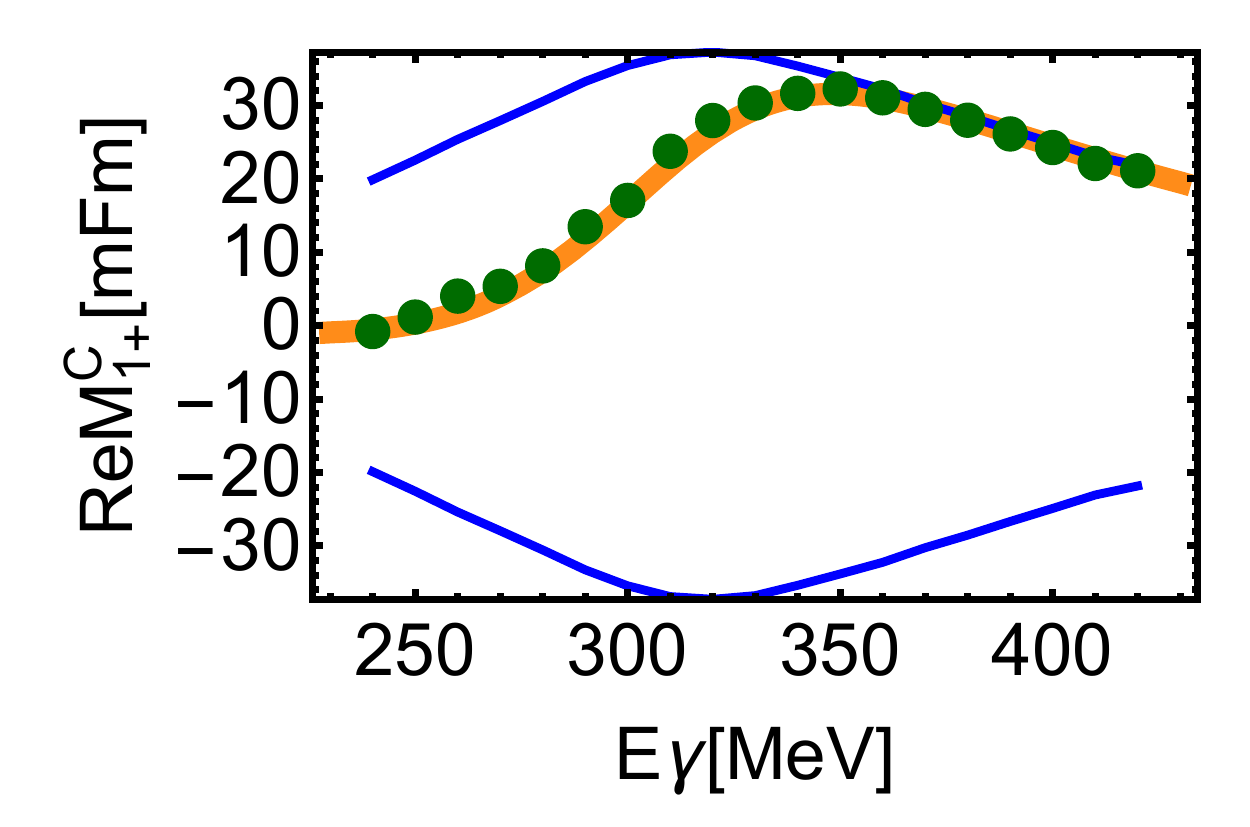}
 \end{overpic}
\begin{overpic}[width=0.325\textwidth]{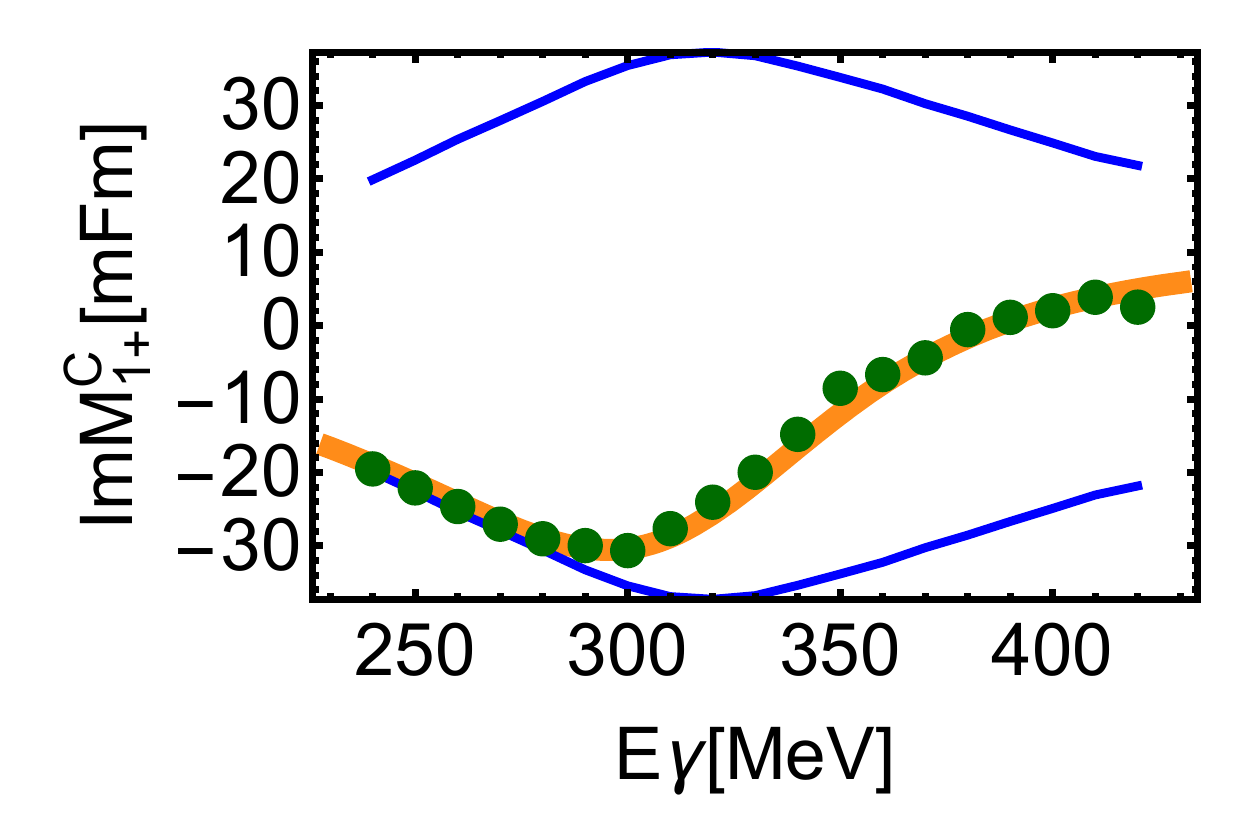}
 \end{overpic}
\begin{overpic}[width=0.325\textwidth]{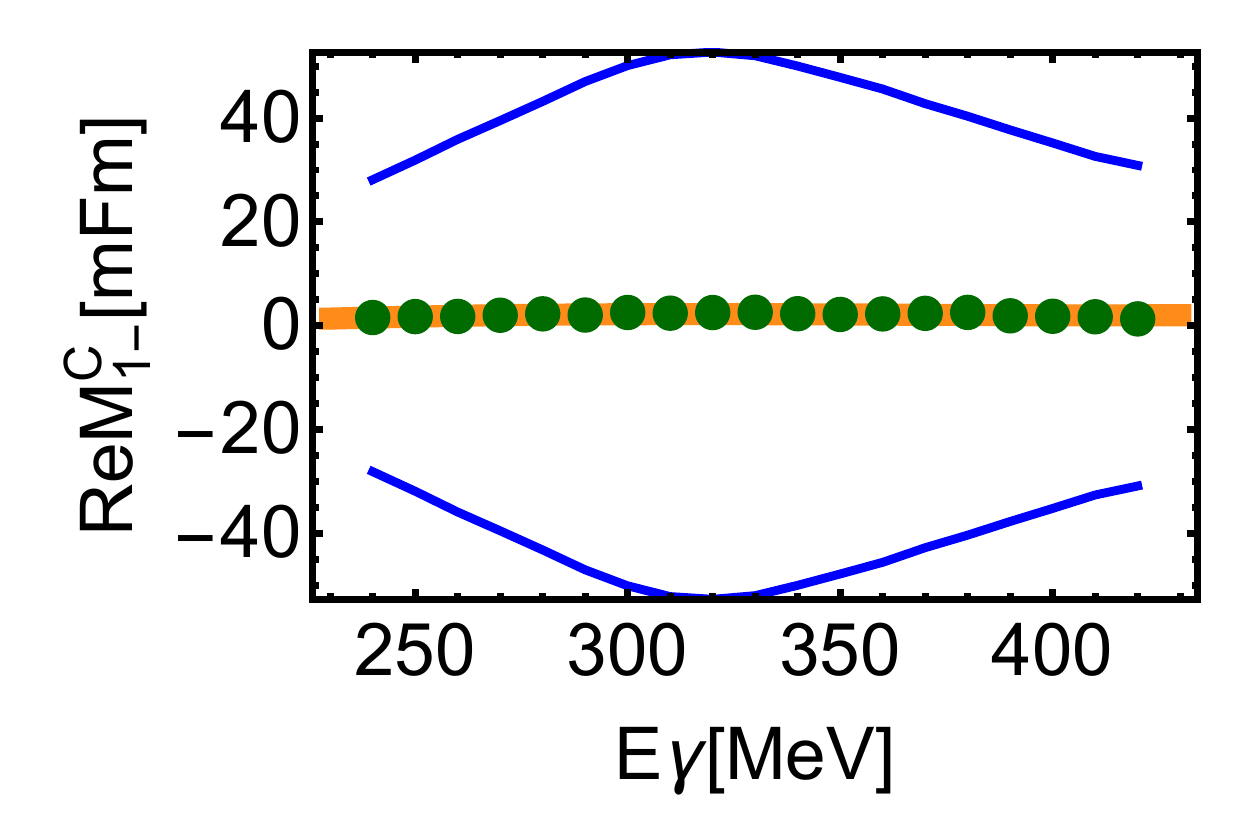}
 \end{overpic} \\
\begin{overpic}[width=0.325\textwidth]{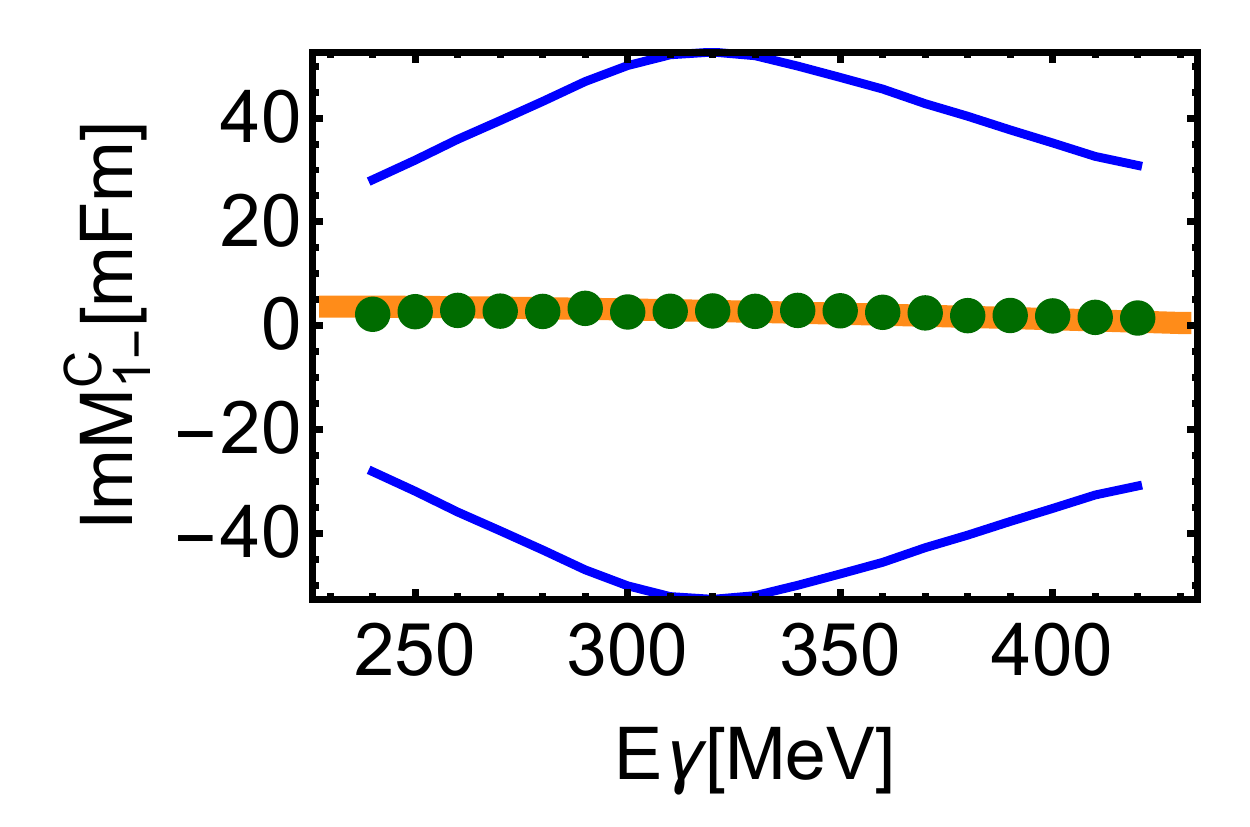}
 \end{overpic}
\caption[Results for the $7$ fit-parameters comprised of the real- and imaginary parts of phase-constrained $S$- and $P$-wave multipoles, as well as values for $\chi^{2} / \mathrm{ndf}$, for a TPWA using $\ell_{\mathrm{max}} = 2$ within the $\Delta$-resonance region. $D$-waves were fixed to the SAID-solution CM12. Five observables $\left\{ \sigma_{0}, \Sigma, T, P, F \right\}$ were fitted, with SAID-pseudodata employed for the recoil-asymmetry $P$.]{Results are summarized for a TPWA-fit with $\ell_{\mathrm{max}} = 2$ to the five observables $\left\{ \sigma_{0}, \Sigma, T, P, F \right\}$ in the $\Delta$-region. Real measured data were used for all observables except for $P$. Pseudodata stemming from the SAID-model were employed in the latter case. The pseudodata have been generated using statistical $5\%$-errors (cf. discussion in the main text). \newline The fits employed a pool of $N_{MC} = 1000$ initial parameter configurations (cf. section \ref{sec:MonteCarloSampling}); $D$-wave multipoles have been locked to the SAID-solution CM12 \cite{WorkmanEtAl2012ChewMPhotoprod}. \newline a.) The obtained values for $\chi^{2} / \mathrm{ndf}$ are plotted against energy for a direct fit to the data (\ref{eq:ChiSquareDirectFitRealDataFitSection}). The global minimum is shown (green dots), as well as other local minima (red dots). From the corresponding chisquare distribution, the mean (red line) as well as the pair of $0.025$- and $0.975$-quantiles (green dashed lines) are shown. The number of degrees of freedom has been estimated to be $\mathrm{ndf} = 110-7 = 103$. \newline b.) The global minimum (green dots) is shown for all fit-parameters (i.e. phase-constrained $S$- and $P$-wave multipoles). The SAID-solution CM12 \cite{WorkmanEtAl2012ChewMPhotoprod, SAID} is shown (solid orange colored curve). For each fit-parameter, the maximal range set by the total cross section (cf. discussion in section \ref{sec:MonteCarloSampling}), as well as its energy-variation, is indicated by the blue solid lines.
}
\label{fig:Lmax2DWavesSAIDFitResultsDeltaRegionPSAID}
\end{figure}

One of the benefits of a higher statistics $P$-dataset would consist of the fact that then the single-energy analysis is not restricted to the $7$ widely spaced energy-bins of the Kharkov-data. In the hypothetical case at hand, it is seen that the above mentioned re-binning procedure, necessary in order to prepare data for the single-energy fit, has to be repeated. However, now the 'statistically weakest' dataset is given by the beam-asymmetry data from Leukel \cite{Leukel:2001,LeukelPhD} (see Figure \ref{fig:DeltaRegionKinematicPlots}). The $\Sigma$-data now dictate the necessary common energy-binning. Thus, the analysis is performed on a grid of $19$ equidistant points from $E_{\gamma} = 240 \hspace*{1.5pt} \mathrm{MeV}$ to $420 \hspace*{1.5pt} \mathrm{MeV}$, with a spacing of $10 \hspace*{1.5pt} \mathrm{MeV}$. This is where the given combination of data coincides. \newline
The Ansatz for the fit of these data mimics the third fit which has been shown above (Figure \ref{fig:Lmax2DWavesSAIDFitResultsDeltaRegionPurelyStat}). This means we only used statistical errors as input to the analysis and performed a direct fit to the data, using a truncation order of $\ell_{\mathrm{max}} = 2$ and adjusting the $D$-waves to SAID CM12 \cite{WorkmanEtAl2012ChewMPhotoprod,SAID}. For the varied $S$- and $P$-wave multipoles, we still did a full Monte Carlo fit according to section \ref{sec:MonteCarloSampling}, applying a pool of $N_{MC} = 1000$ randomly generated initial parameter configurations. The fit to the data in one exemplary energy-bin is shown in Figure \ref{fig:DeltaRegionObsFitAngDistsPSAIDObservable}, while the $\chi^{2}_{\mathrm{data}}/\mathrm{ndf}$ and resulting fit-parameters can be seen for all energies in Figure \ref{fig:Lmax2DWavesSAIDFitResultsDeltaRegionPSAID}. \newline
As one should expect, again a well-separated global minimum is attained. The fit-quality of this minimum is again not great, which may be mainly due to the fact that the systematic uncertainties of the differential cross section have not been taken into account. However, the fit-parameters in the global minimum show excellent agreement with the SAID-model. This is unsurprising again due to the above-mentioned additional model-bias coming from the pseudodata. Still, it is encouraging to see the finer granularity of the energy-binning, resulting in single-energy multipoles which show more subtle details in their energy-variation. \newline

\begin{figure}[h]
\centering
\vspace*{-10pt}
\begin{overpic}[width=0.325\textwidth]{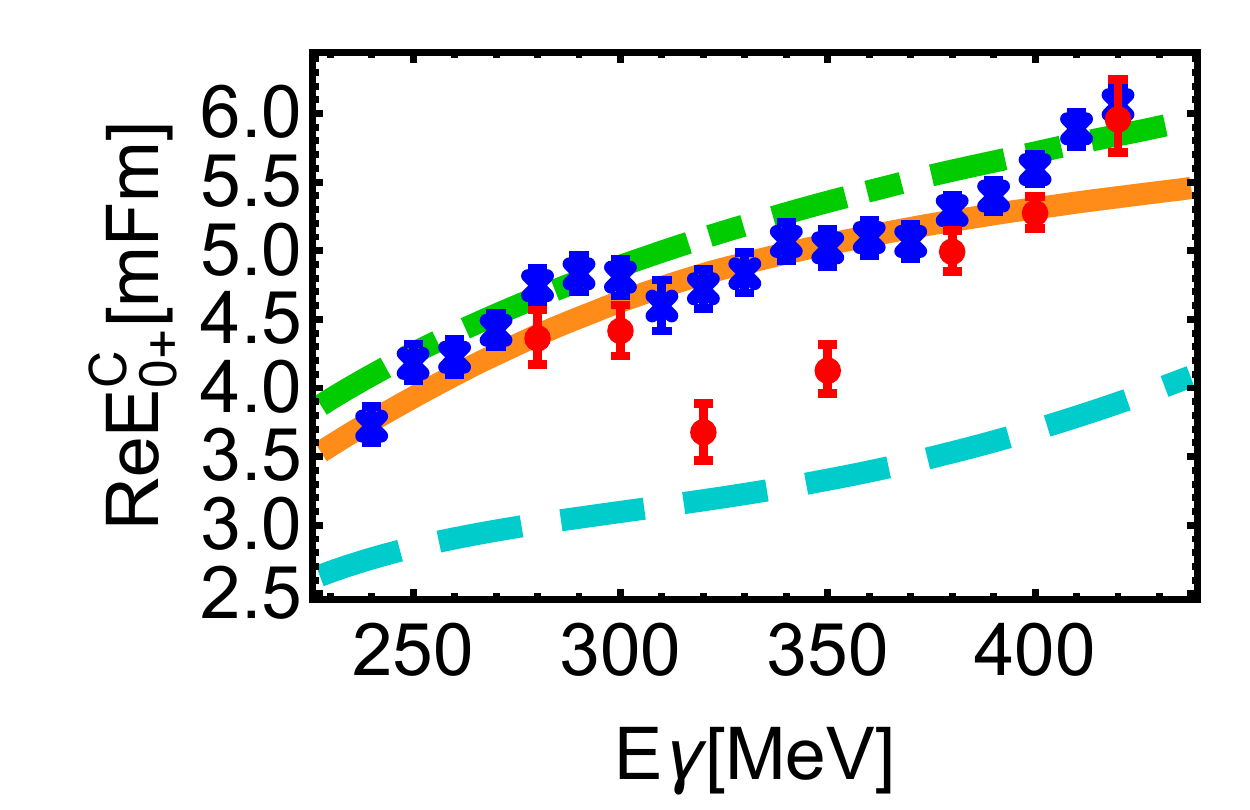}
 \end{overpic}
\begin{overpic}[width=0.325\textwidth]{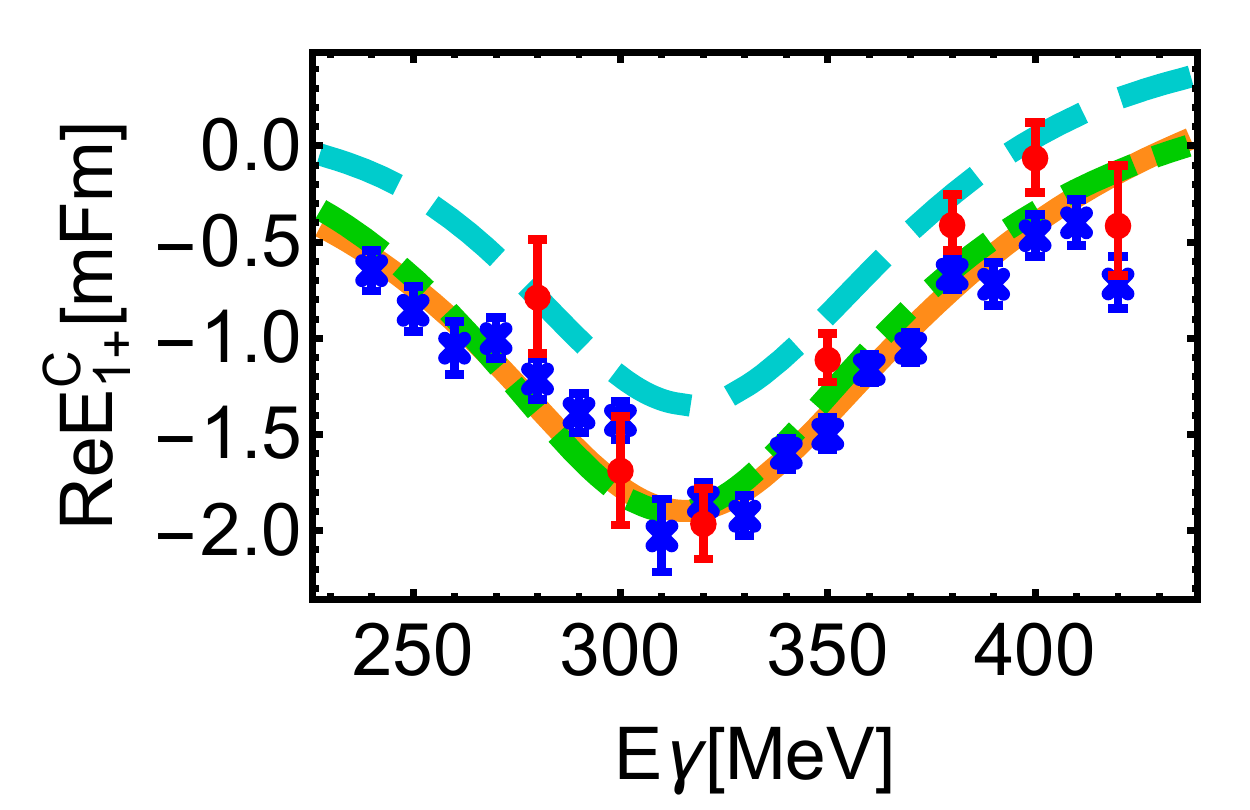}
 \end{overpic}
\begin{overpic}[width=0.325\textwidth]{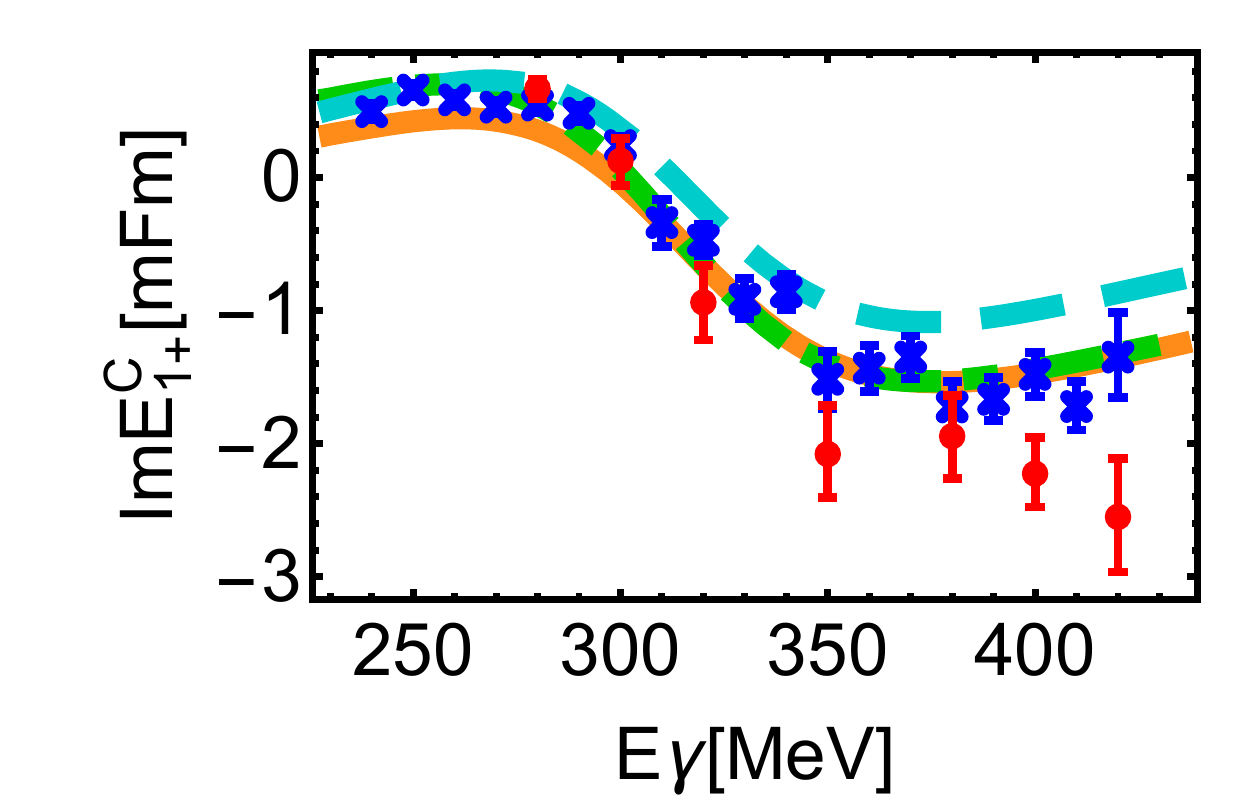}
 \end{overpic} \\
\begin{overpic}[width=0.325\textwidth]{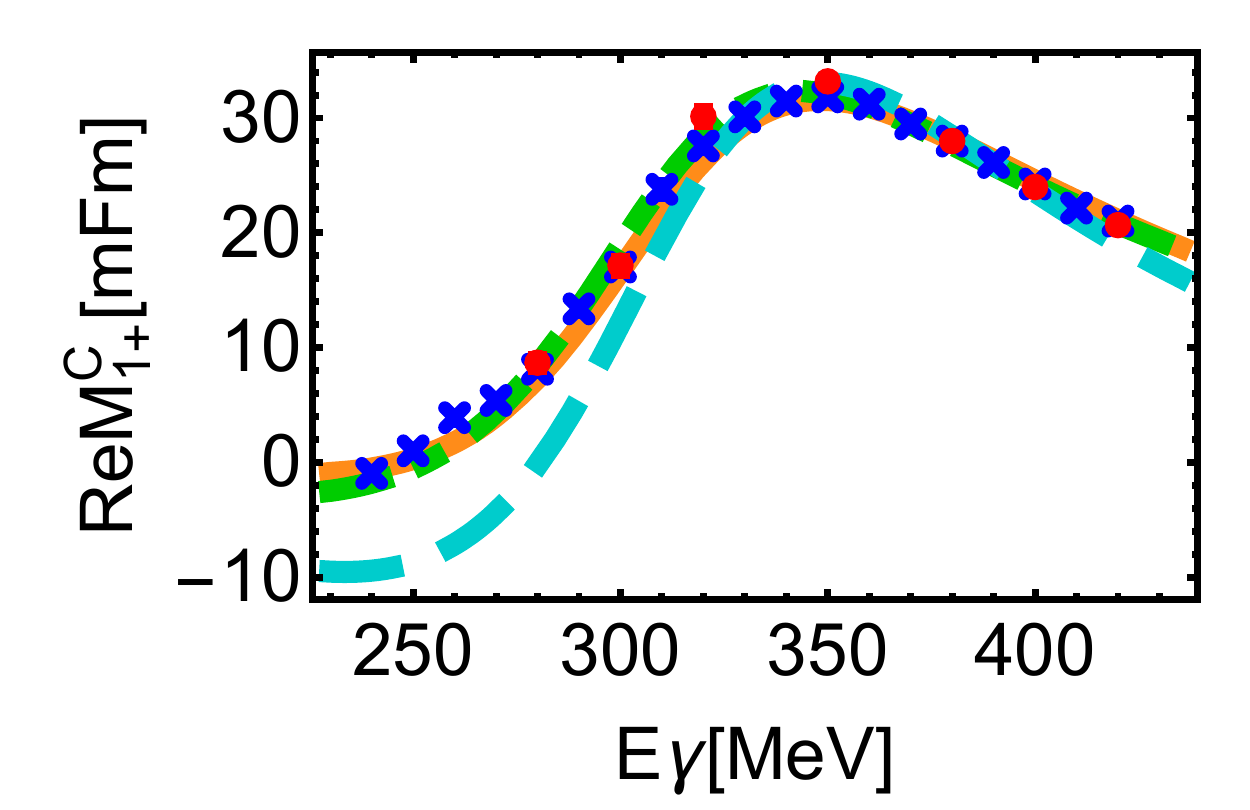}
 \end{overpic}
\begin{overpic}[width=0.325\textwidth]{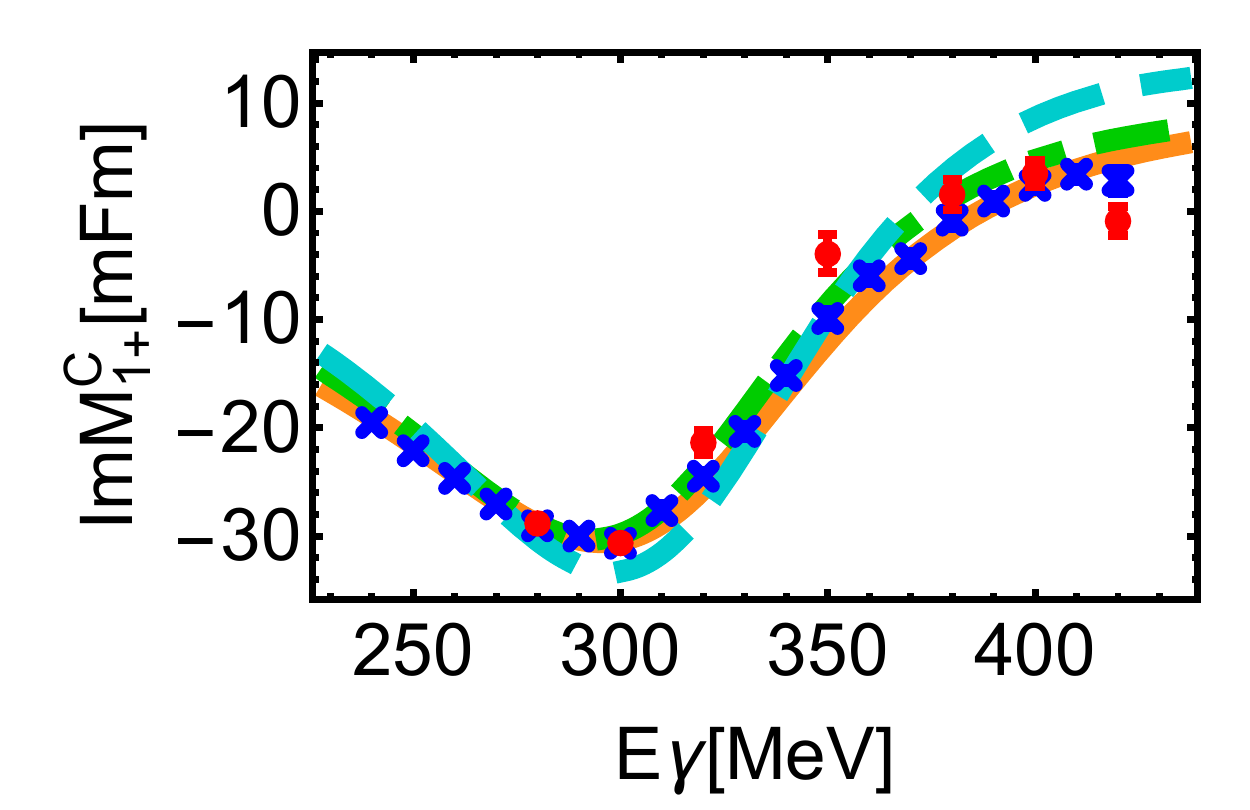}
 \end{overpic}
\begin{overpic}[width=0.325\textwidth]{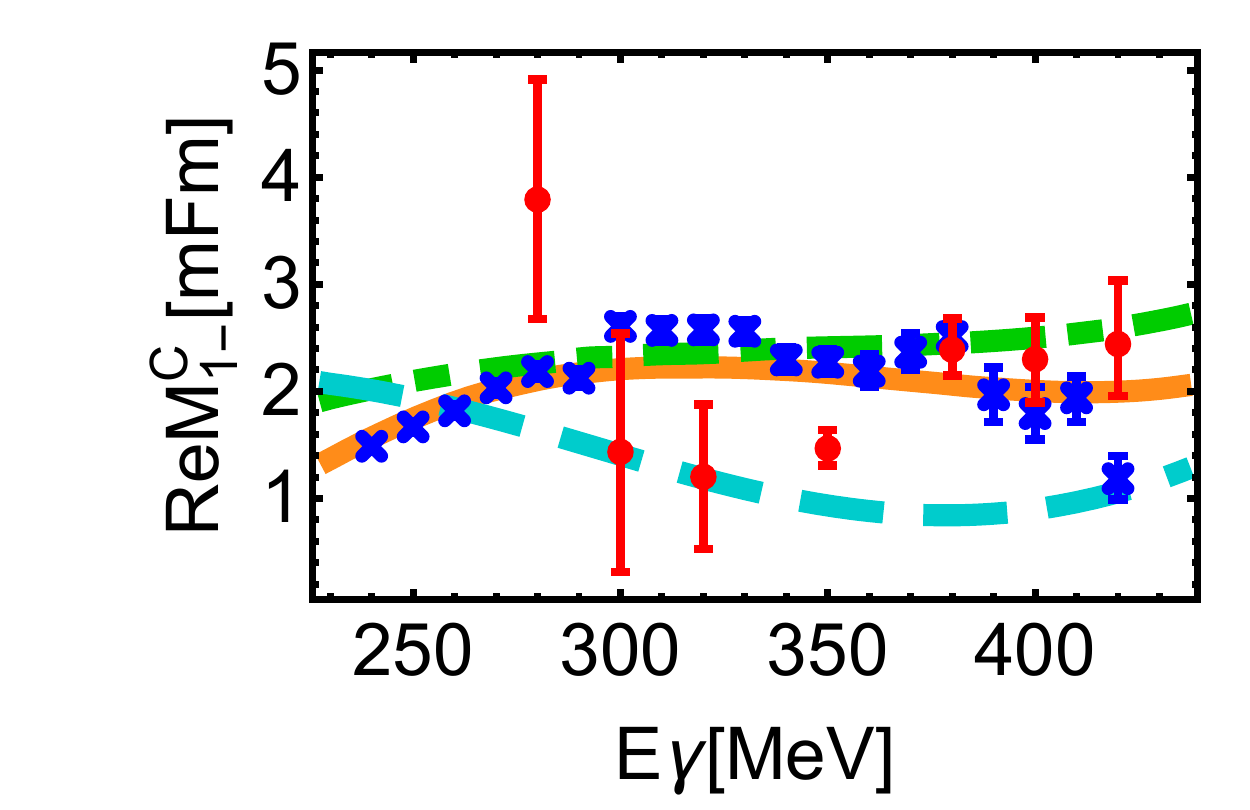}
 \end{overpic} \\
\begin{overpic}[width=0.325\textwidth]{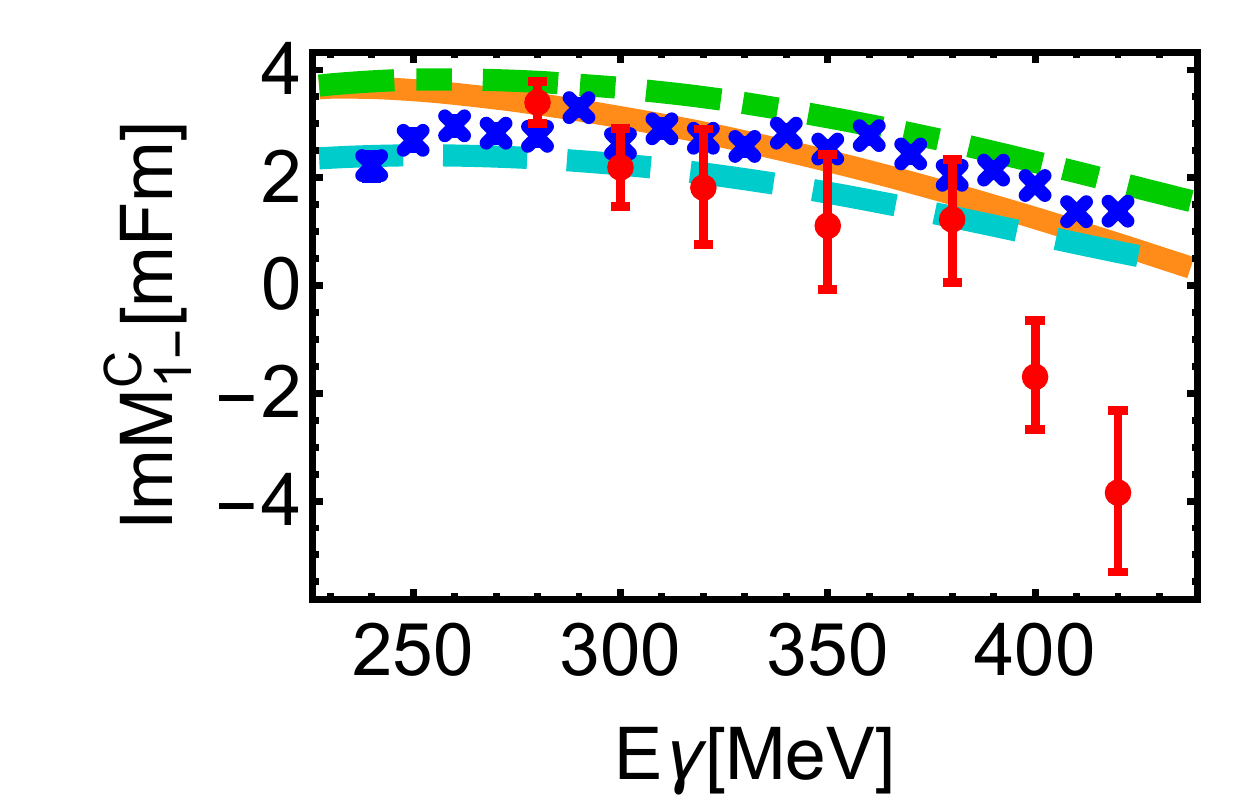}
 \end{overpic}
\vspace*{-5pt}
\caption[A comparison of the results of the bootstrap-analysis performed in the $\Delta$-resonance region, with $D$-waves fixed to SAID CM12. Five observables $\left\{ \sigma_{0}, \Sigma, T, P, F \right\}$ were fitted, with SAID-pseudodata employed for the recoil asymmetry $P$.]{The figures represent a comparison of the results of bootstrap-analyses to energy-dependent PWA-models. Shown are two fits using $\ell_{\mathrm{max}} = 2$, with $D$-waves fixed to the SAID-solution CM12 \cite{WorkmanEtAl2012ChewMPhotoprod,SAID}. The two cases, this time, comprise two different data-scenarios. A fit is shown using only real measured data for the five observables $\left\{ \sigma_{0}, \Sigma, T, P, F \right\}$ (red dots), as well as real data for $\left\{ \sigma_{0}, \Sigma, T, F \right\}$ combined with SAID-pseudodata for $P$ (blue crosses) (see discussion in the main text). Error-bars indicate statistical uncertainties determined from the bootstrapping-procedure (cf. the main text). \newline The results are compared to the PWA-solutions SAID CM12 (orange solid line) \cite{SAID}, BnGa 2014\_02 (cyan dashed line) \cite{BoGa} and MAID2007 (green dash-dotted line) \cite{MAID}.}
\label{fig:BootstrapDeltaRegionResultsPSAIDComparedToPWA}
\end{figure}

The most important point in this paragraph is actually the bootstrap-analysis involving the pseudodata. Again, we did a {\it reduced} bootstrap-TPWA (section \ref{sec:BootstrappingIntroduction}), generating $B=2000$ replicates from the combination of original data and pseudodata and then doing re-fits, each time starting in the global minimum of the fit to the original data (see Figure \ref{fig:Lmax2DWavesSAIDFitResultsDeltaRegionPSAID}). Error-estimates are derived from the bootstrap-distributions and the results are plotted in conjunction with the bootstrap-fit to the purely measured data (i.e. including Kharkov-data for $P$), in Figure \ref{fig:BootstrapDeltaRegionResultsPSAIDComparedToPWA}. In addition, the energy-dependent model-solutions MAID2007 \cite{MAID2007}, SAID CM12 \cite{SAID} and Bonn-Gatchina 2014\_02 \cite{BoGa} are drawn for comparison.

\clearpage

\clearpage

A decrease in the size of the error-bars can be seen in all multipoles. However, an effect is seen most prominently in $E_{1+}$ and $M_{1-}$. In the latter case, the change is actually quite extreme. The relative precision in the determination of $M_{1+}$ however does not get improved substantially, at least when considering Figure \ref{fig:BootstrapDeltaRegionResultsPSAIDComparedToPWA}. \newline
The single-energy fits with and without pseudodata for $P$ do not agree within error-bars in many cases. This occurs definitely due to the additional model-dependence in the data which is present in the former case. As a further consequence of this additional model-bias, the agreement of the resulting single-energy multipoles to the SAID- and MAID-model, which are almost identical, is best. On the other hand, the fit is drawn away from the Bonn-Gatchina model. \newline
Interestingly, the 'dip-like' structure seen in the single-energy results for $E_{0+}$ has disappeared for the fit {\it with} pseudodata. However, at the moment it seems hard to tell whether or not this effect came to pass due to the additional model-dependence in the (pseudo-) data, or simply because the precision in the pseudodata was good enough in order to better determine certain $\left< S,D \right>$- and $\left<P,D\right>$-interference terms, which are themselves necessary in order to correct the values of the $E_{0+}$-multipole.

\clearpage

\subsubsection{$\gamma p \rightarrow \pi^{0} p$ in the $2^{\mathrm{nd}}$ resonance region} \label{subsec:2ndResRegionDataFits}

\textbf{Description of the datasets} \newline

\begin{figure}[h]
 \centering
 \begin{overpic}[width=0.475\textwidth]{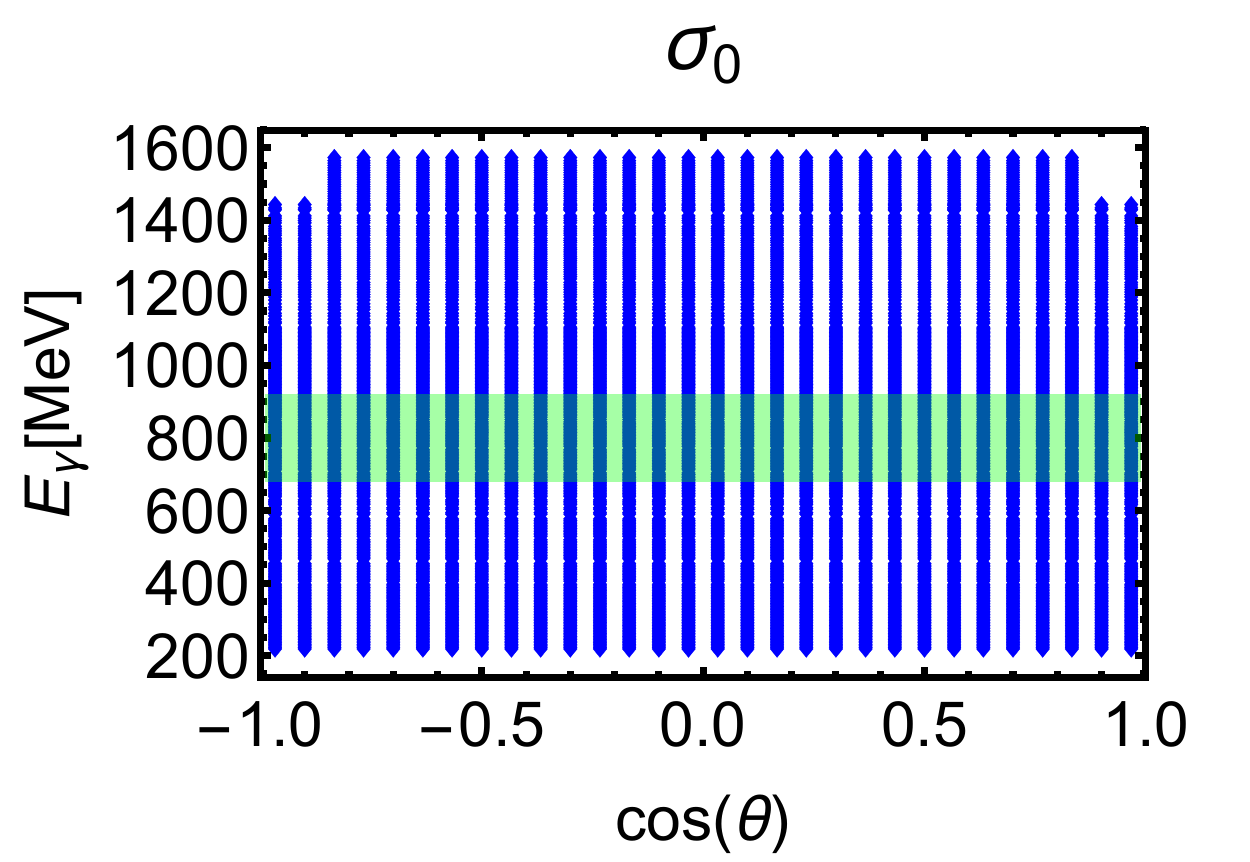}
 \end{overpic}
  \begin{overpic}[width=0.475\textwidth]{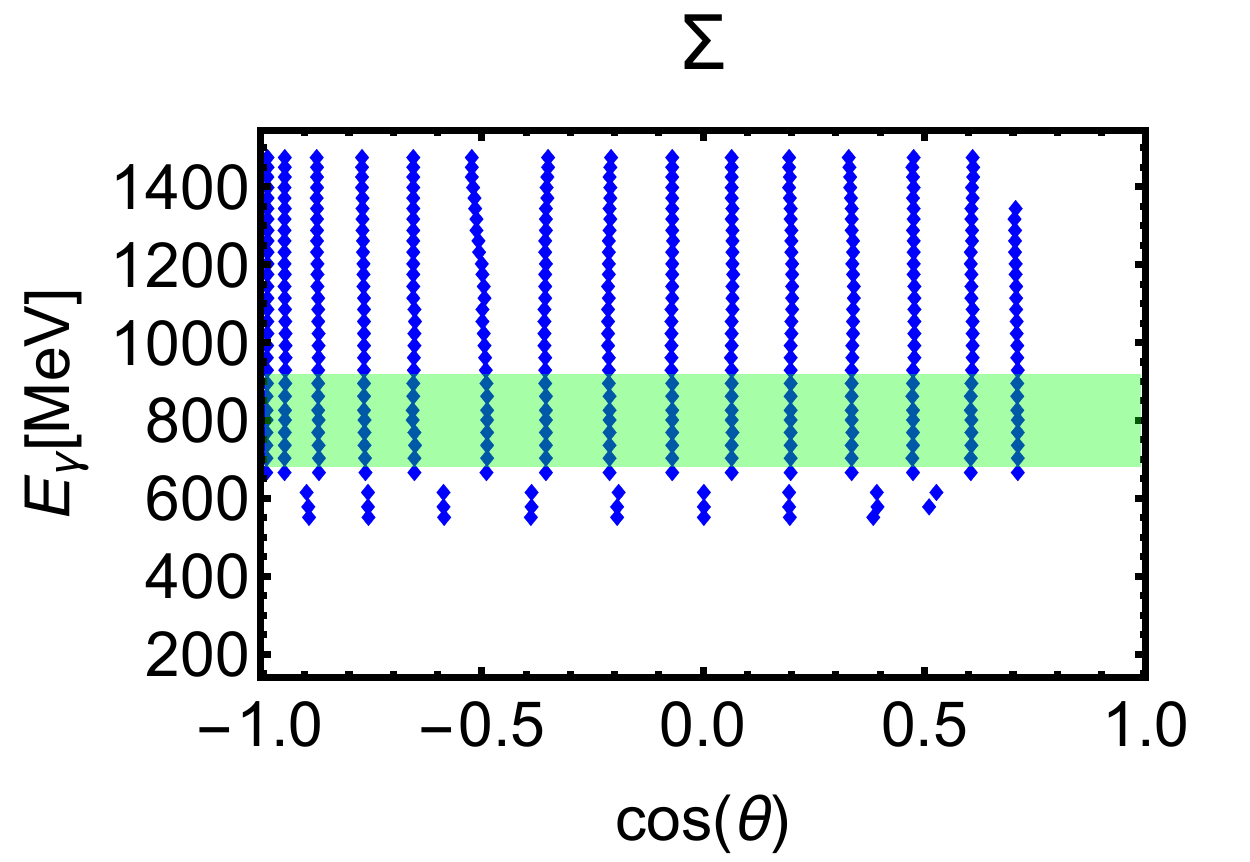}
 \end{overpic} \\
 \begin{overpic}[width=0.475\textwidth]{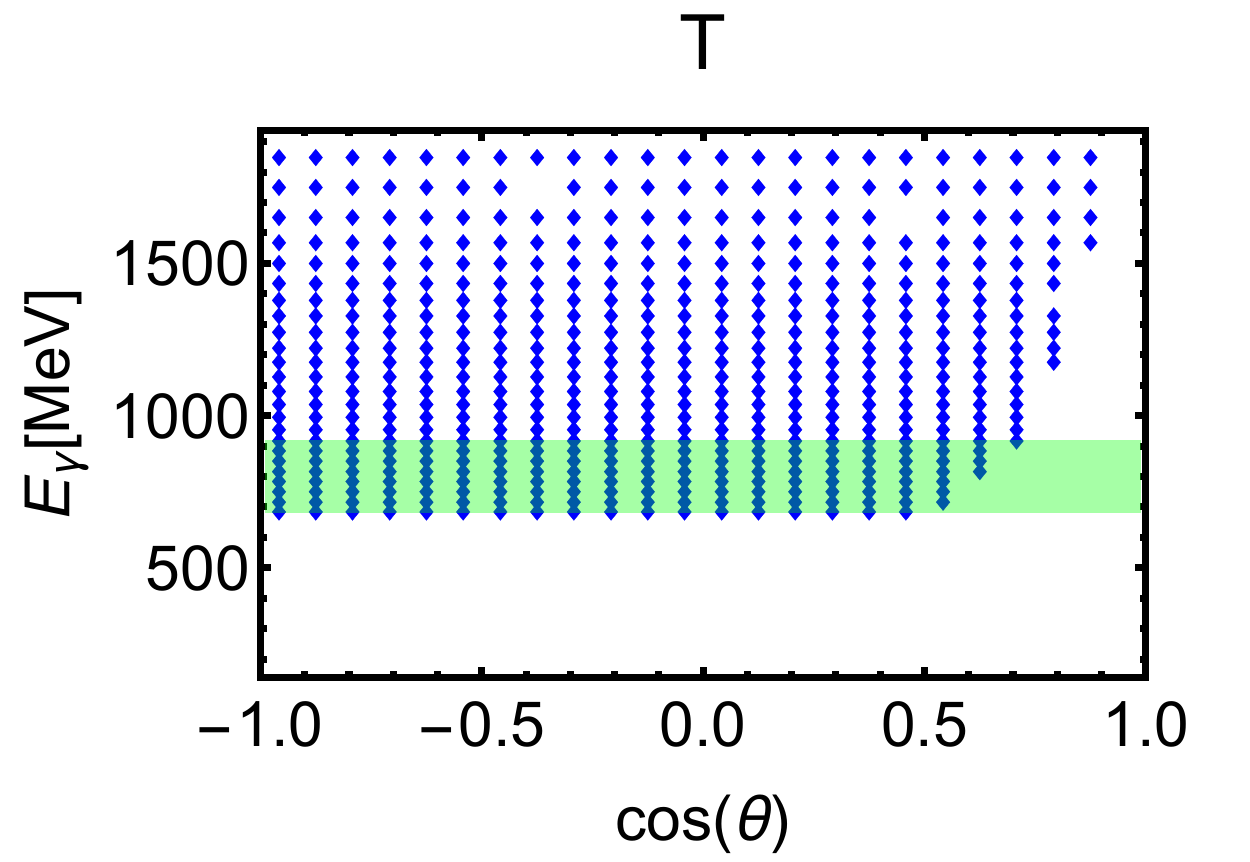}
 \end{overpic}
  \begin{overpic}[width=0.475\textwidth]{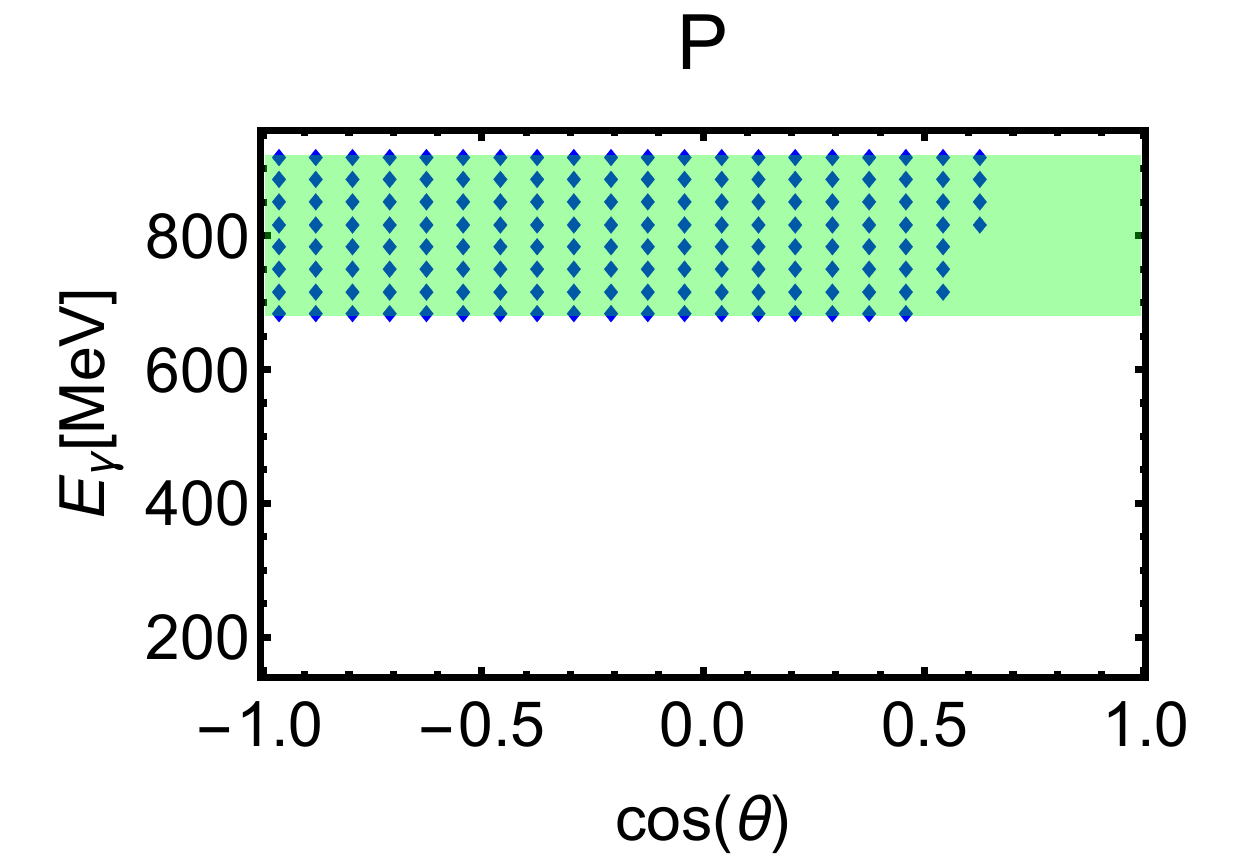}
 \end{overpic} \\
 \begin{overpic}[width=0.475\textwidth]{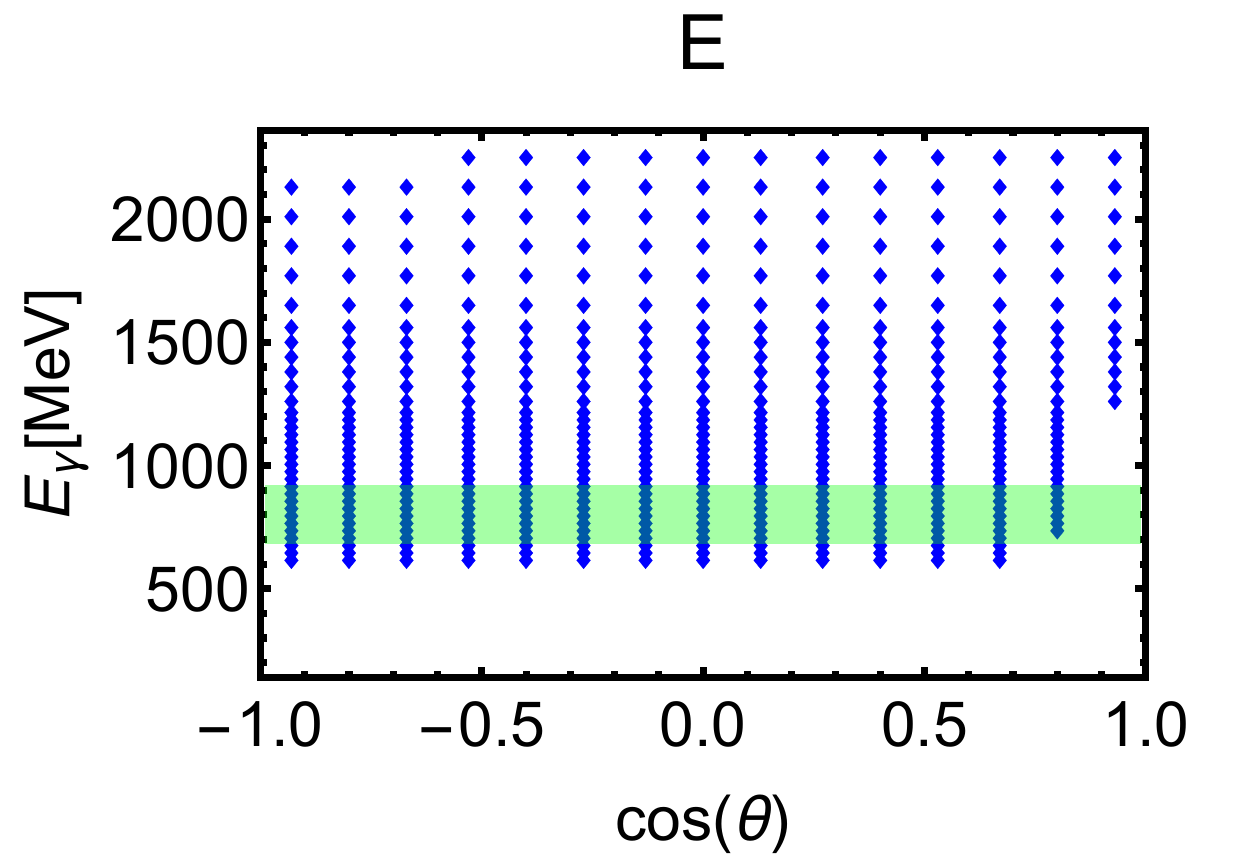}
 \end{overpic}
 \begin{overpic}[width=0.475\textwidth]{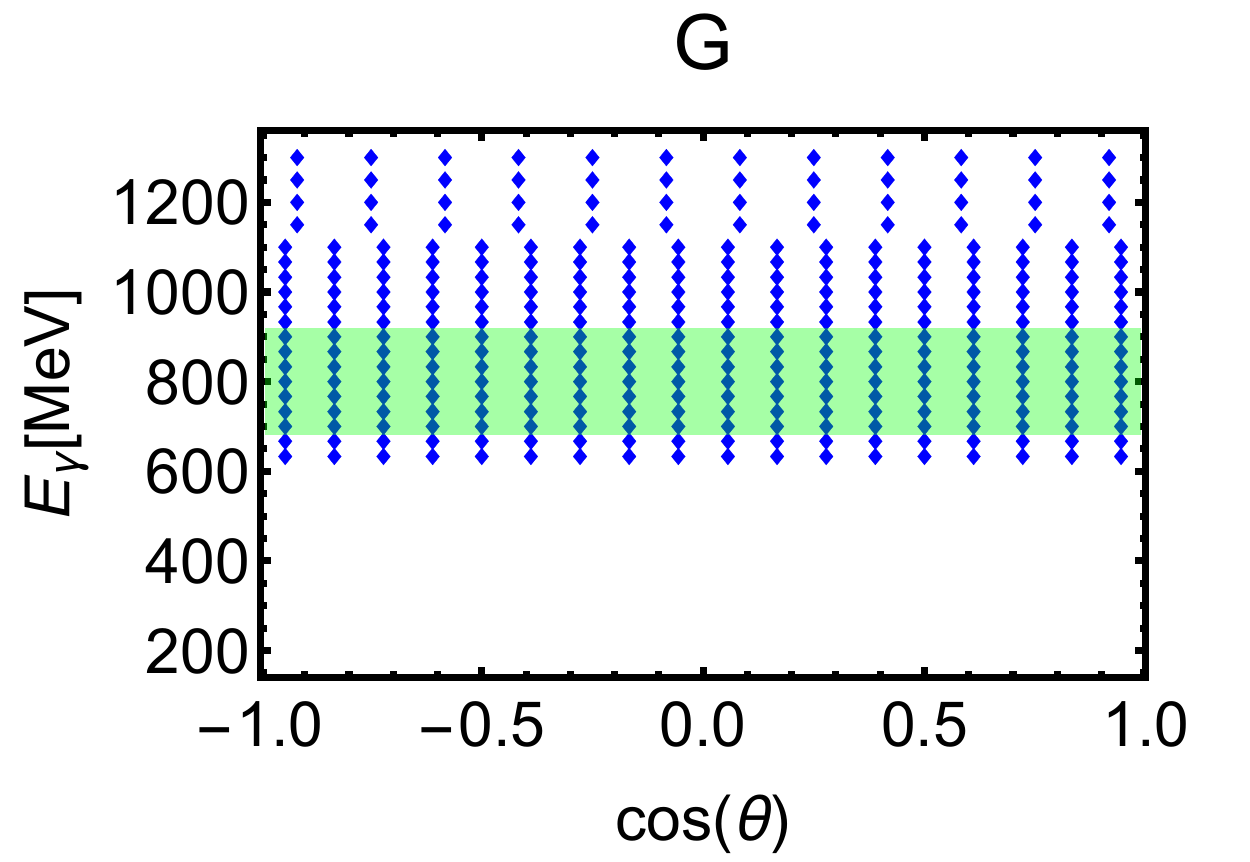}
 \end{overpic} \\
 \begin{overpic}[width=0.475\textwidth]{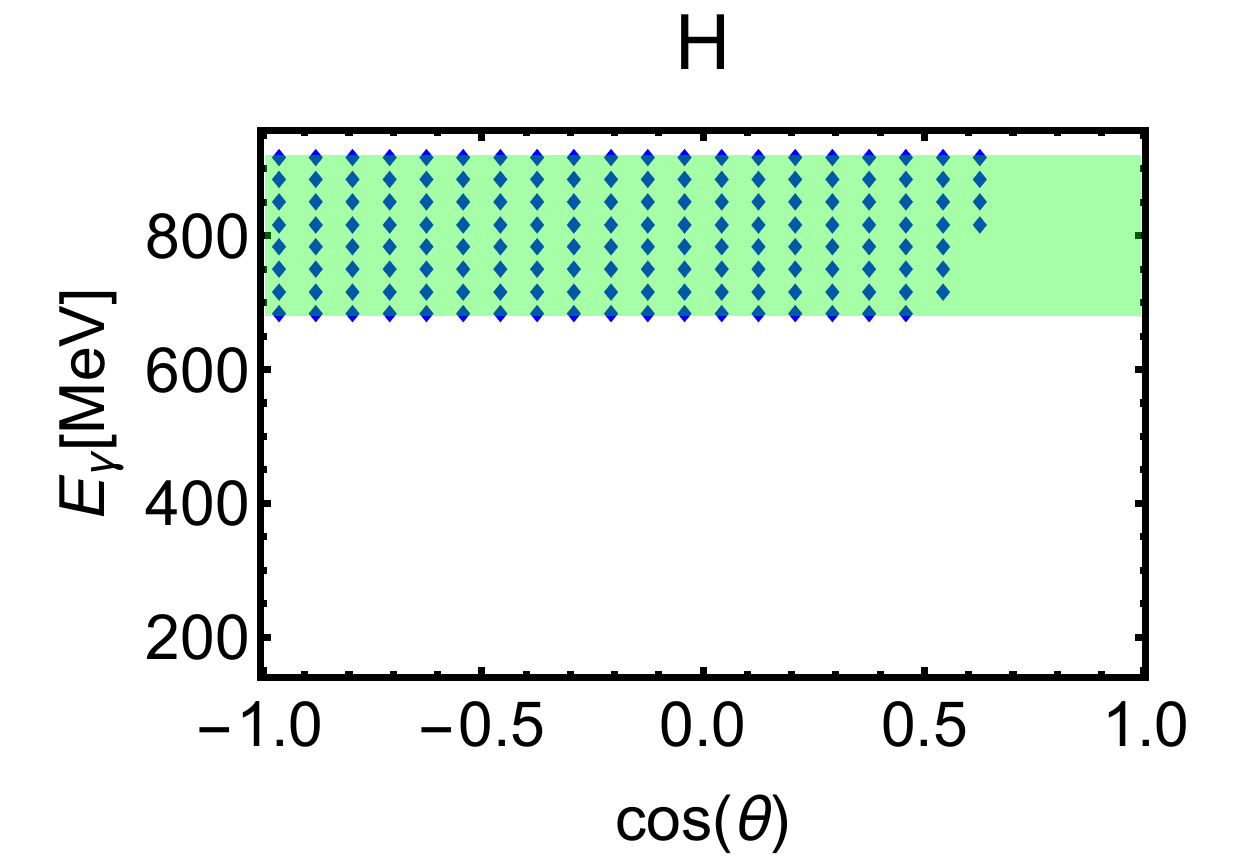}
 \end{overpic}
  \caption[The regions in phase space $(E_{\gamma}, \cos (\theta))$ covered by the seven polarization-datasets analyzed in the $2^{\mathrm{nd}}$ resonance region.]{The plots illustrate the regions in phase space $(E_{\gamma}, \cos (\theta))$ covered by the datasets for the cross section $\sigma_{0}$ \cite{Adlarson:2015} and the dimensionless asymmetries $\Sigma$, $T$, $P$, $E$, $G$ and $H$ \cite{GRAAL, Hartmann:2014, Hartmann:2015, Thiel:2012, Thiel:2016, Gottschall:2014, Gottschall:2015}. Blue markers give the location of individual datapoints. The green-shaded region illustrates the energy-region covered by the $P$- and $H$-datasets, i.e. the area on which the TPWA can be performed (cf. the main text).}
 \label{fig:2ndResRegionKinematicPlots}
\end{figure}

The phase-space coverages of all datasets fitted within the second resonance region are plotted in Figure \ref{fig:2ndResRegionKinematicPlots}. The datasets for $P$ and $H$ published by Hartmann et al. \cite{Hartmann:2014,Hartmann:2015} in this case have the smallest covered energy-region, with largest energy-spacings. Data are given between $E_{\gamma} = 683.5 \hspace*{1pt} \mathrm{MeV}$ and $916.66 \hspace*{1pt} \mathrm{MeV}$, in $8$ energy bins of roughly equal spacing $\Delta E_{\gamma} = 33 \hspace*{1pt} \mathrm{MeV}$. Otherwise, the kinematic grids of both datasets are exactly equal. Angular distributions have $18$ to $20$ datapoints and cover all angles except for roughly one fourth of the full interval, in the forward region. \newline
Since profile functions will have to be calculated from dimensionless asymmetries, the backbone of the analysis will again be formed by the dataset for the differential cross section $\sigma_{0}$. For the second resonance region, we pick here the recently published measurement by Adlarson et al. \cite{Adlarson:2015}. Since we are considering data above the $\Delta$-region, the problems this cross section shows for low energies (cf. section \ref{chap:LFits}) need not be of interest here\footnote{I.e., for the higher energies considered here, the fact that the systematic error of the cross section dominates the statistical error does not lead to consequences as severe as those described for the Adlarson-data \cite{Adlarson:2015} in section \ref{sec:LFitsPaper}, or for the Hornidge-data \cite{Hornidge:2013} in section \ref{subsec:DeltaRegionDataFits}. However, still we decided to add statistical and systematic errors of $\sigma_{0}$ in quadrature for this analysis.}. Nonetheless, the Adlarson-data constitute a cross section-measurement of unprecedented statistics and kinematic coverage. They cover an energy-interval from $E_{\gamma} = 218.37 \hspace*{1pt} \mathrm{MeV}$ to $1572.89 \hspace*{1pt} \mathrm{MeV}$ in roughly equidistant bins of spacing $\Delta E_{\gamma} = 4.4 \hspace*{1pt} \mathrm{MeV}$, thus amounting to $269$ energy-points. Angular distributions cover the full interval in $26$ up to $30$ equally spaced points. Statistical errors are tiny, making up only a few percent of the cross-sections. \newline
The single-polarization observables are completed by measurements for $\Sigma$ by Bartalini et al. (GRAAL) \cite{GRAAL} and $T$ by Hartmann et al. \cite{Hartmann:2014, Hartmann:2015}. The $T$-data cover energies from $E_{\gamma} = 683.5 \hspace*{1pt} \mathrm{MeV}$ to $1847.58 \hspace*{1pt} \mathrm{MeV}$ in $24$ non-equidistant $E_{\gamma}$-bins, while the GRAAL-data for $\Sigma$ are situated between $E_{\gamma} = 551 \hspace*{1pt} \mathrm{MeV}$ and $1475 \hspace*{1pt} \mathrm{MeV}$, in $31$ roughly equidistant $E_{\gamma}$-bins. The angular distributions of the Hartmann-data are covered in $18$ to $23$ equidistant bins in $\cos(\theta)$. However, for the lower energies the forward scattering-region is missing. For the Bartalini-data, between $8$ and $15$ angular bins are given, non-equidistantly in $\cos(\theta)$. Statistical errors of the GRAAL-data are a bit smaller, ranging between $2$ and $10 \hspace*{1pt} \%$ of the asymmetry-values, while for the $T$-dataset errors amount to around $5$ to $20 \hspace*{1pt} \%$. \newline
The set (\ref{eq:2ndResRegionFitObservables}) is completed by data for the double-polarization asymmetries $E$ and $G$. Both datasets used here are first measurements in the respective kinematic regimes, performed by the CBELSA/TAPS-collaboration. The $E$-data by Gottschall et al. \cite{Gottschall:2014,Gottschall:2015} cover a very large energy-region, from $E_{\gamma} = 615 \hspace*{1pt} \mathrm{MeV}$ to $2250 \hspace*{1pt} \mathrm{MeV}$, in $33$ non-equidistant $E_{\gamma}$-bins. Angular distributions are given by $13$ to $15$ points of equal spacing in $\cos(\theta)$ and statistical errors mostly amount to $(8 - 25) \hspace*{1pt} \%$. \newline
In order to remove the double ambiguity, data for the observable $G$ are needed (cf. chapter \ref{chap:Omelaenko}). Within the considered energy region, a first measurement has been published recently by Thiel et al. \cite{Thiel:2012, Thiel:2016}. These data cover an energy region from $E_{\gamma} = 633 \hspace*{1pt} \mathrm{MeV}$ to $1300 \hspace*{1pt} \mathrm{MeV}$ with $19$ energy-bins of non-equidistant spacing. Angular distributions reach over the full interval, with $12$ to $18$ equidistant points in $\cos (\theta)$. The precision of the $G$-measurement is similar to the $E$-data, with statistical errors ranging around $(5 - 20) \hspace*{1pt} \%$. \newline
The aforementioned measurements constitute the data base for the TPWA-fits in the second resonance-region. Table \ref{tab:2ndResRegionSandorfiDataTable} comprises some information for quick reference. \clearpage

\clearpage

\begin{table}[h]
 \centering
 \vspace*{-7.5pt}
\begin{tabular}{rccccr}
\hline
\hline
 Group & Experiment & Obs.'s & $E_{\gamma}$-range [MeV] & $\Delta E_{\gamma}$ [MeV] & av. stat. errors \\
\hline
  &  &  &  &  &  \\
 $1$ & MAMI/CB & $\sigma_{0}$ & $(218.37 - 1572.89)$ & $\sim 4.4$ & $\sim (2 - 4) \%$ \\
  &  &  &  &  &  \\
 $2$ & GRAAL & $\Sigma$ & $(551. - 1475.)$ & $(25. - 31.)$ & $\sim (2. - 10.) \%$ \\
  &  &  &  &  &  \\
 $3$ & CBELSA/TAPS & $\left\{ P, H \right\}$ & $(683.5 - 916.66)$ & $\sim 33.$ & $\sim (8 - 30) \%$ \\
  &  &  &  &  &  \\
 $4$ & CBELSA/TAPS & $T$ & $(683.5 - 1847.58)$ & $(32. - 99.)$ & $\sim (5 - 20) \%$ \\
  &  &  &  &  &  \\
 $5$ & CBELSA/TAPS & $E$ & $(615. - 2250.)$ & $(30. - 120.)$ & $\sim (8 - 25) \%$ \\
  &  &  &  &  &  \\
 $6$ & CBELSA/TAPS & $G$ & $(633. - 1300.)$ & $(34. - 50.)$ & $\sim (5 - 20) \%$ \\
  &  &  &  &  &  \\
\hline
\hline
\end{tabular}
 \caption[Some specifics of the polarization datasets fitted in the $2^{\mathrm{nd}}$-resonance region.]{Some specifics of the polarization datasets fitted in the $2^{\mathrm{nd}}$-resonance region \cite{Adlarson:2015, GRAAL, Hartmann:2014, Hartmann:2015, Thiel:2012, Thiel:2016, Gottschall:2014, Gottschall:2015} are collected. Datasets are divided into groups according to their kinematic binning. Energy-ranges as well as some information on the energy-binning $\Delta E_{\gamma}$ are given. Approximate quantities are marked with a $\sim$. Ranges for binnings as well as averages are provided, whenever appropriate. \newline
 Also, some rough estimate of the precision of the respective datasets is given, by providing an assessment for the average size of the statistical errors, relative to the data, in percent. For similar Tables, see reference \cite{Sandorfi:2010uv}.}
 \label{tab:2ndResRegionSandorfiDataTable}
\end{table}
In total, the datasets provide information on $10 \hspace*{1pt} 018$ datapoints. \newline

\textbf{Kinematic re-binning of data and evaluation of the profile-functions $\check{\Omega}^{\alpha}$} \newline

As in the case of the fit in the $\Delta$-region discussed in the previous section (cf. sec. \ref{subsec:DeltaRegionDataFits}), a central issue consists of the fact that the utilized datasets, even in their shared region of phase-space, do not have the same kinematic binning (Figure \ref{fig:2ndResRegionKinematicPlots}). Therefore, some adjustments are necessary prior to the fitting of multipoles in a TPWA. \newline
Since the CBELSA/TAPS-datasets for $P$ and $H$, both with the exact same binning, cover the minimal energy region common to all data here, they are chosen to dictate the oberall energy-binning for all profile functions $\check{\Omega}^{\alpha}$ prepared for the TPWA. The differential cross section by Adlarson et al. \cite{Adlarson:2015} has by far the highest statistics of all considered datasets. Therefore, it is again no problem to just choose kinematically closest points (both in energy and angle) for the evaluations of the profile functions $\check{\Omega}^{\alpha} = \sigma_{0} \Omega^{\alpha}$. The angular binnings of the individual observables do not have to be adjusted, since they are fitted in the TPWA. \newline
Thus, the TPWA can be performed on the energy-interval between $E_{\gamma} = 683.5 \hspace*{1pt} \mathrm{MeV}$ and $916.66 \hspace*{1pt} \mathrm{MeV}$, as dictated by the $P$- and $H$-measurements. \newline
The kinematic re-binning and the evaluation of the profile-functions leave $1 \hspace*{1pt} 080$ datapoints as input for the multipole analysis. \newline

\textbf{Determination of an $\ell_{\mathrm{max}}$-estimate from angular distributions of the $\check{\Omega}^{\alpha}$} \newline

Just as in the $\Delta$-region analysis, we prepare the multipole analysis itself by extracting

\begin{figure}[h]
 \centering
 \vspace*{20pt}
 \begin{overpic}[width=0.475\textwidth]{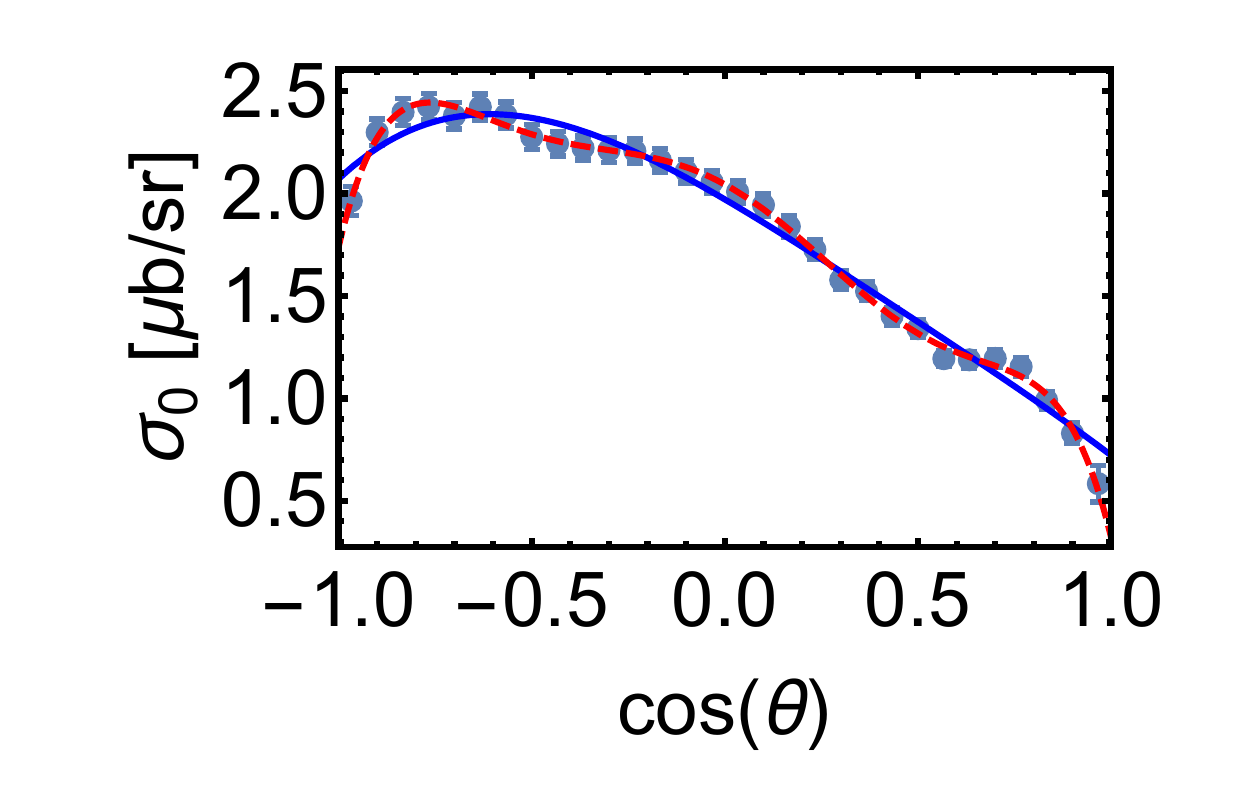}
 \put(81.5,68){\begin{Large}$E_{\gamma} = 884.02 \hspace*{2pt} \mathrm{MeV}$\end{Large}}
 \end{overpic}
 \begin{overpic}[width=0.475\textwidth]{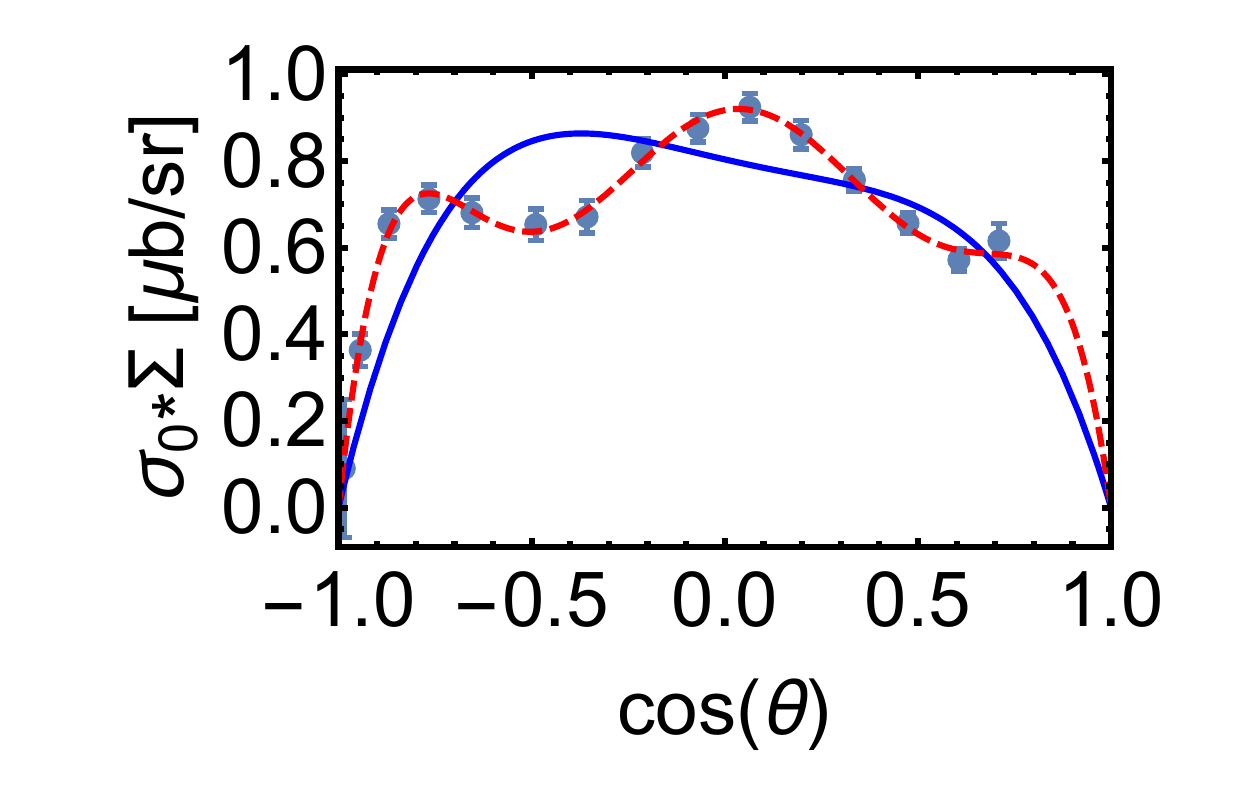}
 \end{overpic} \\
 \begin{overpic}[width=0.475\textwidth]{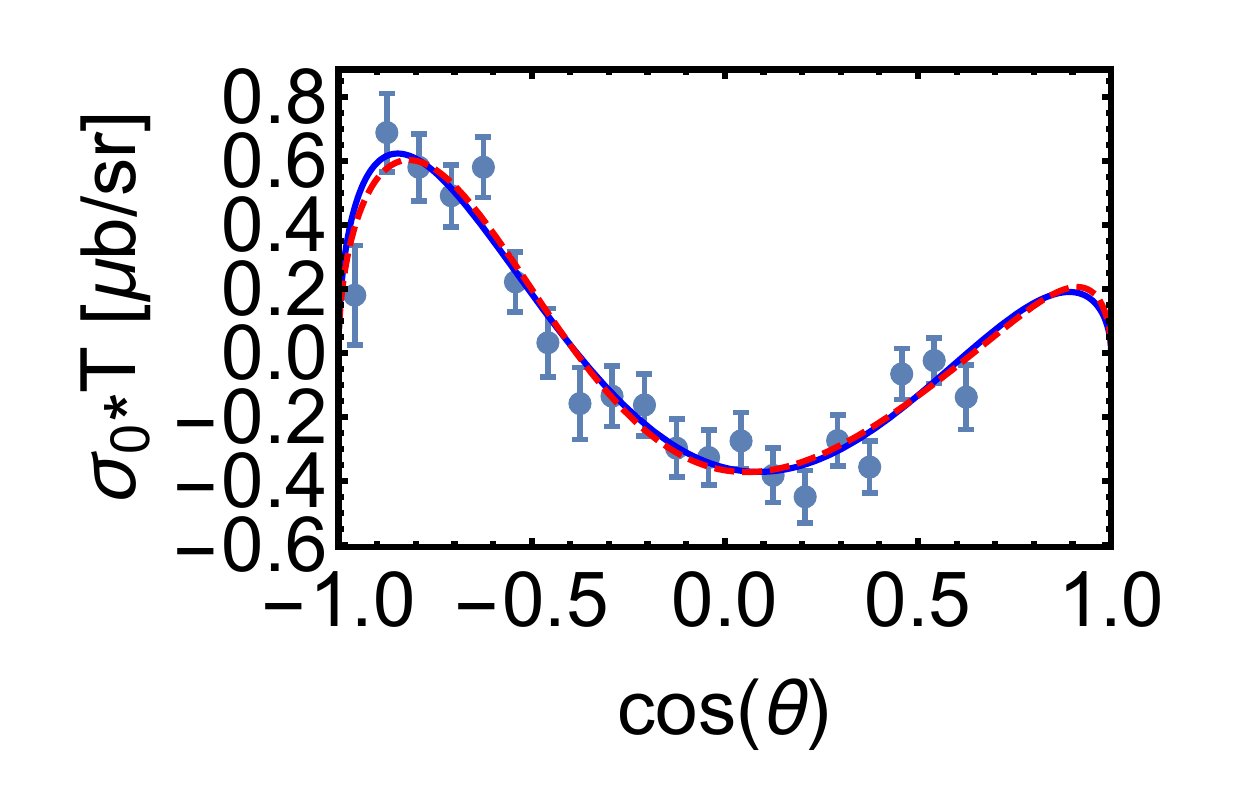}
 \end{overpic}
 \begin{overpic}[width=0.475\textwidth]{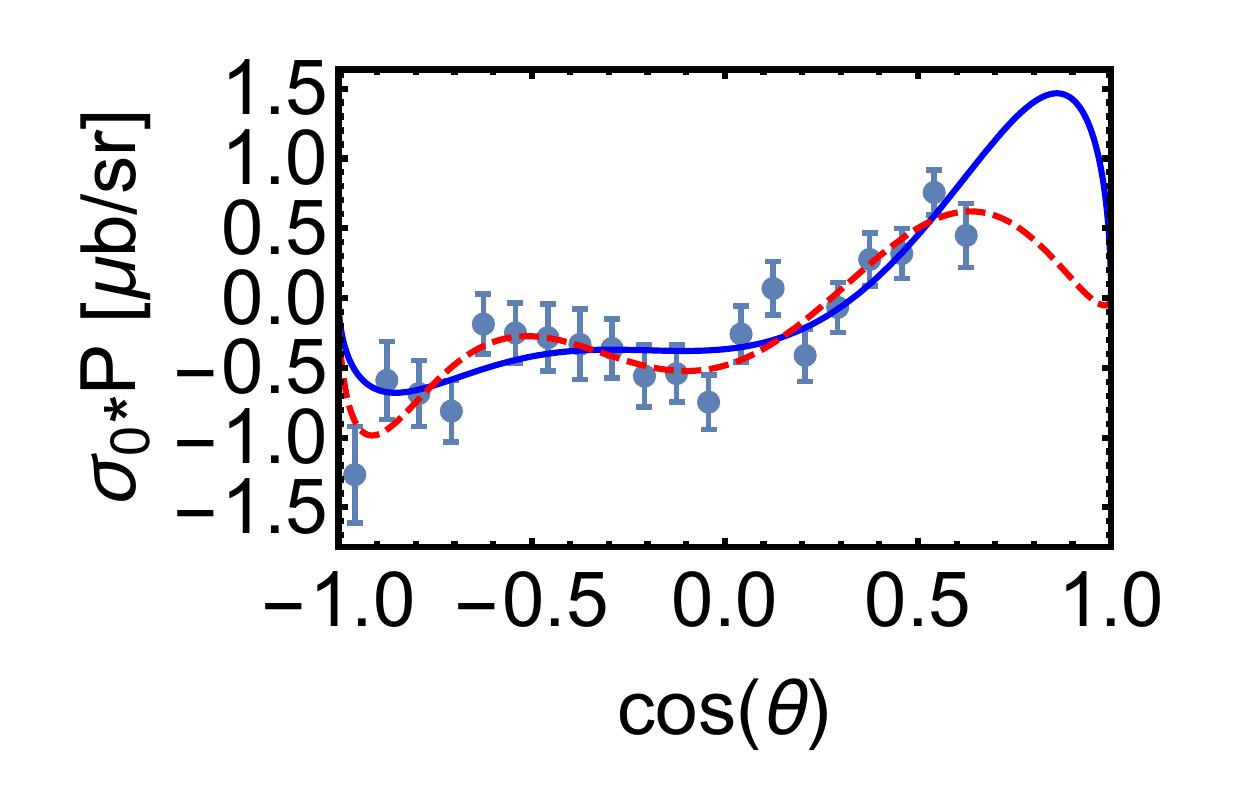}
 \end{overpic} \\
 \begin{overpic}[width=0.475\textwidth]{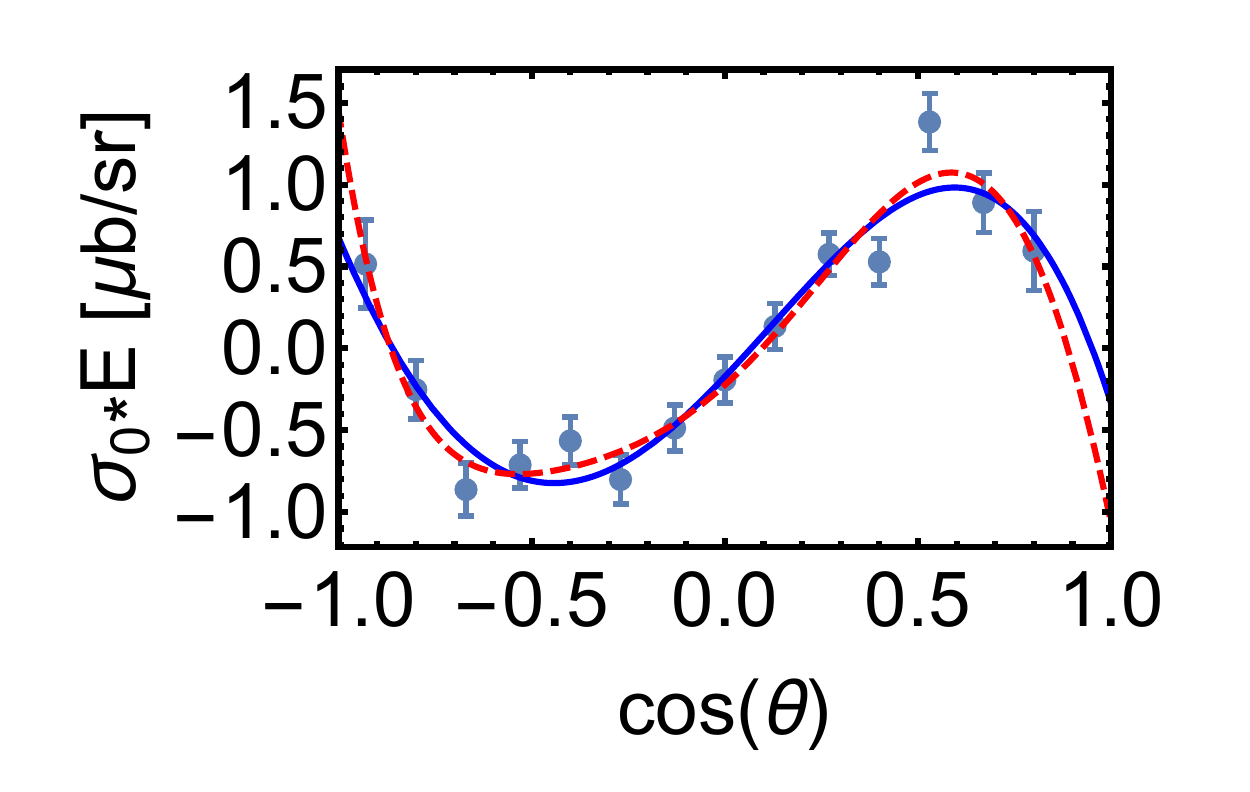}
 \end{overpic}
 \begin{overpic}[width=0.475\textwidth]{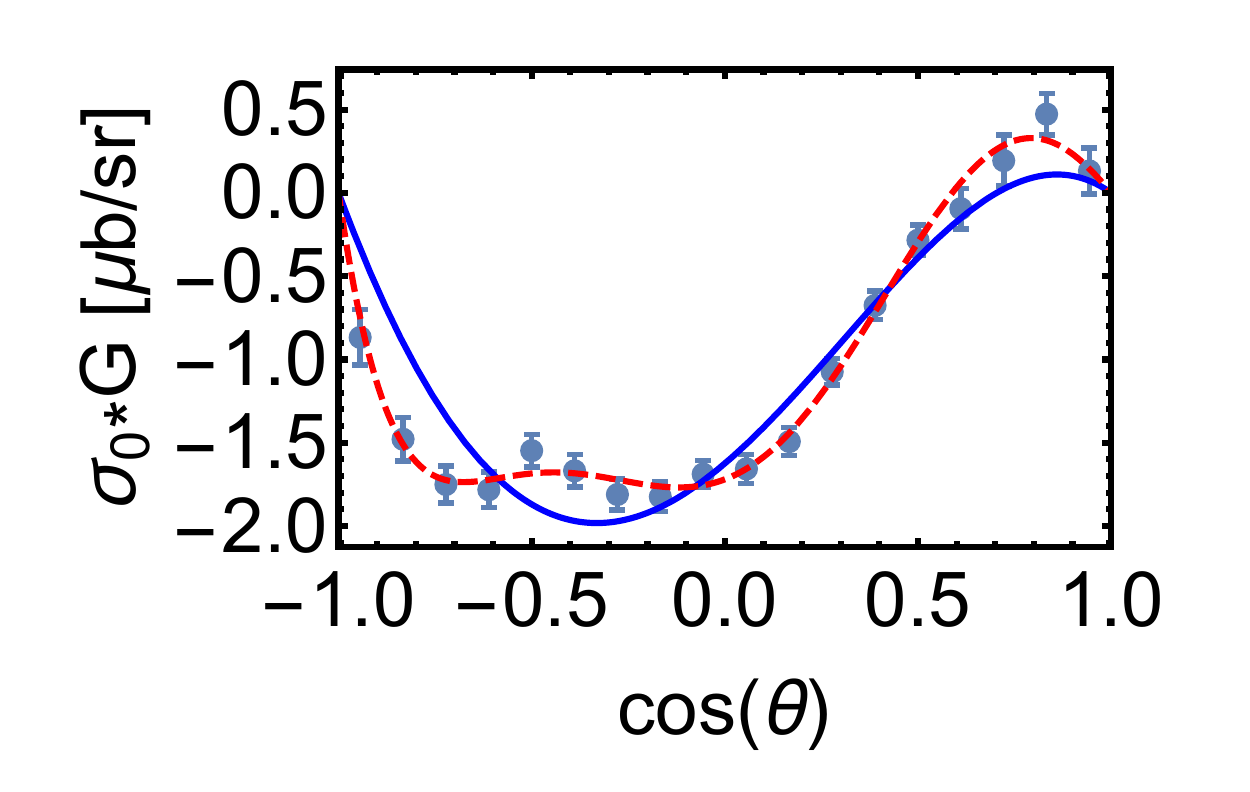}
 \end{overpic} \\
 \begin{overpic}[width=0.475\textwidth]{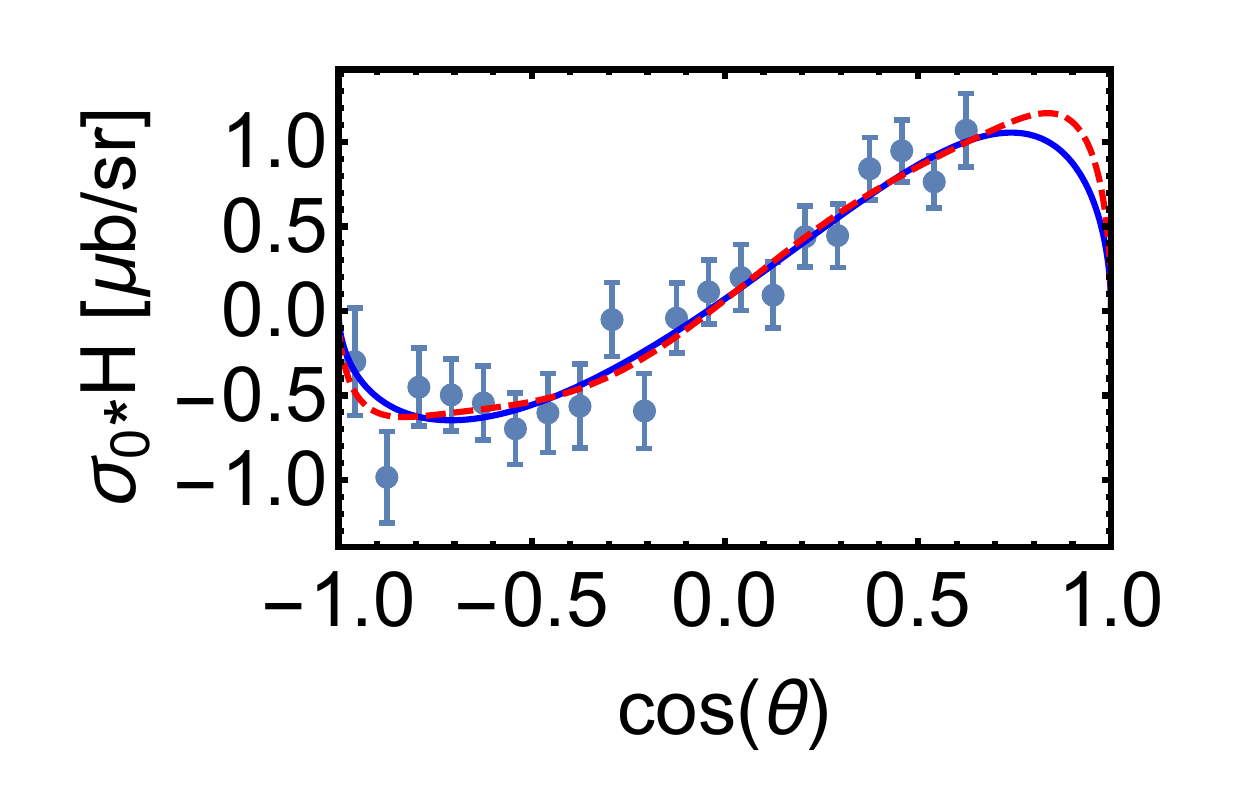}
 \end{overpic}
  \caption[Angular distributions of profile functions for the seven polarization-datasets analyzed in the $2^{\mathrm{nd}}$ resonance region. Data are plotted at a particular energy, $E_{\gamma} = 884.02 \hspace*{1pt} \mathrm{MeV}$.]{Shown here are angular distributions for the differential cross section $\sigma_{0}$ and the profile functions $\check{\Sigma} = \sigma_{0} \Sigma$, $\check{T} = \sigma_{0} T$, $\check{P} = \sigma_{0} P$ , $\check{E} = \sigma_{0} E$ , $\check{G} = \sigma_{0} G$ and $\check{F} = \sigma_{0} F$ analyzed in the second resonance region. Data are plotted at a particular energy, $E_{\gamma} = 884.02 \hspace*{1pt} \mathrm{MeV}$. \newline
  Fits to the angular distributions are also shown, employing the truncation angular momenta $\ell_{\mathrm{max}} = 2$ (blue solid line) and $\ell_{\mathrm{max}} = 3$ (red dashed line). The fit-parametrizations follow the formulas given in equatons (\ref{eq:DCSAngDistChapter4}) to (\ref{eq:PAngDistChapter4}) and (\ref{eq:EAngDistChapter4}) to (\ref{eq:HAngDistChapter4}).}
 \label{fig:2ndResRegionObsFitAngDistPlots}
\end{figure}
\begin{figure}[h]
 \centering
 \begin{overpic}[width=0.45\textwidth]{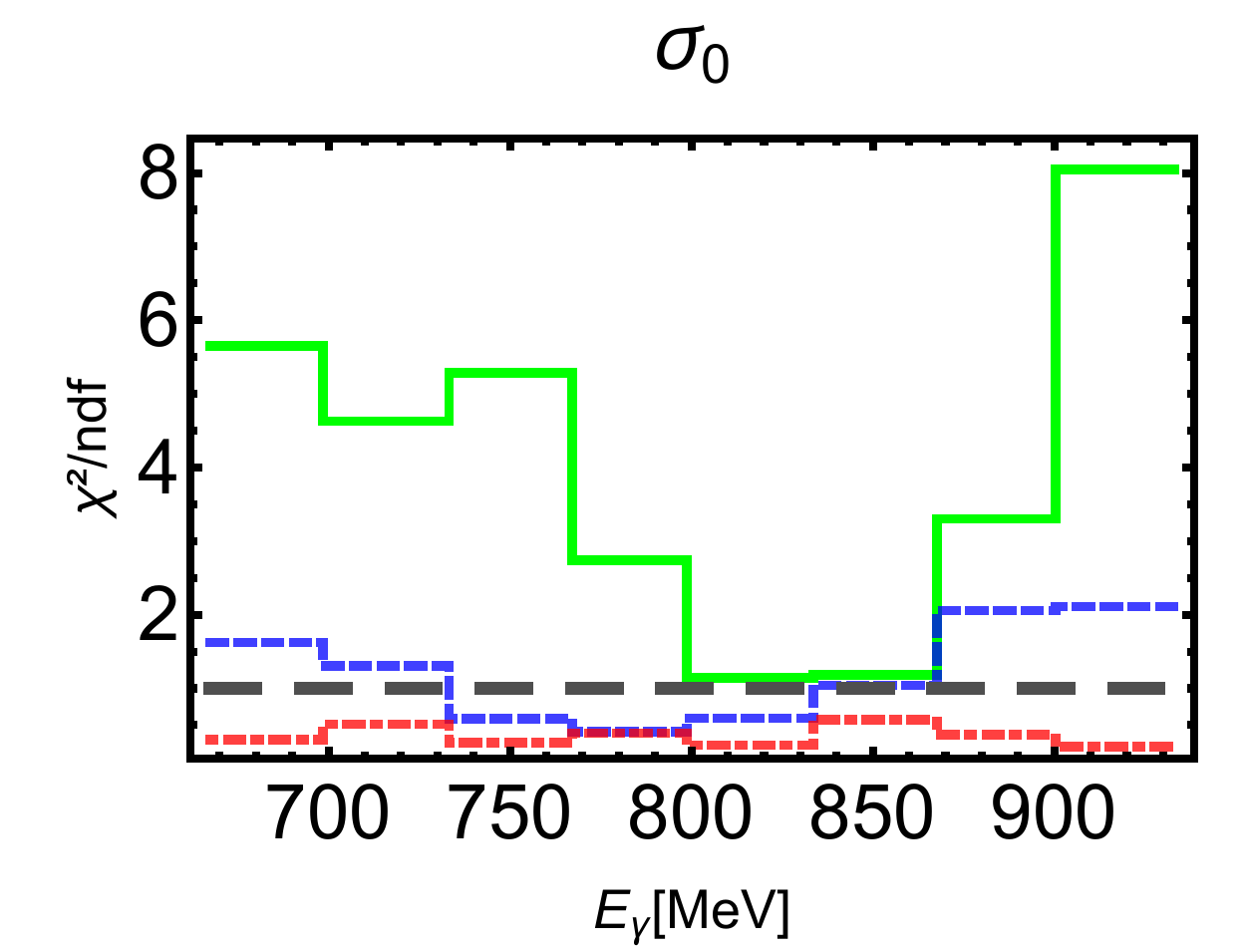}
 \end{overpic}
 \begin{overpic}[width=0.45\textwidth]{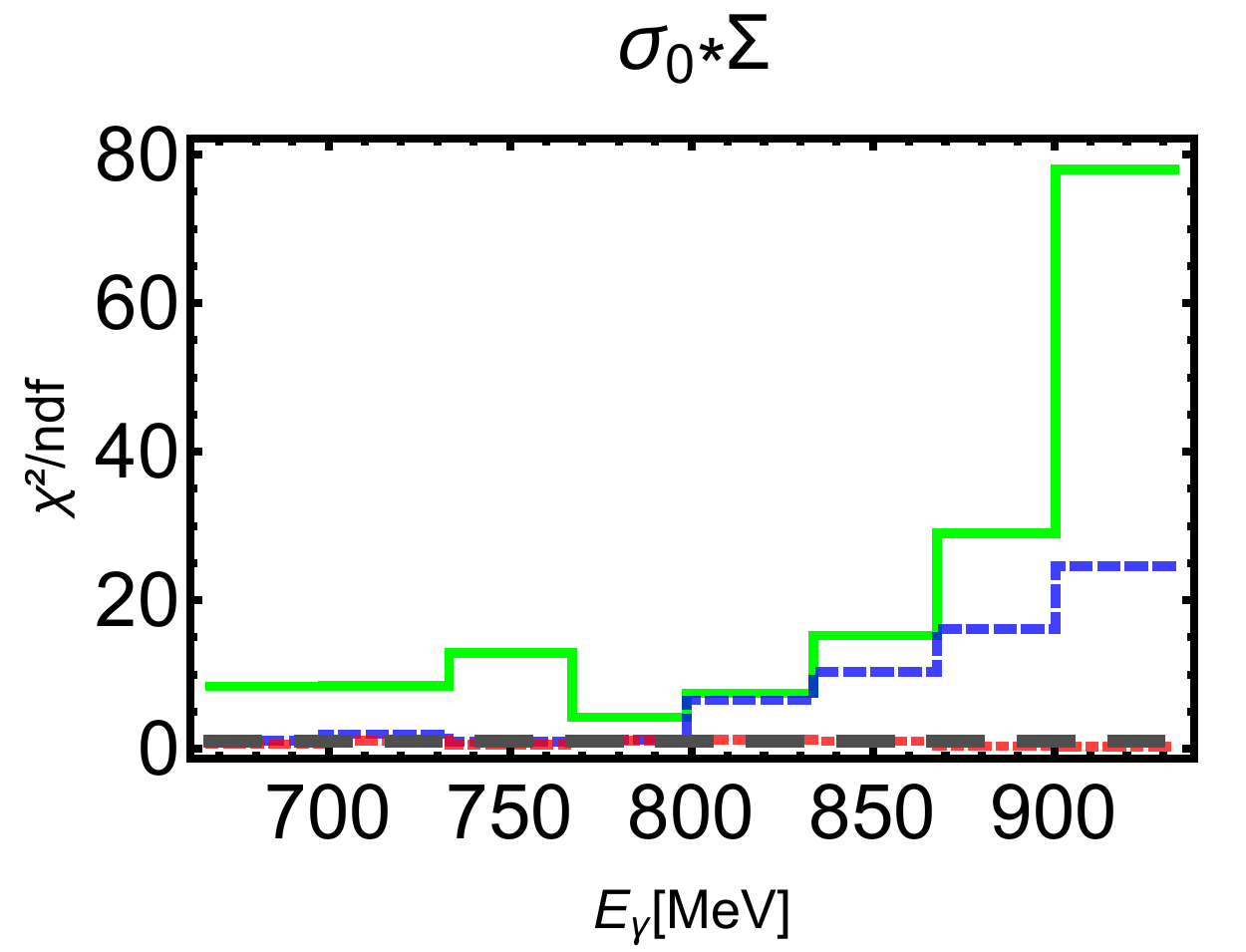}
 \end{overpic} \\
 \begin{overpic}[width=0.45\textwidth]{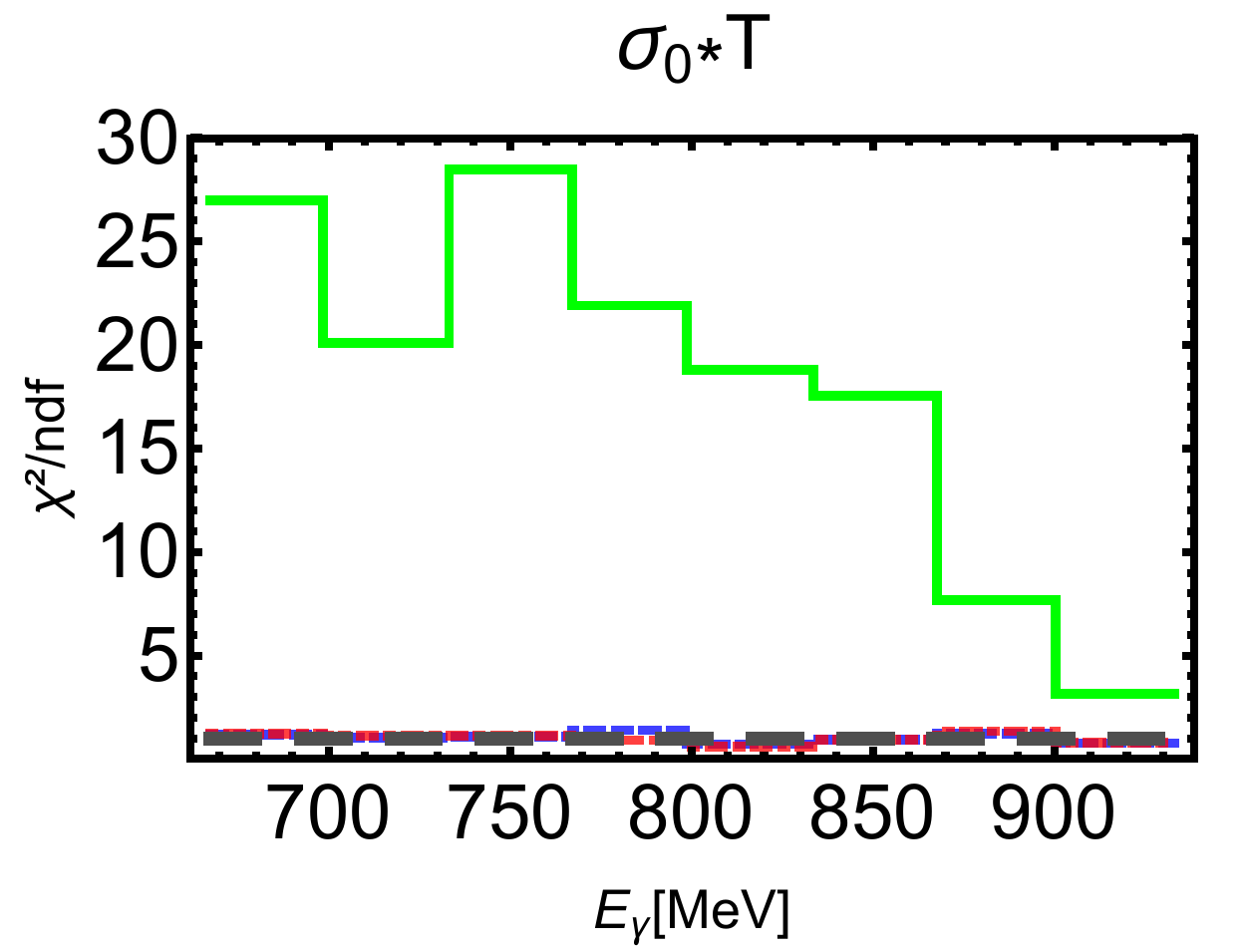}
 \end{overpic}
 \begin{overpic}[width=0.45\textwidth]{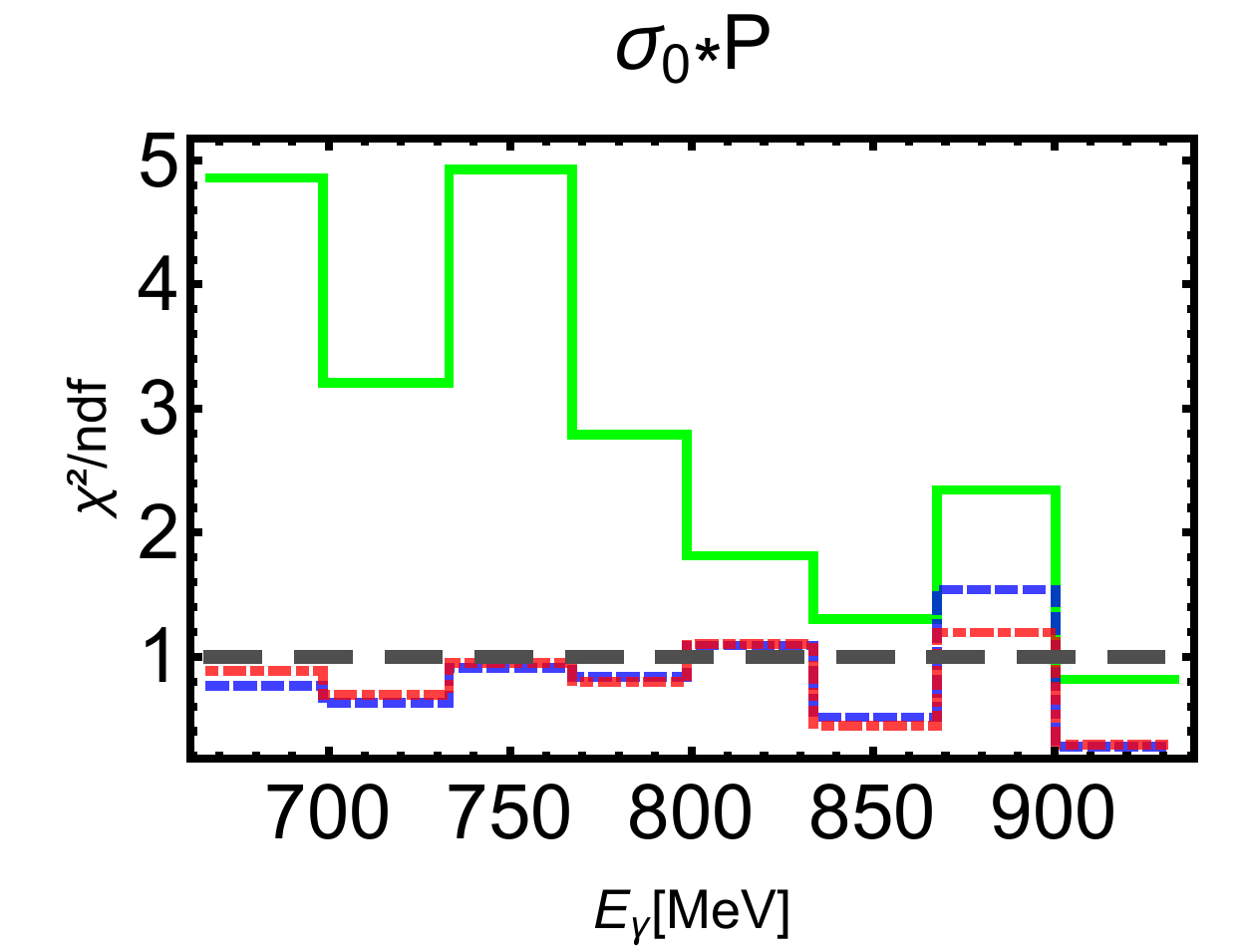}
 \end{overpic} \\
 \begin{overpic}[width=0.45\textwidth]{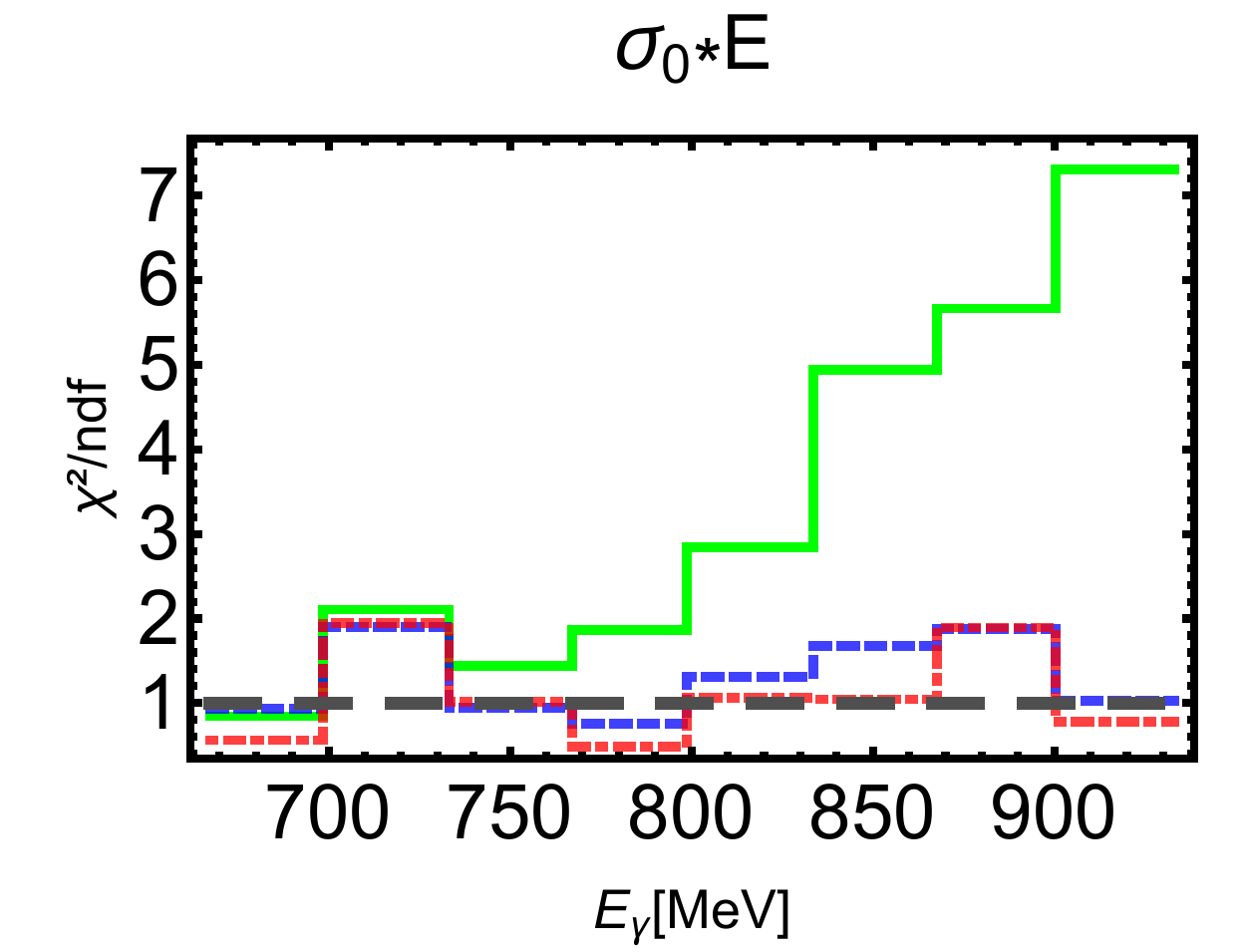}
 \end{overpic}
 \begin{overpic}[width=0.45\textwidth]{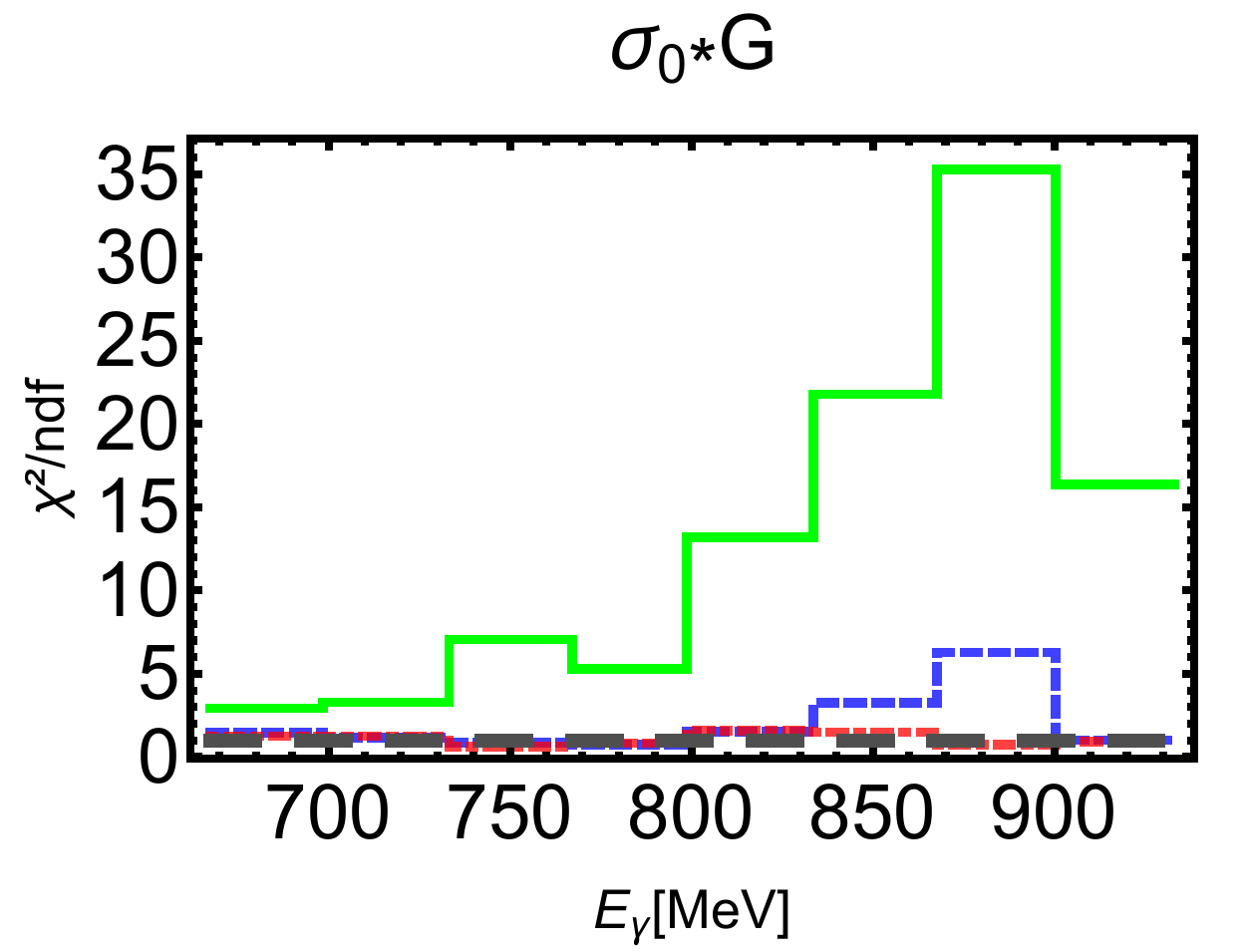}
 \end{overpic} \\
 \begin{overpic}[width=0.45\textwidth]{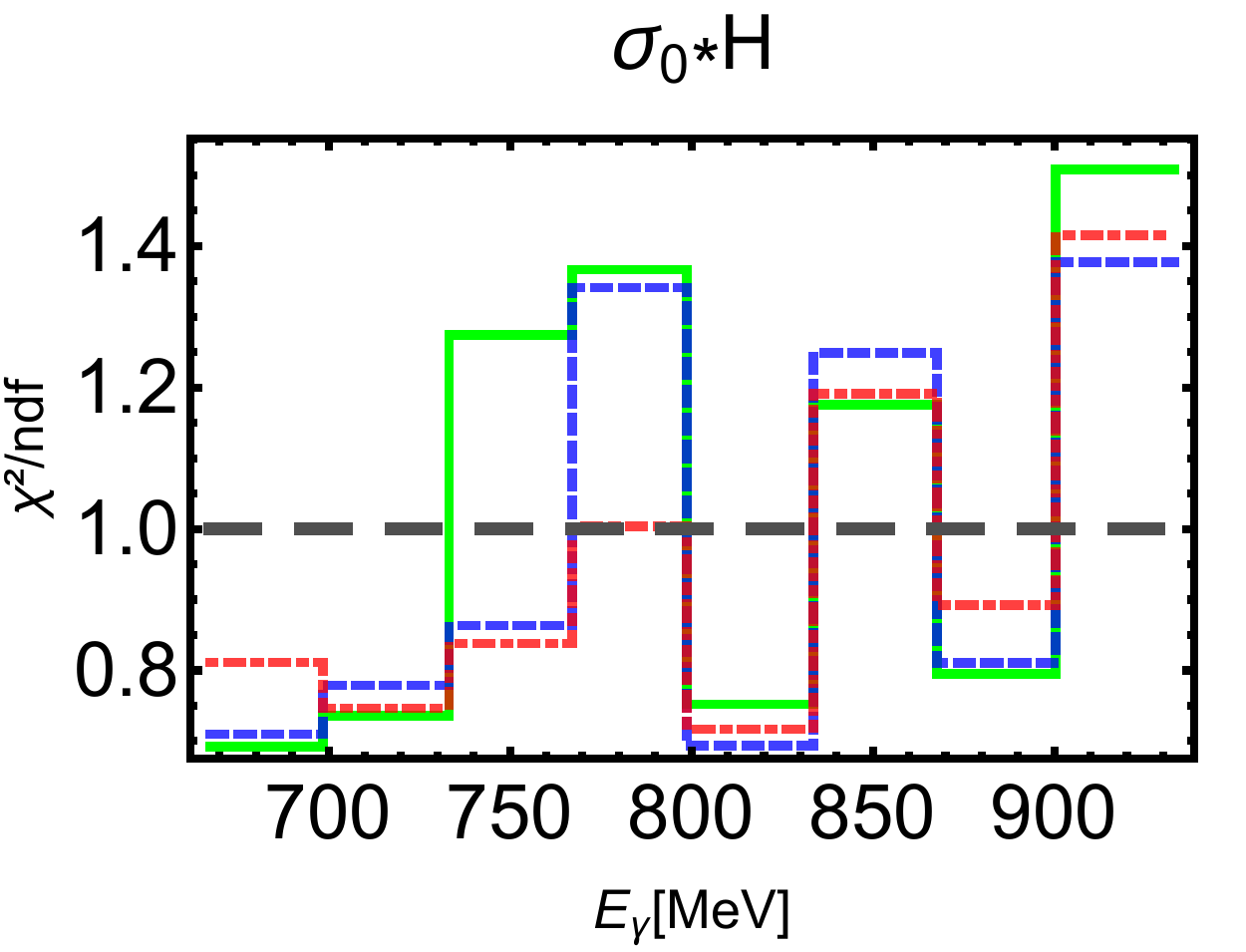}
 \end{overpic}
  \caption[The $\chi^{2}/\mathrm{ndf}$ resulting from fits to the angular distributions of the differential cross section and the six profile functions in the $2^{\mathrm{nd}}$ resonance region, plotted vs. energy.]{The figures show the $\chi^{2}/\mathrm{ndf}$ plotted vs. energy, resulting from fits of the parametrizations (\ref{eq:DCSAngDistChapter4}) to (\ref{eq:PAngDistChapter4}) and (\ref{eq:EAngDistChapter4}) to (\ref{eq:HAngDistChapter4}) to the angular distributions of the differential cross section and the six profile functions in the second resonance region.  Truncations at $\ell_{\mathrm{max}} = 1$ (green solid line), $\ell_{\mathrm{max}} = 2$ (blue dashed line) and $\ell_{\mathrm{max}} = 3$ (red dash-dotted line) are included. The optimal value $\chi^{2}/\mathrm{ndf} = 1$ is indicated by a grey dashed horizontal line.}
 \label{fig:2ndResRegionChisquareLmaxPlots}
\end{figure}

\clearpage

\begin{figure}[h]
 \centering
 \begin{overpic}[width=0.325\textwidth]{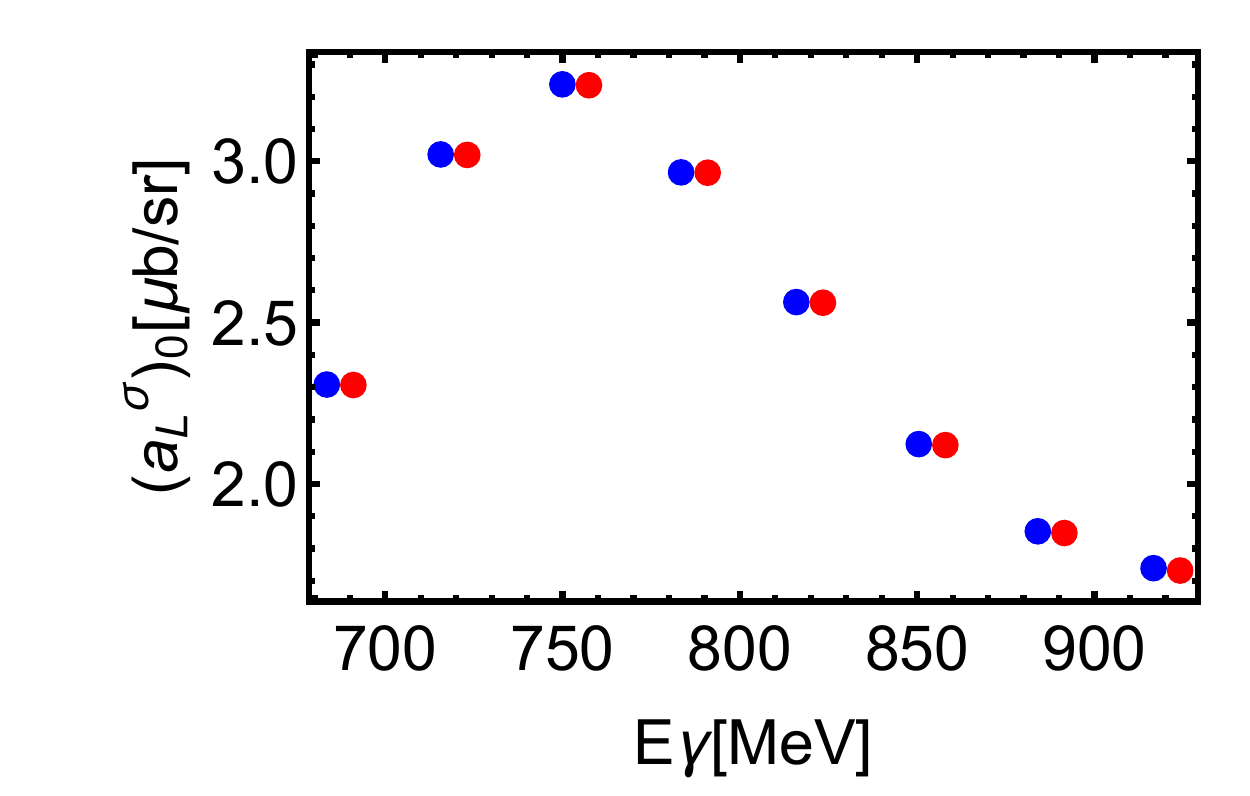}
 \end{overpic}
 \begin{overpic}[width=0.325\textwidth]{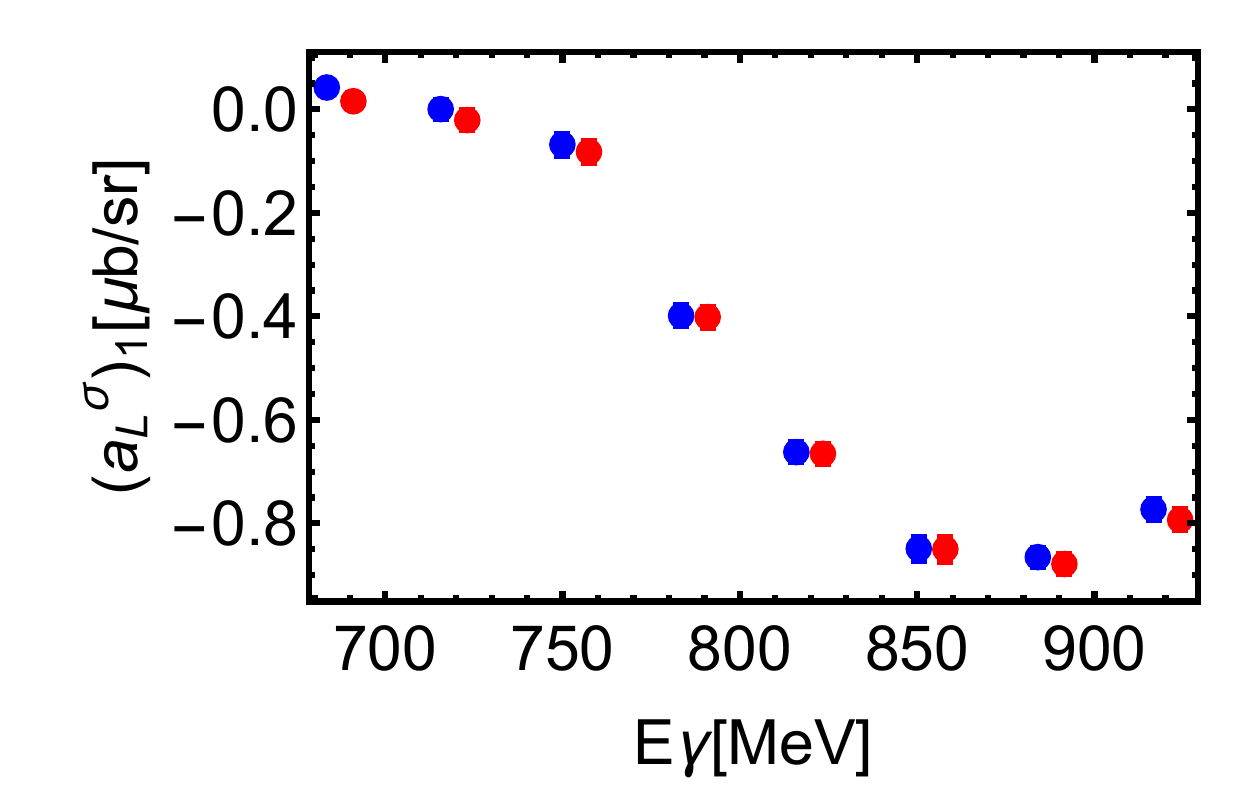}
 \end{overpic}
 \begin{overpic}[width=0.325\textwidth]{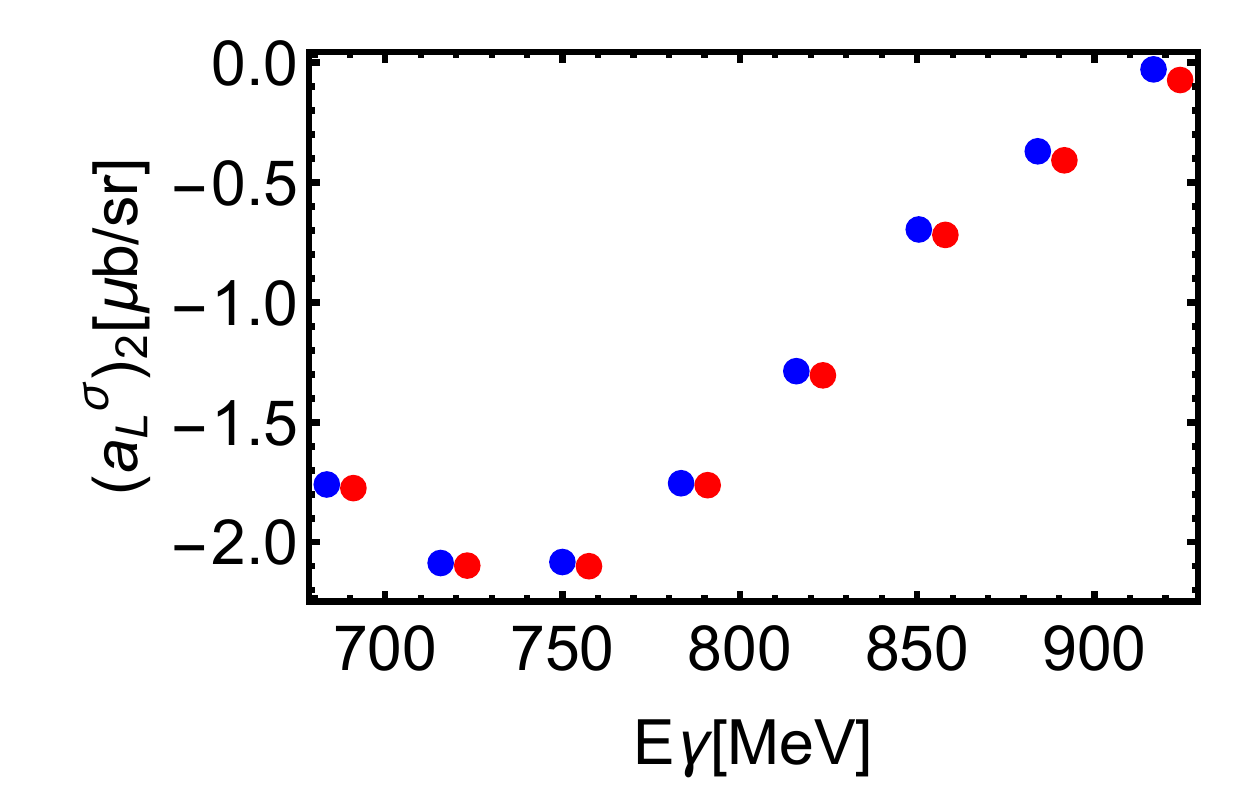}
 \end{overpic} \\
 \begin{overpic}[width=0.325\textwidth]{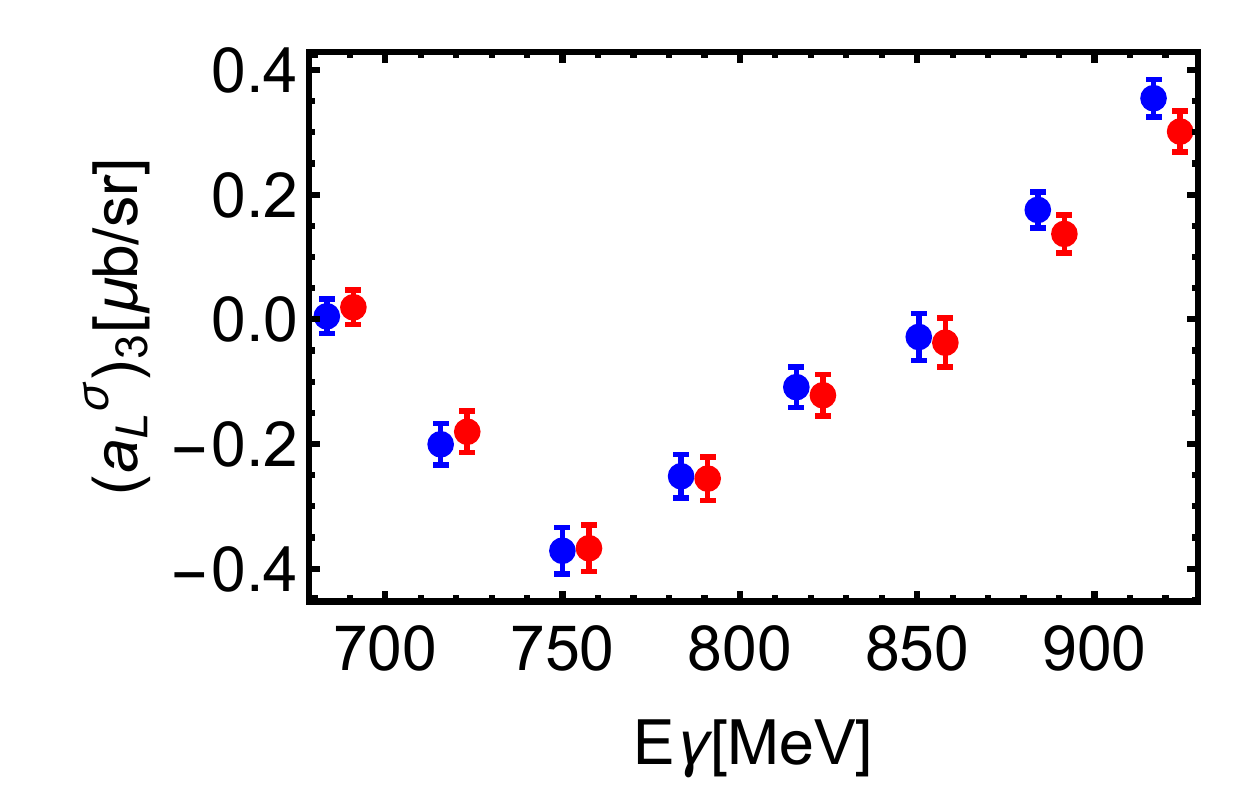}
 \end{overpic}
 \begin{overpic}[width=0.325\textwidth]{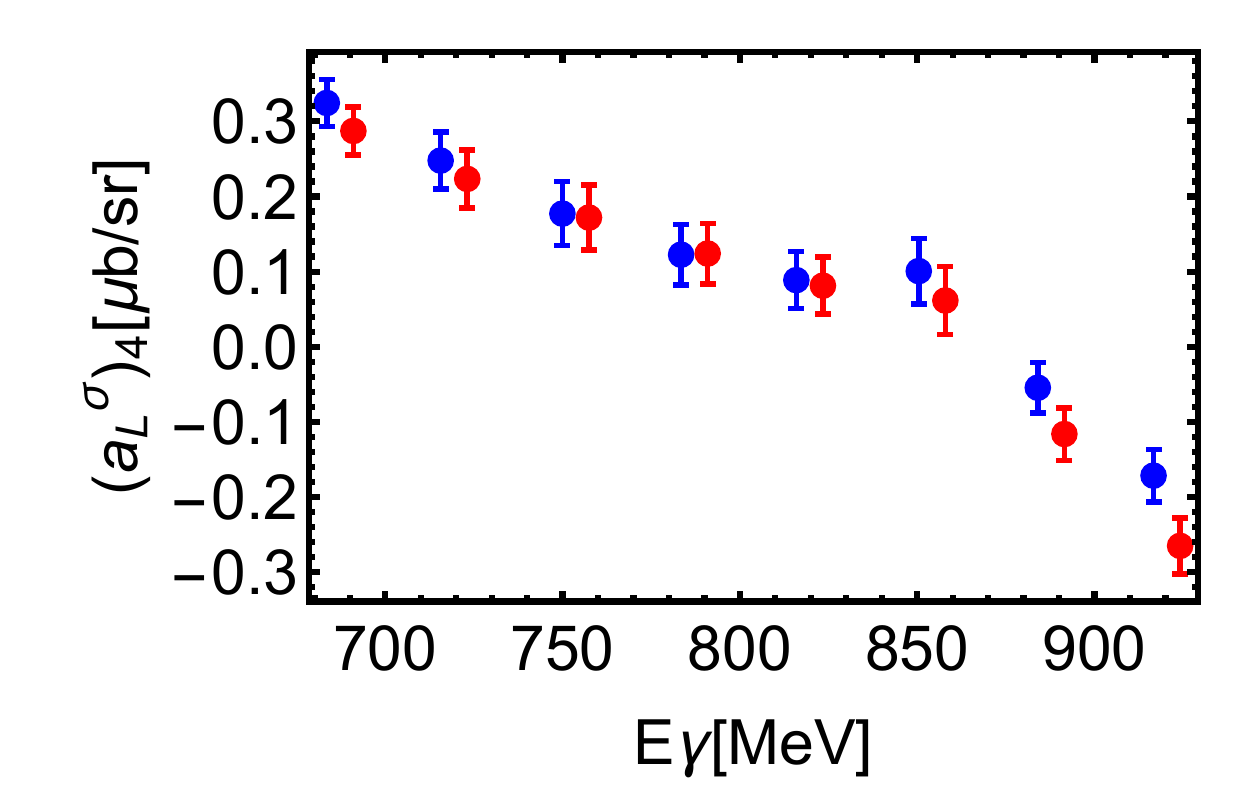}
 \end{overpic}
 \begin{overpic}[width=0.325\textwidth]{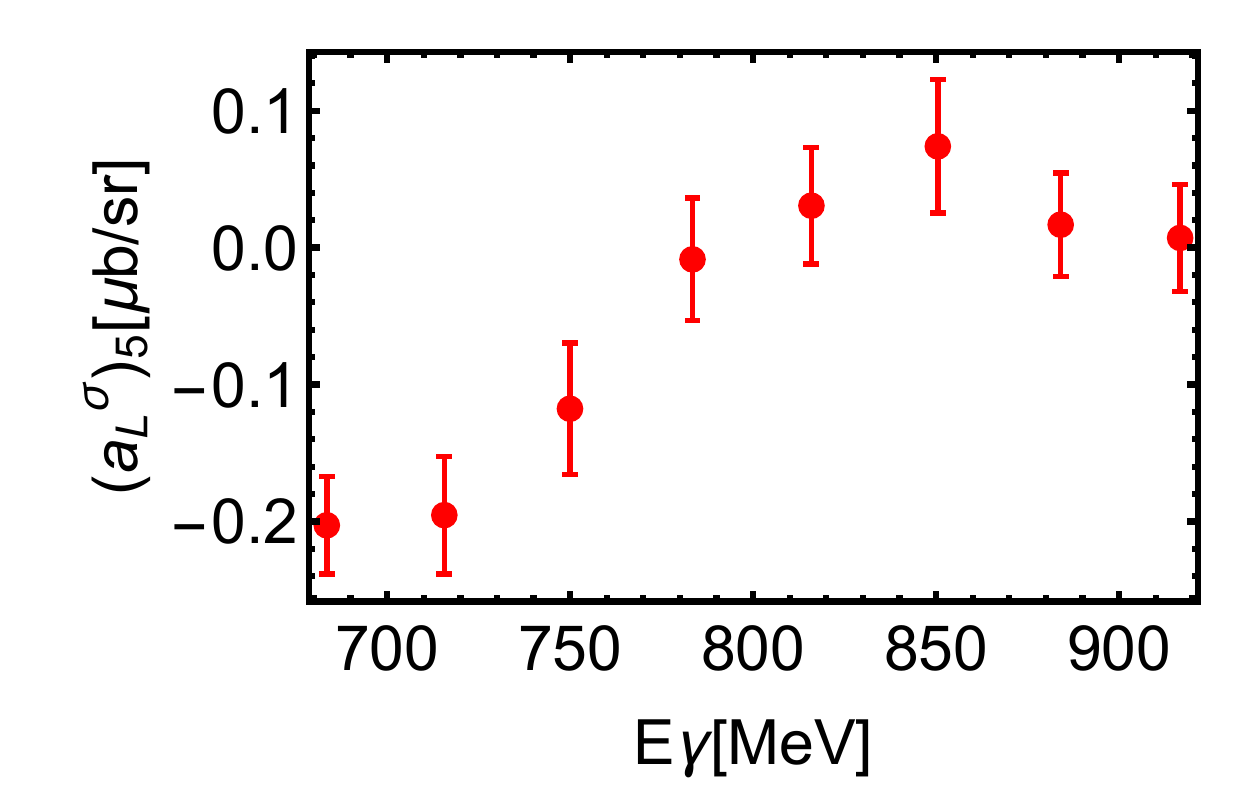}
 \end{overpic} \\

 \begin{overpic}[width=0.325\textwidth]{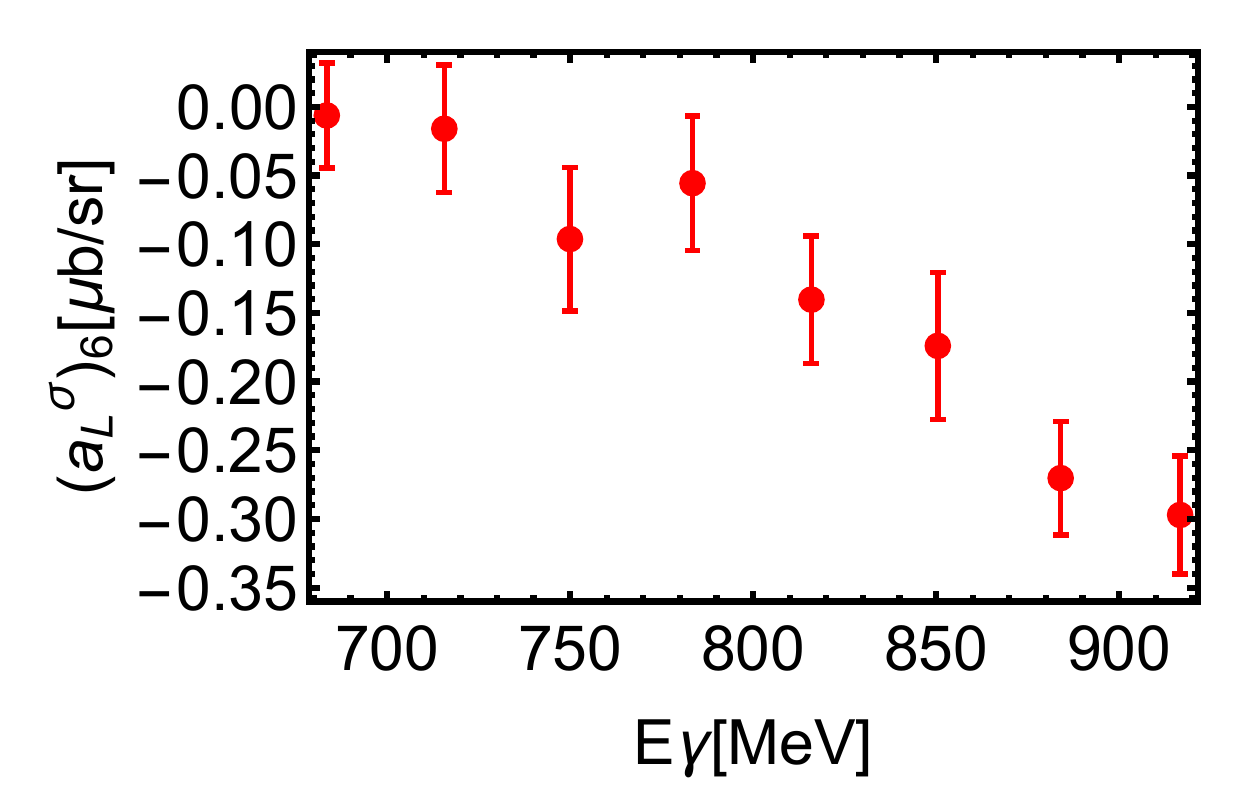}
 \end{overpic} \\
  \begin{overpic}[width=0.325\textwidth]{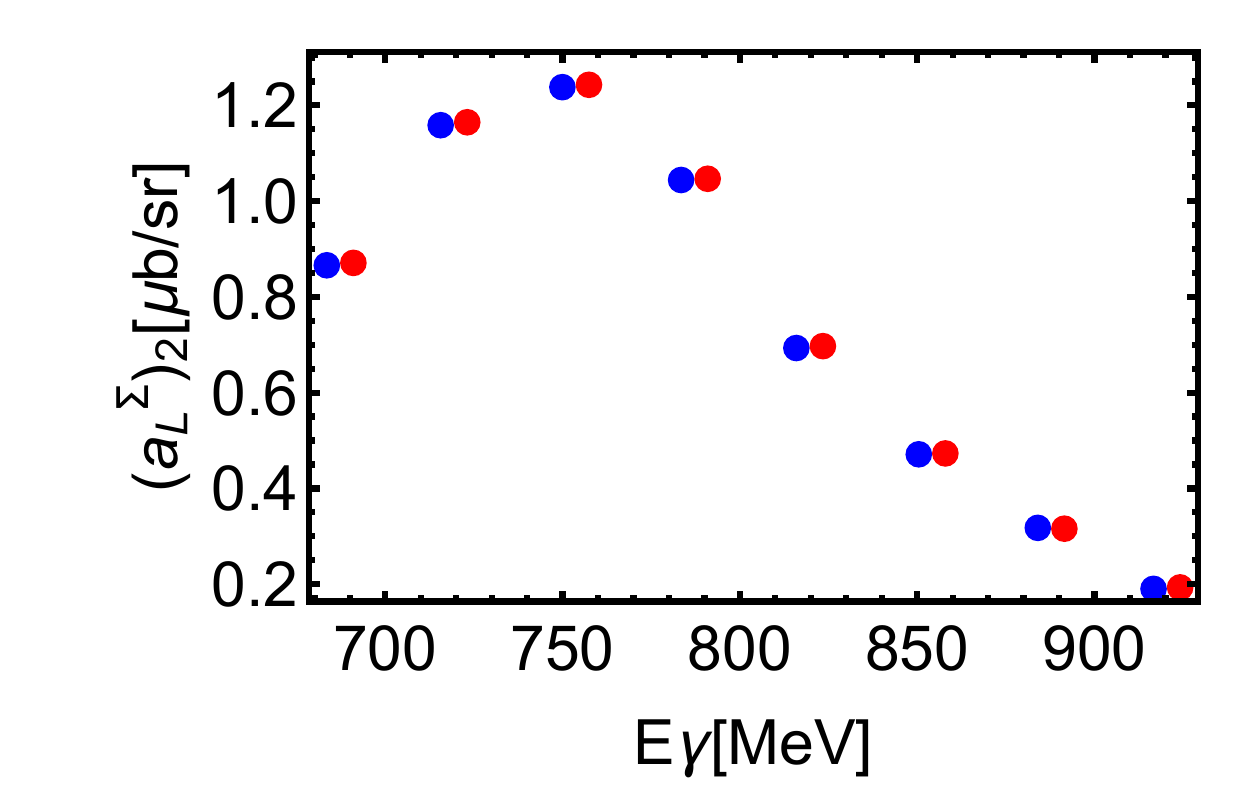}
 \end{overpic}
 \begin{overpic}[width=0.325\textwidth]{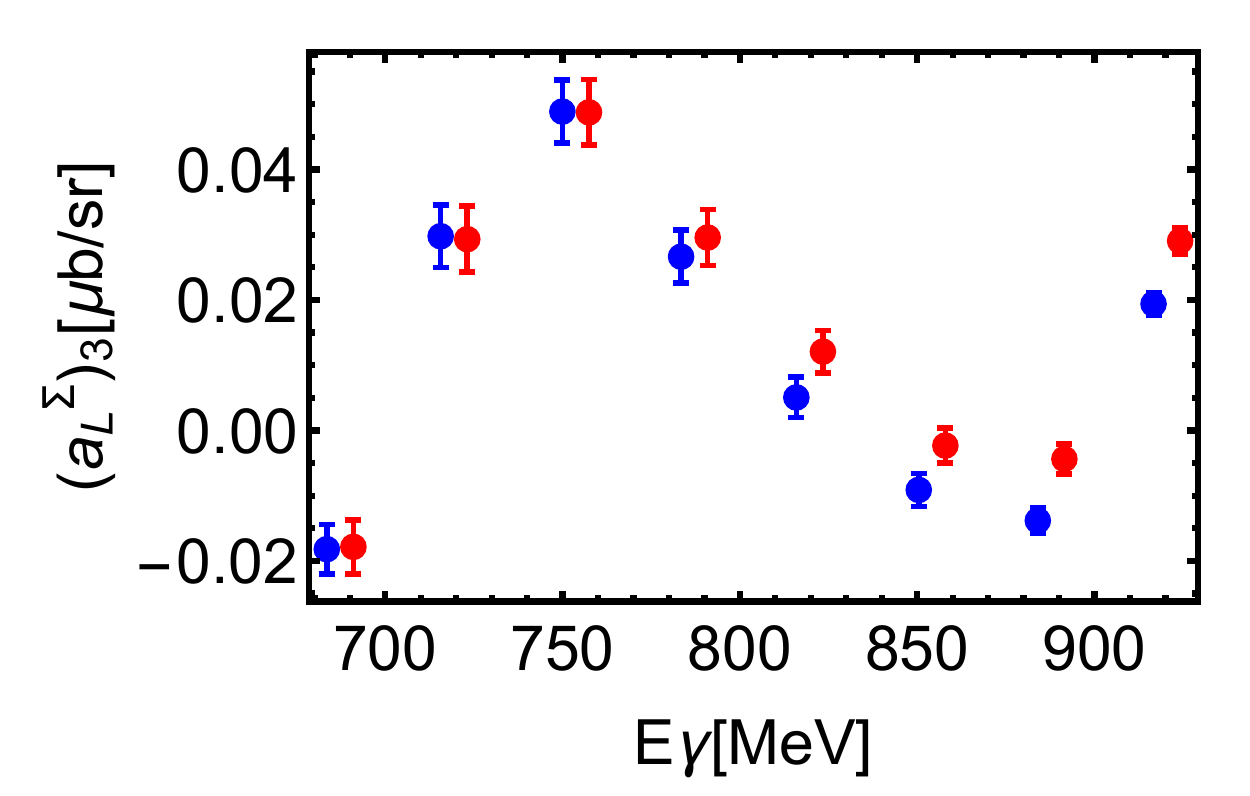}
 \end{overpic}
  \begin{overpic}[width=0.325\textwidth]{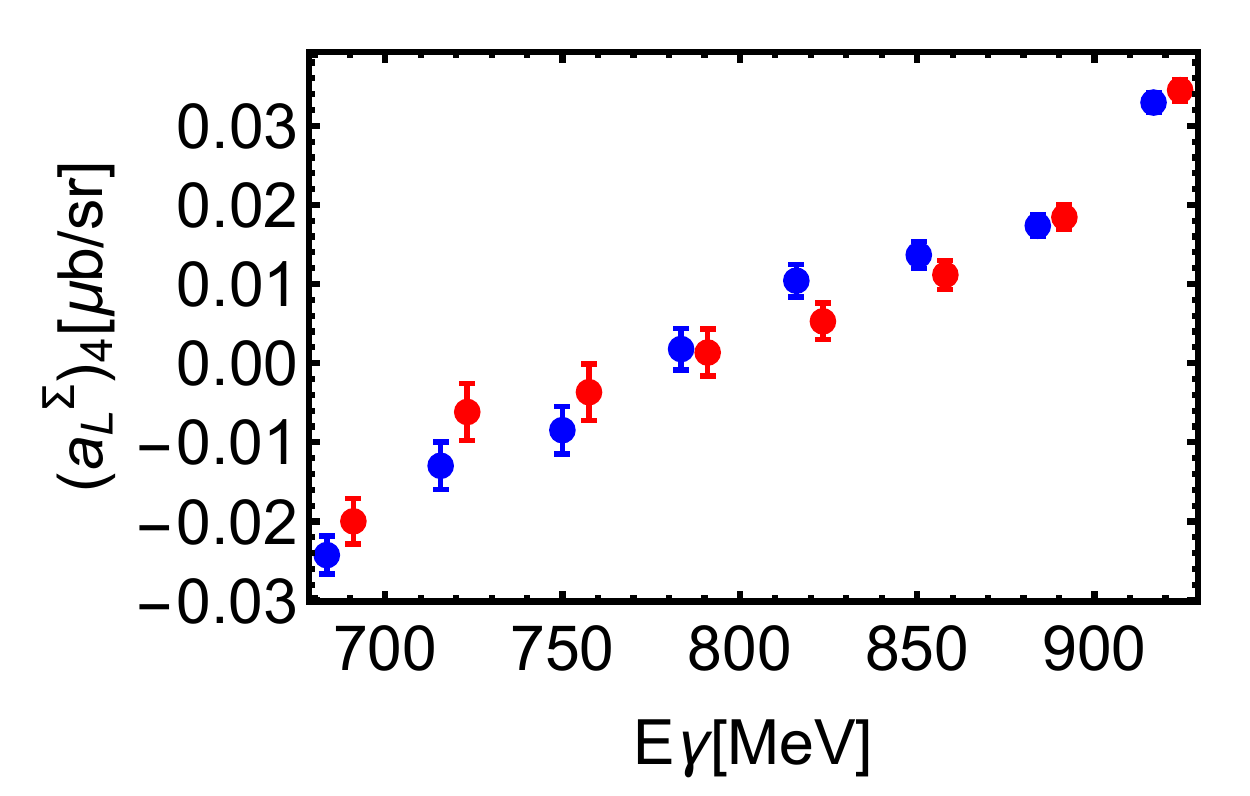}
 \end{overpic} \\
  \begin{overpic}[width=0.325\textwidth]{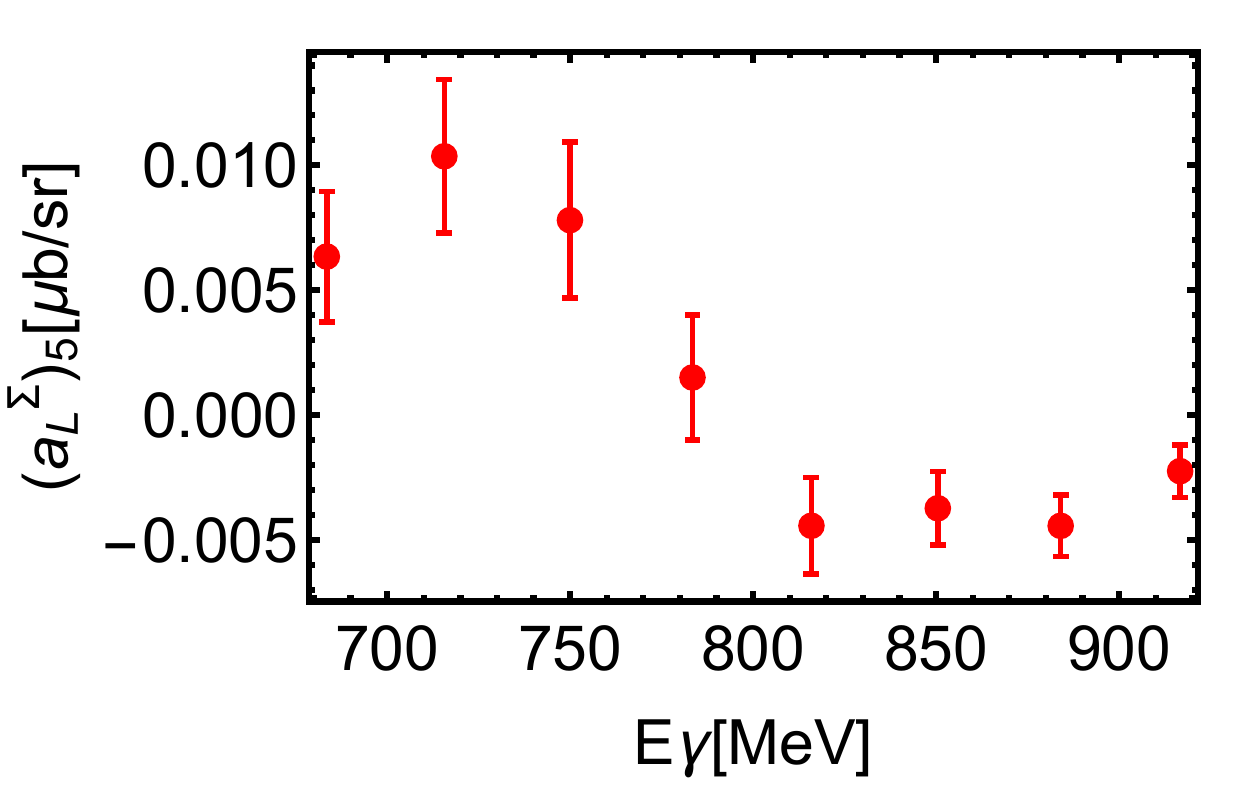}
 \end{overpic}
 \begin{overpic}[width=0.325\textwidth]{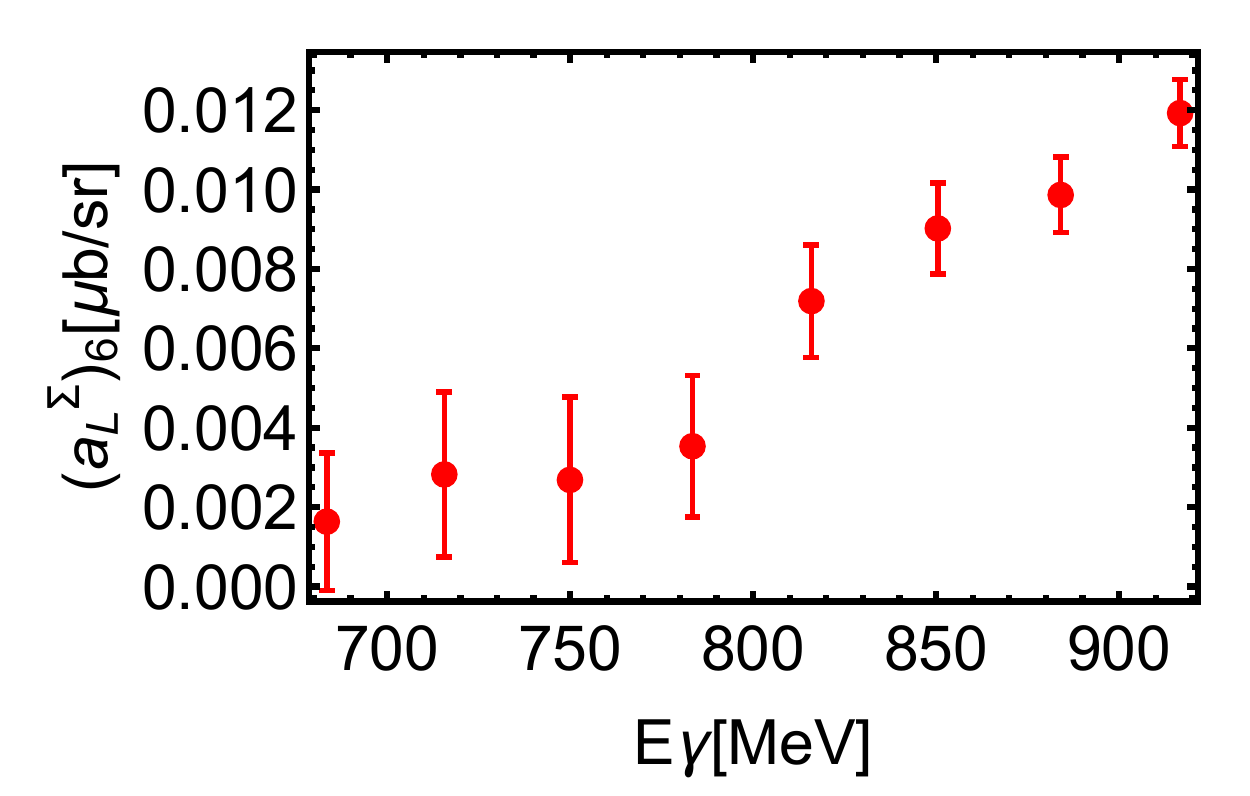}
 \end{overpic} \\
  \begin{overpic}[width=0.325\textwidth]{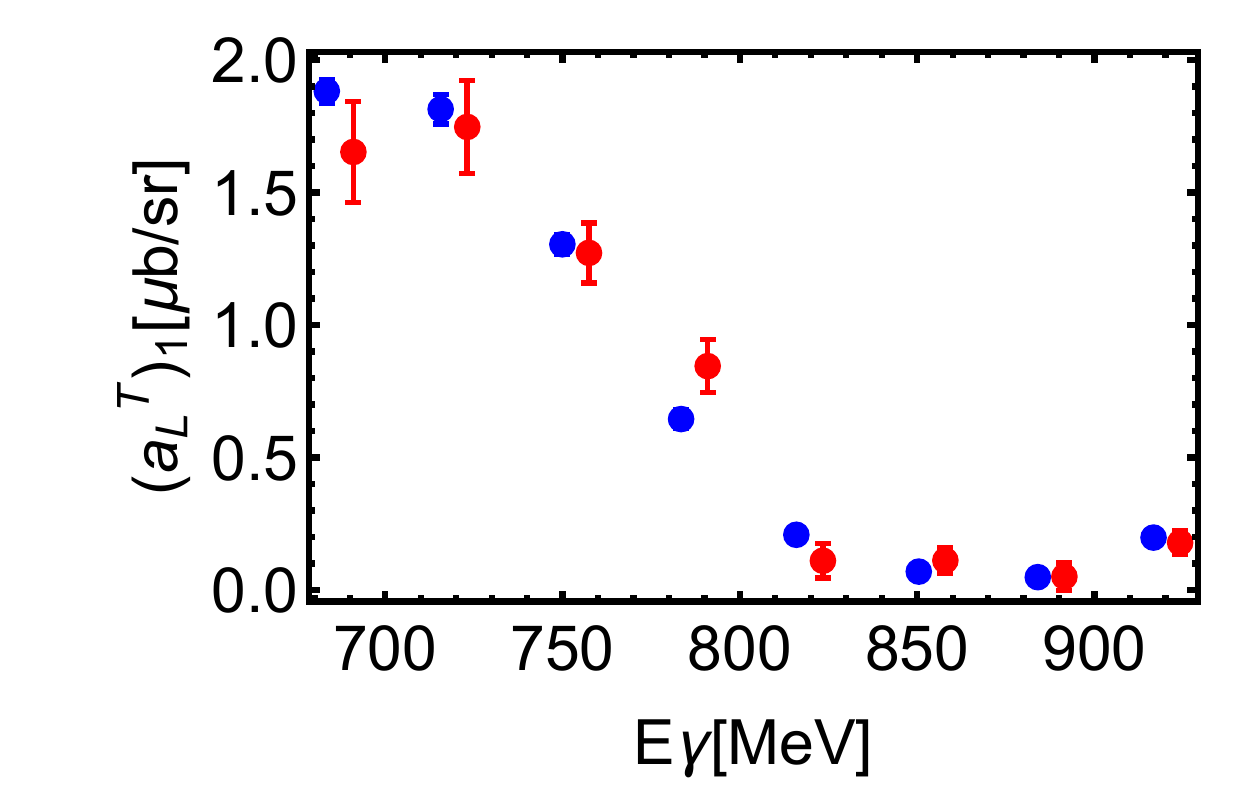}
 \end{overpic}
 \begin{overpic}[width=0.325\textwidth]{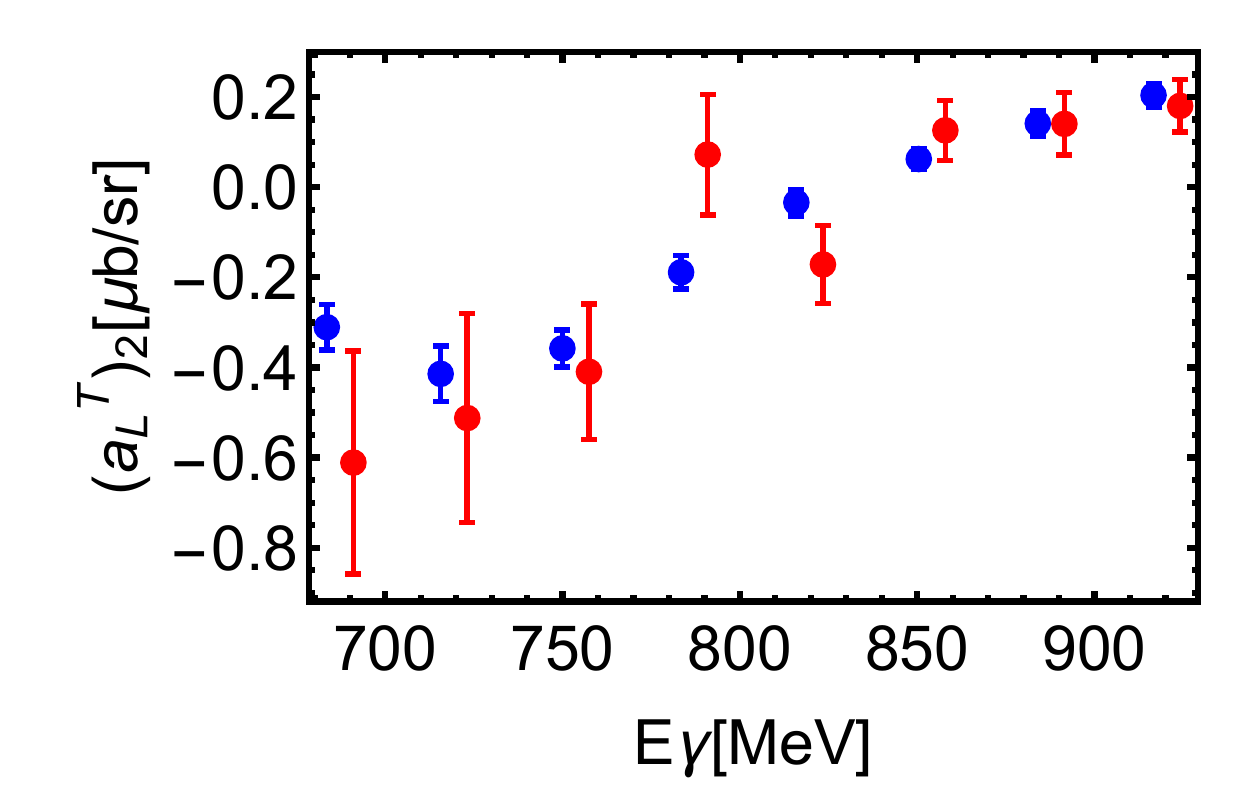}
 \end{overpic}
  \begin{overpic}[width=0.325\textwidth]{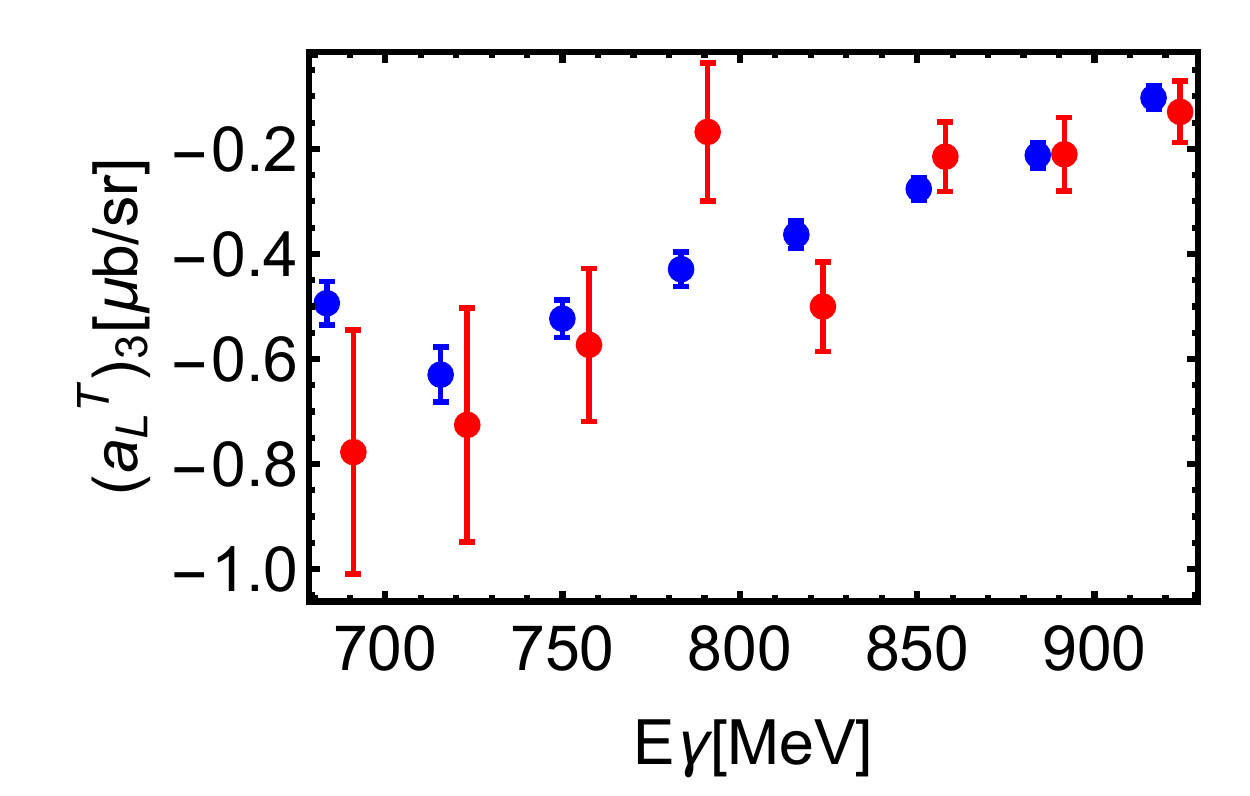}
 \end{overpic} \\
  \begin{overpic}[width=0.325\textwidth]{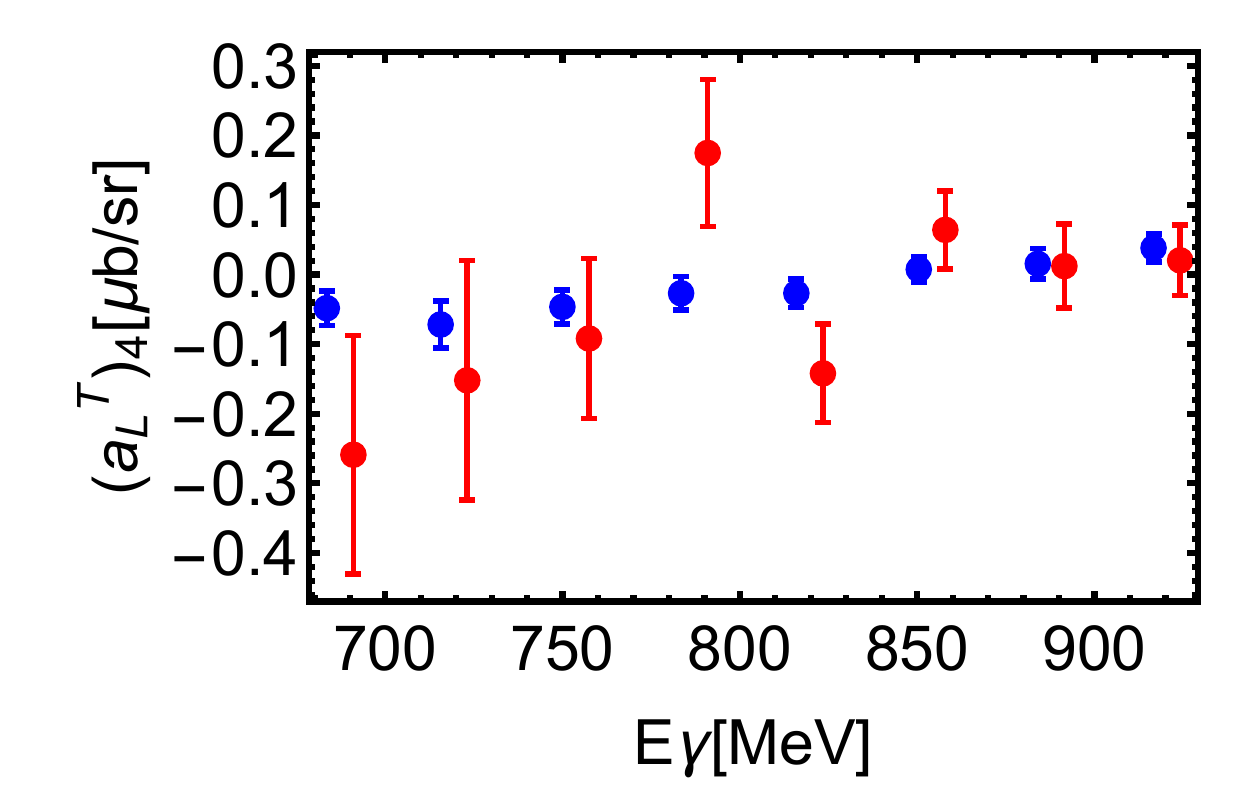}
 \end{overpic}
  \begin{overpic}[width=0.325\textwidth]{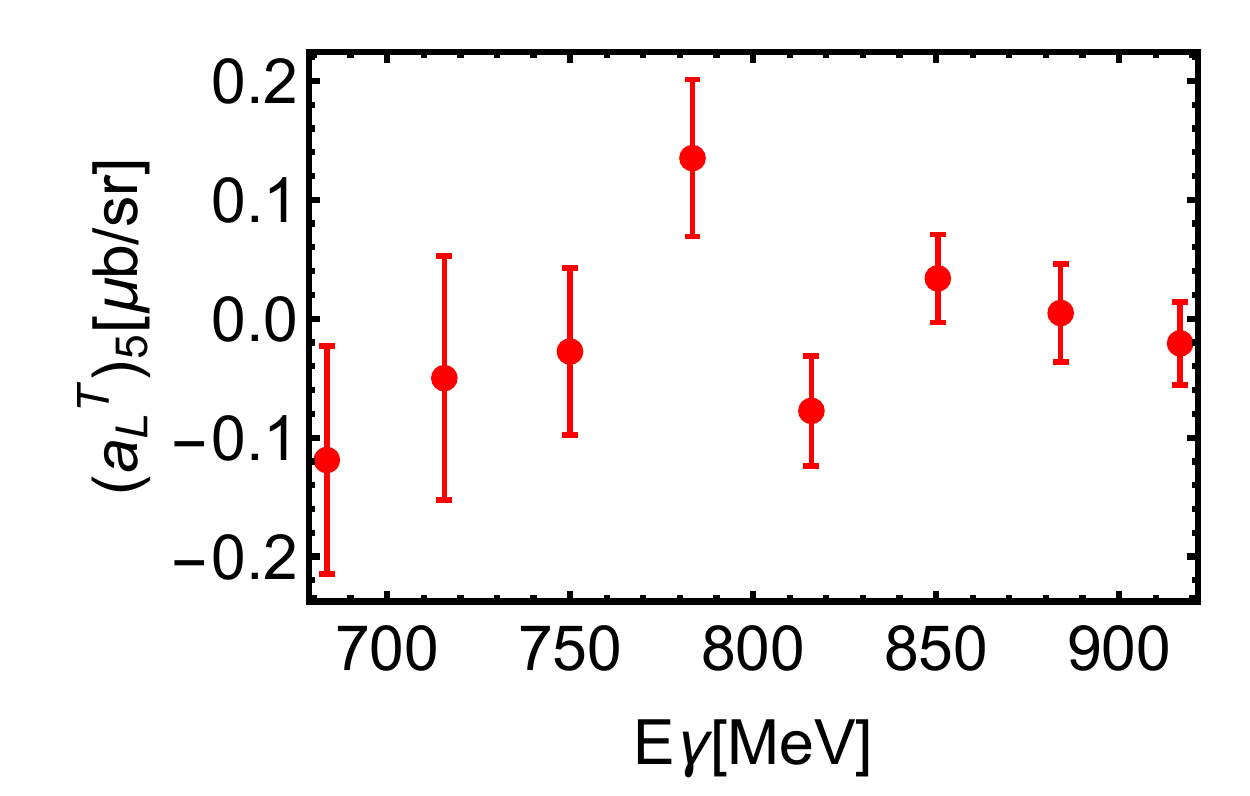}
 \end{overpic}
 \begin{overpic}[width=0.325\textwidth]{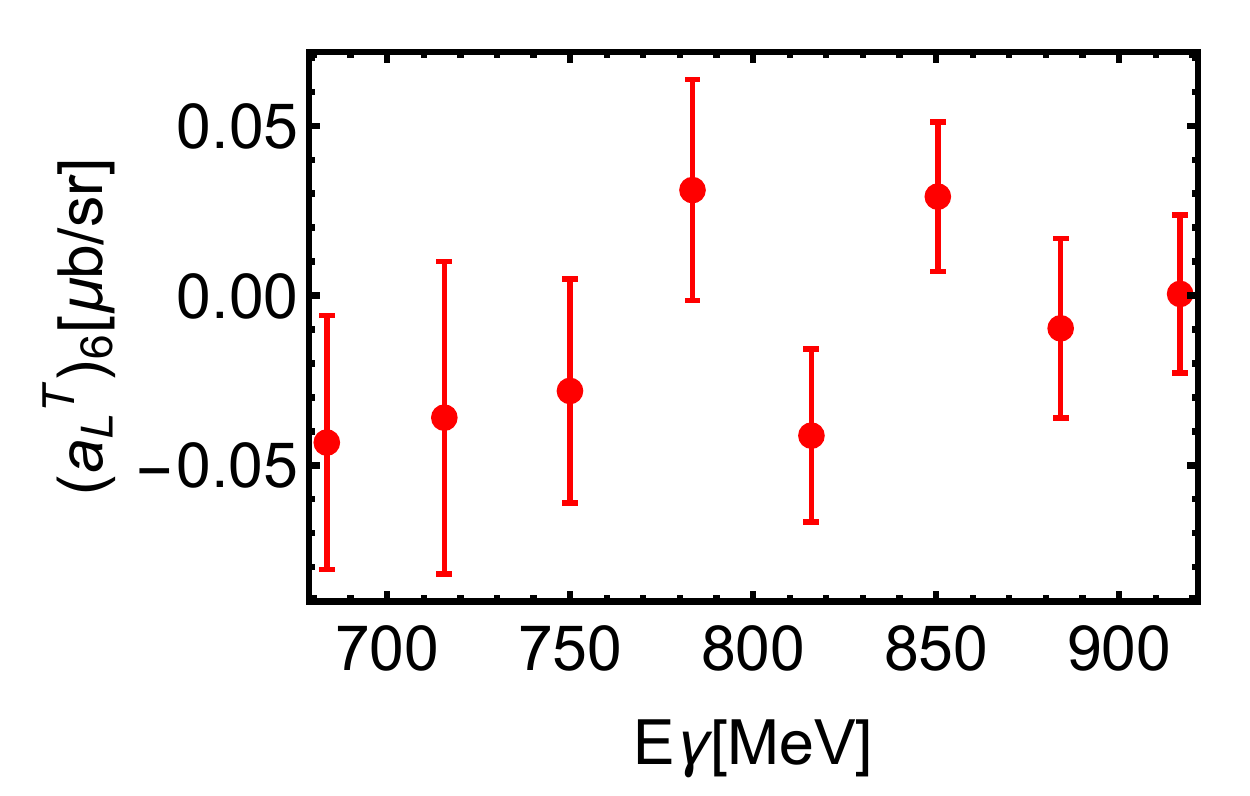}
 \end{overpic} \\
 \caption[Legendre-coefficients for the differential cross section $\sigma_{0}$ and the profile functions $\check{\Sigma}$ and $\check{T}$, extracted by fitting truncations at $\ell_{\mathrm{max}} = 2$ and $\ell_{\mathrm{max}} = 3$, in the $2^{\mathrm{nd}}$ resonance region.]{The pictures show all Legendre-coefficients for $\sigma_{0}$ and the profile functions $\check{\Sigma}$ and $\check{T}$, for truncations at $\ell_{\mathrm{max}} = 2$ (blue points) and $\ell_{\mathrm{max}} = 3$ (red points). In comparison-plots, the results of the $F$-wave truncation have been shifted slightly to higher energies, in order to increase visibility. Errors have been extracted from the fits themselves and are, in this case, not bootstrapped.}
 \label{fig:2ndResRegionFittedLegCoeffsPlotsI}
\end{figure}

\begin{figure}[h]
 \centering
  \begin{overpic}[width=0.325\textwidth]{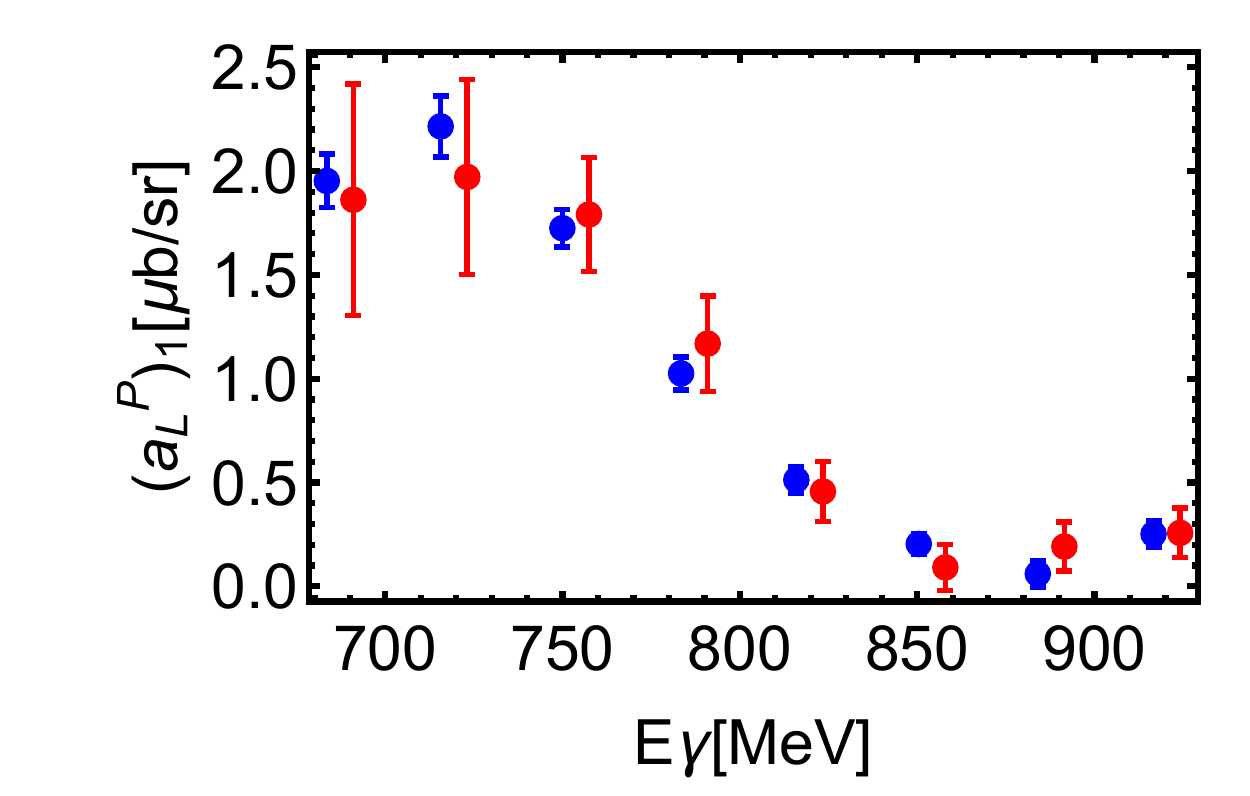}
 \end{overpic}
 \begin{overpic}[width=0.325\textwidth]{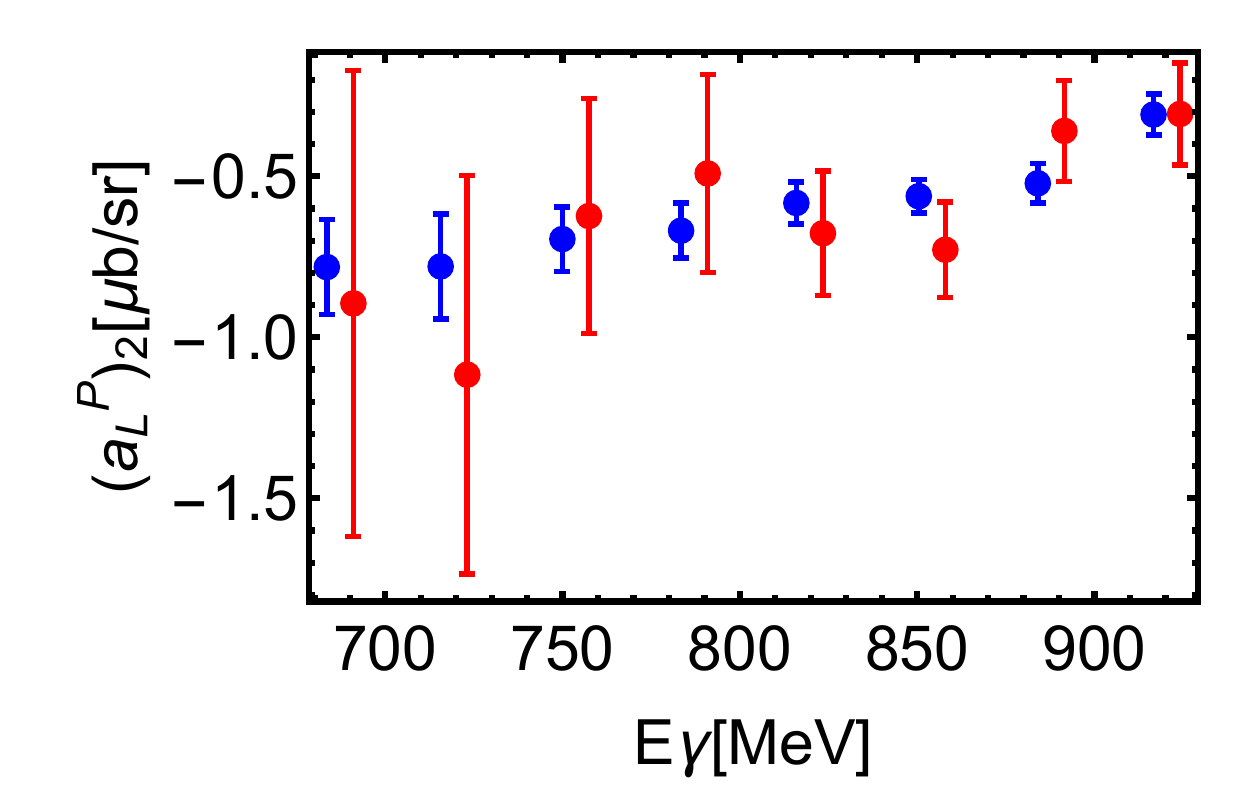}
 \end{overpic}
  \begin{overpic}[width=0.325\textwidth]{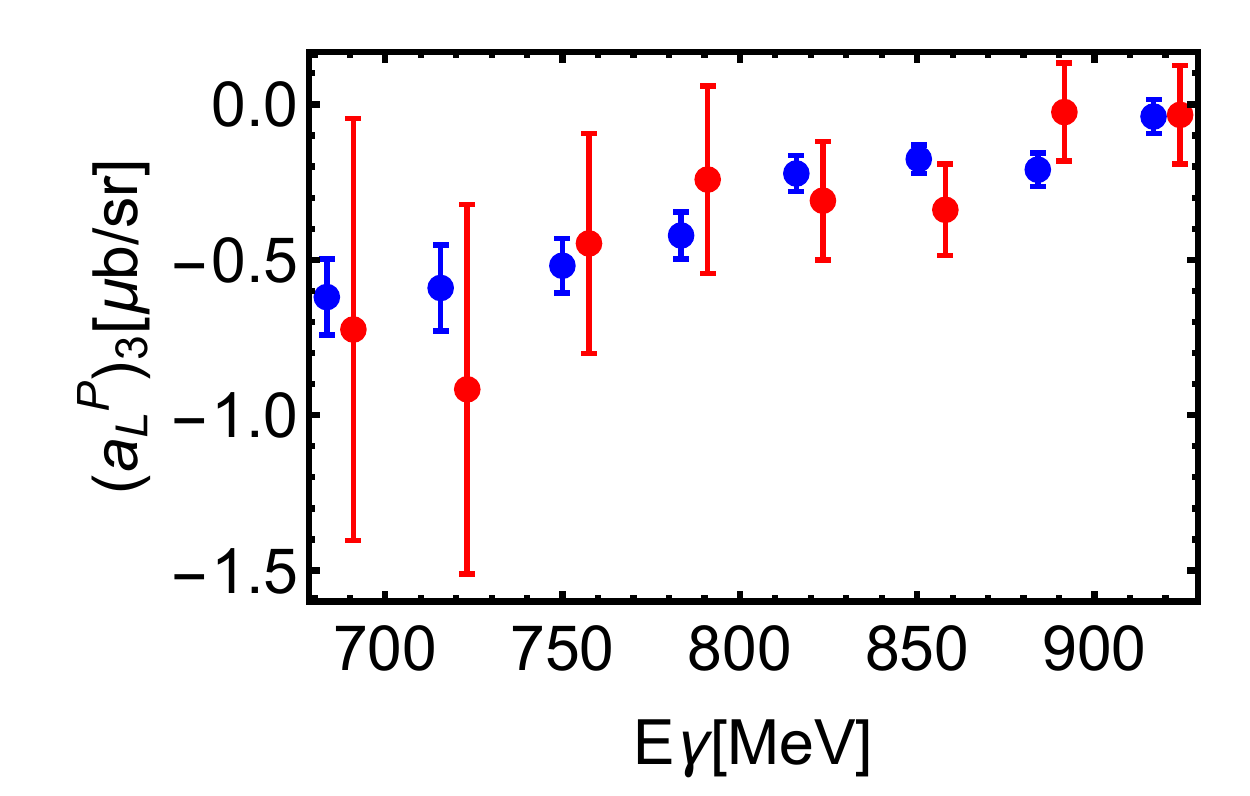}
 \end{overpic} \\
  \begin{overpic}[width=0.325\textwidth]{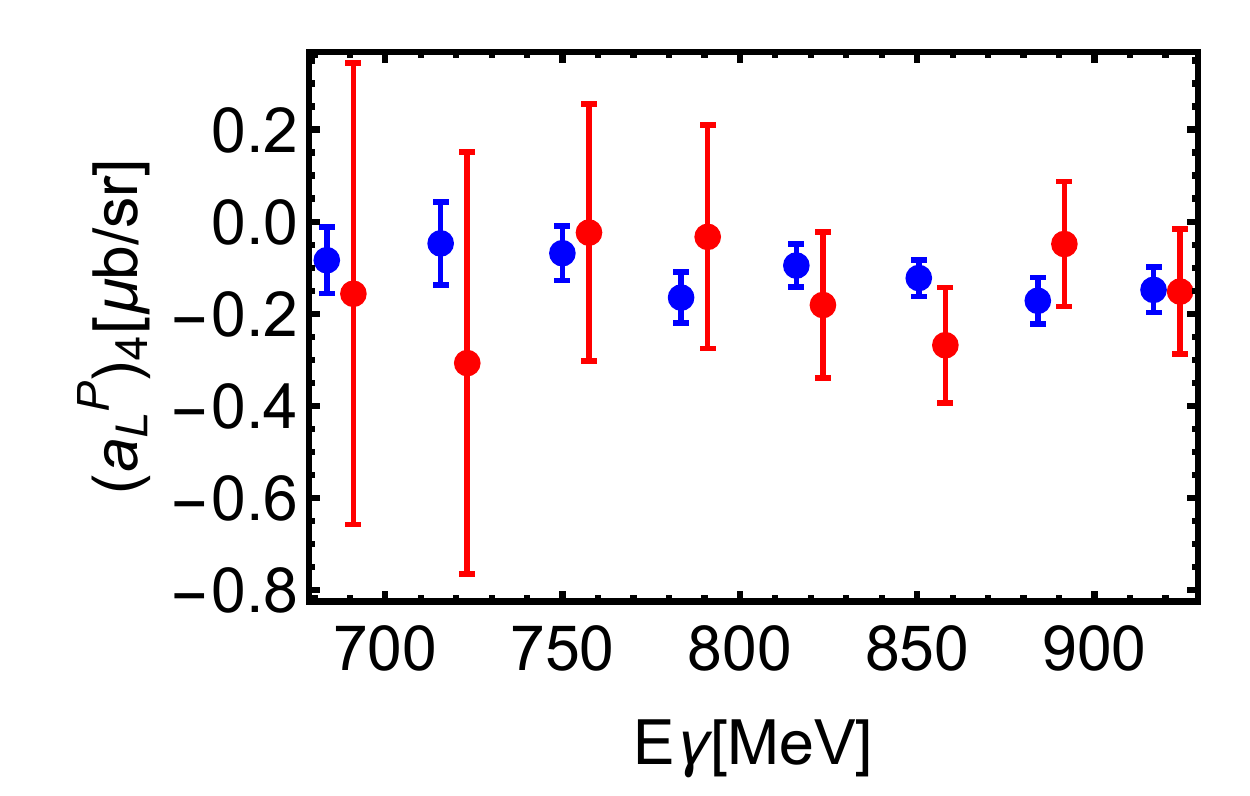}
 \end{overpic}
  \begin{overpic}[width=0.325\textwidth]{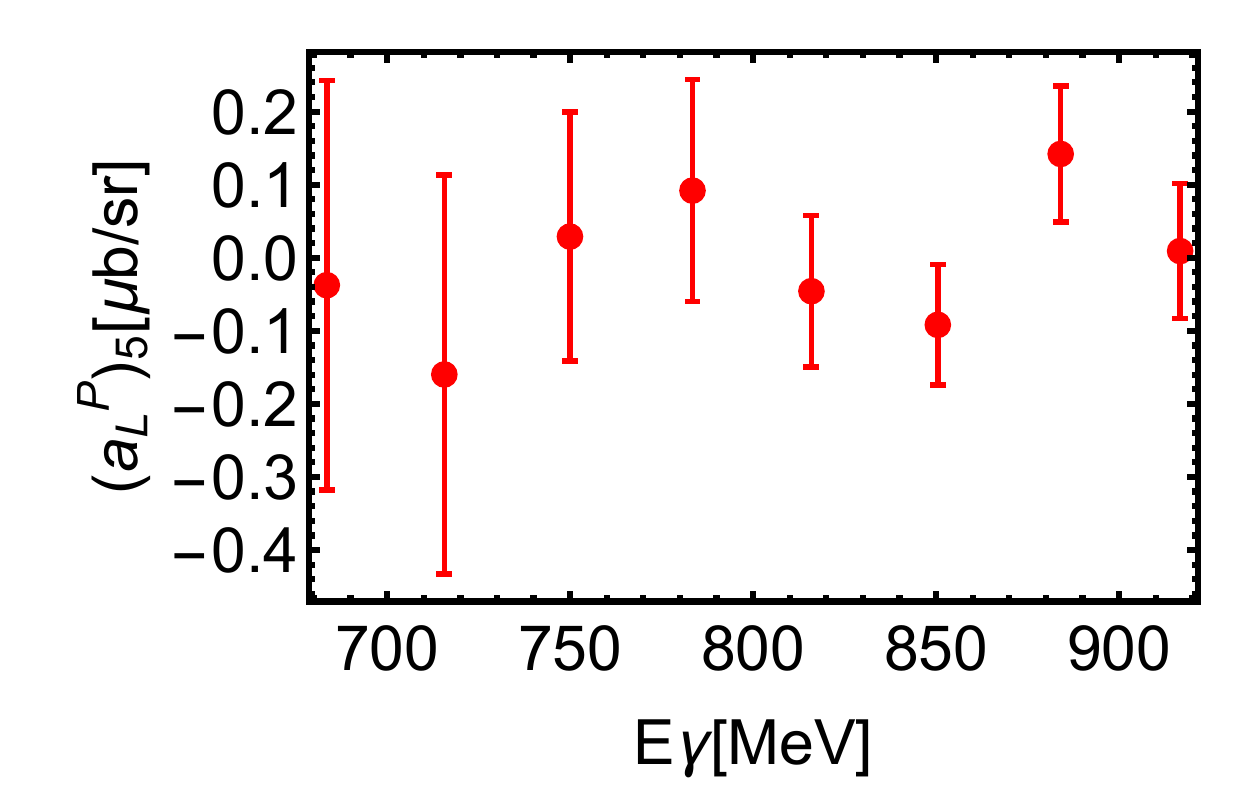}
 \end{overpic}
 \begin{overpic}[width=0.325\textwidth]{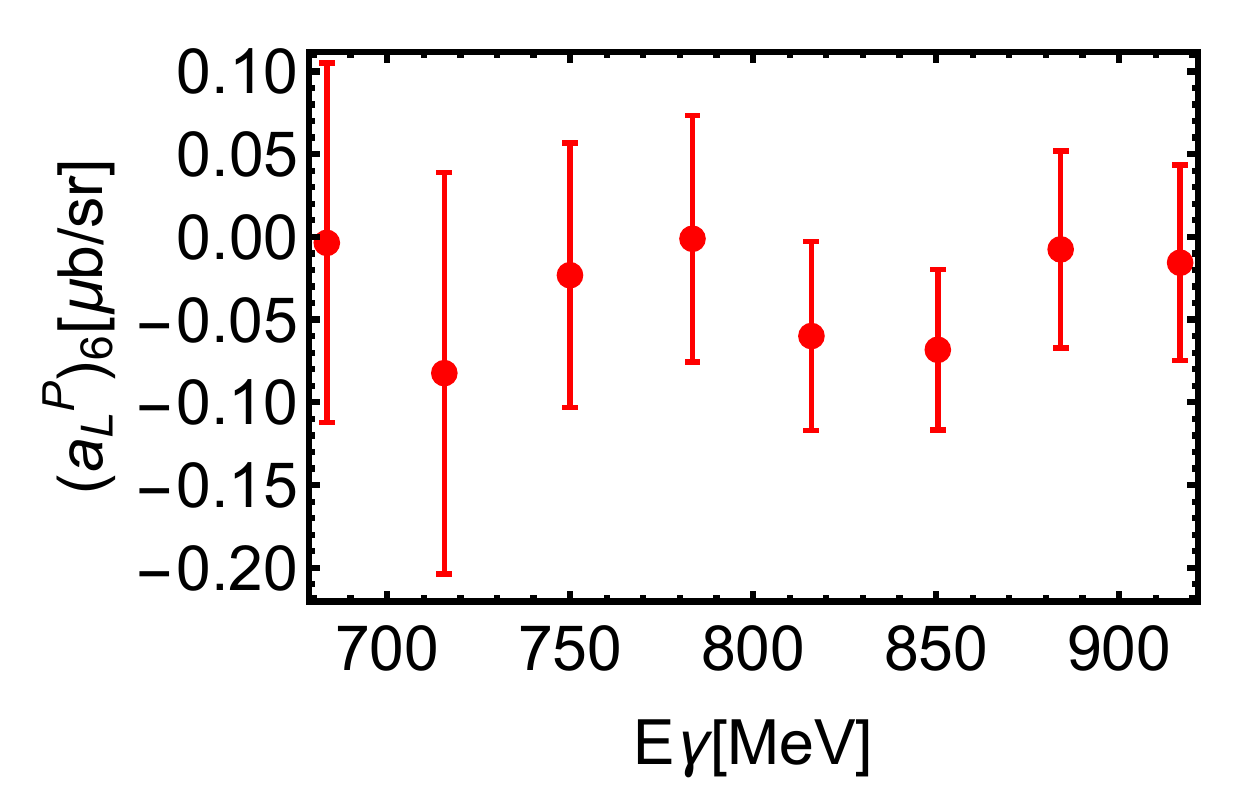}
 \end{overpic} \\
 \begin{overpic}[width=0.325\textwidth]{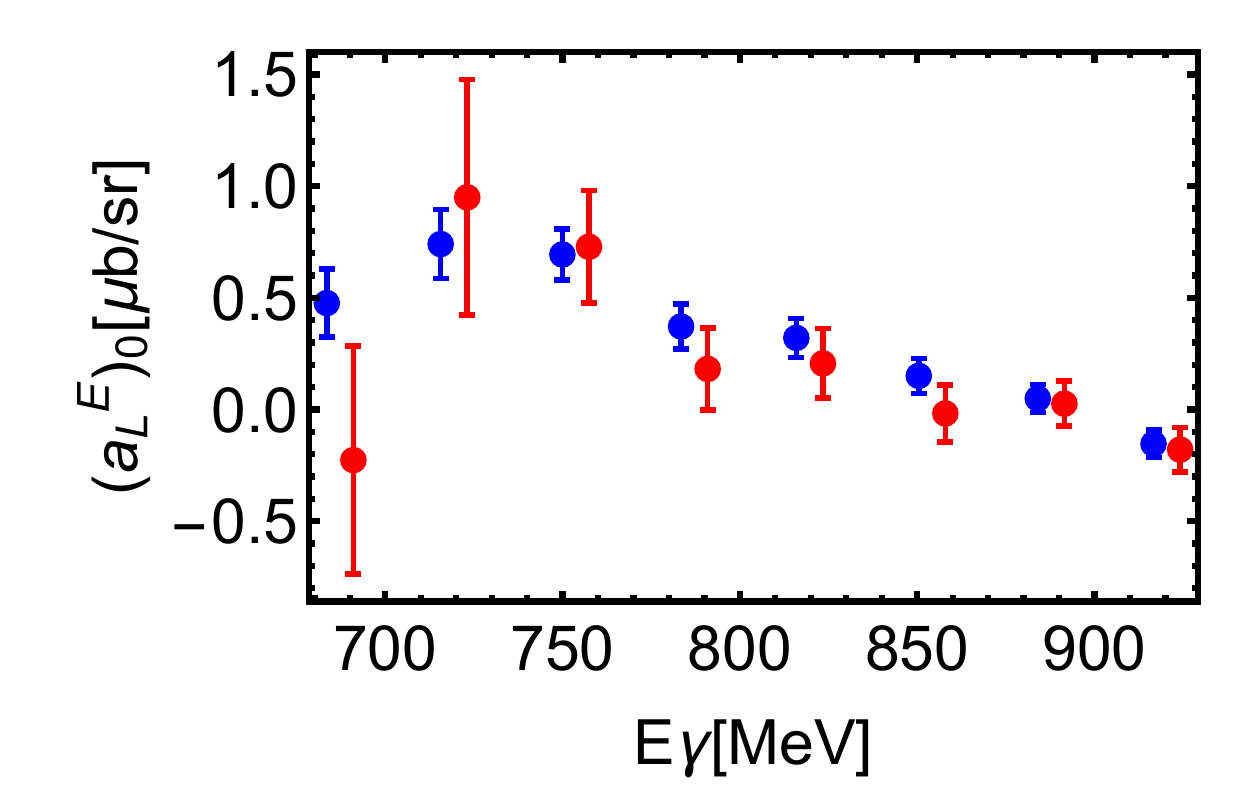}
 \end{overpic}
 \begin{overpic}[width=0.325\textwidth]{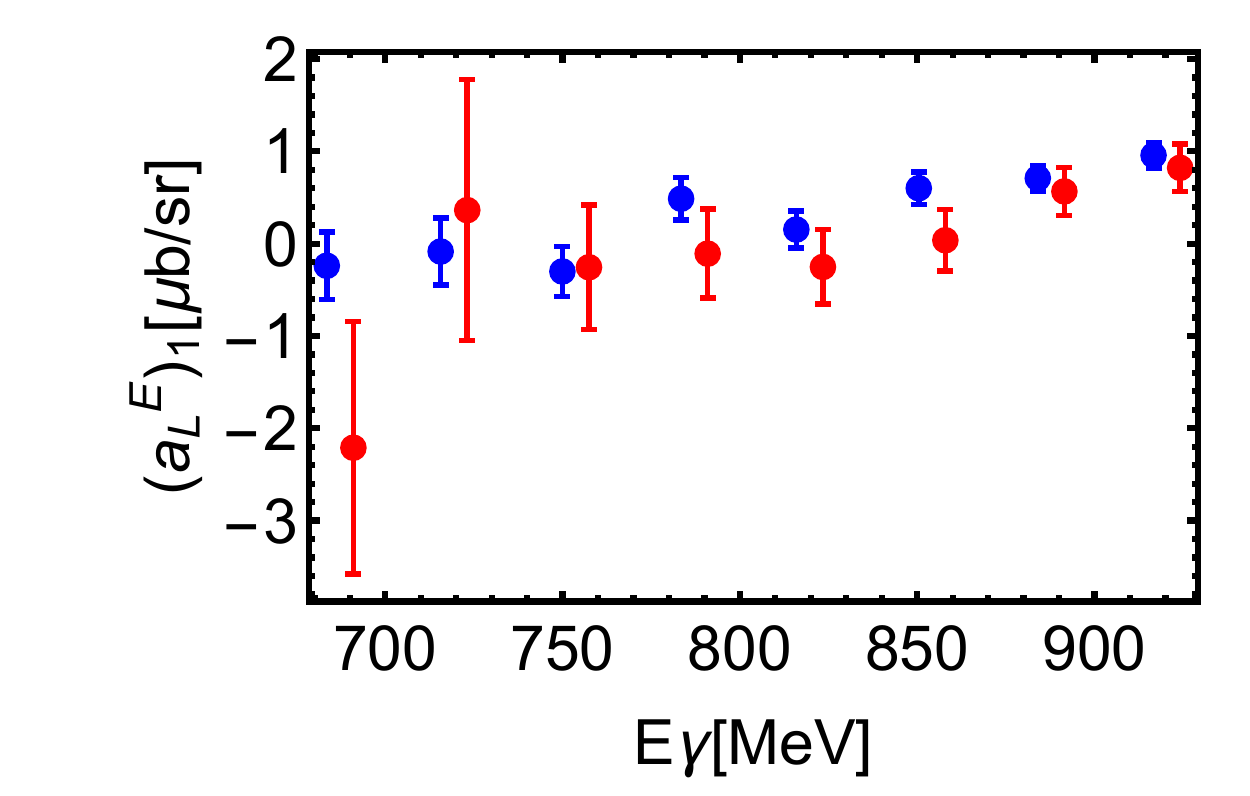}
 \end{overpic}
 \begin{overpic}[width=0.325\textwidth]{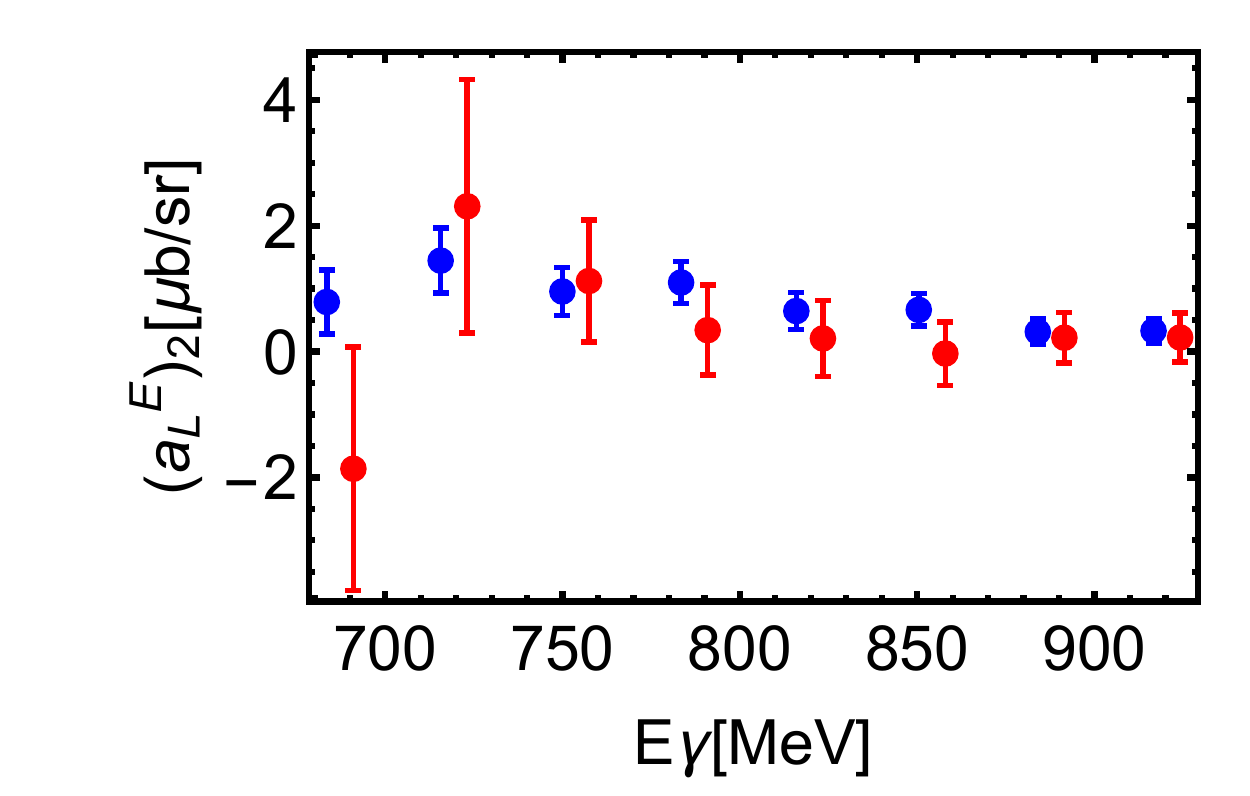}
 \end{overpic} \\
 \begin{overpic}[width=0.325\textwidth]{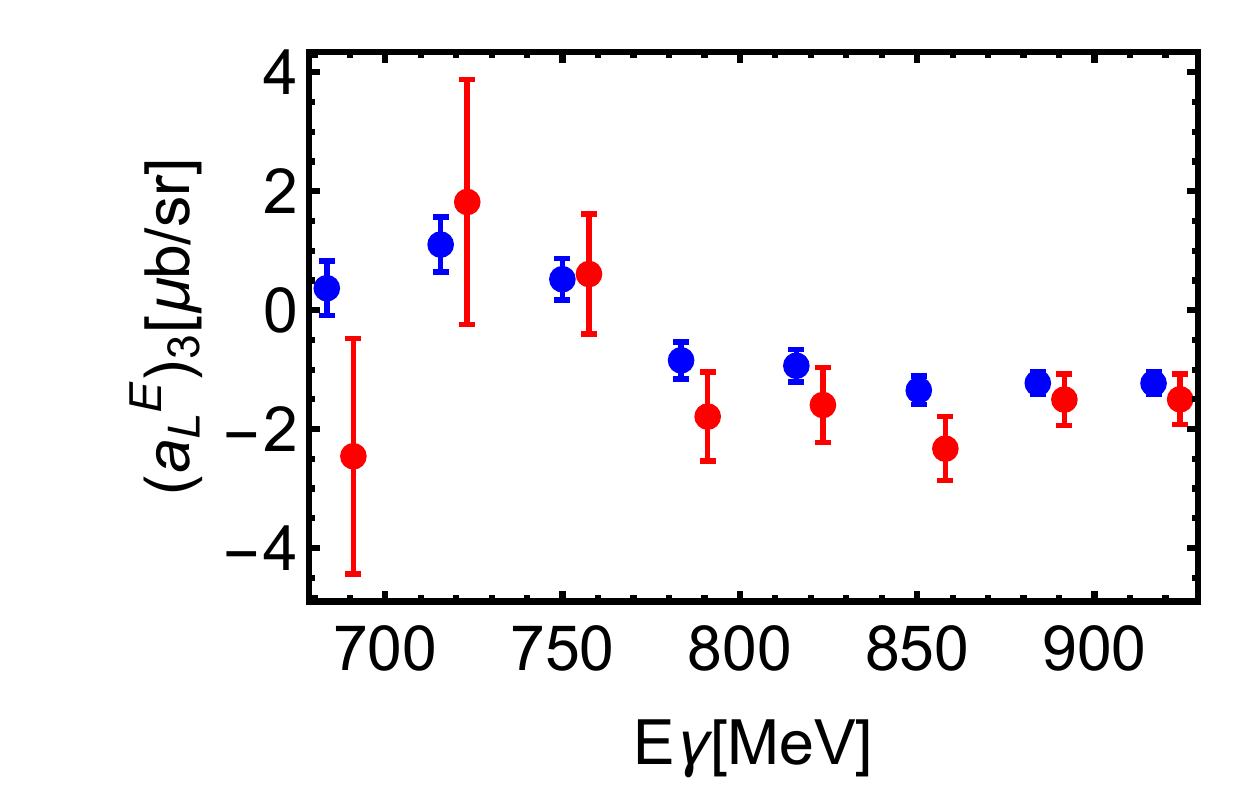}
 \end{overpic}
 \begin{overpic}[width=0.325\textwidth]{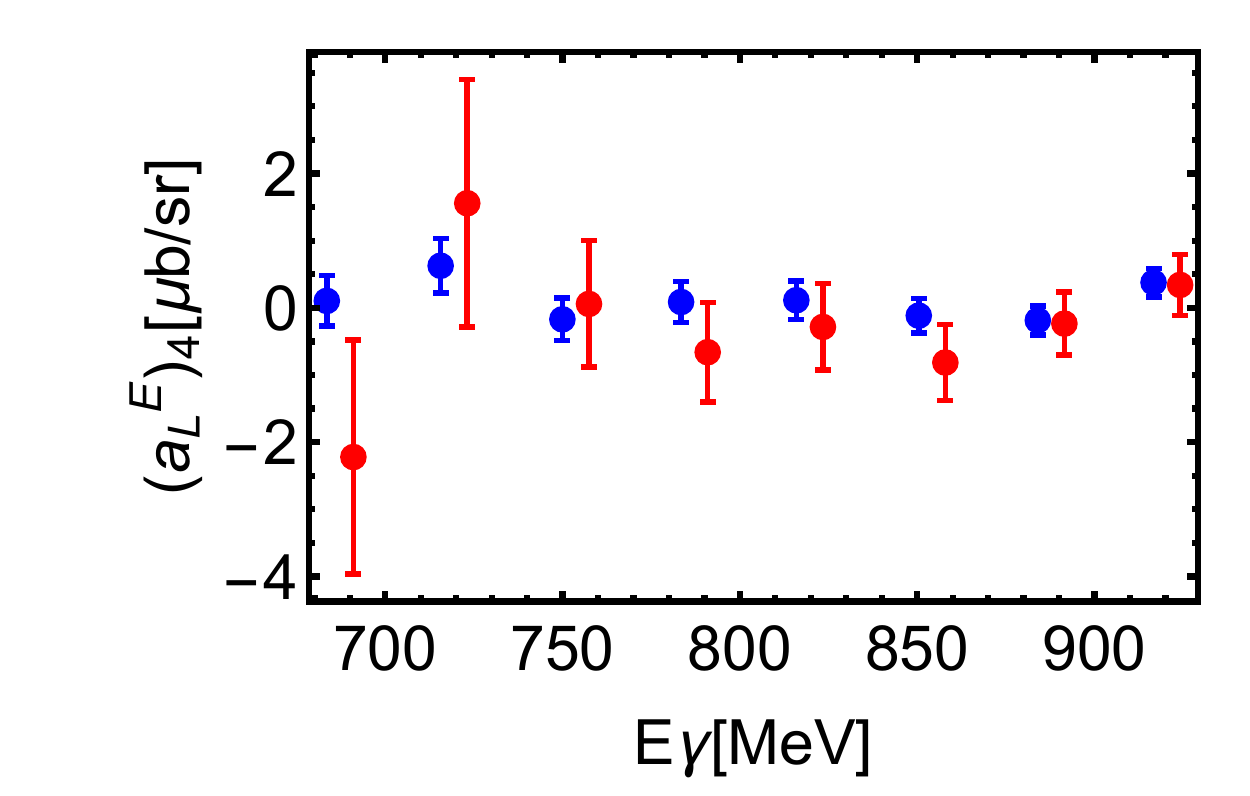}
 \end{overpic}
 \begin{overpic}[width=0.325\textwidth]{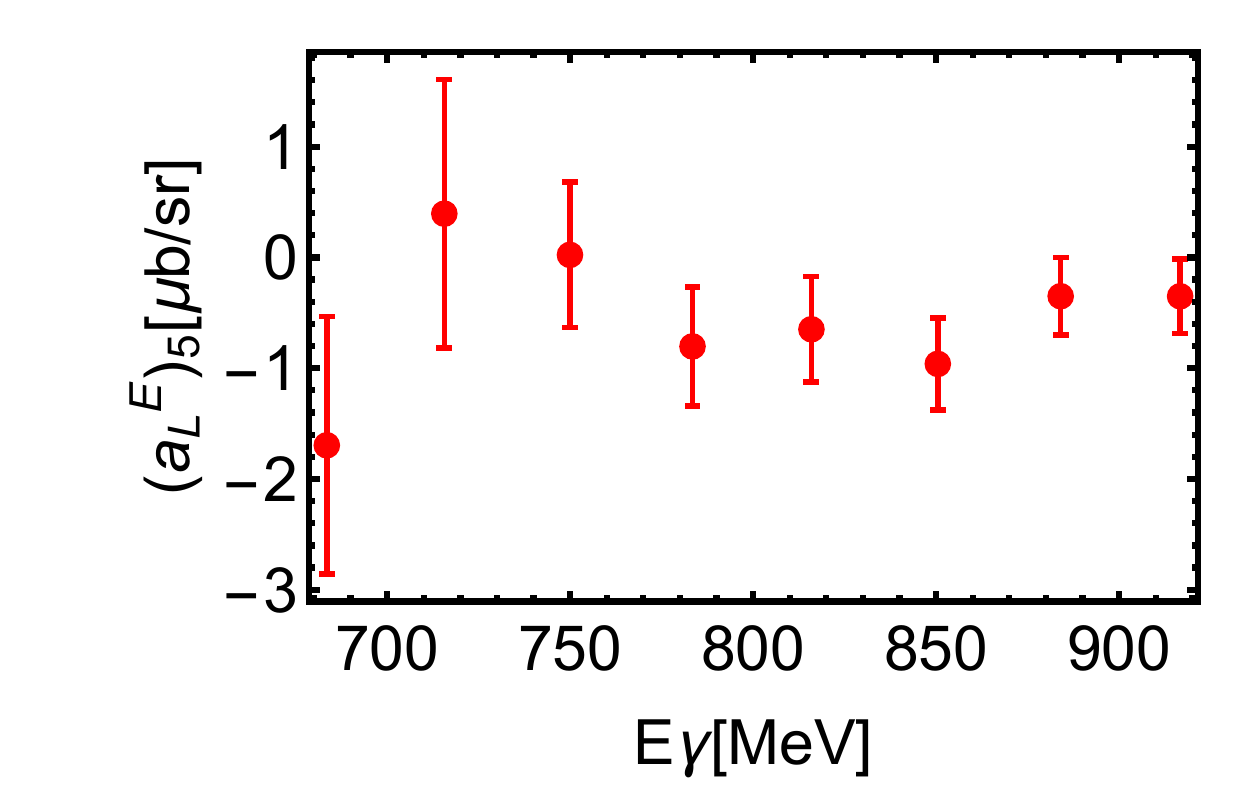}
 \end{overpic} \\
 \begin{overpic}[width=0.325\textwidth]{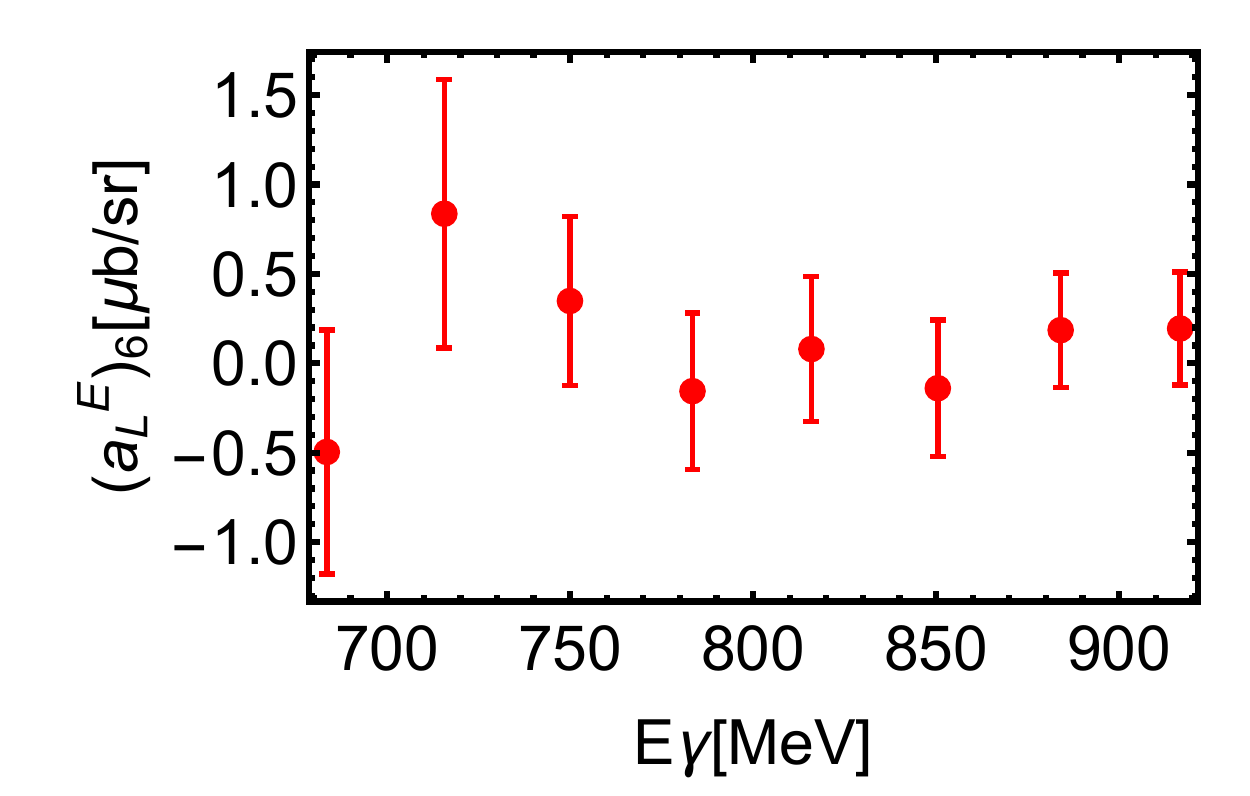}
 \end{overpic} \\
  \begin{overpic}[width=0.325\textwidth]{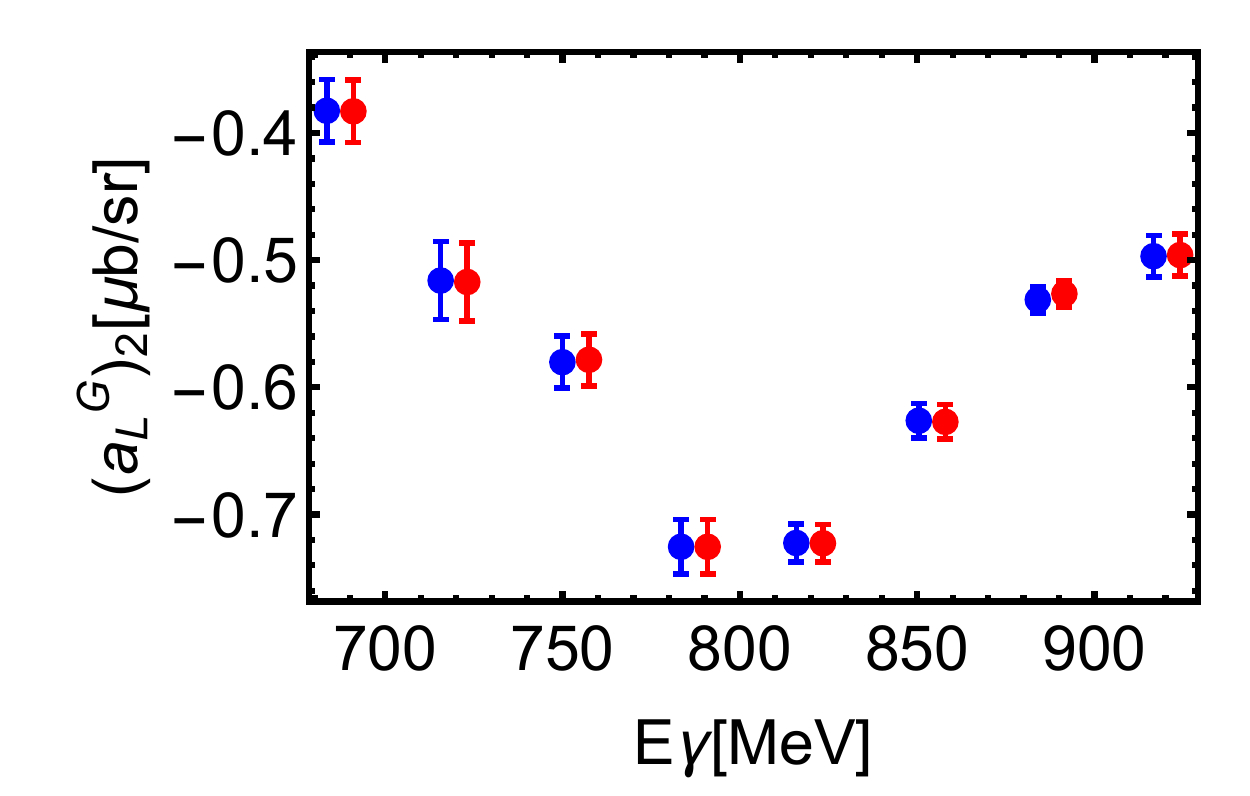}
 \end{overpic}
 \begin{overpic}[width=0.325\textwidth]{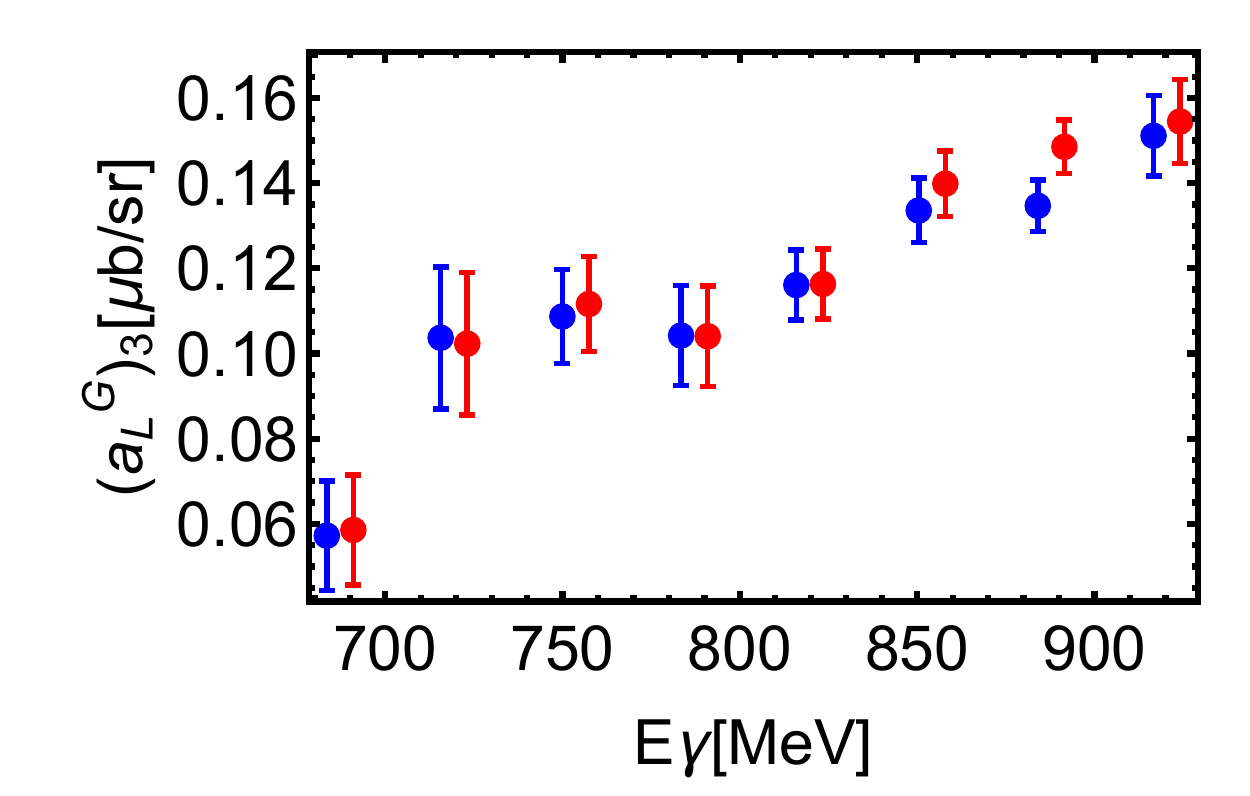}
 \end{overpic}
  \begin{overpic}[width=0.325\textwidth]{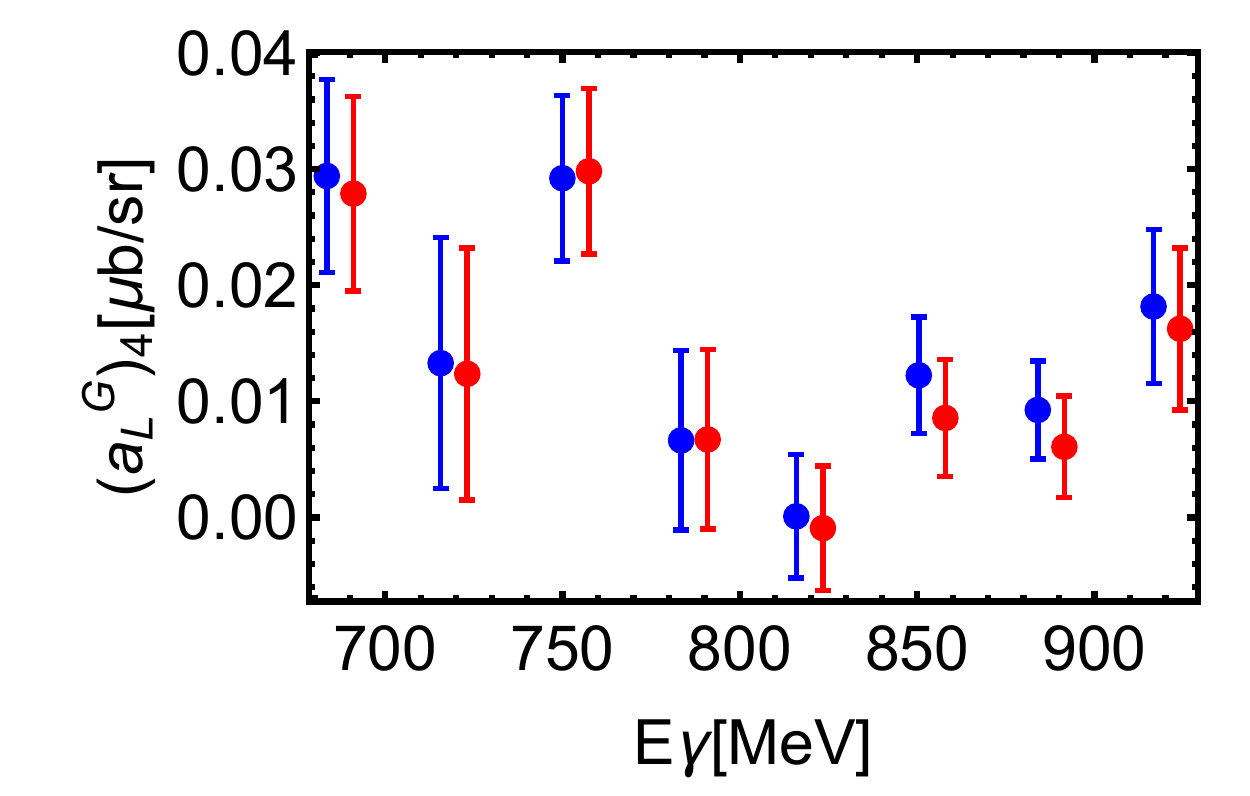}
 \end{overpic} \\
  \begin{overpic}[width=0.325\textwidth]{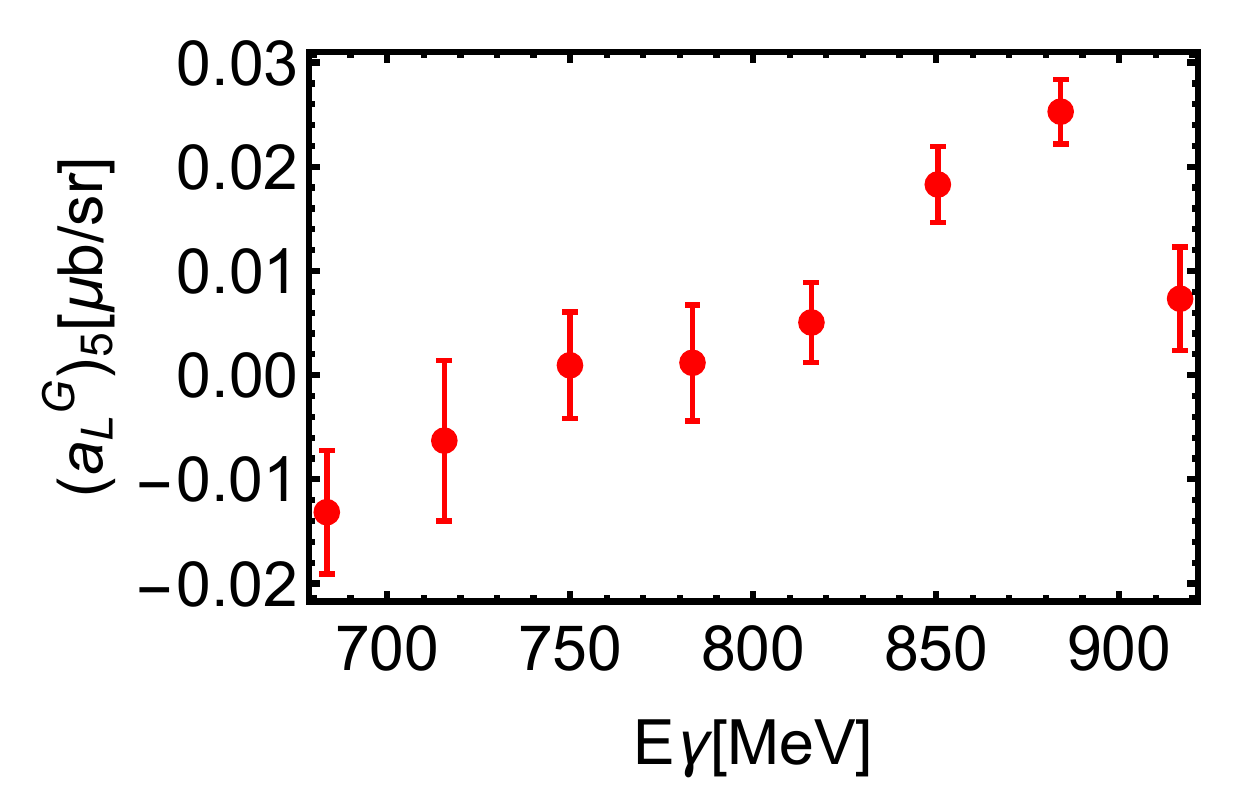}
 \end{overpic}
 \begin{overpic}[width=0.325\textwidth]{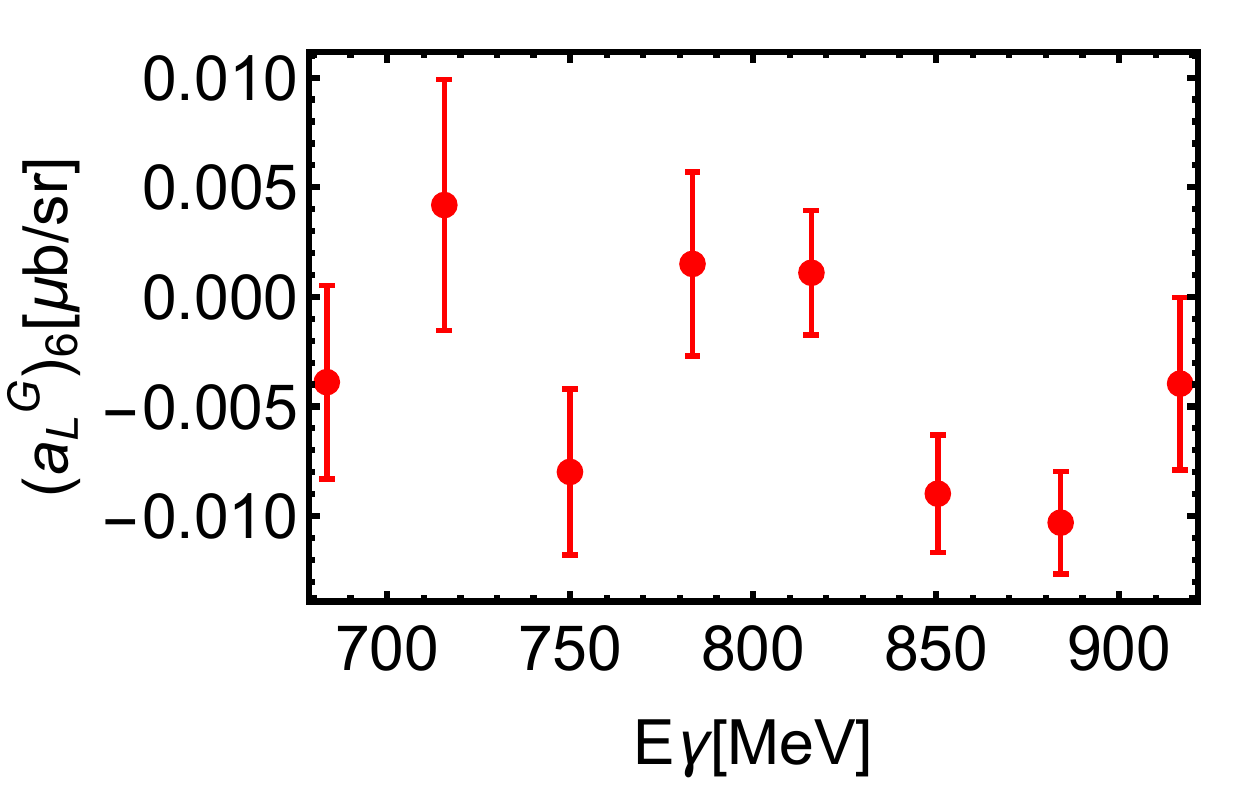}
 \end{overpic} \\
 \caption{The plots represent a continuation of Figure \ref{fig:2ndResRegionFittedLegCoeffsPlotsI}. Shown are the Legendre-coefficients extracted from the profile functions $\check{P}$, $\check{E}$ and $\check{G}$ for $\ell_{\mathrm{max}} = 2$ (blue) and $\ell_{\mathrm{max}} = 3$ (red).}
 \label{fig:2ndResRegionFittedLegCoeffsPlotsII}
\end{figure}

\begin{figure}[h]
 \centering
  \begin{overpic}[width=0.325\textwidth]{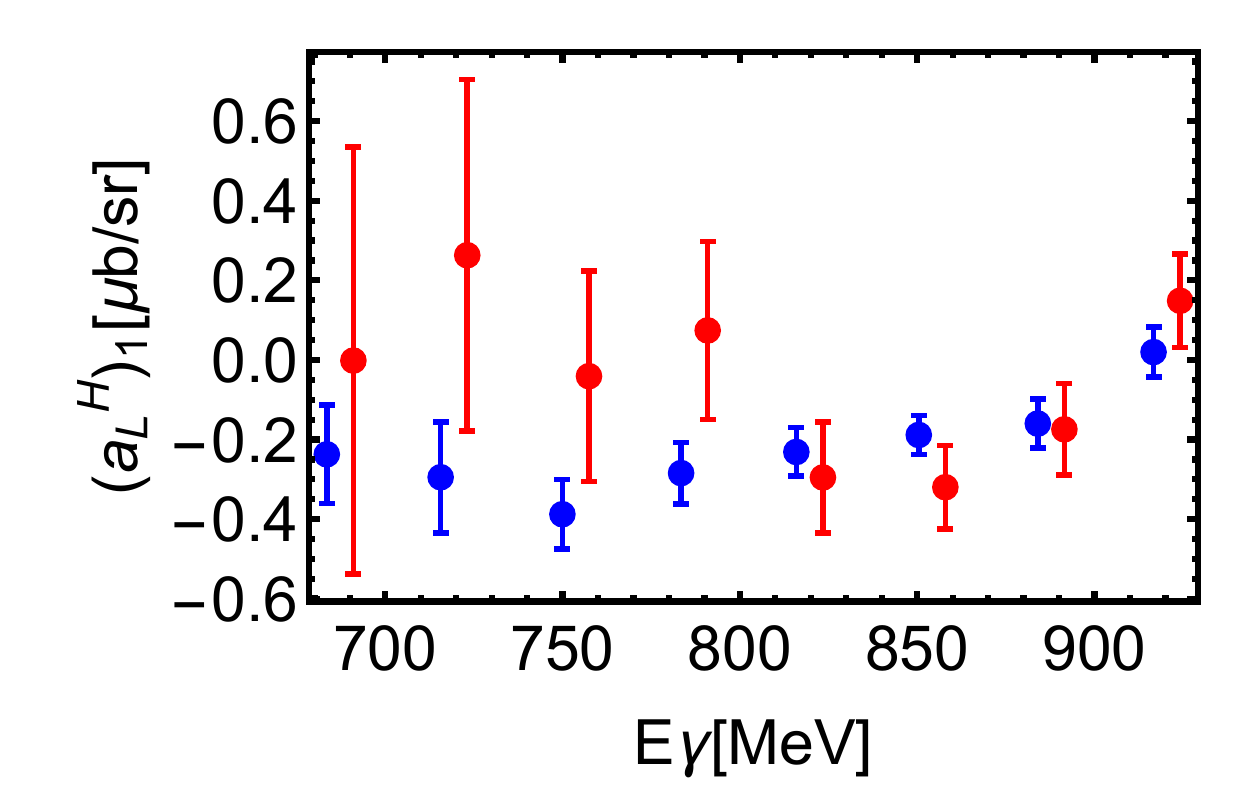}
 \end{overpic}
 \begin{overpic}[width=0.325\textwidth]{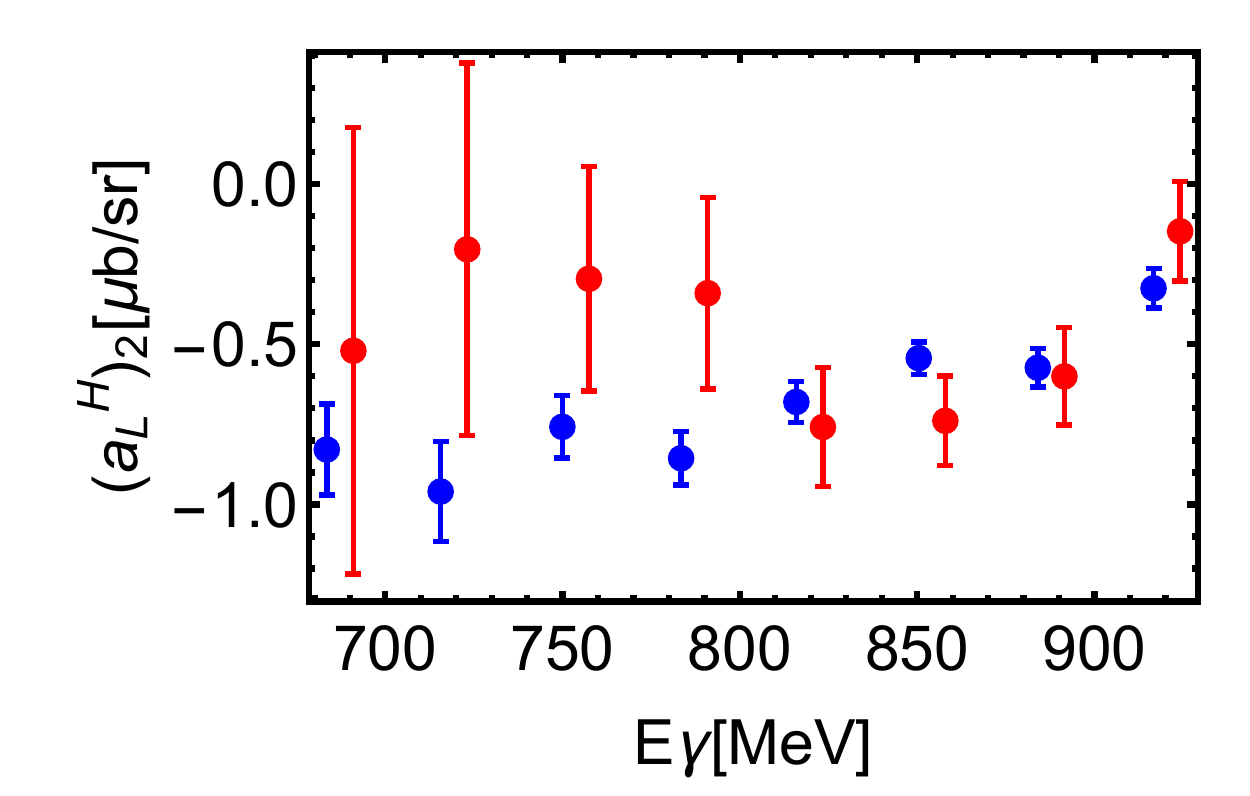}
 \end{overpic}
  \begin{overpic}[width=0.325\textwidth]{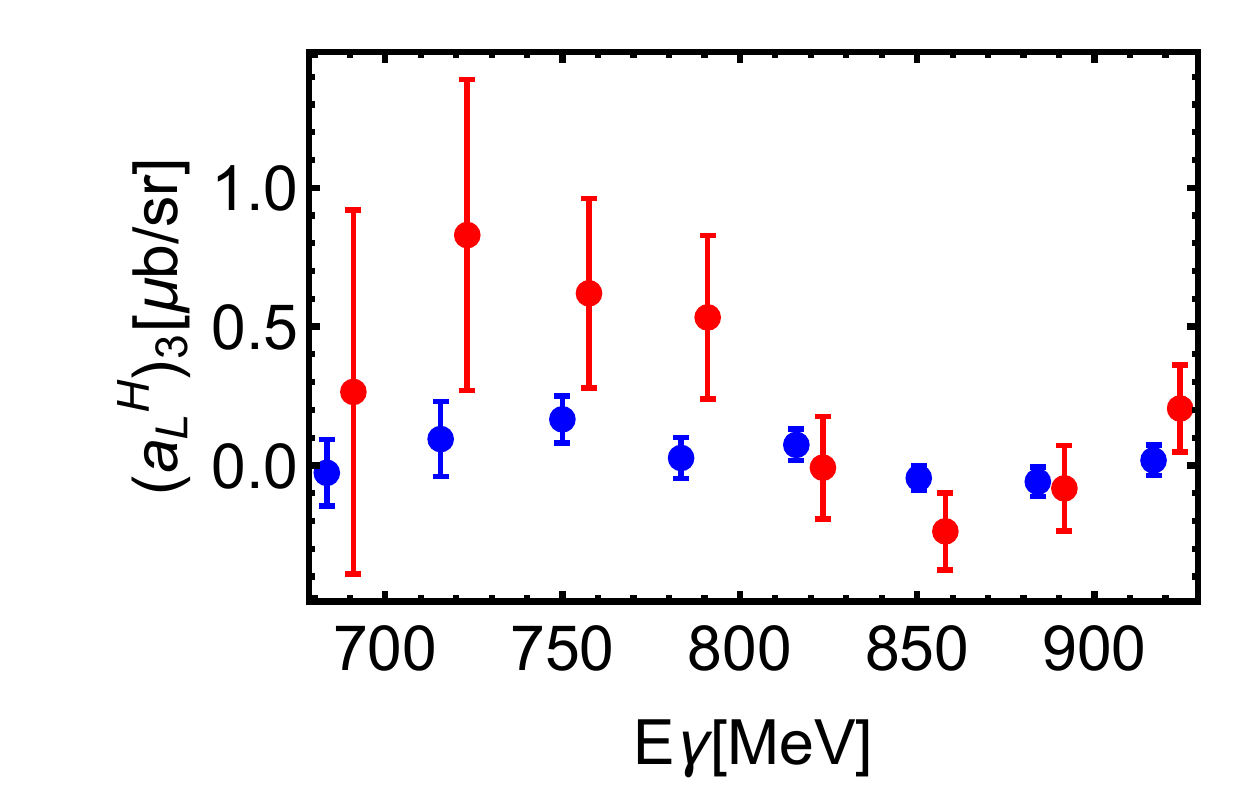}
 \end{overpic} \\
  \begin{overpic}[width=0.325\textwidth]{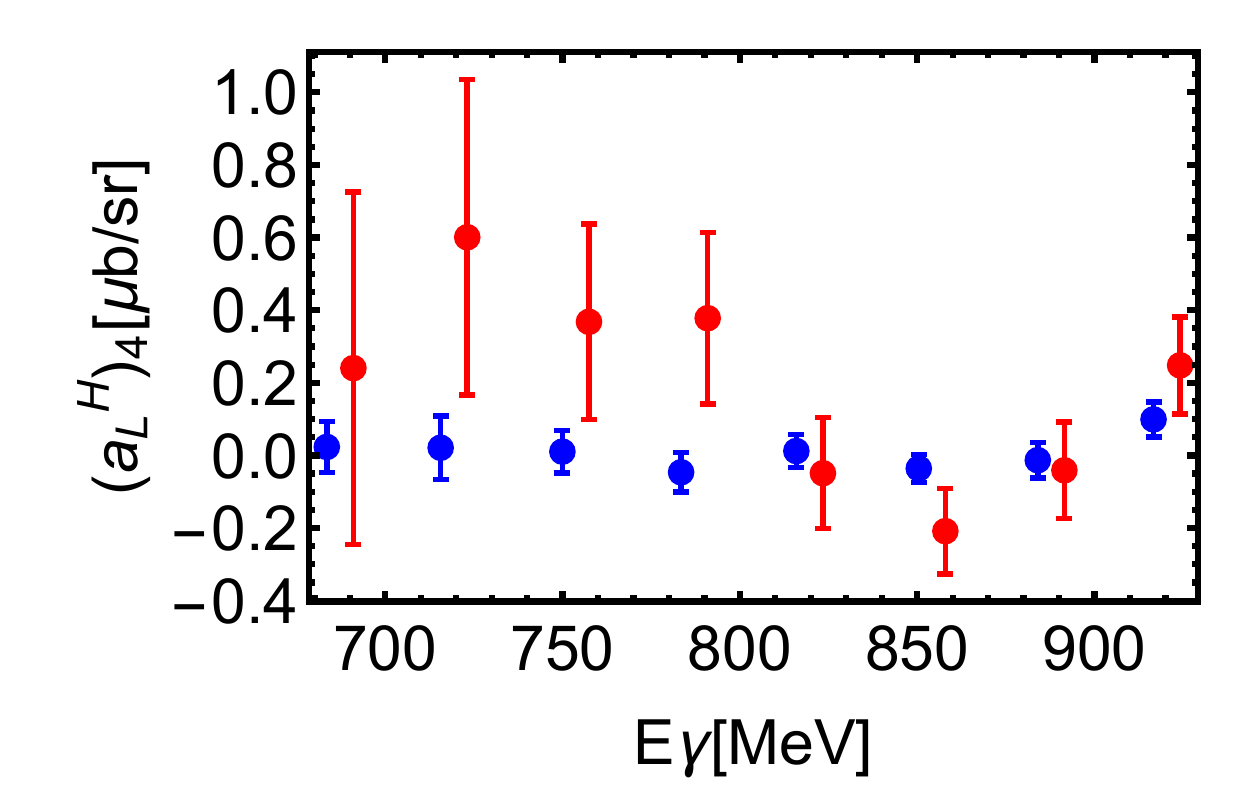}
 \end{overpic}
  \begin{overpic}[width=0.325\textwidth]{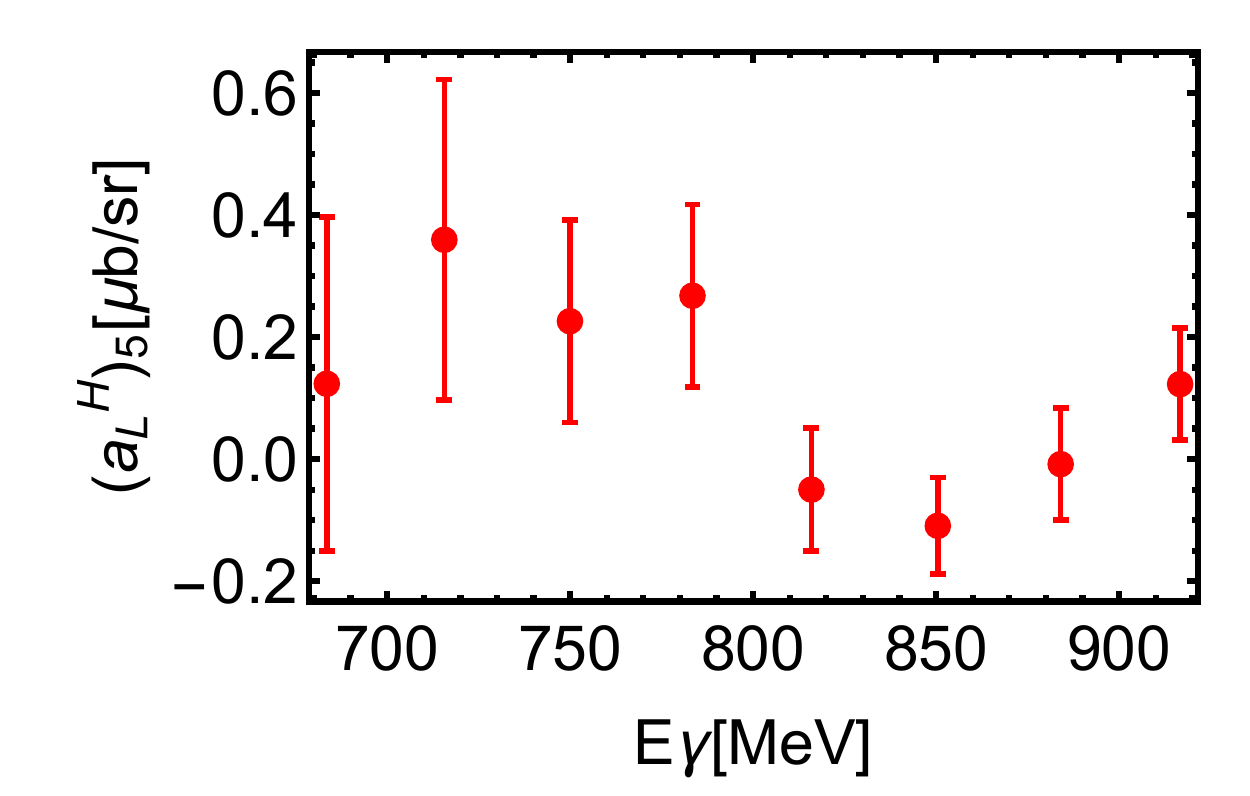}
 \end{overpic}
 \begin{overpic}[width=0.325\textwidth]{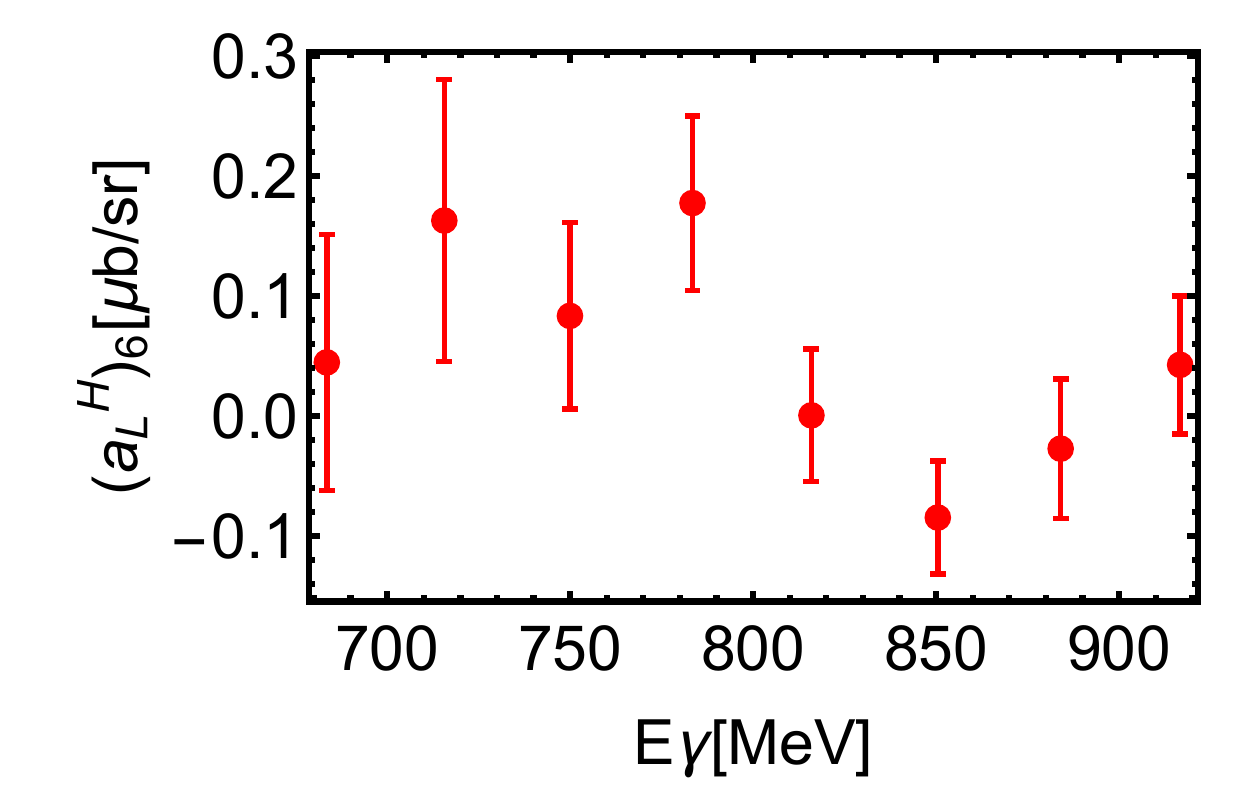}
 \end{overpic}
 \caption{The Figures \ref{fig:2ndResRegionFittedLegCoeffsPlotsI} and \ref{fig:2ndResRegionFittedLegCoeffsPlotsII} are continued. Shown are the Legendre-coefficients extracted from the profile function $\check{H}$ for $\ell_{\mathrm{max}} = 2$ (blue) and $\ell_{\mathrm{max}} = 3$ (red).}
 \label{fig:2ndResRegionFittedLegCoeffsPlotsIII}
\end{figure}

Legendre coefficients in TPWA fit-step $\mathrm{I}$ (cf. section \ref{sec:TPWAFitsIntro}) and using the angular distributions of the profile functions for a simple $\ell_{\mathrm{max}}$-analysis as outlined in detail in chapter \ref{chap:LFits}. The parametrizations for the angular distributions of $\sigma_{0}$, $\check{\Sigma}$, $\check{T}$ and $\check{P}$ can be found in equations (\ref{eq:DCSAngDistChapter4}) to (\ref{eq:PAngDistChapter4}) of section \ref{subsec:DeltaRegionDataFits}. They have to be supplemented here for the three additional beam-traget observables analyzed in the second resonance region. The expressions are quoted here for quick reference (cf. section \ref{sec:LFitsPaper})
\begin{align}
 \check{E} (W,\theta) &= \frac{q}{k} \sum_{k=0}^{2 \ell_{\mathrm{max}}} \left( a_{L} \right)_{k}^{\check{E}} P_{k} (\cos \theta) \mathrm{,} \label{eq:EAngDistChapter4} \\
 \check{G} (W,\theta) &= \frac{q}{k} \sum_{k=2}^{2 \ell_{\mathrm{max}}} \left( a_{L} \right)_{k}^{\check{G}} P^{2}_{k} (\cos \theta) \mathrm{,} \label{eq:GAngDistChapter4} \\
 \check{H} (W,\theta) &= \frac{q}{k} \sum_{k=1}^{2 \ell_{\mathrm{max}}} \left( a_{L} \right)_{k}^{\check{H}} P^{1}_{k} (\cos \theta) \mathrm{.} \label{eq:HAngDistChapter4}
\end{align}
Angular distributions of the differential cross section $\sigma_{0}$, as well as the profile functions for all $6$ polarization observables analyzed in the second resonance region, are shown in Figure \ref{fig:2ndResRegionObsFitAngDistPlots}. Fits of these angular distributions are plotted for $\ell_{\mathrm{max}} = 2$ and $3$. As an example, the seventh overall energy-bin, $E_{\gamma} = 884.02 \hspace*{1pt} \mathrm{MeV}$, has been selected. The quantities $\check{\Sigma}$ and $\check{G}$ are seen to require the $F$-wave truncation for a description of their modulation. Others, like $\check{E}$ or $\check{H}$, are already well-described using just $\ell_{\mathrm{max}} = 2$. The cross section is here a benchmark-dataset, as it is seen to have incredibly small errors. The remaining polarization datasets are seen to be of comparable quality. \newline
In order to infer suitable $\ell_{\mathrm{max}}$-estimates, consideration of the $\chi^{2}/\mathrm{ndf}$, plotted against energy, are useful. The required pictures are shown in Figure \ref{fig:2ndResRegionChisquareLmaxPlots}. Values of $\chi^{2}/\mathrm{ndf}$ have been plotted for $\ell_{\mathrm{max}} = 1$, $2$ and $3$. It can be seen that, as expected, in the second resonance region it is completely impossible to describe the angular distributions satisfactorily using a $P$-wave truncation, i.e. $\ell_{\mathrm{max}} = 1$. This is true for any dataset except for the $H$-data. A truncation at the $D$-waves is seen to describe most of the data well at almost all energies. Exceptions are here the differential cross section $\sigma_{0}$ for the high and low energies, the beam asymmetry $\check{\Sigma}$ at the high energies and the profile function $\check{G}$ for the seventh energy-bin. Once the truncation order is raised to $\ell_{\mathrm{max}} = 3$, the $\chi^{2}/\mathrm{ndf}$ is close to the ideal value of $1$ for all observables and energies. As a result of the investigation of angular distributions, we prefer approaches with $\ell_{\mathrm{max}} = 2$ and $\ell_{\mathrm{max}} = 3$ in the multipole-fits outlined in the ensuing paragraph. \newline
The most important input for fit-step $\mathrm{II}$, i.e. the actual Legendre-coefficients, are shown for all analyzed observables in Figures \ref{fig:2ndResRegionFittedLegCoeffsPlotsI}, \ref{fig:2ndResRegionFittedLegCoeffsPlotsII} and \ref{fig:2ndResRegionFittedLegCoeffsPlotsIII}. The higher Legendre-coefficients, i.e. those newly introduced for $\ell_{\mathrm{max}} = 3$, are small but non-zero and thus show some need for $F$-waves, at least for some observables. Most notably, this is the case for the differential cross section $\sigma_{0}$ and beam-asymmetry $\check{\Sigma}$. The stability of the lower Legendre-coefficients against the increase of $\ell_{\mathrm{max}}$ is quite good in most cases. Some variations can be seen for selected energies of certain observables. The $H$-data seem to be particularly prone to such instabilities in the Legendre-coefficients. Other measurments, like $G$ for instance, are quite stable. \newline
The Legendre coefficients extracted in this paragraph (as well as their standard errors, covariance-matrices, $\ldots$) serve as input for the actual multipole-fits in TPWA step $\mathrm{II}$.

\clearpage

In case of an application of the bootstrap, the Legendre coefficients also have bootstrap-distributions, which are in this case however never shown explicitly (cf. Figure \ref{fig:DeltaRegionFittedLegCoeffsBootstrapHistos} for the $\Delta$-region). \newline

\textbf{Results of TPWA-fits performed to the selected data} \newline

The basis of all fits performed in the second resonance region, be it minimum-searches performed on the original data, or even elaborate bootstrap-analyses, are correlated chisquare-functions just as defined in section \ref{sec:TPWAFitsIntro} (equation (\ref{eq:CorrelatedChisquare}))
\begin{equation}
\chi^{2}_{\mathcal{M}} \left( \left\{ \mathcal{M}_{\ell} \right\} \right) = \sum_{i,j} \Big[ \left(a_{L}^{\mathrm{Fit}}\right)_{i} - \left< \mathcal{M}_{\ell} \right| \left(\mathcal{C}_{L}\right)_{i} \left| \mathcal{M}_{\ell} \right> \Big] \mathrm{\textbf{C}}^{-1}_{ij} \Big[ \left(a_{L}^{\mathrm{Fit}}\right)_{j} - \left< \mathcal{M}_{\ell} \right| \left(\mathcal{C}_{L}\right)_{j} \left| \mathcal{M}_{\ell} \right> \Big] \mathrm{.} \label{eq:CorrelatedChisquareQuotedForAnalysisII}
\end{equation}
Again, the Legendre-coefficients $\left(a_{L}^{\mathrm{Fit}}\right)_{i}$, as well as the covariance matrix $\mathrm{\textbf{C}}_{ij}$ for these coefficients, are imported from the Legendre-fits described in the previous paragraph. The multi-indices $i,j$ comprise information on which observables are fitted and how many Legendre-coefficients are included from every observable. However, since now seven observables are analyzed, the number of fit-coefficients and correpsondingly the range of the multi-indices rise accordingly. For instance, in case a truncation at $\ell_{\mathrm{max}} = 2$ is fitted to the seven observables considered here, the cross section $\sigma_{0}$ yields $5$ coefficients $\left(a_{L}^{\mathrm{Fit}}\right)^{\sigma_{0}}_{0,\ldots,4}$, the observable $E$ the same amount $\left(a_{L}^{\mathrm{Fit}}\right)^{E}_{0,\ldots,4}$, $4$ coefficients would come from the $T$-, $P$- and $H$-asymmetries each, i.e. $\left(a_{L}^{\mathrm{Fit}}\right)^{(T,P,H)}_{1,\ldots,4}$ and $\Sigma$ as well as $G$ each yield $3$ coefficients $\left(a_{L}^{\mathrm{Fit}}\right)^{(\Sigma,G)}_{2,\ldots,4}$. The covariance matrix $\mathrm{\textbf{C}}$ would then grow to become a $28 \times 28$-matrix. \newline
The number of degrees of freedom for each fit is, as before, estimated as the difference between the amount of Legendre coefficients and the number of free parameters in the fit. In case all multipoles are varied in a fully unconstrained analysis, this again becomes the known estimate (see section \ref{sec:MonteCarloSampling})
 \begin{equation}
 \mathrm{ndf} = N_{a^{\alpha}_{k}} - (8 L - 1) \mathrm{.} \label{eq:NDOFFitSecondResReg}
 \end{equation}
\newpage
\begin{figure}[hb]
 \centering
\begin{overpic}[width=0.495\textwidth]{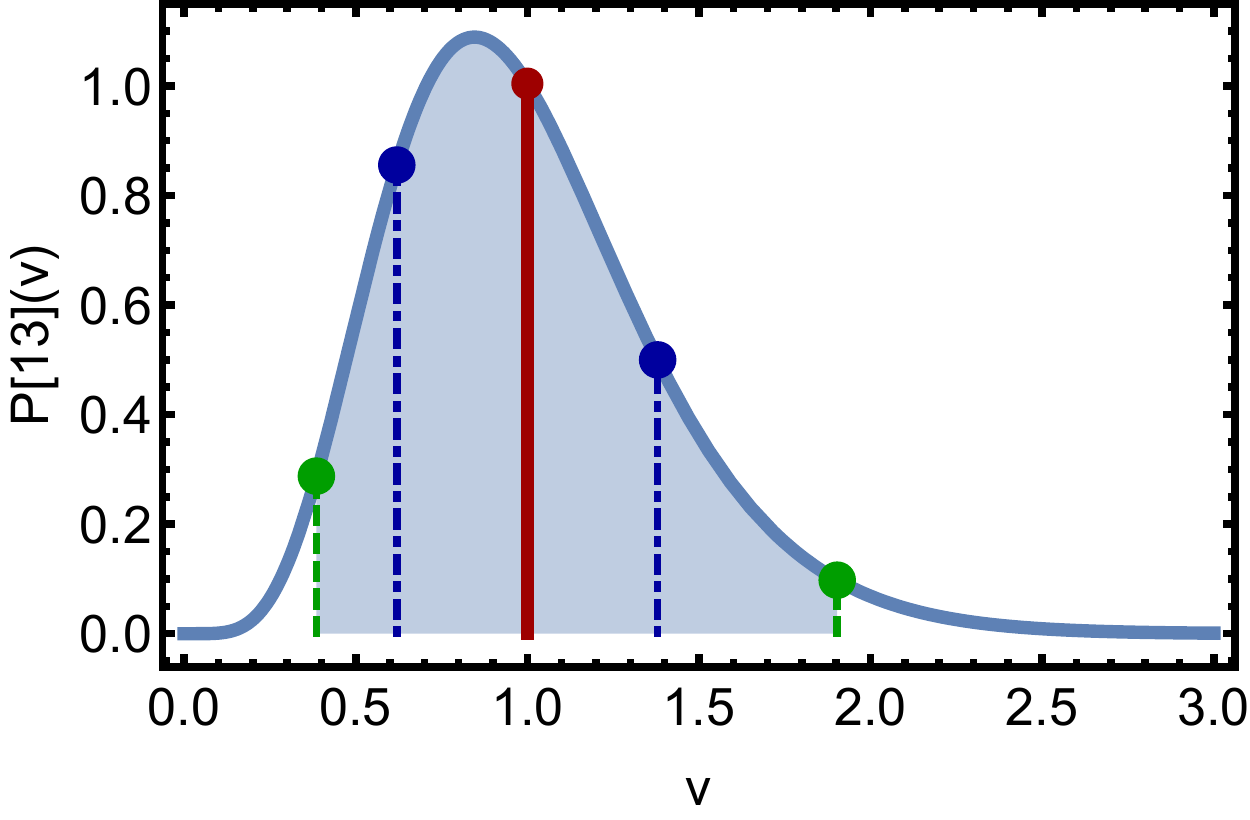}
 \end{overpic}
\begin{overpic}[width=0.495\textwidth]{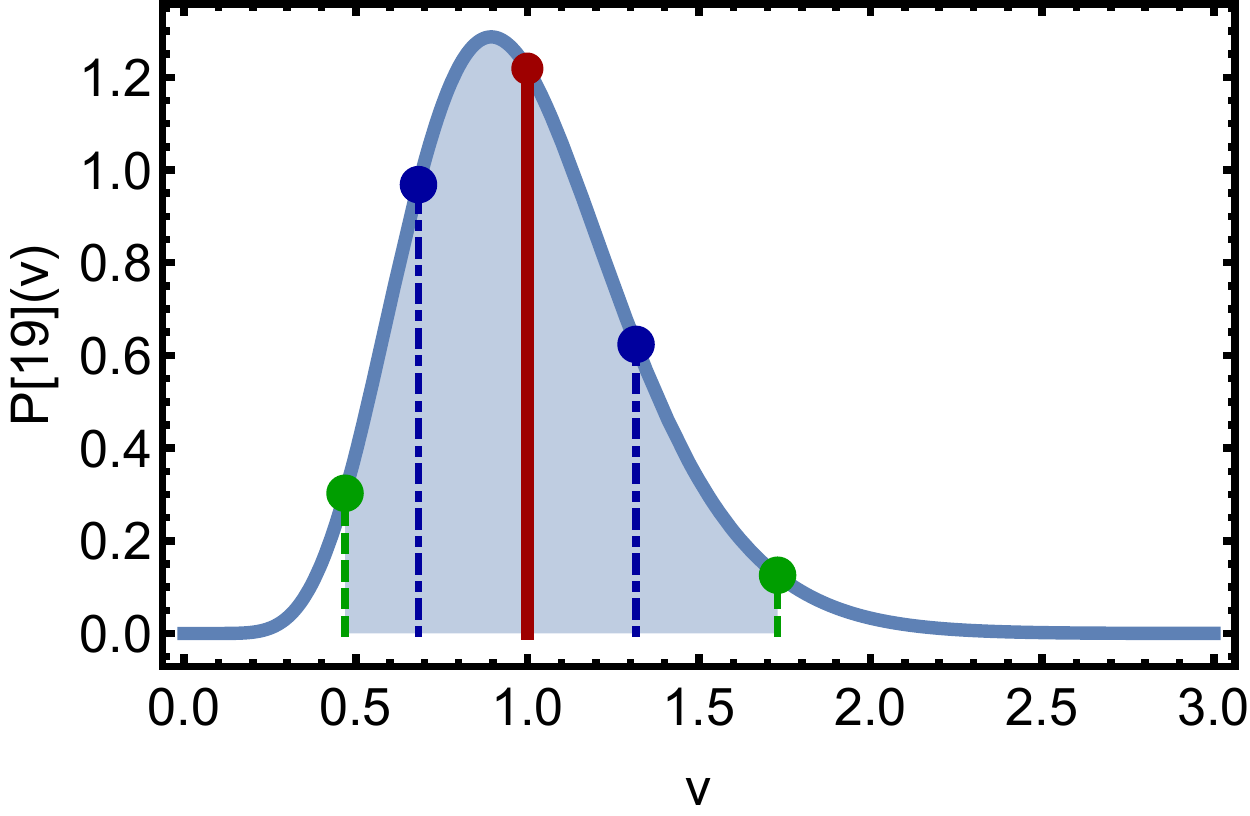}
 \end{overpic}
\caption[Two relevant examples for $\chi^{2}$-distributions in the context of fits within the $2^{\mathrm{nd}}$ resonance region.]{The plots show $\chi^{2}$-distributions for multipole-fits employing $\ell_{\mathrm{max}} = 2$ and $\ell_{\mathrm{max}} = 3$ in the second resonance region, depending on the variable $v$ which takes values of normalized $\chi^{2}$ (cf. equation (\ref{eq:ChisquareDistDefinitionInTermsOfNormalizedChi2})). Due to the fact that now seven observables are analyzed, the estimate for $\mathrm{ndf}$ rises to $r = 13$ in the first case, and $r = 19$ in the second. \newline
The mean is shown as a red solid line, while quantiles are indicated in the same way as in Figure \ref{fig:ExampleChiSquareDistributionDeltaRegion}. The green dashed lines mark the pair of $0.025$- and $0.975$-quantiles, while the blue dash-dotted lines show the $0.16$- and $0.84$-quantiles.}
\label{fig:ExampleChiSquareDistribution2ndResRegion}
\end{figure}
In case of the seven observables considered here, this estimate yields already $\mathrm{ndf} = 13$ in case of $L = \ell_{\mathrm{max}} = 2$ and $\mathrm{ndf} = 19$ for $L = \ell_{\mathrm{max}} = 3$. In Figure \ref{fig:ExampleChiSquareDistribution2ndResRegion}, theoretical chisquare-distributions (cf. equation (\ref{eq:ChisquareDistDefinitionInTermsOfNormalizedChi2}), section \ref{subsec:DeltaRegionDataFits}) are plotted for both cases.  \newline
A comparison to the pertinent chisquare-distributions for the fits in the $\Delta$-region (see Figure \ref{fig:ExampleChiSquareDistributionDeltaRegion} in the previous subsection) shows that the comparatively larger number of degrees of freedom causes the distributions to become more tapered. The quantiles defining our $68\%$- and $95\%$-confidence intervals (which are the pairs or $0.16$- and $0.84$-quantiles, as well as $0.025$- and $0.975$-quantiles) come closer to the mean and thus tighten the intervals upon which solutions of the TPWA-fits are to be accepted. Again, in the pictures comprising the $\chi^{2}/\mathrm{ndf}$-results for actual fits, exactly those quantiles will be shown as horizontal lines. \newline

\begin{figure}[hb]
 \centering
\begin{overpic}[width=0.775\textwidth]{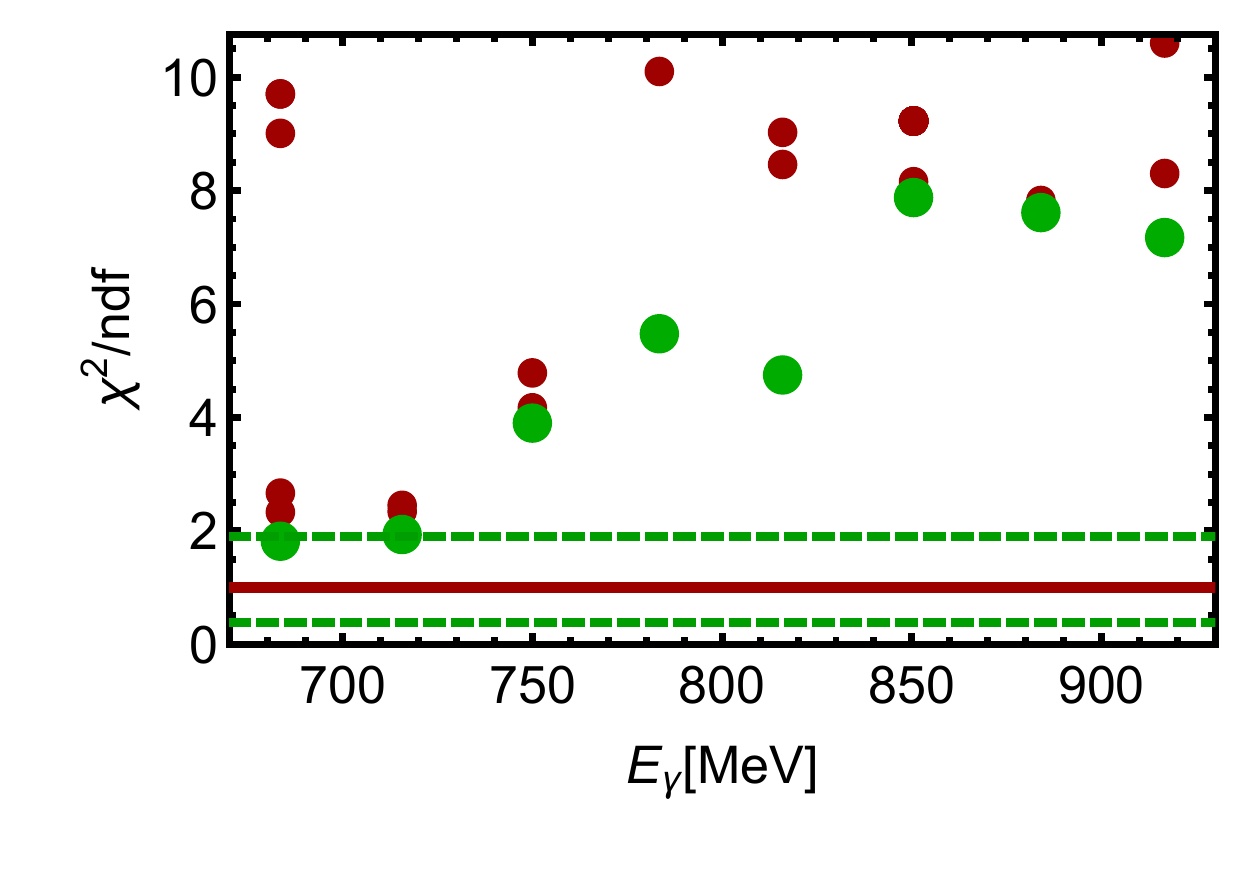}
 \end{overpic}
\vspace*{-5pt}
\caption[The $\chi^{2}/\mathrm{ndf}$ for the best results for the minimum of the correlated chisquare-function, coming from a full Monte Carlo minimum-search applied in the truncation $\ell_{\mathrm{max}} = 2$ within the $2^{\mathrm{nd}}$ resonance region.]{The plot shows the best results for the minimum of the correlated chisquare-function (equation (\ref{eq:CorrelatedChisquareQuotedForAnalysisII})), coming from a full Monte Carlo minimum-search applied in the truncation order $\ell_{\mathrm{max}} = 2$. Shown are the corresponding numbers for $\chi^{2}/\mathrm{ndf}$ which are below or roughly equal to $10$. The results stem from a pool of $N_{MC} = 16000$ initial parameter-configurations. The global minimum is indicated by the big green dots, other local minima are plotted as smaller red-colored dots. \newline
In addition, some parameters of the theoretical chisquare distribution for $\mathrm{ndf} = r = 13$ (which corresponds to $\ell_{\mathrm{max}} = 2$) are included as well. The mean is drawn as a red solid horizontal line, while the pair of $0.025$- and $0.975$-quantiles is indicated by green dashed horizontal lines.}
\label{fig:FirstFitLmax2ChiSquarePlot}
\end{figure}

\begin{figure}[h]
 \centering
\begin{overpic}[width=0.325\textwidth]{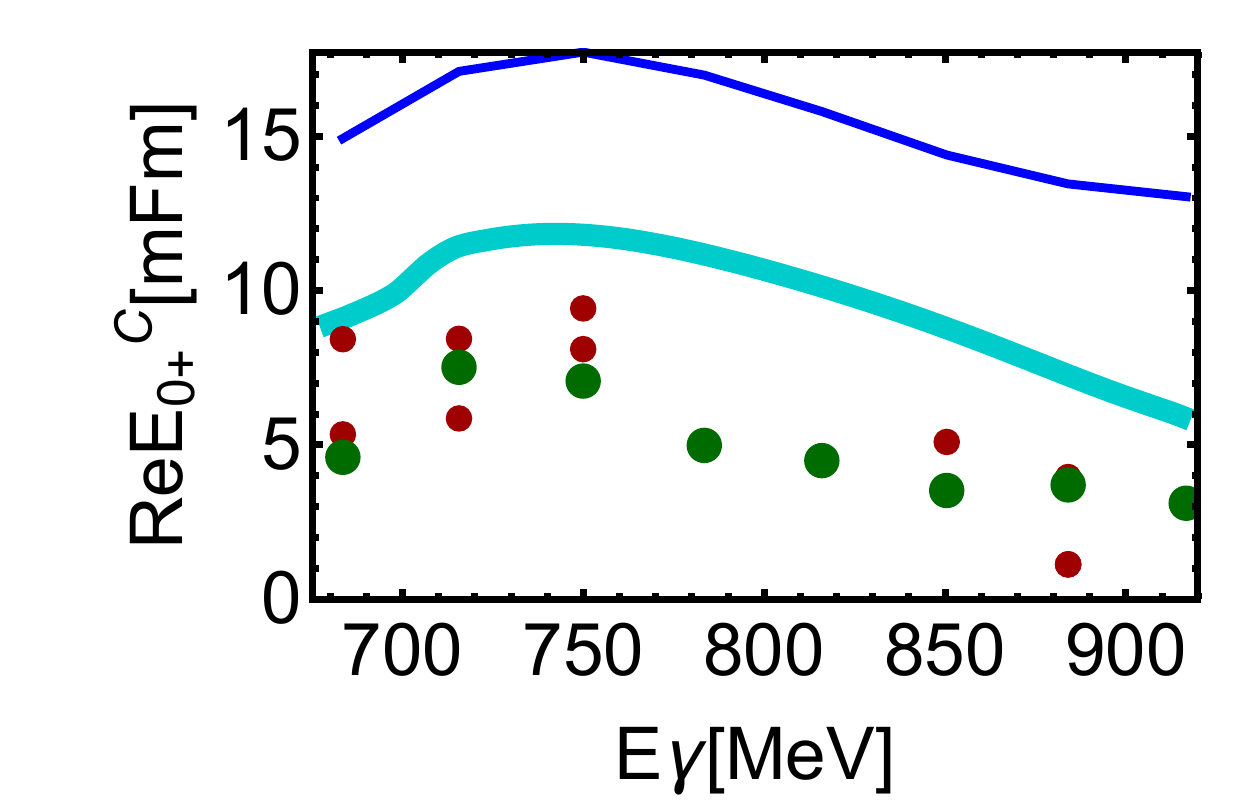}
 \end{overpic}
\begin{overpic}[width=0.325\textwidth]{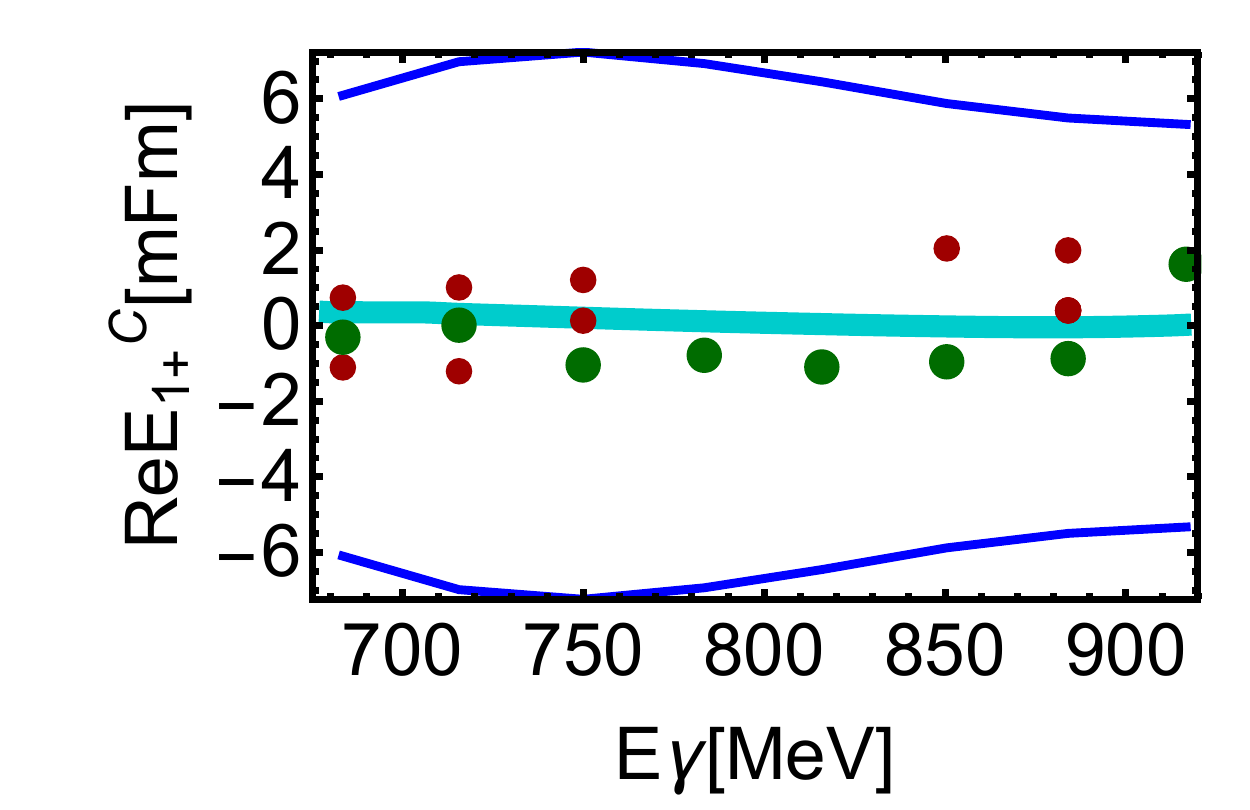}
 \end{overpic}
\begin{overpic}[width=0.325\textwidth]{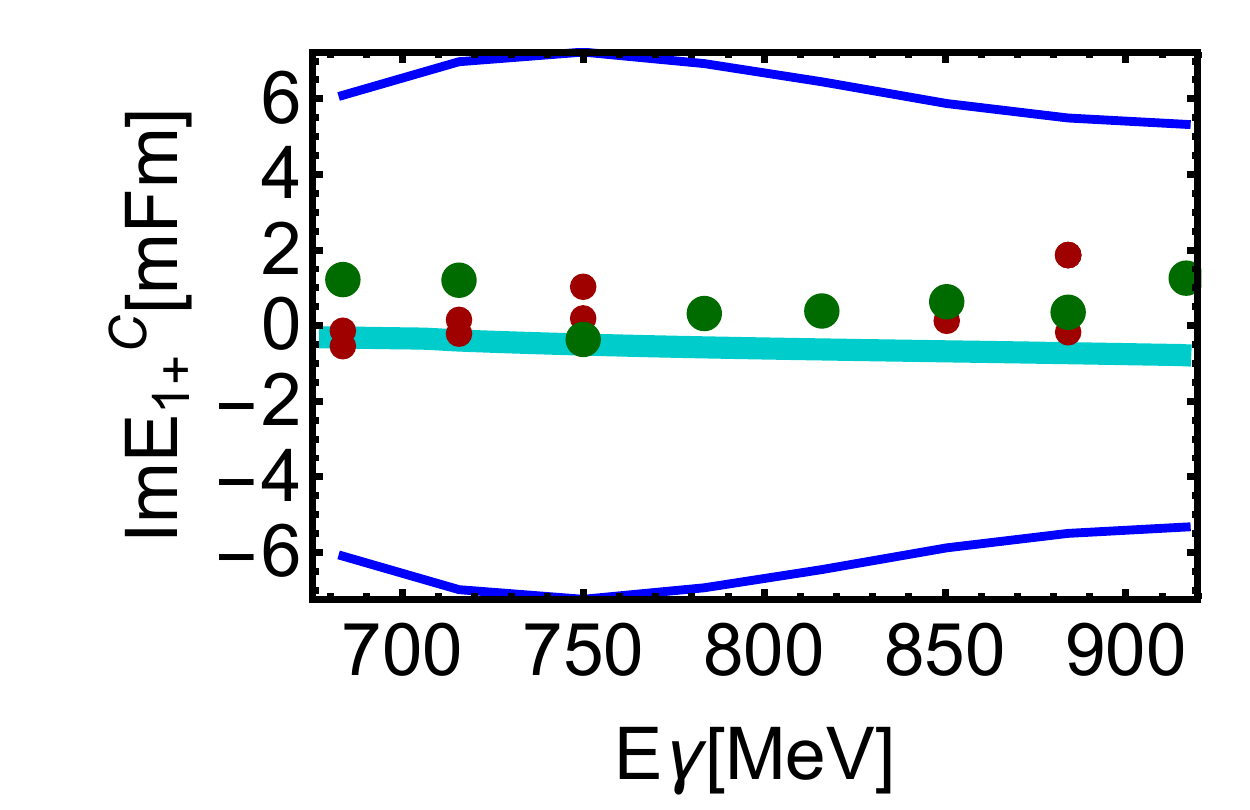}
 \end{overpic} \\
\begin{overpic}[width=0.325\textwidth]{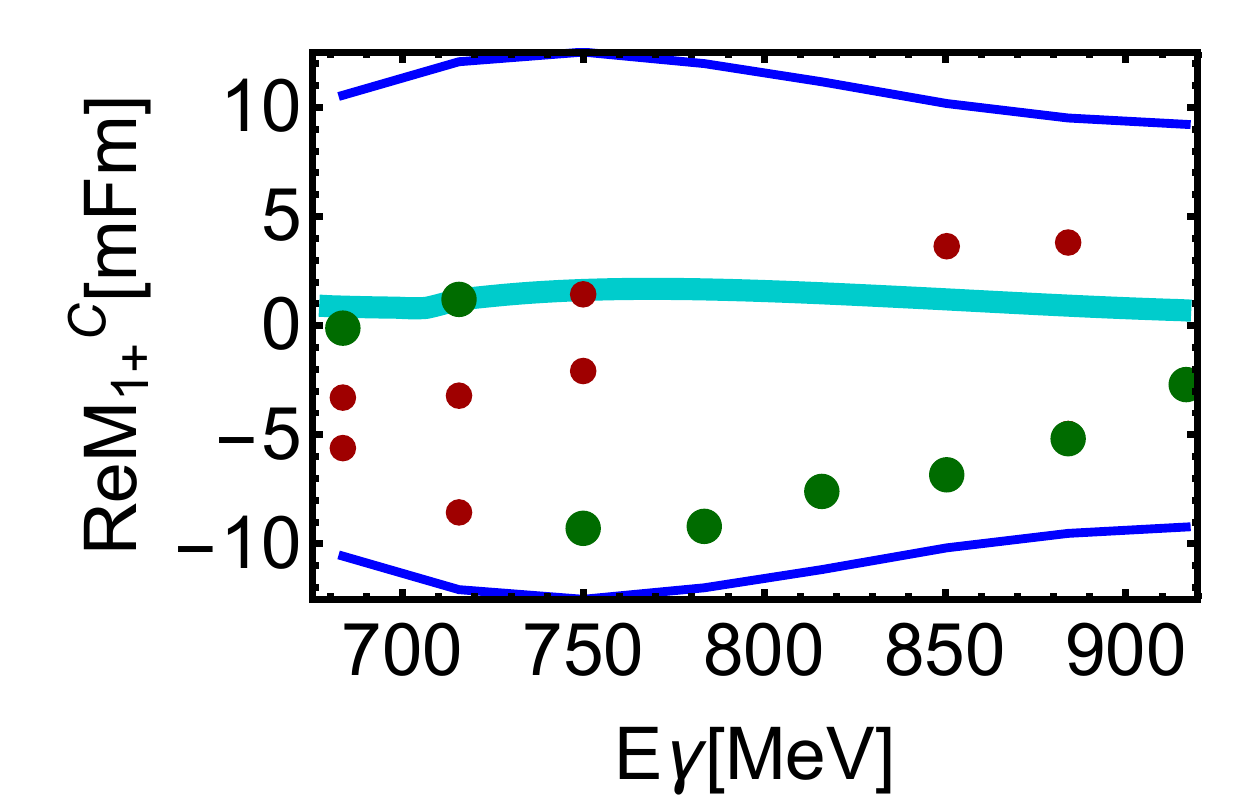}
 \end{overpic}
\begin{overpic}[width=0.325\textwidth]{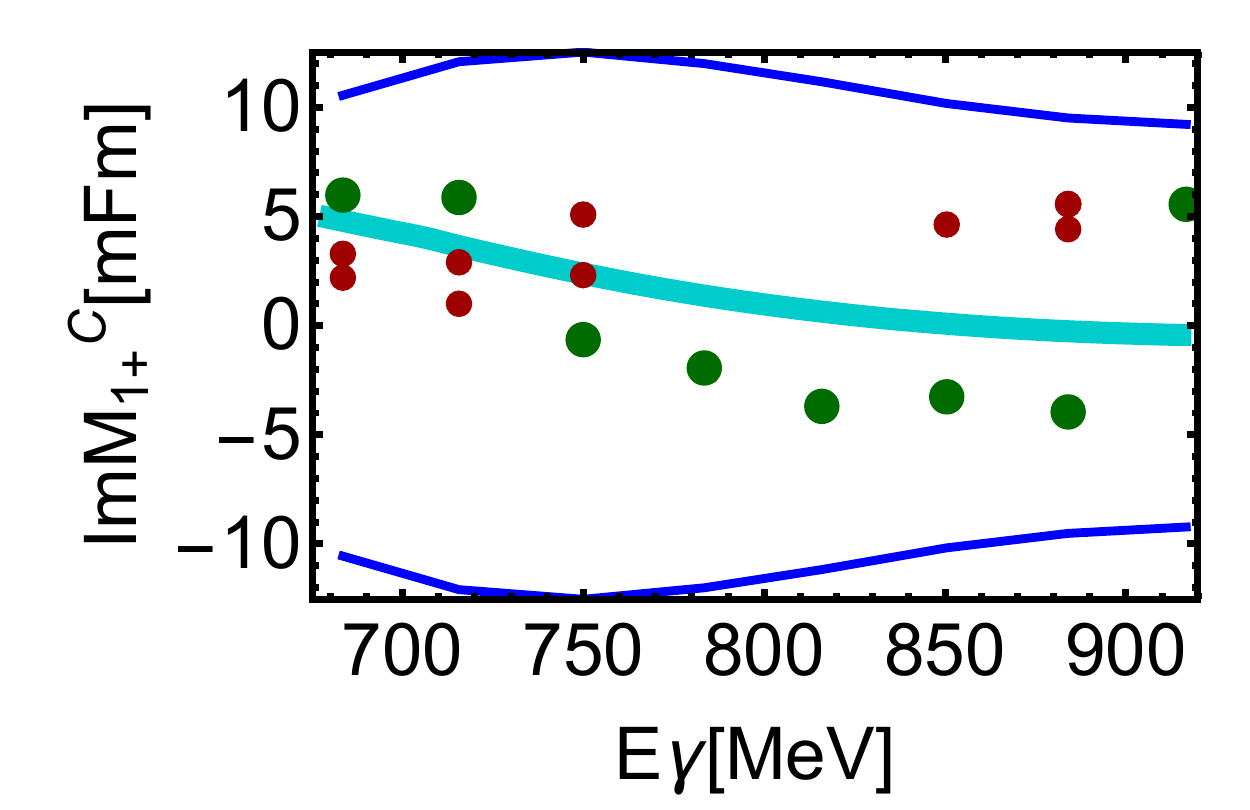}
 \end{overpic}
\begin{overpic}[width=0.325\textwidth]{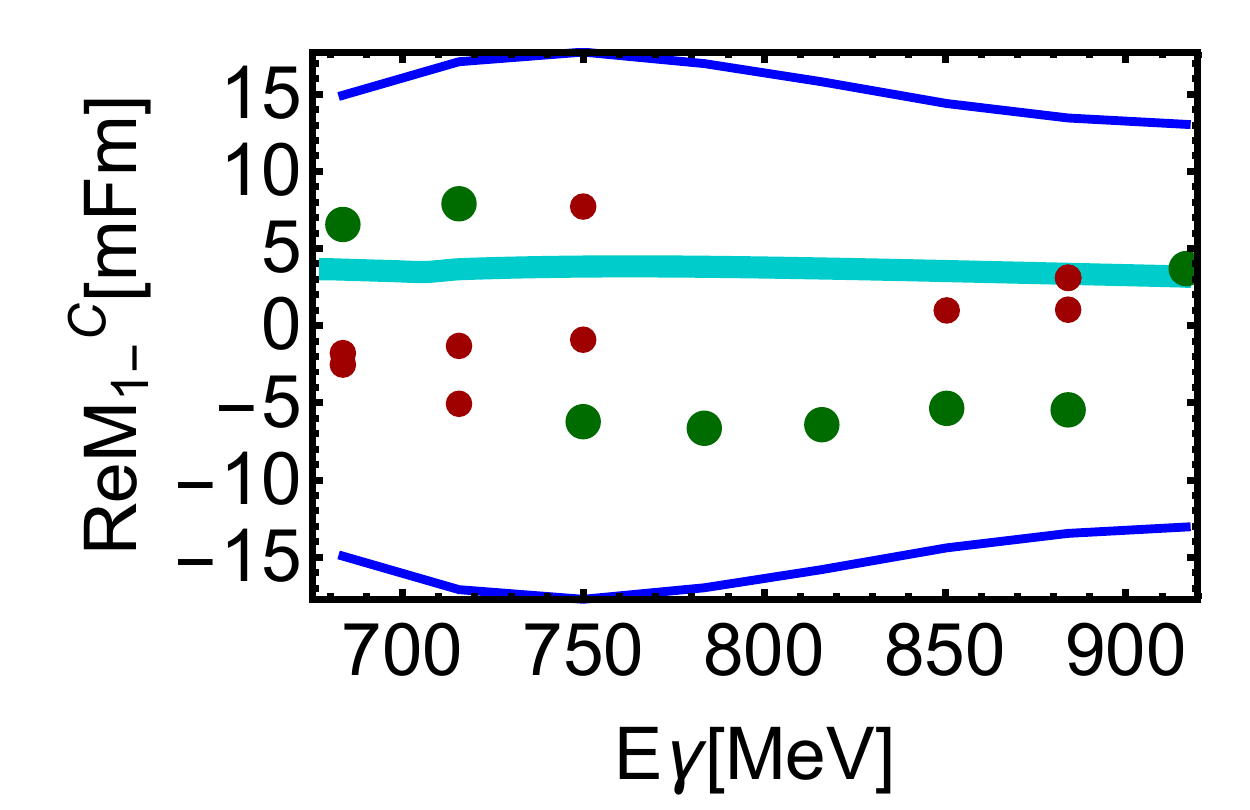}
 \end{overpic} \\
\begin{overpic}[width=0.325\textwidth]{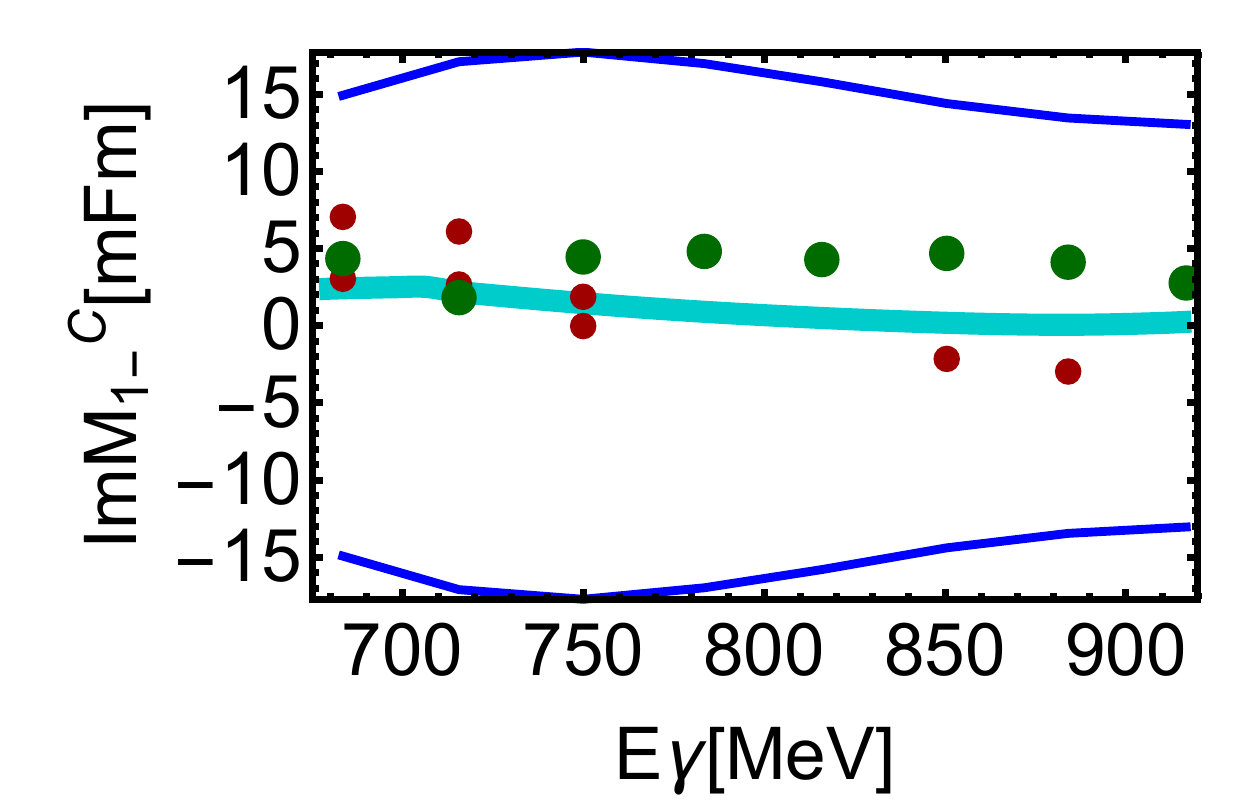}
 \end{overpic}
\begin{overpic}[width=0.325\textwidth]{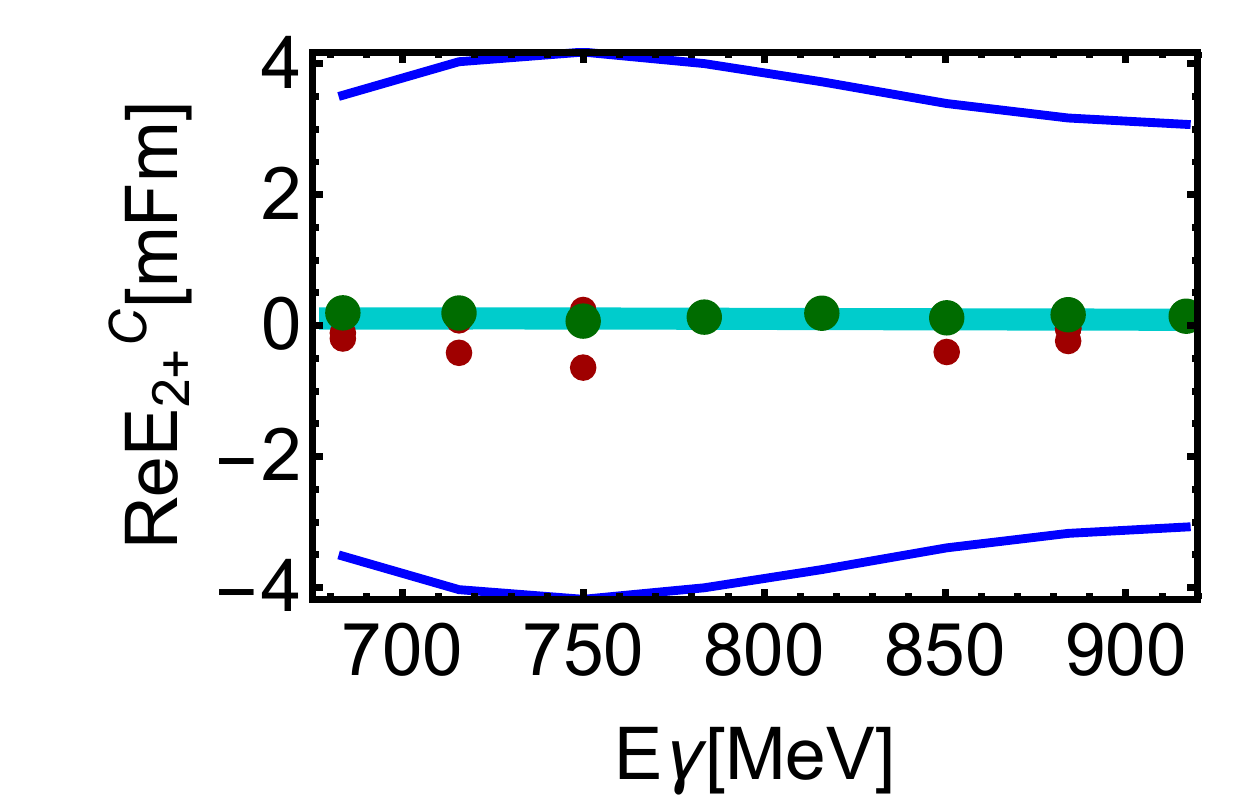}
 \end{overpic}
\begin{overpic}[width=0.325\textwidth]{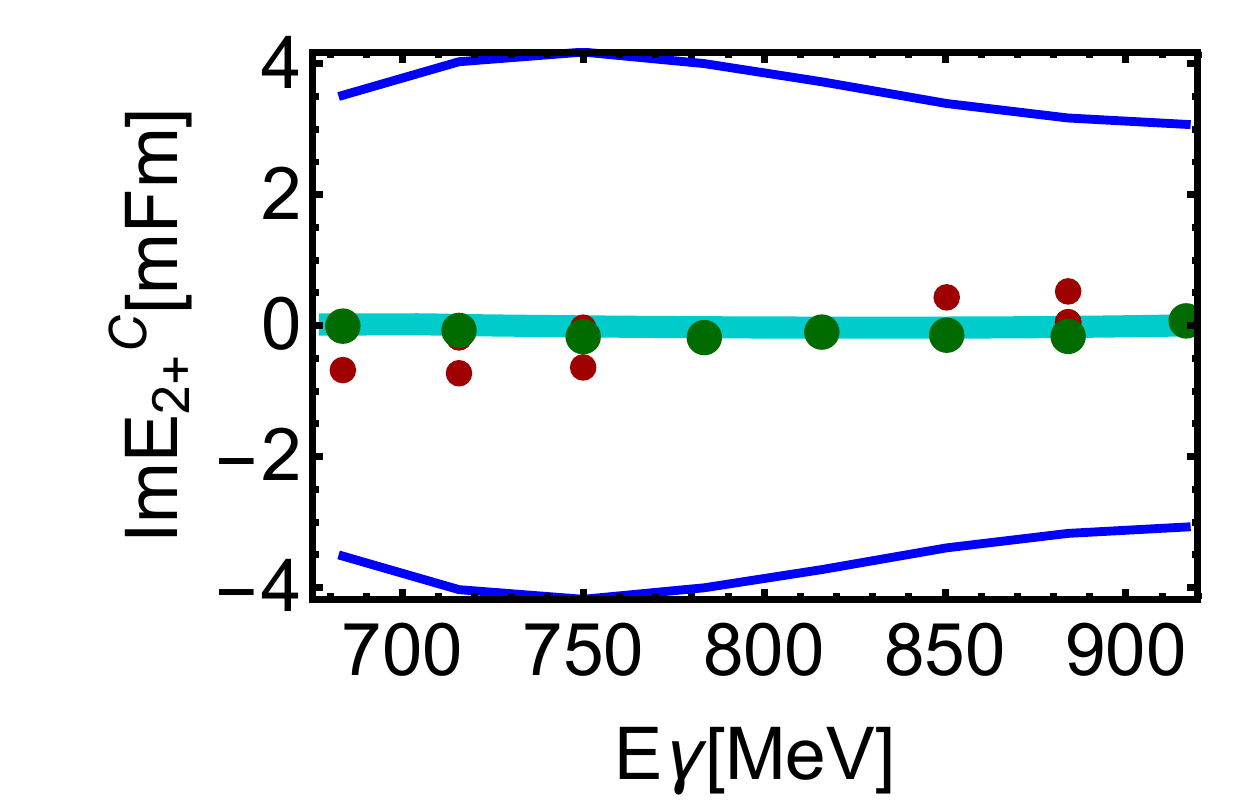}
 \end{overpic} \\
\begin{overpic}[width=0.325\textwidth]{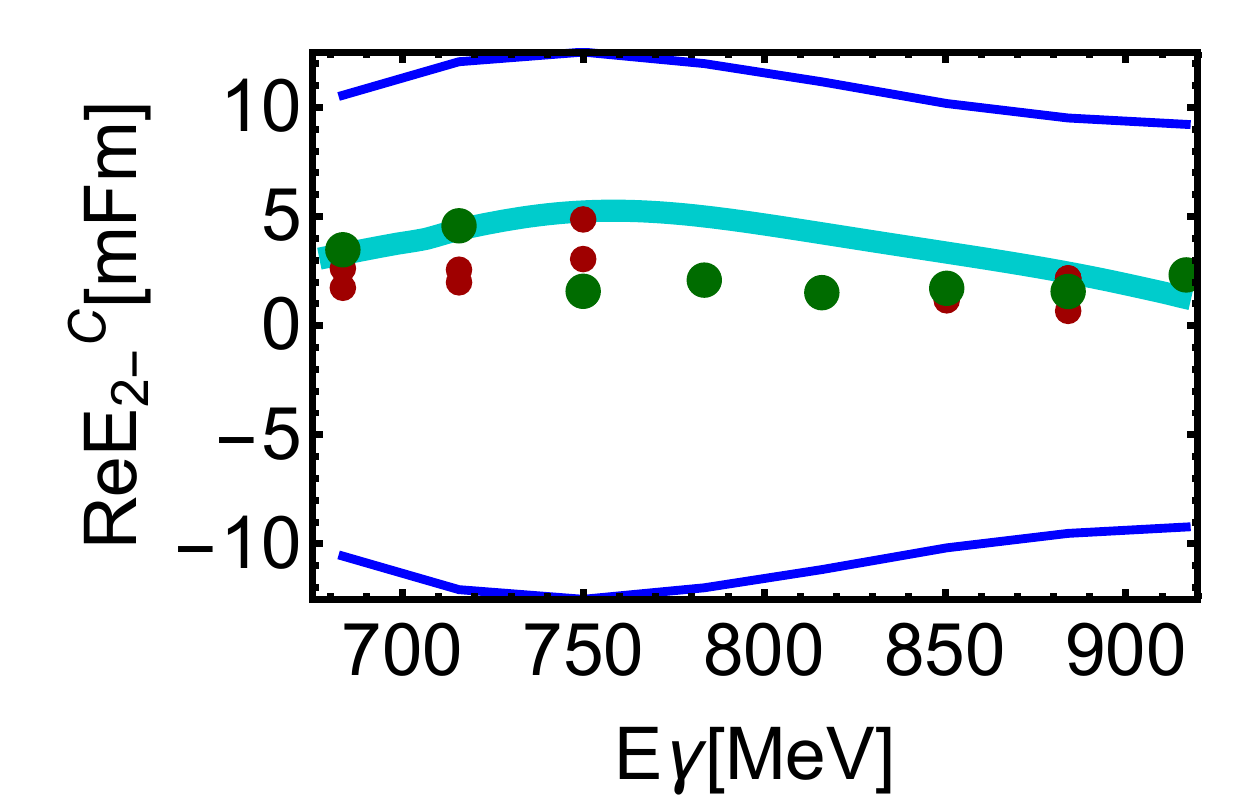}
 \end{overpic}
\begin{overpic}[width=0.325\textwidth]{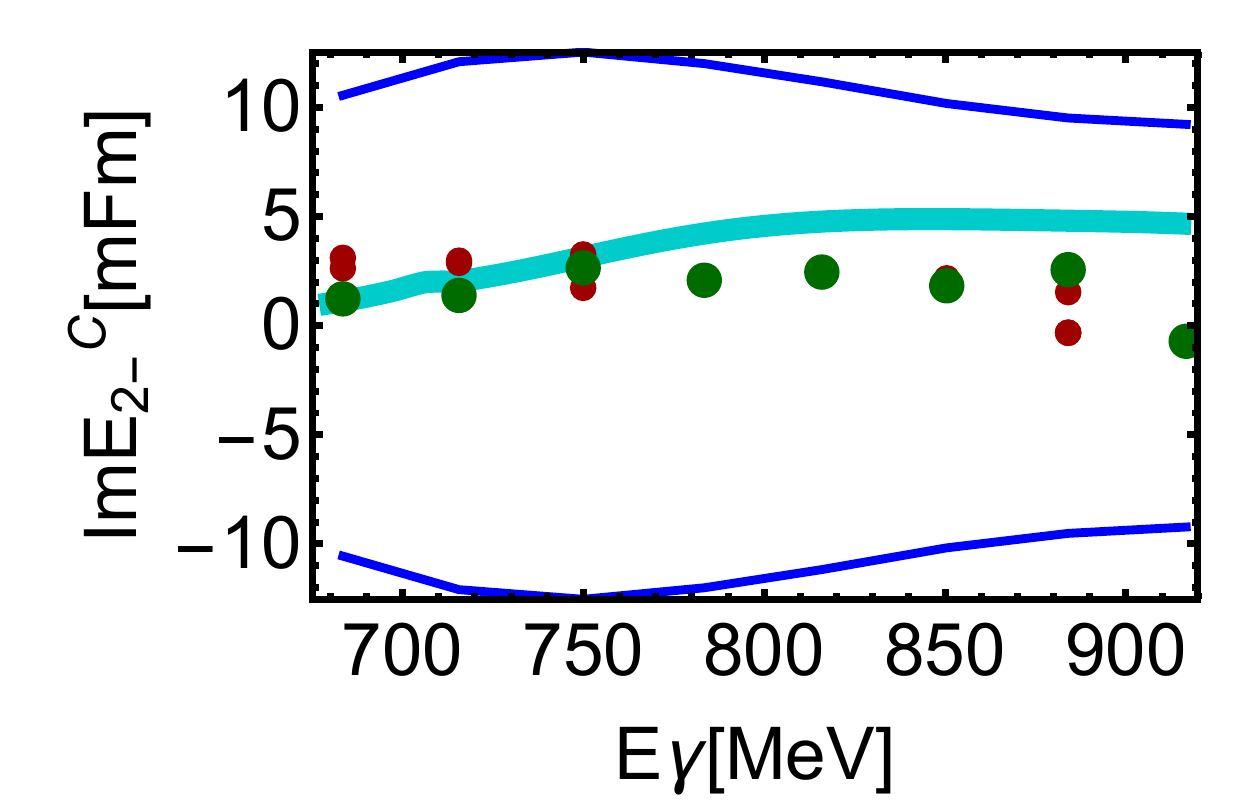}
 \end{overpic}
\begin{overpic}[width=0.325\textwidth]{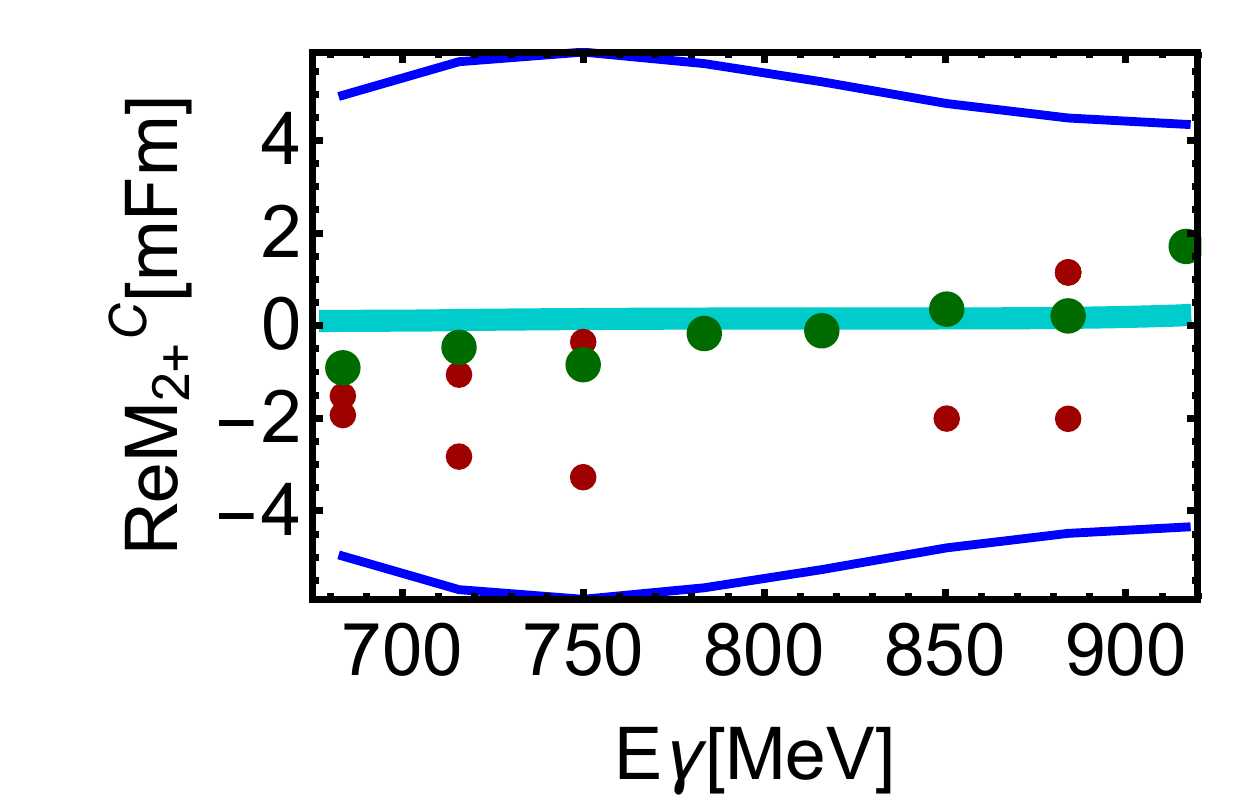}
 \end{overpic} \\
\begin{overpic}[width=0.325\textwidth]{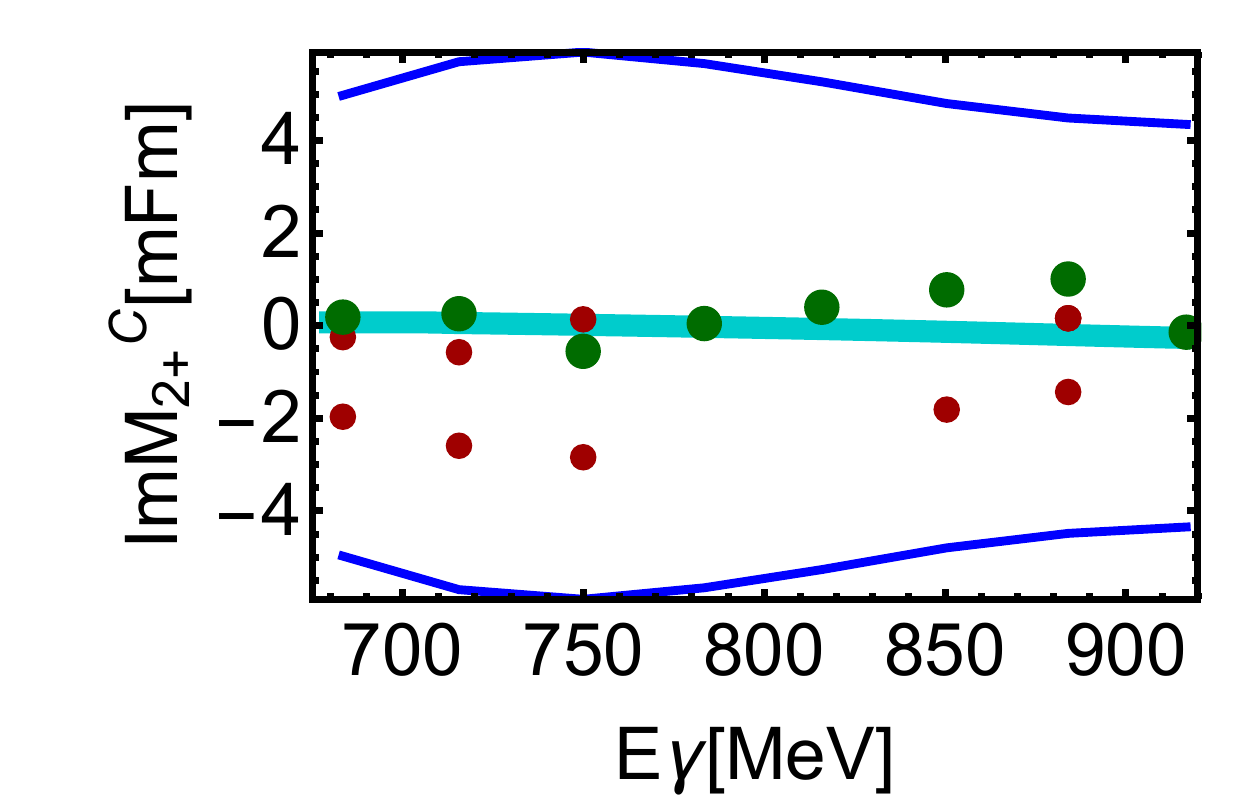}
 \end{overpic}
\begin{overpic}[width=0.325\textwidth]{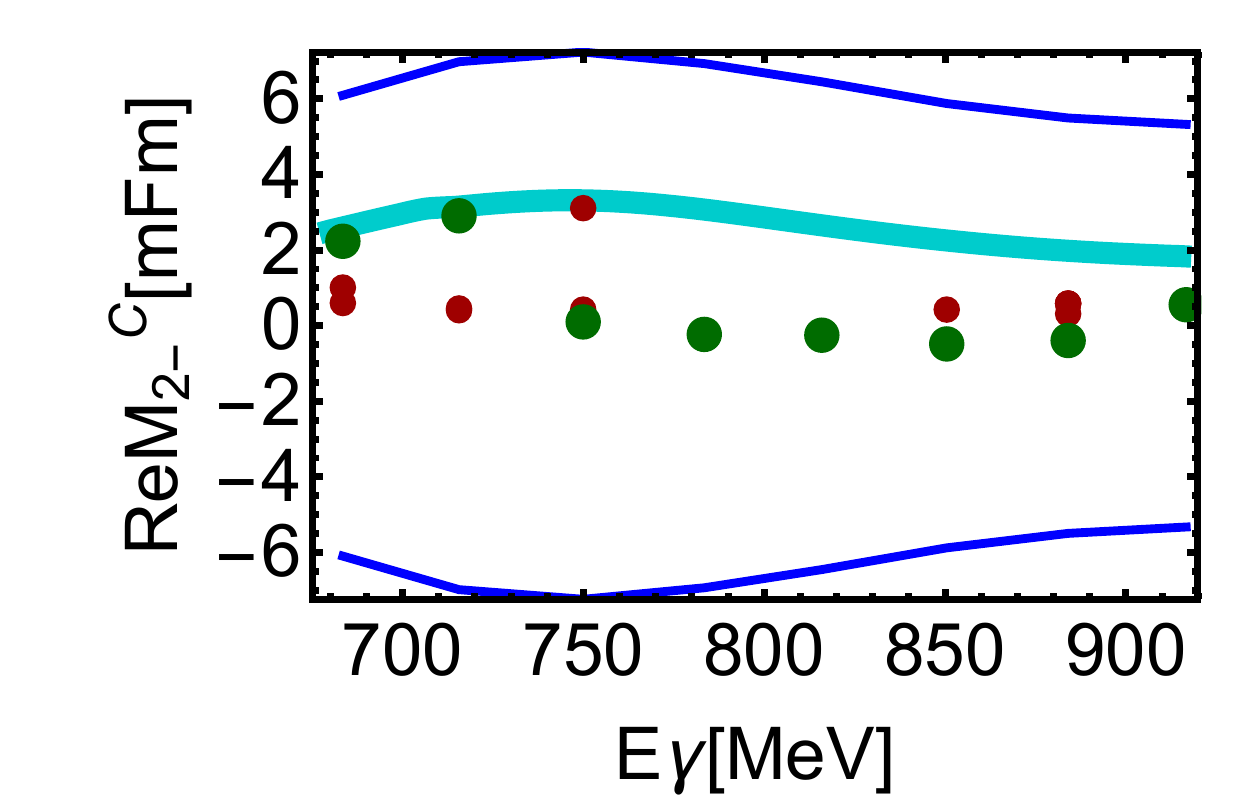}
 \end{overpic}
\begin{overpic}[width=0.325\textwidth]{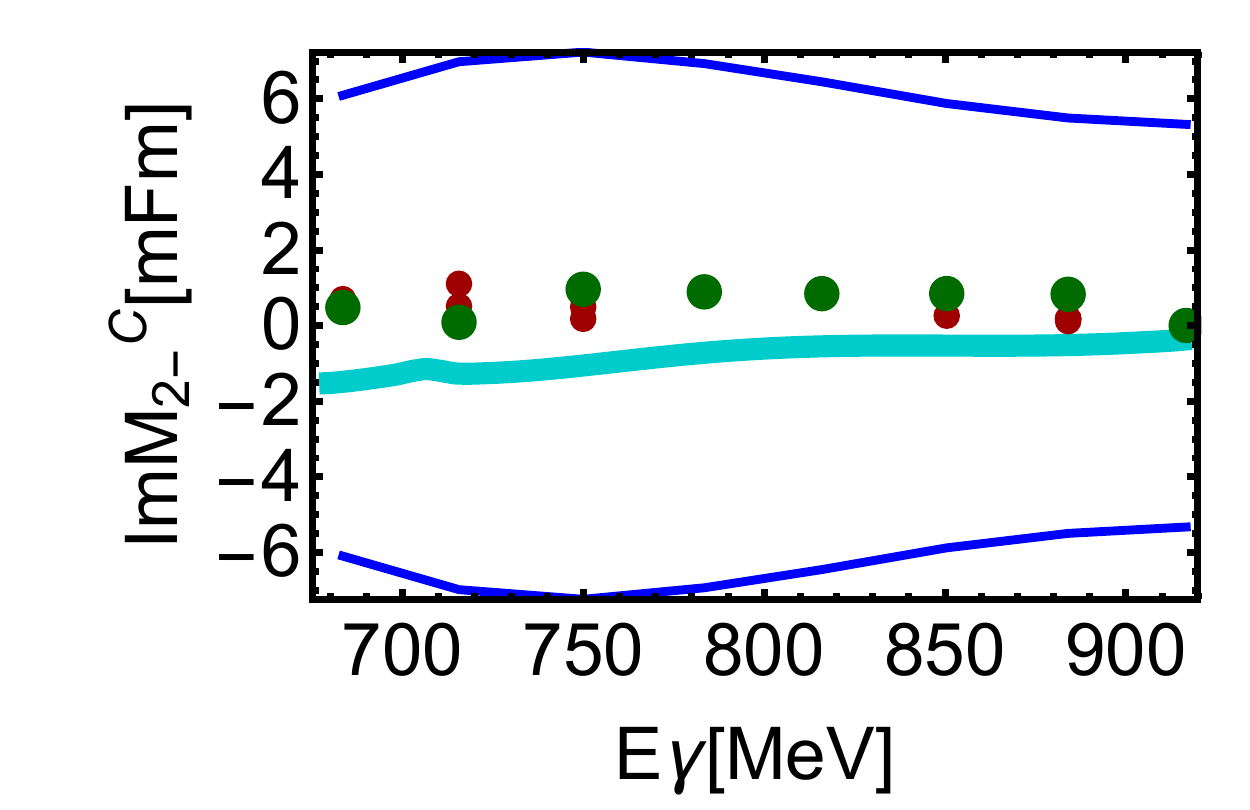}
 \end{overpic}
\caption[Results for the $15$ fit-parameters comprised of the real- and imaginary parts of phase-constrained $S$-, $P$- and $D$-wave multipoles, for a model-independent TPWA with $\ell_{\mathrm{max}} = 2$, within the $2^{\mathrm{nd}}$ resonance region.]{The plots show results for the $15$ parameters comprised of the real- and imaginary parts of the phase-constrained multipoles, for $\ell_{\mathrm{max}} = 2$. Results stem from the full Monte Carlo minimum-search in the second resonance-region, employing $N_{MC} = 16000$ start-configurations. The global minimum is indicated by the big green dots, while local minima are plotted as smaller red dots. Here, all local minima have been included that are within a range of $\chi^{2}_{\mathrm{Best}}/\mathrm{ndf} + 1$ from the global minimum $\chi^{2}_{\mathrm{Best}}$. As a comparison, the Bonn-Gatchina model-solution BnGa 2014\_02 \cite{BoGa} is shown as a thick cyan-colored curve. \newline
Furthermore, the plot-range has been adjusted in order indicate the maximal range for each parameter, constrained by the total cross section alone (cf. the discussion in section \ref{sec:MonteCarloSampling}). This range may of course vary with energy, and the borders of the range are shown a thin blue solid lines. Beyond these, it is impossible to obtain multipole-solutions.}
\label{fig:FirstFitLmax2MultipolesPlots}
\end{figure}

First, the minimal scenario compatible with the $\ell_{\mathrm{max}}$-analyses described above is attempted, i.e. we perform a TPWA-fit for $\ell_{\mathrm{max}} = 2$. Multipoles are fitted under the standard phase-constraint (cf. section \ref{sec:TPWAFitsIntro}), which leads to $(8 * 2 - 1) = 15$ free parameters for the real- and imaginary parts. This is tried in spite of the fact that for some energies, the $\chi^{2}$-values resulting from fits to the angular distributions only (see Figure \ref{fig:2ndResRegionChisquareLmaxPlots}) already suggested a need for $\ell_{\mathrm{max}} = 3$. \newline
For the $8$ energies analyzed here, the correlated chisquare (\ref{eq:CorrelatedChisquareQuotedForAnalysisII}) has been minimized using the Monte Carlo-technique outlined in section \ref{sec:MonteCarloSampling}. For this $D$-wave truncation, a total of $N_{MC} = 16000$ sampling-points have been employed. Then, all non-redundant solutions have been sorted out of the pool of $N_{MC}$ final parameter configurations. Results are shown in Figures \ref{fig:FirstFitLmax2ChiSquarePlot}, where the $\chi^{2}/\mathrm{ndf}$ for the best attained minima is plotted, as well as Figure \ref{fig:FirstFitLmax2MultipolesPlots} which shows the corresponding results for the multipole-parameters. \newline
Using the seven observables in the second resonance region, it has been possible to obtain a global minimum in the fit for $\ell_{\mathrm{max}} = 2$. This is in accordance with the theoretical results of chapter \ref{chap:Omelaenko}. When comparing to the theoretical chisquare distribution suitable for this fit, it is seen that almost none of the obtained global minima are acceptable fits.
\newpage

Figure \ref{fig:FirstFitLmax2ChiSquarePlot} shows that for the lowest two energies, the global minimum is practically touching the theoretical $0.975$-quantile and thus the absolute upper border for acceptable fits. For all the remaining higher energies, the best solution is at least 2 units (in $\chi^{2}/\mathrm{ndf}$) above this quantile. Therefore, while it is a nice feature of this fit to yield global minima, practically all of them have to be rejected for probability-theoretical reasons. Furthermore, for all energies except for the fourth and fifth bin, some local minima exist which are relatively close, in $\chi^{2}$, to the unique best solution. These are most likely remnants of the instabilities caused by accidental ambiguities and can, already for $\ell_{\mathrm{max}}=2$, endanger the uniqueness. \newline
Interesting effects can also be observed when considering the values of the resulting multipole-parameters, Figure \ref{fig:FirstFitLmax2MultipolesPlots}. We plot them here in comparison to the solution BnGa 2014\_02 \cite{BoGa} obtained in an energy-dependent fit by the Bonn-Gatchina group. This already represents an acceptable model in the considered energy region. However, we see that especially for those energies where the global minimum had quite bad $\chi^{2}$, the corresponding fit-parameters completely miss the Bonn-Gatchina model. This is only different in the lowest two energy-bins. Here, some multipoles, for instance $E_{2-}$ or $M_{1+}$, come out not so bad. For all energies, the $S$-wave $E_{0+}$ is clearly missing some strength, such that it can only rise to values approximately half of the BnGa-model. It will soon be seen that problems with missing strength in the $S$-wave can in the end be attributed to the ignorance of higher partial waves. 

%

\clearpage

 Local minima are shown in the plots, but we have confined the attention to those within a range of $\chi^{2}_{\mathrm{Best}}/\mathrm{ndf} + 1$ from the global minimum $\chi^{2}_{\mathrm{Best}}/\mathrm{ndf}$, which occur for all energies except the fourth, fifth and eighth bin. These local minima look, in their energy-dependence, like splitting and later (probably) re-joining branches of potentially ambiguous solutions. The global minimum is then \textit{jumping} among those branches of solutions, as can be seen in the transition from the second to the third energy-bin. Some of these local minima are, in the third energy bin, closer to the Bonn-Gatchina model, although their chisquare is a bit worse than the global minimum. \newline
The rapid decrease in fit-quality for the higher energies visible in Figure \ref{fig:FirstFitLmax2ChiSquarePlot} is caused by the neglect of the $F$-waves in this fit and can be attributed to interferences among higher partial waves, with small modulus and lower dominant waves. This effect has been discussed at length in section \ref{sec:LFitsPaper}. Here, while the $F$-waves were not hinted at strongly in the fits of Legendre-parametrizations to angular distributions (cf. Figure \ref{fig:2ndResRegionChisquareLmaxPlots} in the previous paragraph), the effect of neglecting them is quite extreme in the multipole-fit itself. We therefore choose to investigate the same data in a TPWA with truncation order raised by one. \newline

For the increased truncation order, a fit has been preformed analogously to the previous case. However, for $\ell_{\mathrm{max}} = 3$ the number of free parameters amounts to $(8 * 3 - 1) = 23$. Thus, the number of possible ambiguities rises and therefore forces a need to increase the number $N_{MC}$ of initial parameter-configurations for the minimum-search. Here, we employed $N_{MC} = 32000$.

\vfill

\begin{figure}[hb]
 \centering
\begin{overpic}[width=0.495\textwidth]{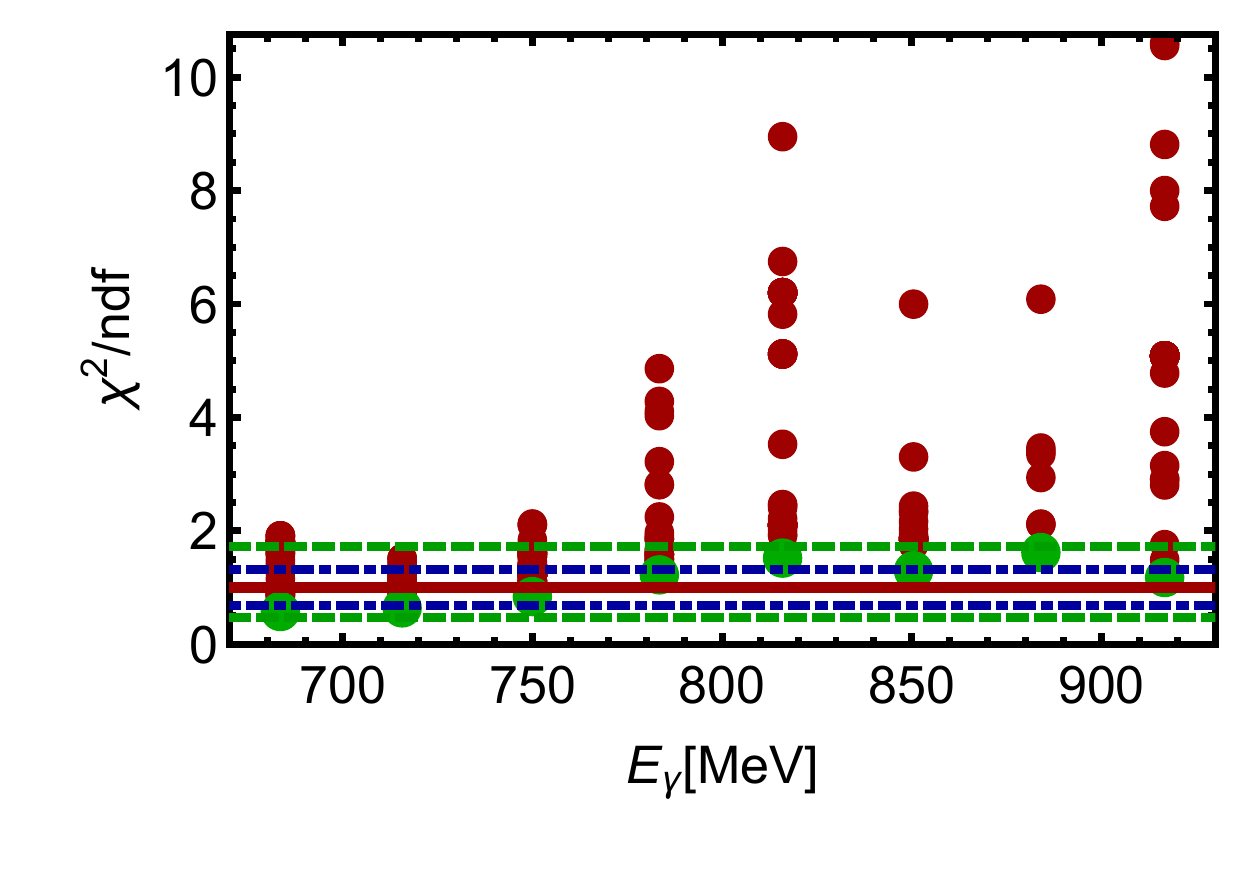}
 \end{overpic}
\begin{overpic}[width=0.495\textwidth]{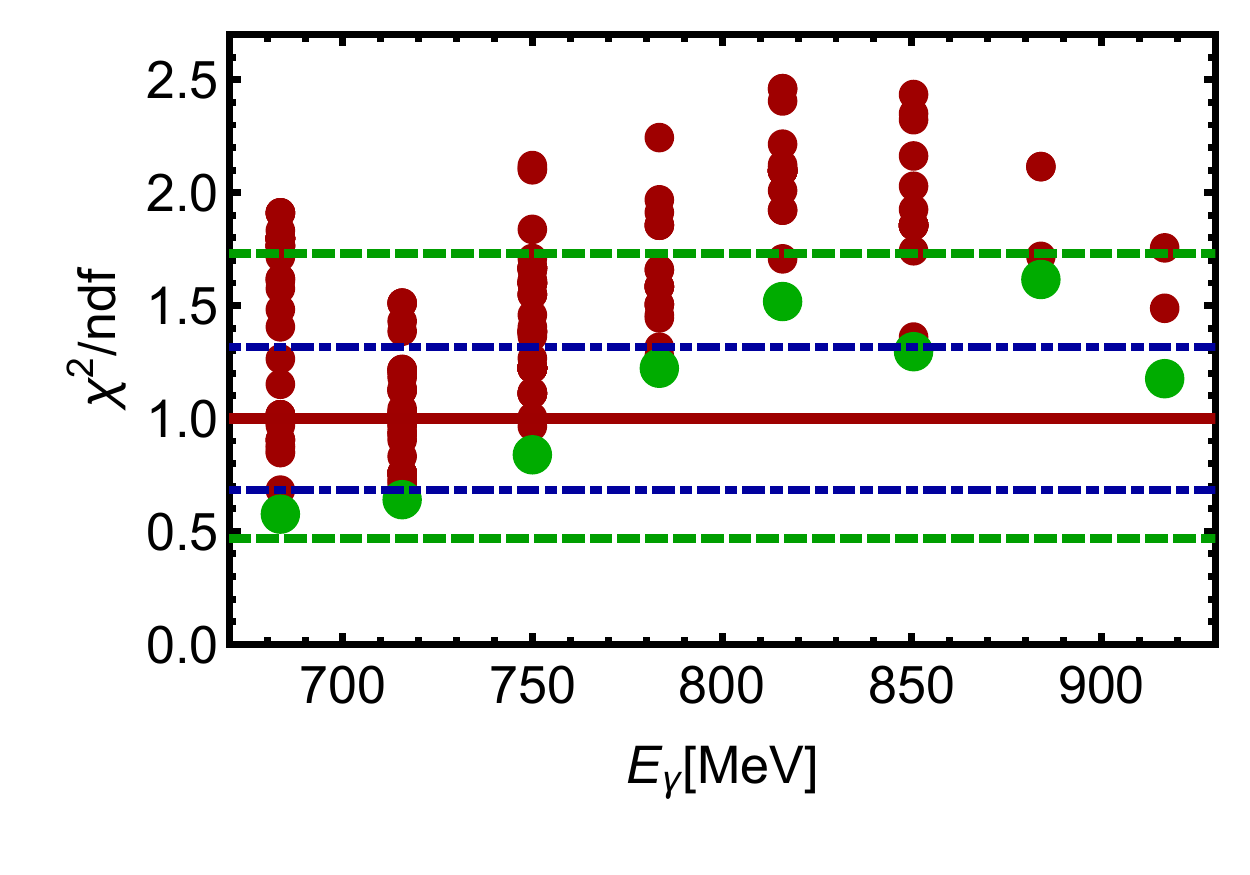}
 \end{overpic}
\vspace*{-15pt}
\caption[The $\chi^{2}/\mathrm{ndf}$ for the best results for the minimum of the correlated chisquare-function, coming from a full Monte Carlo minimum-search applied in the truncation $\ell_{\mathrm{max}} = 3$ within the $2^{\mathrm{nd}}$ resonance region.]{Both plots shows the best results for the full Monte Carlo minimum-search applied in the truncation $\ell_{\mathrm{max}} = 3$, once for a relatively wide plot-range (left) and in a more detailed picture showing only the global minimum and local minima close to it (right). The results stem from a pool of $N_{MC} = 32000$ initial parameter-configurations. The global minimum is indicated by the big green dots, other local minima are plotted as smaller red-colored dots. \newline
In addition, some information on the theoretical chisquare distribution for $\mathrm{ndf} = r = 19$ (which corresponds to $\ell_{\mathrm{max}} = 3$) is included via the horizontal lines. The mean is drawn as a red solid line, while the pair of $0.025$- and $0.975$-quantiles is indicated by green dashed lines and that made of the $0.16$- and $0.84$-quantiles by a blue dash-dotted line.}
\label{fig:SecondFitLmax3ChiSquarePlots}
\end{figure}
\begin{figure}[h]
 \centering
\begin{overpic}[width=0.325\textwidth]{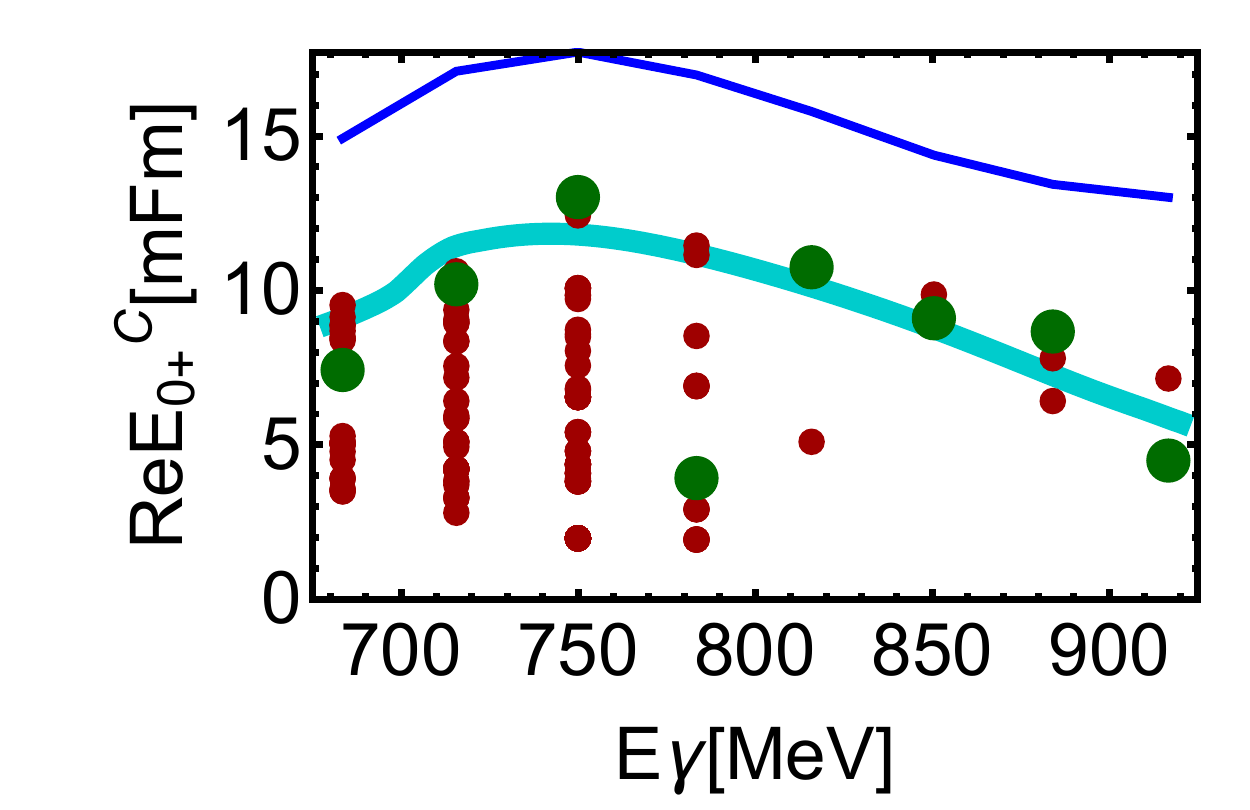}
 \end{overpic}
\begin{overpic}[width=0.325\textwidth]{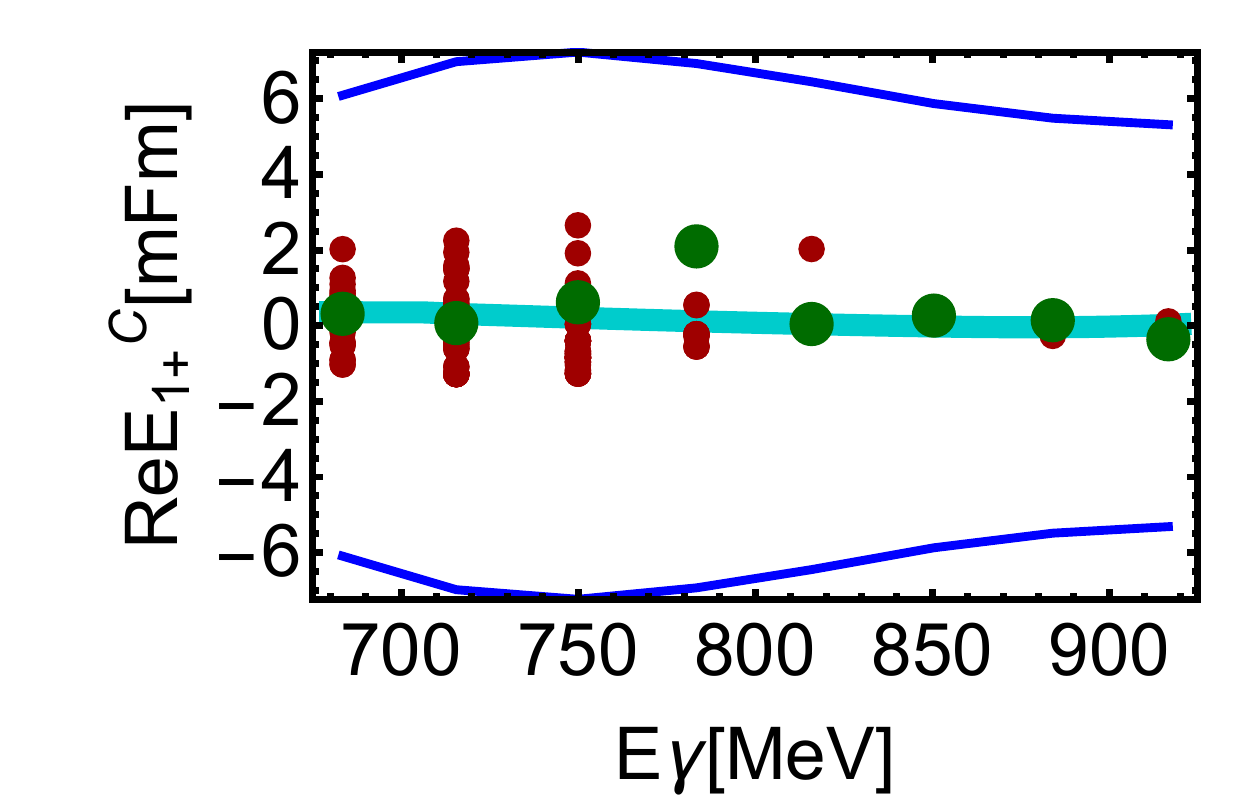}
 \end{overpic}
\begin{overpic}[width=0.325\textwidth]{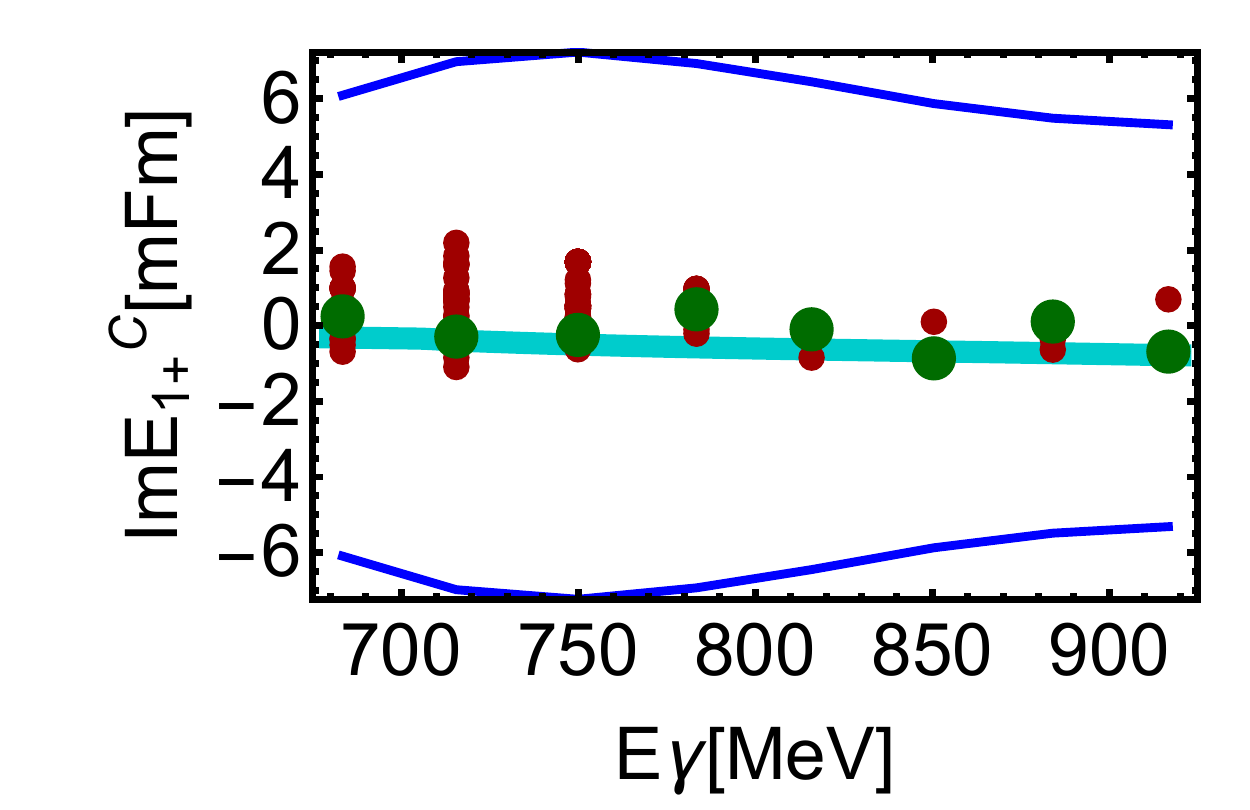}
 \end{overpic} \\
\begin{overpic}[width=0.325\textwidth]{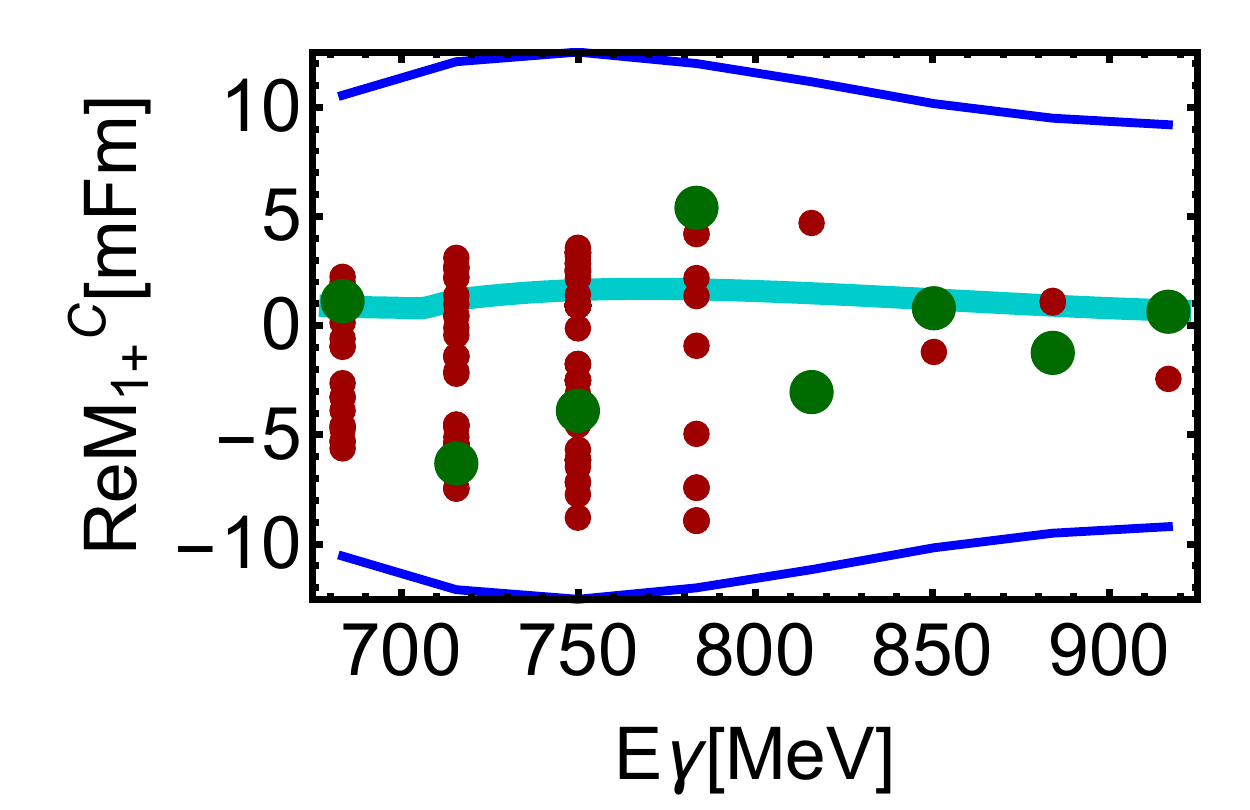}
 \end{overpic}
\begin{overpic}[width=0.325\textwidth]{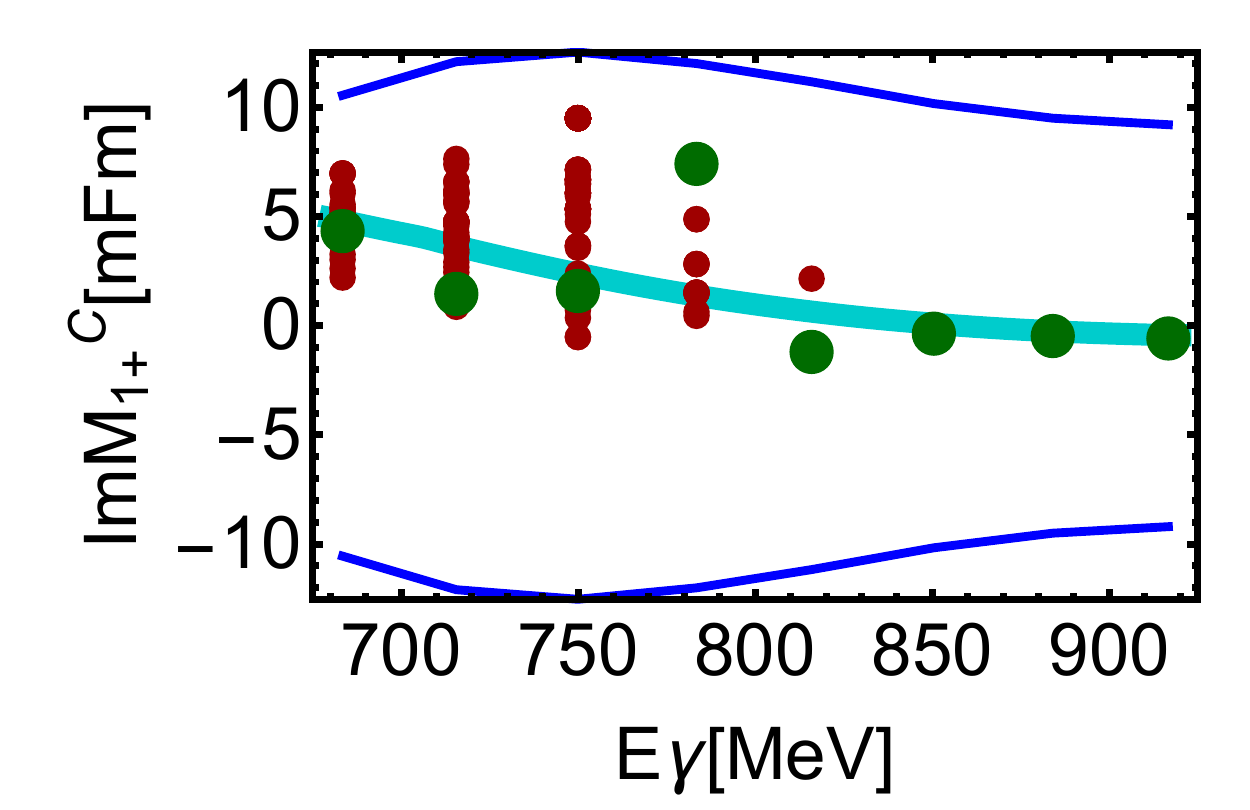}
 \end{overpic}
\begin{overpic}[width=0.325\textwidth]{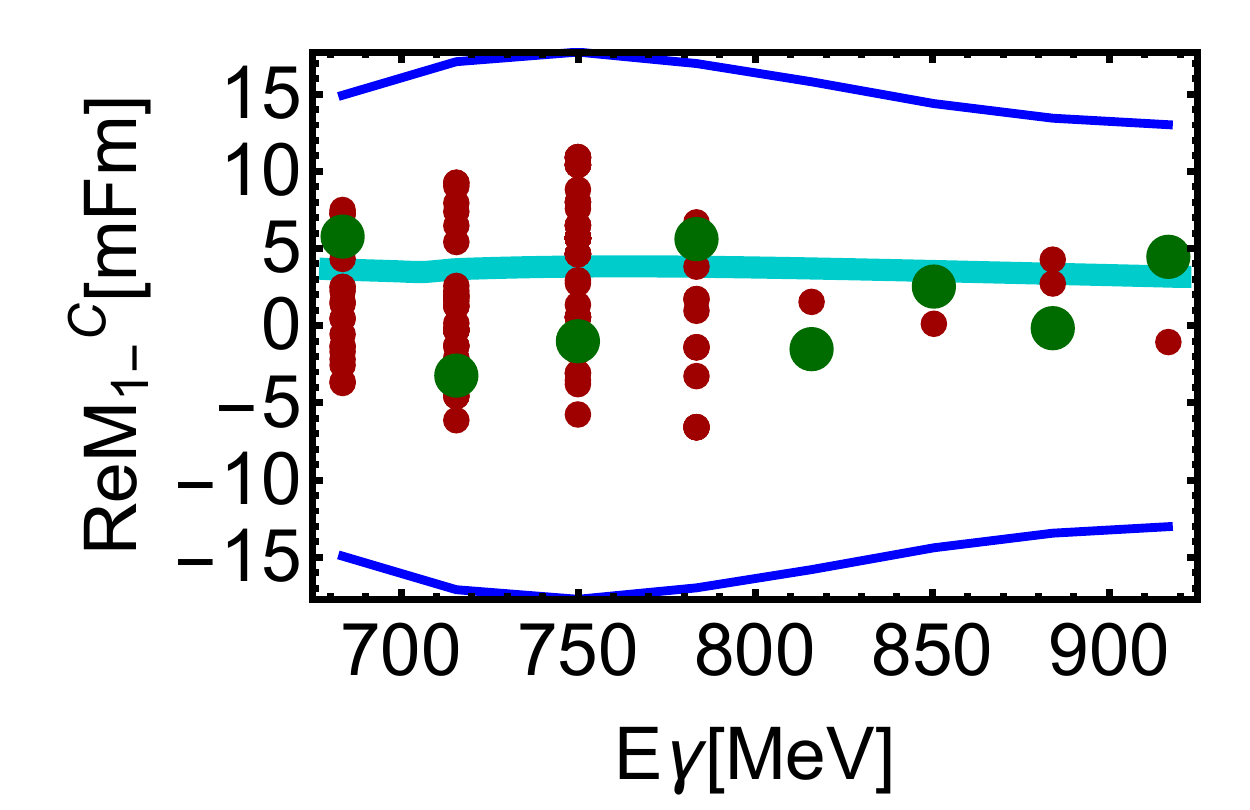}
 \end{overpic} \\
\begin{overpic}[width=0.325\textwidth]{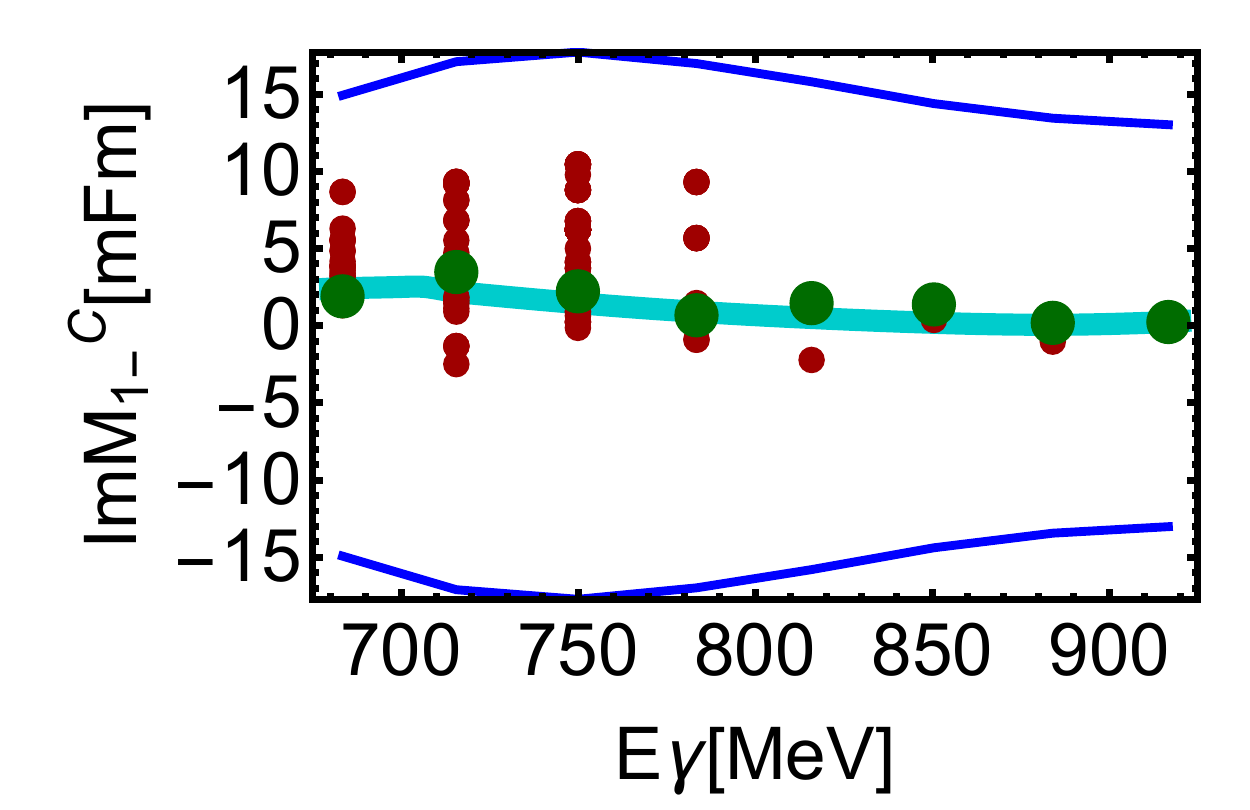}
 \end{overpic}
\begin{overpic}[width=0.325\textwidth]{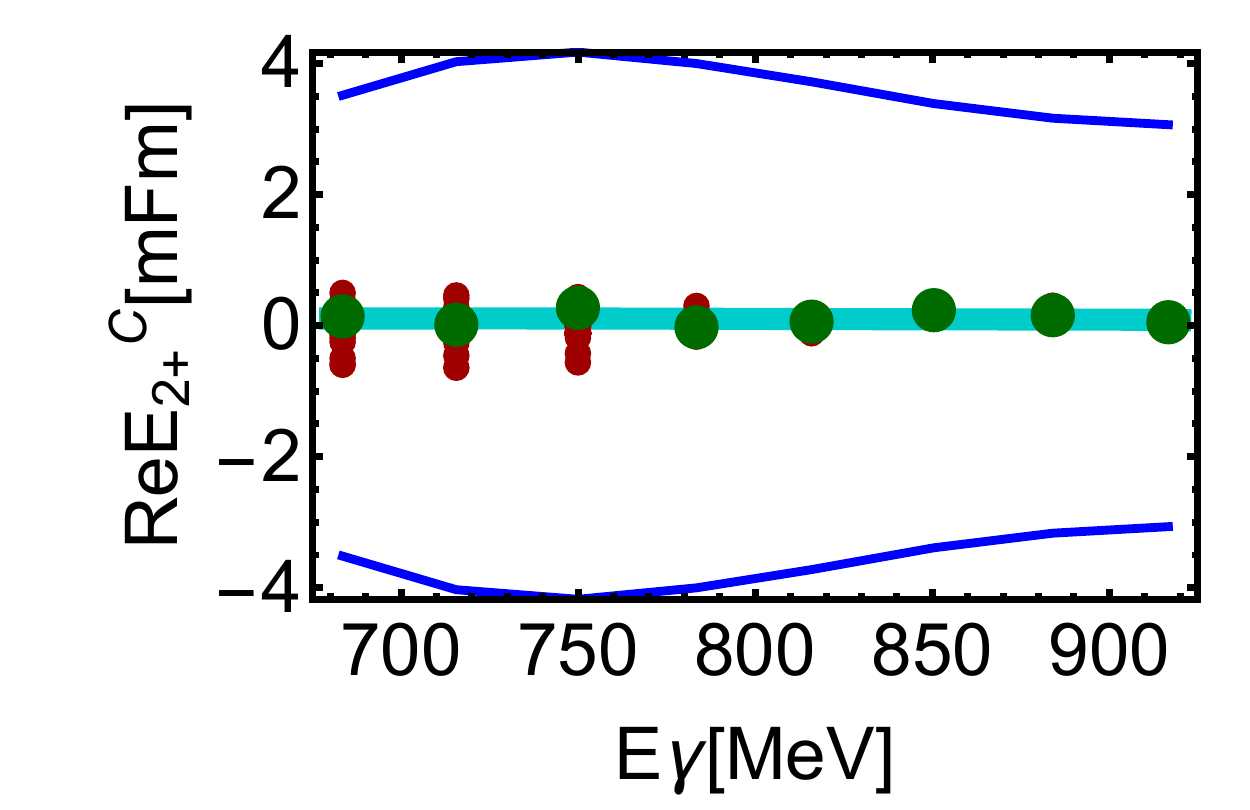}
 \end{overpic}
\begin{overpic}[width=0.325\textwidth]{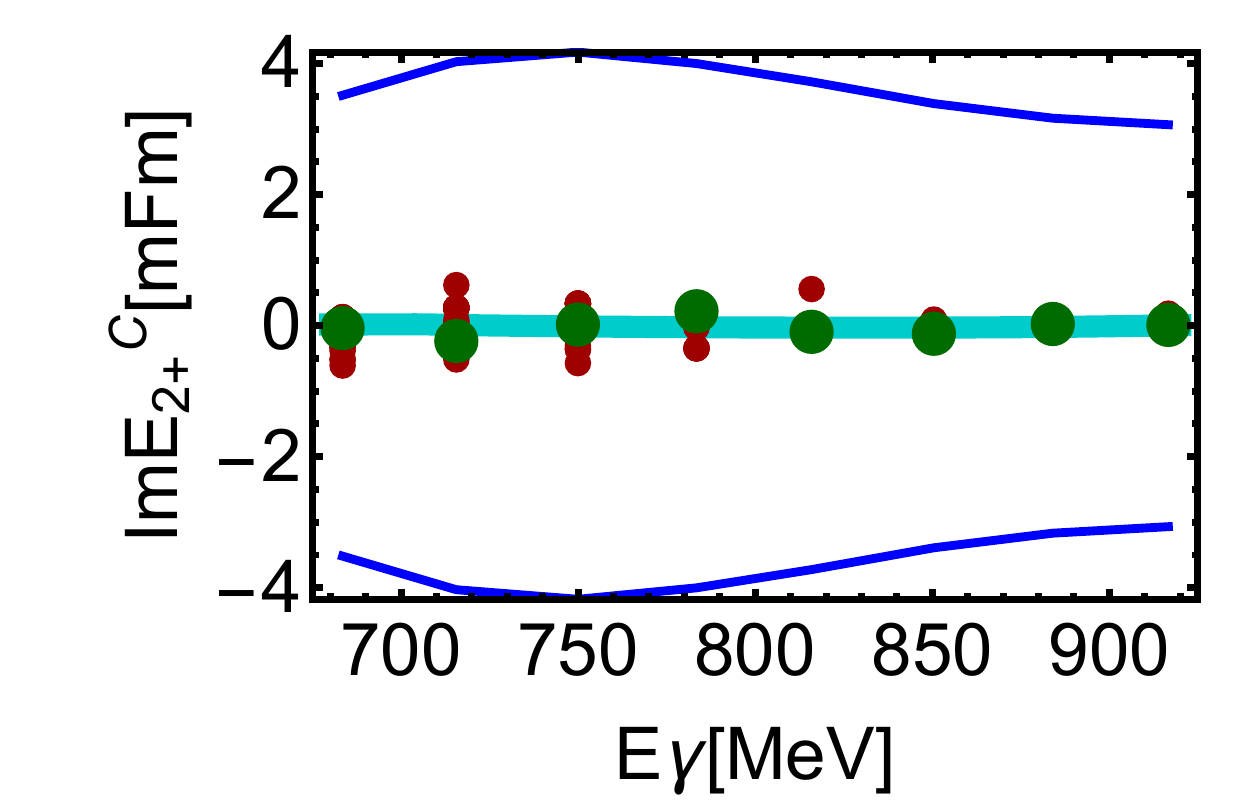}
 \end{overpic} \\
\begin{overpic}[width=0.325\textwidth]{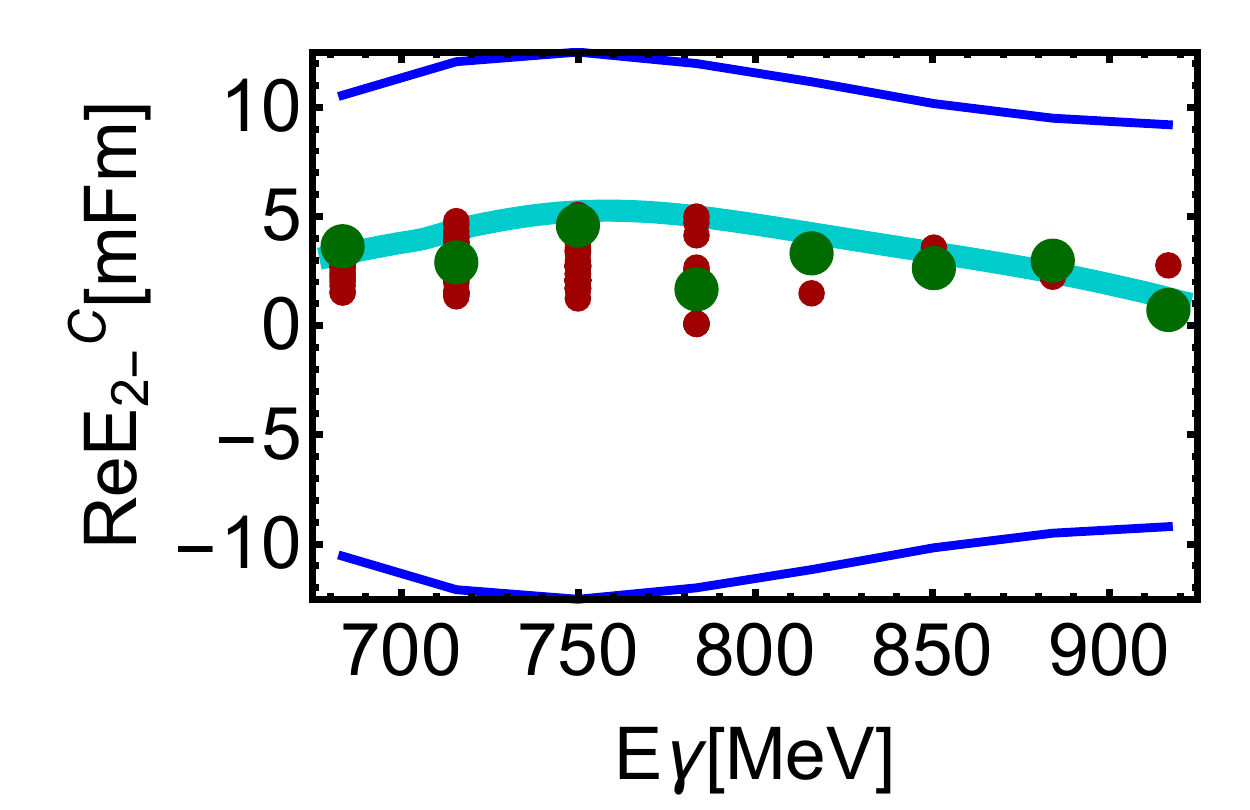}
 \end{overpic}
\begin{overpic}[width=0.325\textwidth]{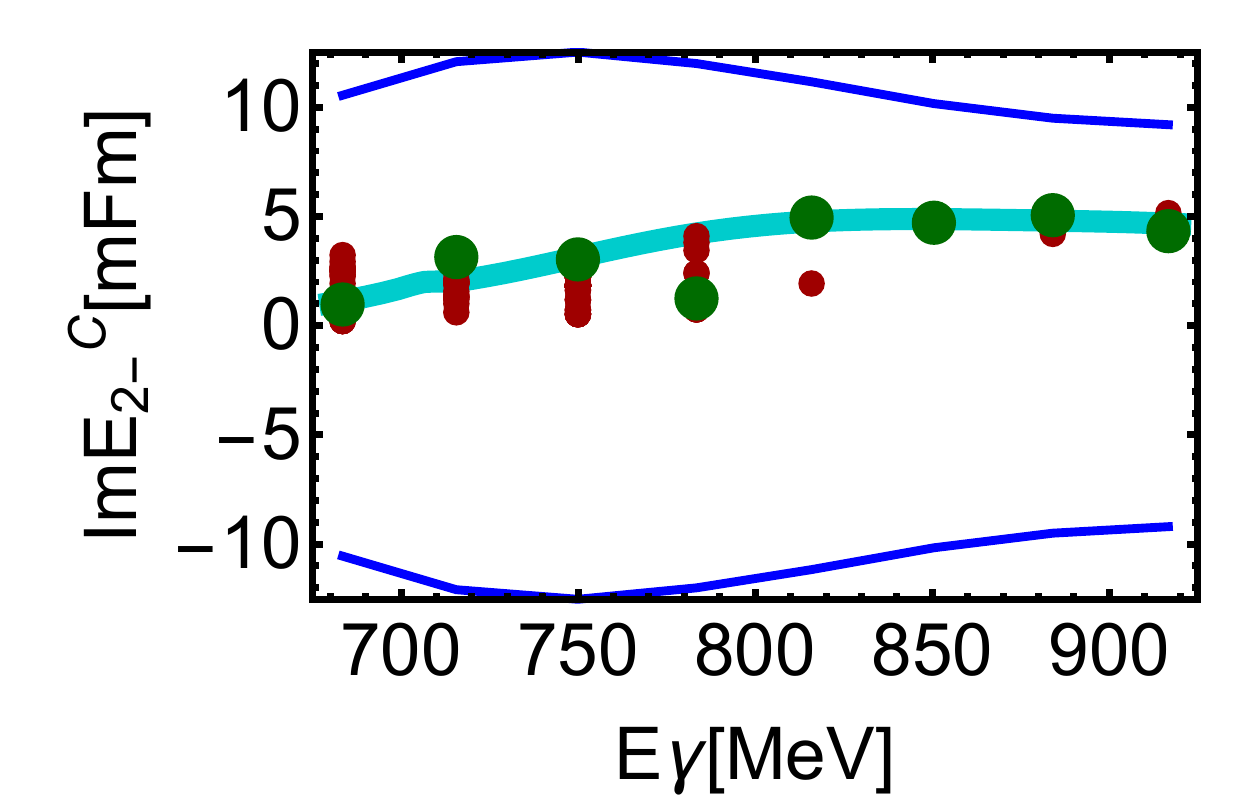}
 \end{overpic}
\begin{overpic}[width=0.325\textwidth]{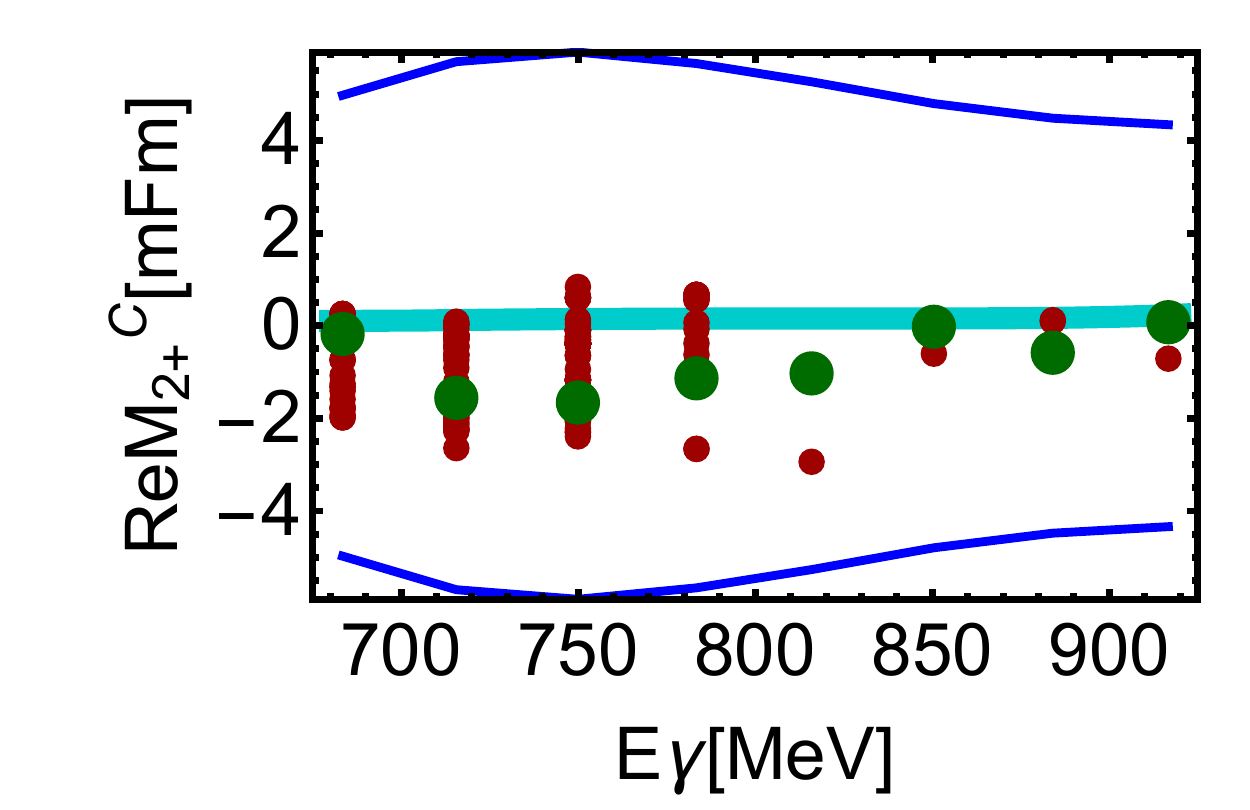}
 \end{overpic} \\
\begin{overpic}[width=0.325\textwidth]{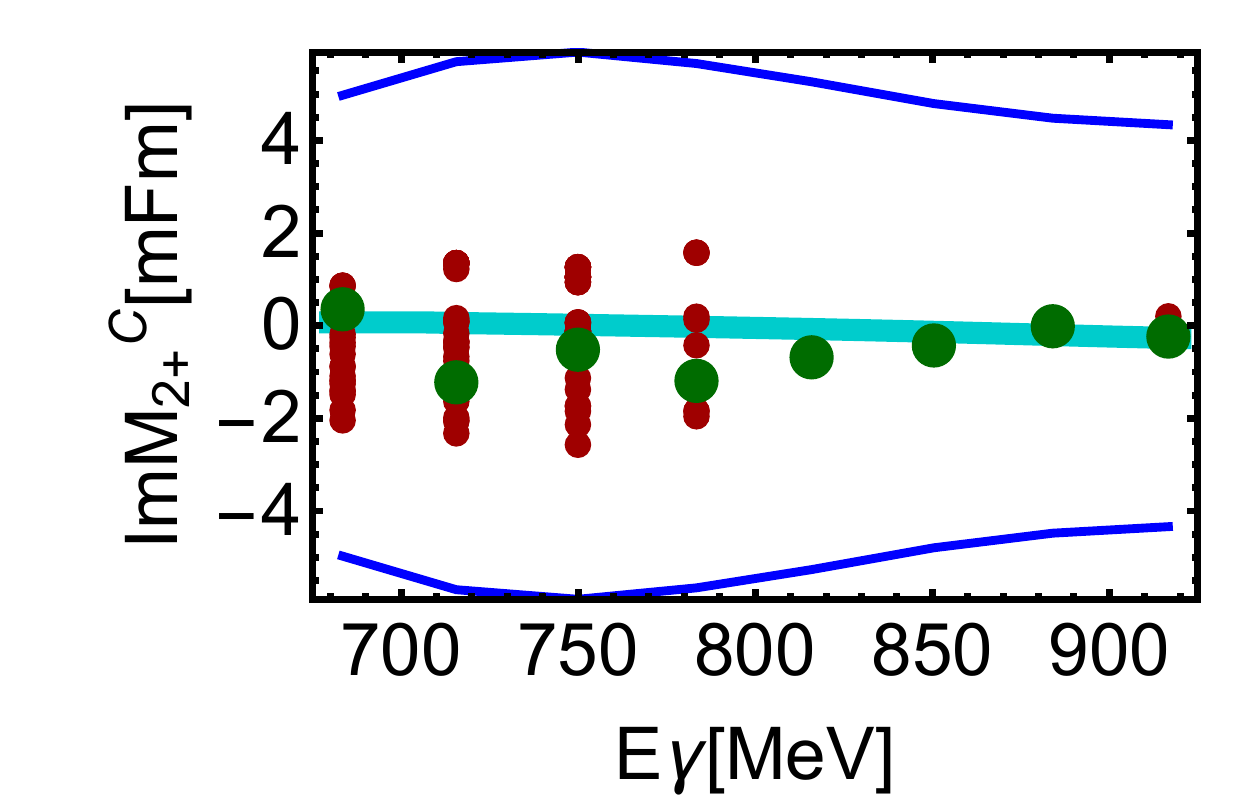}
 \end{overpic}
\begin{overpic}[width=0.325\textwidth]{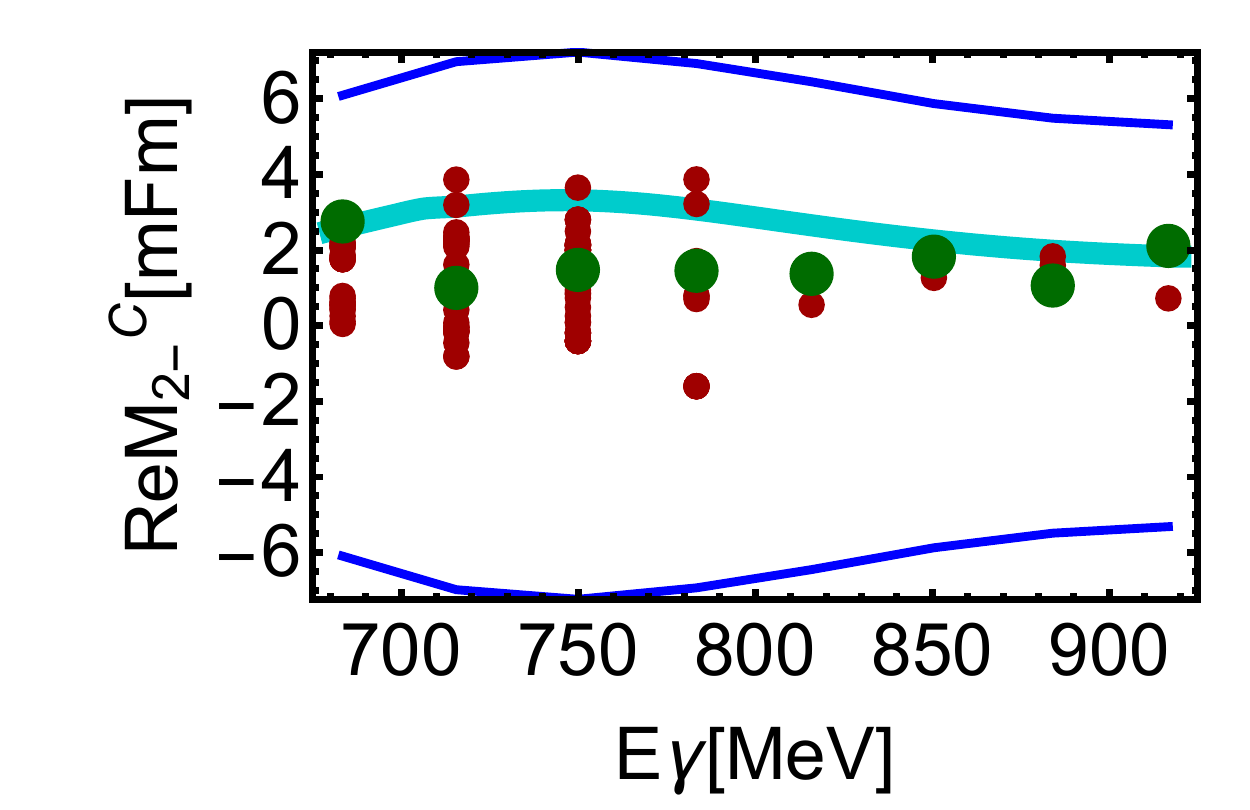}
 \end{overpic}
\begin{overpic}[width=0.325\textwidth]{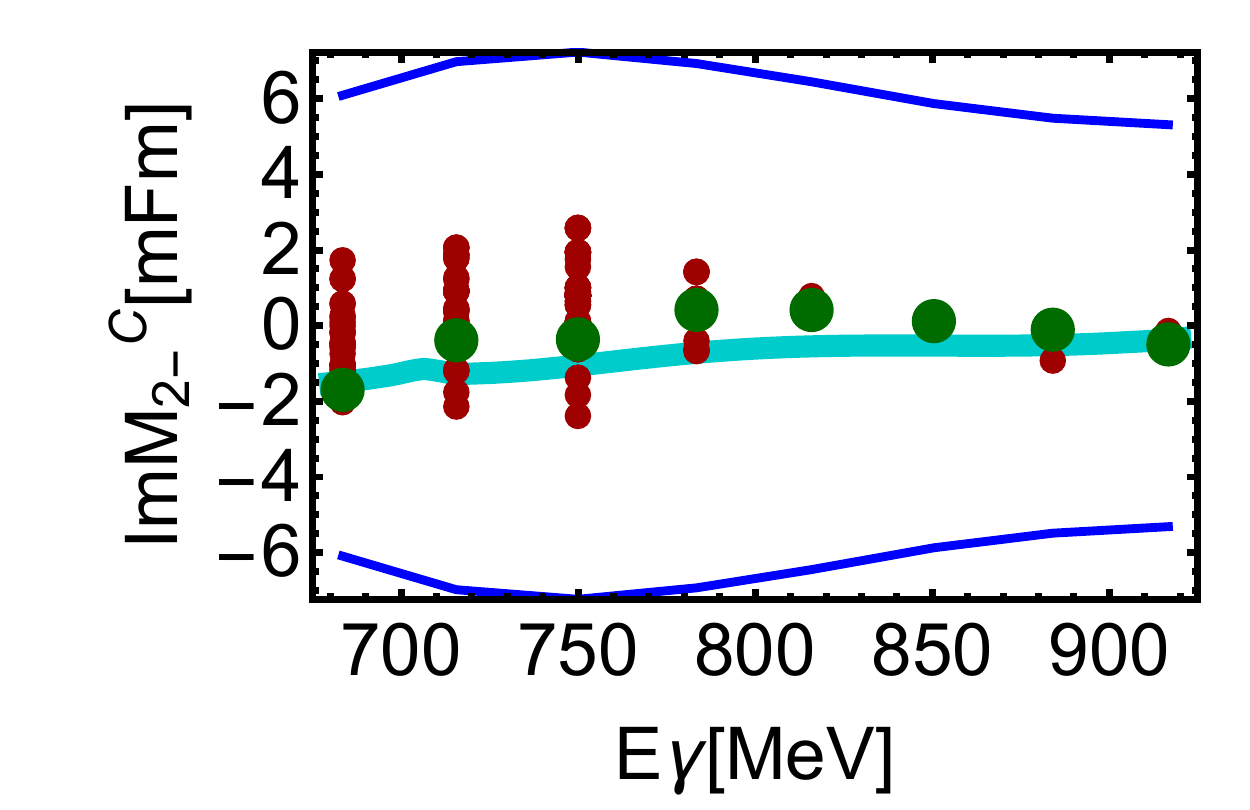}
 \end{overpic} \\
\begin{overpic}[width=0.325\textwidth]{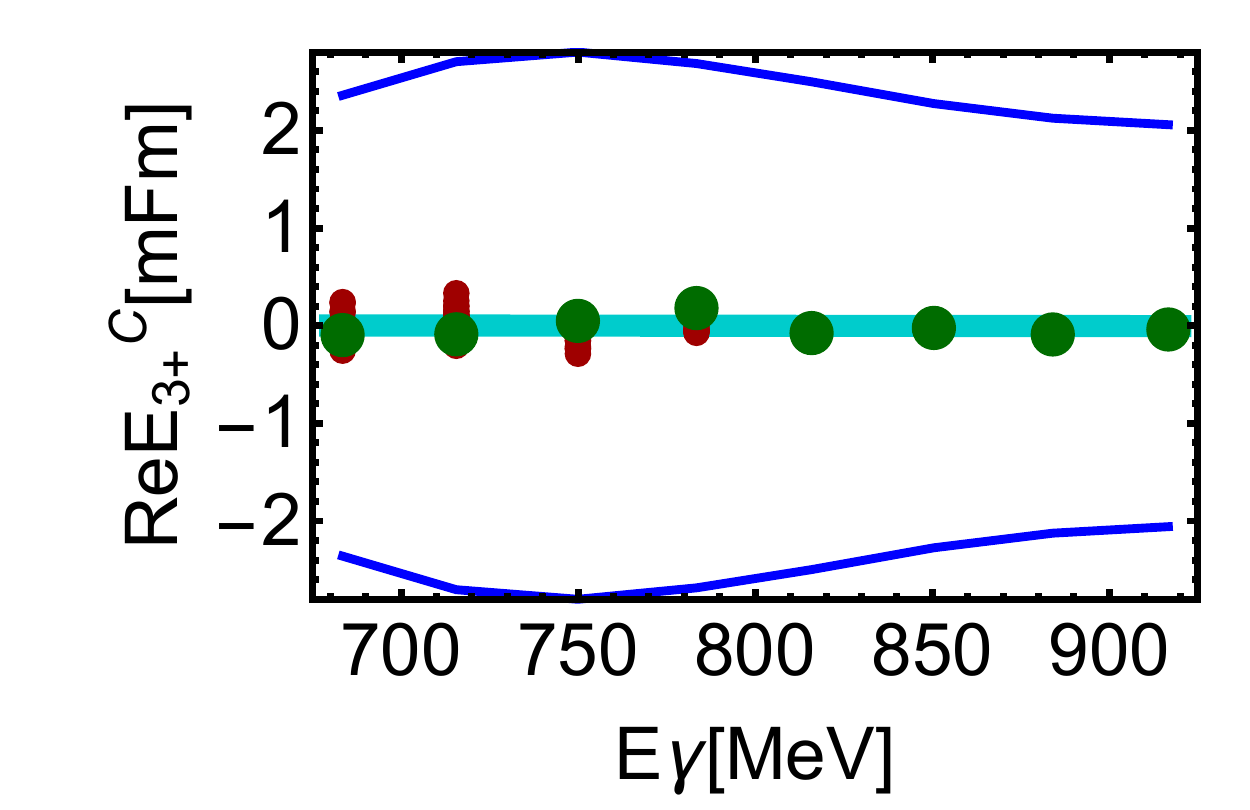}
 \end{overpic}
\begin{overpic}[width=0.325\textwidth]{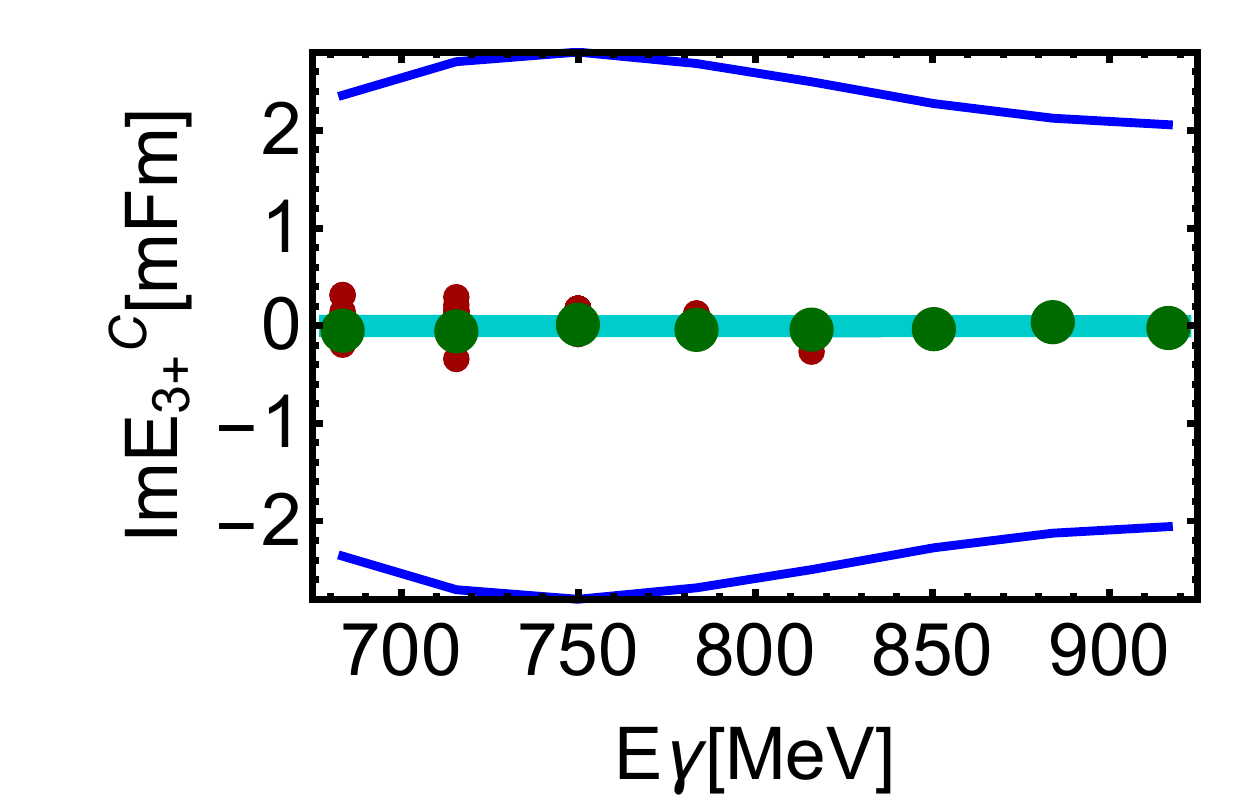}
 \end{overpic}
\begin{overpic}[width=0.325\textwidth]{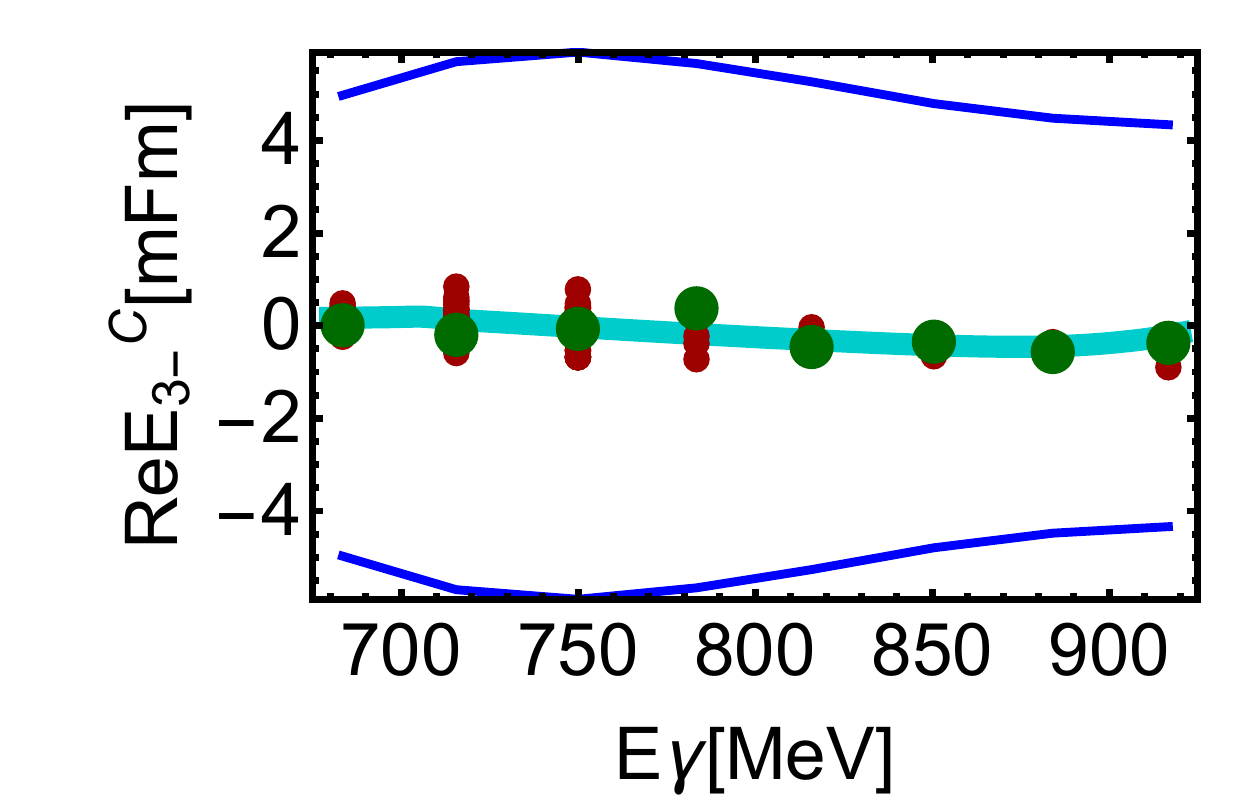}
 \end{overpic} \\
\begin{overpic}[width=0.325\textwidth]{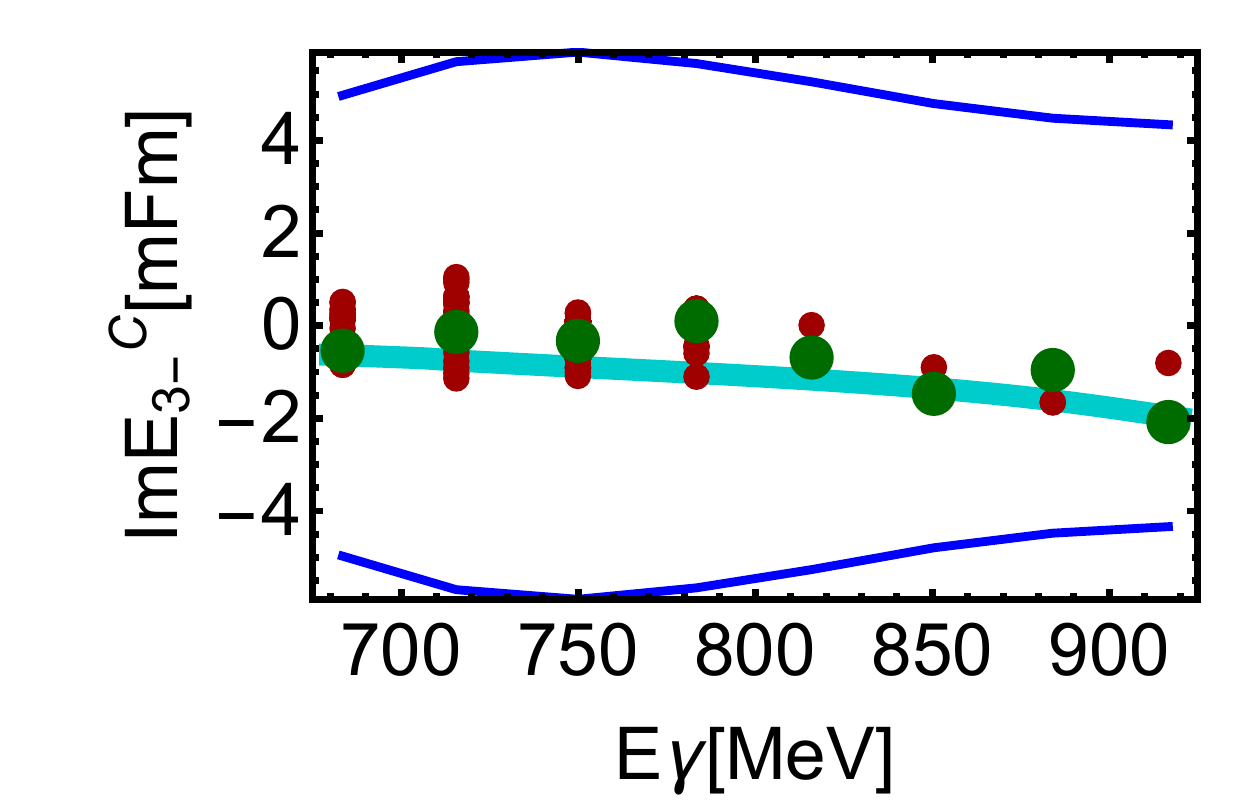}
 \end{overpic}
\begin{overpic}[width=0.325\textwidth]{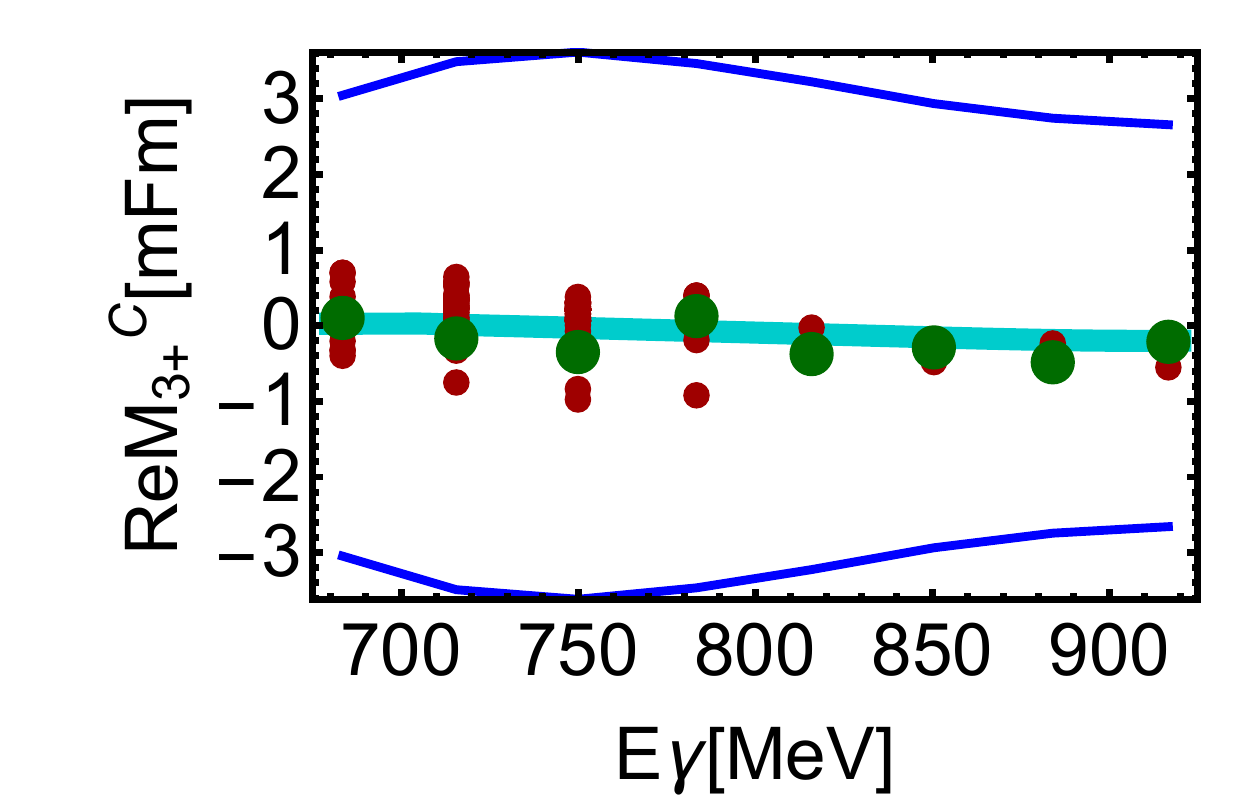}
 \end{overpic}
\begin{overpic}[width=0.325\textwidth]{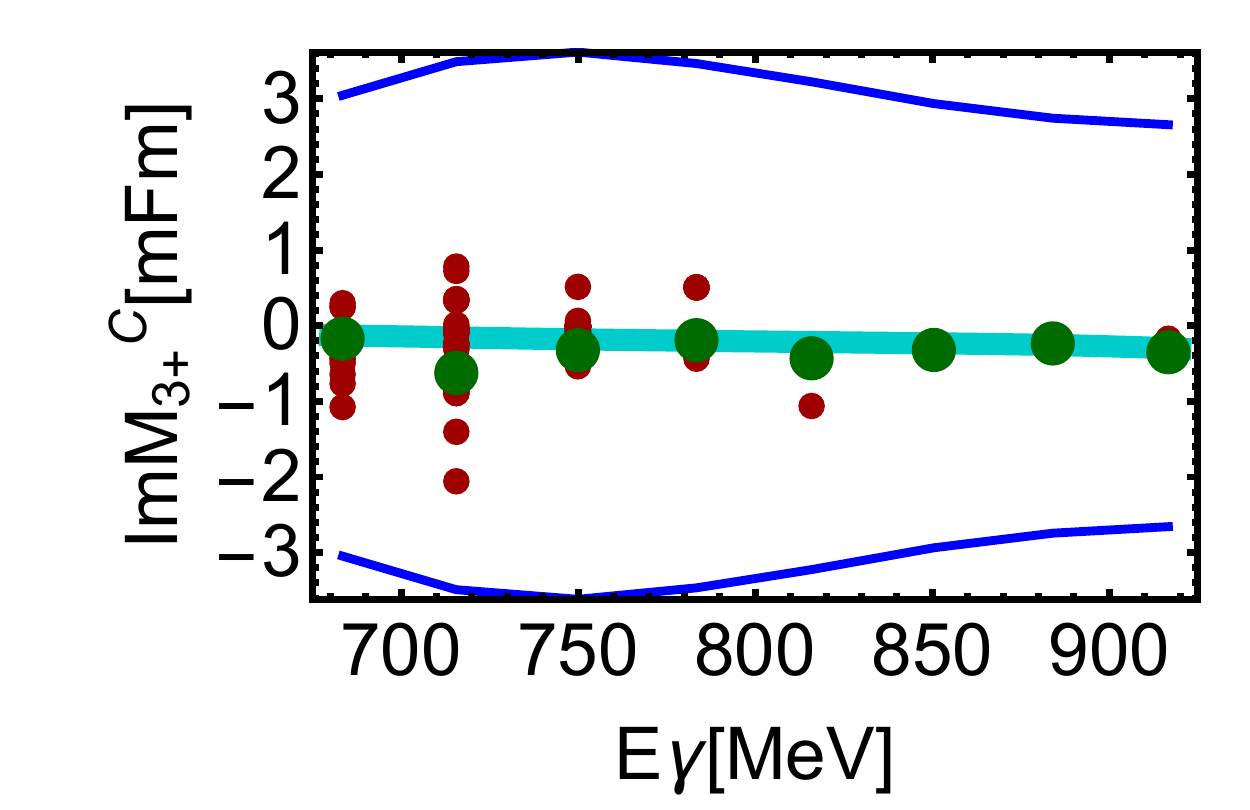}
 \end{overpic}
\caption{The plots show results for the fully unconstrained TPWA at $\ell_{\mathrm{max}} = 3$. They are continued in Figure \ref{fig:SecondFitLmax3MultipolesPlotsII}.}
\label{fig:SecondFitLmax3MultipolesPlotsI}
\end{figure}

\clearpage

\begin{figure}[ht]
 \centering
\begin{overpic}[width=0.325\textwidth]{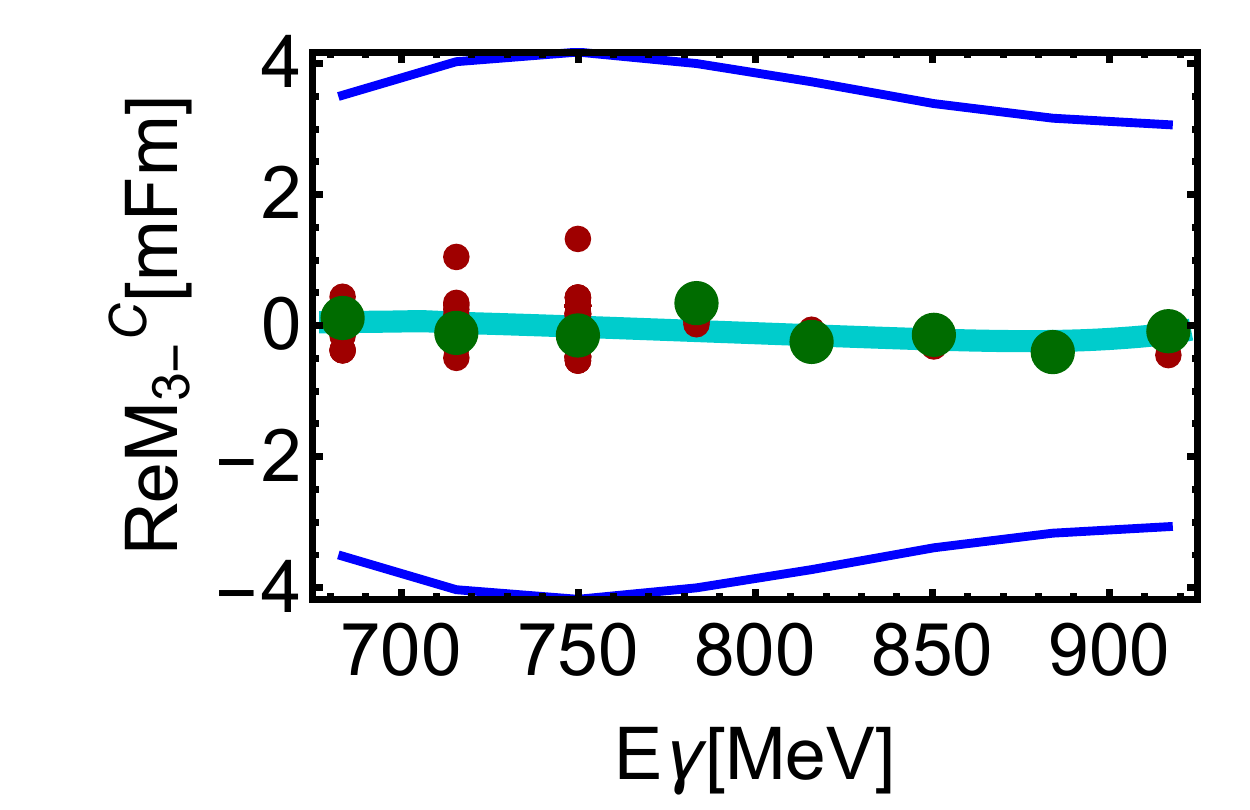}
 \end{overpic}
\begin{overpic}[width=0.325\textwidth]{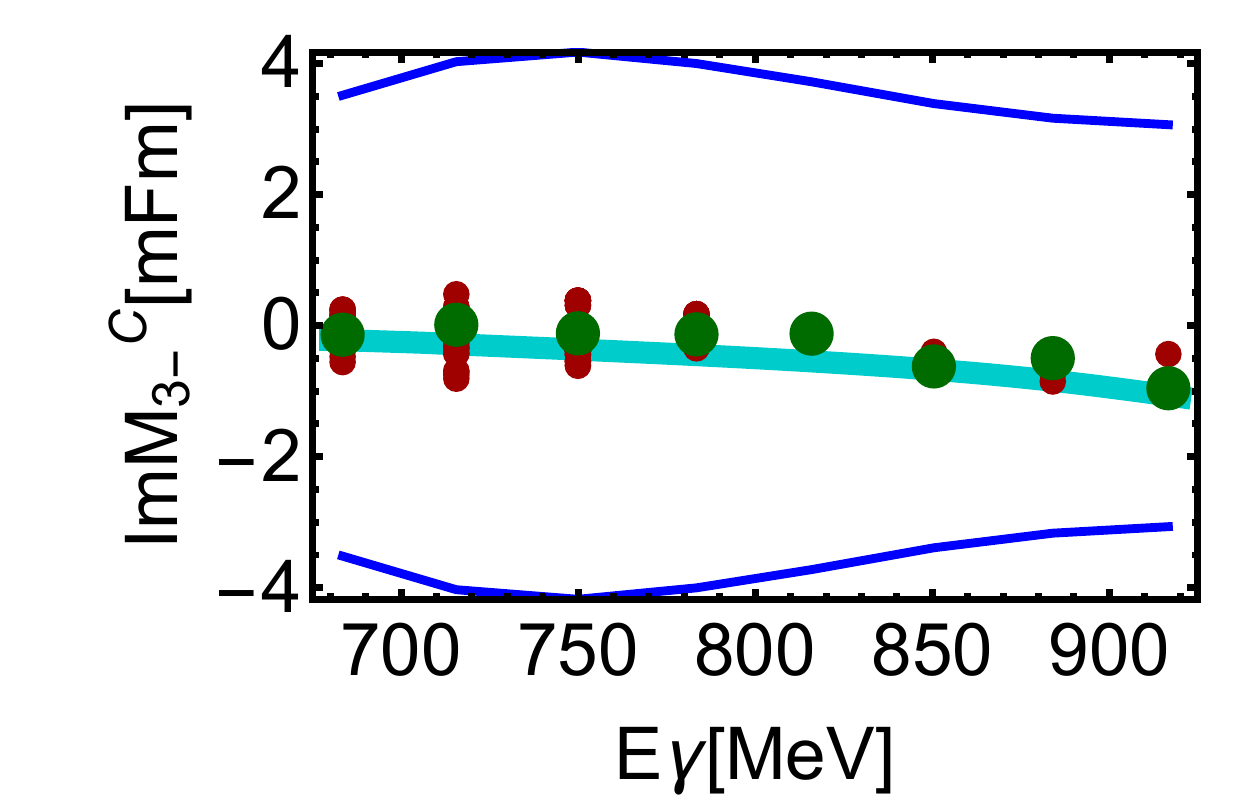}
 \end{overpic}
\caption[Results for the $23$ fit-parameters comprised of the real- and imaginary parts of phase-constrained $S$-, $P$-, $D$- and $F$-wave multipoles, for a model-independent TPWA with $\ell_{\mathrm{max}} = 3$, within the $2^{\mathrm{nd}}$ resonance region.]{The diagrams shown here, combined with those visible in Figure \ref{fig:SecondFitLmax3MultipolesPlotsI}, depict the results for the $23$ fit-parameters comprised of the real- and imaginary parts of phase-constrained multipoles for $\ell_{\mathrm{max}} = 3$. The layout of the plots is analogous to Figure \ref{fig:FirstFitLmax2MultipolesPlots}. Again, the attained global minimum is indicated by big green dots, while local minima are drawn as smaller red dots. However, here we included all local minima below the $0.975$-quantile of the theoretical chisquare-distribution for $\mathrm{ndf} = r = 19$ (cf. Figure \ref{fig:SecondFitLmax3ChiSquarePlots}). The energy-dependent PWA-solution BnGa 2014\_02 \cite{BoGa} is shown as a thick cyan-colored curve.}
\label{fig:SecondFitLmax3MultipolesPlotsII}
\end{figure}

Results for the $\chi^{2} / \mathrm{ndf}$ of the non-redundant final configurations are illustrated in Figure \ref{fig:SecondFitLmax3ChiSquarePlots}, while multipole-solutions are shown in Figures \ref{fig:SecondFitLmax3MultipolesPlotsI} and \ref{fig:SecondFitLmax3MultipolesPlotsII}. \newline
Again, as before, a global minimum is found when fitting the mathematically over-complete set of $7$ observables. However, for the lower $4$ energy-bins, there is an over-abundance of local minima very close in $\chi^{2}$. It is however a pleasant finding that the fit-quality of the global minima has increased as a result of the higher truncation. Most of them lie within the $68\%$-confidence interval, while all of them fall into the $95\%$-confidence interval. Thus, there are no energy-bins where a fit has to be rejected for probabilistic reasons. The abundance of local minima is most extreme in the lower four energy-bins, while the situation clears up a bit for the higher four bins. In the latter case, local minima are still present but a bit better separated from the global minimum. Furthermore, for the highest four energy-bins, most of the local minima fall outside of the $95\%$ confidence-interval. \newline
This situation is reflected in the plots showing the attained multipole-solutions (cf. Figures \ref{fig:SecondFitLmax3MultipolesPlotsI} and \ref{fig:SecondFitLmax3MultipolesPlotsII}). Here, we have included all local minima within the $95\%$ confidence-interval defined by the theoretical chisquare distribution for $\mathrm{ndf} = r = 19$. Especially for  most of the $S$-, $P$- and $D$-wave multipoles, there is quite some scatter in the resulting fit-parameters within the lower half of the considered energy-range. This is most extreme for the $S$- and $P$-waves, especially for $ E_{0+}^{C} $, $ M_{1+}^{C}$ and $M_{1-}^{C}$. Also in the magnetic $D$-wave multipoles $M_{2\pm}^{C}$, this behavior is prominently visible. As a positive side-remark, the $S$-wave $ E_{0+}^{C} $ now has solutions with enough strength to reach the Bonn-Gatchina model-curve. This is different compared to the fit with $\ell_{\mathrm{max}} = 2$ (cf. Figure \ref{fig:FirstFitLmax2MultipolesPlots}) and shows the importance of taking the $F$-waves into consideration. \newline
For the results of the $F$-waves themselves, the scatter in the lower $4$ energy-bins is somewhat reduced. Especially for the electric $F$-wave multipoles, the data constrain them to be quite small for all the obtained solutions, which is consistent with the Bonn-Gatchina model in this energy-region. This constraint seems to be relaxed for the magnetic multipoles, especially for $M_{3+}^{C}$. Still, the range of the scatter in the solutions for the $F$-waves is in all cases less extreme compared to the lower partial waves. \newline
For the higher $4$ energy-bins, the fact that problems with the over-abundance of solutions clear up is reflected in the multipole-solutions. In most cases, the global minimum is located actually quite close to the Bonn-Gatchina solution. If it is not, there exists in most cases another fit-minimum within the $95\%$-confidence interval, which agrees better with Bonn-Gatchina. Also, the number of local minima close to the best solution is not extreme for the higher energies. Mostly, one finds only one or two additional solutions. \newline
Still, when considered over the whole energy-range, the solution-behavior of this fit is not satisfactory. One does not find a well-separated global minimum in agreement with the energy-dependent model, but instead a band of local minima all fairly close (in $\chi^{2}$) to the best solution, which define a band of values for the multipoles. Although the attained global minimum is in some instances quite close to the Bonn-Gatchina model, the scatter of the solution-band can be quite large. However, it in all cases encompasses the Bonn-Gatchina solution. \newline
The situation found here for the photoproduction of $\pi^{0}$-mesons reflects the behavior described in the abstract the paper by Sandorfi, Hoblit, Kamano and Lee \cite{Sandorfi:2010uv}, which reads:
\begin{quotation}
 \textit{``In fitting multipoles, we use a combined Monte Carlo sampling of the amplitude space, with gradient minimization, and find a shallow $\chi^{2}$ valley pitted with a very large number of local minima. This results in broad bands of multipole solutions that are experimentally indistinguishable.''}
\end{quotation}
To be fair, it has to be stated that in this paper, the channel of $K \Lambda$-photoproduction has been analyzed and higher partial waves have been fixed to (phenomenological) Born-terms. Still, the results seem quite similar. Also, the truncation order for the varied multipoles of Sandorfi \textit{et al.} coincides with the one chosen here, i.e. $L=\ell_{\mathrm{max}}=3$. \newline
We re-encounter here a behavior already seen in the analyses of data in the $\Delta$-region: as soon as one tries to fit small higher multipoles out of the data, problems with ambiguities appear. In case one leaves these small multipoles out of the picture while fitting, the situation with ambiguities improves, however it is not possible to obtain a good fit due to disregarded interferences. \newline

As a way out of this issue, again, model-dependence is introduced into the TPWA. Here, we resort to fixing the real- and imaginary parts of the $F$-wave multipoles to the values of the solution BnGa2014\_02 \cite{BoGa}. The $S$-, $P$- and $D$-wave multipoles are varied as free parameters in the minimizations of the $\chi^{2}$-function (\ref{eq:CorrelatedChisquareQuotedForAnalysisII}). For the varied parameters, the accessible amplitude-space is still scanned for solutions using the Monte Carlo-sampling (section \ref{sec:MonteCarloSampling}), employing a pool of $N_{MC} = 16000$ randomly drawn initial conditions. \newline
A few more remarks are in order on details of this fit. First of all, the total cross section $\bar{\sigma}$ used in the Monte Carlo sampling has to be corrected for the fact that the $F$-waves are now fixed to Bonn-Gatchina, i.e. (in analogy to equation (\ref{eq:TCSCorrectedDeltaRegionFit}) section \ref{subsec:DeltaRegionDataFits}, where $D$-waves were fixed to SAID)
\begin{equation}
 \bar{\sigma}_{\bm{c}} := \bar{\sigma} - \Big\{ 80 \left| E^{\mathrm{BnGa}}_{3+} \right|^{2} + 18 \left| E^{\mathrm{BnGa}}_{3 -} \right|^{2} + 48 \left| M^{\mathrm{BnGa}}_{3+} \right|^{2} + 36 \left| M^{\mathrm{BnGa}}_{3-} \right|^{2} \Big\} \mathrm{,} \label{eq:TCSCorrected2ndResRegionFit}
\end{equation}
with $\bar{\sigma}_{\bm{c}}$ now constraining the parameter-space of the fitted multipoles. Secondly, the $F$-wave multipoles have to be implemented obeying the correct constraint for the overall phase. Since the latter is $E^{C}_{0+} = \left| E^{C}_{0+} \right| > 0$, the Bonn-Gatchina $F$-waves have to be rotated with the inverse of the original Bonn-Gatchina $E_{0+}$-phase. Thus, one not only implements information on the $F$-waves into the fit, but implicitly also knowledge about the Bonn-Gatchina $S$-wave. Here, we dispense with explicit plots showing the implemented model-multipoles. They can be considered in the respective plots in Figures \ref{fig:SecondFitLmax3MultipolesPlotsI} and \ref{fig:SecondFitLmax3MultipolesPlotsII}.

\begin{figure}[ht]
 \centering
\begin{overpic}[width=0.495\textwidth]{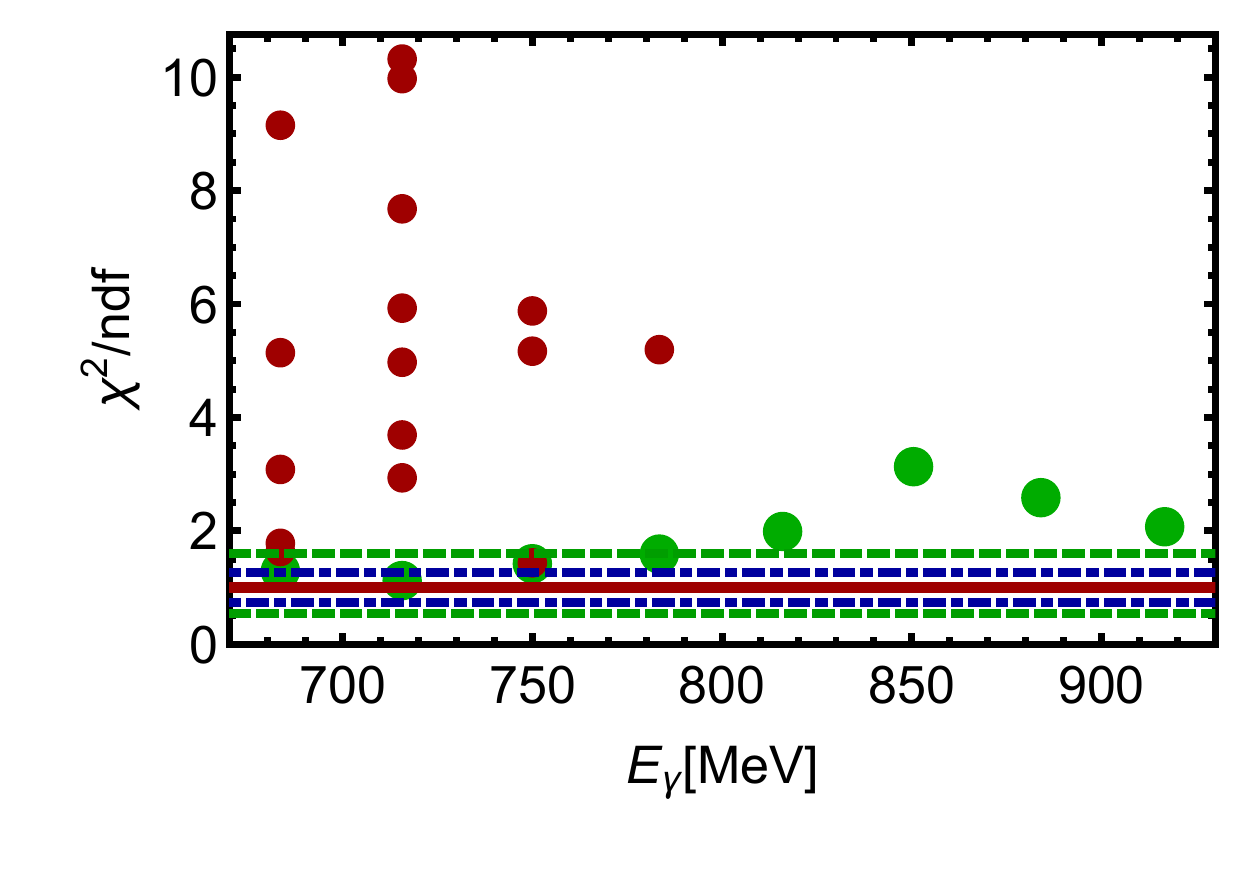}
 \end{overpic}
\begin{overpic}[width=0.495\textwidth]{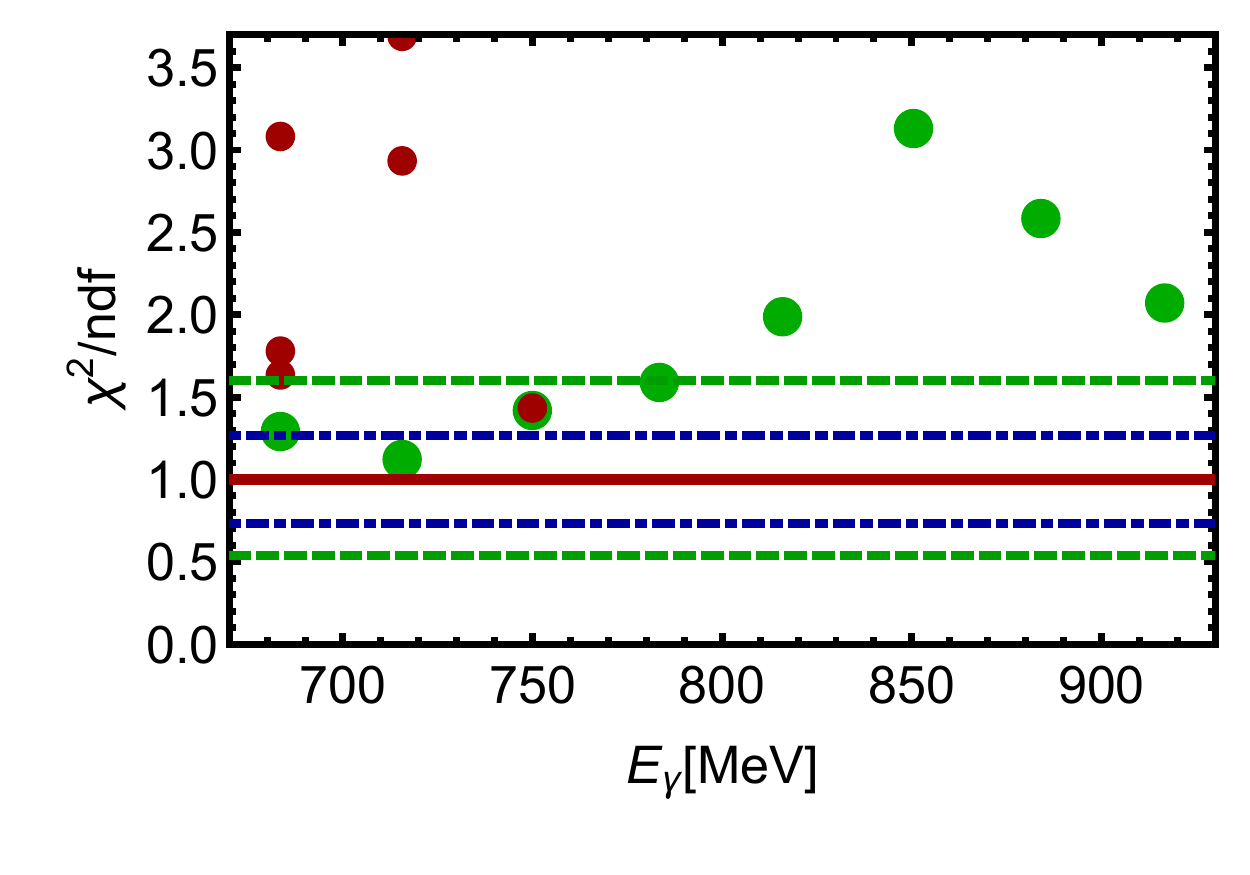}
 \end{overpic}
\vspace*{-15pt}
\caption[The $\chi^{2}/\mathrm{ndf}$ for the best results of the full Monte Carlo minimum-search applied in a truncation at $\ell_{\mathrm{max}} = 3$, with $F$-waves fixed to the Bonn-Gatchina model and $S$-, $P$- and $D$-waves varied, within the $2^{\mathrm{nd}}$ resonance region.]{Best results for the full Monte Carlo minimum-search applied in a truncation at $\ell_{\mathrm{max}} = 3$, with $F$-waves fixed to the Bonn-Gatchina model and $S$-, $P$- and $D$-waves varied, are plotted. A wide plot-range is shown (left) as well as a more detailed picture which includes only the global minimum and local minima close to it (right). All non-redundant solutions coming from a pool of $N_{MC} = 16000$ initial parameter-configurations can be seen. The global minimum is indicated by the big green dots, other local minima are plotted as smaller red-colored dots. \newline
In addition, some information on the theoretical chisquare distribution for the appropriate estimate $\mathrm{ndf} = r = 27$ is included via the horizontal lines. The mean is a red solid line, while the pair of $0.025$- and $0.975$-quantiles is indicated by green dashed lines and that made of the $0.16$- and $0.84$-quantiles by a blue dash-dotted line.}
\label{fig:ThirdFitLmax3FWavesFixedChiSquarePlots}
\end{figure}

\begin{figure}[h]
 \centering
\begin{overpic}[width=0.325\textwidth]{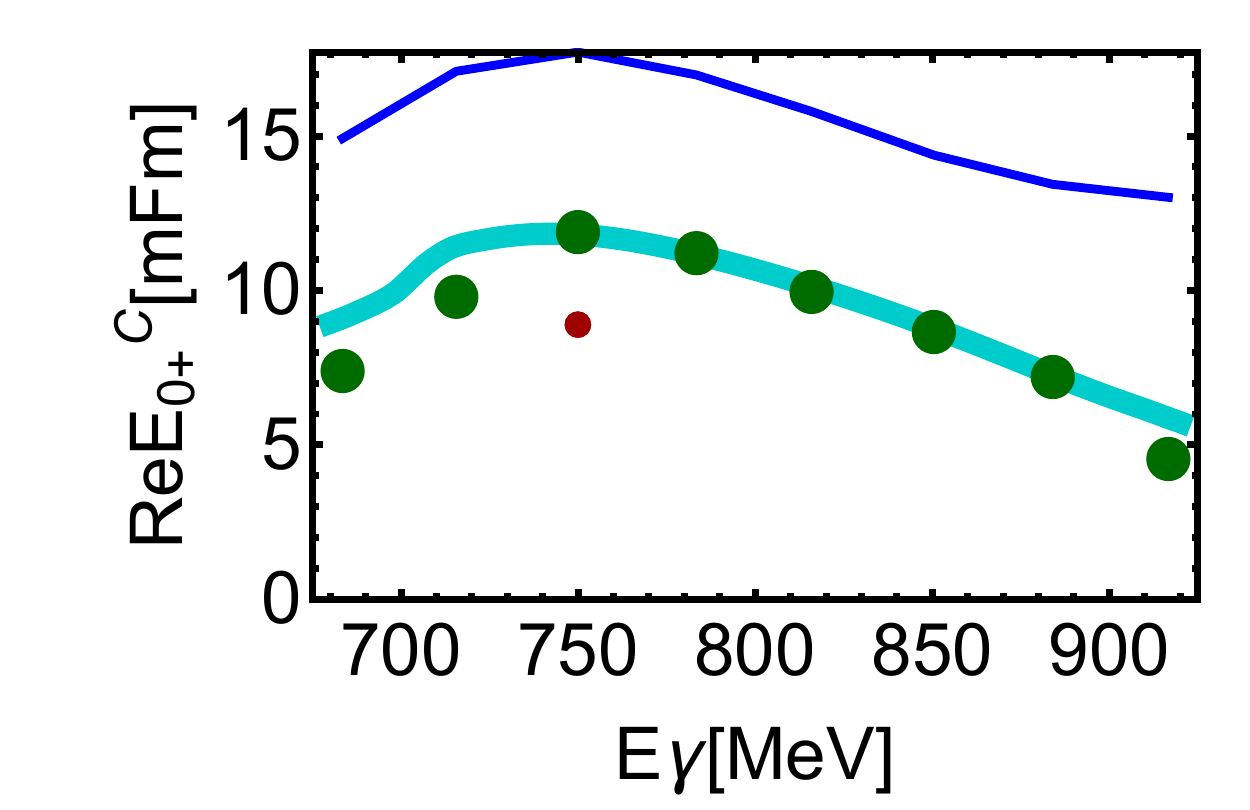}
 \end{overpic}
\begin{overpic}[width=0.325\textwidth]{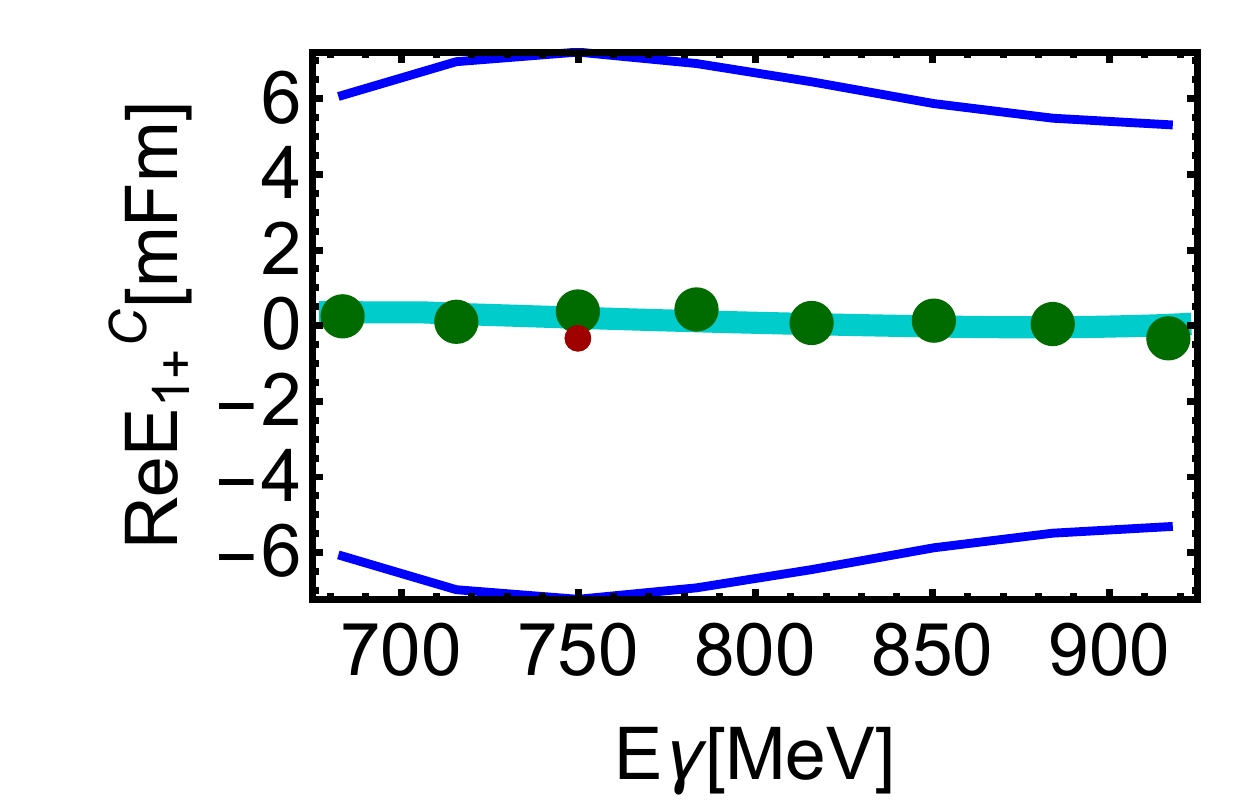}
 \end{overpic}
\begin{overpic}[width=0.325\textwidth]{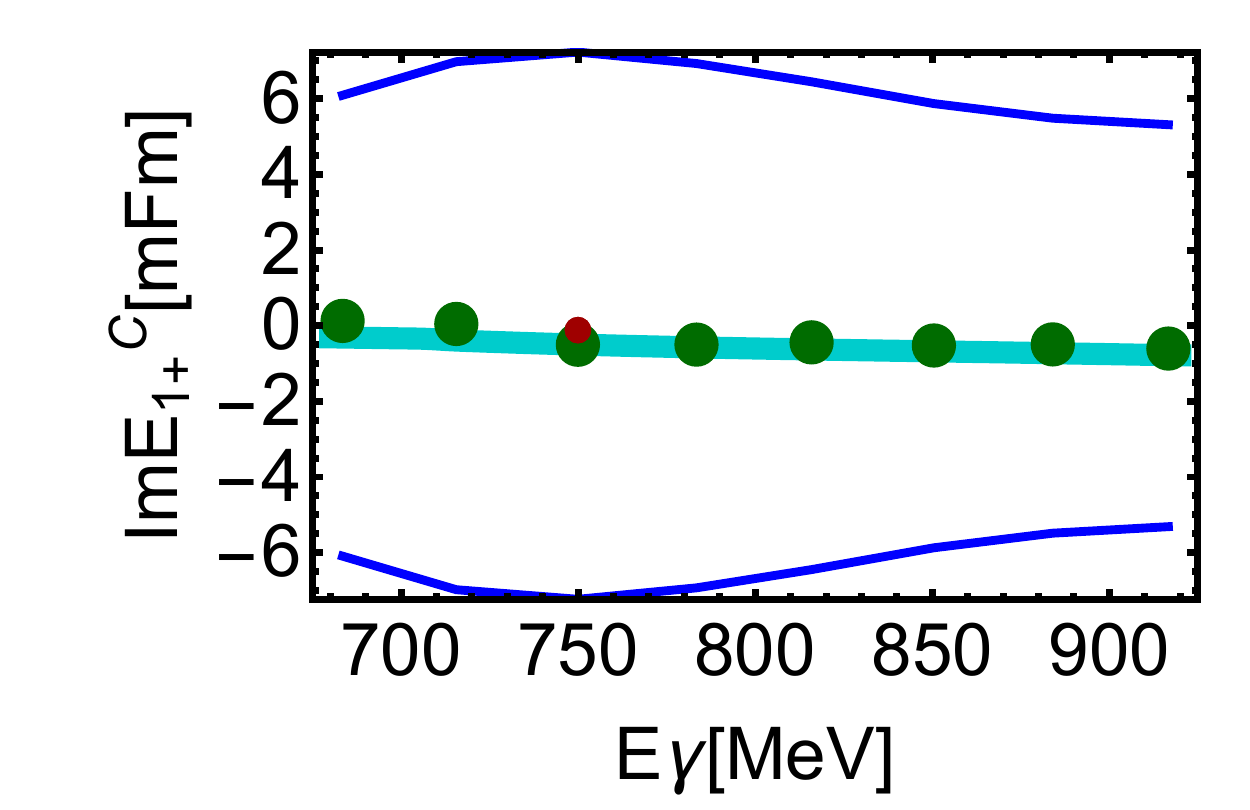}
 \end{overpic} \\
\begin{overpic}[width=0.325\textwidth]{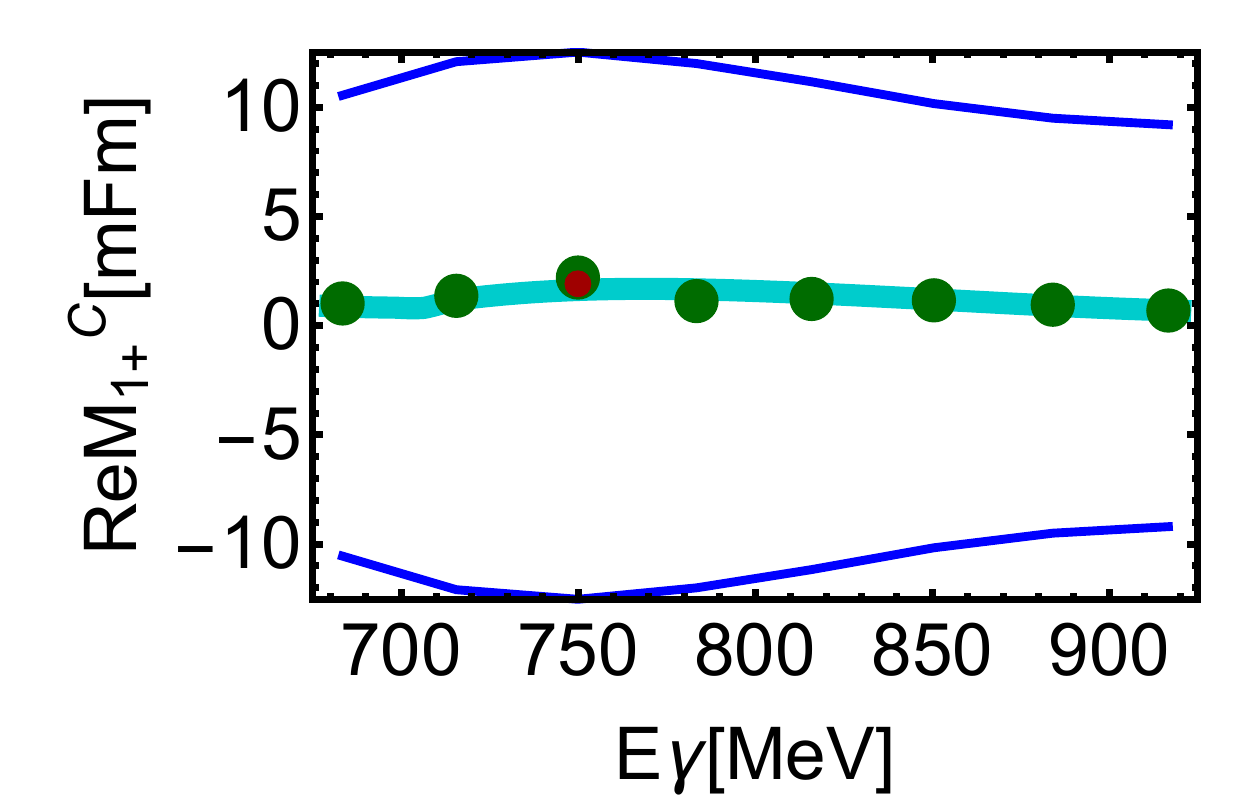}
 \end{overpic}
\begin{overpic}[width=0.325\textwidth]{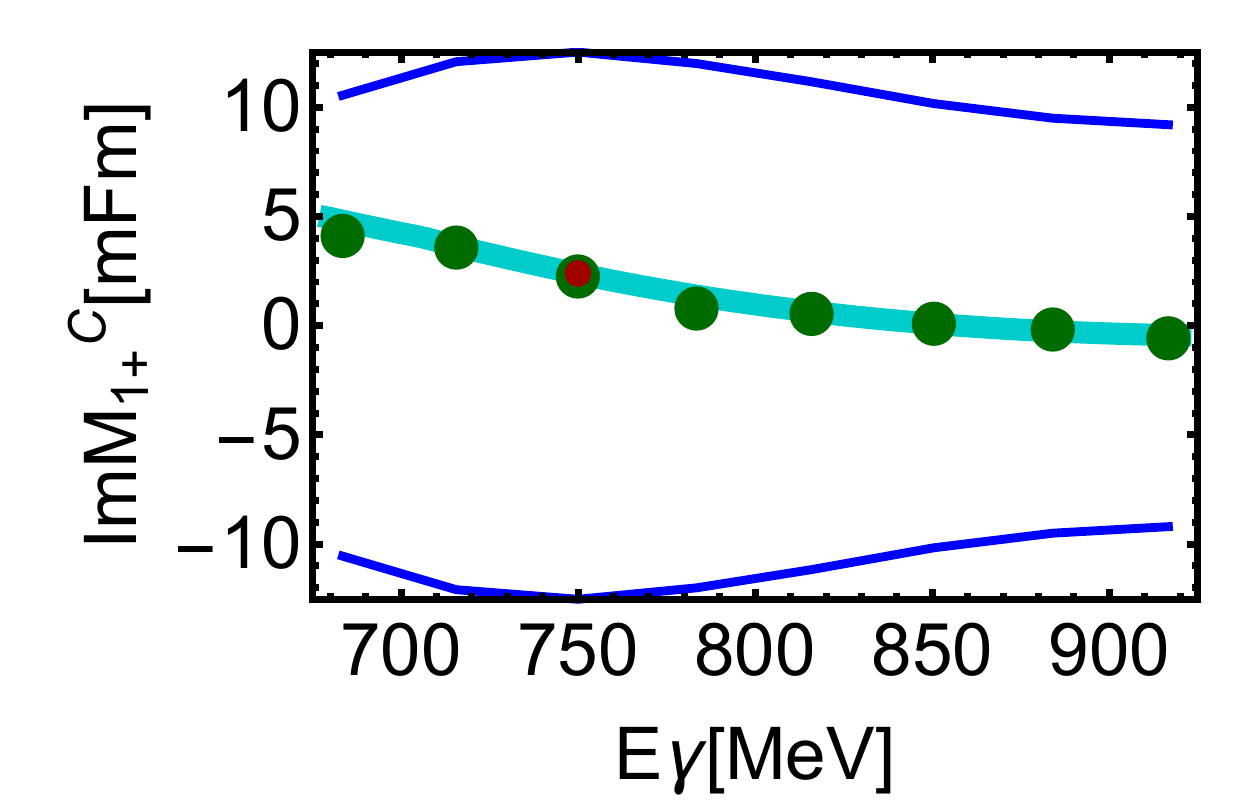}
 \end{overpic}
\begin{overpic}[width=0.325\textwidth]{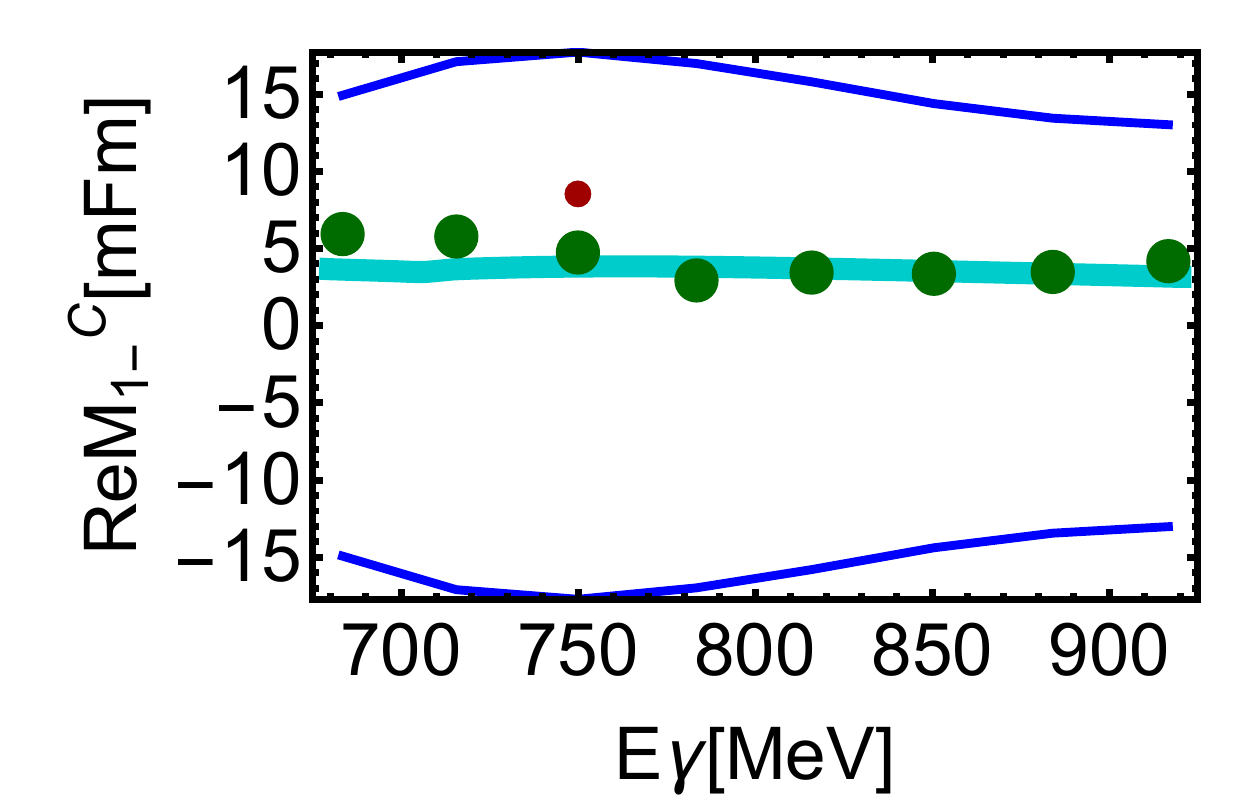}
 \end{overpic} \\
\begin{overpic}[width=0.325\textwidth]{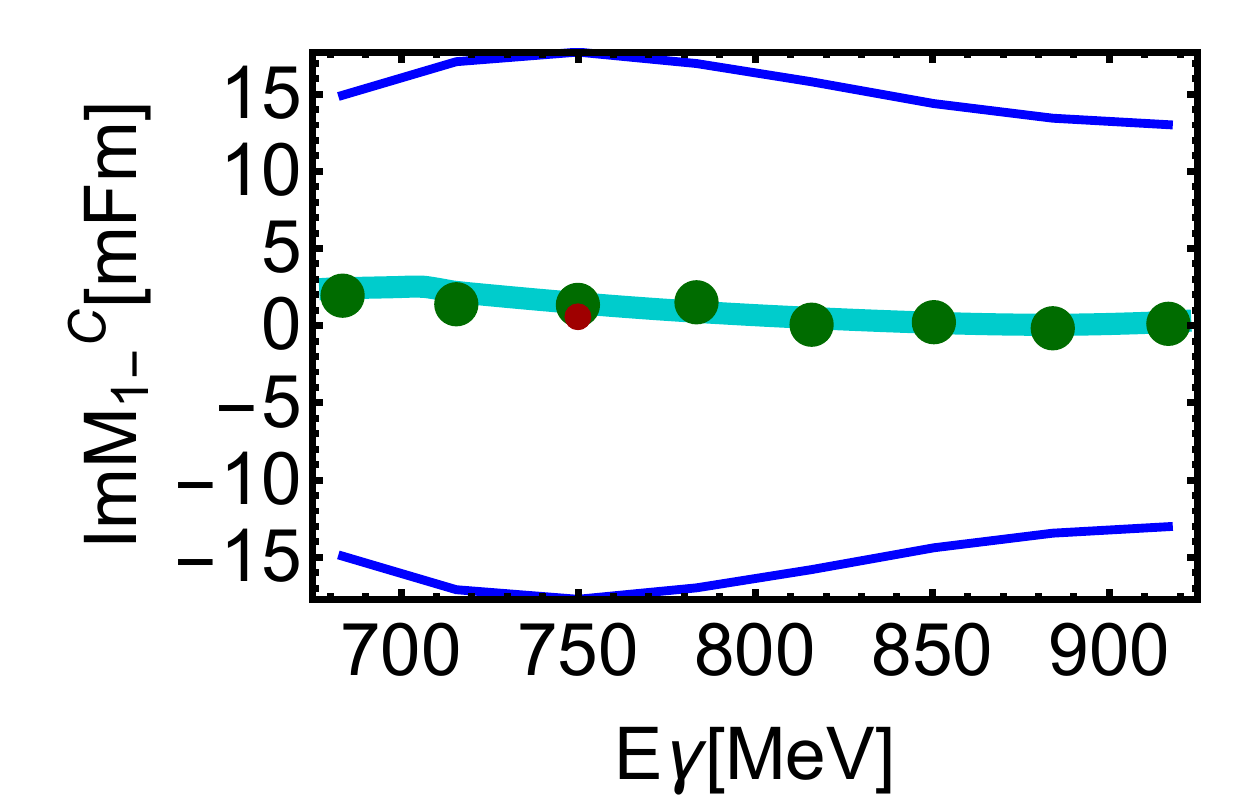}
 \end{overpic}
\begin{overpic}[width=0.325\textwidth]{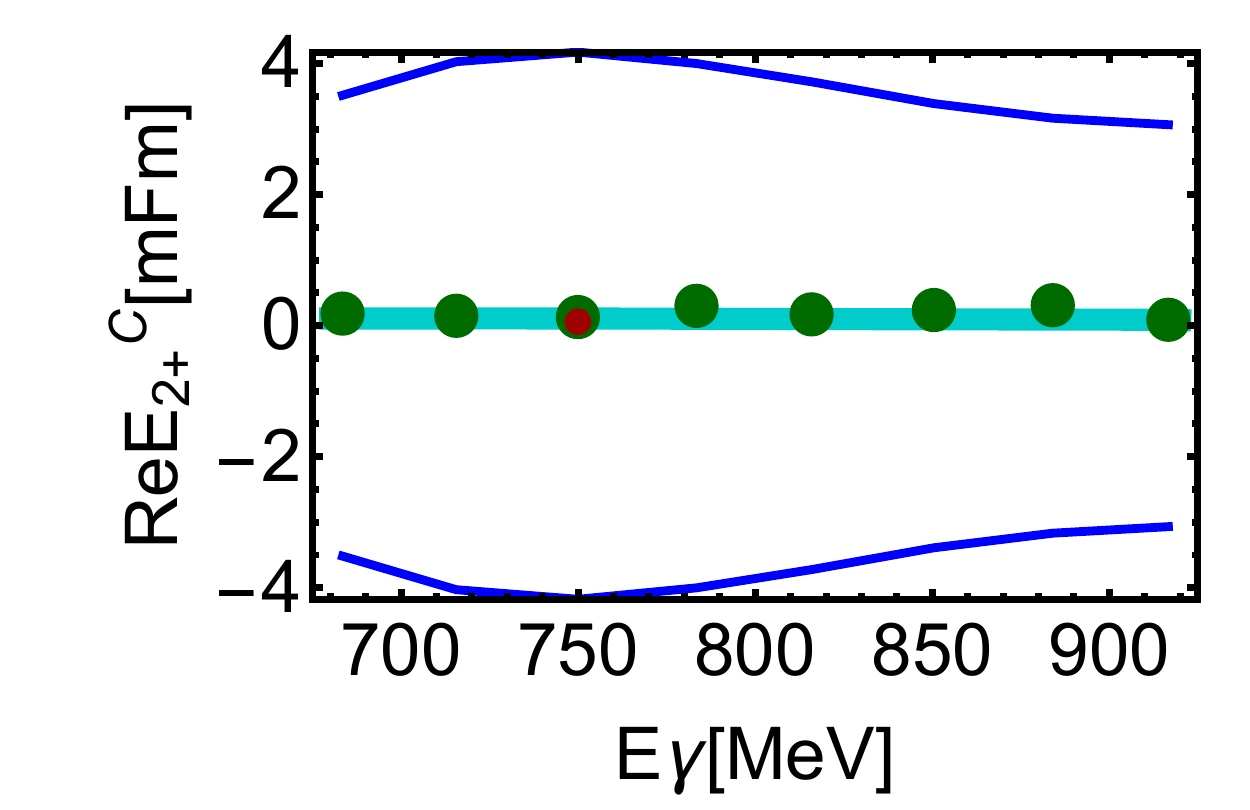}
 \end{overpic}
\begin{overpic}[width=0.325\textwidth]{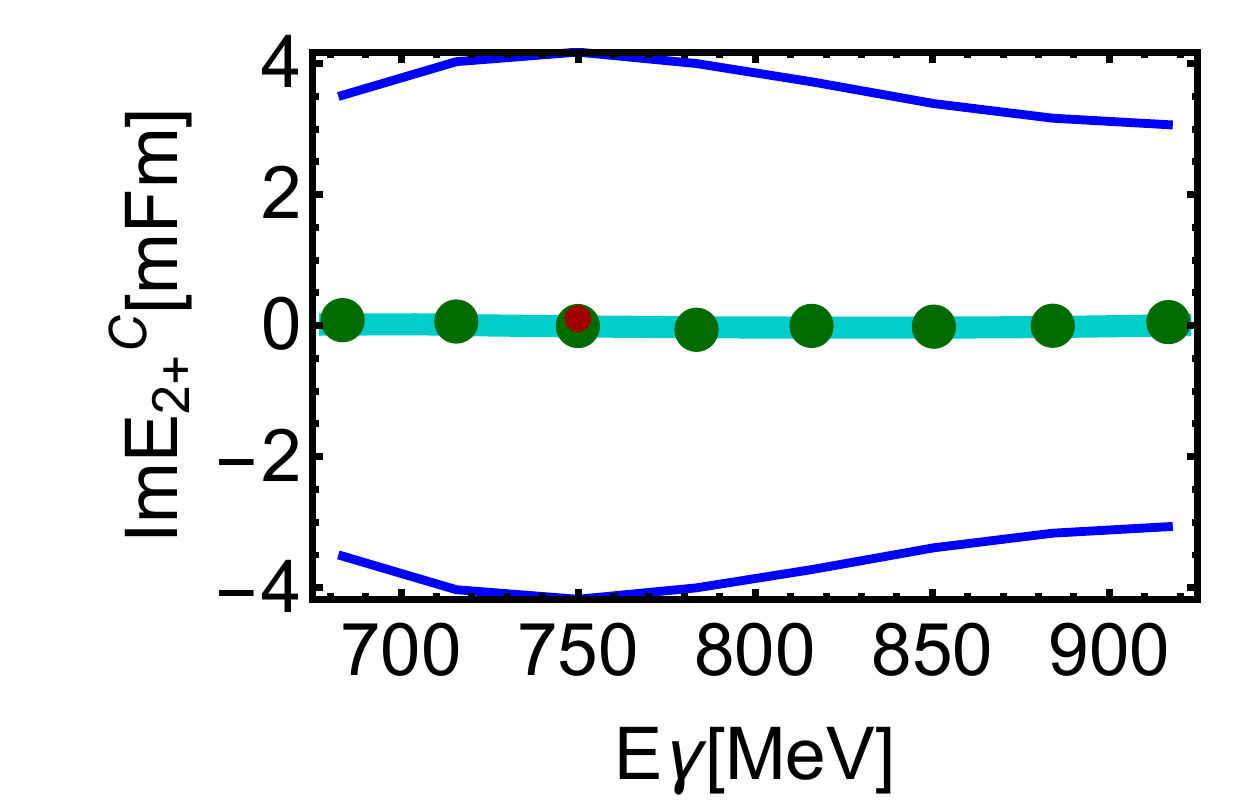}
 \end{overpic} \\
\begin{overpic}[width=0.325\textwidth]{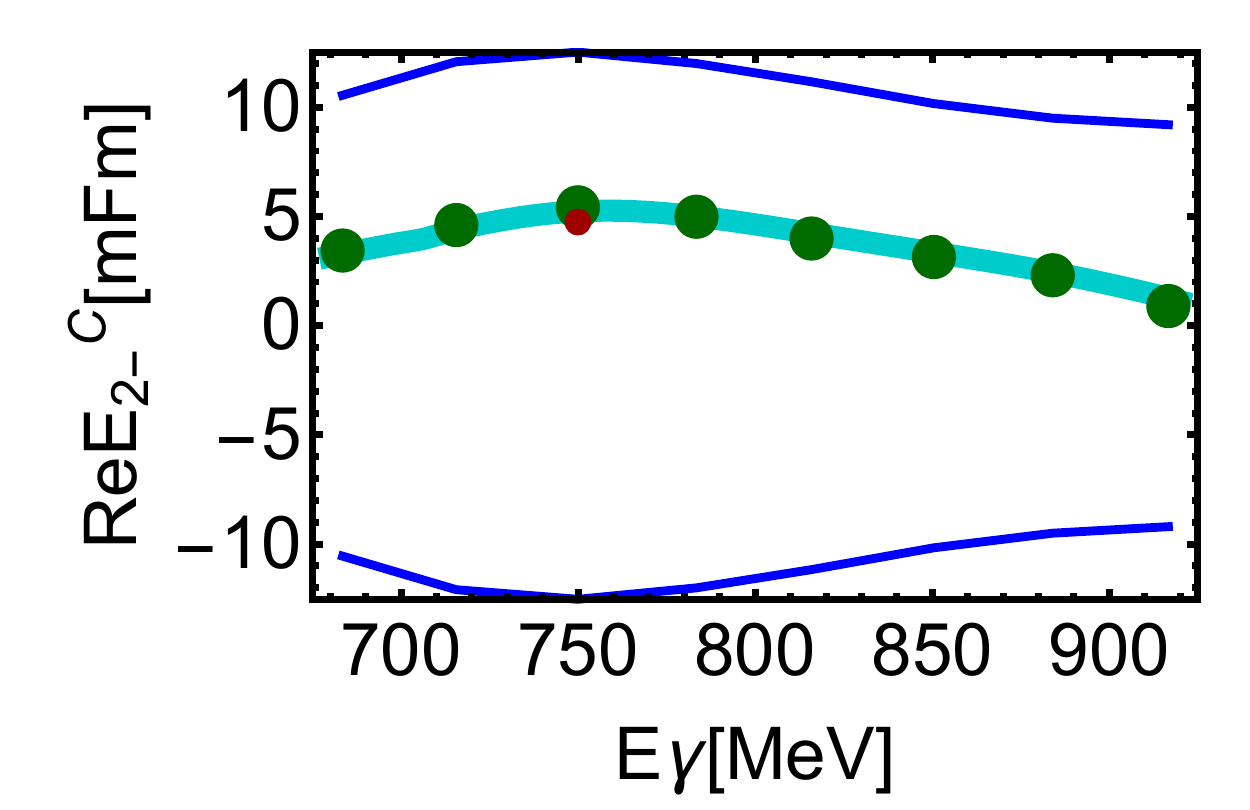}
 \end{overpic}
\begin{overpic}[width=0.325\textwidth]{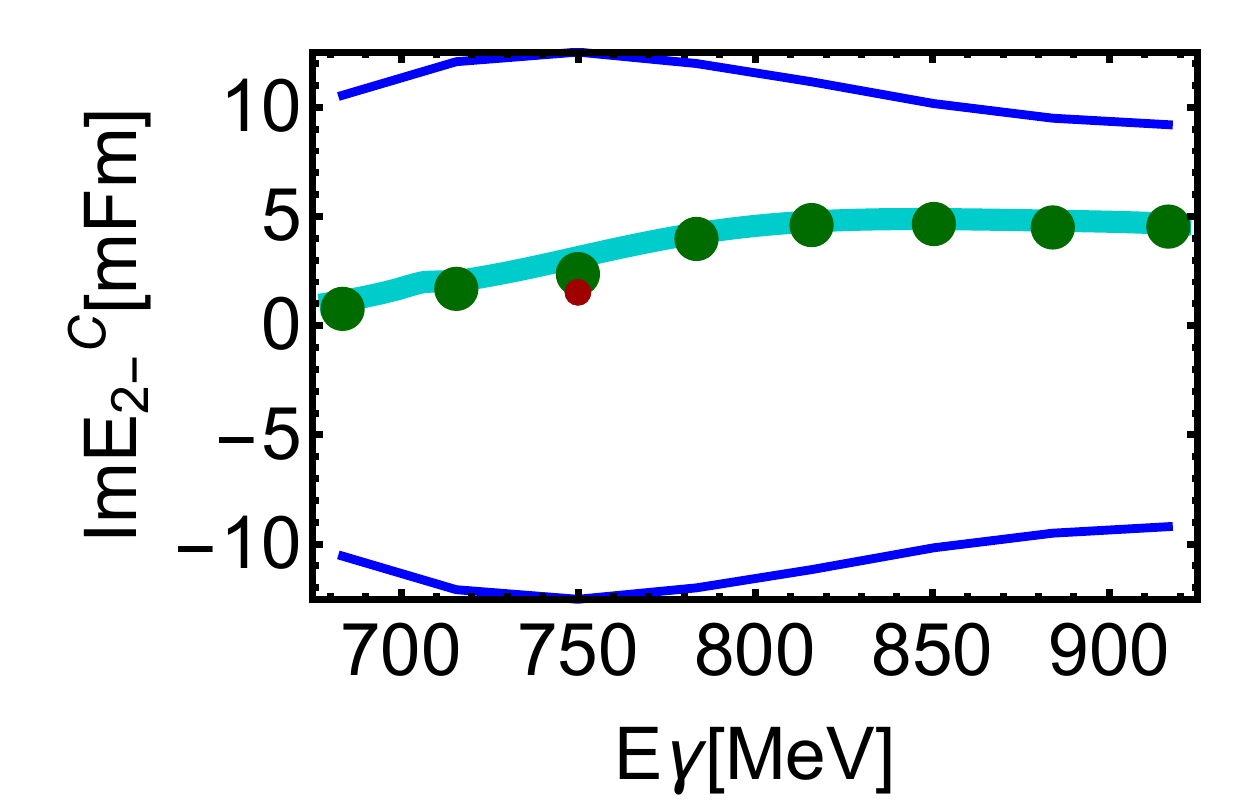}
 \end{overpic}
\begin{overpic}[width=0.325\textwidth]{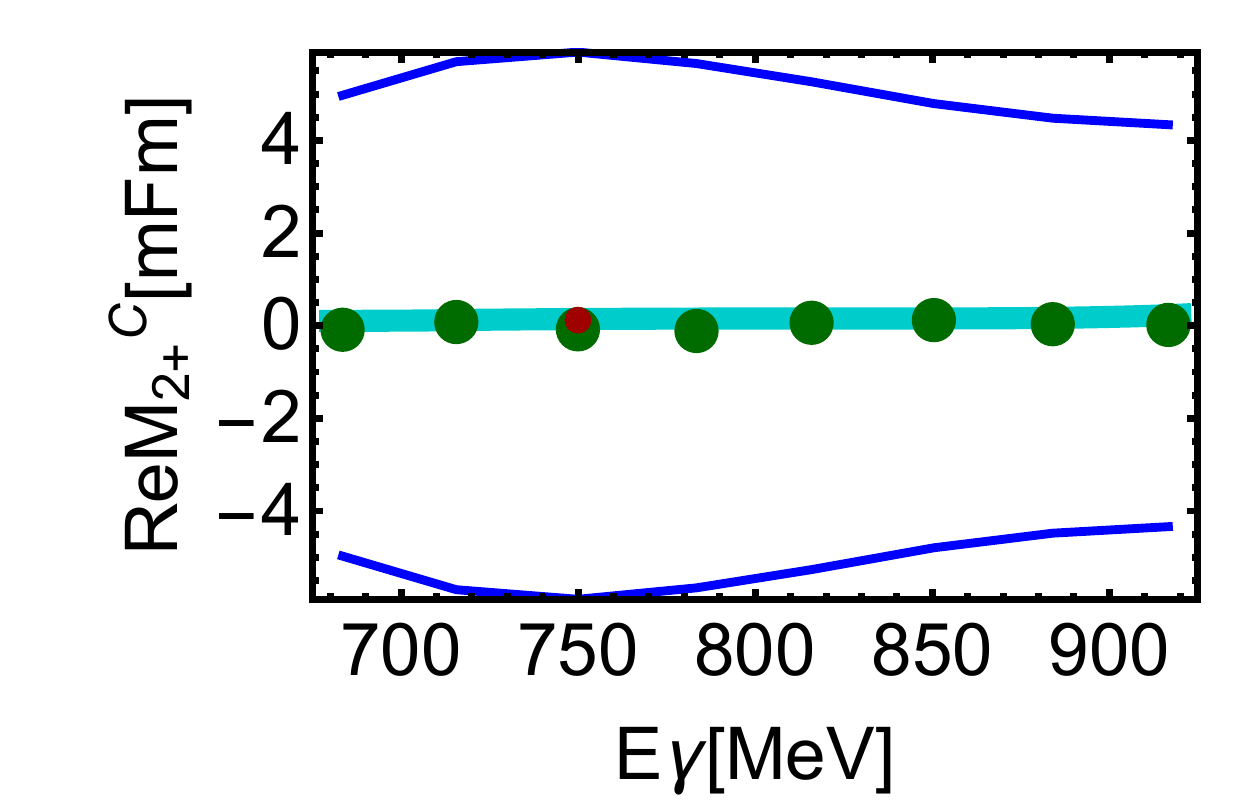}
 \end{overpic} \\
\begin{overpic}[width=0.325\textwidth]{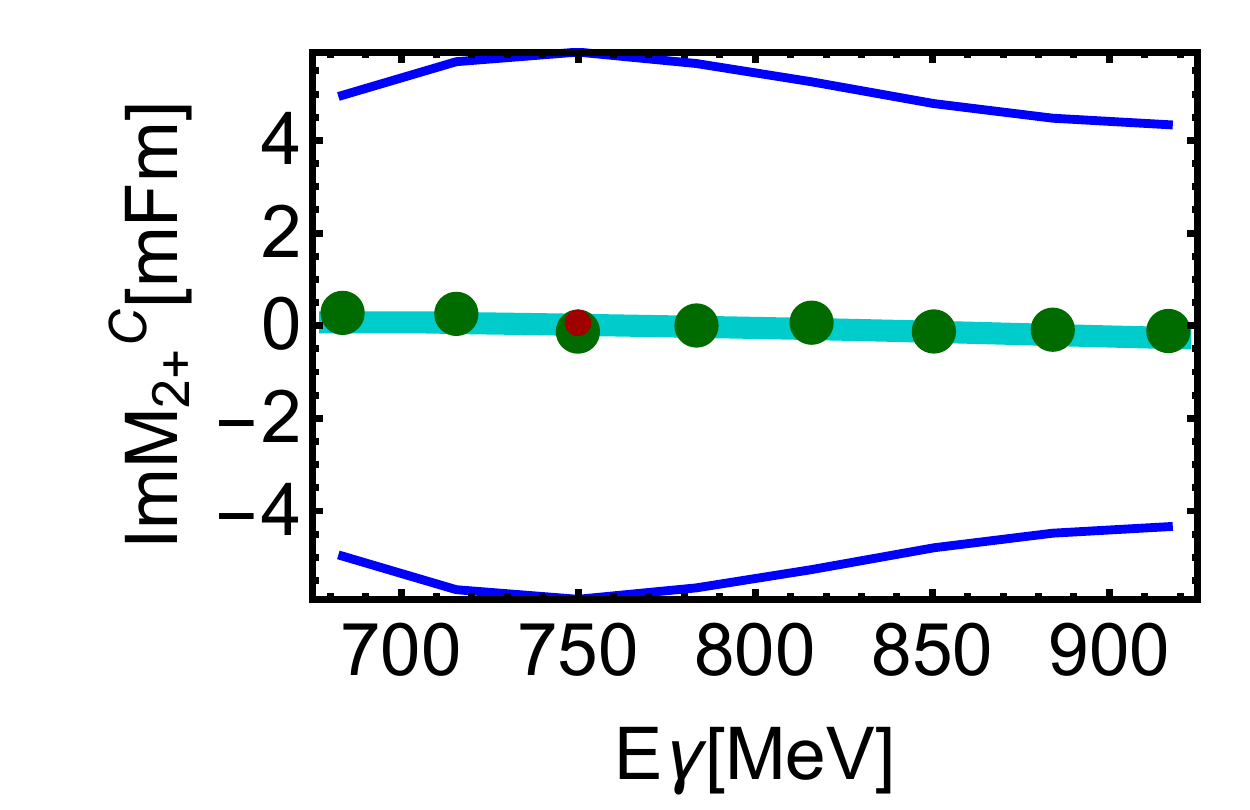}
 \end{overpic}
\begin{overpic}[width=0.325\textwidth]{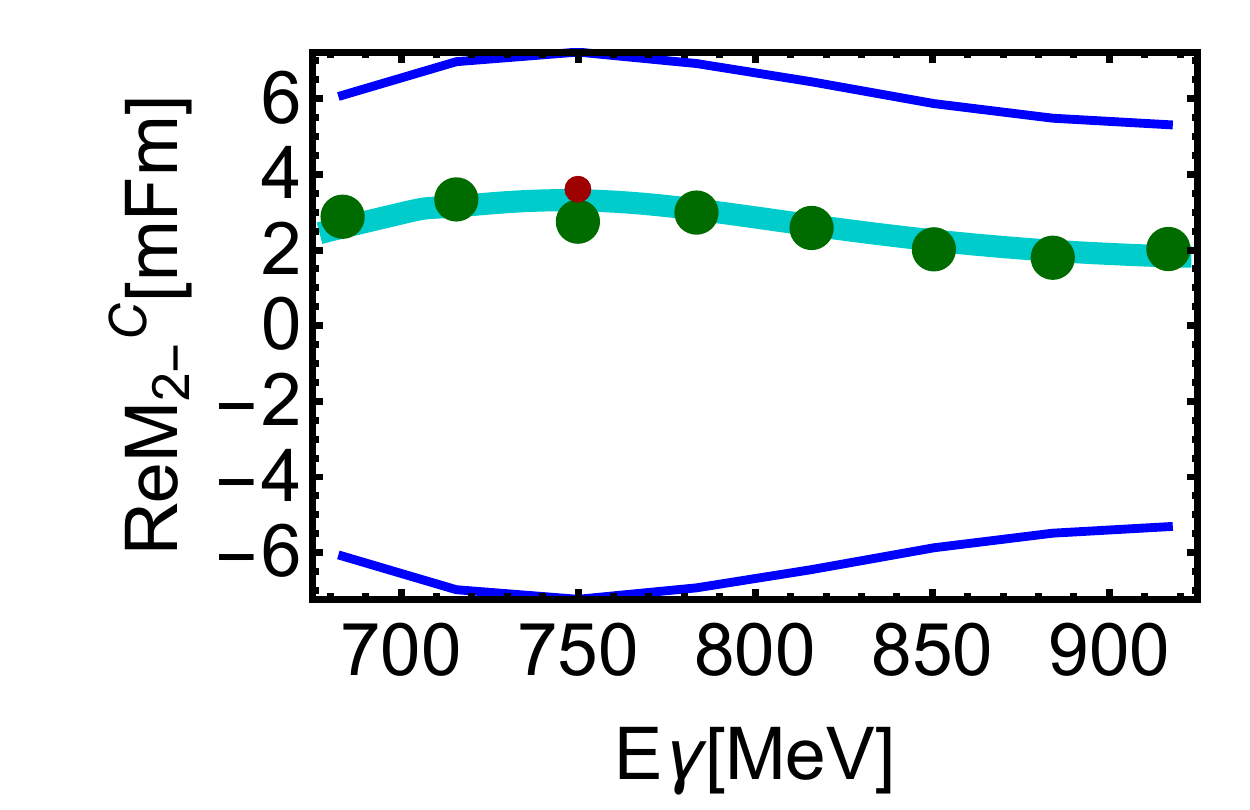}
 \end{overpic}
\begin{overpic}[width=0.325\textwidth]{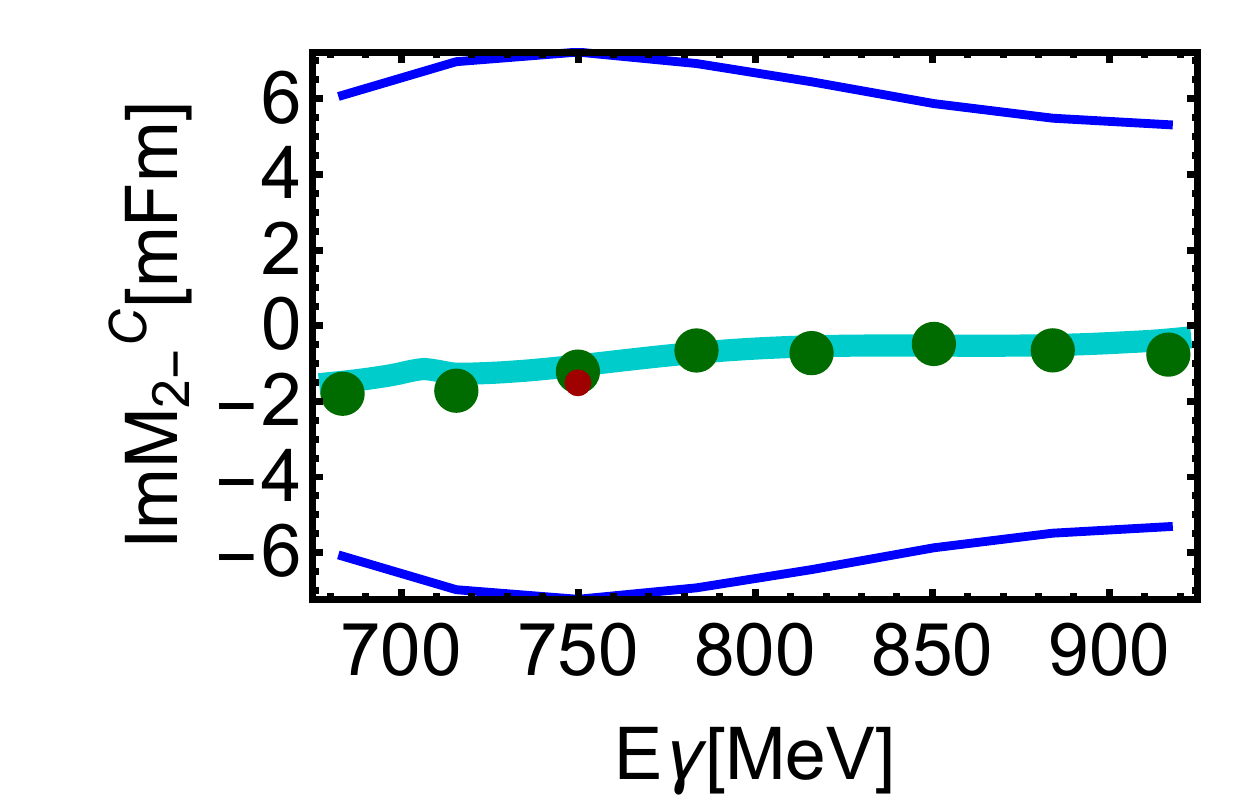}
 \end{overpic}
\caption[Results for the $15$ fit-parameters comprised of the real- and imaginary parts of phase-constrained $S$-, $P$- and $D$-wave multipoles, for a TPWA with $\ell_{\mathrm{max}} = 3$ and $F$-waves fixed to BnGa 2014\_02, within the $2^{\mathrm{nd}}$ resonance region.]{The plots show results for the $15$ parameters comprised of the real- and imaginary parts of the phase-constrained multipoles, for a truncation at $\ell_{\mathrm{max}} = 3$ and all $F$-waves fixed to the Bonn-Gatchina solution BnGa2014\_02 \cite{BoGa}. A Monte Carlo minimum-search has still been done for the fitted parameters, employing a pool of $N_{MC} = 16000$ start-configurations. The attained global minimum is indicated by the big green dots, while local minima are plotted as smaller red dots. All local minima below the $0.975$-quantile of the chisquare-distribution for $\mathrm{ndf} = r = 27$ have been included. The Bonn-Gatchina model-solution BnGa 2014\_02 \cite{BoGa} is shown for the $S$-, $P$- and $D$-waves as a thick cyan-colored curve. \newline
Furthermore, the plot-range has been adjusted in analogy to the Figures \ref{fig:FirstFitLmax2MultipolesPlots}, \ref{fig:SecondFitLmax3MultipolesPlotsI} and \ref{fig:SecondFitLmax3MultipolesPlotsII}.}
\label{fig:ThirdFitLmax3FWavesFixedMultipolesPlots}
\end{figure}

We still estimate the number of degrees of freedom in the fit as the difference between the employed Legendre-coefficients and the varied fit-parameters, which in this case becomes $\mathrm{ndf} = 27$. \newline
Results for the $\chi^{2}/\mathrm{ndf}$ of the non-redundant solutions found in this fit are displayed in Figure \ref{fig:ThirdFitLmax3FWavesFixedChiSquarePlots}. It is seen immediately that a global minimum is found and (almost) all local minima are well-separated from this best solution. The inclusion of the $F$-wave from Bonn-Gatchina really has helped with purging the solution-pool from ambiguities.
An exception is given in the third energy-bin, where a local minimum is very close to the global one. Here, the best solution has $\chi^{2}/\mathrm{ndf}=1.4189$, while the next best local minimum has $\chi^{2}/\mathrm{ndf}=1.4337$. For the first energy-bin, relatively close local minima are present, however they lie above the $0.975$-quantile of the theoretical chisquare-distribution for this fit (see Figure \ref{fig:ThirdFitLmax3FWavesFixedChiSquarePlots}). For the higher $4$ energy-bins, the global minimum is essentially unique, no local optimum is found any more. However, while the global minimum is nicely separated for the highest energies, the fit-quality gets progressively worse starting at the fifth energy. This is most extreme at the sixth energy, where the best solution is located at $\chi^{2}/\mathrm{ndf} \simeq 3$. \newline
The multipole solutions (Figure \ref{fig:ThirdFitLmax3FWavesFixedMultipolesPlots}) show the global minimum in quite good agreement with the Bonn-Gatchina model. The almost-degenerate local minimum in the third energy-bin does not show great deviations from neither the global minimum, nor the Bonn-Gatchina model. Differences are largest here for the real parts of $E_{0+}^{C}$ and $M_{1-}^{C}$, while for all the remaining parameters both solutions look almost identical within the chosen plot-range. \newline
Still, the obtained result is encouraging.

\clearpage

\clearpage

Fixing the $F$-waves to a model has removed the ambiguity-problems of the full fit with $\ell_{\mathrm{max}}=3$ (i.e. all multipoles varied), while in comparison to the case of a strict truncation at $\ell_{\mathrm{max}}=2$, the fit-quality has been improved substantially since important interferences between $F$-waves and lower partial waves are still taken into account. \newline
Therefore, in spite of the loss in fit-quality in the higher energies, we choose the present result as the starting point of a bootstrap-analysis (section \ref{sec:BootstrappingIntroduction}). It remains to be seen whether the single local minimum obtained in the third energy-bin will turn up again in such a resampling-analysis. \newline

The global minimum found in the previous fit serves as a starting point for the resampling, i.e. we will perform a reduced bootstrap-TPWA according to section \ref{sec:BootstrappingIntroduction}. This means we draw bootstrap-replicates $\left\{\check{\Omega}_{\ast}^{\alpha}\right\}$ for the profile function out of normal distributions as $\mathcal{N} \left( \check{\Omega}^{\alpha}, \Delta \check{\Omega}^{\alpha} \right) \longrightarrow \check{\Omega}_{\ast}^{\alpha}$ (cf. section \ref{subsec:DeltaRegionDataFits}, especially equation (\ref{eq:BootstrapDefinitionDeltaRegionFit}), as well as the whole of section \ref{sec:BootstrappingIntroduction}), thus generating an ensemble of $B=2000$ bootstrap-datasets. For each of the replicate-datasets, we omit the full Monte Carlo minimum-search due to numerical tractability and instead do a single minimization of the chisquare-function (\ref{eq:CorrelatedChisquareQuotedForAnalysisII}), starting at the global minimum obtained earlier. Again, $S$-, $P$- and $D$-wave multipoles are varied, with $F$-waves fixed to BnGa 2014\_02 \cite{BoGa}. \newline
In this way, bootstrap-distributions are obtained for the fit-parameters (i.e. the real- and imaginary parts of phase-constrained multipoles). As an example, bootstrap-distributions for the fourth energy-bin can be seen in Figure \ref{fig:BootstrapHistos2ndResRegionEnergy4MainText}, while distributions for all energies are contained in appendix \ref{sec:NumericalTPWAFitResults}. From the bootstrap-distributions of the individual fit-parameters $\hat{\theta}_{i} = \left( \mathcal{M}_{\ell}^{C} \right)_{i}$, the $0.16$- and $0.84$-quantiles define a confidence-interval for the individual parameter, which then defines the upper and lower bootstrap-errors $\Delta_{\pm}$. The global minimum is always quoted as the main result, in conjunction with these asymmetric errors (for more details, see section \ref{subsec:DeltaRegionDataFits}). Furthermore, the mean $\hat{\theta}_{i}^{\ast} (\cdot)$, standard error $\widehat{\mathrm{se}}_{B} \left( \hat{\theta}^{\ast}_{i} \right)$ and bias-estimate $\widehat{\mathrm{bias}}_{B}$ are extracted from the bootstrap-distributions. Lastly, we define and extract a bias test-parameter defined as $\delta_{\mathrm{bias}} := \left| \widehat{\mathrm{bias}}_{B} \right|/\widehat{\mathrm{se}}_{B}$. This makes it possible to check the simple bias rule-of-thumb $\delta_{\mathrm{bias}} < 0.25$ (see section \ref{subsec:DeltaRegionDataFits}, equation (\ref{eq:BiasTestParameterDefinition}) as well as section \ref{sec:BootstrappingIntroduction}). The results stemming from the bootstrap-distribution for the fourth energy-bin can be seen in Table \ref{tab:2ndResRegionResultsFourthEnergyMainText}. Numerical data for all the energies can be found in appendix \ref{sec:NumericalTPWAFitResults}. \newline

The extracted numbers reflect the behavior seen in the histograms. Mostly, quite normal-shaped distributions are observed which are, at least for the fourth energy, mostly quite symmetric. Therefore, the upper and lower bootstrap-errors $\Delta_{\pm}$ do not show great deviances and are both very close to the standard deviation $\widehat{\mathrm{se}}_{B}$. This is not always the case for all the remaining energies, where multipole-parameters can have quite asymmetric gaussian-shaped distributions (see appendix \ref{sec:NumericalTPWAFitResults}). \newline
Furthermore, in the fourth energy-bin, the mean $\hat{\theta}_{i}^{\ast} (\cdot)$ of each parameter-distribution is very close to the global minimum of the fit to the original data, i.e. all parameters have small (almost vanishing) bias, always well below one fourth of the standard error. All these results mark the fourth energy-bin as an example for a quite successful bootstrap-analysis. \newline
Regarding the remaining energy-bins, one also usually encounters (sometimes asymmetric) normal-shaped distributions with mostly small bias, with biases tending to become smaller for the higher energies.
\begin{figure}[h]
\begin{overpic}[width=0.325\textwidth]{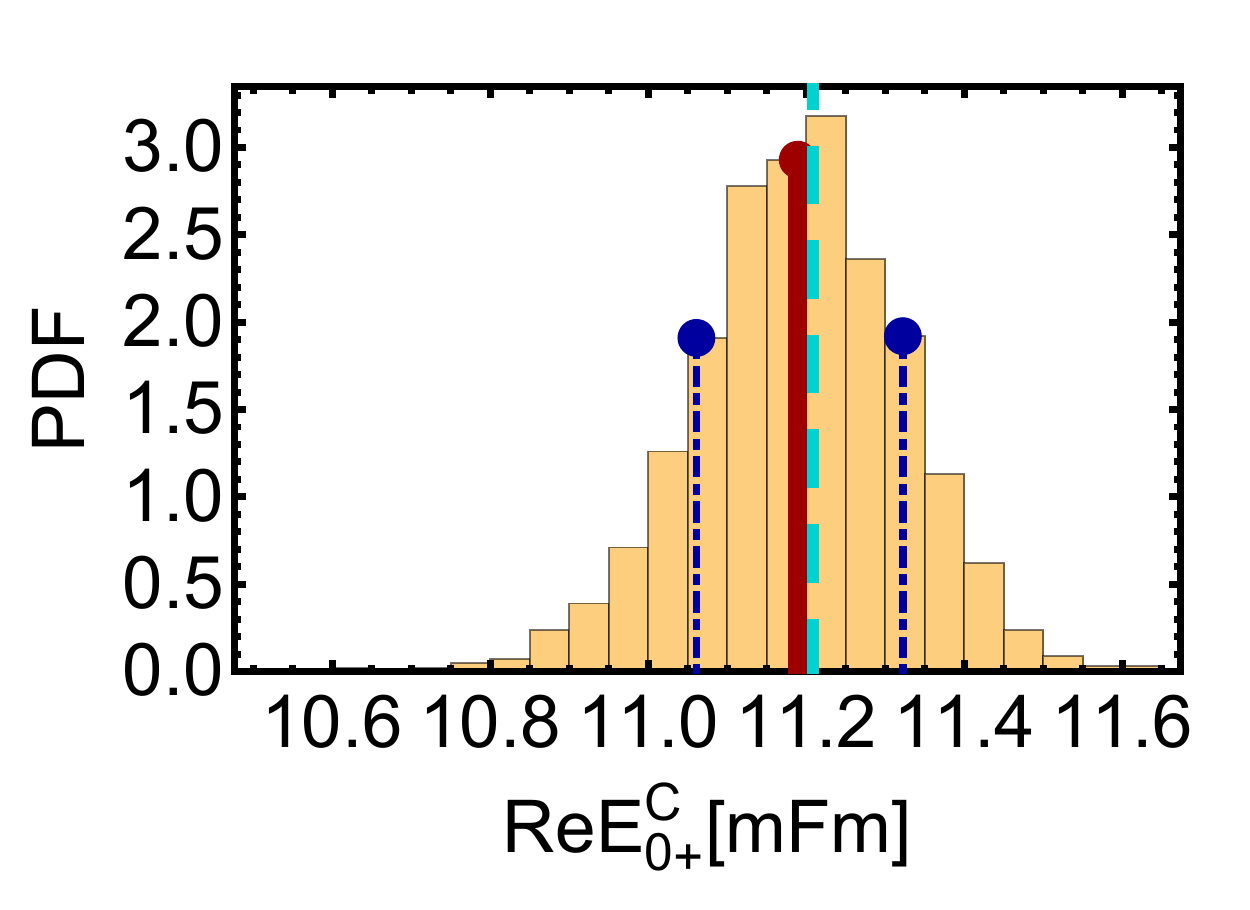}
 \end{overpic}
\begin{overpic}[width=0.325\textwidth]{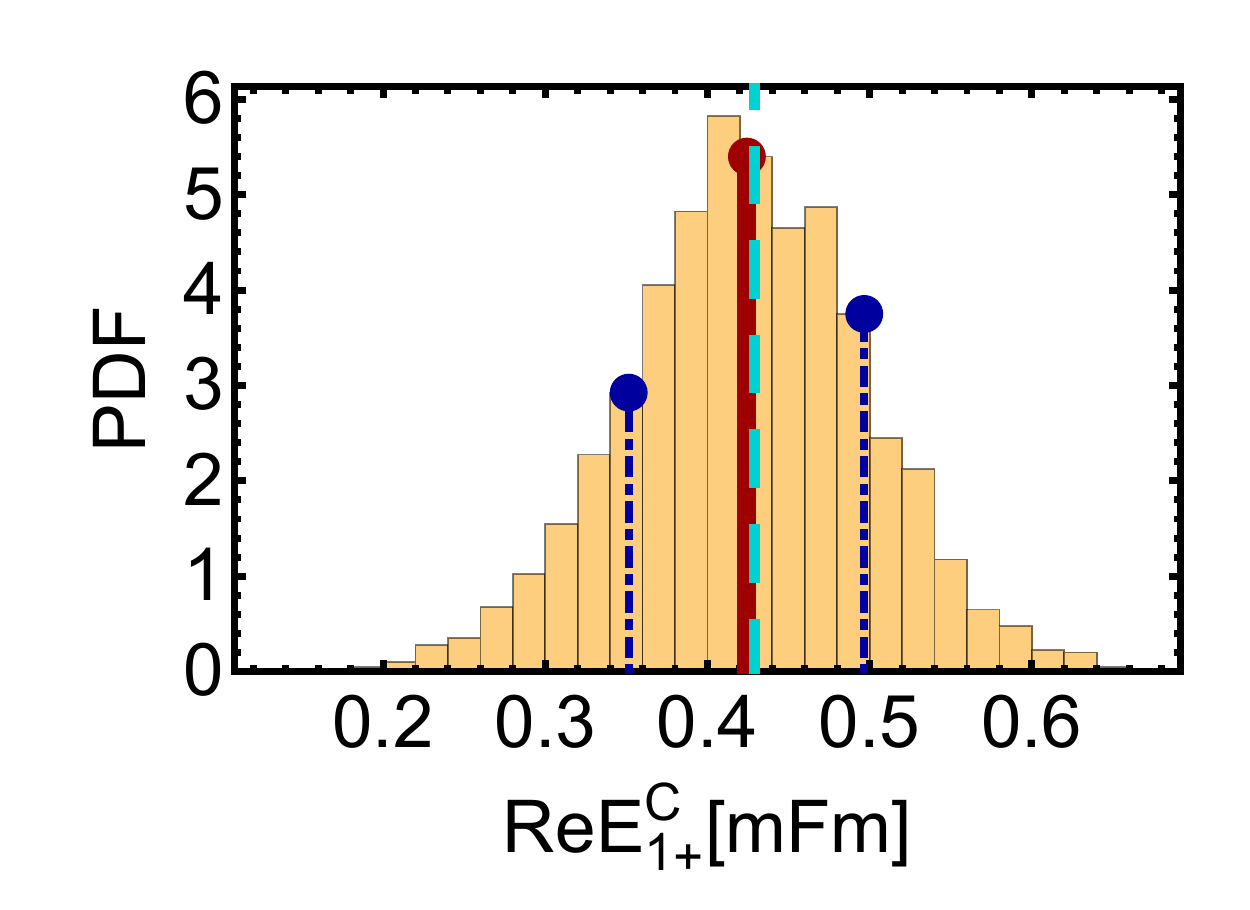}
 \end{overpic}
\begin{overpic}[width=0.325\textwidth]{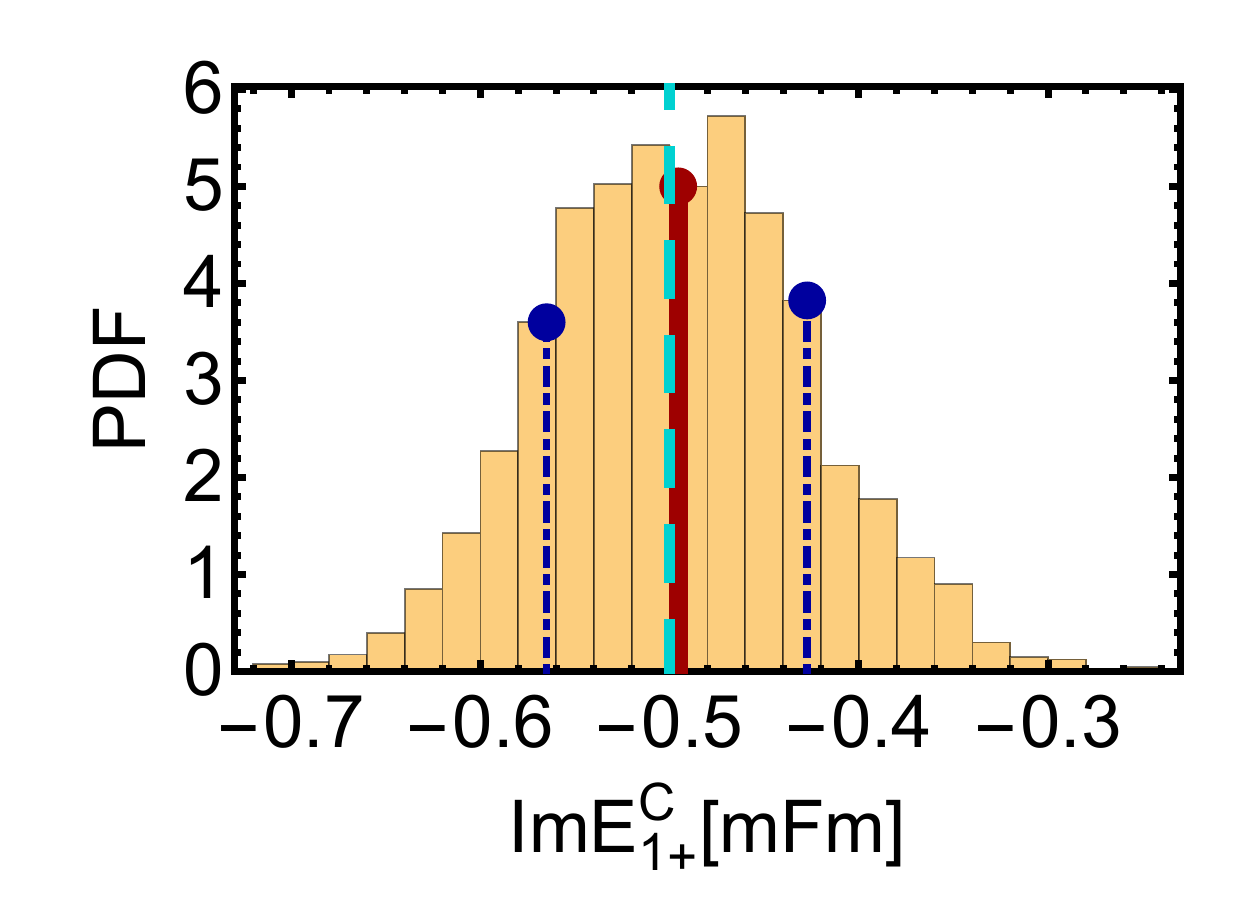}
 \end{overpic} \\
\begin{overpic}[width=0.325\textwidth]{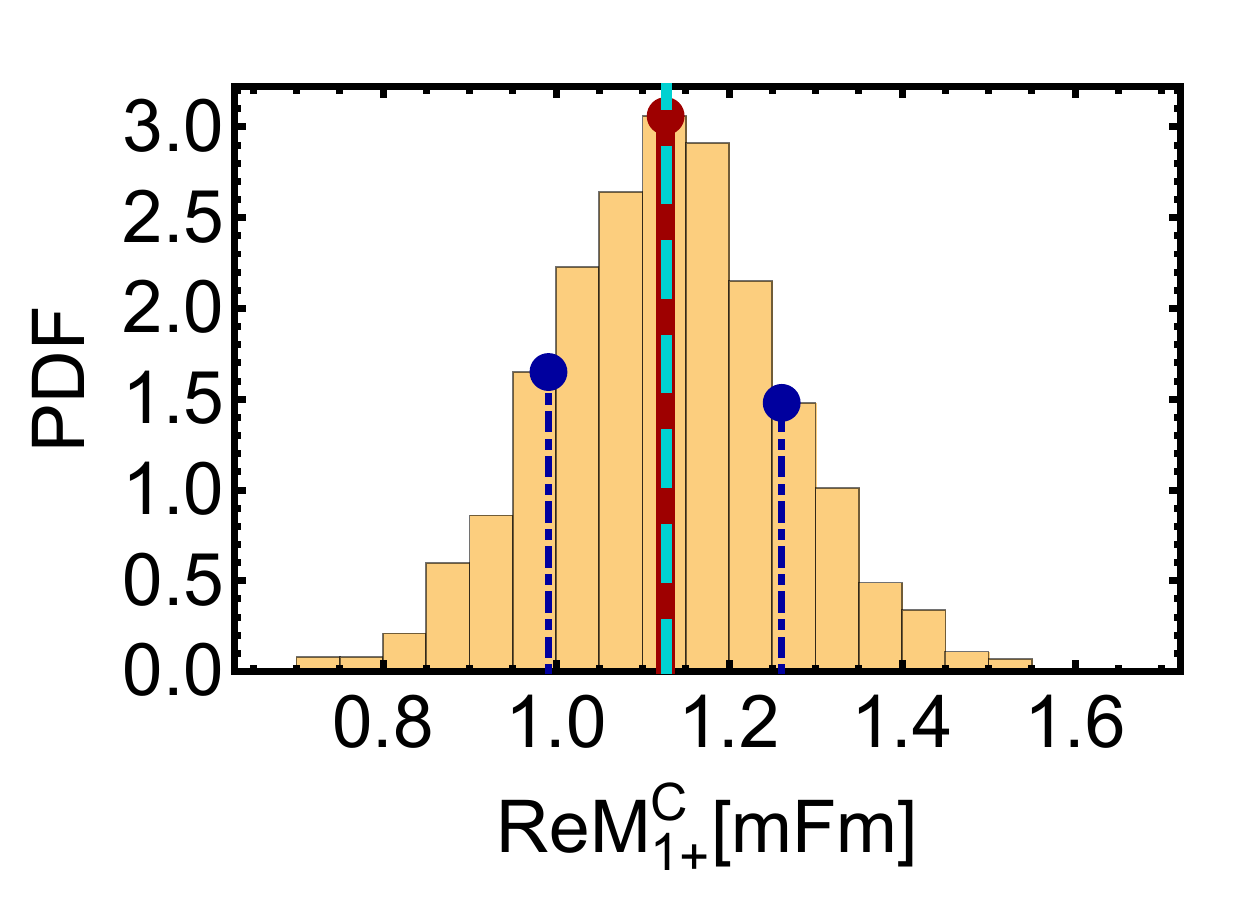}
 \end{overpic}
\begin{overpic}[width=0.325\textwidth]{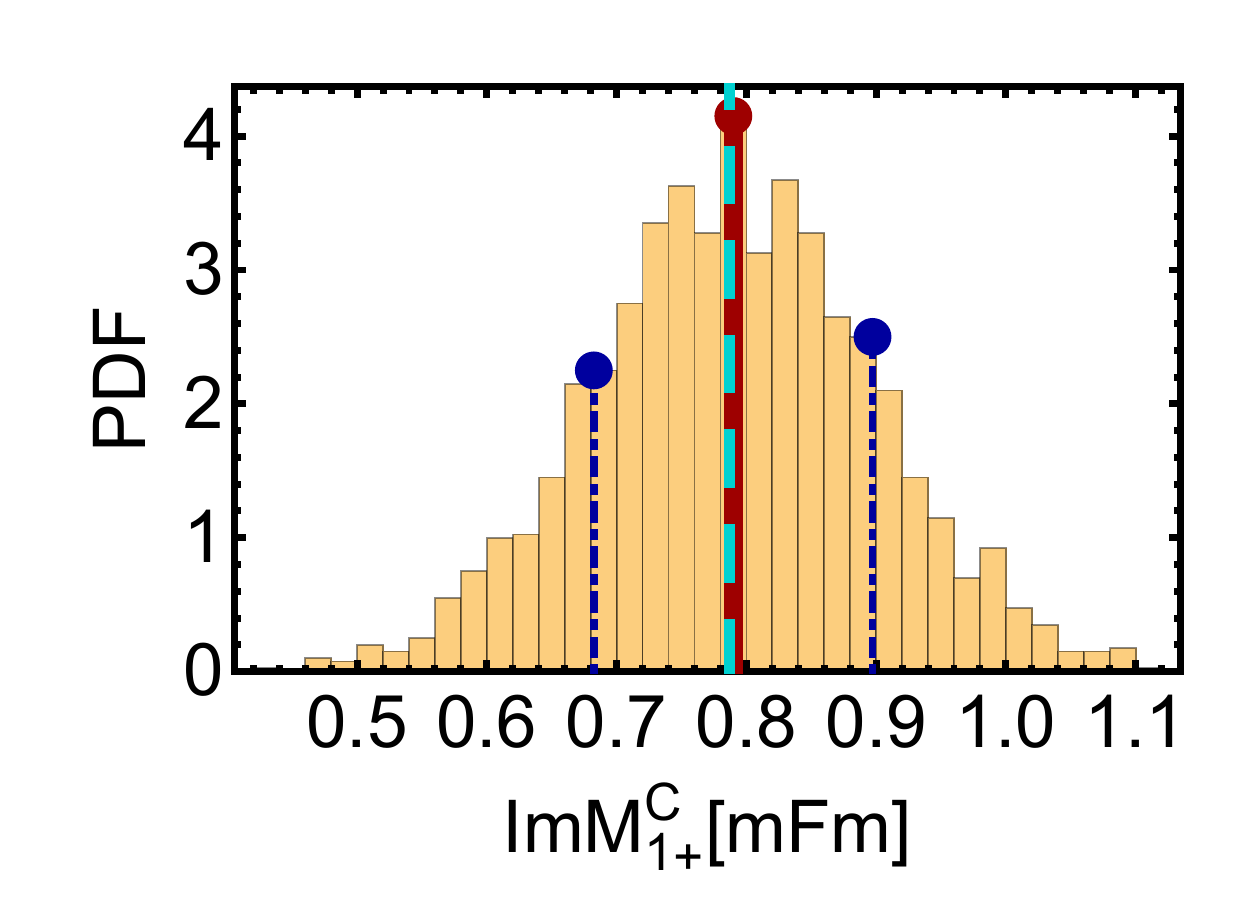}
 \end{overpic}
\begin{overpic}[width=0.325\textwidth]{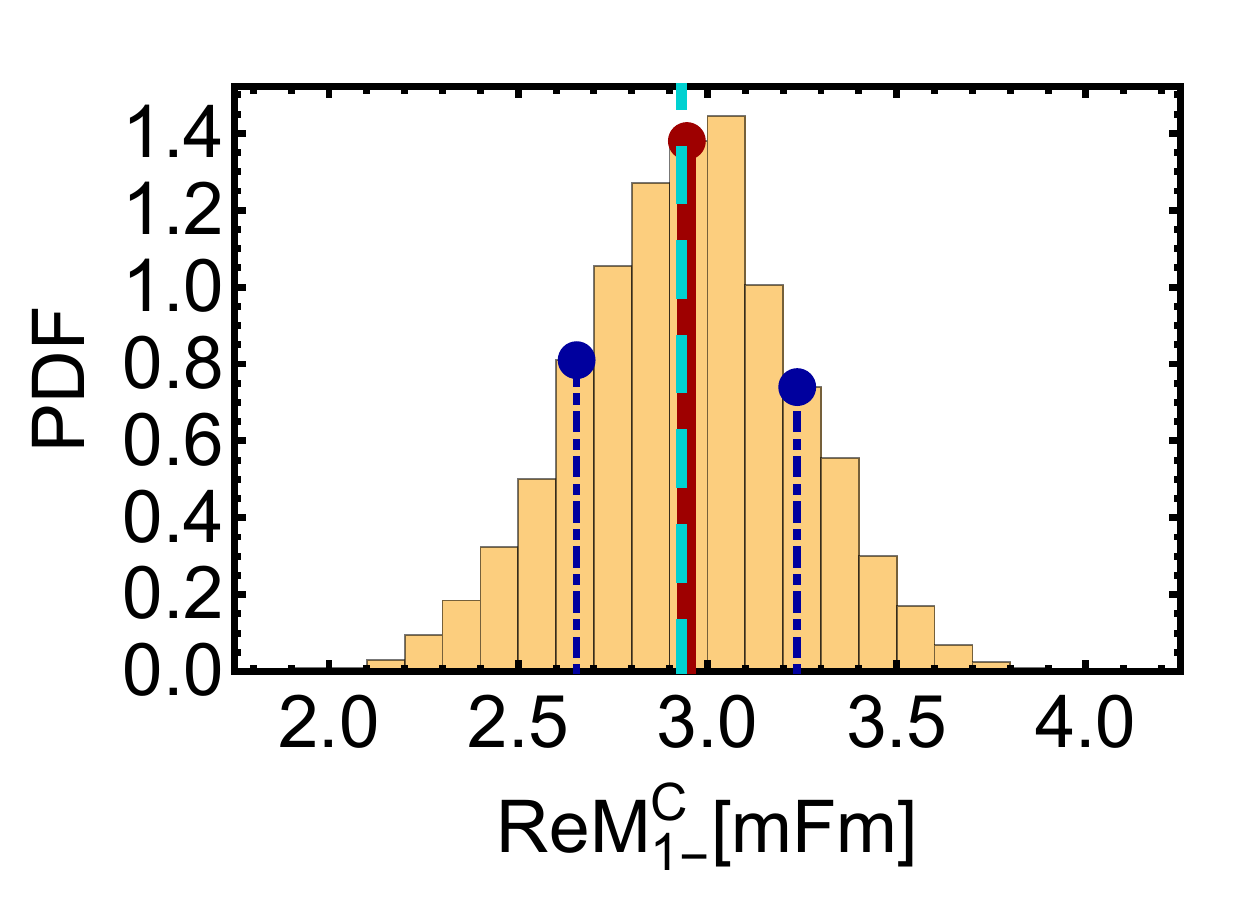}
 \end{overpic} \\
\begin{overpic}[width=0.325\textwidth]{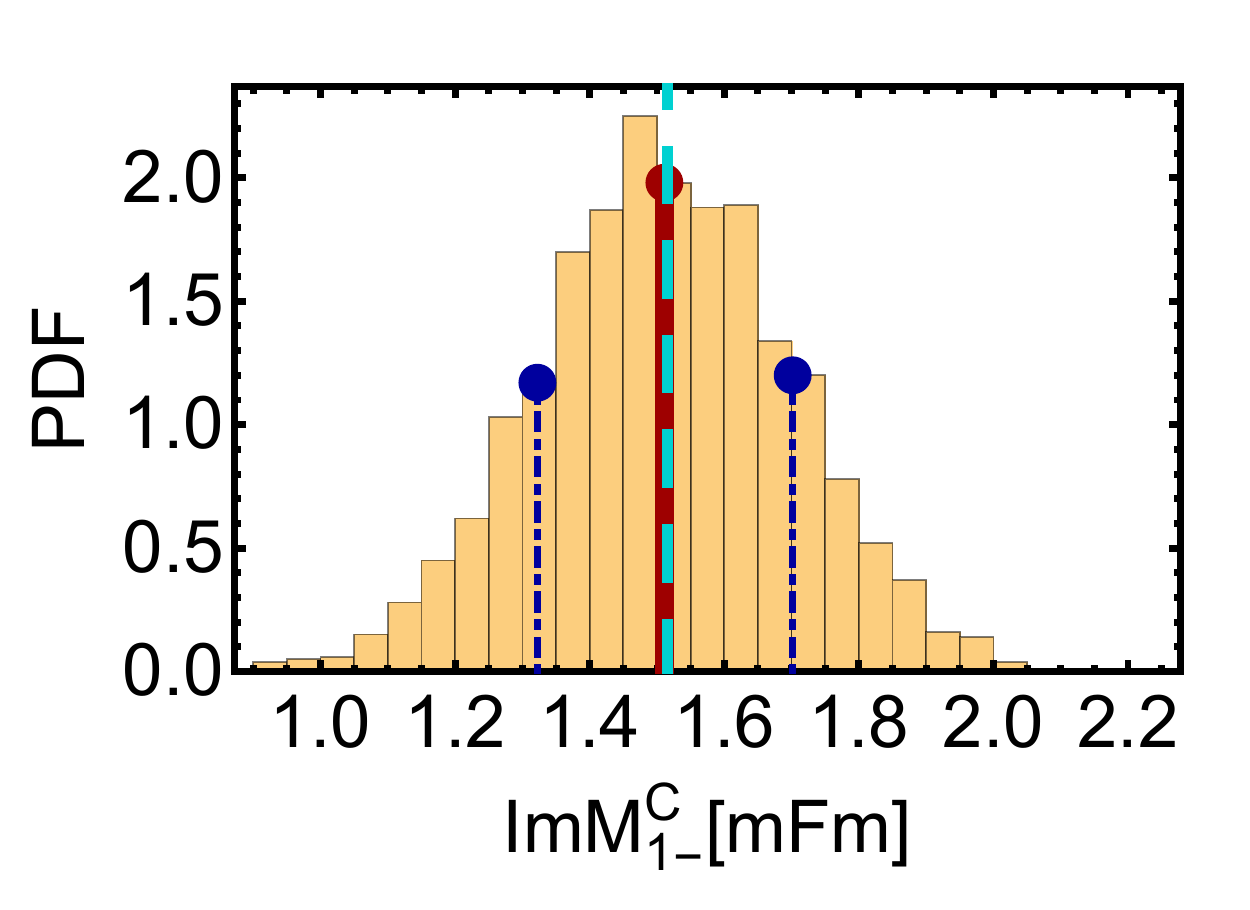}
 \end{overpic}
\begin{overpic}[width=0.325\textwidth]{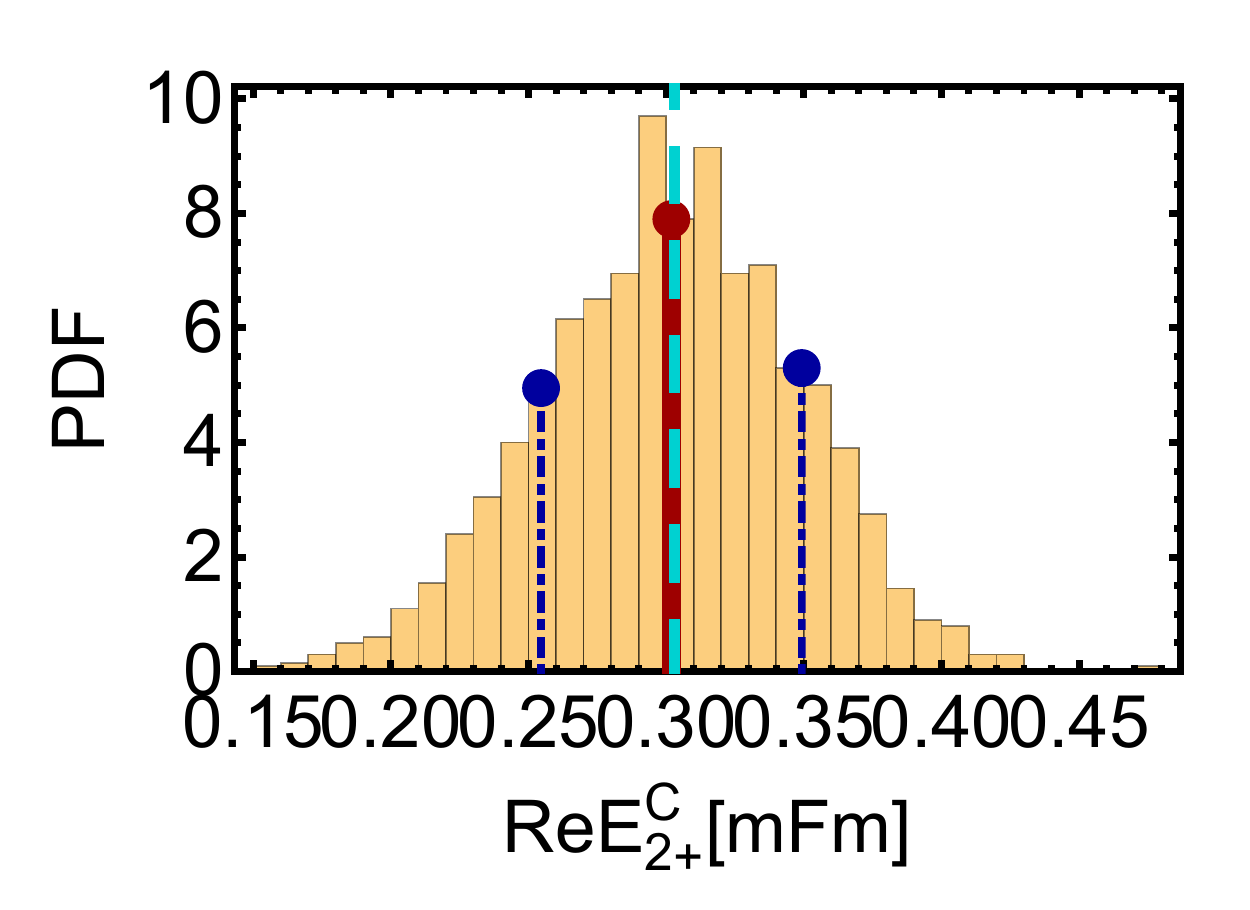}
 \end{overpic}
\begin{overpic}[width=0.325\textwidth]{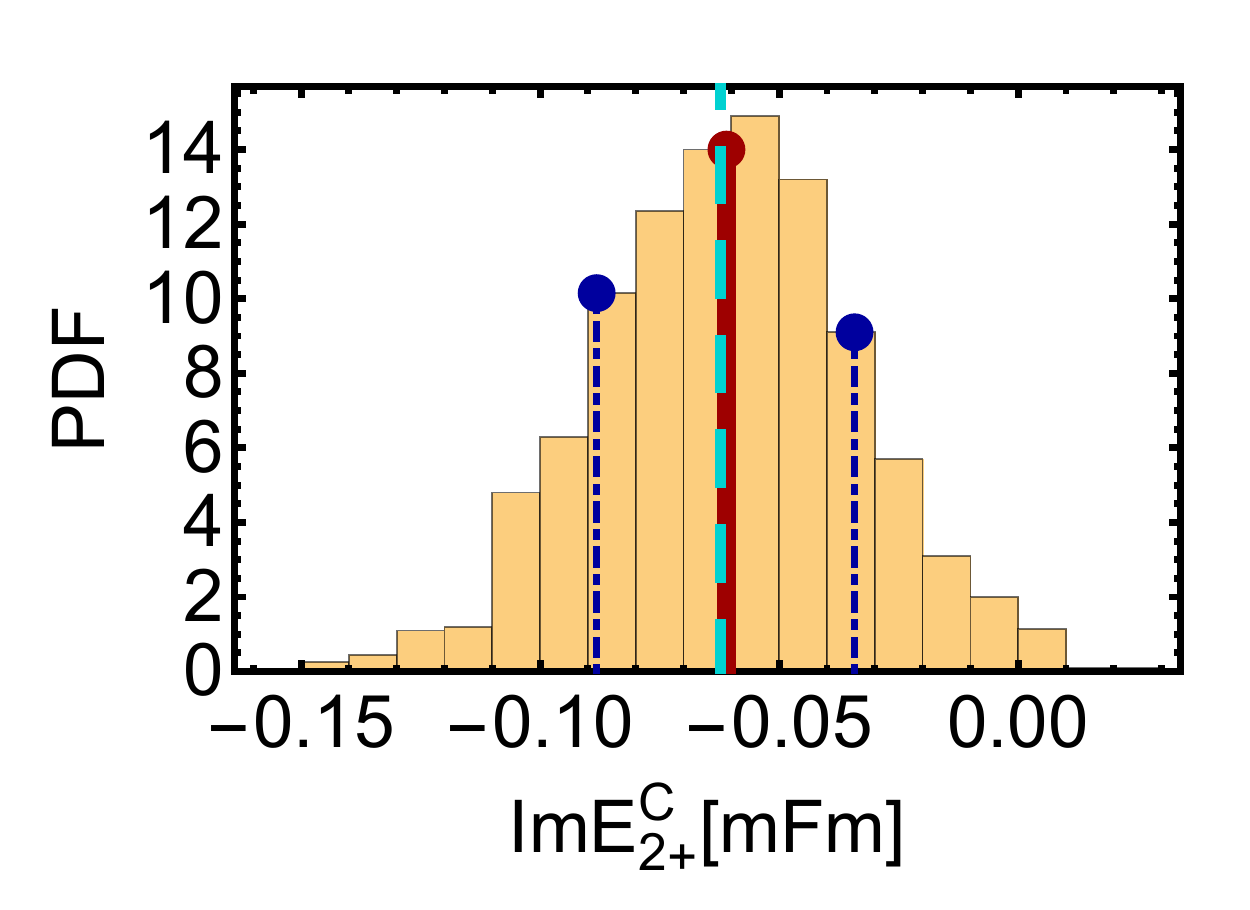}
 \end{overpic} \\
\begin{overpic}[width=0.325\textwidth]{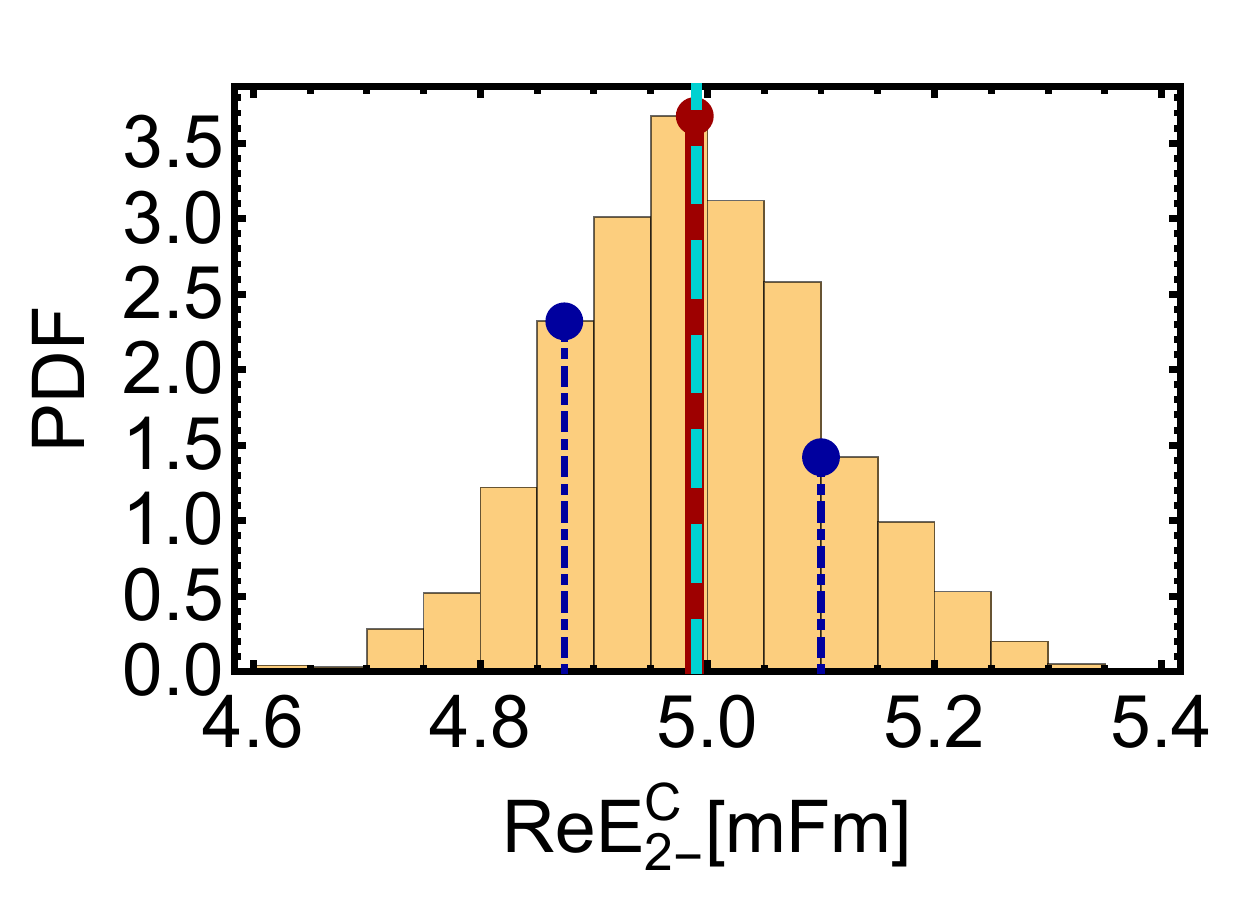}
 \end{overpic}
\begin{overpic}[width=0.325\textwidth]{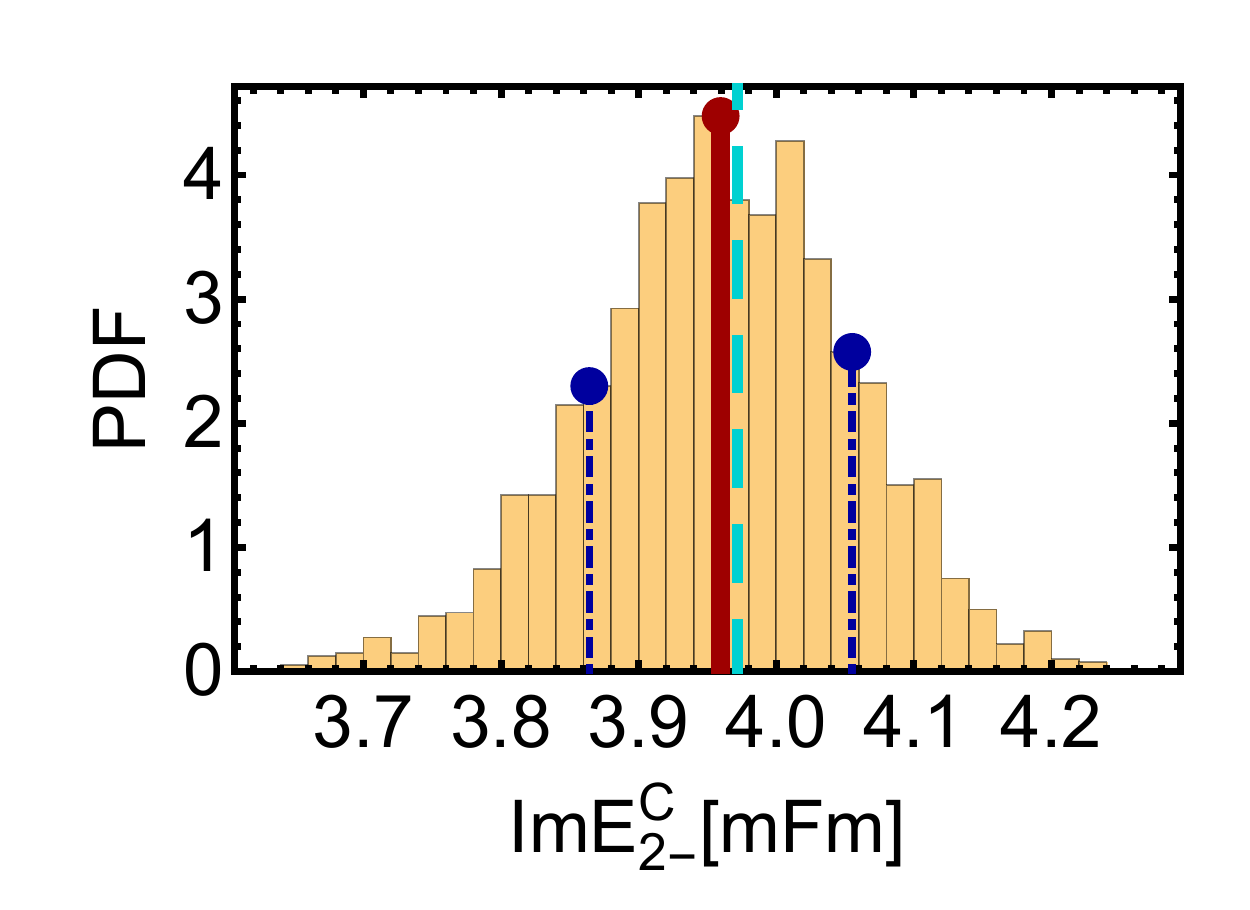}
 \end{overpic}
\begin{overpic}[width=0.325\textwidth]{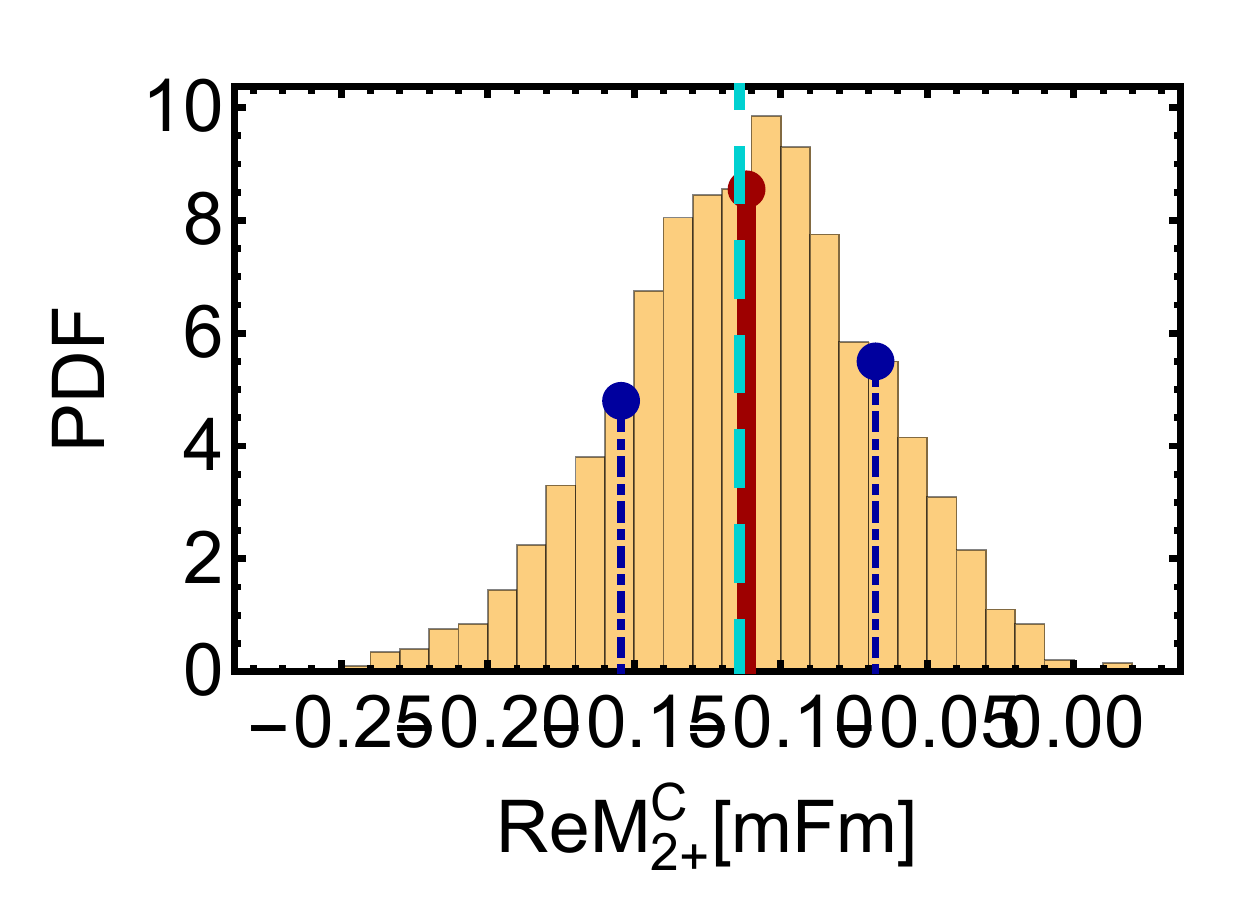}
 \end{overpic} \\
\begin{overpic}[width=0.325\textwidth]{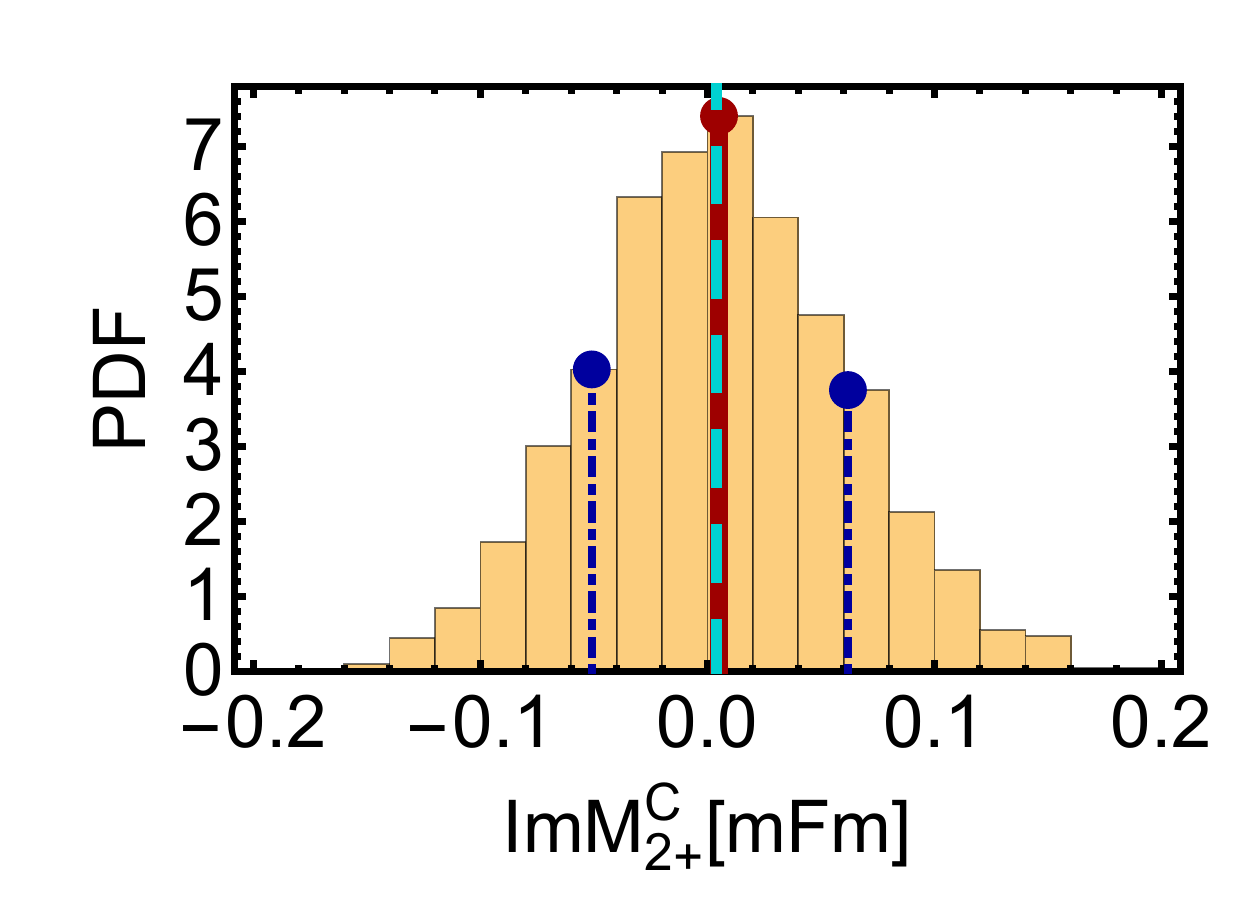}
 \end{overpic}
\begin{overpic}[width=0.325\textwidth]{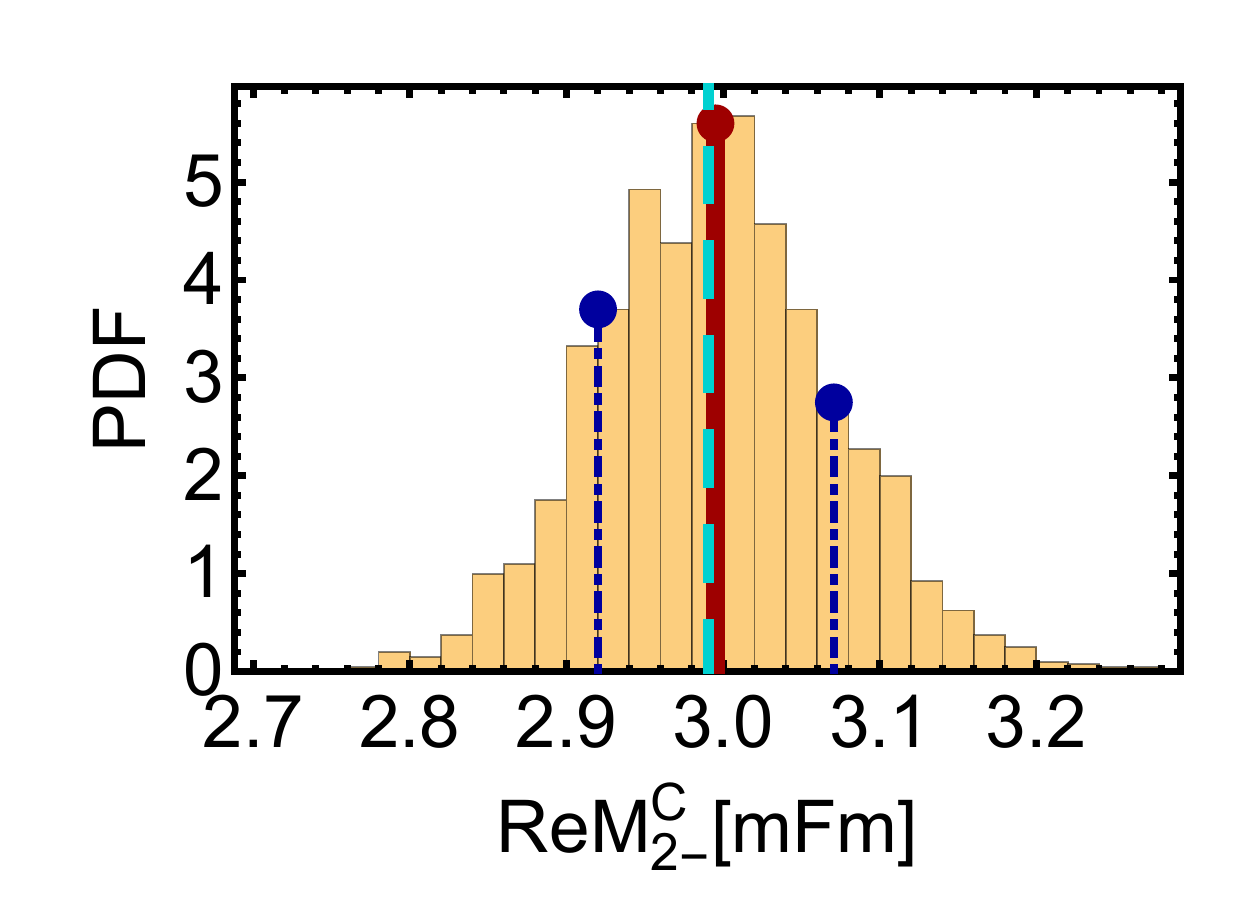}
 \end{overpic}
\begin{overpic}[width=0.325\textwidth]{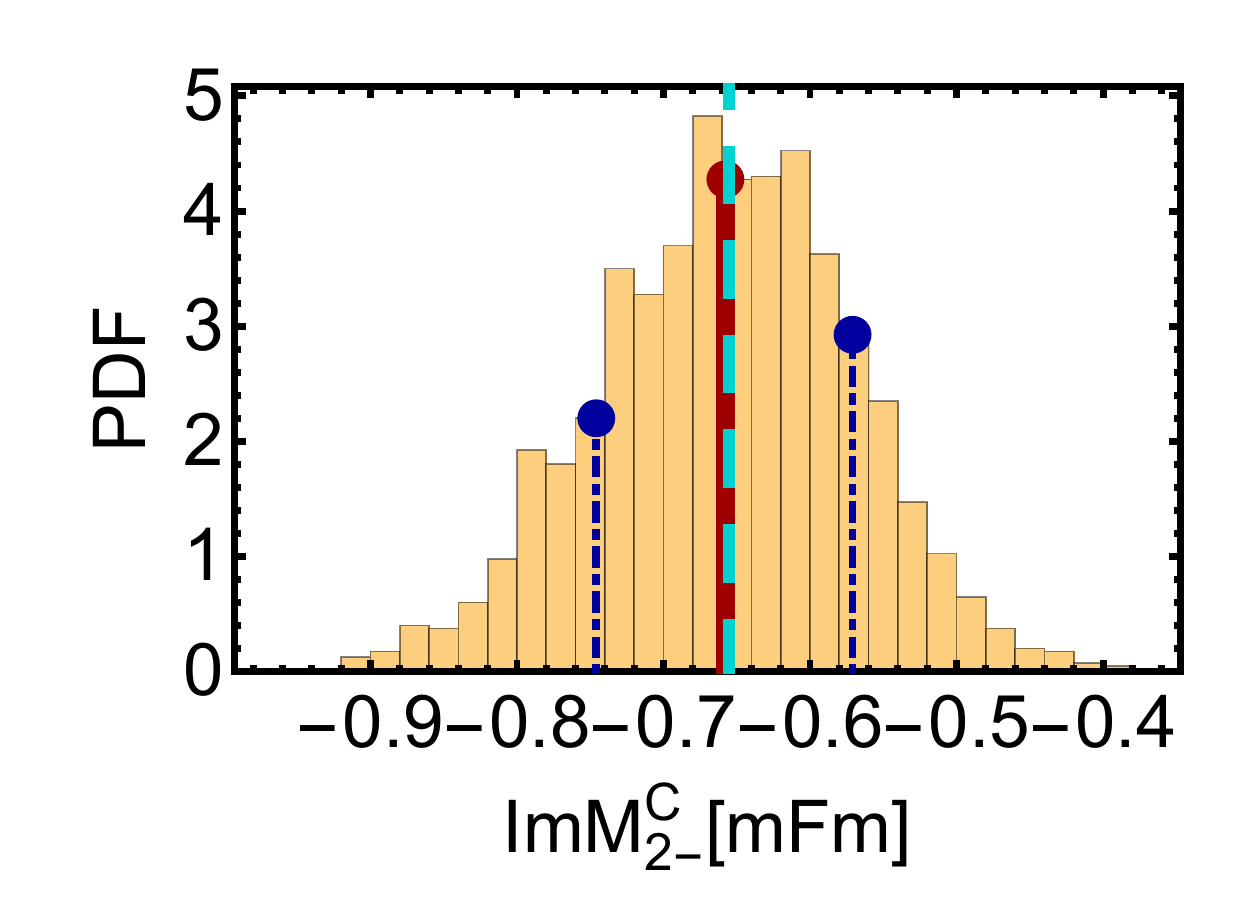}
 \end{overpic}
\caption[Bootstrap-distributions for multipole-fit-parameters in an analysis of photoproduction data on the second resonance region. The fourth energy-bin, \newline $E_{\gamma }\text{ = 783.42 MeV}$, is printed to aid the discussion in the main text.]{The plots show bootstrap-distributions for the real- and imaginary parts of phase-constrained $S$-, $P$- and $D$-wave multipoles. As an example, the fourth energy-bin, $E_{\gamma }\text{ = 783.42 MeV}$, is shown. The distributions result from an ensemble of $B=2000$ bootstrap-replicates. They have been normalized to unit probability via use of the object \textit{HistogramDistribution} in MATHEMATICA \cite{MathematicaLanguage,MathematicaBonnLicense}. Thus, $y$-axes are labeled as \textit{PDF}. \newline
The mean of each distribution is shown as a red solid line, while the $0.16$- and $0.84$-quantiles are indicated by blue dash-dotted lines. The global minimum of the fit to the original data is plotted as a cyan-colored dashed horizontal line. \newline
Distributions for all energies are shown in appendix \ref{sec:NumericalTPWAFitResults}.}
\label{fig:BootstrapHistos2ndResRegionEnergy4MainText}
\end{figure}

\clearpage

\begin{table}[h]
\centering
\begin{tabular}{c|c|c|c|c|c}
\multicolumn{2}{l|}{$ E_{\gamma }\text{ = 783.42 MeV} $} & \multicolumn{2}{c|}{ $ \text{ndf = 27} $} & \multicolumn{2}{c}{ $ \chi ^2\text{/ndf = 1.58912} $ } \\
\hline
\hline
$ \hat{\theta}_{i} = \left( \mathcal{M}_{\ell}^{C} \right)_{i} \text{[mFm]} $ & $ \left(\hat{\theta}_{i}^{\mathrm{Best}}\right)_{- \Delta_{-}}^{+ \Delta_{+}} $ & $ \hat{\theta}_{i}^{\ast} (\cdot) $ & $ \widehat{\mathrm{se}}_{B} \left( \hat{\theta}^{\ast}_{i} \right) $ & $ \widehat{\mathrm{bias}}_{B} $ & $ \delta_{\mathrm{bias}} $\\
\hline
 $ \mathrm{Re} \left[ E_{0+}^{C} \right]  $ & $ 11.2086_{-0.14756}^{+0.11384} $ & $ 11.1893 $ & $ 0.13499 $ & $ -0.01923 $ & $ 0.14246 $ \\
 $ \mathrm{Re} \left[ E_{1+}^{C} \right]  $ & $ 0.42901_{-0.07755}^{+0.06766} $ & $ 0.42432 $ & $ 0.07473 $ & $ -0.00469 $ & $ 0.06272 $ \\
 $ \mathrm{Im} \left[ E_{1+}^{C} \right]  $ & $ -0.50005_{-0.06495}^{+0.07273} $ & $ -0.49548 $ & $ 0.06986 $ & $ 0.00457 $ & $ 0.06541 $ \\
 $ \mathrm{Re} \left[ M_{1+}^{C} \right]  $ & $ 1.12782_{-0.13661}^{+0.13299} $ & $ 1.12668 $ & $ 0.13766 $ & $ -0.00115 $ & $ 0.00834 $ \\
 $ \mathrm{Im} \left[ M_{1+}^{C} \right]  $ & $ 0.78681_{-0.10437}^{+0.11039} $ & $ 0.78994 $ & $ 0.10845 $ & $ 0.00313 $ & $ 0.02882 $ \\
 $ \mathrm{Re} \left[ M_{1-}^{C} \right]  $ & $ 2.93183_{-0.27751}^{+0.30533} $ & $ 2.9456 $ & $ 0.2942 $ & $ 0.01377 $ & $ 0.04681 $ \\
 $ \mathrm{Im} \left[ M_{1-}^{C} \right]  $ & $ 1.51603_{-0.1938}^{+0.18564} $ & $ 1.51098 $ & $ 0.19009 $ & $ -0.00505 $ & $ 0.02657 $ \\
 $ \mathrm{Re} \left[ E_{2+}^{C} \right] $ & $ 0.30299_{-0.04845}^{+0.04625} $ & $ 0.30188 $ & $ 0.04695 $ & $ -0.00111 $ & $ 0.02365 $ \\
 $ \mathrm{Im} \left[ E_{2+}^{C} \right] $ & $ -0.06221_{-0.02596}^{+0.02798} $ & $ -0.06103 $ & $ 0.02742 $ & $ 0.00118 $ & $ 0.04313 $ \\
 $ \mathrm{Re} \left[ E_{2-}^{C} \right] $ & $ 4.9901_{-0.11617}^{+0.1099} $ & $ 4.98882 $ & $ 0.11585 $ & $ -0.00127 $ & $ 0.01097 $ \\
 $ \mathrm{Im} \left[ E_{2-}^{C} \right] $ & $ 3.97206_{-0.10793}^{+0.08308} $ & $ 3.95957 $ & $ 0.09654 $ & $ -0.01249 $ & $ 0.12938 $ \\
 $ \mathrm{Re} \left[ M_{2+}^{C} \right] $ & $ -0.11402_{-0.04045}^{+0.04639} $ & $ -0.11165 $ & $ 0.04363 $ & $ 0.00237 $ & $ 0.05442 $ \\
 $ \mathrm{Im} \left[ M_{2+}^{C} \right] $ & $ 0.00387_{-0.0548}^{+0.058} $ & $ 0.00509 $ & $ 0.05665 $ & $ 0.00122 $ & $ 0.02155 $ \\
 $ \mathrm{Re} \left[ M_{2-}^{C} \right] $ & $ 2.99081_{-0.07067}^{+0.07992} $ & $ 2.99511 $ & $ 0.07605 $ & $ 0.0043 $ & $ 0.0566 $ \\
 $ \mathrm{Im} \left[ M_{2-}^{C} \right] $ & $ -0.65531_{-0.0906}^{+0.08437} $ & $ -0.65777 $ & $ 0.08842 $ & $ -0.00246 $ & $ 0.0278 $ \\
\end{tabular}
\caption[Parameters extracted from the bootstrap-distributions of the fourth energy-bin, $E_{\gamma }\text{ = 783.42 MeV}$.]{Here, we collect results of parameters extracted from the bootstrap-distributions of the fourth energy-bin, $E_{\gamma }\text{ = 783.42 MeV}$, which have been plotted in Figure \ref{fig:BootstrapHistos2ndResRegionEnergy4MainText}. An ensemble of $B = 2000$ bootstrap-replicates has been applied. The Monte Carlo minimum-search described in the main text has found a global minimum with $\chi ^2\text{/ndf = 1.58912}$. All quantities are explained in the main text. \newline
All numbers, except for $\delta_{\mathrm{bias}}$, are given in milli-Fermi. Numerical results for all the remaining energies are contained in appendix \ref{sec:NumericalTPWAFitResults}.}
\label{tab:2ndResRegionResultsFourthEnergyMainText}
\end{table}

For a few selected parameters in the first and second energy-bin, as well as the first two parameters of the seventh energy, the bias rule-of-thumb $\delta_{\mathrm{bias}} < 0.25$ is violated slightly. In all cases where such a violation is met, one observes either a very long tail of the distribution in one particular direction, with some very few fit-results located very far away from the center. Or, sometimes, a small side-bump is seen which is not separated from the primary distribution. However, distributions showing such small bumps are only a few in between. \newline

\begin{figure}[h]
\begin{overpic}[width=0.325\textwidth]{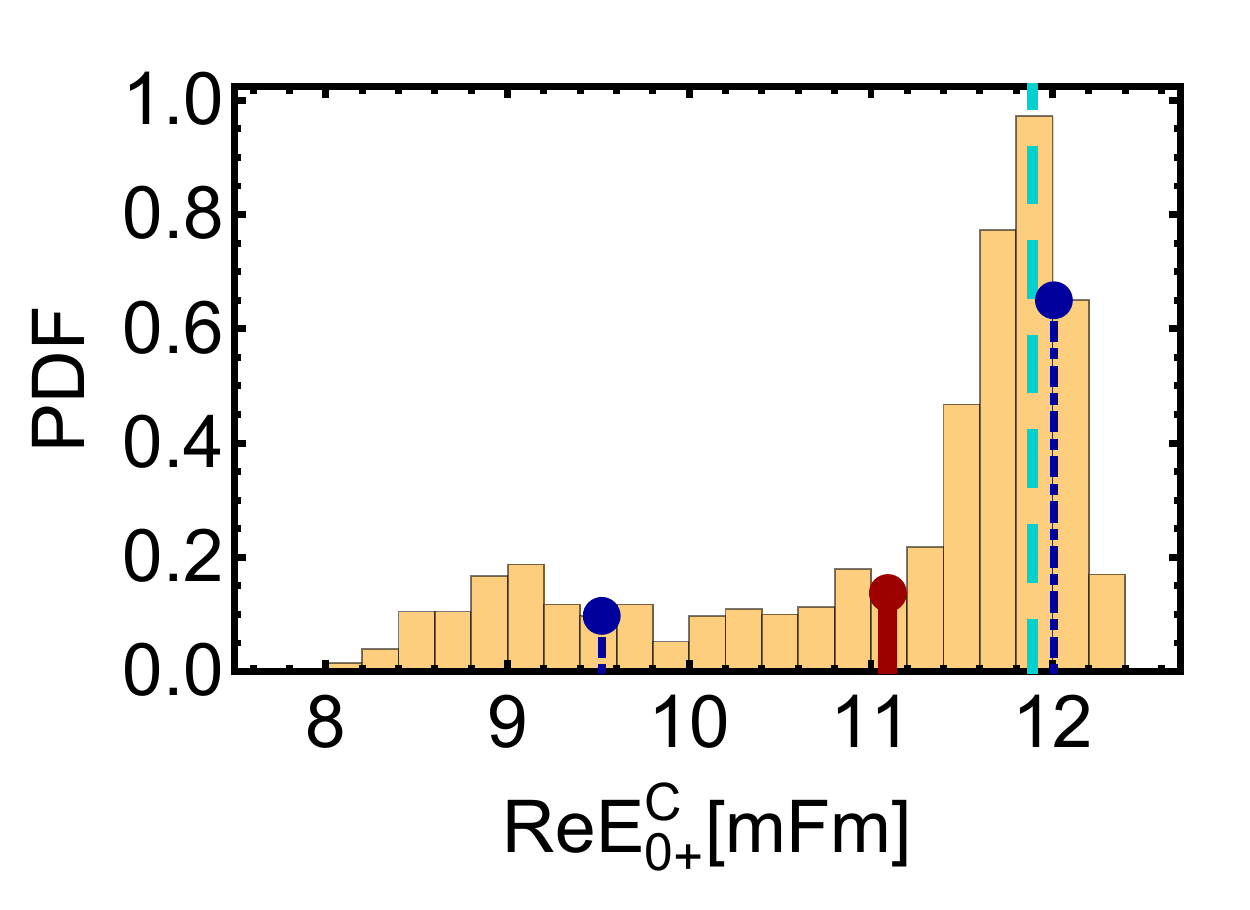}
 \end{overpic}
\begin{overpic}[width=0.325\textwidth]{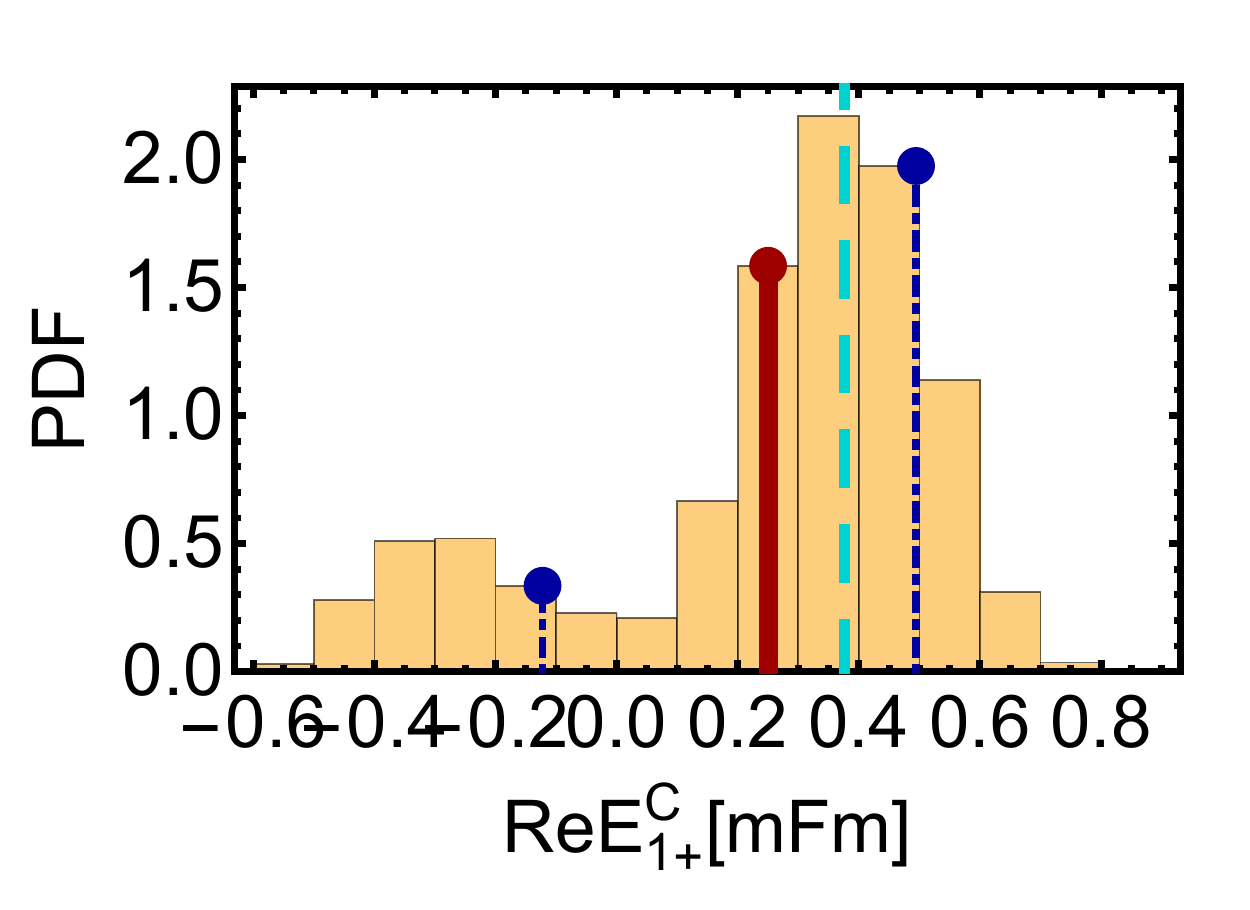}
 \end{overpic}
\begin{overpic}[width=0.325\textwidth]{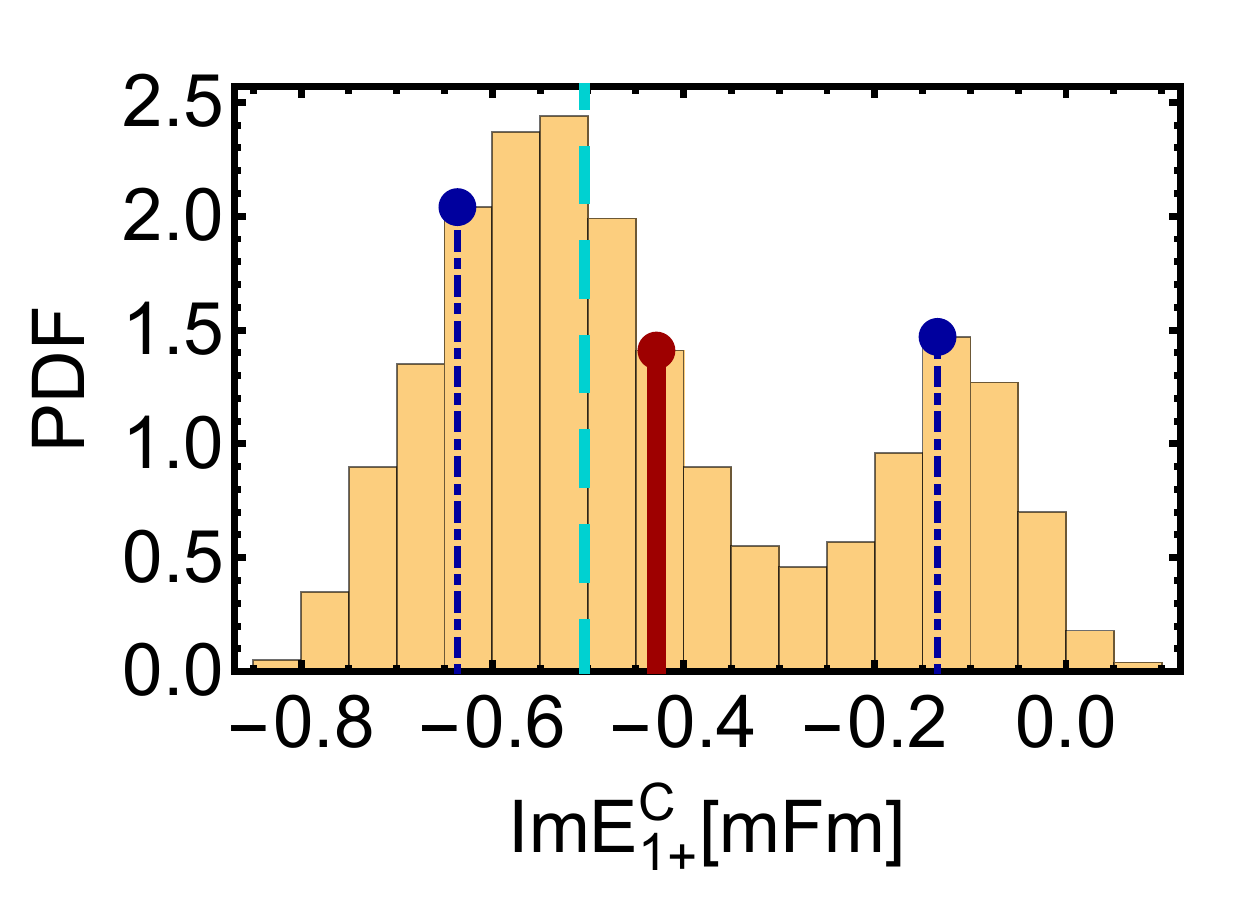}
 \end{overpic} \\
\begin{overpic}[width=0.325\textwidth]{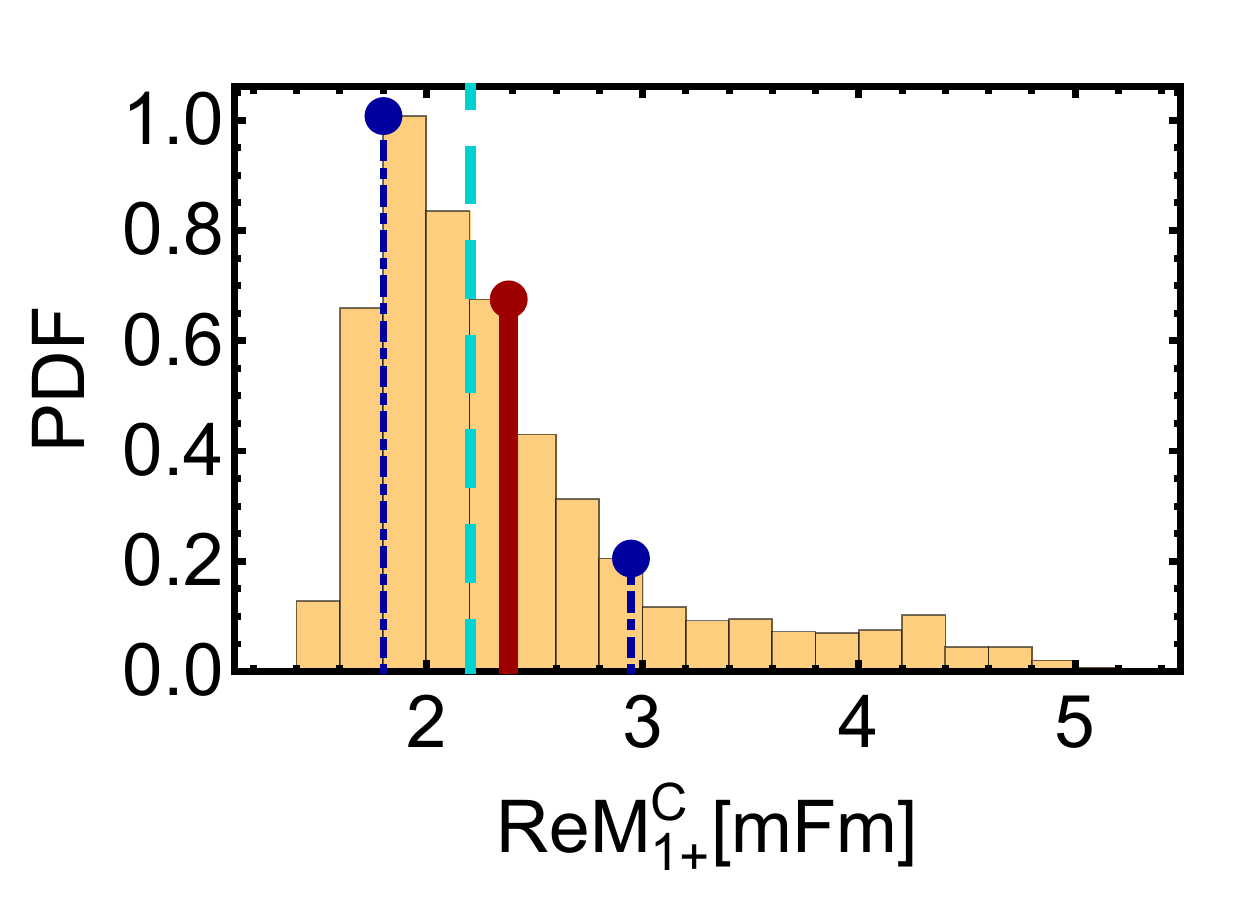}
 \end{overpic}
\begin{overpic}[width=0.325\textwidth]{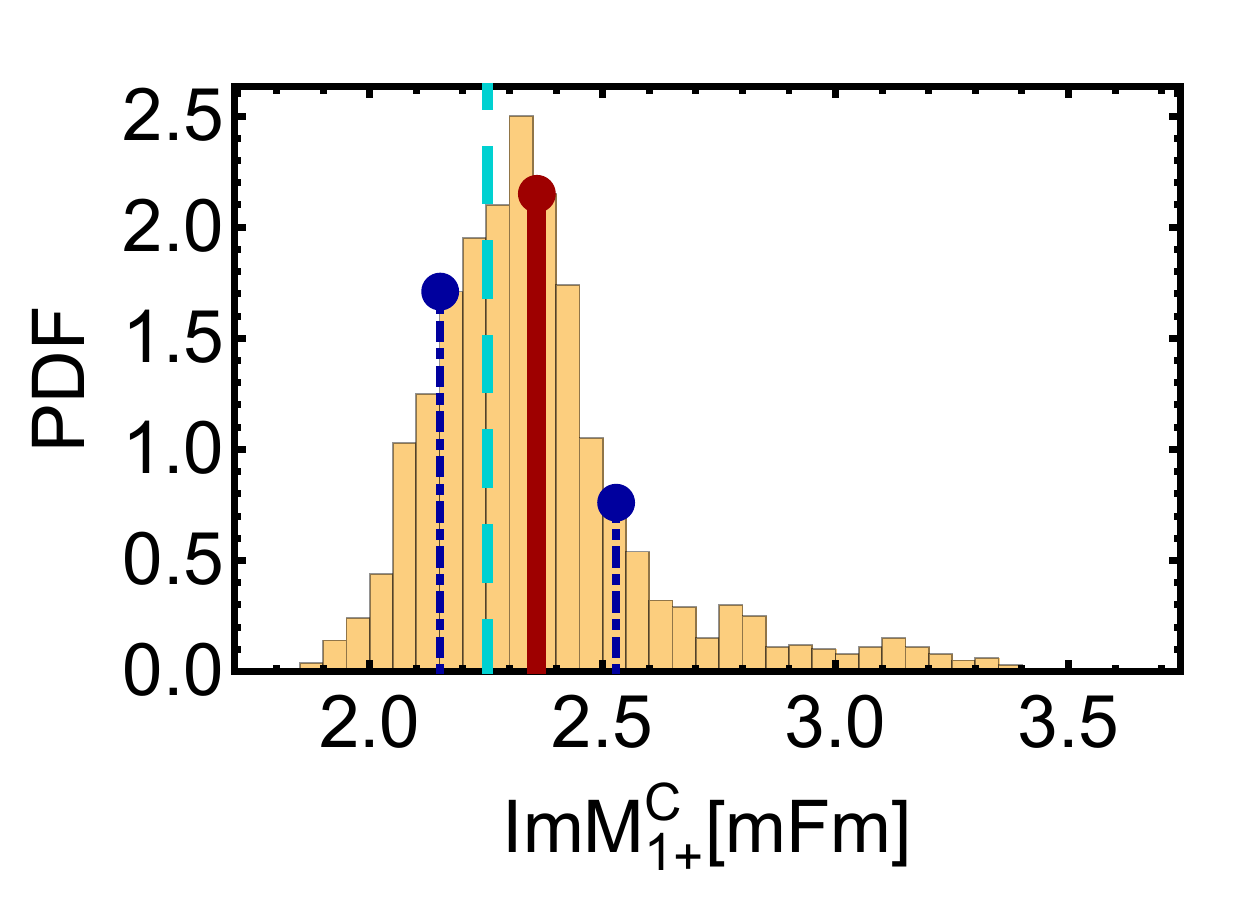}
 \end{overpic}
\begin{overpic}[width=0.325\textwidth]{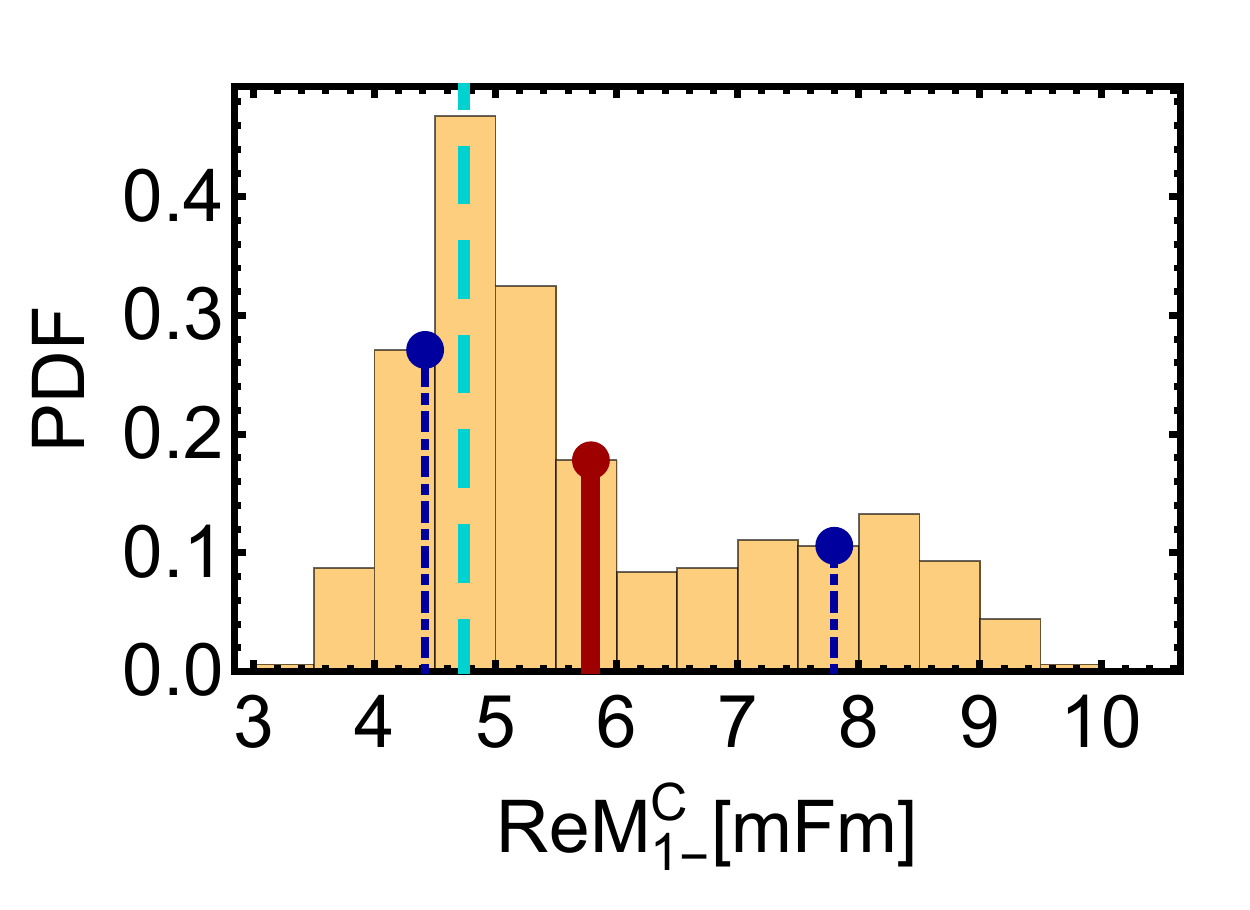}
 \end{overpic} \\
\begin{overpic}[width=0.325\textwidth]{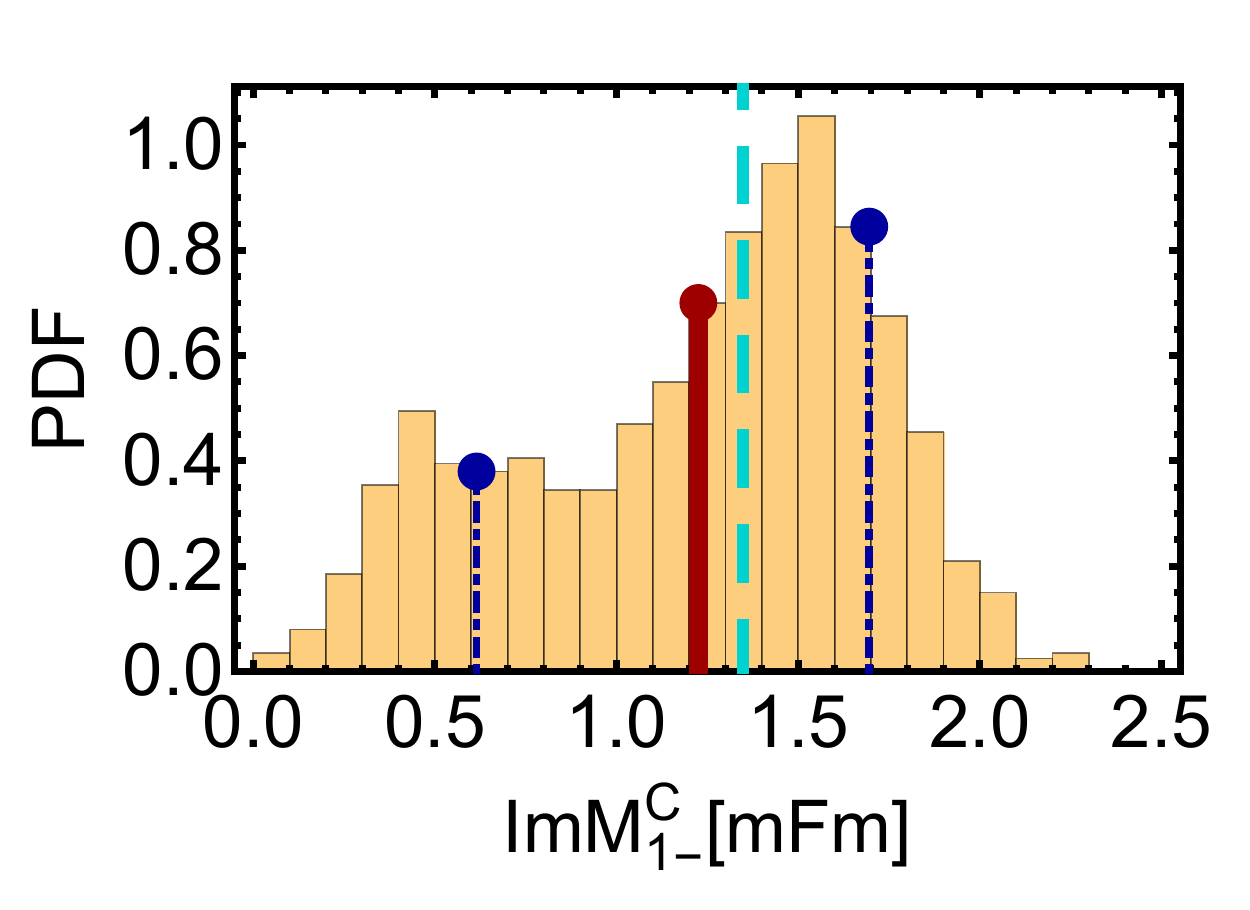}
 \end{overpic}
\begin{overpic}[width=0.325\textwidth]{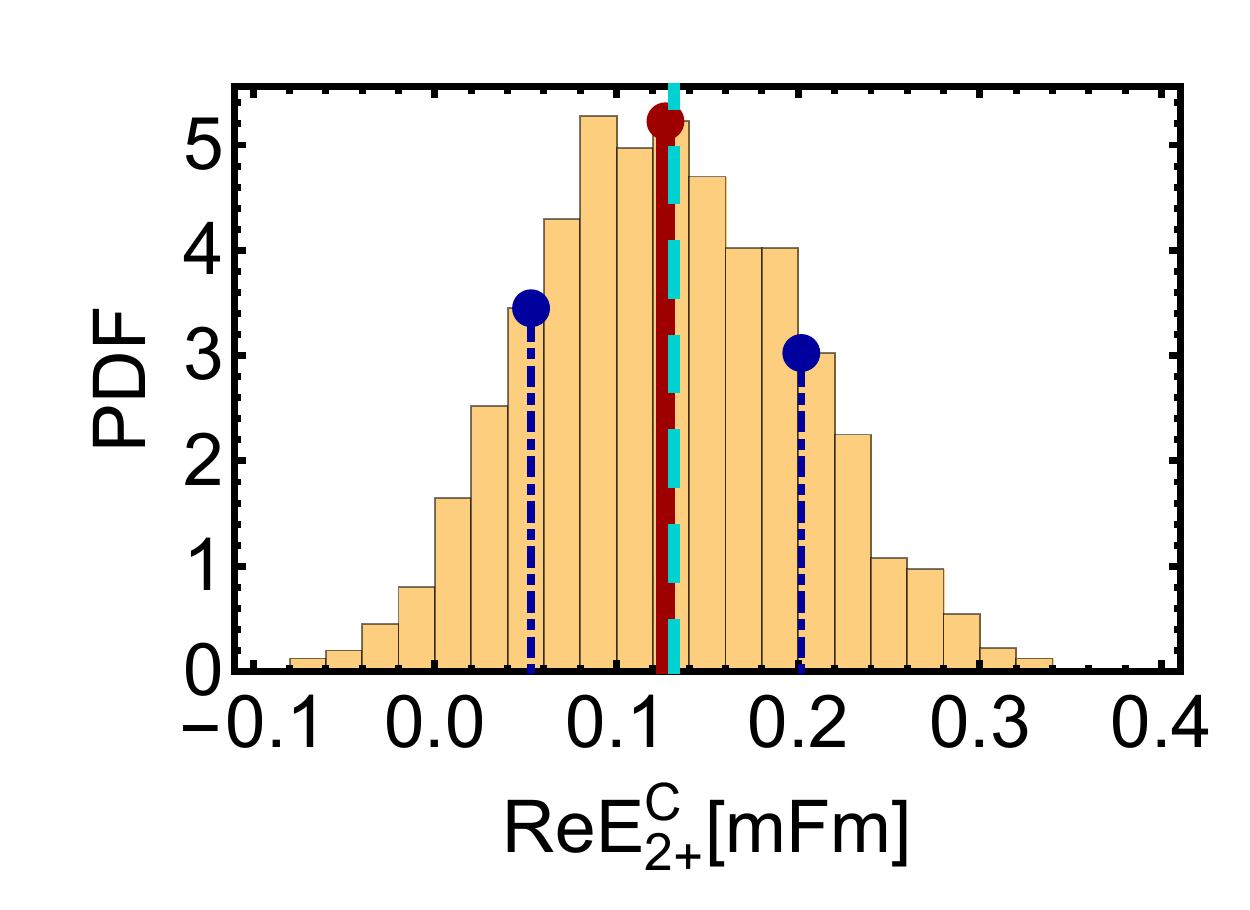}
 \end{overpic}
\begin{overpic}[width=0.325\textwidth]{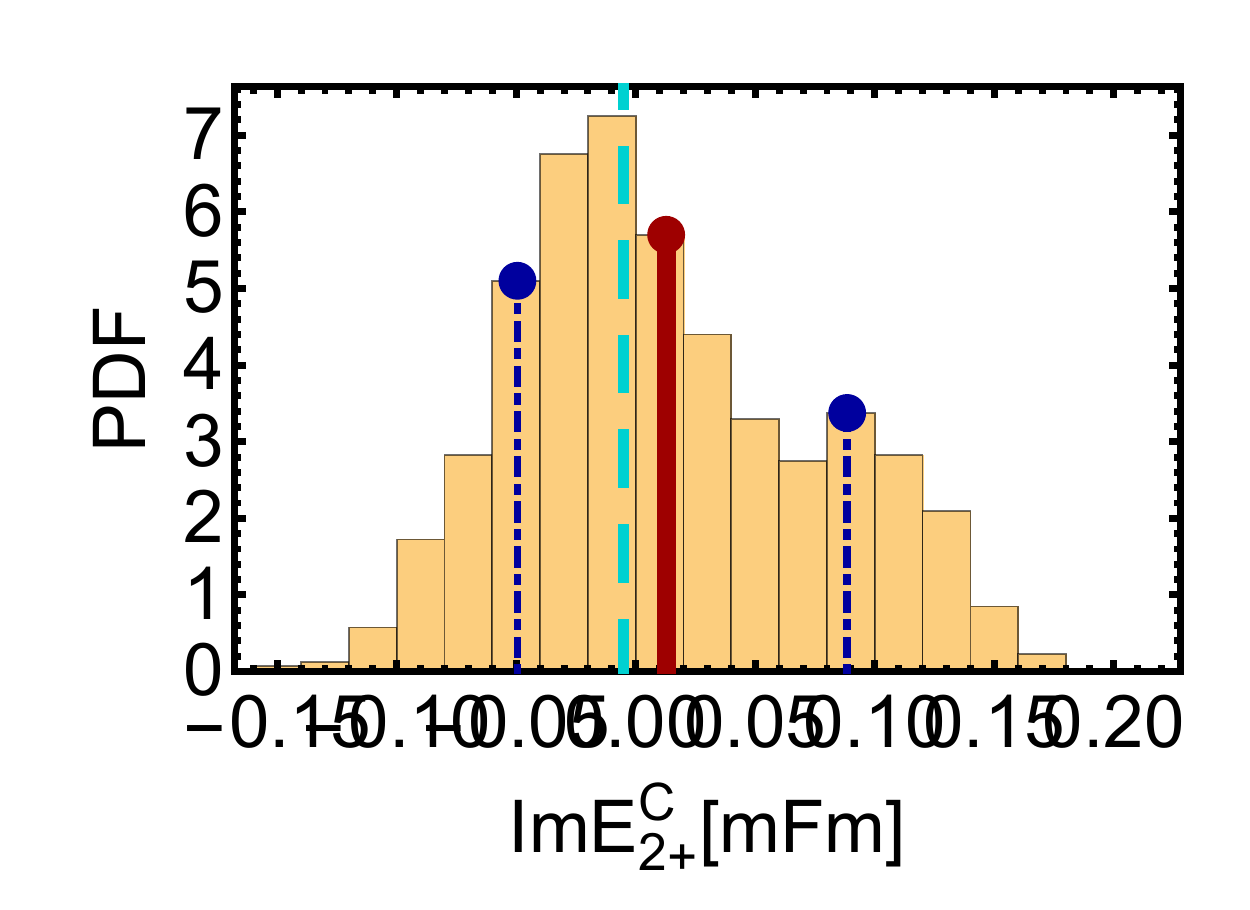}
 \end{overpic} \\
\begin{overpic}[width=0.325\textwidth]{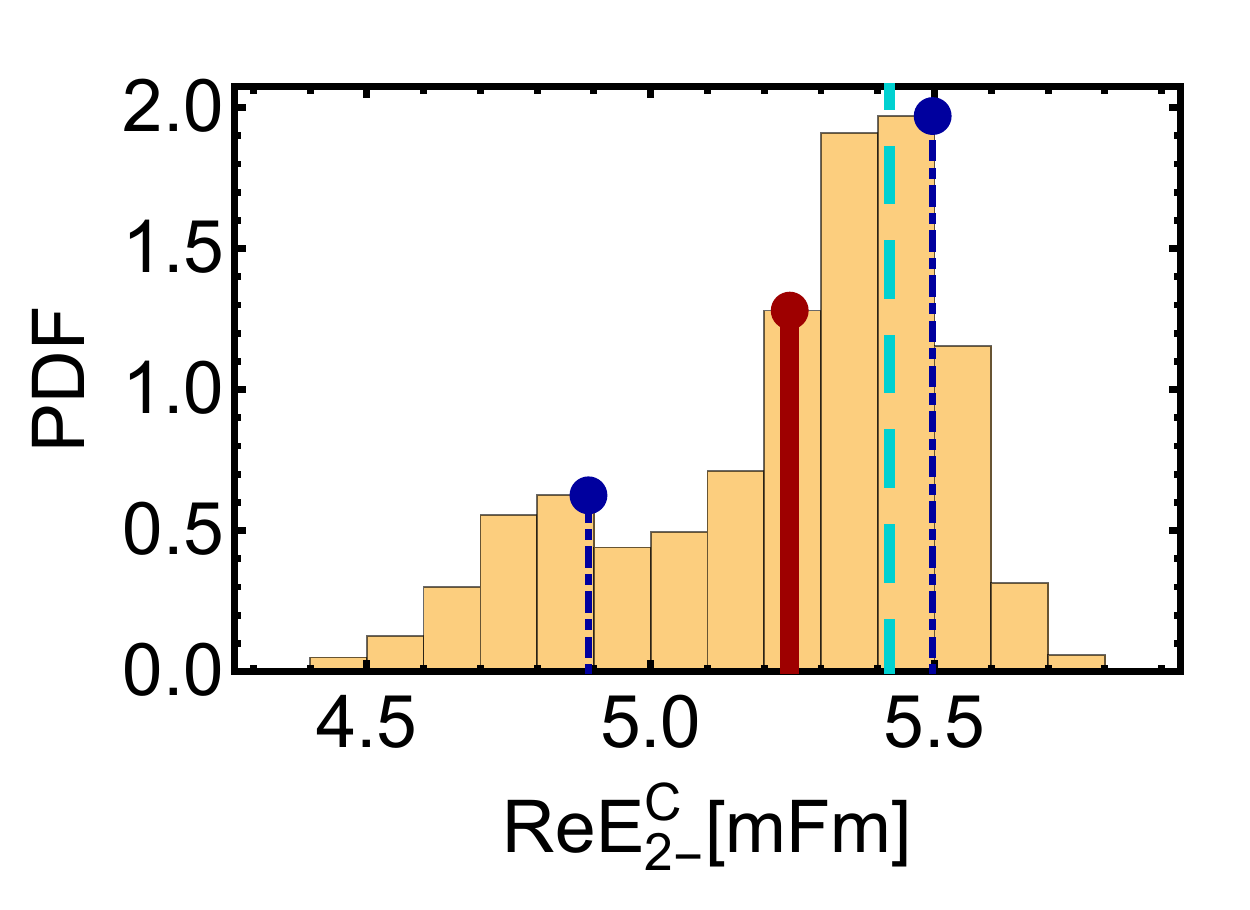}
 \end{overpic}
\begin{overpic}[width=0.325\textwidth]{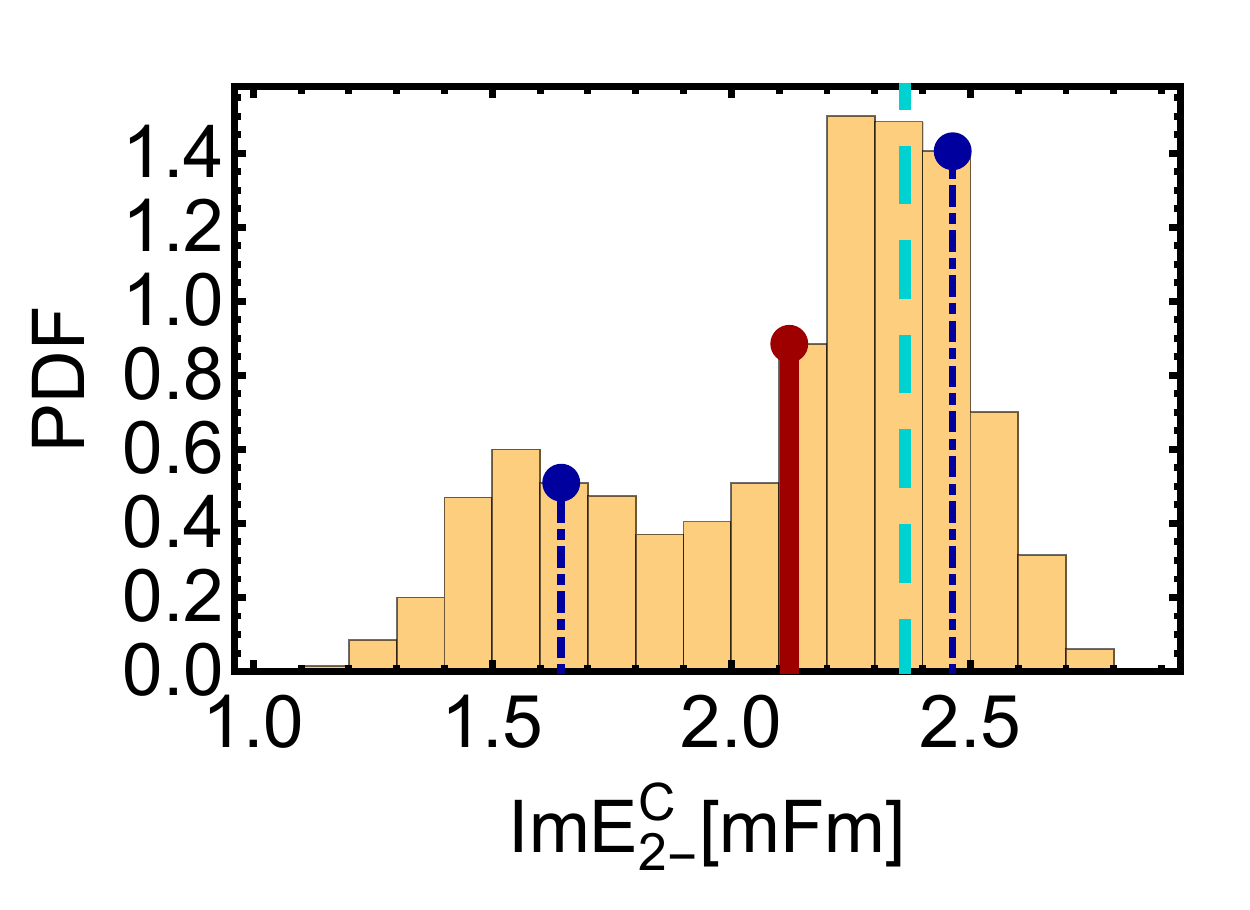}
 \end{overpic}
\begin{overpic}[width=0.325\textwidth]{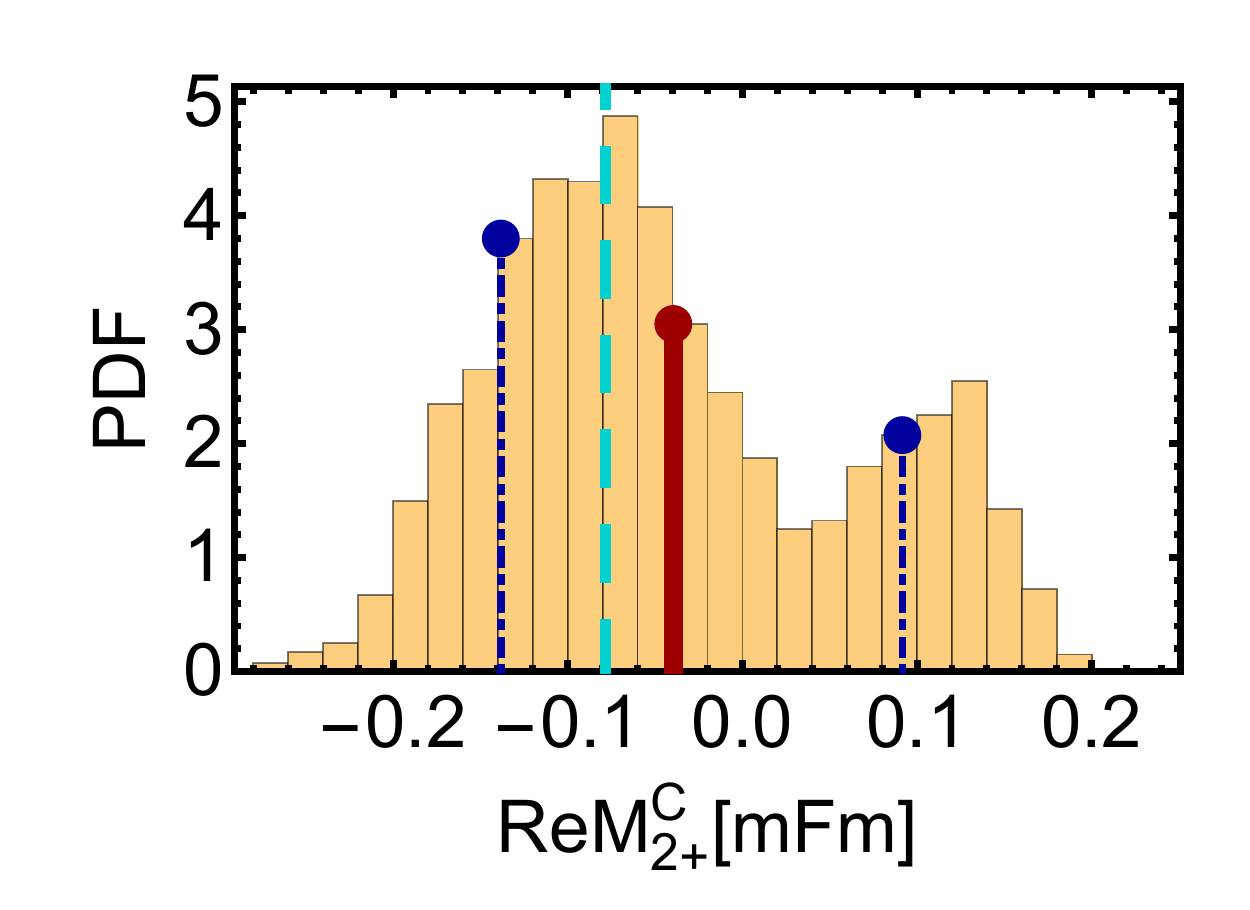}
 \end{overpic} \\
\begin{overpic}[width=0.325\textwidth]{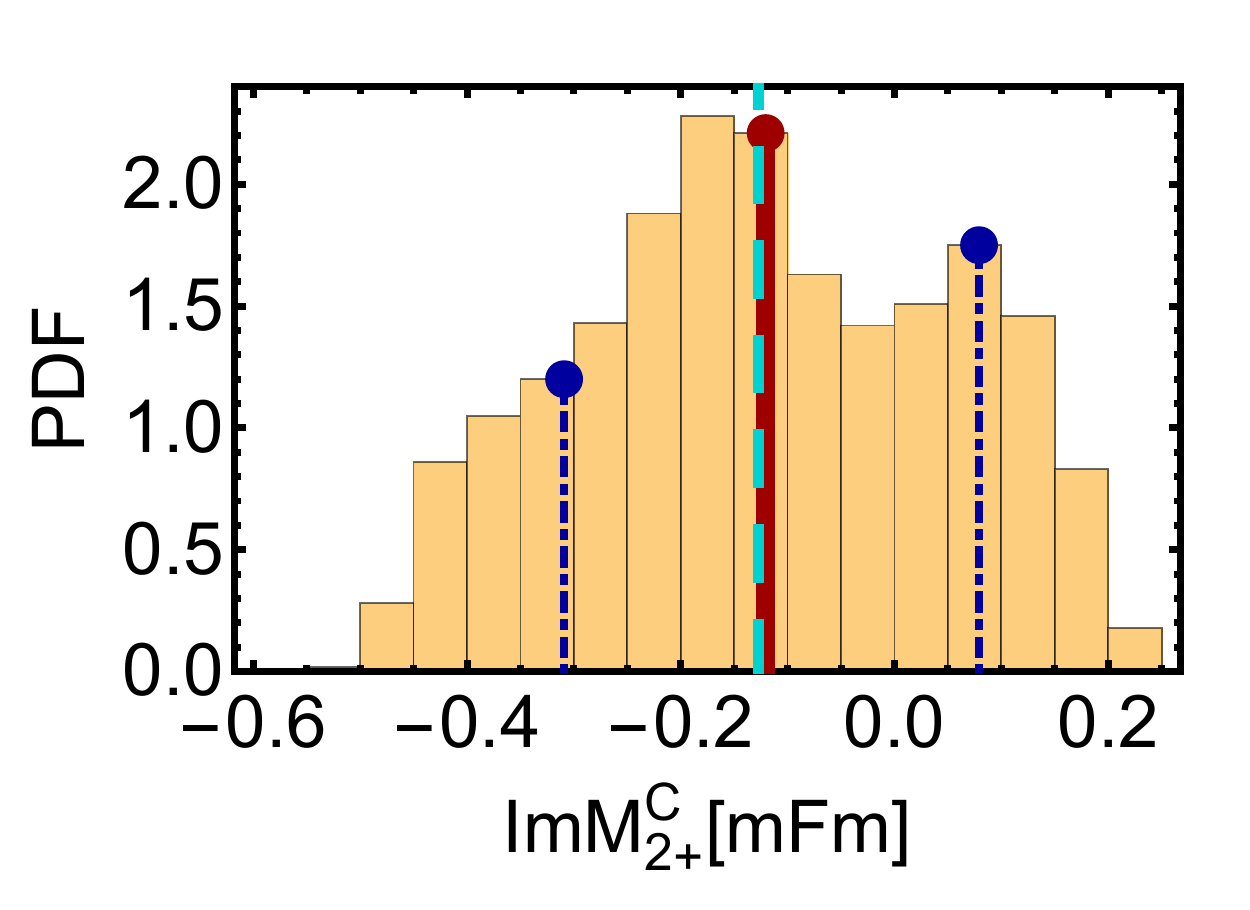}
 \end{overpic}
\begin{overpic}[width=0.325\textwidth]{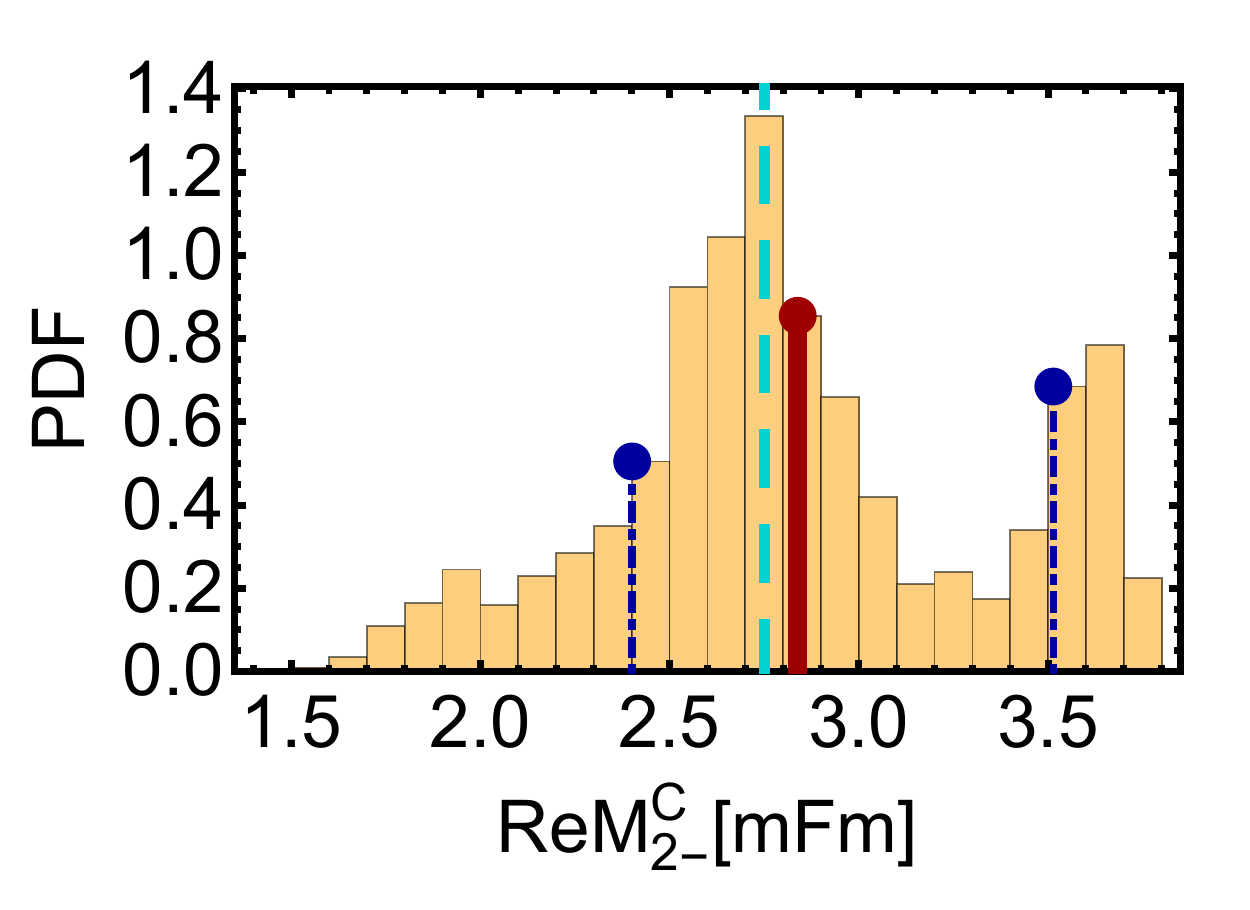}
 \end{overpic}
\begin{overpic}[width=0.325\textwidth]{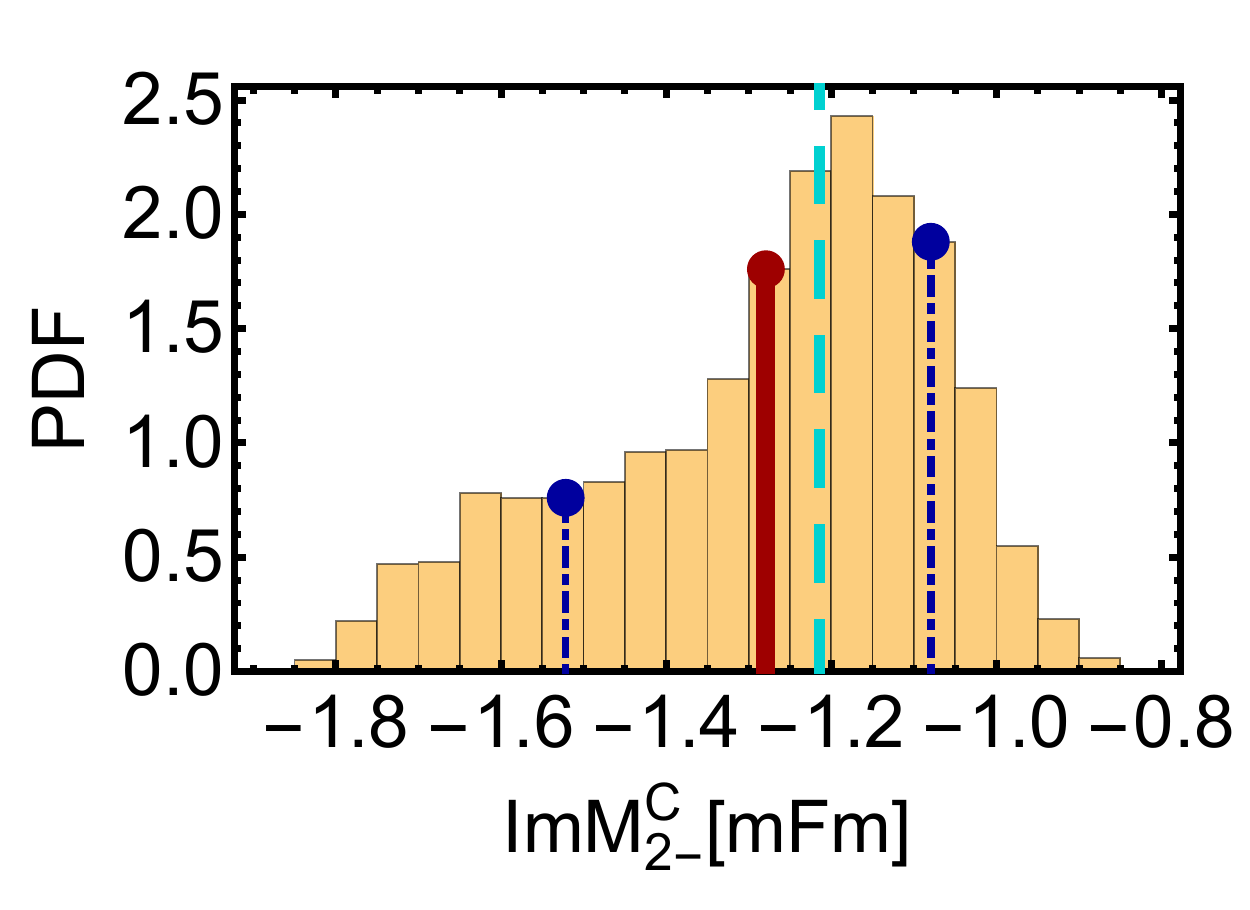}
 \end{overpic}
\caption[Bootstrap-distributions for multipole-fit-parameters in an analysis of photoproduction data on the second resonance region. The third energy-bin, \newline $E_{\gamma }\text{ = 749.94 MeV}$, is printed to aid the discussion in the main text.]{Bootstrap-distributions for the real- and imaginary parts of phase-constrained $S$-, $P$- and $D$-wave multipoles are shown in the same way as in Figure \ref{fig:BootstrapHistos2ndResRegionEnergy4MainText}, but this time for the third energy-bin, $E_{\gamma }\text{ = 749.94 MeV}$. Again, an ensemble of $B=2000$ bootstrap-replicates has been the basis of these results. For more details on the plotting-scheme, see the caption of Figure \ref{fig:BootstrapHistos2ndResRegionEnergy4MainText}. \newline
In most parameter-distributions, a heavy bias is caused by the fact that structures of multiple peaks (modes) are seen. A comparison to Figure \ref{fig:ThirdFitLmax3FWavesFixedMultipolesPlots} shows that here the bootstrap has mapped out distributions merging both the global minimum and the next best local minimum, which have already been found in the Monte Carlo minimum-search applied to the original data (cf. Figures \ref{fig:ThirdFitLmax3FWavesFixedChiSquarePlots} and \ref{fig:ThirdFitLmax3FWavesFixedMultipolesPlots}).}
\label{fig:BootstrapHistos2ndResRegionEnergy3MainText}
\end{figure}

All the statements above apply to each energy-bin, except for the third one. In the latter case, more serious problems are encountered in the bootstrap-analysis. Figure \ref{fig:BootstrapHistos2ndResRegionEnergy3MainText} shows the corresponding distributions, while the numerical data extracted for this energy-bin can be found in appendix \ref{sec:NumericalTPWAFitResults}. Here, large violations of the bias rule-of-thumb occur and the plotted distributions directly tell why: almost every parameter shows here a distribution containing two connected peaks (or modes), both with roughly normal shape. One of both is almost always centered exactly on the global minimum found in the Monte Carlo-fit to the original data. A comparison with Figure \ref{fig:ThirdFitLmax3FWavesFixedMultipolesPlots} tells where the neighboring peaks come from. Their centers always coincide well with the almost degenerate local minimum found in the Monte Carlo minimum-search, using $N_{MC}=16000$, which has been described above (also see the $\chi^{2}$-plots in Figure \ref{fig:ThirdFitLmax3FWavesFixedChiSquarePlots}, where the minimum can be found). Therefore, the bootstrap has mapped out this ambiguity and illustrated the fact that right now, the data are not able to resolve both possible solutions, at least within their current statistical accuracy.

\clearpage

\begin{figure}[h]
\begin{overpic}[width=0.325\textwidth]{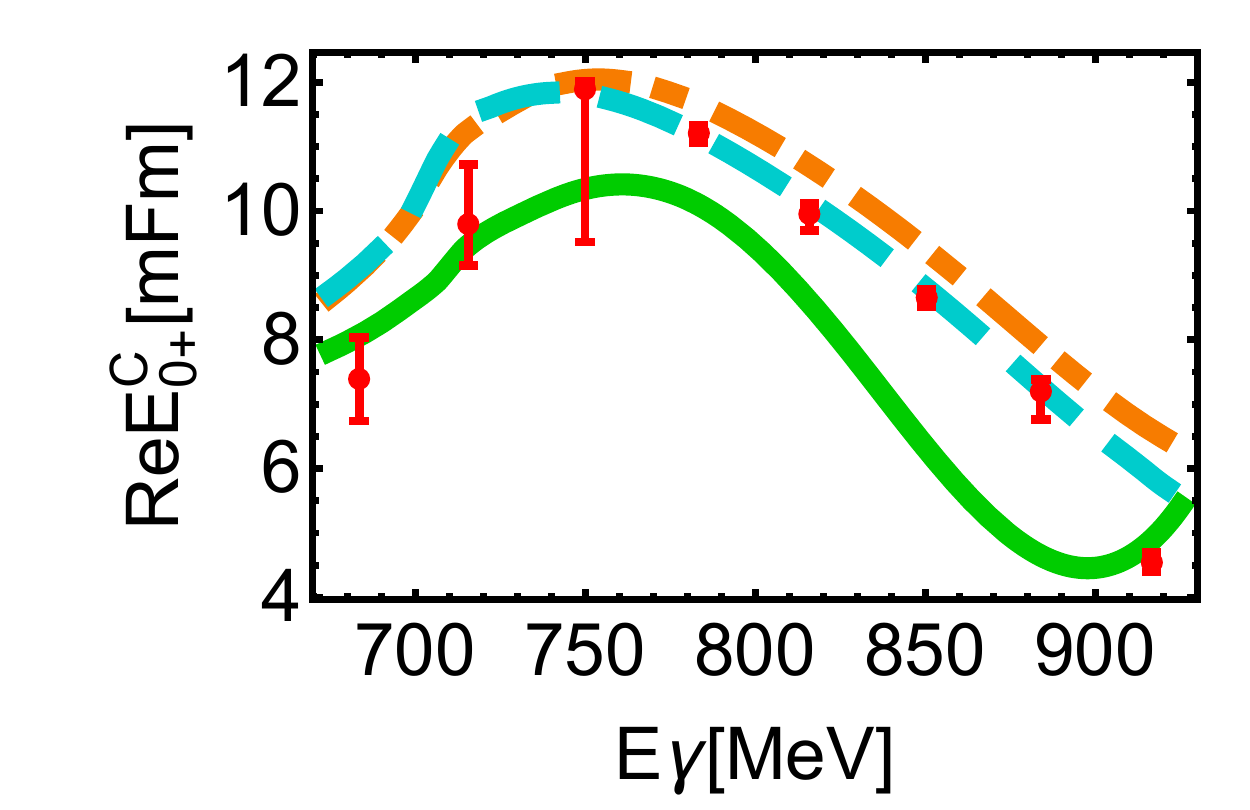}
 \end{overpic}
\begin{overpic}[width=0.325\textwidth]{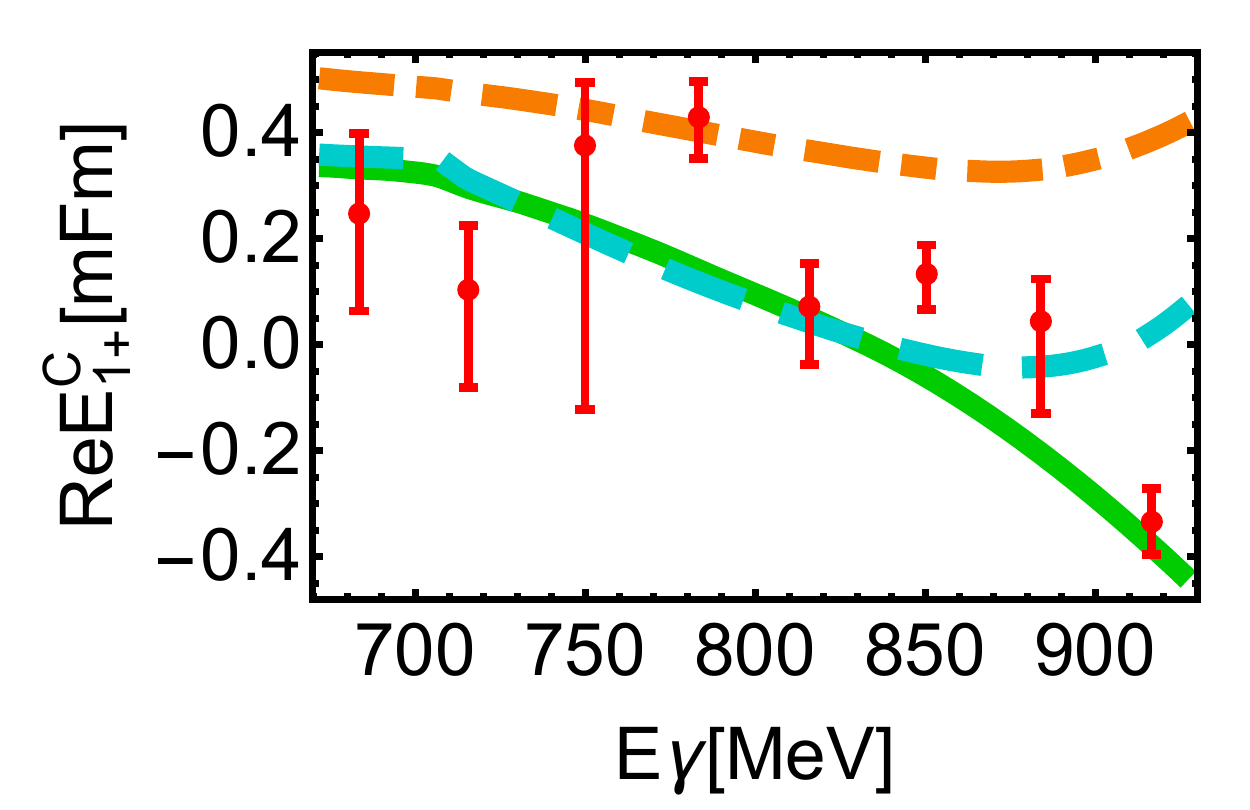}
 \end{overpic}
\begin{overpic}[width=0.325\textwidth]{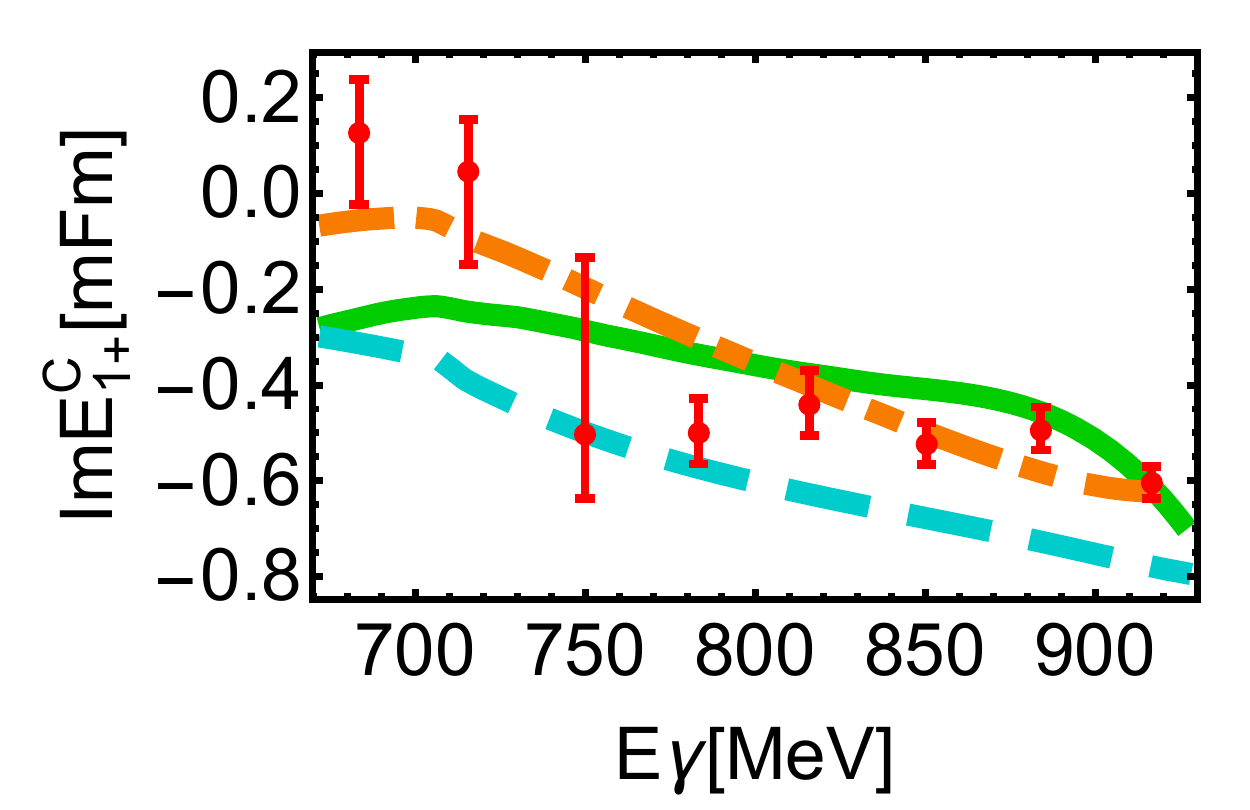}
 \end{overpic} \\
\begin{overpic}[width=0.325\textwidth]{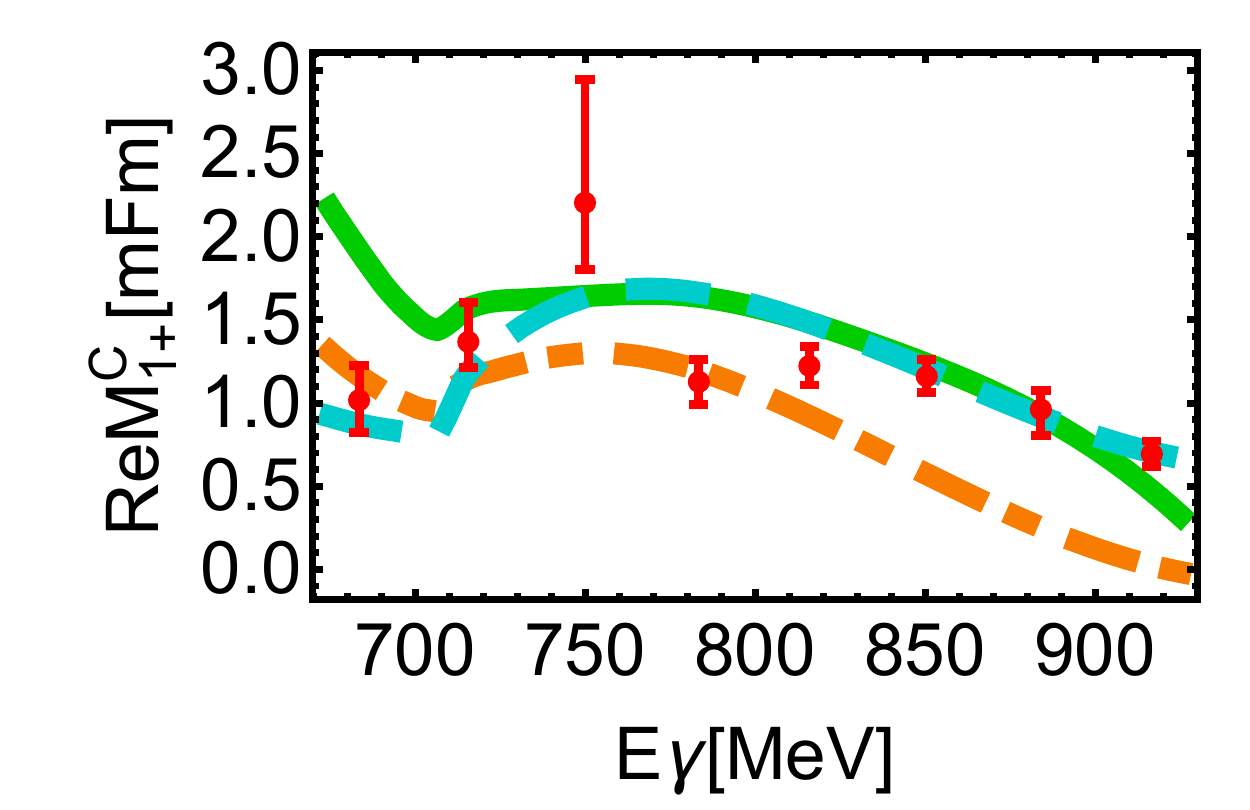}
 \end{overpic}
\begin{overpic}[width=0.325\textwidth]{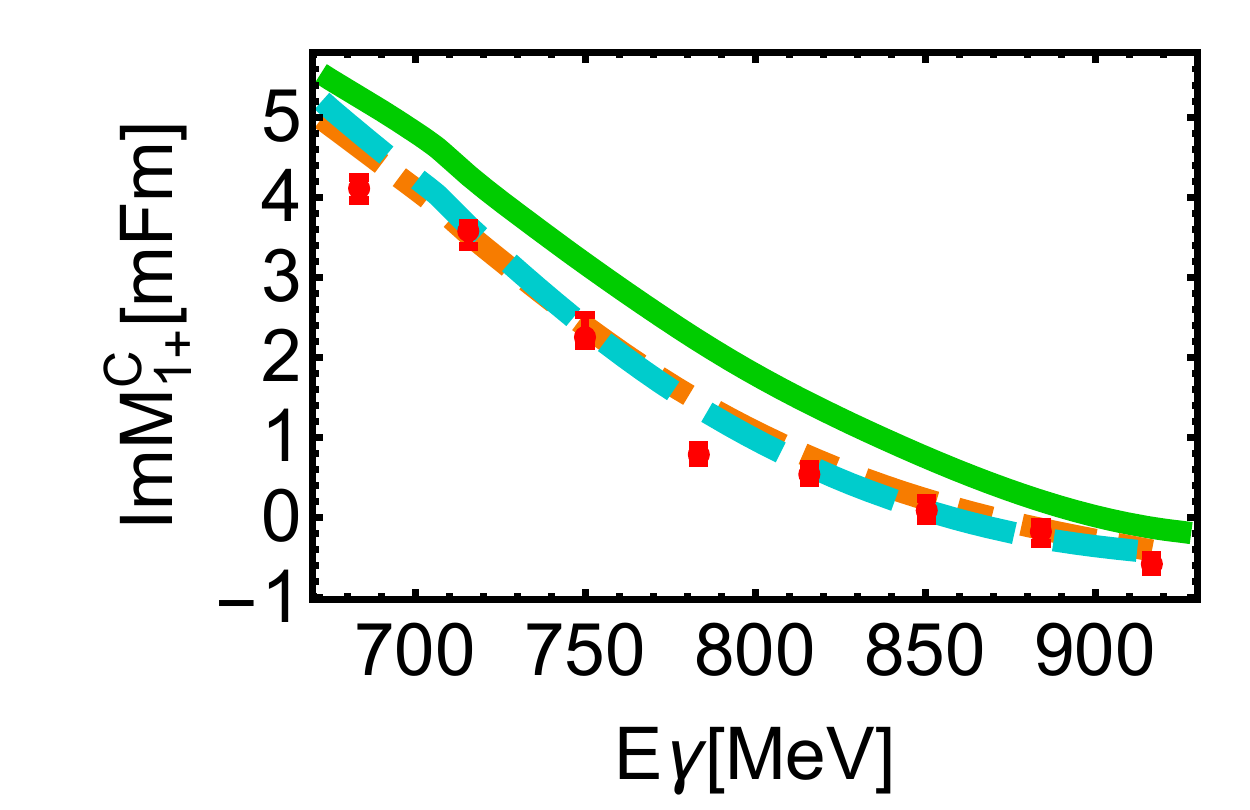}
 \end{overpic}
\begin{overpic}[width=0.325\textwidth]{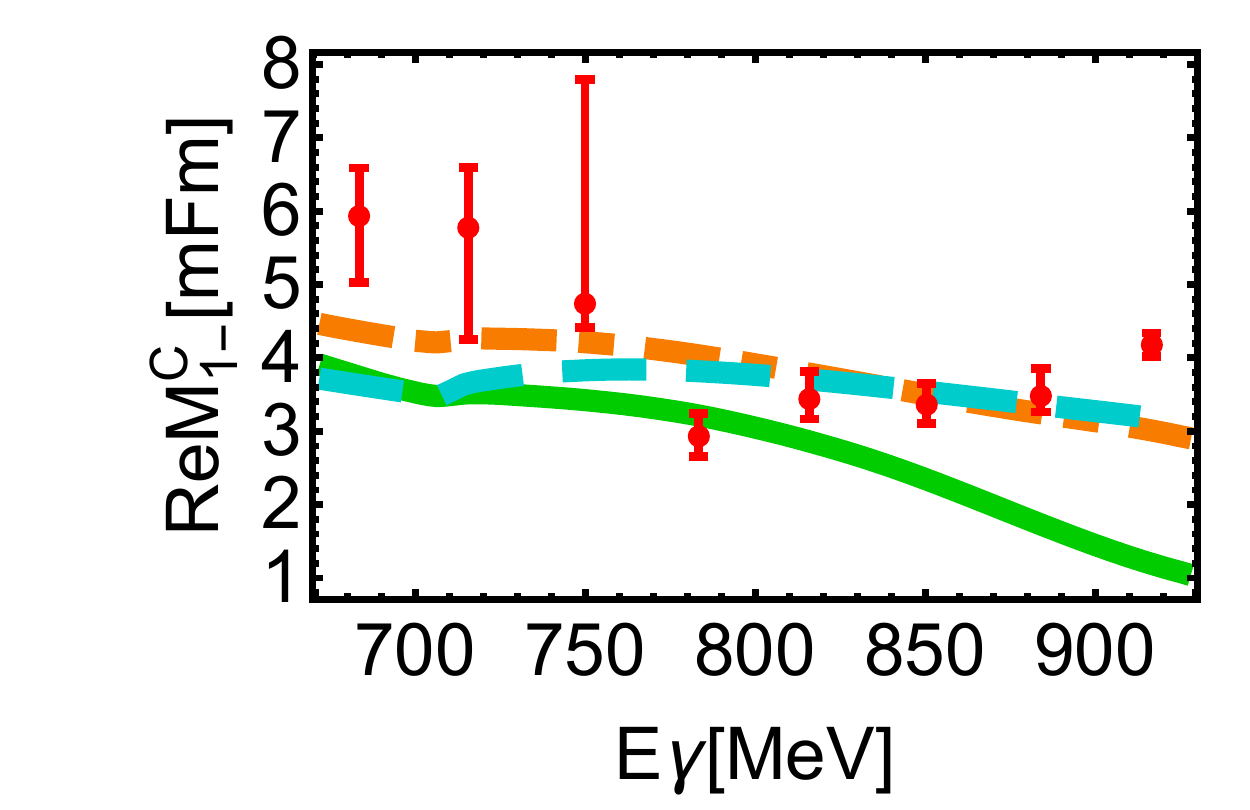}
 \end{overpic} \\
\begin{overpic}[width=0.325\textwidth]{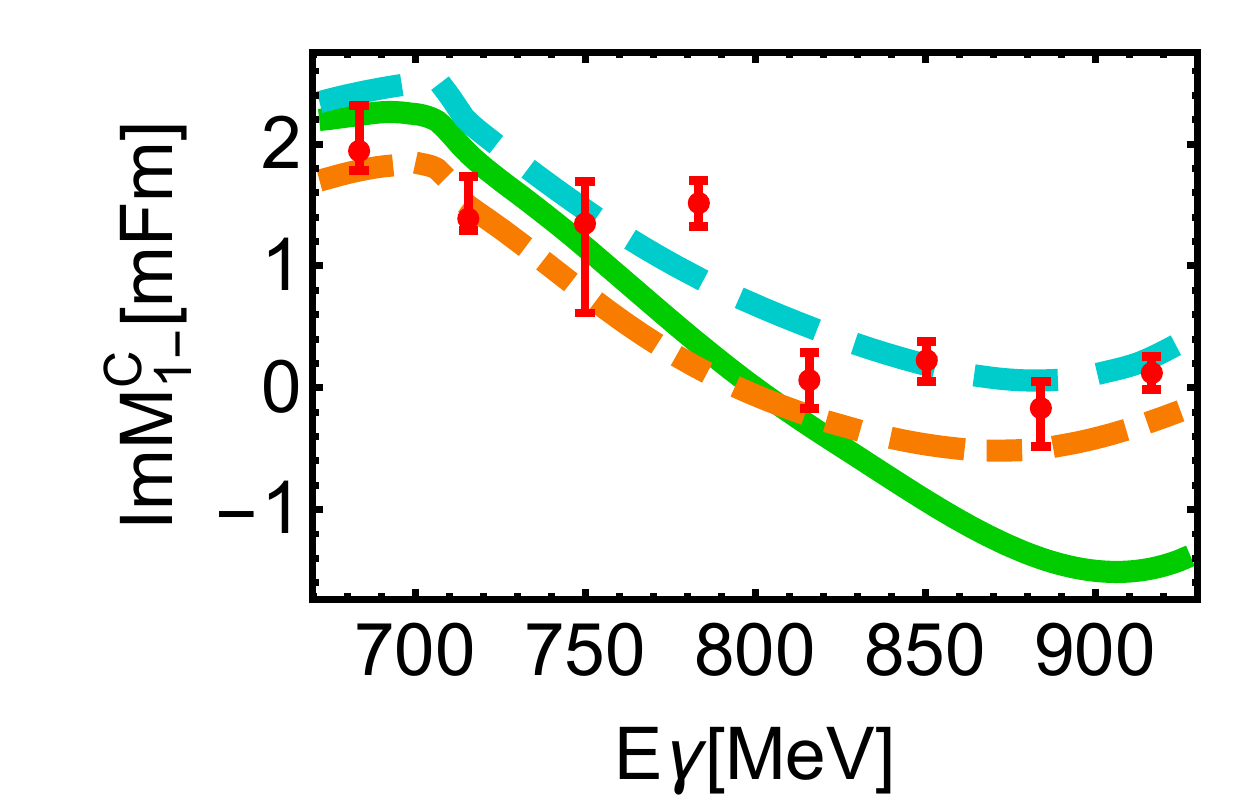}
 \end{overpic}
\begin{overpic}[width=0.325\textwidth]{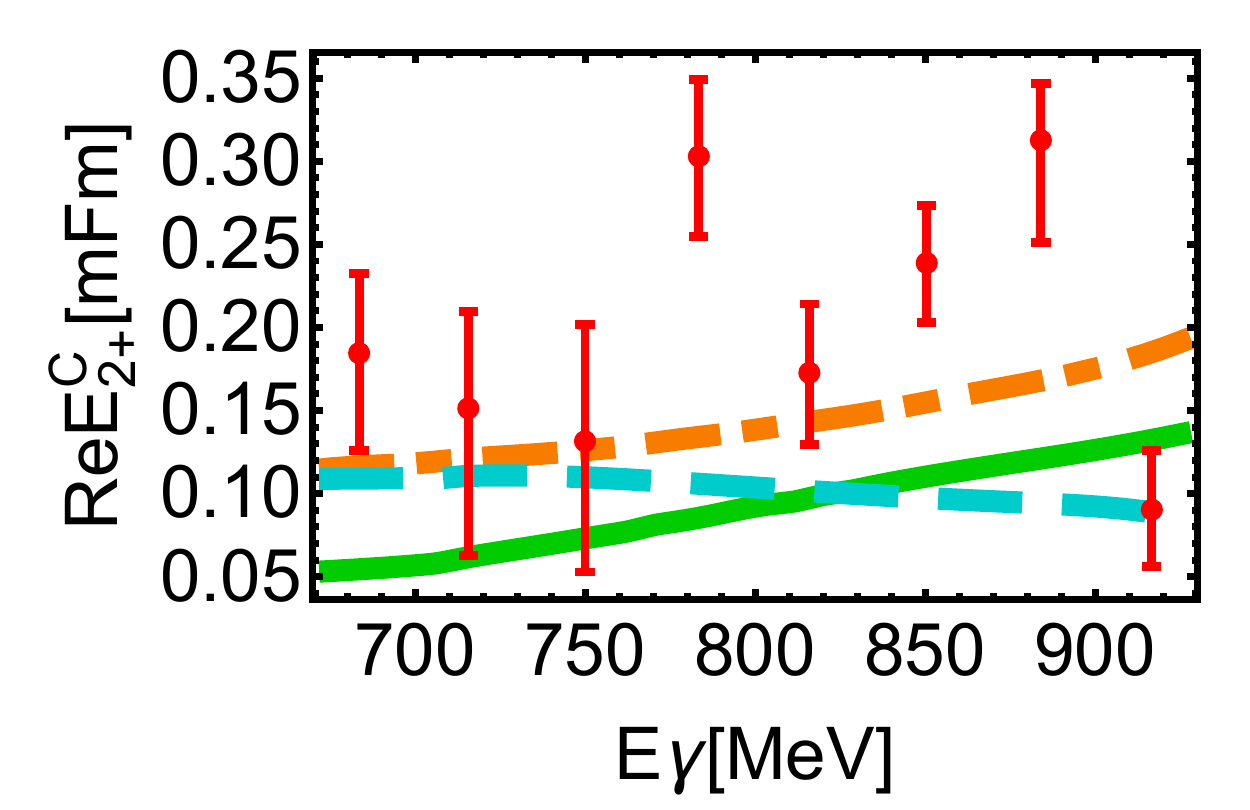}
 \end{overpic}
\begin{overpic}[width=0.325\textwidth]{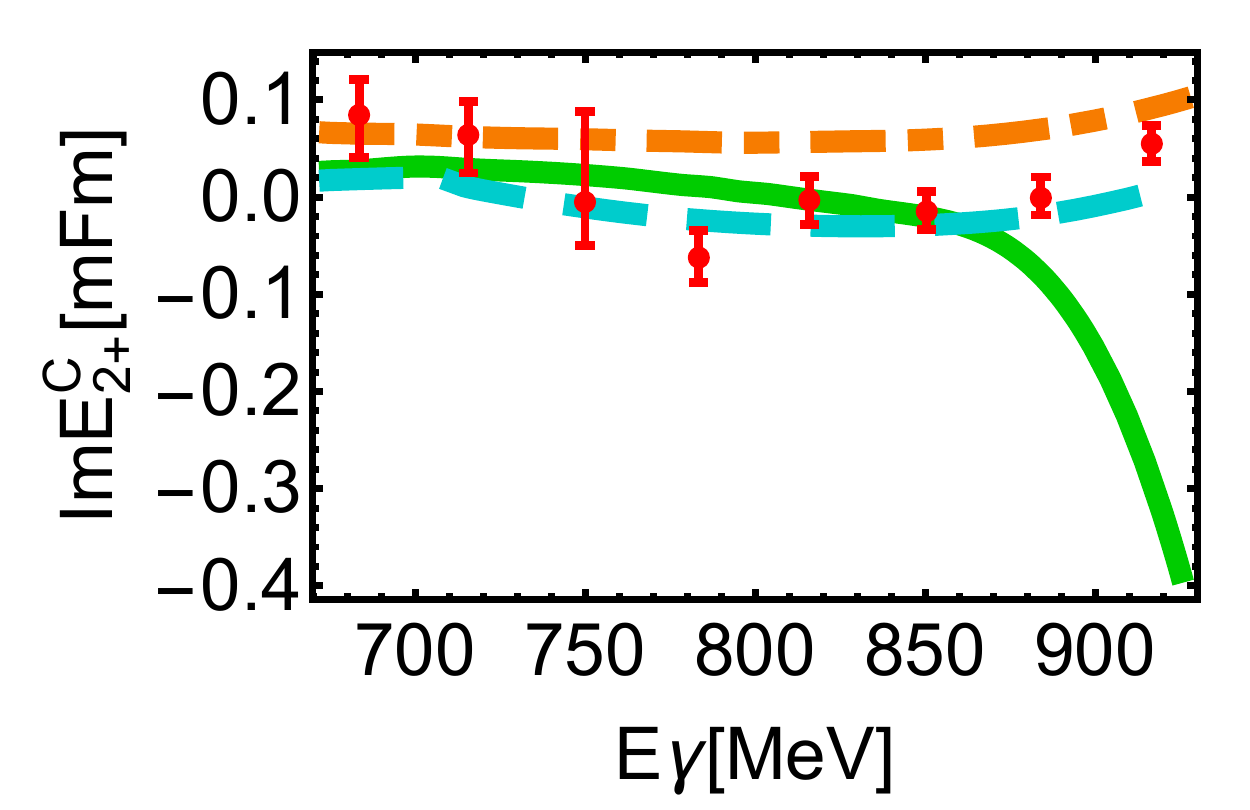}
 \end{overpic} \\
\begin{overpic}[width=0.325\textwidth]{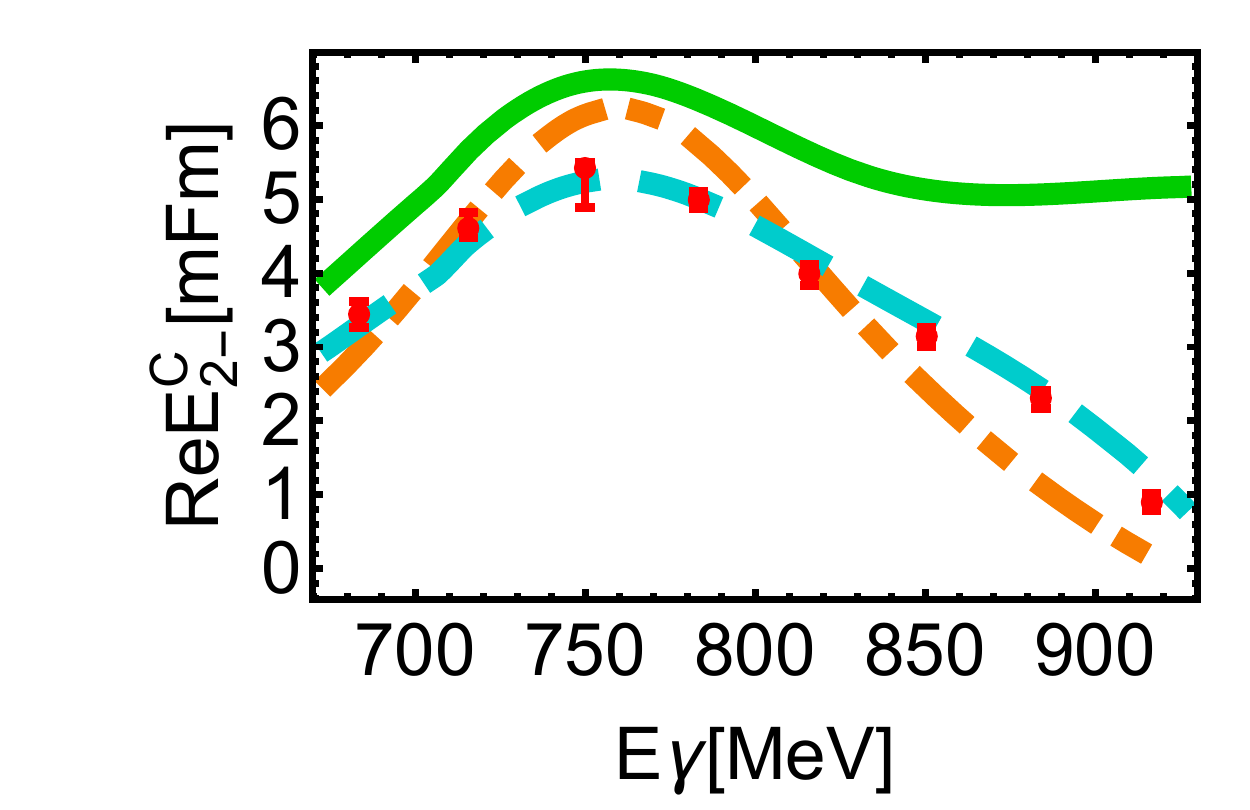}
 \end{overpic}
\begin{overpic}[width=0.325\textwidth]{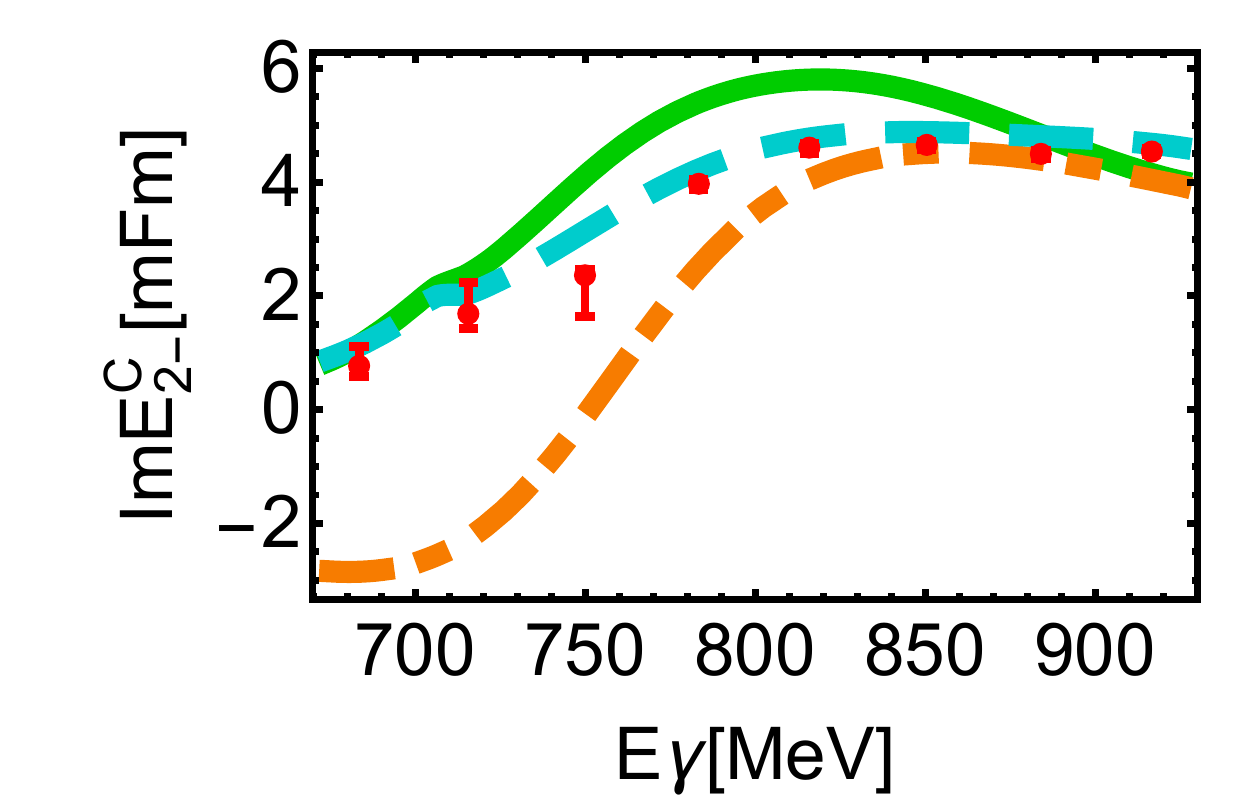}
 \end{overpic}
\begin{overpic}[width=0.325\textwidth]{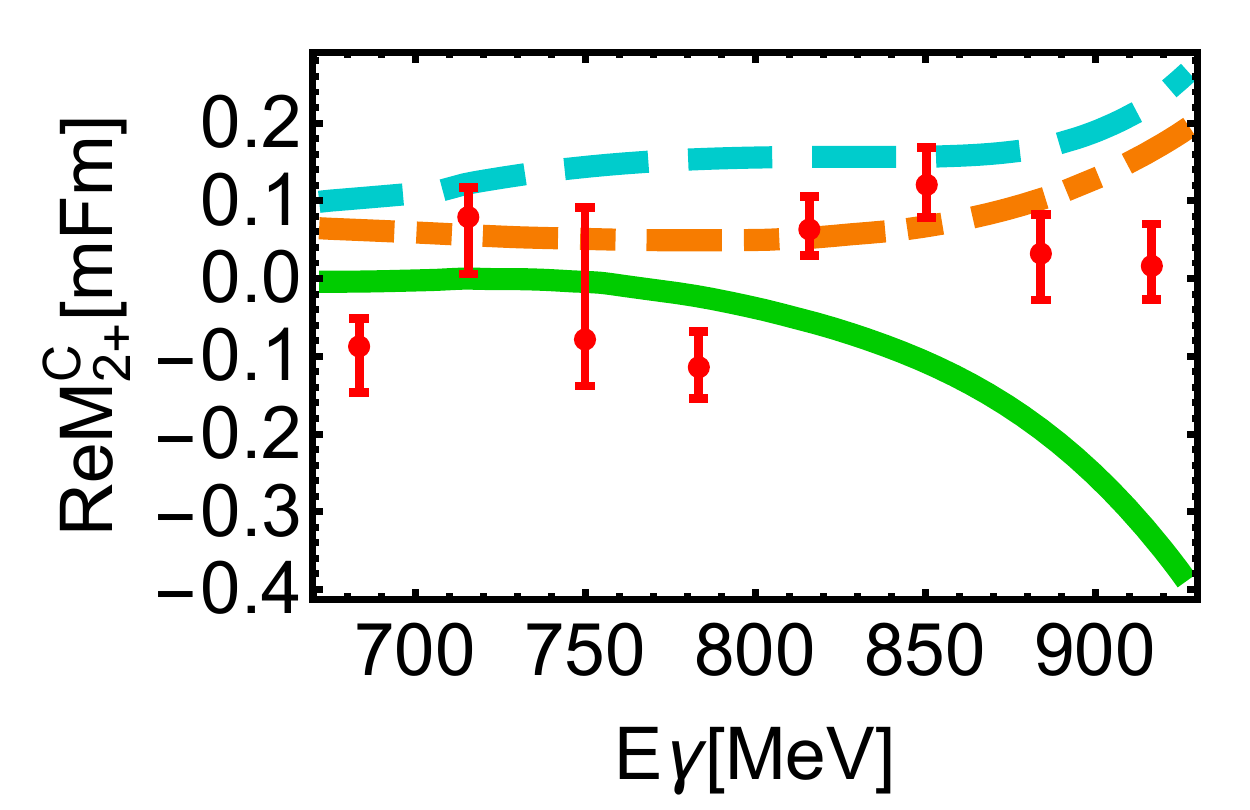}
 \end{overpic} \\
\begin{overpic}[width=0.325\textwidth]{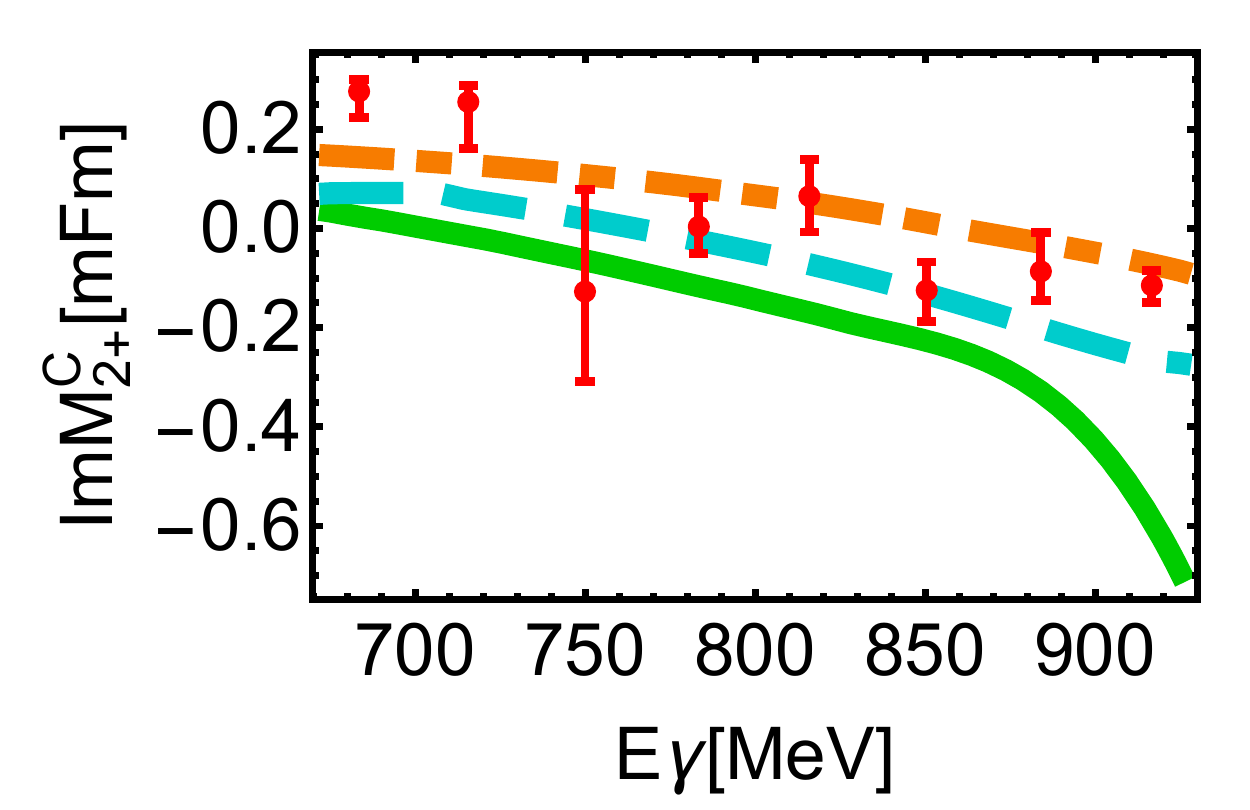}
 \end{overpic}
\begin{overpic}[width=0.325\textwidth]{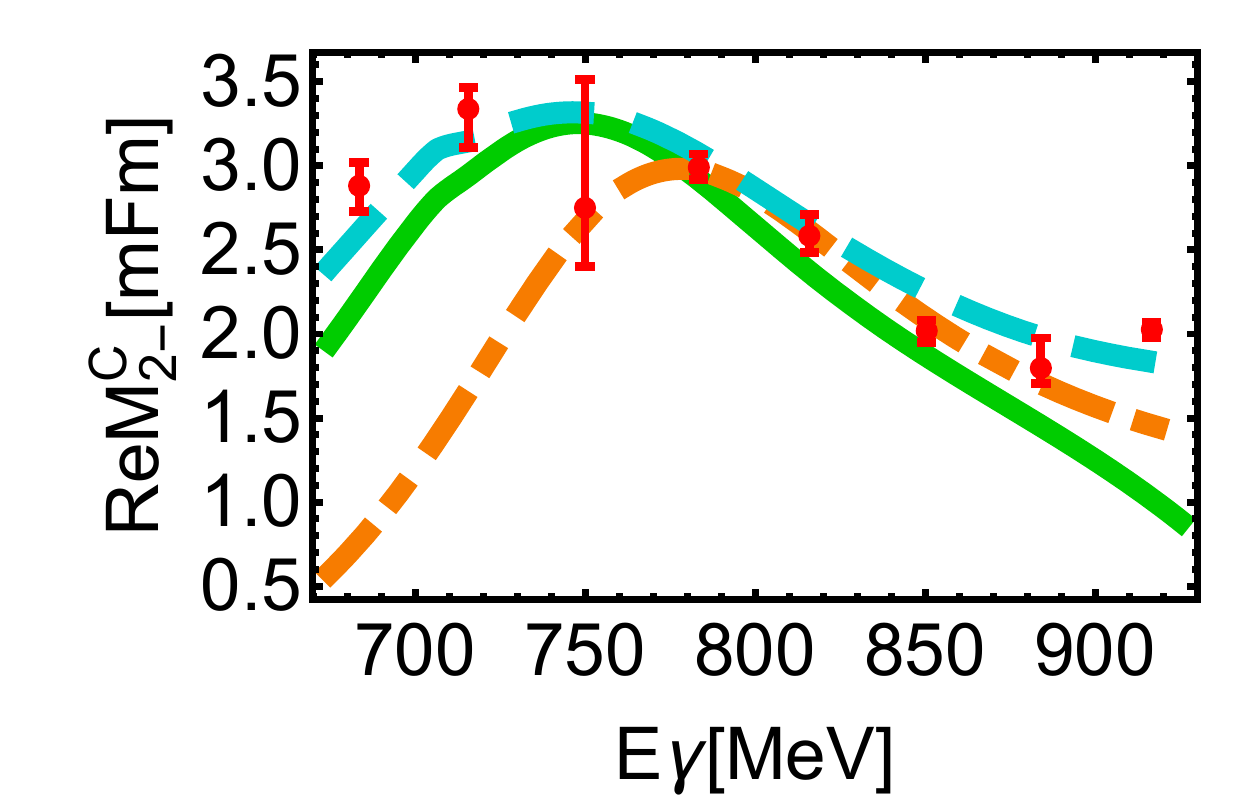}
 \end{overpic}
\begin{overpic}[width=0.325\textwidth]{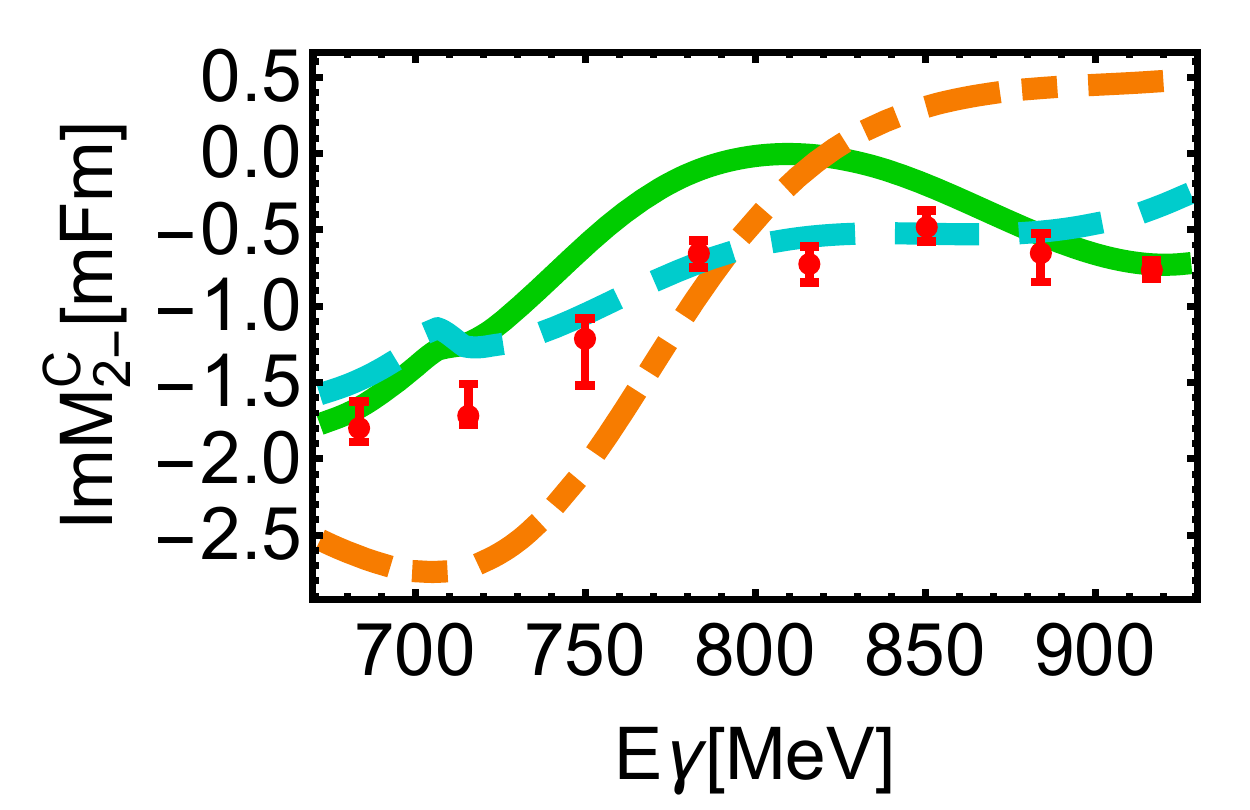}
 \end{overpic}
\caption[A comparison of the results of the bootstrap-analysis performed in the $2^{\mathrm{nd}}$ resonance region, to energy-dependent PWA-models.]{The plots show a comparison of the results of the bootstrap-analysis to energy-dependent PWA-models. Red dots and error-bars represent the global minimum of the fit to the original data, as well as the asymmetric bootstrap-errors $\Delta_{\pm}$. The model-solutions included for comparison are MAID2007 \cite{MAID2007, MAID} (green solid line), BnGa 2014\_02 \cite{BoGa} (cyan dashed line) and SAID CM12 \cite{WorkmanEtAl2012ChewMPhotoprod, SAID} (orange dash-dotted line). \newline
The (sometimes) strongly asymmetric error-bars for the results in the third energy-bin, $E_{\gamma }\text{ = 749.94 MeV}$, result from the fact that here, the bootstrap-distributions encompass two merged ambiguous solutions (see Figure \ref{fig:BootstrapHistos2ndResRegionEnergy3MainText} and the discussion in the main text). Therefore, the error-bars in this particular bin should be considered with extreme care!}
\label{fig:Bootstrap2ndResRegionResultsComparedToPWA}
\end{figure}

\clearpage

Unfortunately, there does not seem to be a unique and simple method to resolve these issues present in the third energy-bin. Nonetheless, we proceed by extracting bootstrap-results for all energies in the same way, most importantly the confidence-interval. In the ensuing discussion, the third bin should always be seen under the huge caveat that the confidence interval extracted from it and correspondingly also the asymmetric bootstrap-errors, encompass a distribution of two merged solutions. \newline
The bootstrap-results for all fitted multipoles and all energies are compared to solutions of energy-dependent PWA-models in Figure \ref{fig:Bootstrap2ndResRegionResultsComparedToPWA}. First of all, and maybe not so surprisingly, the global minimum stays well within the parameter-regions encompassed by the energy-dependent models. The relative size of the bootstrap-errors tends to become smaller for the higher energies. The relative errors for the $S$-wave $E_{0+}^{C}$ are actually quite large for the lower energies and their size decreases drastically only when going to higher energies. Other multipoles with small relative errors are $M_{1+}^{C}$, $E_{2-}^{C}$ and $M_{2-}^{C}$. All these multipoles have in common, that in the considered energy-region, the well-established resonances $N(1520) \frac{3}{2}^{-}$ and $\Delta(1600) \frac{3}{2}^{+}$ can couple to them (see the Table at the end of section \ref{sec:LFitsPaper}). The remaining multipoles, for instance $E_{2+}^{C}$, tend to have larger relative errors. \newline
When comparing to the energy-dependent models, a slight preference of Bonn-Gatchina is found, rooting from the fact that the fit itself already carries dependence on this PWA. However, for some fit-parameters and energy-regions, this can also be different. For instance, for the parameter $\mathrm{Re} \left[ E_{0+}^{C} \right]$, MAID is favored for the lowest energies. As another example, the results for $\mathrm{Re} \left[ M_{2+}^{C} \right]$ seem to favor the SAID-model over Bonn-Gatchina for almost all energies. With these considerations, we conclude the discussion of the bootstrap-analysis. \newline

\vfill

\begin{figure}[hb]
 \centering
\begin{overpic}[width=0.495\textwidth]{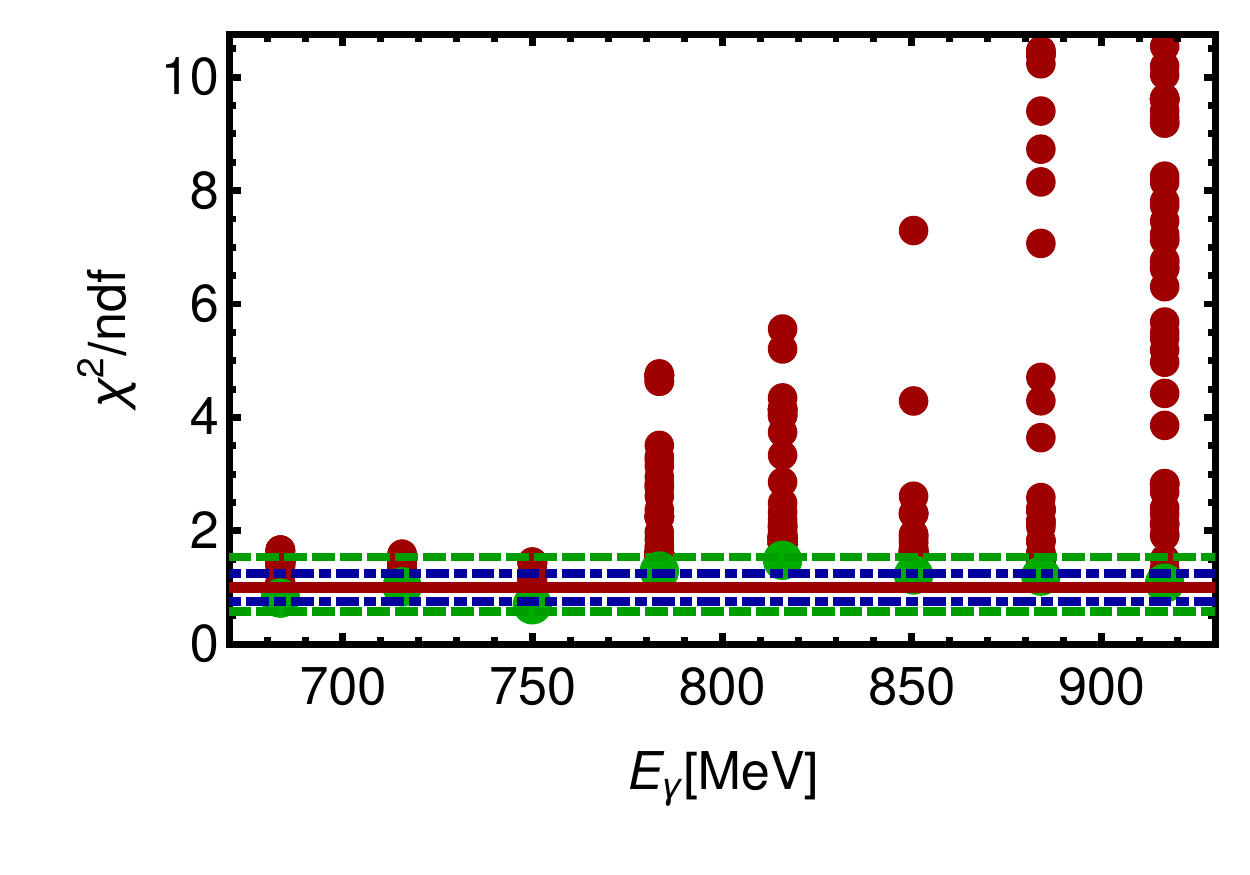}
 \end{overpic}
\begin{overpic}[width=0.495\textwidth]{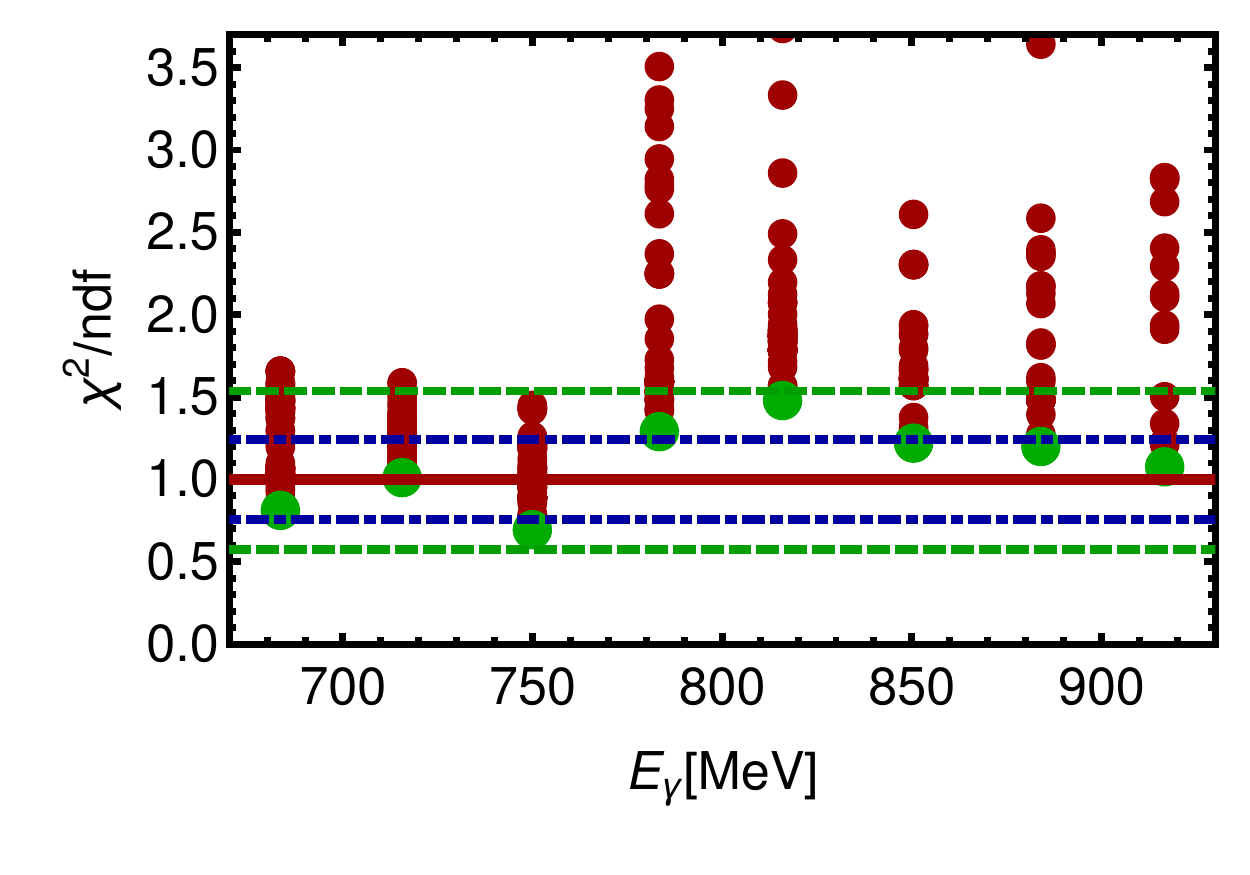}
 \end{overpic}
\vspace*{-15pt}
\caption[The $\chi^{2}/\mathrm{ndf}$ for the best results of the full Monte Carlo minimum-search applied in a truncation at $\ell_{\mathrm{max}} = 4$, with $G$-waves fixed to the Bonn-Gatchina model and $S$-, $P$-, $D$- and $F$-waves varied, within the $2^{\mathrm{nd}}$ resonance region.]{Both plots show the best results for the full Monte Carlo minimum-search applied in the truncation $\ell_{\mathrm{max}} = 4$, with $G$-waves fixed to BnGa 2014\_02 \cite{BoGa} and all remaining mutlipoles varied, once for a relatively wide plot-range (left) and in a more detailed one (right). The results stem from a pool of $N_{MC} = 50000$ initial parameter-configurations. The global minimum is indicated by the big green dots, other local minima are plotted as smaller red-colored dots. \newline
In addition, some information on the theoretical chisquare distribution for the estimate $\mathrm{ndf} = r = 27$ is included via the horizontal lines. The mean is drawn as a red solid line, while the pair of $0.025$- and $0.975$-quantiles is indicated by green dashed lines and that made of the $0.16$- and $0.84$-quantiles by a blue dash-dotted line.}
\label{fig:FourthFitLmax4GWavesBnGaChiSquarePlots}
\end{figure}

\clearpage

\begin{figure}[h]
 \centering
\begin{overpic}[width=0.325\textwidth]{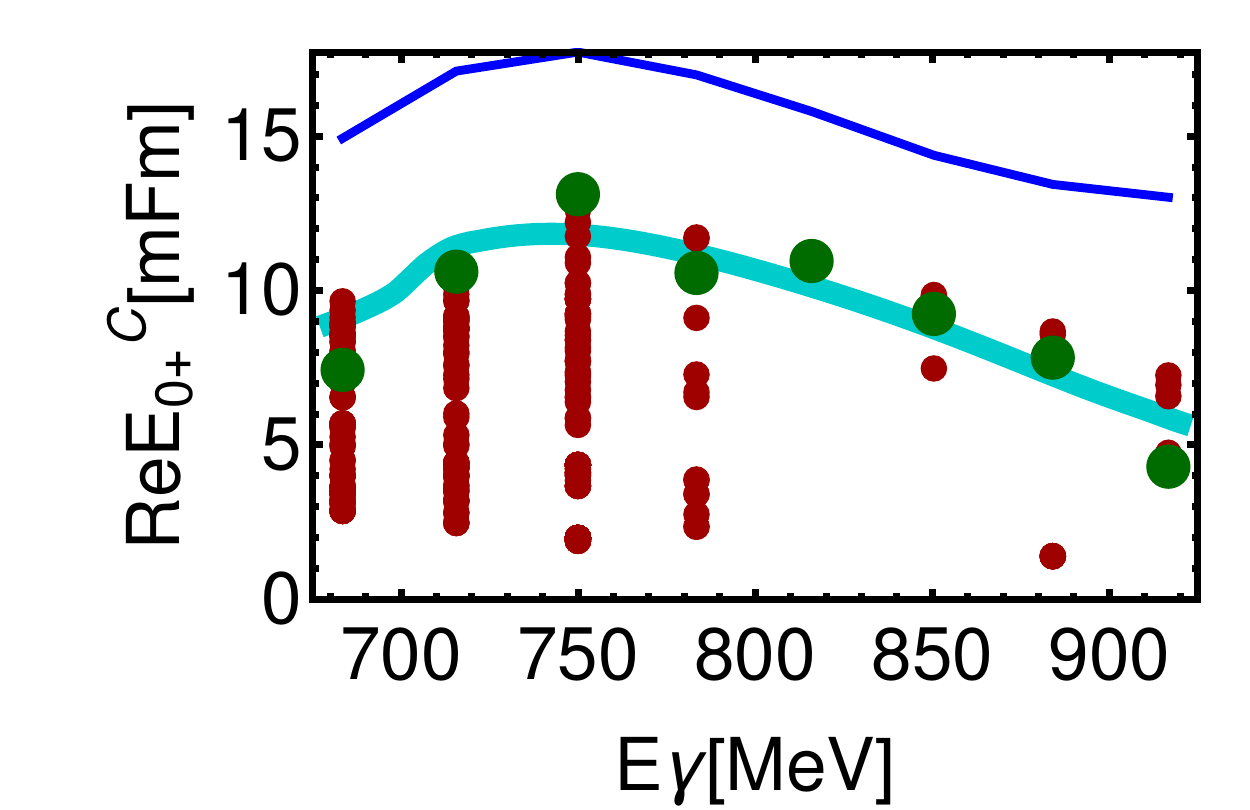}
 \end{overpic}
\begin{overpic}[width=0.325\textwidth]{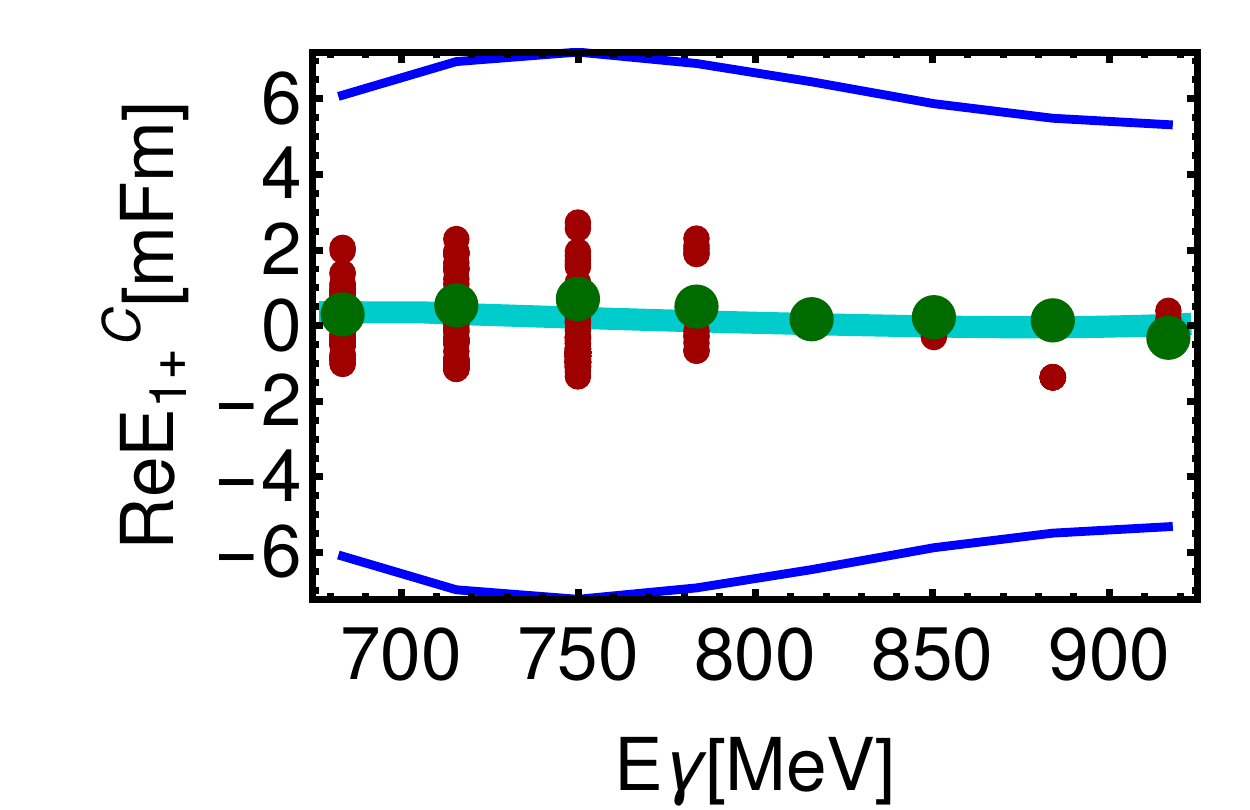}
 \end{overpic}
\begin{overpic}[width=0.325\textwidth]{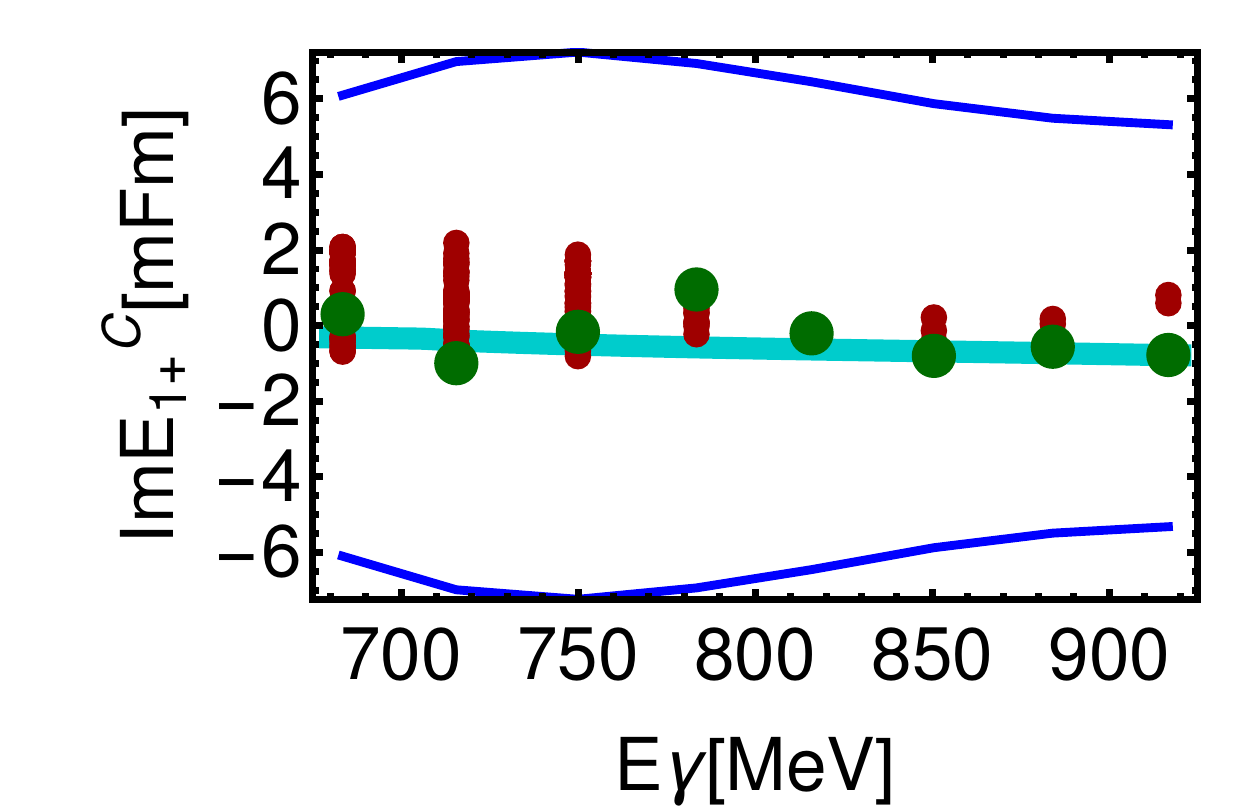}
 \end{overpic} \\
\begin{overpic}[width=0.325\textwidth]{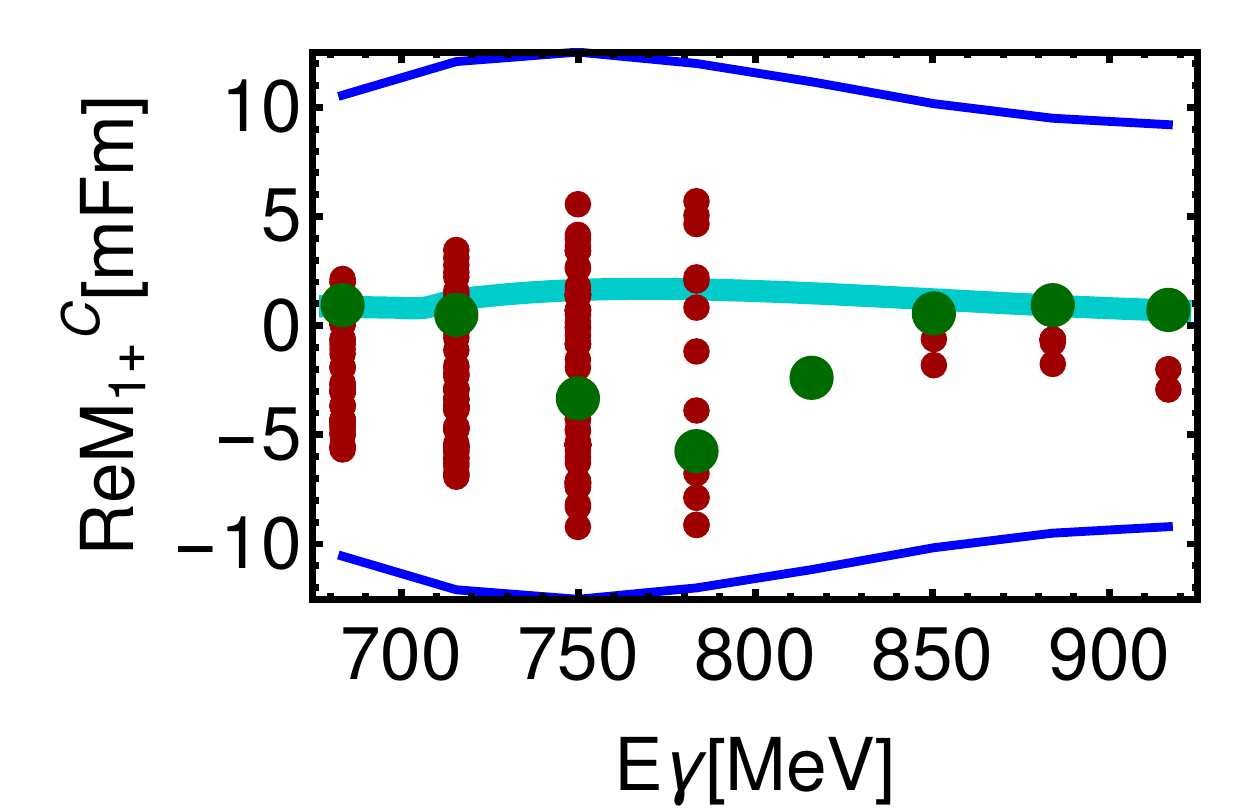}
 \end{overpic}
\begin{overpic}[width=0.325\textwidth]{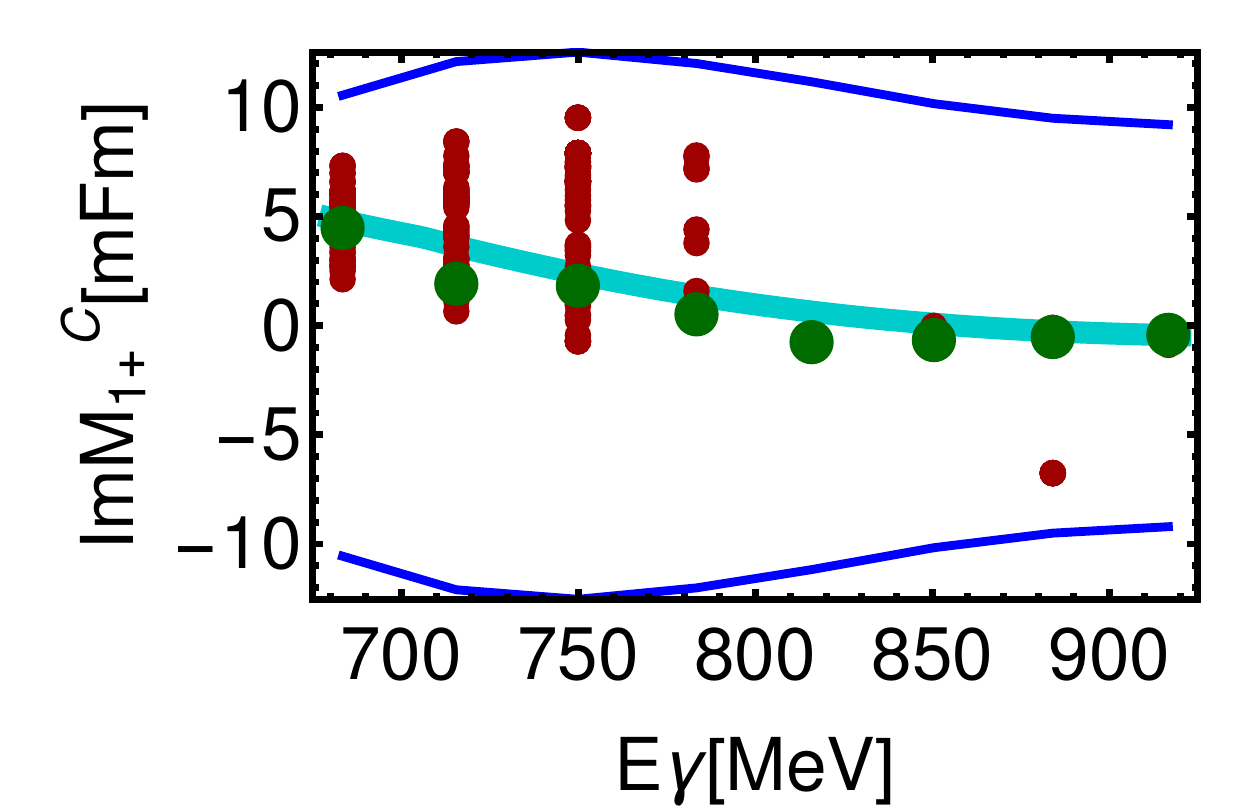}
 \end{overpic}
\begin{overpic}[width=0.325\textwidth]{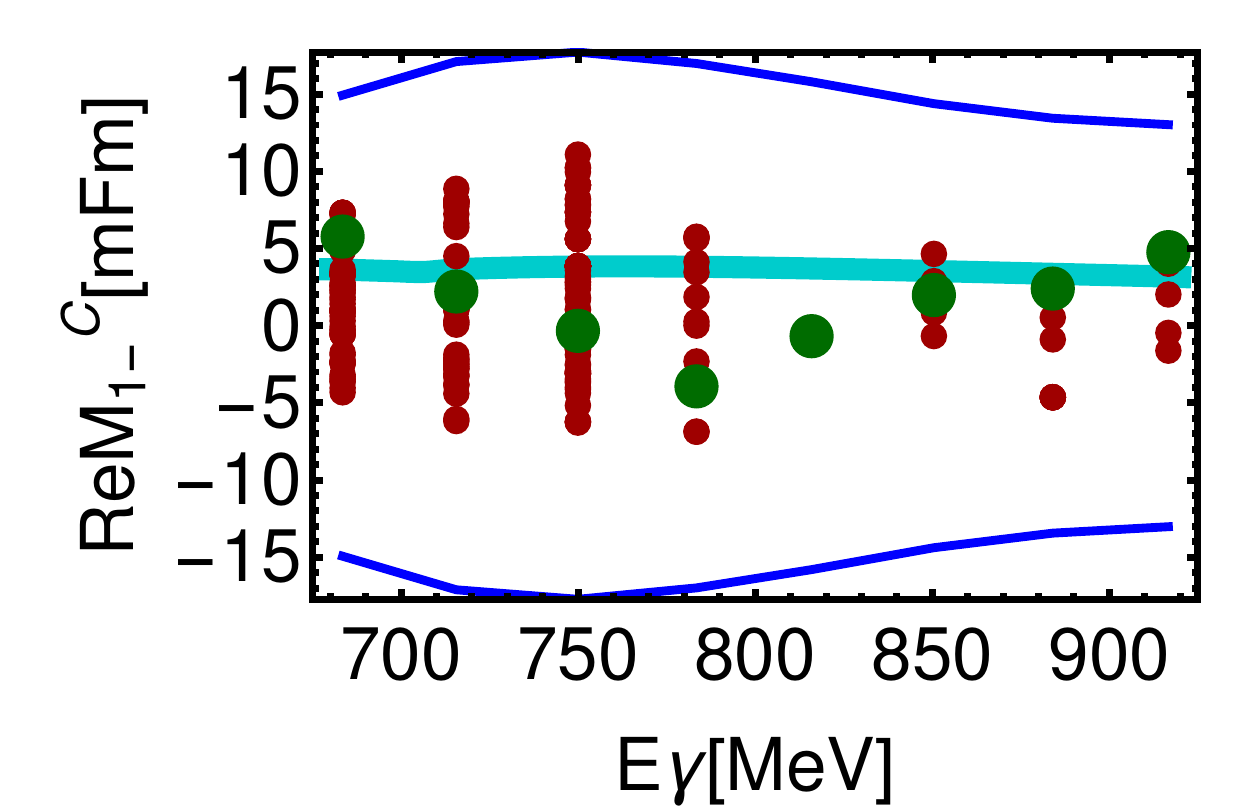}
 \end{overpic} \\
\begin{overpic}[width=0.325\textwidth]{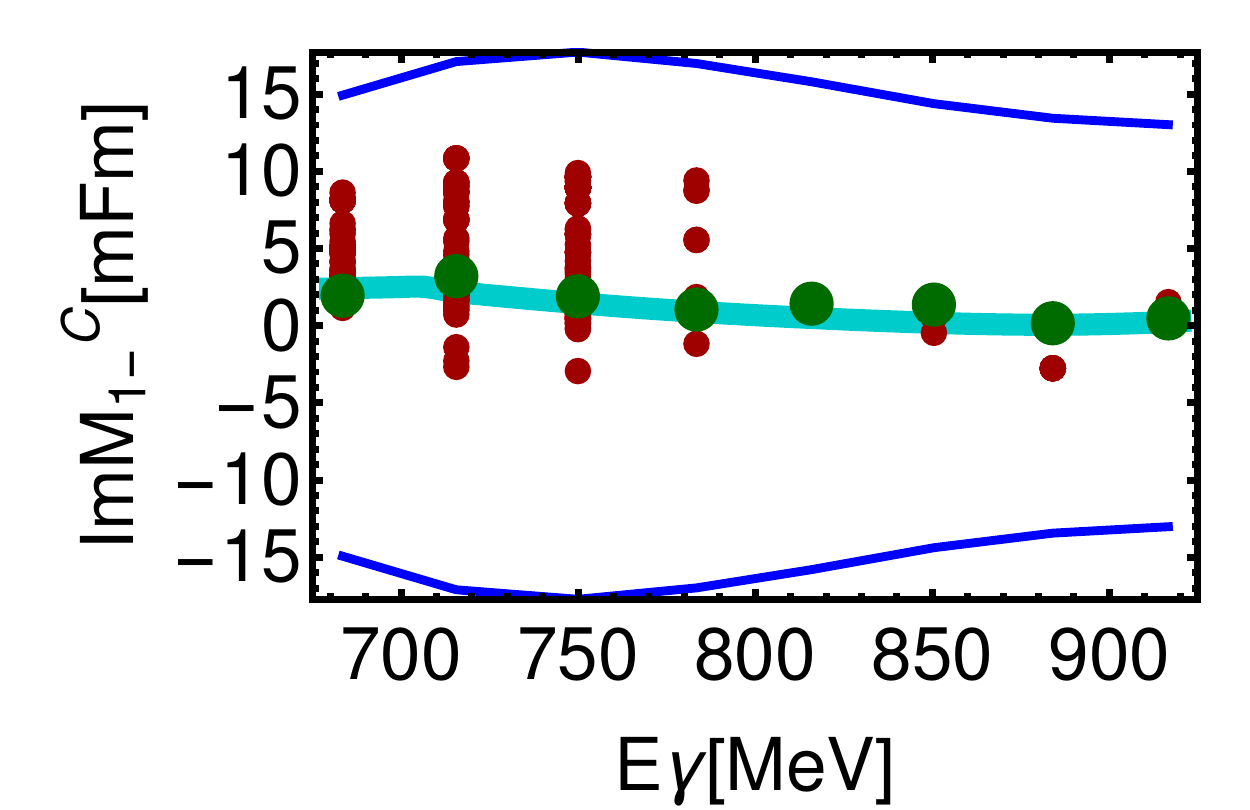}
 \end{overpic}
\begin{overpic}[width=0.325\textwidth]{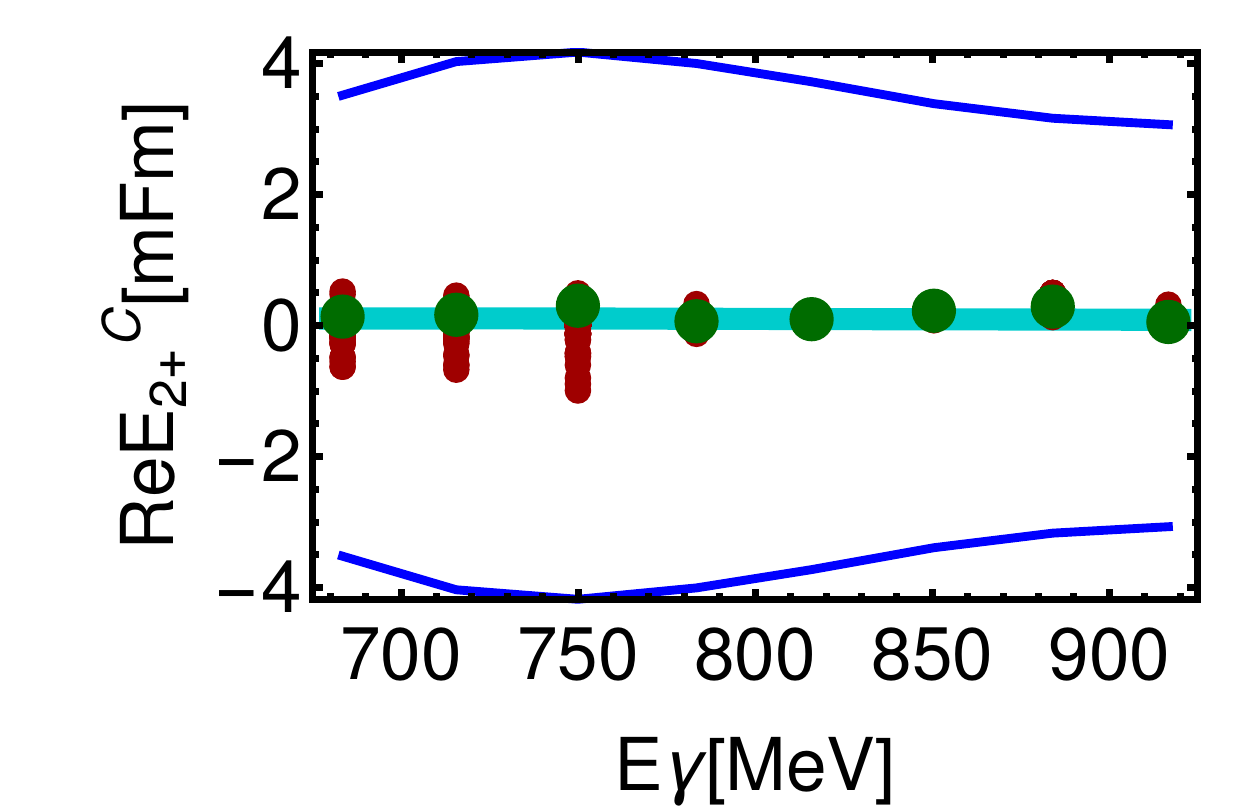}
 \end{overpic}
\begin{overpic}[width=0.325\textwidth]{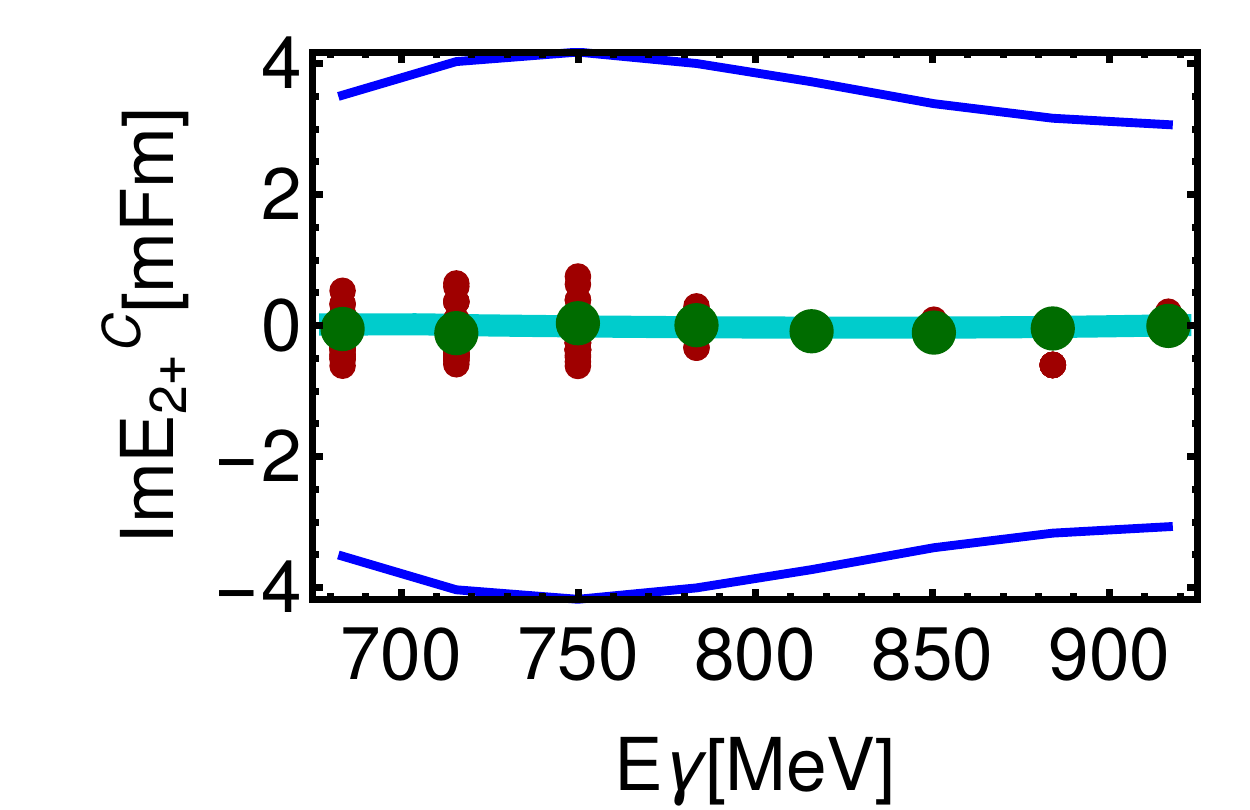}
 \end{overpic} \\
\begin{overpic}[width=0.325\textwidth]{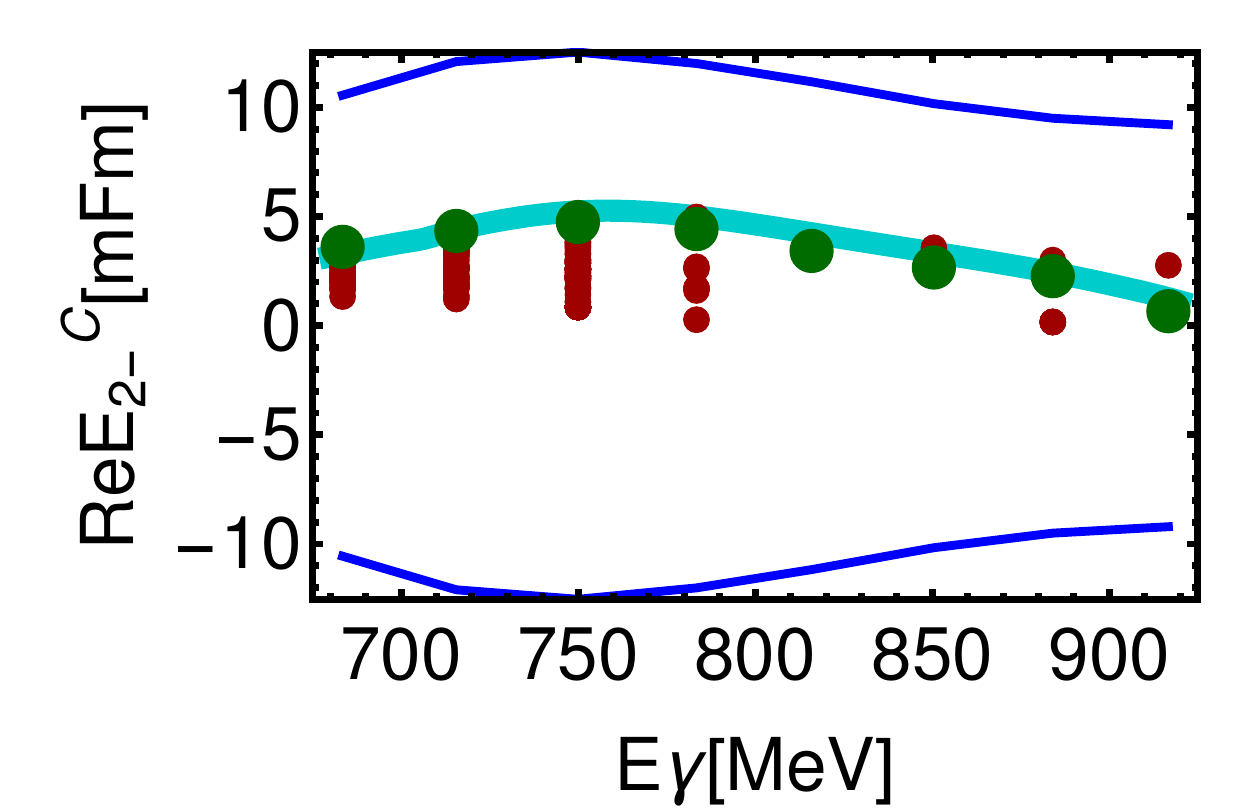}
 \end{overpic}
\begin{overpic}[width=0.325\textwidth]{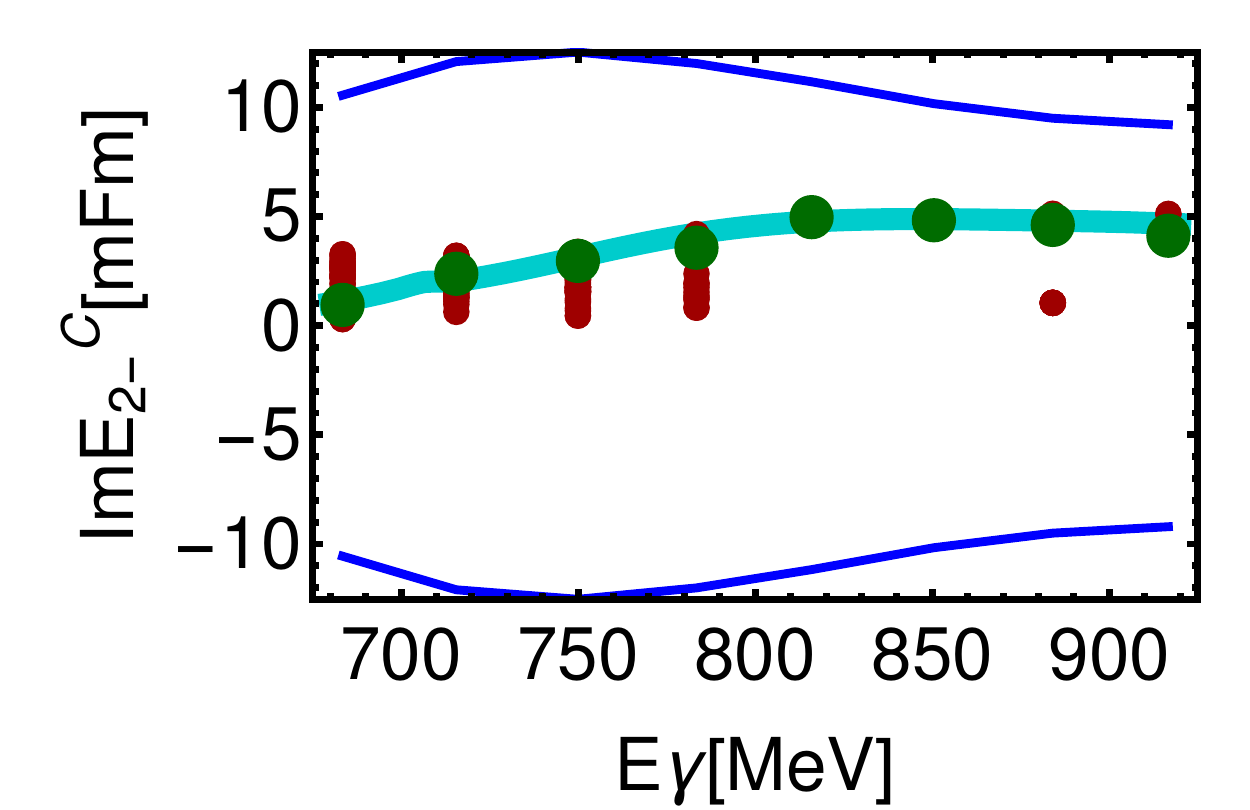}
 \end{overpic}
\begin{overpic}[width=0.325\textwidth]{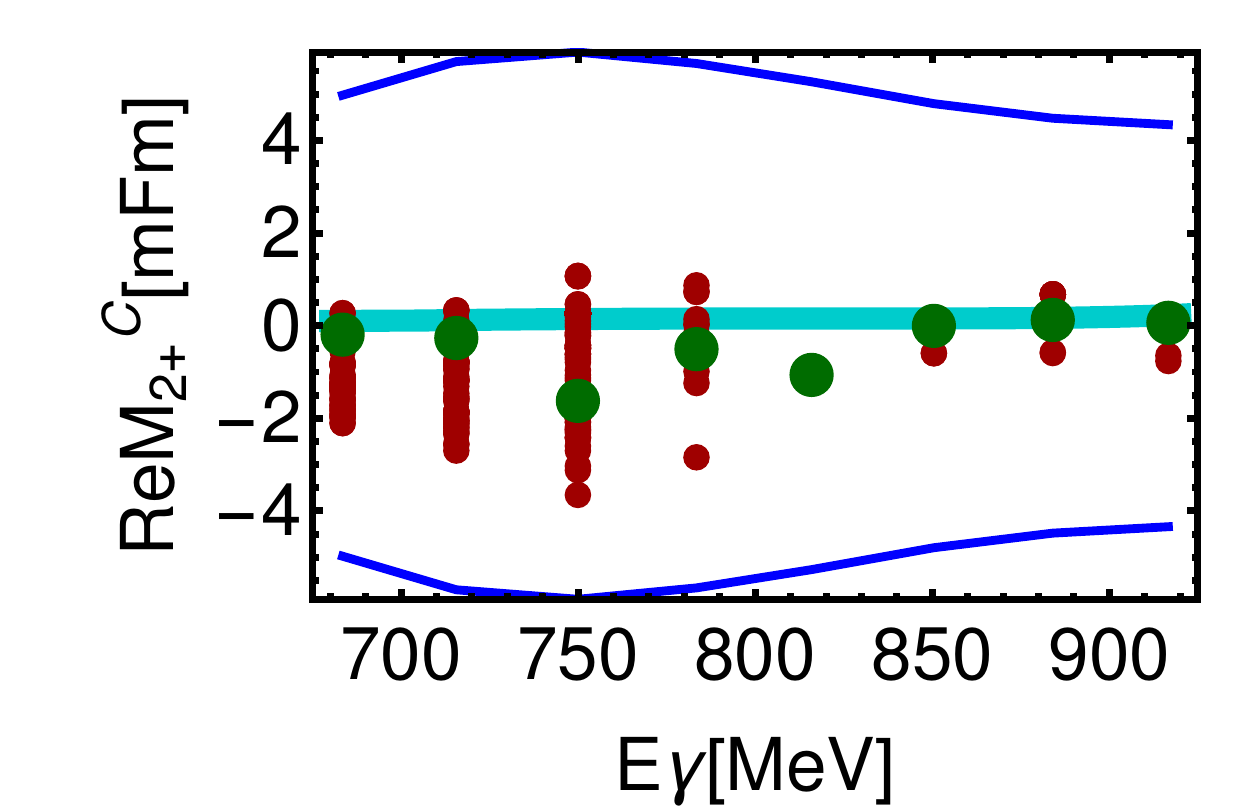}
 \end{overpic} \\
\begin{overpic}[width=0.325\textwidth]{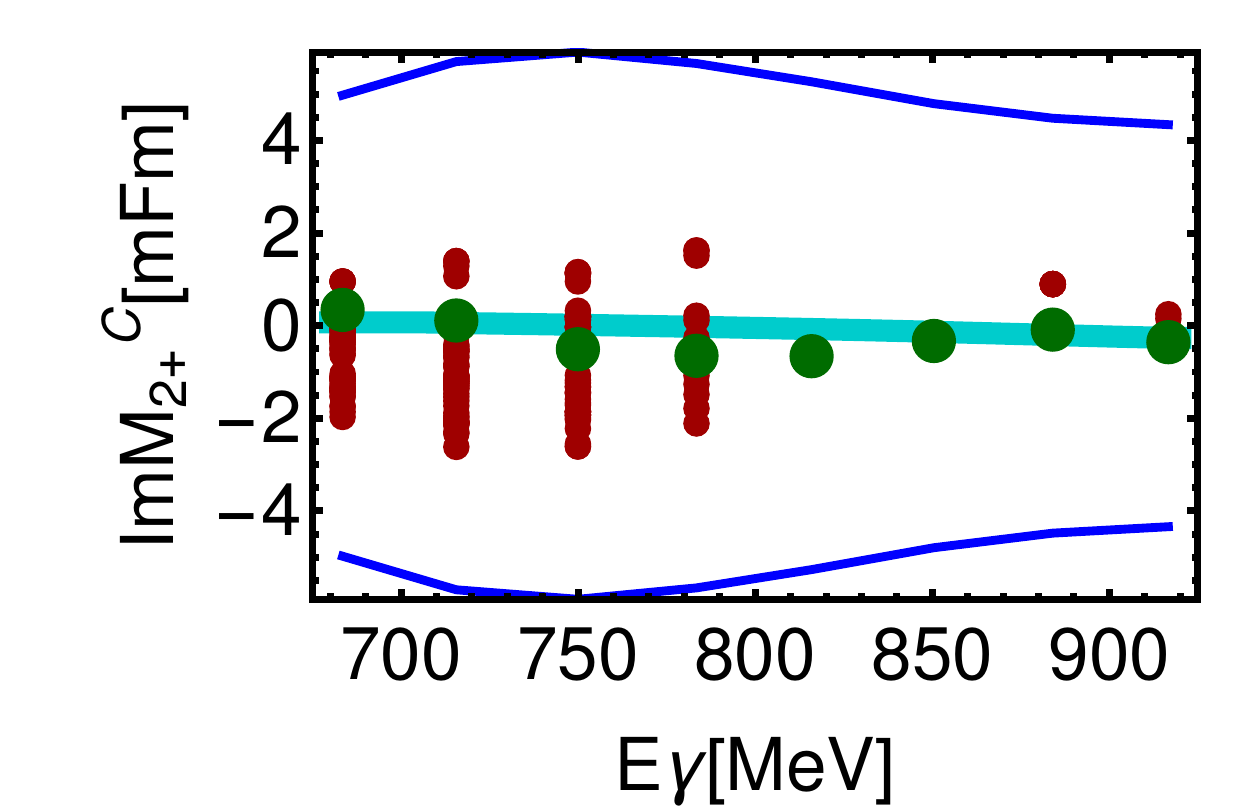}
 \end{overpic}
\begin{overpic}[width=0.325\textwidth]{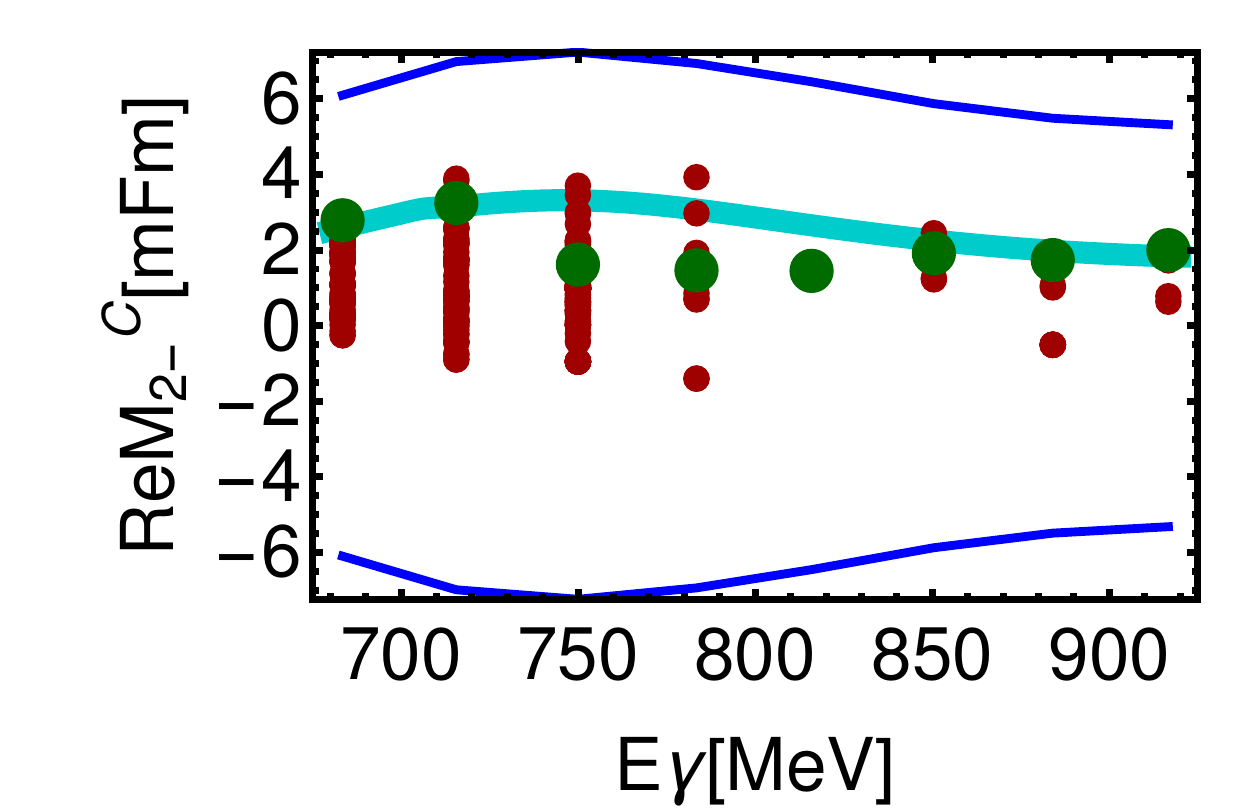}
 \end{overpic}
\begin{overpic}[width=0.325\textwidth]{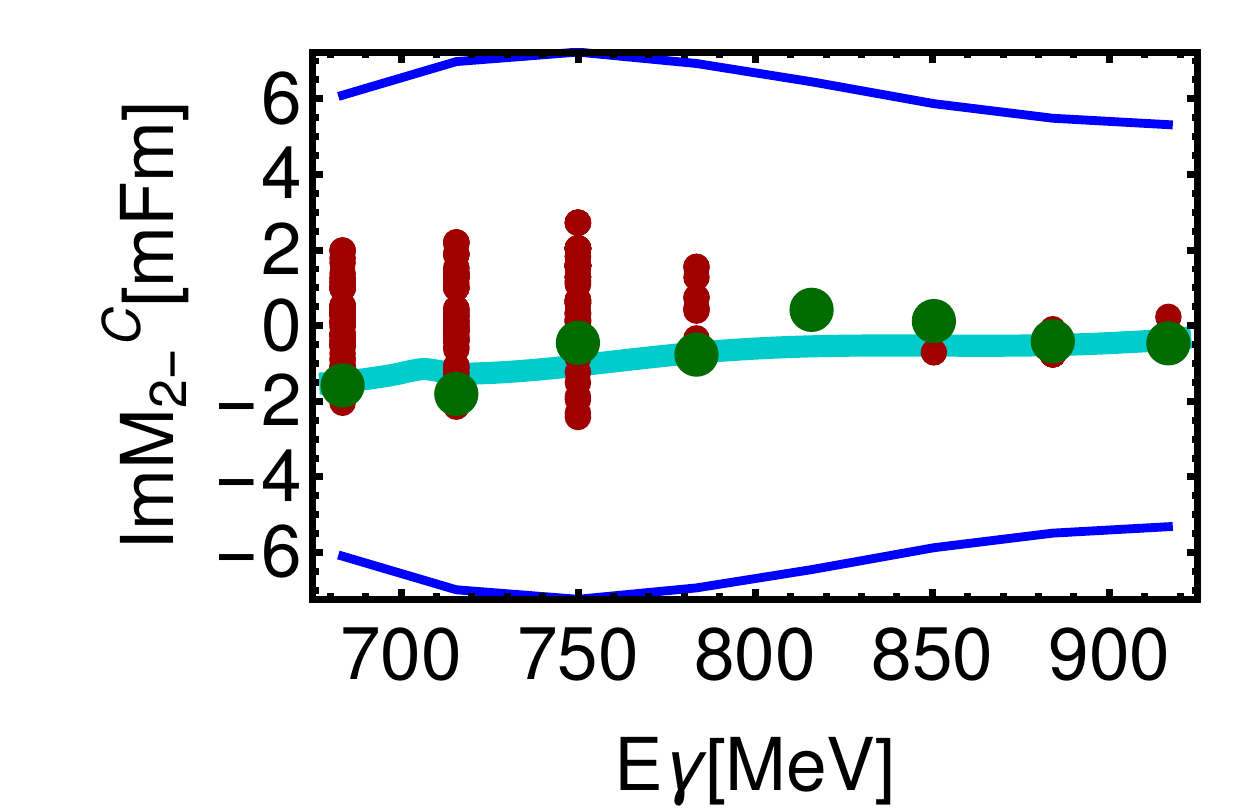}
 \end{overpic} \\
\begin{overpic}[width=0.325\textwidth]{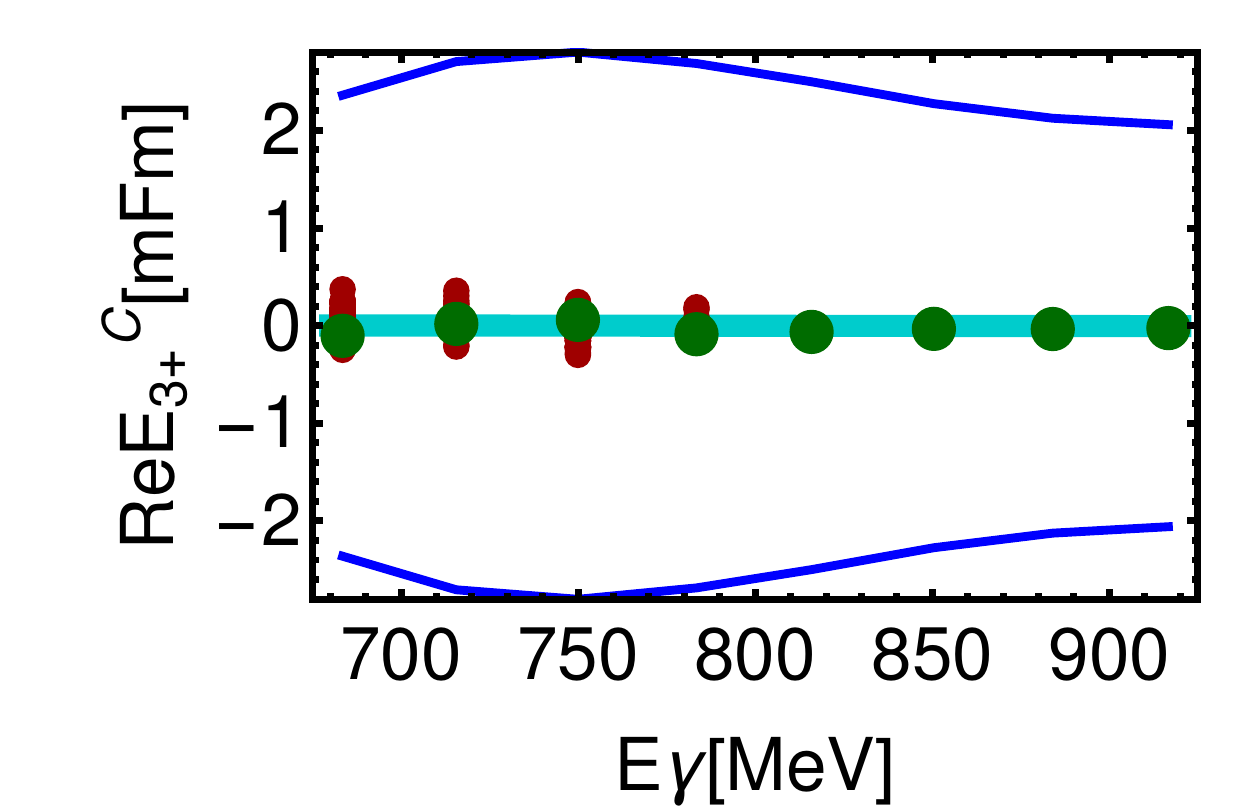}
 \end{overpic}
\begin{overpic}[width=0.325\textwidth]{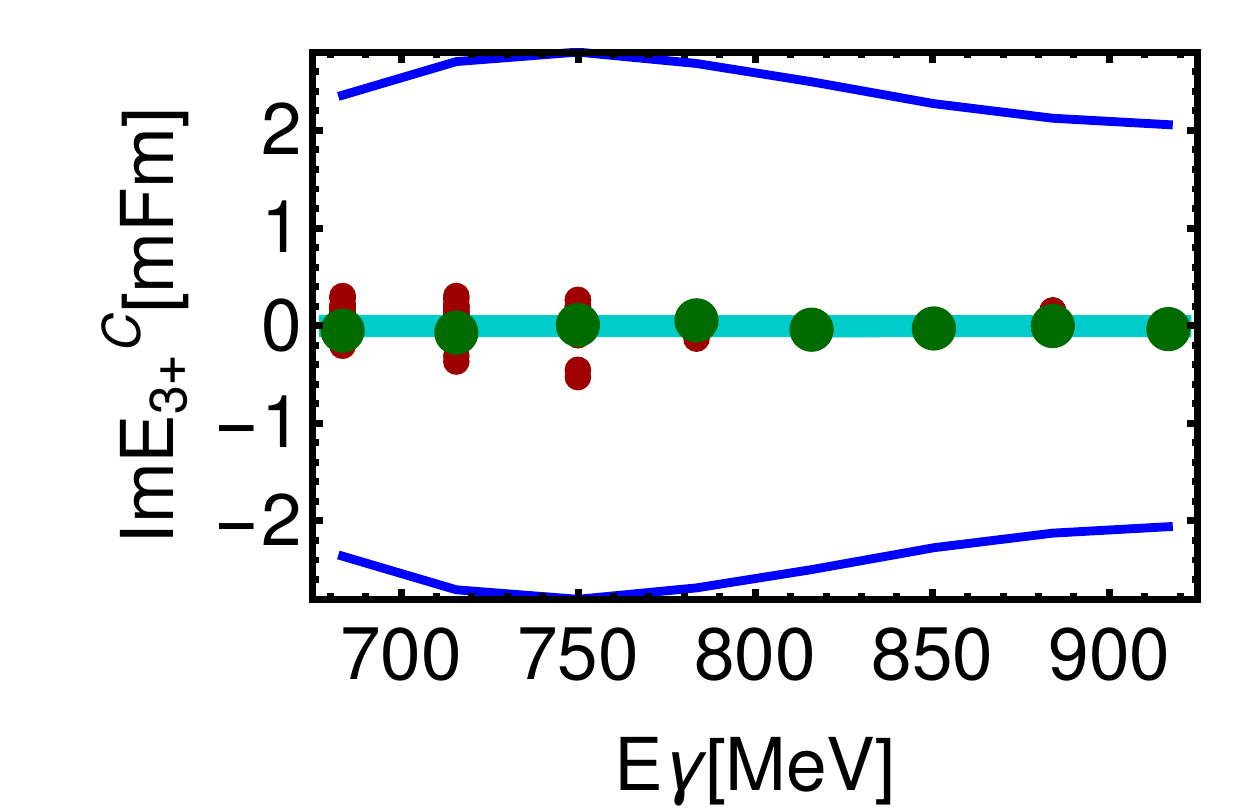}
 \end{overpic}
\begin{overpic}[width=0.325\textwidth]{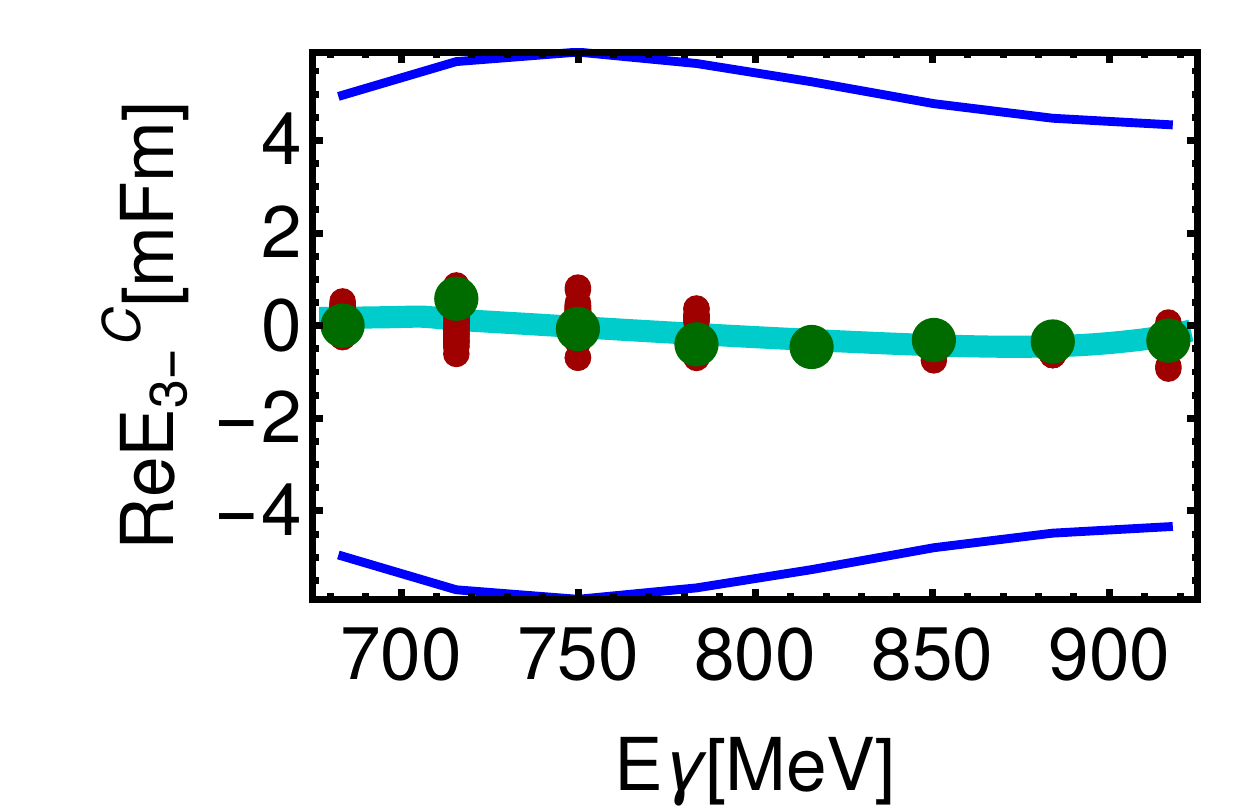}
 \end{overpic} \\
\begin{overpic}[width=0.325\textwidth]{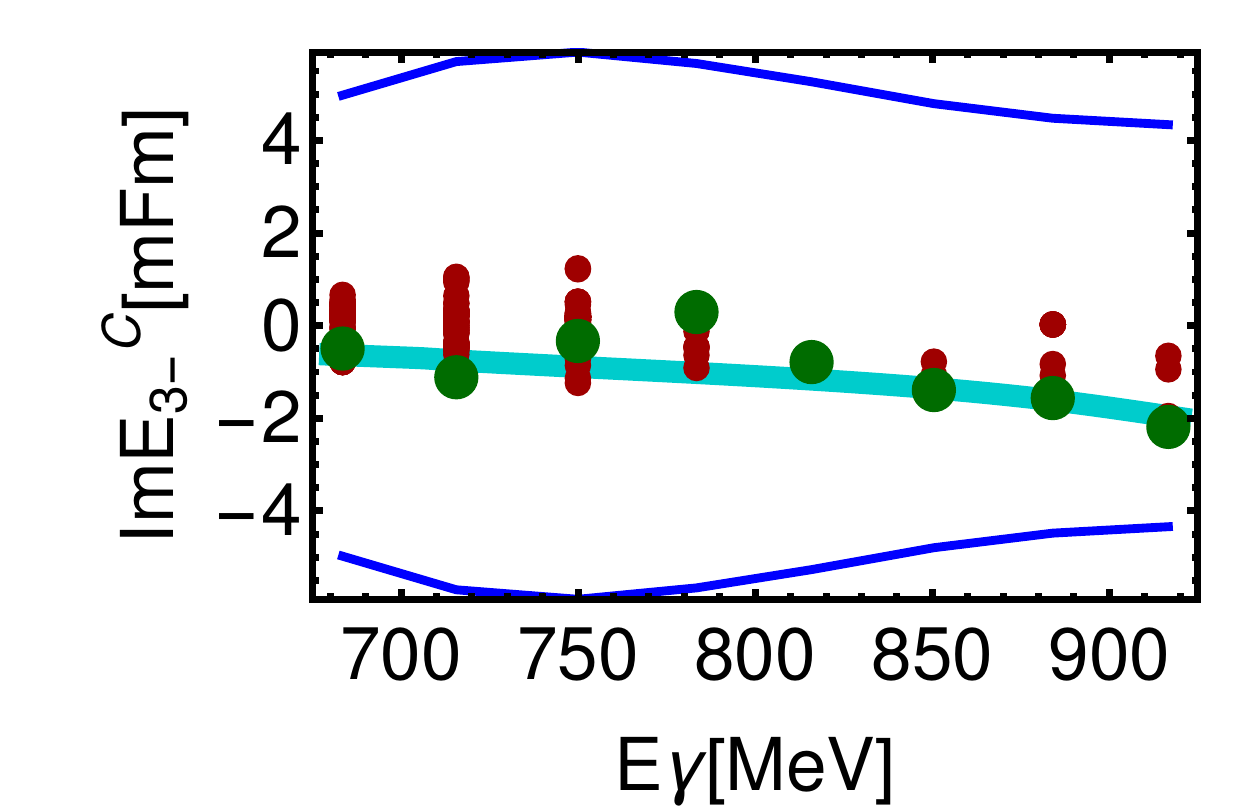}
 \end{overpic}
\begin{overpic}[width=0.325\textwidth]{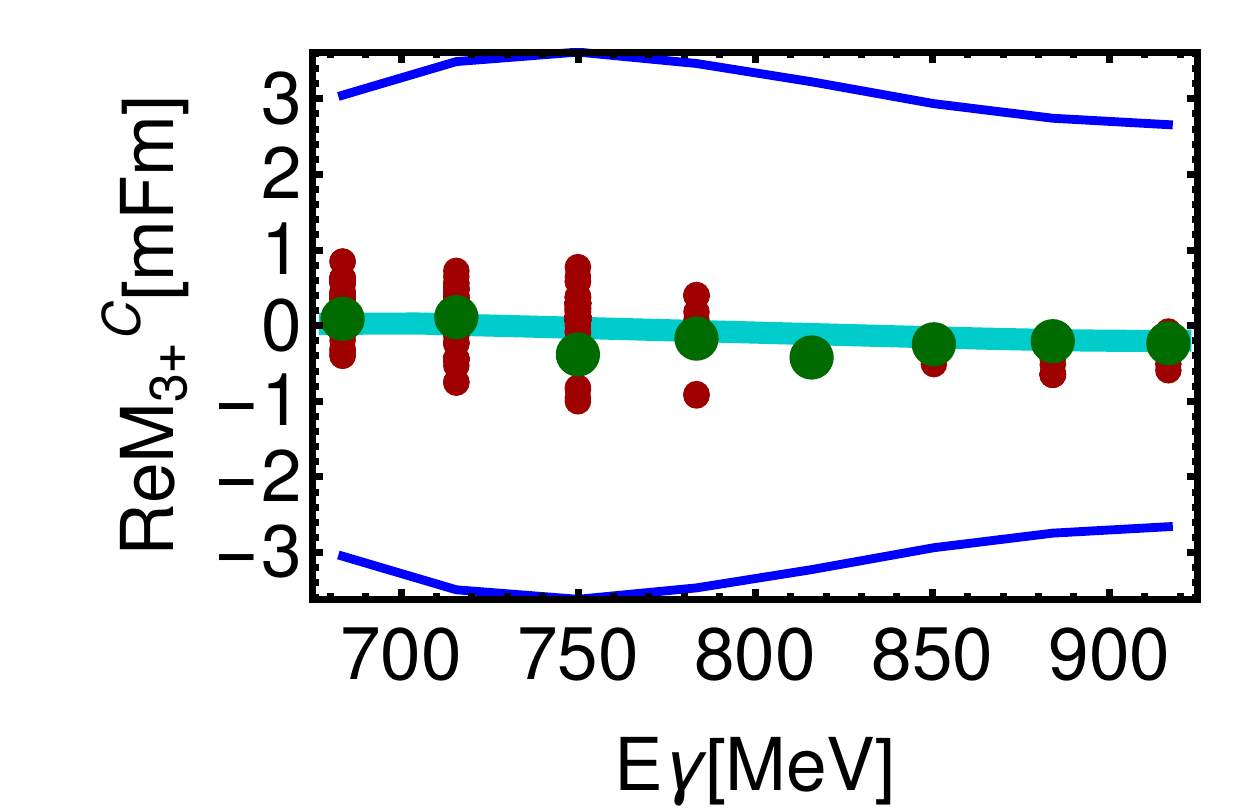}
 \end{overpic}
\begin{overpic}[width=0.325\textwidth]{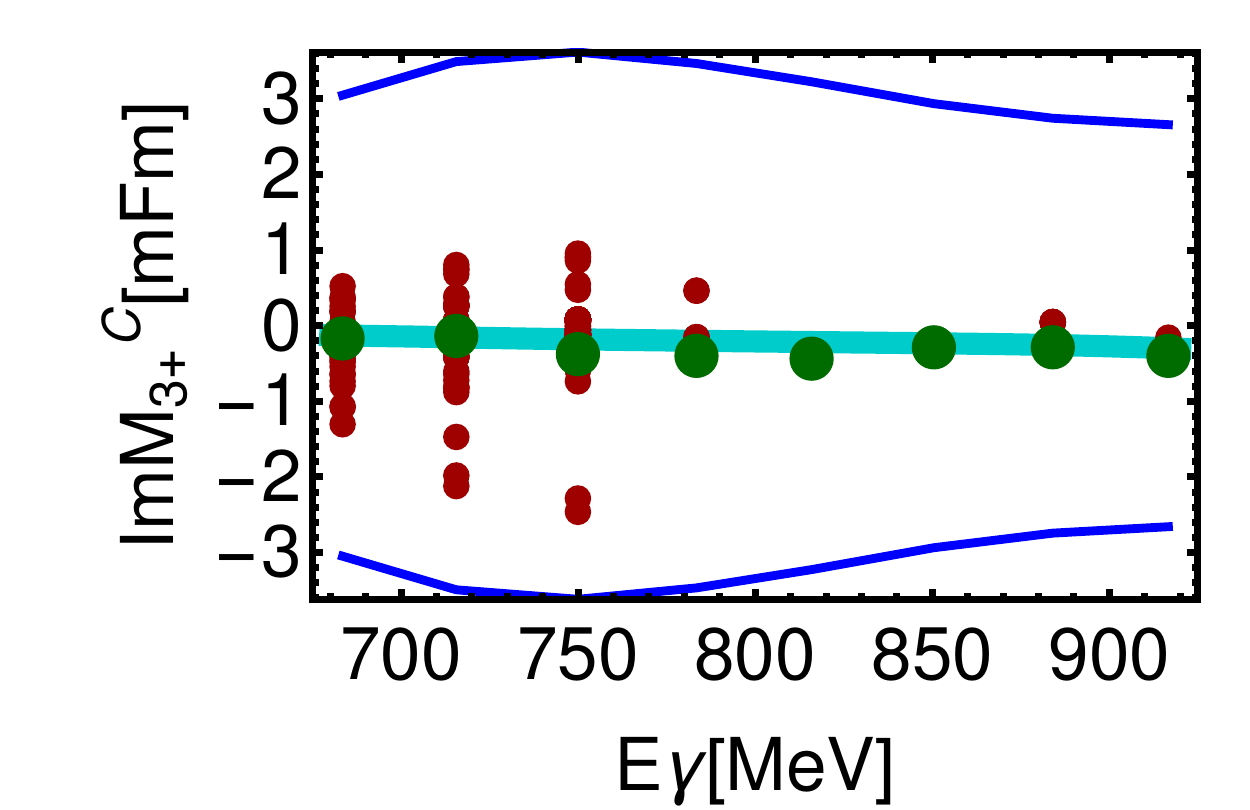}
 \end{overpic}
\caption{The plots show results for the TPWA at $\ell_{\mathrm{max}} = 4$, with $G$-waves fixed to Bonn-Gatchina \cite{BoGa}. They are continued in Figure \ref{fig:FourthFitLmax4GWavesBnGaMultipolesPlotsII}.}
\label{fig:FourthFitLmax4GWavesBnGaMultipolesPlotsI}
\end{figure}

\begin{figure}[ht]
 \centering
\begin{overpic}[width=0.325\textwidth]{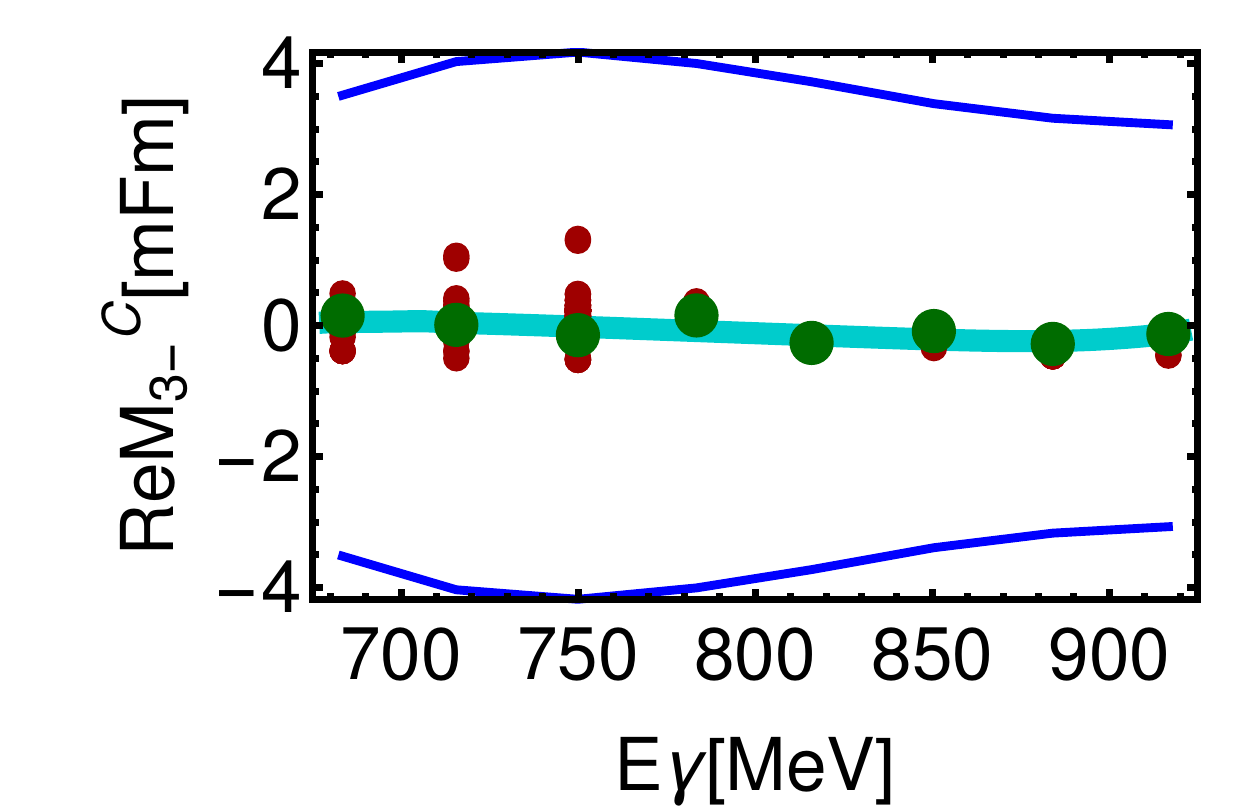}
 \end{overpic}
\begin{overpic}[width=0.325\textwidth]{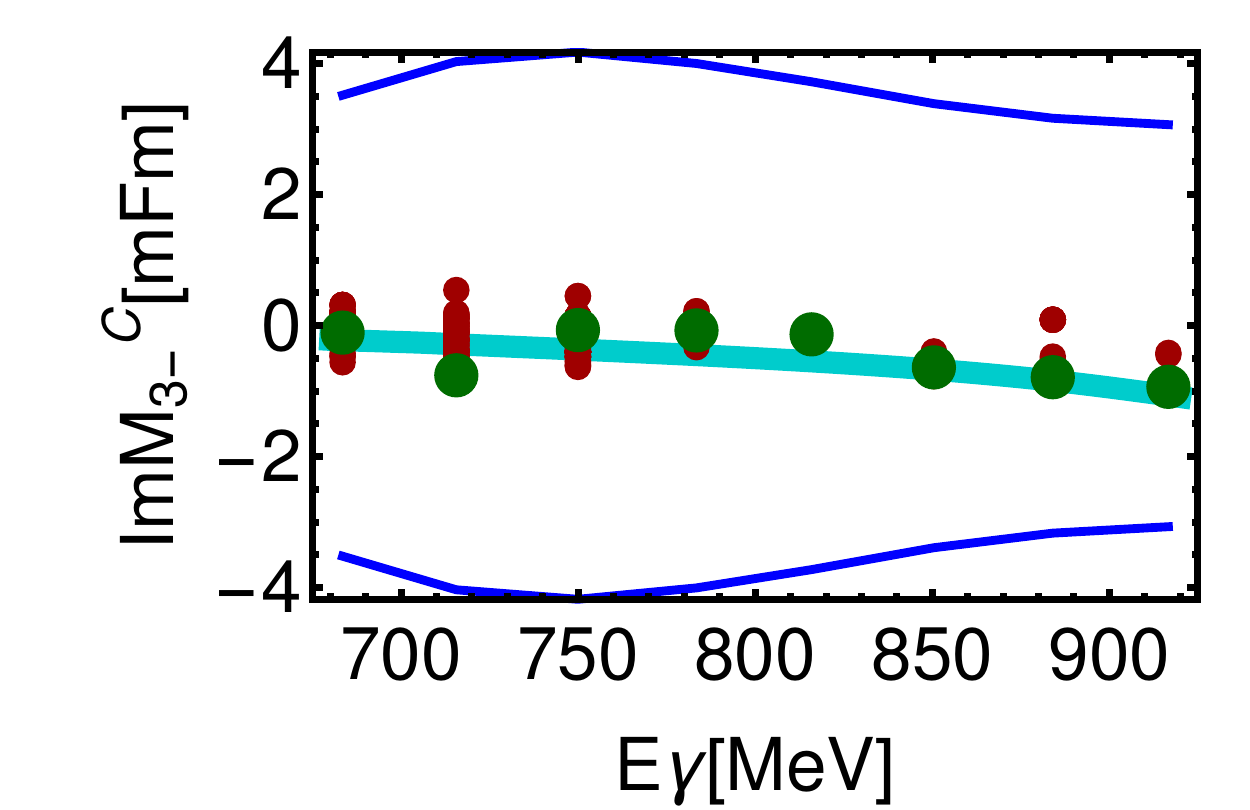}
 \end{overpic}
\caption[Results for the $23$ fit-parameters comprised of the real- and imaginary parts of phase-constrained $S$-, $P$-, $D$- and $F$-wave multipoles, for a TPWA with $\ell_{\mathrm{max}} = 4$ and $G$-waves fixed to BnGa 2014\_02, within the $2^{\mathrm{nd}}$ resonance region.]{The diagrams shown here, combined with those visible in Figure \ref{fig:FourthFitLmax4GWavesBnGaMultipolesPlotsI}, depict the results for the $23$ fit-parameters comprised of the real- and imaginary parts of phase-constrained $S$-, $P$-, $D$- and $F$-wave multipoles. In the corresponding fits, the $G$-waves have been set to values of BnGa2014\_02 \cite{BoGa} (rotated to the standard phase-contraint of a real and potitive $S$-wave). The layout of the plots is analogous to Figure \ref{fig:FirstFitLmax2MultipolesPlots}. Again, the attained global minimum is indicated by big green dots, while local minima are drawn as smaller red dots. Here, we included all local minima below the $0.975$-quantile of the theoretical chisquare-distribution for $\mathrm{ndf} = r = 27$. The energy-dependent PWA-solution BnGa 2014\_02 \cite{BoGa} is shown as a thick cyan-colored curve.}
\label{fig:FourthFitLmax4GWavesBnGaMultipolesPlotsII}
\end{figure}

The single unattractive feature of the previous fit has been the relatively low fit-quality in the higher energy-bins (cf. Figure \ref{fig:ThirdFitLmax3FWavesFixedChiSquarePlots}). To remedy the situation, an analysis has been performed trying the analogous case, but for one truncation order higher. This means, we performed an analysis for $\ell_{\mathrm{max}} = 4$, fixing the $G$-waves to BnGa 2014\_02 \cite{BoGa} and varying the $S$-, $P$-, $D$- and $F$-wave multipoles. \newline
For this analysis, a full Monte Carlo minimum-search (section \ref{sec:MonteCarloSampling}) was done, employing $N_{MC} = 50000$. The estimate for the number of degrees of freedom in this fit amounts to $\mathrm{ndf} = r = 33$. Results for the $\chi^{2}/\mathrm{ndf}$ of the various non-redundant minima are plotted in Figure \ref{fig:FourthFitLmax4GWavesBnGaChiSquarePlots}, while the corresponding results for the multipoles can be inspected in Figures \ref{fig:FourthFitLmax4GWavesBnGaMultipolesPlotsI} and \ref{fig:FourthFitLmax4GWavesBnGaMultipolesPlotsII}.\newline
Indeed, for this Ansatz the fit-quality of the global minimum is improved for the higher energies. Furthermore, as for all fits done before, the seven observables allow for a global minimum and mathematically exactly degenerate solutions are fully absent. However, there exist again, like in the case of the fully unconstrained fit for $\ell_{\mathrm{max}} = 3$ discussed above, substantial issues with local minima situated very close to the best solution in $\chi^{2}$. This is the case especially for the lower half of the considered energy-region, but also for the higher energies, the global minimum is not well-separated. \newline
When looking at the results for the multipoles, it becomes apparent that for most of them, the global minimum gets closer to the Bonn-Gatchina solution when compared to the fully unconstrained $\ell_{\mathrm{max}}=3$-fit (cf. Figures \ref{fig:SecondFitLmax3MultipolesPlotsI} and \ref{fig:SecondFitLmax3MultipolesPlotsII}). Some parameters provide notable exceptions to this statement, namely $\mathrm{Re} \left[ M_{1+}^{C} \right]$, $\mathrm{Re} \left[ M_{1-}^{C} \right]$, $\mathrm{Re} \left[ M_{2+}^{C} \right]$ and $\mathrm{Re} \left[ M_{2-}^{C} \right]$. Apart from the global minimum, Figures \ref{fig:FourthFitLmax4GWavesBnGaMultipolesPlotsI} and \ref{fig:FourthFitLmax4GWavesBnGaMultipolesPlotsII} show all local minima below the $0.975$-quantile of the chisquare-distribution for $\mathrm{ndf} = 33$. These local minima form quite substantial bands of solutions for the lower four energies, while for the higher energies the bands seem to clear up. This again looks quite similar to the fully unconstrained fit. \newline
In summary, the hope that constraining the $G$-waves to Bonn-Gatchina may resolve the ambiguity-issues of the unconstrained $\ell_{\mathrm{max}}=3$-fit did not turn out as true. This fit also does not yield a satisfactory result over the whole energy-range and cannot provide a good basis for a bootstrap-analysis. \newline
Therefore, we are here content with the fact fixing the $F$-wave multipoles to the energy-dependent model yielded a good global minimum for this analysis in the second resonance-region. Also, this result allowed for a sound application of the bootstrap, which yielded reliable uncertainties for the extracted multipoles, except for the third energy-bin where it was still able to unveil a problem with a discrete ambiguity.

\clearpage

\clearpage

\section{Summary, conclusions and outlook} \label{chap:Conclusions}

\allowdisplaybreaks

This section will be used in order to highlight some important results found in the course of this thesis. Also, prospects for future research will be given, building on the results and methods elaborated in this work.

\subsection{Selected results} \label{sec:KinematicsDerivations}

The beginning of this thesis was marked by a mathematical investigation of the posed problem, namely the uniqueness of the extraction of electric and magnetic photoproduction  multipoles $\left\{ E_{\ell \pm}, M_{\ell \pm} \right\}$ from complete sets of polarization observables $\check{\Omega}^{\alpha}$. This extraction had to be done in a truncated partial wave analysis (TPWA), employing a maximal orbital angular momentum $L=\ell_{\mathrm{max}}$. For the mathematical investigation, the academic version of the problem was considered where all higher partial waves above $L$ vanish exactly. \newline
Once the angular distributions of the polarization observables have been decomposed into associated Legendre polynomials $P_{\ell}^{m} (\cos \theta)$, the above stated problem can be rephrased as a uniqueness problem in the solution of certain bilinear equation systems, built up out of the compositions of the extracted Legendre-coefficients $\left(a_{L}\right)_{k}^{\check{\Omega}^{\alpha}}$ (in other contexts referred to as 'moments' \cite{MikhasenkoProceeding2014}) in terms of multipoles. Such equation systems are defined by hermitean matrices $\left( \mathcal{C}_{L}\right)_{k}^{\check{\Omega}^{\alpha}}$ which themselves depend on the observables in question, on the form of the multipole expansion of the photoproduction amplitude and (in their dimension) on the truncation order $L$. The equation systems read
\begin{equation}
\left(a_{L}\right)_{k}^{\check{\Omega}^{\alpha}} = \left< \mathcal{M}_{\ell} \right| \left( \mathcal{C}_{L}\right)_{k}^{\check{\Omega}^{\alpha}} \left| \mathcal{M}_{\ell} \right> \mathrm{.} \label{eq:BilinearEqSystemConclusions}
\end{equation}
Here, multipoles have been collected into the vector $\left| \mathcal{M}_{\ell} \right> = \left[ E_{0+}, E_{1+}, \ldots, M_{L-} \right]^{T}$. The matrices $ \left( \mathcal{C}_{L}\right)_{k}^{\check{\Omega}^{\alpha}}$ have been worked out and shown up to $L=5$ in this work (Appendix \ref{sec:TPWAFormulae}). \newline
Rather than working with the bilinear equations (\ref{eq:BilinearEqSystemConclusions}) directly, it has turned out as more advantageous to use the elegant Ansatz by Omelaenko \cite{Omelaenko} in order to investigate the uniqueness of, or equivalently the appearance of ambiguities in, a TPWA. \newline
The idea is to employ a basis for the $4$ spin-amplitudes of photoproduction which diagonalizes $4$ polarization observables. Famously \cite{ChTab, Omelaenko}, the profile functions of the $4$ group $\mathcal{S}$ observables $\left\{\check{\Omega}^{\alpha_{S}}\right\} = \left\{ \sigma_{0}, \Sigma, T, P \right\}$ are \textit{diagonal} when written in terms of transversity-amplitudes $b_{i}$, i.e. they are just sums of moduli squared:
\begin{equation}
\check{\Omega}^{\alpha_{S}} = \frac{1}{2} \left( \pm \left|b_{1}\right|^{2} \pm \left|b_{2}\right|^{2} \pm \left|b_{3}\right|^{2} + \left|b_{4}\right|^{2}\right) \mathrm{.} \label{eq:GroupSIsDiagonalConclusions}
\end{equation}
These quantities therefore serve as a starting point in Omelaenko's analysis. Other bases for the spin-observables may also be chosen, with different combinations of $4$ observables diagonal\footnote{In this work, an exhaustive list has been given for these alternative possibilities, defined by amplitudes connected to the transversity amplitudes $b_{i}$ by unitary transformations. Further details on this issue have however been moved to Appendix \ref{subsec:AmbiguitiesSimDiagObservablesI}.}. Thus, one is not forced to start at the group $\mathcal{S}$ observables, but they are a natural choice due to physical reasons. \newline
Due to the appearance of factors of $\sin \theta$ in the definitions of the $b_{i}$, it is not possible to discuss polynomial amplitudes in photoproduction using the standard angular variable $\cos \theta$. An elegant mathematical way out, first employed by Gersten \cite{Gersten}, is to make use of $t = \tan \frac{\theta}{2}$. Upon change of variables, it is seen that the transversity amplitudes become, in a finite truncation $L$, linear factor decompositions of finite polynomials, times a kinematic pole in $t^{2}$. Further factors are given by an angle-dependent phase, a complex normalization $a_{2L}$ and a convention-dependent factor $\mathcal{C}$. A consistent final form for the amplitudes is, with explicit reference to the energy $W$ suppressed from now on
\begin{align}
 b_{1} \left(\theta\right) = - \hspace*{1pt} \mathcal{C} \hspace*{1pt} a_{2L} \hspace*{1pt} \frac{\exp \left(- i \frac{\theta}{2}\right)}{\left( 1 + t^{2} \right)^{L}} \hspace*{1pt} \prod_{k = 1}^{2L} \left( t + \beta_{k} \right) &\mathrm{,} \hspace*{5pt} b_{2} \left(\theta\right) = - \hspace*{1pt} \mathcal{C} \hspace*{1pt} a_{2L} \hspace*{1pt} \frac{\exp \left(i \frac{\theta}{2}\right)}{\left( 1 + t^{2} \right)^{L}} \hspace*{1pt} \prod_{k = 1}^{2L} \left( t - \beta_{k} \right)  \mathrm{,} \label{eq:b1and2LinFactDecomp} \\
  b_{3} \left(\theta\right) = \mathcal{C} \hspace*{1pt} a_{2L} \hspace*{1pt} \frac{\exp \left(- i \frac{\theta}{2}\right)}{\left( 1 + t^{2} \right)^{L}} \hspace*{1pt} \prod_{k = 1}^{2L} \left( t + \alpha_{k} \right) &\mathrm{,} \hspace*{5pt} b_{4} \left(\theta\right) = \mathcal{C} \hspace*{1pt} a_{2L} \hspace*{1pt} \frac{\exp \left(i \frac{\theta}{2}\right)}{\left( 1 + t^{2} \right)^{L}} \hspace*{1pt} \prod_{k = 1}^{2L} \left( t - \alpha_{k} \right)  \mathrm{.} \label{eq:b3and4LinFactDecomp}
\end{align}
Here, the complex variables given by the normalization coefficient $a_{2L}$ and the polynomial-roots $\left\{ \alpha_{i}, \beta_{j} \right\}$ are equivalent to the complex multipoles $\left\{E_{\ell \pm}, M_{\ell \pm}\right\}$, but are a lot better suited to discuss the ambiguity problem. In this work, consistent expressions and routines have been worked out to pass from multipoles to roots, which were not easily obtainable from the literature. \newline
Another peculiarity of photoproduction is the fact that the $\alpha$- and $\beta$-roots are not completely independent. Rather, it can be shown that the polynomials represented by the products over linear factors in the expressions for $b_{2}$ and $b_{4}$ ((\ref{eq:b1and2LinFactDecomp}) and (\ref{eq:b3and4LinFactDecomp})) have to have identical free terms (i.e. they are equal at $\theta=0$). Thus, one can derive the following multiplicative constraint among the roots
\begin{equation}
 \boxed{\prod_{j = 1}^{2 L} \alpha_{j} = \prod_{k = 1}^{2 L} \beta_{k} \mathrm{.}} \label{eq:OmelMultConstraintConclusions}
\end{equation}
This expression, though looking very simple, has turned out to be very important in the discussion of the ambiguity problem in photoproduction. The reason is that each ambiguity, while also leaving the group $\mathcal{S}$ observables (\ref{eq:GroupSIsDiagonalConclusions}) invariant, has to fulfill this constraint. \newline
Inspection of the definitions (\ref{eq:GroupSIsDiagonalConclusions}) tells that the only way to compose ambiguous roots out of an already existing solution $\left\{ \alpha_{i}, \beta_{j} \right\}$ is to either complex conjugate all the roots, or only subsets of them. The conjugation of all roots
\begin{equation}
 \alpha_{j} \longrightarrow \alpha_{j}^{\ast} \mathrm{,} \hspace*{2.5pt} \beta_{k} \longrightarrow \beta_{k}^{\ast} \mathrm{,} \hspace*{2.5pt} \forall j, k = 1, \ldots, 2L \mathrm{,} \label{eq:DoubleAmbDefConclusions}
\end{equation}
has been called \textit{double ambiguity} and it is seen quickly that it respects (\ref{eq:OmelMultConstraintConclusions}). All remaining combinatorially possible conjugations of subsets of the roots have been termed \textit{accidental ambiguities}. These possible solutions may (approximately) fulfill the constraint (\ref{eq:OmelMultConstraintConclusions}) by accident and may therefore turn up as additional solutions, or they may not\footnote{A more precise mathematical formalization of the different kinds of ambiguities has been carried out in the course of this work as well. This is elaborated further in Appendix \ref{sec:AccidentalAmbProofs}.}. Both the double ambiguity and the accidental ambiguities fall under the collective name of so-called \textit{discrete ambiguities}. \newline
For the illustration of the peculiar phenomenon of accidental ambiguities in photoproduction, as well as to get a feeling for their frequency of appearance, so-called \textit{ambiguity diagrams} have been useful. Here, one has to introduce the phases $\varphi_{j}$ and $\psi_{k}$ of the complex roots $\alpha_{j} = \left| \alpha_{j} \right| e^{i \varphi_{j}}$ and $\beta_{k} = \left| \beta_{k} \right| e^{i \psi_{k}}$. Then, the constraint (\ref{eq:OmelMultConstraintConclusions}) becomes in terms of the phases
\begin{equation}
 \varphi_{1} + \ldots + \varphi_{2L} = \psi_{1} + \ldots + \psi_{2L} \mathrm{.} \label{eq:OmelAddConstraintConclusions}
\end{equation}
\begin{figure}[h]
 \centering
 \begin{overpic}[width=0.955\textwidth]{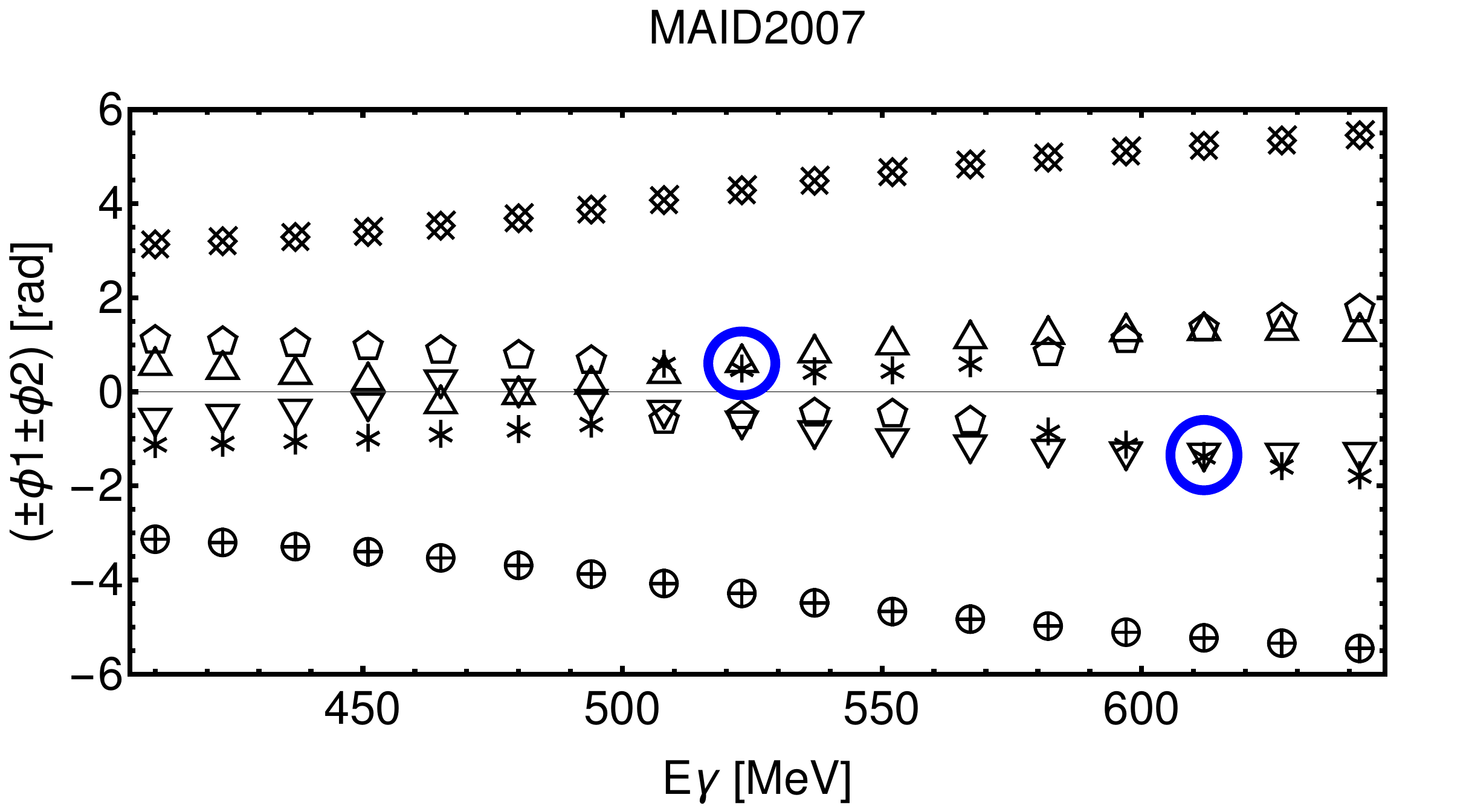}
 \end{overpic}
 \caption[Ambiguity diagram generated from the $S$- and $P$-waves of the model-solution MAID2007 for $\pi^{0}$-photoproduction.]{Shown here is an ambiguity diagram generated from the $S$- and $P$-waves of the model-solution MAID2007 \cite{MAID2007,MAID} for $\pi^{0}$-photoproduction.  Different sign-choices for the linear combinations of the phases $\left\{\varphi_{1}, \varphi_{2}\right\}$ and $\left\{\psi_{1}, \psi_{2}\right\}$ are plotted against energy $E_{\gamma}$. The labeling scheme for the different linear combinations is: $\circ ( \varphi_{1} + \varphi_{2})$, $\bigtriangleup (\varphi_{1} - \varphi_{2})$, $\bigtriangledown (- \varphi_{1} + \varphi_{2})$, $\diamond (- \varphi_{1} - \varphi_{2})$,
 $+ (\psi_{1} + \psi_{2})$, ${\Large \ast} (\psi_{1} - \psi_{2})$, $\pentagon (- \psi_{1} + \psi_{2})$, $\times (- \psi_{1} -\psi_{2})$. \newline
 Furthermore, two examples for specific energy bins where an accidental ambiguity occurs are emphasized by blue circles. }
 \label{fig:ConclusionPlotAmbiguityDiagramMAID2017}
\end{figure}
It is now possible to check graphically if one of the combinatorial possibilities to conjugate roots fulfills this constraint (\ref{eq:OmelAddConstraintConclusions}), i.e. if for any combination of signs the equality
\begin{equation}
 \pm \varphi_{1} \pm \ldots \pm \varphi_{2L} = \pm \psi_{1} \pm \ldots \pm \psi_{2L} \mathrm{,} \label{eq:OmelAddConstraintConclusionsII}
\end{equation}
is satisfied exactly or at least approximately. In Figure \ref{fig:ConclusionPlotAmbiguityDiagramMAID2017}, such an ambiguity diagram is plotted which comes from the $S$- and $P$-wave multipoles (i.e. $L=1$) of the model MAID2007 \cite{MAID2007,MAID} for $\pi^{0}$-photoproduction. The sums of phases $\varphi_{1} + \varphi_{2}$ and $\psi_{1} + \psi_{2}$, as well as the combinations $\pm \varphi_{1} \pm \varphi_{2}$ and $\pm \psi_{1} \pm \psi_{2}$, are plotted. As marked in the plot, for this MAID-solutions there exist energies where accidental ambiguities can occur. \newline
Interestingly, it turned out that for academic TPWA-problems which posses exact solutions, i.e. for solutions of synthetic data where all partial waves above $L$ vanish exactly, the accidental ambiguities turn out to be no problem. Due to the discrete nature of the TPWA, with analyses performed at disconnected points in energy individually, there do not exist continuous curves for the linear combinations $\pm \varphi_{1} \pm \varphi_{2}$ and $\pm \psi_{1} \pm \psi_{2}$ (see Figure \ref{fig:ConclusionPlotAmbiguityDiagramMAID2017}) and thus, Omelaenko's constraint (\ref{eq:OmelAddConstraintConclusions}) is never satisfied exactly for the accidental ambiguities. Instead, it is almost certain that a small error remains. As has been worked out in this thesis\footnote{The mathematical details have been moved to appendix \ref{sec:AccidentalAmbProofs}, in particular appendix \ref{subsec:AccidentalAmbProofsII}.}, the consequence of this fact is that accidental ambiguities never appear as mathematical ambiguities, at least in analyses of truncated model data. Therefore, only the double-ambiguity (\ref{eq:DoubleAmbDefConclusions}) needs to be resolved. \newline
In order to do this, observables are needed in addition to the group $\mathcal{S}$, $\left\{\check{\Omega}^{\alpha_{S}}\right\} = \left\{ \sigma_{0}, \Sigma, T, P \right\}$ (equation (\ref{eq:GroupSIsDiagonalConclusions})), which change under the double ambiguity transformation (\ref{eq:DoubleAmbDefConclusions}). An example is the profile function of the $F$-observable
\begin{equation}
 \check{F} \left(\theta\right) = \frac{\sigma_{0} \left(\pi\right)}{2 \left( 1 + t^{2} \right)^{2L}} \mathrm{Im} \left[ - \prod_{k=1}^{2L} \left( t + \alpha_{k}^{\ast} \right) \left( t + \beta_{k} \right) + \prod_{k=1}^{2L} \left( t - \alpha_{k}^{\ast} \right) \left( t - \beta_{k} \right) \right] \mathrm{,} \label{eq:FExplicitFormulaConclusions}
\end{equation}
which just changes its sign. Therefore, using Omelaenko's algebra it has been possible to postulate, at least in idealized cases, the existence of \textit{complete sets of 5 observables} in a TPWA. An example which has been investigated often in this work is 
\begin{equation}
\left\{ \sigma_{0}, \Sigma, T, P, F \right\} \mathrm{.} \label{eq:ExampleCompleteSetConclusions}
\end{equation}
This result has been surprising at first, since it seems to be in contrast to the extraction of the full spin-amplitudes (for instance the $b_{i}$) out of the data at each energy and angle individually. In the latter case, Chiang and Tabakin \cite{ChTab} have pointed out that $8$ observables are needed, including double-polarization measurements with recoil-polarization. However, both problems, i.e. the TPWA and the extraction of the full amplitudes, are quite different in their nature, as has been pointed out in a recent paper \cite{WorkmanEtAl2017}. \newline
Of course, the complete sets proposed just on the basis of studies of ambiguities have to be either confirmed or disconfirmed in an explicit solution of the inverse problem posed by the TPWA. For the problem at hand, such solutions could only be obtained numerically. In this work, extensive analyses have been performed to truncated theory-data stemming from the MAID2007-model \cite{MAID2007,MAID}. The results for these truncated data have turned out fully consistent with the mathematical understanding on the discrete ambiguities, acquired by a detailed study of Omelaenko's Ansatz \cite{Omelaenko}. For fits to the group $\mathcal{S}$ observables, the appearance of the double ambiguity, as well as additional solutions which have been attributed to accidental ambiguities suggested by diagrams such as Figure \ref{fig:ConclusionPlotAmbiguityDiagramMAID2017}, have been fully confirmed. For the proposed complete sets of $5$ such as (\ref{eq:ExampleCompleteSetConclusions}), it was possible to uniquely solve for the generating MAID-solution for data truncated at $L=1$, $2$, $3$ and $4$. \newline
A Monte Carlo-Ansatz has been developed in order to solve such model-TPWA problems for all multipoles up to an overall phase, without imposing any further model-assumptions. The Ansatz is based in the fact that the total cross section, which always arises as a by-product of a TPWA-analysis, already strongly constrains the relevant parameter-space for the real- and imaginary-parts of the multipoles. Since the total cross section $\bar{\sigma}$ is in any truncation order $L$ strictly a sum of modulus-squared multipoles \cite{LeukelPhD}
\begin{equation}
 \bar{\sigma} = 2 \pi \frac{q}{k} \sum_{\ell = 0}^{L} \Big\{ (\ell + 1)^{2} (\ell + 2) \left| E_{\ell+} \right|^{2} + (\ell - 1) \ell^{2} \left| E_{\ell -} \right|^{2}  + \ell (\ell + 1)^{2} \left| M_{\ell+} \right|^{2} + \ell^{2} (\ell + 1) \left| M_{\ell-} \right|^{2} \Big\} \mathrm{,}
\end{equation}
it requires solutions to lie on a higher-dimensional ellipsoid in parameter space, which may thus be sampled prior to fitting. The proposed Monte Carlo-method also later found application in numerical fits to real photoproduction data. While the analyses of truncated model-data fully confirmed the complete sets of $5$ in a TPWA, it has also been pointed out that this completeness may be lost once all partial waves are contributing to the data, or the data themselves have errors. In both cases, the assumption of the existence of an exact solution is not valid any more, and it is in such cases that the accidental ambiguities can become dangerous. \clearpage
One further very interesting mathematical result found by L. Tiator \cite{LotharPrivateComm2016} is the existence of \textit{complete sets of 4 observables} in a TPWA for photoproduction. This has been found from analyses of synthetic data similar to the one's described above and the result has been discovered only very recently, towards the end of this thesis project. Furthermore, up to now it has not been possible to understand the complete sets of $4$ mathematically using Omelaenko's algebraic Ansatz. In total, there exist $196$ distinct possibilities to form a complete set of $4$, which have all been listed. \newline

\begin{figure}[h]
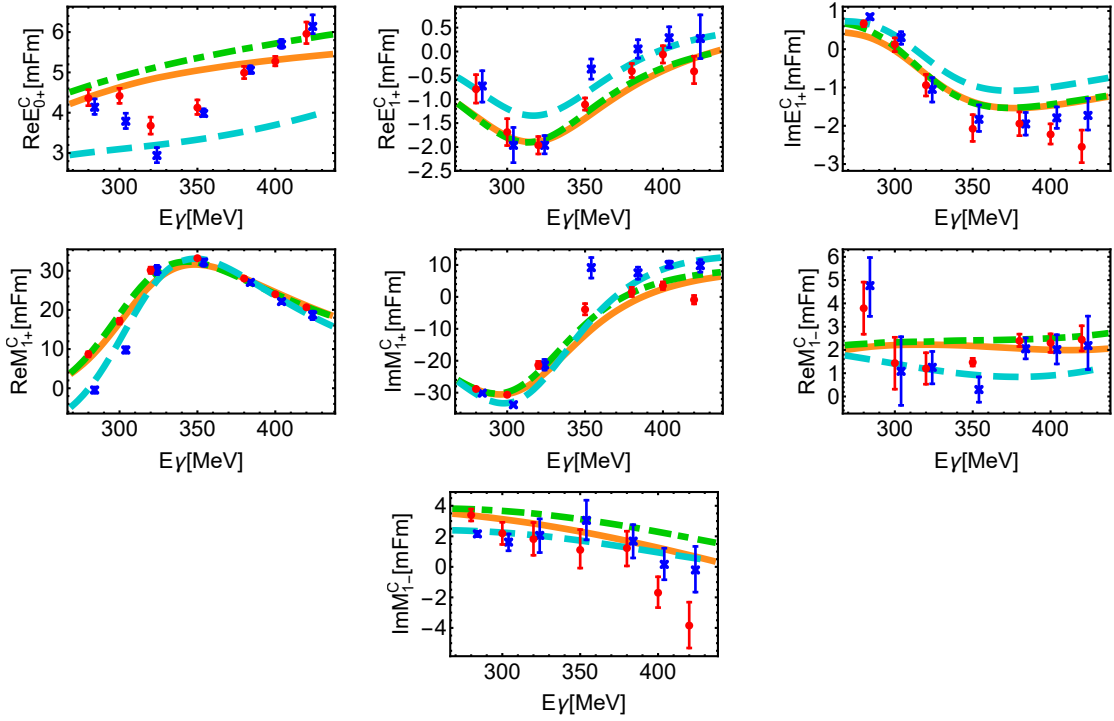

\centering
\vspace*{-8pt}
\begin{overpic}[width=0.325\textwidth]{Lmax2_DWavesSAID_NoDWaves_ResultsCompareToPWA1.pdf}
 \end{overpic}
\begin{overpic}[width=0.325\textwidth]{Lmax2_DWavesSAID_NoDWaves_ResultsCompareToPWA2.pdf}
 \end{overpic}
\begin{overpic}[width=0.325\textwidth]{Lmax2_DWavesSAID_NoDWaves_ResultsCompareToPWA3.pdf}
 \end{overpic} \\
\begin{overpic}[width=0.325\textwidth]{Lmax2_DWavesSAID_NoDWaves_ResultsCompareToPWA4.pdf}
 \end{overpic}
\begin{overpic}[width=0.325\textwidth]{Lmax2_DWavesSAID_NoDWaves_ResultsCompareToPWA5.pdf}
 \end{overpic}
\begin{overpic}[width=0.325\textwidth]{Lmax2_DWavesSAID_NoDWaves_ResultsCompareToPWA6.pdf}
 \end{overpic} \\
\begin{overpic}[width=0.325\textwidth]{Lmax2_DWavesSAID_NoDWaves_ResultsCompareToPWA7.pdf}
 \end{overpic} \vspace*{-5pt}
 \caption[Results for a bootstrap-analysis of 5 observables $\left\{\sigma_{0}, \Sigma, T, P, F\right\}$ in the $\Delta$-resonance region, with a strict truncation at $L=1$, as well as a fit with $D$-waves fixed to the SAID-model.]{Red dots denote results of a TPWA-fit performed to the observables $\left\{\sigma_{0}, \Sigma, T, P, F\right\}$ \cite{Hornidge:2013,LeukelPhD,Leukel:2001,Belyaev:1983,Schumann:2015} in the $\Delta$-resonance region, with a truncation at $\ell_{\mathrm{max}} = 2$ and $D$-waves fixed to the model-solution SAID CM12 \cite{WorkmanEtAl2012ChewMPhotoprod, SAID}. Blue crosses (slightly shifted to the right) indicate a fit to the same data, but using a strict truncation at $\ell_{\mathrm{max}} = 1$ and no model-dependence. Errors have been determined using the bootstrap. \newline Results are compared to the PWA-solutions SAID CM12 (orange solid line) \cite{SAID}, BnGa 2014\_02 (cyan dashed line) \cite{BoGa} and MAID2007 (green dash-dotted line) \cite{MAID}.}
 \label{fig:ConclusionPlotUniqueSolDeltaRegionLmax1}
\end{figure}

Regarding numerical calculations, more has been done than the analyses of synthetic model-data mentioned above. In order to properly account for the effects of statistical uncertainties in the data, i.e. the propagation of errors to the multipole fit-parameters as well as the possible generation of ambiguities, a well-known resampling-analysis technique
known as the \textit{bootstrap} \cite{EfronOriginal, EfronTibshiraniBook, DavisonHinkley} has been applied to the TPWA-problem. Bootstrap-TPWA analyses have been illustrated briefly on pseudodata, then also applied in analyses of real photoproduction data. Data have been analyzed solely for the reaction of $\pi^{0}$-photoproduction
\begin{equation}
 \gamma p \longrightarrow \pi^{0} p \mathrm{.} \label{eq:Pi0ChannelConclusions}
\end{equation}
A useful preparatory step in combined analyses of photoproduction data has been the extraction of Legendre-coefficients $\left(a_{L}\right)_{k}^{\check{\Omega}^{\alpha}}$ out of the angular distributions of the profile functions $\check{\Omega}^{\alpha} = \sigma_{0} \Omega^{\alpha}$. Such fits can already provide useful estimates for the optimal truncation order $L$ in the truncated partial wave analysis. Furthermore, the extracted Legendre-coefficients have turned out to be quantities well-suited for the comparison to energy-dependent PWA-models, since already some statements can then be made on the importance of certain partial-wave interferences in the data.
Therefore, a comprehensive survey of recent polarization-data has been published in the course of this work, describing and showing results of the application and of this useful Legendre-fit method. \newline
The next step in the TPWA of real data consists of the extraction of estimates for the real- and imaginary parts of multiples out of the fitted Legendre-coefficients, using the defining equations (\ref{eq:BilinearEqSystemConclusions}). This analysis has been performed for two combined sets of polarization measurements for the $\pi^{0}$-production (\ref{eq:Pi0ChannelConclusions}), overlapping in different energy-regions. In the $\Delta$-region, i.e. for laboratory-energies from threshold up to $E_{\gamma} = 500 \hspace*{1pt} \mathrm{MeV}$, the mathematically complete set $\left\{ \sigma_{0}, \Sigma, T, P, F  \right\}$ \cite{Hornidge:2013,LeukelPhD,Leukel:2001,Belyaev:1983,Schumann:2015} has been analyzed. Within the so-called second resonance-region, reaching from $E_{\gamma} = 500 \hspace*{1pt} \mathrm{MeV}$ up to roughly $E_{\gamma} = 900 \hspace*{1pt} \mathrm{MeV}$, a combined set of seven polarization observables has been fitted which already form a mathematically over-complete set: $\left\{ \sigma_{0}, \Sigma, T, P, E, G, H  \right\}$ \cite{Adlarson:2015,GRAAL,Hartmann:2014,Hartmann:2015,Gottschall:2014,Thiel:2012,Thiel:2016}. \newline
Within the $\Delta$-region, for the first time a unique solution for the $S$- and $P$-wave multipoles has been extracted from the observables (\ref{eq:ExampleCompleteSetConclusions}) in a completely model-independent way, using a strict truncation at $L=1$. Errors have been determined using the bootstrap. The results are shown and compared to energy-dependent PWA-models in Figure \ref{fig:ConclusionPlotUniqueSolDeltaRegionLmax1}. Within the 1980s, Grushin and collaborators \cite{Grushin} have also obtained quite model-independent single-energy multipoles in the $\Delta$-region. However, their datasets differed from those analyzed in this thesis. Furthermore, slight model-assumptions have been made for the initial parameter-configurations in the Grushin-fits. The Monte Carlo-method used in this work is completely free of such assumptions. \newline
Judging from the results of the initial Legendre-fits, as well as the $\chi^{2}/\mathrm{ndf}$ of the actual multipole-fits, some small influence of the $D$-wave was inferred even in the $\Delta$-region.
\clearpage

\begin{figure}[h]
\centering
\begin{overpic}[width=0.325\textwidth]{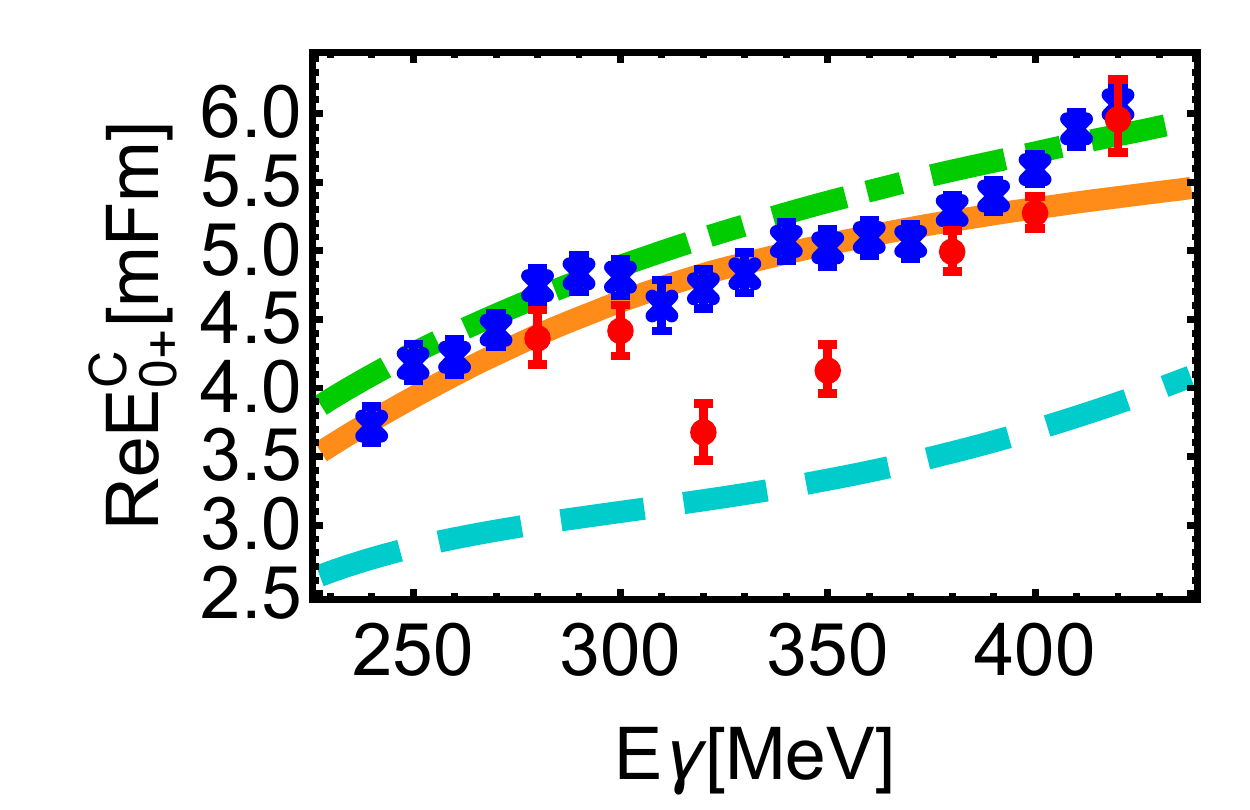}
 \end{overpic}
\begin{overpic}[width=0.325\textwidth]{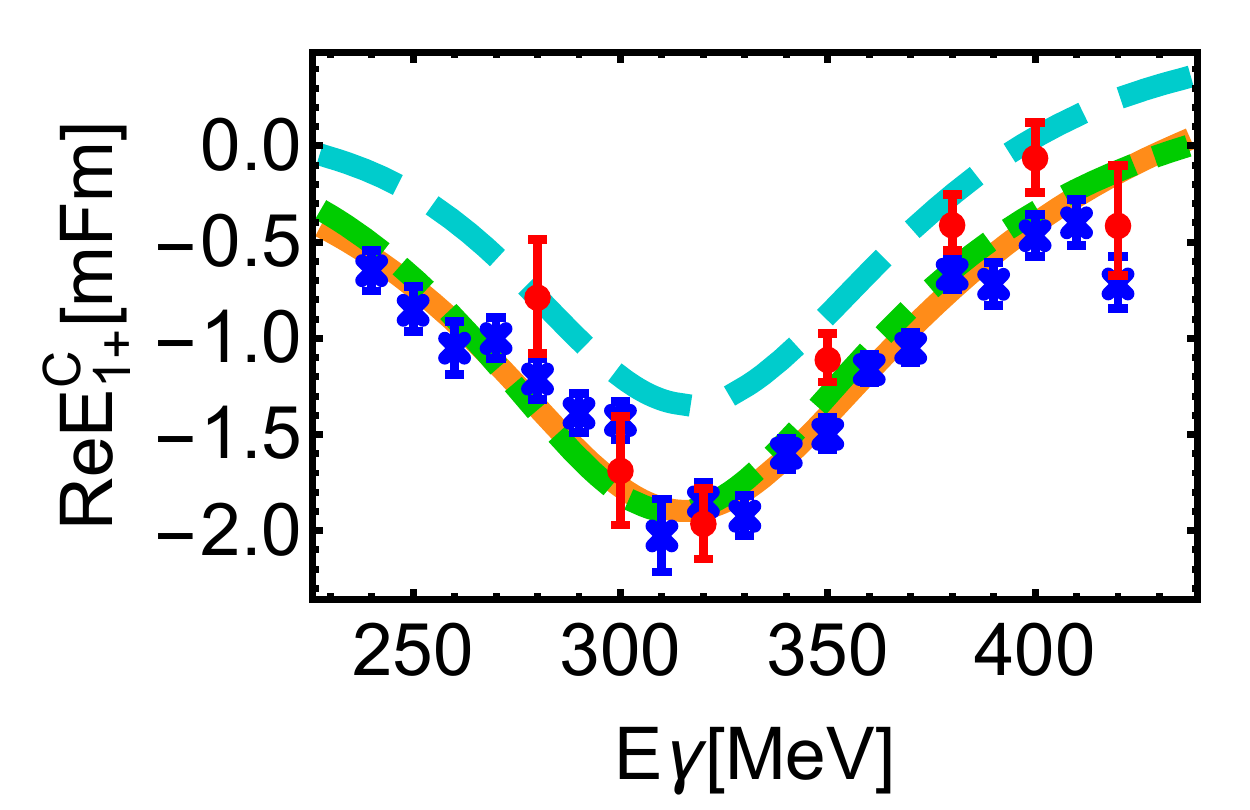}
 \end{overpic}
\begin{overpic}[width=0.325\textwidth]{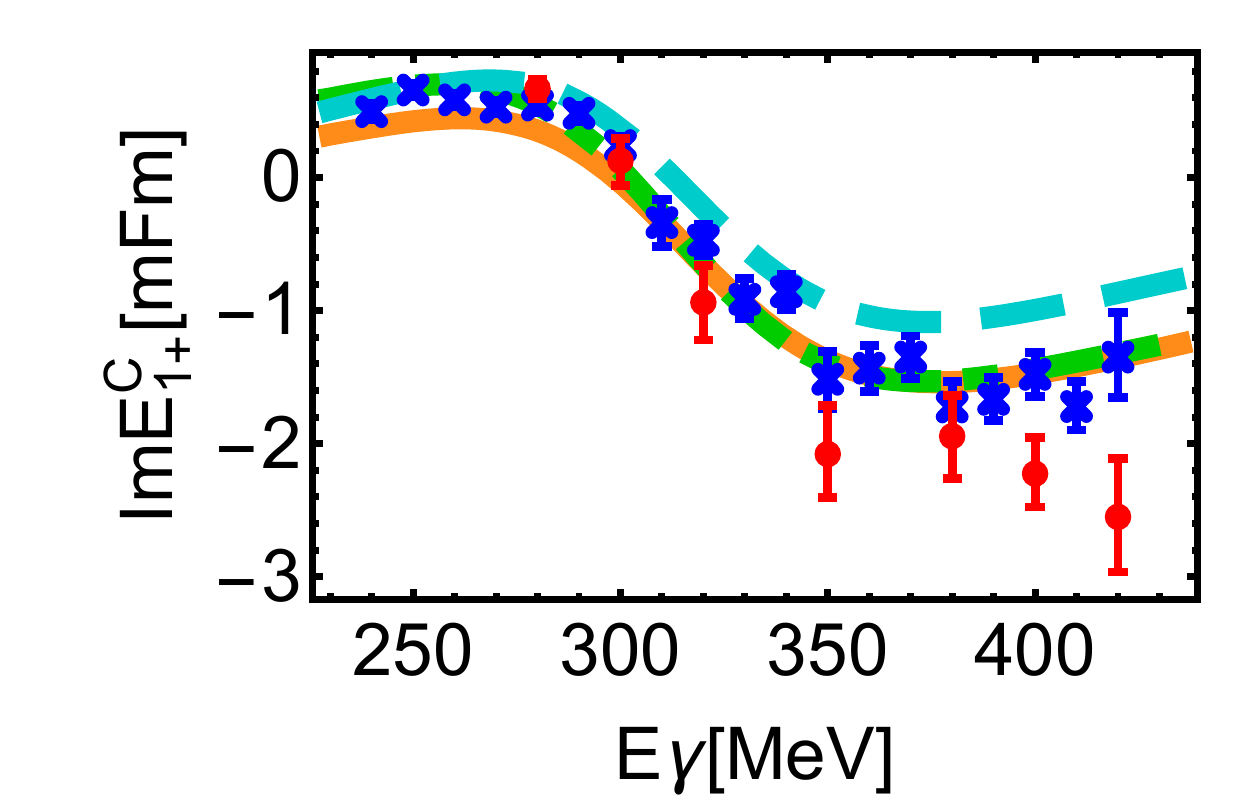}
 \end{overpic} \\
\begin{overpic}[width=0.325\textwidth]{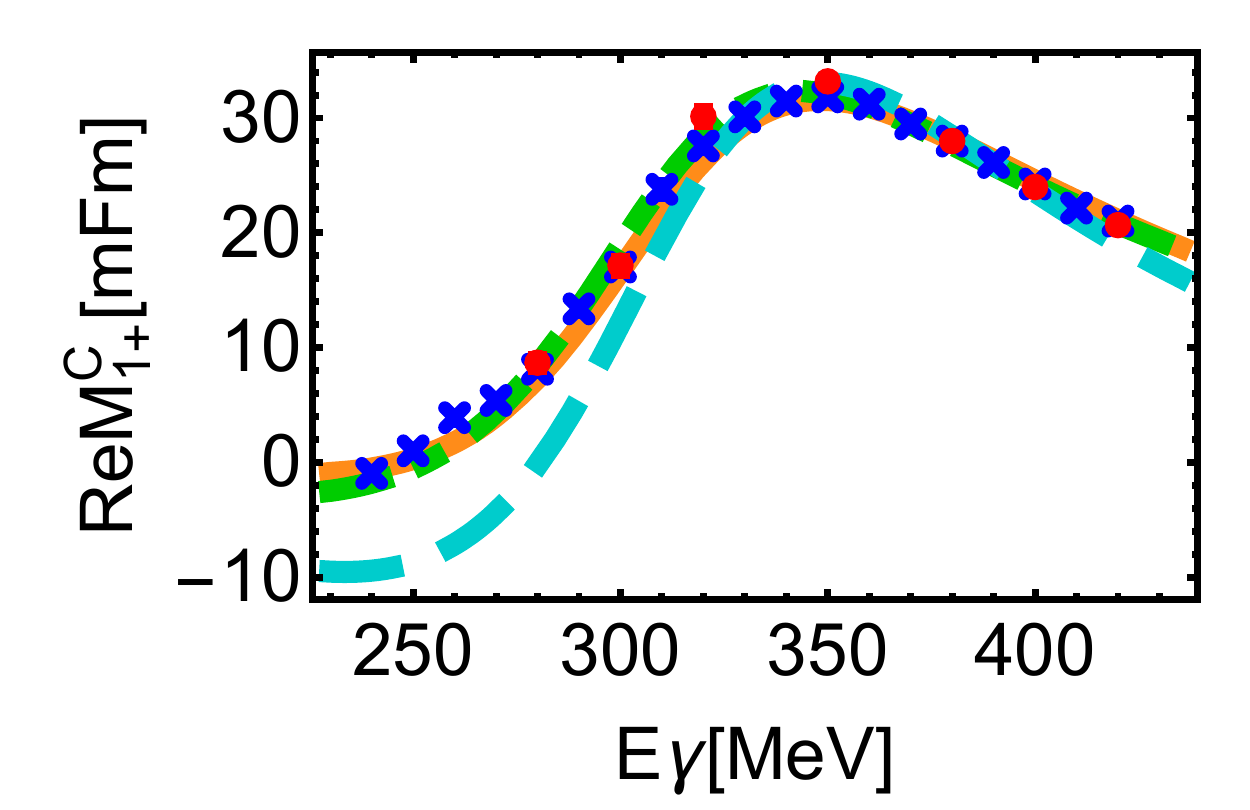}
 \end{overpic}
\begin{overpic}[width=0.325\textwidth]{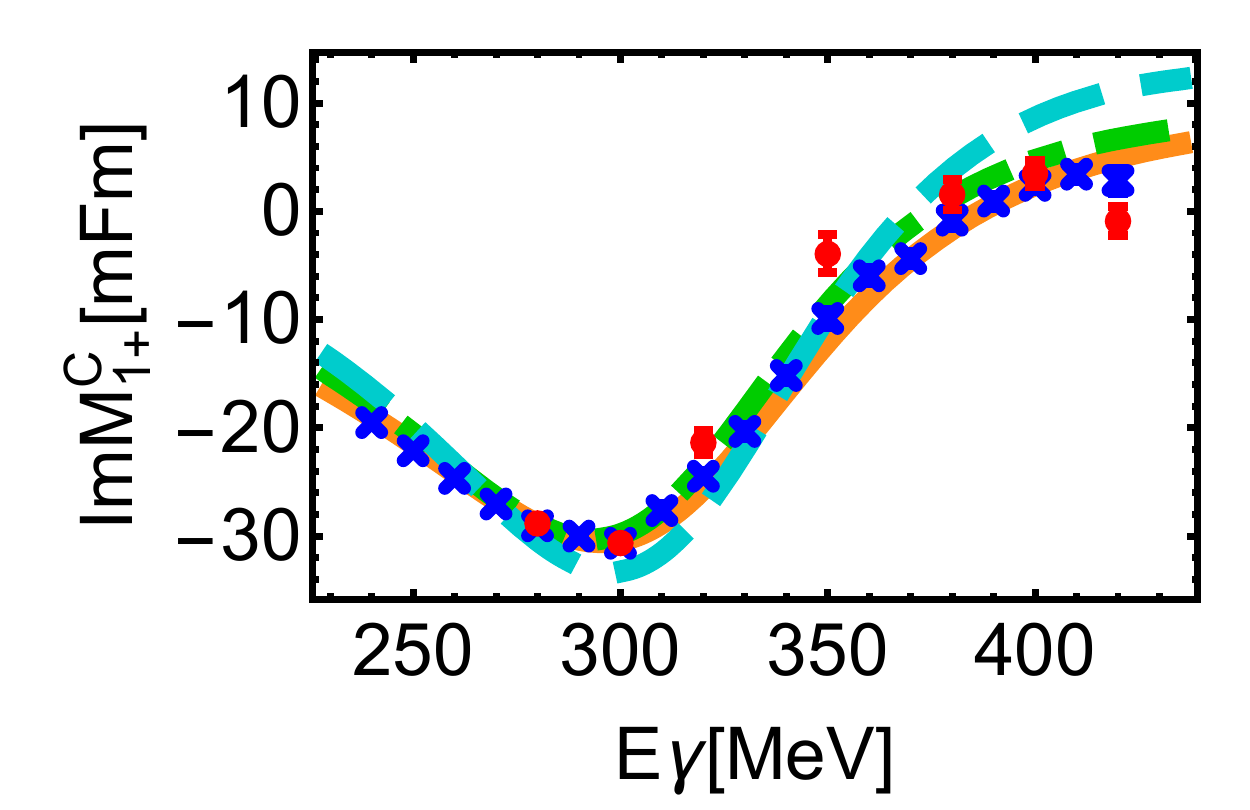}
 \end{overpic}
\begin{overpic}[width=0.325\textwidth]{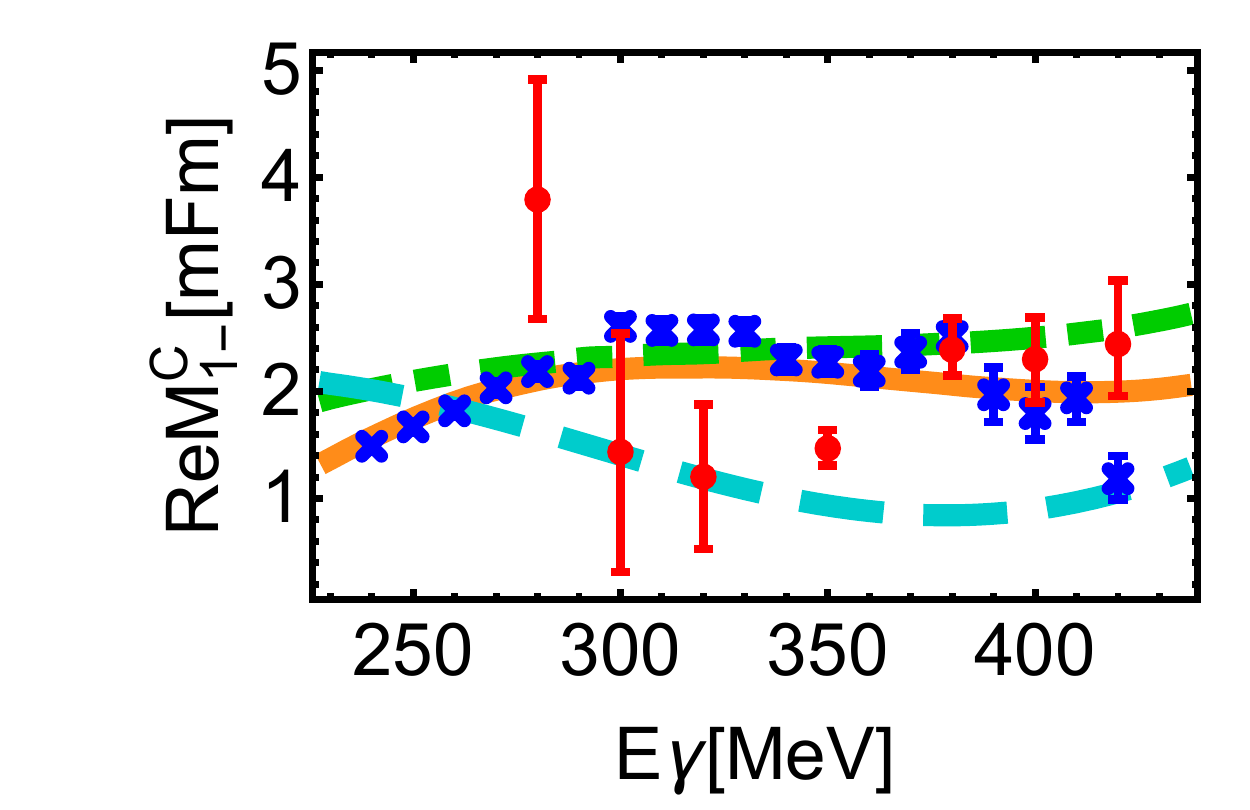}
 \end{overpic} \\
\begin{overpic}[width=0.325\textwidth]{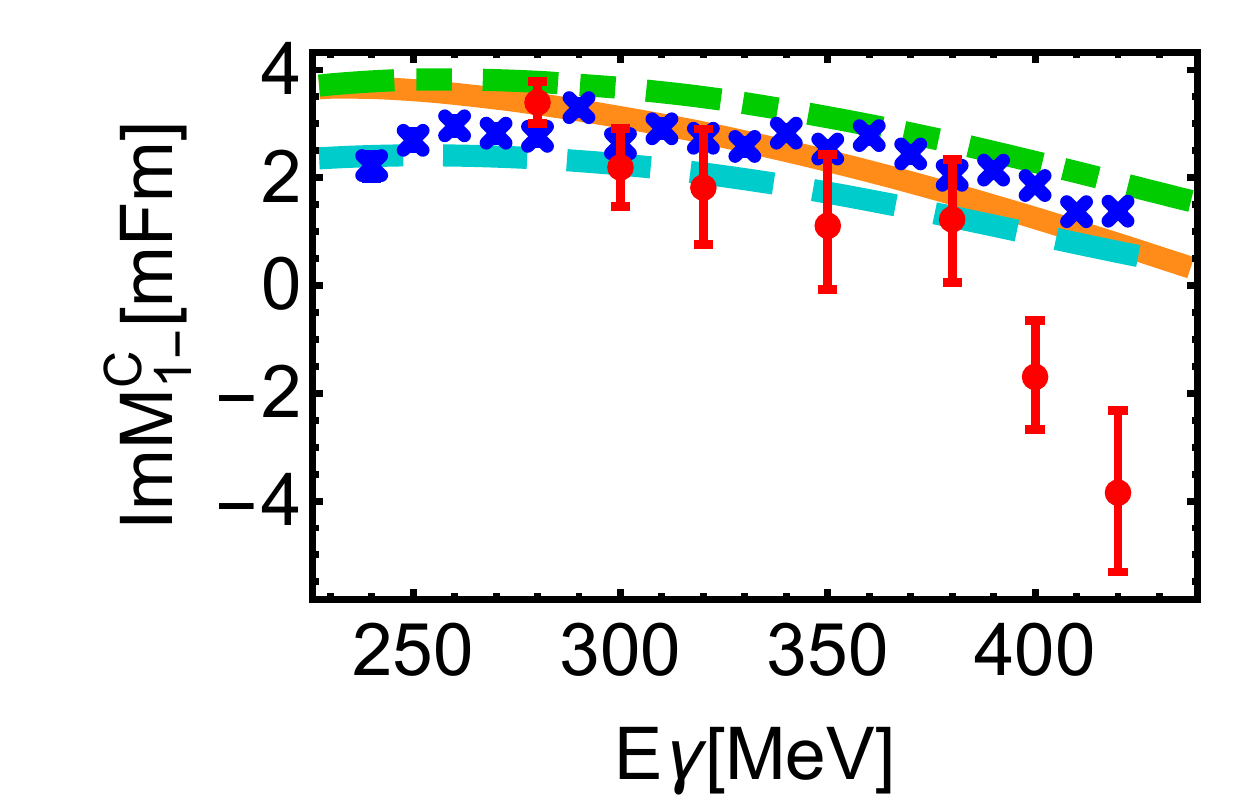}
 \end{overpic}
 \caption[A comparison of the results of the bootstrap-analysis performed in the $\Delta$-resonance region, with $D$-waves fixed to SAID CM12. Five observables $\left\{ \sigma_{0}, \Sigma, T, P, F \right\}$ were fitted, with SAID-pseudodata employed for the recoil asymmetry $P$. This is a re-printing for the conclusions-section.]{Shown are two fits using $\ell_{\mathrm{max}} = 2$, with $D$-waves fixed to the SAID-solution CM12 \cite{WorkmanEtAl2012ChewMPhotoprod,SAID}. The two cases comprise two different data-scenarios. A fit is shown using only real measured data for the five observables $\left\{ \sigma_{0}, \Sigma, T, P, F \right\}$ \cite{Hornidge:2013,LeukelPhD,Leukel:2001,Belyaev:1983,Schumann:2015} (red dots), as well as real data for $\left\{ \sigma_{0}, \Sigma, T, F \right\}$ \cite{Hornidge:2013,LeukelPhD,Leukel:2001,Schumann:2015} combined with SAID-pseudodata for $P$ \cite{RonPrivComm,WorkmanEtAl2012ChewMPhotoprod} (blue crosses). Error-bars indicate statistical uncertainties determined from the bootstrap. \newline The results are compared to the PWA-solutions SAID CM12 (orange solid line) \cite{SAID}, BnGa 2014\_02 (cyan dashed line) \cite{BoGa} and MAID2007 (green dash-dotted line) \cite{MAID}.}
 \label{fig:ConclusionPlotUniqueSolDeltaRegionPSAID}
\end{figure}

However, a completely model-independent TPWA-fit varying all multipoles up to $L=2$ did not result in a well-separated global minimum any more. Instead, $D$-waves have been fixed to values coming from the energy-dependent PWA-solution SAID CM12 (and also, for comparison, to other models). \newline
This inclusion of the $D$-waves improved the $\chi^{2}$ slightly and also showed some modifications of the lower, fitted partial waves. Such modifications are caused by the interferences between high and low partial waves in the data. The results are shown in Figure \ref{fig:ConclusionPlotUniqueSolDeltaRegionLmax1}. \newline
Concerning the analysis of the $5$ observables in the $\Delta$-resonance region, it has to be mentioned that the data for the observable $P$ \cite{Belyaev:1983} had poor statistics compared to the other measurements. This resulted in a relatively limited energy-binning for the single-energy fits shown in Figure \ref{fig:ConclusionPlotUniqueSolDeltaRegionLmax1}, as well as in relatively large statistical error-bars for the fitted multipoles, especially for $M_{1-}$. Therefore, a hypothetical scenario has been investigated where SAID-pseudodata \cite{RonPrivComm,WorkmanEtAl2012ChewMPhotoprod} with $5\%$-errors for the observable $P$ have been fitted in combination with the measured data for the remaining observables. The improvement on the energy-binning and the statistical errors facilitated by more precise $P$-data is immense and it is illustrated by a comparison shown in Figure \ref{fig:ConclusionPlotUniqueSolDeltaRegionPSAID}. \clearpage
However, also in this case it has not been possible to fit out the $D$-waves model-independently and instead they were again adjusted to the SAID-model \cite{SAID}. \newline
The following phenomenon, here seen in the $\Delta$-region, has been a repeated feature of realistic TPWA-fits. In case one fits one order in $L$ too low, missing some interference-contributions caused by small partial waves at $L+1$, which are however still important, a global minimum can mostly be obtained. Introducing then the next higher order into the TPWA and doing a completely model-independent analysis, varying the (small) higher partial waves freely, often results in the loss of uniqueness. The only reasonable way out then consists of introducing mild model-dependences into the analysis, for instance by fixing some of the higher (or sometimes lower) partial waves to values from energy-dependent models. In this case, mathematically (over-) complete sets of observables can very well fix all the remaining multipoles reliably. \newline
The exact same thing happened for the multipole-analysis of the set $\left\{ \sigma_{0}, \Sigma, T, P, E, G, H  \right\}$ \cite{Adlarson:2015,GRAAL,Hartmann:2014,Hartmann:2015,Gottschall:2014,Thiel:2012,Thiel:2016} in the second resonance-region. A completely model-independent TPWA ,varying all multipoles up to $L=2$, resulted in a global minimum, but not in a satisfactory fit quality. The $F$-waves had to be fixed to the model-solution BnGa 2014\_02 \cite{BoGa}, in order to arrive at a unique solution for the varied lower multipoles. \newline

In summary, this thesis features a treatment of the problem of discrete ambiguities in a TPWA for pseudoscalar meson photoproduction, reviewing and further elaborating upon the algebraic Ansatz by Omelaenko \cite{Omelaenko}. The mathematical considerations have lead to the postulation of complete sets of 5 observables. These purely algebraic results have been fully confirmed in numerical analyses of synthetic, truncated MAID model-data \cite{MAID2007,MAID}. The possibility to obtain completeness in a TPWA with even just $4$ observables has been discovered numerically by Tiator towards the end of this thesis-project, but has not yet been understood mathematically. \newline
For the analysis of real photoproduction data, the bootstrap has been applied to TPWA-fitting. For specific examples of data in $\pi^{0}$-photoproduction, useful results have been obtained. These consisted either of obtaining unique solutions for multipoles up to a phase in a completely model-independent way, or under mild model-assumptions for some higher multipoles.

\subsection{Suggestions for future research} \label{sec:SuggestionsForFuture}

This thesis, as in any scientific work, has raised new questions in the course of the solution of the initially proposed problem. Here, we list a few branches that could be followed, building on the foundation already laid out.

\begin{itemize}
 \item[1.] \underline{Mathematical understanding of the complete sets of 4} \newline
 Attempts have already been made to understand why complete sets of $4$ observables can be found numerically, but have not yet been lead to a conclusive answer. It is not even clear that a parametrization of the amplitudes in terms of Omelaenko-roots \cite{Omelaenko} is really the correct one to answer this question. The problem consists of the fact that none of the $196$ complete sets of $4$ is diagonal in the transversity-basis. Furthermore, none of the $196$ complete sets of $4$ seem to be simultaneously diagonalizable at all. Thus, following a logic similar to Omelaenko's may be an entirely false approach. \newline
 In case one investigates the Omelaenko-parametrizations of the complete sets of $4$ however, it is interesting to see that subsets of the $4$ observables are now invariant under a richer variety of symmetry-transformations acting on the roots, compared to just complex conjugations. Some possibilities are just sign-reflections
\begin{equation}
 \alpha_{i} \longrightarrow - \alpha_{i} \mathrm{,} \label{eq:SignChangeSymmetry}
\end{equation}
or maybe even exchange-type symmetries
\begin{equation}
 \alpha_{i} \longrightarrow \beta_{i} \mathrm{,} \hspace*{5pt} \beta_{j} \longrightarrow \alpha_{j} \mathrm{.}  \label{eq:ExChangeSymmetry}
\end{equation}
 It is then interesting to see that for specific examples of complete sets of $4$, it is often the case that $2$ or even $3$ observables are invariant under richer symmetries such as (\ref{eq:SignChangeSymmetry}) or (\ref{eq:ExChangeSymmetry}), while the fourth one usually resolves them, as one should expect. \newline
 However, this observation has been made only for a few hand-selected examples and is far from a satisfying solution to this potentially quite complicated problem.
 \item[2.] \underline{Obtaining a deeper mathematical understanding of ambiguities in PWA} \newline
 As mentioned in the introduction to chapter \ref{chap:Omelaenko}, during the development of this thesis, investigations have been made \cite{MyPhasePaper} on the nature of ambiguities in truncated partial wave analyses of scalar scattering processes, described by one amplitude $A(W,\theta)$. Here, the differential cross section $\sigma_{0} = \left| A (W,\theta) \right|^{2}$ is unchanged with respect to an energy- and angle-dependent rotation of the amplitude
 \begin{equation}
 A(W,\theta) \rightarrow \tilde{A}(W,\theta) := e^{i \phi (W, \theta)} A(W,\theta) \mathrm{.} \label{eq:ContAmbConclusions}
 \end{equation}
 This is called continuum ambiguity. Once however a truncation of the partial wave series for $A(W,\theta)$ is fitted to data, so called discrete ambiguities, analogues of the Omelaenko-ambiguities in photoproduction, are known to appear. It has been found that any transformation of the form (\ref{eq:ContAmbConclusions}) re-mixes the partial waves linearly, but generally as an infinite series \cite{AlfredPhasePaper,MyPhasePaper}. Discrete ambiguities have been identified as a kind of sub-class in the general rotations (\ref{eq:ContAmbConclusions}), with the property that they exhaust all possibilities to rotate a truncated amplitude again into a truncated one \cite{MyPhasePaper}. \newline
 This connection has been worked out using a few new ideas for numerical techniques and it would be interesting in which way the findings for the scalar amplitudes generalize to a spin-reaction such as photoproduction, and in which way they may not.
 \item[3.] \underline{Application of analysis-techniques to further datasets} \newline
 Real photoproduction data have been investigated in this work for the example of the $\pi^{0}$-photoproduction channel. This has a reason, since data in this channel usually have the best statistics and common phase-space coverage. \newline
 However, there exist data on many more reactions, such as production of charged poins $\gamma p \rightarrow \pi^{+} n$, eta mesons $\gamma N \rightarrow \eta N$, or kaons $\gamma p \rightarrow K^{+} \Lambda$, which can be analyzed as well. In particular, the $K \Lambda$-channel seems attractive since the $\Lambda$ is self-analyzing \cite{Sandorfi:2010uv}. \newline
 Also, new and improved data on the pion-channels may of course improve the results shown in this work.
 \item[4.] \underline{Numeric optimization of bootstrap-TPWA fits} \newline
 The fit-methods employed in this work are focusing heavily on sampling, either in the Monte-Carlo routine for the generation of initial multipole-parameters, as well as in the application of the bootstrap, which is based on sampling from the data itself. First tests of the methods have been done successfully with MATHEMATICA \cite{Mathematica8,Mathematica11,MathematicaLanguage,MathematicaBonnLicense}, but it has to be clear that once the statistics and phase-space coverage of the data as well as the truncation-order $L$ rise, the analyses will become more and more numerically demanding. Then, it might become necessary to adapt the full analysis, or just parts of it, to a different programming language which makes it possible to port the whole method to cluster-computing. What is helpful is that both the bootstrap and the applied Monte Carlo-method for sampling initial parameters, are highly suitable for parallelization. This fact has also been exploited in the MATHEMATICA-routines. 
\end{itemize}

\clearpage
\thispagestyle{empty}
\textcolor{white}{Hallo Welt :-)}
\clearpage

\appendix
\appendixpage
\addappheadtotoc
\renewcommand{\thesubsection}{\Alph{subsection}}


\subsection{Matrix representations of polarization observables}\label{sec:ObservableAlgebra}

In this appendix, the most important properties as well as explicit representations of the $4 \times 4$-matrices defining the $16$ polarization observables of pseudoscalar meson
photoproduction are collected. Matrix-representations for observables written in helicity-, transversity- and CGLN-amplitudes are given (cf. section \ref{subsec:PhotoproductionObs}). The bulk of this material has already been collected in the appendices of the thesis \cite{MyDiplomaThesis}.

\subsubsection{Representations for helicity- and transversity amplitudes} \label{subsec:HelTrGammaReps}

As described in section \ref{subsec:PhotoproductionObs}, the $16$ polarization observables of pseudoscalar meson photoproduction take a particularly simple form once the helicity- or transversity basis is chosen for spin-quantization. For instance, using transversity amplitudes they can be written as bilinear forms
\begin{equation}
 \check{\Omega}^{\alpha} = \frac{1}{2} \left< H \right| \Gamma^{\alpha} \left| H \right> = \frac{1}{2} \sum_{i,j=1}^{4} H_{i}^{\ast} \Gamma^{\alpha}_{ij} H_{j} \mathrm{,} \hspace*{10pt} \alpha = 1,\ldots,16 \mathrm{,} \label{eq:BilHelPrFormAppendix}
\end{equation}
using the $16$ Dirac $\Gamma$-matrices (observables in the transversity basis are written using $\tilde{\Gamma}$-matrices). We use this section to mention the representation-independent properties of these matrices which are important for the complete experiment and also to list the explicit representations used by Chiang/Tabakin \cite{ChTab} as well as this work. \newline
The 16 $ 4 \times 4 $ $ \Gamma $ matrices can be defined \cite{ChTab} as hermitean versions of the known Dirac matrices:

\begin{equation}
\Gamma^{\alpha = 1 \mathrm{,} \ldots \mathrm{,} 16} = \mathbbm{1} \mathrm{,} \gamma^{0} \mathrm{,} \vec{\gamma} \mathrm{,} i \sigma^{0x} \mathrm{,} i \sigma^{0y} \mathrm{,} i \sigma^{0z} \mathrm{,} i \sigma^{xy} \mathrm{,} i \sigma^{xz}  \mathrm{,} i \sigma^{zy} \mathrm{,} i \gamma^{5} \gamma^{0} \mathrm{,} i \gamma^{5} \vec{\gamma} \mathrm{,} \gamma^{5} \mathrm{.} \label{eq:GammaDef}
\end{equation}

Here, $ \vec{\gamma} $ refers to $ \gamma $ matrices with spatial indices $ \left( \gamma^{1} \mathrm{,} \gamma^{2} \mathrm{,} \gamma^{3} \right) = \left( \gamma^{x} \mathrm{,} \gamma^{y} \mathrm{,} \gamma^{z} \right) $ and $ \gamma^{5} = i \gamma^{0} \gamma^{x} \gamma^{y} \gamma^{z} $. For definitions of the antisymmetric tensor forms $ \sigma^{\mu \nu} = \frac{i}{2} \left[\gamma^{\mu} \mathrm{,} \gamma^{\nu} \right] $ we refer to the pertinent literature on relativistic quantum mechanics. \newline
We fix the convention that greek indices $ ( \alpha, \beta, \ldots ) $ run from 1 to 16, while latin indices $ (a,b, \ldots ) $ (generally denoting explicit matrix elements) take values from 1 to 4. The general properties of the $ \Gamma $ matrices are (cf. \cite{ChTab}):

\begin{description}

\item {\textbf{(a)}} $ \Gamma^{\alpha} $ are hermitean and unitary, i.e. 
\begin{equation}
(\Gamma^{\alpha})^{\dagger} = \Gamma^{\alpha} \quad \& \quad (\Gamma^{\alpha})^{\dagger} \Gamma^{\alpha} = \mathbbm{1} \quad \alpha = 1, \ldots ,16 \mathrm{.} \label{eq:HermiteanUnitary}
\end{equation}
(The hermitean conjugate $ (\Gamma^{\alpha})^{\dagger} $ is defined as $ (\Gamma^{\alpha})_{ij}^{\dagger}  = (\Gamma^{\alpha}_{ji})^{\ast} $.)

\item {\textbf{(b)}} $ \Gamma^{\alpha} $ are orthogonal under the trace operation:
\begin{equation}
\mathrm{Tr}\left[\Gamma^{\alpha} \Gamma^{\beta}\right] = 4 \delta_{\alpha \beta} \mathrm{.} \label{eq:GammaOrthogonal}
\end{equation}
Here, $\delta_{\alpha \beta}$ is the usual Kronecker-symbol $ \delta_{\alpha \beta} = ( 1, \alpha = \beta \wedge 0, \alpha \not = \beta ) $.

\item {\textbf{(c)}} The $ \Gamma^{\alpha} $ are linearly independent. In combination with \textbf{(b)}, it is seen that they form an orthogonal basis for the vectorspace of complex $ 4 \times 4 $ matrices $ M_{4}(\mathbb{C}) $. Using the orthogonality property, every $ X \in M_{4}(\mathbb{C}) $ can be expanded as
\begin{equation}
X = \sum \limits_{\alpha = 1}^{16}C_{\alpha} \Gamma^{\alpha} \mathrm{,} \label{eq:ExpandMatrix}
\end{equation}

with $ C_{\alpha} = \frac{1}{4} \mathrm{Tr}\left[X \Gamma^{\alpha}\right] $.

\item {\textbf{(d)}} An equivalent statement of the basis-properties expressed in \textbf{(c)} is the validity of the \textit{completeness relation}:
\begin{equation}
\sum \limits_{\alpha =1}^{16} \Gamma^{\alpha}_{ba} \Gamma^{\alpha}_{st} = 4 \delta_{as} \delta_{bt} \mathrm{.} \label{eq:GammaCompleteness}
\end{equation}

Expressing the expansion (\ref{eq:ExpandMatrix}) explicitly in terms of matrix-elements quickly yields this relation.

\item {\textbf{(e)}} Applying the properties \textbf{(c)} and \textbf{(d)} to a product of matrix-elements of arbitrary matrices $A$, $B \in M_{4}(\mathbb{C})$, it can be seen that the $ \Gamma^{\alpha} $ have to satisfy the \textit{Fierz identities} (see \cite{ChTab}):
\begin{equation}
\Gamma^{\alpha}_{ij} \Gamma^{\beta}_{st} = \sum \limits_{\delta, \eta}^{} C_{\delta \eta}^{\alpha \beta} \Gamma^{\delta}_{it} \Gamma^{\eta}_{sj} \quad \mathrm{where} \quad C_{\delta \eta}^{\alpha \beta} = \frac{1}{16} \mathrm{Tr} \left[ \Gamma^{\delta} \Gamma^{\alpha} \Gamma^{\eta} \Gamma^{\beta} \right] \mathrm{.} \label{eq:GammaFierz}
\end{equation}
A proof of these identities can be found in many places in the literature. Reference \cite{MyDiplomaThesis} contains a proof by Nishi \cite{Nishi}. \newline
The Fierz identities play a vital role in the solution of the complete experiment problem for the full photoproduction amplitudes by Chiang and Tabakin (cf. section \ref{sec:CompExpsFullAmp} and reference \cite{ChTab}).
\end{description}

It is very important to note that all of the properties above remain untouched under a unitary change of basis (ref. \cite{ChTab})

\begin{equation}
\Gamma^{\alpha} \quad \longrightarrow \quad U^{\dagger} \Gamma^{\alpha} U \mathrm{,} \label{eq:UnitaryTr} 
\end{equation}

using a unitary matrix $ U $. The transversity transformation $ U^{(4)} $ defined in sec \ref{subsec:PhotoproductionAmpl} is unitary and leads to $ \tilde{\Gamma} $ matrices of the form (\ref{eq:UnitaryTr}) (with $ U = (U^{(4)})^{\dagger} $). Thus the $ \Gamma $- and $ \tilde{\Gamma} $ matrices satisfy the same properties \textbf{(a)} to \textbf{(d)}.

Chiang and Tabakin list explicit $ \tilde{\Gamma} $ matrices already transformed by means of (\ref{eq:UnitaryTr}), using $ U =  (U^{(4)})^{\dagger} $ (see App. A of ref. \cite{ChTab}). We reverse this $ U^{(4)} $ transformation in order to obtain our consistent expressions for the $\Gamma$-matrices. We list the explicit matrices used in this work in the remainder of this section, in equations (\ref{eq:GammaD}) to (\ref{eq:GammaTildePL}). \newline
It can be seen in both cases of $\Gamma$- and $\tilde{\Gamma}$-matrices that the resulting matrices fall into four classes with four members each corresponding to their shape (cf. \cite{ChTab}). The shapes are diagonal (D), right-parallelogram (PR), antidiagonal (AD) and left-parallelogram (PL). We indicate this fact in the used notation (equations (\ref{eq:GammaD}) to (\ref{eq:GammaTildePL})).

\clearpage
\newpage

\begin{align}
\Gamma_{\mathrm{D}} = 
\left[ \begin{array}{cccc} 
a & 0 & 0 & 0 \\
0 & b & 0 & 0 \\
0 & 0 & c & 0 \\
0 & 0 & 0 & d \end{array} \right] 
\qquad \mathrm{,}& \qquad 
\begin{array}{ccccc}
\mathrm{ } & a & b & c & d \\
\hline
\Gamma^{1} & +1 & +1 & +1 & +1 \\
\Gamma^{9} & +1 & -1 & +1 & -1 \\
\Gamma^{2} & +1 & +1 & -1 & -1 \\
\Gamma^{15} & -1 & +1 & +1 & -1 \\
\end{array} \label{eq:GammaD}
\\
\Gamma_{\mathrm{PR}} =
\left[ \begin{array}{cccc} 
0 & 0 & a & 0 \\
0 & 0 & 0 & b \\
c & 0 & 0 & 0 \\
0 & d & 0 & 0 \end{array} \right] 
\qquad \mathrm{,}& \qquad 
\begin{array}{ccccc}
\mathrm{ } & a & b & c & d \\
\hline
\Gamma^{12} & -i & -i & +i & +i \\
\Gamma^{5} & +i & -i & -i & +i \\
\Gamma^{8} & -1 & +1 & -1 & +1 \\
\Gamma^{16} & +1 & +1 & +1 & +1 \\
\end{array} \label{eq:GammaPR}
\\
\Gamma_{\mathrm{AD}} =
\left[ \begin{array}{cccc} 
0 & 0 & 0 & a \\
0 & 0 & b & 0 \\
0 & c & 0 & 0 \\
d & 0 & 0 & 0 \end{array} \right] 
\qquad \mathrm{,}& \qquad 
\begin{array}{ccccc}
\mathrm{ } & a & b & c & d \\
\hline
\Gamma^{4} & +1 & -1 & -1 & +1 \\
\Gamma^{3} & +i & +i & -i & -i \\
\Gamma^{7} & +i & -i & +i & -i \\
\Gamma^{6} & -1 & -1 & -1 & -1 \\
\end{array} \label{eq:GammaAD}
\\
\Gamma_{\mathrm{PL}} =
\left[ \begin{array}{cccc} 
0 & a & 0 & 0 \\
b & 0 & 0 & 0 \\
0 & 0 & 0 & c \\
0 & 0 & d & 0 \end{array} \right] 
\qquad \mathrm{,}& \qquad 
\begin{array}{ccccc}
\mathrm{ } & a & b & c & d \\
\hline
\Gamma^{10} & +i & -i & +i & -i \\
\Gamma^{11} & -1 & -1 & -1 & -1 \\
\Gamma^{14} & +i & -i & -i & +i \\
\Gamma^{13} & -1 & -1 & +1 & +1 \\
\end{array} \label{eq:GammaPL}
\end{align}

\begin{align}
\tilde \Gamma_{\mathrm{D}} = \tilde \Gamma_{\mathrm{S}} = 
\left[ \begin{array}{cccc} 
a & 0 & 0 & 0 \\
0 & b & 0 & 0 \\
0 & 0 & c & 0 \\
0 & 0 & 0 & d \end{array} \right] 
\qquad \mathrm{,}& \qquad 
\begin{array}{ccccc}
\mathrm{ } & a & b & c & d \\
\hline
\tilde \Gamma^{1} & +1 & +1 & +1 & +1 \\
\tilde \Gamma^{4} & +1 & +1 & -1 & -1 \\
\tilde \Gamma^{10} & -1 & +1 & +1 & -1 \\
\tilde \Gamma^{12} & -1 & +1 & -1 & +1 \\
\end{array} \label{eq:GammaTildeD}
\\
\tilde \Gamma_{\mathrm{PR}} = \tilde \Gamma_{\mathrm{BT}} = 
\left[ \begin{array}{cccc} 
0 & 0 & a & 0 \\
0 & 0 & 0 & b \\
c & 0 & 0 & 0 \\
0 & d & 0 & 0 \end{array} \right] 
\qquad \mathrm{,}& \qquad 
\begin{array}{ccccc}
\mathrm{ } & a & b & c & d \\
\hline
\tilde \Gamma^{3} & -i & -i & + i & +i \\
\tilde \Gamma^{5} & +1 & -1 & +1 & -1 \\
\tilde \Gamma^{9} & +1 & +1 & +1 & +1 \\
\tilde \Gamma^{11} & +i & -i & -i & +i \\
\end{array} \label{eq:GammaTildePR}
\\
\tilde \Gamma_{\mathrm{AD}} = \tilde \Gamma_{\mathrm{BR}} = 
\left[ \begin{array}{cccc} 
0 & 0 & 0 & a \\
0 & 0 & b & 0 \\
0 & c & 0 & 0 \\
d & 0 & 0 & 0 \end{array} \right] 
\qquad \mathrm{,}& \qquad 
\begin{array}{ccccc}
\mathrm{ } & a & b & c & d \\
\hline
\tilde \Gamma^{14} & -1 & +1 & +1 & -1 \\
\tilde \Gamma^{7} & -i & -i & +i & +i \\
\tilde \Gamma^{16} & +i & -i & +i & -i \\
\tilde \Gamma^{2} & +1 & +1 & +1 & +1 \\
\end{array} \label{eq:GammaTildeAD}
\\
\tilde \Gamma_{\mathrm{PL}} = \tilde \Gamma_{\mathrm{TR}} = 
\left[ \begin{array}{cccc} 
0 & a & 0 & 0 \\
b & 0 & 0 & 0 \\
0 & 0 & 0 & c \\
0 & 0 & d & 0 \end{array} \right] 
\qquad \mathrm{,}& \qquad 
\begin{array}{ccccc}
\mathrm{ } & a & b & c & d \\
\hline
\tilde \Gamma^{6} & -1 & -1 & +1 & +1 \\
\tilde \Gamma^{13} & +i & -i & -i & +i \\
\tilde \Gamma^{8} & -i & +i & -i & +i \\
\tilde \Gamma^{15} & -1 & -1 & -1 & -1 \\
\end{array} \label{eq:GammaTildePL}
\end{align}

\clearpage

\subsubsection{Representations for CGLN-amplitudes} \label{subsec:CGLNMatrixReps}

Within section \ref{subsec:PhotoproductionObs} in the main text, quite involved formulas were mentioned for the $16$ polarization observables written in terms of CGLN-amplitudes $\left\{F_{i} (W,\theta), i = 1,\ldots,4\right\}$. However, these longer expressions somewhat disguise the fact that when written in the CGLN-basis, observables are also just bilinear
hermitean forms
\begin{equation}
 \check{\Omega}^{\alpha} = \frac{1}{2} \left< F \right| \hat{A}^{\alpha} \left| F \right> = \frac{1}{2} \sum_{i,j=1}^{4} F_{i}^{\ast} \hat{A}^{\alpha}_{ij} F_{j} \mathrm{,} \hspace*{10pt} \alpha = 1,\ldots,16 \mathrm{,} \label{eq:BilCGLNPrFormAppendix}
\end{equation}
defined in terms of a set of $16$ generally angular-dependent matrices $\hat{A}^{\alpha} (\theta)$. The $\hat{A}^{\alpha}$ can be reached from the algebra of $\Gamma$-matrices described in the previous section by means of the transformation $\hat{T}_{H \rightarrow F}$ connecting helicity- and CGLN-amplitudes, i.e. $\left|H\right> = \hat{T}_{H \rightarrow F} \left| F \right>$ with
\begin{small}
\begin{equation}
\hat{T}_{H \rightarrow F} = \left[ \begin{array}{c|c|c|c}
0 & 0 & \frac{i}{\sqrt{2}} \sin (\theta) \sin \left( \frac{\theta}{2} \right) & - \frac{i}{\sqrt{2}} \sin (\theta) \sin \left( \frac{\theta}{2} \right) \\
\hline
- i \sqrt{2} \sin \left( \frac{\theta}{2} \right) & - i \sqrt{2} \sin \left( \frac{\theta}{2} \right) & - i \sqrt{2} \sin (\frac{\theta}{2}) \cos^{2} \left( \frac{\theta}{2} \right) & - i \sqrt{2} \sin (\frac{\theta}{2}) \cos^{2} \left( \frac{\theta}{2} \right) \\
\hline
0 & 0 & \frac{i}{\sqrt{2}} \sin (\theta) \cos \left( \frac{\theta}{2} \right) & \frac{i}{\sqrt{2}} \sin (\theta) \cos \left( \frac{\theta}{2} \right) \\
\hline
- i \sqrt{2} \cos \left( \frac{\theta}{2} \right) & i \sqrt{2} \cos \left( \frac{\theta}{2} \right) & i \sqrt{2} \cos (\frac{\theta}{2}) \sin^{2} \left( \frac{\theta}{2} \right) & - i \sqrt{2} \cos (\frac{\theta}{2}) \sin^{2} \left( \frac{\theta}{2} \right) \\
       \end{array} \right] \mathrm{.} \label{eq:HelCGLNTrafoAppendix}
\end{equation}
\end{small}
A representation change from the helicity to the CGLN-basis can thus be accomplished via
\begin{equation}
 \Gamma^{\alpha} \rightarrow \hat{A}^{\alpha} = \hat{T}_{H \rightarrow F}^{\dagger} \Gamma^{\alpha} \hat{T}_{H \rightarrow F} \mathrm{.} \label{eq:RepChangeToCGLNBasisAppendices}
\end{equation}
Since $\hat{T}_{H \rightarrow F}$ is invertible, the $\hat{A}^{\alpha}$ are also a basis of $M_{4}(\mathbb{C})$. However, some convenient properties of the $\Gamma$-matrices, most notably their orthogonality (\ref{eq:GammaOrthogonal}) and the fulfillment of the completeness relation (\ref{eq:GammaCompleteness}), are lost once one transforms to the $\hat{A}^{\alpha}$. This is caused by the fact that $\hat{T}_{H \rightarrow F}$ is \textit{not} unitary. \newline
The Fierz relations are in some sense preserved, i.e. the $\hat{A}$-matrices satisfy
\begin{equation}
\hat{A}^{\alpha}_{ij} \hat{A}^{\beta}_{st} = \sum \limits_{\delta, \eta}^{} C_{\delta \eta}^{\alpha \beta} \hat{A}^{\delta}_{it} \hat{A}^{\eta}_{sj} \mathrm{,} \label{eq:AFierz}
\end{equation}
with the coefficients $C_{\delta \eta}^{\alpha \beta}$ still evaluated using the $\Gamma$-matrices as in equation (\ref{eq:GammaFierz}). This directly implies that the Fierz relations for \textit{observables}, which arise by contracting the free matrix-indices appropriately with amplitude-vectors, are fully re\-pre\-sen\-ta\-tion-independent, as they should (cf. references \cite{MyDiplomaThesis}, \cite{ChTab} for listings of these relations). \newline
In Tables \ref{tab:A1toA8} and \ref{tab:A9toA16} we give a set of consistent $\hat{A}$-matrices used in this work. The numbering of the $\hat{A}^{\alpha}$-matrices corresponds to Table \ref{tab:ChTabHelTrObs} and the $\Gamma$-matrices given in the preceding section. We repeat it here for convenience
\begin{align}
 \check{\Omega}^{1} = \sigma_{0} \mathrm{,} \hspace*{5pt} \check{\Omega}^{2} = - \check{C}_{z^{\prime}} \mathrm{,} \hspace*{5pt} \check{\Omega}^{3} = \check{G} &\mathrm{,} \hspace*{5pt} \check{\Omega}^{4} = - \check{\Sigma} \mathrm{,} \hspace*{5pt} \check{\Omega}^{5} = \check{H} \mathrm{,} \hspace*{5pt} \check{\Omega}^{6} = - \check{T}_{x^{\prime}} \mathrm{,} \label{eq:Omega1To6Observables} \\
 \check{\Omega}^{7} = - \check{O}_{z^{\prime}} \mathrm{,} \hspace*{5pt} \check{\Omega}^{8} = \check{L}_{x^{\prime}} \mathrm{,} \hspace*{5pt} \check{\Omega}^{9} = \check{E} &\mathrm{,} \hspace*{5pt} \check{\Omega}^{10} = - \check{T} \mathrm{,} \hspace*{5pt} \check{\Omega}^{11} = \check{F} \mathrm{,} \hspace*{5pt} \check{\Omega}^{12} = \check{P} \mathrm{,} \label{eq:Omega7To12Observables} \\
 \check{\Omega}^{13} = - \check{T}_{z^{\prime}} \mathrm{,} \hspace*{5pt} \check{\Omega}^{14} = \check{O}_{x^{\prime}} &\mathrm{,} \hspace*{5pt} \check{\Omega}^{15} = \check{L}_{z^{\prime}} \mathrm{,} \hspace*{5pt} \check{\Omega}^{16} = - \check{C}_{x^{\prime}} \mathrm{.} \label{eq:Omega13To16Observables}
\end{align}
The algebra described in this appendix, in conjunction with the multipole expansion defined by equations (\ref{eq:MultExpF1}) to (\ref{eq:MultExpF4}) in section \ref{subsec:PhotoproductionAmpl}, forms the backbone of all considerations in this work.

\clearpage

\begin{sidewaystable}
\begin{small}
\hspace*{10pt}
\begin{align}
\hspace*{-30pt}\hat{A}^{1} = \left[ \begin{array}{cccc}
 2 & -2 \cos \left(\theta\right) & 0 & \sin \left(\theta\right)^{2} \\
 -2 \cos \left(\theta\right) & 2 & \sin \left(\theta\right)^{2} & 0 \\
 0 & \sin \left(\theta\right)^{2} & \sin \left(\theta\right)^{2} & \cos \left(\theta\right) \sin \left(\theta\right)^{2} \\
 \sin \left(\theta\right)^{2} & 0 & \cos \left(\theta\right) \sin \left(\theta\right)^{2} & \sin \left(\theta\right)^{2} \end{array} \right]
 \quad &\mathrm{,} \quad 
\hat{A}^{2} = \left[ \begin{array}{cccc}
 -2 \cos \left(\theta\right) & 2 & \sin \left(\theta\right)^{2} & 0 \\
 2 & -2 \cos \left(\theta\right) & 0 & \sin \left(\theta\right)^{2} \\
 \sin \left(\theta\right)^{2} & 0 & 0 & 0 \\
 0 & \sin \left(\theta\right)^{2} & 0 & 0 \end{array} \right] \nonumber \\
 \quad \nonumber \\
 \quad \nonumber \\
\hspace*{-30pt}\hat{A}^{3} = \left[ \begin{array}{cccc}
 0 & 0 & 0 & - i  \sin \left(\theta\right)^{2} \\
 0 & 0 & - i  \sin \left(\theta\right)^{2} & 0 \\
 0 & i  \sin \left(\theta\right)^{2} & 0 & 0 \\
 i  \sin \left(\theta\right)^{2} & 0 & 0 & 0 \end{array} \right]
 \quad &\mathrm{,} \quad 
\hat{A}^{4} = \left[ \begin{array}{cccc}
 0 & 0 & 0 & \sin \left(\theta\right)^{2} \\
 0 & 0 & \sin \left(\theta\right)^{2} & 0 \\
 0 & \sin \left(\theta\right)^{2} & \sin \left(\theta\right)^{2} & \cos \left(\theta\right) \sin \left(\theta\right)^{2} \\
 \sin \left(\theta\right)^{2} & 0 & \cos \left(\theta\right) \sin \left(\theta\right)^{2} & \sin \left(\theta\right)^{2} \end{array} \right] \nonumber \\
 \quad \nonumber \\
 \quad \nonumber \\
\hspace*{-30pt}\hat{A}^{5} = \left[ \begin{array}{cccc}
 0 & 2  i  \sin \left(\theta\right) &  i  \sin \left(\theta\right) & i  \cos \left(\theta\right) \sin \left(\theta\right) \\
 -2  i  \sin \left(\theta\right) & 0 & - i  \cos \left(\theta\right) \sin \left(\theta\right) & - i  \sin \left(\theta\right) \\
 - i  \sin \left(\theta\right) & i  \cos \left(\theta\right) \sin \left(\theta\right) & 0 & 0 \\
 - i  \cos \left(\theta\right) \sin \left(\theta\right) & i  \sin \left(\theta\right) & 0 & 0 \end{array} \right]
 \quad &\mathrm{,} \quad 
\hat{A}^{6} = \left[ \begin{array}{cccc}
 0 & 0 & \sin \left(\theta\right)^{2} & 0 \\
 0 & 0 & 0 & \sin \left(\theta\right)^{2} \\
 \sin \left(\theta\right)^{2} & 0 & \cos \left(\theta\right) \sin \left(\theta\right)^{2} & \sin \left(\theta\right)^{2} \\
 0 & \sin \left(\theta\right)^{2} & \sin \left(\theta\right)^{2} & \cos \left(\theta\right) \sin \left(\theta\right)^{2} \end{array} \right] \nonumber \\
 \quad \nonumber \\
 \quad \nonumber \\
\hspace*{-30pt}\hat{A}^{7} = \left[ \begin{array}{cccc}
 0 & 0 &  i  \sin \left(\theta\right)^{2} & 0 \\
 0 & 0 & 0 & i  \sin \left(\theta\right)^{2} \\
 - i  \sin \left(\theta\right)^{2} & 0 & 0 & 0 \\
 0 & - i  \sin \left(\theta\right)^{2} & 0 & 0 \end{array} \right]
 \quad &\mathrm{,} \quad 
\hat{A}^{8} = \left[ \begin{array}{cccc}
 2 \sin \left(\theta\right) & 0 & \cos \left(\theta\right) \sin \left(\theta\right) & \sin \left(\theta\right) \\
 0 & -2 \sin \left(\theta\right) & -\sin \left(\theta\right) & -\cos \left(\theta\right) \sin \left( \theta\right) \\
 \cos \left(\theta\right) \sin \left(\theta\right) & -\sin \left(\theta\right) & -\sin \left(\theta\right)^{3} & 0 \\
 \sin \left(\theta\right) & -\cos \left(\theta\right) \sin \left(\theta\right) & 0 & \sin \left(\theta\right)^{3} \end{array} \right] \nonumber
\end{align}
\end{small}
\caption{Matrices $ \hat{A}^{\alpha} $ ($ \alpha = 1, \ldots, 8 $), representing observables written in the bilinear-CGLN-product-form $ \check{\Omega}^{\alpha} = \frac{1}{2} \left< F \right| \hat{A}^{\alpha} \left| F \right> $ (cf. \cite{MyDiplomaThesis}).}
\label{tab:A1toA8}
\end{sidewaystable}

\begin{sidewaystable}
\begin{small}
\hspace*{-20pt}
\begin{align}
\hat{A}^{9} = \left[ \begin{array}{cccc}
 -2 & 2 \cos \left(\theta\right) & 0 & -\sin \left(\theta\right)^{2} \\
 2 \cos \left(\theta\right) & -2 & -\sin \left(\theta\right)^{2} & 0 \\
 0 & -\sin \left(\theta\right)^{2} & 0 & 0 \\
 -\sin \left(\theta\right)^{2} & 0 & 0 & 0 \end{array} \right]
 \quad &\mathrm{,} \quad
\hat{A}^{10} = \left[ \begin{array}{cccc}
 0 & 0 &  i  \sin \left(\theta\right) &  i  \cos \left(\theta\right) \sin \left(\theta\right) \\
 0 & 0 & - i  \cos \left(\theta\right) \sin \left(\theta\right) & - i  \sin \left(\theta\right) \\
 - i  \sin \left(\theta\right) & i  \cos \left(\theta\right) \sin \left(\theta\right) & 0 & - i  \sin \left(\theta\right)^{3} \\
 - i  \cos \left(\theta\right) \sin \left(\theta\right) &  i  \sin \left(\theta\right) & i  \sin \left(\theta\right)^{3} & 0 \end{array} \right] \nonumber \\
 \quad \nonumber \\
 \quad \nonumber \\
\hat{A}^{11} = \left[ \begin{array}{cccc}
 0 & 0 & \sin \left(\theta\right) & \cos \left(\theta\right) \sin \left(\theta\right) \\
 0 & 0 & -\cos \left(\theta\right) \sin \left(\theta\right) & -\sin \left(\theta\right) \\
 \sin \left(\theta\right) & -\cos \left(\theta\right) \sin \left(\theta\right) & 0 & 0 \\
 \cos \left(\theta\right) \sin \left(\theta\right) & -\sin \left(\theta\right) & 0 & 0 \end{array} \right]
 \quad &\mathrm{,} \quad
\hat{A}^{12} = \left[ \begin{array}{cccc}
 0 & 2  i  \sin \left(\theta\right) &  i  \sin \left(\theta\right) & i  \cos \left(\theta\right) \sin \left(\theta\right) \\
 -2  i  \sin \left(\theta\right) & 0 & - i  \cos \left(\theta\right) \sin \left(\theta\right) & - i  \sin \left(\theta\right) \\
 - i  \sin \left(\theta\right) & i  \cos \left(\theta\right) \sin \left(\theta\right) & 0 & - i  \sin \left(\theta\right)^{3} \\
 - i  \cos \left(\theta\right) \sin \left(\theta\right) &  i  \sin \left(\theta\right) & i  \sin \left(\theta\right)^{3} & 0 \end{array} \right] \nonumber \\
 \quad \nonumber \\
 \quad \nonumber \\
\hat{A}^{13} = \left[ \begin{array}{cccc}
 0 & 0 & -\cos \left(\theta\right) \sin \left(\theta\right) & -\sin \left(\theta\right) \\
 0 & 0 & \sin \left(\theta\right) & \cos \left(\theta\right) \sin \left(\theta\right) \\
 -\cos \left(\theta\right) \sin \left(\theta\right) & \sin \left(\theta\right) & \sin \left(\theta\right)^{3} & 0 \\
 -\sin \left(\theta\right) & \cos \left(\theta\right) \sin \left(\theta\right) & 0 & -\sin \left(\theta\right)^{3} \end{array} \right]
 \quad &\mathrm{,} \quad
\hat{A}^{14} = \left[ \begin{array}{cccc}
 0 & 0 & - i  \cos \left(\theta\right) \sin \left(\theta\right) & - i  \sin \left(\theta\right) \\
 0 & 0 &  i  \sin \left(\theta\right) & i  \cos \left(\theta\right) \sin \left(\theta\right) \\
 i \cos \left(\theta\right) \sin \left(\theta\right) & - i  \sin \left(\theta\right) & 0 & 0 \\
 i \sin \left(\theta\right) & - i  \cos \left(\theta\right) \sin \left(\theta\right) & 0 & 0 \end{array} \right] \nonumber \\
 \quad \nonumber \\
 \quad \nonumber \\
\hat{A}^{15} = \left[ \begin{array}{cccc}
 -2 \cos \left(\theta\right) & 2 & \sin \left(\theta\right)^{2} & 0 \\
 2 & -2 \cos \left(\theta\right) & 0 & \sin \left(\theta\right)^{2} \\
 \sin \left(\theta\right)^{2} & 0 & \cos \left(\theta\right) \sin \left(\theta\right)^{2} & \sin \left(\theta\right)^{2} \\
 0 & \sin \left(\theta\right)^{2} & \sin \left(\theta\right)^{2} & \cos \left(\theta\right) \sin \left(\theta\right)^{2} \end{array} \right]
 \quad &\mathrm{,} \quad
\hat{A}^{16} = \left[ \begin{array}{cccc}
 2 \sin \left(\theta\right) & 0 & \cos \left(\theta\right) \sin \left(\theta\right) & \sin \left(\theta\right) \\
 0 & -2 \sin \left(\theta\right) & -\sin \left(\theta\right) & -\cos \left(\theta\right) \sin \left(\theta\right) \\
 \cos \left(\theta\right) \sin \left(\theta\right) & -\sin \left(\theta\right) & 0 & 0 \\
 \sin \left(\theta\right) & -\cos \left(\theta\right) \sin \left(\theta\right) & 0 & 0 \end{array} \right] \nonumber
\end{align}
\end{small}
\caption{Matrices $ \hat{A}^{\alpha} $ ($ \alpha = 9, \ldots, 16 $), representing observables written in the bilinear-CGLN-product-form $ \check{\Omega}^{\alpha} = \frac{1}{2} \left< F \right| \hat{A}^{\alpha} \left| F \right> $ (cf. \cite{MyDiplomaThesis}).}
\label{tab:A9toA16}
\end{sidewaystable}

\clearpage

\subsubsection{The Fierz identities} \label{subsec:FierzIdentities}

In this appendix, the Fierz identities described in point \textbf{(e)} of Appendix \ref{subsec:HelTrGammaReps} are listed. It is useful to have a quick reference to these identities, since they play an important role in the solution of the complete experiment according to Chiang and Tabakin \cite{ChTab} (see also section \ref{sec:CompExpsFullAmp}) and furthermore, sometime useful (in-) equalities among the observables can be derived from them. \newline
It can be verified that the algebra of observables shown in equations (\ref{eq:GammaD}) to (\ref{eq:GammaTildePL}) (or equivalently also in Tables \ref{tab:A1toA8} and  \ref{tab:A9toA16}) fulfills the Fierz identities (\ref{eq:GammaFierz}). When written in terms of observables $\check{\Omega}^{\alpha} = \frac{1}{2} \left< H \right| \Gamma^{\alpha} \left| H \right> = \frac{1}{2} \left< F \right| \hat{A}^{\alpha} \left| F \right>$, these identities read
\begin{equation}
\check{\Omega}^{\alpha} \check{\Omega}^{\beta} = \sum \limits_{\delta, \eta}^{} C_{\delta \eta}^{\alpha \beta} \check{\Omega}^{\delta} \check{\Omega}^{\eta} \quad \mathrm{with} \quad C_{\delta \eta}^{\alpha \beta} = \frac{1}{16} \mathrm{Tr} \left[ \Gamma^{\delta} \Gamma^{\alpha} \Gamma^{\eta} \Gamma^{\beta} \right] \mathrm{.} \label{eq:ObservableFierz}
\end{equation}
Chiang and Tabakin \cite{ChTab} find a set of $37$ distinct Fierz identities out of this compact form of the equation. In the course of this work, their identities were checked and found to be consistent by use of a MATHEMATICA program \cite{Mathematica8,Mathematica11,MathematicaLanguage,MathematicaBonnLicense} that computes all the possible traces $C_{\delta \eta}^{\alpha \beta}$, using the matrices listed in (\ref{eq:GammaD}) to (\ref{eq:GammaPL}). \newline
Below, we list all the $37$ identities. We do not formulate the equations in terms of the indexed profile functions $\check{\Omega}^{\alpha}$ (i.e. in the way they are found in reference \cite{ChTab}), but instead replace them with the actual names of the observables, employing the relations (\ref{eq:Omega1To6Observables}) to (\ref{eq:Omega13To16Observables}). \newline

\textbf{Linear-quadratic relations:}

\begin{align}
\sigma_{0}^{2}  &= \frac{1}{4} \Big( \sigma_{0}^{2} + \check{\Sigma}^{2} + \check{T}^{2} + \check{P}^{2} + \check{E}^{2} + \check{G}^{2} + \check{H}^{2} + \check{F}^{2} \nonumber \\
 & \hspace*{22.5pt} + \check{O}_{x^{\prime}}^{2} + \check{O}_{z^{\prime}}^{2} + \check{C}_{x^{\prime}}^{2} + \check{C}_{z^{\prime}}^{2} + \check{T}_{x^{\prime}}^{2} + \check{T}_{z^{\prime}}^{2} + \check{L}_{x^{\prime}}^{2} + \check{L}_{z^{\prime}}^{2}  \Big) \mathrm{,} \tag{L.0} \label{eq:L.0}
\end{align}
{\allowdisplaybreaks
\begin{align}
\sigma_{0} \check{\Sigma}  &= \check{T} \check{P} + \check{T}_{x^{\prime}} \check{L}_{z^{\prime}} - \check{L}_{x^{\prime}} \check{T}_{z^{\prime}} \mathrm{,} \tag{L.tr} \label{eq:L.tr} \\
\sigma_{0} \check{T}  &= \check{\Sigma} \check{P} + \check{C}_{z^{\prime}} \check{O}_{x^{\prime}} - \check{O}_{z^{\prime}} \check{C}_{x^{\prime}} \mathrm{,} \tag{L.br} \label{eq:L.br} \\
\sigma_{0} \check{P}  &= \check{\Sigma} \check{T} + \check{G} \check{F} - \check{H} \check{E} \mathrm{,} \tag{L.bt} \label{eq:L.bt} \\
\sigma_{0} \check{G}  &= \check{F} \check{P} + \check{O}_{z^{\prime}} \check{L}_{z^{\prime}} + \check{O}_{x^{\prime}} \check{L}_{x^{\prime}} \mathrm{,} \tag{L.1} \label{eq:L.1} \\
\sigma_{0} \check{H}  &= - \check{E} \check{P} + \check{O}_{z^{\prime}} \check{T}_{z^{\prime}} + \check{O}_{x^{\prime}} \check{T}_{x^{\prime}} \mathrm{,} \tag{L.2} \label{eq:L.2} \\
\sigma_{0} \check{E}  &= - \check{H} \check{P} + \check{C}_{z^{\prime}} \check{L}_{z^{\prime}} + \check{C}_{x^{\prime}} \check{L}_{x^{\prime}} \mathrm{,} \tag{L.3} \label{eq:L.3} \\
\sigma_{0} \check{F}  &= \check{G} \check{P} + \check{C}_{z^{\prime}} \check{T}_{z^{\prime}} + \check{C}_{x^{\prime}} \check{T}_{x^{\prime}} \mathrm{,} \tag{L.4} \label{eq:L.4} \\
\sigma_{0} \check{O}_{x^{\prime}}  &= \check{C}_{z^{\prime}} \check{T} + \check{G} \check{L}_{x^{\prime}} + \check{H} \check{T}_{x^{\prime}} \mathrm{,} \tag{L.5} \label{eq:L.5} \\
\sigma_{0} \check{O}_{z^{\prime}}  &= - \check{C}_{x^{\prime}} \check{T} + \check{G} \check{L}_{z^{\prime}} + \check{H} \check{T}_{z^{\prime}} \mathrm{,} \tag{L.6} \label{eq:L.6} \\
\sigma_{0} \check{C}_{x^{\prime}}  &= - \check{O}_{z^{\prime}} \check{T} + \check{E} \check{L}_{x^{\prime}} + \check{F} \check{T}_{x^{\prime}} \mathrm{,} \tag{L.7} \label{eq:L.7} \\
\sigma_{0} \check{C}_{z^{\prime}}  &= \check{O}_{x^{\prime}} \check{T} + \check{E} \check{L}_{z^{\prime}} + \check{F} \check{T}_{z^{\prime}} \mathrm{,} \tag{L.8} \label{eq:L.8} \\
\sigma_{0} \check{T}_{x^{\prime}}  &= \check{L}_{z^{\prime}} \check{\Sigma} + \check{H} \check{O}_{x^{\prime}} + \check{F} \check{C}_{x^{\prime}} \mathrm{,} \tag{L.9} \label{eq:L.9} \\
\sigma_{0} \check{T}_{z^{\prime}}  &= - \check{L}_{x^{\prime}} \check{\Sigma} + \check{H} \check{O}_{z^{\prime}} + \check{F} \check{C}_{z^{\prime}} \mathrm{,} \tag{L.10} \label{eq:L.10} \\
\sigma_{0} \check{L}_{x^{\prime}}  &= - \check{T}_{z^{\prime}} \check{\Sigma} + \check{G} \check{O}_{x^{\prime}} + \check{E} \check{C}_{x^{\prime}} \mathrm{,} \tag{L.11} \label{eq:L.11} \\
\sigma_{0} \check{L}_{z^{\prime}}  &= \check{T}_{x^{\prime}} \check{\Sigma} + \check{G} \check{O}_{z^{\prime}} + \check{E} \check{C}_{z^{\prime}} \mathrm{.} \tag{L.12} \label{eq:L.12}
\end{align}
}

\textbf{Quadratic relations:}
{\allowdisplaybreaks
\begin{align}
\check{C}_{z^{\prime}} \check{O}_{z^{\prime}} + \check{O}_{x^{\prime}} \check{C}_{x^{\prime}} - \check{G} \check{E} - \check{H} \check{F} &= 0 \mathrm{,} \tag{Q.b} \label{eq:Q.b} \\
\check{G} \check{H} + \check{E} \check{F} - \check{T}_{x^{\prime}} \check{L}_{x^{\prime}} - \check{T}_{z^{\prime}} \check{L}_{z^{\prime}} &= 0 \mathrm{,} \tag{Q.t} \label{eq:Q.t} \\
\check{C}_{z^{\prime}} \check{C}_{x^{\prime}} + \check{O}_{z^{\prime}} \check{O}_{x^{\prime}} - \check{T}_{x^{\prime}} \check{T}_{z^{\prime}} - \check{L}_{x^{\prime}} \check{L}_{z^{\prime}} &= 0 \mathrm{,} \tag{Q.r} \label{eq:Q.r} \\
- \check{\Sigma} \check{G} + \check{T} \check{F} + \check{O}_{z^{\prime}} \check{T}_{x^{\prime}} - \check{O}_{x^{\prime}} \check{T}_{z^{\prime}} &= 0 \mathrm{,} \tag{Q.bt.1} \label{eq:Q.bt.1} \\
- \check{\Sigma} \check{H} - \check{T} \check{E} - \check{O}_{z^{\prime}} \check{L}_{x^{\prime}} + \check{O}_{x^{\prime}} \check{L}_{z^{\prime}} &= 0 \mathrm{,} \tag{Q.bt.2} \label{eq:Q.bt.2} \\
- \check{\Sigma} \check{E} - \check{T} \check{H} + \check{C}_{z^{\prime}} \check{T}_{x^{\prime}} - \check{C}_{x^{\prime}} \check{T}_{z^{\prime}} &= 0 \mathrm{,} \tag{Q.bt.3} \label{eq:Q.bt.3} \\
- \check{\Sigma} \check{F} + \check{T} \check{G} - \check{C}_{z^{\prime}} \check{L}_{x^{\prime}} + \check{C}_{x^{\prime}} \check{L}_{z^{\prime}} &= 0 \mathrm{,} \tag{Q.bt.4} \label{eq:Q.bt.4} \\
- \check{\Sigma} \check{O}_{x^{\prime}} + \check{P} \check{C}_{z^{\prime}} - \check{G} \check{T}_{z^{\prime}} + \check{H} \check{L}_{z^{\prime}} &= 0 \mathrm{,} \tag{Q.br.1} \label{eq:Q.br.1} \\
\check{\Sigma} \check{O}_{z^{\prime}} + \check{P} \check{C}_{x^{\prime}} - \check{G} \check{T}_{x^{\prime}} + \check{H} \check{L}_{x^{\prime}} &= 0 \mathrm{,} \tag{Q.br.2} \label{eq:Q.br.2} \\
\check{\Sigma} \check{C}_{x^{\prime}} + \check{P} \check{O}_{z^{\prime}} + \check{E} \check{T}_{z^{\prime}} - \check{F} \check{L}_{z^{\prime}} &= 0 \mathrm{,} \tag{Q.br.3} \label{eq:Q.br.3} \\
\check{\Sigma} \check{C}_{z^{\prime}} - \check{P} \check{O}_{x^{\prime}} - \check{E} \check{T}_{x^{\prime}} + \check{F} \check{L}_{x^{\prime}} &= 0 \mathrm{,} \tag{Q.br.4} \label{eq:Q.br.4} \\
\check{T} \check{T}_{x^{\prime}} - \check{P} \check{L}_{z^{\prime}} - \check{H} \check{C}_{z^{\prime}} + \check{F} \check{O}_{z^{\prime}} &= 0 \mathrm{,} \tag{Q.tr.1} \label{eq:Q.tr.1} \\
\check{T} \check{T}_{z^{\prime}} + \check{P} \check{L}_{x^{\prime}} + \check{H} \check{C}_{x^{\prime}} - \check{F} \check{O}_{x^{\prime}} &= 0 \mathrm{,} \tag{Q.tr.2} \label{eq:Q.tr.2} \\
- \check{T} \check{L}_{x^{\prime}} - \check{P} \check{T}_{z^{\prime}} + \check{G} \check{C}_{z^{\prime}} - \check{E} \check{O}_{z^{\prime}} &= 0 \mathrm{,} \tag{Q.tr.3} \label{eq:Q.tr.3} \\
- \check{T} \check{L}_{z^{\prime}} + \check{P} \check{T}_{x^{\prime}} - \check{G} \check{C}_{x^{\prime}} + \check{E} \check{O}_{x^{\prime}} &= 0 \mathrm{.} \tag{Q.tr.4} \label{eq:Q.tr.4}
\end{align}
}
\textbf{Square relations:}
{\allowdisplaybreaks
\begin{align}
 \check{G}^{2} +  \check{H}^{2} +  \check{E}^{2} +  \check{F}^{2}
 &=    \sigma_{0}^{2} - \check{\Sigma}^{2} - \check{T}^{2} +  \check{P}^{2} \mathrm{,} \tag{S.bt} \label{eq:S.bt} \\
 \check{O}_{x^{\prime}}^{2} + \check{O}_{z^{\prime}}^{2} + \check{C}_{x^{\prime}}^{2} + \check{C}_{z^{\prime}}^{2}
 &=    \sigma_{0}^{2} - \check{\Sigma}^{2} + \check{T}^{2} -  \check{P}^{2} \mathrm{,} \tag{S.br} \label{eq:S.br} \\
\check{T}_{x^{\prime}}^{2} + \check{T}_{z^{\prime}}^{2} +  \check{L}_{x^{\prime}}^{2} +  \check{L}_{z^{\prime}}^{2}
 &=    \sigma_{0}^{2} + \check{\Sigma}^{2} - \check{T}^{2} -  \check{P}^{2} \mathrm{,} \tag{S.tr} \label{eq:S.tr} \\
 \check{G}^{2} +  \check{H}^{2} -  \check{E}^{2} -  \check{F}^{2}
 &=    \check{O}_{x^{\prime}}^{2} + \check{O}_{z^{\prime}}^{2} - \check{C}_{x^{\prime}}^{2} - \check{C}_{z^{\prime}}^{2} \mathrm{,} \tag{S.b} \label{eq:S.b} \\
-  \check{G}^{2} +  \check{H}^{2} -  \check{E}^{2} +  \check{F}^{2}
 &=   \check{T}_{x^{\prime}}^{2} + \check{T}_{z^{\prime}}^{2} -  \check{L}_{x^{\prime}}^{2} -  \check{L}_{z^{\prime}}^{2} \mathrm{,} \tag{S.t} \label{eq:S.t} \\
 \check{O}_{x^{\prime}}^{2} - \check{O}_{z^{\prime}}^{2} + \check{C}_{x^{\prime}}^{2} - \check{C}_{z^{\prime}}^{2}
 &=   \check{T}_{x^{\prime}}^{2} - \check{T}_{z^{\prime}}^{2} +  \check{L}_{x^{\prime}}^{2} -  \check{L}_{z^{\prime}}^{2} \mathrm{.} \tag{S.r} \label{eq:S.r}
\end{align}
}
\clearpage

\subsection{Explicit expressions for observables in a TPWA} \label{sec:TPWAFormulae}

In this appendix section, the matrices $\left(\mathcal{C}_{L}\right)^{\alpha}_{k}$ that define the Legendre-coefficients in the TPWA standard form (cf. equations (\ref{eq:LowEAssocLegStandardParametrization1}) and (\ref{eq:LowEAssocLegStandardParametrization2}) in section \ref{sec:CompExpsTPWA}), i.e.
\begin{equation}
\left(a_{L}\right)_{k}^{\check{\Omega}^{\alpha}} \equiv \left(a_{L}\right)_{k}^{\alpha} = \left< \mathcal{M}_{\ell} \right| \left( \mathcal{C}_{L}\right)_{k}^{\check{\Omega}^{\alpha}} \left| \mathcal{M}_{\ell} \right> \mathrm{,} \label{eq:BilinearEqSystemAppendix}
\end{equation}
are derived and listed for a sufficiently high truncation. The number $\alpha$ here runs over all observables, $\alpha = 1,\ldots,16$. These coefficients appear in the TPWA-form given in equations (\ref{eq:LowEAssocLegStandardParametrization1}) and (\ref{eq:LowEAssocLegStandardParametrization2}) of section \ref{sec:CompExpsTPWA}, where we repeat the angular distribution of an arbitraty polarization observable $\check{\Omega}^{\alpha}$ here for convenience
\begin{equation}
\check{\Omega}^{\alpha} \left( W, \theta \right) = \rho \hspace*{3pt} \sum \limits_{k = \beta_{\alpha}}^{2 \ell_{\mathrm{max}} + \beta_{\alpha} + \gamma_{\alpha}} \left(a_{L}\right)_{k}^{\check{\Omega}^{\alpha}} \left( W \right) P^{\beta_{\alpha}}_{k} \left( \cos \theta \right) \mathrm{.}  \label{eq:LowEAssocLegStandardParametrization1Appendix} 
\end{equation}
The numbers $\beta_{\alpha}$ and $\gamma_{\alpha}$ that define the angular form and number of coefficients for each observable are collected in Table \ref{tab:AngularDistributionsParameters} in section \ref{sec:CompExpsTPWA}, but shall also be given here again for quick reference (see Table \ref{tab:AngularDistributionsParametersAppendix} below). \newline
The quantities (\ref{eq:BilinearEqSystemAppendix}) appear in the brief notation given above in a lot of places in the main text, whenever a numerical solution of exact theory data, or a fit of (pseudo-) data is described (cf. the end of appendix \ref{subsec:AccidentalAmbProofsIII} or the explanations in section \ref{sec:TPWAFitsIntro}). With the information given in this appendix section, the corresponding minimized functions called $\Phi$ (in case of theory-data) or $\chi^{2}$ (real data) can be understood in detail. \newline
For all investigations done in the main text, a description of the $\left(\mathcal{C}_{L}\right)^{\alpha}_{k}$ for $L ? \ell_{\mathrm{max}} \leq 5$ is sufficient. For this reason, as well as for the fact that the resulting marices can become quite large, we confine here to a truncation at $L = 5$. The expressions for lower truncvation orders can be deduced from the matrices given here quite easily, as shall be explained later. \newline
Now, suppose one has a finite expansion for a particular observable at $L=5$, for instance the unpolarized differential cross section
\begin{table}[h]
\centering
\begin{tabular}{cc|cccccc||cc|cccccccc}
\hline
\hline
  &  &  &  &  &  &  &  &  &  &  &  &  &  &  &  &  \\
Type & $ \check{\Omega}^{\alpha} $ &  & $ \beta_{\alpha} $ &  & $ \gamma_{\alpha} $ &  & $ \delta_{\alpha} $ & Type & $ \check{\Omega}^{\alpha} $ &  & $ \beta_{\alpha} $ &  & $ \gamma_{\alpha} $ &  & $ \delta_{\alpha} $ &   \\
\hline
  &  &  &  &  &  &  &  &  &  &  &  &  &  &  &  &  \\
  & $ I \left( \theta \right) $ &  & $ 0 $ &  & $ 0 $ &  & $ -2 $ &  & $ \check{O}_{x'} $ &  & $ 1 $ &  & $ 0 $ &  & $ -1 $ &   \\
 $\mathcal{S}$ & $ \check{\Sigma} $ &  & $ 2 $ &  & $ -2 $ &  & $ -2 $ & $\mathcal{BR}$ & $ \check{O}_{z'} $ &  & $ 2 $ &  & $ -1 $ &  & $ -1 $ &   \\
  & $ \check{T} $ &  & $ 1 $ &  & $ -1 $ &  & $ -1 $ &  & $ \check{C}_{x'} $ &  & $ 1 $ &  & $ 0 $ &  & $ -1 $ &   \\
  & $ \check{P} $ &  & $ 1 $ &  & $ -1 $ &  & $ -1 $ &  & $ \check{C}_{z'} $ &  & $ 0 $ &  & $ +1 $ &  & $ -1 $ &   \\
\hline
  &  &  &  &  &  &  &  &  &  &  &  &  &  &  &  &  \\
  & $ \check{E} $ &  & $ 0 $ &  & $ 0 $ &  & $ -1 $ &  & $ \check{T}_{x'} $ &  & $ 2 $ &  & $ -1 $ &  & $ -2 $ &   \\
 $\mathcal{BT}$ & $ \check{G} $ &  & $ 2 $ &  & $ -2 $ &  & $ -1 $ & $\mathcal{TR}$ & $ \check{T}_{z'} $ &  & $ 1 $ &  & $ 0 $ &  & $ -2 $ &   \\
  & $ \check{H} $ &  & $ 1 $ &  & $ -1 $ &  & $ -1 $ &  & $ \check{L}_{x'} $ &  & $ 1 $ &  & $ 0 $ &  & $ -2 $ &   \\
  & $ \check{F} $ &  & $ 1 $ &  & $ -1 $ &  & $ -1 $ &  & $ \check{L}_{z'} $ &  & $ 0 $ &  & $ +1 $ &  & $ -2 $ &   \\
\hline
\hline
\end{tabular}
\caption[Definition of the angular parametrization in a TPWA, for all 16 observables. Re-print for appendix \ref{sec:TPWAFormulae}.]{Description of the angular parametrizations (cf. Table \ref{tab:AngularDistributionsParameters} in the main text).}
\label{tab:AngularDistributionsParametersAppendix}
\end{table}
\clearpage
{\allowdisplaybreaks
\begin{align}
 \sigma_{0} &= \rho \hspace*{3pt} \Big( \left(a_{5}\right)^{\sigma_{0}}_{0} \hspace*{1.5pt} P_{0} ( \cos \theta ) + \left(a_{5}\right)^{\sigma_{0}}_{1} \hspace*{1.5pt} P_{1} ( \cos \theta ) + \left(a_{5}\right)^{\sigma_{0}}_{2} \hspace*{1.5pt} P_{2} ( \cos \theta ) + \left(a_{5}\right)^{\sigma_{0}}_{3} \hspace*{1.5pt} P_{3} ( \cos \theta )  \nonumber \\
  & \hspace*{22.5pt} + \left(a_{5}\right)^{\sigma_{0}}_{4} \hspace*{1.5pt} P_{4} ( \cos \theta ) + \left(a_{5}\right)^{\sigma_{0}}_{5} \hspace*{1.5pt} P_{5} ( \cos \theta ) + \left(a_{5}\right)^{\sigma_{0}}_{6} \hspace*{1.5pt} P_{6} ( \cos \theta ) + \left(a_{5}\right)^{\sigma_{0}}_{7} \hspace*{1.5pt} P_{7} ( \cos \theta )  \nonumber \\
  & \hspace*{22.5pt} + \left(a_{5}\right)^{\sigma_{0}}_{8} \hspace*{1.5pt} P_{8} ( \cos \theta ) + \left(a_{5}\right)^{\sigma_{0}}_{9} \hspace*{1.5pt} P_{9} ( \cos \theta ) + \left(a_{5}\right)^{\sigma_{0}}_{10} \hspace*{1.5pt} P_{10} ( \cos \theta ) \Big) \mathrm{.} \label{eq:ExpDCSIntoLegLmax5}
\end{align}
}
The question now is how to calculate the matrices $\left( \mathcal{C}_{L}\right)_{k}^{\check{\Omega}^{\alpha}}$, in this case $\left( \mathcal{C}_{5}\right)_{k}^{\sigma_{0}}$, defining the Legendre coefficients according to equation (\ref{eq:BilinearEqSystemAppendix}). \newline
In order to implement an automable routine for this purpose, one has to employ the orthogonality of the associated Legendre polynomials $P_{\ell}^{m} (x)$ in the lower index, with upper indices equal:
\begin{equation}
 \int_{-1}^{1} dx \hspace{3pt} P^{m}_{\ell} (x) P^{m}_{k} (x) = \frac{2}{2 \ell + 1} \frac{(\ell + m)!}{(\ell - m)!} \delta_{\ell k} \mathrm{.} \label{eq:OrthogonalityAssocLegPolys}
\end{equation}
Next it has to be noted that while the full multipole series defined in equations (\ref{eq:MultExpF1}) to (\ref{eq:MultExpF4}) is an infinite series in the angular variables, rendering the connection between CGLN amplitudes and multipoles non-trivial, it breaks down to a finite linear connection in case one truncates it for any finite order $L = \ell_{\mathrm{max}} \geq 1$. Generally, for finite truncations, the step of building CGLN amplitudes out of multipoles 
\begin{equation}
 \mathrm{multipoles} \hspace*{2.5pt} \mathcal{M}_{\ell} \longrightarrow \mathrm{CGLN-amplitudes} \hspace*{2.5pt} F_{i} \mathrm{,} \label{eq:ConnectionMultCGLNNonInvertible}
\end{equation}
proceeds by equations that are linear and non-invertible for any $L \geq 1$. One can formalize a finite truncation of the infinite series (\ref{eq:MultExpF1}) to (\ref{eq:MultExpF4}) by introducing a $\left( 4 \times \left[ 4L \right] \right)$-matrix $R (\theta)$ and writing

\begin{equation}
 F_{i} (W,\theta) = \sum_{j=1}^{4L} R_{ij} (\theta) \left( \mathcal{M}_{\ell} \right)_{j} \mathrm{,} \hspace*{2.5pt} \mathrm{or} \hspace*{2.5pt} \left| F \right> = R (\theta) \left| \mathcal{M}_{\ell} \right> \mathrm{.} \label{eq:LinearConnectionCGLNToMult}
\end{equation}
The object $\left| \mathcal{M}_{\ell} \right>$ here denotes the vector which collects all the multipoles in a finite truncation according to the convention fixed everywhere in this work, i.e.
\begin{equation}
 \left| \mathcal{M}_{\ell} \right> = \left[ E_{0+}, E_{1+}, M_{1+}, M_{1-}, E_{2+}, E_{2-}, M_{2+}, M_{2-}, \ldots  \right]^{T} \mathrm{.} \label{eq:MultipoleVectorAppendix}
\end{equation}
The adjoint of this vector, which is necessary for the evaluation of the bilinear forms, is then defined as
\begin{equation}
 \left< \mathcal{M}_{\ell} \right| = \left[ E^{\ast}_{0+}, E^{\ast}_{1+}, M^{\ast}_{1+}, M^{\ast}_{1-}, E^{\ast}_{2+}, E^{\ast}_{2-}, M^{\ast}_{2+}, M^{\ast}_{2-}, \ldots  \right] \mathrm{.} \label{eq:AdjointMultipoleVectorAppendix}
\end{equation}
Furthermore, the matrix $R$ depends on $\theta$ through the Legendre-polynomials. It is instructive to consider the expression (\ref{eq:LinearConnectionCGLNToMult}) explicitly for the simple special case $L = 1$, where $R$ is a singular $(4 \times 4)$-matrix. The multipole expansions (\ref{eq:MultExpF1}) to (\ref{eq:MultExpF4}) then yield
\begin{equation}
 \left[ \begin{array}{c} F_{1} \\ F_{2} \\ F_{3} \\ F_{4}  \end{array} \right] = \left[ \begin{array}{cccc} 1 & 3 \cos \theta & 3 \cos \theta & 0 \\
                                                                                                            0 & 0 & 2 & 1 \\
                                                                                                            0 & 3 & - 3 & 0 \\
                                                                                                            0 & 0 & 0 & 0 \end{array} \right]
                                                                                                            \left[ \begin{array}{c} E_{0+} \\ E_{1+} \\ M_{1+} \\ M_{1-}  \end{array} \right] \mathrm{.} \label{eq:Lmax1ExplicitLinearRelationCGLNToMult}
\end{equation}
For any higher truncation order $L$, the corresponding matrix $R$ can similarly be read from the truncated versions of the infinite multipole series (\ref{eq:MultExpF1}) to (\ref{eq:MultExpF4}). \newline
Next, one has to utilize the formalized TPWA of the CGLN amplitudes (\ref{eq:LinearConnectionCGLNToMult}) and combine it with a consistent algebra of matrices that defines the polarization observables. Such an algebra is provided for example by the matrices $\hat{A}^{\alpha}$ listed in the Tables \ref{tab:A1toA8} and \ref{tab:A9toA16} of appendix \ref{subsec:CGLNMatrixReps}, which yield the $16$ polarization observables $\check{\Omega}^{\alpha}$ in terms of CGLN amplitudes. Then, one obtains the observables in a TPWA truncated at $L$ as
\begin{equation}
 \check{\Omega}^{\alpha} = \frac{1}{2} \left< F \right| \hat{A}^{\alpha} \left| F \right> \equiv \left< \mathcal{M}_{\ell} \right| \bar{A}_{\mathcal{M}}^{\alpha} (\theta) \left| \mathcal{M}_{\ell} \right> \mathrm{,} \label{eq:ObsAsMultipoleBilinearForm}
\end{equation}
where the dependence on the multipoles is now defined in terms of a $\theta$-dependent $\left( \left( \left[ 4 L \right] \times \left[ 4L \right] \right) \right)$-matrix $\bar{A}_{\mathcal{M}}^{\alpha} (\theta)$. The latter can be obtained from the defining matrix $\hat{A}^{\alpha}$ by means of the identity
\begin{equation}
 \bar{A}_{\mathcal{M}}^{\alpha} (\theta) = \left[ R (\theta) \right]^{\dagger} \hat{A}^{\alpha} R (\theta) \mathrm{.} \label{eq:DefOfBarAMatrix}
\end{equation}
Now, the expression for the general angular distribution (\ref{eq:LowEAssocLegStandardParametrization1Appendix}) has to be compared with the general orthogonality relation for the associated Legendre-polynomials (\ref{eq:OrthogonalityAssocLegPolys}) in order to find the appropriate projection formula that yields the coefficients $\left(a_{L}\right)_{k}^{\alpha}$. It reads
\begin{equation}
 \left(a_{L}\right)_{k}^{\alpha} = \frac{2 k + 1}{2} \frac{(k - \beta_{\alpha})!}{(k + \beta_{\alpha})!} \int_{-1}^{1} d \cos \theta \hspace{3.75pt} P^{\beta_{\alpha}}_{k} \left( \cos \theta \right) \hspace{3pt} \check{\Omega}^{\alpha} \left( W, \cos \theta \right) \mathrm{.} \label{eq:LegCoeffProjectionFromObs}
\end{equation}
Equivalently, one can determine the matrix $\left(\mathcal{C}_{L}\right)^{\alpha}_{k}$ defining a particular Legendre coefficient according to equation (\ref{eq:BilinearEqSystemAppendix}) using the following relation (which is true due to the fact that multipoles carry no angular dependence)
\begin{equation}
 \left(\mathcal{C}_{L}\right)_{k}^{\alpha} = \frac{2 k + 1}{2} \frac{(k - \beta_{\alpha})!}{(k + \beta_{\alpha})!} \int_{-1}^{1} d \cos \theta \hspace{3.75pt} P^{\beta_{\alpha}}_{k} \left( \cos \theta \right) \hspace{3pt} \bar{A}_{\mathcal{M}}^{\alpha} (\theta) \mathrm{.} \label{eq:LegCoeffMatrixComponentProjection}
\end{equation}
This equation should be understood as a projection that proceeds in each matrix-component individually. It has a form that is particularly suitable for implementing it using computer algebra systems, in this case MATHEMATICA \cite{Mathematica8,Mathematica11,MathematicaLanguage,MathematicaBonnLicense}, in order to automatically evaluate the $\left(\mathcal{C}_{L}\right)^{\alpha}_{k}$ for any finite order $L$. Once the expressions have been evaluated once, they can be stored on disk and then used for all the numerical minimizations described in the main text. \newline
We use the Tables \ref{tab:DCSMatrix0} to \ref{tab:FObsMatrix10} contained in this appendix section in order to show all the matrices $\left(\mathcal{C}_{L}\right)^{\alpha}_{k}$ defining the Legendre coefficients $\left(a_{L}\right)_{k}^{\alpha}$ appearing in a TPWA with $L = 5$ for the group $\mathcal{S}$ and $\mathcal{BT}$-observables. These matrices are the same as the ones used in the publication in section \ref{sec:LFitsPaper}, where they have been the foundation for all interpretations and are also shown as color-plots. \newline
Similarly as in section \ref{sec:LFitsPaper}, horizontal and vertical lines have been drawn into the matrices in order to aid the readability. These lines distinguish contributions coming from partial-waves of different order $L = 0, 1, 2, \ldots$, or $S, P, D, \ldots$-waves, which can come from the multipole vector (\ref{eq:MultipoleVectorAppendix}) or its adjoint (\ref{eq:AdjointMultipoleVectorAppendix}) in the definition of the bilinear form (\ref{eq:BilinearEqSystemAppendix}). For instance, one line separates the first row and first column of every matrix from the rest of it. The numbers contained in those separated entries define all interference-terms in the coefficient that either have a factor of the single $S$-wave $E_{0+}$ and/or of the complex conjugate $E_{0+}^{\ast}$ in them. As another example, the rows $2$, $3$ and $4$ as well as the columns $2$, $3$ and $4$ are separated by lines. These define all interference terms containing factors of the three $P$-waves $\left\{ E_{1+}, M_{1+}, M_{1-} \right\}$ and/or their complex conjugates. This notation-scheme for the matrices proceeds analogously to higher orders, effectively dividing the matrices into blocks defining partial-wave interferences between certain orders of multipoles only. More details can be found in section \ref{sec:LFitsPaper}. \newline
In particular, it is now also apparent how to obtain the coefficient-matrices $\left(\mathcal{C}_{L}\right)_{k}^{\alpha}$ for truncation-orders $L < 5$. All that has to be done for example to get the matrices for $L = 4$ is to strip the ones printed here from their last $4$ rows and last $4$ columns. More generally matrices for any $L < 5$ are defined by the sub-matrices composed out of the first $4L$ rows and first $4L$ columns of the ones printed in the following Tables. \newline

\vspace*{50pt}

\begin{table}[hb]
\centering
\begin{small}
$
\left[
\begin{array}{c|ccc|cccc|cccc|cccc|cccc}
 1 & 0 & 0 & 0 & 0 & 0 & 0 & 0 & 0 & 0 & 0 & 0 & 0 & 0 & 0 & 0 & 0 & 0 & 0 & 0 \\
\hline
 0 & 6 & 0 & 0 & 0 & 0 & 0 & 0 & 0 & 0 & 0 & 0 & 0 & 0 & 0 & 0 & 0 & 0 & 0 & 0 \\
 0 & 0 & 2 & 0 & 0 & 0 & 0 & 0 & 0 & 0 & 0 & 0 & 0 & 0 & 0 & 0 & 0 & 0 & 0 & 0 \\
 0 & 0 & 0 & 1 & 0 & 0 & 0 & 0 & 0 & 0 & 0 & 0 & 0 & 0 & 0 & 0 & 0 & 0 & 0 & 0 \\
\hline
 0 & 0 & 0 & 0 & 18 & 0 & 0 & 0 & 0 & 0 & 0 & 0 & 0 & 0 & 0 & 0 & 0 & 0 & 0 & 0 \\
 0 & 0 & 0 & 0 & 0 & 2 & 0 & 0 & 0 & 0 & 0 & 0 & 0 & 0 & 0 & 0 & 0 & 0 & 0 & 0 \\
 0 & 0 & 0 & 0 & 0 & 0 & 9 & 0 & 0 & 0 & 0 & 0 & 0 & 0 & 0 & 0 & 0 & 0 & 0 & 0 \\
 0 & 0 & 0 & 0 & 0 & 0 & 0 & 6 & 0 & 0 & 0 & 0 & 0 & 0 & 0 & 0 & 0 & 0 & 0 & 0 \\
\hline
 0 & 0 & 0 & 0 & 0 & 0 & 0 & 0 & 40 & 0 & 0 & 0 & 0 & 0 & 0 & 0 & 0 & 0 & 0 & 0 \\
 0 & 0 & 0 & 0 & 0 & 0 & 0 & 0 & 0 & 9 & 0 & 0 & 0 & 0 & 0 & 0 & 0 & 0 & 0 & 0 \\
 0 & 0 & 0 & 0 & 0 & 0 & 0 & 0 & 0 & 0 & 24 & 0 & 0 & 0 & 0 & 0 & 0 & 0 & 0 & 0 \\
 0 & 0 & 0 & 0 & 0 & 0 & 0 & 0 & 0 & 0 & 0 & 18 & 0 & 0 & 0 & 0 & 0 & 0 & 0 & 0 \\
\hline
 0 & 0 & 0 & 0 & 0 & 0 & 0 & 0 & 0 & 0 & 0 & 0 & 75 & 0 & 0 & 0 & 0 & 0 & 0 & 0 \\
 0 & 0 & 0 & 0 & 0 & 0 & 0 & 0 & 0 & 0 & 0 & 0 & 0 & 24 & 0 & 0 & 0 & 0 & 0 & 0 \\
 0 & 0 & 0 & 0 & 0 & 0 & 0 & 0 & 0 & 0 & 0 & 0 & 0 & 0 & 50 & 0 & 0 & 0 & 0 & 0 \\
 0 & 0 & 0 & 0 & 0 & 0 & 0 & 0 & 0 & 0 & 0 & 0 & 0 & 0 & 0 & 40 & 0 & 0 & 0 & 0 \\
\hline
 0 & 0 & 0 & 0 & 0 & 0 & 0 & 0 & 0 & 0 & 0 & 0 & 0 & 0 & 0 & 0 & 126 & 0 & 0 & 0 \\
 0 & 0 & 0 & 0 & 0 & 0 & 0 & 0 & 0 & 0 & 0 & 0 & 0 & 0 & 0 & 0 & 0 & 50 & 0 & 0 \\
 0 & 0 & 0 & 0 & 0 & 0 & 0 & 0 & 0 & 0 & 0 & 0 & 0 & 0 & 0 & 0 & 0 & 0 & 90 & 0 \\
 0 & 0 & 0 & 0 & 0 & 0 & 0 & 0 & 0 & 0 & 0 & 0 & 0 & 0 & 0 & 0 & 0 & 0 & 0 & 75 \\
\end{array}
\right]
$
\end{small}
\caption{This is the matrix $\left( \mathcal{C}_{5}\right)_{0}^{\sigma_{0}}$, which defines the coefficient $\left(a_{5}\right)_{0}^{\sigma_{0}} = \left< \mathcal{M}_{\ell} \right| \left( \mathcal{C}_{5}\right)_{0}^{\sigma_{0}} \left| \mathcal{M}_{\ell} \right>$ that appears in the observable $\sigma_{0}$ for a truncation at $L=5$.}
\label{tab:DCSMatrix0}
\end{table}

\clearpage

\begin{table}[h]
\centering
\scalebox{.75}{
$
\left[
\begin{array}{c|ccc|cccc|cccc|cccc|cccc}
 0 & 3 & 1 & -1 & 0 & 0 & 0 & 0 & 0 & 0 & 0 & 0 & 0 & 0 & 0 & 0 & 0 & 0 & 0 & 0 \\
\hline
 3 & 0 & 0 & 0 & \frac{72}{5} & -\frac{3}{5} & \frac{9}{5} & -\frac{9}{5} & 0 & 0 & 0 & 0 & 0 & 0 & 0 & 0 & 0 & 0 & 0 & 0 \\
 1 & 0 & 0 & 0 & 0 & 1 & \frac{27}{5} & \frac{3}{5} & 0 & 0 & 0 & 0 & 0 & 0 & 0 & 0 & 0 & 0 & 0 & 0 \\
 -1 & 0 & 0 & 0 & 0 & -1 & 0 & 3 & 0 & 0 & 0 & 0 & 0 & 0 & 0 & 0 & 0 & 0 & 0 & 0 \\
\hline
 0 & \frac{72}{5} & 0 & 0 & 0 & 0 & 0 & 0 & \frac{270}{7} & -\frac{36}{35} & \frac{18}{7} & -\frac{18}{7} & 0 & 0 & 0 & 0 & 0 & 0 & 0 & 0 \\
 0 & -\frac{3}{5} & 1 & -1 & 0 & 0 & 0 & 0 & 0 & \frac{27}{5} & 0 & 0 & 0 & 0 & 0 & 0 & 0 & 0 & 0 & 0 \\
 0 & \frac{9}{5} & \frac{27}{5} & 0 & 0 & 0 & 0 & 0 & 0 & \frac{9}{5} & \frac{144}{7} & \frac{36}{35} & 0 & 0 & 0 & 0 & 0 & 0 & 0 & 0 \\
 0 & -\frac{9}{5} & \frac{3}{5} & 3 & 0 & 0 & 0 & 0 & 0 & -\frac{9}{5} & 0 & \frac{72}{5} & 0 & 0 & 0 & 0 & 0 & 0 & 0 & 0 \\
\end{array}
\right]
$
}
\caption{This is the matrix $\left( \mathcal{C}_{5}\right)_{1}^{\sigma_{0}}$, which defines the coefficient $\left(a_{5}\right)_{1}^{\sigma_{0}} = \left< \mathcal{M}_{\ell} \right| \left( \mathcal{C}_{5}\right)_{1}^{\sigma_{0}} \left| \mathcal{M}_{\ell} \right>$ that appears in the observable $\sigma_{0}$ for a truncation at $L=5$.}
\label{tab:DCSMatrix1}
\end{table}

\begin{table}[h]
\centering
\scalebox{.68}{
$
\left[
\begin{array}{c|ccc|cccc|cccc|cccc|cccc}
 0 & 0 & 0 & 0 & 6 & 1 & 3 & -3 & 0 & 0 & 0 & 0 & 0 & 0 & 0 & 0 & 0 & 0 & 0 & 0 \\
\hline
 0 & 3 & 3 & -3 & 0 & 0 & 0 & 0 & \frac{180}{7} & -\frac{9}{7} & \frac{36}{7} & -\frac{36}{7} & 0 & 0 & 0 & 0 & 0 & 0 & 0 & 0 \\
 0 & 3 & -1 & -1 & 0 & 0 & 0 & 0 & 0 & 3 & \frac{72}{7} & \frac{12}{7} & 0 & 0 & 0 & 0 & 0 & 0 & 0 & 0 \\
 0 & -3 & -1 & 0 & 0 & 0 & 0 & 0 & 0 & -3 & 0 & 6 & 0 & 0 & 0 & 0 & 0 & 0 & 0 & 0 \\
\hline
 6 & 0 & 0 & 0 & \frac{108}{7} & -\frac{12}{7} & \frac{36}{7} & -\frac{36}{7} & 0 & 0 & 0 & 0 & \frac{450}{7} & -\frac{18}{7} & \frac{50}{7} & -\frac{50}{7} & 0 & 0 & 0 & 0 \\
 1 & 0 & 0 & 0 & -\frac{12}{7} & -1 & 3 & -3 & 0 & 0 & 0 & 0 & 0 & \frac{72}{7} & 0 & 0 & 0 & 0 & 0 & 0 \\
 3 & 0 & 0 & 0 & \frac{36}{7} & 3 & \frac{36}{7} & \frac{9}{7} & 0 & 0 & 0 & 0 & 0 & \frac{36}{7} & \frac{250}{7} & \frac{20}{7} & 0 & 0 & 0 & 0 \\
 -3 & 0 & 0 & 0 & -\frac{36}{7} & -3 & \frac{9}{7} & 3 & 0 & 0 & 0 & 0 & 0 & -\frac{36}{7} & 0 & \frac{180}{7} & 0 & 0 & 0 & 0 \\
\end{array}
\right]
$
}
\caption{This is the matrix $\left( \mathcal{C}_{5}\right)_{2}^{\sigma_{0}}$, which defines the coefficient $\left(a_{5}\right)_{2}^{\sigma_{0}} = \left< \mathcal{M}_{\ell} \right| \left( \mathcal{C}_{5}\right)_{2}^{\sigma_{0}} \left| \mathcal{M}_{\ell} \right>$ that appears in the observable $\sigma_{0}$ for a truncation at $L=5$.}
\label{tab:DCSMatrix2}
\end{table}

\clearpage

\begin{table}[h]
\centering
\scalebox{.66}{
$
\left[
\begin{array}{c|ccc|cccc|cccc|cccc|cccc}
 0 & 0 & 0 & 0 & 0 & 0 & 0 & 0 & 10 & 3 & 6 & -6 & 0 & 0 & 0 & 0 & 0 & 0 & 0 & 0 \\
\hline
 0 & 0 & 0 & 0 & \frac{18}{5} & \frac{18}{5} & \frac{36}{5} & -\frac{36}{5} & 0 & 0 & 0 & 0 & 40 & -2 & 10 & -10 & 0 & 0 & 0 & 0 \\
 0 & 0 & 0 & 0 & 6 & 0 & -\frac{12}{5} & -\frac{18}{5} & 0 & 0 & 0 & 0 & 0 & 6 & \frac{50}{3} & \frac{10}{3} & 0 & 0 & 0 & 0 \\
 0 & 0 & 0 & 0 & -6 & 0 & -3 & 0 & 0 & 0 & 0 & 0 & 0 & -6 & 0 & 10 & 0 & 0 & 0 & 0 \\
\hline
 0 & \frac{18}{5} & 6 & -6 & 0 & 0 & 0 & 0 & 20 & -\frac{12}{5} & 12 & -12 & 0 & 0 & 0 & 0 & \frac{1050}{11} & -\frac{50}{11} & \frac{150}{11} & -\frac{150}{11} \\
 0 & \frac{18}{5} & 0 & 0 & 0 & 0 & 0 & 0 & -\frac{10}{3} & -\frac{12}{5} & 6 & -6 & 0 & 0 & 0 & 0 & 0 & \frac{50}{3} & 0 & 0 \\
 0 & \frac{36}{5} & -\frac{12}{5} & -3 & 0 & 0 & 0 & 0 & 10 & \frac{36}{5} & 6 & \frac{12}{5} & 0 & 0 & 0 & 0 & 0 & 10 & \frac{600}{11} & \frac{60}{11} \\
 0 & -\frac{36}{5} & -\frac{18}{5} & 0 & 0 & 0 & 0 & 0 & -10 & -\frac{36}{5} & 2 & \frac{18}{5} & 0 & 0 & 0 & 0 & 0 & -10 & 0 & 40 \\
\end{array}
\right]
$
}
\caption{This is the matrix $\left( \mathcal{C}_{5}\right)_{3}^{\sigma_{0}}$, which defines the coefficient $\left(a_{5}\right)_{3}^{\sigma_{0}} = \left< \mathcal{M}_{\ell} \right| \left( \mathcal{C}_{5}\right)_{3}^{\sigma_{0}} \left| \mathcal{M}_{\ell} \right>$ that appears in the observable $\sigma_{0}$ for a truncation at $L=5$.}
\label{tab:DCSMatrix3}
\end{table}

\begin{table}[h]
\centering
\scalebox{.615}{
$
\left[
\begin{array}{c|ccc|cccc|cccc|cccc|cccc}
 0 & 0 & 0 & 0 & 0 & 0 & 0 & 0 & 0 & 0 & 0 & 0 & 15 & 6 & 10 & -10 & 0 & 0 & 0 & 0 \\
\hline
 0 & 0 & 0 & 0 & 0 & 0 & 0 & 0 & \frac{30}{7} & \frac{72}{7} & \frac{90}{7} & -\frac{90}{7} & 0 & 0 & 0 & 0 & \frac{630}{11} & -\frac{30}{11} & \frac{180}{11} & -\frac{180}{11} \\
 0 & 0 & 0 & 0 & 0 & 0 & 0 & 0 & 10 & 0 & -\frac{30}{7} & -\frac{54}{7} & 0 & 0 & 0 & 0 & 0 & 10 & \frac{270}{11} & \frac{60}{11} \\
 0 & 0 & 0 & 0 & 0 & 0 & 0 & 0 & -10 & 0 & -6 & 0 & 0 & 0 & 0 & 0 & 0 & -10 & 0 & 15 \\
\hline
 0 & 0 & 0 & 0 & \frac{18}{7} & \frac{54}{7} & \frac{90}{7} & -\frac{90}{7} & 0 & 0 & 0 & 0 & \frac{180}{7} & -\frac{180}{77} & \frac{1620}{77} & -\frac{1620}{77} & 0 & 0 & 0 & 0 \\
 0 & 0 & 0 & 0 & \frac{54}{7} & 0 & 0 & 0 & 0 & 0 & 0 & 0 & -\frac{60}{11} & -\frac{30}{7} & 10 & -10 & 0 & 0 & 0 & 0 \\
 0 & 0 & 0 & 0 & \frac{90}{7} & 0 & -\frac{36}{7} & -\frac{72}{7} & 0 & 0 & 0 & 0 & \frac{180}{11} & \frac{90}{7} & \frac{540}{77} & \frac{270}{77} & 0 & 0 & 0 & 0 \\
 0 & 0 & 0 & 0 & -\frac{90}{7} & 0 & -\frac{72}{7} & 0 & 0 & 0 & 0 & 0 & -\frac{180}{11} & -\frac{90}{7} & \frac{30}{11} & \frac{30}{7} & 0 & 0 & 0 & 0 \\
\end{array}
\right]
$
}
\caption{This is the matrix $\left( \mathcal{C}_{5}\right)_{4}^{\sigma_{0}}$, which defines the coefficient $\left(a_{5}\right)_{4}^{\sigma_{0}} = \left< \mathcal{M}_{\ell} \right| \left( \mathcal{C}_{5}\right)_{4}^{\sigma_{0}} \left| \mathcal{M}_{\ell} \right>$ that appears in the observable $\sigma_{0}$ for a truncation at $L=5$.}
\label{tab:DCSMatrix4}
\end{table}

\clearpage

\begin{table}[h]
\centering
\scalebox{.655}{
$
\left[
\begin{array}{c|ccc|cccc|cccc|cccc|cccc}
 0 & 0 & 0 & 0 & 0 & 0 & 0 & 0 & 0 & 0 & 0 & 0 & 0 & 0 & 0 & 0 & 21 & 10 & 15 & -15 \\
\hline
 0 & 0 & 0 & 0 & 0 & 0 & 0 & 0 & 0 & 0 & 0 & 0 & 5 & 20 & 20 & -20 & 0 & 0 & 0 & 0 \\
 0 & 0 & 0 & 0 & 0 & 0 & 0 & 0 & 0 & 0 & 0 & 0 & 15 & 0 & -\frac{20}{3} & -\frac{40}{3} & 0 & 0 & 0 & 0 \\
 0 & 0 & 0 & 0 & 0 & 0 & 0 & 0 & 0 & 0 & 0 & 0 & -15 & 0 & -10 & 0 & 0 & 0 & 0 & 0 \\
\hline
 0 & 0 & 0 & 0 & 0 & 0 & 0 & 0 & \frac{10}{7} & \frac{150}{7} & \frac{150}{7} & -\frac{150}{7} & 0 & 0 & 0 & 0 & \frac{420}{13} & -\frac{20}{13} & \frac{420}{13} & -\frac{420}{13} \\
 0 & 0 & 0 & 0 & 0 & 0 & 0 & 0 & \frac{40}{3} & 0 & 0 & 0 & 0 & 0 & 0 & 0 & -\frac{105}{13} & -\frac{20}{3} & 15 & -15 \\
 0 & 0 & 0 & 0 & 0 & 0 & 0 & 0 & 20 & 0 & -\frac{60}{7} & -\frac{150}{7} & 0 & 0 & 0 & 0 & \frac{315}{13} & 20 & \frac{105}{13} & \frac{60}{13} \\
 0 & 0 & 0 & 0 & 0 & 0 & 0 & 0 & -20 & 0 & -20 & 0 & 0 & 0 & 0 & 0 & -\frac{315}{13} & -20 & \frac{45}{13} & 5 \\
\end{array}
\right]
$
}
\caption{This is the matrix $\left( \mathcal{C}_{5}\right)_{5}^{\sigma_{0}}$, which defines the coefficient $\left(a_{5}\right)_{5}^{\sigma_{0}} = \left< \mathcal{M}_{\ell} \right| \left( \mathcal{C}_{5}\right)_{5}^{\sigma_{0}} \left| \mathcal{M}_{\ell} \right>$ that appears in the observable $\sigma_{0}$ for a truncation at $L=5$.}
\label{tab:DCSMatrix5}
\end{table}

\begin{table}[h]
\centering
\scalebox{.63}{
$
\left[
\begin{array}{c|ccc|cccc|cccc|cccc|cccc}
 0 & 0 & 0 & 0 & 0 & 0 & 0 & 0 & 0 & 0 & 0 & 0 & 0 & 0 & 0 & 0 & 0 & 0 & 0 & 0 \\
\hline
 0 & 0 & 0 & 0 & 0 & 0 & 0 & 0 & 0 & 0 & 0 & 0 & 0 & 0 & 0 & 0 & \frac{63}{11} & \frac{360}{11} & \frac{315}{11} & -\frac{315}{11} \\
 0 & 0 & 0 & 0 & 0 & 0 & 0 & 0 & 0 & 0 & 0 & 0 & 0 & 0 & 0 & 0 & 21 & 0 & -\frac{105}{11} & -\frac{225}{11} \\
 0 & 0 & 0 & 0 & 0 & 0 & 0 & 0 & 0 & 0 & 0 & 0 & 0 & 0 & 0 & 0 & -21 & 0 & -15 & 0 \\
\hline
 0 & 0 & 0 & 0 & 0 & 0 & 0 & 0 & 0 & 0 & 0 & 0 & 0 & \frac{450}{11} & \frac{350}{11} & -\frac{350}{11} & 0 & 0 & 0 & 0 \\
 0 & 0 & 0 & 0 & 0 & 0 & 0 & 0 & 0 & 0 & 0 & 0 & \frac{225}{11} & 0 & 0 & 0 & 0 & 0 & 0 & 0 \\
 0 & 0 & 0 & 0 & 0 & 0 & 0 & 0 & 0 & 0 & 0 & 0 & \frac{315}{11} & 0 & -\frac{140}{11} & -\frac{400}{11} & 0 & 0 & 0 & 0 \\
 0 & 0 & 0 & 0 & 0 & 0 & 0 & 0 & 0 & 0 & 0 & 0 & -\frac{315}{11} & 0 & -\frac{360}{11} & 0 & 0 & 0 & 0 & 0 \\
\end{array}
\right]
$
}
\caption{This is the matrix $\left( \mathcal{C}_{5}\right)_{6}^{\sigma_{0}}$, which defines the coefficient $\left(a_{5}\right)_{6}^{\sigma_{0}} = \left< \mathcal{M}_{\ell} \right| \left( \mathcal{C}_{5}\right)_{6}^{\sigma_{0}} \left| \mathcal{M}_{\ell} \right>$ that appears in the observable $\sigma_{0}$ for a truncation at $L=5$.}
\label{tab:DCSMatrix6}
\end{table}

\clearpage

\begin{table}[h]
\centering
\scalebox{.607}{
$
\left[
\begin{array}{c|ccc|cccc|cccc|cccc|cccc}
 0 & 0 & 0 & 0 & 0 & 0 & 0 & 0 & 0 & 0 & 0 & 0 & 0 & 0 & 0 & 0 & 0 & 0 & 0 & 0 \\
\hline
 0 & 0 & 0 & 0 & 0 & 0 & 0 & 0 & 0 & 0 & 0 & 0 & 0 & 0 & 0 & 0 & 0 & 0 & 0 & 0 \\
 0 & 0 & 0 & 0 & 0 & 0 & 0 & 0 & 0 & 0 & 0 & 0 & 0 & 0 & 0 & 0 & 0 & 0 & 0 & 0 \\
 0 & 0 & 0 & 0 & 0 & 0 & 0 & 0 & 0 & 0 & 0 & 0 & 0 & 0 & 0 & 0 & 0 & 0 & 0 & 0 \\
\hline
 0 & 0 & 0 & 0 & 0 & 0 & 0 & 0 & 0 & 0 & 0 & 0 & 0 & 0 & 0 & 0 & -\frac{252}{143} & \frac{9450}{143} & \frac{6300}{143} & -\frac{6300}{143} \\
 0 & 0 & 0 & 0 & 0 & 0 & 0 & 0 & 0 & 0 & 0 & 0 & 0 & 0 & 0 & 0 & \frac{378}{13} & 0 & 0 & 0 \\
 0 & 0 & 0 & 0 & 0 & 0 & 0 & 0 & 0 & 0 & 0 & 0 & 0 & 0 & 0 & 0 & \frac{504}{13} & 0 & -\frac{2520}{143} & -\frac{7875}{143} \\
 0 & 0 & 0 & 0 & 0 & 0 & 0 & 0 & 0 & 0 & 0 & 0 & 0 & 0 & 0 & 0 & -\frac{504}{13} & 0 & -\frac{630}{13} & 0 \\
\end{array}
\right]
$
}
\caption{This is the matrix $\left( \mathcal{C}_{5}\right)_{7}^{\sigma_{0}}$, which defines the coefficient $\left(a_{5}\right)_{7}^{\sigma_{0}} = \left< \mathcal{M}_{\ell} \right| \left( \mathcal{C}_{5}\right)_{7}^{\sigma_{0}} \left| \mathcal{M}_{\ell} \right>$ that appears in the observable $\sigma_{0}$ for a truncation at $L=5$.}
\label{tab:DCSMatrix7}
\end{table}

\begin{table}[h]
\centering
\scalebox{.695}{
$
\left[
\begin{array}{c|ccc|cccc|cccc|cccc|cccc}
 0 & 0 & 0 & 0 & 0 & 0 & 0 & 0 & 0 & 0 & 0 & 0 & 0 & 0 & 0 & 0 & 0 & 0 & 0 & 0 \\
\hline
 0 & 0 & 0 & 0 & 0 & 0 & 0 & 0 & 0 & 0 & 0 & 0 & 0 & 0 & 0 & 0 & 0 & 0 & 0 & 0 \\
 0 & 0 & 0 & 0 & 0 & 0 & 0 & 0 & 0 & 0 & 0 & 0 & 0 & 0 & 0 & 0 & 0 & 0 & 0 & 0 \\
 0 & 0 & 0 & 0 & 0 & 0 & 0 & 0 & 0 & 0 & 0 & 0 & 0 & 0 & 0 & 0 & 0 & 0 & 0 & 0 \\
\hline
 0 & 0 & 0 & 0 & 0 & 0 & 0 & 0 & 0 & 0 & 0 & 0 & 0 & 0 & 0 & 0 & 0 & 0 & 0 & 0 \\
 0 & 0 & 0 & 0 & 0 & 0 & 0 & 0 & 0 & 0 & 0 & 0 & 0 & 0 & 0 & 0 & 0 & 0 & 0 & 0 \\
 0 & 0 & 0 & 0 & 0 & 0 & 0 & 0 & 0 & 0 & 0 & 0 & 0 & 0 & 0 & 0 & 0 & 0 & 0 & 0 \\
 0 & 0 & 0 & 0 & 0 & 0 & 0 & 0 & 0 & 0 & 0 & 0 & 0 & 0 & 0 & 0 & 0 & 0 & 0 & 0 \\
\end{array}
\right]
$
}
\caption{This is the matrix $\left( \mathcal{C}_{5}\right)_{8}^{\sigma_{0}}$, which defines the coefficient $\left(a_{5}\right)_{8}^{\sigma_{0}} = \left< \mathcal{M}_{\ell} \right| \left( \mathcal{C}_{5}\right)_{8}^{\sigma_{0}} \left| \mathcal{M}_{\ell} \right>$ that appears in the observable $\sigma_{0}$ for a truncation at $L=5$.}
\label{tab:DCSMatrix8}
\end{table}

\clearpage

\begin{table}[h]
\centering
\scalebox{.7415}{
$
\left[
\begin{array}{c|ccc|cccc|cccc|cccc|cccc}
 0 & 0 & 0 & 0 & 0 & 0 & 0 & 0 & 0 & 0 & 0 & 0 & 0 & 0 & 0 & 0 & 0 & 0 & 0 & 0 \\
\hline
 0 & 0 & 0 & 0 & 0 & 0 & 0 & 0 & 0 & 0 & 0 & 0 & 0 & 0 & 0 & 0 & 0 & 0 & 0 & 0 \\
 0 & 0 & 0 & 0 & 0 & 0 & 0 & 0 & 0 & 0 & 0 & 0 & 0 & 0 & 0 & 0 & 0 & 0 & 0 & 0 \\
 0 & 0 & 0 & 0 & 0 & 0 & 0 & 0 & 0 & 0 & 0 & 0 & 0 & 0 & 0 & 0 & 0 & 0 & 0 & 0 \\
\hline
 0 & 0 & 0 & 0 & 0 & 0 & 0 & 0 & 0 & 0 & 0 & 0 & 0 & 0 & 0 & 0 & 0 & 0 & 0 & 0 \\
 0 & 0 & 0 & 0 & 0 & 0 & 0 & 0 & 0 & 0 & 0 & 0 & 0 & 0 & 0 & 0 & 0 & 0 & 0 & 0 \\
 0 & 0 & 0 & 0 & 0 & 0 & 0 & 0 & 0 & 0 & 0 & 0 & 0 & 0 & 0 & 0 & 0 & 0 & 0 & 0 \\
 0 & 0 & 0 & 0 & 0 & 0 & 0 & 0 & 0 & 0 & 0 & 0 & 0 & 0 & 0 & 0 & 0 & 0 & 0 & 0 \\
\end{array}
\right]
$
}
\caption{This is the matrix $\left( \mathcal{C}_{5}\right)_{9}^{\sigma_{0}}$, which defines the coefficient $\left(a_{5}\right)_{9}^{\sigma_{0}} = \left< \mathcal{M}_{\ell} \right| \left( \mathcal{C}_{5}\right)_{9}^{\sigma_{0}} \left| \mathcal{M}_{\ell} \right>$ that appears in the observable $\sigma_{0}$ for a truncation at $L=5$.}
\label{tab:DCSMatrix9}
\end{table}

\begin{table}[h]
\centering
\scalebox{.7415}{
$
\left[
\begin{array}{c|ccc|cccc|cccc|cccc|cccc}
 0 & 0 & 0 & 0 & 0 & 0 & 0 & 0 & 0 & 0 & 0 & 0 & 0 & 0 & 0 & 0 & 0 & 0 & 0 & 0 \\
\hline
 0 & 0 & 0 & 0 & 0 & 0 & 0 & 0 & 0 & 0 & 0 & 0 & 0 & 0 & 0 & 0 & 0 & 0 & 0 & 0 \\
 0 & 0 & 0 & 0 & 0 & 0 & 0 & 0 & 0 & 0 & 0 & 0 & 0 & 0 & 0 & 0 & 0 & 0 & 0 & 0 \\
 0 & 0 & 0 & 0 & 0 & 0 & 0 & 0 & 0 & 0 & 0 & 0 & 0 & 0 & 0 & 0 & 0 & 0 & 0 & 0 \\
\hline
 0 & 0 & 0 & 0 & 0 & 0 & 0 & 0 & 0 & 0 & 0 & 0 & 0 & 0 & 0 & 0 & 0 & 0 & 0 & 0 \\
 0 & 0 & 0 & 0 & 0 & 0 & 0 & 0 & 0 & 0 & 0 & 0 & 0 & 0 & 0 & 0 & 0 & 0 & 0 & 0 \\
 0 & 0 & 0 & 0 & 0 & 0 & 0 & 0 & 0 & 0 & 0 & 0 & 0 & 0 & 0 & 0 & 0 & 0 & 0 & 0 \\
 0 & 0 & 0 & 0 & 0 & 0 & 0 & 0 & 0 & 0 & 0 & 0 & 0 & 0 & 0 & 0 & 0 & 0 & 0 & 0 \\
\hline
 0 & 0 & 0 & 0 & 0 & 0 & 0 & 0 & 0 & 0 & 0 & 0 & 0 & 0 & 0 & 0 & 0 & 0 & 0 & 0 \\
 0 & 0 & 0 & 0 & 0 & 0 & 0 & 0 & 0 & 0 & 0 & 0 & 0 & 0 & 0 & 0 & 0 & 0 & 0 & 0 \\
 0 & 0 & 0 & 0 & 0 & 0 & 0 & 0 & 0 & 0 & 0 & 0 & 0 & 0 & 0 & 0 & 0 & 0 & 0 & 0 \\
 0 & 0 & 0 & 0 & 0 & 0 & 0 & 0 & 0 & 0 & 0 & 0 & 0 & 0 & 0 & 0 & 0 & 0 & 0 & 0 \\
\hline
 0 & 0 & 0 & 0 & 0 & 0 & 0 & 0 & 0 & 0 & 0 & 0 & 0 & 0 & 0 & 0 & 0 & 0 & 0 & 0 \\
 0 & 0 & 0 & 0 & 0 & 0 & 0 & 0 & 0 & 0 & 0 & 0 & 0 & 0 & 0 & 0 & 0 & 0 & 0 & 0 \\
 0 & 0 & 0 & 0 & 0 & 0 & 0 & 0 & 0 & 0 & 0 & 0 & 0 & 0 & 0 & 0 & 0 & 0 & 0 & 0 \\
 0 & 0 & 0 & 0 & 0 & 0 & 0 & 0 & 0 & 0 & 0 & 0 & 0 & 0 & 0 & 0 & 0 & 0 & 0 & 0 \\
\hline
 0 & 0 & 0 & 0 & 0 & 0 & 0 & 0 & 0 & 0 & 0 & 0 & 0 & 0 & 0 & 0 & -\frac{7938}{323} & \frac{952560}{4199} & \frac{436590}{4199} & -\frac{436590}{4199} \\
 0 & 0 & 0 & 0 & 0 & 0 & 0 & 0 & 0 & 0 & 0 & 0 & 0 & 0 & 0 & 0 & \frac{952560}{4199} & 0 & 0 & 0 \\
 0 & 0 & 0 & 0 & 0 & 0 & 0 & 0 & 0 & 0 & 0 & 0 & 0 & 0 & 0 & 0 & \frac{436590}{4199} & 0 & -\frac{198450}{4199} & -\frac{992250}{4199} \\
 0 & 0 & 0 & 0 & 0 & 0 & 0 & 0 & 0 & 0 & 0 & 0 & 0 & 0 & 0 & 0 & -\frac{436590}{4199} & 0 & -\frac{992250}{4199} & 0 \\
\end{array}
\right]
$
}
\caption{This is the matrix $\left( \mathcal{C}_{5}\right)_{10}^{\sigma_{0}}$, which defines the coefficient $\left(a_{5}\right)_{10}^{\sigma_{0}} = \left< \mathcal{M}_{\ell} \right| \left( \mathcal{C}_{5}\right)_{10}^{\sigma_{0}} \left| \mathcal{M}_{\ell} \right>$ that appears in the observable $\sigma_{0}$ for a truncation at $L=5$.}
\label{tab:DCSMatrix10}
\end{table}

\clearpage

\begin{table}[h]
\centering
\scalebox{.6825}{
$
\left[
\begin{array}{c|ccc|cccc|cccc|cccc|cccc}
 0 & 0 & 0 & 0 & \frac{1}{2} & \frac{1}{2} & -\frac{1}{2} & \frac{1}{2} & 0 & 0 & 0 & 0 & 0 & 0 & 0 & 0 & 0 & 0 & 0 & 0 \\
\hline
 0 & -\frac{3}{2} & \frac{1}{2} & -\frac{1}{2} & 0 & 0 & 0 & 0 & \frac{15}{7} & \frac{9}{14} & -\frac{9}{7} & \frac{9}{7} & 0 & 0 & 0 & 0 & 0 & 0 & 0 & 0 \\
 0 & \frac{1}{2} & \frac{1}{2} & \frac{1}{2} & 0 & 0 & 0 & 0 & 0 & \frac{1}{2} & -\frac{6}{7} & -\frac{1}{7} & 0 & 0 & 0 & 0 & 0 & 0 & 0 & 0 \\
 0 & -\frac{1}{2} & \frac{1}{2} & 0 & 0 & 0 & 0 & 0 & 0 & -\frac{1}{2} & 0 & -\frac{1}{2} & 0 & 0 & 0 & 0 & 0 & 0 & 0 & 0 \\
\hline
 \frac{1}{2} & 0 & 0 & 0 & -\frac{36}{7} & -\frac{1}{7} & \frac{9}{7} & -\frac{9}{7} & 0 & 0 & 0 & 0 & \frac{75}{14} & \frac{6}{7} & -\frac{50}{21} & \frac{50}{21} & 0 & 0 & 0 & 0 \\
 \frac{1}{2} & 0 & 0 & 0 & -\frac{1}{7} & -\frac{1}{2} & -\frac{1}{2} & \frac{1}{2} & 0 & 0 & 0 & 0 & 0 & \frac{6}{7} & 0 & 0 & 0 & 0 & 0 & 0 \\
 -\frac{1}{2} & 0 & 0 & 0 & \frac{9}{7} & -\frac{1}{2} & \frac{18}{7} & \frac{9}{14} & 0 & 0 & 0 & 0 & 0 & \frac{9}{7} & -\frac{125}{42} & -\frac{5}{21} & 0 & 0 & 0 & 0 \\
 \frac{1}{2} & 0 & 0 & 0 & -\frac{9}{7} & \frac{1}{2} & \frac{9}{14} & \frac{3}{2} & 0 & 0 & 0 & 0 & 0 & -\frac{9}{7} & 0 & -\frac{15}{7} & 0 & 0 & 0 & 0 \\
\end{array}
\right]
$
}
\caption{This is the matrix $\left( \mathcal{C}_{5}\right)_{2}^{\check{\Sigma}}$, which defines the coefficient $\left(a_{5}\right)_{2}^{\check{\Sigma}} = \left< \mathcal{M}_{\ell} \right| \left( \mathcal{C}_{5}\right)_{2}^{\check{\Sigma}} \left| \mathcal{M}_{\ell} \right>$ that appears in the observable $\check{\Sigma}$ for a truncation at $L=5$.}
\label{tab:SigmaMatrix2}
\end{table}

\begin{table}[h]
\centering
\scalebox{.71}{
$
\left[
\begin{array}{c|ccc|cccc|cccc|cccc|cccc}
 0 & 0 & 0 & 0 & 0 & 0 & 0 & 0 & \frac{1}{2} & \frac{1}{2} & -\frac{1}{2} & \frac{1}{2} & 0 & 0 & 0 & 0 & 0 & 0 & 0 & 0 \\
\hline
 0 & 0 & 0 & 0 & -\frac{9}{10} & \frac{3}{5} & 0 & 0 & 0 & 0 & 0 & 0 & 2 & \frac{1}{2} & -\frac{7}{6} & \frac{7}{6} & 0 & 0 & 0 & 0 \\
 0 & 0 & 0 & 0 & \frac{1}{2} & 0 & \frac{2}{5} & \frac{3}{5} & 0 & 0 & 0 & 0 & 0 & \frac{1}{2} & -\frac{5}{6} & -\frac{1}{6} & 0 & 0 & 0 & 0 \\
 0 & 0 & 0 & 0 & -\frac{1}{2} & 0 & \frac{1}{2} & 0 & 0 & 0 & 0 & 0 & 0 & -\frac{1}{2} & 0 & -\frac{1}{2} & 0 & 0 & 0 & 0 \\
\hline
 0 & -\frac{9}{10} & \frac{1}{2} & -\frac{1}{2} & 0 & 0 & 0 & 0 & -\frac{8}{3} & \frac{3}{5} & 0 & 0 & 0 & 0 & 0 & 0 & \frac{105}{22} & \frac{20}{33} & -\frac{45}{22} & \frac{45}{22} \\
 0 & \frac{3}{5} & 0 & 0 & 0 & 0 & 0 & 0 & -\frac{1}{6} & -\frac{2}{5} & -\frac{1}{2} & \frac{1}{2} & 0 & 0 & 0 & 0 & 0 & \frac{5}{6} & 0 & 0 \\
 0 & 0 & \frac{2}{5} & \frac{1}{2} & 0 & 0 & 0 & 0 & \frac{7}{6} & 0 & \frac{3}{2} & \frac{3}{5} & 0 & 0 & 0 & 0 & 0 & \frac{7}{6} & -\frac{30}{11} & -\frac{3}{11} \\
 0 & 0 & \frac{3}{5} & 0 & 0 & 0 & 0 & 0 & -\frac{7}{6} & 0 & \frac{1}{2} & \frac{9}{10} & 0 & 0 & 0 & 0 & 0 & -\frac{7}{6} & 0 & -2 \\
\end{array}
\right]
$
}
\caption{This is the matrix $\left( \mathcal{C}_{5}\right)_{3}^{\check{\Sigma}}$, which defines the coefficient $\left(a_{5}\right)_{3}^{\check{\Sigma}} = \left< \mathcal{M}_{\ell} \right| \left( \mathcal{C}_{5}\right)_{3}^{\check{\Sigma}} \left| \mathcal{M}_{\ell} \right>$ that appears in the observable $\check{\Sigma}$ for a truncation at $L=5$.}
\label{tab:SigmaMatrix3}
\end{table}

\clearpage

\begin{table}[h]
\centering
\scalebox{.67}{
$
\left[
\begin{array}{c|ccc|cccc|cccc|cccc|cccc}
 0 & 0 & 0 & 0 & 0 & 0 & 0 & 0 & 0 & 0 & 0 & 0 & \frac{1}{2} & \frac{1}{2} & -\frac{1}{2} & \frac{1}{2} & 0 & 0 & 0 & 0 \\
\hline
 0 & 0 & 0 & 0 & 0 & 0 & 0 & 0 & -\frac{9}{14} & \frac{6}{7} & -\frac{3}{14} & \frac{3}{14} & 0 & 0 & 0 & 0 & \frac{21}{11} & \frac{9}{22} & -\frac{12}{11} & \frac{12}{11} \\
 0 & 0 & 0 & 0 & 0 & 0 & 0 & 0 & \frac{1}{2} & 0 & \frac{5}{14} & \frac{9}{14} & 0 & 0 & 0 & 0 & 0 & \frac{1}{2} & -\frac{9}{11} & -\frac{2}{11} \\
 0 & 0 & 0 & 0 & 0 & 0 & 0 & 0 & -\frac{1}{2} & 0 & \frac{1}{2} & 0 & 0 & 0 & 0 & 0 & 0 & -\frac{1}{2} & 0 & -\frac{1}{2} \\
\hline
 0 & 0 & 0 & 0 & -\frac{6}{7} & \frac{9}{14} & \frac{3}{14} & -\frac{3}{14} & 0 & 0 & 0 & 0 & -\frac{129}{77} & \frac{60}{77} & -\frac{36}{77} & \frac{36}{77} & 0 & 0 & 0 & 0 \\
 0 & 0 & 0 & 0 & \frac{9}{14} & 0 & 0 & 0 & 0 & 0 & 0 & 0 & -\frac{2}{11} & -\frac{5}{14} & -\frac{1}{2} & \frac{1}{2} & 0 & 0 & 0 & 0 \\
 0 & 0 & 0 & 0 & \frac{3}{14} & 0 & \frac{3}{7} & \frac{6}{7} & 0 & 0 & 0 & 0 & \frac{12}{11} & \frac{3}{14} & \frac{81}{77} & \frac{81}{154} & 0 & 0 & 0 & 0 \\
 0 & 0 & 0 & 0 & -\frac{3}{14} & 0 & \frac{6}{7} & 0 & 0 & 0 & 0 & 0 & -\frac{12}{11} & -\frac{3}{14} & \frac{9}{22} & \frac{9}{14} & 0 & 0 & 0 & 0 \\
\end{array}
\right]
$
}
\caption{This is the matrix $\left( \mathcal{C}_{5}\right)_{4}^{\check{\Sigma}}$, which defines the coefficient $\left(a_{5}\right)_{4}^{\check{\Sigma}} = \left< \mathcal{M}_{\ell} \right| \left( \mathcal{C}_{5}\right)_{4}^{\check{\Sigma}} \left| \mathcal{M}_{\ell} \right>$ that appears in the observable $\check{\Sigma}$ for a truncation at $L=5$.}
\label{tab:SigmaMatrix4}
\end{table}

\begin{table}[h]
\centering
\scalebox{.79}{
$
\left[
\begin{array}{c|ccc|cccc|cccc|cccc|cccc}
 0 & 0 & 0 & 0 & 0 & 0 & 0 & 0 & 0 & 0 & 0 & 0 & 0 & 0 & 0 & 0 & \frac{1}{2} & \frac{1}{2} & -\frac{1}{2} & \frac{1}{2} \\
\hline
 0 & 0 & 0 & 0 & 0 & 0 & 0 & 0 & 0 & 0 & 0 & 0 & -\frac{1}{2} & 1 & -\frac{1}{3} & \frac{1}{3} & 0 & 0 & 0 & 0 \\
 0 & 0 & 0 & 0 & 0 & 0 & 0 & 0 & 0 & 0 & 0 & 0 & \frac{1}{2} & 0 & \frac{1}{3} & \frac{2}{3} & 0 & 0 & 0 & 0 \\
 0 & 0 & 0 & 0 & 0 & 0 & 0 & 0 & 0 & 0 & 0 & 0 & -\frac{1}{2} & 0 & \frac{1}{2} & 0 & 0 & 0 & 0 & 0 \\
\hline
 0 & 0 & 0 & 0 & 0 & 0 & 0 & 0 & -\frac{16}{21} & \frac{15}{14} & 0 & 0 & 0 & 0 & 0 & 0 & -\frac{15}{13} & \frac{32}{39} & -\frac{9}{13} & \frac{9}{13} \\
 0 & 0 & 0 & 0 & 0 & 0 & 0 & 0 & \frac{2}{3} & 0 & 0 & 0 & 0 & 0 & 0 & 0 & -\frac{5}{26} & -\frac{1}{3} & -\frac{1}{2} & \frac{1}{2} \\
 0 & 0 & 0 & 0 & 0 & 0 & 0 & 0 & \frac{1}{3} & 0 & \frac{3}{7} & \frac{15}{14} & 0 & 0 & 0 & 0 & \frac{27}{26} & \frac{1}{3} & \frac{21}{26} & \frac{6}{13} \\
 0 & 0 & 0 & 0 & 0 & 0 & 0 & 0 & -\frac{1}{3} & 0 & 1 & 0 & 0 & 0 & 0 & 0 & -\frac{27}{26} & -\frac{1}{3} & \frac{9}{26} & \frac{1}{2} \\
\end{array}
\right]
$
}
\caption{This is the matrix $\left( \mathcal{C}_{5}\right)_{5}^{\check{\Sigma}}$, which defines the coefficient $\left(a_{5}\right)_{5}^{\check{\Sigma}} = \left< \mathcal{M}_{\ell} \right| \left( \mathcal{C}_{5}\right)_{5}^{\check{\Sigma}} \left| \mathcal{M}_{\ell} \right>$ that appears in the observable $\check{\Sigma}$ for a truncation at $L=5$.}
\label{tab:SigmaMatrix5}
\end{table}

\clearpage

\begin{table}[h]
\centering
\scalebox{.73}{
$
\left[
\begin{array}{c|ccc|cccc|cccc|cccc|cccc}
 0 & 0 & 0 & 0 & 0 & 0 & 0 & 0 & 0 & 0 & 0 & 0 & 0 & 0 & 0 & 0 & 0 & 0 & 0 & 0 \\
\hline
 0 & 0 & 0 & 0 & 0 & 0 & 0 & 0 & 0 & 0 & 0 & 0 & 0 & 0 & 0 & 0 & -\frac{9}{22} & \frac{12}{11} & -\frac{9}{22} & \frac{9}{22} \\
 0 & 0 & 0 & 0 & 0 & 0 & 0 & 0 & 0 & 0 & 0 & 0 & 0 & 0 & 0 & 0 & \frac{1}{2} & 0 & \frac{7}{22} & \frac{15}{22} \\
 0 & 0 & 0 & 0 & 0 & 0 & 0 & 0 & 0 & 0 & 0 & 0 & 0 & 0 & 0 & 0 & -\frac{1}{2} & 0 & \frac{1}{2} & 0 \\
\hline
 0 & 0 & 0 & 0 & 0 & 0 & 0 & 0 & 0 & 0 & 0 & 0 & -\frac{15}{22} & \frac{15}{11} & -\frac{5}{33} & \frac{5}{33} & 0 & 0 & 0 & 0 \\
 0 & 0 & 0 & 0 & 0 & 0 & 0 & 0 & 0 & 0 & 0 & 0 & \frac{15}{22} & 0 & 0 & 0 & 0 & 0 & 0 & 0 \\
 0 & 0 & 0 & 0 & 0 & 0 & 0 & 0 & 0 & 0 & 0 & 0 & \frac{9}{22} & 0 & \frac{14}{33} & \frac{40}{33} & 0 & 0 & 0 & 0 \\
 0 & 0 & 0 & 0 & 0 & 0 & 0 & 0 & 0 & 0 & 0 & 0 & -\frac{9}{22} & 0 & \frac{12}{11} & 0 & 0 & 0 & 0 & 0 \\
\end{array}
\right]
$
}
\caption{This is the matrix $\left( \mathcal{C}_{5}\right)_{6}^{\check{\Sigma}}$, which defines the coefficient $\left(a_{5}\right)_{6}^{\check{\Sigma}} = \left< \mathcal{M}_{\ell} \right| \left( \mathcal{C}_{5}\right)_{6}^{\check{\Sigma}} \left| \mathcal{M}_{\ell} \right>$ that appears in the observable $\check{\Sigma}$ for a truncation at $L=5$.}
\label{tab:SigmaMatrix6}
\end{table}

\begin{table}[h]
\centering
\scalebox{.73}{
$
\left[
\begin{array}{c|ccc|cccc|cccc|cccc|cccc}
 0 & 0 & 0 & 0 & 0 & 0 & 0 & 0 & 0 & 0 & 0 & 0 & 0 & 0 & 0 & 0 & 0 & 0 & 0 & 0 \\
\hline
 0 & 0 & 0 & 0 & 0 & 0 & 0 & 0 & 0 & 0 & 0 & 0 & 0 & 0 & 0 & 0 & 0 & 0 & 0 & 0 \\
 0 & 0 & 0 & 0 & 0 & 0 & 0 & 0 & 0 & 0 & 0 & 0 & 0 & 0 & 0 & 0 & 0 & 0 & 0 & 0 \\
 0 & 0 & 0 & 0 & 0 & 0 & 0 & 0 & 0 & 0 & 0 & 0 & 0 & 0 & 0 & 0 & 0 & 0 & 0 & 0 \\
\hline
 0 & 0 & 0 & 0 & 0 & 0 & 0 & 0 & 0 & 0 & 0 & 0 & 0 & 0 & 0 & 0 & -\frac{177}{286} & \frac{225}{143} & -\frac{75}{286} & \frac{75}{286} \\
 0 & 0 & 0 & 0 & 0 & 0 & 0 & 0 & 0 & 0 & 0 & 0 & 0 & 0 & 0 & 0 & \frac{9}{13} & 0 & 0 & 0 \\
 0 & 0 & 0 & 0 & 0 & 0 & 0 & 0 & 0 & 0 & 0 & 0 & 0 & 0 & 0 & 0 & \frac{6}{13} & 0 & \frac{60}{143} & \frac{375}{286} \\
 0 & 0 & 0 & 0 & 0 & 0 & 0 & 0 & 0 & 0 & 0 & 0 & 0 & 0 & 0 & 0 & -\frac{6}{13} & 0 & \frac{15}{13} & 0 \\
\end{array}
\right]
$
}
\caption{This is the matrix $\left( \mathcal{C}_{5}\right)_{7}^{\check{\Sigma}}$, which defines the coefficient $\left(a_{5}\right)_{7}^{\check{\Sigma}} = \left< \mathcal{M}_{\ell} \right| \left( \mathcal{C}_{5}\right)_{7}^{\check{\Sigma}} \left| \mathcal{M}_{\ell} \right>$ that appears in the observable $\check{\Sigma}$ for a truncation at $L=5$.}
\label{tab:SigmaMatrix7}
\end{table}

\clearpage

\begin{table}[h]
\centering
\scalebox{.72}{
$
\left[
\begin{array}{c|ccc|cccc|cccc|cccc|cccc}
 0 & 0 & 0 & 0 & 0 & 0 & 0 & 0 & 0 & 0 & 0 & 0 & 0 & 0 & 0 & 0 & 0 & 0 & 0 & 0 \\
\hline
 0 & 0 & 0 & 0 & 0 & 0 & 0 & 0 & 0 & 0 & 0 & 0 & 0 & 0 & 0 & 0 & 0 & 0 & 0 & 0 \\
 0 & 0 & 0 & 0 & 0 & 0 & 0 & 0 & 0 & 0 & 0 & 0 & 0 & 0 & 0 & 0 & 0 & 0 & 0 & 0 \\
 0 & 0 & 0 & 0 & 0 & 0 & 0 & 0 & 0 & 0 & 0 & 0 & 0 & 0 & 0 & 0 & 0 & 0 & 0 & 0 \\
\hline
 0 & 0 & 0 & 0 & 0 & 0 & 0 & 0 & 0 & 0 & 0 & 0 & 0 & 0 & 0 & 0 & 0 & 0 & 0 & 0 \\
 0 & 0 & 0 & 0 & 0 & 0 & 0 & 0 & 0 & 0 & 0 & 0 & 0 & 0 & 0 & 0 & 0 & 0 & 0 & 0 \\
 0 & 0 & 0 & 0 & 0 & 0 & 0 & 0 & 0 & 0 & 0 & 0 & 0 & 0 & 0 & 0 & 0 & 0 & 0 & 0 \\
 0 & 0 & 0 & 0 & 0 & 0 & 0 & 0 & 0 & 0 & 0 & 0 & 0 & 0 & 0 & 0 & 0 & 0 & 0 & 0 \\
\end{array}
\right]
$
}
\caption{This is the matrix $\left( \mathcal{C}_{5}\right)_{8}^{\check{\Sigma}}$, which defines the coefficient $\left(a_{5}\right)_{8}^{\check{\Sigma}} = \left< \mathcal{M}_{\ell} \right| \left( \mathcal{C}_{5}\right)_{8}^{\check{\Sigma}} \left| \mathcal{M}_{\ell} \right>$ that appears in the observable $\check{\Sigma}$ for a truncation at $L=5$.}
\label{tab:SigmaMatrix8}
\end{table}

\begin{table}[h]
\centering
\scalebox{.74}{
$
\left[
\begin{array}{c|ccc|cccc|cccc|cccc|cccc}
 0 & 0 & 0 & 0 & 0 & 0 & 0 & 0 & 0 & 0 & 0 & 0 & 0 & 0 & 0 & 0 & 0 & 0 & 0 & 0 \\
\hline
 0 & 0 & 0 & 0 & 0 & 0 & 0 & 0 & 0 & 0 & 0 & 0 & 0 & 0 & 0 & 0 & 0 & 0 & 0 & 0 \\
 0 & 0 & 0 & 0 & 0 & 0 & 0 & 0 & 0 & 0 & 0 & 0 & 0 & 0 & 0 & 0 & 0 & 0 & 0 & 0 \\
 0 & 0 & 0 & 0 & 0 & 0 & 0 & 0 & 0 & 0 & 0 & 0 & 0 & 0 & 0 & 0 & 0 & 0 & 0 & 0 \\
\hline
 0 & 0 & 0 & 0 & 0 & 0 & 0 & 0 & 0 & 0 & 0 & 0 & 0 & 0 & 0 & 0 & 0 & 0 & 0 & 0 \\
 0 & 0 & 0 & 0 & 0 & 0 & 0 & 0 & 0 & 0 & 0 & 0 & 0 & 0 & 0 & 0 & 0 & 0 & 0 & 0 \\
 0 & 0 & 0 & 0 & 0 & 0 & 0 & 0 & 0 & 0 & 0 & 0 & 0 & 0 & 0 & 0 & 0 & 0 & 0 & 0 \\
 0 & 0 & 0 & 0 & 0 & 0 & 0 & 0 & 0 & 0 & 0 & 0 & 0 & 0 & 0 & 0 & 0 & 0 & 0 & 0 \\
\hline
 0 & 0 & 0 & 0 & 0 & 0 & 0 & 0 & 0 & 0 & 0 & 0 & 0 & 0 & 0 & 0 & 0 & 0 & 0 & 0 \\
 0 & 0 & 0 & 0 & 0 & 0 & 0 & 0 & 0 & 0 & 0 & 0 & 0 & 0 & 0 & 0 & 0 & 0 & 0 & 0 \\
 0 & 0 & 0 & 0 & 0 & 0 & 0 & 0 & 0 & 0 & 0 & 0 & 0 & 0 & 0 & 0 & 0 & 0 & 0 & 0 \\
 0 & 0 & 0 & 0 & 0 & 0 & 0 & 0 & 0 & 0 & 0 & 0 & 0 & 0 & 0 & 0 & 0 & 0 & 0 & 0 \\
\hline
 0 & 0 & 0 & 0 & 0 & 0 & 0 & 0 & 0 & 0 & 0 & 0 & 0 & 0 & 0 & 0 & -\frac{1764}{2431} & \frac{11025}{4862} & 0 & 0 \\
 0 & 0 & 0 & 0 & 0 & 0 & 0 & 0 & 0 & 0 & 0 & 0 & 0 & 0 & 0 & 0 & \frac{441}{221} & 0 & 0 & 0 \\
 0 & 0 & 0 & 0 & 0 & 0 & 0 & 0 & 0 & 0 & 0 & 0 & 0 & 0 & 0 & 0 & \frac{49}{221} & 0 & \frac{1225}{2431} & \frac{11025}{4862} \\
 0 & 0 & 0 & 0 & 0 & 0 & 0 & 0 & 0 & 0 & 0 & 0 & 0 & 0 & 0 & 0 & -\frac{49}{221} & 0 & \frac{490}{221} & 0 \\
\hline
 0 & 0 & 0 & 0 & 0 & 0 & 0 & 0 & 0 & 0 & 0 & 0 & -\frac{1764}{2431} & \frac{441}{221} & \frac{49}{221} & -\frac{49}{221} & 0 & 0 & 0 & 0 \\
 0 & 0 & 0 & 0 & 0 & 0 & 0 & 0 & 0 & 0 & 0 & 0 & \frac{11025}{4862} & 0 & 0 & 0 & 0 & 0 & 0 & 0 \\
 0 & 0 & 0 & 0 & 0 & 0 & 0 & 0 & 0 & 0 & 0 & 0 & 0 & 0 & \frac{1225}{2431} & \frac{490}{221} & 0 & 0 & 0 & 0 \\
 0 & 0 & 0 & 0 & 0 & 0 & 0 & 0 & 0 & 0 & 0 & 0 & 0 & 0 & \frac{11025}{4862} & 0 & 0 & 0 & 0 & 0 \\
\end{array}
\right]
$
}
\caption{This is the matrix $\left( \mathcal{C}_{5}\right)_{9}^{\check{\Sigma}}$, which defines the coefficient $\left(a_{5}\right)_{9}^{\check{\Sigma}} = \left< \mathcal{M}_{\ell} \right| \left( \mathcal{C}_{5}\right)_{9}^{\check{\Sigma}} \left| \mathcal{M}_{\ell} \right>$ that appears in the observable $\check{\Sigma}$ for a truncation at $L=5$.}
\label{tab:SigmaMatrix9}
\end{table}

\clearpage

\begin{table}[h]
\centering
\scalebox{.78}{
$
\left[
\begin{array}{c|ccc|cccc|cccc|cccc|cccc}
 0 & 0 & 0 & 0 & 0 & 0 & 0 & 0 & 0 & 0 & 0 & 0 & 0 & 0 & 0 & 0 & 0 & 0 & 0 & 0 \\
\hline
 0 & 0 & 0 & 0 & 0 & 0 & 0 & 0 & 0 & 0 & 0 & 0 & 0 & 0 & 0 & 0 & 0 & 0 & 0 & 0 \\
 0 & 0 & 0 & 0 & 0 & 0 & 0 & 0 & 0 & 0 & 0 & 0 & 0 & 0 & 0 & 0 & 0 & 0 & 0 & 0 \\
 0 & 0 & 0 & 0 & 0 & 0 & 0 & 0 & 0 & 0 & 0 & 0 & 0 & 0 & 0 & 0 & 0 & 0 & 0 & 0 \\
\hline
 0 & 0 & 0 & 0 & 0 & 0 & 0 & 0 & 0 & 0 & 0 & 0 & 0 & 0 & 0 & 0 & 0 & 0 & 0 & 0 \\
 0 & 0 & 0 & 0 & 0 & 0 & 0 & 0 & 0 & 0 & 0 & 0 & 0 & 0 & 0 & 0 & 0 & 0 & 0 & 0 \\
 0 & 0 & 0 & 0 & 0 & 0 & 0 & 0 & 0 & 0 & 0 & 0 & 0 & 0 & 0 & 0 & 0 & 0 & 0 & 0 \\
 0 & 0 & 0 & 0 & 0 & 0 & 0 & 0 & 0 & 0 & 0 & 0 & 0 & 0 & 0 & 0 & 0 & 0 & 0 & 0 \\
\hline
 0 & 0 & 0 & 0 & 0 & 0 & 0 & 0 & 0 & 0 & 0 & 0 & 0 & 0 & 0 & 0 & 0 & 0 & 0 & 0 \\
 0 & 0 & 0 & 0 & 0 & 0 & 0 & 0 & 0 & 0 & 0 & 0 & 0 & 0 & 0 & 0 & 0 & 0 & 0 & 0 \\
 0 & 0 & 0 & 0 & 0 & 0 & 0 & 0 & 0 & 0 & 0 & 0 & 0 & 0 & 0 & 0 & 0 & 0 & 0 & 0 \\
 0 & 0 & 0 & 0 & 0 & 0 & 0 & 0 & 0 & 0 & 0 & 0 & 0 & 0 & 0 & 0 & 0 & 0 & 0 & 0 \\
\hline
 0 & 0 & 0 & 0 & 0 & 0 & 0 & 0 & 0 & 0 & 0 & 0 & 0 & 0 & 0 & 0 & 0 & 0 & 0 & 0 \\
 0 & 0 & 0 & 0 & 0 & 0 & 0 & 0 & 0 & 0 & 0 & 0 & 0 & 0 & 0 & 0 & 0 & 0 & 0 & 0 \\
 0 & 0 & 0 & 0 & 0 & 0 & 0 & 0 & 0 & 0 & 0 & 0 & 0 & 0 & 0 & 0 & 0 & 0 & 0 & 0 \\
 0 & 0 & 0 & 0 & 0 & 0 & 0 & 0 & 0 & 0 & 0 & 0 & 0 & 0 & 0 & 0 & 0 & 0 & 0 & 0 \\
\hline
 0 & 0 & 0 & 0 & 0 & 0 & 0 & 0 & 0 & 0 & 0 & 0 & 0 & 0 & 0 & 0 & -\frac{3087}{4199} & \frac{10584}{4199} & \frac{441}{4199} & -\frac{441}{4199} \\
 0 & 0 & 0 & 0 & 0 & 0 & 0 & 0 & 0 & 0 & 0 & 0 & 0 & 0 & 0 & 0 & \frac{10584}{4199} & 0 & 0 & 0 \\
 0 & 0 & 0 & 0 & 0 & 0 & 0 & 0 & 0 & 0 & 0 & 0 & 0 & 0 & 0 & 0 & \frac{441}{4199} & 0 & \frac{2205}{4199} & \frac{11025}{4199} \\
 0 & 0 & 0 & 0 & 0 & 0 & 0 & 0 & 0 & 0 & 0 & 0 & 0 & 0 & 0 & 0 & -\frac{441}{4199} & 0 & \frac{11025}{4199} & 0 \\
\end{array}
\right]
$
}
\caption{This is the matrix $\left( \mathcal{C}_{5}\right)_{10}^{\check{\Sigma}}$, which defines the coefficient $\left(a_{5}\right)_{10}^{\check{\Sigma}} = \left< \mathcal{M}_{\ell} \right| \left( \mathcal{C}_{5}\right)_{10}^{\check{\Sigma}} \left| \mathcal{M}_{\ell} \right>$ that appears in the observable $\check{\Sigma}$ for a truncation at $L=5$.}
\label{tab:SigmaMatrix10}
\end{table}

\begin{table}[h]
\centering
\scalebox{.64}{
$
\left[
\begin{array}{c|ccc|cccc|cccc|cccc|cccc}
 0 & \frac{3i}{2} & -\frac{3i}{2} & 0 & 0 & 0 & 0 & 0 & 0 & 0 & 0 & 0 & 0 & 0 & 0 & 0 & 0 & 0 & 0 & 0 \\
\hline
 -\frac{3i}{2} & 0 & 0 & 0 & \frac{36i}{5} & \frac{6i}{5} & -\frac{9i}{2} & 0 & 0 & 0 & 0 & 0 & 0 & 0 & 0 & 0 & 0 & 0 & 0 & 0 \\
 \frac{3i}{2} & 0 & 0 & 0 & 0 & 0 & -\frac{27i}{10} & \frac{6i}{5} & 0 & 0 & 0 & 0 & 0 & 0 & 0 & 0 & 0 & 0 & 0 & 0 \\
 0 & 0 & 0 & 0 & 0 & \frac{3i}{2} & 0 & \frac{3i}{2} & 0 & 0 & 0 & 0 & 0 & 0 & 0 & 0 & 0 & 0 & 0 & 0 \\
\hline
 0 & -\frac{36i}{5} & 0 & 0 & 0 & 0 & 0 & 0 & \frac{135i}{7} & \frac{108i}{35} & -9 i & 0 & 0 & 0 & 0 & 0 & 0 & 0 & 0 & 0 \\
 0 & -\frac{6i}{5} & 0 & -\frac{3i}{2} & 0 & 0 & 0 & 0 & 0 & -\frac{27i}{10} & 0 & 0 & 0 & 0 & 0 & 0 & 0 & 0 & 0 & 0 \\
 0 & \frac{9i}{2} & \frac{27i}{10} & 0 & 0 & 0 & 0 & 0 & 0 & 0 & -\frac{72i}{7} & \frac{108i}{35} & 0 & 0 & 0 & 0 & 0 & 0 & 0 & 0 \\
 0 & 0 & -\frac{6i}{5} & -\frac{3i}{2} & 0 & 0 & 0 & 0 & 0 & \frac{9i}{2} & 0 & \frac{36i}{5} & 0 & 0 & 0 & 0 & 0 & 0 & 0 & 0 \\
\end{array}
\right]
$
}
\caption{This is the matrix $\left( \mathcal{C}_{5}\right)_{1}^{\check{T}}$, which defines the coefficient $\left(a_{5}\right)_{1}^{\check{T}} = \left< \mathcal{M}_{\ell} \right| \left( \mathcal{C}_{5}\right)_{1}^{\check{T}} \left| \mathcal{M}_{\ell} \right>$ that appears in the observable $\check{T}$ for a truncation at $L=5$.}
\label{tab:TObsMatrix1}
\end{table}

\clearpage

\begin{table}[h]
\centering
\scalebox{.62}{
$
\left[
\begin{array}{c|ccc|cccc|cccc|cccc|cccc}
 0 & 0 & 0 & 0 & 2 i & -\frac{ i }{2} & -2 i & -\frac{ i }{2} & 0 & 0 & 0 & 0 & 0 & 0 & 0 & 0 & 0 & 0 & 0 & 0 \\
\hline
 0 & 0 & -2 i & \frac{ i }{2} & 0 & 0 & 0 & 0 & \frac{60 i }{7} & \frac{15 i }{14} & -\frac{36 i }{7} & -\frac{6 i }{7} & 0 & 0 & 0 & 0 & 0 & 0 & 0 & 0 \\
 0 & 2 i & 0 & -\frac{ i }{2} & 0 & 0 & 0 & 0 & 0 & -\frac{ i }{2} & -\frac{24 i }{7} & \frac{10 i }{7} & 0 & 0 & 0 & 0 & 0 & 0 & 0 & 0 \\
 0 & -\frac{ i }{2} & \frac{ i }{2} & 0 & 0 & 0 & 0 & 0 & 0 & 2 i & 0 & 2 i & 0 & 0 & 0 & 0 & 0 & 0 & 0 & 0 \\
\hline
 -2 i & 0 & 0 & 0 & 0 & \frac{10 i }{7} & -\frac{36 i }{7} & \frac{6 i }{7} & 0 & 0 & 0 & 0 & \frac{150 i }{7} & 3 i & -\frac{200 i }{21} & -\frac{25 i }{21} & 0 & 0 & 0 & 0 \\
 \frac{ i }{2} & 0 & 0 & 0 & -\frac{10 i }{7} & 0 & -\frac{ i }{2} & -2 i & 0 & 0 & 0 & 0 & 0 & -\frac{24 i }{7} & 0 & 0 & 0 & 0 & 0 & 0 \\
 2 i & 0 & 0 & 0 & \frac{36 i }{7} & \frac{ i }{2} & 0 & \frac{15 i }{14} & 0 & 0 & 0 & 0 & 0 & -\frac{6 i }{7} & -\frac{250 i }{21} & \frac{10 i }{3} & 0 & 0 & 0 & 0 \\
 \frac{ i }{2} & 0 & 0 & 0 & -\frac{6 i }{7} & 2 i & -\frac{15 i }{14} & 0 & 0 & 0 & 0 & 0 & 0 & \frac{36 i }{7} & 0 & \frac{60 i }{7} & 0 & 0 & 0 & 0 \\
\end{array}
\right]
$
}
\caption{This is the matrix $\left( \mathcal{C}_{5}\right)_{2}^{\check{T}}$, which defines the coefficient $\left(a_{5}\right)_{2}^{\check{T}} = \left< \mathcal{M}_{\ell} \right| \left( \mathcal{C}_{5}\right)_{2}^{\check{T}} \left| \mathcal{M}_{\ell} \right>$ that appears in the observable $\check{T}$ for a truncation at $L=5$.}
\label{tab:TObsMatrix2}
\end{table}

\begin{table}[h]
\centering
\scalebox{.635}{
$
\left[
\begin{array}{c|ccc|cccc|cccc|cccc|cccc}
 0 & 0 & 0 & 0 & 0 & 0 & 0 & 0 & \frac{5i}{2} & -i & -\frac{5i}{2} & -i & 0 & 0 & 0 & 0 & 0 & 0 & 0 & 0 \\
\hline
 0 & 0 & 0 & 0 & \frac{3i}{10} & -\frac{6i}{5} & -3 i & 0 & 0 & 0 & 0 & 0 & 10 i & i & -\frac{35i}{6} & -\frac{5i}{3} & 0 & 0 & 0 & 0 \\
 0 & 0 & 0 & 0 & \frac{5i}{2} & 0 & \frac{i}{5} & -\frac{6i}{5} & 0 & 0 & 0 & 0 & 0 & -i & -\frac{25i}{6} & \frac{5i}{3} & 0 & 0 & 0 & 0 \\
 0 & 0 & 0 & 0 & -i & 0 & i & 0 & 0 & 0 & 0 & 0 & 0 & \frac{5i}{2} & 0 & \frac{5i}{2} & 0 & 0 & 0 & 0 \\
\hline
 0 & -\frac{3i}{10} & -\frac{5i}{2} & i & 0 & 0 & 0 & 0 & \frac{5i}{3} & \frac{6i}{5} & -7 i & 0 & 0 & 0 & 0 & 0 & \frac{525i}{22} & \frac{100i}{33} & -\frac{225i}{22} & -\frac{25i}{11} \\
 0 & \frac{6i}{5} & 0 & 0 & 0 & 0 & 0 & 0 & -\frac{5i}{3} & \frac{i}{5} & -i & -\frac{5i}{2} & 0 & 0 & 0 & 0 & 0 & -\frac{25i}{6} & 0 & 0 \\
 0 & 3 i & -\frac{i}{5} & -i & 0 & 0 & 0 & 0 & \frac{35i}{6} & 0 & -\frac{i}{2} & \frac{6i}{5} & 0 & 0 & 0 & 0 & 0 & -\frac{5i}{3} & -\frac{150i}{11} & \frac{40i}{11} \\
 0 & 0 & \frac{6i}{5} & 0 & 0 & 0 & 0 & 0 & -\frac{5i}{3} & 3 i & -i & \frac{3i}{10} & 0 & 0 & 0 & 0 & 0 & \frac{35i}{6} & 0 & 10 i \\
\end{array}
\right]
$
}
\caption{This is the matrix $\left( \mathcal{C}_{5}\right)_{3}^{\check{T}}$, which defines the coefficient $\left(a_{5}\right)_{3}^{\check{T}} = \left< \mathcal{M}_{\ell} \right| \left( \mathcal{C}_{5}\right)_{3}^{\check{T}} \left| \mathcal{M}_{\ell} \right>$ that appears in the observable $\check{T}$ for a truncation at $L=5$.}
\label{tab:TObsMatrix3}
\end{table}

\clearpage

\begin{table}[h]
\centering
\scalebox{.605}{
$
\left[
\begin{array}{c|ccc|cccc|cccc|cccc|cccc}
 0 & 0 & 0 & 0 & 0 & 0 & 0 & 0 & 0 & 0 & 0 & 0 & 3 i & -\frac{3i}{2} & -3 i & -\frac{3i}{2} & 0 & 0 & 0 & 0 \\
\hline
 0 & 0 & 0 & 0 & 0 & 0 & 0 & 0 & \frac{3i}{7} & -\frac{18i}{7} & -\frac{27i}{7} & -\frac{9i}{14} & 0 & 0 & 0 & 0 & \frac{126i}{11} & \frac{21i}{22} & -\frac{72i}{11} & -\frac{27i}{11} \\
 0 & 0 & 0 & 0 & 0 & 0 & 0 & 0 & 3 i & 0 & \frac{3i}{7} & -\frac{27i}{14} & 0 & 0 & 0 & 0 & 0 & -\frac{3i}{2} & -\frac{54i}{11} & \frac{21i}{11} \\
 0 & 0 & 0 & 0 & 0 & 0 & 0 & 0 & -\frac{3i}{2} & 0 & \frac{3i}{2} & 0 & 0 & 0 & 0 & 0 & 0 & 3 i & 0 & 3 i \\
\hline
 0 & 0 & 0 & 0 & 0 & -\frac{27i}{14} & -\frac{27i}{7} & \frac{9i}{14} & 0 & 0 & 0 & 0 & \frac{18i}{7} & \frac{9i}{11} & -\frac{648i}{77} & -\frac{81i}{77} & 0 & 0 & 0 & 0 \\
 0 & 0 & 0 & 0 & \frac{27i}{14} & 0 & 0 & 0 & 0 & 0 & 0 & 0 & -\frac{21i}{11} & \frac{3i}{7} & -\frac{3i}{2} & -3 i & 0 & 0 & 0 & 0 \\
 0 & 0 & 0 & 0 & \frac{27i}{7} & 0 & 0 & -\frac{18i}{7} & 0 & 0 & 0 & 0 & \frac{72i}{11} & -\frac{9i}{14} & -\frac{54i}{77} & \frac{27i}{22} & 0 & 0 & 0 & 0 \\
 0 & 0 & 0 & 0 & -\frac{9i}{14} & 0 & \frac{18i}{7} & 0 & 0 & 0 & 0 & 0 & -\frac{27i}{11} & \frac{27i}{7} & -\frac{21i}{22} & \frac{3i}{7} & 0 & 0 & 0 & 0 \\
\end{array}
\right]
$
}
\caption{This is the matrix $\left( \mathcal{C}_{5}\right)_{4}^{\check{T}}$, which defines the coefficient $\left(a_{5}\right)_{4}^{\check{T}} = \left< \mathcal{M}_{\ell} \right| \left( \mathcal{C}_{5}\right)_{4}^{\check{T}} \left| \mathcal{M}_{\ell} \right>$ that appears in the observable $\check{T}$ for a truncation at $L=5$.}
\label{tab:TObsMatrix4}
\end{table}

\begin{table}[h]
\centering
\scalebox{.655}{
$
\left[
\begin{array}{c|ccc|cccc|cccc|cccc|cccc}
 0 & 0 & 0 & 0 & 0 & 0 & 0 & 0 & 0 & 0 & 0 & 0 & 0 & 0 & 0 & 0 & \frac{7i}{2} & -2 i & -\frac{7i}{2} & -2 i \\
\hline
 0 & 0 & 0 & 0 & 0 & 0 & 0 & 0 & 0 & 0 & 0 & 0 & \frac{i}{2} & -4 i & -\frac{14i}{3} & -\frac{4i}{3} & 0 & 0 & 0 & 0 \\
 0 & 0 & 0 & 0 & 0 & 0 & 0 & 0 & 0 & 0 & 0 & 0 & \frac{7i}{2} & 0 & \frac{2i}{3} & -\frac{8i}{3} & 0 & 0 & 0 & 0 \\
 0 & 0 & 0 & 0 & 0 & 0 & 0 & 0 & 0 & 0 & 0 & 0 & -2 i & 0 & 2 i & 0 & 0 & 0 & 0 & 0 \\
\hline
 0 & 0 & 0 & 0 & 0 & 0 & 0 & 0 & \frac{i}{21} & -\frac{30i}{7} & -5 i & 0 & 0 & 0 & 0 & 0 & \frac{42i}{13} & \frac{16i}{39} & -\frac{126i}{13} & -\frac{28i}{13} \\
 0 & 0 & 0 & 0 & 0 & 0 & 0 & 0 & \frac{8i}{3} & 0 & 0 & 0 & 0 & 0 & 0 & 0 & -\frac{28i}{13} & \frac{2i}{3} & -2 i & -\frac{7i}{2} \\
 0 & 0 & 0 & 0 & 0 & 0 & 0 & 0 & \frac{14i}{3} & 0 & \frac{2i}{7} & -\frac{30i}{7} & 0 & 0 & 0 & 0 & \frac{189i}{26} & -\frac{4i}{3} & -\frac{21i}{26} & \frac{16i}{13} \\
 0 & 0 & 0 & 0 & 0 & 0 & 0 & 0 & -\frac{4i}{3} & 0 & 4 i & 0 & 0 & 0 & 0 & 0 & -\frac{42i}{13} & \frac{14i}{3} & -\frac{12i}{13} & \frac{i}{2} \\
\end{array}
\right]
$
}
\caption{This is the matrix $\left( \mathcal{C}_{5}\right)_{5}^{\check{T}}$, which defines the coefficient $\left(a_{5}\right)_{5}^{\check{T}} = \left< \mathcal{M}_{\ell} \right| \left( \mathcal{C}_{5}\right)_{5}^{\check{T}} \left| \mathcal{M}_{\ell} \right>$ that appears in the observable $\check{T}$ for a truncation at $L=5$.}
\label{tab:TObsMatrix5}
\end{table}

\clearpage

\begin{table}[h]
\centering
\scalebox{.642}{
$
\left[
\begin{array}{c|ccc|cccc|cccc|cccc|cccc}
 0 & 0 & 0 & 0 & 0 & 0 & 0 & 0 & 0 & 0 & 0 & 0 & 0 & 0 & 0 & 0 & 0 & 0 & 0 & 0 \\
\hline
 0 & 0 & 0 & 0 & 0 & 0 & 0 & 0 & 0 & 0 & 0 & 0 & 0 & 0 & 0 & 0 & \frac{6i}{11} & -\frac{60i}{11} & -\frac{60i}{11} & -\frac{45i}{22} \\
 0 & 0 & 0 & 0 & 0 & 0 & 0 & 0 & 0 & 0 & 0 & 0 & 0 & 0 & 0 & 0 & 4 i & 0 & \frac{10i}{11} & -\frac{75i}{22} \\
 0 & 0 & 0 & 0 & 0 & 0 & 0 & 0 & 0 & 0 & 0 & 0 & 0 & 0 & 0 & 0 & -\frac{5i}{2} & 0 & \frac{5i}{2} & 0 \\
\hline
 0 & 0 & 0 & 0 & 0 & 0 & 0 & 0 & 0 & 0 & 0 & 0 & 0 & -\frac{75i}{11} & -\frac{200i}{33} & -\frac{25i}{33} & 0 & 0 & 0 & 0 \\
 0 & 0 & 0 & 0 & 0 & 0 & 0 & 0 & 0 & 0 & 0 & 0 & \frac{75i}{22} & 0 & 0 & 0 & 0 & 0 & 0 & 0 \\
 0 & 0 & 0 & 0 & 0 & 0 & 0 & 0 & 0 & 0 & 0 & 0 & \frac{60i}{11} & 0 & \frac{20i}{33} & -\frac{200i}{33} & 0 & 0 & 0 & 0 \\
 0 & 0 & 0 & 0 & 0 & 0 & 0 & 0 & 0 & 0 & 0 & 0 & -\frac{45i}{22} & 0 & \frac{60i}{11} & 0 & 0 & 0 & 0 & 0 \\
\end{array}
\right]
$
}
\caption{This is the matrix $\left( \mathcal{C}_{5}\right)_{6}^{\check{T}}$, which defines the coefficient $\left(a_{5}\right)_{6}^{\check{T}} = \left< \mathcal{M}_{\ell} \right| \left( \mathcal{C}_{5}\right)_{6}^{\check{T}} \left| \mathcal{M}_{\ell} \right>$ that appears in the observable $\check{T}$ for a truncation at $L=5$.}
\label{tab:TObsMatrix6}
\end{table}

\begin{table}[h]
\centering
\scalebox{.627}{
$
\left[
\begin{array}{c|ccc|cccc|cccc|cccc|cccc}
 0 & 0 & 0 & 0 & 0 & 0 & 0 & 0 & 0 & 0 & 0 & 0 & 0 & 0 & 0 & 0 & 0 & 0 & 0 & 0 \\
\hline
 0 & 0 & 0 & 0 & 0 & 0 & 0 & 0 & 0 & 0 & 0 & 0 & 0 & 0 & 0 & 0 & 0 & 0 & 0 & 0 \\
 0 & 0 & 0 & 0 & 0 & 0 & 0 & 0 & 0 & 0 & 0 & 0 & 0 & 0 & 0 & 0 & 0 & 0 & 0 & 0 \\
 0 & 0 & 0 & 0 & 0 & 0 & 0 & 0 & 0 & 0 & 0 & 0 & 0 & 0 & 0 & 0 & 0 & 0 & 0 & 0 \\
\hline
 0 & 0 & 0 & 0 & 0 & 0 & 0 & 0 & 0 & 0 & 0 & 0 & 0 & 0 & 0 & 0 & -\frac{27i}{286} & -\frac{1350i}{143} & -\frac{2025i}{286} & -\frac{225i}{143} \\
 0 & 0 & 0 & 0 & 0 & 0 & 0 & 0 & 0 & 0 & 0 & 0 & 0 & 0 & 0 & 0 & \frac{54i}{13} & 0 & 0 & 0 \\
 0 & 0 & 0 & 0 & 0 & 0 & 0 & 0 & 0 & 0 & 0 & 0 & 0 & 0 & 0 & 0 & \frac{81i}{13} & 0 & \frac{135i}{143} & -\frac{1125i}{143} \\
 0 & 0 & 0 & 0 & 0 & 0 & 0 & 0 & 0 & 0 & 0 & 0 & 0 & 0 & 0 & 0 & -\frac{36i}{13} & 0 & \frac{90i}{13} & 0 \\
\end{array}
\right]
$
}
\caption{This is the matrix $\left( \mathcal{C}_{5}\right)_{7}^{\check{T}}$, which defines the coefficient $\left(a_{5}\right)_{7}^{\check{T}} = \left< \mathcal{M}_{\ell} \right| \left( \mathcal{C}_{5}\right)_{7}^{\check{T}} \left| \mathcal{M}_{\ell} \right>$ that appears in the observable $\check{T}$ for a truncation at $L=5$.}
\label{tab:TObsMatrix7}
\end{table}

\clearpage

\begin{table}[h]
\centering
\scalebox{.71}{
$
\left[
\begin{array}{c|ccc|cccc|cccc|cccc|cccc}
 0 & 0 & 0 & 0 & 0 & 0 & 0 & 0 & 0 & 0 & 0 & 0 & 0 & 0 & 0 & 0 & 0 & 0 & 0 & 0 \\
\hline
 0 & 0 & 0 & 0 & 0 & 0 & 0 & 0 & 0 & 0 & 0 & 0 & 0 & 0 & 0 & 0 & 0 & 0 & 0 & 0 \\
 0 & 0 & 0 & 0 & 0 & 0 & 0 & 0 & 0 & 0 & 0 & 0 & 0 & 0 & 0 & 0 & 0 & 0 & 0 & 0 \\
 0 & 0 & 0 & 0 & 0 & 0 & 0 & 0 & 0 & 0 & 0 & 0 & 0 & 0 & 0 & 0 & 0 & 0 & 0 & 0 \\
\hline
 0 & 0 & 0 & 0 & 0 & 0 & 0 & 0 & 0 & 0 & 0 & 0 & 0 & 0 & 0 & 0 & 0 & 0 & 0 & 0 \\
 0 & 0 & 0 & 0 & 0 & 0 & 0 & 0 & 0 & 0 & 0 & 0 & 0 & 0 & 0 & 0 & 0 & 0 & 0 & 0 \\
 0 & 0 & 0 & 0 & 0 & 0 & 0 & 0 & 0 & 0 & 0 & 0 & 0 & 0 & 0 & 0 & 0 & 0 & 0 & 0 \\
 0 & 0 & 0 & 0 & 0 & 0 & 0 & 0 & 0 & 0 & 0 & 0 & 0 & 0 & 0 & 0 & 0 & 0 & 0 & 0 \\
\end{array}
\right]
$
}
\caption{This is the matrix $\left( \mathcal{C}_{5}\right)_{8}^{\check{T}}$, which defines the coefficient $\left(a_{5}\right)_{8}^{\check{T}} = \left< \mathcal{M}_{\ell} \right| \left( \mathcal{C}_{5}\right)_{8}^{\check{T}} \left| \mathcal{M}_{\ell} \right>$ that appears in the observable $\check{T}$ for a truncation at $L=5$.}
\label{tab:TObsMatrix8}
\end{table}

\begin{table}[h]
\centering
\scalebox{.7425}{
$
\left[
\begin{array}{c|ccc|cccc|cccc|cccc|cccc}
 0 & 0 & 0 & 0 & 0 & 0 & 0 & 0 & 0 & 0 & 0 & 0 & 0 & 0 & 0 & 0 & 0 & 0 & 0 & 0 \\
\hline
 0 & 0 & 0 & 0 & 0 & 0 & 0 & 0 & 0 & 0 & 0 & 0 & 0 & 0 & 0 & 0 & 0 & 0 & 0 & 0 \\
 0 & 0 & 0 & 0 & 0 & 0 & 0 & 0 & 0 & 0 & 0 & 0 & 0 & 0 & 0 & 0 & 0 & 0 & 0 & 0 \\
 0 & 0 & 0 & 0 & 0 & 0 & 0 & 0 & 0 & 0 & 0 & 0 & 0 & 0 & 0 & 0 & 0 & 0 & 0 & 0 \\
\hline
 0 & 0 & 0 & 0 & 0 & 0 & 0 & 0 & 0 & 0 & 0 & 0 & 0 & 0 & 0 & 0 & 0 & 0 & 0 & 0 \\
 0 & 0 & 0 & 0 & 0 & 0 & 0 & 0 & 0 & 0 & 0 & 0 & 0 & 0 & 0 & 0 & 0 & 0 & 0 & 0 \\
 0 & 0 & 0 & 0 & 0 & 0 & 0 & 0 & 0 & 0 & 0 & 0 & 0 & 0 & 0 & 0 & 0 & 0 & 0 & 0 \\
 0 & 0 & 0 & 0 & 0 & 0 & 0 & 0 & 0 & 0 & 0 & 0 & 0 & 0 & 0 & 0 & 0 & 0 & 0 & 0 \\
\end{array}
\right]
$
}
\caption{This is the matrix $\left( \mathcal{C}_{5}\right)_{9}^{\check{T}}$, which defines the coefficient $\left(a_{5}\right)_{9}^{\check{T}} = \left< \mathcal{M}_{\ell} \right| \left( \mathcal{C}_{5}\right)_{9}^{\check{T}} \left| \mathcal{M}_{\ell} \right>$ that appears in the observable $\check{T}$ for a truncation at $L=5$.}
\label{tab:TObsMatrix9}
\end{table}

\clearpage

\begin{table}[h]
\centering
\scalebox{.75}{
$
\left[
\begin{array}{c|ccc|cccc|cccc|cccc|cccc}
 0 & 0 & 0 & 0 & 0 & 0 & 0 & 0 & 0 & 0 & 0 & 0 & 0 & 0 & 0 & 0 & 0 & 0 & 0 & 0 \\
\hline
 0 & 0 & 0 & 0 & 0 & 0 & 0 & 0 & 0 & 0 & 0 & 0 & 0 & 0 & 0 & 0 & 0 & 0 & 0 & 0 \\
 0 & 0 & 0 & 0 & 0 & 0 & 0 & 0 & 0 & 0 & 0 & 0 & 0 & 0 & 0 & 0 & 0 & 0 & 0 & 0 \\
 0 & 0 & 0 & 0 & 0 & 0 & 0 & 0 & 0 & 0 & 0 & 0 & 0 & 0 & 0 & 0 & 0 & 0 & 0 & 0 \\
\hline
 0 & 0 & 0 & 0 & 0 & 0 & 0 & 0 & 0 & 0 & 0 & 0 & 0 & 0 & 0 & 0 & 0 & 0 & 0 & 0 \\
 0 & 0 & 0 & 0 & 0 & 0 & 0 & 0 & 0 & 0 & 0 & 0 & 0 & 0 & 0 & 0 & 0 & 0 & 0 & 0 \\
 0 & 0 & 0 & 0 & 0 & 0 & 0 & 0 & 0 & 0 & 0 & 0 & 0 & 0 & 0 & 0 & 0 & 0 & 0 & 0 \\
 0 & 0 & 0 & 0 & 0 & 0 & 0 & 0 & 0 & 0 & 0 & 0 & 0 & 0 & 0 & 0 & 0 & 0 & 0 & 0 \\
\hline
 0 & 0 & 0 & 0 & 0 & 0 & 0 & 0 & 0 & 0 & 0 & 0 & 0 & 0 & 0 & 0 & 0 & 0 & 0 & 0 \\
 0 & 0 & 0 & 0 & 0 & 0 & 0 & 0 & 0 & 0 & 0 & 0 & 0 & 0 & 0 & 0 & 0 & 0 & 0 & 0 \\
 0 & 0 & 0 & 0 & 0 & 0 & 0 & 0 & 0 & 0 & 0 & 0 & 0 & 0 & 0 & 0 & 0 & 0 & 0 & 0 \\
 0 & 0 & 0 & 0 & 0 & 0 & 0 & 0 & 0 & 0 & 0 & 0 & 0 & 0 & 0 & 0 & 0 & 0 & 0 & 0 \\
\hline
 0 & 0 & 0 & 0 & 0 & 0 & 0 & 0 & 0 & 0 & 0 & 0 & 0 & 0 & 0 & 0 & 0 & 0 & 0 & 0 \\
 0 & 0 & 0 & 0 & 0 & 0 & 0 & 0 & 0 & 0 & 0 & 0 & 0 & 0 & 0 & 0 & 0 & 0 & 0 & 0 \\
 0 & 0 & 0 & 0 & 0 & 0 & 0 & 0 & 0 & 0 & 0 & 0 & 0 & 0 & 0 & 0 & 0 & 0 & 0 & 0 \\
 0 & 0 & 0 & 0 & 0 & 0 & 0 & 0 & 0 & 0 & 0 & 0 & 0 & 0 & 0 & 0 & 0 & 0 & 0 & 0 \\
\hline
 0 & 0 & 0 & 0 & 0 & 0 & 0 & 0 & 0 & 0 & 0 & 0 & 0 & 0 & 0 & 0 & 0 & -\frac{95256i}{4199} & -\frac{47628i}{4199} & \frac{3969i}{4199} \\
 0 & 0 & 0 & 0 & 0 & 0 & 0 & 0 & 0 & 0 & 0 & 0 & 0 & 0 & 0 & 0 & \frac{95256i}{4199} & 0 & 0 & 0 \\
 0 & 0 & 0 & 0 & 0 & 0 & 0 & 0 & 0 & 0 & 0 & 0 & 0 & 0 & 0 & 0 & \frac{47628i}{4199} & 0 & 0 & -\frac{99225i}{4199} \\
 0 & 0 & 0 & 0 & 0 & 0 & 0 & 0 & 0 & 0 & 0 & 0 & 0 & 0 & 0 & 0 & -\frac{3969i}{4199} & 0 & \frac{99225i}{4199} & 0 \\
\end{array}
\right]
$
}
\caption{This is the matrix $\left( \mathcal{C}_{5}\right)_{10}^{\check{T}}$, which defines the coefficient $\left(a_{5}\right)_{10}^{\check{T}} = \left< \mathcal{M}_{\ell} \right| \left( \mathcal{C}_{5}\right)_{10}^{\check{T}} \left| \mathcal{M}_{\ell} \right>$ that appears in the observable $\check{T}$ for a truncation at $L=5$.}
\label{tab:TObsMatrix10}
\end{table}

\begin{table}[h]
\centering
\scalebox{.635}{
$
\left[
\begin{array}{c|ccc|cccc|cccc|cccc|cccc}
 0 & -\frac{3i}{2} & -\frac{i}{2} & -i & 0 & 0 & 0 & 0 & 0 & 0 & 0 & 0 & 0 & 0 & 0 & 0 & 0 & 0 & 0 & 0 \\
\hline
 \frac{3i}{2} & 0 & 0 & 0 & -\frac{36i}{5} & -\frac{6i}{5} & -\frac{9i}{10} & -\frac{18i}{5} & 0 & 0 & 0 & 0 & 0 & 0 & 0 & 0 & 0 & 0 & 0 & 0 \\
 \frac{i}{2} & 0 & 0 & 0 & 0 & 2 i & -\frac{27i}{10} & \frac{6i}{5} & 0 & 0 & 0 & 0 & 0 & 0 & 0 & 0 & 0 & 0 & 0 & 0 \\
 i & 0 & 0 & 0 & 0 & -\frac{i}{2} & 0 & \frac{3i}{2} & 0 & 0 & 0 & 0 & 0 & 0 & 0 & 0 & 0 & 0 & 0 & 0 \\
\hline
 0 & \frac{36i}{5} & 0 & 0 & 0 & 0 & 0 & 0 & -\frac{135i}{7} & -\frac{108i}{35} & -\frac{9i}{7} & -\frac{54i}{7} & 0 & 0 & 0 & 0 & 0 & 0 & 0 & 0 \\
 0 & \frac{6i}{5} & -2 i & \frac{i}{2} & 0 & 0 & 0 & 0 & 0 & \frac{27i}{10} & 0 & 0 & 0 & 0 & 0 & 0 & 0 & 0 & 0 & 0 \\
 0 & \frac{9i}{10} & \frac{27i}{10} & 0 & 0 & 0 & 0 & 0 & 0 & \frac{27i}{5} & -\frac{72i}{7} & \frac{108i}{35} & 0 & 0 & 0 & 0 & 0 & 0 & 0 & 0 \\
 0 & \frac{18i}{5} & -\frac{6i}{5} & -\frac{3i}{2} & 0 & 0 & 0 & 0 & 0 & -\frac{9i}{10} & 0 & \frac{36i}{5} & 0 & 0 & 0 & 0 & 0 & 0 & 0 & 0 \\
\end{array}
\right]
$
}
\caption{This is the matrix $\left( \mathcal{C}_{5}\right)_{1}^{\check{P}}$, which defines the coefficient $\left(a_{5}\right)_{1}^{\check{P}} = \left< \mathcal{M}_{\ell} \right| \left( \mathcal{C}_{5}\right)_{1}^{\check{P}} \left| \mathcal{M}_{\ell} \right>$ that appears in the observable $\check{P}$ for a truncation at $L=5$.}
\label{tab:PObsMatrix1}
\end{table}

\clearpage

\begin{table}[h]
\centering
\scalebox{.6265}{
$
\left[
\begin{array}{c|ccc|cccc|cccc|cccc|cccc}
 0 & 0 & 0 & 0 & -2 i & \frac{i}{2} & -i & -\frac{3i}{2} & 0 & 0 & 0 & 0 & 0 & 0 & 0 & 0 & 0 & 0 & 0 & 0 \\
\hline
 0 & 0 & 0 & -\frac{3i}{2} & 0 & 0 & 0 & 0 & -\frac{60i}{7} & -\frac{15i}{14} & -\frac{12i}{7} & -\frac{30i}{7} & 0 & 0 & 0 & 0 & 0 & 0 & 0 & 0 \\
 0 & 0 & 0 & -\frac{i}{2} & 0 & 0 & 0 & 0 & 0 & \frac{5i}{2} & -\frac{24i}{7} & \frac{10i}{7} & 0 & 0 & 0 & 0 & 0 & 0 & 0 & 0 \\
 0 & \frac{3i}{2} & \frac{i}{2} & 0 & 0 & 0 & 0 & 0 & 0 & -i & 0 & 2 i & 0 & 0 & 0 & 0 & 0 & 0 & 0 & 0 \\
\hline
 2 i & 0 & 0 & 0 & 0 & -\frac{10i}{7} & 0 & -\frac{30i}{7} & 0 & 0 & 0 & 0 & -\frac{150i}{7} & -3 i & -\frac{50i}{21} & -\frac{25i}{3} & 0 & 0 & 0 & 0 \\
 -\frac{i}{2} & 0 & 0 & 0 & \frac{10i}{7} & 0 & -\frac{5i}{2} & 0 & 0 & 0 & 0 & 0 & 0 & \frac{24i}{7} & 0 & 0 & 0 & 0 & 0 & 0 \\
 i & 0 & 0 & 0 & 0 & \frac{5i}{2} & 0 & \frac{15i}{14} & 0 & 0 & 0 & 0 & 0 & 6 i & -\frac{250i}{21} & \frac{10i}{3} & 0 & 0 & 0 & 0 \\
 \frac{3i}{2} & 0 & 0 & 0 & \frac{30i}{7} & 0 & -\frac{15i}{14} & 0 & 0 & 0 & 0 & 0 & 0 & -\frac{12i}{7} & 0 & \frac{60i}{7} & 0 & 0 & 0 & 0 \\
\end{array}
\right]
$
}
\caption{This is the matrix $\left( \mathcal{C}_{5}\right)_{2}^{\check{P}}$, which defines the coefficient $\left(a_{5}\right)_{2}^{\check{P}} = \left< \mathcal{M}_{\ell} \right| \left( \mathcal{C}_{5}\right)_{2}^{\check{P}} \left| \mathcal{M}_{\ell} \right>$ that appears in the observable $\check{P}$ for a truncation at $L=5$.}
\label{tab:PObsMatrix2}
\end{table}

\begin{table}[h]
\centering
\scalebox{.63}{
$
\left[
\begin{array}{c|ccc|cccc|cccc|cccc|cccc}
 0 & 0 & 0 & 0 & 0 & 0 & 0 & 0 & -\frac{5i}{2} & i & -\frac{3i}{2} & -2 i & 0 & 0 & 0 & 0 & 0 & 0 & 0 & 0 \\
\hline
 0 & 0 & 0 & 0 & -\frac{3i}{10} & \frac{6i}{5} & -\frac{3i}{5} & -\frac{12i}{5} & 0 & 0 & 0 & 0 & -10 i & -i & -\frac{5i}{2} & -5 i & 0 & 0 & 0 & 0 \\
 0 & 0 & 0 & 0 & -\frac{i}{2} & 0 & \frac{i}{5} & -\frac{6i}{5} & 0 & 0 & 0 & 0 & 0 & 3 i & -\frac{25i}{6} & \frac{5i}{3} & 0 & 0 & 0 & 0 \\
 0 & 0 & 0 & 0 & 2 i & 0 & i & 0 & 0 & 0 & 0 & 0 & 0 & -\frac{3i}{2} & 0 & \frac{5i}{2} & 0 & 0 & 0 & 0 \\
\hline
 0 & \frac{3i}{10} & \frac{i}{2} & -2 i & 0 & 0 & 0 & 0 & -\frac{5i}{3} & -\frac{6i}{5} & -i & -6 i & 0 & 0 & 0 & 0 & -\frac{525i}{22} & -\frac{100i}{33} & -\frac{75i}{22} & -\frac{100i}{11} \\
 0 & -\frac{6i}{5} & 0 & 0 & 0 & 0 & 0 & 0 & \frac{5i}{3} & -\frac{i}{5} & -3 i & -\frac{i}{2} & 0 & 0 & 0 & 0 & 0 & \frac{25i}{6} & 0 & 0 \\
 0 & \frac{3i}{5} & -\frac{i}{5} & -i & 0 & 0 & 0 & 0 & -\frac{5i}{6} & \frac{18i}{5} & -\frac{i}{2} & \frac{6i}{5} & 0 & 0 & 0 & 0 & 0 & \frac{20i}{3} & -\frac{150i}{11} & \frac{40i}{11} \\
 0 & \frac{12i}{5} & \frac{6i}{5} & 0 & 0 & 0 & 0 & 0 & 5 i & -\frac{3i}{5} & -i & \frac{3i}{10} & 0 & 0 & 0 & 0 & 0 & -\frac{5i}{2} & 0 & 10 i \\
\end{array}
\right]
$
}
\caption{This is the matrix $\left( \mathcal{C}_{5}\right)_{3}^{\check{P}}$, which defines the coefficient $\left(a_{5}\right)_{3}^{\check{P}} = \left< \mathcal{M}_{\ell} \right| \left( \mathcal{C}_{5}\right)_{3}^{\check{P}} \left| \mathcal{M}_{\ell} \right>$ that appears in the observable $\check{P}$ for a truncation at $L=5$.}
\label{tab:PObsMatrix3}
\end{table}

\clearpage

\begin{table}[h]
\centering
\scalebox{.6075}{
$
\left[
\begin{array}{c|ccc|cccc|cccc|cccc|cccc}
 0 & 0 & 0 & 0 & 0 & 0 & 0 & 0 & 0 & 0 & 0 & 0 & -3 i & \frac{3i}{2} & -2 i & -\frac{5i}{2} & 0 & 0 & 0 & 0 \\
\hline
 0 & 0 & 0 & 0 & 0 & 0 & 0 & 0 & -\frac{3i}{7} & \frac{18i}{7} & -\frac{9i}{7} & -\frac{45i}{14} & 0 & 0 & 0 & 0 & -\frac{126i}{11} & -\frac{2i}{22} & -\frac{36i}{11} & -\frac{63i}{11} \\
 0 & 0 & 0 & 0 & 0 & 0 & 0 & 0 & -i & 0 & \frac{3i}{7} & -\frac{27i}{14} & 0 & 0 & 0 & 0 & 0 & \frac{7i}{2} & -\frac{54i}{11} & \frac{2i}{11} \\
 0 & 0 & 0 & 0 & 0 & 0 & 0 & 0 & \frac{5i}{2} & 0 & \frac{3i}{2} & 0 & 0 & 0 & 0 & 0 & 0 & -2 i & 0 & 3 i \\
\hline
 0 & 0 & 0 & 0 & 0 & \frac{27i}{14} & 0 & -\frac{45i}{14} & 0 & 0 & 0 & 0 & -\frac{18i}{7} & -\frac{9i}{11} & -\frac{162i}{77} & -\frac{8i}{11} & 0 & 0 & 0 & 0 \\
 0 & 0 & 0 & 0 & -\frac{27i}{14} & 0 & 0 & 0 & 0 & 0 & 0 & 0 & \frac{2i}{11} & -\frac{3i}{7} & -\frac{7i}{2} & -i & 0 & 0 & 0 & 0 \\
 0 & 0 & 0 & 0 & 0 & 0 & 0 & -\frac{18i}{7} & 0 & 0 & 0 & 0 & -\frac{18i}{11} & \frac{9i}{2} & -\frac{54i}{77} & \frac{27i}{22} & 0 & 0 & 0 & 0 \\
 0 & 0 & 0 & 0 & \frac{45i}{14} & 0 & \frac{18i}{7} & 0 & 0 & 0 & 0 & 0 & \frac{63i}{11} & -\frac{9i}{7} & -\frac{2i}{22} & \frac{3i}{7} & 0 & 0 & 0 & 0 \\
\end{array}
\right]
$
}
\caption{This is the matrix $\left( \mathcal{C}_{5}\right)_{4}^{\check{P}}$, which defines the coefficient $\left(a_{5}\right)_{4}^{\check{P}} = \left< \mathcal{M}_{\ell} \right| \left( \mathcal{C}_{5}\right)_{4}^{\check{P}} \left| \mathcal{M}_{\ell} \right>$ that appears in the observable $\check{P}$ for a truncation at $L=5$.}
\label{tab:PObsMatrix4}
\end{table}

\begin{table}[h]
\centering
\scalebox{.6525}{
$
\left[
\begin{array}{c|ccc|cccc|cccc|cccc|cccc}
 0 & 0 & 0 & 0 & 0 & 0 & 0 & 0 & 0 & 0 & 0 & 0 & 0 & 0 & 0 & 0 & -\frac{7i}{2} & 2 i & -\frac{5i}{2} & -3 i \\
\hline
 0 & 0 & 0 & 0 & 0 & 0 & 0 & 0 & 0 & 0 & 0 & 0 & -\frac{i}{2} & 4 i & -2 i & -4 i & 0 & 0 & 0 & 0 \\
 0 & 0 & 0 & 0 & 0 & 0 & 0 & 0 & 0 & 0 & 0 & 0 & -\frac{3i}{2} & 0 & \frac{2i}{3} & -\frac{8i}{3} & 0 & 0 & 0 & 0 \\
 0 & 0 & 0 & 0 & 0 & 0 & 0 & 0 & 0 & 0 & 0 & 0 & 3 i & 0 & 2 i & 0 & 0 & 0 & 0 & 0 \\
\hline
 0 & 0 & 0 & 0 & 0 & 0 & 0 & 0 & -\frac{i}{21} & \frac{30i}{7} & -\frac{5i}{7} & -\frac{30i}{7} & 0 & 0 & 0 & 0 & -\frac{42i}{13} & -\frac{16i}{39} & -\frac{42i}{13} & -\frac{112i}{13} \\
 0 & 0 & 0 & 0 & 0 & 0 & 0 & 0 & -\frac{8i}{3} & 0 & 0 & 0 & 0 & 0 & 0 & 0 & \frac{28i}{13} & -\frac{2i}{3} & -4 i & -\frac{3i}{2} \\
 0 & 0 & 0 & 0 & 0 & 0 & 0 & 0 & -\frac{2i}{3} & 0 & \frac{2i}{7} & -\frac{30i}{7} & 0 & 0 & 0 & 0 & -\frac{63i}{26} & \frac{16i}{3} & -\frac{2i}{26} & \frac{16i}{13} \\
 0 & 0 & 0 & 0 & 0 & 0 & 0 & 0 & 4 i & 0 & 4 i & 0 & 0 & 0 & 0 & 0 & \frac{84i}{13} & -2 i & -\frac{12i}{13} & \frac{i}{2} \\
\end{array}
\right]
$
}
\caption{This is the matrix $\left( \mathcal{C}_{5}\right)_{5}^{\check{P}}$, which defines the coefficient $\left(a_{5}\right)_{5}^{\check{P}} = \left< \mathcal{M}_{\ell} \right| \left( \mathcal{C}_{5}\right)_{5}^{\check{P}} \left| \mathcal{M}_{\ell} \right>$ that appears in the observable $\check{P}$ for a truncation at $L=5$.}
\label{tab:PObsMatrix5}
\end{table}

\clearpage

\begin{table}[h]
\centering
\scalebox{.659}{
$
\left[
\begin{array}{c|ccc|cccc|cccc|cccc|cccc}
 0 & 0 & 0 & 0 & 0 & 0 & 0 & 0 & 0 & 0 & 0 & 0 & 0 & 0 & 0 & 0 & 0 & 0 & 0 & 0 \\
\hline
 0 & 0 & 0 & 0 & 0 & 0 & 0 & 0 & 0 & 0 & 0 & 0 & 0 & 0 & 0 & 0 & -\frac{6i}{11} & \frac{60i}{11} & -\frac{30i}{11} & -\frac{105i}{22} \\
 0 & 0 & 0 & 0 & 0 & 0 & 0 & 0 & 0 & 0 & 0 & 0 & 0 & 0 & 0 & 0 & -2 i & 0 & \frac{10i}{11} & -\frac{75i}{22} \\
 0 & 0 & 0 & 0 & 0 & 0 & 0 & 0 & 0 & 0 & 0 & 0 & 0 & 0 & 0 & 0 & \frac{7i}{2} & 0 & \frac{5i}{2} & 0 \\
\hline
 0 & 0 & 0 & 0 & 0 & 0 & 0 & 0 & 0 & 0 & 0 & 0 & 0 & \frac{75i}{11} & -\frac{50i}{33} & -\frac{175i}{33} & 0 & 0 & 0 & 0 \\
 0 & 0 & 0 & 0 & 0 & 0 & 0 & 0 & 0 & 0 & 0 & 0 & -\frac{75i}{22} & 0 & 0 & 0 & 0 & 0 & 0 & 0 \\
 0 & 0 & 0 & 0 & 0 & 0 & 0 & 0 & 0 & 0 & 0 & 0 & -\frac{15i}{11} & 0 & \frac{20i}{33} & -\frac{200i}{33} & 0 & 0 & 0 & 0 \\
 0 & 0 & 0 & 0 & 0 & 0 & 0 & 0 & 0 & 0 & 0 & 0 & \frac{105i}{22} & 0 & \frac{60i}{11} & 0 & 0 & 0 & 0 & 0 \\
\end{array}
\right]
$
}
\caption{This is the matrix $\left( \mathcal{C}_{5}\right)_{6}^{\check{P}}$, which defines the coefficient $\left(a_{5}\right)_{6}^{\check{P}} = \left< \mathcal{M}_{\ell} \right| \left( \mathcal{C}_{5}\right)_{6}^{\check{P}} \left| \mathcal{M}_{\ell} \right>$ that appears in the observable $\check{P}$ for a truncation at $L=5$.}
\label{tab:PObsMatrix6}
\end{table}

\begin{table}[h]
\centering
\scalebox{.635}{
$
\left[
\begin{array}{c|ccc|cccc|cccc|cccc|cccc}
 0 & 0 & 0 & 0 & 0 & 0 & 0 & 0 & 0 & 0 & 0 & 0 & 0 & 0 & 0 & 0 & 0 & 0 & 0 & 0 \\
\hline
 0 & 0 & 0 & 0 & 0 & 0 & 0 & 0 & 0 & 0 & 0 & 0 & 0 & 0 & 0 & 0 & 0 & 0 & 0 & 0 \\
 0 & 0 & 0 & 0 & 0 & 0 & 0 & 0 & 0 & 0 & 0 & 0 & 0 & 0 & 0 & 0 & 0 & 0 & 0 & 0 \\
 0 & 0 & 0 & 0 & 0 & 0 & 0 & 0 & 0 & 0 & 0 & 0 & 0 & 0 & 0 & 0 & 0 & 0 & 0 & 0 \\
\hline
 0 & 0 & 0 & 0 & 0 & 0 & 0 & 0 & 0 & 0 & 0 & 0 & 0 & 0 & 0 & 0 & \frac{27i}{286} & \frac{1350i}{143} & -\frac{675i}{286} & -\frac{900i}{143} \\
 0 & 0 & 0 & 0 & 0 & 0 & 0 & 0 & 0 & 0 & 0 & 0 & 0 & 0 & 0 & 0 & -\frac{54i}{13} & 0 & 0 & 0 \\
 0 & 0 & 0 & 0 & 0 & 0 & 0 & 0 & 0 & 0 & 0 & 0 & 0 & 0 & 0 & 0 & -\frac{27i}{13} & 0 & \frac{135i}{143} & -\frac{1125i}{143} \\
 0 & 0 & 0 & 0 & 0 & 0 & 0 & 0 & 0 & 0 & 0 & 0 & 0 & 0 & 0 & 0 & \frac{72i}{13} & 0 & \frac{90i}{13} & 0 \\
\end{array}
\right]
$
}
\caption{This is the matrix $\left( \mathcal{C}_{5}\right)_{7}^{\check{P}}$, which defines the coefficient $\left(a_{5}\right)_{7}^{\check{P}} = \left< \mathcal{M}_{\ell} \right| \left( \mathcal{C}_{5}\right)_{7}^{\check{P}} \left| \mathcal{M}_{\ell} \right>$ that appears in the observable $\check{P}$ for a truncation at $L=5$.}
\label{tab:PObsMatrix7}
\end{table}

\clearpage

\begin{table}[h]
\centering
\scalebox{.72}{
$
\left[
\begin{array}{c|ccc|cccc|cccc|cccc|cccc}
 0 & 0 & 0 & 0 & 0 & 0 & 0 & 0 & 0 & 0 & 0 & 0 & 0 & 0 & 0 & 0 & 0 & 0 & 0 & 0 \\
\hline
 0 & 0 & 0 & 0 & 0 & 0 & 0 & 0 & 0 & 0 & 0 & 0 & 0 & 0 & 0 & 0 & 0 & 0 & 0 & 0 \\
 0 & 0 & 0 & 0 & 0 & 0 & 0 & 0 & 0 & 0 & 0 & 0 & 0 & 0 & 0 & 0 & 0 & 0 & 0 & 0 \\
 0 & 0 & 0 & 0 & 0 & 0 & 0 & 0 & 0 & 0 & 0 & 0 & 0 & 0 & 0 & 0 & 0 & 0 & 0 & 0 \\
\hline
 0 & 0 & 0 & 0 & 0 & 0 & 0 & 0 & 0 & 0 & 0 & 0 & 0 & 0 & 0 & 0 & 0 & 0 & 0 & 0 \\
 0 & 0 & 0 & 0 & 0 & 0 & 0 & 0 & 0 & 0 & 0 & 0 & 0 & 0 & 0 & 0 & 0 & 0 & 0 & 0 \\
 0 & 0 & 0 & 0 & 0 & 0 & 0 & 0 & 0 & 0 & 0 & 0 & 0 & 0 & 0 & 0 & 0 & 0 & 0 & 0 \\
 0 & 0 & 0 & 0 & 0 & 0 & 0 & 0 & 0 & 0 & 0 & 0 & 0 & 0 & 0 & 0 & 0 & 0 & 0 & 0 \\
\end{array}
\right]
$
}
\caption{This is the matrix $\left( \mathcal{C}_{5}\right)_{8}^{\check{P}}$, which defines the coefficient $\left(a_{5}\right)_{8}^{\check{P}} = \left< \mathcal{M}_{\ell} \right| \left( \mathcal{C}_{5}\right)_{8}^{\check{P}} \left| \mathcal{M}_{\ell} \right>$ that appears in the observable $\check{P}$ for a truncation at $L=5$.}
\label{tab:PObsMatrix8}
\end{table}

\begin{table}[h]
\centering
\scalebox{.73}{
$
\left[
\begin{array}{c|ccc|cccc|cccc|cccc|cccc}
 0 & 0 & 0 & 0 & 0 & 0 & 0 & 0 & 0 & 0 & 0 & 0 & 0 & 0 & 0 & 0 & 0 & 0 & 0 & 0 \\
\hline
 0 & 0 & 0 & 0 & 0 & 0 & 0 & 0 & 0 & 0 & 0 & 0 & 0 & 0 & 0 & 0 & 0 & 0 & 0 & 0 \\
 0 & 0 & 0 & 0 & 0 & 0 & 0 & 0 & 0 & 0 & 0 & 0 & 0 & 0 & 0 & 0 & 0 & 0 & 0 & 0 \\
 0 & 0 & 0 & 0 & 0 & 0 & 0 & 0 & 0 & 0 & 0 & 0 & 0 & 0 & 0 & 0 & 0 & 0 & 0 & 0 \\
\hline
 0 & 0 & 0 & 0 & 0 & 0 & 0 & 0 & 0 & 0 & 0 & 0 & 0 & 0 & 0 & 0 & 0 & 0 & 0 & 0 \\
 0 & 0 & 0 & 0 & 0 & 0 & 0 & 0 & 0 & 0 & 0 & 0 & 0 & 0 & 0 & 0 & 0 & 0 & 0 & 0 \\
 0 & 0 & 0 & 0 & 0 & 0 & 0 & 0 & 0 & 0 & 0 & 0 & 0 & 0 & 0 & 0 & 0 & 0 & 0 & 0 \\
 0 & 0 & 0 & 0 & 0 & 0 & 0 & 0 & 0 & 0 & 0 & 0 & 0 & 0 & 0 & 0 & 0 & 0 & 0 & 0 \\
\end{array}%
\right]
$
}
\caption{This is the matrix $\left( \mathcal{C}_{5}\right)_{0}^{\check{E}}$, which defines the coefficient $\left(a_{5}\right)_{0}^{\check{E}} = \left< \mathcal{M}_{\ell} \right| \left( \mathcal{C}_{5}\right)_{0}^{\check{E}} \left| \mathcal{M}_{\ell} \right>$ that appears in the observable $\check{E}$ for a truncation at $L=5$.}
\label{tab:EObsMatrix0}
\end{table}

\clearpage

\begin{table}[h]
\centering
\scalebox{.724}{
$
\left[
%
\right]
$
}
\caption{This is the matrix $\left( \mathcal{C}_{5}\right)_{1}^{\check{E}}$, which defines the coefficient $\left(a_{5}\right)_{1}^{\check{E}} = \left< \mathcal{M}_{\ell} \right| \left( \mathcal{C}_{5}\right)_{1}^{\check{E}} \left| \mathcal{M}_{\ell} \right>$ that appears in the observable $\check{E}$ for a truncation at $L=5$.}
\label{tab:EObsMatrix1}
\end{table}

\begin{table}[h]
\centering
\scalebox{.681}{
$
\left[
%
\right]
$
}
\caption{This is the matrix $\left( \mathcal{C}_{5}\right)_{2}^{\check{E}}$, which defines the coefficient $\left(a_{5}\right)_{2}^{\check{E}} = \left< \mathcal{M}_{\ell} \right| \left( \mathcal{C}_{5}\right)_{2}^{\check{E}} \left| \mathcal{M}_{\ell} \right>$ that appears in the observable $\check{E}$ for a truncation at $L=5$.}
\label{tab:EObsMatrix2}
\end{table}

\clearpage

\begin{table}[h]
\centering
\scalebox{.668}{
$
\left[
%
\right]
$
}
\caption{This is the matrix $\left( \mathcal{C}_{5}\right)_{3}^{\check{E}}$, which defines the coefficient $\left(a_{5}\right)_{3}^{\check{E}} = \left< \mathcal{M}_{\ell} \right| \left( \mathcal{C}_{5}\right)_{3}^{\check{E}} \left| \mathcal{M}_{\ell} \right>$ that appears in the observable $\check{E}$ for a truncation at $L=5$.}
\label{tab:EObsMatrix3}
\end{table}

\begin{table}[h]
\centering
\scalebox{.591}{
$
\left[
%
\right]
$
}
\caption{This is the matrix $\left( \mathcal{C}_{5}\right)_{4}^{\check{E}}$, which defines the coefficient $\left(a_{5}\right)_{4}^{\check{E}} = \left< \mathcal{M}_{\ell} \right| \left( \mathcal{C}_{5}\right)_{4}^{\check{E}} \left| \mathcal{M}_{\ell} \right>$ that appears in the observable $\check{E}$ for a truncation at $L=5$.}
\label{tab:EObsMatrix4}
\end{table}

\clearpage

\begin{table}[h]
\centering
\scalebox{.646}{
$
\left[
%
\right]
$
}
\caption{This is the matrix $\left( \mathcal{C}_{5}\right)_{5}^{\check{E}}$, which defines the coefficient $\left(a_{5}\right)_{5}^{\check{E}} = \left< \mathcal{M}_{\ell} \right| \left( \mathcal{C}_{5}\right)_{5}^{\check{E}} \left| \mathcal{M}_{\ell} \right>$ that appears in the observable $\check{E}$ for a truncation at $L=5$.}
\label{tab:EObsMatrix5}
\end{table}

\begin{table}[h]
\centering
\scalebox{.641}{
$
\left[
%
\right]
$
}
\caption{This is the matrix $\left( \mathcal{C}_{5}\right)_{2}^{\check{G}}$, which defines the coefficient $\left(a_{5}\right)_{2}^{\check{G}} = \left< \mathcal{M}_{\ell} \right| \left( \mathcal{C}_{5}\right)_{2}^{\check{G}} \left| \mathcal{M}_{\ell} \right>$ that appears in the observable $\check{G}$ for a truncation at $L=5$.}
\label{tab:GObsMatrix2}
\end{table}

\begin{table}[h]
\centering
\scalebox{.634}{
$
\left[
%
\right]
$
}
\caption{This is the matrix $\left( \mathcal{C}_{5}\right)_{3}^{\check{G}}$, which defines the coefficient $\left(a_{5}\right)_{3}^{\check{G}} = \left< \mathcal{M}_{\ell} \right| \left( \mathcal{C}_{5}\right)_{3}^{\check{G}} \left| \mathcal{M}_{\ell} \right>$ that appears in the observable $\check{G}$ for a truncation at $L=5$.}
\label{tab:GObsMatrix3}
\end{table}

\clearpage

\begin{table}[h]
\centering
\scalebox{.6115}{
$
\left[
%
\right]
$
}
\caption{This is the matrix $\left( \mathcal{C}_{5}\right)_{4}^{\check{G}}$, which defines the coefficient $\left(a_{5}\right)_{4}^{\check{G}} = \left< \mathcal{M}_{\ell} \right| \left( \mathcal{C}_{5}\right)_{4}^{\check{G}} \left| \mathcal{M}_{\ell} \right>$ that appears in the observable $\check{G}$ for a truncation at $L=5$.}
\label{tab:GObsMatrix4}
\end{table}

\begin{table}[h]
\centering
\scalebox{.685}{
$
\left[
%
\right]
$
}
\caption{This is the matrix $\left( \mathcal{C}_{5}\right)_{5}^{\check{G}}$, which defines the coefficient $\left(a_{5}\right)_{5}^{\check{G}} = \left< \mathcal{M}_{\ell} \right| \left( \mathcal{C}_{5}\right)_{5}^{\check{G}} \left| \mathcal{M}_{\ell} \right>$ that appears in the observable $\check{G}$ for a truncation at $L=5$.}
\label{tab:GObsMatrix5}
\end{table}

\clearpage

\begin{table}[h]
\centering
\scalebox{.7}{
$
\left[
%
\right]
$
}
\caption{This is the matrix $\left( \mathcal{C}_{5}\right)_{6}^{\check{G}}$, which defines the coefficient $\left(a_{5}\right)_{6}^{\check{G}} = \left< \mathcal{M}_{\ell} \right| \left( \mathcal{C}_{5}\right)_{6}^{\check{G}} \left| \mathcal{M}_{\ell} \right>$ that appears in the observable $\check{G}$ for a truncation at $L=5$.}
\label{tab:GObsMatrix6}
\end{table}

\begin{table}[h]
\centering
\scalebox{.659}{
$
\left[
%
\right]
$
}
\caption{This is the matrix $\left( \mathcal{C}_{5}\right)_{7}^{\check{G}}$, which defines the coefficient $\left(a_{5}\right)_{7}^{\check{G}} = \left< \mathcal{M}_{\ell} \right| \left( \mathcal{C}_{5}\right)_{7}^{\check{G}} \left| \mathcal{M}_{\ell} \right>$ that appears in the observable $\check{G}$ for a truncation at $L=5$.}
\label{tab:GObsMatrix7}
\end{table}

\clearpage

\begin{table}[h]
\centering
\scalebox{.72}{
$
\left[
\begin{array}{c|ccc|cccc|cccc|cccc|cccc}
 0 & 0 & 0 & 0 & 0 & 0 & 0 & 0 & 0 & 0 & 0 & 0 & 0 & 0 & 0 & 0 & 0 & 0 & 0 & 0 \\
\hline
 0 & 0 & 0 & 0 & 0 & 0 & 0 & 0 & 0 & 0 & 0 & 0 & 0 & 0 & 0 & 0 & 0 & 0 & 0 & 0 \\
 0 & 0 & 0 & 0 & 0 & 0 & 0 & 0 & 0 & 0 & 0 & 0 & 0 & 0 & 0 & 0 & 0 & 0 & 0 & 0 \\
 0 & 0 & 0 & 0 & 0 & 0 & 0 & 0 & 0 & 0 & 0 & 0 & 0 & 0 & 0 & 0 & 0 & 0 & 0 & 0 \\
\hline
 0 & 0 & 0 & 0 & 0 & 0 & 0 & 0 & 0 & 0 & 0 & 0 & 0 & 0 & 0 & 0 & 0 & 0 & 0 & 0 \\
 0 & 0 & 0 & 0 & 0 & 0 & 0 & 0 & 0 & 0 & 0 & 0 & 0 & 0 & 0 & 0 & 0 & 0 & 0 & 0 \\
 0 & 0 & 0 & 0 & 0 & 0 & 0 & 0 & 0 & 0 & 0 & 0 & 0 & 0 & 0 & 0 & 0 & 0 & 0 & 0 \\
 0 & 0 & 0 & 0 & 0 & 0 & 0 & 0 & 0 & 0 & 0 & 0 & 0 & 0 & 0 & 0 & 0 & 0 & 0 & 0 \\
\end{array}%
\right]
$
}
\caption{This is the matrix $\left( \mathcal{C}_{5}\right)_{1}^{\check{H}}$, which defines the coefficient $\left(a_{5}\right)_{1}^{\check{H}} = \left< \mathcal{M}_{\ell} \right| \left( \mathcal{C}_{5}\right)_{1}^{\check{H}} \left| \mathcal{M}_{\ell} \right>$ that appears in the observable $\check{H}$ for a truncation at $L=5$.}
\label{tab:HObsMatrix1}
\end{table}

\clearpage

\begin{table}[h]
\centering
\scalebox{.637}{
$
\left[
%
\right]
$
}
\caption{This is the matrix $\left( \mathcal{C}_{5}\right)_{2}^{\check{H}}$, which defines the coefficient $\left(a_{5}\right)_{2}^{\check{H}} = \left< \mathcal{M}_{\ell} \right| \left( \mathcal{C}_{5}\right)_{2}^{\check{H}} \left| \mathcal{M}_{\ell} \right>$ that appears in the observable $\check{H}$ for a truncation at $L=5$.}
\label{tab:HObsMatrix2}
\end{table}

\begin{table}[h]
\centering
\scalebox{.621}{
$
\left[
%
\right]
$
}
\caption{This is the matrix $\left( \mathcal{C}_{5}\right)_{3}^{\check{H}}$, which defines the coefficient $\left(a_{5}\right)_{3}^{\check{H}} = \left< \mathcal{M}_{\ell} \right| \left( \mathcal{C}_{5}\right)_{3}^{\check{H}} \left| \mathcal{M}_{\ell} \right>$ that appears in the observable $\check{H}$ for a truncation at $L=5$.}
\label{tab:HObsMatrix3}
\end{table}

\clearpage

\begin{table}[h]
\centering
\scalebox{.58}{
$
\left[
%
\right]
$
}
\caption{This is the matrix $\left( \mathcal{C}_{5}\right)_{4}^{\check{H}}$, which defines the coefficient $\left(a_{5}\right)_{4}^{\check{H}} = \left< \mathcal{M}_{\ell} \right| \left( \mathcal{C}_{5}\right)_{4}^{\check{H}} \left| \mathcal{M}_{\ell} \right>$ that appears in the observable $\check{H}$ for a truncation at $L=5$.}
\label{tab:HObsMatrix4}
\end{table}

\begin{table}[h]
\centering
\scalebox{.6424}{
$
\left[
%
\right]
$
}
\caption{This is the matrix $\left( \mathcal{C}_{5}\right)_{5}^{\check{H}}$, which defines the coefficient $\left(a_{5}\right)_{5}^{\check{H}} = \left< \mathcal{M}_{\ell} \right| \left( \mathcal{C}_{5}\right)_{5}^{\check{H}} \left| \mathcal{M}_{\ell} \right>$ that appears in the observable $\check{H}$ for a truncation at $L=5$.}
\label{tab:HObsMatrix5}
\end{table}

\clearpage

\begin{table}[h]
\centering
\scalebox{.661}{
$
\left[
%
\right]
$
}
\caption{This is the matrix $\left( \mathcal{C}_{5}\right)_{6}^{\check{H}}$, which defines the coefficient $\left(a_{5}\right)_{6}^{\check{H}} = \left< \mathcal{M}_{\ell} \right| \left( \mathcal{C}_{5}\right)_{6}^{\check{H}} \left| \mathcal{M}_{\ell} \right>$ that appears in the observable $\check{H}$ for a truncation at $L=5$.}
\label{tab:HObsMatrix6}
\end{table}

\begin{table}[h]
\centering
\scalebox{.627}{
$
\left[
%
\right]
$
}
\caption{This is the matrix $\left( \mathcal{C}_{5}\right)_{1}^{\check{F}}$, which defines the coefficient $\left(a_{5}\right)_{1}^{\check{F}} = \left< \mathcal{M}_{\ell} \right| \left( \mathcal{C}_{5}\right)_{1}^{\check{F}} \left| \mathcal{M}_{\ell} \right>$ that appears in the observable $\check{F}$ for a truncation at $L=5$.}
\label{tab:FObsMatrix1}
\end{table}

\clearpage

\begin{table}[h]
\centering
\scalebox{.629}{
$
\left[
%
\right]
$
}
\caption{This is the matrix $\left( \mathcal{C}_{5}\right)_{2}^{\check{F}}$, which defines the coefficient $\left(a_{5}\right)_{2}^{\check{F}} = \left< \mathcal{M}_{\ell} \right| \left( \mathcal{C}_{5}\right)_{2}^{\check{F}} \left| \mathcal{M}_{\ell} \right>$ that appears in the observable $\check{F}$ for a truncation at $L=5$.}
\label{tab:FObsMatrix2}
\end{table}

\begin{table}[h]
\centering
\scalebox{.66}{
$
\left[
%
\right]
$
}
\caption{This is the matrix $\left( \mathcal{C}_{5}\right)_{3}^{\check{F}}$, which defines the coefficient $\left(a_{5}\right)_{3}^{\check{F}} = \left< \mathcal{M}_{\ell} \right| \left( \mathcal{C}_{5}\right)_{3}^{\check{F}} \left| \mathcal{M}_{\ell} \right>$ that appears in the observable $\check{F}$ for a truncation at $L=5$.}
\label{tab:FObsMatrix3}
\end{table}

\clearpage

\begin{table}[h]
\centering
\scalebox{.617}{
$
\left[
%
\right]
$
}
\caption{This is the matrix $\left( \mathcal{C}_{5}\right)_{10}^{\check{F}}$, which defines the coefficient $\left(a_{5}\right)_{10}^{\check{F}} = \left< \mathcal{M}_{\ell} \right| \left( \mathcal{C}_{5}\right)_{10}^{\check{F}} \left| \mathcal{M}_{\ell} \right>$ that appears in the observable $\check{F}$ for a truncation at $L=5$.}
\label{tab:FObsMatrix10}
\end{table}

\clearpage

\subsection{Additional material for chapter \ref{chap:Omelaenko}} \label{sec:AdditionsChapterII}

\subsubsection{The double ambiguity transformation acting on the transversity amplitudes}\label{sec:DoubleAmbiguityTrafoActingOnBi}

In the following, the action the double ambiguity transformation (introduced in section \ref{sec:WBTpaper}) has on the photoproduction transversity amplitudes $\left\{b_{1}, \hspace*{1.5pt} b_{2}, \hspace*{1.5pt} b_{3}, \hspace*{1.5pt} b_{4}\right\}$ shall be derived and discussed. \newline
Fot this purpose, it is useful to recall that upon using the angular variable $t = \tan \theta/2$, one can write the following linear factor decompositions for the transversity amplitudes $b_{2}$ and $b_{4}$ in a truncation of the multipole expansion at $L = \ell_{\mathrm{max}}$ (cf. equations (38) and (39) of section \ref{sec:WBTpaper}):
\begin{align}
 b_{2} \left(\theta\right) &= - \hspace*{1pt} \mathcal{C} \hspace*{1pt} a_{2L} \hspace*{1pt} \frac{\exp \left(i \frac{\theta}{2}\right)}{\left( 1 + t^{2} \right)^{L}} \hspace*{1pt} \prod_{k = 1}^{2L} \left( t - \beta_{k} \right)  \mathrm{,} \label{eq:b2LinFactDecomp} \\
  b_{4} \left(\theta\right) &= \mathcal{C} \hspace*{1pt} a_{2L} \hspace*{1pt} \frac{\exp \left(i \frac{\theta}{2}\right)}{\left( 1 + t^{2} \right)^{L}} \hspace*{1pt} \prod_{k = 1}^{2L} \left( t - \alpha_{k} \right)  \mathrm{,} \label{eq:b4LinFactDecomp}
\end{align}
where again explicit dependencies on the CMS energy $W$ were suppressed. The amplitudes $b_{1}$ and $b_{3}$ can be obtained by using equation (13) of section \ref{sec:WBTpaper}:
\begin{equation}
 b_{1} \left(\theta\right) = b_{2} \left(-\theta\right)\,, \quad
b_{3} \left(\theta\right) = b_{4} \left(-\theta\right) \,.
\label{eq:TransversityAmplitudeAngleRel}
\end{equation}
The profile functions of the observables of the group $\mathcal{S}$, $\left\{\sigma_{0}, \check{\Sigma}, \check{T}, \check{P}\right\}$, are in the transversity basis represented by diagonal matrices (cf. appendix \ref{sec:ObservableAlgebra}) and are therefore strictly the sum of squared moduli of the $b_{i}$, i.e. (omitting the phase-space factor $\rho = q / k$)
\begin{equation}
\check{\Omega}^{\alpha_{S}} \left(W, \theta\right) = \frac{1}{2} \left( \pm \left|b_{1}\right|^{2} \pm \left|b_{2}\right|^{2} \pm \left|b_{3}\right|^{2} + \left|b_{4}\right|^{2}\right) \mathrm{.} \label{eq:GroupSIsDiagonal}
\end{equation}
The complex conjugation of all roots appearing in (\ref{eq:b2LinFactDecomp}) and (\ref{eq:b4LinFactDecomp}) leaves all the squared moduli in (\ref{eq:GroupSIsDiagonal}) invariant and therefore also any group $\mathcal{S}$ observable. In section \ref{sec:WBTpaper}, this transformation was was called the \textit{double ambiguity}:
\begin{equation}
\alpha_{i} \rightarrow \alpha_{i}^{\ast}\,,\quad  \beta_{j}
\rightarrow \beta_{j}^{\ast} \,, \quad \forall i,j = 1,\ldots,2L \,.
\label{eq:DoubleAmbiguityDetail}
\end{equation}
The question is now: what happens to $\left\{b_{1}  \ldots ,b_{4}\right\}$ under this transformation? \newline
In order to answer this question, one further useful thing that should be done is to express the complex number $\mathcal{C}$, which depends on the convention chosen in the definition of the $b_{i}$ and is therefore fixed, by polar coordinates:
\begin{equation}
\mathcal{C} = \left| \mathcal{C} \right| e^{i \varphi_{\mathcal{C}}} \mathrm{.} \label{eq:ConventionConstantPolarCoordinates}
\end{equation}
Now, using the assumption that a truncation at $L = \ell_{\mathrm{max}}$ describes the amplitude exactly and therefore also that the linear factor decompositions (\ref{eq:b2LinFactDecomp}), (\ref{eq:b4LinFactDecomp}) and (\ref{eq:TransversityAmplitudeAngleRel}) are exact, the behaviour of the amplitude $b_{4}$ shall be investigated as an example case. If all roots are conjugated under the double ambiguity transfomation (\ref{eq:DoubleAmbiguityDetail}), where for $b_{4}$ only the transformation of the roots $\alpha_{k} \rightarrow \alpha_{k}^{\ast}$ is important, one sees that
\begin{align}
 b_{4} \left(\theta\right) \overset{\mathrm{(\ref{eq:DoubleAmbiguityDetail})}}{\longrightarrow} b_{4}^{\mathrm{D.A.}} \left(\theta\right) &= \mathcal{C} \hspace*{1pt} a_{2L} \hspace*{1pt} \frac{\exp \left(i \frac{\theta}{2}\right)}{\left( 1 + t^{2} \right)^{L}} \hspace*{1pt} \prod_{k = 1}^{2L} \left( t - \alpha_{k}^{\ast} \right) \nonumber \\
&= \mathcal{C} \hspace*{1pt} a_{2L} \hspace*{1pt} \frac{\exp \left(i \frac{\theta}{2}\right)}{\left( 1 + t^{2} \right)^{L}} \hspace*{1pt} \left[ \prod_{k = 1}^{2L} \left( t - \alpha_{k} \right) \right]^{\ast} \mathrm{.} \label{eq:b4DAmbDerivStep1}
\end{align}
Equation (\ref{eq:ConventionConstantPolarCoordinates}) implies $\mathcal{C}^{\ast} = \mathcal{C} e^{- 2 i \varphi_{\mathcal{C}}}$, such that one can derive for the complex conjugate of $b_{4}$:
\begin{align}
  b_{4}^{\ast} \left(\theta\right) &= \mathcal{C}^{\ast} \hspace*{1pt} a_{2L}^{\ast} \hspace*{1pt} \frac{\exp \left(- i \frac{\theta}{2}\right)}{\left[\left( 1 + t^{2} \right)^{L}\right]^{\ast}} \hspace*{1pt} \left[ \prod_{k = 1}^{2L} \left( t - \alpha_{k} \right) \right]^{\ast} \nonumber \\
&= \mathcal{C}^{\ast} \hspace*{1pt} a_{2L} \hspace*{1pt} \frac{\exp \left(- i \frac{\theta}{2}\right)}{\left( 1 + t^{2} \right)^{L}} \hspace*{1pt} \left[ \prod_{k = 1}^{2L} \left( t - \alpha_{k} \right) \right]^{\ast} \nonumber \\
&= e^{- 2 i \varphi_{\mathcal{C}}} e^{- i \theta} \mathcal{C} \hspace*{1pt} a_{2L} \hspace*{1pt} \frac{\exp \left(i \frac{\theta}{2}\right)}{\left( 1 + t^{2} \right)^{L}} \hspace*{1pt} \left[ \prod_{k = 1}^{2L} \left( t - \alpha_{k} \right) \right]^{\ast} = e^{- 2 i \varphi_{\mathcal{C}}} e^{- i \theta} b_{4}^{\mathrm{D.A.}} \left(\theta\right) \mathrm{,} \label{eq:b4DAmbDerivStep2}
\end{align}
where the fact that $t$ and $a_{2L}$ are real was employed ($a_{2L}$ can always be assumed to be real, since this coefficient encodes the information on the unknown overall phase in a single energy multipole analysis, cf. section \ref{sec:WBTpaper}). One therefore arrives at the relation:
\begin{equation}
  b_{4}^{\mathrm{D.A.}} \left(\theta\right) = e^{2 i \varphi_{\mathcal{C}}} e^{i \theta} b_{4}^{\ast} \left(\theta\right) \mathrm{.} \label{eq:b4DAmbTrafoRule}
\end{equation}
The transformation rules for the remaining transversity amplitudes can be derived in the same way. The results are:
\begin{align}
 b_{1}^{\mathrm{D.A.}} \left(\theta\right) &= e^{2 i \varphi_{\mathcal{C}}} e^{- i \theta} b_{1}^{\ast} \left(\theta\right) \mathrm{,} \label{eq:b1DAmbTrafoRule} \\
 b_{2}^{\mathrm{D.A.}} \left(\theta\right) &= e^{2 i \varphi_{\mathcal{C}}} e^{i \theta} b_{2}^{\ast} \left(\theta\right) \mathrm{,} \label{eq:b2DAmbTrafoRule} \\
 b_{3}^{\mathrm{D.A.}} \left(\theta\right) &= e^{2 i \varphi_{\mathcal{C}}} e^{- i \theta} b_{3}^{\ast} \left(\theta\right) \mathrm{.} \label{eq:b3DAmbTrafoRule}
\end{align}
Upon combination of equations (\ref{eq:b4DAmbTrafoRule}), (\ref{eq:b1DAmbTrafoRule}), (\ref{eq:b2DAmbTrafoRule}) and (\ref{eq:b3DAmbTrafoRule}), it is seen that the double ambiguity transformation acts on the transversity amplitudes as an antilinear ambiguity transformation 
\begin{equation}
b_{i} \left(\theta\right) \rightarrow b^{\mathrm{D.A.}}_{i} \left(\theta\right) = \sum_{j} \left[\mathcal{A}\left(\theta\right)\right]_{ij} b^{\ast}_{j}\left(\theta\right) \label{eq:DoubleAmbTrafoForm} \mathrm{,}
\end{equation}
with the $\theta$-dependent transformation matrix:
\begin{equation}
\mathcal{A} \left(\theta\right) = e^{2 i \varphi_{\mathcal{C}}} \left[ \begin{array}{cccc}
 e^{- i \theta} & 0 & 0 & 0 \\
 0 & e^{i \theta} & 0 & 0 \\
 0 & 0 & e^{- i \theta} & 0 \\
 0 & 0 & 0 & e^{i \theta} \\
\end{array} \right] \mathrm{.} \label{eq:DoubleAmbTrafoMatrix}
\end{equation}
The fact that the double ambiguity transformation acts in the above given way is known and was published by Keaton and Workman, reference \cite{KeatonWorkman2}. \newline
It is now instructive to apply the formalism on the discrete ambiguities of polarization observables given by Chiang and Tabakin \cite{ChTab} (cf. section \ref{sec:CompExpsFullAmp}), especially their study of antilinear ambiguities, to the double ambiguity. In case one considers a generic observable (more precisely, a profile function)
\begin{equation}
 \check{\Omega}^{\alpha} = \frac{1}{2} \sum_{i,j} b_{i}^{\ast} \tilde{\Gamma}^{\alpha}_{ij} b_{j} \mathrm{,} \label{eq:PolObsGenericChapter23}
\end{equation}
and investigates the behaviour of this observable under an antilinear tranformation acting on the transversity amplitudes (with now a general matrix $A$)
\begin{equation}
 b_{i} \rightarrow \bar{b}_{i} = \sum_{j} A_{ij} b^{\ast}_{j} \label{eq:AntilinearTrafoGeneral}
\end{equation}
it can be seen that \cite{ChTab}
\begin{align}
 \bar{\check{\Omega}}^{\alpha} &= \frac{1}{2} \sum_{i,j} \bar{b}_{i}^{\ast} \tilde{\Gamma}^{\alpha}_{ij} \bar{b}_{j} = \frac{1}{2} \sum_{i,j,k,l} A^{\ast}_{ik} b_{k} \tilde{\Gamma}^{\alpha}_{ij} A_{jl} b^{\ast}_{k^{\prime}} \mathrm{,} \nonumber \\
&= \frac{1}{2} \sum_{i,j,k,l} b^{\ast}_{k^{\prime}} b_{k}  A^{\dagger}_{ki}  \tilde{\Gamma}^{\alpha}_{ij} A_{jl} = \frac{1}{2} \sum_{l,k} b^{\ast}_{k^{\prime}} \left( A^{\dagger}  \tilde{\Gamma}^{\alpha} A \right)^{T}_{lk} b_{k} \mathrm{.} \label{eq:AntilinearTrafoActingOnObs}
\end{align}
Therefore, the observable $\check{\Omega}^{\alpha}$ is invariant under the antilinear transformation $A$ if and only if the condition
\begin{equation}
 \left( A^{\dagger}  \tilde{\Gamma}^{\alpha} A \right)^{T} = \tilde{\Gamma}^{\alpha} \mathrm{,} \label{eq:ConditionInvAntilinAmb}
\end{equation}
is fulfilled. In the course of this work, the invariance of all 16 observables under the double ambiguity transformation (\ref{eq:DoubleAmbTrafoMatrix}) was tested using the invariance condition (\ref{eq:ConditionInvAntilinAmb}). All algebraic calculations were performed using MATHEMATICA \cite{Mathematica8,Mathematica11,MathematicaLanguage,MathematicaBonnLicense}. The results are collected in Tables \ref{tab:GroupSDoubleAmbiguityMatricesI}, \ref{tab:GroupSDoubleAmbiguityMatricesII} and \ref{tab:GroupSDoubleAmbiguityMatricesIII}. \newline
The most important results of chapter IV in section \ref{sec:WBTpaper}, where the response of the observables to the double ambiguity transformation was discussed solely on the footing of the root-functions $f \left(\theta, \alpha\right)$ and $f \left(\theta, \beta\right)$, are recovered. Furthermore, additional interesting facts about the behaviour of the $\mathcal{BR}$- and $\mathcal{TR}$ observables can be observed. In summary:
\begin{itemize}
 \item[(i)] The group $\mathcal{S}$ observables $\left\{\sigma_{0}, \check{\Sigma}, \check{T}, \check{P}\right\}$ as well as the BT observables $\check{E}$ and $\check{H}$ are invariant under the double ambiguity transformation $\mathcal{A} \left(\theta\right)$ (\ref{eq:DoubleAmbTrafoMatrix}) and therefore cannot resolve any partial wave ambiguities originating from it.
 \item[(ii)] The two BT observables $\check{G}$ and $\check{F}$ change sign under the double ambiguity transformation $\mathcal{A} \left(\theta\right)$ and can resolve the corresponding ambiguities.
 \item[(iii)] All $\mathcal{BR}$- and $\mathcal{TR}$ observables are also generally not invariant under $\mathcal{A} \left(\theta\right)$. Interestingly, they show a rotational mixing pattern, with two observables rotated into each other in pairs. This pattern shall be elaborated more in the following.
\end{itemize}
The left hand side of the invariance condition (\ref{eq:ConditionInvAntilinAmb}), which defines the transformed observable (\ref{eq:AntilinearTrafoActingOnObs}) under an antilinear transformation, is in case of the $\mathcal{BR}$- and $\mathcal{TR}$ observables, Tables \ref{tab:GroupSDoubleAmbiguityMatricesII} and \ref{tab:GroupSDoubleAmbiguityMatricesIII}, seen to be $\theta$-dependent. This is very different from the group $\mathcal{S}$ and BT-observables (Table \ref{tab:GroupSDoubleAmbiguityMatricesI}), where one has either invariance or a sign change. Once the resulting $\theta$-dependent matrices that represent the transformed $\mathcal{BR}$- and $\mathcal{TR}$ observables are expanded into the orthonormal basis of the $\tilde{\Gamma}$-matrices, one observes the above mentioned rotational pattern, which is already included in Tables \ref{tab:GroupSDoubleAmbiguityMatricesII} and \ref{tab:GroupSDoubleAmbiguityMatricesIII}. \newline
Once the replacement rules in the left columns of the Tables are used, the rotational transformations can be written in terms of profile functions of observables. The results of this step are given in equations (\ref{eq:BrTrDAmbRotationsI}) to (\ref{eq:BrTrDAmbRotationsIV}):
\begin{align}
\left[\begin{array}{c} \check{O}_{x^{\prime}} \\ \check{O}_{z^{\prime}} \end{array} \right] &\overset{\mathrm{D.A.}}{\longrightarrow} 
\left[\begin{array}{cc} \cos \left(2 \theta\right) & \sin \left(2 \theta\right) \\
 \sin \left(2 \theta\right) & -\cos \left(2 \theta\right) \end{array} \right] \left[\begin{array}{c} \check{O}_{x^{\prime}} \\ \check{O}_{z^{\prime}} \end{array} \right] \mathrm{,} \label{eq:BrTrDAmbRotationsI} \\
\left[\begin{array}{c} \check{C}_{x^{\prime}} \\ \check{C}_{z^{\prime}} \end{array} \right] &\overset{\mathrm{D.A.}}{\longrightarrow} 
\left[\begin{array}{cc} - \cos \left(2 \theta\right) & - \sin \left(2 \theta\right) \\
 - \sin \left(2 \theta\right) & \cos \left(2 \theta\right) \end{array} \right] \left[\begin{array}{c} \check{C}_{x^{\prime}} \\ \check{C}_{z^{\prime}} \end{array} \right] \mathrm{,} \label{eq:BrTrDAmbRotationsII} \\
\left[\begin{array}{c} \check{T}_{x^{\prime}} \\ \check{T}_{z^{\prime}} \end{array} \right] &\overset{\mathrm{D.A.}}{\longrightarrow} 
\left[\begin{array}{cc} \cos \left(2 \theta\right) & \sin \left(2 \theta\right) \\
 \sin \left(2 \theta\right) & -\cos \left(2 \theta\right) \end{array} \right] \left[\begin{array}{c} \check{T}_{x^{\prime}} \\ \check{T}_{z^{\prime}} \end{array} \right] \mathrm{,} \label{eq:BrTrDAmbRotationsIII} \\
\left[\begin{array}{c} \check{L}_{x^{\prime}} \\ \check{L}_{z^{\prime}} \end{array} \right] &\overset{\mathrm{D.A.}}{\longrightarrow} 
\left[\begin{array}{cc} - \cos \left(2 \theta\right) & - \sin \left(2 \theta\right) \\
 - \sin \left(2 \theta\right) & \cos \left(2 \theta\right) \end{array} \right] \left[\begin{array}{c} \check{L}_{x^{\prime}} \\ \check{L}_{z^{\prime}} \end{array} \right] \mathrm{.} \label{eq:BrTrDAmbRotationsIV}
\end{align}
All matrices appearing in these transformations are orthogonal ($M^{T} M = \mathbbm{1}$) and have determinant (-1). Therefore, they are strictly speaking not pure rotations but rotations plus mirror reflections (pseudo-rotations) in two-observable space. \newline
Another iteresting fact that arises from the $\theta$-dependence of the tranformed $\mathcal{BR}$- and $\mathcal{TR}$ observables is that there are preferred single values of the scattering angle where these observables are invariant. From expressions (\ref{eq:BrTrDAmbRotationsI}) to (\ref{eq:BrTrDAmbRotationsIV}), one can infer the values of $\theta$ for which the respective observable is unchanged. The results are given in Table \ref{tab:BrTrInvarianceThetas}. The angular point for which an observable can remain invariant is either $\theta = 0$ or $\theta = (\pi/2)$. Since in a truncated partial wave analysis one is always, ideally, using data that are given over the whole angular range $\theta \in \left[0,\pi\right]$, these single points of invariance can however not lead to any discrete partial wave ambiguities. \newline
We close this appendix section with a clarification of the vocabulary that surrounds the term \textit{double ambiguity}, which can be sometimes used confusingly. \newline
On the one hand, there exists the double ambiguity \textit{transformation}, either written in terms of roots as in (\ref{eq:DoubleAmbiguityDetail}), or as a transformation on the full transversity amplitudes as in (\ref{eq:DoubleAmbTrafoForm}) and (\ref{eq:DoubleAmbTrafoMatrix}). This is a discrete transformation on the amplitudes which relates different ambiguous solutions of the group $\mathcal{S}$ observables (and furthermore $E$ and $H$) to each other. \newline
On the other hand, one can speak of a multipole solution which originates from another one by the antilinear transformation (\ref{eq:DoubleAmbTrafoMatrix}) as the double ambiguity \textit{of another partial wave solution}. For instance, if in a truncation at $L=1$ the Omelaenko roots $\left\{\alpha_{1}, \alpha_{2}, \beta_{1}, \beta_{2}\right\}$ represent the true (or physical) solution, then the set of roots $\left\{\alpha_{1}^{\ast}, \alpha_{2}^{\ast}, \beta_{1}^{\ast}, \beta_{2}^{\ast}\right\}$
is the double ambiguity of the true solution. If however the set of roots $\left\{\alpha_{1}^{\ast}, \alpha_{2}, \beta_{1}, \beta_{2}^{\ast}\right\}$ fulfills Omelaenko's constraint, equation (40) in section \ref{sec:WBTpaper}, either exactly or approximately, then it is a valid candidate for an accidental ambiguity. In this case, the set of roots $\left\{\alpha_{1}, \alpha_{2}^{\ast}, \beta_{1}^{\ast}, \beta_{2}\right\}$ defines the double ambiguity of the aforementioned accidental symmetry. \newline
For both meanings described above, the word double ambiguity is used in section \ref{sec:WBTpaper} as well as the remainder of this work. However, it should always be clear from the context what is actually meant.

\begin{table}[h]
\centering
\begin{tabular}{c|cccccccc}
\hline
\hline
Observable & $O_{x^{\prime}}$ & $O_{z^{\prime}}$ & $C_{x^{\prime}}$ & $C_{z^{\prime}}$ & $T_{x^{\prime}}$ & $T_{z^{\prime}}$ & $L_{x^{\prime}}$ & $L_{z^{\prime}}$ \\
$\theta_{\mathrm{inv.}}$ $\left[\mathrm{rad}\right]$ & $0$ & $\left(\pi/2\right)$ & $\left(\pi/2\right)$ & $0$ & $0$ & $\left(\pi/2\right)$ & $\left(\pi/2\right)$ & $0$ \\
\hline
\hline
\end{tabular}
\caption[Angles for which specific $\mathcal{BR}$- and $\mathcal{TR}$-observables are invariant under the double ambiguity transformation.]{This Table summarizes the angles for which specific $\mathcal{BR}$- and $\mathcal{TR}$-observables are invariant under the double ambiguity transformation (\ref{eq:DoubleAmbTrafoMatrix}).}
\label{tab:BrTrInvarianceThetas}
\end{table}

\clearpage

\begin{table}[h]
 \centering
\begin{tabular*}{\linewidth}{c@{\extracolsep\fill}cc}
\hline 
\hline \\
 Observable $\check{\Omega}^{\alpha}$ & $\tilde{\Gamma}^{\alpha}$ & $\left( \mathcal{A} \left(\theta\right)^{\dagger} \tilde{\Gamma}^{\alpha} \mathcal{A} \left(\theta\right) \right)^{T}$ \\ \\
\hline \\
 $\check{\Omega}^{1} = \sigma_{0}$ & $ \tilde{\Gamma}^{1} = \left[
\begin{array}{cccc}
 1 & 0 & 0 & 0 \\
 0 & 1 & 0 & 0 \\
 0 & 0 & 1 & 0 \\
 0 & 0 & 0 & 1 \\
\end{array}
\right] $ & $ \left[
\begin{array}{cccc}
 1 & 0 & 0 & 0 \\
 0 & 1 & 0 & 0 \\
 0 & 0 & 1 & 0 \\
 0 & 0 & 0 & 1 \\
\end{array}
\right] = \tilde{\Gamma}^{1} $ \\ \\
 $\check{\Omega}^{4} = - \check{\Sigma}$ & $ \tilde{\Gamma}^{4} = \left[
\begin{array}{cccc}
 1 & 0 & 0 & 0 \\
 0 & 1 & 0 & 0 \\
 0 & 0 & -1 & 0 \\
 0 & 0 & 0 & -1 \\
\end{array}
\right] $ & $ \left[
\begin{array}{cccc}
 1 & 0 & 0 & 0 \\
 0 & 1 & 0 & 0 \\
 0 & 0 & -1 & 0 \\
 0 & 0 & 0 & -1 \\
\end{array}
\right] =\tilde{\Gamma}^{4} $ \\ \\
 $\check{\Omega}^{10} = - \check{T}$ & $ \tilde{\Gamma}^{10} = \left[
\begin{array}{cccc}
 -1 & 0 & 0 & 0 \\
 0 & 1 & 0 & 0 \\
 0 & 0 & 1 & 0 \\
 0 & 0 & 0 & -1 \\
\end{array}
\right] $ & $  \left[
\begin{array}{cccc}
 -1 & 0 & 0 & 0 \\
 0 & 1 & 0 & 0 \\
 0 & 0 & 1 & 0 \\
 0 & 0 & 0 & -1 \\
\end{array}
\right] = \tilde{\Gamma}^{10} $ \\ \\
 $\check{\Omega}^{12} = \check{P}$ & $ \tilde{\Gamma}^{12} = \left[
\begin{array}{cccc}
 -1 & 0 & 0 & 0 \\
 0 & 1 & 0 & 0 \\
 0 & 0 & -1 & 0 \\
 0 & 0 & 0 & 1 \\
\end{array}
\right] $ & $ \left[
\begin{array}{cccc}
 -1 & 0 & 0 & 0 \\
 0 & 1 & 0 & 0 \\
 0 & 0 & -1 & 0 \\
 0 & 0 & 0 & 1 \\
\end{array}
\right] = \tilde{\Gamma}^{12} $ \\ \\
 $\check{\Omega}^{3} = \check{G}$ & $ \tilde{\Gamma}^{3} = \left[
\begin{array}{cccc}
 0 & 0 & -i & 0 \\
 0 & 0 & 0 & -i \\
 i & 0 & 0 & 0 \\
 0 & i & 0 & 0 \\
\end{array}
\right] $ & $ \left[
\begin{array}{cccc}
 0 & 0 & i & 0 \\
 0 & 0 & 0 & i \\
 -i & 0 & 0 & 0 \\
 0 & -i & 0 & 0 \\
\end{array}
\right] = - \tilde{\Gamma}^{3} $ \\ \\
 $\check{\Omega}^{5} = \check{H}$ & $ \tilde{\Gamma}^{5} = \left[
\begin{array}{cccc}
 0 & 0 & 1 & 0 \\
 0 & 0 & 0 & -1 \\
 1 & 0 & 0 & 0 \\
 0 & -1 & 0 & 0 \\
\end{array}
\right] $ & $ \left[
\begin{array}{cccc}
 0 & 0 & 1 & 0 \\
 0 & 0 & 0 & -1 \\
 1 & 0 & 0 & 0 \\
 0 & -1 & 0 & 0 \\
\end{array}
\right] = \tilde{\Gamma}^{5} $ \\ \\
 $\check{\Omega}^{9} = \check{E}$ & $ \tilde{\Gamma}^{9} = \left[
\begin{array}{cccc}
 0 & 0 & 1 & 0 \\
 0 & 0 & 0 & 1 \\
 1 & 0 & 0 & 0 \\
 0 & 1 & 0 & 0 \\
\end{array}
\right] $ & $ \left[
\begin{array}{cccc}
 0 & 0 & 1 & 0 \\
 0 & 0 & 0 & 1 \\
 1 & 0 & 0 & 0 \\
 0 & 1 & 0 & 0 \\
\end{array}
\right] = \tilde{\Gamma}^{9} $ \\ \\
 $\check{\Omega}^{11} = \check{F}$ & $ \tilde{\Gamma}^{11} = \left[
\begin{array}{cccc}
 0 & 0 & i & 0 \\
 0 & 0 & 0 & -i \\
 -i & 0 & 0 & 0 \\
 0 & i & 0 & 0 \\
\end{array}
\right] $ & $ \left[
\begin{array}{cccc}
 0 & 0 & -i & 0 \\
 0 & 0 & 0 & i \\
 i & 0 & 0 & 0 \\
 0 & -i & 0 & 0 \\
\end{array}
\right] = - \tilde{\Gamma}^{11} $ \\ \\
\hline
\hline
\end{tabular*}
\caption[Effect of the double ambiguity transformation on the group $\mathcal{S}$ and BT observables.]{This Table shows the effect of the double ambiguity transformation ((\ref{eq:DoubleAmbTrafoForm}) $\&$ (\ref{eq:DoubleAmbTrafoMatrix})) on the group $\mathcal{S}$ and $\mathcal{BT}$ observables. The left column lists the observables as well as the numbering of the definition $\check{\Omega}^{\alpha}$ in terms of the bilinear hermiten form (\ref{eq:PolObsGenericChapter23}). The column in the middle shows the corresponding $\tilde{\Gamma}^{\alpha}$-matrix and on the right the result of the ambiguity test, equation (\ref{eq:ConditionInvAntilinAmb}), is shown.}
\label{tab:GroupSDoubleAmbiguityMatricesI}
\end{table}
\begin{sidewaystable}[h]
 \centering
\begin{tabular*}{\linewidth}{c@{\extracolsep\fill}cc}
\hline 
\hline \\
 Observable $\check{\Omega}^{\alpha}$ & $\tilde{\Gamma}^{\alpha}$ & $\left( \mathcal{A} \left(\theta\right)^{\dagger} \tilde{\Gamma}^{\alpha} \mathcal{A} \left(\theta\right) \right)^{T}$ \\ \\
\hline \\
 $\check{\Omega}^{14} = \check{O}_{x^{\prime}}$ & $ \tilde{\Gamma}^{14} = \left[
\begin{array}{cccc}
 0 & 0 & 0 & -1 \\
 0 & 0 & 1 & 0 \\
 0 & 1 & 0 & 0 \\
 -1 & 0 & 0 & 0 \\
\end{array}
\right]$ & $ \left[
\begin{array}{cccc}
 0 & 0 & 0 & -e^{-2 i \theta } \\
 0 & 0 & e^{2 i \theta } & 0 \\
 0 & e^{-2 i \theta } & 0 & 0 \\
 -e^{2 i \theta } & 0 & 0 & 0 \\
\end{array}
\right] = -\sin(2 \theta ) \tilde{\Gamma}^{7} + \cos(2 \theta ) \tilde{\Gamma}^{14} $ \\ \\
 $\check{\Omega}^{7} = - \check{O}_{z^{\prime}}$ & $ \tilde{\Gamma}^{7} = \left[
\begin{array}{cccc}
 0 & 0 & 0 & -i \\
 0 & 0 & -i & 0 \\
 0 & i & 0 & 0 \\
 i & 0 & 0 & 0 \\
\end{array}
\right] $ & $ \left[
\begin{array}{cccc}
 0 & 0 & 0 & i e^{-2 i \theta } \\
 0 & 0 & i e^{2 i \theta } & 0 \\
 0 & -i e^{-2 i \theta } & 0 & 0 \\
 -i e^{2 i \theta } & 0 & 0 & 0 \\
\end{array}
\right]  = -\cos(2 \theta ) \tilde{\Gamma}^{7} - \sin(2 \theta ) \tilde{\Gamma}^{14} $ \\ \\
 $\check{\Omega}^{16} = - \check{C}_{x^{\prime}}$ & $ \tilde{\Gamma}^{16} = \left[
\begin{array}{cccc}
 0 & 0 & 0 & i \\
 0 & 0 & -i & 0 \\
 0 & i & 0 & 0 \\
 -i & 0 & 0 & 0 \\
\end{array}
\right] $ & $ \left[
\begin{array}{cccc}
 0 & 0 & 0 & -i e^{-2 i \theta } \\
 0 & 0 & i e^{2 i \theta } & 0 \\
 0 & -i e^{-2 i \theta } & 0 & 0 \\
 i e^{2 i \theta } & 0 & 0 & 0 \\
\end{array}
\right] = -\sin(2 \theta ) \tilde{\Gamma}^{2} - \cos(2 \theta ) \tilde{\Gamma}^{16} $ \\ \\
 $\check{\Omega}^{2} = - \check{C}_{z^{\prime}}$ & $ \tilde{\Gamma}^{2} = \left[
\begin{array}{cccc}
 0 & 0 & 0 & 1 \\
 0 & 0 & 1 & 0 \\
 0 & 1 & 0 & 0 \\
 1 & 0 & 0 & 0 \\
\end{array}
\right] $ & $ \left[
\begin{array}{cccc}
 0 & 0 & 0 & e^{-2 i \theta } \\
 0 & 0 & e^{2 i \theta } & 0 \\
 0 & e^{-2 i \theta } & 0 & 0 \\
 e^{2 i \theta } & 0 & 0 & 0 \\
\end{array}
\right] = \cos(2 \theta ) \tilde{\Gamma}^{2} - \sin(2 \theta ) \tilde{\Gamma}^{16} $ \\ \\
\hline
\hline
\end{tabular*}
\caption[Effect of the double ambiguity transformation on the $\mathcal{BR}$ observables.]{This Table is the continuation of Table \ref{tab:GroupSDoubleAmbiguityMatricesI} for the $\mathcal{BR}$-observables.}
\label{tab:GroupSDoubleAmbiguityMatricesII}
\end{sidewaystable}
\begin{sidewaystable}[h]
 \centering
\begin{tabular*}{\linewidth}{c@{\extracolsep\fill}cc}
\hline 
\hline \\
 Observable $\check{\Omega}^{\alpha}$ & $\tilde{\Gamma}^{\alpha}$ & $\left( \mathcal{A} \left(\theta\right)^{\dagger} \tilde{\Gamma}^{\alpha} \mathcal{A} \left(\theta\right) \right)^{T}$ \\ \\
\hline \\
 $\check{\Omega}^{6} = - \check{T}_{x^{\prime}}$ & $ \tilde{\Gamma}^{6} = \left[
\begin{array}{cccc}
 0 & -1 & 0 & 0 \\
 -1 & 0 & 0 & 0 \\
 0 & 0 & 0 & 1 \\
 0 & 0 & 1 & 0 \\
\end{array}
\right] $ & $ \left[
\begin{array}{cccc}
 0 & -e^{-2 i \theta } & 0 & 0 \\
 -e^{2 i \theta } & 0 & 0 & 0 \\
 0 & 0 & 0 & e^{-2 i \theta } \\
 0 & 0 & e^{2 i \theta } & 0 \\
\end{array}
\right] = \cos(2 \theta ) \tilde{\Gamma}^{6} + \sin(2 \theta ) \tilde{\Gamma}^{13} $ \\ \\
 $\check{\Omega}^{13} = - \check{T}_{z^{\prime}}$ & $ \tilde{\Gamma}^{13} = \left[
\begin{array}{cccc}
 0 & i & 0 & 0 \\
 -i & 0 & 0 & 0 \\
 0 & 0 & 0 & -i \\
 0 & 0 & i & 0 \\
\end{array}
\right] $ & $ \left[
\begin{array}{cccc}
 0 & -i e^{-2 i \theta } & 0 & 0 \\
 i e^{2 i \theta } & 0 & 0 & 0 \\
 0 & 0 & 0 & i e^{-2 i \theta } \\
 0 & 0 & -i e^{2 i \theta } & 0 \\
\end{array}
\right] = \sin(2 \theta ) \tilde{\Gamma}^{6} - \cos(2 \theta ) \tilde{\Gamma}^{13} $ \\ \\
 $\check{\Omega}^{8} = \check{L}_{x^{\prime}}$ & $ \tilde{\Gamma}^{8} = \left[
\begin{array}{cccc}
 0 & -i & 0 & 0 \\
 i & 0 & 0 & 0 \\
 0 & 0 & 0 & -i \\
 0 & 0 & i & 0 \\
\end{array}
\right] $ & $ \left[
\begin{array}{cccc}
 0 & i e^{-2 i \theta } & 0 & 0 \\
 -i e^{2 i \theta } & 0 & 0 & 0 \\
 0 & 0 & 0 & i e^{-2 i \theta } \\
 0 & 0 & -i e^{2 i \theta } & 0 \\
\end{array}
\right] = -\cos(2 \theta ) \tilde{\Gamma}^{8} - \sin(2 \theta ) \tilde{\Gamma}^{15} $ \\ \\
 $\check{\Omega}^{15} = \check{L}_{z^{\prime}}$ & $ \tilde{\Gamma}^{15} = \left[
\begin{array}{cccc}
 0 & -1 & 0 & 0 \\
 -1 & 0 & 0 & 0 \\
 0 & 0 & 0 & -1 \\
 0 & 0 & -1 & 0 \\
\end{array}
\right] $ & $ \left[
\begin{array}{cccc}
 0 & -e^{-2 i \theta } & 0 & 0 \\
 -e^{2 i \theta } & 0 & 0 & 0 \\
 0 & 0 & 0 & -e^{-2 i \theta } \\
 0 & 0 & -e^{2 i \theta } & 0 \\
\end{array}
\right] = -\sin(2 \theta ) \tilde{\Gamma}^{8} + \cos(2 \theta ) \tilde{\Gamma}^{15} $ \\ \\
\hline
\hline
\end{tabular*}
\caption[Effect of the double ambiguity transformation on the $\mathcal{TR}$ observables.]{This Table is the continuation of Table \ref{tab:GroupSDoubleAmbiguityMatricesI} for the $\mathcal{TR}$-observables.}
\label{tab:GroupSDoubleAmbiguityMatricesIII}
\end{sidewaystable}

\clearpage

\subsubsection{Further results on accidental ambiguities} \label{sec:AccidentalAmbProofs}

In section \ref{sec:WBTpaper} and appendix \ref{sec:DoubleAmbiguityTrafoActingOnBi}, it was shown that the linear factor decompositions of the transversity amplitudes $b_{i}$ in a TPWA truncated at $L \equiv \ell_{\mathrm{max}}$ (using the angular variable $t = \tan \frac{\theta}{2}$)
\begin{align}
 b_{2} \left(\theta\right) &= - \hspace*{1pt} \mathcal{C} \hspace*{1pt} a_{2L} \hspace*{1pt} \frac{\exp \left(i \frac{\theta}{2}\right)}{\left( 1 + t^{2} \right)^{L}} \hspace*{1pt} \prod_{k = 1}^{2L} \left( t - \beta_{k} \right)  \mathrm{,} \label{eq:b2LinFactDecompSecAccAmb} \\
  b_{4} \left(\theta\right) &= \mathcal{C} \hspace*{1pt} a_{2L} \hspace*{1pt} \frac{\exp \left(i \frac{\theta}{2}\right)}{\left( 1 + t^{2} \right)^{L}} \hspace*{1pt} \prod_{k = 1}^{2L} \left( t - \alpha_{k} \right)  \mathrm{,} \label{eq:b4LinFactDecompSecAccAmb} \\
 b_{1} \left(\theta\right) &= b_{2} \left(- \theta\right) \mathrm{,} \hspace*{5pt} b_{3} \left(\theta\right) = b_{4} \left(- \theta\right) \mathrm{,} \label{eq:b1b3LinFactDecompositionsSecAccAmb}
\end{align}
were fundamental for the derivation of the discrete partial wave ambiguities of the group $\mathcal{S}$ observables $\left\{\sigma_{0}, \check{\Sigma}, \check{T}, \check{P}\right\}$, the latter being linear combinations of moduli-squares of the $b_{i}$: $\check{\Omega}^{\alpha_{S}} = \frac{1}{2} \left( \pm \left|b_{1}\right|^{2} \pm \left|b_{2}\right|^{2} \pm \left|b_{3}\right|^{2} + \left|b_{4}\right|^{2} \right)$. \newline
Once the moduli of the decompositions given above are squared, the Omelaenko-roots (e.g. $\alpha_{k}$) always occur in the combination $\left( t - \alpha_{k}^{\ast} \right)\left( t - \alpha_{k} \right)$, where is was assumed that $t$ is real. Therefore, complex conjugation of either all roots $\left\{\alpha_{k}, \beta_{k}\right\}$, or only subsets of them leaves all group $\mathcal{S}$ observables invariant. The first possibility was referred to as the \textit{double ambiguity}, cf. appendix \ref{sec:DoubleAmbiguityTrafoActingOnBi}, while the partial wave ambiguities originating from the second possibility are called \textit{accidental ambiguities}. \newline \newline
Furthermore, it turns out that the complex roots appearing in the group $\mathcal{S}$ observables (\ref{eq:b2LinFactDecompSecAccAmb}) to (\ref{eq:b1b3LinFactDecompositionsSecAccAmb}) are not independent. Rather, the TPWA itself produces the following multiplicative constraint among the $\left\{\alpha_{k}, \beta_{k}\right\}$
\begin{equation}
 \prod_{k = 1}^{2L} \alpha_{k} = \prod_{k = 1}^{2L} \beta_{k} \mathrm{,} \label{eq:ConsistencyRelationAccAmb}
\end{equation}
which was therefore termed the \textit{consistency relation} in section \ref{sec:WBTpaper}. Generally it is true that once a TPWA is fitted in the standard form (\ref{eq:LowEAssocLegStandardParametrization1}) and (\ref{eq:LowEAssocLegStandardParametrization2}), with real and imaginary parts of the multipoles as relevant variables, is has the constraint (\ref{eq:ConsistencyRelationAccAmb}) built into it implicitly. Therefore it can never violate it. This fact shall be motivated in more detail later. \newline
In the analyses of mathematically exactly solvable theoretical model-data performed in this thesis (cf. sec. \ref{sec:TheoryDataFits}), it turned out that the accidental ambiguities never exactly fulfill the constraint (\ref{eq:ConsistencyRelationAccAmb}), but only satisfy it up to a small numerical error. This can also be seen in the ambiguity diagram shown in Figure 2 of section \ref{sec:WBTpaper}. \newline
Once accidental symmetries of the group $\mathcal{S}$ are present which only satisfy the consistency relation approximately, interesting mathematical consequences can be deduced for the multipole-solutions corresponding to them. These consequences then furthermore fortify the statement that, at least in mathematically exactly solvable truncated partial wave analyses, complete experiments can be found that contain only 5 observables. \newline
Before the above mentioned derivation can be done, the accidental ambiguities have to be defined in a mathematically more rigorous way. Furthermore, the notion of ``approximate fulfillment'' of the constraint (\ref{eq:ConsistencyRelationAccAmb}) also has to be given an explicit formal guise.
\clearpage

\paragraph{Mathematical preliminaries} \label{subsec:AccidentalAmbProofsI} \textcolor{white}{:-)} \newline

In the beginning, we assume that an exact \textit{true} or \textit{physical} solution to the TPWA problem exists and that it may be represented by a set of complex Omelaenko-roots $\left\{ \alpha_{k}, \beta_{k} \right\}$ (this is a standard assumption which was also made in reference \cite{Omelaenko} and section \ref{sec:WBTpaper}). \newline
Furthermore, it is useful to introduce a discrete set $\mathbbm{U}_{\mathcal{R}} \subset \mathbbm{C}^{4 L}$ as the subset of points on which the original TPWA-solution $\left\{ \alpha_{k}, \beta_{k} \right\}$ as well as all of the discrete ambiguities of the group $\mathcal{S}$ shall exist. Therefore, one can generally introduce the root-vector
\begin{equation}
 \left( \alpha_{1},\ldots,\alpha_{2L},\beta_{1},\ldots,\beta_{2L}\right) \in \mathbbm{U}_{\mathcal{R}} \mathrm{.} \label{eq:DefRootVectorInU}
\end{equation}
There is of course the isomorphism $\mathbbm{C}^{4 L} \sim \mathbbm{R}^{8 L}$, such that the root vector may equivalently be represented by an $8L$-dimensional real vector. \newline
It is now necessary to mathematically formalize all the different possibilities that exist for forming complex conjugates for all possible subsets of the roots $\left\{ \alpha_{k}, \beta_{k} \right\}$. In order to describe this finite amount of possibilities, we introduce a finite set of maps $\bm{\uppi}_{\hspace*{0.035cm}n} \hspace*{2pt} : \hspace*{2pt} \mathbbm{U}_{\mathcal{R}} \rightarrow \mathbbm{U}_{\mathcal{R}}$ that act on the root vector (\ref{eq:DefRootVectorInU}) as:
\begin{equation}
 \left( \alpha_{1},\ldots,\alpha_{2L},\beta_{1},\ldots,\beta_{2L}\right) \rightarrow \left( \bm{\uppi}_{\hspace*{0.035cm}n} \left(\alpha_{1}\right),\ldots, \bm{\uppi}_{\hspace*{0.035cm}n} \left(\alpha_{2L}\right), \bm{\uppi}_{\hspace*{0.035cm}n} \left(\beta_{1}\right),\ldots, \bm{\uppi}_{\hspace*{0.035cm}n} \left(\beta_{2L}\right)\right) \mathrm{.} \label{eq:PiNMapMathAction}
\end{equation}
In total, since there are $4L$ Omelaenko-roots for an arbitrary truncation order $L$, there exist $2^{4L}$ different possibilities to form complex conjugates of subsets of them (a proof of this fact is given at the end of this appendix). \newline
We introduce here the elegant numbering scheme of Gersten (ref. \cite{Gersten}) in order to uniquely assign to each index $n$ a particular map $\bm{\uppi}_{\hspace*{0.035cm}n}$. Therefore, we define the index $n$ to run over the set of numbers $\left\{0,1,\ldots,2^{4L} - 1\right\}$, such that any $n$ has a unique binary representation expressed by the formula
\begin{equation}
 n = \sum_{k = 1}^{2L} \mu_{k} \left( n \right) 2^{(k-1)} + \sum_{k^{\prime} = 1}^{2L} \nu_{k^{\prime}} \left( n \right) 2^{(2 L + k^{\prime} - 1)} \mathrm{.} \label{eq:BinaryRepresentationOfN}
\end{equation}
The binary expansion coefficients $\mu_{k} \left( n \right)$ and $\nu_{k^{\prime}} \left( n \right)$ can take the values $0$ or $1$.
Utilizing this expansion into powers of $2$, one can accomplish the above mentioned one-to-one correspondence of indices $n$ to maps $\bm{\uppi}_{\hspace*{0.035cm}n}$ by the following definition
\begin{equation}
 \bm{\uppi}_{\hspace*{0.035cm}n} \left(\alpha_{k}\right) := \begin{cases}
                    \alpha_{k} &\mathrm{,} \hspace*{3pt} \mu_{k} \left(n\right) = 0 \\
                    \alpha_{k}^{\ast} &\mathrm{,} \hspace*{3pt} \mu_{k} \left(n\right) = 1 
                   \end{cases}  \hspace*{5pt} \mathrm{and} \hspace*{5pt}
 \bm{\uppi}_{\hspace*{0.035cm}n} \left(\beta_{k^{\prime}}\right) := \begin{cases}
                    \beta_{k^{\prime}} &\mathrm{,} \hspace*{3pt} \nu_{k^{\prime}} \left(n\right) = 0 \\
                    \beta_{k^{\prime}}^{\ast} &\mathrm{,} \hspace*{3pt} \nu_{k^{\prime}} \left(n\right) = 1 
                   \end{cases} \mathrm{.} \label{eq:PiNMapsDefinition}
\end{equation}
Since effectively the $\bm{\uppi}_{\hspace*{0.035cm}n}$ just conjugate particular roots while leaving the remaining ones as they are, it is equivalent to define the action of these maps on the phases of $\alpha_{k} = \left| \alpha_{k} \right| e^{i \varphi_{k}}$ and $\beta_{k^{\prime}} = \left| \beta_{k^{\prime}} \right| e^{i \psi_{k^{\prime}}}$ as
\begin{equation}
 \bm{\uppi}_{\hspace*{0.035cm}n} \left(\varphi_{k}\right) := \begin{cases}
                    + \varphi_{k} &\mathrm{,} \hspace*{3pt} \mu_{k} \left(n\right) = 0 \\
                    - \varphi_{k} &\mathrm{,} \hspace*{3pt} \mu_{k} \left(n\right) = 1 
                   \end{cases}  \hspace*{5pt} \mathrm{and} \hspace*{5pt}
 \bm{\uppi}_{\hspace*{0.035cm}n} \left(\psi_{k^{\prime}}\right) := \begin{cases}
                    + \psi_{k^{\prime}} &\mathrm{,} \hspace*{3pt} \nu_{k^{\prime}} \left(n\right) = 0 \\
                    - \psi_{k^{\prime}} &\mathrm{,} \hspace*{3pt} \nu_{k^{\prime}} \left(n\right) = 1 
                   \end{cases} \mathrm{.} \label{eq:PiNMapsDefinitionActingOnPhases}
\end{equation}
In order to illustrate the definitions given up to now, Table \ref{tab:AllCCPossibilitiesLEquals1Case} lists all correspondences between the map index $n$ and different combinations of sign changes as defined in (\ref{eq:PiNMapsDefinitionActingOnPhases}) for the lowest non-trivial truncation angular momentum $L = 1$. \newpage
\begin{table}[h]
\centering
\begin{tabular}{c|cccc||c|cccc}
\hline
\hline
$n$ of $\bm{\uppi}_{\hspace*{0.035cm}n}$ & $\varphi_{1}$ & $\varphi_{2}$ & $\psi_{1}$ & $\psi_{2}$ & $n$ of $\bm{\uppi}_{\hspace*{0.035cm}n}$ & $\varphi_{1}$ & $\varphi_{2}$ & $\psi_{1}$ & $\psi_{2}$  \\
\hline
 $0$ & $+$ & $+$ & $+$ & $+$ & $8$ & $+$ & $+$ & $+$ & $-$ \\
 $1$ & $-$ & $+$ & $+$ & $+$ & $9$ & $-$ & $+$ & $+$ & $-$ \\
 $2$ & $+$ & $-$ & $+$ & $+$ & $10$ & $+$ & $-$ & $+$ & $-$ \\
 $3$ & $-$ & $-$ & $+$ & $+$ & $11$ & $-$ & $-$ & $+$ & $-$ \\
\hline
 $4$ & $+$ & $+$ & $-$ & $+$ & $12$ & $+$ & $+$ & $-$ & $-$ \\
 $5$ & $-$ & $+$ & $-$ & $+$ & $13$ & $-$ & $+$ & $-$ & $-$ \\
 $6$ & $+$ & $-$ & $-$ & $+$ & $14$ & $+$ & $-$ & $-$ & $-$ \\
 $7$ & $-$ & $-$ & $-$ & $+$ & $15$ & $-$ & $-$ & $-$ & $-$ \\
\hline
\hline
\end{tabular}
\caption[All possibilities for complex conjugation of Omelaenko-roots, for $L=1$.]{All possibilities for complex conjugation of roots, numbered according to the binary representation (\ref{eq:BinaryRepresentationOfN}) and the rule (\ref{eq:PiNMapsDefinitionActingOnPhases}), for a truncation at $L = 1$.}
\label{tab:AllCCPossibilitiesLEquals1Case}
\end{table}
The set of all possible maps $\bm{\uppi}_{\hspace*{0.035cm}n}$, corresponding to all possible ways of forming discrete partial wave ambiguities out of the $\left\{ \alpha_{k}, \beta_{k} \right\}$, shall be defined as
\begin{equation}
 \mathcal{P} := \left\{ \bm{\uppi}_{\hspace*{0.035cm}n} : \mathbbm{U}_{\mathcal{R}} \rightarrow \mathbbm{U}_{\mathcal{R}} \hspace*{3pt} | \hspace*{3pt} n = 0,\ldots,2^{4L} - 1 \right\} \mathrm{.} \label{eq:DefAllPermutations}
\end{equation}
At this point, it is important to note that the full set $\mathbbm{U}_{\mathcal{R}}$ can be mapped out by acting on any $\bm{\uppi}_{\hspace*{0.035cm}n} \left( \alpha_{k}, \beta_{k} \right)$ (which is itself reached from the true solution $\left\{\alpha_{k},\beta_{k}\right\}$) with the full set $\mathcal{P}$. For example, for the cases listed in Table \ref{tab:AllCCPossibilitiesLEquals1Case}, it is true that $\bm{\uppi}_{\hspace*{0.035cm}9} \left( \alpha_{k}, \beta_{k} \right) = \bm{\uppi}_{\hspace*{0.035cm}1} \left[ \bm{\uppi}_{\hspace*{0.035cm}8} \left( \alpha_{k}, \beta_{k} \right) \right]$. Furthermore, the prescription
\begin{equation}
 \bm{\uppi}_{\hspace*{0.035cm}n} \left( \alpha_{k}, \beta_{k} \right) = \begin{cases}
                    \bm{\uppi}_{\hspace*{0.035cm} (n+8) } \left[ \bm{\uppi}_{\hspace*{0.035cm}8} \left( \alpha_{k}, \beta_{k} \right) \right] &\mathrm{,} \hspace*{3pt} n \leq \frac{1}{2} 2^{4L} - 1 \\
                    \bm{\uppi}_{\hspace*{0.035cm} (n-8) } \left[ \bm{\uppi}_{\hspace*{0.035cm}8} \left( \alpha_{k}, \beta_{k} \right) \right] &\mathrm{,} \hspace*{3pt} n > \frac{1}{2} 2^{4L} - 1
                   \end{cases} \mathrm{,} \label{eq:LawAnyOtherMapOutOfPi8}
\end{equation}
generates the full set of ambiguity roots out of $\bm{\uppi}_{\hspace*{0.035cm}8} \left( \alpha_{k}, \beta_{k} \right)$. Similar laws may be deduced by starting from any other set of already transformed roots. \newline
Therefore, one does not even have to start at the true solution $\left\{\alpha_{k},\beta_{k}\right\}$ in order to map out all discrete ambiguities. Any other solution could also be used for this purpose, provided that it is an exact solution of the group $\mathcal{S}$ observables $\left\{\sigma_{0}, \check{\Sigma}, \check{T}, \check{P}\right\}$ (As shall be shown later, this is generically only the case for the double ambiguity of the true solution.). \newline
However, since the discrete ambiguities of the group $\mathcal{S}$ observables shall be classified here and the latter are assumed to have the true solution $\left\{\alpha_{k},\beta_{k}\right\}$, we shall always start at this particular true set of roots. \newline
Furthermore, it should be mentioned here that the maps $\bm{\uppi}_{\hspace*{0.035cm}n}$ themselves and the results $\bm{\uppi}_{\hspace*{0.035cm}n} \left( \alpha_{k}, \beta_{k} \right)$ of these maps applied to a particular set of roots are different kinds of objects, since the latter are just points in $\mathbbm{U}_{\mathcal{R}}$. Still the vocabulary utilized in the following discussion may use the same words for these different concepts and the context should always clarify what is meant. \newline \newline
Now we continue by clarifying which maps can generate the accidental ambiguities. The definition in the introduction of this appendix section (app. \ref{sec:AccidentalAmbProofs}) tells that these are all maps contained in $\mathcal{P}$, except for the identity $\bm{\uppi}_{\hspace*{0.035cm}0} = \mathbbm{1}$ and the double ambiguity transformation $\bm{\uppi}_{\hspace*{0.035cm}\left( 2^{4L} - 1 \right)} = \mathrm{D.A.}$ which returns the double ambiguity $\left\{\alpha_{k}^{\ast},\beta_{k}^{\ast}\right\}$ of the true
solution. Therefore, the relevant set of maps for the discussion of the accidental ambiguities is identified as
\begin{equation}
 \hat{\mathcal{P}} := \left\{ \bm{\uppi}_{\hspace*{0.035cm}n} : \mathbbm{U}_{\mathcal{R}} \rightarrow \mathbbm{U}_{\mathcal{R}} \hspace*{3pt} | \hspace*{3pt} n = 1,\ldots,2^{4L} - 2 \right\} \equiv \mathcal{P} \setminus \left\{ \bm{\uppi}_{\hspace*{0.035cm} 0 }, \bm{\uppi}_{\hspace*{0.035cm}\left( 2^{4L} - 1 \right)} \right\} \mathrm{.} \label{eq:DefAccAmbSubsetOfAllPermutations}
\end{equation}
The total number of candidates for accidental ambiguities, which is at the same time an upper bound for the number of actually occurring ambiguities of that kind, amounts to
\begin{equation}
  N_{\mathrm{AC}}^{\mathrm{total}} = 2^{4 L} - 2 \mathrm{.} \label{eq:AccAmbCountingTotalN}
\end{equation}
It is a fact that, as already noted by Omelaenko \cite{Omelaenko}, the accidental ambiguities are not fully independent. Rather they occur in pairs, with the double ambiguity transformation relating two solutions according to
\begin{equation}
   \left( \bm{\uppi}_{\hspace*{0.035cm}n} \left( \alpha_{k}\right), \bm{\uppi}_{\hspace*{0.035cm}n} \left( \beta_{k} \right) \right) \overset{\mathrm{D.A.}}{\longrightarrow} \left( \bm{\uppi}_{\hspace*{0.035cm}n} \left( \alpha_{k}\right)^{\ast}, \bm{\uppi}_{\hspace*{0.035cm}n} \left( \beta_{k} \right)^{\ast} \right) \mathrm{.} \label{eq:DoubleAmbOfPiNAMbiguity}
\end{equation}
Furthermore, the general numbering scheme given by (\ref{eq:BinaryRepresentationOfN}), (\ref{eq:PiNMapsDefinition}) and (\ref{eq:PiNMapsDefinitionActingOnPhases}) facilitates the reformulation of the transformation (\ref{eq:DoubleAmbOfPiNAMbiguity}) in a more elegant way. The investigation of the example case shown in Table \ref{tab:AllCCPossibilitiesLEquals1Case} allows for the deduction of the following formula, which also holds true for 
the higher truncation angular momenta:
\begin{equation}
  \bm{\uppi}_{\hspace*{0.035cm}n} \left( \alpha_{k}, \beta_{k} \right) \overset{\mathrm{D.A.}}{\longrightarrow} \bm{\uppi}_{\hspace*{0.035cm}\left(2^{4 L} - n - 1\right)} \left( \alpha_{k}, \beta_{k} \right) \mathrm{.} \label{eq:DoubleAmbOfPiNAMbiguityShort}
\end{equation}
Since ambiguities can always be linked in pairs in the above given way, it is seen that the upper bound for the non-redundant accidental ambiguities is just given by the half of equation (\ref{eq:AccAmbCountingTotalN})
\begin{equation}
 N_{\mathrm{AC}} = \frac{1}{2} \left( 2^{4 L} - 2 \right) \mathrm{.} \label{eq:AccAmbCounting}
\end{equation}
In order to provide an intuition for the rapid growth of the power laws (\ref{eq:AccAmbCountingTotalN}) and (\ref{eq:AccAmbCounting}), the resulting numbers are listed in Table \ref{tab:AccAmbPossibilityNumber}. If multipoles are fitted, it is not so easy any more to recognize which pairs of solutions are linked in the way expressed in equation (\ref{eq:DoubleAmbOfPiNAMbiguityShort}). Therefore it was chosen to list both numbers for the upper bound of the non-redundant 
accidental symmetries (\ref{eq:AccAmbCounting}), as well as the upper bound for actually encountered multipole solutions (\ref{eq:AccAmbCountingTotalN}). In both cases, the exponential nature of the law is seen to cause an extremely fast growth. For the non-redundant ambiguities, the number of possibilities reaches the millions for $L = 6$, while for the number of multipole solutions this order of magnitude is already reached at $L = 5$.

\begin{table}[h]
\centering
\begin{tabular}{c|r|r}
\hline
\hline
 $L$ & $N_{\mathrm{AC}}$ & $N_{\mathrm{AC}}^{\mathrm{total}}$ \\
\hline
 $1$ & $7$ & $14$ \\
 $2$ & $127$ & $254$ \\
 $3$ & $2\hspace*{1pt}047$ & $4\hspace*{1pt}094$ \\
 $4$ & $32\hspace*{1pt}767$ & $65\hspace*{1pt}534$ \\
 $5$ & $524\hspace*{1pt}287$ & $1\hspace*{1pt}048\hspace*{1pt}574$ \\
 $6$ & $8\hspace*{1pt}388\hspace*{1pt}607$ & $16\hspace*{1pt}777\hspace*{1pt}214$ \\
\hline
\hline
\end{tabular}
\caption[Maximally possible number of ambiguities for the lowest truncation angular momenta.]{This Table lists evaluations of the power-laws (\ref{eq:AccAmbCountingTotalN}) and (\ref{eq:AccAmbCounting}) for the lowest non-trivial truncation angular momenta.}
\label{tab:AccAmbPossibilityNumber}
\end{table}
\newpage
Having formalized what the accidental ambiguities actually are as well as established that the upper bound for their number grows exponentially fast, it is now time to focus on the one condition in the TPWA-formalism that is capable of removing such ambiguities, namely the consistency relation (\ref{eq:ConsistencyRelationAccAmb}). Expressed in terms of the root-phases $\varphi_{k}$ and $\psi_{k^{\prime}}$, this constraint reads
\begin{equation}
 \varphi_{1} + \ldots + \varphi_{2L} = \psi_{1} + \ldots + \psi_{2L} \mathrm{.} \label{eq:ConsistencyRelationRoots}
\end{equation}
If an accidental ambiguity given by acting on the phases as defined in (\ref{eq:PiNMapsDefinitionActingOnPhases}) exactly fulfills this constraint, it is a valid ambiguity of the group $\mathcal{S}$ observables. In case exact fulfillment of (\ref{eq:ConsistencyRelationRoots}) is not given, the ambiguity is in principle ruled out. \newline
However, it is interesting to consider the case that some particular $\bm{\uppi}_{\hspace*{0.035cm}n} \in \hat{\mathcal{P}}$ violates (\ref{eq:ConsistencyRelationRoots}) only by a small numerical error, or stated equivalently, fulfills is approximately.
This situation is exactly what happens generically in analyses of theoretical model data. Therefore a mathematical formulation of this case is desirable. Assuming that some $\bm{\uppi}_{\hspace*{0.035cm}n} \in \hat{\mathcal{P}}$ fulfills (\ref{eq:ConsistencyRelationRoots}) approximately, one could just express this as
\begin{equation}
 \bm{\uppi}_{\hspace*{0.035cm}n} \left(\varphi_{1}\right) + \ldots + \bm{\uppi}_{\hspace*{0.035cm}n} \left(\varphi_{2L}\right) \simeq \bm{\uppi}_{\hspace*{0.035cm}n} \left(\psi_{1}\right) + \ldots + \bm{\uppi}_{\hspace*{0.035cm}n} \left(\psi_{2L}\right) \mathrm{.} \label{eq:ConsistencyRelationApproxFulfillmentRoots}
\end{equation}
However, for further derivations it is more advantageous to have a statement given as an exact equality, instead of using the sign $\simeq$. It is possible to obtain such a statement by introducing a real parameter that measures the violation of the constraint in equation (\ref{eq:ConsistencyRelationApproxFulfillmentRoots}). \newline
There has to exist a real quantity $\epsilon_{\bm{\uppi}_{\hspace*{0.015cm}n}}$ associated to the particular ambiguity $\bm{\uppi}_{\hspace*{0.035cm}n}$ under consideration such that the validity of the relation (\ref{eq:ConsistencyRelationApproxFulfillmentRoots}) is fully equivalent to
\begin{equation}
 \bm{\uppi}_{\hspace*{0.035cm}n} \left(\varphi_{1}\right) + \ldots + \bm{\uppi}_{\hspace*{0.035cm}n} \left(\varphi_{2L}\right) = \bm{\uppi}_{\hspace*{0.035cm}n} \left(\psi_{1}\right) + \ldots + \bm{\uppi}_{\hspace*{0.035cm}n} \left(\psi_{2L}\right) + \epsilon_{\bm{\uppi}_{\hspace*{0.015cm}n}} \mathrm{.} \label{eq:ConsistencyRelationEpsilonFormulationRoots}
\end{equation}
By assumption, the quantity $\epsilon_{\bm{\uppi}_{\hspace*{0.015cm}n}}$ has an infinitesimally small modulus: $\left| \epsilon_{\bm{\uppi}_{\hspace*{0.015cm}n}} \right| \ll 1$. Furthermore, it is important to see that in case this small parameter would be negative, the $\epsilon$-parameter corresponding to the double ambiguity $\bm{\uppi}_{\hspace*{0.035cm}\left(2^{4L} - 1 - n\right)}$ of
$\bm{\uppi}_{\hspace*{0.035cm}n}$ would be positive:
\begin{align}
 \epsilon_{\bm{\uppi}_{\hspace*{0.035cm}\left(2^{4L} - 1 - n\right)}} &= \bm{\uppi}_{\hspace*{0.035cm}\left(2^{4L} - 1 - n\right)} \left(\varphi_{1}\right) + \ldots + \bm{\uppi}_{\hspace*{0.035cm}\left(2^{4L} - 1 - n\right)} \left(\varphi_{2L}\right) - \bm{\uppi}_{\hspace*{0.035cm}\left(2^{4L} - 1 - n\right)} \left(\psi_{1}\right)  \nonumber \\
  &\hspace*{12.5pt} - \ldots -  \bm{\uppi}_{\hspace*{0.035cm}\left(2^{4L} - 1 - n\right)} \left(\psi_{2L}\right) \nonumber \\
  &= - \bm{\uppi}_{\hspace*{0.035cm}n} \left(\varphi_{1}\right) - \ldots - \bm{\uppi}_{\hspace*{0.035cm}n} \left(\varphi_{2L}\right) + \bm{\uppi}_{\hspace*{0.035cm}n} \left(\psi_{1}\right) + \ldots + \bm{\uppi}_{\hspace*{0.035cm}n} \left(\psi_{2L}\right) \nonumber \\
  &= - \epsilon_{\bm{\uppi}_{\hspace*{0.015cm}n}} \mathrm{.} \label{eq:EpsilonDoubleAmbRelationProof}
\end{align}
For this reason, in subsequent derivations when a generic accidental ambiguity $\bm{\uppi}_{\hspace*{0.035cm}n} \in \hat{\mathcal{P}}$ is considered, it will always be possible to assume that without loss of generality $\epsilon_{\bm{\uppi}_{\hspace*{0.015cm}n}} > 0$ is valid. \newline
When stated in terms of full root variables and not phases, the equation (\ref{eq:ConsistencyRelationEpsilonFormulationRoots}) reads
\begin{equation}
\bm{\uppi}_{\hspace*{0.035cm}n} \left(\alpha_{1}\right) \ast \ldots \ast \bm{\uppi}_{\hspace*{0.035cm}n} \left(\alpha_{2L}\right) = \bm{\uppi}_{\hspace*{0.035cm}n} \left(\beta_{1}\right) \ast \ldots \ast \bm{\uppi}_{\hspace*{0.035cm}n} \left(\beta_{2L}\right) \ast e^{i \epsilon_{\bm{\uppi}_{\hspace*{0.0025cm}n}}} \mathrm{.} \label{eq:GeneralizedViolatedCRWithNIndex}
\end{equation}
This is the form in which it shall be used during the ensuing appendix section. \newline
Before closing this section, the meaning of the quantities $\epsilon_{\bm{\uppi}_{\hspace*{0.015cm}n}}$ shall be illustrated by a particular example.

\begin{sidewaysfigure}[h]
 \centering
  \begin{overpic}[width=0.99\textwidth]{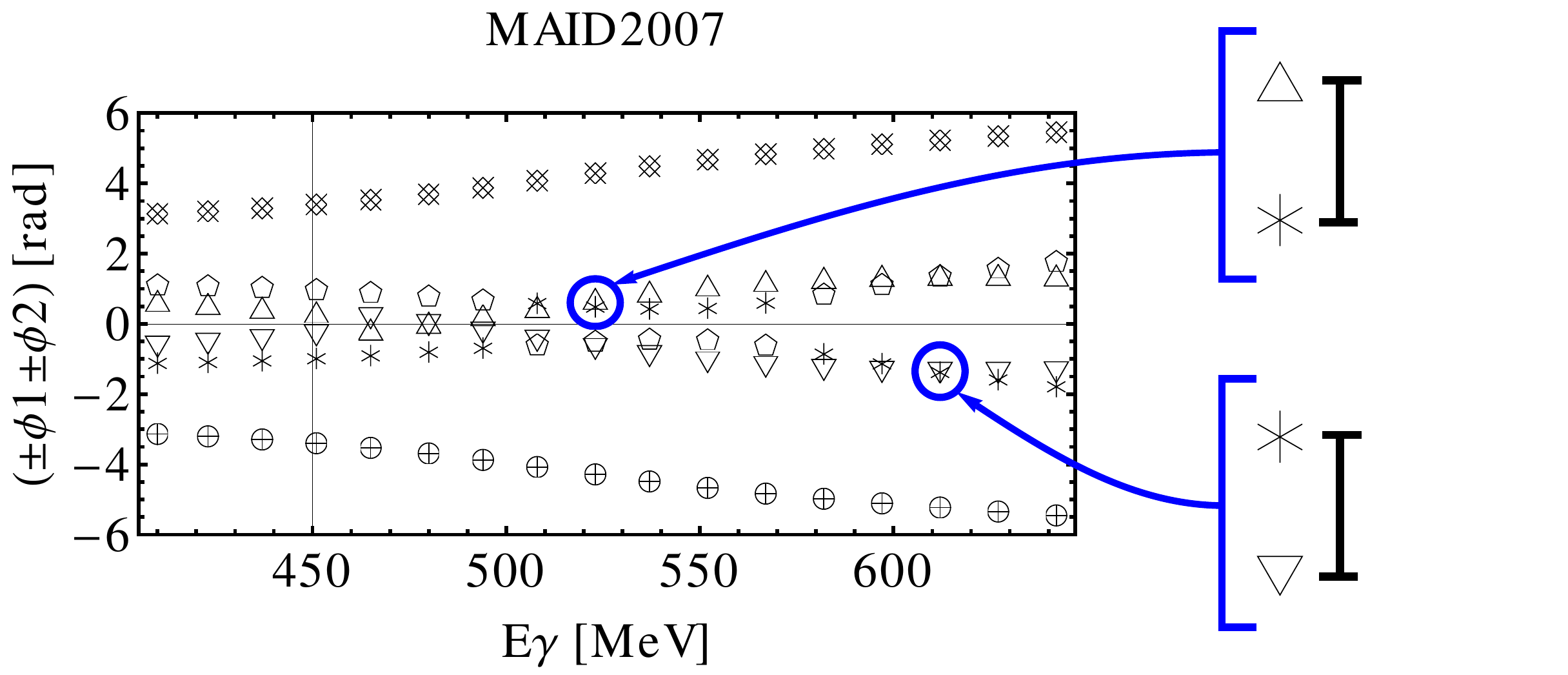}
 \put(88.25,32.3){\begin{Huge}$= \hspace*{2.5pt} \epsilon_{\bm{\uppi}_{\hspace*{0.015cm}10}}$\end{Huge}}
 \put(88.25,9.70){\begin{Huge}$= \hspace*{2.5pt} \epsilon_{\bm{\uppi}_{\hspace*{0.015cm}9}}$\end{Huge}}
  \end{overpic}
\caption[Excerpt of the ambiguity diagram for the $S$- and $P$-wave multipoles ($L = 1$) of the MAID2007 solution for $\pi^{0}$ photoproduction. Accidental ambiguities are illustrated explicitly.]{This picture contains an excerpt of the ambiguity diagram for the $S$- and $P$-wave multipoles ($L = 1$) of the MAID2007 solution \cite{MAID2007,MAID} for $\pi^{0}$ photoproduction, which is given in Figure 2 of section \ref{sec:WBTpaper}. Different sign-choices for the linear combinations of the phases $\left\{\varphi_{1}, \varphi_{2}\right\}$ and $\left\{\psi_{1}, \psi_{2}\right\}$ are plotted against energy $E_{\gamma}$. Furthermore, two examples for specific energy bins where an accidental ambiguity $\bm{\uppi}_{\hspace*{0.035cm}n} \in \hat{\mathcal{P}}$ occurs are emphasized by the blue circles. The two cases are zoomed out in order to clarify their corresponding parameters $\epsilon_{\bm{\uppi}_{\hspace*{0.015cm}n}}$. \newline The labeling scheme for the different linear combinations is: $\circ ( \varphi_{1} + \varphi_{2})$, $\bigtriangleup (\varphi_{1} - \varphi_{2})$, $\bigtriangledown (- \varphi_{1} + \varphi_{2})$, $\diamond (- \varphi_{1} - \varphi_{2})$,
 $+ (\psi_{1} + \psi_{2})$, ${\Large \ast} (\psi_{1} - \psi_{2})$, $\pentagon (- \psi_{1} + \psi_{2})$, $\times (- \psi_{1} -\psi_{2})$.}
\label{fig:ExampleEpsilonPiParametersInAmbiguityDiagram}
\end{sidewaysfigure}
\clearpage

In Figure \ref{fig:ExampleEpsilonPiParametersInAmbiguityDiagram}, an extract of the ambiguity diagram given in Figure 2 of section \ref{sec:WBTpaper} is shown. Here, the energy region was restricted to the interval from $400$ to $650$ MeV. \newline \newline
The consistency relation holds for the phases $\left\{\varphi_{k}, \psi_{k}\right\}$ of the original solution, i.e.
\begin{equation}
 \varphi_{1} + \varphi_{2} = \psi_{1} + \psi_{2} \mathrm{.} \label{eq:CRLmax1Example}
\end{equation}
The symbols in Figure \ref{fig:ExampleEpsilonPiParametersInAmbiguityDiagram} correspond to different $\bm{\uppi}_{\hspace*{0.015cm}n} \in \mathcal{P}$ acting on the left- and right hand side of equation (\ref{eq:CRLmax1Example}), respectively. Different linear combinations $\bm{\uppi}_{\hspace*{0.015cm}n} \left(\varphi_{1}\right) + \bm{\uppi}_{\hspace*{0.015cm}n} \left(\varphi_{2}\right)$ and $\bm{\uppi}_{\hspace*{0.015cm}n} \left(\psi_{1}\right) + \bm{\uppi}_{\hspace*{0.015cm}n} \left(\psi_{2}\right)$ are represented by different symbols as explained in the Figure caption. \newline
The symbols $\circ$ and $+$ at the bottom of the plot correspond to the identity-map $\bm{\uppi}_{\hspace*{0.015cm}0} \left(\varphi_{k}, \psi_{k}\right)$. They lie exactly on top of each other as a results of the validity of equation (\ref{eq:CRLmax1Example}). The symbols $\diamond$ and $\times$ at the top of Figure \ref{fig:ExampleEpsilonPiParametersInAmbiguityDiagram} represent the double ambiguity of the original solution, i.e. $\bm{\uppi}_{\hspace*{0.015cm}15} \left(\varphi_{k}, \psi_{k}\right)$. \newline
The approximate validity of the consistency relation for accidental ambiguities, which is generally expressed by equation (\ref{eq:ConsistencyRelationApproxFulfillmentRoots}), reads for this special case
\begin{equation}
 \bm{\uppi}_{\hspace*{0.015cm}n} \left(\varphi_{1}\right) + \bm{\uppi}_{\hspace*{0.015cm}n} \left(\varphi_{2}\right) \simeq \bm{\uppi}_{\hspace*{0.015cm}n} \left(\psi_{1}\right) + \bm{\uppi}_{\hspace*{0.015cm}n} \left(\psi_{2}\right) \mathrm{.} \label{eq:ConsistencyRelationApproxFulfillmentRootsLmax1}
\end{equation}
Two cases are highlighted in Figure \ref{fig:ExampleEpsilonPiParametersInAmbiguityDiagram} where the approximate fulfillment expressed by (\ref{eq:ConsistencyRelationApproxFulfillmentRootsLmax1}), for accidental ambiguities $\bm{\uppi}_{\hspace*{0.015cm}n} \in \hat{\mathcal{P}}$, is given. \newline
The first one occurs at $E_{\gamma} \simeq 520 \hspace*{2pt} \mathrm{MeV}$, where (\ref{eq:ConsistencyRelationApproxFulfillmentRootsLmax1}) is fulfilled for the map $\bm{\uppi}_{\hspace*{0.015cm}10}$ (i.e. $\varphi_{1} - \varphi_{2} \simeq \psi_{1} - \psi_{2}$). Here, the finite separation of the corresponding symbols, which is expressed by the violation parameter $\epsilon_{\bm{\uppi}_{\hspace*{0.015cm}10}}$, is still visible for the plot scale of the original ambiguity diagram. Regardless, the meaning of the parameter $\epsilon_{\bm{\uppi}_{\hspace*{0.015cm}10}}$ is further explained by a zoom in Figure \ref{fig:ExampleEpsilonPiParametersInAmbiguityDiagram}. \newline
The second example is situated at the energy $E_{\gamma} \simeq 610 \hspace*{2pt} \mathrm{MeV}$. The symbols here correspond to the validity of (\ref{eq:ConsistencyRelationApproxFulfillmentRootsLmax1}) for the accidental ambiguity $\bm{\uppi}_{\hspace*{0.015cm}9}$ (in this case: $- \varphi_{1} + \varphi_{2} \simeq \psi_{1} - \psi_{2}$). The parameter $\epsilon_{\bm{\uppi}_{\hspace*{0.015cm}9}}$ really needs to be emphazised by a zoom, since for the scale and size of the ambiguity diagram it is not visible. The violation parameter $\epsilon_{\bm{\uppi}_{\hspace*{0.015cm}9}}$ is still not vanishing, but it is one order of magnitude smaller than $\epsilon_{\bm{\uppi}_{\hspace*{0.015cm}10}}$ from the first example. Both examples mentioned here will be met again when the ambiguities occurring in numerical analyses of MAID theory data are discussed (cf. section \ref{sec:TheoryDataFits}). \newline
These examples conclude the illustration of the $\epsilon_{\bm{\uppi}_{\hspace*{0.015cm}n}}$-parameters, and thereby also this appendix section. We continue with the deduction of mathematical consequences for the multipole solutions corresponding to accidental ambiguities that fulfill the consistency relation approximately, i.e. for which (\ref{eq:ConsistencyRelationEpsilonFormulationRoots}) is valid. \newline

\textit{Appendix: Proof of the fact that for N complex numbers, $2^{N}$ combinations of complex conjugates and non-conjugates exist:} \newline \newline
In order to show the statement for any set of $N$ variables $\left(z_{1},\ldots,z_{N}\right)$, the complete induction \cite{ForsterI} first of all has to be anchored:
\newpage

\underline{$N = 1$:} \newline

Consider the complex variable $z_{1} \in \mathbbm{C}$. Evidently, there exist two possible conjugation maps, also limiting the number of the above mentioned subsets:
\begin{align}
z_{1} &\overset{1}{\longrightarrow} z_{1} =: \bm{\uppi}_{\hspace*{0.035cm}0} \left( z_{1} \right) \mathrm{,} \label{eq:NEquals1Map1} \\
z_{1} &\overset{2}{\longrightarrow} z_{1}^{\ast} =: \bm{\uppi}_{\hspace*{0.035cm}1} \left( z_{1} \right) \mathrm{.} \label{eq:NEquals1Map2}
\end{align}
Now, the inductive step can be performed. \newline

\underline{$N \rightarrow N+1$:} \newline

Assume that for $N$ complex variables $\left(z_{1},\ldots,z_{N}\right)$, the $2^{N}$ possibilities are encompassed by the maps
\begin{equation}
 \left(z_{1},\ldots,z_{N}\right) \longrightarrow \left(\bm{\uppi}_{\hspace*{0.035cm}n}\left(z_{1}\right),\ldots,\bm{\uppi}_{\hspace*{0.035cm}n}\left(z_{N}\right)\right) \mathrm{,} \hspace*{5pt} n = 0,\ldots,2^{N} - 1 \mathrm{.} \label{eq:NCompVarAllMaps}
\end{equation}
Then, for each $\bm{\uppi}_{\hspace*{0.035cm}n}$, there exist again $2$ possibilities in case one extends the above mentioned set by one complex variable:
\begin{align}
 \left(z_{1},\ldots,z_{N}, z_{N+1}\right) &\overset{1}{\longrightarrow} \left(\bm{\uppi}_{\hspace*{0.035cm}n}\left(z_{1}\right),\ldots,\bm{\uppi}_{\hspace*{0.035cm}n}\left(z_{N}\right), z_{N+1}\right) \mathrm{,} \label{eq:NPlus1CompVarMap1} \\
 \left(z_{1},\ldots,z_{N},z_{N+1}\right) &\overset{2}{\longrightarrow} \left(\bm{\uppi}_{\hspace*{0.035cm}n}\left(z_{1}\right),\ldots,\bm{\uppi}_{\hspace*{0.035cm}n}\left(z_{N}\right), z_{N+1}^{\ast}\right) \mathrm{.} \label{eq:NPlus1CompVarMap2}
\end{align}
In conclusion, if the set consisting of $N$ complex variables is enlarged by the further one $z_{N+1}$, there exist $2 \times 2^{N} = 2^{N + 1}$ possibilities.
\begin{flushright}
\textbf{QED.}
\end{flushright}

\paragraph{Invariance statements for accidental ambiguities} \label{subsec:AccidentalAmbProofsII}  \textcolor{white}{:-)} \newline

\begin{figure}[h]
\centering
 \begin{overpic}[width=0.6\textwidth]{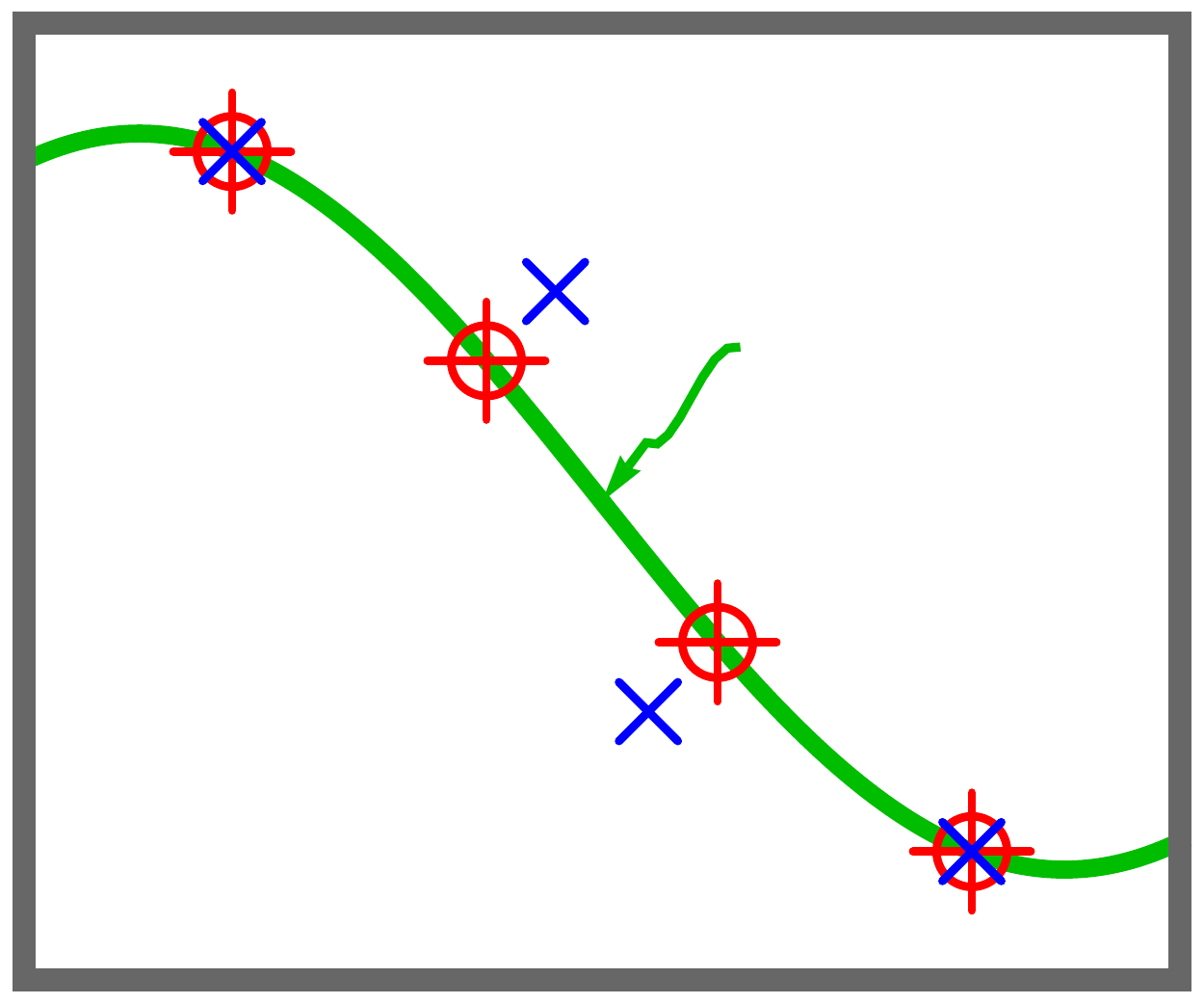}
\put(62.95,53.3){\textcolor{ForestGreen}{$\prod_{k} \alpha_{k} = \prod_{k^{\prime}} \beta_{k^{\prime}}$}}
\put(12.25,55.30){\begin{small}\textcolor{blue}{$\left(\alpha_{k}, \beta_{k}\right)$}\end{small}}
\put(73.47,25.30){\begin{small}\textcolor{blue}{$\left(\alpha_{k}^{\ast}, \beta_{k}^{\ast}\right)$}\end{small}}
\put(32.75,70.30){\begin{small}\textcolor{blue}{$\left(\bm{\uppi}\left[\alpha_{k}\right], \bm{\uppi} \left[\beta_{k}\right]\right)$}\end{small}}
\put(38.38,8.75){\begin{small}\textcolor{blue}{$\left(\bm{\uppi}\left[\alpha_{k}\right]^{\ast}, \bm{\uppi} \left[\beta_{k}\right]^{\ast}\right)$}\end{small}}
\put(18.00,60.80){\begin{large}\textcolor{blue}{$\uparrow$}\end{large}}
\put(52.45,15.25){\begin{large}\textcolor{blue}{$\uparrow$}\end{large}}
\put(44.85,63.95){\begin{large}\textcolor{blue}{$\downarrow$}\end{large}}
\put(79.45,19.8){\begin{large}\textcolor{blue}{$\downarrow$}\end{large}}
\put(-7.85,39.5){$\beta_{2L}$}
\put(24.05,-3.75){$\left(\alpha_{1},\ldots,\alpha_{2L},\beta_{1},\ldots,\beta_{2L-1}\right)$}
 \end{overpic}
\vspace*{5pt}
\caption[The schematic shows a $2$-dimensional representation of the $4L$-dimensional complex root space.]{This schematic shows a $2$-dimensional representation of the $4L$-dimensional complex root space. Possible configurations of roots correspond to blue crosses. Multipole solutions (or more precisely roots equivalent to them) are shown as red crosshairs. The latter have to lie on the green solid line, which represents the sub-manifold of the root-space where the consistency relation is fulfilled.}
\label{fig:RootSpaceCartoon}
\end{figure}
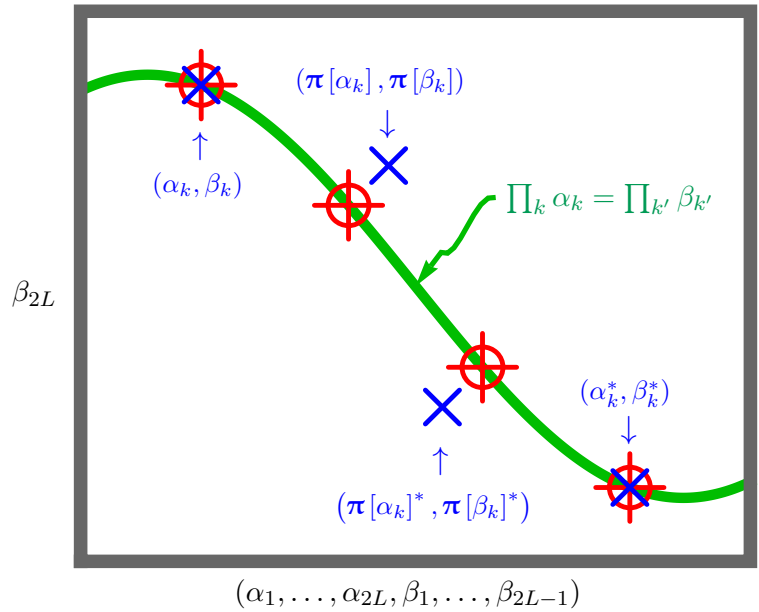

We begin by considering a generic accidental ambiguity $\bm{\uppi} \in \hat{\mathcal{P}}$ (the subscript $n$ shall be dropped in the following, since is is not important for the further discussion what index the ambiguity is carrying), which is assumed to fulfill the consistency relation approximately, i.e.
\begin{equation}
\bm{\uppi} \left(\alpha_{1}\right) \ast \ldots \ast \bm{\uppi} \left(\alpha_{2L}\right) = \bm{\uppi} \left(\beta_{1}\right) \ast \ldots \ast \bm{\uppi} \left(\beta_{2L}\right) \ast e^{i \epsilon_{\bm{\uppi}}} \mathrm{.} \label{eq:GeneralizedViolatedCR}
\end{equation}
Here, the violation parameter $\epsilon_{\bm{\uppi}}$ is assumed to be positive and infinitesimally small, i.e. $\epsilon_{\bm{\uppi}} > 0$ and $\epsilon_{\bm{\uppi}} \ll 1$. Therefore, one has $\left| e^{i \epsilon_{\bm{\uppi}}} - 1 \right| \simeq \left| 1 + i \epsilon_{\bm{\uppi}} - 1 \right| = \epsilon_{\bm{\uppi}} \ll 1$ and the exponential very close to unity. \newline \newline
If (\ref{eq:GeneralizedViolatedCR}) is true with an infinitesimally small $\epsilon_{\bm{\uppi}}$ for some $\bm{\uppi} \in \hat{\mathcal{P}}$, then it is reasonable to assume that a multipole solution $\left\{ \mathcal{M}_{\ell} \right\} = \left\{ E_{\ell\pm}, M_{\ell\pm} \right\}$ of the TPWA standard form (\ref{eq:LowEAssocLegStandardParametrization1}) and (\ref{eq:LowEAssocLegStandardParametrization2}) exists, which is
fully equivalent to some roots $\bm{\tilde{\uppi}} \left( \alpha_{k}, \beta_{k} \right)$, such that the  $\bm{\tilde{\uppi}} \left( \alpha_{k}, \beta_{k} \right)$ and  $\bm{\uppi} \left( \alpha_{k}, \beta_{k} \right)$ are close to each other in root space:
\begin{align}
 d \left[ \bm{\tilde{\uppi}}, \bm{\uppi} \right] &:= \sqrt{ \sum_{k} \left| \bm{\tilde{\uppi}} \left(\alpha_{k} \right) - \bm{\uppi} \left(\alpha_{k} \right) \right|^{2} + \sum_{k^{\prime}} \left| \bm{\tilde{\uppi}} \left(\beta_{k^{\prime}} \right) - \bm{\uppi} \left(\beta_{k^{\prime}} \right) \right|^{2}}  \ll 1 \mathrm{.} \label{eq:RestoredSolSmallDistanceToAccAmb}
\end{align}
The map $\bm{\tilde{\uppi}}$ is then a modified version of $\bm{\uppi} \in \hat{\mathcal{P}}$. The set of multipole parameters $\left\{ E_{\ell\pm}, M_{\ell\pm} \right\}$, or equivalently the roots $\bm{\tilde{\uppi}} \left( \alpha_{k}, \beta_{k} \right)$, a called a multipole solution that is \textit{related to} $\bm{\uppi} \left( \alpha_{k}, \beta_{k} \right)$. \newline
We have to state that up to know the existence of such solutions did not seem provable. Therefore it has to be postulated. Solutions of the above assumed kind are however always found in numerical studies. \newline
Before proceeding with the derivation of the properties of the modified map $\bm{\tilde{\uppi}}$, the fact that the corresponding roots have to exactly fulfill the consistency relation (\ref{eq:ConsistencyRelationAccAmb}) shall be motivated in more detail. In order to see this fact, it is useful to consider two different ways along which the TPWA standard form can be derived. \newline
Firstly, as already mentioned in section \ref{sec:CompExpsTPWA}, it is always possible to assume a truncation of the multipole expansion ((\ref{eq:MultExpF1}) to (\ref{eq:MultExpF4})) at some $L = \ell_{\mathrm{max}}$ and insert this truncation into the bilinear CGLN form $ \check{\Omega}^{\alpha} = \frac{1}{2} \left< F \right| \hat{A}^{\alpha} \left| F \right>$ of the observables.
This way one directly arrives at the standard form given in (\ref{eq:LowEAssocLegStandardParametrization1}) and (\ref{eq:LowEAssocLegStandardParametrization2}), with the consistency relation never showing up explicitly. \newline
However, one could also choose the detour of first exchanging the $F_{i}$ for transversity amplitudes $b_{i}$, truncating at $L$ and then also switching from the angular variable $\cos \theta$ to $\tan \frac{\theta}{2}$. It was shown in section \ref{sec:WBTpaper} that then the consistency relation is always seen to be fulfilled explicitly. If now the TPWA decomposition of the $b_{i}$
is substituted back into the observables in bilinear transversity form $\check{\Omega}^{\alpha} = \frac{1}{2} \left< b \right| \tilde{\Gamma}^{\alpha} \left| b \right>$ and furthermore, one would re-introduce $\cos \theta$ as the angular variable, the \textit{same} standard form would emerge as derived by the more direct way described above. \newline
Therefore it is seen that once a TPWA is fitted, it already has the consistency relation built into it implicitly by construction. By the same argument, configurations in root space that are fully equivalent to multipole solutions have to exactly satisfy the consistency relation. This situation is illustrated in Figure \ref{fig:RootSpaceCartoon}. The picture also indicates the above introduced
multipole solution as the point with the lowest possible distance to $\bm{\uppi} \left( \alpha_{k}, \beta_{k} \right)$ in root space, that still does not voilate the consistency relation. \newline

\begin{figure}[h]
\centering
\hspace*{5pt}
 \begin{overpic}[width=0.95\textwidth]{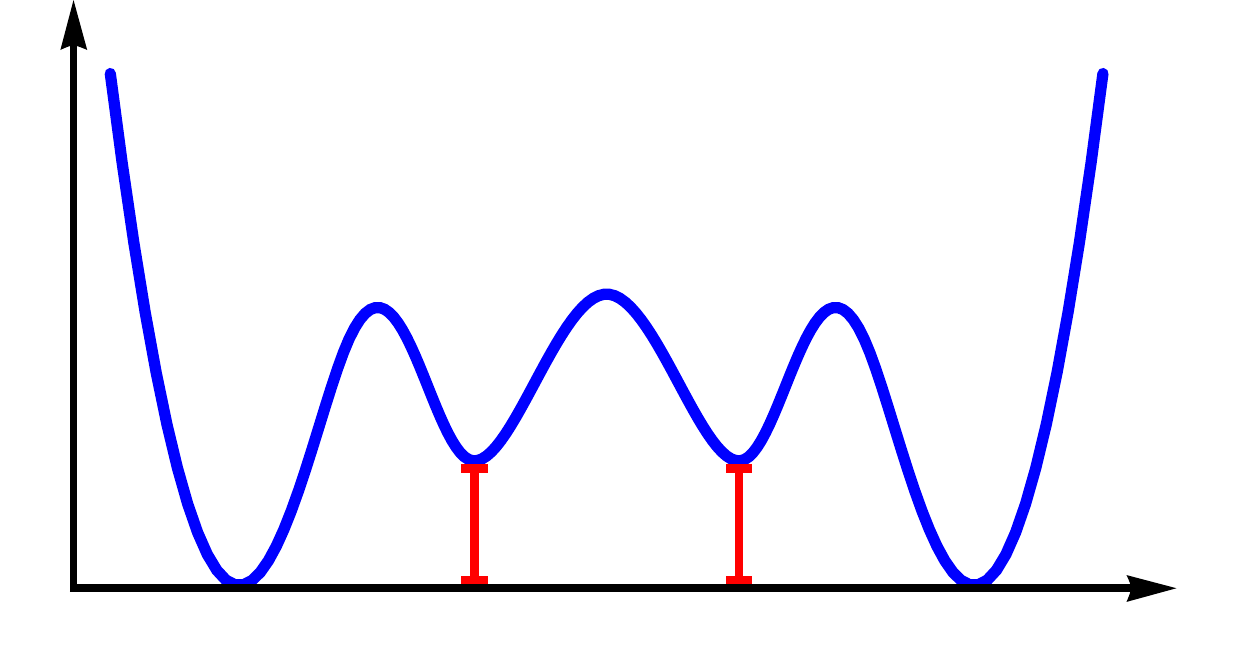}
\put(0.5,50.3){\begin{Large}$\Phi$\end{Large}}
\put(90,1.0){\begin{Large}$\mathcal{M}_{\ell}$\end{Large}}
\put(41.5,10.05){\begin{Large}\textcolor{red}{$\leftarrow$ $\delta \Phi$ $\rightarrow$}\end{Large}}
\put(18.33,1.1){\begin{LARGE}\textcolor{ForestGreen}{$\uparrow$}\end{LARGE}}
\put(14.5,-2.8){\textcolor{ForestGreen}{$\left(\alpha_{k}, \beta_{k}\right)$}}
\put(36.85,1.1){\begin{LARGE}\textcolor{ForestGreen}{$\uparrow$}\end{LARGE}}
\put(28.75,-2.8){\textcolor{ForestGreen}{$\left(\bm{\tilde{\uppi}}\left[\alpha_{k}\right], \bm{\tilde{\uppi}} \left[\beta_{k}\right]\right)$}}
\put(58.05,1.1){\begin{LARGE}\textcolor{ForestGreen}{$\uparrow$}\end{LARGE}}
\put(48.6,-2.8){\textcolor{ForestGreen}{$\left(\bm{\tilde{\uppi}}\left[\alpha_{k}\right]^{\ast}, \bm{\tilde{\uppi}} \left[\beta_{k}\right]^{\ast}\right)$}}
\put(77.04,1.1){\begin{LARGE}\textcolor{ForestGreen}{$\uparrow$}\end{LARGE}}
\put(73.25,-2.8){\textcolor{ForestGreen}{$\left(\alpha_{k}^{\ast}, \beta_{k}^{\ast}\right)$}}
 \end{overpic}
\vspace*{15pt}
\caption[Schematic depicting the behaviour of the discrepancy function (\ref{eq:DefDiscrFunct}), provided that an accidental ambiguity $\bm{\uppi} \in \hat{\mathcal{P}}$ is present which fulfills the consistency relation approximately.]{This picture schematically depicts the behaviour of the discrepancy function (\ref{eq:DefDiscrFunct}) defined in the main text, provided that an accidental ambiguity $\bm{\uppi} \in \hat{\mathcal{P}}$ is present which fulfills the consistency relation approximately. The true solution $\left(\alpha_{k},\beta_{k}\right)$ and its double ambiguity $\left(\alpha_{k}^{\ast},\beta_{k}^{\ast}\right)$ are indicated as exact zeros of $\Phi$.
Furthermore, for the accidential ambiguity $\left(\bm{\tilde{\uppi}}\left[\alpha_{k}\right], \bm{\tilde{\uppi}} \left[\beta_{k}\right]\right)$ and the solution $\left(\bm{\tilde{\uppi}}\left[\alpha_{k}\right]^{\ast}, \bm{\tilde{\uppi}} \left[\beta_{k}\right]^{\ast}\right)$ that is reached from it by applying the double ambiguity transformation, there exists a finite correction $\delta \Phi$. The scale of the correction is greatly exaggerated.}
\label{fig:ChiSquareValleyCartoon}
\end{figure}
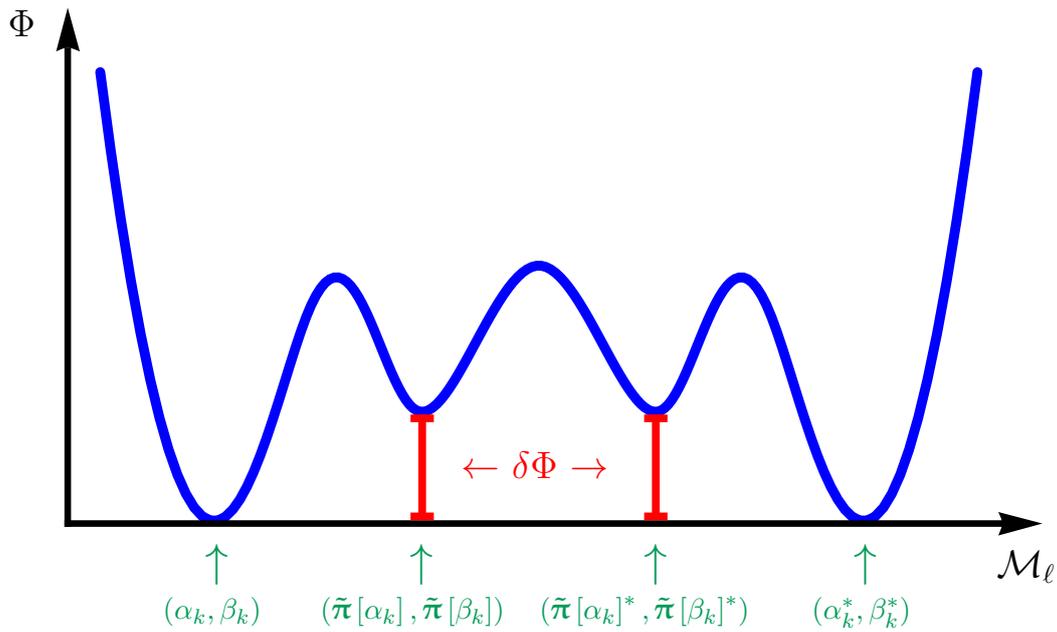

By assumption, the roots obtained by the modified map $\bm{\tilde{\uppi}}$ have to satisfy the consistency constraint exactly, i.e.
\begin{equation}
\prod_{k} \bm{\tilde{\uppi}} \left(\alpha_{k}\right) = \prod_{k^{\prime}} \bm{\tilde{\uppi}} \left(\beta_{k^{\prime}}\right) \mathrm{.} \label{eq:ModifiedRootsFulfillCR}
\end{equation}
Moreover, the requirement that the roots $\bm{\uppi} \left(\alpha_{k},\beta_{k}\right)$ and $\bm{\tilde{\uppi}} \left(\alpha_{k},\beta_{k}\right)$ are close to each other in root space can be formulated equivalently as
\begin{equation}
 \bm{\tilde{\uppi}} \left(\alpha_{k}\right) \simeq \bm{\uppi} \left(\alpha_{k}\right) \mathrm{,} \hspace*{5pt}  \bm{\tilde{\uppi}} \left(\beta_{k^{\prime}}\right) \simeq \bm{\uppi} \left(\beta_{k^{\prime}}\right) \mathrm{,} \hspace*{5pt} k, k^{\prime} = 1,\ldots,2L \mathrm{.} \label{eq:ClosenessOfRootsExpressed1}
\end{equation}
In the following, in order to reformulate the closeness condition (\ref{eq:ClosenessOfRootsExpressed1}) in a more usable way, the assumption has to be made that the moduli of all the transformed roots $\bm{\uppi} \left(\alpha_{k},\beta_{k}\right)$ and $\bm{\tilde{\uppi}} \left(\alpha_{k},\beta_{k}\right)$ by far exceed the distance between the two transformations $\bm{\uppi}$ and $\bm{\tilde{\uppi}}$ in root space. For the $\alpha$-roots, this condition reads
\begin{equation}
 \left| \bm{\uppi} \left( \alpha_{k} \right) \right| \gg \left| \bm{\tilde{\uppi}} \left( \alpha_{k} \right) - \bm{\uppi} \left( \alpha_{k} \right) \right| \mathrm{,} \hspace*{4pt} \left| \bm{\tilde{\uppi}} \left( \alpha_{k} \right) \right| \gg \left| \bm{\tilde{\uppi}} \left( \alpha_{k} \right) - \bm{\uppi} \left( \alpha_{k} \right) \right| \mathrm{,} \hspace*{3pt} k = 1,\ldots,2L \mathrm{,} \label{eq:AssumptionOnModuliOfTransformedRoots1}
\end{equation}
and for the $\beta$-roots
\begin{equation}
 \left| \bm{\uppi} \left( \beta_{k^{\prime}} \right) \right| \gg \left| \bm{\tilde{\uppi}} \left( \beta_{k^{\prime}} \right) - \bm{\uppi} \left( \beta_{k^{\prime}} \right) \right| \mathrm{,} \hspace*{4pt} \left| \bm{\tilde{\uppi}} \left( \beta_{k^{\prime}} \right) \right| \gg \left| \bm{\tilde{\uppi}} \left( \beta_{k^{\prime}} \right) - \bm{\uppi} \left( \beta_{k^{\prime}} \right) \right| \mathrm{,} \hspace*{3pt} k^{\prime} = 1,\ldots,2L \mathrm{.} \label{eq:AssumptionOnModuliOfTransformedRoots2}
\end{equation}
By this, the assumption is also made implicitly that none of the roots transformed under $\bm{\uppi}$ or $\bm{\tilde{\uppi}}$ vanish. The conditions (\ref{eq:AssumptionOnModuliOfTransformedRoots1}) and (\ref{eq:AssumptionOnModuliOfTransformedRoots2}) are typically fulfilled in analyses of mathematically exactly solvable model data. \newline 
In case the above made assumptions are true, it is permissible to reformulate (\ref{eq:ClosenessOfRootsExpressed1}) as
\begin{equation}
 \frac{\bm{\tilde{\uppi}} \left(\alpha_{k}\right)}{\bm{\uppi} \left(\alpha_{k}\right)} \simeq 1 \mathrm{,} \hspace*{5pt}  \frac{\bm{\tilde{\uppi}} \left(\beta_{k^{\prime}}\right)}{\bm{\uppi} \left(\beta_{k^{\prime}}\right)} \simeq 1 \mathrm{,} \hspace*{5pt} k, k^{\prime} = 1,\ldots,2L \mathrm{.} \label{eq:ClosenessOfRootsExpressed2}
\end{equation}
Since all transformed roots in the above given fractions are generally complex variables, the violation of unity in the conditions (\ref{eq:ClosenessOfRootsExpressed2}) can occur into any direction in the complex plane. Therefore, the validity of (\ref{eq:ClosenessOfRootsExpressed2}) is generally equivalent to the existence of sets of complex quantities $\left\{ \xi_{k} \in \mathbbm{C} \hspace*{2pt} | \hspace*{2pt} k = 1,\ldots,2L \right\}$ and $\left\{ \zeta_{k} \in \mathbbm{C} \hspace*{2pt} | \hspace*{2pt} k = 1,\ldots,2L \right\}$ which are infinitesimally small, i.e.
\begin{equation}
 \left|\xi_{k}\right| \ll 1 \mathrm{,} \hspace*{2pt} \left|\zeta_{k^{\prime}}\right| \ll 1 \mathrm{,} \hspace*{2pt} \forall k \mathrm{,} \hspace*{1pt} k^{\prime} = 1 \mathrm{,} \ldots \mathrm{,} 2L \mathrm{,} \label{eq:XiKZetaKAreInfinitesimal}
\end{equation}
and that parametrize the violation of unity in equation (\ref{eq:ClosenessOfRootsExpressed2}) in the following way
\begin{equation}
 \frac{\bm{\tilde{\uppi}} \left(\alpha_{k}\right)}{\bm{\uppi} \left(\alpha_{k}\right)} = e^{\xi_{k}} \mathrm{,} \hspace*{5pt}  \frac{\bm{\tilde{\uppi}} \left(\beta_{k^{\prime}}\right)}{\bm{\uppi} \left(\beta_{k^{\prime}}\right)} = e^{\zeta_{k^{\prime}}} \mathrm{,} \hspace*{5pt} k, k^{\prime} = 1,\ldots,2L \mathrm{.} \label{eq:ClosenessOfRootsExpressed3}
\end{equation}
It has to be said here that it seems not to be possible to calculate the $\xi_{k}$ and $\zeta_{k}$ analytically without knowing both the $\bm{\uppi} \left( \alpha_{k}, \beta_{k} \right)$ \textit{and} $\bm{\tilde{\uppi}} \left( \alpha_{k}, \beta_{k} \right)$. However, in numerical analyses they can be found (see section \ref{sec:TheoryDataFits}). \newline
With the relations given above, a transformation law which turns the roots $\left\{\alpha_{k},\beta_{k}\right\}$ of the true solution into roots for the above mentioned multipole solution $\bm{\tilde{\uppi}} \left(\alpha_{k},\beta_{k}\right)$ has already been found, namely
\begin{align}
 &\alpha_{k} \rightarrow \bm{\tilde{\uppi}} \left(\alpha_{k}\right) = \bm{\uppi} \left(\alpha_{k}\right) \ast e^{\xi_{k}} \hspace*{10pt} \mathrm{\&} \hspace*{10pt} \beta_{k^{\prime}} \rightarrow \bm{\tilde{\uppi}} \left(\beta_{k^{\prime}}\right) = \bm{\uppi} \left(\beta_{k^{\prime}}\right) \ast e^{\zeta_{k^{\prime}}} \mathrm{,} \hspace*{2pt} \nonumber \\
 &\forall k \mathrm{,} \hspace*{1pt} k^{\prime} = 1 \mathrm{,} \ldots \mathrm{,} 2L \mathrm{.}  \label{eq:ModifiedAmbiguityTrafo}
\end{align}
Furthermore, the moduli of the variables $\xi_{k}$ and $\zeta_{k}$ measure the closeness of the two sets of transformed roots $\bm{\uppi} \left(\alpha_{k},\beta_{k}\right)$ and $\bm{\tilde{\uppi}} \left(\alpha_{k},\beta_{k}\right)$ in root space. To see this, one can approximate the euclidean distance between both sets of roots as
\begin{align}
 d \left[ \bm{\tilde{\uppi}}, \bm{\uppi} \right] &= \sqrt{ \sum_{k} \left| \bm{\tilde{\uppi}} \left(\alpha_{k} \right) - \bm{\uppi} \left(\alpha_{k} \right) \right|^{2} + \sum_{k^{\prime}} \left| \bm{\tilde{\uppi}} \left(\beta_{k^{\prime}} \right) - \bm{\uppi} \left(\beta_{k^{\prime}} \right) \right|^{2}} \nonumber \\
 &\simeq \sqrt{ \sum_{k} \left| \bm{\uppi} \left(\alpha_{k} \right) \ast \left( 1 + \xi_{k} \right) - \bm{\uppi} \left(\alpha_{k} \right) \right|^{2} + \sum_{k^{\prime}} \left| \bm{\uppi} \left(\beta_{k^{\prime}} \right) \ast \left( 1 + \zeta_{k^{\prime}} \right) - \bm{\uppi} \left(\beta_{k^{\prime}} \right) \right|^{2}} \nonumber \\
 &= \sqrt{ \sum_{k} \left| \xi_{k} \right|^{2} \left| \bm{\uppi} \left(\alpha_{k} \right) \right|^{2} + \sum_{k^{\prime}} \left| \zeta_{k^{\prime}} \right|^{2} \left| \bm{\uppi} \left( \beta_{k^{\prime}} \right) \right|^{2}} \mathrm{.} \label{eq:DistanceRestoredSolToAmb}
\end{align}
Therefore, the smaller the moduli of the $\xi_{k}$ and $\zeta_{k}$, the closer both configurations of roots are situated in root space. \newline
The assumption that the modified roots $\bm{\tilde{\uppi}} \left(\alpha_{k},\beta_{k}\right)$ respect the consistency relation (\ref{eq:ModifiedRootsFulfillCR}) can be turned into an additive constraint for the infinitesimally small complex variables $\xi_{k}$ and $\zeta_{k}$. To do this, one has to start with
\begin{equation}
 \prod_{k=1}^{2L} \left( \bm{\uppi} \left(\alpha_{k}\right) \ast e^{\xi_{k}} \right) = \prod_{k^{\prime}=1}^{2L} \left( \bm{\uppi} \left(\beta_{k^{\prime}}\right) \ast e^{+ \zeta_{k^{\prime}}} \right) \mathrm{,} \label{eq:RestoringTrafoRespectsCR}
\end{equation}
and reformulate this equation according to
\begin{align}
 1 &= \prod_{k=1}^{2L} \left( \bm{\uppi} \left(\alpha_{k}\right) \ast e^{\xi_{k}} \right) \ast \prod_{k^{\prime}=1}^{2L} \left( \bm{\uppi} \left(\beta_{k^{\prime}}\right) \ast e^{+ \zeta_{k^{\prime}}} \right)^{-1} \nonumber \\
 &= \left(\prod_{i = 1}^{2L} \bm{\uppi} \left(\alpha_{i}\right) \ast \prod_{j=1}^{2L} \bm{\uppi} \left(\beta_{j}\right)^{-1}\right) \ast e^{\sum_{k} \xi_{k}} \ast e^{-\sum_{k^{\prime}} \zeta_{k^{\prime}}} \nonumber \\
&= \left(\prod_{i = 1}^{2L} \bm{\uppi} \left(\alpha_{i}\right) \ast \prod_{j=1}^{2L} \bm{\uppi} \left(\beta_{j}\right)^{-1}\right) \ast e^{\sum_{k} \left( \xi_{k} - \zeta_{k}\right)} \mathrm{.} \label{eq:RestoringTrafoDerivation}
\end{align}
The assumed approximate fulfillment of the consistency relation, equation (\ref{eq:GeneralizedViolatedCR}), can also be reformulated as
\begin{equation}
\left(\prod_{i = 1}^{2L} \bm{\uppi} \left(\alpha_{i}\right) \ast \prod_{j=1}^{2L} \bm{\uppi} \left(\beta_{j}\right)^{-1}\right) \ast e^{- i \epsilon_{\bm{\uppi}}} = 1 \mathrm{.} \label{eq:GeneralizedViolatedCRIII}
\end{equation}
The results (\ref{eq:RestoringTrafoDerivation}) and (\ref{eq:GeneralizedViolatedCRIII}) can be equal only if the quantities $\xi_{k}$ and $\zeta_{k^{\prime}}$ fulfill the following requirement
\begin{equation}
 \sum_{k = 1}^{2L} \left( - \xi_{k} \right) + \sum_{k^{\prime}=1}^{2L} \zeta_{k^{\prime}} = i \epsilon_{\bm{\uppi}} \mathrm{.} \label{eq:RestoringParametersConstraint}
\end{equation}
The modified root transformations (\ref{eq:ModifiedAmbiguityTrafo}), combined with the requirement that the complex parameters $\xi_{k}$ and $\zeta_{k^{\prime}}$ are small and that they fulfill the additive constraint (\ref{eq:RestoringParametersConstraint}), define the most general family of transformations that manage to restore the validity of the consistency relation (i.e. equation (\ref{eq:ModifiedRootsFulfillCR})) out of the initially assumed slightly violated case (\ref{eq:GeneralizedViolatedCR}). If the TPWA problem is solved numerically for an exactly solvable case and the multipole solution closest to $\bm{\uppi} \left(\alpha_{k},\beta_{k}\right)$ is found, then all the applied minimization algorithm does is to select one particular element out of this family. \newline \newline
Now we can proceed with the actual purpose of this appendix, namely to investigate the behaviour of the group $\mathcal{S}$ observables $\left\{\sigma_{0}, \Sigma, T, P\right\}$ under the modified accidental ambiguity transformation (\ref{eq:ModifiedAmbiguityTrafo}). We therefore assume that roots $\bm{\tilde{\uppi}} \left(\alpha_{k},\beta_{k}\right)$ of a multipole solution corresponding to the accidental ambiguity $\bm{\uppi} \in \hat{\mathcal{P}}$ have been found. Since the complex variables $\xi_{k}$ and $\zeta_{k^{\prime}}$ are assumed to have an infinitesimally small modulus (equation (\ref{eq:XiKZetaKAreInfinitesimal})), it is allowed to linearize the transformation law (\ref{eq:ModifiedAmbiguityTrafo})
\begin{align}
 \alpha_{k} \rightarrow \bm{\uppi} \left(\alpha_{k}\right) \ast (1 + \xi_{k}) \hspace*{10pt} \mathrm{\&} \hspace*{10pt} \beta_{k^{\prime}} \rightarrow \bm{\uppi} \left(\beta_{k^{\prime}}\right) \ast (1 + \zeta_{k^{\prime}}) \mathrm{,} \hspace*{5pt} k,k^{\prime}=1\mathrm{,} \ldots \mathrm{,} 2L \mathrm{.}  \label{eq:ModifiedAmbiguityTrafoII}
\end{align}
In the same way, terms that are of higher than linear order in the $\left\{\xi_{k},\zeta_{k}\right\}$ are neglected in all the following calculations. It is assumed that due to the smallness of these parameters, one always obtains a good approximation in this way. \newline
Since the group $\mathcal{S}$ observables are just linear combinations of squared moduli of the transversity amplitudes $b_{i}$, it is for a start sufficient to consider the behaviour of one transversity amplitude squared, e.g. $\left| b_{4} \right|^{2}$, under the transformation (\ref{eq:ModifiedAmbiguityTrafoII}):
\begin{align}
 \left| b_{4} \right|^{2} &= \frac{\left| \mathcal{C} \right|^{2} \left| a_{2L} \right|^{2}}{\left( 1 + t^{2} \right)^{2L}} \prod_{k=1}^{2L} \left( t - \alpha_{k}^{\ast} \right) \left( t - \alpha_{k} \right) \nonumber \\
 &\rightarrow \left| \tilde{b}_{4} \right|^{2} := \frac{\left| \mathcal{C} \right|^{2} \left| a_{2L} \right|^{2}}{\left( 1 + t^{2} \right)^{2L}} \prod_{k=1}^{2L} \left[ t - \left( \bm{\uppi} \left(\alpha_{k}\right) \left[1 + \xi_{k}\right] \right)^{\ast} \right]  \left[ t - \bm{\uppi} \left(\alpha_{k}\right) \left(1 + \xi_{k}\right) \right] \mathrm{.} \label{eq:B4ModSquareModifiedAmbTrafoI}
\end{align}
Neglecting the pre-factors for a moment, the tranformed modulus-squared $\left| \tilde{b}_{4} \right|^{2}$ can now be simplified further
\begin{align}
 \left| \tilde{b}_{4} \right|^{2} &\propto \prod_{k=1}^{2L} \left[ t - \left( \bm{\uppi} \left(\alpha_{k}\right) \left[1 + \xi_{k}\right] \right)^{\ast} \right]  \left[ t - \bm{\uppi} \left(\alpha_{k}\right) \left(1 + \xi_{k}\right) \right] \nonumber \\
&= \prod_{k=1}^{2L} \left[ t - \bm{\uppi} \left(\alpha_{k}\right)^{\ast} - \bm{\uppi} \left(\alpha_{k}\right)^{\ast} \xi_{k}^{\ast} \right]  \left[  t - \bm{\uppi} \left(\alpha_{k}\right) - \bm{\uppi} \left(\alpha_{k}\right) \xi_{k} \right] \nonumber \\
&= \prod_{k=1}^{2L} \Big( \left[ t - \bm{\uppi} \left(\alpha_{k}\right)^{\ast} \right]  \left[ t - \bm{\uppi} \left(\alpha_{k}\right) \right] - \bm{\uppi} \left(\alpha_{k}\right)^{\ast} \xi_{k}^{\ast} \left[ t - \bm{\uppi} \left(\alpha_{k}\right) \right] \nonumber \\
& \hspace*{28.5pt} -  \bm{\uppi} \left(\alpha_{k}\right) \xi_{k} \left[ t - \bm{\uppi} \left(\alpha_{k}\right)^{\ast} \right] + \mathcal{O}\left(\xi_{k}^{2}\right) \Big) \nonumber \\
&\simeq \prod_{k=1}^{2L} \Big( \left[ t - \bm{\uppi} \left(\alpha_{k}\right)^{\ast} \right]  \left[ t - \bm{\uppi} \left(\alpha_{k}\right) \right] - 2 \mathrm{Re} \left[ \bm{\uppi} \left(\alpha_{k}\right) \xi_{k} \left( t - \bm{\uppi} \left(\alpha_{k}\right)^{\ast} \right) \right] \Big) \mathrm{,} \label{eq:B4SquareCorrectionDerivationI}
\end{align}
where the definition of the real part was used and all terms of order $\mathcal{O}\left(\xi_{k}^{2}\right)$ have been dropped. \newline
In order to further evaluate each individual term of the product, two distinct cases have to be considered for all $k \in \left\{ 1\mathrm{,}\ldots \mathrm{,} 2L \right\}$. The first possibility is that the root $\alpha_{k}$ is not conjugated under $\bm{\uppi} \in \hat{\mathcal{P}}$,
\begin{align}
 &\left(1.\right) \bm{\uppi} \left(\alpha_{k}\right) = \alpha_{k} \mathrm{:} \nonumber \\
 &\hspace*{16.5pt}\Rightarrow \left[ t - \bm{\uppi} \left(\alpha_{k}\right)^{\ast} \right]  \left[ t - \bm{\uppi} \left(\alpha_{k}\right) \right] - 2 \mathrm{Re} \left[ \bm{\uppi} \left(\alpha_{k}\right) \xi_{k} \left( t - \bm{\uppi} \left(\alpha_{k}\right)^{\ast} \right) \right] \nonumber \\
 & \hspace*{31.5pt} = \left[ t - \alpha_{k}^{\ast} \right]  \left[ t - \alpha_{k} \right] - 2 \mathrm{Re} \left[ \alpha_{k} \xi_{k} \left( t - \alpha_{k}^{\ast} \right) \right] \mathrm{,} \label{eq:CorrectionTermCases1}
\end{align}
while in the second case, $\alpha_{k}$ gets conjugated:
\begin{align}
 &\left(2.\right) \bm{\uppi} \left(\alpha_{k}\right) = \alpha_{k}^{\ast} \mathrm{:} \nonumber \\
 &\hspace*{16.5pt}\Rightarrow \left[ t - \bm{\uppi} \left(\alpha_{k}\right)^{\ast} \right]  \left[ t - \bm{\uppi} \left(\alpha_{k}\right) \right] - 2 \mathrm{Re} \left[ \bm{\uppi} \left(\alpha_{k}\right) \xi_{k} \left( t - \bm{\uppi} \left(\alpha_{k}\right)^{\ast} \right) \right] \nonumber \\
 & \hspace*{31.5pt} = \left[ t - \alpha_{k} \right] \left[ t - \alpha_{k}^{\ast} \right] - 2 \mathrm{Re} \left[ \alpha_{k}^{\ast} \xi_{k} \left( t - \alpha_{k} \right) \right] \nonumber \\
& \hspace*{31.5pt} = \left[ t - \alpha_{k}^{\ast} \right]  \left[ t - \alpha_{k} \right] - 2 \mathrm{Re} \left[ \alpha_{k} \xi_{k}^{\ast} \left( t - \alpha_{k}^{\ast} \right) \right] \mathrm{.} \label{eq:CorrectionTermCases2}
\end{align}
The behaviour already mentioned in the introduction of this appendix (app. \ref{sec:AccidentalAmbProofs}) can be observed here, namely that each product $\left[ t - \alpha_{k}^{\ast} \right]  \left[ t - \alpha_{k} \right]$ indeed does not change under the accidental ambiguity $\bm{\uppi} \in \hat{\mathcal{P}}$. The second term defined by a real part in each case however exists due to the modification in the transformation law (\ref{eq:ModifiedAmbiguityTrafo}). Since this term has a very similar form in equations (\ref{eq:CorrectionTermCases1}) and (\ref{eq:CorrectionTermCases2}), the following definition is strongly motivated:
\begin{equation}
 \xi_{k}^{\bm{\uppi}} := \begin{cases}
                    \xi_{k} &\mathrm{,} \hspace*{2pt} \bm{\uppi} \left(\alpha_{k}\right) = \alpha_{k} \\
                    \xi_{k}^{\ast} &\mathrm{,} \hspace*{2pt} \bm{\uppi} \left(\alpha_{k}\right) = \alpha_{k}^{\ast}
                   \end{cases} \mathrm{.} \label{eq:DeltaPKDefinition}
\end{equation}
With this, again dropping terms of order $\mathcal{O} \left( \xi_{k}^{2} \right)$ in each factor of the product over $k$, the correction (\ref{eq:B4ModSquareModifiedAmbTrafoI}) can be written as
\begin{equation}
\left| \tilde{b}_{4} \right|^{2} \simeq \frac{\left| \mathcal{C} \right|^{2} \left| a_{2L} \right|^{2}}{\left( 1 + t^{2} \right)^{2L}} \prod_{k=1}^{2L} \left( \left[ t - \alpha_{k}^{\ast} \right] \left[ t - \alpha_{k} \right] - 2 \mathrm{Re} \left[ \alpha_{k} \xi_{k}^{\bm{\uppi}} \left( t - \alpha_{k}^{\ast} \right) \right] \right) \mathrm{.} \label{eq:B4ModSquareCorrectionIntermediateStep}
\end{equation}
Upon introducting polar coordinates for the violation parameters $\xi_{k}$, the expansion (\ref{eq:B4ModSquareCorrectionIntermediateStep}) can be reduced further. Using $\xi_{k} = \left| \xi_{k} \right| e^{i \rho_{k}}$, it is seen directly that equation (\ref{eq:DeltaPKDefinition}) can be expressed by $\xi_{k}^{\bm{\uppi}} = \left| \xi_{k} \right| e^{i \rho_{k}^{\bm{\uppi}}} $ with the phase $\rho_{k}^{\bm{\uppi}}$ defined as
\begin{equation}
 \rho_{k}^{\bm{\uppi}} := \begin{cases}
                    \rho_{k} &\mathrm{,} \hspace*{2pt} \bm{\uppi} \left(\alpha_{k}\right) = \alpha_{k} \\
                    \left(2 \pi - \rho_{k}\right) &\mathrm{,} \hspace*{2pt} \bm{\uppi} \left(\alpha_{k}\right) = \alpha_{k}^{\ast}
                   \end{cases} \mathrm{.} \label{eq:RhoPKDefinition}
\end{equation}
The modulus $\left| \xi_{k} \right|$ can now be pulled out of the real part and the latter can then be defined as a new function (this is just done in order to shorten the ensuing expressions)
\begin{equation}
 \kappa \left( \alpha_{n}, \rho_{n}^{\bm{\uppi}}, t \right) \equiv \kappa_{n}^{\bm{\uppi}} := \mathrm{Re} \left[ \alpha_{n} e^{i \rho_{n}^{\bm{\uppi}}} \left( t - \alpha_{n}^{\ast} \right) \right] \mathrm{.} \label{eq:KappaFunctionDefQuote2}
\end{equation}
Equation (\ref{eq:B4ModSquareCorrectionIntermediateStep}) can now be written even shorter, while keeping the important (infinitesimally small) moduli $\left| \xi_{k} \right|$ explicit:
\begin{align}
\left| \tilde{b}_{4} \right|^{2} &\simeq \frac{\left| \mathcal{C} \right|^{2} \left| a_{2L} \right|^{2}}{\left( 1 + t^{2} \right)^{2L}} \prod_{k = 1}^{2L}  \left[ \left( t - \alpha_{k}^{\ast} \right)  \left( t - \alpha_{k} \right) - 2 \left| \xi_{k} \right| \kappa^{\bm{\uppi}}_{k} \right] \mathrm{.} \label{eq:NonFullyExpandedB4ModSquare}
\end{align}
All that remains to be done now is to expand the product in equation (\ref{eq:NonFullyExpandedB4ModSquare}). Again, only terms linear in the $\left| \xi_{k} \right|$ shall be taken into account explicitly. The result is given in the following expression
\begin{align}
\left| \tilde{b}_{4} \right|^{2} &= \frac{\left| \mathcal{C} \right|^{2} \left| a_{2L} \right|^{2}}{\left( 1 + t^{2} \right)^{2L}}  \left[ \prod_{k = 1}^{2L} \left( t - \alpha_{k}^{\ast} \right)  \left( t - \alpha_{k} \right) - 2 \sum_{n=1}^{2L} \left( \left| \xi_{n} \right| \kappa_{n}^{\bm{\uppi}} \prod_{m \neq n}^{2L} \left[ t - \alpha_{m}^{\ast} \right]  \left[ t - \alpha_{m} \right] \right) \right] \nonumber \\
 & \hspace*{12pt} + \mathcal{O}\left(\left|\xi_{n}\right|^{2}\right) \mathrm{,} \label{eq:ExpandedB4ModSquare}
\end{align}
The validity of this expansion is non-trivial to see on a first sight. Therefore, equation (\ref{eq:ExpandedB4ModSquare}) is proven for any finite $L$ at the end of this appendix section. \newline
The important point to note is know that the squared modulus of the amplitude $b_{4}$ is not invariant anymore under the modified ambiguity transformation (\ref{eq:ModifiedAmbiguityTrafo}). Generally, one can write
\begin{equation}
 \left| \tilde{b}_{4} \right|^{2} = \left| b_{4} \right|^{2} + \delta \left| b_{4} \right|^{2} \mathrm{.}
\end{equation}
The correction $\delta \left| b_{4} \right|^{2}$ has been evaluated to linear order in the quantities $\xi_{k}$, with the result
\begin{equation}
 \delta  \left| b_{4} \right|^{2} \left(t\right) = (-2) \frac{\left| \mathcal{C} \right|^{2} \left| a_{2L} \right|^{2}}{\left( 1 + t^{2} \right)^{2L}} \sum_{n=1}^{2L} \left( \left| \xi_{n} \right| \kappa_{n}^{\bm{\uppi}} \prod_{m \neq n}^{2L} \left[ t - \alpha_{m}^{\ast} \right]  \left[ t - \alpha_{m} \right] \right) \mathrm{.} \label{eq:B4SquaredCorrFunct}
\end{equation}
Corrections in the orders $\mathcal{O} \left(\left|\xi_{k}\right|^{n}\right)$ for any $n > 1$ could of course also be derived systematically, though the algebraic effort would increase drastically. The most important point to note is however that already in the lowest order, the correction $\delta \left| b_{4} \right|^{2}$ is generally clearly non-vanishing. \newline
All the calculations done above for the case of the amplitude $b_{4}$ can of course also be repeated for the remaining transversity amplitudes. The result generally reads
\begin{equation}
 \left| b_{i} \right|^{2} \rightarrow \left| \tilde{b}_{i} \right|^{2} = \left| b_{i} \right|^{2} + \delta \left| b_{i} \right|^{2} \mathrm{,} \hspace*{5pt} i = 1,\ldots,4 \mathrm{.} \label{eq:BiSquaredAccAmbCorrectionLaw}
\end{equation}
The correction functions $\delta \left| b_{1} \right|^{2}$, $\delta \left| b_{2} \right|^{2}$ and $\delta \left| b_{3} \right|^{2}$ shall be deduced in the following, also up to linear order in the quantities $\xi_{k}$ and $\zeta_{k^{\prime}}$. In order to do this, one can start from the expression (\ref{eq:B4SquaredCorrFunct}) and use only a few replacements. \newline
First of all, since the third transversity amplitude is just given by $b_{3} \left(\theta\right) = b_{4} \left(- \theta\right)$ (cf. equation (\ref{eq:b1b3LinFactDecompositionsSecAccAmb})), it is sufficient to replace all roots $\alpha_{k}$ with $\left(- \alpha_{k}\right)$ in equation (\ref{eq:B4SquaredCorrFunct}). Also, the form of the $\kappa$-function (\ref{eq:KappaFunctionDefQuote2}) changes, such that a new one has to be introduced. The result for the correction $\delta \left| b_{3} \right|^{2}$ can be summarized as
\begin{align}
 \delta  \left| b_{3} \right|^{2} \left(t\right) &= 2 \frac{\left| \mathcal{C} \right|^{2} \left| a_{2L} \right|^{2}}{\left( 1 + t^{2} \right)^{2L}} \sum_{n=1}^{2L} \left( \left| \xi_{n} \right| \bar{\kappa}_{n}^{\bm{\uppi}} \prod_{m \neq n}^{2L} \left[ t + \alpha_{m}^{\ast} \right]  \left[ t + \alpha_{m} \right] \right) \mathrm{,} \label{eq:B3SquaredCorrFunct1} \\
 \bar{\kappa} \left( \alpha_{n}, \rho_{n}^{\bm{\uppi}}, t \right) &\equiv \bar{\kappa}_{n}^{\bm{\uppi}} := \mathrm{Re} \left[ \alpha_{n} e^{i \rho_{n}^{\bm{\uppi}}} \left( t + \alpha_{n}^{\ast} \right) \right] \mathrm{.} \label{eq:B3SquaredCorrFunct2}
\end{align}
The correction $\delta \left| b_{2} \right|^{2}$ has the same formal appearance as the expression (\ref{eq:B4SquaredCorrFunct}) for the case of $b_{4}$. The only two differences are that firstly, the roots $\beta_{k}$ have to take the place of the $\alpha_{k}$ ($b_{2}$ is decomposed in terms of $\beta$-roots, cf. (\ref{eq:b2LinFactDecompSecAccAmb})) and secondly the modification of the original accidental ambiguity transformation $\bm{\uppi}$ is here accomplished by the parameters $\zeta_{k}$ (cf. equation (\ref{eq:ModifiedAmbiguityTrafo})). Upon introducting polar coordinates $\zeta_{k} = \left| \zeta_{k} \right| e^{i \eta_{k}}$, again a distinction of cases has to be made exactly as in expression (\ref{eq:RhoPKDefinition}). The latter is done by defining
\begin{equation}
 \eta_{k}^{\bm{\uppi}} := \begin{cases}
                    \eta_{k} &\mathrm{,} \hspace*{2pt} \bm{\uppi} \left(\beta_{k}\right) = \beta_{k} \\
                    \left(2 \pi - \eta_{k}\right) &\mathrm{,} \hspace*{2pt} \bm{\uppi} \left(\beta_{k}\right) = \beta_{k}^{\ast}
                   \end{cases} \mathrm{.} \label{eq:EtaPKDefinition}
\end{equation}
Therefore, the correction function $\delta  \left| b_{2} \right|^{2}$ becomes
\begin{equation}
 \delta  \left| b_{2} \right|^{2} \left(t\right) = (-2) \frac{\left| \mathcal{C} \right|^{2} \left| a_{2L} \right|^{2}}{\left( 1 + t^{2} \right)^{2L}} \sum_{n=1}^{2L} \left( \left| \zeta_{n} \right| \kappa_{n}^{\bm{\uppi}} \left( \beta_{n}, \eta_{n}^{\bm{\uppi}}, t \right) \prod_{m \neq n}^{2L} \left[ t - \beta_{m}^{\ast} \right]  \left[ t - \beta_{m} \right] \right) \mathrm{,} \label{eq:B2SquaredCorrFunct}
\end{equation}
where it is important to make explicit that the kappa-function is now evaluated for new roots $\beta_{k}$ and phases $\eta_{k}^{\bm{\uppi}}$. \newline
Using the fact that $b_{1} \left(\theta\right) = b_{2} \left(- \theta\right)$ (again cf. (\ref{eq:b1b3LinFactDecompositionsSecAccAmb})), the final missing correction $\delta  \left| b_{1} \right|^{2}$ can be quickly found from equation (\ref{eq:B2SquaredCorrFunct}). The result is
\begin{equation}
 \delta  \left| b_{1} \right|^{2} \left(t\right) = 2 \frac{\left| \mathcal{C} \right|^{2} \left| a_{2L} \right|^{2}}{\left( 1 + t^{2} \right)^{2L}} \sum_{n=1}^{2L} \left( \left| \zeta_{n} \right| \bar{\kappa}_{n}^{\bm{\uppi}} \left( \beta_{n}, \eta_{n}^{\bm{\uppi}}, t \right) \prod_{m \neq n}^{2L} \left[ t + \beta_{m}^{\ast} \right]  \left[ t + \beta_{m} \right] \right) \mathrm{.} \label{eq:B1SquaredCorrFunct}
\end{equation}
Since the corrections to the squared moduli of all the transversity amplitudes have been found, everything is assembled to achieve the final goal, namely to derive an expression for the correction to the group $\mathcal{S}$ observables $\left\{\sigma_{0}, \Sigma, T, P\right\}$. Since the latter are just given as $\check{\Omega}^{\alpha_{S}} = \frac{1}{2} \left( \pm \left|b_{1}\right|^{2} \pm \left|b_{2}\right|^{2} \pm \left|b_{3}\right|^{2} + \left|b_{4}\right|^{2} \right)$, it is directly seen from (\ref{eq:BiSquaredAccAmbCorrectionLaw}) that generally the resulting correction can be written as
\begin{equation}
 \check{\Omega}^{\alpha_{S}} \rightarrow \tilde{\check{\Omega}}^{\alpha_{S}} = \check{\Omega}^{\alpha_{S}} + \delta \check{\Omega}^{\alpha_{S}} \mathrm{,} \label{eq:GroupSObsTrafoRuleAccAmb}
\end{equation}
where $\delta \check{\Omega}^{\alpha_{S}}$ is composed of the $\delta  \left| b_{i} \right|^{2}$ in the following way
\begin{equation}
 \delta \check{\Omega}^{\alpha_{S}} = \frac{1}{2} \left( \pm \delta  \left| b_{1} \right|^{2} \pm \delta  \left| b_{2} \right|^{2} \pm \delta  \left| b_{3} \right|^{2} + \delta  \left| b_{4} \right|^{2} \right) \mathrm{.} \label{eq:GroupSObsCorrForm}
\end{equation}
By inserting the results (\ref{eq:B4SquaredCorrFunct}), (\ref{eq:B3SquaredCorrFunct1}), (\ref{eq:B2SquaredCorrFunct}) and (\ref{eq:B1SquaredCorrFunct}), the final explicit expression for the correction $\delta \check{\Omega}^{\alpha_{S}}$ can be assembled, while only keeping the linear order $\mathcal{O} \left(\left|\xi_{k}\right|, \left|\zeta_{k}\right|\right)$ explicit:
\begin{align}
 \delta \check{\Omega}^{\alpha_{S}} &= \frac{\left| \mathcal{C} \right|^{2} \left| a_{2L} \right|^{2}}{\left( 1 + t^{2} \right)^{2L}} \sum_{n=1}^{2L} \Bigg( \pm \left| \zeta_{n} \right| \bar{\kappa}_{n}^{\bm{\uppi}} \left( \beta_{n}, \eta_{n}^{\bm{\uppi}}, t \right) \prod_{m \neq n}^{2L} \left[ t + \beta_{m}^{\ast} \right]  \left[ t + \beta_{m} \right] \nonumber \\
 & \hspace*{12.5pt} \mp \left| \zeta_{n} \right| \kappa_{n}^{\bm{\uppi}} \left( \beta_{n}, \eta_{n}^{\bm{\uppi}}, t \right) \prod_{m \neq n}^{2L} \left[ t - \beta_{m}^{\ast} \right]  \left[ t - \beta_{m} \right] \pm \left| \xi_{n} \right| \bar{\kappa}_{n}^{\bm{\uppi}} \left( \alpha_{n}, \rho_{n}^{\bm{\uppi}}, t \right) \nonumber \\
 & \hspace*{12.5pt} \times \prod_{m \neq n}^{2L} \left[ t + \alpha_{m}^{\ast} \right]  \left[ t + \alpha_{m} \right] - \left| \xi_{n} \right| \kappa_{n}^{\bm{\uppi}} \left( \alpha_{n}, \rho_{n}^{\bm{\uppi}}, t \right) \prod_{m \neq n}^{2L} \left[ t - \alpha_{m}^{\ast} \right]  \left[ t - \alpha_{m} \right] \Bigg) \nonumber \\
 & \hspace*{12.5pt} + \mathcal{O}\left(\left|\xi_{k}\right|^{2}, \left|\zeta_{k}\right|^{2}\right) \mathrm{.} \label{eq:GroupSObsCorrComplete}
\end{align}
Considering the above given expression, it is clear that $\delta \check{\Omega}^{\alpha_{S}}$ can only generally vanish in a limit where $\left| \xi_{k} \right|$ and $\left| \zeta_{k^{\prime}} \right|$ approach zero. Then, it can directly be seen by means of the transformation law (\ref{eq:ModifiedAmbiguityTrafo}) that the modified ambiguity map $\bm{\tilde{\uppi}} \rightarrow \bm{\uppi} \in \hat{\mathcal{P}}$. More importantly, the constraint (\ref{eq:RestoringParametersConstraint}) directly demands that $\epsilon_{\bm{\uppi}} \rightarrow 0$. \newline
In reverse, if $\epsilon_{\bm{\uppi}}$ approaches zero, then $\bm{\uppi} \in \hat{\mathcal{P}}$ satisfies the consistency relation exactly, cf. (\ref{eq:GeneralizedViolatedCR}). Then there would be no need to define a modified map $\bm{\tilde{\uppi}}$ in order to find a multipole solution that is related to $\bm{\uppi}$. In other words, in the modified transformation law (\ref{eq:ModifiedAmbiguityTrafo}) it is possible to choose $\xi_{k} = \zeta_{k^{\prime}} = 0$. In this case, it is directly seen that (\ref{eq:GroupSObsCorrComplete}) vanishes. \newline
Therefore, the correction $\delta \check{\Omega}^{\alpha_{S}}$ generally is only zero if and only if the accidental ambiguity $\bm{\uppi} \left(\alpha_{k},\beta_{k}\right)$ happens to respect the consistency relation exactly. In case only a tiny violation $\epsilon_{\bm{\uppi}} > 0$ is present, $\delta \check{\Omega}^{\alpha_{S}}$ is non-zero. \newline
In has to be said that while the case $\epsilon_{\bm{\uppi}} \rightarrow 0$ cannot be excluded from the outset for a general accidental ambiguity $\bm{\uppi} \in \hat{\mathcal{P}}$, in the analysis of model data performed in this work, a non-vanishing $\epsilon_{\bm{\uppi}} > 0$ was always found. Assuming that this is always the case, only the double ambiguity of the true solution satisfies the consistency relation exactly while leaving the group $\mathcal{S}$ observables invariant. The assumed generic $\bm{\uppi} \in \hat{\mathcal{P}}$ is still an exact symmetry of the group $\mathcal{S}$, while violating the consistency constraint. The latter can be restored by the modified map $\bm{\tilde{\uppi}}$, at the price of loosing the exact symmetry of $\left\{\sigma_{0}, \Sigma, T, P\right\}$. The resulting generic cases for different types of ambiguities of the group $\mathcal{S}$ observables therefore amount to three, assuming that an $\epsilon_{\bm{\uppi}} > 0$ is always present for all $\bm{\uppi} \in \hat{\mathcal{P}}$. They are summarized in Table \ref{tab:ThreeCases}. \newline \newline
\begin{table}[h]
 \centering
\begin{tabular*}{\linewidth}{m{0.75cm}|m{4.0cm}|m{3.5cm}|m{4.5cm}}
\hline
\hline
 & & & \\
 Case & Invariance of group $\mathcal{S}$ & Consistency relation & Type of ambiguity \\
\hline
 & & & \\
 I & Exact symmetry & Exactly fulfilled & \hspace*{7.5pt} True solution $\left(\alpha_{k},\beta_{k}\right)$ \newline \hspace*{50pt} $\updownarrow$ \newline Double Ambiguity $\left(\alpha_{k}^{\ast},\beta_{k}^{\ast}\right)$ \\
 & & & \\
\hline
 & & & \\
 II & Exact symmetry & Slightly broken & Generic accidental ambiguity $\left(\alpha_{k},\beta_{k}\right) \rightarrow \bm{\uppi} \left(\alpha_{k},\beta_{k}\right)$ \\
 & & & \\
\hline
 & & & \\
 III & Slightly broken & Exactly fulfilled & Modified accidental ambiguity $\left(\alpha_{k},\beta_{k}\right) \rightarrow \bm{\tilde{\uppi}} \left(\alpha_{k},\beta_{k}\right)$ \\
 & & & \\
\hline
\hline
\end{tabular*}
\caption[Summary and properties of different types of ambiguities in a TPWA.]{This Table summarizes the different cases of on the one hand, exact or approximate symmetry of the group $\mathcal{S}$ observables $\left\{\sigma_{0}, \Sigma, T, P\right\}$ and on the other hand, exact or approximate fulfillment of the consistency relation. Shown are only the scenarios described in the main text. The type of ambiguity that generically shows the mentioned behaviour is also indicated.}
\label{tab:ThreeCases}
\end{table}
\newpage
Finally we consider the consequences the correction (\ref{eq:GroupSObsCorrComplete}) has for the numeric solution of exactly solvable model data for the group $\mathcal{S}$ observables by minimization of a suitably defined discrepancy function. \newline
The latter can be defined as a function of the real and imaginary parts $\mathcal{M}_{\ell}$ of phase-constrained multipoles in the following way (note that this object is not yet a statistical $\chi^{2}$)
\begin{equation}
\Phi \left( \left\{ \mathcal{M}_{\ell} \right\} \right) := \sum_{\alpha_{S}, t_{k}} \left[ \check{\Omega}^{\alpha_{S}}_{\mathrm{Data}} \left(t_{k}\right) - \check{\Omega}^{\alpha_{S}}_{\mathrm{Fit}} \left(t_{k}, \left\{ \mathcal{M}_{\ell} \right\}\right) \right]^{2} \mathrm{.} \label{eq:DefDiscrFunct}
\end{equation}
The sum runs over the group $\mathcal{S}$ indices $\alpha_{S}$ and a finite set of angular points $t_{k}$ where the datapoints $\check{\Omega}^{\alpha_{S}}_{\mathrm{Data}} \left(t_{k}\right)$ exist. The functions $\check{\Omega}^{\alpha_{S}}_{\mathrm{Fit}} \left(t_{k}, \left\{ \mathcal{M}_{\ell} \right\}\right)$ denote the multipole decompositions of the grous S observables for a TPWA truncated at $L$ (cf. appendix \ref{sec:TPWAFormulae}). \newline
In case the fit parameters $\left\{ \mathcal{M}_{\ell} \right\}$ approach the true solution, i.e. multipoles equivalent to the roots $\left(\alpha_{k}, \beta_{k}\right)$, then $\Phi$ is precisely zero. \newline
If however the parameters $\left\{ \mathcal{M}_{\ell} \right\}$ tend towards a solution that corresponds to the transformed roots $\bm{\tilde{\uppi}} \left(\alpha_{k}, \beta_{k}\right)$, $\Phi$ becomes finite and does generally not vanish. The correction to the discrepancy function can be found by replacing $\check{\Omega}^{\alpha_{S}}_{\mathrm{Fit}} \left(t_{k}, \left\{ \mathcal{M}_{\ell} \right\}\right)$ with $\tilde{\check{\Omega}}^{\alpha_{S}}$ from (\ref{eq:GroupSObsTrafoRuleAccAmb}) in the definition (\ref{eq:DefDiscrFunct}):
\begin{equation}
\delta \Phi = \sum_{\alpha_{S}, t_{k}} \left( \check{\Omega}^{\alpha_{S}} \left(t_{k}\right) - \tilde{\check{\Omega}}^{\alpha_{S}} \left(t_{k}\right) \right)^{2} = \sum_{\alpha_{S}, t_{k}} \left[\delta \check{\Omega}^{\alpha_{S}} \left(t_{k}\right) \right]^{2} \mathrm{.} \label{eq:DeltaDiscrFunct}
\end{equation}
This result is represented as a schematic in Figure \ref{fig:ChiSquareValleyCartoon}. The numerical analyses of model data (see section \ref{sec:TheoryDataFits}) exactly verify the situation that is represented in the picture in a simplified way. \newline \newline
The detailed discussion of the clear definition of accidental ambiguities, as well as the invariance properties of the group $\mathcal{S}$ observables under such ambiguities is now finished. In the next appendix section, we turn to the question about the circumstances under which the accidental ambiguities become important, or in other words, when they can endanger the unique solvability of a particular set of observables. We will be particularly concerned with sets of observables that were called complete in section \ref{sec:WBTpaper}, i.e. that contain only one observable in addition to the group $\mathcal{S}$ that is capable of resolving the double ambiguity.

\newpage

\textit{Appendix: Proof of equation (\ref{eq:ExpandedB4ModSquare}):}
\newline \newline
In order to show that the expansion (\ref{eq:ExpandedB4ModSquare}) is valid for arbitrary $L$, we choose induction \cite{ForsterI} and therefore have to anchor the latter procedure by considering the following simple case: 
\newline \newline
\underline{$L = 1$:} \newline
\newline
Here, the expansion of the product over $k$ in the important term of the initial expression (\ref{eq:NonFullyExpandedB4ModSquare}) is quickly done and yields
\allowdisplaybreaks
\begin{align}
 & \hspace*{12.5pt} \prod_{k = 1}^{2 \ast 1} \left( \left[ t - \alpha_{k}^{\ast} \right] \left[ t - \alpha_{k} \right] - 2 \left|\xi_{k}\right| \kappa^{\bm{\uppi}}_{k} \right) \nonumber \\
&= \left( \left[ t - \alpha_{1}^{\ast} \right] \left[ t - \alpha_{1} \right] - 2 \left|\xi_{1}\right| \kappa^{\bm{\uppi}}_{1} \right) \ast \left( \left[ t - \alpha_{2}^{\ast} \right] \left[ t - \alpha_{2} \right] - 2 \left|\xi_{2} \right| \kappa^{\bm{\uppi}}_{2} \right) \nonumber \\
&= \left( \prod_{k = 1}^{2} \left[ t - \alpha_{k}^{\ast} \right] \left[ t - \alpha_{k} \right] \right) - 2 \left| \xi_{1} \right| \kappa^{\bm{\uppi}}_{1} \left[ t - \alpha_{2}^{\ast} \right] \left[ t - \alpha_{2} \right] - 2 \left| \xi_{2} \right| \kappa^{\bm{\uppi}}_{2} \left[ t - \alpha_{1}^{\ast} \right] \left[ t - \alpha_{1} \right] \nonumber \\
& \hspace*{12.5pt} + \mathcal{O}\left(\left|\xi_{k}\right|^{2}\right) \nonumber \\
&= \prod_{k = 1}^{2} \left[ t - \alpha_{k}^{\ast} \right] \left[ t - \alpha_{k} \right] - 2 \sum_{n=1}^{2} \left( \left| \xi_{n} \right| \kappa_{n}^{\bm{\uppi}} \prod_{m \neq n}^{2} \left[ t - \alpha_{m}^{\ast} \right] \left[ t - \alpha_{m} \right] \right) + \mathcal{O}\left(\left|\xi_{k}\right|^{2}\right) \mathrm{.} \label{eq:InductiveProof1}
\end{align}
Therefore, for the case $L = 1$, equation (\ref{eq:ExpandedB4ModSquare}) is seen to be valid. Next, one has to do the inductive step:
\newline \newline
\underline{$L \rightarrow L+1$:} \newline
\newline
The first move consists of writing down the product in the non-expanded form of $\left| b_{4} \right|^{2}$ (eq. (\ref{eq:NonFullyExpandedB4ModSquare})) for the truncation angular momentum $L+1$ and then splitting the whole product into factors coming from $k$-values up to $2L$, as well as the two coming from $k = 2L+1$ and $k = 2L+2$:
\begin{align}
 & \hspace*{7.5pt} \prod_{k = 1}^{2 (L+1)} \left( \left[ t - \alpha_{k}^{\ast} \right] \left[ t - \alpha_{k} \right] - 2 \left| \xi_{k} \right| \kappa^{\bm{\uppi}}_{k} \right) \nonumber \\
&= \prod_{k = 1}^{2L} \left( \left[ t - \alpha_{k}^{\ast} \right] \left[ t - \alpha_{k} \right] - 2 \left| \xi_{k} \right| \kappa^{\bm{\uppi}}_{k} \right) \ast \prod_{k = 2L+1}^{2L+2} \left( \left[ t - \alpha_{k}^{\ast} \right] \left[ t - \alpha_{k} \right] - 2 \left| \xi_{k} \right| \kappa^{\bm{\uppi}}_{k} \right) \mathrm{.} \label{eq:InductiveProof2step1}
\end{align}
Inserting the expression (\ref{eq:ExpandedB4ModSquare}) (up to prefactors) for the first factor and fully expanding the second factor (up to terms $\mathcal{O}\left( \left| \xi_{k} \right|^{2} \right)$), one arrives at an expression which facilitates the completion of the proof:
\allowdisplaybreaks
\begin{align}
 & \hspace*{7.5pt} \Bigg[ \prod_{k = 1}^{2L} \left( t - \alpha_{k}^{\ast} \right)  \left( t - \alpha_{k} \right) - 2 \sum_{n=1}^{2L} \left( \left| \xi_{n} \right| \kappa_{n}^{\bm{\uppi}} \prod_{m \neq n}^{2L} \left[ t - \alpha_{m}^{\ast} \right]  \left[ t - \alpha_{m} \right] \right) + \mathcal{O}\left(\left|\xi_{k}\right|^{2}\right) \Bigg] \nonumber \\
& \ast \Bigg[ \prod_{k = 2L+1}^{2L+2} \left( t - \alpha_{k}^{\ast} \right)  \left( t - \alpha_{k} \right) - 2 \left| \xi_{2L+1} \right| \kappa^{\bm{\uppi}}_{2L+1} \left( t - \alpha_{2L+2}^{\ast} \right)  \left( t - \alpha_{2L+2} \right)  \nonumber \\
 & \hspace*{9.5pt} - 2 \left| \xi_{2L+2} \right| \kappa^{\bm{\uppi}}_{2L+2} \left( t - \alpha_{2L+1}^{\ast} \right)  \left( t - \alpha_{2L+1} \right) + \mathcal{O}\left(\left|\xi_{k}\right|^{2}\right) \Bigg] \nonumber \\
&= \prod_{k = 1}^{2(L+1)} \left( t - \alpha_{k}^{\ast} \right)  \left( t - \alpha_{k} \right) \nonumber \\
& \hspace{12pt} - \prod_{k = 1}^{2L} \left( t - \alpha_{k}^{\ast} \right)  \left( t - \alpha_{k} \right) \ast 2 \sum_{n = 2L+1}^{2L+2} \left| \xi_{n} \right| \kappa_{n}^{\bm{\uppi}}
 \prod_{\substack{m=2L+1 , \\ m \neq n}}^{2L+2} \left( t - \alpha_{m}^{\ast} \right)  \left( t - \alpha_{m} \right) \nonumber \\
& \hspace{12pt} - 2 \left( \sum_{n=1}^{2L} \left| \xi_{n} \right| \kappa_{n}^{\bm{\uppi}} \prod_{m \neq n}^{2L} \left[ t - \alpha_{m}^{\ast} \right]  \left[ t - \alpha_{m} \right] \right) \prod_{k = 2L+1}^{2L+2} \left( t - \alpha_{k}^{\ast} \right)  \left( t - \alpha_{k} \right) + \mathcal{O}\left(\left|\xi_{k}\right|^{2}\right) \nonumber \\
&= \prod_{k = 1}^{2(L+1)} \left( t - \alpha_{k}^{\ast} \right)  \left( t - \alpha_{k} \right) - 2 \sum_{n = 1}^{2(L+1)} \left| \xi_{n} \right| \kappa_{n}^{\bm{\uppi}} \prod_{m \neq n}^{2(L+1)} \left( t - \alpha_{m}^{\ast} \right)  \left( t - \alpha_{m} \right) \nonumber \\
& \hspace*{12pt} + \mathcal{O}\left(\left|\xi_{k}\right|^{2}\right) \mathrm{.}
\end{align}
Utilizing the principle of complete induction \cite{ForsterI}, one then sees that expression (\ref{eq:ExpandedB4ModSquare}) is valid for any $L$. 
\begin{flushright}
\textbf{QED.}
\end{flushright}

\paragraph{Are accidental ambiguities important?} \label{subsec:AccidentalAmbProofsIII}  \textcolor{white}{:-)} \newline

Up to now the assumption was made that in case of a TPWA truncated at some finite $L$, which is assumed to possess a mathematically exact solution, the accidental symmetries of the group $\mathcal{S}$ observables $\left\{ \sigma_{0}, \Sigma, T, P \right\}$ were effectively negligible. Then, only one observable needed to be added to the group $\mathcal{S}$ which is capable of removing the only exact symmetry, namely the double ambiguity. In this way, it was possible to postulate complete sets of polarization observables that contain just 5 quantities as opposed to the 8 of Chiang and Tabakin \cite{ChTab}. \newline
Now, we will elaborate under which circumstances this completeness can be threatened. Assuming that no other discrete ambiguities apart from the double ambiguity and the accidental ambiguities exist, this question is closely tied to the issue of the conditions under which the latter become relevant. We organize the discussion according to the three most important scenarios. \newline

\textbf{Probability for the appearance of exact accidental ambiguities with $\epsilon_{\bm{\uppi}} = 0$.} \newline

As always stated before, the possibility of having an exact accidental ambiguity $\bm{\uppi} \in \hat{\mathcal{P}}$, i.e. one with $\epsilon_{\bm{\uppi}} = 0$, was assumed to be ruled out. But is this assumption truly justified? In case such an exact accidental ambiguity exists, it can by way of the arguments made in appendix \ref{subsec:AccidentalAmbProofsII} be an exact symmetry of the group $\mathcal{S}$ observables. Then, one may think of a scenarion where the completeness with five observables would be lost, provided that e.g. the $F$ or $G$ observable would be incapable of resolving this ambiguity. It can be seen from their definition (cf. equations (64) and (65) of section \ref{sec:WBTpaper})
\begin{align}
 \check{F} \left(\theta\right) &= \frac{\sigma_{0} \left(\pi\right)}{2 \left( 1 + t^{2} \right)^{2L}} \mathrm{Im} \left[ - \prod_{k=1}^{2L} \left( t + \alpha_{k}^{\ast} \right) \left( t + \beta_{k} \right) + \prod_{k=1}^{2L} \left( t - \alpha_{k}^{\ast} \right) \left( t - \beta_{k} \right) \right] \mathrm{,} \label{eq:FExplicitFormula} \\
 \check{G} \left(\theta\right) &= \frac{\sigma_{0} \left(\pi\right)}{2 \left( 1 + t^{2} \right)^{2L}} \mathrm{Im} \left[ \prod_{k=1}^{2L} \left( t + \alpha_{k}^{\ast} \right) \left( t + \beta_{k} \right) + \prod_{k=1}^{2L} \left( t - \alpha_{k}^{\ast} \right) \left( t - \beta_{k} \right) \right] \mathrm{,} \label{eq:GExplicitFormula}
\end{align}
that those two observables may still very well be capable of resolving the additional ambiguity coming from $\bm{\uppi}$. Still it is interesting to investigate the scenario of such an exact accidental ambiguity. \newline
Numerical constellations of phases $\varphi_{k}$ and $\psi_{k^{\prime}}$ of the Omelaenko-roots $\alpha_{k}$ and $\beta_{k^{\prime}}$ can be easily constructed that fulfill an accidental ambiguity exactly. One can consider for example the case $L=1$ with the four values of the phases
\begin{equation}
 \varphi_{1} = \frac{\pi}{2} \mathrm{,} \hspace*{2pt}  \varphi_{2} = \frac{\pi}{3} \mathrm{,} \hspace*{2pt}  \psi_{1} = \frac{\pi}{2} \mathrm{,} \hspace*{2pt}  \psi_{2} = \frac{\pi}{3} \mathrm{.}  \label{eq:ExactAccAmbPhasesExample}
\end{equation}
In this example the ambiguity $\bm{\uppi}_{10} = (+,-,+,-)$ is exactly fulfilled. Therefore, it is clearly seen that in principle numerical solutions that generate exact accidental ambiguities exist and that they cannot be disregarded from the outset. \newline
In the following, we will give some mathematical (or better said, geometrical) arguments to show that while such special numerical configurations can in fact exist, in case of an exactly solvable TPWA they are highly unlikely to occur. \newline \newline
First, there is again the assumption that an exact true solution to the TPWA problem exists, which is represented by the $4L$ complex roots $\left(\alpha_{k},\beta_{k}\right)$. The phases of these roots then have to fulfill the consistency relation exactly
\begin{equation}
 \varphi_{1} + \ldots + \varphi_{2L} = \psi_{1} + \ldots + \psi_{2L} \mathrm{.} \label{eq:ConsRelExactlyMathDiscussion}
\end{equation}
The validity of this constraint means that for example the phase $\psi_{2L}$ can be evaluated in terms of the other phases in a differentiable way. Effectively, the space on which (\ref{eq:ConsRelExactlyMathDiscussion}) is fulfilled has one dimension less than the full $4L$ dimensional space of the Omelaenko-phases. In mathematical terms, (\ref{eq:ConsRelExactlyMathDiscussion}) defines an $\left(4L-1\right)$-dimensional hypersurface (or sub-manifold) $M_{\mathcal{CR}}^{\left(4L-1\right)}$ in the space spanned by all the phases $\left(\varphi_{k},\psi_{k}\right)$. By convention, we assume all these phases to take values in the interval $\left[ - \pi, \pi \right]$, such that the full volume of interest in the parameter space is a higher dimensional cube $\left[ - \pi, \pi \right]^{4L}$. \newline
Equivalently, the configurations of phases that correspond to the accidental ambiguity $\bm{\uppi} \in \hat{\mathcal{P}}$, i.e. for which
\begin{equation}
 \bm{\uppi} \left( \varphi_{1} \right) + \ldots + \bm{\uppi} \left( \varphi_{2L} \right) = \bm{\uppi} \left( \psi_{1} \right) + \ldots + \bm{\uppi} \left( \psi_{2L} \right) \mathrm{.} \label{eq:ConsRelExactlyMathDiscussionAccAmb}
\end{equation}
is valid, define a $\left(4L - 1\right)$-dimensional hypersurface $M_{\bm{\uppi}}^{\left(4L-1\right)}$. All numerical configurations for which an exact accidental ambiguity can occur have to fulfill (\ref{eq:ConsRelExactlyMathDiscussion}) \textit{and} (\ref{eq:ConsRelExactlyMathDiscussionAccAmb}) at the same time. Therefore, they have to lie on the intersection $ M_{\bm{\uppi}}^{\left(4L-1\right)} \cap M_{\mathcal{CR}}^{\left(4L-1\right)} $. \newline
We now argue that the probability for this to occur is infinitesimally small. In order to do this, it will prove fruitful to define a sub-volume of $\left[ - \pi, \pi \right]^{4L}$ as the set of points for which the accidental ambiguity $\bm{\uppi} \in \hat{\mathcal{P}}$ is not exactly fulfilled as in equation (\ref{eq:ConsRelExactlyMathDiscussionAccAmb}), but is allowed to be violated by a small numerical error. The maximum possible value of this error will be called $\epsilon$. One has to note that there is a direct connection between this $\epsilon$ and the parameters $\epsilon_{\bm{\uppi}}$ we defined in appendix \ref{subsec:AccidentalAmbProofsI} and that measured the violation of the consistency relation for a particular accidental ambiguity $\bm{\uppi}$. The volume that is going to be defined encapsulates all parameter configurations that are possible for ambiguities $\bm{\uppi} \in \hat{\mathcal{P}}$ with violation parameter $\epsilon_{\bm{\uppi}}$ smaller than or equal to the $\epsilon$ that defines the volume. \newline
It will be very helpful for the ensuing discussion to introduce an appropriately defined Heaviside $\theta$-function belonging to the above mentioned volume according to
\begin{equation}
 \theta_{\bm{\uppi}} \left( \varphi_{1}, \ldots, \varphi_{2L}, \psi_{1}, \ldots, \psi_{2L}; \epsilon \right) := \theta \left[ \epsilon^{2} - \left( \sum_{k=1}^{2L} \bm{\uppi} \left(\varphi_{k}\right) - \sum_{k^{\prime}=1}^{2L} \bm{\uppi} \left(\psi_{k^{\prime}}\right) \right) \right] \mathrm{.} \label{eq:EpsilonVolThetaFunction}
\end{equation}
This function can be used to define the set of points we want to introduce, namely
\begin{equation}
 V^{4L}_{\bm{\uppi}} \left( \epsilon \right) := \left\{ \left( \varphi_{1}, \ldots, \varphi_{2L}, \psi_{1}, \ldots, \psi_{2L} \right) \in \left[ - \pi, \pi \right]^{4L} \hspace*{2.5pt} | \hspace*{2.5pt} \theta_{\bm{\uppi}} \left( \varphi_{k}, \psi_{k}; \epsilon \right) = 1 \right\} \mathrm{.} \label{eq:EpsilonVolDefinition}
\end{equation}
The real focus of interest in the following discussion however will be set on an $(4L-1)$-dimensional hypersurface in the cube $\left[ - \pi, \pi \right]^{4L}$, which is defined as the intersection of the volume (\ref{eq:EpsilonVolDefinition}) with the sub-manifold $M^{(4L-1)}_{\mathcal{CR}}$. Therefore, we are interested in
\begin{equation}
 \tilde{M}^{4L-1}_{\bm{\uppi}} \left( \epsilon \right) := V^{4L}_{\bm{\uppi}} \left( \epsilon \right) \cap M_{\mathcal{CR}}^{\left(4L-1\right)} \mathrm{,} \label{eq:InterestingIntersection}
\end{equation}
and even more importantly, the evaluation of the higher dimensional area this manifold has, as measured \textit{on the hypersurface} $M^{(4L-1)}_{\mathcal{CR}}$. \newline
The points contained in the intersection $\tilde{M}^{4L-1}_{\bm{\uppi}} \left( \epsilon \right)$ are just those numerical configurations of the Omelaenko-phases $\left(\varphi_{k},\psi_{k}\right)$ that satisfy the consistency relation of the true solution exactly, while being allowed to break the constraint (\ref{eq:ConsRelExactlyMathDiscussionAccAmb}) by a small numerical error smaller or equal to $\epsilon$. It is a sub-surface of the manifold $M^{(4L-1)}_{\mathcal{CR}}$ and the $(4L-1)$-dimensional volume of this sub-set, as compared to the full volume of $M^{(4L-1)}_{\mathcal{CR}}$, will tell us something about the probability of the corresponding ambiguity to occurr. \newline
In order to evaluate the needed volumes (or areas), some results from the general theory of integration on sub-manifolds of $\mathbbm{R}^{n}$ will be needed (see e.g. \cite{ForsterIII}). \newline
Suppose there is a function defined on some open subset $U \subset \mathbbm{R}^{n}$ and the precise operation of which can be expressed as
\begin{equation}
 f : U \rightarrow \mathbbm{R}^{n} ; \hspace*{2.5pt} \left( x_{1}, \ldots, x_{n} \right) \mapsto f \left( x_{1}, \ldots, x_{n} \right) \mathrm{.} \label{eq:GeneralHigherDimFunction}
\end{equation}
This function shall now be integrated over an $(n-1)$-dimensional manifold (or hypersurface) $M^{(n-1)}$. Furthermore it shall be assumed that, upon some renumbering of the $x$-variables, it is always possible to parametrize the manifold in the following way
\begin{equation}
 \left( x_{1}, \ldots, x_{n-1} \right) \mapsto \left( x_{1}, \ldots, x_{n-1}, F \left( x_{1}, \ldots, x_{n-1} \right) \right) \mathrm{,} \label{eq:SubmanifoldParametrization}
\end{equation}
where $F \left( x_{1}, \ldots, x_{n-1} \right)$ is some differentiable function. In case this is possible, the integral of the function $f$ over the hypersurface $M^{(n-1)}$ is declared as
\allowdisplaybreaks
\begin{align}
 &\int_{M^{n-1}} f \left( x_{1}, \ldots, x_{n} \right) d S \left( x_{1}, \ldots, x_{n} \right) \nonumber \\
 &:= \int_{x_{1}^{\mathrm{I}}}^{x_{1}^{\mathrm{II}}} dx_{1} \ldots \int_{x_{n-1}^{\mathrm{I}}}^{x_{n-1}^{\mathrm{II}}} dx_{n-1} \sqrt{1+ \left| \mathrm{grad} F \left( x_{1}, \ldots, x_{n-1} \right) \right|^{2}} \nonumber \\
 & \hspace*{134pt} \times f \left( x_{1}, \ldots, x_{n-1}, F \left( x_{1}, \ldots, x_{n-1} \right) \right) \nonumber \\
 &\equiv \int_{c_{1}^{\mathrm{I}}}^{c_{1}^{\mathrm{II}}} dx_{1} \ldots \int_{c_{n}^{\mathrm{I}}}^{c_{n}^{\mathrm{II}}} dx_{n} \sqrt{1+ \left| \mathrm{grad} F \left( x_{1}, \ldots, x_{n-1} \right) \right|^{2}} \nonumber \\
 & \hspace*{115pt} \times \delta \left( F\left[ x_{1}, \ldots, x_{n-1} \right] - x_{n} \right) f \left( x_{1}, \ldots, x_{n} \right) \mathrm{.} \label{eq:DeltaFctGeneralDefIntHypersurface}
\end{align}
The factor $\sqrt{1 + \left| \mathrm{grad} F \right|^{2}}$ stems from the determinant of the Gram measure tensor (again, see \cite{ForsterIII}). The integral is performed within the limits $\left[ x_{i}^{\mathrm{I}}, x_{i}^{\mathrm{II}} \right]$, which should directly follow from the sub-manifold under consideration. This however also means that, in order for the integral to only cover the point-set $M^{n-1}$, the integration boundaries $\left[ x_{i}^{\mathrm{I}}, x_{i}^{\mathrm{II}} \right]$ can become quite complicated to find and in particular, boundaries with larger indices can depend on the integration variables with respect to which one integrates subsequently. For example, the boundaries $\left[ x_{n-1}^{\mathrm{I}}, x_{n-1}^{\mathrm{II}} \right]$ are generally not constant but may depend on all variables $(x_{1}, \ldots, x_{n-2})$. All these issues are in some sense circumvented by restoring the $n$-th integration-dimension and introducing the Dirac $\delta$-function in the last step of equation (\ref{eq:DeltaFctGeneralDefIntHypersurface}). The integration boundaries $\left[ c_{i}^{\mathrm{I}}, c_{i}^{\mathrm{II}} \right]$ are here just constants that define the maximal range of each integration variable. \newline
Furthermore, it is seen as a corollary that the $(n-1)$-dimensional volume of the manifold $M^{(n-1)}$ can be calculated by just applying the integral (\ref{eq:DeltaFctGeneralDefIntHypersurface}) to the unit function, i.e.
\begin{align}
 \mathcal{V}_{M}^{(n-1)} &:= \int_{M^{n-1}} d S \left( x_{1}, \ldots, x_{n} \right) \nonumber \\
  &= \int_{x_{1}^{\mathrm{I}}}^{x_{1}^{\mathrm{II}}} dx_{1} \ldots \int_{x_{n-1}^{\mathrm{I}}}^{x_{n-1}^{\mathrm{II}}} dx_{n-1} \sqrt{1+ \left| \mathrm{grad} F \left( x_{1}, \ldots, x_{n-1} \right) \right|^{2}} \nonumber \\
  &= \int_{c_{1}^{\mathrm{I}}}^{c_{1}^{\mathrm{II}}} dx_{1} \ldots \int_{c_{n}^{\mathrm{I}}}^{c_{n}^{\mathrm{II}}} dx_{n} \sqrt{1+ \left| \mathrm{grad} F \left( x_{1}, \ldots, x_{n-1} \right) \right|^{2}} \nonumber \\
 & \hspace*{115pt} \times \delta \left( F\left[ x_{1}, \ldots, x_{n-1} \right] - x_{n} \right) \mathrm{.} \label{eq:VolumeSubManifoldDefinition}
\end{align}
We now apply the introduced concepts to the situation in the higher-dimensional cube $\left[ - \pi, \pi \right]^{4L}$ of the Omelaenko-phases. First of all, a function $F$ is needed that defines the manifold of the true solution $M^{(4L-1)}_{\mathcal{CR}}$. To do this, we choose the convention of expressing the Omelaenko-phase $\psi_{2L}$ in terms of the remaining phases by use of the consistency relation (\ref{eq:ConsRelExactlyMathDiscussion})
\begin{equation}
 \psi_{2L} \equiv F \left( \varphi_{1}, \ldots, \varphi_{2L}, \psi_{1}, \ldots, \psi_{2L-1} \right) := \sum_{k=1}^{2L} \varphi_{k} - \sum_{k^{\prime}=1}^{2L-1} \psi_{k^{\prime}} \mathrm{.} \label{eq:FFunctionCRManifoldDef}
\end{equation}
The gradient of this function is quickly evaluated to be the following vector with $(4L-1)$ components:
\begin{equation}
 \mathrm{grad} F \left( \varphi_{1} \ldots, \psi_{2L-1} \right) = \left( 1,\ldots,1,-1,\ldots,-1 \right)^{T} \mathrm{.} \label{eq:GradFCRResult}
\end{equation}
Since the resulting gradient does not depend on any phase variable any more, the Gram determinant very conveniently evaluates to
\begin{equation}
 \sqrt{ 1 + \left| \mathrm{grad} F \right|^{2}} = \sqrt{1 + \left( 4 L - 1 \right)} = 2 \sqrt{L} \mathrm{.} \label{eq:CRGramMeasureFactor}
\end{equation}
The volume of the full manifold $M_{\mathcal{CR}}^{(4L-1)}$ can still be calculated relatively conveniently. Utilizing the formula (\ref{eq:VolumeSubManifoldDefinition}) as well as everything already evaluated above, one obtains
\allowdisplaybreaks
\begin{align}
\mathcal{V}^{(4L-1)}_{\mathcal{CR}} &= \int_{M^{(4L-1)}_{\mathcal{CR}}} d S \left(\varphi_{1},\ldots,\varphi_{2L},\psi_{1},\ldots,\psi_{2L}\right) \nonumber \\
 &= \int_{\varphi_{1}^{\mathrm{I}}}^{\varphi_{1}^{\mathrm{II}}} d \varphi_{1} \ldots \int_{\varphi_{2L}^{\mathrm{I}}}^{\varphi_{2L}^{\mathrm{II}}} d \varphi_{2L} \int_{\psi_{1}^{\mathrm{I}}}^{\psi_{1}^{\mathrm{II}}} d \psi_{1} \ldots \int_{\psi_{2L-1}^{\mathrm{I}}}^{\psi_{2L-1}^{\mathrm{II}}} d \psi_{2L-1} \sqrt{1 + \left| \mathrm{grad} F \right|^{2}} \nonumber \\
 &= 2 \sqrt{L} \int_{-\pi}^{\pi} d \varphi_{1} \ldots \int_{-\pi}^{\pi} d \varphi_{2L} \int_{-\pi}^{\pi} d \psi_{1} \ldots \int_{-\pi}^{\pi} d \psi_{2L} \hspace*{2.5pt} \delta \left( \sum_{k=1}^{2L} \varphi_{k} - \sum_{k^{\prime}=1}^{2L} \psi_{k^{\prime}} \right) \mathrm{.} \label{eq:CRVolumeMostGeneralEvaluation}
\end{align}
A closed expression for this integral can be derived for arbitrary truncation angular momenta $L$. For the derivation see appendix \ref{sec:ProjectionIntegrals}, where all the integrals introduced here are evaluated in some detail. The result reads
\begin{equation}
 \mathcal{V}^{(4L-1)}_{\mathcal{CR}} = 2 \sqrt{L} \frac{\pi^{4L - 1}}{\left(4L-1\right)!} \sum_{k = 0}^{\left( 2 L - 1 \right)} (-)^{k} \binom{4L}{k} \left(4L-2k\right)^{\left(4L-1\right)} \mathrm{,} \label{eq:CRVolumeMostGeneralExpression}
\end{equation}
where the binomial coefficients
\begin{equation}
 \binom{n}{k} = \frac{n!}{k! (n-k)!} \mathrm{,} \label{eq:BinCoeffDefinitionMainText}
\end{equation}
have been introduced. The formula (\ref{eq:CRVolumeMostGeneralExpression}) actually not only holds for the hypersurface $M_{\mathcal{CR}}^{(4L-1)}$, corresponding to the exact validity of the consistency relation (\ref{eq:ConsRelExactlyMathDiscussion}) for the true solution, but furthermore it does so for any $M_{\bm{\uppi}}^{(4L-1)}$ which is defined by the formula (\ref{eq:ConsRelExactlyMathDiscussionAccAmb}) for any $\bm{\uppi} \in \hat{\mathcal{P}}$. All that has to be done to see this, is to define a function $F_{\bm{\uppi}}$ in much the same way as in (\ref{eq:FFunctionCRManifoldDef})
\begin{equation}
 \psi_{2L} \equiv F_{\bm{\uppi}} \left( \varphi_{1}, \ldots, \varphi_{2L}, \psi_{1}, \ldots, \psi_{2L-1} \right) := \sum_{k=1}^{2L} \bm{\uppi} \left(\varphi_{k}\right) - \sum_{k^{\prime}=1}^{2L-1} \bm{\uppi} \left( \psi_{k^{\prime}} \right) \mathrm{.} \label{eq:FFunctionAccAmbManifoldDef}
\end{equation}
The gradient of $F_{\bm{\uppi}}$ is again quite simple. It is the $(4L-1)$-dimensional constant vector
\begin{equation}
 \mathrm{grad} F_{\bm{\uppi}} \left( \varphi_{1} \ldots, \psi_{2L-1} \right) = \left( \pm 1,\ldots, \pm 1,\mp1,\ldots,\mp1 \right)^{T} \mathrm{,} \label{eq:GradFAccAmbResult}
\end{equation}
It can now be seen quickly that the Gram determinant evaluated for this gradient is the same as (\ref{eq:CRGramMeasureFactor}). Following the steps described in the beginning of appendix \ref{sec:ProjectionIntegrals}, it is seen that the result of the integral (\ref{eq:CRVolumeMostGeneralEvaluation}) does not change as well. \newline
Now it is time to formally define the sought after $(4L-1)$-dimensional volume of the intersection point set (\ref{eq:InterestingIntersection}). This is just the area of the projection of the $4L$-dimensional volume $V^{4L}_{\bm{\uppi}} \left( \epsilon \right)$ on the hypersurface $M_{\mathcal{CR}}^{(4L-1)}$. It can be obtained by integrating the Heaviside function (\ref{eq:EpsilonVolThetaFunction}) over $M_{\mathcal{CR}}^{(4L-1)}$ via the definition (\ref{eq:DeltaFctGeneralDefIntHypersurface}), i.e.
\allowdisplaybreaks
\begin{align}
\tilde{\mathcal{V}}^{(4L-1)}_{\bm{\uppi}} \left(\epsilon\right) &= \int_{M^{(4L-1)}_{\mathcal{CR}}} \theta_{\bm{\uppi}} \left(\varphi_{1},\ldots,\varphi_{2L},\psi_{1},\ldots,\psi_{2L};\epsilon\right) d S \left(\varphi_{1},\ldots,\varphi_{2L},\psi_{1},\ldots,\psi_{2L}\right) \nonumber \\
 &= \int_{\varphi_{1}^{\mathrm{I}}}^{\varphi_{1}^{\mathrm{II}}} d \varphi_{1} \ldots \int_{\varphi_{2L}^{\mathrm{I}}}^{\varphi_{2L}^{\mathrm{II}}} d \varphi_{2L} \int_{\psi_{1}^{\mathrm{I}}}^{\psi_{1}^{\mathrm{II}}} d \psi_{1} \ldots \int_{\psi_{2L-1}^{\mathrm{I}}}^{\psi_{2L-1}^{\mathrm{II}}} d \psi_{2L-1}  \sqrt{ 1 + \left| \mathrm{grad} F \right|^{2}} \nonumber \\
 &\hspace*{15pt} \times \theta \Bigg[ \epsilon^{2} - \left( \sum_{k=1}^{2L} \bm{\uppi} \left(\varphi_{k}\right) - \sum_{k^{\prime}=1}^{2L-1} \bm{\uppi} \left(\psi_{k^{\prime}}\right) + \bm{\uppi} \left[ \sum_{k=1}^{2L} \varphi_{k} - \sum_{k^{\prime}=1}^{2L-1} \psi_{k^{\prime}} \right] \right)^{2} \Bigg] \nonumber \\
 &= 2 \sqrt{L} \int_{-\pi}^{\pi} d \varphi_{1} \ldots \int_{-\pi}^{\pi} d \varphi_{2L} \int_{-\pi}^{\pi} d \psi_{1} \ldots \int_{-\pi}^{\pi} d \psi_{2L} \hspace*{2.5pt} \delta \left( \sum_{k=1}^{2L} \varphi_{k} - \sum_{k^{\prime}=1}^{2L} \psi_{k^{\prime}} \right) \nonumber \\
 &\hspace*{15pt} \times \theta \Bigg[ \epsilon^{2} - \left( \sum_{k=1}^{2L}  \bm{\uppi} \left(\varphi_{k}\right) - \sum_{k^{\prime}=1}^{2L} \bm{\uppi} \left(\psi_{k^{\prime}}\right) \right)^{2} \Bigg] \mathrm{.} \label{eq:PiNVolumeProjectionGeneralExpression}
\end{align}
It is quickly recognizable that once this expression is evaluated for a particular accidental ambiguity, it automatically yields also an expression for a second ambiguity which is connected to the first one via the Double Ambiguity transformation. In this case, the term in the squared bracket within the $\theta$-function changes its sign, which has no effect on the result. Generally, configurations connected by the Double Ambiguity live on the same hypersurfaces, so the volume (\ref{eq:PiNVolumeProjectionGeneralExpression}) just stays the same. Actually, there is some further reduction of the possible projection volume formulas in a particular order $L$, which is discussed below. \newline
The derivation of a closed expression for this integral for arbitrary $L$ seems like a highly formidable task. Surprisingly, it is possible to get quite far towards this goal, but the calculations are lengthy. A detailed treatment of the reduction of the integral (\ref{eq:PiNVolumeProjectionGeneralExpression}) is given in appendix \ref{sec:ProjectionIntegrals}, while here we only quote the main results. \newline

As it turns out, the quantity (\ref{eq:PiNVolumeProjectionGeneralExpression}) can be generally written as
\allowdisplaybreaks
\begin{align}
 \tilde{\mathcal{V}}^{(4L-1)}_{\bm{\uppi}} \left(\epsilon\right) &= \mathcal{V}^{(4L-1)}_{\mathcal{CR}} - 2 \sqrt{L} \int_{\bm{C}}^{\epsilon} d \epsilon^{\prime} \left[  \frac{d \mathrm{I}_{1}}{d \epsilon^{\prime}} \left(\epsilon^{\prime}\right) + \frac{d \mathrm{I}_{2}}{d \epsilon^{\prime}} \left(\epsilon^{\prime}\right) \right] \mathrm{,} \label{eq:FormalFinalExpressionProjectionVolumeMainText}
\end{align}
with the general expressions for the derivatives $\mathrm{I}_{1}^{\prime} (\epsilon)$ and $\mathrm{I}_{2}^{\prime} (\epsilon)$ being
\allowdisplaybreaks
\begin{align}
 \hspace*{-8.5pt} \frac{d \mathrm{I}_{1}}{d \epsilon} &= \frac{(-) \pi^{(4L-2)}}{2^{4L-1} \left( 4L - 2 \right)!} \left[ \sum_{k=0}^{n_{1}}  \sum_{k^{\prime}=0}^{n_{2}} \right]_{(k + k^{\prime}) > 2L} \nonumber \\
 & \hspace*{12.5pt} \times \Bigg\{ \left( 1 - \theta\left[ \epsilon - 2 \pi \left( n_{2} - 2 k^{\prime} \right) \right] \right) \frac{(-)^{(k + k^{\prime})}}{\left( n_{1} - 1 \right)!} \binom{n_{1}}{k}  \binom{n_{2}}{k^{\prime}} \nonumber \\
  & \hspace*{35.5pt} \times \sum_{r = (4L - n_{1} - 1)}^{(4L - 2)} \frac{(4L-2)!}{r!} \hspace*{2pt} \bm{\upkappa}_{(4L - r - 2)}^{\left( L, n_{1}, k, k^{\prime} \right)} \left[ i \left( - 4 k^{\prime} + 2 n_{2} - \frac{\epsilon}{\pi} \right) \right]^{r} \nonumber \\
 & \hspace*{23.25pt} - \left( 1 - \theta\left[ \epsilon - 2 \pi \left( 2 k - n_{1} \right) \right] \right) \frac{(-)^{(k + k^{\prime})}}{\left( n_{2} - 1 \right)!} \binom{n_{1}}{k}  \binom{n_{2}}{k^{\prime}}  \nonumber \\
  & \hspace*{35.5pt} \times \sum_{r = (4L - n_{2} - 1)}^{(4L - 2)} (-)^{r} \frac{(4L-2)!}{r!} \hspace*{2pt} \bm{\upkappa}_{(4L - r - 2)}^{\left( L, n_{2}, k, k^{\prime} \right)} \left[ i \left( 4 k - 2 n_{1} - \frac{\epsilon}{\pi} \right) \right]^{r} \Bigg\} \mathrm{,} \label{eq:2ndIntegralFinalResultReformulatedMainText}
\end{align}
and
\begin{equation}
 \frac{d \mathrm{I}_{2}}{d \epsilon} \left(\epsilon\right) = \frac{d \mathrm{I}_{1}}{d \epsilon} \left(\epsilon\right) \Big|_{n_{1} \leftrightarrow n_{2}} \mathrm{.} \label{eq:I2FromI1MainText}
\end{equation}
The square-brackets around the double-sum in equation (\ref{eq:2ndIntegralFinalResultReformulatedMainText}) have been written in order to explicitly indicate that this sum is restricted to those terms fulfilling $(k + k^{\prime}) > 2L$. The coefficients $\bm{\upkappa}_{j}^{\left( L, n, k, k^{\prime} \right)}$ that appear here are defined as
\begin{equation}
 \bm{\upkappa}_{j}^{\left( L, n, k, k^{\prime} \right)} := (-)^{(n - 1)} \binom{n - 1}{j} \frac{\Gamma (4L - j - 1)}{\Gamma (4L - n)} \left[ 2 i (2 [k + k^{\prime}] - 4L) \right]^{j} \mathrm{.} \label{eq:DefinitionKappasMainText}
\end{equation}
Furthermore, the index $n_{1}$ counts the number of Omelaenko-phases that do not change their sign under the ambiguity $\bm{\uppi} \in \hat{\mathcal{P}}$, while $n_{2}$ is equal to the number of phases that \textit{do} change sign. Interestingly, this is the only information about the ambiguity $\bm{\uppi}$ the projection volume formula is sensitive to. In particular, the knowledge about which phases precisely change their sign is not important, but only how many do this (see appendix \ref{sec:ProjectionIntegrals} for more details). Naturally, both indices have to fulfill the constraint $n_{1} + n_{2} = 4 L$, since one always has a total number of $4L$ Omelaenko-phases at each order in $L$. \newline
Finally, equation (\ref{eq:FormalFinalExpressionProjectionVolumeMainText}) contains an a priori unknown integration constant $\bm{C}$ which can always be fixed by imposing consistent boundary conditions for (\ref{eq:FormalFinalExpressionProjectionVolumeMainText}), e.g. $\tilde{\mathcal{V}}^{(4L-1)}_{\bm{\uppi}} \left(\epsilon \to \infty \right) \equiv \mathcal{V}^{(4L-1)}_{\mathcal{CR}}$. However, investigations of the results for (\ref{eq:FormalFinalExpressionProjectionVolumeMainText}) and specifically (\ref{eq:2ndIntegralFinalResultReformulatedMainText}) for some specific values of $L$, $n_{1}$ and $n_{2}$ have hinted heavily at the fact that choosing $\bm{C} = n^{<} \ast 2 \pi$, with $n^{<}$ defined as $n^{<} = \mathrm{min} (n_{1},n_{2})$, is the universally correct choice. This latter point was until now however not provable. \newline \newline
Below the definition (\ref{eq:PiNVolumeProjectionGeneralExpression}) above, it was breifly mentioned that the number of distinct projection volume formulas gets reduced further beyound the fact that, for ambiguities related via the Double Ambiguity transformation, results have to be the same. The reduction originates from the above mentioned circumstance that the quantity (\ref{eq:PiNVolumeProjectionGeneralExpression}) is insensitive to the information about which phases precisely switch their sign. Only the numbers $n_{1}$ and $n_{2}$ are important. Out of this reason, in conjunction with the fact that phases related by the D.A.-transformation live on the same hypersurfaces, one is lead to the result that the $(4^{2L} - 2)$ possible accidental ambiguities (for arbitrary $L$), leading to $(4^{2L}/2 - 1)$ generally distinct sub-manifolds of $\left[ - \pi, \pi\right]^{4L}$, generate only $2L$ different expressions for the projection volume (\ref{eq:PiNVolumeProjectionGeneralExpression}) (this fact may be verified by complete induction \cite{ForsterI}). \newline

We now start to draw the connection of the up to now almost purely geometrical discussion, to a probabilistic argument which expresses the fact that while the numerical configurations for exact accidental ambiguities indeed exist, they are very unlikely to occur. In order to do this, we assume that the dynamics of pseudoscalar meson photoproduction distribute the surface $M_{\mathcal{CR}}^{\left(4L-1\right)}$ with random numbers having a flat probability distribution function (PDF). This assumption is clearly not fully correct in the sense that photoproduction does not distribute uniformly over $M_{\mathcal{CR}}^{\left(4L-1\right)}$ but rather instead has some smooth PDF that favors certain regions. The introduction of a PDF seems quite \textit{ad hoc} at this point. However, when real data are analyzed, it makes more sense to think about the result of a fit as some smooth likelihood function in amplitude space, rather than to expect the existence of exact solutions. In case the existence of one exact solution is postulated as done above, this one solution would in the language of probability distributions correspond to a Dirac $\delta$-peak. Still, for the argument to be made, the rough assumption of a flat PDF shall at first suffice. \newline
Then, the probability for the ambiguity $\bm{\uppi} \in \hat{\mathcal{P}}$ to be an accidental symmetry with maximal numerical error $\epsilon$ is given by the fraction formed by dividing the intersection volume $\tilde{\mathcal{V}}^{(4L-1)}_{\bm{\uppi}} \left( \epsilon \right)$ by the full volume $\mathcal{V}^{(4L-1)}_{\mathcal{CR}}$ of $M_{\mathcal{CR}}^{\left(4L-1\right)}$. The fact that this ratio just gives the correct probability can be illustrated by a simpler example. If one considers the box region defined in the two dimensional plane as
\begin{equation}
 C := \left\{ (x,y) \in \mathbbm{R}^{2} \hspace*{2pt} | \hspace*{2pt} x \in \left[-1,1\right] \hspace*{2pt} \& \hspace*{2pt} y \in \left[-1,1\right] \right\} \mathrm{,} \label{eq:BoxRegionFullDef}
\end{equation}
and generates random numbers over $C$ with a flat distribution, then the chance to hit the upper right quadrant
\begin{equation}
 \tilde{C} := \left\{ (x,y) \in \mathbbm{R}^{2} \hspace*{2pt} | \hspace*{2pt} x \in \left[0,1\right] \hspace*{2pt} \& \hspace*{2pt} y \in \left[0,1\right] \right\} \mathrm{,} \label{eq:BoxRegionUpRightDef}
\end{equation}
is given by the ratio of the two areas, in this case $1/4$. \newline
In the more complicated case of the phases $\varphi_{k}$ and $\psi_{k}$ of the Omelaenko-roots, it is (using the above made assumptions) possible to generally define the probability
\begin{equation}
 P_{\bm{\uppi}} \left( \epsilon \right) := \frac{\tilde{\mathcal{V}}_{\bm{\uppi}}^{(4L-1)} \left( \epsilon \right)}{\mathcal{V}_{\mathcal{CR}}^{\left(4L-1\right)}} \mathrm{.} \label{eq:ProbDef}
\end{equation}
In order to generalize this definition, the existence of a smooth PDF has to be postulated which is defined on the hypersurface $M^{(4L-1)}_{\mathcal{CR}}$, namely
\begin{equation}
 \Pi \left( \varphi, \psi \right) = \Pi \left( \varphi_{1}, \ldots, \varphi_{2L}, \psi_{1}, \ldots, \psi_{2L-1} \right) \mathrm{,} \label{eq:DefGeneralPDFonMCR}
\end{equation}
and furthermore that this function is normalized according to
\begin{equation}
 \int_{M^{(4L-1)}_{\mathcal{CR}}} d^{2L} \varphi \hspace*{2pt} d^{2L-1} \psi \hspace*{2pt} \Pi \left( \varphi, \psi \right) = 1 \mathrm{.} \label{eq:NormalizationGeneralPDFonMCR}
\end{equation}
Then, the proper generalization of the probability (\ref{eq:ProbDef}) is given by
\allowdisplaybreaks
\begin{align}
 P_{\bm{\uppi}} \left( \epsilon \right) :=& \int_{\tilde{M}^{4L-1}_{\bm{\uppi}} \left( \epsilon \right) } d^{2L} \varphi \hspace*{2pt} d^{2L-1} \psi \hspace*{2pt} \Pi \left( \varphi, \psi \right) \nonumber \\
 =& \int_{M^{(4L-1)}_{\mathcal{CR}}} d^{2L} \varphi \hspace*{2pt} d^{2L-1} \psi \hspace*{2pt} \theta_{\bm{\uppi}} \left(\varphi,\psi;\epsilon\right) \hspace*{2pt} \Pi \left( \varphi, \psi \right) \mathrm{.} \label{eq:DefGeneralProbability}
\end{align}
The following argument generally holds (under some weak assumptions) for the probabilities (\ref{eq:ProbDef}) as well as the more general definition (\ref{eq:DefGeneralProbability}), for it mainly draws from the geometric properties of the projection-manifolds $\tilde{M}^{4L-1}_{\bm{\uppi}} (\epsilon)$ which enter both definitions. \newline
It is the central claim of this appendix section that in the limit $\epsilon \rightarrow 0$, i.e. for the exact fulfillment of the constraint (\ref{eq:ConsRelExactlyMathDiscussionAccAmb}) and therefore the case of an exact accidental ambiguity, the probability $P_{\bm{\uppi}} \left( \epsilon \right)$ approaches zero
\begin{equation}
  \lim\limits_{\epsilon \to 0}{P_{\bm{\uppi}} \left( \epsilon \right)} = 0 \mathrm{.} \label{eq:LimitConjecture}
\end{equation}
In a geometric sense, when $\epsilon$ approaches zero and the consistency relation is assumed to be valid exactly again for the accidental ambiguity $\bm{\uppi} \in \hat{\mathcal{P}}$, then the $(4L-1)$-dimensional intersection volume $\tilde{M}^{4L-1}_{\bm{\uppi}} \left( \epsilon \right)$ approaches the strict intersection $M^{(4L-1)}_{\bm{\uppi}} \cap M_{\mathcal{CR}}^{(4L-1)}$. Since the latter set is formed by intersecting two $(4L-1)$-dimensional non-equal hypersurfaces, it has itself the dimension $(4L-2)$. Therefore, once the limit $\epsilon \rightarrow 0$ is taken, the set $\tilde{M}^{4L-1}_{\bm{\uppi}} \left( \epsilon \right)$ ``looses a dimension'' compared to $M_{\mathcal{CR}}^{(4L-1)}$. During this process, it becomes a \textit{set of measure zero} in the volume units that are valid on the hypersurface $M_{\mathcal{CR}}^{(4L-1)}$. The fact that a lower dimensional subset of a manifold has zero measure is a standard result from calculus-textbooks, see for example \cite{ForsterIII}. This argument from calculus also directly forces the general integral (\ref{eq:PiNVolumeProjectionGeneralExpression}) to vanish for $\epsilon \rightarrow 0$. \newline
Using the above made simple assumption about a flat PDF valid on $M_{\mathcal{CR}}^{(4L-1)}$, it becomes clear that then by means of the definition (\ref{eq:ProbDef}) also the probability for the exact accidental symmetry to occur becomes zero. The statements made here should generalize to any smooth PDF valid in amplitude space and should therefore also hold for the more general probability (\ref{eq:DefGeneralProbability}), provided $\Pi \left( \varphi, \psi \right)$ is not a Dirac $\delta$-distribution centered directly on the intersection $M^{(4L-1)}_{\bm{\uppi}} \cap M_{\mathcal{CR}}^{(4L-1)}$. \newline

\begin{figure}[h]
 \centering
 \begin{overpic}[width=0.95\textwidth]{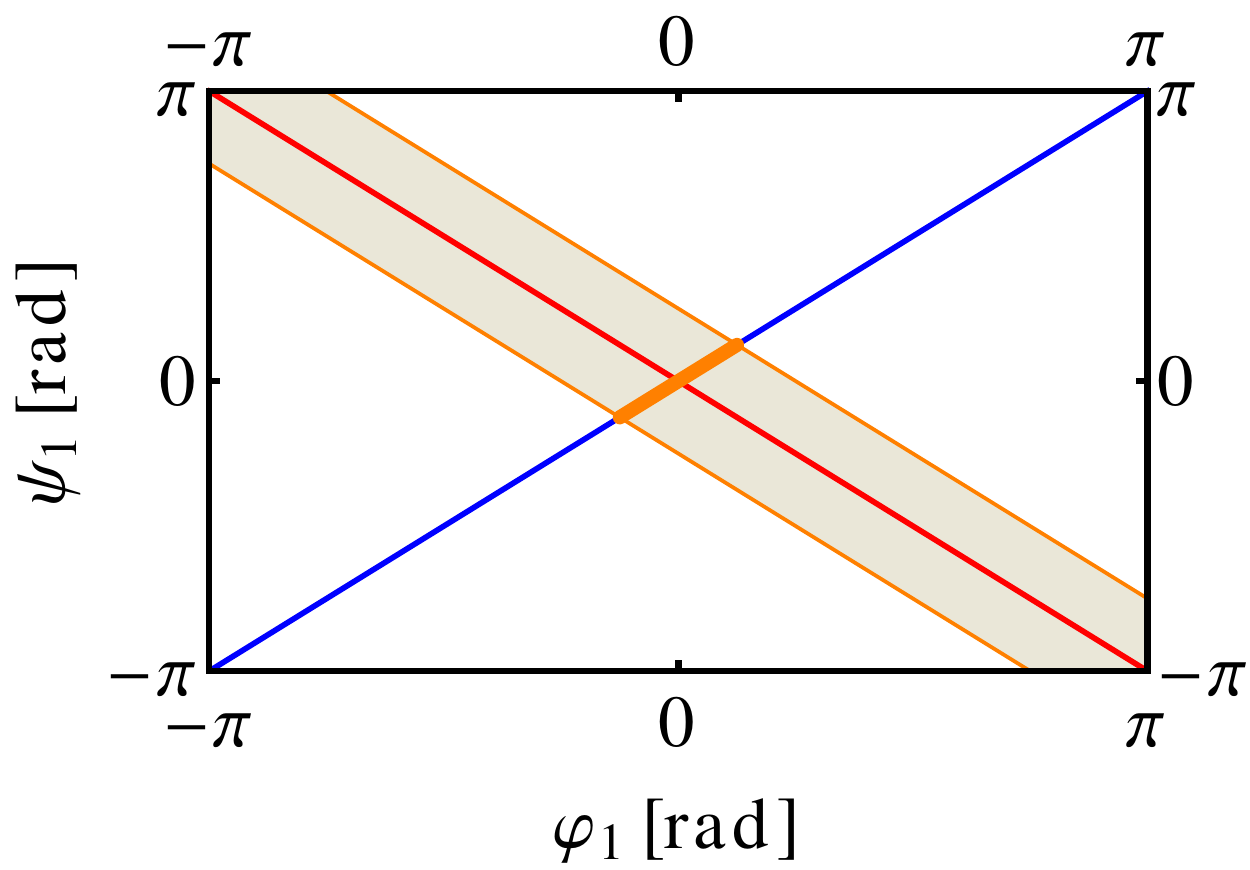}
 \end{overpic}
\vspace*{10pt}
\caption[Illustration of the geometric discussion on Omelaenko-phases given in the main text, for the example of one Omelaenko $\varphi$- and $\psi$-phase each.]{This picture illustrates the geometric discussion of the main text for the example of one Omelaenko $\varphi$- and $\psi$-phase each. Drawn is the parameter space $\left[ - \pi, \pi \right]^{2}$ of these two phases. The blue line represents the hypersurface $M_{\mathcal{CR}}^{1}$ of exact validity of the consistency relation $\varphi_{1} = \psi_{1}$, while the red line indicates points $M_{\bm{\uppi}}^{1}$ of the only ambiguity, $\varphi_{1} = - \psi_{1}$. \newline The orange shaded area denotes the volume $V^{2}_{\bm{\uppi}} \left( \epsilon \right)$ for $\epsilon = \frac{\pi}{4}$. The thick orange line represents the projection surface $\tilde{M}^{1}_{\bm{\uppi}} \left( \epsilon \right)$. For $\epsilon \rightarrow 0$, the one dimensional projection surface would shrink to a point in this example.}
\label{fig:PedagogicalPhaseVolumeExample}
\end{figure}

We will use the rest of this appendix to illustrate the above made abstract statements by particular examples. The first of those shall be given by just two Omelaenko-phases, $\varphi_{1}$ and $\psi_{1}$. This case never occurs in a TPWA, however it has the advantage of being very illustrative for the basic concepts described in this appendix. \newline
A graphical representation of the phase-parameter space $\left[ - \pi, \pi \right]^{2}$ in this case as well as of all the important point sets introducted above is shown in Figure \ref{fig:PedagogicalPhaseVolumeExample}. The (in this case 1-dimensional) hypersurface $M_{\mathcal{CR}}^{1}$ which corresponds to the validity of the consistency relation for the true solution
\begin{equation}
 \varphi_{1} = \psi_{1} \mathrm{,} \label{eq:CRTrivialPedagogicalCase}
\end{equation}
is drawn as the blue straight line, passing through the origin and having the slope 1. The only non-redundant accidental ambiguity is given by for example switching the sign of the $\psi_{1}$-phase, i.e. $\bm{\uppi} \left( \varphi_{1}, \psi_{1} \right) = \left( \varphi_{1}, - \psi_{1} \right)$. The hypersurface $M_{\bm{\uppi}}^{1}$ upon which the consistency relation
\begin{equation}
 \varphi_{1} = - \psi_{1} \mathrm{,} \label{eq:CRAccAmbTrivialPedagogicalCase}
\end{equation}
for $\bm{\uppi}$ is valid, is represented by the red line of slope $-1$ that passes through the origin as well. The volume $V^{2}_{\bm{\uppi}} \left( \epsilon \right)$ is indicated as the orange colored area around $M_{\bm{\uppi}}^{1}$, where in order to enhance the visibility, the quite large value of $\epsilon = \frac{\pi}{4}$ was chosen.

The interesting region of intersection $\tilde{M}^{1}_{\bm{\uppi}} \left( \epsilon \right)$ is indicated as a thick orange colored line segment of the blue colored hypersurface $M_{\mathcal{CR}}^{1}$. The length (or 1-dimensional volume) of the full hypersurface $M_{\mathcal{CR}}^{1}$ can be obtained by either continuing the formula (\ref{eq:CRVolumeMostGeneralExpression}) to $L = \frac{1}{2}$, or by computing the hypotenuse of a right-angled triangle whose cathetuses are both $2 \pi$, which in both cases yields $\mathcal{V}_{\mathcal{CR}} = \sqrt{2} \hspace*{2pt} 2 \pi$. Furthermore, the straightforward application of the formula (\ref{eq:PiNVolumeProjectionGeneralExpression}) yields a closed expression for the projection volume $\tilde{\mathcal{V}}_{\bm{\uppi}} \left(\epsilon\right)$, i.e.
\allowdisplaybreaks
\begin{align}
\tilde{\mathcal{V}}_{\bm{\uppi}} \left(\epsilon\right) &= \int_{M^{1}_{\mathcal{CR}}} \theta\left(\varphi_{1},\psi_{1};\epsilon\right) d S \left(\varphi_{1},\psi_{1}\right) \nonumber \\
 &= \int_{-\pi}^{\pi} d \varphi_{1} \int_{-\pi}^{\pi} d \psi_{1} \sqrt{1 + \left| \mathrm{grad} F \left[\varphi_{1}\right] \right|^{2}} \hspace*{3pt} \delta \left( \varphi_{1} - \psi_{1} \right) \hspace*{3pt} \theta\left( \varphi_{1},\psi_{1};\epsilon \right) \nonumber \\
 &= \int_{-\pi}^{\pi} d \varphi_{1} \sqrt{1 + \left| \mathrm{grad} F \left[\varphi_{1}\right] \right|^{2}} \hspace*{3pt} \theta\left( \varphi_{1},\varphi_{1};\epsilon \right) \nonumber \\
 &= \sqrt{2} \int_{-\pi}^{\pi} d \varphi_{1} \theta \left( \epsilon^{2} - \left[ 2 \varphi_{1} \right]^{2} \right) \nonumber \\
 &= \sqrt{2} \left( \epsilon +(2 \pi -\epsilon ) \theta[ \epsilon -2 \pi ] \right) \mathrm{.} \label{eq:LowerDimExampleVolumeCalculation}
\end{align}
%
This is just the length of the thick orange colored line in Figure \ref{fig:PedagogicalPhaseVolumeExample}. The obtained expression is seen to satisfy the correct anticipated boundary conditions. For the maximum reasonable value of $\epsilon$, $\epsilon = 2 \pi$, it approaches the full volume $\mathcal{V}_{\mathcal{CR}} = \sqrt{2} \hspace*{2pt} 2 \pi$. In the limit $\epsilon \rightarrow 0$, the formula yields zero exactly as postulated for arbitrary projection volumes before. This behaviour is directly reflected in Figure \ref{fig:PedagogicalPhaseVolumeExample}. For $\epsilon \rightarrow 2 \pi$ the orange colored area covers the full parameter space $\left[ - \pi, \pi \right]^{2}$, thereby also reaching over the full hypersurface $M_{\mathcal{CR}}^{1}$ (represented by the blue line). For $\epsilon \rightarrow 0$ on the other hand, the shaded area approaches $M_{\bm{\uppi}}^{1}$ (red line). At the same time, the thick orange colored line, i.e. $\tilde{M}^{1}_{\bm{\uppi}} \left( \epsilon \right)$, shrinks to a single point exactly at the origin. In the limiting process, the 1-dimensional point set $\tilde{M}^{1}_{\bm{\uppi}} \left( \epsilon \right)$ becomes a zero-dimensional one, i.e. just a point. When measured in the 1-dimensional volume units valid on the hypersurface $M_{\mathcal{CR}}^{1}$, this point is just a set of measure zero. \newline

\begin{figure}[H]
 \centering
 \begin{overpic}[width=0.475\textwidth]{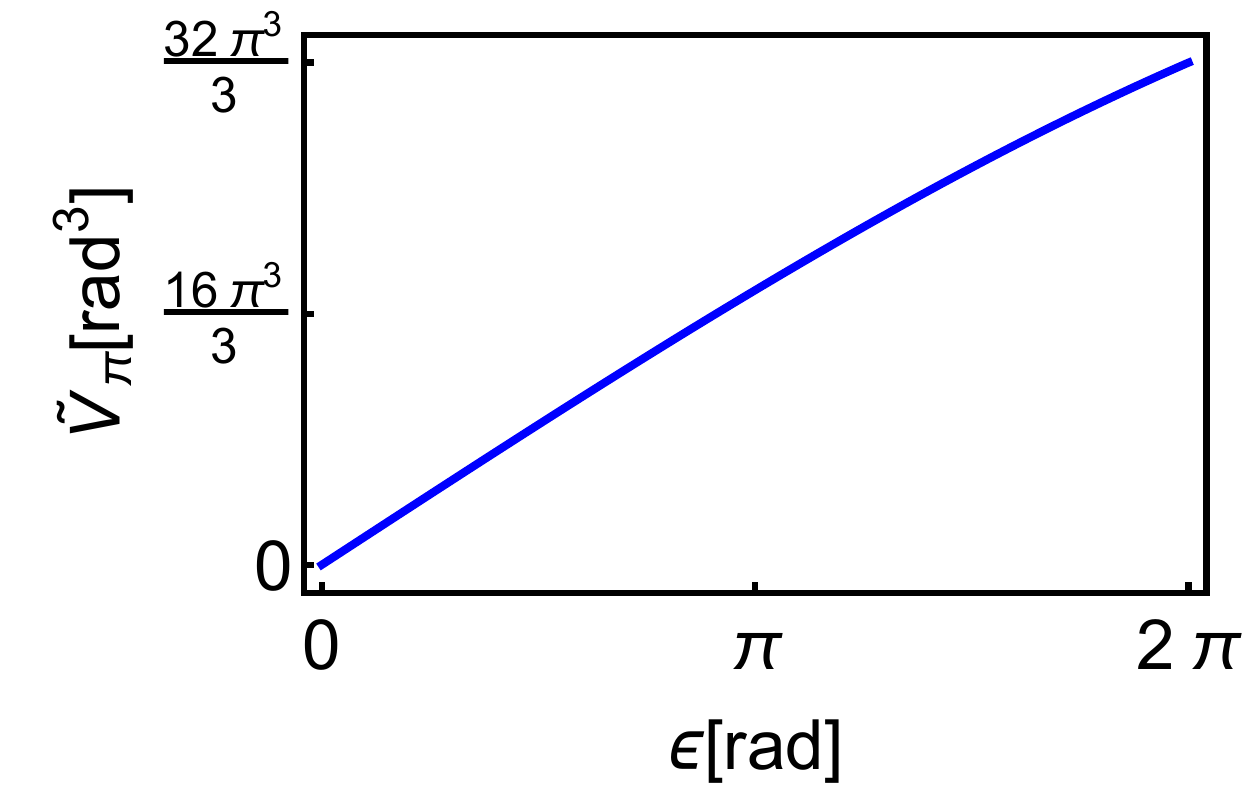}
 \put(27,52.5){a.)}
 \end{overpic}
 \begin{overpic}[width=0.475\textwidth]{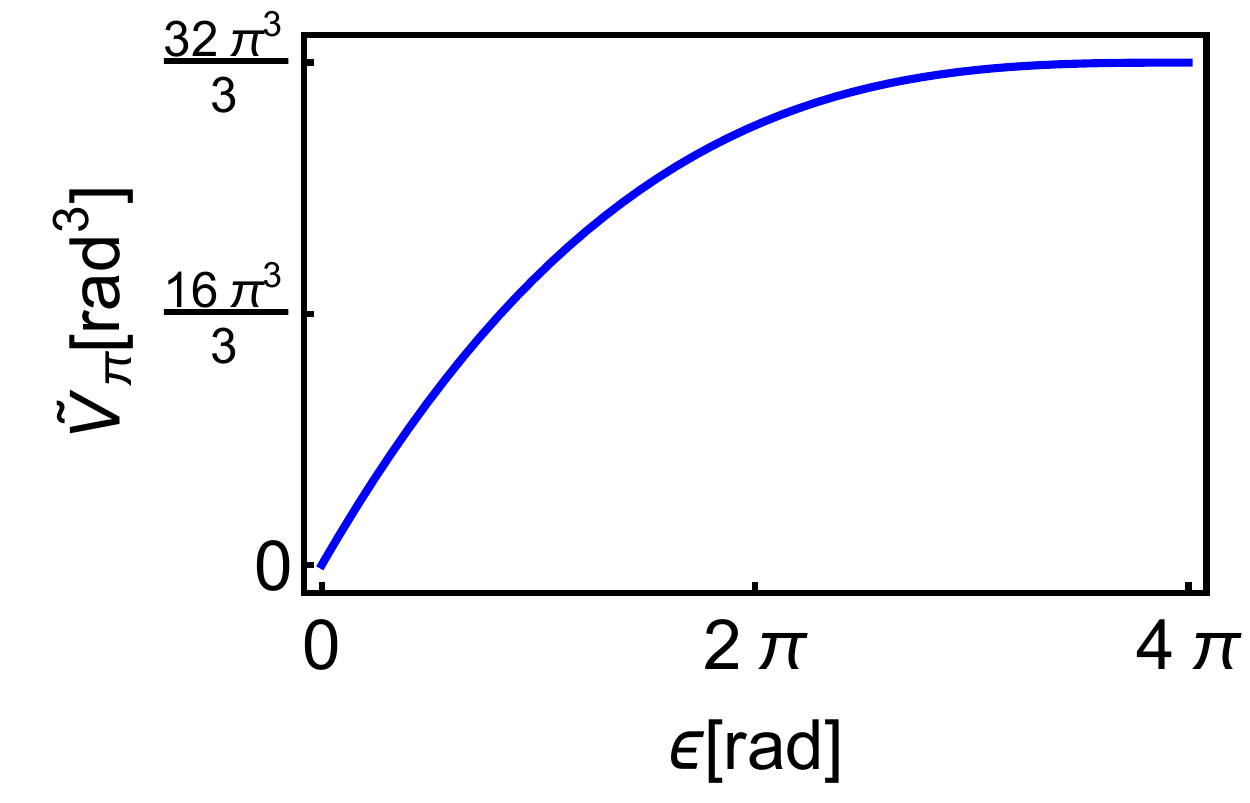}
 \put(27,52.5){b.)}
 \end{overpic}
\caption[All relevant projection volumes for the case $L = 1$ are shown as functions of $\epsilon$.]{The relevant projection volumes for $L = 1$ are shown as functions of $\epsilon$. Expressions for the functions are collected in Table \ref{tab:Lmax1VolumesTable}. The cases refer to different accidental ambiguities as: a.) $\bm{\uppi}_{8}$, $\bm{\uppi}_{11}$, $\bm{\uppi}_{13}$ and $\bm{\uppi}_{14}$; b.) $\bm{\uppi}_{9}$, $\bm{\uppi}_{10}$ and $\bm{\uppi}_{12}$.}
\label{fig:VolumeProjectionIntegralsPlots}
\end{figure}

The usefulness of the previously discussed example consists of the fact that all cases encountered in a realistic TPWA are in some sense just higher-dimensional generalizations of the situation depicted in Figure \ref{fig:PedagogicalPhaseVolumeExample}. The simplest possible realistic case, a truncation at $L = 1$, shall also be discussed. The results obtained will also lead to useful estimates about the percentage of actually dangerous accidental ambiguities appearing in a TPWA. \newline
In this case, all possible discrete ambiguities for the Omelaenko-phases $( \varphi_{1}, \varphi_{2}, \psi_{1}, \psi_{2} )$, i.e. $\left\{ \bm{\uppi}_{0} , \ldots, \bm{\uppi}_{15} \right\}$, are listed in Table \ref{tab:AllCCPossibilitiesLEquals1Case} of appendix \ref{subsec:AccidentalAmbProofsI}. As mentioned underneath of expression (\ref{eq:PiNVolumeProjectionGeneralExpression}), all accidental ambiguities that have to be considered for the calculation have to be non-relatable by the Double Ambiguity transformation, for example those that act on the phase $\psi_{2}$ as $\bm{\uppi}_{\hspace*{0.015cm}n} \left( \psi_{2} \right) = - \psi_{2}$. In this way, the ambiguity transformations $\bm{\uppi}_{\left( 8, \ldots, 14 \right)}$ are selected. As an example, we explicitly show here the expression of the projection volume of the ambiguity $\bm{\uppi}_{10}$ as a function of $\epsilon$. This specific ambiguity acts as $\bm{\uppi}_{10} ( \varphi_{1}, \varphi_{2}, \psi_{1}, \psi_{2} ) = ( \varphi_{1}, - \varphi_{2}, \psi_{1}, - \psi_{2} )$. Therefore, the volume $V^{3}_{\bm{\uppi}_{10}}$ is described by the Heaviside function $\theta_{\bm{\uppi}_{10}} \left( \varphi_{1}, \varphi_{2}, \psi_{1}, \psi_{2}; \epsilon \right) = \theta \left[ \epsilon^{2} - \left( \varphi_{1} - \varphi_{2} - \psi_{1} + \psi_{2} \right)^{2} \right]$. With this, the evaluation of the volume $\tilde{\mathcal{V}}^{3}_{\bm{\uppi}_{10}} \left(\epsilon\right)$ via the general projection integral (\ref{eq:PiNVolumeProjectionGeneralExpression}) reduces to
\allowdisplaybreaks
\begin{align}
\tilde{\mathcal{V}}^{3}_{\bm{\uppi}_{10}} \left(\epsilon\right) &= \int_{M^{3}_{\mathcal{CR}}} \theta_{\bm{\uppi}_{10}} \left(\varphi_{1},\varphi_{2},\psi_{1},\psi_{2};\epsilon\right) d S \left(\varphi_{1},\varphi_{2},\psi_{1},\psi_{2}\right) \nonumber \\
 &= \int_{-\pi}^{\pi} d \varphi_{1} \int_{-\pi}^{\pi} d \varphi_{2} \int_{-\pi}^{\pi} d \psi_{1} \int_{-\pi}^{\pi} d \psi_{2} \sqrt{1 + \left| \mathrm{grad} F \left[\varphi_{1},\varphi_{2},\psi_{1}\right] \right|^{2}} \nonumber \\
 & \hspace*{45pt} \times \delta \left( \varphi_{1} + \varphi_{2} - \psi_{1} - \psi_{2} \right) \theta \left[\epsilon^{2} - \left( \varphi_{1} - \varphi_{2} - \psi_{1} + \psi_{2} \right)^{2}\right] \nonumber \\
 &= \frac{32 \pi^{3}}{3} - \frac{1}{6} (4 \pi - \epsilon)^{3} \theta \left( 4 \pi - \epsilon \right) \mathrm{,} \label{eq:Pi10VolumeCalculation}
\end{align}
where the final closed expression was obtained using the results of appendix \ref{sec:ProjectionIntegrals} as well as MATHEMATICA \cite{Mathematica8,Mathematica11,MathematicaLanguage,MathematicaBonnLicense}. This formula is seen, as in the previous example (\ref{eq:LowerDimExampleVolumeCalculation}), to respect the correct behaviour at the boundaries. For $\epsilon \rightarrow 4 \pi$ (and more generally for $\epsilon \rightarrow \infty$) it is equal to the full volume of $M^{3}_{\mathcal{CR}}$, which in this case is (cf. (\ref{eq:CRVolumeMostGeneralExpression})) $\mathcal{V}^{3}_{\mathcal{CR}} = (32/3) \pi^{3}$. For $\epsilon \rightarrow 0$, the result of equation (\ref{eq:Pi10VolumeCalculation}) tends to zero. \newline \newline
Going beyond the particular example (\ref{eq:Pi10VolumeCalculation}), it is seen that all relevant cases for non-redundant ambiguities $\bm{\uppi}_{\hspace*{0.015cm}n} \in \hat{\mathcal{P}}$ can be classified into two different groups where the integral $\tilde{\mathcal{V}}^{3}_{\bm{\uppi}_{\hspace*{0.015cm}n}} \left(\epsilon\right)$ is the same for all elements of the respective group. The result for $\tilde{\mathcal{V}}^{3}_{\bm{\uppi}_{\hspace*{0.015cm}n}} \left(\epsilon\right)$ depends in any case just on the number $n_{2}$ of phases that change their sign (or the number $n_{1}$ of phases that do not change, respectively). Combined with the fact that ambiguities related via the D.A.-transformation exist on the same hypersurfaces, one arrives at the two distinct projection volume formulas for $L = 1$. All cases are given in Table \ref{tab:Lmax1VolumesTable}. \newline
The formulas that are found confirm that all relevant integrals satisfy consistent boundary conditions. One generally has
\begin{equation}
 \tilde{\mathcal{V}}^{3}_{\bm{\uppi}_{\hspace*{0.015cm}n}} \left(\epsilon\right) \rightarrow \begin{cases}
                    0 &\mathrm{,} \hspace*{2pt} \epsilon \rightarrow 0 \\
                    \frac{32}{3} \pi^{3} &\mathrm{,} \hspace*{2pt} \epsilon \rightarrow \epsilon_{\mathrm{max}}
                   \end{cases} \mathrm{.} \label{eq:ProjVolLmax1BoundaryConditions}
\end{equation}
In principle one could also take the limit $\epsilon \rightarrow \infty$ in the second case, but it is a fact that each projection volume saturates at a certain value $\epsilon_{\mathrm{max}}$ and remains constant for any $\epsilon$ larger than this value. In Figure \ref{fig:VolumeProjectionIntegralsPlots}, all relevant projection integrals for $L = 1$ are plotted as functions of $\epsilon$ within the range $\left[ 0, \epsilon_{\mathrm{max}} \right]$. In particular, all integrals evaluated in this example confirm that the intersections $\tilde{M}^{3}_{\bm{\uppi}_{\hspace*{0.015cm}n}}$ become sets of measure zero if the corresponding accidental ambiguity is assumed to satisfy the consistency relation exactly. \newline

Finally, we provide an estimate for the fraction of all possible accidental ambiguities that do not fulfill the consistency constraint exactly, but are allowed to violate it within a reasonable numerical error. For this error, we choose $\epsilon = 5^{\circ}$, since this value defines the region where the small-angle approximation is valid. In this case, the application of the simplest possible estimate (\ref{eq:ProbDef}) yields that only roughly 1 $\%$ - 2 $\%$ of all possible Omelaenko-phases satisfy the consistency relation for a particular $\bm{\uppi}_{\hspace*{0.015cm}n} \in \hat{\mathcal{P}}$ within this $5^{\circ}$-range. More precise numbers are given in Table \ref{tab:Lmax1VolumesTable}. \newline
This rough estimate however also suggests that only a few percent of the maximally possible accidental ambiguities, listed as $N_{\mathrm{AC}}$ in Table \ref{tab:AccAmbPossibilityNumber} of appendix \ref{subsec:AccidentalAmbProofsI} for some low $L$, can actually turn up as dangerous ambiguous multipole solutions. Furthermore, the few-percent fraction is also a result once the probability $P_{\bm{\uppi}}$ is evaluated for $\epsilon = 5^{\circ}$ in the simplified example case of just two Omelaenko-phases discussed above. Therefore, this result has up to now been more or less invariant under increase of the dimension of the phase-parameter volume $\left[ -\pi, \pi \right]^{4L}$. Whether or not this invariance holds also for the higher dimensions we cannot say at this point. \newline

We finish this discussion by briefly summarizing its main results. It was generally seen that the parameter regions of exact accidental ambiguities are sets of measure zero. This means that they can still be non-empty, which is consistent with the fact that it was generally quite easy to construct explicit parameter configurations that indeed fulfill $\epsilon_{\bm{\uppi}} = 0$ for a particular $\bm{\uppi} \in \hat{\mathcal{P}}$. However, since the dangerous parameter regions have a fully insignificant measure compared to the full volume of the relevant hypersurface $M_{\mathcal{CR}}^{(4L - 1)}$ in phase-parameter space, their elements may be regarded as highly unprobable. \clearpage
\vfill
\begin{table}[H]
 \centering
 \scalebox{.95}{
\begin{tabular}{ccccc}
\hline 
\hline \\
 Ambiguity ${\bm{\uppi}_{\hspace*{0.015cm}n}}$ & $n_{1}$ & $n_{2}$ & Volume \hspace*{0.5pt} $\tilde{\mathcal{V}}^{3}_{\bm{\uppi}_{\hspace*{0.015cm}n}} \hspace*{-2pt} \left(\epsilon\right)$ & $\frac{\tilde{\mathcal{V}}^{3}_{\bm{\uppi}_{\hspace*{0.015cm}n}} \left(\epsilon\right)}{\mathcal{V}^{3}_{\mathcal{CR}}}\Big|_{\epsilon \equiv 5^{\circ}}$ $\left[\%\right]$ \\
\hline \\
$\begin{aligned}\bm{\uppi}_{8} &= \left(+,+,+,-\right) \\ \bm{\uppi}_{11} &= \left(-,-,+,-\right) \\ \bm{\uppi}_{13} &= \left(-,+,-,-\right) \\ \bm{\uppi}_{14} &= \left(+,-,-,-\right) \end{aligned}$ & $\begin{aligned} 3 \\ 1 \\ 1 \\ 1 \end{aligned}$ & $\begin{aligned} 1 \\ 3 \\ 3 \\ 3 \end{aligned}$ & $ \frac{32 \pi^{3}}{3} - \frac{1}{6} \left[ 64 \pi^{3} - 36 \pi^{2} \epsilon + \epsilon^{3} \right] \left[ 1 - \theta (\epsilon - 2 \pi) \right] $ & $\simeq 1.56$ \\ \\
\hline \\
$\begin{aligned} \bm{\uppi}_{9} &= \left(-,+,+,-\right) \\ \bm{\uppi}_{10} &= \left(+,-,+,-\right) \\ \bm{\uppi}_{12} &= \left(+,+,-,-\right) \end{aligned}$ & $\begin{aligned} 2 \\ 2 \\ 2 \end{aligned}$ & $\begin{aligned} 2 \\ 2 \\ 2 \end{aligned}$ & $\frac{32 \pi^{3}}{3} - \frac{1}{6} (4 \pi - \epsilon)^{3} \theta \left( 4 \pi - \epsilon \right)$ & $\simeq 2.07$ \\ \\
\hline \hline
\end{tabular}
}
\caption[Summary of the relevant projection volumes $\tilde{\mathcal{V}}^{(4L-1)}_{\bm{\uppi}_{\hspace*{0.015cm}n}}$ for $L = 1$.]{This Table summarizes all the relevant projection volumes $\tilde{\mathcal{V}}^{(4L-1)}_{\bm{\uppi}_{\hspace*{0.015cm}n}}$ for the case $L = 1$. All the non-redundant accidental ambiguities ${\bm{\uppi}_{\hspace*{0.015cm}n}}$ that act on $\psi_{2L}$ as ${\bm{\uppi}_{\hspace*{0.015cm}n}} \left(\psi_{2L}\right) = - \psi_{2L}$ are listed. All the remaining accidental ambiguities are related to the latter via the double ambiguity transformation and therefore have the same projection volumes. \newline It is found (see appendix \ref{sec:ProjectionIntegrals}) that the final expression for $\mathcal{V}^{3}_{\bm{\uppi}_{\hspace*{0.015cm}n}} \hspace*{-2pt} \left(\epsilon\right)$ depends only on the number of phases that experience a sign change under the ambiguity transformation, but not on which phases precisely are conjugated. Therefore, the ambiguity $\bm{\uppi}_{8}$ has the same projection volume formula as the ambiguities $\bm{\uppi}_{11}$, $\bm{\uppi}_{13}$ and $\bm{\uppi}_{14}$. The explicit $\epsilon$-dependent expression (cf. appendix \ref{sec:ProjectionIntegrals}) is given.}
\label{tab:Lmax1VolumesTable}
\end{table}

\clearpage

\textbf{In realistic descriptions of photoproduction, the multipole series is an infinite series.}  \newline

\begin{figure}[hb]
 \centering
\hspace*{-277.5pt}
 \begin{overpic}[width=0.25\textwidth]{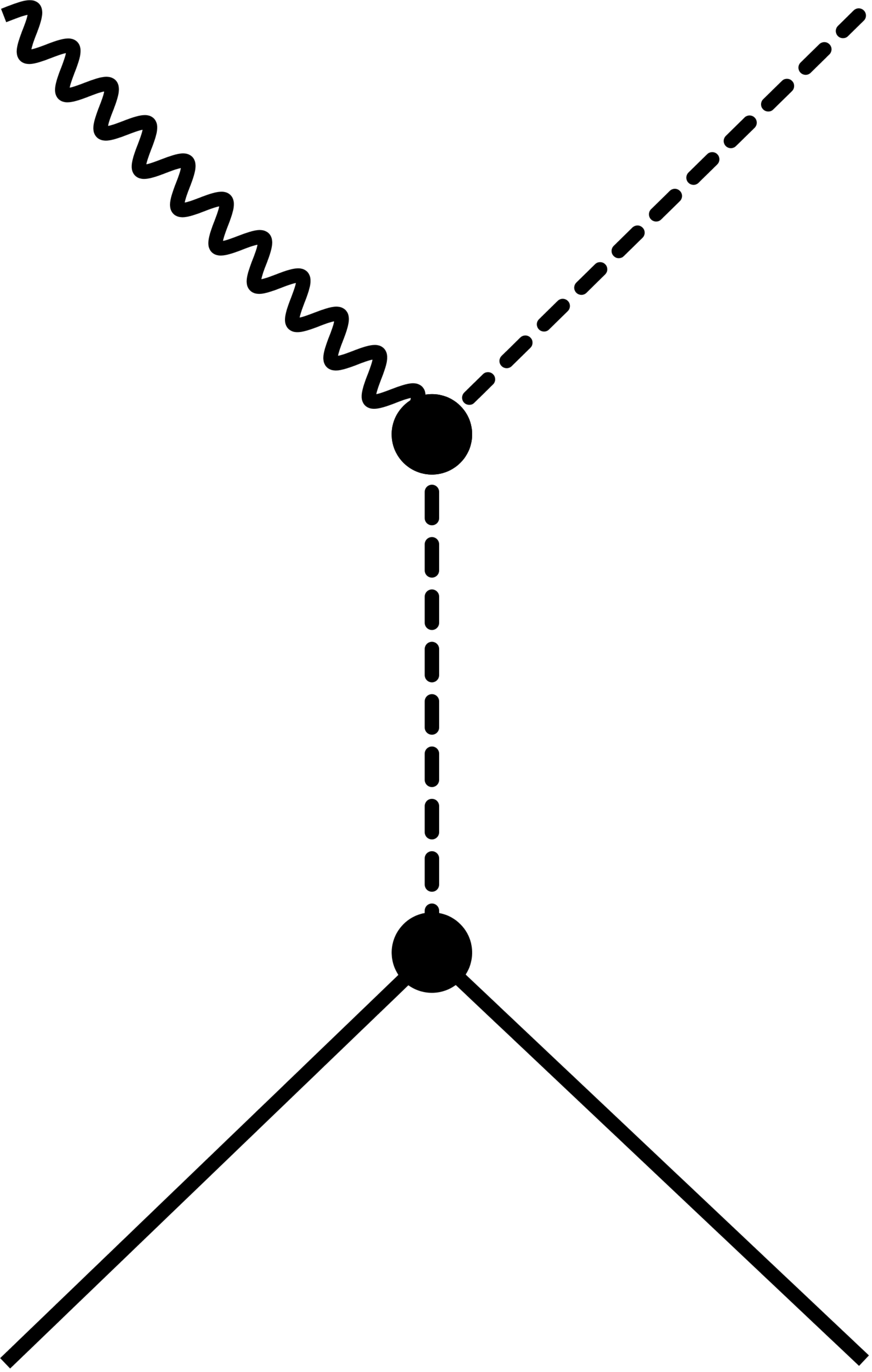}
 \put(9,79){$\gamma$}
 \put(49,79){$\pi^{+}$}
 \put(35,48){$\pi^{+}$}
 \put(9,14.5){$p$}
 \put(50.5,15){$n$}
 \put(65,47.5){\begin{Large}$\implies$ $A_{2}\left(s,t\right) \propto \frac{1}{t - m_{\pi}^{2}} = \frac{1}{2 \left( k q \cos \theta - E_{\gamma} E_{\pi}\right)} $, with:\end{Large}}
 \put(87.5,17.5){\begin{Large}$\frac{1}{t - m_{\pi}^{2}} = \frac{(-)}{4 k q} \sum \limits_{\ell=0}^{\infty} (2 \ell + 1) Q_{\ell} \left( \frac{E_{\gamma} E_{\pi}}{k q} \right) P_{\ell} \left(\cos\theta\right)$\end{Large}}
 \end{overpic}
\vspace*{15pt}
\caption[A pole-contribution in the $t$-channel contributes to an infinite tower of $s$-channel partial waves.]{The diagram shows pion exchange in the $t$-channel (cf. \cite{Grushin}). This process is allowed to contribute to photoproduction of the $n \pi^{+}$ or $p \pi^{-}$ final states. The propagator of the intermediate pion yields a pole contribution to the invariant amplitude $A_{2}(s,t)$. This pole yields generally non-vanishing contributions to an infinite tower of partial waves, since it has the angular variable $\cos \theta$ in the denominator. $Q_{\ell}(y)$ are Legendre functions of the second kind.}
\label{fig:PionPoleInfinityPWaves}
\end{figure}
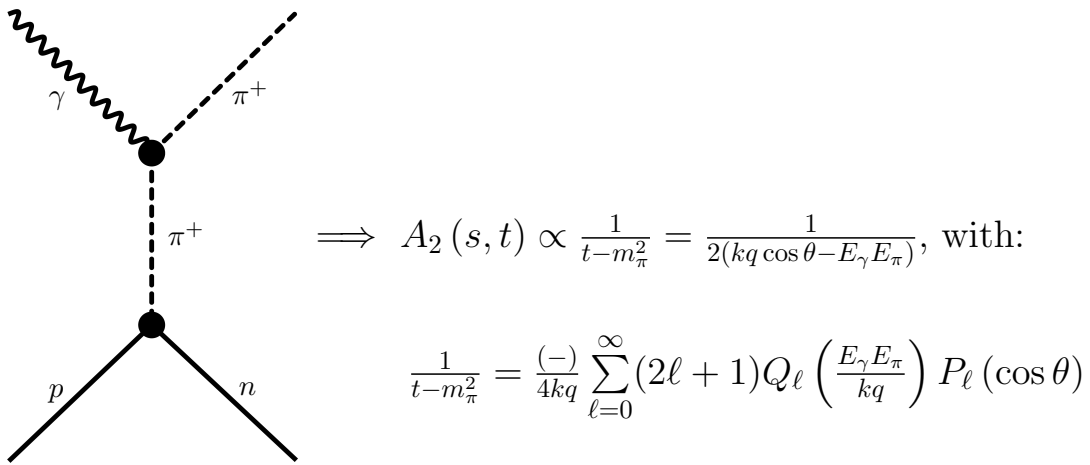

In the whole discussion up to this point, it was assumed that a truncation at some $L = \ell_{\mathrm{max}}$ was exact, i.e. that all higher partial waves with $\ell > L$ are zero and the truncated multipole expansion exactly describes for example the CGLN amplitudes $F_{i}$. Then, it was allowed to assume that an exact, ``true'' solution of the TPWA problem exists. \newline
However, in reality the partial wave series is infinite, caused by certain kinds for Feynman diagrams present in practically all realistic models for photoproduction (\cite{Anisovich:2011fc},\cite{DeborahPhotoProd},\cite{Kamano:2013iva}). An example of such a diagram is the pion pole (cf. e.g. \cite{Grushin}) shown in Figure \ref{fig:PionPoleInfinityPWaves}, which is allowed in the charged channels of pion photoproduction. It yields a pole in $t = m_{\pi}^{2}$ (now the Mandelstam variable $t$) for the resulting full amplitude and therefore is directly seen to contribute to all partial waves. \newline
More generally, all diagrams that feature such a $t$-channel exchange contribute a pole at some value of the Mandelstam variable $t$. More general $t$ channel exchange mechanisms could feature either vector mesons $(\omega, \phi)$ or Reggions. They all generally contribute to all partial waves up to infinity. \newline
Therefore, a truncation of the multipole series at some finite $L$ is itself an approximation with a finite, but possibly small, numerical error. Thereby the assumption that a truncation is exact is generally not correct. \newline \newline \newline \newline

\begin{figure}[hb]
 \centering
\hspace*{5pt}
 \begin{overpic}[width=0.95\textwidth]{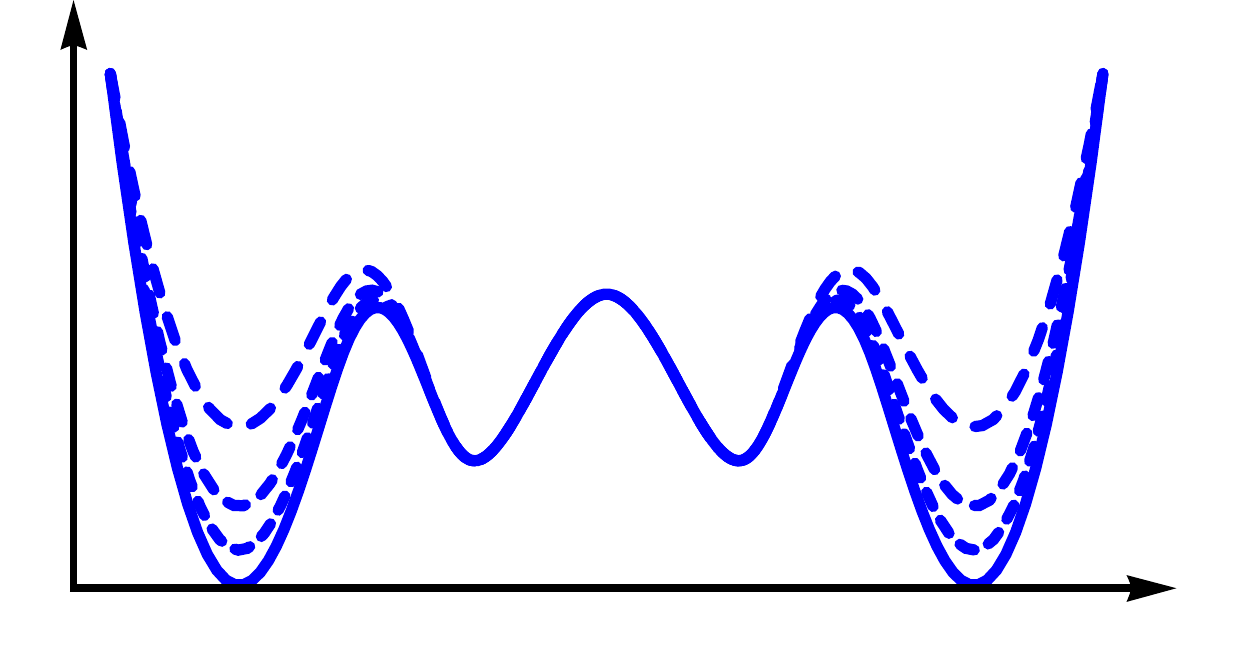}
\put(-3,50.3){\begin{Large}$\Phi$\end{Large}}
\put(90,1.0){\begin{Large}$\mathcal{M}_{\ell}$\end{Large}}
 \end{overpic}
\vspace*{15pt}
\caption[A schematic showing the speculated behavior of a discrepancy function $\Phi$, once the truncation order is raised.]{This picture shows schematically what can be expected to happen to a discrepancy function $\Phi$ corresponding to model data for the group $\mathcal{S}$ with contributions from $L \to \infty$, once the truncation order is raised. The assumed ``true'' solution as well as its double ambiguity become exact zeros of the discrepancy function within the reach of numerical accuracy. For the remaining accidental ambiguities, a finite offset should still exists by way of the arguments made in appendix \ref{subsec:AccidentalAmbProofsII} and the beginning of this appendix section. \newline
What this picture cannot encapsulate is the vast growth of the number of potentially present local minima, which rises exponentially with the truncation angular momentum $L$.}
\label{fig:ChiSquareValleyCartoon2}
\end{figure}

\clearpage
What one however is generally allowed to assume for all physical kinematics, is that the partial wave expansion of e.g. the CGLN amplitudes $F_{i}$ converges to a finite result for the 
full amplitudes $F_{i}^{\infty}$, i.e.
\begin{equation}
 \lim\limits_{L \to \infty}{\sum_{i=1}^{4} \left| F_{i}^{\infty} \left( W, \theta \right) - F_{i}^{\mathrm{trunc.}} \left( W, \theta; L \right) \right|} = 0 \mathrm{.} \label{eq:PWExpConverges}
\end{equation}
The truncation error therefore becomes smaller for larger $L$. \newline
A simple ansatz to restore the existence of in a sense an exact multipole solution would be given by ``overtruncating'' the multipole series in a fit, until the true solution assumed to be provided by nature becomes exact again within the numerical accuracy of the numerical minimization (the idea is illustrated in Figure \ref{fig:ChiSquareValleyCartoon2}). By way of the arguments made above, the probability for the appearance of exact accidental symmetries is not strictly zero anymore, but in case that the numerical accuracy is good enough, it can still be assumed to be quite small. \newline
The only real drawback of this simple idea is that the upper bound of possible accidental ambiguities $N_{\mathrm{AC}} = \frac{1}{2} \left( 2^{4L} - 2 \right)$ (equation (\ref{eq:AccAmbCounting})) rises exponentially with \newline

\clearpage
increasing $L$. The exact solvability (within the region of numerical accuracy) is bought at the price of producing a very large number of candidates for accidental ambiguities. \newline
The pion pole shown in Figure \ref{fig:PionPoleInfinityPWaves} for instance converges quite slowly and can still yield significant contributions for truncation angular momenta as large as $L = 20$. If, based on the considerations made above, we make the rough estimate that only $2 \%$ of all possible accidental symmetries fulfill the consistency relation to within $5^{\circ}$ accuracy, this would still result in roughly $ 0.02 \times \frac{1}{2} \left( 2^{4\times20} - 2 \right) \simeq 1.2 \times 10^{22}$ accidental ambiguities! It is not really clear whether it is still numerically possible to resolve the ``true'' exact minimum out of that many possibilities. In other words, in this case the exponential growth of the candidates for accidental ambiguities can have a catastrophic effect for the solvability. \newline
As a final remark we would like to add that in case a strong pole contribution similar to the example of the pion $t$-channel exchange is present and a good model amplitude for this process is known, one can always just parametrize the $t$-channel pole explicitly in a fit. Fitted are then only those contributions to the multipoles that do not stem from the pole. The multipole series for these remainder parts can then again converge quite quickly (see \cite{Grushin}). \newline
This procedure effectively removes the large amount of ambiguities estimated above. This however comes at the cost of making the TPWA model-dependent. Furthermore, the Omelaenko formalism is not strictly applicable any more once some part of the amplitude is fixed to a rigid model function while the remainder is fitted. \newline

\textbf{Real experimental data are not exactly solvable any more.}  \newline

Now we assume that the scenario described in the previous paragraph, meaning that one has a set of infinitely precise model data which have been generated with all partial waves for $\ell \rightarrow \infty$ while higher partial waves are suppressed, is at hand. Furthermore, we suppose a fit with an "overtruncation" to a complete experiment (e.g. $\left\{\sigma_{0}, \Sigma, T, P, F\right\}$) from this dataset was still capable of yielding the correct physical solution for the multipoles. \newline
The solution can be found for instance by numerical minimization of a least-distance function, which for the purpose of solving ideal model data in a TPWA truncated at $L$ can be defined as (with data given at specific angular points $c_{k_{\alpha}} \equiv \cos \left( \theta_{k_{\alpha}} \right)$, where the $\left\{ k_{\alpha} \right\}$ may be different for each observable)
\allowdisplaybreaks
\begin{align}
\hspace*{-5pt}\Phi \left( \left\{ \mathcal{M}_{\ell} \right\} \right) :=& \sum_{\alpha, c_{k_{\alpha}}} \left[ \check{\Omega}^{\alpha}_{\mathrm{Data}} \left(c_{k_{\alpha}}\right) - \check{\Omega}^{\alpha}_{\mathrm{Fit}} \left(c_{k_{\alpha}}, \left\{ \mathcal{M}_{\ell} \right\}\right) \right]^{2} \nonumber \\
 =& \sum_{\alpha, c_{k_{\alpha}}} \left[ \check{\Omega}^{\alpha}_{\mathrm{Data}} \left(c_{k_{\alpha}}\right) - \rho \sum_{n = \beta_{\alpha}}^{2 L + \beta_{\alpha} + \gamma_{\alpha}} \left< \mathcal{M}_{\ell} \right| \left( \mathcal{C}_{L}\right)_{n}^{\check{\Omega}^{\alpha}}  \left| \mathcal{M}_{\ell} \right> \hspace*{2pt} P^{\beta_{\alpha}}_{n} \left(  c_{k_{\alpha}} \right) \right]^{2} \mathrm{.} \label{eq:DefDiscrFunctII}
\end{align}
In the second step we made the form of the model function in a TPWA explicit, which was also detailed in section \ref{sec:CompExpsTPWA} and appendix \ref{sec:TPWAFormulae}. We note that the quantity (\ref{eq:DefDiscrFunctII}) does not yet have a statistical interpretation. Furthermore, the assumption that the found set of real- and imaginary parts of phase-constrained multipoles $\left\{\mathcal{M}_{\ell}\right\}$ is the true solution of the given ideal model data, is equivalent to the statement that the function $\Phi \left( \left\{ \mathcal{M}_{\ell} \right\} \right)$ vanishes with a sufficient numerical accuracy (for instance $10^{-20}$). We assume that even in this case, with an overtruncated TPWA-fit and model data generated with an infinite tower of partial waves, the convergence of the partial wave expansion still causes the exact solvability using just five observables. Therefore we regard the accidental ambiguities as basically un-important even in this case. The question is now at which point the "exact" solvability is lost once one considers data taken in the real world. \newline

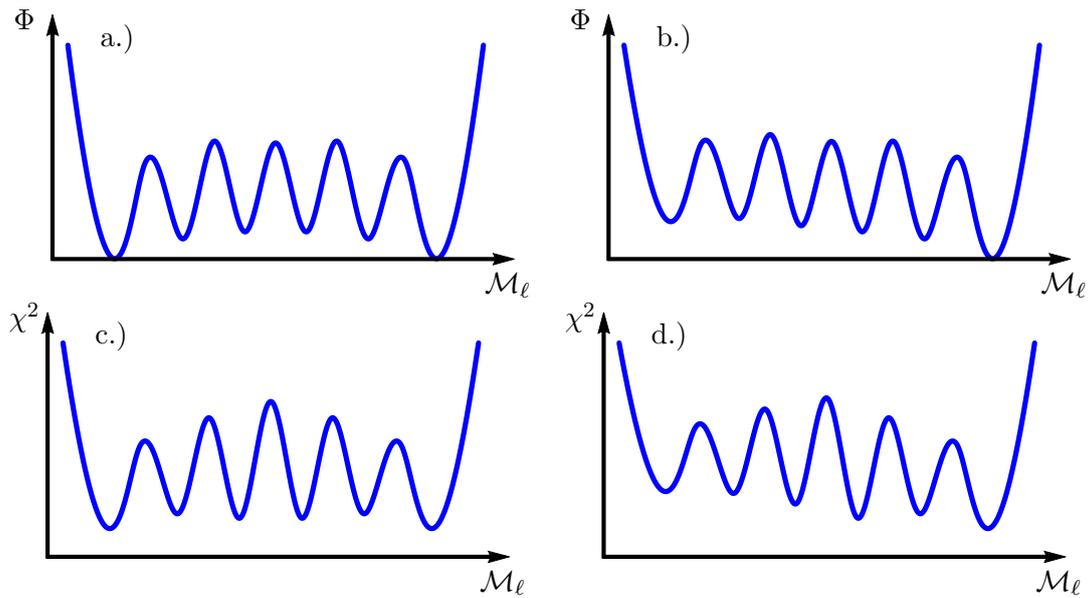
\begin{figure}[ht]
 \centering
\begin{overpic}[width=0.45\textwidth]{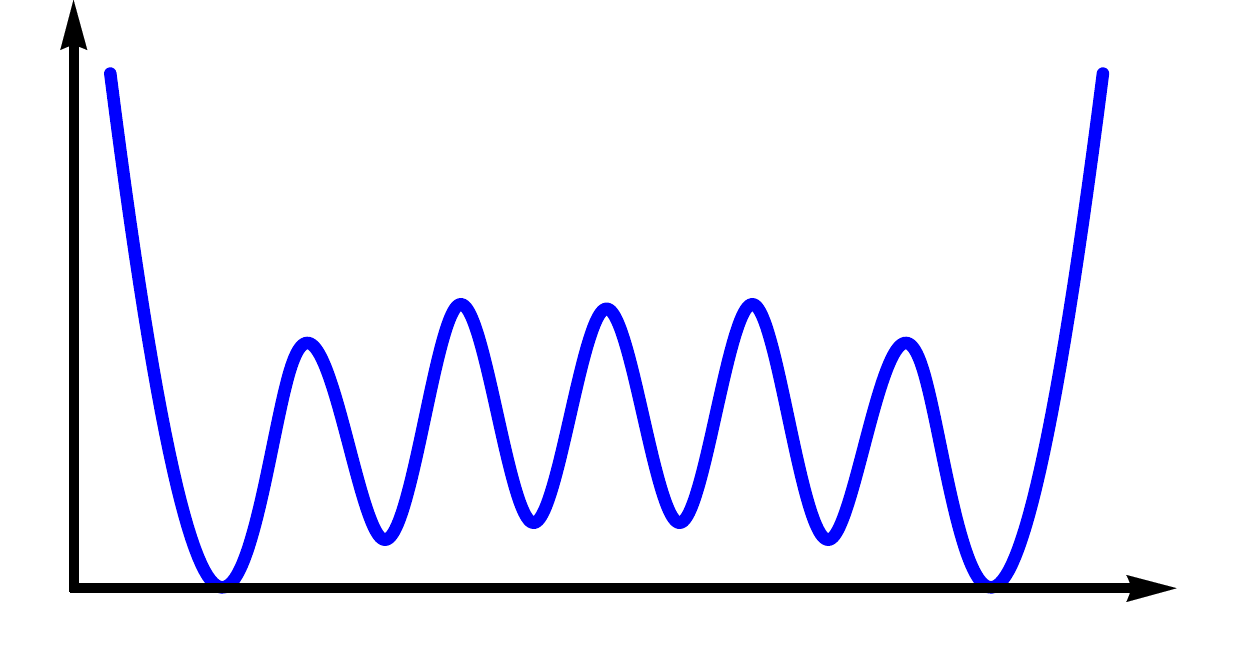}
\put(-1.5,50.3){$\Phi$}
\put(88.5,-0.5){$\mathcal{M}_{\ell}$}
\put(15,47){a.)}
 \end{overpic} \hspace*{5pt}
\begin{overpic}[width=0.45\textwidth]{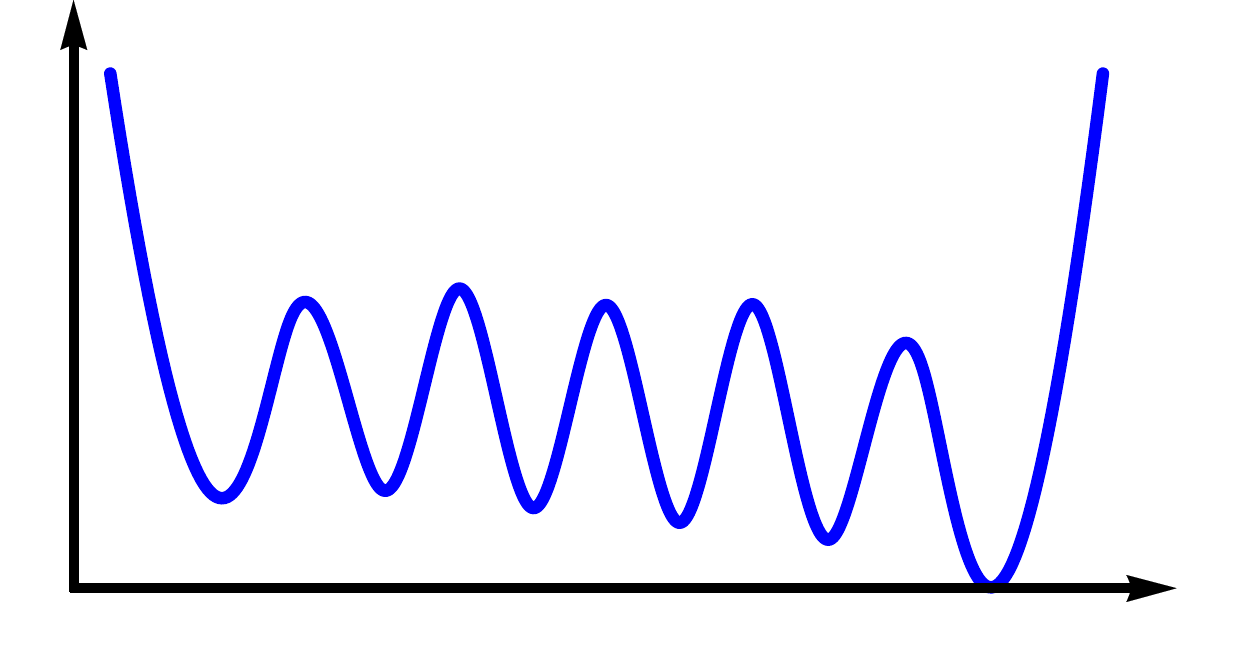}
\put(-1.5,50.3){$\Phi$}
\put(88.5,-0.5){$\mathcal{M}_{\ell}$}
\put(15,47){b.)}
\end{overpic} \\
\vspace*{7.5pt}
\begin{overpic}[width=0.45\textwidth]{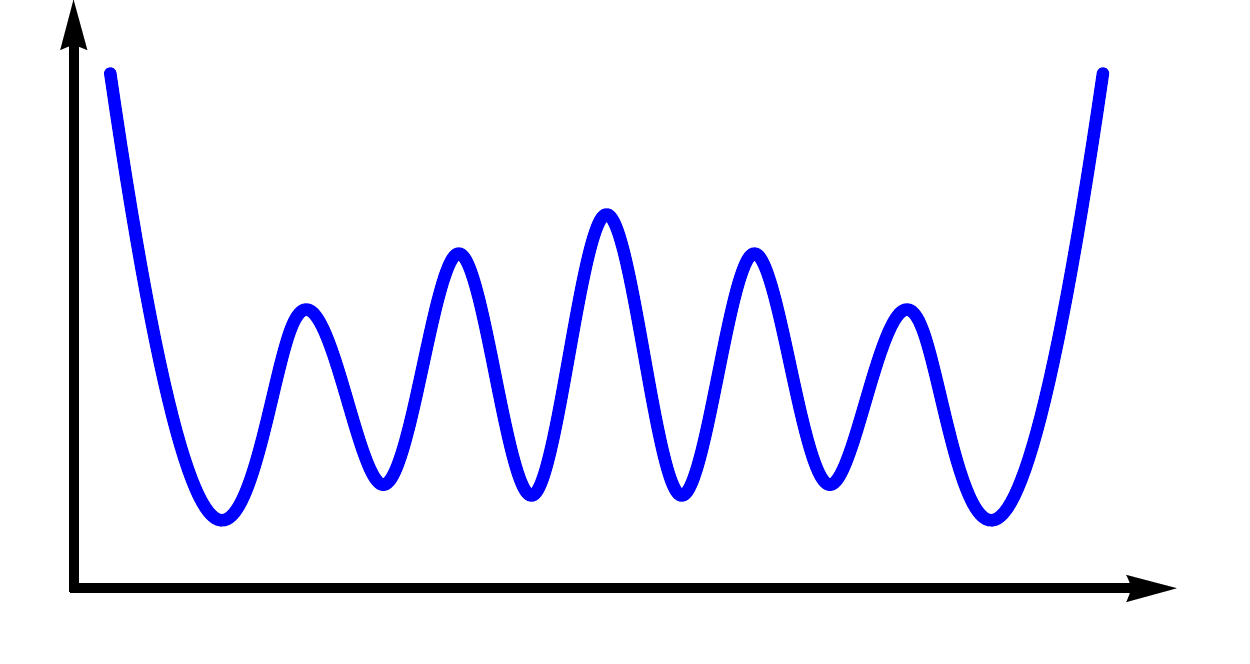}
\put(-1.5,50.3){$\chi^{2}$}
\put(88.5,-0.5){$\mathcal{M}_{\ell}$}
\put(15,47){c.)}
 \end{overpic} \hspace*{5pt}
\begin{overpic}[width=0.45\textwidth]{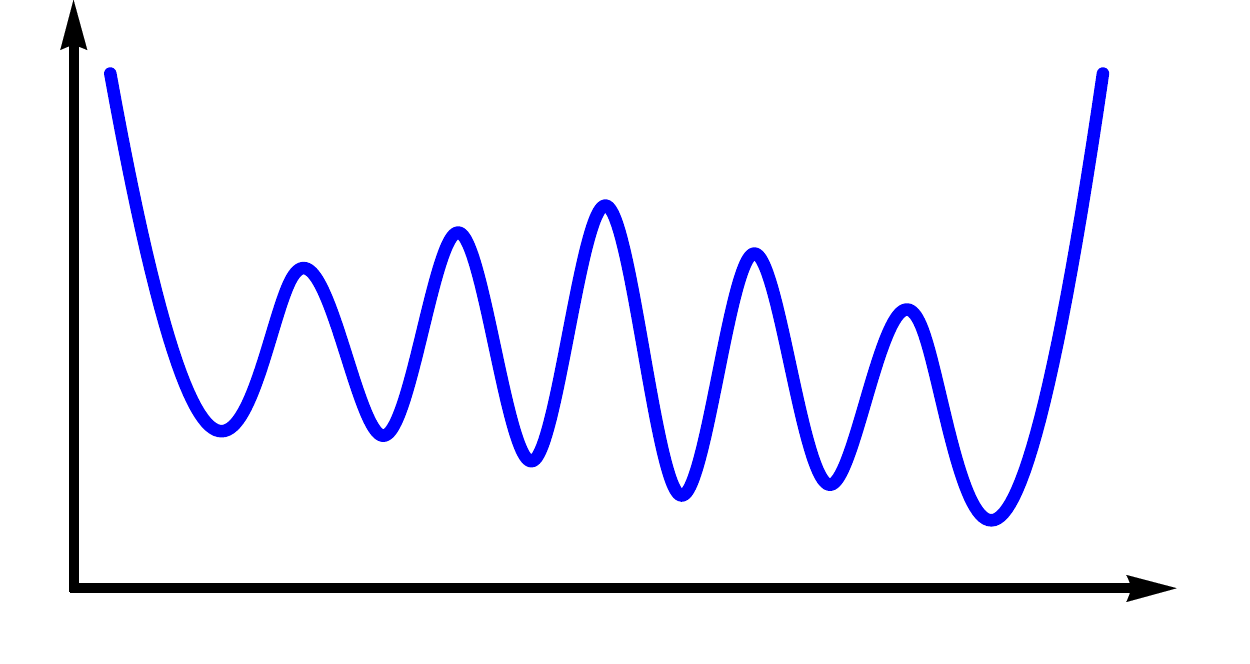}
\put(-1.5,50.3){$\chi^{2}$}
\put(88.5,-0.5){$\mathcal{M}_{\ell}$}
\put(15,47){d.)}
 \end{overpic}
\vspace*{7.5pt}
\caption[Illustration of the effect of compatibility/incompatibility of the equation system (\ref{eq:BilinearEqSystem}) on the minima-structure of the discrepancy function $\Phi_{\mathcal{M}}$.]{The four diagrams illustrate the effect of compatibility/incompatibility of the equation system (\ref{eq:BilinearEqSystem}) on the minima-structure of the minimized function $\Phi$ (equation (\ref{eq:DefDiscrFunctII})) in case exact model data are solved, or the $\chi^{2}$-function (\ref{eq:DefChi2Funct}) for a fit of real data. The figures correspond to the cases: a.) compatible equations $\&$ incomplete experiment (e.g. just group $\mathcal{S}$ observables are solved); b.) compatible equations $\&$ complete experiment; c.) incompatible equations $\&$ incomplete experiment; d.) incompatible equations $\&$ (possibly over-) complete experiment. \newline
Similar pictures have been given by Grushin \cite{Grushin} and also been reprinted in reference \cite{MyDiplomaThesis}.}
\label{fig:CompletenessChi2Cases}
\end{figure}

It is a basic fact of life that every measurement comes with an experimental uncertainty, a statistical and a systematic one. For the moment systematics shall not be regarded. They are expected to only make the below mentioned matters even worse. The existence of statistical uncertainties alone renders the notion of an "exact" solution of a TPWA not applicable any more. One does not look for a mathematical solution of the TPWA, but instead tests a specific truncation statistically by for example minimizing a $\chi^{2}$-function, which for the assumption of uncorrelated data points is only a slight modifcation of equation (\ref{eq:DefDiscrFunctII})
\begin{align}
\hspace*{-5pt}\chi^{2} \left( \left\{ \mathcal{M}_{\ell} \right\} \right) &= \sum_{\alpha, c_{k_{\alpha}}} \left[ \frac{ \check{\Omega}^{\alpha}_{\mathrm{Data}} \left(c_{k_{\alpha}}\right) - \rho \sum_{n = \beta_{\alpha}}^{2 L + \beta_{\alpha} + \gamma_{\alpha}} \left< \mathcal{M}_{\ell} \right| \left( \mathcal{C}_{L}\right)_{n}^{\check{\Omega}^{\alpha}}  \left| \mathcal{M}_{\ell} \right> \hspace*{2pt} P^{\beta_{\alpha}}_{n} \left(  c_{k_{\alpha}} \right)}{\Delta \check{\Omega}^{\alpha}_{\mathrm{Data}} \left(c_{k_{\alpha}}\right) } \right]^{2} \mathrm{.} \label{eq:DefChi2Funct}
\end{align}
Each difference is normalized with respect to the normal-distributed statistical errors $\Delta \check{\Omega}^{\alpha}_{\mathrm{Data}} \left(c_{k_{\alpha}}\right)$ of each data point. The minimization of (\ref{eq:DefChi2Funct}), or fit, now yields an estimate for the parameters $\left\{\mathcal{M}_{\ell}\right\}$ which is considered statistically acceptable if $\chi^{2}$ is equal or close to the number of degrees of freedom $\mathrm{ndf}$ of the fit ($\mathrm{ndf} = N_\mathrm{datapoints} -  N_\mathrm{parameters}$). Therefore the function (\ref{eq:DefChi2Funct}) is non-vanishing even in case a good estimate for the correct, physical multipole-solution is found. \newline
In the language of Grushin \cite{Grushin}, the system of bilinear hermitean forms which enters the numerical minimization in both cases of either (\ref{eq:DefDiscrFunctII}) or (\ref{eq:DefChi2Funct}), i.e. that of Legendre-coefficients expressed in terms of multipole-parameters (see section \ref{sec:CompExpsTPWA} and appendix \ref{sec:TPWAFormulae})
\begin{equation}
\left(a_{L}\right)_{k}^{\check{\Omega}^{\alpha}} = \left< \mathcal{M}_{\ell} \right| \left( \mathcal{C}_{L}\right)_{k}^{\check{\Omega}^{\alpha}} \left| \mathcal{M}_{\ell} \right> \mathrm{,} \label{eq:BilinearEqSystem}
\end{equation}
is not "compatible" any more in case of the fit of real data. Compatibility means that a set of Legendre coefficients $\left(a_{L}\right)_{k}^{\check{\Omega}^{\alpha}}$ is extracted from the data, which constitutes a numerical configuration on the left-hand-side of the equations (\ref{eq:BilinearEqSystem}) such that at least one exact solution for the right-hand-side exists. This assumption was always made in the study of the theoretical discrete ambiguities of the system (\ref{eq:BilinearEqSystem}) described in chapter \ref{chap:Omelaenko} of this work. \newline
However, for a fit of real data with uncertainties, compatibility cannot be assumed any more. Therefore, the experimental uncertainties add an additional layer of complexity on the already, depending on the order $L$, complicated ambiguity structure caused by the non-linearity of the equations (\ref{eq:BilinearEqSystem}). \newline
The completeness or non-completeness of sets of observables both in the academic case of exact data and compatible equation-systems, as well as the fit of real data with generally incompatible systems, is represented in a schematic way in Figure \ref{fig:CompletenessChi2Cases}. \newline \newline
\clearpage
The Figure \ref{fig:CompletenessChi2Cases}.a.) shows the case where an incomplete experiment which however only has discrete ambiguities (like for example the group $\mathcal{S}$ $\left\{\sigma_{0}, \Sigma, T, P\right\}$) is solved in a situation without measurement uncertainties. Two exact solutions exist connected by the double ambiguity transformation (see appendix \ref{sec:DoubleAmbiguityTrafoActingOnBi}), while several accidental ambiguities are also shown which however, according to the arguments made in appendix \ref{subsec:AccidentalAmbProofsII}, should always have a finite offset in $\Phi$ compared to the two exact solutions. In Figure \ref{fig:CompletenessChi2Cases}.b.), the double ambiguity of the true solution, as well as all degeneracies coming from accidental symmetries that are connected via the double ambiguity transformation, have been resolved. In a mathematically exact situation, this could have been achieved by including for instance the $F$- or $G$ observable. \newline
Figure \ref{fig:CompletenessChi2Cases}.c.) depicts the situation for a fit of real data, in case one again has an incomplete set of four observables that admits only discrete ambiguities. A global deepest minimum can be assumed to exist, but it has an exact mathematical double ambiguity. The best minimum is in any case not strictly zero. Further local minima, which are likely to be related to the accidental symmetries, are also indicated and they again occur in degenerate pairs caused by the double ambiguity transformation. In Figure \ref{fig:CompletenessChi2Cases}.d.) finally, the situation is shown where the inclusion of at least one, or possible even more, polarization observables caused the $\chi^{2}$-minimization to adopt a unique global minimum. \newpage
The fact that the $\chi^{2}$ in a fit does not vanish strictly greatly complicates the situation, since one now cannot distinguish the accidental ambiguities any more according to the criteria detailed in appendix \ref{subsec:AccidentalAmbProofsII} as well as the first paragraph of this appendix section. This means that, depending on the truncation order $L$, an exponentially growing number of local minima comes into play which may be related to the accidental ambiguities of the equation system (\ref{eq:BilinearEqSystem}) (in this context, cf. the numbers printed in Table \ref{tab:AccAmbPossibilityNumber}). \newline
The question emerges here about whether or not the accidental symmetries derived under the assumption of an exactly solvable TPWA still exist in case a fit of data is done. It has to be said that even when the left and right hand side of the equation system (\ref{eq:BilinearEqSystem}), or equivalently of the group $\mathcal{S}$ observables written in a linear factor decomposition of the transversity amplitudes $b_{i}$ ((\ref{eq:b2LinFactDecompSecAccAmb}) to (\ref{eq:b1b3LinFactDecompositionsSecAccAmb})), cannot match for any configuration of the varied parameters (either multipoles $\left\{\mathcal{M}_{\ell}\right\}$ or Omelaenko-roots $\left\{\alpha_{k},\beta_{k}\right\}$), it is still possible to perform symmetry transformations on the parameters and therewith leave at least the right hand side invariant. Therefore, while the compatibility was an important assumption in order to discriminate among accidental symmetries and exact solutions, for the mere existence of accidental ambiguities in a fit, it does not need to be assumed. In other words, the symmetries of the equation system (\ref{eq:BilinearEqSystem}) do not care about compatibility and are present in both cases. The non-linear nature of the equation systems present in a TPWA causes the appearance of an exponentially growing number of roots that can be resolved in the academic case of an exactly solvable analysis, but can cause severe problems once a fit of real data is done. \newline
Therefore, in case a mathematically complete set of observables with large errors is fitted in a truncation order $L > 1$, it cannot be assumed that the global minimum of
the $\chi^{2}$-function necessarily yields the correct physical solution. It can very well happen that a minimum corresponding to an accidental symmetry in this case yields the smallest minimum value of $\chi^{2}$. It is seen to be necessary in a TPWA-fit to investigate the global minimum as well as all equally good local minima, e.g. in a range of $\chi^{2}_{\mathrm{min}}/\mathrm{ndf} = \chi^{2}_{\mathrm{glob.}}/\mathrm{ndf} + 1$. Numerically demanding techniques have to be employed in order to be able to find all those minima (see chapter \ref{chap:TPWA}). \newline
There are in principle three ways to improve the situation in case a TPWA-fit of a mathematically complete set comprised of $5$ observables does not yield a well-separated global minimum. These are
\begin{itemize}
 \item[(i)] \underline{The increase of the precision of the already measured $5$ observables:} \newline
 Smaller statistical errors should make the minima of $\chi^{2}$ less shallow and more pronounced. One can even anticipate the dissappearance of some ambiguous solutions. In any case, the fit should become more stabe by increased measurement-precision, in that it becomes easier to numerically map out the minima of $\chi^{2}$. \newline
 However, one cannot assume that the compatibility of (\ref{eq:BilinearEqSystem}) is restored by more precise data. The systematic errors in the data alone will probably still maintain the incompatibility. Furthermore, more precise data may be expected to require higher multipoles to be varied in the fit and as described in previous appendix sections, this causes an exponential rise of the number of possible ambiguities. \newline \newline
 \item[(ii)] \underline{Measurement of additional observables:} \newline
 Another Ansatz for the stabilization of TPWA-fits is to enlarge the mathematically complete sets of $5$ observables postulated in this work by further polarization observables, therefore generating mathematically over-complete experiments. More observables $\check{\Omega}^{\alpha}$ simply generate more Legendre-coefficients $\left(a_{L}\right)^{\alpha}_{k}$ from their angular distributions and therewith also more numbers that are capable of constraining the unknowns $\left\{\mathcal{M}_{\ell}\right\}$. For a detailed description about how many Legendre-coefficients are facilitated by each observable in a truncation order $L = \ell_{\mathrm{max}}$, see the Table \ref{tab:NumberOfLegendreCoefficients} in section \ref{sec:CompExpsTPWA}. \newline
 An advantage is here that one can first pick additional quantities from the class $\mathcal{BT}$ and therefore avoid the less accessible observables of type $\mathcal{BR}$ and $\mathcal{TR}$. However, quantities from the latter two classes yield even more Legendre-coefficients than the beam-target observables (cf. Table \ref{tab:NumberOfLegendreCoefficients}).
 \item[(iii)] \underline{Demand correct analyticity-constraints for the multipoles:} \newline
 Another idea of an additional constraint that does not yet need a model is to fit the multipoles in such a way that their analyticity-properties are respected, regarding them as functions of a complex energy-variable. This constraint is closely related to the fundamental principle of (micro)-causality \cite{PolkinghorneEtAl}. The Mainz-Tuzla collaboration \cite{MainzTuzlaCollaboration} is currently doing work in this direction.
 \item[(iv)] \underline{Gradually introducing model-dependence into the TPWA:} \newline
 If the procedures (i), (ii) or even (iii) did not yield the desired result of a unique best estimate for the multipoles, the only remaining way to still obtain unique parameters is to
 increase the model-dependence. The prerequisite for this of course is the existence of an energy-dependent model-solution for the pho\-to\-pro\-duction-channel under consideration. \newline
 If such a model is at hand, yielding parameters $\mathcal{M}_{\ell}^{\mathrm{mod.}}$ for the multipoles, it is possible to gradually introduce a bias towards this model until a unique best estimate for the multipoles emerges. \newline
A first way for introducing somewhat weak model constraints is to fix the higher partial waves, which in the energy region of consideration are not assumed to contain any resonance-contributions and are known from the model to be small, to the model-values and not vary them any more in the fit. All the remaining multipoles are still running freely in the fit and are not subjected to any constraint. \newline
One way to strengthen the model constraints is to use so-called penalty-term fitting, as outlined in reference \cite{SchumannEtAl}. Here, either all varied multipoles or a subset of them is bound to the energy-dependent solution by modifying the $\chi^{2}$-function of the TPWA-fit. This is done by adding penalty-terms to the function (\ref{eq:DefChi2Funct}). Denoting the latter function now as $\chi^{2}_{\mathrm{Data}}$, the modified definition reads (defined in a slightly different way compared to reference \cite{SchumannEtAl})
\allowdisplaybreaks
\begin{align}
 \chi^{2} \left( \left\{ \mathcal{M}_{\ell} \right\} \right) &= \chi^{2}_{\mathrm{Data}} \left( \left\{ \mathcal{M}_{\ell} \right\} \right) + \sum_{i=1}^{4L} \lambda_{i} \Bigg[ \left( \mathrm{Re} \mathcal{M}_{\ell}^{i} - \mathrm{Re} \left[\mathcal{M}_{\ell}^{\mathrm{mod.}}\right]^{i} \right)^{2} \nonumber \\ 
 & \hspace*{120pt} + \left( \mathrm{Im} \mathcal{M}_{\ell}^{i} - \mathrm{Im} \left[\mathcal{M}_{\ell}^{\mathrm{mod.}}\right]^{i} \right)^{2} \Bigg] \mathrm{,} \label{eq:Chi2WithPenaltyTerm}
\end{align}
with an individual penalty-term for each multipole, multiplied by a parameter $\lambda_{i}$ that determines the strength of each constraint. The $\lambda_{i}$ are in principle free and tuneable parameters. One criterion to restrict this freedom is to choose them in such a way that the minimum of (\ref{eq:Chi2WithPenaltyTerm}) is not too different from that of the original $\chi^{2}_{\mathrm{Data}}$ (see \cite{SchumannEtAl}, where $1 < \chi^{2} / \chi^{2}_{\mathrm{Data}} < 1.05$ is proposed). Though we mention the method of penalty-term fitting here for the sake of completeness, it was never employed in the course of this work. The above mentioned two methods for the introduction of a model-bias are by no means the only one's available. \newline
The disadvantage is now of course that the TPWA is not fully model-in\-de\-pen\-dent any more. But still, also in this model-dependent guise a single-energy fit may be capable to map out structures in the multipoles $\mathcal{M}_{\ell}$ that originate from the data, but are not encapsulated by the energy-dependent model due to overly restrictive constraints.
\end{itemize}
The only case where point (i), i.e. having $5$ observables precise enough to yield a unique solution for the multipoles, worked out was in an analysis of $\pi^{0}$-pho\-to\-pro\-duc\-tion data in the $\Delta$-resonance region. Details on the results, and even studies on the influence of the precision of certain observables, are given in section \ref{subsec:DeltaRegionDataFits}. \newline
A combination of (ii) and (iv), meaning the fit of an overcomplete set with weak model constraints, is outlined in section \ref{subsec:2ndResRegionDataFits}. There, a fit of $7$ polarization observables measured for the process $\gamma p \rightarrow \pi^{0} p$ in the $2^{\mathrm{nd}}$ resonance region was performed.

\clearpage

\paragraph{Evaluation of the projection integrals in the higher dimensional cube of Omelaenko phases} \label{sec:ProjectionIntegrals}  \textcolor{white}{:-)} \newline

We start out with the integral expression for the full $(4L-1)$-dimensional sub-volume of the cube $\left[- \pi, \pi\right]^{4L}$, where the consistency relation is satisfied exactly. This integral was declared in equation (\ref{eq:CRVolumeMostGeneralEvaluation}) of appendix \ref{subsec:AccidentalAmbProofsIII} and it reads:
{\allowdisplaybreaks
\begin{align}
\mathcal{V}^{(4L-1)}_{\mathcal{CR}} &= 2 \sqrt{L} \int_{-\pi}^{\pi} d \varphi_{1} \ldots \int_{-\pi}^{\pi} d \psi_{2L} \hspace*{2.5pt} \delta \left( \sum_{k=1}^{2L} \varphi_{k} - \sum_{k^{\prime}=1}^{2L} \psi_{k^{\prime}} \right) \mathrm{.} \label{eq:CRVolumeMostGeneralEvaluationStep1}
\end{align}
}
\noindent
The correct Ansatz \cite{MishaPrivComm} in order to obtain a closed expression for this integral consists of using the Fourier-representation of the Dirac $\delta$-function
\begin{align}
\delta \left( \sum_{k=1}^{2L} \varphi_{k} - \sum_{k^{\prime}=1}^{2L} \psi_{k^{\prime}} \right) &=  \int_{- \infty}^{\infty} \frac{d p}{2 \pi} e^{i p \left( \sum_{k} \varphi_{k} - \sum_{k^{\prime}} \psi_{k^{\prime}} \right)} \nonumber \\
 &= \int_{- \infty}^{\infty} \frac{d p}{2 \pi} \prod_{k = 1}^{2L} e^{i p \varphi_{k}} \prod_{k^{\prime} = 1}^{2L} e^{- i p \psi_{k^{\prime}}} \mathrm{.} \label{eq:DeltaFctFourierRep}
\end{align}
Upon inserting this expression into (\ref{eq:CRVolumeMostGeneralEvaluationStep1}) and commuting the $p$-integration with the $\varphi$- and $\psi$-integrations, one obtains:
\begin{equation}
 \mathcal{V}^{(4L-1)}_{\mathcal{CR}} = \frac{\sqrt{L}}{\pi} \int_{- \infty}^{\infty} d p  \prod_{k = 1}^{2L} \int_{-\pi}^{\pi} d \varphi_{k} e^{i p \varphi_{k}} \times \prod_{k^{\prime} = 1}^{2L} \int_{-\pi}^{\pi} d \psi_{k^{\prime}} e^{- i p \psi_{k^{\prime}}}  \mathrm{.} \label{eq:CRVolumeMostGeneralEvaluationStep2}
\end{equation}
The $\varphi$-integrations yield
\begin{equation}
 \int_{-\pi}^{\pi} d \varphi_{k} e^{i p \varphi_{k}} = \frac{(-i)}{p} \left( e^{i p \pi} - e^{- i p \pi} \right) \mathrm{,} \label{eq:VarPhiIntegration}
\end{equation}
and it is furthermore quite important to note that the $\psi$-integrations yield exactly the same formula, although there the exponential in the integrand has a different sign:
\begin{equation}
 \int_{-\pi}^{\pi} d \psi_{k^{\prime}} e^{- i p \psi_{k^{\prime}}} = \frac{i}{p} \left( e^{- i p \pi} - e^{ i p \pi} \right) = \frac{(-i)}{p} \left( e^{ i p \pi} - e^{- i p \pi} \right) \mathrm{.} \label{eq:PsiIntegration}
\end{equation}
From this one can see that in fact the integral (\ref{eq:CRVolumeMostGeneralEvaluationStep1}) is the same for the consistency relation of the true solution, i.e. $\sum_{k} \varphi_{k} - \sum_{k^{\prime}} \psi_{k^{\prime}} = 0$, as well as for every other Omelaenko ambiguity $\bm{\uppi} \in \hat{\mathcal{P}}$ where one would have $\sum_{k} \bm{\uppi} \left(\varphi_{k}\right) - \sum_{k^{\prime}} \bm{\uppi}\left(\psi_{k^{\prime}}\right) = 0$ (cf. the definitions in appendix \ref{subsec:AccidentalAmbProofsI}). \newline
Utilizing (\ref{eq:VarPhiIntegration}) and (\ref{eq:PsiIntegration}), the integral (\ref{eq:CRVolumeMostGeneralEvaluationStep2}) can be further reduced as
\begin{align}
 \mathcal{V}^{(4L-1)}_{\mathcal{CR}} &= \frac{\sqrt{L}}{\pi} \left( - i \right)^{4L} \int_{- \infty}^{\infty} d p \left( \frac{e^{ i p \pi} - e^{- i p \pi}}{p} \right)^{4L} \nonumber \\
 &= \sqrt{L} \pi^{4L - 2}  \int_{- \infty}^{\infty} d x \left( \frac{e^{ i x} - e^{- i x}}{x} \right)^{4L} \mathrm{,} \label{eq:CRVolumeMostGeneralEvaluationStep3}
\end{align}
where in the last step, the variable $x := \pi p$ was introduced. For the evaluation of the remaining $x$-integration, the general binomial expansion
\begin{equation}
 \left( x + y \right)^{n} = \sum_{k = 0}^{n} \binom{n}{k} x^{n - k} y^{k} = \sum_{k = 0}^{n} \binom{n}{k} x^{k} y^{n - k} \label{eq:BinExpansionAppendix}
\end{equation}
with binomial coefficients
\begin{equation}
\binom{n}{k} = \frac{n!}{k! (n-k)!} \mathrm{,} \label{eq:BinCoeffAppendices}
\end{equation}
has to be invoked. Applying this expansion to the numerator of the integrand, we obtain
\begin{align}
 \left( e^{i x} - e^{- i x} \right)^{4L} &= \left( e^{i x} + \left[ - e^{- i x} \right] \right)^{4L} = \sum_{k = 0}^{4L} (-)^{k} \binom{4L}{k} \left[ e^{i x} \right]^{4L - k} \left[ e^{-i x}  \right]^{k} \nonumber \\
 &= \sum_{k = 0}^{4L} (-)^{k} \binom{4L}{k} e^{i \left( 4L - 2k \right) x} \mathrm{.} \label{eq:ExpTermBinomialExpansion}
\end{align}
In case of convergence, the full integral can be decomposed upon shifting the pole by replacing $x \rightarrow (x - i \delta)$ $(\delta \ll 1)$ in the denominator, as well as using (\ref{eq:ExpTermBinomialExpansion}). The result becomes:
\begin{equation}
 \mathcal{V}^{(4L-1)}_{\mathcal{CR}} = \sqrt{L} \pi^{4L - 2} \sum_{k = 0}^{4L} (-)^{k} \binom{4L}{k} \int_{- \infty}^{\infty} d x \frac{e^{i \left( 4L - 2k \right) x}}{\left(x - i \delta\right)^{4L}} \mathrm{.} \label{eq:CRVolumeMostGeneralEvaluationStep4}
\end{equation}
The remaining integral under the sum can be evaluated using the residue theorem. However, it is important to note that due to the fact that the pole was shifted to the upper half of the complex $x$-plane, the only integrals that are non-vanishing fulfill the requirement $\left( 4L - 2k \right) > 0$, or $k < 2L$ equivalently. For all terms not fulfilling this constraint, the integration contour used for the integral has to have a giant semi-arc in the lower half of the complex plane (since there, $e^{i \left( 4L - 2k \right) x} = e^{- i \left| 4L - 2k \right| x}$ approaches zero for large negative $\mathrm{Im}\left[x\right]$). Since the pole is not situated in the lower half of the comple plane, the latter integrals (i.e. those for $k \geq 2L$) are all zero. Therefore, our integral can be expressed as:
\begin{equation}
 \mathcal{V}^{(4L-1)}_{\mathcal{CR}} = \sqrt{L} \pi^{4L - 2} \sum_{k = 0}^{2L-1} (-)^{k} \binom{4L}{k} \left( 2 \pi i \right) \mathrm{Res}_{i \delta} \left[ \frac{e^{i \left( 4L - 2k \right) x}}{\left(x - i \delta\right)^{4L}} \right] \mathrm{.} \label{eq:CRVolumeMostGeneralEvaluationStep5}
\end{equation}
The order of the pole at $i \delta$ is $4L$. Therefore, one has to utilize the general expression for the residue of a function $f(z)$ in an $n$-th order pole located at some $z_{0} \in \mathbbm{C}$
\begin{equation}
 \mathrm{Res}_{z_{0}} \left[ f (z) \right] = \frac{1}{(n-1)!} \lim\limits_{z \rightarrow z_{0}}{\frac{d^{n-1}}{d z^{n-1}} \left[ \left( z - z_{0} \right)^{n} f(z) \right]} \mathrm{.} \label{eq:GeneralResidueFormula}
\end{equation}
Therefore, it is seen that the integral in (\ref{eq:CRVolumeMostGeneralEvaluationStep4}) becomes
\begin{align}
 2 \pi i \hspace*{2pt} \mathrm{Res}_{i \delta} \left[ \frac{e^{i \left( 4L - 2k \right) x}}{\left(x - i \delta\right)^{4L}} \right] &= \frac{2 \pi i}{(4L-1)!} \lim\limits_{x \rightarrow i \delta}{\frac{d^{4L-1}}{d x^{4L-1}} \left[ e^{i \left( 4L - 2k \right) x} \right]} \nonumber \\
 &= \frac{2 \pi}{(4L - 1)!} \left[ 4L - 2k \right]^{(4L-1)} e^{- \left[ 4L - 2k \right] \delta} \mathrm{.} \label{eq:RelevantResidueInt1}
\end{align}
Inserting this expression into (\ref{eq:CRVolumeMostGeneralEvaluationStep5}) and taking the limit $\delta \rightarrow 0$ yields the result for the general surface integral (\ref{eq:CRVolumeMostGeneralEvaluationStep1}):
\begin{equation}
 \mathcal{V}^{(4L-1)}_{\mathcal{CR}} = 2 \sqrt{L} \frac{\pi^{4L - 1}}{\left(4L-1\right)!} \sum_{k = 0}^{\left( 2 L - 1 \right)} (-)^{k} \binom{4L}{k} \left(4L-2k\right)^{\left(4L-1\right)} \mathrm{.} \label{eq:CRVolumeMostGeneralEvaluationFinalStep}
\end{equation}
\begin{figure}[ht]
\centering
 \hspace*{0.65pt} \begin{overpic}[width=0.30\textwidth]{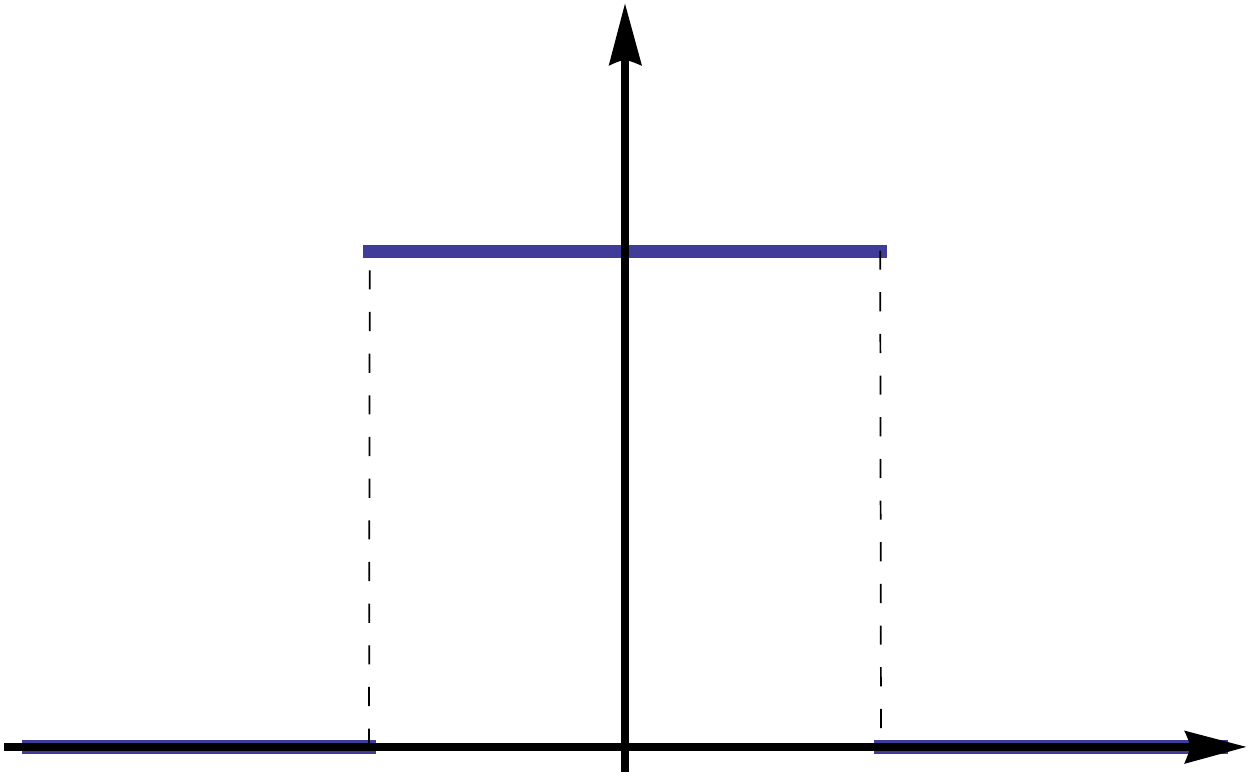}
  \put(43,-17.5){\begin{Huge}$=$\end{Huge}}
 \end{overpic} \vspace*{30pt} \\
\begin{overpic}[width=0.30\textwidth]{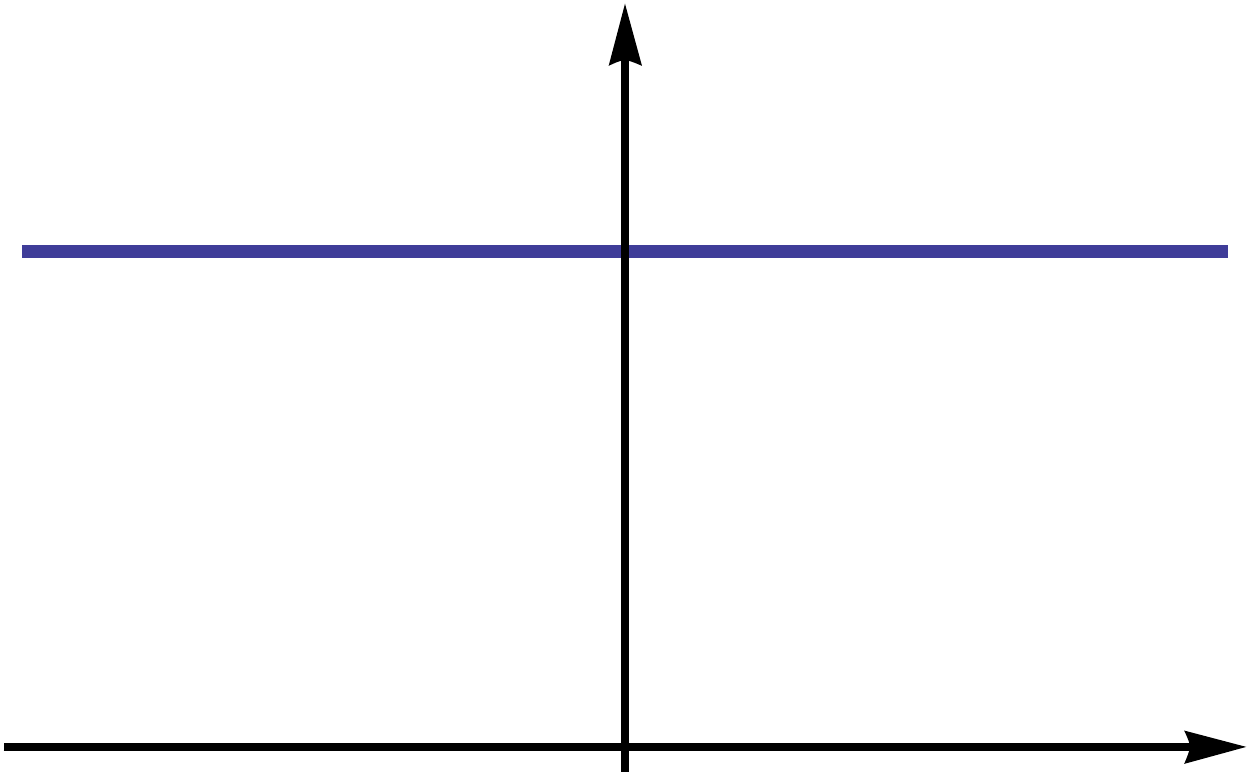}
  \put(100.75,25){\begin{LARGE}$-$\end{LARGE}}
 \end{overpic} \hspace*{8.5pt}
\begin{overpic}[width=0.30\textwidth]{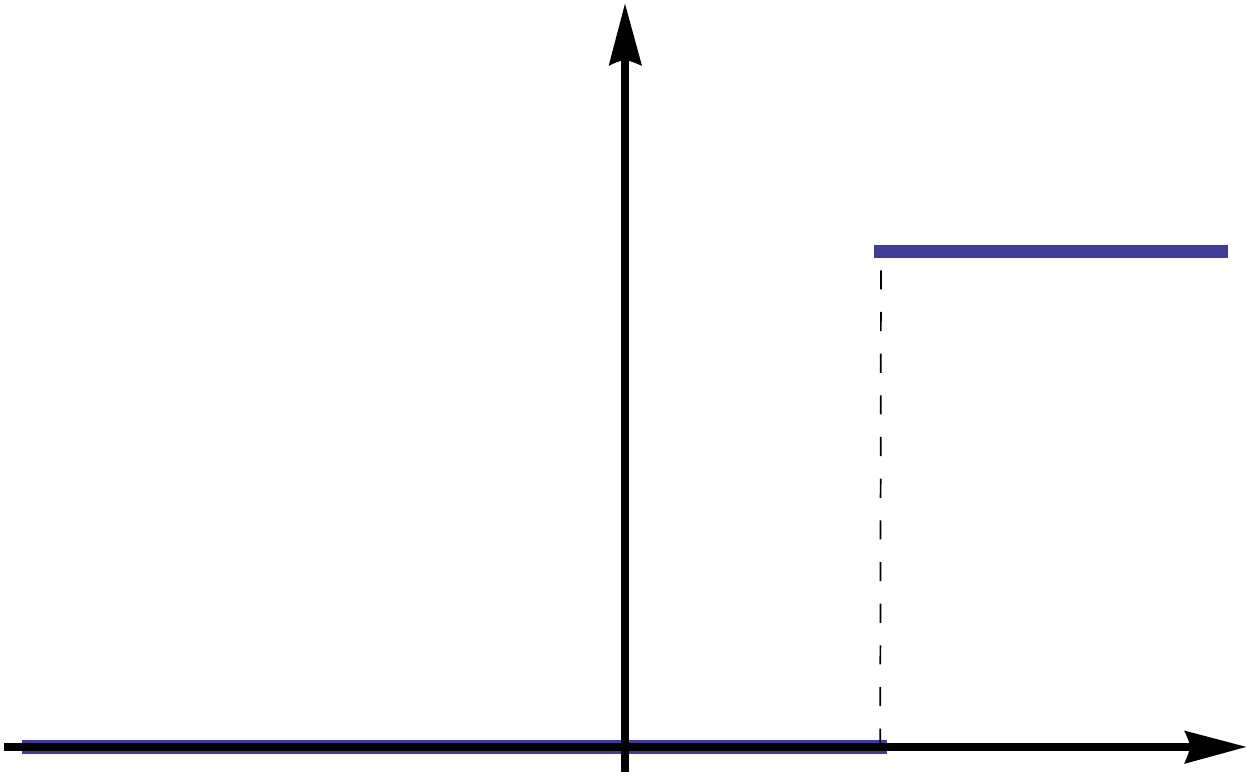}
  \put(100.75,25){\begin{LARGE}$-$\end{LARGE}}
 \end{overpic} \hspace*{8.5pt}
\begin{overpic}[width=0.30\textwidth]{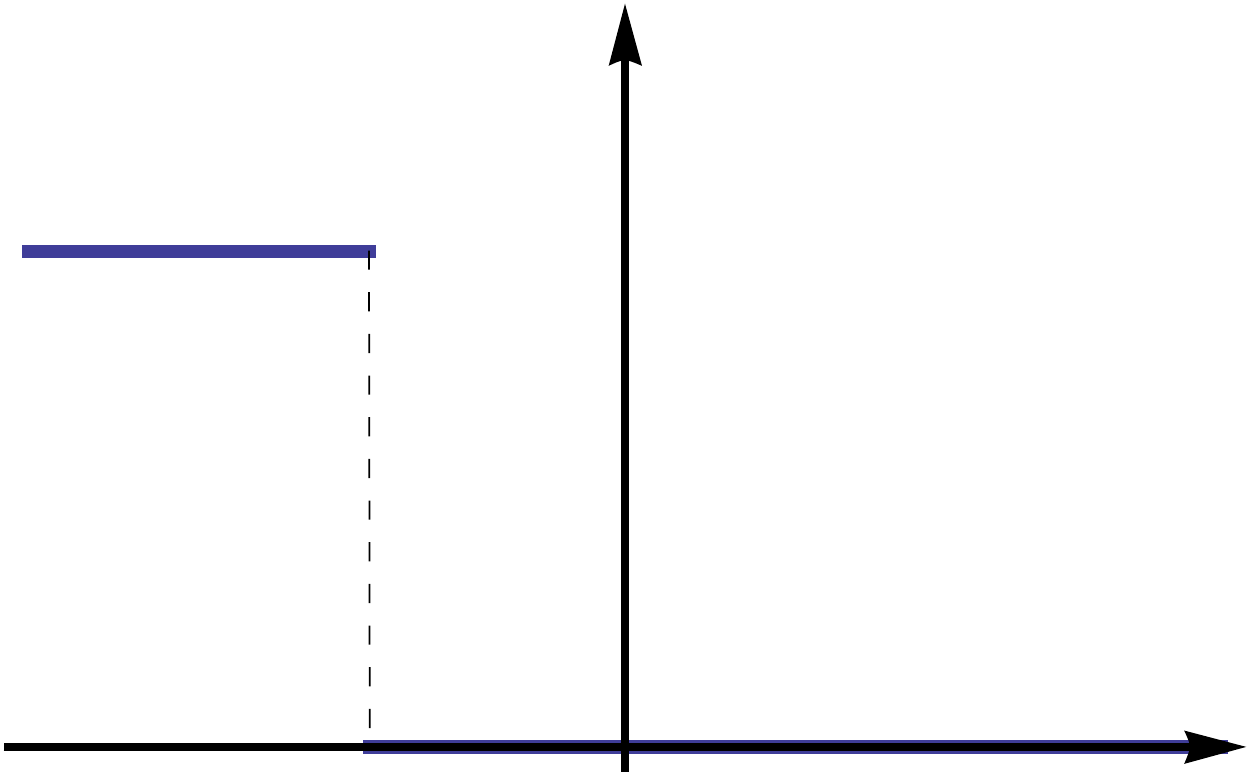}
 \end{overpic}
\caption[Illustration of the identity for the Heaviside $\theta$-functions.]{This figure illustrates the identity for Heaviside $\theta$-functions (\ref{eq:ThetaFunctionIdentity}) used for the reduction of the integral (\ref{eq:PiNVolumeProjectionGeneralExpressionAppendices}) in the main text. This identity is true up to point-sets of vanishing measure.}
\label{fig:ThetaFunctionIdentityFigure}
\end{figure}
Next we turn to the evaluation of the general expression for the projection integral (\ref{eq:PiNVolumeProjectionGeneralExpression}) in appendix \ref{subsec:AccidentalAmbProofsIII}, i.e.
{\allowdisplaybreaks
\begin{align}
\tilde{\mathcal{V}}^{(4L-1)}_{\bm{\uppi}} \left(\epsilon\right) &= 2 \sqrt{L} \int_{-\pi}^{\pi} d \varphi_{1} \ldots \int_{-\pi}^{\pi} d \varphi_{2L} \int_{-\pi}^{\pi} d \psi_{1} \ldots \int_{-\pi}^{\pi} d \psi_{2L} \hspace*{2.5pt} \delta \left( \sum_{k=1}^{2L} \varphi_{k} - \sum_{k^{\prime}=1}^{2L} \psi_{k^{\prime}} \right) \nonumber \\
 &\hspace*{15pt} \hspace*{25pt} \times \theta \Bigg[ \epsilon^{2} - \left( \sum_{k=1}^{2L}  \bm{\uppi} \left(\varphi_{k}\right) - \sum_{k^{\prime}=1}^{2L} \bm{\uppi} \left(\psi_{k^{\prime}}\right) \right)^{2} \Bigg] \mathrm{.} \label{eq:PiNVolumeProjectionGeneralExpressionAppendices}
\end{align}
}
This integral looks very complicated. However, it can be greatly simplified by virtue of the identity
\begin{equation}
 \theta \left( \epsilon^{2} - y^{2} \right) = 1 - \theta \left( - \epsilon + y \right) - \theta \left( - \epsilon - y \right) \mathrm{,} \label{eq:ThetaFunctionIdentity}
\end{equation}
for an arbitrary real quantity $y$ (see Figure \ref{fig:ThetaFunctionIdentityFigure}). Technically speaking, this identity is only true for all points except the ``boundary points'' of the Heaviside $\theta$-function, i.e. those points for which $y = \pm \epsilon$. However, this set of points is again a set of zero measure. Therefore, equation (\ref{eq:ThetaFunctionIdentity}) is correct only if used under an integral. \newline
Applying it to the original integral (\ref{eq:PiNVolumeProjectionGeneralExpressionAppendices}), one gets
{\allowdisplaybreaks
\begin{align}
\tilde{\mathcal{V}}^{(4L-1)}_{\bm{\uppi}} \left(\epsilon\right) &= 2 \sqrt{L} \int_{-\pi}^{\pi} d \varphi_{1} \ldots \int_{-\pi}^{\pi} d \varphi_{2L} \int_{-\pi}^{\pi} d \psi_{1} \ldots \int_{-\pi}^{\pi} d \psi_{2L} \hspace*{2.5pt} \delta \left( \sum_{k=1}^{2L} \varphi_{k} - \sum_{k^{\prime}=1}^{2L} \psi_{k^{\prime}} \right) \nonumber \\
 &\hspace*{12.5pt} - 2 \sqrt{L} \int_{-\pi}^{\pi} d \varphi_{1} \ldots \int_{-\pi}^{\pi} d \varphi_{2L} \int_{-\pi}^{\pi} d \psi_{1} \ldots \int_{-\pi}^{\pi} d \psi_{2L} \hspace*{2.5pt} \delta \left( \sum_{k=1}^{2L} \varphi_{k} - \sum_{k^{\prime}=1}^{2L} \psi_{k^{\prime}} \right) \nonumber \\
 & \hspace*{50pt} \times \theta \Bigg[ - \epsilon + \left( \sum_{k=1}^{2L}  \bm{\uppi} \left(\varphi_{k}\right) - \sum_{k^{\prime}=1}^{2L} \bm{\uppi} \left(\psi_{k^{\prime}}\right) \right) \Bigg] \nonumber \\
 &\hspace*{12.5pt} - 2 \sqrt{L} \int_{-\pi}^{\pi} d \varphi_{1} \ldots \int_{-\pi}^{\pi} d \varphi_{2L} \int_{-\pi}^{\pi} d \psi_{1} \ldots \int_{-\pi}^{\pi} d \psi_{2L} \hspace*{2.5pt} \delta \left( \sum_{k=1}^{2L} \varphi_{k} - \sum_{k^{\prime}=1}^{2L} \psi_{k^{\prime}} \right) \nonumber \\
 & \hspace*{50pt} \times \theta \Bigg[ - \epsilon - \left( \sum_{k=1}^{2L}  \bm{\uppi} \left(\varphi_{k}\right) - \sum_{k^{\prime}=1}^{2L} \bm{\uppi} \left(\psi_{k^{\prime}}\right) \right) \Bigg] \nonumber \\
 &= \mathcal{V}^{(4L-1)}_{\mathcal{CR}} - 2 \sqrt{L} \left(  \mathrm{I}_{1} \left(\epsilon\right) + \mathrm{I}_{2} \left(\epsilon\right) \right) \mathrm{.} \label{eq:PiNVolumeProjectionGeneralExpressionAppendicesII}
\end{align}
}
The first term here, $\mathcal{V}^{(4L-1)}_{\mathcal{CR}}$, is known. Thus the only thing to be done is to evaluate the generally $\epsilon$-dependent integrals $\mathrm{I}_{1}$ and $\mathrm{I}_{2}$. We begin by investigating
\begin{align}
 \mathrm{I}_{1} \left(\epsilon\right) &= \int_{-\pi}^{\pi} d \varphi_{1} \ldots \int_{-\pi}^{\pi} d \varphi_{2L} \int_{-\pi}^{\pi} d \psi_{1} \ldots \int_{-\pi}^{\pi} d \psi_{2L} \hspace*{2.5pt} \delta \left( \sum_{k=1}^{2L} \varphi_{k} - \sum_{k^{\prime}=1}^{2L} \psi_{k^{\prime}} \right) \nonumber \\
 & \hspace*{135pt} \times \theta \Bigg( - \epsilon + \sum_{k=1}^{2L}  \bm{\uppi} \left(\varphi_{k}\right) - \sum_{k^{\prime}=1}^{2L} \bm{\uppi} \left(\psi_{k^{\prime}}\right) \Bigg) \mathrm{.} \label{eq:I1Definition}
\end{align}
The problem is that there is still a $\theta$-function under the integral. Therefore, we choose the detour of first differentiating the expression (\ref{eq:I1Definition}) with respect to $\epsilon$, solving the resulting integral and then in the end integrating the result once, again with respect to $\epsilon$. The derivative is the integral
\begin{align}
 (-) \frac{d \mathrm{I}_{1} \left(\epsilon\right)}{d \epsilon} &= \int_{-\pi}^{\pi} d \varphi_{1} \ldots \int_{-\pi}^{\pi} d \varphi_{2L} \int_{-\pi}^{\pi} d \psi_{1} \ldots \int_{-\pi}^{\pi} d \psi_{2L} \hspace*{2.5pt} \delta \left( \sum_{k=1}^{2L} \varphi_{k} - \sum_{k^{\prime}=1}^{2L} \psi_{k^{\prime}} \right) \nonumber \\
 & \hspace*{135pt} \times \delta \Bigg( - \epsilon + \sum_{k=1}^{2L}  \bm{\uppi} \left(\varphi_{k}\right) - \sum_{k^{\prime}=1}^{2L} \bm{\uppi} \left(\psi_{k^{\prime}}\right) \Bigg) \mathrm{.} \label{eq:DerivativeI1Definition}
\end{align}
We proceed by deriving an expression for this integral using the same strategy that already yielded an expression for $\mathcal{V}^{(4L-1)}_{\mathcal{CR}}$. In order to do this, we employ again the Fourier representation (\ref{eq:DeltaFctFourierRep}) for the first $\delta$-function and represent the second one as
\begin{align}
 \hspace*{-3pt} \delta \Bigg( - \epsilon + \sum_{k=1}^{2L}  \bm{\uppi} \left(\varphi_{k}\right) - \sum_{k^{\prime}=1}^{2L} \bm{\uppi} \left(\psi_{k^{\prime}}\right) \Bigg) &= \int_{- \infty}^{\infty} \frac{d q}{2 \pi} e^{i q \left( \sum_{k} \bm{\uppi} \left( \varphi_{k} \right) - \sum_{k^{\prime}} \bm{\uppi} \left( \psi_{k^{\prime}} \right) - \epsilon \right)} \nonumber \\
&= \int_{- \infty}^{\infty} \frac{d q}{2 \pi} e^{- i q \epsilon} \prod_{k=1}^{2L} e^{i q \hspace*{0.5pt} \bm{\uppi} \left( \varphi_{k} \right)} \prod_{k^{\prime} = 1}^{2L} e^{- i q \hspace*{0.5pt} \bm{\uppi} \left( \psi_{k^{\prime}} \right)} \label{eq:DeltaFunctionFourierRepII}
\end{align}
Inserting the Fourier representations into (\ref{eq:DerivativeI1Definition}) and again commuting integral signs, we get
\begin{align}
 (-) \frac{d \mathrm{I}_{1} \left(\epsilon\right)}{d \epsilon} &= \int_{- \infty}^{\infty} \frac{d p}{2 \pi} \int_{- \infty}^{\infty} \frac{d q}{2 \pi} e^{- i q \epsilon} \prod_{k=1}^{2L} \int_{-\pi}^{\pi} d \varphi_{k} e^{i \varphi_{k} \left( p + \sigma\left[ \bm{\uppi} \left( \varphi_{k} \right) \right] \ast q \right)} \nonumber \\
 & \hspace{102pt} \times \prod_{k^{\prime}=1}^{2L} \int_{-\pi}^{\pi} d \psi_{k^{\prime}} e^{- i \psi_{k^{\prime}} \left( p + \sigma\left[ \bm{\uppi} \left( \psi_{k^{\prime}} \right) \right] \ast q \right)} \mathrm{,} \label{eq:Derivation2ndIntegralStep1}
\end{align}
where the $\sigma$-symbol just extracts the sign of a particular Omelaenko-phase when this phase is transformed under the ambiguity $\bm{\uppi}$, e.g. $\sigma\left[ \bm{\uppi} \left( \varphi_{k} \right) \right] = \sigma\left[ \pm \varphi_{k} \right] = \pm$. \newline
For a general ambiguity transformation $\bm{\uppi} \in \hat{\mathcal{P}}$ there exist, among all transformed phases $\left\{ \bm{\uppi} \left( \varphi_{k} \right) , \bm{\uppi} \left( \psi_{k^{\prime}} \right) \right\}$, $n_{1}$ phases that do not change their sign under this transformation, while the signs of $n_{2}$ phases are flipped. Naturally, since there are $4L$ Omelaenko phases in total for each truncation order $L$, both integers $n_{1}$ and $n_{2}$ introduced here have to satisfy the constraint
\begin{equation}
 n_{1} + n_{2} = 4L \mathrm{.} \label{eq:N1N2Constraint}
\end{equation}
Then, the integral (\ref{eq:Derivation2ndIntegralStep1}) reduces to
{\allowdisplaybreaks
\begin{align}
  (-) \frac{d \mathrm{I}_{1} \left(\epsilon\right)}{d \epsilon} &= (-i)^{(n_{1} + n_{2})} \int_{- \infty}^{\infty} \frac{d p}{2 \pi} \int_{- \infty}^{\infty} \frac{d q}{2 \pi} e^{- i q \epsilon} \left( \frac{e^{i \pi \left( p + q \right)}- e^{- i \pi \left( p+q \right)}}{p+q} \right)^{n_{1}} \nonumber \\
 & \hspace*{12.5pt} \times \left( \frac{e^{i \pi \left( p - q \right)}- e^{- i \pi \left( p - q \right)}}{p - q} \right)^{n_{2}}  \nonumber \\
 &= \frac{\pi^{\left(4L-4\right)}}{4} \int_{- \infty}^{\infty} d x \int_{- \infty}^{\infty} d y e^{- i y \frac{\epsilon}{\pi}} \left( \frac{e^{i \left( x + y \right)}- e^{- i \left( x+y \right)}}{x+y} \right)^{n_{1}} \nonumber \\
 & \hspace*{12.5pt} \times \left( \frac{e^{i  \left( x - y \right)}- e^{- i  \left( x - y \right)}}{x - y} \right)^{n_{2}} \mathrm{,} \label{eq:Derivation2ndIntegralStep3}
\end{align}
}
where $x := \pi p$ and $y := \pi q$ have been defined and in the last step we used the fact that for $L \geq 1$, one always has $(-i)^{n_{1} + n_{2}} = 1$. It is interesting to note that for each ambiguity transformation $\bm{\uppi} \in \hat{\mathcal{P}}$, the information of precisely \textit{which} phases change their sign does not enter the integral (\ref{eq:Derivation2ndIntegralStep3}) and therefore also the sought after projection volume. The only information that fully defines this volume is \textit{how many} phases are conjugated. Therefore, the integral (\ref{eq:Derivation2ndIntegralStep3}) yields the same answer for a large class of different ambigutiy transformations. Furthermore, it can be recognized that once the above given integral and therefore the function $\mathrm{I}_{1} \left(\epsilon\right)$ are fully evaluated, the function $\mathrm{I}_{2} \left(\epsilon\right)$ can be obtained by just interchanging the numbers $n_{1}$ and $n_{2}$, i.e.
\begin{equation}
 \mathrm{I}_{2} \left(\epsilon\right) = \mathrm{I}_{1} \left(\epsilon\right) \Big|_{n_{1} \leftrightarrow n_{2}} \mathrm{.} \label{eq:I2FromI1}
\end{equation}
 The double integral (\ref{eq:Derivation2ndIntegralStep3}) over $x$ and $y$ should, if convergent, not depend in the order in which the integrations are performed. Furthermore, judging by the form of the integrand, it seems promising to do the successive integrations again using \newline
\begin{figure}[h]
\centering
 \begin{overpic}[width=0.975\textwidth]{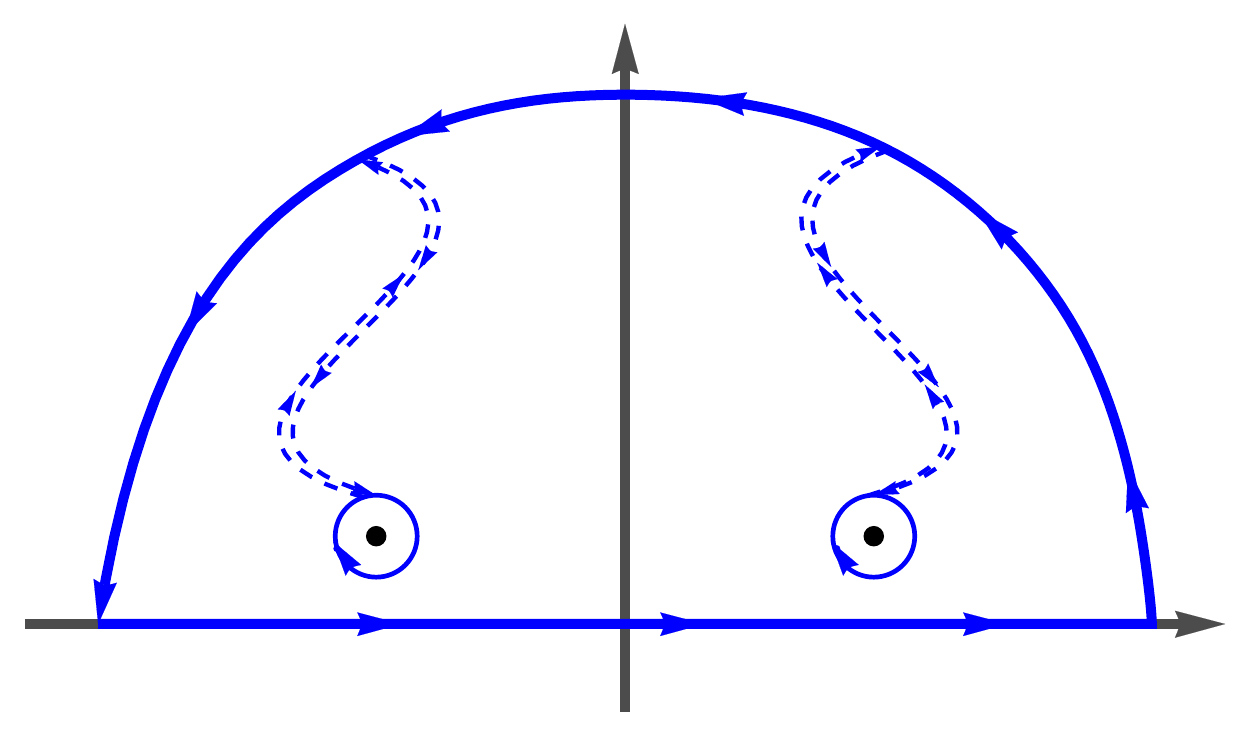}
  \put(52,55.5){\begin{Large}$\mathrm{Im} \left[x\right]$\end{Large}}
  \put(92,4){\begin{Large}$\mathrm{Re} \left[x\right]$\end{Large}}
  \put(53,26.5){$x = y + i \delta$,}
  \put(53,23.5){pole-order $n_{2}$}
  \put(65,21.25){\rotatebox{315}{$\rightarrow$}}
  \put(31,26.5){$x = - y + i \delta$,}
  \put(31,23.5){pole-order $n_{1}$}
  \put(32,22){\rotatebox{225}{$\rightarrow$}}
 \end{overpic}
\caption[Integration-contour used in the application of the residue theorem.]{Application of the residue theorem to the evaluation of the integral (\ref{eq:XIntegral}): both poles in $x = \pm y$ are displaced by a small imaginary part $i \delta$ into the upper half of the complex $x$-plane. The $x$-integral from $-\infty$ to $+\infty$ is transformed into a closed contour in the complex plane. This contour is deformed around the two poles of the integrand, which are now located at $x = y + i \delta$ and $x = - y + i \delta$. The dashed contributions, which are remainders of the contour-deformation, cancel out and do not yield a net contribution to the full integral. The big semi-arc in the upper half of the complex-plane is taken to complex infinity and it is assumed that in this limit its contribution is vanishing (This we do not show, but just assume. It turns out to be true for terms satisfying equation (\ref{eq:SumRestriction1}).). Then the application of Cauchy's integral formula for the closed contour directly yields the statement that the original integral (\ref{eq:XIntegral}) is given as the sum of the residues at the two poles (cf. equation (\ref{eq:XIntegralStep3})).}
\label{fig:ComplexPlanePolesOfIntegrand}
\end{figure}
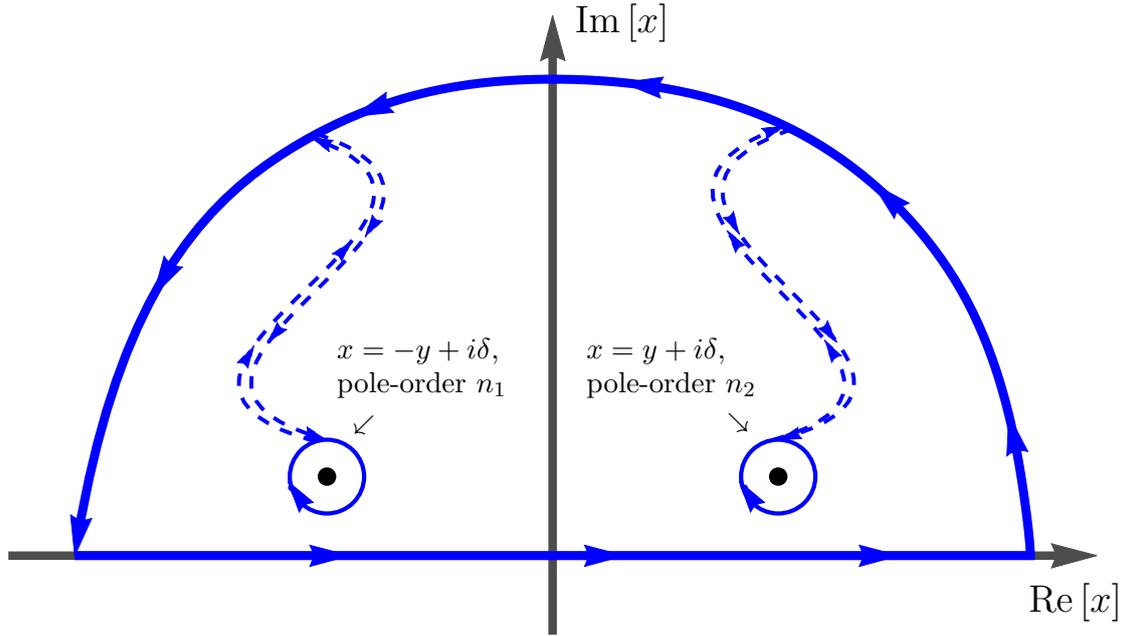
the residue calculus. We choose the route of first performing the $x$-integral, i.e. the evaluation of
\begin{equation}
 \int_{- \infty}^{\infty} d x \left( \frac{e^{i \left( x + y \right)}- e^{- i \left( x+y \right)}}{x+y} \right)^{n_{1}} \left( \frac{e^{i  \left( x - y \right)}- e^{- i  \left( x - y \right)}}{x - y} \right)^{n_{2}} \mathrm{.} \label{eq:XIntegral}
\end{equation}
We choose to replace $x \rightarrow \left(x - i \delta\right)$ with infinitesimally small $\delta$ in the denominators, thereby shifting the poles of the integrand to the upper half of the complex $x$-plane (cf. Figure \ref{fig:ComplexPlanePolesOfIntegrand}). The integrand now has, on the contrary to $\mathcal{V}^{(4L-1)}_{\mathcal{CR}}$ evaluated above, two poles with generally different orders. There is a pole of order $n_{1}$ in $x = - y + i \delta$ and a pole of order $n_{2}$ in $x = y + i \delta$. However, the residue theorem still remains to be usable. Before applying it, we decompose the powers in the numerator using the binomial formula (\ref{eq:BinExpansionAppendix}), which yields
\begin{align}
\left( e^{i (x+y)} - e^{- i (x+y)} \right)^{n_{1}} &= \sum_{k=0}^{n_{1}} \binom{n_{1}}{k} (-)^{n_{1} - k} e^{i \left( 2 k - n_{1} \right) (x+y)} \mathrm{,} \label{eq:FirstPowerBinomDecomp} \\
\left( e^{i (x-y)} - e^{- i (x-y)} \right)^{n_{2}} &= \sum_{k^{\prime}=0}^{n_{2}} \binom{n_{2}}{k^{\prime}} (-)^{n_{2} - k^{\prime}} e^{i ( 2 k^{\prime} - n_{2}) (x-y)} \mathrm{.} \label{eq:SecondPowerBinomDecomp}
\end{align}
Noting that $(-)^{n_{1} - k} \ast (-)^{n_{2} - k^{\prime}} = (-)^{n_{1} + n_{2}} \ast (-)^{- k - k^{\prime}} = (-)^{4L} \ast (-)^{k + k^{\prime}} = (-)^{k + k^{\prime}} $ holds, 
the pole-shifted version of (\ref{eq:XIntegral}) becomes the double-sum
{\allowdisplaybreaks
\begin{align}
 & \hspace*{12.5pt} \sum_{k=0}^{n_{1}} \sum_{k^{\prime}=0}^{n_{2}} (-)^{k + k^{\prime}} \binom{n_{1}}{k} \binom{n_{2}}{k^{\prime}} \int_{- \infty}^{\infty} dx \frac{e^{i \left( 2 k - n_{1} \right) (x+y)} e^{i ( 2 k^{\prime} - n_{2}) (x-y)} }{\left( x + y - i \delta \right)^{n_{1}} \left( x - y - i \delta \right)^{n_{2}}} \nonumber \\
 &= \sum_{k=0}^{n_{1}} \sum_{k^{\prime}=0}^{n_{2}} (-)^{k + k^{\prime}} \binom{n_{1}}{k} \binom{n_{2}}{k^{\prime}} e^{i \left( 2 [ k - k^{\prime} ] - [ n_{1} - n_{2} ] \right) y} \nonumber \\
 & \hspace*{140.5pt} \times \int_{- \infty}^{\infty} dx \frac{e^{i \left( 2 [ k + k^{\prime} ] - 4L \right) x} }{\left( x + y - i \delta \right)^{n_{1}} \left( x - y - i \delta \right)^{n_{2}}} \nonumber \\
 &=: \sum_{k=0}^{n_{1}} \sum_{k^{\prime}=0}^{n_{2}} (-)^{k + k^{\prime}} \binom{n_{1}}{k} \binom{n_{2}}{k^{\prime}} e^{i \left( 2 [ k - k^{\prime} ] - [ n_{1} - n_{2} ] \right) y} \hspace*{5pt} \tilde{\mathbbm{I}}_{(k,k^{\prime})}^{(n_{1},n_{2})} \left(y\right) \mathrm{.} \label{eq:XIntegralStep2}
\end{align}
}
In the integral $\tilde{\mathbbm{I}}_{(k,k^{\prime})}^{(n_{1},n_{2})}$, only terms with $(2 [k + k^{\prime}] - 4L ) > 0$ contribute (cf. remarks made earlier following equation (\ref{eq:CRVolumeMostGeneralEvaluationStep4})), restricting the double sum in front of equation (\ref{eq:XIntegralStep2}) to
\begin{equation}
 (k + k^{\prime}) > 2L \mathrm{.} \label{eq:SumRestriction1}
\end{equation}
For these terms, we get a sum of the residues of the two poles:
\begin{align}
 \tilde{\mathbbm{I}}_{(k,k^{\prime})}^{(n_{1},n_{2})} \left(y\right) &\equiv \int_{- \infty}^{\infty} dx \frac{e^{i \left( 2 [ k + k^{\prime} ] - 4L \right) x} }{\left( x + y - i \delta \right)^{n_{1}} \left( x - y - i \delta \right)^{n_{2}}} \nonumber \\
 &= (2 \pi i) \mathrm{Res}_{(-y + i \delta)} \left[ \frac{e^{i \left( 2 [ k + k^{\prime} ] - 4L \right) x} }{\left( x + y - i \delta \right)^{n_{1}} \left( x - y - i \delta \right)^{n_{2}}} \right] \nonumber \\
& \hspace*{8.25pt} + (2 \pi i) \mathrm{Res}_{(y + i \delta)} \left[ \frac{e^{i \left( 2 [ k + k^{\prime} ] - 4L \right) x} }{\left( x + y - i \delta \right)^{n_{1}} \left( x - y - i \delta \right)^{n_{2}}} \right]  \mathrm{.} \label{eq:XIntegralStep3}
\end{align}
The first residue can be evaluated using equation (\ref{eq:GeneralResidueFormula}) and becomes
\begin{equation}
 \mathrm{Res}_{(-y + i \delta)} \left[ \ldots \right] = \frac{1}{\left( n_{1} - 1 \right)!} \lim\limits_{x \rightarrow (-y + i \delta)}{ \frac{d^{n_{1}-1}}{d x^{n_{1}-1}} \frac{e^{i \left( 2 [ k + k^{\prime} ] - 4L \right) x}}{\left( x - y - i \delta \right)^{n_{2}}} } \mathrm{.} \label{eq:FirstResidue}
\end{equation}
Upon reformulating the exponent of the denominator using the constraint (\ref{eq:N1N2Constraint}), i.e. by setting $n_{2} = 4L - n_{1}$, we see that we need a general formula for a derivative of the general form
\begin{equation}
 \frac{d^{n-1}}{d x^{n-1}} \frac{e^{K x}}{x^{C-n}} \mathrm{,} \label{eq:DerivativeGeneralForm}
\end{equation}
with a complex constant $K$ and integers $n$ and $C$ obeying $C > n$. The expression for this derivative can be found by applying the generalized Leibnitz rule
\begin{equation}
 \frac{d^{N}}{d x^{N}} \left[ u (x) v (x) \right] = \sum_{k=0}^{N} \binom{N}{k} \frac{d^{k} u(x)}{d x^{k}} \frac{d^{N-k} v(x)}{dx^{N-k}} \mathrm{.} \label{eq:GeneralizedLeibnitzRule}
\end{equation}
Using this rule, equation (\ref{eq:DerivativeGeneralForm}) becomes:
\begin{align}
 \frac{d^{n-1}}{d x^{n-1}} \frac{e^{K x}}{(x+\alpha)^{C-n}} &= \sum_{j=0}^{(n-1)} \binom{n-1}{j} \left( \frac{d^{j}}{d x^{j}} e^{K x} \right)  \left( \frac{d^{n-j-1}}{dx^{n-j-1}} (x + \alpha)^{- (C-n)} \right) \nonumber \\
 &= \sum_{j=0}^{(n-1)} \binom{n-1}{j} K^{j} e^{K x} \left[ \prod_{l=1}^{n-j-1} (n - C - l + 1) \right] (x + \alpha)^{j+1-C} \nonumber \\
 &= \frac{e^{Kx}}{(x+\alpha)^{C-1}} \sum_{j=0}^{(n-1)} \binom{n-1}{j} \left[ \prod_{l=1}^{n-j-1} (n - C - l + 1) \right] \left[K (x+\alpha)\right]^{j} \mathrm{.} \label{eq:DerivativeEvaluation}
\end{align}
The product over $l$ in the sum arises because of the fact that $C > n$ and therefore the exponent in the derivative of $x^{-(C-n)}$ is always negative. Thus the prefactor of the resulting derivative cannot be expressed in terms of factorials. Applying the general result (\ref{eq:DerivativeEvaluation}) to the first residue (\ref{eq:FirstResidue}), we obtain the expression
{\allowdisplaybreaks
\begin{align}
 \mathrm{Res}_{(-y + i \delta)} \left[ \ldots \right] &= \frac{1}{\left( n_{1} - 1 \right)!} \lim\limits_{x \rightarrow (-y + i \delta)}{ \frac{d^{n_{1}-1}}{d x^{n_{1}-1}} \frac{e^{i \left( 2 [ k + k^{\prime} ] - 4L \right) x}}{\left( x - y - i \delta \right)^{n_{2}}} } \nonumber \\
 &= \frac{1}{\left( n_{1} - 1 \right)!} \lim\limits_{x \rightarrow (-y + i \delta)} \frac{e^{i \left( 2 [k + k^{\prime}] - 4L \right) x}}{(x - y - i \delta)^{(4L-1)}} \sum_{j=0}^{(n_{1} - 1)} \binom{n_{1}-1}{j} \nonumber \\ 
& \hspace*{12.5pt} \times \left[ \prod_{l=1}^{n_{1}-j-1} (n_{1} - 4L - l + 1) \right] \left[ i (2 [k + k^{\prime}] - 4L) (x - y - i \delta) \right]^{j} \nonumber \\
 &= \frac{1}{\left( n_{1} - 1 \right)!} \frac{e^{- i \left( 2 [k + k^{\prime}] - 4L \right) (y - i \delta)}}{(- 2 y)^{(4L-1)}} \sum_{j=0}^{(n_{1} - 1)} \binom{n_{1}-1}{j} \nonumber \\
 & \hspace*{12.5pt} \times \left[ \prod_{l=1}^{n_{1}-j-1} (n_{1} - 4L - l + 1) \right] (-)^{j} \left[ 2 i (2 [k + k^{\prime}] - 4L) y \right]^{j} \mathrm{.} \label{eq:FirstResidueEvaluated}
\end{align}
}
The contribution to the original integral (\ref{eq:XIntegral}) coming from this term can be found by taking the limit $\delta \rightarrow 0$ and multipilying by $(2 \pi i)$. We obtain:
{\allowdisplaybreaks
\begin{equation}
 (2 \pi i) \lim\limits_{\delta \rightarrow 0}{\mathrm{Res}_{(-y + i \delta)} \left[ \ldots \right]} \equiv (-) \frac{2 \pi i}{\left( n_{1} - 1 \right)!} \frac{e^{- i \left( 2 [k + k^{\prime}] - 4L \right) y}}{(2 y)^{(4L-1)}} P_{(n_{1} - 1)}^{(I)} \left( y \right) \mathrm{,}  \label{eq:FirstResidueFinalResult}
\end{equation}
}
where we have introduced the polynomial of order $(n_{1} - 1)$ in the $y$-variable:
{\allowdisplaybreaks
\begin{align}
  P_{(n_{1} - 1)}^{(I)} \left( y \right) &:= \sum_{j=0}^{(n_{1} - 1)} (-)^{j} \binom{n_{1}-1}{j} \left[ \prod_{l=1}^{n_{1}-j-1} (n_{1} - 4L - l + 1) \right] \left[ 2 i (2 [k + k^{\prime}] - 4L) y \right]^{j} \nonumber \\
 &= \sum_{j=0}^{(n_{1} - 1)} (-) (-)^{j-j+n_{1}} \binom{n_{1}-1}{j} (4L - n_{1})_{(n_{1} - j - 1)} \left[ 2 i (2 [k + k^{\prime}] - 4L) \right]^{j} y^{j} \nonumber \\
 &= \sum_{j=0}^{(n_{1} - 1)} (-)^{(n_{1}-1)} \binom{n_{1}-1}{j} \frac{\Gamma (4L - j - 1)}{\Gamma (4L - n_{1})} \left[ 2 i (2 [k + k^{\prime}] - 4L) \right]^{j} y^{j} \nonumber \\
 &\equiv \sum_{j=0}^{(n_{1} - 1)} \bm{\upkappa}_{j}^{\left( L, n_{1}, k, k^{\prime} \right)} y^{j} \mathrm{.}  \label{eq:FirstResPolynomialDefinition}
\end{align}
}
The rather involved expressions for the coefficients $\bm{\upkappa}_{j}^{\left( L, n_{1}, k, k^{\prime} \right)}$ can just be read off here. The definition of these coefficients can be written naturally either by using the Pochhammer symbols
\begin{equation}
 (a)_{(n)} := a (a+1) (a+2) \ldots (a + n - 1) \mathrm{,} \label{eq:DefPochhammerAppendix}
\end{equation}
or using the Gamma-function $\Gamma (z)$, which comes into play by means of the identity $(a)_{(n)} = \Gamma (a + n) / \Gamma (a)$. In the same way that lead to equation (\ref{eq:FirstResidueFinalResult}), it is possible to derive a closed expression for the second residue contributing to the integral (\ref{eq:XIntegral}), which is
{\allowdisplaybreaks
\begin{equation}
 (2 \pi i) \lim\limits_{\delta \rightarrow 0}{\mathrm{Res}_{(y + i \delta)} \left[ \ldots \right]} = \frac{2 \pi i}{\left( n_{2} - 1 \right)!} \frac{e^{i \left( 2 [k + k^{\prime}] - 4L \right) y}}{(2 y)^{(4L-1)}} P_{(n_{2} - 1)}^{(II)} \left( y \right) \mathrm{,}  \label{eq:SecondResidueFinalResult}
\end{equation}
}
with the polynomial
{\allowdisplaybreaks
\begin{align}
P_{(n_{2} - 1)}^{(II)} \left( y \right) := \sum_{j=0}^{(n_{2} - 1)} (-)^{j} \bm{\upkappa}_{j}^{\left( L, n_{2}, k, k^{\prime} \right)} y^{j} \mathrm{.}  \label{eq:SecondResPolynomialDefinition}
\end{align}
}
Finally, the integral (\ref{eq:XIntegralStep3}) can be brought to the following closed expression, which has a pole of order $(4L - 1)$ in the variable $y$ at the origin:
\begin{equation}
\tilde{\mathbbm{I}}_{(k,k^{\prime})}^{(n_{1},n_{2})} \left(y\right) = \frac{2 \pi i}{(2 y)^{(4L - 1)}} \left( \frac{e^{i \left( 2 [k + k^{\prime}] - 4L \right) y}}{\left( n_{2} - 1 \right)!}  P_{(n_{2} - 1)}^{(II)} \left( y \right) - \frac{e^{- i \left( 2 [k + k^{\prime}] - 4L \right) y}}{\left( n_{1} - 1 \right)!}  P_{(n_{1} - 1)}^{(I)} \left( y \right) \right) \mathrm{.} \label{eq:RelevantXIntegralFinalExpression}
\end{equation}
With the aid of this intermediate result, the original integral (\ref{eq:Derivation2ndIntegralStep3}) becomes
{\allowdisplaybreaks
\begin{align}
 \hspace*{-8.5pt} (+) \frac{d \mathrm{I}_{1}}{d \epsilon} &= \frac{1}{4} \pi^{4 L - 4} \int_{-\infty}^{\infty} dy e^{- i y \frac{\epsilon}{\pi}} \left[ \sum_{k=0}^{n_{1}}  \sum_{k^{\prime}=0}^{n_{2}} \right]_{(k + k^{\prime}) > 2L} (-)^{(k + k^{\prime})} \binom{n_{1}}{k}  \binom{n_{2}}{k^{\prime}}  \nonumber \\
  & \hspace*{85pt} \times e^{i (2 [k - k^{\prime}] - [n_{1} - n_{2}]) y} (-) \tilde{\mathbbm{I}}_{(k,k^{\prime})}^{(n_{1},n_{2})} \left(y\right) \nonumber \\
  &= \frac{1}{4} \pi^{4 L - 4} \frac{2 \pi i}{2^{4L - 1}} \left[ \sum_{k=0}^{n_{1}}  \sum_{k^{\prime}=0}^{n_{2}} \right]_{(k + k^{\prime}) > 2L} (-)^{(k + k^{\prime})} \binom{n_{1}}{k}  \binom{n_{2}}{k^{\prime}} \nonumber \\
  & \hspace*{12.5pt} \times \int_{-\infty}^{\infty} dy e^{- i y \frac{\epsilon}{\pi}}  e^{i (2 [k - k^{\prime}] - [n_{1} - n_{2}]) y} \frac{1}{y^{(4L - 1)}} \nonumber \\
  & \hspace*{12.5pt} \times \left[ \frac{e^{- i \left( 2 [k + k^{\prime}] - 4L \right) y}}{\left( n_{1} - 1 \right)!}  P_{(n_{1} - 1)}^{(I)} \left( y \right) - \frac{e^{i \left( 2 [k + k^{\prime}] - 4L \right) y}}{\left( n_{2} - 1 \right)!}  P_{(n_{2} - 1)}^{(II)} \left( y \right) \right] \nonumber \\
  &= \pi^{4L - 3} \frac{i}{2^{4L}} \left[ \sum_{k=0}^{n_{1}}  \sum_{k^{\prime}=0}^{n_{2}} \right]_{(k + k^{\prime}) > 2L} (-)^{(k + k^{\prime})} \binom{n_{1}}{k}  \binom{n_{2}}{k^{\prime}}  \times  \nonumber \\
  & \hspace*{12.5pt} \int_{-\infty}^{\infty} \frac{dy}{y^{(4L - 1)}} \left[ \frac{e^{ i \left( - 4 k^{\prime} + 2 n_{2} - \frac{\epsilon}{\pi} \right) y}}{\left( n_{1} - 1 \right)!}  P^{(I)} \left( y \right) - \frac{e^{i \left( 4 k - 2 n_{1} - \frac{\epsilon}{\pi} \right) y}}{\left( n_{2} - 1 \right)!}  P^{(II)} \left( y \right) \right] \mathrm{.} \label{eq:2ndIntegralXIntPerformed}
\end{align}
}
Finally, only the $y$-integration needs to be done. We shift the pole according to $y \rightarrow \left( y - i \eta \right)$ in the denominator, with infinitesimally small $\eta \ll 1$. Then, we split the integral in the last line of equation (\ref{eq:2ndIntegralXIntPerformed}) as
{\allowdisplaybreaks
\begin{align}
 \hat{\mathrm{I}} &:= \int_{-\infty}^{\infty} \frac{dy}{(y - i \eta)^{(4L - 1)}} \left[ \frac{e^{ i \left( - 4 k^{\prime} + 2 n_{2} - \frac{\epsilon}{\pi} \right) y}}{\left( n_{1} - 1 \right)!}  P^{(I)} \left( y \right) - \frac{e^{i \left( 4 k - 2 n_{1} - \frac{\epsilon}{\pi} \right) y}}{\left( n_{2} - 1 \right)!}  P^{(II)} \left( y \right) \right] \nonumber \\
  &\equiv \hat{\mathrm{I}}_{1} - \hat{\mathrm{I}}_{2} \mathrm{,} \label{eq:FinalIntegralSplit}
\end{align}
and continue with the evaluation of
\begin{equation}
  \hat{\mathrm{I}}_{1} = \frac{1}{\left( n_{1} - 1 \right)!} \int_{-\infty}^{\infty} dy \frac{e^{ i \left( - 4 k^{\prime} + 2 n_{2} - \frac{\epsilon}{\pi} \right) y}}{(y - i \eta)^{(4L - 1)}}  P^{(I)}_{(n_{1} - 1)} \left( y \right) \mathrm{.} \label{eq:I1FInalYIntegral}
\end{equation}
We note that here, only terms with $(- 4 k^{\prime} + 2 n_{2} - \frac{\epsilon}{\pi}) > 0$ can contribute, which is equivalent to the sum restriction
\begin{equation}
 k^{\prime} < \frac{n_{2}}{2} - \frac{\epsilon}{4 \pi} \mathrm{,} \label{eq:SumRestriction2}
\end{equation}
where again we emphasize that $\epsilon > 0$ has to be valid. The relevant residue becomes
{\allowdisplaybreaks
\begin{align}
 \hat{\mathrm{I}}_{1} &= \frac{2 \pi i}{\left( n_{1} - 1 \right)!} \mathrm{Res}_{i \eta} \left[ \frac{e^{ i \left( - 4 k^{\prime} + 2 n_{2} - \frac{\epsilon}{\pi} \right) y}}{(y - i \eta)^{(4L - 1)}}  P^{(I)}_{(n_{1} - 1)} \left( y \right) \right] \nonumber \\
 &= \frac{2 \pi i}{\left( n_{1} - 1 \right)! \left( 4L - 2 \right)!} \lim\limits_{y \rightarrow i \eta}{ \frac{d^{4L - 2}}{dy^{4L - 2}} e^{ i \left( - 4 k^{\prime} + 2 n_{2} - \frac{\epsilon}{\pi} \right) y} P^{(I)}_{(n_{1} - 1)} \left( y \right) } \mathrm{.} \label{eq:I1ResidueFormula}
\end{align}
}
Again, a general formula for a derivative is needed and the generalized Leibnitz rule (\ref{eq:GeneralizedLeibnitzRule}) quickly yields
\begin{equation}
 \frac{d^{C - 2}}{dy^{C - 2}} e^{K y} P_{(n - 1)}(y) = \sum_{r = 0}^{(C - 2)} \binom{C - 2}{r} K^{r} e^{K y} \frac{d^{C - r - 2} P_{(n - 1)} (y) }{dy^{C - r - 2}}  \mathrm{.} \label{eq:GeneralizedDerivativeII}
\end{equation}
The derivative acting on the polynomial of order $(n-1)$ can only yield non-vanishing contributions if and only if the condition $(C - r - 2) \leq (n - 1)$, or equivalently $r \geq (C - n - 1)$ is fulfilled. Therefore, we obtain:
\begin{equation}
 \frac{d^{C - 2}}{dy^{C - 2}} e^{K y} P_{(n - 1)}(y) = \sum_{r = (C - n - 1)}^{(C - 2)} \binom{C - 2}{r} K^{r} e^{K y}  P^{(C - r - 2)}_{(n - 1)} (y)  \mathrm{,} \label{eq:GeneralizedDerivativeII2}
\end{equation}
where here, the notation $P_{(n-1)}^{(C-r-2)}$ denotes the $(C - r - 2)$-th derivative of the polynomial $P_{(n-1)}$ of order $(n-1)$. With everything assembled until now, the integral $\hat{\mathrm{I}}_{1}$ becomes:
{\allowdisplaybreaks
\begin{align}
 \hat{\mathrm{I}}_{1} &= \frac{2 \pi i}{\left( n_{1} - 1 \right)! \left( 4L - 2 \right)!} \sum_{r = (4L - n_{1} - 1)}^{(4L - 2)} \binom{4L - 2}{r} \left[ i \left( - 4 k^{\prime} + 2 n_{2} - \frac{\epsilon}{\pi} \right) \right]^{r} \nonumber \\
  & \hspace*{166pt} \times e^{ - \left( - 4 k^{\prime} + 2 n_{2} - \frac{\epsilon}{\pi} \right) \eta}  \left[P^{(I)}\right]^{(4L - r - 2)}_{(n_{1} - 1)} (i \eta) \mathrm{.} \label{eq:I1FinalResidueFormula}
\end{align}
}
From the definition of the integral $\hat{\mathrm{I}}_{2}$, i.e.
\begin{equation}
  \hat{\mathrm{I}}_{2} = \frac{1}{\left( n_{2} - 1 \right)!} \int_{-\infty}^{\infty} dy \frac{e^{ i \left( 4 k - 2 n_{1} - \frac{\epsilon}{\pi} \right) y}}{(y - i \eta)^{(4L - 1)}}  P^{(II)}_{(n_{2} - 1)} \left( y \right) \mathrm{,} \label{eq:I2FInalYIntegral}
\end{equation}
we see that the only non-vanishing integrals have to satisfy $(4 k - 2 n_{1} - \frac{\epsilon}{\pi}) > 0$, or 
\begin{equation}
k > \frac{n_{1}}{2} + \frac{\epsilon}{4 \pi} \mathrm{.} \label{eq:SumRestriction3}
\end{equation}
Manipulations similar to those that lead to the expression (\ref{eq:I1FinalResidueFormula}) yield for the integral $\hat{\mathrm{I}}_{2}$:
{\allowdisplaybreaks
\begin{align}
 \hat{\mathrm{I}}_{2} &= \frac{2 \pi i}{\left( n_{2} - 1 \right)! \left( 4L - 2 \right)!} \sum_{r = (4L - n_{2} - 1)}^{(4L - 2)} \binom{4L - 2}{r} \left[ i \left( 4 k - 2 n_{1} - \frac{\epsilon}{\pi} \right) \right]^{r} \nonumber \\
  & \hspace*{166pt} \times e^{ - \left( 4 k - 2 n_{1} - \frac{\epsilon}{\pi} \right) \eta}  \left[P^{(II)}\right]^{(4L - r - 2)}_{(n_{2} - 1)} (i \eta) \mathrm{.} \label{eq:I2FinalResidueFormula}
\end{align}
}
The combination of the results (\ref{eq:I1FinalResidueFormula}) and (\ref{eq:I2FinalResidueFormula}), as well as taking the limit $\eta \rightarrow 0$, yields the following answer for the derivative of $\mathrm{I}_{1} (\epsilon)$ with respect to $\epsilon$
{\allowdisplaybreaks
\begin{align}
 \hspace*{-8.5pt} \frac{d \mathrm{I}_{1}}{d \epsilon} &= \frac{(-) \pi^{(4L-2)}}{2^{4L-1} \left( 4L - 2 \right)!} \nonumber \\
 & \hspace*{12.5pt} \times \Bigg\{ \frac{1}{\left( n_{1} - 1 \right)!} \left[ \sum_{k=0}^{n_{1}}  \sum_{k^{\prime}=0}^{n_{2}} \right]_{(k + k^{\prime}) > 2L}^{k^{\prime} < \frac{n_{2}}{2} - \frac{\epsilon}{4 \pi}} (-)^{(k + k^{\prime})} \binom{n_{1}}{k}  \binom{n_{2}}{k^{\prime}} \sum_{r = (4L - n_{1} - 1)}^{(4L - 2)} \binom{4L - 2}{r} \nonumber \\
  & \hspace*{130.5pt} \times \left[ i \left( - 4 k^{\prime} + 2 n_{2} - \frac{\epsilon}{\pi} \right) \right]^{r}  \left[P^{(I)}\right]^{(4L - r - 2)}_{(n_{1} - 1)} ( 0 ) \nonumber \\
 & \hspace*{23.25pt} - \frac{1}{\left( n_{2} - 1 \right)!} \left[ \sum_{k=0}^{n_{1}}  \sum_{k^{\prime}=0}^{n_{2}} \right]_{(k + k^{\prime}) > 2L}^{k > \frac{n_{1}}{2} + \frac{\epsilon}{4 \pi}} (-)^{(k + k^{\prime})} \binom{n_{1}}{k}  \binom{n_{2}}{k^{\prime}} \sum_{r = (4L - n_{2} - 1)}^{(4L - 2)} \binom{4L - 2}{r} \nonumber \\
  & \hspace*{130.5pt} \times \left[ i \left( 4 k - 2 n_{1} - \frac{\epsilon}{\pi} \right) \right]^{r}  \left[P^{(II)}\right]^{(4L - r - 2)}_{(n_{2} - 1)} ( 0 ) \Bigg\} \mathrm{.} \label{eq:2ndIntegralYIntPerformed}
\end{align}
}
For a standard $n$-th order polynomial
\begin{equation}
 P_{(n)} (z) = \sum_{j = 0}^{n} C_{j} z^{j} \mathrm{,} \label{eq:NthOrderPolynomial}
\end{equation}
the $m$-th order derivative, provided $m \leq n$, reads
\begin{equation}
 P_{(n)}^{(m)} (z) = \sum_{j = 0}^{(n-m)} C_{j+m} \left[ \prod_{k=1}^{m} (j + k) \right] z^{j} \mathrm{.} \label{eq:MthDerivativeOfNthOrderPolynomial}
\end{equation}
Setting $z = 0$ in this derivative, one pulls out the free term, i.e. the coefficient of $z^{0} = 1$
\begin{equation}
 P_{(n)}^{(m)} (0) = C_{m} m! \mathrm{.} \label{eq:MthDerivativeOfNthOrderPolynomialZerothCoeff}
\end{equation}
With this knowledge and furthermore by inspection of the polynomials (\ref{eq:FirstResPolynomialDefinition}) and (\ref{eq:SecondResPolynomialDefinition}), we see that the derivative of $\mathrm{I}_{1}$ (equation (\ref{eq:2ndIntegralYIntPerformed})) becomes
{\allowdisplaybreaks
\begin{align}
 \hspace*{-8.5pt} \frac{d \mathrm{I}_{1}}{d \epsilon} &= \frac{(-) \pi^{(4L-2)}}{2^{4L-1} \left( 4L - 2 \right)!} \nonumber \\
 & \hspace*{12.5pt} \times \Bigg\{ \frac{1}{\left( n_{1} - 1 \right)!} \left[ \sum_{k=0}^{n_{1}}  \sum_{k^{\prime}=0}^{n_{2}} \right]_{(k + k^{\prime}) > 2L}^{k^{\prime} < \frac{n_{2}}{2} - \frac{\epsilon}{4 \pi}} (-)^{(k + k^{\prime})} \binom{n_{1}}{k}  \binom{n_{2}}{k^{\prime}} \sum_{r = (4L - n_{1} - 1)}^{(4L - 2)} \binom{4L - 2}{r} \nonumber \\
  & \hspace*{100.5pt} \times  (4L - r - 2)! \hspace*{2pt} \bm{\upkappa}_{(4L - r - 2)}^{\left( L, n_{1}, k, k^{\prime} \right)} \left[ i \left( - 4 k^{\prime} + 2 n_{2} - \frac{\epsilon}{\pi} \right) \right]^{r} \nonumber \\
 & \hspace*{23.25pt} - \frac{1}{\left( n_{2} - 1 \right)!} \left[ \sum_{k=0}^{n_{1}}  \sum_{k^{\prime}=0}^{n_{2}} \right]_{(k + k^{\prime}) > 2L}^{k > \frac{n_{1}}{2} + \frac{\epsilon}{4 \pi}} (-)^{(k + k^{\prime})} \binom{n_{1}}{k}  \binom{n_{2}}{k^{\prime}} \sum_{r = (4L - n_{2} - 1)}^{(4L - 2)} \binom{4L - 2}{r} \nonumber \\
  & \hspace*{55.5pt} \times (-)^{(4L-r-2)} (4L - r - 2)! \hspace*{2pt} \bm{\upkappa}_{(4L - r - 2)}^{\left( L, n_{2}, k, k^{\prime} \right)} \left[ i \left( 4 k - 2 n_{1} - \frac{\epsilon}{\pi} \right) \right]^{r} \Bigg\} \mathrm{.} \label{eq:2ndIntegralFinalResult}
\end{align}
}
Now, one can get to a final expression for $\mathrm{I}_{1}$ by integrating once with respect to $\epsilon$, provided that the correct borders of integration are chosen. In order to express the integral however, it is better to replace the second restriction of the double sums in equation (\ref{eq:2ndIntegralYIntPerformed}) by suitably defined Heaviside-$\theta$-functions multiplying the summed term. For the first double-sum, the restriction is $k^{\prime} < \frac{n_{2}}{2} - \frac{\epsilon}{4 \pi}$ while for the second sum it reads $k > \frac{n_{1}}{2} + \frac{\epsilon}{4 \pi}$. These two conditions are equivalent to the following requirements for $\epsilon$
\begin{align}
 \epsilon &< 2 \pi \left( n_{2} - 2 k^{\prime} \right) \mathrm{,} \label{eq:FirstRestrictionForEpsilon} \\
 \epsilon &< 2 \pi \left( 2 k - n_{1} \right) \mathrm{.} \label{eq:SecondRestrictionForEpsilon}
\end{align}
Therefore, it is seen that equation (\ref{eq:2ndIntegralYIntPerformed}) can be rewritten as
{\allowdisplaybreaks
\begin{align}
 \hspace*{-8.5pt} \frac{d \mathrm{I}_{1}}{d \epsilon} &= \frac{(-) \pi^{(4L-2)}}{2^{4L-1} \left( 4L - 2 \right)!} \left[ \sum_{k=0}^{n_{1}}  \sum_{k^{\prime}=0}^{n_{2}} \right]_{(k + k^{\prime}) > 2L} \nonumber \\
 & \hspace*{12.5pt} \times \Bigg\{ \left( 1 - \theta\left[ \epsilon - 2 \pi \left( n_{2} - 2 k^{\prime} \right) \right] \right) \frac{(-)^{(k + k^{\prime})}}{\left( n_{1} - 1 \right)!} \binom{n_{1}}{k}  \binom{n_{2}}{k^{\prime}} \nonumber \\
  & \hspace*{35.5pt} \times \sum_{r = (4L - n_{1} - 1)}^{(4L - 2)} \frac{(4L-2)!}{r!} \hspace*{2pt} \bm{\upkappa}_{(4L - r - 2)}^{\left( L, n_{1}, k, k^{\prime} \right)} \left[ i \left( - 4 k^{\prime} + 2 n_{2} - \frac{\epsilon}{\pi} \right) \right]^{r} \nonumber \\
 & \hspace*{23.25pt} - \left( 1 - \theta\left[ \epsilon - 2 \pi \left( 2 k - n_{1} \right) \right] \right) \frac{(-)^{(k + k^{\prime})}}{\left( n_{2} - 1 \right)!} \binom{n_{1}}{k}  \binom{n_{2}}{k^{\prime}}  \nonumber \\
  & \hspace*{35.5pt} \times \sum_{r = (4L - n_{2} - 1)}^{(4L - 2)} (-)^{r} \frac{(4L-2)!}{r!} \hspace*{2pt} \bm{\upkappa}_{(4L - r - 2)}^{\left( L, n_{2}, k, k^{\prime} \right)} \left[ i \left( 4 k - 2 n_{1} - \frac{\epsilon}{\pi} \right) \right]^{r} \Bigg\} \mathrm{,} \label{eq:2ndIntegralFinalResultReformulated}
\end{align}
}
where also some minor simplifications of binomial coefficients and factorials were applied. Furthermore, by invoking the relation (\ref{eq:I2FromI1}) we see that using the formula above for $d \mathrm{I}_{1} / d \epsilon$, it is very simple to find the expression for $d \mathrm{I}_{2} / d \epsilon$, namely by just interchanging the integers $n_{1}$ and $n_{2}$ everywhere in the expression. \newline
Considering the formula (\ref{eq:2ndIntegralFinalResultReformulated}), it is apparent why it is very formidable to find a general closed expression for its integral, i.e. $\mathrm{I}_{1} (\epsilon)$. The integration operator can be pulled past the double-sum, but then there is generally a product of a Heaviside-$\theta$-function and a polynomial coming from the summation over $r$ and it is not clear how to derive a closed expression for this integral, at least for general values of $L$, $n_{1}$ and $n_{2}$. For special cases however, this final integration can be performed explicitly (as shall be shown below). \newline
Furthermore, the integration over $\epsilon$ introduces one a priori unknown integration constant which we call $\bm{C}$. The value of this constant has to be fixed for each special case by invoking consistent boundary conditions. For example, the requirement
\begin{equation}
 \tilde{\mathcal{V}}^{(4L-1)}_{\bm{\uppi}} \left(\epsilon = 0\right) \equiv 0 \mathrm{,} \label{eq:ConsistencyBoundaryCondition}
\end{equation}
fixes the final free integration constant. Bearing this in mind, the final result for $\tilde{\mathcal{V}}^{(4L-1)}_{\bm{\uppi}} \left(\epsilon\right)$ becomes formally (cf. equation (\ref{eq:PiNVolumeProjectionGeneralExpressionAppendicesII}))
{\allowdisplaybreaks
\begin{align}
 \tilde{\mathcal{V}}^{(4L-1)}_{\bm{\uppi}} \left(\epsilon\right) &= \mathcal{V}^{(4L-1)}_{\mathcal{CR}} - 2 \sqrt{L} \int_{\bm{C}}^{\epsilon} d \epsilon^{\prime} \left[  \frac{d \mathrm{I}_{1}}{d \epsilon^{\prime}} \left(\epsilon^{\prime}\right) + \frac{d \mathrm{I}_{2}}{d \epsilon^{\prime}} \left(\epsilon^{\prime}\right) \right] \nonumber \\
 &= \mathcal{V}^{(4L-1)}_{\mathcal{CR}} - 2 \sqrt{L} \int_{\bm{C}}^{\epsilon} d \epsilon^{\prime} \left[  \frac{d \mathrm{I}_{1}}{d \epsilon^{\prime}} \left(\epsilon^ {\prime}\right) + \left( n_{1} \leftrightarrow n_{2} \right) \right] \mathrm{,} \label{eq:FormalFinalExpressionProjectionVolume}
\end{align}
}
where the constant $\bm{C}$ is determined after the fact by invoking for example (\ref{eq:ConsistencyBoundaryCondition}). \newline

We continue with a discussion of some examples for the formulas accumulated until now. First, the artificial case of just one $\varphi$-phase and just one $\psi$-phase shall be discussed, which is also represented in Figure \ref{fig:PedagogicalPhaseVolumeExample} of appendix \ref{subsec:AccidentalAmbProofsIII}. The only interesting ambiguity in this case would correspond to the choice of setting $L = 1/2$ and $n_{1} = n_{2} = 1$ in the expression (\ref{eq:2ndIntegralFinalResultReformulated}). However, one has to bear in mind that now, since $n_{1} = n_{2} = 1$ is given, one gets an additional minus sign from the factor $(-i)^{n_{1} + n_{2}}$ appearing in equation (\ref{eq:Derivation2ndIntegralStep3}). Thus, from equation (\ref{eq:2ndIntegralFinalResultReformulated}) one obtains in this particular special case
\begin{equation}
 \frac{d \mathrm{I}_{1}}{d \epsilon} (\epsilon) = \frac{d \mathrm{I}_{2}}{d \epsilon} (\epsilon) = \frac{(-)}{2} \left( 1 - \theta\left[ \epsilon - 2 \pi \right] \right) \mathrm{.} \label{eq:SimplestDerivativeExampleTwoPhases}
\end{equation}
The general integral with arbitrary constant $\bm{C}$ defined in equation (\ref{eq:FormalFinalExpressionProjectionVolume}) becomes (using MATHEMATICA \cite{Mathematica8,Mathematica11,MathematicaLanguage,MathematicaBonnLicense})
{\allowdisplaybreaks
\begin{align}
 \tilde{\mathcal{V}}^{1}_{\bm{\uppi}} \left(\epsilon\right) &= \sqrt{2} \big[ -\bm{C}+2 \pi +\epsilon +(2 \pi -\epsilon +[\bm{C}-2 \pi ] \theta[\bm{C}-2 \pi ]) \theta[-\bm{C}+\epsilon ] \theta[-2 \pi +\epsilon ] \nonumber \\ 
 & \hspace*{33.5pt} + (\bm{C}-2 \pi +[2 \pi -\epsilon ] \theta[-2 \pi +\epsilon ]) \theta[\bm{C}-2 \pi ,\bm{C}-\epsilon ] \big] \mathrm{,} \label{eq:SimplestExampleGeneralIntegral}
\end{align}
}
where the two-dimensional $\theta$-function is defined by
\begin{equation}
 \theta \left( x, y \right) := \begin{cases}
                    1 &\mathrm{,} \hspace*{2pt} x \geq 0 \hspace*{2pt} \mathrm{and} \hspace*{2pt} y \geq 0 \\
                    0 &\mathrm{,} \hspace*{2pt} \mathrm{else} 
                   \end{cases} \mathrm{.} \label{eq:TwoDimensionalThetaFunction}
\end{equation}
Invoking the boundary condition (\ref{eq:ConsistencyBoundaryCondition}), i.e. by setting $\epsilon = 0$ in the expression (\ref{eq:SimplestExampleGeneralIntegral}) and searching the roots of the resulting equation, we obtain
\begin{equation}
 (\bm{C}-2 \pi ) (-1+\theta[\bm{C}-2 \pi ]) \equiv 0 \mathrm{.} \label{eq:SimplestExampleBoundaryCondition}
\end{equation}
It is seen quickly that this equation is generally fulfilled for $\bm{C} = 2 \pi$ (but the choice of any other $\bm{C} > 2 \pi$ would actually yield the same function). Therefore, our final result becomes
\begin{equation}
 \tilde{\mathcal{V}}^{1}_{\bm{\uppi}} \left(\epsilon\right) = \sqrt{2} \left( \epsilon +(2 \pi -\epsilon ) \theta[-2 \pi +\epsilon ] \right) \mathrm{.} \label{eq:SimplestExampleFinalResult}
\end{equation}
For values of $\epsilon$ respecting $\epsilon < 2 \pi$, this formula reduces to the very simple linear relation
\begin{equation}
 \tilde{\mathcal{V}}^{1}_{\bm{\uppi}} \left(\epsilon\right) = \sqrt{2} \epsilon \mathrm{.} \label{eq:SimplestExampleFinalResultSmallEpsilon}
\end{equation}
Now we treat a more realistic but also more complicated case. We look at $L = 1$, i.e. an $S$- and $P$-wave approximation and consider first the ambiguity case of $n_{1} = n_{2} = 2$ and each ambiguity described by these numbers we call $\bm{\uppi}^{\mathrm{I}} \in \hat{\mathcal{P}}$. Then we of course have $n_{1} + n_{2} = 4L = 4$ as initially required. Setting $L = 1$ and $n_{1} = n_{2} = 2$ in equation (\ref{eq:2ndIntegralFinalResultReformulated}), we get
\begin{equation}
 \frac{d \mathrm{I}_{1}}{d \epsilon} (\epsilon) = \frac{d \mathrm{I}_{2}}{d \epsilon} (\epsilon) = 2 \pi \epsilon + \frac{1}{8} \left[ -(4 \pi + \epsilon)^{2} + (\epsilon - 4 \pi)^{2} \theta \left( \epsilon - 4 \pi \right) \right] \mathrm{.} \label{eq:Example22DerivativeFourPhases}
\end{equation}
The correct constant of integration in this case turns out to be $\bm{C} = 4 \pi$ or any real number larger than this. Settling with $4 \pi$ in this case, we obtain the following expression for the projection volume (\ref{eq:FormalFinalExpressionProjectionVolume}):
\begin{equation}
 \tilde{\mathcal{V}}^{3}_{\bm{\uppi}^{\mathrm{I}}} \left(\epsilon\right) = \frac{32 \pi^{3}}{3} - \frac{1}{6} (4 \pi - \epsilon)^{3} \theta \left( 4 \pi - \epsilon \right) \mathrm{.} \label{eq:Example22DerivativeFinalResult}
\end{equation}
The only distinct projection volume formula for $L = 1$, despite the fact that there are $14$ accidental ambiguities possible in this order, is given by any ambiguity transformation $\bm{\uppi}^{\mathrm{II}} \in \hat{\mathcal{P}}$ for which $n_{1} = 1$ and $n_{2} = 3$ are valid. The master formula for the derivative (\ref{eq:2ndIntegralFinalResultReformulated}) yields in this case
\begin{equation}
 \frac{d \mathrm{I}_{1}}{d \epsilon} (\epsilon) = \frac{d \mathrm{I}_{2}}{d \epsilon} (\epsilon) = \frac{1}{8} \left( \epsilon^{2} - 12 \pi^{2} \right) (1 - \theta [\epsilon -2 \pi ]) \mathrm{.} \label{eq:Example13DerivativeFourPhases}
\end{equation}
Here, the choice $\bm{C} = 2 \pi$ results in the correct boundary behaviour and the projection volume is
\begin{equation}
 \tilde{\mathcal{V}}^{3}_{\bm{\uppi}^{\mathrm{II}}} \left(\epsilon\right) = \frac{32 \pi^{3}}{3} - \frac{1}{6} \left[ 64 \pi^{3} - 36 \pi^{2} \epsilon + \epsilon^{3} \right] \left[ 1 - \theta (\epsilon - 2 \pi) \right] \mathrm{.} \label{eq:Example13DerivativeFinalResult}
\end{equation}
From all the explicit examples for the evaluation of the projection integral (\ref{eq:FormalFinalExpressionProjectionVolume}) investigated in this thesis, we anticipate here without proof that
\begin{equation}
 \bm{C} = n^{<} \ast 2 \pi \mathrm{,} \label{eq:CorrectchoiceforC}
\end{equation}
may always be the correct choice for the integration constant $\bm{C}$, where the index $n^{<}$ is defined as
\begin{equation}
 n^{<} := \mathrm{min} \left( n_{1}, n_{2} \right) \mathrm{.} \label{eq:DefOfNsmaller}
\end{equation}
We note here that for all the considered example cases, even for $L > 1$, the expression resulting from the general formula (\ref{eq:2ndIntegralFinalResultReformulated}) is cut off at the the maximal value $\epsilon = n^{<} \ast 2 \pi$ by the $\theta$-functions. Choosing this cutoff-point for the integration constant $\bm{C}$ seems like a natural decision since with it, another consistent boundary condition which could be exchanged for the condition (\ref{eq:ConsistencyBoundaryCondition}) is fulfilled, namely
\begin{equation}
  \tilde{\mathcal{V}}^{(4L-1)}_{\bm{\uppi}} \left(\epsilon = n^{<} \ast 2 \pi \right) \equiv \mathcal{V}^{(4L-1)}_{\mathcal{CR}} \mathrm{.} \label{eq:ConsistencyBoundaryConditionNoII}
\end{equation}
As a final remark we would like to mention the fact that the treatment of the integrals in this appendix section, leading to the two central results (\ref{eq:CRVolumeMostGeneralEvaluationFinalStep}) and (\ref{eq:2ndIntegralFinalResultReformulated}), was far from being mathematically rigorous. No explicit investigation of the convergence and general well-posedness of the integrals as done, meaning a check of whether or not the displacement of the appearing poles into the upper half of the complex plane really yields the same result as displacing the poles into the lower half. Also, no clean mathematical estimate of the contributions of the giant semi arcs in the contour integrals (see Figure \ref{fig:ComplexPlanePolesOfIntegrand}) was done and the vanishing of those contributions was argued in a very non-rigorous way. It was also not checked whether or not commuting of the $x$- and $y$-integrations that lead to the master formula (\ref{eq:2ndIntegralFinalResultReformulated}) really yields the same result. \newline
We are here content with the fact that it was possible to derive the explicit expressions (\ref{eq:CRVolumeMostGeneralEvaluationFinalStep}) and (\ref{eq:2ndIntegralFinalResultReformulated}) and furthermore, that they passed every numerical cross-check we performed using MATHEMATICA \cite{Mathematica8,Mathematica11,MathematicaLanguage,MathematicaBonnLicense} (i.e. by evaluating the integrals from their definitions (\ref{eq:CRVolumeMostGeneralEvaluationStep1}) and (\ref{eq:PiNVolumeProjectionGeneralExpressionAppendices}) directly).

\subsubsection{Discrete ambiguities of simultaneously diagonalizable sets of observables} \label{sec:AmbiguitiesSimDiagObservables}

An important point in the discussion of the discrete partial wave ambiguities of the group $\mathcal{S}$ observables in section \ref{sec:WBTpaper} and appendix \ref{sec:DoubleAmbiguityTrafoActingOnBi} was the fact that, when written in the transversity basis $\left\{b_{1}, \ldots, b_{4}\right\}$, the four
observables $\left\{\sigma_{0}, \check{\Sigma}, \check{T}, \check{P}\right\} \equiv \left\{\check{\Omega}^{\alpha_{S}}\right\}$ are strictly sums of moduli-squared of
the $b_{i}$:
\begin{equation}
\check{\Omega}^{\alpha_{S}} \left(W, \theta\right) = \frac{1}{2} \left( \pm \left|b_{1}\right|^{2} \pm \left|b_{2}\right|^{2} \pm \left|b_{3}\right|^{2} + \left|b_{4}\right|^{2}\right) \mathrm{.} \label{eq:GroupSIsDiagonalChapter24}
\end{equation}
Therefore, in a TPWA the decomposition of the transversity amplitudes into products over linear factors allowed for the discussion of the discrete ambiguities
of the group $\mathcal{S}$, making use of the appearing roots $\left\{\alpha_{k}, \beta_{k}\right\}$. \newline
A valid question is now: Is there anything special about the 4 observables of the group $\mathcal{S}$ and the transversity amplitudes $b_{i}$, or could one also start from a different set of observables, using a different set of spin-amplitudes? \newline
It is of course known that experimentally, the group $\mathcal{S}$ observables are the most accessible (using a BT double-polarization measurement to extract $P$) and therefore indeed have a special status as compared to the remaining 12 double polarization measurements. This is however not the point that is asked for here, since the discussion shall be mainly concerned with the mathematical properties of the bilinear forms defining the observables. \newline
Mathematically speaking, the group $\mathcal{S}$ observables being sums of squares of transversity amplitudes (equation (\ref{eq:GroupSIsDiagonalChapter24})) is synonymous to the fact that the $\tilde{\Gamma}$-matrices representing them are diagonal (cf. appendix \ref{subsec:HelTrGammaReps}). Therefore a different phrasing of the above posed question would be: do subsets of 4 observables other than the group $\mathcal{S}$ observables exist, that are simultaneously diagonalizable? \newline
In the following, this question shall be answered mathematically for all combinations of 4 observables whose diagonalizing amplitudes are connected to the $\left\{b_{i}\right\}$ by a unitary linear transformation. Furthermore, the derivation of discrete partial wave ambiguities shall be discussed briefly for one example case of a diagonalizable subset.

\paragraph{Subsets of simultaneously diagonalizable polarization observables} \label{subsec:AmbiguitiesSimDiagObservablesI}  \textcolor{white}{:-)} \newline

It is well known (see for example chapter 1 of reference \cite{ShankarQM}), that two general $N \times N$-matrices $A$ $\&$ $B$, with possibly complex entries, can be diagonalized simultaneously if and only if they commute:
\begin{equation}
 \left[A,B\right] \equiv AB - BA = 0 \mathrm{.} \label{eq:KommutatorEqualsZero}
\end{equation}
This mathematical statement can be applied directly in order to search for subsets of 4 polarization observables that can be diagonalized simultaneously. Using MATHEMATICA \cite{Mathematica8,Mathematica11,MathematicaLanguage,MathematicaBonnLicense}, the following procedure was done in the course of this work: for all possible subsets of four ordered non-equal indices, labelling the considered matrices $\tilde{\Gamma}^{\alpha}$ (representing observables $\check{\Omega}^{\alpha}$), i.e.
\begin{equation}
\left\{ \left(\alpha,\beta,\gamma,\delta\right) \hspace*{2pt} | \hspace*{2pt} \alpha,\beta,\gamma,\delta \in 1,\ldots,16 \hspace*{2pt} \& \hspace*{2pt} \alpha < \beta < \gamma < \delta \right\} \mathrm{,} \label{eq:IndicesSetDefinition}
\end{equation}
it was tested whether or not the corresponding four matrices $\left\{\tilde{\Gamma}^{\alpha},\tilde{\Gamma}^{\beta},\tilde{\Gamma}^{\gamma},\tilde{\Gamma}^{\delta}\right\}$ commute among each other, i.e. if the conditions
\begin{equation}
  \left[\tilde{\Gamma}^{\alpha},\tilde{\Gamma}^{\beta}\right] = \left[\tilde{\Gamma}^{\alpha},\tilde{\Gamma}^{\gamma}\right] = \left[\tilde{\Gamma}^{\alpha},\tilde{\Gamma}^{\delta}\right] = \left[\tilde{\Gamma}^{\beta},\tilde{\Gamma}^{\gamma}\right] = \left[\tilde{\Gamma}^{\beta},\tilde{\Gamma}^{\delta}\right] = \left[\tilde{\Gamma}^{\gamma},\tilde{\Gamma}^{\delta}\right] = 0 \mathrm{,} \label{eq:ConditionCheckSimDiagObs}
\end{equation}
are fulfilled. The question may now be asked whether one has to start in the transversity representation in order to check the conditions (\ref{eq:ConditionCheckSimDiagObs}), or if the results are invariant under a change of amplitudes and one could also have investigated the observables in a representation corresponding to different amplitudes and gotten the same results. \newline
In order to investigate this, it has to be noted that the vanishing of a general commutator $\left[A, B\right] = 0$ is invariant if the matrices $A$ and $B$ are transformed with an invertible matrix $\hat{T}$ according to
\begin{equation}
A \rightarrow \hat{T} A \hat{T}^{-1} \mathrm{,} \hspace*{5pt} B \rightarrow \hat{T} B \hat{T}^{-1} \mathrm{.} \label{eq:ABSimilarityTrafo}
\end{equation}
Then, it is evident that
\begin{align}
\left[A, B\right] \rightarrow \left[\hat{T} A \hat{T}^{-1}, \hat{T} B \hat{T}^{-1}\right] &= \hat{T} A \hat{T}^{-1} \hat{T} B \hat{T}^{-1} - \hat{T} B \hat{T}^{-1} \hat{T} A \hat{T}^{-1} \nonumber \\ 
 &= \hat{T} A B \hat{T}^{-1} - \hat{T} B A \hat{T}^{-1} = \hat{T} \left[A, B\right] \hat{T}^{-1} = 0 \mathrm{,} \label{eq:InvCOmmutatorSimTrafo}
\end{align}
holds. However, a change of spin amplitudes generally does not introduce the inverse of the transformation matrix, but the hemitean conjugate. In case transversity amplitudes $\left| b \right>$ are connected to some new set of amplitudes $\left| \hat{b} \right>$ by a transformation matrix $\hat{U}$, i.e. $\left| b \right> = \hat{U} \left| \hat{b} \right>$, it is seen from the definition of profile functions as bilinear forms:
\begin{equation}
\check{\Omega}^{\alpha} = \frac{1}{2} \left< b \right| \tilde{\Gamma}^{\alpha} \left| b \right> = \frac{1}{2} \left< \hat{b} \right| \hat{U}^{\dagger} \tilde{\Gamma}^{\alpha} \hat{U} \left| \hat{b} \right> = \frac{1}{2} \left< \hat{b} \right| \hat{\Gamma}^{\alpha} \left| \hat{b} \right> \mathrm{.} \label{eq:TrafoRuleOfGammaMatricesInduced}
\end{equation}
The transformation $\tilde{\Gamma}^{\alpha} \rightarrow \hat{\Gamma}^{\alpha} = \hat{U}^{\dagger} \tilde{\Gamma}^{\alpha} \hat{U}$ is induced by the amplitude change. Therefore, the vanishing of commutators is generally preserved only if the matrix $\hat{U}$ fulfills $\left(\hat{U}^{\dagger}\right)^{-1} = \hat{U} \leftrightarrow \hat{U}^{\dagger} \hat{U} = \mathbbm{1}$, i.e. if it is unitary. \newline
It is seen that the results of the conditions (\ref{eq:ConditionCheckSimDiagObs}) are the same whether one starts in the representation corresponding to transversity- ($b_{i}$) or helicity-amplitudes ($H_{i}$). However, in case one transforms to CGLN-amplitudes, the commutators do indeed change since the amplitude-transformation-matrix is not unitary in this case. Out of this reason, we can only claim in this appendix that we have found all simultaneously diagonalizable subsets of observables that are unitarily connected to the transversity-representation. If other representations, possibly diagonalizing even different subsets of $4$ observables exist, cannot be said with certainty at this point. \newline
With the above given facts, it is valid to just use the transversity representation, i.e. the $\tilde{\Gamma}^{\alpha}$-matrices, and test for all possible index-combinations (\ref{eq:IndicesSetDefinition}) if the condition (\ref{eq:ConditionCheckSimDiagObs}) is fulfilled. The results are 15 subsets of four polarization observables that are simultaneously diagonalizable. All of them are summarized in Table \ref{tab:SimDiagSetsSummary}. \newline
A few interesting facts have to be noted regarding the found subsets. First of all, the unpolarized cross section $\sigma_{0}$ belongs to all of them. This should come as no surprise, since $\tilde{\Gamma}^{1} = \mathbbm{1}$ commutes with alle other $\tilde{\Gamma}^{\alpha}$-matrices and furthermore, one has $\hat{\Gamma}^{1} = \hat{U} \tilde{\Gamma}^{1} \hat{U}^{\dagger} = \hat{U} \hat{U}^{\dagger} = \mathbbm{1} $ (from now on the convention $\left| b \right> \rightarrow \left| \hat{b} \right>  = \hat{U} \left| b \right> $ shall be chosen). Therefore, $\sigma_{0}$ has to belong to each diagonalizable subset and it has to be represented by the unit-matrix. \newline
Regarding the general structure of the diagonalizable subsets, it is found that they can be separated according to the groups the remaining polarization observables, aside from $\sigma_{0}$, belong to. This is also indicated in Table \ref{tab:SimDiagSetsSummary}. The original diagonalizable set, i.e. the group $\mathcal{S}$ observables, is logically the only one that contains only single-spin asymmetries apart from $\sigma_{0}$. Then, for the possibility of having three additional observables that belong only to two types of polarization measurements, the possible combinations (group $\mathcal{S}$ $\&$ BT), (group $\mathcal{S}$ $\&$ $\mathcal{BR}$), and (group $\mathcal{S}$ $\&$ $\mathcal{TR}$) emerge. For each of these combinations, two subsets exist.
\begin{table}[h]
 \centering
\begin{tabular}{ccc}
\hline 
\hline \\
 Set-Nr. & Observable-groups in the set (aside from $\sigma_{0}$) & Observables \\
\hline \\
$1$ & group $\mathcal{S}$ & $\left\{\sigma_{0}, \Sigma, T, P\right\}$ \\ \\
\hline \\
$\begin{aligned} &2  \\ &3  \end{aligned}$ & group $\mathcal{S}$ $\&$ BT & $\begin{aligned} &\left\{\sigma_{0}, P, G, F\right\} \\ &\left\{\sigma_{0}, P, E, H\right\}  \end{aligned}$ \\ \\
\hline \\
$\begin{aligned} &4  \\ &5  \end{aligned}$ & group $\mathcal{S}$ $\&$ $\mathcal{BR}$ & $\begin{aligned} &\left\{\sigma_{0}, T, O_{x^{\prime}}, C_{z^{\prime}}\right\} \\ &\left\{\sigma_{0}, T, O_{z^{\prime}}, C_{x^{\prime}}\right\}  \end{aligned}$ \\ \\
\hline \\
$\begin{aligned} &6  \\ &7  \end{aligned}$ & group $\mathcal{S}$ $\&$ $\mathcal{TR}$ & $\begin{aligned} &\left\{\sigma_{0}, \Sigma, T_{x^{\prime}}, L_{z^{\prime}}\right\} \\ &\left\{\sigma_{0}, \Sigma, T_{z^{\prime}}, L_{x^{\prime}}\right\}  \end{aligned}$ \\ \\
\hline \\
$\begin{aligned} & \hspace*{6pt} 8  \\ & \hspace*{6pt} 9 \\ & \hspace*{6pt} 10 \\ & \hspace*{6pt} 11 \\ & \hspace*{6pt} 12 \\ & \hspace*{6pt} 13 \\ & \hspace*{6pt} 14 \\ & \hspace*{6pt} 15  \end{aligned}$ & BT $\&$ $\mathcal{BR}$ $\&$ $\mathcal{TR}$ & $\begin{aligned} &\left\{\sigma_{0}, E, C_{x^{\prime}}, L_{x^{\prime}}\right\} \\ &\left\{\sigma_{0}, E, C_{z^{\prime}}, L_{z^{\prime}}\right\} \\ &\left\{\sigma_{0}, G, O_{x^{\prime}}, L_{x^{\prime}}\right\} \\ &\left\{\sigma_{0}, G, O_{z^{\prime}}, L_{z^{\prime}}\right\} \\ &\left\{\sigma_{0}, H, O_{x^{\prime}}, T_{x^{\prime}}\right\} \\ &\left\{\sigma_{0}, H, O_{z^{\prime}}, T_{z^{\prime}}\right\} \\ &\left\{\sigma_{0}, F, C_{x^{\prime}}, T_{x^{\prime}}\right\} \\ &\left\{\sigma_{0}, F, C_{z^{\prime}}, T_{z^{\prime}}\right\} \\ \end{aligned}$ \\ \\
\hline \hline
\end{tabular}
\caption[List of $15$ simultaneously diagonalizable subsets of polarization observables.]{Here, all simultaneously diagonalizable subsets found as described in the main text are listed. In total there are 15 of them. The column in the middle indicates the groups to which all observables in the respective set, apart from $\sigma_{0}$, belong.}
\label{tab:SimDiagSetsSummary}
\end{table}
\clearpage
Lastly, three different classes of polarization measurements are only found for the combination (BT $\&$ $\mathcal{BR}$ $\&$ $\mathcal{TR}$). Eight subsets exist in this case. \newline \newline
Having obtained all possible simultaneously diagonalizable subsets, it is meaningful to ask how the diagonalizing spin amplitudes are connected to the transversity amplitudes $b_{i}$ in each of the 15 cases. As was seen above, the respective transformation matrices have to be unitary. Therefore, we define 15 diagonalizing systems of spin amplitudes according to
\begin{equation}
\hat{b}^{(j)}_{i} := \sum_{k} \hat{U}^{(j)}_{ik} b_{k}\mathrm{,} \hspace*{5pt} \mathrm{or} \hspace*{5pt} \left| \hat{b}^{(j)} \right> := \hat{U}^{(j)} \left| b \right> \mathrm{,} \hspace*{5pt} j = 1,\ldots,15 \mathrm{.} \label{eq:DefOfDiagonalizingAmplitudes}
\end{equation}
The procedure used to determine the unitary matrices $\hat{U}^{(j)}$ is described in the following. For $j = 9$, i.e. the set $\left\{\sigma_{0}, E, C_{z^{\prime}}, L_{z^{\prime}}\right\}$, it is known that the helicity amplitudes $H_{i}$ diagonalize it (cf. \cite{ChTab}, \cite{FTS}) and therefore the matrix $\hat{U}^{(9)}$ is taken over directly from Chiang and Tabakin \cite{ChTab}. For all the other cases, the $\hat{U}^{(j)}$ have to be constructed. \newline
In order to do this, for each ordered index combination (\ref{eq:IndicesSetDefinition}) that corresponds to one of the sets in Table \ref{tab:SimDiagSetsSummary}, the eigenvectors $\left(\bm{v}^{\beta}_{1},\bm{v}^{\beta}_{2},\bm{v}^{\beta}_{3},\bm{v}^{\beta}_{4}\right)$ of $\tilde{\Gamma}^{\beta}$ were used in order to build the diagonalizing transformation:
\begin{equation}
\hat{T}^{\left(\mathrm{I}\right)} := \left[ \begin{array}{c} \left( \bm{v}^{\beta}_{1} \right)^{T} \\ \left( \bm{v}^{\beta}_{2} \right)^{T} \\ \left( \bm{v}^{\beta}_{3} \right)^{T} \\ \left( \bm{v}^{\beta}_{4} \right)^{T}  \end{array} \right] \mathrm{,} \label{eq:DiagTrafoStepI}
\end{equation}
and the transformed $\tilde{\Gamma}^{\alpha_{i}}$-matrices (for $\alpha_{i} \in \left\{\alpha,\beta,\gamma,\delta\right\}$)
\begin{equation}
\tilde{\Gamma}^{\alpha_{i}}_{\left(\mathrm{I}\right)} := \hat{T}^{\left(\mathrm{I}\right)} \tilde{\Gamma}^{\alpha_{i}} \left( \hat{T}^{\left(\mathrm{I}\right)} \right)^{-1} \mathrm{,} \label{eq:TransformedGammasStepI}
\end{equation}
were evaluated. In case the $\tilde{\Gamma}^{\alpha_{i}}_{\left(\mathrm{I}\right)}$ were already diagonal, the matrix $\hat{T}^{\left(\mathrm{I}\right)}$ was used to evaluate $\hat{U}$ (see definition below). If this was not the case, then for one of the non-diagonal $\tilde{\Gamma}^{\alpha_{i}}_{\left(\mathrm{I}\right)}$, e.g. $\tilde{\Gamma}^{\gamma}_{\left(\mathrm{I}\right)}$, the eigenvectors $\left(\left(\bm{v}^{\left(\mathrm{I}\right)}\right)^{\gamma}_{1},\left(\bm{v}^{\left(\mathrm{I}\right)}\right)^{\gamma}_{2},\left(\bm{v}^{\left(\mathrm{I}\right)}\right)^{\gamma}_{3},\left(\bm{v}^{\left(\mathrm{I}\right)}\right)^{\gamma}_{4}\right)$ were used in order to build a second diagonalizing transformation
\begin{equation}
\hat{T}^{\left(\mathrm{II}\right)} := \left[ \begin{array}{c} \left[ \left(\bm{v}^{\left(\mathrm{I}\right)}\right)^{\gamma}_{1} \right]^{T} \\ \left[ \left(\bm{v}^{\left(\mathrm{I}\right)}\right)^{\gamma}_{2} \right]^{T} \\ \left[ \left(\bm{v}^{\left(\mathrm{I}\right)}\right)^{\gamma}_{3} \right]^{T} \\ \left[ \left(\bm{v}^{\left(\mathrm{I}\right)}\right)^{\gamma}_{4} \right]^{T}  \end{array} \right] \mathrm{,} \label{eq:DiagTrafoStepII}
\end{equation}
and transform the $\tilde{\Gamma}^{\alpha_{i}}_{\left(\mathrm{I}\right)}$ to: $\tilde{\Gamma}^{\alpha_{i}}_{\left(\mathrm{II}\right)} := \hat{T}^{\left(\mathrm{II}\right)} \tilde{\Gamma}^{\alpha_{i}}_{\left(\mathrm{I}\right)} \left( \hat{T}^{\left(\mathrm{II}\right)} \right)^{-1}$. \newline
At last, the $\tilde{\Gamma}^{\alpha_{i}}_{\left(\mathrm{II}\right)}$ were diagonal in all condsidered cases. Depending on whether the procedure terminated at step (I) or step (II), the full transformation matrix $\mathbbm{M}$ was either defined as $\mathbbm{M} = \hat{T}^{\left(\mathrm{I}\right)}$, or $\mathbbm{M} = \hat{T}^{\left(\mathrm{II}\right)} \hat{T}^{\left(\mathrm{I}\right)}$. Then, in order to get the unitary amplitude transformation matrix $\hat{U}$, this $\mathbbm{M}$ was normalized according to:
\begin{equation}
\hat{U} := \frac{1}{\left(\left| \mathrm{det} \left[\mathbbm{M}\right] \right|\right)^{1/4}} \mathbbm{M} \mathrm{.} \label{eq:DefinitionOfU}
\end{equation}
With this, in each of the 15 cases a unitary matrix $\hat{U}^{(j)}$ with $\mathrm{det}\left[\hat{U}^{(j)}\right] = \pm 1$ was obtained. \newline
The Tables \ref{tab:SimDiagSetsI} to \ref{tab:SimDiagSetsVIII} collect the results of the procedure described above. The unitary matrices $\hat{U}^{(j)}$ are given for all simultaneously diagonalizable subsets. Furthermore, the form obtained by the then diagonalized observables when written using the amplitudes $\hat{b}^{(j)}_{i}$ defined by (\ref{eq:DefOfDiagonalizingAmplitudes}) is also provided. \newline
A few remarks are in order on the definition of the matrices $\hat{U}^{(j)}$ (cf. equation (\ref{eq:DefinitionOfU})).
\begin{itemize}
 \item[$\ast)$] First of all, the $\hat{U}^{(j)}$ are not determined uniquely. Instead of normalizing $\mathbbm{M}$ as in equation (\ref{eq:DefinitionOfU}), i.e. using the modulus $\left| \mathrm{det} \left[\mathbbm{M}\right] \right|$, one may just divide by $\left( \mathrm{det} \left[\mathbbm{M}\right] \right)^{1/4}$. Then, $\mathrm{det}\left[\hat{U}^{(j)}\right] = 1$ would be valid all the time, at the price of complicating the mathematical form of the matrix-entries of the $\hat{U}^{(j)}$, in case all columns of $\mathbbm{M}$ are normalized by the same factor. The latter complication could be avoided by distributing the phase of $\left( \mathrm{det} \left[\mathbbm{M}\right] \right)^{1/4}$ differently on the columns of the matrix $\mathbbm{M}$ in the normalization process. Since the determinant is multilinear in the columns of its argument-matrix, the result would still satisfy $\mathrm{det}\left[\hat{U}^{(j)}\right] = 1$. However, there is more than one way of distribution of the phase of the normalization-factor, so again it is not possible to uniquely fix the $\hat{U}^{(j)}$ in this way. \newline
It is of course also possible to just change the numbering of the $\hat{b}^{(j)}_{i}$-amplitudes (with respect to $i$), which corresponds to a rearrangement of the rows of $\hat{U}^{(j)}$. But this source of non-uniqueness is trivial.
 \item[$\ast)$] It is not clear whether or not the found amplitudes $\hat{b}^{(j)}_{i}$ and corresponding transformation matrices $\hat{U}^{(j)}$ pertain to any particular choice of spin-quantization scheme. The change from the standard z-axis quantization of the initial and final baryon spins in the photoproduction matrix-element
\begin{center} $\left< m_{s_{f}} \right| \hat{\epsilon}_{c}^{\lambda=+1} \cdot \hat{\vec{J}} \left| m_{s_{i}} \right>$ \end{center}
to either helicity- or transversity quantum numbers, $(\lambda_{i}, \lambda_{f})$ or $(t_{i}, t_{f})$, lead naturally to systems of spin-amplitudes, the $H_{i}$ or $b_{i}$, which diagonalized certain subsets of observables (cf. chapter 2.3 of reference \cite{MyDiplomaThesis}). \newline
However, here the transformed amplitudes $\hat{b}^{(j)}_{i}$ were determined solely for the purpose of diagonalizing observables, while a change of spin-quantization axes was never included in the analysis. The idea suggests itself that probably all found diagonalizable subsets can also be diagonalized by choosing a suitable scheme of spin-quantization. However, this last point was not investigated further in the course of this work.
\end{itemize}
Despite these remarks, it is the central result of this appendix that 15 subsets of simultaneously diagonalizable observables were found, containing 4 observables each. Furthermore, the diagonalizing sets of spin amplitudes $\hat{b}^{(j)}_{i}$ have been introduced via transformations $\hat{U}^{(j)}$ acting on the transversity amplitudes $\left| b \right>$, where a consistent set of unitary matrices $\left\{\hat{U}^{(j)} \mathrm{,} \hspace*{2pt} j=1,\ldots,15\right\}$ was constructed. \newline
This result may now be used to derive the discrete partial wave ambiguities of each of the 15 subsets in a truncated partial wave analysis. In order to do this, one would have to do algebraic investigations similar to the ones originally done by Omelaenko (cf. section \ref{sec:WBTpaper}) for each case. It is however not clear from the beginning whether or not all of the algebra goes through as in the case of the group $\mathcal{S}$ observables. Rather one should explicitly perform a check for each subset under investigation. In the course of this work, not all of the cases were checked. Nonetheless, appendix \ref{subsec:AmbiguitiesSimDiagObservablesII} will treat one example case. \newline \newline
There is of course also the question in which way the results found here are useful for practical partial wave analyses. The usefulness is obvious once one of the subsets listed in Table \ref{tab:SimDiagSetsSummary} is considered. But what if the observables met in an analysis cannot be diagonalized simultaneously? It may be that a lot more subsets of 4 observables have the latter property than the 15 we found that can be diagonalized. \newline
For instance, at the time of this writing the Mainz-Tuzla PWA-collaboration (\cite{MainzTuzlaCollaboration}) is performing a partial wave analysis of the observables $\left\{\sigma_{0}, \Sigma, T, F\right\}$ for the process $\gamma p \longrightarrow \eta p$, with data taken at MAMI. 
Up to now, no sensible proposal can be given on how to treat such non-diagonalizable subsets of observables and derive all their discrete ambiguities in a truncated partial wave analysis.

%
\begin{sidewaystable}[h]
 \centering
\begin{tabular*}{\linewidth}{cc@{\extracolsep\fill}m{7.5cm}c}
\hline 
\hline \\
 Set-Nr. $j$ & Observables & \hspace*{70pt} $\hat{b}_{i}^{(j)}$-representation & Transformation matrix $\hat{U}^{(j)}$ \\
\hline
$1$ & $\left\{ \sigma_{0}, \Sigma, T, P \ \right\}$ & {\begin{align}
               \sigma_{0} &= \frac{1}{2} \left( \left| \hat{b}_{1}^{(1)} \right|^{2} + \left| \hat{b}_{2}^{(1)} \right|^{2} +\left| \hat{b}_{3}^{(1)} \right|^{2} +\left| \hat{b}_{4}^{(1)} \right|^{2} \right) \nonumber \\
               - \check{\Sigma} &= \frac{1}{2} \left( \left| \hat{b}_{1}^{(1)} \right|^{2} + \left| \hat{b}_{2}^{(1)} \right|^{2} - \left| \hat{b}_{3}^{(1)} \right|^{2} - \left| \hat{b}_{4}^{(1)} \right|^{2} \right) \nonumber \\
               - \check{T} &= \frac{1}{2} \left( - \left| \hat{b}_{1}^{(1)} \right|^{2} + \left| \hat{b}_{2}^{(1)} \right|^{2} + \left| \hat{b}_{3}^{(1)} \right|^{2} - \left| \hat{b}_{4}^{(1)} \right|^{2} \right) \nonumber \\
               \check{P} &= \frac{1}{2} \left( - \left| \hat{b}_{1}^{(1)} \right|^{2} + \left| \hat{b}_{2}^{(1)} \right|^{2} - \left| \hat{b}_{3}^{(1)} \right|^{2} + \left| \hat{b}_{4}^{(1)} \right|^{2} \right) \nonumber \end{align}} & $\hat{U}^{(1)} = \left[
\begin{array}{cccc}
 1 & 0 & 0 & 0 \\
 0 & 1 & 0 & 0 \\
 0 & 0 & 1 & 0 \\
 0 & 0 & 0 & 1 \\
\end{array} \right]$ \\
\\
\hline
\hline
\end{tabular*}
\caption[Summary for case 1 of simultaneously diagonalizable observables.]{This Table lists the first, admittedly trivial case of simultaneously diagonalizable polarization observables, the group $\mathcal{S}$ observables (Set 1 of Table \ref{tab:SimDiagSetsSummary}). The set as well as the form of its observables written in the diagonalizing amplitudes is given. Furthermore, the unitary amplitude transformation $\hat{U}^{(1)}$, which in this case is just the identity, is provided.}
\label{tab:SimDiagSetsI}
\end{sidewaystable}
\begin{sidewaystable}[h]
 \centering

\caption[Summary for cases 2 and 3 of simultaneously diagonalizable observables.]{This is the continuation of Table \ref{tab:SimDiagSetsI}. Details on the diagonalizable subsets 2 and 3 listed in the summary Table \ref{tab:SimDiagSetsSummary} are given.}
\label{tab:SimDiagSetsII}
\end{sidewaystable}
\begin{sidewaystable}[h]
 \centering
%
\caption[Summary for cases 4 and 5 of simultaneously diagonalizable observables.]{This is the continuation of Table \ref{tab:SimDiagSetsII}. Details on the diagonalizable subsets 4 and 5 listed in the summary Table \ref{tab:SimDiagSetsSummary} are given.}
\label{tab:SimDiagSetsIII}
\end{sidewaystable}
\begin{sidewaystable}[h]
 \centering
%
\caption[Summary for cases 6 and 7 of simultaneously diagonalizable observables.]{This is the continuation of Table \ref{tab:SimDiagSetsIII}. Details on the diagonalizable subsets 6 and 7 listed in the summary Table \ref{tab:SimDiagSetsSummary} are given.}
\label{tab:SimDiagSetsIV}
\end{sidewaystable}
\begin{sidewaystable}[h]
 \centering
%
\caption[Summary for cases 8 and 9 of simultaneously diagonalizable observables.]{This is the continuation of Table \ref{tab:SimDiagSetsIV}. Details on the diagonalizable subsets 8 and 9 listed in the summary Table \ref{tab:SimDiagSetsSummary} are given.}
\label{tab:SimDiagSetsV}
\end{sidewaystable}
\begin{sidewaystable}[h]
 \centering
%
\caption[Summary for cases 10 and 11 of simultaneously diagonalizable observables.]{This is the continuation of Table \ref{tab:SimDiagSetsV}. Details on the diagonalizable subsets 10 and 11 listed in the summary Table \ref{tab:SimDiagSetsSummary} are given.}
\label{tab:SimDiagSetsVI}
\end{sidewaystable}
\begin{sidewaystable}[h]
 \centering
%
\caption[Summary for cases 12 and 13 of simultaneously diagonalizable observables.]{This is the continuation of Table \ref{tab:SimDiagSetsVI}. Details on the diagonalizable subsets 12 and 13 listed in the summary Table \ref{tab:SimDiagSetsSummary} are given.}
\label{tab:SimDiagSetsVII}
\end{sidewaystable}
\begin{sidewaystable}[h]
 \centering
%
\caption[Summary for cases 14 and 15 of simultaneously diagonalizable observables.]{This is the continuation of Table \ref{tab:SimDiagSetsVII}. Details on the diagonalizable subsets 14 and 15 listed in the summary Table \ref{tab:SimDiagSetsSummary} are given.}
\label{tab:SimDiagSetsVIII}
\end{sidewaystable}

\clearpage

\paragraph{The Omelaenko procedure for one example of a diagonalizable subset} \label{subsec:AmbiguitiesSimDiagObservablesII}  \textcolor{white}{:-)} \newline

In this appendix section, the derivation of the discrete partial wave ambiguities present in a TPWA shall be performed for the example case of the simultaneously diagonalizable subset 2 contained in the summary Table \ref{tab:SimDiagSetsSummary}, appendix \ref{subsec:AmbiguitiesSimDiagObservablesI}. This is one of the two diagonalizable (group $\mathcal{S}$ $\&$ BT)-type subsets and it contains the observables
\begin{equation}
\left\{ \sigma_{0}, P, G, F \right\} \mathrm{.} \label{eq:Set2ToBeDiagonalized}
\end{equation}
These observables are diagonalized by the spin amplitudes $\left| \hat{b} \right> := \left| \hat{b}^{(2)} \right> = \hat{U}^{(2)} \left| b \right>$, where $\left| b \right>$ are the transversity amplitudes (the superscript $(2)$ on amplitudes and matrixes shall be dropped in the following). Written out explicitly, this means
\begin{equation}
 \left[ \begin{array}{c} \hat{b}_{1} \\ \hat{b}_{2} \\ \hat{b}_{3} \\ \hat{b}_{4}\end{array} \right] = \hat{U} \left[ \begin{array}{c} b_{1} \\ b_{2} \\ b_{3} \\ b_{4}\end{array} \right] = \frac{1}{\sqrt{2}} \left[ 
\begin{array}{cccc}
 0 & i & 0 & 1 \\
 i & 0 & 1 & 0 \\
 0 & -i & 0 & 1 \\
 -i & 0 & 1 & 0 \\
\end{array} \right] \left[ \begin{array}{c} b_{1} \\ b_{2} \\ b_{3} \\ b_{4}\end{array} \right] \mathrm{,} \label{eq:Set2DiagTrafoActingOnTransversityBasis}
\end{equation}
where the explicit expression for the matrix $\hat{U}$ can be found in Table \ref{tab:SimDiagSetsII} of appendix \ref{subsec:AmbiguitiesSimDiagObservablesI}. The resulting expressions for the profile functions $\check{\Omega}^{\alpha}$ of all 16 polarization observables when written in the $\hat{b}_{i}$-amplitudes, as well as the corresponding representation matrices $\hat{\Gamma}^{\alpha} = \hat{U} \tilde{\Gamma}^{\alpha} \hat{U}^{\dagger}$, are given in Tables \ref{tab:SetIIDiagonalizedI} and \ref{tab:SetIIDiagonalizedII}. As expected, in this basis the observables (\ref{eq:Set2ToBeDiagonalized}) are sums of moduli squared of the $\hat{b}_{i}$-amplitudes, which is equivalent to their $\hat{\Gamma}$-matrices being diagonal. Therefore, since a diagonal representation of the set (\ref{eq:Set2ToBeDiagonalized}) has been found, the only thing that remains to be done is to derive the linear factor decompositions of a non-redundant subset of the $\hat{b}_{i}$ in a TPWA, which shall be done in the following. \newline
Using the definitions of the $b_{i}$ in terms of CGLN-amplitudes (see equations (8) to (11) in section \ref{sec:WBTpaper}), MATHEMATICA \cite{Mathematica8,Mathematica11,MathematicaLanguage,MathematicaBonnLicense} yields for equation (\ref{eq:Set2DiagTrafoActingOnTransversityBasis})
\begin{align}
 \hat{b}_{1} \left(\theta\right) &= \frac{\mathcal{C}}{\sqrt{2}} e^{i \frac{\theta}{2}} \left[ \left(1 - i\right) \left(F_{1} \left(\theta\right) - e^{- i \theta} F_{2} \left(\theta\right) \right) + \sin\theta \left(F_{3} \left(\theta\right) + e^{- i \theta} F_{4} \left(\theta\right)\right) \right] \mathrm{,} \label{eq:B1HatOutOfMATHEMATICA} \\
 \hat{b}_{2} \left(\theta\right) &= \frac{\mathcal{C}}{\sqrt{2}} e^{- i \frac{\theta}{2}} \left[ \left(1 - i\right) \left(F_{1} \left(\theta\right) - e^{i \theta} F_{2} \left(\theta\right) \right) - \sin\theta \left(F_{3} \left(\theta\right) + e^{i \theta} F_{4} \left(\theta\right)\right) \right] \mathrm{,} \label{eq:B2HatOutOfMATHEMATICA} \\
 \hat{b}_{3} \left(\theta\right) &= \frac{\mathcal{C}}{\sqrt{2}} e^{i \frac{\theta}{2}} \left[ \left(1 + i\right) \left(F_{1} \left(\theta\right) - e^{- i \theta} F_{2} \left(\theta\right) \right) - \sin\theta \left(F_{3} \left(\theta\right) + e^{- i \theta} F_{4} \left(\theta\right)\right) \right] \mathrm{,} \label{eq:B3HatOutOfMATHEMATICA} \\
 \hat{b}_{4} \left(\theta\right) &= \frac{\mathcal{C}}{\sqrt{2}} e^{- i \frac{\theta}{2}} \left[ \left(1 + i\right) \left(F_{1} \left(\theta\right) - e^{i \theta} F_{2} \left(\theta\right) \right) + \sin\theta \left(F_{3} \left(\theta\right) + e^{i \theta} F_{4} \left(\theta\right)\right) \right] \mathrm{.} \label{eq:B4HatOutOfMATHEMATICA}
\end{align}
It is seen that the angular reflection relation that holds for the original transversity amplitudes (cf. equation (13) in section \ref{sec:WBTpaper}) is still valid for the $\hat{b}_{i}$-amplitudes in a truncation at $L = \ell_{\mathrm{max}}$:
\begin{equation}
 \hat{b}_{1} \left(\theta\right) = \hat{b}_{2} \left(- \theta\right) \mathrm{,} \hspace*{5pt} \hat{b}_{3} \left(\theta\right) = \hat{b}_{4} \left(- \theta\right) \mathrm{.} \label{eq:HatBAmplitudesAngularRelation}
\end{equation}
Using the relation $\frac{1}{\sqrt{2}} \left(1 \pm i\right) = e^{\pm i \frac{\pi}{4}}$, as well as some rearrangement of prefactors, the expressions (\ref{eq:B1HatOutOfMATHEMATICA}) to (\ref{eq:B4HatOutOfMATHEMATICA}) can be brought into a form that is a bit more suitable
%

%
\begin{table}[h]
 \centering
\begin{tabular*}{\linewidth}{c@{\extracolsep\fill}cc}
\hline 
\hline \\
 Observable & $\hat{b}_{i}$-representation $\check{\Omega}^{\alpha} = \frac{1}{2} \left< \hat{b} \right| \hat{\Gamma}^{\alpha} \left| \hat{b} \right>$ & Representation matrix $\hat{\Gamma}^{\alpha}$ \\ \\
\hline \\
 $\check{\Omega}^{1} = \sigma_{0}$ & $\frac{1}{2} \left( \left| \hat{b}_{1} \right|^{2} + \left| \hat{b}_{2} \right|^{2} +\left| \hat{b}_{3} \right|^{2} +\left| \hat{b}_{4} \right|^{2} \right)$ & $\left[
\begin{array}{cccc}
 1 & 0 & 0 & 0 \\
 0 & 1 & 0 & 0 \\
 0 & 0 & 1 & 0 \\
 0 & 0 & 0 & 1 \\
\end{array} \right]$ \\ \\
 $\check{\Omega}^{4} = - \check{\Sigma}$ & $ - \mathrm{Re} \left[ \hat{b}_{1} \hat{b}_{3}^{\ast} + \hat{b}_{2} \hat{b}_{4}^{\ast} \right]$ & $\left[
\begin{array}{cccc}
 0 & 0 & -1 & 0 \\
 0 & 0 & 0 & -1 \\
 -1 & 0 & 0 & 0 \\
 0 & -1 & 0 & 0 \\
\end{array} \right]$ \\ \\
 $\check{\Omega}^{10} = - \check{T}$ & $ \mathrm{Re} \left[ \hat{b}_{2} \hat{b}_{4}^{\ast} - \hat{b}_{1} \hat{b}_{3}^{\ast} \right]$ & $\left[
\begin{array}{cccc}
 0 & 0 & -1 & 0 \\
 0 & 0 & 0 & 1 \\
 -1 & 0 & 0 & 0 \\
 0 & 1 & 0 & 0 \\
\end{array} \right]$ \\ \\
 $\check{\Omega}^{12} = \check{P}$ & $ \frac{1}{2} \left( \left| \hat{b}_{1} \right|^{2} - \left| \hat{b}_{2} \right|^{2} + \left| \hat{b}_{3} \right|^{2} - \left| \hat{b}_{4} \right|^{2} \right)$ & $\left[
\begin{array}{cccc}
 1 & 0 & 0 & 0 \\
 0 & -1 & 0 & 0 \\
 0 & 0 & 1 & 0 \\
 0 & 0 & 0 & -1 \\
\end{array} \right]$ \\ \\
 $\check{\Omega}^{3} = \check{G}$ & $ \frac{1}{2} \left( \left| \hat{b}_{1} \right|^{2} + \left| \hat{b}_{2} \right|^{2} - \left| \hat{b}_{3} \right|^{2} - \left| \hat{b}_{4} \right|^{2} \right)$ & $\left[
\begin{array}{cccc}
 1 & 0 & 0 & 0 \\
 0 & 1 & 0 & 0 \\
 0 & 0 & -1 & 0 \\
 0 & 0 & 0 & -1 \\
\end{array} \right]$ \\ \\
 $\check{\Omega}^{5} = \check{H}$ & $ \mathrm{Im} \left[ \hat{b}_{2} \hat{b}_{4}^{\ast} - \hat{b}_{1} \hat{b}_{3}^{\ast} \right] $ & $\left[
\begin{array}{cccc}
 0 & 0 & -i & 0 \\
 0 & 0 & 0 & i \\
 i & 0 & 0 & 0 \\
 0 & -i & 0 & 0 \\
\end{array} \right]$ \\ \\
 $\check{\Omega}^{9} = \check{E}$ & $ \mathrm{Im} \left[ \hat{b}_{1} \hat{b}_{3}^{\ast} + \hat{b}_{2} \hat{b}_{4}^{\ast} \right] $ & $\left[
\begin{array}{cccc}
 0 & 0 & i & 0 \\
 0 & 0 & 0 & i \\
 -i & 0 & 0 & 0 \\
 0 & -i & 0 & 0 \\
\end{array}
 \right]$ \\ \\
 $\check{\Omega}^{11} = \check{F}$ & $ \frac{1}{2} \left( \left| \hat{b}_{1} \right|^{2} - \left| \hat{b}_{2} \right|^{2} - \left| \hat{b}_{3} \right|^{2} + \left| \hat{b}_{4} \right|^{2} \right) $ & $\left[
\begin{array}{cccc}
 1 & 0 & 0 & 0 \\
 0 & -1 & 0 & 0 \\
 0 & 0 & -1 & 0 \\
 0 & 0 & 0 & 1 \\
\end{array} \right]$ \\ \\
\hline
\hline
\end{tabular*}
\caption[Group $\mathcal{S}$- and $\mathcal{BT}$ observables in an amplitude-basis which diagonalizes \newline $\left\{ \sigma_{0}, P, G, F \right\}$.]{This Table lists the group $\mathcal{S}$ and BT observables written in the spin amplitudes $\left| \hat{b} \right> := \left| \hat{b}^{(2)} \right> = \hat{U}^{(2)} \left| b \right> $. Matrices $\hat{\Gamma}^{\alpha}$ that represent the observables via $\check{\Omega}^{\alpha} = \frac{1}{2} \left< \hat{b} \right| \hat{\Gamma}^{\alpha} \left| \hat{b} \right>$ are given in the right column.}
\label{tab:SetIIDiagonalizedI}
\end{table}
\begin{table}[h]
 \centering
\begin{tabular*}{\linewidth}{c@{\extracolsep\fill}cc}
\hline 
\hline \\
 Observable & $\hat{b}_{i}$-representation $\check{\Omega}^{\alpha} = \frac{1}{2} \left< \hat{b} \right| \hat{\Gamma}^{\alpha} \left| \hat{b} \right>$ & Representation matrix $\hat{\Gamma}^{\alpha}$ \\ \\
\hline \\
 $\check{\Omega}^{14} = \check{O}_{x^{\prime}}$ & $ \mathrm{Im} \left[ \hat{b}_{1} \hat{b}_{2}^{\ast} - \hat{b}_{3} \hat{b}_{4}^{\ast} \right] $ & $\left[
\begin{array}{cccc}
 0 & i & 0 & 0 \\
 -i & 0 & 0 & 0 \\
 0 & 0 & 0 & -i \\
 0 & 0 & i & 0 \\
\end{array} \right]$ \\ \\
 $\check{\Omega}^{7} = - \check{O}_{z^{\prime}}$ & $ \mathrm{Re} \left[ \hat{b}_{1} \hat{b}_{2}^{\ast} - \hat{b}_{3} \hat{b}_{4}^{\ast} \right] $ & $\left[
\begin{array}{cccc}
 0 & 1 & 0 & 0 \\
 1 & 0 & 0 & 0 \\
 0 & 0 & 0 & -1 \\
 0 & 0 & -1 & 0 \\
\end{array} \right]$ \\ \\
 $\check{\Omega}^{16} = - \check{C}_{x^{\prime}}$ & $ \mathrm{Re} \left[ \hat{b}_{1} \hat{b}_{4}^{\ast} - \hat{b}_{2} \hat{b}_{3}^{\ast} \right] $ & $\left[
\begin{array}{cccc}
 0 & 0 & 0 & 1 \\
 0 & 0 & -1 & 0 \\
 0 & -1 & 0 & 0 \\
 1 & 0 & 0 & 0 \\
\end{array} \right]$ \\ \\
 $\check{\Omega}^{2} = - \check{C}_{z^{\prime}}$ & $ \mathrm{Im} \left[ \hat{b}_{1} \hat{b}_{4}^{\ast} + \hat{b}_{2} \hat{b}_{3}^{\ast} \right] $ & $\left[
\begin{array}{cccc}
 0 & 0 & 0 & i \\
 0 & 0 & i & 0 \\
 0 & -i & 0 & 0 \\
 -i & 0 & 0 & 0 \\
\end{array} \right]$ \\ \\
 $\check{\Omega}^{6} = - \check{T}_{x^{\prime}}$ & $ \mathrm{Re} \left[ \hat{b}_{1} \hat{b}_{4}^{\ast} + \hat{b}_{2} \hat{b}_{3}^{\ast} \right] $ & $\left[ 
\begin{array}{cccc}
 0 & 0 & 0 & 1 \\
 0 & 0 & 1 & 0 \\
 0 & 1 & 0 & 0 \\
 1 & 0 & 0 & 0 \\
\end{array} \right]$ \\ \\
 $\check{\Omega}^{13} = - \check{T}_{z^{\prime}}$ & $ \mathrm{Im} \left[ \hat{b}_{1} \hat{b}_{4}^{\ast} - \hat{b}_{2} \hat{b}_{3}^{\ast} \right] $ & $\left[ 
\begin{array}{cccc}
 0 & 0 & 0 & i \\
 0 & 0 & -i & 0 \\
 0 & i & 0 & 0 \\
 -i & 0 & 0 & 0 \\
\end{array} \right]$ \\ \\
 $\check{\Omega}^{8} = \check{L}_{x^{\prime}}$ & $ \mathrm{Im} \left[ \hat{b}_{1} \hat{b}_{2}^{\ast} + \hat{b}_{3} \hat{b}_{4}^{\ast} \right] $ & $\left[
\begin{array}{cccc}
 0 & i & 0 & 0 \\
 -i & 0 & 0 & 0 \\
 0 & 0 & 0 & i \\
 0 & 0 & -i & 0 \\
\end{array} \right]$ \\ \\
 $\check{\Omega}^{15} = \check{L}_{z^{\prime}}$ & $ - \mathrm{Re} \left[ \hat{b}_{1} \hat{b}_{2}^{\ast} + \hat{b}_{3} \hat{b}_{4}^{\ast} \right] $ & $\left[
\begin{array}{cccc}
 0 & -1 & 0 & 0 \\
 -1 & 0 & 0 & 0 \\
 0 & 0 & 0 & -1 \\
 0 & 0 & -1 & 0 \\
\end{array} \right]$ \\ \\
\hline
\hline
\end{tabular*}
\caption[$\mathcal{BR}$- and $\mathcal{TR}$- observables in an amplitude-basis which diagonalizes \newline $\left\{ \sigma_{0}, P, G, F \right\}$.]{This is the continuation of Table \ref{tab:SetIIDiagonalizedI}. Listed are the $\mathcal{BR}$- and $\mathcal{TR}$ observables written in the spin amplitudes $\hat{b}_{i}$.}
\label{tab:SetIIDiagonalizedII}
\end{table}
\clearpage
for what is to follow:
\begin{align}
 \hat{b}_{1} \left(\theta\right) &= \mathcal{C} e^{- i \frac{\pi}{4}} e^{i \frac{\theta}{2}} \left[ \left(F_{1} \left(\theta\right) - e^{- i \theta} F_{2} \left(\theta\right) \right) + \frac{e^{i \frac{\pi}{4}}}{\sqrt{2}} \sin\theta \left(F_{3} \left(\theta\right) + e^{- i \theta} F_{4} \left(\theta\right)\right) \right] \mathrm{,} \label{eq:B1HatRearranged} \\
 \hat{b}_{2} \left(\theta\right) &= \mathcal{C} e^{- i \frac{\pi}{4}} e^{- i \frac{\theta}{2}} \left[ \left(F_{1} \left(\theta\right) - e^{i \theta} F_{2} \left(\theta\right) \right) - \frac{e^{i \frac{\pi}{4}}}{\sqrt{2}} \sin\theta \left(F_{3} \left(\theta\right) + e^{i \theta} F_{4} \left(\theta\right)\right) \right] \mathrm{,} \label{eq:B2HatRearranged} \\
 \hat{b}_{3} \left(\theta\right) &= \mathcal{C} e^{i \frac{\pi}{4}} e^{i \frac{\theta}{2}} \left[ \left(F_{1} \left(\theta\right) - e^{- i \theta} F_{2} \left(\theta\right) \right) - \frac{e^{- i \frac{\pi}{4}}}{\sqrt{2}} \sin\theta \left(F_{3} \left(\theta\right) + e^{- i \theta} F_{4} \left(\theta\right)\right) \right] \mathrm{,} \label{eq:B3HatRearranged} \\
 \hat{b}_{4} \left(\theta\right) &= \mathcal{C} e^{i \frac{\pi}{4}} e^{- i \frac{\theta}{2}} \left[ \left(F_{1} \left(\theta\right) - e^{i \theta} F_{2} \left(\theta\right) \right) + \frac{e^{- i \frac{\pi}{4}}}{\sqrt{2}} \sin\theta \left(F_{3} \left(\theta\right) + e^{i \theta} F_{4} \left(\theta\right)\right) \right] \mathrm{.} \label{eq:B4HatRearranged}
\end{align}
Using the angular reflection property (\ref{eq:HatBAmplitudesAngularRelation}), one can choose $\hat{b}_{2}$ and $\hat{b}_{4}$ as the non-redundant spin amplitudes. Considering the equations (\ref{eq:B2HatRearranged}) and (\ref{eq:B4HatRearranged}), it is seen that the CGLN-amplitudes and therefore the angular dependence coming from their standard multipole-expansion, enter in both amplitdes $\hat{b}_{2}$ and $\hat{b}_{4}$ via the same terms $\left(F_{1} \left(\theta\right) - e^{i \theta} F_{2} \left(\theta\right) \right)$ and $\sin\theta \left(F_{3} \left(\theta\right) + e^{i \theta} F_{4} \left(\theta\right)\right)$. It is therefore mandatory to investigate the form of these two terms in a truncation of the multipole expansion at some arbitrary $L \equiv \ell_{\mathrm{max}}$. \newline
The same algebraic steps as performed in appendix A of section \ref{sec:WBTpaper} lead to the insight that, provided the angular variable is changed according to $\cos \theta \rightarrow t = \tan \frac{\theta}{2}$, both terms mentioned above are essentially finite polynomials in $t$, up to angular dependent prefactors:
\begin{align}
\left(F_{1} \left(\theta\right) - e^{i \theta} F_{2} \left(\theta\right) \right) &= \frac{1}{\left(1 + t^{2}\right)^{L}} \times A_{2L} \left(t\right) \mathrm{,} \label{eq:FundamentalAPolynomial} \\
\sin\theta \left(F_{3} \left(\theta\right) + e^{i \theta} F_{4} \left(\theta\right) \right) &= \frac{1}{\left(1 + t^{2}\right)^{L}} \times t \times B_{2L - 2} \left(t\right) \mathrm{,} \label{eq:FundamentalBPolynomial}
\end{align}
with polynomials
\begin{equation}
 A_{2L} \left(t\right) := \sum_{k=0}^{2L} a_{k} t^{k} \mathrm{,} \hspace*{5pt} B_{2L} \left(t\right) := \sum_{n=0}^{2L - 2} b_{n} t^{n} \mathrm{.} \label{eq:ABPolysSetIIDefinitions}
\end{equation}
The intermediate steps of the calculation as well as explicit expressions for the coefficients of $A_{2L}$ and $B_{2L - 2}$ have not been shown here in favor of brevity. The amplitudes $\hat{b}_{2}$ and $\hat{b}_{4}$, when expressed in terms of $A_{2L}$ and $B_{2L - 2}$, read
\begin{align}
 \hat{b}_{2} \left(\theta\right) &= \mathcal{C} \frac{e^{- i \frac{\pi}{4}} e^{- i \frac{\theta}{2}}}{\left(1 + t^{2}\right)^{L}} \left[ A_{2L} \left(t\right) - \frac{e^{i \frac{\pi}{4}}}{\sqrt{2}} \times t \times B_{2 L -2} \left(t\right)  \right] \mathrm{,} \label{eq:B2DecompABPolynomials} \\
\hat{b}_{4} \left(\theta\right) &= \mathcal{C} \frac{e^{i \frac{\pi}{4}} e^{- i \frac{\theta}{2}}}{\left(1 + t^{2}\right)^{L}} \left[ A_{2L} \left(t\right) + \frac{e^{- i \frac{\pi}{4}}}{\sqrt{2}} \times t \times B_{2 L -2} \left(t\right)  \right] \mathrm{.} \label{eq:B4DecompABPolynomials}
\end{align}
This form of the non-redundant spin amplitudes recommends the following definition of new polynomials $C_{2L}$ and $D_{2L}$:
\begin{align}
 C_{2L} \left(t\right) &:= A_{2L} \left(t\right) - \frac{e^{i \frac{\pi}{4}}}{\sqrt{2}} \times t \times B_{2 L -2} \left(t\right) \equiv \sum_{k = 0}^{2L} c_{k} t^{k} \mathrm{,} \label{eq:DefCPolynomial} \\
 D_{2L} \left(t\right) &:= A_{2L} \left(t\right) + \frac{e^{- i \frac{\pi}{4}}}{\sqrt{2}} \times t \times B_{2 L -2} \left(t\right) \equiv \sum_{k = 0}^{2L} d_{k} t^{k} \mathrm{.} \label{eq:DefDPolynomial} 
\end{align}
A structure very similar to the investigation of the discrete partial wave ambiguities of the group $\mathcal{S}$ observables (section \ref{sec:WBTpaper}) appears, namely two polynomials that have the same leading and free complex coefficients (as is apparent from their definitions (\ref{eq:DefCPolynomial}) and (\ref{eq:DefDPolynomial})), i.e.:
\begin{equation}
c_{2L} = d_{2L} \mathrm{,} \hspace*{5pt} C_{2L} \left(t = 0\right) = D_{2L} \left(t = 0\right) \leftrightarrow c_{0} = d_{0} \mathrm{.} \label{eq:LeadingAndFreeTermsOfCDAreEqual}
\end{equation}
It is now possible to define the normalized polynomials
\begin{equation}
\hat{C}_{2L} \left(t\right) := \frac{C_{2L} \left(t\right)}{c_{2L}} = t^{2L} + \sum_{k=0}^{2L - 1} \hat{c}_{k} t^{k} \mathrm{,} \hspace*{5pt} \hat{D}_{2L} \left(t\right) := \frac{D_{2L} \left(t\right)}{c_{2L}} = t^{2L} + \sum_{k=0}^{2L - 1} \hat{d}_{k} t^{k} \mathrm{,} \label{eq:DefOfNormalizedCDPolynomials}
\end{equation}
which by means of the fundamental theorem of algebra are seen to fully decompose into their linear factors
\begin{equation}
 \hat{C}_{2L} \left(t\right) = \prod_{k=1}^{2L} \left( t - \hat{\alpha}_{k} \right) \mathrm{,} \hspace*{5pt} \hat{D}_{2L} \left(t\right) = \prod_{k^{\prime}=1}^{2L} \left( t - \hat{\beta}_{k^{\prime}} \right) \mathrm{.} \label{eq:CDPolynomialsLinFactDecompositions}
\end{equation}
The hat on the roots $\hat{\alpha}_{k}$ and $\hat{\beta}_{k^{\prime}}$ was chosen in accordance with the whole notation for quantities belonging to set 2 (i.e. equation (\ref{eq:Set2ToBeDiagonalized})) in this appendix. With everything done above, one arrives at the following linear factor decompositions of the non-redundant amplitudes $\hat{b}_{2}$ and $\hat{b}_{4}$
\begin{align}
 \hat{b}_{2} \left(\theta\right) &= \mathcal{C} c_{2L} \frac{e^{- i \frac{\pi}{4}} e^{- i \frac{\theta}{2}}}{\left(1 + t^{2}\right)^{L}} \prod_{k=1}^{2L} \left( t - \hat{\alpha}_{k} \right) \mathrm{,} \label{eq:B2LinFactDecomp} \\
\hat{b}_{4} \left(\theta\right) &= \mathcal{C} c_{2L} \frac{e^{i \frac{\pi}{4}} e^{- i \frac{\theta}{2}}}{\left(1 + t^{2}\right)^{L}} \prod_{k^{\prime}=1}^{2L} \left( t - \hat{\beta}_{k^{\prime}} \right) \mathrm{,} \label{eq:B4LinFactDecomp}
\end{align}
while the remaining $\hat{b}$-amplitudes can be reached via $\hat{b}_{1} (t) = \hat{b}_{2} (-t)$ and $\hat{b}_{3} (t) = \hat{b}_{4} (-t)$ (or equivalently, by reversing the signs of the roots $\hat{\alpha}_{k}$ and $\hat{\beta}_{k^{\prime}}$). \newline
The equality of the free terms of $C_{2L}$ and $D_{2L}$ (second relation in (\ref{eq:LeadingAndFreeTermsOfCDAreEqual})) implies a multiplicative constraint among the roots exactly as in Omelaenko's treatment of the group $\mathcal{S}$ observabes:
\begin{equation}
\prod_{k=1}^{2L} \hat{\alpha}_{k} = \prod_{k=1}^{2L} \hat{\beta}_{k} \mathrm{.} \label{eq:ConsistencyRelationSet2}
\end{equation}
This constraint is seen to be always fulfilled only by the double ambiguity of the observables (\ref{eq:Set2ToBeDiagonalized})
\begin{equation}
\hat{\alpha}_{k} \rightarrow \hat{\alpha}_{k}^{\ast} \hspace*{5pt} \& \hspace*{5pt} \hat{\beta}_{k} \rightarrow \hat{\beta}_{k}^{\ast} \hspace*{5pt} \forall k = 1,\ldots,2L \mathrm{.} \label{eq:DoubleAmbTrafoSet2}
\end{equation}
The number of possible further (accidental) ambiguities, arising from the complex conjugation of only a subset of the roots $\left\{\hat{\alpha}_{k},\hat{\beta}_{k}\right\}$,
may again be reduced by the constraint (\ref{eq:ConsistencyRelationSet2}) in the same way as described in appendix \ref{sec:AccidentalAmbProofs} for the group $\mathcal{S}$ observables. \newline
We close this appendix section by a description of the response of the polarization observables $\check{\Omega}^{\alpha}$ on the double ambiguity transformation (\ref{eq:DoubleAmbTrafoSet2}). Following a derivation similar to the one in appendix \ref{sec:DoubleAmbiguityTrafoActingOnBi}, it becomes clear that for the diagonalizable subset (\ref{eq:Set2ToBeDiagonalized}) the double ambiguity transformation (\ref{eq:DoubleAmbTrafoSet2}) again acts as a $\theta$-dependent antilinear transformation on the $\hat{b}_{i}$-amplitudes, i.e.
\begin{align}
\hat{b}_{i} \left(\theta\right) &\rightarrow \hat{b}^{\mathrm{D.A.}}_{i} \left(\theta\right) = \sum_{j} \left[ \hat{\mathcal{A}} \left(\theta\right) \right]_{ij} \hat{b}^{\ast}_{j} \left(\theta\right) \mathrm{,} \hspace*{5pt} \mathrm{with} \label{eq:Set2DoubleAmbGeneral} \\
\hat{\mathcal{A}} \left(\theta\right) &\equiv \hat{\mathcal{A}}^{(2)} \left(\theta\right)  = e^{i \left(2 \varphi_{\mathcal{C}} + \frac{\pi}{2} \right)} \left[ \begin{array}{cccc}
 \left(- \right) e^{i \theta} & 0 & 0 & 0 \\
 0 & \left(- \right) e^{- i \theta} & 0 & 0 \\
 0 & 0 & e^{i \theta} & 0 \\
 0 & 0 & 0 & e^{- i \theta} \\
\end{array} \right] \mathrm{.} \label{eq:Set2DoubleAmbMatrix}
\end{align}
The matrix $\hat{\mathcal{A}} \left(\theta\right) = \hat{\mathcal{A}}^{(2)} \left(\theta\right)$ (for set 2, cf. Table \ref{tab:SimDiagSetsSummary}), attains a form similar to the matrix (\ref{eq:DoubleAmbTrafoMatrix}) that describes the action of the double ambiguity of the group $\mathcal{S}$ observables, expect for the fact that here the elements of the diagonal matrix are rotated by different constant phases and furthermore the signs of the arguments of the exponentials are reversed. \newline
In order to test the invariance of the polarization observables $\check{\Omega}^{\alpha}$ under the transformation defined by (\ref{eq:Set2DoubleAmbGeneral}) and (\ref{eq:Set2DoubleAmbMatrix}), one has to check whether or not the invariance condition analogous to equation (\ref{eq:ConditionInvAntilinAmb}) is fulfilled:
\begin{equation}
 \hat{\Gamma}^{\alpha} = \left(\hat{\mathcal{A}}^{\dagger} \left(\theta\right) \hat{\Gamma}^{\alpha} \hat{\mathcal{A}} \left(\theta\right) \right)^{T} \mathrm{.} \label{eq:SetIIConditionInvAntilinerAmbiguity}
\end{equation}
The responses of the observables to the double ambiguity transformation are collected in Tables \ref{tab:GroupSDoubleAmbiguityMatricesSet2I} to \ref{tab:GroupSDoubleAmbiguityMatricesSet2III}, where the results of the right hand side of equation (\ref{eq:SetIIConditionInvAntilinerAmbiguity}) are listed and compared to the original $\hat{\Gamma}^{\alpha}$-matrices. \newline \newline
To summarize the overall behaviour of the observables:
\begin{itemize}
 \item[(i)] The two group $\mathcal{S}$ observables $\sigma_{0}$ and $\check{P}$ contained in the set (\ref{eq:Set2ToBeDiagonalized}) as well as all BT observables $\left\{\check{G},\check{H},\check{E},\check{F}\right\}$ are seen to be invariant under the double ambituity transformation (\ref{eq:Set2DoubleAmbMatrix}).
 \item[(ii)] Two group $\mathcal{S}$ observables, $\check{\Sigma}$ and $\check{T}$, change their sign under (\ref{eq:Set2DoubleAmbMatrix}).
 \item[(iii)] The remaining observables of the types $\mathcal{BR}$ and $\mathcal{TR}$ show angular dependent mixing patterns similar to the ones found for the double ambiguity of the group $\mathcal{S}$ observables in appendix \ref{sec:DoubleAmbiguityTrafoActingOnBi}.
\end{itemize}
In case one wishes to find mathematically complete sets of observables and therefore, as discussed in appendix \ref{sec:AccidentalAmbProofs}, may be allowed to disregards the accidental ambiguitites, it is again
possible to construct such sets by adding to the diagonalized subset (\ref{eq:Set2ToBeDiagonalized}) one observable that can resolve the double ambiguity (\ref{eq:Set2DoubleAmbMatrix}). \newline
Therefore, it is directly seen to be possible to find new complete sets that arise out of the discussion of the discrete partial wave ambiguities of subset (\ref{eq:Set2ToBeDiagonalized}), the completeness of which is new compared to the 
considerations of ambiguities of the group $\mathcal{S}$, section \ref{sec:WBTpaper}. Examples for these new complete sets would be
\begin{equation}
 \left\{\sigma_{0}, \check{\Sigma}, \check{P}, \check{G}, \check{F}\right\} \hspace*{5pt} \mathrm{and} \hspace*{5pt} \left\{\sigma_{0}, \check{T}, \check{P}, \check{G}, \check{F}\right\} \mathrm{.} \label{eq:NewCompleteSets}
\end{equation}
These two sets again consist out of just 5 observables. Furthermore, the need for any $\mathcal{BR}$- or $\mathcal{TR}$ observable is excluded in order to obtain completeness, at least in the mathematically precise sense. \newline
However, it is also apparent that certain overcomplete sets are similar for the investigations starting at the group $\mathcal{S}$ observables and the diagonalizable subset (\ref{eq:Set2ToBeDiagonalized}), for example
\begin{equation}
 \left\{\sigma_{0}, \check{\Sigma}, \check{T}, \check{P}, \check{G}, \check{F}\right\} \mathrm{.} \label{eq:CommonOvercompleteSet}
\end{equation}
The existence of such common complete sets should come as no surprise, since the solvability criteria of the standard form of the observables in a TPWA ((\ref{eq:LowEAssocLegStandardParametrization1}) and (\ref{eq:LowEAssocLegStandardParametrization2})) have to be independent of the spin amplitudes in terms of which the observables were represented before the truncation of the multipole series was done. Once the equations for the truncated partial wave expansion are brought to their standard form, it should be simply irrelevant which basis for the full amplitudes was used in order to derive it.
\begin{table}[h]
 \centering
\begin{tabular*}{\linewidth}{c@{\extracolsep\fill}cc}
\hline 
\hline \\
 Observable $\check{\Omega}^{\alpha}$ & $\hat{\Gamma}^{\alpha}$ & $\left( \hat{\mathcal{A}} \left(\theta\right)^{\dagger} \hat{\Gamma}^{\alpha} \hat{\mathcal{A}} \left(\theta\right) \right)^{T}$ \\ \\
\hline \\
 $\check{\Omega}^{1} = \sigma_{0}$ & $ \hat{\Gamma}^{1} = \left[
\begin{array}{cccc}
 1 & 0 & 0 & 0 \\
 0 & 1 & 0 & 0 \\
 0 & 0 & 1 & 0 \\
 0 & 0 & 0 & 1 \\
\end{array}
\right] $ & $ \left[
\begin{array}{cccc}
 1 & 0 & 0 & 0 \\
 0 & 1 & 0 & 0 \\
 0 & 0 & 1 & 0 \\
 0 & 0 & 0 & 1 \\
\end{array}
\right] = \hat{\Gamma}^{1} $ \\ \\
 $\check{\Omega}^{4} = - \check{\Sigma}$ & $ \hat{\Gamma}^{4} = \left[
\begin{array}{cccc}
 0 & 0 & -1 & 0 \\
 0 & 0 & 0 & -1 \\
 -1 & 0 & 0 & 0 \\
 0 & -1 & 0 & 0 \\
\end{array}
\right] $ & $ \left[
\begin{array}{cccc}
 0 & 0 & 1 & 0 \\
 0 & 0 & 0 & 1 \\
 1 & 0 & 0 & 0 \\
 0 & 1 & 0 & 0 \\
\end{array}
\right] = - \hat{\Gamma}^{4} $ \\ \\
 $\check{\Omega}^{10} = - \check{T}$ & $ \hat{\Gamma}^{10} = \left[
\begin{array}{cccc}
 0 & 0 & -1 & 0 \\
 0 & 0 & 0 & 1 \\
 -1 & 0 & 0 & 0 \\
 0 & 1 & 0 & 0 \\
\end{array}
\right] $ & $  \left[
\begin{array}{cccc}
 0 & 0 & 1 & 0 \\
 0 & 0 & 0 & -1 \\
 1 & 0 & 0 & 0 \\
 0 & -1 & 0 & 0 \\
\end{array}
\right] = - \hat{\Gamma}^{10} $ \\ \\
 $\check{\Omega}^{12} = \check{P}$ & $ \hat{\Gamma}^{12} = \left[
\begin{array}{cccc}
 1 & 0 & 0 & 0 \\
 0 & -1 & 0 & 0 \\
 0 & 0 & 1 & 0 \\
 0 & 0 & 0 & -1 \\
\end{array}
\right] $ & $ \left[
\begin{array}{cccc}
 1 & 0 & 0 & 0 \\
 0 & -1 & 0 & 0 \\
 0 & 0 & 1 & 0 \\
 0 & 0 & 0 & -1 \\
\end{array}
\right] = \hat{\Gamma}^{12} $ \\ \\
 $\check{\Omega}^{3} = \check{G}$ & $ \hat{\Gamma}^{3} = \left[
\begin{array}{cccc}
 1 & 0 & 0 & 0 \\
 0 & 1 & 0 & 0 \\
 0 & 0 & -1 & 0 \\
 0 & 0 & 0 & -1 \\
\end{array}
\right] $ & $ \left[
\begin{array}{cccc}
 1 & 0 & 0 & 0 \\
 0 & 1 & 0 & 0 \\
 0 & 0 & -1 & 0 \\
 0 & 0 & 0 & -1 \\
\end{array}
\right] = \hat{\Gamma}^{3} $ \\ \\
 $\check{\Omega}^{5} = \check{H}$ & $ \hat{\Gamma}^{5} = \left[
\begin{array}{cccc}
 0 & 0 & -i & 0 \\
 0 & 0 & 0 & i \\
 i & 0 & 0 & 0 \\
 0 & -i & 0 & 0 \\
\end{array}
\right] $ & $ \left[
\begin{array}{cccc}
 0 & 0 & -i & 0 \\
 0 & 0 & 0 & i \\
 i & 0 & 0 & 0 \\
 0 & -i & 0 & 0 \\
\end{array}
\right] = \hat{\Gamma}^{5} $ \\ \\
 $\check{\Omega}^{9} = \check{E}$ & $ \hat{\Gamma}^{9} = \left[
\begin{array}{cccc}
 0 & 0 & i & 0 \\
 0 & 0 & 0 & i \\
 -i & 0 & 0 & 0 \\
 0 & -i & 0 & 0 \\
\end{array}
\right] $ & $ \left[
\begin{array}{cccc}
 0 & 0 & i & 0 \\
 0 & 0 & 0 & i \\
 -i & 0 & 0 & 0 \\
 0 & -i & 0 & 0 \\
\end{array}
\right] = \hat{\Gamma}^{9} $ \\ \\
 $\check{\Omega}^{11} = \check{F}$ & $ \hat{\Gamma}^{11} = \left[
\begin{array}{cccc}
 1 & 0 & 0 & 0 \\
 0 & -1 & 0 & 0 \\
 0 & 0 & -1 & 0 \\
 0 & 0 & 0 & 1 \\
\end{array}
\right] $ & $ \left[
\begin{array}{cccc}
 1 & 0 & 0 & 0 \\
 0 & -1 & 0 & 0 \\
 0 & 0 & -1 & 0 \\
 0 & 0 & 0 & 1 \\
\end{array}
\right] = \hat{\Gamma}^{11} $ \\ \\
\hline
\hline
\end{tabular*}
\caption[Effect of the double ambiguity transformation on the group $\mathcal{S}$- and $\mathcal{BT}$ observables in an alternative spin-amplitude basis.]{This Table shows the response of the group $\mathcal{S}$ and BT observables on the double ambiguity transformation ((\ref{eq:Set2DoubleAmbGeneral}) $\&$ (\ref{eq:Set2DoubleAmbMatrix})). The column in the middle shows the $\hat{\Gamma}^{\alpha}$-matrices corresponding to the observables and on the right the result of the ambiguity test, the right hand side of equation (\ref{eq:SetIIConditionInvAntilinerAmbiguity}), is shown.}
\label{tab:GroupSDoubleAmbiguityMatricesSet2I}
\end{table}
\begin{sidewaystable}[h]
 \centering
\begin{tabular*}{\linewidth}{c@{\extracolsep\fill}cc}
\hline 
\hline \\
 Observable $\check{\Omega}^{\alpha}$ & $\hat{\Gamma}^{\alpha}$ & $\left( \hat{\mathcal{A}} \left(\theta\right)^{\dagger} \hat{\Gamma}^{\alpha} \hat{\mathcal{A}} \left(\theta\right) \right)^{T}$ \\ \\
\hline \\
 $\check{\Omega}^{14} = \check{O}_{x^{\prime}}$ & $ \hat{\Gamma}^{14} = \left[
\begin{array}{cccc}
 0 & i & 0 & 0 \\
 -i & 0 & 0 & 0 \\
 0 & 0 & 0 & -i \\
 0 & 0 & i & 0 \\
\end{array}
\right]$ & $ \left[
\begin{array}{cccc}
 0 & -i e^{2 i \theta } & 0 & 0 \\
 i e^{-2 i \theta } & 0 & 0 & 0 \\
 0 & 0 & 0 & i e^{2 i \theta } \\
 0 & 0 & -i e^{-2 i \theta } & 0 \\
\end{array}
\right] = \sin(2 \theta ) \hat{\Gamma}^{7} - \cos(2 \theta ) \hat{\Gamma}^{14} $ \\ \\
 $\check{\Omega}^{7} = - \check{O}_{z^{\prime}}$ & $ \hat{\Gamma}^{7} = \left[
\begin{array}{cccc}
 0 & 1 & 0 & 0 \\
 1 & 0 & 0 & 0 \\
 0 & 0 & 0 & -1 \\
 0 & 0 & -1 & 0 \\
\end{array}
\right] $ & $ \left[
\begin{array}{cccc}
 0 & e^{2 i \theta } & 0 & 0 \\
 e^{-2 i \theta } & 0 & 0 & 0 \\
 0 & 0 & 0 & -e^{2 i \theta } \\
 0 & 0 & -e^{-2 i \theta } & 0 \\
\end{array}
\right]  = \cos(2 \theta ) \hat{\Gamma}^{7} + \sin(2 \theta ) \hat{\Gamma}^{14} $ \\ \\
 $\check{\Omega}^{16} = - \check{C}_{x^{\prime}}$ & $ \hat{\Gamma}^{16} = \left[
\begin{array}{cccc}
 0 & 0 & 0 & 1 \\
 0 & 0 & -1 & 0 \\
 0 & -1 & 0 & 0 \\
 1 & 0 & 0 & 0 \\
\end{array}
\right] $ & $ \left[
\begin{array}{cccc}
 0 & 0 & 0 & -e^{2 i \theta } \\
 0 & 0 & e^{-2 i \theta } & 0 \\
 0 & e^{2 i \theta } & 0 & 0 \\
 -e^{-2 i \theta } & 0 & 0 & 0 \\
\end{array}
\right] = - \sin(2 \theta ) \hat{\Gamma}^{2} - \cos(2 \theta ) \hat{\Gamma}^{16} $ \\ \\
 $\check{\Omega}^{2} = - \check{C}_{z^{\prime}}$ & $ \hat{\Gamma}^{2} = \left[
\begin{array}{cccc}
 0 & 0 & 0 & i \\
 0 & 0 & i & 0 \\
 0 & -i & 0 & 0 \\
 -i & 0 & 0 & 0 \\
\end{array}
\right] $ & $ \left[
\begin{array}{cccc}
 0 & 0 & 0 & i e^{2 i \theta } \\
 0 & 0 & i e^{-2 i \theta } & 0 \\
 0 & -i e^{2 i \theta } & 0 & 0 \\
 -i e^{-2 i \theta } & 0 & 0 & 0 \\
\end{array}
\right] = \cos(2 \theta ) \hat{\Gamma}^{2} - \sin(2 \theta ) \hat{\Gamma}^{16} $ \\ \\
\hline
\hline
\end{tabular*}
\caption[Effect of the double ambiguity transformation on the $\mathcal{BR}$ observables in an alternative spin-amplitude basis.]{This is the continuation of Table \ref{tab:GroupSDoubleAmbiguityMatricesSet2I} for the $\mathcal{BR}$ observables.}
\label{tab:GroupSDoubleAmbiguityMatricesSet2II}
\end{sidewaystable}
\begin{sidewaystable}[h]
 \centering
\begin{tabular*}{\linewidth}{c@{\extracolsep\fill}cc}
\hline 
\hline \\
 Observable $\check{\Omega}^{\alpha}$ & $\hat{\Gamma}^{\alpha}$ & $\left( \hat{\mathcal{A}} \left(\theta\right)^{\dagger} \hat{\Gamma}^{\alpha} \hat{\mathcal{A}} \left(\theta\right) \right)^{T}$ \\ \\
\hline \\
 $\check{\Omega}^{6} = - \check{T}_{x^{\prime}}$ & $ \hat{\Gamma}^{6} = \left[
\begin{array}{cccc}
 0 & 0 & 0 & 1 \\
 0 & 0 & 1 & 0 \\
 0 & 1 & 0 & 0 \\
 1 & 0 & 0 & 0 \\
\end{array}
\right] $ & $ \left[
\begin{array}{cccc}
 0 & 0 & 0 & -e^{2 i \theta } \\
 0 & 0 & -e^{-2 i \theta } & 0 \\
 0 & -e^{2 i \theta } & 0 & 0 \\
 -e^{-2 i \theta } & 0 & 0 & 0 \\
\end{array}
\right] = - \cos(2 \theta ) \hat{\Gamma}^{6} - \sin(2 \theta ) \hat{\Gamma}^{13} $ \\ \\
 $\check{\Omega}^{13} = - \check{T}_{z^{\prime}}$ & $ \hat{\Gamma}^{13} = \left[
\begin{array}{cccc}
 0 & 0 & 0 & i \\
 0 & 0 & -i & 0 \\
 0 & i & 0 & 0 \\
 -i & 0 & 0 & 0 \\
\end{array}
\right] $ & $ \left[
\begin{array}{cccc}
 0 & 0 & 0 & i e^{2 i \theta } \\
 0 & 0 & -i e^{-2 i \theta } & 0 \\
 0 & i e^{2 i \theta } & 0 & 0 \\
 -i e^{-2 i \theta } & 0 & 0 & 0 \\
\end{array}
\right] = - \sin(2 \theta ) \hat{\Gamma}^{6} + \cos(2 \theta ) \hat{\Gamma}^{13} $ \\ \\
 $\check{\Omega}^{8} = \check{L}_{x^{\prime}}$ & $ \hat{\Gamma}^{8} = \left[
\begin{array}{cccc}
 0 & i & 0 & 0 \\
 -i & 0 & 0 & 0 \\
 0 & 0 & 0 & i \\
 0 & 0 & -i & 0 \\
\end{array}
\right] $ & $ \left[
\begin{array}{cccc}
 0 & -i e^{2 i \theta } & 0 & 0 \\
 i e^{-2 i \theta } & 0 & 0 & 0 \\
 0 & 0 & 0 & -i e^{2 i \theta } \\
 0 & 0 & i e^{-2 i \theta } & 0 \\
\end{array}
\right] = - \cos(2 \theta ) \hat{\Gamma}^{8} - \sin(2 \theta ) \hat{\Gamma}^{15} $ \\ \\
 $\check{\Omega}^{15} = \check{L}_{z^{\prime}}$ & $ \hat{\Gamma}^{15} = \left[
\begin{array}{cccc}
 0 & -1 & 0 & 0 \\
 -1 & 0 & 0 & 0 \\
 0 & 0 & 0 & -1 \\
 0 & 0 & -1 & 0 \\
\end{array}
\right] $ & $ \left[
\begin{array}{cccc}
 0 & -e^{2 i \theta } & 0 & 0 \\
 -e^{-2 i \theta } & 0 & 0 & 0 \\
 0 & 0 & 0 & -e^{2 i \theta } \\
 0 & 0 & -e^{-2 i \theta } & 0 \\
\end{array}
\right] = - \sin(2 \theta ) \hat{\Gamma}^{8} + \cos(2 \theta ) \hat{\Gamma}^{15} $ \\ \\
\hline
\hline
\end{tabular*}
\caption[Effect of the double ambiguity transformation on the $\mathcal{TR}$ observables in an alternative spin-amplitude basis.]{This is the continuation of Table \ref{tab:GroupSDoubleAmbiguityMatricesSet2I} for the $\mathcal{TR}$ observables.}
\label{tab:GroupSDoubleAmbiguityMatricesSet2III}
\end{sidewaystable}

\clearpage

\subsection{Additional material for chapter \ref{chap:TPWA}} \label{sec:AdditionsChapterIV}

\subsubsection{Reduction of the correlated $\chi^{2}$-function by diagonalization of the covariance matrix} \label{sec:CorrelatedChi2Reduction}

The correlated $\chi^{2}$-function from section \ref{sec:TPWAFitsIntro}, equation (\ref{eq:CorrelatedChisquare}), is defined as
\begin{equation}
\chi^{2} = \sum_{i,j} \Big[ \left(a_{L}^{\mathrm{Fit}}\right)_{i} - \left< \mathcal{M}_{\ell} \right| \left(C_{L}\right)_{i} \left| \mathcal{M}_{\ell} \right> \Big] \mathrm{\textbf{C}}^{-1}_{ij} \Big[ \left(a_{L}^{\mathrm{Fit}}\right)_{j} - \left< \mathcal{M}_{\ell} \right| \left(C_{L}\right)_{j} \left| \mathcal{M}_{\ell} \right> \Big] \mathrm{.} \label{eq:CorrelatedChisquareAppendix}
\end{equation}
By way of its definition, the covariance matrix for any vector of random variables $\left( X_{1}, \ldots, X_{j} \right)$ is real and symmetric:
\begin{equation}
 \mathrm{\textbf{C}}_{ij} = \mathrm{E} \left[ \left( X_{i} - \mathrm{E} \left[ X_{i} \right] \right)  \left( X_{j} - \mathrm{E} \left[ X_{j} \right] \right) \right] \mathrm{,} \label{eq:CovMatrixDefAppendix}
\end{equation}
where $\mathrm{E} \left[ X \right]$ denotes the expectation value of the random variable $X$. In order to reduce the expression (\ref{eq:CorrelatedChisquareAppendix}) to an equivalent function that has the mathematical form of an error-weighted $\chi^{2}$ \cite{BlobelLohrmann}, one imposes the standard knowledge from linear algebra \cite{FalkoLorenz2} that every real and symmetric matrix $\mathrm{\textbf{C}}$ can be diagonalized by an orthogonal matrix $\mathrm{O}$, built from the eigenvectors of $\mathrm{\textbf{C}}$, by means of the transformation
\begin{equation}
 \mathrm{O}^{-1} \left( \mathrm{\textbf{C}}^{-1} \right) \mathrm{O} \equiv \mathrm{O}^{T} \left( \mathrm{\textbf{C}}^{-1} \right) \mathrm{O} =: \bm{D} \mathrm{.} \label{eq:DiagCorrMatrixAppendix}
\end{equation}
The diagonal matrix $\bm{D}$ has been defined in the last step. Tellinghuisen \cite{TellinghuisenCorrelatedFit} is another reference for the reduction described here. Defining the residual-vectors
\begin{equation}
 \bm{\Delta}_{i} := \left(a_{L}^{\mathrm{Fit}}\right)_{i} - \left< \mathcal{M}_{\ell} \right| \left(C_{L}\right)_{i} \left| \mathcal{M}_{\ell} \right> \mathrm{,} \label{eq:DefResidualVectorAppendix}
\end{equation}
it can be seen that the original correlated $\chi^{2}$-function (\ref{eq:CorrelatedChisquareAppendix}) is equal to
{\allowdisplaybreaks
\begin{align}
 \chi^{2} &= \bm{\Delta}^{T} \left( \mathrm{\textbf{C}}^{-1} \right) \bm{\Delta} = \bm{\Delta}^{T} \mathbbm{1} \left( \mathrm{\textbf{C}}^{-1} \right) \mathbbm{1} \bm{\Delta} = \bm{\Delta}^{T} \mathrm{O} \mathrm{O}^{T} \left( \mathrm{\textbf{C}}^{-1} \right) \mathrm{O} \mathrm{O}^{T} \bm{\Delta} \nonumber \\
  &= \bm{\Delta}^{T} \mathrm{O} \bm{D} \mathrm{O}^{T} \bm{\Delta} \equiv \bm{\tilde{\Delta}}^{T} \bm{D} \bm{\tilde{\Delta}} \mathrm{,} \label{eq:CorrMatrixDiagonalizationInChiSquare}
\end{align}
}
where the ``rotated'' residual vectors $\bm{\tilde{\Delta}} := \mathrm{O}^{T} \bm{\Delta} $ were defined in the last step. Furthermore, elaborating the final result of the calculation above a bit more, it is seen that the mathematical form of an error weighted $\chi^{2}$ emerges

{\allowdisplaybreaks
\begin{align}
 \chi^{2} &= \sum_{i} \bm{D}_{ii} \left( \bm{\tilde{\Delta}}_{i} \right)^{2} = \sum_{i} \bm{D}_{ii} \left( \sum_{k} \mathrm{O}^{T}_{ik} \left[ \left(a_{L}^{\mathrm{Fit}}\right)_{k} - \left< \mathcal{M}_{\ell} \right| \left(C_{L}\right)_{k} \left| \mathcal{M}_{\ell} \right> \right] \right)^{2} \nonumber \\
 &=  \sum_{i} \bm{D}_{ii} \left( \left[ \sum_{k} \mathrm{O}^{T}_{ik} \left(a_{L}^{\mathrm{Fit}}\right)_{k} \right] - \left< \mathcal{M}_{\ell} \right| \left[ \sum_{k} \mathrm{O}^{T}_{ik}  \left(C_{L}\right)_{k} \right] \left| \mathcal{M}_{\ell} \right>  \right)^{2} \mathrm{,} \label{eq:ReductionToDiagonalChisquare}
\end{align}
}
with weights given by the only non-vanishing matrix elements $\bm{D}_{ii}$ of $\bm{D}$. The weights are therefore also equal to the eigenvalues of $\mathrm{\textbf{C}}$. \newline
The form (\ref{eq:ReductionToDiagonalChisquare}) can now be implemented with every search-algorithm that may have problems with the manifestly non-diagonal function (\ref{eq:CorrelatedChisquareAppendix}). One only has to be careful to employ the rotated Legendre-coefficients and TPWA fit-matrices
\begin{equation}
 \left(a_{L}^{\mathrm{Fit}}\right)_{i}^{R} := \sum_{j} \mathrm{O}^{T}_{ij} \left(a_{L}^{\mathrm{Fit}}\right)_{j} \mathrm{,} \hspace*{5pt} \left(C_{L}\right)_{i}^{R} := \sum_{j} \mathrm{O}^{T}_{ij} \left(C_{L}\right)_{j} \mathrm{.} \label{eq:DefRotatedFitParametersAndMatrices}
\end{equation}
The latter are originating from the expressions listed in Appendix \ref{sec:TPWAFormulae}. It is important to observe that matrix-indices of $\left(C_{L}\right)_{j}$ are suppressed in equation (\ref{eq:DefRotatedFitParametersAndMatrices}). Rather, full matrices are linearly combined using the orthogonal matrix $\mathrm{O}^{T}$.

\clearpage

\subsubsection{Algorithms for the sampling of the $\bar{\sigma}$-ellipsoid, as well as sorting redundant solutions out of the TPWA solution-pool} \label{sec:MCSamplingAlgorithms}

Here, we list two algorithms, mentioned in section \ref{sec:MonteCarloSampling} of the main text, in the form of pseudo-code. Notations for loops and logical inquiries should be self-explanatory. Whenever a variable is (re-) assigned with some value, this is represented by the symbol $\gets$. \newline
Furthermore we should clarify our notation for multipole parameter-vectors since the latter, apart from an index labelling the components, acquire a new index $i$ because of the pool of $N_{MC}$ start-con\-fi\-gu\-ra\-tions (and later solutions) mentioned in the main text. \newline
We label here the pools of start-parameters and the solution-pools by different notations. The initial parameter-vectors for phase-contained multipoles are then denoted as
\begin{equation}
 \left( \mathcal{M}_{\ell}^{C,0} \right)_{i} := \left( \mathrm{Re} \left[ E_{0+}^{C} \right]^{0}_{i}, \mathrm{Re} \left[ E_{1+}^{C} \right]^{0}_{i}, \ldots, \mathrm{Im} \left[ M_{L-}^{C} \right]^{0}_{i} \right)^{T} \mathrm{,} \label{eq:InitialParVectDefinition}
\end{equation}
while $\left[ \left( \mathcal{M}_{\ell}^{C,0} \right)_{i} \right]_{n}$ indicates the individual components. The solutions lying in the pool upon fitting are labelled according to the same scheme, but without the subscript $0$. \newline We now come to the algorithm used to sample the $\bar{\sigma}$-ellipsoid. For the case of a general truncation at $L$, with phase-constrained multipoles fitted and no partial waves assigned to model parameters, it takes the form shown in \textbf{Algorithm \ref{alg:SigmaBarEllipsoidAlgorithm}}. \newline
\begin{algorithm}[h]
\caption{Sampling the $\bar{\sigma}$-ellipsoid}\label{alg:SigmaBarEllipsoidAlgorithm}
\begin{algorithmic}[1]
\State $\bar{\sigma}_{\mathrm{rest}} \gets \bar{\sigma}$ \Comment{Initialize $\bar{\sigma}_{\mathrm{rest}}$ by the number TCS from data}
\For{$i \gets 1, \ldots, N_{MC}$} \Comment{$N_{MC}$-loop} 
\For{$n \gets 2,\ldots,(8 L - 1)$} \Comment{Parameter-loop}
\State $\left[ \left( \mathcal{M}_{\ell}^{C,0} \right)_{i} \right]_{n} \gets \mathrm{RandomReal}\left[ - \sqrt{\frac{\bar{\sigma}_{\mathrm{rest}}}{c_{n}}}, \sqrt{\frac{\bar{\sigma}_{\mathrm{rest}}}{c_{n}}} \right]$ \Comment{Draw parameter}
\State $\bar{\sigma}_{\mathrm{rest}} \gets \bar{\sigma}_{\mathrm{rest}} - \left[ \left( \mathcal{M}_{\ell}^{C,0} \right)_{i} \right]^{2}_{n}$ \Comment{Re-assign $\bar{\sigma}_{\mathrm{rest}}$}
\EndFor
\State $\left[ \left( \mathcal{M}_{\ell}^{C,0} \right)_{i} \right]_{1} \gets + \sqrt{\frac{\bar{\sigma}_{\mathrm{rest}}}{c_{1}}}$ \Comment{Assign remainder to $\mathrm{Re} \left[ E_{0+} \right]^{0}_{i}$}
\EndFor
\end{algorithmic}
\end{algorithm}
All parameters of the multipoles except $\mathrm{Re} \left[ E_{0+} \right]^{0}_{i}$ are drawn from a random number generator operating on a suitably defined interval. In the definition of the latter, the normalizations $c_{n}$ appear which can just be read off the equation (\ref{eq:TCSInTermsOfMultsLZeroExpansion}) (section \ref{sec:MonteCarloSampling}). In the main text, this was done for the example $L=1$ in equations (\ref{eq:SWaveParInt}) to (\ref{eq:MultParIntsLmax1M1Minus}). One can see quickly in simple examples that this algorithm does \textit{not} sample the ellipsoid uniformly. \newline
The second algorithm mentioned in the main text operates on the solution-pool and tries, in this very simple version, to sort all solutions out of the pool that are redundant, i.e. distinct only due to numerical scatter in the minimization routine. Before presenting it, we define the euclidean norm between two parameter vectors
\begin{equation}
 d_{E} \left[ \left( \mathcal{M}_{\ell}^{C} \right)_{i}, \left( \mathcal{M}_{\ell}^{C} \right)_{j} \right] := \sqrt{ \sum_{n = 1}^{(8L-1)} \left( \left[ \left( \mathcal{M}_{\ell}^{C} \right)_{i} \right]_{n} - \left[ \left( \mathcal{M}_{\ell}^{C} \right)_{j} \right]_{n} \right)^{2} } \mathrm{.} \label{eq:EuclideanNormAppChap4}
\end{equation}
Furthermore, a useful concept from the mathematical literature \cite{ForsterII} is a so-called \textit{$\epsilon$-ball}, which generalized to any $\mathbbm{R}^{n}$ and using the euclidean distance is defined as
\begin{equation}
 B_{\epsilon} \left( x \right) := \left\{ y \in \mathbbm{R}^{n} \hspace*{1.75pt} \big| \hspace*{1.75pt} d_{E} \left[x,y\right] < \epsilon \right\} \mathrm{.} \label{eq:EpsilonBallDefinition}
\end{equation}
Our algorithm treats solutions that lie within each others $B_{\epsilon}$ as redundant and tries to get rid of them. The only open problem before running it, is how to fix $\epsilon$ suitably. This is in principle a matter of choice, but it may be helpful to orient it on the numerical precision under which the fit-routine was run before. In \textit{FindMinimum} \cite{MathematicaLanguage} for instance, it is possible to adjust the AccuracyGoal and PrecisionGoal options by integers $a$ and $p$ such that the convergence criterium can be fulfilled for any two iterations $x_{n}$ and $x_{m}$ of each individual parameter, with distance
\begin{equation}
 \left| x_{n} - x_{m} \right| \leq \mathrm{max} \left( 10^{- a}, 10^{- p} \hspace*{1.75pt} \left| x_{n} \right| \right) \mathrm{.} \label{eq:ConvCritMATHDefinition}
\end{equation}
Locking the integer $p$ at the largest possible value (formally $\infty$), the minimizer tries to determine each parameter to an accuracy of $x_{\mathrm{acc.}} \equiv 10^{-a}$. We estimate the minimally allowed coordinate distance for each parameter of two non-redundant solutions to be
\begin{equation}
 \Delta x_{\mathrm{coo.}}^{\mathrm{min.}} = 100 \hspace*{1pt} x_{\mathrm{acc.}} \mathrm{,} \label{eq:EstimateForMinCooDist}
\end{equation}
which is already quite generous. Then, the minimal euclidean distance for $(8L-1)$ multipole-parameters can be evaluated from (\ref{eq:EstimateForMinCooDist}) and is then a good estimate for $\epsilon$
\begin{equation}
 d_{E}^{\mathrm{min}} = \sqrt{\sum_{n=1}^{(8 L - 1)} \Delta x_{\mathrm{coo.}}^{\mathrm{min.}}} = \sqrt{100 \hspace*{1pt} (8L-1)} \sqrt{x_{\mathrm{acc.}}} = \sqrt{(8L-1)} 10^{1-\frac{a}{2}} =: \epsilon \mathrm{.} \label{eq:DefEpsilon}
\end{equation}
Different, even larger choices for $\epsilon$ are also possible. \textbf{Algorithm \ref{alg:SortOutAlgorithm}} listed below has worked in many cases where it correctly sorted redundant solutions out of the solution-pool.
\begin{algorithm}
\caption{Sorting out redundant solutions}\label{alg:SortOutAlgorithm}
\begin{algorithmic}[1]
\State $\epsilon \gets \mathrm{const}$ \Comment{Give suitably chosen constant value to $\epsilon$}
\State $N_{\mathrm{nonred}} \gets 1$
\State $\mathrm{NonRedSols} \left[ N_{\mathrm{nonred}} \right] \gets \left\{ \left( \chi^{2}_{\mathcal{M}} \right)_{1}, \left( \mathcal{M}_{\ell}^{C} \right)_{1} \right\}$
\For{$i \gets 2, \ldots, N_{MC}$} \Comment{Loop through solution-pool}
\State $\mathrm{Flag} \gets 1$
\For{$j \gets 1,\ldots,N_{\mathrm{nonred}}$} \Comment{Loop through non-redundant solutions}
\If{$d_{E} \left[ \mathrm{NonRedSols} \left[ j \right][2], \left( \mathcal{M}_{\ell}^{C} \right)_{i} \right] < \epsilon$}
\State $\mathrm{Flag} \gets 0$
\EndIf
\EndFor
\If{$\mathrm{Flag} = 1$}
\State $N_{\mathrm{nonred}} \gets N_{\mathrm{nonred}} + 1$
\State $\mathrm{NonRedSols} \left[ N_{\mathrm{nonred}} \right] \gets \left\{ \left( \chi^{2}_{\mathcal{M}} \right)_{i}, \left( \mathcal{M}_{\ell}^{C} \right)_{i} \right\}$
\EndIf
\EndFor
\end{algorithmic}
\end{algorithm}

\clearpage

\subsubsection{Comments on the bootstrap-Ansatz chosen in this work} \label{sec:BootstrapAnsatzComments}

In the following, three aspects concerning the chosen probability model (\ref{eq:TPWAEmpDistrFunctDefinition}) and (\ref{eq:NormalDistDef}), as well as the resampling-Ansatz in equations (\ref{eq:TPWABootstrapDataset}) to (\ref{eq:BootstrapFitStepOneAndTwo}) of section \ref{sec:BootstrappingIntroduction}, are elaborated: \newline
\begin{itemize}
 \item[1.)] The profile functions $\check{\Omega}^{\alpha} = \sigma_{0} \hspace*{1pt} \Omega^{\alpha}$ are products of two observables which, as we assume here, correspond to uncorrelated and normal distributed random variables. Then, in the most general case, the product of such random variables does \textit{not} follow a gaussian distribution. \newline
 To see this, one considers two uncorrelated random variables $X_{1}$ and $X_{2}$ drawn from the distribution functions
\begin{equation}
 \mathcal{N} \left( \mu_{1}, \sigma_{1} \right) \longrightarrow X_{1} \mathrm{,} \hspace*{5pt} \mathcal{N} \left( \mu_{2}, \sigma_{2} \right) \longrightarrow X_{2} \mathrm{.} \label{eq:TwoUncorrelatedNormalRandomVars}
\end{equation}
According to well-known methods of probability theory, the probability distribution function of the product random variable $Z = X_{1} X_{2}$ can be formulated as the integral \cite{MishaPrivComm,MDSpringerAlgebraRandomVar}
\begin{equation}
 \mathrm{P}_{X_{1} X_{2}} (u) = \int_{- \infty}^ {\infty} dx \int_{- \infty}^ {\infty} dy \hspace*{1pt} \delta (x y - u) \hspace*{1pt} \frac{\exp \left[{-\frac{\left( x - \mu_{1} \right)^{2}}{2 \sigma_{1}^{2}}}\right]}{\sqrt{2 \pi} \sigma_{1}} \frac{\exp \left[{-\frac{\left( y - \mu_{2} \right)^{2}}{2 \sigma_{2}^{2}}}\right]}{\sqrt{2 \pi} \sigma_{2}} \mathrm{.} \label{eq:ProductPDFGeneralExpression}
\end{equation}
For general values of $\mu_{1}$, $\mu_{2}$, $\sigma_{1}$ and $\sigma_{2}$ this integral is quite involved. In case both means $\mu_{i}$ vanish, the integral becomes a modified Bessel function of the second kind. In both cases, the result is clearly non-gaussian. \newline
Thus, one may wonder under which circumstances the Ansatz (\ref{eq:TPWAEmpDistrFunctDefinition}) is justified and, even approximately, correct. Here, our assumption of a normal distribution for $\check{\Omega}^{\alpha}$ is validated by the fact that, at least for the photoproduction channel $\gamma p \longrightarrow \pi^{0} p$ considered in this work, the modern experiments can determine the unpolarized cross section $\sigma_{0}$ with high statistical precision. Typical values for statistical errors are $\Delta \sigma_{0} \simeq \left(0.1 \hspace*{1pt} ,\ldots, \hspace*{1pt} 0.2 \right) \left[\frac{\mu b}{sr}\right]$ (cf. \cite{Adlarson:2015,Hornidge:2013}). Therefore, except for the threshold-region, where cross sections are small, one generally has the situation that $\Delta \sigma_{0} \ll \sigma_{0}$. \newline
To see that this really helps the approximation (\ref{eq:TPWAEmpDistrFunctDefinition}), we now adapt the formula to the specific case of a polarization measurement. The random variable $X_{1}$ corresponds to $\sigma_{0}$, i.e. it has $\mu_{1} > 0$ and $\sigma_{1} \ll \mu_{1}$ and therefore especially $\sigma_{1} \ll 1 \left[\frac{\mu b}{sr}\right]$. The dimensionless asymmetry $\Omega^{\alpha}$ is represented by $X_{2}$. Thus it may have arbitrary mean $\mu_{2}$ in the interval $\left[ -1, 1 \right]$. Especially, we do not forbid the case $\mu_{2} = 0$. \newline
Formally, it is of course possible to study the limit of $\sigma_{1}$ going to zero. Then, the distribution function of the product, eq. (\ref{eq:ProductPDFGeneralExpression}), becomes
\allowdisplaybreaks
\begin{align}
 \lim\limits_{\sigma_{1} \to 0}{ \mathrm{P}_{X_{1} X_{2}} (u)} &=  \lim\limits_{\sigma_{1} \to 0}{ \int_{- \infty}^ {\infty} dx \int_{- \infty}^ {\infty} dy \hspace*{1pt} \delta (x y - u) \hspace*{1pt} \frac{\exp \left[{-\frac{\left( x - \mu_{1} \right)^{2}}{2 \sigma_{1}^{2}}}\right]}{\sqrt{2 \pi} \sigma_{1}} \frac{\exp \left[{-\frac{\left( y - \mu_{2} \right)^{2}}{2 \sigma_{2}^{2}}}\right]}{\sqrt{2 \pi} \sigma_{2}}} \nonumber \\
  &= \int_{- \infty}^ {\infty} dx \int_{- \infty}^ {\infty} dy \hspace*{1pt} \delta (x y - u) \hspace*{1pt} \lim\limits_{\sigma_{1} \to 0}{ \frac{\exp \left[{-\frac{\left( x - \mu_{1} \right)^{2}}{2 \sigma_{1}^{2}}}\right]}{\sqrt{2 \pi} \sigma_{1}}} \frac{\exp \left[{-\frac{\left( y - \mu_{2} \right)^{2}}{2 \sigma_{2}^{2}}}\right]}{\sqrt{2 \pi} \sigma_{2}} \nonumber \\
  &= \int_{- \infty}^ {\infty} dx \int_{- \infty}^ {\infty} dy \hspace*{1pt} \delta (x y - u) \hspace*{1pt} \delta \left( x - \mu_{1} \right) \frac{\exp \left[{-\frac{\left( y - \mu_{2} \right)^{2}}{2 \sigma_{2}^{2}}}\right]}{\sqrt{2 \pi} \sigma_{2}} \nonumber \\
  &= \int_{- \infty}^ {\infty} dy \hspace*{1pt} \delta (\mu_{1} y - u) \hspace*{1pt} \frac{\exp \left[{-\frac{\left( y - \mu_{2} \right)^{2}}{2 \sigma_{2}^{2}}}\right]}{\sqrt{2 \pi} \sigma_{2}} \nonumber \\
  &= \frac{\exp \left[{-\frac{\left( u - \mu_{1} \mu_{2} \right)^{2}}{2 \left( \left| \mu_{1} \right| \sigma_{2}\right)^{2}}}\right]}{\sqrt{2 \pi} \left| \mu_{1} \right| \sigma_{2}} = \mathcal{N} \left( \mu_{1} \mu_{2}, \mu_{1} \sigma_{2} \right)  \mathrm{,} \label{eq:ProductPDFApproximation}
\end{align}
i.e. one obtains a standard normal distribution with mean $\mu_{1} \mu_{2}$ and standard error $\left| \mu_{1} \right| \sigma_{2} = \mu_{1} \sigma_{2}$, since $\mu_{1} > 0$ was assumed. \newline
For real data, one of course never has $\Delta \sigma_{0} = 0$. However, in case the statistical errors of the cross section  $\sigma_{0}$ as mentioned above, it can very well be that the limiting case (\ref{eq:ProductPDFApproximation}) is fulfilled in an already quite good approximation. \newline
Nonetheless, in practical analyses, one should always check whether the boot\-strap-Ansatz chosen here is justified. This means, one draws synthetic data for $\sigma_{0}$ and $\Omega^{\alpha}$ individually and then investigates the resulting distribution of $\check{\Omega}^{\alpha}$. For any practical dataset fitted in this work (see sections \ref{subsec:DeltaRegionDataFits} and \ref{subsec:2ndResRegionDataFits}), the gaussianity of the latter distributions was given up to a good approximation.
\item[2.)] We attempt to anchor the parametric bootstrap described in section \ref{sec:BootstrappingIntroduction} a little bit more in the available statistical literature. \newline For a fit to some dataset $\left\{ (x_{i}, y_{i}) \big| i = 1,\ldots,n) \right\}$, where each datapoint has a standard error $\sigma_{i}$, the scheme amounts to drawing new datapoints from a normal distribution centered at each individual datapoint
\begin{equation}
 \mathcal{N} \left( y_{i}, \sigma_{i} \right) \longrightarrow y_{i}^{\ast} \mathrm{.} \label{eq:PhysicistsBootstrap}
\end{equation}
The values $x_{i}$ remain fixed in this procedure. In the literature, there exists one variant of the bootstrap that looks very similar to the scheme above. This is the {\it wild bootstrap} first introduced by Wu \cite{WuWildBootstrap}. Here, a fit to the original data first yields fit-function values $\hat{y}_{i} = f (x_{i}; \hat{\beta})$ as well as estimated raw residuals $\hat{\epsilon}_{i}$. Then, a wild bootstrap can be done by generating bootstrap datasets according to the prescription
\begin{equation}
 y_{i}^{\ast} = \hat{y}_{i} + \hat{\epsilon}_{i} \hspace*{1pt} v_{i} \mathrm{.} \label{eq:WildBootstrap1}
\end{equation}
Here, $v_{i}$ denotes a random variable drawn from an a priori arbitrary, discrete or continuous distribution. The only requirement Wu demands this function to fulfill is that it has mean $0$ and variance $1$. A natural choice would be given by drawing from a normal distribution
\begin{equation}
 \mathcal{N} \left( 0 , 1 \right) \longrightarrow v_{i} \mathrm{,} \label{eq:WildBootstrap2}
\end{equation}
but other choices are possible as well. We note here the fact that the parametric bootstrap (\ref{eq:PhysicistsBootstrap}) used in this work can be re-written in the wild bootstrap form (\ref{eq:WildBootstrap1}). However, the random variables $v_{i}$ are in this case drawn from the following distribution function
\begin{equation}
 \mathcal{N} \left( 1 , \frac{\sigma_{i}}{\left| \hat{\epsilon}_{i} \right|} \right) \longrightarrow v_{i} \mathrm{,} \label{eq:PhysicistsWildBootstrap2}
\end{equation}
which can be seen to violate Wu's requirements. Therefore, it is seen that while the parametric bootstrap used in this thesis is quite similar to a wild bootstrap, it is not really quite exactly a wild bootstrap.
\item[3.)] In section \ref{sec:BootstrappingIntroduction} of the main text, it was mentioned that in case correlations exist within the data itself and in case such correlations are known and quantified, that they should be incorporated into the bootstrap probability model $\hat{\mathrm{P}}$ used for the resampling analysis. \newline
In the bootstrap-Ansatz chosen in this work, bootstrap-data are drawn at each angle individually, without taking into account neighboring angles. One may thus ask the question, whether or not any quantifiable statistical correlations between adjacent points ($=$ bins) exist in the angular distributions of polarization observables. \newline
We quote here a recent PhD-thesis \cite{JanPhD}, which studies such statistical angular correlations for polarization data taken with the Crystal-Barrel detector. According to this work, the angular resolutions of such modern detectors is so good, that for standard-binnings of $10$ to $20$ angle-bins, statistical correlations among different datapoints are negligible in a very good approximation.
\end{itemize}

\clearpage

\subsubsection{Results of selected TPWA bootstrap-analyses} \label{sec:NumericalTPWAFitResults}

In this appendix, more details on the results of some selected bootstrap-TPWA fits to real datasets for the reaction $\gamma p \rightarrow \pi^{0} p$ are presented. They provide more information on results mentioned in section \ref{sec:RealWorldDataFits}. Results are sorted according to the considered energy-regions.

\paragraph{$\Delta$-region}  \textcolor{white}{:-)} \newline

Here, results are collected for some specific fits to the set of five polarization observables $\left\{ \sigma_{0}, \Sigma, T, P, F \right\}$ \cite{Hornidge:2013,LeukelPhD,Leukel:2001,Schumann:2015,Belyaev:1983} in the $\Delta$-resonance region, see section \ref{subsec:DeltaRegionDataFits} of the main text. \newline
The results of a fully unconstrained TPWA with all $S$- and $P$-wave multipoles varied in the fit (i.e. $\ell_{\mathrm{max}} = 2$) are shown in Figure \ref{fig:Lmax2UnconstrainedFitMultipoleResultsDeltaRegionPlethora}. These plots are complementary to those shown in Figure \ref{fig:Lmax2UnconstrainedFitResultsDeltaRegion} of the main text. \newline
After that, numerical results (Tables \ref{tab:DeltaRegionResultsFirstEnergy} to \ref{tab:DeltaRegionResultsFifthSixthSeventhEnergy}) and histograms with bootstrap distributions of fit-parameters (Figures \ref{fig:BootstrapHistosDeltaRegionEnergy1} to \ref{fig:BootstrapHistosDeltaRegionEnergies6and7}) are shown for a reduced boostrap-TPWA discussed towards the end of section \ref{subsec:DeltaRegionDataFits}. The starting point in these bootstrap-analyses was the global minimum of the fit shown in Figure \ref{fig:Lmax2DWavesSAIDFitResultsDeltaRegionPurelyStat}, with $D$-waves fixed to the model SAID CM12 \cite{WorkmanEtAl2012ChewMPhotoprod,SAID}.

\begin{figure}[ht]
 \centering
  \vspace*{-10pt}
 \begin{overpic}[width=0.465\textwidth]{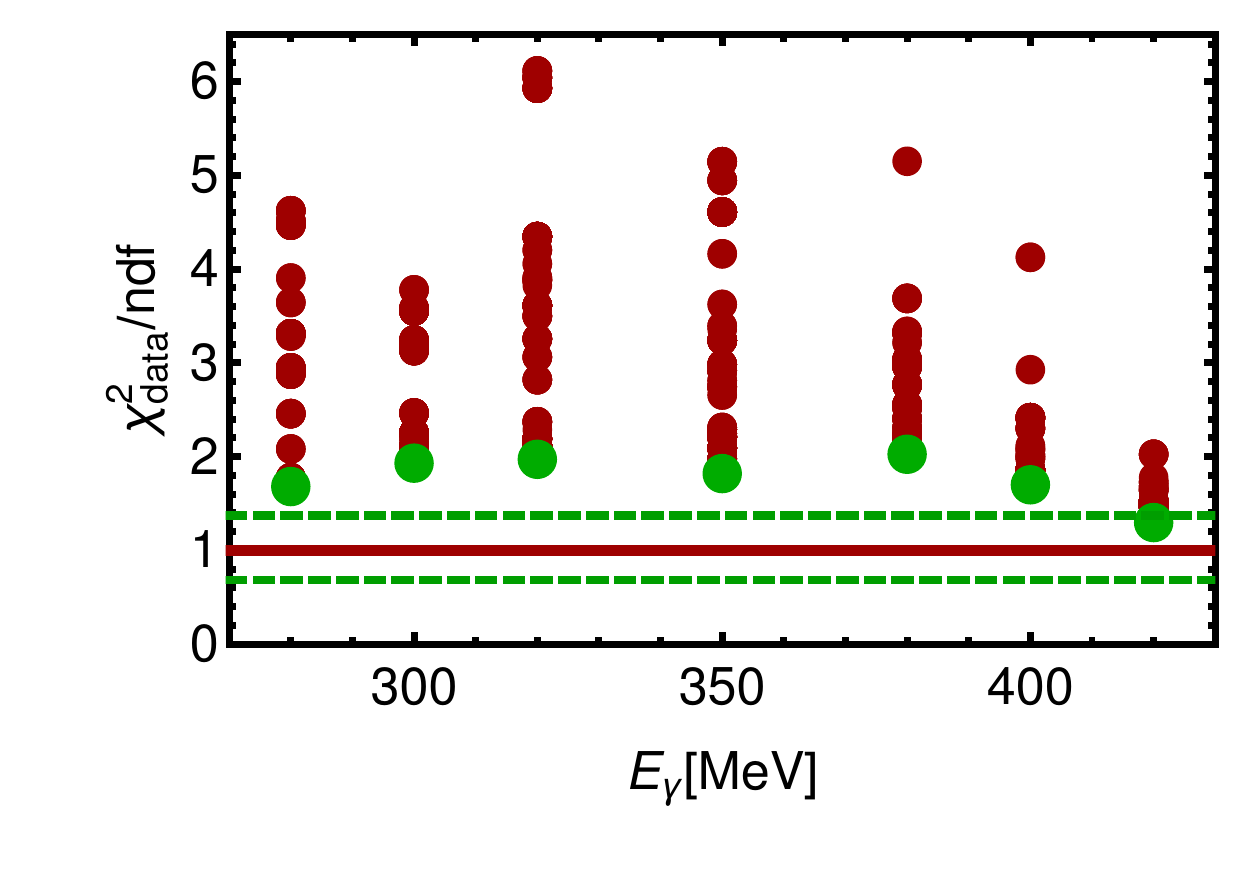}
 \put(-5,64){a.)}
 \end{overpic} \\
 \vspace*{-10pt}
\begin{overpic}[width=0.325\textwidth]{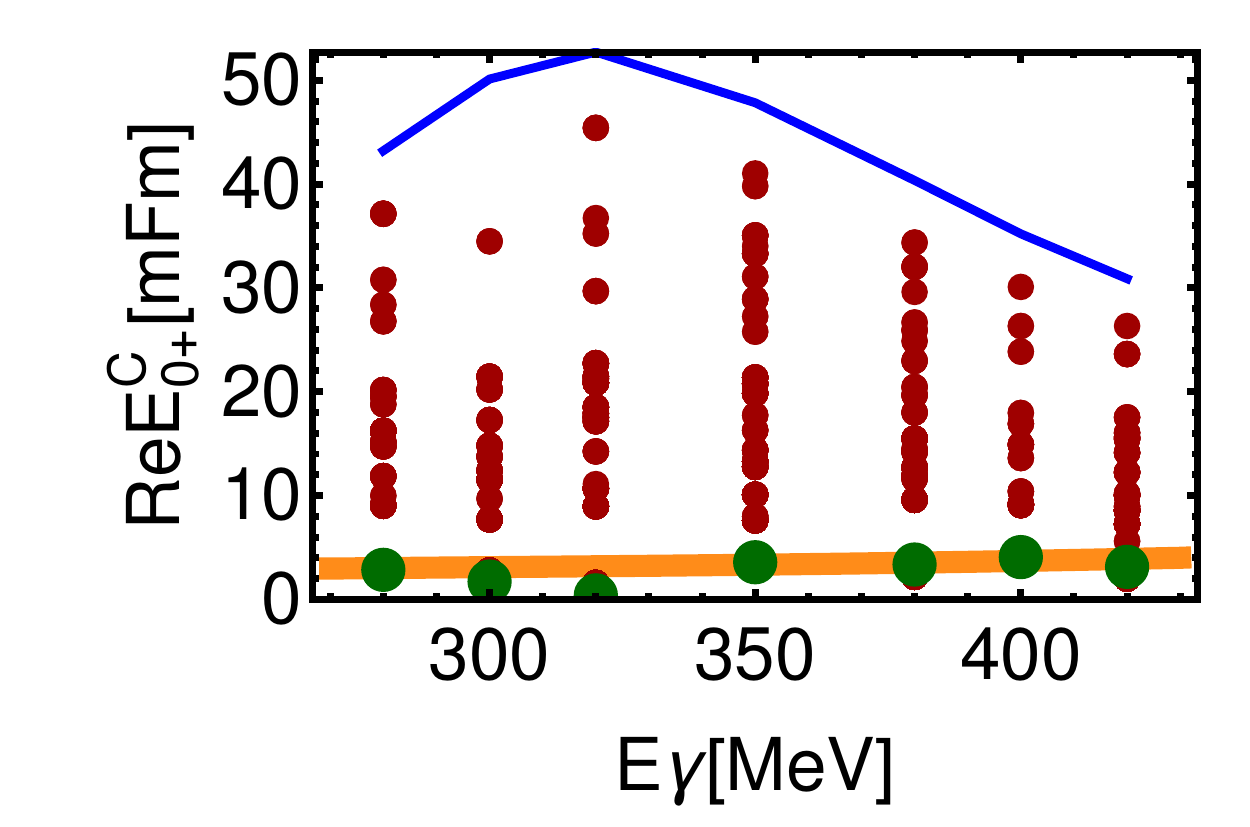}
 \put(0.5,60){b.)}
 \end{overpic}
\begin{overpic}[width=0.325\textwidth]{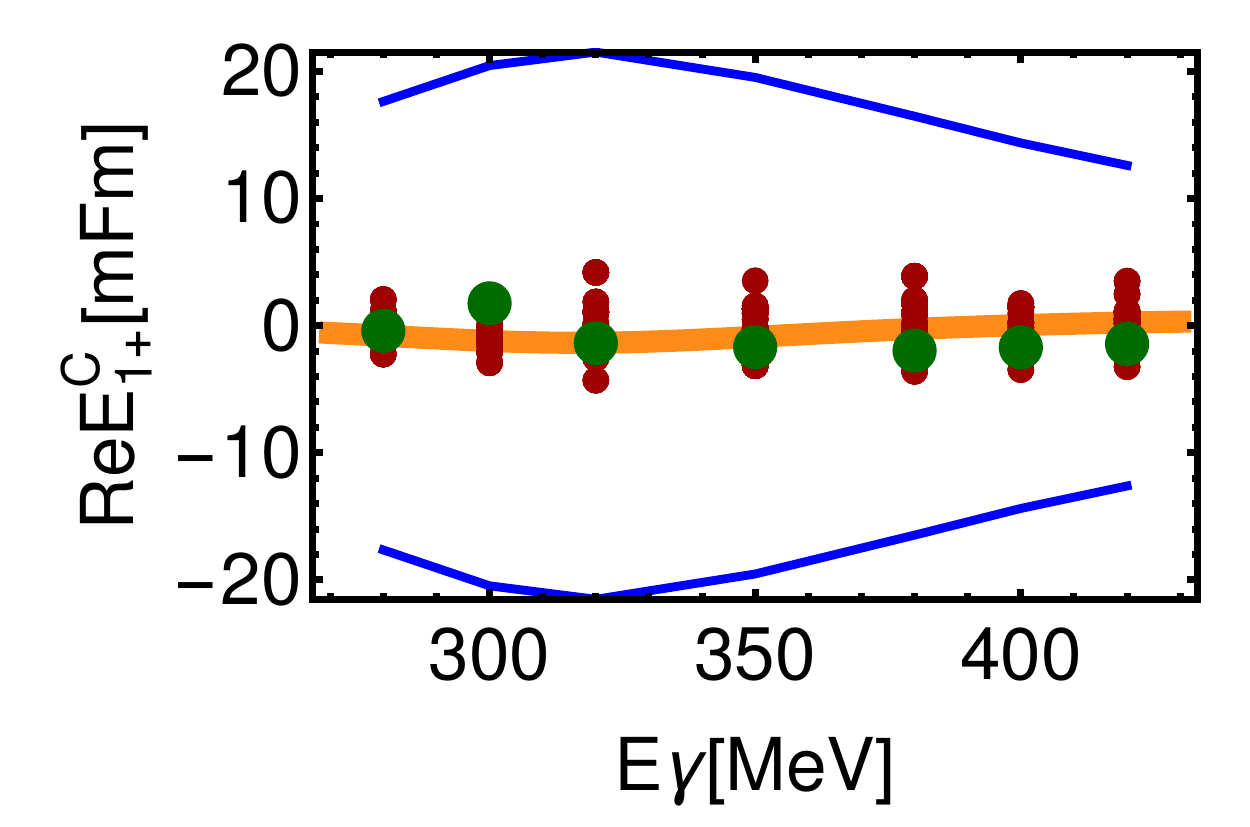}
 \end{overpic}
\begin{overpic}[width=0.325\textwidth]{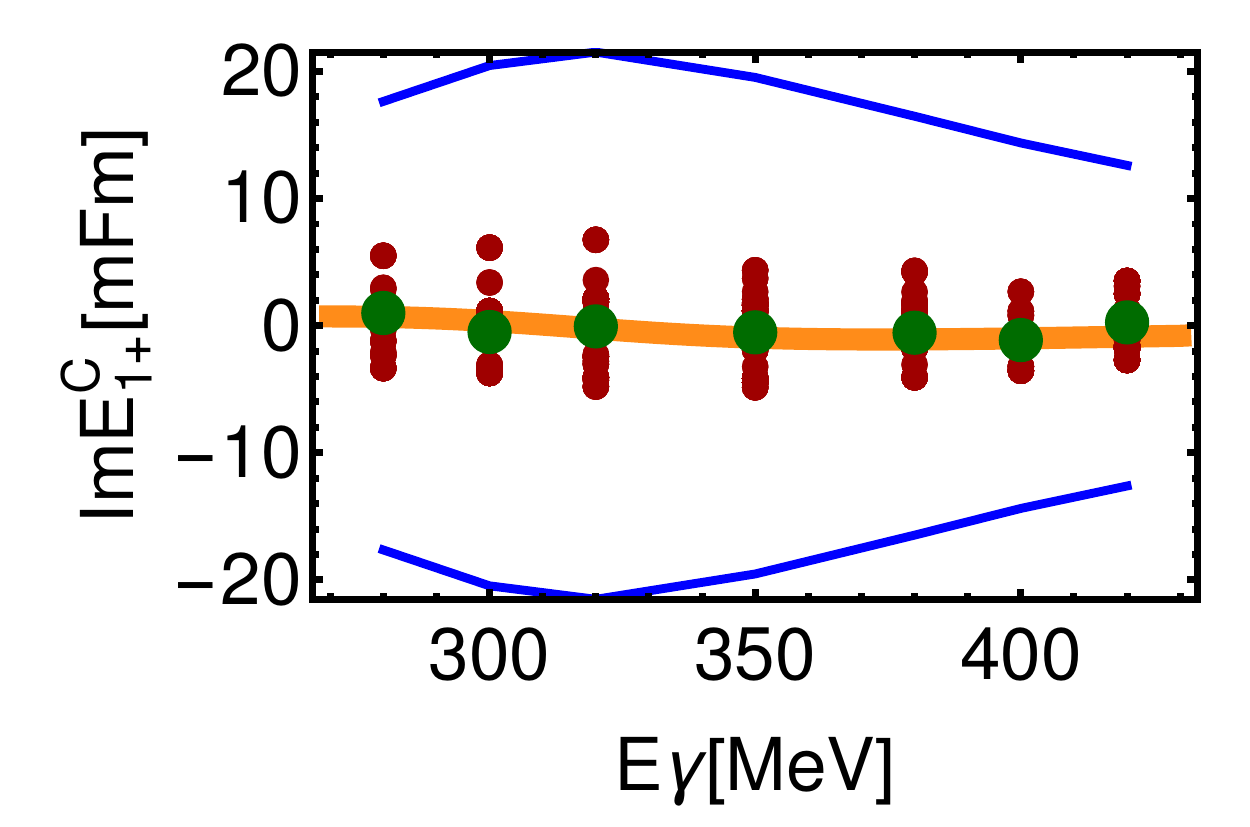}
 \end{overpic} \\
\begin{overpic}[width=0.325\textwidth]{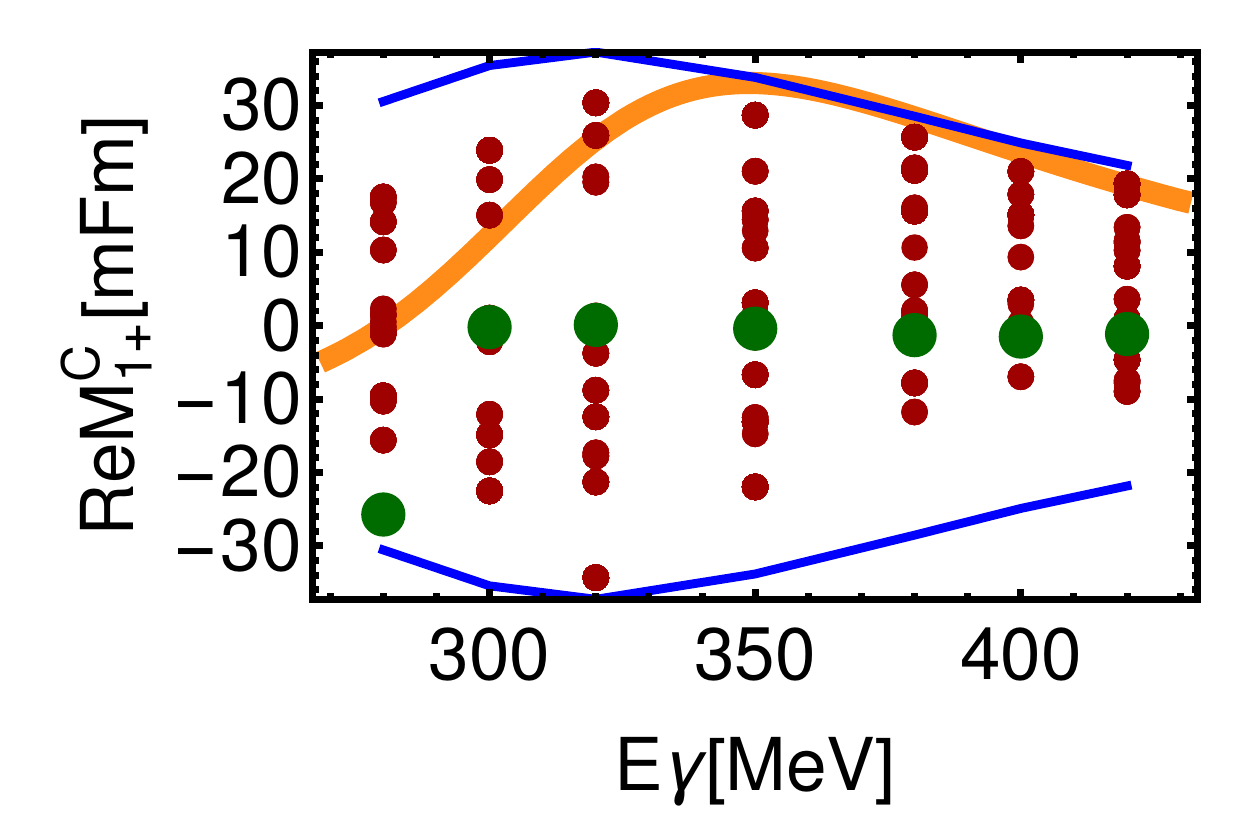}
 \end{overpic}
\begin{overpic}[width=0.325\textwidth]{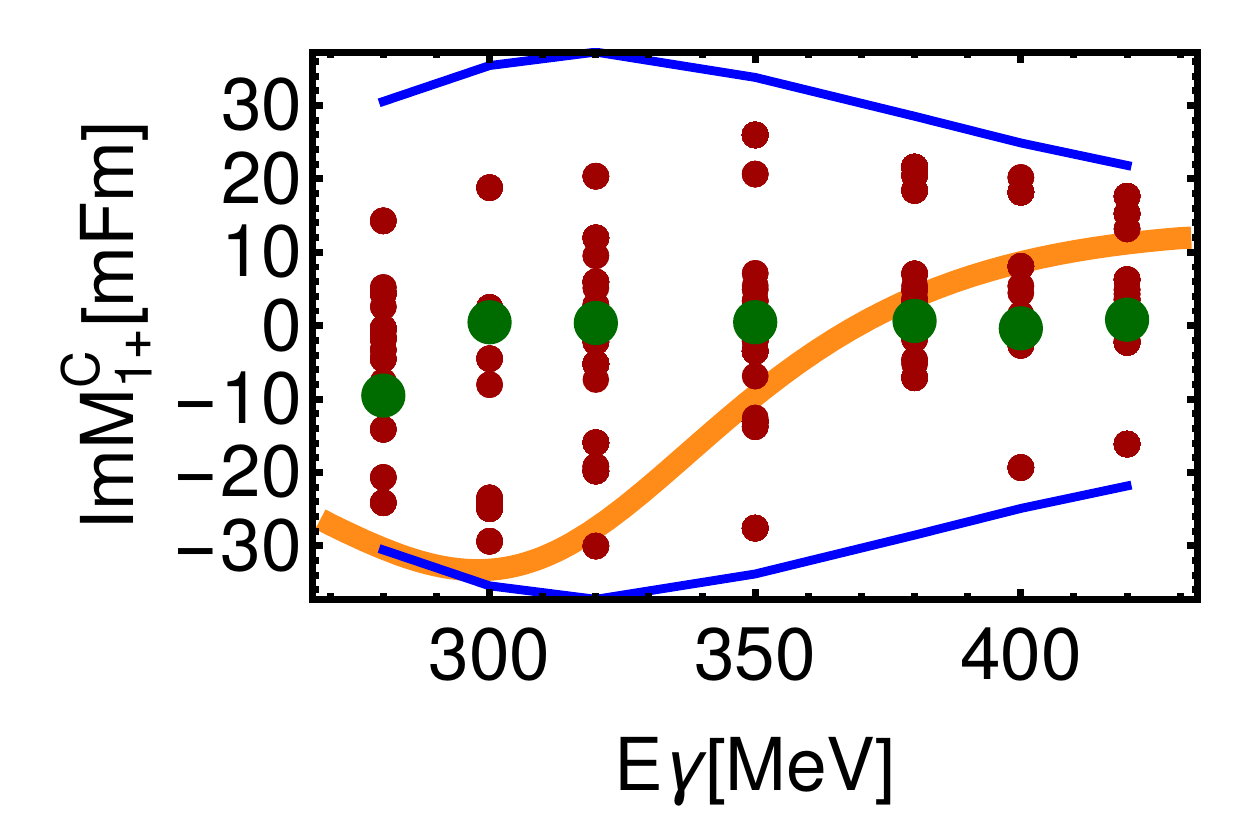}
 \end{overpic}
\begin{overpic}[width=0.325\textwidth]{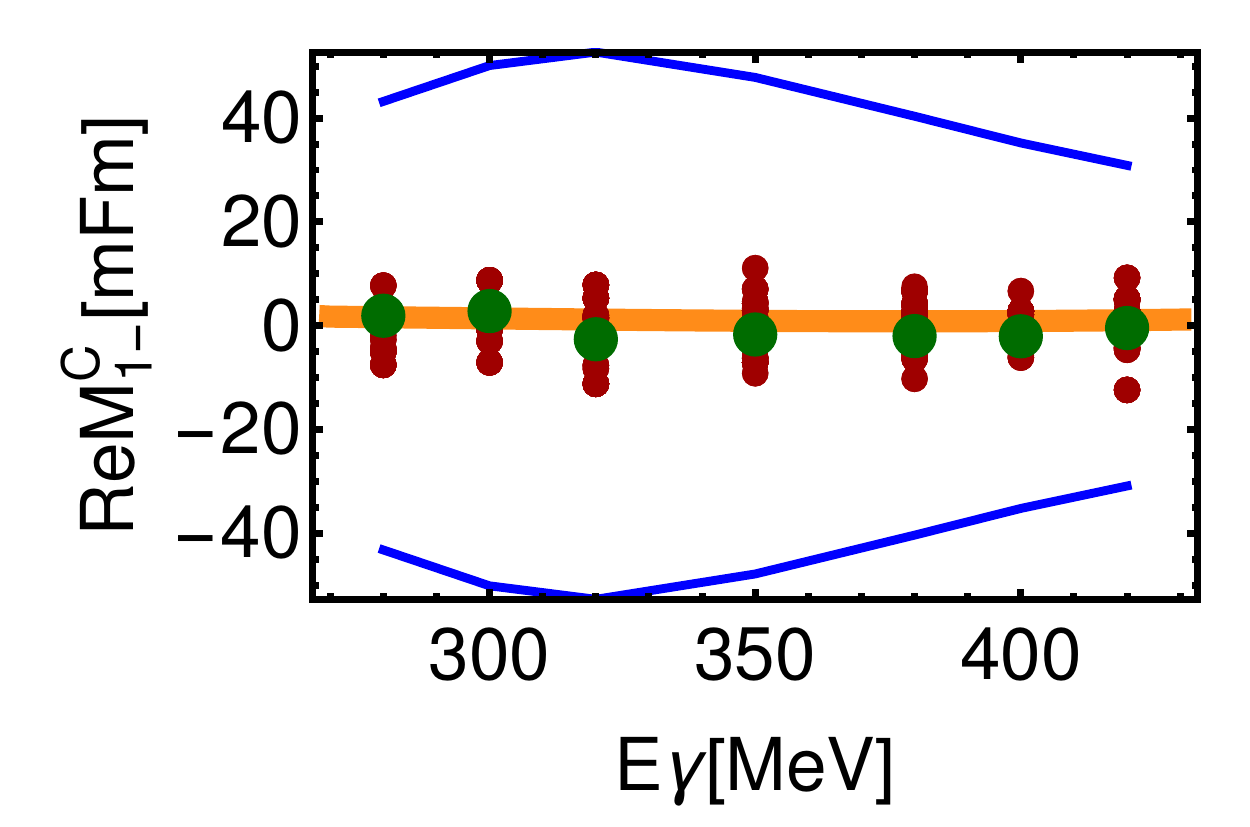}
 \end{overpic} \\
\begin{overpic}[width=0.325\textwidth]{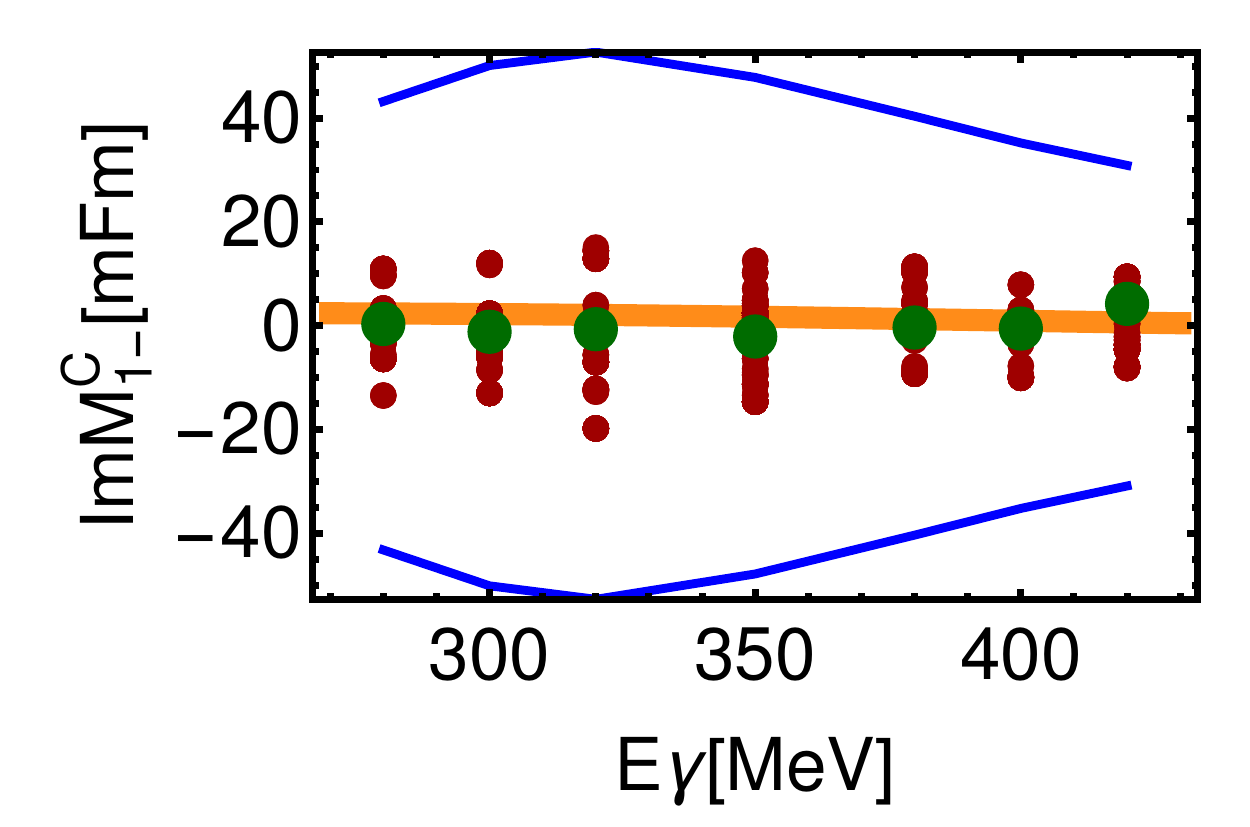}
 \end{overpic}
\begin{overpic}[width=0.325\textwidth]{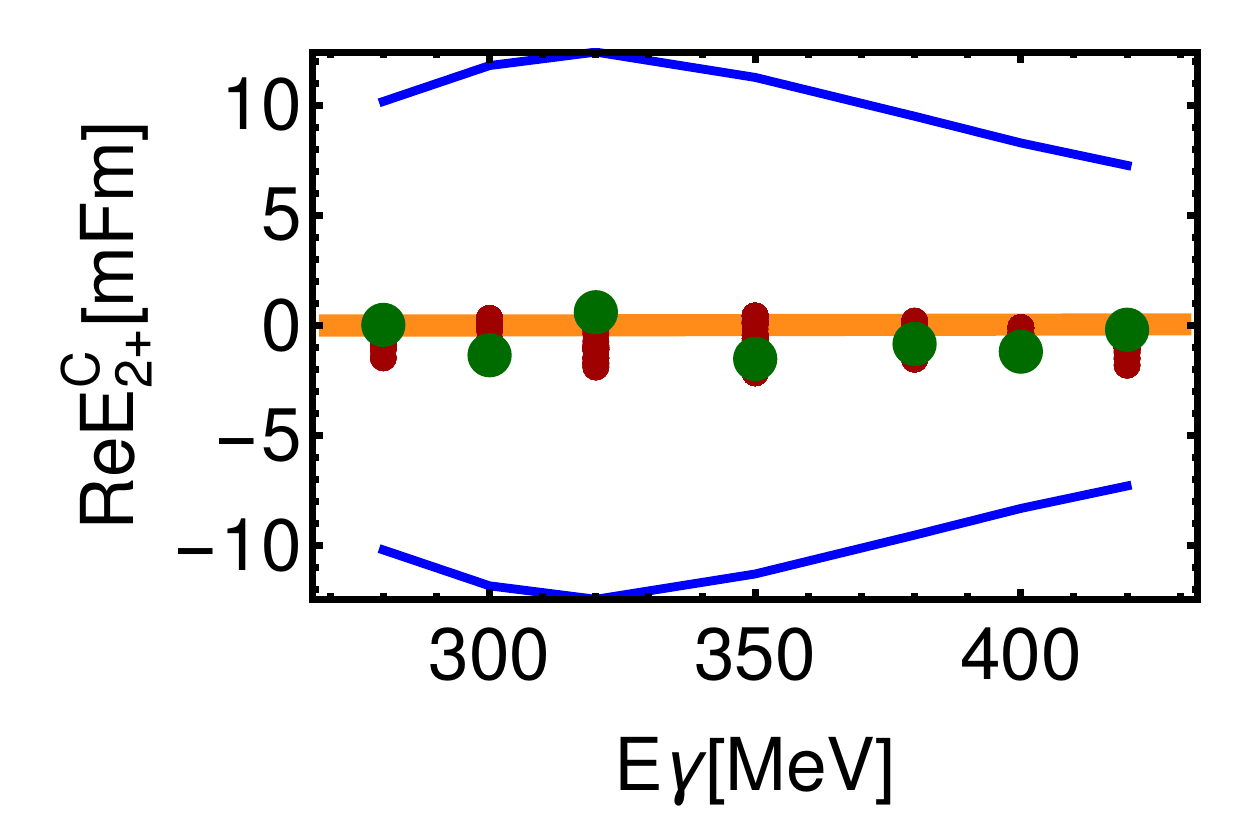}
 \end{overpic}
\begin{overpic}[width=0.325\textwidth]{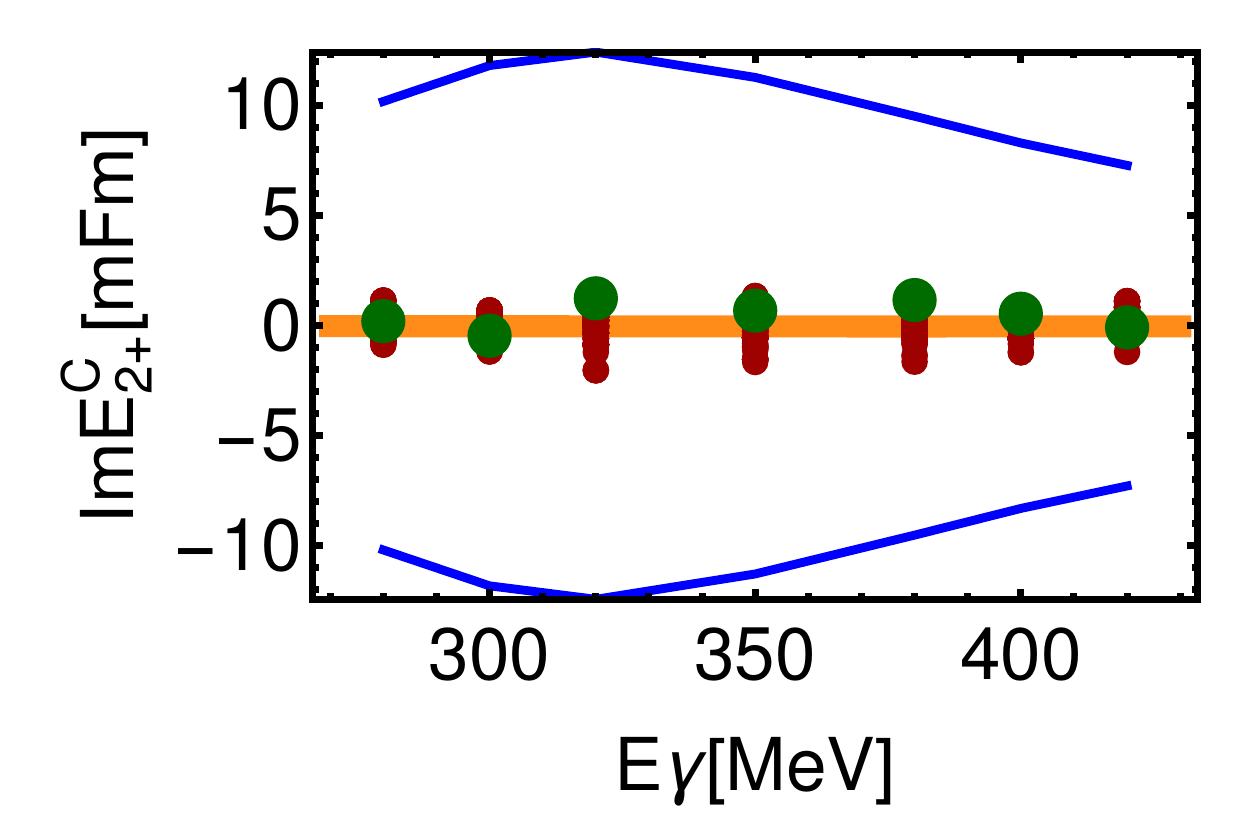}
 \end{overpic} \\
\begin{overpic}[width=0.325\textwidth]{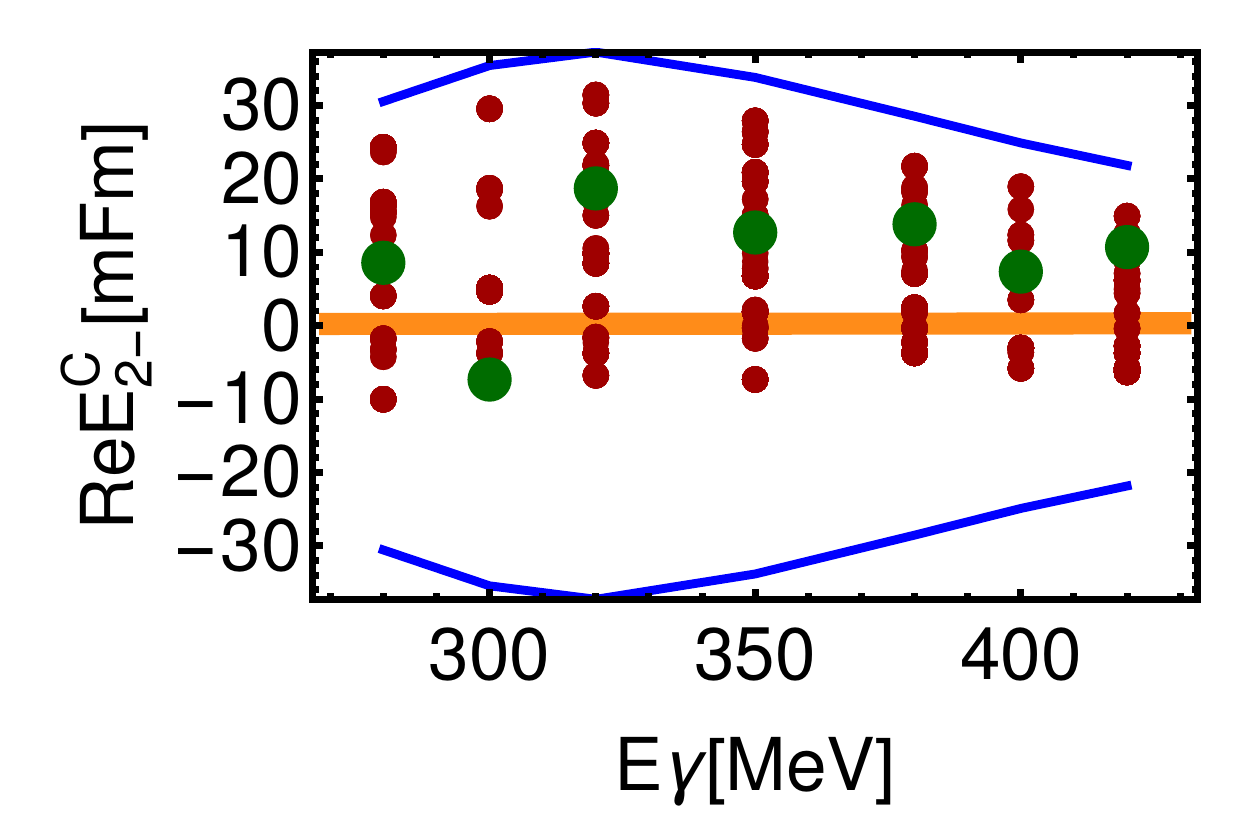}
 \end{overpic}
\begin{overpic}[width=0.325\textwidth]{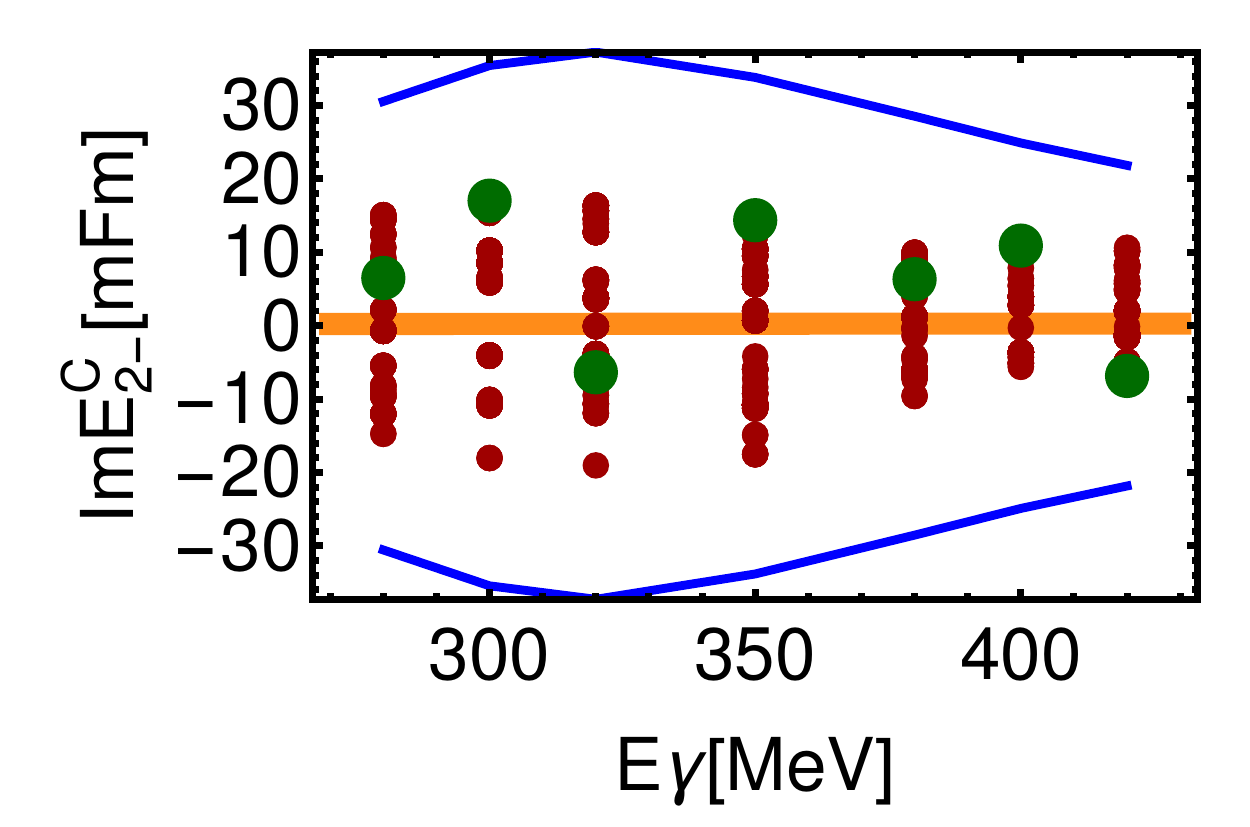}
 \end{overpic}
\begin{overpic}[width=0.325\textwidth]{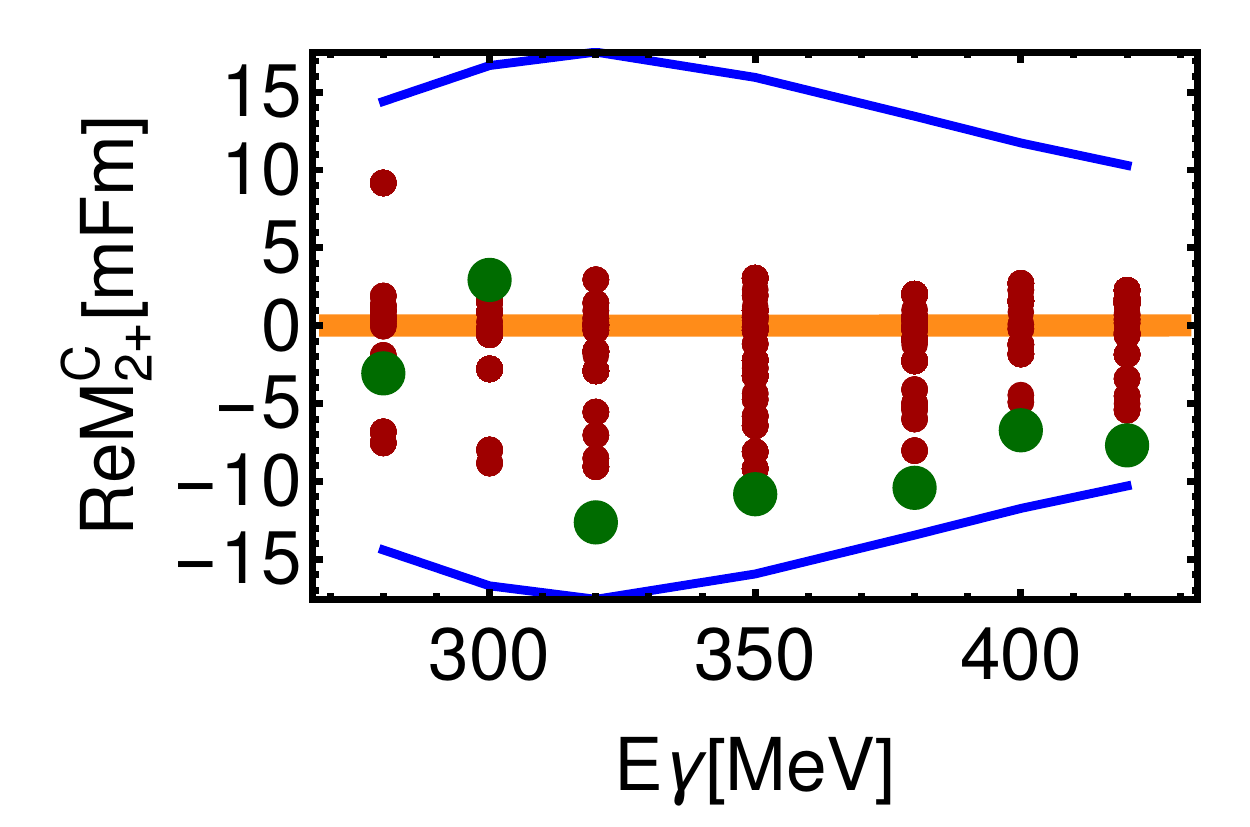}
 \end{overpic} \\
\begin{overpic}[width=0.325\textwidth]{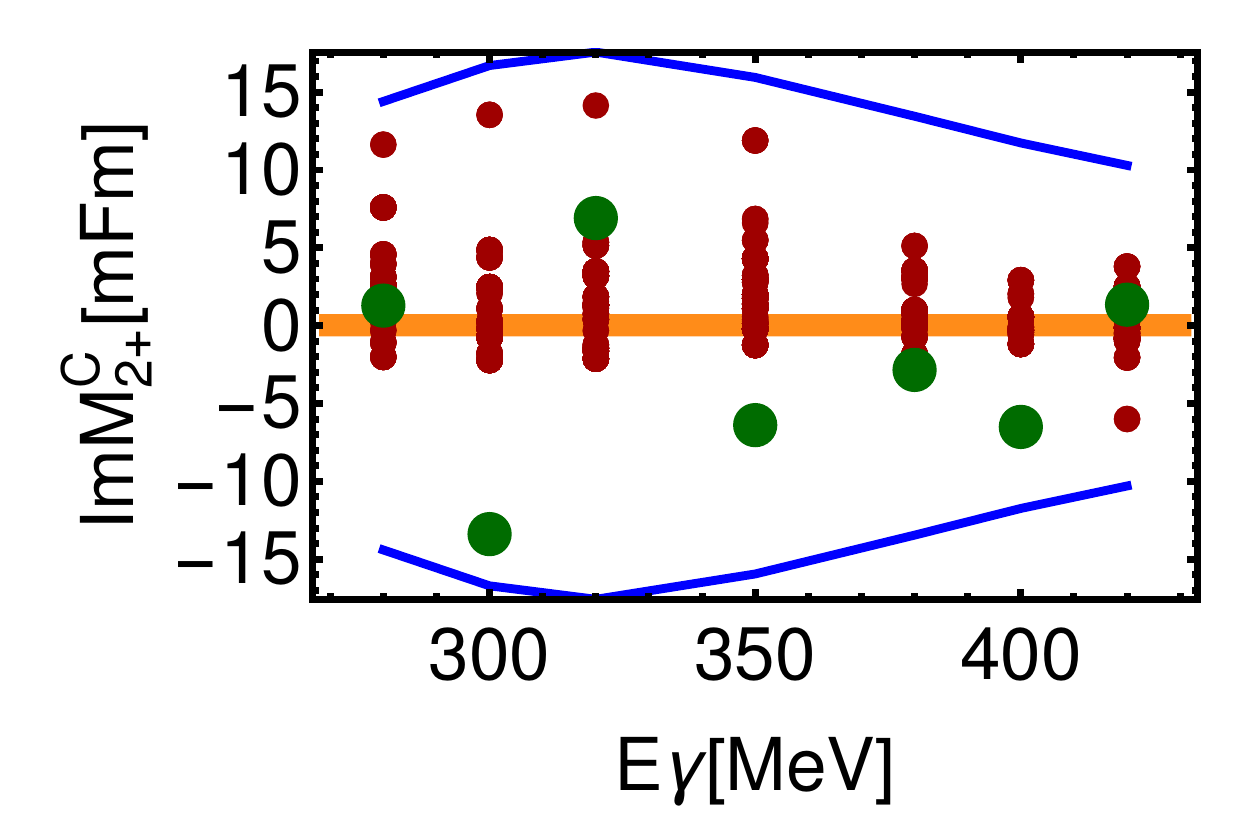}
 \end{overpic}
\begin{overpic}[width=0.325\textwidth]{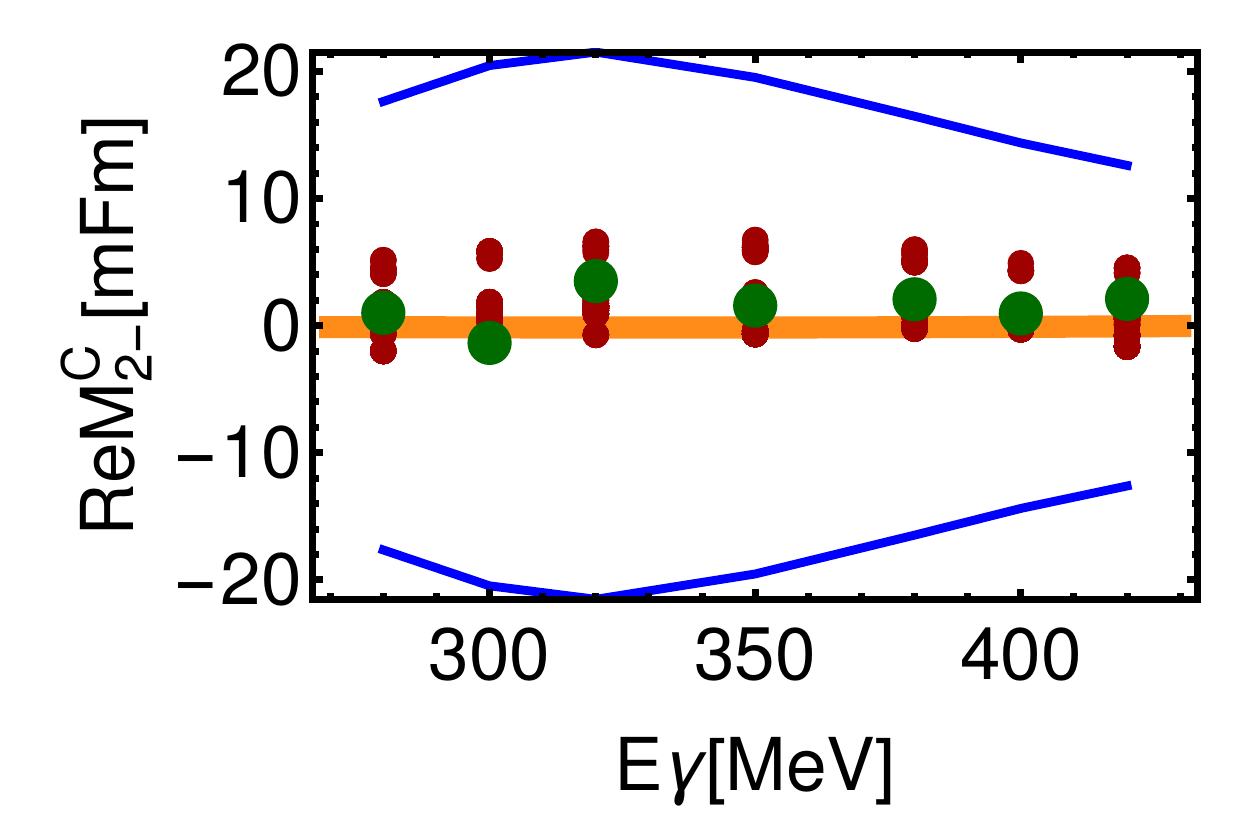}
 \end{overpic}
\begin{overpic}[width=0.325\textwidth]{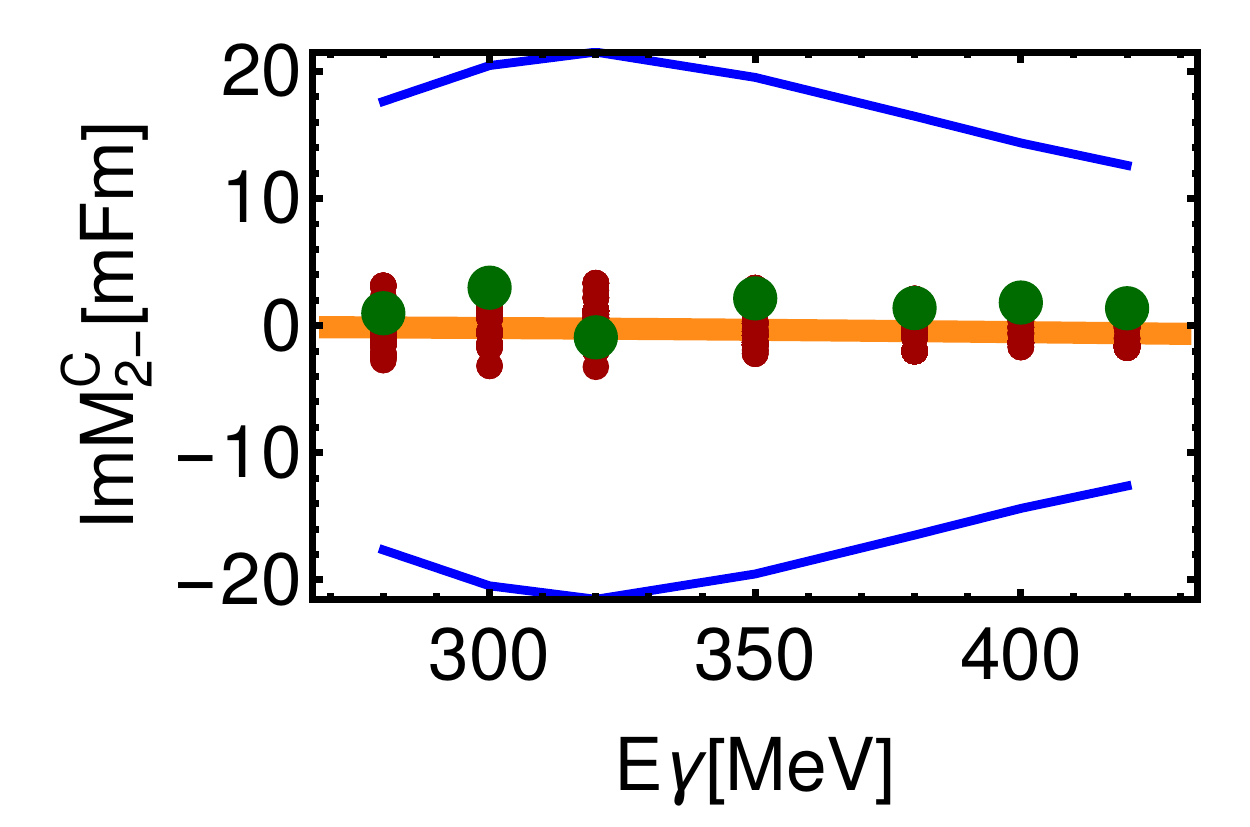}
 \end{overpic} \vspace*{-2pt}
\caption[Results for the multipoles (global and all local minima) are shown for an unconstrained TPWA with $\ell_{\mathrm{max}} = 2$ to a set of $5$ observables $\left\{ \sigma_{0}, \Sigma, \check{T}, P, \check{F} \right\}$ in the $\Delta$-region, using $N_{MC} = 8000$ start configurations.]{Results for an unconstrained TPWA with: $\ell_{\mathrm{max}} = 2$, $N_{MC} = 8000$. Plots have been obtained using a direct fit to the data. \newline a.) The $\chi^{2} / \mathrm{ndf}$ is plotted vs. energy for the global minimum (large green dots), as well as all remaining non-redundant solutions (smaller red dots). \newline b.) Multipole-solutions are shown. The global minimum (green dots) and all non-redundant local minima (red dots) are compared to the SAID CM12-solution \cite{WorkmanEtAl2012ChewMPhotoprod,SAID} (solid orange line).} 
\label{fig:Lmax2UnconstrainedFitMultipoleResultsDeltaRegionPlethora}
\end{figure}

\clearpage

\textbf{Numerical results of fits} \\ \vspace*{10pt}

\begin{table}[h]
\centering
\begin{tabular}{c|c|c|c|c|c}
\multicolumn{2}{l|}{$ E_{\gamma }\text{ = 280.0 MeV} $} & \multicolumn{2}{c|}{ $ \text{ndf = 70} $} & \multicolumn{2}{c}{ $ \chi^{2}_{\mathrm{data}}\text{/ndf = 1.87711} $ } \\
\hline
\hline
$ \hat{\theta}_{i} = \left( \mathcal{M}_{\ell}^{C} \right)_{i} \text{[mFm]} $ & $ \left(\hat{\theta}_{i}^{\mathrm{Best}}\right)_{- \Delta_{-}}^{+ \Delta_{+}} $ & $ \hat{\theta}_{i}^{\ast} (\cdot) $ & $ \widehat{\mathrm{se}}_{B} \left( \hat{\theta}^{\ast}_{i} \right) $ & $ \widehat{\mathrm{bias}}_{B} $ & $ \delta_{\mathrm{bias}} $\\
\hline
$ \mathrm{Re} \left[ E_{0+}^{C} \right] $ & $ 4.36177_{-0.18739}^{+0.20656} $ & $ 4.36952 $ & $ 0.19409 $ & $ 0.00775 $ & $ 0.03995 $\\
$ \mathrm{Re} \left[ E_{1+}^{C} \right] $ & $ -0.78855_{-0.29083}^{+0.30283} $ & $ -0.7799 $ & $ 0.29555 $ & $ 0.00865 $ & $ 0.02926 $\\
$ \mathrm{Im} \left[ E_{1+}^{C} \right] $ & $ 0.66936_{-0.07575}^{+0.06823} $ & $ 0.66525 $ & $ 0.07319 $ & $ -0.00411 $ & $ 0.05612 $\\
$ \mathrm{Re} \left[ M_{1+}^{C} \right] $ & $ 8.69217_{-0.6799}^{+0.60664} $ & $ 8.6689 $ & $ 0.64432 $ & $ -0.02327 $ & $ 0.03611 $\\
$ \mathrm{Im} \left[ M_{1+}^{C} \right] $ & $ -28.831_{-0.24886}^{+0.27283} $ & $ -28.8121 $ & $ 0.26334 $ & $ 0.01894 $ & $ 0.07191 $\\
$ \mathrm{Re} \left[ M_{1-}^{C} \right] $ & $ 3.79078_{-1.11517}^{+1.12208} $ & $ 3.806 $ & $ 1.12481 $ & $ 0.01522 $ & $ 0.01353 $\\
$ \mathrm{Im} \left[ M_{1-}^{C} \right] $ & $ 3.3973_{-0.38767}^{+0.38818} $ & $ 3.39845 $ & $ 0.38769 $ & $ 0.00115 $ & $ 0.00296 $
\end{tabular}
\caption[Numerical results are collected for a bootstrap-analysis for a TPWA-fit of photoproduction data within the $\Delta$-resonance region. The $D$-waves were fixed to SAID CM12 \cite{WorkmanEtAl2012ChewMPhotoprod}. Shown are results for the first energy-bin, \newline $E_{\gamma }\text{ = 280.0 MeV}$.]{Numerical results are collected for a bootstrap-analysis for a TPWA-fit of photoproduction data within the $\Delta$-resonance region, with $S$- and $P$-wave multipoles varied in the fit, while the $D$-waves were fixed to SAID CM12 \cite{WorkmanEtAl2012ChewMPhotoprod} (see section \ref{subsec:DeltaRegionDataFits}). An ensemble of $B = 2000$ bootstrap-replicates has been applied. Shown are results for the first energy-bin, $E_{\gamma }\text{ = 280.0 MeV}$. Here, a global minimum has been found with $\chi^{2}_{\mathrm{data}}\text{/ndf = 1.87711}$. \newline
From the bootstrap-distributions of the fit-parameters, we extract quantiles which then define a confidence-interval for the individual parameter, composed of upper and lower bootstrap-errors $\Delta_{\pm}$. The global minimum is quoted in conjunction with these asymmetric errors (for more details, see the main text). Furthermore, the mean $\hat{\theta}_{i}^{\ast} (\cdot)$, standard error $\widehat{\mathrm{se}}_{B} \left( \hat{\theta}^{\ast}_{i} \right)$ and bias-estimate $\widehat{\mathrm{bias}}_{B}$ are extracted from the bootstrap-distributions. Lastly, we define and extract a bias test-parameter defined as $\delta_{\mathrm{bias}} := \left| \widehat{\mathrm{bias}}_{B} \right|/\widehat{\mathrm{se}}_{B}$. \newline
All numbers are given in milli-Fermi, except for $\delta_{\mathrm{bias}}$ which does not carry dimension.}
\label{tab:DeltaRegionResultsFirstEnergy}
\end{table}

\clearpage

\begin{table}[h]
\centering
\begin{tabular}{c|c|c|c|c|c}
\multicolumn{2}{l|}{$ E_{\gamma }\text{ = 300.0 MeV} $} & \multicolumn{2}{c|}{ $ \text{ndf = 71} $} & \multicolumn{2}{c}{ $ \chi^{2}_{\mathrm{data}}\text{/ndf = 2.25531} $ } \\
\hline
\hline
$ \hat{\theta}_{i} = \left( \mathcal{M}_{\ell}^{C} \right)_{i} \text{[mFm]} $ & $ \left(\hat{\theta}_{i}^{\mathrm{Best}}\right)_{- \Delta_{-}}^{+ \Delta_{+}} $ & $ \hat{\theta}_{i}^{\ast} (\cdot) $ & $ \widehat{\mathrm{se}}_{B} \left( \hat{\theta}^{\ast}_{i} \right) $ & $ \widehat{\mathrm{bias}}_{B} $ & $ \delta_{\mathrm{bias}} $\\
\hline
$ \mathrm{Re} \left[ E_{0+}^{C} \right] $ & $ 4.41611_{-0.18328}^{+0.18863} $ & $ 4.41948 $ & $ 0.18477 $ & $ 0.00336 $ & $ 0.0182 $ \\
$ \mathrm{Re} \left[ E_{1+}^{C} \right] $ & $ -1.68894_{-0.28135}^{+0.28053} $ & $ -1.6865 $ & $ 0.28401 $ & $ 0.00244 $ & $ 0.00858 $ \\
$ \mathrm{Im} \left[ E_{1+}^{C} \right] $ & $ 0.12562_{-0.1873}^{+0.17058} $ & $ 0.11982 $ & $ 0.17648 $ & $ -0.00579 $ & $ 0.03284 $ \\
$ \mathrm{Re} \left[ M_{1+}^{C} \right] $ & $ 17.1527_{-0.75141}^{+0.70903} $ & $ 17.1358 $ & $ 0.72882 $ & $ -0.01694 $ & $ 0.02324 $ \\
$ \mathrm{Im} \left[ M_{1+}^{C} \right] $ & $ -30.6406_{-0.38924}^{+0.43256} $ & $ -30.6189 $ & $ 0.4116 $ & $ 0.02171 $ & $ 0.05275 $ \\
$ \mathrm{Re} \left[ M_{1-}^{C} \right] $ & $ 1.43489_{-1.12082}^{+1.10789} $ & $ 1.44125 $ & $ 1.1055 $ & $ 0.00636 $ & $ 0.00575 $ \\
$ \mathrm{Im} \left[ M_{1-}^{C} \right] $ & $ 2.19197_{-0.72309}^{+0.72336} $ & $ 2.1945 $ & $ 0.71559 $ & $ 0.00253 $ & $ 0.00354 $
\end{tabular}
\caption[Numerical results of a bootstrap-analysis are collected for a TPWA-fit of photoproduction data within the $\Delta$-resonance region. The $D$-waves were fixed to SAID CM12 \cite{WorkmanEtAl2012ChewMPhotoprod}. Shown are results for the second energy-bin, \newline $E_{\gamma }\text{ = 300.0 MeV}$.]{Numerical results are collected for a TPWA bootstrap-analysis of photoproduction data within the $\Delta$-resonance region. Here, the second energy-bin, $E_{\gamma }\text{ = 300.0 MeV}$, is shown. For more details, see the description of Table \ref{tab:DeltaRegionResultsFirstEnergy}.} \bigskip
\begin{tabular}{c|c|c|c|c|c}
\multicolumn{2}{l|}{$ E_{\gamma }\text{ = 320.0 MeV} $} & \multicolumn{2}{c|}{ $ \text{ndf = 71} $} & \multicolumn{2}{c}{ $ \chi^{2}_{\mathrm{data}}\text{/ndf = 2.46793} $ } \\
\hline
\hline
$ \hat{\theta}_{i} = \left( \mathcal{M}_{\ell}^{C} \right)_{i} \text{[mFm]} $ & $ \left(\hat{\theta}_{i}^{\mathrm{Best}}\right)_{- \Delta_{-}}^{+ \Delta_{+}} $ & $ \hat{\theta}_{i}^{\ast} (\cdot) $ & $ \widehat{\mathrm{se}}_{B} \left( \hat{\theta}^{\ast}_{i} \right) $ & $ \widehat{\mathrm{bias}}_{B} $ & $ \delta_{\mathrm{bias}} $\\
\hline
 $ \mathrm{Re} \left[ E_{0+}^{C} \right] $ & $ 3.67725_{-0.20659}^{+0.21267} $ & $ 3.6808 $ & $ 0.20791 $ & $ 0.00355 $ & $ 0.01708 $ \\
 $ \mathrm{Re} \left[ E_{1+}^{C} \right] $ & $ -1.96483_{-0.18186}^{+0.18254} $ & $ -1.9596 $ & $ 0.17913 $ & $ 0.00523 $ & $ 0.02919 $ \\
 $ \mathrm{Im} \left[ E_{1+}^{C} \right] $ & $ -0.93835_{-0.28192}^{+0.27936} $ & $ -0.9394 $ & $ 0.28109 $ & $ -0.00105 $ & $ 0.00373 $ \\
 $ \mathrm{Re} \left[ M_{1+}^{C} \right] $ & $ 30.1383_{-0.81333}^{+0.76587} $ & $ 30.1105 $ & $ 0.79277 $ & $ -0.02783 $ & $ 0.0351 $ \\
 $ \mathrm{Im} \left[ M_{1+}^{C} \right] $ & $ -21.3661_{-1.0677}^{+1.16557} $ & $ -21.3398 $ & $ 1.12889 $ & $ 0.02633 $ & $ 0.02332 $ \\
 $ \mathrm{Re} \left[ M_{1-}^{C} \right] $ & $ 1.20231_{-0.67436}^{+0.67539} $ & $ 1.2006 $ & $ 0.66427 $ & $ -0.00171 $ & $ 0.00257 $ \\
 $ \mathrm{Im} \left[ M_{1-}^{C} \right] $ & $ 1.81728_{-1.05728}^{+1.08716} $ & $ 1.8252 $ & $ 1.06129 $ & $ 0.00792 $ & $ 0.00746 $
\end{tabular}
\caption[Numerical results of a bootstrap-analysis are collected for a TPWA-fit of photoproduction data within the $\Delta$-resonance region. The $D$-waves were fixed to SAID CM12 \cite{WorkmanEtAl2012ChewMPhotoprod}. Shown are results for the third energy-bin, \newline $E_{\gamma }\text{ = 320.0 MeV}$.]{Numerical results are collected for a TPWA bootstrap-analysis of photoproduction data within the $\Delta$-resonance region. Here, the third energy-bin, $E_{\gamma }\text{ = 320.0 MeV}$, is shown. For more details, see the description of Table \ref{tab:DeltaRegionResultsFirstEnergy}.} \bigskip
\begin{tabular}{c|c|c|c|c|c}
\multicolumn{2}{l|}{$ E_{\gamma }\text{ = 350.0 MeV} $} & \multicolumn{2}{c|}{ $ \text{ndf = 72} $} & \multicolumn{2}{c}{ $ \chi^{2}_{\mathrm{data}}\text{/ndf = 2.92216} $ } \\
\hline
\hline
$ \hat{\theta}_{i} = \left( \mathcal{M}_{\ell}^{C} \right)_{i} \text{[mFm]} $ & $ \left(\hat{\theta}_{i}^{\mathrm{Best}}\right)_{- \Delta_{-}}^{+ \Delta_{+}} $ & $ \hat{\theta}_{i}^{\ast} (\cdot) $ & $ \widehat{\mathrm{se}}_{B} \left( \hat{\theta}^{\ast}_{i} \right) $ & $ \widehat{\mathrm{bias}}_{B} $ & $ \delta_{\mathrm{bias}} $\\
\hline
 $ \mathrm{Re} \left[ E_{0+}^{C} \right] $ & $ 4.12632_{-0.16819}^{+0.1907} $ & $ 4.13615 $ & $ 0.18187 $ & $ 0.00983 $ & $ 0.05407 $ \\
 $ \mathrm{Re} \left[ E_{1+}^{C} \right] $ & $ -1.11254_{-0.11417}^{+0.13926} $ & $ -1.09985 $ & $ 0.12748 $ & $ 0.01269 $ & $ 0.09951 $ \\
 $ \mathrm{Im} \left[ E_{1+}^{C} \right] $ & $ -2.07794_{-0.32956}^{+0.36794} $ & $ -2.0628 $ & $ 0.35378 $ & $ 0.01514 $ & $ 0.0428 $ \\
 $ \mathrm{Re} \left[ M_{1+}^{C} \right] $ & $ 33.193_{-0.28003}^{+0.1663} $ & $ 33.1397 $ & $ 0.22645 $ & $ -0.05335 $ & $ 0.2356 $ \\
 $ \mathrm{Im} \left[ M_{1+}^{C} \right] $ & $ -3.93992_{-1.68021}^{+1.82484} $ & $ -3.8695 $ & $ 1.74887 $ & $ 0.07042 $ & $ 0.04026 $ \\
 $ \mathrm{Re} \left[ M_{1-}^{C} \right] $ & $ 1.4676_{-0.15782}^{+0.172} $ & $ 1.4725 $ & $ 0.17162 $ & $ 0.0049 $ & $ 0.02856 $ \\
 $ \mathrm{Im} \left[ M_{1-}^{C} \right] $ & $ 1.10874_{-1.18874}^{+1.3299} $ & $ 1.1655 $ & $ 1.29632 $ & $ 0.05676 $ & $ 0.04379 $
\end{tabular}
\caption[Numerical results of a bootstrap-analysis are collected for a TPWA-fit of photoproduction data within the $\Delta$-resonance region. The $D$-waves were fixed to SAID CM12 \cite{WorkmanEtAl2012ChewMPhotoprod}. Shown are results for the fourth energy-bin, \newline $E_{\gamma }\text{ = 350.0 MeV}$.]{Numerical results are collected for a TPWA bootstrap-analysis of photoproduction data within the $\Delta$-resonance region. Here, the fourth energy-bin, $E_{\gamma }\text{ = 350.0 MeV}$, is shown. For more details, see the description of Table \ref{tab:DeltaRegionResultsFirstEnergy}.}
\label{tab:DeltaRegionResultsSecondThirdFourthEnergy}
\end{table}

\clearpage

\begin{table}[h]
\centering
\begin{tabular}{c|c|c|c|c|c}
\multicolumn{2}{l|}{$ E_{\gamma }\text{ = 380.0 MeV} $} & \multicolumn{2}{c|}{ $ \text{ndf = 72} $} & \multicolumn{2}{c}{ $ \chi^{2}_{\mathrm{data}}\text{/ndf = 2.19302} $ } \\
\hline
\hline
$ \hat{\theta}_{i} = \left( \mathcal{M}_{\ell}^{C} \right)_{i} \text{[mFm]} $ & $ \left(\hat{\theta}_{i}^{\mathrm{Best}}\right)_{- \Delta_{-}}^{+ \Delta_{+}} $ & $ \hat{\theta}_{i}^{\ast} (\cdot) $ & $ \widehat{\mathrm{se}}_{B} \left( \hat{\theta}^{\ast}_{i} \right) $ & $ \widehat{\mathrm{bias}}_{B} $ & $ \delta_{\mathrm{bias}} $\\
\hline
 $ \mathrm{Re} \left[ E_{0+}^{C} \right] $ & $ 4.99243_{-0.146}^{+0.15545} $ & $ 4.99888 $ & $ 0.1535 $ & $ 0.00644 $ & $ 0.04197 $ \\
 $ \mathrm{Re} \left[ E_{1+}^{C} \right] $ & $ -0.41218_{-0.13091}^{+0.15974} $ & $ -0.39863 $ & $ 0.14761 $ & $ 0.01356 $ & $ 0.09185 $ \\
 $ \mathrm{Im} \left[ E_{1+}^{C} \right] $ & $ -1.94375_{-0.31474}^{+0.30514} $ & $ -1.94635 $ & $ 0.31211 $ & $ -0.0026 $ & $ 0.00833 $ \\
 $ \mathrm{Re} \left[ M_{1+}^{C} \right] $ & $ 27.9952_{-0.14201}^{+0.04007} $ & $ 27.9411 $ & $ 0.10771 $ & $ -0.05403 $ & $ 0.50163 $ \\
 $ \mathrm{Im} \left[ M_{1+}^{C} \right] $ & $ 1.5421_{-1.41353}^{+1.418} $ & $ 1.54 $ & $ 1.42293 $ & $ -0.0021 $ & $ 0.00148 $ \\
 $ \mathrm{Re} \left[ M_{1-}^{C} \right] $ & $ 2.38898_{-0.24212}^{+0.2887} $ & $ 2.4097 $ & $ 0.26877 $ & $ 0.02072 $ & $ 0.07709 $ \\
 $ \mathrm{Im} \left[ M_{1-}^{C} \right] $ & $ 1.22644_{-1.16627}^{+1.1069} $ & $ 1.19 $ & $ 1.15162 $ & $ -0.03644 $ & $ 0.03164 $
\end{tabular}
\caption[Numerical results of a bootstrap-analysis are collected for a TPWA-fit of photoproduction data within the $\Delta$-resonance region. The $D$-waves were fixed to SAID CM12 \cite{WorkmanEtAl2012ChewMPhotoprod}. Shown are results for the fifth energy-bin, \newline $E_{\gamma }\text{ = 380.0 MeV}$.]{Numerical results are collected for a TPWA bootstrap-analysis of photoproduction data within the $\Delta$-resonance region. Here, the fifth energy-bin, $E_{\gamma }\text{ = 380.0 MeV}$, is shown. For more details, see the description of Table \ref{tab:DeltaRegionResultsFirstEnergy}.} \bigskip
\begin{tabular}{c|c|c|c|c|c}
\multicolumn{2}{l|}{$ E_{\gamma }\text{ = 400.0 MeV} $} & \multicolumn{2}{c|}{ $ \text{ndf = 72} $} & \multicolumn{2}{c}{ $ \chi^{2}_{\mathrm{data}}\text{/ndf = 2.15544} $ } \\
\hline
\hline
$ \hat{\theta}_{i} = \left( \mathcal{M}_{\ell}^{C} \right)_{i} \text{[mFm]} $ & $ \left(\hat{\theta}_{i}^{\mathrm{Best}}\right)_{- \Delta_{-}}^{+ \Delta_{+}} $ & $ \hat{\theta}_{i}^{\ast} (\cdot) $ & $ \widehat{\mathrm{se}}_{B} \left( \hat{\theta}^{\ast}_{i} \right) $ & $ \widehat{\mathrm{bias}}_{B} $ & $ \delta_{\mathrm{bias}} $\\
\hline
 $ \mathrm{Re} \left[ E_{0+}^{C} \right] $ & $ 5.27353_{-0.11052}^{+0.12334} $ & $ 5.2807 $ & $ 0.11749 $ & $ 0.00717 $ & $ 0.06099 $ \\
 $ \mathrm{Re} \left[ E_{1+}^{C} \right] $ & $ -0.06412_{-0.1757}^{+0.18532} $ & $ -0.05885 $ & $ 0.18453 $ & $ 0.00527 $ & $ 0.02855 $ \\
 $ \mathrm{Im} \left[ E_{1+}^{C} \right] $ & $ -2.22505_{-0.25121}^{+0.27457} $ & $ -2.21228 $ & $ 0.26212 $ & $ 0.01278 $ & $ 0.04875 $ \\
 $ \mathrm{Re} \left[ M_{1+}^{C} \right] $ & $ 23.9895_{-0.21403}^{+0.13164} $ & $ 23.9489 $ & $ 0.17568 $ & $ -0.04056 $ & $ 0.23087 $ \\
 $ \mathrm{Im} \left[ M_{1+}^{C} \right] $ & $ 3.48831_{-1.1693}^{+1.1736} $ & $ 3.48175 $ & $ 1.13325 $ & $ -0.00656 $ & $ 0.00578 $ \\
 $ \mathrm{Re} \left[ M_{1-}^{C} \right] $ & $ 2.29986_{-0.40242}^{+0.38997} $ & $ 2.29975 $ & $ 0.40539 $ & $ -0.00011 $ & $ 0.00027 $ \\
 $ \mathrm{Im} \left[ M_{1-}^{C} \right] $ & $ -1.69618_{-0.97295}^{+1.04385} $ & $ -1.6589 $ & $ 1.00412 $ & $ 0.03728 $ & $ 0.03713 $
\end{tabular}
\caption[Numerical results of a bootstrap-analysis are collected for a TPWA-fit of photoproduction data within the $\Delta$-resonance region. The $D$-waves were fixed to SAID CM12 \cite{WorkmanEtAl2012ChewMPhotoprod}. Shown are results for the sixth energy-bin, \newline $E_{\gamma }\text{ = 400.0 MeV}$.]{Numerical results are collected for a TPWA bootstrap-analysis of photoproduction data within the $\Delta$-resonance region. Here, the sixth energy-bin, $E_{\gamma }\text{ = 400.0 MeV}$, is shown. For more details, see the description of Table \ref{tab:DeltaRegionResultsFirstEnergy}.} \bigskip
\begin{tabular}{c|c|c|c|c|c}
\multicolumn{2}{l|}{$ E_{\gamma }\text{ = 420.0 MeV} $} & \multicolumn{2}{c|}{ $ \text{ndf = 72} $} & \multicolumn{2}{c}{ $ \chi^{2}_{\mathrm{data}}\text{/ndf = 1.67992} $ } \\
\hline
\hline
$ \hat{\theta}_{i} = \left( \mathcal{M}_{\ell}^{C} \right)_{i} \text{[mFm]} $ & $ \left(\hat{\theta}_{i}^{\mathrm{Best}}\right)_{- \Delta_{-}}^{+ \Delta_{+}} $ & $ \hat{\theta}_{i}^{\ast} (\cdot) $ & $ \widehat{\mathrm{se}}_{B} \left( \hat{\theta}^{\ast}_{i} \right) $ & $ \widehat{\mathrm{bias}}_{B} $ & $ \delta_{\mathrm{bias}} $\\
\hline
 $ \mathrm{Re} \left[ E_{0+}^{C} \right] $ & $ 5.95627_{-0.24104}^{+0.29171} $ & $ 5.98035 $ & $ 0.26679 $ & $ 0.02408 $ & $ 0.09027 $ \\
 $ \mathrm{Re} \left[ E_{1+}^{C} \right] $ & $ -0.41556_{-0.25655}^{+0.31556} $ & $ -0.38185 $ & $ 0.29674 $ & $ 0.03371 $ & $ 0.11359 $ \\
 $ \mathrm{Im} \left[ E_{1+}^{C} \right] $ & $ -2.54992_{-0.41387}^{+0.43629} $ & $ -2.5407 $ & $ 0.41727 $ & $ 0.00922 $ & $ 0.0221 $ \\
 $ \mathrm{Re} \left[ M_{1+}^{C} \right] $ & $ 20.6744_{-0.35855}^{+0.18756} $ & $ 20.5854 $ & $ 0.28293 $ & $ -0.08905 $ & $ 0.31476 $ \\
 $ \mathrm{Im} \left[ M_{1+}^{C} \right] $ & $ -0.911_{-1.29552}^{+1.34447} $ & $ -0.86125 $ & $ 1.34502 $ & $ 0.04975 $ & $ 0.03699 $ \\
 $ \mathrm{Re} \left[ M_{1-}^{C} \right] $ & $ 2.44232_{-0.48481}^{+0.5943} $ & $ 2.4896 $ & $ 0.54749 $ & $ 0.04728 $ & $ 0.08636 $ \\
 $ \mathrm{Im} \left[ M_{1-}^{C} \right] $ & $ -3.84469_{-1.4711}^{+1.52973} $ & $ -3.799 $ & $ 1.514 $ & $ 0.04569 $ & $ 0.03018 $
\end{tabular}
\caption[Numerical results of a bootstrap-analysis are collected for a TPWA-fit of photoproduction data within the $\Delta$-resonance region. The $D$-waves were fixed to SAID CM12 \cite{WorkmanEtAl2012ChewMPhotoprod}. Shown are results for the seventh energy-bin, \newline $E_{\gamma }\text{ = 420.0 MeV}$.]{Numerical results are collected for a TPWA bootstrap-analysis of photoproduction data within the $\Delta$-resonance region. Here, the seventh energy-bin, $E_{\gamma }\text{ = 420.0 MeV}$, is shown. For more details, see the description of Table \ref{tab:DeltaRegionResultsFirstEnergy}.}
\label{tab:DeltaRegionResultsFifthSixthSeventhEnergy}
\end{table}

\clearpage

\textbf{Histograms for fit-parameters resulting from application of the bootstrap} \newline

\begin{figure}[h]
\centering
\begin{overpic}[width=0.325\textwidth]{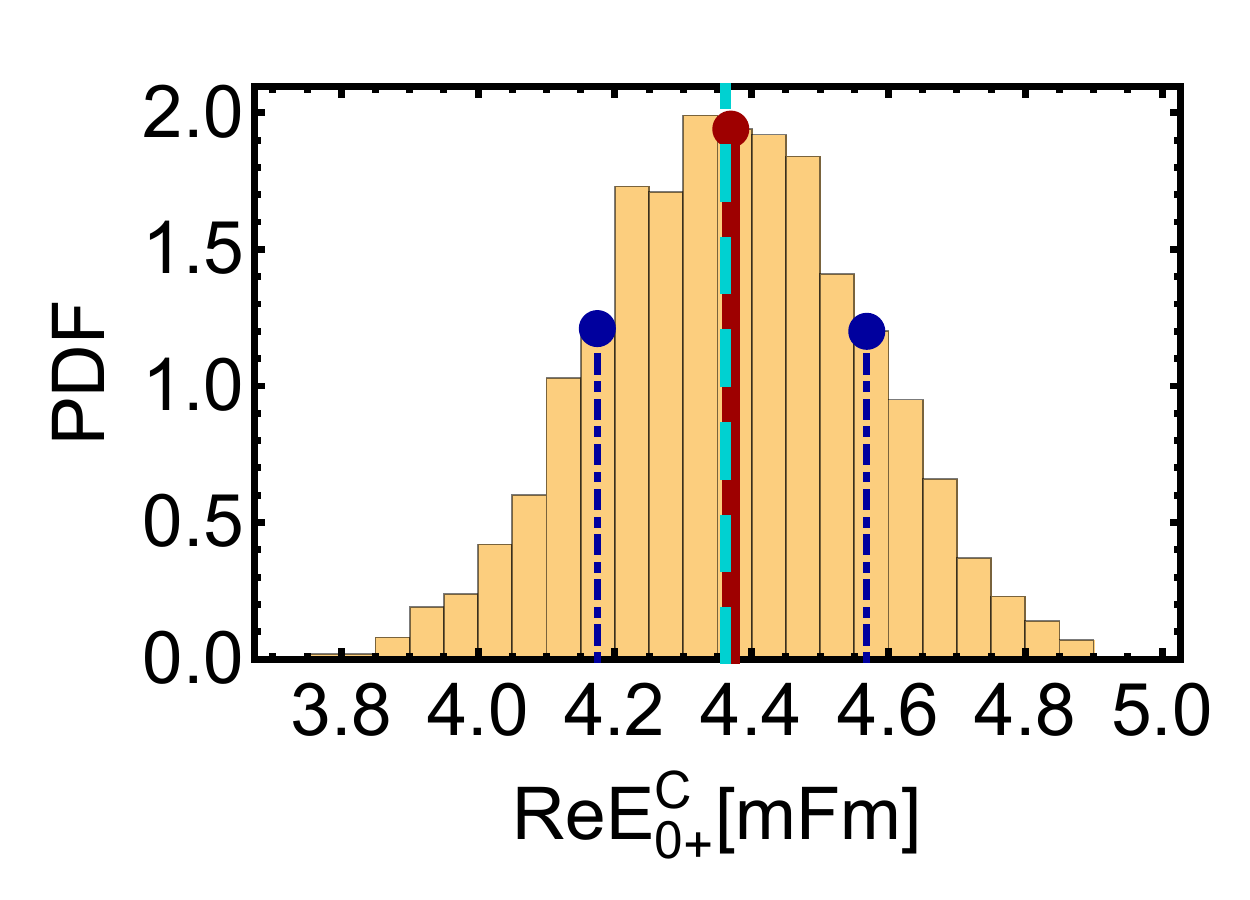}
 \end{overpic}
\begin{overpic}[width=0.325\textwidth]{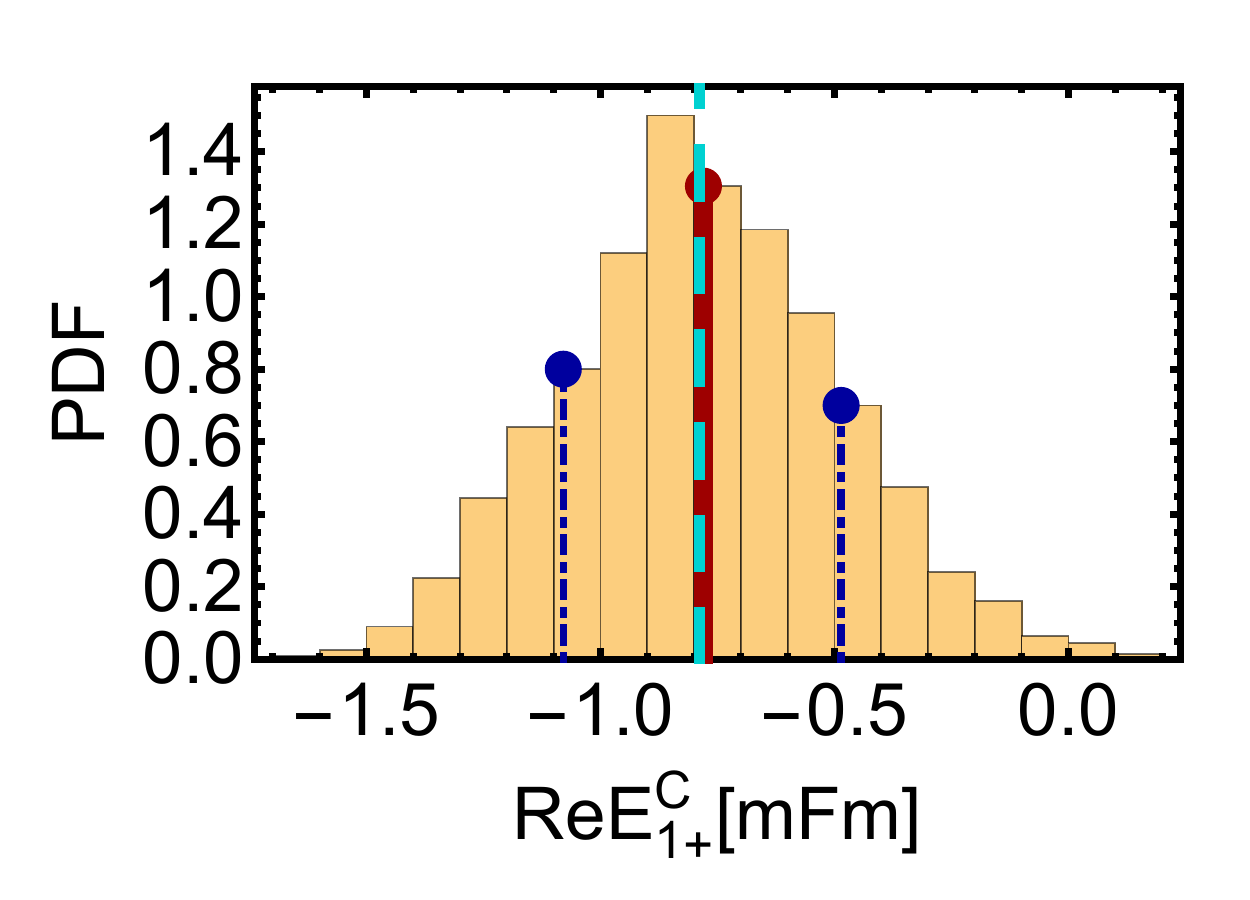}
 \end{overpic}
\begin{overpic}[width=0.325\textwidth]{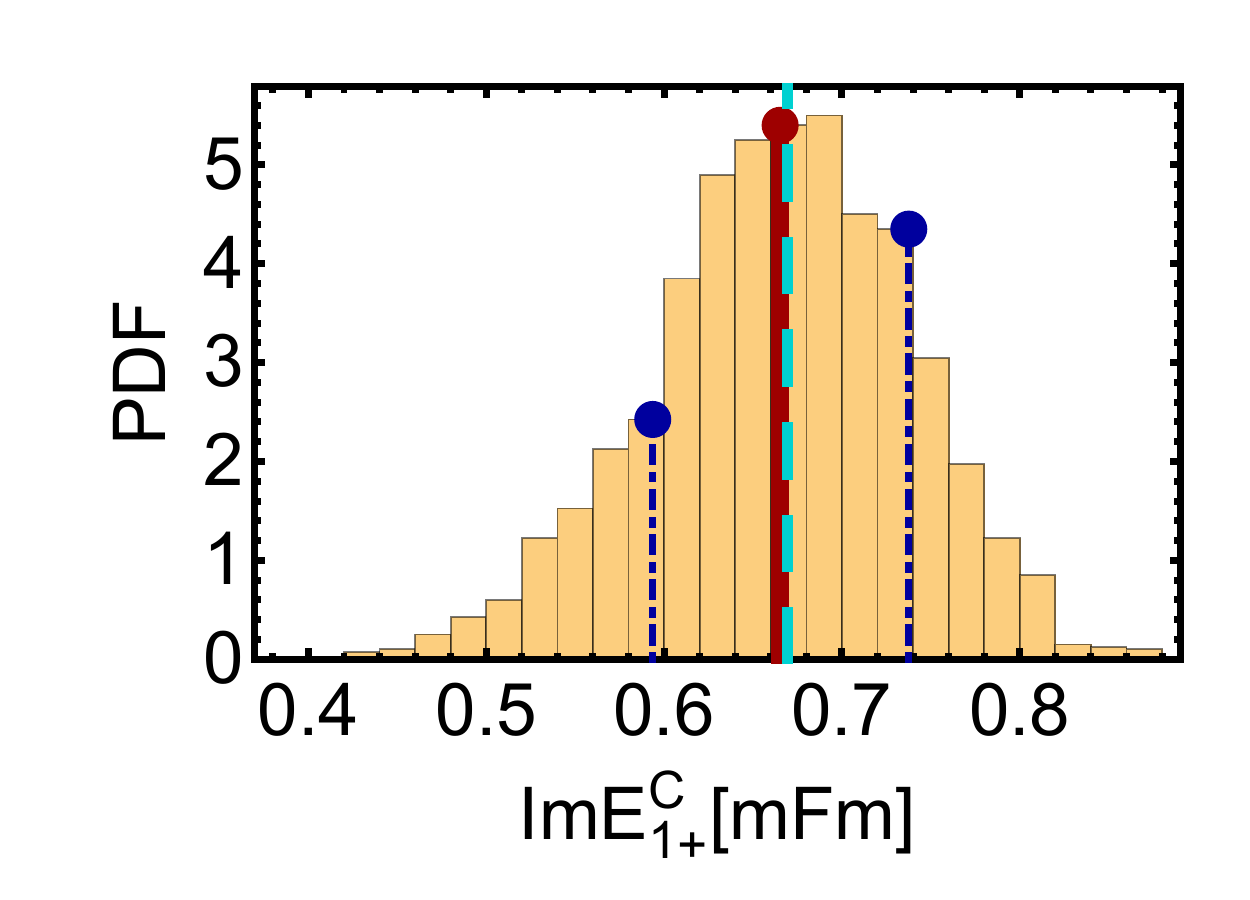}
 \end{overpic} \\
\begin{overpic}[width=0.325\textwidth]{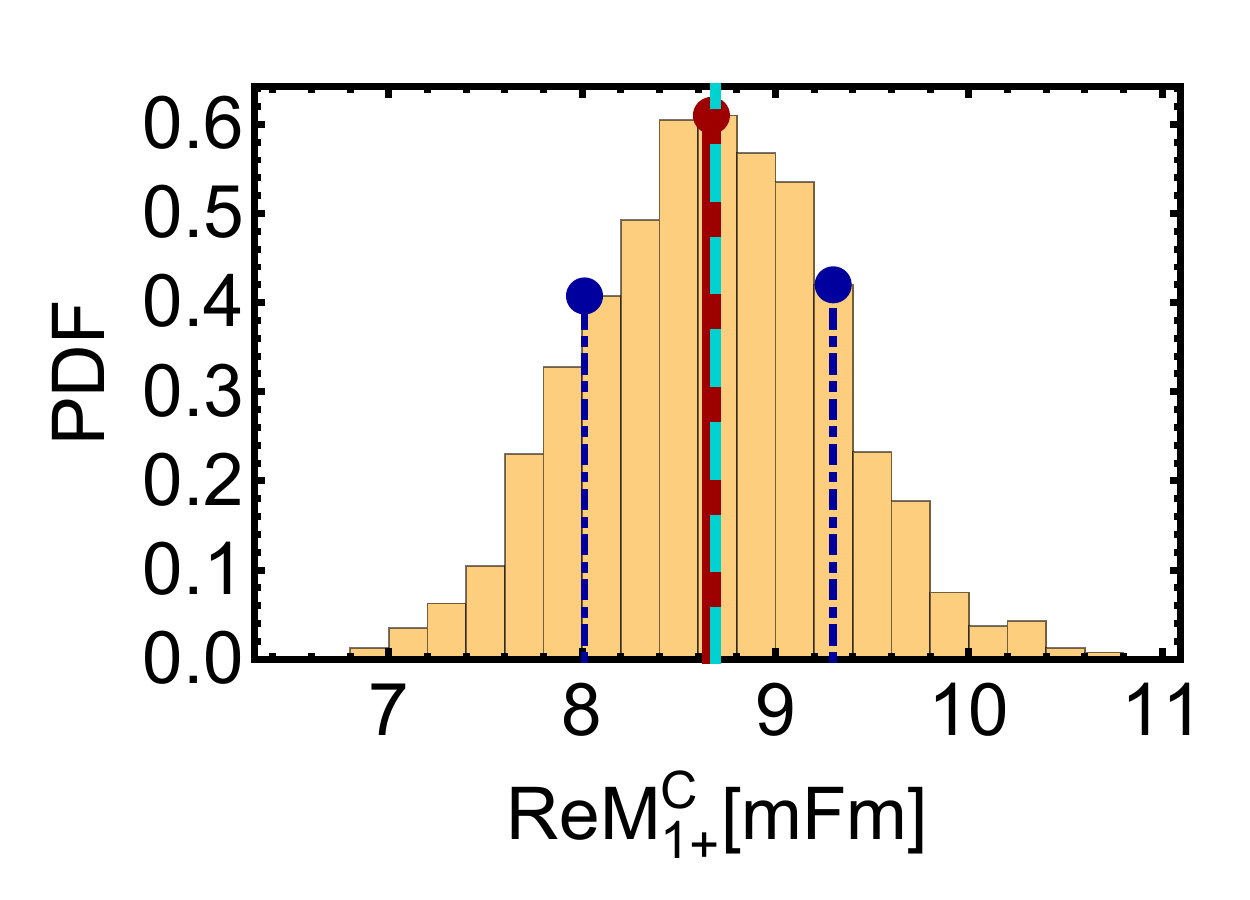}
 \end{overpic}
\begin{overpic}[width=0.325\textwidth]{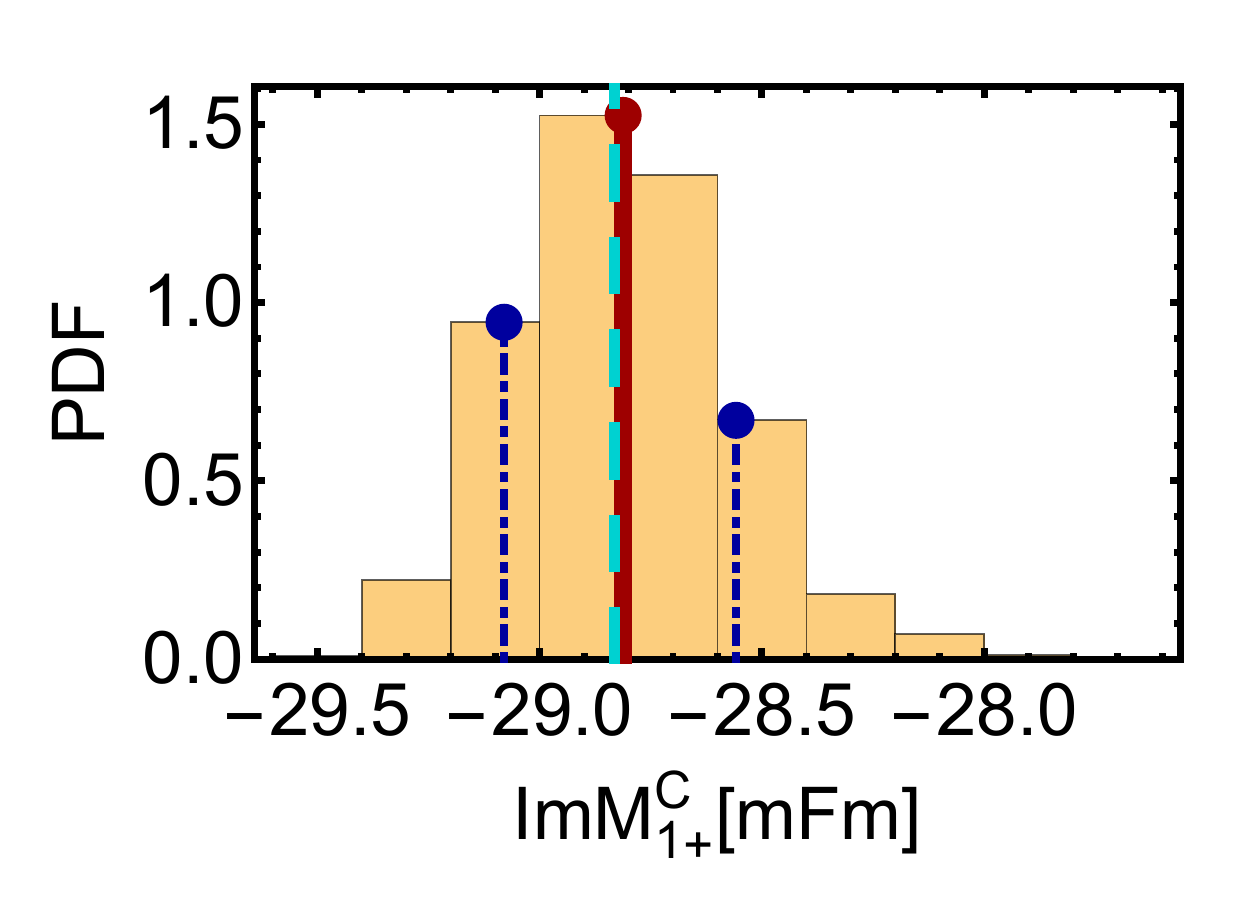}
 \end{overpic}
\begin{overpic}[width=0.325\textwidth]{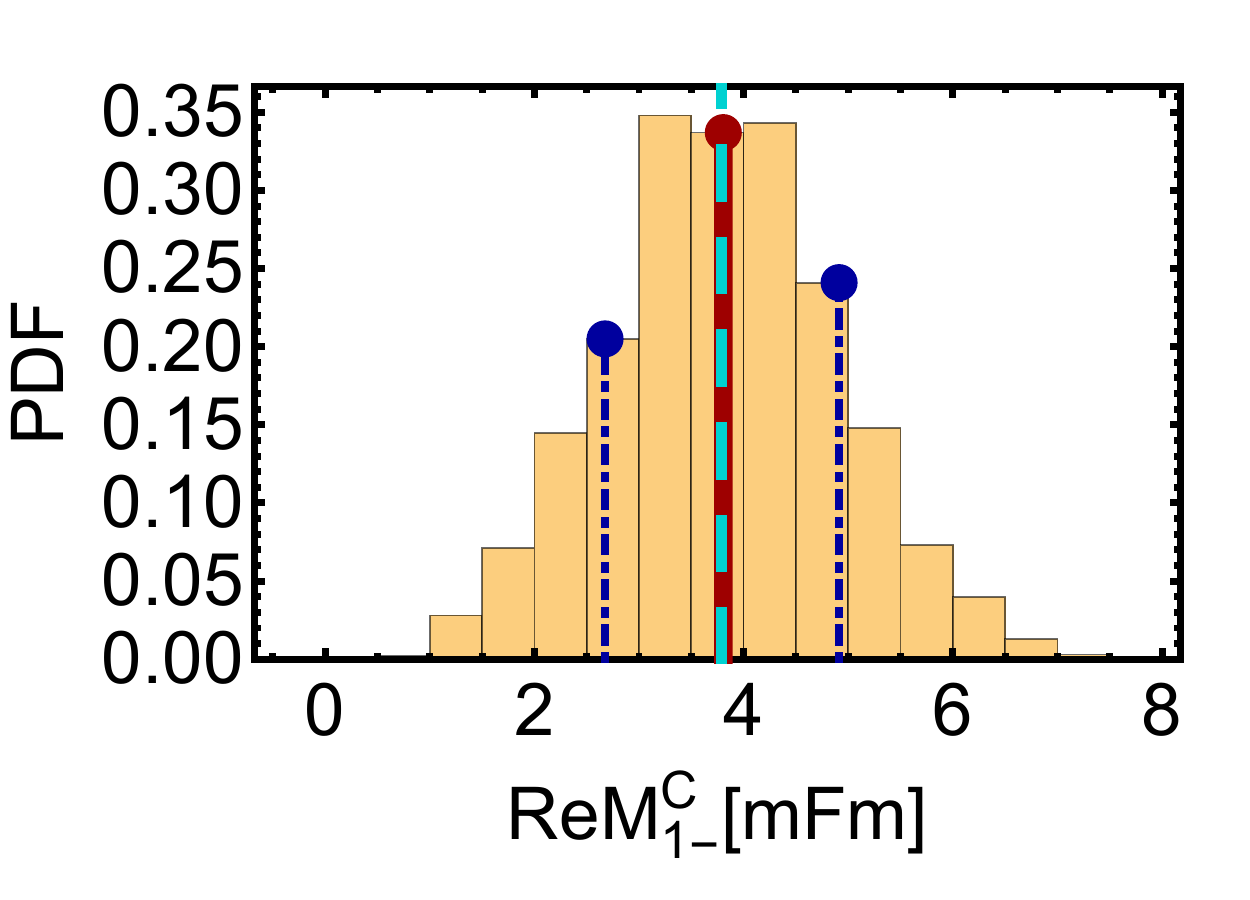}
 \end{overpic} \\
\begin{overpic}[width=0.325\textwidth]{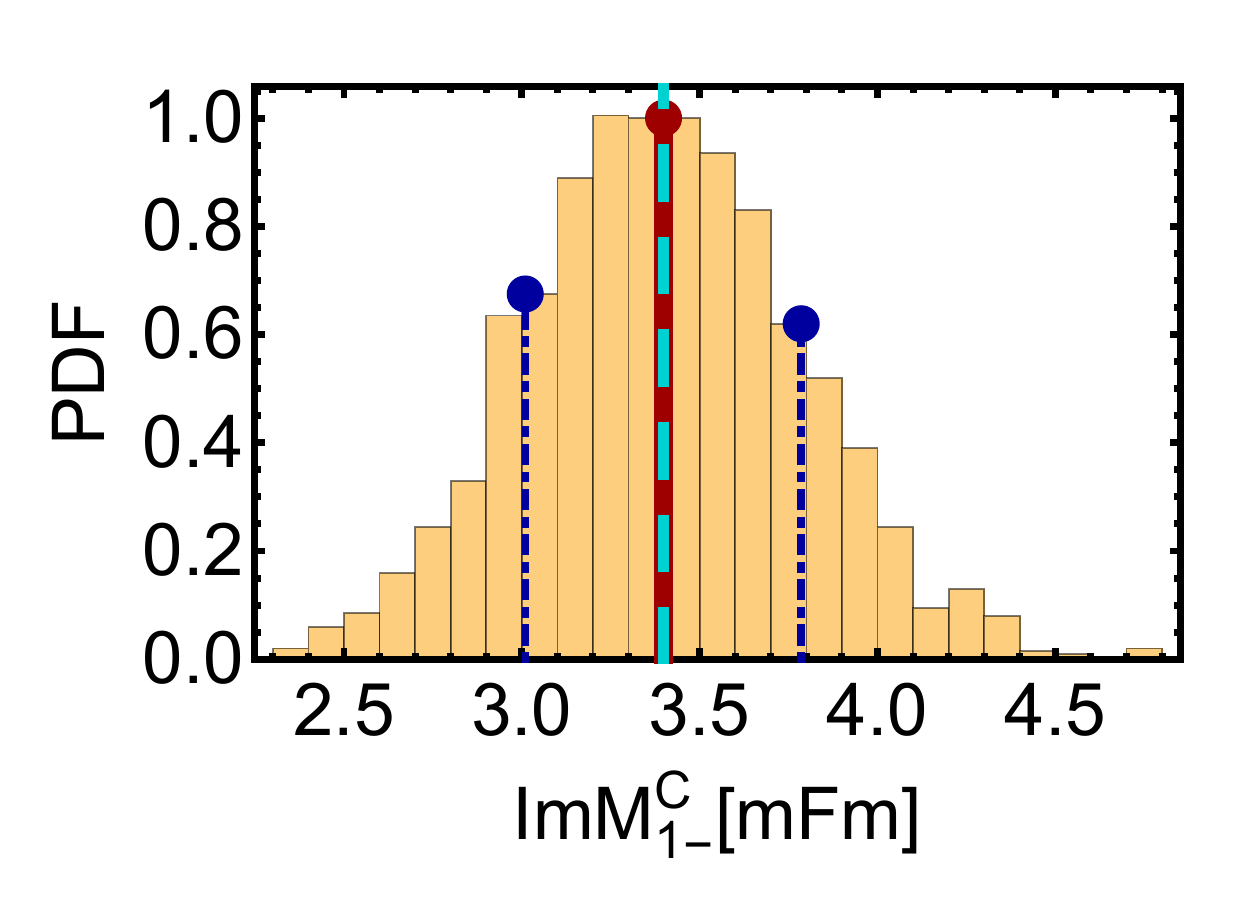}
 \end{overpic}
\caption[Bootstrap-distributions for multipole fit-parameters in an analysis of photoproduction data on the $\Delta$-resonance region. The first energy-bin, \newline $E_{\gamma }\text{ = 280.0 MeV}$, is shown.]{The histograms show bootstrap-distributions for the real- and imaginary parts of phase-constrained $S$- and $P$-wave multipoles, for a TPWA bootstrap-analysis of photoproduction data in the $\Delta$-resonance region (see section \ref{subsec:DeltaRegionDataFits}). The first energy-bin, $E_{\gamma }\text{ = 280.0 MeV}$, is shown. An ensemble of $B=2000$ bootstrap-replicates has been the basis of these results, using solely the statistical errors of the original datasets. The $D$-waves were held fixed to the SAID-solution CM12 \cite{WorkmanEtAl2012ChewMPhotoprod,SAID} during the entire analysis. The fit to the original data is shown in Figure \ref{fig:Lmax2DWavesSAIDFitResultsDeltaRegionPurelyStat} in the main text. \newline
The distributions have been normalized to $1$ via use of the object \textit{HistogramDistribution} in MATHEMATICA \cite{Mathematica8,Mathematica11,MathematicaLanguage,MathematicaBonnLicense}. Thus, $y$-axes are labelled as \textit{PDF}. The mean of each distribution is shown as a red solid line, while the $0.16$- and $0.84$-quantiles are indicated by blue dash-dotted lines. The global minimum of the fit to the original data is plotted as a cyan-colored dashed horizontal line.}
\label{fig:BootstrapHistosDeltaRegionEnergy1}
\end{figure}

\clearpage

\begin{figure}[h]
\centering
\vspace*{-10pt}
\begin{overpic}[width=0.325\textwidth]{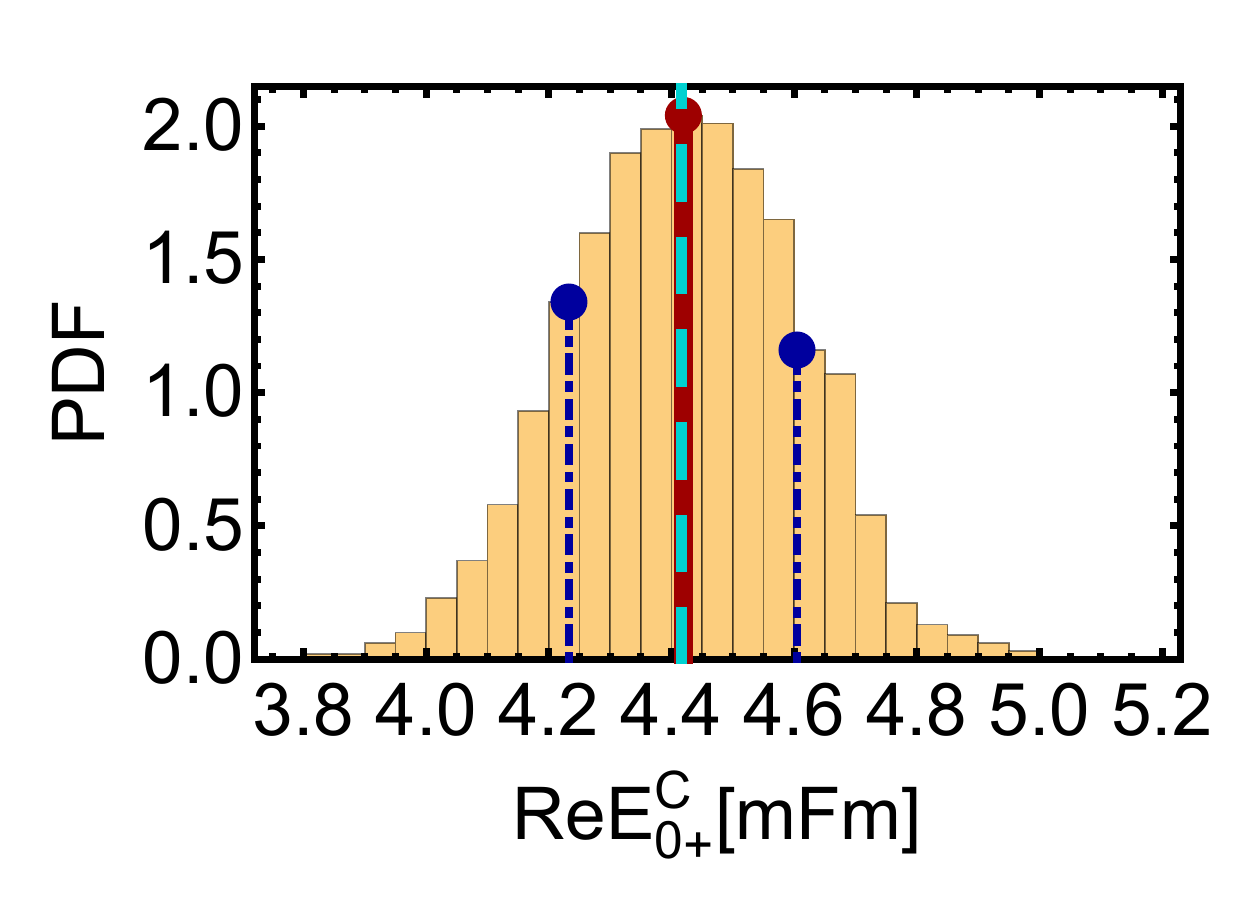}
 \end{overpic}
\begin{overpic}[width=0.325\textwidth]{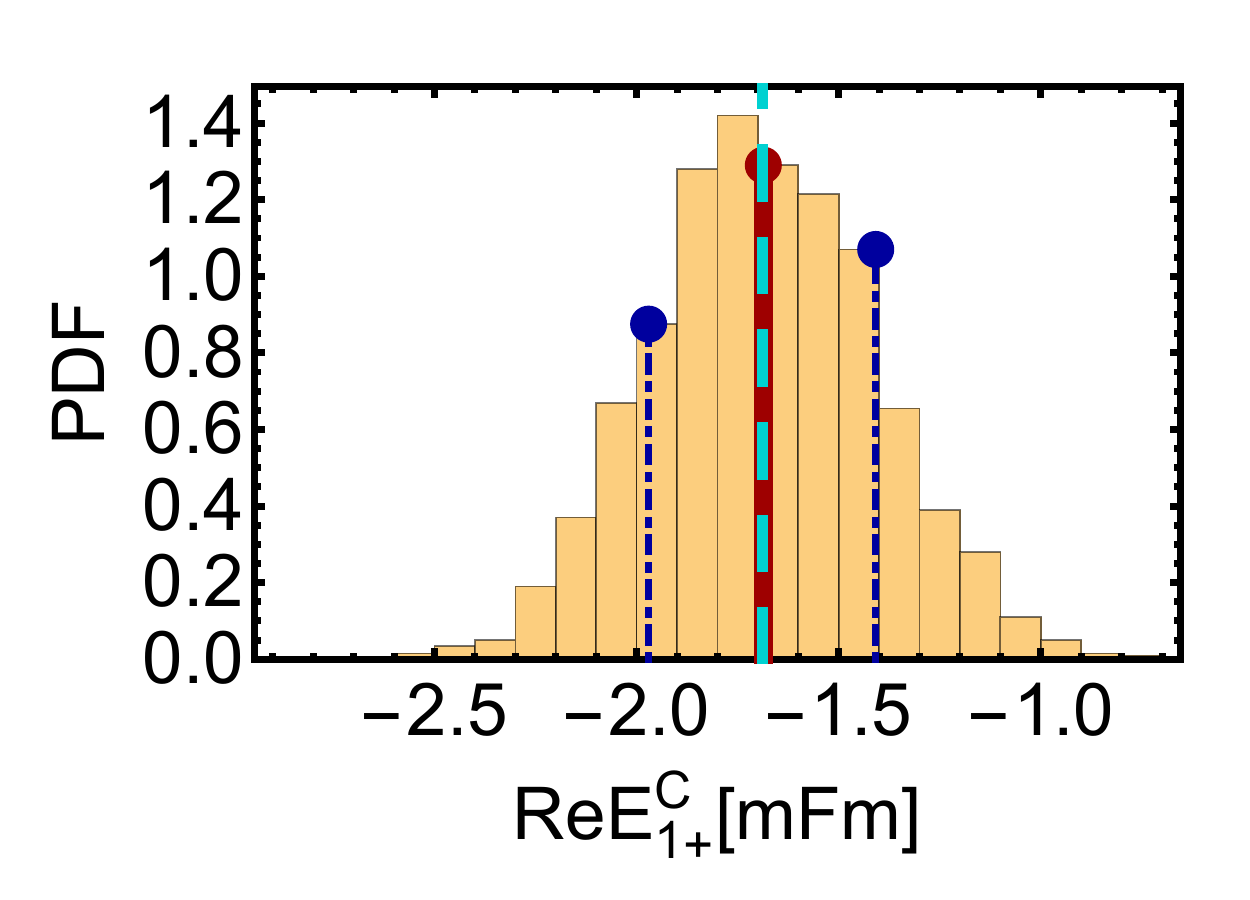}
 \end{overpic}
\begin{overpic}[width=0.325\textwidth]{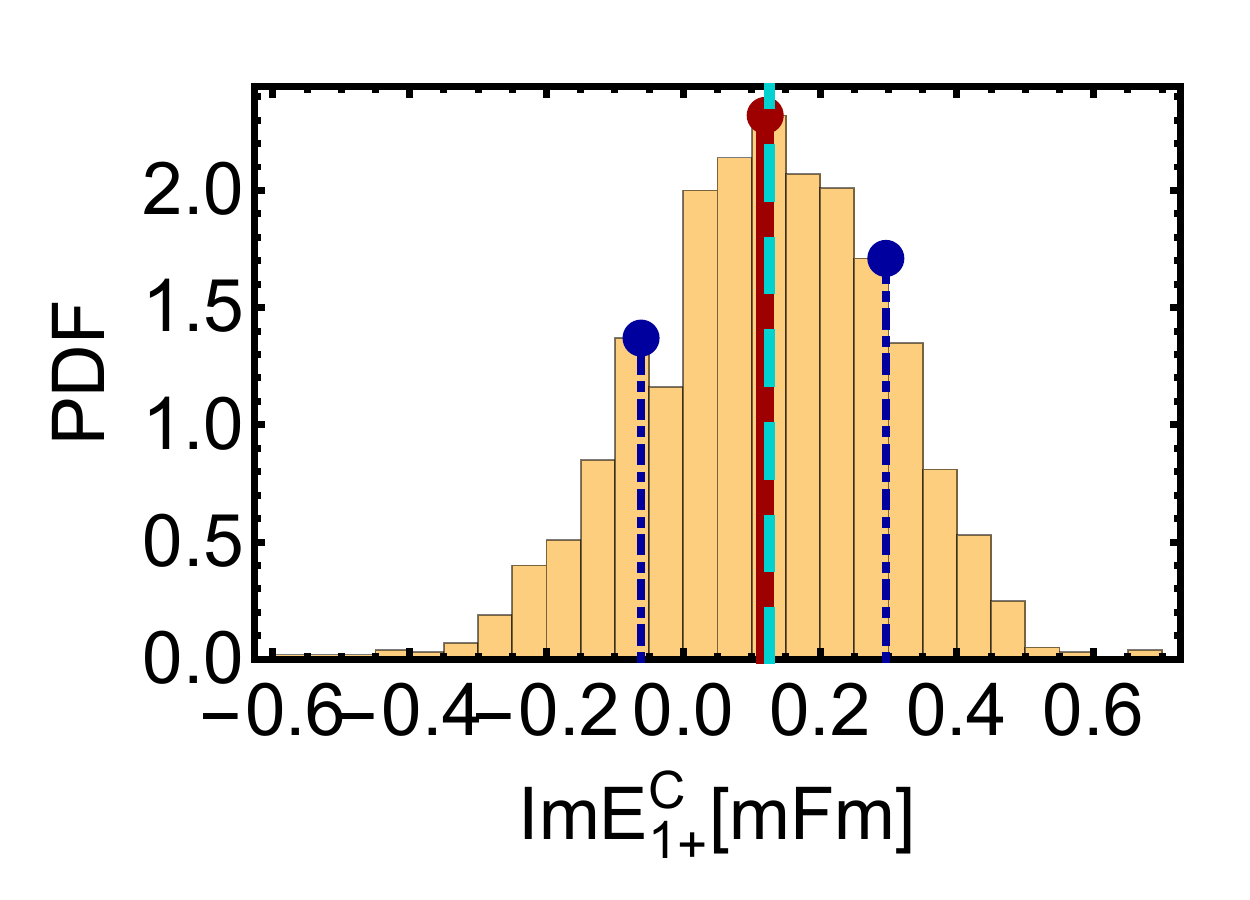}
 \end{overpic} \\ \vspace*{-8pt}
\begin{overpic}[width=0.325\textwidth]{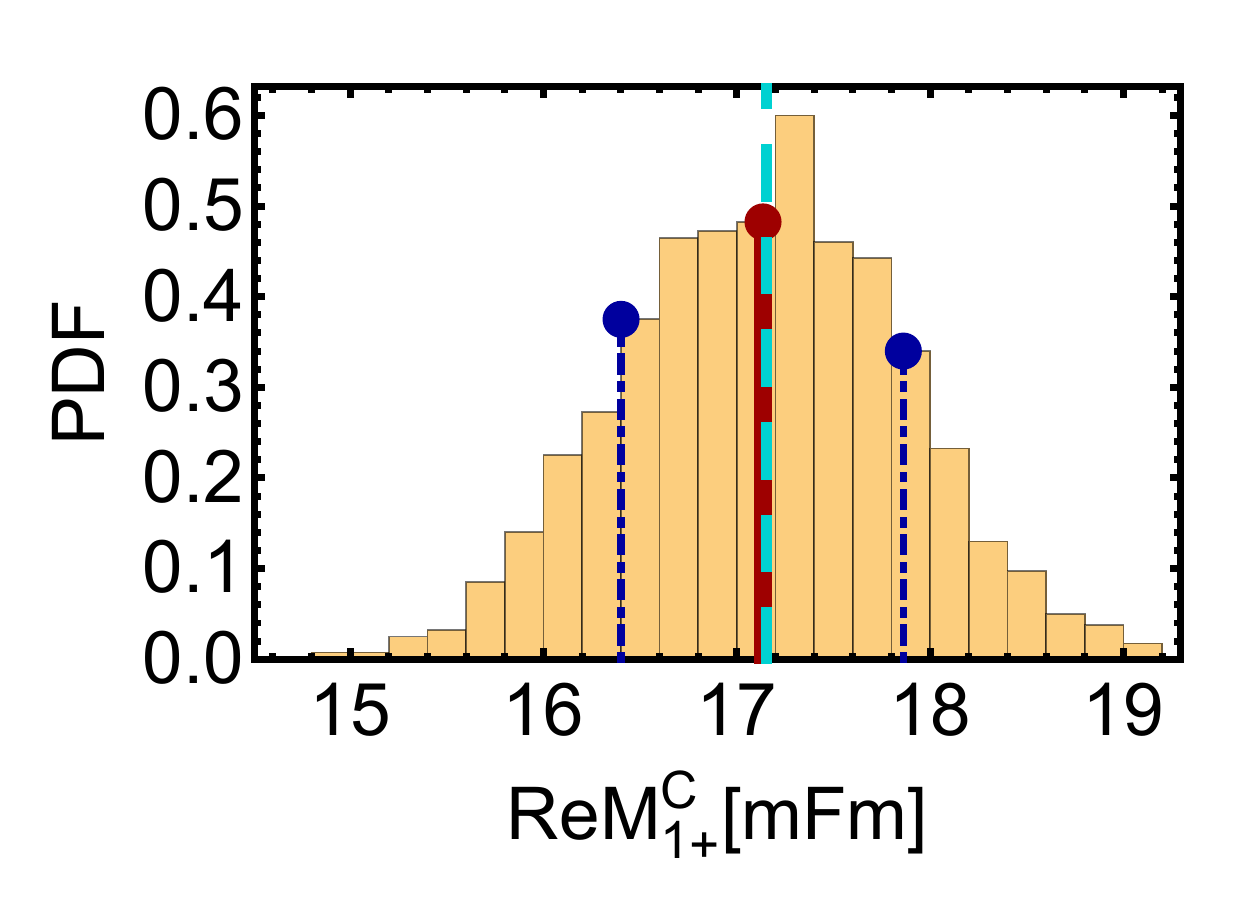}
 \end{overpic}
\begin{overpic}[width=0.325\textwidth]{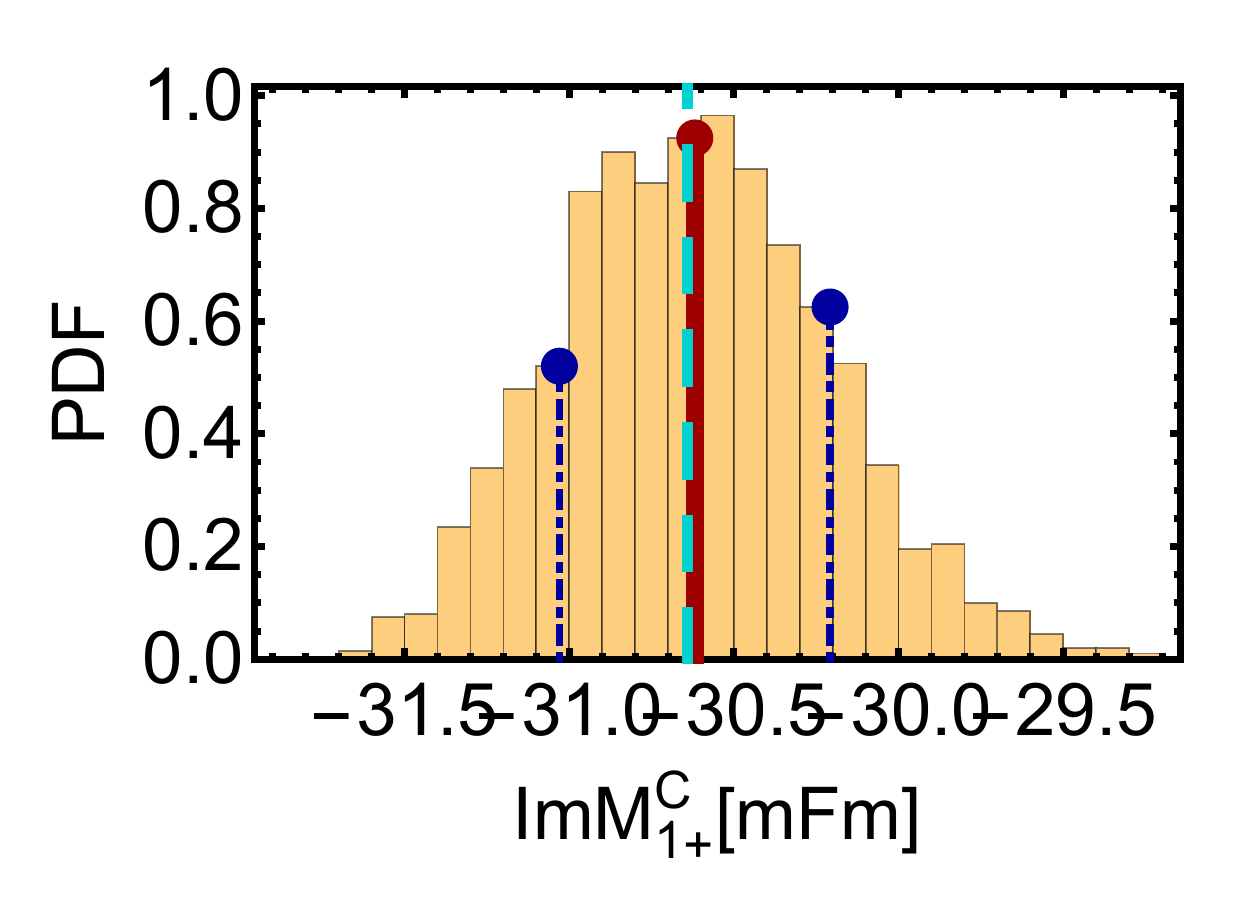}
 \end{overpic}
\begin{overpic}[width=0.325\textwidth]{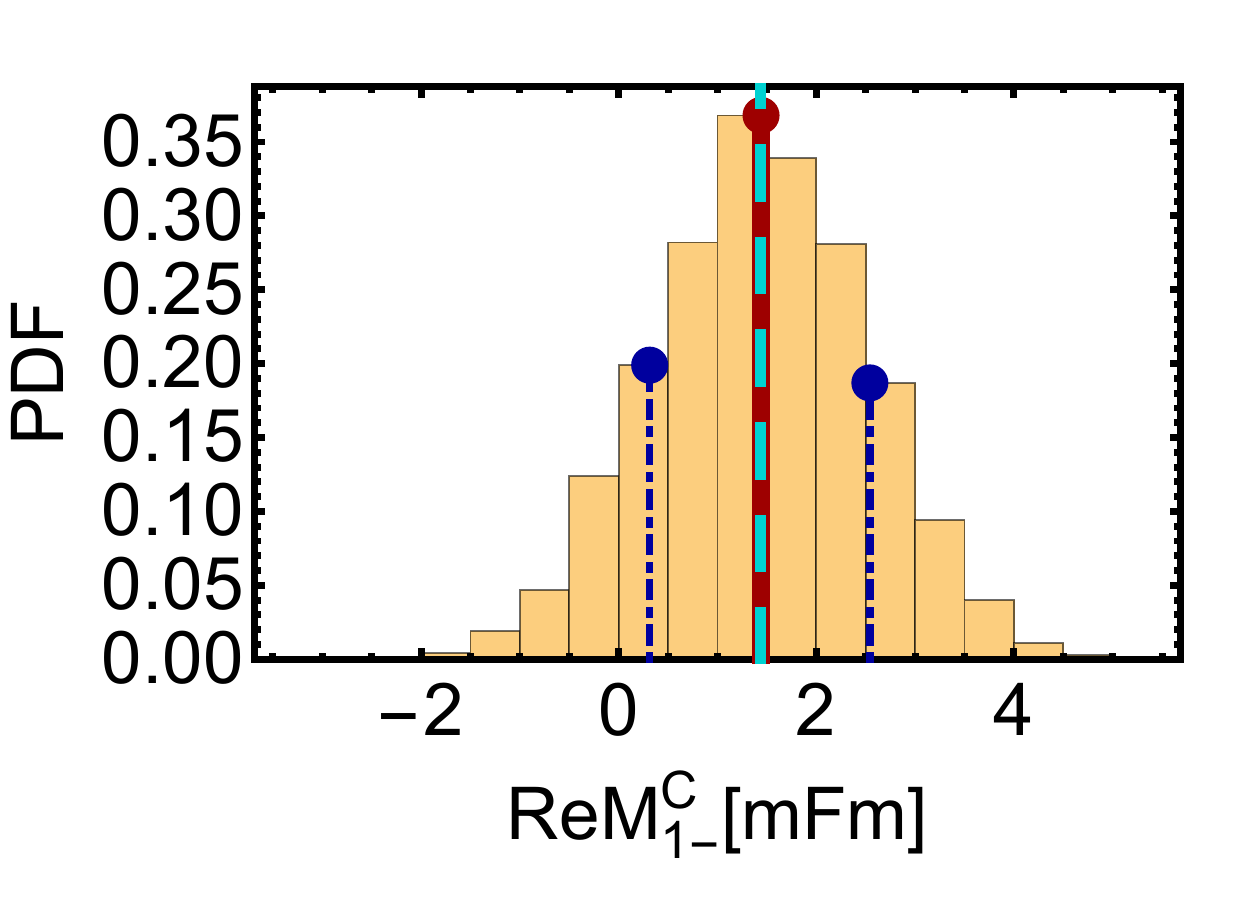}
 \end{overpic} \\ \vspace*{-8pt}
\begin{overpic}[width=0.325\textwidth]{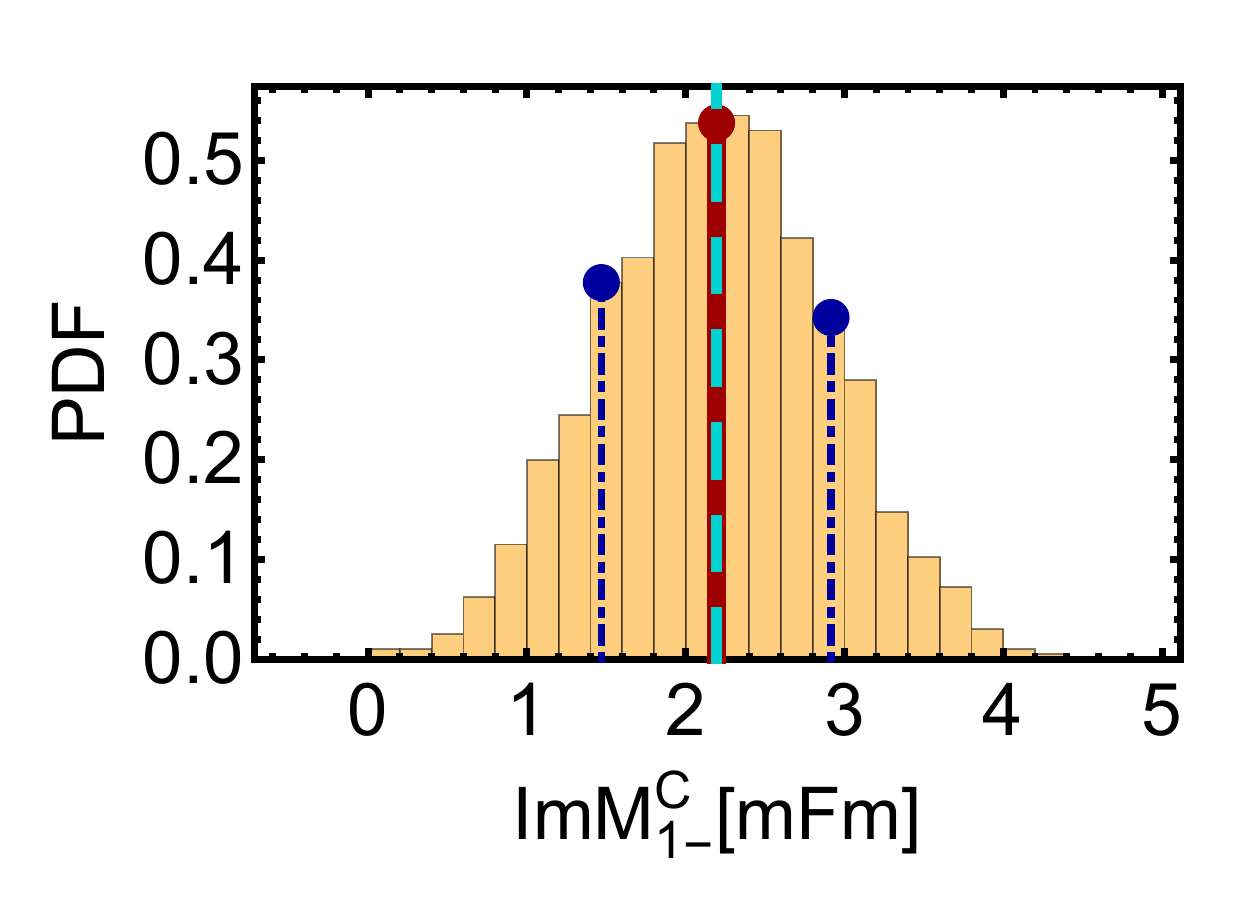}
 \end{overpic}
\caption[Bootstrap-distributions for multipole fit-parameters in an analysis of photoproduction data on the $\Delta$-resonance region. The second energy-bin, \newline $E_{\gamma }\text{ = 300.0 MeV}$, is shown.]{The histograms belong to the same bootstrap-analysis shown in Figure \ref{fig:BootstrapHistosDeltaRegionEnergy1}, but here the second energy-bin, $E_{\gamma }\text{ = 300.0 MeV}$, is shown.} \bigskip
\begin{overpic}[width=0.325\textwidth]{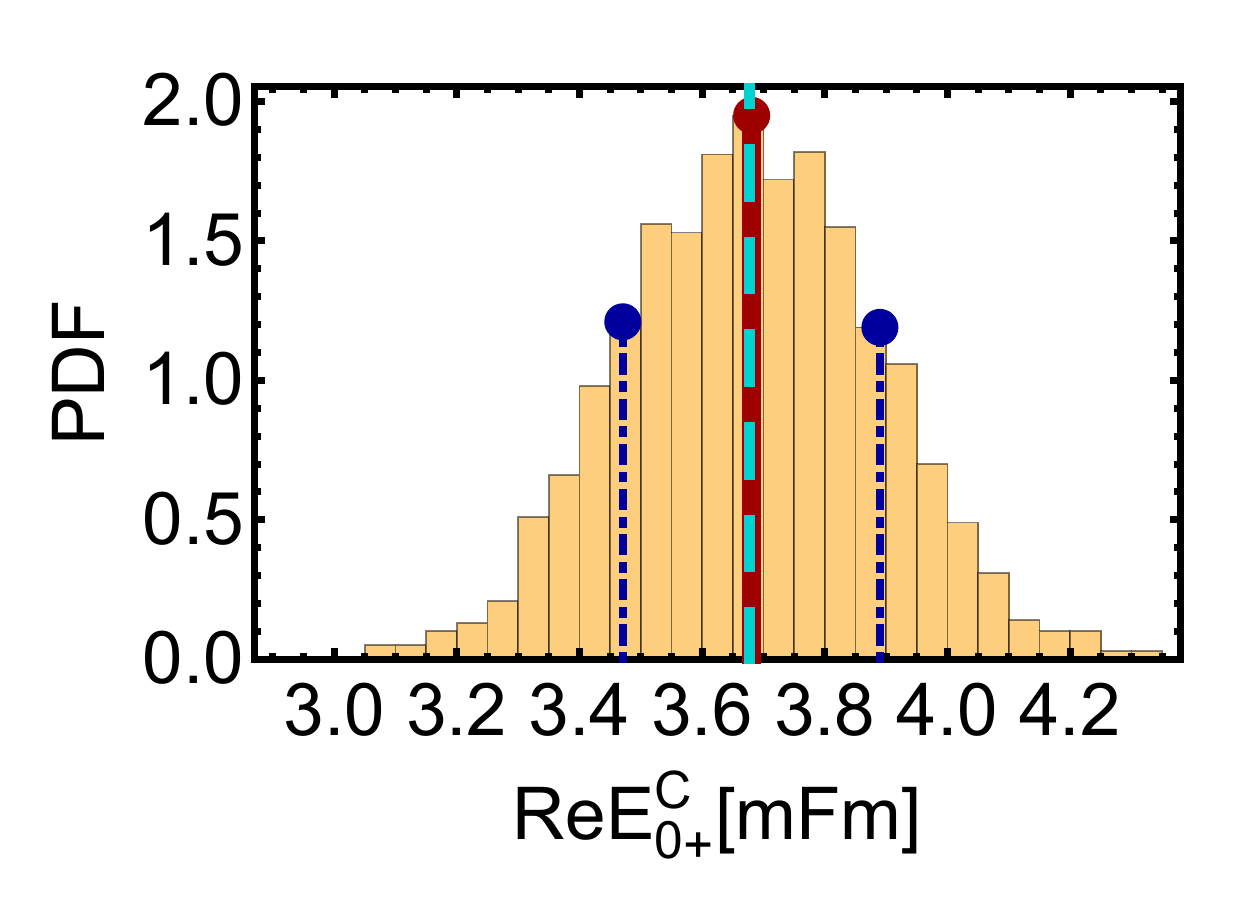}
 \end{overpic}
\begin{overpic}[width=0.325\textwidth]{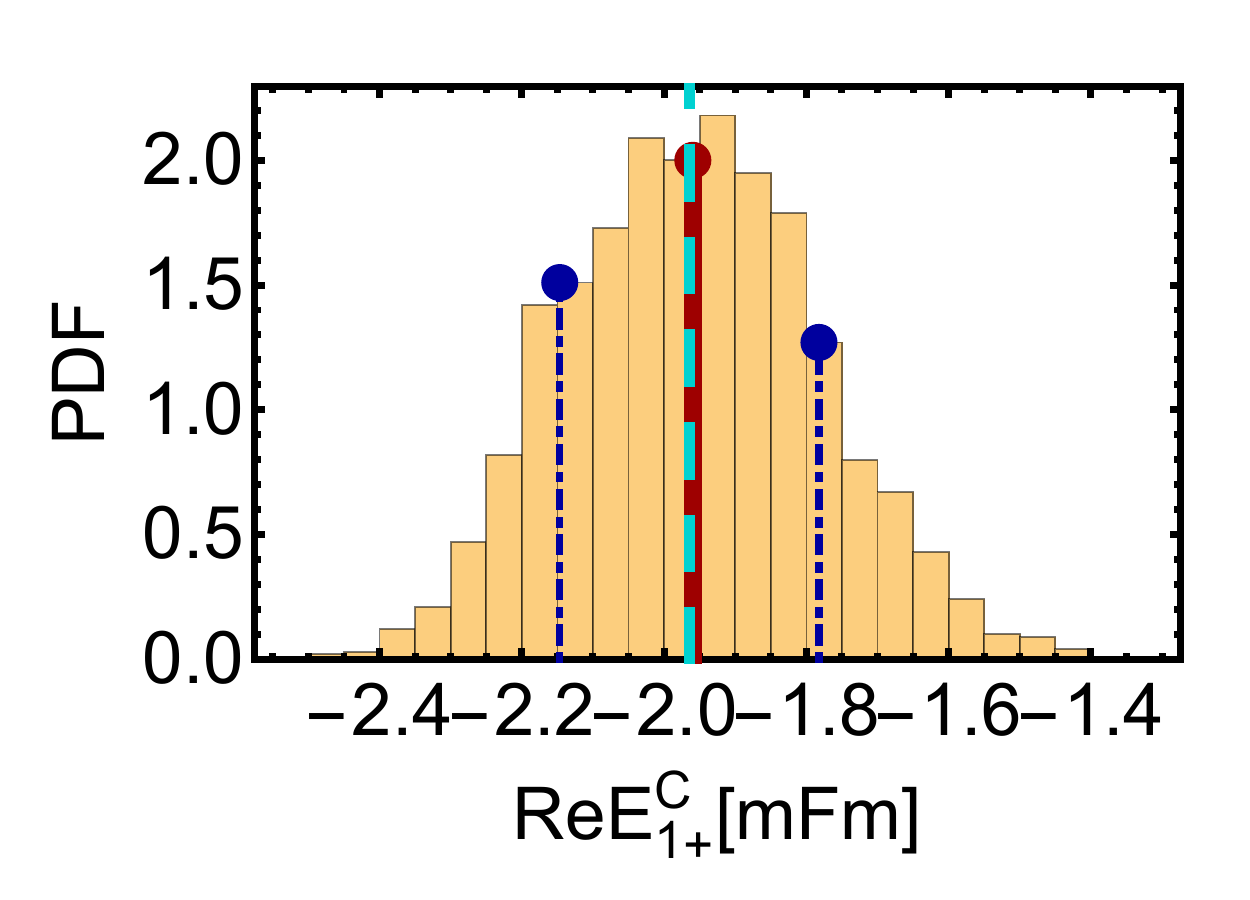}
 \end{overpic}
\begin{overpic}[width=0.325\textwidth]{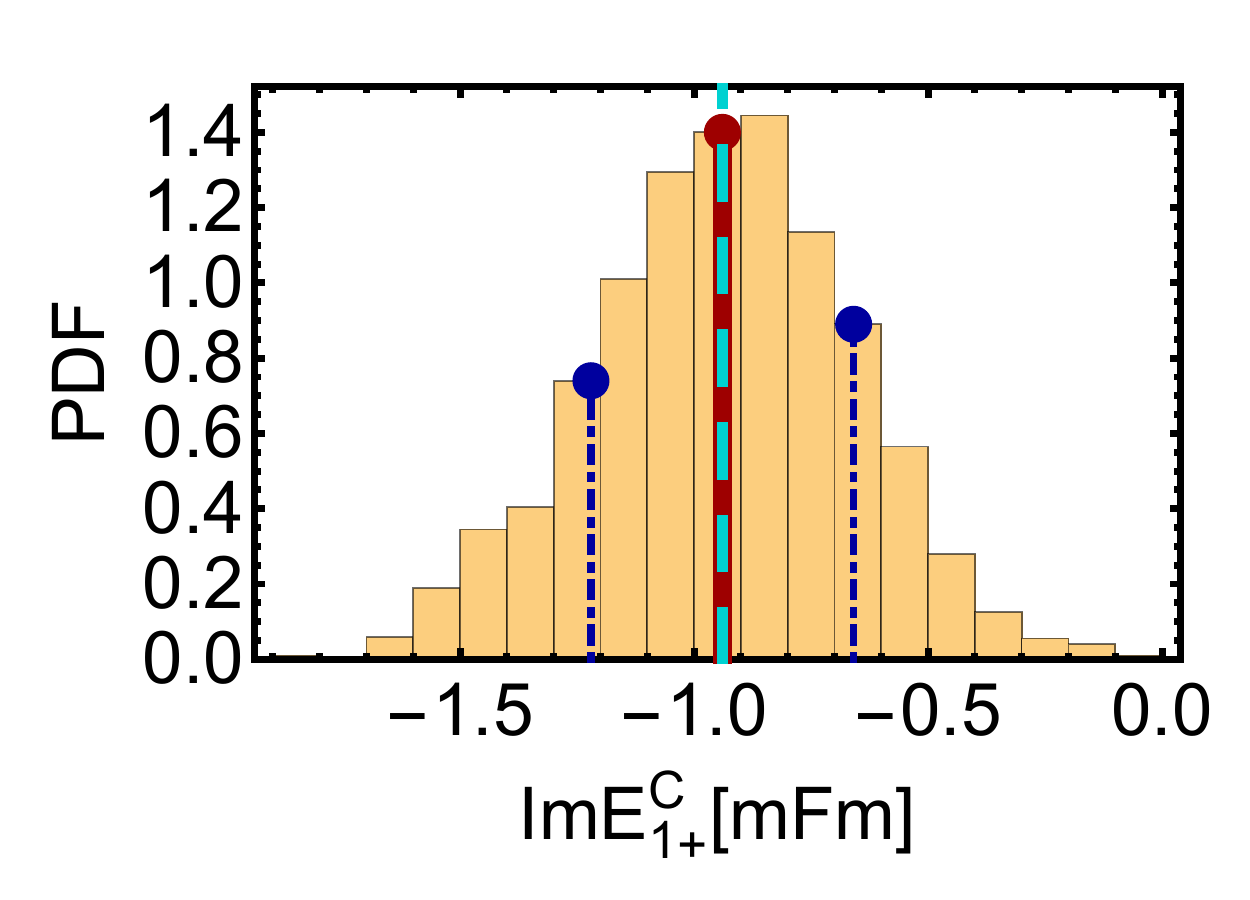}
 \end{overpic} \\ \vspace*{-8pt}
\begin{overpic}[width=0.325\textwidth]{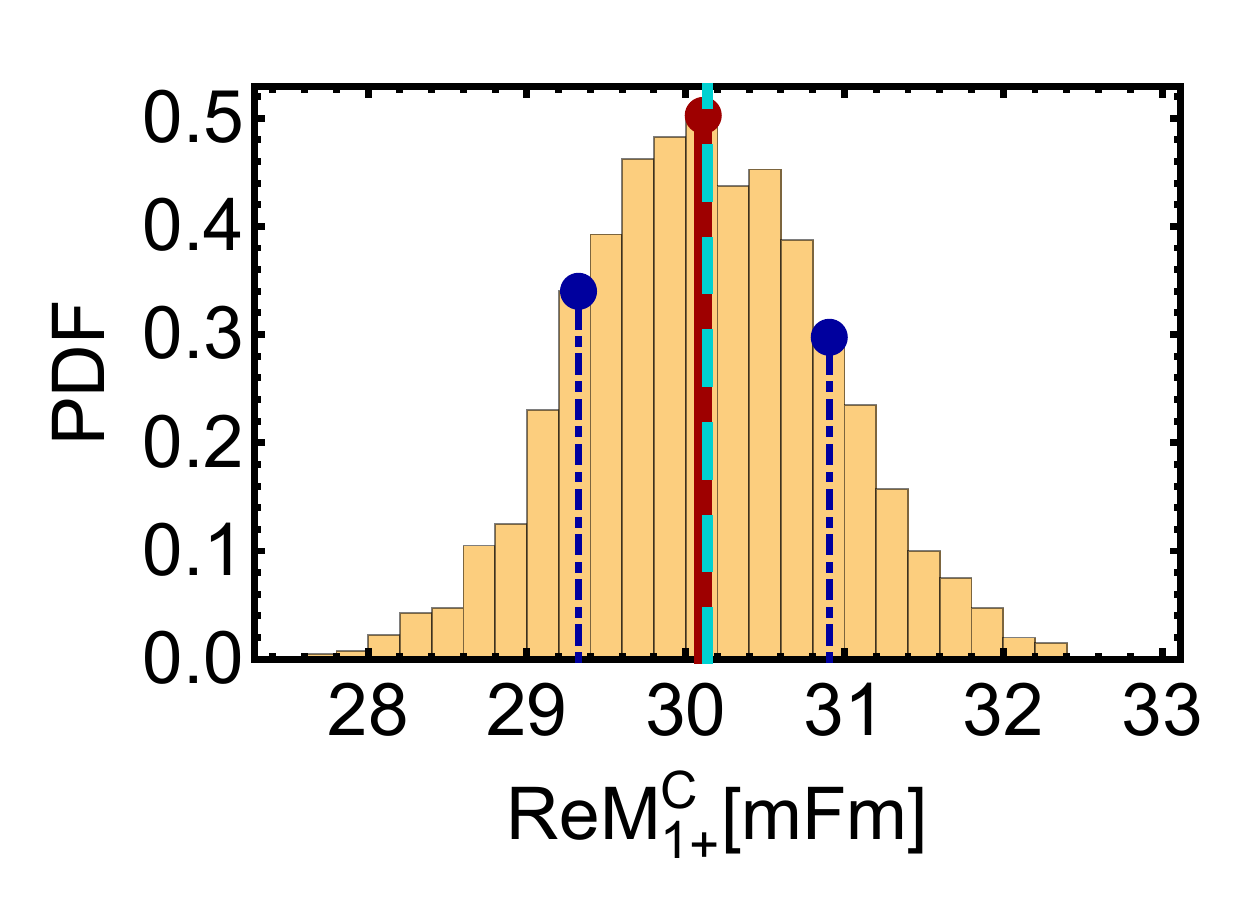}
 \end{overpic}
\begin{overpic}[width=0.325\textwidth]{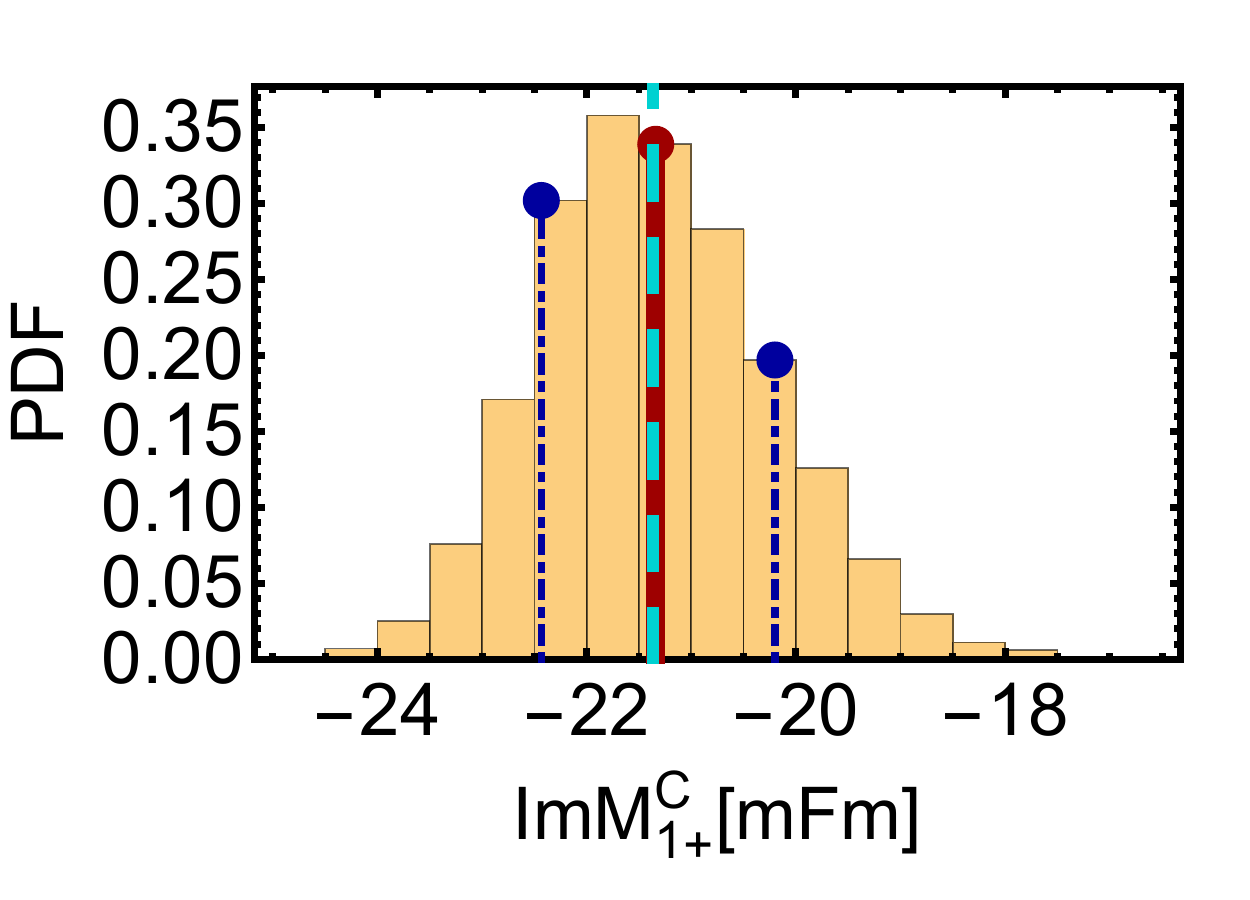}
 \end{overpic}
\begin{overpic}[width=0.325\textwidth]{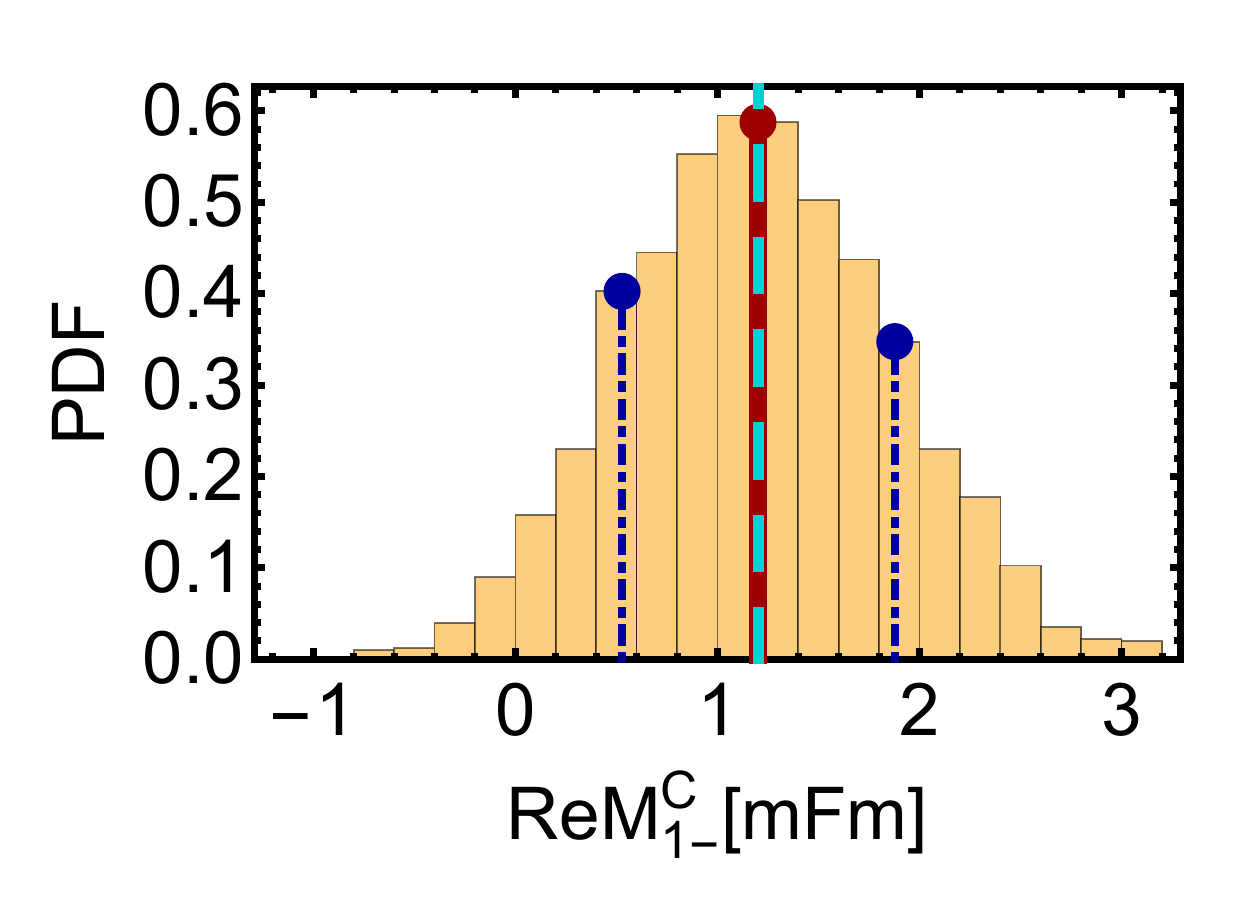}
 \end{overpic} \\ \vspace*{-8pt}
\begin{overpic}[width=0.325\textwidth]{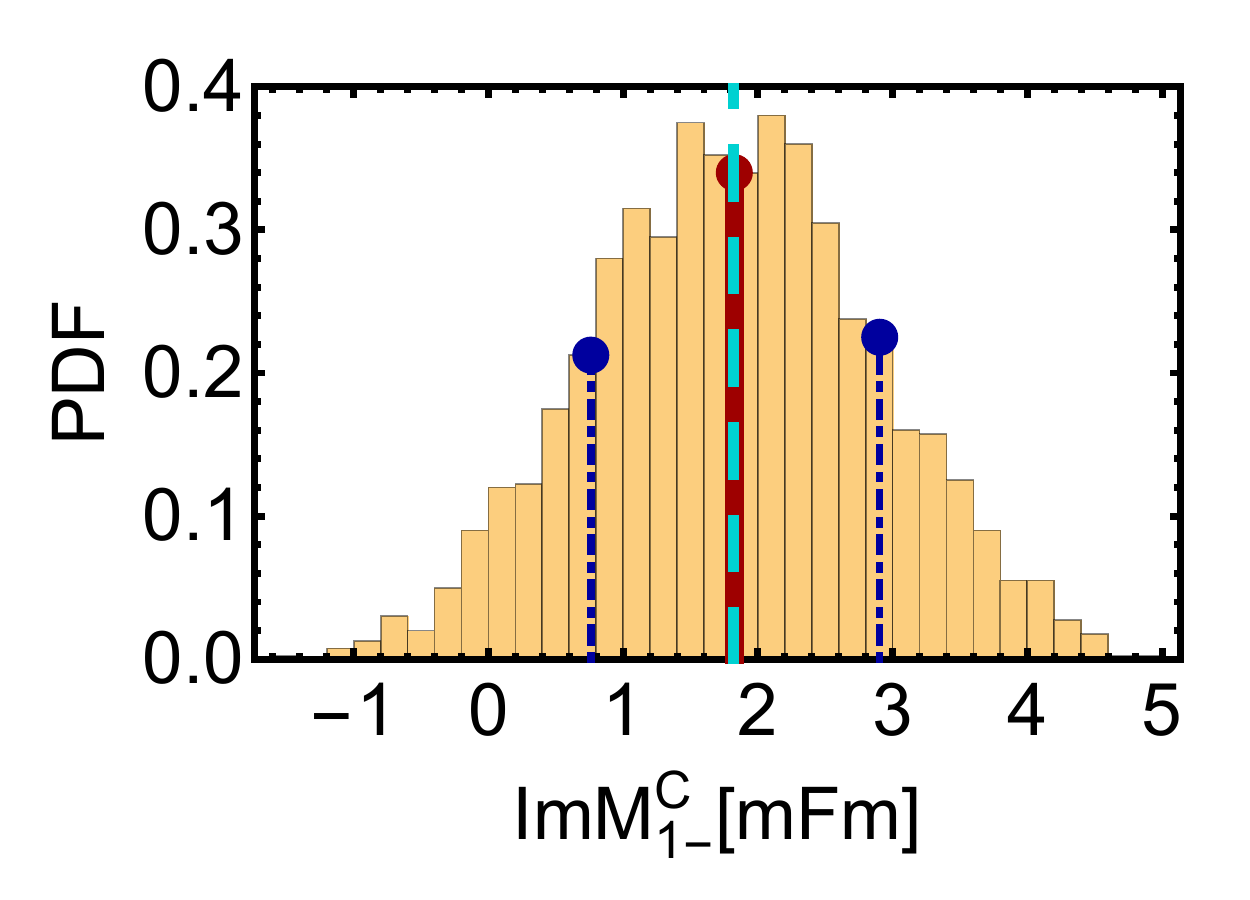}
 \end{overpic}
\caption[Bootstrap-distributions for multipole fit-parameters in an analysis of photoproduction data on the $\Delta$-resonance region. The third energy-bin, \newline $E_{\gamma }\text{ = 320.0 MeV}$, is shown.]{The histograms belong to the same bootstrap-analysis shown in Figure \ref{fig:BootstrapHistosDeltaRegionEnergy1}, but here the third energy-bin, $E_{\gamma }\text{ = 320.0 MeV}$, is shown.}
\label{fig:BootstrapHistosDeltaRegionEnergies2and3}
\end{figure}

\clearpage

\begin{figure}[h]
\centering
\vspace*{-10pt}
\begin{overpic}[width=0.325\textwidth]{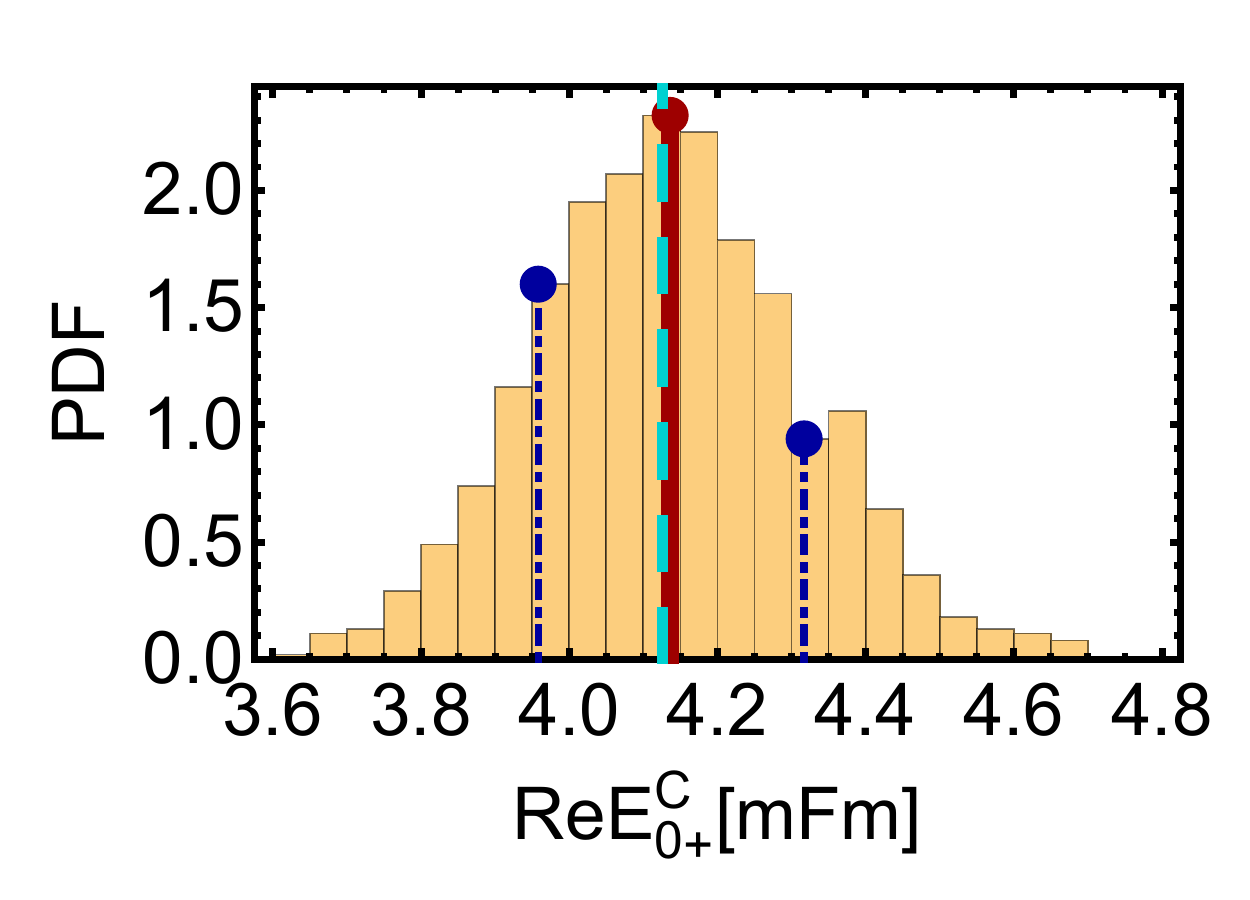}
 \end{overpic}
\begin{overpic}[width=0.325\textwidth]{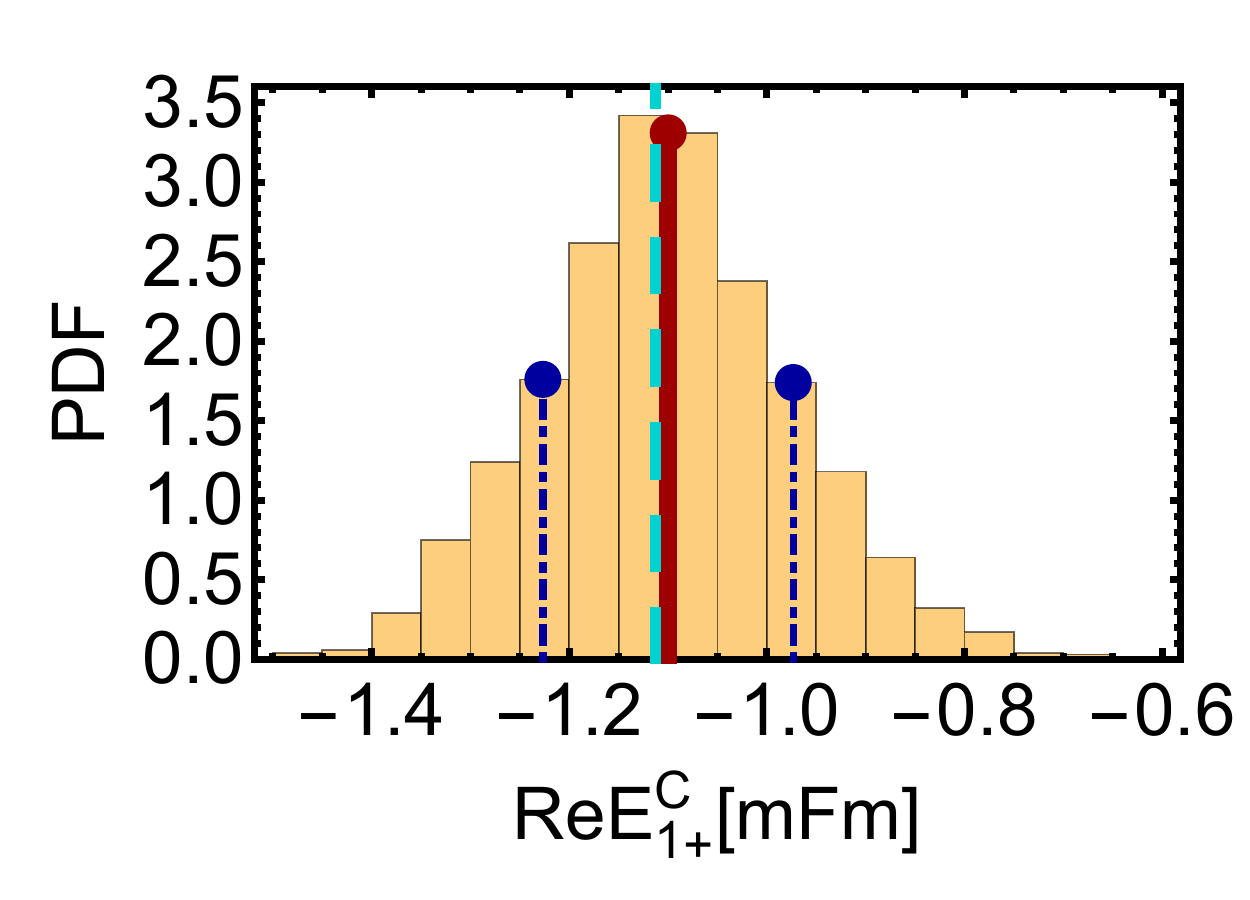}
 \end{overpic}
\begin{overpic}[width=0.325\textwidth]{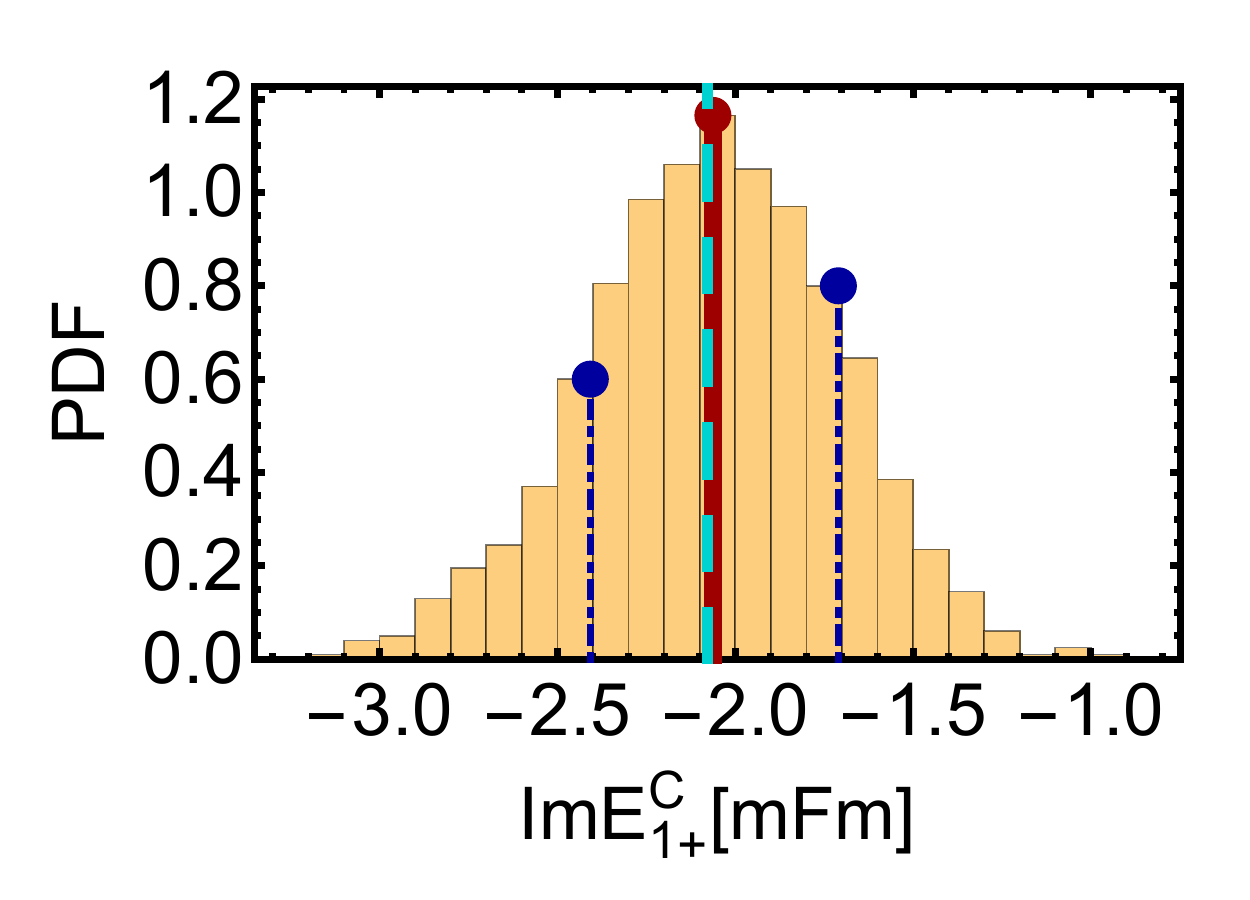}
 \end{overpic} \\ \vspace*{-8pt}
\begin{overpic}[width=0.325\textwidth]{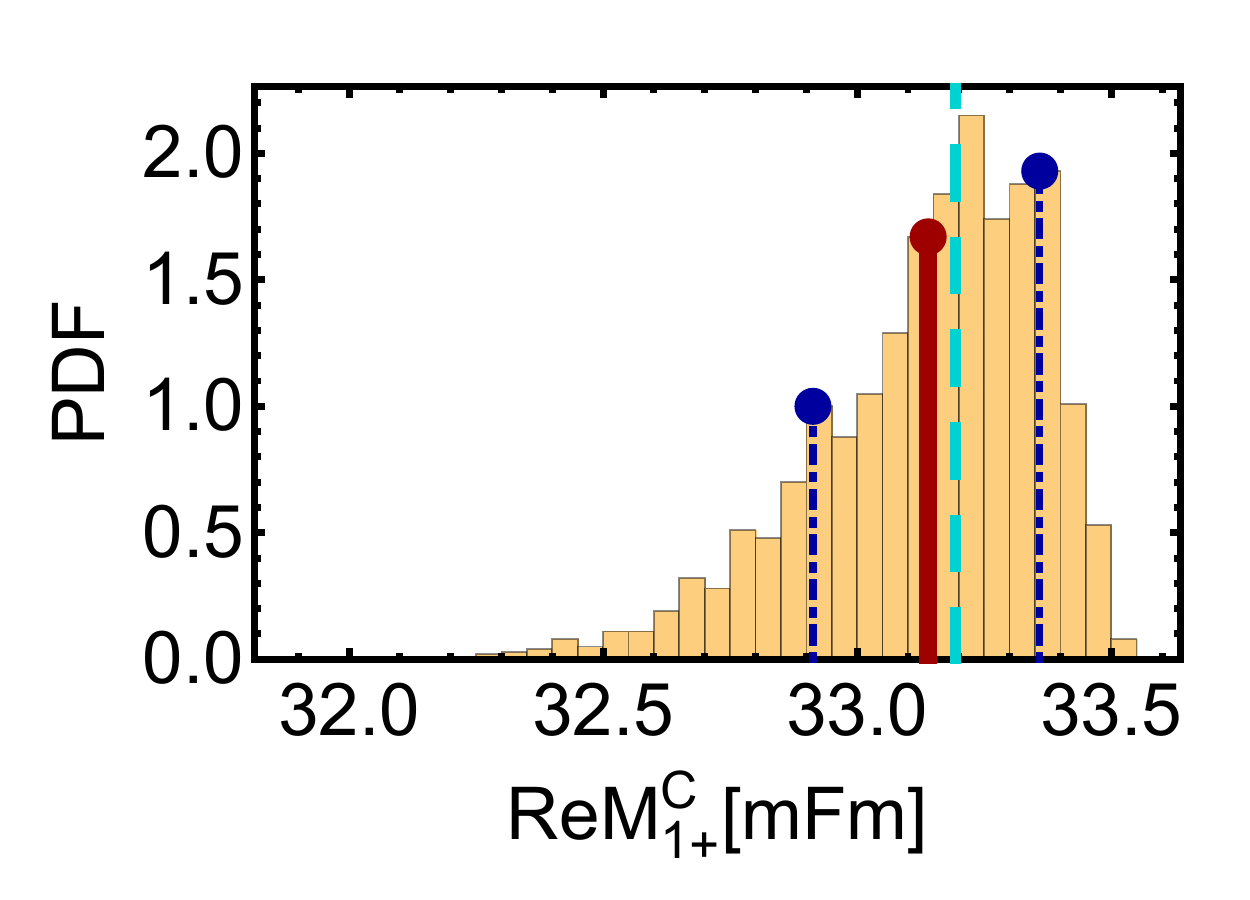}
 \end{overpic}
\begin{overpic}[width=0.325\textwidth]{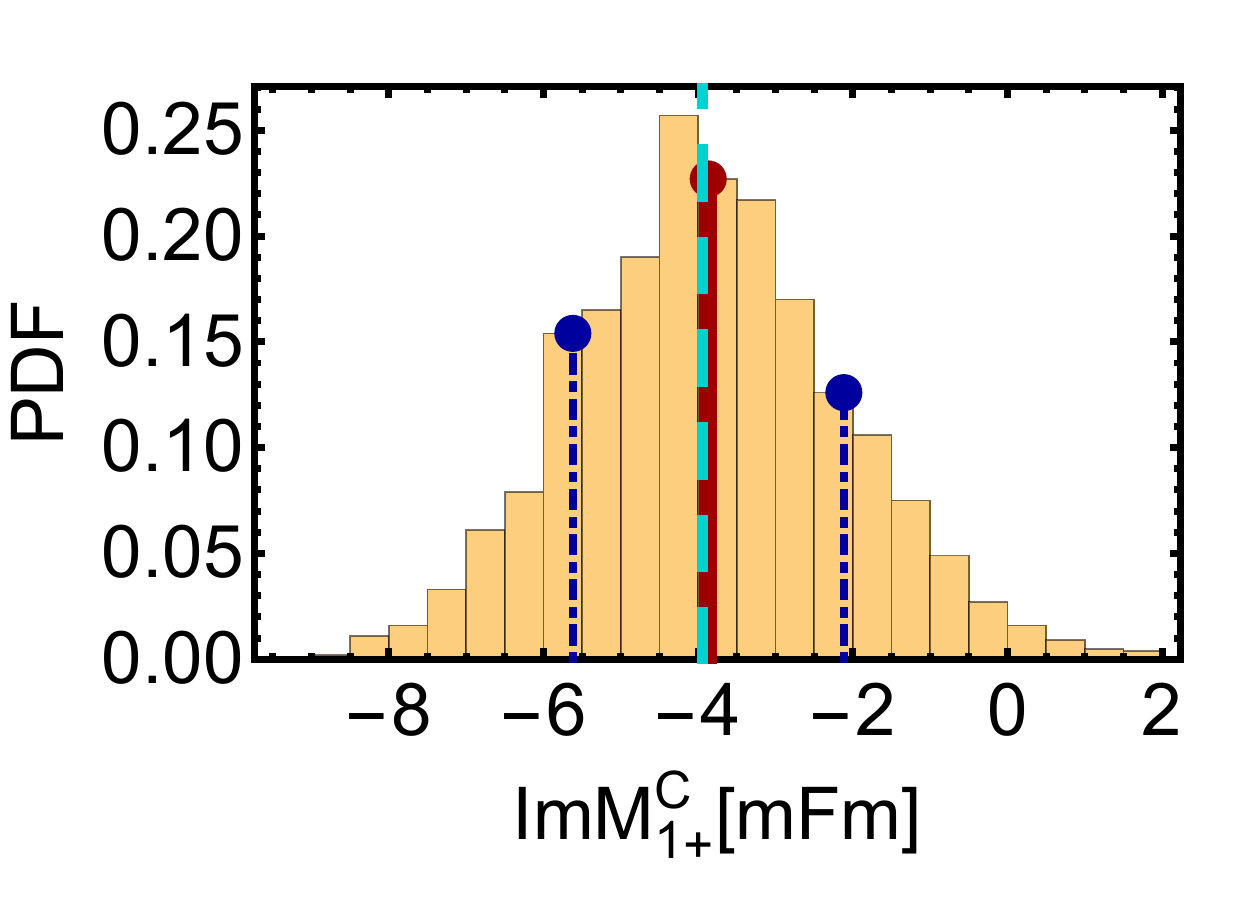}
 \end{overpic}
\begin{overpic}[width=0.325\textwidth]{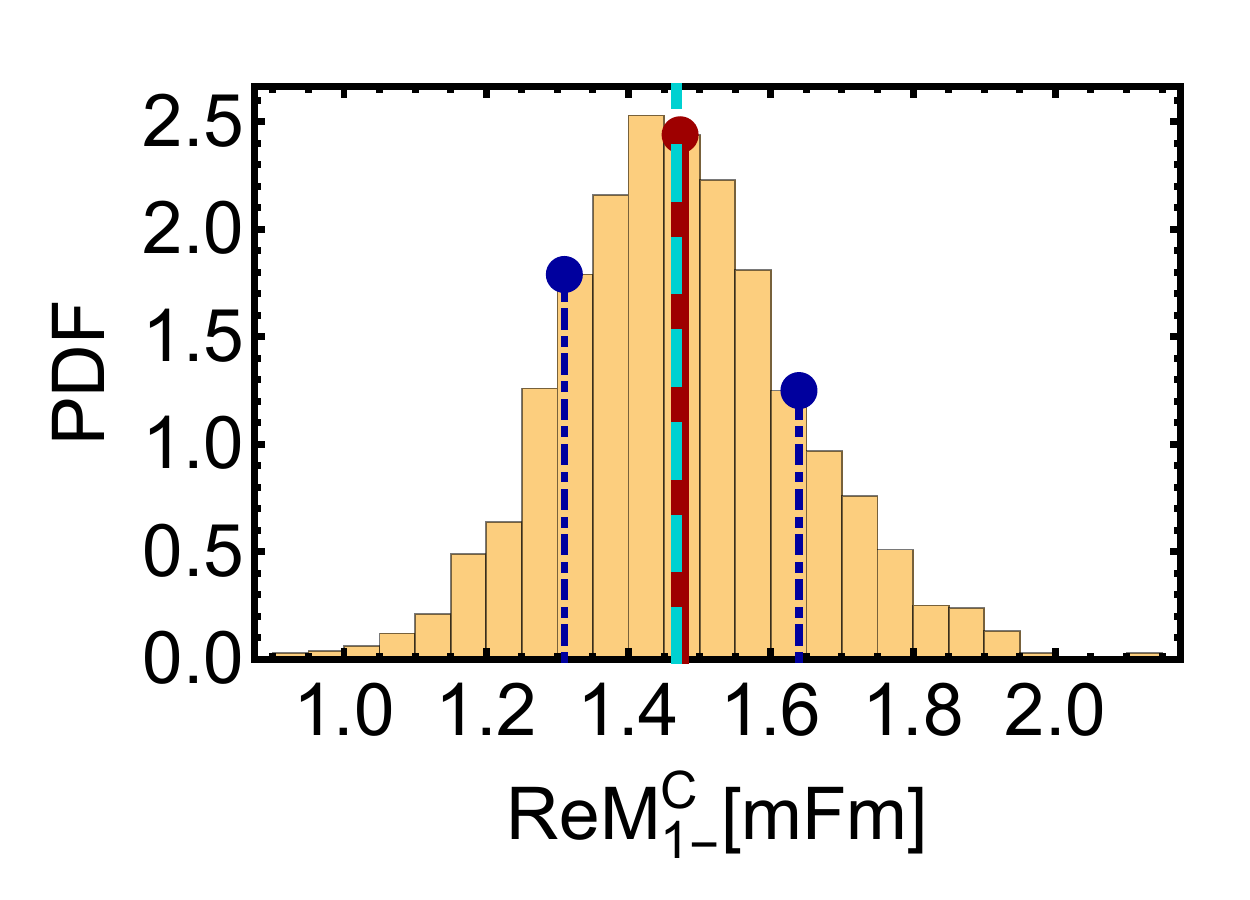}
 \end{overpic} \\ \vspace*{-8pt}
\begin{overpic}[width=0.325\textwidth]{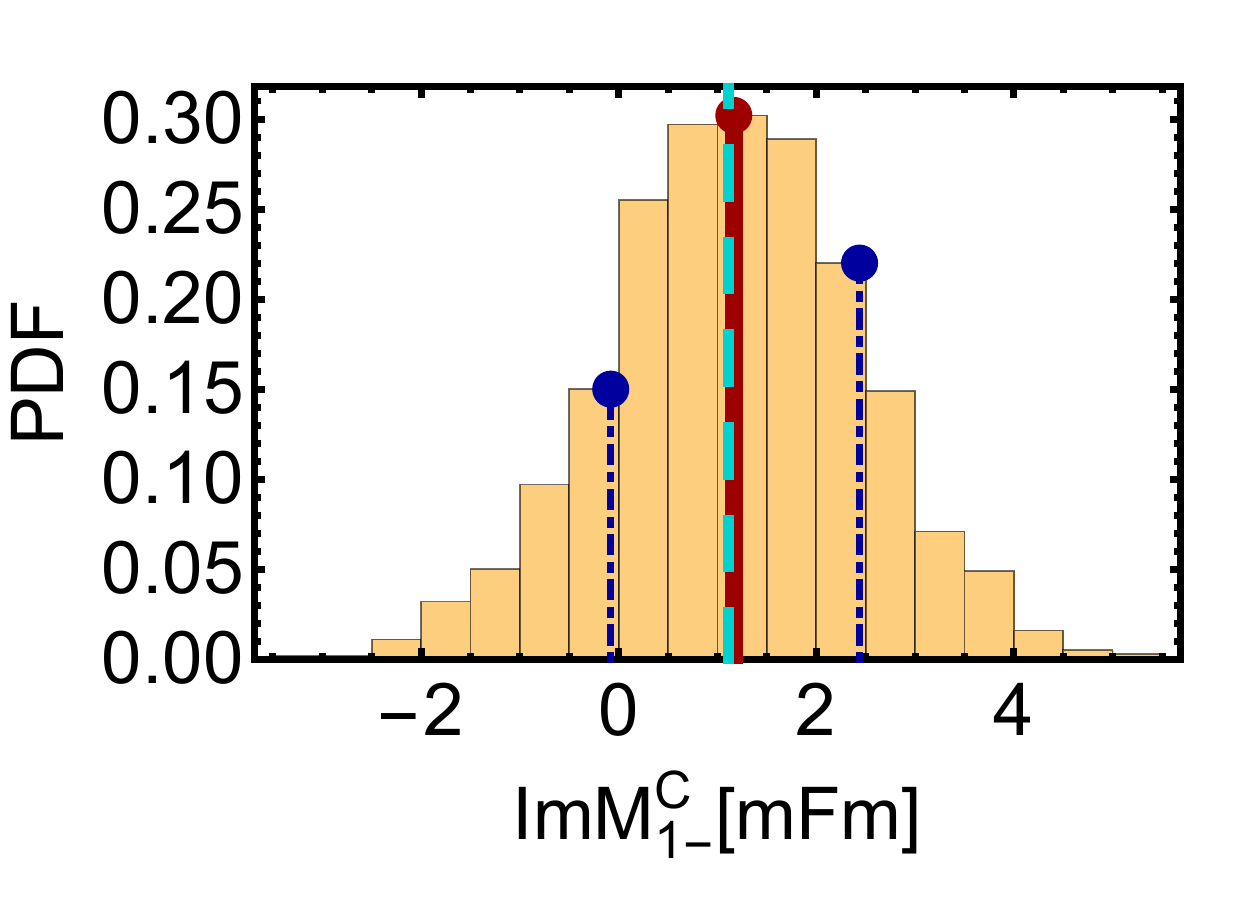}
 \end{overpic}
\caption[Bootstrap-distributions for multipole fit-parameters in an analysis of photoproduction data on the $\Delta$-resonance region. The fourth energy-bin, \newline $E_{\gamma }\text{ = 350.0 MeV}$, is shown.]{The histograms belong to the same bootstrap-analysis shown in Figure \ref{fig:BootstrapHistosDeltaRegionEnergy1}, but here the fourth energy-bin, $E_{\gamma }\text{ = 350.0 MeV}$, is shown.} \bigskip
\begin{overpic}[width=0.325\textwidth]{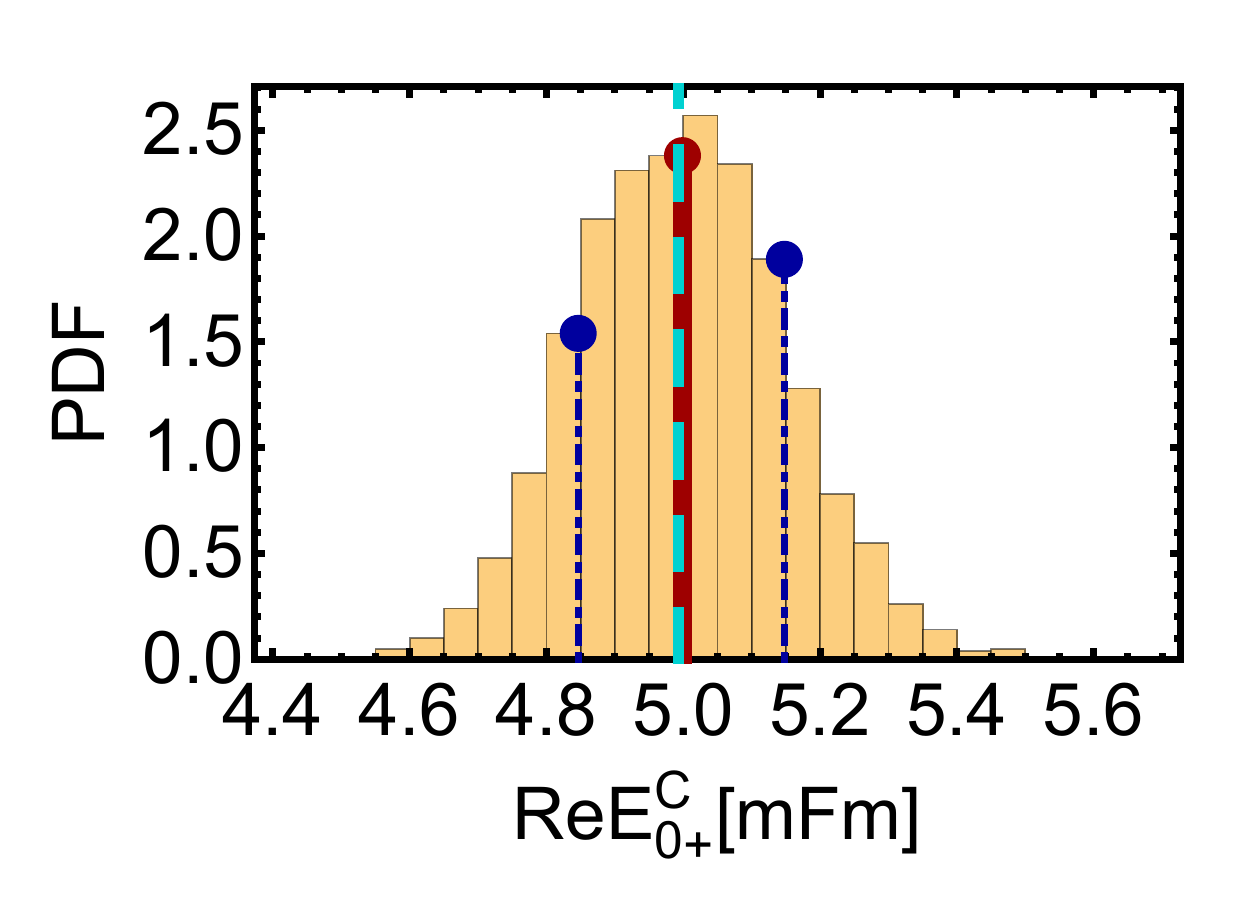}
 \end{overpic}
\begin{overpic}[width=0.325\textwidth]{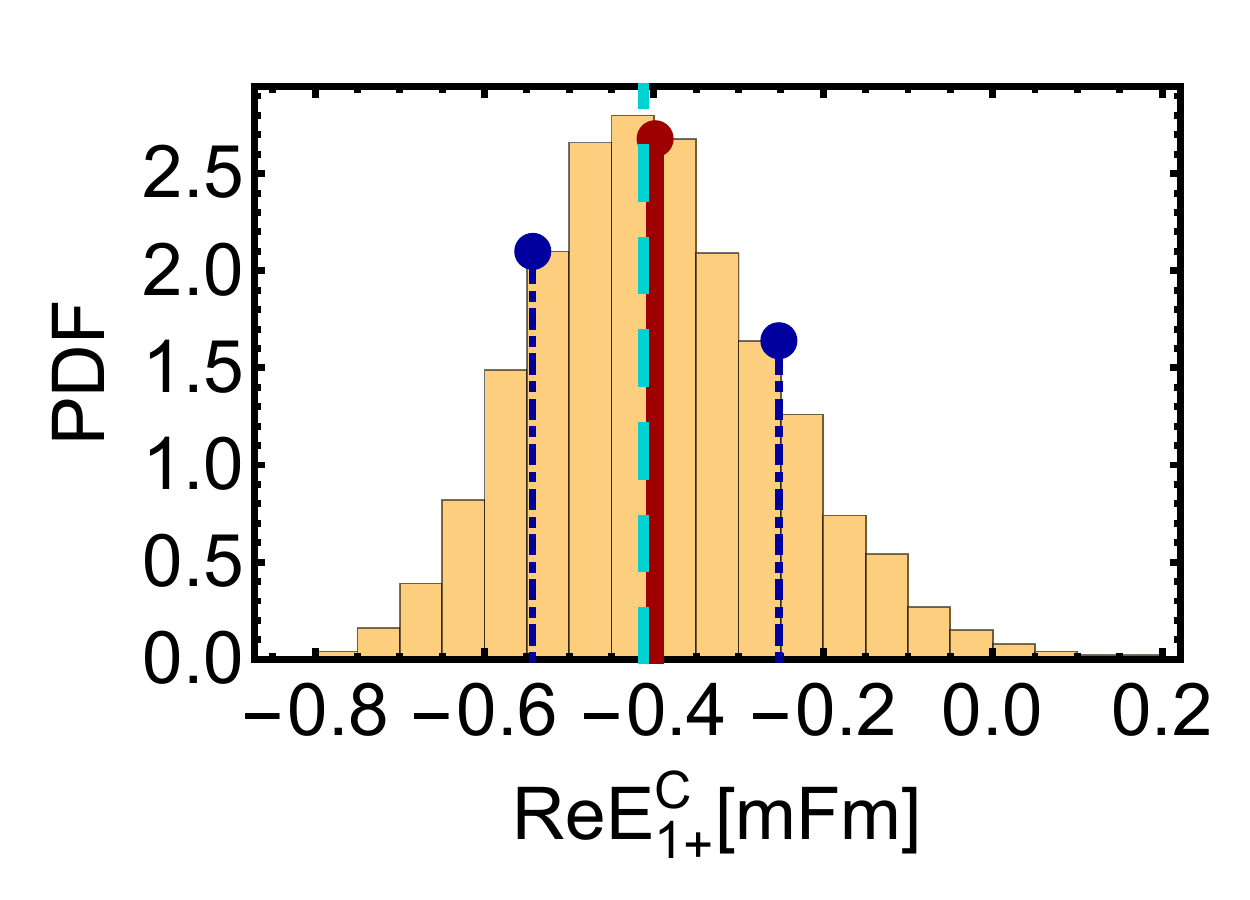}
 \end{overpic}
\begin{overpic}[width=0.325\textwidth]{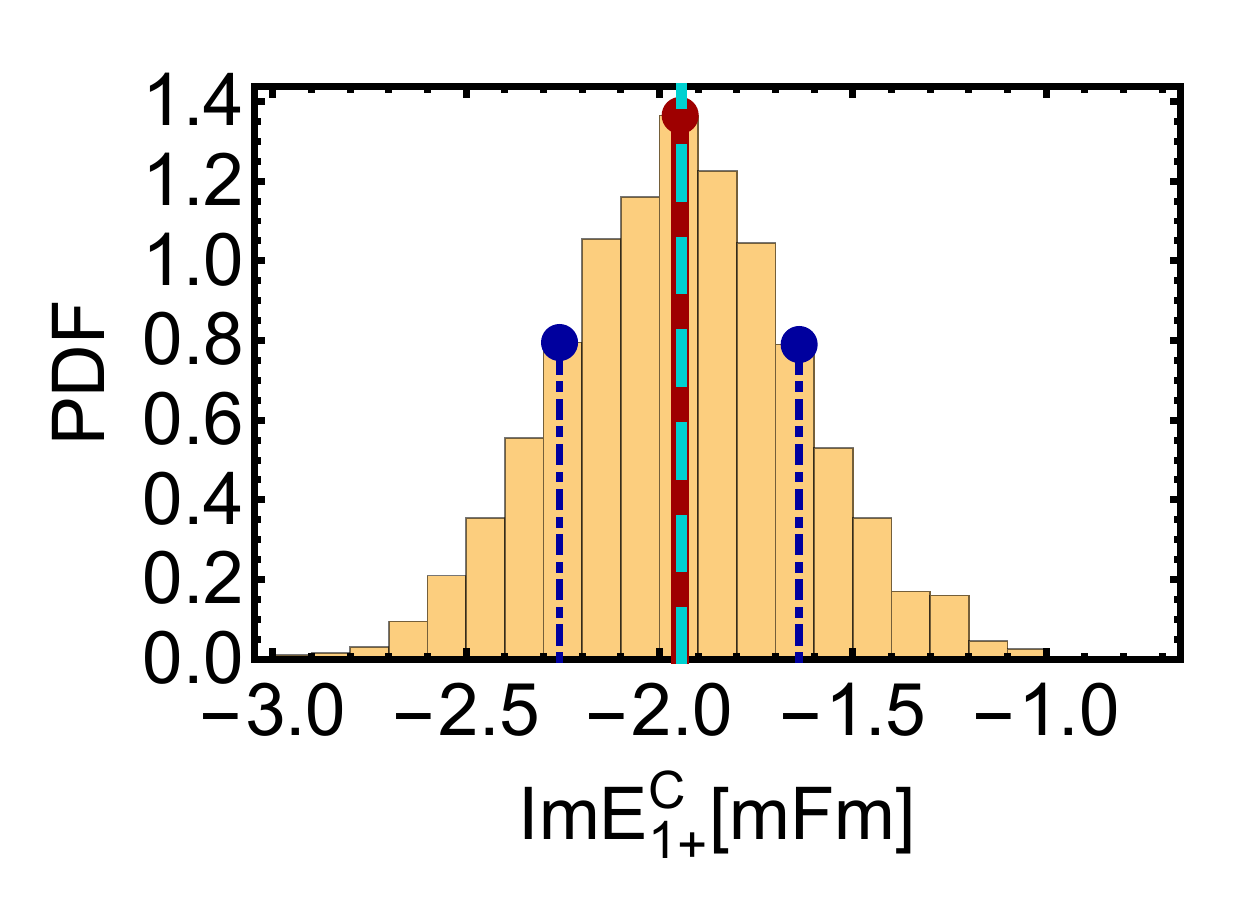}
 \end{overpic} \\ \vspace*{-8pt}
\begin{overpic}[width=0.325\textwidth]{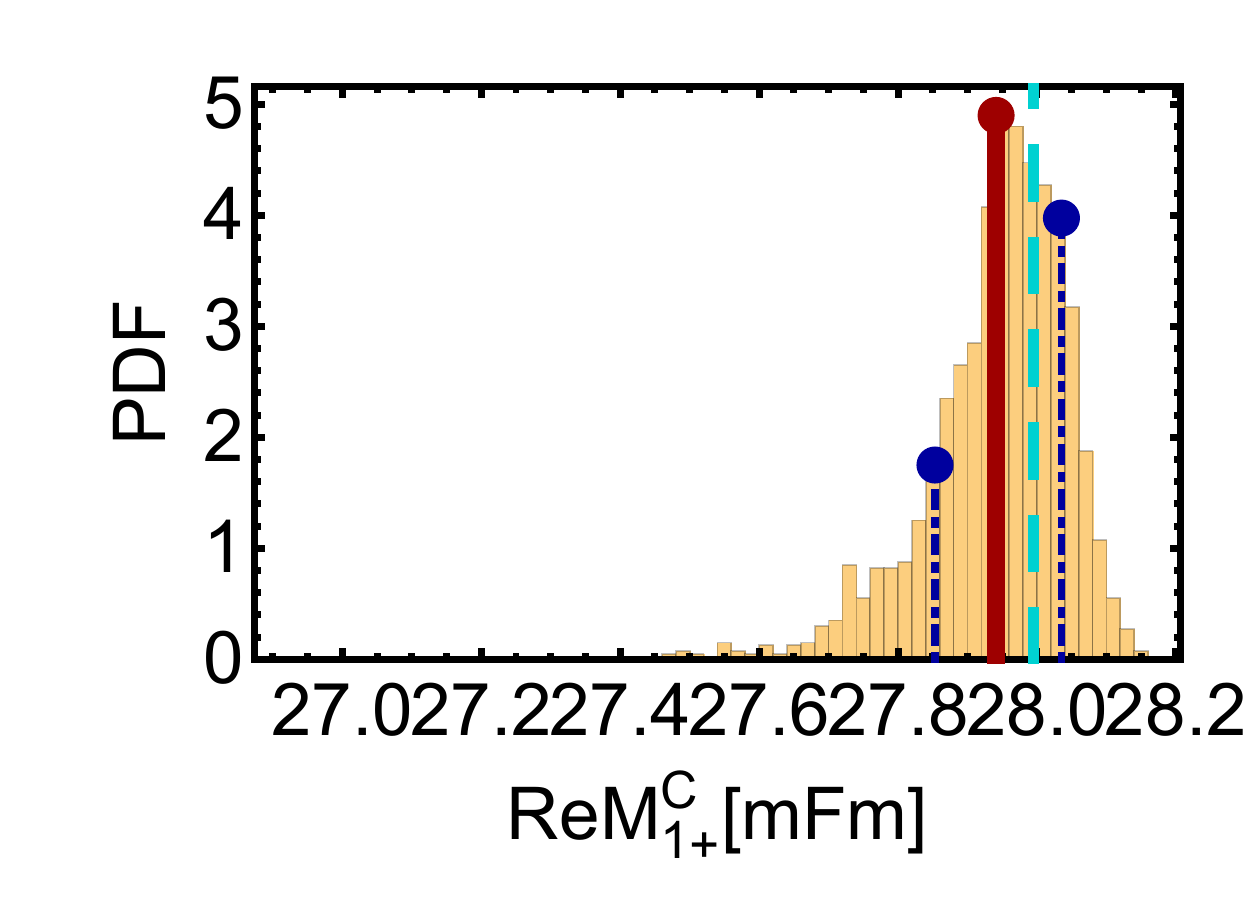}
 \end{overpic}
\begin{overpic}[width=0.325\textwidth]{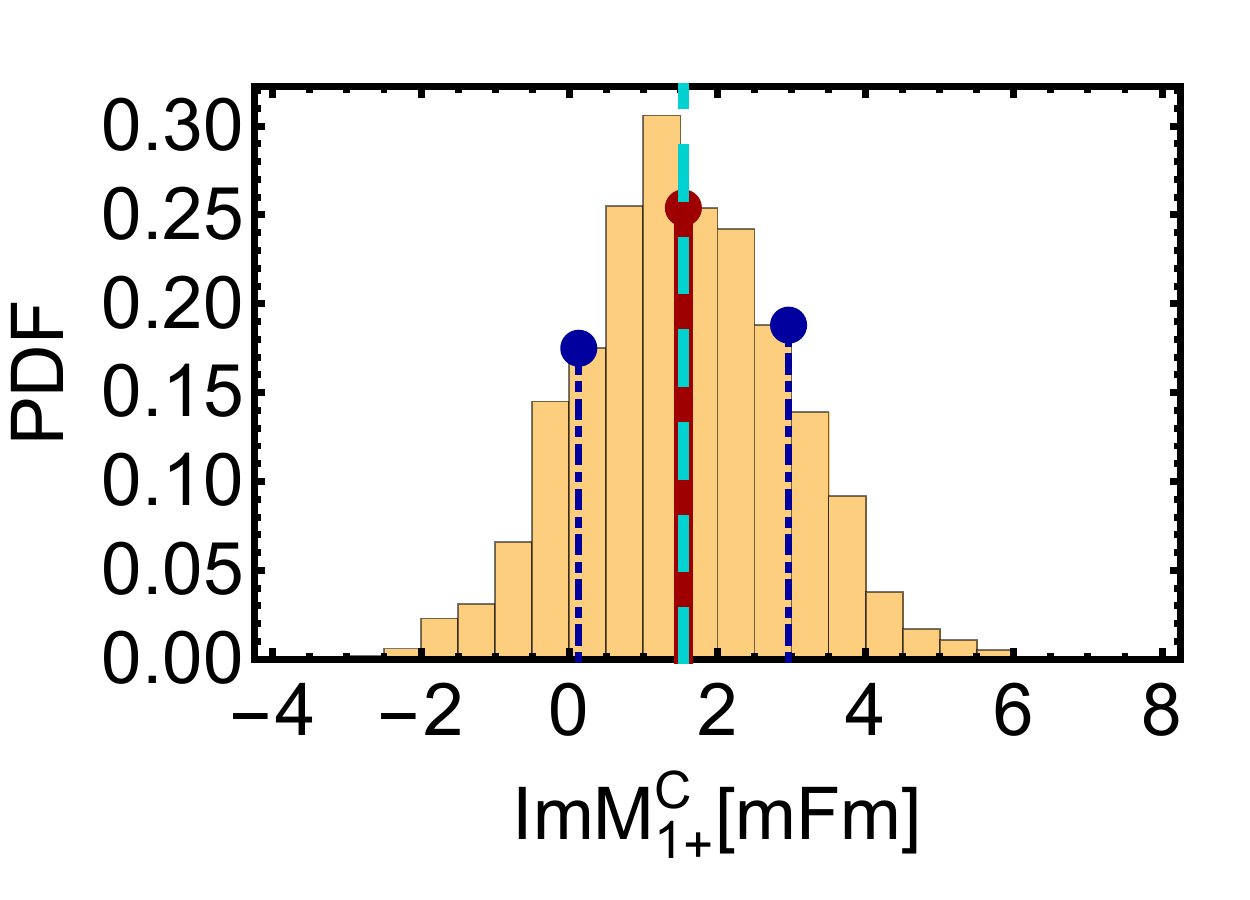}
 \end{overpic}
\begin{overpic}[width=0.325\textwidth]{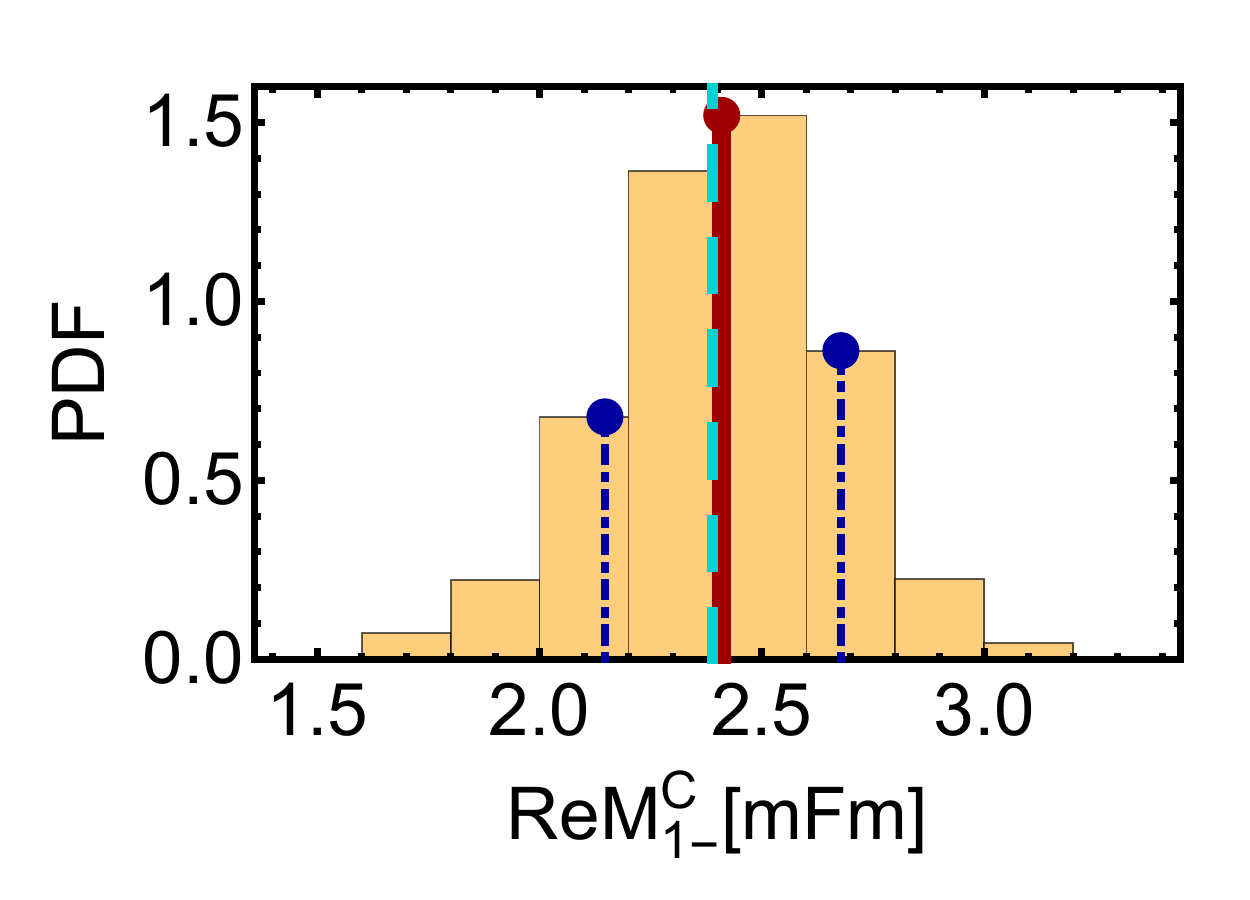}
 \end{overpic} \\ \vspace*{-8pt}
\begin{overpic}[width=0.325\textwidth]{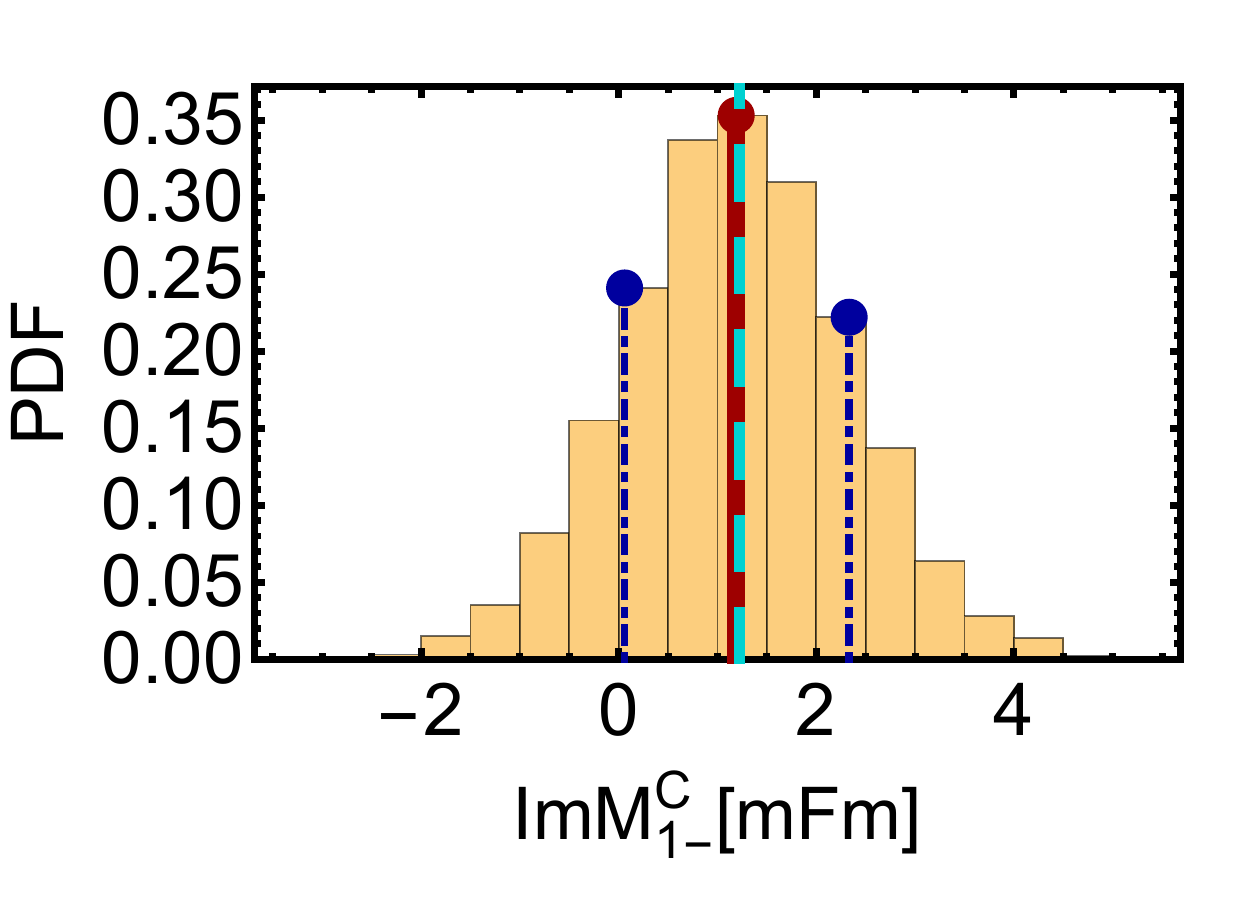}
 \end{overpic}
\caption[Bootstrap-distributions for multipole fit-parameters in an analysis of photoproduction data on the $\Delta$-resonance region. The fifth energy-bin, \newline $E_{\gamma }\text{ = 380.0 MeV}$, is shown.]{The histograms belong to the same bootstrap-analysis shown in Figure \ref{fig:BootstrapHistosDeltaRegionEnergy1}, but here the fifth energy-bin, $E_{\gamma }\text{ = 380.0 MeV}$, is shown.}
\label{fig:BootstrapHistosDeltaRegionEnergies4and5}
\end{figure}

\clearpage

\begin{figure}[h]
\centering
\vspace*{-10pt}
\begin{overpic}[width=0.325\textwidth]{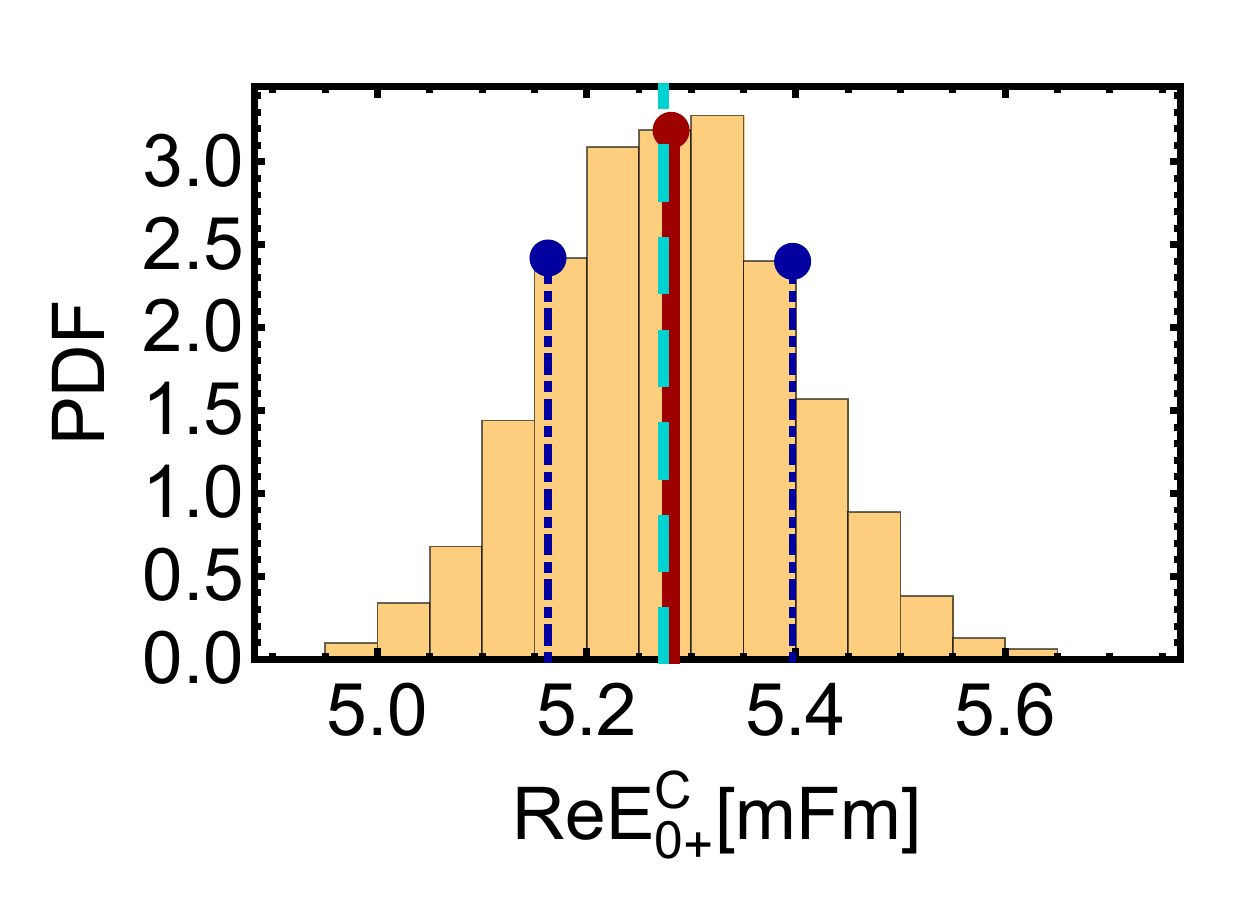}
 \end{overpic}
\begin{overpic}[width=0.325\textwidth]{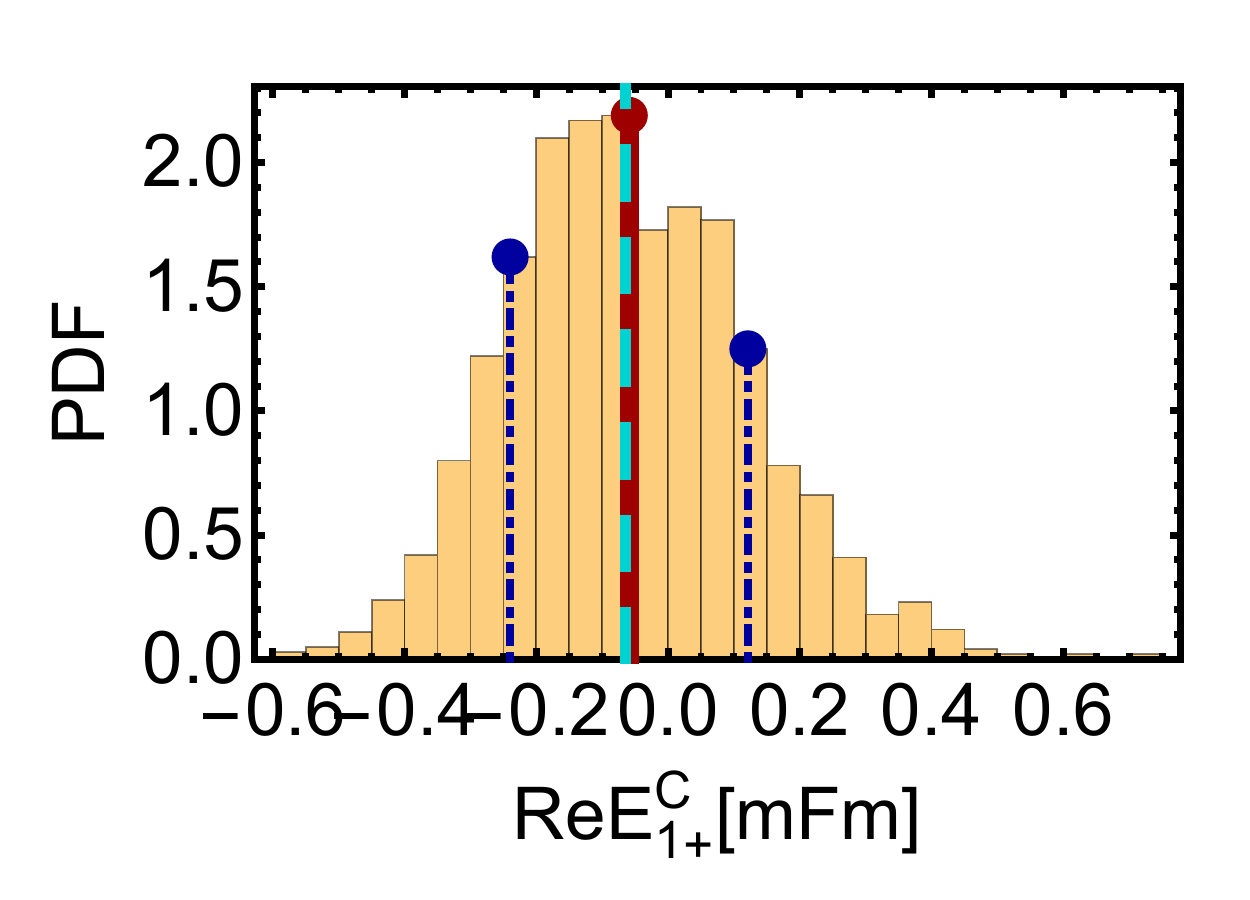}
 \end{overpic}
\begin{overpic}[width=0.325\textwidth]{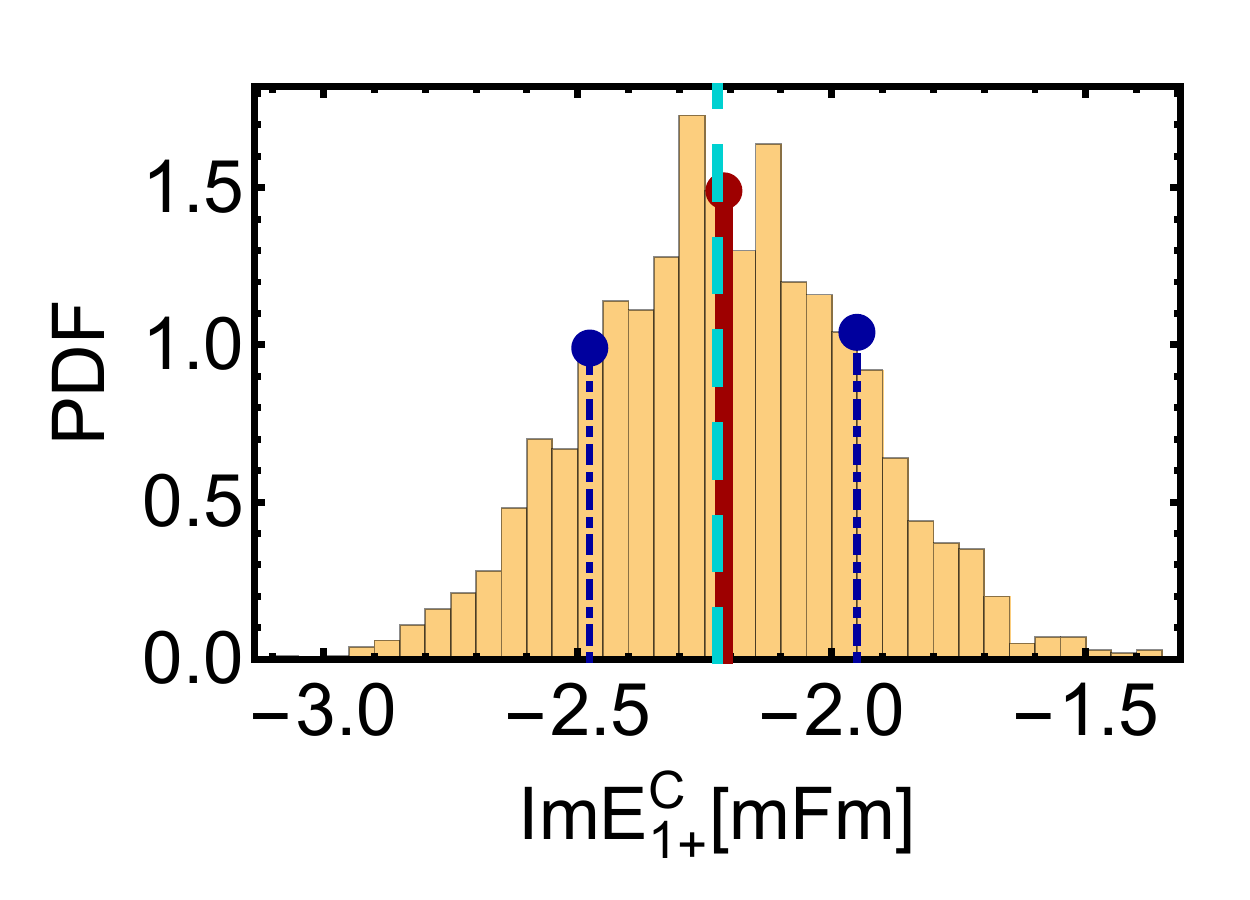}
 \end{overpic} \\ \vspace*{-8pt}
\begin{overpic}[width=0.325\textwidth]{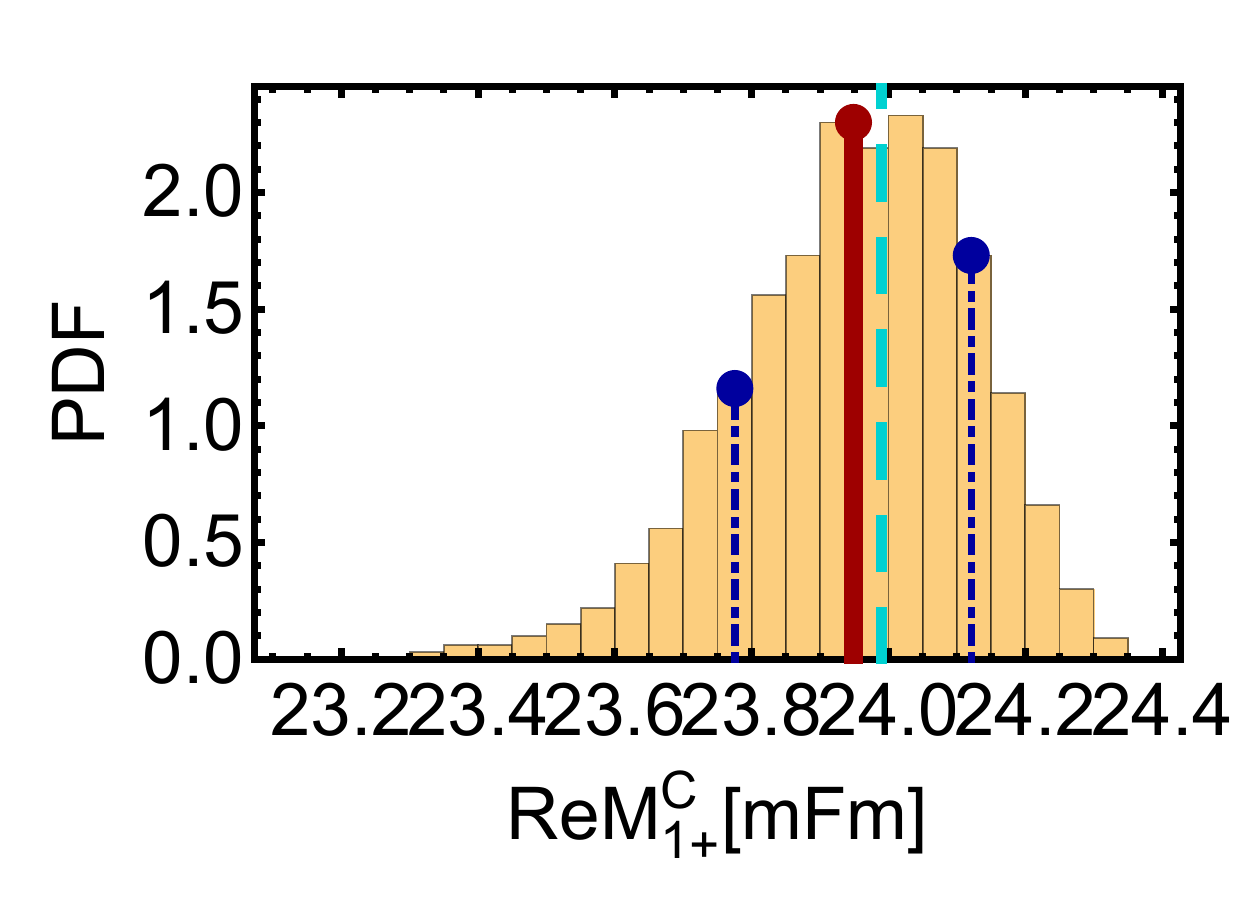}
 \end{overpic}
\begin{overpic}[width=0.325\textwidth]{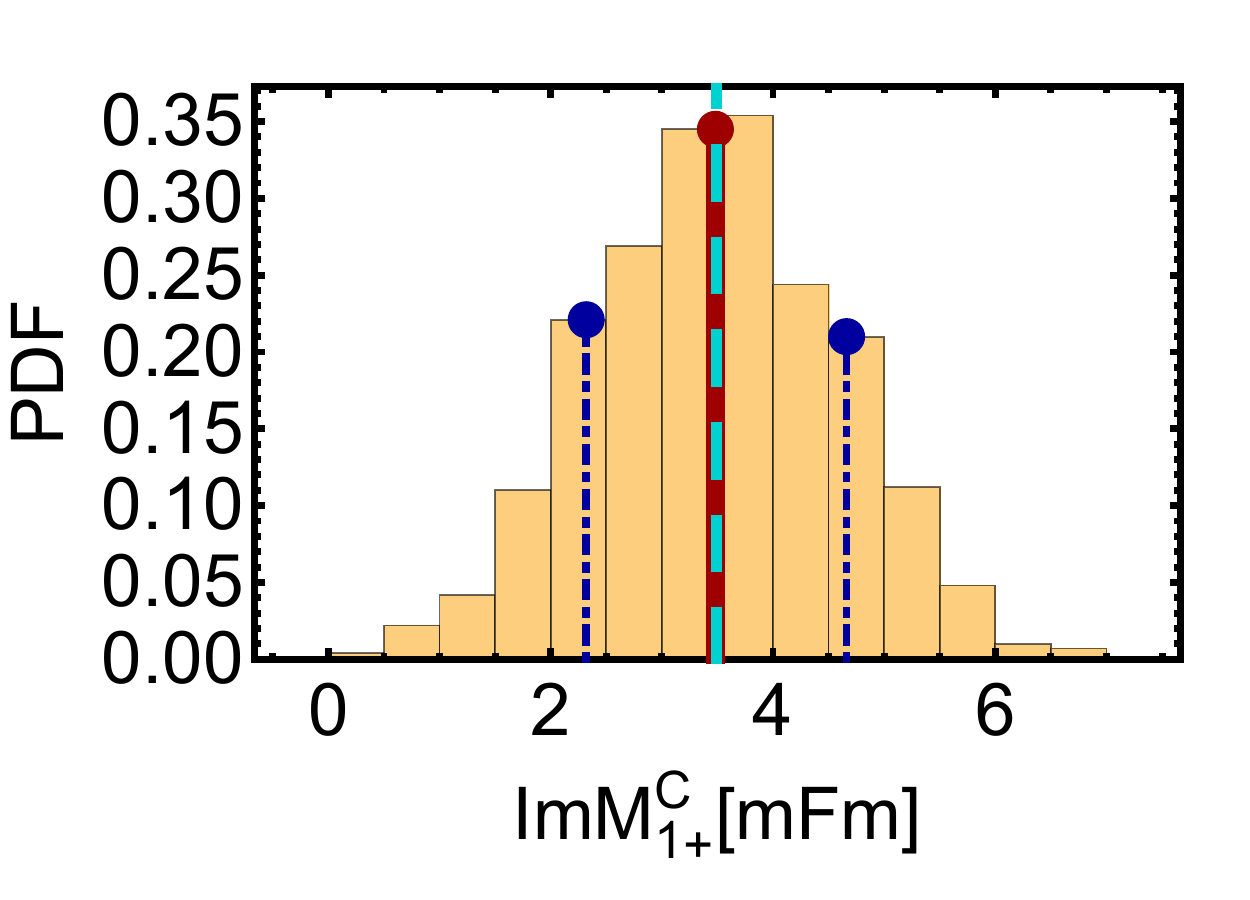}
 \end{overpic}
\begin{overpic}[width=0.325\textwidth]{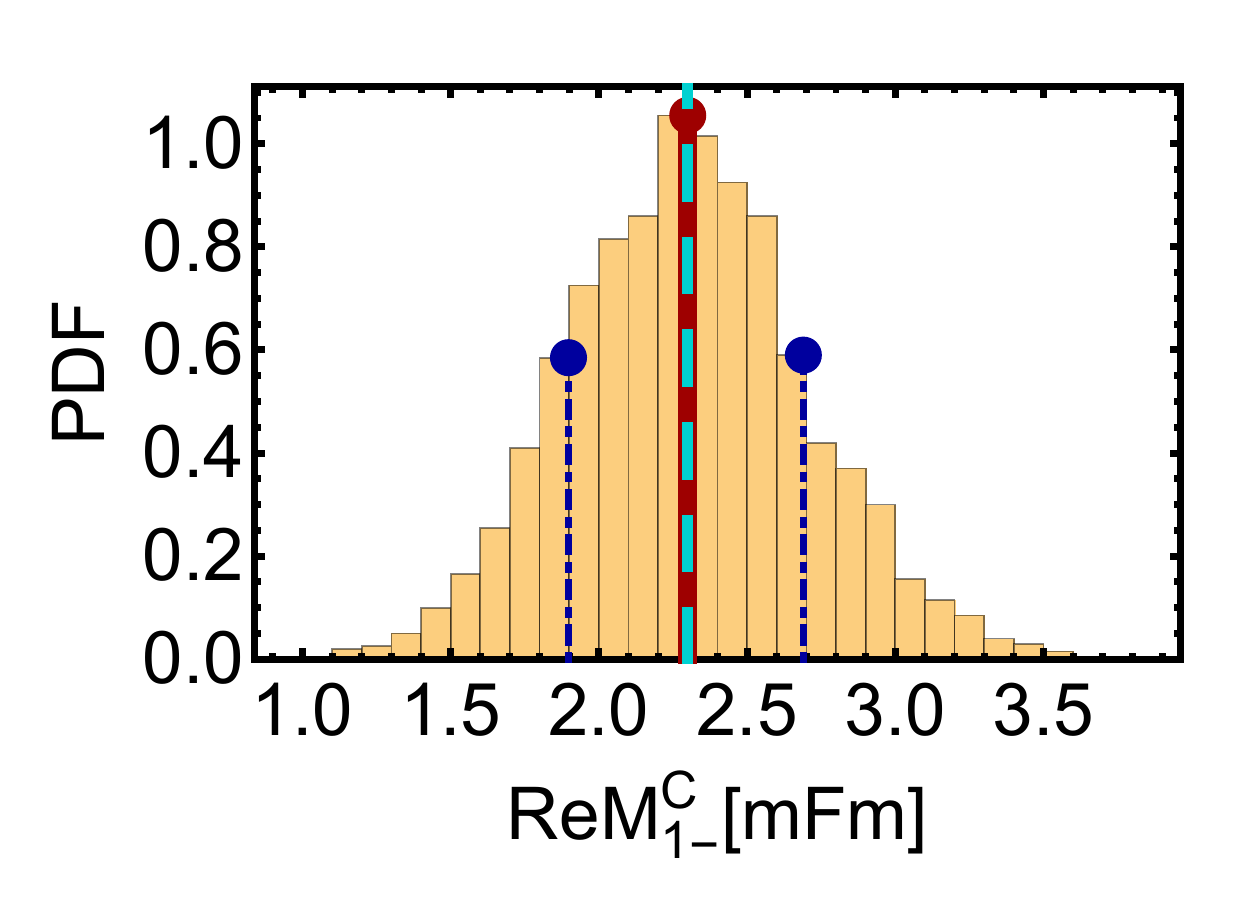}
 \end{overpic} \\ \vspace*{-8pt}
\begin{overpic}[width=0.325\textwidth]{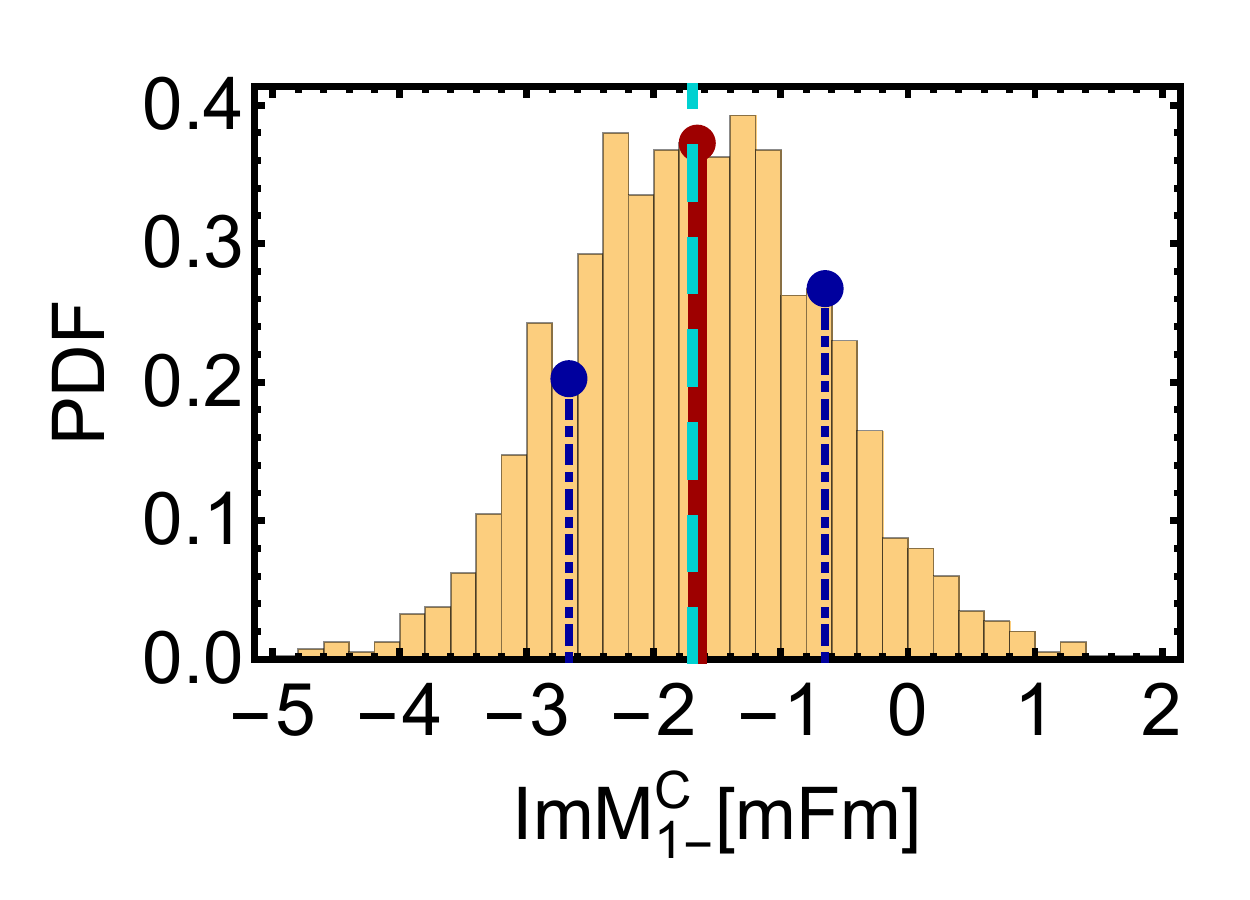}
 \end{overpic}
\caption[Bootstrap-distributions for multipole fit-parameters in an analysis of photoproduction data on the $\Delta$-resonance region. The sixth energy-bin, \newline $E_{\gamma }\text{ = 400.0 MeV}$, is shown.]{The histograms belong to the same bootstrap-analysis shown in Figure \ref{fig:BootstrapHistosDeltaRegionEnergy1}, but here the sixth energy-bin, $E_{\gamma }\text{ = 400.0 MeV}$, is shown.} \bigskip
\begin{overpic}[width=0.325\textwidth]{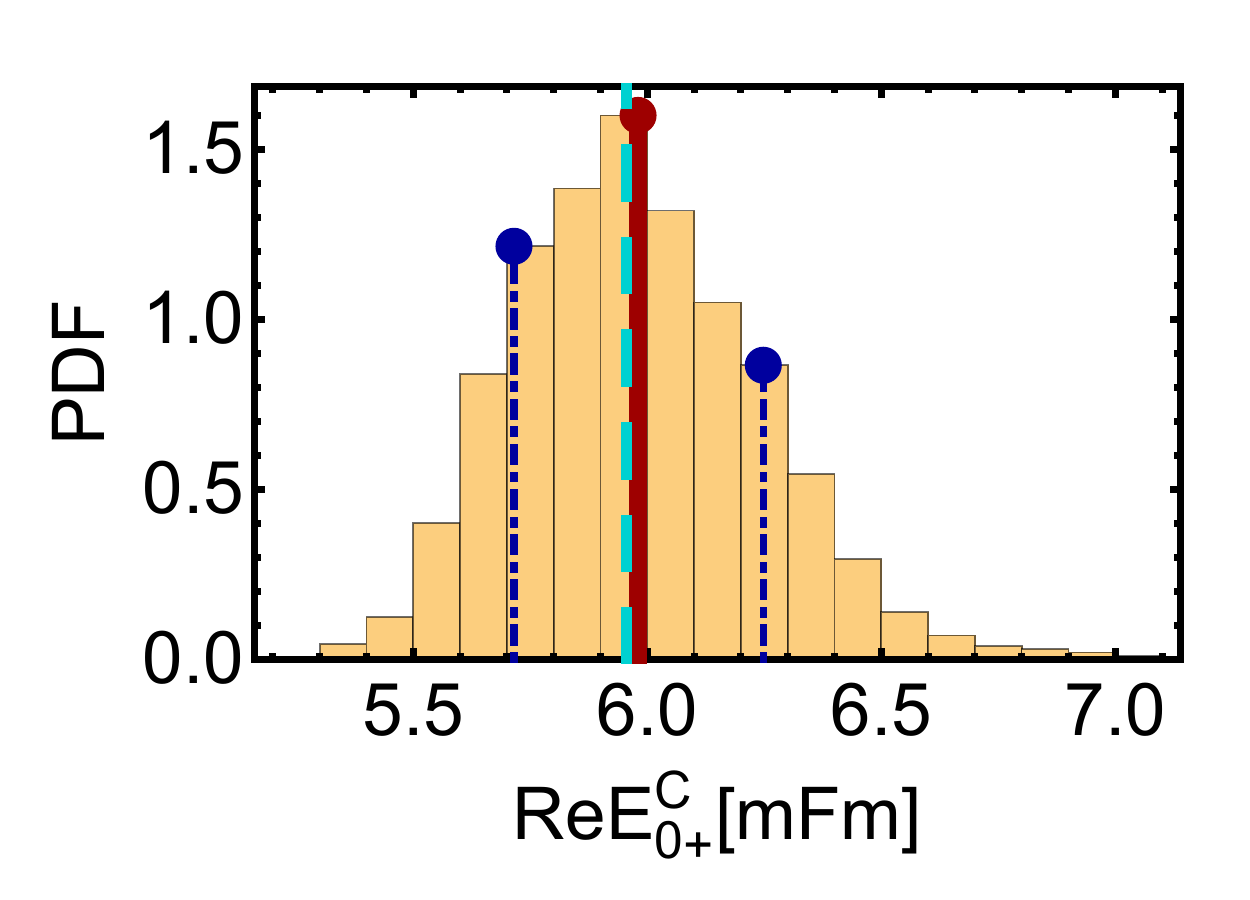}
 \end{overpic}
\begin{overpic}[width=0.325\textwidth]{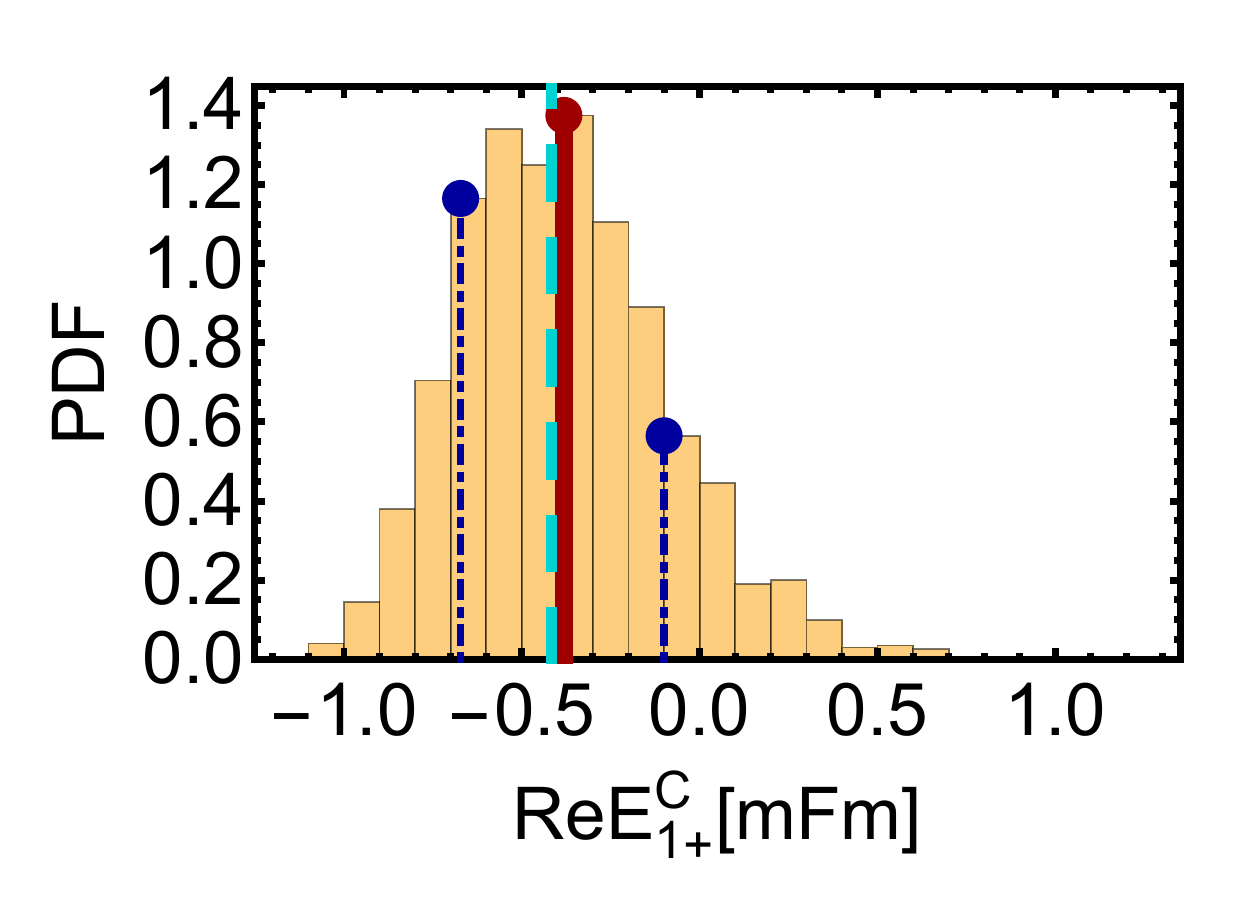}
 \end{overpic}
\begin{overpic}[width=0.325\textwidth]{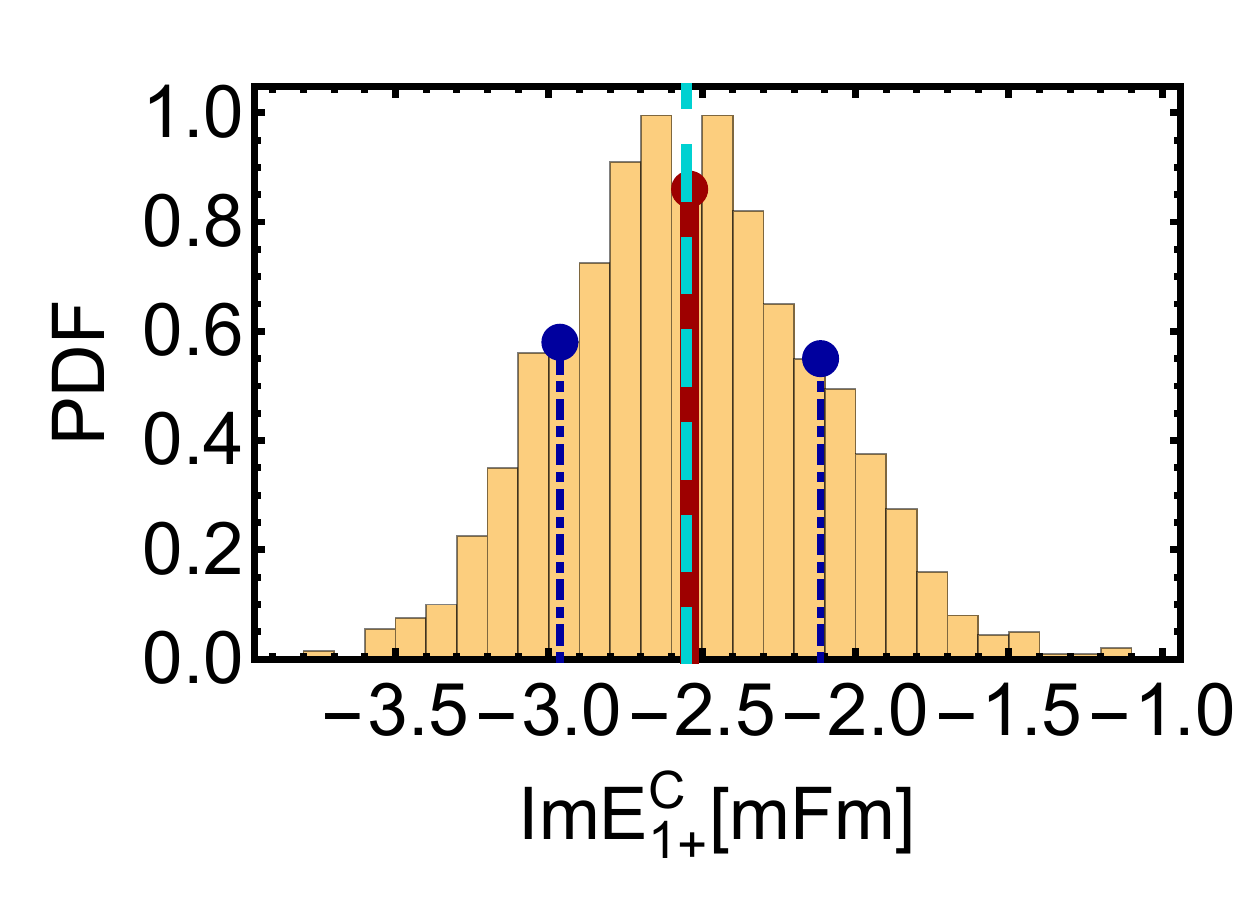}
 \end{overpic} \\ \vspace*{-8pt}
\begin{overpic}[width=0.325\textwidth]{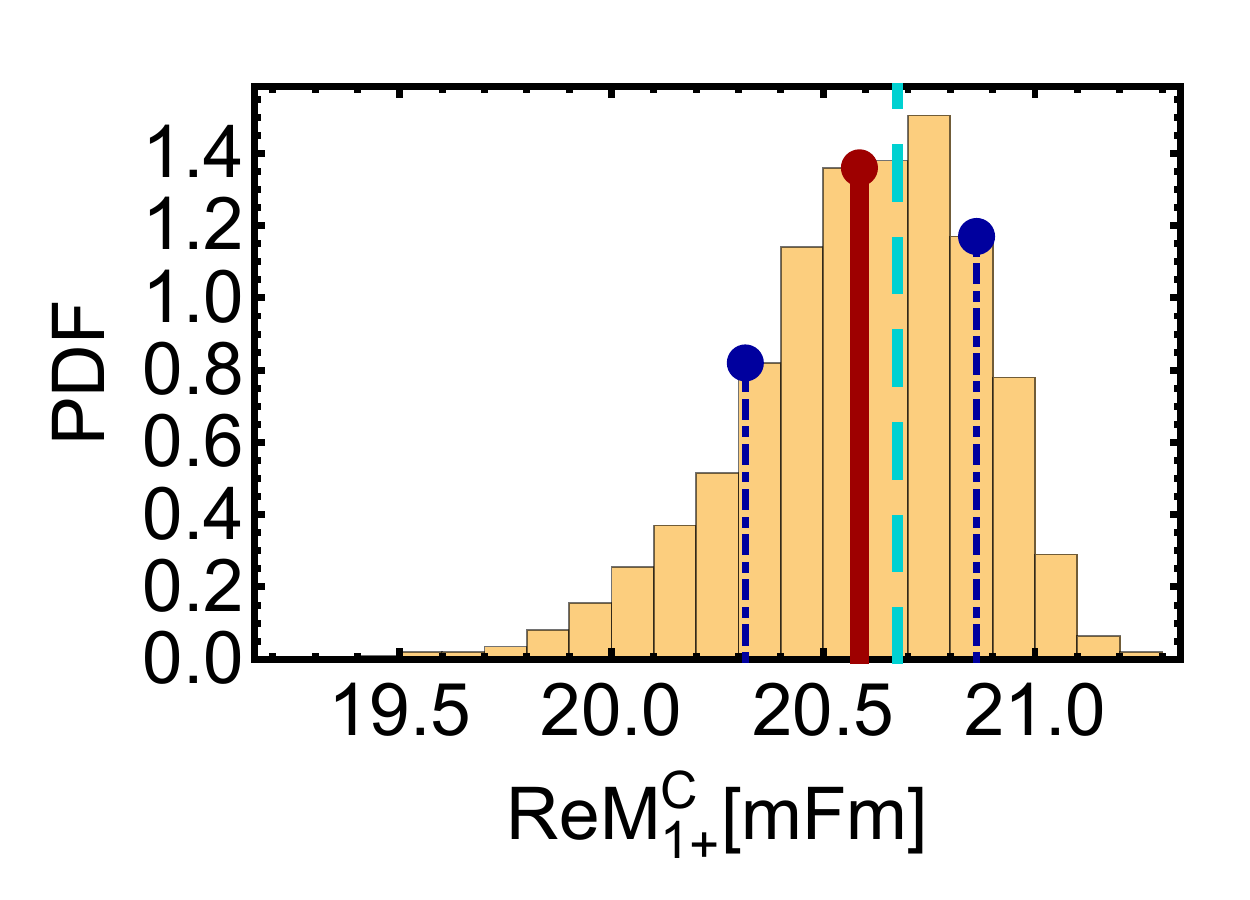}
 \end{overpic}
\begin{overpic}[width=0.325\textwidth]{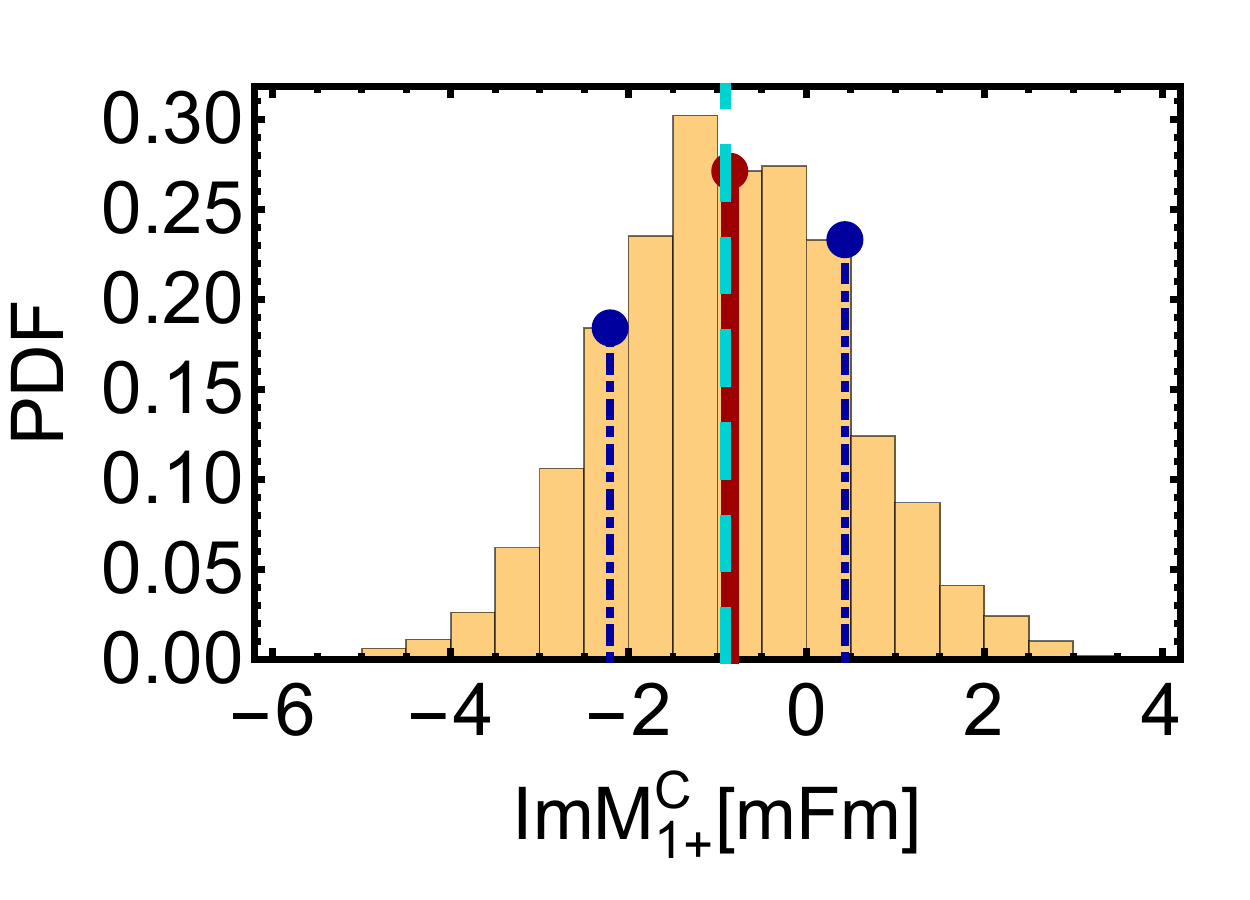}
 \end{overpic}
\begin{overpic}[width=0.325\textwidth]{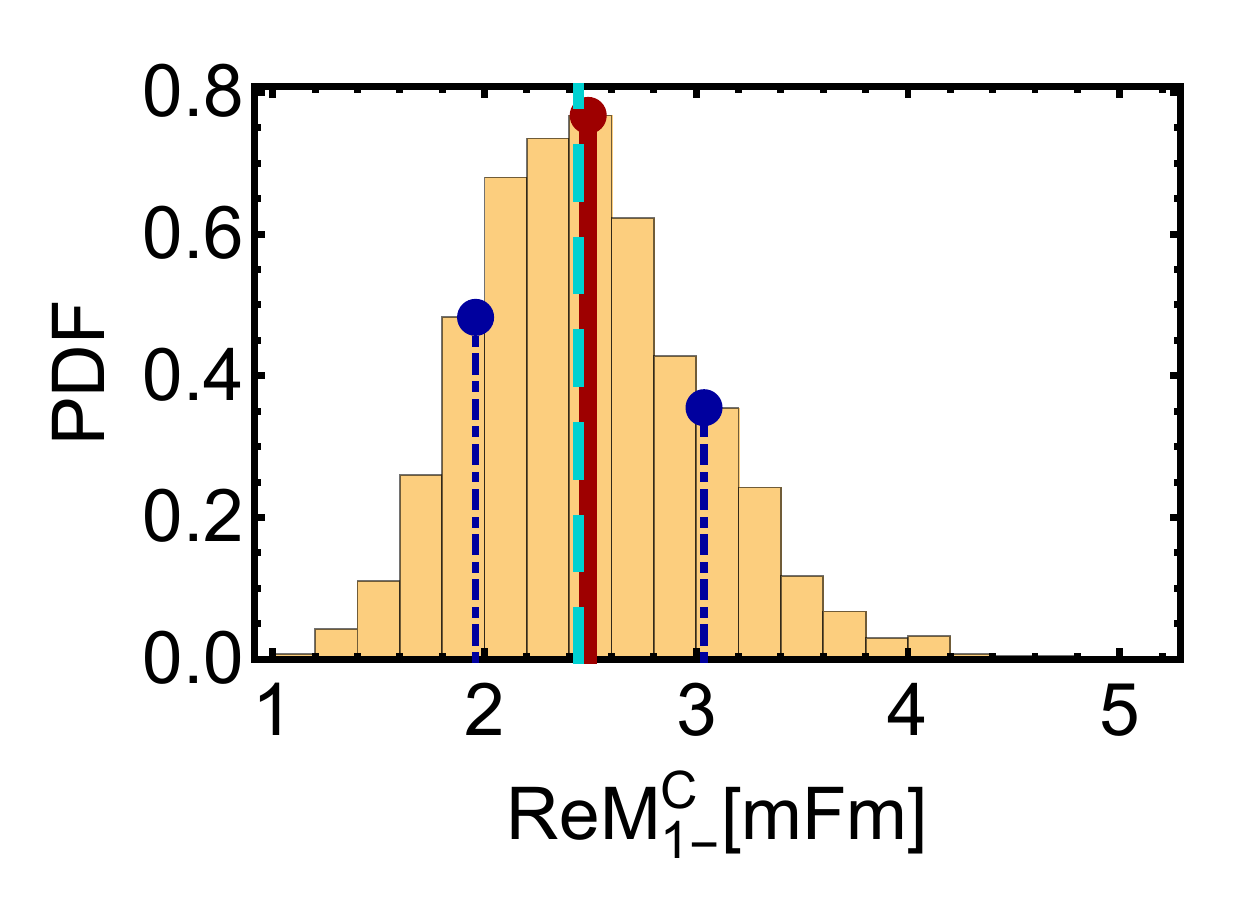}
 \end{overpic} \\ \vspace*{-8pt}
\begin{overpic}[width=0.325\textwidth]{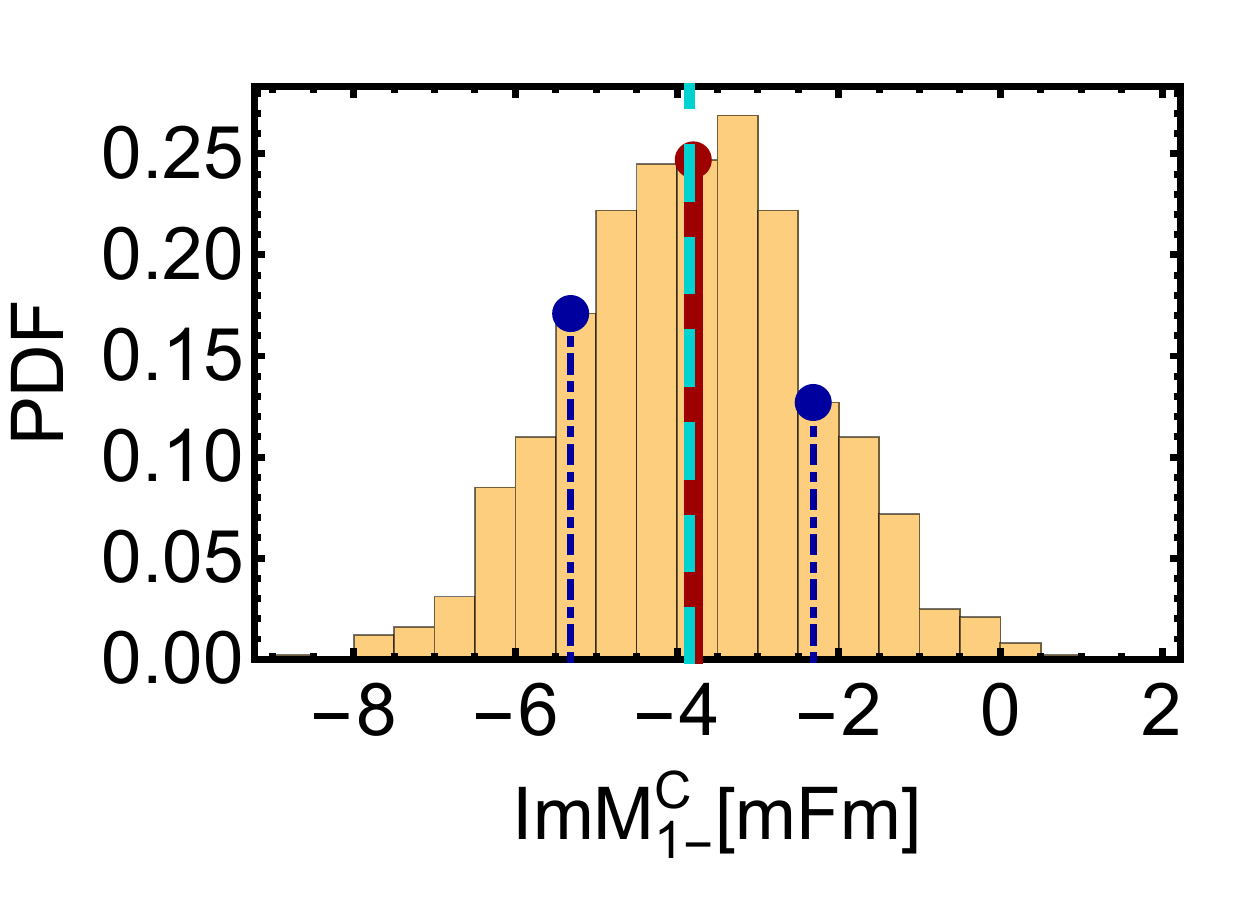}
 \end{overpic}
\caption[Bootstrap-distributions for multipole fit-parameters in an analysis of photoproduction data on the $\Delta$-resonance region. The seventh energy-bin, \newline $E_{\gamma }\text{ = 420.0 MeV}$, is shown.]{The histograms belong to the same bootstrap-analysis shown in Figure \ref{fig:BootstrapHistosDeltaRegionEnergy1}, but here the seventh energy-bin, $E_{\gamma }\text{ = 420.0 MeV}$, is shown.}
\label{fig:BootstrapHistosDeltaRegionEnergies6and7}
\end{figure}

\clearpage

\paragraph{$2^{\mathrm{nd}}$ resonance-region} \label{subsec:MultHistograms} \textcolor{white}{:-)} \newline

From all fits performed in the second resonance-region, only one has been selected as a candidate for a detailed bootstrap-analysis. This has been the TPWA for $\ell_{\mathrm{max}} = 3$, with all $F$-wave multipoles fixed to BnGa2014\_02 \cite{BoGa} and all remaining partial waves varied freely, which yielded a satisfactory global minimum (see section \ref{subsec:2ndResRegionDataFits}). \newline
In the following, numerical results and figures for bootstrap-distributions are listed based for a resampling-analysis performed with an ensemble of $B=2000$ replicate datasets. More details are given in the respective Figure- and Table-captions, as well as the main text. \newline

\textbf{Numerical results of the fit} \newline


\begin{table}[h]
\centering
\begin{tabular}{c|c|c|c|c|c}
\multicolumn{2}{l|}{$ E_{\gamma }\text{ = 683.5 MeV} $} & \multicolumn{2}{c|}{ $ \text{ndf = 27} $} & \multicolumn{2}{c}{ $ \chi ^2\text{/ndf = 1.2907} $ } \\
\hline
\hline
$ \hat{\theta}_{i} = \left( \mathcal{M}_{\ell}^{C} \right)_{i} \text{[mFm]} $ & $ \left(\hat{\theta}_{i}^{\mathrm{Best}}\right)_{- \Delta_{-}}^{+ \Delta_{+}} $ & $ \hat{\theta}_{i}^{\ast} (\cdot) $ & $ \widehat{\mathrm{se}}_{B} \left( \hat{\theta}^{\ast}_{i} \right) $ & $ \widehat{\mathrm{bias}}_{B} $ & $ \delta_{\mathrm{bias}} $\\
\hline
$ \mathrm{Re} \left[ E_{0+}^{C} \right]  $ & $ 7.39249_{-0.65342}^{+0.64931} $ & $ 7.3892 $ & $ 0.70994 $ & $ -0.00329 $ & $ 0.00463 $\\
$ \mathrm{Re} \left[ E_{1+}^{C} \right]  $ & $ 0.24718_{-0.1837}^{+0.152} $ & $ 0.23463 $ & $ 0.17022 $ & $ -0.01255 $ & $ 0.07373 $\\
$ \mathrm{Im} \left[ E_{1+}^{C} \right]  $ & $ 0.12657_{-0.14935}^{+0.11179} $ & $ 0.10295 $ & $ 0.15146 $ & $ -0.02362 $ & $ 0.15593 $\\
$ \mathrm{Re} \left[ M_{1+}^{C} \right]  $ & $ 1.01757_{-0.19226}^{+0.20798} $ & $ 1.03465 $ & $ 0.25834 $ & $ 0.01708 $ & $ 0.06611 $\\
$ \mathrm{Im} \left[ M_{1+}^{C} \right]  $ & $ 4.10966_{-0.15475}^{+0.13874} $ & $ 4.10475 $ & $ 0.15724 $ & $ -0.00491 $ & $ 0.03125 $\\
$ \mathrm{Re} \left[ M_{1-}^{C} \right]  $ & $ 5.93443_{-0.90586}^{+0.65492} $ & $ 5.8043 $ & $ 0.90445 $ & $ -0.13013 $ & $ 0.14388 $\\
$ \mathrm{Im} \left[ M_{1-}^{C} \right]  $ & $ 1.94401_{-0.16103}^{+0.3728} $ & $ 2.068 $ & $ 0.33073 $ & $ 0.12399 $ & $ 0.3749 $\\
$ \mathrm{Re} \left[ E_{2+}^{C} \right] $ & $ 0.18458_{-0.05863}^{+0.04778} $ & $ 0.178 $ & $ 0.05742 $ & $ -0.00658 $ & $ 0.11452 $\\
$ \mathrm{Im} \left[ E_{2+}^{C} \right] $ & $ 0.08452_{-0.04356}^{+0.03655} $ & $ 0.08136 $ & $ 0.04259 $ & $ -0.00316 $ & $ 0.07412 $\\
$ \mathrm{Re} \left[ E_{2-}^{C} \right] $ & $ 3.44245_{-0.17442}^{+0.17232} $ & $ 3.43925 $ & $ 0.17316 $ & $ -0.0032 $ & $ 0.0185 $\\
$ \mathrm{Im} \left[ E_{2-}^{C} \right] $ & $ 0.77078_{-0.19243}^{+0.33339} $ & $ 0.84673 $ & $ 0.32016 $ & $ 0.07595 $ & $ 0.23722 $\\
$ \mathrm{Re} \left[ M_{2+}^{C} \right] $ & $ -0.08731_{-0.05923}^{+0.03563} $ & $ -0.10124 $ & $ 0.05872 $ & $ -0.01393 $ & $ 0.23716 $\\
$ \mathrm{Im} \left[ M_{2+}^{C} \right] $ & $ 0.2766_{-0.05252}^{+0.0241} $ & $ 0.25839 $ & $ 0.0617 $ & $ -0.01821 $ & $ 0.29516 $\\
$ \mathrm{Re} \left[ M_{2-}^{C} \right] $ & $ 2.88131_{-0.15079}^{+0.13816} $ & $ 2.86262 $ & $ 0.18717 $ & $ -0.01869 $ & $ 0.09983 $\\
$ \mathrm{Im} \left[ M_{2-}^{C} \right] $ & $ -1.79922_{-0.09074}^{+0.17665} $ & $ -1.75553 $ & $ 0.14772 $ & $ 0.0437 $ & $ 0.29581 $\\
\end{tabular}
\caption[Numerical results of a bootstrap-analysis are collected for a TPWA-fit of photoproduction data within the second resonance region. The $F$-waves were fixed to BnGa2014\_02. Shown are results for the first energy-bin, \newline $E_{\gamma }\text{ = 683.5 MeV}$.]{Numerical results of a bootstrap-analysis are collected for a TPWA-fit of photoproduction data within the second resonance region, with $S$-, $P$- and $D$-wave multipoles varied in the fit, while the $F$-waves were fixed to BnGa2014\_02 (see section \ref{subsec:2ndResRegionDataFits}). An ensemble of $B = 2000$ bootstrap-replicates has been applied. Shown are results for the first energy-bin, $E_{\gamma }\text{ = 683.5 MeV}$. Here, a global minimum has been found with $\chi ^2\text{/ndf = 1.2907}$. \newline
From the bootstrap-distributions of the fit-parameters, we extract quantiles which then define a confidence-interval for the individual parameter, composed of upper and lower bootstrap-errors $\Delta_{\pm}$. The global minimum is quoted in conjunction with these asymmetric errors (for more details, see the main text). Furthermore, the mean $\hat{\theta}_{i}^{\ast} (\cdot)$, standard error $\widehat{\mathrm{se}}_{B} \left( \hat{\theta}^{\ast}_{i} \right)$ and bias-estimate $\widehat{\mathrm{bias}}_{B}$ are extracted from the bootstrap-distributions. Lastly, we define and extract a bias test-parameter defined as $\delta_{\mathrm{bias}} := \left| \widehat{\mathrm{bias}}_{B} \right|/\widehat{\mathrm{se}}_{B}$. \newline
All numbers are given in milli-Fermi, except for $\delta_{\mathrm{bias}}$ which does not carry dimension.}
\label{tab:2ndResRegionResultsFirstEnergy}
\end{table}

\begin{table}[h]
\centering
\begin{tabular}{c|c|c|c|c|c}
\multicolumn{2}{l|}{$ E_{\gamma }\text{ = 715.61 MeV} $} & \multicolumn{2}{c|}{ $ \text{ndf = 27} $} & \multicolumn{2}{c}{ $ \chi ^2\text{/ndf = 1.1215} $ } \\
\hline
\hline
$ \hat{\theta}_{i} = \left( \mathcal{M}_{\ell}^{C} \right)_{i} \text{[mFm]} $ & $ \left(\hat{\theta}_{i}^{\mathrm{Best}}\right)_{- \Delta_{-}}^{+ \Delta_{+}} $ & $ \hat{\theta}_{i}^{\ast} (\cdot) $ & $ \widehat{\mathrm{se}}_{B} \left( \hat{\theta}^{\ast}_{i} \right) $ & $ \widehat{\mathrm{bias}}_{B} $ & $ \delta_{\mathrm{bias}} $\\
\hline
$ \mathrm{Re} \left[ E_{0+}^{C} \right]  $ & $ 9.79635_{-0.63786}^{+0.91978} $ & $ 9.9298 $ & $ 0.7938 $ & $ 0.13345 $ & $ 0.16811 $ \\
$ \mathrm{Re} \left[ E_{1+}^{C} \right]  $ & $ 0.10347_{-0.18454}^{+0.12137} $ & $ 0.07078 $ & $ 0.1514 $ & $ -0.03269 $ & $ 0.21593 $ \\
$ \mathrm{Im} \left[ E_{1+}^{C} \right]  $ & $ 0.04585_{-0.1934}^{+0.10974} $ & $ 0.0036 $ & $ 0.16937 $ & $ -0.04225 $ & $ 0.24945 $ \\
$ \mathrm{Re} \left[ M_{1+}^{C} \right]  $ & $ 1.36828_{-0.15314}^{+0.23714} $ & $ 1.42105 $ & $ 0.23336 $ & $ 0.05277 $ & $ 0.22613 $ \\
$ \mathrm{Im} \left[ M_{1+}^{C} \right]  $ & $ 3.57637_{-0.19228}^{+0.10127} $ & $ 3.52675 $ & $ 0.15577 $ & $ -0.04962 $ & $ 0.31857 $ \\
$ \mathrm{Re} \left[ M_{1-}^{C} \right]  $ & $ 5.77555_{-1.52274}^{+0.81728} $ & $ 5.4535 $ & $ 1.24255 $ & $ -0.32205 $ & $ 0.25919 $ \\
$ \mathrm{Im} \left[ M_{1-}^{C} \right]  $ & $ 1.38823_{-0.09374}^{+0.34387} $ & $ 1.52583 $ & $ 0.26236 $ & $ 0.13759 $ & $ 0.52444 $ \\
$ \mathrm{Re} \left[ E_{2+}^{C} \right] $ & $ 0.15133_{-0.08847}^{+0.05821} $ & $ 0.13669 $ & $ 0.07607 $ & $ -0.01464 $ & $ 0.19249 $ \\
$ \mathrm{Im} \left[ E_{2+}^{C} \right] $ & $ 0.06397_{-0.03913}^{+0.03415} $ & $ 0.06104 $ & $ 0.03848 $ & $ -0.00293 $ & $ 0.07627 $ \\
$ \mathrm{Re} \left[ E_{2-}^{C} \right] $ & $ 4.60824_{-0.13324}^{+0.20814} $ & $ 4.64353 $ & $ 0.1757 $ & $ 0.03529 $ & $ 0.20084 $ \\
$ \mathrm{Im} \left[ E_{2-}^{C} \right] $ & $ 1.68734_{-0.25991}^{+0.54432} $ & $ 1.81075 $ & $ 0.41591 $ & $ 0.12341 $ & $ 0.29672 $ \\
$ \mathrm{Re} \left[ M_{2+}^{C} \right] $ & $ 0.07912_{-0.0732}^{+0.03785} $ & $ 0.05969 $ & $ 0.06226 $ & $ -0.01943 $ & $ 0.31206 $ \\
$ \mathrm{Im} \left[ M_{2+}^{C} \right] $ & $ 0.25532_{-0.09374}^{+0.03329} $ & $ 0.21817 $ & $ 0.09174 $ & $ -0.03715 $ & $ 0.40491 $ \\
$ \mathrm{Re} \left[ M_{2-}^{C} \right] $ & $ 3.33847_{-0.22718}^{+0.12974} $ & $ 3.28188 $ & $ 0.22021 $ & $ -0.0566 $ & $ 0.25701 $ \\
$ \mathrm{Im} \left[ M_{2-}^{C} \right] $ & $ -1.71871_{-0.06017}^{+0.20854} $ & $ -1.6386 $ & $ 0.1634 $ & $ 0.08011 $ & $ 0.49026 $ \\
\end{tabular}
\caption{Numerical results are collected for a TPWA bootstrap-analysis of photoproduction data within the second resonance region. Here, the second energy-bin, $E_{\gamma }\text{ = 715.61 MeV}$, is shown. For more details, see the description of Table \ref{tab:2ndResRegionResultsFirstEnergy}.}
\label{tab:2ndResRegionResultsSecondEnergy}
\end{table}


\begin{table}[h]
\centering
\begin{tabular}{c|c|c|c|c|c}
\multicolumn{2}{l|}{$ E_{\gamma }\text{ = 749.94 MeV} $} & \multicolumn{2}{c|}{ $ \text{ndf = 27} $} & \multicolumn{2}{c}{ $ \chi ^2\text{/ndf = 1.41885} $ } \\
\hline
\hline
$ \hat{\theta}_{i} = \left( \mathcal{M}_{\ell}^{C} \right)_{i} \text{[mFm]} $ & $ \left(\hat{\theta}_{i}^{\mathrm{Best}}\right)_{- \Delta_{-}}^{+ \Delta_{+}} $ & $ \hat{\theta}_{i}^{\ast} (\cdot) $ & $ \widehat{\mathrm{se}}_{B} \left( \hat{\theta}^{\ast}_{i} \right) $ & $ \widehat{\mathrm{bias}}_{B} $ & $ \delta_{\mathrm{bias}} $\\
\hline
 $ \mathrm{Re} \left[ E_{0+}^{C} \right]  $ & $ 11.8904_{-2.37246}^{+0.11652} $ & $ 11.0932 $ & $ 1.13524 $ & $ -0.79721 $ & $ 0.70224 $ \\
 $ \mathrm{Re} \left[ E_{1+}^{C} \right]  $ & $ 0.37581_{-0.4982}^{+0.11862} $ & $ 0.2501 $ & $ 0.28829 $ & $ -0.12571 $ & $ 0.43607 $ \\
 $ \mathrm{Im} \left[ E_{1+}^{C} \right]  $ & $ -0.50333_{-0.13319}^{+0.36897} $ & $ -0.42838 $ & $ 0.21584 $ & $ 0.07495 $ & $ 0.34725 $ \\
 $ \mathrm{Re} \left[ M_{1+}^{C} \right]  $ & $ 2.20498_{-0.403}^{+0.74136} $ & $ 2.381 $ & $ 0.72297 $ & $ 0.17602 $ & $ 0.24346 $ \\
 $ \mathrm{Im} \left[ M_{1+}^{C} \right]  $ & $ 2.25322_{-0.10205}^{+0.27573} $ & $ 2.35878 $ & $ 0.24778 $ & $ 0.10556 $ & $ 0.42601 $ \\
 $ \mathrm{Re} \left[ M_{1-}^{C} \right]  $ & $ 4.73861_{-0.31979}^{+3.05856} $ & $ 5.78775 $ & $ 1.53111 $ & $ 1.04914 $ & $ 0.68522 $ \\
 $ \mathrm{Im} \left[ M_{1-}^{C} \right]  $ & $ 1.34752_{-0.73305}^{+0.34774} $ & $ 1.22495 $ & $ 0.48557 $ & $ -0.12257 $ & $ 0.25243 $ \\
 $ \mathrm{Re} \left[ E_{2+}^{C} \right] $ & $ 0.13153_{-0.07863}^{+0.07013} $ & $ 0.12687 $ & $ 0.07285 $ & $ -0.00466 $ & $ 0.06392 $ \\
 $ \mathrm{Im} \left[ E_{2+}^{C} \right] $ & $ -0.005_{-0.04451}^{+0.09345} $ & $ 0.01272 $ & $ 0.06365 $ & $ 0.01772 $ & $ 0.2784 $ \\
 $ \mathrm{Re} \left[ E_{2-}^{C} \right] $ & $ 5.42066_{-0.53026}^{+0.07604} $ & $ 5.2449 $ & $ 0.28066 $ & $ -0.17576 $ & $ 0.62624 $ \\
 $ \mathrm{Im} \left[ E_{2-}^{C} \right] $ & $ 2.36311_{-0.71899}^{+0.09988} $ & $ 2.12115 $ & $ 0.36578 $ & $ -0.24196 $ & $ 0.66149 $ \\
 $ \mathrm{Re} \left[ M_{2+}^{C} \right] $ & $ -0.07839_{-0.05989}^{+0.16996} $ & $ -0.03955 $ & $ 0.10132 $ & $ 0.03884 $ & $ 0.38338 $ \\
 $ \mathrm{Im} \left[ M_{2+}^{C} \right] $ & $ -0.12724_{-0.18193}^{+0.20638} $ & $ -0.12053 $ & $ 0.17338 $ & $ 0.00671 $ & $ 0.0387 $ \\
 $ \mathrm{Re} \left[ M_{2-}^{C} \right] $ & $ 2.75003_{-0.34904}^{+0.76384} $ & $ 2.8385 $ & $ 0.49721 $ & $ 0.08847 $ & $ 0.17793 $ \\
 $ \mathrm{Im} \left[ M_{2-}^{C} \right] $ & $ -1.21433_{-0.30738}^{+0.13481} $ & $ -1.27922 $ & $ 0.20582 $ & $ -0.0649 $ & $ 0.31531 $ \\
\end{tabular}
\caption[Numerical results of a bootstrap-analysis are collected for a TPWA-fit of photoproduction data within the second resonance region. The $F$-waves were fixed to BnGa2014\_02. The third energy-bin, $E_{\gamma }\text{ = 749.94 MeV}$, is shown.]{Numerical results are collected for a TPWA bootstrap-analysis of photoproduction data within the second resonance region. Here, the third energy-bin, $E_{\gamma }\text{ = 749.94 MeV}$, is shown. For more details, see the description of Table \ref{tab:2ndResRegionResultsFirstEnergy}.}
\label{tab:2ndResRegionResultsThirdEnergy}
\end{table}

\begin{table}[h]
\centering
\begin{tabular}{c|c|c|c|c|c}
\multicolumn{2}{l|}{$ E_{\gamma }\text{ = 783.42 MeV} $} & \multicolumn{2}{c|}{ $ \text{ndf = 27} $} & \multicolumn{2}{c}{ $ \chi ^2\text{/ndf = 1.58912} $ } \\
\hline
\hline
$ \hat{\theta}_{i} = \left( \mathcal{M}_{\ell}^{C} \right)_{i} \text{[mFm]} $ & $ \left(\hat{\theta}_{i}^{\mathrm{Best}}\right)_{- \Delta_{-}}^{+ \Delta_{+}} $ & $ \hat{\theta}_{i}^{\ast} (\cdot) $ & $ \widehat{\mathrm{se}}_{B} \left( \hat{\theta}^{\ast}_{i} \right) $ & $ \widehat{\mathrm{bias}}_{B} $ & $ \delta_{\mathrm{bias}} $\\
\hline
 $ \mathrm{Re} \left[ E_{0+}^{C} \right]  $ & $ 11.2086_{-0.14756}^{+0.11384} $ & $ 11.1893 $ & $ 0.13499 $ & $ -0.01923 $ & $ 0.14246 $ \\
 $ \mathrm{Re} \left[ E_{1+}^{C} \right]  $ & $ 0.42901_{-0.07755}^{+0.06766} $ & $ 0.42432 $ & $ 0.07473 $ & $ -0.00469 $ & $ 0.06272 $ \\
 $ \mathrm{Im} \left[ E_{1+}^{C} \right]  $ & $ -0.50005_{-0.06495}^{+0.07273} $ & $ -0.49548 $ & $ 0.06986 $ & $ 0.00457 $ & $ 0.06541 $ \\
 $ \mathrm{Re} \left[ M_{1+}^{C} \right]  $ & $ 1.12782_{-0.13661}^{+0.13299} $ & $ 1.12668 $ & $ 0.13766 $ & $ -0.00115 $ & $ 0.00834 $ \\
 $ \mathrm{Im} \left[ M_{1+}^{C} \right]  $ & $ 0.78681_{-0.10437}^{+0.11039} $ & $ 0.78994 $ & $ 0.10845 $ & $ 0.00313 $ & $ 0.02882 $ \\
 $ \mathrm{Re} \left[ M_{1-}^{C} \right]  $ & $ 2.93183_{-0.27751}^{+0.30533} $ & $ 2.9456 $ & $ 0.2942 $ & $ 0.01377 $ & $ 0.04681 $ \\
 $ \mathrm{Im} \left[ M_{1-}^{C} \right]  $ & $ 1.51603_{-0.1938}^{+0.18564} $ & $ 1.51098 $ & $ 0.19009 $ & $ -0.00505 $ & $ 0.02657 $ \\
 $ \mathrm{Re} \left[ E_{2+}^{C} \right] $ & $ 0.30299_{-0.04845}^{+0.04625} $ & $ 0.30188 $ & $ 0.04695 $ & $ -0.00111 $ & $ 0.02365 $ \\
 $ \mathrm{Im} \left[ E_{2+}^{C} \right] $ & $ -0.06221_{-0.02596}^{+0.02798} $ & $ -0.06103 $ & $ 0.02742 $ & $ 0.00118 $ & $ 0.04313 $ \\
 $ \mathrm{Re} \left[ E_{2-}^{C} \right] $ & $ 4.9901_{-0.11617}^{+0.1099} $ & $ 4.98882 $ & $ 0.11585 $ & $ -0.00127 $ & $ 0.01097 $ \\
 $ \mathrm{Im} \left[ E_{2-}^{C} \right] $ & $ 3.97206_{-0.10793}^{+0.08308} $ & $ 3.95957 $ & $ 0.09654 $ & $ -0.01249 $ & $ 0.12938 $ \\
 $ \mathrm{Re} \left[ M_{2+}^{C} \right] $ & $ -0.11402_{-0.04045}^{+0.04639} $ & $ -0.11165 $ & $ 0.04363 $ & $ 0.00237 $ & $ 0.05442 $ \\
 $ \mathrm{Im} \left[ M_{2+}^{C} \right] $ & $ 0.00387_{-0.0548}^{+0.058} $ & $ 0.00509 $ & $ 0.05665 $ & $ 0.00122 $ & $ 0.02155 $ \\
 $ \mathrm{Re} \left[ M_{2-}^{C} \right] $ & $ 2.99081_{-0.07067}^{+0.07992} $ & $ 2.99511 $ & $ 0.07605 $ & $ 0.0043 $ & $ 0.0566 $ \\
 $ \mathrm{Im} \left[ M_{2-}^{C} \right] $ & $ -0.65531_{-0.0906}^{+0.08437} $ & $ -0.65777 $ & $ 0.08842 $ & $ -0.00246 $ & $ 0.0278 $ \\
\end{tabular}
\caption[Numerical results of a bootstrap-analysis are collected for a TPWA-fit of photoproduction data within the second resonance region. The $F$-waves were fixed to BnGa2014\_02. The fourth energy-bin, $E_{\gamma }\text{ = 783.42 MeV}$, is shown.]{Numerical results are collected for a TPWA bootstrap-analysis of photoproduction data within the second resonance region. Here, the fourth energy-bin, $E_{\gamma }\text{ = 783.42 MeV}$, is shown. For more details, see the description of Table \ref{tab:2ndResRegionResultsFirstEnergy}.}
\label{tab:2ndResRegionResultsFourthEnergy}
\end{table}


\begin{table}[h]
\centering
\begin{tabular}{c|c|c|c|c|c}
\multicolumn{2}{l|}{$ E_{\gamma }\text{ = 815.92 MeV} $} & \multicolumn{2}{c|}{ $ \text{ndf = 27} $} & \multicolumn{2}{c}{ $ \chi ^2\text{/ndf = 1.98783} $ } \\
\hline
\hline
$ \hat{\theta}_{i} = \left( \mathcal{M}_{\ell}^{C} \right)_{i} \text{[mFm]} $ & $ \left(\hat{\theta}_{i}^{\mathrm{Best}}\right)_{- \Delta_{-}}^{+ \Delta_{+}} $ & $ \hat{\theta}_{i}^{\ast} (\cdot) $ & $ \widehat{\mathrm{se}}_{B} \left( \hat{\theta}^{\ast}_{i} \right) $ & $ \widehat{\mathrm{bias}}_{B} $ & $ \delta_{\mathrm{bias}} $\\
\hline
 $ \mathrm{Re} \left[ E_{0+}^{C} \right]  $ & $ 9.95371_{-0.25026}^{+0.16162} $ & $ 9.90675 $ & $ 0.23261 $ & $ -0.04696 $ & $ 0.20188 $ \\
 $ \mathrm{Re} \left[ E_{1+}^{C} \right]  $ & $ 0.07167_{-0.10872}^{+0.08166} $ & $ 0.05848 $ & $ 0.09898 $ & $ -0.0132 $ & $ 0.13332 $ \\
 $ \mathrm{Im} \left[ E_{1+}^{C} \right]  $ & $ -0.44091_{-0.06367}^{+0.07112} $ & $ -0.43768 $ & $ 0.06603 $ & $ 0.00323 $ & $ 0.04899 $ \\
 $ \mathrm{Re} \left[ M_{1+}^{C} \right]  $ & $ 1.22389_{-0.11518}^{+0.11741} $ & $ 1.22625 $ & $ 0.11646 $ & $ 0.00236 $ & $ 0.02025 $ \\
 $ \mathrm{Im} \left[ M_{1+}^{C} \right]  $ & $ 0.53953_{-0.10778}^{+0.1188} $ & $ 0.5472 $ & $ 0.11378 $ & $ 0.00767 $ & $ 0.06741 $ \\
 $ \mathrm{Re} \left[ M_{1-}^{C} \right]  $ & $ 3.43973_{-0.27131}^{+0.37058} $ & $ 3.5007 $ & $ 0.34566 $ & $ 0.06097 $ & $ 0.17639 $ \\
 $ \mathrm{Im} \left[ M_{1-}^{C} \right]  $ & $ 0.06262_{-0.23033}^{+0.22757} $ & $ 0.05858 $ & $ 0.23195 $ & $ -0.00405 $ & $ 0.01744 $ \\
 $ \mathrm{Re} \left[ E_{2+}^{C} \right] $ & $ 0.17275_{-0.04304}^{+0.04139} $ & $ 0.17122 $ & $ 0.04289 $ & $ -0.00153 $ & $ 0.03573 $ \\
 $ \mathrm{Im} \left[ E_{2+}^{C} \right] $ & $ -0.00307_{-0.02481}^{+0.02406} $ & $ -0.0031 $ & $ 0.02423 $ & $ -0.00003 $ & $ 0.00125 $ \\
 $ \mathrm{Re} \left[ E_{2-}^{C} \right] $ & $ 3.99203_{-0.15914}^{+0.12023} $ & $ 3.974 $ & $ 0.13893 $ & $ -0.01803 $ & $ 0.12975 $ \\
 $ \mathrm{Im} \left[ E_{2-}^{C} \right] $ & $ 4.60794_{-0.10254}^{+0.07244} $ & $ 4.59192 $ & $ 0.0909 $ & $ -0.01602 $ & $ 0.17626 $ \\
 $ \mathrm{Re} \left[ M_{2+}^{C} \right] $ & $ 0.06324_{-0.03405}^{+0.04213} $ & $ 0.06726 $ & $ 0.03764 $ & $ 0.00402 $ & $ 0.10684 $ \\
 $ \mathrm{Im} \left[ M_{2+}^{C} \right] $ & $ 0.06528_{-0.07235}^{+0.0733} $ & $ 0.06637 $ & $ 0.07269 $ & $ 0.00109 $ & $ 0.01505 $ \\
 $ \mathrm{Re} \left[ M_{2-}^{C} \right] $ & $ 2.58418_{-0.10121}^{+0.12596} $ & $ 2.59795 $ & $ 0.11951 $ & $ 0.01377 $ & $ 0.11526 $ \\
 $ \mathrm{Im} \left[ M_{2-}^{C} \right] $ & $ -0.72475_{-0.11839}^{+0.11456} $ & $ -0.7264 $ & $ 0.11707 $ & $ -0.00165 $ & $ 0.01408 $ \\
\end{tabular}
\caption[Numerical results of a bootstrap-analysis are collected for a TPWA-fit of photoproduction data within the second resonance region. The $F$-waves were fixed to BnGa2014\_02. The fifth energy-bin, $E_{\gamma }\text{ = 815.92 MeV}$, is shown.]{Numerical results are collected for a TPWA bootstrap-analysis of photoproduction data within the second resonance region. Here, the fifth energy-bin, $E_{\gamma }\text{ = 815.92 MeV}$, is shown. For more details, see the description of Table \ref{tab:2ndResRegionResultsFirstEnergy}.}
\label{tab:2ndResRegionResultsFifthEnergy}
\end{table}

\begin{table}[h]
\centering
\begin{tabular}{c|c|c|c|c|c}
\multicolumn{2}{l|}{$ E_{\gamma }\text{ = 850.45 MeV} $} & \multicolumn{2}{c|}{ $ \text{ndf = 27} $} & \multicolumn{2}{c}{ $ \chi ^2\text{/ndf = 3.13052} $ } \\
\hline
\hline
$ \hat{\theta}_{i} = \left( \mathcal{M}_{\ell}^{C} \right)_{i} \text{[mFm]} $ & $ \left(\hat{\theta}_{i}^{\mathrm{Best}}\right)_{- \Delta_{-}}^{+ \Delta_{+}} $ & $ \hat{\theta}_{i}^{\ast} (\cdot) $ & $ \widehat{\mathrm{se}}_{B} \left( \hat{\theta}^{\ast}_{i} \right) $ & $ \widehat{\mathrm{bias}}_{B} $ & $ \delta_{\mathrm{bias}} $\\
\hline
 $ \mathrm{Re} \left[ E_{0+}^{C} \right]  $ & $ 8.65553_{-0.13749}^{+0.13067} $ & $ 8.64948 $ & $ 0.13396 $ & $ -0.00605 $ & $ 0.04518 $ \\
 $ \mathrm{Re} \left[ E_{1+}^{C} \right]  $ & $ 0.13332_{-0.06707}^{+0.05456} $ & $ 0.12825 $ & $ 0.06136 $ & $ -0.00507 $ & $ 0.08265 $ \\
 $ \mathrm{Im} \left[ E_{1+}^{C} \right]  $ & $ -0.52357_{-0.04209}^{+0.04605} $ & $ -0.52131 $ & $ 0.04499 $ & $ 0.00226 $ & $ 0.05033 $ \\
 $ \mathrm{Re} \left[ M_{1+}^{C} \right]  $ & $ 1.16061_{-0.0973}^{+0.10384} $ & $ 1.16266 $ & $ 0.10403 $ & $ 0.00205 $ & $ 0.01974 $ \\
 $ \mathrm{Im} \left[ M_{1+}^{C} \right]  $ & $ 0.08961_{-0.12626}^{+0.14556} $ & $ 0.09975 $ & $ 0.14073 $ & $ 0.01014 $ & $ 0.07206 $ \\
 $ \mathrm{Re} \left[ M_{1-}^{C} \right]  $ & $ 3.36311_{-0.25829}^{+0.28371} $ & $ 3.3786 $ & $ 0.26988 $ & $ 0.01549 $ & $ 0.0574 $ \\
 $ \mathrm{Im} \left[ M_{1-}^{C} \right]  $ & $ 0.22642_{-0.17238}^{+0.15626} $ & $ 0.21867 $ & $ 0.16253 $ & $ -0.00775 $ & $ 0.04767 $ \\
 $ \mathrm{Re} \left[ E_{2+}^{C} \right] $ & $ 0.23876_{-0.03551}^{+0.03476} $ & $ 0.23865 $ & $ 0.03663 $ & $ -0.00012 $ & $ 0.00327 $ \\
 $ \mathrm{Im} \left[ E_{2+}^{C} \right] $ & $ -0.01468_{-0.01876}^{+0.02074} $ & $ -0.01387 $ & $ 0.02044 $ & $ 0.00081 $ & $ 0.03939 $ \\
 $ \mathrm{Re} \left[ E_{2-}^{C} \right] $ & $ 3.14503_{-0.13457}^{+0.12146} $ & $ 3.13748 $ & $ 0.12589 $ & $ -0.00756 $ & $ 0.06003 $ \\
 $ \mathrm{Im} \left[ E_{2-}^{C} \right] $ & $ 4.65642_{-0.08378}^{+0.06008} $ & $ 4.64507 $ & $ 0.07338 $ & $ -0.01135 $ & $ 0.15466 $ \\
 $ \mathrm{Re} \left[ M_{2+}^{C} \right] $ & $ 0.12063_{-0.04156}^{+0.04822} $ & $ 0.12374 $ & $ 0.04626 $ & $ 0.00311 $ & $ 0.06718 $ \\
 $ \mathrm{Im} \left[ M_{2+}^{C} \right] $ & $ -0.12467_{-0.06245}^{+0.05709} $ & $ -0.12733 $ & $ 0.0603 $ & $ -0.00266 $ & $ 0.04414 $ \\
 $ \mathrm{Re} \left[ M_{2-}^{C} \right] $ & $ 2.01845_{-0.07054}^{+0.0648} $ & $ 2.01661 $ & $ 0.06883 $ & $ -0.00184 $ & $ 0.02677 $ \\
 $ \mathrm{Im} \left[ M_{2-}^{C} \right] $ & $ -0.48252_{-0.09571}^{+0.10654} $ & $ -0.4773 $ & $ 0.10115 $ & $ 0.00522 $ & $ 0.05163 $ \\
\end{tabular}
\caption[Numerical results of a bootstrap-analysis are collected for a TPWA-fit of photoproduction data within the second resonance region. The $F$-waves were fixed to BnGa2014\_02. The sixth energy-bin, $E_{\gamma }\text{ = 850.45 MeV}$, is shown.]{Numerical results are collected for a TPWA bootstrap-analysis of photoproduction data within the second resonance region. Here, the sixth energy-bin, $E_{\gamma }\text{ = 850.45 MeV}$, is shown. For more details, see the description of Table \ref{tab:2ndResRegionResultsFirstEnergy}.}
\label{tab:2ndResRegionResultsSixthEnergy}
\end{table}


\begin{table}[h]
\centering
\begin{tabular}{c|c|c|c|c|c}
\multicolumn{2}{l|}{$ E_{\gamma }\text{ = 884.02 MeV} $} & \multicolumn{2}{c|}{ $ \text{ndf = 27} $} & \multicolumn{2}{c}{ $ \chi ^2\text{/ndf = 2.58421} $ } \\
\hline
\hline
$ \hat{\theta}_{i} = \left( \mathcal{M}_{\ell}^{C} \right)_{i} \text{[mFm]} $ & $ \left(\hat{\theta}_{i}^{\mathrm{Best}}\right)_{- \Delta_{-}}^{+ \Delta_{+}} $ & $ \hat{\theta}_{i}^{\ast} (\cdot) $ & $ \widehat{\mathrm{se}}_{B} \left( \hat{\theta}^{\ast}_{i} \right) $ & $ \widehat{\mathrm{bias}}_{B} $ & $ \delta_{\mathrm{bias}} $\\
\hline
 $ \mathrm{Re} \left[ E_{0+}^{C} \right]  $ & $ 7.19845_{-0.44113}^{+0.1828} $ & $ 7.10395 $ & $ 0.30636 $ & $ -0.0945 $ & $ 0.30846 $ \\
 $ \mathrm{Re} \left[ E_{1+}^{C} \right]  $ & $ 0.04392_{-0.17369}^{+0.07985} $ & $ 0.00432 $ & $ 0.12462 $ & $ -0.0396 $ & $ 0.31778 $ \\
 $ \mathrm{Im} \left[ E_{1+}^{C} \right]  $ & $ -0.49482_{-0.04084}^{+0.04912} $ & $ -0.49036 $ & $ 0.04515 $ & $ 0.00446 $ & $ 0.09885 $ \\
 $ \mathrm{Re} \left[ M_{1+}^{C} \right]  $ & $ 0.962_{-0.15687}^{+0.113} $ & $ 0.94348 $ & $ 0.13066 $ & $ -0.01853 $ & $ 0.1418 $ \\
 $ \mathrm{Im} \left[ M_{1+}^{C} \right]  $ & $ -0.17535_{-0.14875}^{+0.11566} $ & $ -0.19222 $ & $ 0.13188 $ & $ -0.01688 $ & $ 0.12798 $ \\
 $ \mathrm{Re} \left[ M_{1-}^{C} \right]  $ & $ 3.48156_{-0.21932}^{+0.37689} $ & $ 3.5546 $ & $ 0.3035 $ & $ 0.07304 $ & $ 0.24066 $ \\
 $ \mathrm{Im} \left[ M_{1-}^{C} \right]  $ & $ -0.16712_{-0.31659}^{+0.21853} $ & $ -0.207 $ & $ 0.25471 $ & $ -0.03988 $ & $ 0.15658 $ \\
 $ \mathrm{Re} \left[ E_{2+}^{C} \right] $ & $ 0.31265_{-0.0614}^{+0.03408} $ & $ 0.29924 $ & $ 0.04802 $ & $ -0.01341 $ & $ 0.27928 $ \\
 $ \mathrm{Im} \left[ E_{2+}^{C} \right] $ & $ -0.00081_{-0.01762}^{+0.02153} $ & $ 0.00118 $ & $ 0.0198 $ & $ 0.00199 $ & $ 0.10043 $ \\
 $ \mathrm{Re} \left[ E_{2-}^{C} \right] $ & $ 2.3085_{-0.14137}^{+0.09768} $ & $ 2.28563 $ & $ 0.12291 $ & $ -0.02287 $ & $ 0.18609 $ \\
 $ \mathrm{Im} \left[ E_{2-}^{C} \right] $ & $ 4.49854_{-0.08711}^{+0.05704} $ & $ 4.48438 $ & $ 0.07086 $ & $ -0.01416 $ & $ 0.19982 $ \\
 $ \mathrm{Re} \left[ M_{2+}^{C} \right] $ & $ 0.03211_{-0.05969}^{+0.05085} $ & $ 0.02834 $ & $ 0.0552 $ & $ -0.00377 $ & $ 0.06821 $ \\
 $ \mathrm{Im} \left[ M_{2+}^{C} \right] $ & $ -0.08653_{-0.05789}^{+0.07798} $ & $ -0.07886 $ & $ 0.06573 $ & $ 0.00767 $ & $ 0.11662 $ \\
 $ \mathrm{Re} \left[ M_{2-}^{C} \right] $ & $ 1.79735_{-0.09298}^{+0.1785} $ & $ 1.82745 $ & $ 0.1304 $ & $ 0.0301 $ & $ 0.23082 $ \\
 $ \mathrm{Im} \left[ M_{2-}^{C} \right] $ & $ -0.65275_{-0.18985}^{+0.12616} $ & $ -0.67472 $ & $ 0.14732 $ & $ -0.02197 $ & $ 0.14914 $ \\
\end{tabular}
\caption[Numerical results of a bootstrap-analysis are collected for a TPWA-fit of photoproduction data within the second resonance region. The $F$-waves were fixed to BnGa2014\_02. The seventh energy-bin, $E_{\gamma }\text{ = 884.02 MeV}$, is shown.]{Numerical results are collected for a TPWA bootstrap-analysis of photoproduction data within the second resonance region. Here, the seventh energy-bin, $E_{\gamma }\text{ = 884.02 MeV}$, is shown. For more details, see the description of Table \ref{tab:2ndResRegionResultsFirstEnergy}.}
\label{tab:2ndResRegionResultsSeventhEnergy}
\end{table}

\begin{table}[h]
\centering
\begin{tabular}{c|c|c|c|c|c}
\multicolumn{2}{l|}{$ E_{\gamma }\text{ = 916.66 MeV} $} & \multicolumn{2}{c|}{ $ \text{ndf = 27} $} & \multicolumn{2}{c}{ $ \chi ^2\text{/ndf = 2.0712} $ } \\
\hline
\hline
$ \hat{\theta}_{i} = \left( \mathcal{M}_{\ell}^{C} \right)_{i} \text{[mFm]} $ & $ \left(\hat{\theta}_{i}^{\mathrm{Best}}\right)_{- \Delta_{-}}^{+ \Delta_{+}} $ & $ \hat{\theta}_{i}^{\ast} (\cdot) $ & $ \widehat{\mathrm{se}}_{B} \left( \hat{\theta}^{\ast}_{i} \right) $ & $ \widehat{\mathrm{bias}}_{B} $ & $ \delta_{\mathrm{bias}} $\\
\hline
 $ \mathrm{Re} \left[ E_{0+}^{C} \right]  $ & $ 4.54308_{-0.14789}^{+0.15163} $ & $ 4.54487 $ & $ 0.1512 $ & $ 0.0018 $ & $ 0.01189 $ \\
 $ \mathrm{Re} \left[ E_{1+}^{C} \right]  $ & $ -0.33427_{-0.06255}^{+0.06341} $ & $ -0.33422 $ & $ 0.06314 $ & $ 0.00005 $ & $ 0.00082 $ \\
 $ \mathrm{Im} \left[ E_{1+}^{C} \right]  $ & $ -0.60443_{-0.03314}^{+0.03393} $ & $ -0.60398 $ & $ 0.03353 $ & $ 0.00045 $ & $ 0.01344 $ \\
 $ \mathrm{Re} \left[ M_{1+}^{C} \right]  $ & $ 0.69371_{-0.07536}^{+0.07819} $ & $ 0.6942 $ & $ 0.07628 $ & $ 0.00049 $ & $ 0.00644 $ \\
 $ \mathrm{Im} \left[ M_{1+}^{C} \right]  $ & $ -0.57962_{-0.09854}^{+0.1118} $ & $ -0.57115 $ & $ 0.10426 $ & $ 0.00847 $ & $ 0.08122 $ \\
 $ \mathrm{Re} \left[ M_{1-}^{C} \right]  $ & $ 4.17838_{-0.15967}^{+0.16289} $ & $ 4.18097 $ & $ 0.16439 $ & $ 0.00259 $ & $ 0.01576 $ \\
 $ \mathrm{Im} \left[ M_{1-}^{C} \right]  $ & $ 0.12173_{-0.13831}^{+0.13791} $ & $ 0.12237 $ & $ 0.13707 $ & $ 0.00064 $ & $ 0.00469 $ \\
 $ \mathrm{Re} \left[ E_{2+}^{C} \right] $ & $ 0.09039_{-0.03427}^{+0.03571} $ & $ 0.0915 $ & $ 0.03515 $ & $ 0.00111 $ & $ 0.03167 $ \\
 $ \mathrm{Im} \left[ E_{2+}^{C} \right] $ & $ 0.05493_{-0.01861}^{+0.0185} $ & $ 0.05483 $ & $ 0.01861 $ & $ -0.00009 $ & $ 0.00501 $ \\
 $ \mathrm{Re} \left[ E_{2-}^{C} \right] $ & $ 0.89961_{-0.11739}^{+0.11945} $ & $ 0.90008 $ & $ 0.11791 $ & $ 0.00046 $ & $ 0.00392 $ \\
 $ \mathrm{Im} \left[ E_{2-}^{C} \right] $ & $ 4.5392_{-0.0602}^{+0.04792} $ & $ 4.53219 $ & $ 0.05418 $ & $ -0.00701 $ & $ 0.12931 $ \\
 $ \mathrm{Re} \left[ M_{2+}^{C} \right] $ & $ 0.01623_{-0.04356}^{+0.054} $ & $ 0.0218 $ & $ 0.04921 $ & $ 0.00557 $ & $ 0.11321 $ \\
 $ \mathrm{Im} \left[ M_{2+}^{C} \right] $ & $ -0.11471_{-0.03451}^{+0.03006} $ & $ -0.11689 $ & $ 0.03232 $ & $ -0.00218 $ & $ 0.06749 $ \\
 $ \mathrm{Re} \left[ M_{2-}^{C} \right] $ & $ 2.02705_{-0.04966}^{+0.04267} $ & $ 2.02286 $ & $ 0.04673 $ & $ -0.00418 $ & $ 0.08951 $ \\
 $ \mathrm{Im} \left[ M_{2-}^{C} \right] $ & $ -0.76426_{-0.0532}^{+0.06502} $ & $ -0.75794 $ & $ 0.06039 $ & $ 0.00632 $ & $ 0.10469 $ \\
\end{tabular}
\caption[Numerical results of a bootstrap-analysis are collected for a TPWA-fit of photoproduction data within the second resonance region. The $F$-waves were fixed to BnGa2014\_02. The eighth energy-bin, $E_{\gamma }\text{ = 916.66 MeV}$, is shown.]{Numerical results are collected for a TPWA bootstrap-analysis of photoproduction data within the second resonance region. Here, the eighth energy-bin, $E_{\gamma }\text{ = 916.66 MeV}$, is shown. For more details, see the description of Table \ref{tab:2ndResRegionResultsFirstEnergy}.}
\label{tab:2ndResRegionResultsEighthEnergy}
\end{table}

\clearpage



\begin{figure}[h]
\vspace*{15pt}
\begin{overpic}[width=0.325\textwidth]{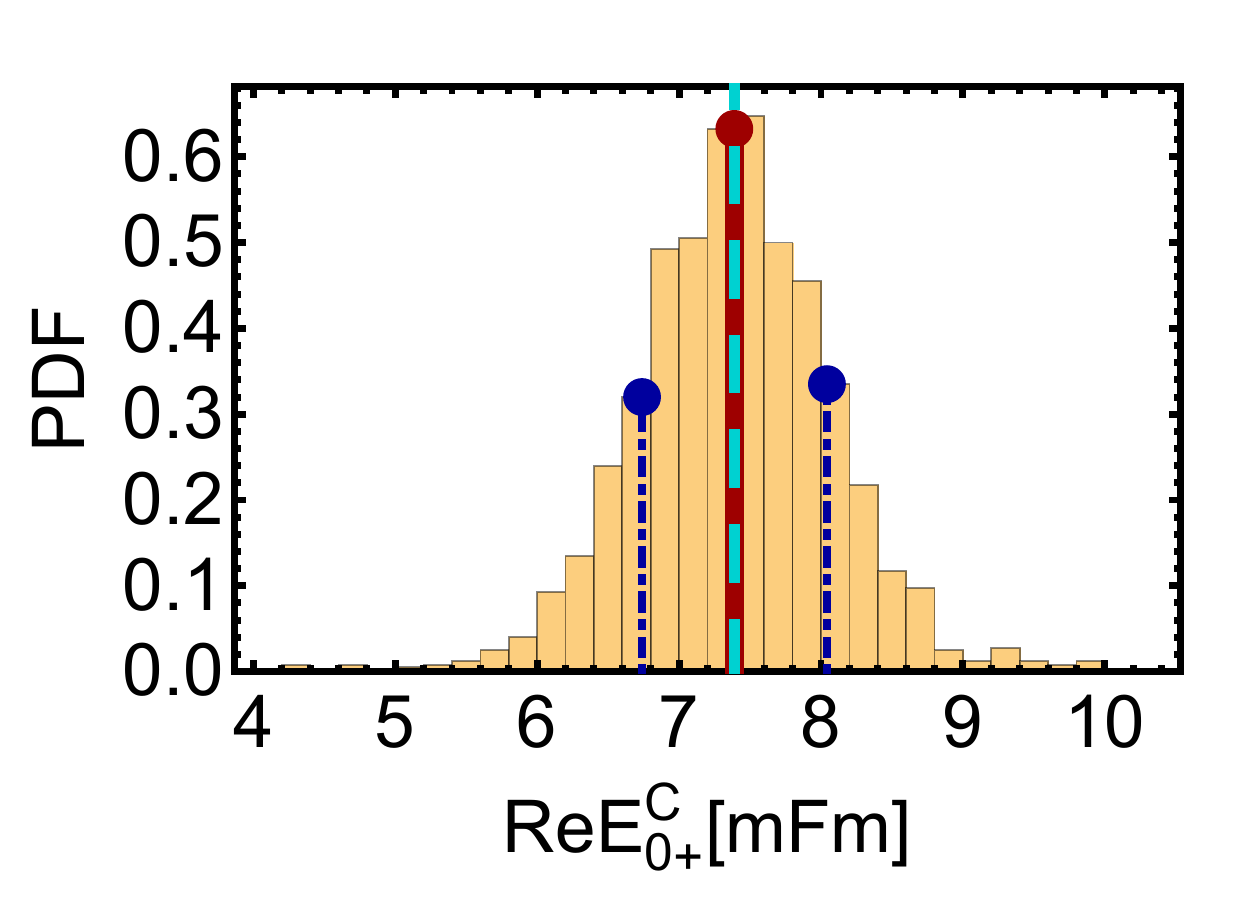}
 \put(1,80){\textbf{Histograms for fit-parameters resulting from the application of the bootstrap}}
 \end{overpic}
\begin{overpic}[width=0.325\textwidth]{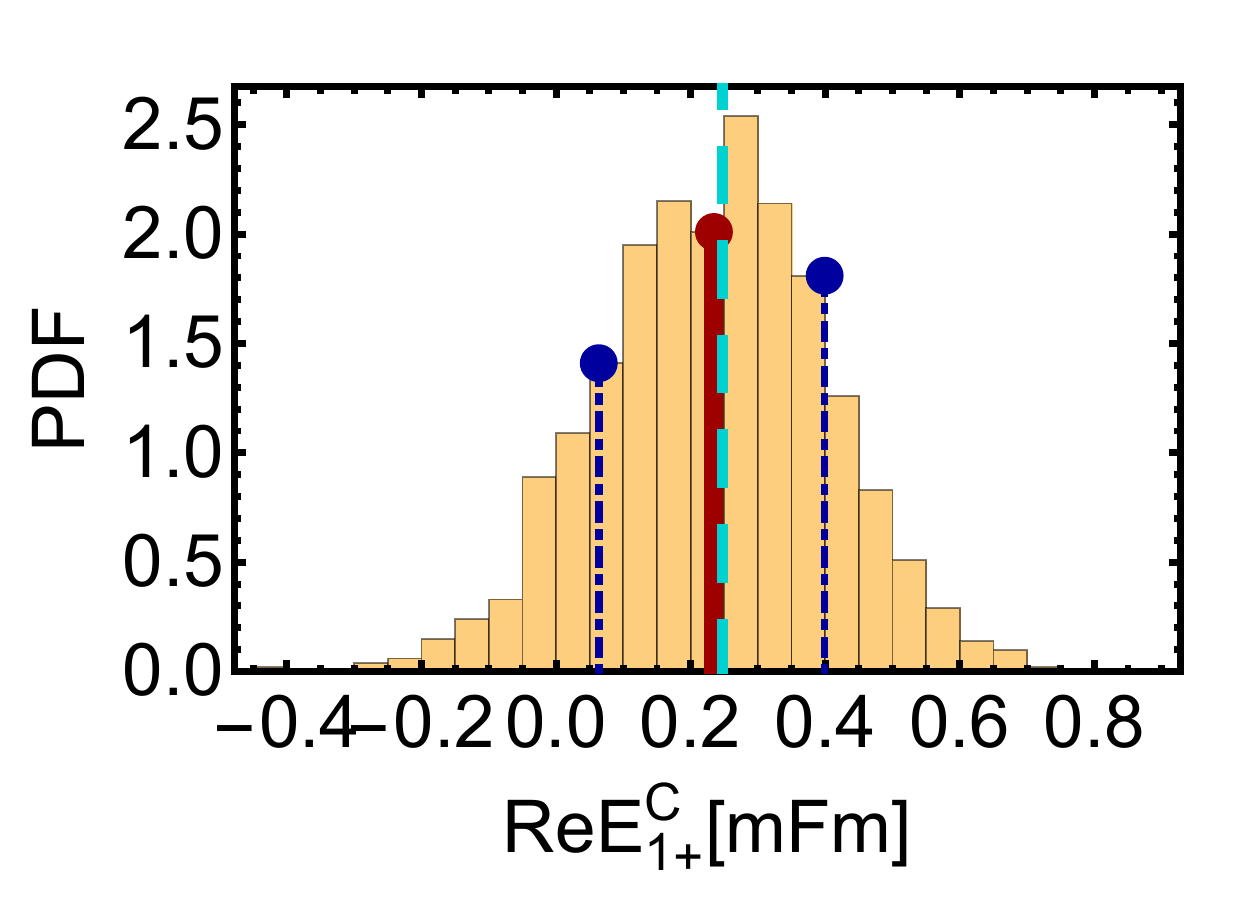}
 \end{overpic}
\begin{overpic}[width=0.325\textwidth]{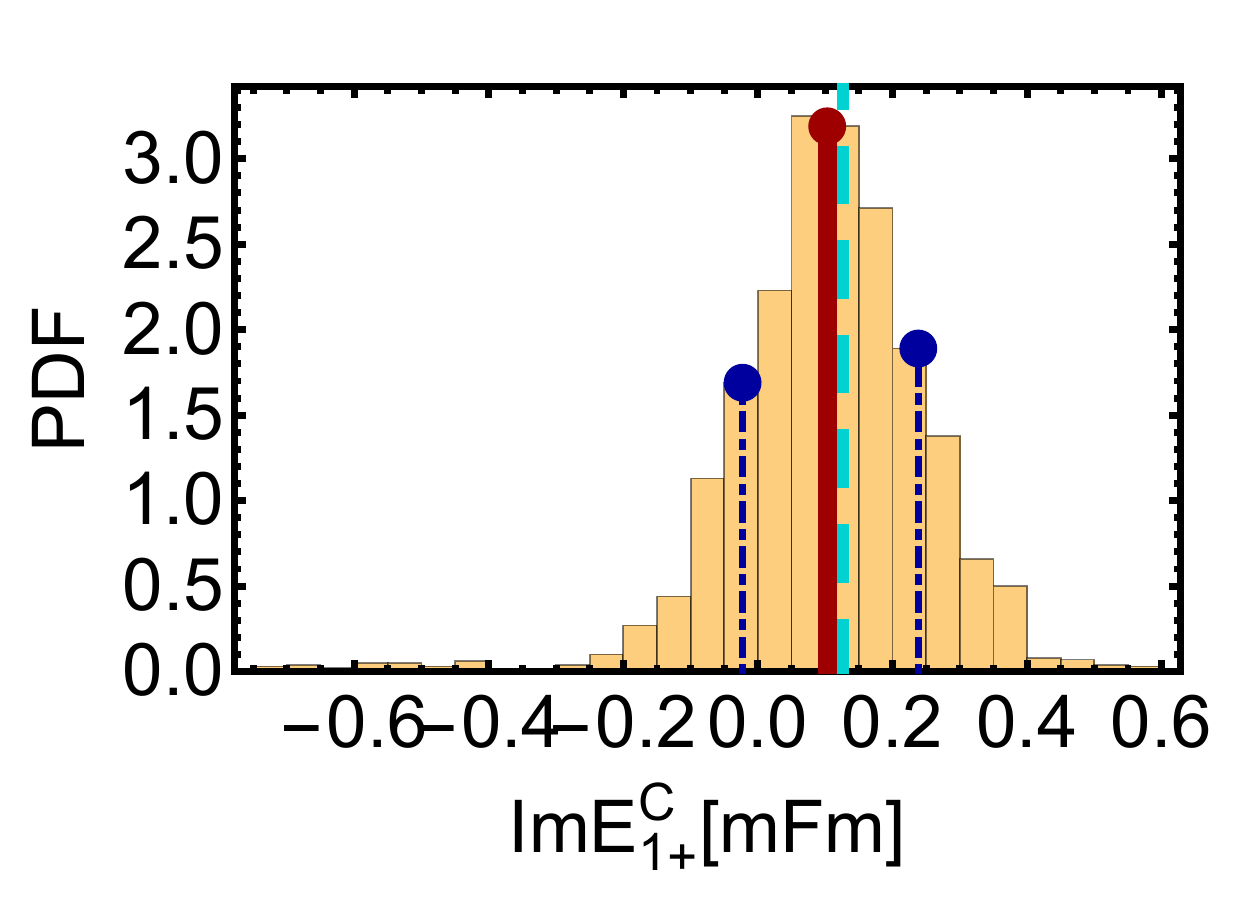}
 \end{overpic} \\
\begin{overpic}[width=0.325\textwidth]{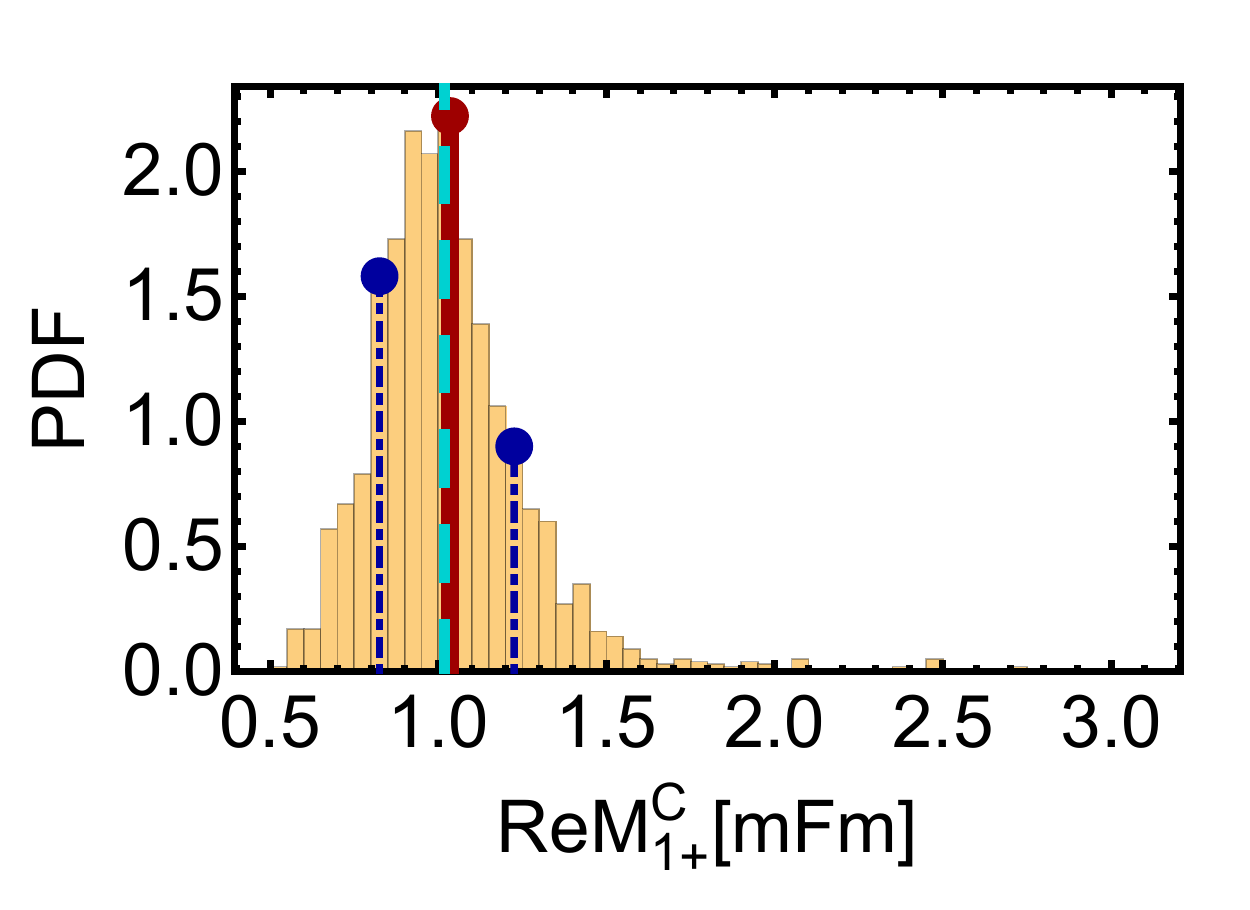}
 \end{overpic}
\begin{overpic}[width=0.325\textwidth]{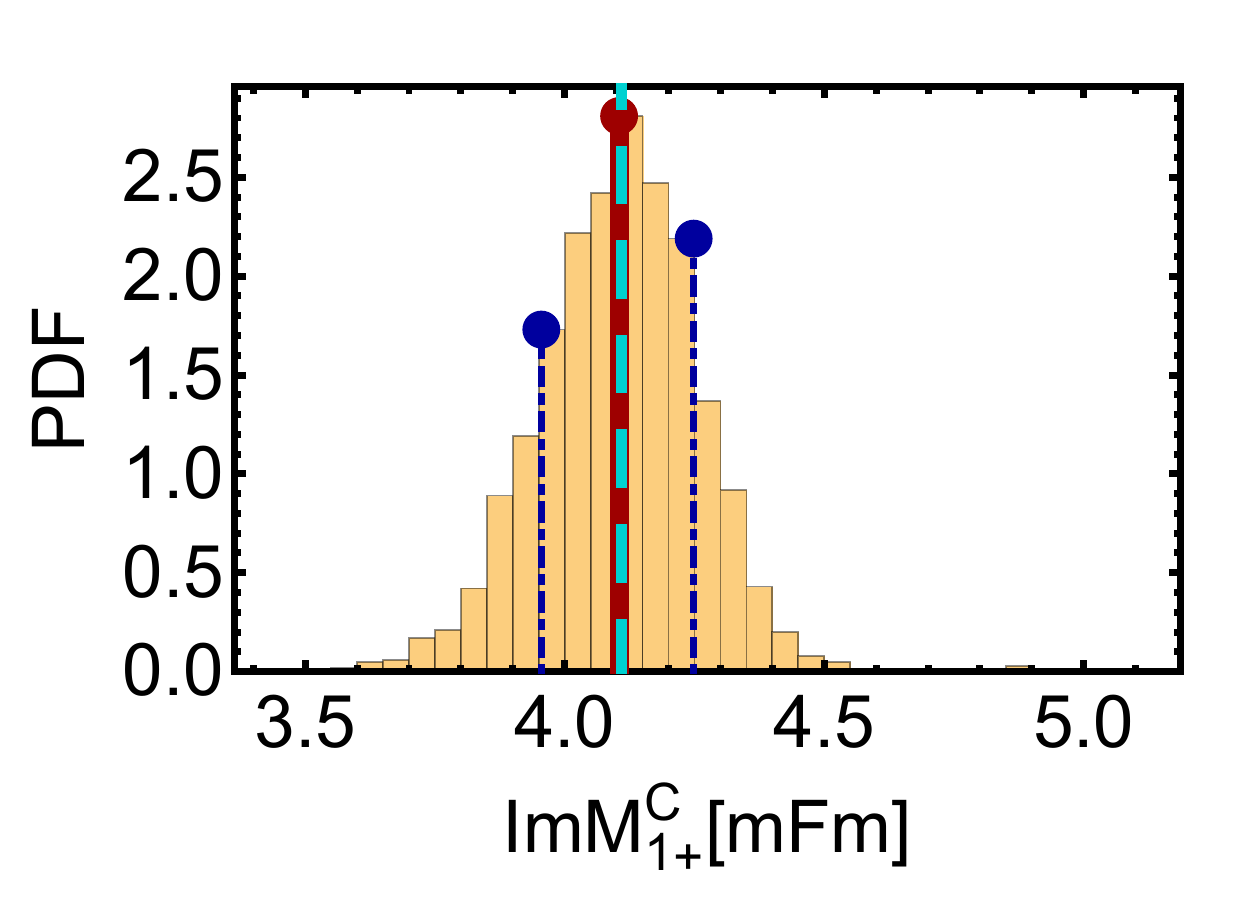}
 \end{overpic}
\begin{overpic}[width=0.325\textwidth]{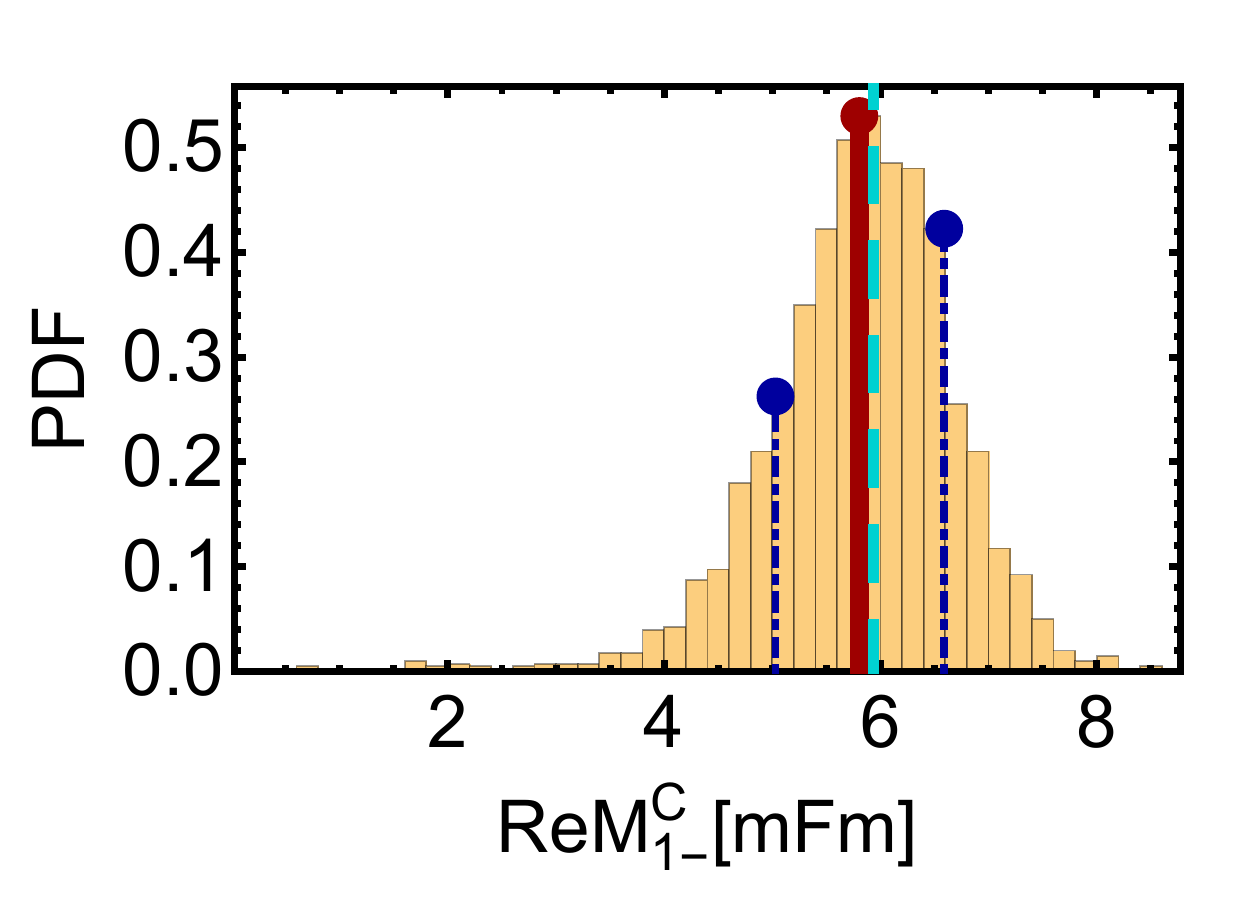}
 \end{overpic} \\
\begin{overpic}[width=0.325\textwidth]{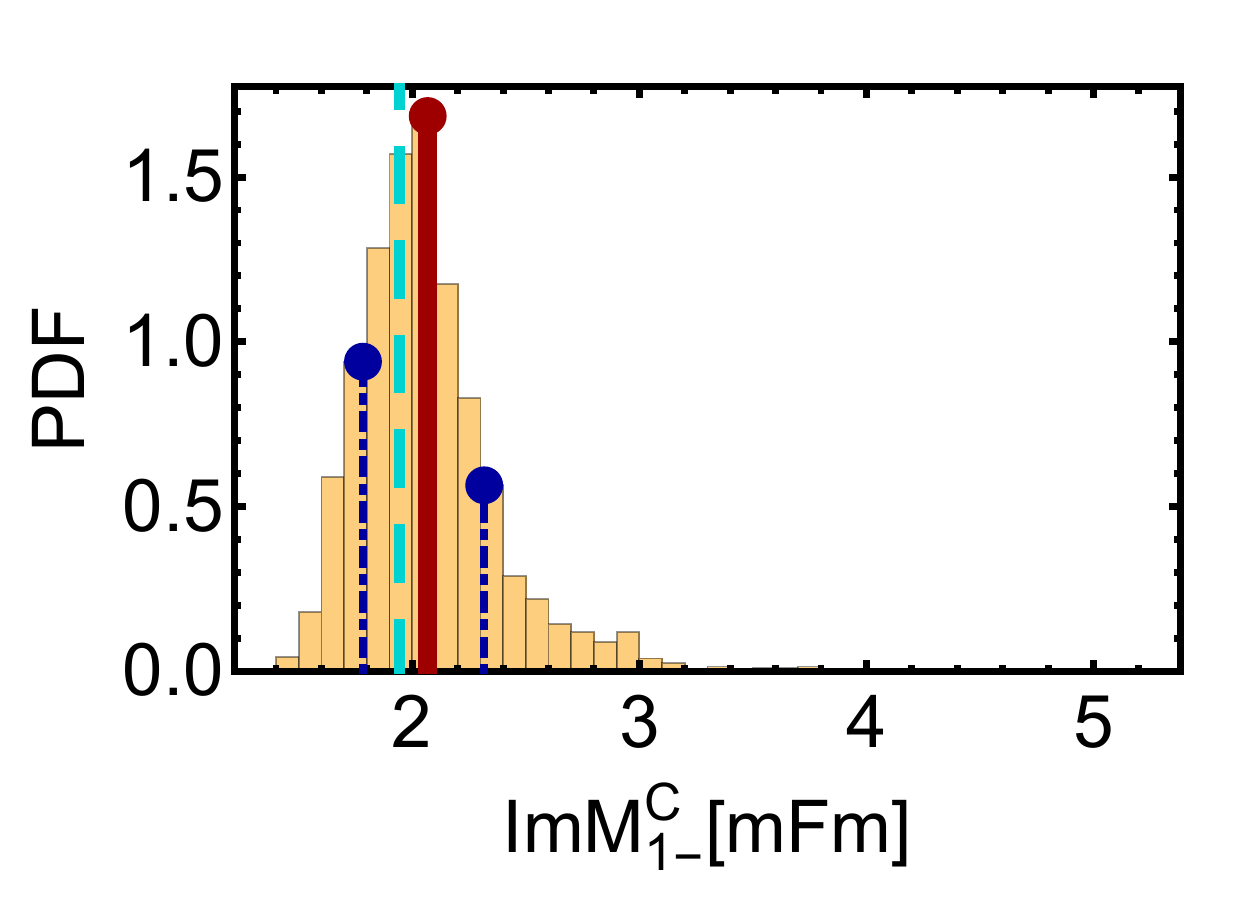}
 \end{overpic}
\begin{overpic}[width=0.325\textwidth]{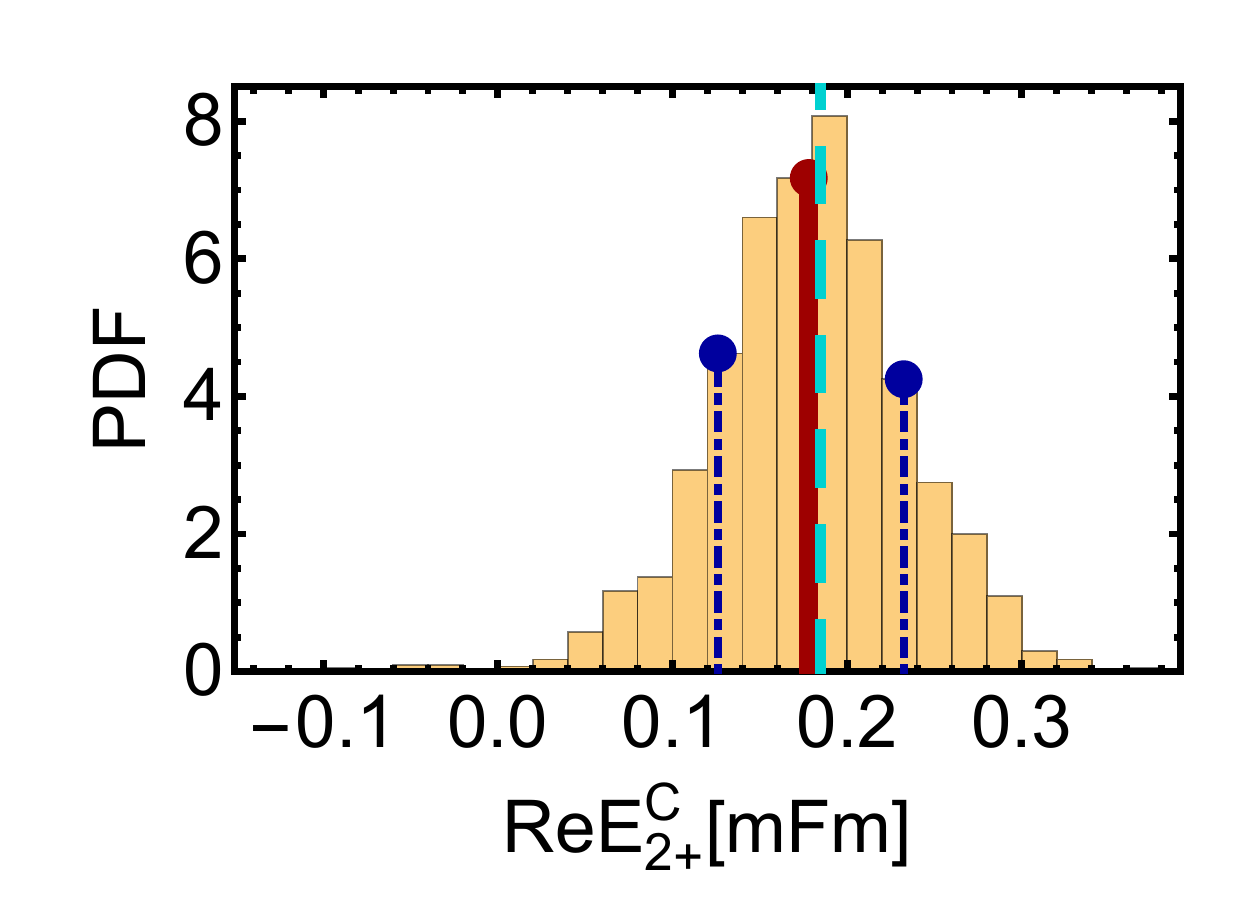}
 \end{overpic}
\begin{overpic}[width=0.325\textwidth]{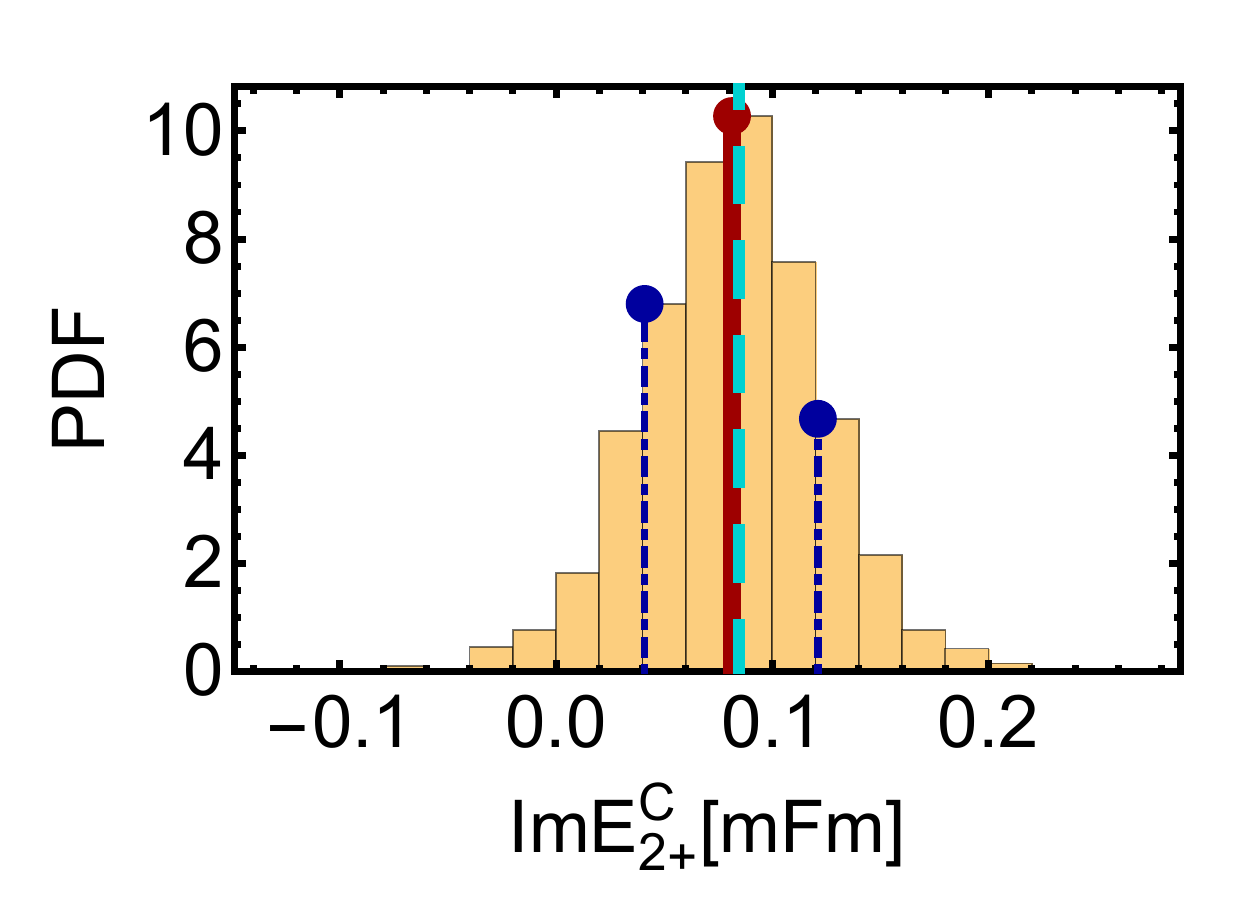}
 \end{overpic} \\
\begin{overpic}[width=0.325\textwidth]{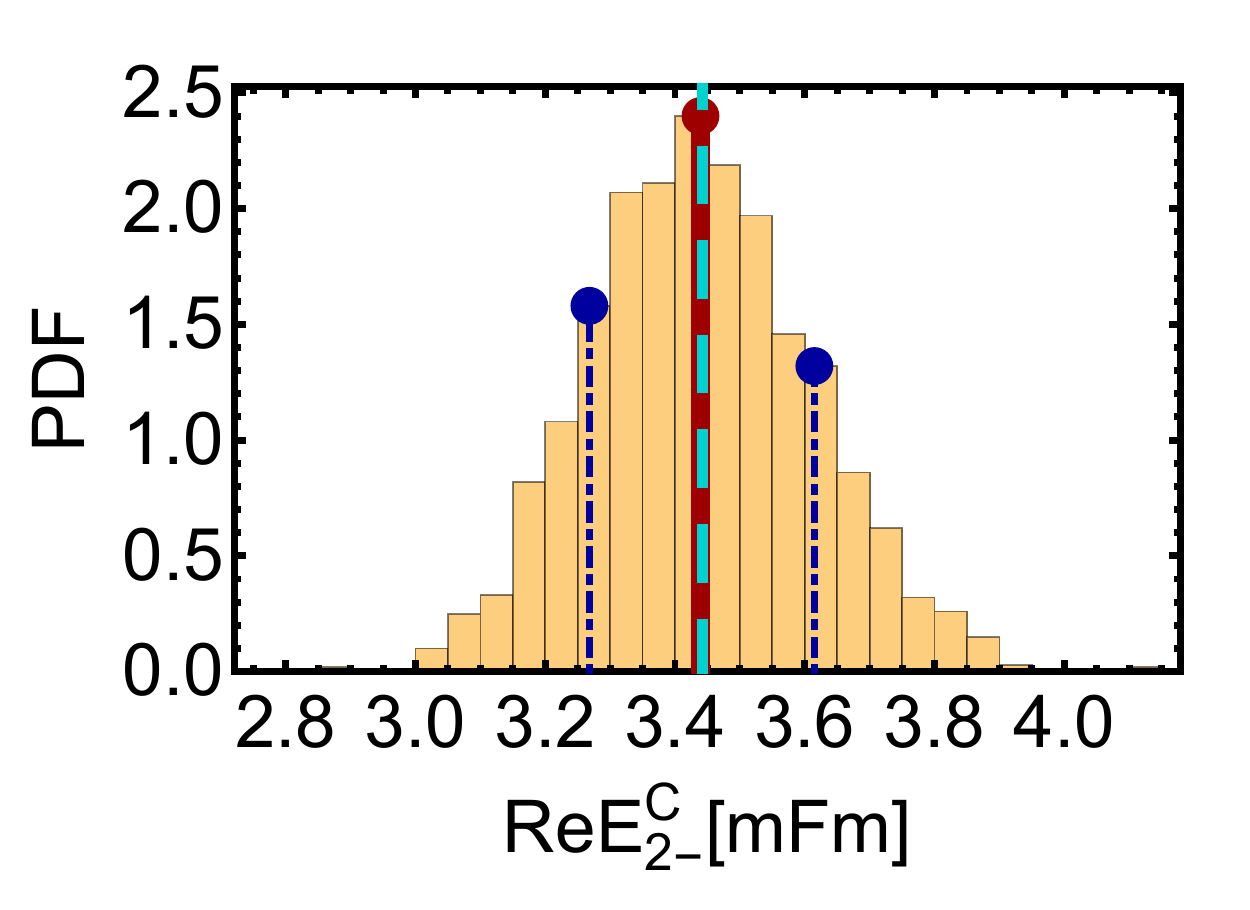}
 \end{overpic}
\begin{overpic}[width=0.325\textwidth]{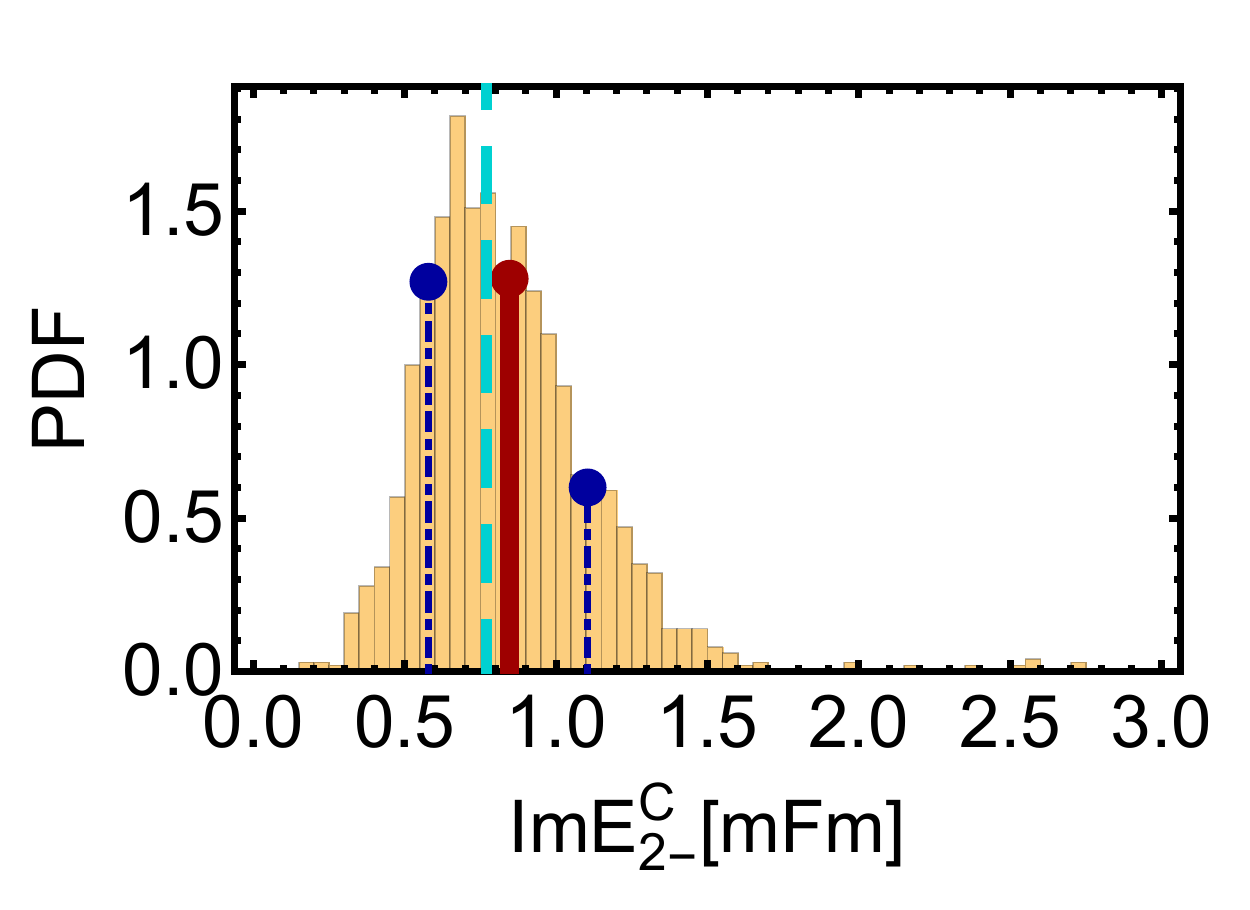}
 \end{overpic}
\begin{overpic}[width=0.325\textwidth]{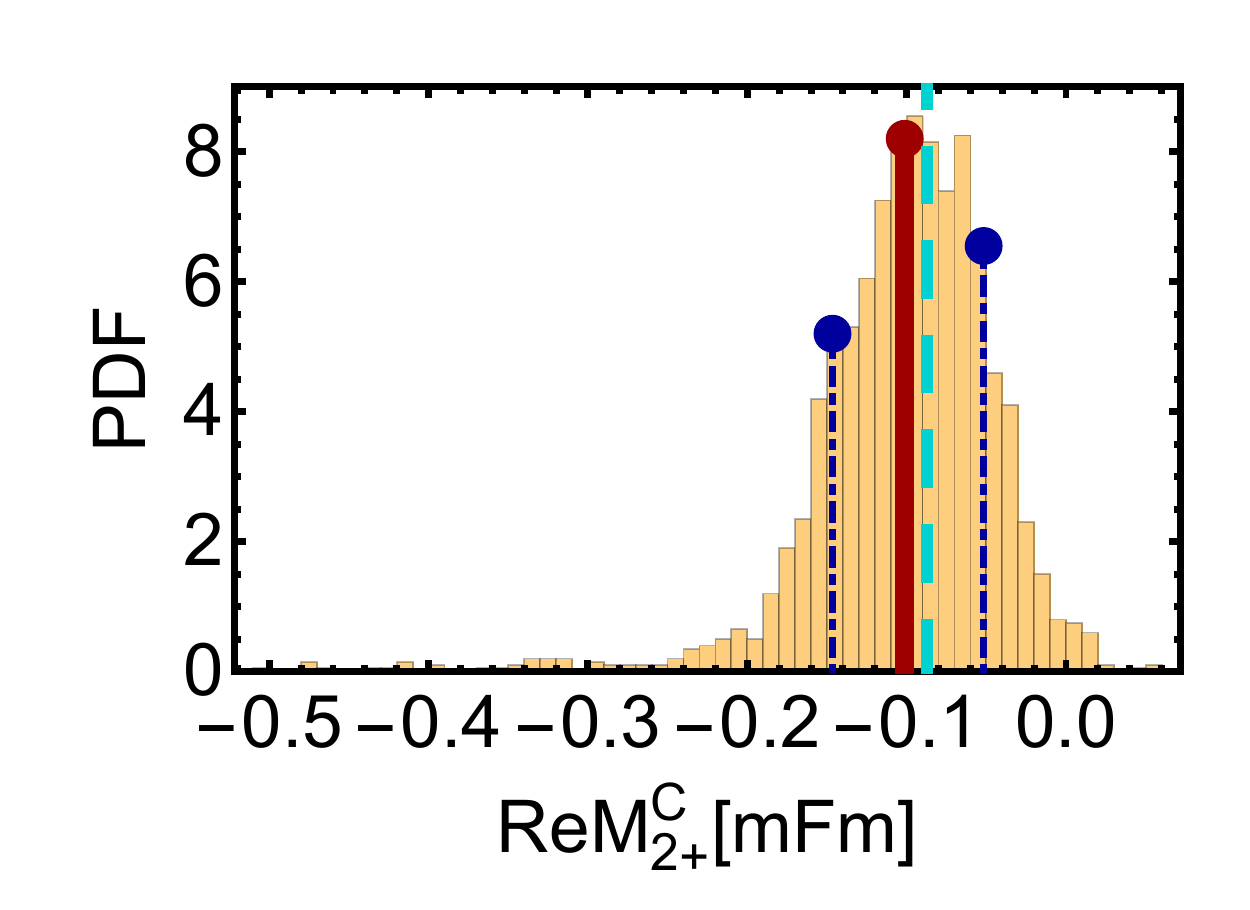}
 \end{overpic} \\
\begin{overpic}[width=0.325\textwidth]{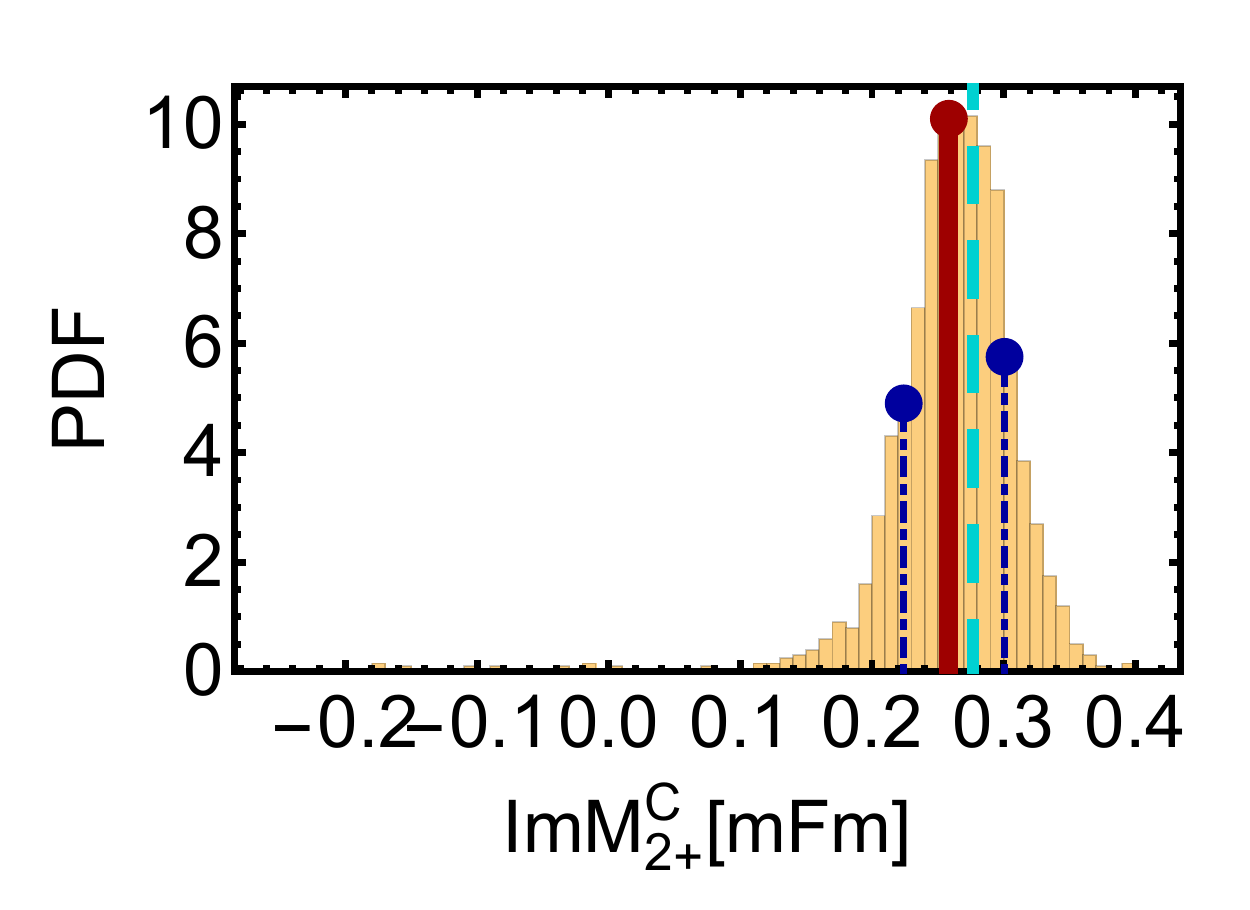}
 \end{overpic}
\begin{overpic}[width=0.325\textwidth]{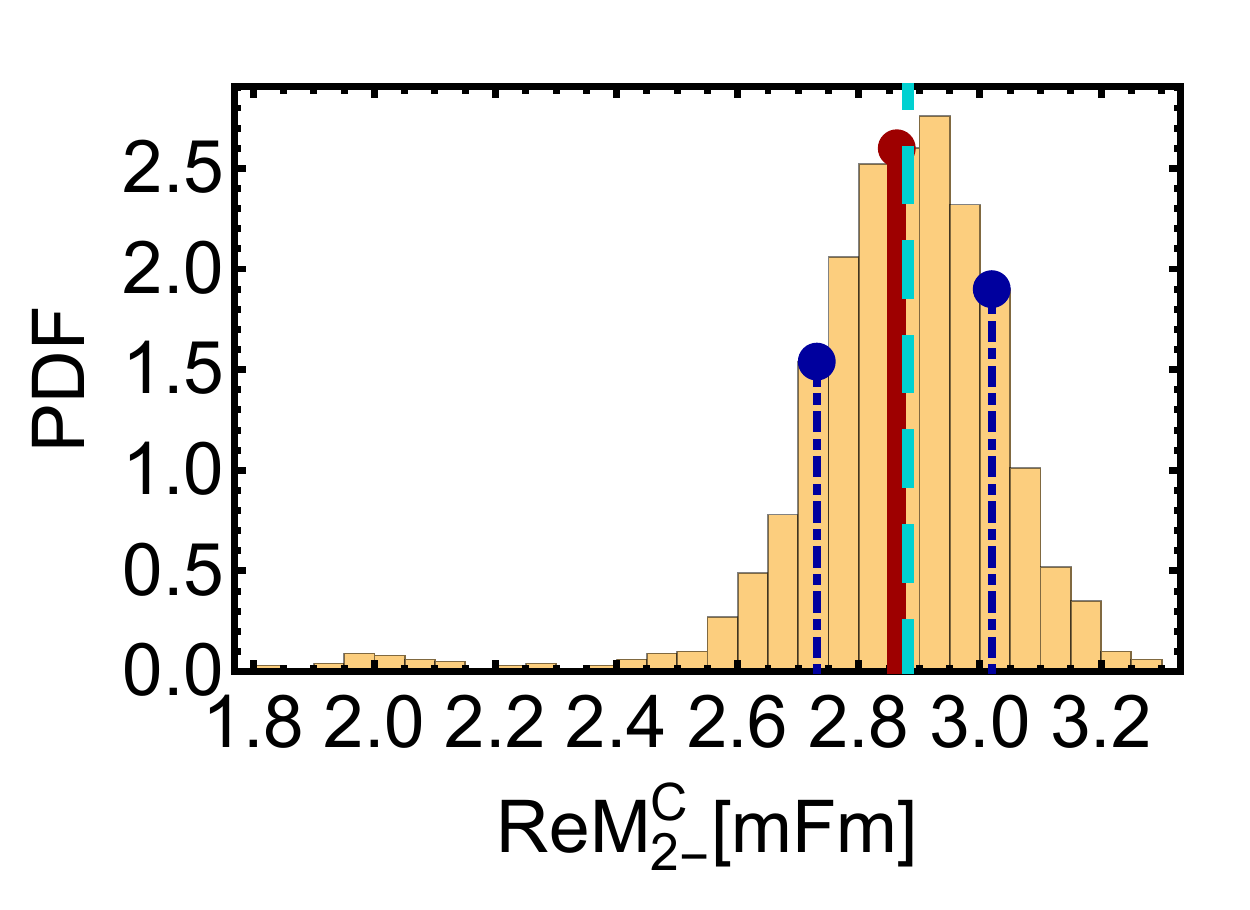}
 \end{overpic}
\begin{overpic}[width=0.325\textwidth]{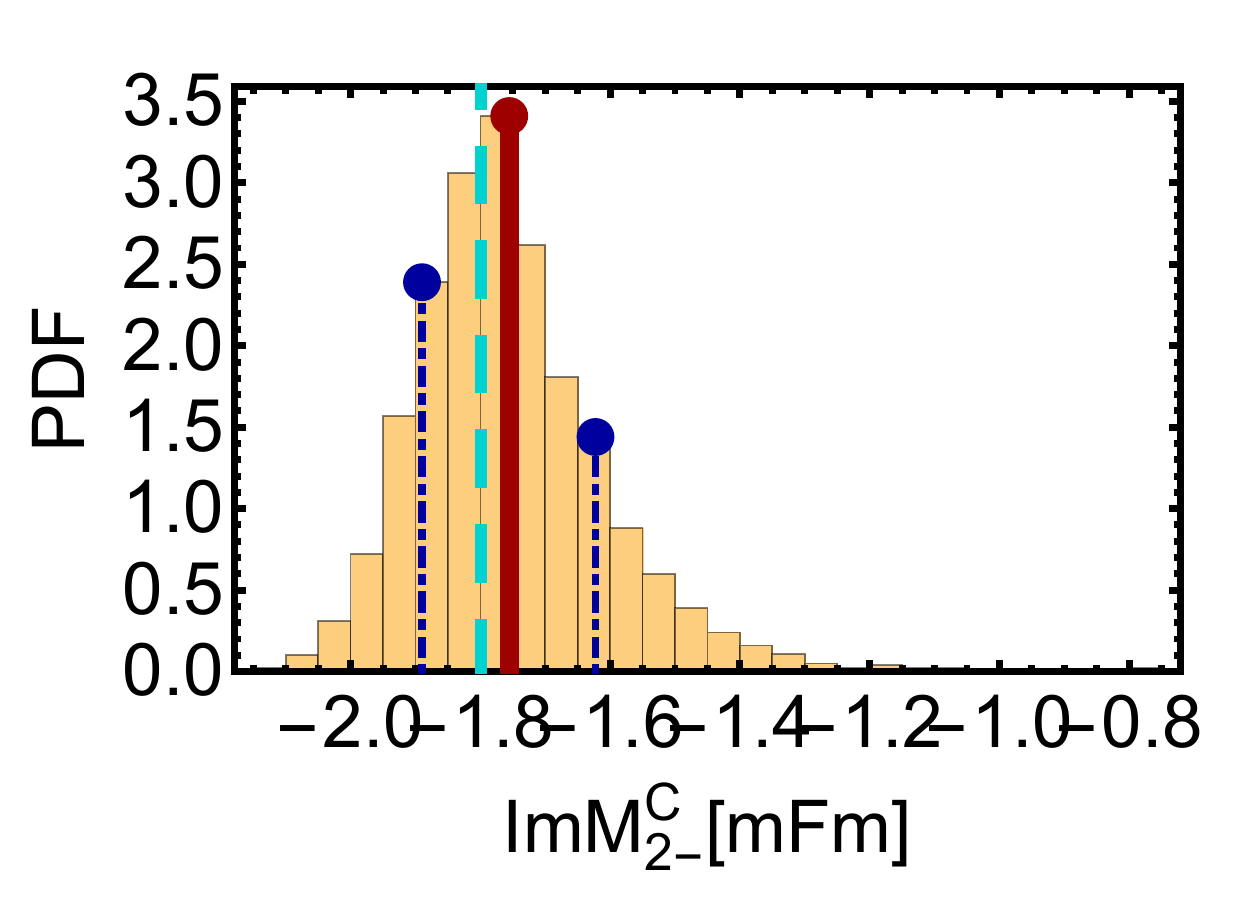}
 \end{overpic}
\caption[Bootstrap-distributions for multipole fit-parameters in an analysis of photoproduction data on the second resonance region. The first energy-bin, \newline $E_{\gamma }\text{ = 683.5 MeV}$, is shown.]{The histograms show bootstrap-distributions for the real- and imaginary parts of phase-constrained $S$-, $P$- and $D$-wave multipoles, for a TPWA bootstrap-analysis of photoproduction data in the second resonance region (see section \ref{subsec:2ndResRegionDataFits}). The first energy-bin, $E_{\gamma }\text{ = 683.5 MeV}$, is shown. An ensemble of $B=2000$ bootstrap-replicates has been the basis of these results. \newline
The distributions have been normalized to $1$ via use of the object \textit{HistogramDistribution} in MATHEMATICA \cite{Mathematica8,Mathematica11,MathematicaLanguage,MathematicaBonnLicense}. Thus, $y$-axes are labelled as \textit{PDF}. The mean of each distribution is shown as a red solid line, while the $0.16$- and $0.84$-quantiles are indicated by blue dash-dotted lines. The global minimum of the fit to the original data is plotted as a cyan-colored dashed horizontal line.}
\label{fig:BootstrapHistos2ndResRegionEnergy1}
\end{figure}

\clearpage

\begin{figure}[h]
\begin{overpic}[width=0.325\textwidth]{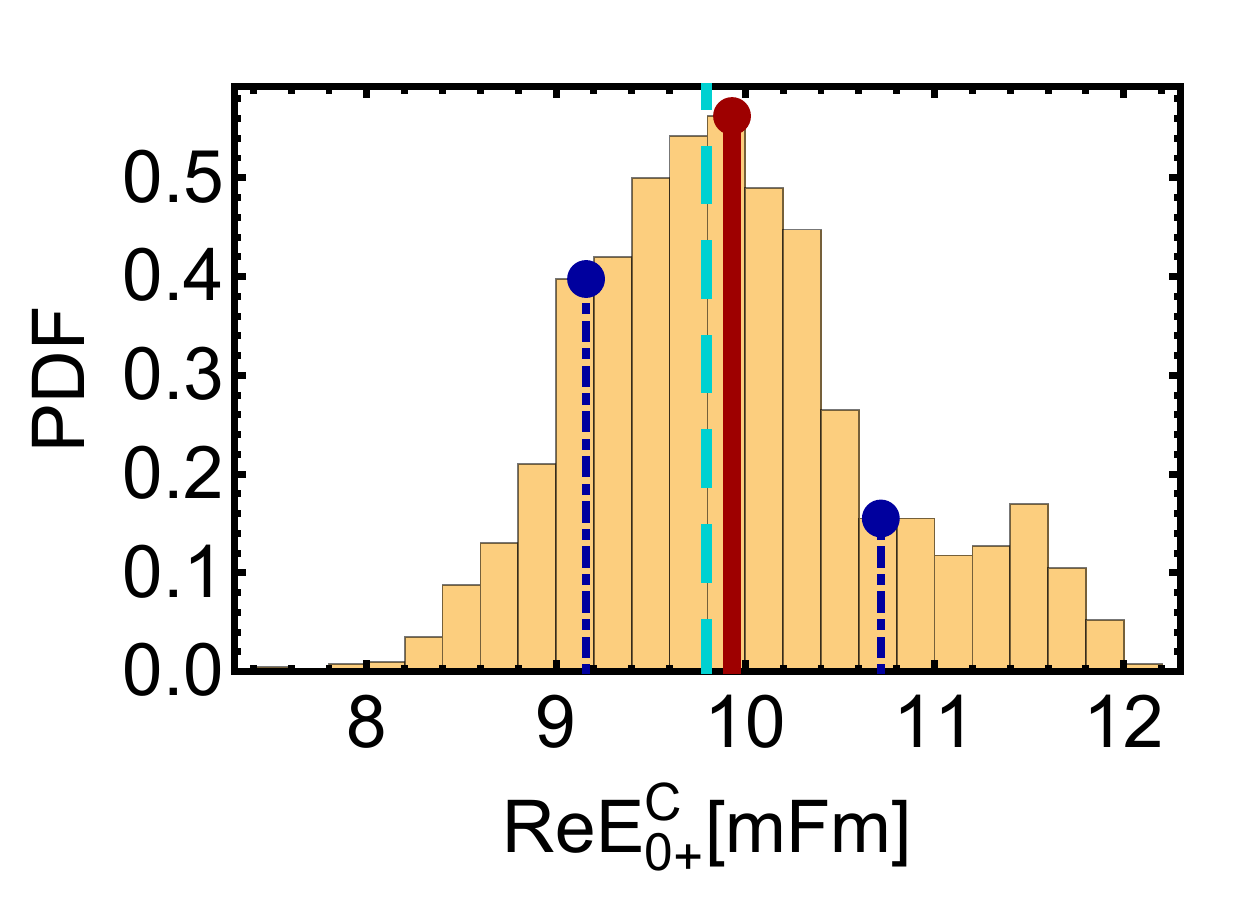}
 \end{overpic}
\begin{overpic}[width=0.325\textwidth]{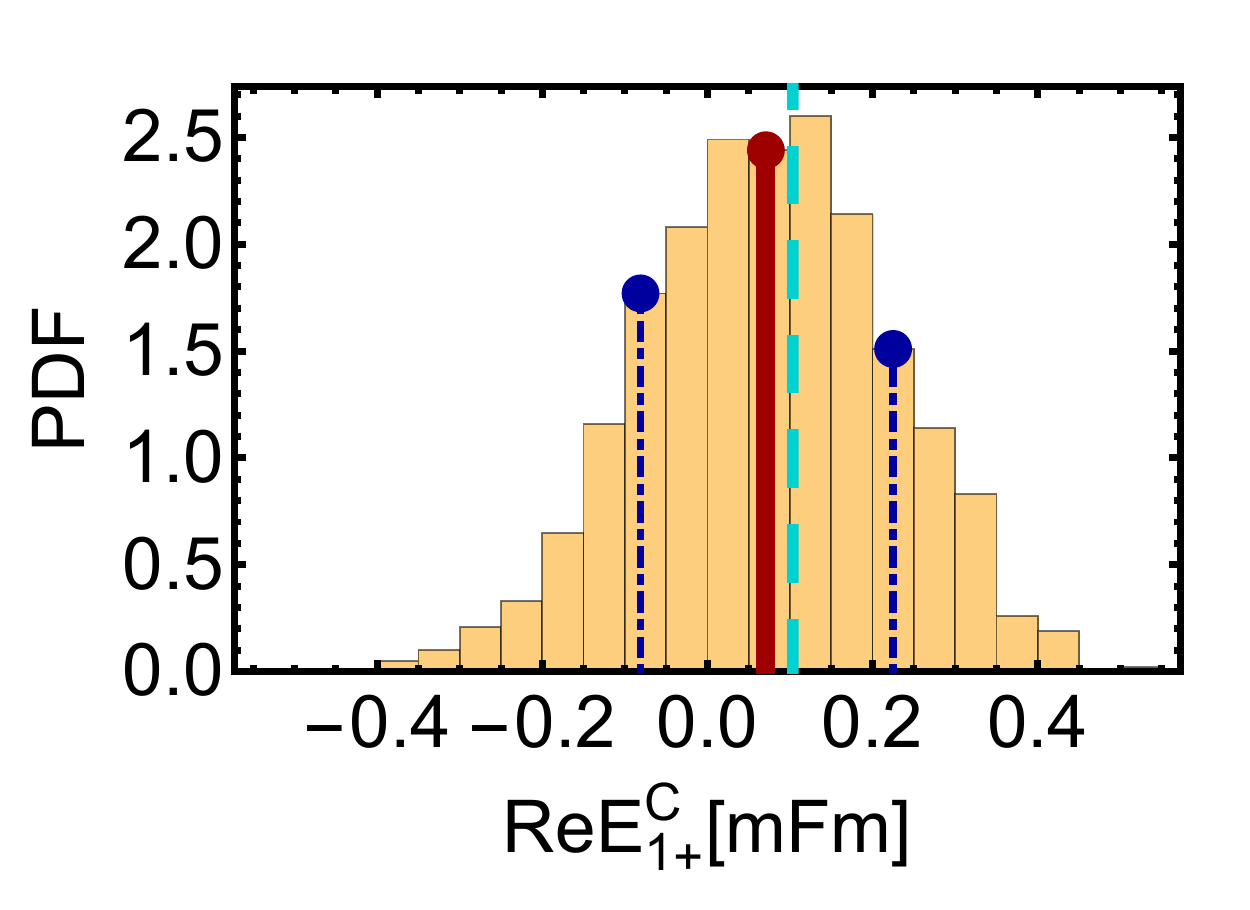}
 \end{overpic}
\begin{overpic}[width=0.325\textwidth]{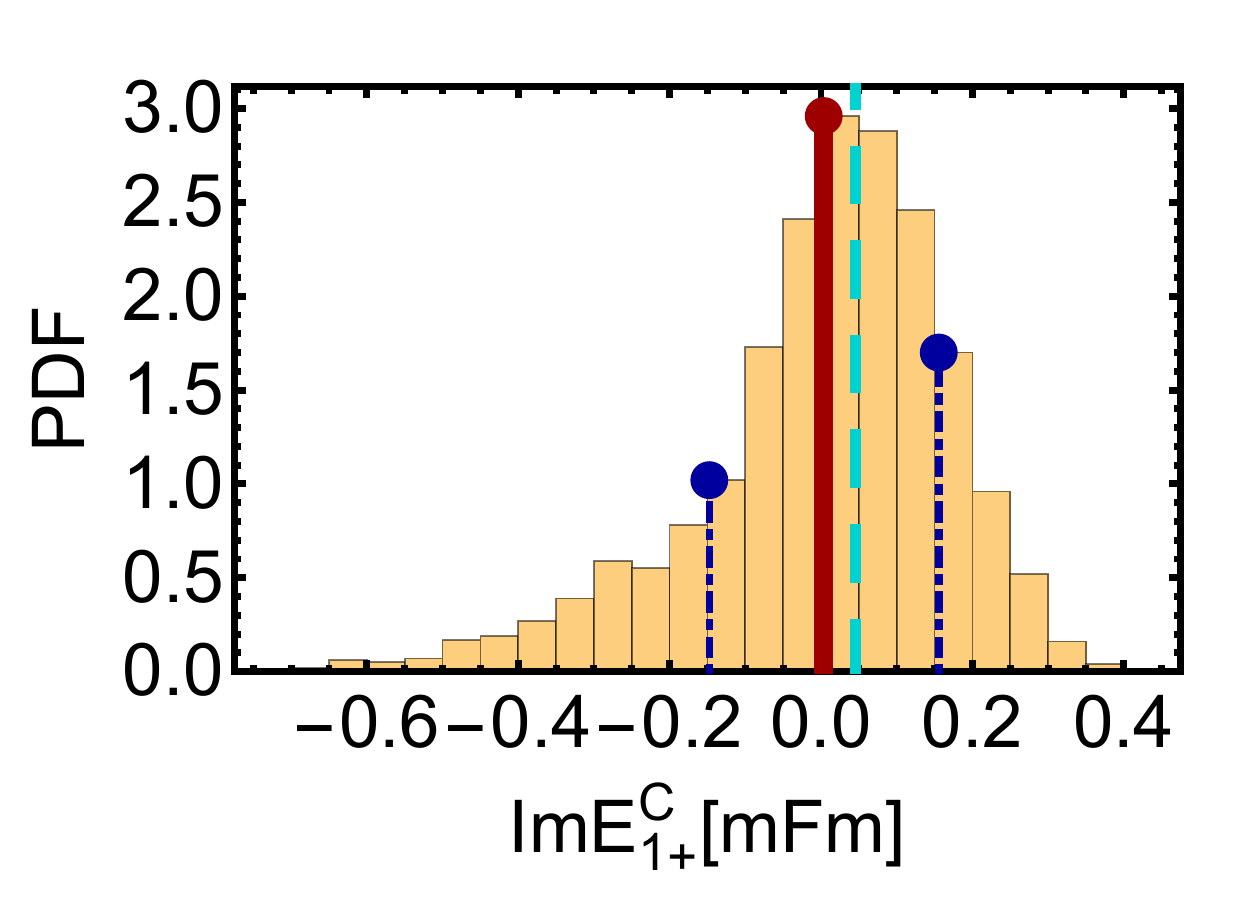}
 \end{overpic} \\
\begin{overpic}[width=0.325\textwidth]{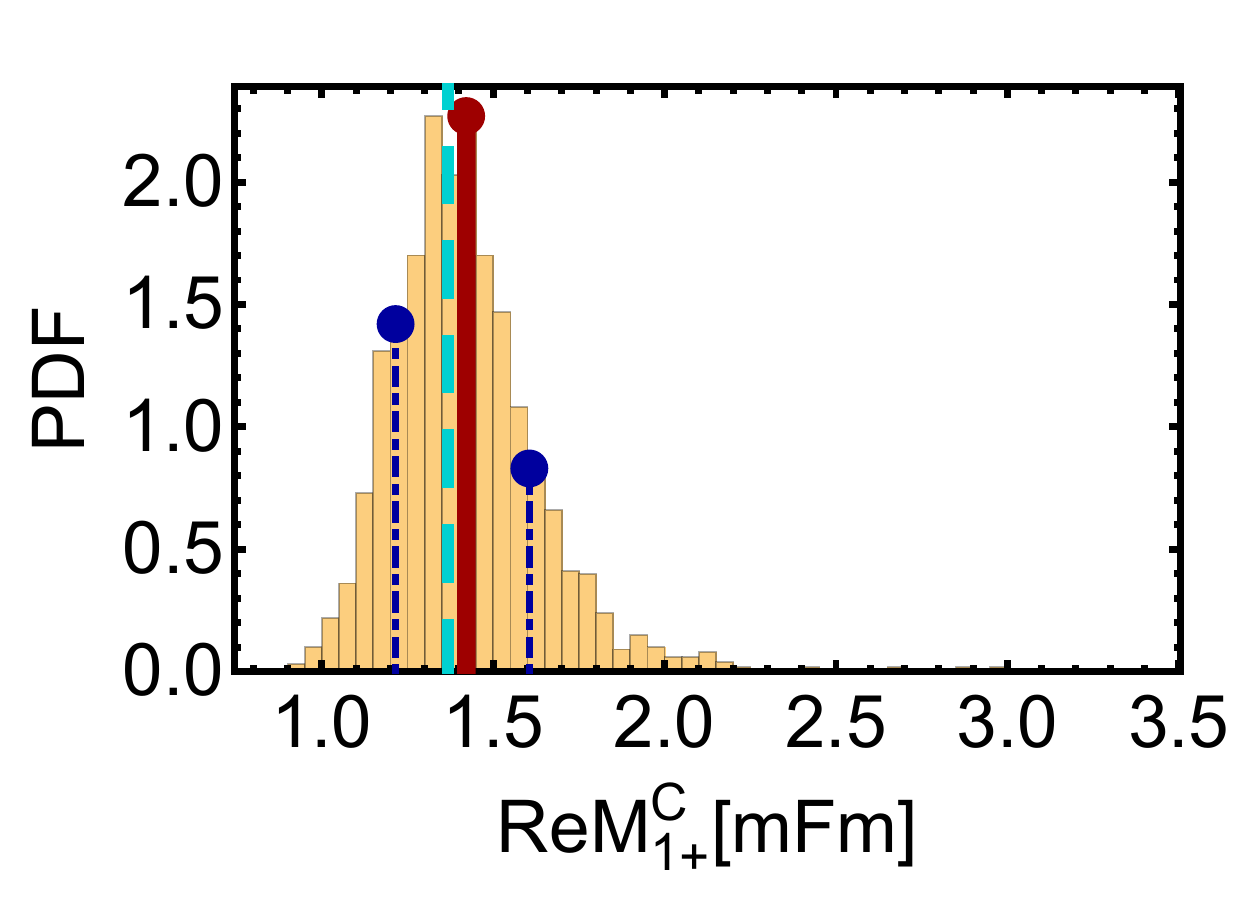}
 \end{overpic}
\begin{overpic}[width=0.325\textwidth]{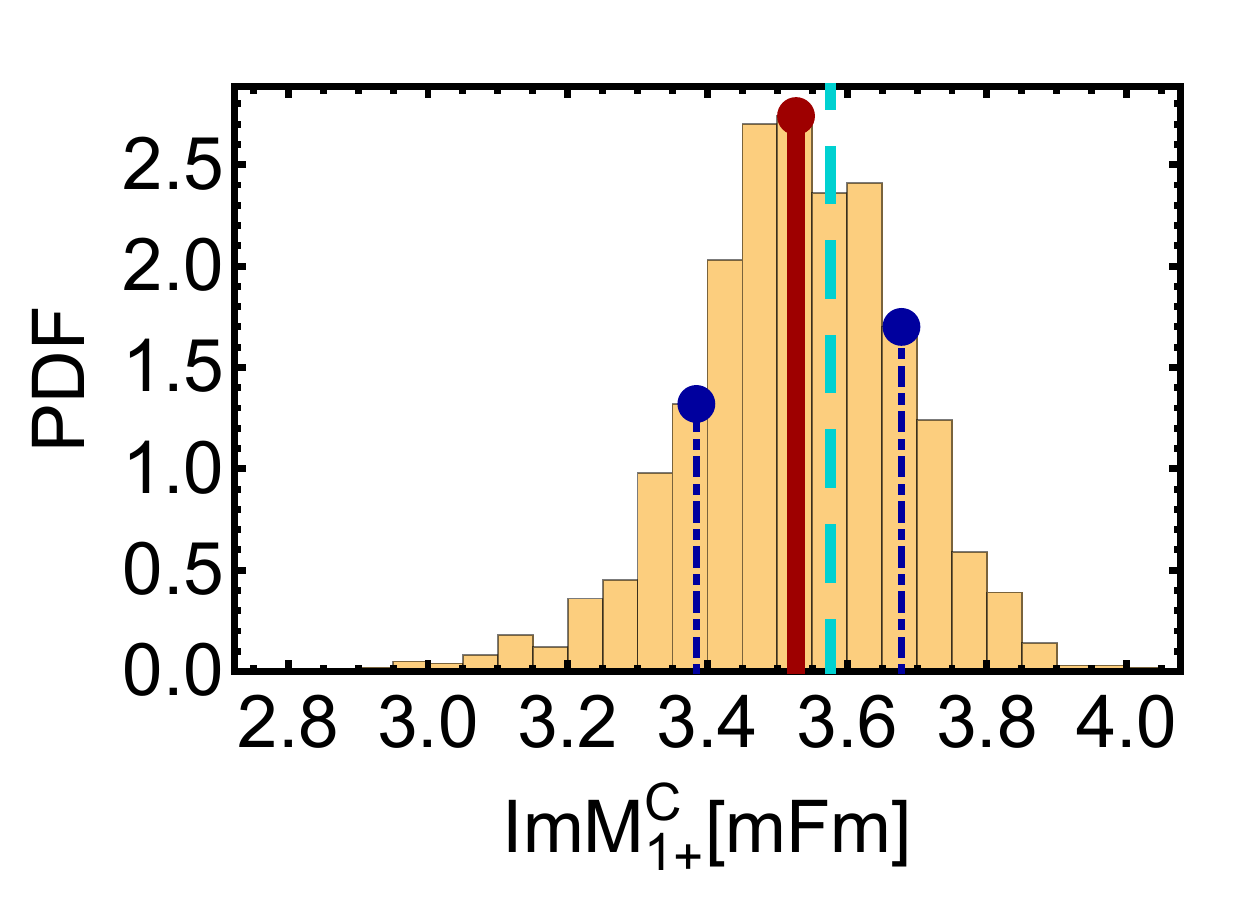}
 \end{overpic}
\begin{overpic}[width=0.325\textwidth]{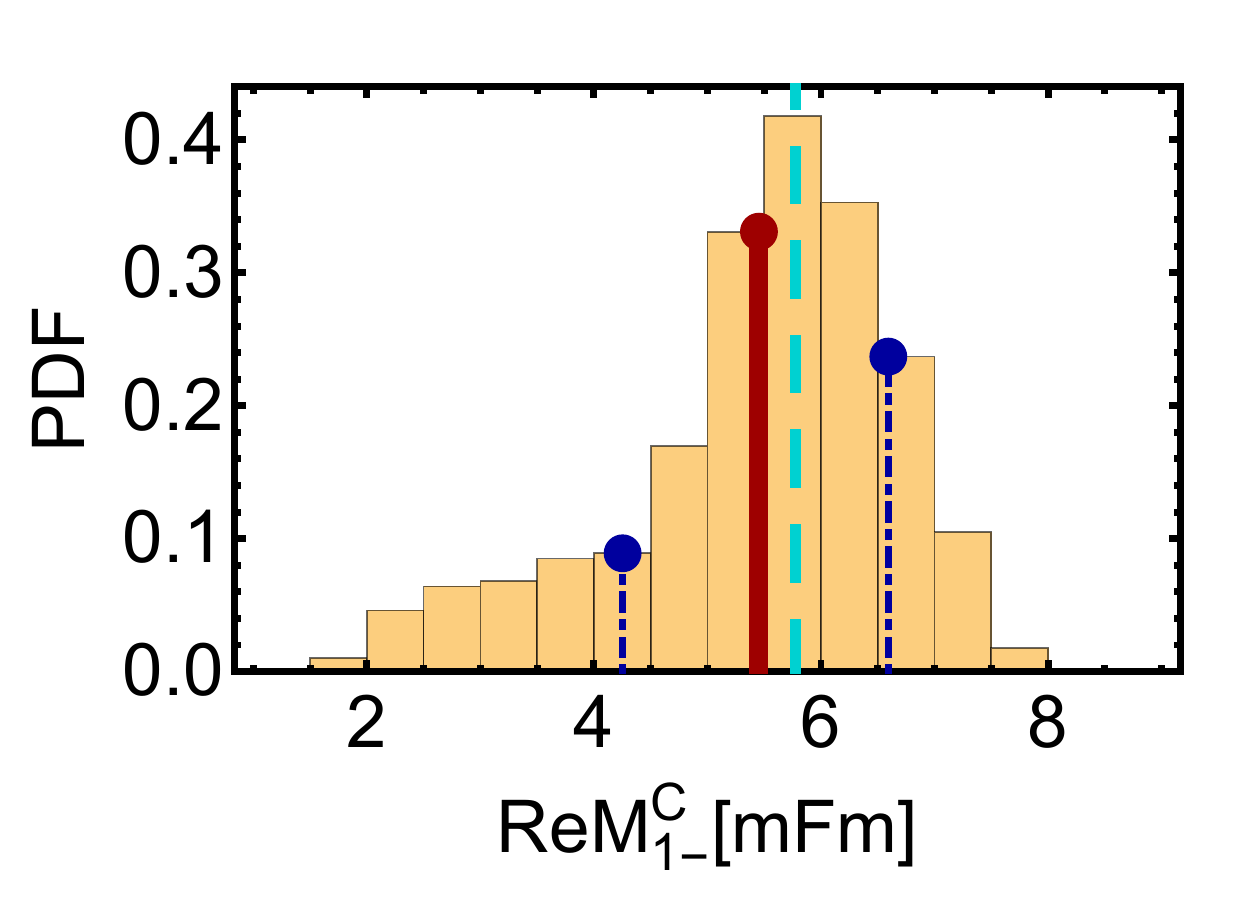}
 \end{overpic} \\
\begin{overpic}[width=0.325\textwidth]{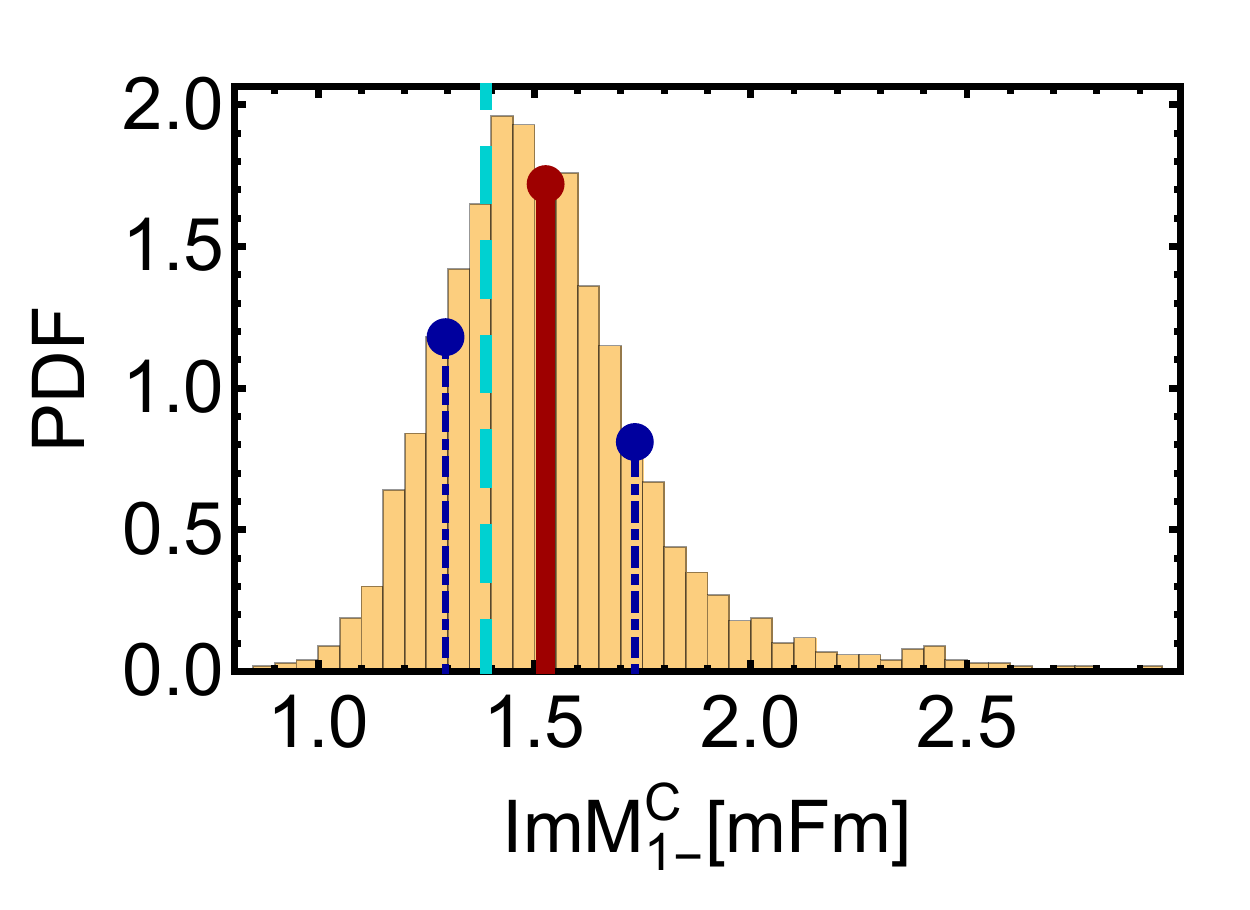}
 \end{overpic}
\begin{overpic}[width=0.325\textwidth]{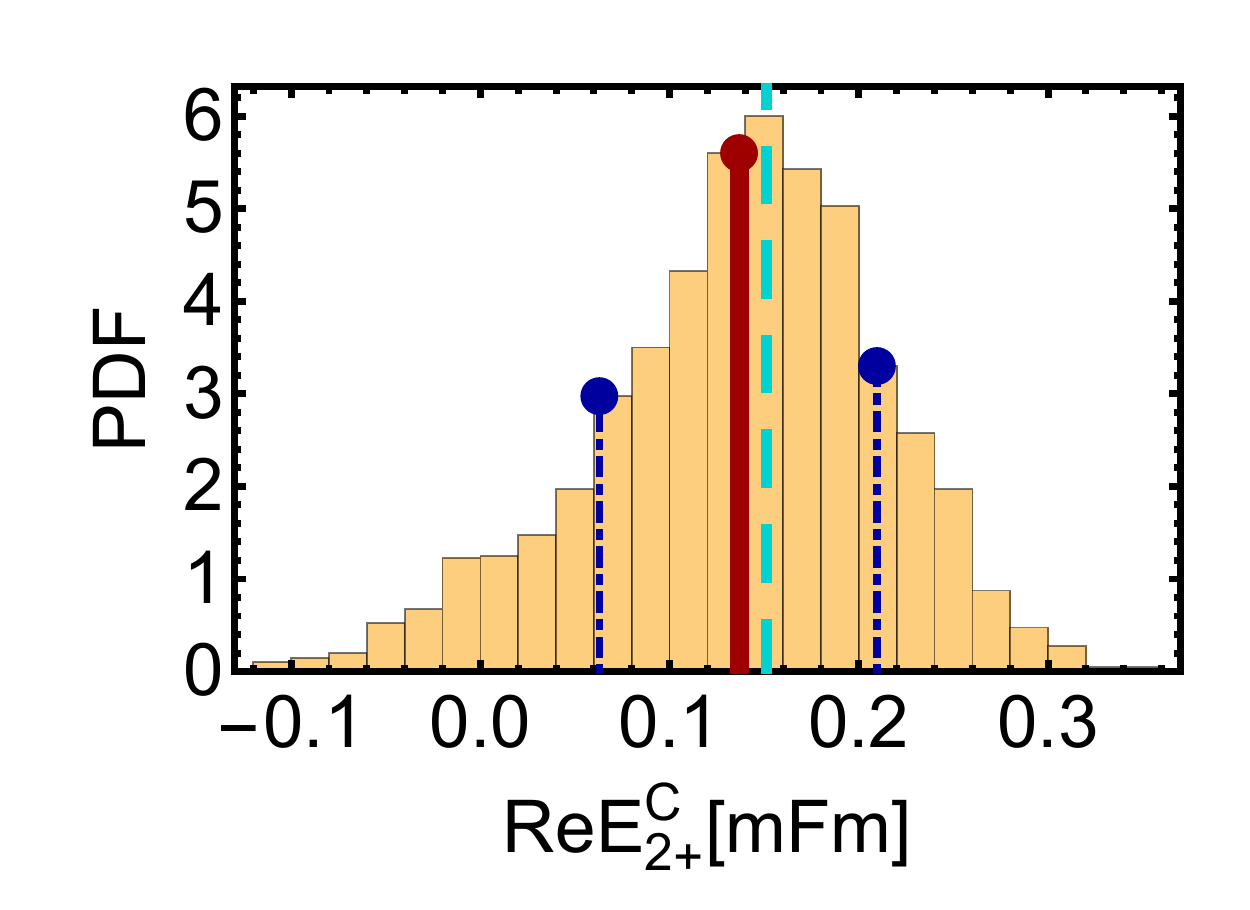}
 \end{overpic}
\begin{overpic}[width=0.325\textwidth]{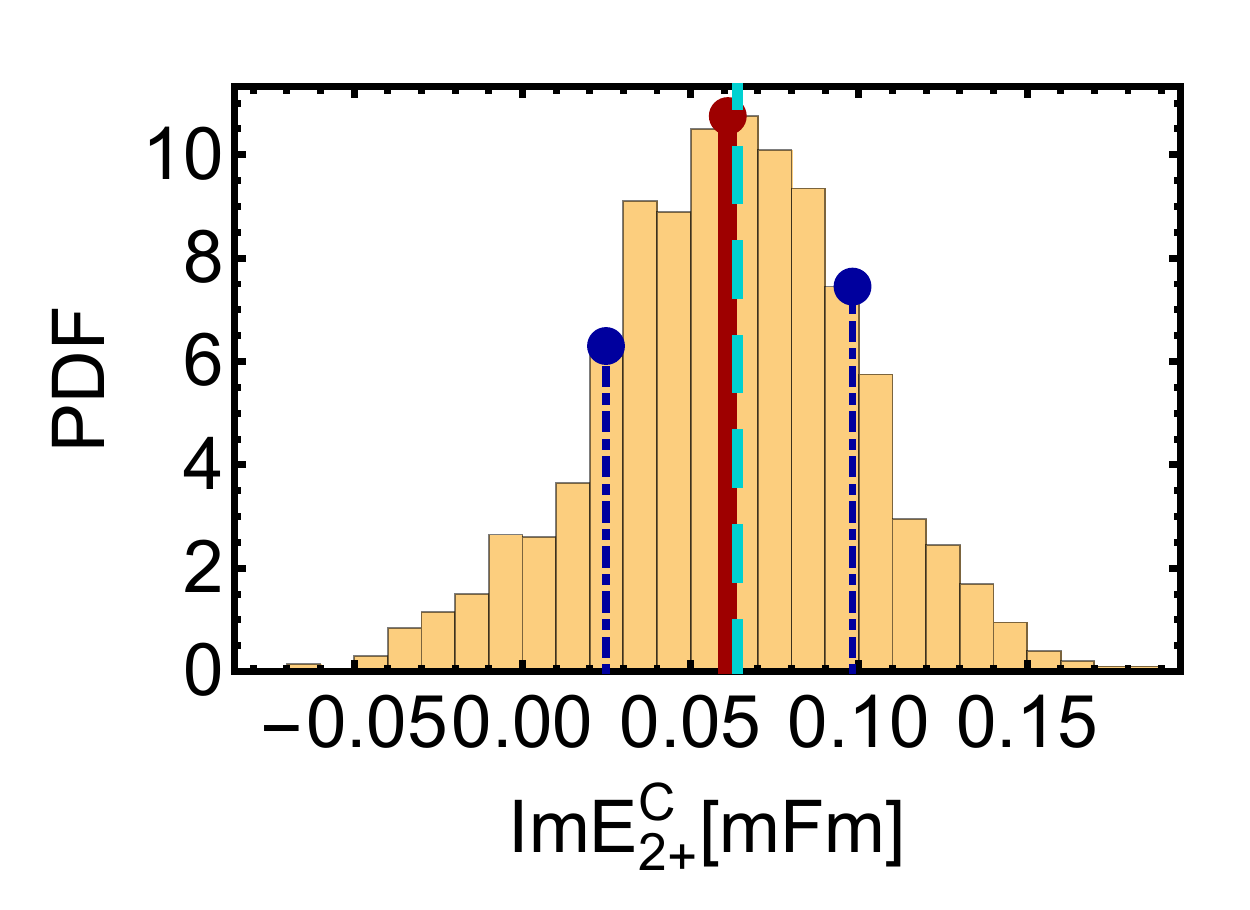}
 \end{overpic} \\
\begin{overpic}[width=0.325\textwidth]{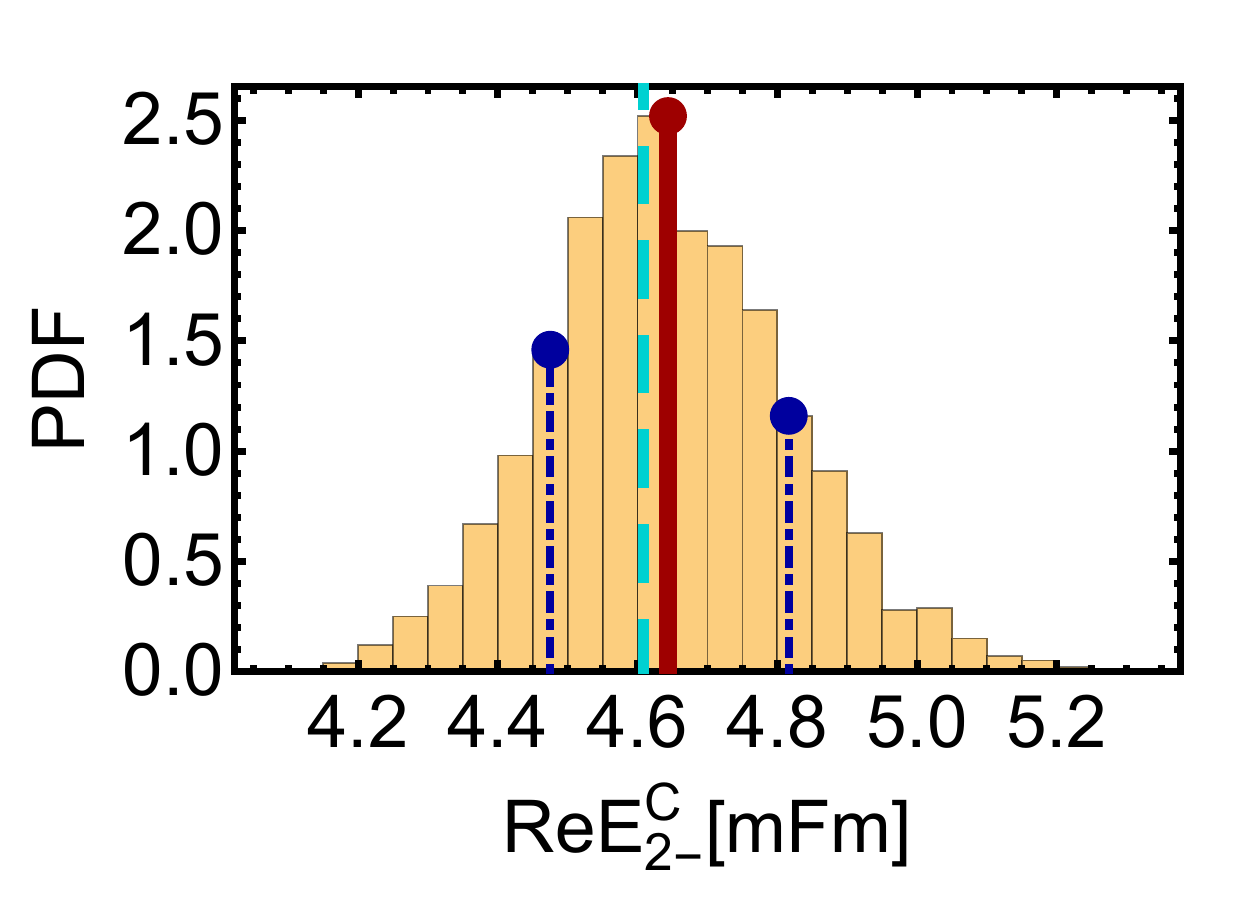}
 \end{overpic}
\begin{overpic}[width=0.325\textwidth]{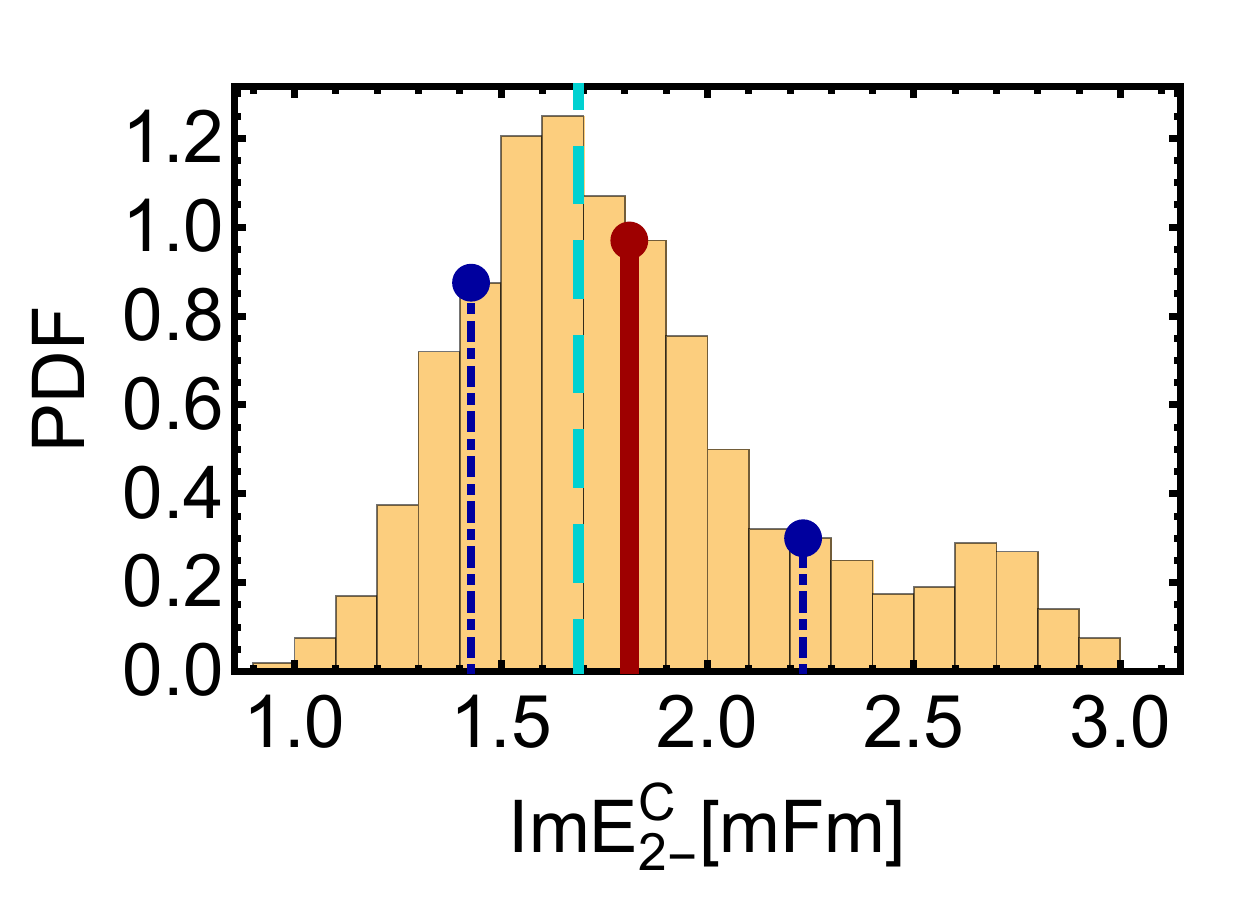}
 \end{overpic}
\begin{overpic}[width=0.325\textwidth]{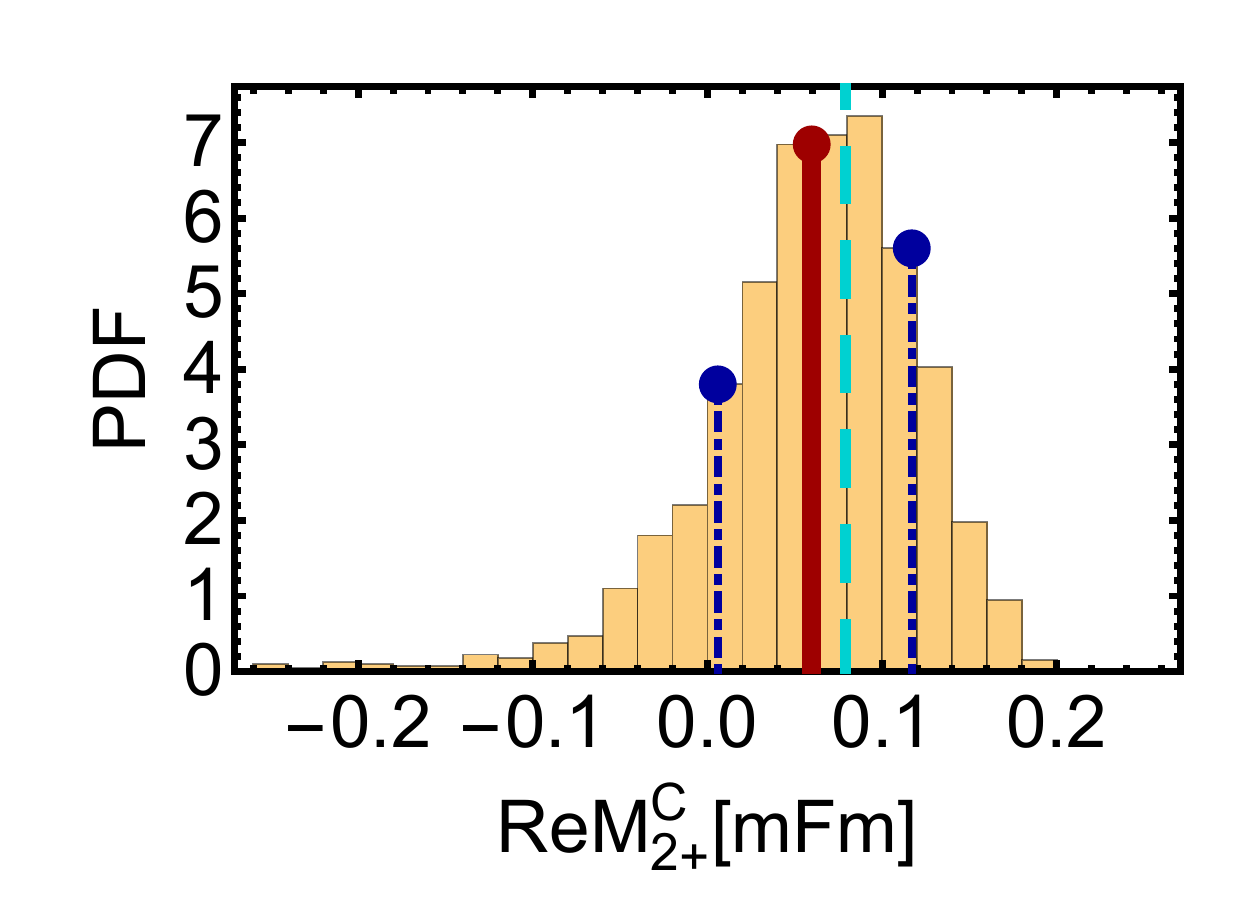}
 \end{overpic} \\
\begin{overpic}[width=0.325\textwidth]{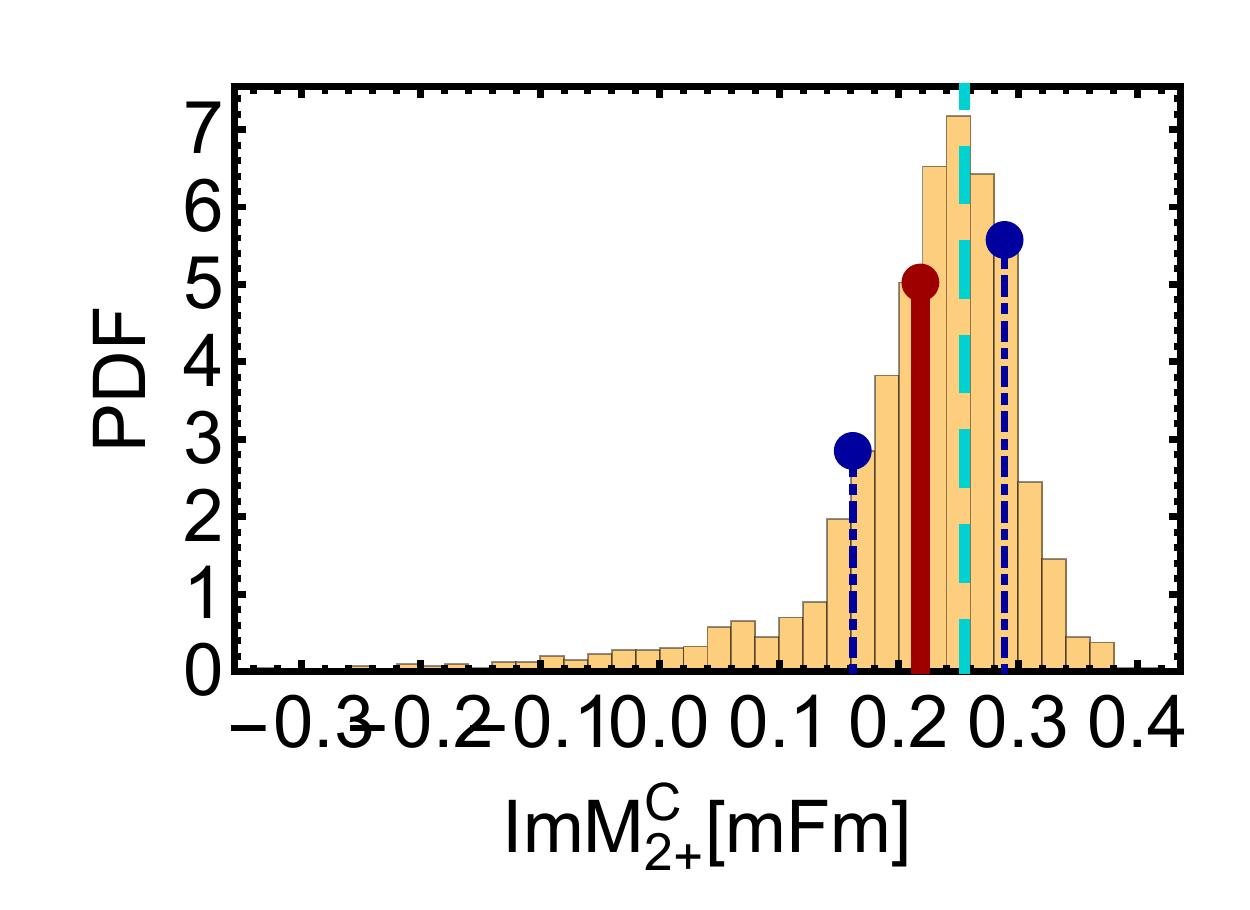}
 \end{overpic}
\begin{overpic}[width=0.325\textwidth]{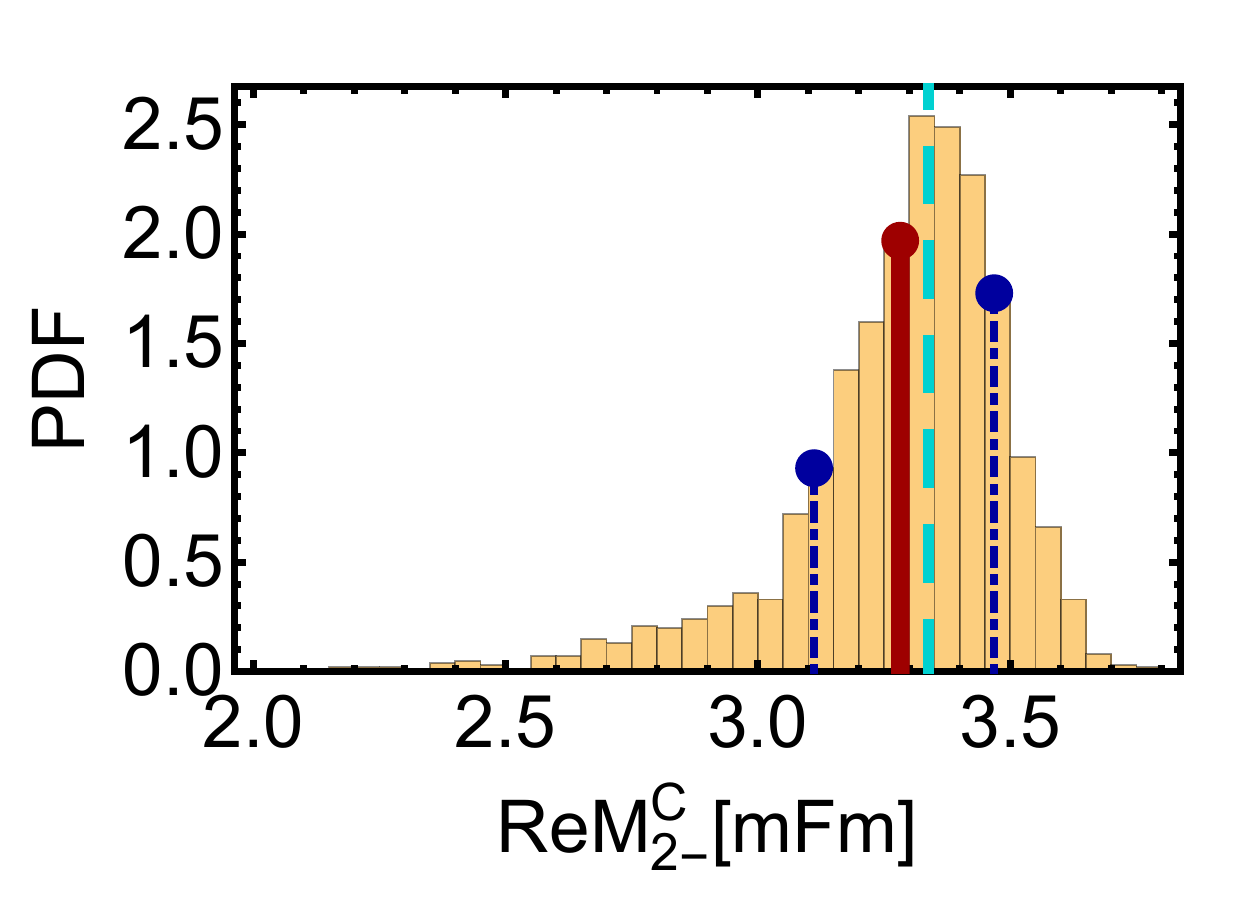}
 \end{overpic}
\begin{overpic}[width=0.325\textwidth]{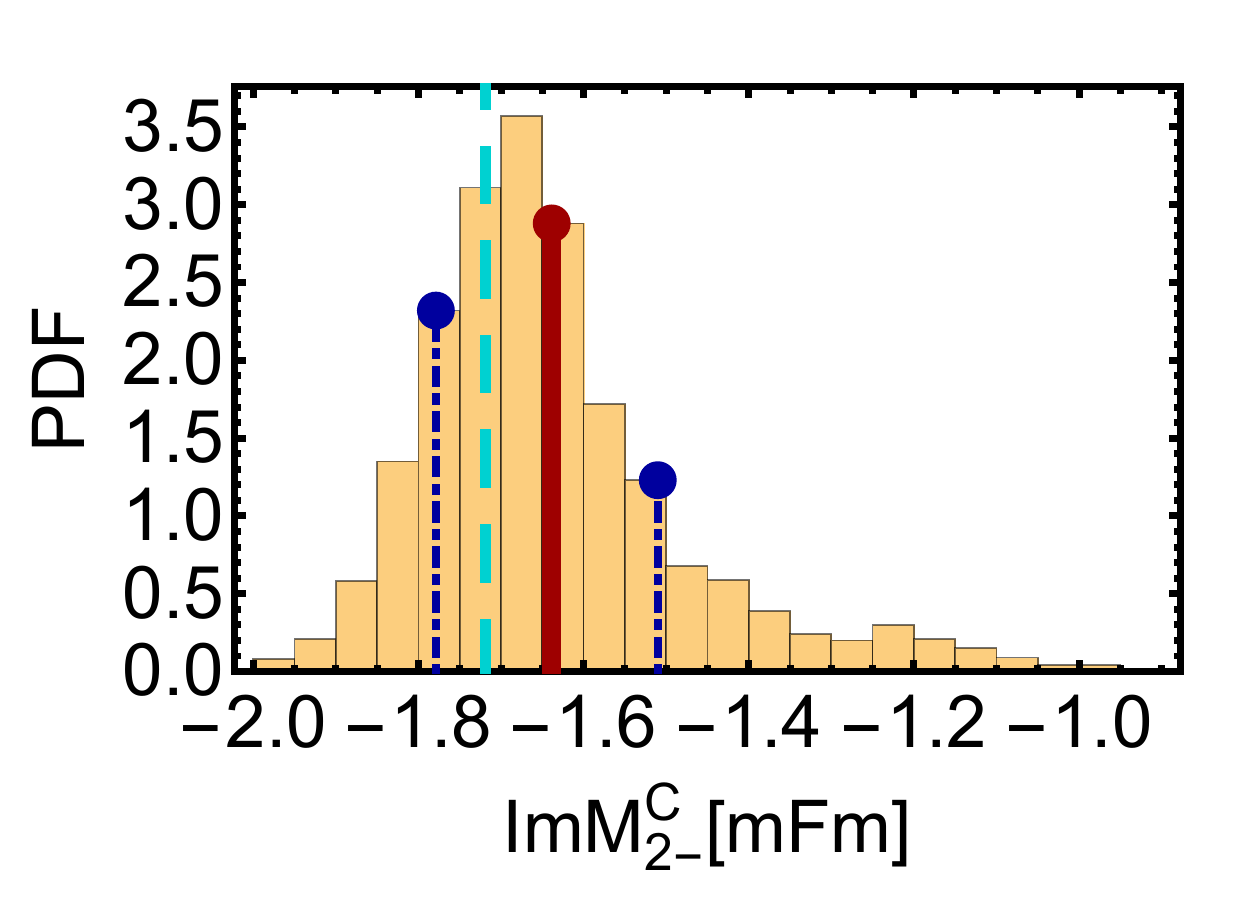}
 \end{overpic}
\caption[Bootstrap-distributions for multipole fit-parameters in an analysis of photoproduction data on the second resonance region. The second energy-bin, \newline $E_{\gamma }\text{ = 715.61 MeV}$, is shown.]{The histograms show bootstrap-distributions for the real- and imaginary parts of phase-constrained $S$-, $P$- and $D$-wave multipoles, for a TPWA bootstrap-analysis of photoproduction data in the second resonance region (see section \ref{subsec:2ndResRegionDataFits}). The second energy-bin, $E_{\gamma }\text{ = 715.61 MeV}$, is shown. An ensemble of $B=2000$ bootstrap-replicates has been the basis of these results. \newline
The distributions have been normalized to $1$ via use of the object \textit{HistogramDistribution} in MATHEMATICA \cite{Mathematica8,Mathematica11,MathematicaLanguage,MathematicaBonnLicense}. Thus, $y$-axes are labelled as \textit{PDF}. The mean of each distribution is shown as a red solid line, while the $0.16$- and $0.84$-quantiles are indicated by blue dash-dotted lines. The global minimum of the fit to the original data is plotted as a cyan-colored dashed horizontal line.}
\label{fig:BootstrapHistos2ndResRegionEnergy2}
\end{figure}

\clearpage

\begin{figure}[h]
\begin{overpic}[width=0.325\textwidth]{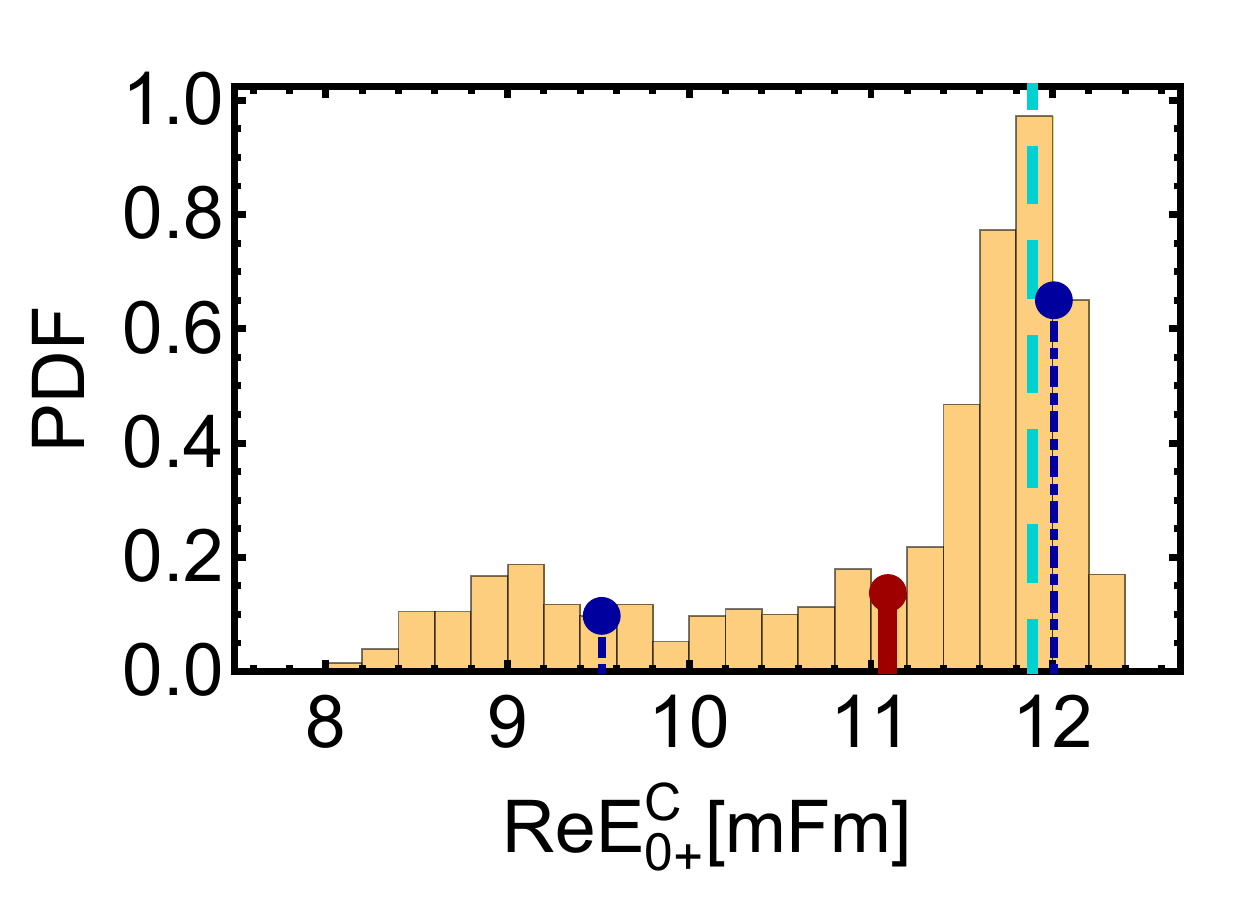}
 \end{overpic}
\begin{overpic}[width=0.325\textwidth]{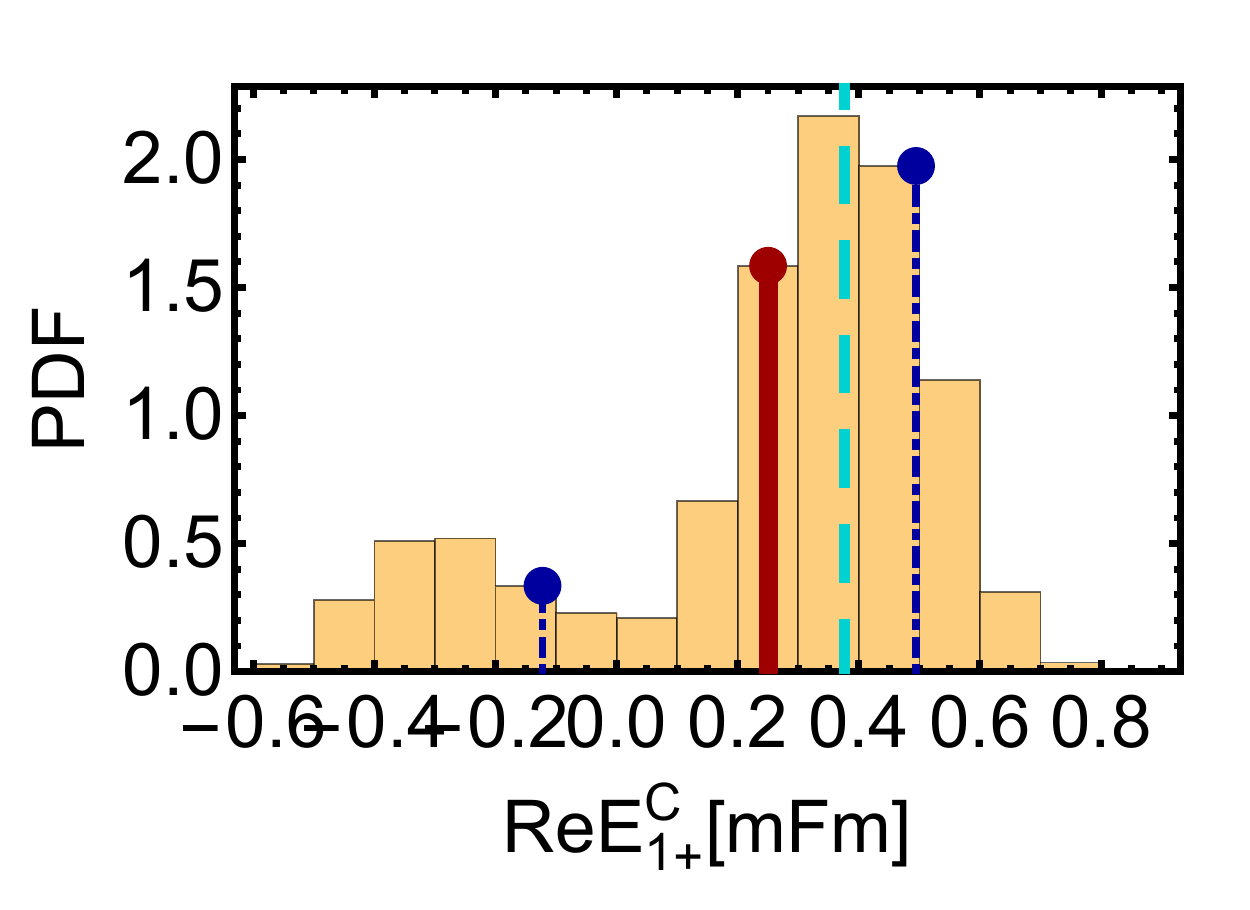}
 \end{overpic}
\begin{overpic}[width=0.325\textwidth]{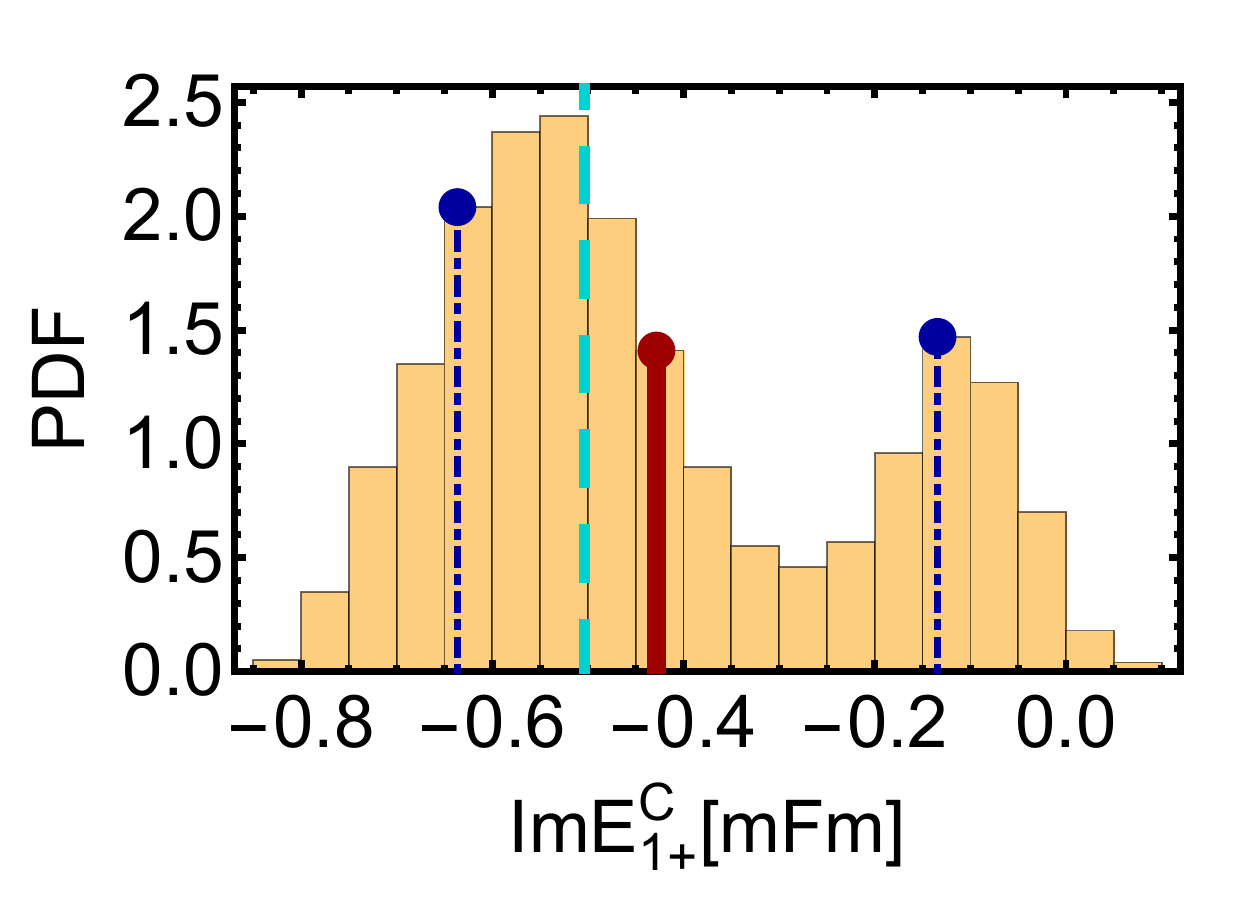}
 \end{overpic} \\
\begin{overpic}[width=0.325\textwidth]{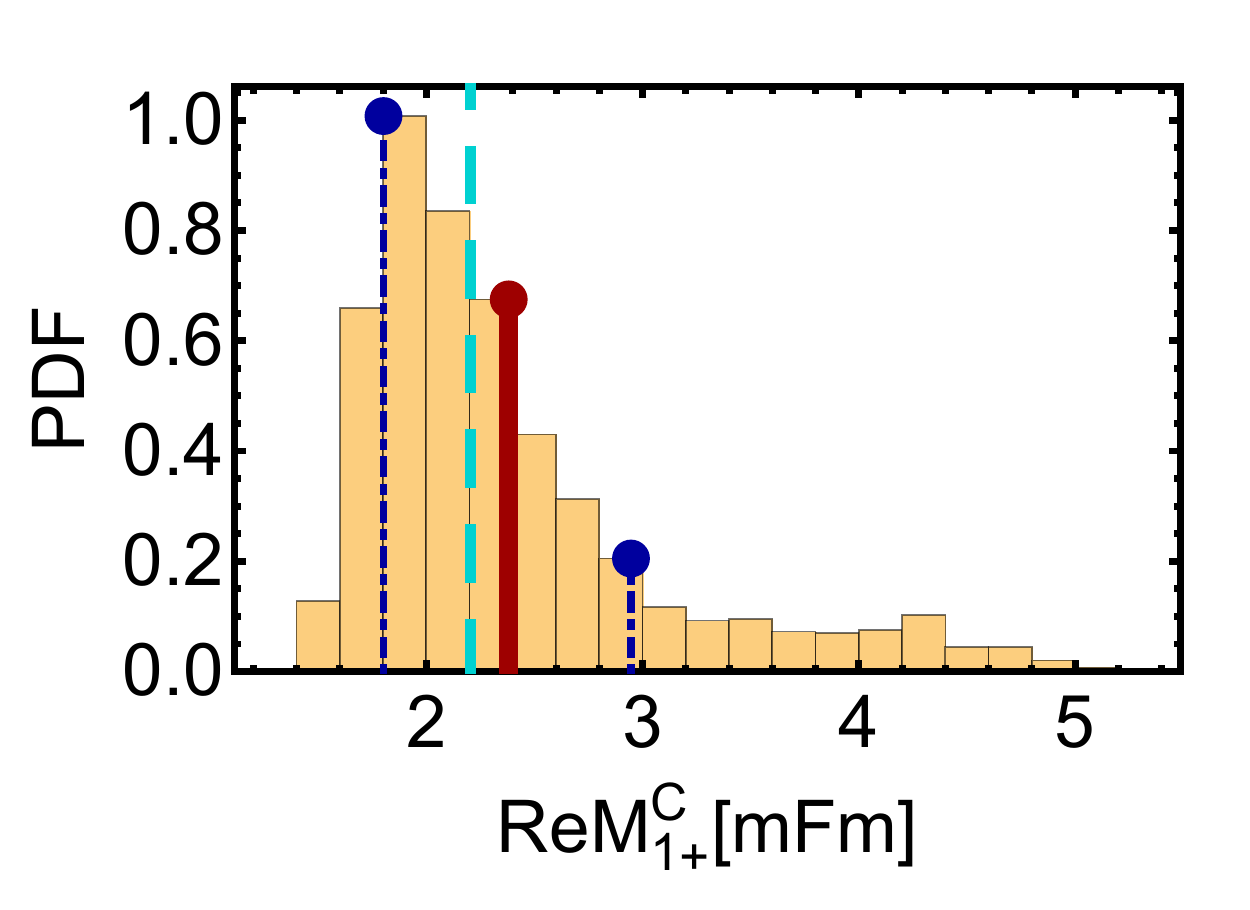}
 \end{overpic}
\begin{overpic}[width=0.325\textwidth]{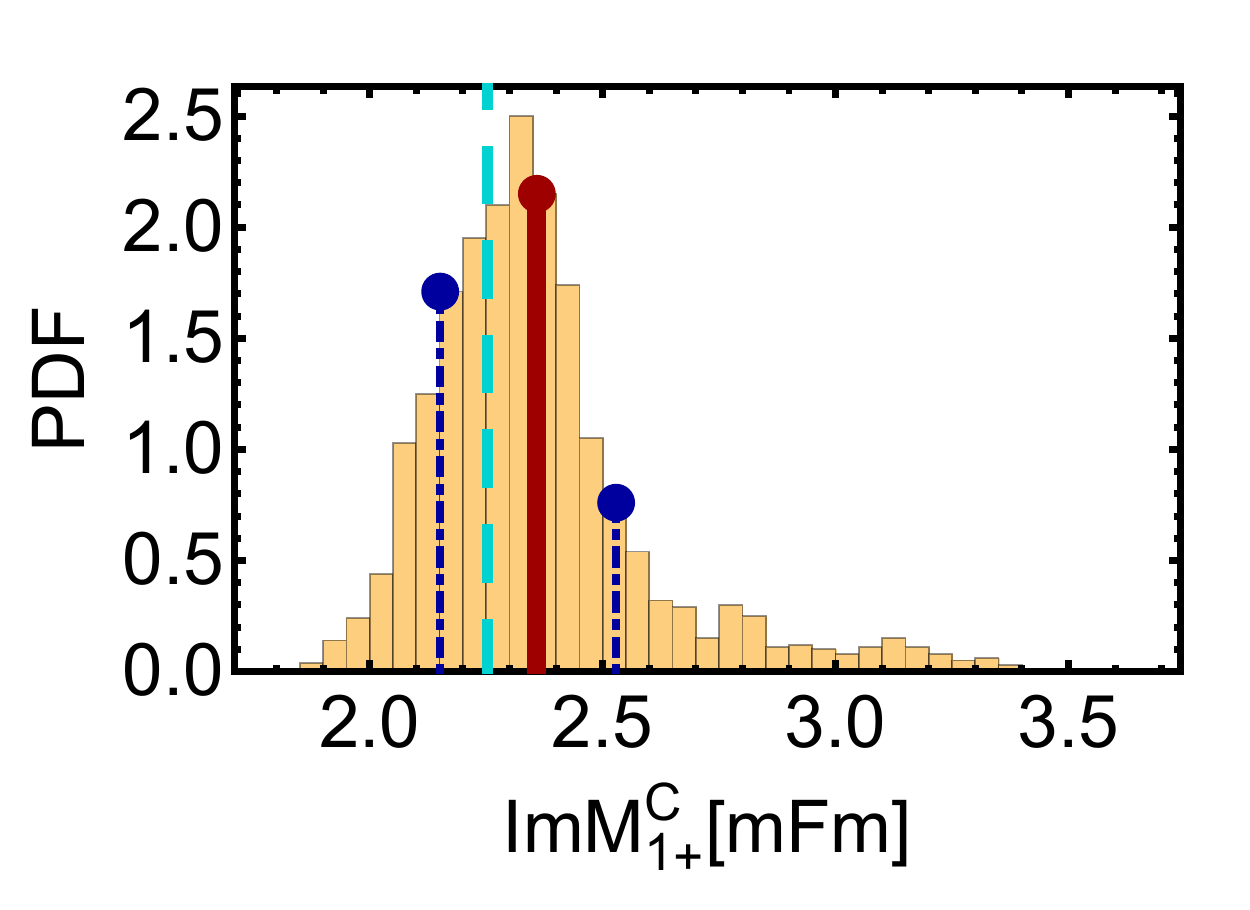}
 \end{overpic}
\begin{overpic}[width=0.325\textwidth]{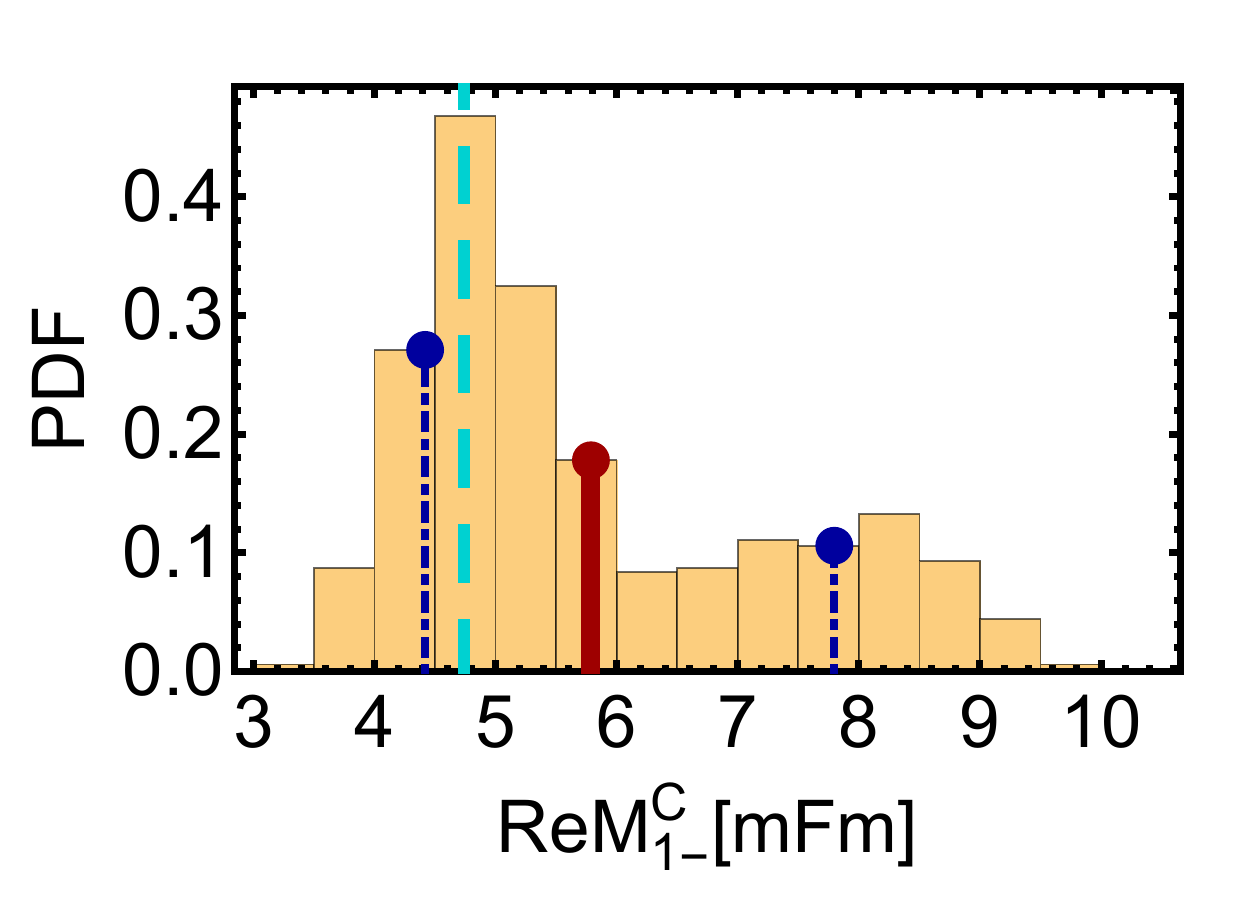}
 \end{overpic} \\
\begin{overpic}[width=0.325\textwidth]{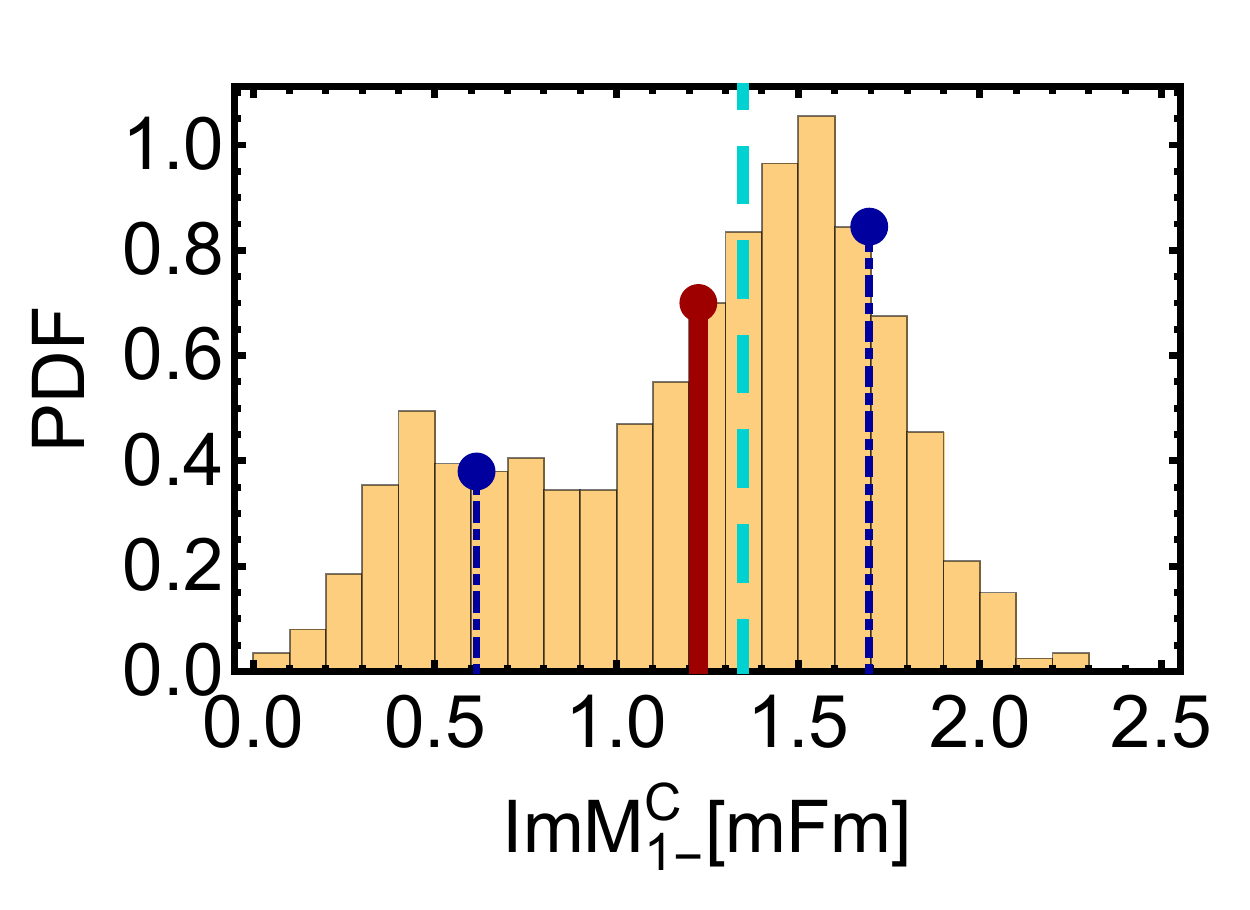}
 \end{overpic}
\begin{overpic}[width=0.325\textwidth]{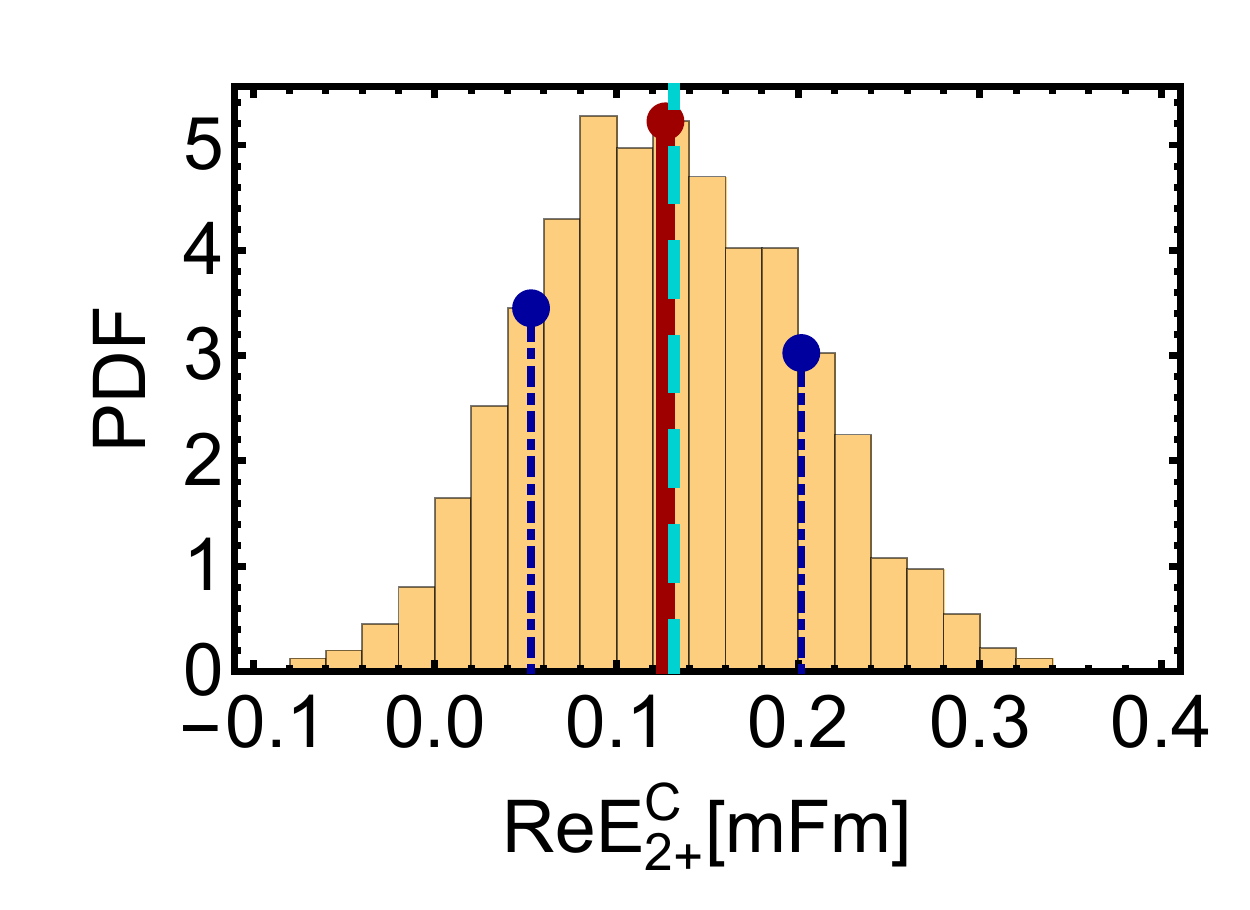}
 \end{overpic}
\begin{overpic}[width=0.325\textwidth]{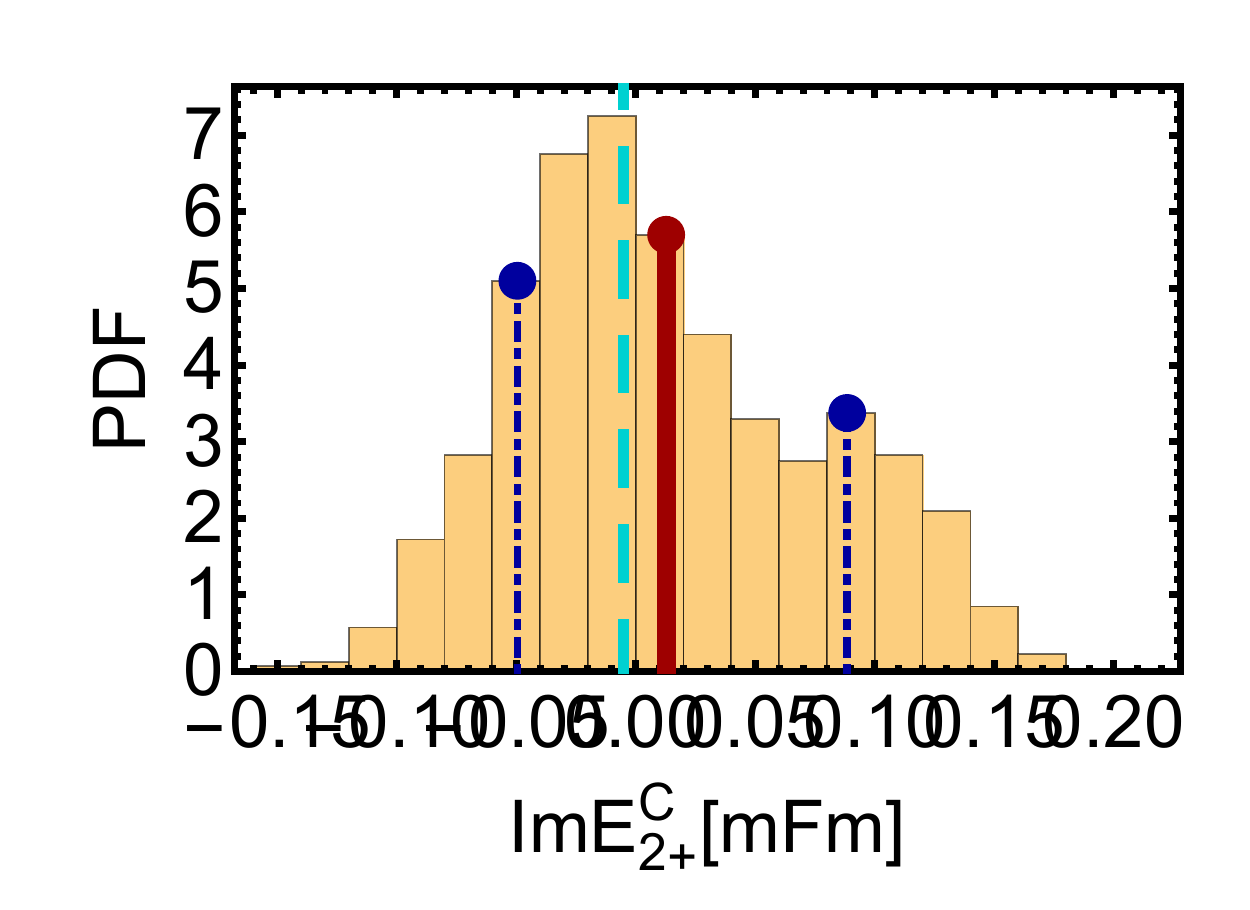}
 \end{overpic} \\
\begin{overpic}[width=0.325\textwidth]{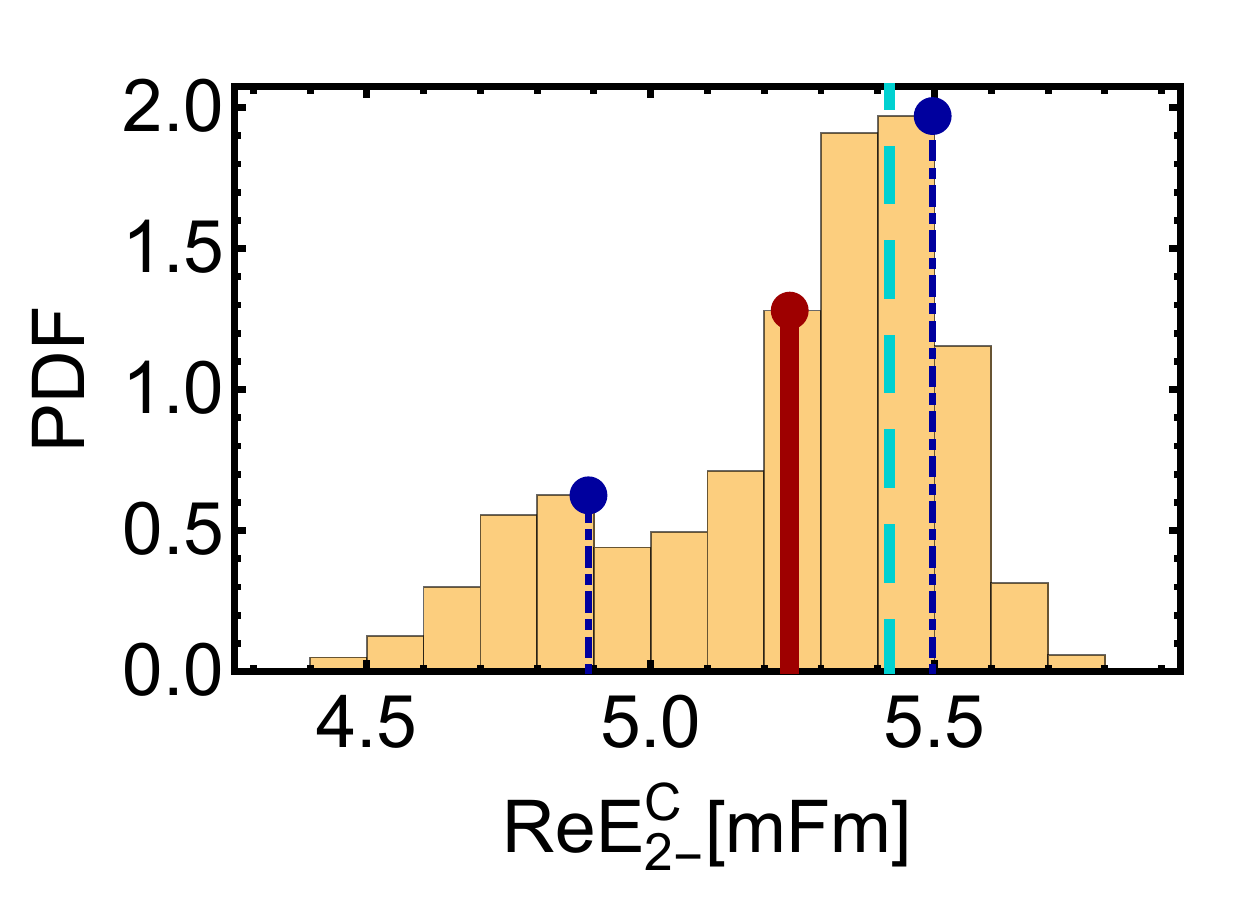}
 \end{overpic}
\begin{overpic}[width=0.325\textwidth]{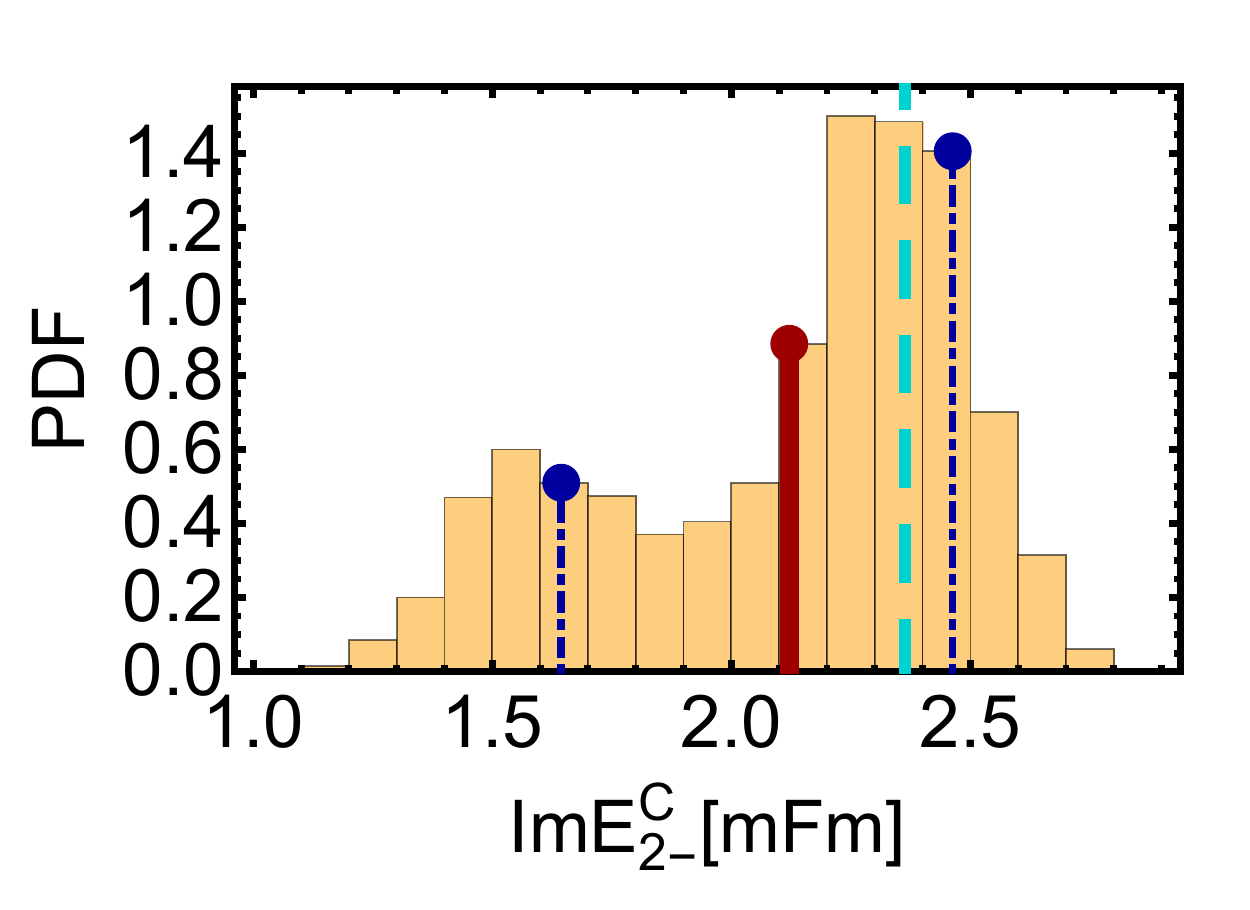}
 \end{overpic}
\begin{overpic}[width=0.325\textwidth]{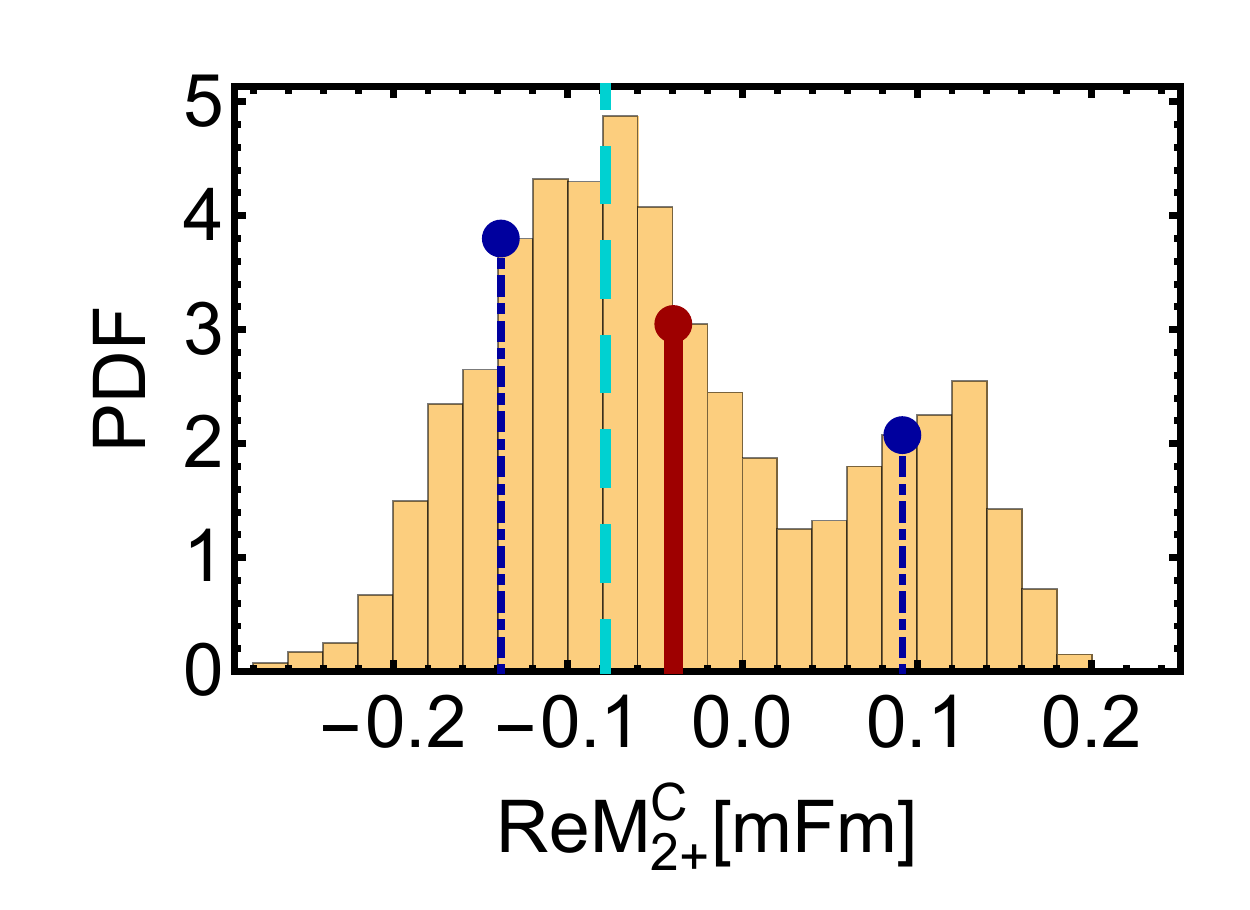}
 \end{overpic} \\
\begin{overpic}[width=0.325\textwidth]{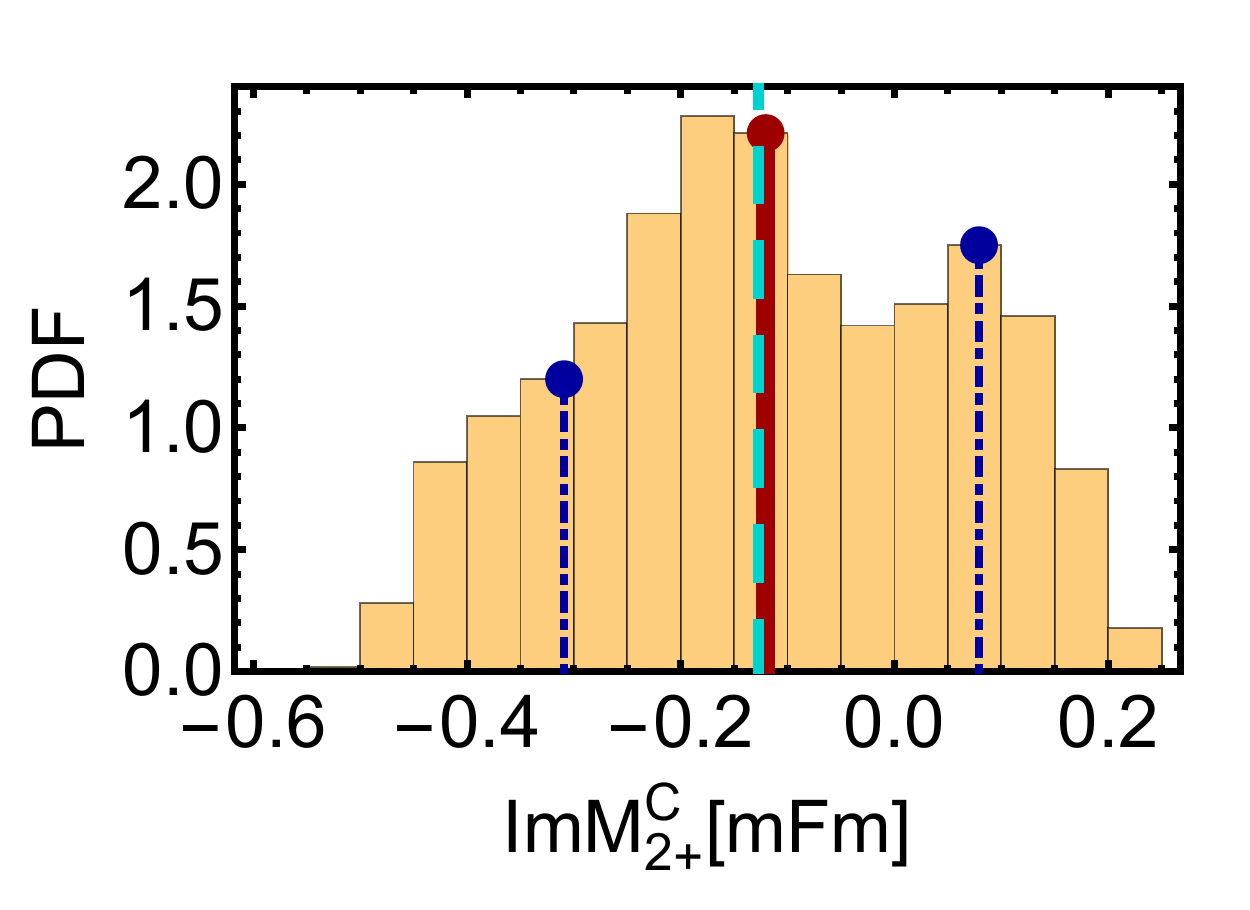}
 \end{overpic}
\begin{overpic}[width=0.325\textwidth]{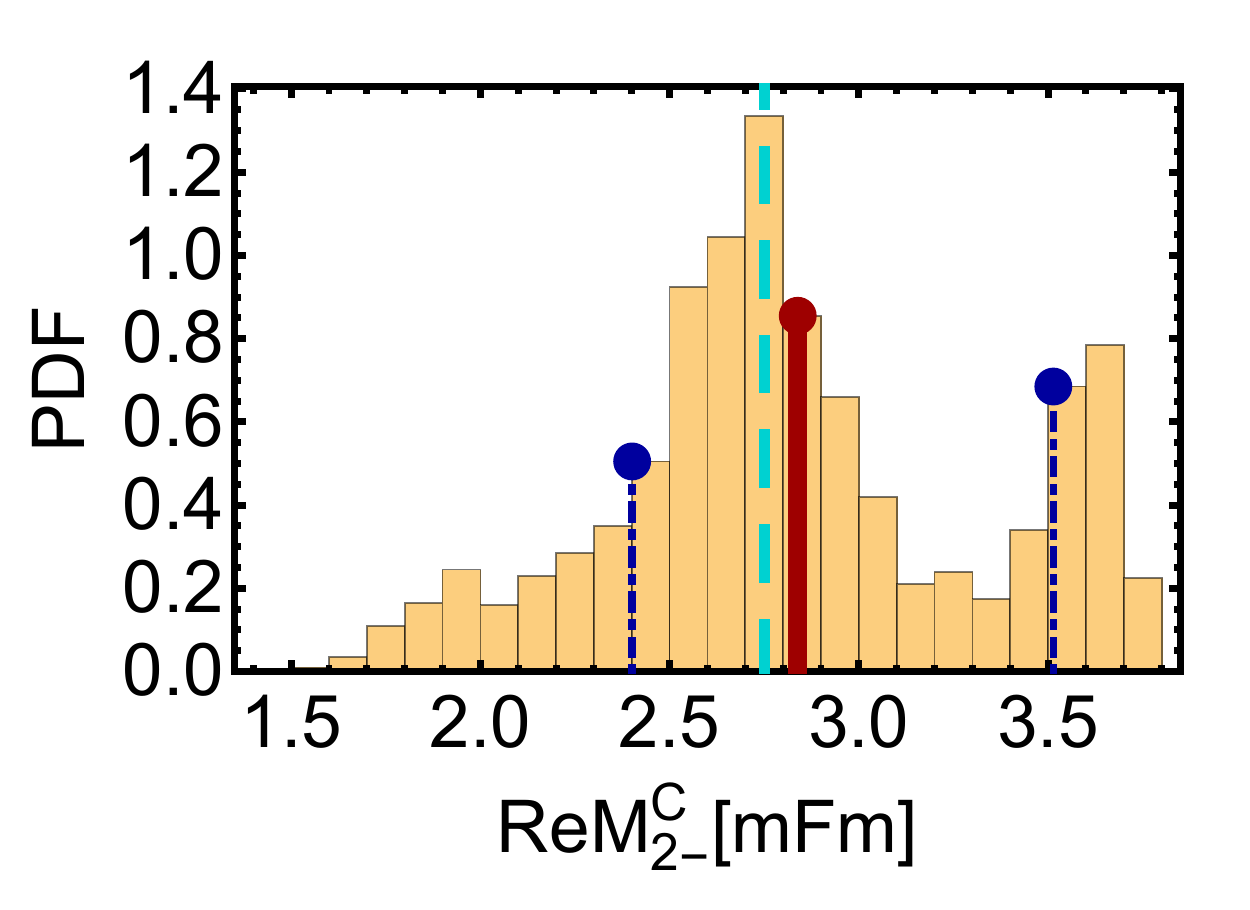}
 \end{overpic}
\begin{overpic}[width=0.325\textwidth]{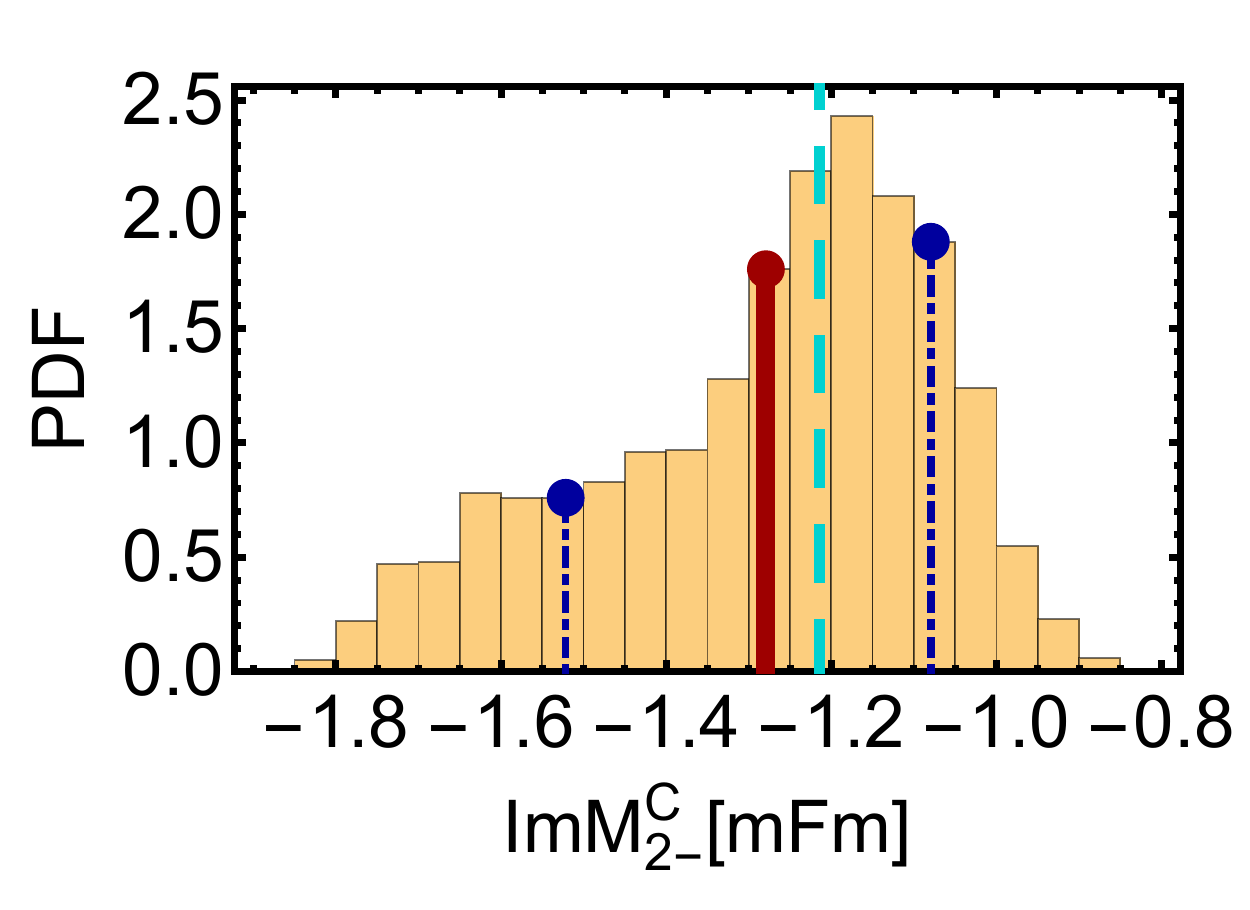}
 \end{overpic}
\caption[Bootstrap-distributions for multipole fit-parameters in an analysis of photoproduction data on the second resonance region. The third energy-bin, \newline $E_{\gamma }\text{ = 749.94 MeV}$, is shown.]{The histograms show bootstrap-distributions for the real- and imaginary parts of phase-constrained $S$-, $P$- and $D$-wave multipoles, for a TPWA bootstrap-analysis of photoproduction data in the second resonance region (see section \ref{subsec:2ndResRegionDataFits}). The third energy-bin, $E_{\gamma }\text{ = 749.94 MeV}$, is shown. An ensemble of $B=2000$ bootstrap-replicates has been the basis of these results. \newline
The distributions have been normalized to $1$ via use of the object \textit{HistogramDistribution} in MATHEMATICA \cite{Mathematica8,Mathematica11,MathematicaLanguage,MathematicaBonnLicense}. Thus, $y$-axes are labelled as \textit{PDF}. The mean of each distribution is shown as a red solid line, while the $0.16$- and $0.84$-quantiles are indicated by blue dash-dotted lines. The global minimum of the fit to the original data is plotted as a cyan-colored dashed horizontal line.}
\label{fig:BootstrapHistos2ndResRegionEnergy3}
\end{figure}

\clearpage

\begin{figure}[h]
\begin{overpic}[width=0.325\textwidth]{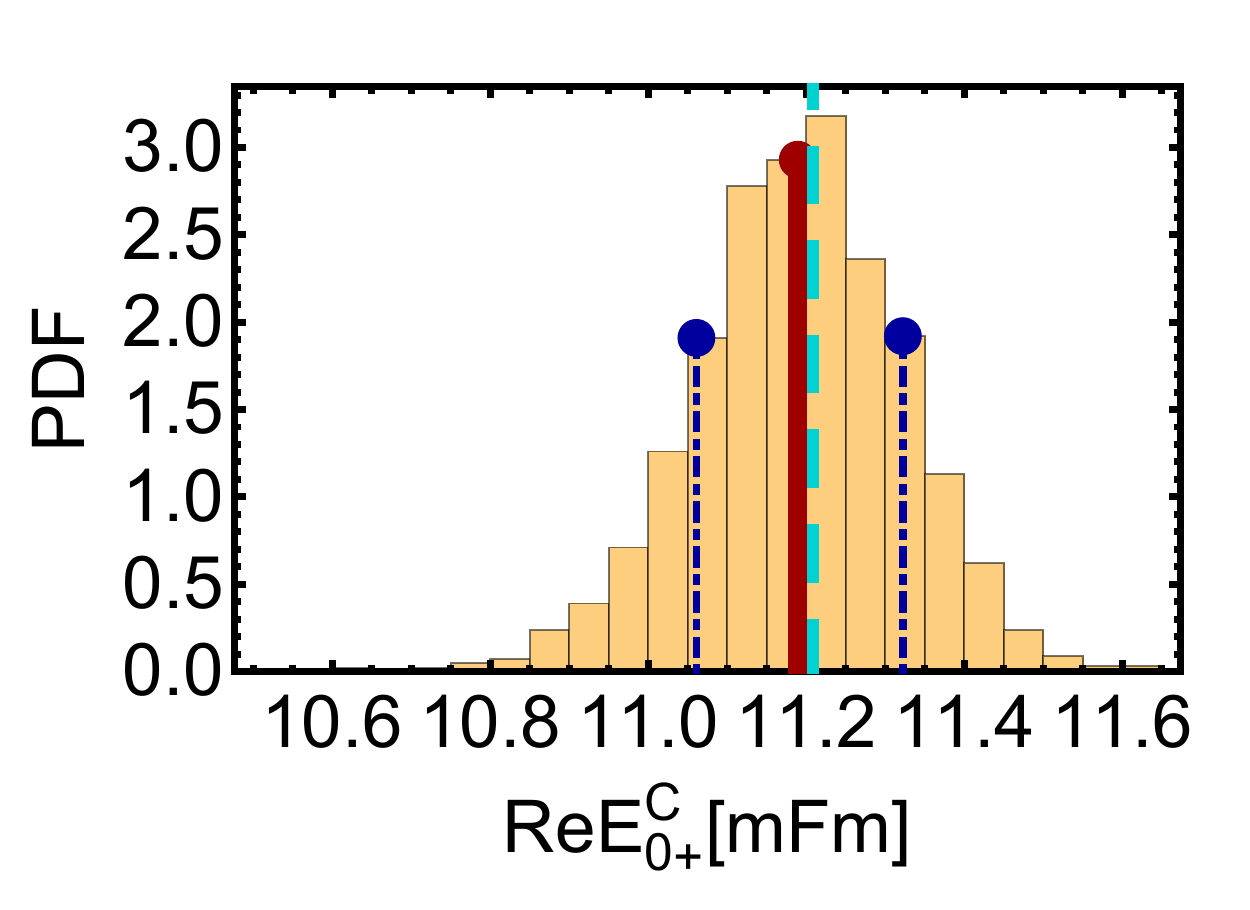}
 \end{overpic}
\begin{overpic}[width=0.325\textwidth]{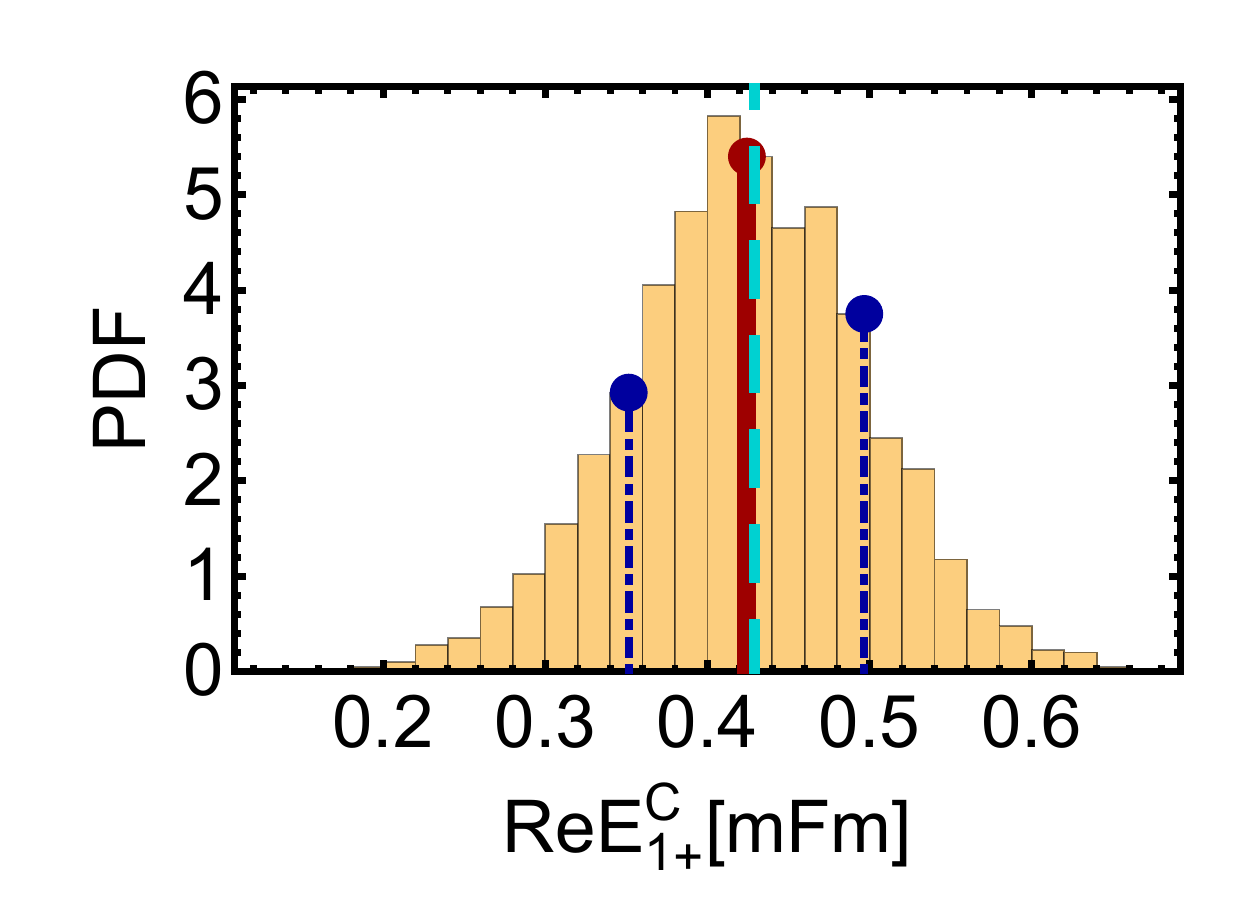}
 \end{overpic}
\begin{overpic}[width=0.325\textwidth]{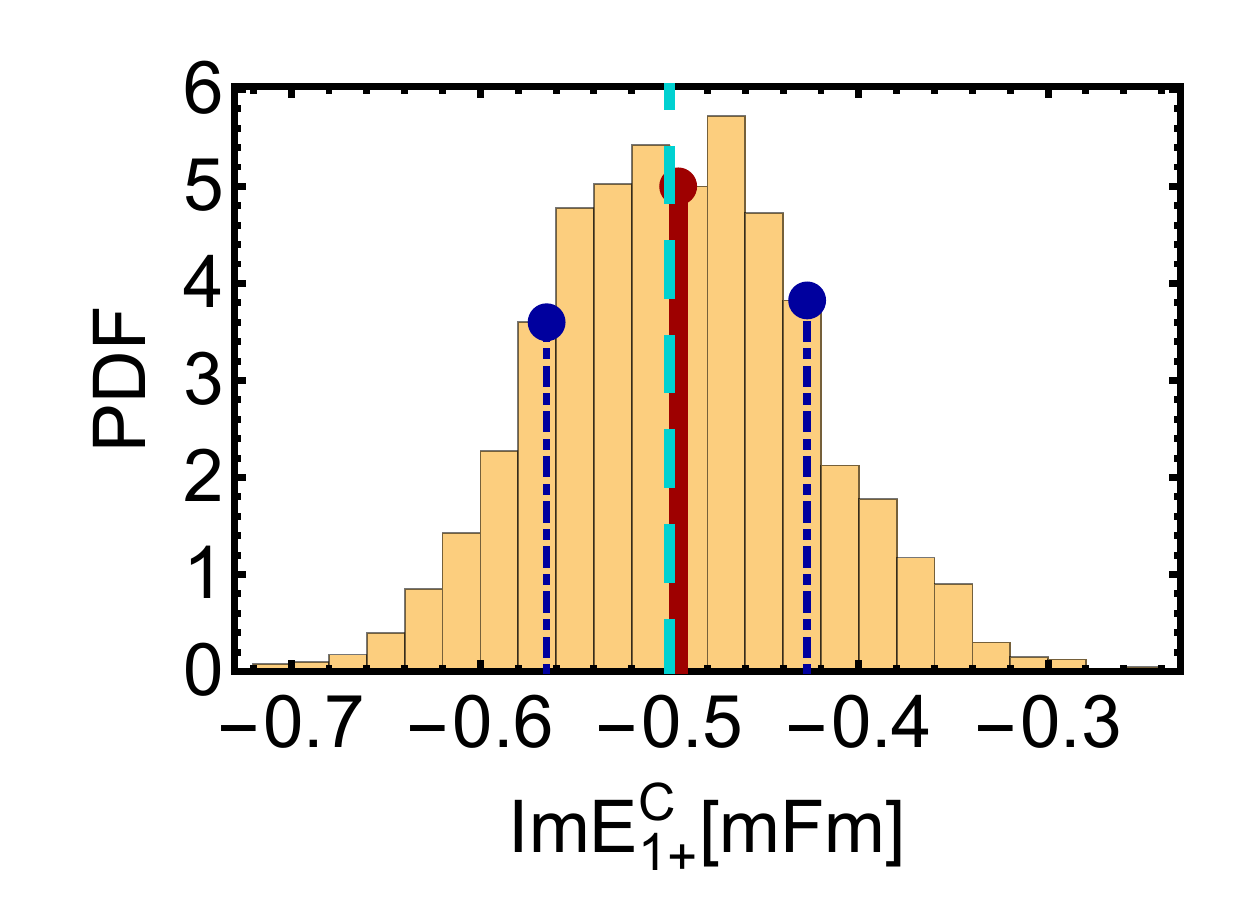}
 \end{overpic} \\
\begin{overpic}[width=0.325\textwidth]{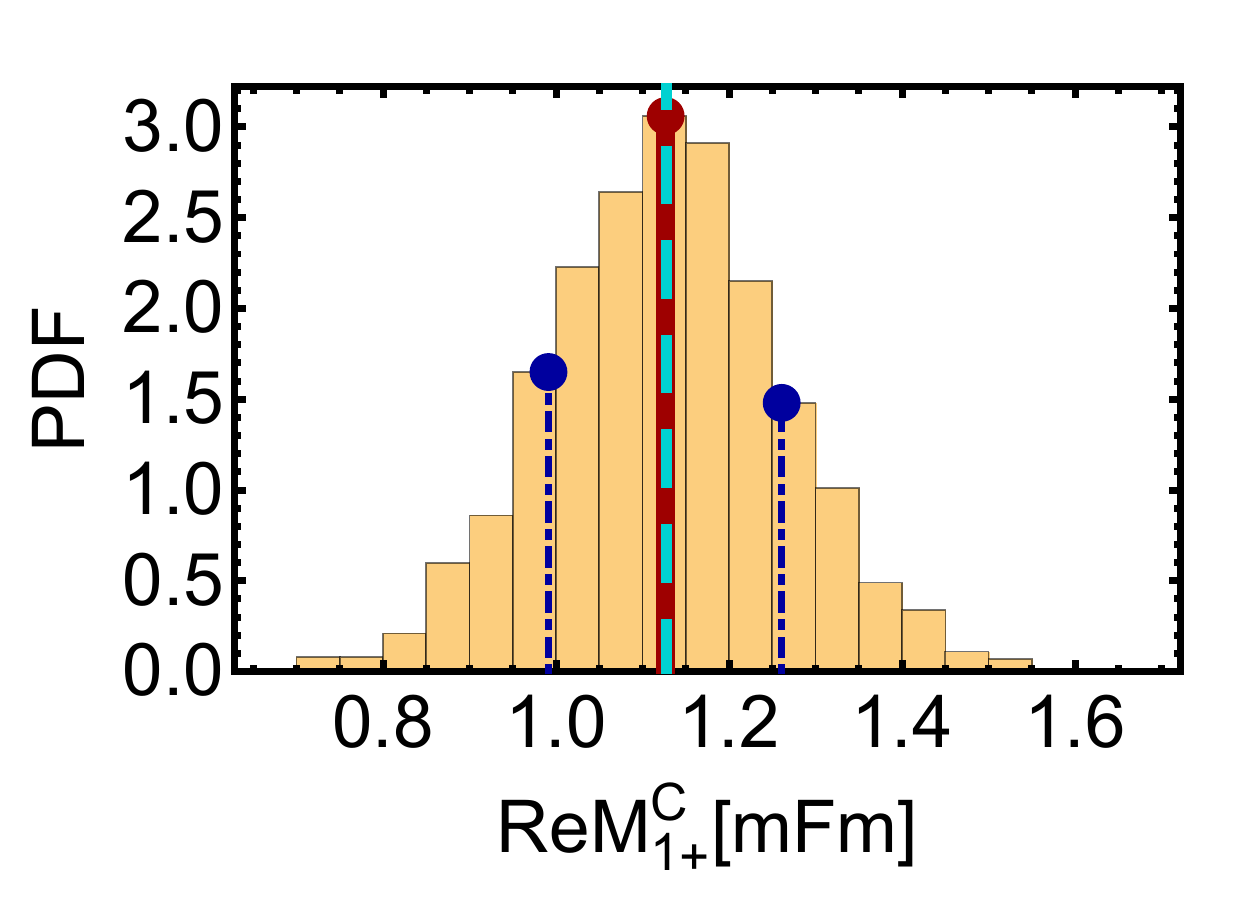}
 \end{overpic}
\begin{overpic}[width=0.325\textwidth]{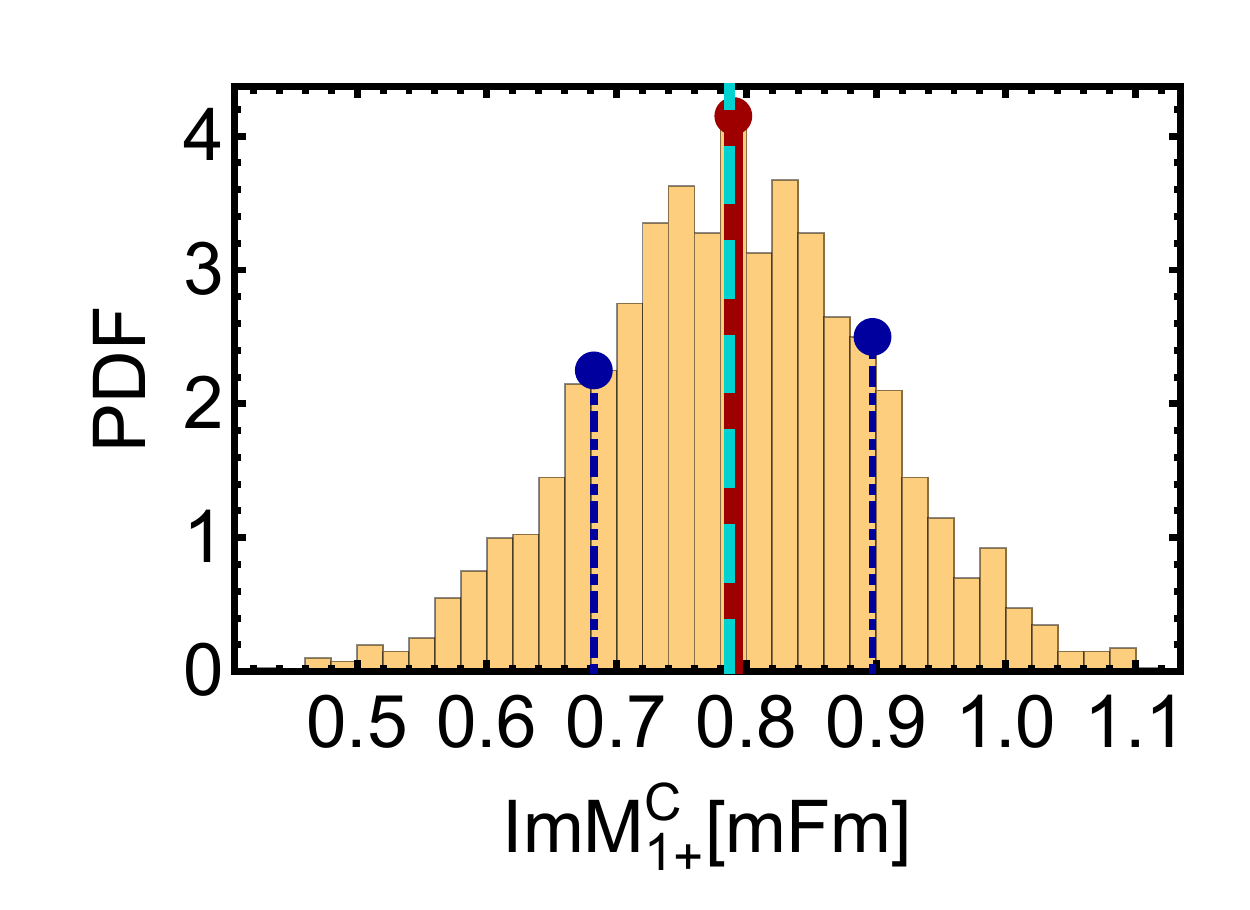}
 \end{overpic}
\begin{overpic}[width=0.325\textwidth]{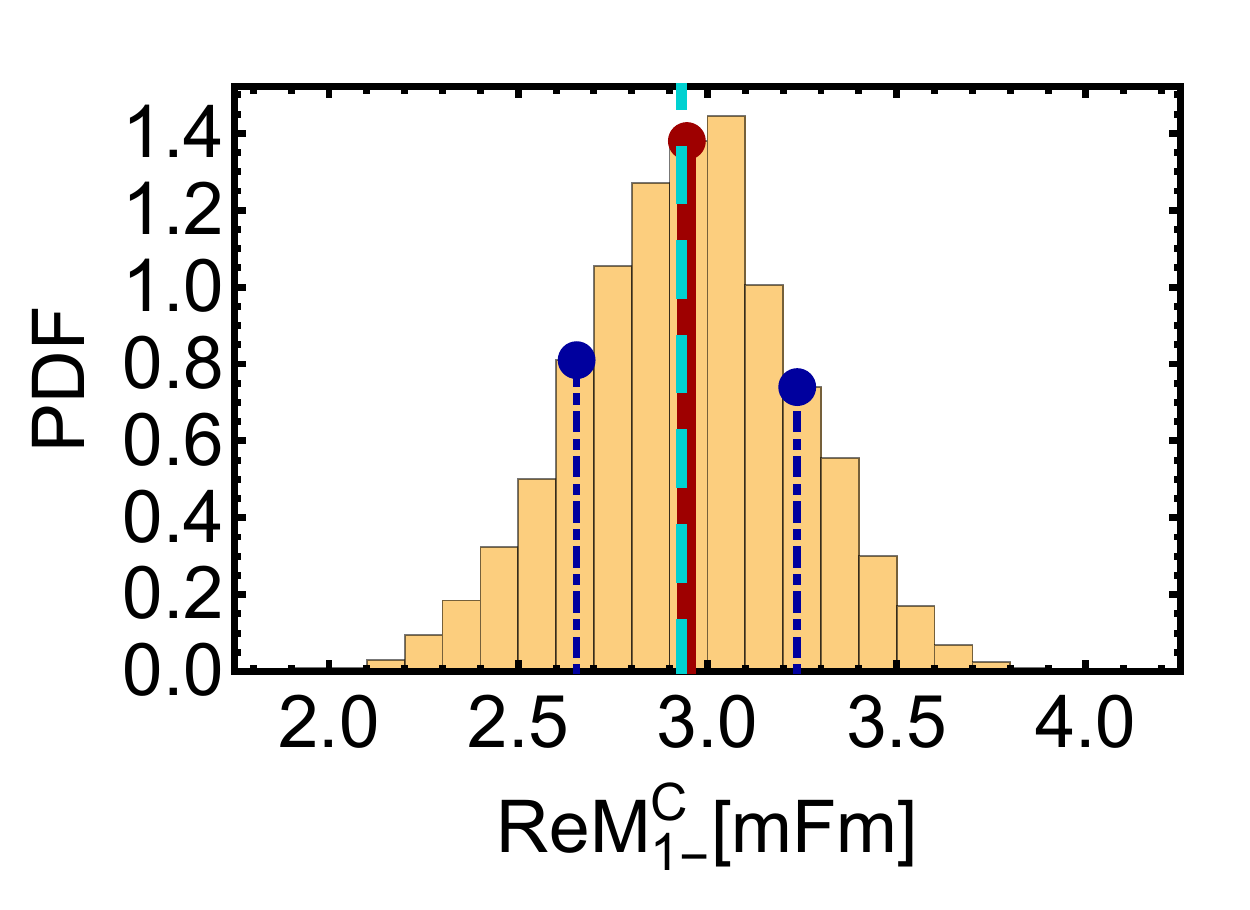}
 \end{overpic} \\
\begin{overpic}[width=0.325\textwidth]{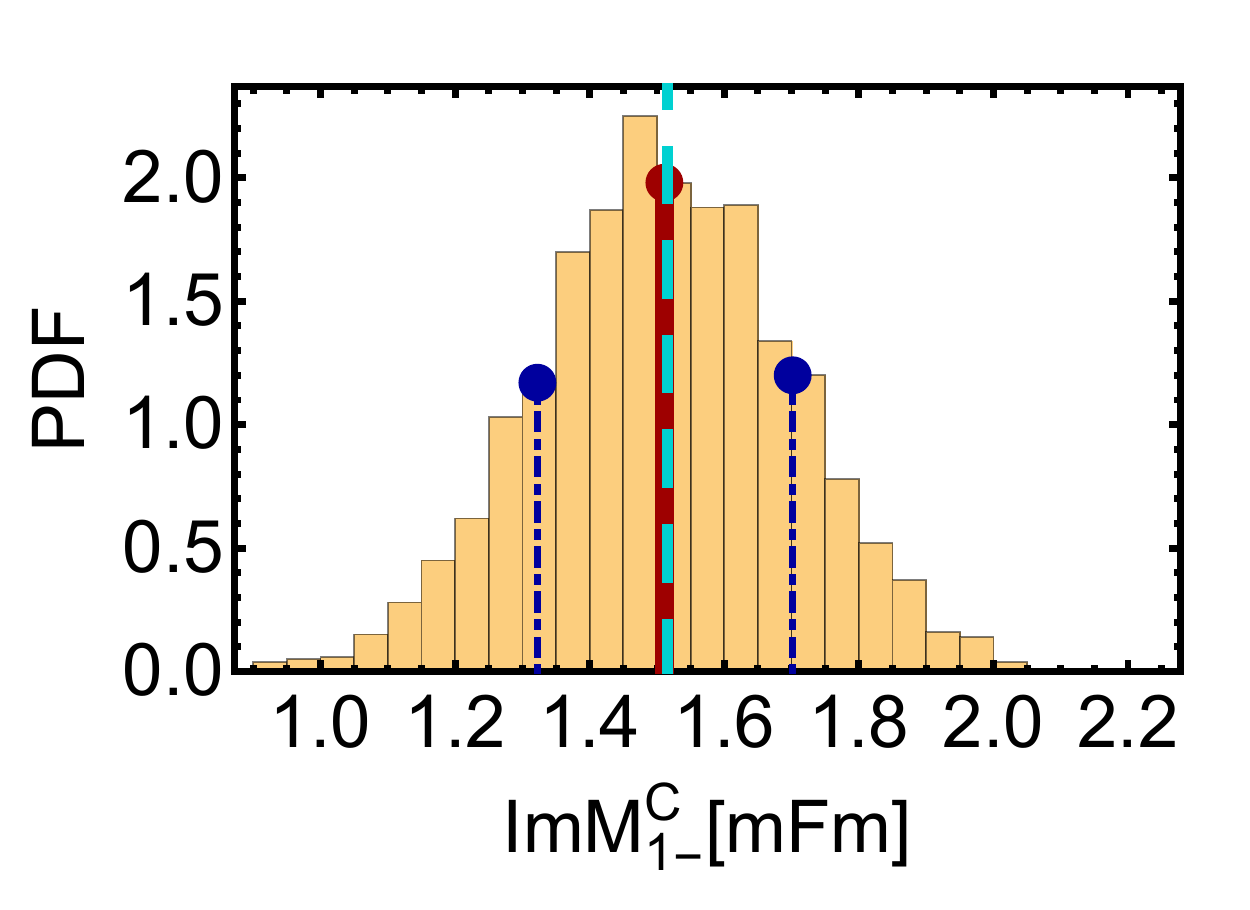}
 \end{overpic}
\begin{overpic}[width=0.325\textwidth]{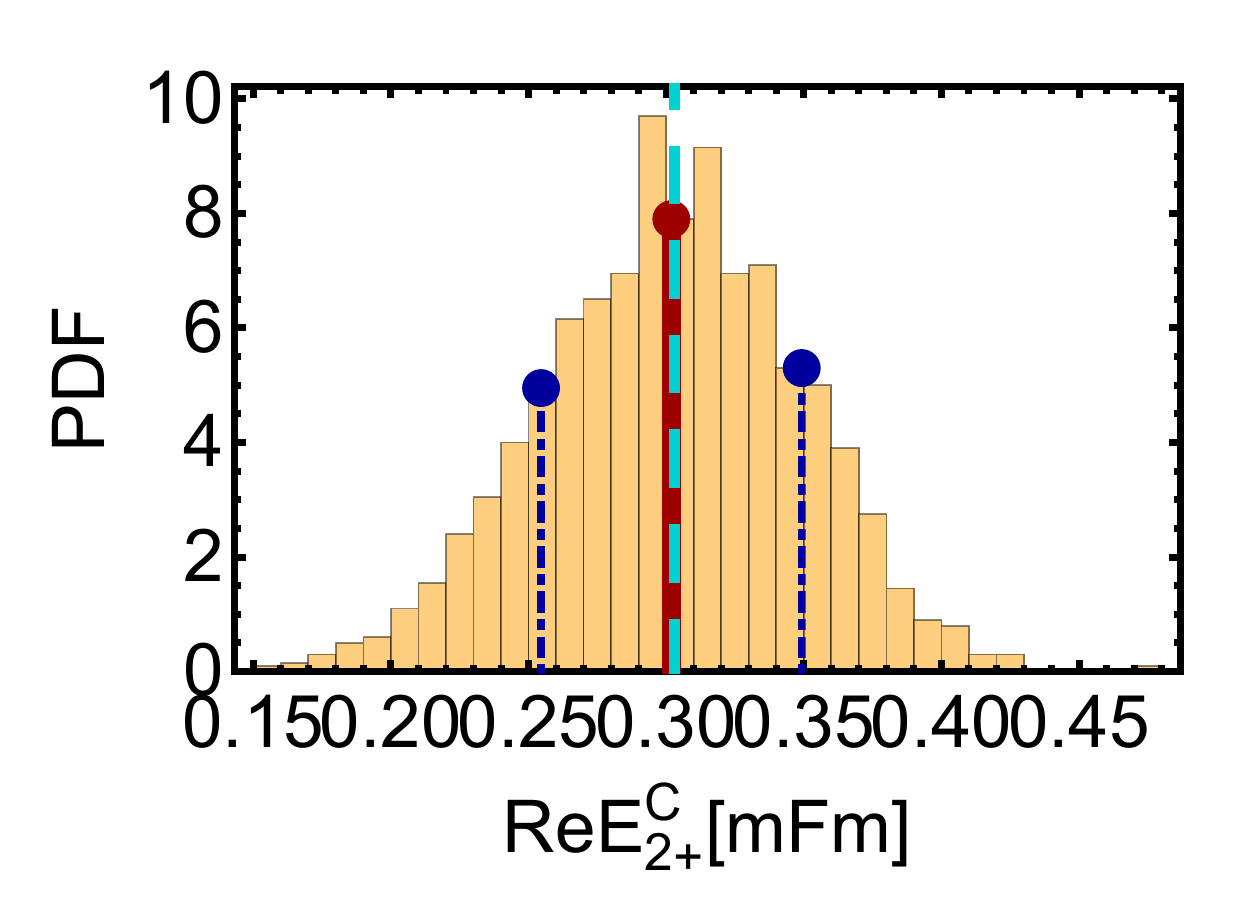}
 \end{overpic}
\begin{overpic}[width=0.325\textwidth]{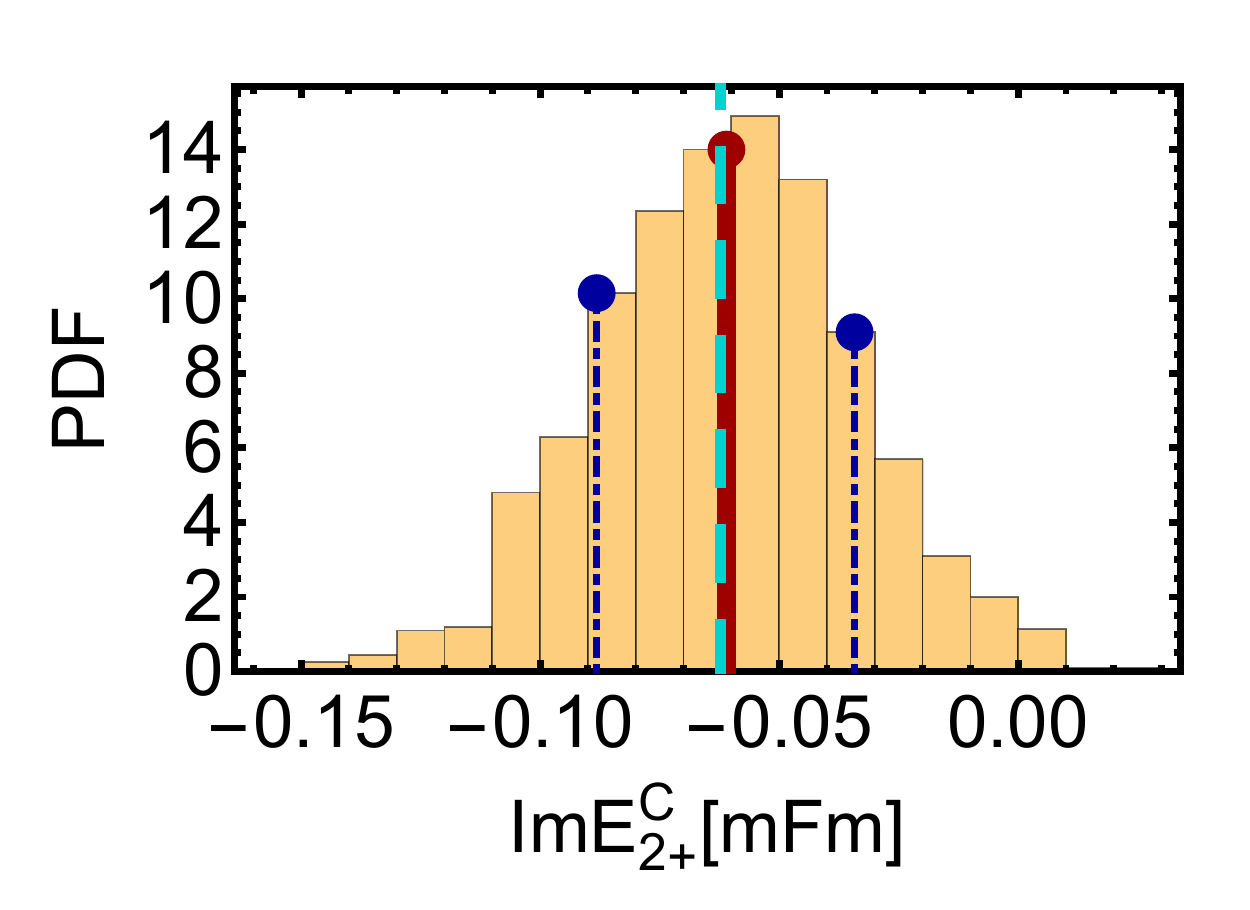}
 \end{overpic} \\
\begin{overpic}[width=0.325\textwidth]{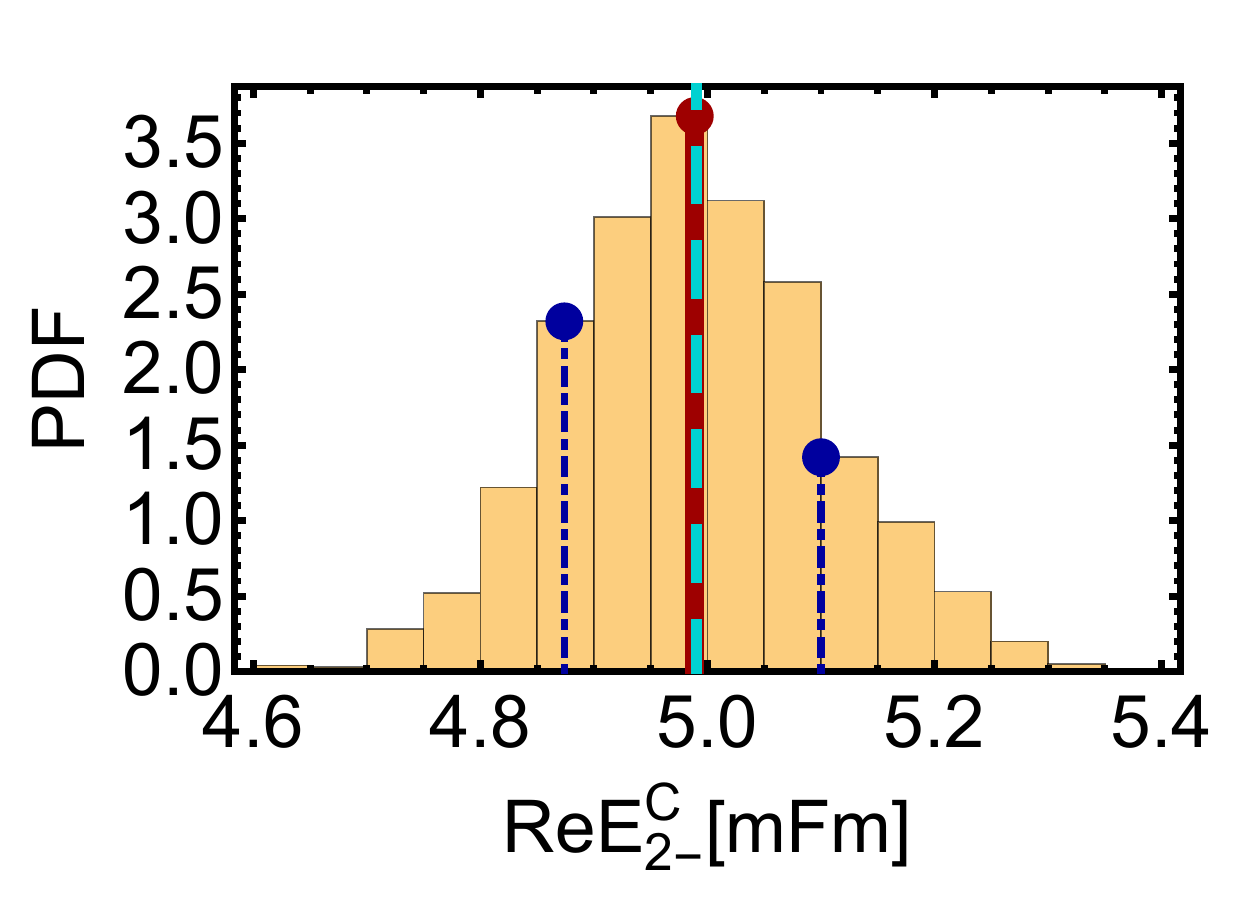}
 \end{overpic}
\begin{overpic}[width=0.325\textwidth]{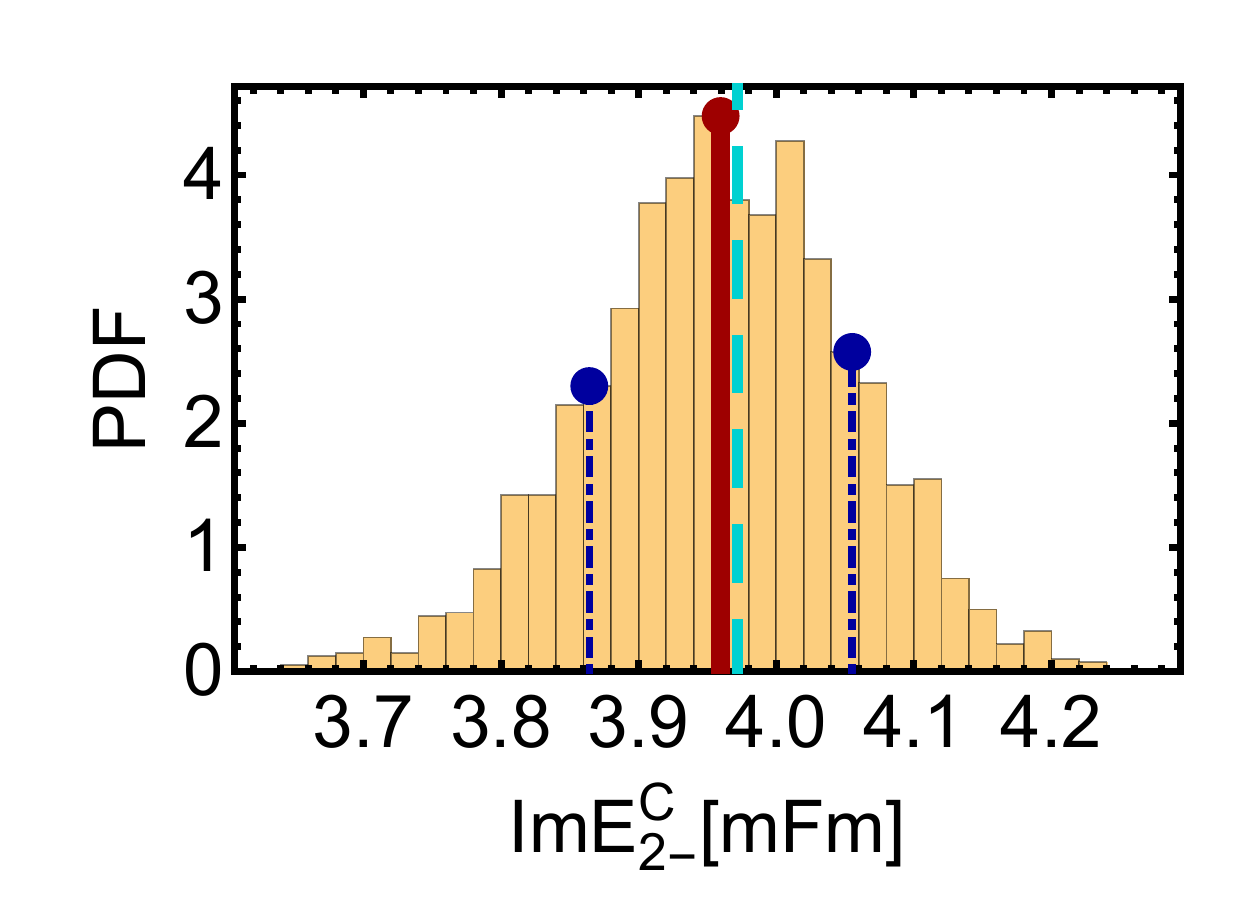}
 \end{overpic}
\begin{overpic}[width=0.325\textwidth]{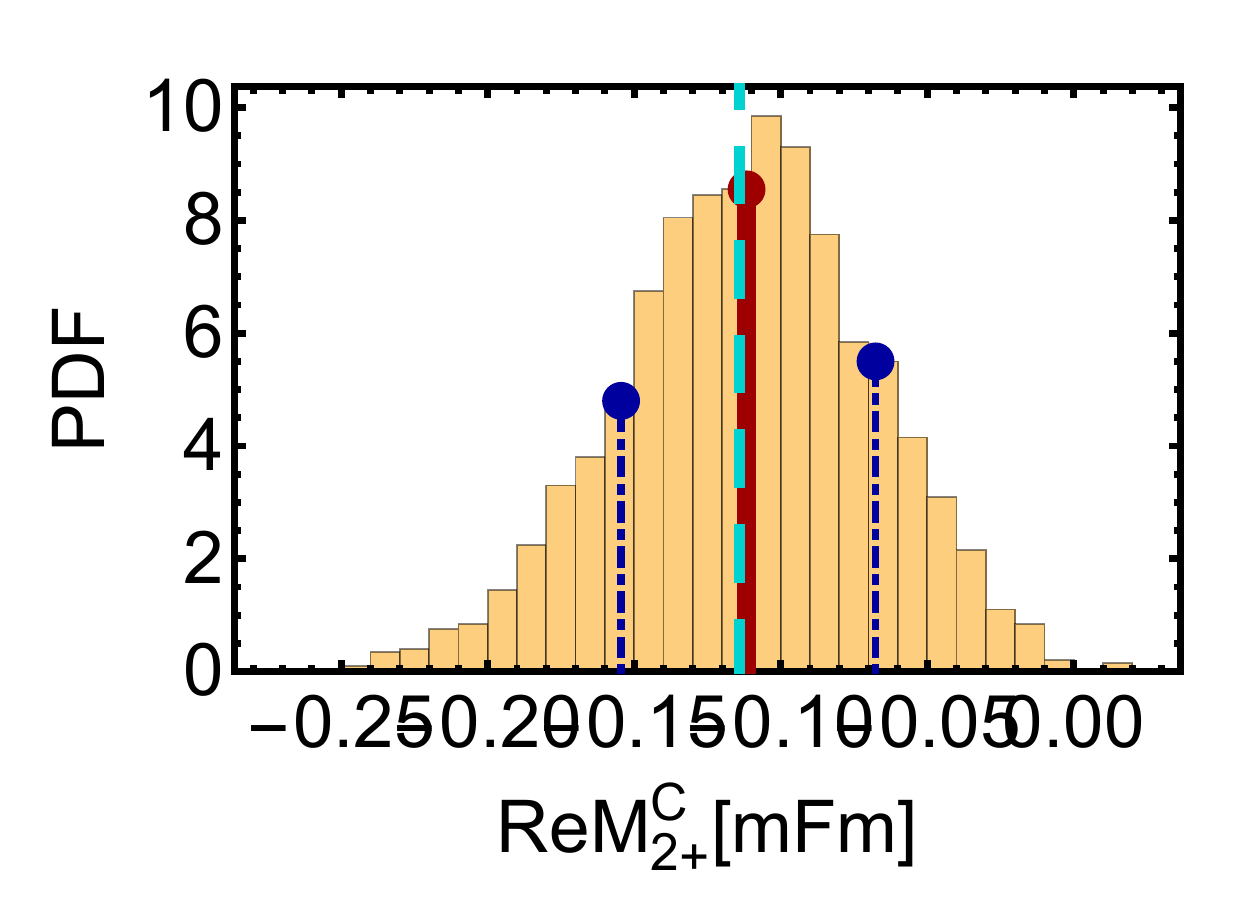}
 \end{overpic} \\
\begin{overpic}[width=0.325\textwidth]{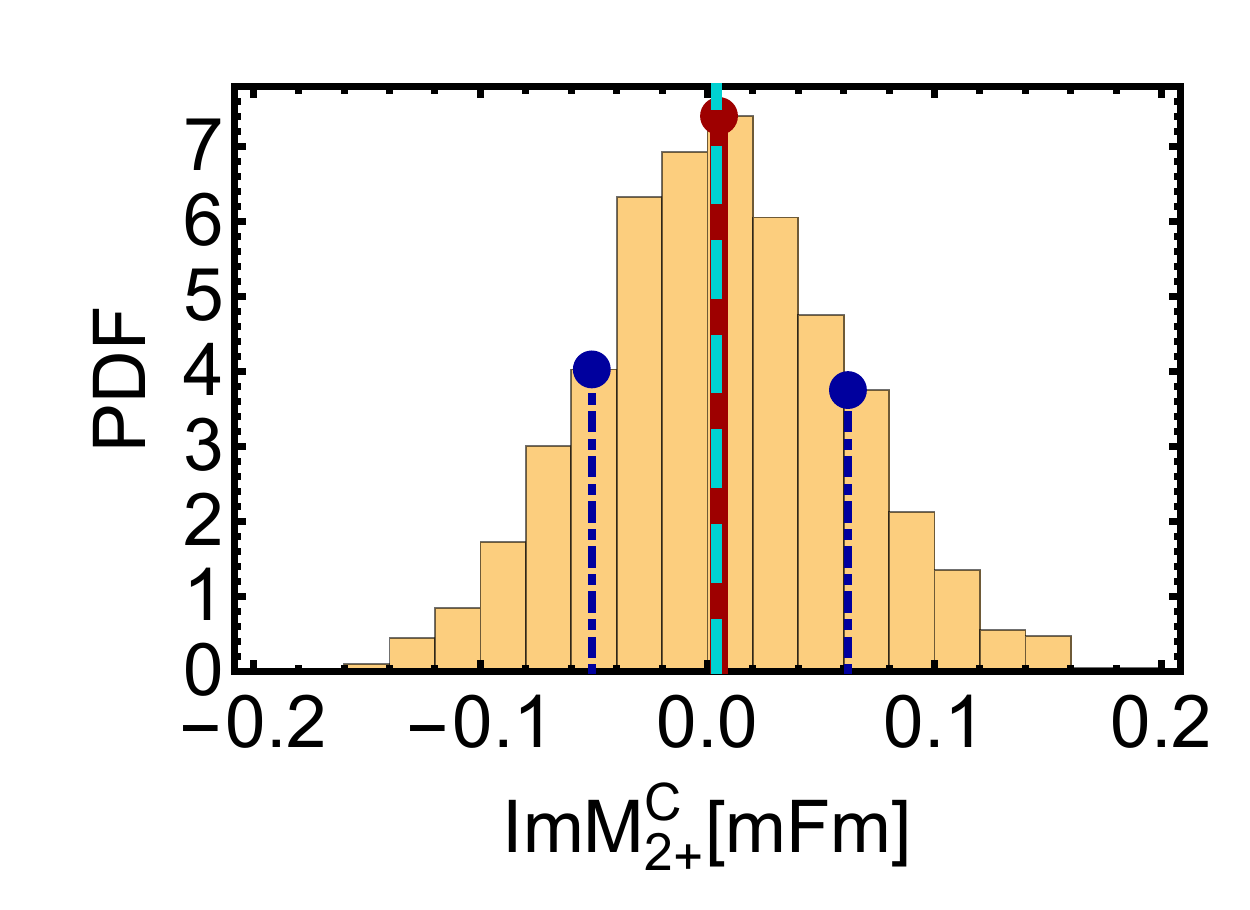}
 \end{overpic}
\begin{overpic}[width=0.325\textwidth]{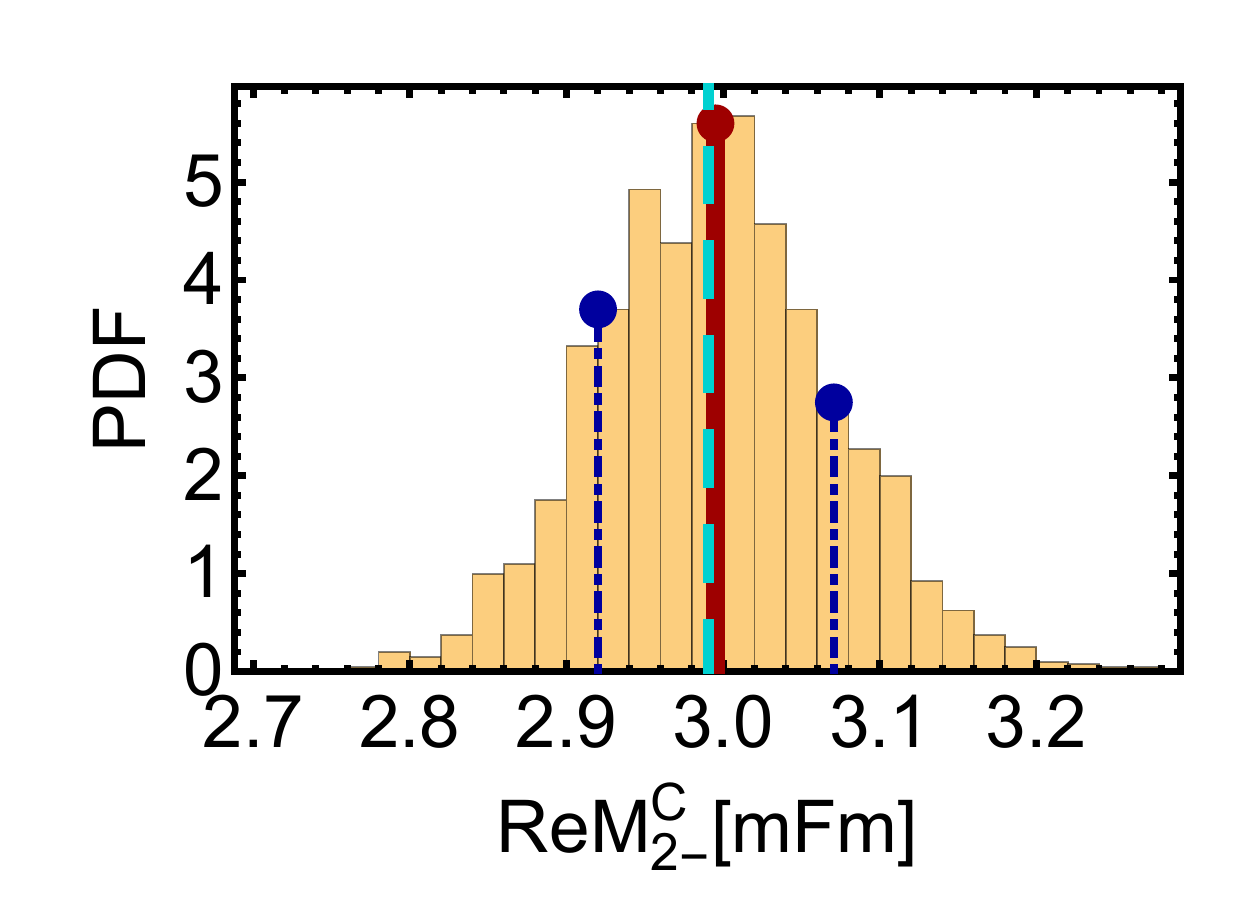}
 \end{overpic}
\begin{overpic}[width=0.325\textwidth]{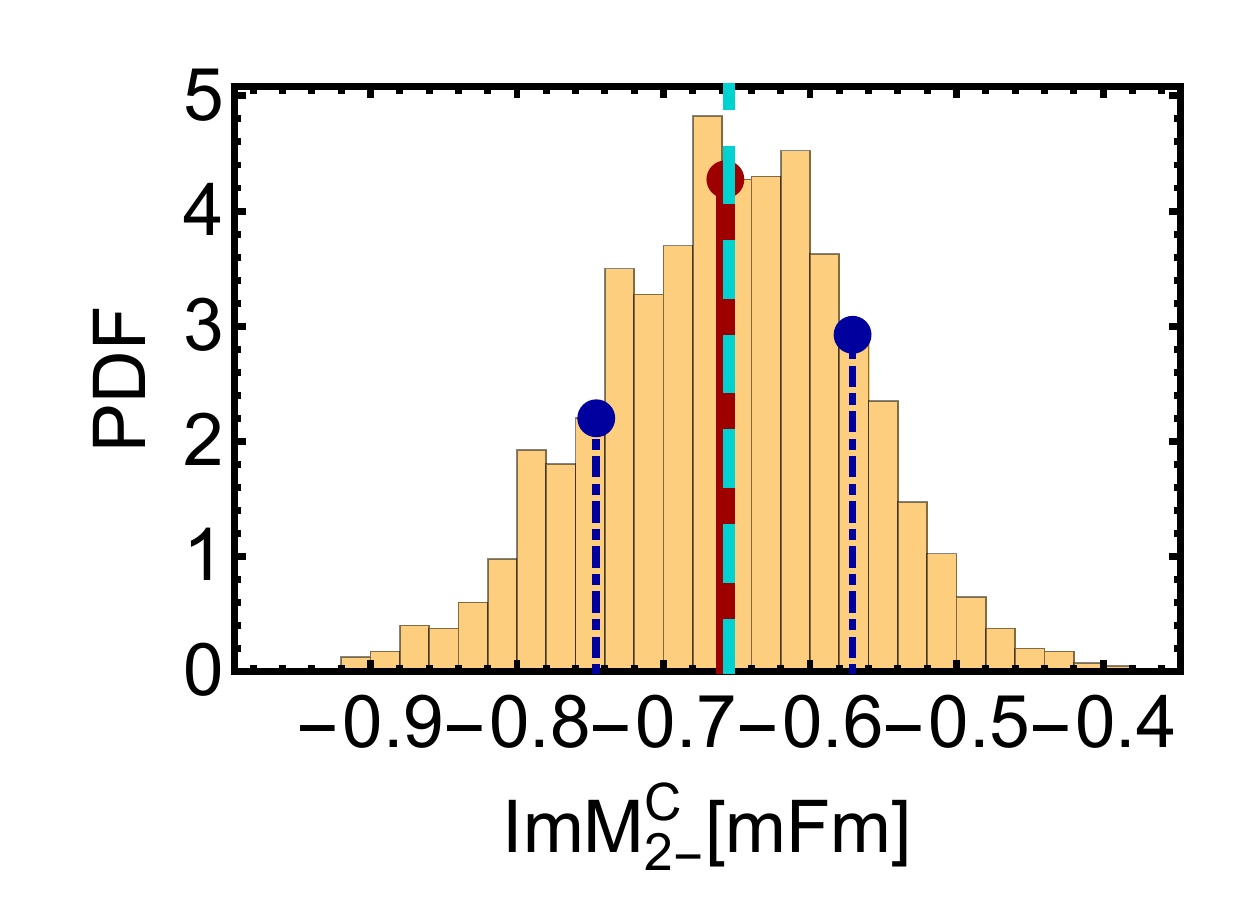}
 \end{overpic}
\caption[Bootstrap-distributions for multipole fit-parameters in an analysis of photoproduction data on the second resonance region. The fourth energy-bin, \newline $E_{\gamma }\text{ = 783.42 MeV}$, is shown.]{The histograms show bootstrap-distributions for the real- and imaginary parts of phase-constrained $S$-, $P$- and $D$-wave multipoles, for a TPWA bootstrap-analysis of photoproduction data in the second resonance region (see section \ref{subsec:2ndResRegionDataFits}). The fourth energy-bin, $E_{\gamma }\text{ = 783.42 MeV}$, is shown. An ensemble of $B=2000$ bootstrap-replicates has been the basis of these results. \newline
The distributions have been normalized to $1$ via use of the object \textit{HistogramDistribution} in MATHEMATICA \cite{Mathematica8,Mathematica11,MathematicaLanguage,MathematicaBonnLicense}. Thus, $y$-axes are labelled as \textit{PDF}. The mean of each distribution is shown as a red solid line, while the $0.16$- and $0.84$-quantiles are indicated by blue dash-dotted lines. The global minimum of the fit to the original data is plotted as a cyan-colored dashed horizontal line.}
\label{fig:BootstrapHistos2ndResRegionEnergy4}
\end{figure}

\clearpage

\begin{figure}[h]
\begin{overpic}[width=0.325\textwidth]{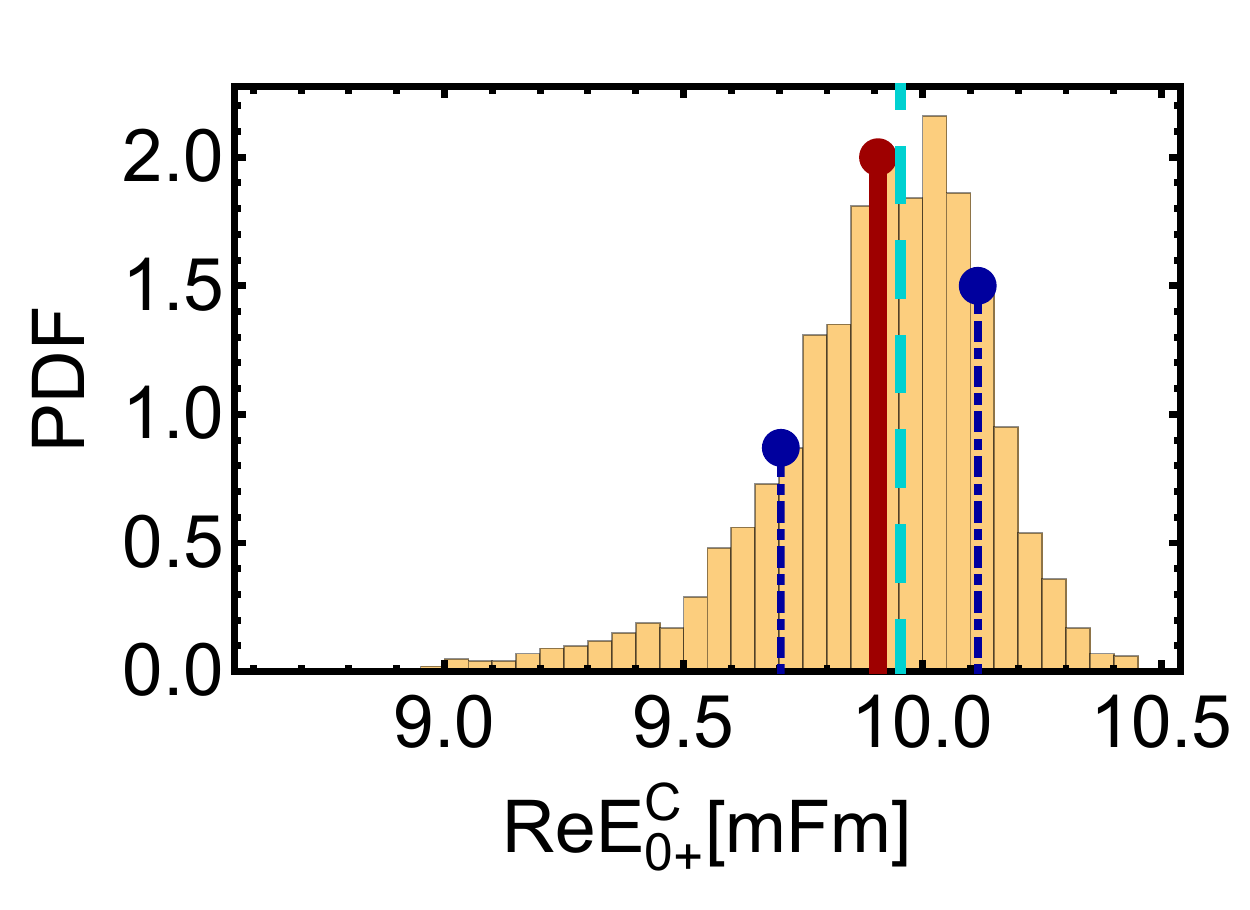}
 \end{overpic}
\begin{overpic}[width=0.325\textwidth]{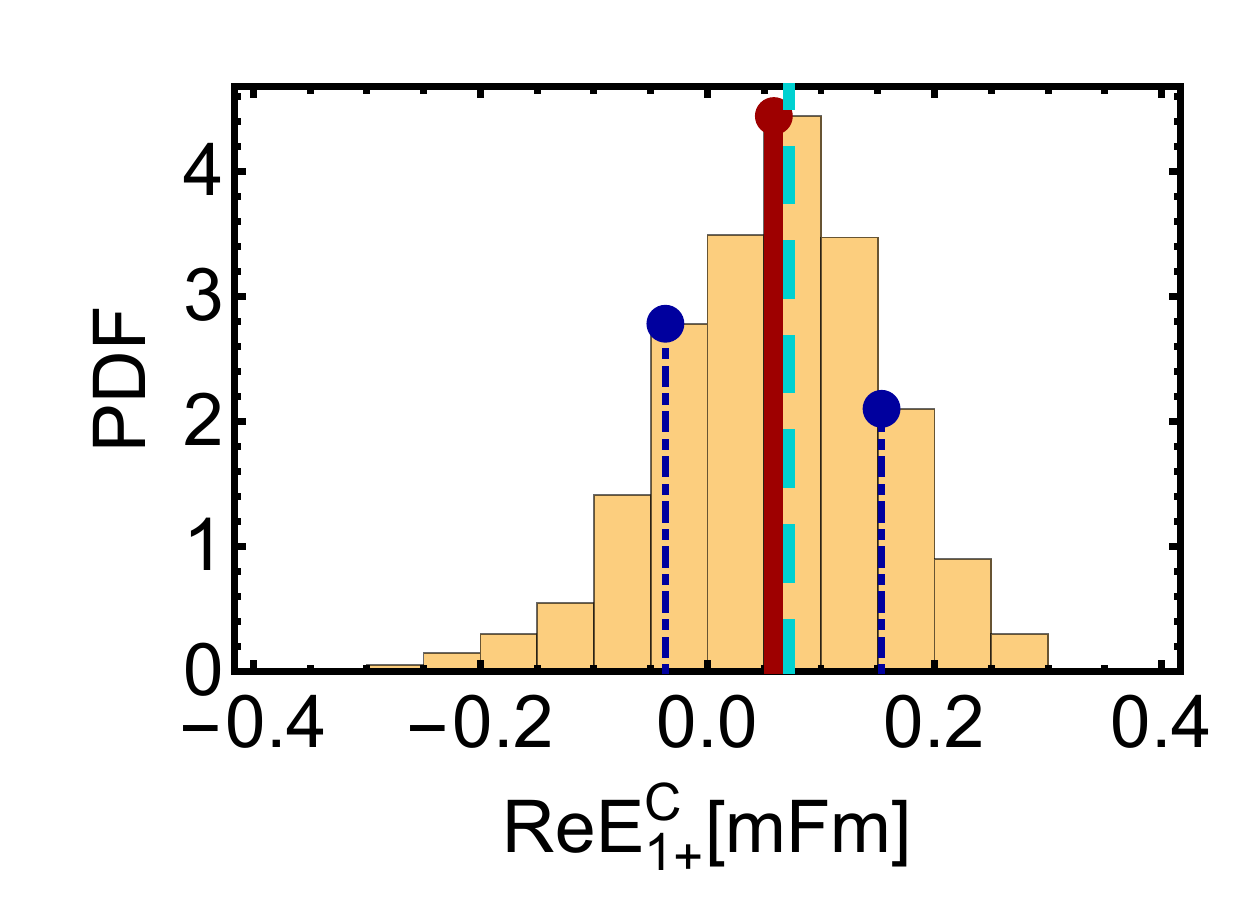}
 \end{overpic}
\begin{overpic}[width=0.325\textwidth]{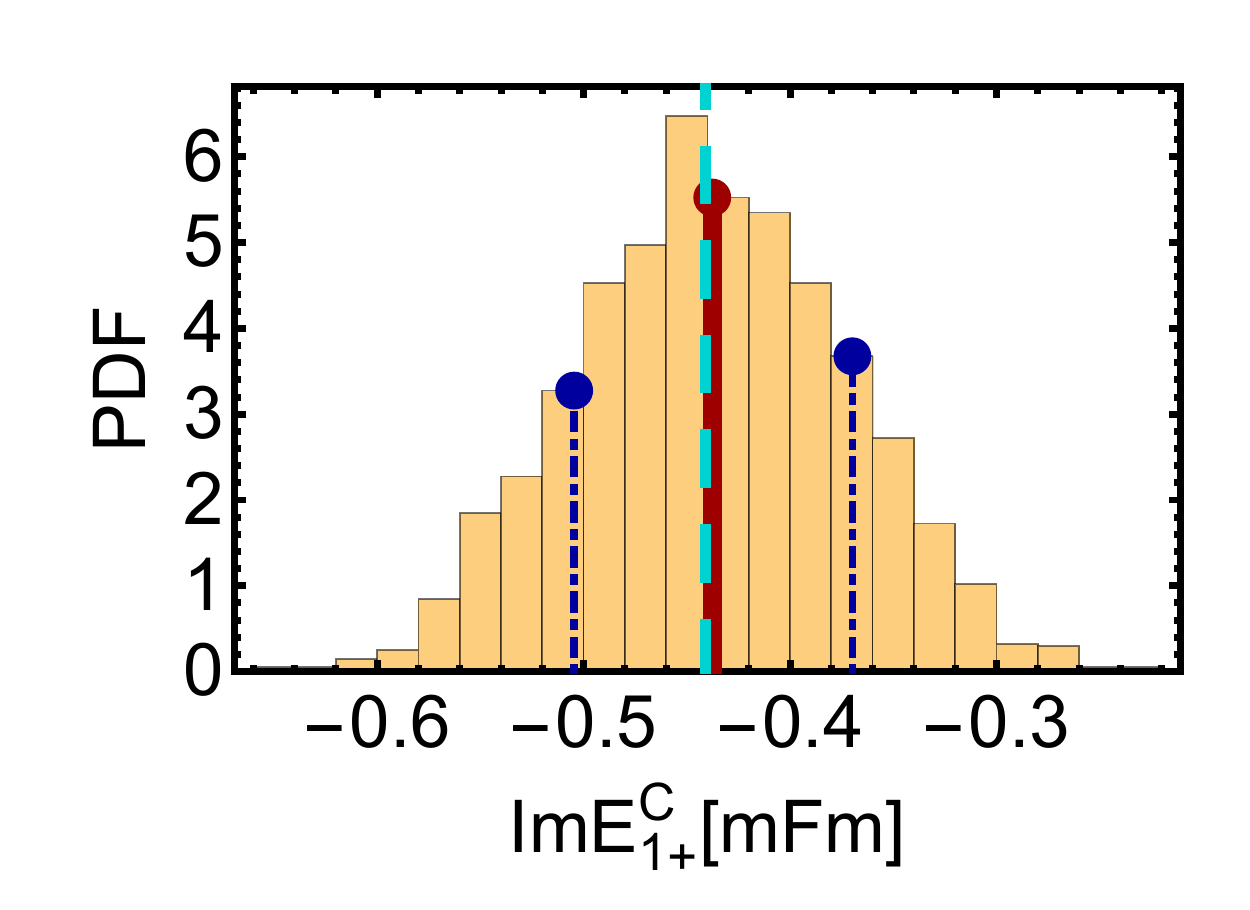}
 \end{overpic} \\
\begin{overpic}[width=0.325\textwidth]{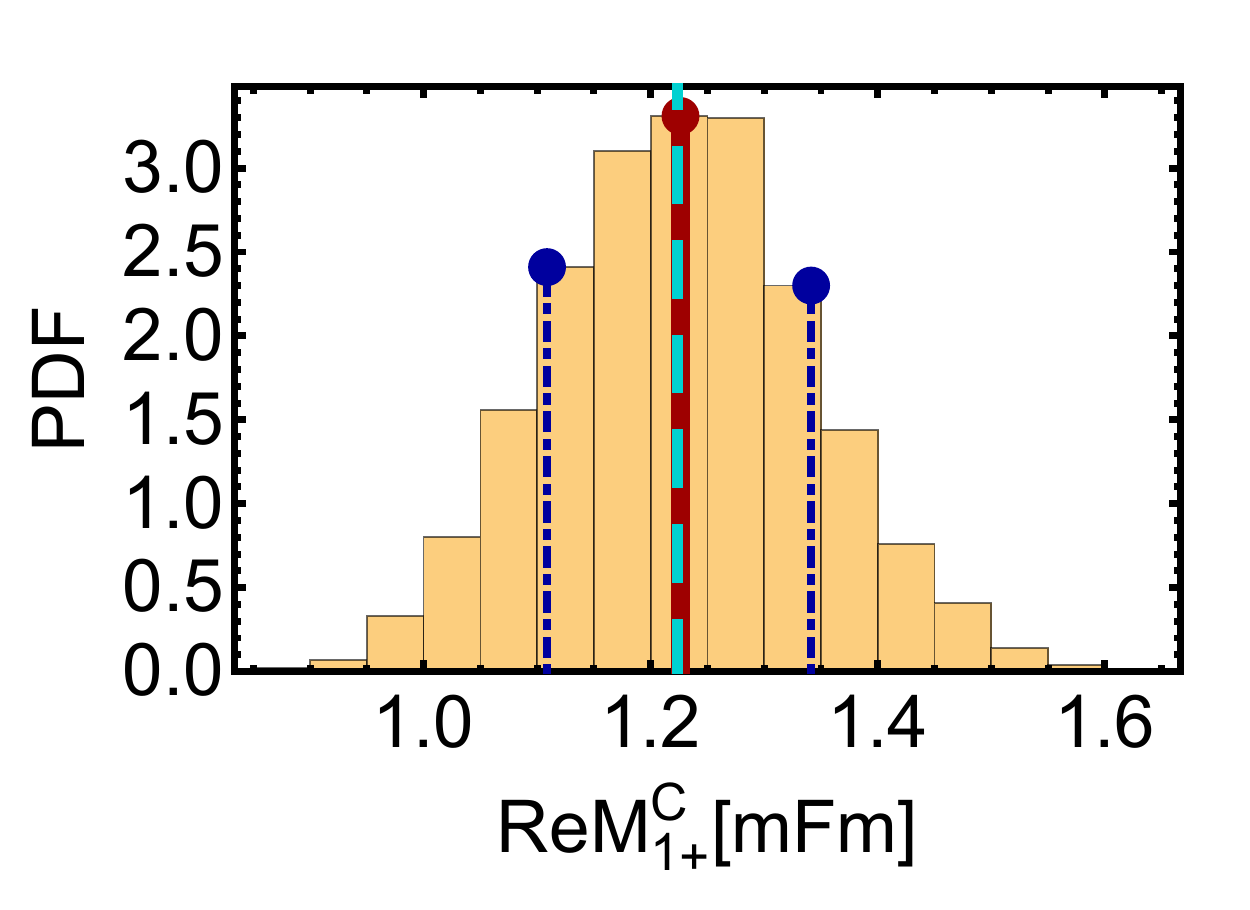}
 \end{overpic}
\begin{overpic}[width=0.325\textwidth]{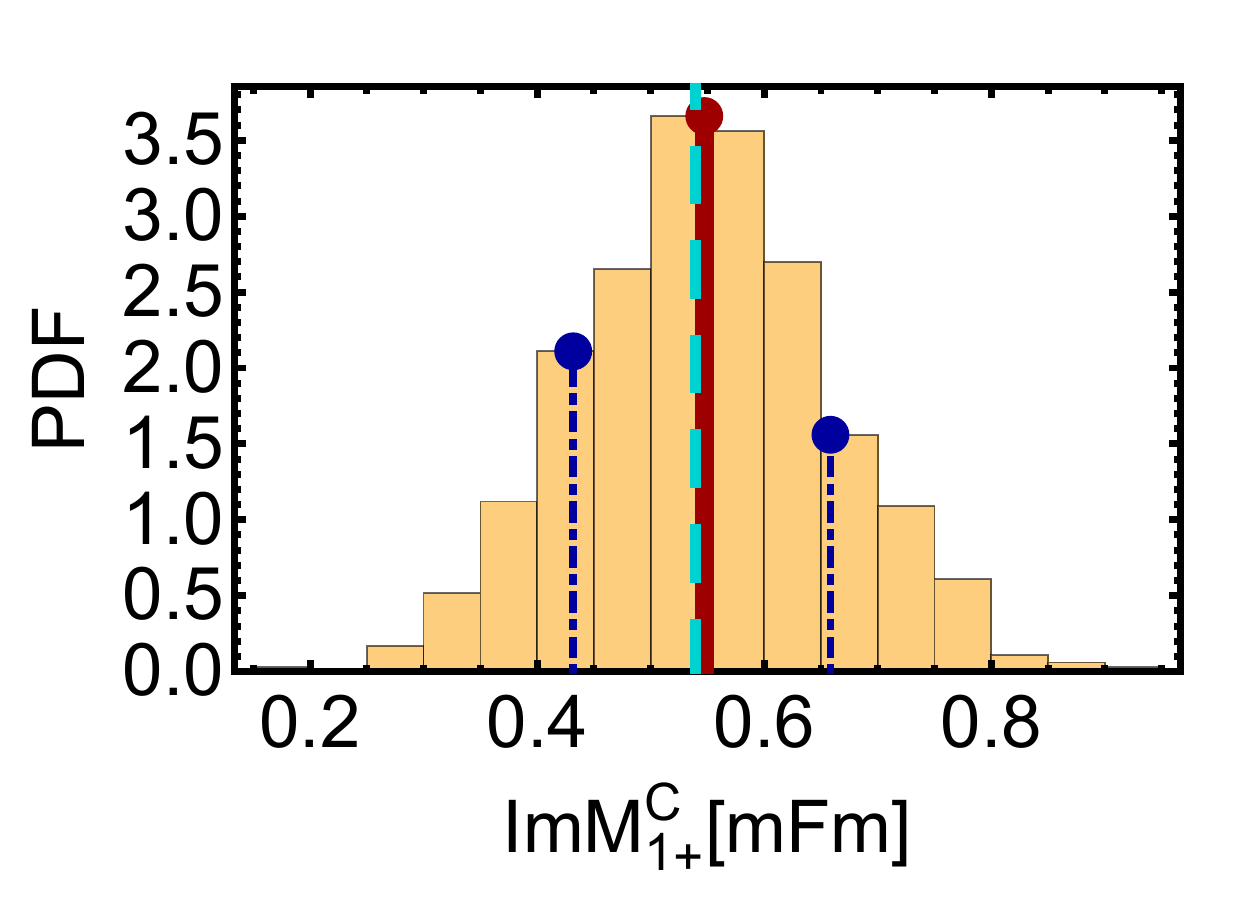}
 \end{overpic}
\begin{overpic}[width=0.325\textwidth]{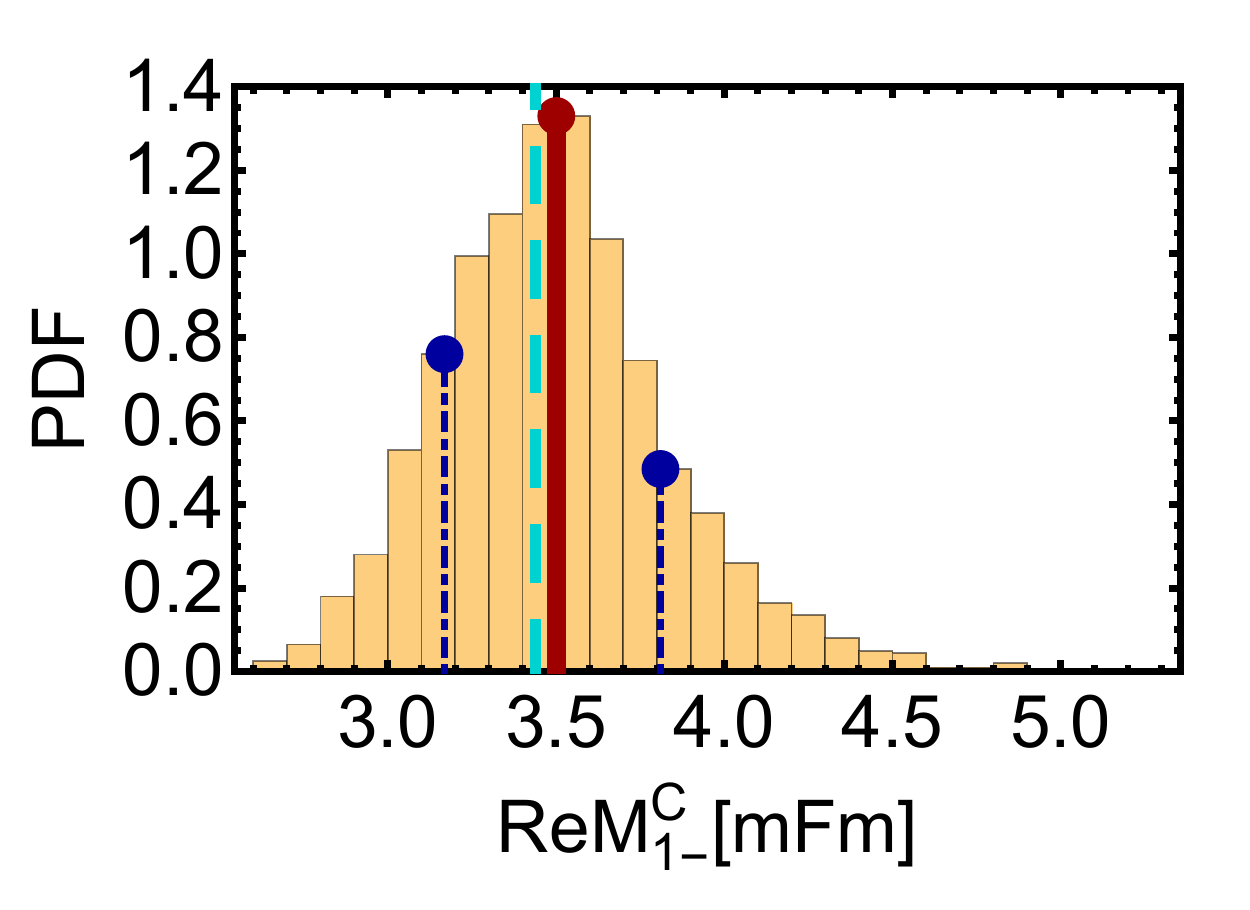}
 \end{overpic} \\
\begin{overpic}[width=0.325\textwidth]{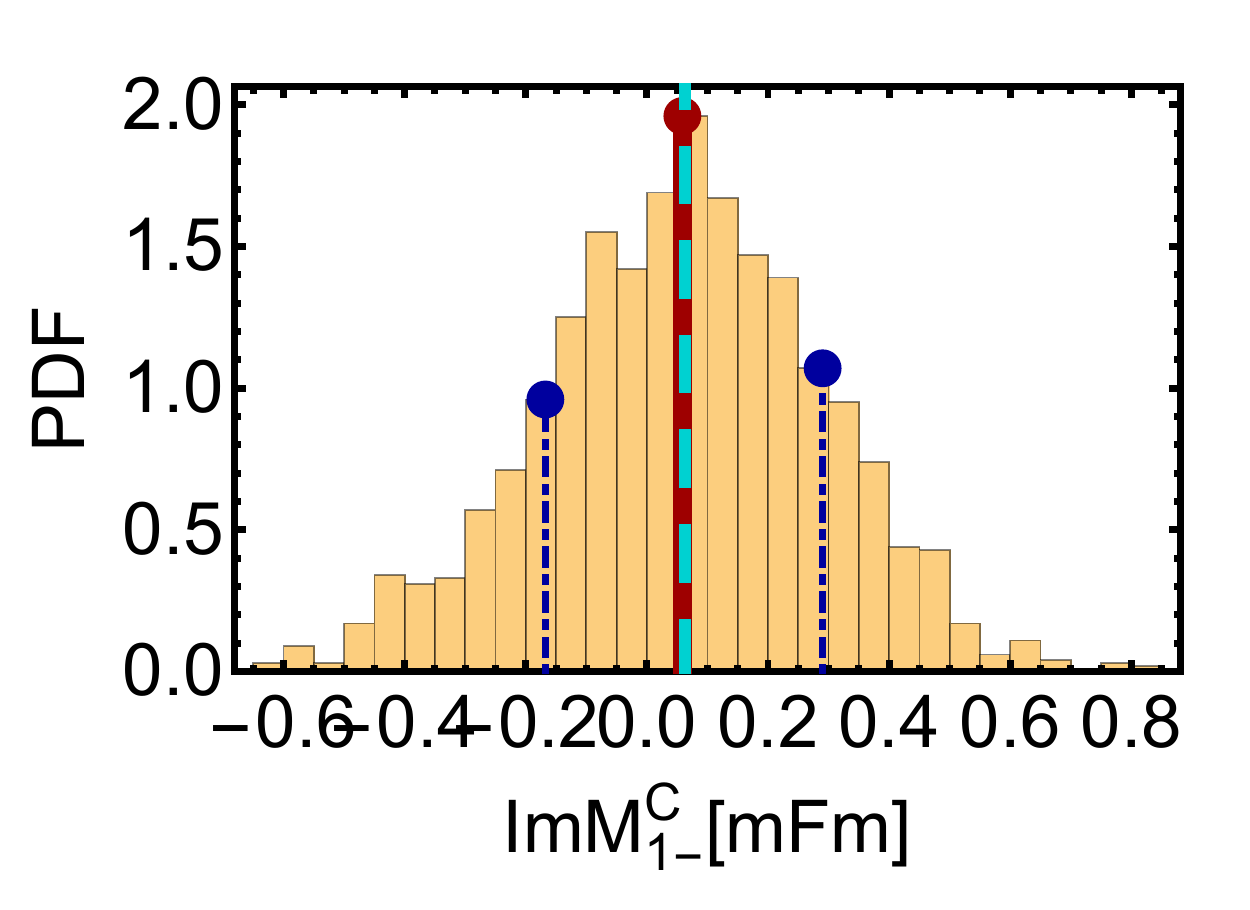}
 \end{overpic}
\begin{overpic}[width=0.325\textwidth]{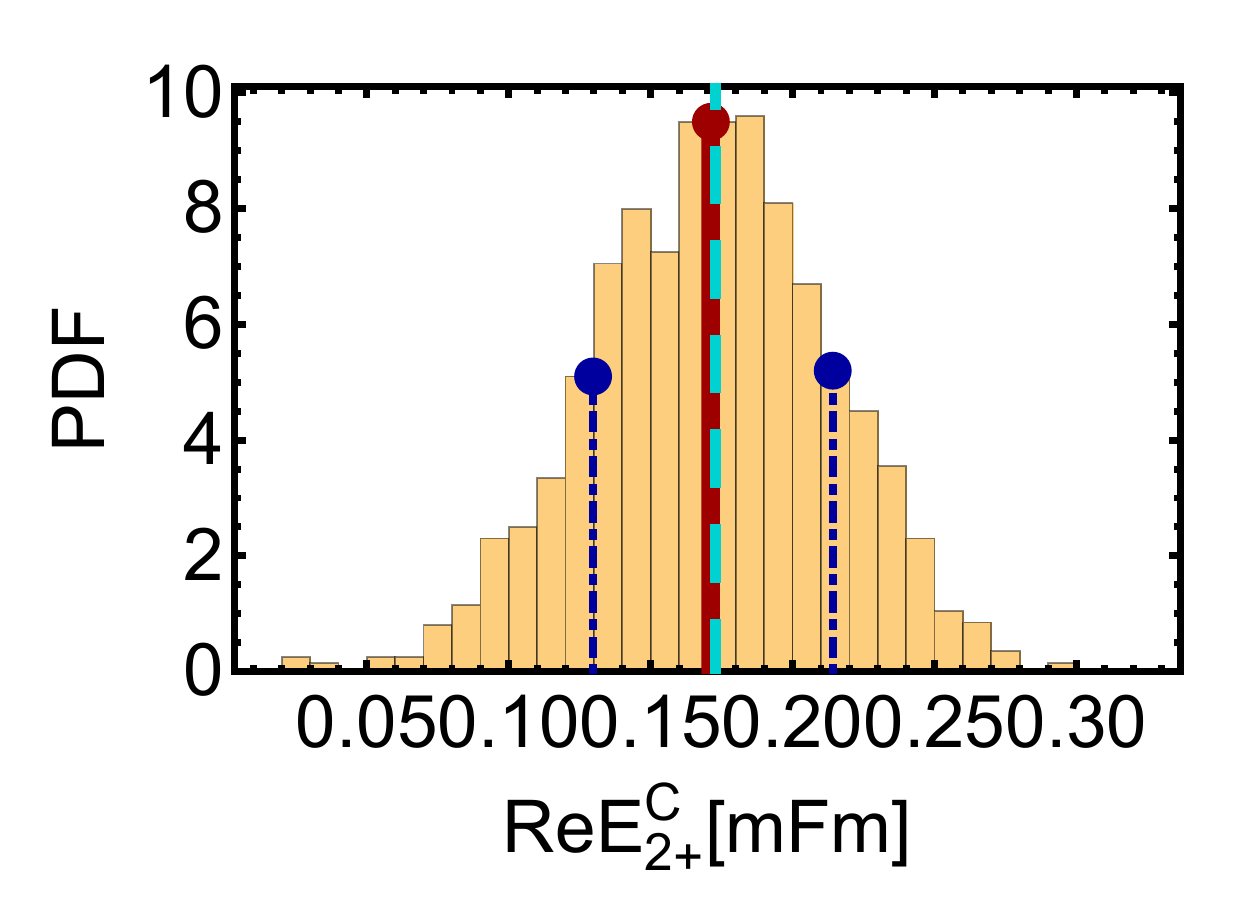}
 \end{overpic}
\begin{overpic}[width=0.325\textwidth]{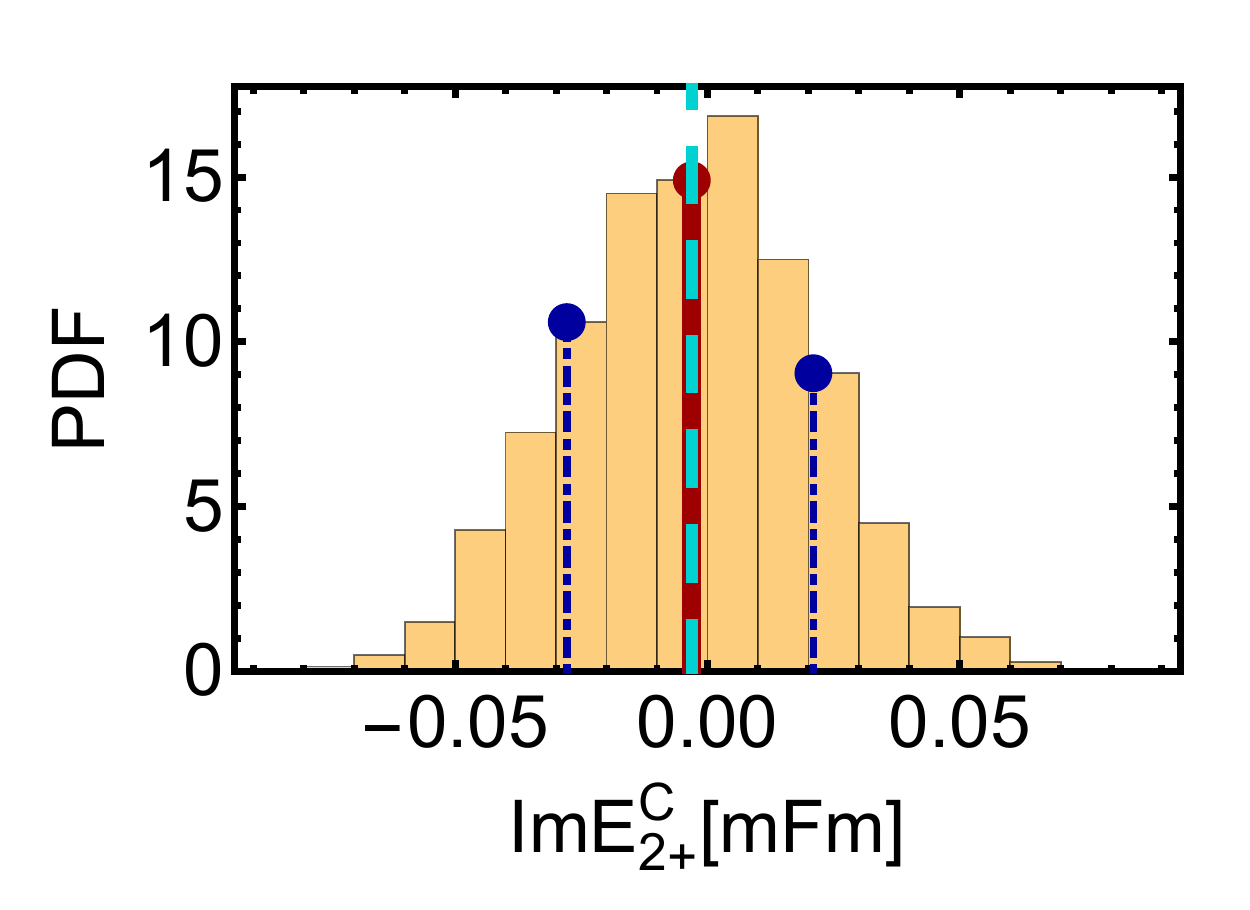}
 \end{overpic} \\
\begin{overpic}[width=0.325\textwidth]{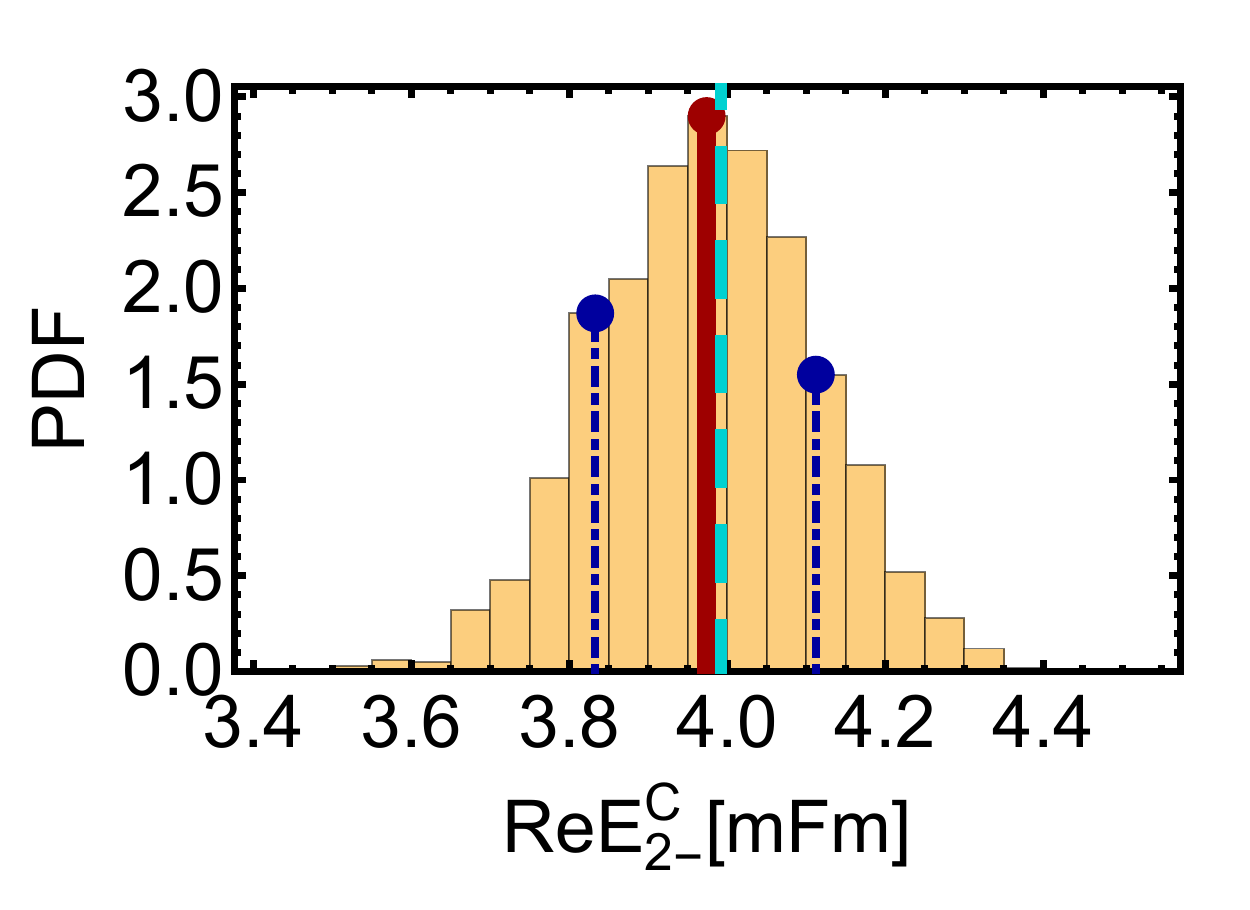}
 \end{overpic}
\begin{overpic}[width=0.325\textwidth]{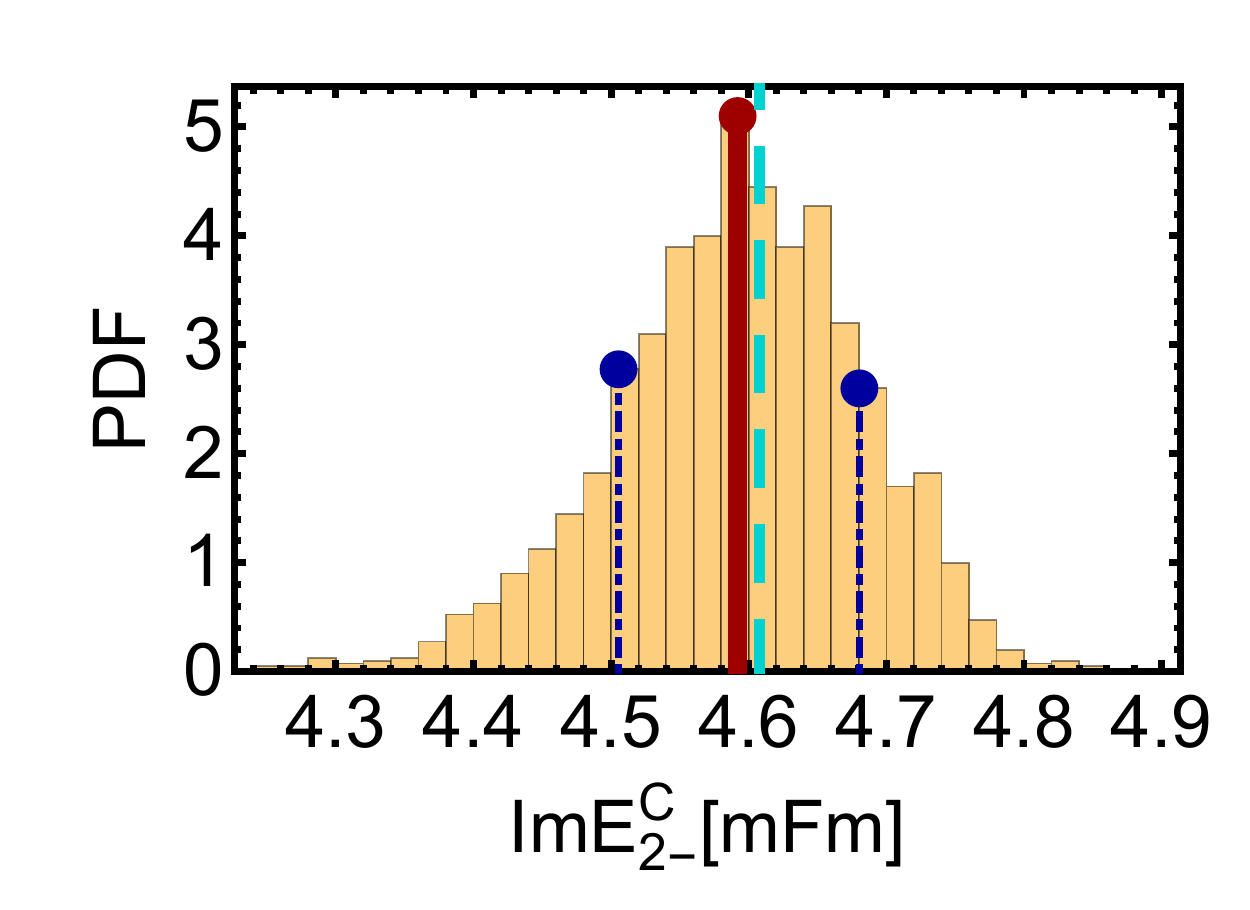}
 \end{overpic}
\begin{overpic}[width=0.325\textwidth]{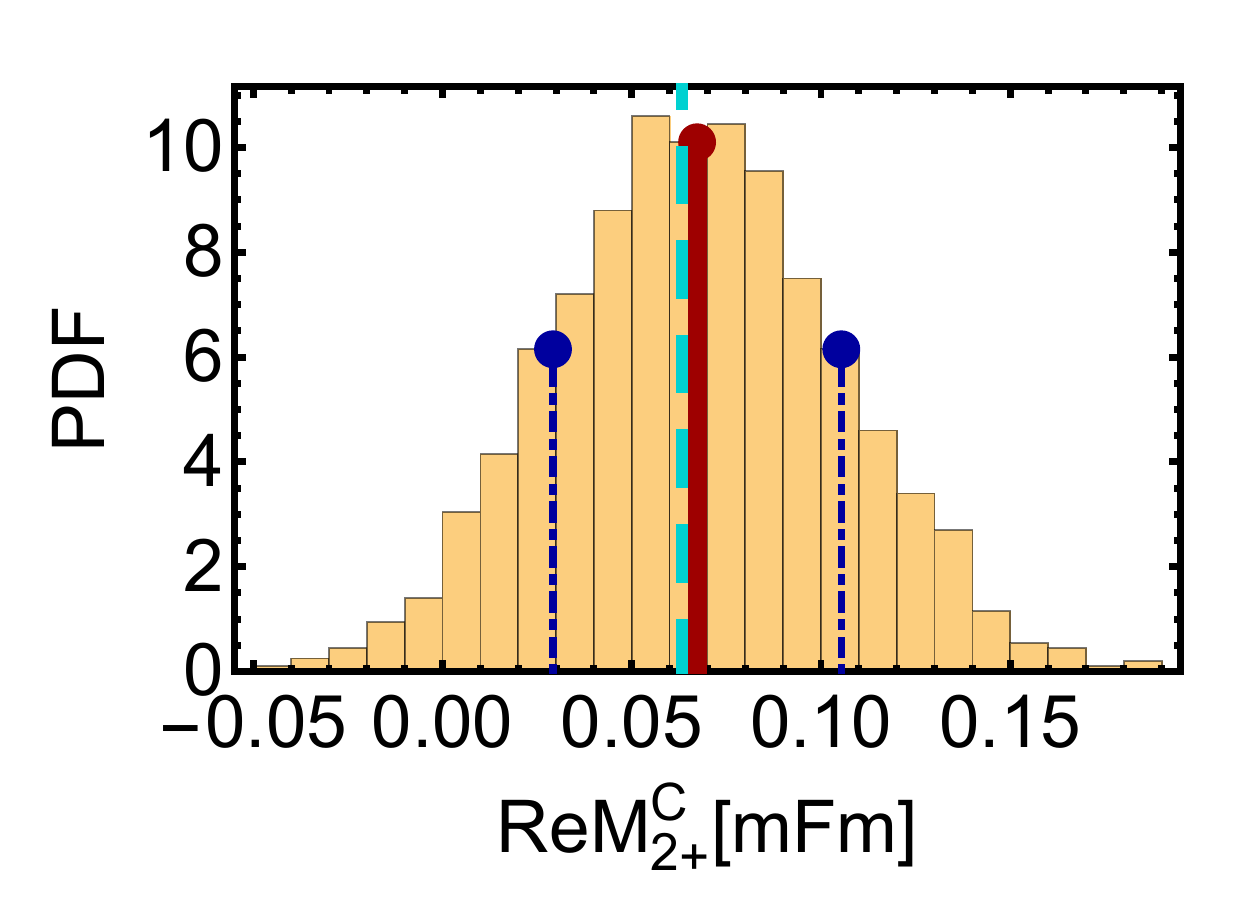}
 \end{overpic} \\
\begin{overpic}[width=0.325\textwidth]{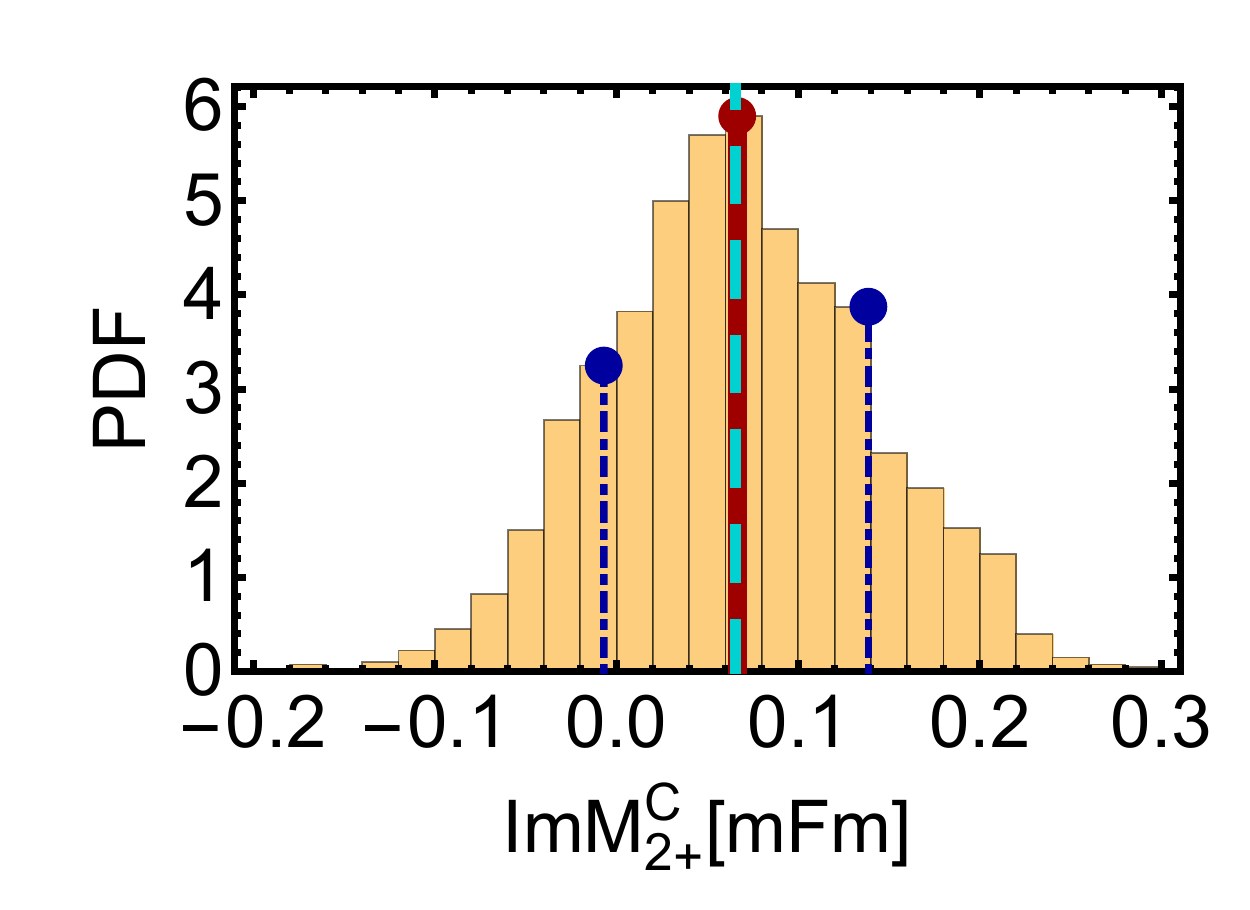}
 \end{overpic}
\begin{overpic}[width=0.325\textwidth]{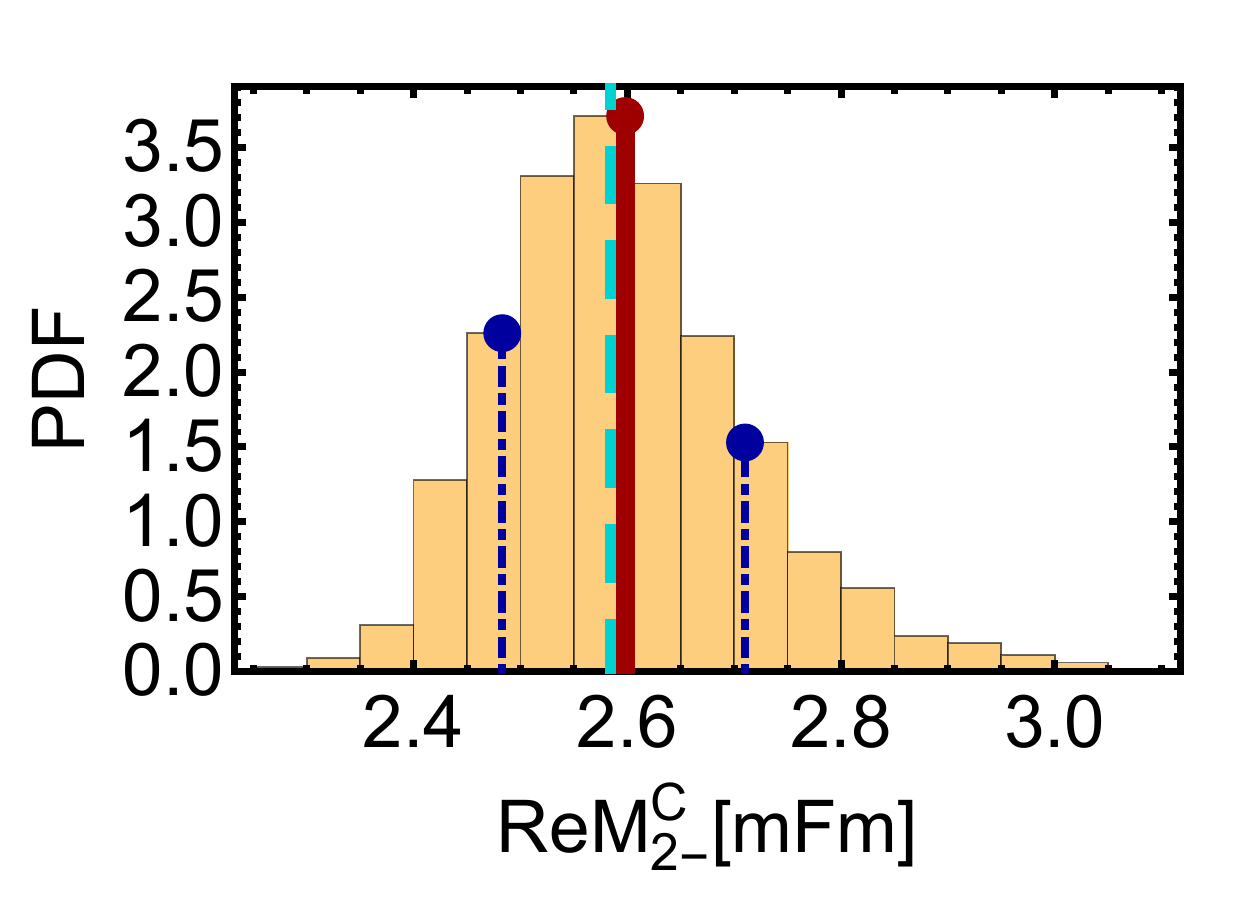}
 \end{overpic}
\begin{overpic}[width=0.325\textwidth]{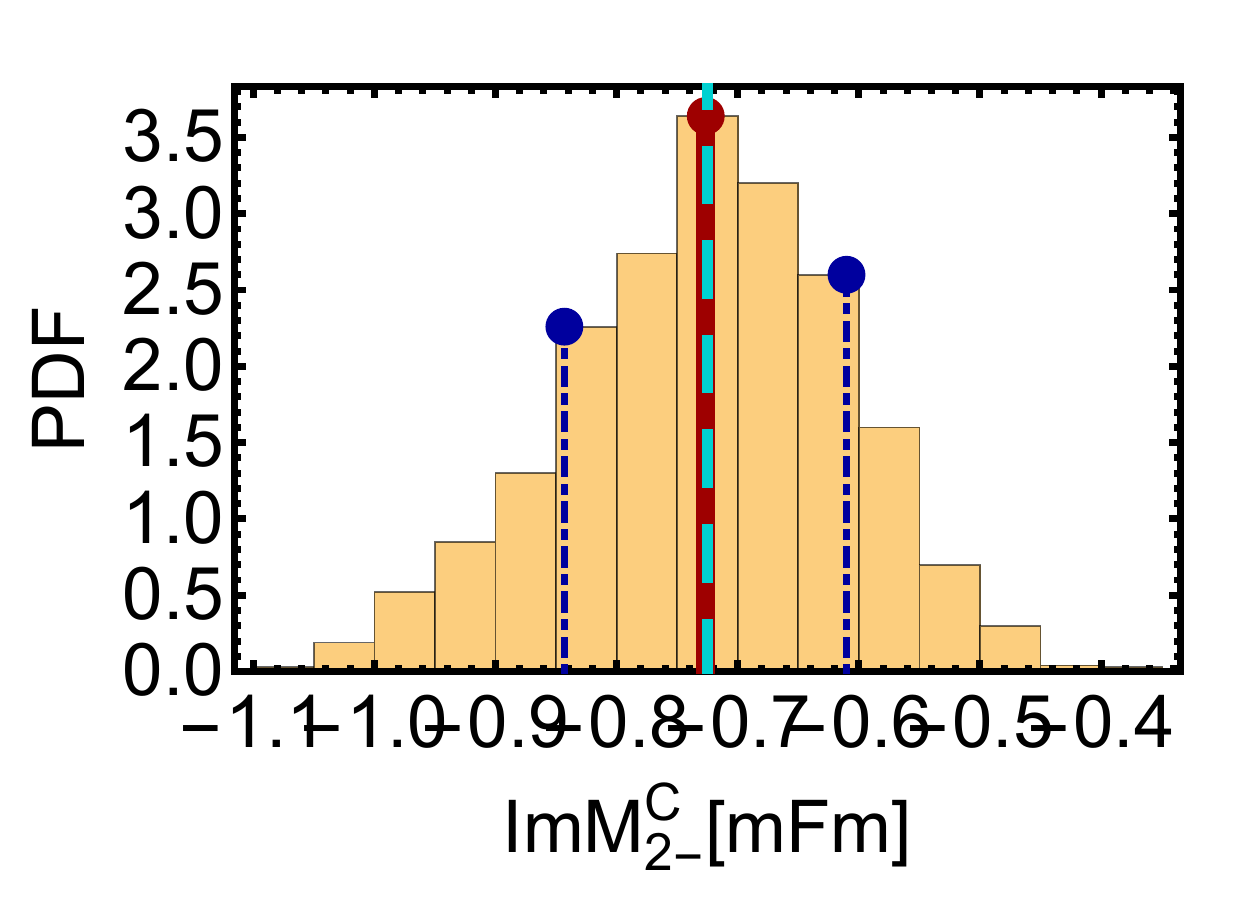}
 \end{overpic}
\caption[Bootstrap-distributions for multipole fit-parameters in an analysis of photoproduction data on the second resonance region. The fifth energy-bin, \newline $E_{\gamma }\text{ = 815.92 MeV}$, is shown.]{The histograms show bootstrap-distributions for the real- and imaginary parts of phase-constrained $S$-, $P$- and $D$-wave multipoles, for a TPWA bootstrap-analysis of photoproduction data in the second resonance region (see section \ref{subsec:2ndResRegionDataFits}). The fifth energy-bin, $E_{\gamma }\text{ = 815.92 MeV}$, is shown. An ensemble of $B=2000$ bootstrap-replicates has been the basis of these results. \newline
The distributions have been normalized to $1$ via use of the object \textit{HistogramDistribution} in MATHEMATICA \cite{Mathematica8,Mathematica11,MathematicaLanguage,MathematicaBonnLicense}. Thus, $y$-axes are labelled as \textit{PDF}. The mean of each distribution is shown as a red solid line, while the $0.16$- and $0.84$-quantiles are indicated by blue dash-dotted lines. The global minimum of the fit to the original data is plotted as a cyan-colored dashed horizontal line.}
\label{fig:BootstrapHistos2ndResRegionEnergy5}
\end{figure}

\clearpage

\begin{figure}[h]
\begin{overpic}[width=0.325\textwidth]{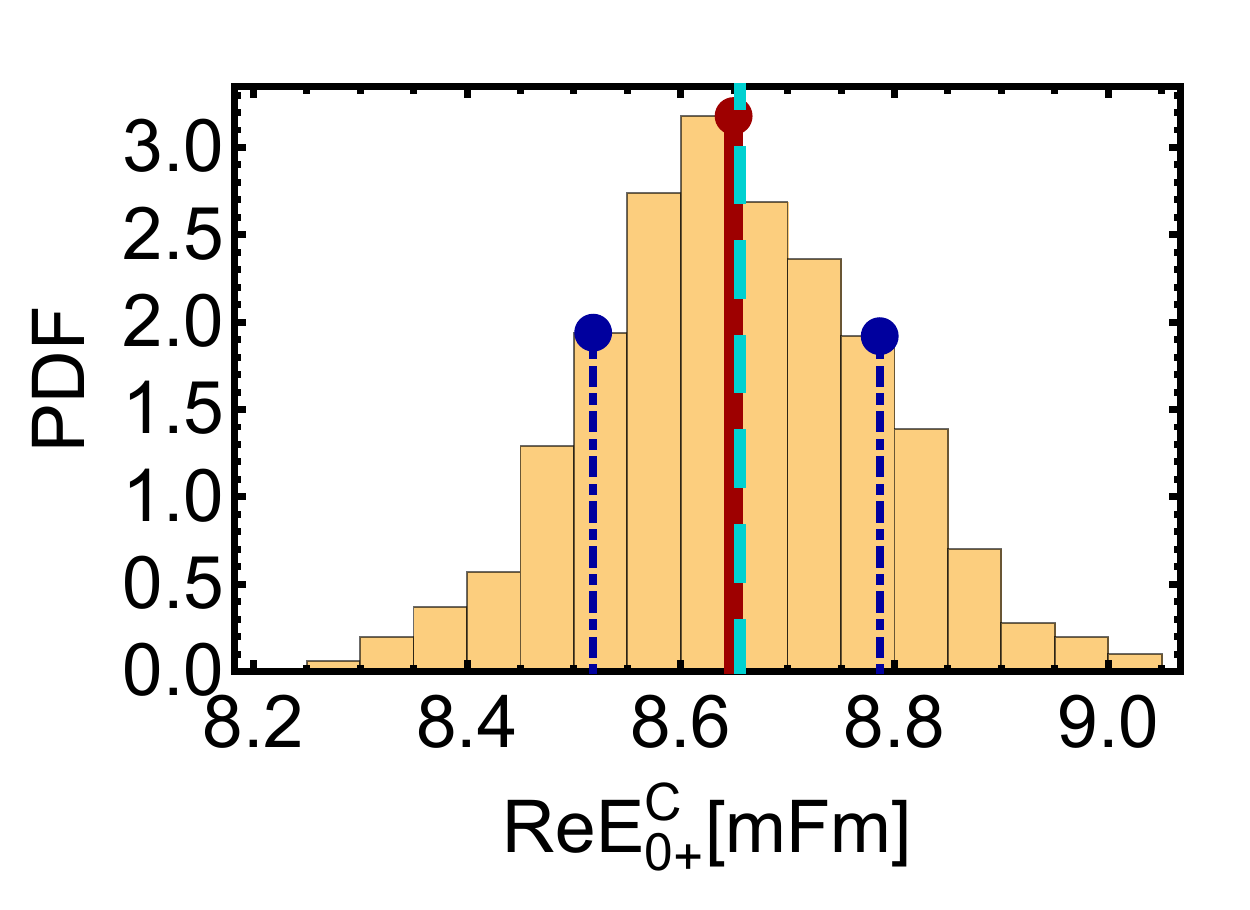}
 \end{overpic}
\begin{overpic}[width=0.325\textwidth]{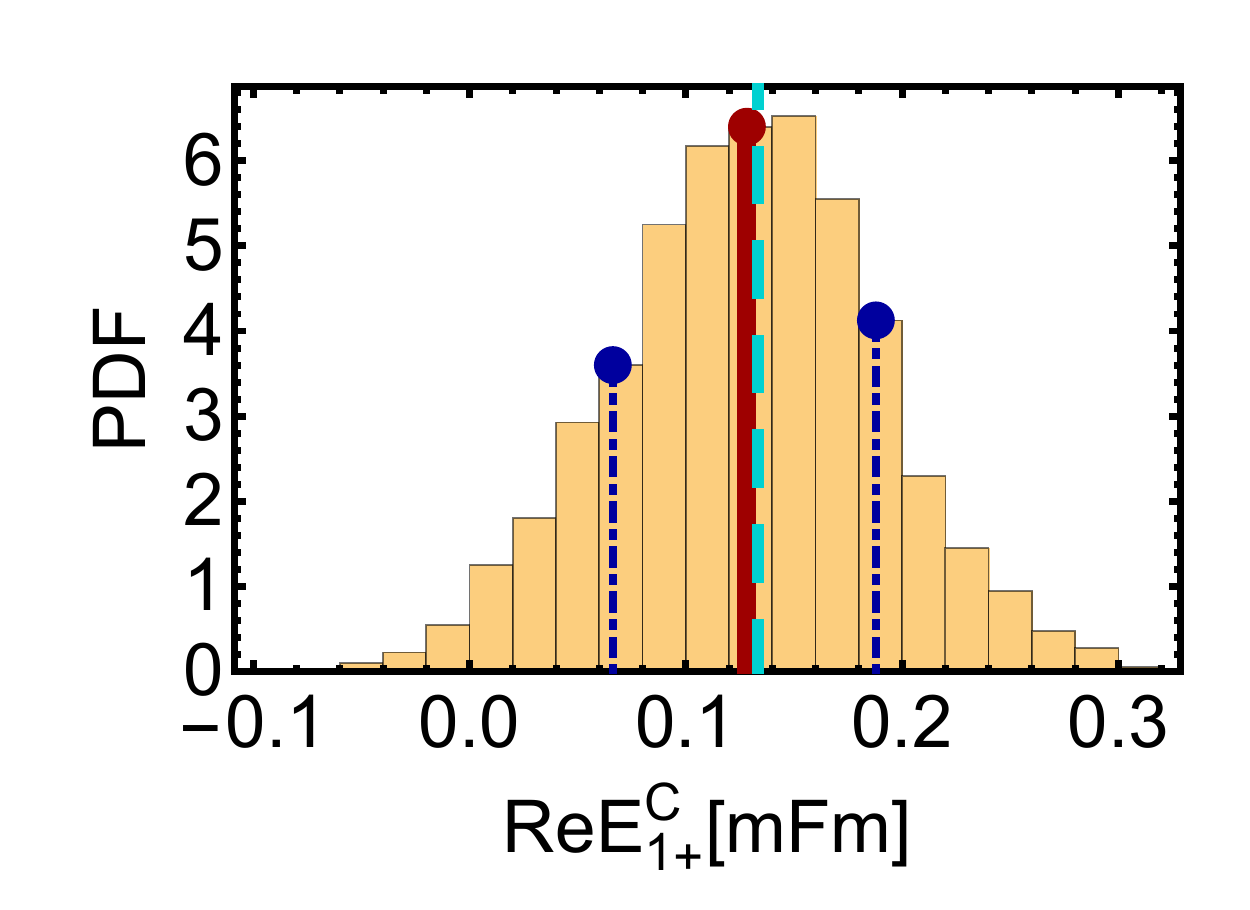}
 \end{overpic}
\begin{overpic}[width=0.325\textwidth]{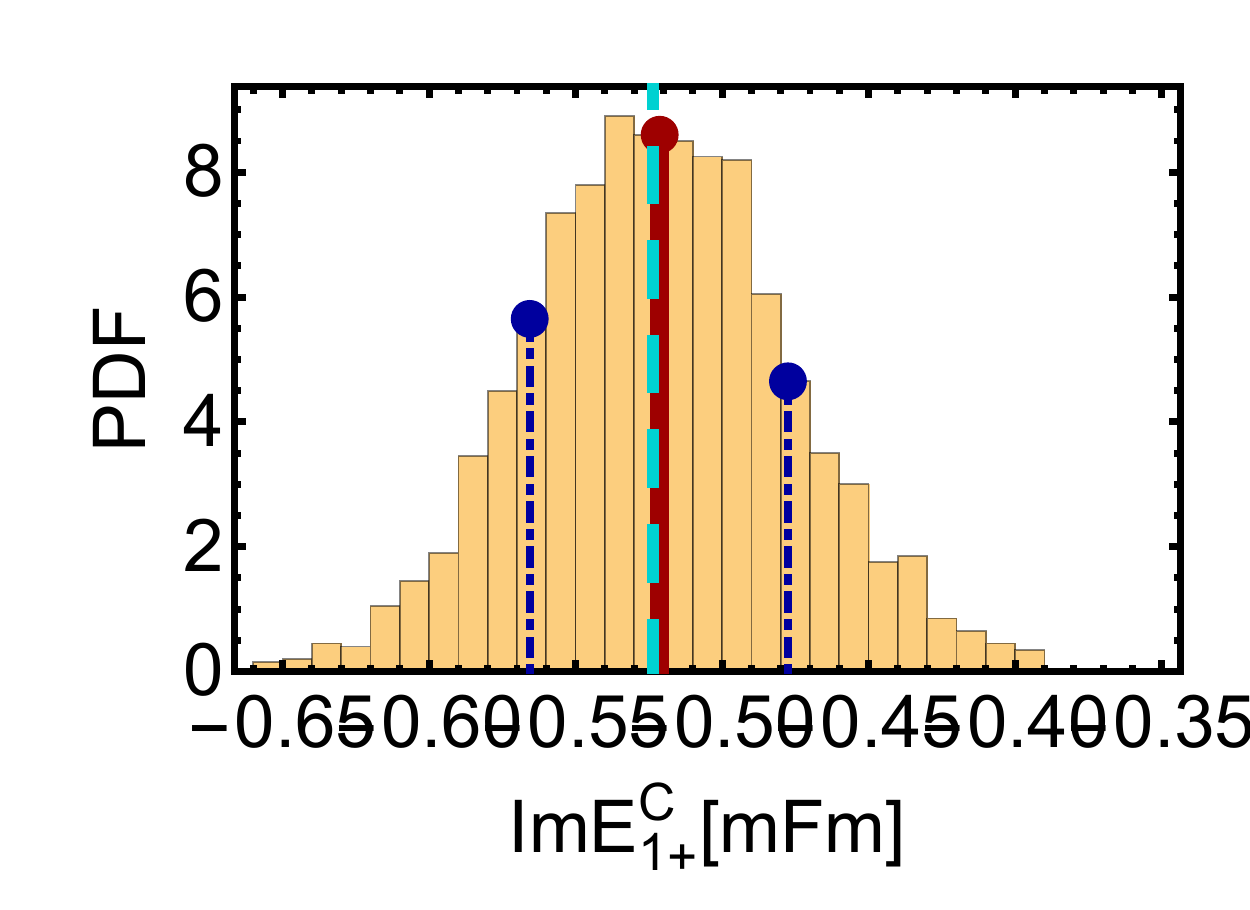}
 \end{overpic} \\
\begin{overpic}[width=0.325\textwidth]{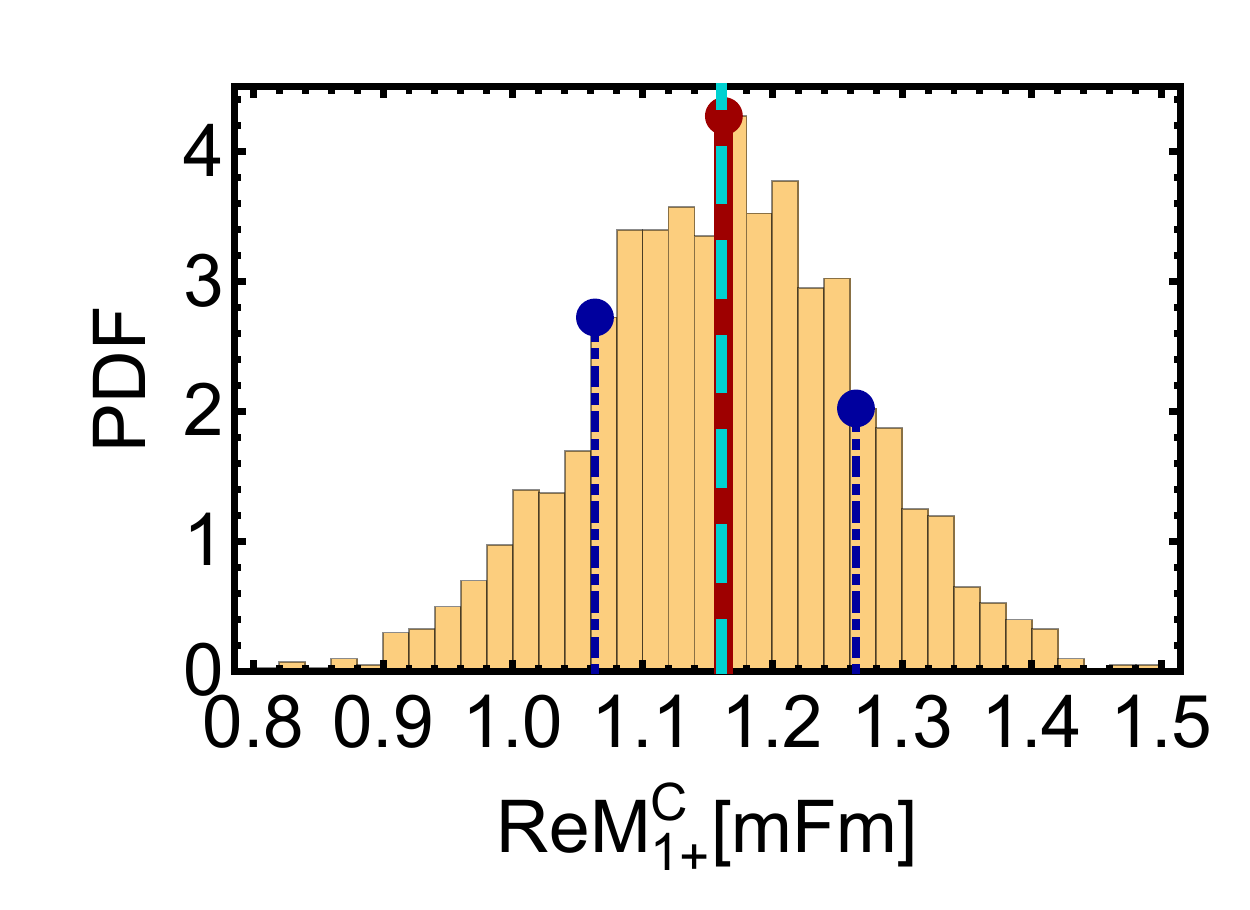}
 \end{overpic}
\begin{overpic}[width=0.325\textwidth]{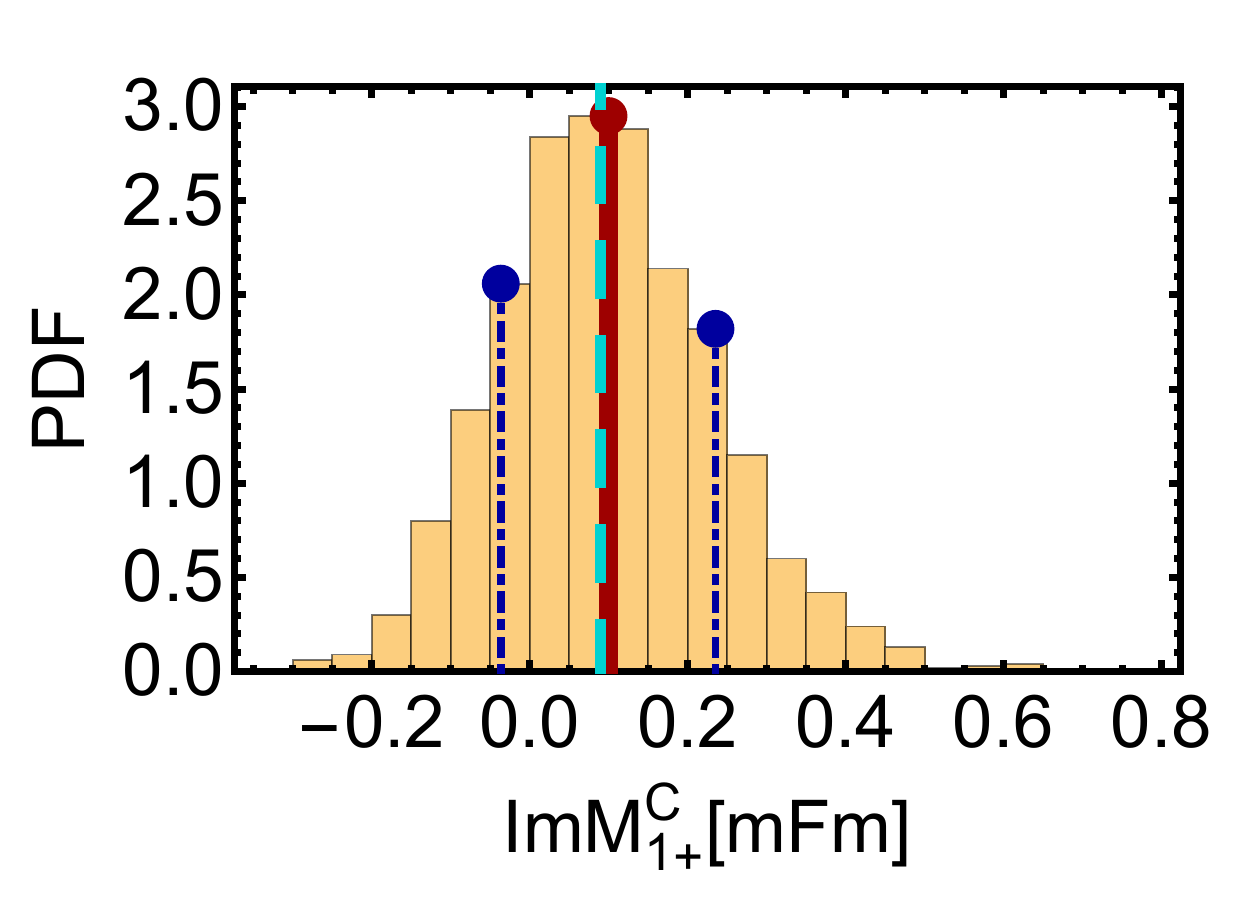}
 \end{overpic}
\begin{overpic}[width=0.325\textwidth]{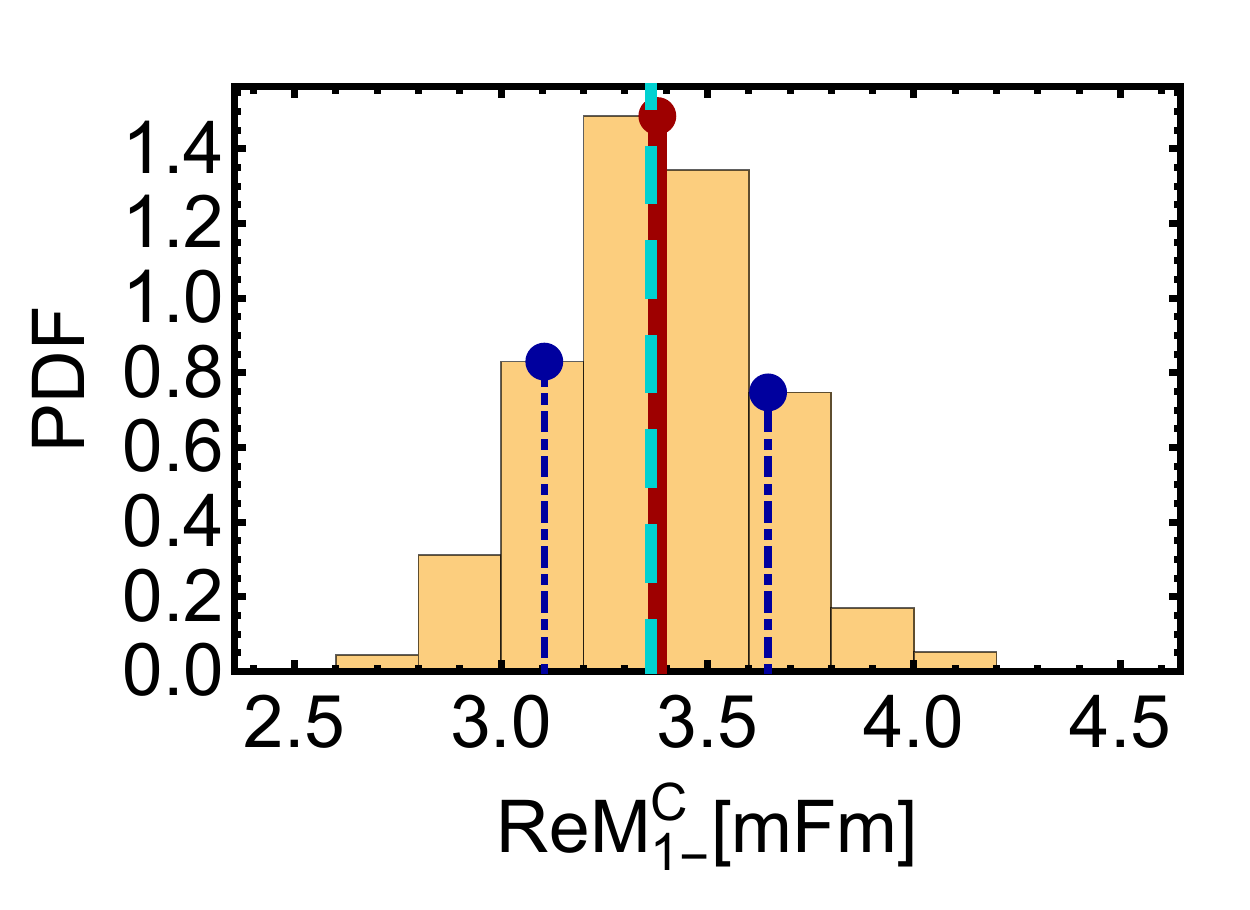}
 \end{overpic} \\
\begin{overpic}[width=0.325\textwidth]{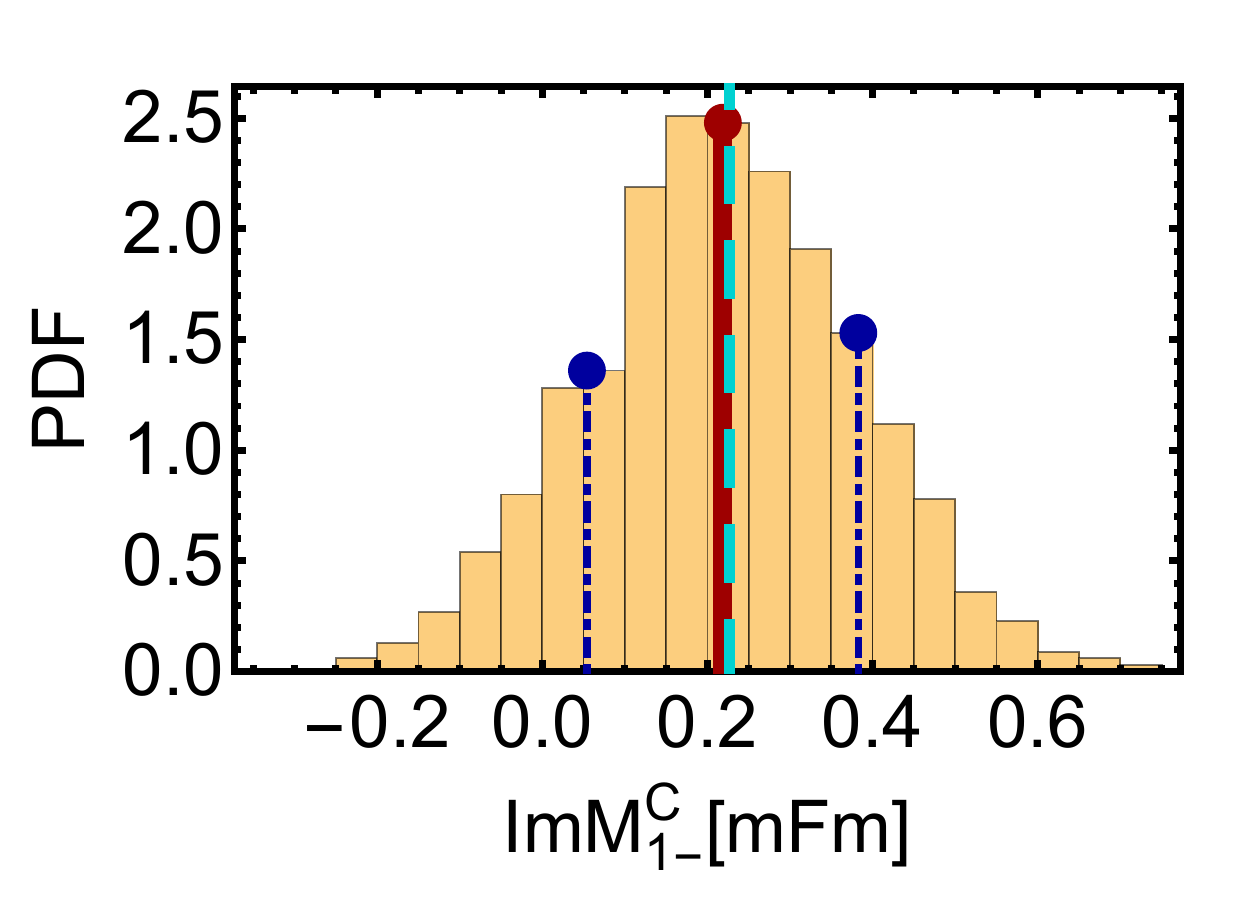}
 \end{overpic}
\begin{overpic}[width=0.325\textwidth]{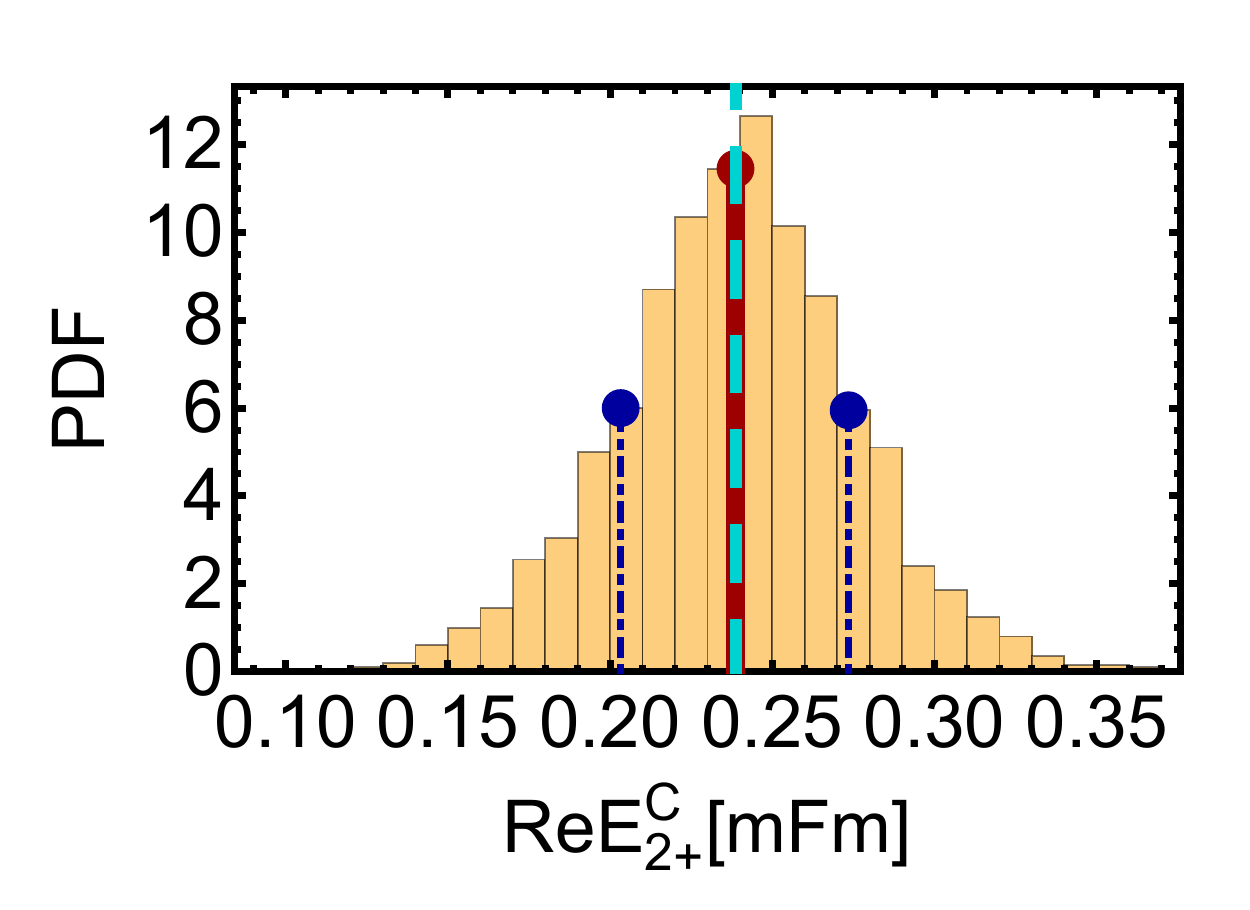}
 \end{overpic}
\begin{overpic}[width=0.325\textwidth]{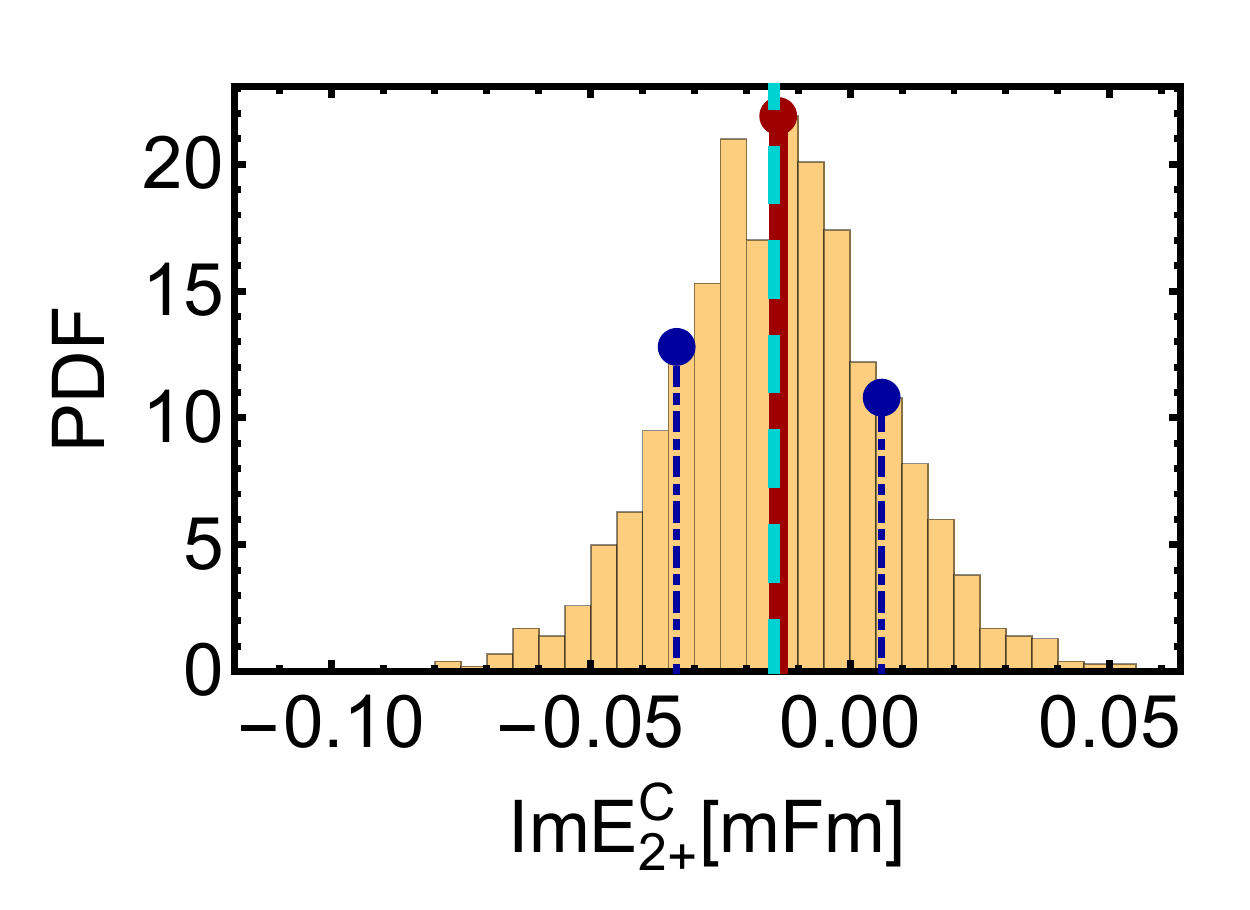}
 \end{overpic} \\
\begin{overpic}[width=0.325\textwidth]{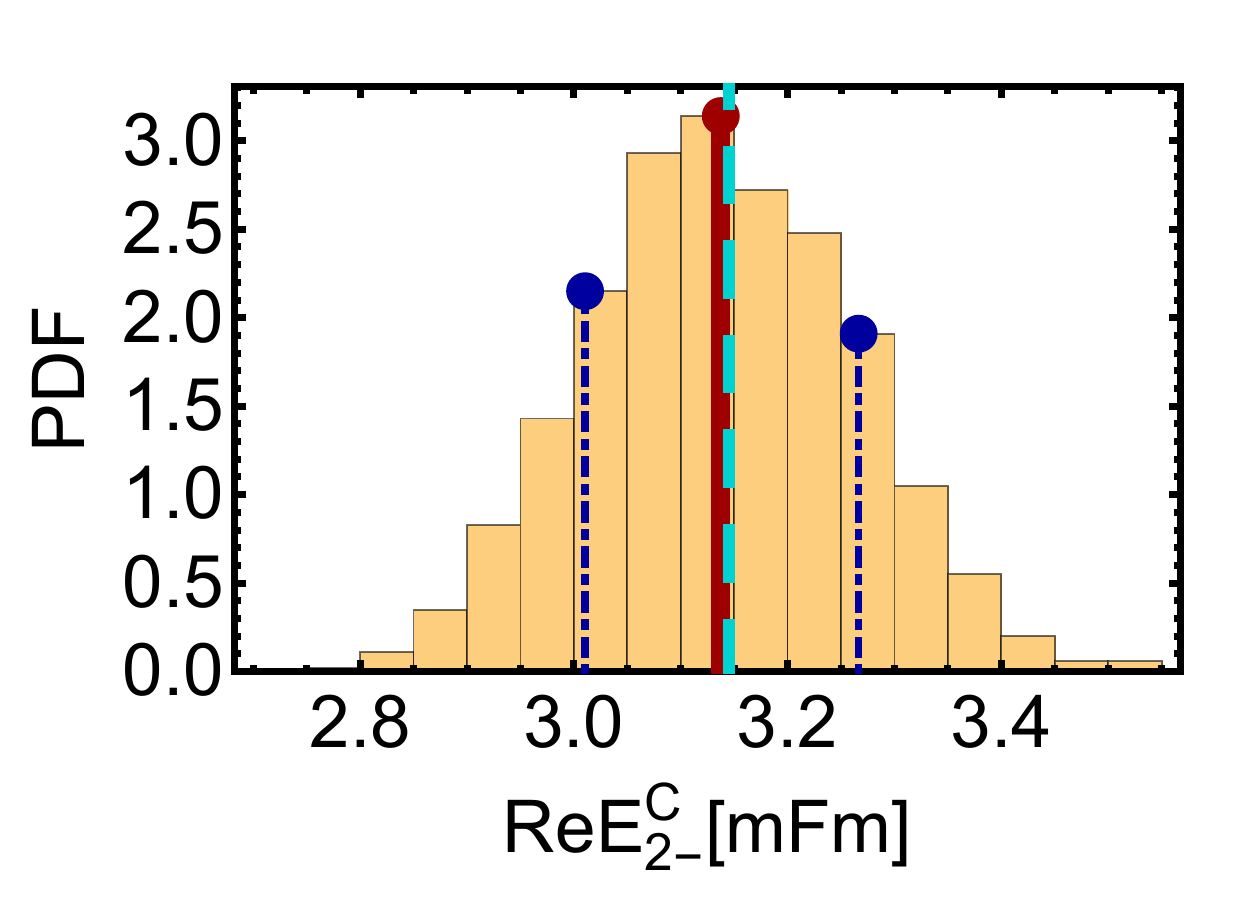}
 \end{overpic}
\begin{overpic}[width=0.325\textwidth]{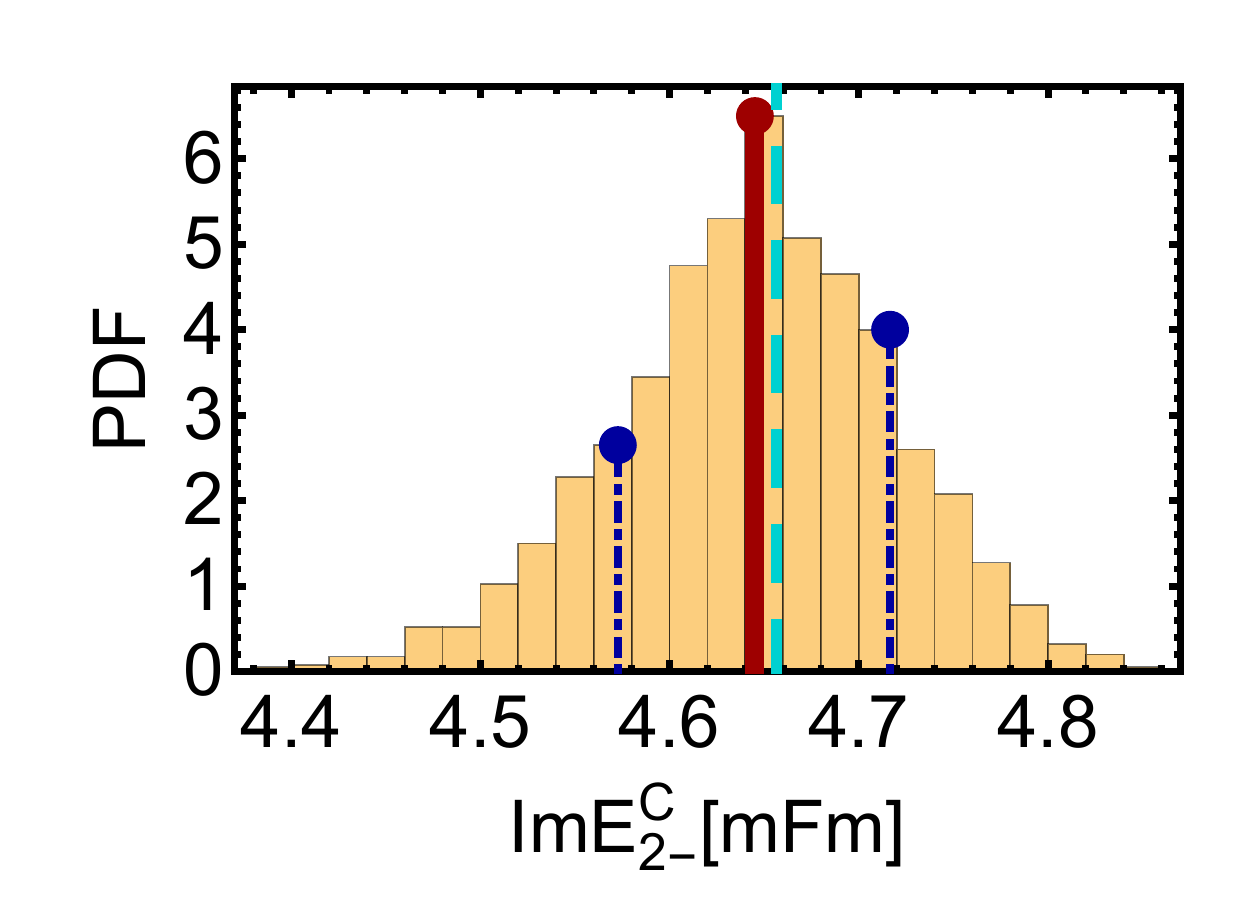}
 \end{overpic}
\begin{overpic}[width=0.325\textwidth]{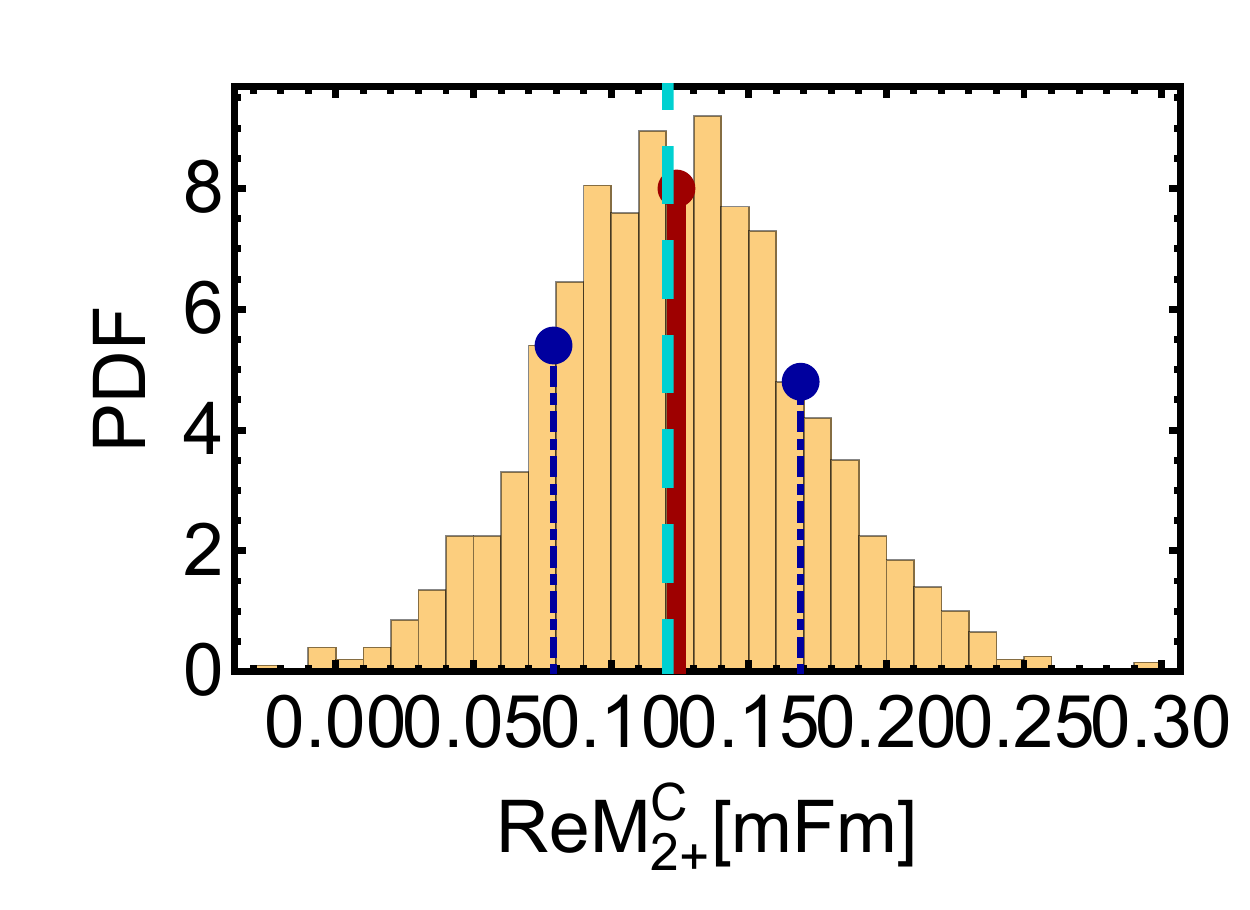}
 \end{overpic} \\
\begin{overpic}[width=0.325\textwidth]{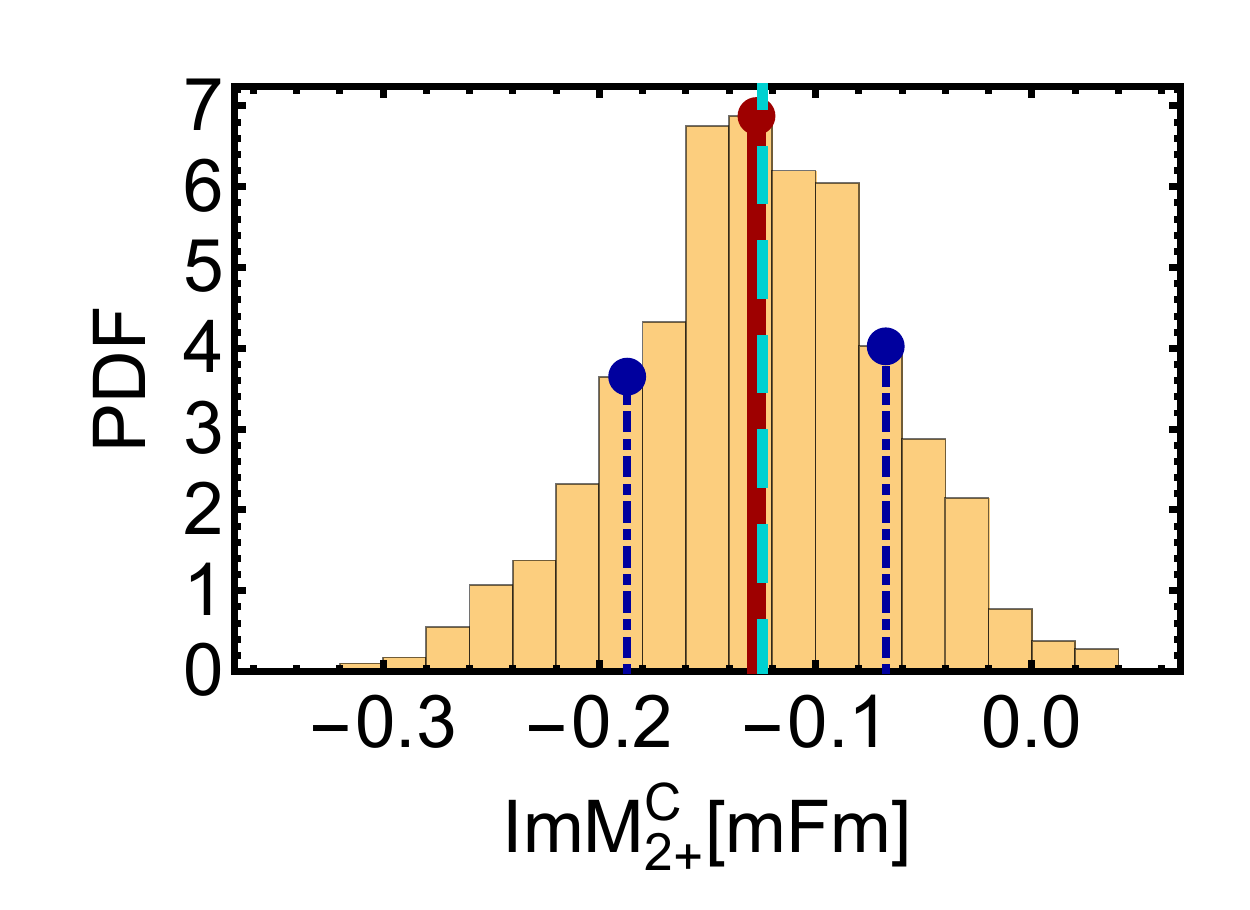}
 \end{overpic}
\begin{overpic}[width=0.325\textwidth]{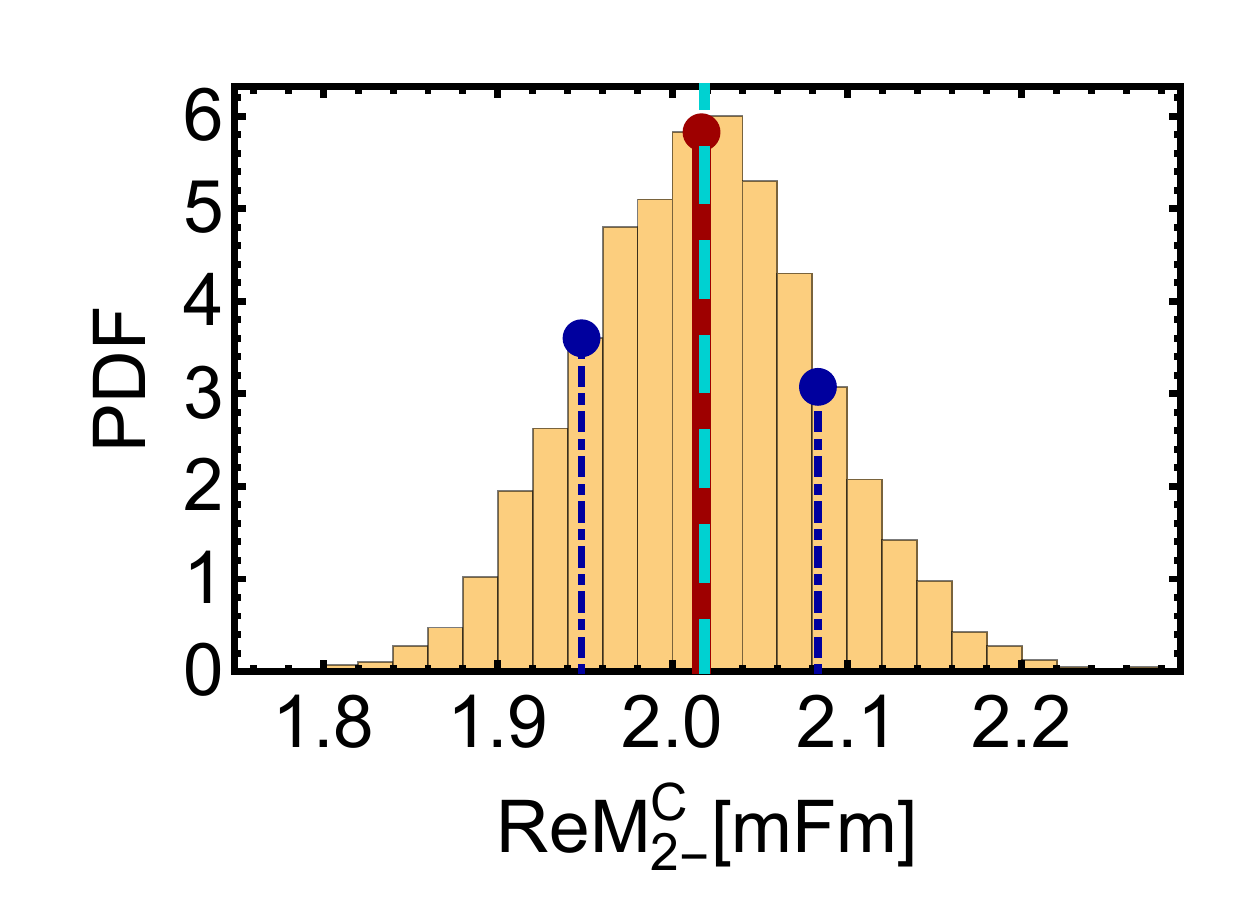}
 \end{overpic}
\begin{overpic}[width=0.325\textwidth]{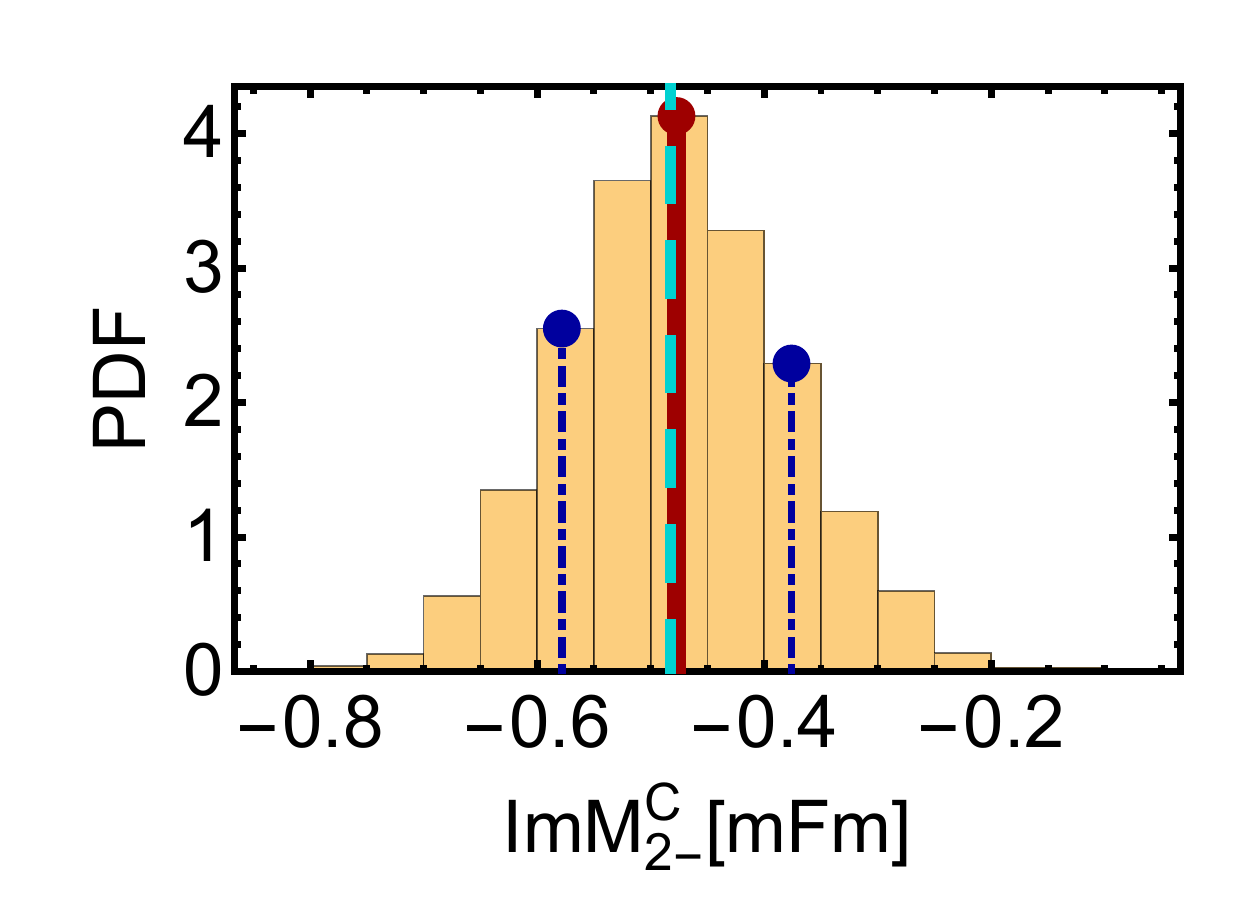}
 \end{overpic}
\caption[Bootstrap-distributions for multipole fit-parameters in an analysis of photoproduction data on the second resonance region. The sixth energy-bin, \newline $E_{\gamma }\text{ = 850.45 MeV}$, is shown.]{The histograms show bootstrap-distributions for the real- and imaginary parts of phase-constrained $S$-, $P$- and $D$-wave multipoles, for a TPWA bootstrap-analysis of photoproduction data in the second resonance region (see section \ref{subsec:2ndResRegionDataFits}). The sixth energy-bin, $E_{\gamma }\text{ = 850.45 MeV}$, is shown. An ensemble of $B=2000$ bootstrap-replicates has been the basis of these results. \newline
The distributions have been normalized to $1$ via use of the object \textit{HistogramDistribution} in MATHEMATICA \cite{Mathematica8,Mathematica11,MathematicaLanguage,MathematicaBonnLicense}. Thus, $y$-axes are labelled as \textit{PDF}. The mean of each distribution is shown as a red solid line, while the $0.16$- and $0.84$-quantiles are indicated by blue dash-dotted lines. The global minimum of the fit to the original data is plotted as a cyan-colored dashed horizontal line.}
\label{fig:BootstrapHistos2ndResRegionEnergy6}
\end{figure}

\clearpage

\begin{figure}[h]
\begin{overpic}[width=0.325\textwidth]{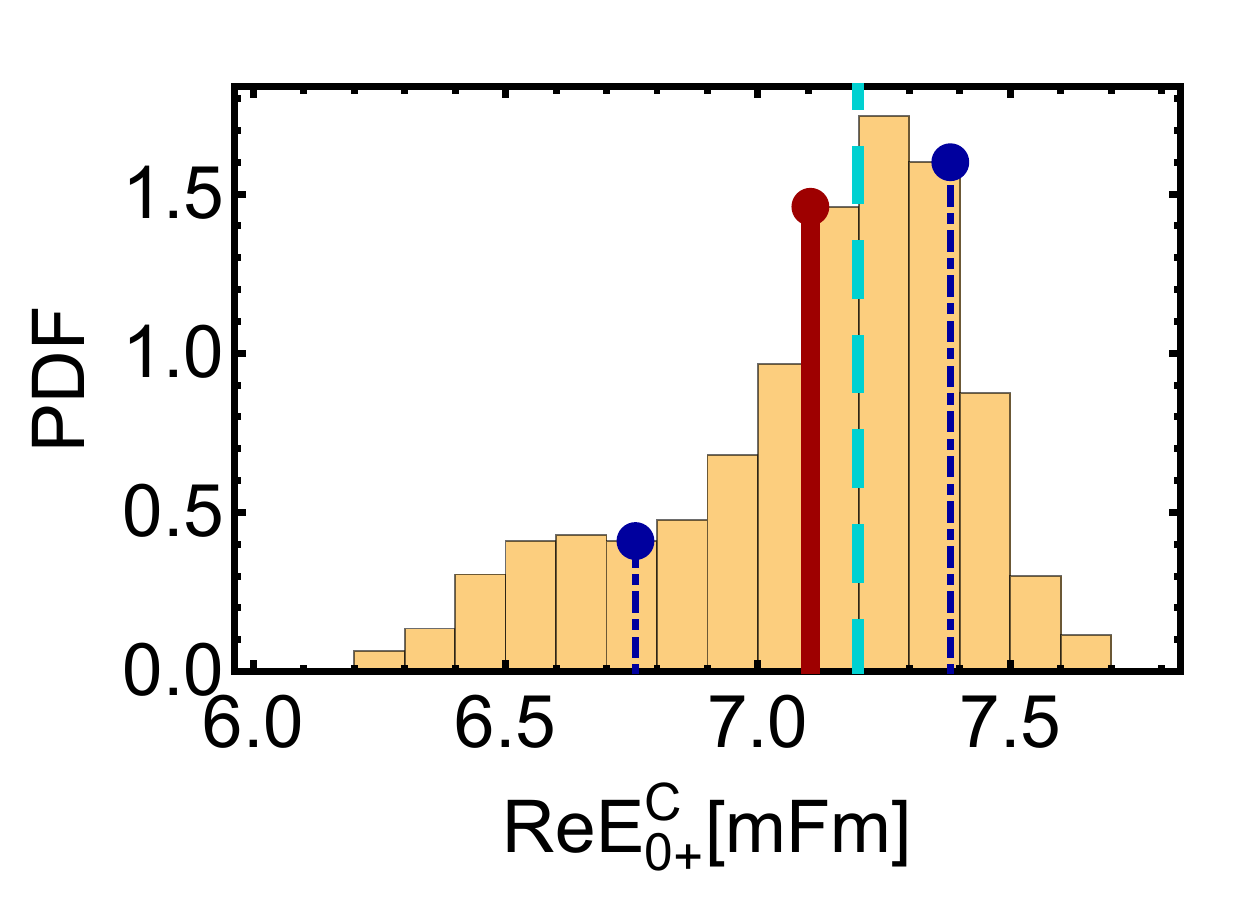}
 \end{overpic}
\begin{overpic}[width=0.325\textwidth]{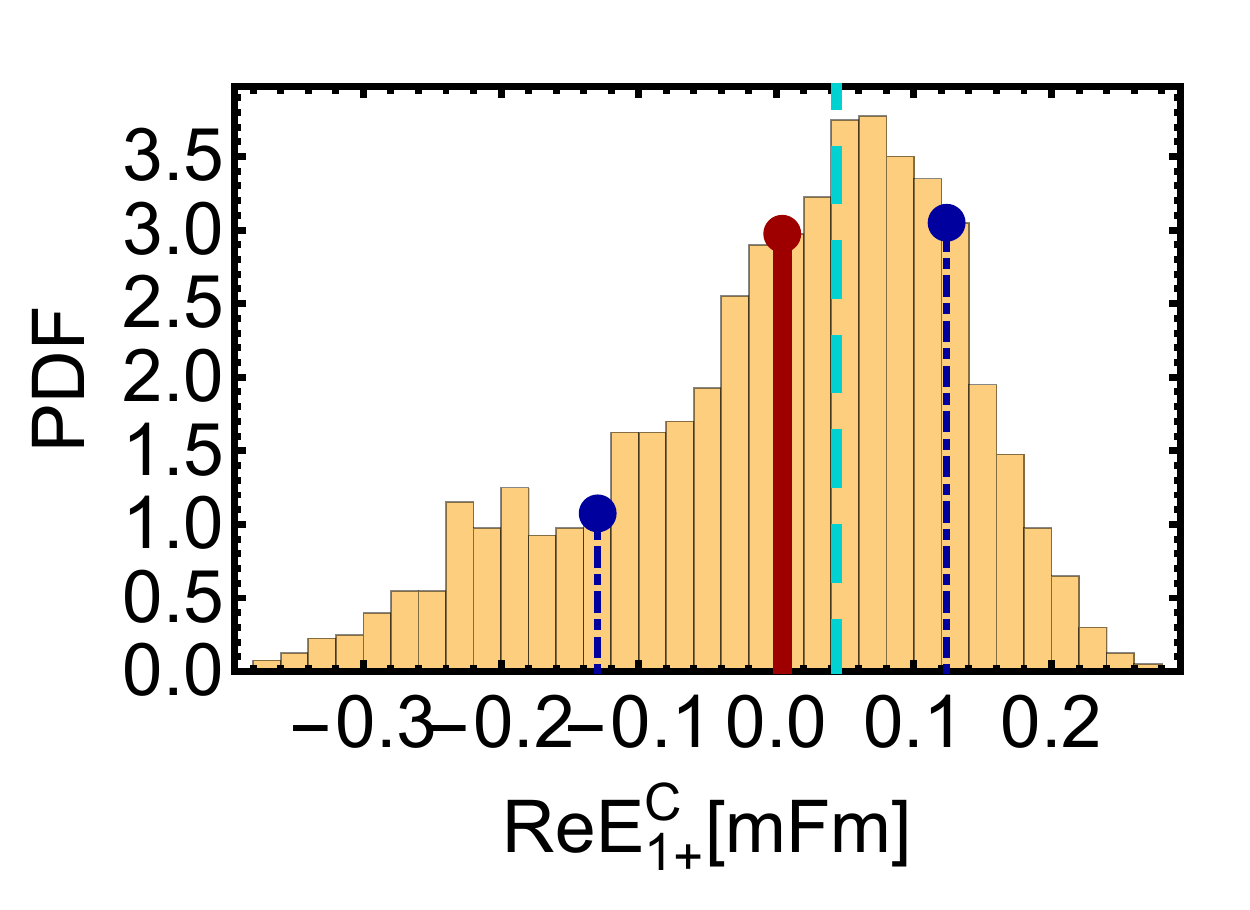}
 \end{overpic}
\begin{overpic}[width=0.325\textwidth]{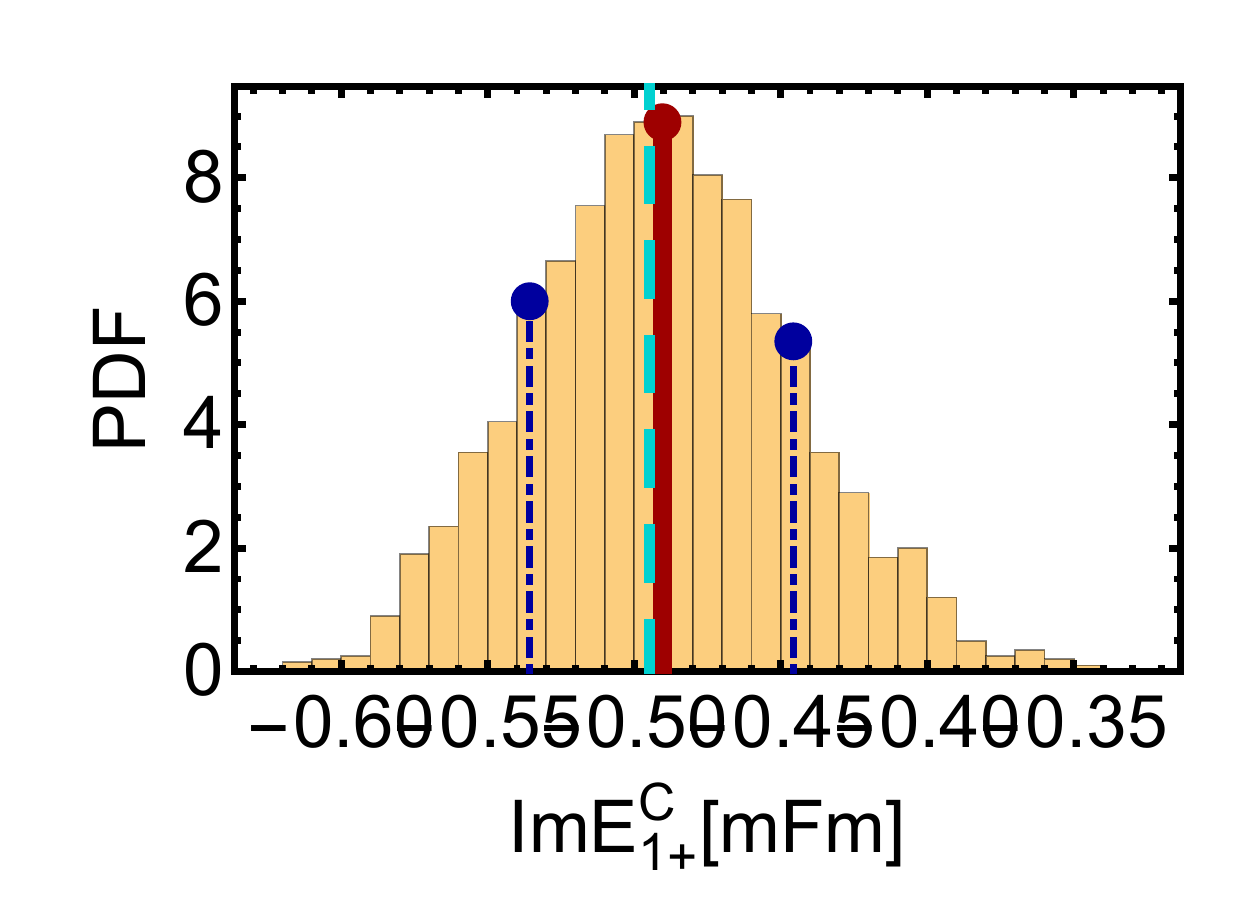}
 \end{overpic} \\
\begin{overpic}[width=0.325\textwidth]{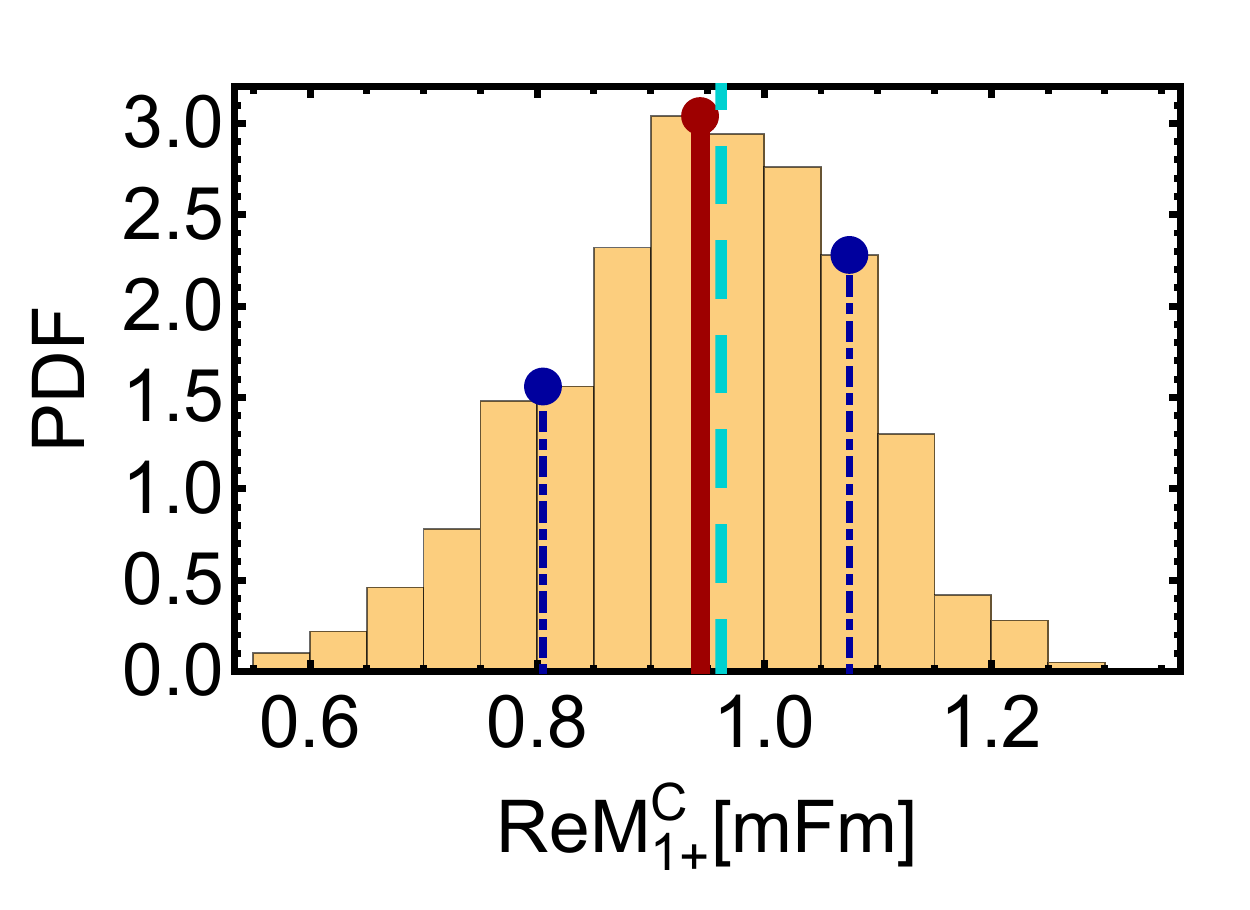}
 \end{overpic}
\begin{overpic}[width=0.325\textwidth]{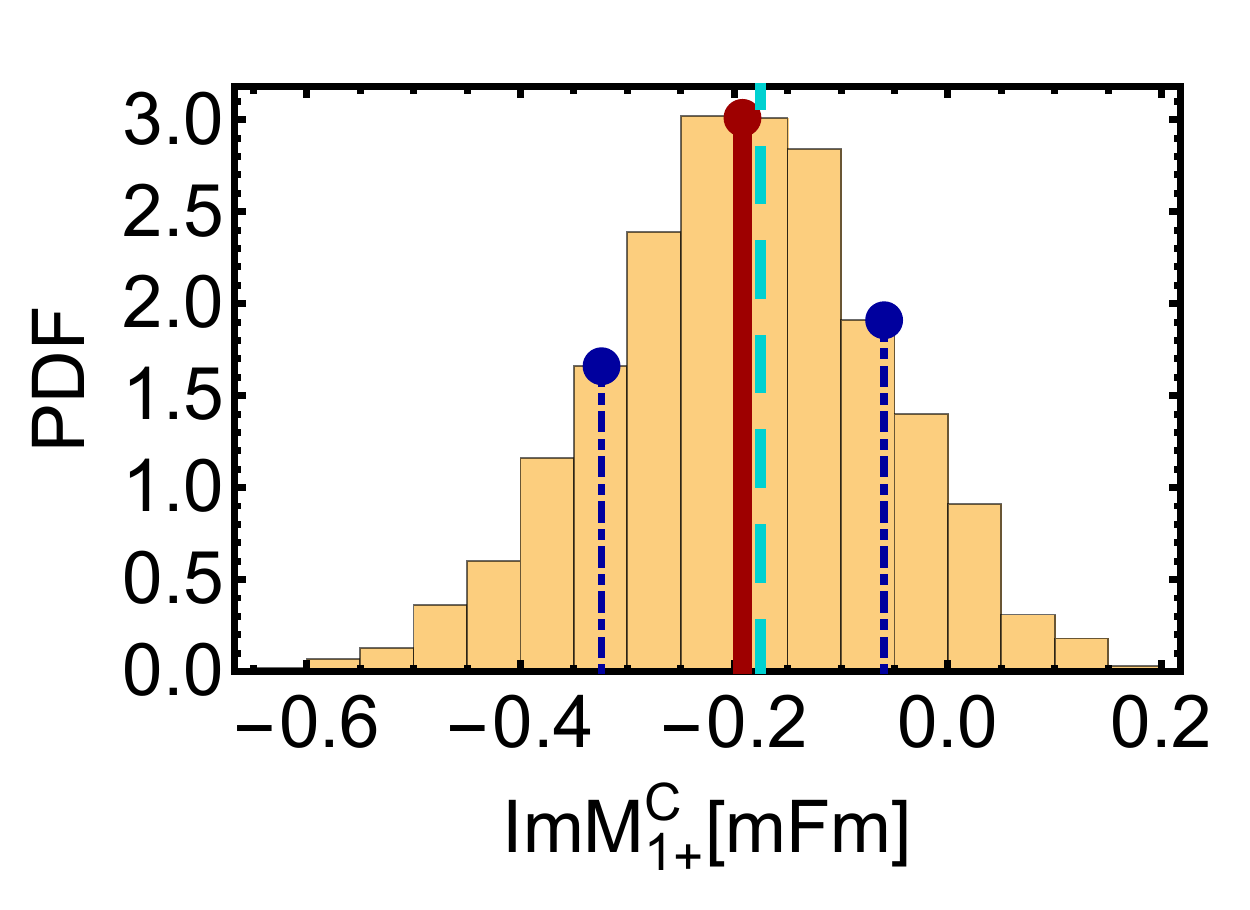}
 \end{overpic}
\begin{overpic}[width=0.325\textwidth]{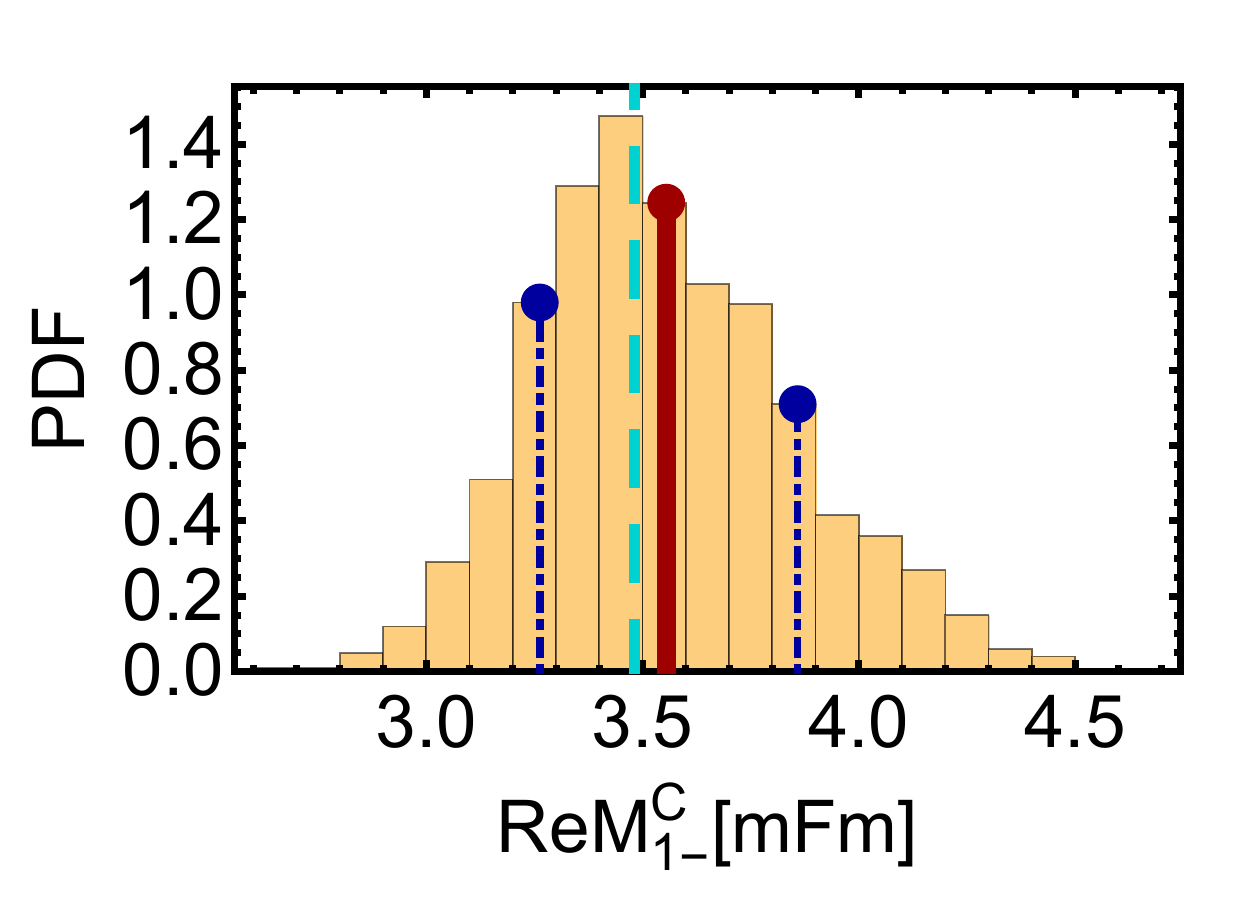}
 \end{overpic} \\
\begin{overpic}[width=0.325\textwidth]{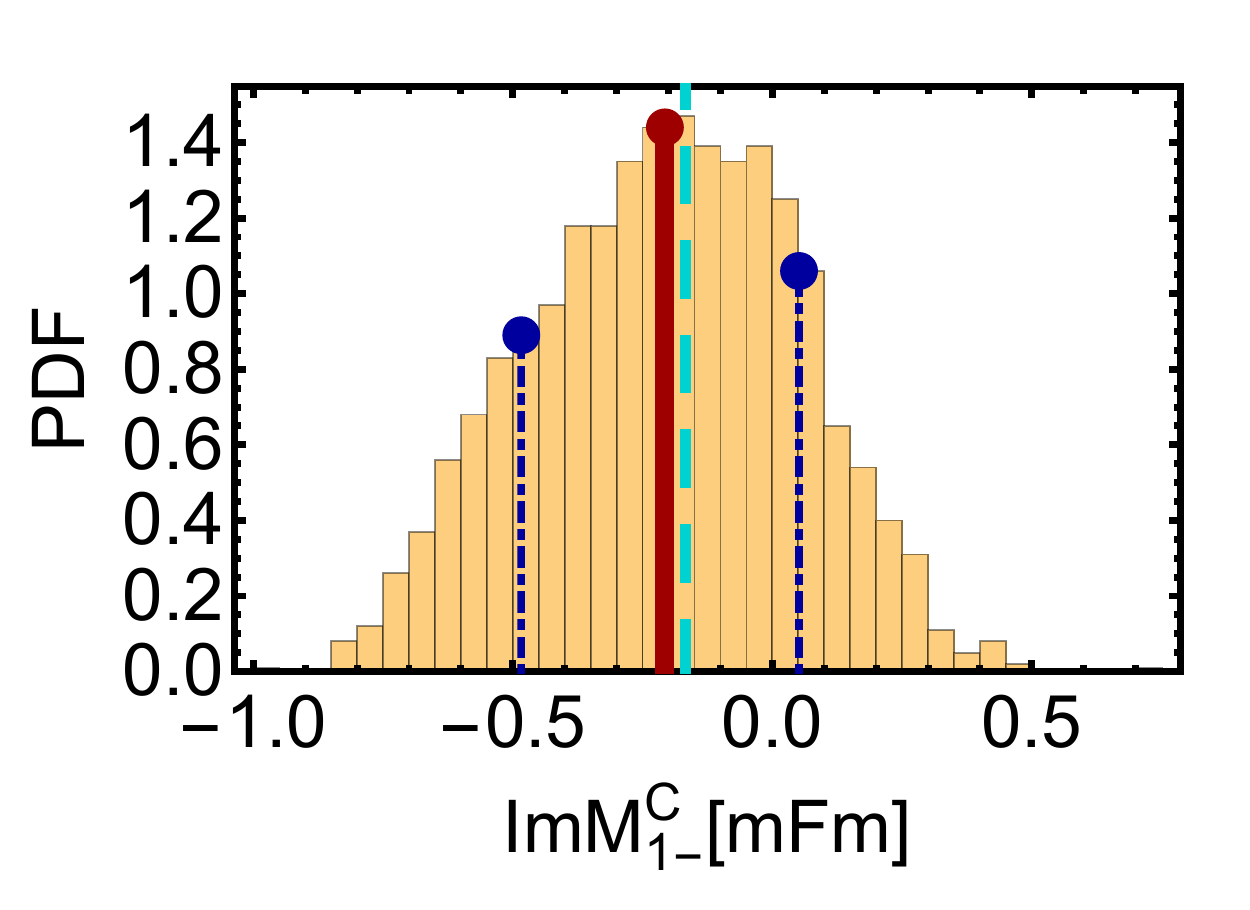}
 \end{overpic}
\begin{overpic}[width=0.325\textwidth]{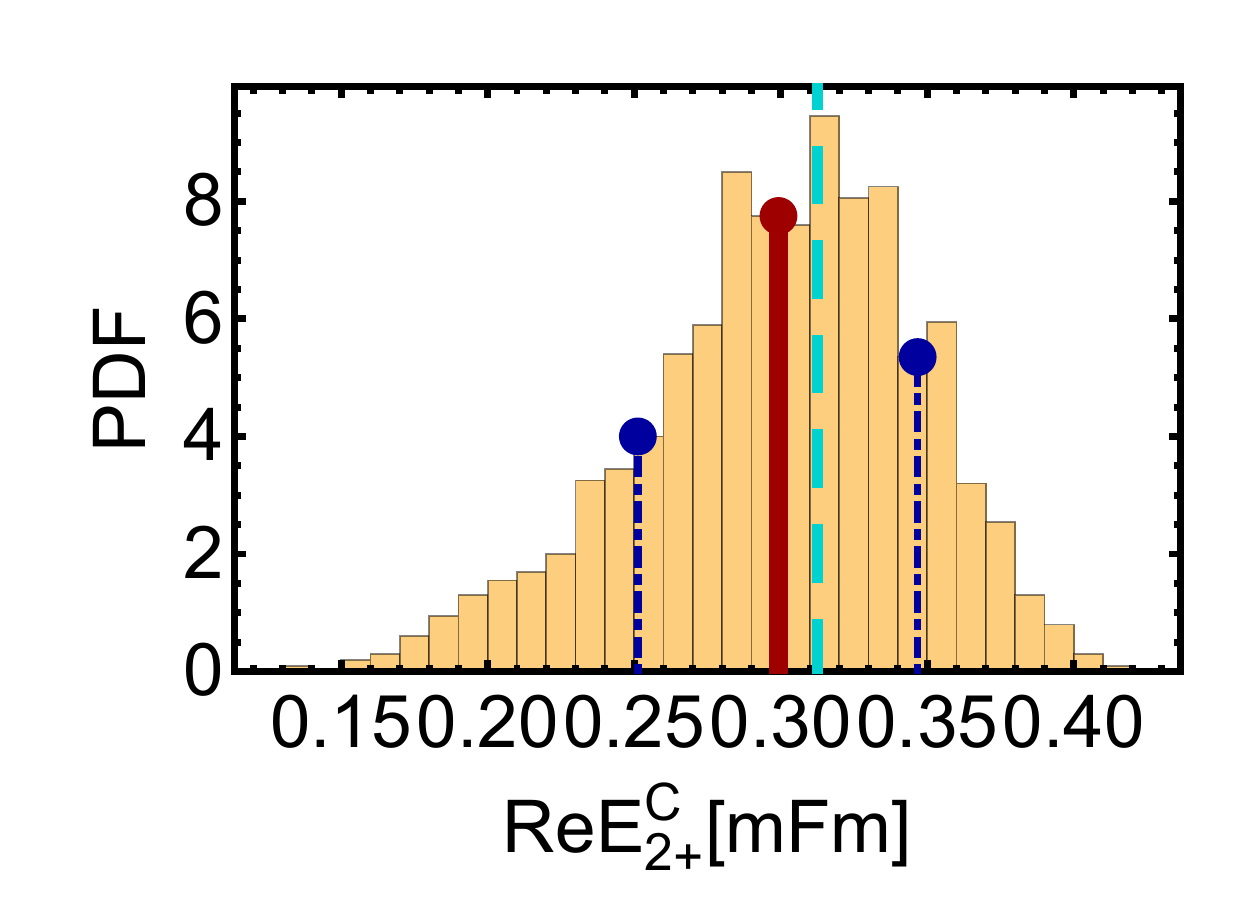}
 \end{overpic}
\begin{overpic}[width=0.325\textwidth]{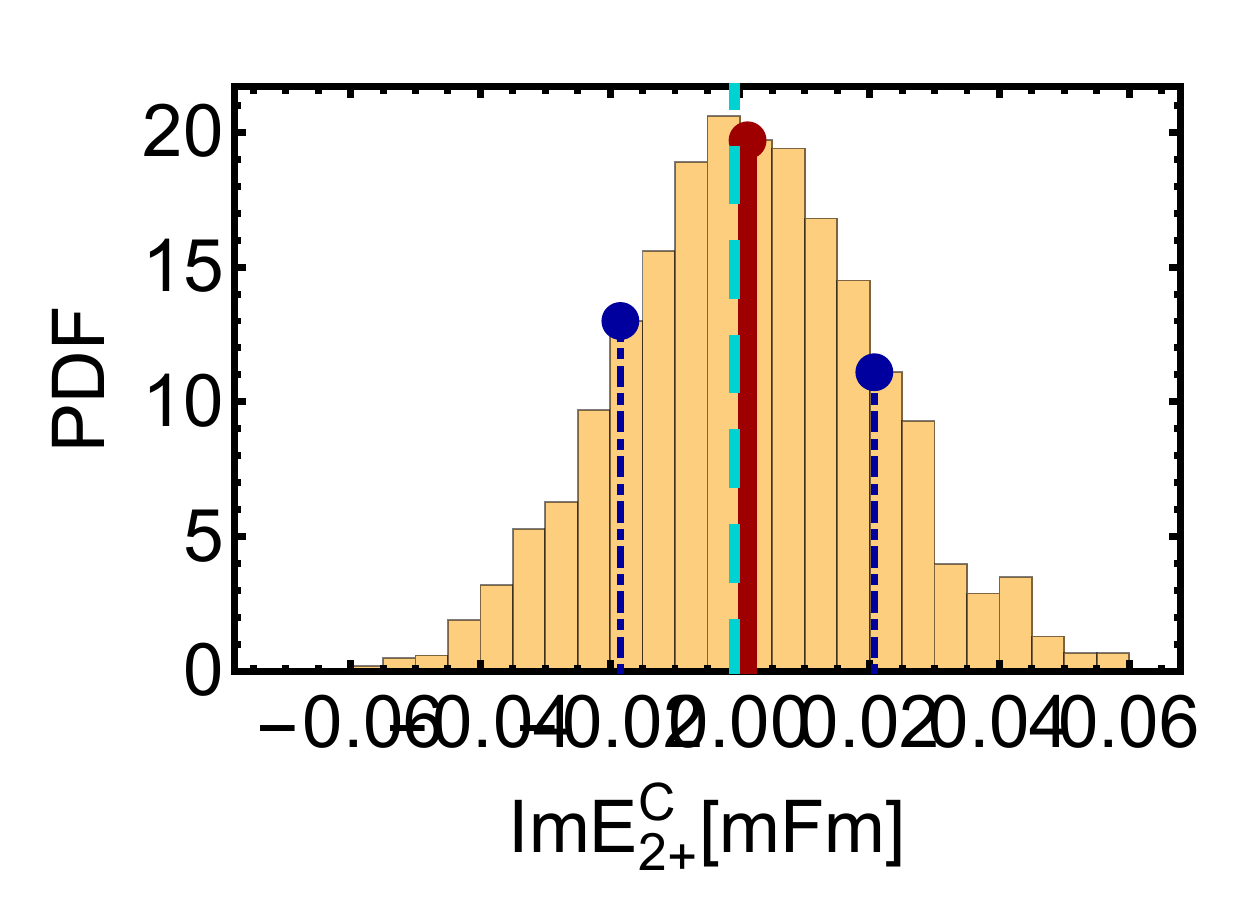}
 \end{overpic} \\
\begin{overpic}[width=0.325\textwidth]{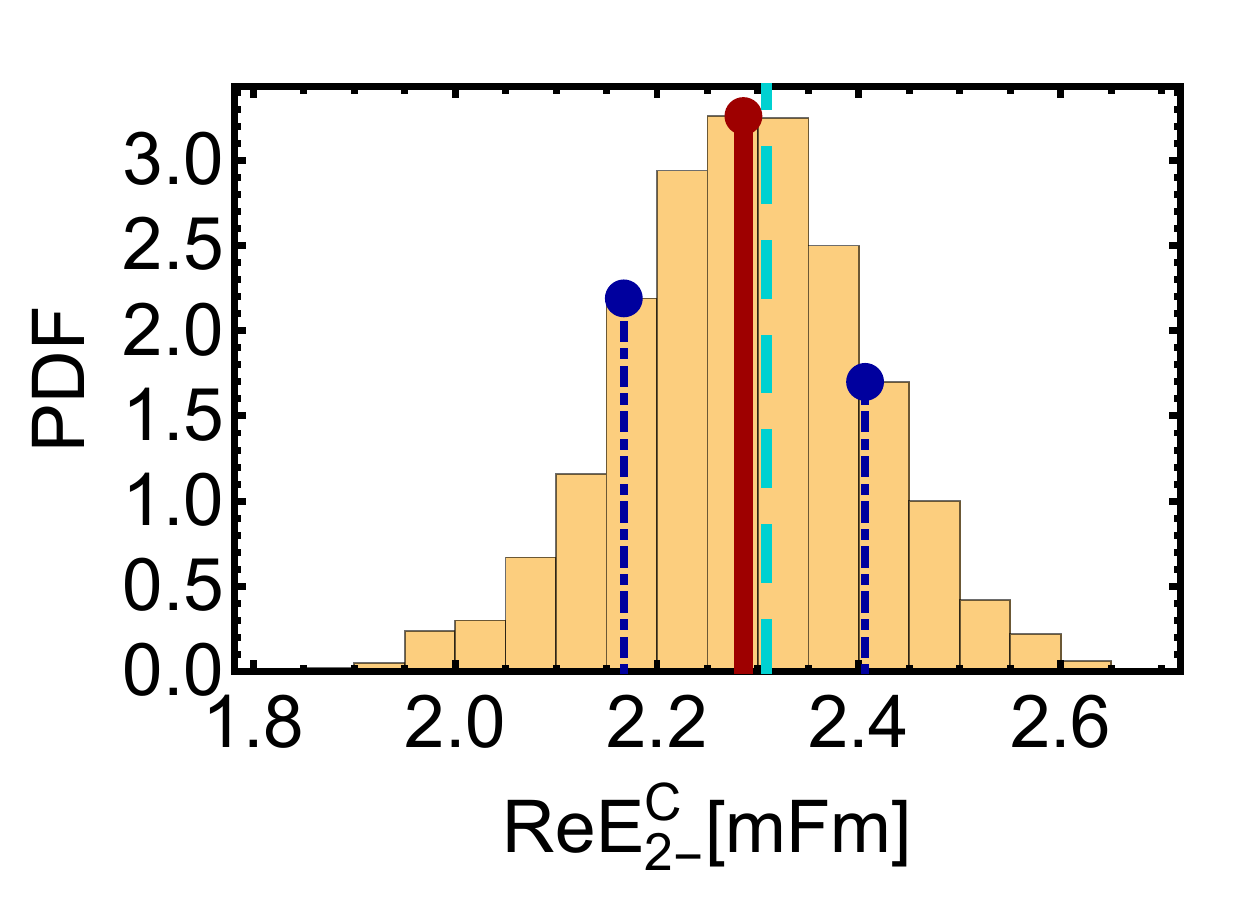}
 \end{overpic}
\begin{overpic}[width=0.325\textwidth]{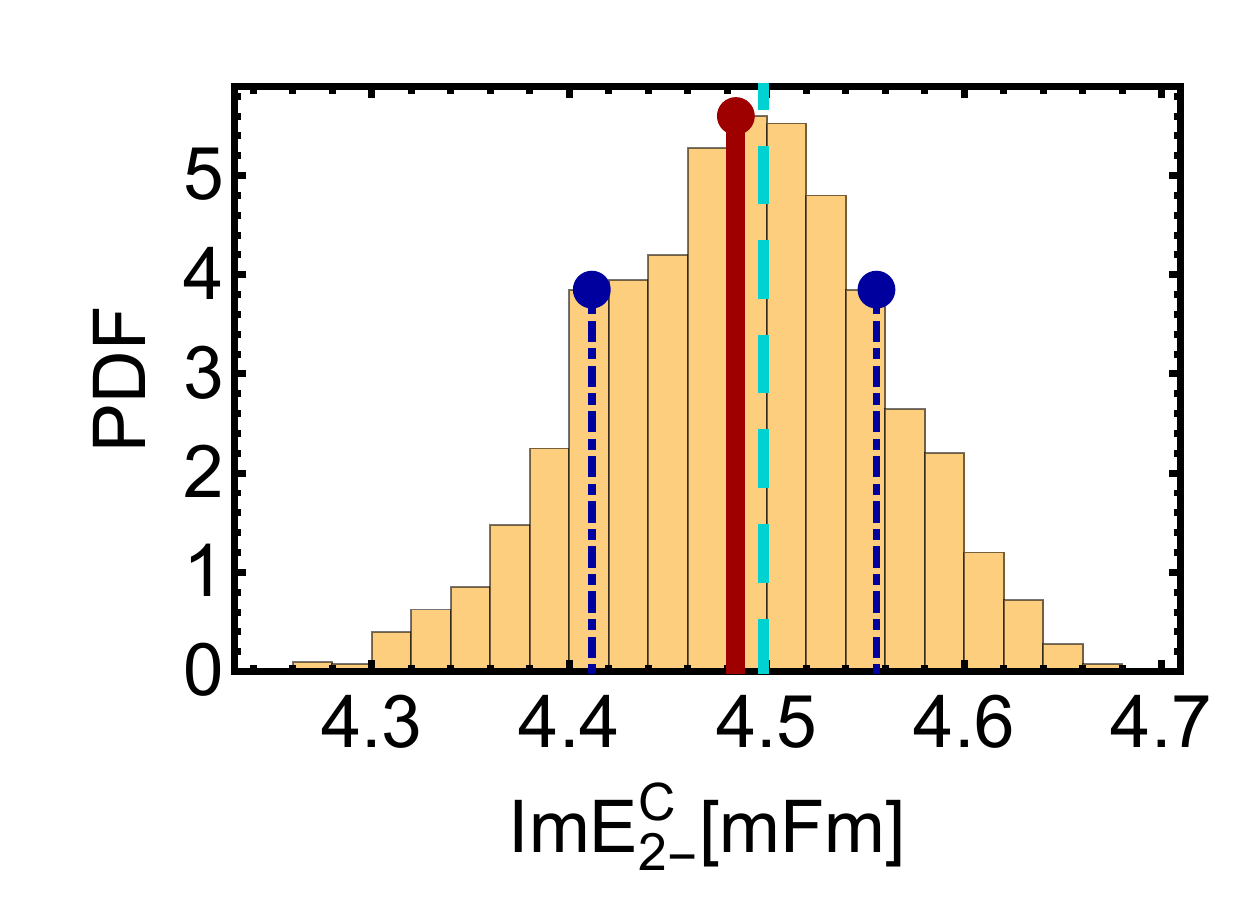}
 \end{overpic}
\begin{overpic}[width=0.325\textwidth]{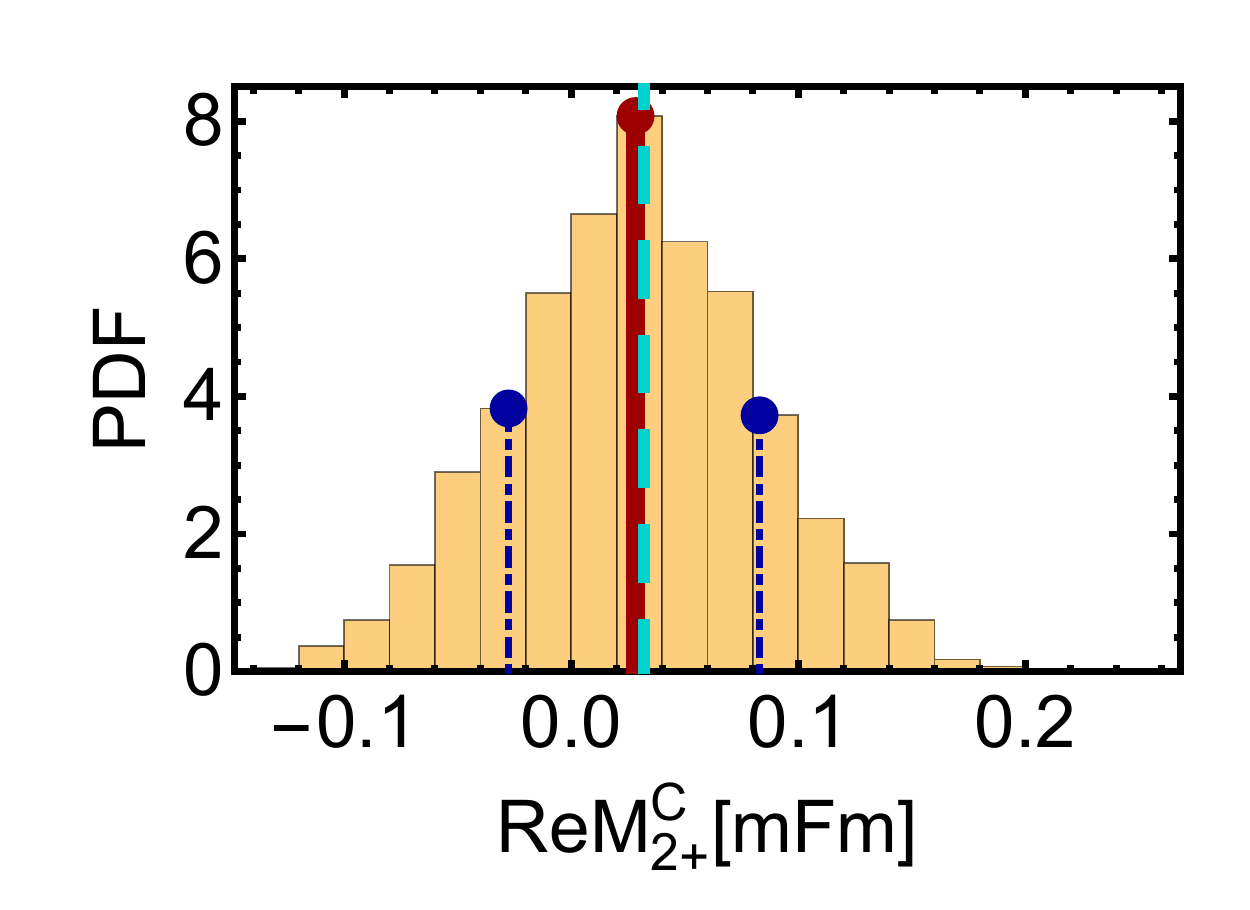}
 \end{overpic} \\
\begin{overpic}[width=0.325\textwidth]{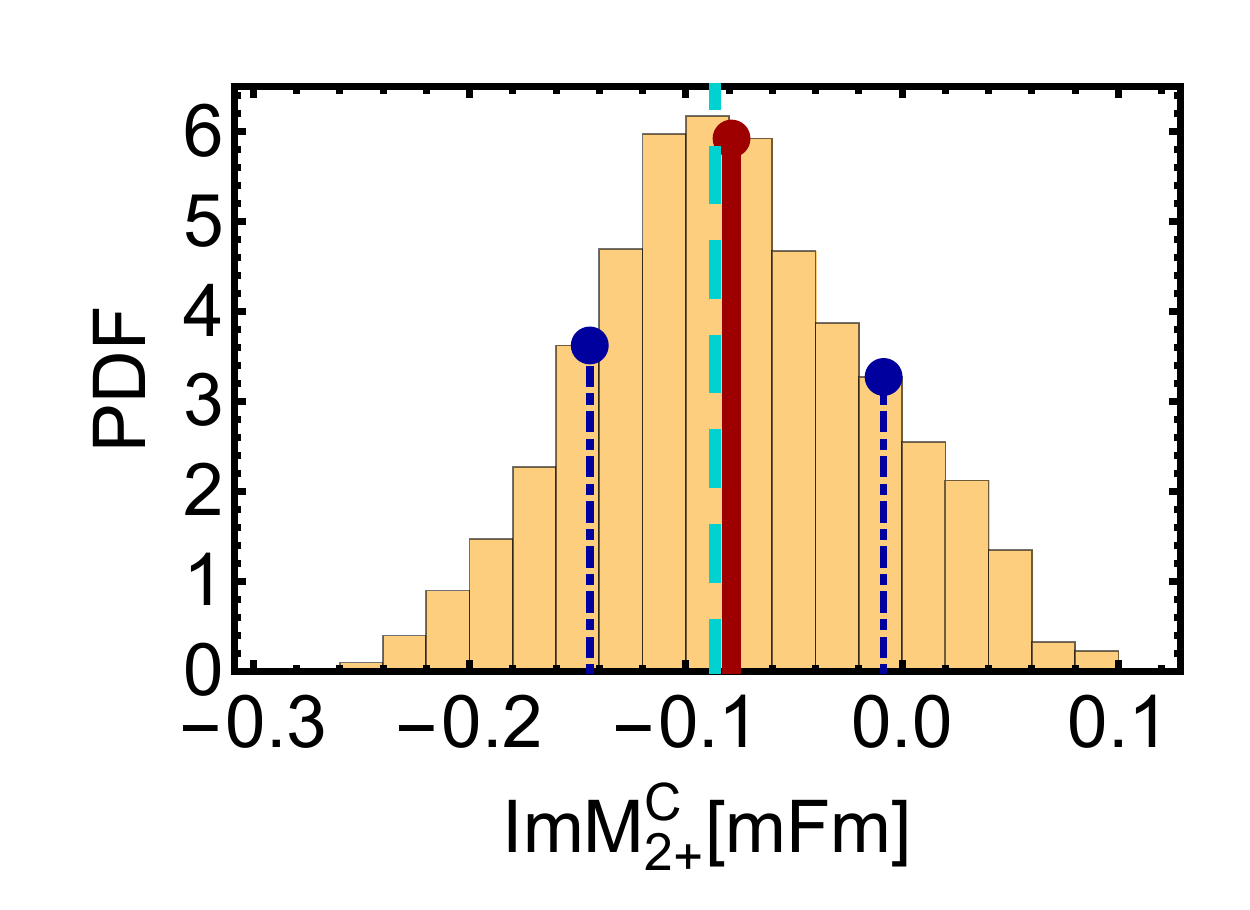}
 \end{overpic}
\begin{overpic}[width=0.325\textwidth]{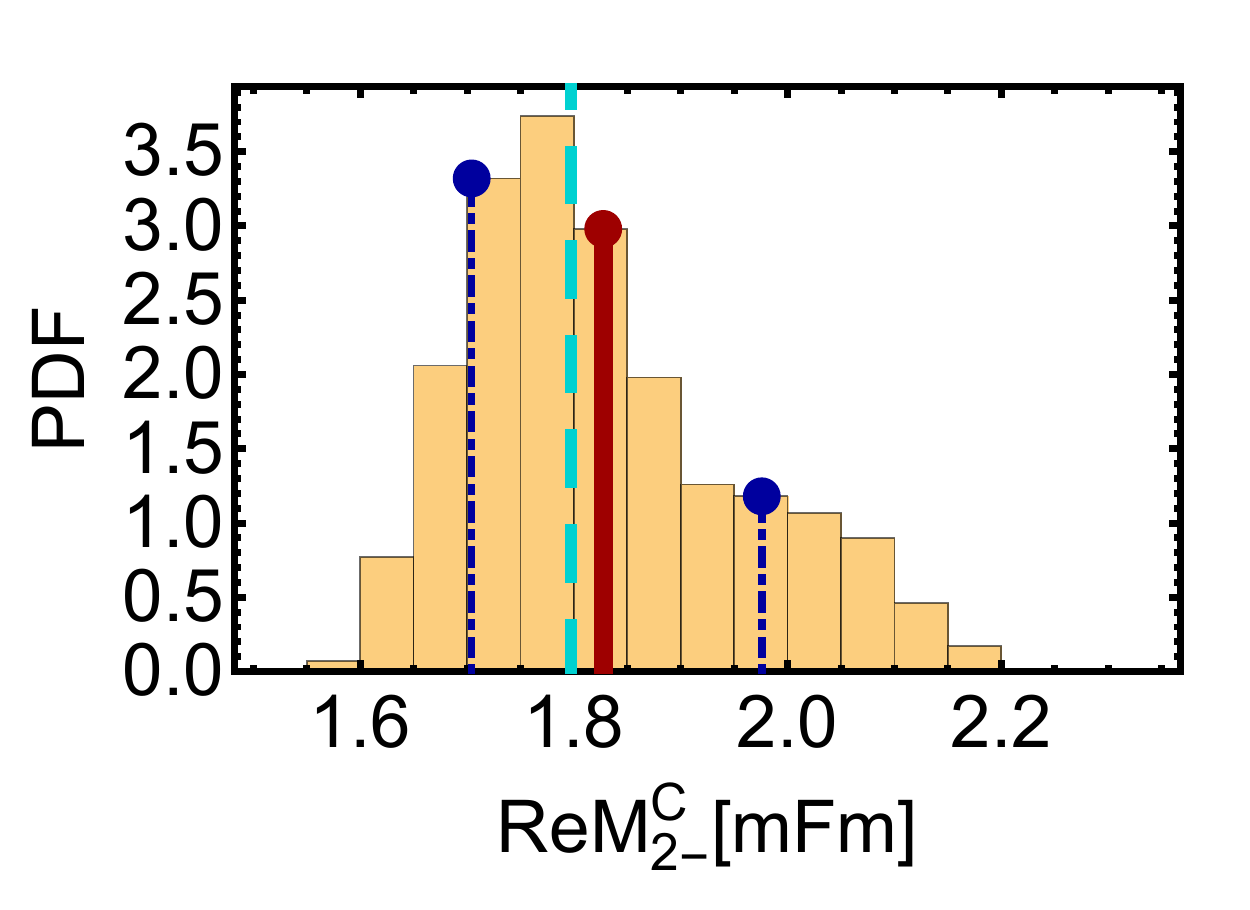}
 \end{overpic}
\begin{overpic}[width=0.325\textwidth]{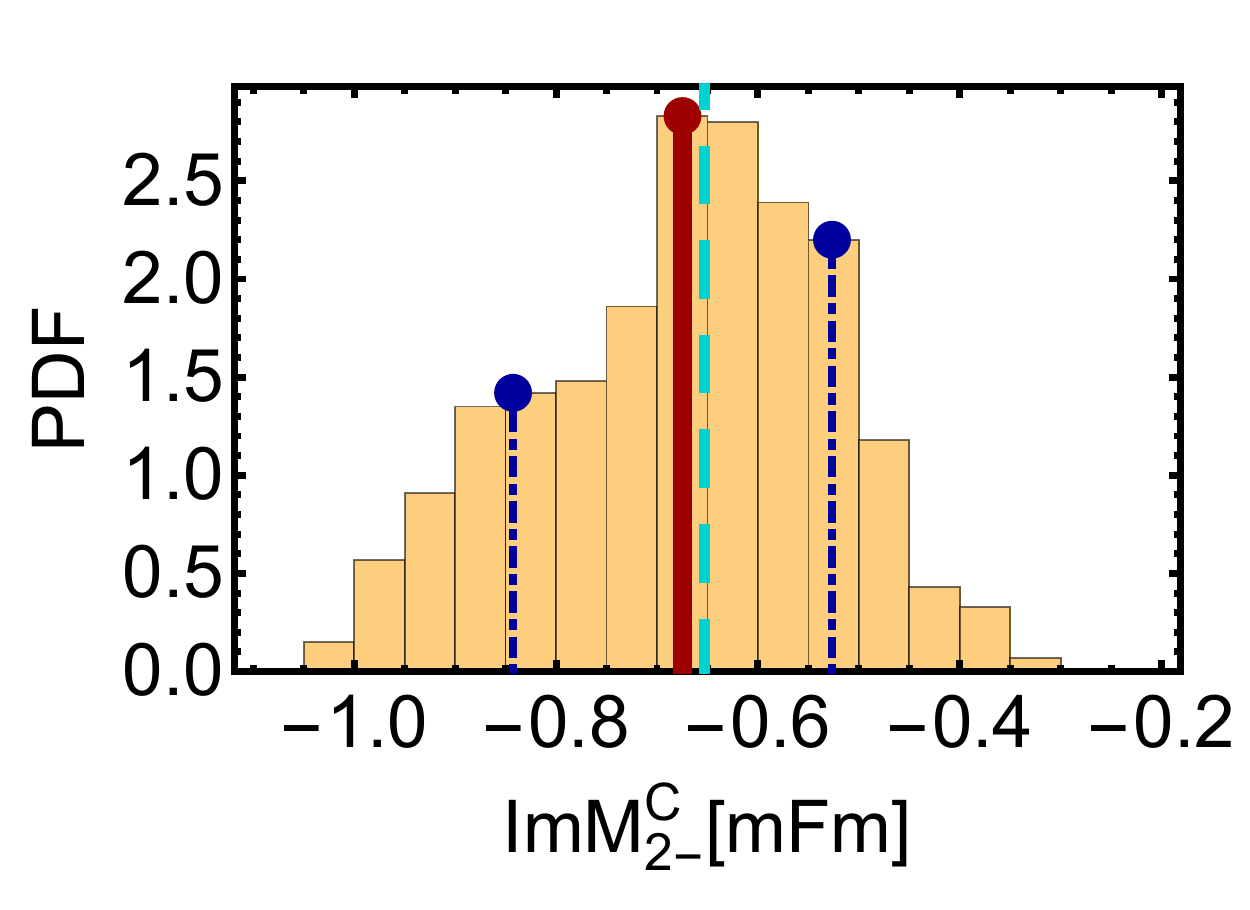}
 \end{overpic}
\caption[Bootstrap-distributions for multipole fit-parameters in an analysis of photoproduction data on the second resonance region. The seventh energy-bin, \newline $E_{\gamma }\text{ = 884.02 MeV}$, is shown.]{The histograms show bootstrap-distributions for the real- and imaginary parts of phase-constrained $S$-, $P$- and $D$-wave multipoles, for a TPWA bootstrap-analysis of photoproduction data in the second resonance region (see section \ref{subsec:2ndResRegionDataFits}). The seventh energy-bin, $E_{\gamma }\text{ = 884.02 MeV}$, is shown. An ensemble of $B=2000$ bootstrap-replicates has been the basis of these results. \newline
The distributions have been normalized to $1$ via use of the object \textit{HistogramDistribution} in MATHEMATICA \cite{Mathematica8,Mathematica11,MathematicaLanguage,MathematicaBonnLicense}. Thus, $y$-axes are labelled as \textit{PDF}. The mean of each distribution is shown as a red solid line, while the $0.16$- and $0.84$-quantiles are indicated by blue dash-dotted lines. The global minimum of the fit to the original data is plotted as a cyan-colored dashed horizontal line.}
\label{fig:BootstrapHistos2ndResRegionEnergy7}
\end{figure}

\clearpage

\begin{figure}[h]
\begin{overpic}[width=0.325\textwidth]{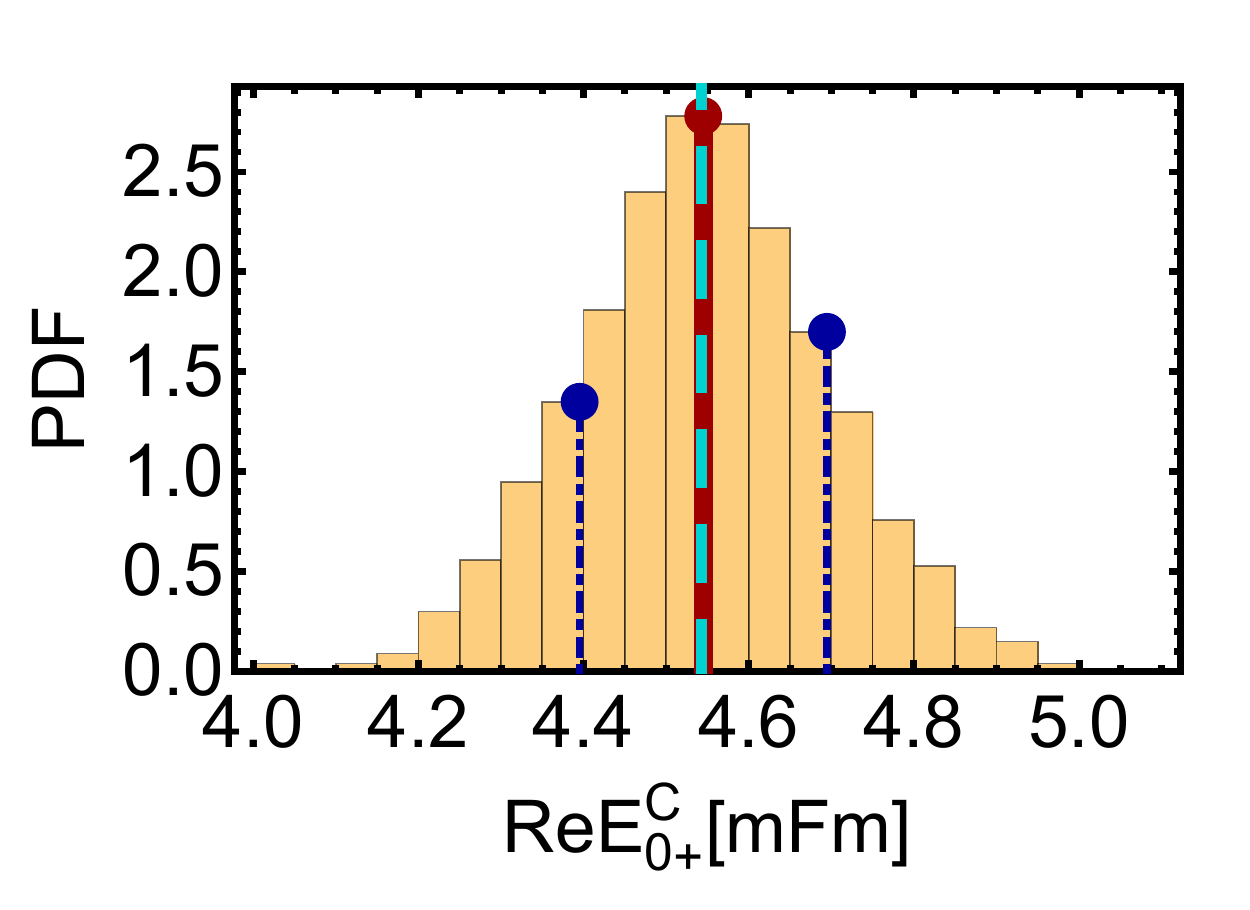}
 \end{overpic}
\begin{overpic}[width=0.325\textwidth]{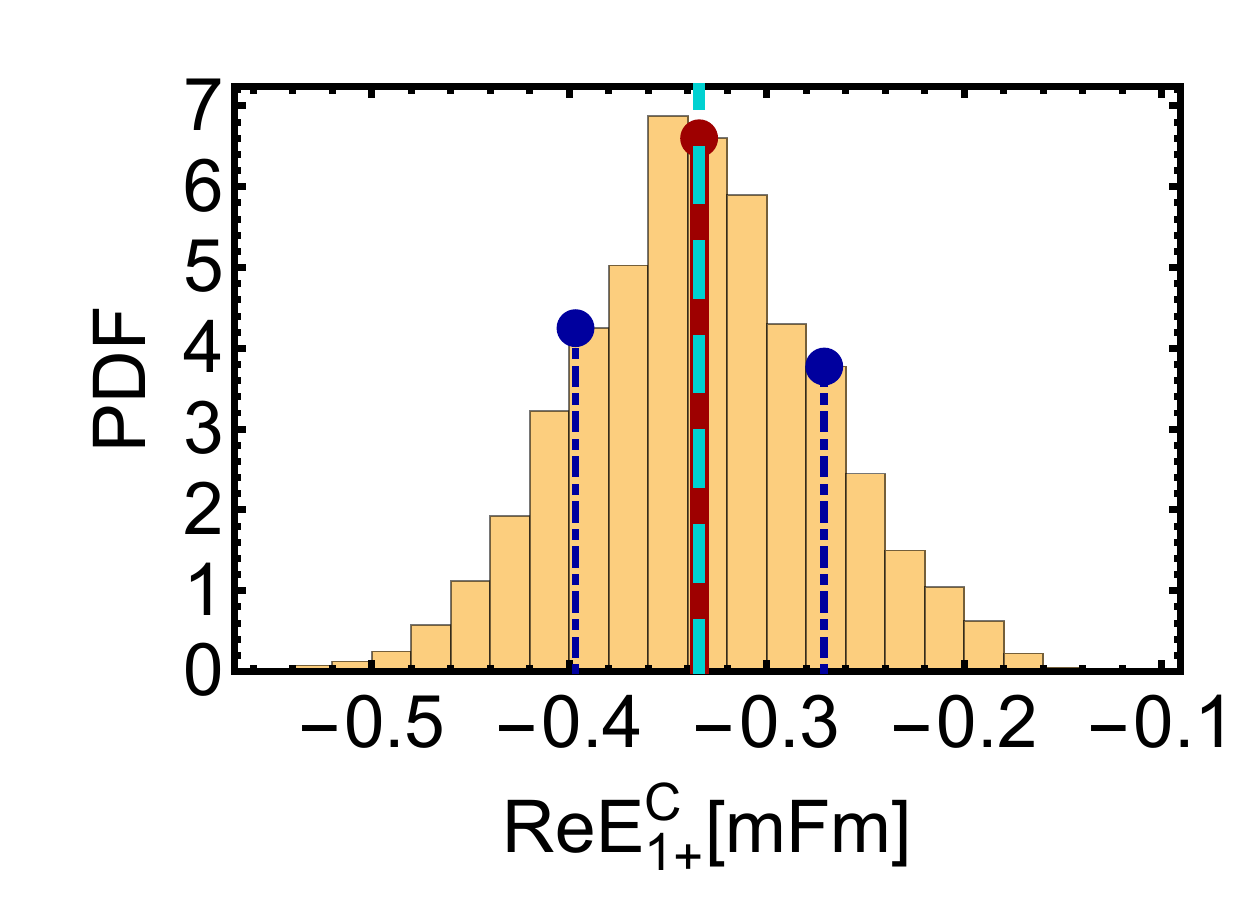}
 \end{overpic}
\begin{overpic}[width=0.325\textwidth]{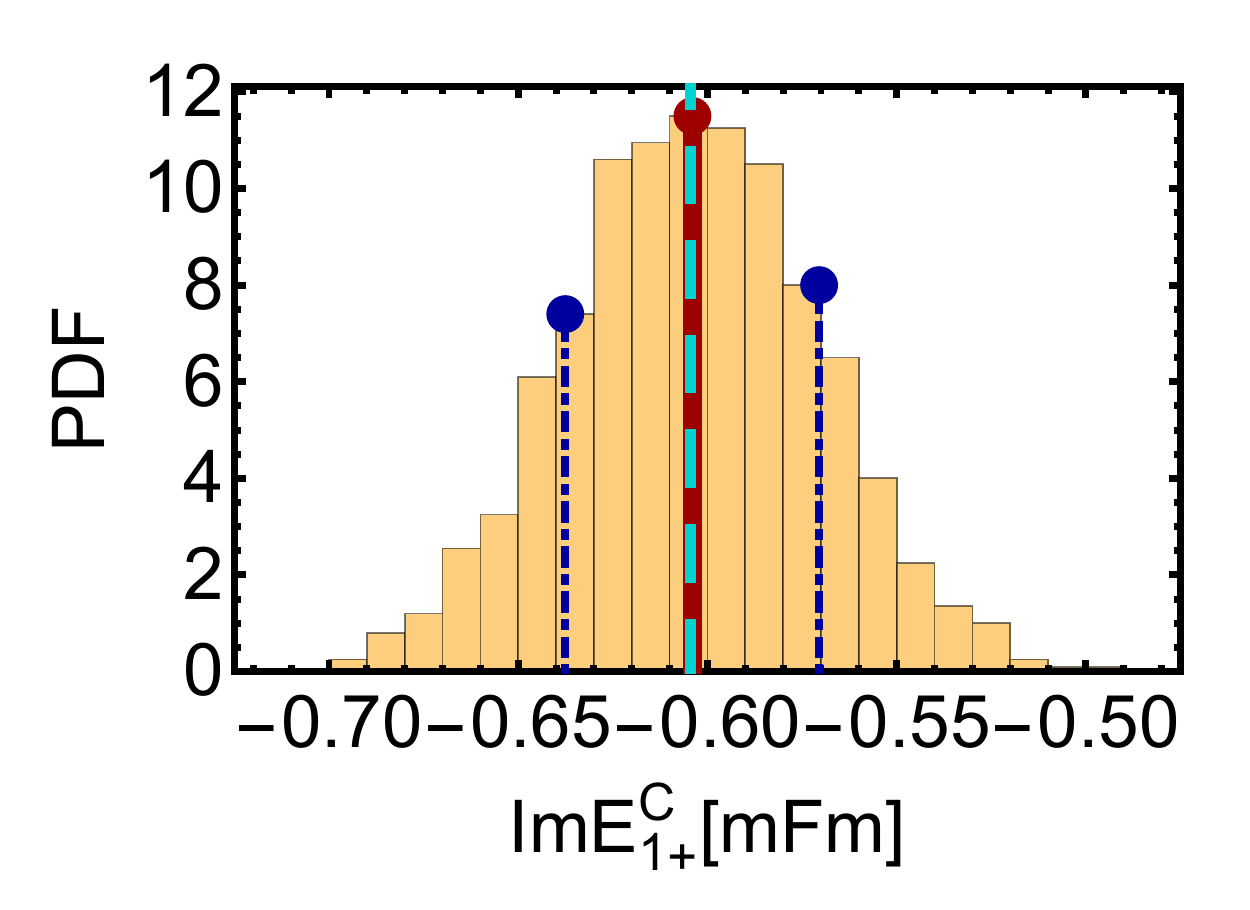}
 \end{overpic} \\
\begin{overpic}[width=0.325\textwidth]{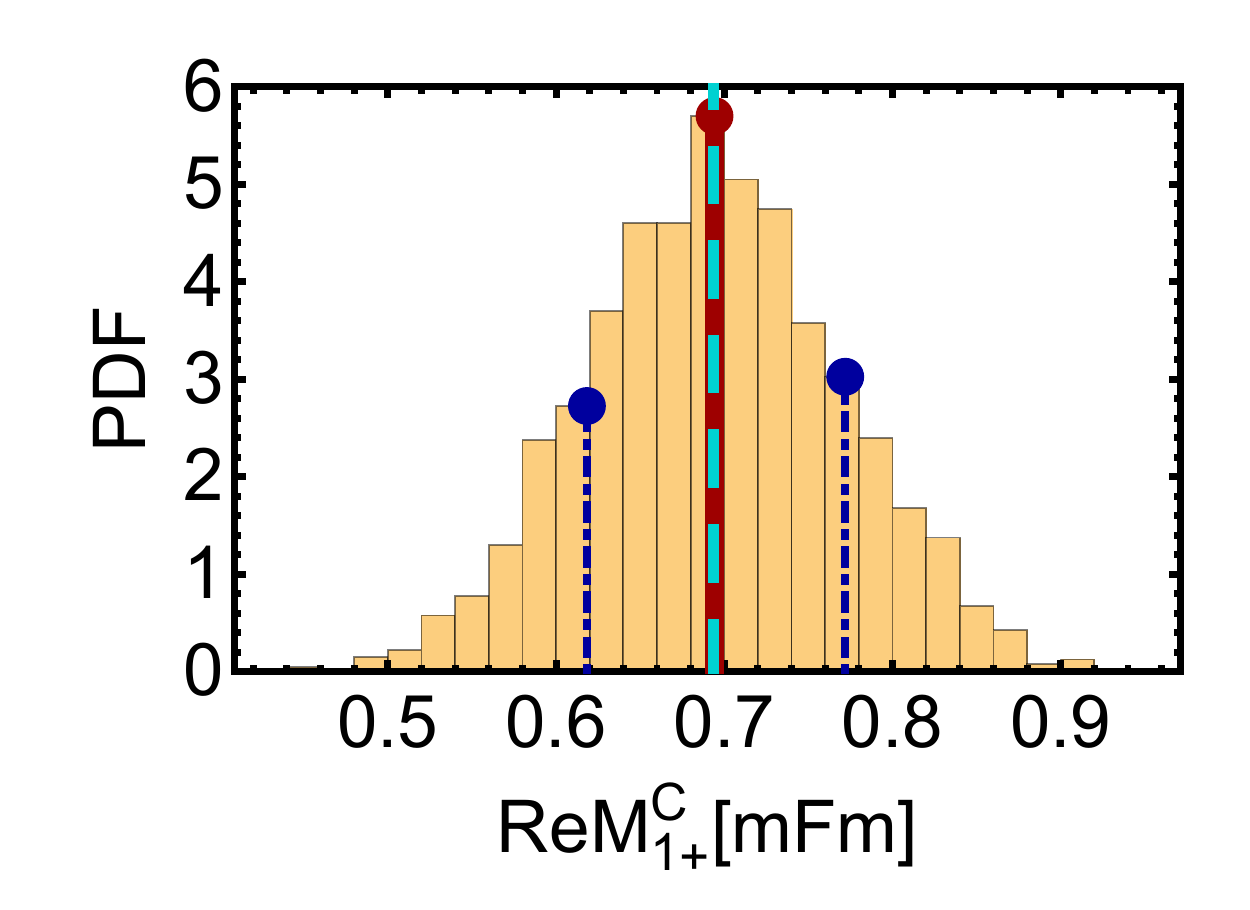}
 \end{overpic}
\begin{overpic}[width=0.325\textwidth]{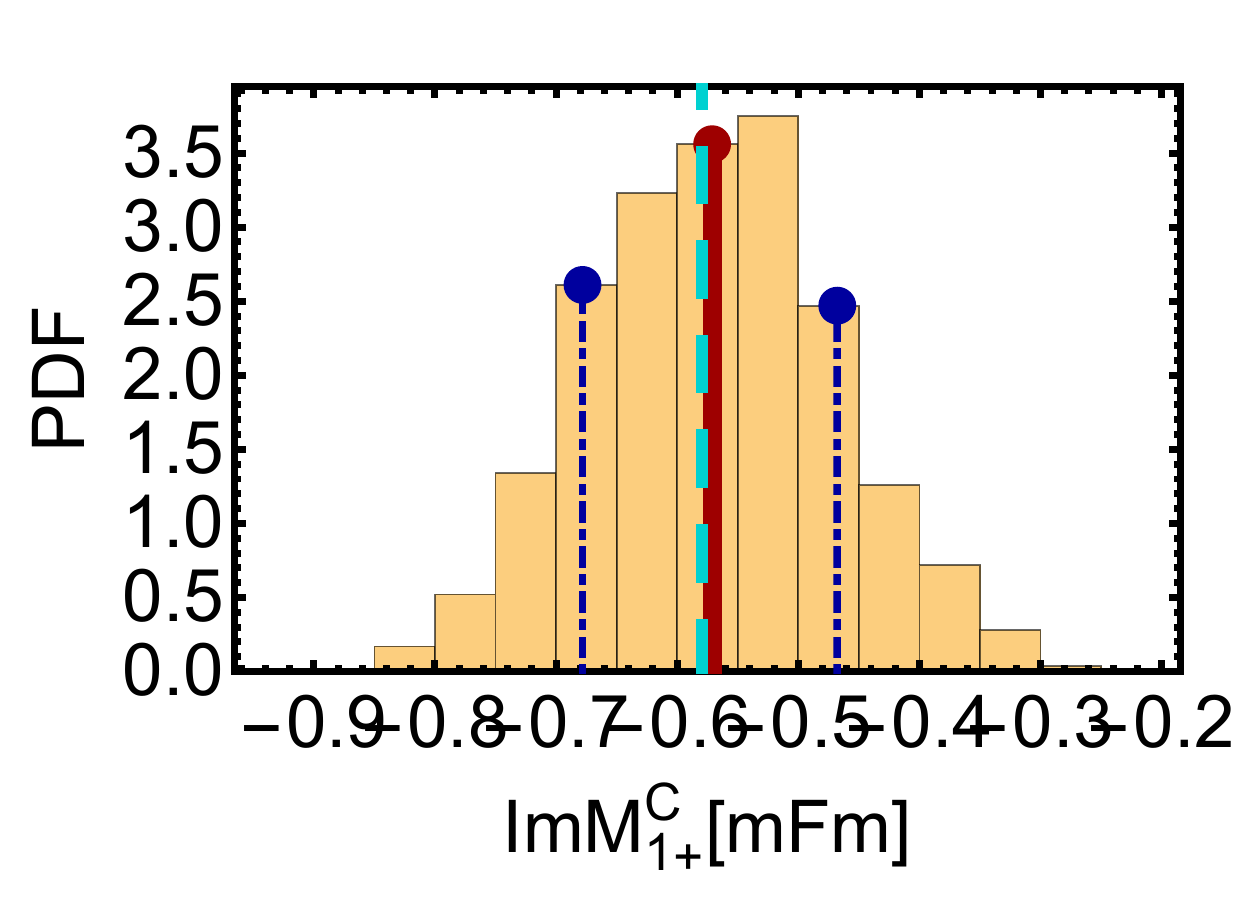}
 \end{overpic}
\begin{overpic}[width=0.325\textwidth]{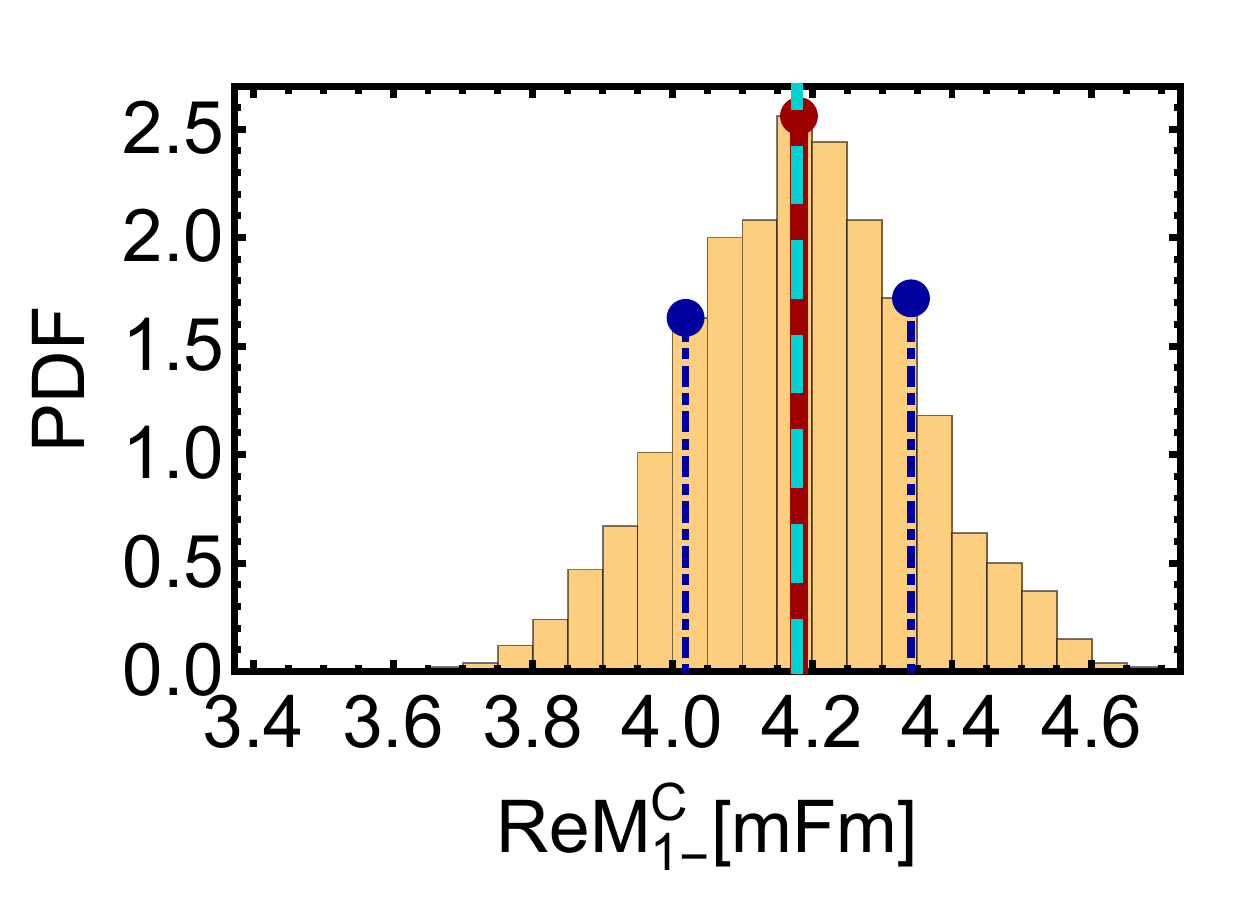}
 \end{overpic} \\
\begin{overpic}[width=0.325\textwidth]{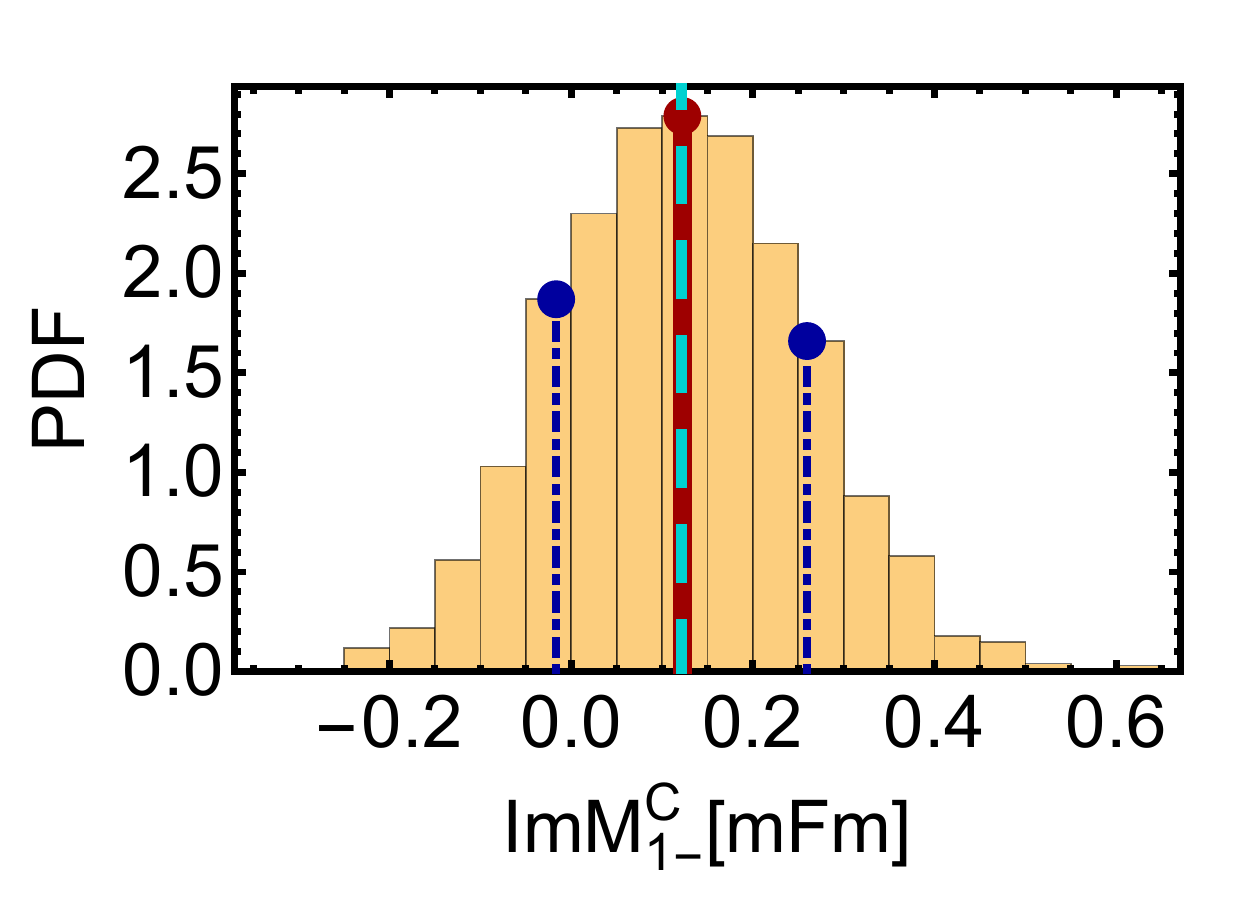}
 \end{overpic}
\begin{overpic}[width=0.325\textwidth]{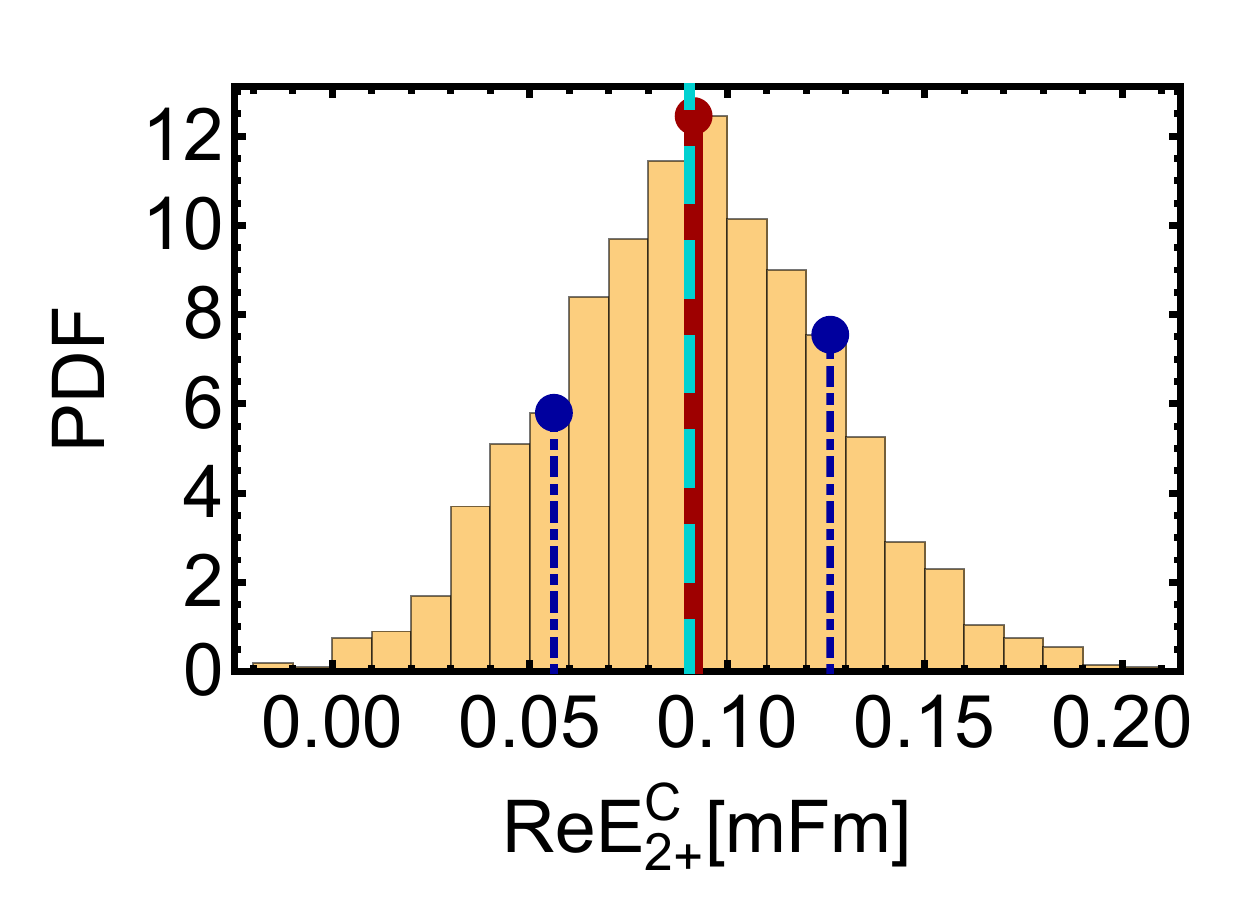}
 \end{overpic}
\begin{overpic}[width=0.325\textwidth]{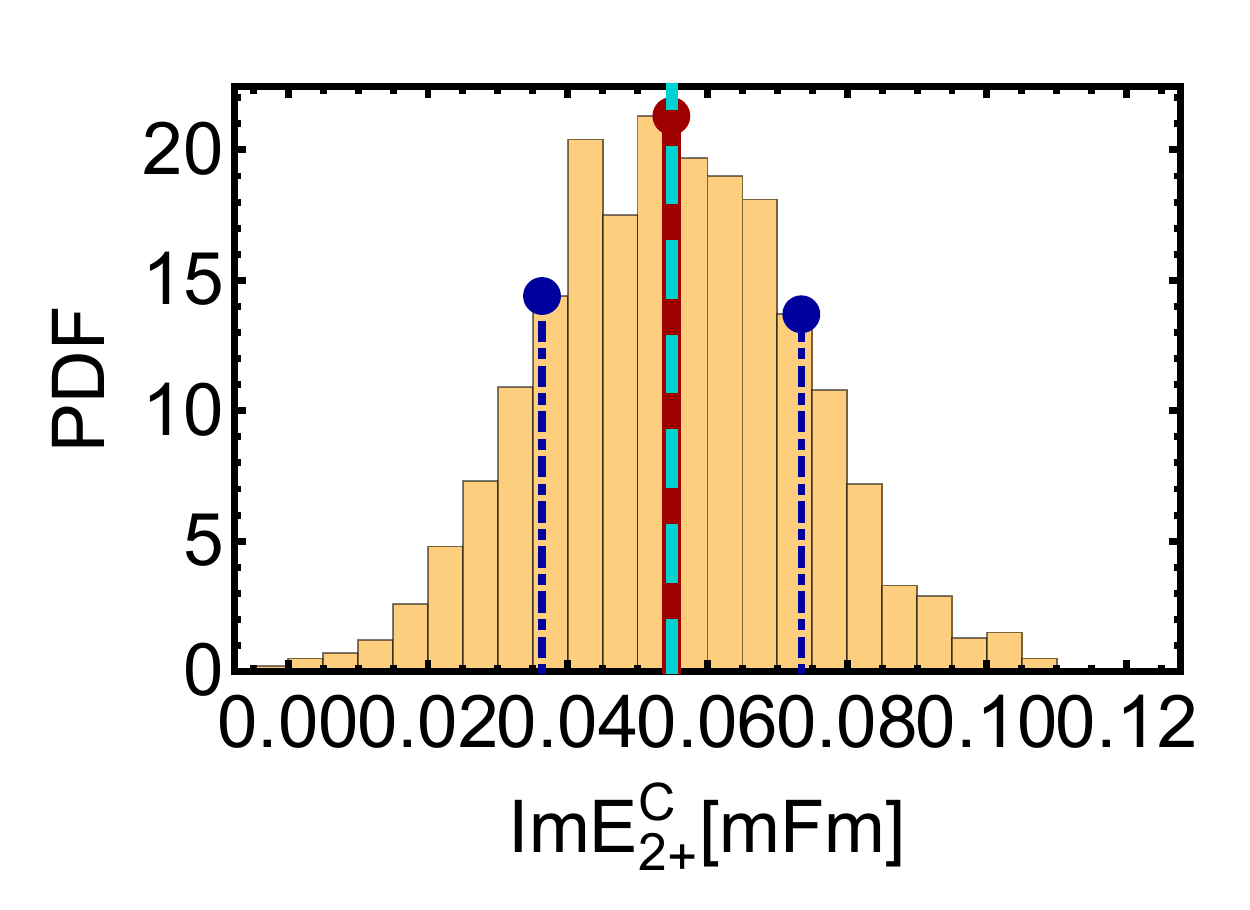}
 \end{overpic} \\
\begin{overpic}[width=0.325\textwidth]{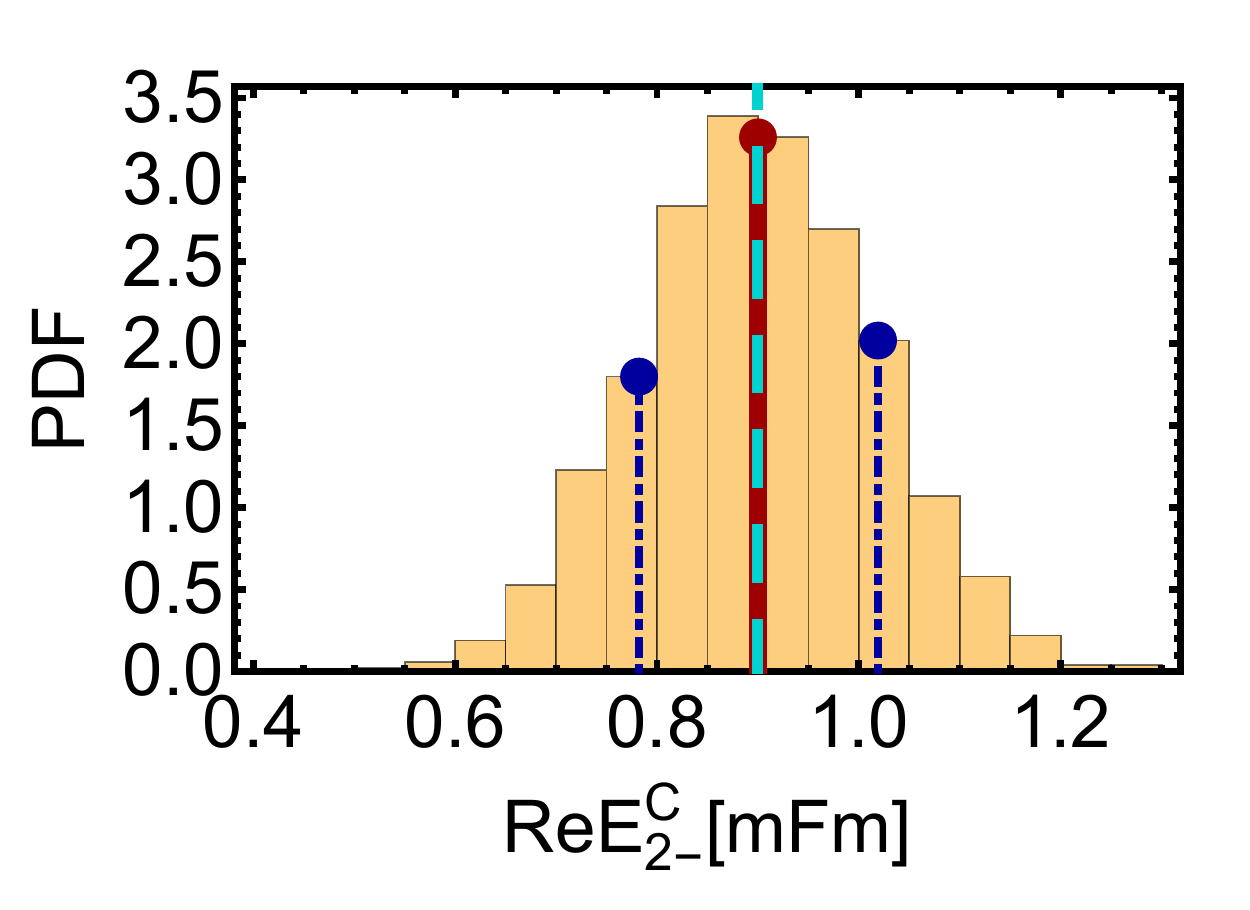}
 \end{overpic}
\begin{overpic}[width=0.325\textwidth]{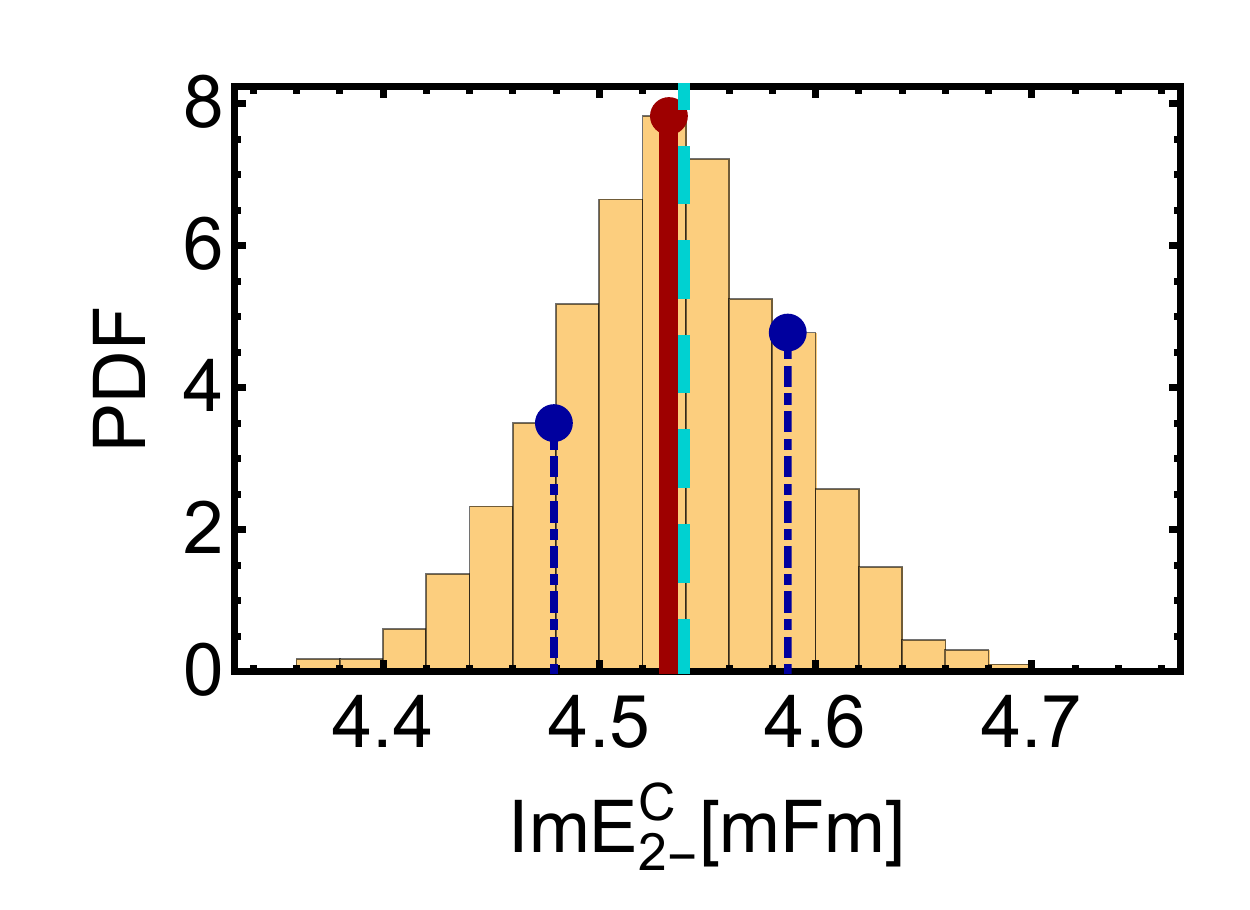}
 \end{overpic}
\begin{overpic}[width=0.325\textwidth]{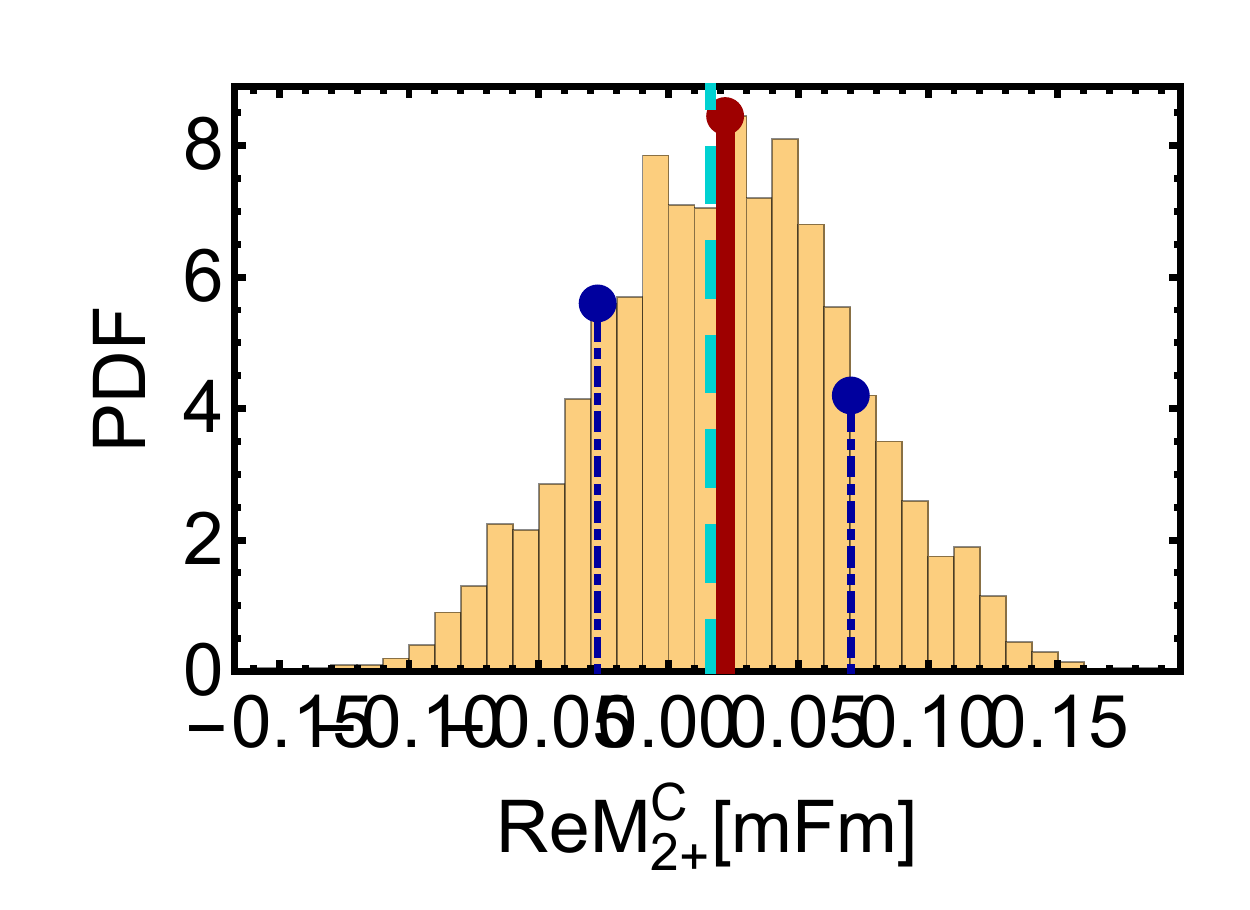}
 \end{overpic} \\
\begin{overpic}[width=0.325\textwidth]{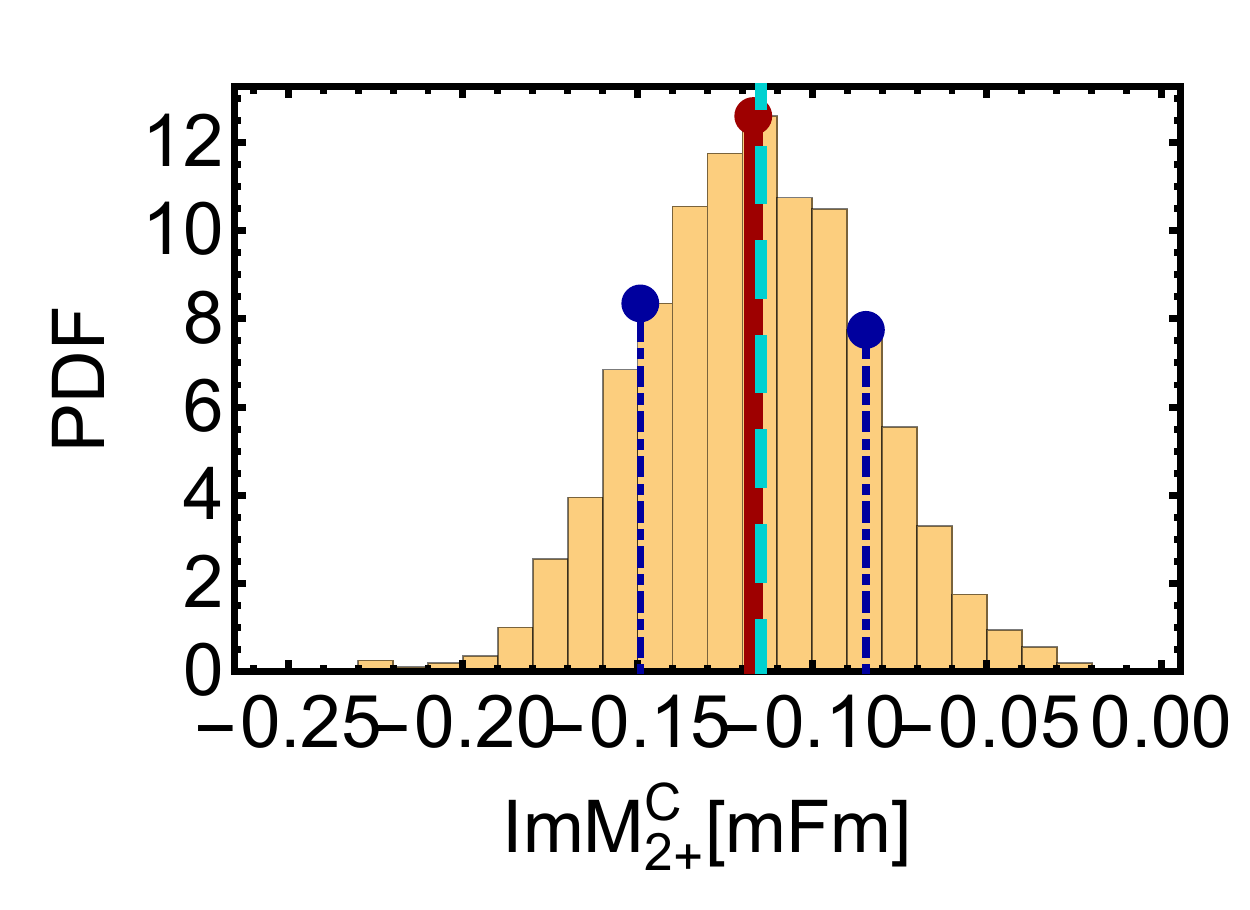}
 \end{overpic}
\begin{overpic}[width=0.325\textwidth]{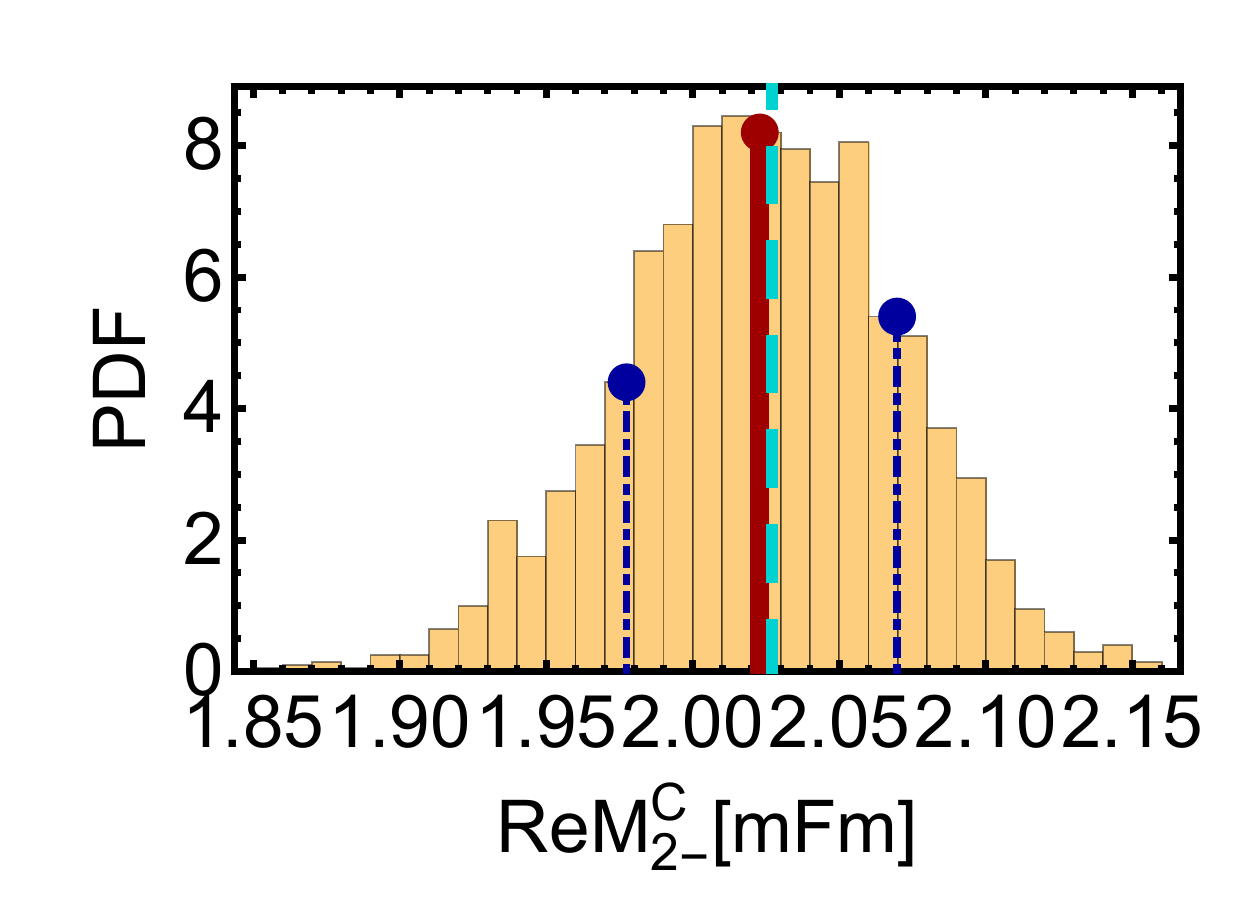}
 \end{overpic}
\begin{overpic}[width=0.325\textwidth]{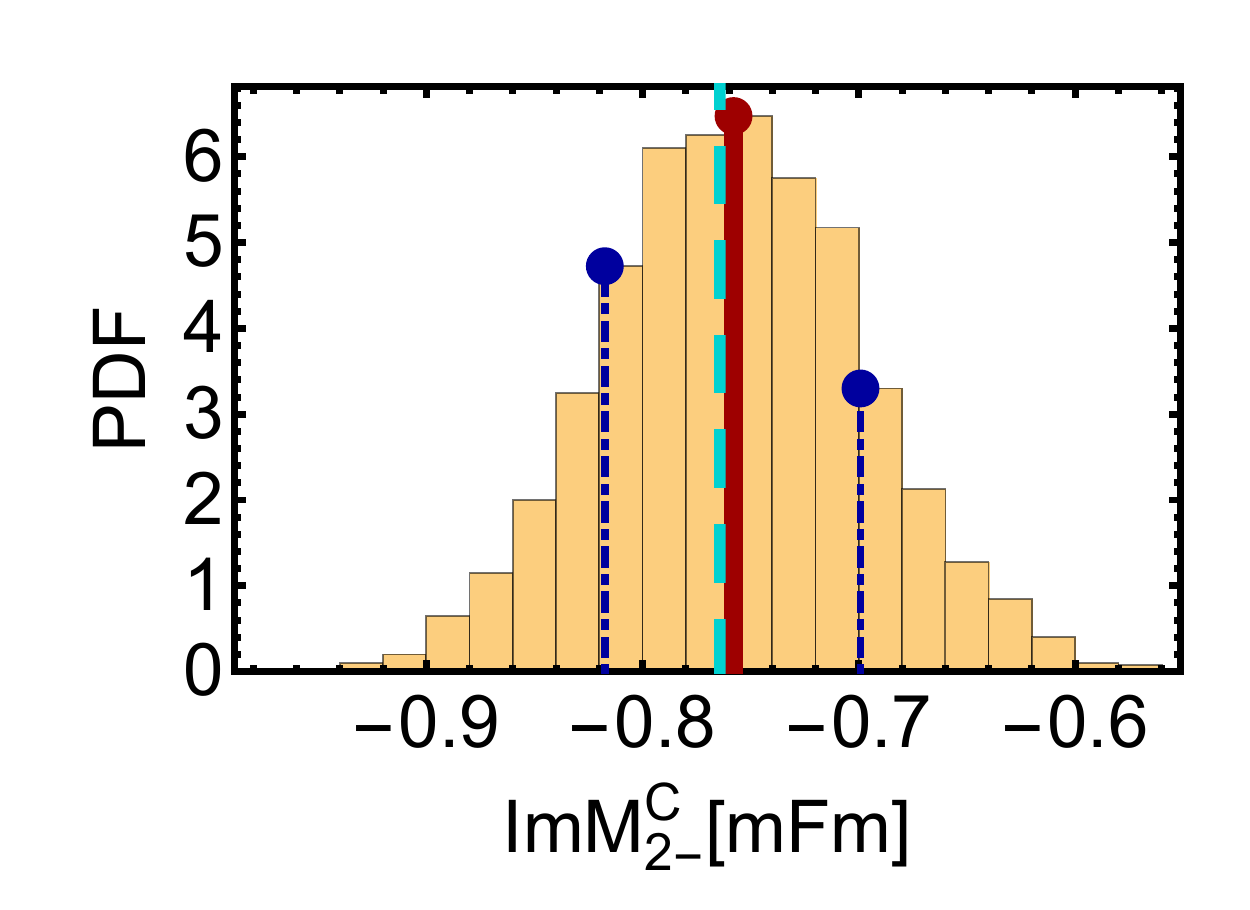}
 \end{overpic}
\caption[Bootstrap-distributions for multipole fit-parameters in an analysis of photoproduction data on the second resonance region. The eighth energy-bin, \newline $E_{\gamma }\text{ = 916.66 MeV}$, is shown.]{The histograms show bootstrap-distributions for the real- and imaginary parts of phase-constrained $S$-, $P$- and $D$-wave multipoles, for a TPWA bootstrap-analysis of photoproduction data in the second resonance region (see section \ref{subsec:2ndResRegionDataFits}). The eighth energy-bin, $E_{\gamma }\text{ = 916.66 MeV}$, is shown. An ensemble of $B=2000$ bootstrap-replicates has been the basis of these results. \newline
The distributions have been normalized to $1$ via use of the object \textit{HistogramDistribution} in MATHEMATICA \cite{Mathematica8,Mathematica11,MathematicaLanguage,MathematicaBonnLicense}. Thus, $y$-axes are labelled as \textit{PDF}. The mean of each distribution is shown as a red solid line, while the $0.16$- and $0.84$-quantiles are indicated by blue dash-dotted lines. The global minimum of the fit to the original data is plotted as a cyan-colored dashed horizontal line.}
\label{fig:BootstrapHistos2ndResRegionEnergy8}
\end{figure}

\clearpage

\phantomsection
\addcontentsline{toc}{section}{List of Tables}
\listoftables

\clearpage

\phantomsection
\addcontentsline{toc}{section}{List of Figures}
\listoffigures

\clearpage

\bibliographystyle{alpha}

\phantomsection
\addcontentsline{toc}{section}{References}
\bibliography{lit}

\clearpage
\thispagestyle{empty}
\textcolor{white}{Hallo Welt :-)}
\clearpage



\newpage \clearpage
\selectlanguage{ngerman}
\thispagestyle{empty}
\section*{Danksagung}

An dieser Stelle bedanke ich mich bei allen, die mich bei meiner Dissertation unterst"utzt und zum Gelingen der Arbeit beigetragen haben. \newline

An erster Stelle geb"uhrt mein Dank meinem Doktorvater, Prof. Dr. Reinhard Beck, f"ur die M"oglichkeit, in seiner Arbeitsgruppe an einem interessanten Thema zu arbeiten, sowie f"ur die gute Betreuung w"ahrend dieser Zeit. F"ur seine Bereitschaft, mir die Teilnahme an vielen Konferenzen und Workshops zu erm"oglichen und somit meinen wissenschaftlichen und pers"onlichen Horizont zu erweitern, bedanke ich mich hier noch einmal ausdr"ucklich. \newline

PD. Dr. Bernard C. Metsch danke ich sehr f"ur die "Ubernahme des Zweitgutachtens. \newline

Prof. Dr. Ulf-G. Mei{\ss}ner und Prof. Dr. Jens Franke spreche ich meinen Dank aus f"ur die Bereitschaft zur Teilnahme an der Promotionskommission. \newline

Die finanzielle Unterst"utzung dieser Arbeit wurde durch die DFG als Teil des SFB/TR16, sowie durch ein Stipendium der Bonn-Cologne Graduate School of Physics and Astronomy sichergestellt. \newline

Den Mitgliedern der Bonn-Gatchina Partialwellenanalyse-Gruppe Dr. Andrei Sarantsev, Dr. Alexei Anisovich, Dr. Viktor Nikonov, Prof. Dr. Ulrike Thoma und Prof. Dr. Eberhard Klempt danke ich f"ur die M"oglichkeit der Zusammenarbeit. \newline

Gro{\ss}er Dank geb"uhrt Dr. Lothar Tiator daf"ur, dass er mir Einsicht in ansonsten sehr schwer zug"angliche wissenschaftliche Quellen erm"oglichte (Omelaenko-Papier) und immer ein offenes Ohr f"ur meine Fragen, Ideen und Ans"atze hatte. Auch f"ur das geduldige Lesen und Beantworten meiner Emails bedanke ich mich. \newline 
Im Laufe dieser Arbeit konnte ich, ebenfalls erm"oglicht durch Dr. Lothar Tiator, an einigen der j"ahrlichen PWA-Meetings in Mainz teilnehmen. Die weiteren Teilnehmer haben zu der guten und produktiven Atmosph"are bei diesen Treffen beigetragen, namentlich: Alfred {\v{S}}varc, Jugoslav Stahov, Hedim Osmanovi{\'{c}}, Mirza Had{\v{z}}imehmedovi{\'{c}}, Rifat Omerovi{\'{c}}, Viktor Ka\-she\-va\-rov, Kirill Nikonov, Michael Ostrick, Jannes Nys und Lefteris Markou. \newline
Insbesondere ergaben sich aus diesen Meetings einige interessante Diskussionen mit Alfred {\v{S}}varc zum allgemeinen Problem der Kontinuumsambiguit"aten, welche in eine Kollaboration m"undeten die dann auch Fr"uchte trug und f"ur die ich mich hier nochmal bedanke. \newline

Dr. Ron Workman von der George Washington University spreche ich f"ur die Bereitstellung der SAID-Modelldaten f"ur meine Analysen in der $\Delta$-Region meinen Dank aus. \newline

Ich danke den Arbeitsgruppen der experimentellen Hadronenphysik von Prof. Beck, Prof. Thoma, Prof. Ketzer und Jun. Prof. Thiel f"ur die angenehme Arbeitsatmosph"are und reichhaltige Un\-ter\-st"ut\-zung w"ahrend der Bearbeitungszeit meiner Dissertation. \newline
Der ehemaligen und aktuellen Besetzung des B"uros $1.004$, Dr. Marcus Gr"uner, Damian-Maria Piontek, Dr. Sabine B"ose, Farah Afzal, Henri St"ubner, Julian G"unther, Gerrit Grut\-zeck, Vitalij Adam und Luca Gottardi, geb"uhrt mein Dank daf"ur, dass sie meinen Arbeitsplatz zu einem Ort gemacht hat, an dem ich gerne und konzentriert arbeiten konnte.

\clearpage
\thispagestyle{empty}

Mein Dank gilt ebenfalls Farah Afzal und Jun. Prof. Dr. Annika Thiel f"ur die allgemeine Unterst"utzung bei wissenschaftlichen Fragen und insbesondere f"ur die Mitarbeit an unserer gemeinsamen Ver"offentlichung. \newline
Gro{\ss}en Teil zur Fertigstellung dieser Arbeit beigetragen hat Dr. Michael Lang, da er stets bereit war, mir bei meinen Anliegen zu Hardware- und Software-Problemen Hilfe zu leisten. \newline 
Ich spreche Dr. Christian Honisch f"ur die interessanten Konversationen w"ahrend unserer gemeinsamen Mittagessen meinen Dank aus, Dr. Jan Hartmann und Karsten Spieker f"ur die gemeinsamen Unternehmungen auf Konferenzreisen. \newline
Die in\-te\-res\-santen Diskussionen mit Mikhail Mikhasenko sowohl zu wissenschaftlichen Themen, als auch dar"uber hinaus waren f"ur mich wertvoll, da mir sein schier unersch"opfliches Wissen des "Ofteren ein Quell der Erleuchtung war. \newline
F"ur das Korrekturlesen meiner Arbeit bedanke ich mich nochmal bei Annika, Christian und Mikhail. \newline
Allen, welche bis hier noch nicht namentlich aufgef"uhrt wurden, u.a. Philipp Hoffmeister, Dr. Martin Urban, Peter Klassen, Dimitri Schaab, Dr. Eric Gutz, Roman Schmitz, Sebastian Ciupka, Eugenia Fix, Peter Pauli, Lena Mikhasenka, Mathias Wagner, Michael H"osgen, Rocio Reyes Ramos, Elizaveta Fotina, Dr. Andrew Wilson, Tobias Seifen, Dr. Harald van Pee, Christian Hammann, Philipp Mahlberg, Matthias Kube, Johannes M"ullers, Georg Urff, Merlin Rossbach, Florian Kalischewski, Jonas M"uller und allen anderen, die ich hier nicht aufgez"ahlt habe, geb"uhrt mein Dank. \newline

Die Unterst"utzung aus dem Kreis meiner Familie und Freunde hat mir die Fertigstellung dieser Arbeit erleichtert. Besonders hervorheben m"ochte ich meine Eltern, Helmut und Irmgard Wunderlich, die mich immer ermuntert haben, meinen Weg in der Wissenschaft zu suchen und an einer Promotion zu arbeiten. Meiner Patentante, Rita Kr"amer, danke ich f"ur viele enorm wertvolle Ratschl"age in allen Lebenslagen. \newline
Der allergr"o"ste Dank jedoch geb"uhrt meiner Frau, {\it Hsu Ning-Hsin} (chin.: \begin{CJK*}{UTF8}{bkai}徐寧昕\end{CJK*}), f"ur die liebevolle Begleitung und Unterst"utzung auf meinem Weg.

\end{document}